## Proceedings of 17th International Meeting on

# Fully 3D
# Image Reconstruction
# in
# Radiology
# and
# Nuclear Medicine

**(Fully3D 2023)**

# Preface

Contained within this volume are the scholarly contributions presented in both oral and poster formats at Fully3D 2023: The 17th International Meeting on Fully Three-Dimensional Image Reconstruction in Radiology and Nuclear Medicine. This conference convened from July 16-21, 2023, at Stony Brook University in New York.

For ease of reference, all papers are organized alphabetically according to the last names of the primary authors.

Our heartfelt appreciation goes out to all participants who took the time to submit, present, and revise their work for inclusion in these proceedings.

Collectively, we would also like to express our profound gratitude to our generous sponsors, detailed in subsequent pages, who have played an instrumental role in offering awards and facilitating the various conference activities.

Additionally, our thanks extend to the diligent reporter who collated invaluable feedback from attendees, which can be found in the pages that follow.

September 7, 2023

Fully3D 2023 Co-Chairs

Jerome Liang
Paul Vaska
Chuan Huang

# Sponsors

**Corporate Level 1**

GE HealthCare

Neusoft Medical Systems

UNITED 联影 IMAGING

**Corporate Level 2**

Canon

CANON MEDICAL RESEARCH, USA INC.

PHILIPS

SYFF

**Academic**

Stony Brook Medicine

Renaissance School of Medicine
Stony Brook University

Department of Radiology

## Stony Brook University Hosts 17th International Fully3D Imaging Conference

By Liza N. Burby

*Worldwide leaders in the field of medical imaging algorithms gathered for a five-day conference in Stony Brook University's Charles B. Wang Center to discuss the latest advancements, highlighting Stony Brook and the Renaissance School of Medicine as a major player in medical imaging science.*

Nearly 200 scientists, students and industry researchers from around the world met for the 17th International Meeting on Fully 3D Image Reconstruction in Radiology and Nuclear Medicine, known as Fully3D 2023, a biennial research conference sponsored for the first time by Stony Brook University (SBU), Stony Brook Medicine and in particular the Department of Radiology. They gathered in the Charles B. Wang Center Sunday, July 16 through Friday, July 21 to present the latest research on the mathematics and theory behind X-ray CT (computed tomography), PET (positron emission tomography) and SPECT (single photon emission computed tomography) imaging.

The focus of the conference is Image reconstruction – the process of taking the raw data from medical imaging systems and generating optimized 3D images that radiologists use to diagnose disease. It brought together the brightest minds in the field, including keynote speakers Dr. Lihong Wang, an inductee in the National Academy of Inventors and the National Academy of Engineering; Dr. Kris Krishna Kandarpa, Director of Research Sciences and Strategic Directions at the National Institute of Biomedical Imaging & Bioengineering; Dr. Bahaa Ghammraoui, a medical imaging scientist at the US Food and Drug Administration; and Dr. Yvonne Lui, Vice Chair for Research at New York University and President of the American Society of Neuroradiology. Their topics ranged from photoacoustic, light-speed and quantum imaging, to recent advancements in medical imaging technologies, including quantitative molecular imaging using PET and SPECT, photon energy-resolving X-ray CT, and artificial intelligence (AI) machine-learning methods. In addition, Stony Brook's own were represented as doctoral candidate Tianyun Zhao gave an oral presentation and five others presented posters at the conference, including Xiaoyu Duan, who won a poster award that was selected by a committee of esteemed scientists independent of Stony Brook.

### A Major Player

The first Fully3D imaging meeting was held in 1991 at Corsendonk, Belgium, where 65 world experts and trainees in the field met on a college campus to exchange their ideas. Fully 3D 2023 had 180 attendees with over half from Europe and Asia. For many it was the first in-person conference they attended since the pandemic. The choice for SBU to serve as host was a natural fit since it has been at the forefront of medical imaging research for decades, serving as the birthplace of MRI technology and diagnostic CT colonography (CTC), and making important contributions in the development of X-ray, CT, PET, SPECT, MRI, optical and multimodality imaging. The new and growing PET Research Center (including the BAHL Molecular Imaging Laboratory and Cyclotron) has imaging capabilities that rival the best in the world and was the site of a tour enjoyed by many attendees at the event.

Fully3D 2023 was co-chaired by three professors doing research in imaging science who are affiliated with the Department of Radiology: Dr. Jerome Z. Liang, Professor of Radiology, Biomedical Engineering, Electric and Computer Engineering, and Computer Science, who has attended nearly all of the previous conferences; Dr. Paul Vaska, Professor of Biomedical Engineering and Radiology; and Dr. Chuan Huang, former Associate Professor of Radiology who is now at Emory University, but returned to work on this conference. All three co-chairs have national and international reputations in the field of medical imaging, particularly in reconstruction of low-dose CT (including CTC), hardware and methods for high-resolution and quantitative PET and SPECT, and advanced AI algorithms for PET and MRI clinical utility.

Liang said that the biennial conference gives attendees a chance to come together to stay on top of the medical imaging field. “Medical imaging hardware has been evolving over the years resulting in better spatial resolution and improved measurements,” Liang said. “This conference was focused on what to do with the raw data from the scanners—image reconstruction, its mathematical process and the different modalities—and turning it into something that radiologists and other medical experts can look at. There are new techniques with AI that are being used, which were highlighted in the sessions, as well as techniques being used to improve the images. This is a very active field of research that’s not just producing new algorithms, but also evaluating them.”

Vaska said the conference enhanced Stony Brook’s status as a major player in medical imaging science. “Attendance exceeded expectations and the audience during the talks was deeply engaged, asking pertinent questions throughout all discussion periods,” he said. “Poster sessions were also well attended and resulted in many lively discussions. There was a strong consensus that the keynote talks covered highly relevant topics, and the speakers themselves said they were impressed with the caliber of the conference. We were also complimented on the venue, campus and staff who were most professional and supportive.”

Keynote speaker Lui said in addition to finding the Stony Brook location an easy trip from the city, she appreciated the interactions. “I enjoyed speaking with the attendees and organizers. I hope my discussion on our work using deep learning to advance MR image reconstruction was useful. I’m a neuroimager by training and a practicing radiologist,” she said. “It was great to interact with imaging scientists and hear the latest advances in reconstruction and consider how these could be useful in practice."

**Practical Applications**

Virtually all the major breakthroughs in image reconstruction have been presented throughout the history of this conference, according to Dr. Grant T. Gullberg from the University of California, San Francisco, who was also a 2023 presenter and has attended nearly every meeting. "This then translates to the technological advances beyond medicine, including the airlines where CT scans can be used to detect a crack in an engine blade," he said.

Dr. Rolf Clackdoyle from the Université Grenoble Alpes in France agreed, adding that while the scientists' main mandate is to publish and produce new research generally in radiology and nuclear medicine, "the receiving end of that is also industry that wants to develop new tools. The homeland security field learns a lot from the medical imaging field and copies the technology and adapts it. The main place our research leads to is medicine, but the same technology is used to scan luggage in the airport."

The Stony Brook poster award winner, Duan, a fifth year PhD student in biomedical engineering and in medical physics, said her work—which is focused on breast imaging when there's a suspicious lesion—is an example of the practical applications of what the conference is all about. "Our clinical motivation is to avoid having a woman called back after an annual screening and make it clear for the radiologist to distinguish if a lesion is benign or malignant, reducing the patient's anxiety and the cost."

Duan added that beyond the industry applications was the opportunity for her to meet colleagues from all over the world and to hear other presentations.

Tianyun Zhao, a third year PhD biomedical engineering student working in Stony Brook's Department of Radiology, said that's what made the conference meaningful for him. Not only did he volunteer, but he also gave a presentation on using anatomical MRI to improve PET imaging performance without the need for large amounts of training data, which led to useful feedback from researchers whose work he's studied.

"This conference allowed me to see a lot of people whose papers I've read before but never had the chance to meet," Zhao said. "I was able to communicate and meet with the person who did the research rather than just interpret from their paper. Networking and exploring this field means I can have better knowledge of the whole imaging field."

Ten-time attendee Dr. Margaret Daube-Witherspoon of the University of Pennsylvania agreed a major benefit of this conference is the ability to network during meals as well as sessions. "This week I've gotten a few ideas on how to do motion correction in our work and some specific ideas that help us reconstruct PET data. It's also just getting up to speed on CT and especially the deep-learning methods," she said. "The importance of the conference overall is to maintain a connection with experts in my field and expand my knowledge base."

**Advanced Details**

Dr. Yuxiang Xing of Tsinghua University in China who earned her PhD in electrical engineering from Stony Brook in 2003, has been attending the conference for more than 15 years and stressed the significance of this supportive scientific community. "With all the cutting-edge science and technology, I learned a lot. This conference is always about the most advanced technology problems. I think Stony Brook should host more of these kinds of conferences. We need this kind of support we've gotten from the radiology department. We have a lot of young investigators, so this is a very good community to help them with their career and interest in this area and their contributions to this work."

For Dr. Emil Sidky from the University of Chicago, who has been attending since 2003, this conference has always been about the details. "For me, this is my favorite conference, because I'm an image reconstruction nerd. This is the one where we can talk about the details of the algorithms and the image formation and the mathematics and all these details that for everybody else their eyes glaze over," he said. "There's a lot of very technical questions also. In a normal radiology conference it wouldn't be like that because you mostly have radiologists or clinical businesses who don't really care that much about what's happening inside the scanner. But for us, it's a formation of the image so we're really focused on a lot of the theoretical details of how that image is formed."

According to Dr. Chuan Huang, the conference allowed for a relaxed, free exchange of ideas. "The feedback we've gotten had been enthusiastically positive. And many of the attendees developed long-term friendship through the professional network opportunity provided by this conference, and some of them have seen each other for the past 20 to 30 years, which is also part of the attraction of this event."

# Committee Members

## Organization Committee Co-Chairs

Jerome Liang • Paul Vaska • Chuan Huang

## Scientific Committee

| | |
|---|---|
| Evren Asma | Canon Medical Research USA |
| Ti Bai | University of Texas Southwestern Medical Center |
| Freek Beekman | Dept. Radiation Science & Technology, Section Biomedical Imaging, TU Delft |
| Yannick Berker | Siemens Healthcare GmbH |
| Richard E. Carson | Yale University |
| Shaojie Chang | Mayo Clinic |
| Guang-Hong Chen | University of Wisconsin in Madison |
| Margaret Daube-Witherspoon | University of Pennsylvania |
| Bruno De Man | GE Research |
| Michel Defrise | VUB |
| Joyita Dutta | University of Massachusetts Amherst |
| Matthias J Ehrhardt | University of Bath |
| Georges El Fakhri | Harvard Medical School, Massachusetts General Hospital |
| Jeffrey A. Fessler | University of Michigan |
| Eric Frey | Johns Hopkins University |
| Grace Gang | University of Pennsylvania |
| Hewei Gao | Tsinghua University |
| Yongfeng Gao | United Imaging Healthcare |
| Kuang Gong | Massachusetts General Hospital |
| Jens Gregor | University of Tennessee |
| Grant T. Gullberg | University of California San Francisco |
| Chuan Huang | Emory University/Stony Brook University |
| Brian Hutton | University College London |
| Xun Jia | Johns Hopkins University |
| Jakob Sauer Jørgensen | Technical University of Denmark (DTU) |
| Marc Kachelriess | German Cancer Research Center (DKFZ) |
| Joel Karp | UPenn |
| Paul Kinahan | University of Washington |
| Michael King | Univ of Mass Chan Medical School |
| Thomas Koehler | Philips Research Hamburg |
| Hiroyuki Kudo | University of Tsukuba |
| Patrick La Riviere | The University of Chicago |
| Tobias Lasser | Technical University of Munich |
| Yusheng Li | UIH America |
| Jerome Zhengrong Liang | Stony Brook University |
| Yihuan Lu | United Imaging Healthcare at Shanghai |
| Hongbing Lu | Fourth Military Medical University |

# Committee Members

| | |
|---|---|
| Jianhua Ma | Southern Medical University |
| Nicole Maass | Siemens Healthineers |
| Samuel Matej | University of Pennsylvania |
| Scott Metzler | University of Pennsylvania |
| Stephen Moore | University of Pennsylvania |
| Xuanqin Mou | Xi'an Jiaotong University |
| Klaus Muller | Stony Brook University |
| Peter Noël | University of Pennsylvania |
| Frederic Noo | University of Utah |
| Johan Nuyts | KU Leuven |
| Jed D. Pack | GE Research |
| Tinsu Pan | The University of Texas, MD, Anderson Cancer Center |
| Xiaochuan Pan | University of Chicago |
| Vladimir Panin | Siemens Medical Solutions USA |
| Amir Pourmorteza | Emory University |
| Jinyi Qi | UC Davis |
| Magdalena Rafecas | University of Luebeck |
| Andrew Reader | King's College London |
| Ahmadreza Rezaei | KU Leuven |
| Cyril Riddell | GE Healthcare |
| Georg Schramm | Stanford University |
| Kuangyu Shi | University of Bern |
| Emil Sidky | The University of Chicago |
| Arkadiusz Sitek | Massachusetts General Hospital |
| Mark Slifstein | Stony Brook University |
| Web Stayman | Johns Hopkins University |
| Karl Stierstorfer | Siemens Healthcare GmbH |
| Suleman Surti | University of Pennsylvania |
| Ken Taguchi | Johns Hopkins University |
| Xiangyang Tang | Emory University School of Medicine |
| Benjamin Tsui | Retired from Johns Hopkins University in May 2020 |
| Paul Vaska | Stony Brook University |
| Dimitris Visvikis | INSERM |
| Jing Wang | UT Southwestern Medical Center |
| Guobao Wang | University of California Davis Health |
| Adam Wang | Stanford University |
| Wenli Wang | Avant Tomography Consulting LLC |
| Ge Wang | RPI |
| Yuxiang Xing | Tsinghua University |
| Wei Xu | Brookhaven National Laboratory |
| Hengyong Yu | UML |
| Zhou Yu | Canon Medical Research USA |
| Larry Zeng | Utah Valley University |
| Hao Zhang | Memorial Sloan Kettering Cancer Center |
| Jian Zhou | Canon Medical Research USA |

## Publication Committee

| | |
|---|---|
| Chuan Huang | Emory University/Stony Brook University |
| Yongfeng Gao | United Imaging Healthcare |
| Amir Pourmorteza | Emory University |
| Shaojie Chang | Mayo Clinic |
| Thomas Wesley Holmes | Emory University |

# Table of Contents

# A Generic Software Design for Computed Tomography in Modern C++

Shiras Abdurahman[1], Robert Frysch[1], Tim Pfeiffer[1], Oliver Beuing[2], and Georg Rose[1]

[1]Institute for Medical Engineering and Research Campus STIMULATE, Otto-von-Guericke-Universität, Magdeburg, Germany
[2]AMEOS Klinikum Bernburg, Bernburg, Germany

**Abstract** In this paper, we propose a flexible software design for CT systems of various detector and acquisition geometries using modern C++. It is based on the generic design of CT data processing algorithms utilizing a high-level abstraction of CT geometry. We also introduce a new toolkit, GCT, to reconstruct images from the parallel-beam (1D and 2D), fan-beam (line and arc-shaped detectors), cone-beam (flat-panel and cylindrical detectors), and rebinned projections using FBP/FDK algorithm demonstrating that the proposed design can be used for the development of scalable and maintainable CT reconstruction software.

## 1 Introduction

Developing novel reconstruction and artifact reduction algorithms is an active area of research in Computed Tomography (CT). To support this effort, flexible, easy-to-use, versatile, and open-source reconstruction software is essential. Although many open-source libraries and toolkits are available, most support one or a handful of detector and projection geometries. In practice, researchers need to reconstruct images from various scanners of distinct CT geometries. Most textbooks discuss the reconstruction of 2D images from one-dimensional parallel and fan-beam (arc- and line-shaped detectors) projections [1]. Their software implementations are widely used for learning, comparative assessment, and the rapid development of novel artifact reduction algorithms. Flat detector-equipped Cone-Beam Computed Tomography (CBCT) systems are ubiquitous for imaging in radiation therapy, interventional radiology (C-arm CBCT), and Non-Destructive Testing (NDT) [2]. Multi-Slice or Multi-Row Detector clinical CT systems (MSCT or MDCT) employ cylindrical detectors for projection acquisition. In addition, most MSCT scanners perform axial rebinning to transform projections from cone-beam to oblique-parallel (cone-parallel) projection geometries [3]. The reconstruction from the rebinned projections is computationally efficient due to the simplified backprojection operation. Besides, it preserves the uniform resolution and noise texture in the axial images [4]. Finally, the reconstruction software must support 2D/3D parallel projection geometry to reconstruct volume from synchrotron projections.

As per the survey of open-source toolkits found in [5], the number of toolkits supporting a wide range of CT geometries is minimal, and the majority are developed for axial cone-beam reconstruction from the flat detector projections. Consequently, researchers have to rely on various toolkits for image reconstruction. They are written in multiple languages (C, C++, MATLAB, and Python) and utilize different geometry conventions and coordinate systems (e.g., DICOM LPS [6], and IEC 61217 [7]). Many of these toolkits employ a single model to describe the projection geometry (e.g., projection matrix for cone-beam projections), and conforming to that model often requires additional effort (e.g., simulation of parallel geometry by placing the X-ray source at a large distance) and computations (e.g., the transformation of projections from cylindrical to flat detector by interpolation). In this scenario, a single but versatile toolkit in which different geometries can be easily configured is very appealing. The most challenging aspect of building such a toolkit is developing software design aligned with modern design guidelines and best practices. Such design should enable the incremental and independent addition of new features (e.g., supporting new detector geometries) while preserving the common interface to algorithms. The design should facilitate the development of reconstruction software that is easy to change, extend and test [8].

Utilizing the generic programming techniques of modern C++, we present a software design supporting a wide range of CT geometries. Rather than merely using templates, the proposed design is based on conceptualizing CT geometry from tiny building blocks and identifying variation and customization points where the changes can be expected and software can be extended [8]. At the same time, we refrain from "over-generalization," and entities and functionalities are differentiated via specialization, making API "hard to be misused." The design eliminates code duplication and enables the development of extendable and maintainable software. We also introduce a reconstruction toolkit GCT (Generic CT) to reconstruct images from the parallel-beam (1D and 2D), fan-beam (line and arc-shaped detectors), cone-beam (flat-panel and cylindrical detectors), and rebinned projections using FBP/FDK algorithm. GCT is implemented in C++ 20 where the numerical computations are accelerated in CUDA C++ (https://gitlab.stimulate.ovgu.de/shiras-abdurahman/gct.git). Though currently, GCT supports only analytical reconstruction from axial scan projections, the generic description of CT geometry discussed in the paper can be used for spiral and iterative reconstructions, CT projection simulation, and artifact correction. The present paper describes only the design and implementation details of CT data processing and reconstruction (relevant to the Fully3D conference). The operational control of CT system components, user interface, and reading, display, and storage of images are beyond the scope of the paper.

## 2 Materials and Methods

The proposed design divides CT geometry into system and rotation angle-specific projection geometries. The system geometry is invariant during the scan, and the projection geometry describes the transformation between the World (WCS) and Detector Coordinate Systems (DCS). The essential component of system geometry is the type of detector with which projections are recorded. To describe how detector elements are arranged and the ray-sampling is performed (type of projection), the following one-dimensional (1D) detectors are defined:

1. DetFanArc: Equi-angular sampled arc-shaped detector for fan-beam projection.
2. DetFanLine: Equi-distant sampled line-shaped detector for fan-beam projection.
3. DetParallel: Equi-distant sampled detector for parallel projection.

It is to be noted that the above detectors are merely abstractions and do not necessarily represent the geometry of the physical detector since detectors of any shape can be used to acquire parallel projections. However, in GCT, DetParallel has to be used to reconstruct images from parallel projections. The type definitions of the 1D detectors can be found in Listing 1 where the modeling of common behavior (information about sampling) is realized by composition.

```
1  struct SamplingInfo1D
2  {
3      unsigned int num_samples;
4      float sampling_width;
5  };
6  struct DetFanArc
7  {
8      SamplingInfo1D sampler;
9  };
10 struct DetFanLine
11 {
12     SamplingInfo1D sampler;
13 };
14 struct DetParallel
15 {
16     SamplingInfo1D sampler;
17 };
```

**Listing 1:** 1D Detectors. The projection geometry and the FBP reconstruction algorithms associated with 1D detectors are described in [1].

As listed in Table. 1, several 2D detectors can be conceptualized as the permutations of two 1D detectors representing detector row (front view) and column (side view) geometries. The generic 1D/2D detector can be defined in modern C++ using variadic templated std::tuple type as shown in Listing 2.

```
1  DetFanLine fl{sampling_info_fl};
2  DetFanArc fa{sampling_info_fa};
3  auto line_det = std::make_tuple(fl);
4  auto cyl_det = std::make_tuple(fa, fl);
```

**Listing 2:** 1D or 2D Detector types.

| DetRow | DetCol | Detector geometry |
|---|---|---|
| DetFanArc | | 1D arc detector |
| DetFanLine | | 1D line detector |
| DetParallel | | 1D parallel detector |
| DetFanArc | DetFanLine | 2D cylindrical detector |
| DetFanLine | DetFanLine | 2D flat detector |
| DetParallel | DetParallel | 2D parallel detector |
| DetParallel | DetFanLine | 2D rebinned detector |
| DetFanArc | DetFanArc | 2D spherical detector |

**Table 1:** The conceptualization of 1D and 2D detectors for parallel, fan-beam, rebinned and cone-beam projections. DetRow and DetCol can be considered as the cross-section of the source-detector pair in the axial and sagittal planes.

To model the unified cone-beam projection geometry (Listing 3) for 2D flat and cylindrical detector projections, we utilize parameters described in [9]. Along with the rotation angle, the cone-beam projection geometry parameters (as a whole or subset of them) could be used for all detectors listed in Table. 1. However, we decided to employ detector-dependent projection geometry parameters (Listing 4) to ensure that the users are not exposed to parameters they do not need and to prevent the passing of wrong parameter values.

```
1  //Vector in World Coordinate System (WCS)
2  struct VecWCS3D
3  {
4      float x, y, z;
5  };
6  struct ConeBeamGeom
7  {
8      float src_det_dist_mm, src_iso_dist_mm;
9      float det_row_cent_pix, det_col_cent_pix;
10     VecWCS3D src_pos;
11     VecWCS3D det_row_dir, det_col_dir,
       src_det_dir;
12 };
```

**Listing 3:** Cone-beam projection parameters. Please refer [9] for the detailed explanations of individual parameters and the geometry computations.

```
1  struct FanBeamGeom
2  {
3      float src_det_dist_mm, src_iso_dist_mm;
4      float det_row_cent_pix;
5      VecWCS2D src_pos;
6      VecWCS2D det_row_dir, src_det_dir;
7  };
8  template<typename DetRow, typename ...DetCol>
9  struct ProjGeom;
10
11 //For 2D flat detector cone-beam projection.
12 template<>
13 struct ProjGeom<DetFanLine, DetFanLine>
14 {
```

```
    ConeBeamGeom pg;
};
//For 2D cylindrical detector cone-beam
   projection.
template<>
struct ProjGeom<DetFanArc, DetFanLine>
{
    ConeBeamGeom pg;
};
//For 2D parallel projection.
template<>
struct ProjGeom<DetParallel, DetParallel>
{
    float det_row_cent_pix, det_col_cent_pix;
    float rot_ang_rad;
};
//For 2D rebinned (cone-parallel) projection.
template<>
struct ProjGeom<DetParallel, DetFanLine>
{
    float src_det_dist_mm, src_iso_dist_mm;
    float det_row_cent_pix, det_col_cent_pix;
    float rot_ang_rad, src_pos_z_mm;
};
//For 1D arc detector fan-beam projection.
template<>
struct ProjGeom<DetFanArc>
{
    FanBeamGeom pg;
};
//For 1D line detector fan-beam projection.
template<>
struct ProjGeom<DetFanLine>
{
    FanBeamGeom pg;
};
//For 1D parallel projection.
template<>
struct ProjGeom<DetParallel>
{
    float det_row_cent_pix, rot_ang_rad;
};
```

**Listing 4:** Detector-dependent projection geometry.

Finally, the complete CT geometry is composed by detector and projection geometries of the entire scan (Listing 5).

```
template<typename DetRow, typename ...DetCol>
struct CTGeom
{
  std::tuple<DetRow, DetCol...> dg;
  std::vector<ProjGeom<DetRow, DetCol...>> pg;
};
```

**Listing 5:** Complete CT geometry. CTGeom is a polymorphic (static) abstraction and can be used to describe the geometry of CT systems of various trajectories and projection geometries.

The piece-wise conceptualization of detector geometry enables the compile-time evaluation of projection and reconstructed image dimensions as shown in Listing 6. Since DetCol is variadic and optional (Listing 5), the presence of the second template argument facilitates the 3D volumetric reconstruction from 2D images, while its absence implies the 2D image reconstruction from 1D projections. This will help to develop a generic type signature for reconstruction function (Listing 7), and the user has to pass appropriate geometry, projection, and image objects to make 2D or 3D reconstruction feasible. Hence, the compiler enforces the correct usage of API without additional documentation and numerous overloaded functions.

```
template<typename ...DetCol>
constexpr size_t projDim()
{
    return sizeof...(DetCol) + 1u;
}
template<typename ...DetCol>
constexpr size_t imgDim()
{
    return sizeof...(DetCol) + 2u;
}
```

**Listing 6:** Compile-time evaluation of projection and reconstructed image dimensions.

```
template<typename PixType, size_t Dim>
struct Proj;
template<typename PixType, size_t Dim>
struct Img;
template<typename DetRow, typename ...DetCol>
void reconFDK(CTGeom<DetRow, DetCol...> const
    &cg, Proj<float, projDim<DetCol...>()>
    const &p, Img<float, imgDim<DetCol...>()>
    &i);
```

**Listing 7:** The FDK reconstruction function.

The design allows the specialized implementations of 2D and 3D data processing algorithms. Although 2D reconstruction is possible using 3D reconstruction toolkits by considering 1D projection as the 2D projection image with a single row, extra effort is needed to set the appropriate geometry parameters. If the back projection operation utilizes in-built 2D interpolation of GPU, the projections should contain at least two rows. GCT eliminates these constraints by having a dedicated implementation to reconstruct 2D images from the 1D projections without any changes to the interface. By avoiding unnecessary computations, specialized 2D processing also enhances computational efficiency.

The design enhances the extendability of the reconstruction software. If the projection geometry can be described by a single model for all projections, detectors of any shape can be integrated. The FDK reconstruction for a new detector type can be realized by adding type and underlying function specializations without any changes to the runFDK function or the implementations specific to existing detector types. The absence of necessary specializations will result in errors during compilation preventing incorrect results or run-time errors. The compiler acts as a guide to the developer by pointing out missing type and function definitions enabling incremental and independent addition of new features without modifications to the existing code. Like runFDK, most data processing functionalities requiring CT geometry can be implemented as high-level meta-functions. Hence, code duplication can be significantly reduced since a small fraction of underlying functionalities needs to be specialized (e.g., computing weight for cosine weighting) without re-writing boilerplate code. The feasibility of template kernels in the CUDA programming model also helps in this regard. Conceptualizing CT geometry from a small number of types also helps to avoid defining multiple concrete classes (e.g., FlatDetectorGeom, LineDetGeom).

On the other hand, the software can be easily changed to meet new requirements. Although we employed the geometry parameters shown in Listing 3, the projection matrix can also be utilized to model the projection geometry of the CBCT system with flat detectors. Changing to projection matrices could have been difficult if we used a single concrete type for projection geometry since they apply only to flat detectors. By making detector-dependent, the projection geometry can be modified at only one place without affecting others (line 15 of Listing 4).

As per the guidelines described in [8], the major single-purpose image processing tasks are implemented as the non-member and non-friend functions (free functions). Exploiting the template argument deduction, static polymorphism is realized by overloading free functions. In addition, the free functions greatly enhance the software testability without introducing artificial dependencies [8]. Unit testing is difficult for a data processing task like backprojection where the input and output are images. On the other hand, the core computations of backprojection are the coordinate transformations involving projective and affine mappings. These functionalities can be extracted, made into free functions, and tested with known projection geometry parameters without unnecessary instantiations. The same transformation functions can be reused while implementing forward projection eliminating code duplication.

The proposed design is based on the selection of geometry at compile-time. In many situations, it can only be known at run-time if the program relies on the user input via command line arguments, GUI, or the DICOM file. We intend to move the decision-making and conditional branching to the "front end" of the reconstruction software. As a result, numerically intensive computations can be freed from branch prediction, thread divergence, and dynamic binding through virtual functions, resulting in improved code readability and performance.

## 3 Results

Fig. 1 displays the reconstructed images from the C-arm CBCT (flat detector), MSCT (cylindrical detector), and synchrotron (2D parallel) projections. The reconstructions were performed using the FDK algorithm of GCT software. Though they are not shown, GCT also contains specialized FDK reconstruction implementations for 2D rebinned, 1D parallel, and 1D fan-beam (arc and line detectors) projections.

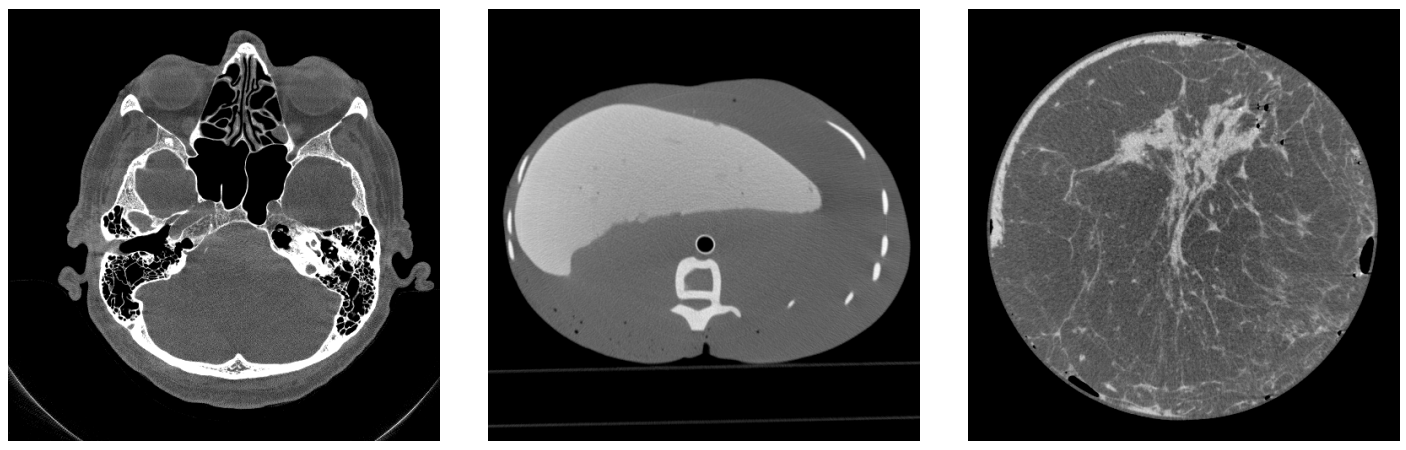

(a) (b) (c)

**Figure 1:** Images reconstructed from: (a) flat detector (C-arm CBCT) projections; (b) cylindrical detector (MSCT) projections; (c) synchrotron (2D parallel) projections.

## 4 Conclusion

We have proposed a software design for CT data processing and reconstruction supporting diverse detector and projection geometries. The design will be valuable for developing commercial and open-source CT reconstruction software. We have also introduced a new reconstruction toolkit GCT to reconstruct images from various axial scan projections. GCT is in active development, and many features will be added in the coming months, including spiral and iterative reconstructions, artifact correction algorithms, and multi-GPU support.

# Effect of Cone-Beam CT Artifacts in the Training Data on the Outcome of Deep Learning Denoising

Andriy Andreyev[1], Parisa Asadi[1], Faguo Yang[1], and Matthew Andrew[1]

[1]Carl Zeiss X-ray Microscopy, Inc, 5300 Central Parkway, Dublin, CA, USA 94568

**Abstract** Deep Learning denoising is increasingly popular to refine the image quality in tomographic imaging. It typically requires training from the images themselves. Very often the images are not artifact-free. Especially, in cone-beam CT with circular scan trajectories, by the nature of cone beam geometry, measured line integrals diverge and the image reconstruction accuracy diminishes very quickly. In this paper, we try to answer the question of whether we should avoid artifact regions when training Deep Learning models, as applied to CT image denoising. We use simulated phantom data as well as real data examples.

## 1 Introduction

Improvement of image quality by deep learning (DL) methods are getting increasingly popular in tomographic imaging [1-6]. Many different ways of applying convolutional neural networks to learn local noise properties and restore the diagnostic quality with the aim of dose reduction and faster imaging throughput have been presented over the years. Cone beam microCT is yet another imaging modality that greatly benefits from DL denoising, often, applied in the image domain after the image has been reconstructed by traditional analytical algorithms. At the same time, an important part of a DL workflow is how training data are sourced. Good quality, unbiased, and representative training data are crucial to the DL algorithms' success. Cone beam CT is quite special in this regard since when using a strictly circular trajectory, only the regions of the object that are very close to the circle plane can be reconstructed exactly [7-8]. Cone beam artifacts commonly result in image blurring and distortion of structural information, increasingly saturating the reconstructed image volume further away from the main circle plane due to missing data.

This paper is trying to answer the practical question of how robust Deep Learning denoising methods are when the training data are contaminated with cone-beam geometry-related artifacts.

## 2 Materials and Methods

Two three-dimensional artificial phantoms: Shepp-Logan and foam phantom [9-10] were employed in this study (Figure 1). The 3D digitized image volume has been forward projected in a cone-beam geometry with a circular acquisition trajectory. Additive Poisson noise has been added to the projection data. In-house implemented FDK image reconstruction algorithm was used for reconstructing the training dataset. For the Shepp-Logan phantom, the same slice of the 2D phantom was duplicated across the entire Z-range of 512 slices. For the foam phantom, two random variations of internal bubble distributions were generated: one for training the models, and another one for inference testing. The input phantom images that were used for forward projection and data creation had 512x512x512 voxel dimensions. The images were reconstructed with matching 512x512x512 voxel dimensions as well. The cone-beam geometry was extended to create a larger cone angle of approximately 45% so that only approximately 250 slices in the middle are considered to be mostly artifact-free. A 180 + fan angular range was used, as the cone beam artifacts are typically more pronounced in this case, with one example shown in Figure *2*, which compares the difference in reconstructed image quality between the central slice (slice index 256) and side slice (slice index 485).

For the DL denoising method, we are using Noise2Noise U-Net architecture network, with standard augmentation parameters, trained over two independent noise realizations of the same dataset [11-12]. As shown in Figure *3*, the training data were truncated to produce a normally defined image volume, matching simulated detector dimensions "Incl. CB", artifact free volume ("Excl. CB"), as well as volume that consisted entirely of artifacts "CB only", as a hypothetical worst-case scenario.

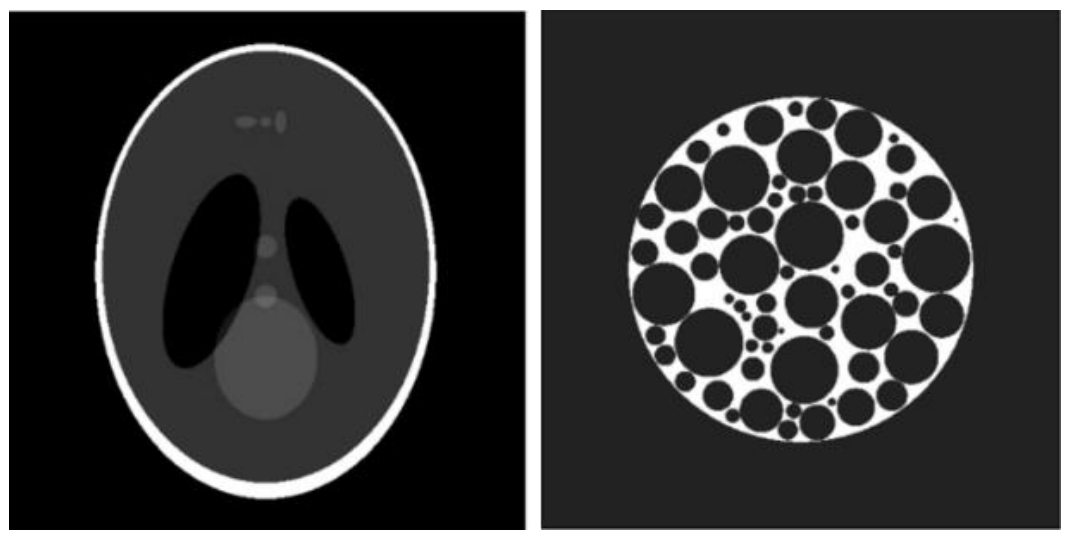

(a) (b)

Figure 1. Digital phantoms used in this study: (a) Shepp-Logan; (b) foam phantom.

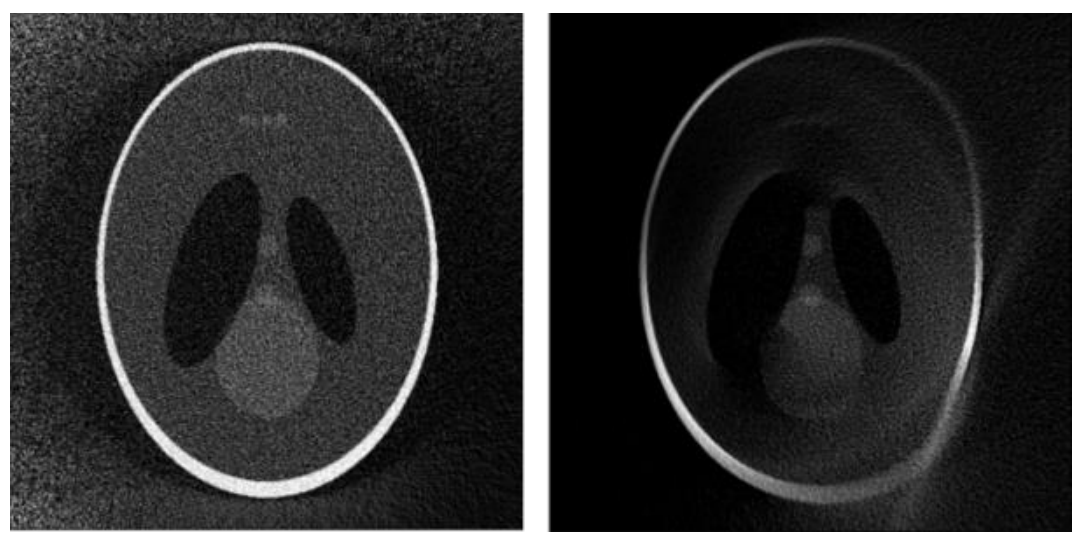

(a) (b)

Figure 2. Shepp-Logan phantom. (a) reconstruction with added noise (central slice), (b) one of the slices in the peripheral range, showing strong cone-beam artifacts.

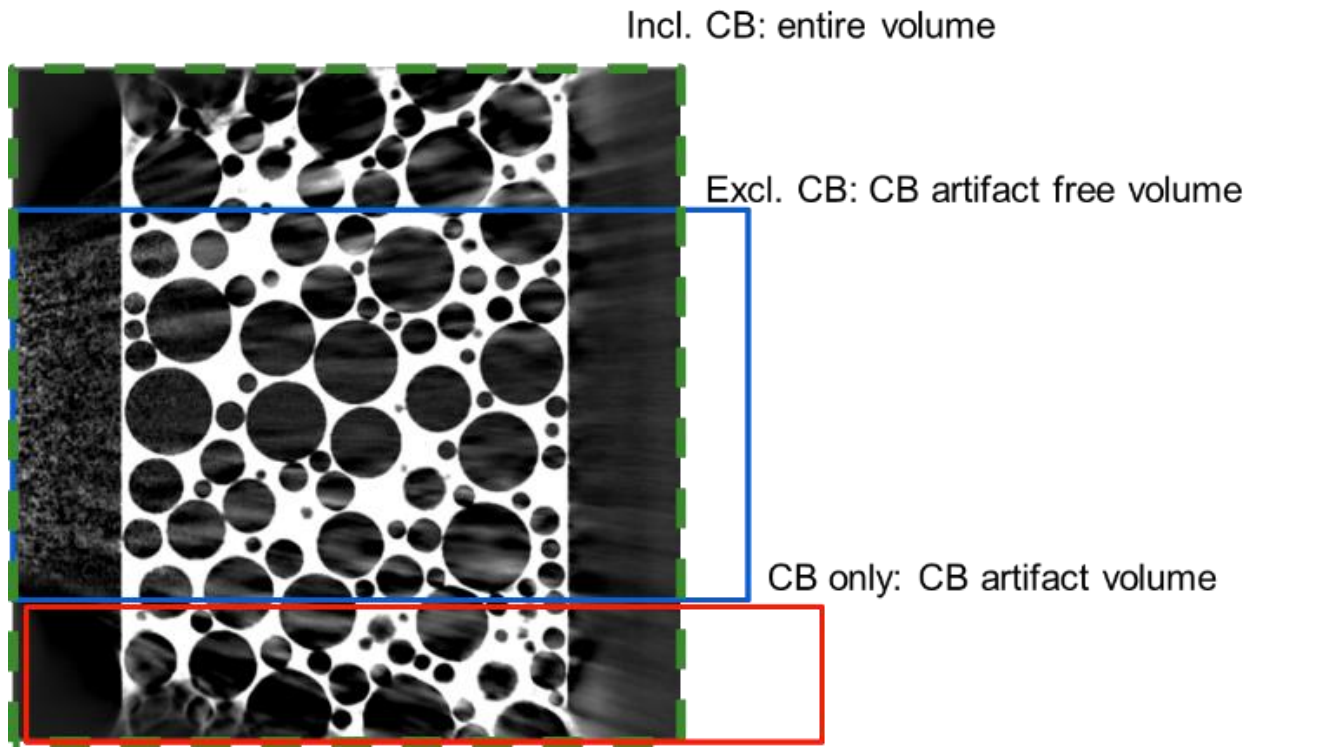

Figure 3. Coronal cross-section through the foam phantom reconstructed dataset, demonstrating the three ROIs that were used in this study.

The images were analyzed visually, as well as the values of MSE and SSIM [12] were calculated from the resulting images.

To supplement the study with some real data results, we used a rock sample dataset, acquired using Carl Zeiss Xradia Versa X-ray Microscope in cone-beam geometry. The dataset dimensions were 1000x1000 pixels with 200 projection views for each noise realization, much lower than normally would be used with a standard reconstruction algorithm.

## 3 Results

In Figure *4*, we show the results for the Shepp-Logan phantom. As one can see, there is a negligible difference between the first two methods to train the denoising model, while CB-only model produced worse image. The results for the foam phantom (Figure *5*) are quite similar, with the CB-only model (c) looking somewhat worse.

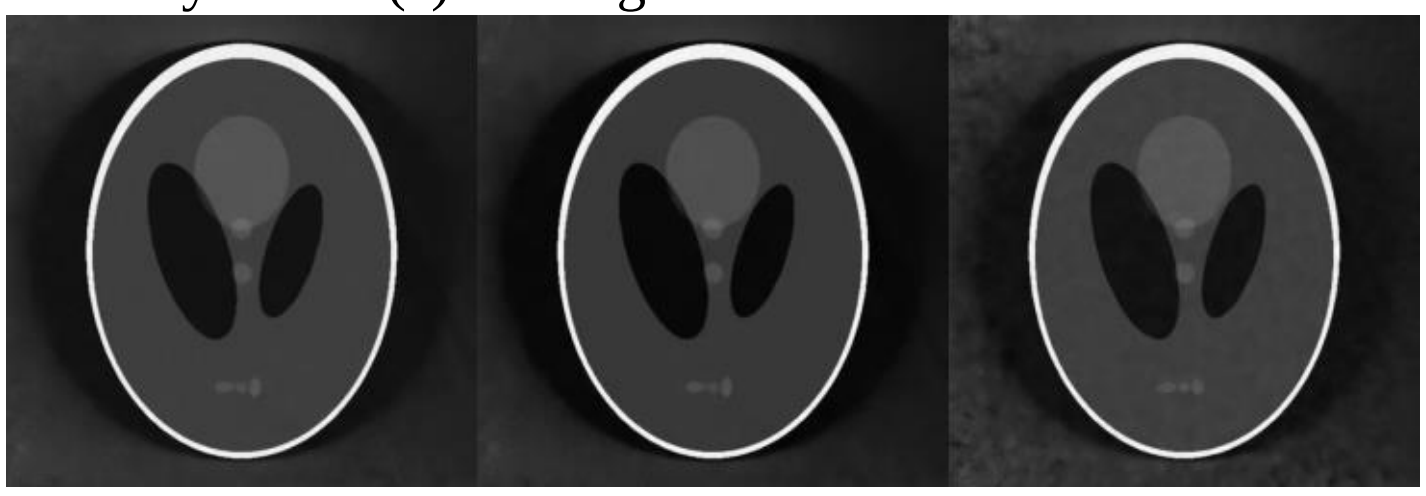

(a) (b) (c)

Figure 4. Shepp-Logan denoising results, central cross-sectional slice 256 shown. (a) denoised using "Incl. CB" trained model, (b) denoised using "Excl. CB" (CB artifact-free); (c) denoised using "CB only" (CB artifact region).

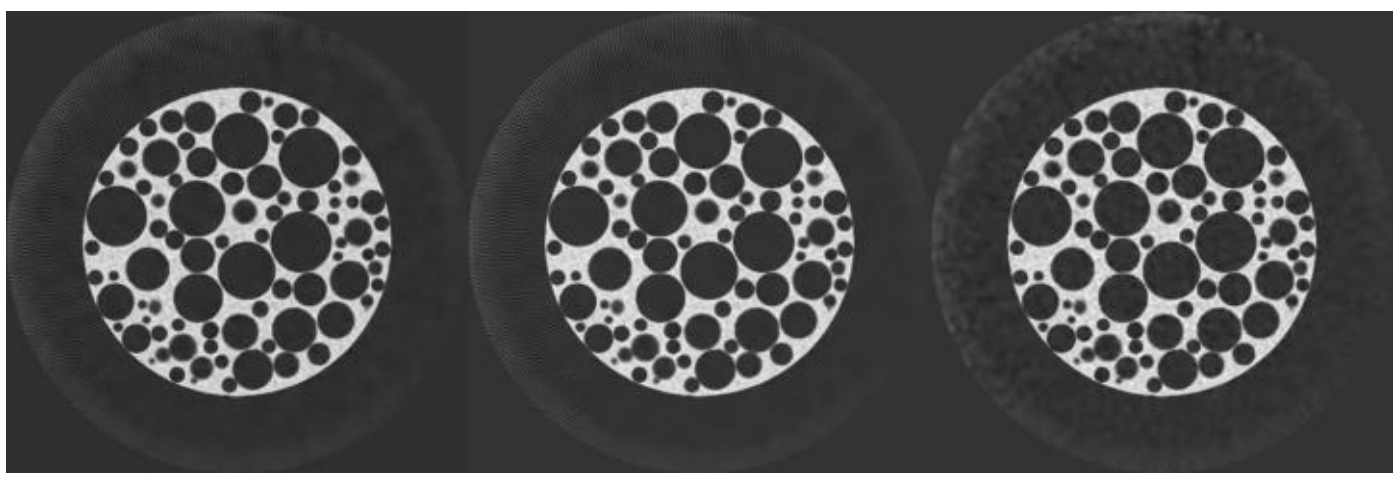

(a) (b) (c)

Figure 5. Foam phantom denoising results, central cross-sectional slice 256 shown. (a) denoised using "Incl. CB" trained model, (b) denoised using "Excl. CB" (CB artifact-free); (c) denoised using "CB only" (CB artifact region).

No significant differences were observed in the training loss curves behavior either (Figure *6*). Mostly minor variations in MSE and SSIM were noted, with CB only trained model showing larger errors in the case of the foam phantom (Table I).

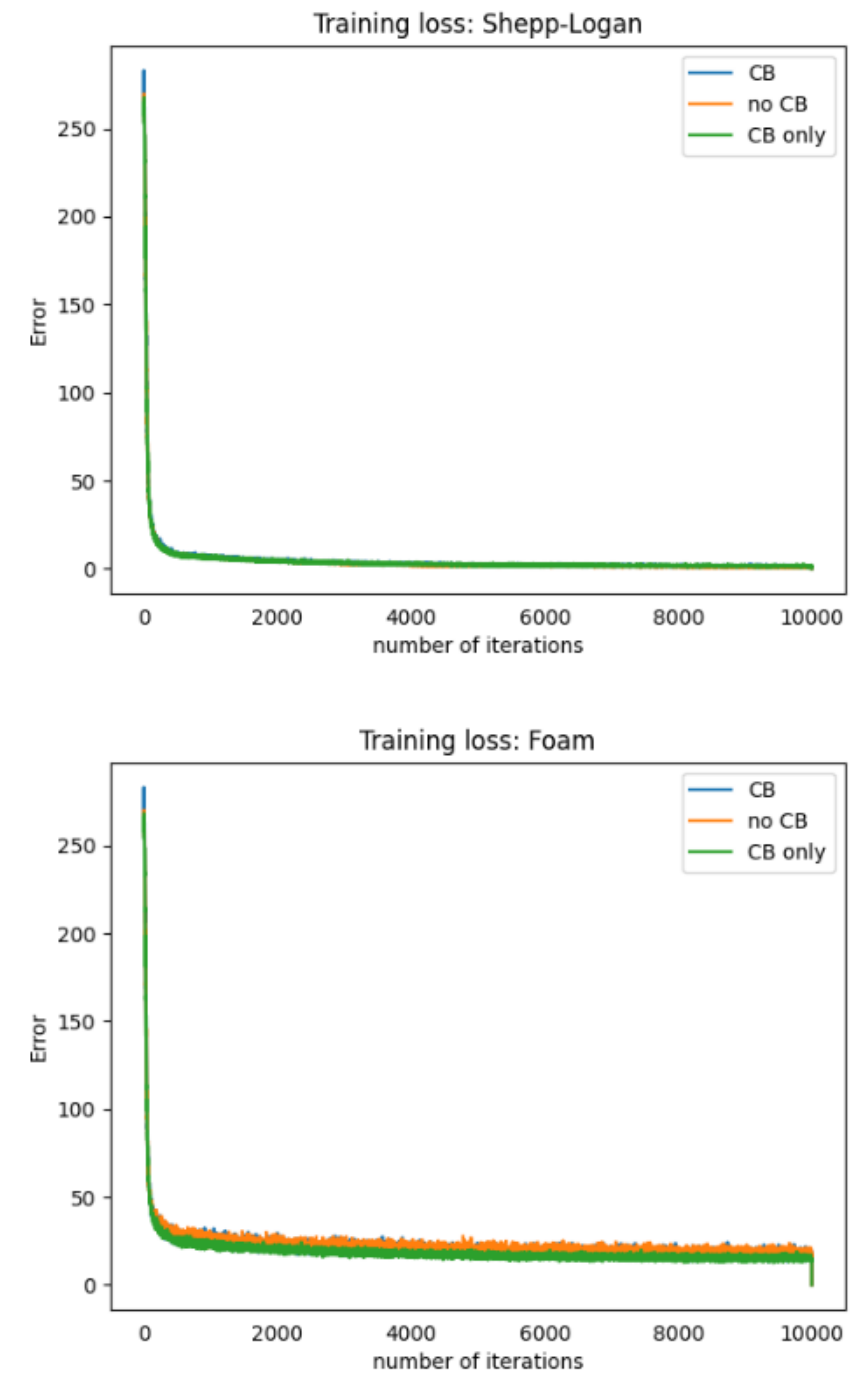

Figure 6. Training loss curves for the Shepp-Logan (top) and foam phantom (bottom).

| | Incl. CB | Excl. CB | CB only |
|---|---|---|---|
| | MSE(SSIM) | MSE(SSIM) | MSE(SSIM) |
| Shepp-Logan | 0.055392 (0.96796) | 0.061462 (0.970223) | 0.060763 (0.966148) |
| Foam phantom | 0.83354 (0.935229) | 0.849188 (0.936431) | 0.933166 (0.928041) |

Table I. Quantitative results showing MSE and SSIM (in parenthesis) values, for both phantoms using the three ways to train the model, as described above.

Finally, in Figure *7* we show the results for the rock sample, reconstructed from real data, 200 projection views, acquired with about 20 degrees cone angle. Overall, consistently with previously observed artificial data, there is no big difference between the model trained over the whole volume and the volume that excluded the CB artifact region. Both models have achieved excellent denoising and image quality improvement over the image reconstructed with the classical algorithm. In this case, we did not use the CB artifact region only for training.

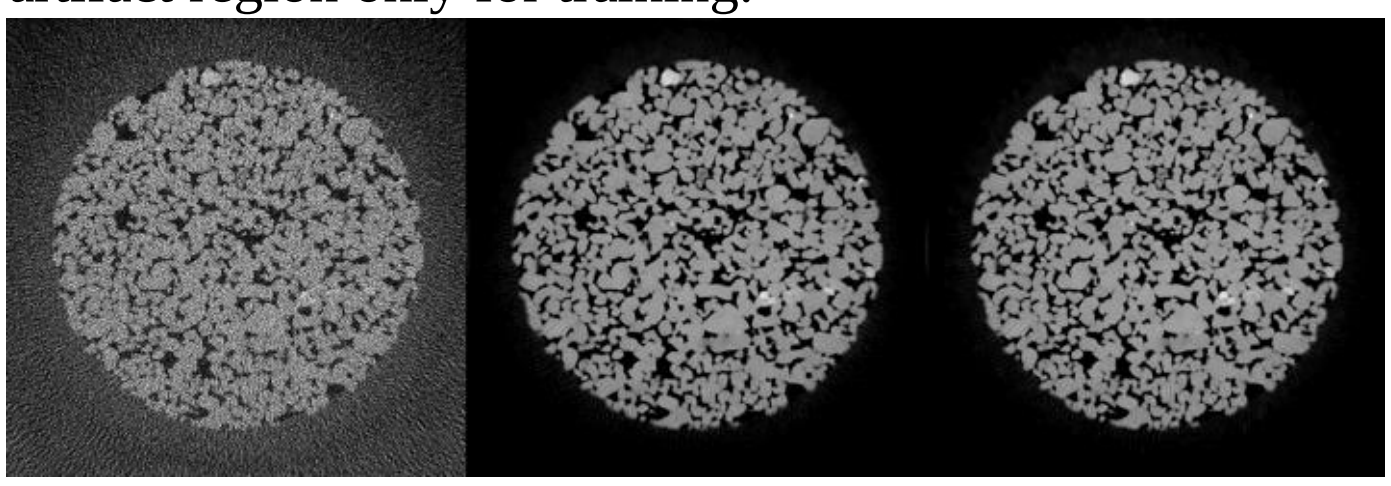

(a) (b) (c)

Figure 7. Reconstructed real data with rock sample: (a) noisy input to the DL denoising algorithm reconstructed with FDK; (b) denoised with DL model trained over the whole volume ("Incl. CB"); (c) denoised with the DL model trained over the volume excluding CB artifacts ("Excl. CB").

## 4 Conclusion

Based on the results presented in this paper, we find that deep learning denoising is quite robust and performs well even when the training data contains a significant amount of CB artifacts. We speculate that cone-beam artifact mainly affects the structural or anatomical information content but not the noise texture distribution, which can still be successfully learned during the training. This is confirmed by both simulated and real data.

Certainly, this finding has been observed only for limited use cases as well as the conclusion could still be dependent on the particular network architecture, data augmentations, hyper-parameters, etc. If the model can still be trained with similar accuracy excluding cone-beam artifact regions, it still makes sense to avoid those regions for the sake of shorter training data preparation and easier data storage requirements. Additionally, there can be a difference in noise appearance as dependent on sampling between central slices and periferal regions, which can bias the network training process.

We still expect that in the case if deep learning algorithms are used not simply for denoising but for tasks that involve knowledge of the structural information - feature identification or segmentation tasks (tumor localization and classification, fractures detection), the conclusion could be different as well, and those algorithms can be more sensitive to the cone beam artifacts in the training data.

# Support Vector Classifier for Metal Detection in CBCT Images

Bernhard Brendel[1], Vithal Trivedi[2], Hans Rosink[2], and Dirk Schäfer[1]

[1]Philips GmbH Innovative Technologies, Research Laboratories Hamburg, Hamburg, Germany
[2]Philips Image Guided Therapy, Advanced Development Mobile Surgery, Best, The Netherlands

**Abstract** Fast classification methods are of interest to decide if computationally expensive processing methods like metal artifact correction (MAC) should be launched automatically to improve interventional C-arm cone beam CT (CBCT) images. Simple threshold-based classification methods are possible if the voxel values of the CBCT image represent HU values. If this is not the case (e.g., to reduce calibration effort) these methods are not reliable. We present a classification method that makes no assumption on the quantitative voxel value scale of an image. It is based on a coarse histogram of min-max normalized voxel values. A linear support vector classifier (SVC) is used to separate the histograms of images without metal from those of image with metal. Tests on datasets of a cadaver study led to very good classification rates of 99-100%.

## 1 Introduction

A major application of mobile C-arm systems is the visualization of anatomical structures and implants during surgical interventions. In many of these interventions the correct placement of metal objects (like screws or plates) with respect to bone structures shall be verified. In some cases, cone beam CT (CBCT) acquisitions are performed, and 3D images are reconstructed, to check the positioning of the metal object in 3D. However, metal objects are often surrounded by strong artifacts in these images, hampering the exact identification of their position. Thus, metal artifact correction (MAC) methods have been put in place [1,2].

MAC methods, especially common 2$^{nd}$ pass methods [1], are computationally expensive. Thus, a computationally cheap decision whether MAC processing is necessary or not (i.e., whether the reconstructed image contains metal or not), is helpful. Also for other purposes, e.g., regarding the interventional workflow, fast methods clarifying whether a metal object has been inserted or not might be of interest.

The simplest method for the classification task detecting whether an image contains metal or not, is to apply a metal segmentation threshold to the image [1]. This requires that the voxel values of the reconstructed image represent HU values. To reconstruct an image on the HU scale, the primary flux intensity during the scan of the object has to be known. Since in some scenarios the primary flux is modulated during the scan to reduce dose, it has to be determined individually for each acquired projection, which requires additional calibration effort.

Sometimes this effort is avoided, since even with a very rough approximation of the primary flux, the quality of the reconstructed images is adequate for diagnostic tasks relating to high contrast structures like bone and metal objects. However, the voxel values in the images do not represent HU values and the threshold-based metal classification mentioned above is not reliable.

In this abstract a computationally cheap method is introduced to classify if a given 3D image contains metal or not. This method does not rely on any relation of the reconstructed voxel values to the HU scale and works robustly as long as the contrast between objects in the image is not compromised and voxel values are not strongly clipped.

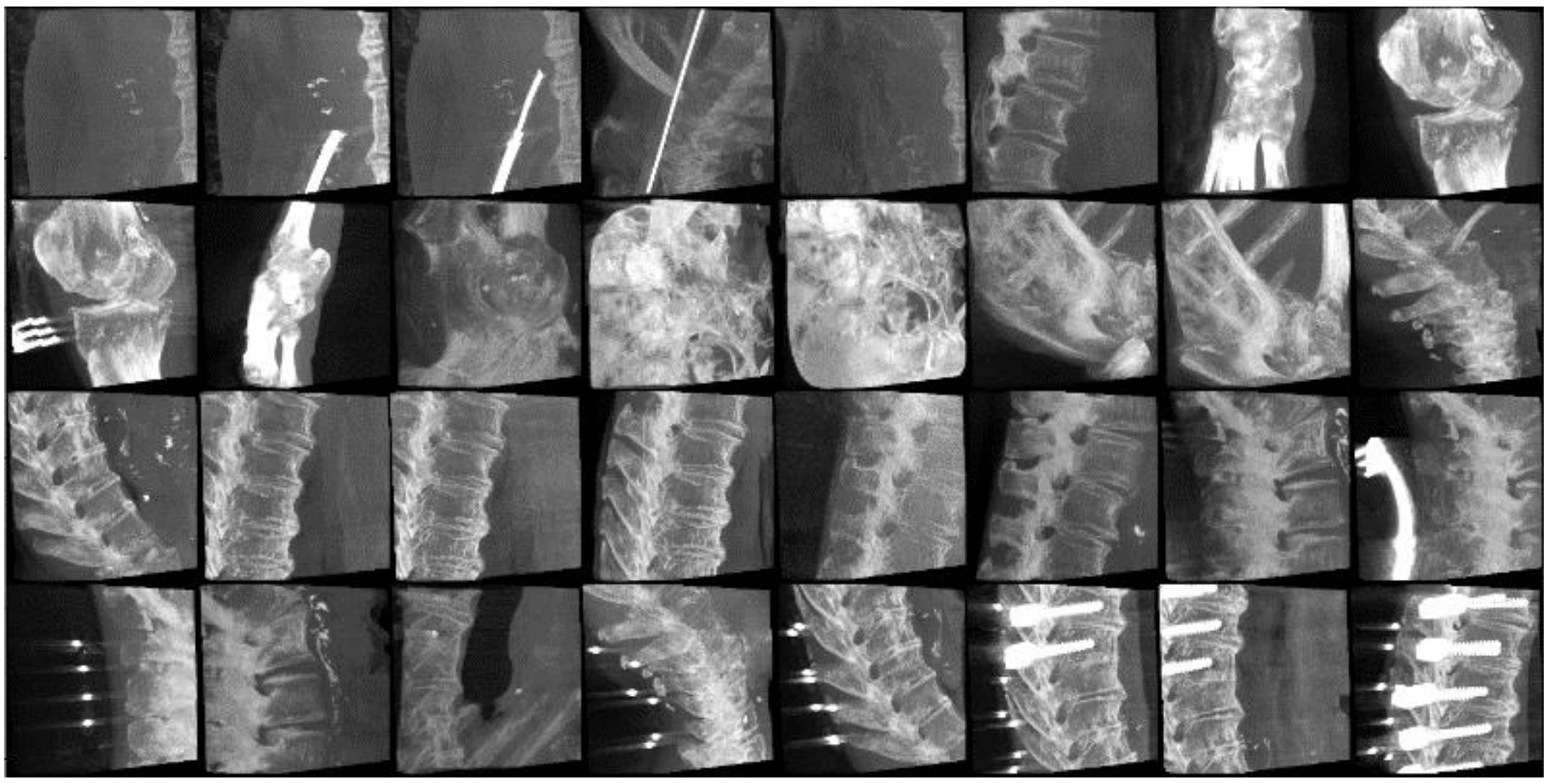

*Figure 1: Maximum intensity projections (MIP) through 3D images of scans representing the different anatomies and configurations with and without metal.*

## 2 Materials and Methods

The following investigations are based on 96 datasets acquired in a cadaver study with a C-arm system. The datasets cover different anatomical regions (ankle, knee, hip, wrist, shoulder, head, thorax, spine) with different metal configurations (no metal, clamps, fiducial markers, wires, bronchoscope, screws). A brief overview is given in Figure 1. These scenarios were scanned with different scan protocols, varying the frame rate and the tube settings (fixed kVp and varying kVp). For each of the 96 scans, 3 reconstructions were performed with different volume sizes ($256^3$, $394^3$, and $512^3$ voxels), resulting in overall 288 images.
The metal detection method introduced here is based on a coarse histogram of an image. It is motivated in the following on two datasets, one without and one with metal (see Figure 2).

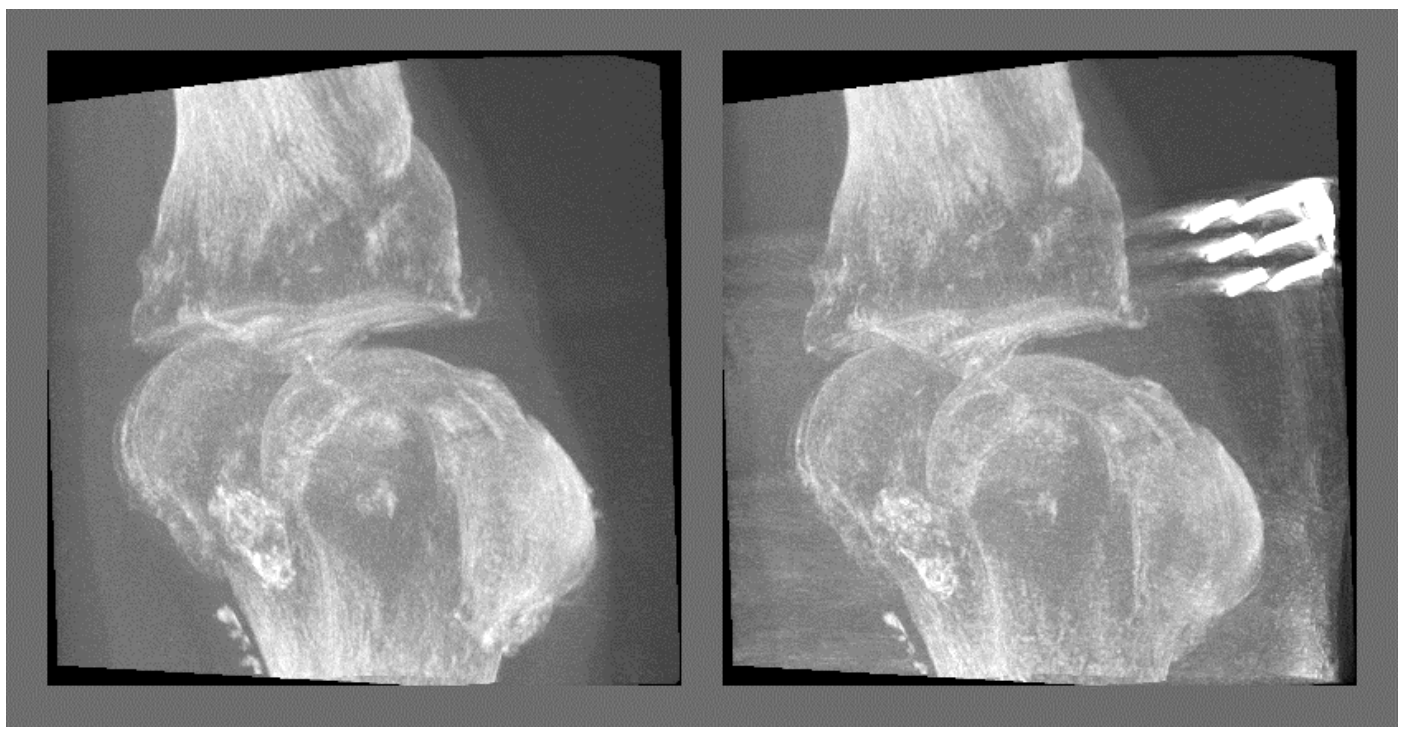

*Figure 2: MIP through 3D images of a knee dataset without (left) and with metal (clamp, right).*

Since it is assumed that the HU scale is unknown and the gray value scale is varying between the images, histograms of min-max normalized voxel values are considered in the following. I.e, the range of voxel values of each image is linearly mapped onto a range from zero to one.
It is expected that the histograms of images without and with metal are different, if the voxel values in dark areas (values lower than air) and of metal are not clipped. This is due to the fact, that voxels representing metal have much higher values than voxel representing air, soft tissue, or bone. Furthermore, metal artifacts lead to voxel values much lower than those of air. The effect on the histograms of normalized voxel values is illustrated in Figure 3. The main peaks (representing the value range of air, soft tissue and bone) in the histograms of images with metal are much narrower and higher than for images without metal due to the much higher range of voxel values before normalization.
The histograms show a strong dependency on the volume size, which can be reduced by normalizing the histogram entries, to represent probabilities instead of number of occurrences (see Figure 4).
The resulting normalized histograms have probabilities very close to zero for voxel values > 0.5, with very low visible differences between images without and with metal.
This can be improved by taking the log of the normalized histograms, as shown in Figure 4, disclosing significant differences in shape for normalized voxel values > 0.5. The different shape is due to the fact, that for cases without metal the voxel value range > 0.5 represent mainly bone, while for cases with metal it represents mainly metal, which on the one hand covers a much smaller volume fraction, and thus results in much lower probabilities, and on the other hand generates a much broader range of voxel values (compared to the air and soft value range) than bone.

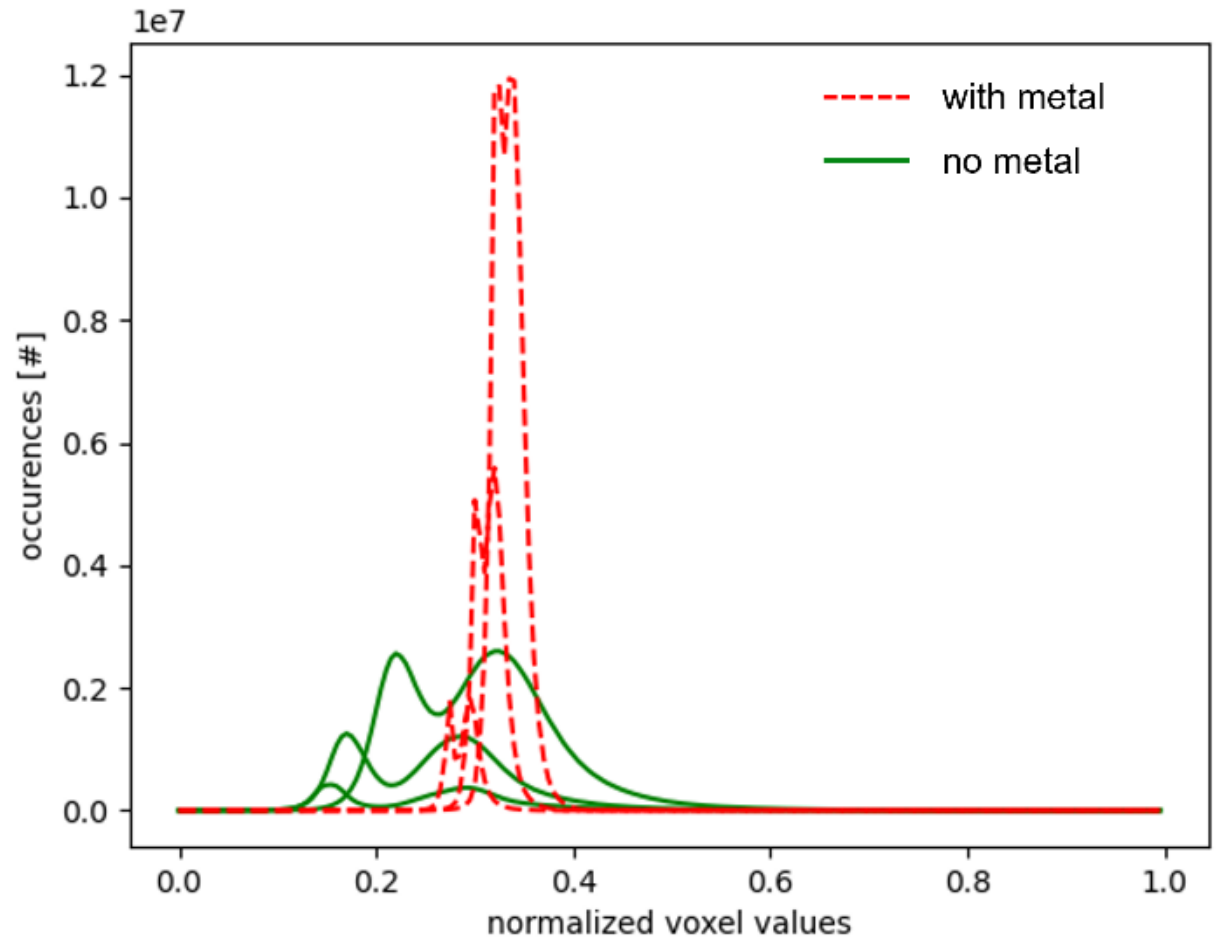

*Figure 3: 200 bin histograms of the two knee dataset shown in Figure 2 for three different volume sizes ($256^3$ / $384^3$ / $512^3$ voxels)*

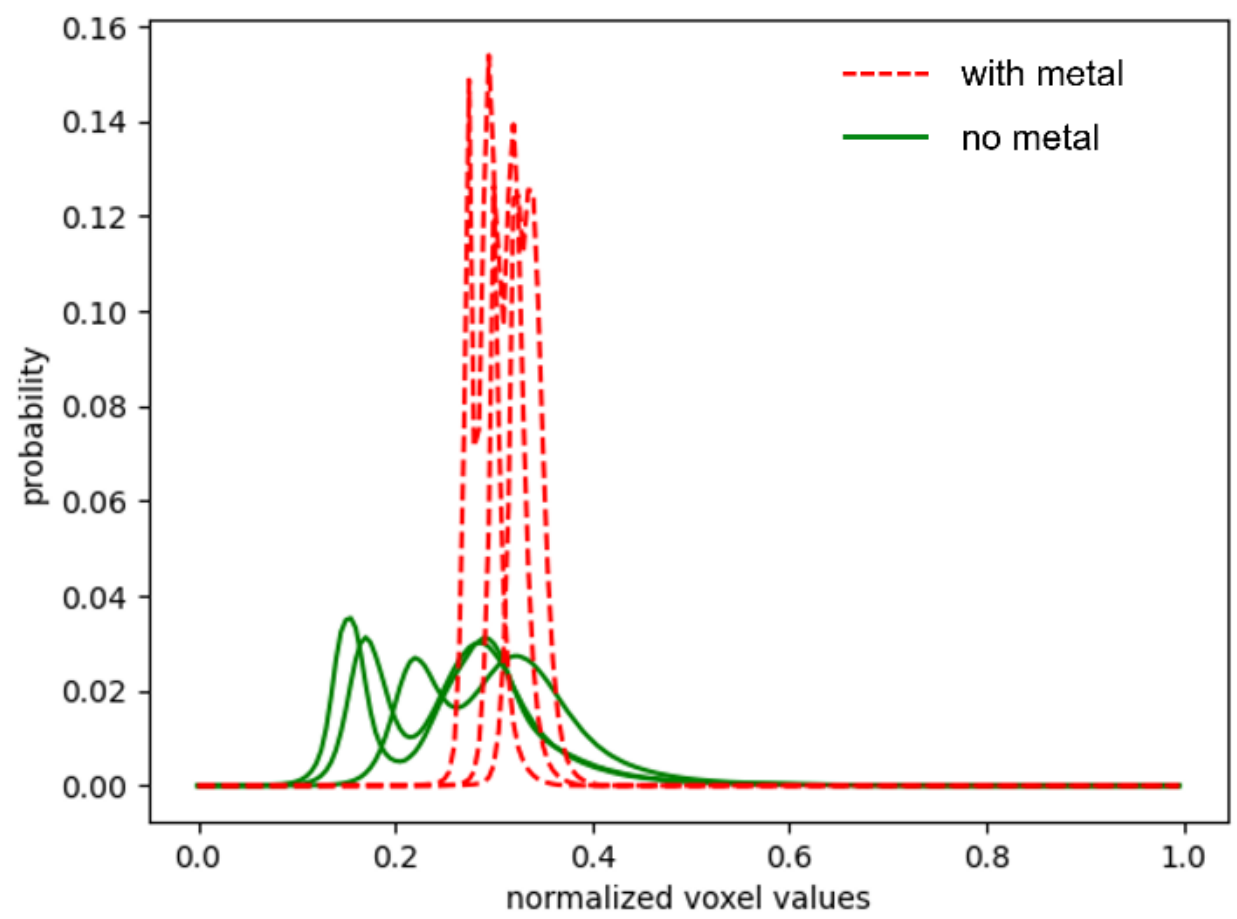

*Figure 4: Normalized 200 bin histograms of the two knee datasets shown in Figure 2 for the three different volume sizes ($256^3$ / $384^3$ / $512^3$ voxels)*

Due to the strong differences illustrated in Figure 4 for cases without and with metal, a histogram with only very few bins can be used as basis for a classifier. In Figure 5 it is shown that even for only 12 bins differences are clearly visible.
The straightforward approach chosen here for a classifier discriminating cases without and with metal is a hyperplane in the 12-dimensional space of the 12 bin histograms that separates these two case classes. The task of finding such a hyperplane is exactly the task solved by a linear Support Vector Classifier (linear SVC) [3]. Advantageously, the linear SVC determines the hyperplane such that the margin between the two classes is maximized, promising good generalization.

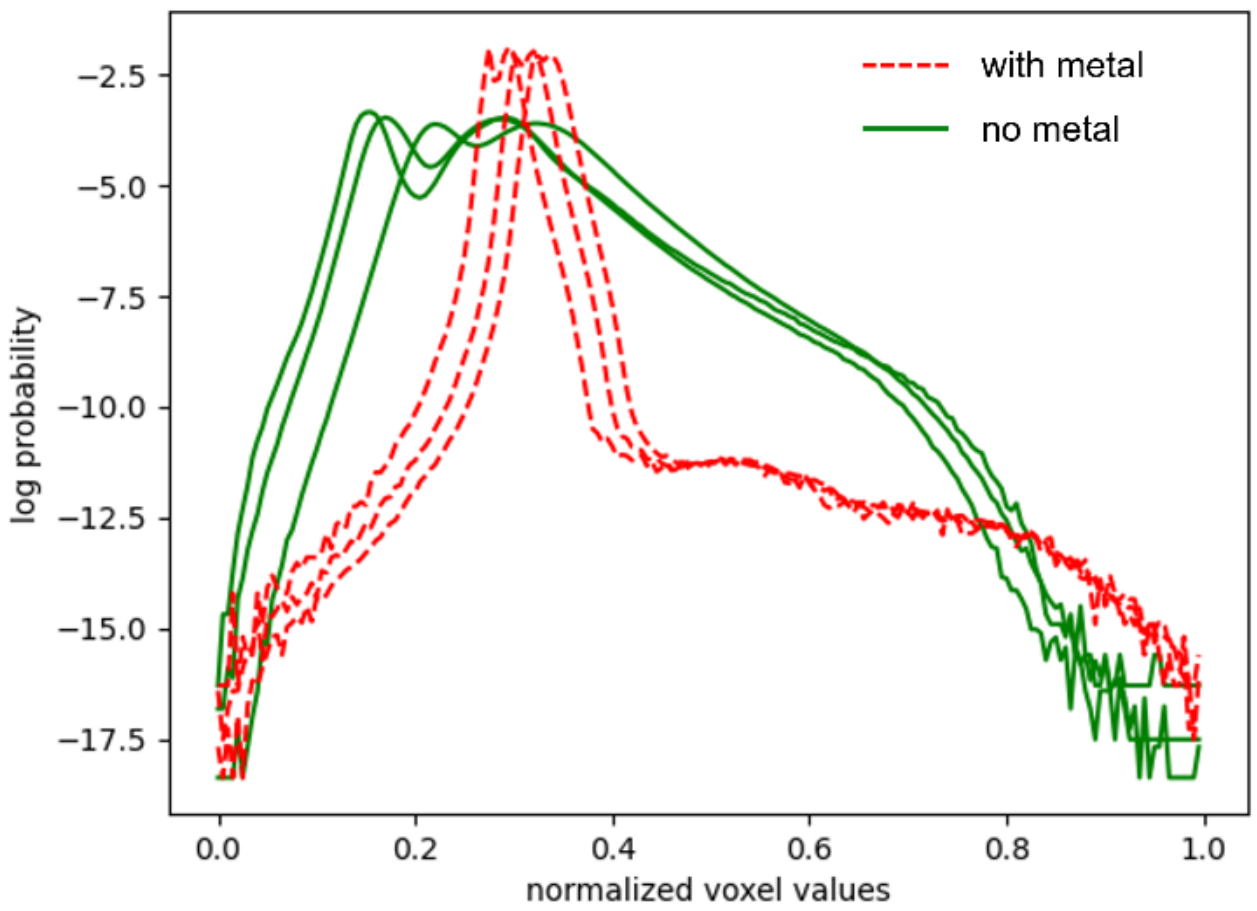

*Figure 4: Logarithm of the normalized 200 bin histograms of the two knee dataset shown in Figure 2 for three different volume sizes (256³ / 384³ / 512³ voxels)*

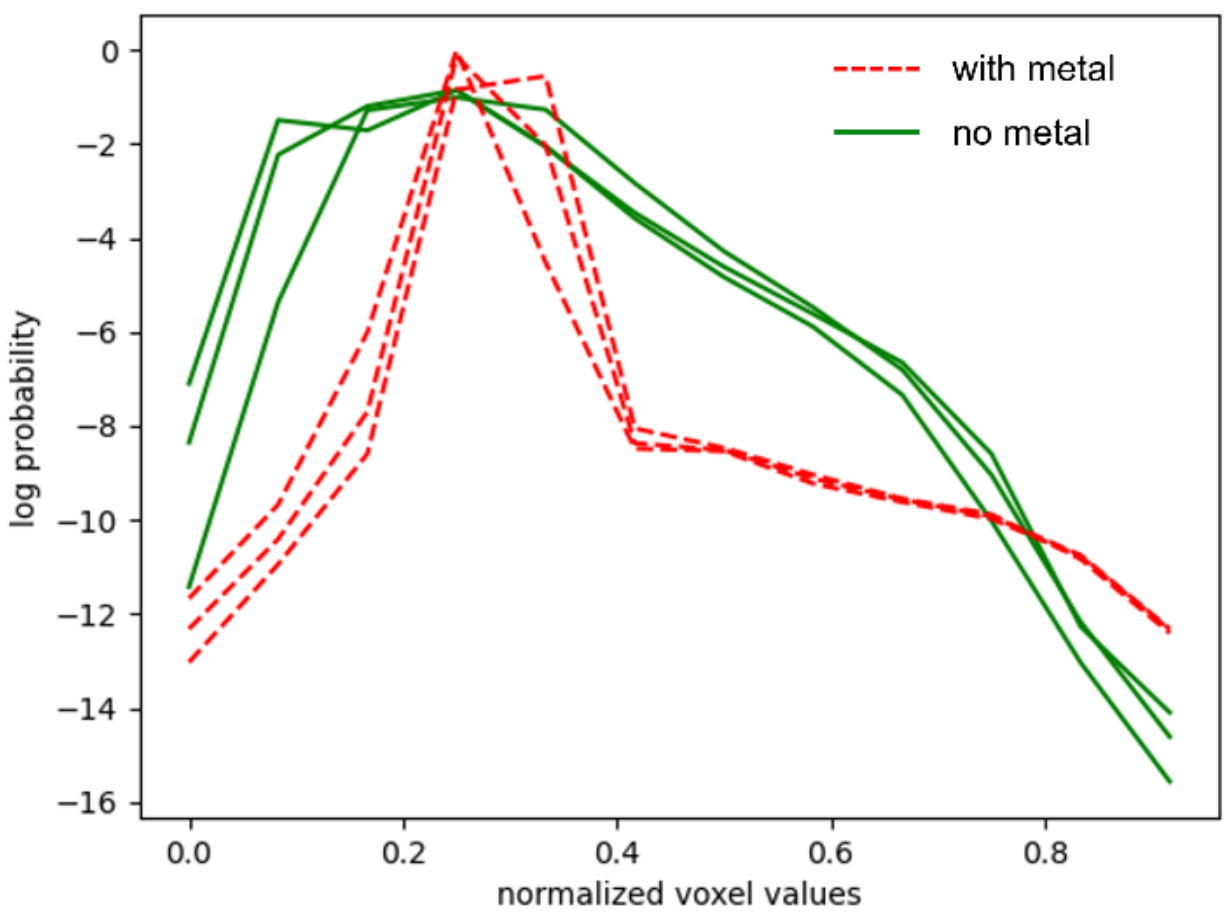

*Figure 5: Logarithm of the normalized 12 bin histograms of the two knee dataset shown in Figure 2 for three different volume sizes (256³ / 384³ / 512³ voxels)*

At runtime, the computation of the linear SVC is very simple. Given the 12 logarithmized histogram values $v_i$ ($i = 1\ldots12$) the classifier output is the logical state of the following inequality $\sum_{i=1}^{12} v_i \cdot c_i + c_0 > 0$ with $c_i$ ($i = 0\ldots12$) being the 13 parameters of the SVC defining the hyperplane.

## 3 Results

The 12 bin histograms of all 96 datasets of this study are visualized in Figure 6 and Figure 7, verifying that the conclusions drawn on the histogram shapes in Figure 4 and Figure 5 generalize well over a wide range of scenraios.

A training of a linear SVC (python sklearn.svm.SVC [5]) with all available 288 images leads to a correct classification rate of 100%. An overfitting is very unlikely since the linear SVC has only 13 free parameters. Still, a leave-one-out test [4] was performed. For this, not only one image was taken out per training, but all three images were taken out that were reconstructed with different volume sizes for one dataset under consideration. Classification rate of the training data is 100% in all 96 leave-one-out tests, and for only one left-out dataset a misclassification occurs. The histograms belonging to the three images reconstructed for this dataset are denoted by the black arrow in Figure 6.

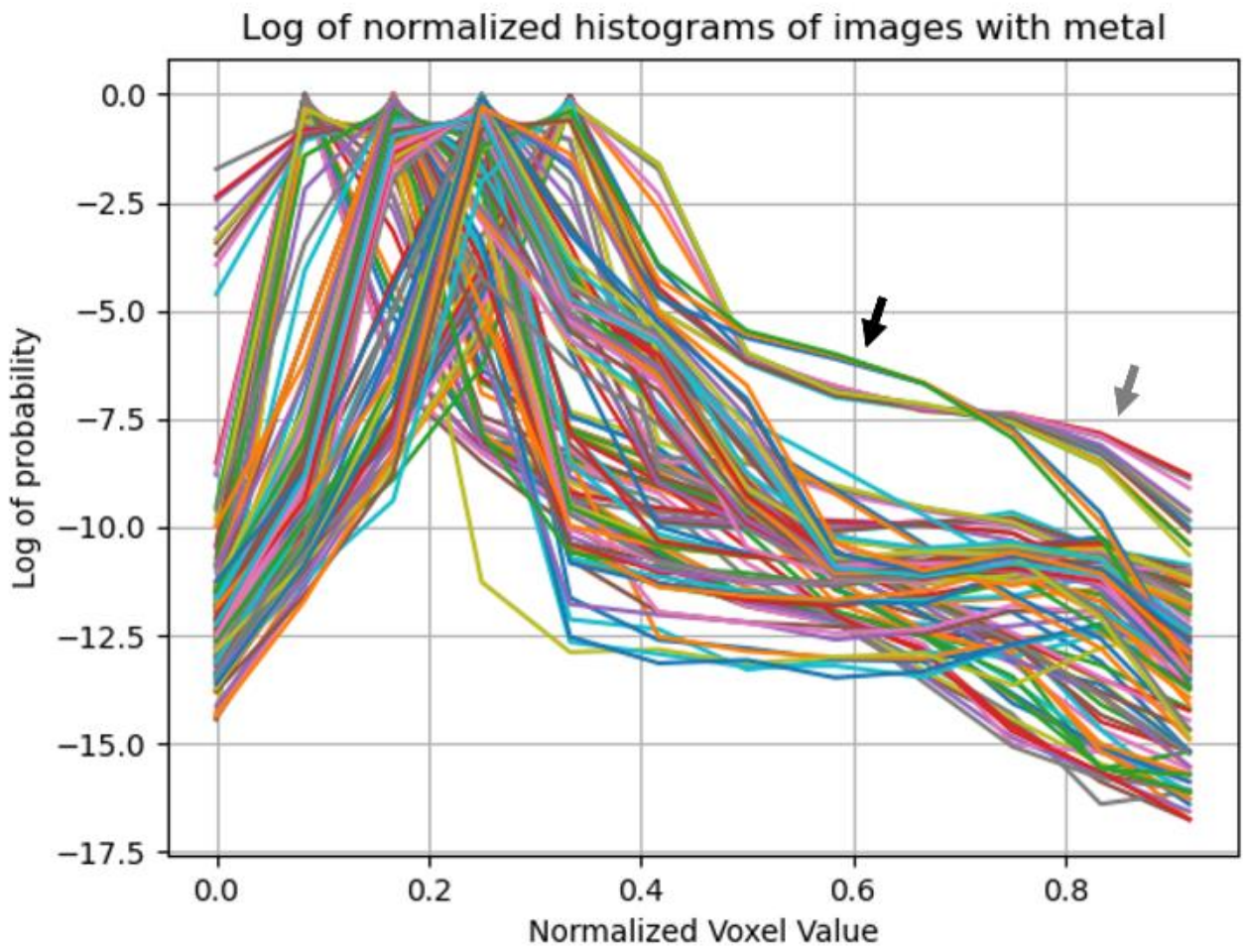

*Figure 6: Logarithm of the normalized 12 bin histograms of all reconstructions of all datasets with metal.*

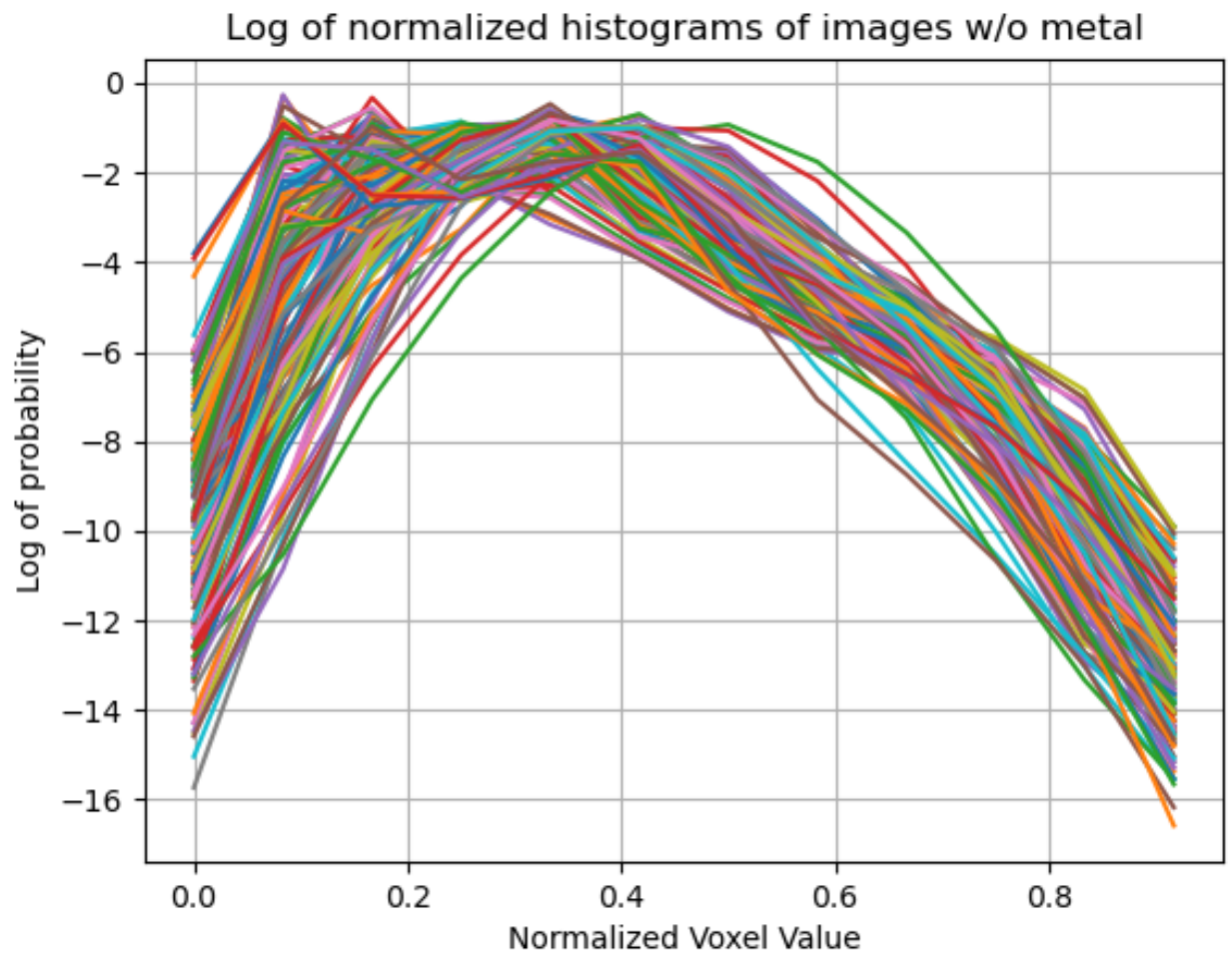

*Figure 7: Logarithm of the normalized 12 bin histograms of all reconstructions of all datasets without metal.*

## 4 Discussion

It has been illustrated that the classification whether a CBCT image contains metal or not is robustly possible with a linear SVC based on a coarse histogram, even if the correlation of the voxel values to the HU scale is completely unknown. To ensure a good performance of the classifier, strong clipping of the voxel values for very low and high values has to be avoided.

As can be seen in Figure 6 (black arrow), the shapes of the histograms for the dataset failing in the leave-one-out test deviate somewhat from the histograms of all other datasets, illustrating that a good representation of all reasonable scan scenarios in the training data is crucial to avoid out-of-distribution errors in the classification, even for this simple model with very low number of parameters. It is however remarkable that the leave-one-test did not fail for the histograms denoted with the gray arrow in Figure 6.

## 5 Conclusion

The computational effort of the classification is very low, and can thus be used to take fast decisions if further processing steps are necessary or helpful, like, e.g., 2nd pass metal artifact correction.

# Quantitative Dual-Energy Spectral CT at Ultra-Low Dose

Kevin M. Brown[1] and Stanislav Žabić[1]

[1]Philips Healthcare, Cleveland, OH, USA, 44122

***Abstract*—We address the clipping-induced bias problem arising from electronic noise in energy-integrating Spectral CT at ultra-low dose, and we show that with proper treatment these kind of systems can deliver quantitative Spectral CT images even at ultra-low dose levels.**

## I. Introduction

Spectral CT imaging with simultaneous multi-energy acquisition can be realized with either energy-integrating [1] or photon-counting detectors. One commonly cited advantage of photon-counting detectors is their lack of electronic noise, and this has been shown to lead to improved HU stability in Spectral CT images at ultra-low dose, compared to energy-integrating detectors [2]. However, a proper attention to the effects of electronic noise at ultra-low dose can lead to a marked improvement in Spectral HU stability for energy-integrating detectors, which we aim to show in the following sections.

### A. Problem Statement

For a monochromatic approximation in energy-integrating (scintilator-based) CT, the measured signal (for each detector) is typically expressed as

$$s = \mathcal{P}(N_\alpha e^{-l}) + \mathcal{D}_\sigma \tag{1}$$

where $\mathcal{P}$ is a Poisson random variable with mean $N_\alpha e^{-l}$, $N_\alpha$ is an initial number of photons at a tube current $\alpha$, $l$ is the line integral of attenuation along a given ray path, and $\mathcal{D}_\sigma$ is a zero-mean Gaussian random variable with standard deviation $\sigma$, that models the electronic noise of the detector [3]. Prior to image reconstruction the measurements $s$ are converted into the "logarithm" or "line integral" domain by

$$p = -\log\left(\frac{s}{N_\alpha}\right) \tag{2}$$

A problem occurs at low dose when the electronic noise is of similar magnitude to the detected number of photons: sometimes the measurement $s$ can be negative, and the logarithm is undefined for values $\leq 0$. Typically negative measurements are clipped to some small positive number, but this clipping introduces a bias to the measurements (as illustrated in Figure 1) that results in image artifacts and inacurate HU values. Here we refer to the clipped measurements as $s_c$, and we can go back and forth between clipped measurements and line-integral values by

$$p_c = -\log\left(\frac{s_c}{N_\alpha}\right) \tag{3}$$

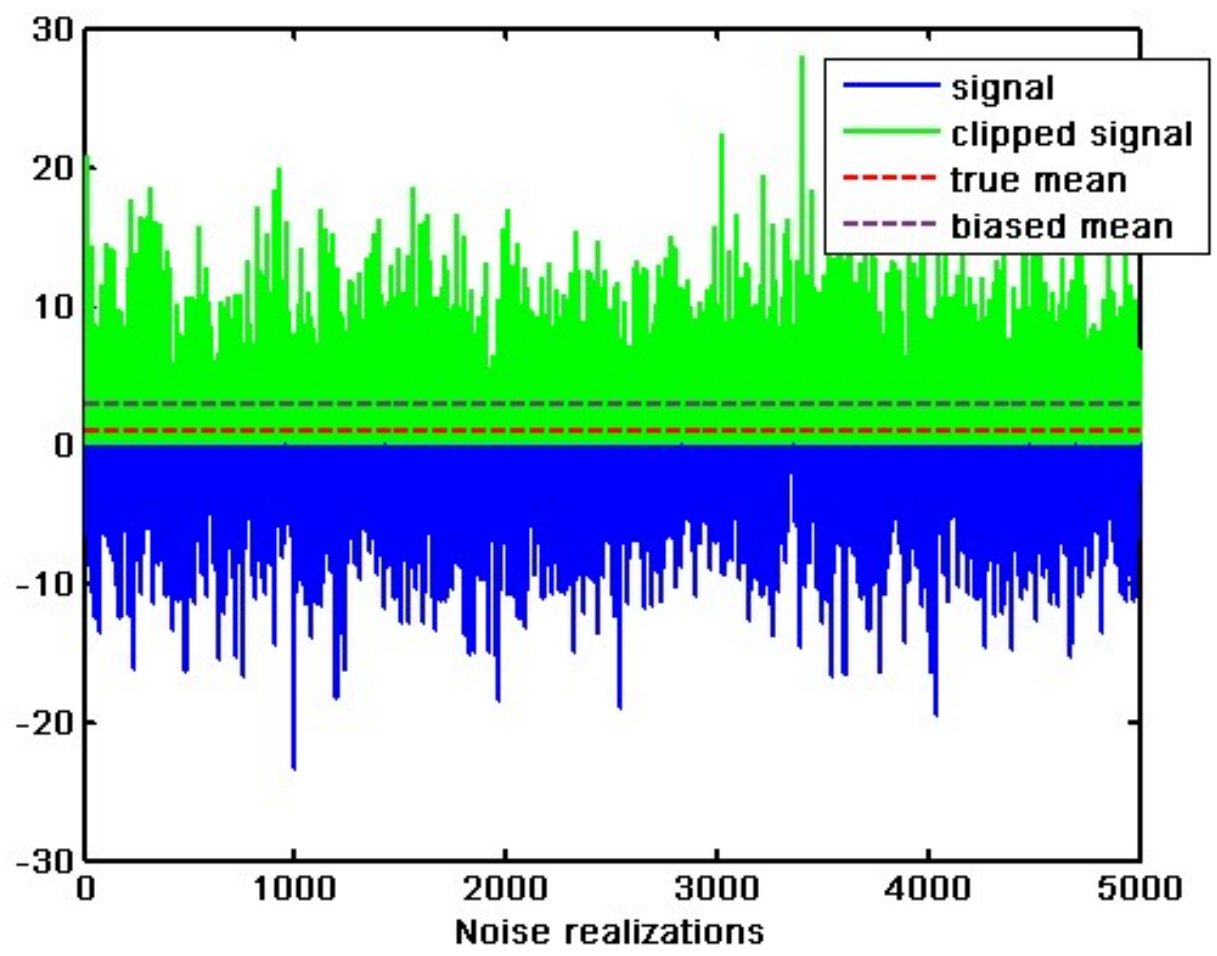

Fig. 1. An illustration of the bias introduced by clipping negative signals.

In a previous publication [4], we demonstrated a method to correct for the clipping-induced bias, even in the case when access to the original un-clipped measurements is lost (this is typically done on many commercial CT scanners because the logarithm operation compresses the dynamic range of the data, thus reducing disk storage requirements). The previous study used only simulated data for the demonstration. This study extends the previous one by applying the method to real measured data from a commercial CT scanner, and also by demonstrating the improvements at ultra-low dose for Spectral CT as well as for conventional images.

## II. Methods

### A. Data Acquistion

All data are acquired using a CT 7500 Spectral Detector scanner (Philips Healthcare) at 120 kVp tube voltage. The Spectral Detector design consists of a dual-layer energy-integrating scintilator which simultaneously measures photons of low and high energies separately. Axial low dose scans are performed at 10 mAs (0.8 mGy), and for comparison ground truth (GT) images we used a scan at 320 mAs (26 mGy). The phantom scanned is the Multi-Energy CT Phantom (from Sun Nuclear) with inserts detailed later below in Figures 3 and 4.

### B. Bias Correction

The method for bias-correction is the same as in our previous abstract [4], repeated here more briefly for the reader's convenience. For electronic noise $\mathcal{D}_\sigma$ with a fixed, known standard deviation $\sigma$, the mean of the true (un-clipped) signal is a monotonic function of the mean of the measured (clipped) signal $s_c$. Thus, if we can estimate the mean of the measured signal $\bar{s_c}$, we can directly find the correct value for the mean of the un-clipped signal $s$. We call this mapping function $\mathcal{F}_\sigma(\bar{s_c})$ (since it depends on the electronic noise level $\sigma$) and note that it can be computed analytically or through Monte-Carlo simulations with Poisson and Gaussian random variables, or by measurement on a real CT system.

We estimate the mean of the clipped signal with a large 3D smoothing filter (9x9x25) applied to the unlogged clipped data $s_c$ to generate smoothed data $s_m$. The smoothed data is then input into the correction function $\mathcal{F}_\sigma$ to generate a sinogram of the estimated mean of the true (unclipped) signal:

$$\bar{s_t} = \mathcal{F}_\sigma(s_m) \tag{4}$$

The final step in the method is to use this (low-res) mean sinogram to correct the measured data. We can't simply replace the measured data by the true mean sinogram $\bar{s_t}$, since it does not have sufficient resolution. We know the mean bias overall by $s_m - \bar{s_t}$, but we cannot simply subtract that bias from each point in the noisy clipped data data $s_c$, because then additional data points would become negative and clipped. This additional clipping can happen even if we subtract the bias from the denoised clipped data $s_{dn}$. To avoid these problems, we make use of the logarithmic identity $\log(a-c) = \log(a)+\log(1-c/a)$ to generate a low-frequency additive correction to the logged raw or denoised data. The final bias correction formula is thus:

$$p_{corr} = -\log\left(\frac{s_{dn}}{N_\alpha}\right) - \log\left(\frac{\bar{s_t}}{s_m}\right) \tag{5}$$

Figure 2-(a) shows an example low-dose logged sinogram ($p_c$), from the low-energy detector layer. White points in the image represent clipped values, where 0 or negative signal was measured, and the logarithm is clipped to some large positive value. Figure 2-(b) shows the estimated bias correction that is derived from this sinogram, using the methods described above. The bias correction is given by the second term in Eq. 5 above.

For the dual-energy applications shown here, the bias correction is computed and applied separately for the low- and high-energy sinograms that we get from the spectral detector acquisition.

### C. Projection Denoising

Even without electronic noise, some method of projection denoising or other approach is needed to tame the noise and bias from the logarithm non-linearity at ultra-low dose [5], [6]. Projection denoising here uses the TV method previously described in [7], delivering denoised data $s_{dn} = TV_{proj}(s_c)$, and both nominal and bias-corrected data use the same denoising method. Besides the bias correction, this is the only non-linear processing applied until the image domain. We reconstruct images at a 5 mm slice thickness using a linear FBP algorithm.

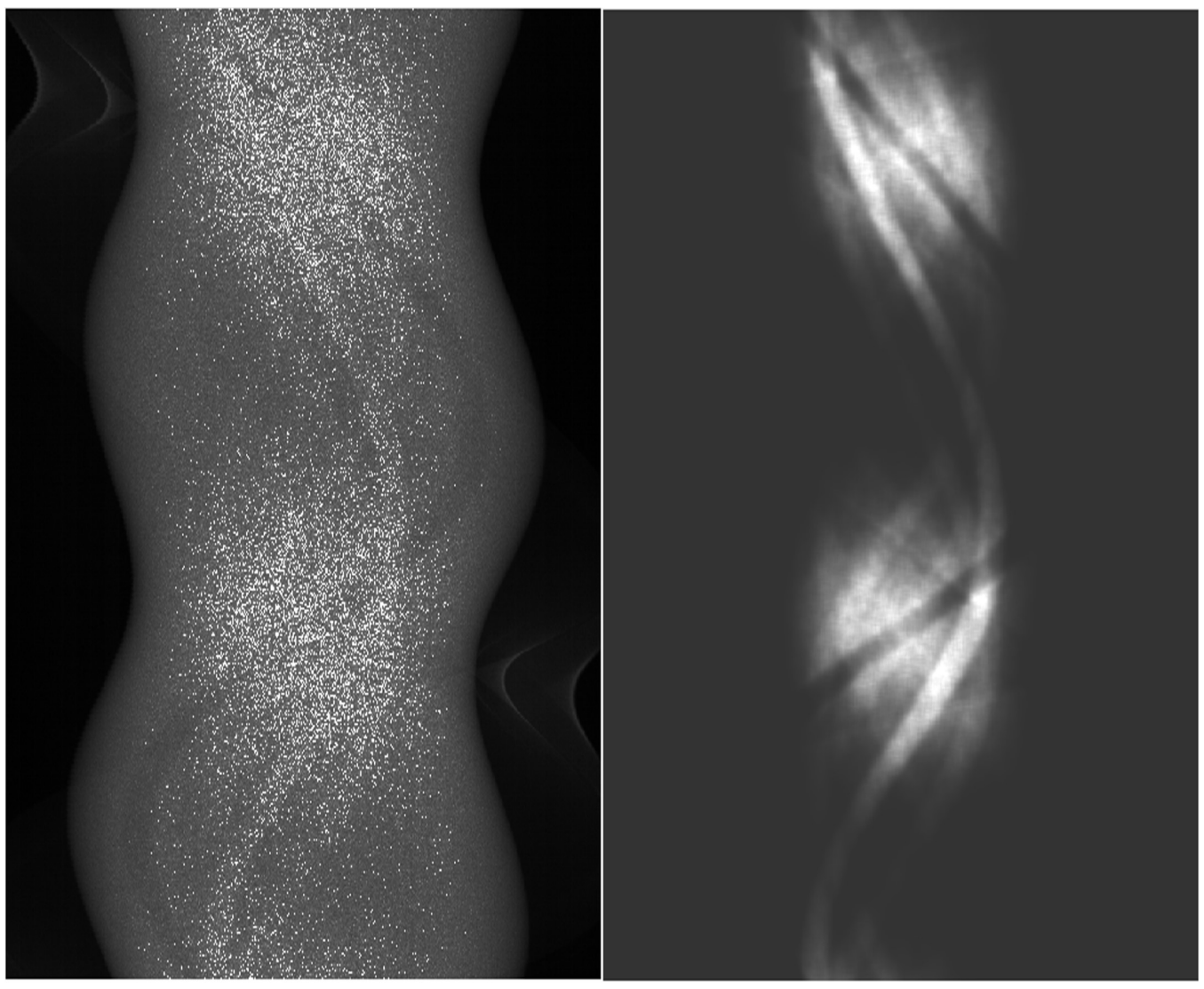

Fig. 2. Left = raw data sinogram of the Low-energy detector layer. Right = computed bias correction by the proposed method (that is, the 2nd term in Eq. 5).

### D. Spectral Decomposition

The simultaneous dual-energy acquistion of the spectral detector permits either projection- or image-domain decomposition. In general it is well known that a projection domain decomposition has advantages over an image-domain decomposition, particularly in suppressing beam-hardening artifacts [8]. For the special case of ultra-low dose scanning investigated here, there are some advantages to an image-domain decomposition, namely the averaging effect of backprojection that reduces the low frequency noise which is challenging to remove from the projection data alone. Thus for this application we use a standard image-domain decomposition defned as:

$$\begin{bmatrix} P_p \\ S_p \end{bmatrix} = \begin{bmatrix} c_{11} & c_{12} \\ c_{21} & c_{22} \end{bmatrix} \times \begin{bmatrix} L_p \\ H_p \end{bmatrix} \tag{6}$$

where $P$ and $S$ are the photo-electric and scatter basis images, where $c_{ij}$ are the coefficients applied pixel-wise for all pixels $p$ in the input low and high-energy images $L$ and $H$.

To deal with and reduce the anti-correlated noise in the resulting Photo and Scatter basis images, we apply the denoising algorithm given in [9], a Huber-penalty minimization that includes an anti-correlated noise model.

## III. Results

Figure 5 shows the reconstructed images of the low- and high-energy detector layers with and without bias correction, for a scan of 10 mAs. With no bias correction, there is a dark shading which can be seen in the center of the phantom, for both the low- and high-energy images. In the difference images shown in the second row, additional strong negative bias can be seen in some of the phantom pins (these are the high-density calcium pins). The bias correction removes both the dark shading in the background and the strong bias in the pins, resulting in difference images that are nearly flat compared to the high-dose images (except for increased noise, which is expected at such low dose). The low- and high-energy images are decomposed to photo/scatter basis images as described above, and these are further used to generate mono-energenic images at 70 and 140 kev, which are shown next in Figure 6. As before, bias correction removes shading in the phantom background as well as restoring errors inside the pins, especially those of higher density.

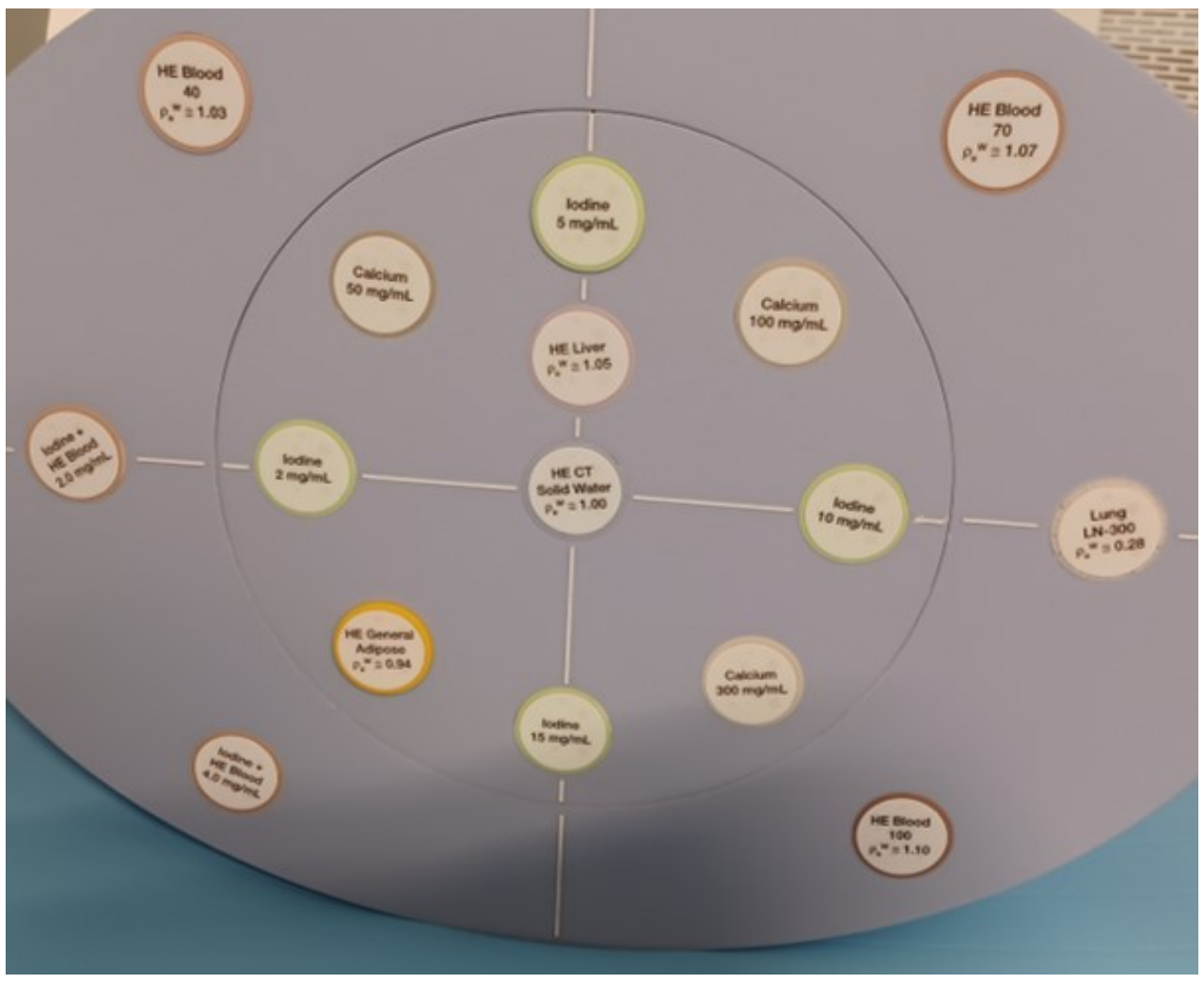

Fig. 3. Position of the material pins inside the phantom.

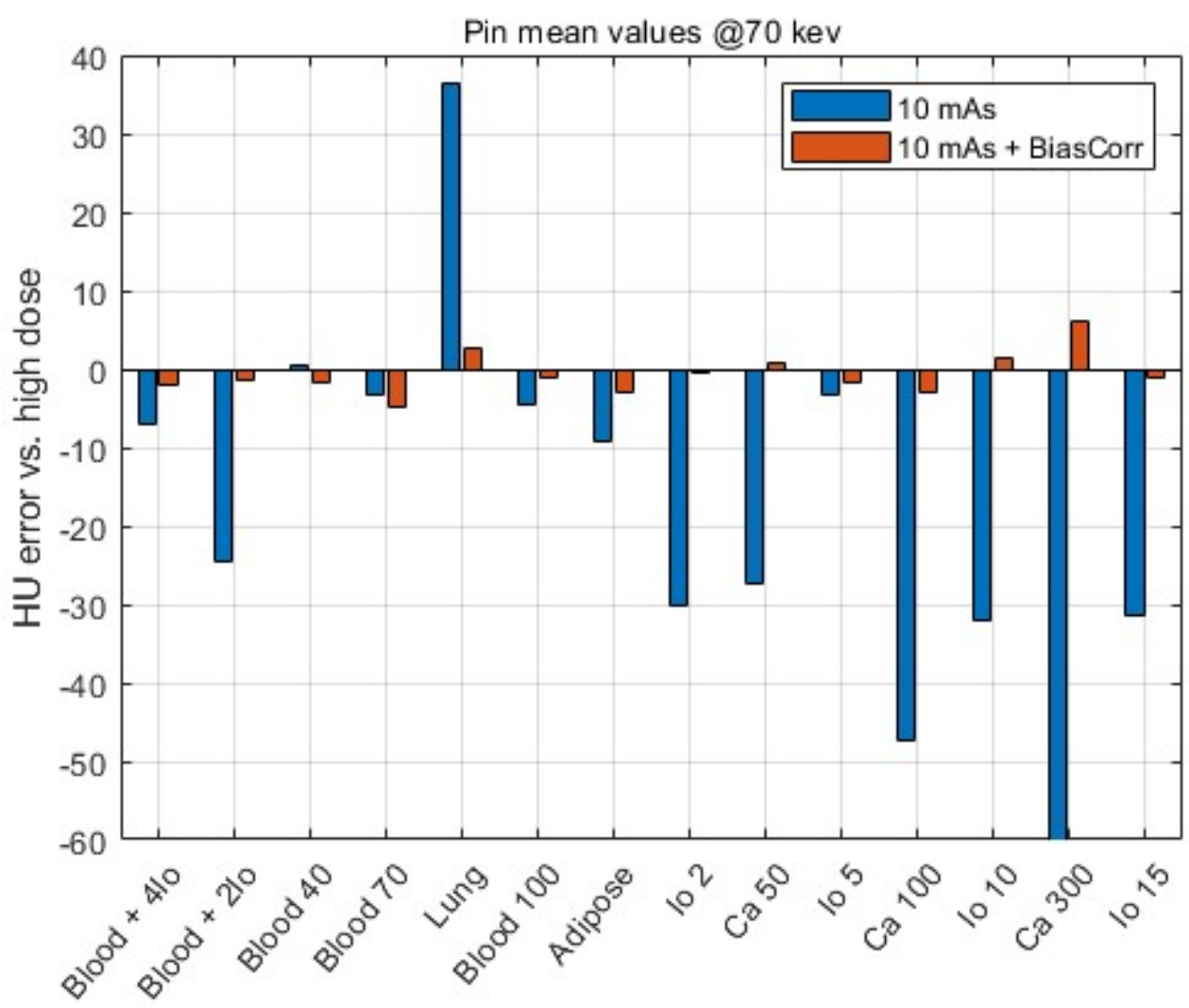

Fig. 4. Error in HU from 320 mAs images measured in each of the pins. The values in each pin (18mm diamter ROI) are averaged over 6 images to achieve these numbers.

### A. Metrics

For a further quantitative comparison, Figure 4 shows the HU error at low dose of the mean values of each material pin in the phantom, at 70 kev mono-E, compared to the high-dose images. Without bias correction, pin values show large errors, depending on the pin composition and pin position within the phantom. With bias correction, all pins show errors of less than 6 HU, with the majority less than 4 HU. Of particular interest is the observation that with the bias correction, the HU stability at low-dose is comparable to that reported recently for a commercial photon-counting detector-based scanner [2].

## IV. Conclusion

A proper treatment of the bias arising from electronic noise enables Dual Energy Spectral CT with energy-integrating detectors to deliver greatly improved quantitative spectral imaging, even at ultra-low dose levels. A more in-depth comparison to CT systems with Photon-Counting detectors is an avenue for future work.

## V. Acknowledgements

The authors gratefully acknowledge Heiner Daerr for assistance with the image domain decomposition coefficients.

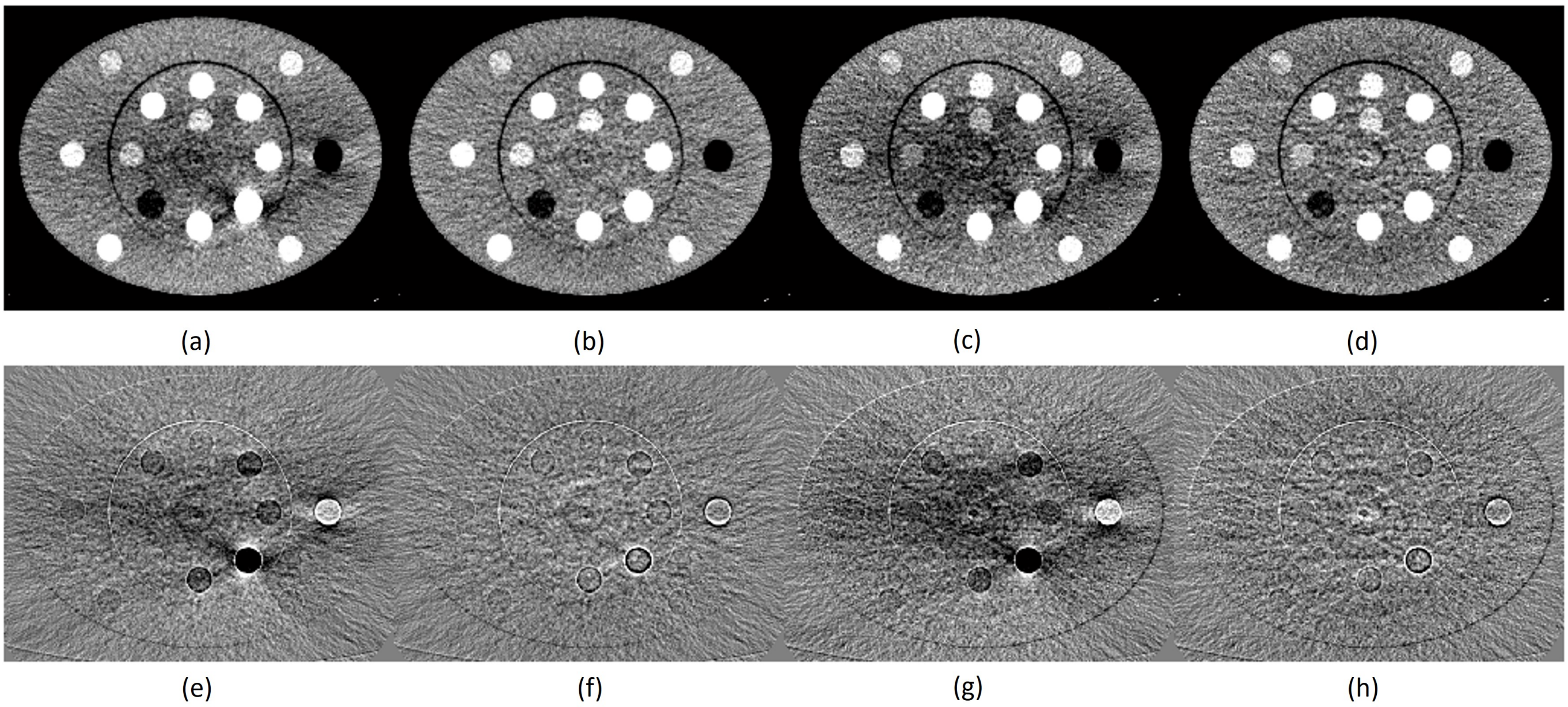

Fig. 5. Bias correction on separate Low and High-energy layer images at 10 mAs. The top row shows: (a) Low-energy image (b) Low-energy image + bias correction (c) High-energy image (d) High-energy image + bias correction. The bottom row shows the difference from the corresponding ground truth 320 mAs images. Four images are avaraged to more easily visualize the bias effects through the noise. Window / Level = 150 / 0 HU.

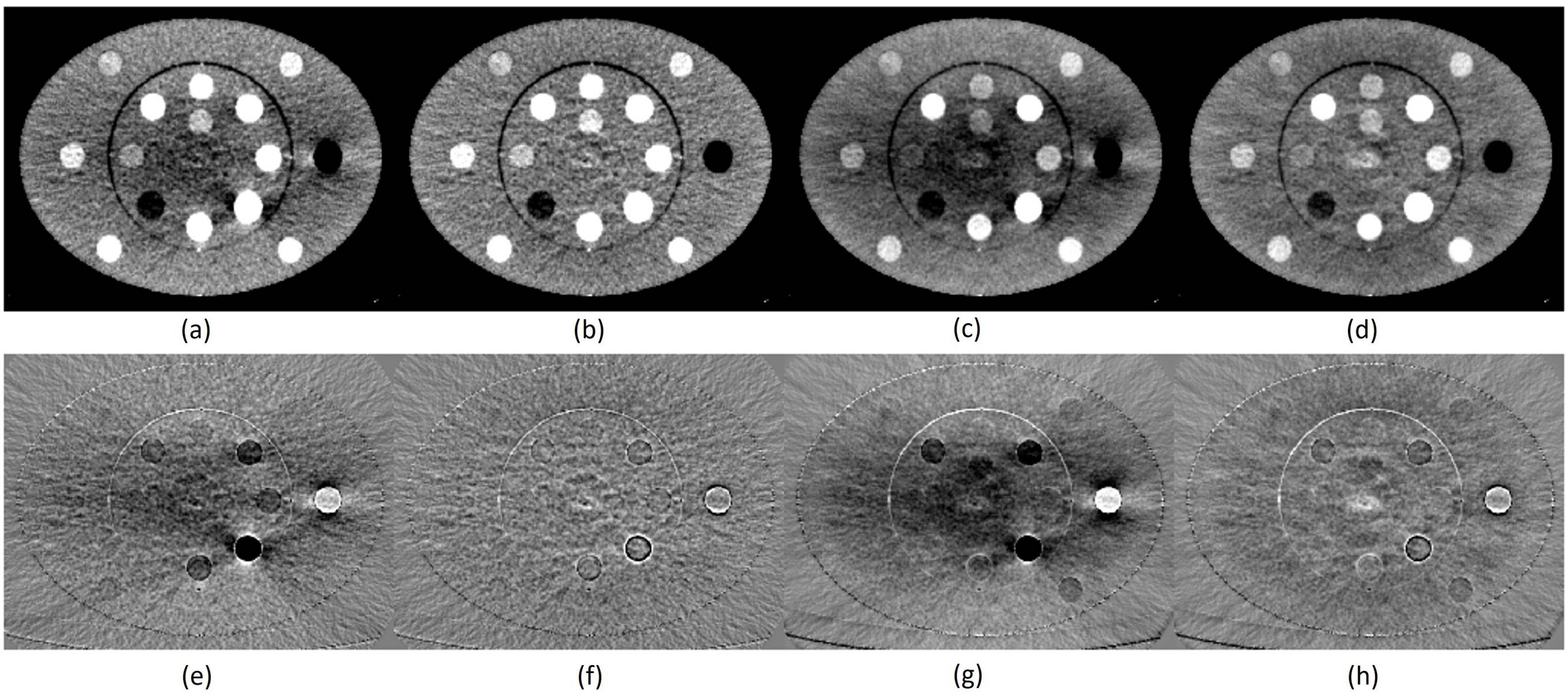

Fig. 6. 70 and 140 keV mono-energetic images. The top row shows 10 mAs images: (a) nominal 70 kev (b) bias-corected 70 kev (c) nominal 140 kev (d) bias-corrected 140 kev. The bottom row shows the corresponding difference images from the GT 320 mAs scan. Window / Level = 150 / 0 HU.

# Convolutional Sparse Coding and Dictionary Learning for improving X-Ray Computed Tomography image quality

Victor Bussy[1], Caroline Vienne[1], Julie Escoda[1], and Valérie Kaftandjian[2]

[1]Paris-Saclay University, CEA, LIST, F-91190, Palaiseau, France
[2]Laboratory of Vibrations and Acoustics, INSA-Lyon, Villeurbanne, France

**Abstract**
This paper evaluates the potential of convolutional sparse coding (CSC) for reducing noise and artefacts in 3DCT images in the case of sparse acquisition configuration. The proposed CSC method is tested on additive-manufactured metallic samples, which present unique challenges for denoising applications due to the fine structure of the material and the wide variety of possible textures in the images. Results indicate that CSC outperforms traditional dictionaries in denoising performance and computation speed, making it a promising method for large-scale tomographic imaging applications.

## 1 Introduction

X-ray Computed Tomography (CT) is a powerful imaging technique for medical and industrial contexts, where it faces the same objective of quickening the acquisition. Even if the reason differs, reducing the dose to the patient in the first case and reducing the inspection time in the second case, the challenge is similar and consists in ensuring the best quality of the reconstructed image from a limited number of projections.
Iterative reconstruction techniques have made it possible to partially solve this problem thanks to regularisation terms and *a priori* knowledge [1]. In recent years, these iterative techniques have been further improved by the theory and development of Compressed Sensing (CS), which allows the reconstruction of high-quality images despite a smaller number of views than the one required for analytic filtered back-projection algorithms [2]. However, conventional CS-based CT reconstructions are computationally expensive because the associated reconstruction methods solve a high-dimensional system with the $\ell_1$ norm. To overcome this problem, *sparse coding*, a particular development of CS, uses a *patch-based* strategy. By only processing small portions of the image, also called *blocks* in 3D, it allows the description of a huge volumetric image by overlapping blocks and thus reduces the demand on digital resources. Many methods using patches have been adapted to tomography in recent years: Non-Local Means (NLM) [3], Block-Matching 4D (BM4D) [4], and in particular sparse coding on redundant dictionaries [5]. The latter has been widely successful thanks to their numerous applications but are still not present in industrial 3D tomography, notably because of the often too long computation time and the artefacts due to the aggregation of patches. Indeed, each block of the image is processed independently and often, for each one, an optimisation must be applied, which makes processing very long. Moreover, the blocks must be overlapped to describe the image well, increasing their number. To overcome these limitations, convolutional sparse coding (CSC), also called shift-invariant sparse coding, proposes a new formalism in which the continuity of the image is taken into account [6]. This property also allows the use of smaller operators. CSC proposes a solution to the problems of numerical resources, computation speed and reconstruction quality. This article will compare dictionary-based and CSC algorithms on industrial data from additive manufacturing for X-ray tomography denoising.

## 2 Methodology

Sparse coding is a widely used technique in signal processing and inverse problems. This technique assumes that a signal $\mathbf{x} \in \mathbb{R}^m$ can be reconstructed for a few elements, called *atoms*, taken from an overcomplete dictionary $\mathbf{D} \in \mathbb{R}^{m \times n}$ [7]. In the sparse coding framework, each block $\mathbf{x}_s$ of a signal $\mathbf{x}$ is encoded using a sparse linear combination $\mathbf{z}_s \in \mathbb{R}^m$ of atoms from the dictionary, such that for each block $\mathbf{x}_s \approx \mathbf{E}_s\mathbf{D}\mathbf{z}_s$, where $\mathbf{E}_s$ represents an operator to extract the block $\mathbf{x}_s$. The problem of finding a sparse representation is called *sparse pursuit* and is the essential point of all sparse coding techniques. Naturally, lots of methods have emerged to solve this problem. The most used are the *matching pursuit* algorithms which will greedily find a solution (OMP [8]), and convex formulations called *basis pursuit* (Iterative soft-shrinkage methods, LARS), which will solve this problem with a $\ell_1$ norm constraint on the representation [9]. With these latter, the objective function to reconstruct the signal becomes the following equation:

$$\hat{\mathbf{z}} = \underset{\mathbf{z},(\mathbf{D})}{\operatorname{argmin}} \frac{1}{2}\sum_s \|\mathbf{D}\mathbf{z}_s - \mathbf{E}_s\mathbf{x}\|_2^2 + \lambda \sum_s \|\mathbf{z}_s\|_1, \tag{1}$$

where $\lambda$ is a parameter that balances the data fidelity and sparsity terms. The dictionary can also be learned simultaneously as the sparse representation. In this case, it becomes a variable of the previous equation, and the problem is called *dictionary learning.*

However, dictionary learning quickly becomes computationally infeasible with long signals and high dimensions. The CSC thus appeared as an evolution of the dictionary-based methods where the global dictionary is, in fact, a banded circulant matrix of a local dictionary. This structure imposes that the blocks are a superposition of the local dictionary's *filters*. The main advantage is that there is no more need to learn a global dictionary made of many atoms but just a small dictionary made of few $m$ filters. Moreover, image continuity is considered, and all translations of the filters are possible. The sparse representation is now obtained with the following:

$$\underset{\mathbf{z},(\mathbf{d})}{\operatorname{argmin}} \frac{1}{2}\left|\left|\sum_{j=1}^{m} \mathbf{d}_j * \mathbf{z}_j - \mathbf{x}\right|\right|_2^2 + \lambda \sum_{j=1}^{m} \|\mathbf{z}_j\|_1, \quad (2)$$

where $\mathbf{d}_j$ represent the filters of the local dictionary, $\mathbf{x}$ represent the image, and $\mathbf{z_j}$ are the *coefficient maps*. In the CSC framework, dictionary learning and basis pursuit are generally done using Alternating Directions Method of Multipliers (ADMM) [10]. By solving the problem in the Fourier domain, [11] proposes a considerable algorithm acceleration.

To compare traditional and convolutional dictionaries, we have selected two denoising methods: unsupervised and supervised.

## 2.1 Denoising with Basis Pursuit

The simplest way to denoise a 3D CT image is to find its sparse representation and to reconstruct the signal in the image basis. This method is straightforward and allows to get rid of the roughness of an image. For traditional dictionary learning, the dictionary is usually learned on another tomographic image, obtained with the same acquisition parameters and reconstructed with the full set of projections.

For CSC, the training volume is first high-pass filtered with a Tikhonov filter. Similarly, the coefficient maps are determined on the high-pass filtered data because, as mentioned in [12], CSC is not well-suited to represent low frequencies. Therefore, high frequencies are denoised separately, and then added back to the low frequencies.

## 2.2 Denoising with joint dictionaries

A second approach to denoise a signal is to use joint dictionaries [13]. Two dictionaries are learned jointly on the sparse and dense reconstructions of the same sample. By forcing the two reconstructions to share the same code or map coefficients but on a different basis, each filter in the created 'sparse' dictionary $D_s$ and 'dense' dictionary $D_d$ are matched. The results of a basis pursuit in the 'sparse' basis can then be used in the 'dense' basis. Using reconstructions of different sizes, [14] even proposes a super-resolution scheme. They also propose an analytical method based on a convex formulation to train both dictionaries simultaneously with classical dictionary learning algorithms. The authors of [13] directly build the two dictionaries by extracting patches at the same positions in the sparse and dense reconstructions. Note that in both methods, to improve the results, the image features in $D_s$ are the voxel values and the first and second-order gradient distributions. $D_d$ is only composed of voxel values.

We propose an intermediate method for the CSC between the analytical and direct methods. Thanks to an ADMM optimiser, a dictionary $D_s$ is learned on the sparse data. Then, each filter is expressed as a linear combination of blocks extracted from the sparse volume. Using the same linear combination and blocks extracted at the same locations in the dense reconstruction, we create a 'dense' version of our initial dictionary. This technique takes advantage of the benefits of the analytical method, which finds the underlying signal structures, but also of the simplicity and computational speed of the direct method. As mentioned previously, the proposed method can also use image features other than voxels. It also has the advantage of requiring few numerical resources, unlike the analytical method, which requires solving a high-dimensional problem. The method is recapped in the Algorithm 1, and Figure 1 shows matched filters in $D_s$ and $D_d$.

**Algorithm 1:** Joint dictionary learning algorithm.

**Data:** Sparse reconstruction $X_s$, Dense reconstruction $X_d$
**Result:** Joint dictionaries $D_s$ and $D_d$
Compute $D_s$ by solving Eq. (2) with $X_s$
Extract blocks $B_s$ and $B_d$ at the same positions in $X_s$ and $X_d$
Concatenate its $1^{st}$ and $2^{nd}$ order derivatives to $B_s$
Solve $B_s Y = D_s$
$D_d \leftarrow B_d Y$

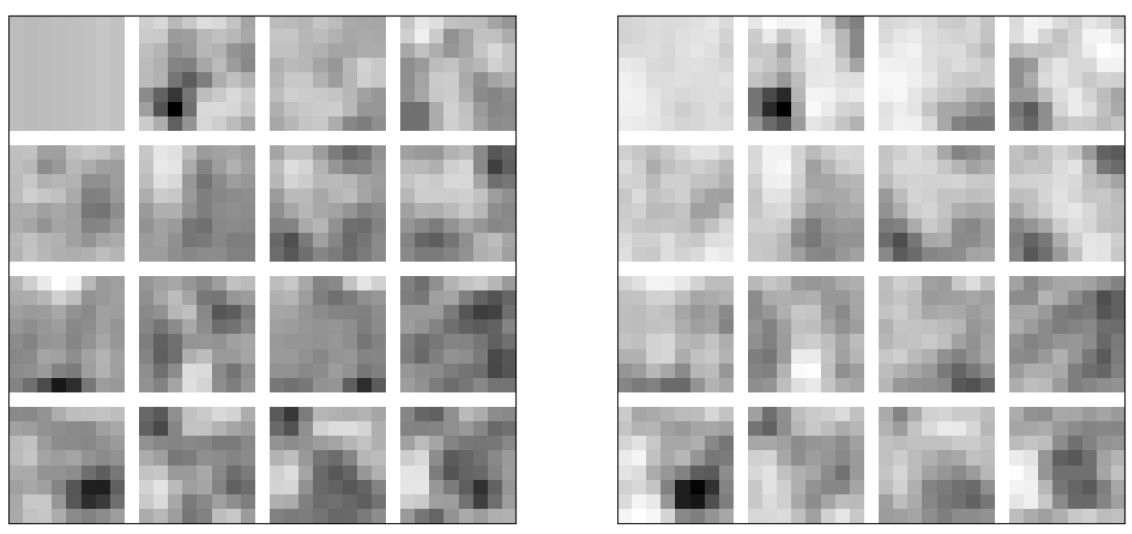

**Figure 1:** Examples of cross-sections from $D_s$ (left) and their match in $D_d$ (right).

## 3 Results

In this section, the previous techniques are tested on three aluminium cubes created by Laser Powder Bed Fusion (L-PBF) additive manufacturing process using different process parameters (laser power and speed). The 3D images are respectively of size 742×601×749, 744×777×723 and 679×670×627 voxels. A fourth cube 500×379×500 is used for training. Voxel size is 20 $\mu m^3$. Figure 2 shows the cross-sections of the different cubes used in this paper. Due to different additive manufacturing processes, the samples have significant texture variations. The sparse and dense images are reconstructed using the FDK algorithm, respectively with 100 and 900 projections regularly acquired around the specimen. The SPORCO package is used for CSC calculations [15].

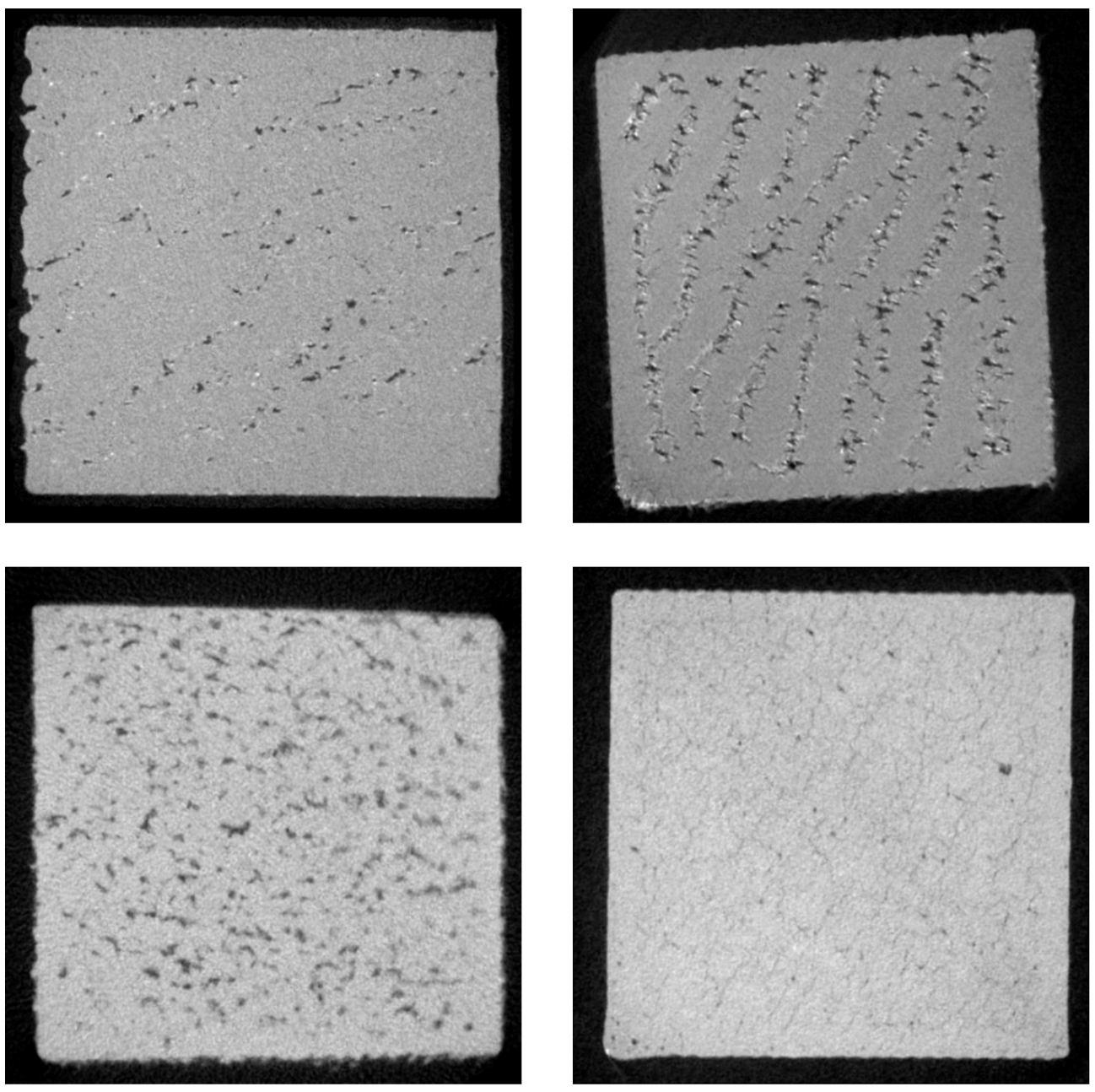

**Figure 2:** Cross-sections of the aluminium cubes. The one on the bottom right is used for training while the others are for tests.

For basis pursuit, the $\lambda$ parameter is set to 0.1. Figure 3 shows that both traditional and convolutional dictionaries allow a better contrast around the cracks and holes. The edges of the image are sharper with traditional dictionaries. With CSC, one can notice some unexpected effects at the edges due to the high-pass filter. The granular aspect of the image is well-corrected between the cracks without losing the image information. The CSC has a stronger smoothing effect.

Then, for the classical joint dictionaries learning problem, we have used the method as described in [14] with the gradient in each direction as additional image features. The dictionary size is $2560 \times 10240$, 512 features are dedicated to the 'dense' blocks, and 2048 are for the sparse blocks and their gradients. The training is made using ADMM. The joint convolutional dictionaries are made of 32 filters of size $8^3$ voxels. 7680 random blocks were extracted to make $B_s$ and $B_d$. The denoised images, shown in Figure 3, appear to be smoothed. The detection of porosities and cracks is much easier. One can notice that CSC appears more textured than traditional dictionaries, but the contrast around little asperities is still good.

In Table 1, we report the Peak Signal-to-Noise Ratio (PSNR) and Structural Similarity Index Measure (SSIM) values for the 100-projections reconstructions of the cubes before and after denoising. The volumes are compared to dense 900-projections reconstructions. NLM (with $\sigma$=0.05, $8^3$ voxels block and $33^3$ voxels search area) and BM4D are also reported for comparison. However, those methods do not improve image quality because the initial image is too textured, and similarities between blocks are hard to find.

**Table 1:** Image quality evaluation of the aluminium cubes with different denoising methods.

| PSNR[dB] / SSIM | Cube 1 | Cube 2 | Cube 3 |
|---|---|---|---|
| Before denoising | 19.16/0.3871 | 22.70/0.3428 | 18.80/0.4561 |
| NLM | 19.12/0.3876 | 22.70./0.3428 | 18.80./0.4562 |
| BM4D | 19.09/0.3912 | 22.71/0.3442 | 18.80/0.4569 |
| Basis Pursuit | 22.56/0.5776 | 25.16/0.6087 | 26.59/0.7188 |
| CSC Basis Pursuit | 25.94/0.6431 | 25.41/0.6427 | 26.77/0.7272 |
| Joint Dictionaries | 24.19/0.5707 | 25.85/0.6802 | 26.67/0.7277 |
| CSC Joint Dictionaries | 25.26/0.5881 | 25.91/0.7051 | 26.71/0.7278 |

Results show the effectiveness of CSC for denoising sparse tomographic reconstructions. CSC outperforms traditional dictionaries in terms of both denoising performance and computation time. CSC takes around 3 minutes to denoise a sample compared to 1 hour for the traditional method, on a CPU Intel Core i9-11950H 2.60GHz.

## 4 Conclusion

We have presented two examples of denoising methods using traditional and convolutional sparse representation for improving the reconstruction quality of additive-manufactured aluminium cubes. Each method has shown, on experimental data, an ability to denoise reconstructions leading to easier analysis and control of the studied samples. CSC has shown a better capacity to denoise tomographic volumes. CSC opens up many possibilities and perspectives for tomographic reconstruction in the Plug-and-Play regularisation framework. Many aspects are still to be treated, including

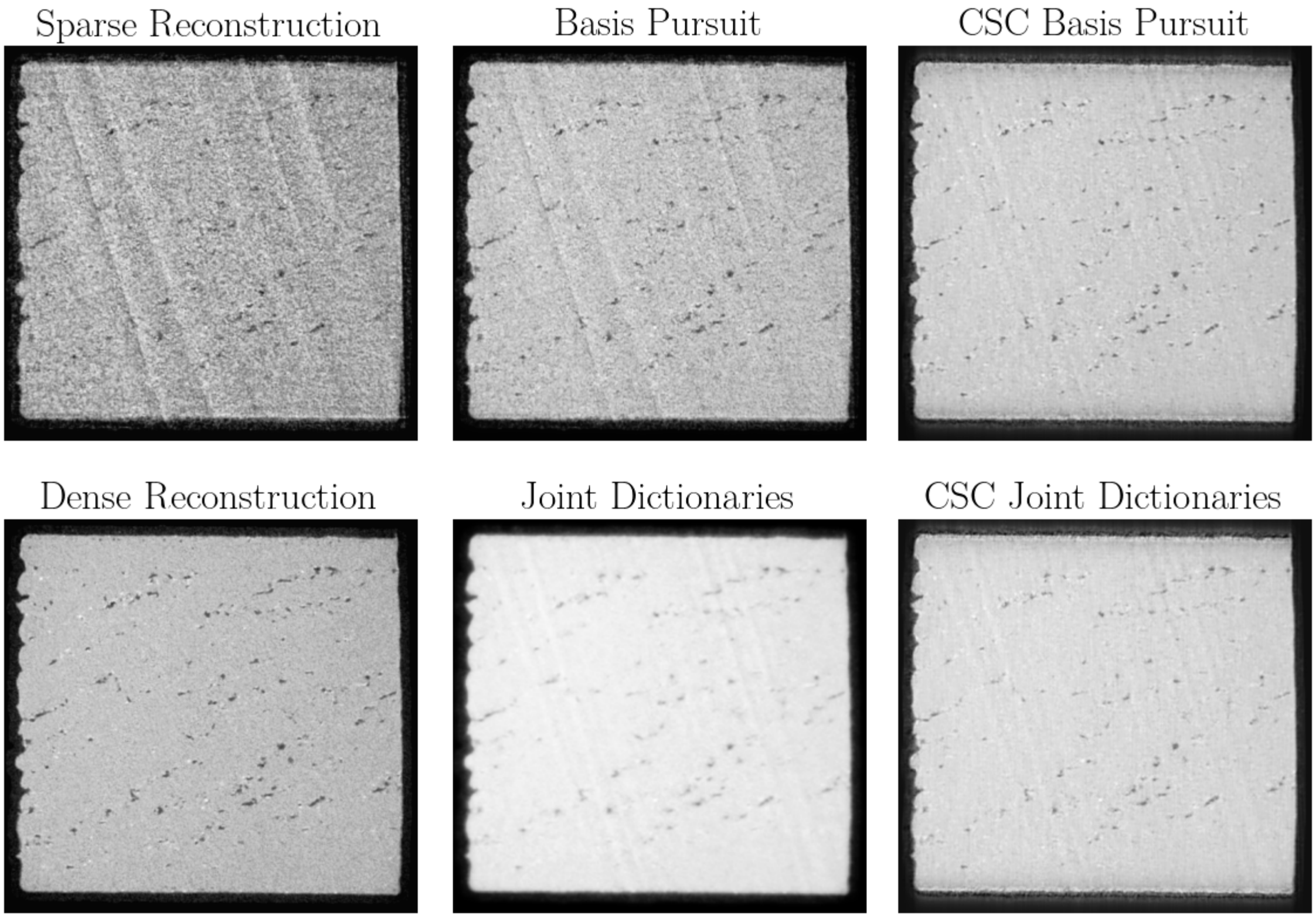

**Figure 3:** Results of the different denoising techniques on the first cube.

classification, segmentation and determining physical properties from map coefficients.

# Robust PET-CT Respiratory-Mismatch Correction Based on Modeled Image Artifact Evaluation and Anatomical Reshaping

Raz Carmi, Nasma Mazzawi, Gali Elkin-Basudo, Lilach Shay Levi, Danielle Ezuz, and Yariv Grobshtein

GE Healthcare Molecular Imaging, Haifa, Israel

**Abstract** Respiratory mismatch in clinical PET-CT is a common source of image artifacts due to inaccurate attenuation-correction, which typically seen as falsely low tracer uptake in the lower lung regions and adjacent organs. Various approaches were suggested to mitigate this problem but achieving practical, efficient, and robust solution is still non-trivial. We propose a clinically practical method to correct respiratory mismatch effects, based on modeled image artifact evaluation and anatomical reshaping. In this method, standard reconstructed PET-CT images are evaluated, and relevant potential artifacts are analyzed. The artifact characteristics control a CT image reshaping model, which is based on predictable respiratory motion pattern, and provides corrected attenuation map for any final PET image reconstruction. Testing the method on numerous and diverse patient studies, in terms of imaging characteristics and clinical protocols, demonstrates accurate and robust artifact elimination.

## 1 Introduction

Attenuation correction in PET medical imaging is an important part of the image reconstruction process. Typically, attenuation correction coefficients are obtained from an associated anatomical image scan with CT or MRI. For achieving high quality functional images and reliable clinical diagnostics, the spatial matching or registration between the two modalities must be accurate, at least where the spatial gradient of the integral attenuation is large. Common sources of image misregistration are sporadic patient movement, and natural respiratory or cardiac organ motion. Although it would be ideal to achieve functional and anatomical image data acquired at the same patient motion phase, it is difficult to accomplish this with typical clinical protocols, where the PET is continuously acquired during free-breathing and the diagnostic CT scan in an arbitrary breath-holding state typically in an inspiration-phase. Therefore, various attempts were conducted to algorithmically correct for this problem within the image reconstruction framework.

In whole-body PET-CT, respiratory-motion functional-anatomical mismatch often causes visible image artifacts in the lower lung regions and the upper abdomen organs below the diaphragm, such as the liver, spleen, and the stomach. In such cases, radiotracer uptake in lesions or other distinctive tissue segments may appear significantly lower or higher than the true uptake values. A common risk is that erroneously low observed uptake might mislead the physician to underestimate true clinical findings.

Several methods were proposed for mitigating the problem. One approach is to use information from gated PET acquisition, either by data-driven or instrumental-based techniques [1-4]. The respiratory motion is estimated, and a derived deformation field is applied on the CT images to generate better adapted attenuation map. A fundamental limitation is that it is not guaranteed that the deformed CT will match any of the PET gated phases since the breath-holding organ positions can be different than any of those in free-breathing. In addition, some approximations are needed to initially reconstruct the gated PET phases without accurate attenuation maps. Another suggestion was to use 4D CT data [5], inevitably with excess radiation dose to the patient. Alternatively, direct 3D image-based PET-CT registration can be utilized to some extent. However, sufficient structural similarity between the relevant functional and anatomical morphologies is not guaranteed, particularly in regions with severe artifacts. Using PET non-attenuation-correction reconstruction to estimate the required attenuation map is also problematic due to the highly inaccurate and inhomogeneous images. Recently, methods utilizing Deep-Learning were also proposed [6-7]. For implementation in a wide clinical use, the chosen solution must be accurate, robust in various imaging conditions, fast, and seamless from the user perspective. To that end we propose and demonstrate a practical approach based on direct modeled artifact evaluation and CT image correction without any gating or dynamic data.

## 2 Materials and Methods

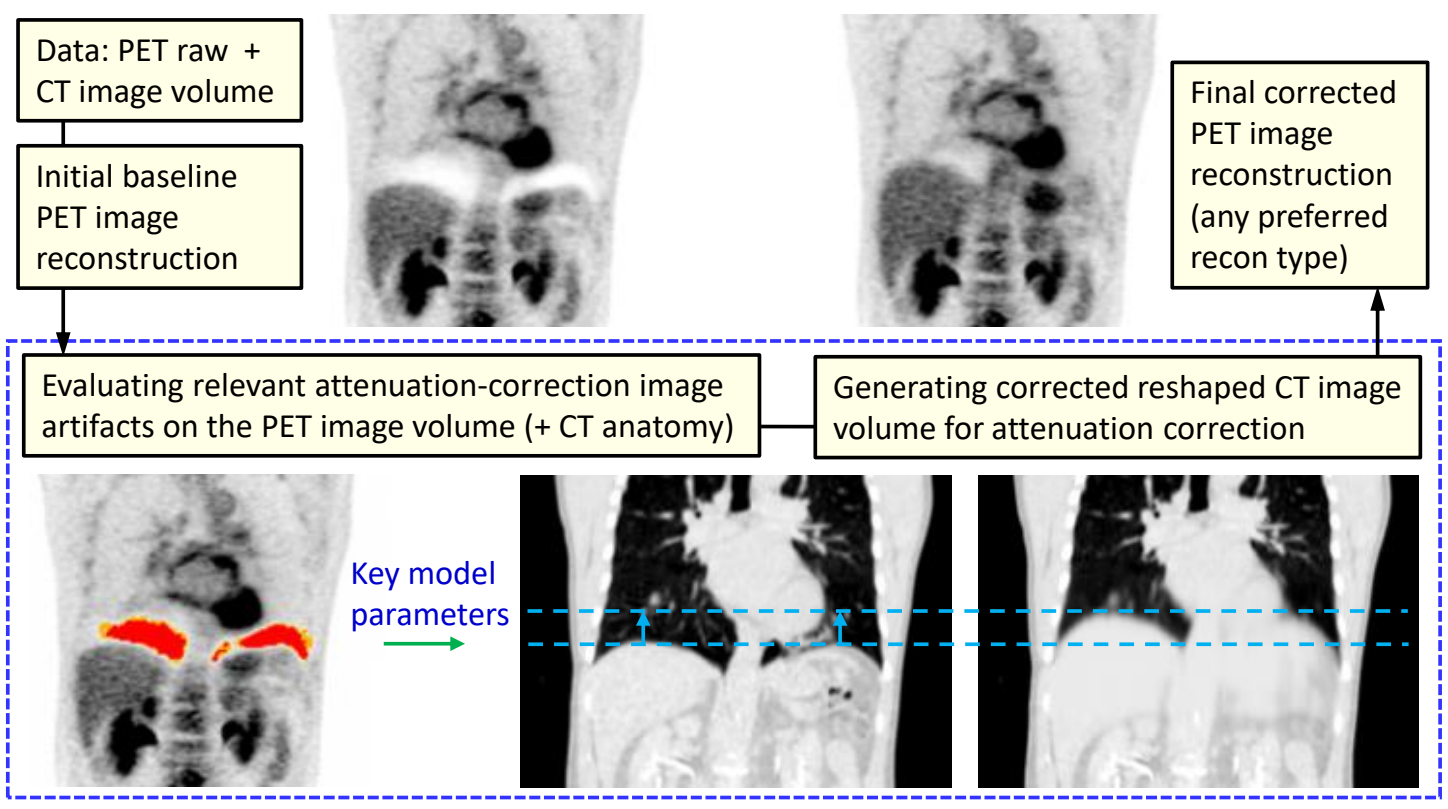

**Fig. 1.** The main concept of the proposed method.

The algorithm first evaluates the PET image artifacts directly from an initial standard PET image volume, and using the original CT. From that, key parameters are derived for controlling a modeled CT image data reshaping based on predictable anatomical respiratory motion pattern. The reshaped CT volume is used to reconstruct corrected PET images. The method is particularly focused on reducing the most dominant respiratory-mismatch artifacts and not on an entire registration of the CT and PET image volumes.

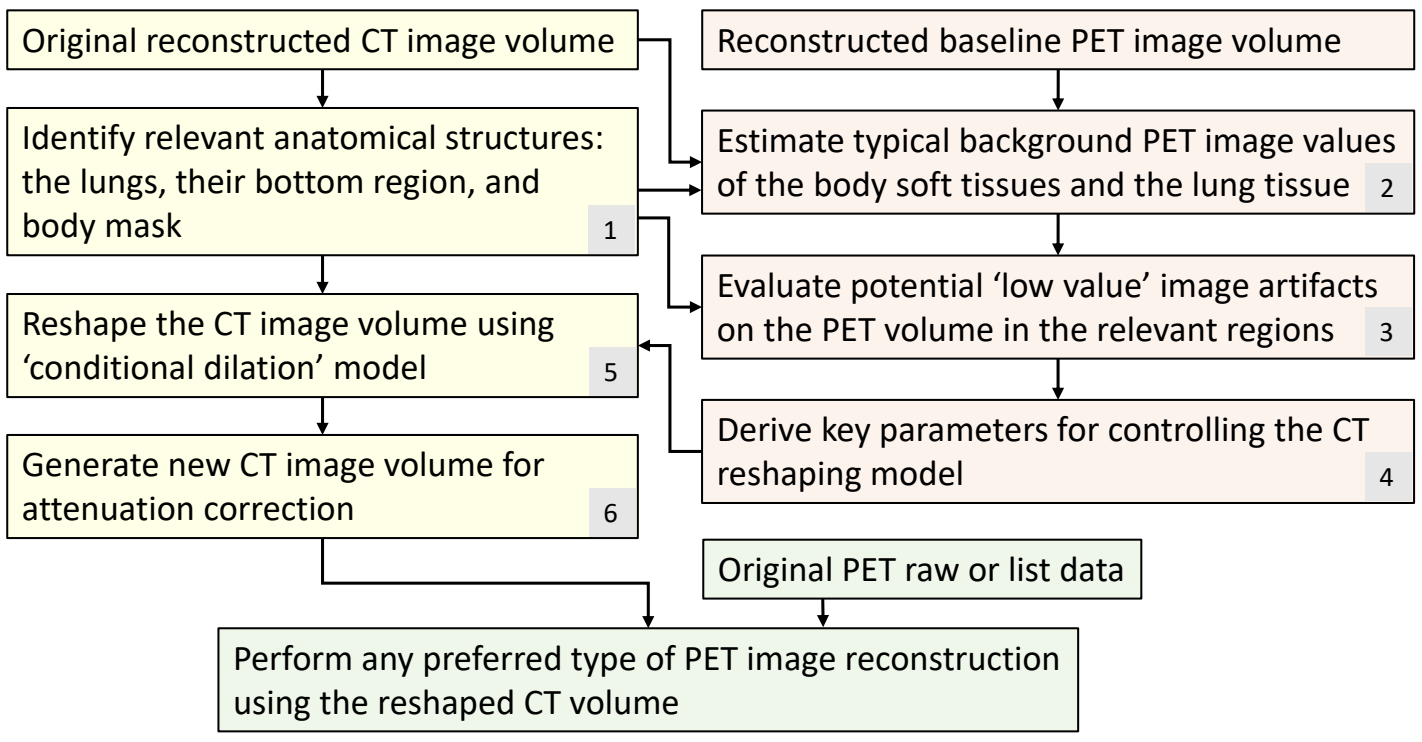

**Fig. 2.** High-level algorithm flowchart. The left blocks are mostly related to the CT data processing, and the right blocks to the PET data. The lower green blocks are final reconstructions for the clinical diagnostics.

Figure 2 shows the main algorithm steps. The input two image volumes are the original breath-holding CT and the single standard (non-gated) PET image volume, reconstructed using the original CT data for attenuation correction (also denoted CTAC). For any PET system type or selected final reconstruction method, a standard baseline PET recon is used for the artifact evaluation step. This, to maintain consistency in diverse conditions.

The first step is identifying the relevant anatomical regions on the CT volume, including the lungs with their bottom region and its contour shape. The patient body mask is extracted as well. This section utilizes conventional image-processing techniques similar to other published methods [8]. The exact region in which the anatomical reshaping model will be applied (fig. 4- A, B) is extracted with emphasis on accurate shape delineation, even in clinical cases of asymmetric or non-homogeneous lungs.

The second step is estimating typical PET 'background' uptake values of the body soft tissues (between fat to soft-bone densities) and of the lung volume. The body tissue background is estimated from a histogram analysis of the PET image values corresponding to the derived body mask, and a limited PET value range around the distribution's median. The lung whole volume background is estimated separately as the median of the non-zero PET image values within the lung mask volume. The two background uptake values are determined in relative units of the PET image scale, independent of any quantitative SUV metric, which in clinical practice may be unreliable or inaccurate.

The evaluation of potential low-value image artifacts on the PET volume in the relevant regions is based on eq. 1.

$$\boldsymbol{M} = \left(\boldsymbol{P} < (B_l \cdot R_l)\right) \,\&\, \left((B_l - \boldsymbol{P}) > (B_b \cdot R_b)\right) \quad (1)$$

where $\boldsymbol{M}$ is the volumetric mask of voxels identified as related to artifacts, $\boldsymbol{P}$ is the volumetric array of PET image voxels which are within the segmented lungs and the determined analyzed region, $B_l$ and $B_b$ are the calculated background uptake values of the lungs and body respectively, and $R_l$ and $R_b$ are optimized ratio parameters. Steps 3 and 4 are illustrated in fig. 3. Using eq. 1, two different artifact severity weights are calculated, subject to two pre-determined $R_l$ values. The weights are summed along the $z$ direction to form a 2D artifact severity map. A high-percentile analysis of the map values determines the required characteristic CT image dilation distance $d$ for the next step. An eccentricity analysis of the artifact map spatial distribution further determines the level of asymmetrical weighting between the right and left lung sides. The logic behind the model is that the whole relevant CT image region under the diaphragm will be dilated upward, as required, in a balanced or nearly homogenous manner depending on a basic shape. This, even if the detected PET image artifacts are more localized to specific areas.

For the CT image conditional dilation process, 2D weights array of local maximal dilation is generated representing a modeled diaphragm motion phase-shift by:

$$\boldsymbol{D} = d \cdot Smooth2D(w_l \cdot \boldsymbol{S}_l + w_r \cdot \boldsymbol{S}_r) \quad (2)$$

where $\boldsymbol{S}_l$ and $\boldsymbol{S}_r$ are the left and right halves of the filled lower lung region contour (fig. 4- B, C). The area's weights are also divided into outer and inner mask shapes based on an observation that the diaphragm motion upward is slighter below the heart and the mediastinum. $d$, $w_l$, and $w_r$ are determined from the artifact analysis step. Note that $\boldsymbol{D}$ is not directly the CT reshaped volume morphology by itself.

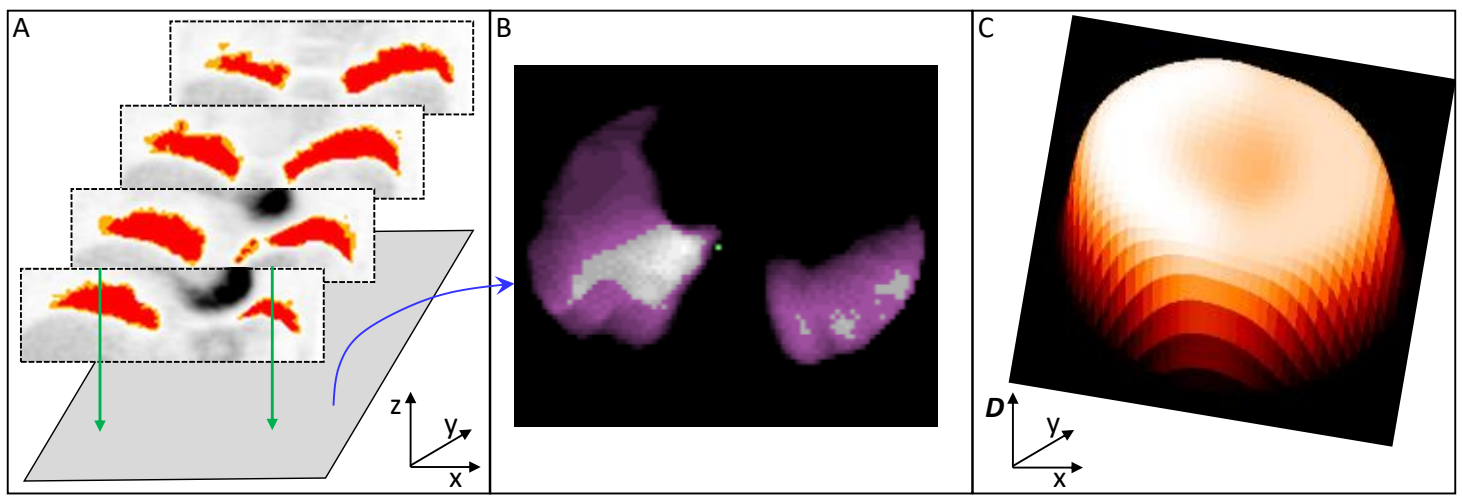

**Fig. 3.** The artifact analysis key steps. A) Identified artifact voxels are assigned with high or low weights (red or yellow resp.) and are summed along the $z$ axis. B) The 2D artifact severity map where grayscale pixels indicate high percentile values. The green dot is the body center. C) The derived modeled shape of maximal dilation weights (see also fig. 4-C).

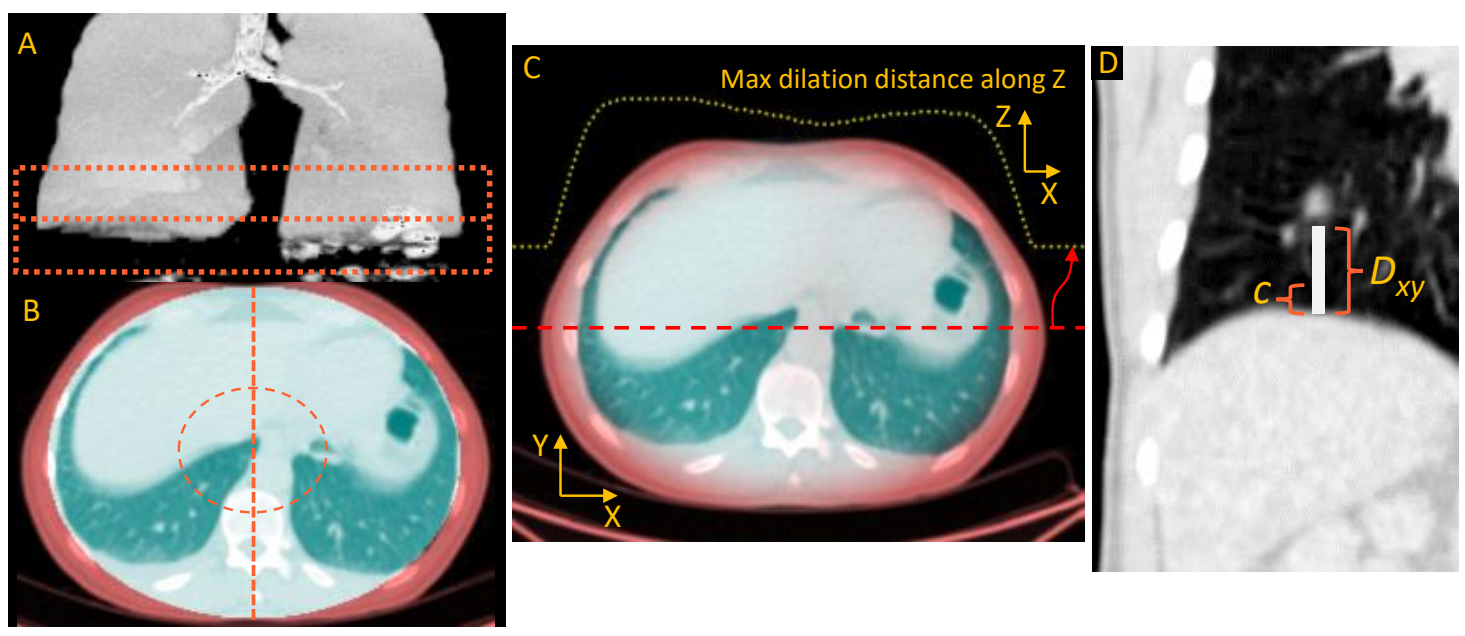

**Fig. 4.** Key steps related to the anatomy from the CT images. A) The automatically identified lower lung region shown on inverse-grayscale MinIP view. B) The identified dilation-intended area (green) shown on the central axial CT slice. The inner contour and the middle line (dotted) are used to optimize the dilation weight model. C) The dilation weights from fig. 3-C shown on top of the CT. D) The two distance values $D_{xy}$ and $c$, used in eq. 3 below, for the conditional dilation process.

The CT image volume is reshaped using 'conditional dilation' model (step 5), schematically shown in eq. 3.

(3)

For each *xy* in a region mask run on *z* along an axial range:

$$\boldsymbol{V}'_{xy}(z) = \max\left(\boldsymbol{V}_{xy}\left[(z-\boldsymbol{D}_{xy}):(z+1)\right]\right),$$

$$if\colon \max\left(\boldsymbol{V}_{xy}\left[(z-\boldsymbol{D}_{xy}):(z-\boldsymbol{D}_{xy}+c+1)\right]\right) < t$$

*stop for this xy*

where $\boldsymbol{V}'$ and $\boldsymbol{V}$ are the corrected and original CT volumes respectively, $\boldsymbol{D}$ is the array of maximal dilation distance for each *xy* in the region mask. The *z* direction is positive toward the patient head. $c$ is a constant distance parameter (except adjustments for margins), and $t$ is a low-HU threshold. The last two parameters enable to identify local transition into lung regions (e.g., with $c = 8$mm and $t = -200$HU). As illustrated in fig. 4-D, the soft tissue HU values fill the lower lung space to a distance determined by the dilation weights until the stopping condition is met.

For the dilation process, the original CT sub-volume initially passes a high-HU thresholding to avoid influence of any local high-density values (e.g., bones, iodine), and these values are returned back after the reshaping. As a final step, a 3D fine smoothing is performed only on the relevant region of the reshaped CT image volume, to avoid any potential sharp edge effects in the reconstruction. An example of the final reshaped CT is shown in fig. 1.

The algorithm's parameters were optimized using 50 PET-CT clinical cases with diverse characteristics and from different systems. Each of the main steps was optimized separately according to its sequential order. The accuracy metric was based on comparative assessment of intermediate steps using dedicatedly built analysis tools. This, since it is impractical to obtain ground truth data without full 4D PET-CT scans. The advantage in our approach is that the whole algorithm can be optimized step by step against intermediate results, and the fully corrected image reconstruction is used only for final verification.

## 3 Results

The correction algorithm was tested on more than 70 PET-CT clinical studies with notable respiratory-mismatch artifacts representing practical diversity of imaging conditions. All cases were acquired and reconstructed using a standard protocol of non-gated free-breathing PET, and a diagnostic CT scan in an arbitrarily breath-holding state typically in inspiration-phase. In fig. 5 (and fig. 1), all cases were reconstructed using standard OSEM method. This, to enable on-par comparison with the image type used for the algorithm's input.

For testing the algorithm robustness, all the cases were processed with the same pre-optimized algorithm setting. The 12 example cases in fig. 5 include different tracers: cases A-I are with the radiotracer $^{18}$F-FDG, J with $^{68}$Ga-DOTATATE, and K, L with $^{18}$F-PSMA. Cases B, C, H and K show lesions in both the liver and lungs, where it can be seen that the liver structures are reasonably corrected while the lung lesions above the artifact areas are unchanged. Cases F and G are with deliberately low administered radiotracer dose leading to high image noise. Case I demonstrates an uncommon scenario where the original attenuation correction led to over correction at the bottom of the lungs (black rims), for which the corrected image still enables to reduce the artifact intensity. Cases J and L are with substantially large mismatch.

In all demonstrated cases the artifact elimination or reduction are very satisfactory. This was also carefully verified on a clinical workstation while reviewing the whole patient volume, and with variable image window.

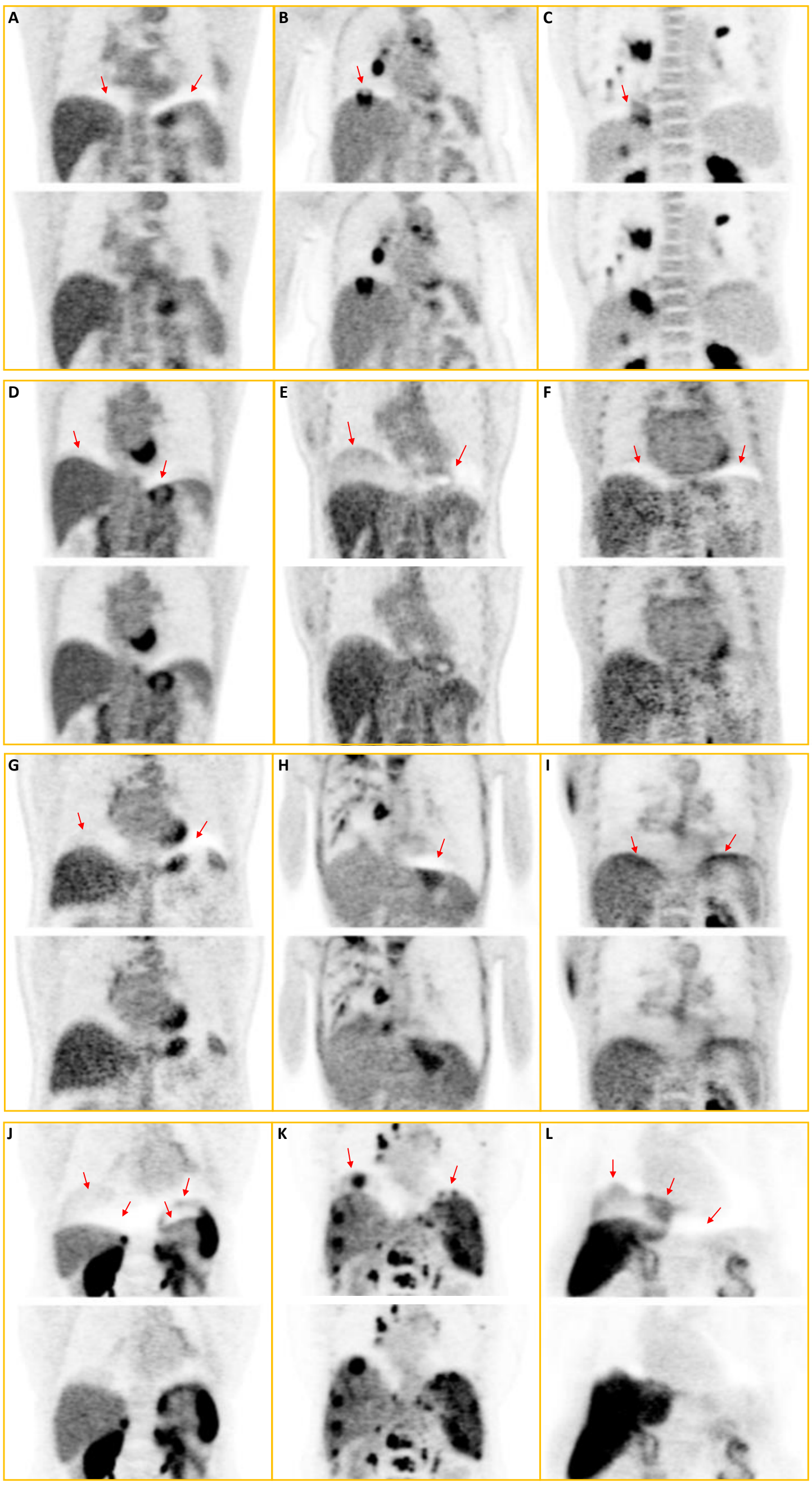

**Fig. 5.** Clinical examples of respiratory-mismatch artifacts. 12 cases (frames A-L) are shown with original vs. corrected reconstructed PET image data, a selected coronal slice of each. On the original cases (upper

image in each frame), red arrows indicate the most visible artifact regions. The lower image in each frame shows the corresponding corrected results. For each case, the grayscale image window was set to best demonstrate the image artifacts (not necessarily as the clinical review window setting). In the example cases, the visualized affected organs are mainly the liver, spleen, stomach, and the spine.

An optional technique to assist clinical reviewing (fig. 6) utilizes the algorithm's analysis results for regional fusion visualization adjustment, in which the CT image volume is automatically shifted rigidly along the axial direction to better match with the PET image volume in the relevant correction regions. The shifting distance is derived from the algorithm's results. Note that the CT and PET images are not morphologically deformed. Therefore, the diagnostic accuracy is fully maintained.

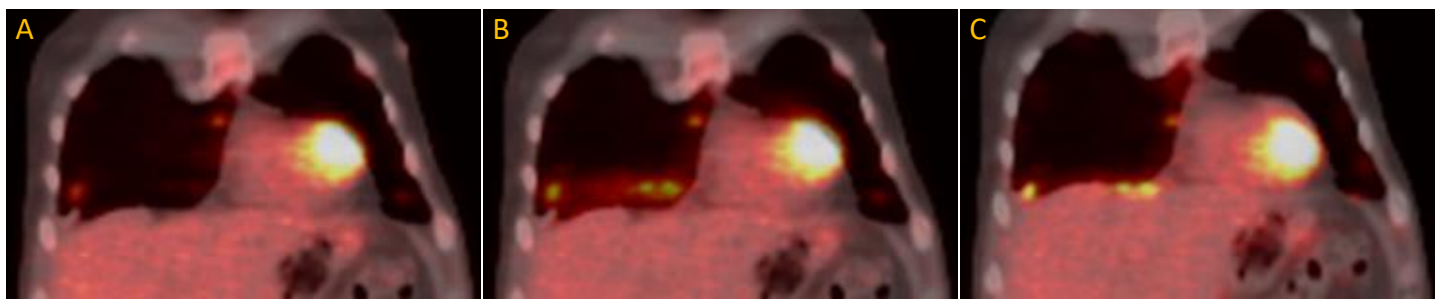

**Fig. 6.** Regional fusion visualization adjustment. A) The original PET image volume having mismatch-related low-value artifacts, fused with the original diagnostic CT. B) The PET recon is corrected using the modified attenuation map, according to the described method. C) The corrected PET recon, fused with the rigidly shifted CT.

In order to test the respiratory mismatch correction in a systematic way, without objective ground-truth references, we developed a verification method based on artificially modifying the CTAC volumetric data for imitating the original cause of respiratory mismatch artifacts (fig. 7). The procedure is as follows: a) A proper PET-CT case without noticeable respiratory-mismatch artifacts is selected; b) An artificial PET-CT spatial mismatch is generated by modifying the CTAC image data using a slab-shifting technique. In this approach, a CT image volume slab is automatically detected around the lower lung regions. The CT image data in the slab is copied, shifted in a constant length downward (e.g., 10 CT pixel-size units), and pasted into the original volume. Although the generated new CT volume is not exactly as it would be with true elastic respiration motion, it enables to mimic the main cause of the relevant artifacts in a simple way; c) The original PET raw data is reconstructed using the slab-shifted CTAC attenuation map, resulting in PET image volume with noticeable mismatch artifacts; d) The full respiratory-mismatch artifact correction is applied using the artifact-induced PET image volume with the slab-shifted CTAC as the inputs. As a result, a corrected CTAC image volume is generated; e) The original PET raw data is reconstructed using the corrected CTAC, and the resultant corrected PET image volume is compared to the original proper PET image volume. Note that this verification approach typically shows a worst-case scenario since additional attenuation artifacts may occur (e.g., near the heart and the ribs) compared to a situation involving true patient respiratory mismatch. In addition to the visualized assessment, we determined a quantitative analysis as follows: a) For each clinical case, the PET image value distribution is measured on a homogenous liver region, and the minimal value change to be considered with clinical significance is defined as $S = FWTM$ of the distribution; b) Calculate voxel-wise on the whole volume the differences $d_1 = abs(C - D)$ and $d_2 = abs(C - E)$ ($C$, $D$, $E$ are the image volumes e.g., as in fig. 7); c) Calculate the array's 'sum of square errors' $e_1 = SSE(d_1)\ for\ d_1 > S$, and $e_2 = SSE(d_2)\ for\ d_2 > S$; d) Determine a relative correction improvement metric $I = 1 - e_2/e_1$. The results on a selected group of 20 clinical cases provided $mean(I)$ = 88 % (CI: 82, 93 %).
The artifact correction accuracy can be also tested using the known technique of synthetic lesion insertion into the PET data and corresponding image reconstruction processes.

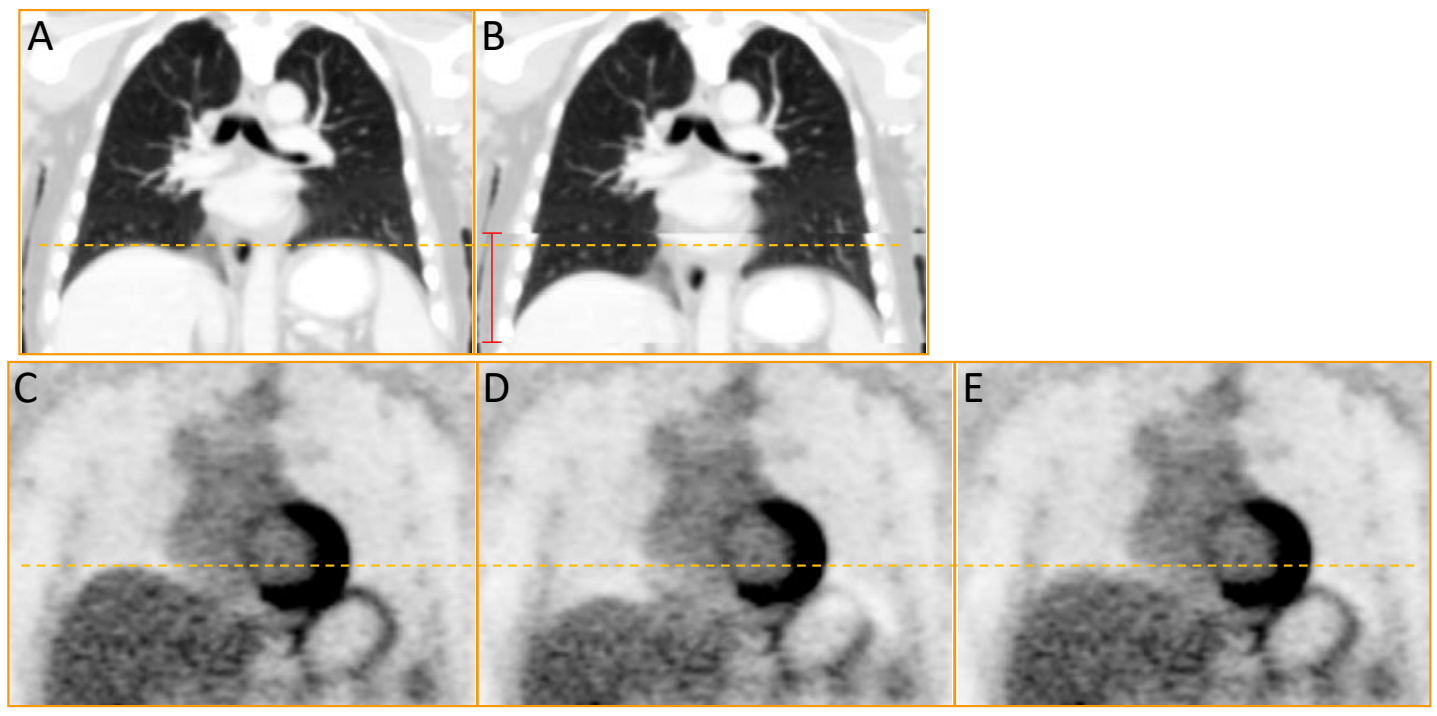

**Fig. 7.** Correction verification method by inducing mismatch artifacts on originally proper PET-CT cases. A) Original CTAC. B) Slab-shifted CTAC (embedded in the red-mark range). C) Original PET recon. D) PET recon using the slab-shifted CTAC. E) Corrected PET recon, where the image data of B and D were used as the algorithm's inputs.

## 4 Discussion and Conclusions

The demonstrated image artifact correction approach has substantial advantages relative to some previously proposed techniques: a) There is no need for gated PET acquisition or reconstruction following motion analysis, which may have their own inaccuracies; b) There is no assumption that the CT breath-hold lung position is identical to the inhale position of the free-breathing PET; c) Only standard PET and CT acquisitions and recons are needed; d) The algorithm optimization is efficiently accomplished by sequentially tuning the parameters of its main steps using special tools, sparing the need for end-to-end process with iteratively reconstructing the PET images multiple times; e) The algorithm is designed to be robust and adapting to various conditions like system type, reconstruction and image parameters, scan time, scanned body range, range of possible respiratory mismatch and artifact severity, tracer type and dose, tracer distribution, clinical conditions, and scan protocol; f) It well fits situations where severe artifacts exist with low structure correlation between the PET and CT images; g) The algorithm is programmatically efficient and integrable within common recon workflows. It is also possible to first process the CT image data, and then applying the PET evaluation and correction steps only on the relevant axial regions after they are acquired; h) The method is inherently suitable for any PET detector and system technologies, and for any final reconstruction algorithm type. This even if the image artifact characteristics can be varied substantially [9-11].

The algorithm's correction strategy utilizes the knowledge that the respiratory mismatch in the lower lung regions corresponds to different phases of the natural diaphragm motion that causes the organs beneath it to change shape and position, mostly up or down against the adjacent lung volume. These morphological changes are typically within predicted limits and constraints. Tissue deformation in additional areas and directions exist as well, but generally they are less pronounced in causing PET-CT reconstruction artifacts due to the smaller integral-attenuation gradients.

Since the algorithm is intended to mitigate the most significant artifacts, and not necessarily to correct image registration in the whole body, the derived few key parameters from the artifact evaluation enable the appropriate CT image reshaping. The generated artificial anatomical volume is much closer to the underlying anatomy corresponding to the PET image volume, than the original CT volume is. This, even if the artificially reshaped CT volume has an approximate shape by itself. As a result, significantly improved attenuation correction and overall better image quality are obtained in all practical clinical situations.

# Dual-energy CT with two-orthogonal-limited-arc scans

Buxin Chen[1], Zheng Zhang[1], Dan Xia[1], Emil Sidky[1], and Xiaochuan Pan[1,2]

[1]Department of Radiology, The University of Chicago, Chicago, USA
[2]Department of Radiation and Cellular Oncology, The University of Chicago, Chicago, USA

**Abstract** In this work, we investigate dual-energy CT with non-overlapping two-orthogonal-limited-arc (TOLA) scans. The TOLA scan configuration consists of two limited-angular-range (LAR) arcs of low- and high-kVp scans whose center lines are orthogonal to each other. Real dual-energy data are collected from a clinical CT scanner in axial mode, and reconstructed from by use of a one-step method for accurate reconstruction with minimal LAR and beam-hardening (BH) artifacts. The method encompasses a constrained optimization problem with directional-total-variation (DTV) constraints on the monochromatic images and the DTV algorithm, as a new instance derived from the non-convex primal-dual (ncPD) algorithm for numerically solving the optimization problem. Qualitative and quantitative evaluations are carried out including the assessment of artifacts reduction, profile plots, and mean pixel values in the basis images. Results suggest that the proposed one-step method can reconstruct from LAR data collected with two orthogonal arcs of as low as 87° each and obtain monochromatic images that are visually and quantitatively close to the reference monochromatic image from full-angular-range data of 360°.

## 1 Introduction

Dual-energy CT (DECT) has been used in clinical and industrial imaging applications, for its improved material differentiation and effective beam-hardening (BH) correction. DECT with limited-angular-range (LAR) data can be potentially used to reduce radiation dose and scanning time and avoid collision between the scanner's moving gantry and the imaged subject. There has been an increasing interest in DECT with LAR data [1–4]. Previous investigations have shown that separate constraints along the image grid's individual axes, such as the directional total variation (DTV) [5], can be effective in reducing, and sometimes eliminating, the LAR artifacts in the reconstruction, while one-step and data-domain-decomposition-based two-step methods have the potential to obtain quantitatively accurate images in DECT from LAR data. Among the two, one-step methods do not require overlapping rays from low- and high-kVp scans, thus enabling flexible scan configuration designs with non-overlapping arcs. This is especially desirable for DECT with LAR scans for increasing the total angular coverage.

In this work, we investigate image reconstruction for DECT with LAR data collected from two-orthogonal-limited-arc (TOLA) scans, which have non-overlapping rays from low- and high-kVp scans. A one-step method is proposed, including an optimization problem considering the non-linear data model and DTV constraints on monochromatic images and a non-convex primal-dual-based algorithm for numerically solving the optimization problem. The proposed method is applied to real data collected from a clinical CT scanner in axial mode, with two full rotations for low- and high-kVp scans. They are referred to as the full-angular-range (FAR) data, and used as references for comparison. LAR data are then extracted from the FAR data to simulate the TOLA scans. Basis images of water and iodine contrast agent are reconstructed and then combined into monochromatic images at 75 keV for visual inspection and also quantitative analysis. Images are also reconstructed by use of the standard method in DECT, i.e., reconstructing kVp images by use of the FBP algorithm followed by an image-domain-based decomposition.

## 2 Materials and Methods

### 2.1 Data collection

Dual-energy data are collected with 80 and 135 kVp using a clinical CT scanner in the axial mode. We extract data from the central row in the 16-row detector, which forms a 2D fan-beam geometry. The source-to-rotation-center distance, source-to-detector distance, and fan angle of a curved detector are 60 cm, 107 cm, and 49°, respectively. The detector has 896 elements. A full rotation of data are collected for each of the low- and high-kVp scans, containing 1200 angular views evenly distributed over 360°.

From the full rotation data, LAR data of TOLA scan configuration, as shown in Fig. 1, are extracted. The TOLA scan consists of two scanning arcs, corresponding to the low- and high-kVp scans, subject to angular ranges $\alpha_1$ and $\alpha_2$, respectively. The center lines of the two arcs, i.e., the line connecting the origin of the $x$-$y$ coordinate system and the center of the scanning arc, are always orthogonal to each other. While the angles of the two center lines w.r.t. the $x$ or $y$ axis can change, i.e., the two lines can rotate in the $x$-$y$ plane around the origin, in the TOLA scan, in this work we focus on the configuration with the two center lines overlapping with the two axes of the $x$-$y$ coordinate system, i.e., $x$ & $y$ axis, as such a design would ensure that the scanning arc, of either the low- or high-kVp scans, is symmetric w.r.t. the $y$- or $x$ axis, along which the DTV constraints are applied. We also select $\alpha_1 = \alpha_2 \equiv \alpha$ and study different LARs with $\alpha = 87°$ and 138°. The TOLA scan configuration can be readily implemented on existing DECT scanners with dual-source or slow-kVp-switching techniques.

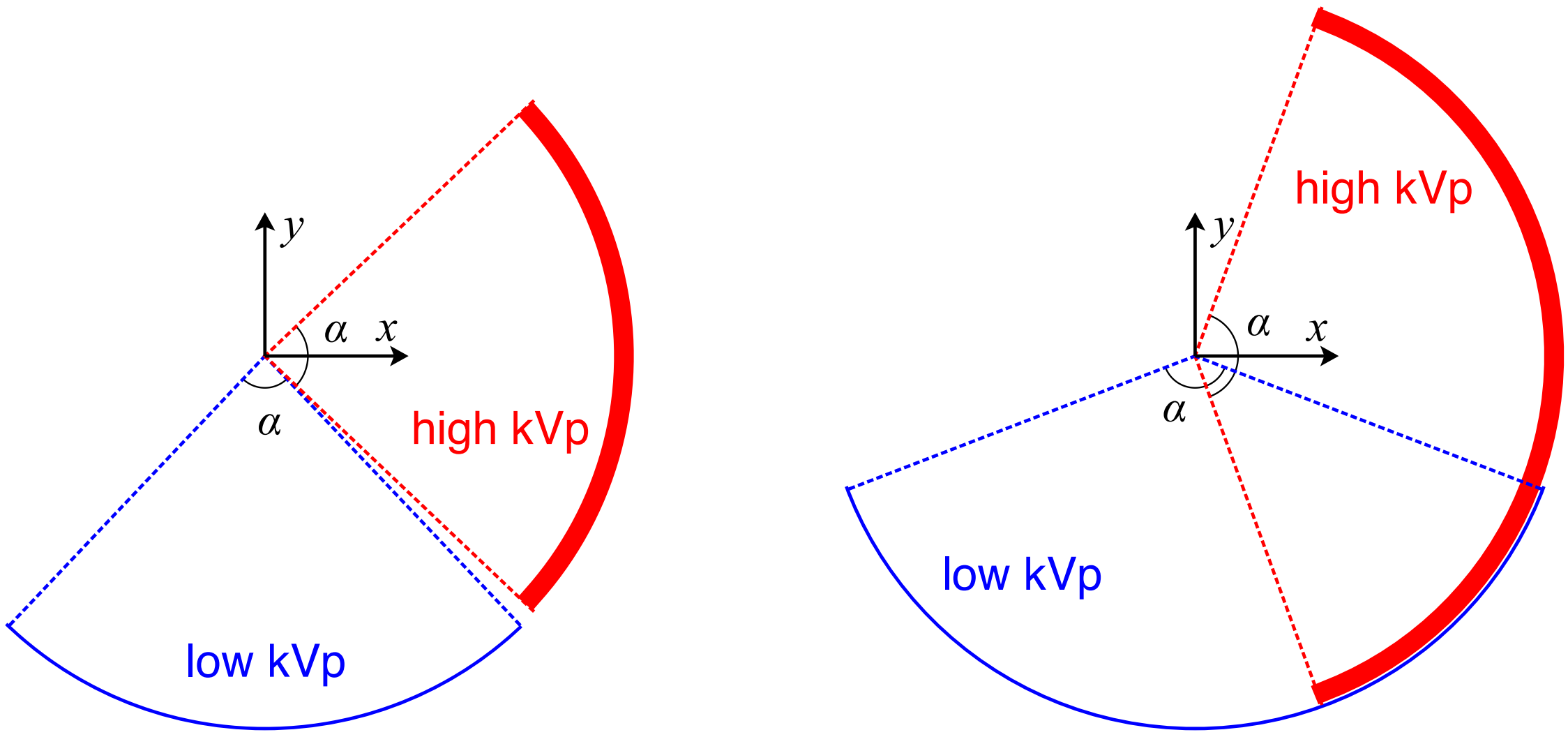

(a) $\alpha = 87°$ (b) $\alpha = 138°$

**Figure 1:** Two-orthogonal-limited-arc (TOLA) scan configuration with (a) non-overlapping ($\alpha = 87°$) and (b) partially overlapping ($\alpha = 138°$) scanning arcs, where the center lines, i.e., the line connecting the origin with the center of the arc, for the low- and high-kVp scanning arcs are orthogonal and overlap with the $x$ and $y$ axis in the 2D coordinate system.

## 2.2 Image reconstruction

Utilizing a material basis decomposition model with the non-linear data model, the following optimization problem is formulated for reconstructing basis images

$$\begin{aligned} \mathbf{b}^* = \arg\min_{\mathbf{b}} \frac{1}{2}||\mathbf{g}(\mathbf{b}) - \mathbf{g}^{[\mathscr{M}]}||_2^2 \\ \text{s.t. } ||\mathscr{D}_x \mathbf{f}_m(\mathbf{b})||_1 \leq t_{mx},\ ||\mathscr{D}_y \mathbf{f}_m(\mathbf{b})||_1 \leq t_{my}, \\ \mathbf{f}_m(\mathbf{b}) \geq 0, \end{aligned} \tag{1}$$

where $\mathbf{b}$ is the congregate basis image vector consisting of two basis images concatenated, $\mathbf{g}(\mathbf{b})$ the non-linear model data for DECT [6], $\mathbf{g}^{[\mathscr{M}]}$ the measured sinogram data,

$$\mathbf{f}_m(\mathbf{b}) = \mu_{m1}\mathbf{b}_1 + \mu_{m2}\mathbf{b}_2 \tag{2}$$

the monochromatic image at energy level $m$, as a linear combination of the two basis images together with their corresponding linear attenuation coefficients, $\mu_{m1}$ and $\mu_{m2}$, at energy level $m$, and $\mathscr{D}_x$ and $\mathscr{D}_y$ partial derivatives, approximated by the two-point difference operators, along the $x$ and $y$ axes. The resulting $\ell_1$ norm of the image partial derivatives are referred to as the DTV of the image, which are upper-bounded by the constraint parameters $t_{mx}$ and $t_{my}$. The objective function and constraints in the optimization problem can take many forms. In this work, the former is selected as the data-$\ell_2$-divergence between the non-linear model data and measured data. The constraints are designed to be DTV and non-negativity constraints on the monochromatic image(s), as they stand for the distribution of linear attenuation coefficients and are often used for visual inspection. We use two different energy levels, 50 and 100 keV, for constraining the basis images and consequently employ a total of 6 constraints, 2 DTV ones and 1 non-negativity one for each energy level.

The optimization problem in Eq. (1) is non-convex, due to the non-linear data model. The non-convex primal-dual (ncPD) algorithm has been previously developed for numerically solving the non-convex optimization problem in DECT based on the non-linear data model [6]. In this work, we derive a new instance of the ncPD algorithm, referred to as the DTV algorithm, incorporating the DTV and non-negativity constraints on the monochromatic images, to numerically solve Eq. (1). The DTV algorithm is directly applied to low- and high-kVp data of FAR and LAR for reconstructing basis images. The standard FBP algorithm is also used to first reconstruct low- and high-kVp images from the data. The pair of kVp images is then decomposed into basis images using an image-domain-based, $2 \times 2$ decomposition matrix. The images reconstructed by use of the FBP algorithm are only for demonstrating the typical LAR artifacts, if not accounted for, under the data conditions studied in the work.

Just like any other algorithms, parameter selection is important in the DTV algorithm. The images are reconstructed onto an image grid of $432 \times 656$ square pixels of 0.8 mm. The low- and high-kVp spectra at 80 and 135 kVp are simulated to best match those from the scanner. The basis materials of water and 20 mg/ml iodine are used, with their linear attenuation coefficients looked up from the NIST database. Finally, the DTV constraint parameters $t_{mx}$ and $t_{my}$ are selected as those yielding the monochromatic images with minimum artifacts. Similarly, a Hanning kernel and a cutoff frequency of 0.5 are selected for the FBP algorithm.

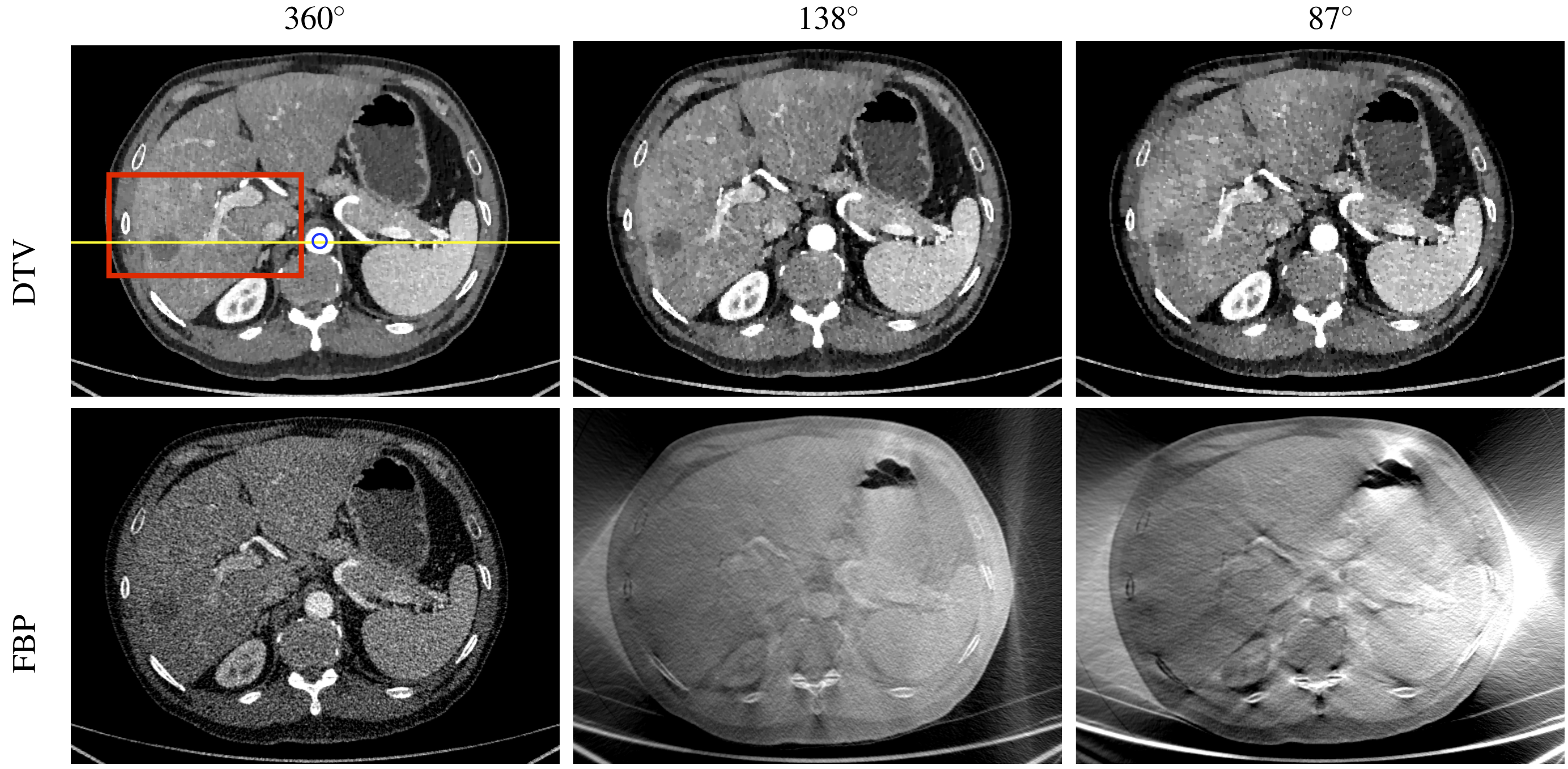

**Figure 2:** Monochromatic images at 75 keV reconstructed by use of the DTV (top row) and FBP (bottom row) algorithms from real data collected with the FAR scan (column 1) and the TOLA scan of different LARs $\alpha = 138°$ (column 2) and $87°$ (column 3). Displaying windows: [-160, 240] HU for top row and column 1 of bottom row, [-1000, 1000] HU for columns 2-3 of bottom row.

## 2.3 Evaluation

With basis images reconstructed, monochromatic images at 75 keV are combined using Eq. (2) and inspected visually. Image profiles are also plotted for quantitative analysis. Further, basis image values within regions of interest, such as aorta, are assessed since they are related to the quantification of iodine concentrations.

# 3 Results

Prior to studies with real data, we have carried out simulation studies with a chest phantom and noiseless DECT data simulated with the TOLA scans. The results have shown that the proposed one-step method can obtain monochromatic images visually identical and quantitatively very close to the truth images of the phantom from data of LAR as low as as $\alpha = 60°$. This sets the performance upper-bound for the method, while the real-data studies, with inconsistencies such as noise, scatter, and decomposition error, are likely to be more challenging. The results from the simulation studies are now shown here to avoid distraction and will be presented at the conference instead.

We first show monochromatic images at 75 keV reconstructed by use of the DTV and FBP algorithms in Fig. 2. It can be observed that the DTV monochromatic images from LAR data are with no or minimal artifacts. The DTV images from LAR data of $\alpha = 138°$ and $87°$ are visually close to the reference DTV image from FAR data. The FBP image from FAR data does not show any LAR artifacts, however it appears darker and with lower contrast than all the DTV images using the same display window, indicating quantitative inaccuracy resulting from uncorrected BH effect. The FBP image from LAR data all show significant LAR artifacts of leakage and distortion, as expected, that deteriorate with decreasing LAR. The observations can be corroborated by the zoomed-in areas in Fig. 3. ROIs of DTV images from LAR data of $\alpha = 138°$ and $87°$ resemble that from FAR data, while ROIs of FBP images show DC shift and obscured anatomic structures due to LAR artifacts.

Next, we show profile plots, along the yellow horizontal line indicated in the top left panel of Fig. 2, of DTV and FBP images from FAR and LAR data in Fig. 4. It can be observed that the profiles of the DTV images from LAR data generally overlap with that from FAR data, indicating effective correction of both LAR and BH, while the profiles of the FBP image from LAR data shows significant deviations from that from FAR data.

Lastly, we compute mean pixel values (MPVs) within the ROI around the aorta, as indicated by the blue circle in the top left panel of Fig. 2, of the 20-mg/ml iodine basis images reconstructed, as they can be related to the quantitative estimation of iodine contrast concentration. The results are shown in Table 1. Using the MPV from the DTV reconstruction from FAR data as the reference, MPVs from the DTV reconstructions from LAR data of $\alpha = 138°$ are close to the reference value from FAR data, while bias exists for LAR data of $\alpha = 87°$. On the other hand, MPVs from FBP reconstructions from LAR data are under-estimated by almost an order of magnitude.

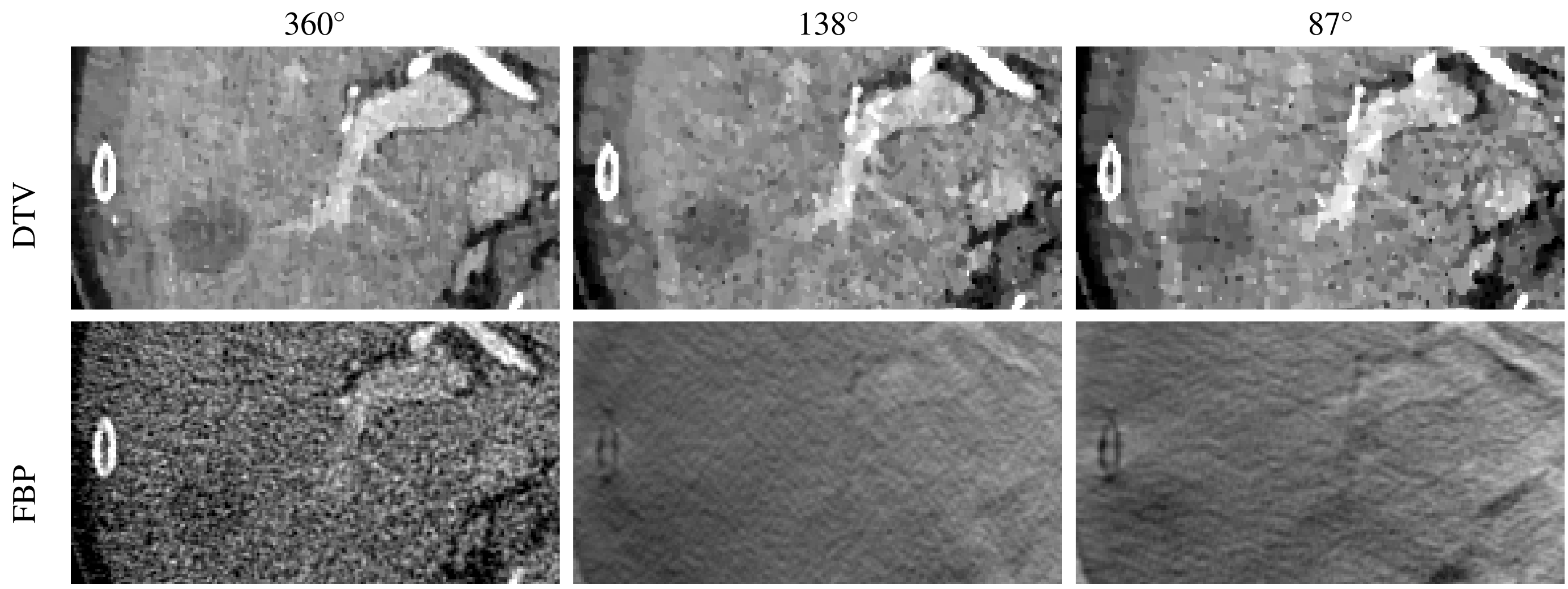

**Figure 3:** Zoomed-in areas, as enclosed by the red rectangular box in the top left panel of Fig. 2, of those monochromatic images at 75 keV in Fig. 2.

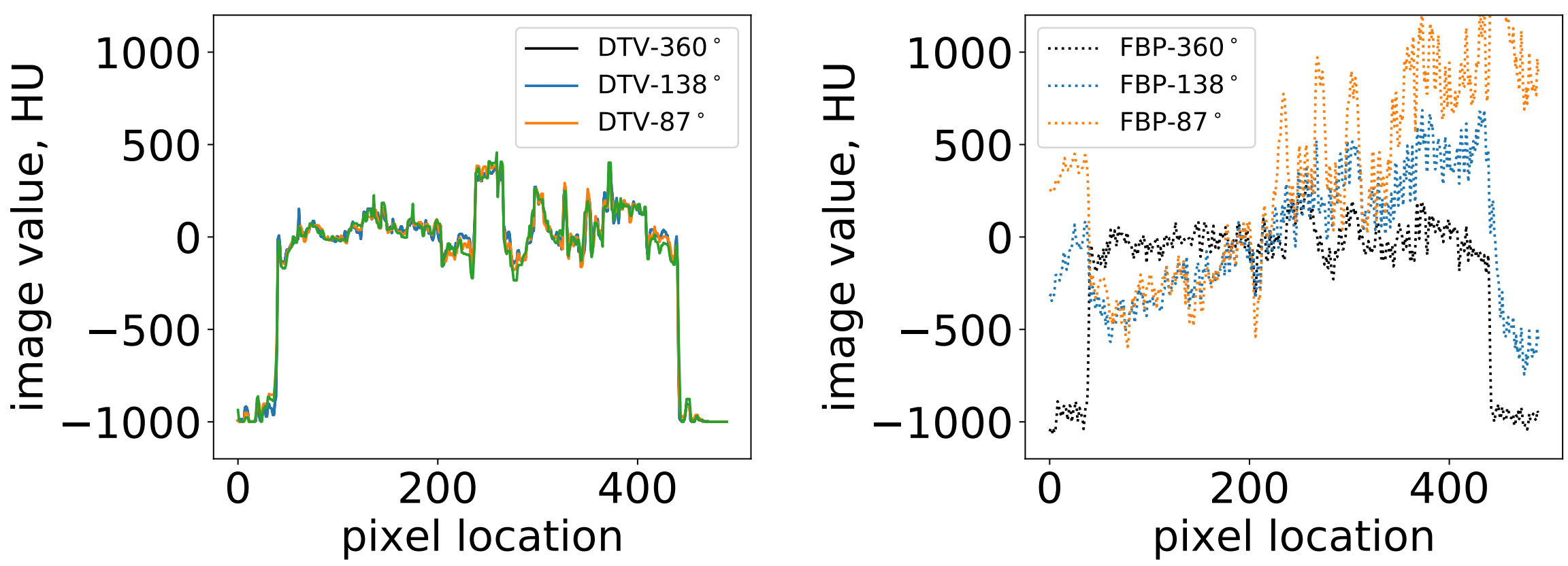

**Figure 4:** Profile plots along the yellow horizontal line indicated in the top left panel of Fig. 2 in the monochromatic images reconstructed by use of the DTV (left) and FBP (right) algorithm from FAR and LAR data.

**Table 1:** Mean pixel values within the ROI around aorta in the 20-mg/ml iodine basis images of the DTV and FBP reconstruction from FAR and LAR data

| | 360° | 138° | 87° |
|---|---|---|---|
| DTV | 0.72 | 0.71 | 0.19 |
| FBP | 0.44 | 0.19 | 0.07 |

## 4 Discussion and conclusion

In this work, we have investigated DECT imaging with TOLA scans. The TOLA scan configuration consists of two scanning arcs of low- and high-kVp scans that are non- or only partially-overlapping. Image reconstruction is thus done by use of a one-step method, including the DTV algorithm that has been derived and tailored to numerically solve the non-convex optimization problem with DTV constraints on the monochromatic images. Real data are collected from a 16-row clinical CT scanner with two full rotations of low- and high-kVp spectra, and qualitative (i.e., visual) and quantitative (i.e., profile plots and MPVs from basis images) evaluations have been carried. The results suggest that the DTV algorithm can reconstruct from data collected with the non-overlapping TOLA scan of as low as $\alpha = 87°$ and obtain monochromatic images that are visually and quantitatively close to the reference image from FAR data.

In this work, the TOLA scan configuration is positioned such that the two center lines of the low- and high-kVp scanning arcs overlap with the $x$ and $y$ axis in the 2D coordinate system. We have also used real data scanned with a single object of human abdomen. The performance of the proposed one-step method for DECT image reconstruction with TOLA scans could be affected by different scanned objects and/or TOLA scan configurations that rotate certain degrees (e.g. 45° so that the centerlines overlap with the two diagonal lines in the coordinate system) w.r.t. the one used in the work. It is of interest to study the impact of these parameters.

The image reconstruction is formulated as a constrained optimization problem, with constraints on monochromatic images. Different designs of constraints are available, such as applying the DTV and non-negativity constraints on the basis images directly, which would have a more direct impact on the basis images and might result in more accurate MPVs in the basis images, to improve the quantitative estimation of iodine concentrations. Such investigations will also be the focus in our future works.

## 5 Acknowledgments

This work is supported in part by NIH Grant Nos. R01EB026282, R01EB023968, and R21CA263660. The contents of this paper are solely the responsibility of the authors and do not necessarily represent the official views of NIH.

# Millisecond CT: a Dual Ring Stationary CT System and its Reconstruction

Changyu Chen[1,2], Yuxiang Xing[1,2], Li Zhang[1,2*], and Zhiqiang Chen[1,2*]

[1]Department of Engineering Physics, Tsinghua University, Beijing, China
[2]Key Laboratory of Particle & Radiation Imaging (Tsinghua University), Ministry of Education, Beijing, China

**Abstract** In this paper, we proposed a new concept of stationary CT system (**M**ulti-**S**egment **D**ual-**R**ing Stationary CT, MSDR-CT). MSDR-CT consists of a source ring and a detector ring that can be implemented conveniently by splicing multiple segments of distributed sources and detectors. With advanced multi-spot X-ray source techniques, X-rays will be fired from different spots sequentially or simultaneously in a programmable manner. Hence, MSDR-CT enables ultra-fast and flexible data acquisition modes which presents great potential in pushing the temporal resolution to the order of millisecond for dynamic imaging. As a proof-of-concept study, we deducted a Hilbert transform analytical reconstruction method in a differentiation-backprojection-filtration format for this system configuration. Furthermore, the discontinuity between segments in the source or detector rings is a protogenetic problem for its practical implementation. A preliminary correction method is proposed to address this problem. Simulated experimental results with the Shepp-Logan phantom confirmed the feasibility of MSDR-CT.

## 1 Introduction

X-ray computed tomography (CT) is one of the most important non-invasive imaging methods in clinical and industrial applications. Most of the current commercial CT scanners take the configuration where a source and a detector module are mounted on a heavy gantry and spin around the object to complete a scan[1]. Due to the mechanical limitation, the temporal resolution of current CT scanners is limited by the rotation speed of gantries, which presents challenges in dynamic imaging for the hearts, lungs, and small animals[2] as well as fast cargo inspection[3].

In recent decades, various new systems have been proposed to accelerate the scanning process. There are mainly two categories: **(1) Multisource rotational CT** decreases the range of rotation in a complete scan by employing multiple X-ray sources and alleviate the limitation of the rotation speed. An initial attempt was conducted by Mayo Clinic known as the dynamic spatial reconstructor (DSR)[4]. Dual-source CT by Siemens halves the scanning time and still serves as a cutting-edge system for cardiac imaging at present. Systems with more X-ray sources[5, 6] were also explored for further improving temporal resolution. **(2) Stationary CT** realizes a full scan by dozens or more sources distributed to cover a big range of angular directions. Hence, mechanical rotation is evitable and the temporal resolution could be extremely high[7]. Electron-beam CT was investigated for cardiac imaging of humans[8] and small animals[9]. But the high cost and limited signal-to-noise ratio of reconstructed images prevented its commercial application[10]. Field emission electron sources are an alternative to thermionic sources for their instantaneous response and electronical controllability where carbon nanotubes (CNT) make ideal field emitters[11]. Incorporated by recent advances of multi-spot technologies, CNT-based stationary CT has drawn a lot of attention[12, 13]. To address the data deficiency problem caused by finite length of source and detector arrays, multi-segment scanning with multiple imaging planes were explored[3, 10]. However, translation through all imaging planes is necessary to complete a scan which is a limitation for the temporal resolution.

In this work, we proposed a new concept of stationary CT (**M**ulti-**S**egment **D**ual-**R**ing Stationary CT, MSDR-CT) where the imaging optical structure is constructed by a source ring and a detector ring. A Hilbert transform reconstruction method and truncation correction methods are derived to validate the feasibility of MSDR-CT.

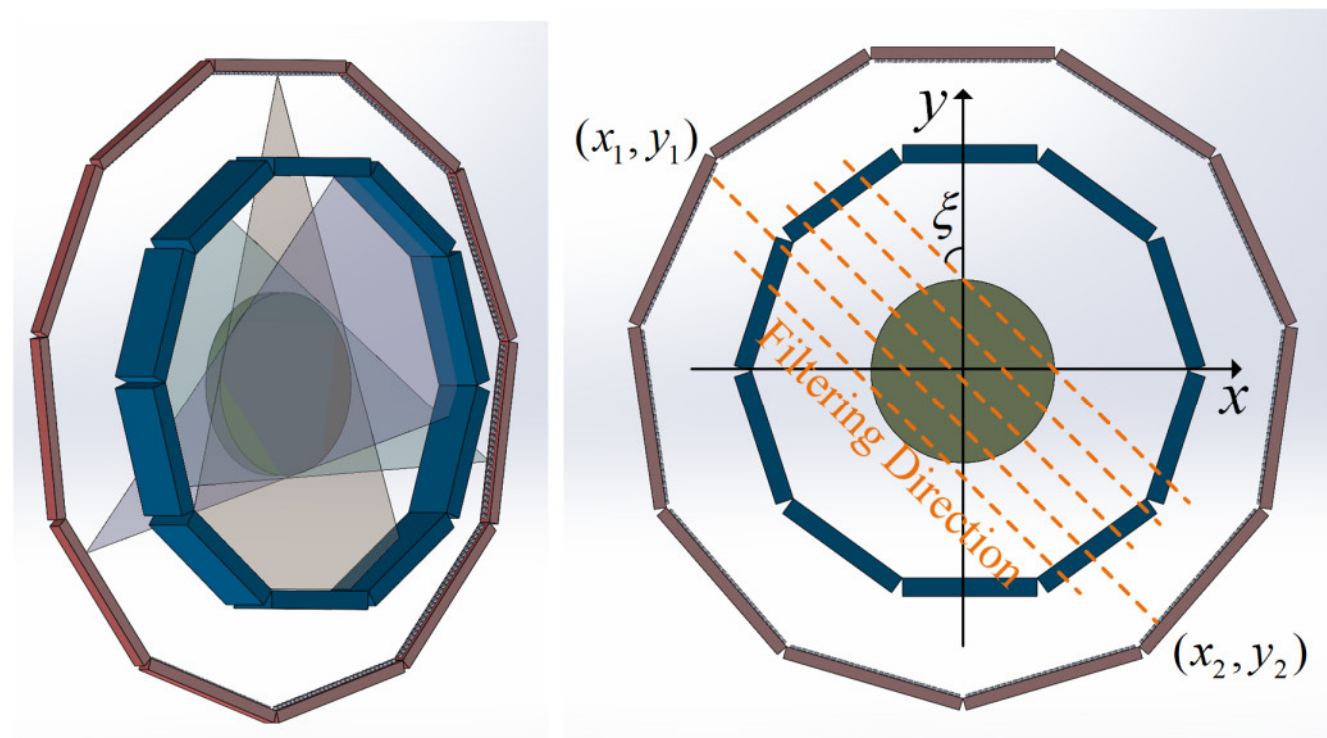

(a) (b)

**Fig. 1** Diagrams of MSDR-CT. (a) Illustration for the system configuration: the red boxes and blue boxes represent the source modules and detector modules, respectively. Each source module contains multiple focal spots (gray circles). (b) Illustration of the system geometry: the orange dotted-lines stand for the direction for Hilbert filtering in the $\xi$ direction.

## 2 Materials and Methods

### 2.1 System Configuration

A general system configuration diagram of MSDR-CT is shown in Fig. 1(a): $N_S$ multi-spot X-ray source modules (the red boxes) and $N_D$ detector modules (the blue boxes) are jointly arranged to form the source and detector rings. The two rings share the same central axis, but with a slight offset in their axial positions to avoid optical obstruction. Enabled by recent advance in X-ray source techniques, each source module contains multiple linearly distributed focal spots which can be fired and switched ultrafast (up to sub-millisecond) under programmable automatic control. As long as signals from different views are separable, multiple

This work was supported in part by the National Natural Science Foundation of China (Grant No. 62031020).
*Corresponding author: Zhiqiang Chen (e-mail: czq@tsinghua.edu.cn); Li Zhang (e-mail: zli@mail.tsinghua.edu.cn).

source modules may emit X-ray simultaneously (three spots are fired in Fig. 1(a)) with individual tube voltage and current setting to accelerate the data acquisition. Hence, the scanning by MSDR-CT can be both rapid and flexible per different scanning strategies to achieve millisecond CT imaging. However, this system configuration adds the complexity of the optical structure since the geometrical relationship between the sources and detectors varies for different focal spots. To support such a design, we present an analytical reconstruction method and examine the reconstruction results in subsequent sections.

## 2.2 Analytical Reconstruction using a Hilbert Transform Method

Because X-rays from each focal spot will be received by detector modules along a polyline, it is problematic to apply an FBP-type algorithm for reconstruction. Inspired by Noo's method for circular trajectories[14], we deducted a Hilbert transform based method for the analytical reconstruction of MSDR-CT. Although the positional relation between the source and detector segments is different from one another (see Fig. 1(b)), the basic components of MSDR-CT could be formulated as a source segment indexed by $i$ and a detector segment indexed by $k$ as depicted in Fig. 2. Assuming the global coordinates attached to the object $f(x,y)$ with the origin being O, the distance from the $i^{\text{th}}$ source segment and $k^{\text{th}}$ detector segment to O are denoted as $D_{\text{SO}}$ and $D_{\text{OD}}$ with the perpendicular foots from O to the segments being $\text{O}_\text{S}$ and $\text{O}_\text{D}$ respectively. $\beta_i$ and $\varphi_k$ represent the rotation angles from the $y$-axis to line $\text{OO}_\text{S}$ and line $\text{OO}_\text{D}$. Each focal spot in the source module is indexed by its offset from $\text{O}_\text{S}$ as $l \in [L_{\text{m-}}, L_{\text{m+}}]$ . A virtual detector (indexed by $u \in [U_{\text{m-}}, U_{\text{m+}}]$) passing O and parallel to the physical detector is introduced for simplicity. Hence, if the ray is emitted by the focal spot $(\beta_i, l)$, the projection collected by the virtual detector cell $(\varphi_k, u)$ is denoted as $p_{\beta_i,\varphi_k}(l,u)$.

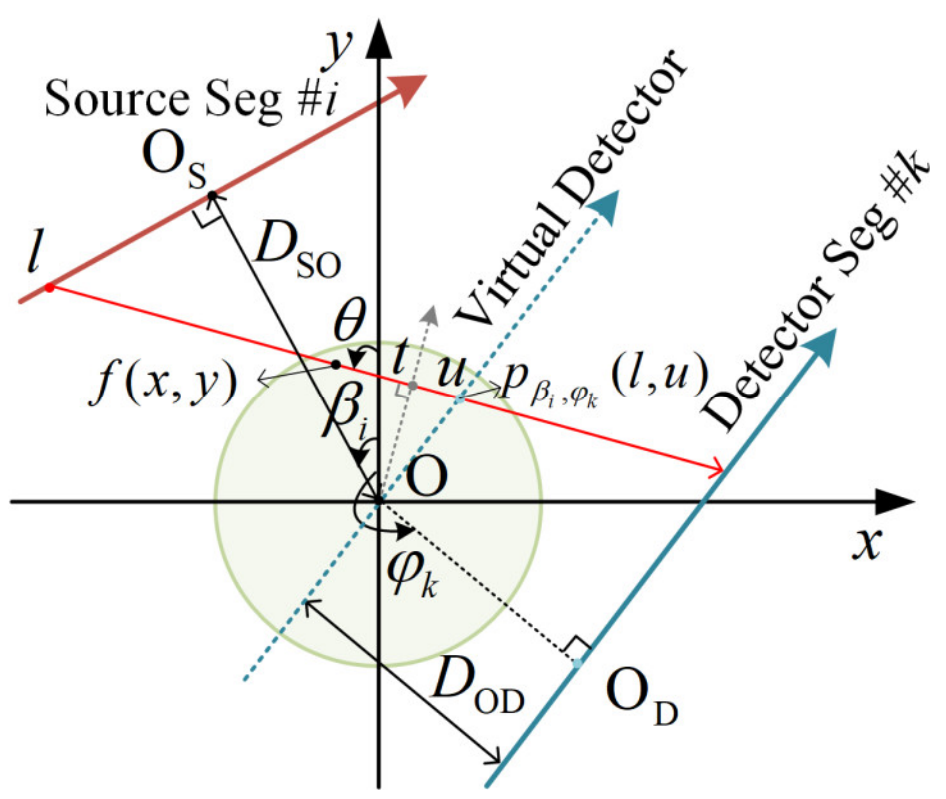

**Fig. 2** Illustration for a general "single source to single detector" geometry of MSDR-CT.

For notation simplicity, the offset between the source and detector rings is ignored here. As in[14], a Hilbert image can be produced by backprojecting the derivative of the projections

$$b_\xi(x,y) = -2\pi H_\xi f(x,y) \tag{1}$$

with $b_\xi(x,y)$ being the DBP image, and $H_\xi f$ the Hilbert transform of $f$ in the $\xi$ direction (see the orange dotted line in Fig. 1(b)). Hence, the reconstruction can be achieved by Hilbert inversion towards a DBP image. More specifically, $b_\xi(x,y)$ can be achieved by the following three steps:

① **Projection weighting.** Cosine weighting is applied according to the distance relationship as

$$\tilde{p}_{\beta_i,\varphi_k}(l,u) = \frac{(D_{\text{SO}} + u\sin\alpha_{i,k}) \cdot p_{\beta_i,\varphi_k}(l,u)}{\left[(l + u\cos\alpha_{i,k})^2 + (D_{\text{SO}} + u\sin\alpha_{i,k})^2\right]^{1/2}} \tag{2}$$

where $\alpha_{i,k} = \varphi_k - \beta_i$ denotes the intersection angle between the $i^{\text{th}}$ source and $k^{\text{th}}$ detector segments.

② **Differentiation.**

$$g_{\beta_i,\varphi_k}(l,u) = \frac{\text{d}}{\text{d}u}\tilde{p}_{\beta_i,\varphi_k}(l,u) \tag{3}$$

③ **Weighted backprojection with two boundary terms.**

$$b_\xi(x,y) = \sum_{m=1}^{2}(-1)^{m-1}\frac{p_{\beta_m,\varphi_{k^*}}(l,u^*)}{\left[(x_m - x)^2 + (y_m - y)^2\right]^{1/2}} + \frac{1}{2}\sum_{i=1}^{N_\text{S}}\int_{L_{\text{m-}}}^{L_{\text{m+}}} \frac{(l\sin\alpha_{i,k} - D_{\text{SO}}\cos\alpha_{i,k})g_{\beta_i,\varphi_k}(l,u)\,\text{sgn}(\sin(\theta_{i,k} - \xi))}{(l\sin\alpha_{i,k} - D_{\text{SO}}\cos\alpha_{i,k} - x\sin\varphi_k + y\cos\varphi_k)^2}\Bigg|_{(k,u)=(k^*,u^*)} \text{d}l \tag{4}$$

where $\theta_{i,k} = \beta_i + \arctan\left(-\frac{l + u\cos\alpha_{i,k}}{D_{\text{SO}} + u\sin\alpha_{i,k}}\right)$ is the projection view defined by the intersection angle between the ray path and the $y$-axis. Generally, one can derive $b_\xi(x,y)$ by backprojecting over $\theta_{i,k} \in [\xi, \xi+\pi)$ with the coordinate of the starting and ending focal spots being $(x_m, y_m)$ (see Fig. 1(b)). As the projection by MSDR-CT is $2\pi$ periodic in $\theta_{i,k}$ , a signum function is introduced to extend the integral range to $\theta_{i,k} \in [\xi, \xi+2\pi)$ . Two boundary terms are used to avoid taking the derivative of the signum function directly. $(k^*, u^*)$ stands for the projection point on the virtual detector coordinate which satisfies $u \in [U_{\text{m-}}, U_{\text{m+}}]$ for

$$u = \frac{(x\cos\beta_i + y\sin\beta_i)D_{\text{SO}} + (x\sin\beta_i - y\cos\beta_i)l}{l\sin\alpha_{i,k} - D_{\text{SO}}\cos\alpha_{i,k} - x\sin\varphi_k + y\cos\varphi_k} \tag{5}$$

Finally, the reconstruction is given by inverting the Hilbert image along $\vec{\tau}^\perp = (-\sin\xi, \cos\xi)$ . Suppose that $f$ is zero for any $t_2 \notin [T_{\text{m-}}, T_{\text{m+}}]$ , the range of inverse Hilbert transform shrinks from $(-\infty, +\infty)$ to $[T_{\text{m-}}^\varepsilon, T_{\text{m+}}^\varepsilon]$

$$f(x,y) = f(t_1\vec{\tau} + t_2\vec{\tau}^\perp) = \frac{-1}{\sqrt{(t_2 - T_{\text{m-}}^\varepsilon)(T_{\text{m+}}^\varepsilon - t_2)}} \times \left(\int_{T_{\text{m-}}^\varepsilon}^{T_{\text{m+}}^\varepsilon}\sqrt{(t_2' - T_{\text{m-}}^\varepsilon)(T_{\text{m+}}^\varepsilon - t_2')}\,\frac{H_\xi f(t_1\vec{\tau} + t_2'\vec{\tau}^\perp)}{\pi(t_2 - t_2')}\text{d}t_2' + C(\xi, t_1)\right) \tag{6}$$

where $\varepsilon$ is a small positive number, $T_{\text{m-}}^\varepsilon = T_{\text{m-}} - \varepsilon$ , and $T_{\text{m+}}^\varepsilon = T_{\text{m+}} + \varepsilon$ . $C(\xi, t_1)$ is a constant to uniquely determine the values of $f$ . Taking the knowledge that $f(t_1\vec{\tau}+\tilde{t}_2\vec{\tau}^{\perp})=0$ over the range $\tilde{t}_2 \in (T^{\varepsilon}_{\text{m-}}, T_{\text{m-}}) \cup (T_{\text{m+}}, T^{\varepsilon}_{\text{m+}})$, $C(\xi, t_1)$ can be obtained by

$$C(\xi,t_1) = -\int_{T^{\varepsilon}_{\text{m-}}}^{T^{\varepsilon}_{\text{m+}}} \sqrt{(t_2'-T^{\varepsilon}_{\text{m-}})(T^{\varepsilon}_{\text{m+}}-t_2')}\,\frac{H_{\xi} f(t_1\vec{\tau}+t_2'\vec{\tau}^{\perp})}{\pi(\tilde{t}_2-t_2')}\mathrm{d}t_2' \quad (7)$$

## 2.3 Truncation Correction for the Discontinuity between Segments

In the previous sections, we assumed seamless connection between the source or detector segments. However, gaps and discontinuity are unavoidable in the mechanical assembly of source and detector rings which poses challenge for the implementation of MSDR-CT.

The gaps in the source ring induce limited-angle problem. Taking the conjugation relationship of projections into consideration, the missing data may be compensated. The radon space of MSDR-CT is illustrated in Fig. 3. The gray region in Fig. 3(a) represents the radon space of each source segment. The "fan angle" of each segment spans over $[-\gamma_2, \gamma_1]$ where $\gamma_1$ and $\gamma_2$ are determined by the half-length of the source array $L_{\text{m}}$ and the radial distance to the origin denoted as $t \in [-R, R]$ (see Fig. 2). For the $i^{\text{th}}$ original segment ( $\beta_i - \beta_1 < \pi$ ), the $n^{\text{th}}$ and $(n-1)^{\text{th}}$ conjugate segments ( $\beta_n - \beta_1 > \pi$ ) provide conjugate projections (shown as the purple region in Fig. 3(b)) with redundant or compensational information. To balance the redundancy and take advantage of the conjugate projection data, a weighting function $W_{\beta_i,\varphi_k}(l,u)$ is proposed.

$$W_{\beta_i,\varphi_k}(l,u) = \begin{cases} \sin^2\left[\dfrac{\pi}{2}\dfrac{\theta_{i,k}-(\beta_i+\gamma_1)}{\beta_n-\pi-\beta_i-2\gamma_1}\right], & \theta_{i,k} \in [\beta_n-\pi-\gamma_1, \beta_i+\gamma_1] \\ 1, & \theta_{i,k} \in (\beta_{n-1}-\pi+\gamma_2, \beta_n-\pi-\gamma_1) \\ \sin^2\left[\dfrac{\pi}{2}\dfrac{\theta_{i,k}-(\beta_i-\gamma_2)}{\beta_{n-1}-\pi-\beta_i+2\gamma_2}\right], & \theta_{i,k} \in [\beta_i-\gamma_2, \beta_{n-1}-\pi+\gamma_2] \end{cases} \quad (8)$$

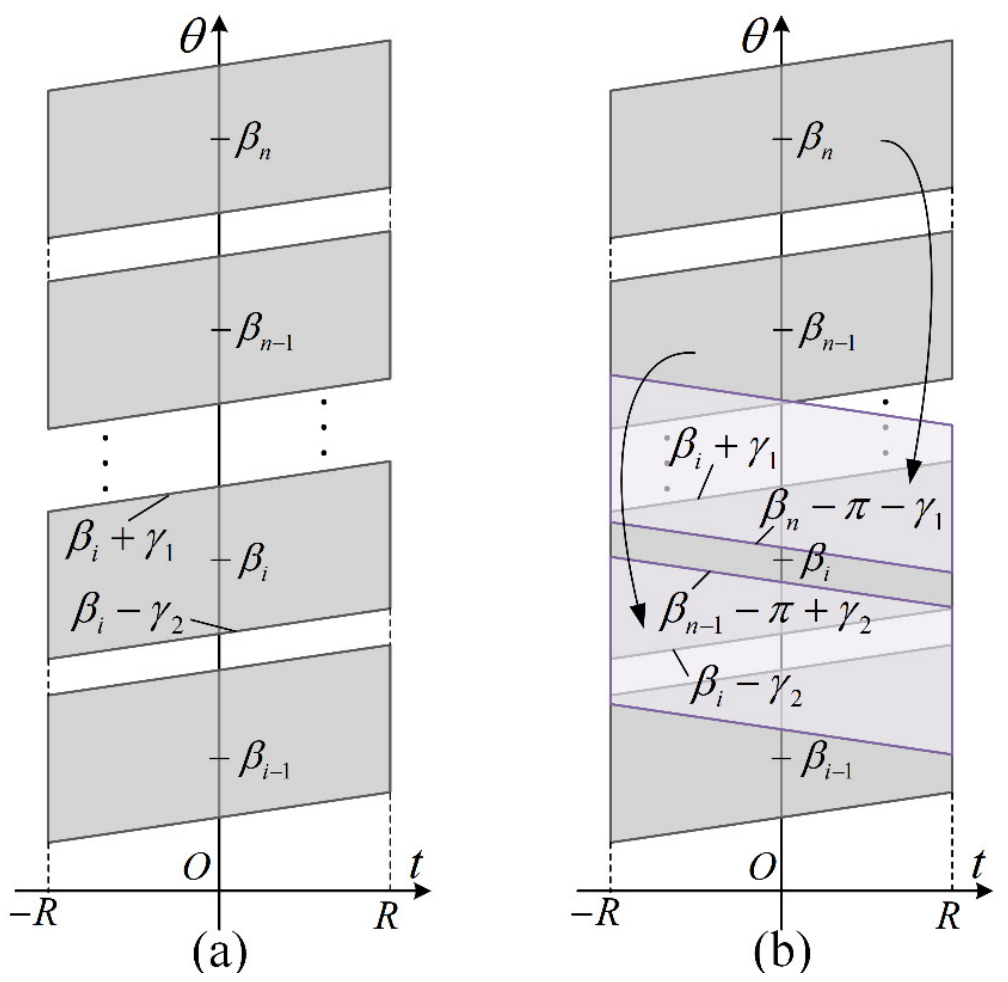

**Fig. 3** (a) Diagram for the radon space of MSDR-CT; (b) Illustration of the conjugation relationship of MSDR-CT.

The gaps between detector segments can be solved by data inter-/extra-polation, iterative reconstruction, and deep learning methods. Here, we simply use linear interpolation since only small gap is considered for now.

## 3 Experimental Results

We evaluated the feasibility of MSDR-CT and the proposed reconstruction method by numerical experiments on a Shepp-Logan phantom. The geometric parameters are listed in Table I. For the reconstruction, the direction of Hilbert filtering is horizontal with $\xi = \pi/2$.

TABLE I
PARAMETERS OF THE MSDR-CT CONFIGURATION

| Parameter | Value |
|---|---|
| Distance between the source and isocenter (mm) | 510.85 |
| Number of source segments | 11 |
| Number of focal spots in each segment | 400 |
| Distance between focal spots (mm) | 0.75 |
| Distance between the detector center and isocenter (mm) | 461.65 |
| Number of detector segments | 10 |
| Number of detector elements in each segment | 600 |
| Size of detector element (mm) | 0.5 |
| Dimension of reconstruction grids (pixels) | 512 |
| Pixel size (mm) | 0.25 |

Preliminary results of the ideal MSDR-CT without discontinuity between segments are demonstrated in Fig. 4, where almost exact reconstruction (Fig. 4(b)) is achieved by the Hilbert transform method. Profiles in Fig. 4(c) quantitatively confirm the consistency of the result to the phantom.

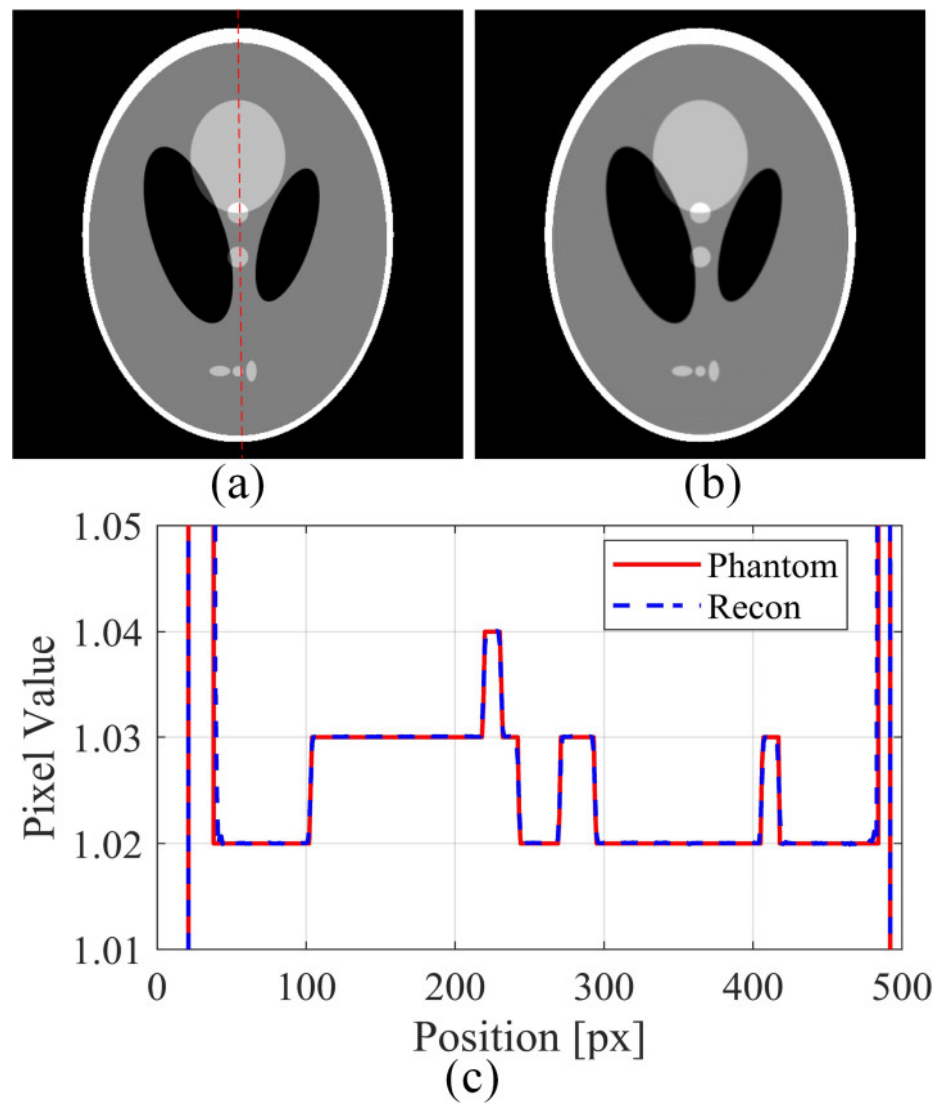

**Fig. 4** Evaluation of the reconstruction by the ideal MSDR-CT. (a) phantom; (b) reconstruction by the proposed method; (c) 1D profiles along the red vertical line in (a). The display window is [1,1.04].

We also evaluated the influence of the gaps in the source /detector rings. **(1) Gaps in the source ring:** The number of focal spots in each source module is decreased from 400 to 350, which induces about 37.5 mm gap between two adjacent segments. Due to the incomplete angular coverage, the reconstruction (see Fig. 5(b)) suffers from obvious bias and streaking artifacts. Compensated by the conjugate projection with the proposed weighting $W_{\beta_i,\varphi_k}(l,u)$ , the artifacts are removed (see Fig. 5(c)). **(2) Gaps in the detector ring:** We decreased the number of detector elements in each module from 600 to 596, which induces a 2 mm gap between adjacent segments. The reconstruction without correction (Fig. 5(d)) is contaminated by artifacts. However, when the missing elements are compensated by the preliminary sinogram interpolation, the reconstruction is well recovered (see Fig. 5(e)).

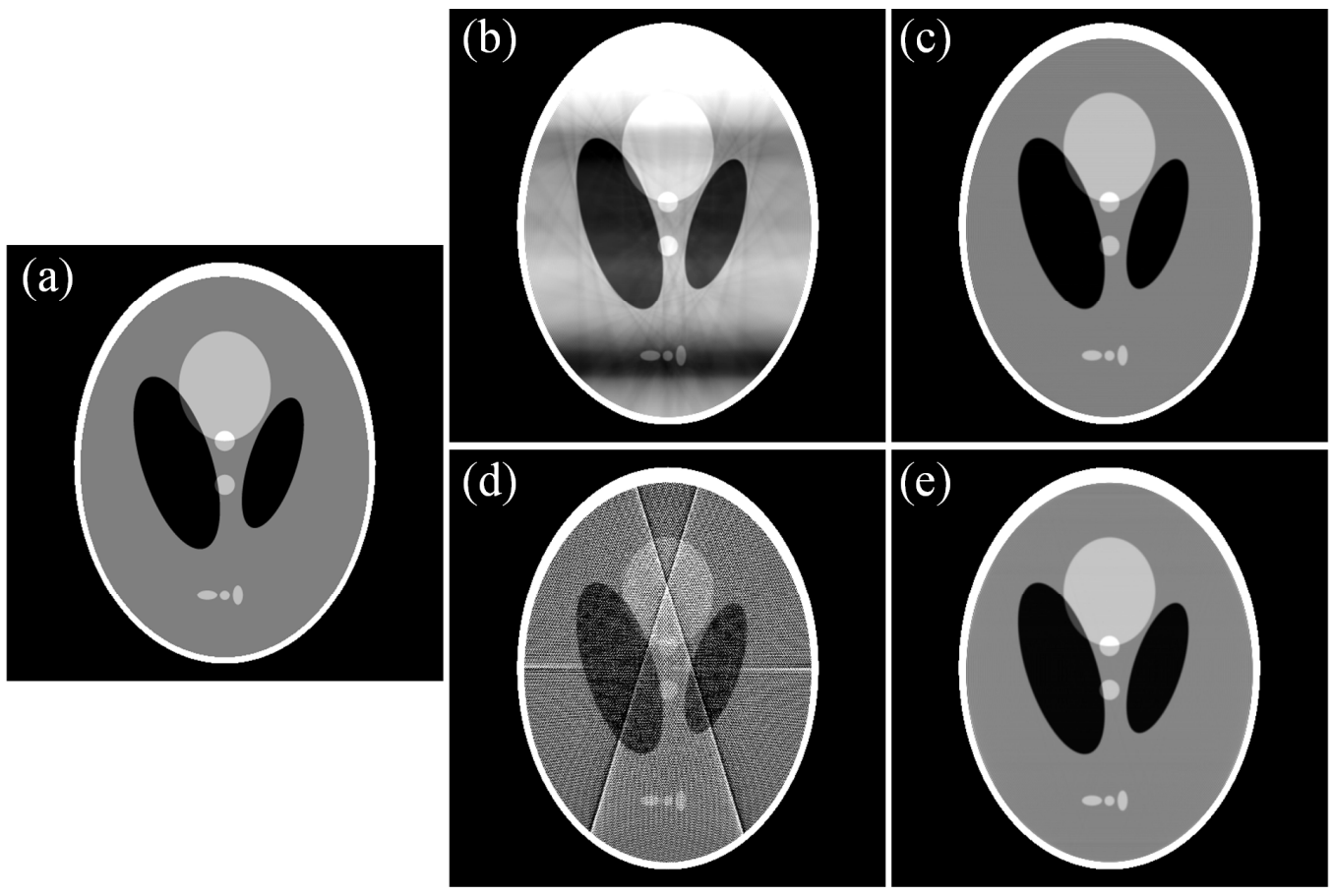

**Fig. 5** Evaluation of the reconstruction by MSDR-CT with discontinuity. (a) phantom; (b)(c) reconstruction w/o and w correction for the gap between source segments; (d)(e) reconstruction w/o and w correction for the gap between detector segments. The display window is [1,1.04].

## 4 Discussion

Stationary CT has great potential in high temporal resolution imaging. We proposed a novel system (MSDR-CT) that is of great potential for millisecond CT imaging. Mainly, it has three advantages: **(1) Ultra-fast and flexible scan protocols.** Different sources in MSDR-CT can be fired and switched ultrafast (up to sub-millisecond) at different technical parameters under programmable control which dramatically accelerates scanning. **(2) Improved angular sampling.** Recent advance in CNT multi-spot sources enables focal spots to densely and linearly distributed in a single source module which will significantly increases the angular sampling for stationary CT. **(3) Easy implementation.** The system could be constructed by splicing ready-made multi-spot source arrays and detector modules to formulate the source and detector rings. However, the optical structure of MSDR-CT is very complicated, which challenges the image reconstruction. We proposed a Hilbert transform method for the analytical reconstruction of MSDR-CT. Results in Section 3 confirm satisfying reconstruction is achievable for MSDR-CT, and validate the feasibility and efficiency of the system. As seamless assembly of multiple modules is impossible in practical implementation, we further explored the influence of discontinuity in the source or detector rings. Due to the gaps between the source or detector segments, reconstruction results will be contaminated by bias or inconsistent artifacts (see Fig. 5(b) and Fig. 5(d)), which indicates the reconstruction is rather sensitive to the continuity of the source and detector rings. To address this problem, preliminary correction methods are proposed. According to the analysis in Fig. 3, conjugate projections could compensate for gaps in the source ring by incorporating a weighting function to balance data redundancy and deficiency. Linear interpolation is a straightforward strategy to alleviate the detector data discontinuity. Simulated experiments verified the efficiency of the correction methods (see Fig. 5(c) (e)), and further validated the feasibility and implementability of MSDR-CT. However, when gaps get bigger, the incompleteness in sampling in the radon space will be severer. We are to research more advanced methods in future works.

## 5 Conclusion

Aiming at ultra-high temporal resolution imaging, we proposed a novel stationary CT MSDR-CT that is convenient to implement and capable of ultra-fast and flexible data acquisition modes per various protocols. As a proof-of-concept study, a Hilbert transform method and truncation correction methods are investigated to address the image reconstruction and discontinuity problems from the principle and practice aspects. Simulated experiments validated the feasibility and implementability of MSDR-CT. MSDR-CT is of great potential to promote real-time imaging of dynamic organs and objects to the order of millisecond. We will present further results of 3D imaging by MSDR-CT in the conference.

# Deep-Projection-Extraction based Reconstruction for Interior Tomography

Changyu Chen[1,2], Li Zhang[1,2], Yuxiang Xing[1,2*], and Zhiqiang Chen[1,2*]

[1]Department of Engineering Physics, Tsinghua University, Beijing, China
[2]Key Laboratory of Particle & Radiation Imaging (Tsinghua University), Ministry of Education, Beijing, China

**Abstract** Interior tomography is a typical strategy for radiation dose reduction in computed tomography (CT), where only a certain region-of-interest (ROI) is scanned. However, given the truncated projection data, ROI reconstruction by conventional analytical algorithms may suffer from severe cupping artifacts. In this paper, we proposed a new **D**eep-**P**rojection-**E**xtraction based **R**econstruction (DPER) for interior tomography. DPER works in dual domains where a sinogram-domain network (SDNet) estimates the contribution of the exterior region to the truncated projection and an image-domain network (IDNet) further mitigates artifacts. Unlike extrapolation-based methods, SDNet is intended to obtain a complete ROI-only sinogram via extraction instead of a fully non-truncated sinogram for both the ROI and exterior regions. We validated DPER with simulation on low-dose CT data. Results indicate that DPER can disclose more reliable structures, and achieve better image quality with better generalization performance than extrapolation-based methods.

## 1 Introduction

Interior tomography is commonly used when only a certain region of the patient is of interest. With X-ray flux collimated to the region-of-interest (ROI), dose exposure to the uninterested exterior region can be reduced. However, projection data will be truncated due to the limited field-of-view (FOV) scan. If filtered back-projection (FBP) algorithm is employed, the truncated filtration will introduce severe cupping artifacts into the reconstructed ROIs which compromises important diagnostic information.

In last decades, many works have gone into interior reconstruction. One straightforward method is sinogram extrapolation under certain assumptions for the exterior region [1, 2]. Other researchers explored exact interior reconstruction with various conditions of known sub-regions[3, 4] and developed reconstruction methods under the differentiated back-projection (DBP) framework[5]. Besides, the sparsity model of ROIs is also integrated into reconstruction[6]. Given an ROI is piecewise constant or polynomial, it can be stably reconstructed via total variation minimization[6, 7]. However, in practical situations, the known subregions are not always available and the sparsity model may be violated.

Recently, deep learning methods have yielded impressive performance in various ill-posed reconstruction problems. Some researchers proposed to use the FBP [8], DBP[9] reconstructions, or direct back-projected projection[10] as the input of U-Net for post-processing. Others focus on sinogram extrapolation [11] or combine it with an image-domain network to form a dual-domain optimization[12]. However, the extrapolation-based methods learn to estimate the undetected projection to reflect the complete measure of both the ROI and the exterior region, which is very challenging especially when the ROI size is of a small portion. Intuitively, the truncated projection data can be viewed as the combination of a complete measure of the ROI and a truncated measure of the exterior region. Hence, exterior reconstruction is highly underdetermined with a big exterior region. Besides, if the scanning is coupled with other nonideal factors such as noise, extrapolation-based methods may not be efficient enough to deal with such distortions inside the truncated projection.

In this work, we proposed a new dual-domain pathway for ROI reconstruction based on Deep-Projection-Extraction (DPER). Specifically, a sinogram-domain network (SDNet) estimates the contribution of the exterior region to the truncated projection, and an image-domain network (IDNet) further mitigates artifacts. Our experiments demonstrate that DPER can disclose more reliable structures and achieve a better image quality with better capability of generalizing to a higher noise level than extrapolation-based methods.

## 2 Materials and Methods

### 2.1 Formulation of Interior Tomography

Let $\mu(x,y)$ being the attenuation map of an object, $\theta \in [0,2\pi)$ the angle of projection views, and $t$ the coordinate of detector elements (see Fig. 1). We can formulate a full-view projection as

$$p_{\mathrm{FULL}}(\theta,t) = \iint \mu(x,y)\delta(\vec{\mathrm{x}}\cdot\vec{\theta} - t)\mathrm{d}x\mathrm{d}y \tag{1}$$

where $\vec{\mathrm{x}} = (x,y)$ and $\vec{\theta} = (\cos\theta, \sin\theta)$.

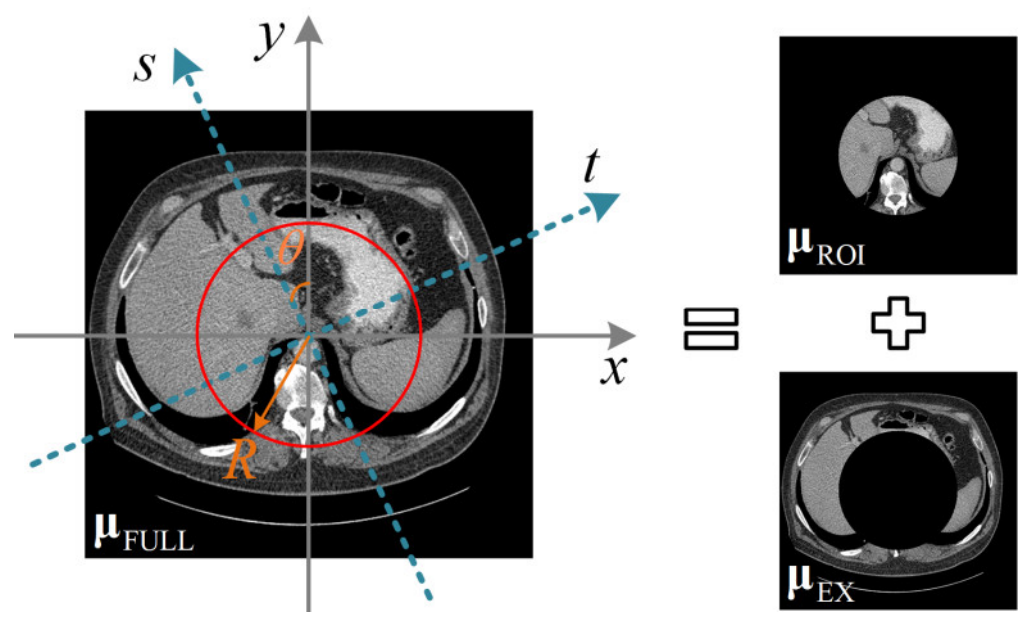

**Fig. 1** Illustration for interior tomography.

For interior tomography, X-ray flux is collimated to fully cover a ROI $\mu_{\mathrm{ROI}} = \mathcal{I}_R(|\vec{\mathrm{x}}|)\mu$ with the FOV restricted to $t \in [-R,R]$ (see the red circle in Fig. 1). $\mathcal{I}_R(\cdot)$ represents the indicator function

$$\mathcal{I}_R(\alpha) = \begin{cases} 1, & |\alpha| \le R \\ 0, & |\alpha| > R \end{cases} \tag{2}$$

This work was supported in part by the National Natural Science Foundation of China (Grant No. 62031020).
*Corresponding author: Zhiqiang Chen (e-mail: czq@tsinghua.edu.cn); Yuxiang Xing (e-mail: xingyx@mail.tsinghua.edu.cn).

Then, a low-dose interior projection can be formulated as

$$p_{\mathrm{IN}}(\theta,t)=\mathcal{I}_R(t)\left(p_{\mathrm{FULL}}(\theta,t)+n_{\mathrm{p}}\right) \tag{3}$$

where $n_{\mathrm{p}}$ denotes the noise. If conventional analytical algorithms (such as FBP) are directly applied to $p_{\mathrm{IN}}$, reconstructions will suffer from obvious cupping artifacts and biased value due to truncated filtration

$$\hat{\mu}_{\mathrm{FBP}}(x,y)=\int_0^{\pi}\mathrm{d}\theta\int_{-R}^{R}p_{\mathrm{IN}}(\theta,t)h(x\cos\theta+y\sin\theta-t)\mathrm{d}t \tag{4}$$

with $h(\cdot)$ being the convolution kernel. It is important to achieve a non-truncated filtration for interior reconstruction.

## 2.2 Existing Extrapolation-based Methods

Intuitively, $\mu_{\mathrm{ROI}}$ can be reconstructed from a non-truncated projection $p_{\mathrm{FULL}}$ (see Fig. 2) with $p_{\mathrm{IN}}$ unchanged inside the original FOV and an additional measure $p_{\mathrm{EX}}$ outside the FOV $t\in(-\infty,-R)\cup(R,+\infty)$:

$$\begin{aligned}p_{\mathrm{EX}}(\theta,t)&=\iint\left(1-\mathcal{I}_R(t)\right)\mu(x,y)\delta(\vec{\mathrm{x}}\cdot\vec{\theta}-t)\mathrm{d}x\mathrm{d}y\\&=\iint\left(1-\mathcal{I}_R(t)\right)\left(1-\mathcal{I}_R(|\vec{\mathrm{x}}|)\right)\mu(x,y)\delta(\vec{\mathrm{x}}\cdot\vec{\theta}-t)\mathrm{d}x\mathrm{d}y\end{aligned} \tag{5}$$

Various methods have been proposed to acquire $p_{\mathrm{EX}}$ via extrapolation either heuristically or in a data-driven manner. FOV extension is intended for the complete measure of $\mu_{\mathrm{EX}}=(1-\mathcal{I}_R(|\vec{\mathrm{x}}|))\mu$. However, compared with $p_{\mathrm{EX}}$, the measured information about $\mu_{\mathrm{EX}}$ in $p_{\mathrm{IN}}$ can be rather limited especially when the ROI is relatively small. Further, when the interior projection is contaminated with other artifacts (such as noise and sparse sampling), extrapolation methods may not be able to efficiently recover the data inside the FOV, hence compromises the image quality.

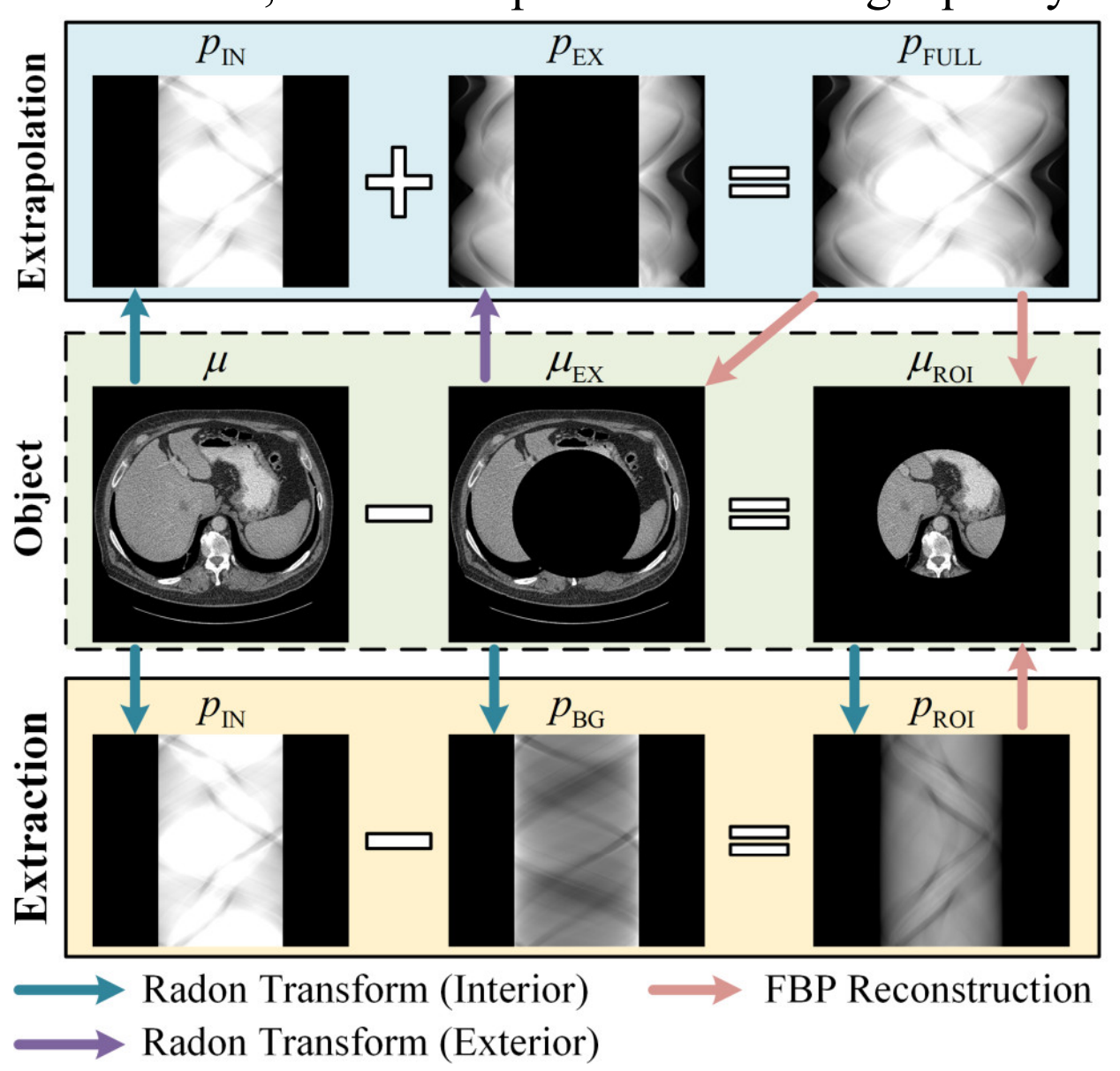

**Fig. 2** Illustration for the extrapolation- and extraction- based pipelines.

## 2.3 Projection Extraction

Besides extrapolation, a non-truncated filtration is also achievable if projections are already complete over $t\in[-R,R]$. Given the rays passing through $\mu_{\mathrm{ROI}}$ only, a complete ROI-only projection $p_{\mathrm{ROI}}$ is generated

$$\begin{aligned}p_{\mathrm{ROI}}(\theta,t)&=\iint\mathcal{I}_R(|\vec{\mathrm{x}}|)\mu(x,y)\delta(\vec{\mathrm{x}}\cdot\vec{\theta}-t)\mathrm{d}x\mathrm{d}y\\&=\iint\mathcal{I}_R(t)\mathcal{I}_R(|\vec{\mathrm{x}}|)\mu(x,y)\delta(\vec{\mathrm{x}}\cdot\vec{\theta}-t)\mathrm{d}x\mathrm{d}y\end{aligned} \tag{6}$$

As $p_{\mathrm{ROI}}(\theta,t)=\mathcal{I}_R(t)p_{\mathrm{ROI}}(\theta,t)$, the filtration of $p_{\mathrm{ROI}}$ within $t\in[-R,R]$ is complete and $\mu_{\mathrm{ROI}}$ can be well reconstructed. Unfortunately, in interior tomography, part of $\mu_{\mathrm{EX}}$ is also scanned which contributes an additional background projection $p_{\mathrm{BG}}$ (see Eq. (7)) to the projection inside the FOV.

$$\begin{aligned}p_{\mathrm{BG}}(\theta,t)&=p_{\mathrm{IN}}(\theta,t)-p_{\mathrm{ROI}}(\theta,t)\\&=\iint\mathcal{I}_R(t)\left(1-\mathcal{I}_R(|\vec{\mathrm{x}}|)\right)\mu(x,y)\delta(\vec{\mathrm{x}}\cdot\vec{\theta}-t)\mathrm{d}x\mathrm{d}y+n_{\mathrm{p}}\end{aligned} \tag{7}$$

Hence, $p_{\mathrm{IN}}$ can be decomposed into two parts: a complete $p_{\mathrm{ROI}}$ from $\mu_{\mathrm{ROI}}$ and a truncated $p_{\mathrm{BG}}$ from $\mu_{\mathrm{EX}}$. In this paper, we introduce a new pathway (DPER) to reconstruct the ROI through extraction (see Fig. 2). Different from extrapolation, DPER only works on data inside the FOV to remove the contribution of $\mu_{\mathrm{EX}}$ and estimate the ROI-only projection $\hat{p}_{\mathrm{ROI}}$ without FOV extension.

## 2.4 Network Architecture and Loss Function

DPER works in dual domains (see Fig. 3): a sinogram-domain network (SDNet) $\varphi_{\mathrm{S}}(\cdot)$ extracts $p_{\mathrm{BG}}$, an FBP layer for domain transformation, and an image-domain network (IDNet) $\varphi_{\mathrm{I}}(\cdot)$ further improves image quality. In practice, SDNet and IDNet employ modified U-Net as the backbone where the number of channels is decreased by half compared to the original version. As the estimation for $p_{\mathrm{BG}}$ requires global properties in the sinogram domain, we enlarge the receptive field (RF) of SDNet by three dilated convolution layers with dilation rate (1,1), (2,1), and (5,1) for each stage of SDNet. The objective function can be formulated as

$$\hat{\vartheta}_{\varphi_{\mathrm{S}}},\hat{\vartheta}_{\varphi_{\mathrm{I}}}=\underset{\vartheta_{\varphi_{\mathrm{S}}},\vartheta_{\varphi_{\mathrm{I}}}}{\arg\min}\left\{\lambda\frac{\left\|\mathcal{R}^{-1}(\varphi_{\mathrm{S}}(p_{\mathrm{IN}}))-\mu_{\mathrm{ROI}}\right\|_2}{\left\|\mu_{\mathrm{ROI}}\right\|_2}+\frac{\left\|\varphi_{\mathrm{I}}(\mathcal{R}^{-1}(\varphi_{\mathrm{S}}(p_{\mathrm{IN}})))-\mu_{\mathrm{ROI}}\right\|_2}{\left\|\mu_{\mathrm{ROI}}\right\|_2}\right\} \tag{8}$$

where $\mathcal{R}^{-1}$ denotes the process of FBP reconstruction with RamLak filtration, $\vartheta_{\varphi_{\mathrm{S}}}$ and $\vartheta_{\varphi_{\mathrm{I}}}$ the network parameters for SDNet and IDNet respectively. $\lambda$ is simply set to 1.

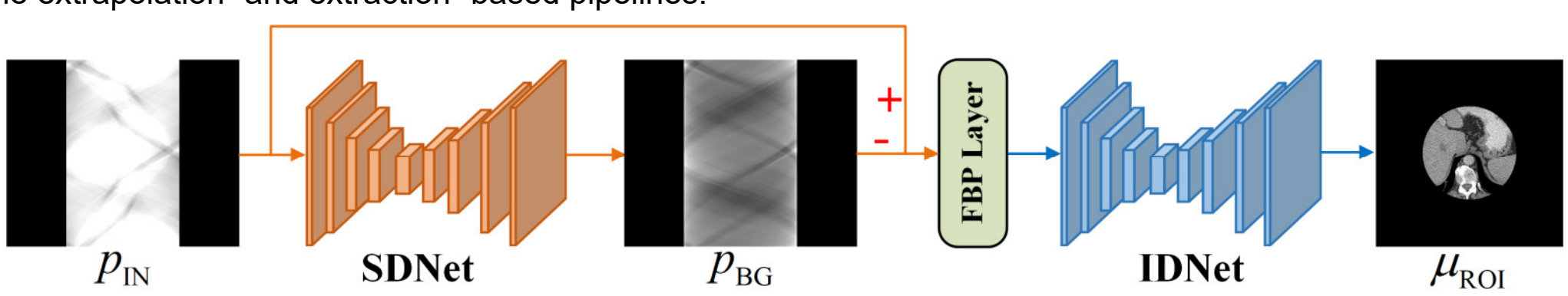

**Fig. 3** Diagram for the network architecture of DPER.

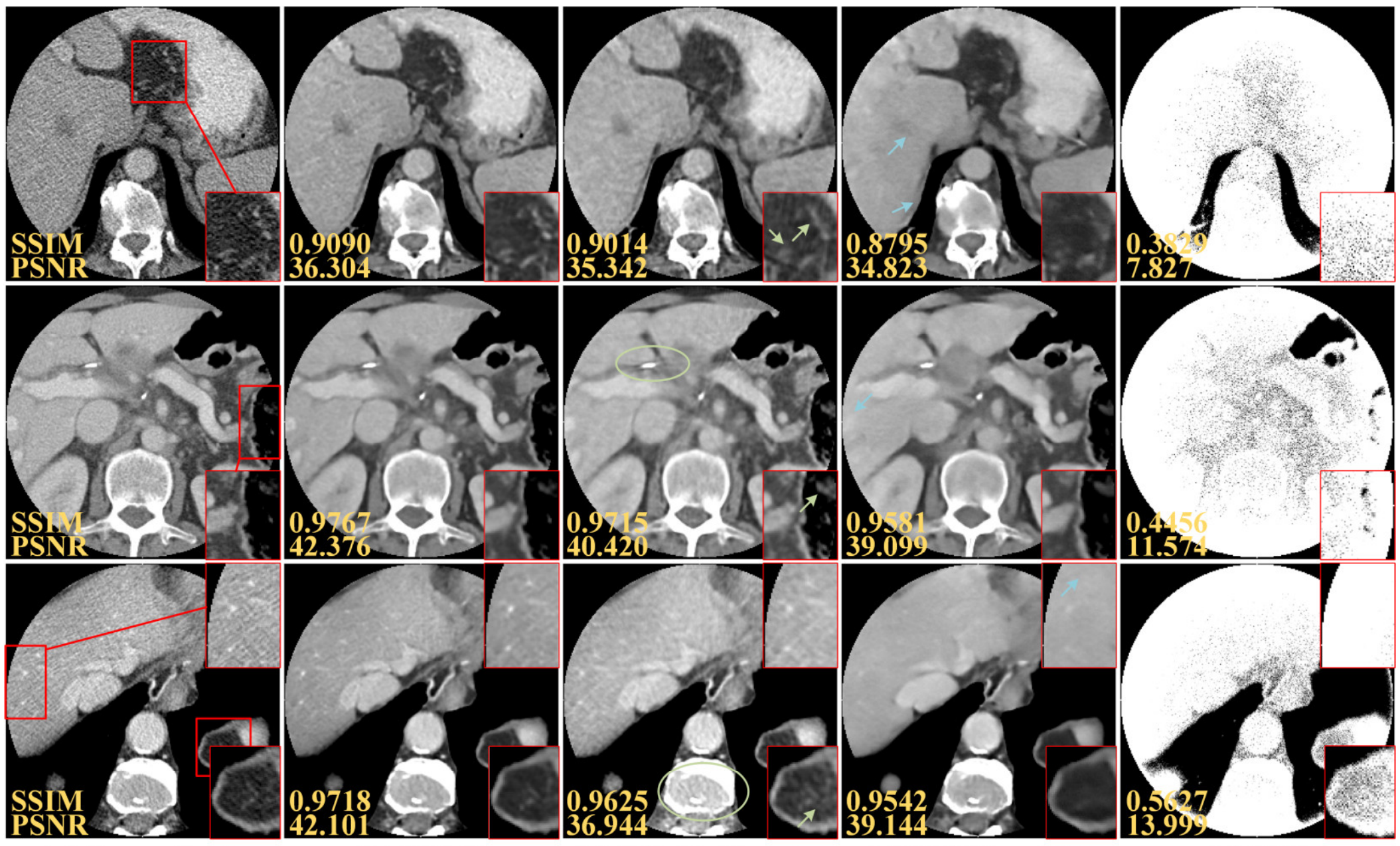

**Fig. 4** Comparison among different methods: ground truth (a), DPER (b), DD_ExDenoise (c), FBPConvNet (d), and FBP (e). The display window is [0.016, 0.024]. SSIM and PSNR are in the bottom-left corner of each image.

## 3 Results

### 3.1 Experimental Set-up

We evaluated our method based on datasets from AAPM Low Dose Grand Challenge and TCIA Low Dose CT Image and Projection Data. ROI scanning is simulated under an equidistant fan-beam configuration and the detailed geometry parameters are listed in Table I. Poisson noise is simulated with $10^5$ incident photons per ray. Structural similarity index (SSIM) and peak signal to noise ratio (PSNR) are used for quantitative evaluation. The mean and coefficient of variation (CV) are employed to evaluate the stability of performance in the whole test set.

TABLE I
PARAMETERS OF THE FAN-BEAM GEOMETRY

| Parameter | Value |
|---|---|
| Distance between the source and isocenter (mm) | 410 |
| Distance between the detector center and isocenter (mm) | 310 |
| Number of detector elements - truncated | 512 |
| Number of detector elements - complete | 1056 |
| Size of detector element (mm) | 0.45 |
| Dimension of reconstruction grids - ROI | 256 |
| Dimension of reconstruction grids - Full | 512 |
| Pixel size (mm) | 0.5 |
| Number of projection views over $2\pi$ | 480 |

We employed three methods as baselines for evaluation: the conventional FBP reconstruction, FBPConvNet, and a dual-domain network (DD_ExDenoise) [12] where the sinogram-domain network employs two heads output for extrapolation outside the FOV and denoising inside the FOV respectively. All the methods were reproduced according to their original implementation.

### 3.2 Experimental Results

Three samples from different patients are displayed in Fig. 4. FBP reconstructions are contaminated by severe bias and noise. FBPConvNet can recover the gray level but the noise cannot be well suppressed. DD_ExDenoise can remove the cupping artifacts and noise to a large extent. But some inconsistent artifacts and false positive structures can be observed (see the green circles and arrows in Fig. 4). Compared with all the methods above, DPER can produce the most reliable structures and consistent image quality (see the zoom-ins in Fig. 4). The quantitative metrics among all test slices are listed in Table II. We can find that DPER achieves the best SSIM and PSNR.

TABLE II
QUANTITATIVE METRICS AMONG ALL TEST SLICES (Photon $10^5$)

| Metric | SSIM | | PSNR | |
|---|---|---|---|---|
| | Mean | CV | Mean | CV |
| **FBP** | 0.4898 | 20.10% | 12.382 | 30.01% |
| **FBPConvNet** | 0.9431 | 3.49% | 37.434 | **6.48%** |
| **DD_ExDenoise** | 0.9605 | 2.80% | 39.771 | 8.76% |
| **DPER** | **0.9680** | **2.50%** | **40.980** | 7.00% |

We further evaluate the methods on the same test set but injected higher noise (corresponding to 5 x $10^4$ photons) in projections. To adapt to the higher noise, models are finetuned on the same dataset from one patient for 10 epochs where mixed noise levels corresponding to 5 x $10^4$ and $10^5$ photons were employed. The same slice is shown in Fig. 5 and the quantitative metrics are listed in Table III. FBPConvNet cannot adapt to the higher noise and produces images with obvious gray level deviation and drastically decreased PSNR. Though DD_ExDenoise may achieve better noise reduction, streaking and inconsistent artifacts are observed. DPER produces images of best quality with fewer artifacts and more identifiable structures than other methods, which validated its capability of generalizing to higher noise levels.

TABLE III
QUANTITATIVE METRICS AMONG ALL TEST SLICES (Photon 5x$10^4$)

| Metric | SSIM | | PSNR | |
|---|---|---|---|---|
| | Mean | CV | Mean | CV |
| **FBP** | 0.4154 | 57.70% | 12.234 | 30.21% |
| **FBPConvNet** | 0.9182 | 3.89% | 32.512 | 6.51% |
| **DD_ExDenoise** | 0.9533 | 2.89% | 37.532 | **6.13%** |
| **DPER** | **0.9587** | **2.71%** | **39.301** | 6.61% |

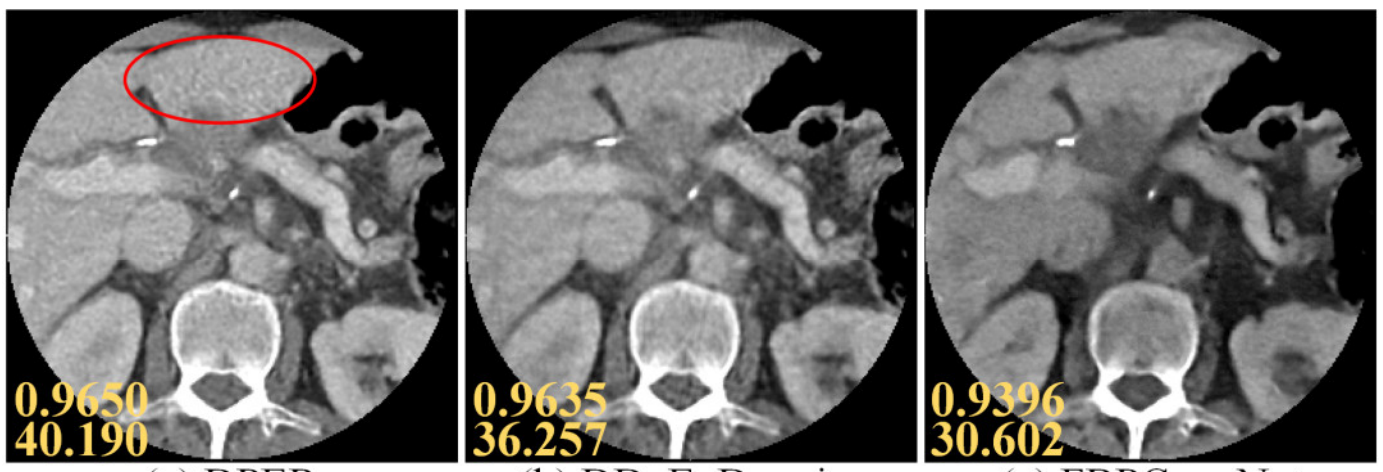

(a) DPER (b) DD_ExDenoise (c) FBPConvNet

**Fig. 5** Comparison among different methods on a higher noise level: DPER (a), DD_ExDenoise (b), and FBPConvNet (c). The display window is [0.016, 0.024]. SSIM and PSNR are in the bottom-left corner of each image.

For the extraction of $p_{\mathrm{BG}}$, dilated convolution is incorporated into SDNet to reach a larger RF. To validate the efficiency of the design, we compared the reconstruction from the dilated network with non-dilated network where SDNet employs the same backbone as IDNet. According to our experiments, both methods achieved similar results in the training set. However, reconstructions by the non-dilated network suffer from severe ring artifacts in the test set (see the red arrows in Fig. 6). From the intermediate result from SDNet (see Fig. 6(b)), we may conclude that the artifact is mainly caused by SDNet though IDNet could subsequently alleviate it to a large extent (see Fig. 6(c)).

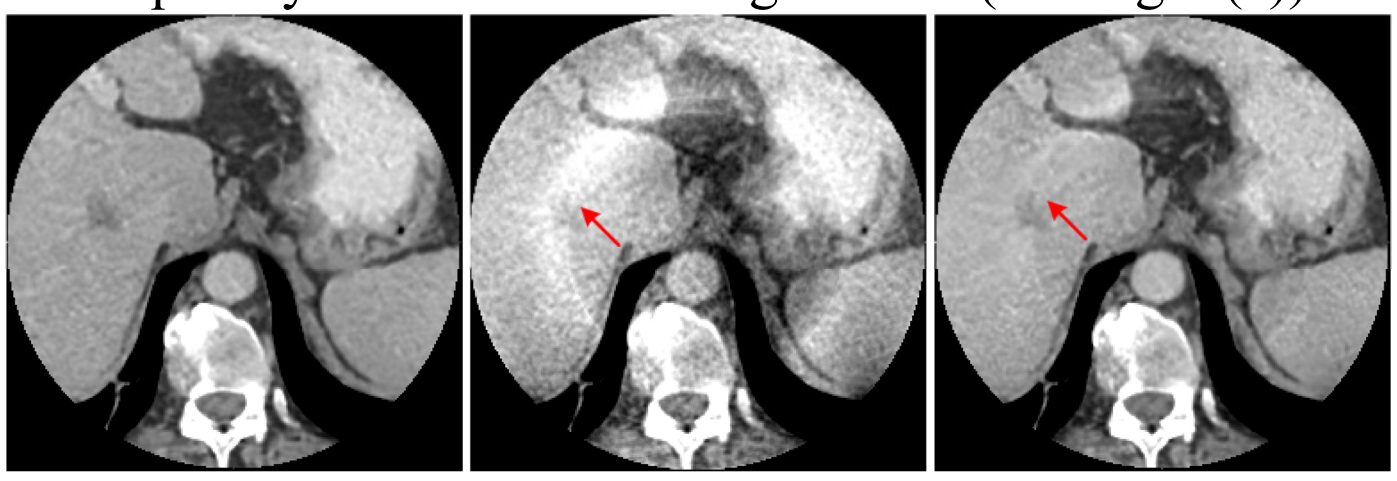

(a) Dilated Network (b) nonDilated_SD (c) nonDilated_ID

**Fig. 6** Comparison of the reconstruction of dilated and nondilated network: dilated network (DPER) (a), SDNet (b) and IDNet (c) reconstruction of the nondilated network. The display window is [0.016, 0.024].

## 4 Discussion

Interior tomography is an ill-posed problem due to the truncated projection data. Reconstructions by conventional algorithms such as FBP suffer from severe cupping artifacts. Although image-domain networks can alleviate most artifacts, the overall image quality is blurry which compromises the recognition of context structures (see the blue arrows in Fig. 4(d)). Compared with image-domain networks, dual-domain methods are more capable of revealing reliable structures with better generalization performance. Under the dual-domain framework, we proposed a new extraction-based reconstruction method DPER. The main difference between DPER and extrapolation-based pathways is how to alleviate the influence of the exterior region which is not fully measured. In the literature, extrapolation-based methods have been proposed to extend the truncated interior projection to acquire a non-truncated projection that contains the full measure of the exterior region. However, DPER learns to estimate the contribution of the exterior region inside the FOV to achieve an ROI-only projection without sinogram extension. DPER requires less prior information about the exterior region which may make the reconstruction more stable. As DPER only works on data inside the FOV, it can be more capable of jointly dealing with other nonideal factors such as noise in the interior projection compared with extrapolation-based methods (see the comparison with DD_ExDenoise in Fig. 4 and Fig. 5). As the contribution of the exterior region is to the whole FOV, the network could not efficiently extract it with limited RF (see Fig. 6(b)). Hence, dilated convolution is employed to expand the RF and promotes the extraction (see Fig. 6(a)). Although DPER has shown promising results for interior reconstruction, slight structural distortions and gray-level error may still happen. Fake structures can be observed when generalize to a different noise level (see the red circles in Fig. 5). We will address these problems in the conference.

## 5 Conclusion

In this paper, we established a new pipeline named DPER for interior tomography that works on extraction. Unlike extrapolation-based methods, DPER only focuses on data inside the FOV to extract the contribution of the exterior region and other artifacts. Experiments on low-dose interior tomography favored DPER in its potential of revealing reliable structures and good generalization.

# Unifying Supervised and Unsupervised Methods for Low-dose CT Reconstruction: a General Framework

Ling Chen[1], Zhishen Huang[2], Yong Long[1], and Saiprasad Ravishankar[3,4]

[1]University of Michigan Shanghai Jiao Tong University Joint Institute, Shanghai Jiao Tong University, Shanghai 200240, China
[2]Amazon, Seattle, WA 98109, USA
[3]Department of Computational Mathematics, Science and Engineering, Michigan State University, East Lansing, MI 48824, USA
[4]Department of Biomedical Engineering, Michigan State University, East Lansing, MI 48824, USA

**Abstract** Recent application of deep learning methods for image reconstruction provides a data-driven approach to address the challenge raised by undersampled measurements or various types of noise. In this work, we propose a general learning framework for X-ray computed tomography (CT) image reconstruction that combines supervised and unsupervised learned models. We leverage both a dictionary learning-based unsupervised solver and supervisedly trained neural network reconstructors in two incarnations of the proposed framework to simulate a fixed-point iteration process. Our experimental results for denoising low-dose CT (LDCT) images demonstrate promising performance of the proposed general framework compared to our recent parallel and cascading SUPER methods for LDCT.

## 1 Introduction

X-ray computed tomography (CT) is widely used in clinical practice to obtain images of bones, blood vessels, and soft tissues inside the human body. Many commercial CT scanners use the filtered-back projection (FBP) technique to produce tomographic images from X-ray measurements. When the X-ray dosage for CT measurements is low, which is desirable to lessen potential harm to patients, the FBP method can lead to compromised image quality. Model-based iterative reconstruction (MBIR) methods [1] were proposed to address such performance degradation in the low-dose X-ray computed tomography (LDCT) setting.

Two classical MBIR formulations are non-adaptively regularized least-square problems and dictionary learning-based optimization problems. Edge-preserving (EP) regularized least-square-based solver is an example for non-adaptive image reconstruction. Dictionary learning-based methods [2] provide improved image reconstruction quality compared to non-adaptive MBIR schemes, but incur expensive computations for sparse encoding. Recent penalized weighted least square (PWLS) methods with regularizers involving learned sparsifying transforms (PWLS-ST [3]) or a union of learned transforms (PWLS-ULTRA [4]) have the advantages of computational efficiency (cheap sparse coding in transform domain) and the representation power of learned models (transforms).

Deep learning-based approaches have also been successfully applied for LDCT image reconstruction (see [5] for a review). FBPConvNet [6] is a CNN-based refinement scheme which operates in the image domain and maps the crude FBP-reconstructed CT images to target images obtained under full-dose setting. Another approach, WavResNet [7], learns a set of filters used for constructing the encoder and decoder of the convolutional framelet denoiser to refine crude LDCT images.

Deep learning methods often require large dataset for training and may have degraded performance when applied to a dataset with different underlying distribution than the training dataset. In comparison, sparsifying transform learning methods only need small training datasets and have shown improved generalization property to new data [4]. Ye et. al. [8] proposed a unified supervised-unsupervised (Serial SUPER) learning method for LDCT image reconstruction that exploits both supervised deep learning models and unsupervised transform learning (ULTRA) solvers. The Serial SUPER method alternates between a neural network-based denoising step and an optimization step with learned dictionary regularization. In a similar spirit, Chen et al. [9] proposed the Parallel SUPER method to simultaneously exploit the supervised module and unsupervised module, where the output of supervised module and unsupervised module are adaptively aggregated and such aggregation blocks iterate to attain convergence to the final refined image.

In this work, we propose a general framework to combine supervised and unsupervised modules for LDCT image reconstruction. The Serial SUPER and the Parallel SUPER can be derived as special instances of this general formulation. We explore two specific structures derived from the general framework, and find that these two structures outperform the standalone supervised module, unsupervised module, the Serial SUPER method and the Parallel SUPER method.

## 2 General Formulation and Algorithms

### 2.1 General formulation

The main idea of the general SUPER framework is to combine the unsupervised module and the supervised module in some form for the purpose of outperforming individual methods, and we iterate such combination blocks to attain convergence. Denote the neural network operator (or some general non-linear operator) as $N(\cdot)$, the unsupervised solver as $T(\cdot)$, and a general aggregation operator as $G(\cdot)$. The general SUPER model in the $n$-th layer can be formulated as:

$$\mathbf{x}_{n+1} = G_n\Big(N_n(\mathbf{z}_n, T(\mathbf{z}_n)), T(\mathbf{x}_n), \mathbf{z}_n\Big), \tag{1}$$

where $\mathbf{x}_n$ is the output of $n$-th layer and $\mathbf{z}$ is the collection of all history of $\mathbf{x}$, i.e., $\mathbf{z}_n = \{\mathbf{x}_{0:n}\}$. The notation $N_n(\mathbf{z}_n, T(\mathbf{z}_n))$ is abused here in the sense that not necessarily the entire history needs to be considered, but instead only partial history can be taken into account. As an example, if we only want to exploit the reconstruction results in the previous three blocks and the output of unsupervised solvers in the previous two blocks, we can set the first argument of $G_n$ to be $N_n(\mathbf{x}_n, \mathbf{x}_{n-1}, \mathbf{x}_{n-2}, T(\mathbf{x}_n), T(\mathbf{x}_{n-1}))$. The same connotation applies for the history input $\mathbf{z}_n$ to $G_n$.

In the formulation (1), the neural networks $N_n$ can take input from previous intermediate output $\mathbf{z}_n$ and output of unsupervised solvers in previous blocks $T(\mathbf{z}_n)$. Using the intermediate output of earlier blocks enables the residual structure. The unsupervised solver only takes in the reconstruction output from previous block without any historical information. This design for $T$ is for simplicity and for accommodating the specific solvers we use in this study. The aggregation $G(\cdot)$ can be nonlinear, piecewise, index-dependent, or identity with respect to some arguments.

### 2.2 Examples of structures

Based on the general formulation (1), the Serial SUPER method can be represented as $x_{n+1} = G_n(N_n(x_n), T(x_n))$, where $G_n(N_n(x_n), T(x_n)) = N_n(x_n)$ when $n$ is odd and $G_n(N_n(x_n), T(x_n)) = T(x_n)$ when $n$ is even. The Parallel SUPER method can be represented in the formulation (1) as $x_{n+1} = \lambda_n N_n(x_n) + (1 - \lambda_n)T(x_n)$, and $G_n(N_n(x_n), T(x_n)) = \lambda_n N_n(x_n) + (1 - \lambda_n)T(x_n)$, where $\lambda_n$ is the combination parameter in the $n$-th layer.

We can also obtain other SUPER models by choosing different aggregation functions and different ways of exploiting history information, for instance the Structure 1 shown in Figure 1 and the Structure 2 shown in Figure 2.

For Structure 1, the final output of the $n$-th layer is from a neural network aggregating the final output of $(n-1)$-th layer and the output of unsupervised solver in $n$-th layer:

$$x_{n+1} = N_n(x_n, T(x_n)). \quad (2)$$

For Structure 2, the difference from structure 1 is that we use older information of the unsupervised solvers in the iteration sequence:

$$x_1 = N_0(x_0), \quad x_{n+1} = N_n(x_n, T(x_{n-1})). \quad (3)$$

#### 2.2.1 Supervised Module

We use FBPConvNet, which is a CNN-based image-domain denoising network, as the supervisedly trained denoiser. FBPConvNet was originally designed for sparse-view CT, while we apply it here to the low-dose CT cases. The FBPConvNet takes low-dose FBP-reconstructed images as network input, and it is trained to map input images to high-quality reference images. FBPConvNet has an architecture similar to the U-Net, and FBPConvNet adopts multichannel filters to increase the capacity of the network.

#### 2.2.2 Unsupervised Module

For the unsupervised module of each layer, we solve the following MBIR problem to reconstruct an image $\mathbf{x} \in \mathbb{R}^{N_p}$ from the corresponding noisy sinogram data $\mathbf{y} \in \mathbb{R}^{N_d}$:

$$\min_{\mathbf{x} \geq 0} J(\mathbf{x}, \mathbf{y}) := \underbrace{\frac{1}{2}\|\mathbf{y} - \mathbf{A}\mathbf{x}\|_{\mathbf{W}}^2}_{:=L(\mathbf{A}\mathbf{x}, \mathbf{y})} + \beta \mathcal{R}(\mathbf{x}), \quad (4)$$

where $\mathbf{W} = \text{diag}\{w_i\} \in \mathbb{R}^{N_d \times N_d}$ is a diagonal weighting matrix with the diagonal elements $w_i$ being the estimated inverse variance of $y_i$, $\mathbf{A} \in \mathbb{R}^{N_d \times N_p}$ is the system matrix of the CT scan, $L(\mathbf{A}\mathbf{x}, \mathbf{y})$ is the data-fidelity term, penalty $\mathcal{R}(\mathbf{x})$ is a (learning-based) regularizer, and the parameter $\beta > 0$ controls the noise and resolution trade-off.

In this work, we use the PWLS-ULTRA method to reconstruct an image $\mathbf{x}$ from noisy sinogram data $\mathbf{y}$ (measurements) with a union of pre-learned transforms $\{\mathbf{\Omega}_k\}_{k=1}^K$. The image reconstruction is done through the following nonconvex optimization problem:

$$T(x_n) = \arg\min_{\mathbf{x}} \Big\{ \frac{1}{2}\|\mathbf{y} - \mathbf{A}\mathbf{x}\|_{\mathbf{W}}^2 + \min_{C_k, \mathbf{z}_j} \sum_{k=1}^{K} \sum_{j \in C_k} \Big( \|\mathbf{\Omega}_k \mathbf{P}_j \mathbf{x} - \mathbf{z}_j\|_2^2 + \gamma^2 \|\mathbf{z}_j\|_0 \Big) \Big\}, \quad (5)$$

where $T(x_n)$ is the reconstructed image by the unsupervised solver *(with initialization $x_n$)* in the $l$-th layer, the operator $\mathbf{P}_j \in \mathbb{R}^{l \times N_p}$ extracts the $j$-th patch of $l$ voxels of image $\mathbf{x}$ as $\mathbf{P}_j\mathbf{x}$, $\mathbf{z}_j$ is the corresponding sparse encoding of the image patch under a matched transform, and $C_k$ denotes the indices of patches grouped into the $k$-th cluster with transform $\mathbf{\Omega}_k$. Minimization over $C_k$ indicates the computation of the cluster assignment of each patch. The regularizer $\mathcal{R}$ includes an encoding error term and an $\ell_0$ sparsity penalty counting the number of non-zero entries with weight $\gamma^2$. The sparse encoding and clustering are computed simultaneously. We leverage the alternating minimization method [4] (with inner iterations for updating $\mathbf{x}$) to solve the optimization problem (5). We also use different (potentially better) initialization in each parallel SUPER layer, which may benefit solving the involved nonconvex optimization problem.

## 3 Experiments

### 3.1 Experiment setup

In our experiments, we use the Mayo Clinic dataset established for the "2016 NIH-AAPM-Mayo Clinic Low Dose CT Grand Challenge" [10]. We chose 520 images from 6 of 10 patients in the dataset, from which 500 slices were used for training and 20 slices were used for validation. We randomly selected 20 images from the remaining 4 patients for testing. We projected the regular dose CT images $\mathbf{x}^\star$ to sinograms $\mathbf{y}$ by adding Poisson and additive Gaussian noise to them as follows:

$$\mathbf{y}_i = -\log\Big(I_0^{-1} \max\big(\text{Poisson}\{I_0 e^{-[\mathbf{A}\mathbf{x}^\star]_i}\} + \mathcal{N}\{0, \sigma^2\}, \varepsilon\big)\Big),$$

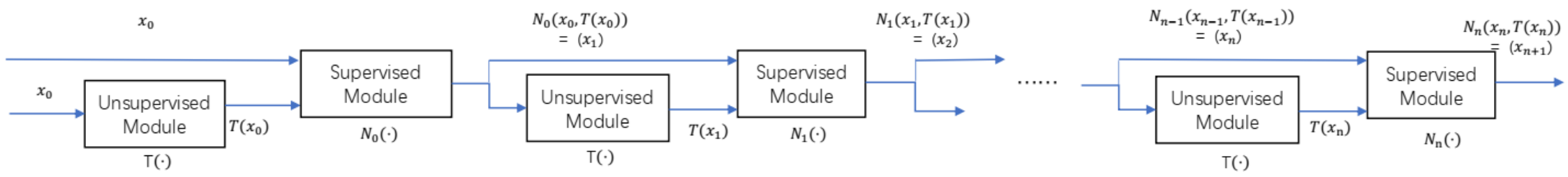

**Figure 1:** The structure 1 of the general SUPER model.

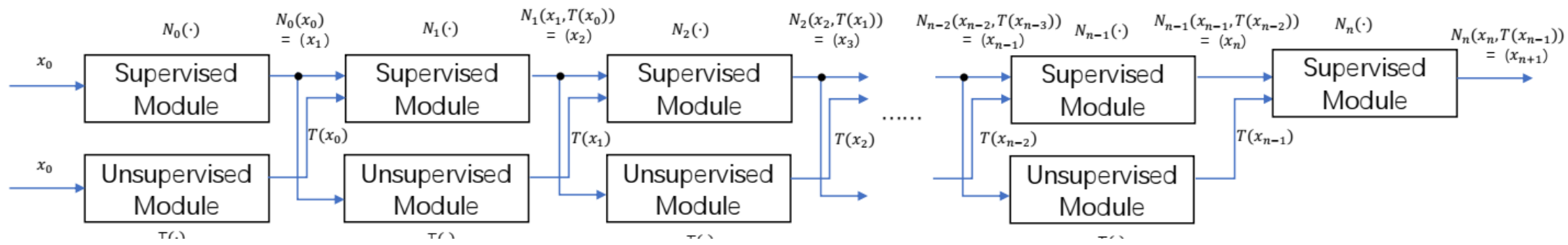

**Figure 2:** The structure 2 of the general SUPER model.

where the original number of incident photons per ray is $I_0 = 10^4$, the Gaussian noise variance is $\sigma^2 = 25$, and $\varepsilon$ is a small positive number to avoid negative measurement data when taking the logarithm [11].

We used the Michigan Image Reconstruction Toolbox to construct fan-beam CT geometry with 736 detectors × 1152 regularly spaced projection views, and a no-scatter mono-energetic source. The width of each detector column is 1.2858 mm, the source to detector distance is 1085.6 mm, and the source to rotation center distance is 595 mm. We reconstructed images of size 512×512 with the pixel size being 0.69 mm × 0.69 mm.

### 3.2 Parameter settings

In both Structure 1 and Structure 2, we use FBPConvNet as the supervised module and PWLS-ULTRA as the unsupervised module. It takes about 10 hours for training a model instance with 10 layers on a GTX Titan GPU graphics processor. During the training of the supervised method, we ran 4 epochs (the amount of epochs is kept small to reduce overfitting risks) with the stochastic gradient descent (SGD) optimizer for the FBPConvNet module in each parallel SUPER layer. The training hyperparameters of FBPConvNet are set as follows: the learning rate decreases logarithmically from 0.001 to 0.0001; the batchsize is set as 1 due to the ability of the GPU; and the momentum parameter is 0.99. The filters were initialized in the various networks during training with i.i.d. random Gaussian entries with zero mean and variance 0.005. For the unsupervised module, we trained a union of 5 sparsifying transforms using 12 slices of regular-dose CT images (which are included in the 500 training slices). Then, we use the pre-learned union of 5 sparsifying transforms to reconstruct images with 5 outer iterations and 5 inner iterations of PWLS-ULTRA. In the training and reconstruction with ULTRA, we set the parameters $\beta = 5\times10^3$ and $\gamma = 20$. PWLS-EP reconstruction is used as the initial $\mathbf{x}_0$ or the input of the networks in the first layer.

We compare the two derived structures from the general supervised-unsupervised integration framework with the unsupervised method (PWLS-EP), standalone supervised module (FBPConvNet), standalone unsupervised module (PWLS-ULTRA), the serial SUPER model and the parallel SUPER method. PWLS-EP is a penalized weighted-least squares reconstruction method with edge-preserving hyperbola regularization. For the unsupervised method (PWLS-EP), we set the parameters $\delta = 20$ and $\beta = 2^{15}$ and ran 100 iterations to obtain convergent results. In the training of the standalone supervised module (FBPConvNet), we ran 100 epochs of training to sufficiently learn the image features with low overfitting risks. In the standalone unsupervised module (PWLS-ULTRA), we used the pre-learned union of 5 sparsifying transforms to reconstruct images. We set the parameters $\beta = 10^4$ and $\gamma = 25$, and ran 1000 alternations with 5 inner iterations to ensure good performance. In the serial SUPER model, we ran 4 epochs of training when learning the supervised modules (FBPConvNet), and we used the pre-learned union of 5 sparsifying transforms and set the parameters $\beta = 5\times10^3$, $\gamma = 20$ and $\mu = 5\times10^5$ to reconstruct images with 20 alternations and 5 inner iterations for the unsupervised module (PWLS-ULTRA). In the parallel SUPER model, we also ran 4 epochs of training when learning the supervised modules (FBPConvNet), and we used the pre-learned union of 5 sparsifying transforms and set the parameters $\beta = 5\times10^3$ and $\gamma = 20$ to reconstruct images with 20 alternations and 5 inner iterations for the unsupervised module (PWLS-ULTRA). We used the optimal combination parameter $\lambda = 0.3$.

We choose root mean square error (RMSE) and structural similarity index measure (SSIM) to quantitatively evaluate the performance of all reconstruction methods. The RMSE in Hounsfield units (HU) is defined as RMSE = $\sqrt{\sum_{j=1}^{N_p}(\hat{\mathbf{x}}_j - \mathbf{x}_j^*)^2/N_p}$, where $\mathbf{x}_j^*$ is the $j$-th pixel of the reference regular-dose image $\mathbf{x}^*$, $\hat{\mathbf{x}}_j$ is the $j$-th pixel of the reconstructed image $\hat{\mathbf{x}}$, and $N_p$ is the number of pixels.

### 3.3 Results

We conducted experiments on 20 test slices (slice 20, slice 50, slice 100, slice 150, and slice 200 of patients L067, L143, L192, and L310) of the Mayo Clinic data. Table 1 shows the averaged image quality of 20 test images with different methods. From Table 1, we observe that the proposed Structure 1 and Structure 2 significantly improve the image quality compared to the standalone FBPConvNet (supervised module) and the standalone PWLS-ULTRA (unsupervised module). They achieve 2.4 HU and 2.3 HU better average RMSE compared with Serial SUPER while their SSIM is comparable with Serial SUPER. They achieved 0.6 HU and 0.5 HU better average RMSE compared with Parallel SUPER while their SSIM is comparable with Parallel SUPER. Fig. 3 shows the reconstructions of L067 (slice 50) using PWLS-EP, PWLS-ULTRA, FBPConvNet, serial SUPER, parallel SUPER, Structure 1, and Structure 2 along with the references (ground truth). The proposed two structures achieved the lowest RMSE and the zoom-in areas show that they can reconstruct image details better.

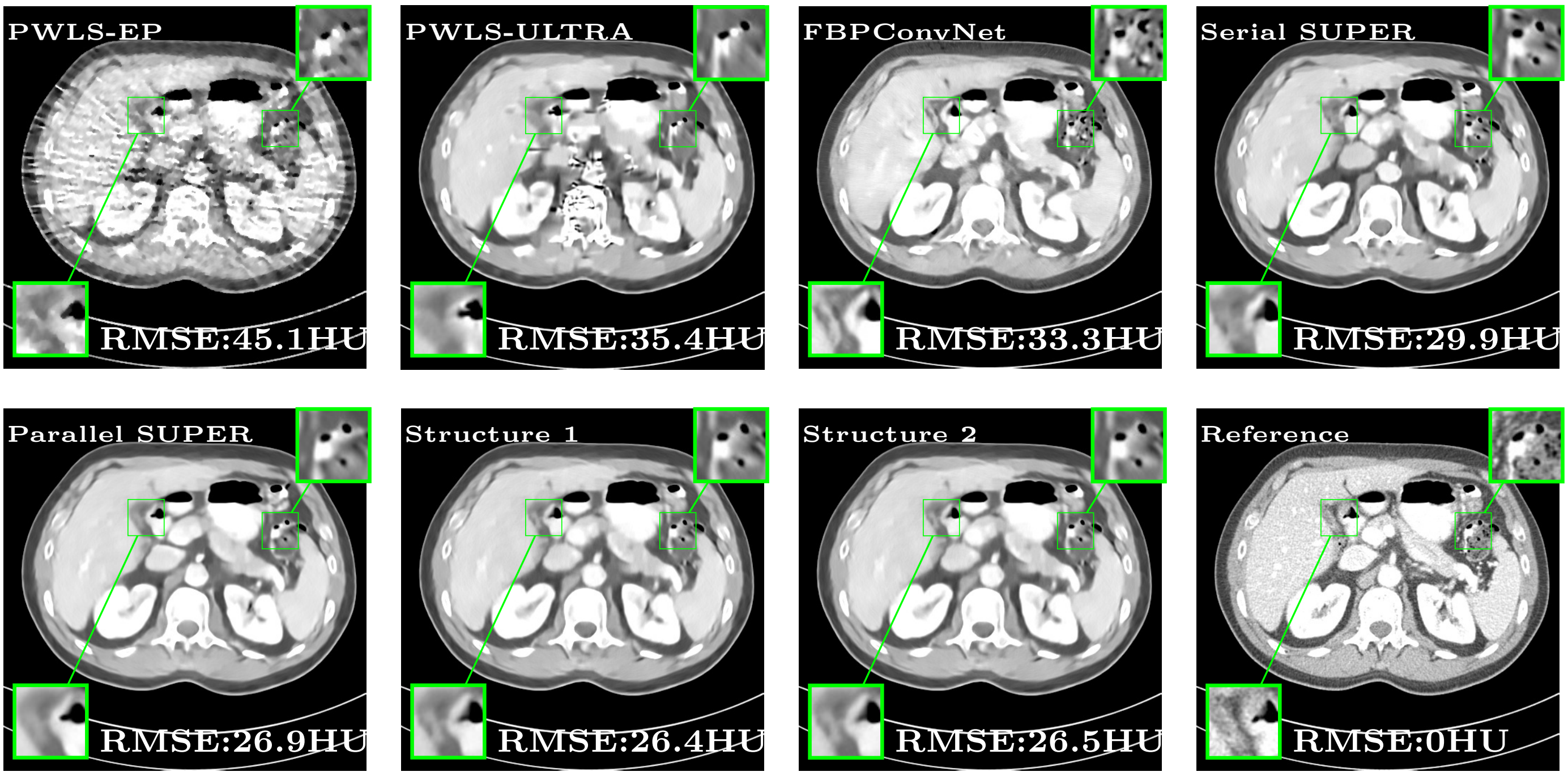

**Figure 3:** Reconstruction of slice 50 from patient L067 using various methods. The display window is [800, 1200] HU.

**Table 1:** Mean RMSE and SSIM of reconstructions of 20 test slices with the PWLS-EP, PWLS-ULTRA, FBPConvNet, Serial SUPER, Parallel SUPER, the proposed Structure 1 and the proposed Structure 2, respectively.

| Method | RMSE (HU) | SSIM |
|---|---|---|
| PWLS-EP | 41.4 | 0.673 |
| PWLS-ULTRA | 32.4 | 0.716 |
| FBPConvNet | 29.2 | 0.688 |
| Serial SUPER | 25.0 | 0.748 |
| Parallel SUPER | 23.2 | 0.751 |
| Structure 1 | **22.6** | 0.745 |
| Structure 2 | **22.7** | 0.747 |

## 4 Conclusion

This paper proposes a general formulation to combine supervised deep learning methods and unsupervised methods for low-dose CT reconstruction. We experiment with two specific integration formats which involve the supervised deep model FBPConvNet and the unsupervised model PWLS-ULTRA. Our framework demonstrates better reconstruction accuracy compared to the individual modules, and the recent serial SUPER and Parallel SUPER methods.

# When iRadonMAP meets Federated Learning: Partially Parameter-sharing Strategy for Robust Low-dose CT Image Reconstruction across Scanners

Shixuan Chen[1,2], Yinda Du[1,2], Boxuan Cao[1,2], Ji He[3], Yaoduo Zhang[3], Zhaoying Bian[1,2], Dong Zeng*[1,2], and Jianhua Ma†[1,2]

[1]School of Biomedical Engineering, Southern Medical University, Guangdong 510515, China
[2]Pazhou Lab (Huangpu) , Guangdong 510000, China
[3]School of Biomedical Engineering, Guangzhou Medical University, Guangdong 511436, China

**Abstract** Image reconstruction from low-dose measurements has been playing an important role in low-dose CT imaging. Deep learning models have been shown to be successful in low-dose CT reconstruction over traditional methods. Recently, a reconstruction framework for Radon inversion with deep learning (i.e., iRadonMAP) is developed to transfer low-dose measurement into CT images directly and efficiently. The iRadonMAP implements the theoretical inverse Radon inversion alongside the image transformation between sinogram and image domais in a network. However, the iRadonMAP is sensitive to protocol-specific perturbations, i.e., geometric model, mAs settings and kVp settings, and it fails to reconstruct the high-fidelity CT images from low-dose measurements across different scanners simultaneously. This might reduce the iRadonMAP generalization. In this work, to address this issue, we constructed a federated learning (FL) framework with iRadonMAP embedded to produce high-quality CT images across different scanners. For simplicity, the proposed framework is denoted as FL-iRadonMAP, i.e., federated learning with iRadonMAP. Specifically, in the FL framework, all clients have normal-dose images/corresponding low-dose sinograms pairs from different scanners to train the iRadonMAP in each local model. Different from traditional FL, the server has a large amount of labeled data, i.e., normal dose/corresponding low-dose CT image pairs, and the labeled data are used to train the global model that enforces the local client to learn different protocol perturbations in each local model. In each client, only the parameters in the image domain of the iRadonMAP are collected and aggregated to the server to obtain global model. We evaluate the presented FL-iRadonMAP on four different CT datasets, and the experimental results confirm that the proposed FL-iRadonMAP has significantly improved performance, preserving the detail texture and suppressing noise-induced artifacts in the CT images across scanners.

## 1 Introduction

Concerns have been raised about the radiation exposure and radiation-induced malignancy risk due to an increasingly large number of CT examinations each year [1]. Methods of reducing radiation dose have been extensively studied. Lowering mAs levels in CT data acquisition protocols is a simple way to reduce radiation dose. However, this might result in an insufficient number of photons detected at detector and hence increase noise-induced artifacts in the final CT images. Various methods have been extensively investigated [2–6]. Among them, the deep learning (DL)-based CT reconstruction methods have been widely developed. Compared with the traditional methods, the DL-based methods not only improve CT image quality but have the advantages of real-time imaging. The DL-based methods can be roughly categorized as model-free methods and model-based methods. The model-free methods learn certain underly mappings completely through end-to-end training in sinogram domain or image domain. It should be noted that the model-free methods have fewer layers than traditional optimization iterations but too many parameters with slow training. In addition, most of these methods have poor generalization with inaccurate solutions. The model-based methods usually combined data consistency and geometry system priors with image features learned from dataset, then unroll an iterative optimization algorithm to a network. For example, He et al. unrolled the model-based iterative reconstruction algorithm in a DL-based framework that had the potential to address both prior knowledge design and parameter selection in an optimization framework [5]. Compared with the model-free methods, the model-based methods relax more constraints and incorporate more domain knowledge in the network to promote the low-dose CT reconstruction performance.

Recently, He et al. developed a reconstruction framework for Radon transformation with deep learning techniques, i.e., iRadonMAP [6]. The architecture of iRadonMAP is similar with FBP algorithm that contains a fully connect filtering layer along the rotation angle direction in the sinogram, a sinusoidal back-projection layer that transforms the filtered sinogram data into the image, and a post processing network to further improve image quality. Meanwhile, it should be noted that the iRadonMAP is geometry system specific, i.e., the iRadonMAP trained on the dataset from one scanner might fails to generalize to new dataset from the other scanner and the iRadonMAP can not be trained on the dataset from different scanner simultaneously due to the different geometry system parameters. Therefore, it is a challenging task for iRadonMAP to reconstruct low-dose CT images from different scanners simultaneously.

Federated learning has become a topic of active research in medical imaging field. The FL enables collaborative training of machine learning models among different organizations and keeping the data private at each local institution, which tackles data heterogeneity from different institutions. Inspired by the FL strategy, we presented a federated learning (FL) framework with iRadonMAP embed-ded (i.e., FL-iRadonMAP) to reconstruct CT images with suppressed noise-induced artifacts across different scanners. Specifically, in the FL framework, there are normal-dose images/corresponding low-dose sinograms pairs from different scanners in each local client, which are used to train the iRadonMAP in each local model. In the server, there has a large amount of high-quality labeled data, i.e., normal dose/corresponding low-dose CT image pairs, which is different from the traditional FL framework. And the labeled data in the server are used to train the global model that enforces the local client to learn different protocol perturbations in each local model. Then, only the parameters in the image domain of the iRadonMAP in each local client are collected and aggregated to the server to obtain global model. The experimental results demonstrate that the presented FL-iRadonMAP is a promising direction to achieve improved CT reconstruction with the iRadonMAP.

*Corresponding author: D. Zeng, zd1989@smu.edu.cn
†Corresponding author: J. Ma, jhma@smu.edu.cn

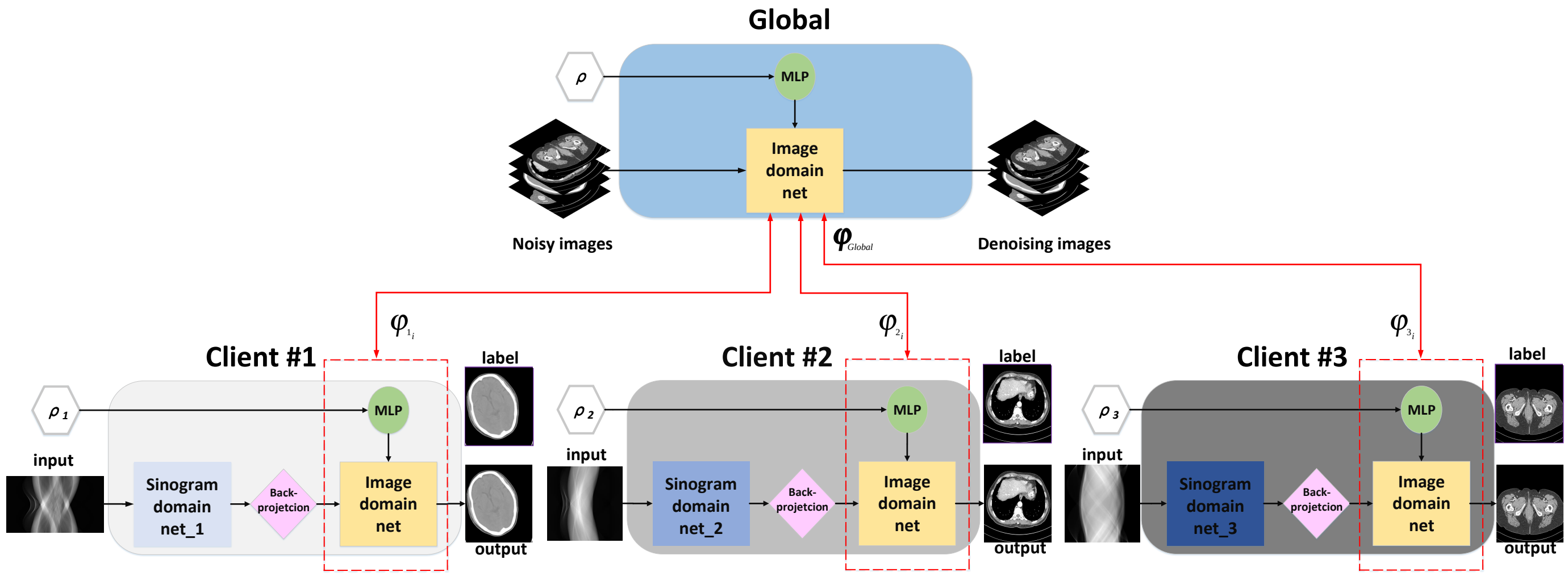

**Figure 1:** The framework of the FL-iRadonMAP.

## 2 Materials and Methods

### 2.1 The iRadonMAP

In previous work [6], the iRadonMAP can be expressed as follows:

$$\mu(i,j) = \sum_{m=1}^{M} \varphi_R \cdot P(n,m)|_{n=INT[i\cos\theta_m + j\sin\theta_m]}, \qquad (1)$$

where, $\mu$ is the reconstructed image, $\varphi_R$ represents the trainable parameters of the back-projection module in iRadonMAP, $P$ is two-dimensional the Radon projection with $\theta$ rotation angle, $m$ is the index of rotation angles, $INT$ denotes the nearest neighbor interpolation, $n$ determines the sinusoid in the sinogram data at a certain location $(i,j)$ in the image domain.

### 2.2 The proposed FL-iRadonMAP

The presented FL-iRadonMAP introduces the iRadonMAP into the FL framework to promote the generalization of iRadonMAP, aiming to reconstruct high-quality CT images across different scanners simultaneously. Specifically, different from the traditional FL framework, the global server in the presented FL-iRadonMAP contains high-quality normal-dose/low-dose CT image pairs to enforce the local client to learn different protocol perturbations in each local model. And the training dataset in each local model are low-dose sinogram data/normal-dose CT images that are from different scanners, which exist great heterogeneity among all the local clients.

#### 2.2.1 Local model training

The loss function in each local client can be expressed as follows:

$$\mathscr{L}(\mu,\mu^*) = \|\mu-\mu^*\|_2^2, \qquad (2)$$

where $\mu^*$ is the reference image, $\|\cdot\|_2^2$ is L2 norm.
In the $t$th round of training the parameters of the $k$th $(k \in L)$ local client can be updated as follows:

$$\boldsymbol{\varphi}_k^{t+1} = \boldsymbol{\varphi}_k^t - \sigma_k \Delta \mathscr{L}^t(\mu,\mu^*), \qquad (3)$$

where $\boldsymbol{\varphi}_k^t(\varphi_p, \varphi_R, \varphi_i)$ represents the parameters at $k$th local model in the $t$th round training, $\varphi_p$ is the parameters of sinogram domain sub-network, $\varphi_i$ is the parameters of image domain sub-network, $\sigma_k$ is the learning rate of client $k$.
It should be noted that the dataset in each client is acquired with different protocols and scanners, which indicates great variations among distribution of the different dataset. In the presented FL-iRadonMAP, the local client trains the iRadonMAP with its own dataset, respectively. Moreover, to preserve the feature specificity of each local model, we collect only the model parameters $\varphi_i$ in the image domain and aggregate to the global model in server for update. Then, the

| | Client #1 | Client #2 | Client #3 |
|---|---|---|---|
| Number of projection angles | 896 | 1152 | 640 |
| Number of detector bins | 1008 | 1104 | 1024 |
| Length of a detector bin(mm) | 0.5480 | 0.5000 | 0.6500 |
| Length of a voxel (mm) | 0.7421 | 0.6641 | 0.7500 |
| Distance between source and detector (mm) | 800.0 | 1085.6 | 750.1 |
| Distance between source and rotation center (mm) | 550.0 | 595.0 | 476.3 |
| Voltage of tube(kVp) | 80 | 140 | 120 |
| Intensity of X-rays | $5\times10^5$ | $1\times10^5$ | $1\times10^5$ |

**Table 1:** Imaging geometries of clients.

update of global model parameters $\boldsymbol{\varphi}_{global}$ can be expressed as follows:

$$\boldsymbol{\varphi}_{global}^{t+1} = \frac{1}{L}\sum_{k=1}^{L} \omega_k \varphi_{k_i}^t, \tag{4}$$

where $\omega_k$ is the weight of $k$th local model in the aggregation. It should be noted that the global model in the server only takes into account the characteristics of the CT images in the local client, but not the characteristics of the sinogram in the local client. This can process the sinogram data across the different scanners in each local client simultaneously within the iRadonMAP, which can promote the generalization of the iRadonMAP and tackle data heterogeneity efficiently.

#### 2.2.2 Global model training

To further improve the iRadonMAP reconstruction performance of the local model, we also collect high-quality normal-dose/low-dose CT image pairs to construct a central dataset in the global server, which is different from the traditional FL framework. The traditional FL only utilizes the global server to aggregate and distribute the corresponding parameters from the network in the local client. In the presented FL-iRadonMAP, the model in the global server can be iteratively updated by the high-quality labeled dataset in each training round, and then used for parameters update in the client. Then Equation 4 can be updated as follows:

$$\boldsymbol{\varphi}_{global}^{t+1} = \frac{1}{L+1}\left(\sum_{k=1}^{L} \omega_k \varphi_{k_i}^t + \omega_{global}\boldsymbol{\varphi}_{global}^t\right), \tag{5}$$

where $\omega_{global}$ is the weight of the global model in parameter aggregation.

## 3 Experiments

### 3.1 Network Architectures

Figure 1 shows the presented FL-iRadonMAP framework. The iRadonMAP in each local client is a modified one in [6] wherein both the sub-network in the sinogram domain and the sub-network in the image domain are composed of convolution residual blocks, and we use one-dimensional convolutional kernels with the size $1\times3$ in the sub-network in the sinogram domain, and the kernels with the size of $3\times3$ in the image domain. To consider the heterogeneity among the dataset in each client from different protocols and scanners, in this work, we utilize the multilayer perceptron (MLP) to map the protocols and geometry parameters into high-dimensional vector $\rho$ as the network input, which represents the source of the CT images, i.e., acquisition protocols and scanners. These vectors can modulate the output feature maps as follows:

$$\widehat{y} = h_1(\rho)y + h_2(\rho), \tag{6}$$

where $y$ is the output feature map of the network layer. $\widehat{y}$ is the modulated feature map. $h_1$, $h_2$ represent the MLP vectors.

### 3.2 Dataset

To validate and evaluate the reconstruction performance of the presented FL-iRadonMAP, with the approval of the Local Institute Research Medical Ethics Committee, four different dataset are collected in the experiment. 100,000 CT images data including head, abdomen and phantom at different kVp, mAs are collected in the global server. The other three CT datasets from different scanners are used in three local clients to represent the heterogeneity among different clients. Specifically, Client #1 contains 1000 head images, Client #2 contains 500 phantom images, and Client #3 contains 1000 abdomen images. To obtain corresponding low-dose measurements, we simulated low-dose CT images/sinogram data from the normal-dose ones based on the previous study [7]. Table 1 lists the geometry parameters and dose levels in the local dataset in the simulation study.

### 3.3 Compared methods and implementation details

In this work, FBP, FedAvg, and iRadonMAP are used as competitive methods. In the presented FL-iRadonMAP the weight of global model and three local client model at parameter aggregation (i.e., $\omega_{global}, \omega_1, \omega_2, \omega_3$) is set to $\frac{7}{10}, \frac{1}{10}, \frac{1}{10}, \frac{1}{10}$, respectively. And the weight of each local client in FedAvg is set to $\frac{1}{3}, \frac{1}{3}, \frac{1}{3}$, respectively. The learning rate of all models is set to $2\times10^{-5}$, and optimized with the RMSProp optimizer. Peak signal-to-noise ratio (PSNR), structural similarity index (SSIM) and root mean square error (RMSE) measures are used to evaluate the reconstruction performance of each competing method.

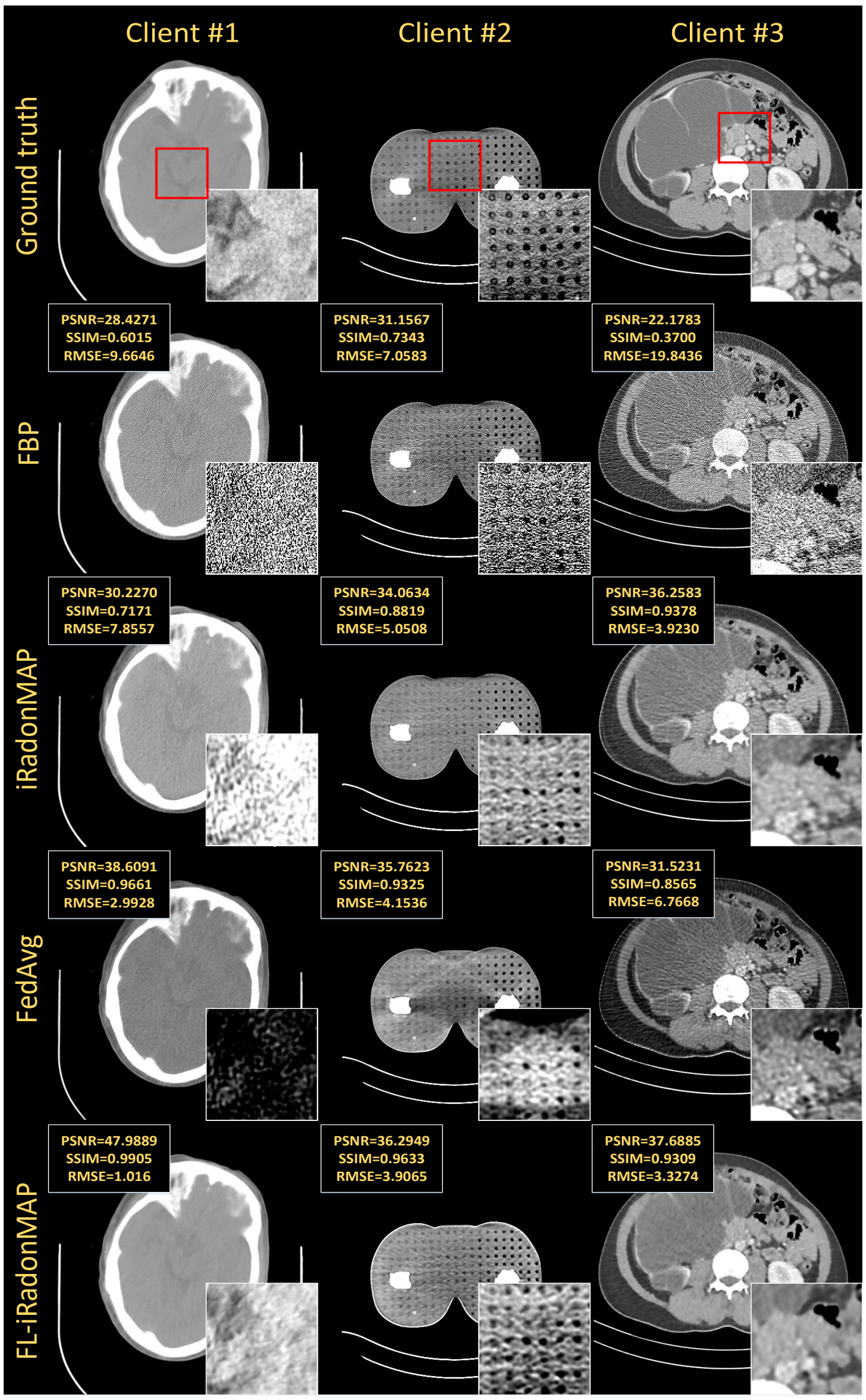

**Figure 2:** Results of FL-iRadonMAP and compared methods.

## 4 Results

Figure 2 shows the qualitative visualization of representative results of the competing methods on three different cases. The normal-dose FBP-reconstructed images are served as the ground-truth. The ROI indicated by the red boxes are zoomed-in for better comparison. From Figure 2, it can be observed that the low-dose FBP-reconstructed images contain severe noise-induced artifacts. The iRadonMAP can effectively remove the noise-induced artifacts but it would result in the loss of structure details as shown in the zoomed-in ROIs. The FedAvg does not work well in resolution preservation since the global server in the FedAvg only aggregates and distributes the parameters during the training process, but fails to consider the variations among the dataset in each client. This might lead to undesired reconstruction results due to the data heterogeneity across scanners. The presented FL-iRadonMAP yields reconstructed CT images with remarkable visual similarity to the ground truth compared to the other competing methods in three datasets with diverse characteristics across scanners. The possible reason is that the dataset in the global server would help tackle data heterogeneity across scanners and promote the reconstruction performance in each client. Moreover, FL-iRadonMAP obtains the best quantitative measurements at all the cases, which is consistent with the visual inspection.

## 5 Conclusion

In this work, we present a FL-iRadonMAP to tackle data heterogeneity among the datasets due to the different protocols and scanners and promote the generalization of the iRadonMAP in the reconstruction tasks. Through extensive experiments on datasets with diverse characteristics, it is demonstrated that the presented FL-iRadonMAP can obtain promising reconstruction results across scanners.

## Acknowledgments

This work was supported in part by the NSFC under Grant U21A6005, and Grant 12226004, and Young Talent Support Project of Guangzhou Association for Science and Technology.

# Learning a Dual-Domain Harmonization Network for Low-dose CT Image Reconstruction across Scanner Changes

Shixuan Chen[1,2], Yinda Du[1,2], Boxuan Cao[1,2], Shengwang Peng[1,2], Ji He[3], Yaoduo Zhang[3], Zhaoying Bian[1,2], Dong Zeng*[1,2], and Jianhua Ma†[1,2]

[1]School of Biomedical Engineering, Southern Medical University, Guangdong 510515, China
[2]Pazhou Lab (Huangpu) , Guangdong 510000, China
[3]School of Biomedical Engineering, Guangzhou Medical University, Guangdong 511436, China

**Abstract** Deep learning (DL) have shown great potential in the low-dose CT imaging field. Existing works show that DL-based CT denosing/reconstruction methods can obtain high-quality CT images at low-dose cases, and outperform traditional model-based methods. However, existing DL-based methods are generally designed based on the dataset from one site, hindering their application to the new dataset from other site with limited generalization performance. The main reason is that there exists variation across scanners, acquisition protocols and patient populations. This could lead to data heterogeneity among different centers and low efficiency in downstream task. To address this issue, in this work, we present a Dual Domain Harmonization Network (DuDoHNet) with consideration of the filter kernel style in the sinogram domain and image style in the image domain. Specifically, multilayer perceptron (MLP) is introduced to characterize difference between the source domain and target domain and PatchGAN is used to model the semantic information between the source domain and target domain. Finally, the presented DuDoHNet allows for reconstructing high-fidelity CT images to reduce confounding data variation and preserve semantic information from different scanners. Experiments on the CT dataset with four different filter kernels from two scanners demonstrate that the presented DuDoHNet outperforms the image-based harmonization networks qualitatively and quantitatively.

## 1 Introduction

X-ray Computed tomography is widely used in the clinics for disease diagnosis and treatment planning. However, it is well-known that ionizing radiation in the CT is harmful to human, which might cause cancerous diseases. Then the as low as reasonable available (ALARA) principle should be followed in the radiology field. However, lowering the X-ray flux by decreasing tube current and exposure time directly would lead to poor CT image quality. Extensive efforts have been made to develop an efficient image denosing/reconstruction method for the clinical used of low-dose CT. Among them, deep learning (DL)-based methods have shown promising results compared to the conventional model-based methods. These DL-based methods leverage large CT image datasets to learn better image representations and produce better image reconstruction results.

Although the DL-based methods offer significant improvement in image quality, it should be noted that most of them are trained with historical dataset and then reconstruct the new unseen dataset, which might lead to limited generalization. The main reason is that there exists heterogeneity of CT dataset among sites. The variation of CT dataset might come from different scanners, different protocols, and different patient population. Pooling data across scanners and sites leads to undesirable increase in non-biological variance [1]. For example, the scanner-induced biases can cause incoherence among the characteristics in the CT images, and the differences in protocol settings can lead to appearance variation of image type. Therefore, it is critical to develop effective harmonization method to reduce data variation and reconstruct high-fidelity CT images from different scanners and protocols.

Various methods have been developed to harmonize/transform CT dataset with varied settings. For example, Yao et al. used a generative model to simultaneously generate energy-resolving CT images at multiple energy bins from existing energy-integrating CT images [2]. D. Kawahara et al. constructed a deep convolutional adversarial network for synthesizing a dual-energy CT image from an equivalent kilovoltage CT image [3]. S. Charyyev et al. presented a residual attention generative adversarial network to synthesize dual-energy CT images from single-energy CT image [4]. These methods take into account the data heterogeneity between different scanning protocols, enable CT images synthesis and harmonization. However, the disadvantage of these methods is that they do not consider the influence of filter kernels on image reconstruction, and there is room for improvement. And then, some DL-based CT kernel transformation methods were proposed to solve this problem. For instance, S. M. Lee et al. developed a convolutional neural network to transform CT images reconstructed with one CT kernel to images with different reconstruction CT filter kernels [5]. Kim et al. reconstructed from different manufacturers through a routable network [6]. These methods consider variability of images with different filter kernels and could achieve image style transformation and uniformity. But the drawback is that they do not mention the influence of scanning protocols on image quality, and all these methods process the data in image domain which lack in the consideration of the features of sinogram domain and result in the limitation of the accuracy of harmonization/transformation.

In this paper, we propose a Dual Domain Harmonization Network (DuDoHNet) by considering the filter kernel style in the sinogram domain and image style in the image domain. Specifically, the presented DuDoHNet can obtain desired sinogram data with well-generated filter kernel via the network, and then reconstruct the final CT images with targeted style. In addition, multilayer perceptron (MLP) is introduced to characterize difference between the source domain and target domain and PatchGAN is used to model the semantic information between the source domain and target domain. Therefore, the presented DuDoHNet can reconstruct high-fidelity CT images to reduce confounding data variation and preserve semantic information from different scanners. Experiments show that DuDoHNet enable to reconstruct and harmonize CT images with various filter kernels from different scanners.

*Corresponding author: D. Zeng, zd1989@smu.edu.cn
†Corresponding author: J. Ma, jhma@smu.edu.cn

## 2 Materials and Methods

The CT image reconstruction procedure can be expressed as follows:

$$I = \mathscr{F}^i\{\mathscr{R}^{-1}[\mathscr{F}^p(P)]\}, \tag{1}$$

where $P$ is the sinogram data, $\mathscr{F}^p$, $\mathscr{F}^i$ are the filtering operation in the sinogram domain and image domain, respectively. $\mathscr{R}^{-1}$ is inverse Radon transformation operator.

To harmonize CT images captured by different scanners, in this work, we construct a deep learning network for CT image harmonization. For example, the measurements (i.e., sinogram) $P_\mathsf{S}$ from scanner $A$ (i.e., source domain $\mathsf{S}$) are passed through the network to reconstruct the CT images $I_\mathsf{S}^\mathsf{T}$ with the style from scanner $B$ (i.e., target domain $\mathsf{T}$). And the expression can be defined as follows:

$$I_\mathsf{S}^\mathsf{T} = \mathbf{M}^i\{\mathscr{R}^{-1}[\mathbf{M}^p(P_\mathsf{S}|\theta^p)|\theta^i]\}, \tag{2}$$

where $\mathbf{M}^p$ and $\mathbf{M}^i$ represent the sub-networks in the sinogram domain and image domain, respectively. $\theta^p$ and $\theta^i$ are the corresponding network parameters. $P_\mathsf{S}$ is sinogram processed by the filter kernel from the source domain, which is $\mathscr{F}_A^p(P_A)$ in Equation 1. Then, we define the dual-domain generator $\mathbf{M}_G$ as the combination of $\mathbf{M}^p$ and $\mathbf{M}^i$ wherein the $\mathbf{M}_G$ can reconstruct sinogram from source domain into the final CT image with style from the target domain. Then Equation 1, 2 can be rewritten as follows:

$$I_\mathsf{S}^\mathsf{T} = \mathbf{M}_G(P_\mathsf{S}|\theta_G). \tag{3}$$

Due to the unpaired data between the source domain and target domain, in this study, WGAN [7] discriminator is introduced, and the loss function can be expressed as follows:

$$L_D = \underset{I_\mathsf{S}^\mathsf{T}\sim P_{\tilde{\mathsf{T}}}}{E}[\mathbf{M}_D(I_\mathsf{S}^\mathsf{T})] - \underset{I_\mathsf{S}\sim P_\mathsf{T}}{E}[\mathbf{M}_D(I_\mathsf{S})], \tag{4}$$

where $P_\mathsf{T}$ is the distribution of the target domain style image, $P_{\tilde{\mathsf{T}}}$ is the distribution of the desired reconstructed image. The loss function of the generator $G$ can be defined as follows:

$$L_G = -\mathbf{M}_D(I_\mathsf{S}^\mathsf{T}). \tag{5}$$

It should be noted that the generative adversarial loss of Equation 5 is unsupervised. To make the network robust, we also introduce a self-transformation loss function to produce the CT images with the style from the source domain (i.e., source-to-source reconstruction), and it is defined as follows:

$$L_{self} = \begin{cases} \|I_\mathsf{S}^\mathsf{S} - I_\mathsf{S}\|_2^2, & while\ self\ transformation \\ 0 & others \end{cases}, \tag{6}$$

where $I_\mathsf{S}^\mathsf{S}$ represents an image with the style from the source domain.

Moreover, we also would like to produce the different filter kernels efficient in the sinogram domain that can produce the CT images with different styles, which can be defined as follows:

$$L_{kernel} = \|P_\mathsf{S} - P_\mathsf{S}^\mathsf{T}\|_1, \tag{7}$$

where $P_\mathsf{S}^\mathsf{T}$ is the sinogram data from the source domain that would be filtered by the filter kernel from the target domain. Therefore, the total loss function can be expressed as follows:

$$L_{total} = \lambda_{GAN}(L_D + L_G) + \lambda_{self}L_{self} + \lambda_{kernel}L_{kernel}, \tag{8}$$

where $\lambda_{GAN}$, $\lambda_{self}$ and $\lambda_{kernel}$ are parameters.

## 3 Experiments

### 3.1 Network Architectures

Figure 1 shows the framework of the presented DuDoHNet. The proposed DuDoHNet consists of a dual-domain generator and a discriminator. The generator network contains sino-network, back-projection module, image-network, and multilayer perceptron (MLP). The sino-network is composed of 8 convolutional residual blocks with channel attention, and the size of convolutional kernel is set to $1\times3$. The back-projection module allows for efficiently processing the gradients. The image-network is composed of 8 convolutional residual blocks with spatial attention, and the size of convolutional kernel is set to $3\times3$. The discriminator is similar with the patchGAN [8]. To consider the geometry parameters from the different scanners, MLP is used to characterize the scanners as follows [9]:

$$v = [f_1, \cdots, f_n], \tag{9}$$

where $v$ is normalized vector, where $f_1, \cdots, f_n$ are normalized CT imaging geometry parameters, the details are shown in Table 1. Furthermore, we concatenate the vector $v_\mathsf{S}$ and the vector $v_\mathsf{T}$ of target domain. The parameter $\rho$ can be defined as follows:

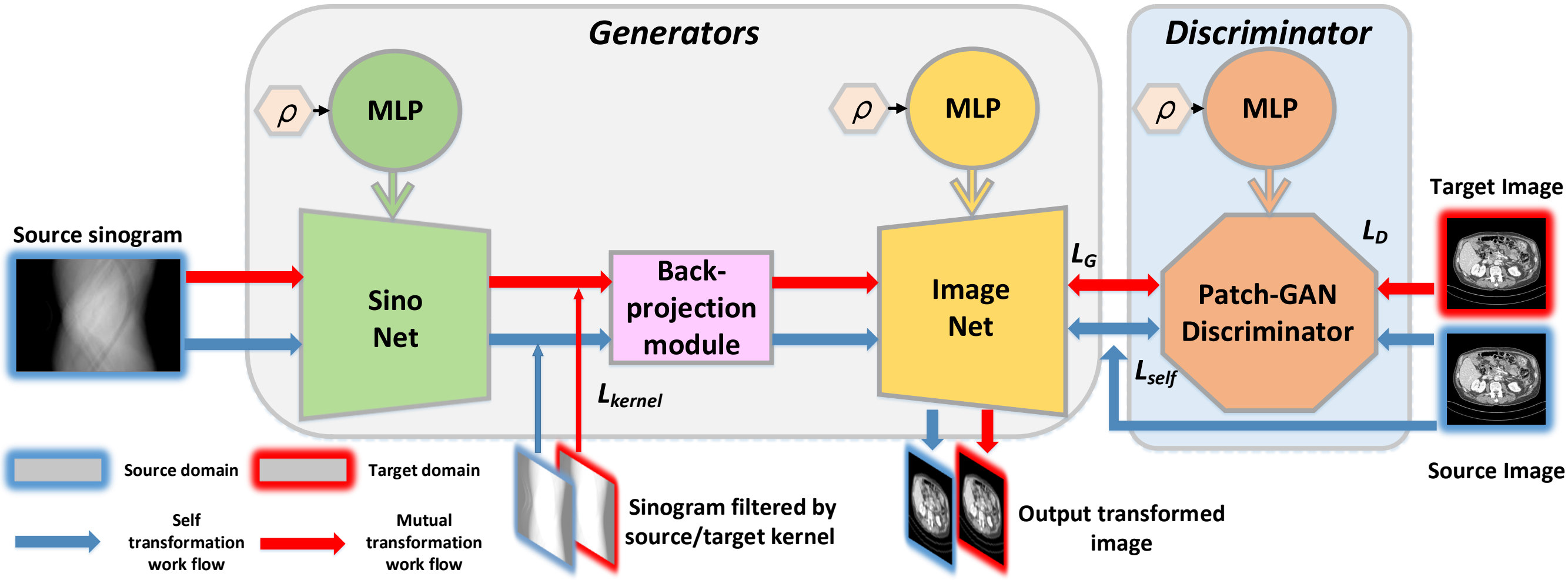

**Figure 1:** Framework of the proppsed DuDoHNet. The blue arrows represent the source-to-source reconstruction procedure, and the red arrows represent the source-to-target reconstruction.

$$\rho = [v_S, v_T]. \tag{10}$$

As shown in Figure 1, this vector is regarded as a part of the network input, and is mapped by three multilayer perceptrons(MLPs) into a high-dimensional modulation vector that modulates the feature map of the middle layer of the network [10].

## 3.2 Dataset

To validate and evaluate the reconstruction performance of the presented DuDoHNet, we collected abdominal CT images from two different scanners, i.e., Scanner #1 and Scanner #2, wherein 1200 images are from Scanner #1 and 1350 images are from Scanner #2. Then we simulated low-dose sinogram data from the two datasets with corresponding geometry parameters. The geometry parameters are listed in Table 1. And then we reconstruct the low-dose siongram from Scanner #1 into CT image with two different filter kernels (i.e., Kernel #1 and Kernel #2), respectively. We reconstruct the low-dose siongram from Scanner #2 into CT image with another two different filter kernels (i.e., Kernel #3 and Kernel #4), respectively. Figure 2 shows the four CT images from two scanners reconstructed with different filter kernels. 90% of the dataset are used for network training, and the remaining are used for experimental validation.

## 3.3 Competing methods and implementation details

We compared the presented DuDoHNet with GAN and cycleGAN that are post-processing methods. Gray-level run-length matrix (GLRLM) is an important feature to measure the higher-order texture features between the reconstructed CT images and the images from target domain, and the consistency correlation coefficient (CCC) of GLRLM is a quantitative harmonization measure. We constructed all models with Pytorch [11] with Adam optimizer, and the learning rate is set to $2 \times 10^{-5}$.

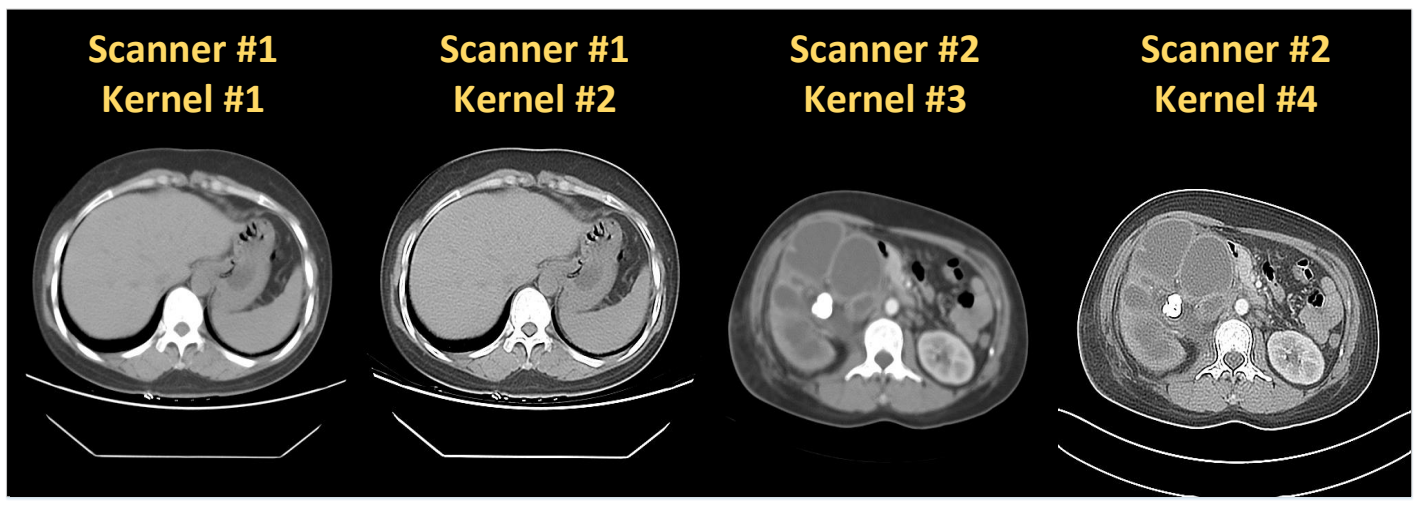

**Figure 2:** Examples from the dataset with different filter kernels from two scanners.

| | Scanner #1 | Scanner #2 |
|---|---|---|
| Number of projection angles | 1152 | 896 |
| Number of detector bins | 1100 | 1008 |
| Length of a detector bin(mm) | 0.5000 | 0.5480 |
| Voltage of tube(kVp) | 120 | 120 |

**Table 1:** Geometry parameters from two scanners used in the simulation study.

## 4 Results

Figure 3 shows the CT images with four filter kernels from two scanners reconstructed by the presented DuDoHNet. It can be seen that the DuDoHNet can produce the CT images from source domain to source domain which are close to those in Figure 2. And the DuDoHNet can produce CT images from source domain to target domain with markedly uniform in terms of structure contrast, over and regional intensity, and noise patterns. For example, from source domain (Scanner #1, Kernel #2) to target domain (Scanner #2, Kernel #3), both the reconstructed CT image (c2) and reconstructed CT image (C3) have similar style. Moreover, CCC measure-ments also indicated that the harmonized CT images are more similar to the reference domain.

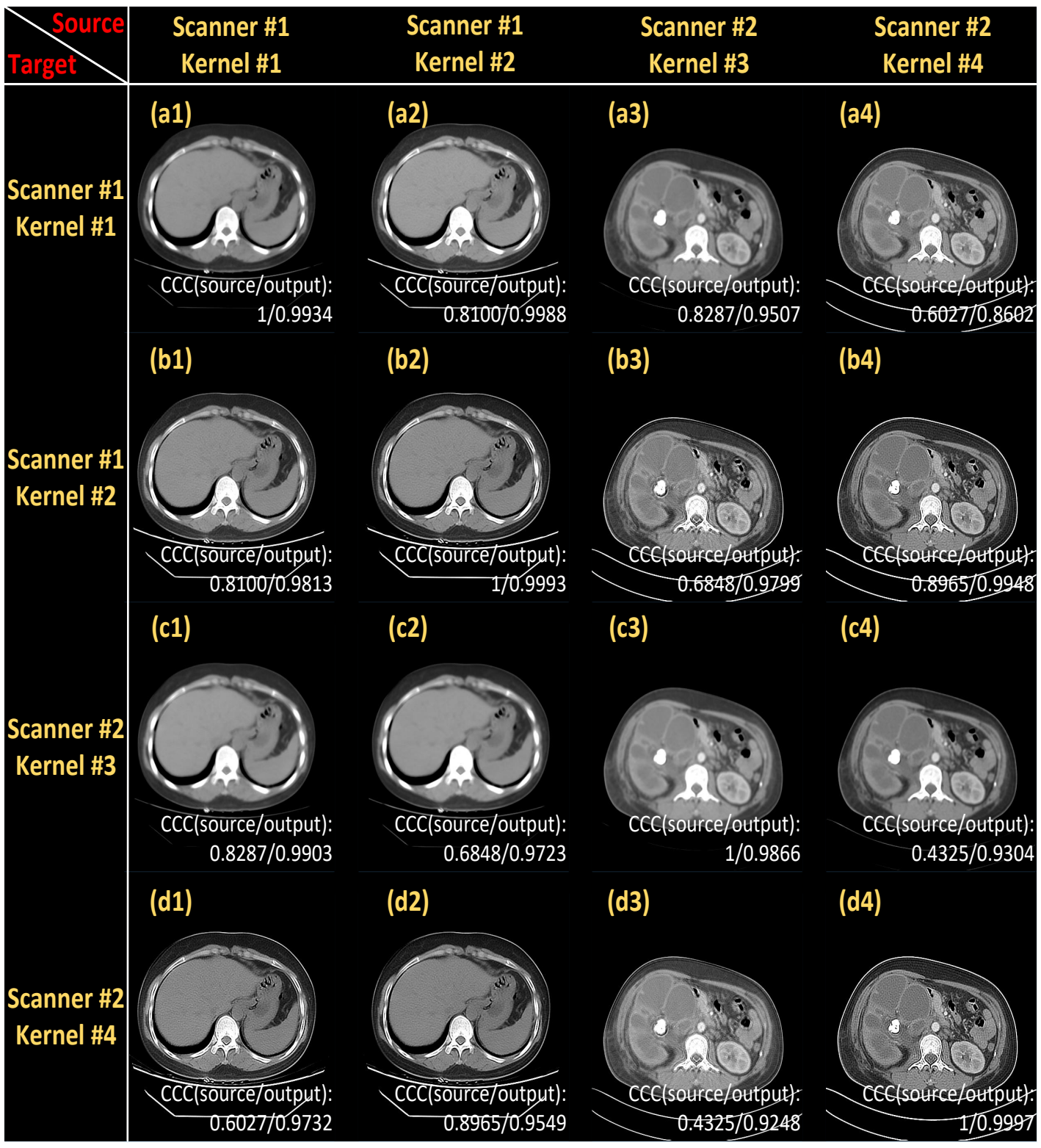

**Figure 3:** Reconstruction results of the presented DuDoHNet.

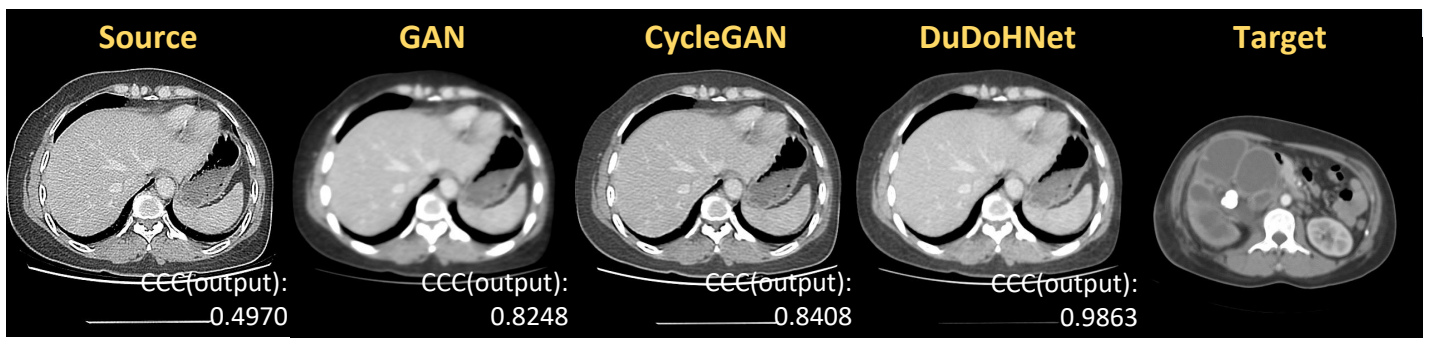

**Figure 4:** Reconstruction results of the competing DuDoHNet.

Figure 4 shows the CT images with kernel #3 reconstructed by the different methods (i.e., from Scanner #1 Kernel #2 to Scanner #2 Kernel #3). It can be observed that the presented DuDoHnet can produce CT images with the style close to the target domain. And the presented DuDoHNet obtains the best CCC performance among all, which is consistent with the visual inspection.

## 5 Conclusion

In this work, we propose a dual domain harmonization network for low-dose reconstruction across scanner changes. Unlike traditional post-processing-based methods, we consider the characteristics in the sinogram to promote the harmonization performance. The experimental results show that the presented DuDoHNet can concurrently model and correct for the site effects across scanner changes while retaining predictive information within scans. This can help facilitate downstream application.

## Acknowledgments

This work was supported in part by the NSFC under Grant U21A6005, and Grant 12226004, and Young Talent Support Project of Guangzhou Association for Science and Technology.

# Protocol Variation Network for Low-dose CT image denoising with Different kVp Settings and Anatomical Regions

Shixuan Chen[1,2], Yinda Du[1,2], Boxuan Cao[1,2], Shengwang Peng[1,2], Zhaoying Bian[1,2], Dong Zeng*[1,2], and Jianhua Ma†[1,2]

[1]School of Biomedical Engineering, Southern Medical University, Guangdong 510515, China
[2]Pazhou Lab (Huangpu) , Guangdong 510000, China

**Abstract** Computed tomography (CT) is a useful diagnostic tool for diseases and injuries detection in clinics. With an increase in the utilization of CT examinations, there is a concern about the general population's radiation exposure. Deep learning (DL)-based methods are efficient to lower radiation dose without sacrificing CT image quality, and have potential to achieve great reconstruction performance. However, these DL-based methods are sensitive to CT protocol selection, i.e., kVp settings and anatomy to be examined. In this work, we propose a protocol variation network (PV-Net) for low-dose CT image denoising to consider the heterogeneity within the different kVp settings and anatomical structures. Specifically, the proposed PV-Net models the characteristics of the CT data, i.e., noise distribution and anatomical structures in the network. Experimental results on the CT datasets demonstrate that the proposed PV-Net leads to improved reconstruction performance with kVp and anatomical structure variations, compared with the other competing methods.

## 1 Introduction

X-ray CT is widely used for various clinical applications due to its advantages of non-invasive, sectional imaging. The potential harmful effects of radiation dose in CT imaging have raised growing concerns, and then low-dose CT imaging with clinically acceptable image quality is desirable. The simple way to lower radiation dose is to reduce mAs or angular sampling per rotation. Meanwhile, these two strategies would lead to severe noise-induced artifacts or view-aliasing artifacts in the filtered back-projection (FBP) reconstructed images. To improve low-dose CT image quality, various methods have been developed. Among them, the deep learning (DL)-based methods have shown excellent reconstruction performance due to their powerful ability to learn deep features and high computational efficiency.

Althouth the existing DL-based methods have potential to suppress artifacts in the CT images, they are usually trained based on the specific dataset acquired with one protocol and could be sensitive to unseen perturbations, i.e., the dataset from the different protocol. This leads to limited generalization and robustness of these DL-based methods. They might fail to process the dataset with large heterogeneity, which degrades the reconstruction performance. Recently Xia et al. [1] proposed a framework for modulating DL model with CT imaging geometric parameters (i.e., PDF) due to the geometry variations, and obtain improved performance. Meanwhile, it should be noted that the crucial parameters in the protocol selection, i.e., kVp settings and anatomical structures, also play a vital role in the final CT image, which can also affects the reconstruction performance of the DL-based methods [2]. In this paper, to consider the protocol variations (i.e., different kVp settings and anatomical structures) in the network, we propose a protocol variation network (PV-Net) for low-dose CT image denoising. The proposed PV-Net takes into account the heterogeneity with the different kVp settings and anatomical structures. Specifically, to model the protocol variations, the proposed PV-Net introduces the characteristics of the CT data, i.e., noise distribution and anatomical structures into the network with well-designed loss function. Using simulation studies based on the CT dataset acquired with different protocols, we compare the proposed PV-Net with conventional FBP, protocol specific network (i.e., PS-Net), and PDF. Experimental results suggest that the proposed PV-Net outperforms the other competing methods, with improved reconstruction performance at the cases of kVp and anatomical structure variations, making the proposed PV-Net a potentially important strategy to reconstruct CT images with protocol variations.

## 2 Materials and Methods

### 2.1 PV-Net

Figure 1 shows the architecture of the proposed PV-Net. The network inputs are the CT images acquired with different protocols, i.e., kVp settings and anatomical regions. Inspired by Xia et al. [1], to modulate the feature maps with different geometries and protocols efficiently in the network, geometry parameters and scanning protocols are parameterized by a normalized vector $v$ as another input of the proposed PV-Net. Then, the proposed PV-Net can be expressed as follows:

$$y = M(x, v|\theta), \tag{1}$$

where $M$ represents the presented PV-Net model, and $\theta$ represents the learnable parameters of $M$. $x, y$ are the input and the output image of the model, respectively. $v = [\rho_1, \cdots, \rho_n]$, $\rho$ is parameterized geometry and protocol in specific CT imaging task. The vector $v$ is mapped by the multilayer perceptron (MLP) to high-dimensional vectors, which modulates the output feature map of the layer with different geometries and protocols efficiently in the network linearly. It can be expressed as follows:

*Corresponding author: D. Zeng, zd1989@smu.edu.cn
†Corresponding author: J. Ma, jhma@smu.edu.cn

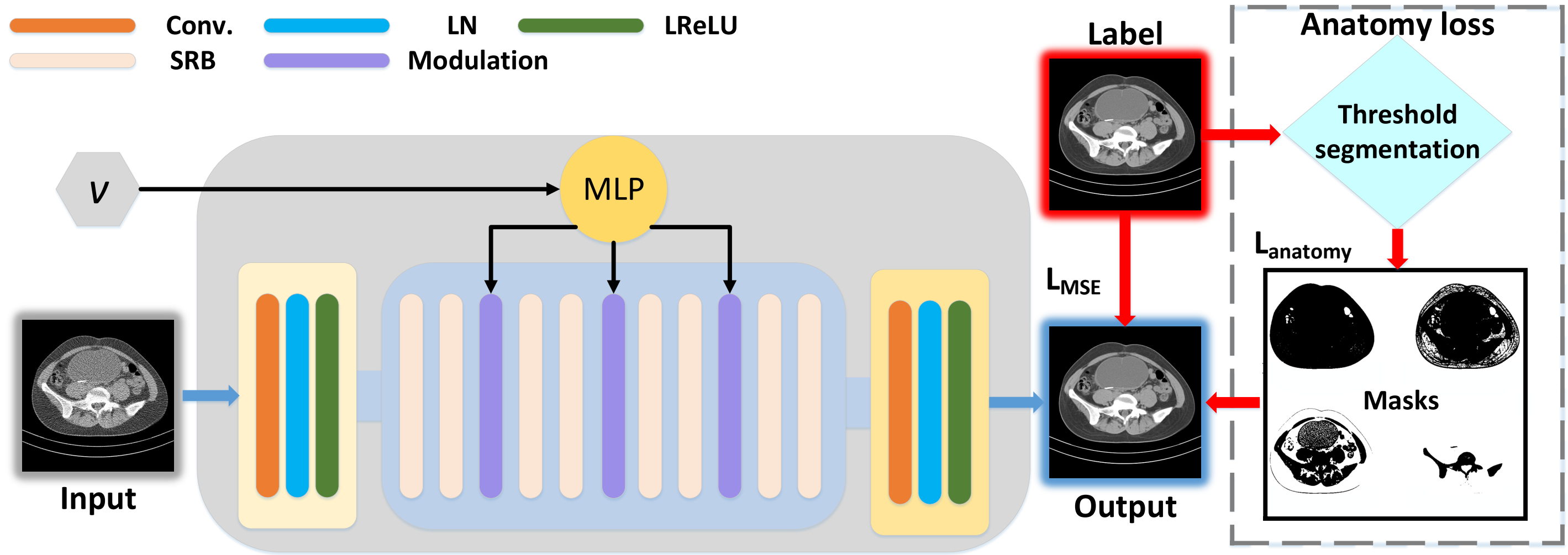

**Figure 1:** Network architecture of the proposed PV-Net.

$$\widehat{f} = h_1(v)f + h_2(v), \tag{2}$$

where $h_1$ and $h_2$ are two different MLP, $f$ is the output feature map of model's layers and $\widehat{f}$ is the modulated feature map. As shown in Figure 1, the firsrt and last part of the network are constructed with a convolutional layer (Conv.), a layer normalization layer (LN) and LeakyReLU (LReLU). The middle of the network is a series of spatial-attention residual blocks (SRB)[3]. The vector $v$ is fed into the MLP and mapped to high-dimensional vectors, then the high-dimensional vectors are transferred into the modulation layers (purple ones) as indicated by Equation 2.

**Algorithm 1:** Anatomy loss function

**Input:** Label image $\mu$, output image of model $\mu^*$, number of segmentation regions $M$
**Output:** Anatomy loss $\mathscr{L}_{anatomy}$
**for** $m$=1:$M$ **do**
  $Mask_m = Threshold\ Segmentation(\mu)$;
  Calculate $\mathbf{S}_m$ and $\mathbf{K}_m$ according to Equation 4, 5;
**end**
**for** $m$=1:$M$ **do**
  Calculate $\mathbf{S}^*_m$ and $\mathbf{K}^*_m$ in $Mask_m(\mu^*)$ according to Equation 4, 5;
**end**
Calculate $\widetilde{\mathbf{S}}$, $\widetilde{\mathbf{S}}^*$, $\widetilde{\mathbf{K}}$ and $\widetilde{\mathbf{K}}^*$ according to Equation 6;
Calculate $\mathscr{L}_{anatomy}$ according to Equation 7;
**return** $\mathscr{L}_{anatomy}$

## 2.2 kVp modulation

It should be noted that same type of CT images but a different kVp setting would affect the DL-based reconstruction performance. The main reason can be attributed to the noise distribution shift between different kVp settings. For example, the noise level on the CT images at lower kVp is higher than that at higher kVp. Therefore, to consider the kVp settings in the proposed PV-Net, the kVp settings can be parameterized as follows:

$$v = [\rho_1, \cdots, \rho_n, \rho_{kVp}], \tag{3}$$

This modulation can guide the network to learn the deep features of CT data at different kVp efficiently and adaptively, and then reconstruct the final CT images with the vector $v$.

## 2.3 Anatomical region modulation

Since different anatomical regions have different anatomical structures and pathological indications, the DL-based methods might have different reconstruction performance. Moreover, it is difficult to modulate different anatomical regions with vector $v$ in Equation 2. Then, to model the characteristics variations among different anatomical regions, in this work, we take advantage of statistical features of the anatomical regions that describe the noise distribution shift among different protocols [4]. In particular, we performed a rough segmentation on the labeled images according to the threshold values in Table 2. Then the statistical features of different anatomical regions are calculated, i.e., skewness and kurtosis, which can be used to model anatomical region variations.

The skewness of different tissues can be obtained through Fisher-Pearson skewness statistic as follows:

$$\mathbf{S} = \frac{n}{(n-1)(n-2)} \frac{\sum_{i=1}^{n}(x_i - \bar{x})^3}{(\frac{1}{n-1}\sum_{i=1}^{n}(x_i - \bar{x})^2)^{\frac{3}{2}}}, \tag{4}$$

where $\bar{x}$ is the mean value of the desired region, and $n$ is the number of pixels in the desired region. The skewness is used to describe the range in which the values in the region are mainly concentrated. Therefore it is used to characterize the structural features in the region.

Similarly, the kurtosis of the different tissues can be difined as follows:

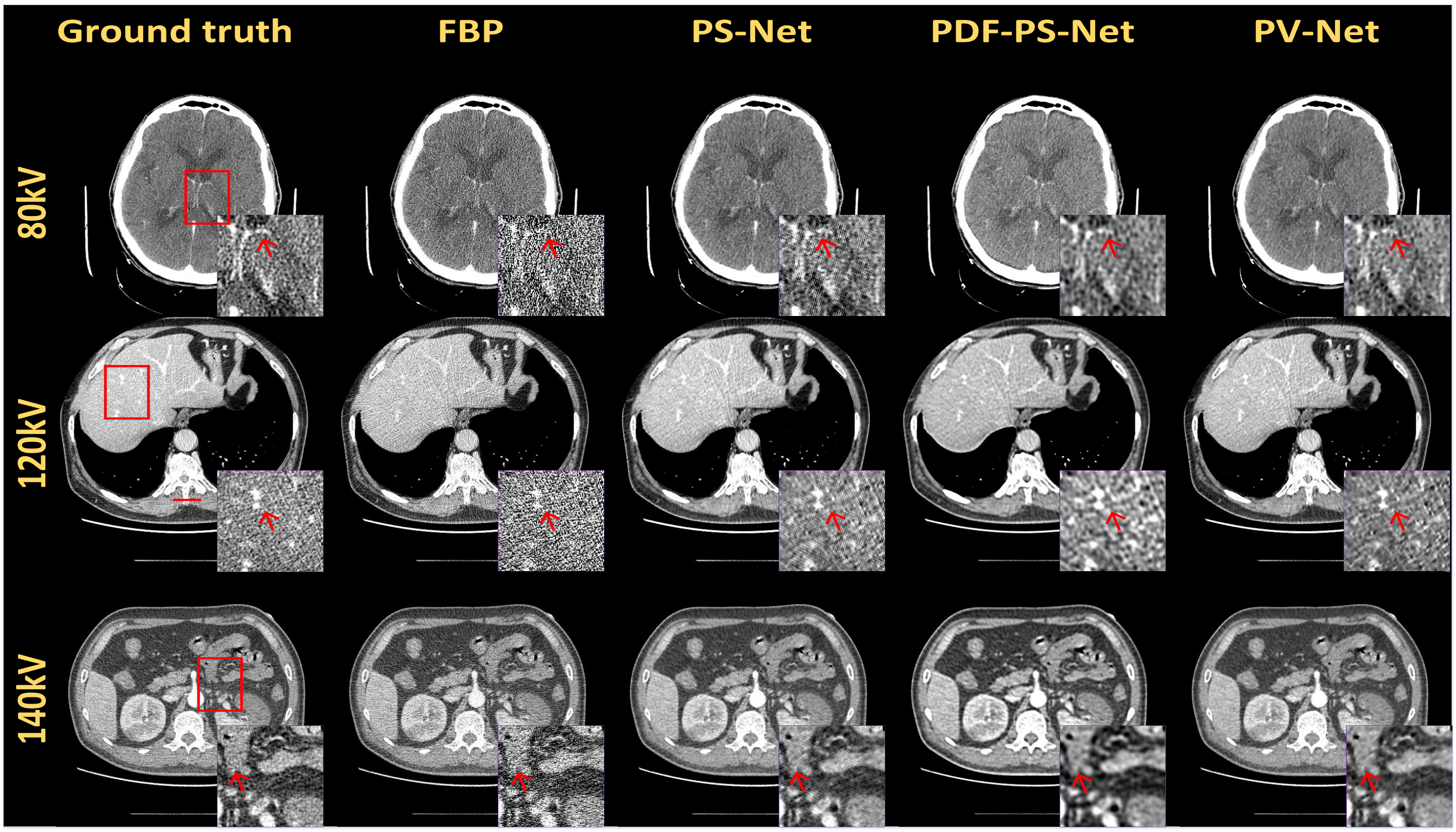

**Figure 2:** Results of PV-Net and comparison methods.

$$\mathbf{K} = \frac{n+1}{(n-1)(n-2)(n-3)} \frac{\frac{1}{n}\sum_{i=1}^{n}(x_i-\overline{x})^4}{(\frac{1}{n-1}\sum_{i=1}^{n}(x_i-\overline{x})^2)^2} - \frac{3(n-1)^2}{(n-2)(n-3)}. \tag{5}$$

This statistic is then used to describe the steepness of the data distribution curve, that is, the degree of deviation of the values within the region. Thus, it mainly characterizes the noise distribution in the region.

With the two statistical features of the anatomical regions, we construct an anatomy loss function for anatomical region modulation. More specifically, as shown in the gray box in Figure 1, we first segment the labeled noise-free image into different regions based on the threshold values in Table 2. Then we calculate the sum of the statistics $\mathbf{S}$ and $\mathbf{K}$ in these regions:

$$\widetilde{\mathbf{S}} = \sum_{m=1}^{M} \mathbf{S}_m, \widetilde{\mathbf{K}} = \sum_{m=1}^{M} \mathbf{K}_m, \tag{6}$$

where $M$ is the number of the segmented regions. Therefore, the anatomy loss function can be expressed as follow:

$$\mathscr{L}_{anatomy} = \frac{1}{M}(\left|\widetilde{\mathbf{S}} - \widetilde{\mathbf{S}}^*\right| + \left|\widetilde{\mathbf{K}} - \widetilde{\mathbf{K}}^*\right|), \tag{7}$$

where * denotes the network output. The anatomy loss function can be summarized in Algorithm 1. Then the total loss function can be expressed as:

$$\mathscr{L} = \mathscr{L}_{MSE} + \lambda \mathscr{L}_{anatomy}. \tag{8}$$

$\lambda$ here is parameter.

## 3 Experiments

### 3.1 Dataset

To validate and evaluate the reconstruction performance of the proposed PV-Net, we collected CT dataset with different scanning protocols (i.e., different kVp settings and anatomical regions). In the dataset 2001 head CT images were scanned at 80 kVp, 1991 chest CT images were scanned at 120 kVp, and 2006 abdominal CT images were scanned at 140 kVp. We perform low-dose simulations based on our previous work [5] to obtain pairing data under different dose level. Ninety percent of the CT images are used for network training, and ten percent are used for network testing.

### 3.2 Compared methods and implementation details

To evaluate the performance of the proposed PV-Net, three competing methods are selected, i.e., FBP, the protocol specific network without modulation layer and anatomy loss function (i.e., PS-Net), and protocol specific network with modulation layer (i.e., PDF-PS-Net). The learning rate of all models is set to $2 \times 10^{-4}$, and optimized with the RMSProp optimizer. Peak signal-to-noise ratio (PSNR), structural similarity index (SSIM) and root mean square error

| Protocol | Methods | PSNR | SSIM | RMSE |
|---|---|---|---|---|
| Brain(80kVp) | FBP | 25.8683±1.5008 | 0.6004±0.0132 | 11.3085±1.4346 |
| | PS-Net | 32.6950±0.7404 | 0.6539±0.0240 | 7.7413±0.6541 |
| | PDF-PS-Net | 35.4417±1.3863 | 0.6841±0.0263 | 5.4918±0.8316 |
| | **PV-Net** | **37.6840±1.1697** | **0.7636±0.0236** | **3.3592±0.4542** |
| Chest(120kVp) | FBP | 23.7044±2.02021 | 0.5304±0.0756 | 17.0905±3.9140 |
| | PS-Net | 27.9415±2.108 | 0.7020±0.0387 | 9.6430±2.9243 |
| | PDF-PS-Net | 29.3819±2.6047 | 0.7140±0.0420 | 7.6680±2.2914 |
| | **PV-Net** | **30.7763±1.7969** | **0.7204±0.0293** | **5.5187±1.3953** |
| Abdomen(140kVp) | FBP | 26.0422±2.5032 | 0.6030±0.0676 | 11.5117±2.9102 |
| | PS-Net | 30.8683 ±2.6608 | 0.7178±0.0492 | 6.9772±2.4951 |
| | PDF-PS-Net | 32.8992 ±2.0179 | 0.7311±0.0353 | 6.2566±2.7775 |
| | **PV-Net** | **33.4222 ±1.8133** | **0.7380±0.0402** | **4.8457±1.6955** |

**Table 1:** Quantitative index of PV-Net and comparison methods.

| Tissue | Hounsfield unit(HU) |
|---|---|
| Air | (-1024,-900) |
| Lung | (-900,-500) |
| Lung parenchyma | (-500,-100) |
| Fat | (-100,-20) |
| Blood | (-20,25) |
| Cerebral white matter | (25,38) |
| Muscle/cerebral gray matter | (38,90) |
| Calcification | (90,400) |
| Bone | (400,1000) |

**Table 2:** Threshold for segmentation.

(RMSE) measures are used to evaluate the reconstruction performance.

## 4 Results

Figure 2 shows the performance of each model on each dataset. To better show the image details, we select a region of interest (shown in the red boxes) on each case and zoom in on the bottom right. As we can see, PS-Net enables to recover a certain image structure, but there are still some remaining noise in the reconstruction results. PDF-PS-Net demonstrates stronger denoising performance, but the reconstruction results at 120 kVp and 140 kVp are slightly over-smooth. Due to the advantage of kVp modulation and unique anatomy loss function, the presented PV-Net has better performance of denoising, superior preservation of anatomical information, and as shown, the reconstruction results of PV-Net higher detail resolution, and lower noise levels. Table 1 shows the quantitative index of each method under each protocol. The results show that the presented PV-Net is able to learn the deep features of data under different protocols, which results in better quantitative metrics of reconstruction results.

## 5 Conclusion

In this work, we propose a protocol variation network to consider the heterogeneity within the different kVp settings and anatomical structures. Extensive experiments on datasets with different protocols demonstrate that the proposal model can improve the performance of deep learning models across multi protocols.

## Acknowledgments

This work was supported in part by the NSFC under Grant U21A6005, and Grant 12226004, and Young Talent Support Project of Guangzhou Association for Science and Technology.

# A Deep Learning Approach to Estimate and Compensate Motion in Non-contrast Head CT scan

Zhennong Chen[1], Quanzheng Li[1], and Dufan Wu[1]

[1]Center for Advanced Medical Computing and Analysis, Massachusetts General Hospital and Harvard Medical School, Boston, USA

Patient's head motion is a major source of image artifacts in the head CT. The motion artifacts can be compensated by accurate motion estimation and compensation during the image reconstruction. Partial angle reconstruction (PAR)-based method is well-known for motion estimation in CT. In this study, we propose the *first* PAR-based deep learning (DL) method to estimate and compensate the motion in the head CT. We designed a DL pipeline to estimate the motion from PARs built by the CT sinogram. In our simulation study with relatively heavy motion, the proposed method achieved good accuracy on the motion parameter estimation, with a mean absolute error (MAE) of 1mm in head translation and 1° in rotation. The MAE of the reconstructed CT images was also reduced from 130 to 54 HU. Simulation results demonstrate that the proposed method has promise to tackle motion problem in clinical head CT.

## 1 Introduction

Head motion in paediatric patients or paitents with head trauma or stroke is a majoy source of image artifacts in head computed tomgoraphy (CT), degrading the image quality and impacting diagnosis[1]. Improved hardware such as faster gantry rotation or dual source designs can reduce motion artifacts but it is challenging to apply these improvements on some systems such as mobile CT and C-arm CT. Another type of harware design is the motion tacking system that measures the motion but the calibration is complicated[2]. Software-based, especially deep learning (DL)-based motion correction has been developed to tackle this problem. Recently, Su et al.[3] proposed a modified U-Net to remove the motion artifacts directly in the image domain. However, the image-domain-based DL method may over-smooth the image, especially when the head motion is large.

Motion estimation is another solution to this problem and it ensures data fidelity. If the true head motion can be correctly estimated, one can use it to compensate the motion during backprojection to correct the motion artifacts. The partial angle reconstruction (PAR)-based method[4] is well-known for an accurate estimation of motion. Concretely, PAR means a series of images of the same object reconstructed from small angular segments of the sinogram, each of which represents the object status in a small time interval. Registration between two PARs reveals the motion between two time frames. To overcome the challenging registration due to the heavy limited-angle artifacts, Maier et al.[5] designed a DL pipeline which takes PARs as the input, predicts the motion and utilizes the spatial transformer module to output the motion-compensated reconstruction in cardiac CT. However, the utility of the PAR-based method to correct the Head CT motion is under-investigated. In this study, we have three major aims: (1) Develop the *first* PAR-based DL method to estimate and compensate motion in non-contrast head CT scans; (2) Design a DL approach to directly output motion parameters from PARs; (3) investigate different DL model designs and compare its performance with the DL model solely relying on image domain[3].

## 2 Materials and Methods

*2.1 Head Motion Model and Simulation*

We modeled the head motion by rigid transformation. To simplify, we only investigated 2D rigid motion in the axial plane in this work, which include three motion parameters: two intra-slice translations, $t_x$ and $t_y$, along the image x- and y- axes and one intra-slice rotation, $\theta_z$, around the z-axis. The motion is modeled according to Jang et al.[6]; the motion at the *i*th view $(t_{x,i}, t_{y,i}, \theta_{z,i})$ is described using a cubic *B*-spline interpolation with 5 control points (CP) equally distributed across time. For example, the translation $t_x$ is described as:

$$t_{x,i} = B_{1D}\left(CP_{tx,0}, \dots CP_{tx,4}, i\right) \qquad (1)$$

where $B_{1D}$ denotes 1-dimensional *B*-spline interpolation, $CP_0$ to $CP_4$ refers to 5 control points. Each motion is described using a different *B*-spline with its own 5 CPs. The first CP ($CP_0$) corresponding to the start of the scan was always set to 0. An example of this motion model can be found in **Figure 1**.

100 CT studies from 37 unique patients without any motion artifacts were retrospectively collected. The pixel size varies from 0.37 to 0.49mm and the slice thickness was resampled to 1.5mm. All images have $512 \times 512 \times 60$ dimension to cover the whole head. We followed existing works for motion simulation: 5 CP values in any *B*-spline were randomly sampled from [-*max*, *max*] where *max* is the maximum magnitude equal to 10mm for translation or 10° for rotation[2,6], representing the normal head motion. We produced 90 motion simulations for each CT study, resulting in 9000 images as our dataset. In each simulation, 3 *B*-splines were generated for $(t_x, t_y, \theta_z)$ respectively and applied to the whole volume. The simulated full-scan (360°, in total 1400 views assumed) sinogram was generated by forward-projecting the moving image with the fan-beam x-ray geometry.

*2.2 Partial Angle Reconstruction (PAR)*

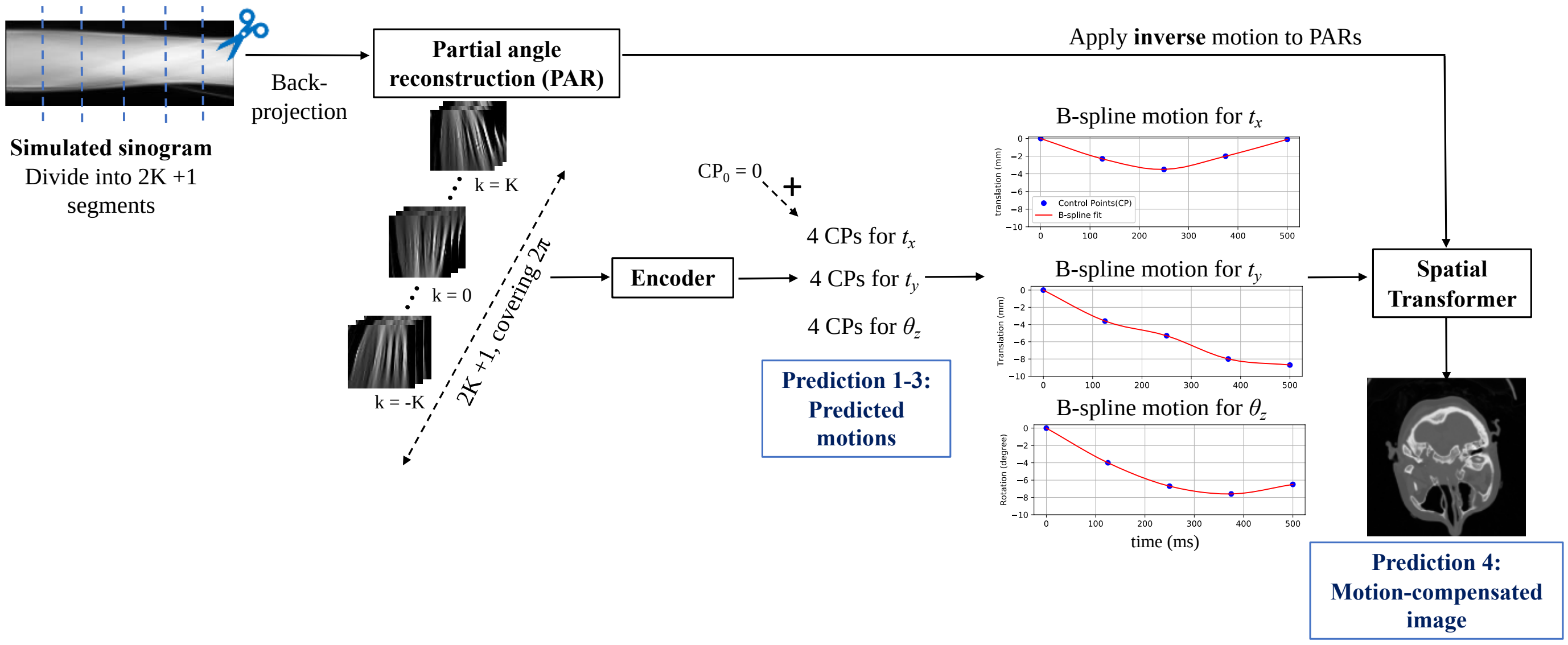

**Figure 1. DL Pipeline**. It shows how we started from the sinogram and predicted four outputs: 3 motion *B*-splines and 1 motion-compensated image. Note in B-spline motion plots, blue dots represent 5 control points, and the x-axis spans the gantry rotation time which is assumed to be 500ms.

The PARs were reconstructed using the simulated sinogram. Concretely, the sinogram was divided equally into $2K+1$ nonoverlapping angular segments. Hence, each segment $k$ $(-K \leq k \leq K)$ covers an angular interval $[\emptyset_k - \frac{\Delta\emptyset}{2}, \emptyset_k + \frac{\Delta\emptyset}{2}]$ where $\emptyset_k = \pi + \Delta\emptyset \times k$ and $\Delta\emptyset = \frac{2\pi}{2K+1}$. Each segment was then filtered-backprojected with the Ram-Lak filter to obtain a PAR. A large value of $K$ =12 was picked to maintain high temporal resolution and reduce the motion artifacts within each PAR. Therefore, each image in our dataset has $2K+1$ = 25 corresponding PARs.

Most CT scanners cannot cover the entire head from top to bottom in one rotation, so we generated PARs using a image patch of the 15 (out of 60) consecutive slices. We investigated two typical patch types: type 1 patch is the 15-slice patch starting from the teeth plane, which contains a variety of structures from the complex maxillofacial bones to the relatively simple cerebrum; type 2 patch is from the middle of the head, which contains mostly the brain tissue surrounded by skull. If not specified, the results we reported corresponds to type 1 patch.

### *2.3 Deep Learning (DL) Pipeline Designs*

A 3D convolutional network is trained to predict the motion CPs. 25 PARs of one image patch were used as the DL model's multi-channel input. These PARs were downsampled to have isotropically 2mm spatial resolution in x and y directions. Similar to Maier et al.[5], an encoder-like (**Figure 2**) network including 3D convolution layers and maxpooling layers extracts image features from the input PARs at different resolutions. Then fully-connected layers are used to decode the 12 motion paramters. Specially, it outputs three sets of 4 CPs (corresponding to $CP_1$ to $CP_4$ ) for $t_x$ , $t_y$ , $\theta_z$ respectively, which are subsequently used to construct three *B*-spline interpolations as the predicted motion. A spatial transformer module applies the *inverse* of predicted motions to PARs, outputing the motion-compensated image. The whole pipeline can be found in **Figure 1**.

To summarize, our DL pipeline takes 25 PARs as the input and outputs 4 predictions: three sets of 4 CPs for $t_x$ (unit: pixel), $t_y$ (pixel), $\theta_z$ (°) respectively and one motion-compensated image. The whole pipeline was trained end-to-end. The mean-absolute-error (MAE), $L$, was calculated by comparing predictions with the ground truth motion CPs and the ground truth image:

$$L = L_{t_x} + L_{t_y} + L_{\theta_z} + w \cdot L_{image} \quad (2)$$

The image loss, $L_{image}$, is highly correlated from the motion CP losses. Hypothetically, addition of the image loss may further increase the model performance by putting a constraint in the image domain; but the necessity of doing so is unkown. Therefore, we investigated two different options of loss weight combination: with the image loss ($w = 1$) and without the image loss ($w = 0$). $w = 1$ was chosen because the normalized image loss and the normalized motion CP loss had the same order of magnitude.

To clarify, since the PARs and the output motion-compensated image were downsampled, the DL image was only used to constrain the model prediction while we did *not* use it as our final image. How we made the final image is elaborated in section 2.4. Our dataset was split on the patient level. 80% of the data was used for training and validation where 5-fold cross validation was performed. The rest 20% was used for independent testing.

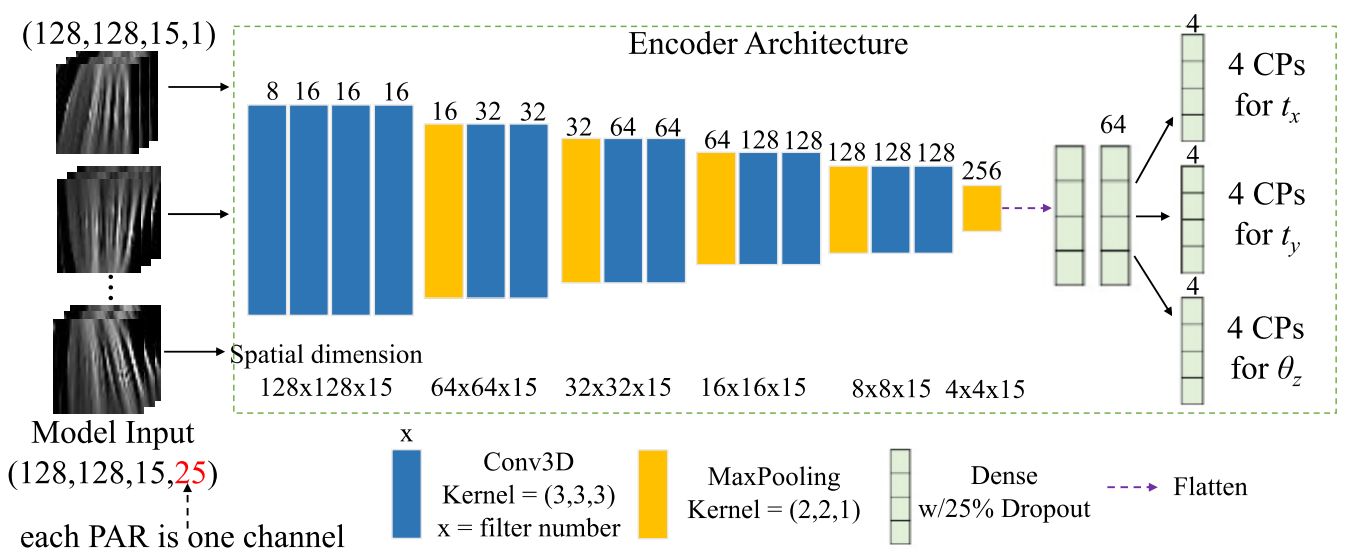

**Figure 2. Encoder architecture.** The input are 25 PARs; each was padded/cropped to (128,128,15).

*2.4 The Final Image Output*

After predicting the motion CPs using downsampled PARs, we interpolated the motion at each time frame using the B-Spline (see **Equation 1**) and applied the *inverse* motion field frame-by-frame during filtered-backprojection. This motion-compensated image, denoted as $I_{DL}$, is the final image output of our proposed method.

*2.5 Quantitaive Evaluation*

The model was evaluated in the testing cohort (with 1800 simulated cases). The predicted motion CPs were averaged across all models from the 5-fold cross-validation. MAE was used to evaluate predicted motion CPs against the ground truth. MAE, root-mean-squared-error (RMSE), and structural similarity index metric (SSIM) were used to evaluate the final motion-corrected image against the ground truth image. All these errors were measured on the the foreground pixels whose values are larger than -10 HU on the ground truth image.

We also applied the image-domain-based DL approach proposed in Su et al.[3] to the testing cohort and generated images denoted as $I_{DL,img}$ , which were evaluated using the same metrics.

The paired t-test was used to determine whether there is a significant difference between the results of two methods. Statistical significance was set to $p \leq 0.05$.

## 3 Results

*3.1 Motion Simulation*

Among the testing cohort, the motion corrupted images $I_m$ had an MAE of 130±33 HU and an RMSE of 286±62 HU. This magnitude level of RMSE indicates that the majority of simulation has the moderate/large motion[2].

*3.2 Investigating Different Loss Weights*

We investigated the influence of the presence of the image loss by setting $w$ in equation (2) to 0 and 1. Two models with and without image loss were trained on the same training group and tested on the testing dataset. **Table 1** shows that the motion CPs errors from two settings are not significantly different ($p$>0.05). Thus, for simplicity, we excluded the image loss in our final model.

*3.3 Motion Prediction*

**Table 1.** MAE of the motion CPs predicted by models trained with $w = 1$ and $w = 0$ and their statistical difference.

| | $w = 0$ | $w = 1$ | $p$ |
|---|---|---|---|
| $t_x$ (mm) | 1.26±0.72 | 1.25±0.70 | 0.83 |
| $t_y$ (mm) | 0.87±0.45 | 0.90±0.49 | 0.38 |
| $\theta_z$ (°) | 1.13±0.49 | 1.13±0.47 | 0.99 |

**Table 2.** MAE of the predicted CPs. The definitions of type 1 and 2 patches can be found in section 2.2.

| | Type 1 patch | Type 2 patch | $p$ |
|---|---|---|---|
| $t_x$ (mm) | 1.00±0.59 | 0.95±0.60 | 0.02 |
| $t_y$ (mm) | 0.70±0.39 | 0.72±0.50 | 0.14 |
| $\theta_z$ (°) | 1.05±0.70 | 1.08±0.94 | 0.89 |

**Table 3.** Evaluations of images. $I_m$ is the motion corrupted image, $I_{DL}$ is from the proposed method, $I_{DL,img}$ is from the image-domain approach[3].

| | $I_m$ | $I_{DL}$ | $I_{DL,img}$ |
|---|---|---|---|
| MAE (HU) | 130±33 | 54±18 | 86±14 |
| RMSE (HU) | 286±62 | 126±40 | 195±25 |
| SSIM | 0.71±0.13 | 0.94±0.05 | 0.83±0.09 |

**Table 2** shows that the MAE between predicted and ground truth motion CPs are as follows: $t_x$ = 1.00±0.59mm, $t_y$ = 0.70±0.39mm, $\theta_z$ = 1.05±0.70°. This ~1mm and ~1° is at the magnitude level of very slight motion according to Kim et al[2]. Furthermore, the model has comparable high performance on type 1 patches from the teeth plane and type 2 patches from the middle of the brain ($p$>0.05 except for $t_x$ ). It demonstrates the robustness of our method on different image features.

*3.4 Performnace on Images*

By applying the inverse of predicted motions, we achieved good performance in the final motion-corrected images. From **Table 3**, we can see across all testing data, the MAE drops from 130±33 in motion-corrupted images $I_m$ to 54±18 HU in motion-corrected images $I_{DL}$; RMSE drops from 286±62 to 126±40 HU; SSIM improves from 0.71±0.13 to 0.94±0.05. The box plots in **Figure 3** show the median and interquartile ranges of each metric. **Figure 4** provides some image results, which show that the motion artifacts are dramatically reduced, and the blurred head bone as well as brain tissues are clear again in the motion-corrected images $I_{DL}$**.**

We also implemented the image-domain method proposed by Su et al.[3], but found relatively poor performance in large motion due to the over-smoothing it introduced. As shown in **Table 3** and **Figure 3**, because most of our simulated motions are moderate to large, the result from the proposed method $I_{DL}$ has significant improvement compared to the image-domain result $I_{DL,img}$ ($p$<0.05, one-tailed t-test). **Figure** 4 also shows that the image-domain method introduces over-smoothing in $I_{DL,img}$. The motion artifacts on bone structures are allievated while the grey matters were over-smoothed.

## 4 Discussions and Conclusion

In this study, we proposed a PAR-based DL method to estimate motion in non-contrast head CT. We evaluated our method in a large simulation study and presented that it accurately predicts the motions (with only ~1mm and ~1° error). This accurate estimation enables us to subsequently use the predicted motions to significantly corrected the motion artifacts in the image (MAE: from 130±33 to 54±18 HU, SSIM: from 0.71±0.13 to 0.94±0.05). We show the better performance of ours in the large head motion compared to the image-domain-based method[3].

This method needs to be further developed, especially with a better head motion model. For example, we shall use 3D motion model and consider different scanning trajectories such as helical trajectory or multi-plane fanbeam. The proposed method should also be further validated with phantom study and real clinical data.

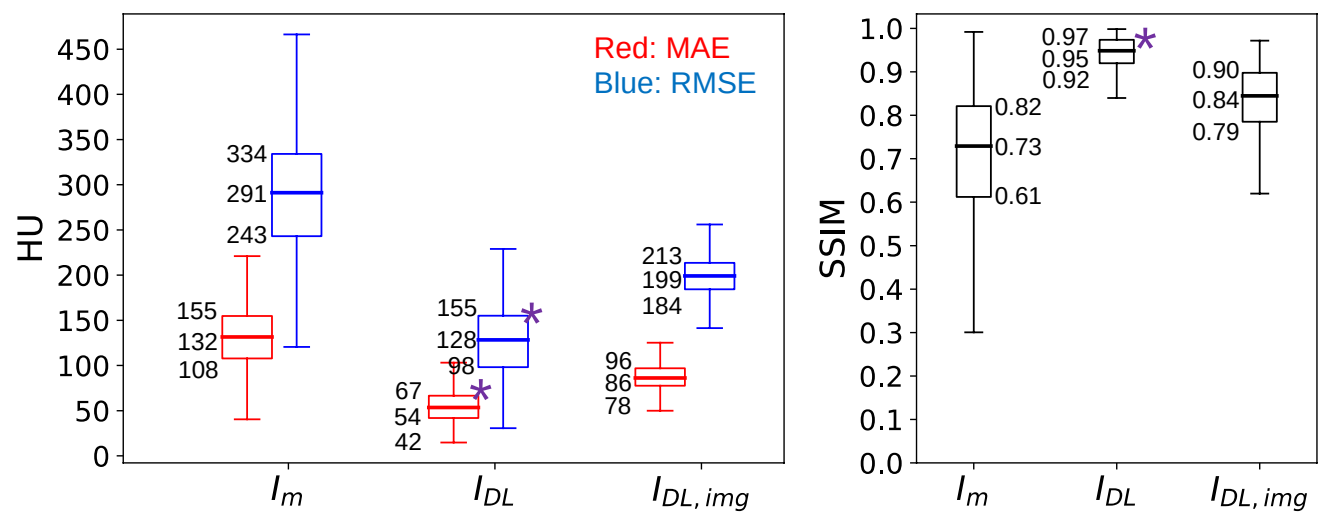

**Figure 3. Box plots of image metrics**. The left image is for MAE (red boxes) and RMSE (blue boxes); the right image is for SSIM. Three lines in each box from top to bottom mean upper quartile, median, lower quartile respectively; their values were provided besides each line. Purple asterisks indicate that $I_{DL}$ has significant improvement over $I_{DL,img}$ for all three metrics.

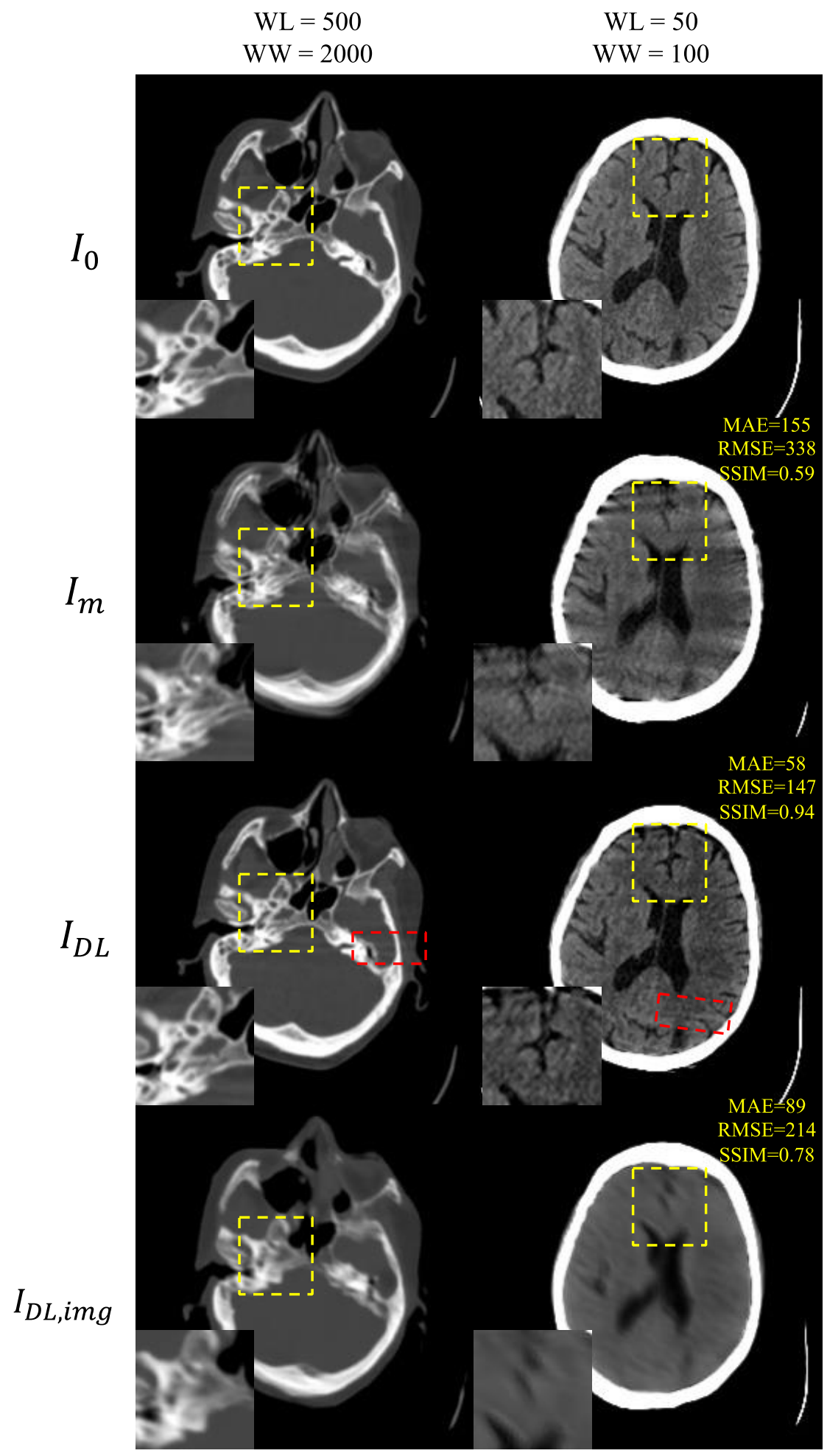

**Figure 4. Image result**. From top to bottom row, we show $I_0$, the static ground truth; $I_m$, the simulated motion-corrupted image; $I_{DL}$, the motion-corrected image by the proposed method; $I_{DL,img}$, the motion-corrected image the image-domain method[3]. The dashed red boxes in the $I_{DL}$ row show that there are still some residual motion artifacts uncorrected due to the error of DL-predicted motion parameters. WL stands for window level. WW stands for window width.

# Reconstruction of miscalibrated sparse-view measurements without geometry information using a transformer model

Theodor Cheslerean-Boghiu[1,2], Franz Pfeiffer[2,3], and Tobias Lasser[1,2]

[1]Department of Computer Science, School of Computation, Information and Technology, Technical University of Munich, Munich, Germany
[2]Munich Institute of Biomedical Engineering, Technical University of Munich, Munich, Germany
[3]Department of Physics, School of Natural Sciences, Technical University of Munich, Munich, Germany

**Abstract** Deep learning based algorithms for X-ray computed tomography reconstruction often rely on domain-transfer modules that use tomographic operators to perform the forward- and backprojection. The used operators restrict the trained network to one single reconstruction geometry. In this manuscript we propose to train a network on sinogram-image pairs without explicitly using a domain-transfer module based on a forward-backward projector. We train the network to implicitly learn the geometry only from image pairs using the attention mechanism, similar to machine translation in natural language processing. We show that the model can partially handle a set of random corruptions in the sinogram to simulate imperfections or miscalibrated angles in the geometry. We run a simple experiment to show that an attention-based architecture is able to implicitly learn geometrical information only from sinogram-image pairs and produce the corresponding reconstruction.

## 1 Introduction

X-ray Computed Tomography (CT) is a non-invasive technique employed for gathering insights into the internal arrangement of a sample or a body part, all without the need for physical dissection. The reconstruction problem of CT is an ill-posed problem for which several categories of algorithms have been developed over the years. Among the original algorithms, which are still used today, we have analytical algorithms like the Filtered Backprojection for 2d CT or FDK for 3d CT [1]. Iterative reconstruction pipelines have been used to overcome the data-limiting issues analytical methods have in sparse-data applications like sparse-view or limited angle CT [2].

More recently, data-driven algorithms have shown promising quality improvement of the reconstruction over analytical and iterative methods [3]. However, most convolution-based deep learning algorithms are unable to encode the geometry information in the learned kernels. Data-driven methods that directly perform the X-ray transform inversion still need to be explored due to the lack of appropriate architectures and the complexity of the operation [2].

The emergence of transformer-based architectures [4] has provided an opportunity to tackle the tomographic reconstruction problem without requiring any forward operators. The nature of the attention mechanism used throughout the transformer architectures overcomes the domain limitations imposed by the convolutional operations or the computational complexity of fully connected layers. Moreover, since the tomographic reconstruction problem represent a domain transfer problem, similarities to machine translation in NLP can be drawn and used to our advantage. A sinogram is no more than a set of measurements, like a sentence is a set of words arranged in some configuration [5]. The reconstruction is also a set of patches arranged correctly, similar to the corresponding sentence in another language. The sinogram and the reconstruction are correlated to one another, similar to two translated sentences in two different languages.

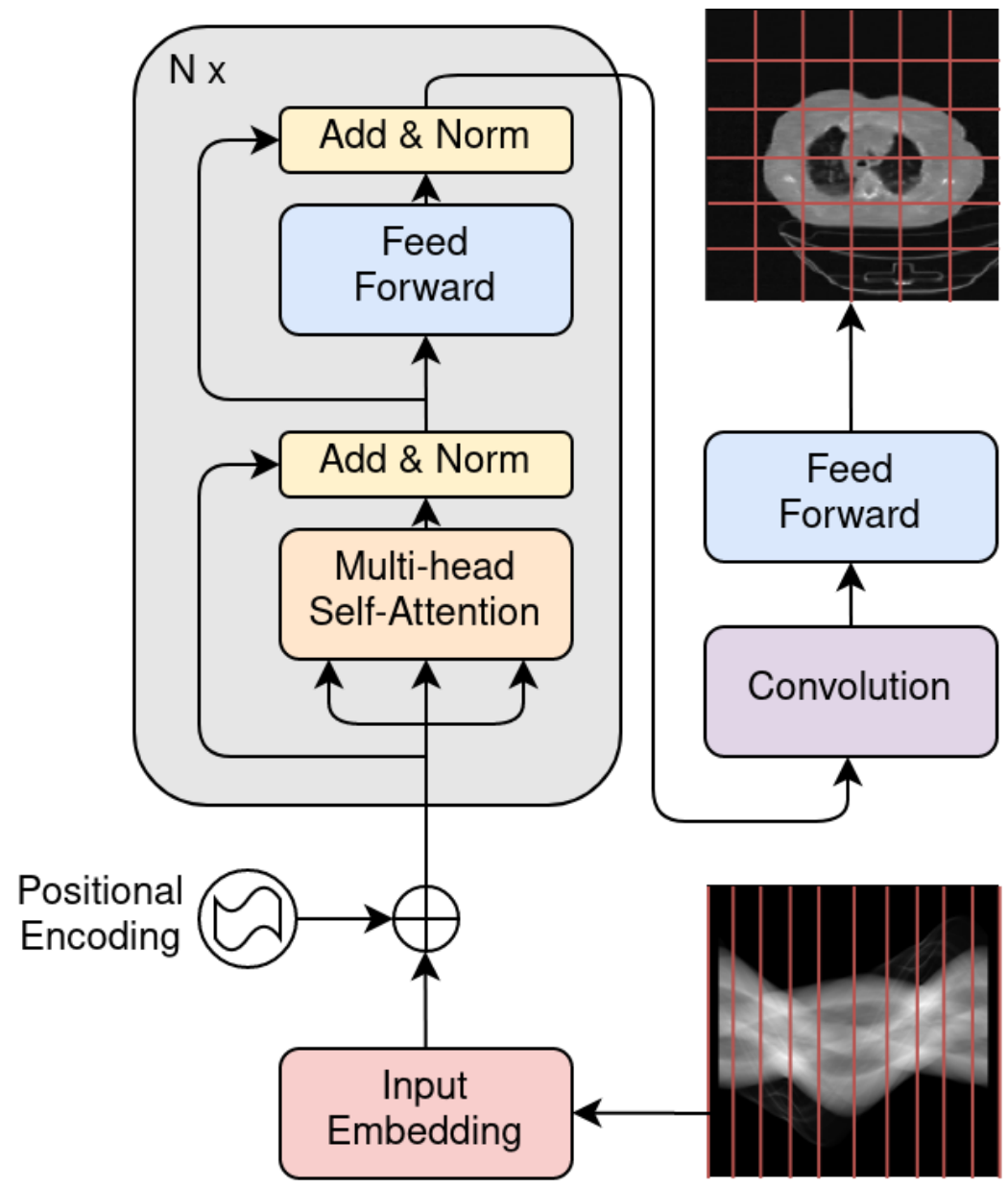

**Figure 1:** Architecture of the proposed reconstruction network. A transformer encoder embeds a set of sinogram measurements (each angle is one token) and a simplified decoder made up of one 1d convolutional layer and one fully connected layer extracts the set of square patches that make up the corresponding reconstruction.

In this manuscript we propose a deep learning algorithm used to perform tomographic reconstruction without requiring any geometrical information about the system. We can summarize our contributions as follows:

- A transformer encoder and one simplified decoder are used to reconstruct fixed-sized sinograms
- We provide insights about the trained model to understand how it implictly learns geometry information.
- We train the model on a set of simulated “acqusition errors” added to the sinogram.
- We show that our proposed model generates results that are comparable in quality to sparse-view FBP.

## 2 Methods

The (discrete) CT reconstruction problem is defined as inverting the X-ray transform of the imaged sample. The X-ray transform is defined as an integral over the space of straight two-dimensional lines $L$ given by $\mathscr{X}f(L) = \int_L f(x)dx$.
For a given slice $x \in \mathbb{R}^{N\times N}$ the discrete X-ray transform generates a set of measurements $y \in \mathbb{R}^{N\times k}$ in a parallel-beam geometry setting. In this configuration we generate $k$-view sinograms ($k = 224$) over a 360 degree arc with 224 pixel detectors from 128-by-128 pixel original volume slices. In this experiment we concentrate on 2d FBP reconstructions for sparse-view chest CT.
We model the inversion operation without providing any geometrical information to the system. We propose a simple deep neural network based on a transformer architecture [4] (seen in Fig. 1) with a simplified decoder head.
The model is trained on pairs of a $k$-view sinogram and the original slice used to generate the sinogram. Our goal is to learn the reconstruction operation without providing the geometrical information. A nice advantage of using the original slices as labels instead of matching the labels to the sinograms (via FBP reconstruction) is the implicit sparse-artifact reduction effect learned by the model.
The transformer architecture enables in each layer the sharing of information between all tokens via the attention mechanism. Compared to conventional convolutional networks, which work on the intrinsic assumption of locality and local invariance imposed by the convolution operation, transformer-based architectures rely on attention layers which allow for both short- and long-range dependencies to be learned by the model.
Vaswani et al. [4] define the attention operation as a function with three input arguments. A dot product is first computed between a "query" $Q \in \mathbb{R}^n$ and a "key" $K \in \mathbb{R}^n$. The result is normalized by the square root of the key's dimension $\sqrt{d_k}$. The resulting weighting factor is then multiplied with the "value" $V \in \mathbb{R}$.

$$\text{Attention}(Q,K,V) = \text{softmax}\left(\frac{QK^T}{\sqrt{d_k}}\right)V \tag{1}$$

In a multihead self-attention scenario the "query", "key", and "value" are embedded from a single input (not three different ones) and the attention operation is replicated on multiple parallel "heads" with different embedding kernels.
A model relying on the attention mechanism splits its input into a set of "units" [1] which are then embedded into **tokens**. The attention operation enables each **token** to attend to every other **token** thus enabling both short- and long-range dependencies in the latent space. Since each **token** represents one single measurement in the input sinogram, measurements taken from complementary angles will be matched together during the training process.

[1] these units can represent anything from image patches to words in a sentence or measurements in a sinogram

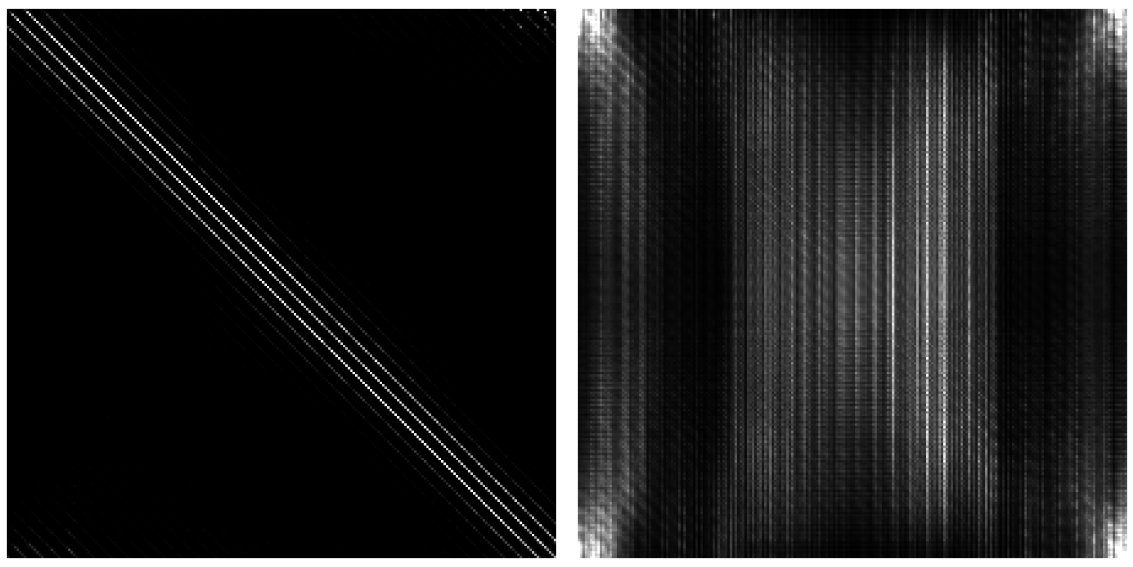

(a) Short-range (b) Long-range

**Figure 2:** 224-by-224 attention maps (softmax$(QK^T/\sqrt{d_k})$) from different attention layer in the trained model. Each row and column correspond to one measurement in the sinogram (e.g. first row and column are the first measurement, last row and column are the k'th measurement). **(a) Short-range.** Second head of the third self-attention layer. Notice the short-range dependencies (around the main diagonal) between relatively close measurements. **(b) Long-range.** Fourth head of the third attention layer. Notice how each measurement (row) attends to a lot more other measurements in the sinogram (columns) based on its position in the geometry.

Unlike the vision transformers approach [6], we don't allow the model to learn any positional encoding of the input measurement tokens. We employ non-trainable additive sinusoidal positional encodings as used in NLP [4].
Since the input geometry is fixed, the network is also limited to one geometry and able to learn said geometry very well.

### 2.1 "Machine Translation"

We can draw similarities between the tomographic reconstruction problem and a machine translation problem from NLP [4]. A set of measurements $y$ is the "sentence" being translated while the set of image tokens $x_{i,j}\,\forall i,j$ is the translation.
A machine translation scenario in NLP consists of two main steps. In the first stage, the sentence to be translated is embedded in an appropriate latent space. Similarly, we do generate latent features for each of the sinogram's measurements $y$ via a transformer encoder block (see Fig. 3-(**Base patches**)). Then, in the second stage, the translation output is generated one token at a time until the transformer decoder predicts an End-Of-Sentence (EOS) token [4]. In our case, the length of the "translation" output is a fixed set of image patches, and we replace the transformer decoder with a sequential application of one convolutional layer and one fully connected layer.
The role of the convolutional layer is to generate the correct number of latent tokens out of the extracted sinogram features. These tokens are then transformed by the fully connected layer into the final reconstruction patches. We show that the transformer encoder learns to generate a patch dictionary from the sinogram, while the convolutional layer performs basically a weighted sum of the dictionary entries

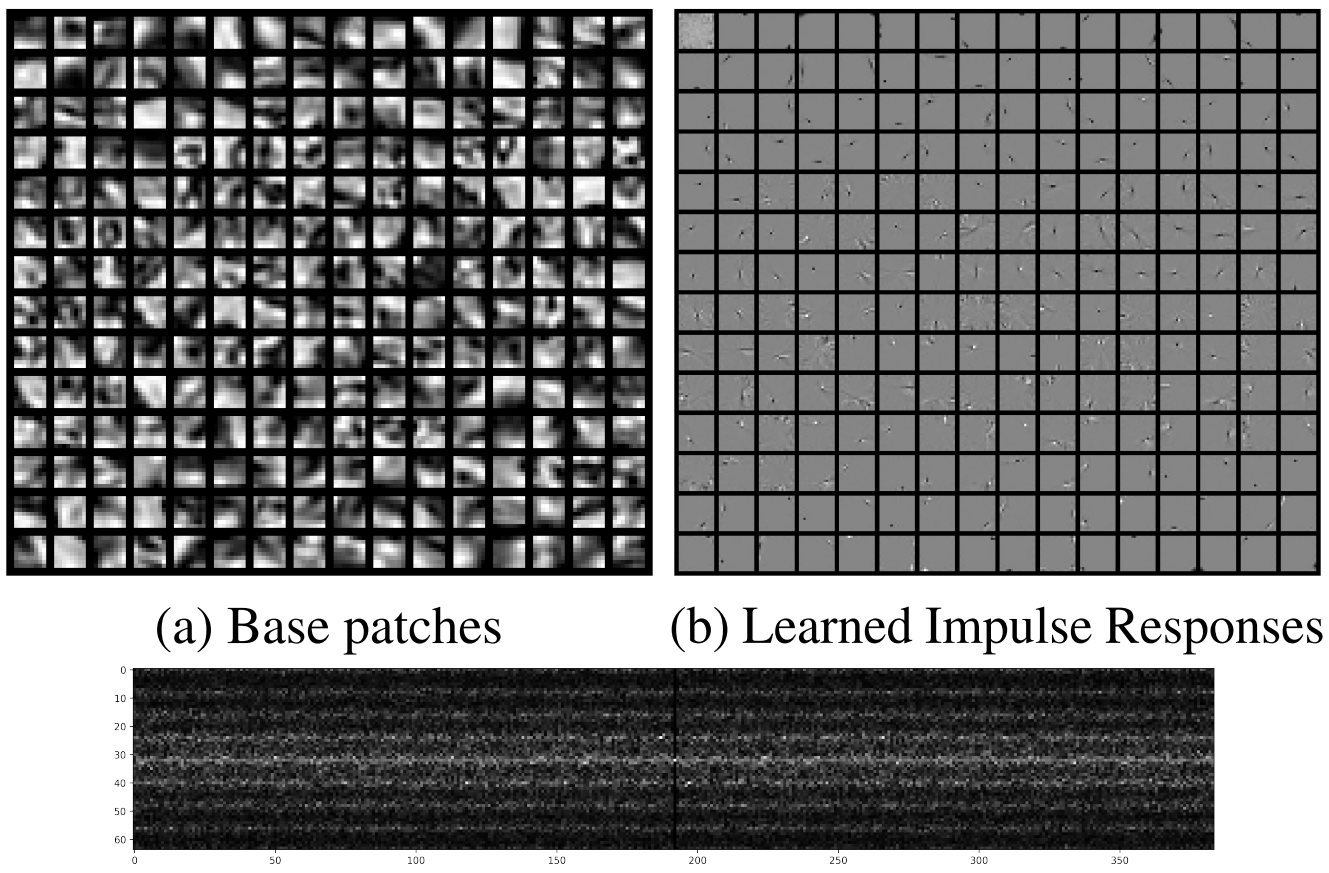

(a) Base patches (b) Learned Impulse Responses

(c) Fully Connected layer weight

**Figure 3:** **(a) Base patches.** The transformer encoder has embedded the 224 measurements (tokens) from an input sinogram into a dictionary of 224 (base) vectors used to generate the reconstruction patches. **(b) Patch dictionary.** 224 16-by-16 pixel maps representing the individual learned kernels of the final convolutional layer for every of the 256 8-by-8 pixel patches that make up the final reconstruction. Each impulse response corresponds to one embedded measurement from **(a)** and provides information about the area in the final reconstruction that is dependent on the corresponding latent patch. **(c) Fully Connected layer weight.** The final fully connected layer takes the embedded tokens and extracts the final 8-by-8 pixel patches.

similar to a inverse cosine transform seen in JPEG compression.

## 3 Experiments

Our training set consists of $\sim$ 15k clinical chest CT slices based on a full-dose protocol in the lung window. Each slice was obtained at a resolution of 512-by-512px and binned down to 128-by-128px. Our test set is composed of $\sim$ 3k clinical chest CT slices.

The feature extractor section is a non-pretrained vision transformer model (ViT) with 12 layers implemented in *timm* [7]. To match the size of the sinogram to the size of the reconstruction we use a convolutional layer to increase the number of tokens from 224 (number of angles in the sinogram) to 256 (number of tokens that make up the final image) and a fully connected layer to reduce the latent representation of the sinogram tokens to the final representation of the image tokens (8-by-8 pixel patches). The final step of the network is to merge all the patches into the final 128-by-128 pixel reconstruction.

Initial experiments without data augmentation have shown that the model has a strong inclination to overfit on the training set. We, therefore, settled on two extensive random data augmentation strategies:

- **Reconstruction-level augmentation**: Rotation (up to 90°), Cropping, Perspective Shift, Brightness and Contrast Change, Vertical Flip, Random Grid Shuffle (7-by-7 grid).
- **Sinogram-level augmentation**: Assuming that one part of the geometry might have been miscalibrated or its corresponding measurements corrupted, we chose to randomly shift up to 32 random measurements by up to 4 pixels in either up or down direction (angular random shift) simulating pixel-level miscalibration, and to disable up to 8 measurements (random 0'ing) to simulate a failure of measurement or storage.

We train the proposed architecture for 6000 epochs with batch size 16 and learning rate $10^{-4}$ and we use *SmoothL1Loss* to the ground truth. The labels are the originally binned slices. The total time for training on 224-view sinograms at a resolution of 128px is around 25 days on one NVIDIA RTX6000 with 24GB of VRAM.

We simulate the sinogram via the *radon* transform implemented in the Python library *skimage*. This is also the bottleneck of the training step.

## 4 Discussion

It is clear that the model does not overfit the training dataset as the testing results are comparable to the FBP reconstructions. Fig. 3 shows the learned convolution kernels and fully connected layer weight. The learned kernels in the convolutional layer contain impulse responses encoding location information for each individual patch of the final reconstruction. Due to the randomness of the weight initialization at the beginning of the training procedure, the set of impulse responses is randomized. The fully connected layer used to extract the final image patches from the latent feature vectors resembles a convolutional operation similar to the FBP kernel.

The ViT-based feature extractor encodes the geometrical information inside its attention layers (Fig. 2), while the employed sinusoidal positional encoding vectors provide the model information about the relative locations of the measurements in the sinogram.

Fig. 4 shows results generated with the trained network for slices from two different upper-body locations. The partial smoothness of the network outputs has been a consistent feature of in the output of deep neural networks trained to perform denoising for sparse-view CT mainly due to the loss of information through the encoding and decoding operation that such networks perform. Macroscopic organ structures like some bones or the heart are reconstructed at an acceptable visual quality while some microscopic structures like lung nodules are also visible in the network output. One advantage of not using a forward-backward operator inside the model is the size of the region-of-interest circle in the network output compared to the circle-limited FBP reconstructions.

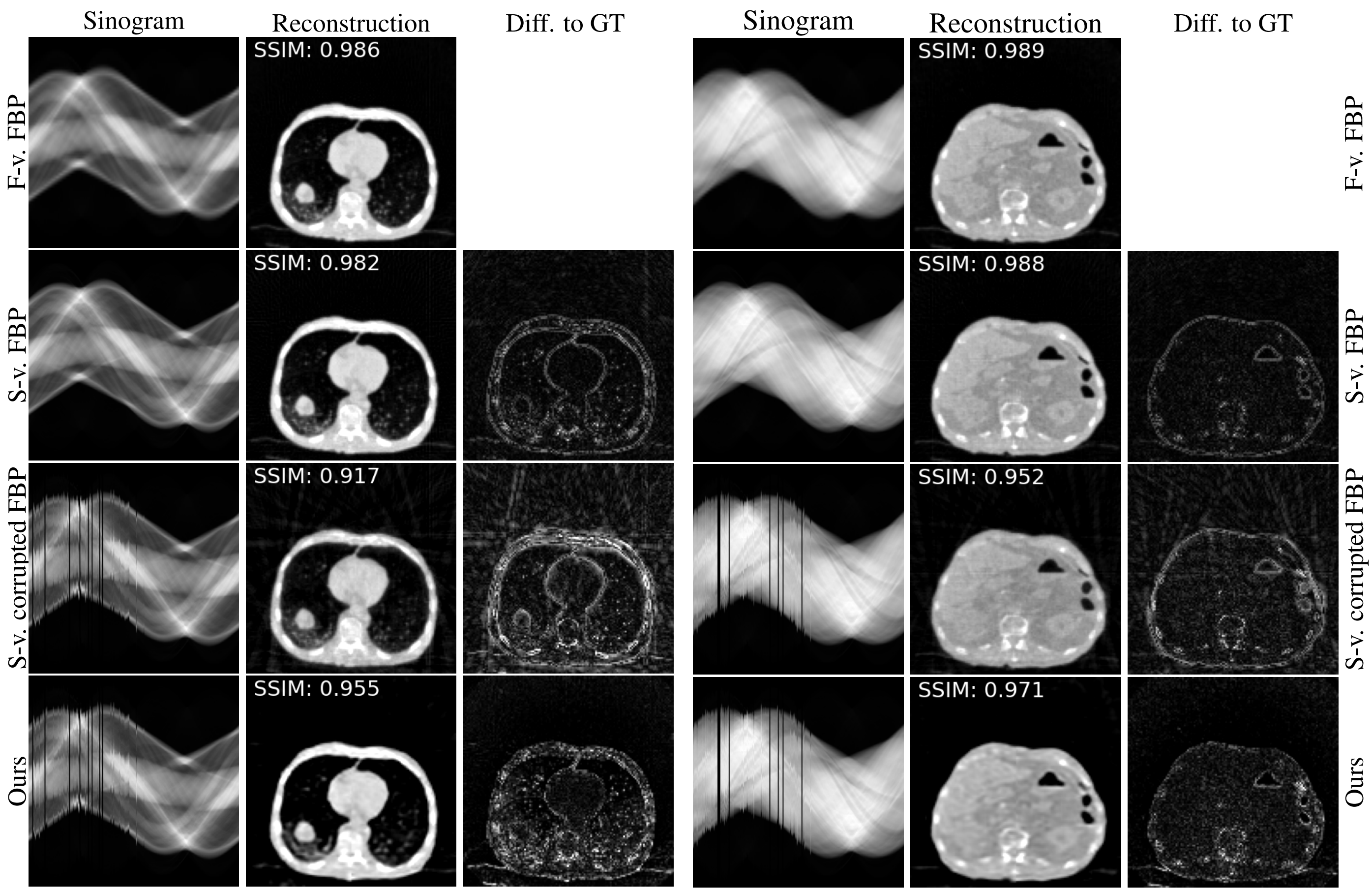

**Figure 4:** Reconstruction comparison between the proposed method (**Ours**) and Ram-Lak filtered FBP for two different CT slices. Each image contains the SSIM value computed against the binned ground truth (GT) (see Sec. 3). (**F-v. FBP**) FBP reconstruction using a 2048-view (**f**ull-**v**iew) sinogram. (**S-v. FBP**) 224-view sinogram (**s**parse-**v**iew) and corresponding FBP reconstruction. (**S-v. corrupted FBP**) 224-view sinogram corrupted with measurement shift and loss (as described in Sec. 3) and corresponding FBP reconstruction. (**Ours**) 224-view sinogram corrupted with measurement shift and loss (as described in Sec. 3) with corresponding reconstruction generated by the proposed network.

## 5 Conclusion

In this manuscript we have shown that given enough data, a complex operation like sparse-view tomographic reconstruction can be learned using a transformer model, and can partially positively improve the quality of the reconstruction over conventional FBP, given the presence of several types of limitations: sparse measurements, miscalibrated measurements, missing angles.
The practicality of such a data-driven solution is limited to the fixed geometry used to generate the input measurements. However, the potential of applying machine translation algorithms towards tomographic reconstruction is demonstrated by our simple experiment and paves the way towards training a full transformer architecture accepting variable-view sinograms to perform tomographic reconstruction.

## Acknowledgement

This work was supported in part by the German Federal Ministry of Health under Grant 2520DAT920 and partially funded by the Technical University of Munich Graduate School.

# Formulation of the ML-EM algorithm based on the continuous-to-continuous data model

Robert Cierniak[1]

[1]Department of Intelligent Computer Systems, Czestochowa University of Technology, Czestochowa, Poland

**Abstract** This abstract introduces a new alternative to the commonly used ML-EM method for positron emission tomography. The conception proposed here is based on a continuous-to-continuous data model, where a forward model used in the reconstruction problem is formulated as a shift-invariant system. The main aim of this report is to show proof that this new conception is based on probabilistic fundamentals. That is important because this reconstruction problem is formulated taking into consideration the statistical properties of signals obtained in the PET technique.

## 1 Introduction

The reconstruction method presented here relates to positron emission tomography (PET). Because the relatively small number of annihilations in a single measurement is observed, the statistical nature of observations has a strong influence on the reconstructed image, and it is necessary to take this fact into account. The commonly used reconstruction method is the maximum likelihood-expectation maximization (ML-EM) algorithm [1, 2]. It must be emphasized that the image processing methodology used in this algorithm is consistent with a discrete-to-discrete (D-D) data model, where the reconstructed image is a priori divided into homogeneous blocks representing pixels. In this conception, individual elements of the system matrix are determined separately for every pixel, and for every annihilation event detected along the given LOR. In this case reconstruction problem is formulated using huge matrices. It makes the reconstruction procedure much more complex than in the case of the analytical methods. We propose a formulation of a statistical iterative reconstruction algorithm based on a continuous-to-continuous data model (C-C). An algorithm of this form was proposed firstly in [3]. The forward model for this problem is presented as a shift-invariant system [4], and the equivalents for the direct measurements are values of the image obtained after the back-projection operation. As a result, most of the drawbacks connected with using the method based on the D-D data model can be avoided, namely: the use of a shift-invariant system allows for the implementation of an FFT algorithm during the most demanding calculations, which as a consequence significantly accelerates performed calculations which are necessary to complete the reconstruction procedure. Unfortunately, there are two serious difficulties: elements of the image obtained after the back-projection operation are strongly correlated, and there are doubts about whether they follow the Poisson probability distribution. In this paper, we show that our approach has statistical fundamentals, and the obtained solutions are consistent with the results of the referential algorithm.

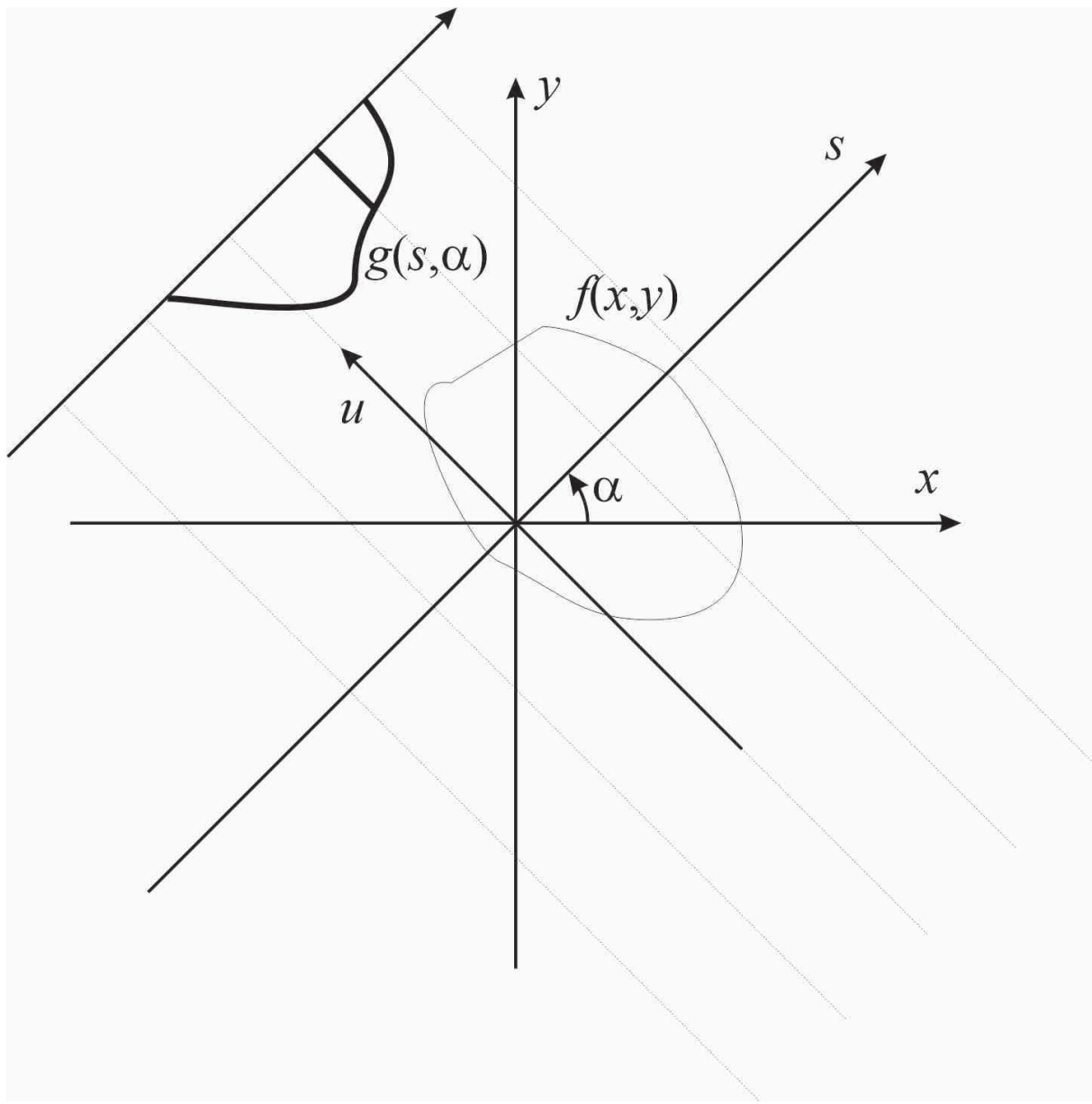

**Figure 1:** Scheme of the measurement system

## 2 Forward Model Formulation

The crucial premise behind the idea presented here is that conceptually, both a measurement system and a reconstructed image are defined in continuous spaces. Let function $f(x,y)$ denote the unknown image representing the distribution of a chosen molecule in space inside the human body in its given cross-section, e.g. distribution of the molecule with a radioactive for cancer diagnostic. Image $f(x,y)$, defined as a function $f:\mathbb{R}^2\rightarrow\mathbb{R}$, will be reconstructed using projections $g(s,\alpha)$ obtained by using the hypothetical measurement system that is presented in the Fig. 1, during scanning process along an axis $s$.
The function $g(s,\alpha)$ means a measurement carried out at a distance $s$ from the origin when a projection is made at a specific angle $\alpha$. It is called the Radon transform and is written mathematically as

$$g(s,\alpha)=\int_{-\infty}^{+\infty}\int_{-\infty}^{+\infty} f(x,y)\cdot\delta(x\cos\alpha+y\sin\alpha-s)\,dxdy. \quad (1)$$

In our approach, the problem of image reconstruction from projections is closely related to the proposed reconstruction method. Firstly, we perform the first step of the reconstruction procedure: the back-projection operation. This operation is described using the following equation:

$$\tilde{f}(x,y) = \int_{-\pi}^{\pi} \int_{-\infty}^{+\infty} g(\bar{s},\alpha)\, int\,(s-\bar{s})\, d\bar{s} d\alpha. \tag{2}$$

It should be underlined presence of the function $int\,(\Delta s)$ in the convolution in the Eq. 2. This continuous function is related to an interpolation whose usage is necessary during the back-projection operation performed in implementations of this reconstruction method. At this stage of consideration, it is enough to assume that we can use a filtration function $int\,(\Delta s)$ over measurements $g(\bar{s},\alpha)$.

All the projections are carried out physically according to eq. (1). Therefore, we can further transform relation (2) into the following form:

$$\tilde{f}(x,y) = \int_{-\pi}^{\pi}\int_{-\infty}^{+\infty}\int_{-\infty}^{+\infty}\int_{-\infty}^{+\infty} f(\bar{\bar{x}},\bar{\bar{y}}) \cdot \delta(\bar{\bar{x}}cos\alpha + \bar{\bar{y}}sin\alpha - \bar{s})\, d\bar{\bar{x}} d\bar{\bar{y}} \cdot \cdot int\,(s-\bar{s})\, d\bar{s} d\alpha. \tag{3}$$

A fundamental conclusion for our further deliberations is that every point of the original image $f(\bar{x},\bar{y})$ makes a contribution to the image $\tilde{f}(x,y)$ obtained after the back-projection operation, depending on the distance $\bar{x}\cos\alpha + \bar{y}\sin\alpha - s$. It means that values of the function $f(\bar{x},\bar{y})$ are projected onto the interpolation function $int$. Bearing this in mind, we can modify relation (**??**) to obtain a more convenient form:

$$\tilde{f}(x,y) = \int_{-\infty}^{+\infty}\int_{-\infty}^{+\infty} f(\bar{x},\bar{y}) \int_{-\pi}^{\pi} int\,(\bar{x}\cos\alpha + \bar{y}\sin\alpha - s)\, d\alpha d\bar{x} d\bar{y}. \tag{4}$$

The above transformation is fundamental for our method because it leads to a shift-invariant system formulation. It is worth emphasizing that the interpolation function $int$ is an essential medium that allows us to create the relation between the original and the image obtained after the back-projection operation. One can say that the interpolation function serves as a medium to project a pixel's influence on other pixels.

It is convenient to present relation (4) in more compact way:

$$\tilde{f}(x,y) = f(x,y) * h(x,y) = \int_{-\infty}^{+\infty}\int_{-\infty}^{+\infty} f(\bar{x},\bar{y})\, h(\bar{x}-x, \bar{y}-y)\, d\bar{x} d\bar{y}, \tag{5}$$

where:

$$h(\Delta x, \Delta y) = \int_{-\pi}^{\pi} int\,((\bar{x}-x)\cos\alpha + (\bar{y}-y)\sin\alpha)\, d\alpha. \tag{6}$$

The back-projection operation represents an accumulation of all the projections which have passed through a given point of the reconstructed image $f(x,y)$, taking into account the blur effect of the LORs by the interpolation operation. The new image obtained in this way $\tilde{f}(x,y)$ includes information about the reconstructed image $f(x,y)$, but this information is strongly blurred.

## 3 Statistical considerations

It has been stated before that a distribution of the registered annihilation events corresponds to the concentration of atoms of a radioactive isotope in a given cross-section of the examined body. It is justified that the number of decays of these nucleons in the reconstructed cross-section, and subsequent annihilation events follow the inhomogeneous Poisson point process $\Pi$. In the case of the proposed reconstruction approach, we will consider the state space $\Omega$ in which this point process $\Pi$ sit as Euclidean space $\mathbb{R}^2$, i.e. $\Omega$ is a measurable space, and the Poisson process $\Pi$ on $\Omega = \mathbb{R}^2$ is a random countable subset $\Pi$ on $\Omega$. Therefore, the probability that in a homogeneous reconstructed plane, $\lambda$ annihilation events are observed is:

$$P\{\Lambda = \lambda\} = e^{\lambda^*} \frac{(\lambda^*)^{\lambda}}{\lambda!} \tag{7}$$

for $\lambda \in \mathbb{N}_0$, where expectation value of the random variable $\Lambda$ is $\lambda^*$. Let us assume that we use the LM method and we maximize the following expression:

Let us assume that we use the ML method and we maximize the following expression:

$$l_1(\lambda) = \ln(P(\Lambda = \lambda)) \approxeq \lambda \ln\frac{\lambda^*}{\lambda} - \lambda^* + \lambda = l_2(f), \tag{8}$$

whereby $\Lambda$ is a random variable with Poisson distribution which represents a number of the observed annihilation events in a given cross-section in a certain time interval.

The same optimal solution as $l_3$ also gives

$$l_3(\lambda) = H\left(\lambda \ln\frac{\lambda^*}{\lambda} - \lambda^* + \lambda\right), \tag{9}$$

wherein constant $H = \int_x \int_y h(x,y)\, dxdy$. Continuing, it is possible to rearrange all three terms in $l_3$ from the Eq. (9), taking into account relations (5) and (6), to the following form:

$$l_4(f) = \int_x \int_y \tilde{f}(x,y) \ln\frac{\tilde{f}^*(x,y)}{\tilde{f}(x,y)} - \tilde{f}^*(x,y) + \tilde{f}(x,y)\, dxdy, \tag{10}$$

wherein:

$$\tilde{f}^*(x,y) = \int_{\bar{x}} \int_{\bar{y}} h(x-\bar{x}, y-\bar{y})\, f^*(\bar{x},\bar{y})\, d\bar{x} d\bar{y}, \tag{11}$$

and

$$\tilde{f}(x,y) = \sum_{k=1}^{\lambda} int\left((x-x_k)\cos\alpha_k + (y-y_k)\sin\alpha_k\right) \quad (12)$$

are points in an image obtained after a back-projection operation. This image can be understood as an equivalent of a set of the direct measurements used in the traditional ML-EM method.

Finally, taking into account results presented by formulas (10)-(12), we can express the reconstruction problem in the following way:

$$f_0(x,y) = \arg\min_{f^*(x,y)} (l_4), \quad (13)$$

It is natural now to use a gradient descent method to find the optimum for $l_4$, i.e. $\frac{\partial l_4}{\partial f^*(x,y)} = 0$. We obtain in this way the main relation, as follows:

$$f^{t+1}(x,y) = f^t(x,y)\frac{1}{H}\int_{\bar{x}}\int_{\bar{y}} \frac{\tilde{f}(\bar{x},\bar{y})}{\int_{\bar{\bar{x}}}\int_{\bar{\bar{y}}} f^t(\bar{\bar{x}},\bar{\bar{y}})\, h_{\Delta x,\Delta y}\, d\bar{\bar{x}}\, d\bar{\bar{y}}} h_{\Delta x,\Delta y}\, d\bar{x}\, d\bar{y} \quad (14)$$

where $\tilde{f}(x,y)$ is an image obtained after the back-projection operation.

That is entirely consistent with the continuous-to-continuous data model. Now, to perform calculations, it is possible to discretize the above formula to the following form:

$$f^{t+1}(x_i,y_j) = f^t(x_i,y_j)\frac{1}{H_{ij}}\sum_{\bar{i}}^{I}\sum_{\bar{j}=1}^{I} \frac{\tilde{f}(x_{\bar{i}},y_{\bar{j}})}{\sum_{\bar{\bar{i}}}\sum_{\bar{\bar{j}}} f^t(x_{\bar{\bar{i}}},y_{\bar{\bar{j}}})\, h_{\Delta i,\Delta j}} h_{\Delta i,\Delta j} \quad (15)$$

wherein $H_{ij}$ is a sum of all coefficients $h_{\Delta i,\Delta j}$ taken into account at the calculation of a given expression $\frac{\tilde{f}(x_{\bar{i}},y_{\bar{j}})}{\sum_{\bar{\bar{i}}}\sum_{\bar{\bar{j}}} f^t(x_{\bar{\bar{i}}},y_{\bar{\bar{j}}}) h_{\Delta i,\Delta j}} h_{\Delta i,\Delta j}$, and $h_{\Delta i,\Delta j}$ are determined according to the formula

$$h_{\Delta i,\Delta j} = \Delta_\alpha \sum_{\psi=0}^{\Psi-1} int\left(\Delta i\cos\psi\Delta_\alpha + \Delta j\sin\psi\Delta_\alpha\right). \quad (16)$$

## 4 Experimental results

In our experiments, we have adapted the well-known Shepp-Logan mathematical phantom of the head (all values divided by $10^{-3}$). We used parallel projections ($L = 512$ virtual detectors on the virtual screen). The number of the parallel views was chosen as $\Psi = 728$ per half-rotation, and the dimension of the processed image was fixed at $I \times I = 512 \times 512$ pixels. After making these assumptions, it is possible to conduct the virtual measurements (with a relatively high degree of noise) and complete all the required parallel projections related to the LORs. Then, through suitable rebinning operations, the back-projection operation can be carried out to obtain an image $\tilde{f}_{ij}$, which is used as a referential image for the reconstruction procedure. In Figures 2 and 3 (at the bottom), the reconstructed image after 1 000 iterations are depicted. All elements $h_{\Delta i,\Delta j}$ were pre-calculated before was started the iterative reconstruction process. The image obtained after the back-projection operation was subjected to an iterative reconstruction process, wherein the convolutions operations were performed in the frequency domain. For comparison, a view of the reconstructed images using a referential reconstruction algorithm based on the D-D data model is presented: Figure 2 for a low level of noise, and Figure 3 for a high level of noise.

## 5 Conclusion

In this paper, it has been shown that a statistical reconstruction method for PET can be formulated which is based on the C-C data model. We have presented a feasible statistical reconstruction ML-EM algorithm. Performed experiments proved that our reconstruction method is relatively fast (in consequence of the FFT use) and gives satisfactory results with suppressed noise. The computational complexity for 2D reconstruction geometries (e.g. parallel rays) is proportional to $I^4$ for each iteration of the D-D reconstruction procedure, but our original approach only needs approximately $8I^2\log_2(2I)$ operations.

## Acknowledgements

The project financed under the program of the Polish Minister of Science and Higher Education under the name "Regional Initiative of Excellence" in the years 2019 - 2023 project number 020/RID/2018/19 the amount of financing PLN 12,000,000.

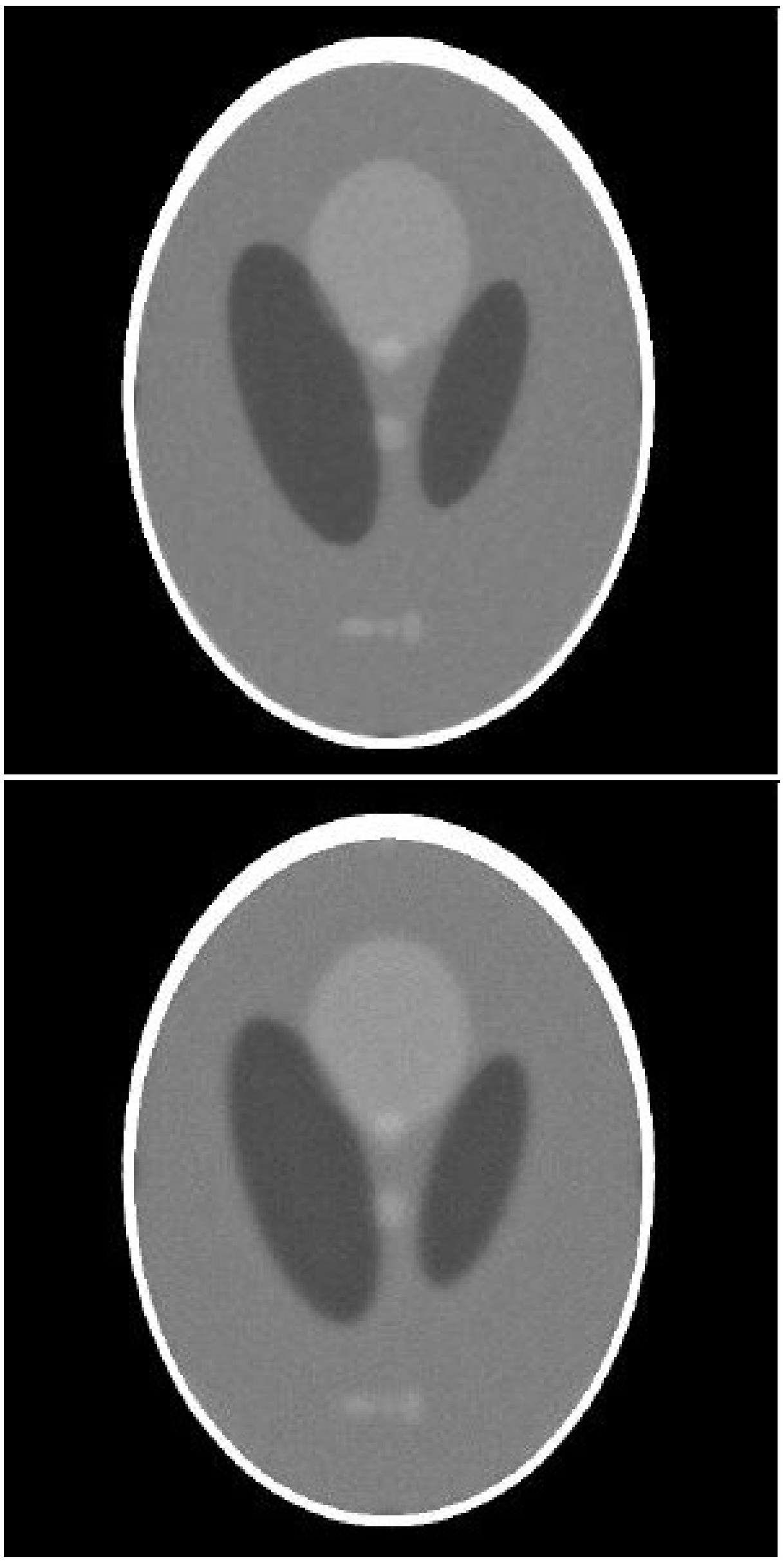

**Figure 2:** Views of the images (window center $C = 1.05 \cdot 10^{-3}$, window width $W = 0.1 \cdot 10^{-3}$): reconstructed image using the referential statistical approach described by (1) after 50 iterations ($MSE = 5.13 \cdot 10^{-7}$) (at the top); reconstructed image using the statistical approach presented in this paper obtained after 1 000 iterations ($MSE = 5.21 \cdot 10^{-7}$) (at the bottom).

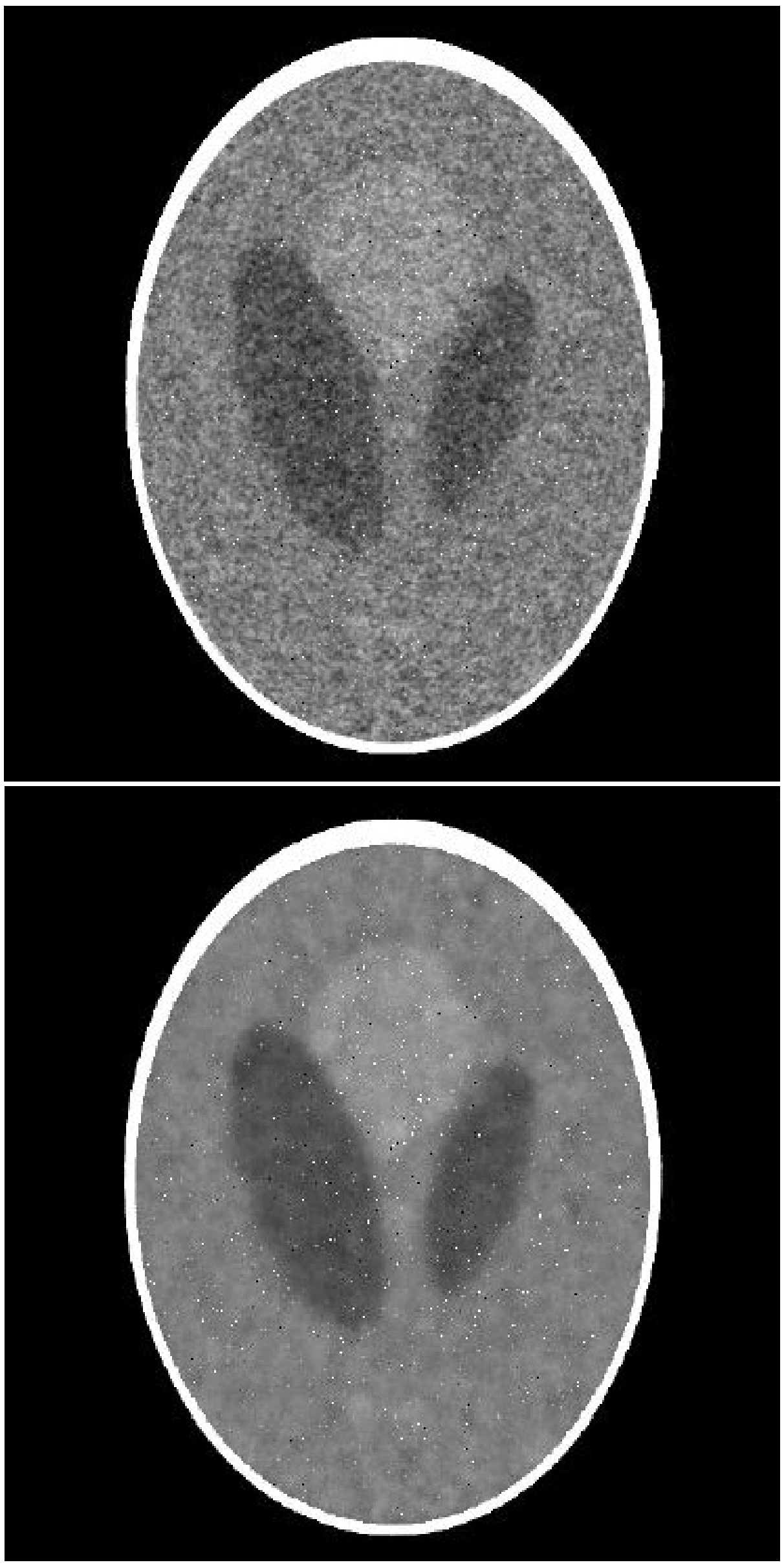

**Figure 3:** Views of the images (window center $C = 1.05 \cdot 10^{-3}$, window width $W = 0.1 \cdot 10^{-3}$): reconstructed image using the referential statistical approach described by (1) after 50 iterations ($MSE = 7.89 \cdot 10^{-7}$) (at the top); reconstructed image using the statistical approach presented in this paper obtained after 1 000 iterations ($MSE = 7.03 \cdot 10^{-7}$) (at the bottom).

# Positron-Range Correction for an On-Chip PET Scanner using Deep Learning

Christoph Clement[1], Gabriele Birindelli[1], Fiammetta Pagano[2, 3], Marco Pizzichemi[2, 3], Marianna Kruithof-De Julio[4], Sibylle Ziegler[5], Axel Rominger[1], Etiennette Auffray[2], and Kuangyu Shi[1]

[1]Department of Nuclear Medicine, Inselspital, University of Bern, Bern, Switzerland
[2]EP Department, CERN, Geneva, Switzerland
[3]Physics Department, University of Milano-Bicocca, Milan, Italy
[4]Department for Biomedical Research, Inselspital, University of Bern, Bern, Switzerland
[5]Department of Nuclear Medicine, University Hospital, Ludwig Maximilian University of Munich, Munich, Germany

**Abstract** Organs-on-Chips (OOCs) are a novel technology that aim to mimic the functions and physiology of human organs in a laboratory setting. Positron Emission Tomography (PET) is a widely-used imaging modality that enables non-invasive monitoring of biological processes in vivo. However, the spatial resolution of current small-scale PET systems is not sufficient for OOC imaging. One of the main factors limiting the spatial resolution of a PET scanner is the positron range, which is the distance that a positron travels before it collides with an electron. In this study, we present a novel Deep Learning (DL)-based approach for correcting the positron-range effect in our previously introduced On-Chip PET scanner. We created a dataset of pairs of non-corrected and corrected images using a Monte-Carlo simulation of a realistic OOC phantom and a fully three-dimensional Maximum-Likelihood Expectation-Maximization (MLEM) iterative reconstruction algorithm. Our results demonstrate the effectiveness of the DL-based positron-range correction algorithm in improving the overall quality of the reconstructed images. This approach has the potential to be a valuable tool for advancing the study of 3D models in radiopharmaceutical research.

## 1 Background

OOCs are a novel technology that aim to mimic the functions and physiology of human organs in a laboratory setting. These devices have gained significant attention in the field of radiopharmaceutical research due to their ability to accurately replicate the physiology of native organs. OOCs contain living human cells that are cultured and perfused to simulate the physiology of native organs. The goal of OOCs is to provide an efficient alternative to traditional in vitro and animal models for drug development, disease modeling, and toxicity testing [1, 2]. PET is a widely-used imaging modality that enables non-invasive monitoring of biological processes in vivo. With the emergence of OOCs as a powerful tool for studying human physiology and drug response, there is a growing need for dedicated PET scanners that can accurately and efficiently image these microfabricated devices. However, the spatial resolution of current small-scale PET systems is not sufficient for OOC imaging. One of the main factors limiting the spatial resolution of a PET scanner is the positron range, which is the distance that a positron travels before it collides with an electron. Positron range is dependent on the energy of the emitted positron and the density of the tissue or material it is traveling through. In PET imaging, the positron range is an important factor to consider when reconstructing the image, as it affects the spatial resolution of the image and the ability to accurately localize the source of the emitted radiation [3].

In our previous work, we introduced a dedicated On-Chip PET scanner that is capable of imaging OOCs [4]. We optimized the design of the system using a Monte-Carlo simulation and predicted the gamma-ray interaction positions with a Convolutional Neural Network (CNN). In this study, we present improved performance results of the scanner achieved by employing a fully three-dimensional MLEM reconstruction algorithm and implementing a deep learning-based positron-range correction algorithm.

The novelties presented in this work are threefold. Firstly, we created a realistic OOC phantom for our Monte-Carlo simulation of the system, providing a more accurate representation of the OOCs in our imaging studies. Secondly, we adapted a fully three-dimensional MLEM iterative reconstruction algorithm to the geometry of our dedicated On-Chip PET scanner. And thirdly, we developed and trained an image-to-image DL model that takes the non-positron-range corrected reconstructed image as input and outputs the positron-range corrected image. Overall, this study demonstrates the potential of using deep learning-based positron-range correction algorithm to improve the performance of a dedicated PET scanner for imaging OOCs.

## 2 Methods

### 2.1 Monte-Carlo Simulation

In order to train the DL-based positron-range correction algorithm, we created a dataset using a Monte-Carlo simulation of the On-Chip PET scanner, which was implemented with the GEANT4 Application for Tomographic Emission (GATE) software [5]. The simulation models the scanner's response to either back-to-back gamma or positron sources placed inside the compartments of a realistic OOC phantom, which is modeled after a commercially available device. The phantom consists of multiple compartments with varying diameters connected by microfluidic channels. In each compartment, a cylindrical source with the same diameter as the compartment is placed to simulate the background radiation and a point source is used for the foreground with a ratio of 8:1. The scanner geometry, as shown in Figure 1, consists of four detectors arranged in a box-like design around the phantom. Each detector is made up of two monolithic Lutetium–yttrium oxyorthosilicate (LYSO) crystals. The back surface of each detector is covered with Silicon photomultipliers (SiPMs) that capture the optical photons emerging from the scintillation process. For more detailed information on the individual steps of the simulation, we refer the reader to our previous work [4].

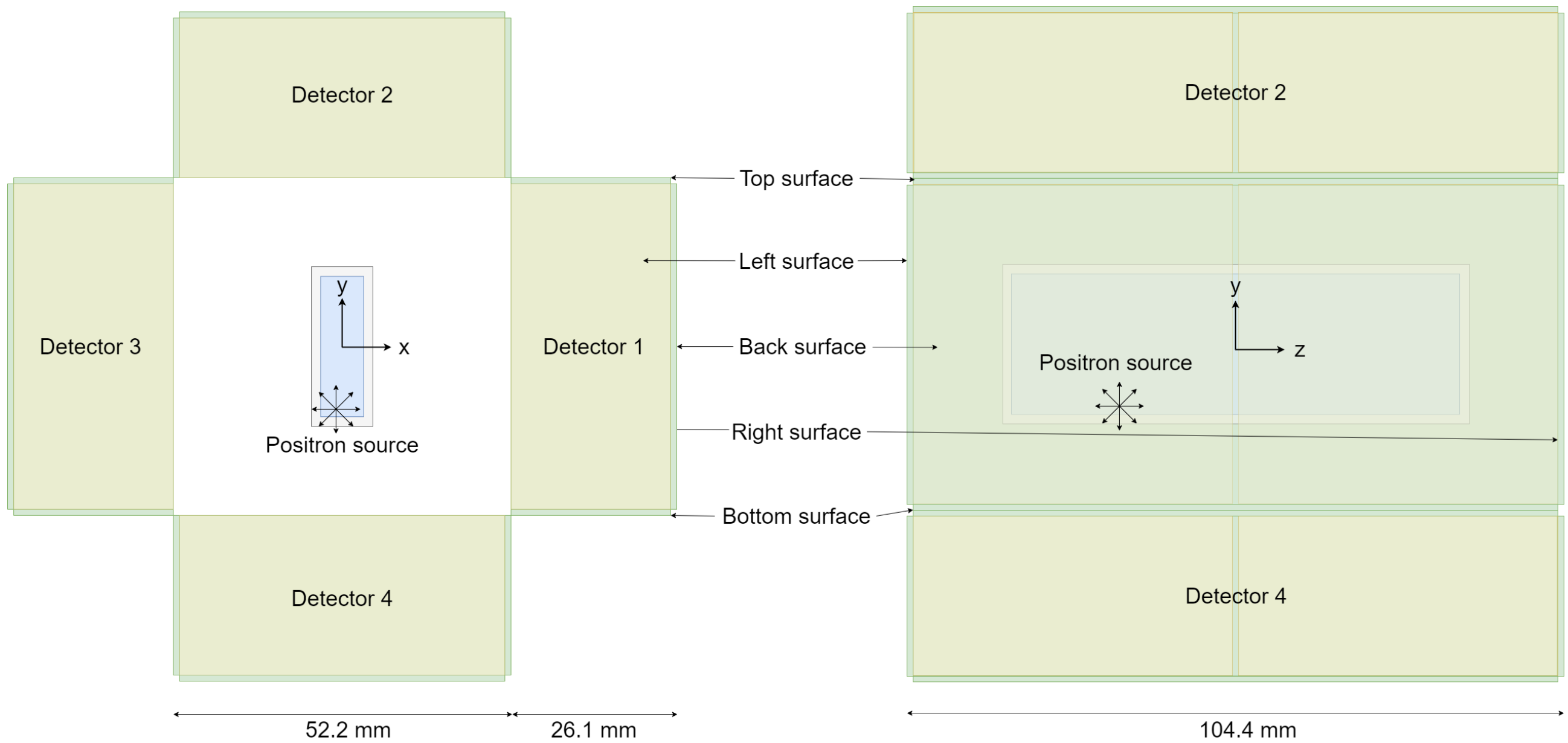

**Figure 1:** Simulation setup viewed from the front and the side. The yellow volumes represent the crystals, the green ones the epoxy layer around the crystal, and the blue one the volume from which the source position is sampled to create the training dataset.

## 2.2 Dataset Creation

The DL-based positron-range correction algorithm should be trained with pairs of reconstructed images, the non-positron-range corrected image as input and the corrected one as output. To do this, we randomly generated 100 placements for the point sources in the compartments of the OOC phantom and ran 200 individual simulations, 100 with F18-positron point sources and 100 with corresponding back-to-back gamma point sources. The simulations with the F18-positrons were used to reconstruct the non-positron-range corrected images, and the simulations with the back-to-back gammas were used to reconstruct the ground truth images. For each simulation, around 800,000 coincidences were recorded.

## 2.3 Reconstruction

To create the pairs of non-corrected and corrected images, we implemented a fully three-dimensional listmode-based MLEM iterative reconstruction algorithm using the Quantitative Emission Tomography Iterative Reconstruction (QETIR) software [6]. We adapted QETIR's reconstruction pipeline to the geometry of the On-Chip PET scanner and converted the GATE coincidences to the listmode-based file format that QETIR requires. The MLEM reconstruction parameters that we used were the following: 200 x 400 x 800 as the image dimensions, 0.1 mm x 0.1 mm x 0.1 mm as the voxel dimensions, five iterations, and four subsets.

## 2.4 Deep Learning-based Positron Range Correction

With the pairs of non-corrected and corrected reconstructed images, we trained an image-to-image model based on the U-Net architecture [7] that predicts the corrected images from the non-corrected one. The dataset of 100 image pairs was randomly split into 80 training and 20 test pairs. We selected a three-dimensional U-Net with leaky ReLU activations, instance normalization, and PixelShuffle [8] upsampling. The rest of the hyperparameters were the following: AdamW [9] as the optimizer, a learning rate of 3e-4 with a cosine-annealing schedule, a weight decay of 1e-6, and pixel-based binary cross-entropy as the loss function. The models were trained with patches of size 64 x 128 x 256 with a batch size of 10 for 1,000 epochs. The training pipeline was implemented with MONAI [10] and PyTorch Lightning [11]. We determined the spatial resolution as the mean Full Width at Half Maximum (FWHM) values of the line profiles drawn through the point sources of the phantom in x-, y-, and z-direction in the reconstructed image. Furthermore, we compared the quality of the predicted images and the ground-truth images using the Peak Signal-to-Noise Ratio (PSNR) and Structural Similarity Index Measure (SSIM) metrics.

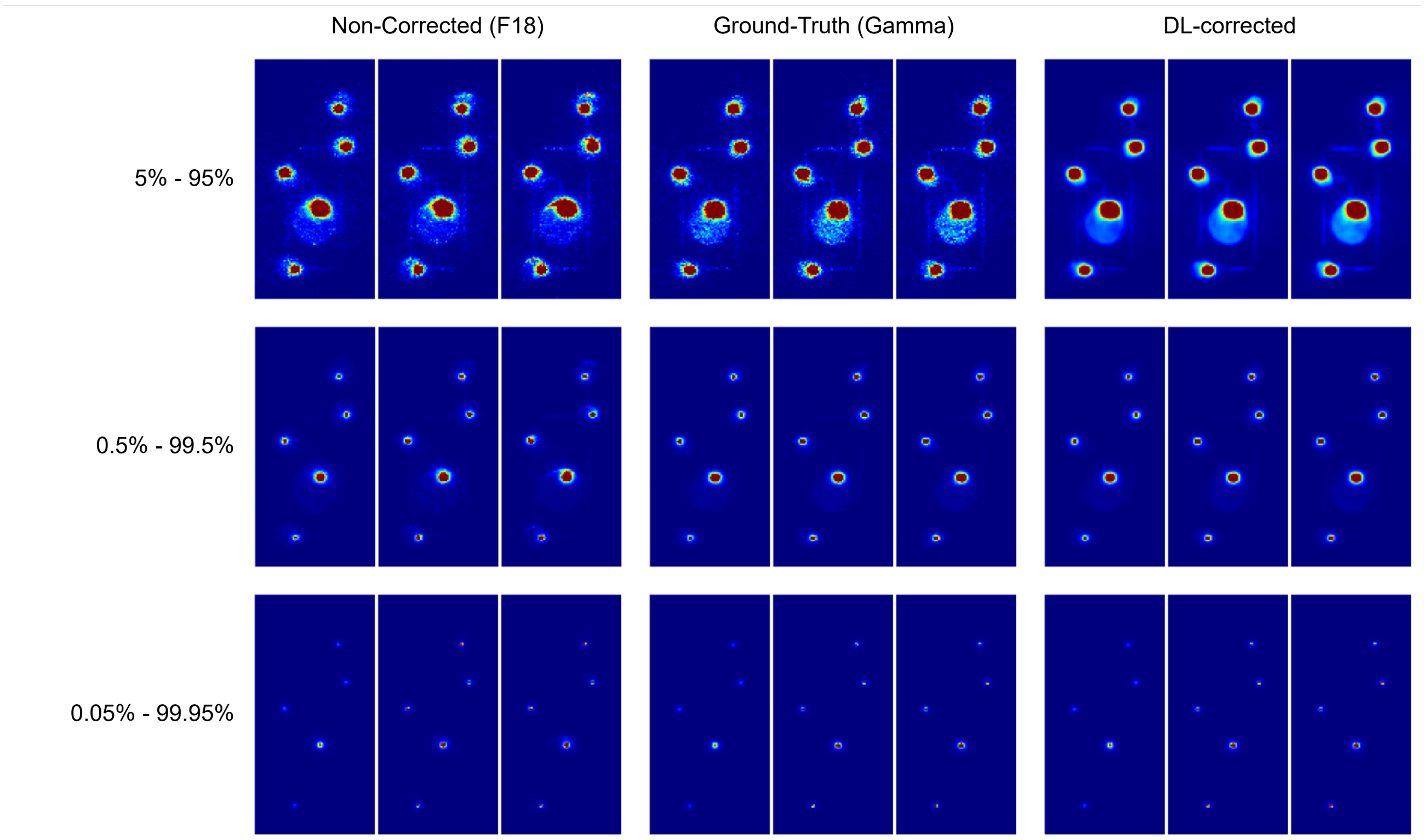

**Figure 2:** Non-positron-range corrected, ground-truth corrected, and predicted reconstructed image for different intensity-percentile ranges. The non-corrected image was created using F18-positron sources, the ground-truth corrected images with back-to-back gamma sources, and the Deep Learning (DL) corrected by the trained model. Each image has dimensions of 200 x 400 x 800 and the middle three slices along the x-dimension are shown. The intensities of the images were scaled to either 5% - 95% , 0.5% - 99.5%, or 0.05% - 99.95% percentile ranges depending on the amount of details that should be visible.

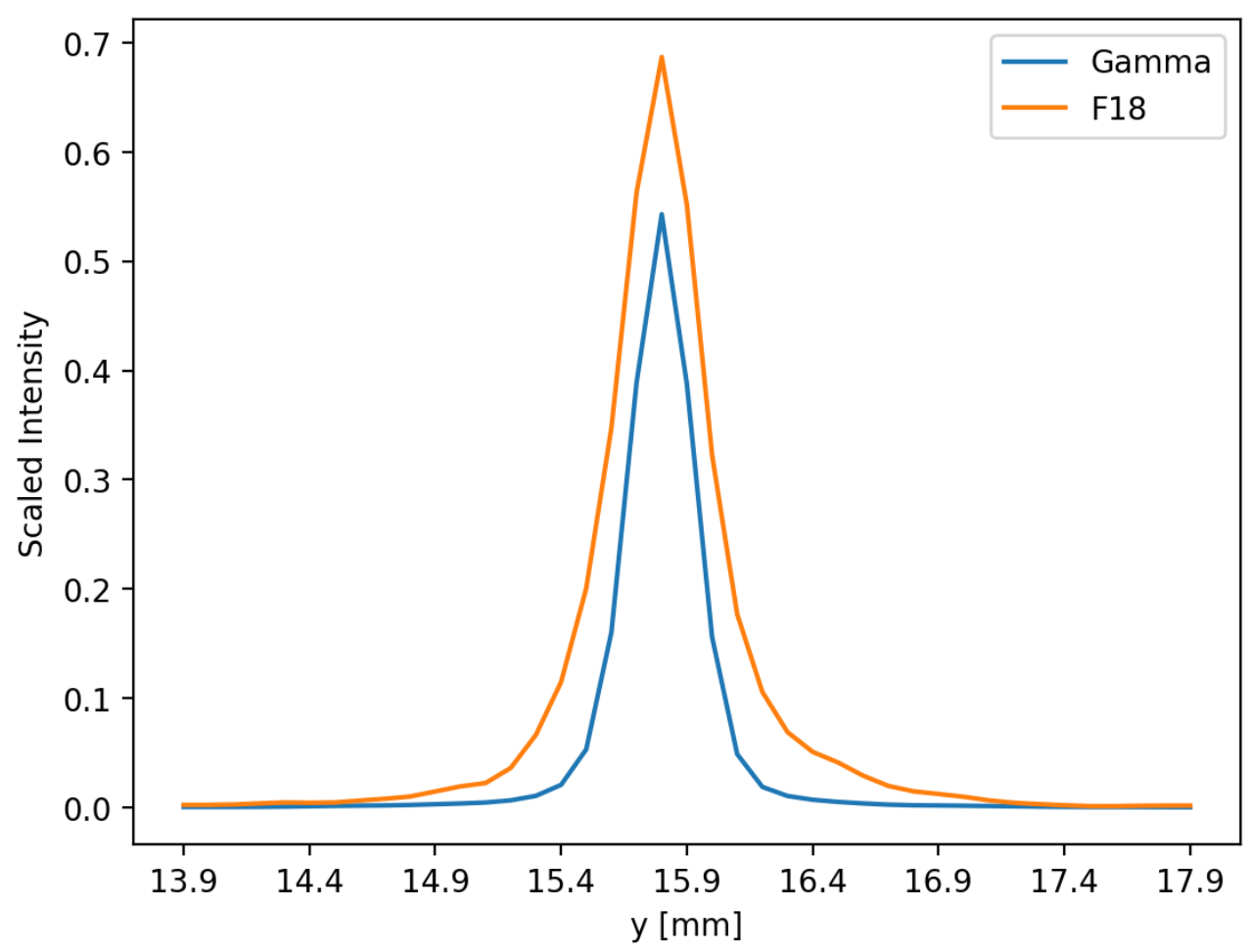

**Figure 3:** Line profiles from a F18-positron source (orange) and back-to-back gama source (blue) drawn through the point source in the large compartment in the transaxial direction. The x-axis depicts the transaxial direction range from 13.9 mm to 17.9 mm. The y-axis represents the scaled intensity.

## 3 Results

In the left column of Figure 2, an example of a non-positron-range corrected reconstructed image from the test dataset is shown, which was created by placing F18-positron point sources inside the compartments of the OOC phantom. In the middle column, the corresponding positron-range corrected image is shown, which was created by using back-to-back gamma point sources instead of the positron ones. The three rows in Figure 2 depict the three different percentile ranges that were used to scale the intensity of the images. When using the range from 5% to 95%, the microfluidic channels between the compartments become visible. For the other two percentile ranges, the focus shifts to details inside the compartment. In the right column of Figure 2, the DL corrected images are shown, which are the direct outputs of the U-Net-like model described in Section 2.4. In Figure 3, two line profiles drawn through the point source in the large compartment in the transaxial direction are shown. The orange one coming from the F18-positron source and the blue one from the gamma point source. Table 1 depicts the mean FWHM values of the non-positron-range corrected, ground truth, and DL-corrected reconstructed images of the test set. They are computed using the line profiles in x-, y-, and z-direction drawn through the point sources in each compartment. In Table 2, the PSNR and SSIM metrics for the three different intensity-scaling percentile ranges of the images in the test set are shown.

## 4 Discussion

The results of this study demonstrate the effectiveness of the DL-based positron-range correction algorithm in improving the overall quality of the reconstructed images. The positron-range effect is clearly visible when comparing the images from the left and middle column of Figure 2 and the line profiles in Figure 3. Especially in the large compartment, but also in some of the smaller ones, a degradation near the borders of the foreground activity can be observed. The DL-based approach improved the spatial resolution of the reconstructed images in the test set from FWHM values of 0.260 mm in the non-corrected images to 0.177 mm in the corrected ones. Our approach improves the spatial resolution by almost 32%, which is more than 91% of the maximal achievable improvement of a FWHM value of 0.169 mm coming from the back-to-back gamma source images. From the right column of Figure 2 showing the DL-corrected images, we can observe that the model is not only able to correct for the positron range effect by removing the resulting artifacts in the large and smaller compartments, but it is also able to reduce the noise level in the images from the small percentile range.

| Non-Corrected | Ground-Truth | DL-corrected |
|---|---|---|
| 0.260 mm | 0.169 mm | 0.177 mm |

**Table 1:** Mean FWHM values of the non-positron-range corrected, ground truth, and DL-corrected reconstructed images of the test set. To compute the FWHM values, line profiles in the three spatial directions are drawn through the point sources in each compartment.

| Percentile Range [%] | Test PSNR [dB] | Test SSIM |
|---|---|---|
| 5 - 95 | 31.2 | 0.852 |
| 0.5 - 99.5 | 48.9 | 0.994 |
| 0.05 - 99.95 | 58.7 | 0.997 |

**Table 2:** Peak Signal-to-Noise Ratio (PSNR) and Structural Similarity Index Measure (SSIM) of images in the test set shown for different intensity-scaling percentile range.

One of the main advantages of our approach is that it is purely data-driven, which means that it is agnostic to the type of radionuclide used. This makes it versatile and applicable to a wide range of imaging scenarios. Additionally, the use of a realistic OOC phantom in the simulation of the dataset and the implementation of a fully three-dimensional MLEM reconstruction algorithm ensure that the dataset is representative of real-world imaging scenarios.

Some limitations of this preliminary work include the small size of the dataset the model was trained on, with only 100 pairs of non-corrected and corrected images in total. To address this issue in future work, we plan to simulate more cases, and also extend the work to include other radionuclides with larger positron ranges. This will add more diversity to the dataset and therefore make it harder for the model to perform the correction. Additionally, it will also be important to test the model on real experimental data in order to validate its performance in a real-world setting.

## 5 Conclusion

In this work, we presented a novel approach for correcting the positron-range effect in an On-Chip PET scanner using a DL-based algorithm. We created a dataset of pairs of non-corrected and corrected images using a Monte-Carlo simulation of a realistic OOC phantom and a fully three-dimensional MLEM iterative reconstruction algorithm. We trained an image-to-image model based on the U-Net architecture to predict the corrected images from the non-corrected ones. Our results demonstrate the effectiveness of the DL-based positron-range correction algorithm in improving the overall quality of the reconstructed images. This approach has the potential to advance the study of 3D models in radiopharmaceutical research and provide a valuable tool for radio pharmacists in the development of radiotheranostics. The overall goal of this work is to provide an imaging device that can enable more detailed and accurate imaging of OOCs, facilitating the advancement of this technology and its applications in drug development and disease modeling.

# An accretion method to regularize digital breast tomosynthesis reconstruction

Leonardo Coito[1], Koen Michielsen[1], and Ioannis Sechopoulos[1, 2, 3]

[1]Department of Medical Imaging, Radboud University Medical Center, Nijmegen, The Netherlands.
[2]Technical Medicine Centre, University of Twente, Enschede, The Netherlands.
[3]Dutch Expert Centre for Screening (LRCB), Nijmegen, The Netherlands.

**Abstract** Digital breast tomosynthesis suffers from limited angle artifacts, which means that the representation of the distribution of fibro-glandular tissue is severely limited in the direction of the missing data. To recover the location of the tissue in the image, we present a regularization method based on the accretion of fibro-glandular tissue that is inspired by the particle accretion phenomenon with attractive interactions. The method is combined with a polychromatic reconstruction algorithm with material decomposition.

The key ingredient of our approach is the correspondence between the fraction of fibro-glandular tissue at each voxel with a number of particles with attractive interactions. In this manner, the particles within one voxel would be attracted to those of a neighbor, generating a redistribution among voxels and the tissue patterns present in breast phantoms.

Reconstruction performance was analyzed quantitative over 65 two-dimensional phantom slices. It was found that the relative error of the reconstructed glandularity was 11.4% on average, ranging between $-48.8\%$ and $+32.4\%$, for glandularity values ranging between 0.07 and 0.62. The Dice similarity coefficient of the fibro-glandular structures in the phantoms after reconstruction was 0.57 on average, ranging between 0.20 and 0.90.

Results indicate that the presented accretion method has shown to improve the results of the iterative reconstruction method by localizing some of the large structures of fibro-glandular tissue in the phantoms studied.

## 1 Introduction

Digital breast tomosynthesis (DBT) overcomes some of the limitations of mammography by acquiring several low-dose planar x-ray projections over a limited angular range. This allows for the reconstruction of a pseudo-3D representation of the breast. However, the sparsity of the sampling and the limited number of projections give rise to artifacts, such as the stretching of features along the direction of the acquisitions in the set of projections. This artifact results in an incomplete separation between overlapping features and in the challenging recovery of their real shape and extent [1, 2].

Iterative algorithms have been developed to alleviate the artifacts by applying constraints over the images, e.g., on the image total variation [3–6] or on the 0-norm of image gradient [7]. Deep learning approaches [8, 9] appear to be able to recover information lost in the limited angle acquisition based on learned prior knowledge on the expected tissue distributions, although they require large amounts of training data and their generalization is not completely clear.

Therefore, accurate image reconstruction from data acquired over limited-angular ranges remains an appealing area of research, and in this work we present a regularization method that is able to recover the distribution of larger fibro-glandular structures by applying a clustering algorithm inspired by planetary accretion mechanics, see e.g. Ref. [10].

## 2 Method

### 2.1 Description

The knowledge that the breast is almost entirely composed of only two tissue types, namely adipose and fibro-glandular tissues, could provide a strong constraint when solving the limited angle reconstruction problem. Directly applying a discrete tomographic reconstruction allowing only these two components is however too limiting due to the presence of skin, calcifications, and possibly masses in a small fraction of the reconstructed volume, in addition to the expected biological variability of the density of adipose and fibro-glandular tissues.

Since such direct assignment of a unique tissue type to each voxel effectively creates clusters of a specific tissue, we were inspired by the phenomenon of planetary accretion to achieve a similar effect in the fibro-glandular tissue as a method to counter the limited angle effect to spread attenuation outside of its actual location. This clustering is then achieved by including attractive forces that exist among individual elements in a system and results in their collision and combination to form larger structures.

To allow straightforward application of such regularization on only the fibro-glandular tissue of the breast, we use a polychromatic iterative method published by Bustamante et al. [11], which optimizes a maximum likelihood cost function with material decomposition that allows for the separation of tissues into a set of base materials, specifically adipose and fibro-glandular tissues for the current application. The forward model is shown in Eq. (1), where $\hat{y}_i$ is the expected value of projection line $i$, $l_{ij}$ the intersection between projection line $i$ and voxel $j$ and $I_0^{i,e}$ the energy fluence source spectrum for the projection. The attenuation of voxel $j$ at energy $e$, $\mu_j^e$, is parameterized by the attenuation of the base materials, $\mu_a^e$, and the fraction of these materials constituting the voxel $w_{a,j}$: $\mu_j^e = \sum_a \mu_a^e w_{a,j}$. Although mathematically there are no specific constraints on the weights $w$, for the reconstruction of breast tissue we expect most values to represent adipose–fibro-glandular

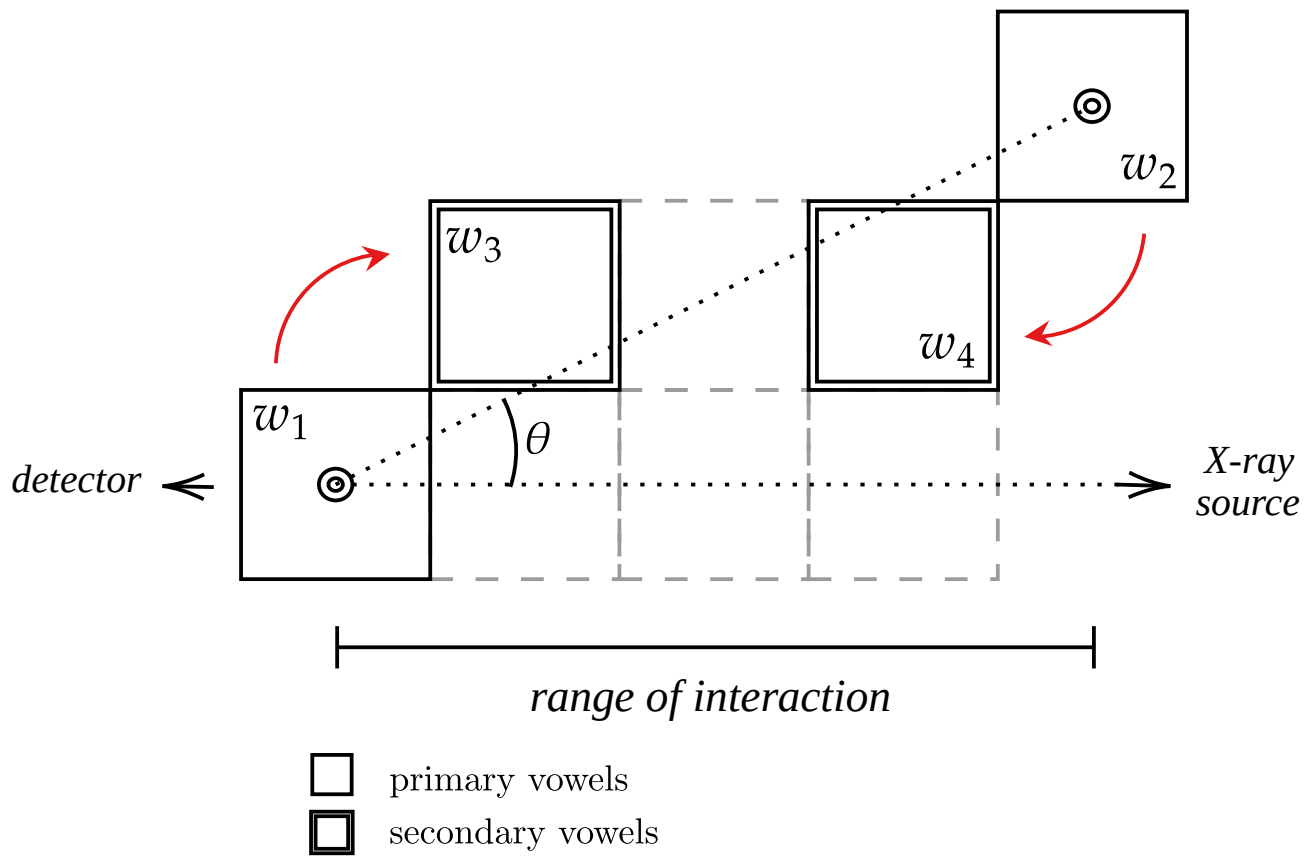

**Figure 1:** Representation of the primary and secondary voxels. The red arrows indicate the direction of the accretion.

mixtures and thus yield values in the interval $[0,1]$.

$$\hat{y}_i(\vec{w}) = \sum_e I_0^{i,e} \exp\left(-\sum_a \mu_a^e \sum_j l_{ij} w_{aj}\right) \tag{1}$$

The main principle in our approach is the representation of the fraction of fibro-glandular tissue at each voxel with a number of particles with attractive interactions. In this manner, the particles of one voxel would be attracted to those of a neighbor, generating redistribution among voxels and, more specifically, accumulation of fibro-glandular and adipose tissues.

The method is applied by alternating updates from the polychromatic iteration and the accretion process as follows. At each iteration, and after the polychromatic reconstruction update, a set of pairs of voxels are chosen to interact, which we dubbed *primary voxels*. First, let us consider one pair of these voxels, named 1 and 2, with fibro-glandular (fg) components $w_{a=fg,j=1,2}$, and the closest neighboring voxels along the line that joins them, 3 and 4, which we dubbed *secondary voxels*, as illustrated in Fig. 1. To simplify notation, here on we omit the material index $a$, as it only correspond to $fg$. Then the accretion operation is driven by the updates

$$\begin{aligned} w_1 &\rightarrow w_1 - \Delta_1, \\ w_2 &\rightarrow w_2 - \Delta_2, \\ w_3 &\rightarrow w_3 + \Delta_1, \\ w_4 &\rightarrow w_4 + \Delta_2, \end{aligned}$$

with

$$\Delta_1 = w_1 w_2 (1 - w_3), \quad \Delta_2 = w_1 w_2 (1 - w_4), \tag{2}$$

The red arrows in the figure represent the particle's flow, given by $\Delta_j$, as their definition constraints the values of $w_j$ to the interval $[0,1]$. Next, for each pair of primary voxels, one is randomly chosen while the other is considered from the parameters of the accretion operation:

| Stage | $n_{iter}$ | $n_{pairs}$ | $\delta_{int}$ (voxels) | $\sigma_\theta$ (°) |
|---|---|---|---|---|
| 1 | $10^4$ | 100 | 20 | 25 |
| 2 | 500 | $10^4$ | 10 | 5 |

**Table 1:** Accretion parameters used in the evaluation.

- **Interacting angle $\theta$:** the angle subtended by the line joining the primary voxels is obtained from a Gaussian distribution with average $\bar{\theta} = 0°$ and standard deviation $\sigma_\theta$. $\theta = 0°$ is the main angle of projections.
- **Range of interaction $\delta_{int}$:** the maximum distance between interacting voxels can slightly define the size of the resulting structures.
- **Number of interacting pairs $n_{pairs}$:** the number of pairs of primary voxels evaluated at each iteration is modified according to the balance between the data-driven and accretion processes.

Each of these accretion parameters can be chosen freely. For the evaluation described in the following section, we employed the values listed in Table 1, where $n_{iter}$ is the number of iterations. A relatively small rate of accretion, $n_{pairs}/n_{iter} \ll 1$, improves the clustering in the correct places as it allows the data-driven adjustments to have an important effect.

## 2.2 Simulation and Evaluation

We evaluated our method using 65 two-dimensional slices extracted from 22 digital phantoms based on segmented breast CT patient scans that where then mechanically compressed using a finite element model [12]. The phantoms have an isotropic pixel size of 0.273 mm. The tissues are classified as either adipose, fibro-glandular, or skin.

Simulated acquisitions were performed for a wide angle breast tomosynthesis geometry with 25 projections over a 50 degrees angular range ($-25$ to 25 degrees), source-detector distance of 647 mm, source-center of rotation distance of 600 mm and detector spacing of 0.1 mm. The acquisition spectra alternated between one with a tube voltage of 30 kV and filtered by 50 $\mu$m of rhodium, and one with tube voltage of 49 kV and filtered by 1 mm of titanium over the 25 projection angles.

We applied our method in a multigrid approach with the application of an average pooling operation in the projection data. The first stage is a low resolution reconstruction (rebinning of 5x5 pixels) and is responsible for the relatively large-sized clusterings of fibro-glandular tissue, while the second stage is in full resolution and develops smaller features. A monotonically decreasing range of interaction is selected along the stages in order to adjust the formation of the specified accretion, $\delta_{int} = 20 \rightarrow 10$ as in Table 1. Furthermore, we focus the interactions in the mean angle of projections with the aim of correcting the stretching artifact at angle $\theta = 0°$. In addition to visual evaluation of the reconstructions, both the amount of fibro-glandular tissue recovered and its location were evaluated. The former by comparison of the reconstructed glandularity to that of the original phantom, and the latter by calculating the Dice similarity coefficient (DSC, Eq. (3)) between the fibro-glandular components of the reconstruction (RE) and the original phantom (GT).

$$DSC = \frac{2(RE \cap GT)}{RE + GT} \tag{3}$$

## 3 Results

Two examples of the evaluated phantoms are shown in Fig. 2, including the ground truth, the reconstruction without accretion for comparison purposes, and the two stages of reconstruction explained in Sec. 2. As can be seen, the accretion in the first stage accounts for the large clustering along the mean projection angles, see Fig. 2c, while the following stage leads to more subtle changes in finding the relatively large areas of fibro-glandular tissue, see Fig. 2d. The difference between the ground truth and the reconstruction is shown in Fig. 2e, where it is possible to observe that the relatively smaller features are missed.

An example where the method performs comparably poorly is shown on the right column of Fig. 2, where not all large structures are encountered and the appearance of some incorrectly placed fibro-glandular structures can be seen.

Quantitative evaluation over 65 two-dimensional slices from 22 phantoms found that the relative error of the reconstructed glandularity was 11.4% on average, ranging between $-48.8\%$ and $+32.4\%$, for glandularity values ranging between 0.07 and 0.62. The Dice similarity coefficient of the fibro-glandular structures in the phantoms after reconstruction was 0.57 on average, ranging between 0.20 and 0.90.

## 4 Discussion & Conclusion

The presented accretion method has shown to visually improve the results of the iterative reconstruction method by localizing some of the large structures of fibro-glandular tissue in the phantoms studied and the total amount of reconstructed fibro-glandular tissue closely matched the ground truth.

However, the output was found to have differences across the features present, with it being not completely successful in reconstructing smaller fibro-glandular features of the phantoms.

In principle, our method is suitable for a direct extension to 3D reconstruction, without much increase in computational cost since the distribution of the set of attracting voxel pairs remains preferentially aligned to the direction of the lines tracing from source to each detector pixel.

We do not expect this method to be able to recover the fine fibro-glandular structures present, however, as we have shown previously [8], being able to localize the bulk of the fibro-glandular tissue in a limited angle reconstruction can be useful for accurate measurements of patient specific breast density and Monte-Carlo dose estimation.

Even though the method has lower accuracy than our learned reconstruction, the fact that no training data is needed, since this is not a learned method, makes this new approach an important tool.

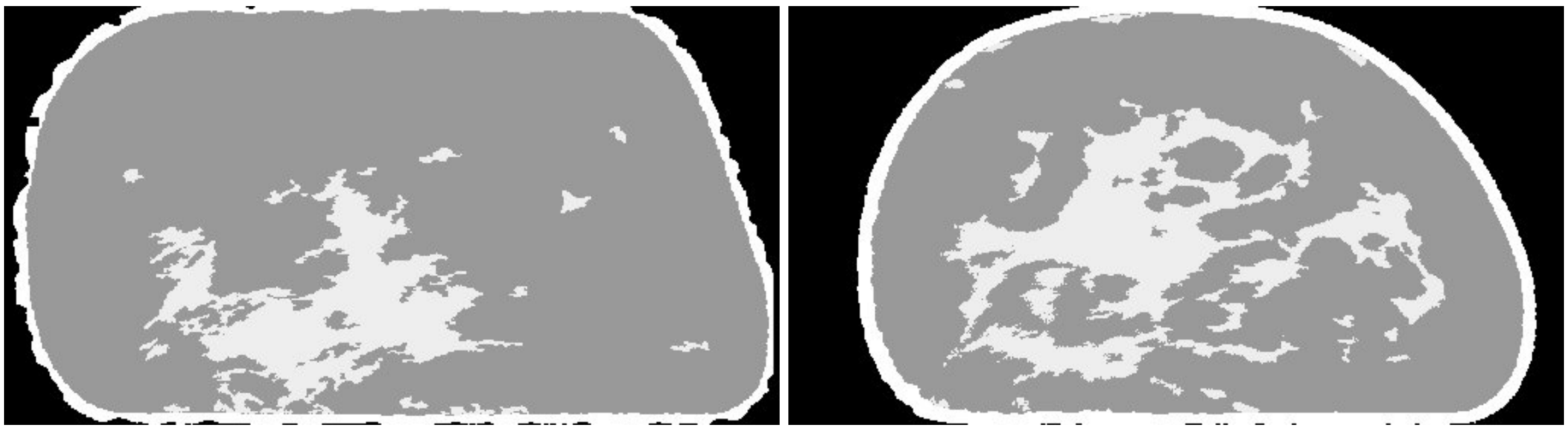

**(a)** Ground truth

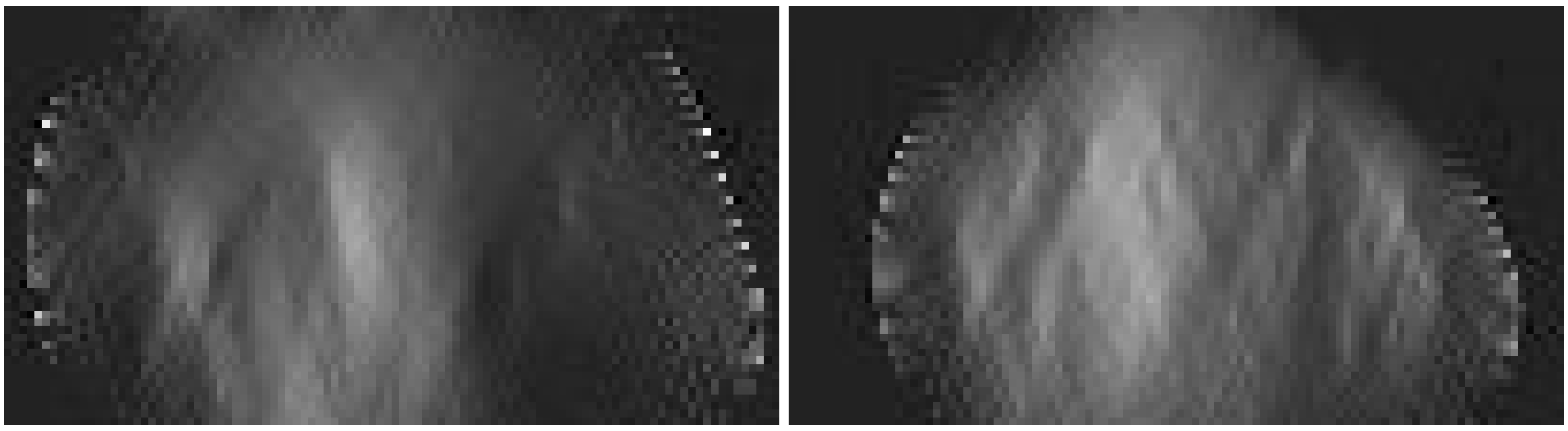

**(b)** No accretion

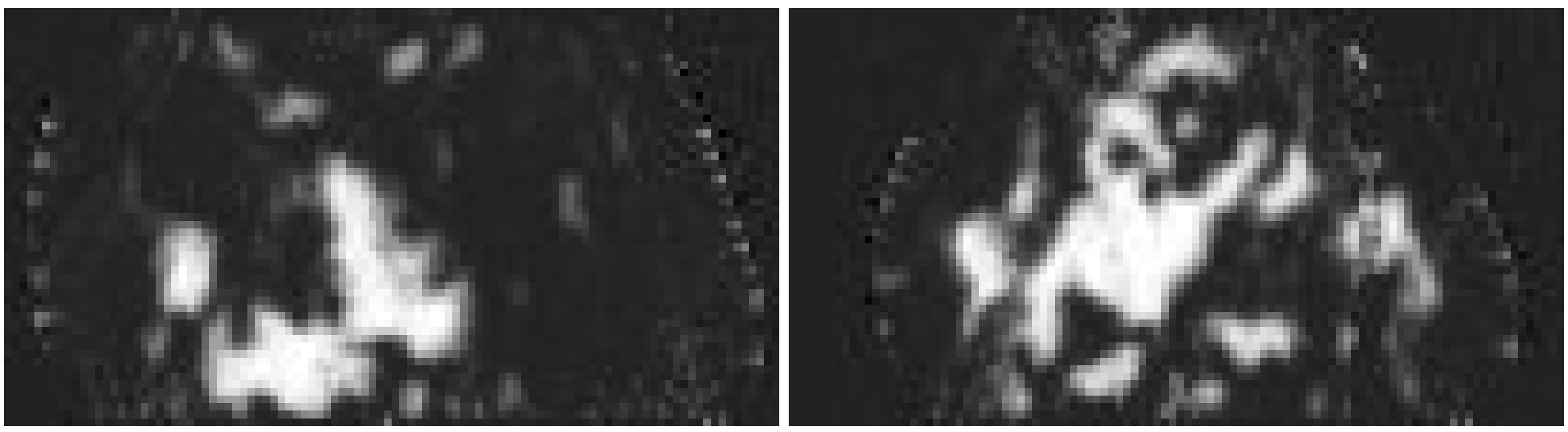

**(c)** First stage

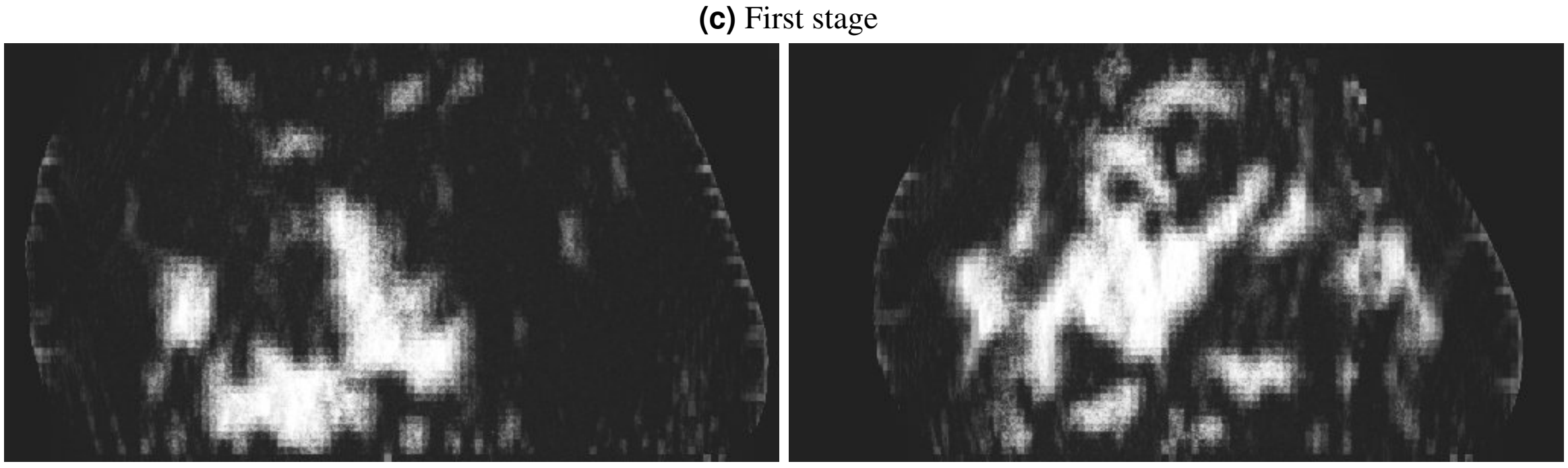

**(d)** Second stage

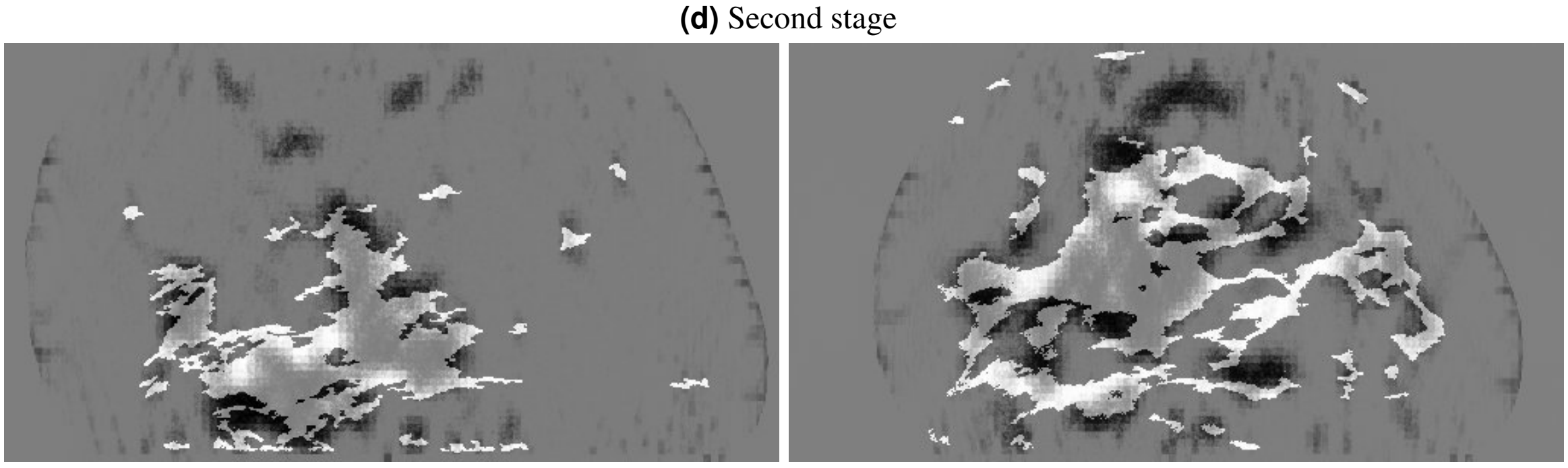

**(e)** Difference between GT and RE

**Figure 2:** Examples of two reconstructions of phantoms with the accretion method. The fraction of fibro-glandular tissue is represented by the white color in Fig. 2a and gray-scale in Figs. 2b, 2c and 2d, in a range from 0 (black) to +1 (white). In Fig. 2e, the range is set from -1 (black) to +1 (white). The DSC values obtained were 0.64 and 0.56 for the left and right columns respectively.

# Deep Learning-based Stopping Power Ratio (SPR) Estimation for Proton Therapy

[1]Wenxiang Cong, [2]Manudeep K. Kalra, [3]Harald Paganetti, [1]Ge Wang
[1]Biomedical Imaging Center, Department of Biomedical Engineering, Rensselaer Polytechnic Institute, Troy, NY 12180
[2]Division of Thoracic Imaging & Intervention, Massachusetts General Hospital, Boston, MA 02114
[3]Department of Radiation Oncology, Massachusetts General Hospital, Boston, MA 02114

**Abstract** Proton therapy is to maximize radiation dose on a tumor site while minimizing radiation effects to surrounding vital organs in a patient. When protons travel through tissues, a pronounced Bragg peak occurs in the energy loss curve of ionizing radiation. The position of the Bragg peak is usually estimated by the stopping power ratio (SPR). The proton therapy is to realize optimal conformity of the high dose deposit with the tumor. The quality of the proton therapy relies on accuracy of the Bragg peak estimation and reliable dose planning. In current clinical practice, the standard stoichiometric calibration method is susceptible to tissue composition variations, and often carries ~3.5% range uncertainty. In this study, we propose a deep learning method to directly map a clinical CT image to corresponding SPR map via an improved ResNet (imResNet) model. Specifically, based on the elemental composition of human tissues, material decomposition is performed from dual-energy CT (DECT) to generate a training dataset to optimize the parameters of imResNet model in the supervised fashion. In this pilot study, we evaluate the proposed method on simulated and clinical CT images, showing that the method achieves accurate SPR calculation.

## 1 Introduction

Proton-matter interaction produces a special dose distribution, releasing most of energy at the Bragg peak, the distal end of the beam range. The proton therapy is to realize optimal conformity of the high dose deposit with the tumor [1]. The quality of proton therapy depends on the accuracy of the Bragg peak prediction, which is estimated by calculating the stopping power ratio (SPR). Inaccurate SPR values cause a range shift of the proton beam, and result in proton dose distribution error and insufficiently treatment. Schneider et al. proposed a stoichiometric calibration method to convert single energy CT (SECT) images to SPR via a calibration curve [2]. However, this calibration method yields an uncertainty of 3–3.5% in the proton range [3]. Virtual monochromatic (VM) images are derived from dual-energy CT (DECT) to reduce beam-hardening artifacts and improve CT number accuracy relative to SECT images [4-6]. To minimize the impact of range uncertainty on treatment plans, dual-energy CT (DECT) was introduced to photon therapy to provide material-specific information by decomposing energy-dependent photoelectric absorption and Compton scattering [7]. Thus, SPR can be calculated on the basis of the Bethe-Bloch formula using the relative electron density and mean excitation energy of tissues [8]. Hudobivnik et al. demonstrated a higher accuracy in SPR derived from DECT than from SECT [9].

It is worth mentioning that SPR estimation via DECT technique is to establish an empirical model to estimate the mean excitation energy in terms of the effective atomic number [7]. The relationship between the mean excitation energy and the effective atomic number is rather complicated and cannot be accurately described by an analytical formula due to different physical characteristics of the two physical quantities [3]. The empirical analytic model often generates considerable errors in the estimated mean excitation energy, compromising the accuracy of SPR estimation.

Emerging deep learning is a powerful technique to perform various types of uncertainty estimation and data modeling through learning and inference in a data-driven fashion [10]. A representative learning-based approach [11] for proton therapy first synthesizes DECT images from SECT images [5], and then computes SPR images from the synthetic DECT image. In this paper, we develop a deep learning approach to directly convert a SECT image to corresponding SPR map. Chemical compositions of human tissues are of importance in calculating the dosimetric distribution in the patient irradiated with radiation [12]. Based on the elemental composition of human tissue, accurate material decompositions are performed using VM images to calculate physical parameters of human tissues, including the electron density, effective atomic number, mean excitation energy, and mass attenuation coefficient of tissues. These physical parameters provide an accurate calculation of SPR and attenuation coefficient of tissues, which can be used to establish training datasets to optimize the parameters of imResNet model in the supervised fashion. Through the imResNet model, clinical SECT images can be converted to corresponding SPR map directly. This method differs significantly from both conventional DECT-based SPR estimation and learning-based synthetic DECT methods [11]. In the next section, we describe our methodology for the SPR estimation. In the third section, we present our representative results. Finally, we discuss relevant issues and conclude the paper in the last section.

## 2 Materials and Methods

**2.1. Convolution Network Architecture:** The convolution neural network (CNN) is a popular architecture for image processing [13]. However, training a deep CNN network often suffers from vanishing/diverging gradients [14]. The residual neural network (ResNet) is an effective architecture, which greatly facilitates extraction of complex and subtle features from data. With use of the shortcut, ResNet allows the network training process to converge more stable and faster than without the shortcut.

For modeling a complicated nonlinear relationship between clinical SECT images and SPR images, here we use an improved ResNet (imResNet) architecture [5]. The imResNet performs convolution for 3D images $\left(1, I, I^2, I^3\right)$ formed from an input 2D CT image $I$ with a filter of 4×1×1 in the first layer to introduce nonlinear terms of input data. The next layers in the imResNet network are 2 residual blocks of 3 convolution layers with 64 filters of 7×7 kernels, followed by 2 residual blocks of 3 convolution layers with 64 filters of 5×5 kernels, and 2 residual blocks of 3 convolution layers with 64 filters of 3×3 kernels. Each residual block performs feed forward processing with shortcut connections skipping 3 layers to implement an identity map. Then, a convolution layer with 64 filters of 3×3 kernels is performed, followed by a convolution layer with 32 filters of 3×3 kernels, and a last layer generates a feature map with a single 3×3 filter as the output. Every layer is done by a ReLU activation function. The network architecture is shown in Figure 1.

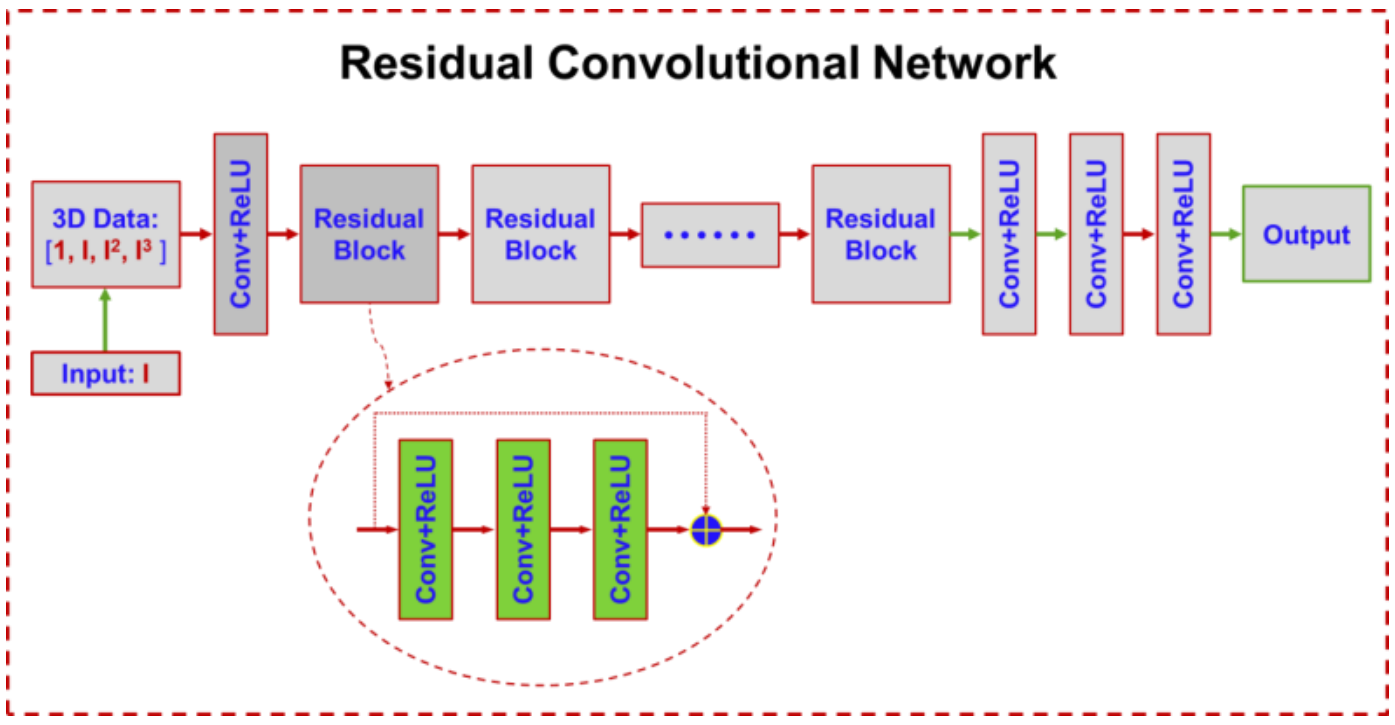

**Fig. 1.** Architecture of the imResNet model with non-linearized input terms.

**2.2. Building training dataset:** Virtual monochromatic (VM) CT images are derived from DECT images to provide quantitative information on tissue composition [15]. At the molecular level, tissues are composed of the basis molecular substances [12, 16]. The elemental composition of body tissue is determined by the relative weight fractions of the basis molecular substances: water ($H_2O$, density of 1.0g/cm$^3$), lipid ($C_{51}H_{98}O_6$, density of 0.9g/cm$^3$), protein ($C_{100}H_{159}N_{26}O_{32}S_{0.7}$, density of 1.34g/cm$^3$), carbohydrate ($C_6H_{10}O_5$, density of 1.52 g/cm$^3$), and mineral ($Ca_3(PO_4)_2$, density of 3.14g/cm$^3$) [12, 16]. The molecular formula of a basis molecular substance contains information about its mass density, atomic number, and atomic weight. Thus, the relative electron density and mean excitation energy of tissues can be accurately estimated from following formulae, thereby calculating the SPR exactly.

The X-ray mass attenuation coefficient of a tissue can be expressed as a composition of basis molecular substances [17]:

$$\begin{aligned}\mu(E)/\rho = {} & w_{\text{water}}\, \mu_{\text{water}}(E)/\rho_{\text{water}} + w_{\text{lipid}}\, \mu_{\text{lipid}}(E)/\rho_{\text{lipid}} \\ & + w_{\text{protein}}\, \mu_{\text{protein}}(E)/\rho_{\text{protein}} + w_{\text{minerals}}\, \mu_{\text{mineral}}(E)/\rho_{\text{mineral}}\end{aligned} \quad (1)$$

The weight fraction of respective components satisfies the normalized condition:

$$w_{\text{water}} + w_{\text{lipid}} + w_{\text{protein}} + w_{\text{mineral}} = 1. \quad (2)$$

VM CT images can be applied to Eq. (1) to form a system of linear equations with respect to weight fractions. The system of linear equations can be solved using an optimization method to determine relative weight fractions of the basis molecular substances [5].

Furthermore, the electron density $\rho_e$ of mixture materials relative to water can be computed from the atomic number and atomic weight of each elemental component [8]:

$$\rho_e = \sum_{i \in mixture} \rho_i \frac{w_i Z_i}{A_i} \Big/ \rho_w \sum_{i \in water} \frac{w_i Z_i}{A_i}, \quad (3)$$

where $\rho_i$, $w_i$, $Z_i$ and $A_i$ are the mass density, elemental weight fraction, atomic number, and atomic weight of element component in a substance, respectively. Thus, the mean excitation energy of tissues is calculated based on the Bragg additivity rule [8]:

$$\ln(I) = \sum_{i \in mixture} \frac{w_i Z_i}{A_i} \ln(I_i) \Big/ \sum_{i \in mixture} \frac{w_i Z_i}{A_i} \quad (4)$$

Therefore, the stopping power ratio (SPR) of intervening tissues can be computed from relative electron density and mean excitation energy using the Bethe-Bloch equation [7]:

$$SPR = \rho_e \frac{\ln\left(2m_e c^2\right) + \ln\left(\beta^2/\left(1-\beta^2\right)\right) - \beta^2 - \ln(I)}{\ln\left(2m_e c^2\right) + \ln\left(\beta^2/\left(1-\beta^2\right)\right) - \beta^2 - \ln(I_w)}, \quad (5)$$

where $m_e c^2$ is the rest mass energy of the electron, $\beta$ is the speed of the proton relative to light speed, $I_w$ is the mean excitation energy of water, $I$ is the mean excitation energy of the tissue, and $\rho_e$ is the relative electron density of the tissue with respect to water.

On the other hand, the linear attenuation coefficient of tissues can be calculated from these physical parameters using the following formula [18, 19]:

$$\mu(E) = \rho \sum_{i \in mixture} \frac{w_i Z_i}{A_i} \begin{pmatrix} k_p \sigma_p(E) Z_i^{3.62} + k_h \sigma_h(E) Z_i^{1.86} \\ + k_c \sigma_c(E) \end{pmatrix}, \quad (6)$$

where $\sigma_P(E)$, $\sigma_h(E)$ and $\sigma_c(E)$ are energy-dependent coefficients describing photoelectric interaction, coherent scattering and Compton scattering, respectively [20, 21].

Based on the elemental composition of human tissue, both linear attenuation coefficients and SPR values can be accurately calculated, generating datasets for the supervised learning to modeling the relationship between two physical quantities.

## 3 Results

The dual-energy CT (DECT) datasets of 10 patients were obtained from Massachusetts General Hospital (MGH, Boston, MA). The DECT scanner was operated at the dual-source scanning mode to acquire two raw datasets at 80kVp and 140kVp respectively. Then, DECT images were reconstructed and further converted into virtual monochromatic (VM) CT images at multiple energy levels from 60keV to 120keV in an increment of 10keV. We generated training dataset using the method described in the preceding section. Training a network was a process of optimizing specific kernels and weights in convolution layers to minimize a loss function, which was defined as $l_1$ norm to evaluate errors between output predictions and ground truth labels on a training dataset. For computational efficiency, the imResNet model was trained using image patches with 64×64. The standard training cycle was followed through the training, validation and testing stages.

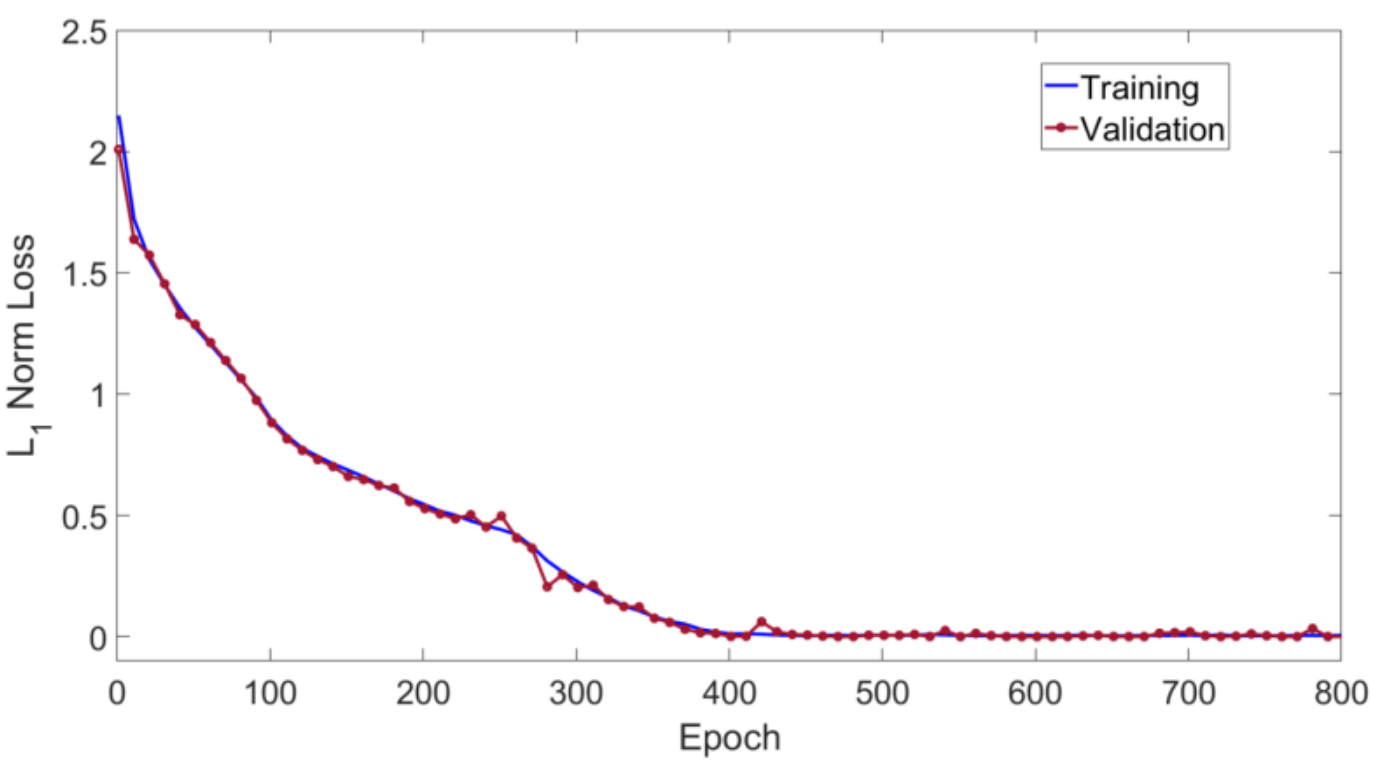

**Fig. 2**. Convergence of the imResNet model training measured with the Manhattan norm loss versus the number of epochs during the training process.

The training procedure of the imResNet model was programmed in Python and the Tensorflow on a PC computer with a NVIDIA Titan XP GPU of 12 GB memory. The network parameters in the convolution kernels were initialized from the Gaussian distribution with zero mean and variance of 0.01. The loss function was minimized using the adaptive moment estimation (ADAM) optimizer with a learning rate of $10^{-4}$ and decay rates of $\beta_1$=0.9 and $\beta_2$=0.999. Data were randomly sampled from the training dataset to maximize the probability of finding the global minimum. The network was trained with 800 epochs, which took about 12 hours. The loss function decreased consistently, showing an excellent convergence and stability. Fig. 2 shows the averaged Manhattan norm loss versus the number of epochs. Fig. 3 presents representative SPR images produced using our trained imResNet model against the ground truth SPR images calculated from the elemental composition data. It can be observed that the trained imResNet delivered high-quality SPR images, with an average relative error of less than 1% in the testing phase. The peak signal-to-noise ratio (PSNR) and structural similarity index measure (SSIM) were used to evaluate the quality of reconstructed SPR images. With the theoretical SPR image as the reference, we calculated PSNR for the reconstructed SPR images, achieving an average PSNR of 55.88±0.125 ($p<0.05$). Also, SSIM was calculated to quantify the similarity between the reconstructed SPR image and the reference image, achieving an average SSIM of 0.9991±0.0018 ($p<0.05$). The results show that the structural information especially texture features are well preserved in the reconstructed SPR images and this method effectively copes tissue composition variations.

## 4 Discussions

DECT images in clinical practice provide material specific information to reflect composition variations of various tissues in the human body. DECT calculates virtual monochromatic (VM) CT images, and extracts electron density and effective atomic number of tissues on CT images pixel-wisely [15]. According to the molecular model of human tissues, the composition of body tissues can be expressed as a linear combination of basis substances. The weight fraction of each basis substance can be determined from VM CT images to calculate elemental composition data of human tissues. Chemical compositions of basis molecular substances can be used to accurately calculate relative electron density and mean excitation energy of tissues for the SPR calculation. Therefore, linear attenuation coefficients and SPR values can be perfectly paired on DECT images pixel-wisely, and generate datasets to train the imResNet model in the supervised learning fashion.

Deep learning-based methods have been introduced for the SPR calculation. A typical learning calculation method of the SPR [11] is to synthesize DECT images, i.e. synthetic low energy CT image and synthetic high energy CT image, from SECT images. The relative electron density and the effective atomic number can be reconstructed from synthetic low energy and high energy CT images. Then, the DECT-based method establishes an empirical model for the estimation of the mean excitation energy in term of the effective atomic number. Furthermore, SPR can be calculated on the basis of the Bethe formula utilizing the relative electron density and mean excitation energy of tissues. However, the relationship between the mean excitation energy and effective atomic number is very complicated and cannot be accurately described by an analytical formula. The empirical analytic model often generates considerable errors in the estimated mean excitation energy, leading to a substantial proton range uncertainty. Significantly different from the learning DECT-based SPR estimation method [11], our deep learning-based approach generates a training dataset from chemical composition data of human tissues, and the deep learning approach directly calculate the stopping power ratio (SPR) map from clinical SECT images via the trained imResNet model. In our data-driven modeling approach, the trained network model is accurate and robust against various types of data perturbations. This deep learning-based approach estimates SPR from regular clinical SECT images directly, solves the problems in the dual-energy stoichiometric method efficiently, and significantly minimizes the range uncertainty of SPR estimation.

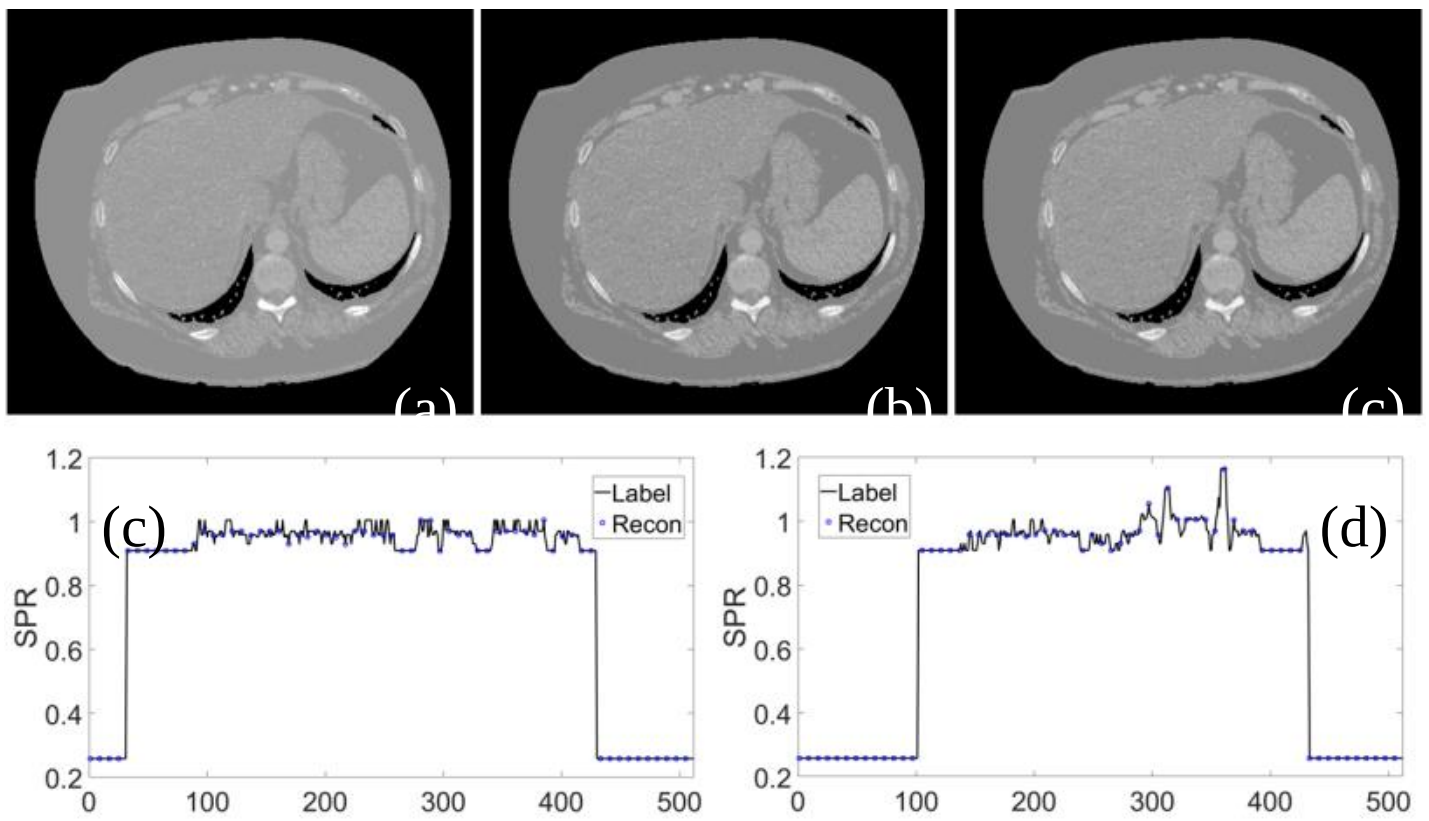

**Fig. 3**. SPR image reconstruction. (a) An input clinical CT image at 120kVp, (b) the SPR image calculated using our trained ImResNet model, (c) the ground truth SPR image calculated in terms of elemental composition of tissues, and (d)-(e) the vertical and horizontal profiles of the SPR image against that of the ground truth.

## 5 Conclusion

We have developed a practical approach to generate a training dataset from chemical composition data of human tissues derived from DECT and VM images, and proposed a deep learning approach to directly calculate the stopping power ratio (SPR) map from clinical SECT images via the trained imResNet model with a novel non-linearized input block and efficient shortcuts. The training of the imResNet model has excellent convergence and stability. We have conducted the quantitative evaluation based on numerical and clinical CT images. From CT images acquired by a SECT scanner, the proposed deep-learning-based approach generates high-quality SPR images with an uncertainty of 1%, thereby significantly minimizing the uncertainty range of SPR estimation for proton therapy. The proposed imResNet model and SPR calculation method have translational potential for the proton therapy.

# Design of Scatter-Decoupled Material Decomposition for Multi-Energy Blended CBCT Using Spectral Modulator with Flying Focal Spot

Yifan Deng[1] and Hewei Gao*[1]

[1]Department of Engineering Physics, Tsinghua University, Beijing 100084, China

**Abstract** In X-ray cone-beam computed tomography (CT), spectral modulator with flying focal spot (SMFFS) technology could be a promising low-cost approach to accurately solving the X-ray scattering problem and physically enabling multi-energy imaging in a unified framework, with no significant misalignment in data sampling of spectral projections. In this work, we advance this technology for spectral cone-beam CT (CBCT) and analyze the design of Scatter-Decoupled Material Decomposition (SDMD) for the blended multi-energy projection and twisted scatter-spectral challenge based on a scatter similarity hypothesis of SMFFS. Physics experiments on a tabletop CBCT system using a GAMMEX multi-energy CT phantom, are carried out to demonstrate the feasibility of our proposed SDMD method for CBCT spectral imaging with SMFFS. In the physics experiments, the mean relative errors in selected ROI for virtual monochromatic image (VMI) are 0.9% for SMFFS, and 5.3% and 16.9% for 80/120 kV dual-energy cone-beam scan with and without scatter correction, respectively.

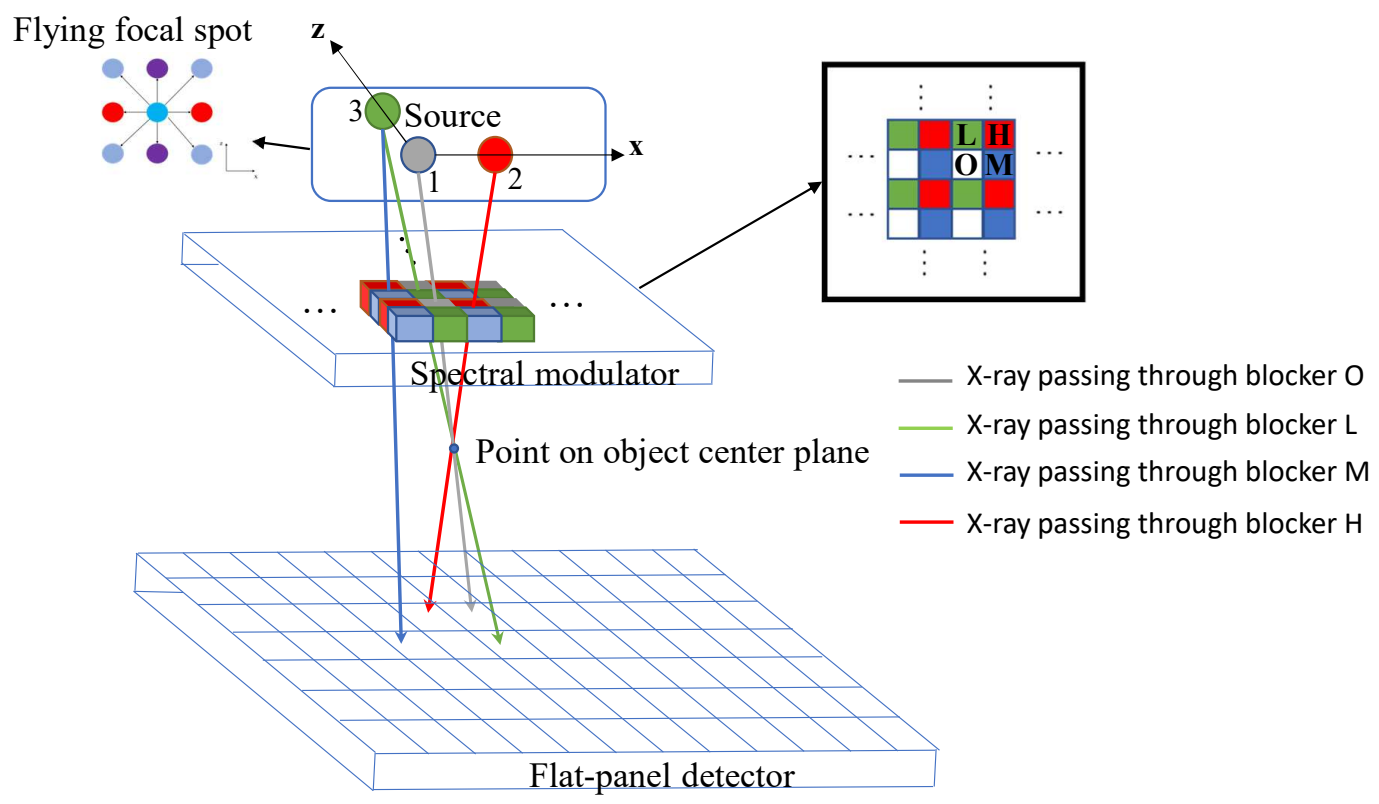

**Figure 1:** An illustrative diagram of the spectral modulator with flying focal spot (SMFFS).

## 1 Introduction

For cone-beam CT, spectral imaging is highly desired and is under active investigation with great progress these days. Numerous spectral CBCT concepts and prototype systems have been developed by using the fast kV-switching and the dual-layer detector technology[1–4]. These investigations have shown the potential of spectral CBCT in diagnostic and interventional radiology and radiotherapy. However, as one of the most important factors affecting spectral CBCT performance, so far in most of the studies in the literature, scatter was either simply estimated and corrected or completely avoided by using a relatively narrowed collimator.

In this work, we advance the SMFFS technology[5] to the multi-energy blended CBCT spectral imaging, and design a novel scatter-decoupled material decomposition method, where scatter correction and spectral imaging are modeled in a unified framework.

## 2 Method

### 2.1 Physical Model of SMFFS

Figure 1 shows an illustration of the SMFFS system, which mainly consists of the X-ray source, the spectral modulator, the scanning object, and the detector. In such a system, the X-ray source should be able to equivalently deflect the focal spot during a CT scan using the flying focal spot technology (or alternatively, with a distributed X-ray source); the spectral modulator consists of partially attenuating blockers and is placed between the X-ray source and the scanning object. In this paper, the spectral modulator was made by overlapping two 1D modulators together. For the SMFFS system with the focal spot at the $k$-th position, in consideration of the scatter intensity $I_s^{(k)}$ received on the detector, the total X-ray intensity measurement with an object in the beam is,

$$I_t^{(k)} = \int_0^{E_{\max}} S^{(k)}(E) e^{-\mu_1(E)L_1^{(k)} - \mu_2(E)L_2^{(k)}} \mathrm{d}E + I_s^{(k)} \tag{1}$$

Here, the superscript $(k)$ corresponds to the different focal spot positions that can minimize the misalignment of spectral projections as shown in Fig. 1; $S^{(k)}(E)$ is the effective spectra passing through the modulator and received by the detector (without object); $\mu_1(E)$ and $\mu_2(E)$ denote the energy-dependent attenuation coefficients of two basis materials, and $L_1^{(k)}$ and $L_2^{(k)}$ represent the corresponding effective path-lengths of two basis materials, respectively. Because the SMFFS system can maintain a good alignment of multi-energy X-rays and very similar scatter distributions among selected focal spot positions [5, 6], we can assume $L_1^{(k)} \approx L_1$, $L_2^{(k)} \approx L_2$, $I_s^{(k)} \approx I_s$. For the simplified problem with three unknowns $L_1, L_2, I_s$, if multiple measurements $I_t^{(k)}$ at different focal spot positions are available, (1) can be mathematically solved in a unified framework.

### 2.2 Analysis

First, we can regard the problem as a three-variable nonlinear problem. As long as we have three measurements at different focal spot positions, mathematically this problem can be solved as a ternary system of linear equations.

However, non-ideal factors occur in the actual situation. One of the most important factors is the penumbra effect. The penumbra effect is caused by the focal spot of the X-ray source with a certain size, and the spectra $S^{(k)}(E)$ corresponding to the X-rays passing through the edge of blockers of the modulator can be very similar and without enough energy separation. In other words, the equation of $k=1$ and $k=2$ may be similar in (1). Therefore, the problem is highly ill-posed in the penumbra area. What's more, for the 2D modulator shown in Fig. 1, the ratio of penumbra area to non-penumbra area is approximately the square of the ratio for the 1D modulator. The valid data for the scatter-decoupled material decomposition can even be 10% in reality. Other factors like the noise and the scatter deviation at different focal spot positions also make the problem difficult to be directly solved. To solve this problem, an iterative method for all data with strong constraints like total variation may be a good choice for the sparse spectral data. But in consideration of the high sparsity and the high complexity, we first conduct a prior analysis of the ill-posedness of the data.

To select the valid data, we use an elimination method to simplify the problem like this,

$$I_t^{(i)} - I_t^{(k)} = \int_0^{E_{\max}} \left(S^{(i)}(E) - S^{(k)}(E)\right) e^{-\sum_i \mu_i(E) L_i} \mathrm{d}E \quad (2)$$

Compared with the ternary nonlinear equations, the simplified equation (2) has the advantages as follows. First, the spectral diversity of the residual spectra like $S^{(i)}(E) - S^{(k)}(E)$ in (2), and the noise level of the residual data like $I_t^{(i)} - I_t^{(k)}$ that follows a Skellam distribution, can better characterize the ill-posedness of the data, which can help us quickly filter out useful data rather than analyze the ill-posedness of the ternary nonlinear equations; Second, the residual spectra can maintain a normal shape as the spectra are generated by different filtration of the 2D modulator. Thus, some typical material decomposition method for dual-energy CT can also be used for this residual data.

As Fig. 1 shows, we can divide the spectral data into different energy levels corresponding to different filtrations. Taking the detector pixels with $I_t^{(1)} = I_{t_O}, I_t^{(2)} = I_{t_L}, I_t^{(3)} = I_{t_H}$ as an example, we can obtain a pair of residual projections without scatter and with enough energy separation as,

$$\begin{aligned} P_{OL} &= -\ln\left(\frac{I_t^{(1)} - I_t^{(2)}}{I_m^{(1)} - I_m^{(2)}}\right) = -\ln\left(\frac{I_{t_O} - I_{t_L}}{I_{m_O} - I_{m_L}}\right) \\ P_{LH} &= -\ln\left(\frac{I_t^{(2)} - I_t^{(3)}}{I_m^{(2)} - I_m^{(3)}}\right) = -\ln\left(\frac{I_{t_L} - I_{t_H}}{I_{m_L} - I_{m_H}}\right) \end{aligned} \quad (3)$$

where, $I_m$ is the X-ray intensity measured without the scanning object in the beam; the subscript of $I_m$ and $I_t$ represents the energy level (O: original spectrum without passing through the modulator, L: low filtration by the modulator, M: middle filtration, H: high filtration), corresponding to different blockers of the modulator.

## 2.3 Implementation

The material decomposition is ill-posed due to the interpolation distortion for the residual data in the penumbra area. Therefore, we propose a two-pass guided material decomposition approach to solving this problem while preserving spatial resolution.

### 2.3.1 Scatter-Decoupled Material Decomposition for Residual Data

The scatter-decoupled material decomposition is similar to the material decomposition in traditional dual-energy CT, but with higher noise and more sparse data. To solve these problems and take advantage of the non-penumbra data with different energy levels, we conduct the guided material decomposition by referring to an iterative similarity-based method mentioned in [7]. The reference image in the material decomposition can be generated by a special combination among the different combinations of residual projections. The special combination $P_{OH} = -\ln\left(\frac{I_{t_O} - I_{t_H}}{I_{m_O} - I_{m_H}}\right)$ has the lowest noise but is not the best choice for energy separation. Therefore, reconstruction from $P_{OH}$ can be used as a guided image for material decomposition using $P_{OL}, P_{LH}$.

### 2.3.2 Scatter Estimation

Taking the decomposition of iodine and water as an example, we use the first-pass, over-smooth iodine image from 2.3.1 to generate the equivalent iodine length $L_{io}$. Then, we generate the iodine-induced spectra $S_{io}^{(k)}(E) = S^{(k)}(E) e^{-\mu_{io} L_{io}}$, hence the total X-ray intensity can be modeled as $I_t^{(k)} = \int_0^{E_{\max}} S_{io}^{(k)}(E) e^{-\mu_{wa} L_{wa}} \mathrm{d}E + I_s$. And the object can be regarded as purely water-equivalent. Therefore, the scatter can be estimated by the residual spectral linearization approach for SMFFS [8]. It should be noted that only data with enough energy separation are used for the scatter estimation. The full-scale scatter distribution is estimated by a regular interpolation. During this processing, the median filter and constraint can also be added to the estimated scatter given its low-frequency property and limited range of scatter-to-primary ratio (SPR) in practice (i.e., $0 < \mathrm{SPR} < 10$).

### 2.3.3 Material Decomposition for Complete Data

After scatter estimation, we utilize the total scatter-corrected data, especially the data in the penumbra area by a blended material decomposition method. In this paper, we use both an analytical method and a one-step model-based material decomposition method (iterative) referring to IFBP[9] as preliminary attempts.

**1) The analytical method**. First, we do preliminary material decomposition using the non-penumbra projection data with enough energy separation to obtain the basis material equivalent projections $P_{m1}, P_{m2}$ by a polynomial fitting method, which is a fast and robust material decomposition method.

Then, we generate virtual monochromatic projections (VMP) based on $P_{m1}, P_{m2}$ and the attenuation coefficients of basis materials at specific energies $E_L, E_H$ as,

$$\begin{aligned} \mathrm{VMP}_{L,np} &= \mu_1(E_L) \cdot P_{m1} + \mu_2(E_L) \cdot P_{m2} \\ \mathrm{VMP}_{H,np} &= \mu_1(E_H) \cdot P_{m1} + \mu_2(E_H) \cdot P_{m2} \end{aligned} \tag{4}$$

where, VMP in the penumbra area can be obtained by a regular interpolation, and $E_L, E_H$ is chosen empirically.

On the other hand, by using the calibrated spectra in the penumbra area and the preliminary iodine result, $P_{m2}$, the VMP in the penumbra area can also be modeled as,

$$\begin{aligned} \mathrm{VMP}_{L,p} &= \sum_{i=0}^{N} \sum_{j=0}^{N-i} c_{p,ij} \cdot P_p^i \cdot P_{m2}^j,\ c_{00} = 0. \\ \mathrm{VMP}_{H,p} &= \sum_{i=0}^{N} \sum_{j=0}^{N-i} d_{p,ij} \cdot P_p^i \cdot P_{m2}^j,\ d_{00} = 0. \end{aligned} \tag{5}$$

where, $c_{p,ij}, d_{p,ij}$ can be generated by the polynomial fitting using a series of sample points of basis material densities in advance; and $P_p$ is the scatter-corrected projection in the penumbra area.

For simplicity, we use a hard threshold to blend the VMPs as (6) shows. And the threshold $t_0$ is empirically chosen as 0.2 in this paper. The VMIs can be directly reconstructed and so do the basis material images after a simple image-domain material decomposition.

$$\mathrm{VMP}_{p,final} = \begin{cases} \mathrm{VMP}_p, & \dfrac{|\mathrm{VMP}_p - \mathrm{VMP}_{np}|}{\mathrm{VMP}_{np}} <= t_0 \\ \mathrm{VMP}_{np}, & \dfrac{|\mathrm{VMP}_p - \mathrm{VMP}_{np}|}{\mathrm{VMP}_{np}} > t_0 \end{cases} \tag{6}$$

**2) The iterative method**. For the iterative method the blended process in each iteration is similar to the polynomial method. First, we do material decomposition based on the one-step iterative formula[9]. Similarly, this step only uses the data with enough energy separation and generates the equivalent basis material projections $P_{m1}, P_{m2}$; then, we implement the steps from equation (4) to (6) as usual, and generate the basis material images by VMPs of two specific energy; and finally, we update the iteration by the basis material images.

Compared to the polynomial method, the iterative method can deal with the slight misalignment of spectral data in SMFFS by using different system matrices in the forward projection step, and has the potential of optimizing the results by adding regularization constraints. However, accurate geometry is difficult to be modeled in reality. The slight misalignment of multi-energy X-rays may be even enhanced by the wrong geometry. Therefore, the analytical method can be a robust and quick method to solve the problem, and the artifacts caused by the slight misalignment are not obvious in a routine noise. It should be noted that in both methods ring fix methods are used for the residual ring artifacts caused by the penumbra effect and the spectra calibration error.

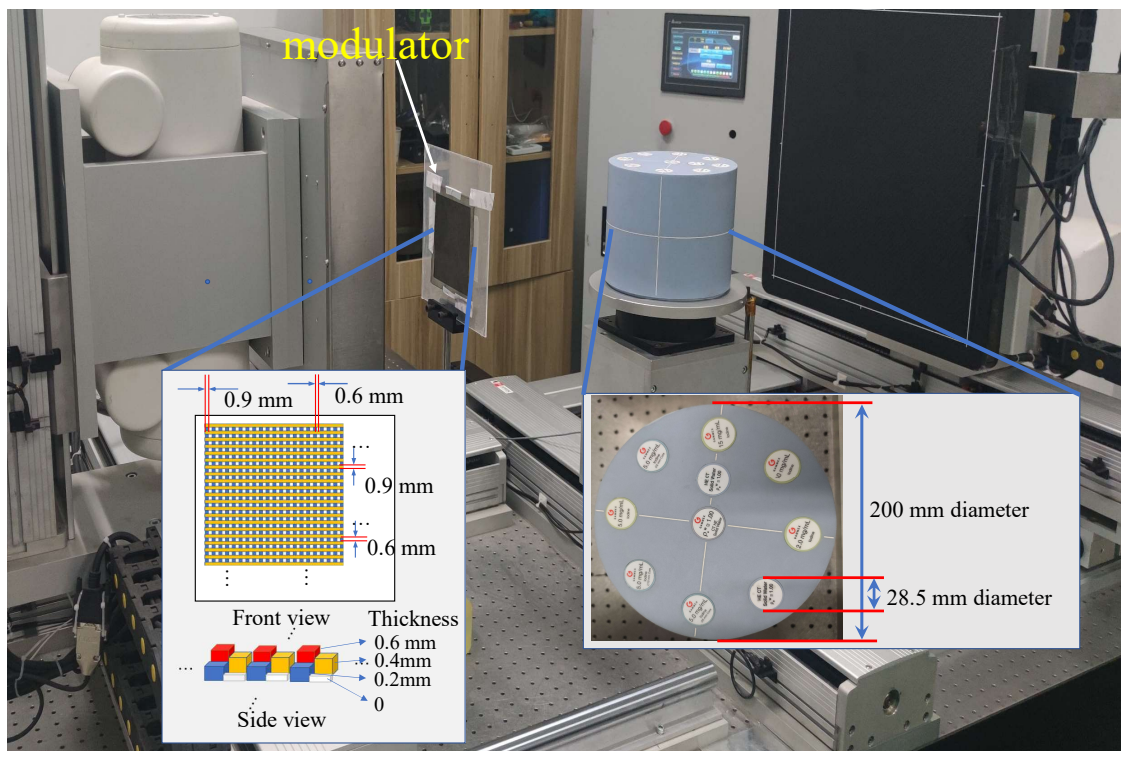

**Figure 2:** The experimental CBCT platform with SMFFS.

## 2.4 Physics Experiments

Physics experiments were conducted on our tabletop CBCT system as shown in Fig. 2. The X-ray source used a Varex G-242 tube with a focal spot of 0.4 mm; the detector was Varex 4030 DX flat-panel detector; the modulator was manufactured by stacking two 1D strip modulators of Moly. with a spacing of 0.6 mm and a period of 1.5 mm. In physics experiments, the source was placed at (0,0,0), (1.32,0,0), and (0,0,1.32) mm sequentially to mimic the flying focal spot deflection, and the source was operated at 120 kVp, 386 mAs in all SMFFS scans; the detector worked in binning 2 mode with $1024 \times 768$ pixels, $0.388 \times 0.388\ mm^2$ per pixel; a GAMMEX multi-energy CT phantom was scanned with collimation wide enough to cover the detector. Besides SMFFS scans, limited by the hardware with the maximum tube voltage of 125 kVp, sequential 80/120 kVp dual-energy scans without a modulator were also collected for comparison, in which the material decomposition was conducted by a conventional polynomial function method.

# 3 Result

Figure 3 shows the spectral imaging results of SMFFS, compared with sequential dual-energy scans of 80 / 120 kVp with a narrowed collimator (fan beam), a wide collimator (cone beam) without and with scatter correction by a kernel-based method fASKS[10] available in the CBCT Software Tools (CST) (Varex Imaging Corporation). Fig. 4 shows the quantitative density results in ROIs of iodine and water images, where their references (dotted-line) are obtained by the user manual of the Multi-Energy CT Phantom. These preliminary spectral imaging results and quantitative analysis above demonstrate that our proposed method can significantly improve the image quality and the quantitative performance of CBCT, even better than that of the sequential dual-energy cone-beam results in our study.

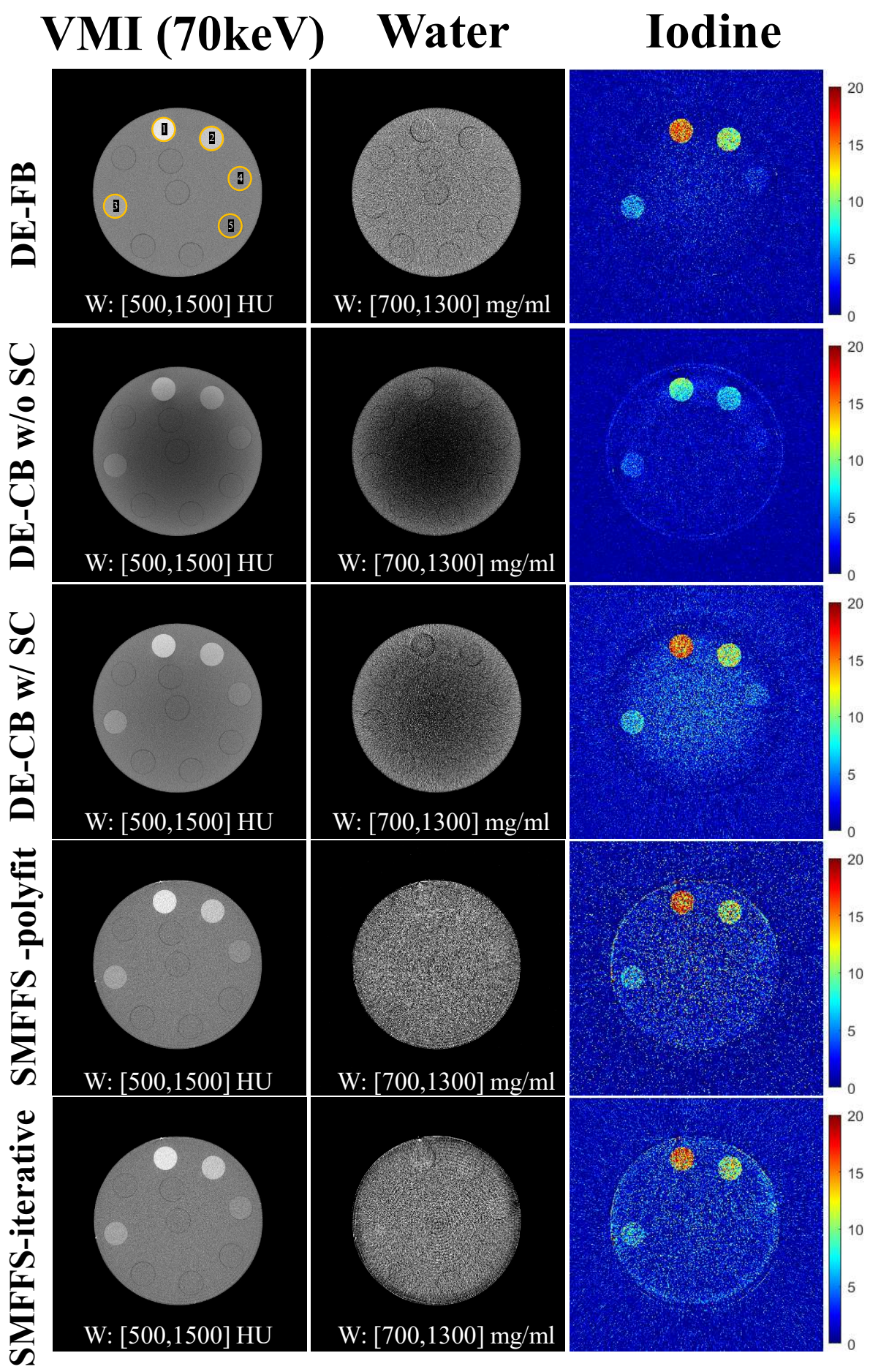

**Figure 3:** Iodine, water, and VMI (70 keV) CT images of SMFFS and the sequential dual-energy scans. Top row: 80 / 120 kVp FB. Second row: 80 / 120 kVp CB without scatter correction. Third row: 80 / 120 kVp CB with scatter correction by fASKS in CST. Fourth row: SMFFS scan with scatter using polynomial fitting for second-pass material decomposition. Bottom row: SMFFS scan with scatter using the iterative method.

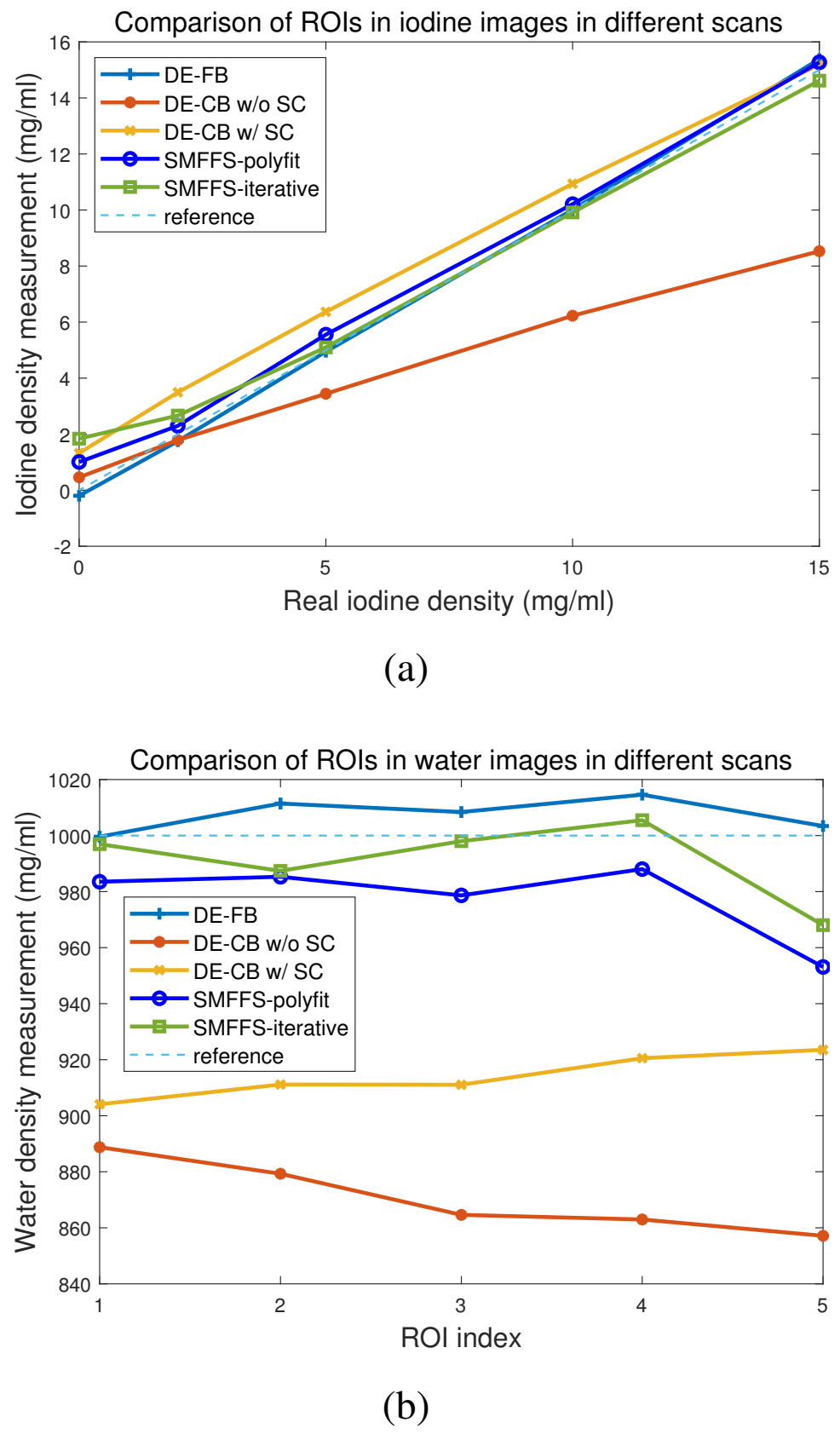

**Figure 4:** Iodine and water averaged densities of ROIs in Fig. 3 and the reference values. (a): averaged densities of ROIs in iodine images; (b): averaged densities of ROIs in water images.

## 4 Conclusions

CBCT has been highly desired to achieve better quantitative performance in recent years, but spectral CBCT is still limited by the accurate scatter correction for dual- or multi-energy data. In this paper, we designed a scatter-decoupled material decomposition (SDMD) method for spectral CBCT imaging with SMFFS and analyze the characteristic of this scatter-decoupled material decomposition. As a preliminary study, physics experiments showed that the scatter-spectral twisted problem in spectral CBCT can be simultaneously solved by our method.

# One-cycle 4D-CT reconstruction with two-level motion field INR

Muge Du[1], Li Zhang[1], Le Shen[1], Yinong Liu[1], Shuo Wang[1] and Yuxiang Xing*[1]

[1]Department of Engineering Physics, Tsinghua University, China

**Abstract**

Background: One-cycle 4D-CT reconstruction can largely reduce scanning time and radiation dose, but the reconstruction can be ill-posed with sparse-view and limited-angle problems.

Method: We proposed a novel implicit neural representation (INR) for one-cycle 4D-CT featured by a basic deformation model combining a two-level motion field INR to a template attenuation field INR, and a shortcut for capturing non-deformable motion. The two-level design in the motion field INR includes global-local division for motion, and pattern-strength modeling for local motion. Besides, a Fourier-Domain-Error rendering loss is proposed to train INR-based CT reconstruction.

Evaluation: The method is evaluated in simulated one-cycle cardiac and lung 4D-CT datasets. It outperforms PICCS with huge advantages and is able to reconstruct high-quality dynamic volumes with rich details.

Novelty and impact: The proposed hierarchical INR structure and FDE loss help reconstruct complex dynamic volumes and solve the sparse-view and limited-angle problems in one-cycle 4D-CT reconstruction, bringing the potential for significant dose reduction by one-cycle 4D-CT.

## 1 Introduction

Traditional 4D-CT reconstruction methods often select several projections inside a time window and reconstruct the phase volume using these projections and prior information[1]. For 4D-CT reconstruction with fewer projections or shorter time window, some methods explore the spatial or temporal correlation between volumes at different phases or prior images[2]. Another type of methods joints the projections at different phases by a shared static volume and the deformation vector field (DVF) or motion field, which is used to deform the static volume into temporal phases[3]. Recently, implicit neural representation (INR), originated from NeRF[4], provided a new methodology in scene representation and reconstruction, and has been applied to medical imaging[5]. Instead of traditional isolated representation based on voxels, INR can construct the scene as a continuous implicit model, composed of a feature embedding for arbitrary location and view query, and neural networks to convert the embedding into the scene prediction. INR is applied to solve the sparse-view[5] and limited-angle problem[6] in CT. Several INR-based methods are proposed to jointly estimate INR for static volume and DVF for dynamic CT[7], [8], but either modeling the DVF explicitly with polynomial-fitting, or relying on preliminary scans for prior DVFs.

One-cycle 4D-CT reduces the dose greatly by scanning for only one motion cycle. However, its reconstruction is ill-posed as the angular coverage in a short time window is often limited. Besides, sparse-view problem may also exist.

In this work, we propose a novel INR for one-cycle 4D-CT reconstruction. It is a combination of a two-level motion field INR and a template attenuation field INR. Experimental study demonstrates that the method can solve the limited-angle and the sparse-view problem in one-cycle 4D-CT effectively and reconstruct rich details with few artefacts. It brings the potential to largely reduce the scanning time and radiation dose needed for 4D-CT.

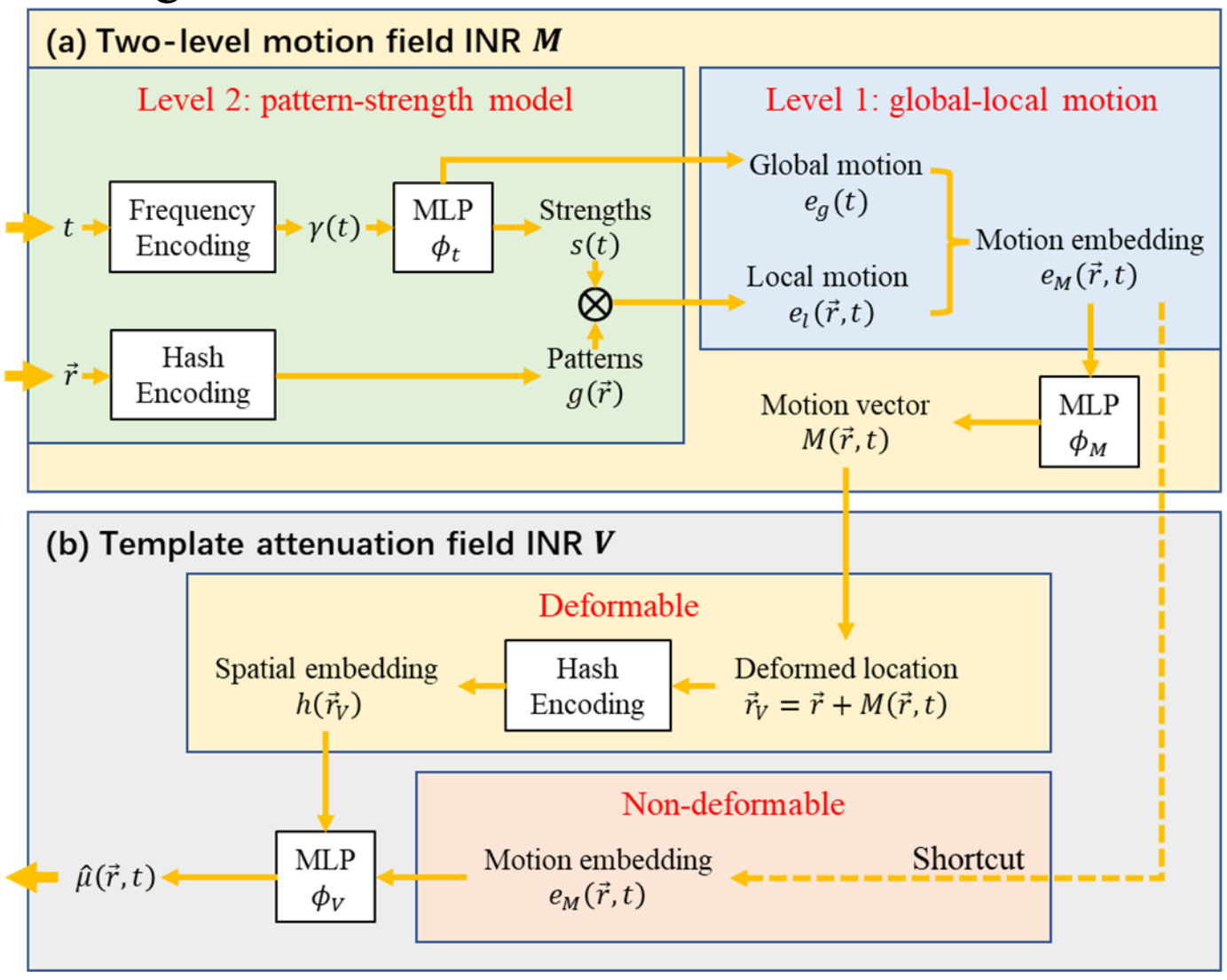

Figure 1 Overview of the proposed INR for one-cycle 4D-CT

## 2 Methods

### 2.1 Method Overview

In 4D CT scanning of the time-variant object, the obtained sinogram $p$ can be written as:

$$p(m,t) = \int_{l(m,t)} \mu(\vec{r},t) d\vec{r} \qquad (1)$$

where $l(m,t)$ stands for the ray casting path for the $m^{th}$ detector at moment $t$, and $\mu(\vec{r},t)$ is the object's attenuation coefficient at spatial location $\vec{r}$ and moment $t$. In this work, we propose an implicit neural representation (INR) for a dynamic object, called DctINR. It is featured by a basic deformation model combining a two-level motion field INR $M$ to a template attenuation field INR $V$, and an additional mechanism to capture non-deformable motion. The overview of DctINR is shown in in Figure 1.

In the basic deformation model, for a dynamic object at any moment $t$ and location $\vec{r}$, its attenuation coefficient $\mu(\vec{r},t)$ is assumed to match the value of $V$ at some deformed location $\vec{r}_V$. The offset from $\vec{r}$ to $\vec{r}_V$ is our motion field denoted by $M(\vec{r},t)$ that is computed from a motion feature embedding $e_M(\vec{r},t)$ sent to an MLP $\phi_M$. In math,

$$M(\vec{r},t) = \phi_M\big(e_M(\vec{r},t)\big) \qquad (2)$$

$$\vec{r}_V = \vec{r} + M(\vec{r},t) \qquad (3)$$

and $\mu(\vec{r},t)$ is computed by querying $V$ with $\vec{r}_V$:

$$\mu(\vec{r},t) = V(\vec{r}_V) \qquad (4)$$

Further, we borrow the concept of residual learning with a shortcut from the motion feature embedding $e_M(\vec{r},t)$ to $V$

This work was supported in part by the National Natural Science Foundation of China (Grant No. 62031020) and the grant from Institute of Precision Medicine, Tsinghua University.

to capture another level's motion modeling non-deformable dynamic components, hence the full method changes from the basic deformation model in Eq. 4 to:

$$\mu(\vec{r},t) = V\big(\vec{r}_V, e_M(\vec{r},t)\big) \quad (5)$$

Next, we introduce the detailed method implementation.

## 2.2 Two-level motion field INR

The motion field INR $M(\vec{r},t)$ is built as a motion embedding $e_M(\vec{r},t)$ fed into an MLP $\phi_M$ as in Eq. 2. Especially, we use a two-level structure for the motion field. The first level is a global-local division. We design the motion embedding as the concatenation of global and local motion components:

$$e_M(\vec{r},t) = \big[e_g(t), e_l(\vec{r},t)\big] \quad (6)$$

The local motion $e_l(\vec{r},t)$is seen as both spatial and time-dependent. The global motion $e_g(t)$ is only time-dependent. For the local motion embedding, the second level is introduced as a pattern-strength model:

$$e_l(\vec{r},t) = p(\vec{r}) \otimes s(t) \quad (7)$$

where the possible motion patterns $p(\vec{r})$ are extracted from spatial information and the strengths $s(t)$ server as a time-dependent scale to linearly combine the patterns, $\otimes$ denotes element-wise multiplication.

As shown in Figure 1-a, we use the learnable multi-resolution hash encoding[9] for the patterns $p(\vec{r})$. Time $t$ is firstly frequency-encoded[4],

$$\gamma(t) = [\sin(2^0 t), \sin(2^1 t), \dots, \sin(2^{L-1} t), \cos(2^0 t), \cos(2^1 t), \dots, \cos(2^{L-1} t)] \quad (8)$$

and then fed into an MLP $\phi_t$, to get $s(t)$ and $e_g(t)$ together:

$$\big[s(t), e_g(t)\big] = \phi_t\big(\gamma(t)\big) \quad (9)$$

## 2.3 Template attenuation field INR

The template attenuation field INR $V$ is implemented as in Figure 1-b. We also use the hash encoding to get a spatial embedding $h(\vec{r}_V)$ for the deformed position $\vec{r}_V$ . Then, $h(\vec{r}_V)$ and the previously computed motion embedding $e_M(\vec{r},t)$ are fed into an MLP $\phi_V$ together to get $\mu(\vec{r},t)$:

$$\mu(\vec{r},t) = V\big(\vec{r}_V, e_M(\vec{r},t)\big) = \phi_V([h(\vec{r}_V), e_M(\vec{r},t)]) \quad (10)$$

## 2.4 Model training

### 2.4.1 Overall training procedure

The training of DctINR follows the tradition in INRs for CT. At each iteration, the sinogram at a random moment $t$ is drawn from the dataset, and $n_r$ rays are randomly sampled from the sinogram for ray casting. For each ray, it is casted towards the light source, and points are sampled uniformly along the intersection of the ray path and the reconstruction region. In total, there are $n_p$ points sampled. For each sampled point, its $(\vec{r},t)$ is sent to DctINR to estimate the attenuation coefficient $\hat{\mu}(\vec{r},t)$. All the estimated values are projected back to render the predicted detector signals $\hat{p}$:

$$\hat{p}(m_i,t) = \int_{l(m_i,t)} \hat{\mu}(\vec{r},t)d\vec{r} \quad i = 0,2,\dots,n-1 \quad (11)$$

### 2.4.2 Fourier-domain-error rendering loss

Like other INRs for CT, DctINR is trained in self-supervision by a rendering loss. A rendering loss is defined as the distance between the predicted and the ground-truth detector signals. A common choice for the distance is MSE or MAE. However, both MSE and MAE loss treat each detector independently, even though the sinogram/ projection itself is semantically meaningful. In this work, we propose a Fourier Domain Error (FDE) loss for CT reconstruction, to effectively capture the long-range dependency across the whole detector panel. For the $n_r$ sampled detectors, Discrete Fourier Transform (DFT) is performed to the detectors' ground-truth signals $p$ and estimated signals $\hat{p}$ . FDE loss is defined as the MAE between $\mathrm{DFT}(p)$ and $\mathrm{DFT}(\hat{p})$:

$$Loss_{FDE}(\hat{p},p) = \frac{1}{n_r}\|\mathrm{DFT}(\hat{p}) - \mathrm{DFT}(p)\|_1 \quad (12)$$

FDE loss measures the rendering error globally among the selected detectors and frequency-wisely. As the lower-frequency error is faster to mitigate during training, FDE loss helps train the network in a coarse-to-fine way.

### 2.4.3 Regularization loss for the motion field

To avoid too magnificent motion vector outputs by the motion field INR and stabilize the training in the early stage, an L2 regularization loss is applied to $M(\vec{r},t)$:

$$Loss_{reg} = \frac{1}{n_p}\|M(\vec{r},\ t)\|_2^2 \quad (13)$$

Hence, the total loss for training DctINR is:

$$Loss_{total} = Loss_{FDE} + \lambda Loss_{reg} \quad (14)$$

## 3 Experimental study results

### 3.1 Dataset

In this work, we evaluate DctINR with a cardiac 4D-CT dataset and a lung 4D-CT dataset, both simulated under cone-beam CT geometry using ground-truth volumes.

For the cardiac dataset, a 20-phase 4D volume covering a whole heartbeat cycle reconstructed from a prospective patient scan is used for 4D-CT projection simulation. For the lung dataset, a 10-phase 4D ground-truth volume provided from SPARE[10]'s Monte Carlo dataset P3\ MC_V_P3_LD_01 is used for 4D-CT projection simulation. In total $N_v$ timestamps across the motion cycle are sampled, each simulated with one projection. For timestamps not centered at provided phases, temporal linear interpolation from two neighboring phases is used to simulate current volume and corresponded CT projection. Therefore, each simulated projection is related to a unique timestamp. The gantry rotation angle at the $i$th timestamp is set as:

$$\frac{2\pi R * (N_v + 1) * i}{{N_v}^2}, i = 0,1,\dots,N_v - 1 \quad (15)$$

where $2\pi R$ is the total angular coverage. This setting ensures the samples are non-overlapped in angle. The default setting of $N_v$ and $R$ is (360, 4) for the cardiac dataset and (180, 2) for the lung dataset. Under this setting, the rotation speed for the cardiac dataset assuming a 60-bpm heart rate is about 0.25s/rot, the rotation speed for the lung dataset assuming a 12-bpm respiratory rate is about 2.5s/rot, and the angular coverage per phase is both $\frac{2\pi}{5}$ in both datasets. Poisson noise with photons of 2e6 is simulated in projection.

## 3.2 Implementations

### 3.2.1 Method training details

DctINR is trained using Adam optimizer, with initial learning rate set as 1e-3 for the cardiac dataset and 3e-3 for the lung dataset. Cosine learning rate scheduler with 0.1 decay ratio is used. A training epoch is defined as the shuffled iteration over the total views, and the epoch number is set as $180000/N_v$, so that DctINR is updated in roughly same times in all experiments. The number of sampled rays $n_r$ for each view is 1024. Loss weight $\lambda$ is 0.1.

### 3.2.2 Comparison methods

The comparison methods include time-averaged FDK using projections from all phases, and PICCS[1] using projections from a short time window with the FDK result as the prior. The time window length is set to three phases, which is the best among all the choices from experiments.

### 3.2.3 Evaluation metrics

RMSE, PSNR and SSIM (using gaussian window with $\sigma = 1.5$) are used for evaluation. All the metrics are computed on attenuation coefficients with unit in $\text{mm}^{-1}$.

The metrics are computed both on the whole volume and within semantic ROIs. For the cardiac dataset, 8 per-phase ROI masks are used. For the lung dataset, 4 phase-averaged ROI masks provided within SPARE dataset are used.

## 3.3 Experimental results

### 3.3.1 One-beat 4D cardiac

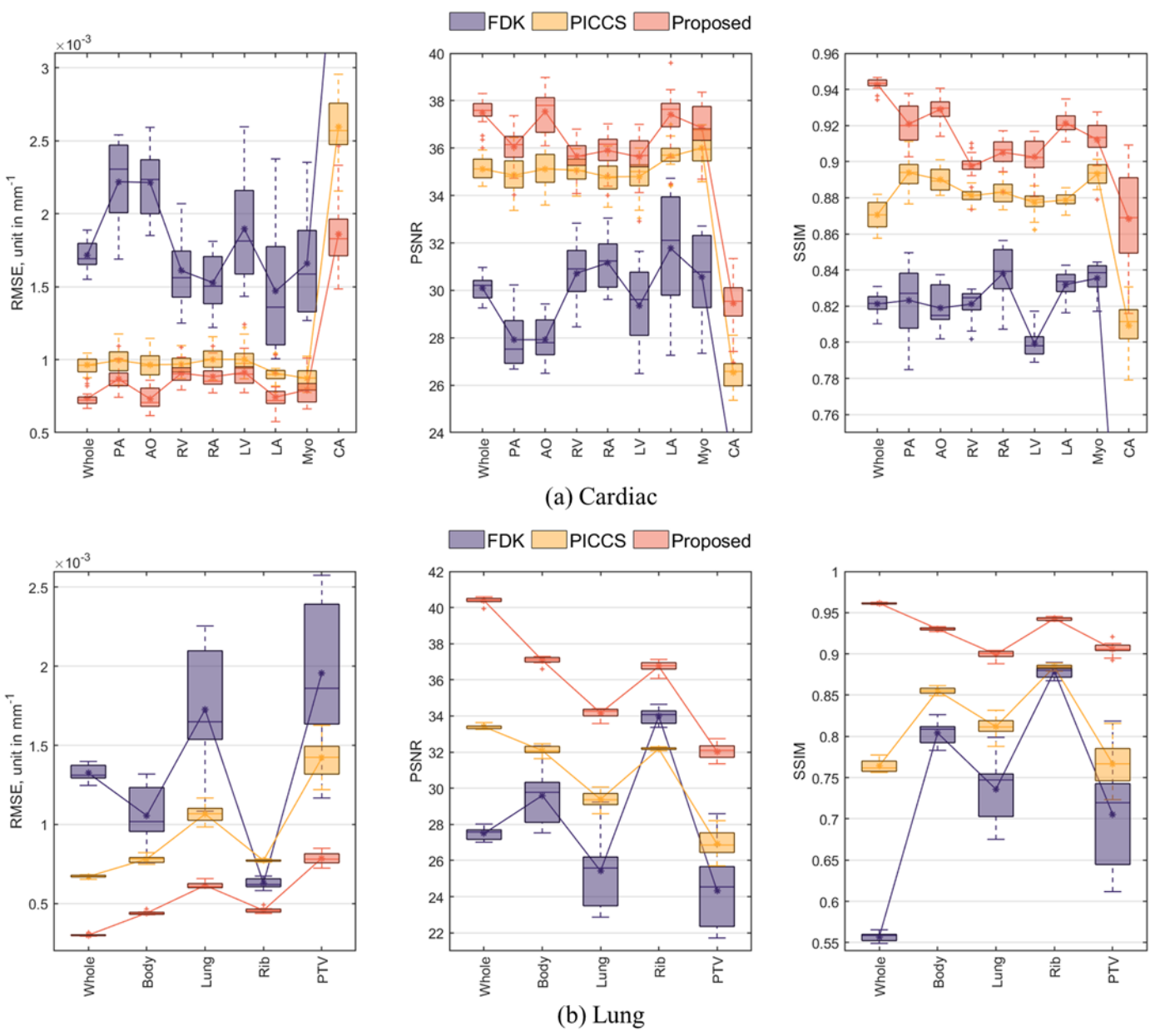

Figure 2. The boxplot and the mean value(solid line) of the per-ROI metrics counted on all phases, for all the methods. For RMSE, the lower the better. For PSNR and SSIM, the higher the better. For cardiac dataset(a), the ROI masks are the pulmonary artery (PA), ascending aorta (AO), right ventricle (RV), right atrium (RA), left ventricle (LV), left atrium (LA), myocardium of LV (Myo), and the coronary artery (CA). For lung dataset (b), PTV is the planning target volume.

Figure 3 shows the example reconstructed coronal and axial views at the 50% phase of the cardiac dataset. DctINR gives the best visual quality with the fewest artefacts, the fewest structural errors and the clearest boundaries. Shown at the zoomed-in ROIs, DctINR can reconstruct high-resolution details, for example, the tiny coronary artery (with 1mm radius) pointed by an arrow in the coronal view's ROI. FDK result is time-averaged, thus contains strong motion artefacts and blurred structures. It also contains streaking artefacts. PICCS can reconstruct dynamic volumes with only a few motion artefacts, but it contains streaking artefacts similar to FDK. Compared to DctINR, PICCS contains more smoothed boundaries, more structural errors and much less details. PICCS fails to reconstruct tiny structures like the coronary arteries shown in the two ROIs.

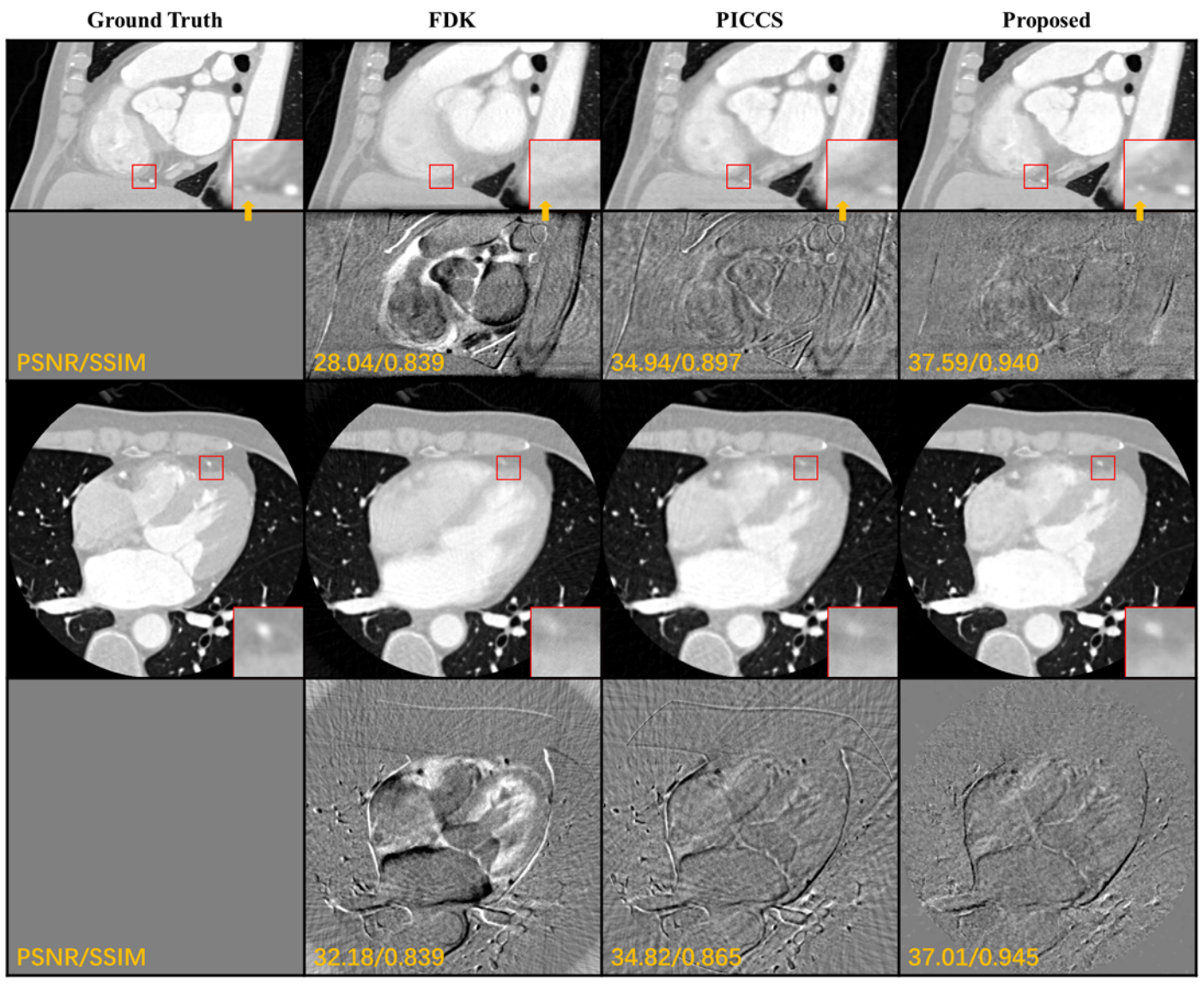

Figure 3. The example reconstructed coronal (upper 2 rows) and axial view (lower 2 rows) at the 50% phase of the cardiac dataset. The display window is [0, 0.03] $\text{mm}^{-1}$ for the reconstructions (1st and 3rd rows) and [-0.005,0.005] $\text{mm}^{-1}$ for the residuals (2nd and 4th rows). The PSNR and SSIM are only for the view.

### 3.3.2 One-breath 4D lung

Figure 2-b shows the quantitative evaluation of all methods on the 4D lung dataset with 180 total views. The lung dataset contains stronger movement and only half number of views compared to the cardiac dataset. Here, the advantage of DctINR over the others is more obvious. It is interesting to notice that DctINR has better performance in the lung dataset than in the cardiac dataset, while the comparison methods perform worse. The time-averaged FDK result indicates strong movement in the lung dataset. These results demonstrate the robust performance of DctINR to strong movement and fewer views.

Figure 4 shows the example reconstructed sagittal and axial views at the 50% phase of the lung dataset. DctINR result is very close to the ground-truth and contains almost no motion artefacts or streaking artefacts. The PTV and the lung nodes can be well reconstructed as shown in the two zoomed-in ROIs. The organ boundaries are also clear. FDK contains strong motion artefacts and streaking artefacts. PICCS is able to compensate some big motions, but the structures are blurred as shown in the zoomed-in ROIs. Besides, the organ boundaries are hard to distinguished in PICCS.

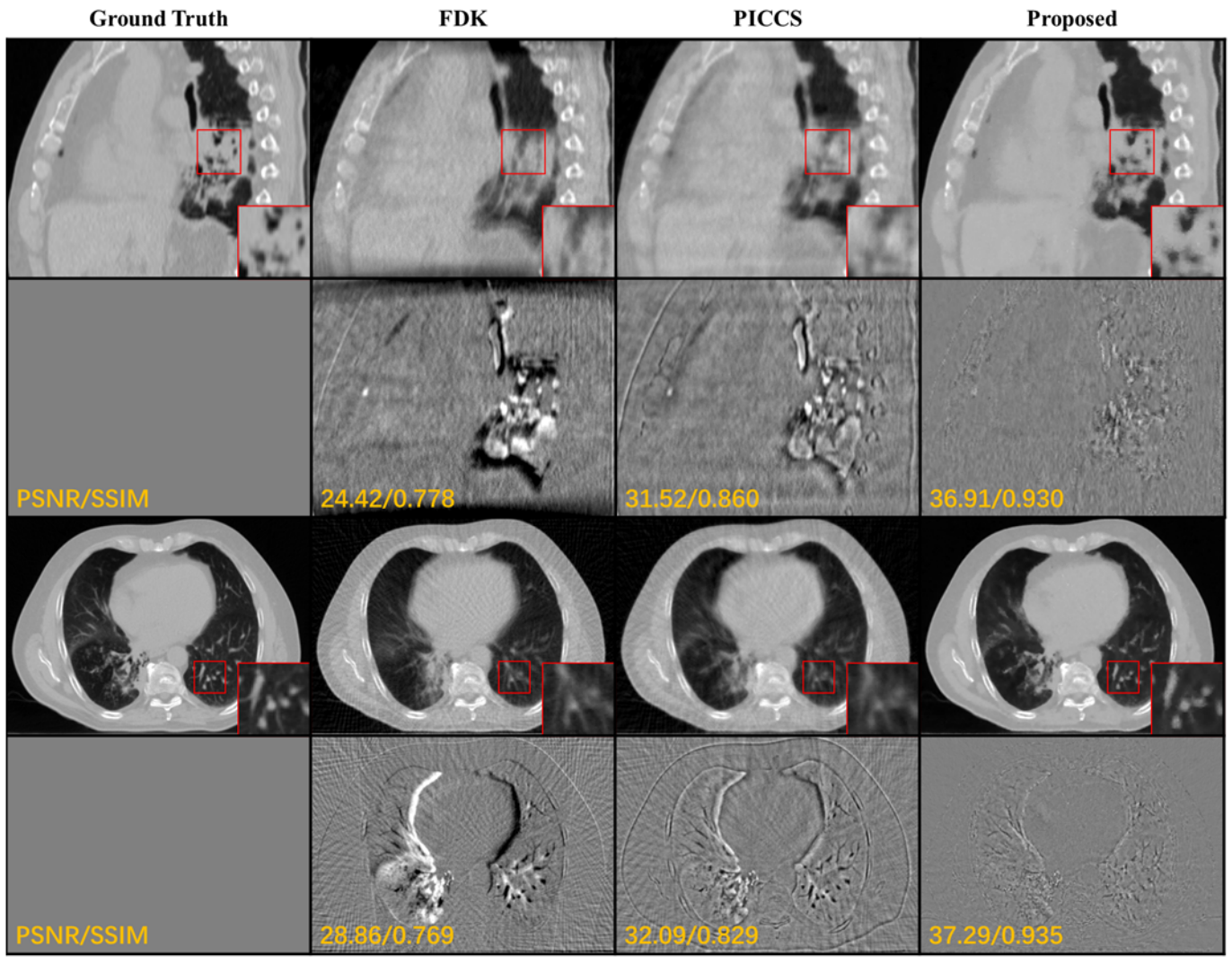

Figure 4. The example reconstructed sagittal (upper 2 rows) and axial view (lower 2 rows) at the 50% phase of the lung dataset. The display window is [0, 0.02] $mm^{-1}$ for the reconstructions (1st and 3rd rows) and [-0.005,0.005] $mm^{-1}$ for the residuals (2nd and 4th rows). The PSNR and SSIM are only for the view.

## 4 Conclusion and Discussion

In this work, we proposed a novel INR that addresses the limited-angle and sparse-view problems in one-cycle 4D-CT reconstruction. DctINR combines a two-level motion field INR to a template attenuation field INR to model complex dynamic objects. Besides, a novel Fourier-Domain-Error rendering loss is proposed for INR-based CT reconstruction. The experimental results in simulated one-cycle cardiac and lung 4D-CT datasets demonstrate DctINR's ability in high-quality 4D-CT reconstruction with rich details. DctINR is beneficial for dose reduction in 4D-CT reconstruction.

In DctINR, the sparse-view problem in one-cycle 4D-CT is solved by model both the motion field and the template attenuation field using INR, which provides a smooth and nearly continuous estimation given sparsely sampled views. Besides, DctINR uses hierarchical designs for modeling complex and realistic dynamic objects. The deformable motion is modeled by deformed location at the template attenuation field, and the non-deformable motion is modeled by an additional shortcut. For the two-level motion field, global and local motion is used in the first level, while the second level views the local motion as a pattern-strength model. In this way, the limited-angle problem in 4D-CT reconstruction can be solved since we do not need to reconstruct an explicit whole 4D volume or motion field, but a hierarchical model instead with basic components only depending on either spatial or temporal information.

There are a few limitations in this work. Firstly, denoising mechanism in method design is absent, which is critical for further dose reduction. Secondly, a comparison to existing DVF-based iterative methods and INR-based methods for 4D-CT reconstruction methods is not performed. Thirdly, the analysis on the sensitivity to different sparse-view and limited-angle settings is absent, which is useful for designing low-dose protocols for DctINR. Fourthly, the proposed method is not evaluated on practical 4D-CT scans, where issues like noises and scatters may limit the performance. We will address these limitations in the future.

# Coded Array Beam X-ray Imaging Based on Spatio-sparsely Distributed Array Sources

Jiayu Duan[1], Yang Li[1], Song Kang[2], Guofu Zhang[2], Jianmei Cai[1], Chengyun Wang[2], Jun Chen*[2] and Xuanqin Mou*[1]
[1] Institute of Intelligence Computing and Data Communication, Xi'an Jiaotong University, Xi'an, China
[2] State Key Laboratory of Optoelectronic Materials and Technologies, Guangdong Province Key Laboratory of Display Material and Technology, School of Electronics and Information Technology, Sun Yat-sen University, Guangzhou, China

**Abstract**
Traditional X-ray tubes and their usage patterns suffer from the problem of overheating which results in limitations of heavy, bulk and expensive. In this paper, we proposed coded array beam X-ray imaging (CABI) method using very low power X-ray array source that consists of a number of spot sources arranged in a line or a plane. Each spot source in the array just covers a small part of the object. A number of coded array beams are radiated for imaging while in each radiation multiple sources simultaneously light. In this manner, the array source works in a very low power state. High density and virtually parallel-beam projections with a number of different rotation angles are reconstructed from the coded projections by the CABI algorithm, which can be used for CT and tomosynthesis imaging. Theoretical analysis proves that the projections from sparse array sources are capable for high density reconstruction since that sparsely spatial information can be restored by densely angular views in tomographic imaging. Hence just a small number of coded lighting modes are needed for imaging and the source power is further decreased. Experiments validate the feasibility of the CABI method. Particularly, with CABI, a flat-panel X-ray array source can work stably for a long time in room temperature without cooling. This study shows a promising modality for the next generation of the X-ray imaging technology.

## 1 Introduction

As the discovery of X-ray, X-ray generators, over 100 years of development, enables to design multiple imaging modalities used as radiography, computed tomography, etc. In most practices, X-ray photons are produced by X-ray tube using high-energy electrons to bombard a high-Z material, e.g., tungsten or molybdenum, where only 1% of energy can be converted into X-rays; the rest appears in the form of heat. At present, X-ray imaging technologies use the tube of a spot source to emit a fan-beam or cone-beam to cover the field of view (FOV) of the object for radiation imaging, which results in a balance problem between the source power and the image quality since the imaging quality is depended on the energy of X-ray photons passing through the object to the detector. Hence, the X-ray tubes are inevitably overheated. The overheating problem limits the use of X-ray apparatus.

Traditionally, a cooling system is applied to ensure the working efficiency of the X-ray tubes, which makes the source heavy, bulky and expensive. When it is used for CT device, the tube is rotated to acquire enough information for reconstruction, which results in complex mechanical architecture and motion artifacts in the image. Cold cathode X-ray source such as carbon nanotube (CNT) has drawn much attention for its compact size and fast switching ability[1]. With these advantages, stationary CT systems are available by using a number of CNTs arranged on a line or ring to complete the rotational scans [2, 3]. To acquire comparable measurements with traditional X-ray source, the CT system based on CNTs needs enough number of CNTs, leading high cost in manufacturing. Moreover, the physical size of the CNT can not be ignored. Thus, CNT-based imaging systems face sparse view problems.

To overcome the limitations of the X-ray source of high heat capacity, a possible way is to introduce the X-ray array source that is consisted of a large number of tightly arranged spot sources wherein each spot source just covers a small part of the object to be imaged. In this way, a single spot source just works in a very low power state. In [2], the authors showed that such high-density array sources are feasible in implementing X-ray imaging function and CT reconstruction. In this study, we suggest further decreasing the power of the array X-ray source by proposing the coded array beam X-ray imaging (CABI) modality that works based on a serial of lighting modes of the array source. In each lighting model, many spot sources, generally half of the array are simultaneously lighted. The detector cell receives mixed projections from different spot sources so that the power of each spot source can be decreased dramatically. Recently developed field-emission cold cathode X-ray source provides possibility of working in this way[1, 4]. However, the received signal cannot be directly used because the signal of each cell combines different line integrals of the object along the rays from different spot sources. To restore each line integral from the mixed signals, we introduce coded apertures to design the lighting models for the restoration. More importantly, in this paper, we explain that projections of spatio-sparsely arranged spot sources indeed contain enough information to restore high density projections of the object. Based on this property the number of lighting models will be few in restoration of high density image and the radiation dose is also saved. Additionally, combined with compressed sensing, the number of lighting models can be decreased further. Hence, the proposed CABI modality would be feasible in practice. We use a flat-panel X-ray array source[5] to simulate the CABI modality and validate its feasibility. It's worth noting that the flat-panel X-ray array source works stably for a long time in room temperature without any cooling apparatus. By our study, the innovation of the CABI method would introduce the following benefits. 1) With the merits of the flat-panel X-ray array source working in room temperature environment, as well as there is a small stand-off distance between the source and the detector since each spot source in the array just needs to cover a small part of the object, traditional bulky X-ray tube might truly become a chip tube, which would greatly decrease the cost and volume of X-ray imaging equipment and CT devices. 2) With the restored high density projections, virtually rotating parallel-beam projections are simultaneously acquired by the stationary array X-ray tube and detector. This fact suggests a new methodology of realizing stationary tomosynthesis and CT devices.

## Materials and Methods

### 2.1 CABI algorithm

The brief process of the CABI method is described in Fig. 1. For simplicity, we take a radial profile to demonstrate the CABI method. The method consists of four steps described as follows: 1) coded array source radiation (left in Fig.1), noting that the fan

---

This work was supported by the National Key Research and Development Program of China (Nos. 2022YFA1204200, 2016YFA0202000)
* Correspondence: xqmou@mail.xjtu.edu.cn (X. Mou), stscjun@mail.sysu.edu.cn (J. Chen)

beam of each spot source does not need to cover entire object. 2) Restoration of fan beam projections of each spot source when data completeness is satisfied (middle in Fig.1). 3) Rebinning for a series of parallel-beam projections at densely rotating angles (right in Fig. 1). 4). A high density projection and tomosynthesis results can be reconstructed from the deconvolved projections.

In the imaging modality of line integral, the discrete CABI process of the flat-panel system can be formulized as:

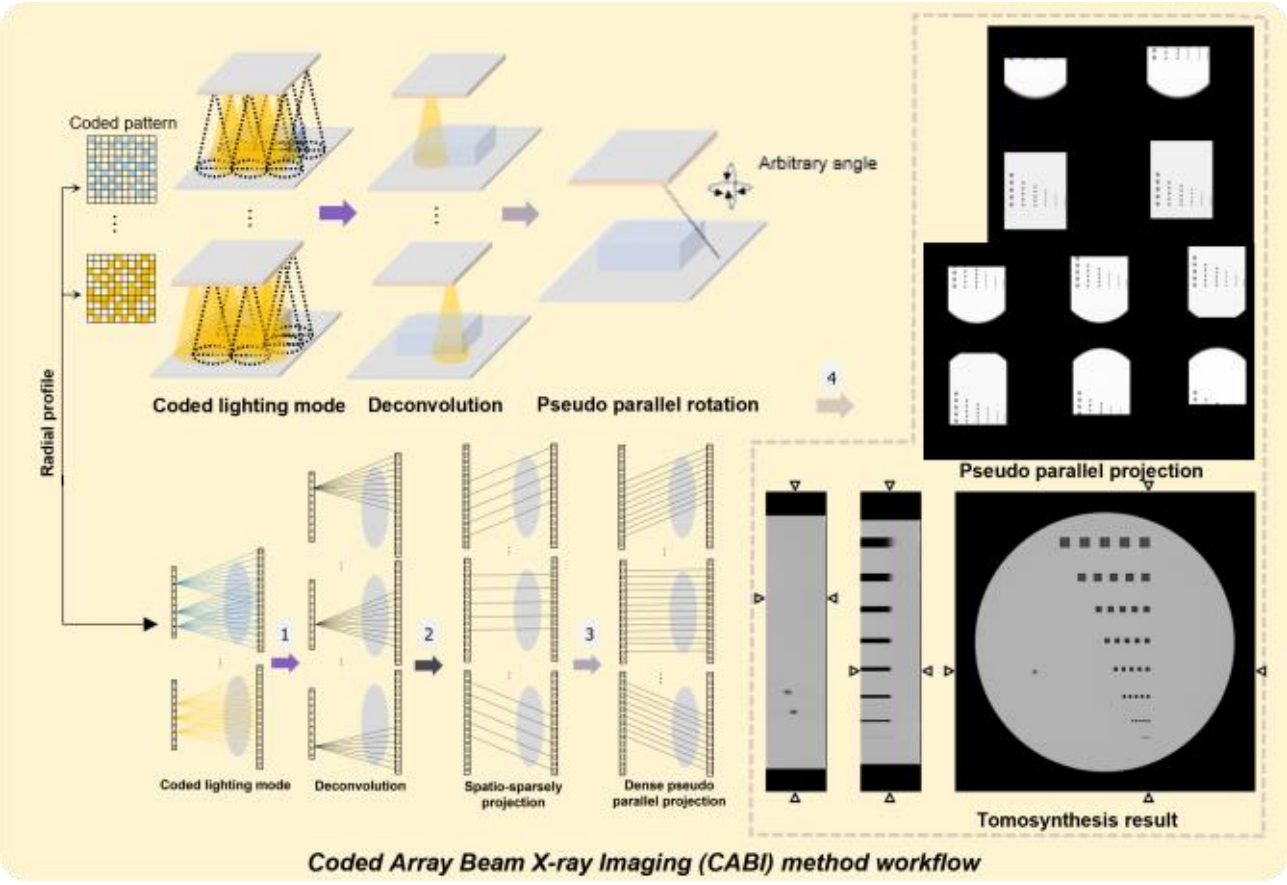

Fig. 1. The diagram of the CABI

$$\mathbf{b}=\mathbf{SAf}=\begin{bmatrix}0&1&\dots&1&1\\1&1&\dots&1&0\\&&\ddots&&\\0&0&\dots&1&1\\1&1&\dots&0&0\end{bmatrix}\begin{bmatrix}\mathbf{P}_{11},\mathbf{P}_{12},\dots,\mathbf{P}_{1N}\\\mathbf{P}_{21},\mathbf{P}_{22},\dots,\mathbf{P}_{2N}\\\dots\\\mathbf{P}_{M1},\mathbf{P}_{M2},\dots,\mathbf{P}_{MN}\end{bmatrix}=\mathbf{SP} \quad (1)$$

where $\mathbf{S}\in\mathbb{R}^{K\times M}$ representing the lighting models, and $M$ is the total number of the flat-panel source. 1 means that the source is lightened up and 0 means off. $\mathbf{A}$ means the system matrix. $\mathbf{P}\in\mathbb{R}^{M\times N}$ denotes the projection data matrix. $\mathbf{P}_{ij}$ means the projection data between i-th source and j-th detector. $\mathbf{f}$ represents the imaging object.

If the rank of (1) is M, we can directly compute each $\mathbf{P}_{ij}$ as:

$$\mathbf{P}=(\mathbf{S}^{\mathrm{T}}\mathbf{S})^{+}\mathbf{S}^{\mathrm{T}}\mathbf{b}\,. \quad (2)$$

However, the direct computation is tedious and ill-posed. Based on the Bayesian rule and medical X-ray imaging physics, this problem can be solved by a statistical iterative reconstruction scheme of maximizing the likelihood function under monochromatic source assumption in real X-ray scan scenarios. Denote $I_{ij}^k$ as the received photons in j-th detector from i-th source in k-th lighting model, which can be formulized as:

$$\bar{I}_{ij}^k=s_i^k I_0 e^{-[Af]_{ij}}\,, \quad (3)$$

where $s_i^k$ is the sampling function. $s_i^k=1$ indicates the i-th source is on in k-th lighting model, while $s_i^k=0$ means that the i-th source is off. For simplicity, we assume each ray has the same incident photons $I_0$. $[Af]_{ij}$ represents the line integral from i-th source to j-th detector.

Remind that the photons from different sources are i.i.d., and can be represented by the Poisson distribution. From the Poisson distribution property, the distribution of total photons in j-th detector also follows Poisson distribution. Thus, we have:

$$\sum_{i=1}^{M}s_i^k I_{ij}^k\sim Poisson(\sum_{i=1}^{M}s_i^k\bar{I}_{ij}^k), \quad (4)$$

where $I_{ij}^k$, $\bar{I}_{ij}^k$ denote the actual and theoretical values of the attenuated X-ray photons, separately. Let $I_j^k=\sum_{i=1}^{M}s_i^k I_{ij}^k$ represents the actual number of the received X-ray photons. $\bar{I}_j^k=\sum_{i=1}^{M}s_i^k\bar{I}_{ij}^k$ is the theoretical value of the total detected photons.

Briefly, the likelihood function $\mathcal{L}(\mathbf{I}\,|\,\mathbf{f})$ can be described as:

$$\mathcal{L}(\mathbf{I}\,|\,\mathbf{f})=\sum_{k=1}^{K}\sum_{j=1}^{N}(I_j^k\log(\bar{I}_j^k)-\bar{I}_j^k)\,, \quad (5)$$

and the optimization function can be optimized by:

$$\mathbf{f}^{z+1}=\left[\mathbf{f}^{z}-\alpha\frac{\sum_{k=1}^{K}\frac{1}{I_j^{k,f^z}}(\sum_{j=1}^{N}(e(\mathbf{f};\mathbf{f}^z)\sum_{i=1}^{M}(Q_3)))}{\sum_{k=1}^{K}\frac{1}{I_j^{k,f^z}}(\sum_{j=1}^{N}(-e(\mathbf{f};\mathbf{f}^z)\sum_{i=1}^{M}(A_{ij}Q_3)+(\sum_{i=1}^{M}Q_3)^2))}\right]_{+}, \quad (6)$$

where $e(\mathbf{f};\mathbf{f}^z)=I_j^k-I_j^{k,f^z}$,

$$I_j^{k,f^z}=\sum_{i=1}^{M}s_i^k I_0\exp(-A_{ij}\mathbf{f}),\quad Q_3=A_{ij}s_i^k I_0\exp(-A_{ij}\mathbf{f})\,.$$

### 2.2 Data completeness for sparse projections

In section 2.1, we have proposed the CABI algorithm, which can restore each spot source projection from mixed measurements using a number of lighting models. In this section, we will explain that the projections from spatio-sparsely arranged spot sources contain the enough information to restore the high density projections.

Recall the Radon transform, let $L(t,\boldsymbol{\theta})$ denote a line in the Euclidean plane, $\boldsymbol{\theta}=(\theta_1,\theta_2)=(\cos\theta,\sin\theta)$ is the angle the normal vector to $L$ makes with the $x_2$ axis, and $t$ is a signed distance of $L$ from the origin:

In [6], the relationship between angle and detector (the signed distance) in the projection can be denoted as:

$$\frac{\partial}{\partial\theta_k}Rf(\boldsymbol{\theta},t)=-\frac{\partial}{\partial t}(R(x_k f))(\boldsymbol{\theta},t),k=1,\dots,\text{space dimension} \quad (7)$$

In 2D parallel-beam geometry, (7) can be rewritten for any direction $\theta$ [7]:

$$\begin{aligned}\frac{\partial}{\partial\theta}Rf(\theta,t)&=\frac{\partial}{\partial\theta_1}Rf(\theta,t)\frac{\partial\theta_1}{\partial\theta}+\frac{\partial}{\partial\theta_2}Rf(\theta,t)\frac{\partial\theta_2}{\partial\theta}\\&=\frac{\partial}{\partial t}(R(x\cdot\boldsymbol{\theta}^{\perp})f))(\boldsymbol{\theta},t)\end{aligned}, \quad (8)$$

where $\boldsymbol{\theta}^{\perp}=(\sin\theta,-\cos\theta)$ is the unit normal vector to $\boldsymbol{\theta}$, $\boldsymbol{\theta}^{\perp}\cdot\boldsymbol{\theta}=0$.

From (8), we can observe that there is an implicit relationship between angle and detector differentials, which is object depended and hence hard to solve. To obtain direct relationship between differentials of the Radon variables, we introduce two pseudo-rotation centers to counteract the influence of weight and transform the implicit relationship between the angle and detector to an explicit relationship, which is shown in Fig. 2 to demonstrate a profile of the 3D Radon transform.

Here, we define three planes parallel to each other: source plane $\xi$, virtual plane $O$, and flat-panel detector plane $\eta$. The distance of $\xi-\eta$ is $H$. $d_i$ represents the distance from $O$ to $\eta$ . Let denote $Rf(O,\theta,d)$ is the Radon transform along line $(\xi-\eta)$ . From Fig. 2, we can observe that $Rf(O_1,\theta,d_1)$ and $Rf(O_2,\theta,d_2)$ denote the same line integral.

Let us denote $T_2$ as the pseudo detector plane parallel to $T_1$. The distance between $T_1$ and $T_2$ is $\tau$ .

Based on $Rf(O_1,\theta,d_1)$ and $Rf(O_2,\theta,d_2)$ , we can calculate the first differential of angle by tilting the line with a slight angle $\Delta\theta$ in $(O_1,d_1)$ and $(O_2,d_2)$ , as the dashed line shown in Fig. 2.

$$\frac{\partial Rf(O_1,\theta,d_1)}{\partial t}\Delta t=\frac{\partial Rf(O_1,\theta,d_1)}{\partial\theta}\Delta\theta-\frac{\partial Rf(O_2,\theta,d_2)}{\partial\theta}\Delta\theta\ . \quad (9)$$

And, it is easy to find that:

$$\Delta t=\tau\tan\Delta\theta\ . \quad (10)$$

When $\Delta\theta$ is small enough, (10) becomes:

$$\Delta t=\tau\Delta\theta\ . \quad (11)$$

Substituting (11) into (9), we get:

$$\frac{\partial Rf(O_1,\theta,d_1)}{\partial t}=\frac{1}{\tau}\left(\frac{\partial Rf(O_1,\theta,d_1)}{\partial\theta}-\frac{\partial Rf(O_2,\theta,d_2)}{\partial\theta}\right), \quad (12)$$

Eq.(12) shows that the differential of $t$ can be precisely recovered by calculating angle differentials of two rotation centers. This formula is independent to the object so that we can easily restore detector information from that of angular views. Covert coordinates in (13) to $(\theta,\eta,o)$, we have,

$$\frac{\partial Rf(O_1,\theta,d_1)}{\partial o}=\frac{\cos^2\theta}{d_1-d_2}\left(\frac{\partial Rf(O_1,\theta,d_1)}{\partial\theta}-\frac{\partial Rf(O_2,\theta,d_2)}{\partial\theta}\right). \quad (13)$$

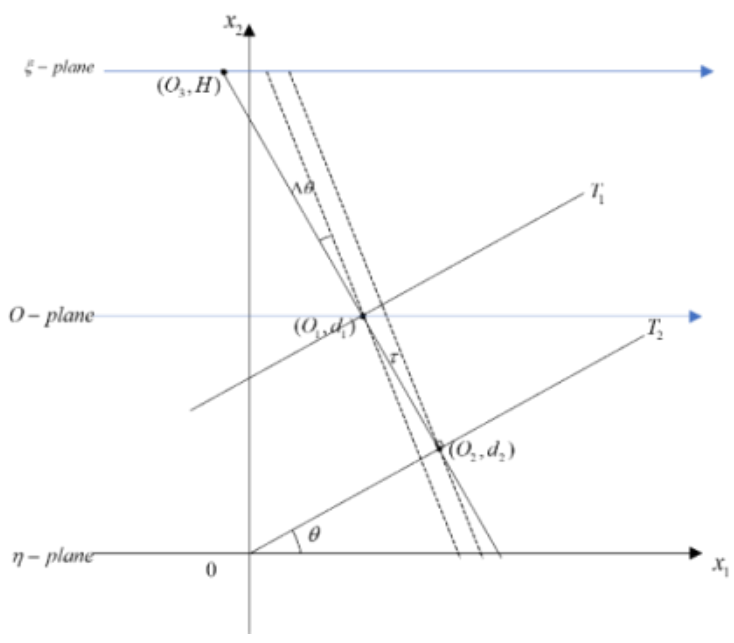

Fig. 2. The pseudo two rotation center geometries

Besides, when we set $d_2=H$ , with a simple derivation, we find:

$$\frac{\partial Rf(O_1,\theta,d_1)}{\partial\eta}=\frac{\cos^2\theta}{H}\frac{\partial R(O_3,\theta,H)}{\partial\theta}\ , \quad (14)$$

and (13) becomes:

$$\frac{\partial Rf(O,\theta,d)}{\partial o}=\frac{\cos^2\theta}{d-H}\left(\frac{\partial Rf(O,\theta,d)}{\partial\theta}-\frac{H}{\cos^2\theta}\frac{\partial Rf(O,\theta,d)}{\partial\eta}\right). \quad (15)$$

Continuing to deduce second-order partial differential equations based on (15), we have:

$$\frac{\partial^2R}{\partial o^2}-\frac{2\cos^2\theta}{(d-H)}\frac{\partial^2R^1}{\partial\theta\partial o}+\frac{\cos^4\theta}{(d-H)^2}\frac{\partial^2R^1}{\partial\theta^2}=\frac{2H\sin\theta\cos\theta}{(d-H)^2}\frac{\partial R}{\partial\eta}+\frac{H^2}{(d-H)^2}\frac{\partial^2R}{\partial\eta^2} \quad (16)$$

The relationship among $\theta,\eta,o$ is described in partial differential equation as (16) if the boundary condition is met. In CABI, when array source is sparse, the rebinned parallel projections have low spatial resolution and dense rotation angles given that the detector is of high resolution. Eqs. (15) and (16) suggest that information in densely angular views can be converted to spatial information for high density restoration, which advocates the similar point with [8]. However, the proposed formulae in the study are more general in describing the relationship. In practice, directly solving the equations is unnecessary since the measured data contain noise and the formulae are sense to noise. Alternatively, we solve the PDE into iterative scheme by introducing compressed sensing [9] and name it as CABI_CS in this paper.

$$\{\mathbf{f}\}=\arg\min_{\mathbf{f}}(F(\mathbf{f})+\beta\|\mathbf{f}\|_{TV})\ , \quad (17)$$

where $F(\mathbf{f})$ is the fidelity term denoted by (5) and solved by (6), and $\|\mathbf{f}\|_{TV}$ the image regularization, respectively. Thus, the proposed CABI_CS algorithm can be described in Table. 1.

**Table. 1 WORKFLOW FOR THE CABI_CS SCHEME**

**Geometry, reconstruction parameters initialization**
**for** each iteration $z=1,...,Z$
—**for** each lighting mode(pattern): $k=1:K$
Compute current received photons:
$$I_j^{k,f^z}=\sum_{i=1}^{M}s_i^kI_0\exp(-A_{ij}\mathbf{f})\ ;$$
Compute $e(\mathbf{f};\mathbf{f}^z)$ ;
Update $\mathbf{f}$ :
$$\mathbf{f}^{z+1}=\left[\mathbf{f}^z-\alpha\frac{\sum_{k=1}^{K}\frac{1}{I_j^{k,f^z}}\left(\sum_{j=1}^{N}(e(\mathbf{f};\mathbf{f}^z)\sum_{i=1}^{M}(Q_3))\right)}{\sum_{k=1}^{K}\frac{1}{I_j^{k,f^z}}\left(\sum_{j=1}^{N}(-e(\mathbf{f};\mathbf{f}^z)\sum_{i=1}^{M}(A_{ij}Q_3)+(\sum_{i=1}^{M}Q_3)^2)\right)}\right]_+$$
—**end**
Total variation operation: $\|\mathbf{f}\|_{TV}$
**end**

2.3 **Materials in simulation and physical experiments**

In simulation, we used a 15mm non-through digital resolution phantom. Eight line-pair structures with square shape and equal width ranging from 0.4 mm to 2.5 mm were listed on a 65mm diameter cylinder. In order to evaluate the Z-axis resolution, we put two balls with 0.5mm diameter in different slices, as shown in third row of Fig.4. The scheme of CABI radiation is same as below real experiment with K=80. Two noise level were tested in the simulation: 5E4 and 1E4 photons.

In the real imaging system, we chose a chicken foot as the imaging object. The thickness of the chicken foot is about 15mm. And the width and length of the chicken foot is about 7.5cmx15cm with package, as shown in Fig. 3(d). The physical experiments were performed in our developed verification platform, as shown in Fig. 3(c). The distance between source and detector is 190mm. The detector size is $11.4cm\times14.6cm$ with $2304\times2940$ bins. The scanning protocol is 36kV in both air scan and object scan. We designed a collimator with slot. The number of array cells is 15x15 with adjacent distance of 3mm. Compared to the resolution of the detector, it is extremely sparse. With simple out and in plugging of coded masks, the coded array beam is realized. In this paper, we chose the simplest orthogonal form. There are a half of lighten cells in each acquisition time. We designed a set of coded apertures depicted in Fig. 3(e). The finest aperture contains two rows of array sources which mean 6mm aperture size at least. With the rotation of apertures, the final coded patterns are 40 in sum.

## 3 Results and Analysis

Fig.4 depicts the reconstruction result of the simulated phantom in two different noise levels. The volume size is set as 0.3mmx0.3mmx0.3mm. From the results, we find that the proposed CABI_CS algorithm can reconstruct sharp images from the overlapping projections in different noise levels. From the line profile, we found that the proposed method can preserve small edges indicated by green line. For Z-plane analysis, we plotted ASF (Artifact spread function) curves of the proposed method, which also shows the efficiency of the CABI_CS method. Fig. 5 shows the physical reconstruction result of the chicken foot with CABI_CS. The volume is set as 0.2mmx0.2mmx0.2mm. During data acquisition, the device temperature is keeping at 60$^o$C due to low power operation. In the experiment, we only performed the constraints along $x_2$ axis. Three different angles are selected to illustrate the process of the pseudo parallel-beam projection, as $\theta = [-10.8°, 0°, 10.8°]$. From the results, we find that the pseudo parallel-beam projections at all angles are clearly reconstructed and the details and structures of the chicken foot are preserved well, even though the coded apertures are larger than 6mm. The results fully prove the effectiveness of the proposed CABI_CS method.

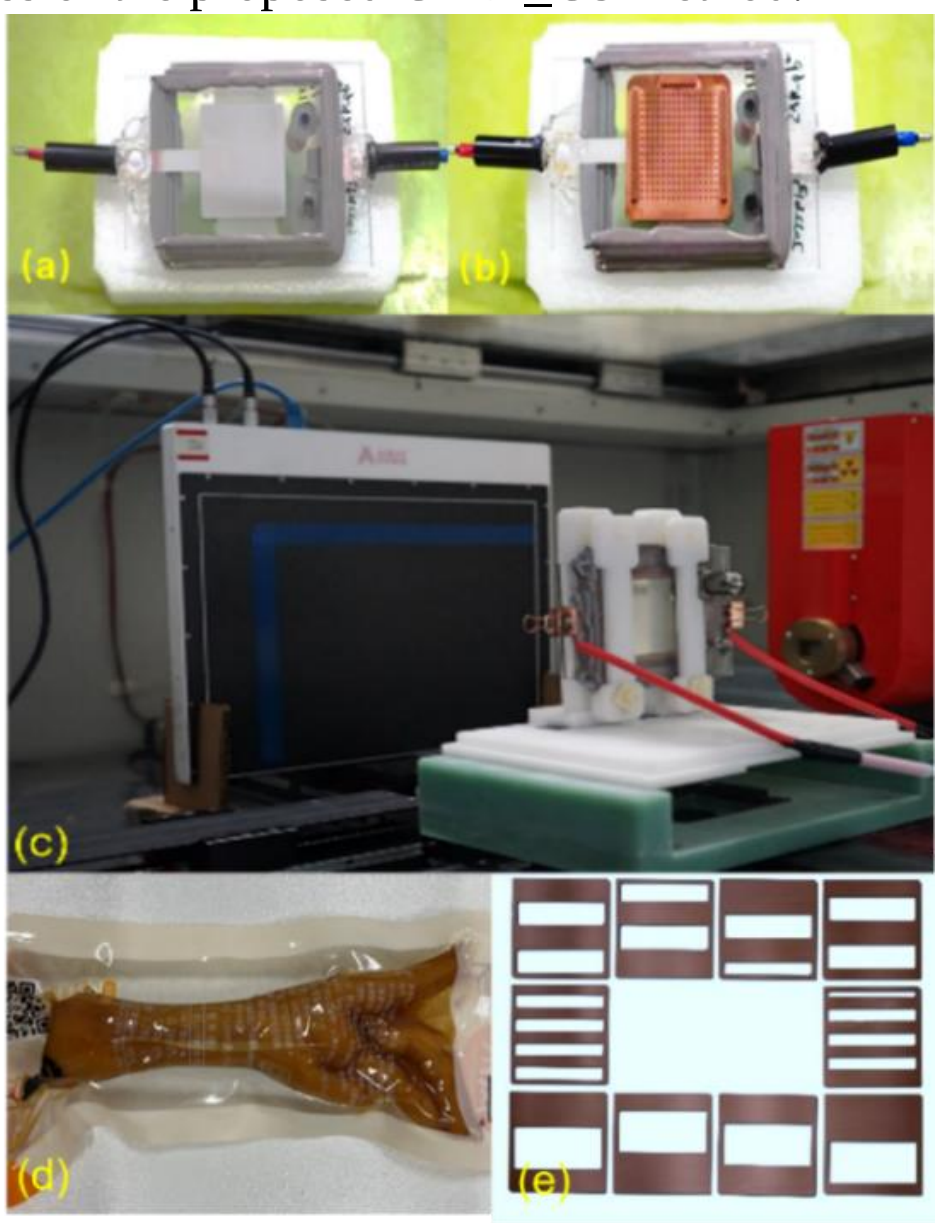

Fig. 3. The physical experiments setting details. (a) shows a flat-panel X-ray array source[4]. (b) is the source covered by a copper collimator. (c) is the geometry setting figuration of the flat-panel source tomosynthesis, where no cooling is applied. (d) shows the chicken foot used in physical experiment. (e) depicts the masks of coded apertures.

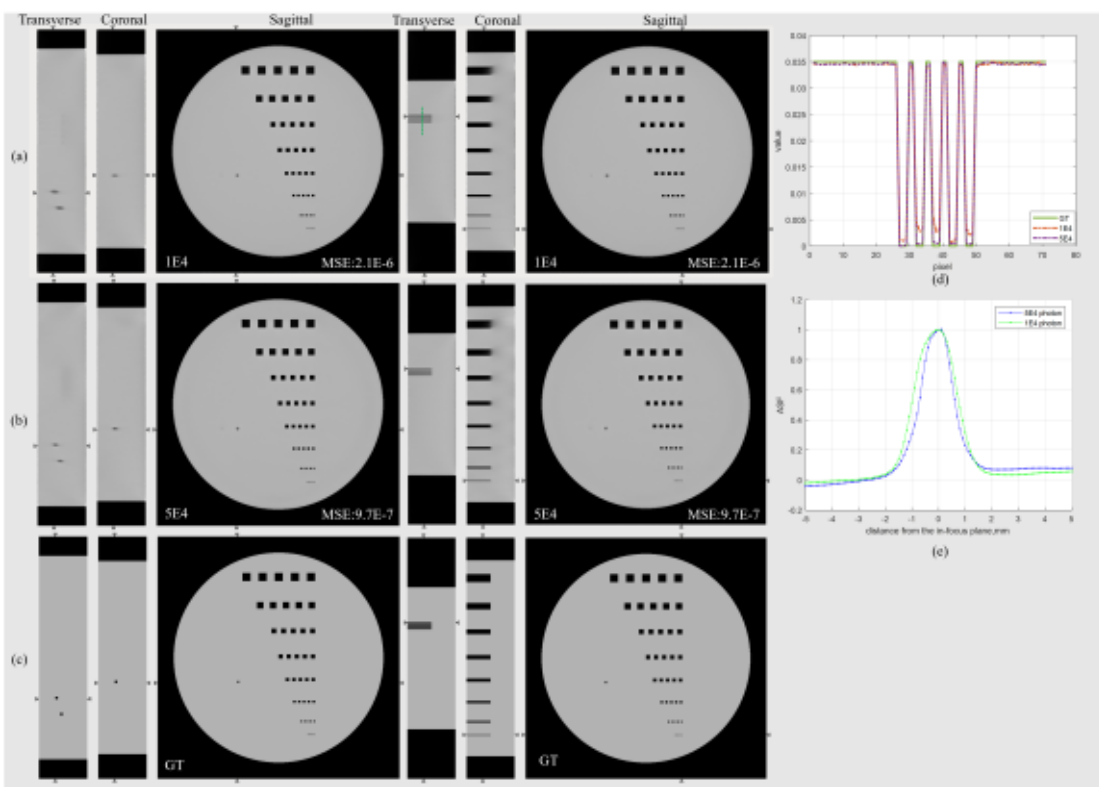

Fig. 4. The reconstruction result (K=80) of the simulation phantom display in [0 0.05], with TV parameter as 0.35.

## 4 Discussion and conclusion

Regardless of the promising results of the CABI_CS algorithm, there are some aspects related to our method which should be discussed. In this paper, we only adopt a simple orthogonal basis of coded apertures. Indeed, how to design coded apertures is on studying, which is one of the crucial topics in developing CABI methods. In the future, we will propose more efficient and compact coded apertures based on specific scenarios. Most importantly, in this study, only TV regularization was applied to CABI calculation although it has achieved remarkable results, clearly, another crucial topic is to further explore reconstruction algorithms with the proposed PDEs to solve the spatio-sparsely projections. In this study, we deduce two dimensional PDEs to describe the relationship between the projections of detector and rotation angles. However, three dimensional PDE, such as John's equation could be used if some coordinate conditions are met[8, 10].

In conclusion, we proposed the CABI method to obtain the high density tomographic image from coded sparse array sources, as well as virtual parallel-beam projections that can be used for stationary CT reconstructions and tomosynthesis imaging. Meanwhile, we theoretically analyze the data completeness in spatio-sparsely array beam. The proposed CABI and coded array X-ray source complement each other toward a new X-ray imaging modality. The simulation and physical results show its possibility of becoming a next generation of the X-ray imaging technologies.

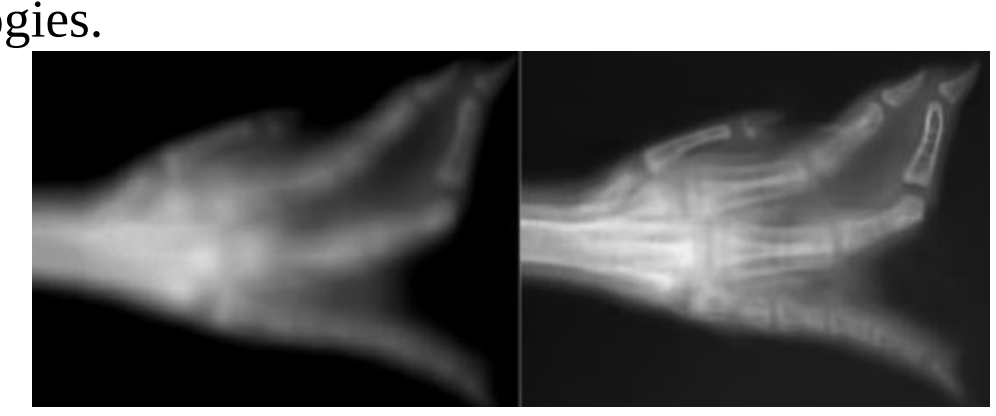

(a)original overlapped projection (b) CABI based reconstruction

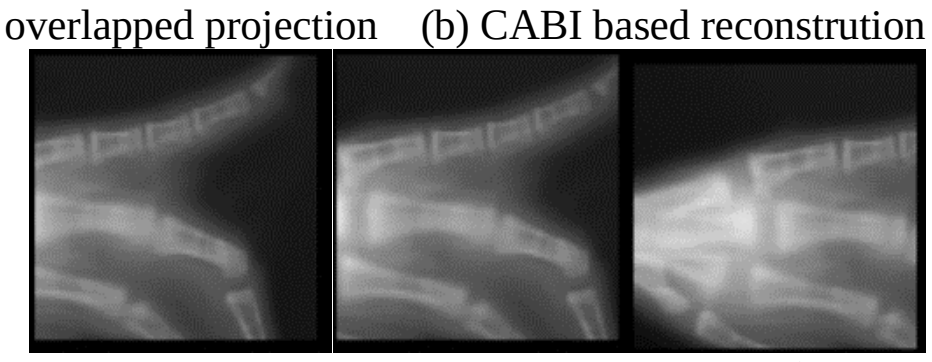

(c) pseudo parallel rotation projection samples of angles -10.8$^0$ (left), 0$^0$ (middle), 10.8$^0$ (right)

Fig. 5. The CABI reconstruction result based on the real tomosynthesis system.

# Characterization of cysts and solid masses using direct-indirect dual-layer flat-panel detector

Xiaoyu Duan, Hailiang Huang, and Wei Zhao

Department of Radiology, Stony Brook Medicine, Stony Brook, NY, United States

**Abstract** Dual-energy (DE) mammography and DE digital breast tomosynthesis provide spectral information, which could be utilized to conduct material decomposition for lesion characterization. Distinguishing solid masses and cysts has the potential to improve diagnostic accuracy and reduce the call back rate for cysts. We proposed a direct-indirect dual-layer flat-panel detector (DLFPD) to acquire low-energy and high-energy images simultaneously, with the benefit of no patient motion between LE and HE exposures. This DLFPD incorporates existing state-of-the-art direct (a-Se) and indirect (CsI) FPDs for breast x-ray imaging, which promises rapid clinical translation. The feasibility of distinguishing solid masses and cysts with direct-indirect DLFPD was validated using projection-based material decomposition method, which decompose masses and cysts into aluminum (Al) and polymethyl methacrylate (PMMA).

## 1 Introduction

Using spectral X-ray imaging and material decomposition to distinguish solid masses and cysts has the potential to improve diagnostic accuracy and reduce the call back rate for cysts.[1-5]

Previous studies have obtained breast spectral information by utilizing systems that incorporate photon counting detectors.[1,2,4] However, these systems are known to be high cost and have limited availability. Alternatively, dual-energy (DE) mammography and DE digital breast tomosynthesis based material decomposition techniques usually acquire low-energy (LE) and high-energy (HE) images separately, by switching the x-ray filter and tube voltage between exposures. This approach, however, can result in patient motion between LE and HE images, which can lead to artifacts in the final image. In contrast, dual-layer detectors enable the simultaneous acquisition of both LE and HE images with a single exposure, eliminating the possibility of motion artifacts and resulting in higher quality images.

We proposed a direct-indirect dual-layer flat-panel detector (DLFPD) and K-edge filter combination, incorporating existing state-of-the-art direct (a-Se) and indirect (CsI) FPDs for breast x-ray imaging, which promises rapid clinical translation.[6] This work distinguishes solid masses and cysts using projection-based material decomposition with direct-indirect DLFPD DE images.

## 2 Materials and Methods

### 2.1 Direct-indirect dual-layer flat-panel detector (DLFPD) for dual energy breast imaging

Our previous spectral simulation study optimized the material and thickness of DLFPD and the x-ray filter as Fig 1 shows.[7] The K-edge filter, Ag (150 μm thickness), shapes x-ray spectrum into two peaks as Fig 2(a) shows. Front-layer (FL) a-Se direct detector with thickness of 200 μm can sufficiently absorb LE photons. Back-layer (BL) CsI indirect detector with thickness of 400 μm enhances the absorption of HE photons, thereby improving energy separation between LE and HE images without added filtration between detectors. Orange and purple curves in Fig 2(b) show analytically calculated photon energy spectra absorbed by the FL a-Se detector (200 μm thickness) and BL CsI detector (400 μm thickness) respectively. The detector pixel pitch for both layers was 85 μm. The 700 μm thick glass between two layers of detectors corresponds to the TFT substrate of the FL detector.

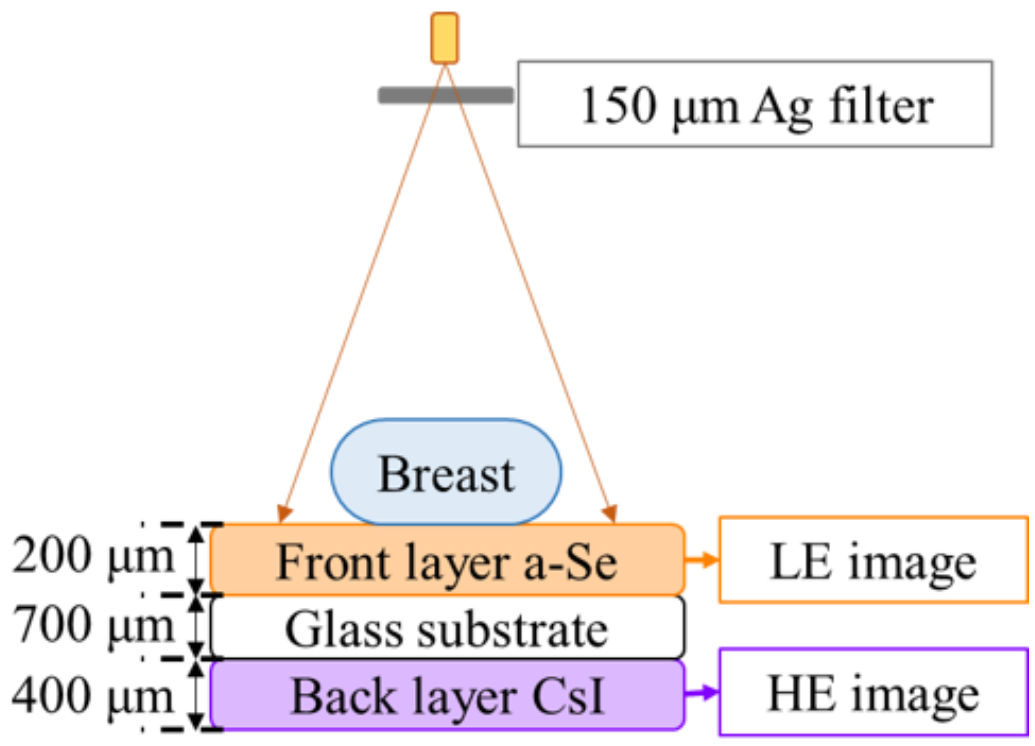

*Fig 1. System design and image geometry for dual-layer detector LE and HE images.*

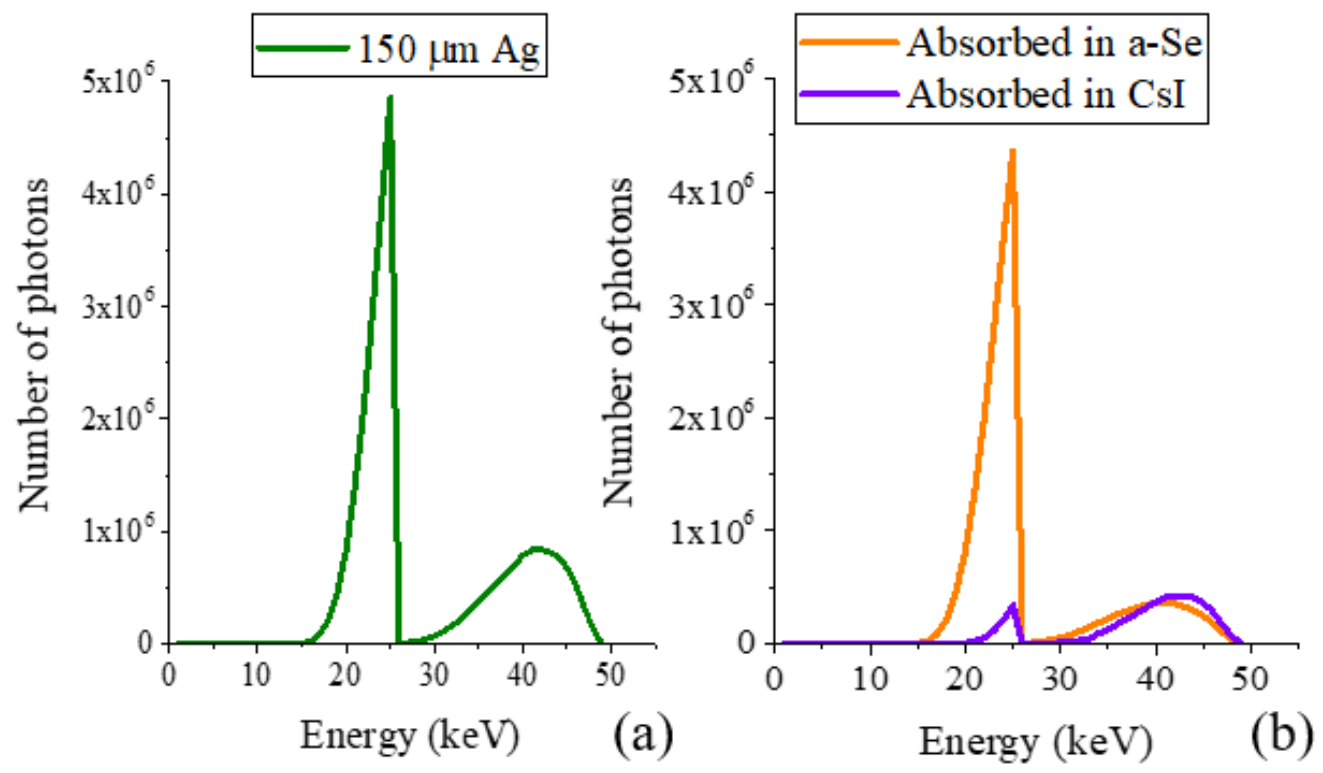

*Fig 2. (a) Energy spectrum of incident photons in after the k-edge filter Ag. (b): Energy spectrum of photons absorbed in each layer of the direct-indirect dual-layer detector. Orange curve: front-layer a-Se detector. Purple curve: back-layer CsI detector.*

### 2.1 Differentiate between solid masses and cysts using material decomposition method

Breast tissue attenuation can be approximated by the linear combination of two reference materials with equivalent thicknesses.[1-3] We used Al and PMMA as the reference materials to differentiate solid mass from cyst. To establish the calibrated lookup table for attenuation of different Al-PMMA thickness calibration combinations, a digital phantom was generated, composed of various Al-PMMA combination blocks, mass, and cyst blocks in different thicknesses (Fig 3).

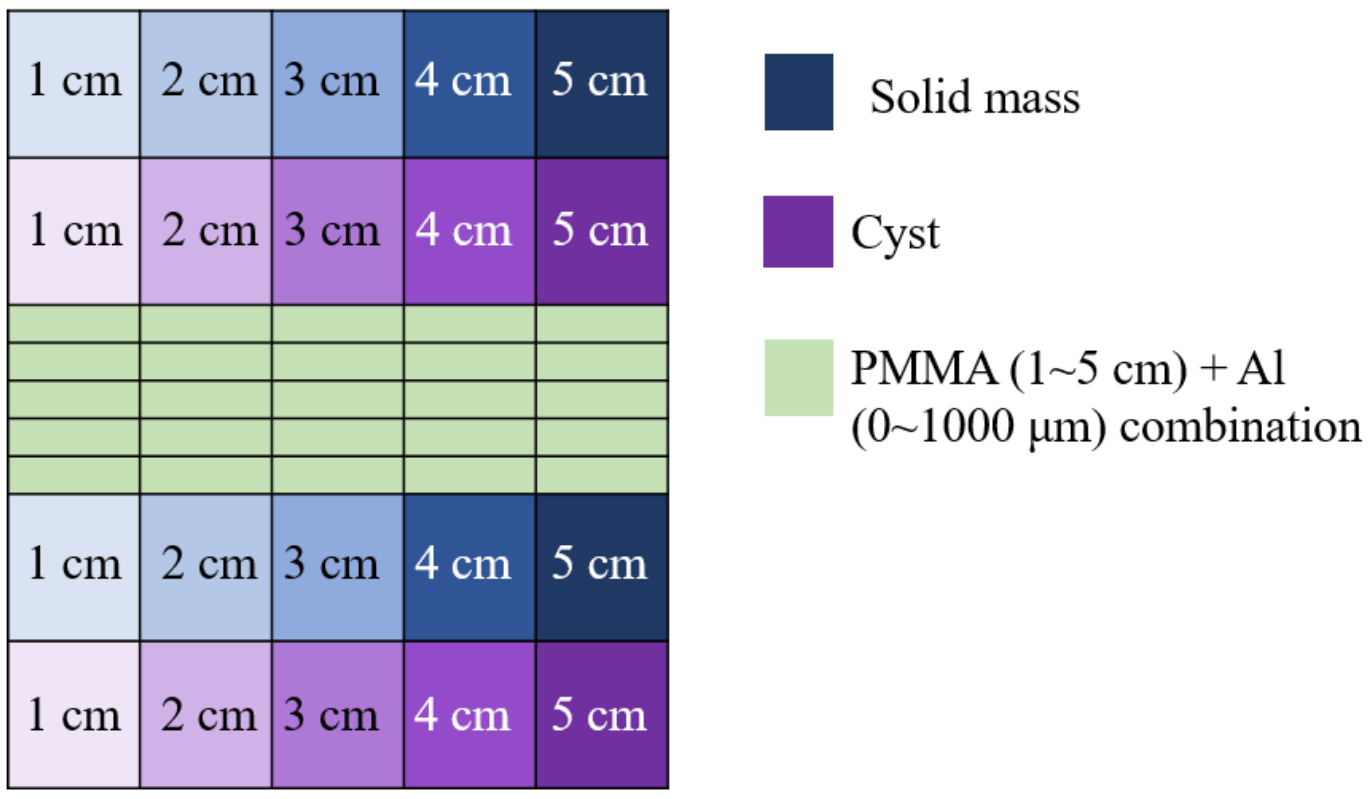

*Fig 3. Cross section of digital calibration phantom for solid mass and cyst classification*

Monte Carlo simulations were conducted to acquire LE and HE images using VICTRE.[8] The attenuations of masses used in the simulation match those for IDC measured by Johns and Yaffe.[9] The attenuations of cysts were generated using Penelope dataset with chemical components provided by Fredenberg et al.[1] Scatter radiation was not involved in this study. The attenuation information of material from virtual FL LE images and BL HE images were calculated as Eq 1 and 2:

$$\boldsymbol{attenuation_{front}} = -\log\left(\frac{\boldsymbol{I_{front}}}{\boldsymbol{I_{0,front}}}\right) \quad (1)$$

$$\boldsymbol{attenuation_{back}} = -\log\left(\frac{\boldsymbol{I_{back}}}{\boldsymbol{I_{0,back}}}\right) \quad (2)$$

$\boldsymbol{I}$ was image intensity after material attenuation. $\boldsymbol{I_0}$ was the un-attenuated signal.

The attenuation of solid masses and cysts were mapped back to corresponding PMMA/Al combination using linear interpolation method. In the equivalent thickness map, the magnitude $\boldsymbol{r}$ and angle $\boldsymbol{\theta}$ of each point were calculated as Eq 3 and 4:

$$\boldsymbol{r} = \sqrt{\boldsymbol{t_{Al}^2} + \boldsymbol{t_{PMMA}^2}} \quad (3)$$

$$\boldsymbol{\theta} = \tan^{-1}\left(\frac{\boldsymbol{t_{Al}}}{\boldsymbol{t_{PMMA}}}\right) \quad (4)$$

Larger value of magnitude represents thicker object. Angle relates to the material effective atomic number.

## 2.2 An analytical model to improve spectral separation with dual-layer detector

As Fig 2(b) shows, the FL absorbs a non-negligible portion of HE photons. This HE photon contamination in FL LE image compromises the energy separation and can cause inaccuracies in material decomposition. We developed a workflow to generate virtual FL LE images that efficiently remove the HE photon contamination (Fig 4). Firstly, FL image could be considered as the sum of a FL LE image and a FL HE image. Because the x-rays that interact with the BL have a similar beam quality to the HE portion of the x-rays that interact with the FL (Fig 2(b)), the attenuation information of a virtual FL HE image was obtained by multiplying a factor $R_{object}$ to attenuation information of the BL as illustrated in Fig 4(b). For the breast phantom, we assumed the adipose and glandular tissue compose the most breast volume and determined the ratio $R_{object}$ from attenuation coefficients of adipose and glandular tissue at the mean energy of FL HE part and BL spectrum (Eq 5, 6, 7)

$$R_{object,ad} = \frac{I_{front,H,blankcorr}}{I_{back,blankcorr}} \sim \exp\left(\left(\mu_{ad}(\overline{E_{back}}) - \mu_{ad}(\overline{E_{front,H}})\right) t_{ad}\right) \quad (5)$$

$$R_{object,gl} = \frac{I_{front,H,blankcorr}}{I_{back,blankcorr}} \sim \exp\left(\left(\mu_{gl}(\overline{E_{back}}) - \mu_{gl}(\overline{E_{front,H}})\right) t_{gl}\right) \quad (6)$$

$$R_{object} = average\left(R_{object,ad}, R_{object,gl}\right) \quad (7)$$

*Tab 1.* $R_{object}$ *of adipose and glandular tissue for 1 cm thickness*

| Material $\boldsymbol{i}$ | $R_{object,i}$ |
|---|---|
| Adipose tissue† | 1.006461002 |
| Glandular tissue† | 1.011316201 |

† Data from ICRP report

Secondly, a virtual FL HE blank scan was estimated from the FL blank scan using the prior knowledge of the x-ray spectrum and detector characteristics (Fig 4(c)). The ratio between blank scan of FL LE and HE was approximated as the ratio of energy fluence of photons incident on the detector ($N \cdot \bar{E}$), absorption coefficient of a-Se detector ($F_{abs,\,Se}$), and the gain of a-Se detector between FL LE and HE spectrum (Eq 8). The ratio $R_{blank}$ was calculated to estimate the virtual FL HE blank scan (Eq 9).

$$\frac{I_{front,L,blank}}{I_{front,H,blank}} = \frac{N_{top,L}}{N_{top,H}} \cdot \frac{\overline{E_{top,L}}}{\overline{E_{top,H}}} \cdot \frac{F_{abs,\,Se,L}}{F_{abs,\,Se,H}} \cdot \frac{W_{Se}}{W_{Se}} \quad (8)$$

$$R_{blank} = \frac{I_{front,H,blank}}{\left(I_{front,L,\,blank} + I_{front,H,\,blank}\right)} \quad (9)$$

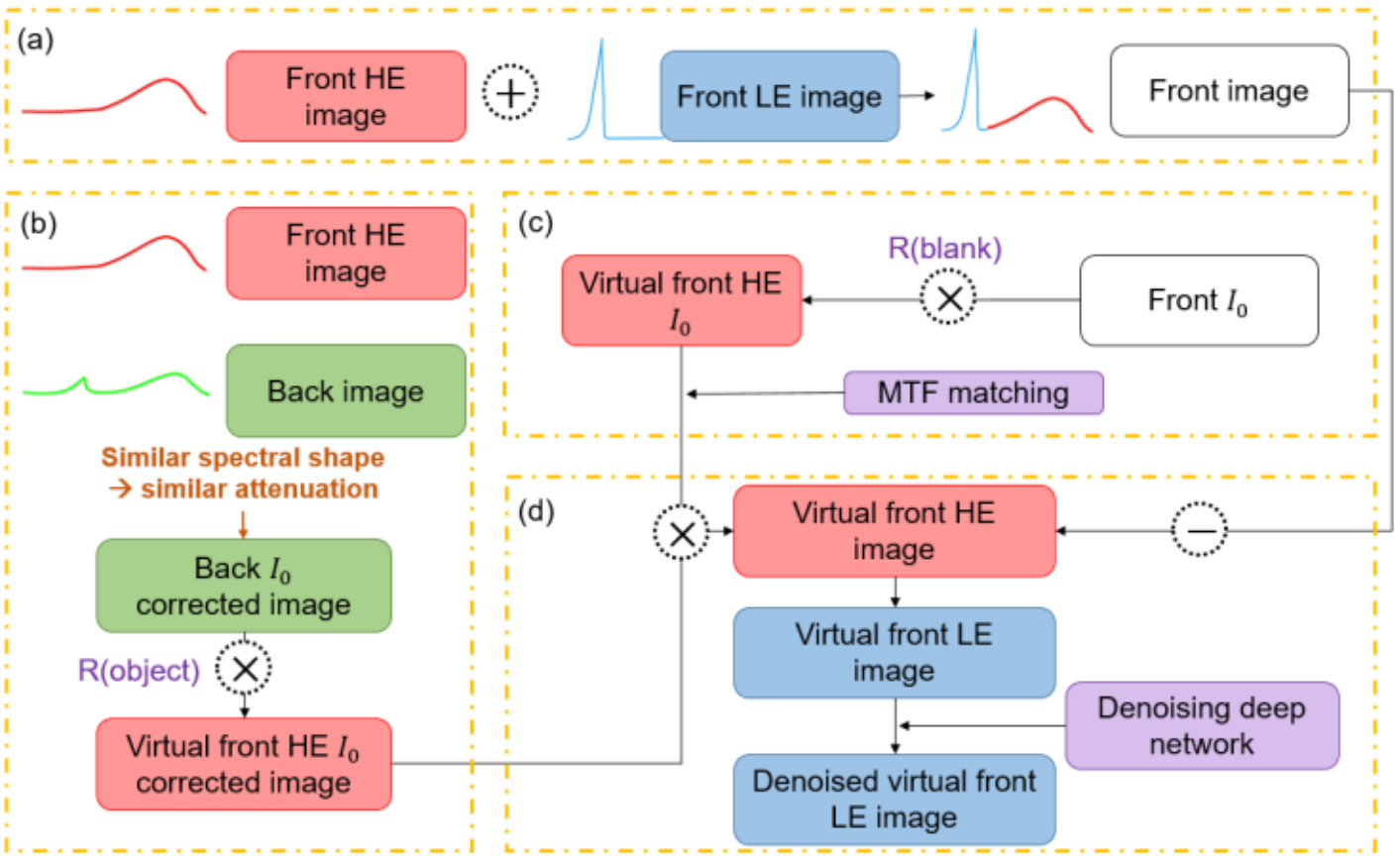

*Fig 4. (a) Photons absorbed by the front-layer detector contain two energy peaks (HE and LE part). (b) The HE part of the front-layer image has similar attenuation information as the back-layer image. We will use a ratio R(object) (calculated from the materials attenuation coefficients) to scale back-layer $I_0$ corrected image to the virtual front HE $I_0$ corrected image. (c) Virtual front HE $I_0$ image will be obtained by scaling the front $I_0$ image with a ratio R(blank). (d) Virtual front LE image can be calculated by subtracting virtual front HE image from the front-layer image. We will also apply a previously developed deep learning denoising neural network to reduce the image noise of the virtual front LE image.*

Finally, as Fig 4(c) shows, the virtual FL HE image was generated by multiplying the virtual FL HE image attenuation and the virtual FL HE blank scan. The virtual FL LE image was generated by subtracting the virtual FL HE part image from the FL image. A deep learning based denoising technique was utilized to reduce the image noise of the virtual FL LE image. [10]

## 3 Results

Black dots in Fig 5(a) and (b) are calibration points of different PMMA/Al thickness combinations using original FL images and virtual FL LE image respectively. When the virtual FL images is used for calibration, the lookup table is linear, which can greatly reduce the error involved in the following interpolation step.

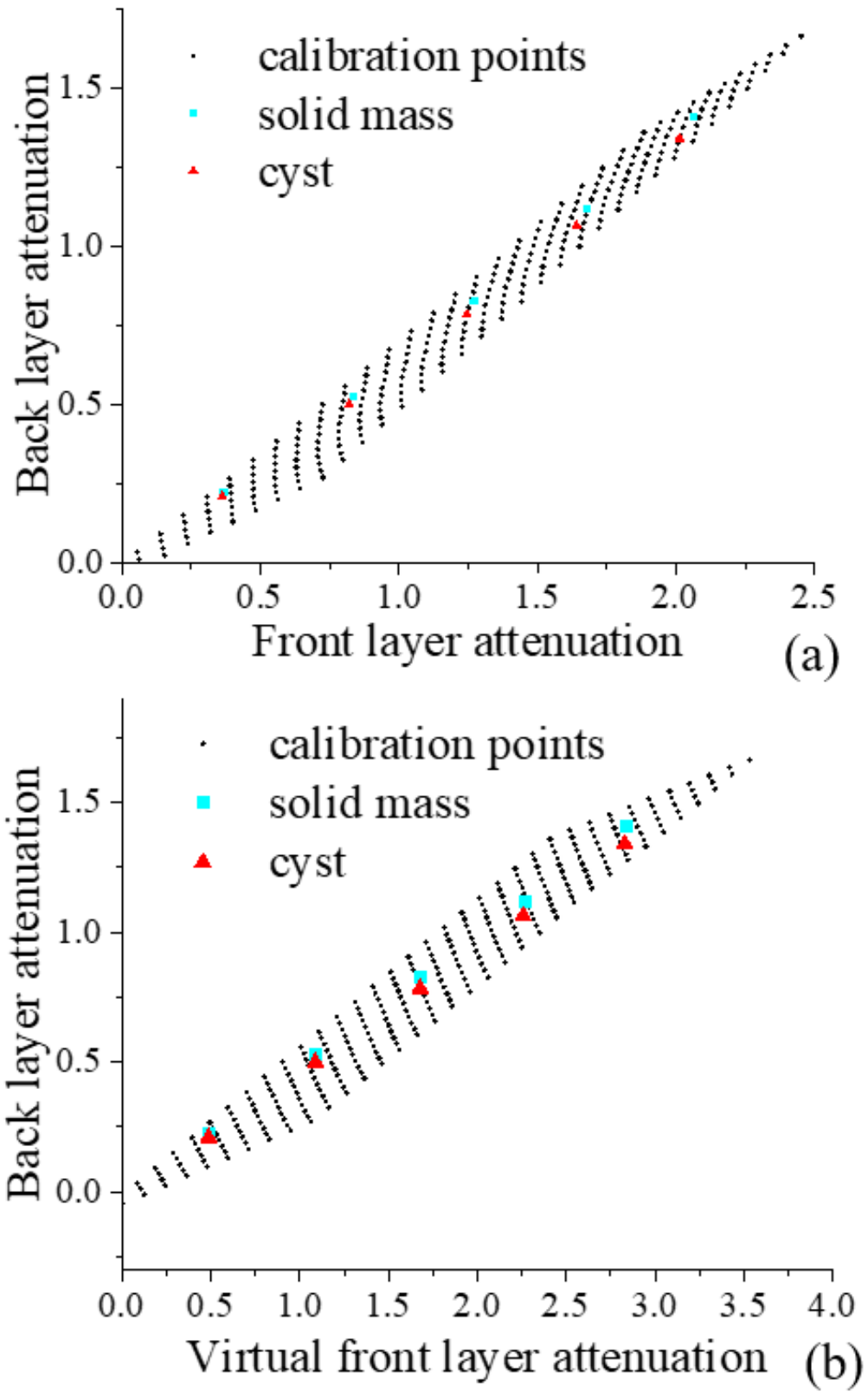

*Fig 5. (a) The calibration look up table of attenuation of front-layer and back-layer images. (b) The calibration look up table of attenuation of virtual front-layer LE and back-layer images.*

Fig 6 shows the interpolated equivalent PMMA/Al thickness for solid masses (with 1 ~ 5 cm thickness) and cysts (with 1 ~ 5 cm thickness), using calibration table in Fig 5(b). The solid masses and cysts were distinguished with different decomposition angles. For solid masses, all five points fall on a linear fitted curve with slope of 36.7. For cysts, points fall on a linear fitted curve with slope of 26.0.

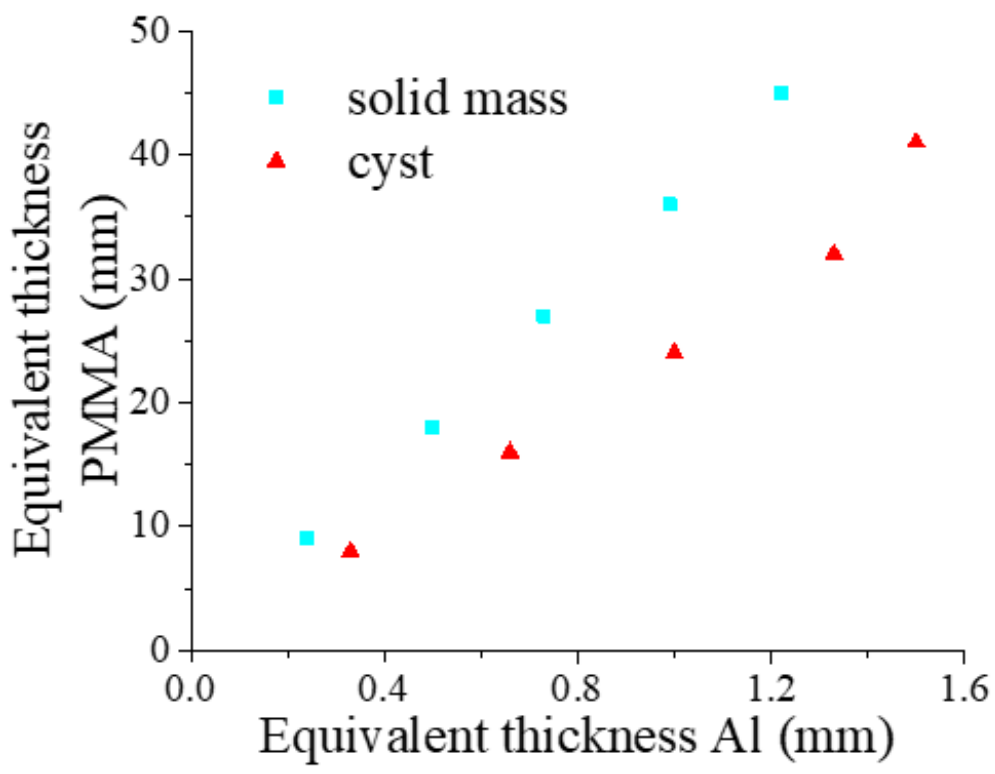

*Fig 6. The equivalent PMMA/Al thickness of solid masses and cysts*

## 4 Discussion

Results of calibration phantom demonstrated the possibility of using two-material (PMMA/Al) decomposition to differentiate between solid mass and cyst. This technique could be developed for lesion characterization in breasts with structured background and thickness roll-off. For this task, breast thickness and volumetric breast density (VBD) estimation are needed as prior knowledge, which could be acquired by our in-house algorithm utilizing dual-energy method decomposition method.[11] VBD values under lesion regions could be approximated as the average of surrounding background region VBD values. Lesion thickness could first be estimated from its measured size parallel to the detector plane, and then be adjusted depending on a DBT volume. The virtual FL and BL attenuation of the targeting lesion could be estimated and used to generate corresponding PMMA/Al equivalent thicknesses. The probability of the lesion being a solid mass or cyst could be determined by a vector calculated as the linear combination of this target lesion's distance to the solid mass and cyst linear fitted curves distances on the Al-PMMA equivalent thickness map.

Our workflow of generating virtual LE reduces the HE photon contamination in the FL image, has the benefit of not only easier material decomposition, but also higher contrast of lesions in contrast-enhanced breast imaging.

In future works, we will incorporate scatter correction for DLFPD images into the workflow.[12,13] By doing so, this workflow would be applicable for conditions without any scatter rejection, such as in DE-DBT.

## 5 Conclusion

We validated the feasibility of distinguishing solid masses and cysts with direct-indirect DLFPD using material decomposition method. We developed a workflow to minimize the contamination of HE photon in FL images, which improved accuracy of material decomposition. This technique could be potentially used in clinical practise to reduce the call back rate for cysts.

# Tomographic Image Reconstruction of Triple Coincidences in PET

Nerea Encina[1], Alejandro López-Montes[2], Jorge Cabello[3], Hasan Sari[4,5], George Prenosil[4], Maurizio Conti[3], and Joaquin L. Herraiz[1*]

[1]Nuclear Physics Group and IPARCOS, University Complutense of Madrid (IdISSC), Madrid, Spain
[2]JHU Department of Biomedical Engineering, Johns Hopkins University, Baltimore, Maryland, United States of America
[3]Siemens Medical Solutions USA, Inc., Knoxville, Tennesse, United States of America
[4]Department of Nuclear Medicine, Bern University Hospital, Bern, Bern, Switzerland
[5]Siemens Healthcare AG, Advanced Clinical Imaging Technology, Lausanne, Switzerland

**Abstract -** Triple coincidences in Positron Emission Tomography (PET) occur when three gamma-rays are detected within the same time coincidence and energy window in a PET scanner. They can be produced by many different causes, and if they are not analyzed separately, they are just reconstructed as two separate double coincidences. As at least one of them will introduce background to the image, and even though this effect may be corrected by a randoms correction, the overall effect is a degradation of the image quality. In this work, we evaluated the performance of different methods to process these triple coincidences, using acquisitions with $^{18}$F, $^{68}$Ga, and $^{124}$I from the Biograph Vision Quadra PET/CT scanner. The contrast and noise of the reconstructed images from triple coincidences were compared with those acquired through the standard method. Our results confirm that images obtained from triples coincidences with a specific treatment have a significant better signal-to-noise ratio  up to a 20% noise reduction) and similar contrast (1% improvement) compared to the current approach. This work shows the importance of proper processing and reconstructing triple coincidences in PET, as they can be as high as randoms in high-sensitivity PET scanners.

## 1 Introduction

Positron Emission Tomography (PET) is based on the detection and image reconstruction of the pairs of annihilation gamma rays from the positrons emitted by a radiotracer. Therefore, triple coincidences in PET (i.e., three gamma-rays detected simultaneously) have been traditionally neglected, despite representing a significant amount of the data acquired in many studies [1]. There are multiple sources of triple coincidences in PET:

**A)** Triples may occur when working with positron-gamma emitters such as $^{124}$I, $^{76}$Br, $^{44}$Sc, $^{86}$Y, $^{52}$Mn, $^{68}$Ga [2] (Fig. 1) as the two annihilation photons may be detected together with a prompt-gamma ray. This type of coincidences may be used to enable dual-isotope PET imaging [3-5] and positronium lifetime analysis [6].

**B)** Even when working with standard pure positron emitters such as $^{18}$F, $^{13}$N, or $^{11}$C, a significant amount of the measured coincidences may be triples:

- They can be due to the inter-detector scatter of one of the gamma rays [7-9], which produces two events (with a total energy of 511 keV) from the same annihilation. These events are rarely recorded in clinical PET scanners due to their narrow 511 keV-centered energy window, but they can account for up to 30% of the coincidences in preclinical studies with worse energy resolution and larger energy windows [7].

- The most common source of triple coincidences in clinical PET scanners is random triples. They are produced when two different positron emissions occur within the same time window, and three or even four out of the four annihilation gamma-rays produced are detected (Fig. 1A). Random triples are becoming more common as the sensitivity and axial FOV of state-of-the-art PET scanners are increasing.

- Another source of triples in PET scanners with Lutetium (Lu)-based scintillators, such as LSO and LYSO, is the emission of an electron and several simultaneous gamma rays from $^{176}$Lu, a long-lived and naturally occurring (2.6% abundance) radioactive element [10].

- Other sources of triples, such as the annihilation of the ortho-positronium into three gamma-rays, have also been studied [11], but in clinical scanners (with a standard energy window) they are far less common compared to the previous cases.

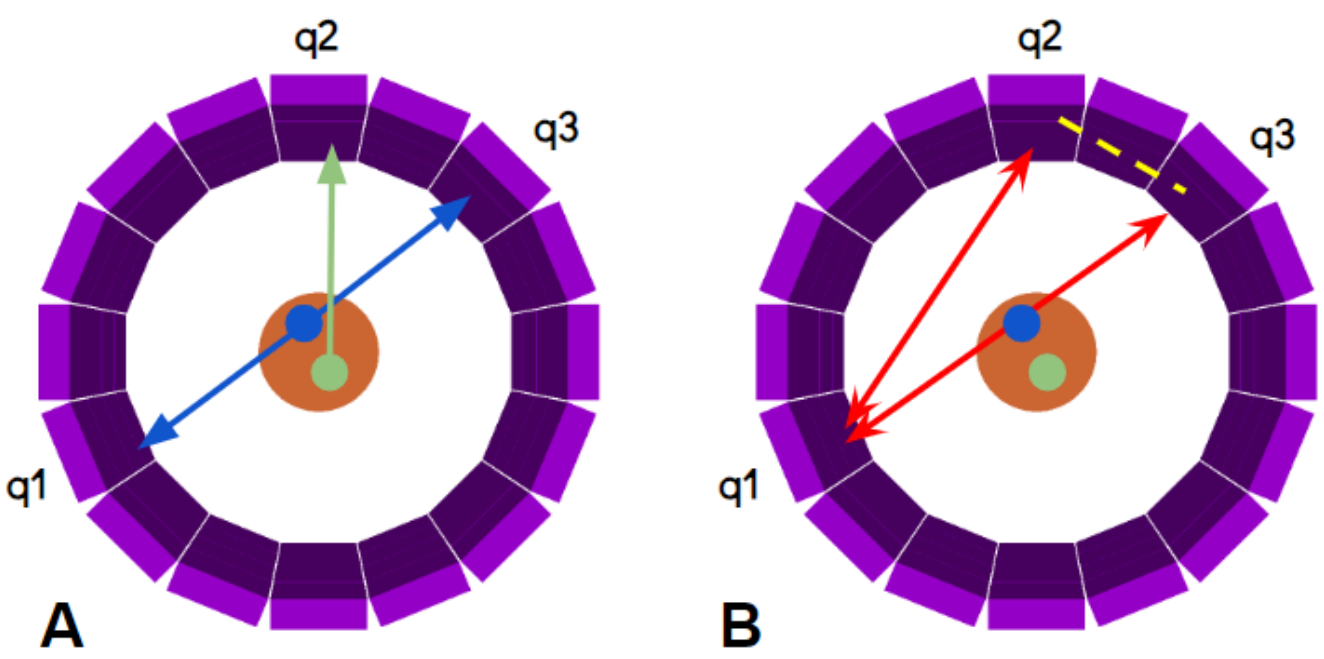

Fig. 1. A) Random event with three detected gamma-rays (i.e. Triple Random). B) The three detection points generate up to three lines of response (LOR) connecting them. Typically, one of them falls out of the FOV (yellow line) and does not need to be considered in the reconstruction. The other two (red lines) can be connected with a V-shape LOR (VLOR), which corresponds to two LORs with a common vertex.

Some new PET scanners are being developed to exploit triple coincidences, such as the J-PET [11] and MI-PET [12] scanners. They provide the possibility to differentiate the nature of each detected gamma (annihilation, prompt-gamma, etc.) based on their energy. However, most commercial PET systems are just optimized for 511 keV gamma-rays detection, and they do not allow the energy differentiation of events detected within the defined energy window. Furthermore, in the case of triple random coincidences (Fig.1A), all gamma rays have the same 511 keV energy, and it is not possible to use the energy information to classify the possible LORs.

From the list-mode data of commercial PET scanners, we can identify triple coincidences as contiguous pairs of double coincidences with a common detector element (Fig. 1B) [5]. Therefore, they can be retrieved from the acquired data and reconstructed separately using a new list-mode file which defines each triple with two LORs. At least one of them will introduce noise to the background, while the other may be the true annihilation pair and contribute to the signal. If they are not properly handled, they will just be reconstructed as two unrelated doubles.

In this work, we studied the amount of triple coincidences in full-body PET acquisitions from Biograph Vision Quadra with commonly used radionuclides and evaluate methods to process them.

## 2 Materials and Methods

### A. Study of Triple Coincidences in real PET Data

The relative amount of triple coincidences with respect to doubles depends on the specific radionuclide used (whether they are pure positron or positron-gamma emitters), the count rate (as it determines the relative amount of random coincidences and $^{176}$Lu background), and the sensitivity and geometry of the scanner.

In this work, we analyzed the triple coincidences present in three different acquisitions from the Biograph Vision Quadra PET/CT scanner at Bern Hospital: a patient from Bern Hospital injected with $^{18}$F-FDG (total activity 328 MBq), a phantom acquisition corresponding to an IQ phantom plus two cylinders placed along the axial FOV, all filled with $^{68}$Ga, and a small vial containing $^{124}$I, respectively. They were considered as acquisitions with high, medium and low activity respectively. The patient data was acquired as part of a study which was approved by the regional ethics committee (KEK 2020/01413) and performed in accordance with the declaration of Helsinki and all relevant national legislation

We also studied the evolution of the triple random coincidences over time, using the patient data ($^{18}$F-FDG). We verified the hypothesis that the majority of the detected triple coincidences correspond to the simultaneous detection of a double coincidence and a single event, i.e. the expected evolution over time of the triple coincidence rate should be proportional to:

$$T(t) = k \cdot D_p(t) \cdot \sqrt{D_r(t)} \quad (1)$$

where $T$ corresponds to the count rate (CR) of triples, $D_p$ and $D_r$ are the CR of double prompt and double randoms respectively, and k is a proportionality constant.

### B. Triples Processing Methods

In this work, we have considered different methods to perform the tomographic image reconstruction of triples:

a) Standard Method: Triples are reconstructed as two unrelated double coincidences. As previously indicated, this may not be optimal, despite currently being the standard procedure.

Proposed Methods: Each of the two LORs of the triple receives a weight ($\omega_1 , \omega_2 ,, \omega_1 + \omega_2 = 1$). The respective weights can be obtained with multiple methods:

A) Using the measured doubles as a reference [7]. This method uses the relative number of doubles measured in each of the LORs to assign a weight to each LOR of the VLORs. It is fast and easy to implement because no prior reconstruction needed [9].

B) Using a projection as a reference. In this case, the weights are obtained from the projections of a previously reconstructed image from the doubles (instead of using the doubles directly as in the previous case). This reduces the noise in the weights' estimation. This approach was successfully used to reduce the artifacts in multiplexed multi-pinhole SPECT [14].

C) Iterative: This approach performs the reconstruction of the triples iteratively using VLORs [5]. The weights of each LOR of the triples are obtained from the relative number of counts of the last projection of the iterative procedure. It is worth noting that VLORs (formed by the union of the 2 LORs of the triple) can be considered as generalized LORs that are bent and about two times larger than standard ones. As Time-of-flight (TOF) PET has shown that reducing the length of the LORs accelerates the convergence of the iterative reconstruction, VLORs reconstruction have a slower convergence.

### C. Evaluation with Patient and IQ Phantom Data

The triples obtained from the list-mode data were processed using the methods previously described and reconstructed with 50 iterations of the ISRA iterative algorithm [16] working in the Histo-Image Space using the TOF information.

The noise, defined as the ratio of the standard deviation and the mean in an expected uniform region was calculated using the liver as the region of interest, and a tumor in the abdominal region was used as the hot region for the quantitative analysis.

## 3 Results

Table 1 shows the count rate (in Mcounts/second) of prompts and random doubles and triples of two different acquisitions. The relative amount of triple coincidences with respect to doubles depends on each case. It is worth noting that due to the high-sensitivity of the Biograph Vision Quadra scanner, the amount of triples may be even higher than the amount of random doubles. This is expected, as in this case, whenever two decays occur "simultaneously", the probability of detecting three out of the four emitted gamma rays (i.e a triple) is even higher than the detection of only two of them (i.e random double)

| Acquisition | Doubles [Mcps] | | Triples [Mcps] | |
|---|---|---|---|---|
| | Prompt-s | Rando-ms | Prompt-s | Rando-ms |
| $^{18}$F-FDG Patient | 3.18 | 1.21 | 1.43 | 0.85 |
| IQ+Cylinders $^{68}$Ga Phantom | 0.90 | 0.33 | 0.50 | 0.29 |

Table 1. Rate of Doubles (Prompts and Randoms) and Triples (Prompts and Randoms) in Millions of counts per second in the 18F-FDG patient and the IQ+cylinders 68Ga phantom acquisition.

The distribution of doubles and triples in the list-mode of the $^{18}$F-FDG acquisition (double prompt: 47.7% ; double random: 18.1%; triple prompt: 21.4%; triple random: 12.7%) indicates that in a regular clinical acquisition more than 30% of the stored LORs may belong to a triple coincidence. This demonstrate that triple events become very relevant in high-sensitivity scanners such as the Biograph Vision Quadra.

It is important to note that triple coincidences may not be stored in the list-mode file as two directly consecutive coincidences. If the count rate is high, it is possible that some other unrelated LORs may have been written in between (even up to 4 events between them as shown in Fig.2). This has to be taken into account in any method based on the retrieval of triples from the list-mode data.

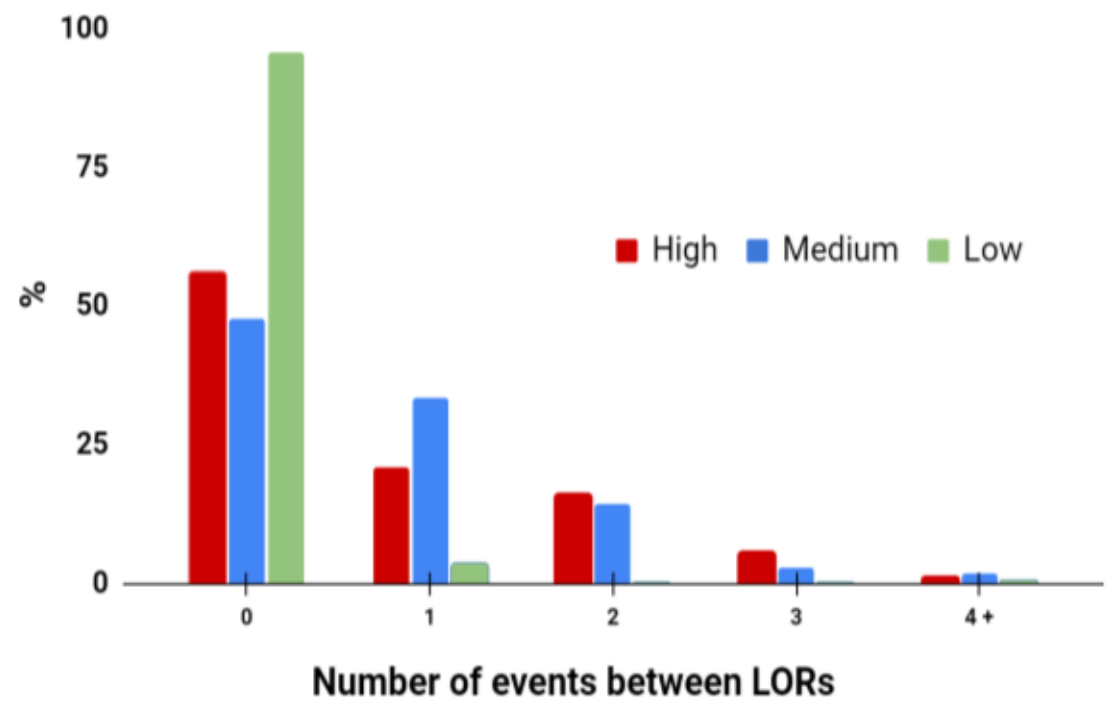

Fig. 2. Number of events in the list-mode file between the two LORs that define a triple coincidences for high, medium and low activity acquisitions.

Figure 3 shows the relative count rate of the $^{18}$F-FDG acquisition with respect to the start of the acquisition. The evolution over time of the total number of the triples prompt coincidences follows the expected behavior of equation (1).

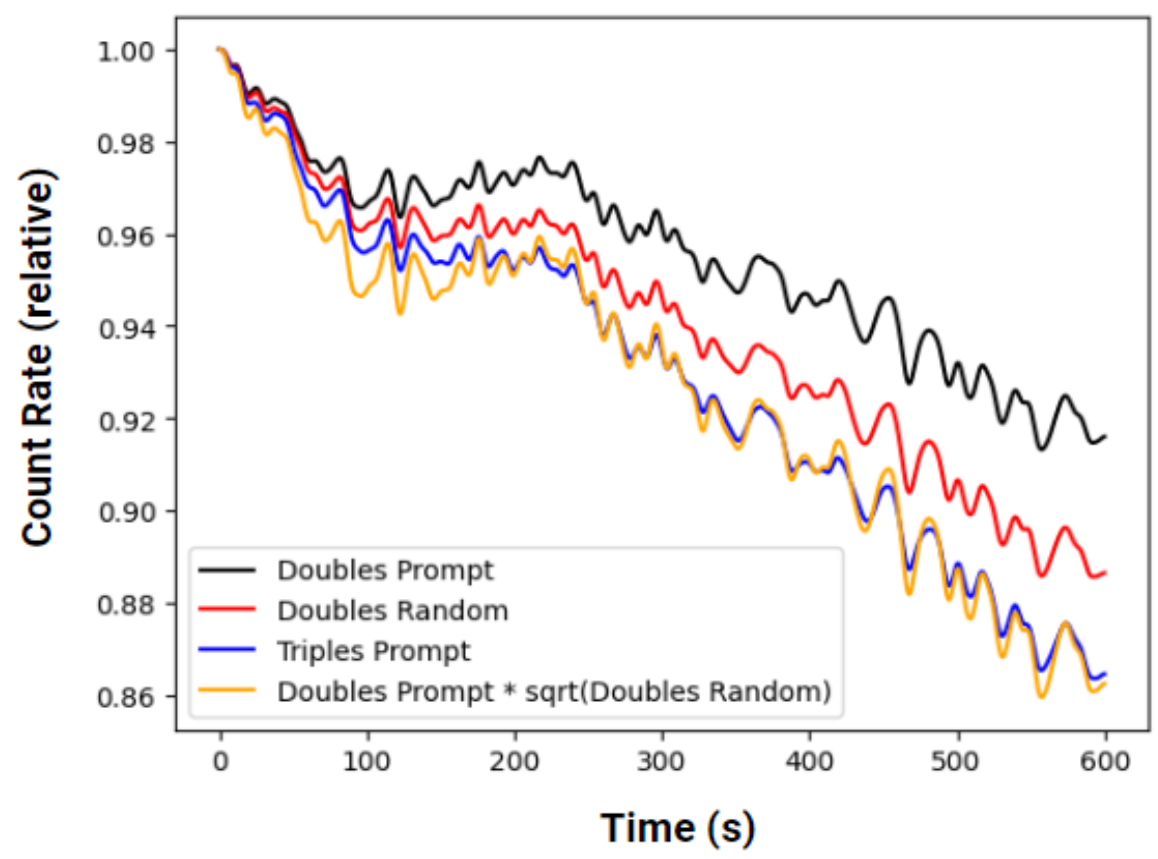

Fig. 3. Count rate (relative to the start of the acquisition) over a 10-min 18F-FDG patient acquisition. The count rate of the triples prompt matches the expected dependence.

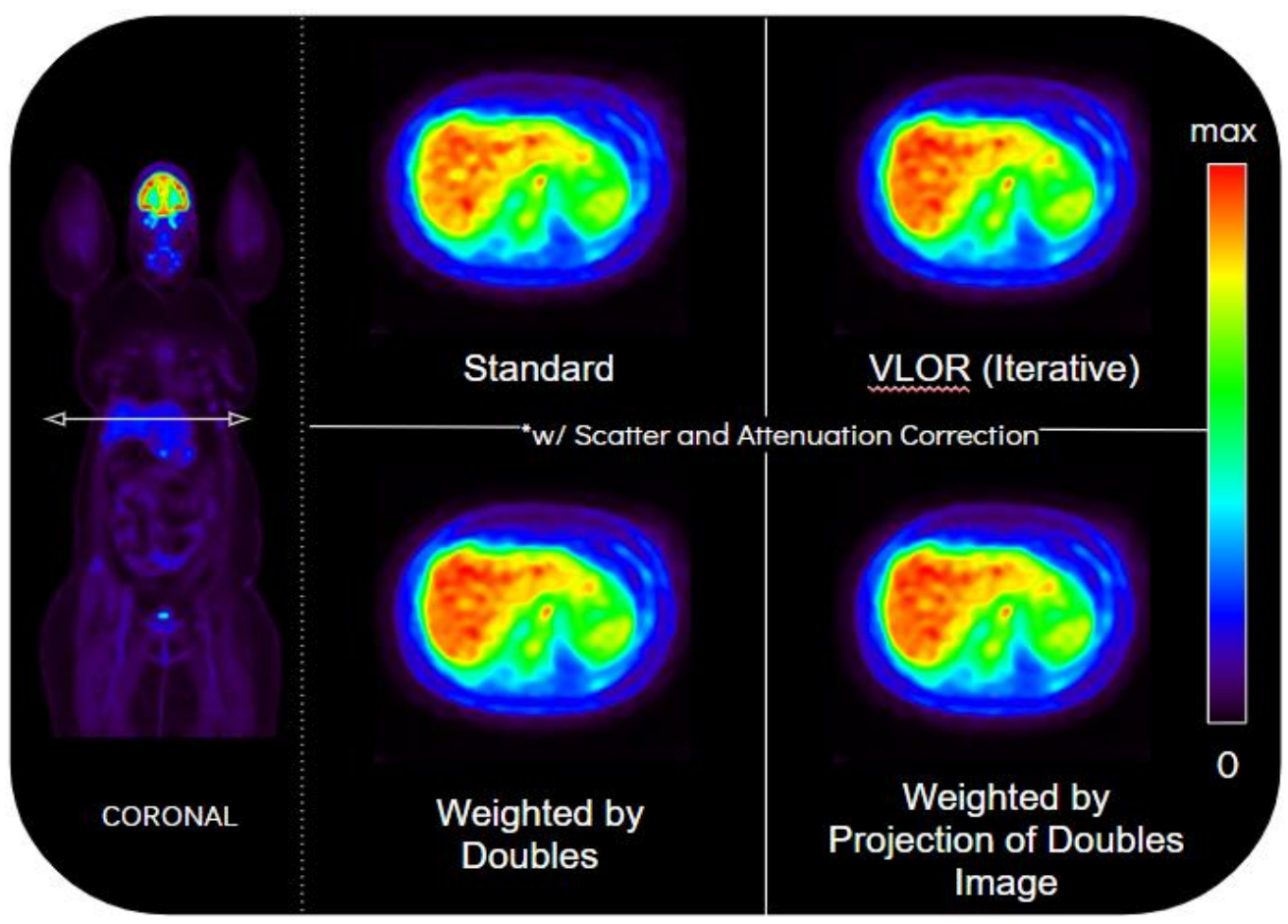

Fig. 4. Reconstructed images from triple coincidences of patient data using the four considered methods. The proposed methods show better results in terms of quality.

Figure 4 shows a coronal and transverse view the reconstructed images from the triples measured in the patient acquisition with $^{18}$F-FDG. It can be observed in the liver region how the standard method yields more noisy images. The quantitative analysis of the noise (using the liver and a tumor as regions of interest) and contrast is shown in Table 2.

| Method | 18F-FDG Patient | | IQ+Cylinders 68Ga Phantom | |
|---|---|---|---|---|
| | Noise (%) | Contrast (%) | Noise (%) | Contrast (%) |
| **Standard** | 5.40 | 6.94 | 11.3 | 3.12 |
| **Doubles as ref.** | 5.37 | 7.00 | 10.0 | 3.15 |
| **Projection as ref.** | 5.13 | 7.01 | 9.2 | 3.16 |
| **Iterative (VLOR)** | 5.08 | 6.84 | 9.1 | 3.08 |

Table 2. Percentage of noise and contrast in the uniform and hot region at patient and extended IQ phantom data acquired.

## 4 Discussion

Although this work is focused on triple coincidences, the analysis could be extended to higher-order coincidences (quadruples, quintuples…). However, the probability of detection of multiple coincidences becomes smaller as we consider additional gammas, and the contribution of quadruples and higher-order coincidences is quite low compared to standard doubles.

Our estimation indicates that up to 30% of the LORs in acquisitions involving standard radionuclides and high-sensitivity scanners like Biograph Vision Quadra belongs to a triple events. These triples can be identified in the list-mode file as consecutive or quasi-consecutive double coincidences that shares a common detector element. In the case of standard radionuclides the detected triples correspond to random events mostly from two decays.

## 5 Conclusion

The amount of triple coincidences present in any PET acquisitions may be very significant, even when working with standard radionuclides such as $^{18}$F. This is especially important in high-sensitivity PET scanners. We have shown how triple coincidences may be retrieved from the list-mode data and properly processed to improve the quality of the reconstructed images with respect of the standard approach of treating triples as two unrelated double coincidences.

A systematic analysis with a larger number of acquisitions, as well as the analysis of the overall impact on the reconstructed images obtained with both doubles and triples coincidences is still needed.

### Acknowlegments

This work has been supported by Siemens Healthineers with a research grant. J.L.H. is currently supported by the Spanish Ministry of Science and Innovation (MCIN) (PID2021-126998OB-I00, and PDC2022-133057-I00/AEI/10.13039/501100011033/ Unión Europea Next GenerationEU/PRTR).

# An update to *elsa* - an elegant framework for tomographic reconstruction

David Frank[1,2], Jonas Jelten[1,2], and Tobias Lasser[1,2]

[1]Department of Computer Science, TUM School of Computation, Information and Technology, Technical University of Munich, Munich, Germany
[2]Munich Institute of Biomedical Engineering, Technical University of Munich, Munich, Germany

**Abstract** In tomographic reconstruction, many different imaging modalities can be expressed using similar mathematical concepts. Our framework *elsa*, which we already presented previously [1], builds on this fact, and provides both a unified mathematical framework and a set of common reconstruction algorithms. These can be applied to various imaging modalities, such as X-ray attenuation computed tomography (CT), Phase-Contrast CT, and (anisotropic) Dark-Field CT. Developed by our team, *elsa* is written in modern C++17, it utilizes the CMake build system for reliability and ease of use. It also provides a Python interface for rapid prototyping. Here, we will present an overview of new features, which include among others new optimization algorithms, the 3D FORBILD phantom, and support for grating-based Phase-Contrast CT to *elsa*. Further, a short demonstration of *elsa*'s capabilities to reconstruct a real-world 3D X-ray attenuation CT dataset is given.

## 1 Introduction

There are many different imaging modalities, for which the following generic optimization problem formulation can be used for solving the inverse problem:

$$\min_x E(x), \quad \text{where } E(x) = D(Ax, m) + \sum_{k=1}^{K} \lambda_k R_k(x) \qquad (1)$$

where $x$ is the unknown quantity to be reconstructed (e.g. the attenuation coefficients for attenuation X-ray CT), $D(Ax,m)$ represents a data fidelity term and $R_k$ are regularization terms with the corresponding regularization parameter $\lambda_k$. The linear operator $A$ represents the physical forward model (such as the Radon Transform for X-ray CT). The data fidelity term expresses the relation of the forward projected quantity $Ax$, and the measured data $m$, via the functional $D$ (e.g. the least squares functional $\frac{1}{2}||Ax-m||_2^2$).
There are multiple different frameworks available that solve specific versions of the optimization problem given in Equation 1, in particular for attenuation X-ray CT. To list some of them, there are the ASTRA Toolbox [2], TIGRE [3, 4], the Core Imaging Library (CIL) [5, 6] and the Operator Discretization Library (ODL) [7].
However, as our research is not solely focused on attenuation X-ray CT with standard geometries, we are looking into more generalized versions of the optimization problem. Specifically, this imposes the requirement for a flexible and extensible operator based framework, which follows the generic form of the optimization problem as given above.
This abstract is structured as follows: The next section gives a brief introduction to *elsa*'s goals, design and the noteworthy additions since its first release. After that, a short walkthrough of a reconstruction of a real-world attenuation X-ray CT dataset is shown using *elsa*'s Python bindings. Finally, a short conclusion and future outlook is given.

## 2 elsa

*elsa* is a reconstruction framework developed at the Computational Imaging and Inverse Problem (CIIP) research group at the Technical University of Munich. It started as a part of a research project, at that time, called *campRecon*. It was later modernized and renamed to *elsa*, and finally open-sourced and released in 2019 [1]. The framework can be found at https://ciip.in.tum.de/elsadocs/. It is available under the Apache 2 open source license.

### 2.1 Goals

The main goals of *elsa* remain unchanged since its first release. However, in the current field of research we have shifted our priorities somewhat.
The primary goal is to deliver a compelling and flexible set of tools for the reconstruction of multiple imaging modalities. This includes a variety of modern iterative reconstruction algorithms, and support to solve a large variety of optimization problems.
Another, important aspect is reproducibility. We strive to support a large variety of reconstruction algorithms, different forward models and regularization techniques, in order to compare and reproduce others research. Further, it is important to us to foster reproducibility of our own research. For this reason, *elsa* is open-source and uses common standards and tools. Of course, we are also hoping for the chance that you find it interesting. If you are considering using *elsa* for anything, consider contacting us, we are happy to help and assist in any way we can.
Finally, we have and will put more effort into interoperability. As mentioned above, there are many great frameworks already out there. And instead of reinventing the wheel again, we want to build on other's work and focus on our strengths.

### 2.2 Design

The design of *elsa* is operator- and optimization-based. This is largely influenced by our need to described multiple problems with a common set of abstractions. It further provides us with a great deal of flexibility. The core abstractions used throughout our framework are represented as C++ classes. The general design is similar to the one given in [1]. However, some important details changed, and hence a short overview of the important concepts is given in the following.

- **DataContainer** This is our core representation of *n*D data. In the mathematical description, this is a column vector. However, for certain operators (e.g. the gradient), this needs to contain more information about dimension, shape, strides and so on. Further, the class can represent both measurement data and the reconstruction volume. The **DataDescriptor** is used to differentiate that, and contains all necessary information about e.g. the geometry, trajectory and detector.

- **LinearOperator** The `LinearOperator` is a base class providing two functions, `apply` and `applyAdjoint`, which compute $Ax$ and $A^T y$ respectively. This is used throughout the framework to implement a representation of a diagonal matrix (`Scaling`), finite differences (`FiniteDifferences`) and the different forward models. Further, vertically and horizontally stacked operators are supported with the `BlockLinearOperator`.

- **Functional** The base class represents a functional $f : x \mapsto f(x) \in \mathbb{K}$, mapping a vector $x \in \mathbb{F}^n$ (where $\mathbb{F} \in \{\mathbb{R}, \mathbb{C}\}$), to a value of the field $\mathbb{K} \in \{\mathbb{R}, \mathbb{C}\}$. In most cases, both the vector and the scalar will be real. There are a number of common norms, that are already implemented, such as $\ell^1$, $\ell^2$, $\ell^\infty$-norm and the Huber loss function. Other functionals implement the transmission and emission log likelihood. If applicable, each functional implements methods to calculate both the gradient and Hessian of the functional.

- **Problem** The `Problem` class acts both as a general optimization problem and a base class for other, more specific problem formulations. It provides a convenient way to compute the gradient and determine the Hessian of the current problem. Derived classes act as a way to express concrete problem formulations such as a (weighted) least squares, (generalized) Tikhonov or LASSO problem.

- **Solver** This is the base class for iterative reconstruction algorithms. If applicable, they accept some (potentially derived) `Problem`, convert it to a form with which the specific algorithm can work, and compute an approximate solution based for the given problem formulation.

### 2.3 Updates

Since, we introduced *elsa* we have been working on many different aspects. For one, a series of new iterative reconstruction algorithms have been introduced. They include (accelerated) proximal gradient descent (also known as ISTA and FISTA [8]), Nesterov's fast gradient methods, order subsets [9], and alternating direction method of multipliers (ADMM) [10]. With the addition of algorithms such as (accelerated) proximal gradient descent and ADMM, we now also support proximal operators.

Another noteworthy addition, is an easy-to-use 3D FORBILD head phantom (see [11] for a discussion on the 2D FORBILD head phantom). To the best of our knowledge, we are one of the only open-source frameworks to provide it. The FORBILD head phantom is more complex than the often used Shepp-Logan phantom. And thus is closer to real-world data, while still preserving the benefits of synthetic data.

For attenuation X-ray CT, many different approximations of the Radon Transform exists. Common examples are Siddon's [12, 13], and Joseph's method [14]. Since the first release of *elsa*, optimized versions of the CPU versions were implemented (see [15, 16]). Further CUDA accelerated implementations of both exist.

The forward model usually does not contain information about the type of detector. *elsa* now also supports curved detectors, not only flat ones. These types of detectors are most often found in medical attenuation X-ray CT scanners.

In the last year, two specific versions of GMRes have been discussed in [17]. Compared to many other solvers, these do not assume that the adjoint of the forward model is mathematical exact. This situation is quite common for efficient implementations of the forward model and its adjoint, which is the case in one of the implementations in *elsa*. Both of these versions of GMRes are implemented in our framework.

As our research is not solely focused on attenuation X-ray CT, other forward models are needed. We have now also full support for X-ray phase-contrast CT, based on grating interferometry. The difficulty in the forward model for phase-contrast CT lies in the need for differentiable basis functions. The typically used pixel-basis function is not differentiable, and hence other basis functions are used. We support both the commonly cited blob [18] and B-Spline basis function [19]. Again, we provide efficient CPU and CUDA implementations.

## 3 Showcase Reconstruction

In this section, we want to present a walkthrough of a reconstruction of the phantoms used in the Helsinki Tomography Challenge (HTC) 2022 [20]. We present how a real-world dataset can be reconstructed using *elsa*'s Python interface. A slightly adapted version of the below code can be found in our GitLab repository (https://gitlab.lrz.de/IP/elsa).

The phantoms from the challenge are acrylic discs with a varying number and shaped holes. They are scanned using a cone-beam set up, with a circular trajectory over the complete circle. A projection is taken at every 0.5°. The data is already preprocessed with black-field and flat-field correction, the center of rotation is aligned, and the negative log transforma-tion has been performed. Hence, no extra preprocessing is necessary for a simple reconstruction. For the purpose of this demonstration, an additional beam hardening correction was performed.

For a reconstruction in *elsa*, we first need to load the data, set up the geometry and trajectory, and then we can reconstruct the data. The dataset is provided as MATLAB `.mat` files. Hence, we can use SciPy's `loadmat` function. We additionally wrap it, such that it returns a Python dictionary. We omit the code for this specific part, and assume we have a function `loadmat` that returns a dictionary containing the meta information and sinogram data.

The next step is to retrieve the necessary information to set up the trajectory. This includes information about the distance from X-ray source to center, and X-ray source to detector, number of angles, and effective detector pixel size. These are provided by the metadata. In the following Python snippet, this information is used to set up a circular trajectory:

```
# filename is the path to the HTC .mat files
mat = loadmat(filename)

# Parameters of the full scan
params = mat["CtDataFull"]["parameters"]

# Distance parameters
ds2c = params["distanceSourceOrigin"]
ds2d = params["distanceSourceDetector"]
dc2d = ds2d - ds2c # Distance from center to detector

# Detector parameters
detector_pixel_size = params["pixelSizePost"]
num_detector_pixels = params["numDetectorsPost"]

# Rough approximation of a volume size
vol_npixels = int(num_detector_pixels / np.sqrt(2))

# Descriptor of the desired volume
vol_desc = elsa.VolumeDescriptor(
    [vol_npixels] * 2, [detector_pixel_size] * 2
)

# Setup circular trajectory given above parameters
sino_desc =
↪   elsa.CircleTrajectoryGenerator.trajectoryFromAngles(
    params["angles"],
    vol_desc,
    ds2c,
    dc2d,
    [0], [0, 0], # no offset of principal point and CoR
    [num_detector_pixels] * 2,
    [detector_pixel_size] * 2,
)
```

Next, we need to create a `DataContainer` with the descriptor and data of the sinogram. The memory layout of the measurements is altered to adhere to our requirements.

```
# Measurement data, converted to single precision
m = mat["CtDataFull"]["sinogram"].astype("float32")

# Adapt memory layout
sino_data = np.flip(m.transpose(1, 0), axis=1)
sinogram = elsa.DataContainer(sino_data, sino_desc)
```

The last step, before the reconstruction, is the choice of approximation of the forward model. In this case, we use *elsa*'s `SiddonsMethod`. However, if you'd like to use GPU acceleration, you can simply use `SiddonsMethodCUDA` or `JosephsMethodCUDA`.

```
# A specific approximation of the forward model
projector = elsa.JosephsMethod(vol_desc, sino_desc)

# Or its CUDA alternative
projector = elsa.JosephsMethodCUDA(vol_desc, sino_desc)
```

You can check whether your present installation supports the CUDA projectors using the function `cudaProjectorsEnabled`. Finally, we set up an iterative reconstruction algorithm:

```
# Setup LASSO problem and FISTA
lasso = elsa.LASSOProblem(projector, sinogram, 7.0)
solver = elsa.FISTA(lasso)

# Run the reconstruction algorithms for 70 iterations
reconstruction = solver.solve(70)
```

Here, we use the least squares problem with $\ell^1$-regularization, which is often referred to as *LASSO*. The regularization parameter (7.0) was chosen by trial and error to give visually the best results. The solver we choose for this specific reconstruction is an implementation of the *accelerated proximal gradient method*, or also knows as *Fast Iterative Shrinkage-Thresholding* (FISTA). Examples of the reconstruction with a selection of the phantoms provided in the challenge can be seen in Figure 1.

## 4 Conclusion and Outlook

*elsa* is still a work in progress. We are porting more and more code of our internal code base to the public and open source repository. This includes support for (anisotropic) X-ray Dark-Field CT [21]. We also work on more advanced solvers, such as Split-Bregman and Primal-Dual Hybrid Gradient. Both of those, will become available in the upcoming months. Currently, a lot of work is put into a better and more general seamless data transfer from Python to C++ and back, and multi-GPU support for the implementations of the forward models.

In summary, *elsa* provides a flexible reconstruction framework, that provides an operator- and optimization based workflow. It is easy to extend, and provides tools ranging from fundamental to advanced.

### Acknowledgments

The authors are currently the core maintainers of the framework, but many others have contributed to *elsa*. As such, we want to thank all the different contributors for their contributions. Specifically, we want to thank the currently actively working contributors David Alkier, Daniel Klitzner, Armin

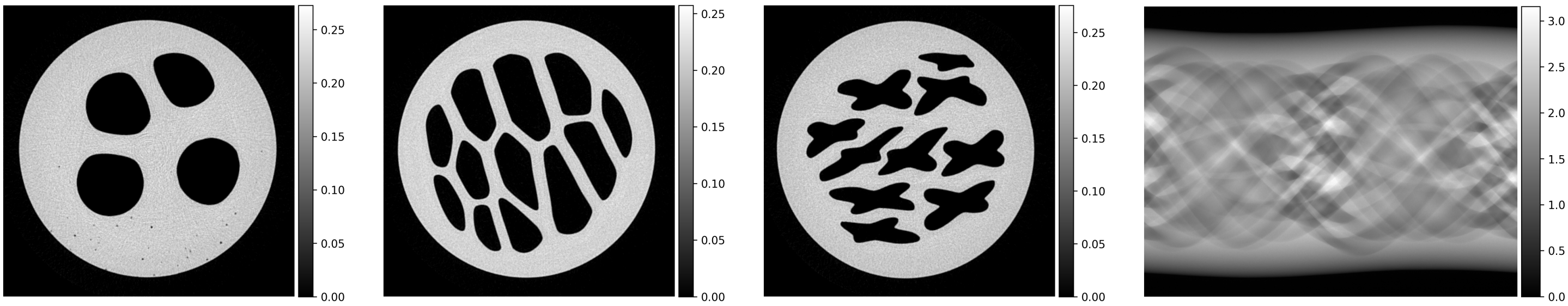

**(a)** Phantom `01c` **(b)** Phantom `04c` **(c)** Phantom `07a` **(d)** Sinogram for phantom `07a`

**Figure 1:** (a)-(c): Reconstruction of some of the test phantoms of the Helsinki Tomography challenge 2022 using all available projections. The problem formulation is the least squares data term with $\ell^1$ regularization, with a heuristically chosen regularization parameter of 7. 70 iterations of FISTA are used to reconstruct the images. (d): sinogram for the phantom shown in (c).

Begert and Noah Dormann. Many thanks for your awesome work!
This work was supported in part by the German Federal Ministry of Health under Grant 2520DAT920.

# Improved resolution on existing CT scanners by utilizing off-center scan regions and Zoom-In Partial Scans (ZIPS)

Lin Fu[1], Eri Haneda[1], Stephen Araujo[1], Ryan Breighner[2], and Bruno De Man[1]

[1]Radiation Imaging, GE Research-Healthcare, Niskayuna, NY, USA
[2]Hospital for Special Surgery, New York, NY, USA

**Abstract** Recently, the Zoom-In Partial Scans (ZIPS) scheme was introduced to boost the spatial resolution of clinical CT scanners. Contrary to the conventional wisdom that a region of interest should be centered in the field of view for the best resolution, in ZIPS the ROI is intentionally placed some distance away from the iso-center to gain higher geometrical magnification, which in turn brings higher resolution. Despite promising simulation results, it remains an open question whether ZIPS is effective on real CT systems, where complicated practical constrains may impact theoretical assumptions. In this paper, we implement the proposed ZIPS scheme on a GE Revolution CT and evaluate image resolution using phantoms and cadaveric bone samples. Clear improvement of the visibility of fine features and trabecular details is observed when comparing ZIPS to the conventional scans at the center. To our knowledge, this is the first demonstration on clinical CT that the resolution of off-center scans can outperform that of a conventional centered scan.

## 1 Introduction

The Zoom-In Partial Scans (ZIPS) is a novel CT scanning scheme that improves the intrinsic spatial resolution of existing clinical multi-slice CT for region-of-interest (ROI) imaging [1]. Unlike a conventional CT scan where the ROI is centered in the field of view, in ZIPS the ROI is intentionally placed some distance away from the iso-center to gain higher geometrical magnification, which in turn improves the imaging resolution for the ROI. Previous simulation studies showed that for an ROI offset of 20 - 30 cm relative to iso-center, the modulation transfer function (MTF) can improve by 30% to 80% when comparing ZIPS to a conventional centered scan. For more details about ZIPS, please refer to [1,2].

Despite the promising simulation results, it remains an open question whether the off-center scanning scheme in ZIPS can be effective on real CT systems, where multiple tradeoff factors can impact resolution in a location-dependent and anisotropic manner, especially in off-center regions. For example, several previous studies observed that image resolution substantially degrades as radial distance increases from the iso-center [3-6]. This is partly due to the continuous rotation of the CT gantry during data acquisition, causing motion blur in the azimuthal direction which increases linearly with the radial distance. The resolution in off-center regions is also impacted by the elongation of the focal spot: because of the angulation of the focal spot on the surface of the x-ray anode, the apparent size of the focal spot increases when viewed along larger fan angles relative to the central ray [5,6].

In this paper, we implement the proposed ZIPS scheme on a GE Revolution CT scanner and evaluate the image resolution using phantoms and cadaveric bone samples. To our knowledge, this is the first demonstration on real gantry-based CT that off-center regions can be utilized to improve resolution over with a conventional centered scan, without a costly upgrade of the scanner hardware. We also acquire two complimentary ZIPS scan of the ROI then register two partial scans to achieve more isotropic improvement of resolution relative to a single off-center scan.

This paper also introduces a simple analytical model to clarify the role of a few major factors in influencing the location- and direction-dependent resolution in CT. The model includes the effect of geometric magnification, detector aperture, focal spot elongation, and azimuthal blur. It may be used to elucidate the rationale of ZIPS, guide the selection of scan parameters, and guide the design of high-resolution image reconstruction algorithms.

## 2. Theory and methods

### Notations

We begin by a theoretical analysis of the location- and direction-dependent resoluiotn in CT. We restrict the analysis to 2-D for simplicity, although the real data results shown in subsequent sections are 3-D.

Let $\boldsymbol{x} = (x, y)^T$ denote the spatial coordinates in the image domain. As shown in Fig. 1, fan-beam projections are measured by moving the x-ray source along a circular trajectory $\boldsymbol{s}(\beta) = (R\cos\beta, R\sin\beta)$, where $\beta$ denotes the angular position of the source and $R$ denotes the source-to-center distance. Here we define angle zero as the positive Y direction. Let $\mathcal{L}(\beta,\theta)$: $\boldsymbol{x} = \boldsymbol{s}(\beta) + t\boldsymbol{\theta}$, $t \in [0, +\infty)$ denote a projection ray originating from the source in the direction of a unit vector $\boldsymbol{\theta} = (\cos\theta, \sin\theta)^T$. The projection data are measured by a curved detector centered at the x-ray source $\boldsymbol{s}$, with the source-to-detector distance $L$.

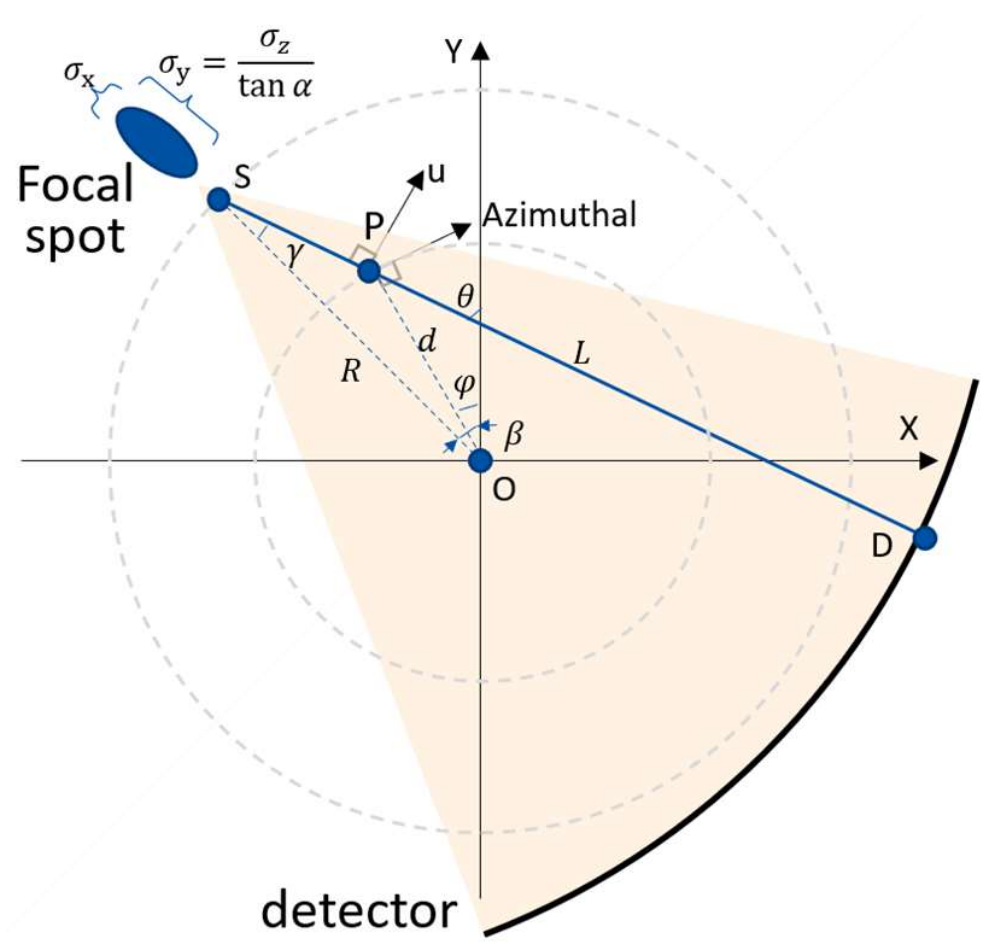

Fig 1. The CT geometry and definition of symbols.

To analyze location- and direction-dependent resolution in off-center regions, we consider a point $P$ represented by the vector $\boldsymbol{d} = (d\cos\varphi,\ d\sin\varphi)^T$, where $d$ is the radial distance relative to the iso-center. We first calculate the geometric magnification factor at point $P$, which will be used in subsequent derivation of the blurring kernels of the detector and source. The magnification factor for $P$ along the projection angle $\theta$ is

$$M(\theta, \boldsymbol{d}) = \frac{L}{|SP|} \geq 1. \tag{1}$$

Because $P$ lies in the ray $\mathcal{L}(\beta, \theta)$, $|SP|$ satisfies

$$\begin{bmatrix} R\cos\beta \\ R\sin\beta \end{bmatrix} = \begin{bmatrix} d\cos\varphi \\ d\sin\varphi \end{bmatrix} + |\mathrm{SP}| \begin{bmatrix} \cos\theta \\ \sin\theta \end{bmatrix}.$$

Eliminating $\beta$ to solve for $|SP|$, and substituting |SP| to (1), we obtain $M$ as a function of $\theta$ and $\boldsymbol{d}$

$$M(\theta, \boldsymbol{d}) = \frac{L}{\sqrt{R^2 - \left(d\sin(\theta - \varphi)\right)^2} - d\cos(\theta - \varphi)}. \tag{2}$$

The fan angle $\gamma$ of the ray going through $P$ satisfies $\sin\gamma / d = \sin(\theta - \varphi) / R$, thus we can express $\gamma$ as a function of $\theta$ and $\boldsymbol{d}$

$$\gamma(\theta, \boldsymbol{d}) = \sin^{-1}\left(\sin(\theta - \varphi)\frac{d}{R}\right).$$

**Overall resolution model**

We focus on analyzing the intrinsic resolution of the projection data in single projection views, excluding the effect the ramp filter kernel used in image reconstruction. We model the location- and view-dependent blurring effect in CT data as the convolution of three factors: detector blur, focal spot (source) blur, and azimuthal blur:

$$B(u; \theta, \boldsymbol{d}) = \det(u; \theta, \boldsymbol{d}) * \mathrm{src}(u; \theta, \boldsymbol{d}) * \mathrm{azm}(u; \theta, \boldsymbol{d}), \tag{3}$$

where $B(u; \theta, \boldsymbol{d})$ denotes the 1-D blurring kernel at point $P$ along the lateral direction $u$, which is perpendicular to the projection ray angle $\theta$; $*$ denotes 1-D convolution.

**Detector blur**

For simplicity, we model the detector response as a rectangular window function, although it is straightforward to substitute other analytical models if needed

$$\det(u; \theta, \boldsymbol{d}) = \mathrm{rect}\left(\frac{M(\theta, \boldsymbol{d})}{\Delta w} u\right) \tag{4}$$

where $\mathrm{rect}(x) = 1$ for $|x| \leq 0.5$ and $\mathrm{rect}(x) = 0$ otherwise. $\Delta w$ is the detector cell pitch, and the factor $M(\theta, \boldsymbol{d})$ shrinks the blurring kernel to compensate for the geometric magnification at point $P$.

**Source blur**

We model the focal spot on the anode surface as a gaussian distribution with the covariance matrix

$$\Sigma = \begin{bmatrix} \sigma_x^2 & & \\ & \sigma_y^2 & \sigma_y\sigma_z \\ & \sigma_y\sigma_z & \sigma_z^2 \end{bmatrix},$$

where the Y and Z components of the distribution are fully correlated, and $\sigma_z = \tan\alpha\, \sigma_y$, $\alpha$ being the anode angle. When viewed along fan angle $\gamma$, the 1-D marginal distribution of the focal spot in the lateral direction ($u$) is still Gaussian, and variance of this 1-D distribution is

$$\begin{aligned} \sigma_u^2(\theta, \boldsymbol{d}) &= \cos^2\gamma\, \sigma_x^2 + \sin^2\gamma\, \sigma_y^2 \\ &= \cos^2\gamma\, \sigma_x^2 + \sin^2\gamma \frac{\sigma_z^2}{\tan^2\alpha}. \end{aligned}$$

Because $\sigma_y$ is usually much larger than $\sigma_x$, the effective size of the focal spot increases as $\gamma$ increases (elongation). After further considering the geometric magnification factor for the focal spot, which is $\frac{1}{1-1/M}$, we model the overall blurring effect of the focal spot at point $P$ as

$$\mathrm{src}(u; \theta, \boldsymbol{d}) = \exp\left(-\frac{1}{2}\left(\frac{u}{\left(1 - 1/M(\theta, \boldsymbol{d})\right)\sigma_u(\theta, \boldsymbol{d})}\right)^2\right) \tag{5}$$

**Azimuthal blur**

We model the azimuthal blur as a rectangular window function

$$\mathrm{azm}(u; \theta, \boldsymbol{d}) = \mathrm{rect}\left(\frac{1}{|\cos(\theta - \varphi)|} \cdot \frac{N_{\mathrm{view}} u}{2\pi d}\right). \tag{6}$$

where $N_{\mathrm{view}}$ is the number of project views per rotation, and $\frac{N_{\mathrm{view}}}{2\pi d}$ is the distance traveled point $P$ relative to the CT gantry in a single projection view. The factor $|\cos(\theta - \varphi)|$ accounts for the angle between the rotation motion and the lateral direction $u$.

**Resolution model for a specific scanner**

Substituting Eqs. 4,5,6 into 3, we obtain an overall model of the angle-dependent resolution at different spatial locations. We evaluated the proposed resolution model using the geometry of a GE Revolution CT scanner. Fig. 2 shows resolution as a function of ray angle $\theta$ for various radial offsets ($d$), focal spot sizes ($\sigma_{\mathrm{x}}$), and number of views per rotation ($N_{\mathrm{view}}$).

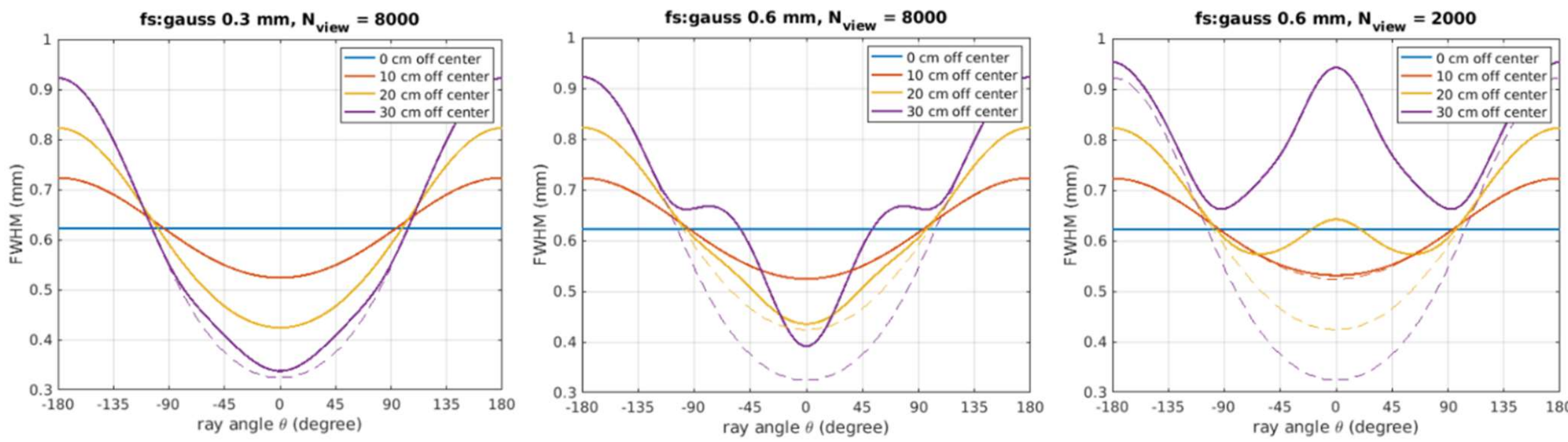

(a) (b) (c)

Fig. 2. Full width at half maximum (FWHM) of the blurring kernel computed by the proposed analytical model. Solid lines - all blurring effects modeled. Dashed lines – only detector blur is modeled, excluding focal spot and azimuthal blur. The subplots show the effect of different focal spot sizes and the number of projection views per rotation. The labeled focal spot sizes (FS) correspond to the full width at 15% maximum of the intensity of focal spot and assume $\sigma_x = \sigma_z$. Without loss of generality, these plots assume $\varphi = 0$.

When $d = 0$ (at iso-center), the resolution is constant in all directions due to symmetry and the resolution is relatively insensitive to focal spot sizes because the resolution is limited by the detector. When the focal spot size and azimuthal blur are sufficiently small (Fig. 2a), resolution substantially improves as the radial distance from the center increases, over a range of projection angles from about -90 to 90 degrees. This is because the bottleneck of detector resolution is overcome by the higher magnification in the off-center regions. When the focal spot size becomes larger (Fig. 2b), the benefit in resolution in off-center regions becomes less, especially at larger radial distances and when $\theta$ is near 45 degrees, where the focal spot elongation becomes the limiting factor of resolution. When the azimuthal blur increases (Fig. 2c), the resolution in off-center regions further degrade especially when $\theta \approx 0$, where the azimuthal blur dominates resolution.

The analysis above shows that the best resolution may be achieved in off-center regions, instead of at the iso-center as in conventional CT scans. Very high resolution may be achieved at large radial offsets (20 – 30 cm), provided that the azimuthal blur and the focal spot size are sufficiently small (Fig. 2a). On the other hand, a relatively small radial offset (10 cm) may still provide noticeable improvement of resolution and the improvement is relatively insensitive to azimuthal blur and focal spot size (10 cm lines, all sub plots in Fig. 2). Depending on the focal spot size and the number of projection views per rotation of a clinical CT system, a proper radial distance may be chosen to achieve the best resolution.

The above analysis also provides some guidance for high-resolution image reconstruction for off-center regions. Because the higher resolution is achieved over a limited angular range of about -90 to 90 degrees, a half-scan reconstruction algorithm should be used to prevent contamination from the lower-resolution conjugate rays.

**Zoom-In Partial Scans (ZIPS)**

As shown in Fig. 2. the improved resolution in off-center regions is anisotropic, especially for higher radial offset. The greatest improvement to resolution is obtained over a limited angular range from about -45 to 45 degrees. To achieve more uniform resolution improvement, ZIPS uses a dual partial scan scheme. Two limited-angle partial scans are performed to collect high-resolution projections of the ROI in complimentary angular ranges, then the partial data are registered and merged into a more isotropic high-resolution image. As shown in Fig. 3, a patient is scanned at two off-center table positions and remains still during each scan. After the first partial scan is completed, the ROI is translated to the second position, where the second partial scan is performed. In Fig. 3, the angle of the radial offset is $\varphi_1 = 45°$ and $\varphi_2 = 135°$ for the bed positions, respectively (angle zero is defined at 12 o'clock position). The exact translation of the ROI between the two partial scans is unknown to the image reconstruction algorithm and will estimated and registered during image reconstruction. For more details, please refer to [1,2].

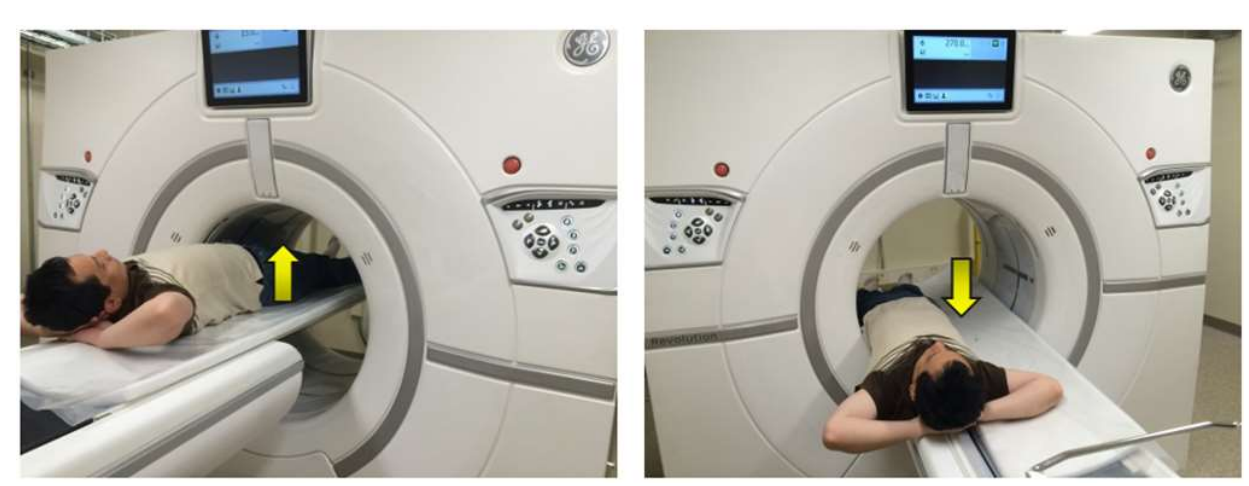

Bed position 1 Bed position 2

Fig. 2. Illustration of the ZIPS scheme. Two limited-angle partial scans are performed to collect high-resolution projections of an ROI in complimentary angular ranges.

## 3. Results

We scanned a resolution phantom and a cadaveric vertebra sample on a GE Revolution CT. The phantom and the sample were placed at the center of the field of view for standard CT, and nominally at 20 cm off center for ZIPS. Each ZIPS acquisition consisted of two scans at two bed positions to cover high resolution data in complimentary angular ranges. All scans used 120 kVp, 500 mAs, 8000 view per rotation, and an extra-small focal spot. No hardware modification was made to the scanner. The CT images were reconstructed on a 512×512 pixel grid over a 5×5 cm square ROI. FBP reconstruction with Edge filter was used for all scans. Parker-weighting was used in off-center reconstructions.

Fig. 3 shows reconstructed images of the resolution phantom. The visual sharpness of features is improved in the off-center scan image compared with the standard CT. The line pair features of 16 lp/cm that are barely distinguishable in the centered scan become clearly visible when scanned at 20 cm off center.

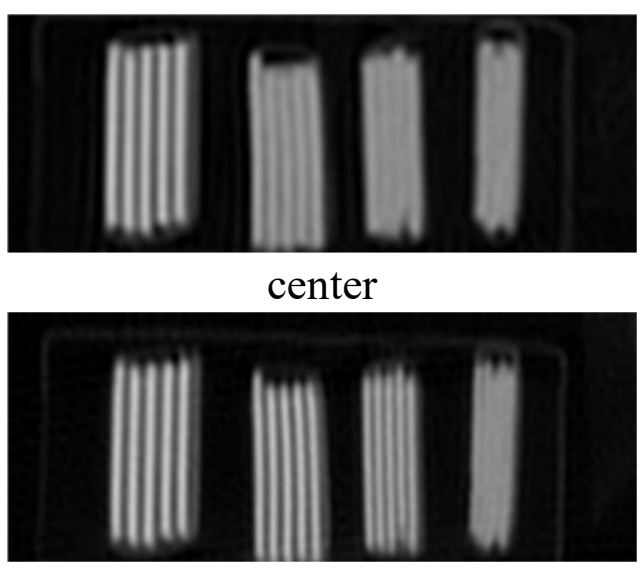

Fig. 3. Comparison of images of a resolution phantom scanned at the center and 20 cm off center. The sizes of the resolution features are 11, 13, 16, and 20.5 lp/cm.

Fig. 4 shows reconstructed images of the bone sample. The visual sharpness of trabecular details is improved in the ZIPS image when compared with the standard CT. The improvement is also seen in the coronal and sagittal reformats. The images from standard CT suffer strong aliasing artifacts especially in coronal and sagittal reformats, which are not present in the ZIPS images.

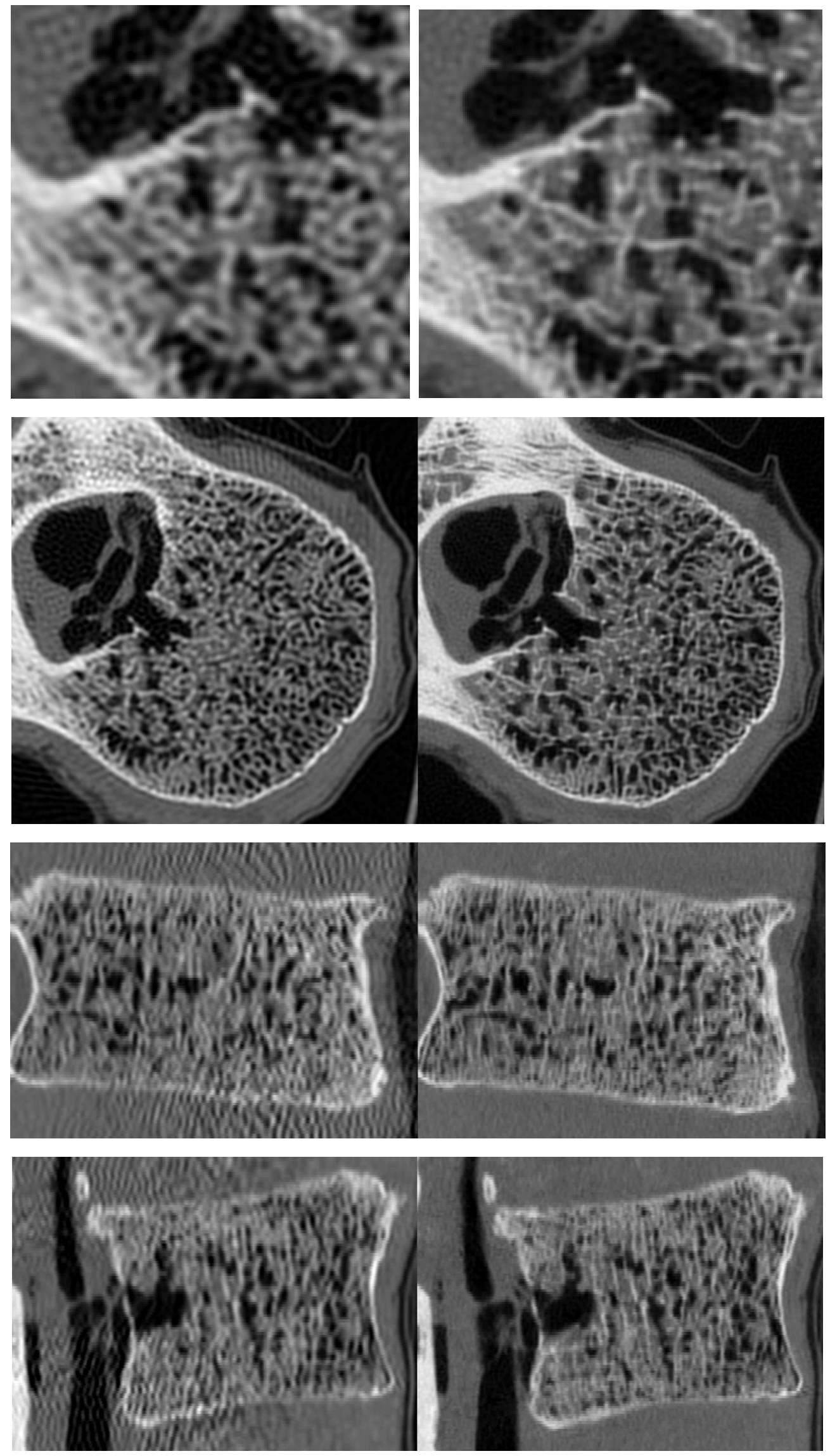

Fig. 4. Comparison of images of a vertebral bone sample scanned at the center (standard CT) and 20 cm off center (ZIPS). From top to bottom: axial (zoom-in), axial (whole image), coronal, sagittal.

## 4. Conclusion

Unlike the conventional assumption that the best resolution is achieved at the center of the field of view, our analytical and experimental results showed that off-center regions can achieve higher spatial resolution than a centered scan, provided that the radial distance, focal spot size, and the angular sampling rates are properly chosen. Clear improvement of visual detectability of fine features was observed with a resolution phantom and a bone sample by using the proposed ZIPS scheme in off-center regions. ZIPS does not require an upgrade of the CT detector array and thus has the potential to be applied to existing clinical CT systems. ZIPS CT is also orthogonal to pure algorithmic resolution-boosting methods and a combination may give further improvement.

**Acknowledgement**

The authors would like to thank Dr. Fernando Quevedo Gonzalez, Hospital for Special Surgery, and Dr. Pengwei Wu, GE Research for valuable discussion. Research reported in this publication was supported by NIBIB of the National Institutes of Health under grant number R01EB028270. The content is solely the responsibility of the authors and does not necessarily represent the official views of the NIH.

# An attenuation field network for cone-beam breast CT with a laterally-shifted detector in short scan

Zhiyang Fu[1], Hsin Wu Tseng[1], and Srinivasan Vedantham[1,2]

[1]Department of Medical Imaging, University of Arizona, Tucson, United States
[2]Department of Biomedical Engineering, University of Arizona, Tucson, United States

**Abstract** In cone-beam computed tomography (CBCT), asymmetrically mounted detectors can enlarge the imaging field of view compared to centered detectors. This geometry can also reduce radiation dose and photon scatter but acquires truncated projections. The truncated projections can be compensated by weighting functions applied in conventional analytic reconstruction methods. We have recently demonstrated the feasibility of asymmetric or laterally-shifted detector geometry in a prone breast CBCT system. To enable the transition of a dedicated breast CT system from prone to upright patient positioning, we propose to use the shifted detector geometry to acquire partial scan (270 degrees) data. A self-supervised attenuation field network (AFN) is proposed to reconstruct such incomplete data. The network learns a continuous mapping from the physical coordinates of imaging volume to its respective attenuation coefficients during training and can render attenuation images or projection data by querying volumetric coordinates during inference. The synthesized projections can be further used in any reconstruction methods. We evaluated the proposed technique using fifty clinical breast CBCT datasets. AFN, when combined with analytical reconstruction methods, obtains visually similar reconstructions, and yields comparable noise variance, contrast, and spatial resolution compared to the reference reconstruction using full-scan data. AFN can potentially enable a partial scan shifted-detector geometry for dedicated breast CBCT imaging.

## 1 Introduction

In cone-beam computed tomography (CBCT) systems, the field of view (FOV) is limited by the x-ray detector size. A laterally-shifted detector or offset-detector is known to extend the FOV without reducing the system magnification. Meanwhile, this "extended FOV" geometry can also reduce photon scatter and radiation dose.[1,2] A recent work from our group demonstrated the feasibility of a shifted detector geometry in dedicated breast CBCT (a 30×30 cm detector positioned with a 5 cm offset compared to a 40×30 cm centered detector).[3,4] In this work, we propose to acquire partial-scan (also known as short scan) data in conjunction with the shifted-detector geometry (referred to as PSSD) to envision a clinical transition from a prone scanner to an upright scanner for dedicated breast CT imaging.

The truncated projection data acquired using a shifted detector can be compensated using weighting functions along the fan-angle direction.[3–8] However, these truncation weightings are based on the data redundancy of fan-beam data and, thus, require full-scan data. We show that direct use of these weighting functions failed to reconstruct the PSSD data and develop an attenuation field network (AFN) instead to approach the goal. AFN adopts the emerging *neural field* paradigm[9,10] in computer vision where a scene is represented as a continuous function of coordinates. In the context of CT, neural fields were learned either in the projection domain[11] or in the image domain[12–15]. Specifically, Sun et al.[11] proposed a projection field network that is suited for sparse-view CT problems since the network cannot extrapolate projection data. Tancik et al.[12] and Zang et al.[13] propose to train neural attenuation field networks yet require the system forward operator for computing training losses in the projection domain. The high memory requirement of the system operator limits the network's application in three-dimensional or dynamic CT imaging. Our proposed learning of a neural attenuation field is inspired by Zha et al.[14] and Rückert et al.[15], where the training reduces to each ray originating from the X-ray source to a detector pixel. This decomposed learning procedure is memory efficient and suited for high-dimensional imaging applications. The rendered attenuation field during inference is regarded as the image reconstruction of the network in these works.[14,15] Differently, we reuse the acquire projections and splice them with the network synthesized projections to form full data for post reconstructions. Here, we demonstrate the feasibility of the unsupervised deep learning method, AFN, for cone-beam breast CT with a shifted detector in a partial scan.

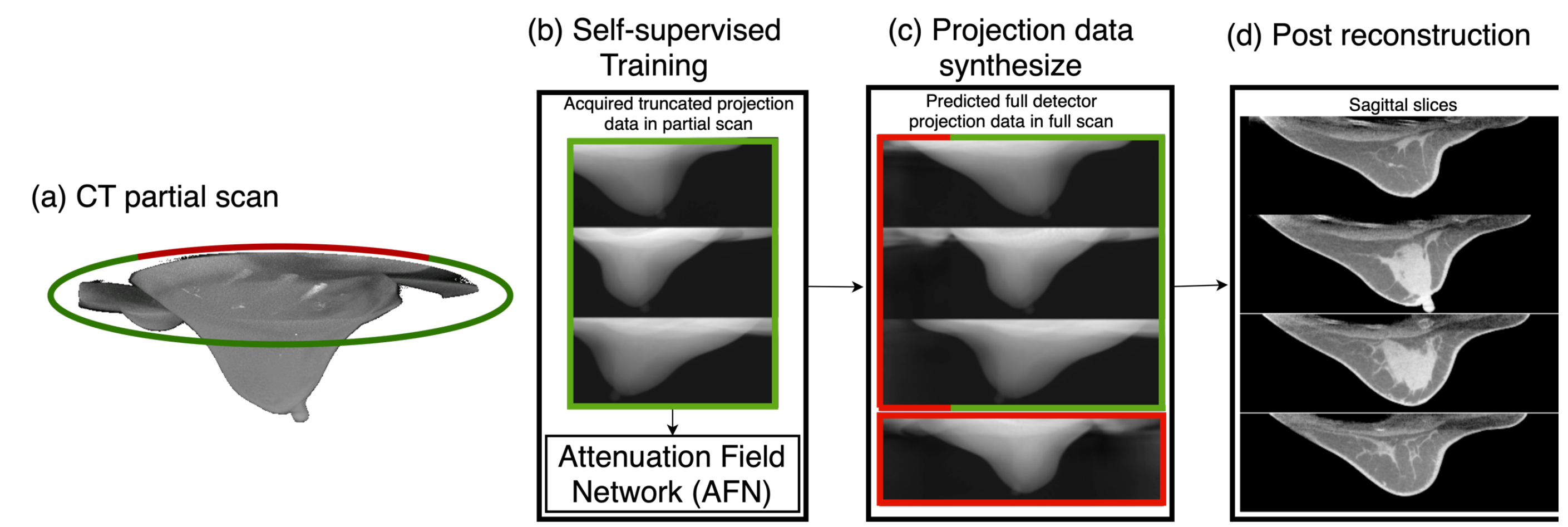

Figure 1. The proposed pipeline for cone beam breast CT data in partial scan using shifted-detector geometry (PSSD). (a) Circular cone beam CT data in partial scan where green arc denote acquired views and red arc denote non-acquired views. (b) An attenuation field network (AFN) is trained using the partial-scan shifted-detector data as indicated by the green box. (c) After self-supervised training, i.e., no labeled data, the network is able to predict both acquired (surrounded by green) and unacquired (surrounded by red) projection data. (d) These synthetic projections along with the original PSSD projections can be used for post reconstructions using methods of user's choice.

## 2 Methods

Figure 1 shows the pipeline of the proposed method. Suppose in a circular cone-beam CT scan, a laterally-shifted detector acquires truncated projection data in a partial scan. The incomplete data is fed to an attenuation field network (AFN) for self-supervised training. The trained AFN can synthesize both acquired and non-acquired projection data as indicated by the green and red bounding boxes. The synthesized projections along with the acquired projections are used for post-reconstructions using analytical methods, iterative methods, or deep learning methods.

Figure 2 illustrates the training procedure of AFN, a fully-connected network whose input is a single continuous three-dimensional (3D) coordinate $\boldsymbol{r} = (x, y, z)$ and whose output is the respective attenuation coefficient $\mu(\boldsymbol{r})$. A minimum training sample of AFN is a ray propagating from the X-ray source $\boldsymbol{s} = (s_x, s_y, s_z)$ to a detector pixel $\boldsymbol{d} = (d_x, d_y, d_z)$. Along the ray $\overrightarrow{\boldsymbol{sd}}$, multiple 3D coordinates denoted as $\boldsymbol{t}_i = \boldsymbol{s} + \alpha_i(\boldsymbol{d} - \boldsymbol{s})$, $0 < \alpha_i < 1$, are randomly sampled. The attenuation coefficients of these sampled coordinates are queried and then integrated, according to the Beer-Lambert's law, to render the projection intensity at the ray end. Since the ray propagation is fully differentiable, we can optimize this model by minimizing the error between the acquired projection and the rendered projection. This error serves as the training loss and is written as $\|\sum_i h(t_i)|\boldsymbol{t}_{i+1} - \boldsymbol{t}_i| - p(\beta, u, v)\|$, where $h(\cdot)$ denotes the fully-connected network, $p$ denotes the acquired projection data (after taking the log), $\beta$ is the view angle, and $(u, v)$ specify a detector pixel. After the network iterates through all the "acquired rays", it can query the entire attenuation field by simply inputting the volumetric coordinates of the image; it may also re-project the attenuation field to obtain projections that are unacquired.

Our network consists of three full-connected layers with a feature dimension of 64. Prior to the input layer, we employed a hash grid encoding[10] to accelerate training as well as enhance the learning of high-frequency features. At the output layer, we used the exponential function with its gradient clipped as the activation to enforce the non-negativeness of attenuations. ReLU activations were used for other layers. For each breast PSSD data, we trained AFN using ADAM optimizer[16] with a learning rate of 1E-3 for 250 epochs. The learning rate decays by one-third every 50 epochs. In each iteration, we randomly selected 2048 rays from an acquired view angle and sampled 512 points per ray to roughly match the reconstruction voxel pitch. The effective batch size is 2048×512=1024 Ki, which amounts to approximately 20 GB GPU memory usage. The training took about two hours, and the rendering of 300-view full-scan data took about 20 minutes on an NVIDIA RTX A6000 graphic card.

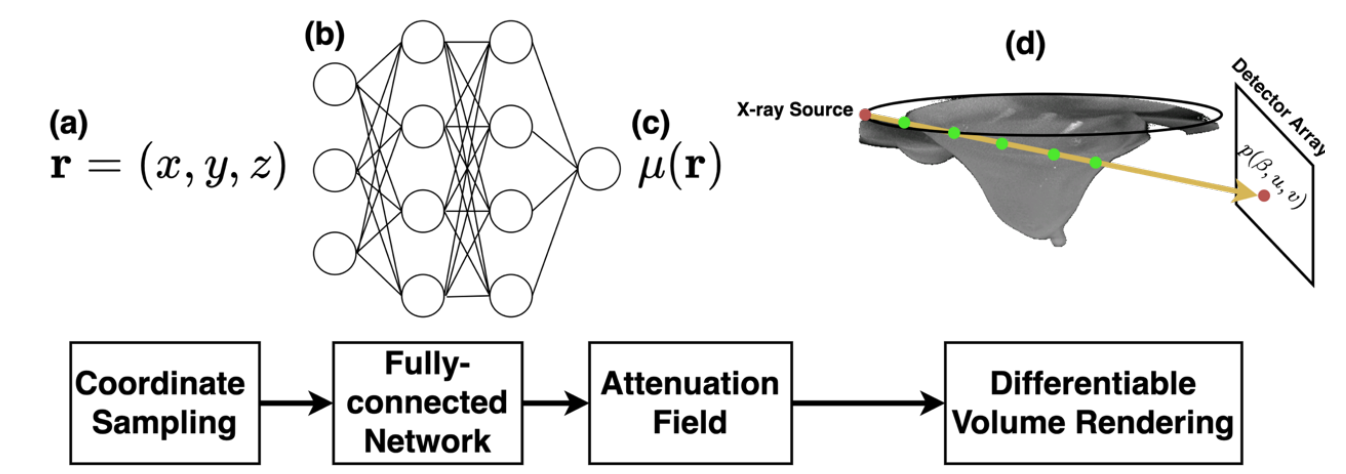

Figure 2. AFN training procedure. Our attenuation field network serves as a continuous function that maps a vector of three-dimensional coordinates (a), through a fully connected network (b), to the respective attenuation coefficient (c). The training procedure of AFN coincides with the conventional ray-tracing algorithm, where we sample points along a ray from the X-ray source to a detector pixel. According to the Beer-Lambert's law, the attenuations of these points are integrated to estimate the projection intensity at the detector. This volume rendering is fully differentiable (d) to allow us to optimize our attenuation representations by minimizing the error between the estimated and the acquired projection.

We used fifty breast CT datasets (HIPPA-compliant, de-identified, IRB approved) of BIRAD 4/5 women acquired on a prone, clinical-prototype scanner. Among the fifty datasets, 26 contain calcifications. The chest wall diameters of the breasts range from 8.5 cm to 18.5 cm, with an average of 13.0 cm. The scanner used a centered detector with 40 cm × 30 cm FOV in 2 × 2 binning mode, yielding a resolution 0.388 mm × 0.388 mm with its matrix size 1024 × 768. To retrospectively obtain PSSD data, we selected 225 views out of 300 views corresponding to 270 degrees ranging from -135 degree to 135 degree where each projection view was truncated 256 out of 1024 on the left. This in total results in an undersampling rate of $\frac{3}{4} \times \frac{3}{4} = \frac{9}{16}$. The full-scan data were reconstructed using the Feldkamp-Davis-Kress (FDK) algorithm[17] (0.273 mm voxel) as reference. Three reconstruction methods were evaluated: 1) AFN was trained using the PSSD data and used to predict missing data such that a full-scan full-size detector (FSFD) data can be formed along with the acquired PSSD data. The spliced FSFD data was reconstructed using FDK with a shifted-detector weight.[8] This additional weighting is applied mainly to suppress the truncation artifacts due to slight inconsistency between AFN synthesized projection and the acquired projection. 2) The PSSD data were reconstructed using FDK with a modified Parker weight[18] to only account for the partial scan. The unacquired ¼ data in the fan-angle dimension was zero-filled. 3) The PSSD data were reconstructed using FDK with a shifted-detector weight[8] to only account for truncated projections. The reconstructed image volume was scaled by the ratio of the number of views in the full scan to the number of views in the partial scan. Quantitative metrics including noise variance estimated in the adipose region, signal difference to noise ratio (SDNR) between adipose and fibroglandualr tissues, and the full-width at half maximum (FWHM) of calcifications along the mediolateral direction and the superior-inferior direction were used.[4]

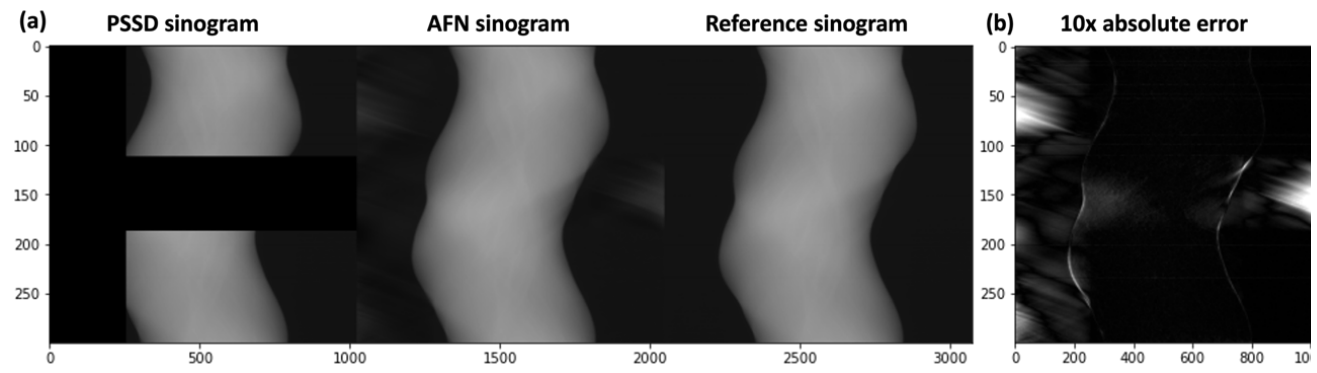

Figure 3. Sinogram prediction using AFN. (a) While AFN is trained on the partial scan shifted-detector (PSSD) sinogram, it can generate a high-fidelity full sinogram. (b) The 10× absolute error between the AFN sinogram and reference sinogram is displayed using the same dynamic range as (a). AFN yields minor errors in the top right and bottom right regions where the data is acquired and yields larger errors elsewhere.

## 3 Results

Figure 3 shows the breast sinogram upper towards the chest wall. Using the partial scan shifted-detector (PSSD) data,

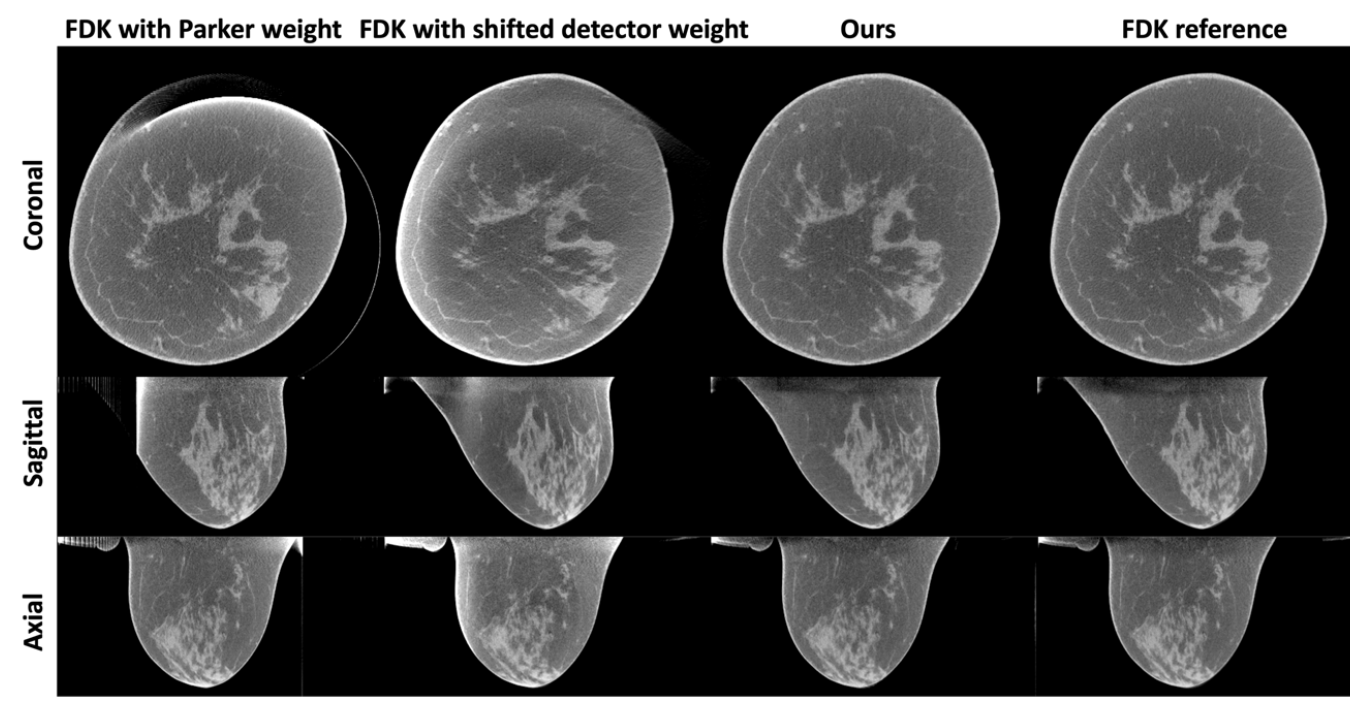

Figure 4. Reconstructions of a medium-size breast (14.5 cm chest wall) in coronal, sagittal, and axial views. Partial scan shifted-detector (PSSD) data were used in the three competing methods, including FDK with Parker weight, FDK with shifted detector weight, and our proposed AFN method. Full scan full-detector (FSFD) data were reconstructed using FDK to obtain the reference on the right. FDK with weighting functions can either account for partial scan or shifted-detector geometry, leading to truncation artifacts (first column) or missing wedge artifacts (second column) as expected. In contrast, AFN predicts the missing projections in the fan-angle dimension and the view angle dimension to yield FSFD data for post reconstructions (third column). Our proposed method delivers artifact-free and visually similar images as the reference. The display window is [0.15, 0.35] cm$^{-1}$.

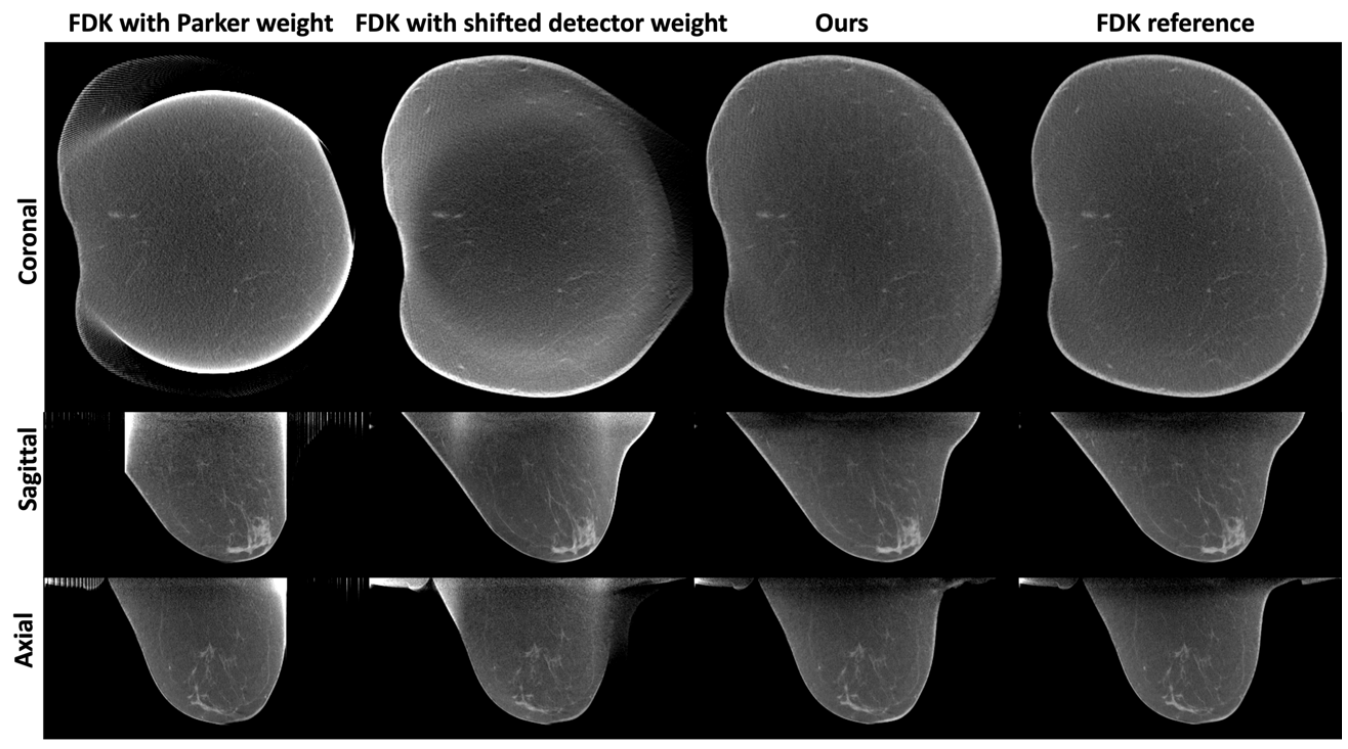

Figure 5. Reconstructions of a large-size breast (18.3 cm chest wall) in coronal, sagittal, and axial views. Partial scan shifted-detector (PSSD) data were used in the three competing methods, including FDK with Parker weight, FDK with shifted detector weight, and our proposed AFN method. Full scan full-detector (FSFD) data were reconstructed using FDK to obtain the reference on the right. FDK with weighting functions can either account for partial scan or shifted-detector geometry, leading to truncation artifacts (first column) or missing wedge artifacts (second column) as expected. In contrast, AFN predicts the missing projections in the fan-angle dimension and the view angle dimension to yield FSFD data for post reconstructions (third column). Our proposed method generates visually similar images as the reference except that some breast skin structures were not fully reconstructed, which is best seen near the chest wall of the axial view. The display window is [0.15, 0.35] cm$^{-1}$.

AFN can generate a high-fidelity sinogram compared to the full-scan reference. As specifically shown in Figure 3(b), other than the large errors outside the sinogram, AFN only makes small errors within the sinogram and marginal errors where the data is acquired (top right and bottom right regions). Figure 4 shows the image reconstructions of a medium-sized breast (chest wall diameter of 14.5 cm). The image reconstructed using FDK with Parker weight exhibits truncation artifacts due to shifted detector, whereas the image obtained using FDK with shifted detector weight manifests the missing wedge artifacts in the coronal plane due to partial scan and inhomogenous attenuations across all three planes. In contrast, our post-FDK reconstruction using AFN compensated projection data is free of artifacts and yields similar visual quality as the FDK reference. Figure 5 shows the reconstruction comparison on a large-size breast (18.3 cm chest wall diameter). Similar observations can be made except that the breast skin structures near the chest wall (due to the larger diameter) are not fully reconstructed in our method as can be seen in the coronal and axial images. Note that these regions correspond to the area where the projection data are most scarce. Table 1 compares the proposed AFN and references FDK methods. Noise variance and SDNR metrics suggest AFN yields comparable noise variance and image contrast as the reference, respectively. Two FWHM metrics indicate AFN maintains a similar spatial resolution of calcifications as the reference.

Table 1. Quantitative comparison between the proposed AFN using PSSD data and the FDK reference using full scan data. Fifty breast cases were evaluated, among which 26 cases contained calcifications. The noise variance is estimated in the adipose region. Signal difference to noise ratio (SDNR) is calculated between the adipose and fibroglandular tissues. The full width at the half maximum (FWHM) of calcifications was computed along two orthogonal directions (ML: mediolateral; SI: superior-inferior). AFN yields comparable quantities as the reference.

| | AFN | FDK reference |
|---|---|---|
| Noise Variance ($\times 10^{-5}$ cm$^{-2}$) | 7.55 ± 0.30 | 7.69 ± 0.35 |
| SDNR | 6.36 ± 1.65 | 6.35 ± 1.83 |
| FWHM-ML (mm) | 1.68 ± 0.72 | 1.61 ± 0.75 |
| FWHM-SI (mm) | 1.55 ± 0.64 | 1.69 ± 0.63 |

## 4 Discussion and Conclusion

In this work, we presented a novel neural field-based method for cone-beam breast CT reconstruction with a simultaneous partial scan and shifted detector geometry. The proposed AFN reduces the learning to each propagating ray such that implicit representations of the attenuations can be learned with high efficacy and memory efficiency. We have shown that the learned AFN can generate high-fidelity projections in areas where data are acquired or unacquired. Nevertheless, we also observed that the post-FDK reconstructions using AFN synthesized projections require further refinement for large-size breasts. We believe this can be achieved in a future study by either using additional constraints/regularizations during network training or using advanced post reconstruction methods such as iterative algorithms or deep learning techniques.

## Ackonwledgments

This work was supported in part by the National Cancer Institute (NCI) of the National Institutes of Health (NIH) grants R01CA241709, R01CA199044, and R21CA134128. The contents are solely the responsibility of the authors and do not represent the official views of the NIH or the NCI.

# Using Translational Stable Diffusion Probabilistic Model (TranSDPM) to Improve the Longitudinal Resolution in Computed Tomography

Yongfeng Gao[1], Liyi Zhao[2], Wenjing Cao[2], Yuan Bao[2], Jian Xu[1], and Guotao Quan[2]

[1] UIH America, Inc., Houston, Texas, USA
[2] United Imaging Healthcare, Shanghai, China

**Abstract** The ability of computed tomography (CT) to clearly distinguish between structures along the z-axis, referred to as the longitudinal resolution, can be limited in some scenarios. Super Resolution (SR) is a crucial challenge in the field of computer vision and image processing, where the objective is to generate a high-quality, high-resolution (HR) image from a low-resolution (LR) input. This task is challenging due to the loss of information and degradation that occurs during the process of downscaling the image, including blurring and noise. In this study, we introduce an innovative solution using a Translational Stable Diffusion Probabilistic Model (TranSDPM) to address the SR challenge in CT images. The diffusion denoise model is built on the concept of non-linear diffusion, which enables the propagation of information from the LR image to the HR image by modeling the degradation process through a Markov chain. This approach has the advantage of preserving the structural information in the LR image, leading to enhanced super-resolution performance and suppressing the noise at the same time. The effectiveness of the proposed method is rigorously evaluated on publicly available datasets at the full dose and a quarter dose demonstrating its superiority and robustness.

## 1 Introduction

Computed Tomography (CT) is a commonly used medical imaging technique that provides high-quality cross-sectional images of the body. However, the ability to clearly distinguish between structures along the z-axis, referred to as the longitudinal resolution, can be limited in some scenarios, such as high thickness reconstruction or high pitch imaging [1,2]. To address this challenge, researchers have employed various techniques to improve the longitudinal resolution of CT scans.

In the broader context, the process of enhancing image resolution is commonly referred to as super-resolution (SR). SR is a critical challenge in the field of computer vision and image processing, where the objective is to generate a high-quality, high-resolution (HR) image from a low-resolution (LR) input. This task is challenging due to the loss of information and degradation that occurs during the downscaling process, including blurring and noise. In recent years, deep learning methods have shown particularly promising results for SR, with various proposed methods in the literature [3,4].

Among deep learning methods, Stable Diffusion Probabilistic Models (SDPMs) have become increasingly popular in recent years due to their ability to handle large and complex data sets and provide robust and accurate predictions in a variety of applications [5-8]. SDPMs are a class of models used in statistical analysis and machine learning for capturing the evolution of systems over time or iterations. SDPMs are based on the idea of stable diffusions, which are continuous-time Markov processes characterized by their long-term behavior. They have demonstrated great success in many tasks, such as image generation, object recognition, and image segmentation.

In this study, we introduce an innovative solution using a Translational Stable Diffusion Probabilistic Model (TranSDPM) to address the super-resolution challenge in CT images. The proposed TranSDPM is built on the concept of non-linear diffusion, which enables the propagation of information from the LR image to the HR image by modeling the degradation process. Unlike traditional image SR techniques that rely on simplistic interpolation methods, TranSDPM generates images through a probabilistic framework that can learn the underlying distribution of high-resolution images. This approach enables the model to produce more accurate and detailed images, reducing the occurrence of artifacts and other issues commonly associated with traditional image SR methods. TranSDPM preserves the structural information in the LR image, leading to enhanced SR performance. The effectiveness of the proposed method is rigorously evaluated on publicly available datasets at full and quarter doses, demonstrating its superiority and robustness.

## 2 Materials and Methods

### *2.1 Dataset*

This study utilized the publicly available clinical dataset, “Clinics for the 2016 NIH-AAPM-Mayo Clinic Low Dose CT Grand Challenge,” which features a comprehensive collection of CT images with a range of dose levels and longitudinal resolutions. Specifically, the dataset includes four image sets, consisting of two dose levels and two longitudinal resolutions (i.e., full dose and quarter dose at 1mm and 3mm resolutions). In addition, two sets of 4mm resolution data were simulated from the 1mm resolution data via a rebin operation for the downsampling.

To enhance the longitudinal resolution of the images, the 2D images were extracted from the coronal and sagittal views, resulting in a total of 6240 images obtained from 10 patients. The image size was 512 x 512, with 8 patients’ images used for training and the remaining 2 patients’ images utilized for testing. Notably, there was no overlap between the training and testing data sets, ensuring the integrity of the results obtained.

### *2.2 Method Description*

Fig. 1 shows the procedure of our TranSDPM method. Training is performed in two stages. In the first stage shown in panel A, an auto-encoder-decoder (highlighted by red outline) is trained to convert images to the latent domain. In the second stage shown in panel B, a noise estimator (highlighted by red outline) is trained to estimate noises in the shape of a latent domain image conditioned by a latent domain low resolution image. This is the core part of the proposed TranSDPM model. As with infinitesimal steps, the diffusion process is reversable so that noise can be removed with the noise estimator. The inference process or image translation from low resolution domain to high resolution domain is illustrated in panel C. The SR latent is initiated as pure Gaussian noise, which is used as input to the noise estimator together with the LR latent. The estimated noise is used to reduce noise from the SR latent by a fixed reverse diffusion routine for one time step. This will not remove all the noise. Instead, this moves the high-dimensional latent vector slightly away from the noise manifold and towards the data manifold. The fidelity preservation task is also performed by the noise estimator during the LR latent conditioned noise estimation process. This one step denoising is repeated for certain times and the final SR latent is translated back to image domain by the decoder. The auto-encoder is CNN based and has a 8×64×64 bottle neck (latent shape). The noise estimator is a CNN based UNet.

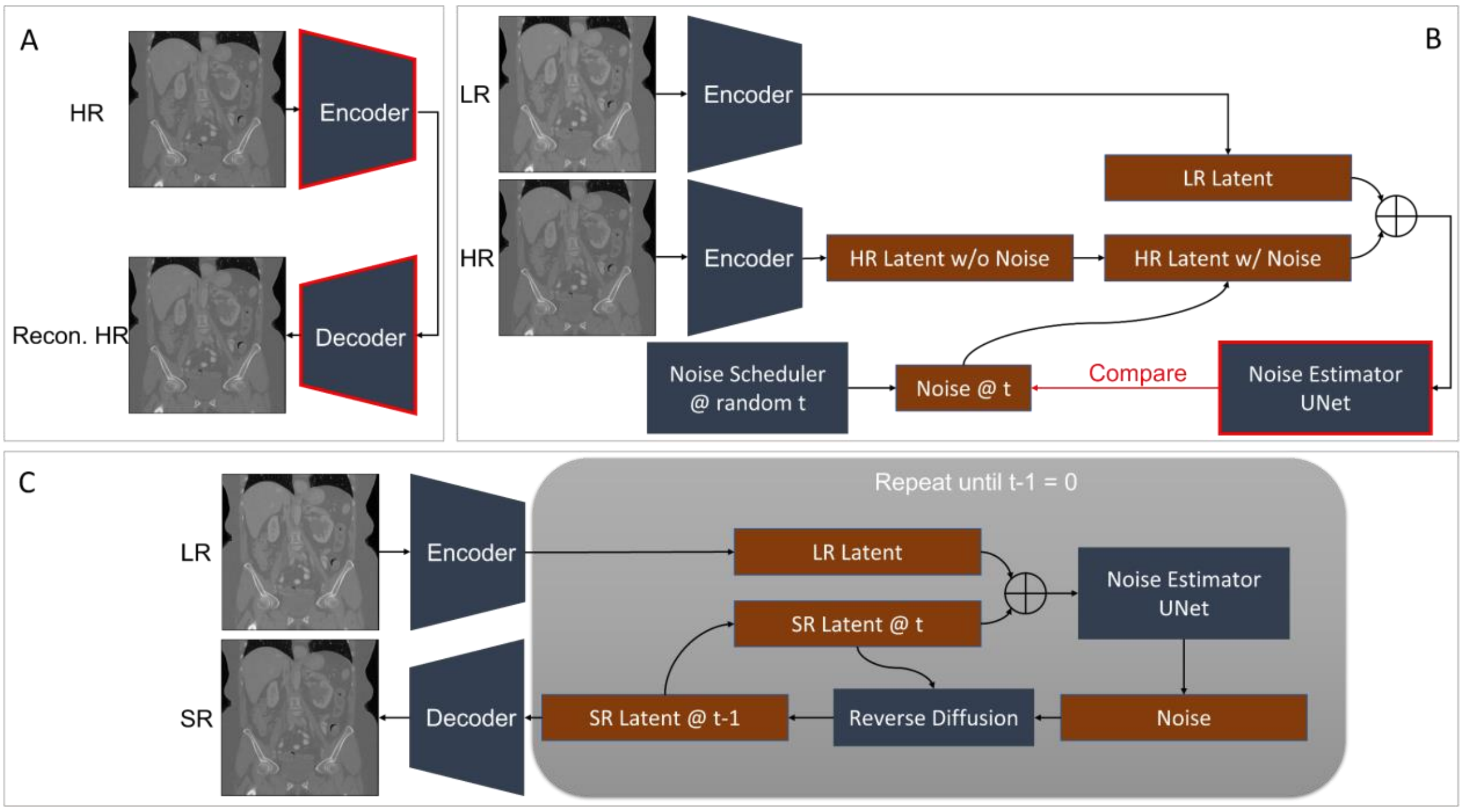

Fig. 1 The Architecture of the proposed TransDPM for longitudinal resolution enhancement. "LR" stands for the low resolution. "HR" stands for the high-resolution images used in the training stage. "SR" stands for the super resolution obtained from the TransDPM model.

## 3 Results

### *3.1 Results with the Simulated LR Data*

The initial phase of the study was focused on assessing the efficacy of the proposed model in enhancing the resolution of low-resolution images, utilizing a simulated dataset. Both the HR and simulated LR data sets underwent voxel-level registration to ensure precise alignment so that we can have a more direct comparison through their difference images. In this study, a four-fold resolution enhancement was used as the task. In the future, we are interested in investigating the potential of TranSDPM to enhance LR images of varying degrees. The evaluation of the simulated data set enabled the identification of potential discrepancies between the enhanced and original images, with the results presented in Figures 2, 3, and 4. It is noted that all the models were trained solely on the full dose data, and the evaluation of the quarter dose was performed without retraining.

Fig. 2 presents an illustrative example of the enhanced resolution in the abdominal area, where the images are displayed in the soft tissue window, and the difference images are displayed in the window with a level of zero for better visualization. From left to right, the images comprise the HR image (1mm), the bicubic-generated image, the TranSDPM generated image, the absolute difference image between the bicubic image and the HR image, and the absolute difference image between the TranSDPM image and the HR image.

The bicubic-generated images exhibit notable smear like appearance, with certain details such as the liver structure not clearly visible, as indicated by the red arrow. In contrast, the TranSDPM images effectively improve this appearance and provide enhanced details of the liver structure. This improvement is also evident in the difference images. The values in the bicubic difference image primarily arise from the structure, whereas those in the TranSDPM difference image are primarily due to noise.

Interestingly, it should be noted that the HR images themselves contain considerable noise, which can obscure structures, especially in low dose scenarios, as illustrated by the left image in the bottom row. By suppressing noise while recovering structure details, the proposed TransDPM approach enables the visualization of more detailed anatomical features in the TranSDPM images compared to HR images, as shown by the areas indicated by the red arrow in the bottom row. Overall, these results demonstrate the effectiveness of the proposed model in enhancing the resolution and clarity of LR medical images.

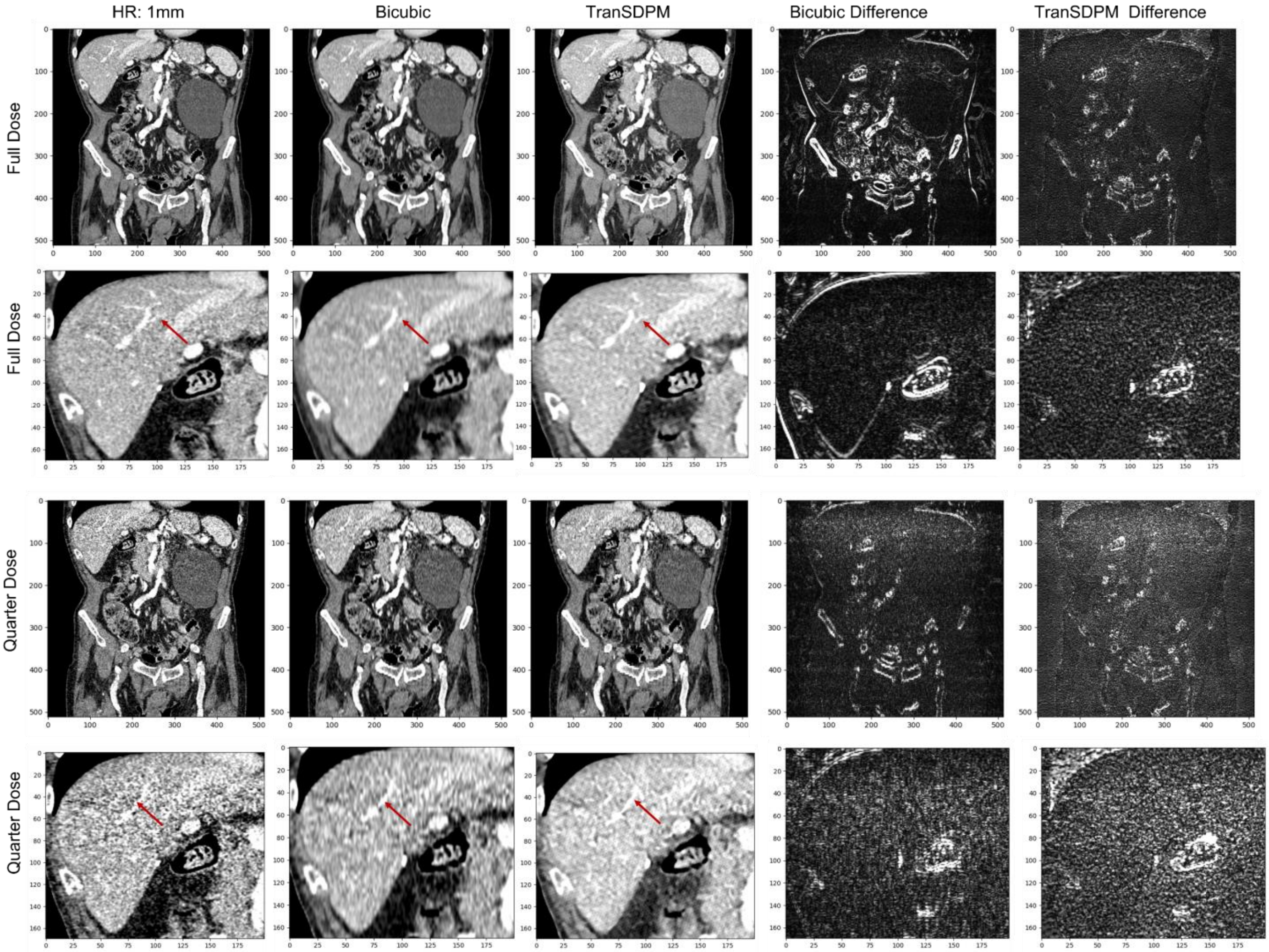

Fig. 2. The results with simulated low-resolution data. It is noted that TranSDPM is only trained with the full dose data and tested on the low dose data directly. The original images are displayed in the soft-tissue window [-125, 225] HU. The difference images are displayed in the window [0 175] for better visualization.

Fig. 3 and Fig. 4 demonstrate the effectiveness of the proposed model in enhancing the resolution of bone and lung images, respectively, with the display windows adjusted accordingly. Similar observations to those in Fig. 2 were also noted in Fig. 3 and Fig. 4. Fig. 3 reveals that the contour surrounding the bone region appears less distinct with some blocky segments in certain areas in the bicubic image. However, the boundaries appear more clear and natural in the TranSDPM images. Meanwhile, Fig. 4 shows that the details of the pulmonary airways are unclear in the bicubic images but can be clearly visualized in the TranSDPM images, highlighting the proposed model's capacity to enhance the details and structures in low-resolution medical images. Through exploring the performance of the TranSDPM model in various applications, its effectiveness and robustness have been demonstrated.

### *3.2 Results with Real Clinical LR Data*

The effectiveness of TranSDPM was further evaluated using real clinical LR data, and the results are presented in Fig. 5. In this evaluation, the model trained on the simulated data will be directly used for the inference. In this instance, the HR and LR images lack pixel-level alignment, so only the original images will be used for evaluation. The results demonstrate that the proposed method can significantly enhance the resolution of 3mm LR images, producing images with resolution comparable to those of the 1mm images. Furthermore, TranSDPM not only enhances the resolution, but also effectively suppresses noise and improves structural details, as shown by the comparison with HR images. These findings indicate the robustness and versatility of TranSDPM, which was trained on a four-fold resolution enhancement and tested on a three-fold resolution enhancement, demonstrating its potential for application in various contexts.

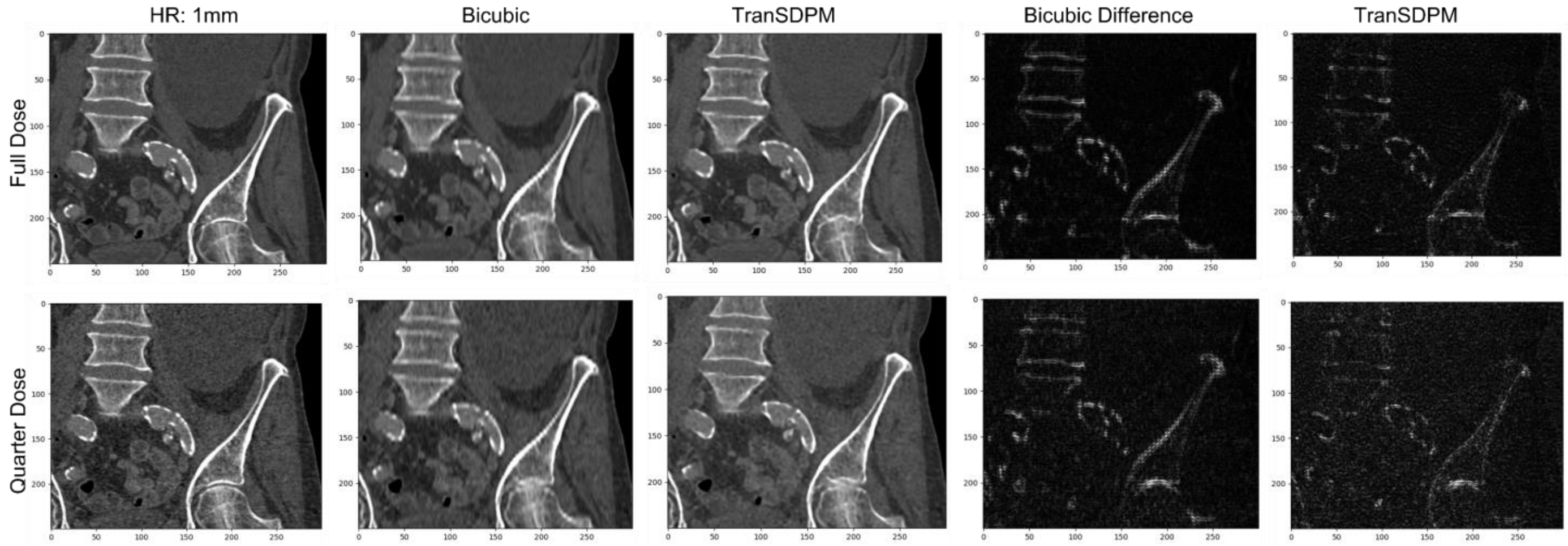

Fig 3. The results with simulated LR data. The original images are displayed in the bone window [-250, 750] HU. The difference images are displayed in the window [0 500] for better visualization.

## 4 Discussion

When designing TranSDPM, we also considered using the LR latent as one step in the reverse diffusion process instead of the condition for the fidelity preservation. However, the desing using the LR as one step failed to generate reasonable image. One possible reason for this is that the noise distribution of the LR data differs from the Gaussian distribution. For the fidelity preservation, based on our current tests, we have not observed any unusual or unexpected structures arising from using the LR data as a condition in the reverse diffusion process.

The proposed model was trained on high-resolution images with a resolution of 1mm. However, as demonstrated in the results section, these HR images still contain noise. This raises the question of how to construct a "noise-free" HR image to serve as a reference standard. Addressing this challenge will be an area of future research interest, as it has the potential to further enhance the accuracy and effectiveness of the proposed model in enhancing the resolution of low-dose medical images.

As described in the Method section, the proposed model was trained exclusively on full dose CT data, yet it was directly tested in the low-dose scenario. Furthermore, during training, the model was trained on a four-fold resolution enhancement and tested on a three-fold resolution enhancement. Both results were highly promising, indicating the model's robustness and potential to handle varying levels of noise. To expand upon these findings, future research may involve training the model on low dose CT data and evaluating it across various folds resolution enhancement scenarios.

## 5 Conclusion

This study proposed a TranSDPM to address the super-resolution challenge in CT images. The feasibility and effectiveness of the proposed approach is demonstrated through rigorous evaluations on publicly available datasets at both full and quarter doses, utilizing both the simulated and real clinial data showcasing its superiority and robustness. Further evaluations will be conducted to continue testing the performance of TranSDPM.

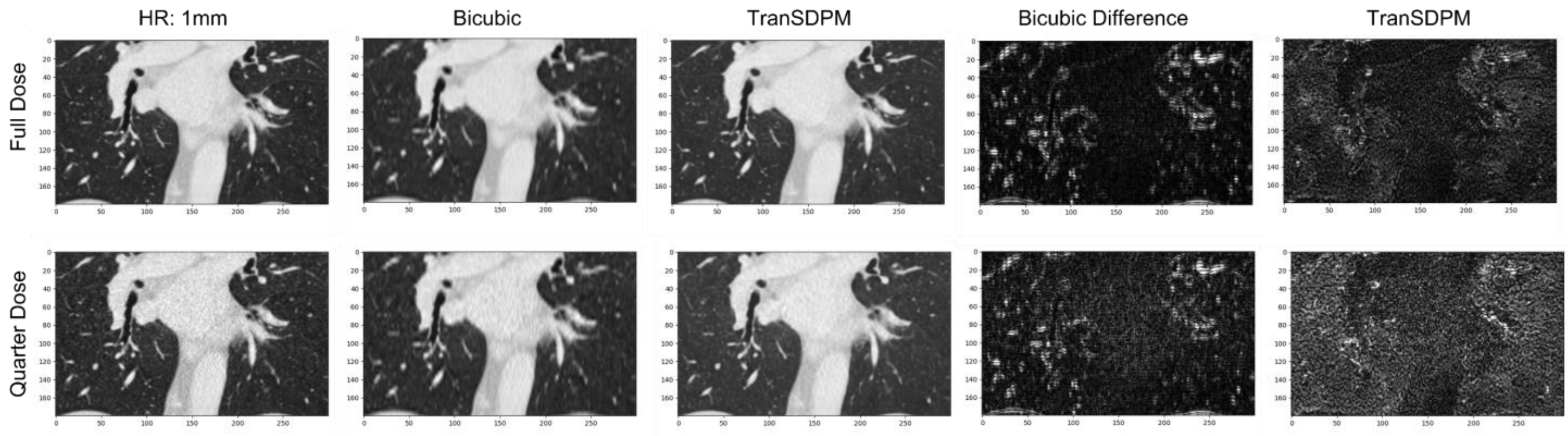

Fig 4. The results with simulated LR data. The original images are displayed in the lung window [-1150, 350] HU. The difference images are displayed in the window [0 300] for better visualization.

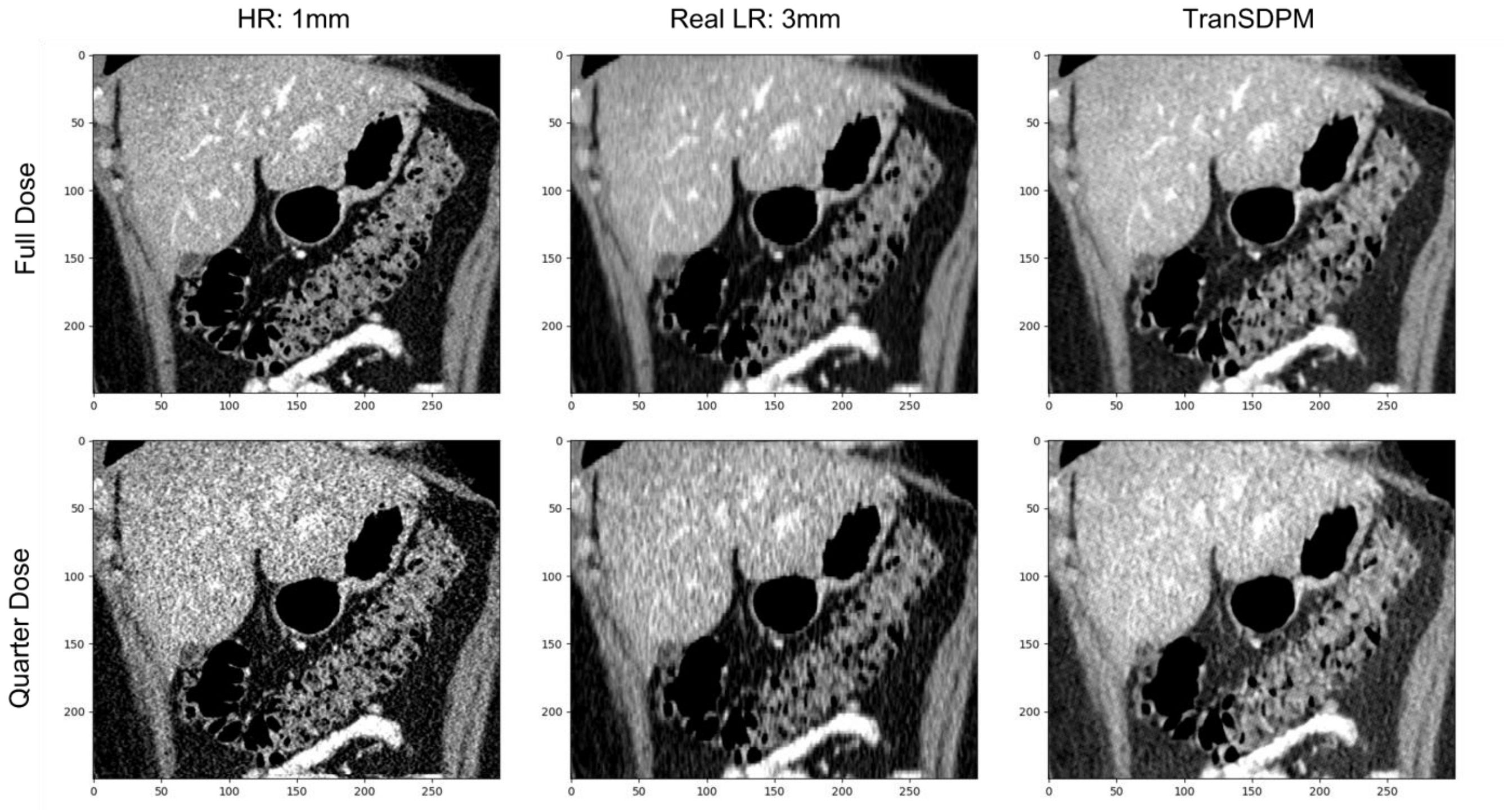

Fig 5. The results with real clinical LR data. All images are displayed in the soft tissue window [-125, 225] HU.

# Motion Estimation in Parallel-Beam Linogram Geometry Using Data Consistency Conditions

Sasha Gasquet[1,2], Laurent Desbat[1], and Pierre-Yves Solane[2]

[1]Univ. Grenoble Alpes, CNRS, UMR 5525, VetAgro Sup, Grenoble INP, TIMC, 38000 Grenoble, France
[2]TIAMA, 215 chemin du Grand Revoyet, F-69230 Saint-Genis-Laval, France

**Abstract** Data consistency conditions (DCCs) express the redundancy in the projections. In X-ray computed tomography, the most common conditions are expressed pairwise on the projections or as equality between projection-based moments and polynomials. The latter is better known in the parallel-beam geometry as the Helgason-Ludwig consistency conditions (HLCCs). The DCCs are often used to self-calibrate radiography systems. In this paper, we adjust data consistency conditions to a time-dependent model of the data in the parallel linogram geometry. We show that it is not possible to estimate the parameters of a uniform motion of a translating object using the DCCs. However, we show that we can estimate the average speed with prior information on the object's center of mass. Then, we model and estimate the parameters of a periodical variation of the motion. Finally, we run simulations to assess the performances of our method.

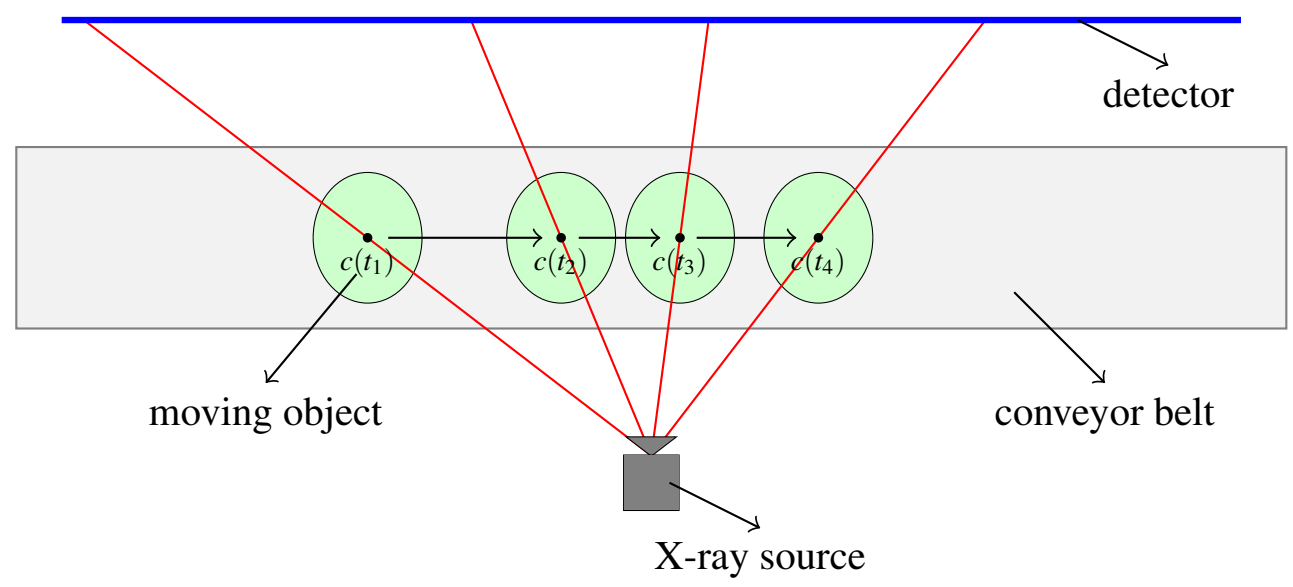

**Figure 1:** The radiography system is constituted of a X-ray source, a horizontal linear detector and a conveyor belt. Everything is stationary, except for the object translating on the conveyor belt.

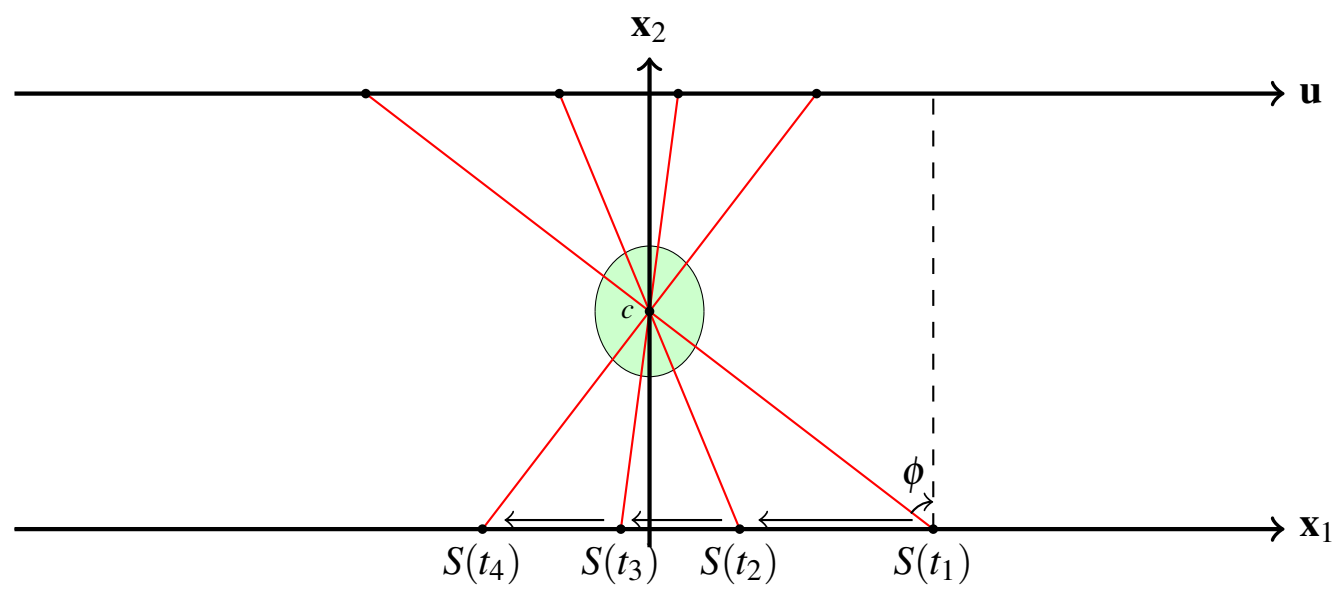

**Figure 2:** The equivalent radiography system. The source and the detector are moving at the same speed in the opposite direction to the translation of the conveyor belt in the Fig. 1. The object is stationary.

## 1 Introduction

In X-ray computed tomography, the data consistency conditions (DCCs) give information on the behavior of the radiography system based on the redundancy of the projections. If some changes occur in the system, the conditions are no longer satisfied. The changes can be detected or even estimated with proper modeling.

In the literature, some conditions are derived from the Helgason-Ludwig consistency conditions (HLCCs) [1][2]. These conditions have been used to estimate the motion of a moving object in the fan-beam geometry with a circular trajectory of the source and in the parallel geometry [3]. A more suitable representation of the data in the geometry with the source on a line is the linogram. The HLCCs have been expressed in the linogram geometry [4]. Results on the estimation of the source position and motion have been published for the fan-beam linogram geometry [5][6][7].

In this work, we consider a radiography system composed of a X-ray source, a horizontal linear detector and an object translating on a conveyor belt. The source and detector are supposed stationary. The object position is defined by its center of mass $c(t)$. The system is represented in the Fig. 1. Equivalently, this system can be considered as a system with a stationary object and a translating source and detector. The translation is the same as in the original radiography system but in the opposite direction. The equivalent system is represented in Fig. 2. At a fixed viewing angle $\phi$, all the X-rays are parallel. The source position and the projection on the detector are supposed point-like. The X-rays are defined in the coordinate system $(\mathbf{x}_1, \mathbf{x}_2)$ as segment from the source $S(t) = (x(t), 0)$ to the detector point with the direction $\gamma_\phi = (\sin\phi, \cos\phi) \in S^1$, where $S^1$ is the unit sphere.

We suppose that the measured object $\mu : \mathbb{R}^2 \to \mathbb{R}$ has a compact support. The projections are modeled by the Beer-Lambert absorption law which makes the link between the object $\mu$, the initial intensity $I_0$ of the X-rays and the intensity $I$ acquired by the detector:

$$I = I_0 \exp\left(-\int_{\mathbb{R}} \mu\left(S(t) + r\gamma_\phi\right) dr\right) \quad (1)$$

A logarithm transform leads to the classical projection form:

$$p(\phi, x(t)) = \int_{\mathbb{R}} \mu\left((x(t), 0) + r\gamma_\phi\right) dr \quad (2)$$

In the following, we first define the parallel linogram geometry and recall the parallel linogram consistency conditions derived from the HLCCs to our geometry. We show that we cannot estimate the parameters of a uniform motion. Then, we model and estimate the parameters of a non uniform mo-tion with the DCCs. Finally, we run simulations to evaluate the accuracy of the method.

## 2 Theory

### 2.1 Parallel linogram geometry

In an equivalent radiography system, the object is considered stationary. The source and the detector are moving at the same speed on two parallel lines separated by a distance $D > 0$. The parallel beam X-rays in the linogram geometry are parallel segments from a source at $S(t) = (x(t), 0)$ on the $x_2 = 0$ axis to a detector at $(x(t) + u, D)$ on the parallel axis $x_2 = D$ where $u \in \mathbb{R}$ and the distance $D$ is fixed. The object position is defined by its center of mass $c = (c_1, c_2)$. We assume $x'(t) < 0, \forall t \in \mathbb{R}$. The offset $u$ on the $\mathbf{x}_1$ axis between the source at $(x(t), 0)$ and the detector point at $(x(t) + u, D)$ is bijectively linked to the projection angle $\phi$ with $u = D\tan(\phi)$, $\phi \in ]-\pi/2, \pi/2[$, $u \in \mathbb{R}$, or equivalently $\phi_u = \arctan(u/D)$. This system is represented in Fig. 3.
The parallel-beam linogram $l$ is defined by:

$$l(u,x) = \int_{\mathbb{R}} \mu((x,0) + r(u,D))dr \tag{3}$$

In our geometry, the X-rays are indexed by the time $t$ at which they are measured. Thus, we define the parallel linogram $\bar{\bar{l}}$ and the parallel linogram operator $\bar{\bar{\mathscr{L}}}$ as:

$$\bar{\bar{\mathscr{L}}}\,\mu\,(u,t) = \bar{\bar{l}}\,(u,t) \tag{4}$$

$$= l(u, x(t)) \tag{5}$$

$$= \int_{\mathbb{R}} \mu((x(t),0) + r(u,D))dr \tag{6}$$

The linogram $\bar{\bar{l}}$ is a weighted linogram. Using the change of variable $r' = r\sqrt{u^2 + D^2}$, we have:

$$\bar{\bar{l}}\,(u,t) = \frac{1}{\sqrt{u^2+D^2}} \int_{\mathbb{R}} \mu((x(t),0) + r'\gamma_{\phi_u})dr' \tag{7}$$

$$= \frac{1}{\sqrt{u^2+D^2}} p(\phi_u, x(t)) \tag{8}$$

where $\gamma_{\phi_u} = (\sin(\phi_u), \cos(\phi_u)) = \frac{1}{\sqrt{u^2+D^2}}(u,D)$.

### 2.2 Helgason-Ludwig Consistency Conditions

In the parallel-beam linogram geometry, the order $n \in \mathbb{N}$ moment of the projections is defined by:

$$J_n(u) = \int_{\mathbb{R}} l(u,x)x^n dx \tag{9}$$

For the Radon transform, the Helgason-Ludwig theorem states that the order $n$ moment is an homogeneous polynomial of order $n$ in $\cos\phi$ and $\sin\phi$ [8]. Such DCCs can be derived for the parallel linogram geometry [4]. We adjust these conditions to our geometry. Using a change of variable, we define the order $n$ moment of the projections as:

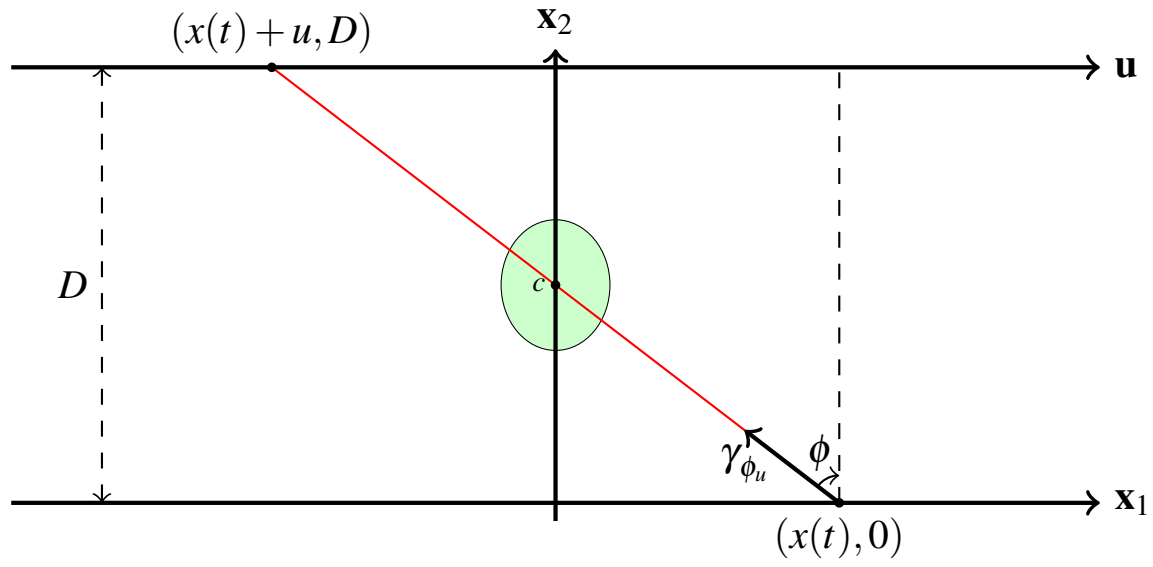

**Figure 3:** The parallel linogram. The object is supposed stationary and its center of mass is denoted by $c$. For a viewing angle $\phi$, the offset $u = D\tan\phi$ between the source and the projection point on the detector is constant. The position of the source and the projection point associated to $\phi$ are respectively $(x(t), 0)$ and $(x(t) + u, D)$.

$$\bar{\bar{J}}_n\,(u) = \int_{\mathbb{R}} \bar{\bar{l}}\,(u,t)x^n(t)|x'(t)|dt \tag{10}$$

**Proposition 1** *(from [4]) The data $\bar{\bar{l}}$ are consistent, i.e. $\bar{\bar{l}}$ is in the range of $\bar{\bar{\mathscr{L}}}$, if and only if*

$$\bar{\bar{J}}_n\,(u) = \sum_{k=0}^{n} c_{n,k}u^k \tag{11}$$

From the data $\bar{\bar{l}}$, we can only compute the time related moment $\tilde{J}_n(u)$ defined as follow:

$$\tilde{J}_n(u) = \int_{\mathbb{R}} \bar{\bar{l}}\,(u,t)t^n dt \tag{12}$$

In the following sub-sections, we use the proposition 1 to estimate parameters related to the motion $x(t)$.

### 2.3 Uniform motion

We first model the source position using 2 real parameters $x_0$ and $v_0$.

$$x(t) = x_0 + v_0 t \tag{13}$$

The parameter $x_0$ cannot be estimated using DCCs [9]. We arbitrarily set $x_0 = 0$. Thus, $\bar{\bar{J}}_n\,(u)$ can be rewritten as:

$$\bar{\bar{J}}_n\,(u) = |v_0|v_0^n \int_{\mathbb{R}} \bar{\bar{l}}\,(u,t)t^n dt \tag{14}$$

$$= \text{sgn}(v_0)v_0^{n+1}\tilde{J}_n(u) \tag{15}$$

We want to estimate $v_0$ using the Eq. (15). Since $\bar{\bar{J}}_n\,(u)$ is defined relatively to $v_0$, we use the proposition 1. Therefore, in addition to $v_0$, we need to estimate the parameters $c_{n,k}$ for $k = 0, \ldots, n$. Let's now consider $u_1, \ldots, u_{n_a}$ where $n_a$ is the number of projections. We get a non-linear system of equations from the Eq. (15).

$$\begin{cases} \operatorname{sgn}(v_0)\displaystyle\sum_{k=0}^{n} c_{n,k}u_1^k - v_0^{n+1}\tilde{J}_n(u_1) = 0 \\ \qquad\qquad\qquad\vdots \\ \operatorname{sgn}(v_0)\displaystyle\sum_{k=0}^{n} c_{n,k}u_{n_a}^k - v_0^{n+1}\tilde{J}_n(u_{n_a}) = 0 \end{cases} \quad \forall n \in \mathbb{N} \tag{16}$$

The system of Eqs. (16) has an infinity of solutions: if $\{(c_{n,k}, v_0),\ \forall n \in \mathbb{N},\ k = 0, \ldots, n\}$ is a solution then $\left\{(\lambda^{n+1}c_{n,k}, \lambda v_0),\ \forall n \in \mathbb{N},\ k = 0, \ldots, n\right\}$ is a solution for any $\lambda \in \mathbb{R}$. Thus, *we cannot determine $v_0$ from the DCCs.*

### 2.4 Estimating $v_0$ from a center of mass property

The parameters of the uniform motion cannot be estimated using the DCCs *only*. However, we can use DCCs with a calibration object to estimate $v_0$. The DCCs of order 0 and 1 are related to the center of mass of an object (This property can be used for misalignment correction of the projections) [9]. We show in this sub-section that we can use the center of mass coordinates to estimate the average velocity $v_0$ of the source from two different projections. Using the definition of $\bar{\bar{J}}_n(u)$ and $x(t)$ in the Eqs. (10) and (13), we get the following:

$$\frac{\bar{\bar{J}}_1(u)}{\bar{\bar{J}}_0(u)} = \frac{\displaystyle\int_{\mathbb{R}} \bar{\bar{l}}(u,t)(x_0 + v_0 t)v_0 dt}{\displaystyle\int_{\mathbb{R}} \bar{\bar{l}}(u,t)v_0 dt} \tag{17}$$

$$= x_0 + v_0 t_c(u) \tag{18}$$

where $t_c(u) = \tilde{J}_1(u)/\tilde{J}_0(u)$ is the temporal center of mass of the projection $u$. Now, using the Eqs. (4) and (10), we get:

$$\frac{\bar{\bar{J}}_1(u)}{\bar{\bar{J}}_0(u)} = \frac{\displaystyle\int_{\mathbb{R}}\int_{\mathbb{R}} \mu(x(t) + ru, rD)x(t)x'(t)drdt}{\displaystyle\int_{\mathbb{R}}\int_{\mathbb{R}} \mu(x(t) + ru, rD)x'(t)drdt} \tag{19}$$

We make the following change of variables:

$$\begin{cases} x_1 = x(t) + ru \\ x_2 = rD \end{cases} \tag{20}$$

Additionally, we have $dx_1dx_2 = |-Dx'(t)|\,drdt$. We recall that $x'(t) < 0$, $\forall t \in \mathbb{R}$. Then, we have $dx_1dx_2 = -Dx'(t)drdt$. Hence, applying the change of variables, we get:

$$\frac{\bar{\bar{J}}_1(u)}{\bar{\bar{J}}_0(u)} = \frac{\displaystyle\int_{\mathbb{R}}\int_{\mathbb{R}} \mu(x_1, x_2)\left(x_1 - \frac{x_2}{D}u\right)dx_1dx_2}{\displaystyle\int_{\mathbb{R}}\int_{\mathbb{R}} \mu(x_1, x_2)dx_1dx_2} \tag{21}$$

$$= c_1 - \frac{u}{D}c_2 \tag{22}$$

where $c = (c_1, c_2)$ is the center of mass of the calibration object $\mu$. From the Eqs. (18) and (22), we have:

$$c_1 - \frac{u}{D}c_2 = x_0 + v_0 t_c(u) \tag{23}$$

For two different projections $u_1$ and $u_2$, we can write the following formula using a linear combination of the Eq. (23).

$$v_0 = -\frac{u_1 - u_2}{D(t_c(u_1) - t_c(u_2))}c_2 \tag{24}$$

We do not need to know $c_1$ nor $x_0$ here but only $c_2$.

### 2.5 Non uniform motion estimation

We now assume the conveyor belt has a non uniform motion due to mechanical instabilities. We model the variations with the time dependent function $\delta(t)$. The position of the source is then defined by:

$$x(t) = x_0 + v_0 t + \delta(t) \tag{25}$$

The motion $\delta$ is assumed to be periodic.

$$\delta(t) = A\sin(\omega t + \psi) \tag{26}$$

We assume $x'(t) < 0$, $\forall t \in \mathbb{R}$. The DCCs can be rewritten using Eqs. (25) and (26) as:

$$\begin{aligned} \bar{\bar{J}}_n(u) = -\int_{\mathbb{R}} \bar{\bar{l}}(u,t)(x_0 + v_0 t + A\sin(\omega t + \psi))^n \times \\ (v_0 + A\omega\cos(\omega t + \psi))dt \end{aligned} \tag{27}$$

Eq. (27) is non-linear in $A$, $\omega$, $\psi$ for all $n \in \mathbb{N}$. We apply the proposition 1 as in the subsection 2.3. We use the 0-order condition to estimate the parameters $A$, $\omega$, $\psi$, $c_{0,0}$ by solving a non-linear system of equations using the Gauss-Newton algorithm based on:

$$-v_0\tilde{J}_0(u) = c_{0,0} + A\omega\int_{\mathbb{R}} \bar{\bar{l}}(u,t)\cos(\omega t + \psi)dt \tag{28}$$

## 3 Simulations

The mean velocity $v_0$ is assumed to be known and $\delta(t)$ is estimated by solving the Eq. (28). The simulation are done using the library RTK [10]. Our phantom is composed of two cylinders respectively with a radius of 45mm and 50mm, and of density $-0.2$ and 0.2. It is placed midway between the source and the detector. The source to detector distance is $D = 480$mm. The linear detector is composed of 500 pixels with a pitch of 0.4mm. The leftmost pixel is the pixel 0. Thus, we set its origin at $u = -200$mm. The moving source position is defined by $x(t) = x_0 + v_0 t + A\sin(\omega t + \psi)$ where we fix $x_0 = 0$mm, $v_0 = -1000$mm/s, $A = 2$, $\omega = 40$, $\psi = \pi/4$. We acquire the projection at a rate of 2000Hz within the interval $[-T/2, T/2]$ where we set $T = 0.6$s. Gaussian noise is added to the projections. The standard deviation of the noise is defined for each pixel as a percentage of its value.

| Noise | 0% | 1% | 3% |
|---|---|---|---|
| $A$ | 1.988 | $1.976 \pm 0.030$ | $1.959 \pm 0.104$ |
| $\omega$ | 39.966 | $39.886 \pm 0.181$ | $39.731 \pm 0.688$ |
| $\psi$ | 0.785 | $0.783 \pm 0.007$ | $0.775 \pm 0.028$ |

**Table 1:** Results of 50 simulations with Gaussian noise added to the projections. The parameter values are $A = 2$, $\omega = 40$, $\psi = \pi/4 \approx 0.785$.

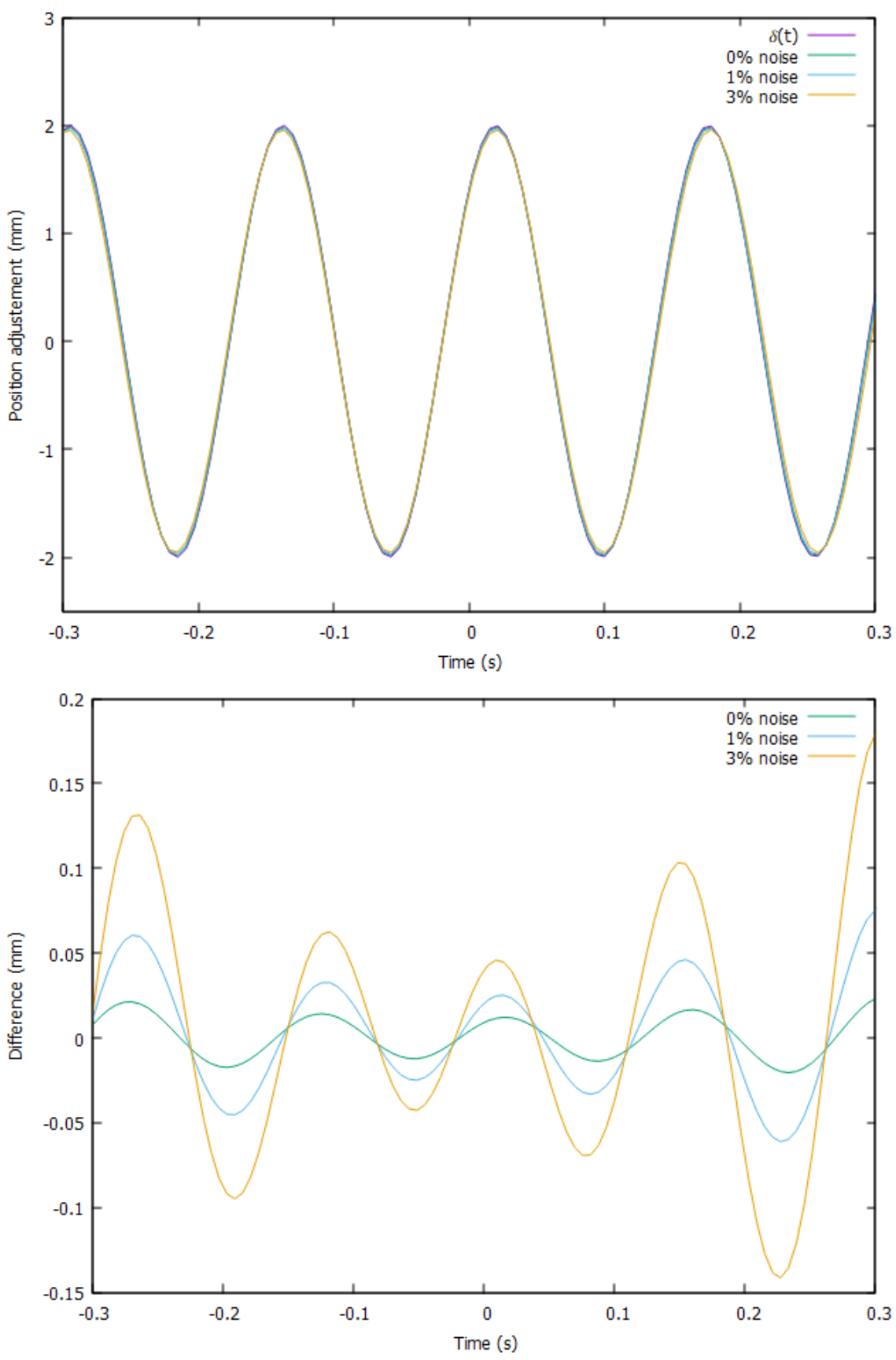

**Figure 4:** Estimation of the motion variation $\delta(t) = A\sin(\omega t + \psi)$ with 0%, 1% and 3% Gaussian noise. *Top:* the function $\delta(t)$ and its estimates. *Bottom:* difference between the theoretical value of $\delta(t)$ and its estimates.

The initialization of the Gauss-Newton algorithm is set close to the real solution. Often, the convergence of the Gauss-Newton algorithm is local. The parameters are therefore initially set to $A = 3$, $\omega = 42$, $\psi = \pi/8$, $c_{0,0} = 0$. The results are given in the table 1. Noticing that $\pi/4 \approx 0.785$, we see that the estimates are quite good. In the Fig. 4, we show the estimation of the function $\delta(t)$ and the difference between the theoretical value of $\delta(t)$ and its estimates. The differences are respectively up to 25$\mu$m and 80$\mu$m for the simulations with 0% and 1% noise. Except for $c_{0,0}$, all the parameters can be roughly estimated a priori using external tools. With different set of initial solution, we could see that the most sensitive parameter is $\omega$. It's worthwhile noticing that the solution is not unique. Indeed, it depends on the definition interval of $\psi$. We have $A\sin(\omega t + \psi + k\pi) = (-1)^k A\sin(\omega t + \psi)$ with $k \in \mathbb{Z}$.

## 4 Conclusion

We have adjusted the Helgason-Ludwig consistency conditions expressed in the parallel linogram geometry to a time-dependent self-calibration problem. We have proven that we cannot estimate the mean velocity $v_0$ using the DCCs. However, we have shown that we can estimate $v_0$ using a priori information on center of mass of a calibration object. We have modeled a non uniform motion with a periodical function and proposed a method to estimate the motion based on 0-order DCC. As in [5], we experimented that higher order moments ($n \geq 1$, cf Eq. (27)) do not provide significant improvements. Moreover, the results can easily be extended to the 3D using multiple 2D plane as done by Nguyen *et al.* [7]. The redundancy in the data will most likely help to get more robust estimates. Nonetheless, the results obtained in the simulations are already good enough for our needs.

# VAE constrained MR guided PET reconstruction

Valentin Gautier[1], Claude Comtat[2], Florent Sureau[2], Alexandre Bousse[3], Louise Friot–Giroux[1], Voichita Maxim[1], and Bruno Sixou[1]

[1]Université de Lyon, INSA-Lyon, UCBL 1, UJM-Saint Etienne, CNRS, Inserm, CREATIS UMR 5220, U1294, F-69621, LYON, France.
[2]BioMaps, Université Paris-Saclay, CEA, CNRS, Inserm, SHFJ, 91401 Orsay, France.
[3]LaTIM, INSERM U1101, Université de Bretagne Occidentale, 29238 Brest, France

**Abstract** In this work, we investigate a deep learning PET-MR joint reconstruction method based on the ADMM algorithm. The a priori information to regularize the inverse problem is obtained with a VAE trained with high-quality images. Adaptive choice of the Lagrangian parameter ensures good convergence properties of the method. The proposed approach is tested on simple cases. It outperforms the classical MLEM for high noise levels.

## 1 Introduction

Dual imaging positron emission tomography (PET)-magnetic resonance imaging (MRI) scanners have been investigated recently as an imaging modality offering both functional and anatomical information. With this hybrid imaging technique, the PET and MRI data are simultaneously acquired. Several works have investigated the synergistic reconstruction of PET and MRI data in order to improve the reconstruction results obtained with conventional independent reconstruction approaches. The idea is to exploit the common features and similarities between PET and MR images.

Variational methods have been investigated by Ehrhardt, Thielemans, Pizarro, et al. [1]. Structural similarity, joint sparsity and alignment of the gradients of the two images is promoted through a Total Variation (TV) prior. Improvements have been proposed to overcome cross-talk artifacts [2]. Recently, Mehranian, Belzunce, Prieto, et al. [3] have proposed a non-convex joint sparsity prior generalizing the joint TV to promote common boundaries while preserving modality-unique features. Their reconstruction framework is based on the augmented Lagrangian method with a scaling to take into account the dependence of the prior on the magnitude of the PET and MR images gradients. The performance of the algorithm is highly dependent on the PET-MR initialization and on the selection rules of hyper-parameters.

Deep learning methods have opened a new area of research for medical image reconstruction and they have allowed substantial improvements over state-of-the-art conventional methods in terms both accuracy and execution time. More specifically, for synergistic PET-MR reconstruction, deep learning approaches have been studied to overcome the limitations of these variational methods [4]. These methods are based on unrolling techniques that leverage the classical iterative algorithms used for image reconstruction. The proposed synergistic PET-MR reconstruction algorithm interconnects two networks to guide one modality with the other. Generative modeling has also been used for PET image denoising with MR images [5].

In this work we propose the deep latent reconstruction method (DLR) for synergistic PET-MR images which leverages the ADMM iterative method of Mehranian, Belzunce, Prieto, et al. The Total Variation regularization is replaced by a learned constraint obtained with a Variational Auto-Encoder (VAE) [6] trained with high-quality PET-MR images. The latent variable is used to represent the common information shared by the two imaging modalities. The proposed algorithm could be used for synergistic reconstruction although we focus on MR guided PET reconstruction in this paper.

The work is structured as follows. In the first section, we summarize the ADMM algorithm and we present the VAE we used as well as our dataset. We then present and discuss preliminary results showing that the proposed method outperforms the classical MLEM algorithm, showing promising for guided reconstruction as well as mutlimodal reconstruction.

## 2 Materials and Methods

### 2.1 Forward imaging models and ADMM approach for synergistic reconstruction

We denote $M$ the number of PET lines of response and $N$ the number of image voxels. The unknown vector $x_{\text{pet}} \in \mathbb{R}^N$ is the radioactive tracer distribution and $P \in \mathbb{R}^{M\times N}$ the detection probability matrix. The forward model considers the data $y_{\text{pet}} \in \mathbb{R}^M$ as random independent Poisson random variables with expected counts $\tilde{y}_{\text{pet}} = Px_{\text{pet}} + r + s$ where $r$ and $s$ the expected number of randoms and scatters. The PET data fidelity is given by the negative Poisson log-likelihood which reads:

$$D_{\text{pet}}(y_{\text{pet}}, x_{\text{pet}}) = \sum_{i=1}^{M} \left( [\tilde{y}_{\text{pet}}]_i - [y_{\text{pet}}]_i \log([\tilde{y}_{\text{pet}}]_i) + \log([y_{\text{pet}}]_i!) \right) . \tag{1}$$

The Magnetic Resonance (MR) imaging model is $\tilde{y}_{\text{mr}} = Ex_{\text{mr}}$ where $x_{\text{mr}} \in \mathbb{R}^N$ is the MR image, $\tilde{y}_{\text{mr}}$ and $y_{\text{mr}} \in \mathbb{R}^{M_v}$ are the expected k-space data and the measurements respectively, $E \in \mathbb{R}^{M_v\times N}$ is the Fourier encoding matrix consisting of the product of the discrete Fourier transform $F$ and subsampling k-space operator, $M_v$ and $N$ are respectively the number of k-space samples and MR image voxels. In the following, we will use fully sampled spectra. We also assume the measurements are corrupted by Gaussian noise, and we define the

MR data fidelity term :

$$D_{\mathrm{mr}}(y_{\mathrm{mr}}, x_{\mathrm{mr}}) = \frac{1}{2}\|Ex_{\mathrm{mr}} - y_{\mathrm{mr}}\|_2^2. \tag{2}$$

Our new method of synergistic reconstruction is inspired from the method of Mehranian, Belzunce, Prieto, et al. [3] and from the approach investigated by Xie, Li, Zhang, et al. [5] for guided mono-modal reconstruction. It is well-known that variational autoencoders (VAEs) allow to reliably represent complex data in a lower dimensional space. Our hypothesis in this work is that by training a VAE to represent both modalities with a single latent variable, it should learn more about the mutual information between them. Thus, we assume that the images are the output of the decoder part of a VAE with a single input $z$,

$$(x_{\mathrm{pet}}, x_{\mathrm{mr}}) = \mathrm{Decoder}(z) \tag{3}$$

where Decoder is the decoder part of the VAE and the latent variable $z$ is used for the low-dimensional representation of the PET and MR images. Our aim is to find the PET-MR solution $(\hat{x}_{\mathrm{pet}}, \hat{x}_{\mathrm{rm}}) \in \mathbb{R}^N \times \mathbb{R}^N$ of the following minimization problem:

$$\begin{aligned}(\hat{x}_{\mathrm{pet}}, \hat{x}_{\mathrm{mr}}, \hat{z}) = \underset{x_{\mathrm{pet}}, x_{\mathrm{mr}}, z}{\arg\min}\ & D_{\mathrm{pet}}(y_{\mathrm{pet}}, x_{\mathrm{pet}}) + D_{mr}(y_{\mathrm{mr}}, x_{\mathrm{mr}}) \\ \text{s.t.}\quad & (x_{\mathrm{pet}}, x_{\mathrm{mr}}) = \mathrm{Decoder}(z)\end{aligned} \tag{4}$$

We apply the augmented Lagrangian method to the constrained optimization problem. We used the ADMM [7] algorithm to solve (4). Denoting $\mu$ the Lagrange multiplier and $\rho = (\rho_{\mathrm{mr}}, \rho_{\mathrm{pet}})$ the Lagrangian hyperparameter, the ADMM iterations can be rewritten:

$$\begin{aligned} x_{\mathrm{pet}}^{n+1} &= \underset{x_{\mathrm{pet}}}{\arg\min} D_{\mathrm{pet}}(y_{\mathrm{pet}}, x_{\mathrm{pet}}) && (5)\\ &\quad + \frac{\rho_{\mathrm{pet}}}{2}\|x_{\mathrm{pet}} - \mathrm{Decoder}(z^n)_{\mathrm{pet}} + \mu_{\mathrm{pet}}^n\|^2 \\ x_{\mathrm{mr}}^{n+1} &= \underset{x_{\mathrm{mr}}}{\arg\min} D_{\mathrm{mr}}(y_{\mathrm{mr}}, x_{\mathrm{mr}}) && (6)\\ &\quad + \frac{\rho_{\mathrm{mr}}}{2}\|x_{\mathrm{mr}} - \mathrm{Decoder}(z^n)_{\mathrm{mr}} + \mu_{\mathrm{mr}}^n\|^2 \\ z^{n+1} &= \underset{z}{\arg\min} \|\mathrm{Decoder}(z) - (x^{n+1} + \mu^n)\|^2 && (7)\\ \mu^{n+1} &= \mu^n + x^{n+1} - \mathrm{Decoder}(z^{n+1}) && (8)\end{aligned}$$

where we denoted $x^n = (x_{\mathrm{pet}}^n, x_{\mathrm{mr}}^n)$, $\mu^n = (\mu_{\mathrm{pet}}^n, \mu_{\mathrm{mr}}^n)$ and $\mathrm{Decoder}(z^n) = (\mathrm{Decoder}(z^n)_{\mathrm{pet}}, \mathrm{Decoder}(z^n)_{\mathrm{mr}})$ . The minimization problem of Eq. (5) is solved with optimization transfer and convex surrogate function similar to the one of the classical MLEM algorithm [5]. We used the following update formula for $x_{\mathrm{pet}}^{n+1}$ at each pixel $j$:

$$\begin{aligned}\left[x_{\mathrm{pet}}^{n+1}\right]_j = \frac{1}{2}\Bigg( & [\mathrm{Decoder}(z^n)_{\mathrm{pet}}]_j - [\mu_{\mathrm{pet}}^n]_j - \frac{p_j}{\rho_{\mathrm{pet}}} \\ & + \sqrt{\left(\mathrm{Decoder}(z^n)_{\mathrm{pet}\,j} - [\mu_{\mathrm{pet}}^n]_j - \frac{p_j}{\rho_{\mathrm{pet}}}\right)^2 + \frac{4p_j[x_{\mathrm{pet,em}}^{n+1}]_j}{\rho_{\mathrm{pet}}}}\Bigg)\end{aligned} \tag{9}$$

where $p_j = \sum_i P_{i,j}$ and $x_{\mathrm{pet,em}}^{n+1}$ is obtained by doing one MLEM step:

$$[x_{\mathrm{pet,em}}^{n+1}]_j = \frac{[x_{\mathrm{pet}}^n]_j}{p_j}\sum_i P_{ij}\frac{[y_{\mathrm{pet}}]_i}{[Px_{\mathrm{pet}}^n]_i + [r]_i + [s]_i} \tag{10}$$

The update of the latent variable in Eq. (7) is obtained with a simple gradient descent with a gradient step $S$ which is easily performed using Tensorflow's Gradient API [8]. This step size S has to be tuned depending on the input data.
The scaling of the Decoder output is also an important issue. It is common practice to train the decoder on normalized data but the image that we are trying to reconstruct is not necessarily normalized. We thus chose to rescale the output of the decoder by using a scaling factor equal to the mean of the current reconstructed image.
The additional ADMM penalty parameter $\rho$ has a strong influence on the convergence rate and is often chosen empirically based on some validation data. In this work, we implemented the adaptive update scheme proposed recently in [9]. Its principle is to balance the relative primal and dual residuals while taking into account the scaling properties of the ADMM problem. We have also implemented the stopping criterion proposed in the same paper with $\varepsilon = 0.02$.

## 2.2 Dataset

The data used for training the VAE and testing our method consists in 840 co-registered 2D brain [$^{18}$F]FDG PET images (20 minutes acquisition) and T1-weighted MR images extracted from 44 acquisitions on a clinical hybrid PET/MR scanner (Signa PET/MR, GE Healthcare) of patients with dementia or epilepsy. These images are of shape $256 \times 256$ and are considered our references. The data used for the reconstructions are generated from the images $x_{pet}$ according to the following:

$$y_{pet} = \mathrm{Poisson}\left(\frac{\alpha}{\|x_{\mathrm{pet}}\|_1}(Px_{\mathrm{pet}} + r + s)\right) \tag{11}$$

where $\alpha$ is a factor used to control the noise level . The signal to noise ratio is lower for lower values of $\alpha$. We add $r$ (random) and $s$ (scatter) events so that they correspond to 1% of the total number of observed events . The resulting sinograms are of shape 256 (number of bins) $\times$ 60 (number of angles) and are then used as the input data for our reconstruction method.

## 2.3 VAE structure and training

In this work, we use VAEs to generate PET and MR images. A VAE is a latent space generative model based on the variational Bayesian inference originally proposed by Kingma and Welling [6]. The structure of our VAE is displayed in figure 1. Each convolution and transposed convolution is described with the format (number of channels, filter size, stride) and the dense layers are described by their number of neurons.

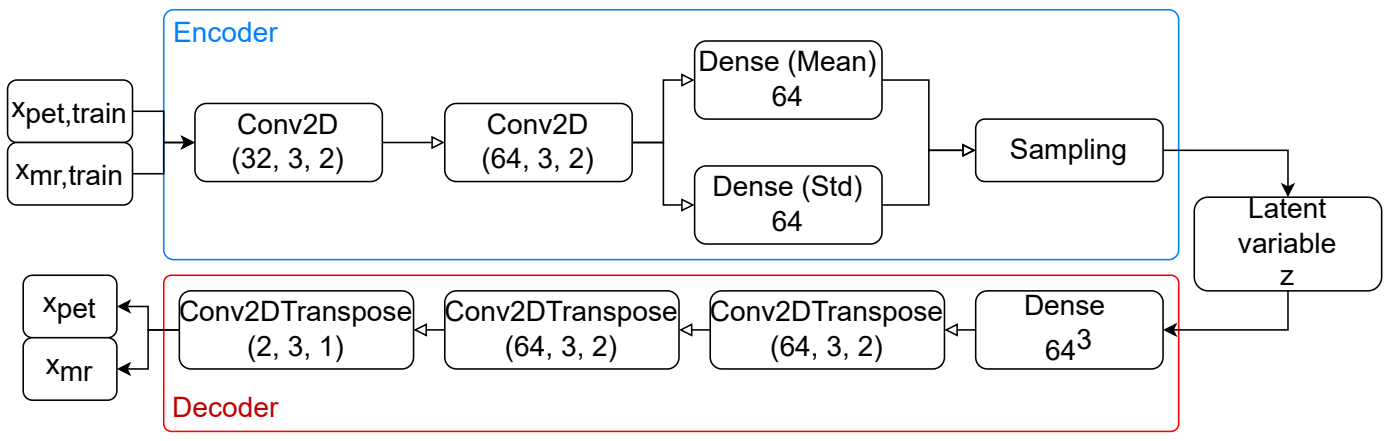

**Figure 1:** Architecture of the VAE.

The Sampling layer is used to perform the reparametrization trick [6] and takes a mean vector and a standard deviation vector to sample a latent variable $z$. We use a multichannel input CNN, which treats PET/MR images as two separate input channels. The VAE we considered in our work is a $\beta$-VAE [10] which allows us to balance the training of the model between latent space regularization and data fidelity. On top of that, we use a $L_2$ loss for the data fidelity term with a weighting parameter to balance the two modalities' contributions to the loss. The training was implemented using the open-source library Keras 2.2.5 with Tensorflow backbone and performed on an NVIDIA RTX A2000 mobile. The network is trained for 500 epochs using the Adam optimizer with a learning rate of $10^{-3}$ and a batch size of 32. The images from our dataset were used as the high-quality references and the VAE was trained on them. The dataset was split into 3 parts: one for training, one for validation (20% of the data) and one for testing (10% of the data).

### 2.4 Experiments

For preliminary tests, we have fixed the MR image to the reference image and only reconstructed the PET image. We then compare the reconstruction results to the ones obtained with the classical MLEM algorithm [11].
We initialize the algorithm with the 10th iteration of MLEM and initialize the latent variable by using the encoder part of the VAE on the initial PET image and the reference MR image. The forward and backward projections are handled by the ASTRA toolbox [12] with a parallel geometry. The test reconstructions were performed on 10 slices from the test set.

## 3 Results

Figure 2 shows a slice from the test set reconstructed with our approach together with the ground truth image and the reconstruction given by the MLEM algorithm . Qualitatively, the DLR approach outperforms MLEM on these very deteriorated data.
Figure 3 shows the evolution of the mean squared error (MSE) and of the constraint $||\text{Decoder}(z^n)_{\text{pet}} - x^n_{\text{pet}}||$ as a function of the ADMM iterations for the slice shown in figure 1 for the proposed method during one reconstruction. We also show the best NRMSE obtained by MLEM run for 30 iterations for comparison: with the adaptive update of the

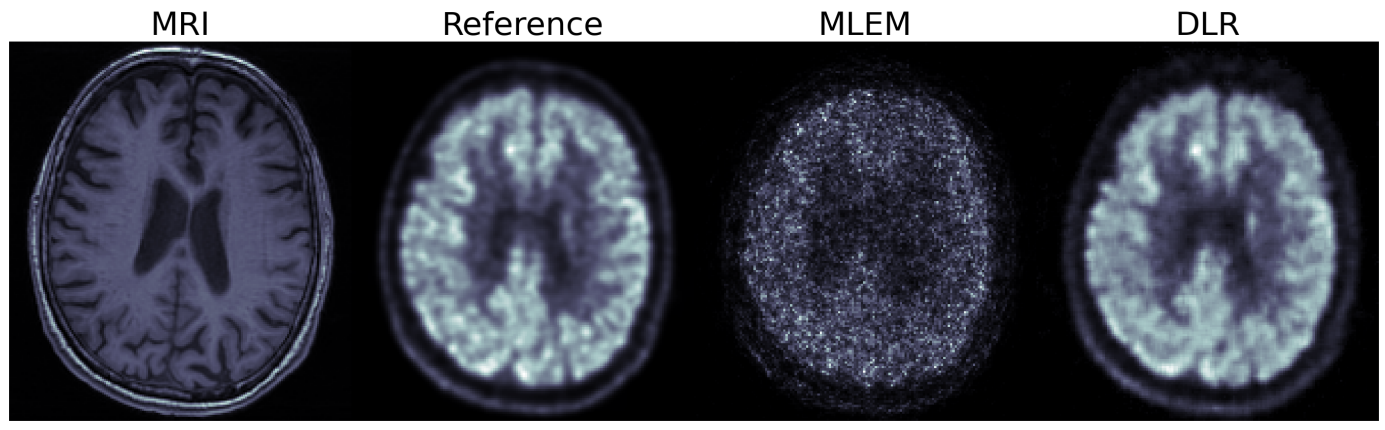

**Figure 2:** Ground truth image and reconstructed images obtained with MLEM and the deep latent reconstruction approach for $\alpha = 10^5$.

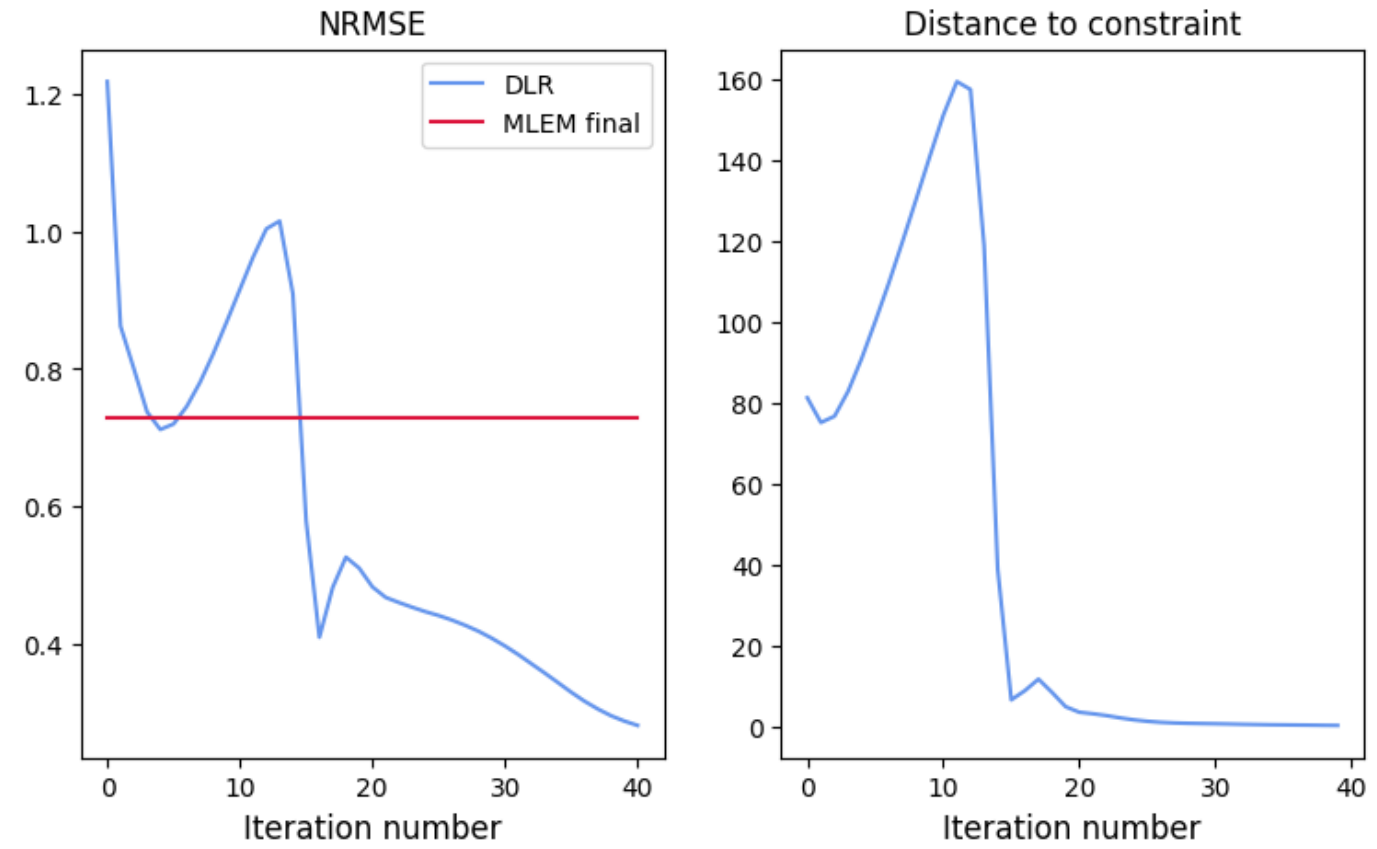

**Figure 3:** Evolution of the NRMSE, SSIM, data fidelity terms and constraint as a function of iterations for $\alpha = 10^5$. The MLEM NRMSE is displayed for comparison.

Lagrangian parameter $\rho$, the behavior of the error metrics is highly nonlinear. The regularizing effect of the constraint is obtained after a few iterations when $\rho$ increases significantly. The large decrease of the constraint corresponding to the VAE is concomitant with the MSE decrease. It should be noted that simultaneously the data fidelity term increases which means that once a good latent vairable has been found, the improvement is due to the constraint. The projection on the image manifold learned with the autoencoder is thus efficient to reduce the noise and the artifacts on the reconstructed image.
The quality of the reconstructions was evaluated quantitatively using the the normalised root mean squared error (NRMSE) and SSIM for several noise levels. The results obtained for several Poisson noise levels are displayed in table 1 together with the value of the gradient step $S$ chosen empirically. The deep latent reconstruction method clearly outperforms the MLEM approach for lower signal to noise ratio.

## 4 Discussion

In this work, we have presented preliminary results obtained with a new deep latent reconstruction approach. The VAE constrained reconstruction framework achieves better performance compared with MLEM for the various noise levels investigated. The main advantage of the method is that the regularizing effect is learned and not based on penality terms like TV or joint TV regularization. By using a VAE, we get a latent variable that sums up the mutual information between the two modalities. With enough data it should be possible to get a latent space from which we can generate any PET image. Moreover, the approach is based on a limited number of hyper-parameters thanks to the automatic update method adopted for the ADMM algorithm. It should be noted that the known convergence properties of the classical ADMM algorithm are not guaranteed since the optimization method uses both a convex surrogate and a non linear constraint.
For the presented MR-aided reconstruction, several aspects will be further investigated. First, we would like to get more data and use more realistic simulations to evaluate the method in a clinically accurate context.
One current limitation of the method is that the latent variable $z$ update from Eq.7 is based on a simple gradient descent with a rough estimate of the gradient step. This sometimes leads to updates outside of the known latent space and may cause the algorithm's divergence. Additional constraints could be studied to improve the search of the optimal latent variable. The simple VAE used could also be improved to generate less blurred images. The VAE that was used here is known to produce blurry images, which is good enough to handle PET images but not enough for MR images. With some improvement, we could also handle MR reconstruction and improve on the PET one. Variants like VAE GAN, InfoVAE [13] or even diffusion models [14] exist in the literature and could lead to improving the quality of the generated images as well as a better use of the mutual information. In the end, the proposed framework is highly flexible and each of its components can be improved individually.

| $\alpha$ | S | MLEM | | DLR | |
|---|---|---|---|---|---|
| | | NRMSE | SSIM | NRMSE | SSIM |
| $10^5$ | $10^{-2}$ | 1.47 | 0.79 | 0.88 | 0.83 |
| $2\times10^5$ | $5\times10^{-3}$ | 1.1 | 0.82 | 0.7 | 0.86 |
| $5\times10^5$ | $10^{-3}$ | 0.68 | 0.86 | 0.65 | 0.87 |

**Table 1:** Comparison of the mean for NRMSE and SSIM for the MLEM method and the proposed DLR method.

## 5 Conclusion

Our aim in this paper is to improve the fusion of the complementary information in PET/MR images. We have investigated a network-constrainted image reconstruction method where a pre-trained multi-channel input VAE trained with high-quality images is used to represent feasible PET and MR images. We show that, using an MR image for guidance, we can find a suitable latent variable to represent our denoised PET data. In future work, we will consider the joint reconstruction of PET and MR images and compare our results with other deep learning based reconstructions approaches such as unrolling.

## 6 Acknowledgment

We acknowledge financial support form the French National Research Agency (ANR) under grant ANR-20-CE45-0020 (ANR MULTIRECON).

# Evaluating Spectral Performance for Quantitative Contrast-Enhanced Breast CT with a GaAs Photon-Counting Detector: A Simulation Approach

Bahaa Ghammraoui, Muhammad Ghani, Andreu Badal, and Stephen J. Glick

Division of Imaging, Diagnostics and Software Reliability (DIDSR), Office of Science and Engineering Laboratories (OSEL), FDA, Silver Spring, MD

**Abstract** Quantitative contrast-enhanced breast computed tomography (CT) has the potential to enhance the diagnosis and management of breast cancer. The traditional methods involve dual-exposure images with different incident spectra to obtain two spectrally separated measurements, which can be used for material discrimination, comes at the expense of increased patient dose and susceptibility to motion artifacts. An alternative approach, using Photon Counting Detectors (PCD), allows for acquisition of multiple energy levels in a single exposure, reducing these issues. GaAs is a particularly promising material for breast PCD-CT due to its high quantum efficiencies and reduction of fluorescence X-rays escaping the pixel within the breast imaging energy range.

This simulation study evaluated the spectral performance of a GaAs photon-counting detector (PCD) for quantitative iodine contrast-enhanced breast CT. Utilizing both projection-based and image-based material decomposition methods, the study produced material-specific images of the breast and used the iodine component images to estimate iodine intake. The accuracy and precision of the method for estimating iodine concentration in breast CT images were assessed for different material decomposition methods, incident spectra, and mean glandular dose (MGD).

The results showed that the GaAs PCD had comparable performance to an ideal PCD in terms of Root Mean Squared Error (RMSE), precision, and accuracy of estimating the iodine intake in the breast. Furthermore, the results demonstrated the effectiveness of both material decomposition methods (projection-and image-based) in making accurate and precise iodine concentration predictions using a GaAs-based photon counting breast CT system, with better performance when applying the projection-based material decomposition approach. The study highlights the potential of GaAs-based photon counting breast CT systems as a viable alternative to traditional imaging methods in terms of material decomposition and iodine concentration estimation.

## 1 Introduction

Quantitative contrast-enhanced breast computed tomography (CT) is a diagnostic technique that uses a contrast agent, such as iodine, to examine the breast tissue for signs of cancer. The contrast agent is injected into the patient's bloodstream and allows for the visualization of blood vessels in the breast tissue. The traditional methods involve dual-exposure images with different incident spectra to obtain two spectrally separated measurements, which can be used for material discrimination. This approach comes at the expense of increased patient dose and susceptibility to motion artifacts. An alternative approach, using Photon Counting Detectors (PCD), allows for acquisition of multiple energy levels in a single exposure, reducing these issues. Photon counting CT detectors (PCD-CT) are a rapidly developing technology in medical imaging. Many major CT companies have recently been developing prototype PCD-CT systems, and the FDA has cleared the first CT system incorporating a PCD. These systems have several advantages over traditional CT systems, including reduced electronic noise, improved resolution, and reduced radiation dose and scanning time. The current development of PCD spectral imaging systems is driven by recent technological advancements, making flat-panel CT PCDs dedicated to breast imaging now possible. Several semiconductor-based X-ray PCD technologies have been developed, including those using silicon (Si), cadmium telluride (CdTe), and gallium arsenide (GaAs). GaAs is a particularly promising material for PCD-CT, as it can achieve high quantum efficiencies with a thickness of 0.5mm, and it is especially useful within the breast imaging energy range (12 to 55 KeV) as its characteristic K-edges lie below the relevant energies (9, 10, and 12 keV), thus reducing the probability of fluorescence X-rays escaping the pixel. This study evaluates the spectral performance of a GaAs photon-counting detector for quantitative breast CT using a simulation approach.

## 2 Methods

In this study, the objective was to evaluate the effects of various factors on the accuracy and precision of the quantitative methods for iodinated contrast-enhanced dedicated breast photon counting CT when using GaAs detector. A numerical breast phantom composed of a 50/50 ratio of glandular and adipose tissue in a cylindrical shape (10 cm diameter) was used in this study. The phantom included iodine targets at concentrations of 0.5, 1.0, and 2.0 mg/cm$^3$, chosen to cover the clinical range of iodine concentrations. Three targets were placed at different locations for each concentration to assess the impact of beam hardening on iodine quantification, resulting in a total of nine targets (see figure 1). The phantom also included a target of 100% glandular tissue equivalent material to test iodine separation from the background over a wide range of densities. The phantom was assumed to be mounted on a rotation stage situated in front of the collimator, with a source-object distance of 80 cm and an object-detector distance of 20 cm. Simulations were performed at different exposure levels to test the effect of dose on accuracy and precision. TIGRE CT open-source software and the publicly available Photon Counting Toolkit (PcTK) [1] were used to generate material-based sinograms and noisy energy-dependent projection data. The detector simulation used 100 $\mu$m-pixel size and 500 $\mu$m-thickness of GaAs with a density of 5.32 g/cm3. The charge cloud size and electronic noise simulation parameters were set at $r$= 11 $\mu$m and $\sigma$ = 2.1 keV, respectively. These values were chosen from a separate validation study and were found to give the best agreement between the simulation software and a prototype GaAs PCD [1]. A PCD with two energy thresholds was simulated, and the low-energy and high-energy thresholds were set at 15 and 32 keV, respectively. Varying levels of mean glandular doses, which correspond to different noise realizations, were generated by simulating a Poisson process for the number of photons recorded in each detector energy bin.

## 2.1 Iodine concentration estimation and image reconstruction

In this work, we tested two frameworks for material decomposition, which were presented in detail elsewhere [2, 3]. The process is briefly described here:

### 2.1.1 Image-Based Decomposition

First, the low and high energy-dependent phantom measurements ($y_l$, $y_h$) at each pixel are converted or scaled to the original linear attenuation coefficient measurement:

$$\mu_i = -\log\left(\frac{y_i}{y_i^0}\right), \tag{1}$$

where $i$ refers to low or high energy levels and $y_i^0$ corresponds to the open beam measurements.

Second, the reconstructed images undergo a typical water precorrection to reduce the cupping artifacts effect resulting from beam hardening. This is done by reconstructing images of a uniform water phantom and normalizing it to have a maximum value of 1. Then, the projection reconstructed images $\mu_i^r$ of the test phantom at each pixel are rescaled using the following equation:

$$\mu_i = \frac{\mu_i^r}{I_i^w}, \tag{2}$$

where $I_i^w$ corresponds to the normalized reconstructed water image.

Finally, the rescaled images are decomposed into three materials: adipose, glandular, and iodine, to determine the local concentration or volumetric portion $t_i$ of each material:

$$\begin{bmatrix}\mu_l\\ \mu_h\end{bmatrix} = \begin{bmatrix}\mu_l^{ad} & \mu_l^{gd} & \mu_l^{id}\\ \mu_h^{ad} & \mu_h^{gd} & \mu_h^{id}\end{bmatrix} \times \begin{bmatrix}t_{ad}\\ t_{gd}\\ t_{id}\end{bmatrix} \tag{3}$$

where $l$ and $h$ indicate the low and high energy bins, and the subscripts ad, gd, and id represent the three-material basis of adipose, glandular, and iodine, respectively. To solve the system of two equations with three unknowns, we applied a simple conservation of volume, the sum-to-one constraint ($t_{ad} + t_{gd} = 1$). Therefore, we can rewrite Equation 3 as follows:

$$\begin{bmatrix}\mu_l - \mu_l^{ad}\\ \mu_h - \mu_h^{ad}\end{bmatrix} = \begin{bmatrix}\mu_l^{gd} - \mu_l^{ad} & \mu_l^{id}\\ \mu_h^{gd} - \mu_h^{ad} & \mu_h^{id}\end{bmatrix} \times \begin{bmatrix}t_{gd}\\ t_{id}\end{bmatrix} \tag{4}$$

The basis material matrix elements can be empirically determined through a calibration procedure using known concentrations of the basis materials of interest, adipose, iodine, and glandular. The scan of the calibration phantom should be performed under the same acquisition parameters as those in the test scan, including kV, threshold configuration, and reconstruction kernel. The empirically determined material matrix is then used to estimate each basis material concentration within a mixture of materials.

### 2.1.2 Projection-Based Decomposition

In this framework, we performed material decomposition for every pixel of the detector using only two basis materials (water and iodine) since we simulated a detector with two energy bins. Under this assumption, the PCD measurements yield the projection values:

$$M_i = -log\left(\frac{\int_0^\infty I_0(E)e^{-(\mu_w(E)t_w+\mu_{id}(E)t_{id})}B_i(E)dE}{\int_0^\infty I_0(E)B_i(E)dE}\right) \tag{5}$$

Where $B_i(E)$ is the energy bin sensitivity, $t_w$ and $t_{id}$ represent the propagation path lengths through water and iodine, respectively, and $\mu_w$ and $\mu_{id}$ are their corresponding linear attenuation coefficients. An empirical method to solve these equations was proposed by Cardinal and Fenster [3] and adopted in this study. This method consists of modeling the solution ($t_w$ and $t_{id}$) by a polynomial function (of order p = 3) of low and high energy measurements ($M_1$ and $M_2$) using a least square fitting method from a calibration set. With this model, $t_w$ and $t_{id}$can be estimated as:

$$\begin{aligned} t_w(M_1,M_2) &= \sum_{m=0}^{P} a_{mn}^{w} M_1^m M_2^n \\ t_{id}(M_1,M_2) &= \sum_{m=0}^{P} a_{mn}^{id} M_1^m M_2^n \end{aligned} \tag{6}$$

The calibration measurements were acquired using a set of water and iodine solutions. The water thicknesses ranged between 0 and 100 mm with a 10 mm step, while iodine concentrations ranged between 0 to 5 mg/cm$^3$ with 1 mg/cm$^3$ step. This led to two material-based sinograms, which were followed by standard filtered back projection algorithm with unapodized ramp filter, allowing for water- and iodine-based images of the phantom ($g_w$ and $g_{id}$). Since the test phantom is made of three materials (adipose, glandular, and water), a post-processing step was needed to quantify these three materials for the two basis images. For that, the same calibration phantom and method described in Equation 4 was used, but with $g_w$ and $g_{id}$ instead of $\mu_l$ and $\mu_h$.

## 2.2 Figure of Merits

The following metrics were used to evaluate the performance of the quantitative methods under study:

- Root-mean-square (RMS) error between the estimated iodine concentration $C_i$ obtained from the dual energy

decomposition and the known values $C_i^{true}$ of the inserted iodine discs:

$$RMSE = \sqrt{\frac{\sum_{i=1}^{9}(C_i - C_i^{true})^2}{9}} \qquad (7)$$

where $C_i$ was calculated at each disk using circular regions of interest (ROIs).

- The correlation $C_r$ between the measured and known values is measured using the normalized correlation factor, calculated between $C_i$ and $C_i^{true}$ using the MATLAB built-in function 'normxcorr2'.
- Precision of the iodine estimation, as measured by the population standard deviation ($\sigma_{ci}$) across different realizations and the standard deviation between locations ($\Sigma_{ci}$). The latter is useful for assessing the effects of beam hardening and cupping artifacts on the estimated values of iodine concentration:

$$\sigma = \sqrt{\frac{\sum_{i=1}^{Nr}(C_i - C_i^{mr})^2}{Nr}}$$
$$\Sigma = \sqrt{\frac{\sum_{i=1}^{9}(C_i - C_i^{ml})^2}{9}} \qquad (8)$$

Where $C_i^{mr}$ represents the mean values of the estimated iodine concentration across different realizations and $C_i^{ml}$ represents the mean values across the three different locations of the disks with the same known concentration.

## 3 Results and Discussions

Figure 1 shows the material-based reconstructed images at 10 mGy for both material decomposition methods and simulated detectors under 55 kVp tungsten anode incident spectrum with 1 mm of added aluminum. The resulting RGB images are also displayed in Figure 2.

The results of our study are summarized in Figure 3 and indicate that the GaAs system with both material decomposition methods performs well in terms of accuracy and precision compared to an ideal PCD system. As expected, better performance was achieved with higher exposure doses and when using the projection-based material decomposition algorithm. On average, the root mean square error (RMSE) values for the ideal and GaAs simulated PCDs differed by less than 0.18 mg/ml, demonstrating that the GaAs PCD-CT system is capable of making accurate predictions.

Both material decomposition methods also performed well in terms of accuracy. At higher doses with the GaAs PCD, the minimum accuracy value Σ (variation across target locations) of the projection-based material decomposition was 0.25 mg/ml, only 0.1 lower than the minimum accuracy value of the image-based method (0.34 mg/ml). This suggests that the image-based material decomposition method performs well in minimizing beam hardening or cupping artifacts on the quantification methods.

A correlation factor of 0.98 or higher was observed in all methods. This suggests that more sophisticated methods can be applied and used to better estimate the iodine concentration in a breast. To further illustrate this point, Figure 4 presents the linear relationship between the measured iodine concentration and the true values at various locations within the phantom. A more sophisticated calibration process that takes into account the location of the targets could also lead to better performance. This finding highlights that the image-based material decomposition method may be favored due to its simplicity compared to the projection-based method, which requires extensive calibration measurements.

It is also noteworthy that the 55 kVp spectrum showed the highest performance at a fixed absorbed dose in terms of RMSE, precision, accuracy, and the correction between the measured and true values. However, it should be noted that these results may not be applicable to all cases, as different breast sizes and variations in tissue composition may lead to optimal results with different spectrums.

In summary, the GaAs PCD demonstrated comparable results to an ideal PCD in terms of Root Mean Squared Error (RMSE), precision, and accuracy. These figures of merit show the effectiveness of both material decomposition methods using a GaAs-based photon counting breast CT system in accurately and precisely estimating iodine concentration.

## 4 Conclusions

In this study, the spectral performance of a GaAs photon-counting detector was evaluated for quantitative breast CT using a simulation approach. Two different methods for producing material-specific images of the breast were tested, and the iodine component image was used to estimate iodine intake in the breast. The results of the simulation study revealed comparable performance between the GaAs detector and an ideal PCD in terms of Root Mean Squared Error (RMSE), precision, and accuracy of estimating iodine intake in the breast. The figures of merit indicate the effectiveness of both projection-based and image-based material decomposition methods in making accurate and precise predictions of iodine concentration using a GaAs-based photon counting breast CT system, with better performance when applying the projection-based material decomposition approach.

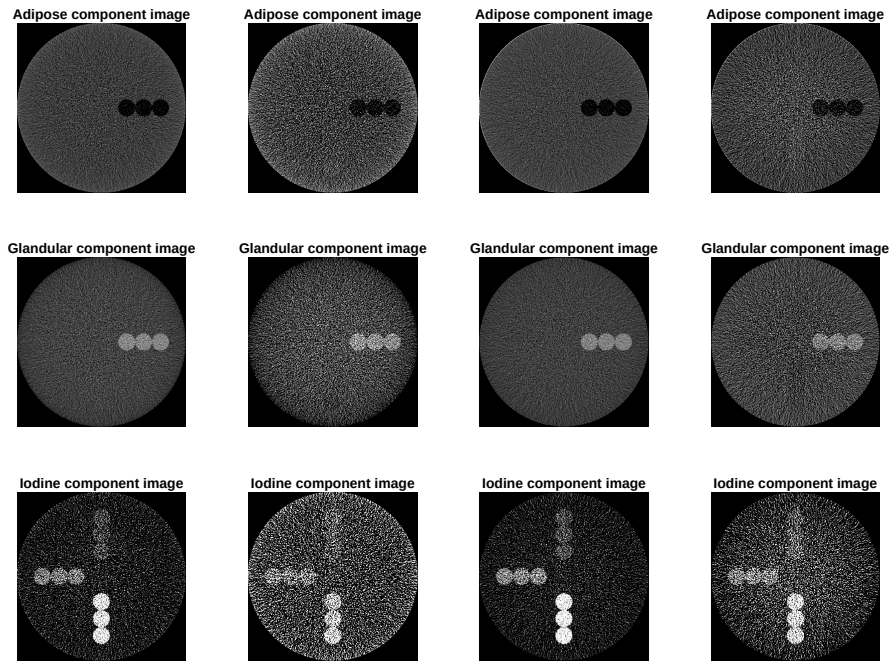

(a) (b) (c) (d)

**Figure 1:** Component-based reconstructed images from projections taken at 55 kVp tungsten anode incident spectrum and a mean glandular dose of 10 mGy. (a) and (b) depict the results from the image-based material decomposition method using an ideal PCD and a GaAs PCD, respectively. (c) and (d) depict the results from the projection-based material decomposition method using an ideal PCD and a GaAs PCD, respectively.

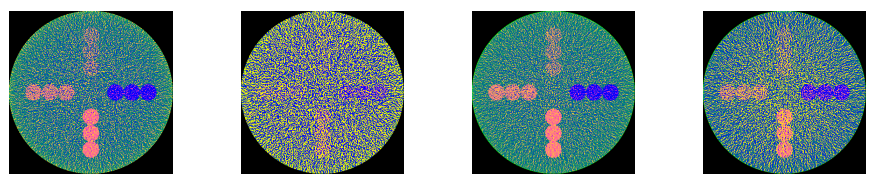

(a) (b) (c) (d)

**Figure 2:** RGB color images derived from figures in 1 mapping adipose fraction to red, glandular to green and iodine to blue.

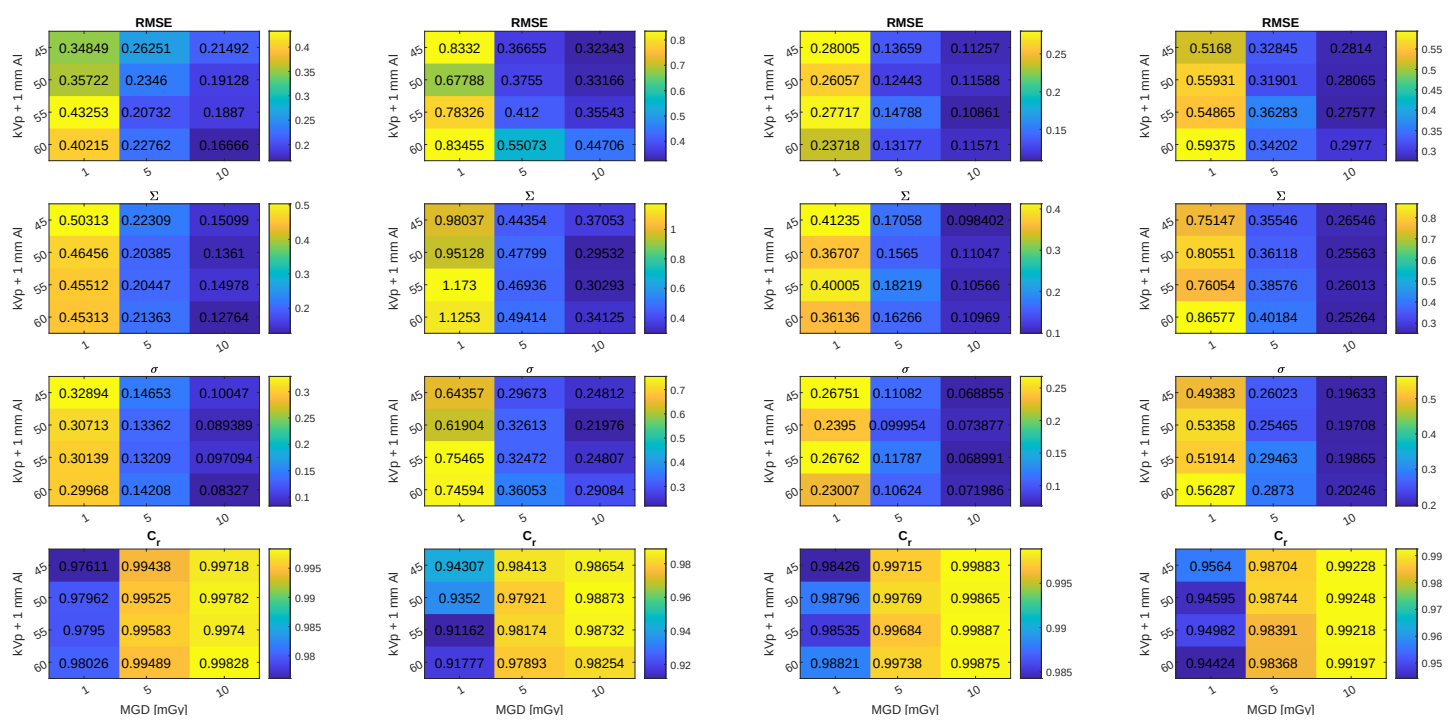

(a) (b) (c) (d)

**Figure 3:** Performance metrics for the evaluation of the imaging methods. The table shows the Root-mean-square error (RMSE), correlation factor, and precision values for the image-based and projection-based material decomposition methods using an ideal PCD and a GaAs PCD.(a) and (b) depict the results from the image-based material decomposition method using an ideal PCD and a GaAs PCD, respectively. (c) and (d) depict the results from the projection-based material decomposition method using an ideal PCD and a GaAs PCD, respectively.

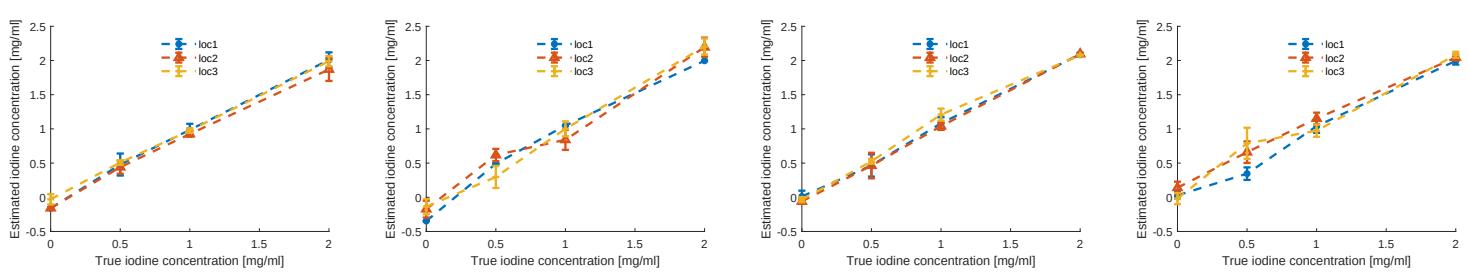

(a) (b) (c) (d)

**Figure 4:** Measured iodine concentration and the true values at various locations within the phantom. The standard variations within the circular ROIs are also presented. (a) and (b) depict the results from the image-based material decomposition method using an ideal PCD and a GaAs PCD, respectively. (c) and (d) depict the results from the projection-based material decomposition method using an ideal PCD and a GaAs PCD, respectively.

# Assessing the Mechanical Properties of Acutely Injured Lungs: A Comparative Study of Traditional and Novel Ventilation Techniques Utilizing Computed Tomographic (CT) Imaging

Jian Gao[1], Emmanuel A. Akor[2], Andrea F. Cruz[3], Bing Han[3], Junfeng Guo[4], Monica L. Hawley[5], Sarah E. Gerard[2], Jacob Herrmann[2,5], Eric A. Hoffman[4], David W. Kaczka[2,3,4,5]

[1]Carver College of Medicine, The University of Iowa, Iowa City, USA
[2]Roy J. Carver Department of Biomedical Engineering, The University of Iowa, Iowa City, USA
[3]Department of Anesthesia, The University of Iowa, Iowa City, USA
[4]Department of Radiology, The University of Iowa, Iowa City, USA
[5]OscillaVent Inc., USA

**Abstract** Acute Respiratory Distress Syndrome (ARDS) is a major cause of respiratory failure, often leading to significant morbidity and mortality. Mechanical ventilation, the primary supportive strategy, can cause ventilator-induced lung injury (VILI) due to uneven strain distribution and heterogeneous gas exchange. In this study, we employed quantitative computed tomographic (qCT) imaging, a novel and impactful advancement in medical imaging, to explore the pathophysiology of ARDS in a porcine model, and to examine its response to two ventilatory strategies: conventional mechanical ventilation (CMV) and multi-frequency ventilation (MFV). We evaluated baseline conditions and subsequent responses to oleic acid-induced lung injury over a 9-hour ventilation period, using biomarkers and static CT imaging to quantify regional and global lung textures and aeration. The MFV group demonstrated a significantly lower respiratory rate and superior oxygenation ($PaO_2$: $FiO_2$ ratio) compared to the CMV group. CT texture analysis and lung aeration assessment indicated that the MFV group had a larger proportion of normal lung tissue and normal-aerated lung volumes, and less consolidated tissue. These findings suggest that MFV could be a more effective ventilation strategy in ARDS, potentially mitigating the risk of VILI and improving oxygenation. The utilization of qCT in this study underscores the transformative impact of this advanced medical imaging technique, and its vital role in deepening our understanding of complex conditions like ARDS.

## 1 Introduction

ARDS is a life-threatening condition characterized by severe respiratory failure, noncardiogenic pulmonary edema, and arterial hypoxemia. This complex syndrome can be triggered by a range of factors including pneumonia, sepsis, severe trauma, aspiration of gastric contents, blood product transfusion, inhalation injury, or burns. Accounting for ~10% of ICU admissions and with a mortality rate of 30-40%,[1] ARDS remains a significant challenge for clinicians worldwide, indicating an urgent need for innovative therapeutic approaches and refined ventilation strategies.

Mechanical ventilation, a cornerstone of ARDS treatments, supports gas exchange and helps manage ARDS symptoms. However, the structural heterogeneity of the lungs can make achieving uniform gas exchange challenging, especially during supportive mechanical ventilation.[2] Furthermore, prolonged or inappropriate use of mechanical ventilation can inadvertently cause VILI,[3] exacerbating the already compromised pulmonary function in ARDS patients. The recently developed MFV technique by Kaczka and co-workers shows potential in addressing these issues.[4,5] MFV employs multiple simultaneous oscillatory frequencies, generating much lower regional strains and potentially reducing lung injury compared with traditional ventilation approaches. For MFV to be used in clinical trials in adult patients, technical modifications and rigorous preclinical testing are necessary.

In this context, qCT imaging, which offers detailed insights into regional lung textures and aeration, is a valuable tool for assessing lung function and pathology in ARDS. Our study utilizes this technology in a porcine model, widely adopted in ARDS research due to its similarity to human lung physiology, to investigate the pathophysiological changes in ARDS. Our aim is to evaluate the effectiveness of two different ventilation strategies, CMV and MFV, in mitigating lung injury and optimizing gas exchange during ARDS. We assess baseline lung conditions and subsequent responses to oleic acid-induced lung injury over a 9-hour ventilation period, using qCT imaging and a range of biomarkers including heart rates, blood pressures, respiratory rates, pH, blood gas levels, and the ratio of partial pressure arterial oxygen and fraction of inspired oxygen ($PaO_2$: $FiO_2$). The outcome of our research could significantly refine ARDS ventilation practices and enhance patient outcomes. By elucidating the effects of different ventilation modalities on lung function, we aim to facilitate the development of personalized, evidence-based treatment plans for ARDS patients and pave the way for novel therapeutic interventions.

## 2 Materials and Methods

Large pigs (n=12, weight of 36.7-70.0 kg) were pre-anesthetized and randomly assigned to either the CMV Group (n=6; weight of 51.1 ± 13.8 kg) or the MFV Group (n=6; weight of 49.1 ± 9.2 kg). An unpaired two-tailed t-test was performed to confirm there was no significant difference for the body weight between the two groups (p-value=0.76, with p-value<0.05 considered statistically significant). To establish baseline data, we collected lung injury-associated biomarkers (heart rates, blood pressures, respiratory rates, pH, blood gases, $PaO_2$: $FiO_2$, and serum cytokines) and performed CT scans on both groups prior to the induction of lung injury. Subsequently, oleic acid was administered to elicit a lung injury resembling ARDS. Different ventilation strategies, specifically CMV and MFV, were then employed in the respective groups. Following a 9-hour ventilation period, we collected another set of lung injury-associated biomarkers and conducted CT scans for both groups. Figure 1 illustrates the experimental setup, and Figure 2 displays the different gas flow rates used in the CMV and MFV groups. Numerous parameters, including oxygen saturation ($SpO_2$), temperature (T, °C), heart rate (HR), systemic arterial pressure (SAP), pulmonary arterial pressure (PAP), central venous pressure (CVP), respiratory rate (RR) and expiratory $CO_2$ levels, were monitored during the experiment.

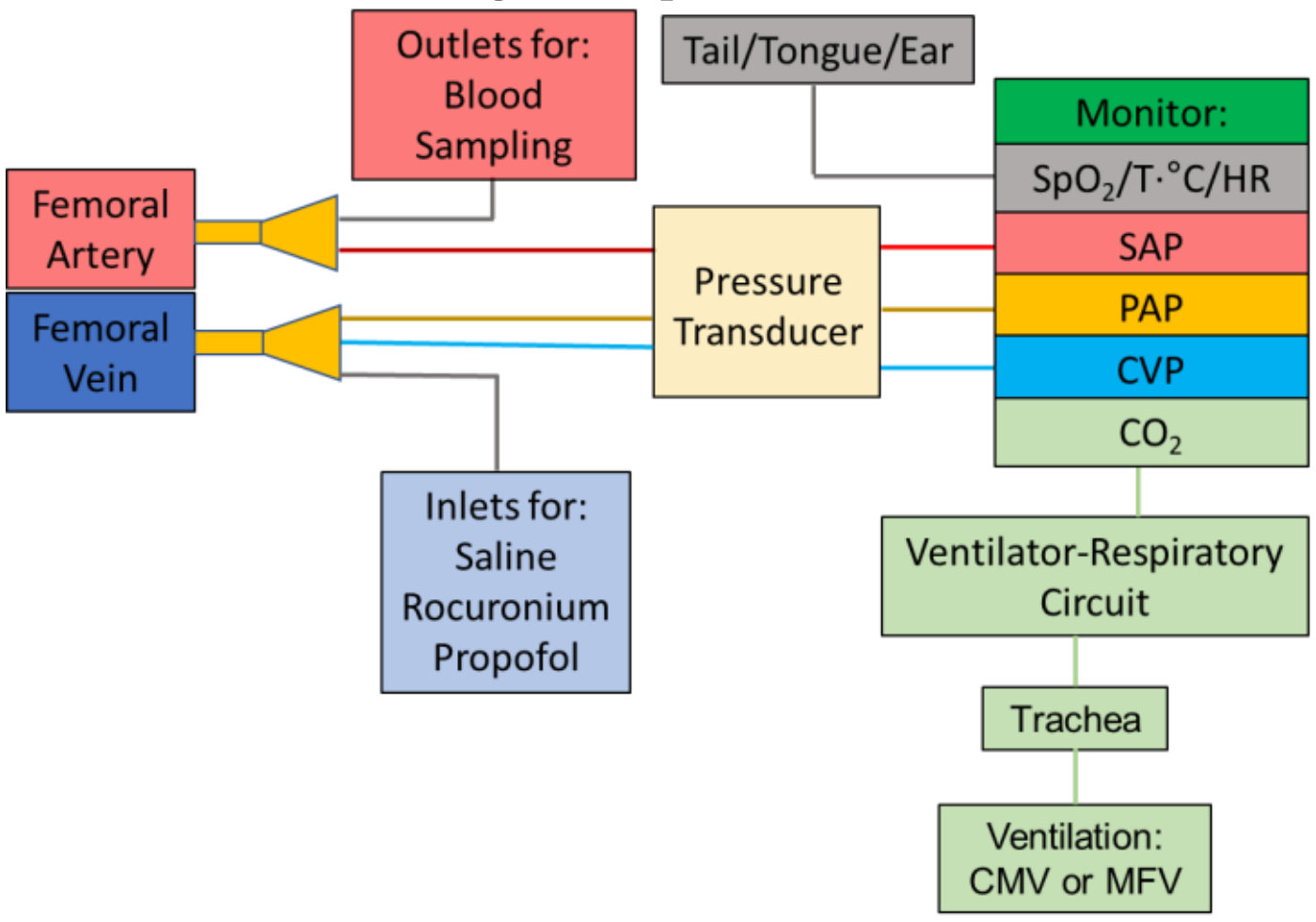

**Figure 1.** Schematic diagram of the experimental setup.

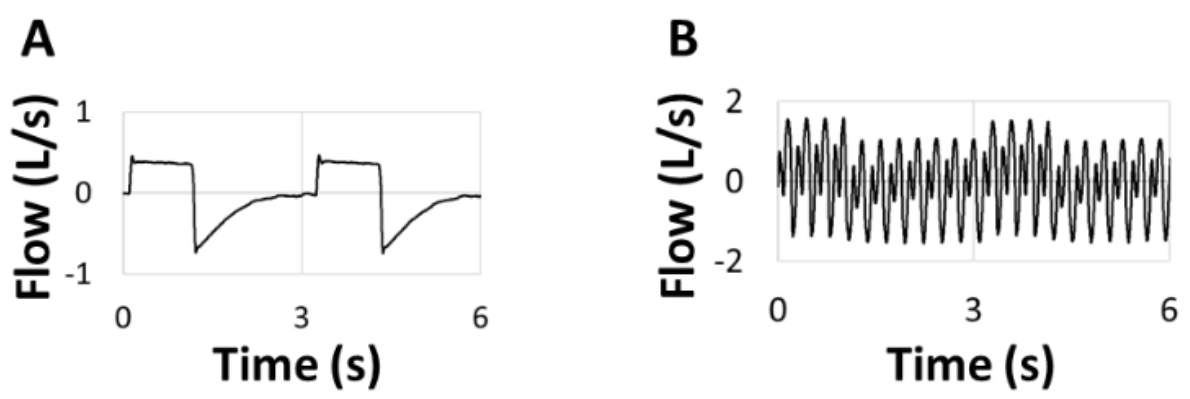

**Figure 2.** Flow rates in the CMV (Panel A) and MFV (Panel B) ventilation strategies.

We obtained 3D-CT image scans of the pigs in the CMV and MFV groups at a constant airway pressure of 30cm$H_2O$ using a Siemens SOMATOM Force scanner (Siemens Healthineers, Forchheim, Germany). These CT image scans were processed by a convolutional neural network for automatic lung parenchyma segmentation.[6] As shown in Figure 3, the mask data files obtained from the segmentation, along with the original CT image scans, were employed by the Pulmonary Analysis Software Suite (PASS)[7] to acquire qCT data and to visualize changes in volume and texture type. Volume changes were based on the total number of voxel volumes for each segmented lung. Texture analysis employed the adaptive multiple feature method, which relies on a standardized training set and labeling of normal and diseased lung regions. Depending on the intensity of the voxels that comprise the images, volumes were partitioned into hyper-aerated (less than -900 HU), normal-aerated (between -900 HU and -500 HU), poor-aerated (between -500 HU and -100 HU), and non-aerated (higher than -100 HU) compartments. The texture types evaluated included normal, ground glass, emphysema, ground glass reticular, bronchovascular, honeycomb, and consolidated. An unpaired two-tailed t-test was performed for the qCT data comparison between the CMV and MFV groups.

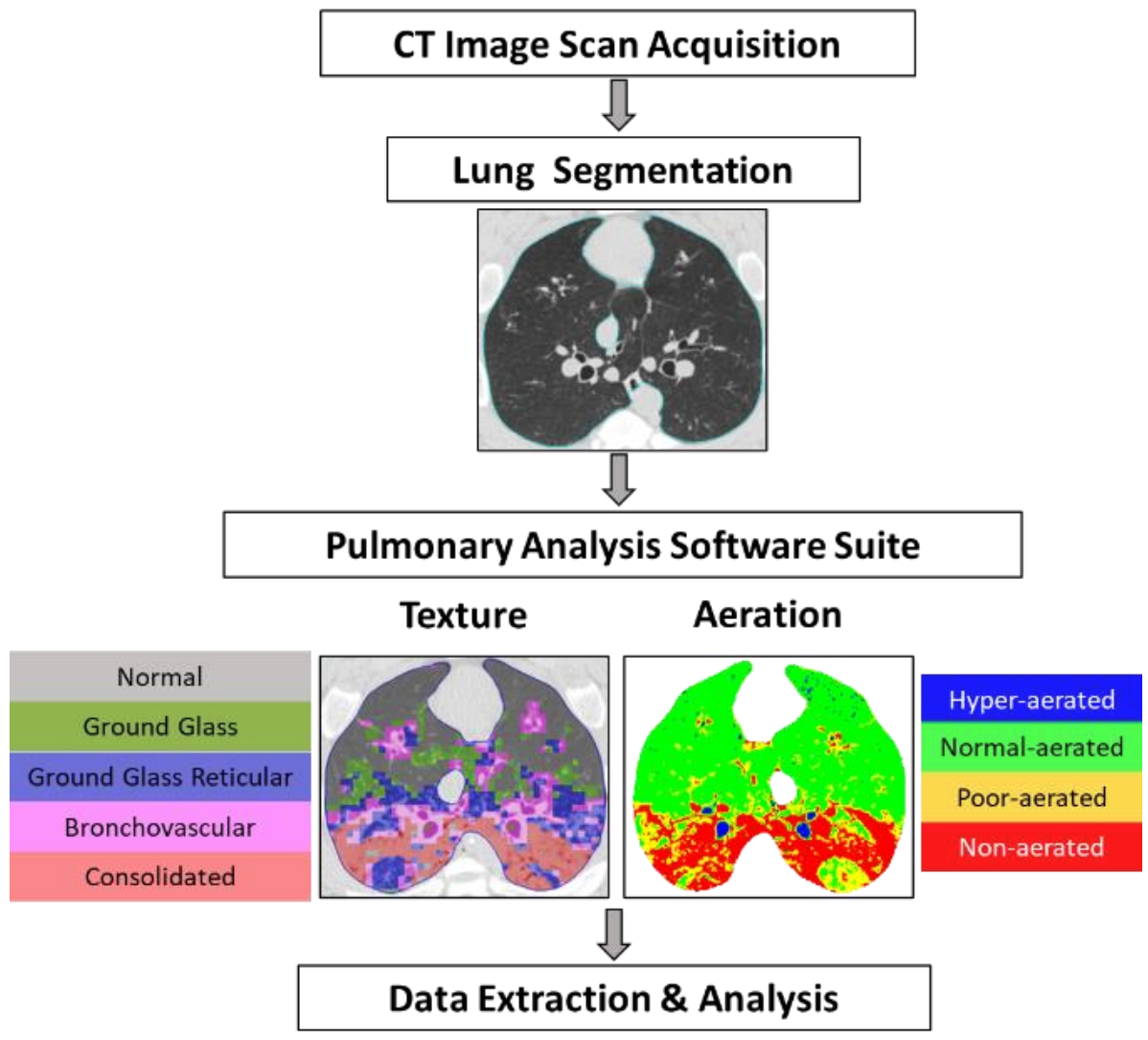

**Figure 3.** Workflow for CT image analysis.

## 3 Results

As outlined in Table 1, there were no significant differences between the CMV and MFV groups at baseline in terms of HR, SAP, PAP, RR, pH, $PaCO_2$, and $PaO_2$: $FiO_2$ ratio. However, after a 9-hour ventilation period following the induction of lung injury with oleic acid, the MFV group demonstrated notable physiological changes. Specifically, there was a significant decrease in RR and a marked increase in the $PaO_2$: $FiO_2$ ratio compared to the CMV group. These results suggest that MFV could potentially be a more effective ventilation strategy for optimal gas exchange in the context of lung injury.

| | Baseline | | 9-hour Ventilation after Injury | |
|---|---|---|---|---|
| | CMV | MFV | CMV | MFV |
| HR ($min^{-1}$) | 101.8 ± 9.3 | 104.0 ± 24.4 | 122.0 ± 20.4 | 133.0 ± 30.4 |
| SAP (mmHg) | 106.5 ± 14.8 | 90.7 ± 10.2 | 98.0 ± 5.9 | 96.0 ± 17.5 |
| PAP (mmHg) | 26.5 ± 6.3 | 16.2 ±9.5 | 36.0 ± 10.1 | 24.0 ± 14.2 |
| RR ($s^{-1}$) | 0.32 ± 0.00 | 0.33 ± 0.07 | **0.48 ± 0.04** | **0.38 ± 0.09*** |
| pH | 7.5 ± 0.1 | 7.5 ± 0.0 | 7.4 ± 0.1 | 7.5 ± 0.1 |
| $PaCO_2$ (mmHg) | 43.8 ± 5.5 | 39.4 ± 5.0 | 47.6 ± 10.1 | 42.2 ± 5.1 |
| $PaO_2$: $FiO_2$ (mmHg) | 401.8 ± 53.3 | 472.1 ± 69.2 | **167.3 ± 82.4** | **310.6 ± 68.2**** |

* p-Value <0.02; ** p-Value <0.008.

**Table 1.** Comparison of baseline and 9-hour post-injury physiological parameters between the CMV and MFV groups.

Figure 4 presents a visual comparison of the CT images, lung textures, and aeration distributions between the CMV and MFV groups at both baseline and after 9 hours of ventilation following injury. No significant differences were observed between the CMV and MFV groups in these visualizations at baseline. After 9 hours of ventilation post-lung injury, the MFV group exhibited less consolidation and a more even distribution of aeration in the images, suggesting a better recovery process under the MFV ventilation strategy.

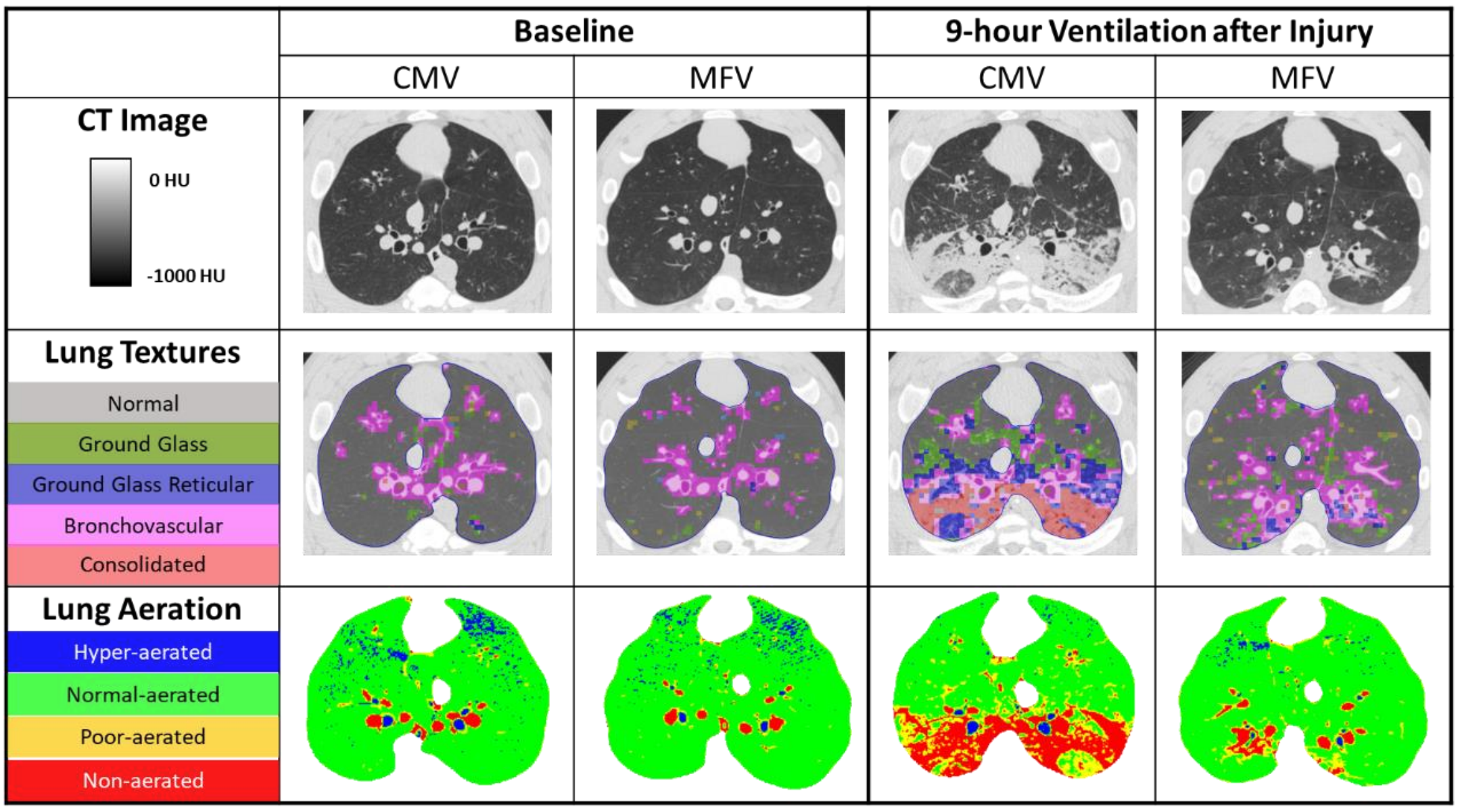

**Figure 4.** Visual comparison of CT images, lung textures, and aeration distributions between the CMV and MFV groups at baseline and after 9 hours of ventilation following injury.

We performed a quantitative analysis of the lung textures in both the CMV and MFV groups at baseline and after 9 hours of ventilation following injury. The texture types of emphysema and honeycomb were excluded from the plots in Figure 5 due to their negligible representation (<1%). At baseline, the CMV and MFV groups showed similar volume percentages for normal, ground glass, ground glass reticular, bronchovascular, and consolidated lung textures. Following 9 hours of ventilation after injury, the MFV group demonstrated a significantly higher volume percentage of normal lung textures (~20% higher in total lung volume) and considerably less consolidation (~14% lower in total lung volume) compared with the CMV group.

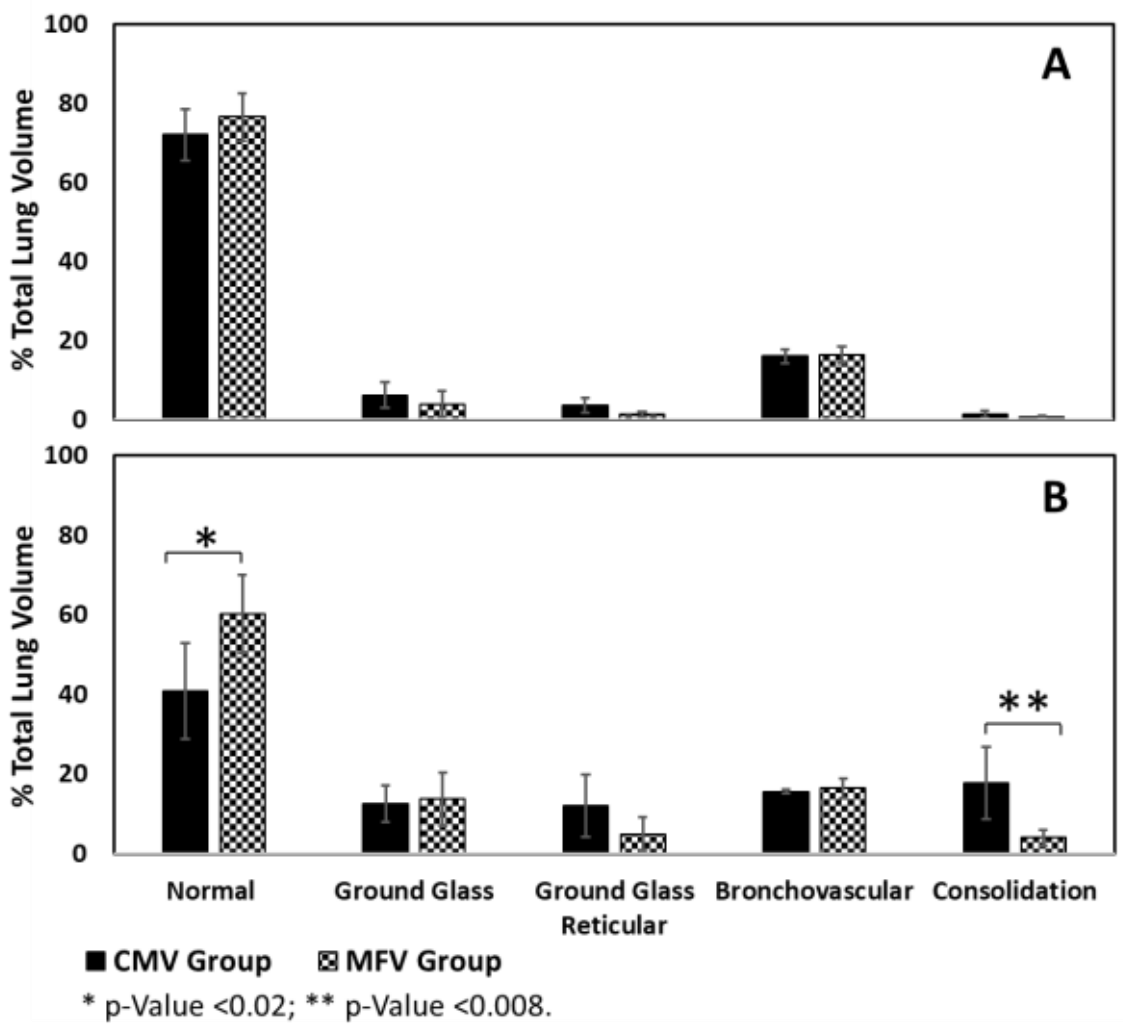

**Figure 5.** Lung texture analysis for the CMV and MFV groups at baseline (Panel A) and after 9-hour ventilation following injury (Panel B).

Lung aeration volumes from both the CMV and MFV groups were further classified into hyper-aeration, normal-aeration, poor-aeration, and non-aeration based on the intensity of the CT images. At baseline, the CMV and MFV groups exhibited a high degree of similarity. Following 9 hours of ventilation post-injury, the MFV group displayed a significantly higher volume percentage of normal-aerated compartments (~20% higher in total lung volume) and significantly less non-aerated tissue (~15% lower in total lung volume) compared to the CMV group. These results align with the lung texture analysis, indicating that the MFV strategy potentially aids in the recovery process by reducing consolidation and restoring normal lung function.

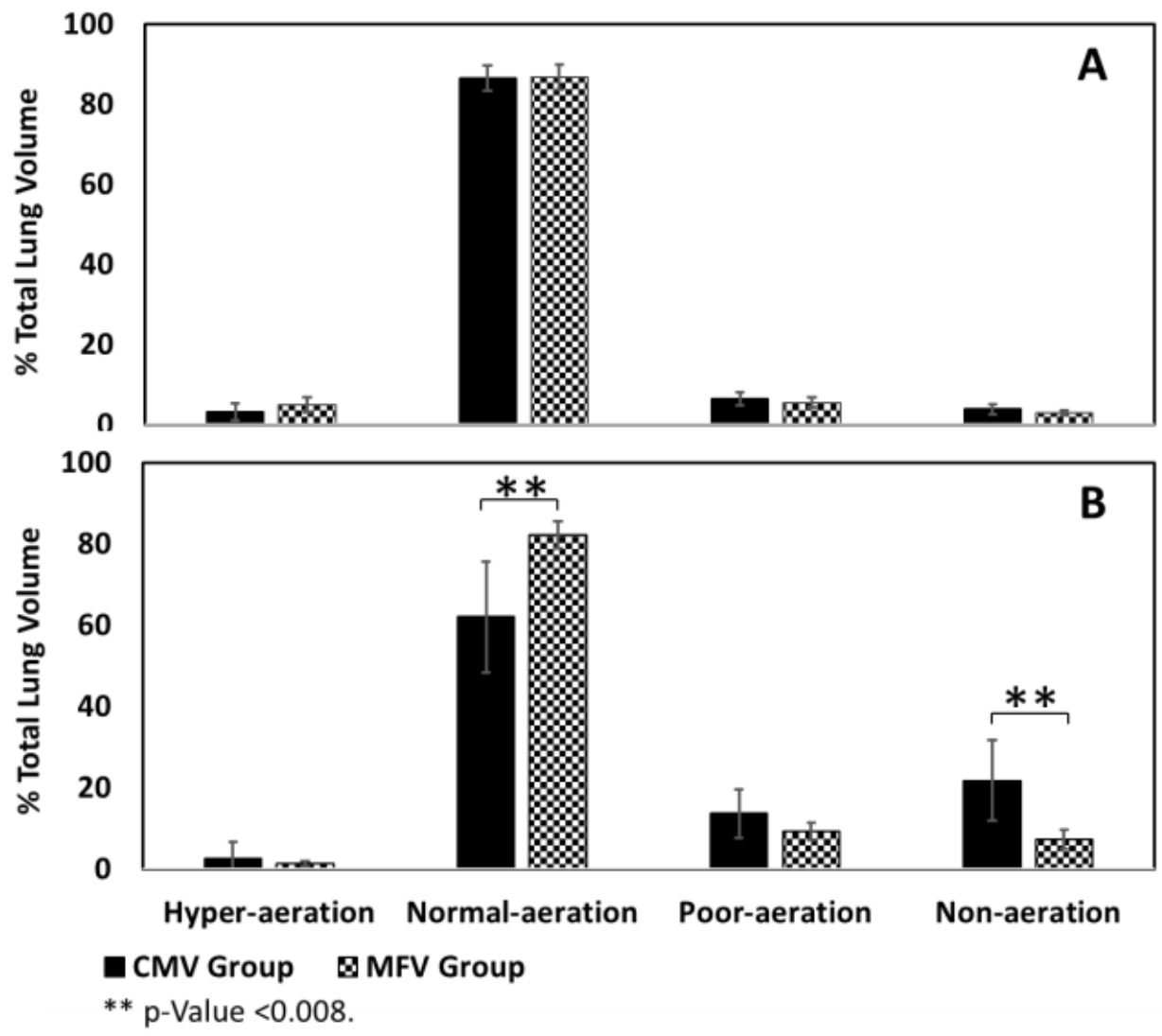

**Figure 6.** Lung aeration analysis for the CMV and MFV groups at baseline (Panel A) and after 9-hour ventilation following injury (Panel B).

To analyze the changes in lung density distribution from baseline to post-injury after 9 hours of ventilation, we plotted the average number of voxels with a CT intensity ranging from -1000 HU to 150 HU for both groups (Figure 7). Following 9 hours of ventilation post-lung injury, the CMV group still demonstrated a significantly larger non-aerated compartment (defined as higher than -100 HU) compared to both its baseline and the MFV group.

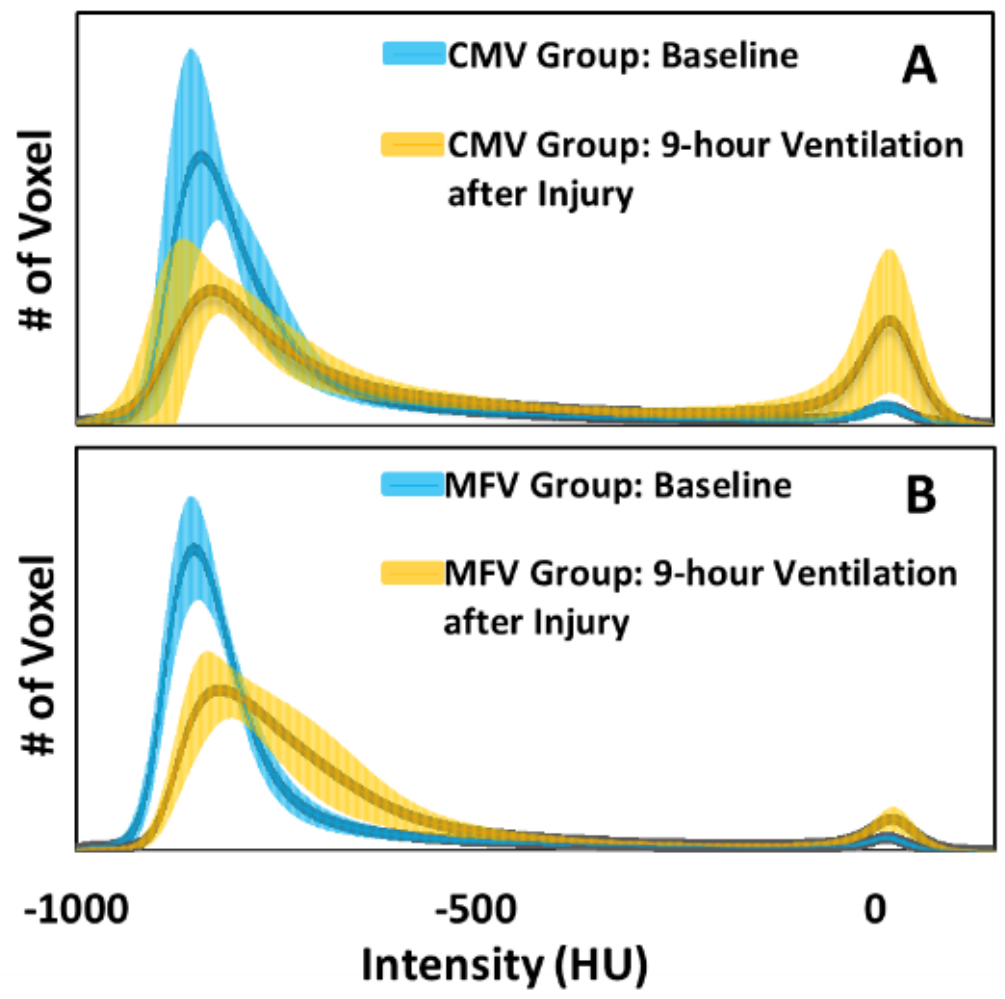

**Figure 7.** Lung density distribution for the CMV group (Panel A) and MFV group (Panel B) at baseline and after 9 hours of ventilation following injury. The error bars are represented as shadows of the lines.

## 4 Discussion

In this study, we undertook a comprehensive investigation into the pathophysiology of ARDS using a porcine model, specifically contrasting the effects of CMV and MFV strategies. Our study utilized qCT imaging to precisely visualize and evaluate the impacts of these ventilation techniques. Notably, the MFV group demonstrated a slower RR, suggesting reduced respiratory stress, and a higher $PaO_2$: $FiO_2$ ratio, indicative of superior gas exchange efficiency, compared to the CMV group. This implies that MFV could potentially enhance oxygenation in ARDS patients, thereby contributing to improved clinical outcomes.

Furthermore, we conducted a thorough comparative study of the lung texture and aeration status for the CMV and MFV groups, both at baseline and 9 hours post-ventilation following injury. The MFV group presented a larger percentage of normal lung tissue and a smaller amount of consolidated tissue after 9 hours of ventilation, compared to the CMV group. This observation suggests more effective lung tissue preservation with MFV, which might help minimize ARDS progression and promote a more favorable prognosis. In addition, our aeration analysis indicated that the MFV strategy was associated with a higher percentage of normal-aerated lung volume. This could potentially improve gas exchange and mitigate the detrimental effects of hypoxia, a frequent complication in ARDS patients. Overall, our findings highlight the potential benefits of the MFV ventilation strategy in managing ARDS, including preservation of lung tissue, enhanced oxygenation, and reduced respiratory stress.

## 5 Conclusion

Through detailed analysis of CT imaging, lung texture, and aeration, our study emphasizes the potential advantages of MFV over CMV in managing ARDS. The data suggest that MFV might facilitate improved oxygenation, a reduced RR, and minimized lung injury. These findings provide valuable insights into our understanding of ARDS and suggest potential improvements in ventilation strategies. Further research is needed to evaluate the effectiveness of MFV in human ARDS patients and to translate these promising results into clinical practice. This study represents a crucial step towards improving patient outcomes in ARDS, a challenging condition for clinicians worldwide. Moreover, this study reinforces the importance of CT imaging reconstruction in the study of lung diseases.

## Acknowledgements

This study is supported by the Carver College of Medicine-Iowa Medical Student Research Program (IMSRP), Department of Defense Grant W81XWH-21-1-0507, and National Institutes of Health Grant S10 OD018526.

## Disclosures

DWK and JH are co-founders and shareholders of OscillaVent, Inc., and are co-inventors on a patent involving multi-frequency oscillatory ventilation. DWK and JH also receive research support from ZOLL Medical Corporation. EAH is co-founder and shareholder of VIDA Diagnostics, a company commercializing lung image analysis software developed, in part, at the University of Iowa. JG is a shareholder of VIDA Diagnostics.

# C-arm CT imaging of the head with the sine-spin trajectory: evaluation of cone-beam artifacts from computer-simulated data of voxelized patient models

Zijia Guo[1], Michael Manhart[2], Philipp Bernhardt[2], Bernd Schreiber[2], Julie DiNitto[3], and Frédéric Noo[1]

[1]Department of Radiology and Imaging Sciences, University of Utah, Salt Lake City, USA
[2]Siemens Healthcare GmbH, Forchheim, Germany
[3]Siemens Medical Solutions, USA

**Abstract** Three-dimensional cone-beam CT imaging of the head is valuable in neuro-interventional radiology, particularly to detect brain hemorrhages and visualize regions of ischemia. To improve image quality, a new data acquisition geometry, called here the sine-spin trajectory, was recently shown to be clinically feasible and meritorious over the classical circular short-scan. In this work, we assess the performance of the sine-spin trajectory in terms of cone-beam artifacts only, using voxelized patient models to retain the complexity of the human head anatomy and the task of interest (low contrast brain imaging). Our results show strong superiority of the sine-spin trajectory over the circular short-scan. They also show strong robustness to relative positioning between sine-spin and head, and to variations in skull thickness, making altogether the sine-spin trajectory highly attractive for head imaging in neuro-interventional radiology.

## 1 Introduction

Three-dimensional cone-beam CT imaging of the head is valuable in neuro-interventional radiology, particularly to detect brain hemorrhages and visualize regions of ischemia. For this reason, the last two decades have seen a continuous stream of efforts dedicated to improving the image quality achieved with this technology, with the ultimate goal of producing images of quality comparable to diagnostic CT.

Initially, a circular short-scan was used for data acquisition, and image reconstruction relied on using the FDK algorithm with Parker weights. One important milestone regarding image reconstruction was to change the algorithm in favor of using rigorous principles of cone-beam tomography [1–3]. The benefits of using these principles for circular short-scan reconstruction were, for example, shown in [4]. Such changes, facilitated by the increase in computer power, can produce images of higher quality given that the cone-beam artifacts induced by the Parker weights are avoided. However, cone-beam artifacts are still present as the data acquisition trajectory is strongly incomplete in terms of Tuy's condition [5].

One obvious but not straightforward approach to avoid cone-beam artifacts is to use a trajectory that is complete. For example, we recently showed that multi-axis robotic systems can be reprogrammed to perform data acquisition with the Line-Ellipse-Line trajectory [6]. Unfortunately, the biplane C-arm systems that are most commonly used in neuro-interventional radiology do not offer such flexibility in motion. Nevertheless, recent developments showed that a nearly complete trajectory, called here the sine-spin trajectory, is clinically feasible [7–9].

Besides cone-beam artifacts, image quality in cone-beam CT is affected by many physical effects including beam-hardening, scatter, and patient motion. In the aforementioned clinical studies [7–9], image quality was assessed using real patient scans, hence cone-beam artifacts were entangled with all other sources of errors. In this work, we aim to assess the performance of the sine-spin trajectory in terms of cone-beam artifacts only, using voxelized patient models to retain the complexity of the human head anatomy and the task of interest (low contrast brain imaging).

## 2 Materials and Methods

### 2.1 Sine-spin trajectory

The sine-spin trajectory differs from the conventional circular short-scan trajectory in that a sinusoidal motion is applied in the cranio-caudal direction during the rotation, as illustrated in Figure 1 (top row). The resulting trajectory lies on a sphere where the sinusoidal motion can be visualized as a smooth variation in the elevation angle. The amplitude of the motion is 10-degree. In Figure 1 (bottom row), it can be seen that the projection of the sine-spin trajectory on the $(x, y)$-plane is identical to the circular short-scan trajectory, while the projection on the $(x, z)$-plane shows that the sine-spin trajectory covers view angles above and below the plane of the short-scan, which leads to improvements in terms of Tuy's condition. However, the sine-spin trajectory is not complete. Hence, cone-beam artifacts can be expected, with magnitude and location dependent on relative positioning between head and trajectory.

### 2.2 Voxelized phantoms

To assess performance of the sine-spin trajectory in terms of cone-beam artifacts only, we used voxelized models of the human head from the BrainWeb database. Specifically, we used subjects 04 and 06 from this database, which respectively correspond to a thin skull and a thick skull. The models associate each voxel with one of twelve tissue indices. We used data available on the NIST website to convert the tissue indices in values of the linear attenuation coefficient at photon energy of 56.5 keV. The selected energy is the mean energy of the spectral x-ray beam typically used in the clinic.

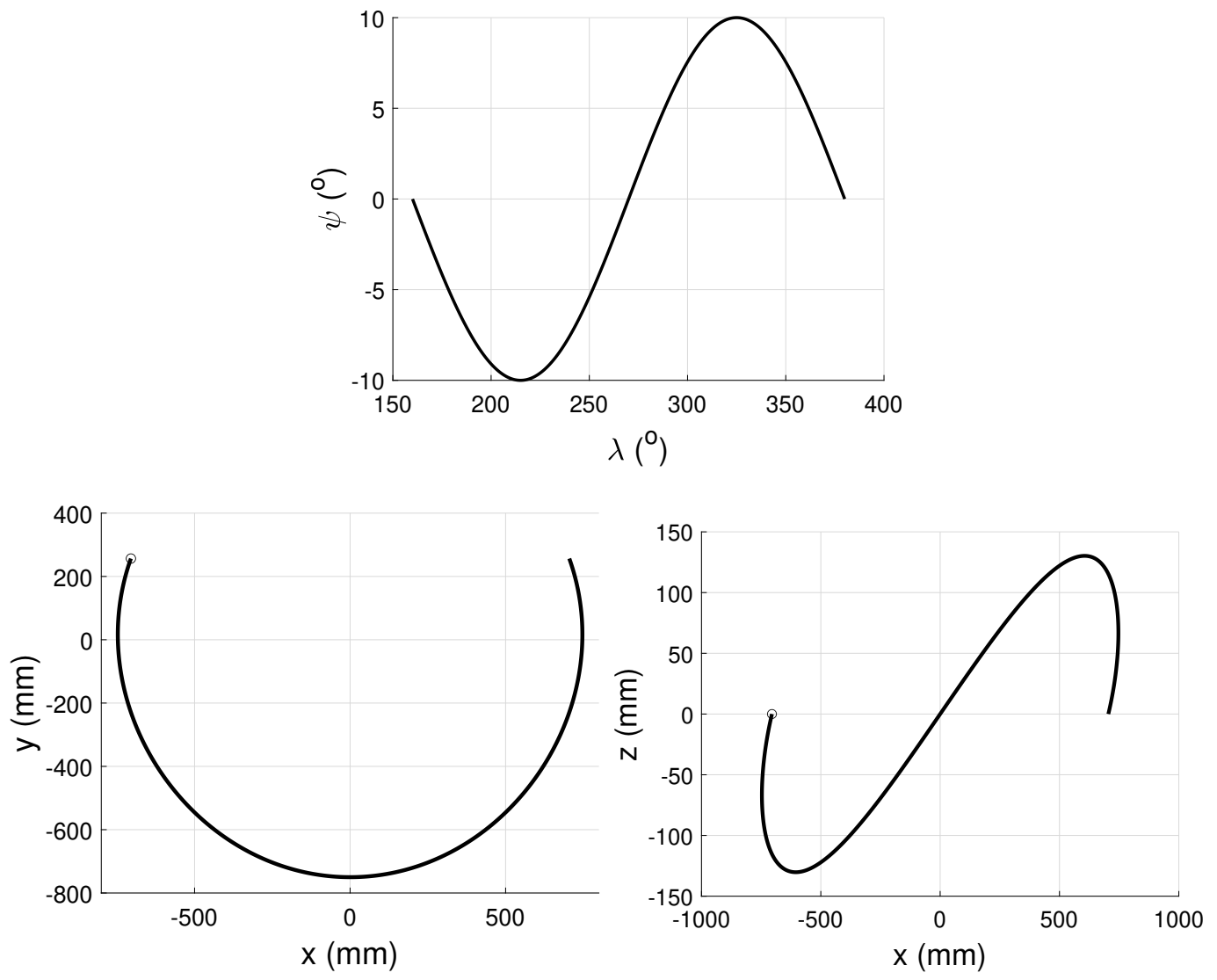

**Figure 1:** Sine-spin trajectory. Top: elevation (cranio-caudal) angle as a function of the rotation angle used for the circular short-scan. Bottom: orthogonal projection of the trajectory on the $(x,y)$-plane and the $(x,z)$-plane; the small circle shows the start position of the trajectory.

### 2.3 Data simulation and image reconstruction

Cone-beam data were simulated using forward projection through the voxelized models, with $2 \times 2$ sub-lines for each detector pixel. The source-to-rotation axis distance was 750 mm, the source-to-detector distance was 1200 mm, the number of views was 551, and the detector pixel size was $0.64 \times 0.64\,\text{mm}^2$, as in [7, 8]. For experiments with noise, we used Poisson statistics such that the noise level in the reconstruction is similar to clinical practice. Image reconstruction (for both sine-spin and circular short-scan) follows the principles of [2, 3], with cubic voxels of side-length equal to 0.5 mm.

To study how relative positioning between sine-spin and head impacts cone-beam artifacts, we repeated the data simulation using variations in positioning of the sine-spin trajectory. We considered four variations in axial positioning, corresponding to centering the trajectory on $z \in \{-10,-5,5,10\}$ mm instead of $z = 0$ mm; and we considered two variations for the start angle: 140° and 180° instead of 160°.

To study the effect of scan range and head anatomy, we used subjects 04 and 06, and considered scan ranges of 210° and 240° instead of the 220° used for all other experiments.

## 3 Results

### 3.1 Comparison with the circular short-scan

Figure 2 shows results obtained for subject 04 using the circular short-scan versus the sine-spin trajectory. While the circular short-scan quickly yields undesirable cone-beam artifacts away from the plane of the source trajectory, the sine-spin trajectory produces high accuracy almost everywhere. For a more quantitative assessment, we also computed the root mean squared deviation (RMSD) from the ground truth in the regions of interest marked in Figure 2. The results, displayed in Table 1 and 2, closely supports the visual assessment.

Figure 3 shows reconstructions with noise added to the data, with slice thickness of 2.5 mm. As can be seen, the effect of data noise is similar between circular short-scan and sine-spin, conveying that the improvement in cone-beam artifacts is not associated with a higher sensitivity to data noise.

### 3.2 Effect of relative positioning

Figure 4 shows the RMSD between the ground truth and the reconstruction as a function of the slice index for the various variations in positioning of the sine-spin trajectory. The RMSD computation is restricted to voxels containing brain tissue. On one hand, variations in the start angle have negligible impact on the slice-by-slice magnitude of cone-beam artifacts. On the other hand, variations in the $z$ positioning has some impact for the bottom and top regions of the brain: as $z$ varies from $z = 10$ mm towards $z = -10$ mm, the magnitude of cone-beam artifacts in the pons and cerebellum regions progressively decreases while increasing for the upper portion of the brain.

### 3.3 Effect of scan length and skull thickness

Figure 5 shows results for subjects 04 versus subject 06 as obtained when varying the scan length. The results are again shown in terms of RSMD through brain pixels, on a slice-by-slice basis. Negligible improvements result from increasing the scan length while reducing it increases the magnitude of cone-beam artifacts in the bottom and top regions of the brain. We also see strong similarity between the two subjects.

## 4 Discussion and Conclusion

Since the sine-spin trajectory is not complete, cone-beam artifacts can be expected. We investigated the magnitude of these artifacts using voxelized patient models. The results showed strong superiority over the circular short-scan. The results also showed strong robustness to relative positioning between sine-spin and head, and to variations in skull thickness, making altogether the sine-spin trajectory highly attractive for head imaging in neuro-interventional radiology.

| | #1 | #2 | #3 | #4 | #5 |
|---|---|---|---|---|---|
| circular | 14.37 | 2.27 | 3.72 | 34.80 | 50.15 |
| sine-spin | 3.62 | 2.25 | 3.58 | 3.37 | 10.97 |

**Table 1:** RMSDs for the 5 regions of interest marked in the coronal slice of Figure 2, expressed in HU.

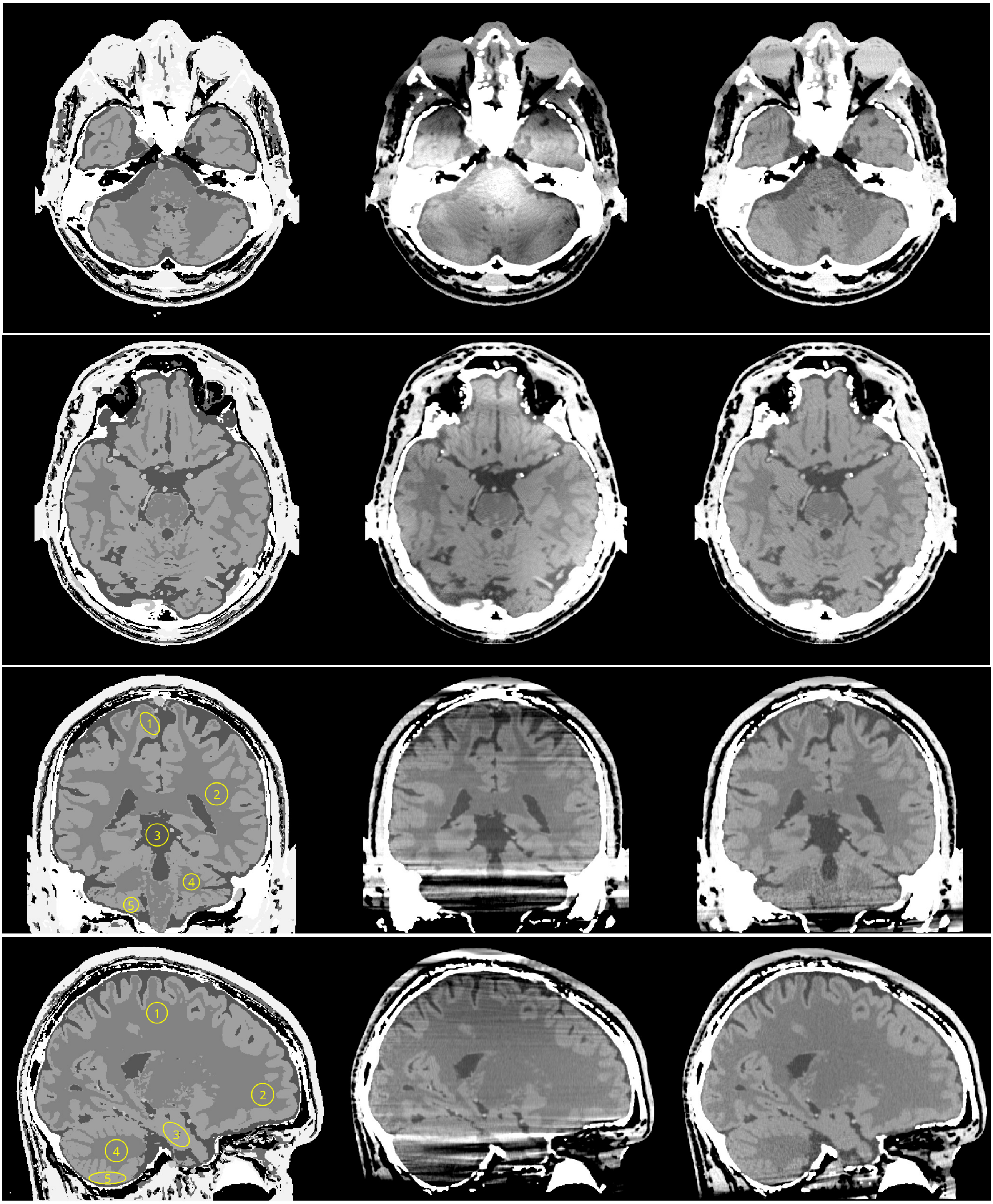

**Figure 2:** Image reconstruction comparison between the sine-spin and the circular short-scan using subject 04. From left to right: ground truth, circular short-scan, and sine-spin. Window: $[0, 80]$ HU.

| | #1 | #2 | #3 | #4 | #5 |
|---|---|---|---|---|---|
| circular | 11.10 | 1.78 | 21.59 | 25.54 | 50.22 |
| sine-spin | 3.24 | 1.66 | 3.43 | 2.83 | 10.78 |

**Table 2:** RMSDs for the 4 regions of interest marked in the sagittal slice of Figure 2, expressed in HU

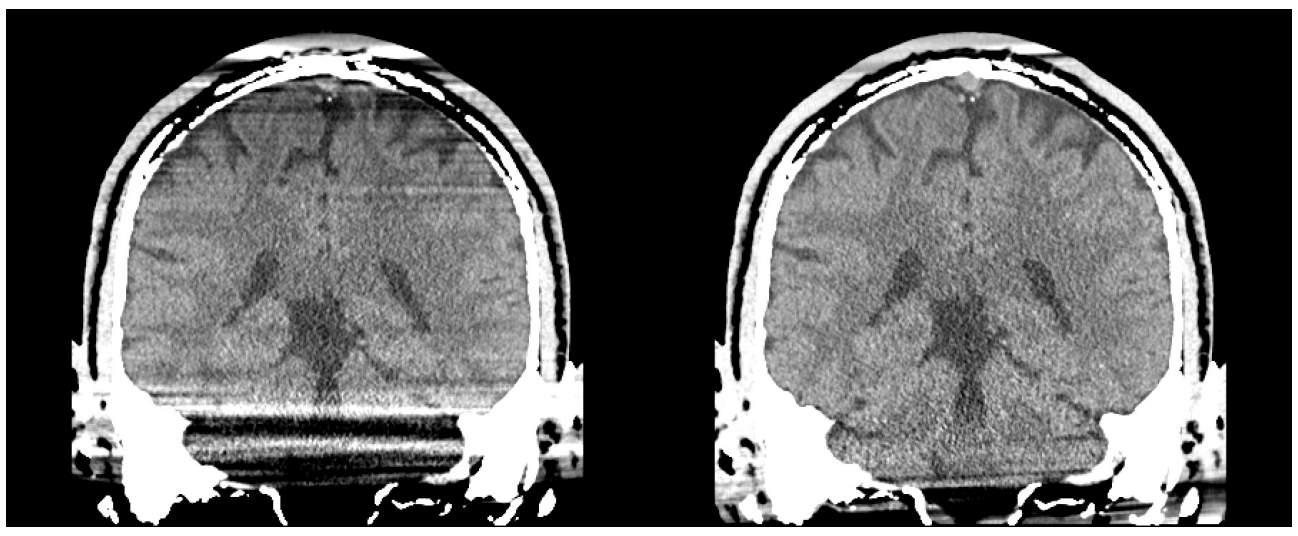

**Figure 3:** Image reconstruction of subject 04 from noisy data. Left: circular short-scan. Right: sine-spin. Window: $[0, 80]$ HU.

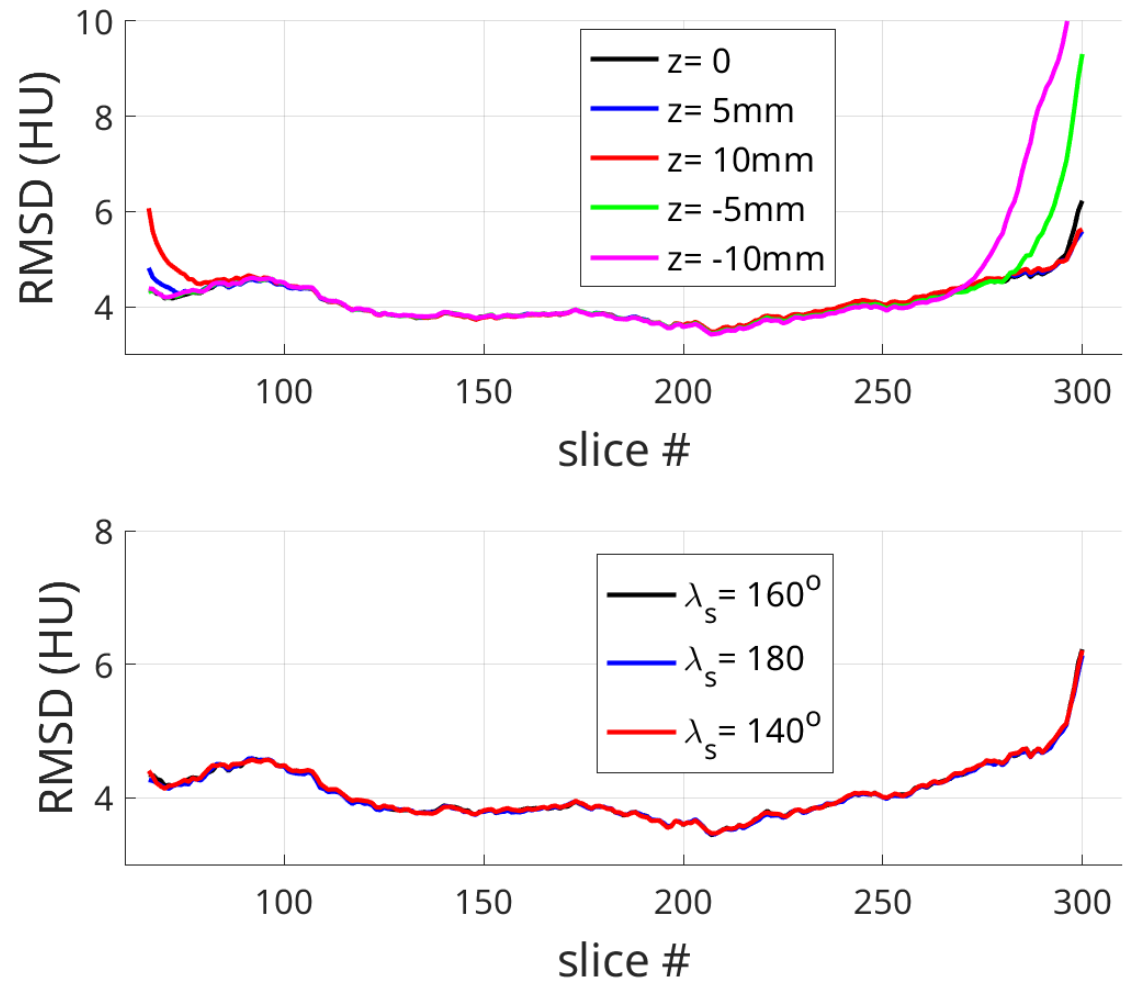

**Figure 4:** Effect of relative positioning between the sine-spin trajectory and the head, quantified using the RMSD over brain pixels and reported on a slice-by-slice basis. Top: variations in axial positioning. Bottom: variation in the start angle.

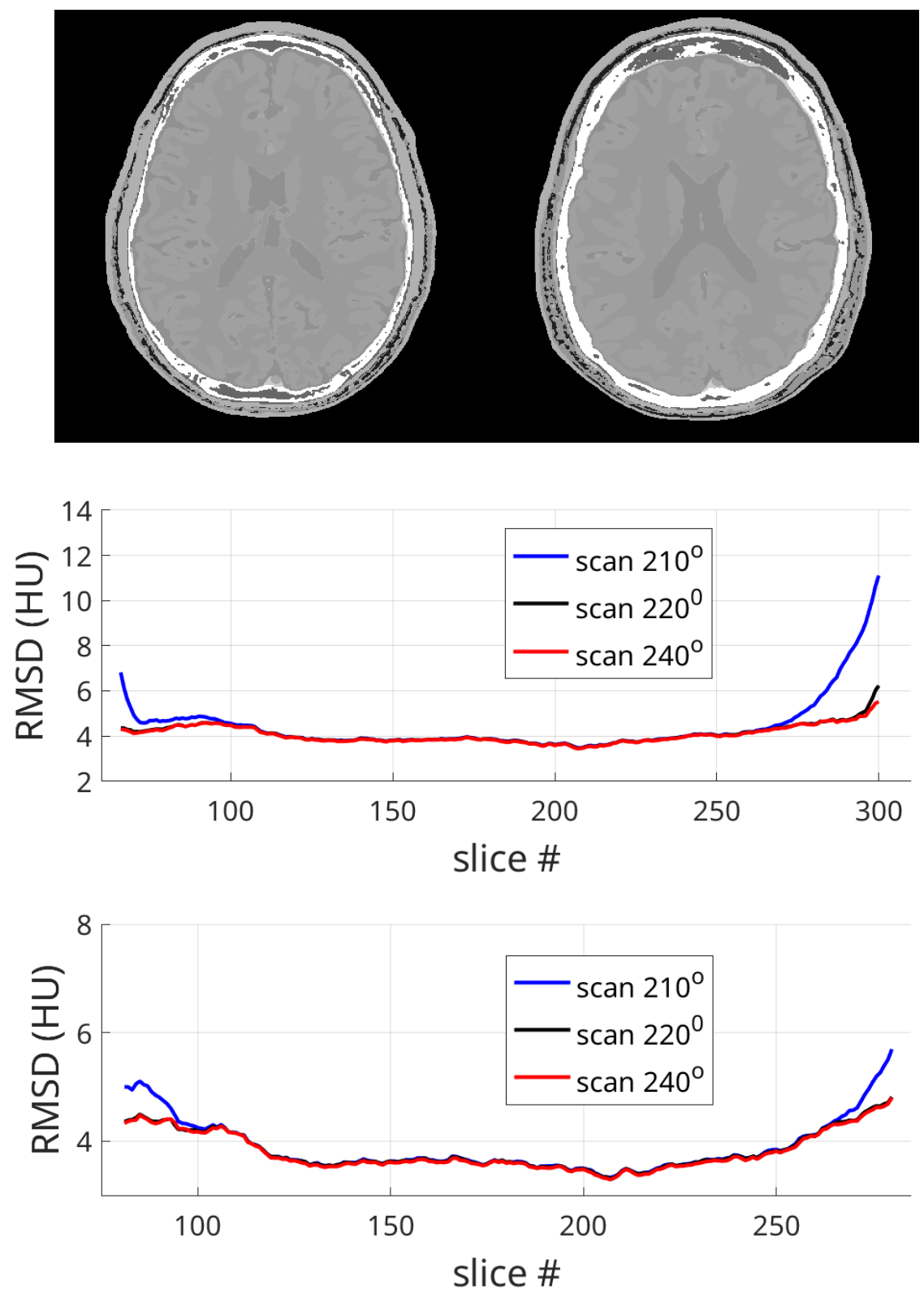

**Figure 5:** Effect of scan length and head anatomy. Top: subjects 04 and 06 strongly differ in skull thickness. Bottom: plots of RMSD over brain pixels reported on a slice-by-slice basis for (top) subject 04 and (bottom) subject 06.

# X-ray Dark-Field Imaging at the Human Scale: Helical Computed Tomography and Surview Imaging

Jakob Haeusele[1], Clemens Schmid[1], Manuel Viermetz[1], Nikolai Gustschin[1], Tobias Lasser[1,2], Thomas Koehler[3,4], and Franz Pfeiffer[1,4,5]

[1]Munich Institute of Biomedical Engineering, Technical University of Munich, 85748 Garching, Germany
[2]Computational Imaging and Inverse Problems, Department of Informatics, Technical University of Munich, 85748 Garching, Germany
[3]Philips Research, 22335 Hamburg, Germany
[4]Institute for Advanced Study, Technical University of Munich, 85748 Garching, Germany
[5]Department of Diagnostic and Interventional Radiology, School of Medicine and Klinikum rechts der Isar, Technical University of Munich, 81675 München, Germany

**Abstract** X-ray imaging is a widely adopted diagnostic tool that uses the attenuation as contrast generating mechanism. Grating-based X-ray dark-field imaging is an interferometric approach that unlocks an additional contrast mechanism. Unlike conventional X-ray techniques, dark-field imaging is also capable of measuring the so-called linear diffusion coefficient. This measure of a sample's small-angle scattering strength yields additional information about a sample's microstructure, which otherwise could not be resolved directly. While initial clinical studies are already being carried out for human chest dark-field radiography, dark-field computed tomography (CT), which is capable of yielding unobstructed 3D views, has only recently been brought to the human scale: With a first prototype system, we demonstrated the feasibility to implement a Talbot-Lau interferometer on a clinical CT system to perform dark-field imaging with clinical acquisition time and field of view. Until now, this prototype was limited to axial scans. In this work, we present our advancements in extending the setups capabilities to also support other modes of acquisition, namely surview and helical scans. The new capabilities of the updated dark-field CT scanner are demonstrated using an anthropomorphic thorax phantom.

## 1 Introduction

X-ray computed tomography is a common medical imaging technique that produces unobstructed 3D-views of a patient's anatomy and allows for fast acquisition and quantitative measurements. However, traditional CT imaging relies solely on attenuation-based contrast and does not utilize the wave-like properties of X-rays. Access to this additional information and novel contrast modalities can be achieved, for example, through techniques such as Talbot-Lau interferometry [1–3], speckle [4], or edge-illumination-based imaging [5].

In Talbot-Lau interferometry, three gratings are introduced into the X-ray beam, which modulate the wavefront and create an interference pattern on the detector [1, 3]. When a sample is inserted into the X-ray beam, it distorts the observed pattern in three ways: it decreases the overall intensity due to attenuation, it causes small-angle scattering that blurs the interference pattern decreasing the fringe amplitude, and it shifts the pattern due to the induced phase shift. These three physical effects give rise to the three contrast modalities obtained via grating-based imaging: the attenuation, dark-field, and phase contrast, respectively. Proper analysis of the interference pattern's distortion allows for the separation of the three contrasts channels. This process is known as *phase retrieval*.

The so-called dark-field signal generated by the effect of small-angle scattering is a promising new tool in the diagnosis of pulmonary diseases as it reveals microstructural changes in the lung parenchyma that are too small to be resolved directly [6]. As a first step in translating dark-field imaging from the bench to clinical usage, a prototype human scale dark-field thorax radiography system is already in use in the university hospital München rechts der Isar. First clinical studies demonstrated the potential of dark-field imaging for the diagnosis of chronic obstructive pulmonary disease (COPD) and COVID-19 [7]. However, this system is limited to acquiring chest radiographs and lacks 3D information.

As a next step, we recently presented a dark-field CT prototype based on a clinical system that overcomes two of the main challenges in bringing dark-field CT imaging into clinical use: the dynamic environment of a clinical CT gantry that introduces perturbations of the grating interferometer, and the continuous data acquisition that prohibits conventional stepping-based phase retrieval methods. The prototype scanner is a clinical Philips Brilliance iCT scanner that was upgraded with a Talbot-Lau interferometer [8, 9].

Until now, this prototype was limited to axial scans and did not support other standard acquisition protocols. In this work, we present our advancements in extending the setups capabilities to also support other acquisitions, namely surview and helical scans.

## 2 Materials and Methods

### 2.1 Talbot-Lau X-ray Dark-Field Imaging

Grating-based Talbot-Lau X-ray dark-field imaging with incoherent sources employs three gratings that are introduced in the beam path to generate an interference pattern on the detector. Fig. 1 shows a schematic of the used Talbot-Lau interferometer used in the dark-field CT prototype. In good approximation the fringe pattern recorded at the detector can be modeled as a truncated Fourier series [2]

$$y = TI + TIDV\cos(\Phi + \varphi), \tag{1}$$

where the state of the interferometer is described by the parameters $I$, $V$, and $\varphi$ which denote the mean intensity,

the fringe amplitude (visibility), and the phase of the fringe pattern, respectively. A sample in the beam attenuates, small-angle scatters, and phase-shifts the incident wave, which alters the measured fringe pattern. These changes are modeled by three sample parameters, which represent the method's different contrast modalities: the sample transmission $T$, the dark-field $D$, and the (differential) phase $\Phi$. These three sample parameters are related to the sample by

$$T = \exp\left[-\int_{L_1}^{L_2} \mu(\vec{S}+l\vec{n})\mathrm{d}l\right], \tag{2}$$

$$D = \exp\left[-\int_{L_1}^{L_2} s_D(l)\varepsilon(\vec{S}+l\vec{n})\mathrm{d}l\right], \tag{3}$$

$$\Phi = \partial_y \int_{L_1}^{L_2} s_\Phi(l)\delta(\vec{S}+l\vec{n})\mathrm{d}l, \tag{4}$$

where the linear attenuation coefficient $\mu$, the refractive index decrement $\delta$, and the linear diffusion coefficient $\varepsilon$ are material specific parameters of the sample. Line integrals are computed from the actual source position $\vec{S}$ in the direction $\vec{n}$ towards the detector pixel, where the object is known to be located between the gratings $G_1$ and $G_2$, which have distances $L_1$ and $L_2$ from the source, respectively. Finally, $s_D$ and $s_\Phi$ are setup geometry specific sensitivities, which are 0 at $L_2$ and grow linearly towards $L_1$, see e.g. [10] for more details. The partial derivative $\partial_y$ is taken perpendicular to the grating lamellae in the projection domain. The grand goal of grating-based phase-contrast and dark-field CT imaging is to determine the sample projection parameters $T$, $D$, and $\Phi$, and reconstruct the spatial distribution of $\mu$, $\varepsilon$, and $\delta$.

### 2.2 Sliding-Window-Based Phase Retrieval

An essential step in processing the data of the dark-field CT scanner is the so-called phase retrieval step. Here, the sample parameters $T$, $D$, and $\Phi$ are extracted from the measured signal, which implies fitting the forward model Eq. (1) to the acquired data sinogram. Therefore, the interferometer parameters $I$, $V$, and $\varphi$ have to be determined via a prior processing step. For this, an approach that models vibrations and distortions of the interferometer occurring in the highly dynamic environment of the CT gantry is used [11].

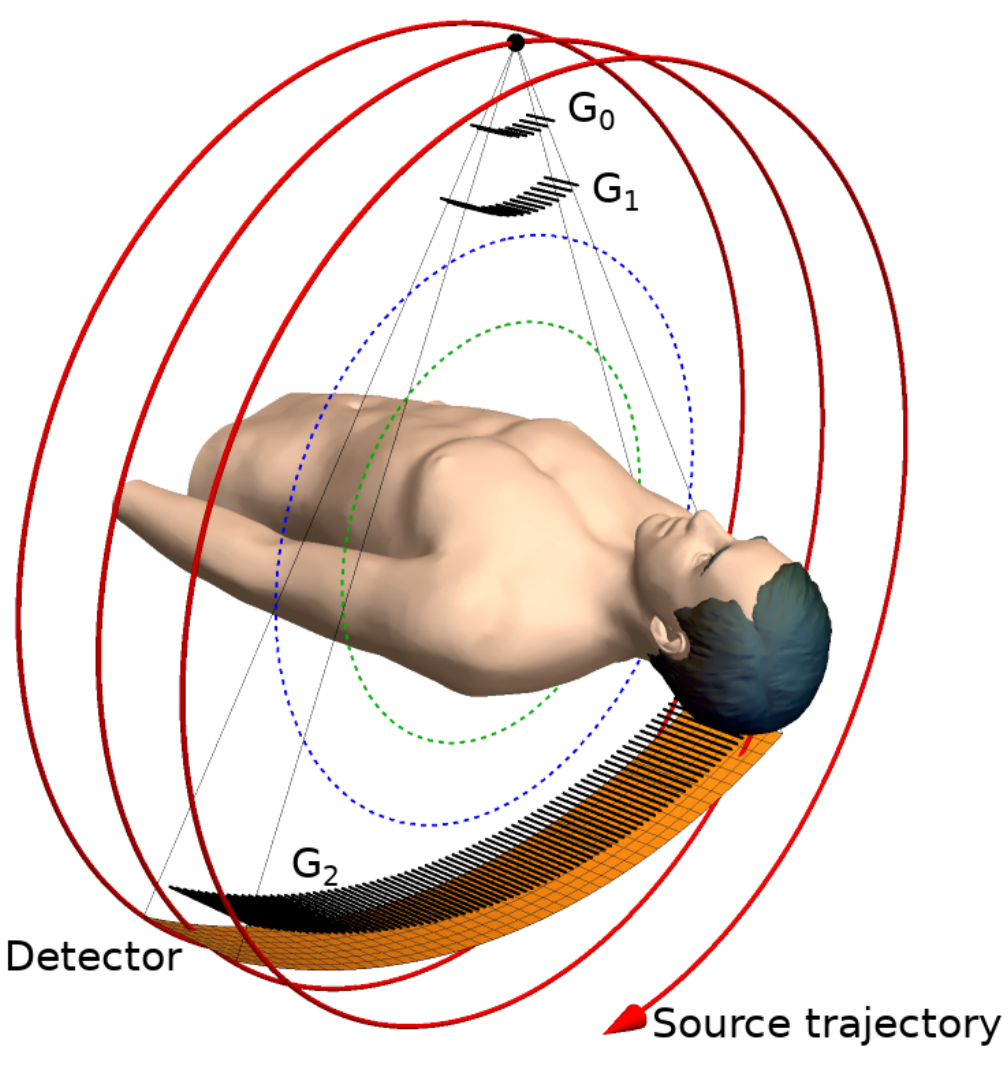

**Figure 1:** Sketch of the Talbot-Lau interferometer in the dark-field CT prototype's gantry. It consists of three cylindrically bent gratings $G_0$, $G_1$, and $G_2$.

For the phase retrieval step, the interferometer phase $\varphi$ is of particular importance as it determines the sampling quality of the sinusoidal signal and thus the accuracy of the fit result. Contrary to conventional lab setups, the necessary phase sampling is in our prototype not generated by a controlled stepping of the gratings [12]: We generate a spatio-temporal modulation of the fringe pattern by exploiting system inherent vibrations that move the gratings, and additionally tune a fine Moiré fringe pattern on the detector that results in a spatial gradient in the interferometer phase.

For our prototype scanner, we use an advanced sliding-window-based phase retrieval algorithm, that is capable of exploiting both the temporal and the spatial modulation of the designed fringe pattern. Due to the continuous data acquisition, we do not have access to multiple shots acquired under the same gantry angle with different grating positions. Our sliding-window algorithm thus groups together adjacent pixels and shots in the data sinogram and performs phase retrieval in each patch. The movement of the sample's projection within the patch is modeled by polynomials and updated by multiple processing passes [13].

### 2.3 Dark-Field Surview Imaging

For surview imaging the beam is collimated to only the central four detector rows. Furthermore, the detector position is fixed during the scan and only the patient couch is moved during the measurement. Detector readings are acquired roughly every 60 µm, providing a fine temporal sampling for phase retrieval. In our prototype scanner, the gantry is supported on an air cushion, which decouples it from its surroundings. Therefore, the couch movement does not introduce additional perturbations to the interferometer. Conversely, the lack of the gantry rotation removes the centrifugal force acting on the gratings that has to be considered for axial scans. Thus low frequency vibrations are excluded from the surview vibration model. For surview scans, we therefore only include the vibrations generated by components on the gantry into our vibration model, which simplifies the employed perturbation model. The main source of vibrations in our setup is the anode drive that induces a grating vibration of approximately 177 Hz [12].

In practice, surview imaging is performed by first acquiring an air surview scan that uses the same scan parameters as the sample scan and next performing the sample surview scan. Based on both scans, the simplified perturbation model is fitted to the data and the interferometer state is determined. Next, the sliding-window-based phase retrieval is performed to extract the transmission $T$ and the dark-field $D$. To access the line integrals, $-\log(\cdot)$ is applied. Finally, post-processing is applied to the data. It involves binning the data in scanning direction and averaging the four detector rows. Furthermore, a beam hardening correction is applied to the dark-field channel [12]. In the attenuation channel, a Laplace-filtered version is mixed in with the original image to enhance edges.

### 2.4 Helical Dark-Field Computed Tomography

For helical acquisitions, the gantry is rotating while the patient couch is moved and the measurement is performed. Therefore, until the reconstruction step, each individual rotation of the gantry can be treated during processing like an axial scan. For the reference scan, it is thus sufficient to use an axial air scan with the same flux, exposure, and rotation time. Using this reference scan, we then determine the interferometer state of the sample scan for each rotation separately. The main difference to the axial case is the couch movement, which introduces an additional movement of the sample projections in $z$-direction. This movement is considered in the phase retrieval step by employing the dual pass method described in [13], and approximating movement within the demodulation window in the transmission signal by a first order polynomial.
Once the sample information has been retrieved in the projection domain, it is reconstructed using a dedicated helical FBP algorithm for the dark-field channel, that accounts for the sensitivity gradient along the rays (see Eq. 3 and 4). This has been described in [10] for axial CT and a different grating geometry where the object is located between $G_0$ and $G_1$ and the idea is adapted here to the new scenario.
The basic idea is that the linear sensitivity gradient along the ray can be eliminated in 2D fan-beam geometry if a direct and a complementary ray are averaged, see Fig. 2. Then, the only change in a 2D FBP is a pre-weighting to compensate a fan-angle dependency of the mean sensitivity after averaging. In helical cone-beam CT, however, there is usually no real complementary ray since the cone-angles of the rays from $\vec{S}_1$ and $\vec{S}_2$ differ. Thus, the averaging of direct and complementary rays is postponed to the back-projection step, where direct and complementary rays are identified as rays passing in opposing directions (i.e. only differing by their cone-angle) through a certain voxel.
The reconstruction algorithm can be summarized as follows:

- pre-weighting the data in native geometry with fan-angle $\alpha$ dependent weight $c(\alpha) = \frac{L_2 - L_1}{L_2 - R\cos\alpha}$, where $R$ is the distance from the source to the iso-center. The reason for the fan-angle dependency is that $\vec{S}_1$ and $\vec{S}_2$ get closer to each other with increasing fan-angle, see Fig. 2.
- re-binning the data into wedge-geometry [14].
- weighting the data with the cosine of the cone-angle and ramp-filtering.

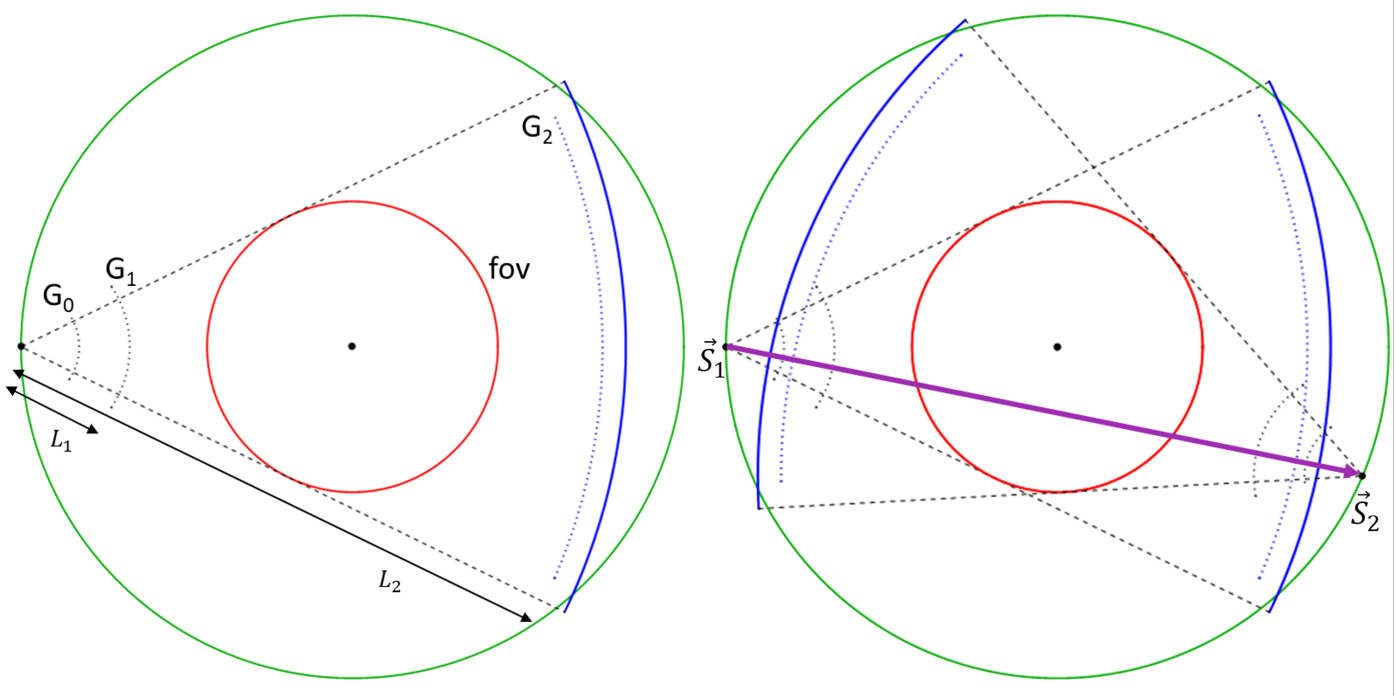

**Figure 2:** Sketch of the system geometry (left) for a certain source position $\vec{S}_1$. On the right, an example direct ray measured from $\vec{S}_1$ is shown together with the complementary source position $\vec{S}_2$.

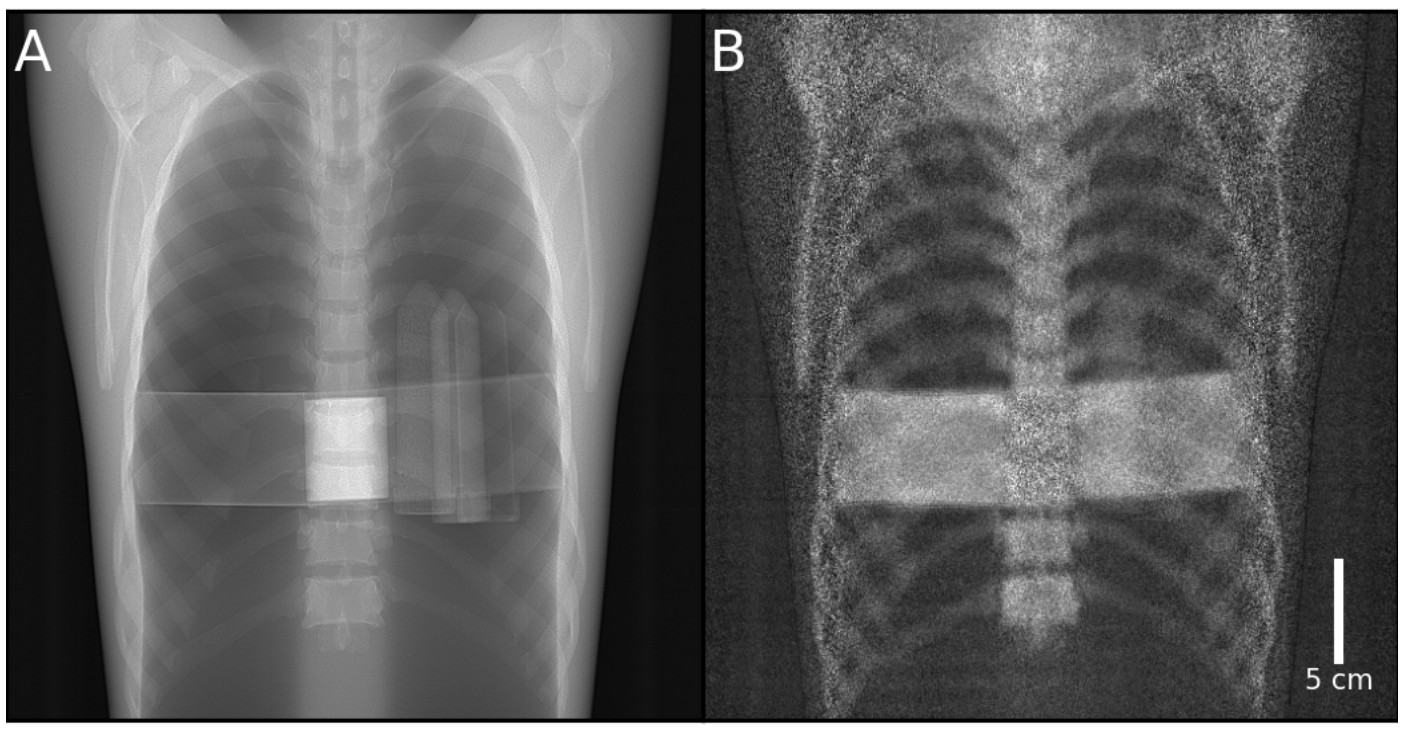

**Figure 3:** Surview scan of a human thorax phantom acquired with the dark-field CT prototype. **A** shows the conventional attenuation-based surview, while **B** depicts the dark-field modality. By the methods design, both channels are perfectly registered.

- For each voxel $\vec{x}$ to reconstruct, back-projection from all projection angles $\beta$ is done using an aperture weighting scheme as described in [15] with the modification that normalization of weights is done over $2\pi$ partners instead of $\pi$ partners, i.e., $w_{\text{all}}(\beta, \vec{x}) = \frac{1}{2} \frac{w_{\text{ap}}(\beta, \vec{x})}{\sum_n w_{\text{ap}}(\beta + 2\pi n, \vec{x})}$. This is analogue to the normalization concept for high-resolution scans as described in [14].

For post-processing, we applied the same polyoxymethylene-based beam hardening correction to the dark-field channel that we use for axial scans, which is described in [12].

## 3 Results

To demonstrate the capabilities of the dark-field CT prototype for surview and helical imaging, two test measurements were performed with a human thorax phantom that was filled with a neoprene foam material to simulate the porous microstructure of lung parenchyma. Additionally, plastic tubes filled with water, cotton, and powdered sugar were inserted into the phantom, as well as a polyoxymethylene cylinder.
First, a dark-field surview scan was performed with a tube voltage of 80 kVp and a current of 200 mA. During scanning, 7006 shots were acquired in 3.2 s to cover a total scan length of 45 cm. The resulting surview scan is depicted in Fig. 3.

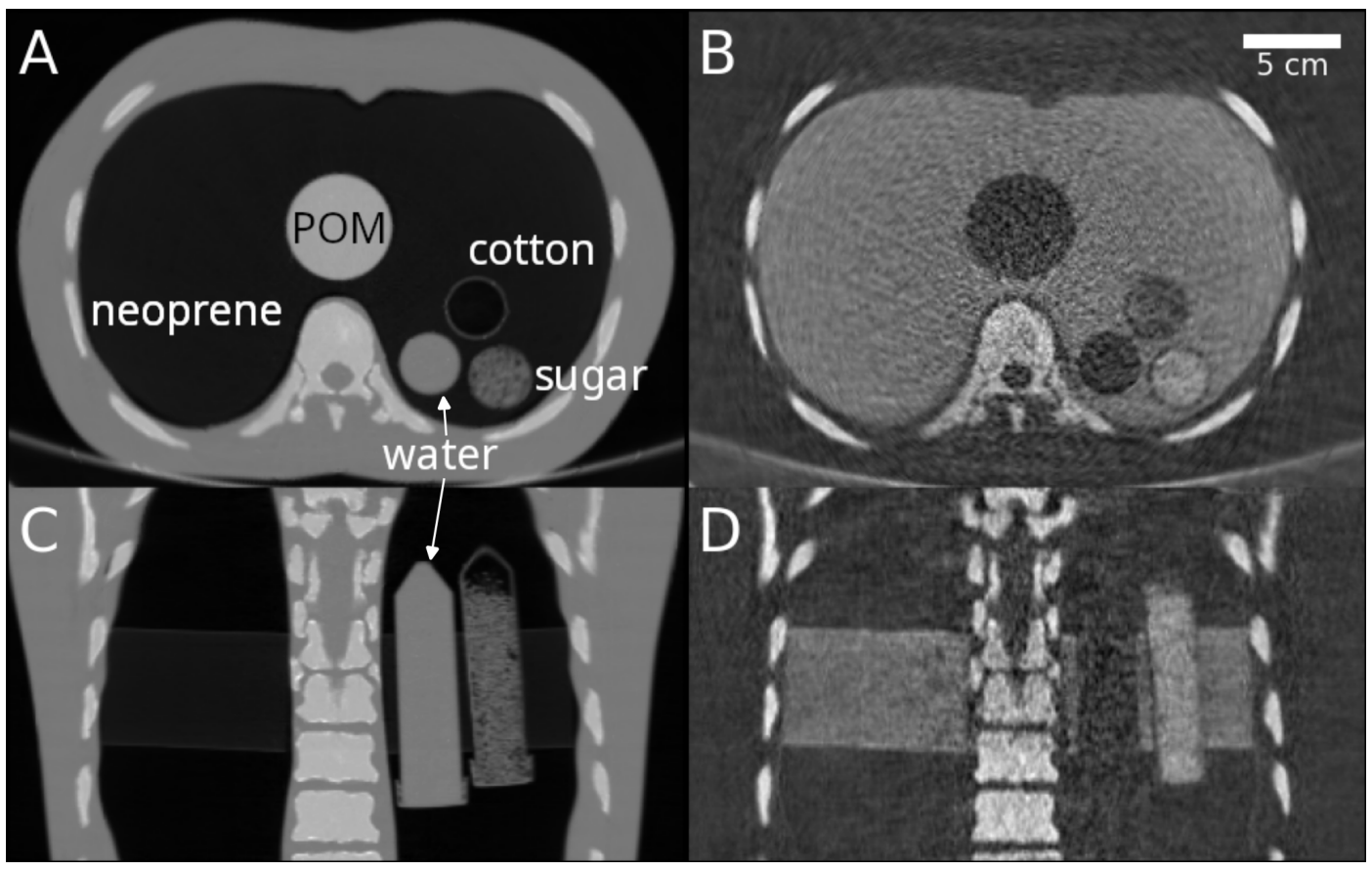

**Figure 4:** Helical CT scan of an anthropomorphic thorax phantom acquired with the dark-field CT prototype. **A** shows a conventional attenuation-based axial slice with window/level 1660/-170 HU, while **B** depicts the dark-field modality. **C** and **D** show respective coronal slices at height of the spine.

Second, a helical dark-field CT measurement was acquired using a current of 540 mA, tube voltage of 80 kVp, and a rotation time of 1.5 s. Using 32 detector rows, 35208 shots were acquired to cover a length of 18 cm in $z$-direction with a pitch of 0.6. The total acquisition time was 22.8 s. The reconstructed helical scan is shown in Fig. 4.
In both the surview and the helical scan, it can be observed that the porous materials – the neoprene insert, dry wool, powdered sugar, and bone surrogate – exhibit a strong dark-field signal, while the polyoxymethylene and water cylinders as well as the soft tissue surrounding the ribs vanish fully in the dark-field. This demonstrates the system's capability to differentiate between porous and homogeneous materials.

## 4 Conclusion

In this work, we presented new results using a dark-field CT prototype based on a clinical system. With our updated scanner, we could extend its imaging capabilities to perform dark-field surview as well as dark-field helical CT measurements. This development brings dark-field CT an important step closer to clinical usage as it demonstrates as a first proof of concept the feasibility of all three standard acquisition modes of clinical CT scanners (axial, helical, and surview) for human-sized dark-field imaging.

## Acknowledgment

This work was supported by the Karlsruhe Nano Micro Facility, a Helmholtz Research Infrastructure at Karlsruhe Institute of Technology. We acknowledge support of the TUM Institute for Advanced Study, funded by the German Excellence Initiative, the European Research Council (H2020, AdG 695045) and Philips GmbH Market DACH.

# Deep Learning Model for SPECT of Thyroid Cancer Using a Compton Camera

Shuo Han[1], Yongshun Xu[1], Dayang Wang[1], Ge Wang[2], and Hengyong Yu[1,*]

[1]Department of Electrical and Computer Engineering, University of Massachusetts Lowell, Lowell, MA, 01854, USA
[2]Department of Biomedical Engineering, Rensselaer Polytechnic Institute, Troy, NY, 12180, USA
[*]Corresponding author, email: hengyong-yu@ieee.org

**Abstract** Thyroid cancer is a malignant tumor which is the most common malignant tumor in the head and neck region. Early diagnosis is crucial for optimal prognosis. Compton camera-based single photon emission computed tomography (SPECT) is a new avenue to study organ functions and pathologies in this context. Compared to gamma camera-based SPECT, Compton camera requires no mechanical collimation and in principle rejects no gamma ray photon. Hence, radiation dose efficiency and/or signal-to-noise ratio will be improved by orders of magnitude for screening and follow-up scans of patients. However, image reconstruction for Compton camera-bases SPECT is challenging, and there is no analytic reconstruction approach. To address this issue, here we propose a deep learning approach to obtain high-resolution image by deblurring weighted backprojection image. Our simulation results show that the proposed framework can effectively deblur backprojection images and produce high-resolution SPECT images.

## 1. Introduction

Thyroid is a very important organ of the human body and serves as the largest endocrine gland. Thyroid gland controls energy production, protein synthesis, and human body regulation by secreting thyroid hormones. Thyroid hormones can promote growth, development, and metabolism. Therefore, when the thyroid function is abnormal, all organs of the human body will be affected, resulting in easy fatigue, depression or persistent hyperactivity, anxiety, insomnia, and severe cases will lead to death [1]. Besides, the 2020 Global Cancer Observatory survey reported that thyroid cancer is responsible for 586,000 cases around the world [2]. Early diagnosis is important for recovery from thyroid cancer. Single photon emission computed tomography (SPECT) can image the anatomical structures such as the location, shape, and size of organs and lesions. And it can also provide information on blood flow, function, and drainage of organs and lesions [3]. This is helpful for early diagnosis of thyroid cancers. In recent years, there has been a growing interest in the development of SPECT imaging technologies and applications. Recent studies have focused on improving image resolution, reducing radiation exposure to the patient, and increasing the accuracy and reliability of SPECT imaging for a variety of medical conditions [4], [5]. Compared with the traditional SPECT imaging system like gamma camera, the advantages of high efficiency and wide energy ranges of the Compton camera make it very attractive in the real-time monitoring [6] and functional imaging.

Among different types of Compton cameras, the design type based on the read-out chip Timepix3 [7] holds great potential for thyroid imaging. Because the Timepix3 does not require any collimator, it can offer a superior detection sensitivity over a broad field of view. Especially in the context of SPECT imaging, it improves the image quality and helps reduce the radiation exposure of the patient. Timepix3 has the capability to promptly transmit each hit pixel to a readout and simultaneously record the time-of-arrival and energy of an incoming gamma rays in the pixel. By possessing information on both the energy and position of the events in the sensor, it is feasible to reconstruct the gamma source's image. However, there is no theoretically exact reconstruction algorithm available yet, particularly for 3D case.

In the recent years, deep learning methods have played an important role in medical image reconstruction [8]. Since there is no theoretically exact image reconstruction algorithm available for Compton camera-based SPECT, in this study, we propose a deep learning framework to solve this problem. To reconstruct high-quality SPECT images from data collected on a Timepix3 camera, we design a deep learning approach to approximate the deconvolution kernel for the 3D SPECT backprojection images. Extensive numerical simulations are performed to train our network. Our preliminary results demonstrate the merits of the proposed approach.

## 2. Materials and Methods

### *2.1. Compton Imaging*

The principle of Compton imaging technology is based on the Compton scatter effect, which refers to the process where an incident photon collides with an extranuclear electrons causing a change in direction and energy loss. In this event, the incident photon transfers a part of its energy to the extranuclear electron, forcing it to escape from the atomic nucleus as a recoil electron. Concurrently, the incident photon itself undergoes energy loss and becomes a scattered photon that is emitted in a different direction.

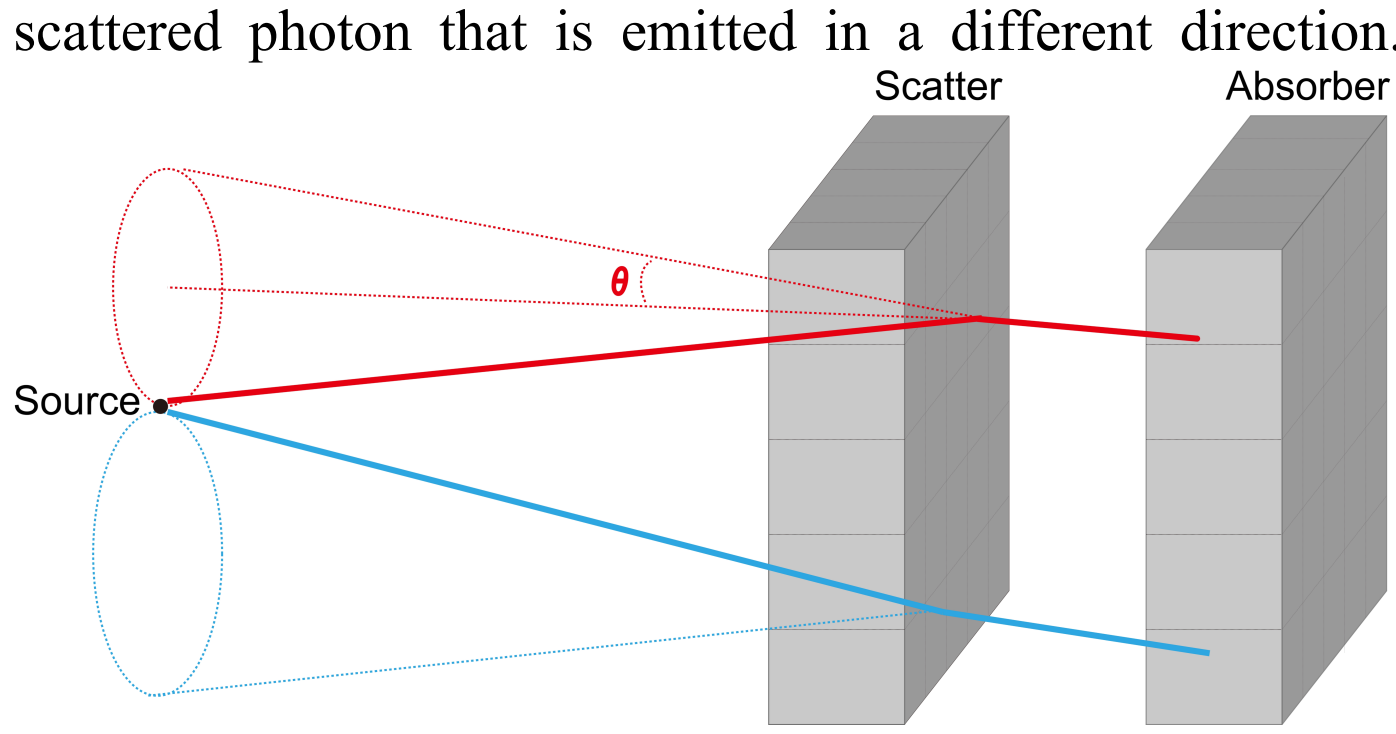

Fig. 1 A typical Compton camera and its imaging event diagram.

During the Compton scatter process, by energy and momentum conservation, the energies of recoil and scattered photons, and the relationship between scattering angle and recoil angle can be derived as follows:

$$E_1 = \frac{E_0^2 \cdot (1-\cos\theta)}{m_0c^2 + E_0 \cdot (1-\cos\theta)}, \quad (1)$$

$$E_m = \frac{E_0}{1+\frac{E_0}{m_0c^2}(1-\cos\theta)}, \quad (2)$$

$$\cot\varphi = \left(1+\frac{E_0}{m_0c^2}\right)\cdot \tan\frac{\theta}{2}. \quad (3)$$

where $E_0$, $E_1$ and $E_m$ are the energies of incident photons, recoil electrons and scattered photons respectively, $\theta$ and $\varphi$ are the scattering and recoil angles respectively, and $m_0c^2 = 0.511$ MeV is the rest mass energy of a free electron.

Generally speaking, the interaction that occurs in the front detector is Compton scattering, while the interaction that occurs in the rear detector is photoelectric absorption. Therefore, for a typical double-layer Compton camera, there is only one Compton scattering. In the case of a single Compton imaging event, the physical process can be characterized as follows: the gamma rays emitted from the radioactive source directly enter the front detector without undergoing scattered. After a single Compton scattering in the front detector, the scattered photon is totally absorbed via photoelectric absorption in the rear detector. Then, the front detector could measure the position of the Compton scattering and the energy of the recoiled electrons, while the rear detector can determine the outgoing direction and energy of the scattered photon. Using Eqs. (1) and (2), the angle of Compton scattering can be estimated as

$$\cos\theta = 1 - \frac{m_0c^2 \cdot E_1}{(E_0 - E_1)\cdot E_0}. \quad (4)$$

### *2.2. Weighted Backprojection*

In this section, we present our reconstruction framework. First, a 3D weighted backprojection (WBP) algorithm in our previous study is used to backproject the obtained projection data into the image domain [9]. In the Compton cameras, a recorded Compton event could be from any position on a Compton cone surface. To reconstruct the underlying source distribution, the WBP algorithm directly backprojects all the events to the corresponding Compton cone surfaces in the 3D image domain. In a 2D image slice, it will trace an elliptical trajectory. In the entire 3D space, the intersection of the cone surfaces yields an initial 3D image of the radioactive source. However, the WBP image contains substantial blurrings. Inspired by the ramp filtering in CT, a 3D ramp filter is applied to the backprojected image for deblurring in our previous work [9]. In this study, the 3D ramp filter is replaced by a 3D Gaussian high pass filter to improve the quality of the image. The 3D Gaussian high-pass filter in the Fourier domain can be represented by:

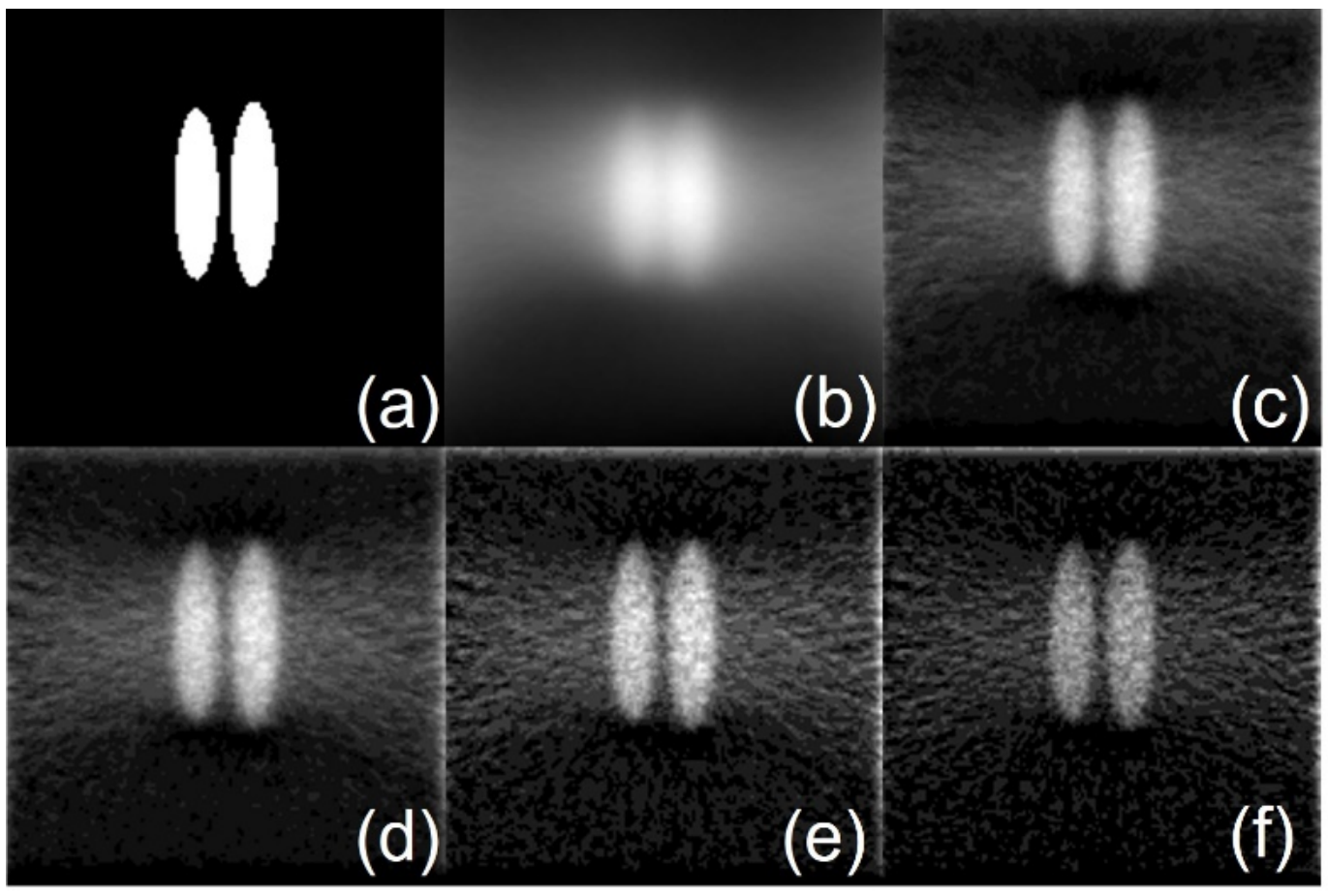

Fig. 2 Representative sagittal views of the Gaussian filtered images. (a) the ground truth, and (b)-(f) the results of d = 1, 2, 4, 6 and 8 respectively.

H(u, v, w) = 1 - e^(-d$^2$ * (u$^2$ + v$^2$ + w$^2$) / (2 * σ$^2$)), where (u, v, w) are variables in the (x, y, z) directions, σ is the standard deviation of the Gaussian function, and d is the cutoff frequency. The cutoff frequency in a Gaussian high-pass filter determines the range in which the filter effectively reject low-frequency signals while passing high-frequency signals outside. Clearly, the cutoff frequency affects the selectivity, transition band and sharpness of a Gaussian high-pass filter. Fig. 2 shows how different cutoff frequencies affect the quality of deblurred images.

### *2.3. EU-Net*

In the medical imaging field, the detection of edges is a critical aspect of image analysis. These edges provide valuable information on boundaries between anatomical structures and tissues. It is essential for diagnosis and treatment planning under various medical conditions. This information is crucial in identifying anomalies, calculating dimensions and volumes, and detecting disease or injury. Inspired by the success of U-Net [10] and MLEFGN [11], here we adapt the Edge-Net from MLEFGN to construct our EU-Net to improve Compton camera-based SPECT image quality (see Fig. 3).

The proposed EU-Net is described as follows. First, Edge-Net to obtain the edge prior of SPECT images. The Edge-Net consists of three modules including the convolution layer, residual block and dense block. It should be noted that Edge-Net is trained with U-Net simultaneously. The input and output of the Edge-Net are blurry SPECT images and clear edges respectively, to learn the capacity to produce clear edges from blurry SPECT images. Then, we use a skip connection to link the backprojected image to the U-Net. Finally, the U-Net takes both the Edge-Net output and the backprojected image as input to generate the recovery image.

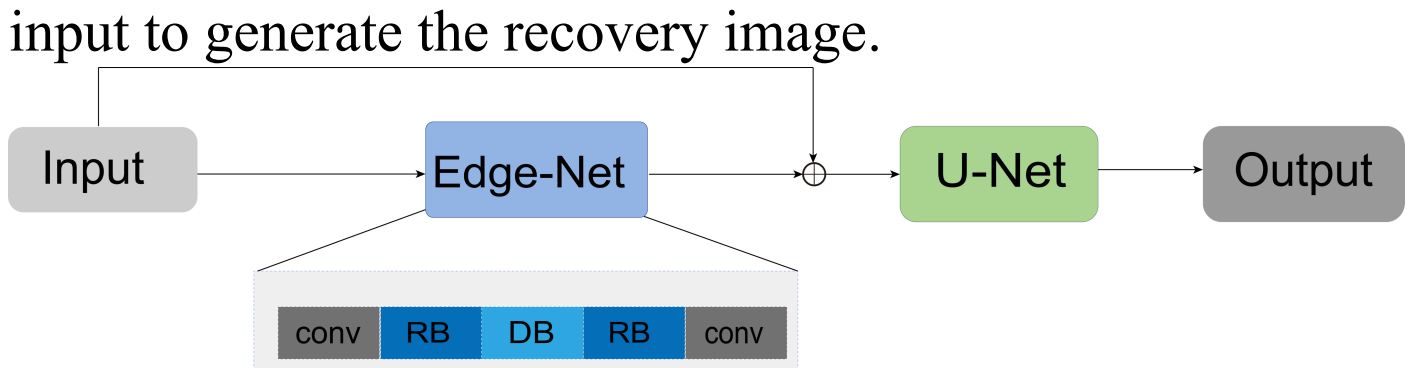

Fig. 3 Architecture of the Compton camera-based SPECT image reconstruction network model EU-Net.

Since the L1 norm is proven to be effective for image recovery [12], both the U-Net and Edge-Net use the L1 norm to enforce the correctness of low-frequency structures. During the training process, the U-Net loss and edge loss are calculated as

$$\mathcal{L}_{\text{U-Net}} = \left\| I'_{\text{clear}} - I_{\text{clear}} \right\|_1, \quad (5)$$

$$\mathcal{L}_{\text{edge}} = \left\| I'_{\text{edge}} - I_{\text{edge}} \right\|_1. \quad (6)$$

where $I'_{\text{clear}}$ represents a predicted image, $I'_{\text{edge}}$ is the predicted edge image, $I_{\text{clear}}$ is the reference image, and $I_{\text{edge}}$ the reference edge image.

## 3. Results

### *3.1. Experimental Setup*

To assess the proficiency and aptitude of the proposed EU-Net, two 3D numerical phantoms are designed for training. Specifically, it is assumed that technetium-99m and cesium-137 are distributed uniformly within these two phantoms, respectively. The gamma rays have energies of 140keV and 662keV. The volume of the two phantoms includes 150x150x150 voxels. The number of Compton events for training datasets are $2*10^5$, $2*10^6$, and $1*10^5$, respectively. Finally, a total of 900 pairs of images are utilized in the training phase. Additionally, a well-designed thyroid phantom used in our previous study [9], utilizing $2*10^6$ events, is employed to test the EU-Net. Both the training and testing dataset are applied by the 3D Gaussian filter, and the cutoff frequency is set to 2. The experiments are conducted on a workstation computer featuring an Intel i9-9920X CPU running at 3.50 GHz, and an NVIDIA GeForce 2080TI graphics processing unit (GPU). The EU-Net is implemented using the PyTorch framework and optimized by the Adam optimizer. The minibatch size is set at 16, and the network is trained for 100 epochs. Finally, 3D rendering images are generated using MATLAB.

### *3.2. Results and Analysis*

Fig. 4 illustrates three different views of the reconstructed SPECT images. The first column represents the ground truths, the second column depicts the WBP results, the third column presents the Gaussian high-pass filtering results, and the fourth column displays the results from the EU-Net. It can be observed that the WBP results are blurry, and it is difficult to distinguish the edges of the simulated thyroid lobes. If the unfiltered results are utilized directly as the input to the network, regardless of the duration of the training process, we cannot obtain satisfactory outcomes. While undergoing Gaussian high-pass filtering, after the high frequency noise removed, the image edges become clearer. However, there still exists significant amount of noise. After further processing by our proposed network, the blurring is efficiently eliminated, and the final results are very close to the ground truths.

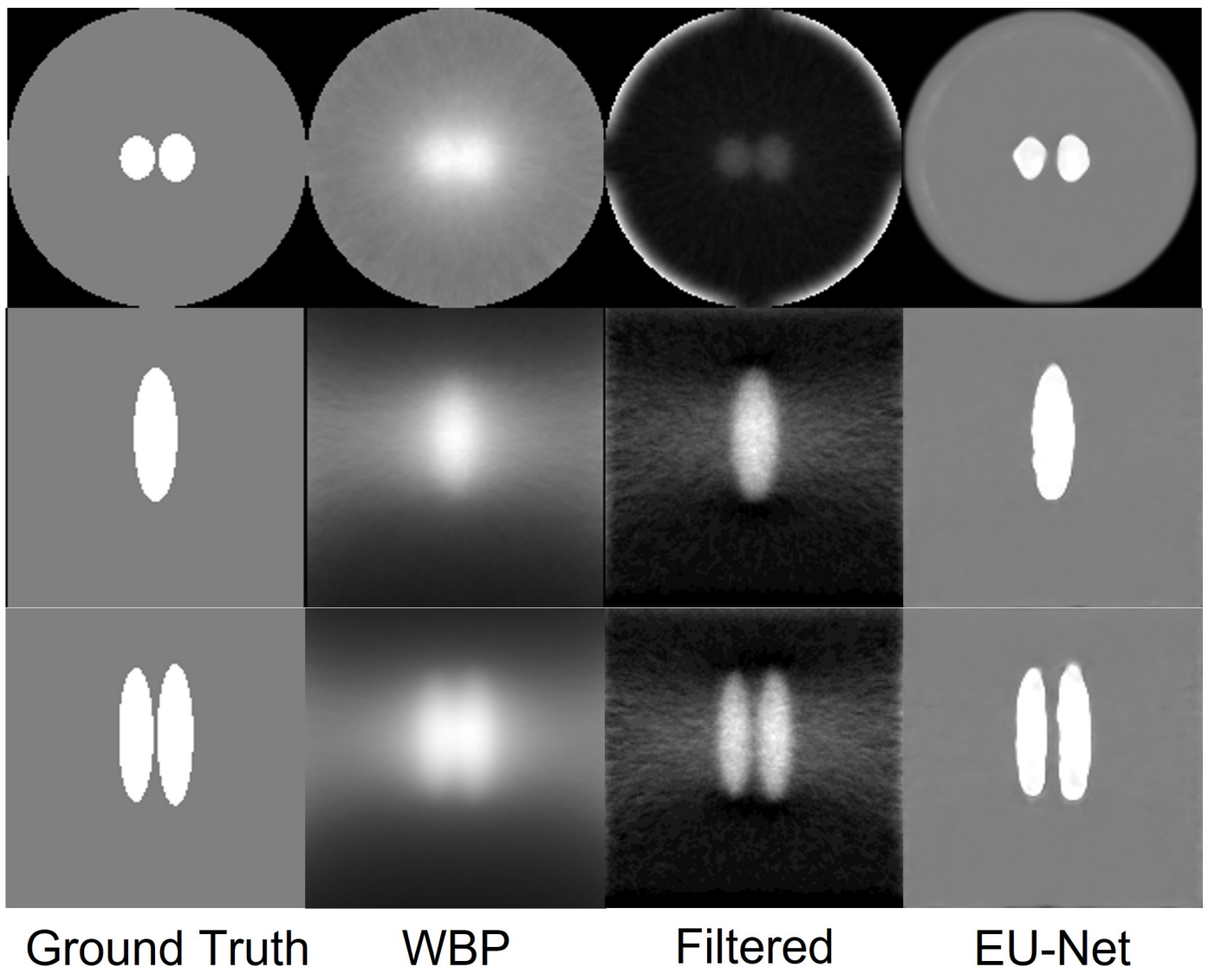

Fig. 4 Three different view of the reconstructed SPECT image from simulated Compton camera data.

Fig. 5 presents 3D rendering results based on different algorithms. They are generated using MATLAB (Version R2021b, MathWorks, Natick, MA) Volume View App via maximum intensity projection. Those results further validate the conclusions observed in the 2D images. It can be observed that the result without Gaussian filtering is very blurry, and the two thyroid lobes are almost connected which makes it difficult for accurately diagnosis. After the Gaussian filtering, the two thyroid lobes are much clear, but they are still a little blurry. However, the performance of our framework remains satisfactory, and two lobes are clearly distinguishable.

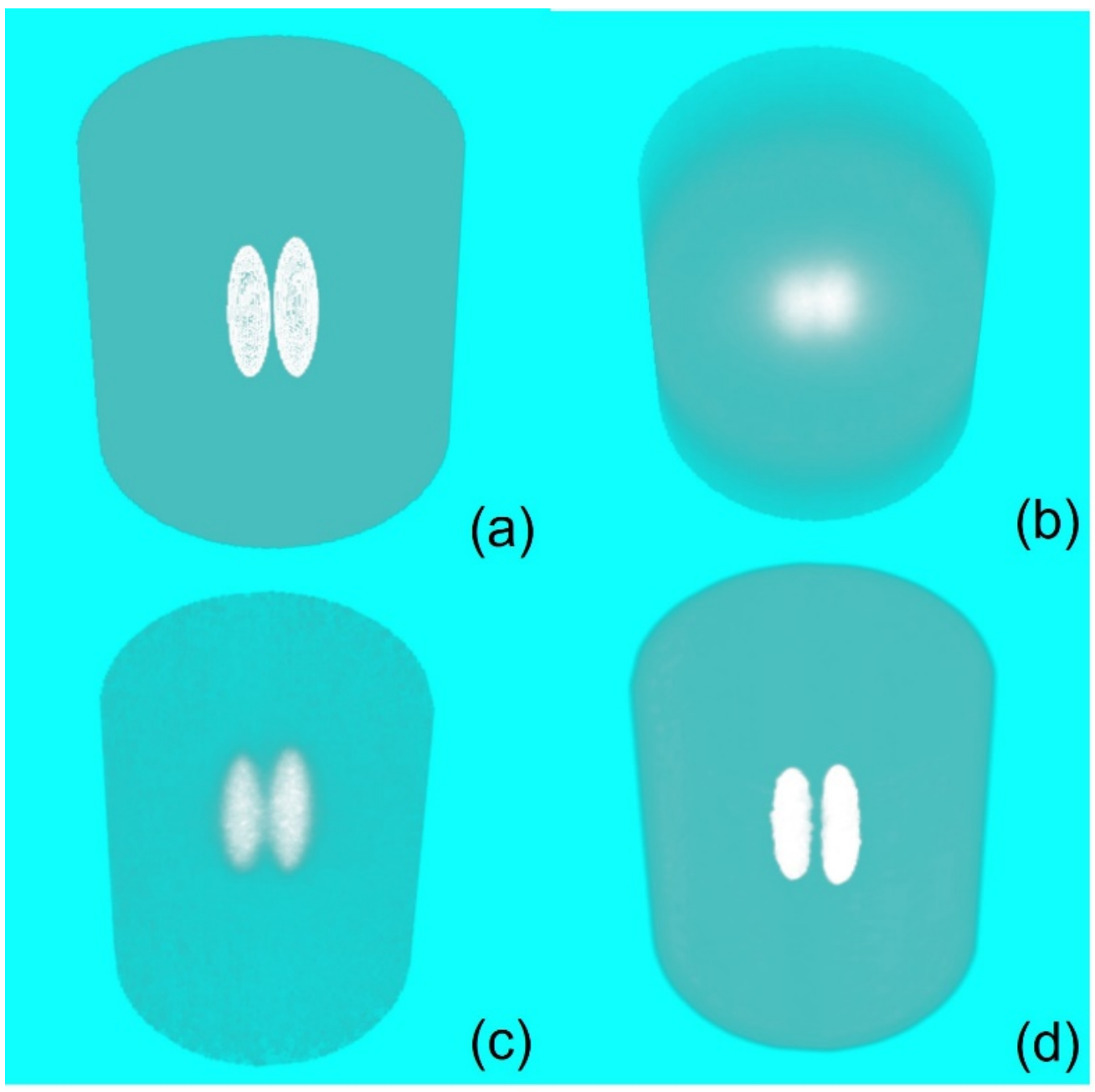

Fig. 5 Three-dimension rendering of the whole volumetric images using the maximum intensity projection. (a) is the ground truth, and (b)-(d) are using WBP, Gaussian filter, and EU-Net, respectively.

## 4. Discussions

Although the Compton camera can be used in many types of cancers, thyroid cancer is well suited for detection using a Compton Camera. There are several reasons as following. First, thyroid cancer is one of the most common endocrine cancers, and its incidence is increasing. Early and accurate diagnosis is crucial for effective treatment, and it can provide patient with the best opportunity for a positive health outcome. Furthermore, traditional imaging modalities such as ultrasound and CT scans have limitations in detecting small and benign thyroid nodules. The use of a Compton Camera, as a nuclear imaging modality, provides a more sensitive and specific alternative for thyroid cancer diagnosis. The high uptake of radioactive isotopes such as iodine by the thyroid gland makes thyroid cancer an ideal target for nuclear imaging, which can effectively capture and analyze the gamma rays emitted to produce high-quality images of the thyroid. Additionally, due to small size of the current Timepix3 Compton camera detector, it is suitable to effectively image small organs such as the thyroid.

Currently, we just use the U-Net as our backbone network to deblur the SPECT image. However, it has some disadvantages. First, the lack of interpretability associated with the model raises questions regarding its ability to provide insights into the underlying mechanism of the deblurring process. Furthermore, the limited generalization of deep learning model presents challenges when applying the model to new and diverse data. Additionally, the risk of overfitting must also be considered when training these models on small or noisy datasets, as it can result in poor performance on unseen data. As can be seen in our results, the quality of our final SPECT images still needs for further improvement. Compressive Sensing (CS) is a highly efficient method for denoising images, and it is totally interpretable. Therefore, by combining the traditional model-based CS method and data-driven deep learning techniques may is the next generation in the field of image reconstruction [13].

## 5 Conclusion

A novel deep learning-based approach is presented in this paper for enhancing the quality of thyroid SPECT images obtained through a Compton camera. In this study, we take the advantages of the WBP and make a modification to this algorithm to generate the pre-backprojection SPECT images. And, then we use the EU-Net to obtain the deblurring reconstruction images. For the inputs with 150 $\times$150 images, the prediction time is about 1.2s. The results of numerical simulations reveal that the proposed network is capable of effectively eliminating blurry and enhance image quality. But, there still exist some challenges at present, including deformations that may arise in the images processed by the network. In the future, we will incorporate more prior knowledge into the neural network model to enhance its scalability and extensibility. Additionally, acquiring some real experimental data to confirm the scalability of the proposed network is under our consideration.

**Acknowledgement:** Research reported in this paper was supported in part by NIH/NCI under award number R21 CA264772.

# High-resolution CT using super-short Zoom-In Partial Scans (ZIPS) and cross-correction of missing data

Eri Haneda[1], Bruno De Man[1], and Lin Fu[1]

[1]Radiation Imaging, GE Research-Healthcare, Niskayuna, NY, USA

**Abstract** The Zoom-In Partial Scans (ZIPS) is a recently introduced CT scanning scheme that utilizes the high geometric magnification in off-center scanning regions to boost the spatial resolution of clinical CT. ZIPS performs *two* complimentary partial scans of a region of interest, then merges the partial data into a high-resolution image. In this paper, we extend ZIPS to more dose-efficient and practical scanning scenarios. First, to minimize patient dose, each partial scan is limited to a 90-degree scan instead of a conventional 180+fan scan. Second, we account for lateral truncation due to the off-center position and the limited size of the CT detector array. Lastly, to fully realize the resolution capability of ZIPS, the two partial scans need to be accurately registered to compensate for the table/patient translation between the two partial scans. Results show the effectiveness of ZIPS in these challenging conditions and much improved resolution relative to a conventional centered CT scan.

## 1 Introduction

The Zoom-In Partial Scans (ZIPS) is a novel CT scanning scheme that improves the spatial resolution of existing clinical multi-slice CT for region-of-interest (ROI) imaging [1]. Unlike a conventional CT scan where the ROI is centered in the field-of-view, in ZIPS the ROI is placed off center, leading to higher geometrical magnification and improving spatial resolution locally when combined with a small focal spot size. For an ROI offset of 20 - 30 cm relative to iso-center, the modulation transfer function (MTF) can improve by 30% to 80% when comparing ZIPS to a conventional centered scan [1,2].

However, there are a few limitations of the existing ZIPS method. (1) Previously, ZIPS was demonstrated in scenarios where each partial scan covers half-scan data (180 +fan angle) [1,2]. This is sub-optimal in terms of dose-efficiency due to a high level of redundancy between the two partial scans. To keep the radiation dose as low as possible, each partial scan may be ideally limited to a super-short quarter-rotation scan (90° fan-beam) while the two scans together can still cover the complete Radon space (180° parallel-beam) of the off-center ROI. However, the limited-angle 90° fan-beam data cannot be directly handled by a standard filtered-backprojection (FBP) algorithm. One may rebin the two limited-angle sinograms into a full sinogram, but the interpolation approximations during rebinning may degrade the resolution of data, especially when considering extension to cone-beam data. More advanced algorithms such as the Katsevich derivative method and the differentiated backprojection method [6-9] may handle limited-angle fan-beam data, but they are not readily avaialble on existing clinical scanners. (2) Lateral truncation is another challenge to ZIPS. Due to the off-center ROI and the increased magnification, the CT detector array may not be large enough to measure the full projection of the patient at some angles. The truncation boundary requires proper handeling to prevent artifacts from propagating into the reconstructed ROI image. (3) Although ZIPS is intended for imaging locally rigid anatomies such as bones, the ROI translation between two partial scans may be impacted by involuntary patient motion or imprecise table translation. We previously used a data-driven image alignment metric to register the two partial scans [2], but in the limited-angle and truncated senarios the feasiblity of such registration remains to be tested due to the less overlap between the partial data.

In this paper, we extend ZIPS imaging to more dose-efficient and practical scanning senarios that entail limited-angle and truncated data. We introduce a cross-correction method to extend the partial sinograms into full sinograms while preserving the high resolution of the original measurements. We further iteratively correct for the missing data and compensate for the actual ROI translation to refine the final reconstructed image.

## 2 Theory

### Notations

We restrict this study to 2-D, although extension to 3-D is conceivable. Let $f(\boldsymbol{x})$ denote the 2-D ROI object we would like to reconstruct, where $\boldsymbol{x} = (x_1, x_2)^T$. The fan-beam projections are measured by moving the x-ray source along a circular trajectory $\boldsymbol{s}(\beta) = (R\cos\beta, R\sin\beta)$, where $\beta$ denotes the angular position of the source and $R$ denotes the source-to-center distance. Let $\mathcal{L}(\beta,\theta)$: $\boldsymbol{x} = \boldsymbol{s}(\beta) + t\boldsymbol{\theta}$, $t \in [0, +\infty)$ denote a projection ray originating from the source in the direction of a unit vector $\boldsymbol{\theta} = (\cos\theta, \sin\theta)^T$. The fan angle of the ray is $\gamma = \beta - \theta$. We denote the fan-beam transform of $f(\boldsymbol{x})$ by

$$p(\beta,\theta) = \int_0^\infty f(\boldsymbol{s}(\beta) + t\boldsymbol{\theta})\mathrm{d}t.$$

### Zoom-in partial scans (ZIPS)

ZIPS utilizes a dual partial scan scheme illustrated in Fig. 1(a). The ROI is scanned at two off-center table positions and remains still during each scan. After the first partial scan is completed, the ROI is translated to the second position, where the second partial scan is performed. Let $\boldsymbol{d}_k = (d_k\cos\varphi_k,\ d_k\sin\varphi_k)^T$ denote the position of the ROI relative to the iso-center in the $k$th partial scan. In the example in Fig. 1(a), $d_1 = d_2$, and $\varphi_1 = 45°$, $\varphi_2 = 135°$ (angle zero is defined at 12 o'clock position). Because of the higher geometric magnification in off-center regions, high-resolution projections of the ROI are measured by the illustrated trajectory arcs. More details can be found in Ref [1].

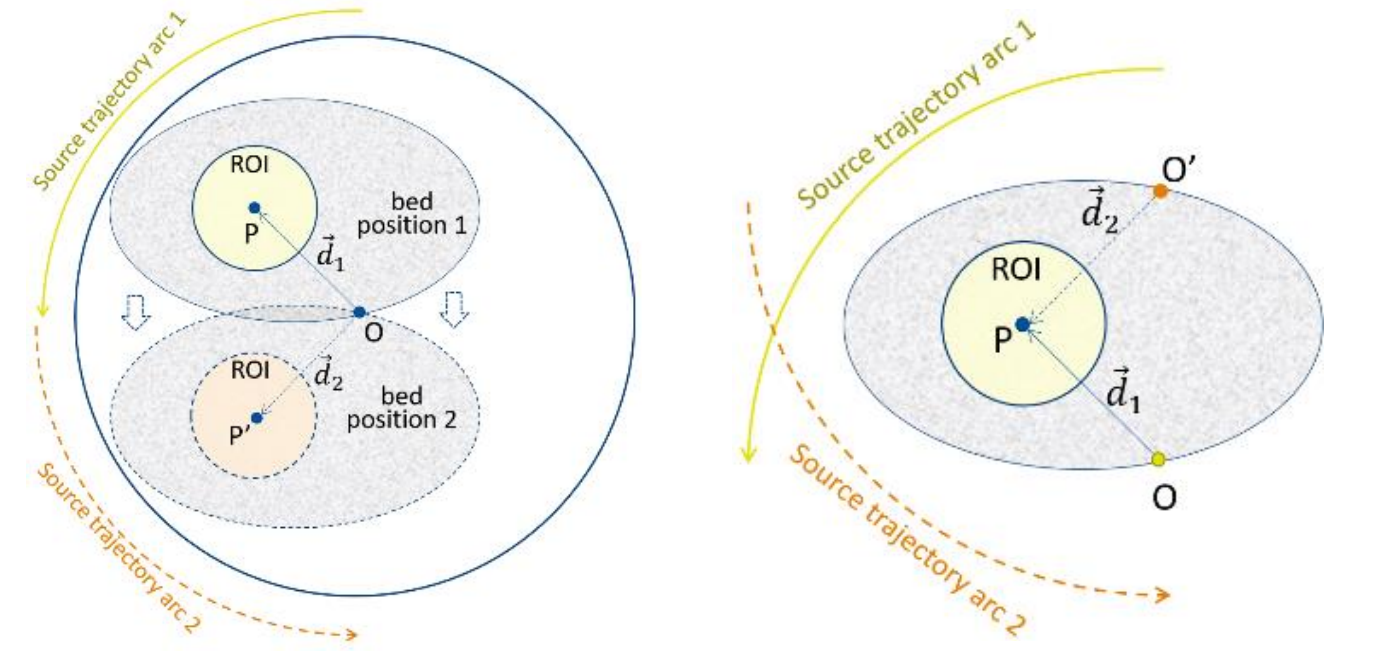

(a) (b)

Fig 1. Illustration of the ZIPS technique. (a) Two limited-angle partial scans are performed to collect high-resolution projections of the ROI in complimentary angular ranges. (b) To help confirm the completeness of data, the same source trajectories are viewed in the ROI's local reference frame, i.e., the ROI is treated as stationary, but the CT scanner virtually translates from $O$ to $O'$.

In the $k$th partial scan, the fan-beam transform of the off-center ROI object is

$$p_k(\beta_k, \theta_k) = \int_0^{\infty} f(\boldsymbol{s}(\beta_k) + t\boldsymbol{\theta}_k - \boldsymbol{d}_k)\mathrm{d}t\,, k = 1, 2.$$

Because of the super-short trajectories and lateral truncation, the measured data is restricted to an interval

$$\Lambda_k = \left\{(\beta_k, \theta_k): |\beta_k - \varphi_k| \le \frac{A_{\max}}{2} \text{ and } |\beta_k - \theta_k| \le \frac{\gamma_{\max}}{2}\right\},$$

where $A_{\max}$ denotes the angle of gantry rotation with respect to the iso-center in each partial scan, and $\gamma_{\max}$ denotes the fan angle of the x-ray beam. In this study, we choose $A_{\max} = \frac{\pi}{2}$ (quarter rotation). Because of the short distance between the x-ray source and the off-center ROI, the quarter rotation is sufficient to cover more than 90° parallel-beam data for the ROI, hence the two partial scans together cover the complete 180° parallel-beam data of the ROI. As shown in Fig. 1(b), any ray through the ROI intersects with at least one of the source trajectories. Depending on the ROI size and offset, it is possible to use $A_{\max} < \frac{\pi}{2}$ to further reduce patient dose while still covering the complete Radon space of the ROI.

### Collinear rays

Although each partial scan suffers from substantial missing data thus cannot be directly handled by a standard FBP algorithm, the two partial scans together cover the complete Radon projection space of the ROI. Thus, if a ray $(\beta_1, \theta_1)$ is not measured by one of the trajectories, it can be interpolated from a collinear ray $(\beta_2, \theta_2)$ from the other trajectory and their conjugate rays.

Let $\mathcal{L}_k(\beta_k, \theta_k)$ denote a ray measured by the $k$th scan. In the ROI's local reference frame, the ray is

$$\mathcal{L}_k(\beta_k, \theta_k): \boldsymbol{x} = \boldsymbol{s}(\beta_k) + t\boldsymbol{\theta}_k - \boldsymbol{d}_k$$

Rays $\mathcal{L}_1$ and $\mathcal{L}_2$ are collinear when

$$\theta_2 = \theta_1 \tag{1}$$

and

$$\text{signed distance}(\mathcal{L}_1, P) = \text{signed distance}(\mathcal{L}_2, P),$$

where $P$ is any point. For simplicity, choose $P = (0,0)^T$, i.e., the center of the ROI, which gives

$$\begin{aligned} &\text{signed distance}(\mathcal{L}_1, \mathbf{0}) = \text{signed distance}(\mathcal{L}_2, \mathbf{0}) \\ &= |\boldsymbol{\theta}_k, \boldsymbol{s}(\beta_k) - \boldsymbol{d}_k| \\ &= R\sin(\beta_k - \theta_k) - d_k \sin(\theta_k - \varphi_k) \end{aligned} \tag{2}$$

where $|\cdot|$ denotes matrix determinant. Eq. 1 & 2 give a condition by which the rays $(\beta_2, \theta_2)$ and $(\beta_1, \theta_1)$ are collinear.

### Cross correction between partial scans

Using the above collinearity condition, we can extend partial sinograms into full sinograms by cross correction between $p_1$ and $p_2$.

$$\hat{p}_k(\beta_k, \theta_k) = \begin{cases} p_k(\beta_k, \theta_k), & \text{if } (\beta_k, \theta_k) \in \Lambda_k \\ p_{\tilde{k}}(\beta_{\tilde{k}}, \theta_{\tilde{k}}), & \text{if } (\beta_k, \theta_k) \notin \Lambda_k \text{ and } (\beta_{\tilde{k}}, \theta_{\tilde{k}}) \in \Lambda_{\tilde{k}} \\ \text{extrapolation}, & \text{otherwise} \end{cases} \tag{3}$$

where $\hat{p}_k$ is the extended $k$th sinogram; $k = 1,2$; $\tilde{k} = 3 - k$; $(\beta_{\tilde{k}}, \theta_{\tilde{k}})$ is the collinear ray of $(\beta_k, \theta_k)$ found by Eq. 1 & 2. In the extended sinogram, $\beta_k \in [0,2\pi)$ and $\gamma_{\max}$ is increased to cover the full projection of the ROI. The cross-correction method addresses both the limited-angle and truncated data. If any missing data cannot be interpolated by the cross correction, we perform linear extrapolation along the lateral direction of the sinogram to ensure smooth transition at the boundary of the available data. The resulting full sinograms can be directly processed by a standard FBP algorithm. It should be noted that the contribution of the interpolated data will be filtered out in subsequent processing steps and will not degrade the resolution of the final reconstructed image.

### Image reconstruction and registration

We reconstruct intermediate images from the extended sinograms by Parker-weighted FBP reconstruction

$$\hat{f}_k(\boldsymbol{x} - \boldsymbol{d}_k) = \left(\mathcal{F}^{-1} m_{\varphi_k}(\boldsymbol{\rho}) \mathcal{F}\right)\mathcal{R}[\hat{p}_k(\beta, \theta)], \tag{4}$$

where $\hat{f}_k$ is a partial image reconstructed from $\hat{p}_k$; $\boldsymbol{x} - \boldsymbol{d}_k$ accounts for the off-center location of the ROI; $\mathcal{R}$ denotes the Parker-weighted FBP operator; $\mathcal{F}$ and $\mathcal{F}^{-1}$ denote the Fourier transform and its inverse; $m_{\varphi_k}(\boldsymbol{\rho}) =$

$$m_{\varphi_k}(\rho\cos\zeta, \rho\sin\zeta) = \begin{cases} 1, & \text{when } |\zeta - \varphi_k + n\pi| \le \frac{\pi}{4} \\ 0, & \text{otherwise} \end{cases}$$

is a Fourier-domain filter that selects a 90° subset of spatial frequencies from the image. The Fourier mask is aligned to the direction $\varphi_k$ of the ROI offset thus it extracts a 90° range of angular frequencies only measured by the $k$th partial scan and, in the meantime, suppresses other angular frequencies such as from the interpolated data. With the Fourier masking, $\hat{f}_1$ and $\hat{f}_2$ contains complementary angular frequencies.

To generate a final output image $\hat{f}_{\text{zips}}$, the partial images $\hat{f}_1$ and $\hat{f}_2$ are registered and merged in the image domain

$$\hat{f}_{\text{zips}}(\boldsymbol{x}; \Delta\boldsymbol{d}, \Delta\theta) = \hat{f}_1(\boldsymbol{x}) + \hat{f}_2(\mathbf{T}_{\Delta\theta}(\boldsymbol{x} - \Delta\boldsymbol{d})),$$

where $\Delta\boldsymbol{d} = (\Delta x, \Delta y)^T$ and $\Delta\theta$ denote the relative translation and rotation between $\hat{f}_1$ and $\hat{f}_2$. $\mathbf{T}_{\Delta\theta} = \begin{bmatrix} \cos\Delta\theta & -\sin\Delta\theta \\ \sin\Delta\theta & \cos\Delta\theta \end{bmatrix}$ is a rotation matrix. The relative motion is not known exactly because of imprecise table translation and involuntary patient motion. The motion vectors are estimated by minimizing the entropy of the output image:

$$\Delta\boldsymbol{d}, \Delta\theta = \underset{\Delta\boldsymbol{d}, \Delta\theta}{\operatorname{argmin}} J_{\text{entropy}}\left(\hat{f}_{\text{zips}}(\boldsymbol{x}; \Delta\boldsymbol{d}, \Delta\theta)\right). \quad (5)$$

For more details on image registration, please refer to [2].

**Iterative correction**

Given the estimated motion vectors $\Delta\boldsymbol{d}$ and $\Delta\theta$, the collinearity condition (Eq. 1 & 2) should be adjusted and a new round of extended sinograms may be generated. The updated collinearity condition including the effect of the motion vectors is

$$\theta_2 = \theta_1 + \Delta\theta,$$
$$|\boldsymbol{\theta}_2, \boldsymbol{s}(\beta_2) - (\boldsymbol{d}_2 + \Delta\boldsymbol{d})| = |\boldsymbol{\theta}_1, \boldsymbol{s}(\beta_1) - \boldsymbol{d}_1|,$$

where $|\cdot|$ denotes matrix determinant. The image reconstruction steps above, including the cross correction, the FBP reconstruction, and the image registration (Eq. 3, 4 & 5) are repeated iteratively to refine the quality of the final ZIPS reconstruction $\hat{f}_{\text{zips}}$.

## 3 Results

The new ZIPS imaging senarios and the corresponding image reconstruction algorithm were evaluated in a CatSim environment [3]. A 22 cm diameter phantom containing resolution features was scanned with a GE Revolution CT with 832 detector pixels, 1.1 mm detector cell pitch, 626 mm source-to-isocenter distance, 1098 mm source-to-detector distance, and 80 cm diameter bore opening. The phantom was positioned at the isocenter for standard CT, and nominally at 20 and 30 cm off center for ZIPS. Each ZIPS acquisition consisted of two 90° gantry rotations. Lateral truncation of projection data was incurred in both 20 cm and 30 cm offset cases. To emulate the effect of involuntary patient motion between the two partial scans, the position of the phantom in the second partial scan was perturbed by $[\Delta x, \Delta y, \Delta\theta]$ = [3 mm, 6 mm, 12°] for the 20 cm offset case, and $[\Delta x, \Delta y, \Delta\theta]$ = [3 mm, 6mm, 4°] for the 30 cm offset case.

All scans used a 70 keV monochromatic x-ray. The standard CT used a flux of 800 mAs. ZIPS used a total flux (including both partial scans) of 800 mAs for the 20 cm offset case, and 1600 mAs for the 30 cm offset case, respectively. The higher flux for the 30 cm offset case was to ensure the higher image resolution was not limited by statistical noise. We evaluated the effect of three angular sampling rates (1000 views/rot, 4000 views/rot, and 8000 views/rot) and four focal spot sizes (0.8 mm, 0.4 mm, 0.3 mm, and 0.1 mm). A 10° anode angle was modeled in the simulation.

The CT images were reconstructed on a 1024×1024 pixel grid over a 10×10 cm and 30×30 cm square ROI. Shepp-Logan filter was used in all cases. The perturbation of the phantom was unknown to the image reconstruction algorithm. Four iterations of cross-correction and registration were used for ZIPS reconstruction.

Fig. 2 shows the measured partial sinograms and algorithmically extended full sinograms after four iterations of correction of missing data and registration. The majority of the missing rays were filled by the cross-correction between the two partial sinograms. The remaining missing rays were extrapolated by linear extrapolation.

Fig. 3 shows the ZIPS reconstructed image in comparison with the standard CT image. The visual sharpness and contrast of features are improved in the ZIPS image compared with the standard CT. The 400 µm features not distinguishable in the standard CT images are clearly revealed by the ZIPS images. On the other hand, Disabling the various correction components in the ZIPS reconstruction algorithm resulted in biased, saturated, or distorted images (Fig. 3(c), (d), (e)). The residual errors of the motion vectors estimated by the proposed method were no greater than 0.06 mm in translation and 0.08° in rotation.

Fig. 4 shows the MTFs of various imaging techniques. Improvement of the MTF by up to 80% (MTF@10%) is observed when comparing ZIPS to conventional CT.

## 4 Conclusion

This paper improved the dose-efficiency and practicality of ZIPS by introducing super-short quarter-rotation partial scans and effectively correcting for missing data and inter-scan ROI motion duirng image reconstruction. Clear improvement of visual detectability of fine features and up to 80% improvedment of MTF was achieved by ZIPS relative to standard CT. ZIPS does not require an upgrade of the CT detector array and thus has the potential to be applied to existing clinical CT systems. ZIPS CT is also orthogonal to pure algorithmic resolution-boosting methods and a combination may give further improvement.

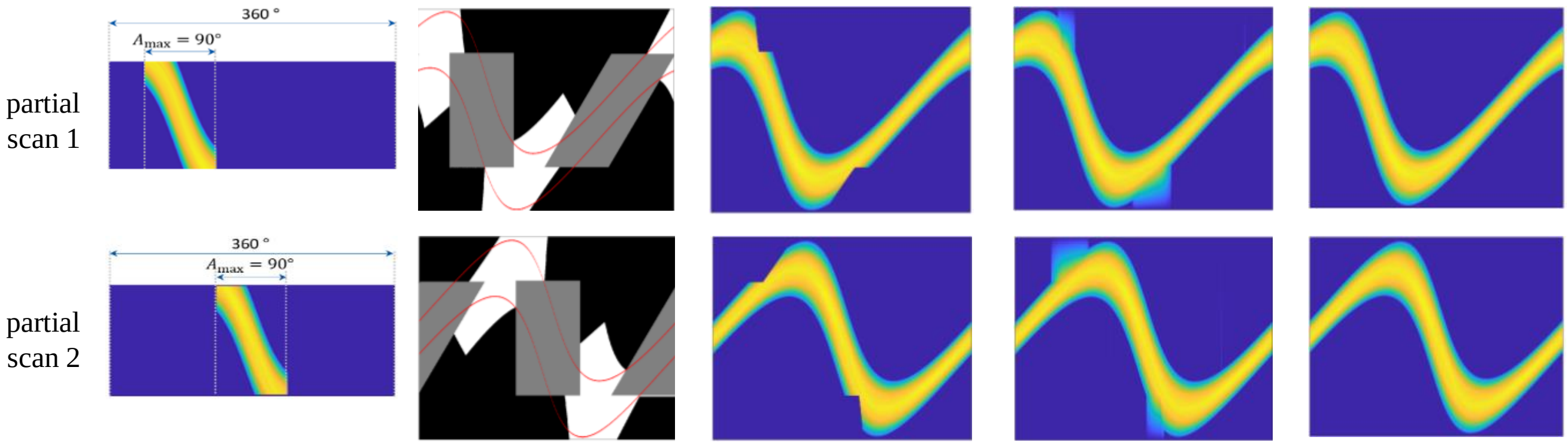

(a) original limited-angle and truncated sinograms (b) maps of data availability (c) extended sinograms with cross-correction (d) extended sinograms with cross-correction and linear extrapolation (e) ground truth

Fig. 2. Correction of missing data in the sinogram domain. In the map of data availability (column b), the gray region represents the measured rays and their conjugate rays; the white region represents the rays from the complimentary partial scan; the black region represents rays not measured by either partial scan. The sinograms are from a 22 cm diameter phantom scanned with a 30 cm offset. The contours of the partial sinograms (column a) are visually similar because the phantom is circular, but their internal details are different.

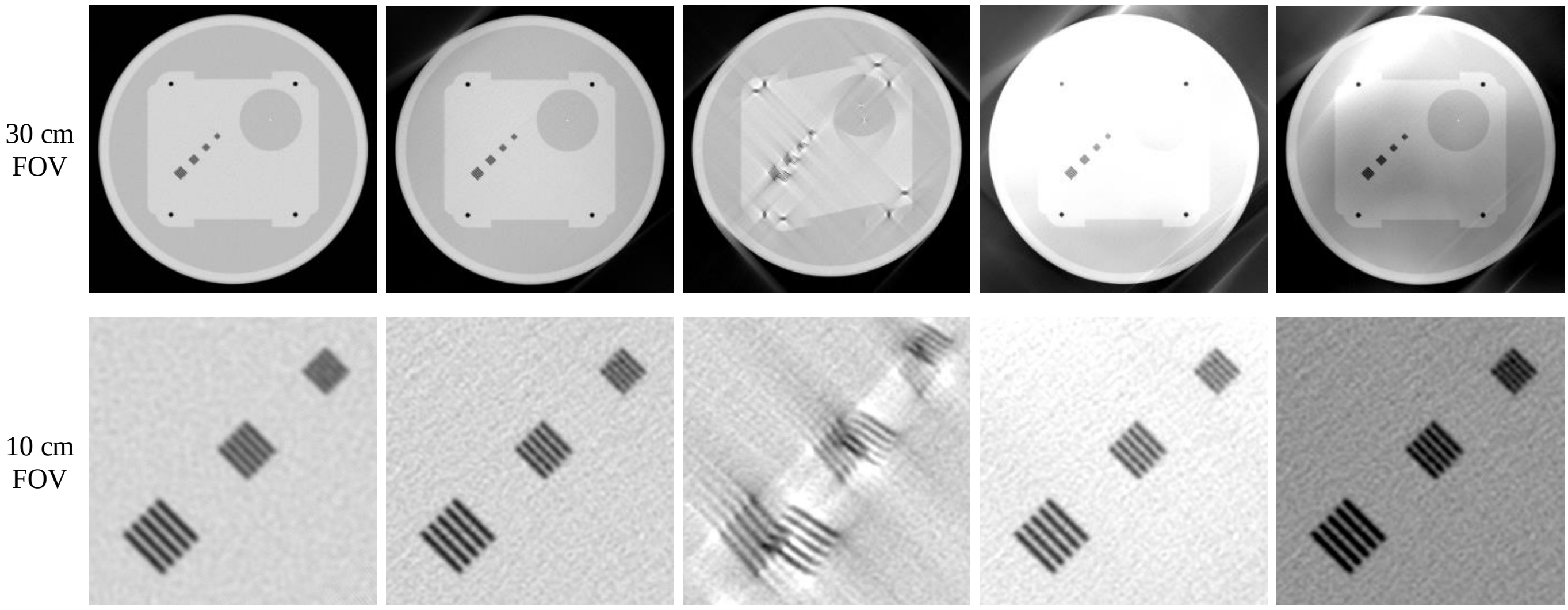

(a) Standard CT (b) ZIPS with all corrections (c) ZIPS without registration (d) ZIPS without truncation correction (e) ZIPS without limited-angle correction

Fig. 3. Reconstructed images of different scanning and reconstruction techniques. The ZIPS image with all corrections shows superior resolution relative to the standard CT image. The widths of individual resolution bars in the 10 cm FOV images are 600, 500, and 400 µm. All cases used 0.3 mm focal spot size and 8000 views/rot. The ZIPS images used 20 cm offset. The display window is [-1000, 400] HU.

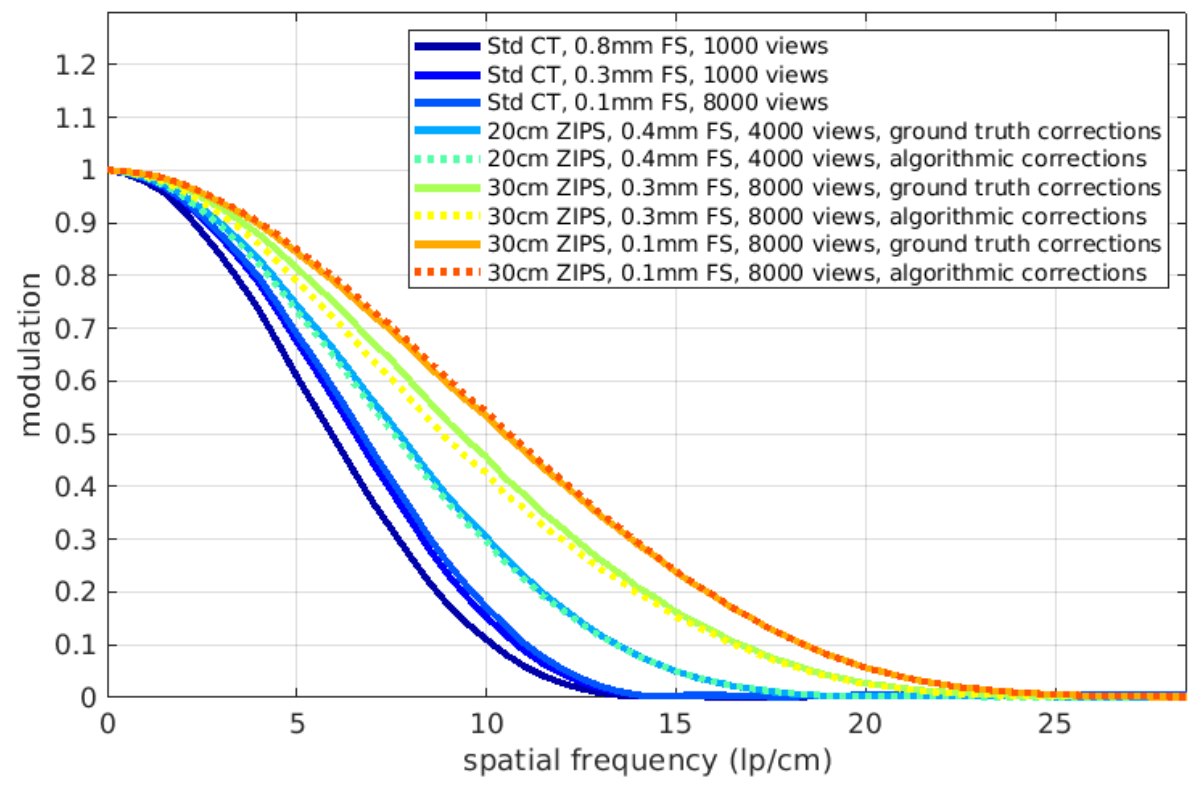

Fig. 4. MTFs of various CT techniques measured with a 50 µm diameter wire. "FS" denotes focal spot size.

## Acknowledgements

The authors would like to thank Dr. Jed Pack, GE Research - Healthcare, and Drs. Ryan Breighner and Fernando Quevedo Gonzalez, Hospital for Special Surgery, for valuable discussions. Research reported in this publication was supported by NIBIB of the National Institutes of Health under grant number R01EB028270. The content is solely the responsibility of the authors and does not necessarily represent the official views of the NIH.

# Dynamic contrast peak estimation by gamma-variate-convolution model for autonomous cardiac CT triggering

Eri Haneda[1], Pengwei Wu[1], Isabelle M. Heukensfeldt Jansen[1], Jed Pack[1], Albert Hsiao[2], Elliot McVeigh[3], and Bruno De Man[1]

[1]GE Research-Healthcare, Niskayuna, NY, USA
[2]Department of Radiology at University of California San Diego, 9500 Gilman Dr, La Jolla, CA, USA
[3]Departments of Bioengineering, Cardiology and Radiology at University of California San Diego, 9500 Gilman Dr, La Jolla, CA, USA

**Abstract** In cardiac CT, it is important to time the scan when the target chambers and vessels are near their peak contrast enhancement. In traditional bolus tracking, this timing is based on the increasing phase of observed time-intensity curve measured in a single region-of-interest (typically in the left heart) and is done empirically by imposing a diagnostic delay after the intensity reaches a pre-defined threshold. We are interested in more accurately predicting contrast peak arrival time of the left heart by analyzing the early contrast dynamics in both the right and left heart regions. We define a gamma-variate-convolution (GVC) model and we fit this model to the increasing phase up to the peak of the intensity curves in the right heart and the beginning phase of the left heart to predict the peak enhancement time of the left chambers. Myocardial perfusion CT datasets were analyzed to demonstrate that the model fits real patient CT contrast dynamics well. The same data was used to demonstrate prediction of the bolus peak enhancement time based on the early observation of bolus time-intensity curves of the heart halves. These times were compared with the optimal time, as defined by the bolus peak prediction time using full intensity measurements. The error was smaller than 1.1 sec in all cases.

## 1 Introduction

Intravenous contrast medium is often injected into a vein in the arm for better visualization of heart structure and blood vessels in cardiac CT. Since the concentration of injected contrast medium is dispersed throughout vessels and cardiac chambers after injection, it is not trivial to predict the respective contrast dynamics. In cardiac CT exams, scan timing needs to be carefully selected so that the contrast bolus is at its peak enhancement in the target vessels and chambers [1]. Timing the CT scan to coincide with the peak contrast concentration is traditionally done either with a 'timing bolus' or with 'bolus tracking'. With a timing bolus, a small volume of contrast is injected in a patient during a test session and repeated single-slice axial scans are performed to track the time-intensity curve and to estimate the delay between the start of the injection and the peak enhancement. After this, the diagnostic cardiac CT exam is performed with the full contrast bolus and the CT scan is started after the estimated delay. With bolus tracking, there is no test session: the full contrast bolus volume is injected, and single-slice axial scans are performed until the CT number in a region-of-interest reaches a predefined threshold. Then, the diagnostic scan will start after the 'diagnostic delay' of several seconds. During the diagnostic delay, the scan table is repositioned, breath hold instructions are delivered, and the scanner collimation is reconfigured. Both approaches require highly trained operators to achieve consistent bolus enhancement.

Our goal is to develop a smart cardiac CT scanner that autonomously triggers the scan by estimating bolus peak time from analyzing time-intensity curves of multiple cardiac chambers [2, 3]. For example, the bolus curve at the left heart is more dispersed than the right heart with a time delay, hence the bolus peak time of the left heart may be estimated from the right heart bolus curve if we can analytically model the relationship. One could train an end-to-end neural network that directly estimates the bolus peak, but it may be challenging to collect sufficient data to train a robust network for this task. We here present a gamma-variate-convolution (GVC) analytical model to formulate the bolus dynamics and to help estimate the peak arrival time. We demonstrate that the model can match to actual measurements from myocardial perfusion CT by comparing it with other reference models. Furthermore, we demonstrate that this simple model can predict the future bolus peak arrival time for a selected left cardiac chamber from the increasing phase up to the peak of time-intensity measurements of contrast dynamics in the right chamber and the beginning phase of left chamber.

Many mathematical models have been proposed for the time-intensity curves of contrast medium injected to human body. Bae et. al. successfully established a simulated model of aortic and hepatic CT contrast enhancement in a patient based on a compartment kinetic model of a whole body using more than 100 differential equations [4]. Although the compartment kinetic model is controlled by physiologically meaningful parameters and gives a rigorous estimation of time-attenuation curves, people often use more tractable and simpler analytical models. Analytical expressions for time-intensity curves are often found in indicator dilution studies [5]. The goal of indicator dilution is to characterize hemodynamic parameters such as patient cardiac output and blood flow by infusing an indicator intravenously [6]. Dye (indicator) is injected to a patient upstream, and the concentration of dye at specific location of vessel downstream is measured to obtain a time-intensity curve. Several methods are available for the measurements, ranging from the classic invasive methods such as dye or thermodilution, to different imaging modalities such as ultrasound, magnetic resonance imaging, and computed tomography [5, 8]. Strouthos et. al. reviewed popular analytical models [5], including a lognormal function, a gamma-variate function, and a lagged normal function. The gamma-variate is probably the most commonly used mathematical function in this context. Although lognormal function and gamma-variate models were empirically defined based on their resemblance with measured curves [8], some papers try to provide a physical interpretation for these mathematical models. For example, for the gamma-variate model, flow in a blood vessel is modeled as a series of mixing compartments [7]. By solving the differential equations that describes the concentration change of each compartment, the gamma-variate time-intensity curve $I(t)$ can be described by:

$$I(t) = \frac{(t-t_0)^{\alpha} e^{\left(-\frac{t-t_0}{\beta}\right)}}{\beta^{(\alpha+1)}\Gamma(\alpha+1)} \delta(t > t_0) \quad (1)$$

$$\Gamma(\alpha+1) = \int_0^{\infty} x^{\alpha} e^{-x} dx, \quad (2)$$

where $\Gamma(\cdot)$ is a gamma function, $\alpha$ designates the number of theoretical mixing compartments, which in turn reflects the degree of turbulance in the flow, and $\beta$ is the ratio of the volume of a mixing compartment to the flow rate.

According to [5], all analytical models show very similar fast wash-in and slow wash-out trends [5]. However, we have observed slightly different trends in the myocardiacl perfusion data, where the wash-in slows down before the peak and the wash-out accelerates after the peak, especially in the right cardiac chambers. For this reason, we use the *gamma-variate-convolution* (GVC) model, where the gamma-variate is used as an impulse response of a dilution system, as was proposed by Mischi et al [8]. Although the gamma-variate function is typically used as a model of the actual output time-intensity curve, Mischi et. al stated that the gamma-variate model can be interpreted as the impulse response of a dilution system using a stochastic derivation, and the time-intensity curve at the site of interest is given by the convolution of the injection function and the gamma-variate function [7, 8].

An advantage of the GVC model is that we can simultaneously model the right and left cardiac chambers by tuning a small set of physiologically meaningful parameters. We will compare the GVC model to traditional anlytical models (i.e. lognormal and original gamma-variate) by fitting each model to actual measurement data from myocardial perfusion CT. We will also demonstrate that the GVC model can predict future peak arrival time in the left cardiac chamber from the increasing phase up to the peak of time-intensity measurements in the right chamber and the beginning phase of left cardiac chamber.

## 2 Materials and Methods

### Gamma-variate-convolution (GVC) model

Our gamma variate convolution model originates from gamma-variate function where a blood pool is modeled as a series of mixing compartments, each completely stirred and of equal volume. Instead of using the model to describe a final output, we will use it as an impulse response of the dilution system. To simplify the data fitting process, we first replaced the scaling factor of gamma-variate equation with a single parameter $A$ by

$$f(t) = A(t-t_0)^{\alpha} e^{\left(-\frac{t-t_0}{\beta}\right)} \delta(t > t_0) \quad (3)$$

where the other physiological parameters were preserved. This function is convolved with an injection function. In our study, a simple rectangular function with injection duration $p$ is used,

$$h(t) = f(t) * rect(p). \quad (4)$$

Figure 1 shows examples of time-intensity curves of the GVC model with different parameter settings. While $\alpha$ and $\beta$ control dispersion level uniformly (Fig. 1 (a)(b)), the parameter $p$ changes the shape of the wash-in and wash-out in a specific way. For smaller $\alpha$ (Fig. 1 (c)) this shape change is more obvious than for larger $\alpha$ (Fig. 1 (d)). This behavior exactly matches our observations in right versus left cardiac chambers, as shown later in the result section. In this study, $A$ and $p$ values were estimated for each patient but do not change between right and left heart, whereas $\alpha$ and time delay $t_0$ were re-estimated for each heart half. $\beta$ was empirically chosen and fixed to 3.5 across all patients.

### Model parameter estimation

The parameters of the GVC model (and other reference models) were optimized by minimizing the mean squared error between the model prediction and measurements. MATLAB's constrained optimization tool with Simplex method was used for the error minimization [9, 10]. Since a positive constant value was observed at the static state, the constant value (50HU) was added to all models. In general, it is known that bolus intensity converges slowly with secondary peaks due to the blood recirculation. For our application only the first pass is of interest, thus we exclude all data after the peak below the threshold (150 HU) from the optimization.

### Bolus peak estimation

Traditionally, scan timing is performed based on the contrast in the left heart, such as the left ventricle and the ascending aorta. Here we use time-intensity data both from the pulmonary artery (PA) and the left atrium (LA) for a time interval that includes the peak of the PA as well as the initial rise of the LA. First model parameters $A, p, t_0^{PA}$ and $\alpha^{PA}$ are estimated from the increasing phase of PA data up to the peak. Then, $A$ and $p$ are fixed and a new $t_0^{LA}$ and $\alpha^{LA}$ are estimated from the partially observed LA data. Based on these estimated parameters, the complete LA time-intensity curve is identified, and its peak time can be computed.

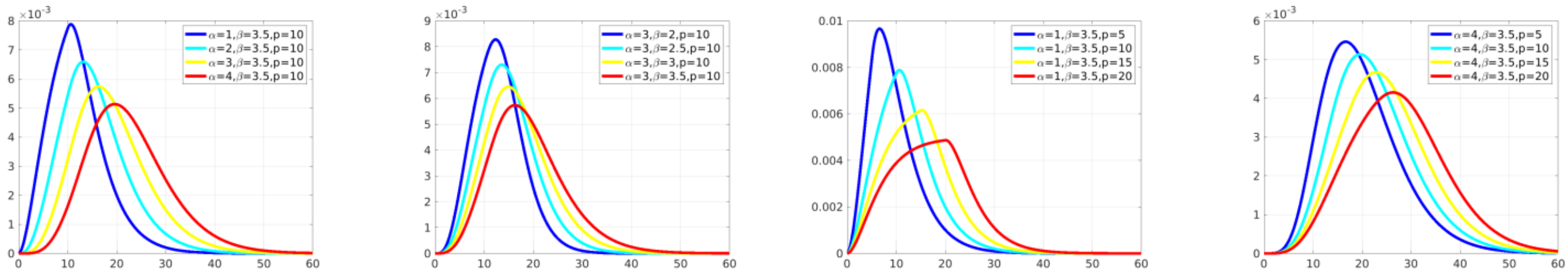

(a) (b) (c) (d)

Figure 1: Time-intensity curves generated by our gamma-variate convolution (GVC) model with different values for three parameters.

## 3 Results

We used five myocardial perfusion datasets acquired at University of California San Diego using a GE Revolution CT scanner. Iodine contrast agent (bolus) was administered to each patient, and about 26 ECG-gated CT scans were acquired for each exam, with 1.5-3.0 sec intervals and at 100 kVp. Each gated data was reconstructed by 512x512x128 voxels covering a 32-cm-diameter field-of-view and 16 cm longitudinal coverage. Scan times for all scans were recorded.

Three examples of time-intensity curves from myocardial perfusion datasets are shown in Figure 2. We first manually found slabs, each visually best represents right ventricle (RV), pulmonary artery (PA), left atrium (LA), left ventricle (LV), ascending aorta (AA), and descending aorta (DA). Then, we drew a circular region-of-interest (ROI) and averaged the values in the ROI over seven neighboring slices to measure the mean intensity. The ROI was selected for each scan to make sure the ROI is within each chamber. The time-intensity curve for superior vena cava (SVC) was excluded from the plot due to the spatial nonuniformity and extreme HU values. This is because the SVC is closer to the injection site and the bolus is not well mixed. The curve of right atrium (RA) could be noisy from the same reason. Therefore, for RA, larger non-circular ROIs were manually segmented to average the signal intensity to compensate for occasional poor mixing. At other locations, we observed a well mixed bolus with uniform values.

Figure 2 shows that the shape of three curves for the right cardiac chambers (RA, RV, and PA) look similar, except for the relative delay. Similarly, the four curves for the left cardiac chambers (LA, LV, AA, and DA) look very similar. However, the curves for the left chamber are more dispersed than the curves for the right chambers due to the effect of the lung circulation. Figure 3 shows three models (GVC, log normal and gamma-variate) fitted to the myocardial perfusion PA and LA using all measurement data, for the three example patient datasets (a, b, and c). The estimated parameters for the GVC are listed in Table 1. For the PA, the GVC model visually better represents the shape of the wash-in and wash-out curves compared to the lognormal and gamma-variate functions. Even though there are only

Table1: Estimated parameters of GVC model

| | $A$ | $p$ | $t_0$ (PA/LA) | $\alpha$ (PA/LA) |
|---|---|---|---|---|
| ID26689 | 629.28 | 17.83 | -0.26/6.29 | 1.29/3.32 |
| ID26499 | 838.99 | 16.21 | -4.05/2.34 | 2.53/4.65 |
| ID27001 | 924.60 | 15.10 | 3.15/8.34 | 1.05/4.51 |

two free parameters for estimating LA given the two fixed parameters, the GVC curves visually fit quite well to both PA and LA measurement data. Quantitatively, the RMSE for lognormal, gamma-variate, and GVC were 37.3, 37.0, and 24.6 HU for PA, 16.3, 17.8, and 14.0 HU for LA over 5 patients, confirming that GVC best approximates the measurements.

In order to predict the scan start time, it is necessary to fit data observed over a partial time before the peak of the left heart. The bolus peak estimation of LA from the increasing phase up to the peak of PA and the beginning phase of LA was implemented and evaluated only using the GVC model. Specifically, we truncated the PA/LA data by including only two more data points after the peak of the PA data. We then fitted the GVC model to the partially observed measurement data to estimate the model parameters. Figure 4 shows three examples of peak time estimation using the GVC model. Starting from the incomplete measurement data (red for PA, dark blue for LA), the parameters $A$, $p$, $\alpha$, and $t_0$ were first estimated for PA in GVC model (orange), and then only $\alpha$ and $t_0$ were tuned for fitting the model to LA given the fixed $A$ and $p$ (light blue). Then, from the complete time-intensity curve estimation of LA, the peak time was found (vertical pink lines). The peak time obtained from the fully observed measurements is also shown (vertical purple lines). The error in peak time estimate between the two ranged between 0.3s and 1.1s. The complete measurement data was also plotted as the ground truth (green). Qualitatively, the GVC model predicts the LA curve well from the incomplete measurements. In this perfusion data, the data sampling interval includes multiple heart beats. Data samples in higher temporal resolution, such as those obtained with pulsed mode acquisitions with constant intervals, will likely improve the peak estimation.

## 4 Conclusion

We demonstrated that gamma-variate-convolution (GVC) model shows better fitting to the time-intensity curves (especially for right cardiac chambers such as right atrium, right ventricle, and pulmonary artery) from myocardial perfusion CT data than two reference mathematical models. The GVC well models the shape of the wash-in and wash-out by convolving the gamma-variate function with an injection function. More importantly, we demonstrated that this simple GVC model can be used for bolus peak arrival time estimation from a sequence of measurements ending 5-8 seconds before the peak with an error smaller than 1.1s.

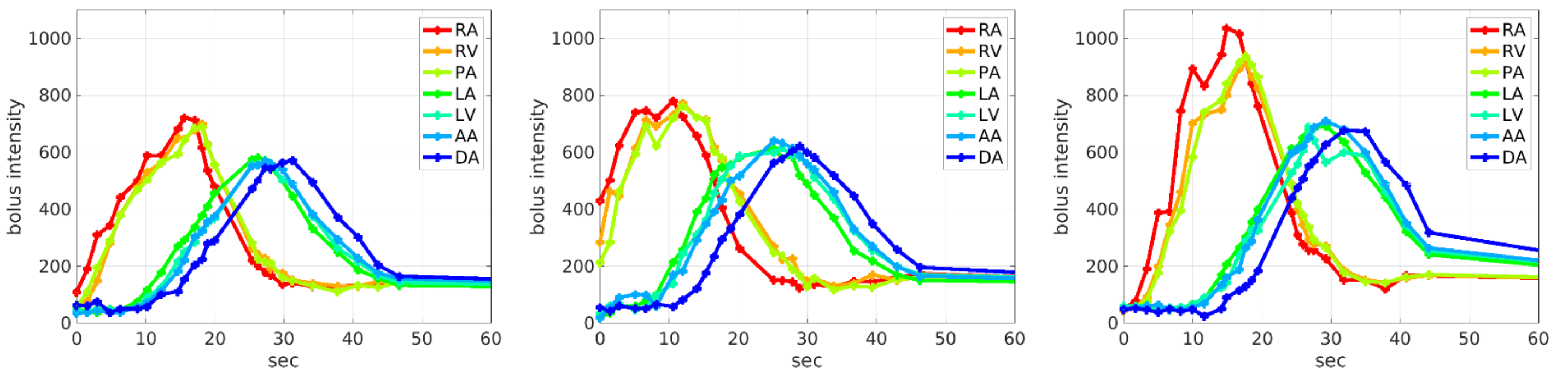

(a) Patient ID 26689 (b) Patient ID 26499 (c) Patient ID 27001

Figure 2: Time-intensity curves are plotted for three myocardial perfusion CT datasets for right atrium (RA), right ventricle (RV), pulmonary artery (PA), left atrium (LA), left ventricle (LV), ascending aorta (AA), and descending aorta (DA).

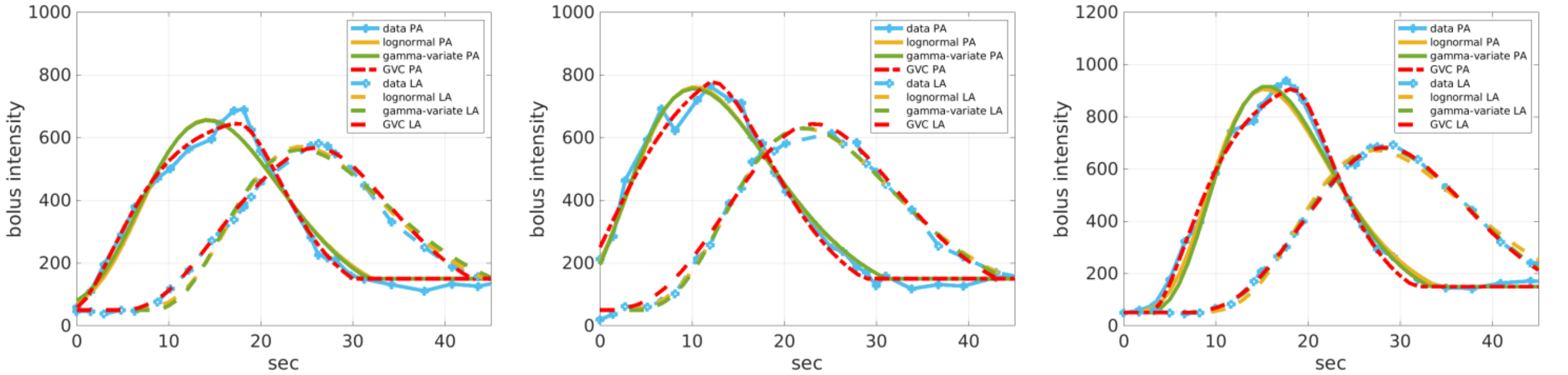

(a) Patient ID 26689 (b) Patient ID 26499 (c) Patient ID 27001

Figure 3: Three different models are fitted to time-intensity curves from three myocardial perfusion data for pulmonary artery (PA) and left atrium (LA): lognormal function, gamma-variate, and our gamma-variate-convolution (GVC) model.

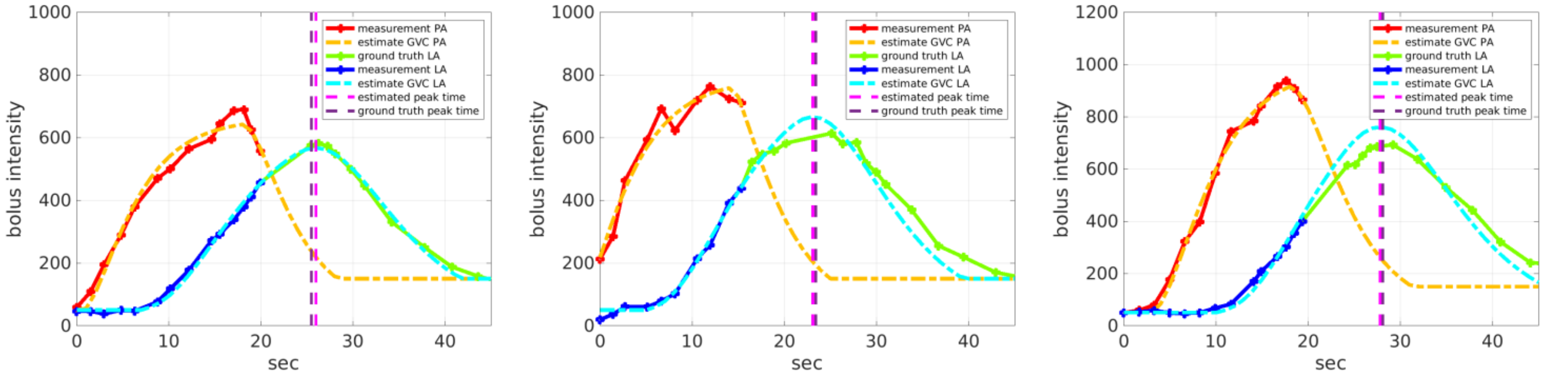

(a) Patient ID 26689 (b) Patient ID 26499 (c) Patient ID 27001

Figure 4: Peak estimation of left atrium (LA) from incomplete measurement data using GVC model. Assuming that we observe two extra data points after the PA peak (red and dark blue), the parameters $A$, $p$, and $\alpha$ are first estimated for PA and fitted (orange), and then only $\alpha$ and $t_0$ were adjusted for LA data fitting. The estimated LA curve is shown in light blue, and the estimated bolus peak time is shown in pink at the maximum.

## Acknowledgement

Research reported in this publication was supported by NHLBI of the National Institutes of Health under grant number R01HL153250. The content is solely the responsibility of the authors and does not necessarily represent the official views of the NIH.

# High Performance Spherical CNN for Medical Image Reconstruction and Denoising

Amirreza Hashemi[1], Yuemeng Feng[1], and Hamid Sabet[1]

[1] Department of Radiology, Massachusetts General Hospital & Harvard Medical School, Boston, MA, USA

**Abstract** Convolutional neural network (CNN) has shown tremendous success in post-processing of a variety of tomography medical imaging systems. CNNs are reliable in dealing with feature learning, handling noise, and handling non-linearity and high dimensional data. However, the efficiency of the conventional CNN largely depends on the undiminished and proper training set. To address this issue, in this work, we aim to investigate the potential for equivariant networks in order to reduce the dependency of CNN on the respective training set. We examine the equivariant CNN on spherical signals for tomographic medical imaging problems. We show higher quality and the computational efficiency of spherical-CNN on denoising and reconstruction of benchmark problems, e.g., we report an order of magnitude difference in the computational cost of same quality image reconstruction using spherical-CNN compared with CNN. Also, we discuss the possibilities of such a network for broader tomography applications, especially applications with omnidirectional representation.

## 1 Introduction

It is well established that artificial intelligence (AI) has immense potential to improve the quality of medical images in several ways such as enhancing spatial resolution, noise reduction, and lowering acquisition time. Among the AI tools, convolutional neural networks (CNN) are distinctly recognized as a powerful tool for image reconstruction and denoising in tomography. They can learn features from the 2D images and use them to reconstruct high-quality 3D images. They can also be trained for image denoising while preserving the details of the images. CNN usage in tomography can improve image quality and make the image reconstruction process more efficient.[1-3] This approach has been shown to produce high-quality images that are comparable to or better than images produced by traditional reconstruction algorithms like as maximum likelihood expectation maximization (MLEM) [4, 5]. However, the enhancement of conventional CNNs has been hampered by many existing limitations that include: overfitting, limited interpretability, limited ability to handle non-Euclidean spaces (e.g. image on the sphere) and missing or insufficient data, and being computationaly expensive.

Also, our work was motivated by recent studies [6-8] on AI image reconstruction of PET and SPECT which showed the inclusion of the equivariant rotated training data would enhance the image quality and lower the computational cost.

Here, we aim to investigate the potential of equivariant spherical CNNs (S-CNNs) for medical imaging applications and particularly for problems where the representative domain is spherical such as brain image. We will discuss the efficiency of S-CNNs for denoising and reconstruction of a benchmark brain image.

## 2 Methods

S-CNNs were first introduced by [9-11] as a specialized type of CNNs that are designed to work with data defined on the signals on a sphere. The sphere is a non-Euclidean space and traditional CNNs are not well-suited to handle data defined in non-Euclidean spaces. One of the key properties of S-CNNs is equivariance, which means that they are able to maintain the symmetry of the input data. For example, if an image of the earth is rotated, an S-CNN will be able to recognize that the image is still an image of the earth, even though it has been rotated. This is beneficial because it allows the network to learn features that are invariant to certain transformations, such as rotation. Another important property of S-CNNs is that they are able to handle data that is defined in a spherical space. This is beneficial because the sphere is a natural space for many types of data, such as images of the earth or the sky. In addition, the sphere has a constant positive curvature, which means that the distance between any two points on the sphere is always less than the distance between the same two points on a flat surface. This property allows S-CNNs to be more efficient at processing data than traditional CNNs, which are designed for flat, Euclidean spaces.

Our S-CNN is built of only three layers that include input, hidden, and output layers with 8 rotational actions on the group space. Each layer is connected to the inner batch normalization and rectified linear unit (ReLU) activation. Input and hidden layers have regular representations with a kernal size of 3 and the output layer has an irreducible representation with one kernel (for the details of S-CNN descriptions and more information on the representative functions, see [12]). For the sake of comparison, a deep learning conventional CNN is chosen with 5 layers (3 hidden layers), 64 channels, and a ReLU activation.

We utilized the pytorch framework (https://pytorch.org/) [13] and the *escnn* package where the details are provided in [12, 14]. The training data/image is a brain image from a brain library of scikit-learn. Simulations were done on Linux Ubuntu v20.04 with an Intel Xeon E5-2687W 3.1GHz CPU, 128GB RAM, and an NVIDIA TITAN RTX GPU card with 24GB of memory.

## 3 Results

To examine the performance of our S-CNN model, we compare the S-CNN results with the conventional CNN for denoising of a noisy brain image and reconstruction from a sinogram of the brain image. Figures 1a and 1b show the true and noisy images respectively. The noise type is Poisson in this case. Both CNN and S-CNN are run for 2000 epochs and the results of the denoised images are shown in figures 1c and 1d at 2000 epoch and figures 1e and 1f at 500 epoch respectively.

The loss function is estimated based on the mean square error (L2 norm) and the loss function vs epoch for SCNN and CNN are shown in Figures 1g and 1h, respectively.

For denoising case, both CNN and S-CNN run on the GPU resulted in similar computational times for 2000 epoch iterations: 104 sec and 106 sec, respectively.

For image reconstruction comparison, the sinogram of the brain image is generated by Radon transform and the forward propagation model is incorporated into the loss function of neural networks.

Results of reconstruction for CNN and SCNN are shown for two epochs at 5000 and 500 iterations and the loss function vs epoch plots are shown in Figures 2c and 2f. Also, the loss function vs epoch plot is zoomed in for lower iteration in Figure 1g to show a clearer view of the result. The minimum loss function values for SCNN and CNN converge to 0.015 and 0.14, respectively. The computational times of CNN and SCNN for 5000 epochs are 766 and 784 seconds, respectively, and the computational time for the SCNN at 500 epochs is estimated to be 87 seconds.

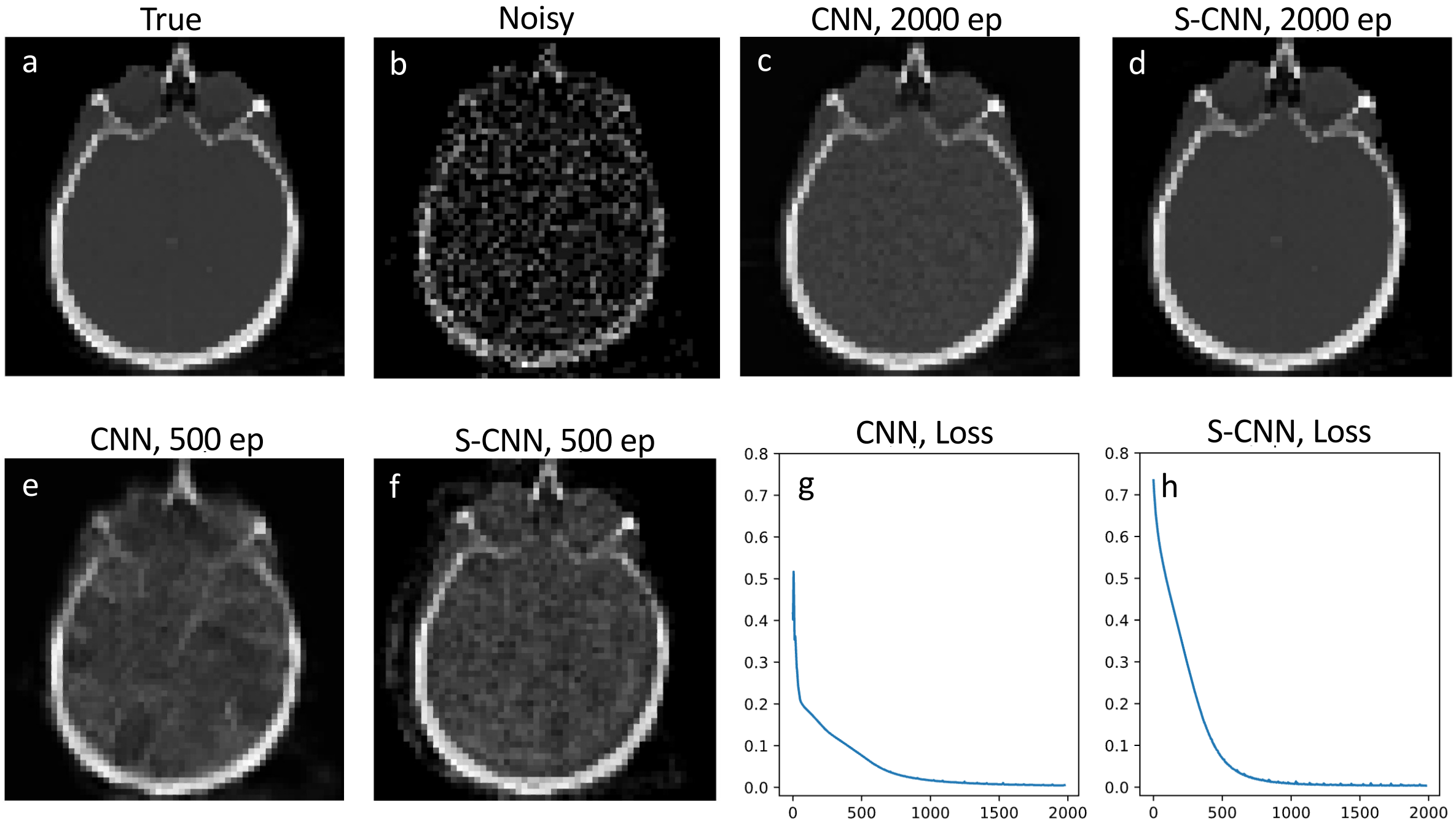

**Figure 1:** Denoising results of brain image. Plots a and b show true and noisy images, comparison between CNN and S-CNN are shown in plots c-f for 2000 and 500 epochs. The loss functions for CNN and S-CNN are shown in plots g and h respectively.

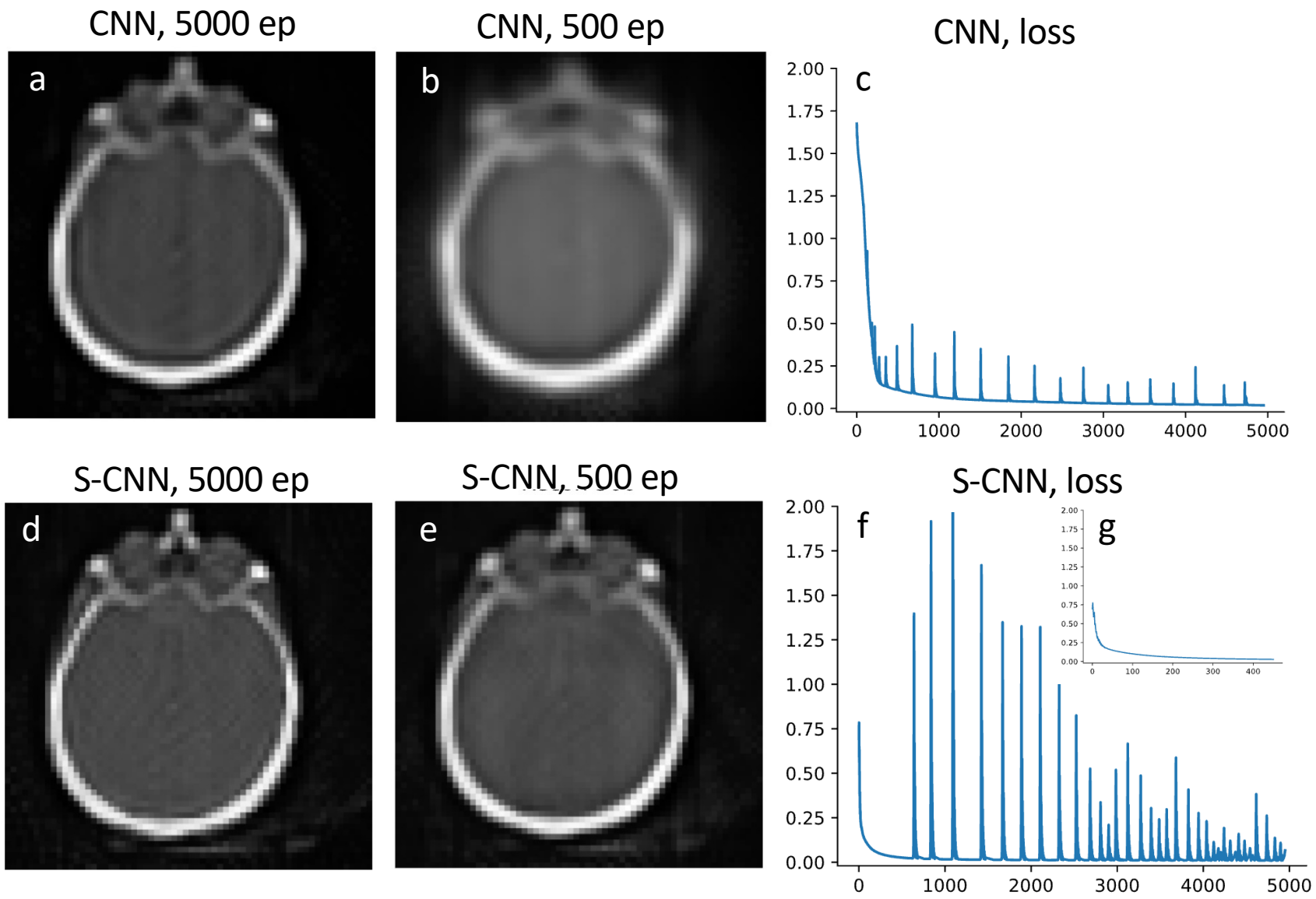

**Figure 2:** Reconstruction results. Comparison between the reconstructed image at 5000 and 500 epochs are shown in plots a and b for CNN and plots d and e for S-CNN. Plots c and f are loss function vs epoch for CNN and S-CNN respectively, subplot g shows the zoom-in view of loss function result for S-CNN.

## 4 Discussion

S-CNN results indicate higher accuracy images in both denoising and reconstruction cases which inherently is related to the inclusion of an equivariant dataset as input for S-CNN and that is not present in conventional CNN. To clarify this point in Figure 3, we show 8 equivariant rotational representative actions for denoising cases over one epoch iteration of S-CNN, and consequently, each training input is transformed into the group rotational space of representative actions. This suggests that the inclusion of equivariant representative actions lowers the dependency of S-CNN on the completeness of its training set and will result in higher accuracy of output. Also, S-CNN convergence is obtained in fewer iterations as the loss function sharply declines to near zero value. In this regard, for the denoising case, the S-CNN loss value at 1000 epoch is nearly equivalent to the loss value of CNN at 2000 epoch and for the reconstruction case, the S-CNN loss value converges to a lower value (0.02) at 500 epoch while the CNN loss value converges to a higher value (0.14) at 5000 epoch. Comparison between Figure 1a and 1e shows that the S-CNN reaches higher quality image reconstruction ten times faster in terms of the number of epochs and nearly nine times faster in terms of computational time.

Additionally, we observe the loss function monotonically decreases with minimal instability up to the minimum loss value which is the main reason for the quick convergence of the S-CNN model. Beyond reaching the minimum, the S-CNN loss function shows oscillatory behavior which is a consequence of the overfitting limitation that occurs for very small errors. On the other hand, for the case of CNN, the loss function shows some instabilities at the early epochs which relate to the main limitations of conventional CNN on dealing with non-Euclidean space and that is apparent in denoising of the noisy image.

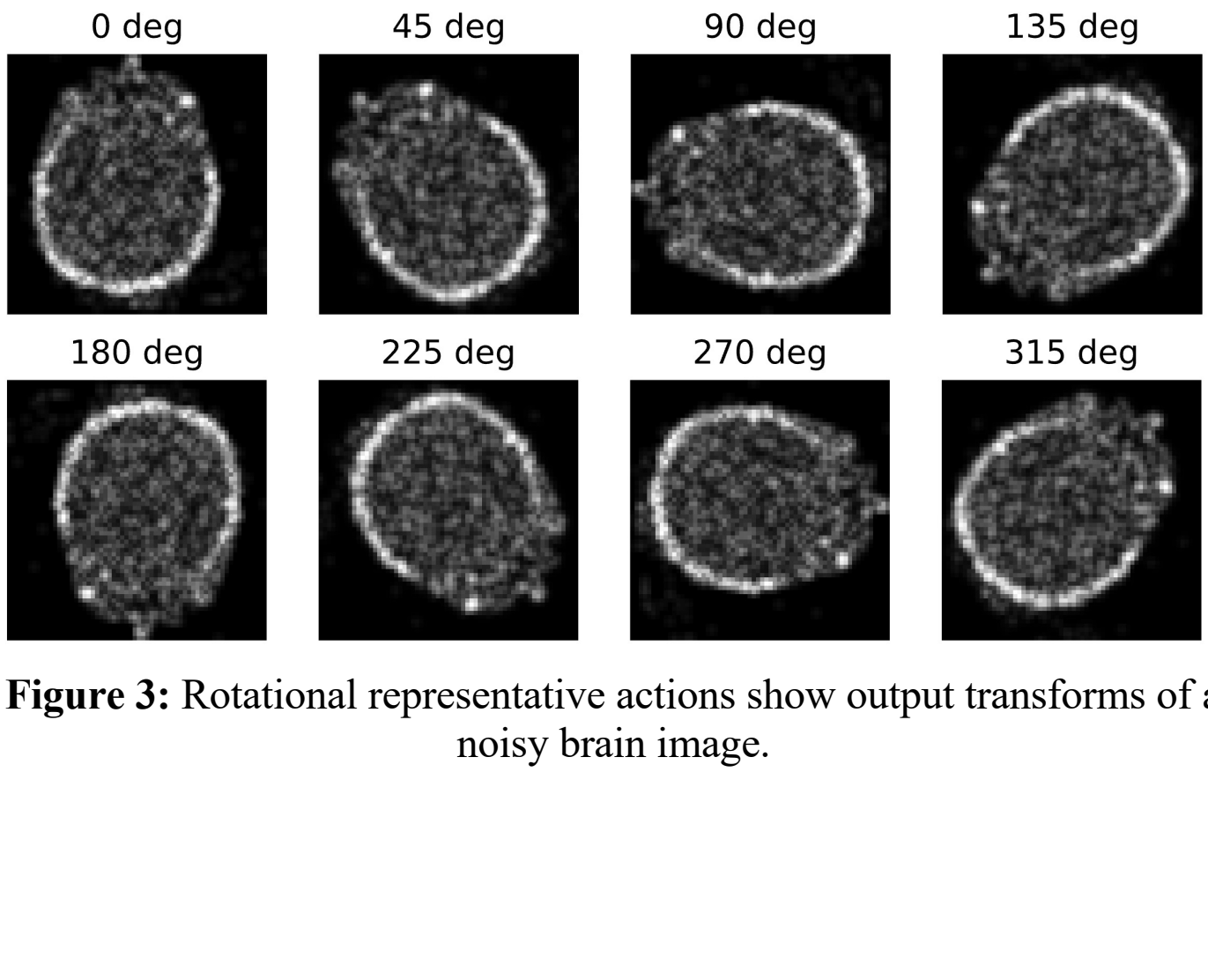

**Figure 3:** Rotational representative actions show output transforms of a noisy brain image.

## 5 Conclusion

In this work, we showed that equivariant S-CNNs improve the image quality for denoising and the reconstruction of a benchmark brain image while the computational cost is considerably lower than the conventional CNNs. The results indicate that the S-CNNs are viable AI tools for broader medical imaging applications, particularly for imaging modalities with omnidirectional outputs such as brain PET or cardiac dedicated scanners. Hence, we plan to further investigate the S-CNN's applicability for these types of problems in the future.

# PACformer: A Locally-Enhanced Efficient Vision Transformer for Sparse-view PAT Restoration

Li He[1, 2, 3], Li Ma[1, 2, 3], Xu Cao[1, 2, 3], Shouping Zhu[1, 2, 3, *], and Yihan Wang[1, 2, 3, *]

[1] Engineering Research Center of Molecular and Neuro Imaging, Ministry of Education & School of Life Science and Technology, Xidian University, Xi'an, Shaanxi 710126, China

[2]Xi'an Key Laboratory of Intelligent Sensing and Regulation of Trans-Scale Life Information & International Joint Research Center for Advanced Medical Imaging and Intelligent Diagnosis and Treatment, School of Life Science and Technology, Xidian University, Xi'an, Shaanxi, 710126, China|

[3] Innovation Center for Advanced Medical Imaging and Intelligent Medicine, Guangzhou Institute of Technology, Xidian University, Guangzhou, Guangdong, 51055, China

**Abstract** Balancing computational cost and imaging quality, convolutional neural networks (CNNs) have been widely studied and applied in sparse-view photoacoustic tomography (PAT) restoration. However, it is difficult for CNNs to extract the long-range dependencies that exist in the continuous fine structures across the entire PAT images as CNNs are naturally localized. Vision Transformer (ViT), a promising deep vision model computing self-attention traversing the whole image, possesses the ability to extract such dependencies. Moreover, the restoration of detailed information, where the classical ViT model has limited utility, is equally important in PAT images. High computational cost also limits its direct application. In this study, an efficient locally-enhanced ViT—PACformer is proposed to restore the continuous fine structures as well as local details in sparse-view PAT. To avoid the excessive computational cost, reducing channel convolution (RCC) is proposed to save algorithmic complexity. On the mice in vivo dataset, PACformer significantly improves the reconstruction quality (~50% improvement on SSIM and ~10dB improvement on PSNR) compared with the classical Universal back-projection (UBP), compared with other existing deep networks, PACformer has a significant advantage in terms of computational complexity and has the best performance on SSIM and PSNR.

.

## 1 Introduction

As a hybrid imaging modality combining optical excitation and ultrasound detection, photoacoustic tomography (PAT) overcomes the disadvantages of low contrast in ultrasound imaging and shallow depth in optical imaging [1]. Current limitations of PAT in system throughput and computational power require using a sparse sampling scheme, leading to degraded images [2], [3]. The consistent goal of both theoretical research and engineering is to efficiently reconstruct high-quality PAT images while using sparse-view sampling. Model-based iterative reconstructions have been used to improve the quality of sparse-view PAT images [4], however, its optimization requires iterative minimization of the penalty function with high computational complexity. When enough paired images are available, supervised deep learning [5] yields excellent results while also considering time efficiency. It is currently one of the most extensively researched techniques for PAT reconstruction in conjunction with deep learning.

Restoration networks generally use convolutional layers (Convs) for feature extraction but struggle with long-range dependencies [6]. Vision Transformer (ViT) based on multi-head self-attentions (MSAs) can model long-range dependencies, making it a feasible solution for sparse-view PAT reconstruction. Shifted-windows (Swin) Transformer [7] achieves excellent performance in computer vision tasks while maintaining efficient computation.

While Swin has achieved success, it still faces challenges when applied to sparse-view PAT restoration. Research [8] indicates that MSAs and Convs have opposite properties in certain aspects. Research [10] has shown that local convolution is beneficial in image restoration since it can leverage the neighborhood of a degraded pixel to restore its clean version. It is crucial to construct MSAs-Convs tandem models that take advantage of their respective strengths. Additionally, the attention score is computed as the dot product of the query and key vectors in the original Swin Transformer. The learned attention maps for certain blocks and heads are often dominated by a few query-key pairs in large models. Finally, U-shaped networks have a wide range of applications in image restoration [9], [10]. The decoding side of this structure performs feature dimension doubling, which is computationally expensive, especially for MSAs.

To address these issues, this study introduces an efficient MSAs-Convs tandem model PACformer for sparse-view PAT restoration. PACformer replaces the feedforward network (FFN) layer with an effective local convolution layer Moved-up Invrted Bottleneck (MIB) for better feature extraction. It also optimizes attention score calculation to prevent any single pixel from dominating the process. Additionally, reducing channel convolution (RCC) is used for reducing complexity. The results show that PACformer outperforms three popular convolutional networks on the open-source sparse-view reconstruction PAT dataset [9]. In most experiments, the proposed method performs better than Uformer despite having only 47% of its algorithmic complexity and 65% of its parameter size.

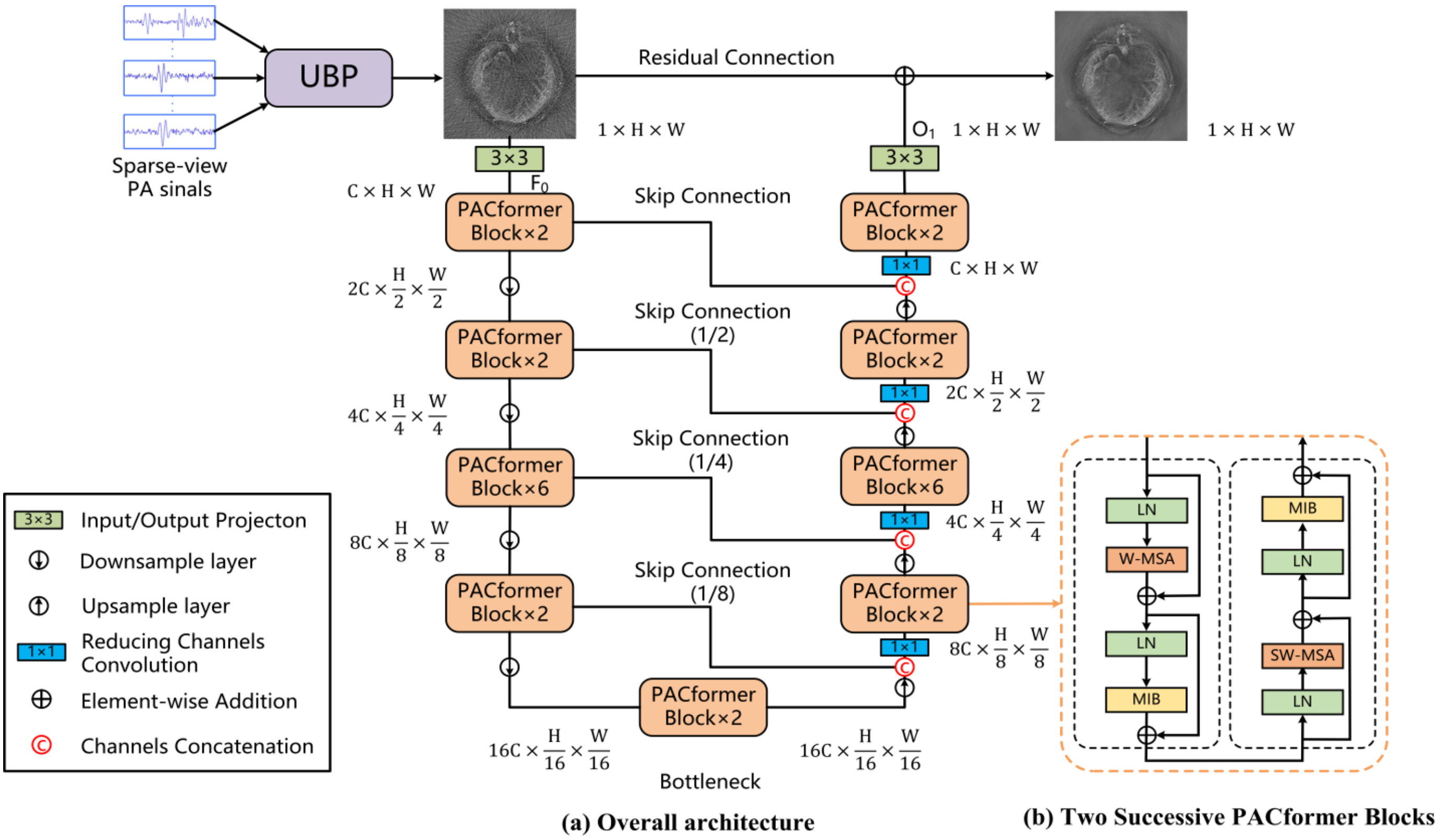

**Fig. 1.** Overall architecture of the proposed network.

## 2 Materials and Methods

### Dataset and Network Implementation

The dataset used is from the open-source sparse-view PAT dataset [9]. It consists of 98 groups of the vascular phantom and 274 groups of mice in vivo. Each of the t datasets consists of 8, 16, 32, 64, 128 sparse views reconstructed PAT images and 512 full-view reconstructed Ground Truth.

The network in this study was implemented via pytorch framework (python 3.8, torch1.11). The loss function is Charbonnier loss $l(I', \hat{I}) = \sqrt{||I' - \hat{I}||^2 + \epsilon^2}$, $\epsilon = 10^{-3}$, where $\hat{I}$ is the ground-truth image, and $\epsilon$ is a constant in all the experiments, the optimizer is AdamW, with initial learning rate $1\times10^{-4}$ and decreasing to $1\times10^{-6}$ with cosine annealing strategy, batch size is set as 8, and the number of iteration epochs is set as 200.

### Overall Architecture

The framework of the proposed network in this study is shown in Fig. 1(a), it is a U-shaped end-to-end network. On the coding side, given an input degraded image $I \in R^{1\times h\times w}$, PACformer first uses 3×3 convolution to obtain the multidimensional low-level features $F_0$. Subsequently, $F_0$ is fed into a 4-stage coding block consisting of the proposed PACformer Blocks, shown in Fig. 1(b). On the decoding side, the output features of the bottleneck are passed into the decoding end of the network. Notice that after concatenation, the number of feature channels will be doubled, we use RCC to halve the dimensionality of the concatenated features before feeding them to the decoder.

### PACformer Blocks

(1). Replace the computation of the self-attention score from the scaled dot production with scaled cosine similarity. As shown in Fig. 2(a), in the classical self-attention computation, the attention score is designed as a scaled dot production t of queries (Q) and keys (K) vectors. However, this leads to a learned attention map for some blocks and heads dominated by a few pixel pairs. To mitigate this problem, we compute the attention score using scaled cosine attention.

(2). Considering the importance of detail information recovery in PAT images, as shown in Fig. 2(b), the FFN layer in the standard Swin Transformer block is replaced by an efficient local convolutional layer MIB, which is close to the algorithmic complexity of both compared with FFN.

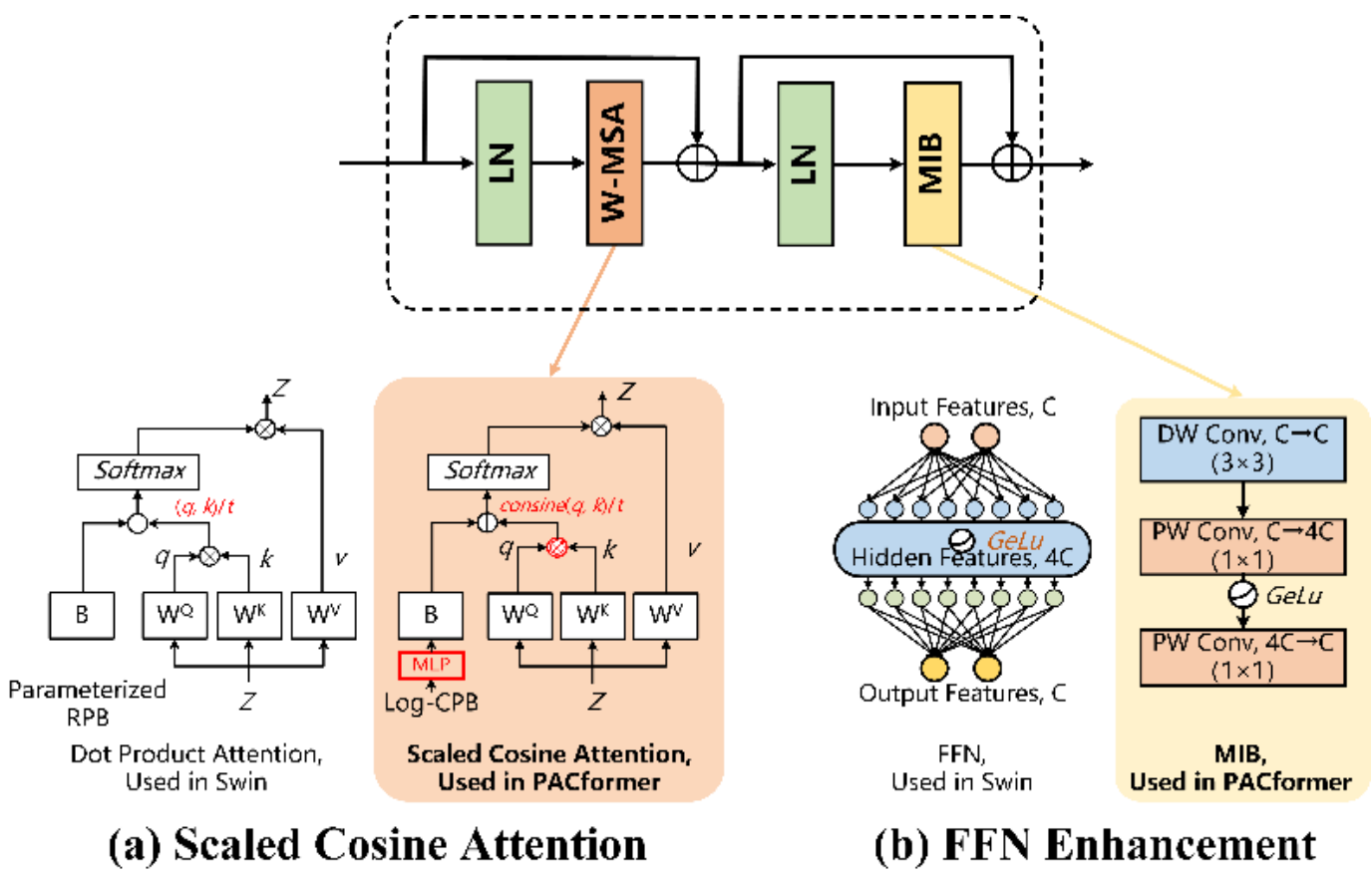

**Fig. 2.** Proposed PACformer Blocks and its enhancements.

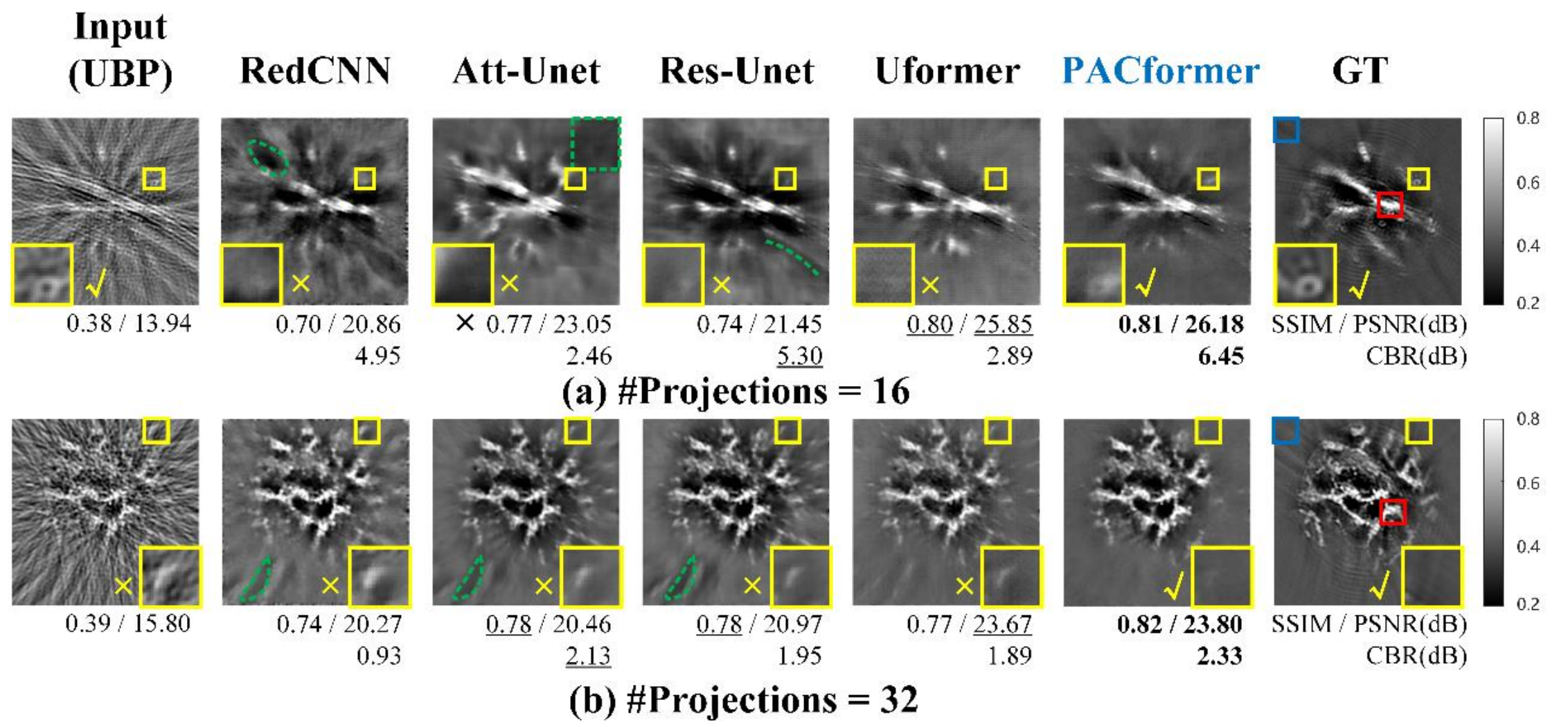

**Fig. 3.** Comparison of the quality of reconstruction of vessel phantom. (a) Reconstructed PAT images from 16 views. (b) Reconstructed PAT images from 32 views. The best and second-best results are underlined and marked in bold, respectively. The red box is the signal region selected for the calculation of CBR, the blue box is the background region, and the yellow box is the zoomed-in ROI region, the green dashed lines indicate artifacts.

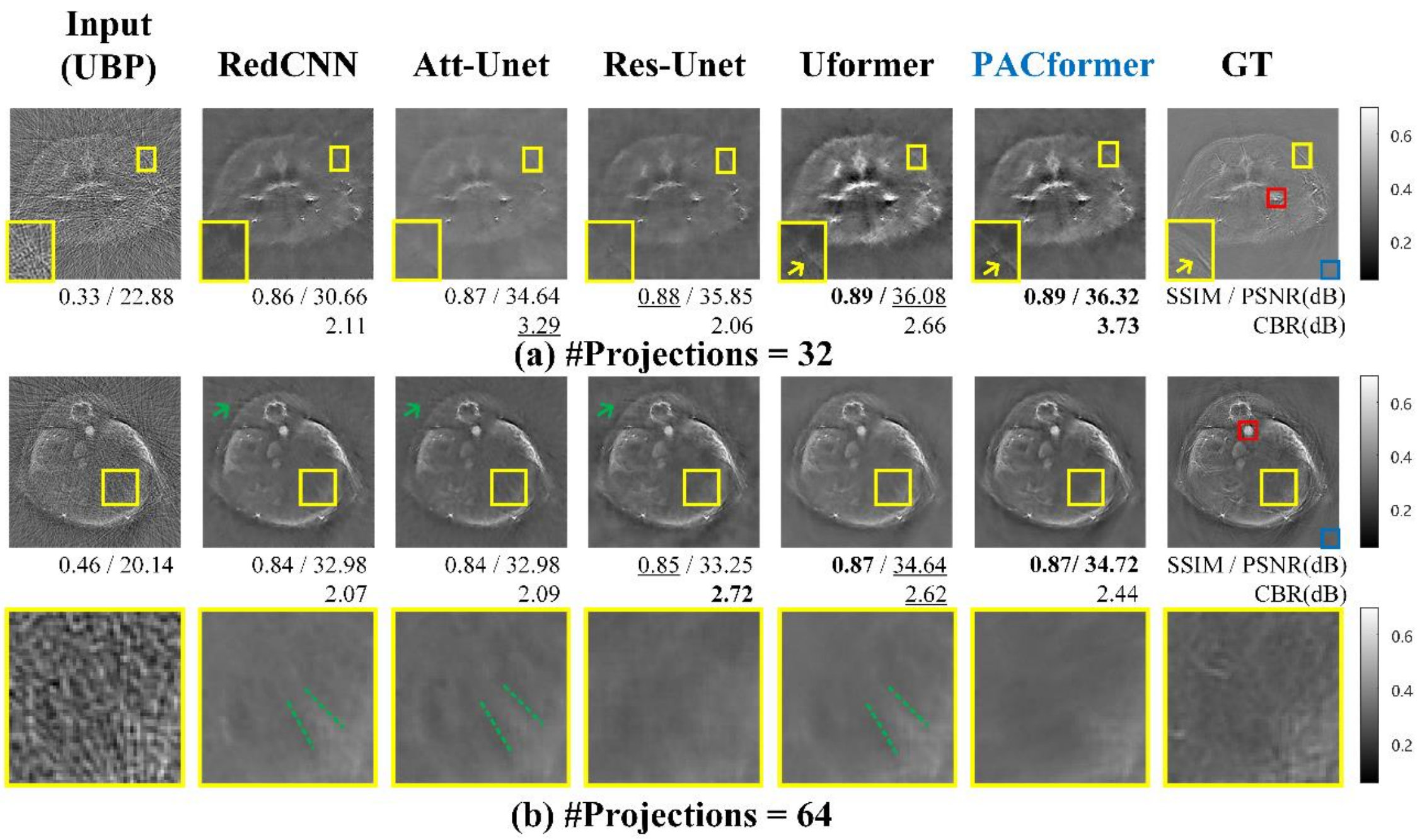

**Fig. 4.** Comparison of the quality of reconstruction of mice in vivo. (a) Reconstructed PAT images from 32 views. (b) Reconstructed PAT images from 64 views. The best and second-best results are underlined and marked in bold, respectively. The red box is the signal region selected for the calculation of CBR, the blue box is the background region, and the yellow box is the zoomed-in ROI region, the green dashed lines indicate artifacts.

## Compared Methods

In order to evaluate the performance of the proposed method, we compare several representative deep networks for quality enhancement. These include RedCNN [11], a classical approach in sparse-view CT reconstruction, two representative works of the U-Net family, Att-Unet [12], Res-Unet [13], and Uformer [10] which enjoys a high capability for capturing both local and global dependencies for image restoration.

## 3 Results

As shown in Fig. 3(a), all the convolutional networks perform poorly in the suppression of background artifacts, the predicted images of RedCNN and Att-Unet show low-value regions that are inconsistent with the surrounding background in the dashed green area. Observing the upper-right yellow ROI region of all images, it can be found that PACformer accurately recovers the signal compared with

**Table I.** Comparison of model parameter sizes and algorithm complexity.

| Item | Uformer | Res-Unet | RedCNN | Att-Unet | **PACformer** |
|---|---|---|---|---|---|
| FLOPs | 60.90 G | 61.55 G | 53.90 G | 59.91 G | **28.31 G** |

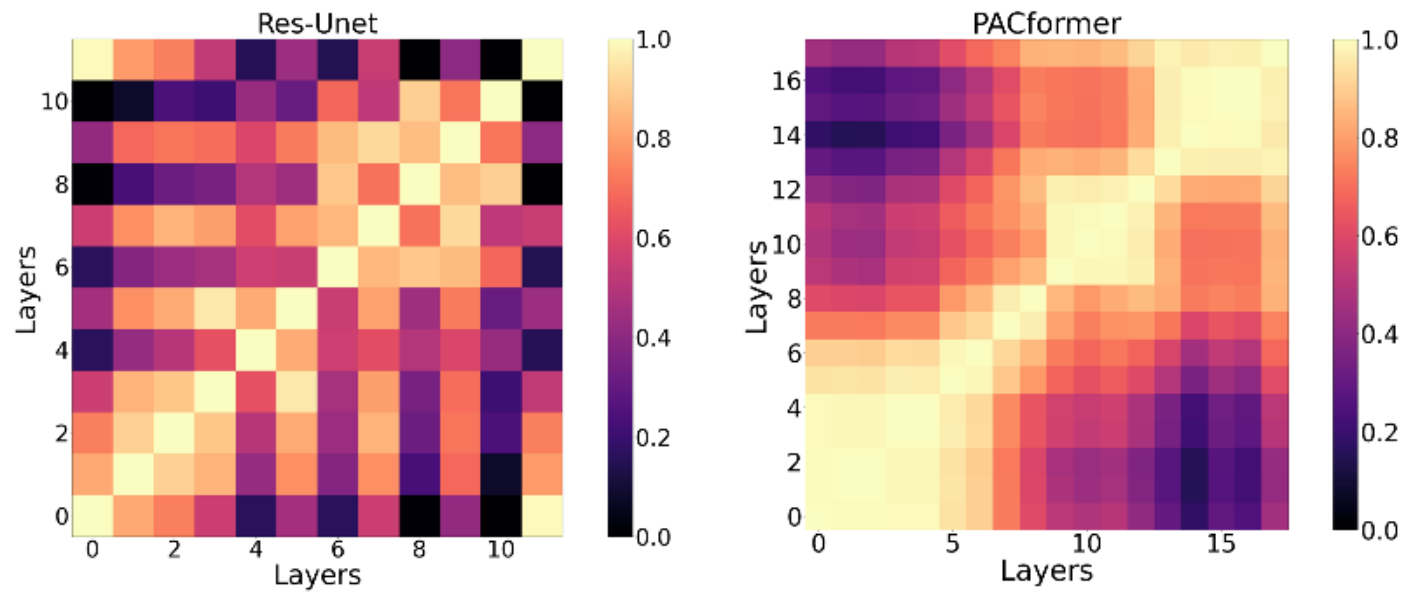

**Fig. 5.** Representation structure of Res-Unet and PACformer show significant differences, with PACformer having highly similar representations throughout the model.

Ufomer, while other convolutional networks appear false negative predictions. Similarly, convolutional methods in Fig. 3(b) show obvious residual artifacts in the background region. Uformer and the proposed method have the performance of a more uniform background region. Observing the zoomed-in area of the ROI region in the yellow box. The proposed method not only can better suppress the positive reconstruction artifacts but also has obvious advantages in detail retention and recovery ability. As shown in Fig. 4(a), PACformer has the best performance in all three metrics. It can be seen that there is a diagonally distributed signal region in the ROI amplification region of the yellow box from Ground Truth, which can be recovered by Uformer and PACformer, while convolutional methods can hardly recover it. At the denser 64 views in Fig. 4(b), the image quality of several methods is significantly improved, but similar to the previous ones, the three convolutional methods perform poorly in the suppression of the background barring artifacts, with residual barring artifacts suggested by the green arrows.

## 4 Discussion

In this study, an efficient ViT model PACformer is proposed for sparse-view PAT restoration. PACformer shows more significant PSNR enhancement relative to the Convs-based methods in Fig. 3 and Fig. 4, the zoomed-in ROI region indicates that MSAs have better recovery ability for this continuous long-range distributed signal region, which may be related to the ability of MSAs to capture the remote dependencies. As Fig.5. shown the proposed method learns global long-range dependency features similar to the features of the intermediate bottleneck layer at the shallow layer of the network. This may be a source of superiority of the proposed method.

Thanks to the parameter-saving RCC design, PACformer has a significant advantage in computational complexity (shown in Table. I). Compared to the three convolutional networks, PACformer achieves consistently better performance in sparse-view PAT reconstruction as shown in Fig. 3 and Fig. 4. Uformer, which is also based on the MSAs, achieves a comparable performance to the proposed method. This indicates the superiority of the MSAs models over the compared convolutional networks.

## 5 Conclusion

The proposed PACformer outperforms three representative convolutional networks as well as the same MSA-based Uformer on sparse-view PAT restoration. Thanks to the RCC and the effective improvements in PACformer blocks, the proposed method maintains the state-of-the-art while having significantly lower computational complexity.

## Acknowledgments

This work was partly supported by the Key Research and Development Program of Shaanxi Province under Grant No. 2023-YBSF-204, the Natural Science Foundation of Chongqing under Grant No. CSTB2023NSCQ-MSX0955, the Fundamental Research Funds for the Central Universities under Grant No. ZYTS23186, and the Science and Technology Program of Guangzhou under Grant No. 2023B03J1255.

# Generation of photon-counting spectral CT images using a score-based diffusion model

Dennis Hein[1] and Mats Persson[1]

[1]Department of Physics, KTH Royal Institute of Technology, Stockholm, Sweden and MedTechLabs, BioClinicum, Karolinska University Hospital, Solna, Sweden.

**Abstract** Deep learning is playing an increasingly important role in medical imaging. One important factor for the development and evaluation of high performing and robust networks is the availability of large and diverse datasets. However, this availability is lacking for novel technologies such as photon-counting spectral CT. One way of generating synthetic data is using score-based diffusion models, a novel class of generative models that have recently shown to perform on par with, or outperform, generative adversarial networks. This paper explores the possibility of utilizing a score-based diffusion model to generate photon-counting spectral CT images. We train a network to generate a pair of 70 and 100 keV virtual monoenergetic images from which we can subsequently recover material basis images via a simple linear transformation. Our results are very encouraging as the resulting network is able to generate realistic output with limited data and training time and with minimal hyperparameter tuning.

## 1 Introduction

Deep neural networks are applied in medical imaging for a large variety of different tasks. However, the performance and robustness of these data driven methods is highly dependent on the availability of large and diverse datasets. If these are not readily available, as is the case for photon-counting spectral CT, then we might want to turn to deep generative models to augment our existing dataset. Generative adversarial networks (GANs) [1] have proven capable of generating amazing results in a myriad of different task. However, due to their adversarial nature, they are notoriously difficult to train. Recently, score-based [2] and denoising diffusion probabilistic models [3] have proved to outperformed GANs in various data generation tasks. Using stochastic calculus, [4] unifies these two classes of models into one framework, which we may call score-based diffusion models [5]. Song et al. [4] models the diffusion process from target data to a prior noise distribution as the solution to a stochastic differential equation (SDE). Generating samples then involves the reverse-SDE, which is also a diffusion process. Notably, this reverse-SDE only depends on the gradients of the data distribution, that is, the score function [4]. Estimating the score function with a neural network gives us a complete framework to generate samples from noise. In this paper, we apply the framework in [4] to the problem of generating photon-counting spectral CT images. Instead of directly generating material basis images, we opt to generate a pair of 70 and 100 keV virtual monoenergetic images and then recover material basis images via a linear transformation. We do this as material basis images in photon-counting spectral CT are usually exceedingly noisy.

## 2 Materials and Methods

### 2.1 Score-based modeling via stochastic differential equations

In this paper we smoothly diffuse the data into noise using a SDE as suggested in [4]. One can subsequently generate samples from noise via the reverse-time SDE once we have an estimate of the score function. More formally, suppose we have i.i.d. samples from the $D$-dimensional data distribution $\boldsymbol{x}(0) \sim p_0$. Let $\{\boldsymbol{x}(t)\}_{t\in[0,T]}$ denote the process that smoothly diffuses the data to a prior distribution $\boldsymbol{x}(T) \sim p_T$. This diffusion process can be obtained as the solution to the Itô SDE

$$d\boldsymbol{x} = \boldsymbol{f}(\boldsymbol{x},t)dt + g(t)d\boldsymbol{w} \tag{1}$$

where $\boldsymbol{w} \in \mathbb{R}^D$ is a standard Wiener process, $f(\cdot,t) : \mathbb{R}^D \to \mathbb{R}^D$ is the drift coefficient, and $g(t) \in \mathbb{R}$ is the diffusion coefficient. Note that we can handcraft the prior by choosing $f(\boldsymbol{x},t), g(t)$ and $T$ s.t. the diffusion process approaches some tractable prior distribution $\pi(\boldsymbol{x}) \approx p_T(\boldsymbol{x})$ at $t = T$. In particular, [4] presents three families of SDEs for this task: Variance Exploding (VE), Variance Preserving (VP), and subVP SDEs[1]. In this paper we consider the VE SDE. It is now possible to generate samples from the data distribution $\boldsymbol{x}(0) \sim p_0$ by sampling $\boldsymbol{x}(T) \sim p_T$ from the prior distribution and reversing the process in (1). Fortunately, a result from [6] states that this SDE has a corresponding reverse-time SDE which is also a diffusion process on the form

$$d\boldsymbol{x} = [\boldsymbol{f}(\boldsymbol{x},t) - g(t)^2\nabla_{\boldsymbol{x}} \log p_t(\boldsymbol{x})]dt + g(t)d\bar{\boldsymbol{w}}, \tag{2}$$

where $\bar{\boldsymbol{w}}$ is a standard Wiener process that is running in reverse time, from $T$ to 0, and $dt$ is an infinitesimal, negative, time step. The forward and reverse-time SDEs are illustrated in Fig. 1. Now, if we know the time-dependent score function $\nabla_{\boldsymbol{x}} \log p_t(\boldsymbol{x})$ then we can generate a sample $\boldsymbol{x}(0) \sim p_0$ from via (2). Let $\boldsymbol{s}_\theta(\boldsymbol{x},t)$ be a neural network parametrized by $\boldsymbol{\theta}$. We then learn $\boldsymbol{\theta}$ s.t. $\boldsymbol{s}_\theta(\boldsymbol{x},t) \approx \nabla_{\boldsymbol{x}} \log p_t(\boldsymbol{x})$ via score-matching. In particular,

$$\begin{aligned}\boldsymbol{\theta}^* = \arg\min_{\boldsymbol{\theta}} \mathbb{E}_t\{\lambda(t)\mathbb{E}_{\boldsymbol{x}(0)}\mathbb{E}_{\boldsymbol{x}(t)|\boldsymbol{x}(0)}[||\boldsymbol{s}_\theta(\boldsymbol{x},t) \\ - \nabla_{\boldsymbol{x}(t)} \log p_{0t}(\boldsymbol{x}(t)|\boldsymbol{x}(0))||_2^2]\},\end{aligned} \tag{3}$$

where $t \sim U(0,T)$, $\lambda : [0,T] \to \mathbb{R}_+$ is a weighting function, $\boldsymbol{x}(0) \sim p_0(\boldsymbol{x}), \boldsymbol{x}(t) \sim p_{0t}(\boldsymbol{x}(t)|\boldsymbol{x}(0))$ and $p_{st}(\boldsymbol{x}(t)|\boldsymbol{x}(s))$ is the

[1]For more details see [4] and relevant appendices therein.

transition kernel from $\boldsymbol{x}(t)$ to $\boldsymbol{x}(s)$ [4]. Typically $\lambda(t) \propto [||\nabla_{\boldsymbol{x}(t)} \log p_{0t}(\boldsymbol{x}(t)|\boldsymbol{x}(0))||_2^2]^{-1}$.

### 2.2 Data

We train the network on a simulated photon-counting spectral CT dataset. These data were obtained by first generating numerical basis phantoms by thresholding CT images from the KiTS19 dataset [7]. We subsequently simulated photon-counting imaging by using the `fanbeam` function in Matlab and a spectral response model of a photon-counting silicon detector [8] with $0.5 \times 0.5$ mm$^2$ pixels for 120 kVp and 200 mAs with 2000 detector pixels and 2000 view angles. We then simulated Poisson noise and used the maximum likelihood method to decompose the simulated energy bin sinograms into soft tissue and bone material basis sinograms. Finally, images were reconstructed on a $1024 \times 1024$ pixel grid using FBP. Two examples slices from this simulated dataset are illustrated in Fig. 2.

### 2.3 Training details

A total of 2025 slices are split into a training (1575 slices) and a test set (450 slices). We trained for 100k iterations[2] on an NVIDIA A6000 GPU using Adam [9] and a learning rate of $2 \times 10^4$. We use a batch size of 32. Images were resized to $512 \times 512$ to reduce graphics memory constraints. For network, we use a version of NCSN++, where we have omitted the anti-aliasing filter and the progressively growing architecture, from [4]. We used channel dimension 16, channel multipliers $[1, 2, 4, 8, 16, 32, 32, 32]$, one ResBlock per resolution, and self-attention layers at the $16 \times 16$ resolution. We augment the training data by randomly flipping the images horizontally. This means that the network will learn to generate images where the anatomy has been mirrored, which might actually be the case for some patients.[3] To solve the reverse-time SDE we use the predictor-corrector sampler from [4] with 1000 steps.

## 3 Results

Qualitative results are available in Fig. 3. We can see directly that the network does a fairly good job of producing realistic images. The size, shape, and anatomy seems reasonable. We note again that since we used random horizontal flipping during training, a mirrored anatomy is not made up by the network and therefore not considered an deviation from realism. Considering the soft tissue basis images in the top two rows, we can see that the network has produced realistic lungs. In addition, the position and shape of the aorta seems fairly reasonable. In the third and fourth rows we can see that the network is able to produce realistic kidneys and intestines. Shape, position, size, and small details all seem reasonable. Finally, in the last row, we can see that the network has learned to produce artifacts resembling what you expect to see from photon starvation when imaging the pelvis.

[2]This was a limitation imposed by time constraints for this preliminary study. We note however that network would most likely benefit from more training.

[3]Situs inversus is a rare congential condition where organs are mirrored from their normal (situs solitus) locations.

## 4 Discussion

Although data generation is a very important and interesting issue in computer vision, data generation in medical imaging is often a means to an end rather than an end by itself. For instance, we might be interested in data generation to augmented our existing datasets. Another interesting application is virtual clinical trials. Note that how realistic the generated data is required to be depends on the application. For instance, in the case of data augmentation, the only requirement is that training on an augmented data set improves performance and robustness of a given network. For virtual clinical trials the bar might be significantly higher. In future work, we will explore how the results in this paper can be used for data augmentation. In addition, we will compare, quantitatively and qualitatively, score-based diffusion models with other deep generative networks, for instance GANs, for data generation in photon-counting spectral CT. Moreover, we note that score-based diffusion models are very well suited for solving inverse problems [10]. Hence, there are a myriad of interesting applications of score-based diffusion models in medical imaging beyond data generation.

## 5 Conclusion

This paper presents some preliminary results on using score-based diffusion models to generate photon-counting spectral CT images. The results are very encouraging as we are able to get good performance despite data and training time limitations and with very little hyperparameter tuning.

## 6 Acknowledgements

This study was financially supported by MedTechLabs, by the Swedish Research council (grant no. 2021-05103) and the Göran Gustafsson foundation (grant no. 2114). Mats Persson and Dennis Hein disclose research collaboration with GE Healthcare.

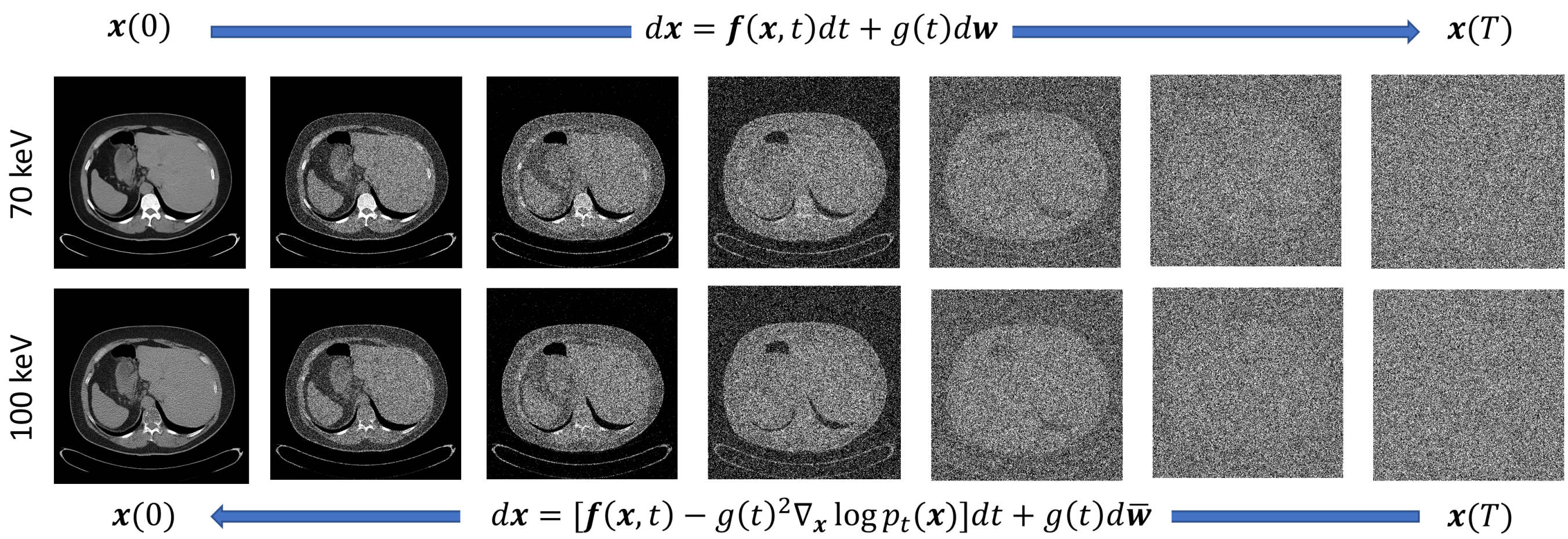

**Figure 1:** Illustration of forward and reverse-time SDEs. Based on Fig. 1 in [4].

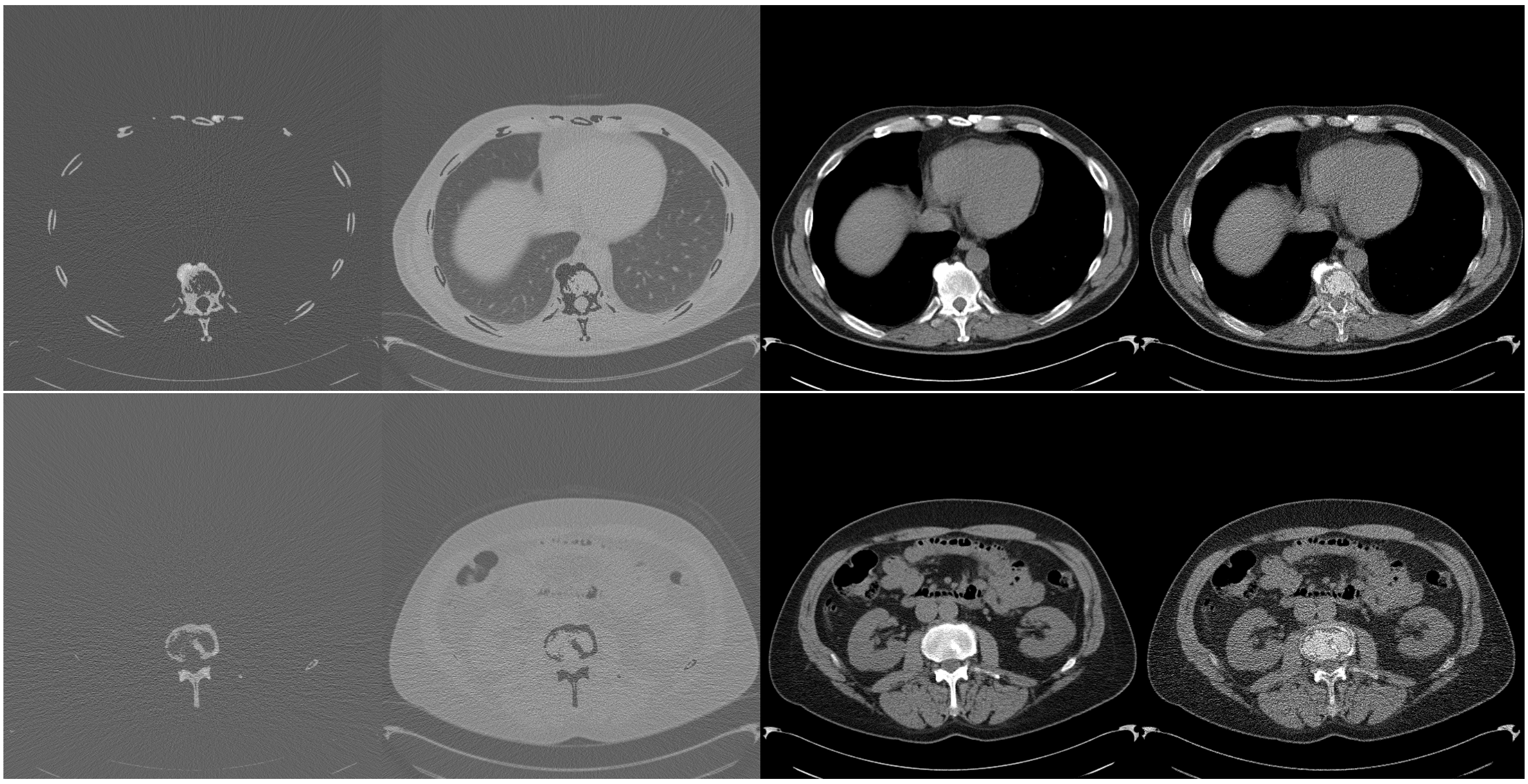

**Figure 2:** Examples from the simulated photon-counting spectral CT dataset based on KiTS19. These data are treated as "real" in this study. Note that these images have been resized to $512 \times 512$. From left to right: bone, soft tissue, 70 keV, and 100 keV.

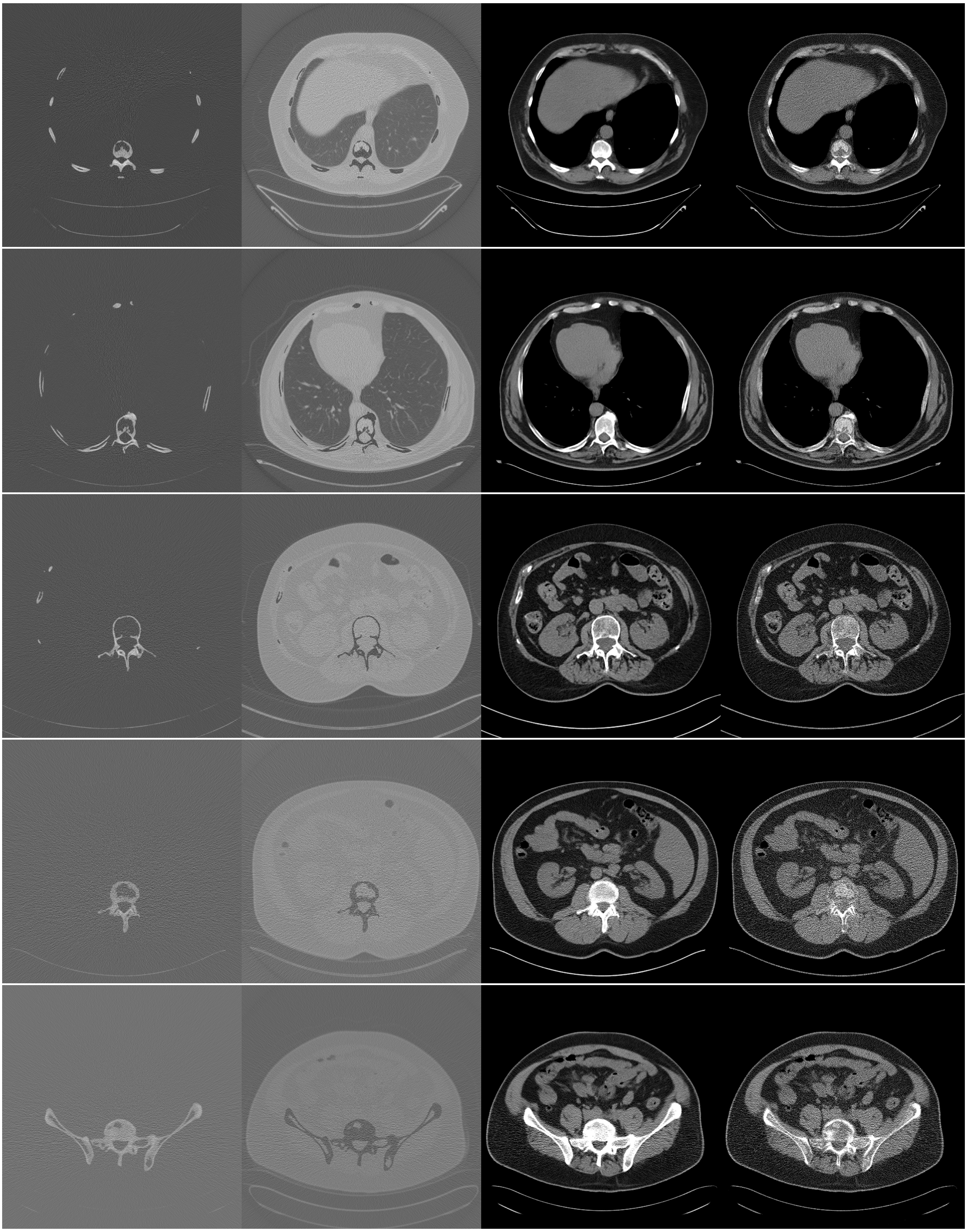

**Figure 3:** Examples of generated data. The virtual monoenergetic images are the output from the network and the material basis images are recovered via a linear transformation. From left to right: bone, soft tissue, 70 keV, and 100 keV.

# Abdominal organ segmentation based on VVBP-Tensor in sparse-view CT imaging

Zixuan Hong[1,3], Chenglin Ning[1], Zhaoying Bian[1,2], Dong Zeng*[1,2], and Jianhua Ma[1,2]

[1]School of Biomedical Engineering, Southern Medical University, Guangdong 510515, China
[2]Pazhou Lab (Huangpu) , Guangdong 510000, China
[3]Department of Radiotherapy, Sun Yat-sen Memorial Hospital, Sun Yat-sen University, Guangzhou, Guangdong 510120, China

**Abstract** Abdominal organ segmentation is critical in abdomen lesion diagnosis, radiotherapy, and follow-up. It is time-consuming and expensive for oncologists to delineate abdominal organs efficiently and accurately. Deep learning (DL)-based strategies have been shown great potential for abdominal organ segmentation and then manual delineation efforts reduction. It should be noted that most of the DL-based segmentation methods are constructed via full-view CT images which are reconstructed through multiple views projections at normal dose. Meanwhile, the radiation dose effect is a major concern in the CT imaging, especially for the full-view CT acquisition. Then, the accumulated radiation dose of the training CT dataset is also serious. Lowering radiation dose, i.e., sparse view sampling, is an effective way in the CT imaging. It is challenging to delineate the abdominal organs in the sparse-view CT images due to the artifacts. In this work, to achieve this goal, we add the effort to segment organ directly from the sparse-view CT measurements. Specifically, we first construct the VVBP-tensor from the sparse-view measurements. Based on the above dataset, we construct a VVBP-Tensor segmentation network (VVBP-SegNet). Experimental results on simulated dataset demonstrate that the proposed VVBP-SegNet obtains best segmentation results than the other competing methods in qualitative and quantitative assessments. This can provide a new insightful strategy for abdominal organ segmentation in the sparse view CT imaging.

## 1 Introduction

Abdominal organ segmentation is a fundamental image analysis task that supports many clinical applications, i.e., organ size quantification, disease diagnosing, radiotherapy and follow up. Manual segmentation of abdominal images is time-consuming and may result in inter- and intra-operator variability. Various segmentation methods have been developed [1]. Among them, deep learning based segmentation methods have achieved remarkable results in medical imaging field [2, 3]. For example, E. Gibson et al. [3] introduced dense connection in each encoder block to segment multiple organs on abdominal CT.

Although these DL-based methods have shown great potential, they still have intrinsic limitations. On the one hand, most of these methods are trained based on full-view CT images which are reconstructed through multiple views projections at normal dose. The radiation burden associated with CT imaging has been a major concern in the wide applications. Reducing radiation dose (i.e., lowering mAs, sparse-view sampling) directly without any treatments would lead to severe noise-induced artifacts or streak artifacts in the images. The DL-based strategies are generally used to improve low-dose CT image quality, for example, iRadon-MAP [4], SLSR-Net [5], VVBP-Net [6], etc. Furthermore, it is challenging to delineate abdominal organ region in the sparse-view CT images with streak artifacts directly.

On the other hand, most of the existing DL-based segmentation methods are designed based on image domain, failing to consider the characteristics latent in the projection/sinogram domain. At the same time, some researches are also developed directly for organ segmentation from the projection data [7, 8]. For example, D. Edmunds et al. segment diaphragm tissue for localizing lung tumor in individual projections [7]. These methods directly delineate the organ in the projection domain, avoiding being influenced by the artifact effect on the reconstructed image. But they neglect the structure characteristics, region size and shape in the image, which leaves a large room for segmentation performance promotion. Thus, the DL-based method constructed based on both projection and image domain is an alternative strategy in the abdominal organ segmentation task, which is also the goal in this study.

In this work, based on the previous study [9], we develop an abdominal organ segmentation network based on the view-by-view backprojection tensor (VVBP-Tensor) measurements which considers the characteristics in both sinogram and image domains in sparse-view CT imaging. For simplicity, the proposed network is termed as VVBP-Tensor segmentation network (VVBP-SegNet). In the proposed VVBP-SegNet, the sparse-view samplings are backprojected into 3D data before summation, and the 3D data are transformed into VVBP-Tensor measurements after applying a sorting operation. The VVBP-Tensor measurements have the potential to provide lossless information for processing and preserve fine details of image. Then we construct the VVBP-SegNet with the VVBP-Tensor measurements in the network. Experiments were conducted using simulation studies with three sparse-view sampling cases. The results demonstrate that the proposed VVBP-SegNet obtains accurate segmentation results with higher Dice coefficient (DSC) and lower 95% Hausdorff distance (HD) in both single-organ and multi-organ segmentation tasks, outperforming the competing method with sparse-view images.

*Corresponding author: D. Zeng, zd1989@smu.edu.cn

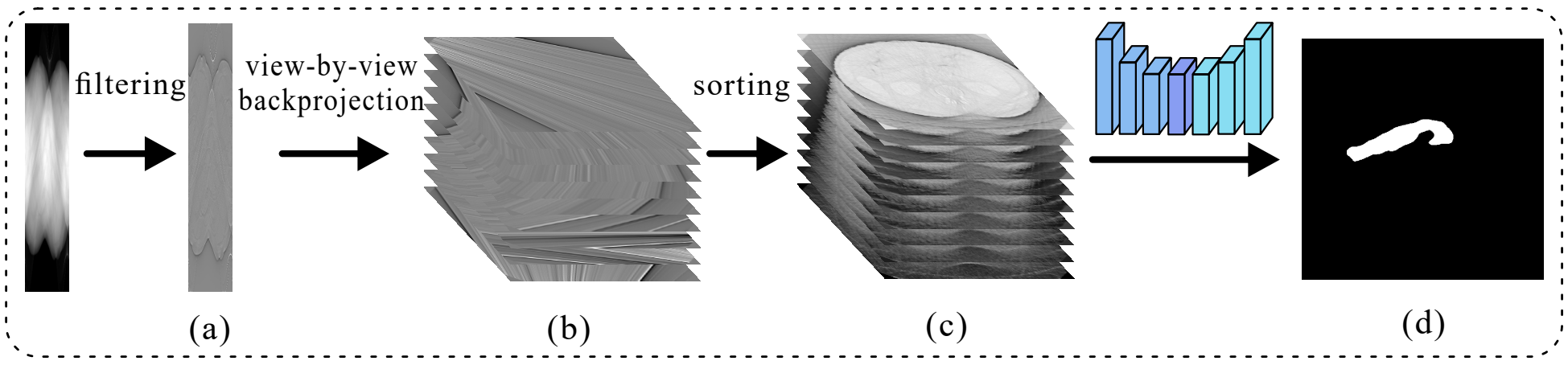

**Figure 1:** Framework of the proposed VVBP-SegNet.

## 2 Materials and Methods

### 2.1 VVBP-SegNet

Figure **??** shows the architecture of the proposed VVBP-SegNet. Based on the previous study [9], the conventional FBP algorithm can be technically decomposed into three steps: (a) filtering, (b) view-by-view backprojection (VVBP), and (c) summing along the z direction (view direction). The filtered sinogram is shown in Figure **??** (a), and the data following VVBP can be treated as an 3-order tensor (VVBP-Tensor) as shown in Figure **??** (b). Then, each mode-3 vector of the VVBP-Tensor is independently sorted according to its pixel values. After sorting, the VVBP-Tensor contains structures similar to those objects as shown in Figure **??** (c), indicating that it has the same organ shape and size as the FBP reconstructed image. Moreover, VVBP-Tensor can provide more information on abdominal organs than the conventional CT images due to the tensor measurements.

As shown in Figure **??** (d), VVBP-SegNet treats the VVBP-Tensor as the network input, and produce the desired abdominal organ segmentation results directly from the network. More, we observe that the structures in different slices of the tensor are highly correlated. We only select 10 middle slices of the tensor as the input to promote network efficiency. Besides, middle slices of VVBP-Tensor have fewer artifacts, which is beneficial for segmentation.

### 2.2 Network structure and loss function

In the proposed VVBP-SegNet, nnU-Net [10] is selected as the backbone network due to its high segmentation performance The nnU-Net is based on the U-Net with down- and up-sampling convolution levels to extract features. The network loss function is the sum of dice loss and binary cross-entropy loss (DC-BCE loss) as follows:

$$L_{DC} = \frac{-2\sum_i \hat{y}_i y_i}{\sum_i y_i + \sum_i \hat{y}_i}, \tag{1}$$

$$L_{BCE} = \sum_i y_i log \hat{y}_i + (1 - y_i)\, log\,(1 - \hat{y}_i), \tag{2}$$

$$L = L_{DC} + L_{BCE}. \tag{3}$$

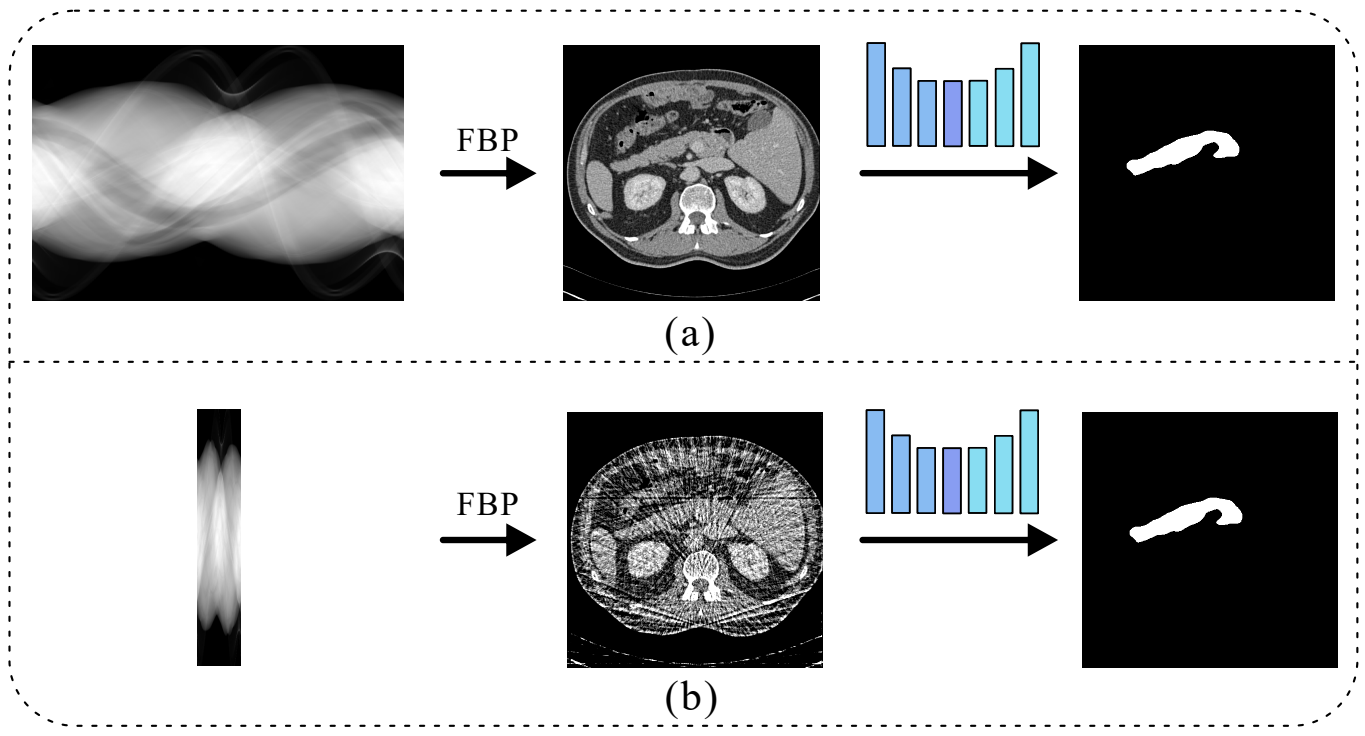

**Figure 2:** Frameworks of the other two competing methods, (a) F-SegNet; (b) S-SegNet.

### 2.3 Dataset

To validate and evaluate the segmentation performance of the proposed VVBP-SegNet, the AbdomenCT-1k dataset is used. The Abdomen-1k dataset is shared online about abdomen CT images for four-organ segmentation, containing the liver, kidney, spleen, and pancreas [11]. We choose about 3000 images of 60 patients, divided into 8/2 for training/validating with some additional cases for testing. The sparse-view CT images and VVBP-Tensor data are obtained based on the previous study [9, 12]. In the experiment, sparse-view sampling with 60 views, 90 views, and 120 views are simulated.

### 2.4 Compared methods and training details

As shown in Figure **??**, two strategies are chosen as the competing methods, i.e., F-SegNet and S-SegNet. Specifically, the F-SegNet is constructed based on the CT images acquired with full-view measurements, and S-SegNet is constructed based on sparse-view CT images with no treatments. The number of epochs for all methods is 1000 when the training and validation loss curves converge. The optimizer is SGD with 0.99 momentum, and the attenuation coefficient is $3e^{-5}$.

## 3 Results

### 3.1 Single-organ segmentation results

Figures **??** and **??** show the single-organ segmentation results obtained by all methods at three different sampling views. The F-SegNet is able to obtain segmentation results of the pancreas closest to the ground truth. Both the presented

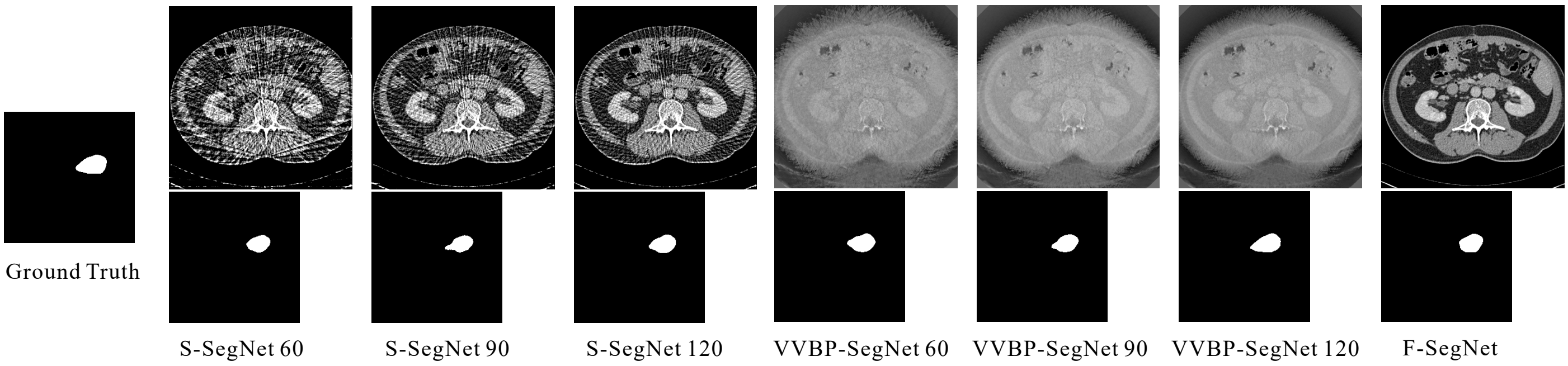

**Figure 3:** The segmentation comparisons of S-SegNet, VVBP-SegNet and F-SegNet in case 1.

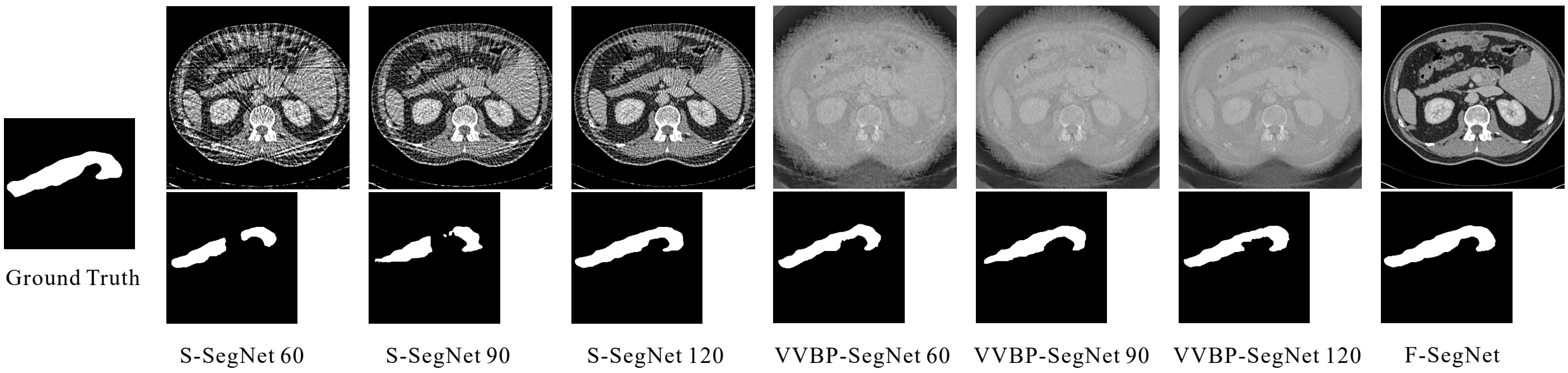

**Figure 4:** The segmentation comparisons of S-SegNet, VVBP-SegNet and F-SegNet in case 2.

VVBP-SegNet and S-SegNet methods can obtain similar segmentation performance with the pancreas with bean-shaped at different sampling views, as shown in Figure **??**. In Figure **??**, S-SegNet fails to delineate the elongated pancreas at 60 and 90 views due to the severe artifacts in the CT image. The proposed VVBP-SegNet achieves robust pancreas segmentation results in all cases, indicating its segmentation ability in the sparse-view imaging. The main reason is that the VVBP-Tensor measurements have the potential to provide lossless information for processing and preserve fine details of image. The numerical comparisons of the average Dice score (DSC) and 95% Hausdorff distance (HD) of testing cases are shown in Table **??**. The VVBP-SegNet has higher Dice scores and lower HD than the S-SegNet algorithm, and is closer to the F-SegNet.

| | Number of views | DSC | 95% HD |
|---|---|---|---|
| F-SegNet | | 0.835 | 3.651 |
| S-SegNet | 60-views | 0.665 | 4.789 |
| | 90-views | 0.729 | 4.500 |
| | 120-views | 0.757 | 4.266 |
| VVBP-SegNet | 60-views | 0.746 | 4.238 |
| | 90-views | 0.800 | 3.954 |
| | 120-views | 0.803 | 3.937 |

**Table 1:** The average Dice coefficient (DSC) and 95% Hausdorff distance (HD) of three methods for single-organ segmentation.

### 3.2 Multi-organ segmentation results

Figure **??** shows the multi-organ segmentation results obtained by the different methods at three different sampling views. From the results, the S-SegNet fails to segment the pancreas with elongated shape at 60 and 90 views, as the same as Figure **??** above. Our proposed VVBP-SegNet predicts shapes of four organs, having similar segmentation results with the F-SegNet in all cases. Table **??** summarized the quantitative measurements on the competing methods in all cases, and the results suggest that the proposed VVBP-SegNet performs better than the S-SegNet, which is consistent with those in the single-organ segmentation task.

| | Number of views | DSC | 95% HD |
|---|---|---|---|
| F-SegNet | | 0.959 | 15.904 |
| S-SegNet | 60-views | 0.937 | 20.486 |
| | 90-views | 0.951 | 17.970 |
| | 120-views | 0.950 | 17.789 |
| VVBP-SegNet | 60-views | 0.938 | 18.457 |
| | 90-views | 0.952 | 17.900 |
| | 120-views | 0.953 | 17.327 |

**Table 2:** The average DSC and 95% HD of three methods for multi-organ segmentation.

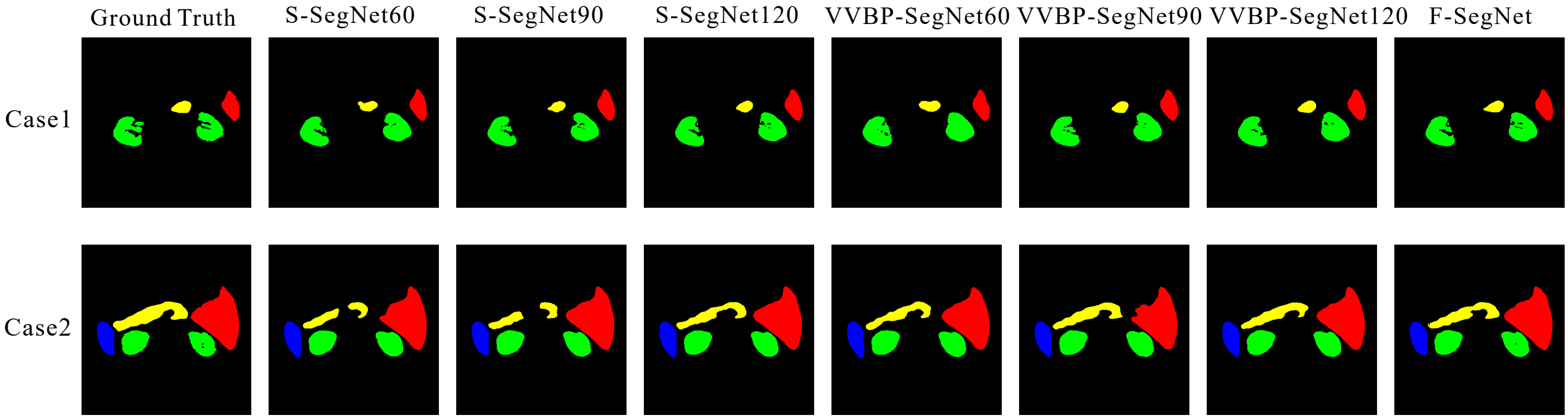

**Figure 5:** The multi-organ segmentation comparisons among different segmentation methods. The liver is red, kidneys are green, spleen is blue, and pancreas are yellow.

## 4 Discussion

The point of the proposed method is identifying the three-dimensional morphology of internal organs with a smaller number of scanning angles. Therefore, the clinical translation of our method would consider the situations of reducing radiation of patients, for example, calculating the volume of organs and nodules with sparse-view CT.
The methods of restoring sparse-view CT image followed by segmentation involves two steps. Compared to our proposed method, the workload of these methods is double. These methods need to suppress artifacts carefully, in case new artifacts introduced further affect subsequent segmentation. In addition, there is still a lot of difficulty for segmentation task due to low contrast and blurry details in restored image. In the proposed method, VVBP-Tensor measurements after sorting reconstruct the pixel values, and exactly highlight the contours of various organs, which is a solution to the above problems.

## 5 Conclusion

In this work, we develop a VVBP-SegNet for abdominal organ segmentation in the sparse-view CT imaging through the newly developed VVBP-Tensor measurements which considers the characteristics in both sinogram and image domains. The experimental results on the simulated sparse-view CT images at the different sampling views demonstrate that the proposed VVBP-SegNet can provide a similar segmentation performance with the F-SegNet, compared with the ground-truth. In the future, more studies will be conducted to demonstrate the efficiency of the proposed VVBP-SegNet.

## 6 Acknowledge

This work was supported in part by the NSFC under Grant U21A6005, and Grant 12226004, and Young Talent Support Project of Guangzhou Association for Science and Technology.

# On the optimal selection of energy thresholds for quantification of gold concentration in photon-counting-based CT

Xiaoyu Hu[1,3], Yuncheng Zhong[3], Kai Yang[2], and Xun Jia[1]

[1]Department of Radiation Oncology and Molecular Radiation Science, Johns Hopkins University School of Medicine, Baltimore, MD 21231, USA
[2]Division of Diagnostic Imaging Physics, Department of Radiology, Massachusetts General Hospital, Boston, MA 02114, USA
[3]Department of Radiation Oncology, University of Texas Southwestern Medical Center, Dallas, TX 75390, USA

**Abstract** Quantitative material decomposition is one of the advantages offered by photon-counting-based dual-energy CT (DECT) or multi-energy CT (MECT). For this task, it is important to determine proper thresholds that separate energy channels to minimize the decomposition error of the material of interest. In this study, we developed an analytical expression for the error that considered projection noise level and mathematical properties of the material decomposition matrix and proposed to use that to determine optimal energy thresholds. Comprehensive simulations were performed to verify this idea in an example problem of quantifying gold concentration. We simulated x-ray projections of a phantom with various inserts containing gold solutions under a 140 kVp x-ray tube voltage, a range of photon counts from 4000 to 128000 per pixel, and enumerated possible energy thresholds. We calculated gold concentration in each setting. Simulation results showed that the optimal energy channels that minimize the error of the gold image were $(30,42]$ and $(42,140]$ keV for DECT, and $(30,42]$, $(42,62]$and $(62,140]$ keV for MECT with three channels. The optimal thresholds estimated by the proposed method matched the simulation results for both DECT and MECT cases.

## 1 Introduction

Material decomposition is one of the main applications of dual-energy or multi-energy computed tomography (DECT or MECT). The benefit of material decomposition is not only to enhance the visibility of the region of interest but also to provide quantitative information of the target material. While most DECT and MECT approaches for material decomposition have employed conventional energy-integrating detectors [1], recent advances in photon-counting detectors (PCD) have offered a new direction for this problem with technological advantages, such as better spectrum separation [2–5].

PCD often operates in a mode acquiring photon counts above user-specified energy thresholds to allow separating photons into different energy channels. Therefore, selecting appropriate thresholds is of critical importance for the achievement of accurate material decomposition and quantification of the spatial distribution of the material of interest. Over the years, great efforts have been devoted to investigating PCD-based material decomposition [4, 5]. Yet, there is a lack of studies to address the relationship between energy threshold selection and resulting accuracy. When K-edge material is one of the base materials, although it is well-accepted that the K-edge should be included in one of the thresholds, other quantities, such as the noise in the projection data, can also affect the decomposition accuracy. In addition, material decomposition usually generates two or more material images, while only one of them may be of interest. In this study, we address the energy threshold selection problem in PCD-based DECT and MECT with an example problem of a quantitative assessment of gold contrast agent concentration. In recent years, there has been increasing interest in using gold nanoparticles for imaging and therapy applications [6]. Due to toxicity concerns, there is a strong need to optimize the imaging system's sensitivity to gold to enable visualization and quantification of gold distribution at low concentration levels.

The contribution and significance of this study are threefold. First, we derived an expression for the numerical accuracy of the decomposition result of the material of interest and proposed to optimize energy thresholds based on this quantity. Second, we found that the optimal selection of thresholds is governed by the balanced consideration of projection data noise and numerical properties of the material decomposition matrix. Third, we performed numerical simulations to validate our theoretical derivations.

## 2 Methods

### 2.1 Theory

Let us define $m \in \mathbb{R}^{N_v \times N_m}$ as the material images to be determined, $P \in \mathbb{R}^{N_p N_d \times N_v}$ as the x-ray projection system matrix, $A \in \mathbb{R}^{N_m \times N_e}$ as the material decomposition matrix, and $u \in \mathbb{R}^{N_p N_d \times N_e}$ as the projection data acquired in a PCD-based CT system after logarithm transform. Assuming that the decomposition matrix $A$ is well-calibrated, the model of material decomposition can be expressed as

$$PmA = u, \tag{1}$$

where $N_p$ is the number of projections, $N_d$ is the number of detector pixels, $N_v$ is the number of voxels, $N_e$ is the number of energy channels, and $N_m$ is the number of base materials. For an actual CT scan, the acquired data $\hat{u} = u + \delta u$ due to errors such as noise, the decomposed result becomes $\hat{m} = m + \delta m$. We have

$$P\delta mA = \delta u. \tag{2}$$

Suppose the first column of $m$ is the material image of interest, we abstracted this image by multiplying a vector $S$ to the right of $m$, namely $m_1 = mS$, where $S \equiv [1,0,..,0]^\top \in \mathbb{R}^{N_m \times 1}$. Thus

$$Pm_1 = PmS = uA^+S, \tag{3}$$

$$\delta m_1 = \delta m S = P^+ \delta u A^+ S. \tag{4}$$

where $\cdot^+$ denotes matrix pseudo-inverse. Note that $\|S\|_p = 1$ for all $p$. We have

$$\|P\|\|m_1\| \geq \|Pm_1\| = \|uA^+S\|, \tag{5}$$

$$\|\delta m_1\| \leq \|P^+\|\|\delta uA^+S\|. \tag{6}$$

With these two equations, we arrive at

$$\frac{\|\delta m_1\|}{\|m_1\|} \leq \kappa_P \frac{\|\delta uA^+S\|}{\|uA^+S\|} \equiv \kappa_P C(T). \tag{7}$$

In Eq. (7), $\kappa_P$ is the condition number of $P$, reflecting the error propagation during the CT reconstruction process. This fact is not affected by energy threshold selection. In the second term, as we adjust the energy threshold $T$, both $u$ and $A$ will be affected. Here we defined the term $C(T)$ and explicitly indicated its dependence on $T$. The threshold will impact the noise level of $u$, as the photon counts in each energy channel are changed. It will also modify the properties of the material decomposition matrix $A$ because basis vectors of different materials in the space spanned by energy channels depend on the threshold $T$. With this expression, we propose to select the optimal energy threshold $T$ to minimize $C(T)$, in order to minimize the relative error of the concentration of the decomposed gold material.

## 2.2 Simulation study

We considered the setup for an in-house PCD-based CT system for small-animal imaging and image guidance in preclinical radiation experiments [5]. The source-to-isocenter distance was 30.49 cm, and the source-to-detector distance was 44.21 cm. The CdTe PCD had 512 pixels with a 0.1 mm pixel size and 0.75 mm width.

Filtered by a 0.25 mm copper sheet, a 140 kVp x-ray spectrum $S(E)$ was used (Figure 1(a)). While 140 kVp is likely too high for small-animal applications, we chose this to achieve enough photon counts above the K-edge of gold at 80 keV. The lowest threshold was fixed at 30 keV to eliminate the electronic noise. To select proper thresholds, we generated a pool of candidate energy thresholds ranging from 32 to 110 keV with a 2 keV step. For DECT, the goal was to select one threshold $T$ from all the candidates to form energy channels: $(30, T]$ and $(T, 140]$ keV. For MECT with three energy channels, we selected two thresholds $T_1$ and $T_2$ for energy channels $(30, T_1]$, $(T_1, T_2]$ and $(T_2, 140]$ keV.

The unattenuated photon counts $C_0$ ranged from 4000 to 128000 per pixel to cover the typical CT dose in small animal imaging settings. For a $C_0$, we repeated the following calculations for each threshold $T$ in the DECT case, or threshold combinations $T = \{T_1, T_2\}$ for the MECT case to compute $C(T)$ and hence decide the optimal selection.

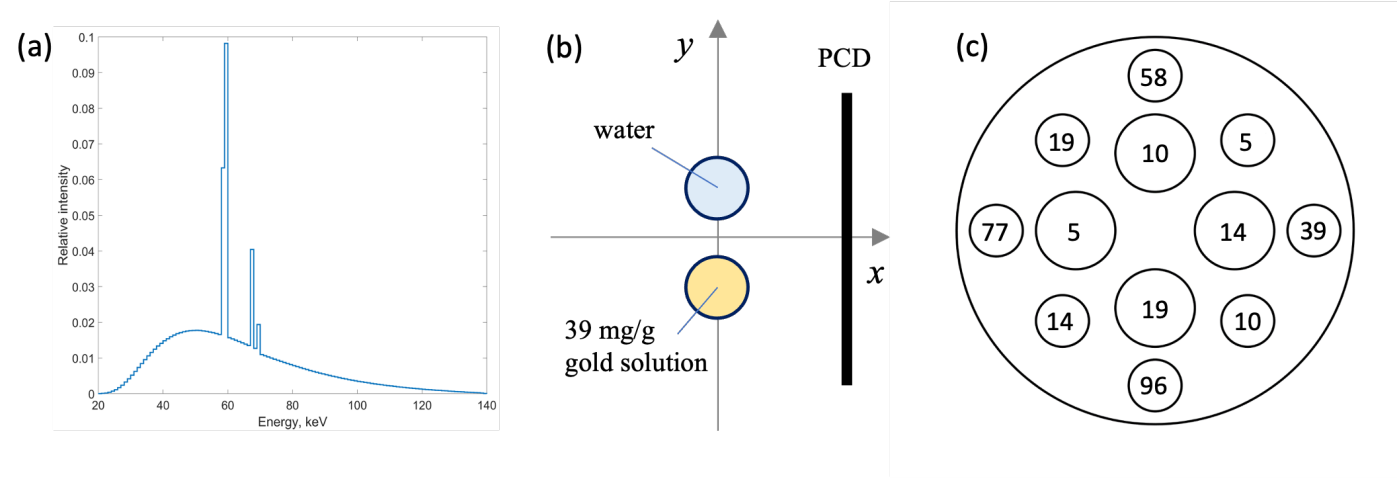

**Figure 1:** (a) X-ray spectrum at 140 kVp; (b) Calibration setup; (c) The phantom.

### 2.2.1 Calibration

The calibration setup is depicted in Figure 1(b). Two vials contained pure water and 39 mg/g gold solution, which mimicked the calibration configuration used in our lab. We simulated detector photon counts for a given energy channel defined by energy threshold $T_L$ and $T_H$:

$$\int_{T_L}^{T_H} S(E)C_0 \exp\left[-\int_L \mu(x,E)\mathrm{d}l\right] \mathrm{d}E, \tag{8}$$

where $L$ is the line connecting the x-ray source to each detector, and $\mu(x,E)$ is the x-ray attenuation coefficient at $x$ for energy $E$, computed based on known material composition in the calibration configuration. The noise was not included in the simulation, because in calibration we do not have dose constraint, and it is possible to acquire sufficient data to mitigate noise.

The fundamental equation for the calibration is the same as Eq. (1), but we formulated it as $BA = g$, where $B \equiv Pm_{gt}$ is the summation of path length times the density of every material along the x-ray line to a detector pixel, $m_{gt}$ is the known material density map for the calibration configuration, and $g$ is the projection data computed above after logarithm transform. Solving this equation with respect to $A$ yielded the decomposition matrix.

### 2.2.2 Material decomposition

We generated a digital phantom with a 3 cm diameter (Figure 1(c)) consisting of gold solutions of various concentrations in a water background. We computed projection data for a DECT/MECT scan with 600 projections uniformly distributed over a $2\pi$ angular range. For each detector bin, the projection data were calculated using the same equation as Eq. (8) except that Poisson noise was generated according to the calculated photon counts. With the computed projection data, we first reconstructed CT images of each energy channel with $312 \times 312$ voxels with 0.1 mm voxel size. We then computed the material images through $\hat{m} = \hat{f}A^+$, where $\hat{f} = P^+\hat{u}$ is the DECT/MECT images, and $P^+$ was implemented via the filtered back projection algorithm[7].

### 2.2.3 Evaluation

Using the decomposition results, we calculated the relative error norm $\|\delta m_1\|/\|m_1\|$. We enumerated all the energy threshold values $T$ and searched for the value that minimizes the relative error norm as the optimal threshold. Meanwhile, Eq. (7) provided the upper bound for the decomposition error. We evaluated the upper bound without the condition number of $P$, namely $C(T)$, which was compared with $\|\delta m_1\|/\|m_1\|$ to evaluate the consistency between the simulation and the theoretical calculations. The condition number of $P$ was not computed because it is a great challenge to estimate the condition number for large matrices. Since it does not depend on the energy threshold, we ignore this scalar when comparing theory and simulation results.

## 3 Results

### 3.1 DECT case

For DECT, the threshold $T$ splits the entire spectrum into two energy channels. The decomposition errors in the gold image computed with simulations and its counterparts $C(T)$ estimated by our theory with respect to different thresholds at $C_0 = 4000$ are plotted in Figure 2(a). Because the relative error term and $C(T)$ differ by a factor $\kappa_P$, it is expected that the two curves share the same trend. While $T = 80$ keV, the K-edge of gold, appeared to be a local minima, the actual minima existed at 42 keV for both simulation and theory. This indicated that different from the common setup that uses the K-edge as the threshold to define energy channels, the optimal choice of $T$ actually is much lower.

To look at this behavior in more detail, we plotted several other curves. Specifically, the first one is the dependence of the relative uncertainty of projection data $\|\delta u\|/\|u\|$ on the threshold $T$ and photon counts $C_0$, as in Figure 2(b). As expected, $\|\delta u\|/\|u\|$ increased as $C_0$ was reduced due to amplified counting noise. As for the dependence on $T$, when one of the energy channels is too narrow, the noise in that channel will dominate the error. Therefore, there must be a threshold (58 keV in our case) in the middle to balance the number of photons at two energy channels.

The second curve is the dependence of the condition number of $A$ matrix ($\kappa_A$) on $T$, as in Figure 2(c). This quantity governs the propagation of noise to the decomposed image during the material decomposition process by solving a linear model. Essentially, the two columns of $A$ contained effective attenuation coefficients of gold and water in different channels. They are linear representations of the gold and water materials in the linear space defined by the two energy channels. The more linear independence between the two, the more favorable it is in terms of the accuracy of material decomposition. The choice $T = 80$ keV at the gold K-edge gave a local minima of $\kappa_A$, as the sharp jump of the x-ray attenuation coefficient of gold at the K-edge differentiates gold from water, when having the two energy channels above and below the K-edge. As moving away from the K-edge, the condition number started to increase. But after a certain energy range, it started to decrease again. As the threshold $T$ further deviates from the K-edge, the two effective attenuation coefficients of gold in the two channels are sufficiently different from those of water, yielding even lower condition numbers of $A$ than that when $T$ equals the K-edge. Note that since we assume that $A$ was well-calibrated, $\kappa_A$ does not depend on $C_0$.

While $\|\delta u\|/\|u\|$ and $\kappa_A$ are not directly related to $C(T)$ due to its complex form, they provided some insights about the overall behavior of $C(T)$ as a function of $T$. It is the combination of the two effects, namely projection noise level affected by photon counts and decomposition matrix $A$'s properties, that led to the overall behavior of the decomposition accuracy observed in Figure 2(a).

Finally, to directly look at the decomposition results, we show in Figure 3 the decomposed gold images with the threshold at 42, 54, 58, and 80 keV respectively. 54 keV and 58 keV are the minima from $\kappa_A$ and $\|\delta u\|/\|u\|$, respectively, 42 keV is the optimal threshold selected by our method, and 80 keV is gold's K-edge. Clearly, the gold image with the threshold at 42 keV is the least noisy. Quantitatively, we computed the decomposition results of gold density for each insert at $C_0 = 4000$ and 128000, as seen in Figure 4(a). With increasing $C_0$, the result approached the ground truth and the uncertainty became smaller. $T = 42$ keV gave the most accurate results compared to other thresholds for all the gold inserts.

### 3.2 MECT case

For MECT with three energy channels and $C_0 = 4000$, Figure 5(a) depicts the relative error of simulation results with the minimal value appearing at $T = \{T_1, T_2\} = \{42, 62\}$ keV, and the function $C(T)$ with the minimal at the same energy thresholds. In contrast, $\kappa_A$ had the minima at $T = \{78, 80\}$ keV, and the relative noise error of $u$ reached its minimum at $T = \{54, 66\}$ keV. The gold material decomposition results are shown in Figure 5(b) for the optimally selected thresholds by the simulation and theory, and for the thresholds at the minimum of $\kappa_A$ and $\|\delta u\|/\|u\|$. The derived gold concentrations are presented in Figure 4(b). Again, the case with $T = \{42, 62\}$ keV outperformed all other cases.

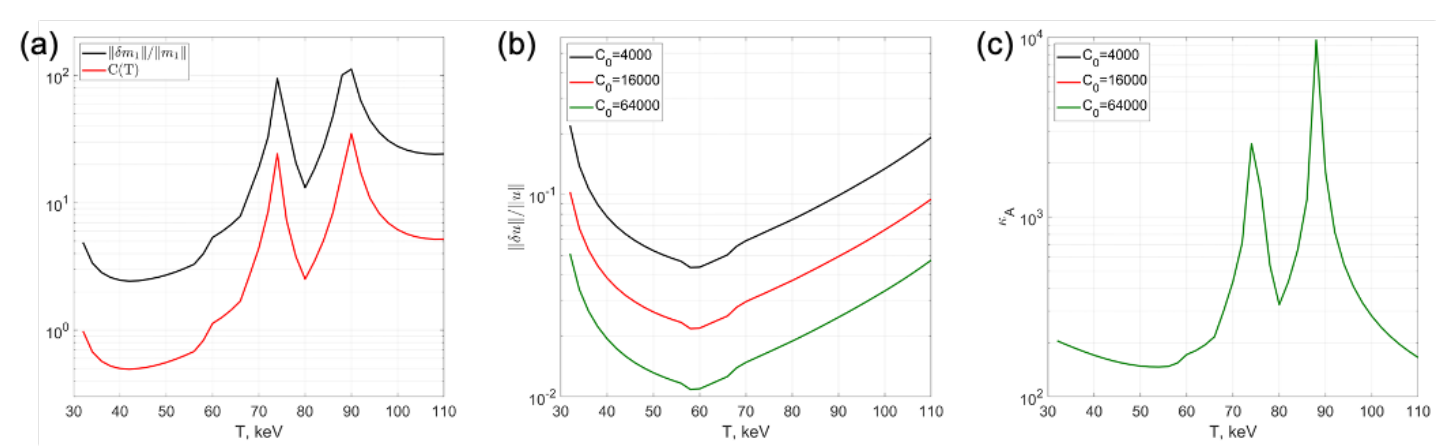

**Figure 2:** (a) $\|\delta m_1\|/\|m_1\|$ and $C(T)$; (b) $\|\delta u\|/\|u\|$ versus $T$ and $C_0$; (c) $\kappa_A$ versus $T$ and $C_0$.

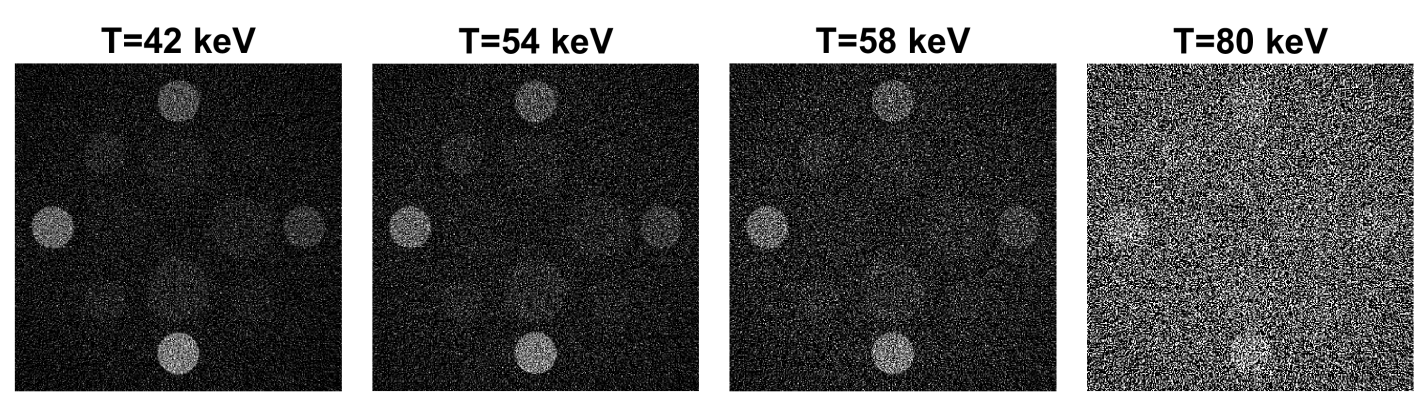

**Figure 3:** Gold images with $C_0 = 4000$ at thresholds at 42, 54, 58, and 80 keV respectively. Display window: $[0, 0.2]$ g/cm$^3$.

## 4 Discussion

The selection criteria $C(T)$ included terms related to the projection data. Hence, the result is actually case-dependent. In practice, it is necessary to decide the optimal thresholds using the proposed method and evaluate $C(T)$ with a phantom most representative of the actual experiments, e.g. with similar phantom size, the material of interest, and material concentrations, etc. We used a phantom with a 3 cm diameter and a range of gold concentrations to illustrate our proposed approach in a setup intended for an experiment currently ongoing in our lab using gold nanoparticle as a probe for kidney damage. For applications in clinical settings, a large phantom mimicking a human should be used. While the method is still expected to be valid, the selected thresholds will be likely different. Additionally, realistic issues such as beam hardening and scattering should be considered.

This study only considered a simple linear reconstruction and decomposition model. It is well-known that employing a regularization approach, classical or recent deep learning-based could further suppress noises in material decomposition problems. Once a regularization term is used, the proposed threshold selection method would not be applicable, and it would be difficult to derive theoretical guidance in these setups. However, we expect that the analysis here based on the linear model could still provide insights for those more complicated decomposition models, because the decomposition results would be still largely determined by the linear model rather than the regularization terms.

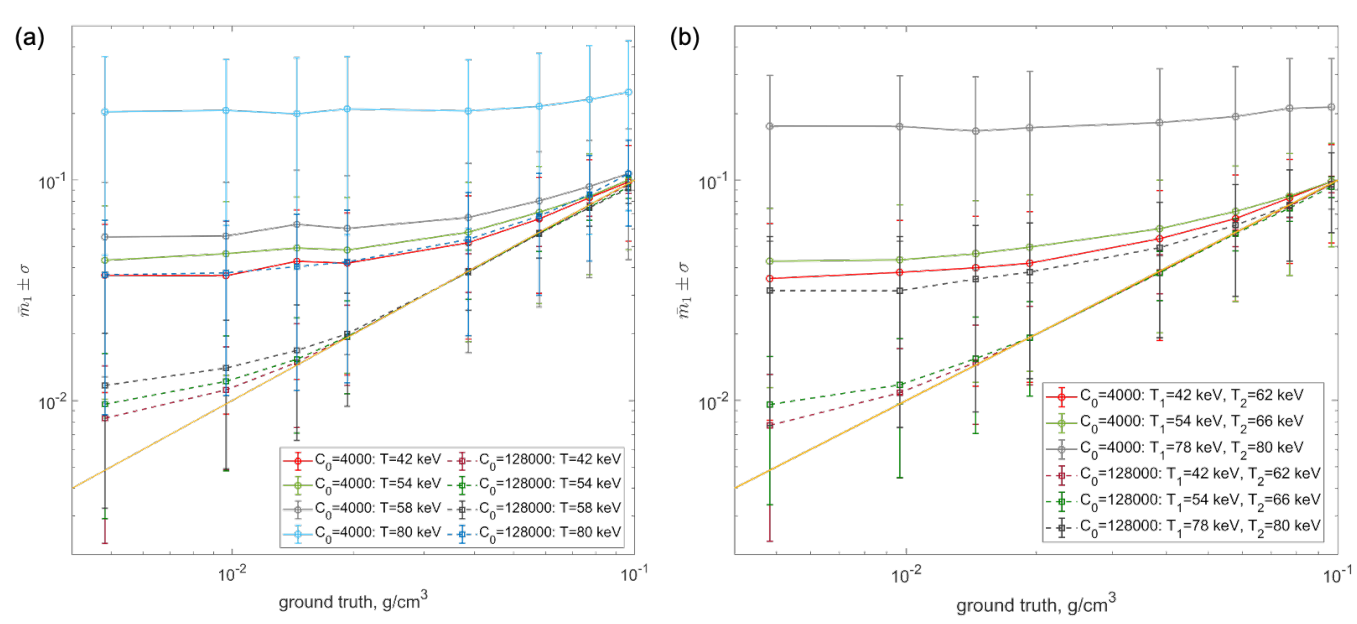

**Figure 4:** Gold quantification results at each inset for different $C_0$ at (a) DECT and (b) MECT. Yellow lines indicate that the mean equals the ground truth.

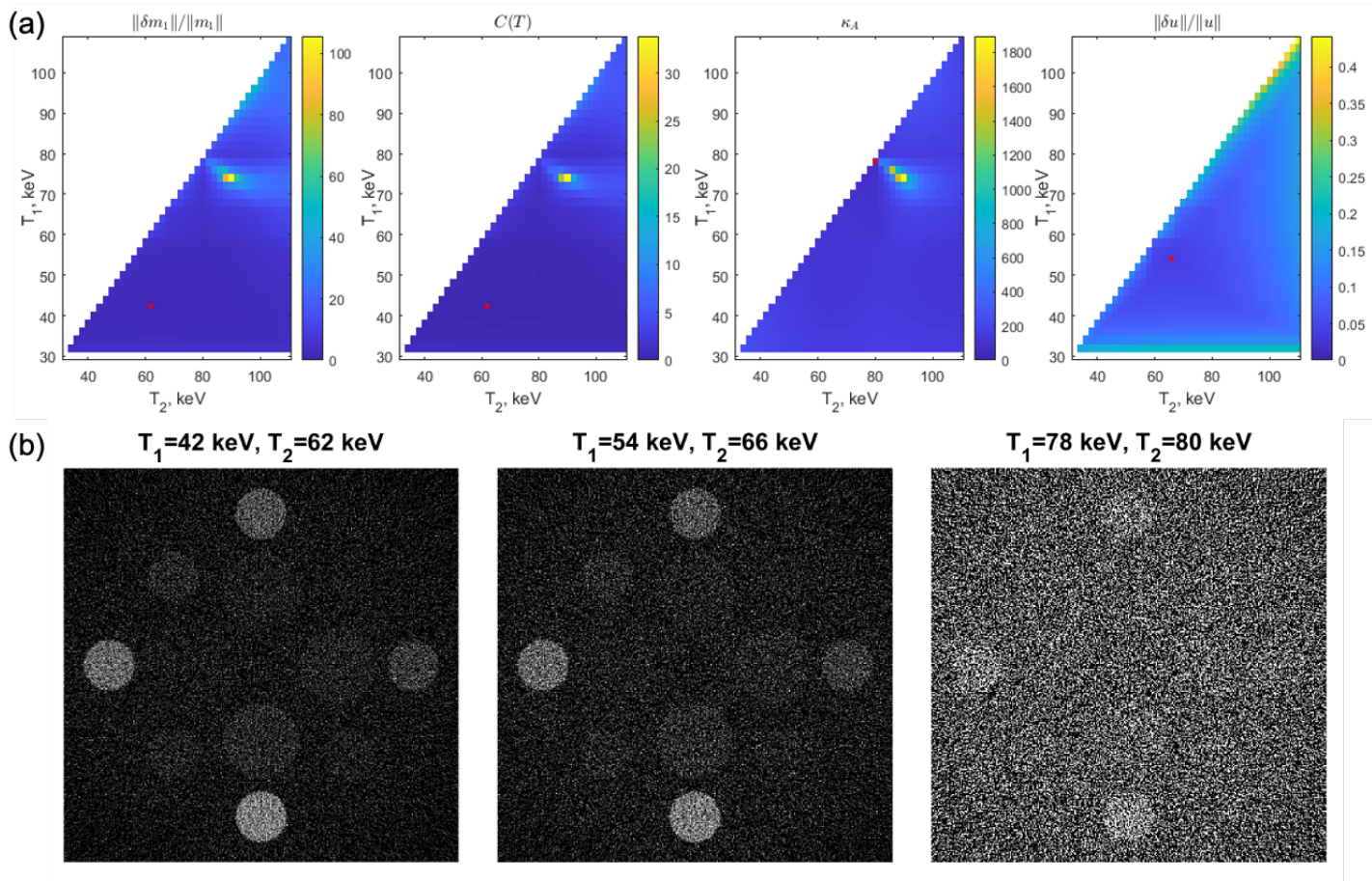

**Figure 5:** (a) $\|\delta m_1\|/\|m_1\|$, $C(T)$, $\kappa_A$ and $\|\delta u\|/\|u\|$ versus $T = \{T_1, T_2\}$; The red stars indicate the minima locations; (b) Gold images with $C_0 = 4000$ at $T$=$\{T_1, T_2\}$=$\{42, 62\}$, $\{54, 66\}$, and $\{78, 80\}$ keV respectively. Display window: $[0, 0.2]$ g/cm$^3$.

## 5 Conclusion

In this study, we derived an expression to characterize the dependence of material decomposition accuracy in PCD-based DECT/MECT on the energy thresholds and used that to guide threshold selection. We focused on the accuracy of only the material of interest. The selection criteria balanced considerations of photon counting noise levels and the numerical properties of the decomposition matrix. Our simulation validated the proposed method in DECT and 3-channel MECT cases.

# kV scattered x-ray imaging for real-time imaging and tumor tracking in lung cancer radiation therapy

Xiaoyu Hu[1,3], Yuncheng Zhong[3], Kai Yang[2], and Xun Jia[1]

[1]Department of Radiation Oncology and Molecular Radiation Science, Johns Hopkins University School of Medicine, Baltimore, MD 21231, USA
[2]Division of Diagnostic Imaging Physics, Department of Radiology, Massachusetts General Hospital, Boston, MA 02114, USA
[3]Department of Radiation Oncology, University of Texas Southwestern Medical Center, Dallas, TX 75390, USA

**Abstract** Tumor tracking is an important task in image-guided radiation therapy for lung cancer. This study investigated the feasibility of Kilo-voltage Real-time Imaging with Scatter Photons (KRISP) using a photon-counting detector (PCD) to measure Compton-scattered x-ray photon signals for real-time volumetric image and tumor tracking during treatment delivery. We used a 120 kV x-ray slice beam to irradiate a slice containing the tumor. We acquired the photons scattered off this plane using a PCD with a parallel-hole collimator. Using a prior CT image as prior information, we formulated the image reconstruction problem as an optimization problem with respect to the deformation vector field between the prior image and the real-time image. The problem was solved via a forward-backward splitting (FBS) algorithm. We conducted an initial evaluation on KRISP using a CIRS lung phantom. It was found that the reconstructed image can capture tumor motion information with a root mean square error of 1.2 mm. Radiation exposure and interference from scattered photons from the mega-voltage therapeutic beam were also investigated.

## 1 Introduction

In lung cancer radiotherapy, real-time image tumor tracking is of critical significance for the management of respiration-induced tumor motion during radiation delivery. Pre-treatment imaging modalities, such as respiratory phase resolved 4D cone beam CT, are available to image the patient's internal anatomy. Yet it is still important to image the anatomical motion on-the-fly during treatment delivery, which offers valuable information for managing variations of motion amplitude and phase, as well as baseline drift from the motion observed prior to the treatment[1].

To date, there is no approach that can provide safe, accurate, reliable, and low-cost real-time imaging and tumor tracking functions. 3D/4D-cone beam CT or tomosynthesis cannot achieve this goal due to the long data acquisition time. X-ray projection-based methods suffer from poor tumor contrast because of the projection of 3D anatomy to a 2D plane. Indirectly estimating tumor position via anatomy surrogates is not reliable. While methods tracking implanted fiducial markers using orthogonal x-rays are accurate and reliable, the invasive procedure to implant markers increases patient risks. Recent advancement in MRI-linear accelerator enables MRI-based tumor tracking. Yet the high cost and hardware complexity likely hinder its clinical penetrations.

The challenge of real-time imaging is to acquire adequate relevant data within a short period of time, e.g. 0.2 sec, to derive motion information. In this study, we investigate the feasibility of an innovative solution Kilo-voltage Real-time Imaging with Scatter Photons (KRISP) that measures scattered x-ray photons in a unique imaging geometry with a photon-counting detector (PCD) and derives images in real-time assisted by patient-specific prior information. In previous studies, we proposed the KRISP idea and studied its feasibility via numerical simulation [2] and initial experimental studies [3]. The current study performs comprehensive experimental validations on this method to characterize its performance, evaluate its accuracy in tumor tracking, and investigate several practical issues for clinical applications.

## 2 Methods

### 2.1 KRISP setup

The basic setup of KRISP is illustrated in Figure 1. Using the x-ray tube of cone beam CT on a typical medical linear accelerator, a planar x-ray beam is delivered to the patient. Different from the typical fan-beam geometry in a CT scan that delivers the x-ray to an axial plan, the planar x-ray in KRISP intersects with the body at a plane of interest (POI) containing the iso-center and parallel to the superior-inferior direction. To image anatomy in the POI, a 2D detector that is oblique to both the x-ray beam plane and the mega-voltage (MV) therapeutic beam plane captures photons scattered out of this POI. The detector has a parallel-hole collimator, so that each pixel receives photons primarily from the direction normal to the detector plane. With this signal, it is possible to form an image capturing anatomical information on the POI. The output of KRISP is a real-time 3D volumetric image with tumor motion information, which is reconstructed by solving an optimization problem that ensures fidelity to the measured scatter signal, aided by a patient-specific prior image. KRISP targets at a temporal resolution of 0.2 sec, estimated based on typical tumor motion speed under respiration and the targeted geometry accuracy.

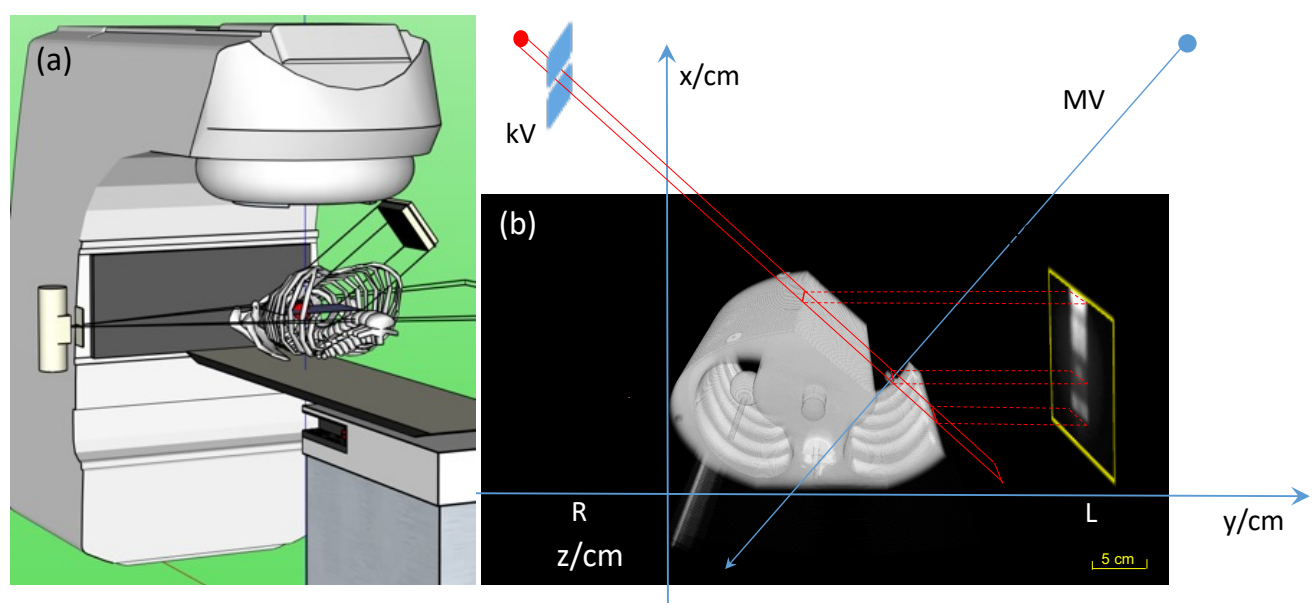

**Figure 1:** (A) Geometry configuration of the proposed system on LINAC. (B) An illustration of the geometry and imaging process.

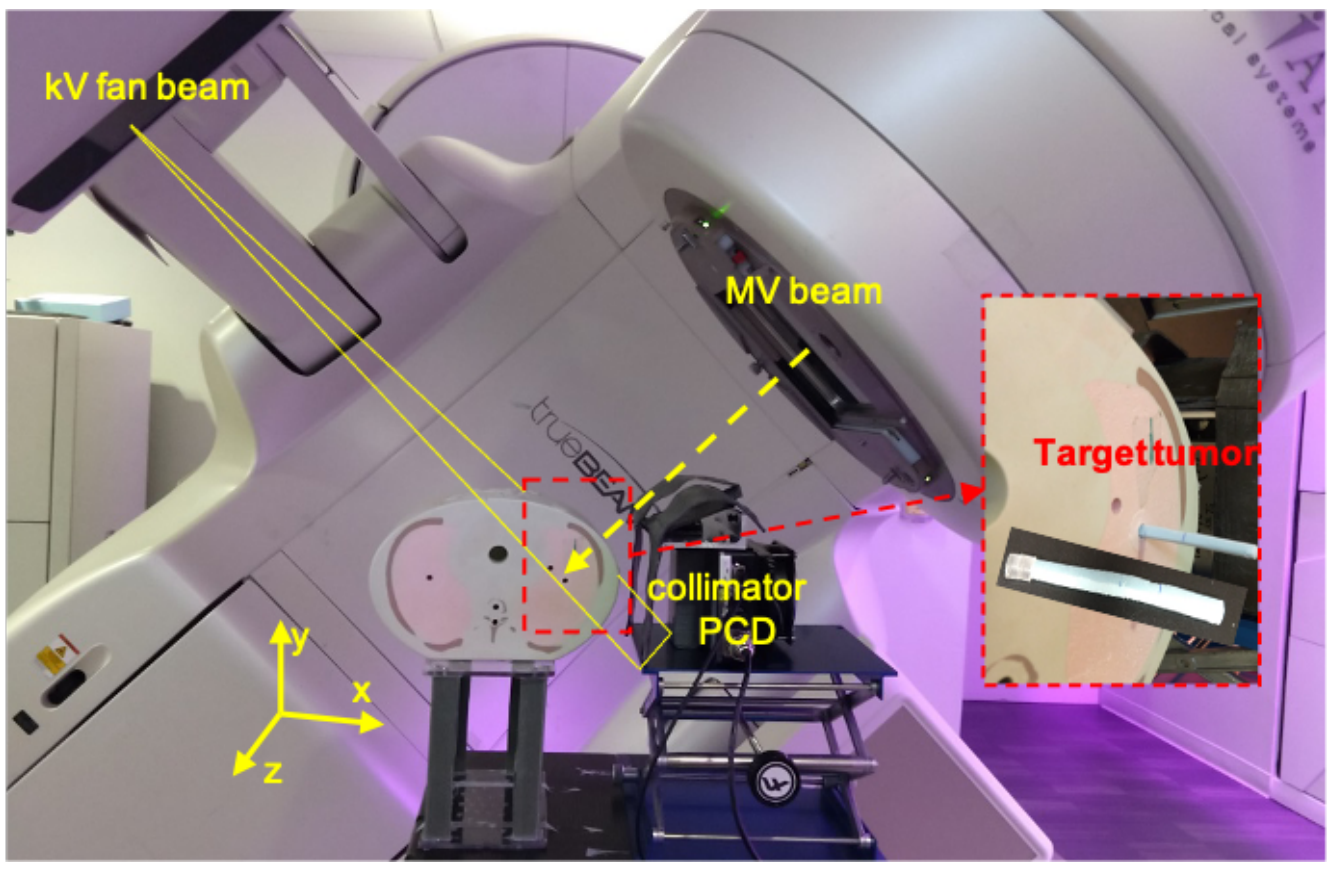

**Figure 2:** Experimental setup on a Varian TrueBeam medical linear accelerator.

## 2.2 Experimental study

The experiment was performed on a Varian TrueBeam machine (Varian Medical System, Palo Alto, CA) with a CIRS lung phantom (Sun Nuclear, Melbourne, FL). The geometry and the experimental setup are illustrated in Figure 2. The on-board x-ray tube was rotated to the gantry angle of 45° location. The on-board kV flat panel detector was retracted and the THOR photon-counting detector (PCD) (Varex Imaging, Salt Lake City, UT) assembly was installed at 135° gantry angle position along the horizontal direction. The detector pixel pitch is 100 $\mu$m and the sensitive area is $5 \times 10$ cm$^2$. A low-energy high-resolution (LEHR) parallel hole x-ray collimator (NuclearFields, Des Plaines, IL) was placed against the front surface of the PCD. The collimator was made from lead with 0.2 mm septa, 1.5 mm hole size, and 35 mm height. The kV fan beam emitted at 120 kVp (half value layer (HVL) of 4.9 mm Al) with 20 mAs (200 mA, 100 ms). The beam size was $4 \times 0.5$ cm$^2$ at iso-center. We measured scattered photon signals with the PCD using an energy threshold of 24 keV to capture photons above this level.

The CIRS phantom contained a spherical tumor with a diameter of 2 cm in the left lung. The tumor was placed at the isocenter as a steady tumor for reference. Next to it was a tunnel along the superior-inferior direction. We inserted an acrylic cylinder as the tumor under motion as shown in the red box in Figure 2. The acrylic cylinder was 12 mm long with a 6.5 mm diameter. We manually moved the target tumor in and out along the tunnel to mimic tumor motion, and recorded the relative distance to the reference tumor during the experiment at various positions as the ground truth tumor position used to study tumor tracking accuracy by KRISP.

## 2.3 Image reconstruction and motion tracking

Due to the low count of scattered photons, it is difficult to form a high-quality image of the POI directly using the measured scatter signal. The left subfigure in Figure 3 illustrates a typical measurement. We employed a patient-specific prior image, e.g. the right subfigure in Figure 3, to facilitate the reconstruction of the real-time image. As such, we formulated the reconstruction task as an optimization problem with respect to a motion vector field between the prior image $f_p(x)$ and the image to be reconstructed:

$$\begin{aligned} v &= \arg\min_v E[v] \\ &= \arg\min_v \frac{1}{2}\|K \otimes f_p(x+v(x)) - Cg\|_F^2 + \frac{\lambda}{2}\|\nabla v\|_F^2, \end{aligned} \tag{1}$$

where $K$ is the blurring kernel under the parallel-hole collimator, $g$ is the corrected measured signal, $\otimes$ is the convolution operator. $f_p$ is the 2D prior CT image extracted from the volumetric CT image at the incoming kV fan-beam plane. The second term enforces solution smoothness, and $\lambda$ is a parameter empirically chosen to balance the importance of the two terms. The intensity of CT images was assumed to be proportional to the x-ray attenuation coefficient dominated by Compton scattering. The correction on $g$ from the raw measured photon counts was applied to manage factors such as x-ray attenuation along the beam path connecting the x-ray source to each pixel. A constant $C$ was introduced in Eq. (1) to manage the intensity difference between the two terms, which was obtained via a normalization step. An example image of corrected measurement $g$ and the prior image $f_p$ are shown in Figure 3.

Similar to a SPECT setup, the blurring kernel $K$ depends on factors such as collimator hole size, septa height, the distance between the detector and scatter location, etc. Because of the imaging geometry, the distance between each detector pixel and scatter location varies among pixels, and hence the kernel is pixel dependent. In this study, we ignored this variation and used one position-independent blurring kernel $K$ corresponding to the one at the isocenter.

We employed a forward-backward splitting (FBS) algorithm [4] to solve the optimization problem. As such, we consid-

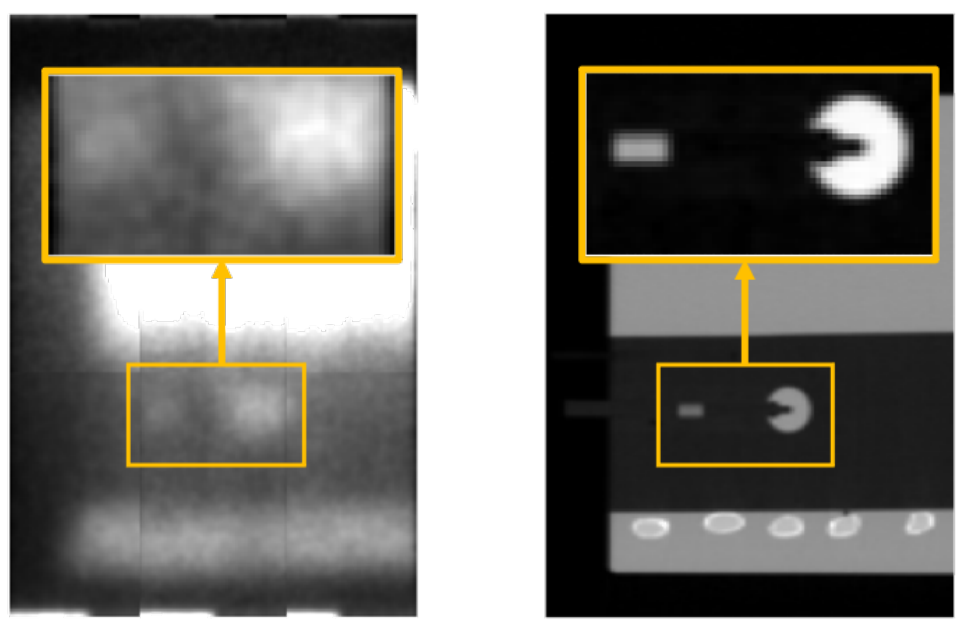

**Figure 3:** Left: experimental image and right: CT-reconstructed image of the slice of interest at phase 1. Display window: [0,1].

ered the optimality condition with respect to $v$ of $E[v]$:

$$\frac{\delta E[v]}{\delta v} = \nabla f_p K \otimes (K \otimes f_p - Cg) - \lambda \Delta v \ni 0. \quad (2)$$

Let us introduce an auxiliary image $s$ and re-write Eq. (2) by adding two additional terms:

$$\begin{aligned} \frac{\delta E[v]}{\delta v} &= \nabla f_p K \otimes (K \otimes f_p - Cg) - \lambda \Delta v \\ &+ \nabla f_p (f_p - s) - \nabla f_p (f_p - s) \ni 0. \end{aligned} \quad (3)$$

We split Eq. (3) into:

$$\begin{aligned} \nabla f_p K \otimes (K \otimes f_p - Cg) - \nabla f_p (f_p - s) = 0, \\ \nabla f_p (f_p - s) - \lambda \Delta v = 0. \end{aligned} \quad (4)$$

Assuming $\nabla f_p \neq 0$, the FBS algorithm can be implemented as an iterative process with the following three steps:

(1) Calculating $s = f_p - K \otimes (K \otimes f_p - Cg)$.;

(2) Solving $v \leftarrow \arg\min_v \frac{1}{2}\|f_p - s\|_F^2 + \frac{\lambda}{2}\|\nabla v\|_F^2$

(3) Updating $f_p$ with $v$.

Note that the second step corresponds to an image registration problem between $f_p$ and $s$, which was implemented using the Demons algorithm.
Once the solution $v$ was obtained, we derived the reconstructed image by deforming the prior image $f_p$ with this vector field. Tumor location was calculated based on the known tumor position in the prior image and the deformation vector field at the tumor position.

### 2.4 Evaluations

We evaluated KRISP from the following perspective. For imaging capability and tumor tracking, in addition to visually inspecting the measured scattered photon images, we performed quantitative measurements on tumor tracking accuracy. As such, we measured the derived tumor position from the reconstruction results and compared that with the ground truth tumor position. We evaluated the agreement between the two using Root Mean Square Error (RMSE).
Regarding its practicality, we evaluated two aspects: radiation exposure and interference from the therapeutic MV beam.

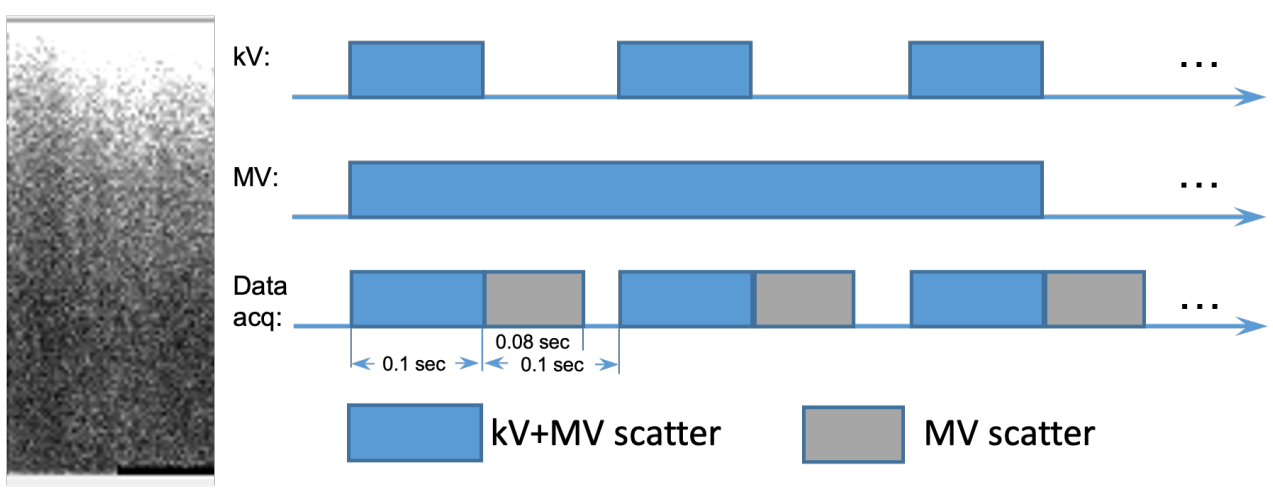

**Figure 4:** Left: Measured images with both kV and MV beams on. Right: Time structure for data acquisition and MV scatter correction.

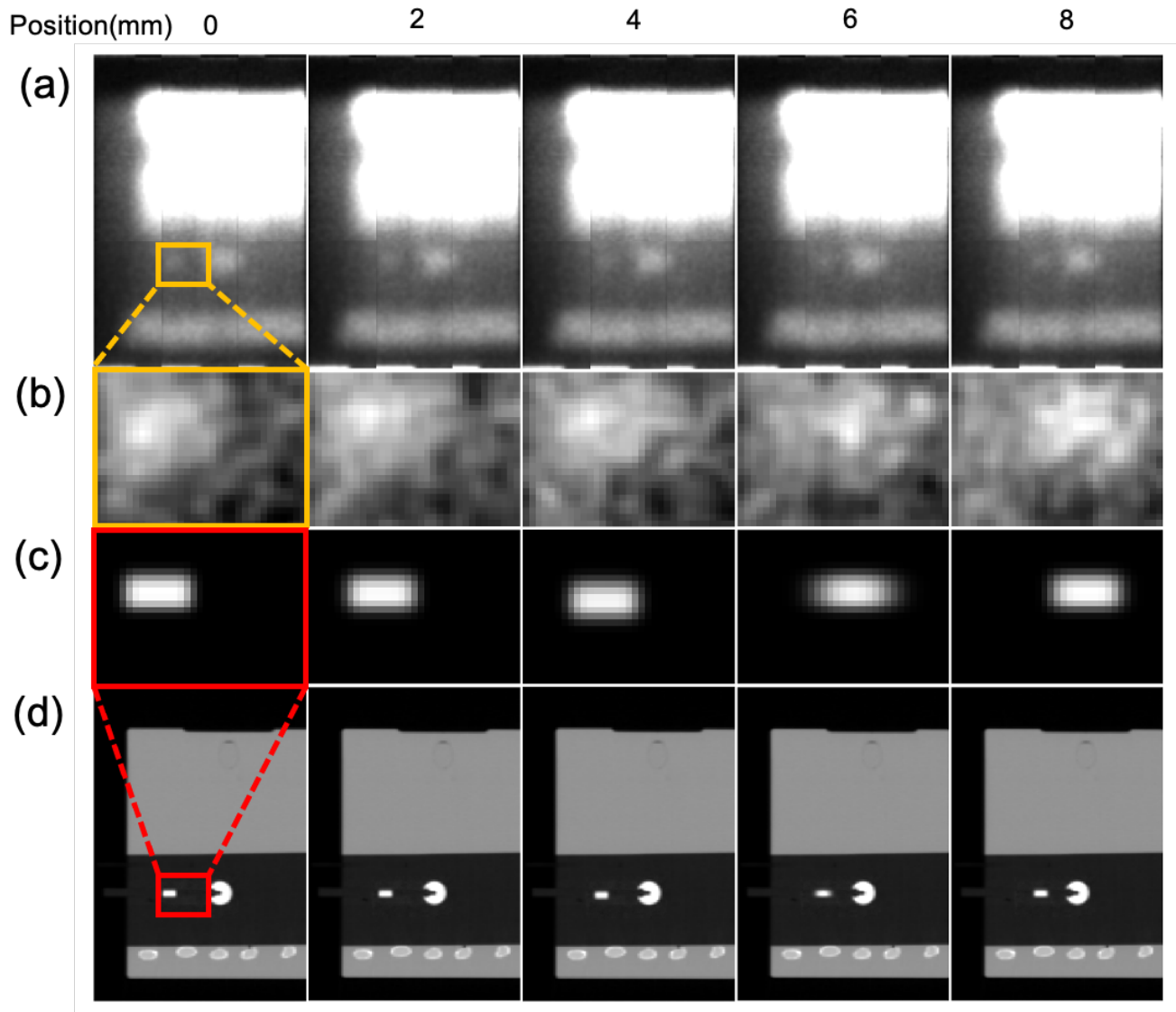

**Figure 5:** (a) Slice images from the experiment. (b) Zoomed in view of the measurement in a region of interest containing the moving tumor. (c)-(d) Zoomed-in and full view of reconstructed images assisted by the prior image. Display window: [0,1].

Specifically, we measured the x-ray exposure to compute the effective dose and peak entrance skin dose. Since KRISP will be used for intra-fractional tumor tracking during treatment delivery, interference of scattered photons from the MV therapeutic beam is a major concern. We investigated this issue experimentally in the same setup as in Figure 2) with the clinically realistic condition of a 600 MU/min 6 MV beam dose rate. As expected, the strong scattered MV beam had a strong contribution to the measured scatter signal, masking the kV image signal (Figure 4). To overcome this challenge, we propose the following approach to remove the impacts of MV photons. As KRISP targets temporal resolution of 0.2 sec with data acquisition duration 0.1 sec, we acquired data during the 0.1 sec interval with kV beam off for a certain length (e.g. 0.08 sec as in Figure 4), which contained solely scattered signals from the MV beam. This MV scattered signal can be scaled (e.g. 0.1/0.08 in this case) to account for the time duration difference and removed from the measured kV+MV signal. As subtraction cannot remove noise, we subtracted the denoised MV signal using a denoising algorithm considering Poisson noise statistics, and denoised the final result one more time. To demonstrate the validity of this approach, we acquired the kV+MV signal and the MV signal separately with the specific time duration, and investigated the resulting image quality using the stationary tumor in the phantom for simplicity. Note that the actual time structure in Figure 4 was not implemented in this proof-of-principle study.

## 3 Results

The measured scattered x-ray photon images from the experiment at the different tumor positions are plotted in Figure 5(a).

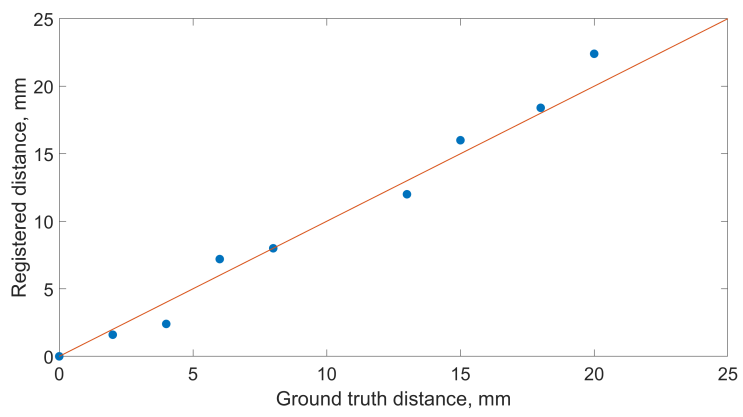

**Figure 6:** Measured tumor position.

The image has a lower intensity at the bottom due to photon fluence attenuation as propagating in the phantom. Within the lung region, we observed that the fixed tumor was visible, but the target (moving) tumor was much weaker due to its small size and blurring effects. Figure 5(b) shows the images of the ROI. In Figure 5(c), we plotted the registered images $f_p$ for the 9 phases. The target tumor motion was clearly reflected in these images, and matched well with the experiment results. Finally, we reassembled the full slice image with the registered image $f_p$, as shown in Figure 5(d).
As for the tumor tracking task, Figure 6 presents the tumor position derived from the motion vector field, as compared to the ground truth tumor position. The RMSE of the calculated tumor position was 1.2 mm.
Based on measured x-ray exposure, the effective dose rate was 0.27 mSv/min and the peak entrance skin dose rate was 240 mGy/min under the very small illumination area (slit beam). In radiotherapy, the gantry rotates to multiple positions, smearing out the skin dose. Fundamentally, our approach is very similar to a cardiology intervention procedure, with a comparable peak skin dose level but a much smaller x-ray beam area and lower effective dose.
Figure 7 presents the images using the scheme proposed in Figure 4 with the MV interference signal removed. The images were acquired at a few phantom positions, with the large tumor clearly visualized. The images were more blurry than those measured for the kV scattered signal only (e.g. Figure 5(a)) due to the application of denoising operations to suppress noise after subtracting MV signal from kV+MV measurement.

## 4 Discussion

In the present study, only the planar image on the plane of x-ray illumination was reconstructed, as the measurement data did not contain information outside this plane. However, the imaging geometry was determined in such a way shown in Figure 1, such that this plane is of the most importance for tumor tracking in radiotherapy. Specifically, the plane is perpendicular to the therapeutic MV beam. Motion in this plane moves the tumor out of the therapeutic beam and hence may substantially underdose the tumor, as well as move normal tissues into the beam and overdose them. In contrast, the impact of motion perpendicular to this plane on the delivered dose is characterized by beam depth dose, which is much more gradual (3mm change in depth causes 1% dose change for a 6MV therapeutic photon beam. Meanwhile, it is well known that tumor motion follows a patient-specific tumor model [5]. Adding this model and volumetric prior image to the image reconstruction process can potentially allow us to achieve volumetric real-time imaging and 3D tumor tracking. For simplicity, we used Matlab to perform the image processing tasks. The computation time cannot meet the requirement for real-time tumor tracking. It is our future work to develop GPU-acceleration technologies and more computationally efficient algorithms to support real-time applications.

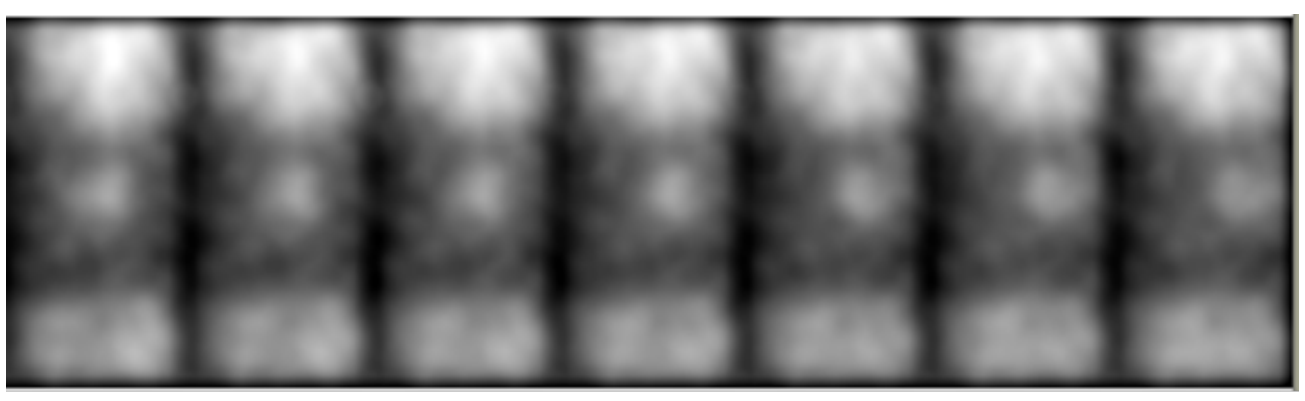

**Figure 7:** Scatter image with MV interference removed for a few phantom positions.

## 5 Conclusion

In this study, to address the task of tumor tracking in image-guided radiation therapy for lung cancer, we investigated the feasibility of KRISP that employs a PCD to measure Compton-scattered x-ray photon signal for real-time volumetric image and tumor tracking during treatment delivery. We performed phantom experiments using a 120 kV x-ray slice beam to irradiate a slice containing the tumor and acquired the photons scattered off this plane using the PCD with a parallel-hole collimator. Aided by a prior CT image, we successfully reconstructed the image for the plane of x-ray illumination and derived the tumor position with an RMSE of 1.2 mm. Radiation exposure and interference from the therapeutic beam were also investigated.

# Fast Reconstruction of Positronium Lifetime Image by the Method of Moments

Bangyan Huang[1] and Jinyi Qi[1]

[1]Department of Biomedical Engineering, University of California, Davis, Davis, CA, USA

**Abstract** The lifetime of ortho-positroniums can be influenced by the microstructure and the concentration of bio-active molecules in human tissue, thereby providing valuable information for better understanding of disease progression and treatment response. There is currently a lack of efficient lifetime image reconstruction methods. Existing methods are either computationally intensive or have poor spatial resolution. This paper presents a fast lifetime image reconstruction method called SIMPLE-Moment, which can reconstruct ortho-positronium lifetime images with a computational time equivalent to that of reconstructing three standard PET activity images. The implementation of this method requires minimal modification to the conventional ordered subset expectation maximization (OSEM) algorithm. A Monte Carlo simulation study using GATE demonstrates that the proposed method can reconstruct high-resolution lifetime images using a PET scanner with an existing time-of-flight (TOF) resolution.

***Novelty and impact: An efficient positronium lifetime image reconstruction method that utilizes the moments of lifetime is developed. This method provides a fast alternative to the previously proposed reconstruction methods.***

## 1 Introduction

Recently, there has been an increasing interest in investigating the lifetime of positrons before they annihilate with electrons in biological tissue using positron emission tomography (PET). Prior to the annihilation, about 40% of positrons form positroniums (Ps) in human tissue. There are two types of Ps, namely ortho-positronium (o-Ps) and para-positronium (p-Ps), with o-Ps occurring 75% of the time and p-Ps occurring 25% of the time. The lifetime of p-Ps is relatively short (125 ps) and is unlikely to be affected by the surrounding micro-environment. However, the lifetime of o-Ps can vary in biological tissue due to two effects—pick-off annihilation and spin-exchange interaction—which result from the interaction between the o-Ps and its surrounding microenvironment. Pick-off annihilation occurs when the positron of the o-Ps annihilates with a foreign electron, while spin-exchange is induced when the surrounding molecules possess unpaired electrons. Therefore, the o-Ps lifetime is dependent on the size of intermolecular voids and the concentration of bio-active molecules in biological materials [1]. It has been reported that the o-Ps lifetime in human adipose tissue is about 0.7 ns longer than that in myxoma tissue [2]. A study also showed that the o-Ps lifetime can be affected by the oxygen concentration in water [3]. Therefore, estimating the o-Ps lifetime *in vivo* can provide useful information for a better understanding of disease progression and treatment response.

Currently there is a lack of efficient image reconstruction methods for positronium lifetime imaging (PLI). The time-of-flight (TOF) direct backprojection method, which positions each event based on the TOF information, is fast but has very poor spatial resolution. We previously introduced a penalized maximum likelihood (PML) method [4], which can produce high-resolution lifetime images but suffers from high computational cost and uses a monoexponential decay model that is inadequate for real-world lifetime distributions. Another method that we developed, known as SPLIT (Statistical Positronium Lifetime Image reconstruction via time-Thresholding) [5], can perform 3D reconstruction and correct random events. This method leverages existing reconstruction algorithms to reconstruct a threshold-activity curve for each voxel and then estimate the lifetime image from these curves. While it has much lower computational cost than the PML method, it still requires reconstruction of tens of activity images to form the threshold-activity curves and perform curve fitting for each voxel. To further reduce the computational cost and eliminate the need for curve fitting, we proposed an alternative method called SIMPLE (Statistical IMage reconstruction of Positron Lifetime via time-wEighting) that can reconstruct images of the average lifetime of all the interaction pathways [6]. The computational cost of this method is equivalent to two standard activity image reconstructions. However, this method cannot estimate the o-Ps lifetime directly.

To overcome this limitation, here we extend the SIMPLE method to include higher orders of moments so that the o-Ps lifetime can be estimated directly. The method of moments has been previously used in optical imaging to estimate fluorescence lifetime [7] and in dynamic PET for kinetic parameter estimation [8]. The proposed SIMPLE-Moment method has the advantage of fast computation speed without the need for curve fitting. This method represents a significant improvement over our previous methods for PLI, as it allows for direct estimation of the o-Ps lifetime while maintaining computational efficiency.

## 2 Materials and Methods

A. Lifetime Event Model in PLI

In this paper, a lifetime event is defined as a tri-coincidence consisting of two annihilation photons and a prompt gamma. It is represented by a line of response (LOR) $i_k$ determined by the detection positions of the two annihilation photons and a time delay between the emission of prompt gamma and the annihilation of positron. The distribution modeling the time delay of the events originated from voxel $j$ can be represented as a summation of multiple exponential decays convolved by a Gaussian function $g(\tau)$ that is characterized by the time resolution of the PET scanner:

$$f\left(\tau|A_{l,j},\lambda_{l,j},l\in\{o,p,d\}\right)= g(\tau)*\sum_{l\in\{o,p,d\}}A_{l,j}\lambda_{l,j}\exp\left(-\lambda_{l,j}\tau\right)u(\tau), \quad (1)$$

where $\lambda_{l,j}$ is the decay rate (inverse of lifetime), $A_{l,j}$ is the intensity of the $l^{\text{th}}$ pathway of positron annihilation, and $u(\tau)$ is a unit step function. We consider three pathways of annihilation including forming o-Ps, p-Ps and direct annihilation, labeled by the subscripts $o, p,$ and $d$, respectively.

B. Lifetime Estimation via the Method of Moments

Using the above PLI event model, the $n^{th}$ moment of the lifetime is written as

$$m_j^n=E[\tau^n]=\int_{-\infty}^{\infty}\tau^n f\left(\tau|A_{l,j},\lambda_{l,j},l\in\{o,p,d\}\right)d\tau, \quad (2)$$

which can be reduced into the following form:

$$m_j^n=\sum_{s=0}^{n}\frac{n!}{(n-s)!}\mu_{n-s}G_{s,j}, \quad (3)$$

where $\mu_k=E_{g(\tau)}\left[\tau^k\right]$ is the $k^{\text{th}}$ moment of the Gaussian distribution $g(\tau)$ and

$$G_{s,j}=\sum_{l\in\{o,p,d\}}\frac{A_{l,j}}{\lambda_{l,j}^s}. \quad (4)$$

With equations (3) and (4), we can estimate the lifetimes and intensities in two steps: first, we calculate an empirical estimate of the moments $m_j^n$ and then find $G_{s,j}$ using equation (3); second, we estimate the lifetimes and intensities using equation (4).

When there is no prior information about the lifetimes and intensities, we need six moment equations (zeroth to fifth) for the six unknowns ( $A_l, \lambda_l, l\in\{o,p,d\}$ ). Fortunately, we know the ratio of the occurrence between o-Ps and p-Ps ($\frac{A_o}{A_p}=3$). Furthermore, we can assume that the lifetimes of p-Ps and direct annihilation are known and fixed at 0.125 ns and 0.4 ns, respectively. Therefore, only three moment equations (zeroth to second) are required and the closed-form solution of $\lambda_o$ is given by

$$\lambda_o=\frac{-B+\sqrt{B^2-4AC}}{2A}, \quad (5)$$

where

$$A=G_0\lambda_d-G_1\lambda_d^2-G_0\lambda_p+G_2\lambda_p\lambda_d^2+4G_1\lambda_p^2 -4G_2\lambda_p^2\lambda_d,$$
$$B=-3G_0\lambda_p^2+3G_2\lambda_p^2\lambda_d^2,$$
$$C=3G_0\lambda_p^2\lambda_d-3G_1\lambda_p^2\lambda_d^2.$$

C. Reconstruction of Moment Images

In order to estimate positronium lifetime images using the method of moments, we must first obtain moment images from the measured lifetime events. Direct reconstruction of the moment images is difficult, so we instead reconstruct an intensity-weighted moment image $w_j^n=x_j m_j^n$ and divide it by the intensity image $\boldsymbol{x}$, which is reconstructed using the two annihilation photons of all the tri-coincidence events.

To reconstruct the intensity-weighted moment image $\boldsymbol{w}^n$, we need to find a projection whose expectation is the forward projection of $\boldsymbol{w}^n$,

$$\overline{z^n}_i=\sum_j H_{ij}w_j^n=\sum_j H_{ij}x_j m_j^n=\sum_j E[y_{ij}]m_j^n, \quad (6)$$

where $\boldsymbol{H}$ is a standard PET system matrix and $y_{ij}$ is the number of events originated in voxel $j$ and detected in LOR $i$. An empirical estimate of $m_j^n$ is

$$\widehat{m_j^n}=\frac{1}{y_{ij}}\sum_{k\in K_{ij}}\tau_k^n, \quad (7)$$

where $K_{ij}$ denotes the set of list-mode indices of the events originated in voxel $j$ and detected in LOR $i$, and $K_i=\bigcup_j K_{ij}$. Thus, we have

$$\sum_j H_{ij}w_j^n=\sum_j E\left[\sum_{k\in K_{ij}}\tau_k^n\right]=E\left[\sum_{k\in K_i}\tau_k^n\right]. \quad (8)$$

Equation (8) indicates that the intensity-weighted moment image can be estimated from time-delay-weighted projection data

$$z_i^n=\sum_{k\in K_i}\tau_k^n. \quad (9)$$

We used a list-mode OSEM algorithm to reconstruct the intensity-weighted moment image $\boldsymbol{w}^n$ with the following updating equation:

$$w^{n}{}_j^{(r+1)}=\frac{w^{n}{}_j^{(r)}}{\sum_i H_{ij}}\sum_{k\in S_r}\frac{H_{i_k j}\tau_{i_k}^n}{\sum_j H_{i_k j}w^{n}{}_j^{(r)}}, \quad (10)$$

where $S_r$ denotes the $r^{th}$ subset of the list-mode event indices. The moment image $\boldsymbol{m}^n$ is then obtained by taking the voxel-wise ratio of the image $\boldsymbol{w}^n$ over the activity image $\boldsymbol{x}$.

D. Simulation Study

We performed a Monte Carlo simulation to validate the proposed method. A rodent phantom was simulated in GATE with an average activity concentration of 20 kBq/cc and a scan duration of 30 minutes. The activity ratio was set to 10 : 15 : 2 : 1 for the lesion, kidneys, liver, and body background to mimic $^{44}$Sc-PSMA uptake. The o-Ps lifetime was set to 2.0 ns inside the lesion and 2.5 ns elsewhere; the lifetimes of p-Ps and direct annihilation were set to 0.125 ns and 0.4 ns, respectively, throughout the body. The occurrence of Ps was set to 40%. The simulated PET scanner was the Neuro-EXPLORER scanner with a ring diameter of 52 cm, an axial length of 49 cm, and a TOF resolution of 250 ps. Prior to the reconstruction, the travel time difference between the prompt gamma and annihilation photons in each PLI event was corrected. To obtain the travel time for each photon, its travel distance was calculated as the distance between the detector and the most likely annihilation position determined by the TOF information along the LOR. The proposed method was compared with the previous SPLIT method. OSEM with 3 subsets and 2 iterations was used to reconstruct the activity and intensity-weighted moment images for the SIMPLE-Moment method and the lifetime-thresholded activity images for the SPLIT method. The reconstruction field-of-view (FOV) was 50 × 50 × 180 mm$^3$ and the reconstruction voxel size was 0.8 × 0.8 × 1.6 mm$^3$.

## 3 Results

The reconstructed first and second moment images are shown in Fig. 1 in comparison with the ground truth images. The biases (standard deviations) in four regions of interest (ROIs) are listed in Table 1. Both 1$^{st}$ and 2$^{nd}$ moment images are accurate in terms of the mean and the variance is increased with the higher moment image. Fig. 2 displays the reconstructed activity and lifetime images, with the corresponding lifetimes in the four ROIs listed in Table 2. In the displayed images, the voxels with activity lower than 50% of the activity in the body background had their lifetime set to zero. The proposed SIMPLE-Moment method slightly under-performed in the lesion as compared to the SPLIT, yielding an average lifetime value that is 0.8% higher than the ground truth. In the normal region, the SIMPLE-Moment method has a higher accuracy and a slightly better variance as compared to the SPLIT method.

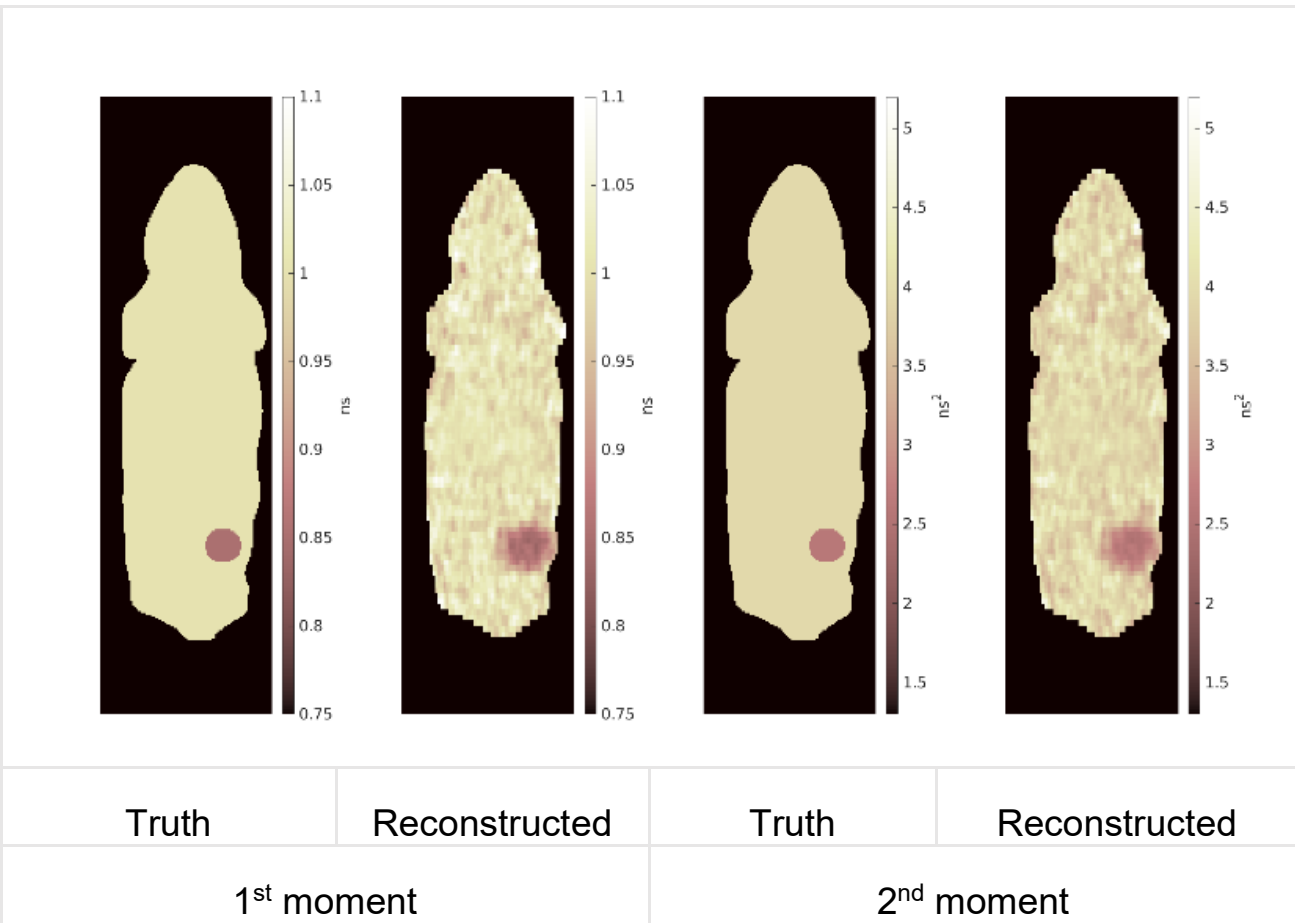

Fig 1. Reconstructed first and second moment images and their ground truths.

Table 1. The biases (standard deviations) of the estimated moments in the lesion, kidney, liver, and body background (in ns for the first moment; ns$^2$ for the second moment).

| | | Lesion | Kidney | Liver | BG |
|---|---|---|---|---|---|
| 1$^{st}$ | Truth | 0.8525 | 1.0025 | 1.0025 | 1.0025 |
| | Recon | 0.002(12) | 0.001(13) | 0.000(20) | 0.000(32) |
| 2$^{nd}$ | Truth | 2.6036 | 3.9536 | 3.9536 | 3.9536 |
| | Recon | 0.02(9) | 0.00(12) | 0.00(19) | 0.00(29) |

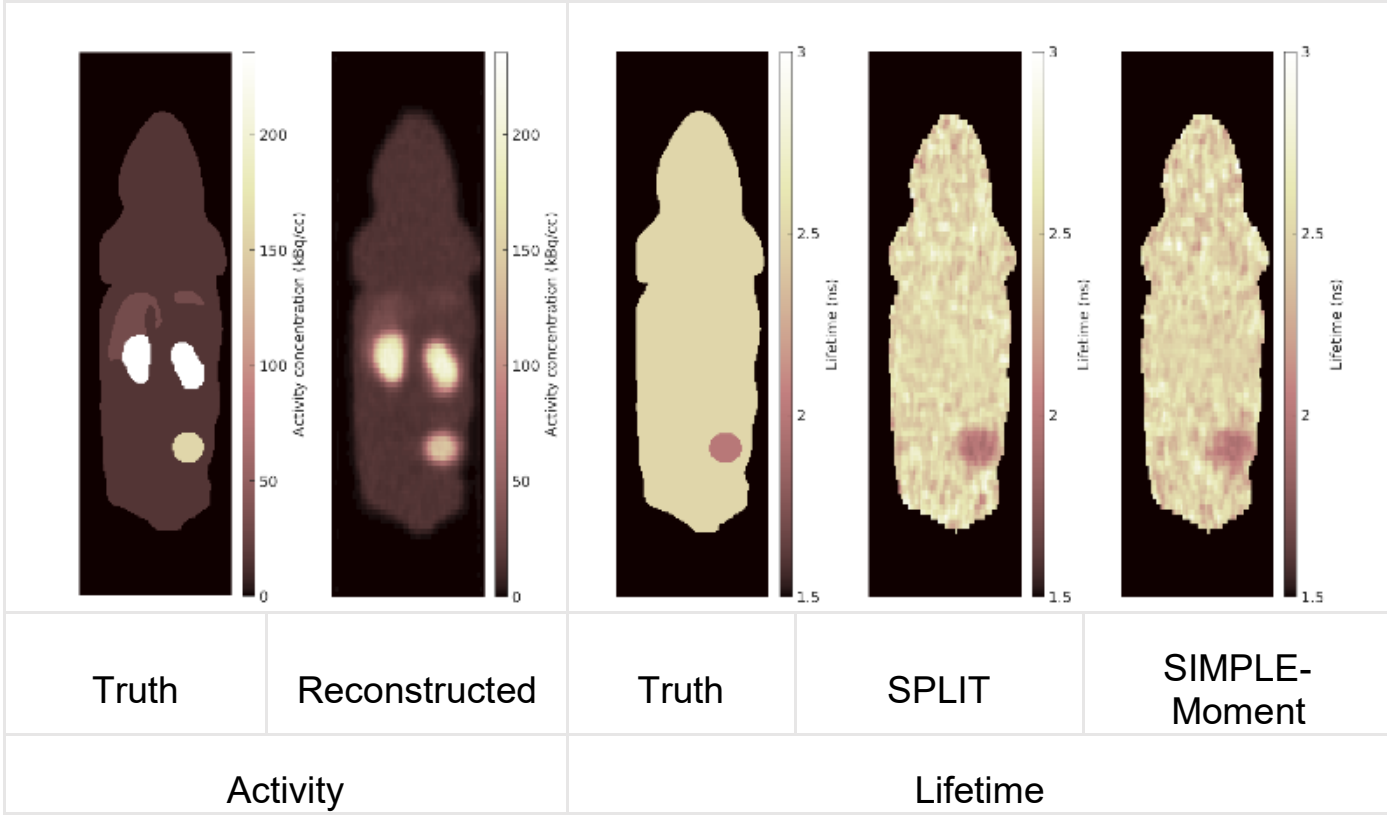

Fig 2. Reconstructed activity and lifetime images and their ground truths.

Table 2. The biases (standard deviations) of reconstructed lifetimes in the lesion, kidney, liver and body background (in ns).

| | Lesion | Kidney | Liver | BG |
|---|---|---|---|---|
| Truth | 2.0 | 2.5 | 2.5 | 2.5 |
| SPLIT | 0.007(54) | 0.028(59) | 0.035(91) | 0.014(136) |
| SIMPLE-Moment | 0.015(56) | -0.006(55) | -0.002(87) | -0.003(132) |

## 4 Discussion

The major advantage of the proposed SIMPLE-Moment method is its low computational cost, which depends on the number of intensity-weighted moment images to be reconstructed. In this study, two intensity-weighted moment images and one activity images were reconstructed, resulting in a 3-fold computational cost compared to a standard activity reconstruction. In comparison, the SPLIT method in this study reconstructed 53 lifetime-thresholded activity images. The computational time of the SIMPLE-Moment method was approximately 10.5 minutes, while the SPLIT reconstruction took around 119 minutes, including 95 minutes for the OSEM reconstruction and 24 minutes for fitting threshold-activity curves. All computations were performed on a Dell computer with dual Intel Xeon E5-2630 v3 2.4 GHz CPUs.

In the absence of prior knowledge about lifetime, $2n - 1$ moment images are required to solve a lifetime model with $n$ interaction pathways. For a typical three-pathway model, this means the first to the fifth moments. However, since the occurrence ratio of o-Ps to p-Ps is known, it is reasonable to reduce one moment image. To obtain a simple closed-form solution to equation (3) and (4) in this study, we further assumed that the lifetimes of p-Ps and direct annihilation are fixed and known in biological tissue. This assumption may be strong and requires further validation through experimental studies.

## 5 Conclusion

We have proposed the SIMPLE-Moment method, a fast and curve-fitting-free approach for reconstructing o-Ps lifetime images based on intensity-weighted moment images. The proposed method has a computational cost comparable to that of three standard activity image reconstructions. The simulation study showed that the proposed SIMPLE-Moment method can accurately reconstruct lifetime images with low variance. Future work will apply the SIMPLE-Moment method to real data.

# Volumetric Breast Density Measurement using Dual-Energy Digital Breast Tomosynthesis with Dual-Shot and Dual-Layer Technique

Hailiang Huang[1], Xiaoyu Duan[1], and Wei Zhao[1]

[1]Department of Radiology, Stony Brook University, Stony Brook, USA

**Abstract** Volumetric breast density (VBD) has been conventionally measured on mammogram or digital breast tomosynthesis (DBT) projection image acquired using single x-ray energy spectrum, which could be greatly impacted by different breast compressions. We aim to develop a method for reproducible VBD measurement using dual-energy (DE) material decomposition with DE DBT, which uses both conventional dual-shot (DS) technique for image acquisition and a dual-layer (DL) detector to minimize patient motion in clinical practice. In-silico experiments were conducted using a virtual clinical trial software (VICTRE) with digital breast phantom and image simulations. The breast attenuations in high- and low-energy image were calibrated using analytical calculations for DS and DL technique. Our results show consistent VBD measurements among all projection angles and reproducible measurements for breast under different compressions. For DL technique, a correction method was applied to reduce the uncertainty in the decomposed thickness map and VBD map.

Novelty and impact: We developed a method to measure VBD using DE DBT and applied it to DS and DL image acquisition technique, which allows reproducible VBD measurement for masking risk and breast cancer risk assessment for dense breast patients.

## 1 Introduction

Breast density describes the fraction of fibroglandular tissue (FGT) perceived radiologically in the breast, and has been used to assess the risk of masking breast lesions and as a biomarker for breast cancer risk.[1] Methods have been developed to measure volumetric breast density (VBD) using conventional mammogram or digital breast tomosynthesis (DBT) projection image acquired using a single x-ray energy spectrum.[2] This single-energy method typically assumes a simple shape for compressed breast that contains fatty tissue only in the periphery, which introduces errors in measured VBD that is greatly impacted by different breast compressions.

Dual-energy (DE) material decomposition has been used to measure VBD from DE mammograms.[3] It measures breast attenuations from high-energy (HE) and low-energy (LE) images and directly solves for breast thickness and density without assumptions made in single-energy method, and it has shown accurate measurement of VBD. Conventional DE image acquisition is based on dual-shot (DS) technique where two images are acquired separately, which is prone to patient motion in clinical practice. This will introduce errors in the quantification for VBD. DE imaging with dual-layer (DL) detector has been used in radiography and allows DE images acquired simultaneously with one x-ray exposure to minimize motion artifacts.[4]

This study aims to develop methods for VBD measurement using DE DBT and material decomposition. We evaluated the performance for DE images acquired using DS and DL techniques by in-silico experiments with digital breast phantoms and image simulations.

## 2 Materials and Methods

The experiment was conducted using a virtual clinical trial software (VICTRE) based on Monte-Carlo simulation and developed by FDA.[5] The image acquisition for DS and DL technique is illustrated in Figure 1. For DS technique, the geometry of Siemens Mammomat Inspiration DBT system was modeled. Twenty five projections were generated in an angular range of 50 degrees. For DL technique, the central projection images were generated. The simulated images included quantum noise and electronic noise and did not include scatter radiation.

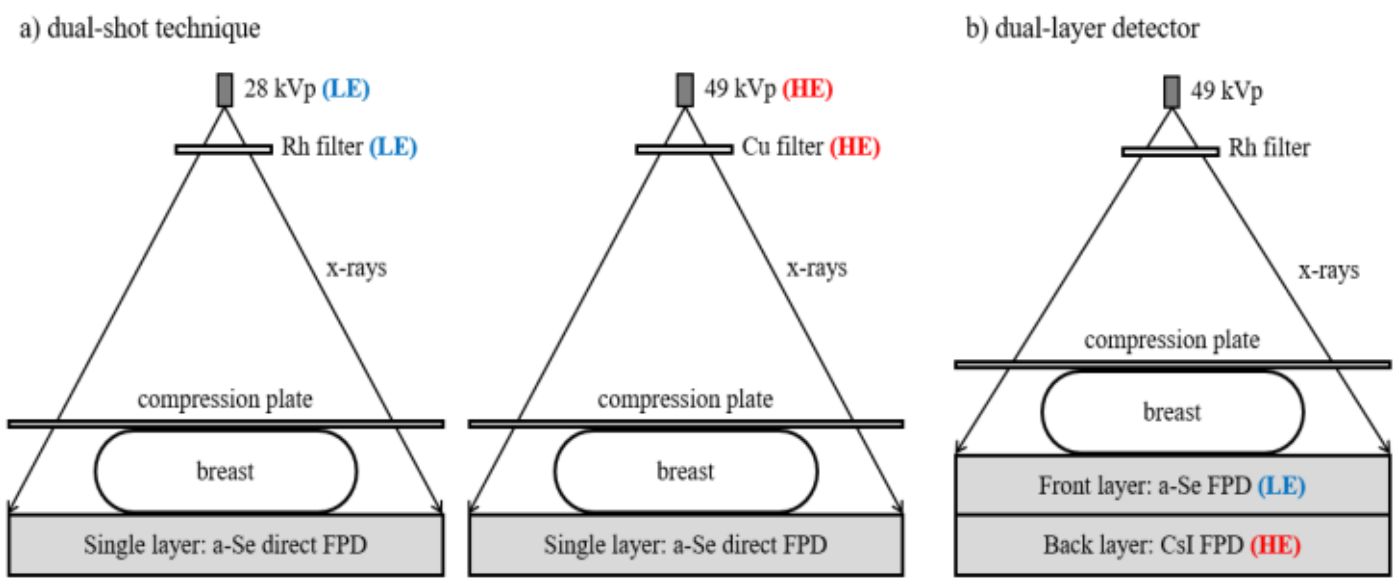

Figure.1 Image acquisition for dual-shot and dual-layer technique

Anthropomorphic breast phantoms were generated with pre-defined VBD and compressed to clinical-relevant breast thickness using VICTRE. To evaluate the reproducibility of our method for repeated measurement, each breast phantom was compressed to different breast thicknesses to simulate a breast under different breast compressions in clinical practice. A slice in a digital breast phantom was shown in Figure 2. The voxels in the phantom volume were labelled to determine their tissue type. In the breast volume, the tissue types other than fatty tissue, skin, and nipple are classified as FGT for two-material decomposition.

The breast attenuations in HE and LE image for a given breast thickness and VBD were calibrated by analytical calculations using the Lambert-Beer's law, which includes spectral simulation, attenuations of fatty tissue and FGT, and the detector response. Figure 3 shows a wide range of calibrated data with each point in the graph (black dot) recording the breast attenuation in HE image ($S_{HE}$) and LE image ($S_{LE}$) for a given breast thickness and VBD. During image decomposition, the breast attenuations measured from the images for each pixel location can be similarly plotted in the graph (red dot). The nearest calibrated point to the measurement was selected to determine the breast thickness and VBD for that pixel location.

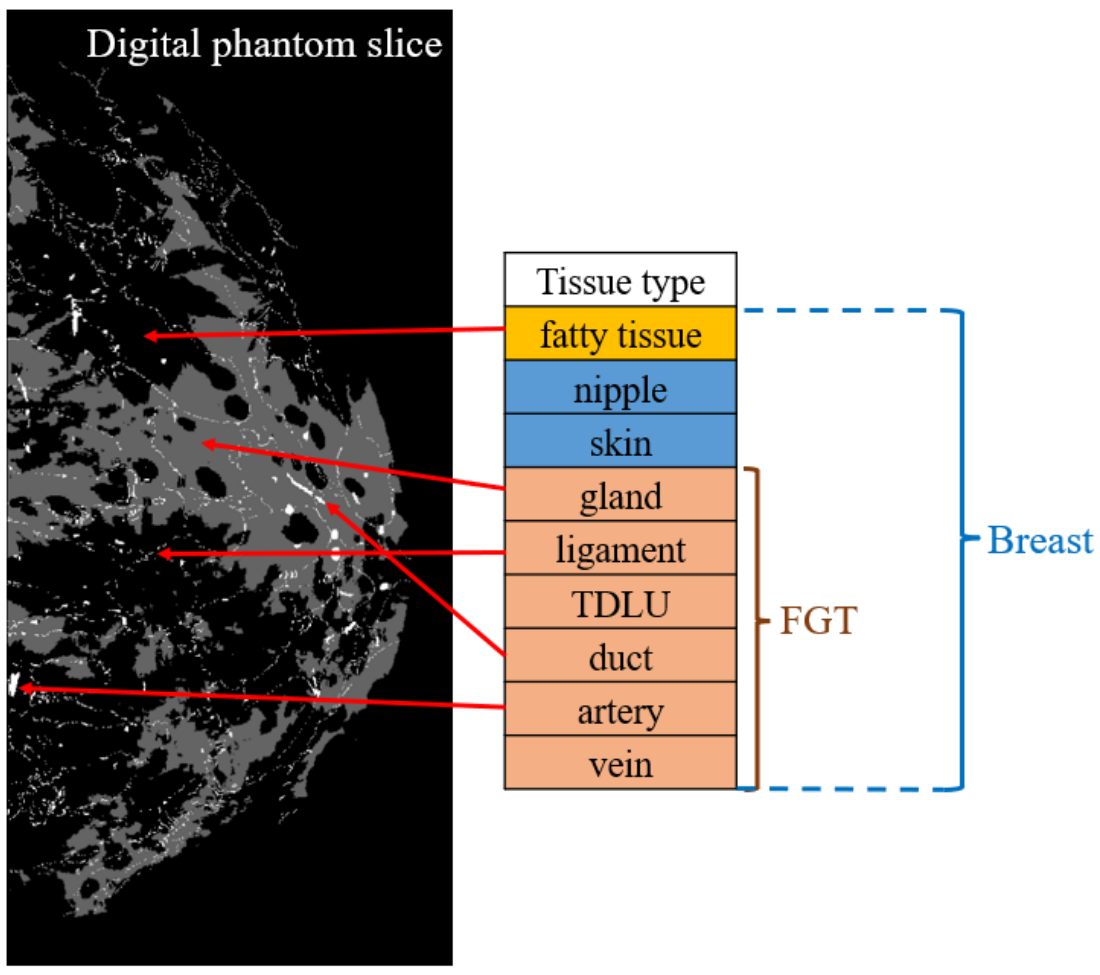

Figure.2 digital phantom slice showing simulated breast tissues

The FGT thickness for each pixel location was calculated as

$$t_{FGT} = t \times VBD_{pixel}$$

where t (total breast thickness) and $VBD_{pixel}$ were obtained from the decomposition.

The total breast volume $V_{breast}$ and the total FGT volume $V_{FGT}$ measured from the HE and LE image pair at oblique angle θ were calculated separately as

$$V = \sum_i t_i \times A \times \cos(\theta)$$

where A is the pixel size and $t_i$ is the value of pixel i on the breast thickness map or the FGT thickness map, and it sums up all pixels in the breast region.

The VBD was calculated as

$$VBD = \frac{V_{FGT}}{V_{breast}} \times 100\%$$

For DL technique, the decomposed thickness map was corrected to reduce uncertainty by estimating the iso-thickness contour on the map. 2D interpolation between contours was used to create a smooth thickness map that decreases continuously in the breast peripheral region. The corrected thickness was used as a prior knowledge and constraint in the decomposition to determine the VBD.

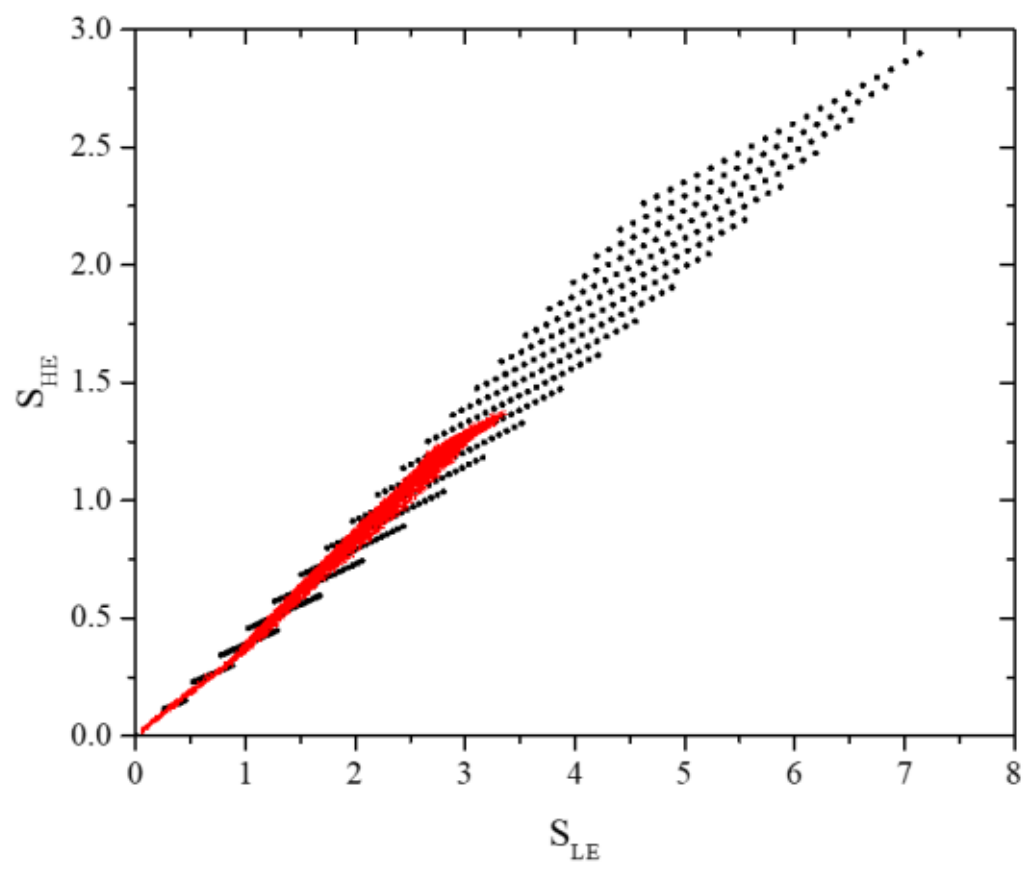

Figure.3 calibrated (black) and measured (red) breast attenuations

## 3 Results

### 3.1 Dual-shot technique

Figure 4 shows the material decomposition using DE projection images of a digital breast phantom with an average thickness (4.6cm) and high density (25.9%). The VBD and breast thickness is quantified throughout the entire breast region including the periphery with decreasing thickness. The method is applicable to central and oblique projection views.

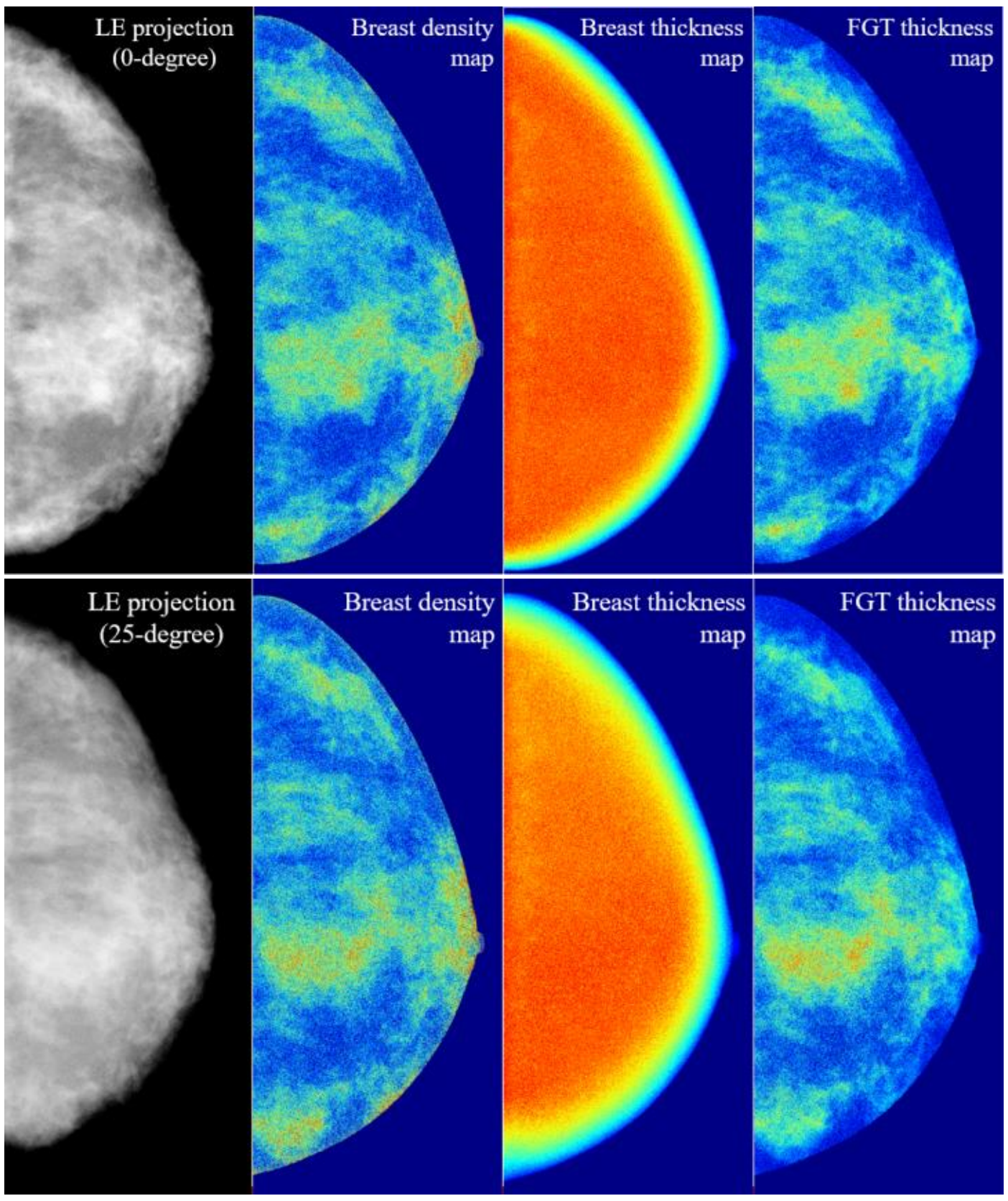

Figure.4 Simulated LE DBT projection images, breast thickness and density maps from material decomposition, and the calculated FGT thickness map for the central (0-degree) and the most oblique (25-degree) projection angle.

Figure 5 shows that the measurement for breast volumes and VBD is consistent among all projection angles.

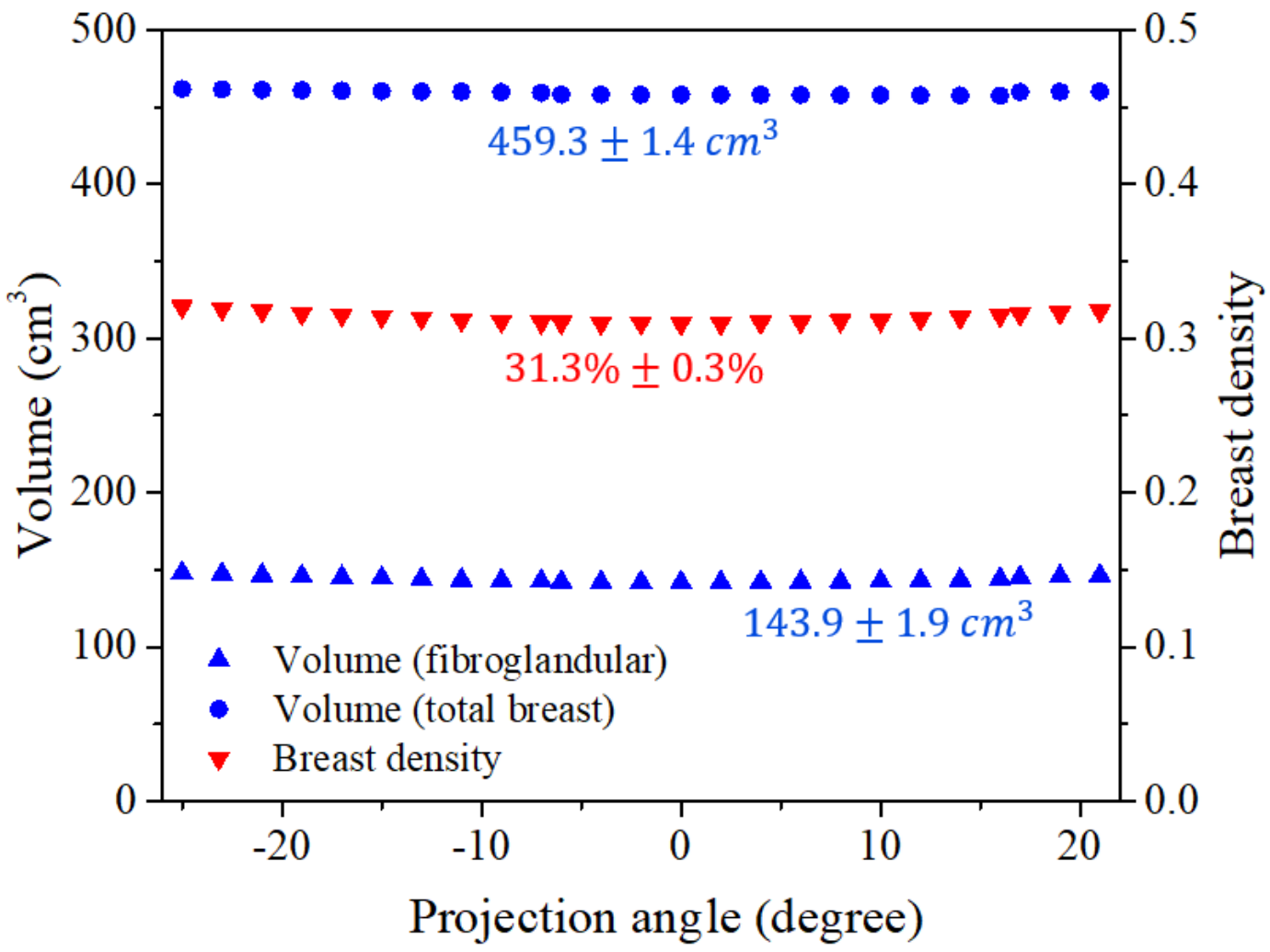

Figure.5 Measurements from DE projection images for all angles.

Table 1 summarizes the simulated digital breast phantoms in different sizes compressed to various clinical-relevant thicknesses. Average (~15%) and high (~30%) density were simulated. The changes in VBD for compressed phantoms with different thicknesses are negligible, which allows evaluation for the reproducibility of measurements.

Table.1 Breast phantoms for evaluation of reproducibility.

| | Breast size in diameter (cm) | Compressed thickness (mm) | VBD calculated from the compressed phantom (%) | | |
|---|---|---|---|---|---|
| | | | 40 mm | 50 mm | 60 mm |
| Phantom 1 | 15.2 | 40, 50, 60 | 14.9 | 14.8 | 14.7 |
| Phantom 2 | 12.1 | 40, 50 | 15.6 | 15.5 | ---- |
| Phantom 3 | 16.1 | 50, 60 | ---- | 32.2 | 32.1 |

Figure 6 shows the reproducibility for the proposed method. The average absolute discrepancy between two repeated measurements for all phantoms is 2.3% ± 1.1%. Largest discrepancy is 3.9% for phantom 1 with 20 mm difference in compressed thickness (40 vs. 60 mm) representing a large variation in breast compression in clinical practice.

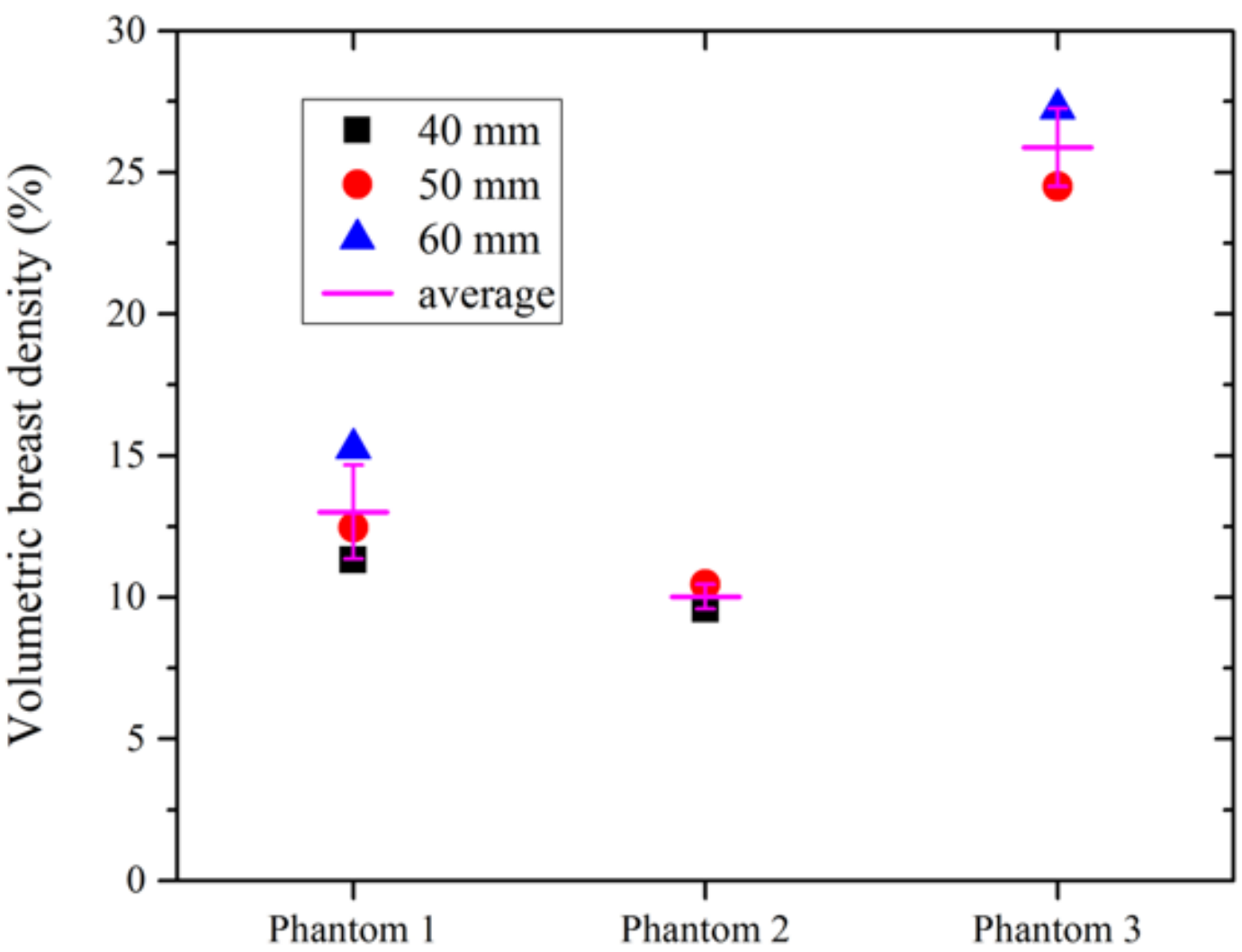

Figure.6 Repeated measurements of VBD for breast phantoms.

### 3.2 Dual-layer technique

Figure 7 shows the decomposed breast thickness and density maps of a digital breast phantom using DE projection images acquired with DL detector. Without correction, the decomposed maps show high image noise. The corrected thickness map shows uniform thickness in the breast center and continuous decreases in the periphery. The noise in the corrected density map using corrected thickness map as constraint for decomposition is much reduced.

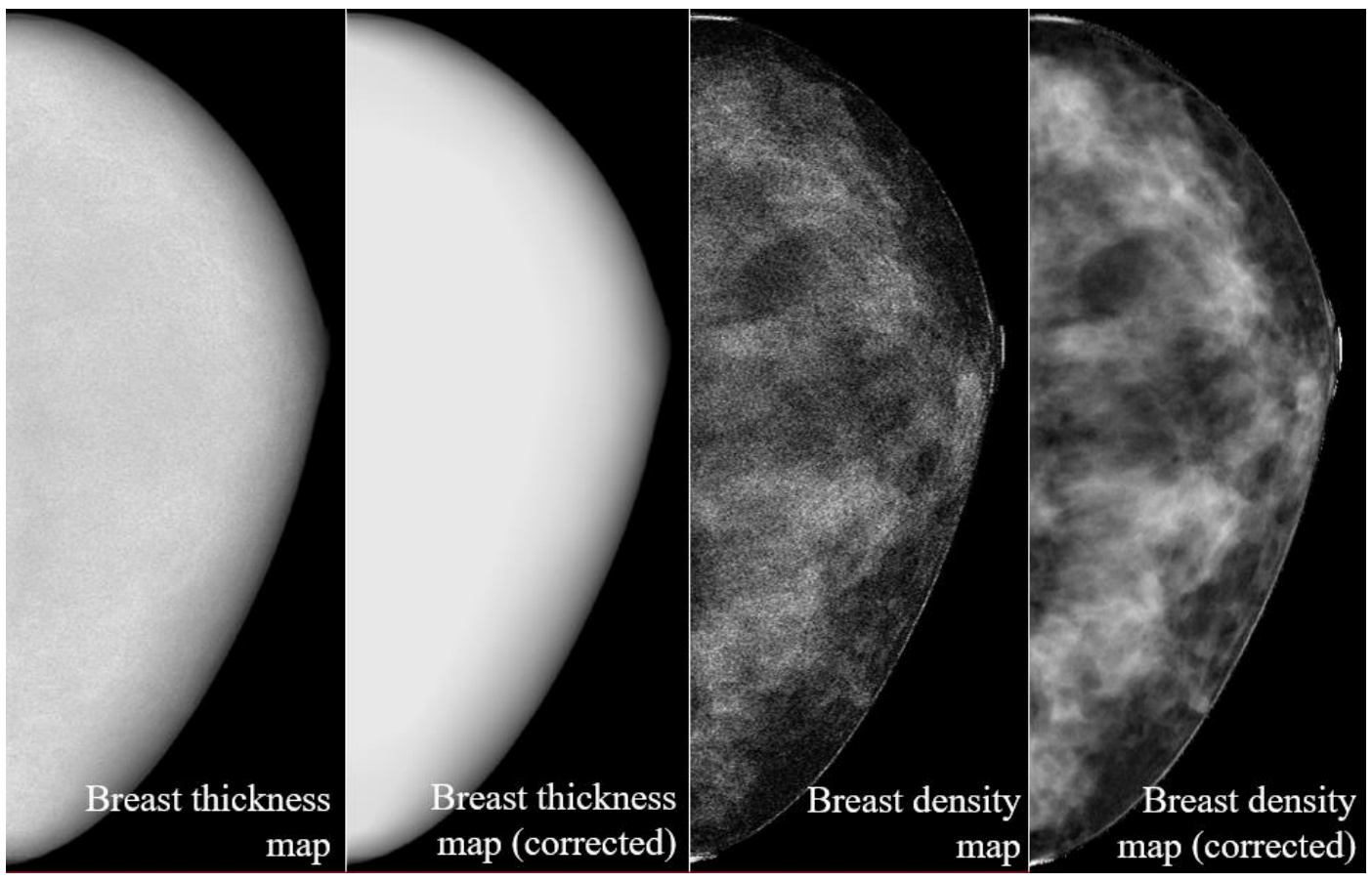

Figure.7 Decomposed breast thickness and density maps before and after correction using DE images acquired from DL detector.

## 4 Discussion

The analytical model used for calculating calibration data could deviate from the physical model in VICTRE used for image simulation, which could introduce systematic errors to the measurement by DE decomposition. The FGT defined includes other tissue types such as blood vessels and ligaments which have x-ray attenuations larger than the FGT. The skin layers on the top and bottom surface of the compressed breast add additional attenuation to the x-ray passing through the breast and are not excluded.

The accuracy of breast attenuations measured from the DE images will be reduced by scatter radiations. Since anti-scatter grid is usually not implemented in DBT systems, an effective image-based scatter removal technique will be required.

Future work includes comparing the VBD measured from DE DBT with those obtained using single-energy methods, and correlating the results with MRI.

## 5 Conclusion

We developed a method to measure VBD using DE DBT with material decomposition and applied it to DS and DL image acquisition technique. Our results show consistent VBD measurements among all DBT projection angles and reproducible measurements for breasts under different breast compressions. The proposed method could allow reproducible VBD measurement for masking risk and breast cancer risk assessment for dense breast patients.

# Pairwise Data Consistency Conditions for the Exponential Fanbeam Transform

Richard Huber[1], Rolf Clackdoyle[1], and Laurent Desbat[1]

[1]Univ. Grenoble Alpes, CNRS, Grenoble INP, TIMC, 38000 Grenoble, France.

**Abstract** Data consistency conditions (DCCs) for projection operators have been of great relevance in the field of tomography, as they allow the determination of measured data's feasibility prior to reconstruction. Particularly useful are DCCs comparing two projections, accordingly called pairwise DCCs (PDCCs). Such conditions compute certain linear functionals dependent on individual projections, whose values must coincide for consistency to hold. For many projection operators, such PDCCs are known, but for the exponential fanbeam transform – which is relevant for Single Photon Emission Tomography (SPECT) – they are not. We show mathematically that no condition of this type can exist for a pair of exponential fanbeam projections. Moreover, we present a novel class of pairwise data consistency conditions, requiring that the difference between certain linear functionals of two projections lie in a specified interval, instead of coinciding as they would in classical PDCCs. This new condition is substantiated by numerical experiments (simulation study) on some phantoms.

## 1 Introduction

Tomographic techniques have become a vital tool in medicine, allowing doctors to observe patients' interior features. A mathematical operator (projection operator) models the underlying physics of the measurement process. The data is usually structured in so-called projections, referring to the data obtained during a specific measurement step. The choice of projection operator depends on the measurement setup used. The capability to determine whether measured data is consistent with the model/projection operator has found broad applications, such as identification/correction of corrupted data, geometric calibration, parameter identification, and motion detection [1–6]. Particularly useful are conditions capable of finding inconsistencies from just two projections, because small collections of arbitrarily oriented projections can be tested using such conditions. We refer to them as pairwise data consistency conditions (PDCCs).

A typical example is parallel-beam Computed Tomography (CT), which is modeled by the Radon transform

$$[\mathscr{R} f](\psi,s) := \int_{\mathbb{R}} f(s\vartheta_\psi^\perp + t\vartheta_\psi)\,\mathrm{d}t \tag{1}$$

for $\psi \in [0,\pi[$, $s \in \mathbb{R}$ and $f \in \mathscr{C}_c^\infty(\mathbb{R}^2)$ (smooth function with compact support), where $\vartheta_\psi = (\cos\psi, \sin\psi)^T$ and $\vartheta_\psi^\perp = (-\sin\psi, \cos\psi)^T$ are the projection directions with the angle $\psi$. In this case, the PDCCs are well-known [7]. For $\psi_1, \psi_2 \in [0,\pi[$ and all $f \in \mathscr{C}_c^\infty(\mathbb{R}^2)$, they take the form

$$\int_{\mathbb{R}} [\mathscr{R} f](\psi_1,s)\,\mathrm{d}s = \int_{\mathbb{R}} [\mathscr{R} f](\psi_2,s)\,\mathrm{d}s. \tag{2}$$

We denote by $\mathscr{R}^\Lambda$ the operator $\mathscr{R}$ only containing two projections $\Lambda = (\psi_1,\psi_2)$; we call it the pairwise Radon transform.

For several other projection operators, PDCCs have been found, two examples being the fanbeam operator [8] and the parallel-beam exponential operator [9]. One noticeable projection operator whose PDCC is yet unknown, is the (pairwise) exponential fanbeam transform

$$[\mathscr{E}_\mu^\Lambda f](\lambda,\phi) := \int_{\mathbb{R}^+} f(\lambda + t\vartheta_\phi)e^{\mu t}\,\mathrm{d}t \tag{3}$$

for $\lambda \in \Lambda = (\lambda_1,\lambda_2) \in \mathbb{R}^2 \times \mathbb{R}^2$ (the fan vertex positions), $\phi \in [-\pi,\pi[$ and $f \in \mathscr{C}_c^\infty(\Omega)$ with constant attenuation parameter $\mu \in \mathbb{R}$. Here, $\Omega$ is an open, connected set whose compact closure does not intersect with the line containing $\lambda_1$ and $\lambda_2$. This operator finds applications in pinhole SPECT imaging and corresponding PDCCs could find applications in the alignment of SPECT/CT data [10]. Here, a projection corresponds to all rays converging on $\lambda$ from any direction. The exponential term models constant attenuation processes; more general attenuation can – under certain assumptions – be converted to this exponential model, making this a mild restriction for many applications. Figure 1 illustrates the associated geometry. In this work, we explore PDCCs for the exponential fanbeam operator. For the conventional fanbeam transform ($\mu = 0$), the PDDC has the form

$$\int_{-\pi}^{\pi} \frac{[\mathscr{E}_0^\Lambda f](\lambda_1,\phi)}{\vartheta_\phi^\perp \cdot \Delta}\,\mathrm{d}\phi = \int_{-\pi}^{\pi} \frac{[\mathscr{E}_0^\Lambda f](\lambda_2,\phi)}{\vartheta_\phi^\perp \cdot \Delta}\,\mathrm{d}\phi \tag{4}$$

with $\Delta = \lambda_2 - \lambda_1$ [8]. Mathematically, (4) states that the function $\left(1/(\vartheta_\phi^\perp \cdot \Delta), -1/(\vartheta_\phi^\perp \cdot \Delta)\right)$ is orthogonal to $\mathrm{Rg}(\mathscr{E}_0^\Lambda)$, which corresponds to the backprojections of $1/(\vartheta_\phi^\perp \cdot \Delta)$ for both projections being equal. Other mentioned PDCCs are of similar mathematical structure. Hence, one might aim to find a pair of functions $\mathrm{G}_{\lambda_1,\lambda_2}, \mathrm{G}_{\lambda_2,\lambda_1}$ such that

$$\int_{-\pi}^{\pi} [\mathscr{E}_\mu^\Lambda f](\lambda_1,\phi)\,\mathrm{G}_{\lambda_1,\lambda_2}(\phi)\,\mathrm{d}\phi = \int_{-\pi}^{\pi} [\mathscr{E}_\mu^\Lambda f](\lambda_2,\phi)\,\mathrm{G}_{\lambda_2,\lambda_1}(\phi)\,\mathrm{d}\phi \tag{5}$$

for all $f \in \mathscr{C}_c^\infty(\Omega)$. As we will show, there are no PDDCs of this form for the exponential fanbeam transform, i.e., (5) does not possess a solution.

## 2 Nonexistence of PDCCs

We first set some relevant notation. Since the map $(\phi,t) \mapsto \lambda + t\vartheta_\phi$ present in (3) is a diffeomorphism for any $\lambda \in \mathbb{R}^2$,

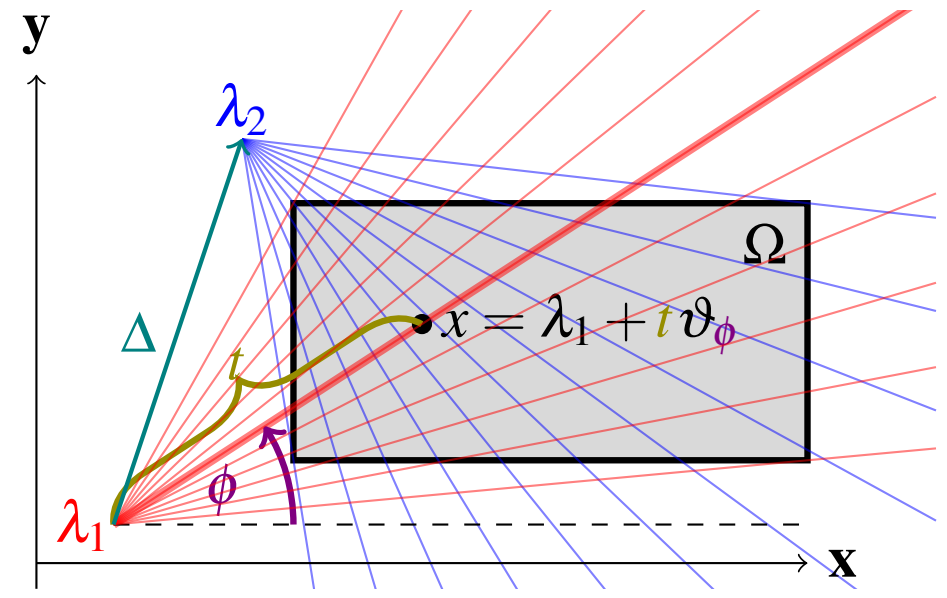

**Figure 1:** Illustration of the geometry of the fanbeam transform with two projections. The black point $x$ on the bold red line and its parametrization in fanbeam coordinates with respect to the projection $\lambda_1$ is highlighted.

we define the inverse parametrizations

$$\phi_i(x) = \arg(x - \lambda_i) \qquad \text{and} \qquad t_i(x) = |x - \lambda_i| \quad (6)$$

for $i \in \{1,2\}$. Further, we define $\Omega' = (\{\lambda_1\} \times \Phi_1) \cup (\{\lambda_2\} \times \Phi_2)$ with $\Phi_i = \phi_i(\Omega)$ for $i \in \{1,2\}$; so $\Omega'$ is essentially a 2-projection sinogram domain.

**Theorem 1.** *There is no pair of non-zero functions* $\mathrm{G}_{\lambda_1,\lambda_2} \in L^1_{loc}(\Phi_1)$ *(functions absolutely integrable on any compact subset) and* $\mathrm{G}_{\lambda_2,\lambda_1} \in L^1_{loc}(\Phi_2)$ *such that* (5) *is satisfied for all* $f \in \mathscr{C}_c^\infty(\Omega)$ *when* $\mu \neq 0$.

*Sketch of proof.* We assume $\mathrm{G}_{\lambda_1,\lambda_2}$ and $\mathrm{G}_{\lambda_2,\lambda_1}$ non-zero satisfying (5) were to exist. Plugging the definition of the exponential fanbeam transform into (5) implies

$$\int_{\Phi_1}\int_{\mathbb{R}^+} f(\lambda_1 + t\vartheta_\phi)e^{\mu t}\,\mathrm{d}t\,\mathrm{G}_{\lambda_1,\lambda_2}(\phi)\,\mathrm{d}\phi = \int_{\Phi_2}\int_{\mathbb{R}^+} f(\lambda_2 + t\vartheta_\phi)e^{\mu t}\,\mathrm{d}t\,\mathrm{G}_{\lambda_2,\lambda_1}(\phi)\,\mathrm{d}\phi. \quad (7)$$

Substituting $x = \lambda_i + t\vartheta_\phi$ for $i \in \{1,2\}$, this is equivalent to

$$\int_\Omega f(x)\frac{\mathrm{G}_{\lambda_1,\lambda_2}(\phi_1(x))}{t_1(x)}e^{\mu t_1(x)}\,\mathrm{d}x = \int_\Omega f(x)\frac{\mathrm{G}_{\lambda_2,\lambda_1}(\phi_2(x))}{t_2(x)}e^{\mu t_2(x)}\,\mathrm{d}x. \quad (8)$$

Since this statement is assumed to be true for all $f \in \mathscr{C}_c^\infty(\Omega)$, the fundamental lemma of variational calculus implies

$$\frac{\mathrm{G}_{\lambda_1,\lambda_2}(\phi_1(x))}{t_1(x)}e^{\mu t_1(x)} = \frac{\mathrm{G}_{\lambda_2,\lambda_1}(\phi_2(x))}{t_2(x)}e^{\mu t_2(x)} \quad (9)$$

for almost all $x \in \Omega$. We show that this equation does not possess a solution (thus contradicting our original assumption) by setting

$$\mathrm{h}_{\lambda_1,\lambda_2}(\phi) = \mathrm{h}_{\lambda_2,\lambda_1}(\phi) = \frac{1}{\Delta \cdot \vartheta_\phi^\perp} \neq 0, \quad (10)$$

which, for all $x \in \Omega$, satisfies

$$\frac{\mathrm{h}_{\lambda_1,\lambda_2}(\phi_1(x))}{t_1(x)} = \frac{\mathrm{h}_{\lambda_2,\lambda_1}(\phi_2(x))}{t_2(x)}. \quad (11)$$

Hence, we set $\mathrm{g}_{\lambda_1,\lambda_2} := \ln(\mathrm{G}_{\lambda_1,\lambda_2}/\mathrm{h}_{\lambda_1,\lambda_2})$ and $\mathrm{g}_{\lambda_2,\lambda_1} := \ln(\mathrm{G}_{\lambda_2,\lambda_1}/\mathrm{h}_{\lambda_2,\lambda_1})$ which (combining (9), (11)) satisfies

$$\mathrm{g}_{\lambda_1,\lambda_2}(\phi_1(x))) - \mathrm{g}_{\lambda_2,\lambda_1}(\phi_2(x)) = \mu(t_2(x) - t_1(x)). \quad (12)$$

Moreover, expressing $t_1(x) - t_2(x)$ explicitly via basic computation – e.g., using the Law of Sines – shows that

$$\mathrm{g}_{\lambda_1,\lambda_2}(\phi_1) - \mathrm{g}_{\lambda_2,\lambda_1}(\phi_2) = \mu\frac{\Delta\cdot(\vartheta_{\phi_1}^\perp - \vartheta_{\phi_2}^\perp)}{\vartheta_{\phi_1}^\perp \cdot \vartheta_{\phi_2}} \quad (13)$$

for all $\phi_1 \in \Phi_1$ and $\phi_2 \in \Phi_2$. It seems dubious that the right-hand side of (13) can be the sum of functions only depending on $\phi_1$ or $\phi_2$, respectively, as required by the left-hand side. Proving that is slightly technical, so we do not go into detail here, but roughly speaking it involves observing that the left-hand side of

$$\mathrm{g}_{\lambda_1,\lambda_2}(\phi_1) - \mathrm{g}_{\lambda_1,\lambda_2}(\widetilde{\phi}_1) = \mu\frac{\Delta\cdot(\vartheta_{\phi_1}^\perp - \vartheta_{\phi_2}^\perp)}{\vartheta_{\phi_1}^\perp \cdot \vartheta_{\phi_2}} - \mu\frac{\Delta\cdot(\vartheta_{\widetilde{\phi}_1}^\perp - \vartheta_{\phi_2}^\perp)}{\vartheta_{\widetilde{\phi}_1}^\perp \cdot \vartheta_{\phi_2}} \quad (14)$$

does not depend on $\phi_2$, but the right-hand side does, as can be verified via evaluation for certain values $\phi_1, \widetilde{\phi}_1, \phi_2$ when $\mu \neq 0$. Since the function in (14) is analytical when the denominator is not zero, the equation is wrong for almost every tuple, particularly the ones representing $\Omega$. So (13) has no solution, and consequently neither does (9), as required to prove the theorem. □

Many of the steps in the proof of Theorem 1 are not specific to the exponential fanbeam transform, but also apply to other projection pair operators. Thus the approach can be used to check if a projection pair operator possesses a PDCC, and also gives a method for identifying them if they do.

Not only range conditions of the specific form (5) cannot exist, but in fact, no $L^2(\Omega')$-continuous conditions of any kind can; even if they were non-linear.

**Theorem 2.** *When* $\mu \neq 0$, *there is no continuous function* $F\colon L^2(\Omega') \to \mathbb{R}$ *which satisfies*

1. $F(g) = 0$ *when* $g \in \mathrm{Rg}(\mathscr{E}_\mu^\Lambda)$,
2. *There is* $g \in L^2(\Omega')$ *with* $F(g) \neq 0$.

*Sketch of proof.* As a direct consequence of Theorem 1, the range of $\mathscr{E}_\mu^\Lambda$ must be dense in $L^2(\Omega')$ as no non-trivial orthogonal vector exists. If $F$ as described were to exist, it would be zero on a dense set, and by continuity zero everywhere, contradicting the second point. □

Real world data are always imperfect – and therefore inconsistent. A DCC which is continuous will be nearly satisfied for very mildly inconsistent data. Thus, a continuous DCC is favoured in practice, even though it can only identify the closure of the range. It is not yet known, whether the range of the exponential fanbeam transform is closed.

## 3 Alternative consistency conditions

That no PDCCs exist, does not imply that there is no overlapping information whatsoever. There might still be a large class of functions for which some kind of consistency criteria are possible.

Evidently, the ratio for $\lambda_1$ to $\lambda_2$ of the exponential weight applied by $\mathscr{E}_\mu^\Lambda$ to any point $x$ is $e^{\mu(t_1(x)-t_2(x))}$, and aside from this factor, the measurements behave like conventional fanbeam data ($\mu = 0$). Hence, we define

$$\overline{\delta}_\mu = \max_{x\in\Omega}\Big(\mu\big(t_1(x)-t_2(x)\big)\Big), \quad \underline{\delta}_\mu = \min_{x\in\Omega}\Big(\mu\big(t_1(x)-t_2(x)\big)\Big).$$

**Theorem 3.** *Let $f \in \mathscr{C}_c^\infty(\Omega)$ with $f \geq 0$ not constantly zero. Then*

$$\left(\frac{\int_{-\pi}^{\pi}[\mathscr{E}_\mu^\Lambda f](\lambda_1,\phi)\frac{1}{\vartheta_\phi^\perp\cdot\Delta}\,\mathrm{d}\phi}{\int_{-\pi}^{\pi}[\mathscr{E}_\mu^\Lambda f](\lambda_2,\phi)\frac{1}{\vartheta_\phi^\perp\cdot\Delta}\,\mathrm{d}\phi}\right) \in \left[e^{\underline{\delta}_\mu}, e^{\overline{\delta}_\mu}\right]. \qquad (15)$$

Note that the left-hand side of (15) (henceforth called consistency quotient) would be 1 for the conventional fanbeam transform (see (4)). So this condition states that the exponential fanbeam transform satisfies the conventional fanbeam transform's PDCC within a certain margin dependent on the geometry (the extent of $\Omega$ and positions $\lambda_1, \lambda_2$) and the attenuation coefficient.

Satisfying (15) does not guarantee consistency, as mildly inconsistent data might still be within the bounds. However, violating the condition implies inconsistency with certainty.

This result does not contradict Theorem 2 since the consistency quotient in (15) is not continuous at $f = 0$. For $f$ giving a non-zero denominator in (15), the denominator does not become zero when sufficiently small noise is added, making the condition stable under low noise conditions.

*Sketch of proof.* We define the function $\widetilde{f}(x) = f(x)e^{\mu(t_1(x)-t_2(x))}$ and the conventional fanbeam's PDCC (4) for $f(x)e^{\mu t_1(x)}$ implies

$$\int_{-\pi}^{\pi}[\mathscr{E}_\mu^\Lambda f](\lambda_1,\phi)\frac{1}{\vartheta_\phi^\perp\cdot\Delta}\,\mathrm{d}\phi = \int_{-\pi}^{\pi}[\mathscr{E}_\mu^\Lambda \widetilde{f}](\lambda_2,\phi)\frac{1}{\vartheta_\phi^\perp\cdot\Delta}\,\mathrm{d}\phi. \qquad (16)$$

Moreover, using the mean value theorem, it is easy to see that

$$\frac{[\mathscr{E}_\mu^\Lambda \widetilde{f}](\lambda_2,\phi)}{[\mathscr{E}_\mu^\Lambda f](\lambda_2,\phi)} \in \left[e^{\underline{\delta}_\mu}, e^{\overline{\delta}_\mu}\right]. \qquad (17)$$

Combining the right-hand side of (16) with (17) and basic integration properties, Theorem 3 follows directly. □

## 4 Numerical experiments

We conducted numerical experiments to corroborate Theorem 3's result and to show that the interval proposed in (15) is not chosen needlessly big, but is violated by inconsistent enough data. To that end, we considered the following setup: We had fan vertex positions $\lambda = 80\vartheta_\varphi$ for $\varphi = \{-90°, -88°, \ldots, 88°, 90°\}$. The imaging domain was a 40 cm × 40 cm square centered at the origin, and the attenuation parameter $\mu = -0.154$ (the attenuation per cm of water). We chose $\Omega = [-20\text{ cm}, 20\text{ cm}] \times [-14\text{ cm}, 8\text{ cm}]$ reflecting a box with width and height to encompass the activity. We computed the consistency quotient for two activity distributions designed to highlight specific aspects of Theorem 3. Those activities were the NCAT phantom [11] and a (Hot Source) phantom with constant activity in a circle with radius 1 cm at the bottom right extremity of $\Omega$. Those phantoms were digitally represented on an array of $N_x \times N_x$ pixels with $N_x = 400$ and were transformed into sinograms with $N_s = 200$ detector pixels positioned equispaced on a flat detector covering the relative angular range of $[-\arctan\left(\frac{5}{12}\right), \arctan\left(\frac{5}{12}\right)]$; see Figure 2. The exponential fanbeam transform was executed with Gratopy [12] employing a pixel-driven approach. We simulated motion inconsistencies by shifting the phantom for projections with $\varphi > 4°$ by 4 cm in the x-direction. This corresponded to an abrupt movement of the patient at 8 minutes after the beginning of a 15-minutes scan.

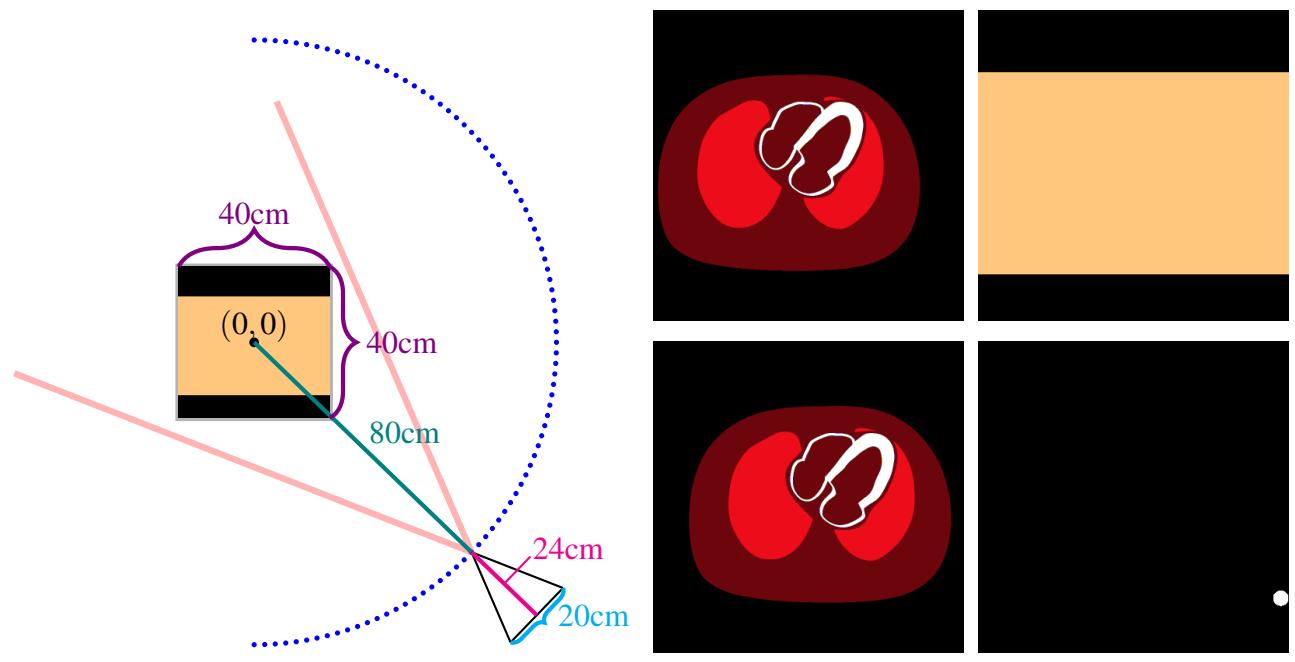

**Figure 2:** Illustration of the geometry used for numerical experiments, with the blue dots representing the fan vertex positions. The central 40 cm × 40 cm box represents the imaging domain containing activity overlayed with an illustration of our choice of $\Omega$. On the right: (upper left) the NCAT phantom and its shifted form (lower left), (upper right) the rectangular domain $\Omega$ and (lower right) the second phantom consisting of a hot source located in the bottom right corner of $\Omega$.

Figure 3 depicts the evaluation of condition (15) for the two scenarios by showing the consistency quotient for the middle projection (associated with $\varphi = 0$) paired with any other $\lambda$, and the corresponding bounds (the right-hand side of (15)). For the NCAT phantom without motion (consistent projections) the consistency quotient remained close to one, and stayed well within the bounds given by Theorem 3. For the NCAT phantom with motion, the consistency quotient behaved similarly, but did violate the consistency bounds for projections near the reference projection. For the second phantom, the data were consistent (no motion) and, as expected, the consistency quotient remained within the bounds, but closely followed the upper right arm of the bounds, which supports the notion that these bounds cannot be improved when using this particular consistency quotient. Simulations with the hot source in the other three corners of the rectangular support region result in tracing the other three arms of the bounds.

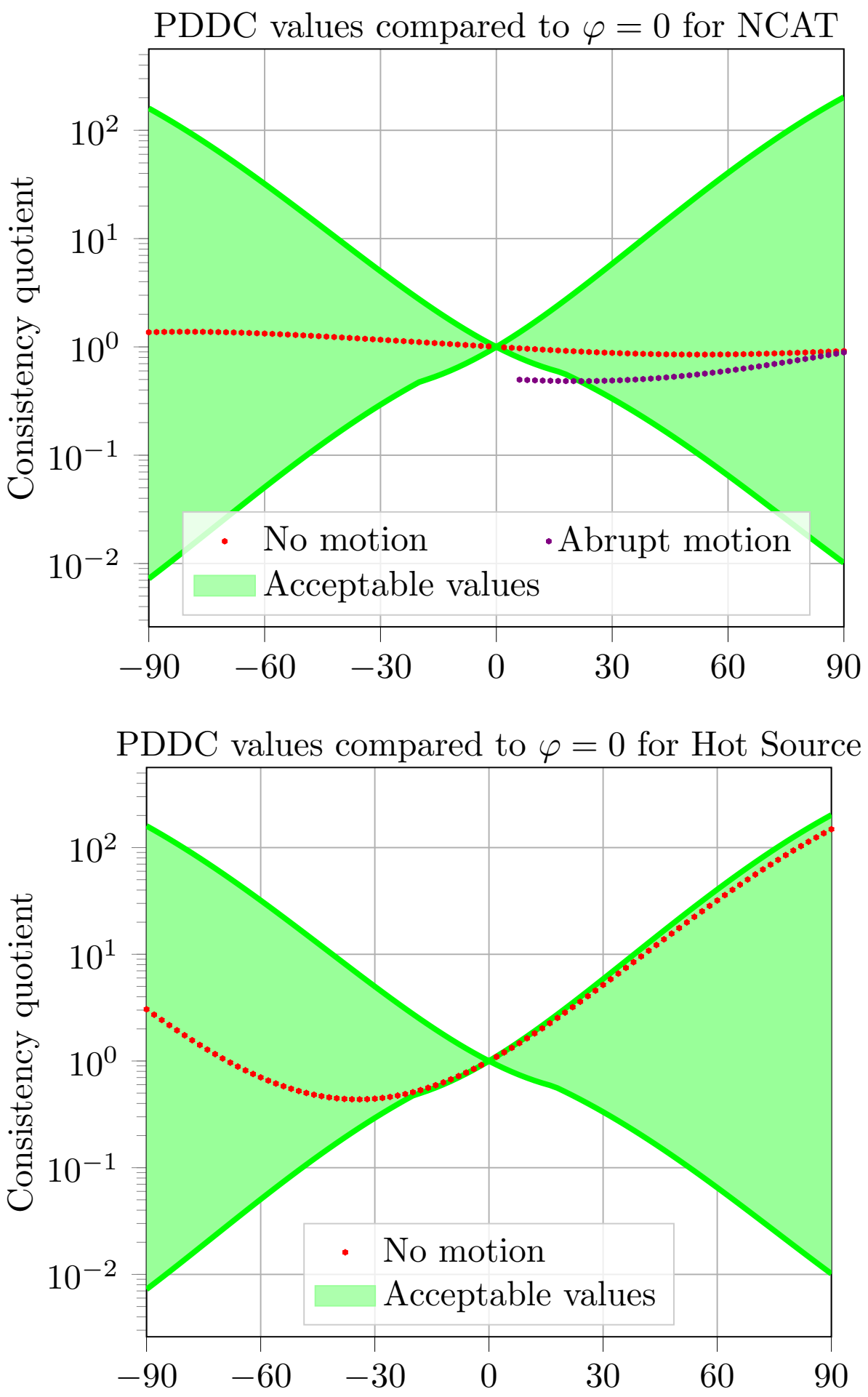

**Figure 3:** Evaluation of (15) in the numerical experiments, on the top for the NCAT phantom, below the second phantom. The consistency quotient of the consistent measurements and the inconsistent measurments with motion at the 8-minute mark are depicted. These observations are made for $\lambda_1 = \vartheta_\varphi$ with the angle $\varphi = 0$, while the angles of $\lambda_2$ are on the $\mathbf{x}$-axis.

## 5 Summary and Outlook

This paper discussed data consistency conditions for the exponential fanbeam transform, showing that (classical) pairwise data consistency conditions cannot exist for this transform. As an alternative, Theorem 3 provides a weak form of PDCC whereby a certain expression, the consistency quotient, must lie within a defined interval if the two projections are consistent. The NCAT simulations illustrated three consequences of Theorem 3, namely (i) that if the projections are consistent, the consistency quotient will lie within the bounds defined by the interval; (ii) equivalently, values lying outside the bounds definitely indicate inconsistency; and (iii) values inside the bounds do not provide information on consistency. For similar projections (fanbeam vertex positions fairly close to each other) the bounds are reasonably small and effectively detected inconsistency. For distant projections though, the bounds can be very broad, which limits their usefulness. However, the functional in (15) can probably be improved upon, to achieve tighter bounds and a consequently stronger PDCC. This might be the topic of future work.

We have illustrated the potential for patient motion identification, but other applications, such as detector sensitivity response or identification of pinhole positions, are conceivable.

**Acknowledgement:** This work was supported by the ANR grant ANR-21-CE45-0026 ‘SPECT-Motion-eDCC’.

# Data-driven approach for metal artifact reduction in dental cone-beam CT with an extra-condition of intra-oral scan data

Chang Min Hyun[1], Kiwan Jeon[2], and Hyoung Suk Park[2]

[1]School of Mathematics and Computing (Computational Science and Engineering), Yonsei University, Seoul, Republic of Korea
[2]National Institute for Mathematical Sciences, Daejun, Republic of Korea

**Abstract** In dental cone-beam computed tomography (CBCT), the presence of metallic inserts such as implants, crowns, and dental fillings causes severe streaking and shading artifacts in a CT image, which consequently prevent the accurate restoration of teeth and bones. Existing metal artifact reduction (MAR) methods have limitations in their ability to preserve morphological structures of teeth in reconstructed CT images. This study presents a novel data-driven approach for MAR in dental CBCT that takes advantage of radiation-free intra-oral scan (IOS) data as an extra-condition for CT image reconstruction. The IOS data play a crucial role by providing sophisticated knowledge for teeth morphological structure and dentition in such a way of acting as an appropriate condition during MAR. We adopt a conditional generative adversarial network framework, where a metal artifact corrector is trained to generate a metal artifact-free image sampled from the target distribution (i.e., the distribution of metal-free CBCT images) under the condition of IOS data. Realistic simulations were performed to show the usefulness and effectiveness of the proposed MAR method.

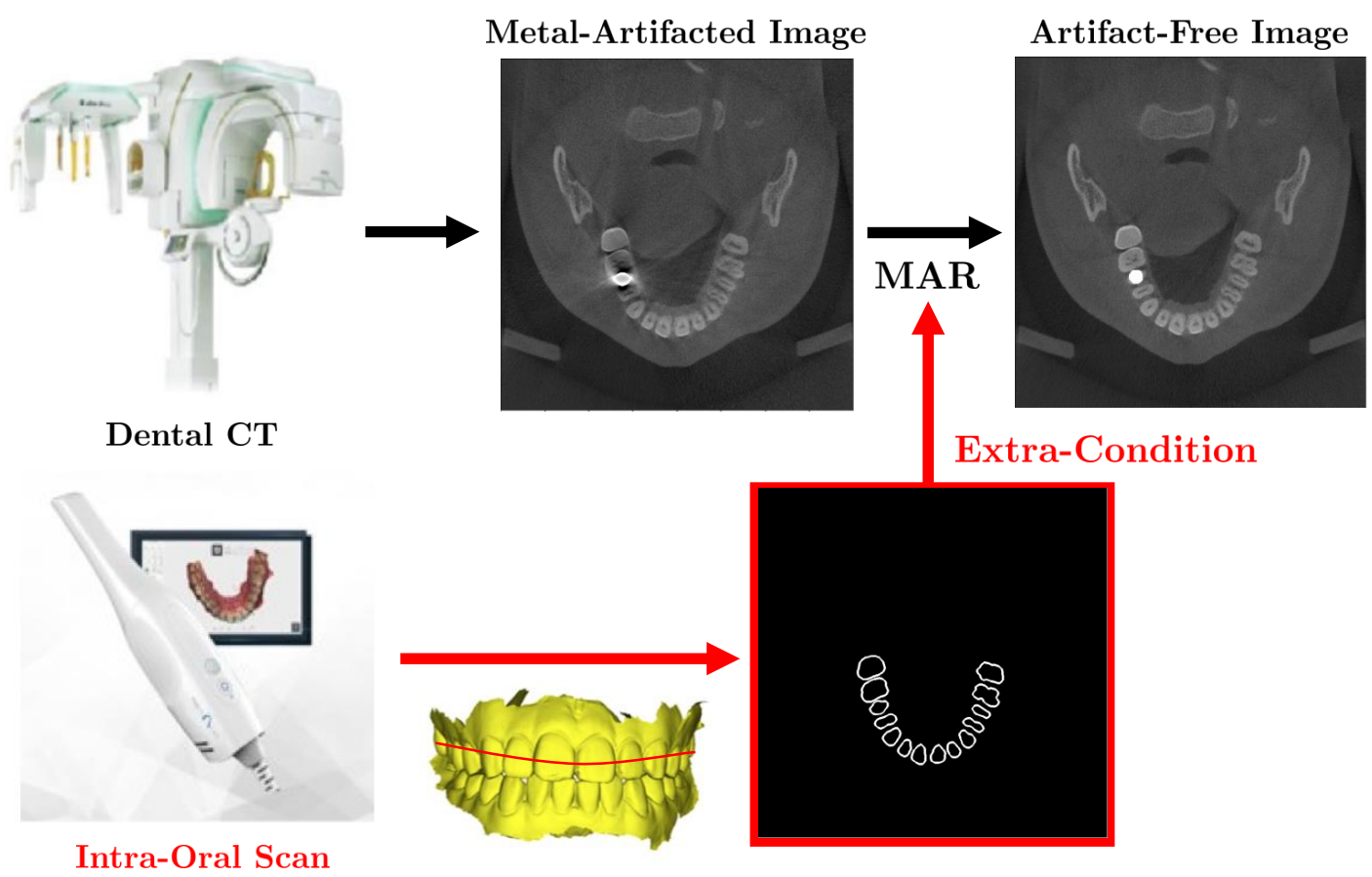

**Figure 1:** Schematic representation of the proposed approach for metal artifact reduction in dental cone-beam CT with an extra-condition of intra-oral scan data.

## 1 Introduction

In various fields of clinical dentistry, dental cone-beam computed tomography (CBCT) has been widely utilized to understand complicated anatomical structures of mandible, maxilla, or skeleton [**Sukovic2003**, **Yun2022**]. However, in the presence of metallic inserts (e.g., implants, crowns, and dental fillings), a CBCT image is deteriorated by metal artifacts, making it difficult to accurately perform downstream tasks such as bone and teeth segmentation [**Gateno2007**]. Hence, reducing metal-related artifacts has drawn increased attention in digital dentistry workflows.

Metal artifacts are caused by several physical factors, such as beam hardening, scattering, noise, nonlinear partial volume, and photon starvation [**Lee2019**, **Park2017**]. Streaking and shadowing artifacts are created, resulting in a significant loss of anatomical details such as teeth morphological structures [**Bayaraa2020**]. Because multiple and large metallic objects commonly present in dental CT scans, it is very difficult to effectively deal with metal artifacts due to strong contamination. What is more, in the practical dental CBCT imaging environment, offset detection, FOV truncation, low radiation dose, and 3D reconstruction characteristics attribute to make metal artifacts more severe [**Park2022A**, **Hyun2022**].

There have been numerous studies on metal artifact reduction (MAR), which include sinogram inpainting-based correction [**Meyer2010**], statistical iterative correction [**DeMan2001**], and dual-energy reconstruction [**Lehmann1981**]. However, existing MAR methods are not fully satisfactory for clinical use. The inpainting-based correction can generate secondary artifacts owing to inaccurate interpolation along the metal trace. The statistical iterative correction and dual-energy approaches require a large computational cost and an additional radiation dose, respectively.

Recent advances in deep learning technology have been being progressing in various medical imaging fields including MAR [**Hyun2021**]. Existing data-driven MAR methods [**Zhang2018**, **Gjesteby2019**, **Park2018**, **Lin2019**, **Yu2020**] have shown great potential to improve the overall image quality, whereas it is still arduous to effectively restore significantly corrupted or missing morphological structures. It appears to be caused by using already severely corrupted projection data only as explicit base knowledge for recovering a target artifact-free image [**Hyun2022**].

Along with the rapid development of intra-oral scan (IOS) technologies, 3D teeth morphological information can be acquired in a digital form, whose accuracy is in a similar level to the conventional physical impression [**Albayrak2021**, **Roig2020**]. Moreover, its data acquisition is irrelevant to radiation threaten. IOS data are timely available as a new means for assisting MAR by providing sophisticated knowledge for teeth morphological structure and dentition. A schematic representation is provided in Figure 1.

In this study, we present a novel data-driven approach for MAR in dental CBCT that takes advantage of radiation-free IOS data as an extra-condition for CT image reconstruction. We adopt a conditional generative adversarial network (GAN) framework [**Mirza2014**, **Gauthier2014**] so that IOS data can be leveraged as an extra-condition during MAR. A reconstruction map is trained to generate a metal artifact-free image sampled from the target distribution (i.e., the distribution of metal-free CBCT images) under the condition of IOS data. The training is performed using a realistic triplet training dataset [**Zhang2018**, **Park2018**], where metal-affected images are simulated according to the CT physics model and IOS data are using teeth segmentation from metal-free CBCT scans.

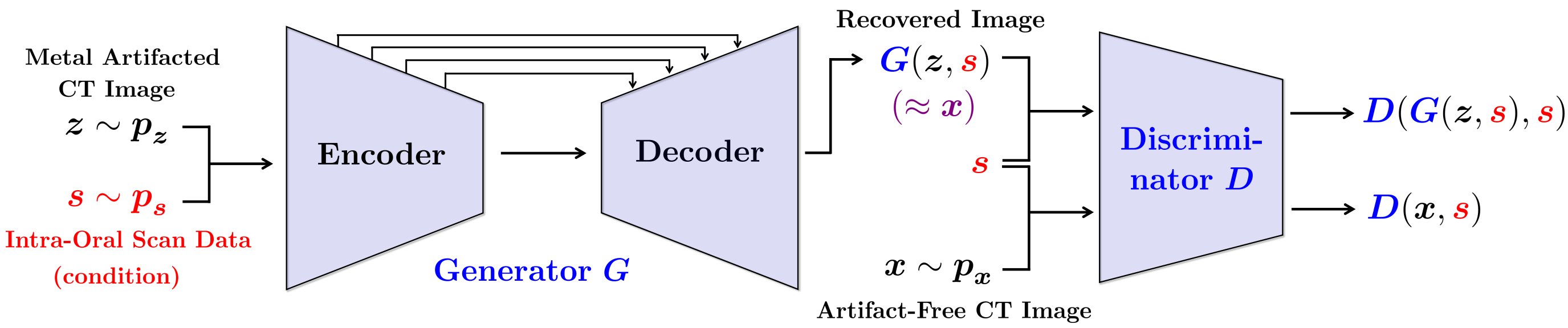

**Figure 2:** Data-driven approach for metal artifact reduction in dental cone-beam CT with an extra-condition of intra-oral scan data.

Experiments using clinical CT and simulated IOS data validated the usefulness and effectiveness of the proposed MAR method.

## 2 Method

Let $p$ be a metal-artfacted CBCT projection, which can be represented by

$$p = \mathscr{S}_{\text{ub}}(-\ln \int_E \eta(E) \exp(-A\mu_E) dE + n), \qquad (1)$$

where $\mathscr{S}_{\text{ub}}$ is a subsampling operator depending on the detector arrangement, $\eta$ represents a normalized energy distribution of a X-ray source, $A$ is a cone beam forward projection operator, $\mu_E$ is a 3D attenuation coefficient distribution of a human body at an energy level $E$, and $n$ is CT noise. A CBCT image is reconstructed by $z = A^\dagger p$, where $A^\dagger$ is the FDK algorithm [**Feldkamp1984**]. In the presence of metallic objects, $z$ is severely contaminated by metal artifacts.

Let $x$ be the artifact-free CBCT image corresponding to $z$. A MA corrector $G : z \to x$ can be learned through a variational model by solving the following minimization problem [**Park2022B**]: For $\lambda > 0$,

$$G = \underset{G}{\text{argmin}}\ d(p_G, p_x) + \lambda \|G(z) - x\|^2_{\ell 2}, \qquad (2)$$

where $d(p_G, p_x)$ is a metric that measures a distance between $p_G$ and $p_x$, and $\|G(z) - x\|^2_{\ell 2}$ represents a pixel-wise loss that constrains to preserving image information in $z$. Here, $p_G$ and $p_x$ are probability distributions for $G(z)$ and $x$, respectively. However, in the presence of metallic objects, existing methods have shown limited performance for estimating $p_x$;

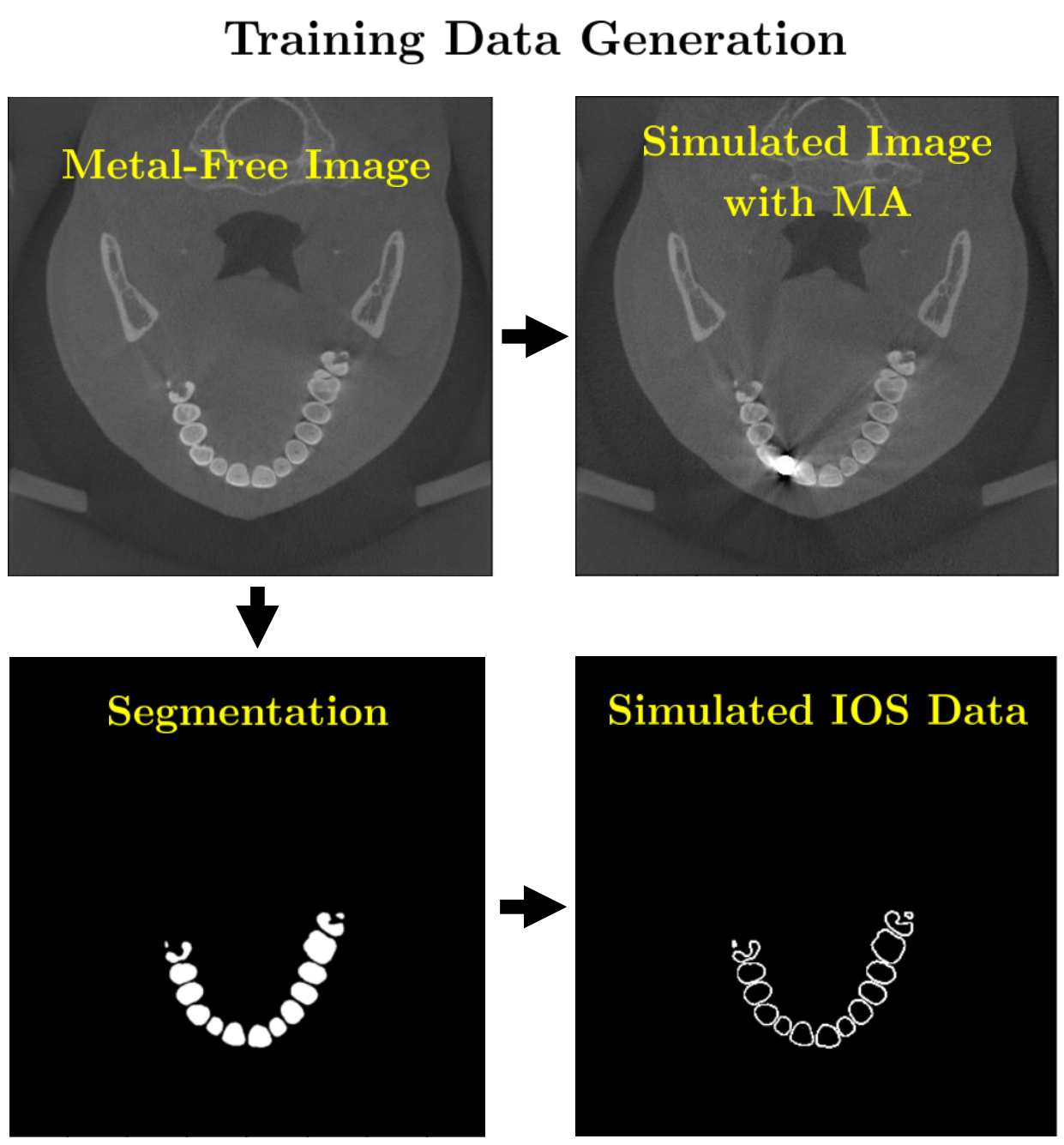

**Figure 3:** Training data generation for metal affected image and IOS data.

so that images in $p_G$ tend to lose anatomical details such as teeth boundaries.

The main objective of this study is to learn the corrector $G$ with an extra-condition of IOS data as a tool for preventing the loss or corruption of teeth boundaries. The overall process is illustrated in Figure 2. Denoting the corresponding IOS data to $z$ by $s$, we seek to learn a corrector $G : (z, s) \to x$ such that

$$G = \underset{G}{\text{argmin}}\ d(p_G, p_{x,s}) + \lambda \|G(z, s) - x\|^2_{\ell 2}, \qquad (3)$$

where $p_{x,s}$ is a joint probability distribution for $x$ and $s$. Note that, in the case when the distance $d$ is given by the Pearson divergence, the minimization for $dist(p_G, p_{x,s})$ can be alternatively achieved using the conditional least-squares GAN framework [**Goodfellow2020**, **Gauthier2014**, **Mao2017**].

To be precise, we assume that $N$ triplet training samples $\{(z^{(i)}, x^{(i)}, s^{(i)})\}_{i=1}^N$ are given. The proposed method finds an optimal $G$ by the aid of a discriminator $D$ through the following minimization: For $\gamma > 0$,

$$\begin{cases} \underset{G}{\operatorname{argmin}} \frac{1}{N}\sum_{i=1}^{N} \|D(G(z^{(i)},s^{(i)}),s^{(i)})\|_{\ell 2}^{2} + \gamma\|G(z^{(i)},s^{(i)}) - x^{(i)}\|_{\ell 2}^{2} \\ \underset{D}{\operatorname{argmin}} \frac{1}{N}\sum_{i=1}^{N} \|D(G(z^{(i)},s^{(i)}),s^{(i)}) + 1\|_{\ell 2}^{2} + \|D(x^{(i)},s^{(i)}) - 1\|_{\ell 2}^{2} \end{cases}, \tag{4}$$

which comprises the mean squared loss under the condition $s$ and conditional least-squares adversarial loss related to the discrepancy between learned and target distributions under the condition $s$.

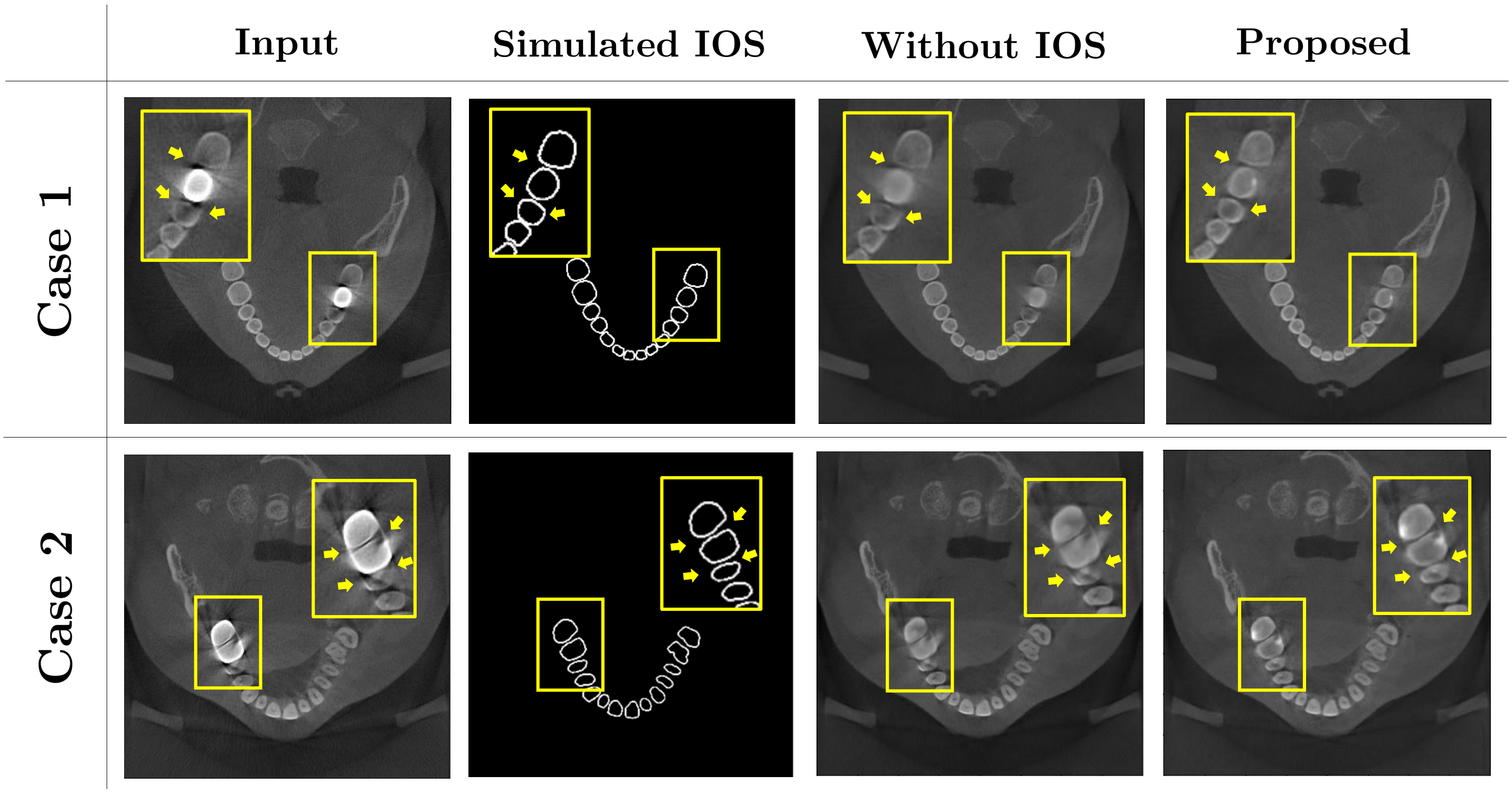

**Figure 4:** Qualitative evaluation in two test cases using clinical CBCT and simulated intra-oral scan data.

## 3 Result

Metal artifact-free scans were obtained from 20 subjects by a commercial CT device (Q-FACE, HDXWILL) with a tube voltage of 85 keV. A voxel size is $800 \times 800 \times 400$ with a resolution of $0.2 \times 0.2 \times 0.2$ mm$^3$. For network training, metal affected and IOS data were generated. The CT physics model in (1) were used for the generation of metal artifacts. Once tooth segmentation was performed in metal artifact-free images, simulated IOS data were obtained by the followed Laplacian filtering, image erosion, and binarization. See Figure 3.

For the generator $G$, the U-shape FCN [**Ronneberger2015**] was used as a backbone network architecture. For the discriminator $D$, we applied three $4 \times 4$ and one $1 \times 1$ convolution operations with the stride of 2 and 1, respectively. The former three operations were performed with an activation function of Leaky ReLU and the last of hyperbolic tangent. The training was based on Adam optimizer, batch normalization, and $200 \times 200$ image patches for reducing a total computational cost [**Hyun2020**].

All experiments were conducted in a computing environment equipped with four GTX3080ti GPUs, two Xeon CPUs E5-2630 v4, and 128GB DDR4 RAM.

We tested the proposed method by using clinical (not simulated) CBCT data with metal artifacts and simulated IOS data. Qualitative evaluations are provided in Figure 4. We compared the proposed method with the U-shape FCN that was trained by data pairs of $z$ and $x$ without IOS data. The experimental results showed that the proposed method can considerably improve fine details of teeth

## 4 Discussion and Conclusion

This study investigated the usefulness of IOS data as extra information for MAR in dental CBCT. IOS data containing sophisticated knowledge for teeth morphological structures can be leveraged as an image prior in such a way of acting as an extra-condition in MAR. To show the effectiveness and potential of IOS data in MAR, we conducted experiments using clinical CBCT and simulated IOS data.

Our future research subjects include the following two topics. The first is to test the proposed method by using real IOS data. The second is to apply our algorithm in a unsupervised learning setup.

For the application using real IOS data, the registration to dental CBCT is prerequisite. Since manual registration is time-consuming and labor-intensive, it is recommended to use an automatic registration method. Fortunately, several deep learning algorithms for dental CBCT-IOS registration [**Chung2020**, **Jang2022**] were recently proposed, seemingly promising high performance even in the circumstance that severe CT artifacts present. These methods can be potentially combined with the proposed method toward imaging in the practical environment.

In recent digital dentistry workflows, the integrated use of IOS, facial soft-tissue scanner, and dental CBCT has been becoming common [**Shujaat2021**]. Similar to IOS data, surface knowledge contained in facial soft-tissue scan data might be utilized as another constraint for MAR.

# Multi-material Decomposition with Triple Layer Flat-Panel Detector CBCT using Model-based and Deep Learning Approaches

Xiao Jiang[1], Xiaoxuan Zhang[2], J. Webster Stayman[1], Grace J. Gang[2]

[1]Department of Biomedical Engineering, Johns Hopkins University, Baltimore, USA
[2]Department of Radiology, University of Pennsylvania, Philadelphia, USA

**Abstract** Spectral CT has been investigated widely for a range of diagnostic applications with increasing potential interest for cone-beam CT (CBCT) applications. Current CBCT technology has largely focused on flat-panel detectors due to their relatively small form factor and ease of integration within a compact gantry that fits well in an interventional suite. The recent commercial availability of triple-layer flat panel detectors has provided a new avenue for spectral CBCT. In particular, while many spectral systems are limited to two channels with different energy sensitivity (e.g. dual layer detectors, kV-switching systems, etc.), a triple-layer system has the potential to be able to perform three material decomposition without additional constraints. Unfortunately, the spectral separation of a triple-layer panel is modest leading to a relatively ill-conditioned material decomposition problem (which consequently can be highly noise magnifying). In this work, we explore the possibility of three material decomposition and CBCT using a triple-layer panel and two sophisticated processing approaches: 1) model-based projection-domain material decomposition and 2) deep-learning-based projection-domain decomposition. Both approaches use simple filtered-backprojection of material line integral estimates to form 3D material maps. A simulation study with realistic measurement models is conducted using anthropomorphic phantoms and three material bases (water, calcium, and exogenous gadolinium contrast agent). A preliminary performance evaluation of reconstructed phantom data is provided to illustrate the potential of spectral CBCT using triple-layer detectors.

## 1 Introduction

Spectral x-ray imaging has the potential to enable a number of novel clinical diagnostics and CT image quality improvements [1] including non-contrast-enhanced image synthesis, virtual monoenergetic images, beam hardening and metal artifact reduction. The capability of spectral method to provide material separation similarly has the potential to enhance various clinical tasks like artery calcification detection and visualization, uric acid characterization/quantification, etc. Such applications are increasingly being developed and translated using diagnostic CT scanners. Research into the application of spectral imaging is also being investigated in cone-beam CT[1]. Cone-beam CT (CBCT) applications often target specific interventional procedures and their associated diagnostics. Applications include hemorrhage detection (including contrast agent extravasation), revascularization assessments, and vascular lumen characterization[2].

Tube voltage switching[3] and multi-layer flat-panel detector[4] are two approaches that have been investigated for spectral cone-beam CT data acquisitions. Tube voltage switching can provide large spectral separation and is compatible with the conventional flat-panel detectors (FPDs). However, this strategy requires a more complex x-ray generator and may have increased sensitivity to patient and gantry motion due to projection mismatches between different energy channels. In contrast, recently available multi-layer flat-panel detectors provide projection data where spectral channels are collected simultaneously, minimizing cross-channel geometry mismatches and motion-induced artifacts. Dual-layer flat-panel detector [4] has been explored for radiography and interventional imaging systems. Strictly speaking, such two channel systems only permit differentiation between two materials. While a volume constraint could be added to permit a third material estimate, such a constraint does not hold universally (e.g. in the lungs).

Three-material decomposition is particularly important for separation of exogenous contrast agents and from anatomy[5]. Applications include angiography [6], perfusion studies, and lesion enhancement.

Recently available triple-layer flat panel detectors provide one potential avenue to produce three-material decompositions without an explicit volume constraint. However, such technology is challenged by the relatively poor spectral separability of these detectors (and consequent ill-conditioning of the decomposition problem). In this work, we conducted a study to investigate the potential of creating accurate water, calcium, and contrast density images using triple-layer CBCT. We developed and evaluated two processing schemes to produce 3D material estimates: 1) a model-based iterative approach; and 2) a deep-learning-based decomposition. Both approaches consider a projection domain decomposition followed by FDK reconstruction. Preliminary evaluations comparing the two approaches are provided.

## 2 Materials and Methods

The two material decomposition approaches are introduced in the following sections.

### 2.1. Model-based projection-domain decomposition

The general forward model for a multi-layer flat-panel detector can be written in matrix notation as [7]:

$$\bar{y}(q) = \mathbf{BS}\exp(-\mathbf{QA}\rho) = \mathbf{BS}\exp(-\mathbf{Q}l) \qquad (1)$$

where $l \in \mathbb{R}^{jk}$ is the vectorized material density projection (i.e., the physical density $\rho$ of each basis material forward projected by operator $\mathbf{A}$ ) with $j$ pixels and $k$ basis materials, $\mathbf{Q}$ stacks the mass attenuation coefficients of each basis material, and $\exp(-\mathbf{Q}l)$ represents the total attenuation in each energy bin. The matrix $\mathbf{S}$ characterizes the overall spectral sensitivities of the system and **B** models the (potentially) layer-dependent blur kernels.

Assuming measurements $y$ follow a multivariate Gaussian distribution, $y \sim \mathcal{N}(\bar{y}, \mathbf{\Sigma})$, the regularized likelihood objective function for material decomposition can be written as:

$$\hat{l} = \text{argmax}_l(\Phi(l; y)) \quad (2)$$

$$\text{where } \Phi(l) = (y - \bar{y}(l))^T \mathbf{\Sigma}^{-1}(y - \bar{y}(l)) + \beta l^T \mathbf{R} l$$

A quadratic regularization functin ($l^T \mathbf{R} l$) is used in this work that penalizes the differences between the 4-nearest neighboring pixels within each material map. We applied Newton's method to solve the objective. Within the n[th] iteration, the material maps are updated according to:

$$l_{n+1} = l_n - \alpha(\nabla_l^2 \Phi)^{-1} \nabla_l \Phi \quad (3)$$

where the gradient, $\nabla_q \Phi$, and Hessian, $\nabla_q^2 \Phi$, of the objective are given by:

$$\nabla_q \Phi = \mathbf{Q}^\mathbf{T} \mathbf{D_1} \mathbf{S}^\mathbf{T} \mathbf{B}^\mathbf{T} \mathbf{\Sigma}^{-1}(\mathbf{BS}\exp(-\mathbf{Q}l) - y) + \beta \mathbf{R} l \quad (4)$$

$$\nabla_q^2 \Phi \approx \mathbf{Q}^\mathbf{T} \mathbf{D_1} \mathbf{S}^\mathbf{T} \mathbf{B}^\mathbf{T} \mathbf{\Sigma}^{-1} \mathbf{B} \mathbf{S} \mathbf{D_1} \mathbf{Q} + \beta \mathbf{R} \quad (5)$$

where $\mathbf{D_1} = \text{diag}\{\exp(-\mathbf{Q}l)\}$. The update step size, $\alpha$, is empirically chosen to be 0.5.

We additionally adopted the following strategies to model additional physical effects, improve performance, and accelerate convergence: 1) The middle and bottom layer projections were registered to that of the top layer using an affine transformation to account for geometry mismatch and pixel grid misalignment among the layers; 2) The projections were pre-processed by deconvolving the blurs **B** using the Richardson–Lucy method [8], thus making the pixels approximately separable; 3) An initial material decomposition was first performed on 8x downsampled projection data, upsampled to the full resolution, and then used as initialization for the Newton update. The latter strategy increased robustness against local minima induced by image noise.

The application in this work focuses on contrast-enhanced studies using gadolinium as the contrast media. The three basis materials were therefore chosen to be water, calcium, and gadolinium. The material-specific regularization strength parameters, $\beta$, were chosen to minimize the error in the gadolinium images. The values are set to $2 \times 10^{-6}, 2 \times 10^{-5}$, and $2 \times 10^{-4}$ for water, calcium, and gadolinium, respectively.

## 2.2. Deep-learning projection-domain decomposition

We have previously developed a deep learning network capable of performing three material decomposition using simulated data generated from a realistic forward model of the triple layer detector (Eq.1). The network architecture is shown in Fig. 1 The input to the network consists of a three-channel input formed by concatenating the three projections from the triple-layer detector, while the output consists of three basis material maps. Material decomposition is performed by a network following the ResUnet architecture with 9 residual blocks (ResBlock) [9]. Each block consists of two 3×3 convolution layers separated by a LeakyReLU activation layer, one residual connection adding the block input to output, and a final LeakyReLU activation layer. Each ResBlock in the encoder portion of the network reduces the dimension of the input by half and doubles the number of feature channels, while each ResBlock in the decoder portion does the opposite. A final 1×1 convolution layer is applied to reduce the number of channels to three - the number of basis materials.

For training, we adopted the following loss function:

$$\mathcal{L} = \mathcal{L}_p + \lambda_1 \mathcal{L}_{edge} + \lambda_2 \mathcal{L}_{consistency} \quad (6)$$

The first term, $\mathcal{L}_p(l, \hat{l}) = \|l - \hat{l}\|_2^2$, is a typical MSE loss which quantifies the difference between the predicted material density projections, $\hat{l}$, and the ground truth, $l$. We additionally included a gradient-based loss term, $\mathcal{L}_{edge}(l, \hat{l}) = \|\nabla l - \nabla \hat{l}\|_1$, which has been explored to preserve spatial resolution[10]. The third term is a data consistency loss that penalized differences in the measurement ($y$) domain rather than the material density line integral ( $l$ ) domain, i.e., $\mathcal{L}_{consistency}(y, \hat{l}) = \|\mathbf{BS}\exp(-\mathbf{Q}\hat{l}) - y\|_2^2$. Such loss functions have been investigated in previous work to incorporate a physics-driven constraint to the network output[11]. The scalars, $\lambda_1, \lambda_2$, control the relative weight of each term. Different combinations of $\lambda_1, \lambda_2$ were investigated for their impact on imaging performance. The optimal weighting was selected to minimize the MSE over all three material maps ($\lambda_1 = 10$ and $\lambda_2 = 1$). The network was implemented in PyTorch. We used the Adam optimizer with a batch size of 4 and terminated after 400 epochs based on empirical observation. The loss function typically decreased by <0.1% in the last 100 iterations.

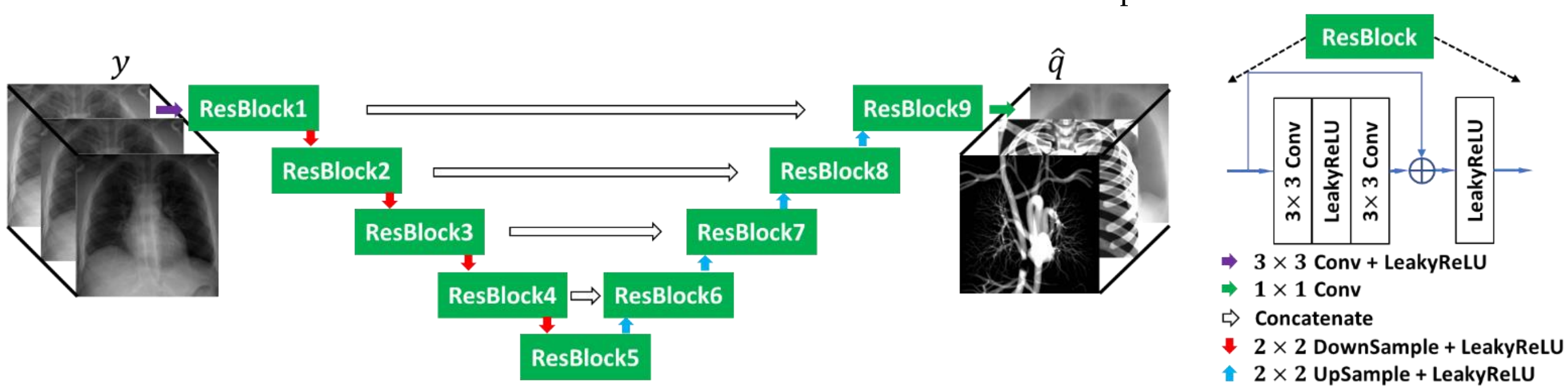

Figure 1: ResUnet network architecture with triple layer projection data ($y$) as an input, and predicted material projections ($\hat{q}$) as the output.

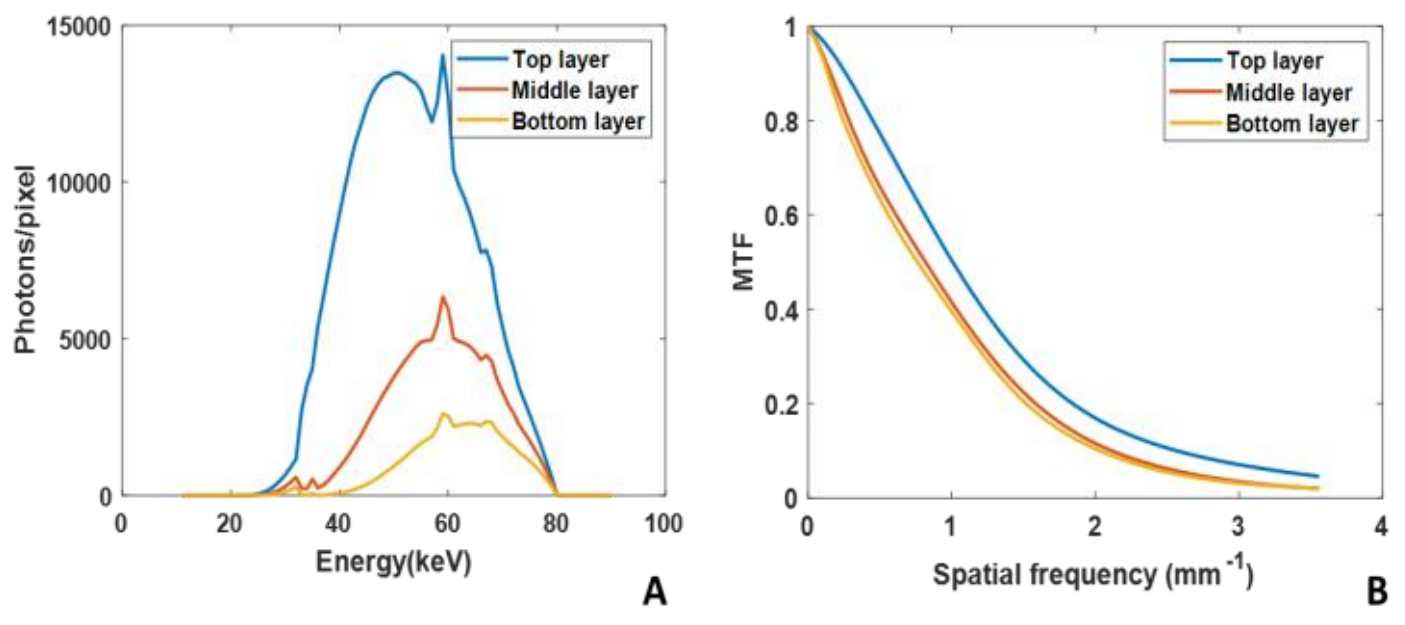

Figure 2: Spectral sensitivities (A) and MTFs for each channel (B).

### 2.3 Phantoms and data simulation

Imaigng phantoms in this work were generated using the XCAT package[12]. High resolution volumetric chest phantoms were genereated at a voxel size of 0.4 mm×0.4 mm×0.4 mm. The ground truth density of water and calcium are obtained using an image-domain decomposition method applied to two attenuation images generated at 60 keV and 100 keV. Vessels are uniformly assigned to 20 mg/ml of gadolinium. The ventricles and atria in the heart were intentionally not enhanced for better visualization of overalying vessels.

To generate simulated projection data for both the model- and learning-based methods, we used the detector blurs, **B**, and spectral sensitivities, **S**, shown in Fig.2. These characteristics are intended to model a realistic triple-layer panel comprised of three stacked indirect flat-panel detectors with ~250 µm of CsI in each layer. Projection data were simulated according to Eq.1 using a pixel size of 0.28mm×0.28mm. Noisy projections was obtained by adding independent Poisson noise before applying **B**.

To generate training data for the learning-based decomposition algorithm, we used 24 XCAT patient models to obtain different realizations of chest phantoms. We simulated 100 projections evenly distributed over 360° for each phantom to obtain 2400 projections, of which 2000 were used for training while the remaining 400 were used for validation. We randomly extracted 32 patches (256 pixels × 256 pixels) from each projection, and randomly applied horizontal or vertical flips for data augmentation.

### 2.4 Reconstruction

The model-based and learning-based algorithms were applied to 360 projections uniformly distributed over 360°. The resulting estimated material density line integrals $\hat{l}$ were reconstructed using an in-house FDK algorithm to obtain the 3D density distribution ($\rho$).

### 2.5. Evaluation

We evaluted the structural similarity index measure (SSIM) and root-mean-square error (RMSE) of the estimated 3D density distribution with ground truth. Note that the ground truth here is the FDK reconstruction of the ground truth mateiral density line integrals from the XCAT phantom. We further compared line profiles through anatomical structures for spatial resolution comparison.

## 3 Results

Figure 3 shows the 3D density distribution of water, calcium, and gadolinium following the model- and learning-based decomposition methods.

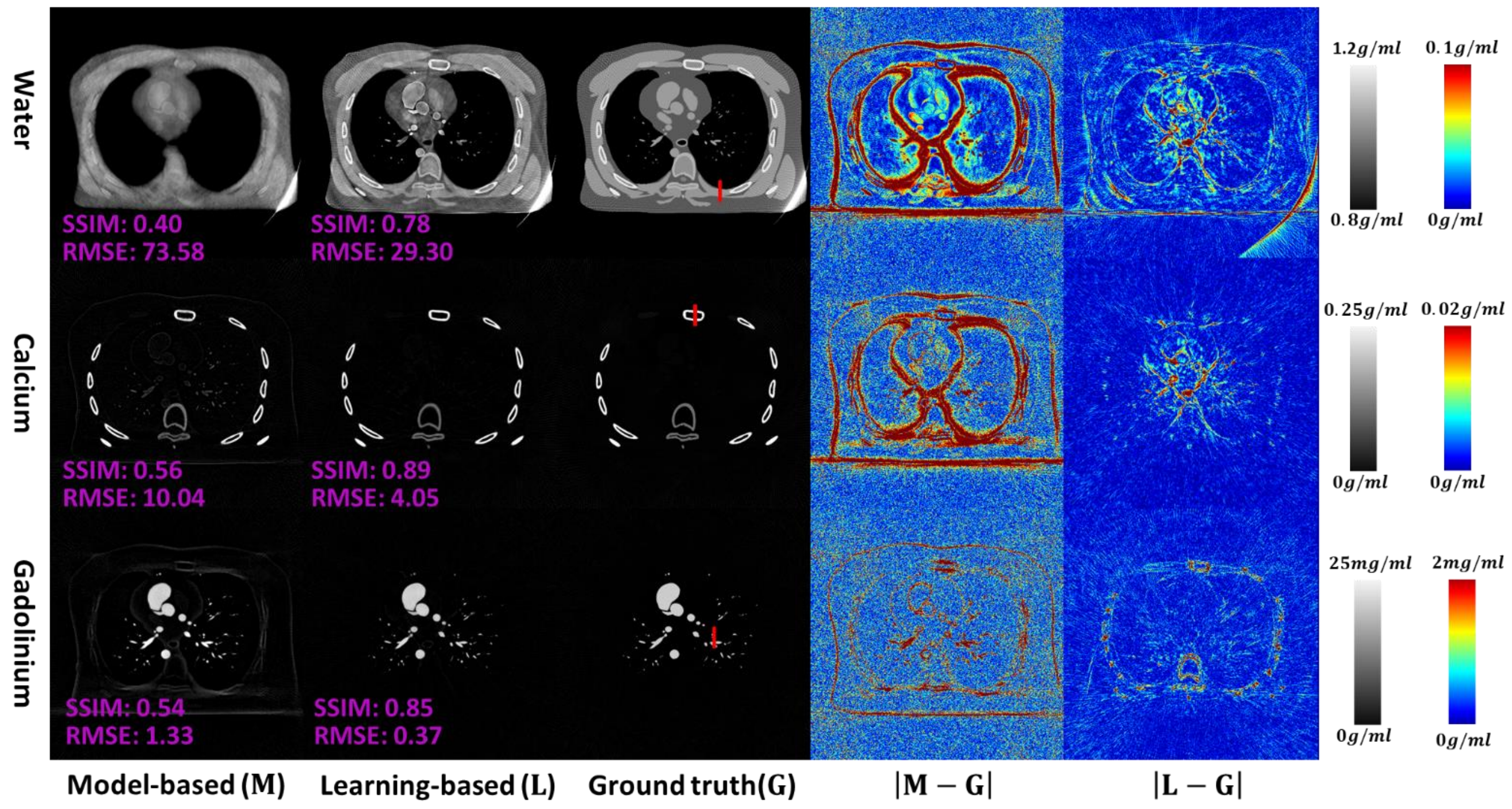

Figure 3: FDK reconstruction of material projections obtained by model-based decomposition, learning-based decomposition, and ground truth projections. The color images display the pixel-wise absolute error. RMSE unit: mg/ml.

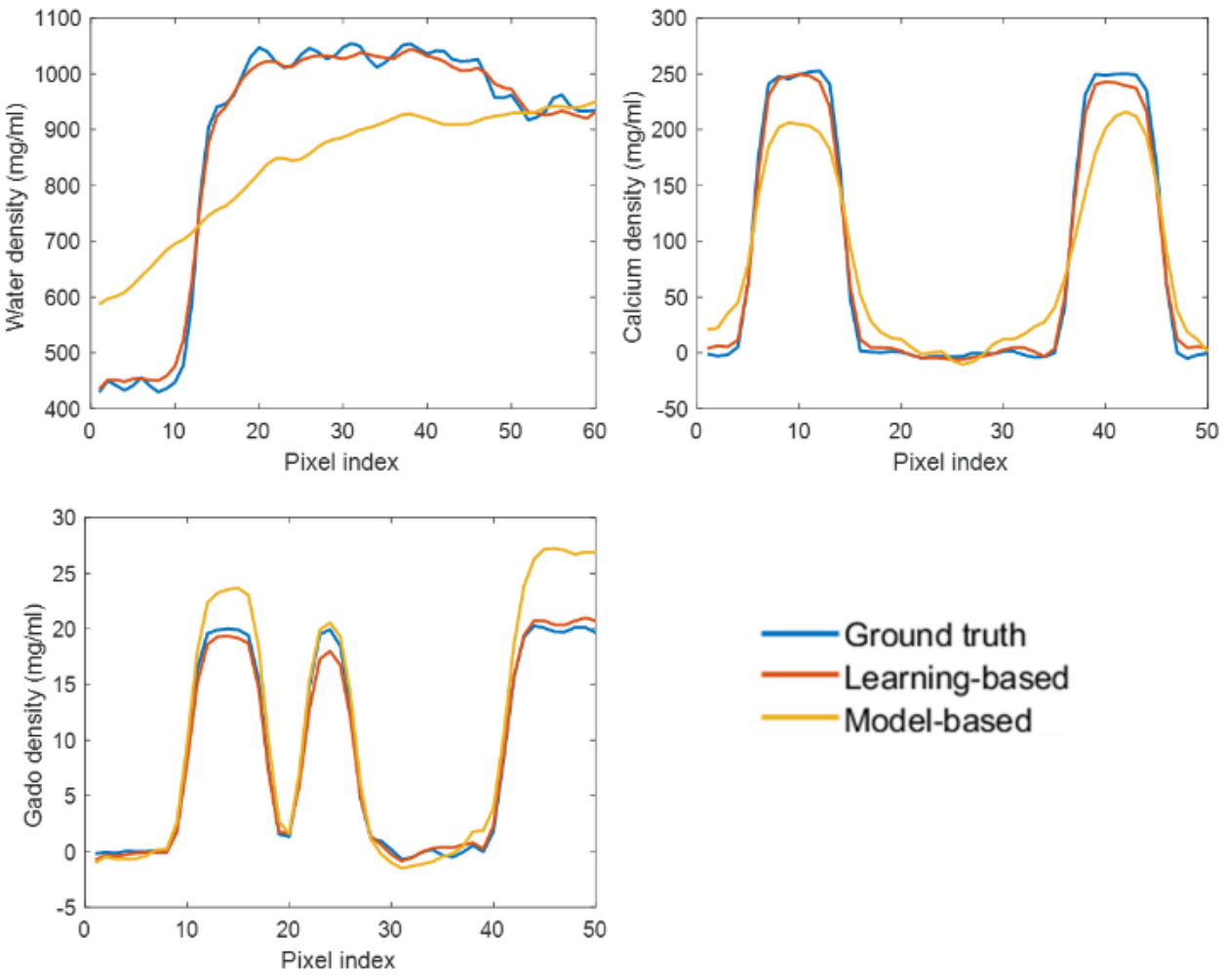

Figure 4: 1D profiles of the line segments in Fig. 3

From visual observation, the model-based method exhibits significant bias in all three material images – the water image appears overly smooth; lung parenchyma is absent in the water images but present on the calcium image; chest wall and patient boundary are present in the gadolium image. Larger errors are observed around tissue boundaries, consistent with previous observations of cross talk effects amongst material maps. Careful tuning and design of regularization can potentially mitigate such biases[13].

The learning-based method, on the other hand, is able to achieve low bias in the calcium and gadolinium images. The water image presents significant artifacts, likely from the propagation of different estimation bias from individual projections. The advantage of the learning-based method over model-based method is also reflected in the higher SSIM and lower RMSE in all three material images.

To further evaluate algorithm performance, we plotted the line profiles across anatomical structures in each material image. Consistent with visual observation, the model-based method produced blurred water and calcium images. The gadolinium image has comparable spatial resolution with the ground truth but exhibits an over-estimation of concentration by ~25%. The learning-based method shows better performance, with line profiles closely resembles those from the ground truth in both spatial resolution and density estimates.

## 4 Discussion and Conclusion

We have presented a simulation study to explore the potential of three material decomposition using a triple-layer panel. Such a system would fit within the current interventional imaging paradigm that largely uses flat-panel detectors for imaging. In our preliminary studies, we find that both model-based and deep-learning methods can provide three material estimates in a water-calcium-gadolinium decomposition problem, though imaging performance is significantly improved with the deep learning approach. We note that the deep learning approach has a significantly faster processing speed making it more appropriate for the workflow of an interventional suite.

A number of limitations of this preliminary work are noted including the focus on gadolinium contrast (as opposed to the more popular iodine contrast agent) due to the greater potential for separation from calcium; a limited investigation of regularization strategies for the model-based approach; and a limited phantom evaluation. Despite these limitations, the technology appears to hold promise and we seek to continue to address these limitations in ongoing simulation studies and physical experiments.

# An investigation on the detection task performance of deep learning-based streak artifacts reduction methods

Hojin Jung[1], Minwoo Yu[1], and Jongduk Baek[1]

[1]Department of Artificial Intelligence, College of Computing, Yonsei University, Seoul, South Korea

**Abstract** In X-ray computed tomography(CT), sparse-view CT has been proposed as a way to reduce a radiation dose. However, sparse-view CT generates streak artifacts that degrade image quality (IQ). To address this problem, various deep learning (DL)-based methods for streak artifacts reduction have been developed. To assess the performance of streak artifacts reduction methods, root mean square error (RMSE) and structural similarity index measure (SSIM) are commonly adopted. However, these two assessments cannot guarantee superiority in terms of diagnostic task performance. In this work, we performed a signal detection task using a convolutional neural network (CNN)-based ideal observer (IO) to evaluate the diagnostic IQ of streak artifacts reduction methods. We compared the performances of three DL-based streak artifacts methods with two differences: the domain where the input of CNN was defined and where the loss function was computed. Our result shows that the methods utilizing CNN in the image domain outperform the methods utilizing CNN in the sinogram domain in terms of SSIM, and RMSE. On the contrary, the methods utilizing CNN in the sinogram domain performed better IO performance of the detection task than the methods utilizing CNN in the image domain.

## 1 Introduction

To reduce radiation exposure during computed tomography (CT) scans, sparse-view sampling using a smaller number of projection views can be adopted [1]. However, when CT images are reconstructed from sparse-view sinograms via filtered back-projection (FBP), streak artifacts are generated, which degrades image quality (IQ). To address this problem, various methods such as iterative reconstruction and deep learning (DL)-based methods have been proposed [2, 3], and shown superior performances in terms of full reference image quality (FR-IQ) metrics such as root mean square error (RMSE) and structural similarity index measure (SSIM) [4]. Based on these FR-IQ results, previous studies [3, 5] concluded that DL-based methods are effective for streak artifacts reduction.

However, these FR-IQ metrics cannot guarantee the superiority of diagnostic tasks such as detection tasks and other diagnosis-related tasks, because the FR-IQ metrics focus on comparing the corrected image with original full-view images. To address this limitation of FR-IQ metrics, performances of the detection task can be employed as task-based IQ (Task-IQ) metrics, and the numerical observers are commonly used to evaluate detection task performances [6]. Among them, an ideal observer (IO) can be used to evaluate the amount of restored information related to the signal detection task. However, the calculation of IO performance is intractable in complex background images such as CT images. Therefore, convolutional neural network (CNN)-based techniques [7, 8] have been proposed to approximate IO performance.

By using CNN-based IO, Zhang et al. and Li et al. [9, 10] evaluated the IO performances of image restoration methods and have confirmed that image domain-based methods cannot improve IO performance. These results are in agreement with the data processing inequality (DPI) theorem [11] which means post-processing cannot restore the degraded information. However, previous studies are limited to image domain-based methods and did not examine sinogram domain-based methods for streak artifact reduction in sparse-view CT. Sinogram domain-based methods can reduce the additional information loss during the image reconstruction process by increasing the number of projection views of the sinogram, thus IO performance can be improved compared to sparse-view CT images.

Therefore, this study aims to compare the detection performance of image domain-based and sinogram domain-based streak artifacts reduction methods.

## 2 Method

### 2.1 Implementation of DL-based streak artifacts reduction methods

In this study, we evaluated three streak artifacts reduction methods, and the network architectures are based on the enhanced deep super-resolution network (EDSR) [12]. EDSR is a network designed for super-resolution of the natural image that ensures high performance through multiple residual connections and an upsampling module with pixel-shuffle layers [13] as shown in Figure 1. Furthermore, we designed the methods with two differences: 1) the domain in which the EDSR is applied and 2) the domain in which the loss function is calculated.

First, we evaluated the method of performing streak artifacts reduction trained by calculating the loss between the output image and the full-view image, referred to as image domain processing-image domain loss method (IPIL). Since the resolution of the input and target images are the same, the upsampling module in the EDSR is removed in IPIL.

Next, we evaluated the method of obtaining upsampled sinograms from sparse-view sinograms trained by calculating the loss between the output sinogram and the full-view sinogram, referred to as sinogram domain processing - sinogram domain loss method (SPSL). The upsampling modules of the EDSR in SPSL used 1D pixel-shuffle layers because it performs 1D upsampling along the projection views unlike the EDSR designed for super-resolution of 2D images. The output sinograms are reconstructed into images via full-view FBP to compare results with other methods.
Last, we evaluated the method of obtaining upsampled sinograms in the same way as SPSL, but the loss is calculated in the image domain after FBP. This method is referred to as sinogram domain processing - image domain loss method (SPIL). Descriptions of three DL-based streak artifacts reduction methods are illustrated in Figure 2.

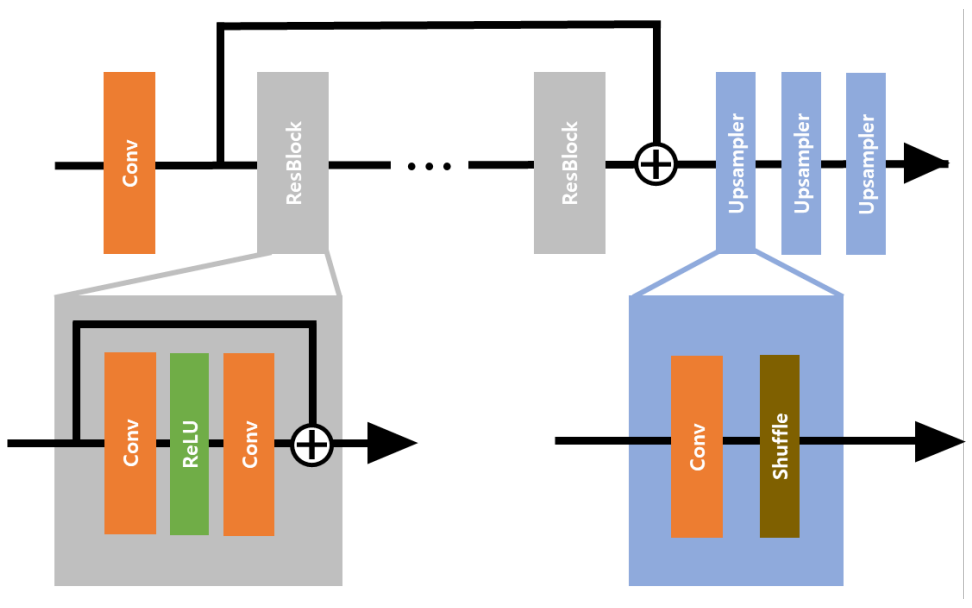

**Figure 1:** Architecture of EDSR

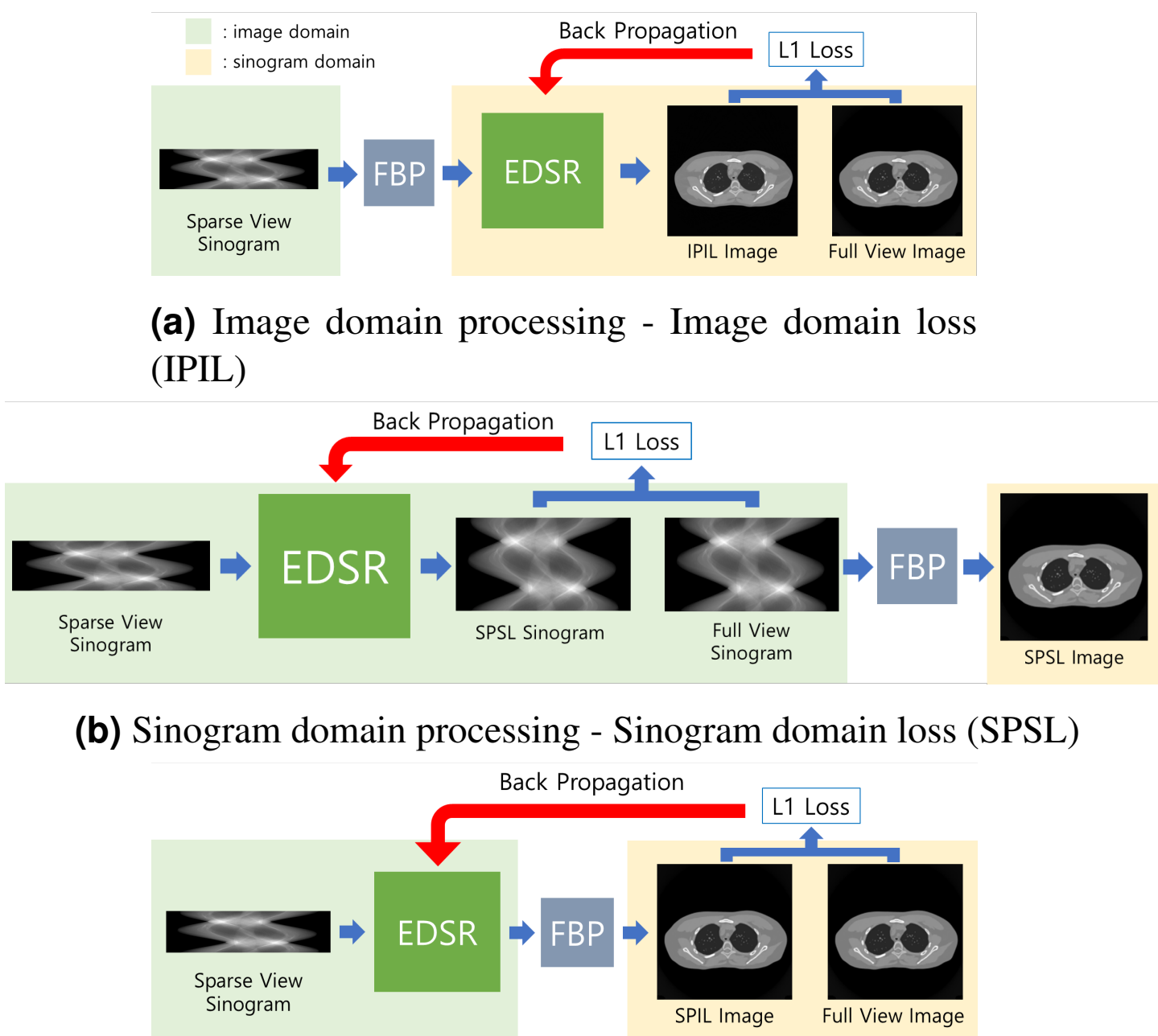

**(a)** Image domain processing - Image domain loss (IPIL)

**(b)** Sinogram domain processing - Sinogram domain loss (SPSL)

**(c)** Sinogram domain processing - Image domain loss (SPIL)

**Figure 2:** Descriptions of streak artifacts reduction methods

## 2.2 Formulation of Signal Detection Task

To evaluate the three streak artifacts reduction methods in terms of Task-IQ, we formulated a signal detection task designed as signal-known-exactly (SKE) and background-unknown conditions. A signal detection task classifies whether a given image satisfies signal absent hypothesis $H_0$ or signal present hypothesis $H_1$. In detail, the signal detection task can be expressed for a given image $g$ as follows:

$$\begin{aligned} H_0 &: g = \mathscr{H}(FBP(\mathscr{N}(s_b))) \\ H_1 &: g = \mathscr{H}(FBP(\mathscr{N}(s_b + s_s))) \end{aligned} \tag{1}$$

where $s_b$ denotes sinogram of background, and $s_s$ denotes sinogram of signal. $\mathscr{H}$ and $\mathscr{N}$ denote image domain processing and sinogram domain processing operator, respectively. Note that the operator $\mathscr{H}$ and $\mathscr{N}$ are dependent on the specific method used for streak artifacts reduction. For instance, in the case of SPIL, $\mathscr{H}$ and $\mathscr{N}$ represent the identity operator and the EDSR-based sinogram domain upsampler, respectively.
We trained a CNN-based IO to classify the signal present case and signal absent case. As a CNN-based IO, we used a CNN architecture that contains residual connections [14], and it is denoted as ResNet-IO.
The architecture of ResNet-IO is shown in Figure 3. To improve the performance of ResNet-IO closer to IO, determining the optimal depth to avoid underfitting or overfitting is important. Therefore, we trained the network using different numbers of convolution layers with residual connections, including 4, 6, 8, 10, and 12 layers and selected the network with the best validation loss.

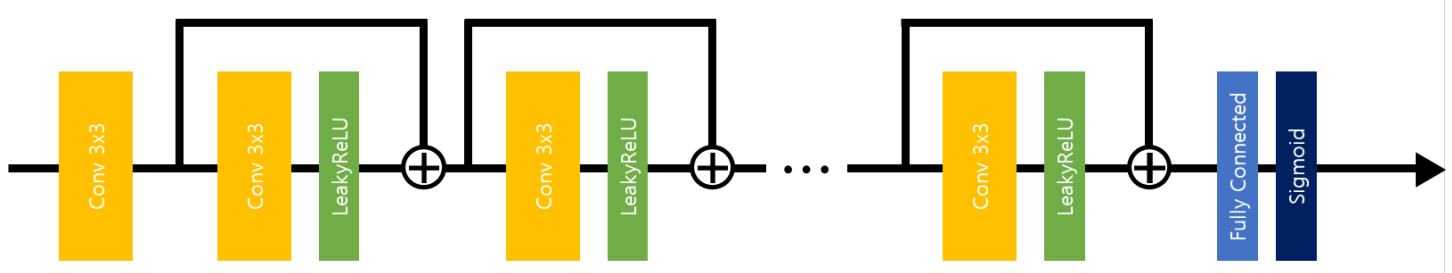

**Figure 3:** Architecture of ResNet-IO

## 2.3 Dataset preparation and training details

We used the "2016 NIH-AAPM-Mayo Clinic Low-Dose CT Grand Challenge" dataset from the Mayo Clinic. This dataset consists of normal-dose data and low-dose data, and we used datasets of 10 patients in normal-dose data to generate paired full-view and sparse-view datasets. We generated full-view sinogram data by using 512-view projections through Siddon's algorithm [15] on a fan-beam geometry system. After that, sparse-view sinograms were generated by down-sampling full-view sinograms into 64-view projections, and sparse-view and full-view CT images were reconstructed through FBP. The fan-beam geometry parameters are summarized in Table 1.
For training streak artifacts reduction networks, we used paired datasets from 8 patients: 6 patients for the training, 1 patient for the validation, and 1 patient for the test. We used the L1 loss function with the initial learning rate of $1 \times 10^{-4}$. In addition, we used Adam optimizer [16], and $\beta_1$ and $\beta_2$ were set as 0.9 and 0.999.
For the dataset of the signal detection task, 2 patients' data were used. We generated Gaussian signals for the signal detection task and varied the size and intensity of the signals: 1) fixing the peak signal intensity to 80 HU and changing the standard deviation of Gaussian signal as $\{1.5, 2, 2.5, 3\}$ pixels, 2) fixing the standard deviation of Gaussian signal as 2 pixels and changing the peak signal intensity within $\{40, 60, 80, 100\}$ HU. Then, 512-view and 64-view sinograms of Gaussian signals were obtained and inserted into the full-view and sparse-view background sinograms, respectively. These signal-inserted sinograms were reconstructed into images by each reconstruction method: 64-view FBP, 512-view FBP, IPIL, SPSL, and SPIL. The reconstructed CT images were cropped into patches of $64 \times 64$ pixels where the signal is located in the center of each patch. We randomly cropped a total of 66000 patches for each method to train a ResNet-IO model, excluding the part where the background of the patch is air. From 66000 patches, 50000 patches were used as training data, 6000 patches as validation data, and the remaining 10000 as test data. We used the binary cross entropy (BCE) loss function with the initial learning rate of $3 \times 10^{-5}$. In addition, we used Adam optimizer [16], and $\beta_1$ and $\beta_2$ were set as 0.9 and 0.999.

**Table 1:** Parameters of fan-beam CT geometry

| | |
|---|---|
| Distance between source and detector center (mm) | 1085.6 |
| Distance between source and isocenter (mm) | 595 |
| Number of detector elements | 512 |
| size of each detector element (mm) | 1.6 |
| Reconstructed Image resolution (pixels) | 0.668 |
| The pixel size of the reconstructed image | 512×512 |

### 2.4 Image quality assessments

For FR-IQ metrics, we used SSIM and RMSE. For the Task-IQ metric, we calculated the $P_c$ value from the result of ResNet-IO. The $P_c$ value is a measure of the accuracy of predicted values. By comparing the $P_c$ values, it is possible to compare the amount of restored information related to the detection task of the three streak artifacts reduction methods relatively.

## 3 Result

### 3.1 FR-IQ evaluation of streak artifacts reduction methods

we evaluated streak artifacts reduction methods in terms of FR-IQ metrics. Table 2 presents the average SSIM and PSNR values from the test dataset. Three DL-based streak artifacts reduction methods show improved values compared to the 64-view FBP images in terms of FR-IQ metrics. Among the three DL-based methods, IPIL shows the highest performance, and SPIL shows better values than SPSL.These results confirm that utilizing CNN in the image domain is superior to utilizing CNN in the sinogram domain in terms

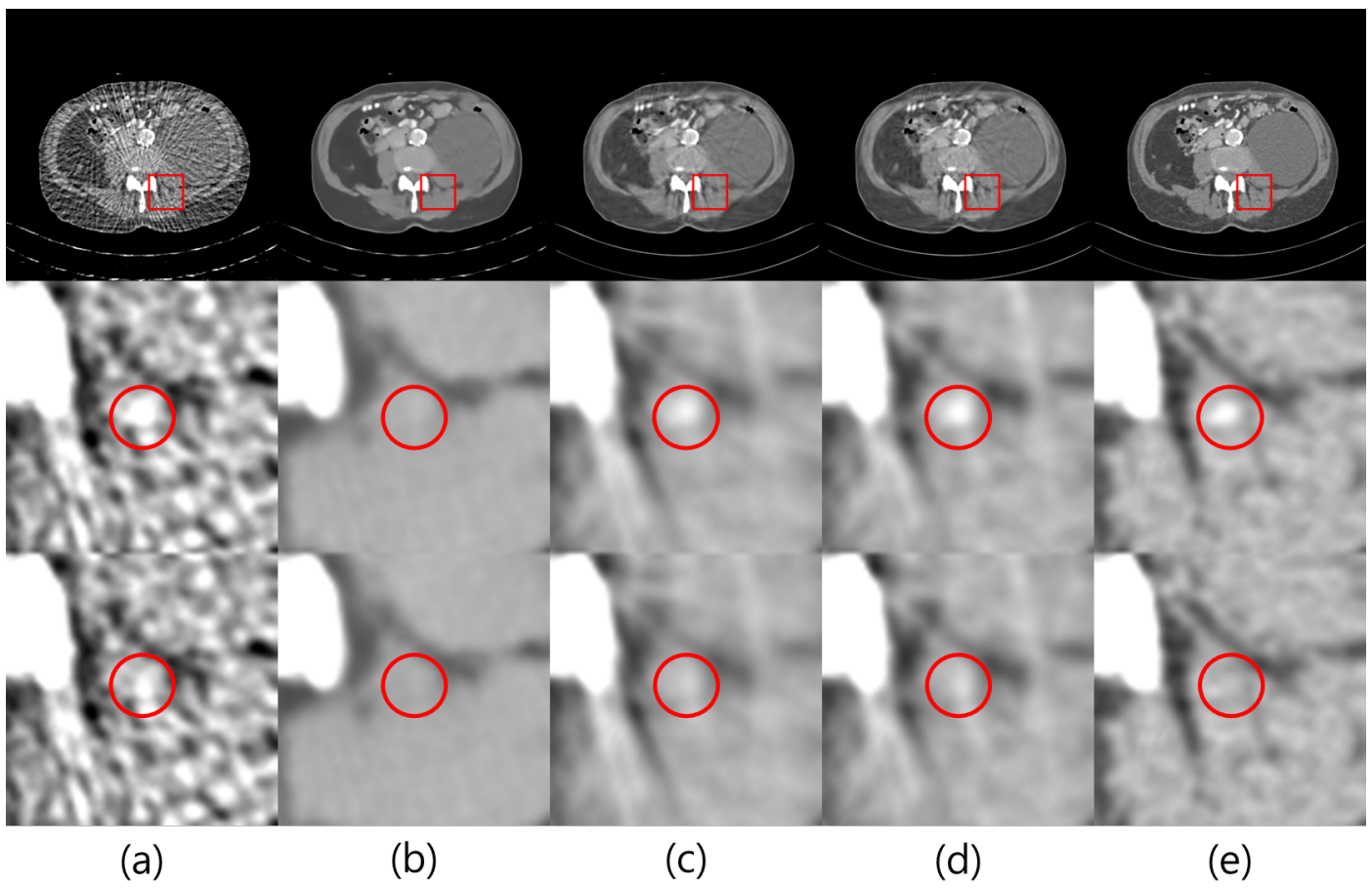

**Figure 4:** Application results of (a) 64-view FBP, (b) IPIL, (c) SPSL, (d) SPIL, and (e) 512-view FBP, respectively. $2^{nd}$ and $3^{rd}$ rows show enlarged patches emphasized in $1^{st}$ row for two cases, each signal present and signal absent. The display windows are $[-250\ 300]HU$ for $1^{st}$ row and $[-200\ 150]HU$ for $2^{nd}$ and $3^{rd}$ rows.

**Table 2:** Average SSIM and RMSE calculated over all 533 test CT images.

| | 64 view FBP | IPIL | SPSL | SPIL |
|---|---|---|---|---|
| SSIM | 0.367±0.0433 | 0.8753±0.0233 | 0.8715±0.024 | 0.8739±0.0239 |
| RMSE(HU) | 103.9112±13.1631 | 21.5347±1.6883 | 22.3133±2.4603 | 22.0533±2.39 |

of FR-IQ metrics.

### 3.2 Task-IQ evaluation of streak artifacts reduction networks

Task-IQ evaluation was performed using the signal detection task. Figure 4 shows the example patches of the signal present and signal absent case. In 64-view FBP images, it is difficult to classify the signal present case and signal absent case because of streak artifacts. Similarly, in the case of IPIL, it is difficult to classify the signal present case and signal absent case even though streak artifacts are removed. In contrast, the signal present and signal absent cases can be classified clearly in the case of SPSL and SPIL.
These observations are also confirmed in the evaluation using $P_c$ values shown in Figure 5. The SPSL and SPIL methods show significantly higher $P_c$ values compared to the IPIL method. The results of the $P_c$ evaluation demonstrate that the application of CNN in the sinogram domain is more effective for improving IO performance enhancement of signal detection task than the application of CNN in the image domain.

## 4 Conclusion

In this study, we compared the detection task performance of three DL-based streak artifacts methods that have differences in the domain where the CNN is utilized and the loss function is computed. We employed the ResNet-IO performance as a Task-IQ metric to evaluate ability to restore information related to the detection task. Our study showed that SPSL and SPIL significantly outperformed IPIL in terms of the ResNet-IO performance of the signal detection task. This investigation suggests that sinogram domain processing is one of the key factors for improving the detection task performance in the field of sparse-view CT.

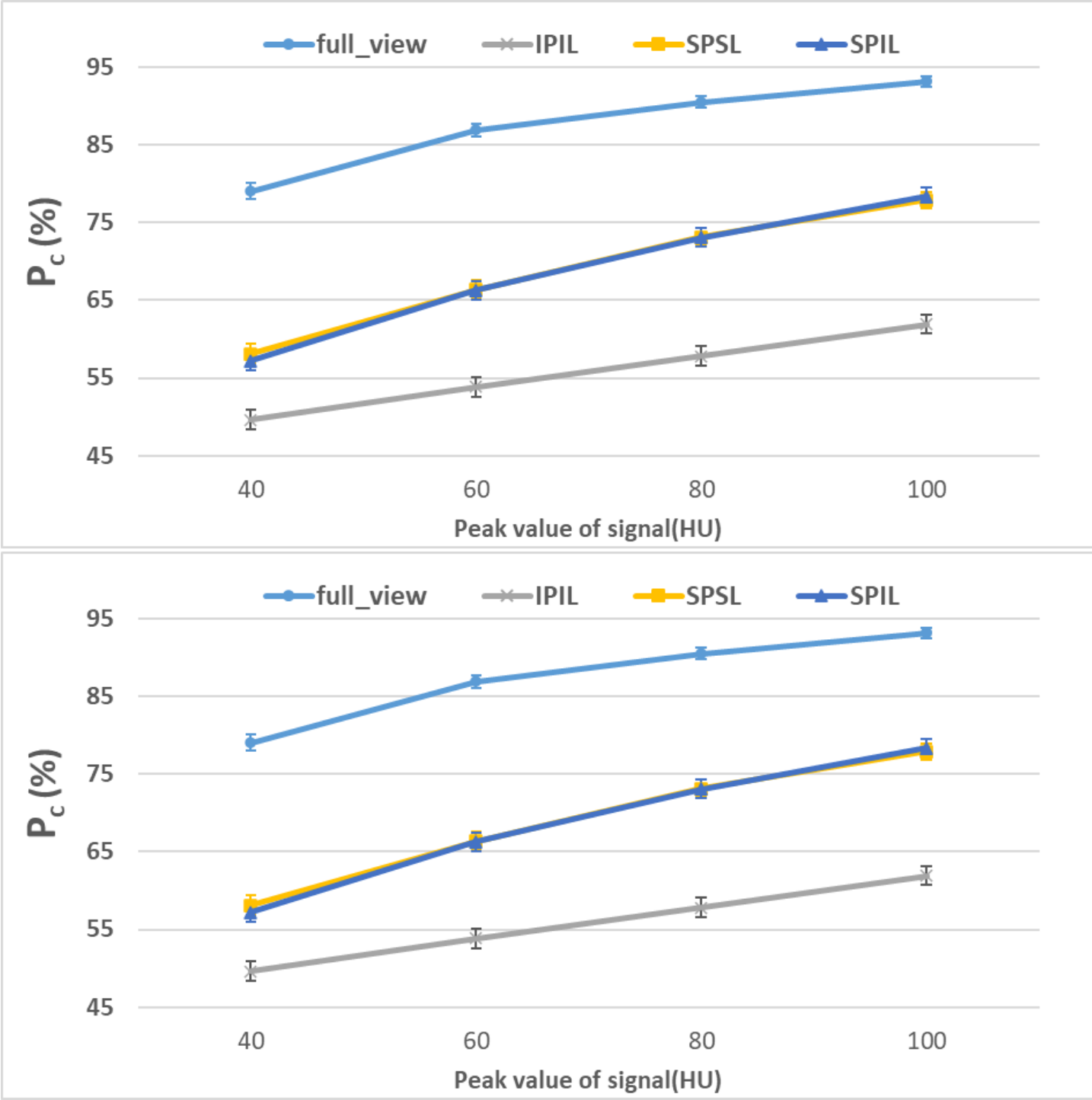

**Figure 5:** Plots of $P_c$ values and its 95% confident intervals: (a) Peak value of the signal is changed, (b) Size of the signal is changed.

## 5 Acknowledgement

This work was supported in part by the National Research Foundation of Korea (NRF) Grant funded by the Korean Government through the Ministry of Science and ICT (MSIT) under Grant RS-2022-00144336, and Institute of Information & Communications Technology Planning & Evaluation (IITP) Grant funded by MSIT (No. 2020-0-01361, Artificial Intelligence Graduate School Program (Yonsei University)). This research was financially supported by the Ministry of Trade, Industry, and Energy (MOTIE), Korea, under the "Project for Research and Development with Middle Markets Enterprises and DNA (Data, Network, AI) Universities" (Technology of intraoperative C-arm CT using low dose X-ray source based on AI) (P0021346) supervised by the Korea Institute for Advancement of Technology (KIAT).

# Monte Carlo-free Deep Scatter Estimation (DSE) with a Linear Boltzmann Transport Solver for Image Guidance in Radiation Therapy

Fabian Jäger[1,2], Joscha Maier[1], Pascal Paysan[3], Michał Walczak[3], and Marc Kachelrieß[1,4]

[1]Division of X-Ray Imaging and Computed Tomography, German Cancer Research Center, Heidelberg, Germany
[2]Department of Physics and Astronomy, Ruprecht-Karls-University Heidelberg, Heidelberg, Germany
[3]Varian Medical Systems Imaging Lab, GmbH, Baden-Dättwil, Switzerland
[4]Medical Faculty, Ruprecht-Karls-University Heidelberg, Heidelberg, Germany

**Abstract** Scatter correction methods currently in clinical use often require first-pass reconstruction and subjected to artifacts and truncation. We use the deep scatter estimation (DSE) introduced in [1] to predict the scatter signal of an on-board CBCT system in the projection domain. To demonstrate that DSE can be trained with scatter estimates other than those from Monte Carlo simulations. The projections and the scatter distribution for training was calculated with a fast-linear Boltzmann transport equation (LBTE) solver. The simulated CBCT system has an anti-scatter grid, a bowtie filter, a titanium prefilter as well as a shifted detector. Training and validation of the neural network used simulated data, while testing was done on simulations as well as phantom measurements. We benchmarked DSE against a kernel-based scatter correction and the LBTE which was used to generate the training data. For the simulated scans, DSE outperforms the kernel and LBTE methods using the mean absolute error compared to a scatter-free reconstruction. For the measurements no scatter free scan is available, again, the DSE shows less visible scatter artifacts.

## 1 Introduction

With cancer being one of the leading causes for death worldwide [2], the effort to extend and improve cancer treatment is significant. One of the principal treatments is radiation therapy, which uses ionizing radiation to deactivate cancer cells. The type of particle used depends on availability, tumor type, and location, but for all particles, accurate representation of the patient's anatomy is required to create a good treatment plan. Better imaging results in less damage to healthy tissue. Therefore, imaging techniques are an important part of radiation therapy. Today's treatment systems, such as Varian Medical Systems' Ethos™, feature an on-board CBCT in the gantry of a linear accelerator (LINAC). This gives the possibility to adapt the treatment plan and help to position the patient for each treatment session.
Current clinical practice in radiotherapy is to acquire a planning CT of the region of interest (ROI). For this, clinical CTs are used, due to the limited image quality of on-board CBCTs. Inferior image quality is caused by artifacts such as scatter or beam hardening, which always occur when the physics of CT data acquisition are not modeled appropriately. In the case of scattering, the reconstruction algorithms assume that the detected photons travel in a straight line through the scanned object, when in fact photons passing through the tissue are scattered, reducing their energy and changing their direction. Thus, the scattered photons cause a secondary signal at the detector, distorting the actual signal of the unscattered photons. In the reconstruction, the scatter is visible as streaks, dark areas and cupping. This results in inaccurate CT values, which are important for calculating the treatment plan.
In conjunction with scatter suppression methods, such as anti-scatter grids or collimators [3], various scatter estimation methods can be used to correct for the artifacts. The gold standard among these methods is Monte Carlo (MC) simulation, which has the disadvantage of very high computation time. For a clinical application, faster methods such as a deterministic solver of the linear Boltzmann transport equation (LBTE) [4, 5] or so-called kernel-based approaches [6] are used. More recently, deep learning (DL) methods such as neural networks are investigated [1, 7, 8]. Most DL-based approaches have the benefit to estimate the scatter in the projection domain and do not rely on a prior reconstruction.
In this work we investigate, the possibility to train DSE with scatter estimated by the LBTE solver Acuros® CTS, which was introduced in [4, 5]. However, it has the downside of relying on a prior reconstruction. Thus, training DSE with scatter estimates can give better results in edge cases such as truncation, for which no complete prior can be reconstructed. In addition, if DSE shows promising results, it should be possible to adapt it to other systems rather easily when they already use Acuros as scatter correction method.

## 2 Materials and Methods

### 2.1 Scatter Estimation

#### Deep Scatter Estimation (DSE)

The deep scatter estimation (DSE) was introduced in reference [1] for scatter estimation in industrial CT, and it was shown that it can be trained by MC simulations for CBCTs with different tube voltages, noise levels, and anatomical regions [7]. More recently, it also showed promising results for the cross and forward scatter estimation in clinical CTs [9]. DSE is a deep convolutional neural network based on the U-net [10]. The encoding (downward) path consists of seven encoding blocks, each of which has three convolutional layers with a $3 \times 3$ kernel followed by a rectified linear unit (ReLU) activation. For pooling, a $3 \times 3$ convolution with a stride of $2 \times 2$ was used and the number of feature channels is doubled, starting with 32 going up to 2048. DSE was implemented using PyTorch Lightning 1.7.7 and PyTorch 1.12.1. Training was performed on an Nvidia RTX A5000 with an AdamW [11] optimizer, an improved version of the Adam [12] optimizer. The weights were initialized using a uniform initialization as proposed in reference [13] and the biases with zeroes. The loss function was the mean absolute percentage error (MAPE) scaled to the scatter-to-primary ratio, the so called SPMAPE as it was introduced in reference [9]:

$$\text{SPMAPE} = \frac{1}{N}\sum\left|\frac{S_{\text{BT}} - S_{\text{DSE}}}{S_{\text{BT}}}\frac{S_{\text{BT}}}{I_{\text{primary, BT}}}\right| \tag{1}$$

$$= \frac{1}{N}\sum\left|\frac{S_{\text{BT}} - S_{\text{DSE}}}{I_{\text{primary, BT}}}\right|, \tag{2}$$

where $S$ is the scatter intensity estimated by a Boltzmann transport (BT) equation solver or DSE and $I_{\text{primary, BT}}$ the primary intensity as calculated by the BT solver. The initial learning rate was set to $5 \cdot 10^{-5}$ and a plateau scheduler was used to decrease the learning rate by half if the validation loss did not decrease for 20 epochs. In addition, an early stopping was used if the validation loss did not decrease for 35 epochs, otherwise the training stopped after 200 epochs. Scatter distributions are known to be of low frequency. Therefore, DSE was not applied to the full projection, but to a downsampled version with $320 \times 320$ pixels. As input, the so-called *pep* function ([7]) was used, which is the scaled projection $p$ and given by:

$$pep = pe^{-p}, \tag{3}$$

with $p$ as:

$$p = -\ln\left(\frac{I_{\text{primary, BT}} + S_{\text{BT}}}{I_0}\right), \tag{4}$$

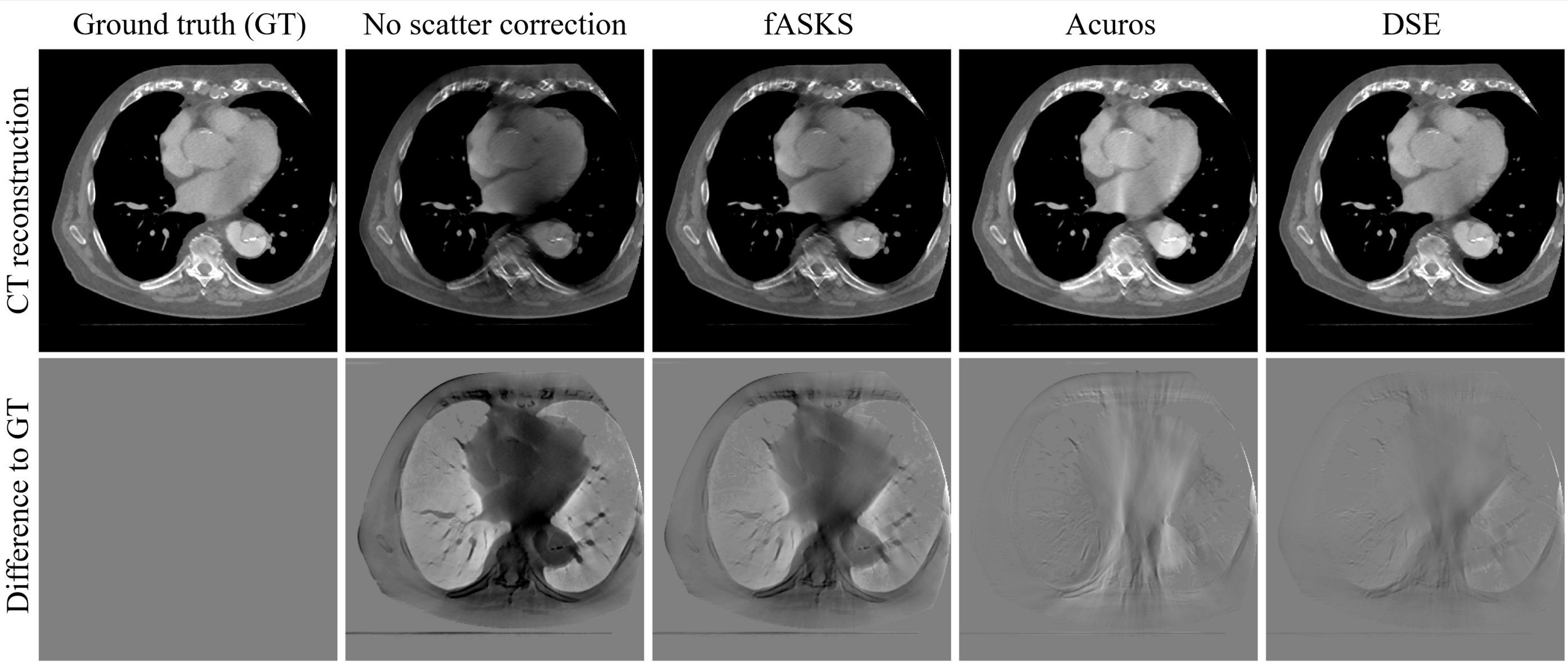

**Figure 1:** CT reconstructions of a simulated and truncated patient scan with and without scatter correction. The ground truth reconstruction used the scatter free forward projections. $C = 0$ HU, $W = 1000$ HU

### Linear Boltzmann Transport

The work process of the deterministic LBTE solver Acuros can be summarized into three steps. First, the photons are traced from the source to all voxels, in a second step the scattering and absorption is calculated in an iterative way, starting with the distribution of unscattered photons. From there, the scatter flux from each voxel to every other voxel can be calculated giving the flux of first order scattered photons at each voxel. This is repeated until a convergence criteria is reached. In the last step, the total scattering flux from all voxels is traced to the detector pixels [4, 5].

## 2.2 Dataset

### Geometry

The scanner geometry used in the simulation corresponds to the Ethos™ by Varian Medical Systems. It includes a titanium pre-filter and bowtie filter, a one-dimensional anti scatter grid, a tube voltage of 125 kV and has a detector shifted by 175 mm to extend the field of measurement. The detector has $1280 \times 1280$ pixels with a $1 \times 4$ binning, resulting in a projection size of $1280 \times 320$.

### Simulation

For training and validation of the proposed DSE several thousand projections were simulated with the LBTE solver Acuros® CTS. For the simulation, 50 clinical CTs from a Siemens Somatom Force (tube voltage of 70 kV) were used as patient prior. Since the scans were from neck to abdomen, several simulations were done at different z position. To simulate realistic projections, the z positions were chosen in a way, that the primary forward projection did not contain the upper or lower edge of the prior CT. To train the network on as many different anatomies as possible, a circular scan was simulated every 34.8 cm. This corresponds to the maximal possible reconstruction size in z. As explained above, the patient position has an influence on the scatter distribution, therefore different x and y positions of the patient were simulated for every z position by randomly shifting the patient 5 times between ±5 cm. For each of these positions, 47 projections were simulated uniformly distributed over 360°. In total, 32,195 projections were simulated and 40 patients were used for training while 10 were used for validation. For testing, several patients from the validation set were fully simulated (840 projections per scan) and iterative reconstructed with a slice thickness of one millimeter. For DSE every 25$^{\text{th}}$ projection was scatter corrected and interpolated for the projections in between. The estimated scatter intensity was clipped to 95% of the uncorrected projection value and smoothed with a Gaussian with a sigma of $\sigma_{x,y} = 5$ px.

#### 2.2.1 Measurements

In addition for testing, a Multipurpose Chest Phantom N1 "LUNGMAN" by Kyoto Kagaku® was measured with different couch positions. As mentioned above, the simulation mimic the Ethos by Varian Medical Systems, thus the measurements were done on said scanner. One scan consisted out of 896 projections and was iterative reconstructed with a slice thickness of one mm. The clipping and the smoothing was applied as described for the full simulated scans.

## 3 Results

### 3.1 Simulation Results

The DSE scatter corrected reconstructions are compared to the ground truth, the scatter free reconstruction of the simulated forward projection, and the different scatter correction methods applicable at the CBCT system, the LBTE solver Acuros and a kernel based approach called fast adaptive scatter kernel superposition (fASKS). Table 1 shows the results for the different scatter correction methods. As introduced above, the scatter correction with the LBTE solver relies on a prior reconstruction, which was scatter corrected with fASKS for better results. On truncated scans as seen in Figure 1 the deviation to the ground truth, the reconstruction of the scatter free forward projections, for the DSE appears to be smaller than Acuros. As mentioned earlier, this is to be expected since LBTE methods rely on a first-pass reconstruction that is not completely artifact-free or even truncated. This leads to errors in the estimated scatter.

### 3.2 Phantom Measurement

For the phantom measurements, DSE and fASKS were compared to Acuros, since it is the Varian preferred standard and there is no scatter free ground truth. Again DSE performs better than fASKS (2), and as it can be seen in Figure 2 the dark streaks from the scatter of the rib bone, seems less pronounced for the DSE corrected scans compared to the other methods.

| Correction Method | MAE |
|---|---|
| Kernel | 52.3 HU |
| BT | 11.1 HU |
| DSE | 6.7 HU |
| Uncorrected | 104.3 HU |

**Table 1:** Mean absolute error (MAE) at the patient between the reconstruction of the simulated primary projection without scatter and the reconstruction of the scatter-corrected projection for the different methods and without any correction.

| Correction Method | MAE |
|---|---|
| Kernel−BT | 26.1 HU |
| Uncorrected−BT | 73.3 HU |
| DSE−BT | 20.6 HU |

**Table 2:** Mean absolute error (MAE) at the patient of the reconstructed measurements of the Lungman phantom between the projections corrected with the linear Boltzmann equation solver (BT) and the projections corrected with DSE or a kernel based approach.

## 4 Discussion

DSE shows good results for the simulated scans. The mean absolute error to the ground truth, the scatter free reconstruction, is smaller for DSE than for the other two correction methods, kernel based and BT solver. The patient position relative to the rotation center appears to be important for the scatter estimation, highlighted by the fact, that the performance of DSE improves if additional input about the patient position are supplied for the scatter estimation. On measured data, no slit scan was available, therefore the BT corrected reconstruction was used for the comparison, since it is the baseline scatter correction by Varian on the used scanner. Again, DSE outperforms the Kernel based approach. In addition, the dark streaks caused by the scatter of the rib bone, seems less pronounced in the DSE reconstructed image, compared to BT corrected image.

## 5 Conclusions

DSE can be trained on scatter estimations of a deterministic LBTE solver and shows good performance on real scans. Supplying the network with additional information about the patient position increases the performance.

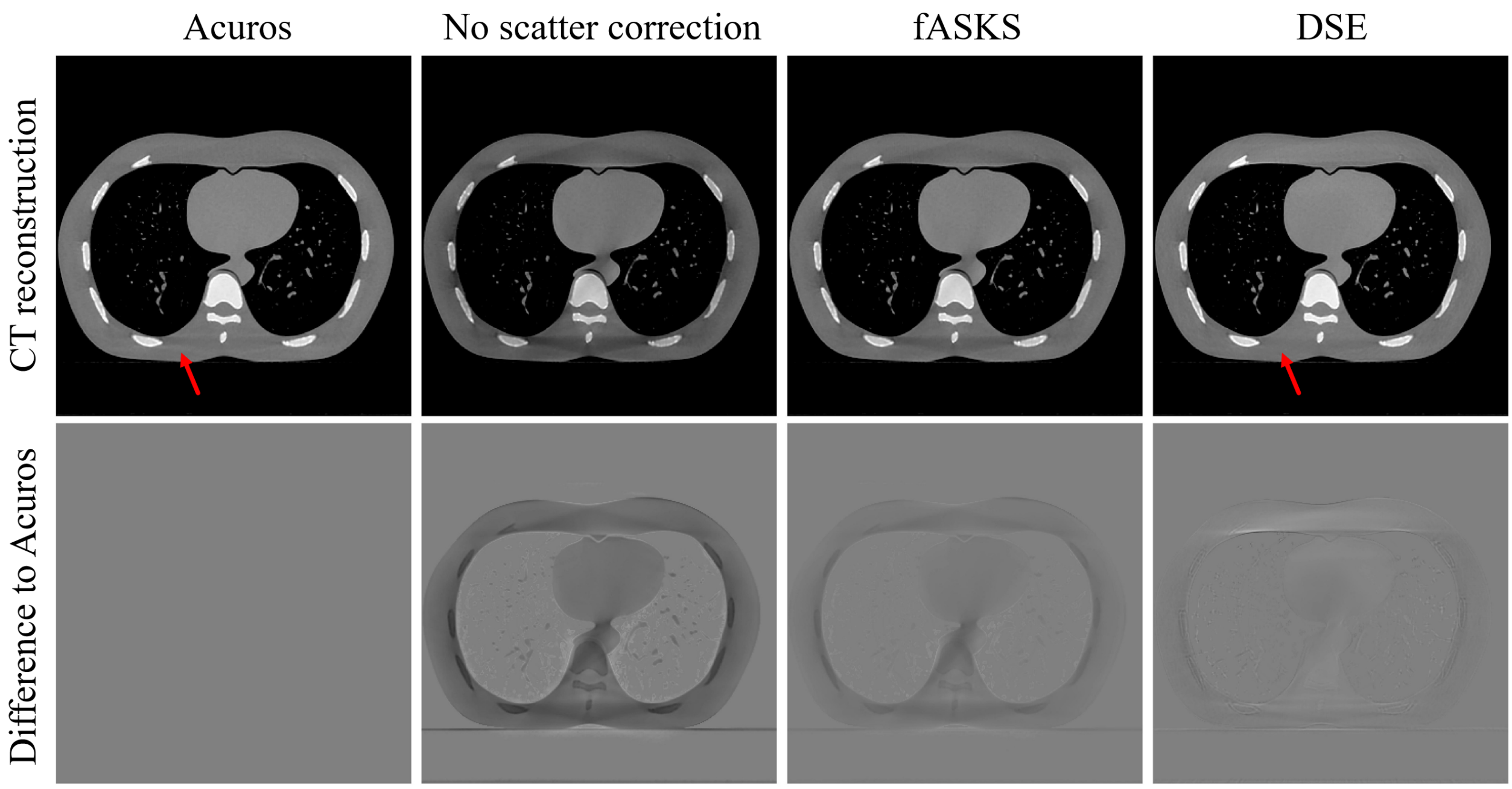

**Figure 2:** CT reconstructions of a Lungman measurement with and without scatter correction. The difference to Acuros is chosen, because it is the baseline scatter correction for the Ethos system. The red arrow show a dark streak which appears to be less pronounced in the DSE corrected reconstruction, compared to the other methods. $C = 0$ HU, $W = 1000$ HU

## 6 Acknowledgements

Part of the code was provided by RayConStruct® GmbH, Nürnberg, Germany. The project was funded by Varian Medical Systems.

# Towards Deep-Learning Partial Volume Correction for SPECT

Théo Kaprelian[1], Ane Etxebeste[1], and David Sarrut[1]

[1]Université de Lyon, CREATIS; CNRS UMR5220; Inserm U1044; INSA-Lyon; Université Lyon 1; Centre Léon Bérard, France.

**Abstract** Partial Volume Effect impacts the spatial resolution of SPECT images. We investigated the feasibility of a deep learning-based Partial Volume Correction method (PVCNet) that compensates for the effect of collimator blurring on 2D projections, before reconstruction. A large dataset containing 600,000 pairs of synthetic projections was generated and used to train two consecutive UNets (one for denoising, one for PVC). Scatter and attenuation were not yet considered in the database. Our proposed PVCNet method achieves 12.8% NMAE reduction compared to conventional Resolution Modeling on the IEC phantom but Recovery Coefficients were not always better for smallest spheres.

## 1 Introduction

Single-Photon Emission Computed Tomography (SPECT) images are impacted by several physical effects that need to be compensated in order to achieve a reasonable image quality [1]: photon attenuation, photon scatter and Partial Volume Effect (PVE). Several works have been able to effectively reduce the impact of the three effects [2–4], but PVE still remains the main limiting factor, leading to inaccurate quantification of the radioactive tracer uptake [5]. PVE is defined as the apparent underestimation of activity in an object of interest due to limited spatial resolution. It is particularly an issue for objects whose size is smaller than the system's resolution volume [6], approximately defined as twice the Full-Width at Half Maximum (FWHM) of the Point Spread Function (PSF) obtained by imaging a point source located at the center of the Field Of View (FOV). Several elements contribute to degrade the spatial resolution: detector and electronics response, collimator septal penetration, collimator scatter and collimator geometric response. The main one is the geometrical response of the collimator that depends on the source-to-collimator distance and the characteristic of the collimator such as hole diameter, septal thickness, length and material. Typical values for the FWHM at 10 cm from the collimator front surface range between 5-15 mm according to the collimator type.

A widely used Partial Volume Correction (PVC) is the Resolution Modeling (RM) method [7] applied in the system's matrix used for forward and back-projections of the Ordered Subset Expectation Maximisation (OSEM) reconstruction algorithm. This method has good noise reduction properties and theoretically converges to the true activity distribution but the resolution gain achieved in practice is limited because of the loss of high frequency information leading to Gibbs artefacts as the number of iteration increases [4]. For this reason, a regularization can be applied, e.g. [8], resulting in a smoother image and requiring additional parameter tuning.

Other PVC techniques [4] include image deconvolution [9] or region-based correction [10]. A drawback of deconvolution methods is that they tend to amplify noise. Region-based corrections rely on a segmentation mask of Regions of Interest (ROI) which may not be easy to define.

Recently, Deep Learning methods have shown promising results in various tasks in nuclear medicine [11]. In SPECT, recent works showed that some neural networks architectures were able to perform scatter correction [12], image reconstruction [13] or projection interpolation [14]. However, to our knowledge, only very few works investigated deep learning-based PVC, e.g. in [15, 16] the net was trained with small datasets and with ground truth images obtained with conventional PVC.

In this work, we propose a deep learning framework trained to compensate the effect of the PSF due to the collimator on the 2D projections, before 3D reconstruction. A large training dataset is first generated by simulation and contains 600,000 pairs of corresponding projections with and without PVE+noise. Our PVC networks are two consecutive UNets henceforth denoted as PVCNet.

## 2 Materials and Methods

### 2.1 Database

We generated a large dataset of simulated pairs of corresponding *input* and *target* 2D SPECT projections. Input projections are the realistic ones with PVE and noise, while target ones are artefact-free.

We first created 3D sources of $^{99m}$Tc made of a large elliptic cylinder background with variable axis size (90-260 mm) with several hot sources (between 1 and 8) of ellipsoidal shapes (8-128 mm axis) randomly oriented and located within the background. Hot source to background activity ratios between 1/1000 and 1/8 were considered. A total of 5 000 3D voxelized ($256^3$ voxels of 2 mm size) activity sources were randomly generated. Then, 2D projections were obtained by forward-projecting each one of the 5 000 sources with ray-tracing using RTK [17] in two different ways: once without resolution modeling ($P_{\mathrm{noPVE}}$ projection) and once with resolution modeling ($P_{\mathrm{PVE}}$ projection). Resolution modeling was performed during forward-projection operator by applying depth-dependant Gaussian convolutional kernel [7] whose parameters were derived from the dimensions of the Siemens-Intevo LEHR collimator following the analytical analysis provided by [6]. We obtain $\mathrm{FWHM}(d) = 0.048 + 1.11d$ where d is the distance from the source to the collimator front face. For $P_{noPVE}$, the same operator was applied but with $\mathrm{FWHM}(d) = 0$, so that the simulated projection is the one that would have been obtained without collimator blurring. $P_{noPVE}$ will thus serve as target projections. Each projection contains $256^2$ pixels with a size of $2.3976^2$ mm. Sources were simulated in air to avoid attenuation and scattering, and the 140 keV photo-peak window of $^{99m}$Tc was considered.
Poisson noise was applied to $P_{PVE}$ to roughly mimic the data detection process. The resulting projection was denoted $P_{PVE,noisy}$. For each of the 5000 sources, we applied the same projection process for 120 evenly distributed angles between 0° and 360°, resulting in 120 triplets ($P^i_{PVE,noisy}$, $P^i_{PVE}$, $P^i_{noPVE}$), for $i = 1, ..., 120$. The source distributions were randomly scaled so that the total number of counts in each projection was comprised between 5,000 and 500,000 such as in realistic clinical applications.

## 2.2 Networks and training

The database described in the previous section was employed to train simultaneously two neural networks: a *Denoiser* and a *PVC* network. The *Denoiser* network was trained to take $P_{PVE,noisy}$ as inputs and to output projections close to the corresponding unnoisy $P_{PVE}$ projections. Input and output of the *Denoiser* have the same number of channels. The *PVC* network then takes as input the output of the *Denoiser* and is trained with $P_{noPVE}$ as target to perform PVC with one projection angle as output. The idea behind this is that since the database generation is completely analytical, we have access to useful intermediate information that can be used to divide training into these two supervised tasks.
Both networks were UNets with 3 encoding/decoding residual blocks with skip-connections. The first layer was a Conv2d expanding the number of channels to 32. Then, each encoding (resp. decoding) block was composed by a sequence of Conv2d (resp. TransposeConv2d)-InstNorm-LeakyRelu-Conv2d-LeakyRelu-InstNorm-MaxPooling(resp. Conv2d). Both networks end by a final convolution layer that outputs the needed number of channels. All kernels were (3,3) convolutions.
The input of the *Denoiser* network was extended to consider several $P_{PVE,noisy}$ projections corresponding to different projections angles of the same source. Considering that the projection to be corrected is at angle $i$, the *Denoiser* takes as input projections of angles : $(i°, i-3°, i+3°, i+90°, i+180°, i+270°)$, *i.e.* the projection to be corrected, two adjacent angles, two orthogonal and opposite ones. Moreover, we further enrich the input of the *Denoiser* to take an additional channel previously obtained by using the full sinogram (120 angles) $P_{PVE,noisy}$ to reconstruct a coarse volume with one iteration of OSEM and RM and then forwardprojecting this volume (without RM) on the same 120 angles to obtain one additional channel per angle. The input/output of *Denoiser* then have 7 channels whereas the *PVC* network has 7 channels as input and outputs only one channel (i.e. the estimated projection $\hat{P}^i_{noPVE}$). The idea here was to exploit additional information contained in the data to help solving this ill-posed inverse problem (different noise realisation, source depth, RM) and to ensure a continuity in the corrected projections. Parameters of both networks were optimized to minimize a L1 loss functions. Networks were trained during 100 epochs with Adam optimizer, with 4 GPUs, a batch size of 256 per GPU and a learning rate of $10^{-4}$ halved every 20 epochs.

## 2.3 Evaluation data and metrics

Performance of the proposed method was evaluated with three experiments. First, we considered an analytical version of the standard NEMA IEC phantom composed of six spheres with increasing diameters 10-37 mm, with 1/40 background ratio and projections obtained with RTK like for the training database. Several reconstructed images were compared: the images reconstructed from $P_{noPVE}$ projections (noPVE-noPVC), from $P_{PVE,noisy}$ projections and RM (PVE-RM), from projections corrected by the networks (PVE-PVCNet) and from projections without any PVC (PVE-noPVC). Then, we evaluated the performance of our proposed method on real acquisition of the NEMA IEC phantom, obtained with a 1/10 background-to-source ratio and the Siemens-Intevo SPECT/CT system. PVCNet was applied to the primary energy window and scatter window independently before scatter correction with the DEW was applied (k=1.1). Finally, we tested our proposed network on real patient data.
All reconstructions were performed with OSEM with 8 subsets and 5 iterations (except for PVE-RM for which 20 iterations were needed), scatter and attenuation correction. Resulting image resolutions were compared by computing the hot sphere contrast Recovery Coefficient (RC) [18] for each sphere and correction method. We also computed the Normalized Root Mean Square Error (NRMSE), Normalized Mean Absolute Error (NMAE), Peak Signal-to-Noise Ratio (PSNR) and Structural Similarity Index (SSIM) of each image.

# 3 Results

Database generation took 3 hours using one hundred parallel CPUs. Training took 50 hours for 100 epochs using 4 GPUs. The proposed PVCNet method was compared to the ground-truth image (noPVE-noPVC), to the widely used Resolution Modeling (RM) method and to the un-corrected one (PVE-noPVC). From now on, FWHM refers to the collimator resolution value at a distance of 28 cm, which corresponds to the employed isocenter-collimator distance to generate the simulated projections and for the acquisition. Visualisation and RC results for the analytical IEC phantom are shown in Figure 1 and show promising results in terms of activity recovery, Gibbs artefact reduction and error reduction. NMAE was divided by two and PVCNet was significantly better in term of all global metrics shown in Figure 1c. However, small spheres (size<FWHM) were completely lost by the network.

Reconstructed images and RC curves for the real IEC acquisition are shown in Figure 2. Figure 2b underlines that we only achieved better correction than RM on the largest sphere (of size 37 mm) in terms of RC. Similarly to the previous experiment, the network struggles to correct PVE on small spheres and some distortion artefacts are visible on the corrected spheres. On the other side, Figure 1a shows that for sphere with size>FWHM, the homogeneity was better retrieved with PVE-PVCNet than PVE-RM and regarding the other studied criteria, PVE-PVCNet outperfomed PVE-RM (Figure 2c).

Real patient reconstructed images are shown in Figure 3 for visual assessment only, as no reference was available.

## 4 Discussion and conclusion

For the first time, this work investigated the feasibility of training a network from simulated projections to compensate the effect of the PSF and to denoise projections, before reconstruction. We showed that building such a database is feasible, and we designed an adapted deep learning architecture to correct both noise and PVE. On simple test cases, PVCNet reduced Partial Volume Effect compared to standard RM method, while requiring less iterations and no regularization. However, on real data acquisition, while NRMSE, NMAE, PSNR and SSIM were better than the values obtained with RM, RC was not. Small spheres were not well recovered. We now envision to improve the realism of the training database by using tumor-like source shapes, heterogeneous activities and projections generated by (fast) Monte Carlo simulations [19, 20]. Finally, considering more projection angles as input could be useful to increase source-depth information.

## Acknowledgement

This research was funded, in part, by MOCAMED (ANR-20-CE45-0025), LYriCAN (INCa-INSERM-DGOS-12563), LABEX PRIMES (ANR-11-LABX-0063, ANR- 11-IDEX-0007), POPEYE (ANR-19-PERM-0007-04). A CC-BY public copyright license has been applied by the authors to the present document and will be applied to all subsequent versions up to the Author Accepted Manuscript arising from this submission, in accordance with the grant's open access conditions. We gratefully acknowledge the support of NVIDIA Corporation with the donation of the Titan Xp GPU used for this research. This work was granted access to the HPC resources of IDRIS under the allocation 2019-101203 made by GENCI (Jean Zay computing center).

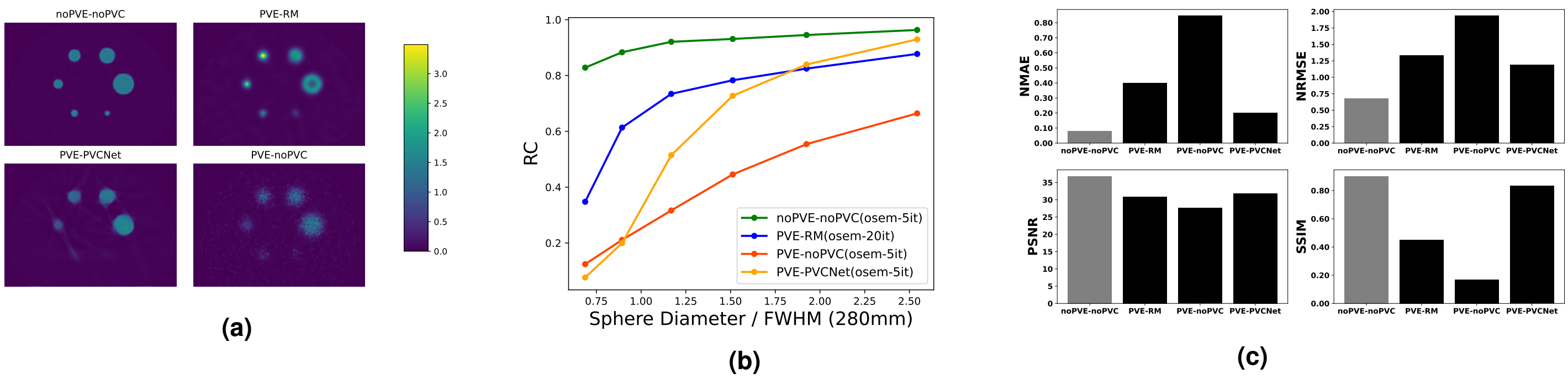

**Figure 1:** (a) Slice of the four reconstructed images (noPVE-noPVC, PVE-RM, PVE-noPVC, PVE-PVCNet), (b) the corresponding Recovery Coefficient (RC) for each reconstructed sphere with respect to the ratio sphere diameter / FWHM (c) comparison of NRMSE, NMAE, PSNR and SSIM. The reference image was the initial voxelised 3D source used for forward projection.

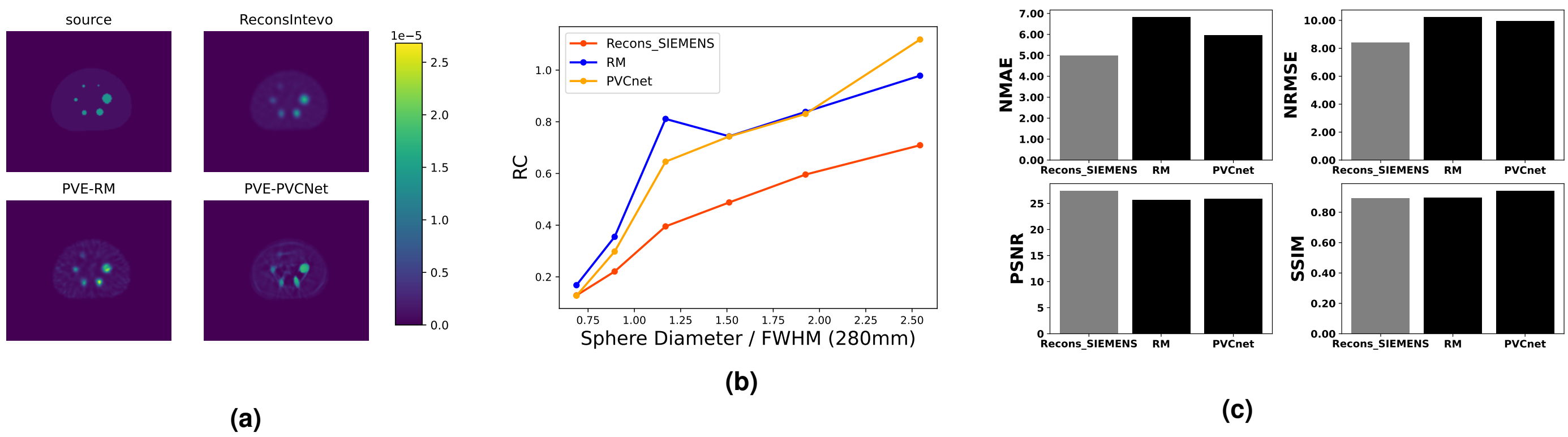

**Figure 2:** (a) Slice of the four images (source, Reconstruction by the INTEVO SPECT system, PVE-RM, PVE-PVCNet) divided by the total number of counts in each image (b) the corresponding Recovery Coefficient (RC) for each reconstructed sphere with respect to the ratio sphere diameter / FWHM (c) NRMSE, NMAE, PSNR and SSIM with a manually contoured reference image knowing the injected activity concentrations in spheres/background

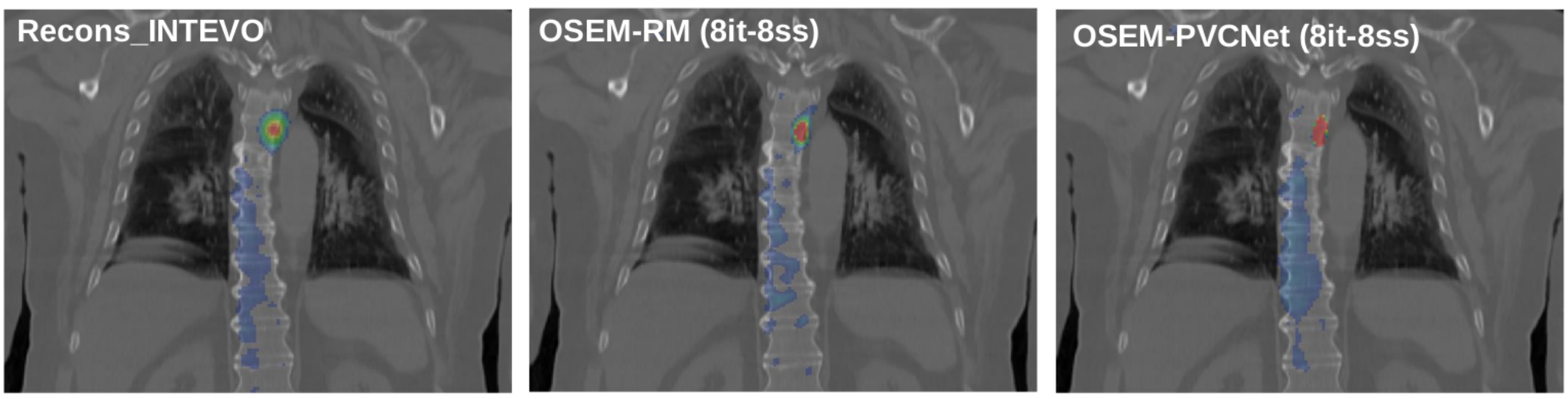

**Figure 3:** Visual comparison of patient SPECT/CT images with different correction methods. All three images were reconstructed with OSEM algorithm (20 iterations, 4 subsets), attenuation and scatter correction.

# Dynamic Spatiotemporal Clustering for Factor Analysis in Dynamic Structures

Valerie Kobzarenko[1], Rostyslav Boutchko, Uttam M. Shrestha[2], Grant T. Gullberg[2], Youngho Seo[2], Debasis Mitra[1]

[1]Department of Computer Science, Florida Institute of Technology, Melbourne, FL, USA
[2]Department of Radiology and Biomedical Imaging, University of California, San Francisco, CA, USA

**Abstract**: **We propose a dynamic PET analysis methodology that improves on top of previously established non-matrix factorization algorithms. Our primary innovation is the improved factor initialization approach and subsequent data driven spatial masking. The initialization is based on time series k-means which detects subsets of voxels whose time-activity curves captures essential features of underlying dynamics. The dynamic binary mask developed based on the initialization work further removes portions of background in order to unmix the factors prior to performing factor analysis of dynamic structures (FADS). We provide the initial validation and simple illustration of our approach using a simulated dynamic dataset. Then, the proposed workflow is applied to clinical dynamic cardiac PET data reconstructed using straightforward filtered back projection (FBP) algorithm, a very challenging type of PET data to work with because of low counts and high noise. In both simulated and real patient data, our algorithm is efficient in identifying different tissues based on inherent tracer dynamics. We first show the importance of a dynamic-targeted initialization, and then improve upon the initialization with our spatiotemporal clustering. The extension of this methodology can enhance the relevance and applicability of dynamic imaging in the clinical domain.**

## 1 Introduction

Dynamic Positron Emission Tomography (dPET) is an approach to PET where a series of spatial distribution of time series is acquired instead of a single static 3D distribution of the injected radiotracer, thereby producing a 4D dataset. Dynamic imaging has been applied to other cases of nuclear imaging and has shown to provide a more succinct representation of the tracer kinetics [1]. A principal advantage of dynamic PET is its capacity to detect differences in tracer exchange rates. This detection then translates into the ability to resolve fine differences in tissue physiology which may have not been captured by the static scan [2].

The primary challenge of dynamic PET comes from the requirements imposed on the acquisition protocols and data processing functionality. The classic approach to dynamic imaging solves for the compartmental model parameters of the tracer kinetics. However, this approach requires data to be acquired over a long time from tracer injection to steady state [3]. Multiple approximation models such as the reference tissue model, have been proposed which can decrease the time used [4]; however, they stringently require imaging during steady state, and can suffer from approximation errors of varying severity. An inadvertent consequence of these acquisition, and processing issues is that the information content from dynamic PET can be limited. This paradox creates technical obstacles which prevent the widespread dynamic PET adoption, while the lack of adoption halts technological progress in dynamic PET. We think that by developing techniques for applying dynamic processing to data acquired using the same (or essentially similar) protocols as used in clinic now, the adoption of the methodology would be more feasible. Therefore, even a small practical improvement achieved in this area can herald significant down-the-line progress [5].

Non-compartment modelling techniques aided by machine learning have also been applied to dynamic PET processing. A dynamic image processing approach that avoids some of the known challenges can use data driven techniques to separate different tissues based solely on their curve shape/time activity. Earlier, our team reported using non-negative matrix factorization method called cluster-initialized factor analysis in dynamic brain PET [6][7]. In this paper, we further improve this method, and apply it to clinical dynamic cardiac PET images. In light of pursuing adoption-oriented goals, we would like to show how one can resolve tracer dynamics, and the corresponding tissue distributions from data acquired, and reconstructed using a common type of scanner, and a generic clinical protocol.

Filtered backprojection (FBP) is a reconstruction algorithm that is widely applied in earlier x-ray computed tomography (CT), PET, and other common tomographic modalities. Since it is an analytic inverse problem solution to the tomographic model, it is numerically correct, and capable of generating excellent images from properly sampled projections that are not overly contaminated by noise. However, in nuclear medicine, FBP is highly suboptimal, both because the angular sampling may be too coarse, and, most importantly, because filtering amplifies structual and random noise, while the backprojection propagates it through to the image (Fig.1). Despite this, due to its speed, FBP is still used in some diagnostic scenarios. For example, the patient data we will use for our algorithm's performance evaluation: the data are FBP-reconstructed PET images of the human heart.

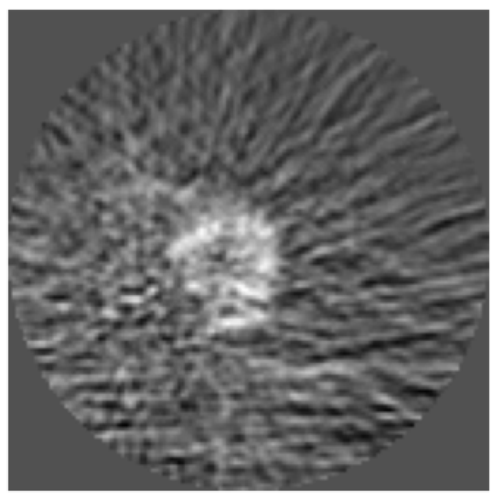
**Figure 1**. Original FBP reconstruction of cardiac PET, a single time frame.

At the submission stage of the paper, we describe the details of the basic processing algorithm, and show examples of its successful application to simulated dynamic data and real patient cardiac dynamic PET data.

## 2 Materials and Methods

The principal data processing approach used in the presented work is Factor Analysis of Dynamic Structures or FADS. This is a non-negative matrix factorization method which represents all time-dependent behavior in the imaged volume as a linear superposition:

$$D(\mathbf{r}, t) = S_j\, C_j(\mathbf{r}) \times f_j(t) \qquad (1)$$

Here, $D$ denotes the 4D distribution, dependent on spatial position $\mathbf{r}$ and time $t$. The factors are denoted by index $j$; usually no more than $J$=3~4 factors are used. Factor choice $J$ is often determined by *a priori* of the tissue imaged and is based on the region imaged (*i.e* in this case we expect myocardium, left and right ventricle, and background). Equation (1) is often presented in a discrete form and solved for $J$ factors or factor curves $f$ and the corresponding factor images $C$.

Even though we are not aware of direct proof of physiological meaningfulness for such representation, the factors curves will correspond to commonly recognizable types of radiotracer dynamics such as blood concentration, specific-binding, and non-specific-binding tissue curves. Consequently, the factor images represent spatial distributions of blood vessels of said tissues. The key steps of FADS and other initialization techniques have been developed, and presented in the past. The present implementation has been presented in [6].

Establishing the right-hand-side terms in (1) is a complex optimization problem with multiple methodological, and numerical challenges such as non-uniqueness, local minima, tissue overlap, etc. Our team overcomes some of these problems by applying additional clustering techniques during the factorization steps. As noted, non-uniqueness is a prevalent issue within the factorization implementation. Yet, a meaningful initialization can aid in this scenario as it drives the decomposition in the correct direction.

*A. Initialization*

An integral part to our initialization approach is the ability to use data-driven initialization without the need for region of interest (ROI) segmentation. We first sample a random large subset of $N$ voxels within the data to get a noisy estimate of the dynamic activity occurring in different regions. Then, we apply time series k-means clustering (2) to the sampled voxels and find $J$ clusters with corresponding centers $f_j$. We treat the $f_j$ curves as the initialization for the following four distinct dynamic regions (pertinent to cardiac imaging): background, right ventricle, left ventricle, and myocardium:

$$\arg\min \sum_{j=1}^{J} \sum_{x \in s_j} \left\| x - \mu_j \right\|^2 \quad . \qquad (2)$$

At time-resolution of some dynamic scans, the temporal behavior of the myocardium, and the background can be similar in shape despite having different spatial distributions. The indicated effect drives the optimization step of FADS to maximally separate the curves despite their inherent temporal similarity, and therefore obscures the underlying representation. This behavior was a predominant target of our efforts, and led to the exploration, and development of the proposed methodology described below. Additionally, as a byproduct of the Gaussian filtering used on the noisy FBP dataset, the background, and myocardium images overlap significanly, more than partial volume observed in images reconstructed using iterative statistical algorithms.

Despite the effective initialization used improving FADS results, in some datasets, FBP amplified the noise to the extent where the improvements in the initialization may be insufficient to allow FADS to extract the correct images. In that case, we apply additional masking, and preprocessing as described below.

*B. Dynamic Spatial Masking*

Factorization of the myocardium ROI can be obstructed by the background. Manual masking of different tissue regions is time consuming, and ineffective. To combat this, we appy the cluster-based improvements described below. Prior to factorization, all voxels in the data are clustered by considering their distance from each of the cluster centers $f_j$. Background cluster is identified by using the curve-shape, and the coefficient distribution.Voxels that have been assigned into the background cluster are subsequently removed from the dataset using a dynamic binary mask (3):

$$M_j = \begin{cases} 0, & \textit{if background cluster} \\ 1, & \textit{otherwise.} \end{cases} \qquad (3)$$

After removal the background, only $J$–1 factor curves remain: the myocardium, and the left and right ventricles.

*C. Clustering updates during factorization*

The newly masked dataset is factorized using FADS [6]. Further improvement of the method comes from employing additional clustering techniques near FADS convergence. We use Density Based Clustering Applications with Noise (DBSCAN) and apply it to the factor images to determine which voxels are within the coefficients ROI, and which ones are obfuscating the decomposition [8]. This update to the methodology is performed for each of the $J$–1 coefficients. Clusters that are not central to the region of interest shown in Fig. 2 are masked out, for example (b); whereas clusters within the ROI are kept, such as (a). This is applied to the last 5-8 iterations.

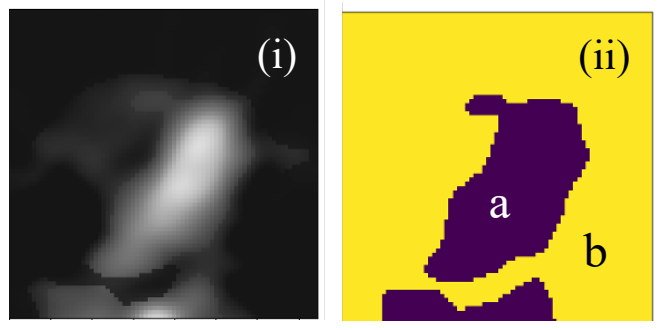

**Figure 2**: (i) 2D slice example. (ii) Final DBSCAN result.

Fig. 2 shows the result at the very last step prior to convergence. The area marked by (a) is a part of this regions desired spatial cluster, while the area marked by (b) is outside of this region, and therefore should not be considered for this $J_i$ coefficient.

## 3 Results

### *A. Simulation*

The work was divided in two stages: demonstrating the validity of the initialization using simulated dynamic data, and subsequently applying the initialization, and update steps on real data. A 128×128 dynamic dataset was created with four regions of different tissues, Fig 3.ii. These regions overlapped at the edges to simulate partial-volume effect or blurring due to filtering. Regions were designed to have different shapes, and levels of overlap. Tissue texture was added using random small variations. Each of the four regions was characterized by distinct temporal tracer dynamics (which can be edited to be more or less similar). Fig.3.i shows the known ground truth dynamic curves.

The dataset was tested with the proposed initialization algorithmic update, and compared with another effective NMF initialization technique, Non-negative Double Single Value Decomposition (NNDSVD) [9]. NNDSVD has been shown to be an effective initialization technique for matrix factorization. This method is also data-based, and is used here as a comparison to show how important a dynamic-data-specific initialization is. Each initialization method selected the known number of *J-1* factors and revealed the resulting factor curves, and factor coefficients were compared. Fig.3.c and Fig.3.d illustrate the results of the reconstruction.

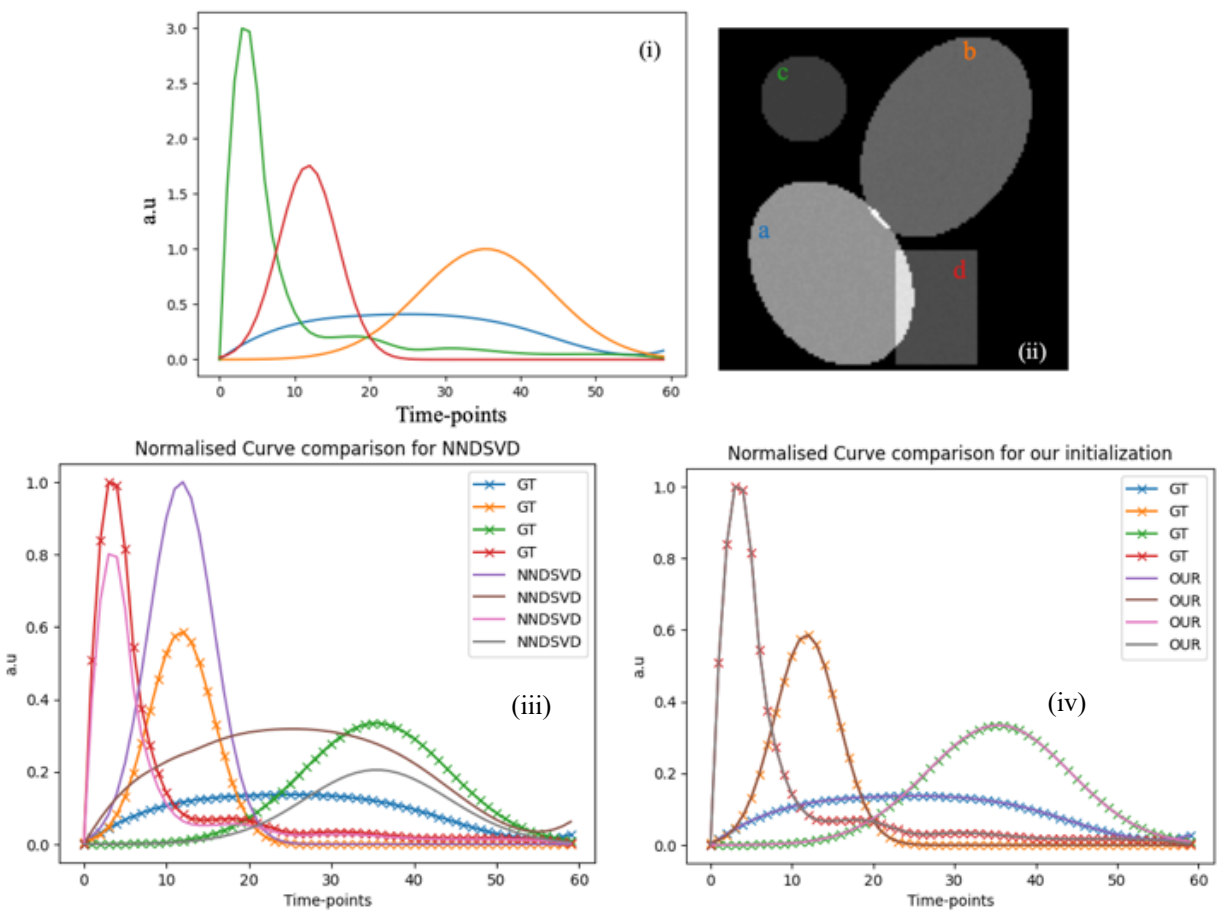

**Figure 3**: (i) Simulated phantom ground truth with four distinct regions (ii) (a is blue, b is orange, c is green, and d is red). (iii) is the normalized NNDSVD result plotted against normalized ground-truth. (iv) is the normalized result for our proposed dynamic-targeted initialization plotted against the normalized ground-truth.

Results of the simulated dataset reconstruction demonstrate that despite NNDSVD's ability to successfully drive factorization in many other applications, it may struggle to reveal the temporal nature of tissues in the dynamic cardiac PET modality. Sampling and clustering voxels, therefore, drives the factorization toward the global minimum, and is better suited for dynamic datasets. The proposed initialization's effectiveness (Fig.3.iv) is expressed as a closer approximation of the factorized curve shape to the known ground truth (Fig.3.i), since after normalization of values to be in the [0,1] range, the (Fig.3.iv) curves overlap the known ground-truth curves, demonstrating that the curve shape is revealed using the dynamic-targeted method. The units on y-axis are arbitrary, the key aspect is each curve's relative shape to one-another. Outfitted with an effective start for factorization we then focused on improving the final factorized results using imbedded clustering.

### *B. Dynamic $NH_3$ PET Results*

We used dynamic $^{13}$N-ammonia cardiac dPET scans acquired at UCSF using a PET/CT scanner (Discovery VCT, GE Healthcare). Each dataset consisted of 25 or 40 images acquired between 0 and 15 minutes post-tracer injection. Each timeframe was reconstructed on the scanner using vendor provided FBP. In order to mitigate intense backprojection noise (see Fig. 1), a (*sigma* = 2.27x, 2.27y, 1z) 3D Gaussian filter was applied to each data frame. A smaller window around the ROI was selected approximately 70x70x30 voxels. Factorization was performed on the dataset, with the proposed initialization, one with and one without the dynamic masking. Fig.4.d-f shows the factorization without removing background clustered pixels. Fig.4.a-c shows our applied methodology of removing the pixels clustered as *background* and factorizing for three factors and successively updating the factors using density based clustering. Fig.4.(i) shows the factor curves resulting from our final factorization. In these curves we can clearly see the expected temporal behaviour of the myocardium, the right ventricle and the left ventricle. Furthermore, spatially, the myocardium is more pronounced in our applied methodology: Fig.4.a-c, as compared to Fig.4.d-f, where restricting the decomposition only to a better initialization still resulted in a ROI obscured by the presence of the background factor. No clear myocardium factor was observed (Fig.4.d) versus the clear myocardium in (Fig.4.a) from our method.

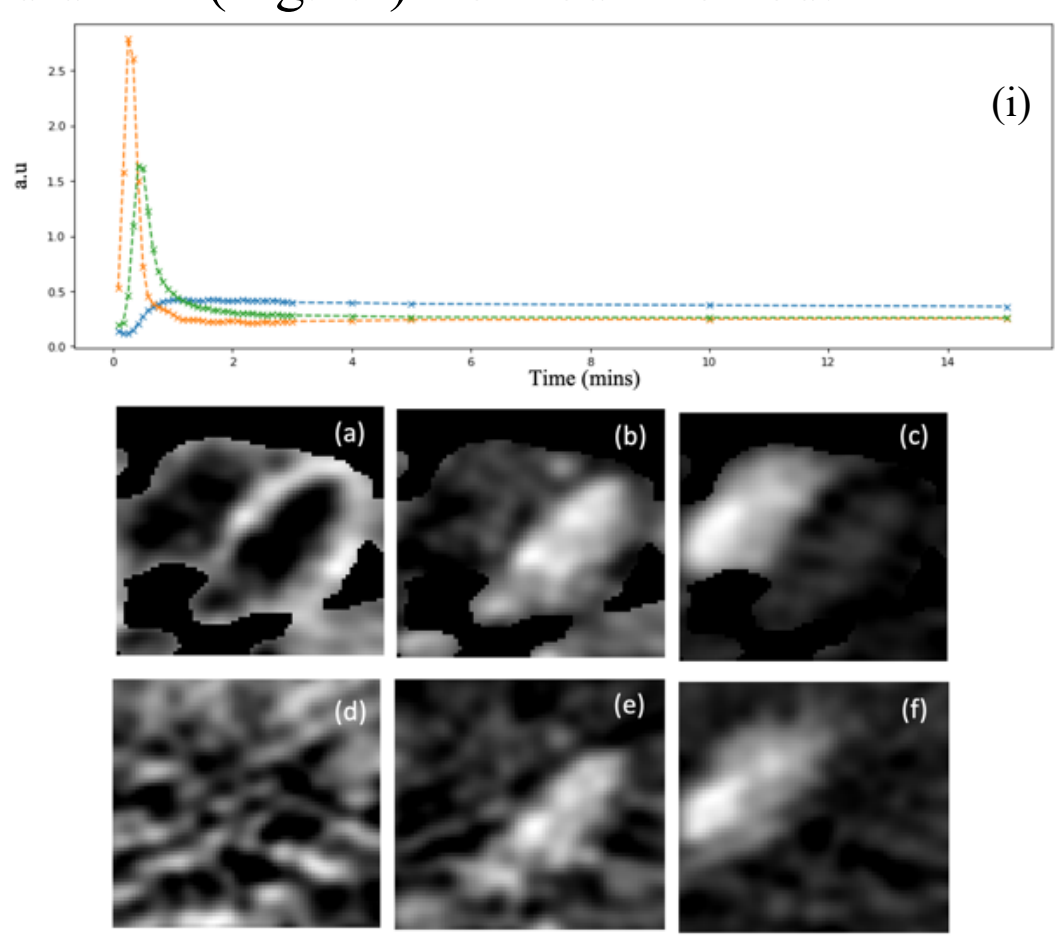

**Figure 4:** (i) Our final curves. Blue is (a), orange is (b), green is (c). (a-c) 2D slice of our result coefficient images. (d-f) coefficient images with the improved factorization, but no spatial clustering.

## 4 Discussion

As seen in the results, processing dynamic images is extremely sensitive to the initialization of factor curves. The initial curves are the driving force for a better decomposition which can mean the difference between the noisy decomposition, and one that reveals the ROIs, and inherent tracer dynamics. Additionally, the nature of this noisy data type, and the usage of FBP reconstruction further complicates the factorization which is where the proposed methodology demonstrates advantages over the traditional method. Even in the simulation we can observe cases when factorization intermediation, such as the clustering proposed, is needed to successfully reveal the underlying factor coefficient ROIs, and the factor curves.

By using voxel TACs allocated into several clusters, initialization becomes more data-driven, and consequently, patient-specific, and data type independent. This methodology better estimates the temporal behavior, and can reveal important aspects of the data such as the expected number of clusters. Yet, as seen in the real data, subsampled voxels' TACs are often noisy, and can be unresolvable by simple factorization despite having a ROI which is clearly seen within the dataset. Thus, for these cases the application of our method's dynamic cluster masking improves the factorization result as the desired region of interest is more pronounced, and no longer has temporal behavior mixed with the background. The methodology therefore expands the application of factorization to more nuanced data, and further increases the relevance of dynamic imaging in clinical settings where the acquisition method may be suboptimal. Similarly, the application of Density-based clustering is an important addition to the factorization steps, and itself aids in improving the results since it considers the spatial domain, and aids in spatial ROI unmixing.

## 5 Conclusion

We hope that this methodology will find applications in datasets where the factorization may be obstructed by the inherent noise from the background, and hope that the methodology can be further applied to other dynamic imaging modalities which suffer from the same drawbacks of factor mixing. Throughout this work we have presented the difference that an improved initialization technique has on the factorization. Conjointly, we also showed that data driven dynamic masking effectively aids in factorization of datasets where the temporal behavior of more than one region is similar. The case presented here was a dataset which failed to have a meaningful factorization result prior to the application of the presented workflow. Our effort now is focused on deriving better quantitation parameters from dynamic data, clustering differentiation, and extending the methodology to be completely unsupervised. As curve shape improves from unmixing with our method, we expect down-the-line improvement in values calculated from the curves. In the future, we would like to validate our work with kinetic parameters.

## Acknowledgements

We would like to acknowledge support from NIH (R01HL135490, R01EB026331 and R15EB030807).

# Hybrid Geometric Calibration in 3D Cone-Beam with Sources on a Plane

Anastasia Konik[1], Laurent Desbat[1], and Yannick Grondin[2]

[1]TIMC-GMCAO, Université Grenoble Alpes, Grenoble, France
[2]SurgiQual Institute, Meylan, France

**Abstract** In this work, we describe a new self-calibration algorithm in the case of 3D cone-beam geometry with sources on a plane parallel to the detector plane. This algorithm is hybrid, so it combines data consistency conditions (DCCs) and the partial knowledge on the pattern of the calibration cage. We explain how we build this algorithm with the modelling of calibration markers by Diracs and the generalization of existing DCCs for this geometry to distributions. This new method can work with truncated projections if the marker projections are not truncated. With this hybrid approach, we build an analytical self-calibration algorithm based on DCCs, robust to projection truncations. We show numerical experiments.

## 1 Introduction

In the recent work [1] a hybrid approach to calibrate a cone-beam system was proposed. This method combines data consistency conditions (DCCs), special equations on projection data, and detected anatomical markers. We continued the development of similar methods and presented hybrid methods in 2D fan-beam geometry and 3D cone-beam geometry, both with sources on a line, see [2, 3]. In our approaches we modelled the calibration markers by Dirac distributions and exploited DCCs on distributions. Compared with [1], our calibration algorithms with such DCCs are analytical, thus we don't need to solve numerical optimization procedures. Moreover, because we use DCCs only on markers, the projection data may be truncated. Only the marker projections must be non-truncated. Thus, moment conditions of DCCs on the set of Diracs yields the extension of existing calibration procedures based on DCCs to truncated projections.
In [4] DCCs for functions were proposed for the 3D cone-beam with sources on a plane parallel to the detector plane. In this work, we show that the generalisation of these DCCs to distributions helps to overcome difficulties in the construction of the analytical calibration procedure for this geometry and allows the hybrid geometric self-calibration with truncated data.

## 2 Calibration problem

3D divergent X-ray systems lead to the cone-beam projection data defined by the following transform:

**Definition 2.1.** *The cone-beam transform of the function $f$ of compact support describing an object is*

$$\mathfrak{D}f(\vec{s}_\lambda, \vec{\zeta}) := \int_0^{+\infty} f(\vec{s}_\lambda + l\vec{\zeta})dl, \quad (1)$$

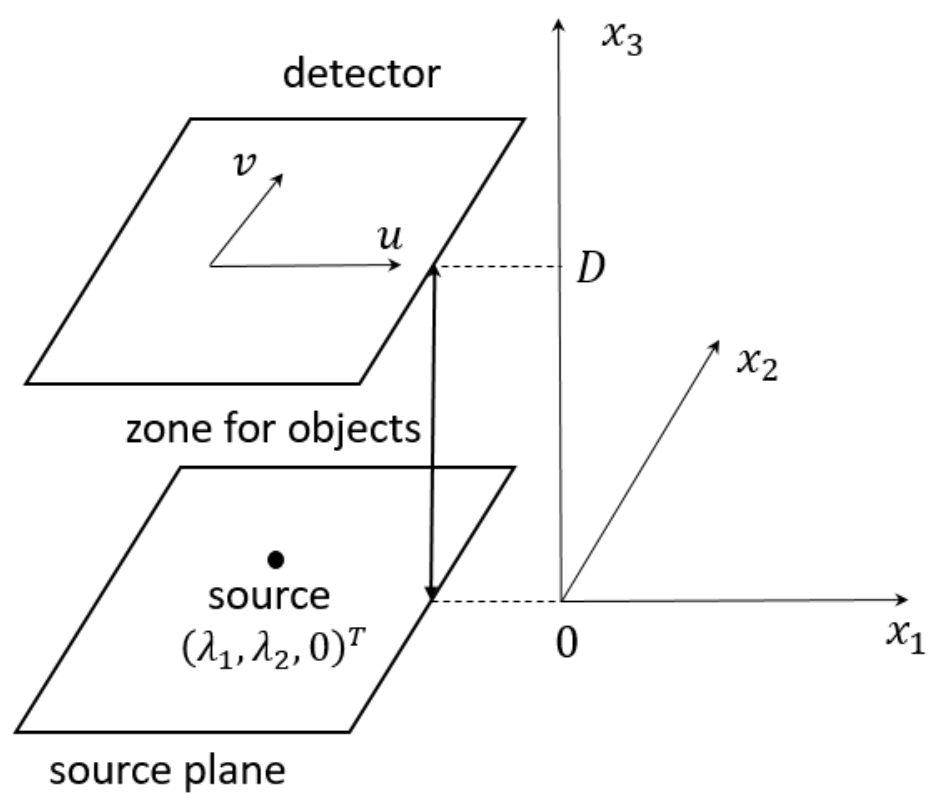

**Figure 1:** The cone-beam geometry with sources on a plane parallel to the detector.

*where $\lambda \in \mathbb{R}$ is the trajectory parameter of the source $\vec{s}_\lambda \in \mathbb{R}^3$ and the (usually unit) vector $\vec{\zeta}$ is the direction of the integration line.*

For the cone-beam transform with sources on a plane parallel to the detector, we work with the geometry presented in the Figure 1, see also [4]. Here we have sources moving in a plane parallel to the detector plane: the source trajectory is $\vec{s}_\lambda = (\lambda_1, \lambda_2, 0)^T$, the detector is in $x_3 = D$, the *non-unit* direction of the integration line $\vec{\zeta} = (u, v, D)^T - (\lambda_1, \lambda_2, 0)^T$, $u$ and $v$ are parameters of the detector. In this case we can rewrite our data as:

**Definition 2.2.** *The cone-beam transform with sources on a plane parallel to the detector plane of a function $f$ of compact support between the source plane $x_3 = 0$ and the detector plane $x_3 = D$ is*

$$\mathfrak{D}f(\lambda_1, \lambda_2, u, v) := \int_0^{+\infty} f(\lambda_1 + l(u - \lambda_1), \lambda_2 + l(v - \lambda_2), lD)dl. \quad (2)$$

We can denote the cone-beam transform for fixed $\lambda_1$ and $\lambda_2$ with $\mathfrak{D}_{\lambda_1,\lambda_2}f(u,v) := \mathfrak{D}f(\lambda_1, \lambda_2, u, v)$ for $f$ with support in $Y_3 = \mathbb{R}^2 \times (D_1, D_2)$, $0 < D_1 < D_2 < D$. Then we consider $\mathfrak{D}_{\lambda_1,\lambda_2}f$ as a function of two variables.
Suppose that we work with a lattice of $u, v$, but the system is moving. So, we don't know exactly:

- source positions $\lambda_{1i}$ and $\lambda_{2i}$ for $P$ projections, $i \in [\![0, P-1]\!]$,
- detector shifts $u_i$, $v_i$ for each source position $i$.

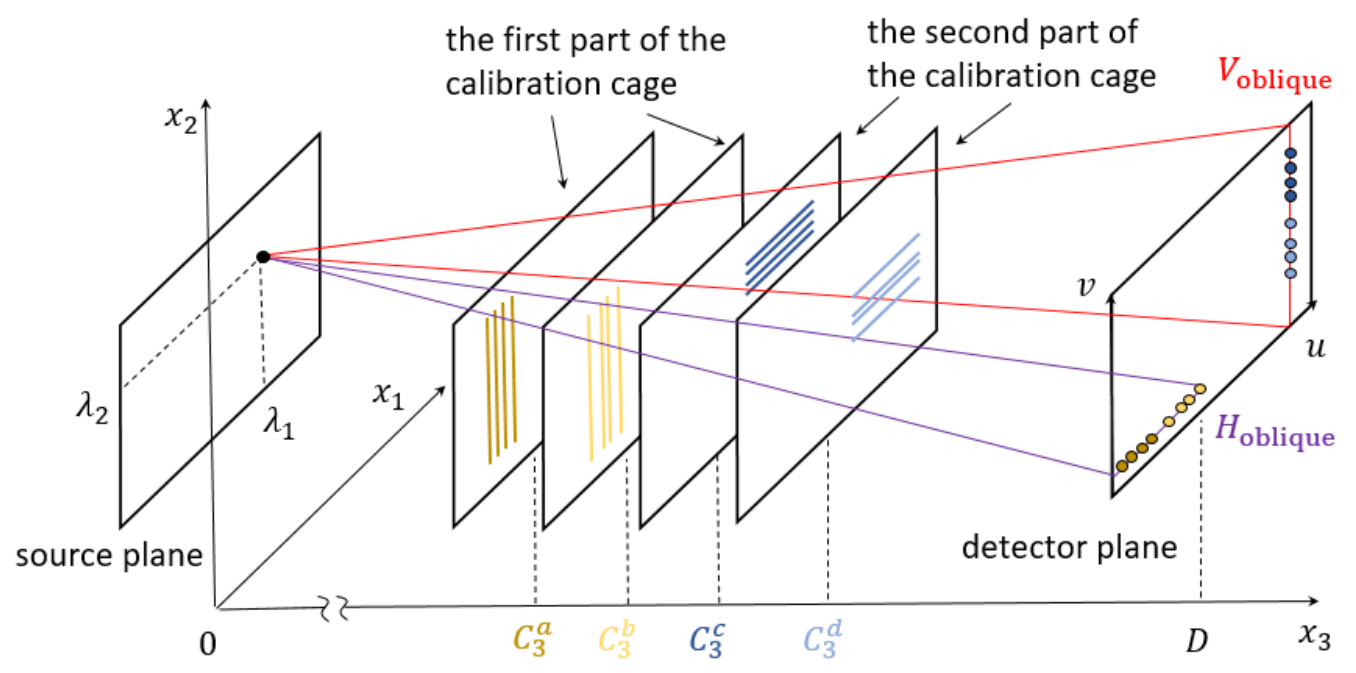

**Figure 2:** The 3D cone-beam geometry with sources on a plane with the calibration cage of two groups of 8 parallel sticks each.

We know $m_i(u,v)$, where $m_i(u,v) = \mathfrak{D}_{\lambda_{1i},\lambda_{2i}} f(u-u_i, v-v_i)$. We need to estimate the geometric calibration parameters $\lambda_{1i}$, $\lambda_{2i}$, $u_i$, $v_i$, $i \in [\![0, P-1]\!]$. We want to do it with the specific calibration cage presented in the Figure 2. So, we add some calibration object with unknown position. We want to use DCCs on projections of this object to perform the full analytical calibration.

Our task can be separated into two independent tasks to find the couple of $\lambda_{1i}$ and $u_i$ for the first task with the help of the first part of the calibration cage (8 vertical sticks) and $\lambda_{2i}$ and $v_i$ for the second task with the second part of the calibration cage (8 horizontal sticks).

We place these sticks as close as possible to the detector and two groups of sticks are separated such that it's possible to select one horizontal detector line for the vertical sticks and one vertical detector line for horizontal sticks containing only projections of one part of our calibration cage. Thus, it's possible to work separately with the projection of vertical sticks for one fixed detector line and the projection of horizontal sticks for another fixed detector column. So, we can say that we need to select two appropriate detector lines (one horizontal and one vertical) for the current source position or two oblique planes (one horizontal $H_{\text{oblique}}$ and one vertical $V_{\text{oblique}}$) passing through each detector line and the current source position. Let us consider the intersection of sticks with one selected oblique plane and points of the calibration cage in this intersection as Dirac distributions.

For each group of 8 sticks we have a known pattern, the same as we used in our previous works [2, 3]. Let us consider only the first vertical group of sticks (the same can be written for the second group). Each subgroup of 4 sticks belongs to one plane. Let us use the superscript $l = a$ or $l = b$ for each group of 4 sticks. We have more unknowns in our calibration problem:

- the first 4 sticks are in the unknown plane $x_3 = C_3^a$, the second 4 sticks are in the unknown plane $x_3 = C_3^b$,
- we suppose that sticks are perpendicular to the $x_1$-axis, then points in the intersection of sticks and the fixed oblique plane $H_{\text{oblique}}$ are $c_{11}^a = p_a - k_1 L$, $c_{21}^a = p_a - L$, $c_{31}^a = p_a + L$, $c_{41}^a = p_a + k_1 L$, $c_{11}^b = p_b - k_2 L$, $c_{21}^b = p_b - k_3 L$, $c_{31}^b = p_b + k_3 L$, $c_{41}^b = p_b + k_2 L$, where $p_a$, $p_b$ are unknown and define the $x_1$-position of the center of mass of each subgroup, $L$, $k_1 > 0$, $k_2 > 0$, $k_3 > 0$ are known; we also assume that $D$ is known.

Thus, by knowing the abscissas $q_{ij}^l$ ($i \in [\![0, P-1]\!]$, $j \in [\![1, 4]\!]$, $l \in \{a, b\}$), the detected projections of the stick intersections with the oblique plane $H_{\text{oblique}}$, for one fixed detector line and some parameters of the pattern of the calibration cage (see previous paragraph), we want to identify the geometrical calibration parameters $\lambda_{1i}$, $u_i$, $i \in [\![0, P-1]\!]$, and the unknown position of the calibration cage $p_a$, $p_b$, $C_3^a$, $C_3^b$. With essentially the same procedure (see the next section), but the vertical oblique plane and 8 horizontal sticks, we identify $\lambda_{2i}$, $v_i$, $i \in [\![0, P-1]\!]$, $p_c$, $p_d$, $C_3^c$, $C_3^d$.

## 3 Hybrid solution

The mathematical theory of the cone-beam transform on distributions with sources on a plane parallel to the detector allowed to build the similar hybrid algorithm that we used in our previous works [2, 3], see section 4.

Let us define $r_a = \frac{D - C_3^a}{C_3^a}$, $r_b = \frac{D - C_3^b}{C_3^b}$. It was possible to derive analytical formulas to calibrate with DCCs in this case (for the first task, the similar algorithm can be constructed for the second task):

1. $r_a$ and $r_b$ can be uniquely estimated from

$$\begin{cases} \sum_{j=1}^{4} (q_{ij}^a)^2 = (1 + r_a)^2 (2 + 2k_1^2) L^2 + K_a(i) \\ \sum_{j=1}^{4} (q_{ij}^b)^2 = (1 + r_b)^2 (2k_2^2 + 2k_3^2) L^2 + K_b(i), \end{cases} \tag{3}$$

where

$$K_l(i) = \frac{1}{4} \left[ \sum_{j=1}^{4} q_{0j}^l \right]^2 + 2\Delta\tilde{M}_1^l(i) \sum_{j=1}^{4} q_{0j}^l + 4[\Delta\tilde{M}_1^l(i)]^2,$$

$$\Delta\tilde{M}_1^l(i) = \frac{1}{4} \left( \sum_{j=1}^{4} q_{ij}^l - \sum_{j=1}^{4} q_{0j}^l \right),$$

and then we can deduce $C_3^a$ and $C_3^b$ from $r_a$ and $r_b$,

2. $p_a$ and $p_b$ can be estimated from

$$\sum_{j=1}^{4} q_{0j}^l = (1 + r_l) 4 p_l, \qquad l \in \{a, b\}, \tag{4}$$

3. from the linear system

$$u_i - \lambda_{1i} r_l = \Delta\tilde{M}_1^l(i) \tag{5}$$

we compute $u_i$ and $\lambda_{1i}$ for each projection: (5) gives us 2 equations with 2 unknowns, since $l \in \{a, b\}$.

## 4 Mathematical basis

In this section we describe the theory that we built and used in the derivation of our algorithm. Firstly, we generalize the definition of the cone-beam transform from the second section to distributions. Then we provide the generalization of known DCCs given in [4] to distributions.

**Definition.** Denote for any open set $\Omega_N \subset \mathbb{R}^N$ the spaces of compactly supported smooth functions $\mathscr{D}(\Omega_N)$, the spaces of smooth functions $\mathscr{E}(\Omega_N)$, $N \in \{2,3\}$. Then $\mathscr{D}'(\Omega_N)$ and $\mathscr{E}'(\Omega_N)$ state for the sets of corresponding distributions.

We need to define the dual operator $\mathfrak{D}^*_{\lambda_1,\lambda_2}$ of $\mathfrak{D}_{\lambda_1,\lambda_2}$. For $f \in \mathscr{D}(\mathbb{R}^3)$ and $\phi \in \mathscr{E}(\mathbb{R}^2)$:

$$(\mathfrak{D}_{\lambda_1,\lambda_2} f, \phi) = \int_{\mathbb{R}^2} \mathfrak{D}_{\lambda_1,\lambda_2} f(u,v)\phi(u,v)dudv$$
$$= \int_{\mathbb{R}^2}\int_0^{+\infty} f(\lambda_1 + l(u-\lambda_1), \lambda_2 + l(v-\lambda_2), lD)dl\phi(u,v)dudv$$
$$= \frac{1}{D}\int_{\mathbb{R}^2}\int_0^{+\infty} f\left(\lambda_1 + \frac{t_3}{D}(u-\lambda_1), \lambda_2 + \frac{t_3}{D}(v-\lambda_2), t_3\right)dt_3$$
$$\times \phi(u,v)dudv = \int_0^{+\infty}\int_{\mathbb{R}^2} f(t_1,t_2,t_3)$$
$$\times \phi\left(\frac{Dt_1 - \lambda_1(D-t_3)}{t_3}, \frac{Dt_2 - \lambda_2(D-t_3)}{t_3}\right)\frac{D}{t_3^2}dt_1dt_2dt_3$$
$$= \langle f, \mathfrak{D}^*_{\lambda_1,\lambda_2}\phi\rangle, \quad (6)$$

where $(\cdot,\cdot)$ is the scalar product in $L^2(\mathbb{R}^2)$, $\langle\cdot,\cdot\rangle$ is the scalar product in $L^2(Y_3)$, we used the change of variables $t_3 = lD$, $dl = \frac{dt_3}{D}$; $t_1 = \lambda_1 + \frac{t_3}{D}(u-\lambda_1)$, $du = \frac{D}{t_3}dt_1$; $t_2 = \lambda_2 + \frac{t_3}{D}(v-\lambda_2)$, $dv = \frac{D}{t_3}dt_2$.

We can define the dual operator for functions from $\mathscr{E}(\mathbb{R}^2)$:

$$\mathfrak{D}^*_{\lambda_1,\lambda_2}\phi(\vec{x}) := \frac{D}{x_3^2}\phi\left(\frac{Dx_1 - \lambda_1(D-x_3)}{x_3}, \frac{Dx_2 - \lambda_2(D-x_3)}{x_3}\right). \quad (7)$$

**Definition 4.1.** *The cone-beam transform on a plane at fixed $\lambda_1$ and $\lambda_2$ of a compactly supported distribution $f \in \mathscr{E}'(Y_3)$ is a distribution from $\mathscr{E}'(\mathbb{R}^2)$ defined by the dual equality*

$$(\mathfrak{D}_{\lambda_1,\lambda_2} f, \phi) = \langle f, \mathfrak{D}^*_{\lambda_1,\lambda_2}\phi\rangle \quad (8)$$

*with the dual operator from* (7).

Since we model the intersection at $\vec{c}$ of a projection line with an opaque stick by a Dirac distribution $\delta_{\vec{c}}$, then

$$(\mathfrak{D}_{\lambda_1,\lambda_2}\delta_{\vec{c}}(u,v), \phi(u,v)) = \langle \delta_{\vec{c}}(\vec{x}), \mathfrak{D}^*_{\lambda_1,\lambda_2}\phi(\vec{x})\rangle$$
$$= \left\langle \delta_{\vec{c}}(\vec{x}), \frac{D}{x_3^2}\phi\left(\frac{Dx_1 - \lambda_1(D-x_3)}{x_3}, \frac{Dx_2 - \lambda_2(D-x_3)}{x_3}\right)\right\rangle$$
$$= \frac{D}{c_3^2}\phi\left(\frac{Dc_1 - \lambda_1(D-c_3)}{c_3}, \frac{Dc_2 - \lambda_2(D-c_3)}{c_3}\right) = \frac{D}{c_3^2}\delta_{\tilde{c}}(\phi),$$
$$\text{where } \tilde{c} = \left(\frac{Dc_1 - \lambda_1(D-c_3)}{c_3}, \frac{Dc_2 - \lambda_2(D-c_3)}{c_3}\right). \quad (9)$$

It's easy to see that $\tilde{c}$ is the perspective projection of $\vec{c}$ in the geometry of $\mathfrak{D}_{\lambda_1,\lambda_2}$.

For the sum of Diracs $f = \sum_{j=1}^n \delta_{\vec{c}_j}$

$$\mathfrak{D}_{\lambda_1,\lambda_2} f = \sum_{j=1}^n \frac{D}{c_{j3}^2}\delta_{\tilde{c}_j},$$
$$\tilde{c}_j = \left(\frac{Dc_{j1} - \lambda_1(D-c_{j3})}{c_{j3}}, \frac{Dc_{j2} - \lambda_2(D-c_{j3})}{c_{j3}}\right). \quad (10)$$

**DCCs.** The DCCs for functions from [4] state:

**Theorem 4.1.** *Define*

$$J_k(\lambda_1,\lambda_2,U,V) = \int_{-\infty}^{+\infty} g(\lambda_1,\lambda_2,u,v)(uU+vV)^k dudv \quad (11)$$

*for all $k = 0,1,2,\dots$ Then $J_k(\lambda_1,\lambda_2,U,V) = \mathscr{P}_k(U,V,-\lambda_1U-\lambda_2V)$, $\mathscr{P}_k(U,V,W)$ is a homogeneous polynomial of degree $k$ and $g(\lambda_1,\lambda_2,\cdot,\cdot)$ has a compact support for all $(\lambda_1,\lambda_2)$ if and only if $g = \mathfrak{D}f$ with compactly supported $f$ in $z > 0$.*

We can generalize the necessary part of these DCCs to distributions of compact support:

**Theorem 4.2.** *If $f \in \mathscr{E}'(Y_3)$, $g_{\lambda_1,\lambda_2} = \mathfrak{D}_{\lambda_1,\lambda_2} f$ is the cone-beam transform on a plane of $f$ for fixed $\lambda_1$, $\lambda_2$, then:*

1. $g_{\lambda_1,\lambda_2} \in \mathscr{E}'(\mathbb{R}^2)$,
2. *for $k = 0,1,2,\dots$ we have the moment conditions:*

$$(g_{\lambda_1,\lambda_2}(u,v), (uU+vV)^k) = \mathscr{P}_k(U,V,-\lambda_1U-\lambda_2V), \quad (12)$$

   *where $\mathscr{P}_k(U,V,W)$ is a homogeneous polynomial of degree $k$.*

*Proof.* Let us prove here the moment conditions that we plan to use. Obviously $(u,v) \longmapsto (uU+vV)^k \in \mathscr{E}(\mathbb{R}^2)$, then

$$(\mathfrak{D}_{\lambda_1,\lambda_2} f(u,v), (uU+vV)^k) = \langle f(\vec{x}), \mathfrak{D}^*_{\lambda_1,\lambda_2}((uU+vV)^k)(\vec{x})\rangle$$
$$= \left\langle f(\vec{x}), \frac{D}{x_3^2}\left(\frac{Dx_1 - \lambda_1(D-x_3)}{x_3}U + \frac{Dx_2 - \lambda_2(D-x_3)}{x_3}V\right)^k\right\rangle$$
$$= \left\langle f(\vec{x}), \frac{D}{x_3^{k+2}}(Dx_1U + Dx_2V + (D-x_3)(-\lambda_1U-\lambda_2V))^k\right\rangle$$
$$= \left\langle f(\vec{x}), \frac{D}{x_3^{k+2}}\sum_{\substack{i,j,l\\ i+j+l=k}}\frac{k!}{i!j!l!}(Dx_1U)^i(Dx_2V)^j\right.$$
$$\left.\times((D-x_3)(-\lambda_1U-\lambda_2V))^l\right\rangle = \sum_{\substack{i,j,l\\ i+j+l=k}}\frac{k!}{i!j!l!}U^iV^j$$
$$\times(-\lambda_1U-\lambda_2V)^l\left\langle f(\vec{x}), \frac{D}{x_3^{k+2}}(Dx_1)^i(Dx_2)^j(D-x_3)^l\right\rangle$$
$$= \mathscr{P}_k(U,V,-\lambda_1U-\lambda_2V).$$

□

**Non-uniqueness of the solution.** Let $f_{M,\vec{t}}(\vec{x}) := f(M\vec{x}+\vec{t})$ with $M = \begin{pmatrix} 1 & 0 & -(u'+\lambda_1')/D \\ 0 & 1 & -(v'+\lambda_2')/D \\ 0 & 0 & 1 \end{pmatrix}$ and $\vec{t} = (\lambda_1', \lambda_2', 0)^T$, then it can be shown for functions

$$\mathfrak{D} f_{M,\vec{t}}(\lambda_1, \lambda_2, u, v) = \mathfrak{D} f(\lambda_1+\lambda_1', \lambda_2+\lambda_2', u-u', v-v'). \tag{13}$$

It can be generalized to distributions $f = \delta_{\vec{c}} \in \mathscr{E}'(Y_3)$. Let us define $f_{M,\vec{t}} \in \mathscr{E}'(Y_3)$ as $\langle f_{M,\vec{t}}(\vec{x}), \phi(\vec{x})\rangle = \langle f(\vec{x}), \phi(M^{-1}(\vec{x}-\vec{t}))\rangle$, where $M^{-1} = \begin{pmatrix} 1 & 0 & (u'+\lambda_1')/D \\ 0 & 1 & (v'+\lambda_2')/D \\ 0 & 0 & 1 \end{pmatrix}$, thus $M^{-1}(\vec{x}-\vec{t}) = \begin{pmatrix} x_1 - \lambda_1' + \frac{u'+\lambda_1'}{D}x_3 \\ x_2 - \lambda_2' + \frac{v'+\lambda_2'}{D}x_3 \\ x_3 \end{pmatrix}$. Then $(\delta_{\vec{c}})_{M,\vec{t}}$ is the distribution $\delta_{M^{-1}(\vec{c}-\vec{t})} \in \mathscr{E}'(Y_3)$. Then it's easy to show

$$\begin{aligned}(\mathfrak{D}_{\lambda_1,\lambda_2}(\delta_{\vec{c}})_{M,\vec{t}}(u,v), \phi(u,v)) &= \\ &= (\mathfrak{D}_{\lambda_1+\lambda_1',\lambda_2+\lambda_2'}\delta_{\vec{c}}(u-u', v-v'), \phi(u,v)).\end{aligned} \tag{14}$$

If we shift the detector by $u'$, $v'$ and the source positions by $-\lambda_1'$, $-\lambda_2'$, then there exists another object with the same projection data from the original source and detector positions. Thus, the source positions cannot be identified better than up to a global shift (idem for the detector shifts) from the data only.

**Derivation of the algorithm.** If we write projection data as

$$m_i^l(u,v) = \mathfrak{D}_{\lambda_{1i},\lambda_{2i}} f^l(u-u_i, v-v_i), \tag{15}$$

$f^l = \sum_{j=1}^4 \delta_{\vec{c}_j^l}$, then from the moments of order 1 of the type $M_1^l(i) = (m_i^l(u,v), u)$ and moments of order 2 of the type $M_2^l(i) = (m_i^l(u,v), u^2)$ we can derive formulas (3), (4), (5). Note that along with the direct calculation of moments for (15), we used that we can compute the same moments with the detected points $q_{ij}^l$ as $M_1^l(i) = \frac{D}{(C_3^l)^2}\sum_{j=1}^4 q_{ij}^l$, $M_2^l(i) = \frac{D}{(C_3^l)^2}\sum_{j=1}^4 (q_{ij}^l)^2$.

## 5 Numerical results

For numerical simulations we launched our algorithm twice: for the first part of the calibration task to find $\lambda_{1i}$, $u_i$ and for the second task to find $\lambda_{2i}$, $v_i$. All values of parameters are given in cm:

1. The known parameters of the calibration cage pattern: $L = 0.4$, $k_1 = 3$, $k_2 = 1$, $k_3 = 2$. We used the same pattern for the group of vertical sticks and for the group of horizontal sticks.

2. The true positions of sticks: $p_a = 5$, $p_b = 8.2$, $p_c = 4$, $p_d = 7.2$, $C_3^a = 8$, $C_3^b = 9.5$, $C_3^c = 8$, $C_3^d = 9.5$.

3. The true calibration parameters: we randomly selected $P = 30$ values for source positions in $[0, 10]$ and fixed $\lambda_{10} = 0$. We chose the grid on $u \in [0, 10]$ with the sampling step 0.01. The detector jitters $u_i$ were generated as random uniform noise on the interval $[-0.05, 0.05]$, $u_0 = 0$. The same was done for the sets of $\lambda_{2i}$, $v_i$.

4. The source-detector distance is fixed $D = 10$.

| Noise level | Noise std | MAE for $\lambda_{1i}$, $\lambda_{2i}$ | MAE for $u_i$, $v_i$ | MAE for $p_l$ | MAE for $C_3^l$ |
|---|---|---|---|---|---|
| 0% | 0 | $3.86E-13$ | $9.67E-14$ | $6.50E-14$ | $7.65E-14$ |
| 10% | 0.001 | $1.80E-2$ | $3.79E-3$ | $3.26E-3$ | $4.42E-3$ |
| 50% | 0.005 | $1.02E-1$ | $2.08E-2$ | $1.66E-2$ | $2.25E-2$ |
| 100% | 0.01 | $1.89E-1$ | $3.76E-2$ | $2.82E-2$ | $4.37E-2$ |
| 200% | 0.02 | $3.87E-1$ | $7.97E-2$ | $6.47E-2$ | $8.91E-2$ |

**Table 1:** Mean absolute errors (MAE) for calibration parameters and positions of the markers; all errors are in cm.

In the Table 1 we present the results of our calibration algorithm from two oblique planes. To simulate detection errors, we added to $q_{ij}^l$ realisations of the Gaussian noise $N(0, \sigma)$, $\sigma = 0.01 \cdot n_l$, where $n_l$ is the noise level, 0.01 is the pixel size of the initial image.

## 6 Conclusion

We have presented a hybrid approach to calibrate cone-beam projections with sources on a plane parallel to the detector with a marker set of partially known geometry and DCCs generalized to compactly supported distributions. We used DCCs on geometric projections of the spherical markers. Thus, DCCs can be computed if projections of the marker set are non-truncated (the rest of the object can be truncated). This is the main advantage of the approach. One disadvantage of our method is the placement of the calibration cage: it has to be parallel to the source and detector planes. Moreover, we see in the Table 1 that our algorithm is sensitive to detection errors. This method requires further numerical simulations and comparisons with other self-calibration methods.

# Improving CT Image Segmentation Accuracy Using StyleGAN Driven Data Augmentation

Soham Bhosale[1], Arjun Krishna[1], Ge Wang[2], and Klaus Mueller[1]

[1]Department of Computer Science, Stony Brook University, Stony Brook, NY, USA
[2]Biomedical Imaging Center, School of Engineering, Rensselaer Polytechnic Institute, Troy, NY, USA

**Abstract** Medical Image Segmentation is a useful application for medical image analysis including detecting diseases and abnormalities in imaging modalities such as MRI, CT etc. Deep learning has proven to be promising for this task but usually has a low accuracy because of the lack of appropriate publicly available annotated or segmented medical datasets. In addition, the datasets that are available may have a different texture because of different dosage values or scanner properties than the images that need to be segmented. This paper presents a StyleGAN-driven approach for segmenting publicly available large medical datasets by using readily available extremely small annotated datasets in similar modalities. The approach involves augmenting the small segmented dataset and eliminating texture differences between the two datasets. The dataset is augmented by being passed through six different StyleGANs that are trained on six different style images taken from the large non-annotated dataset we want to segment. Specifically, style transfer is used to augment the training dataset. The annotations of the training dataset are hence combined with the textures of the non-annotated dataset to generate new anatomically sound images. The augmented dataset is then used to train a U-Net segmentation network which displays a significant improvement in the segmentation accuracy in segmenting the large non-annotated dataset.

## 1 Introduction

CT Image Segmentation is a hallmark of Computer-Aided Diagnosis (CAD). Medical professionals often use it to isolate specific areas of interest in a medical image. This allows them to conduct effective medical analyses for detecting abnormalities such as lung cancer and positioning implants [1]. However, manual segmentation can become time-consuming and complicated depending on what part of a body is being analyzed. Deep learning can make this task significantly more efficient, affordable, and accessible. Nevertheless, this approach is currently very ineffective due to the low availability of large labeled and annotated training datasets. The reasons for this shortage include privacy concerns and the high cost of labeling CT datasets by human experts.

In addition, the added challenge with segmenting a CT dataset is that there is a lot of variability between the publicly available datasets. For example, the images frequently have different dosages which is the difference in amount of x-ray radiation used to synthesize CT images or they could be generated from several different machines each having different parameters for CT reconstruction, which for example could mean the thickness of tissue that each image slice represents may vary. Hence, if we train a segmentation network with a dataset that has images of different dosage and scanner properties than the set of images we intend to segment then the deep network will have low accuracy.

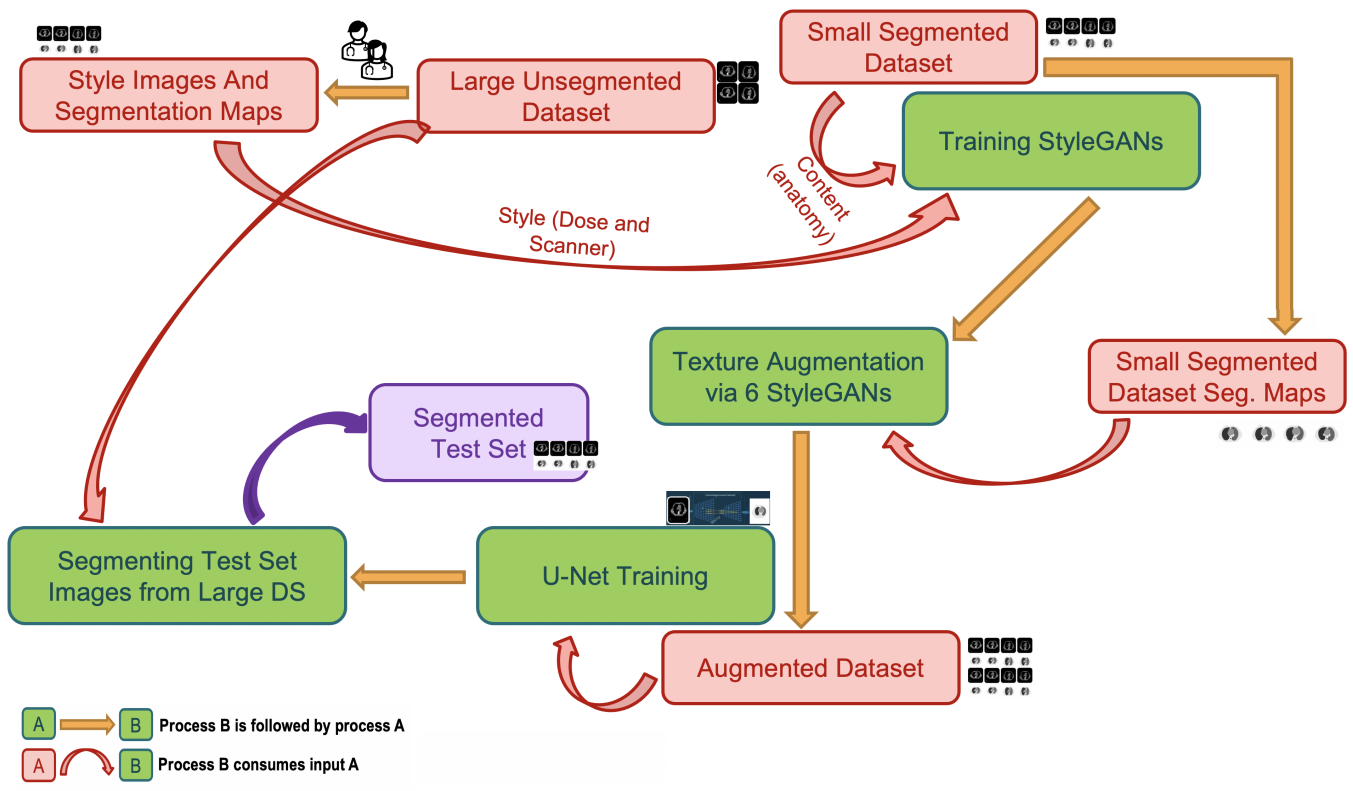

**Figure 1:** Flow starts at the top right corner with the two datasets - a small segmented and a large unsegmented dataset. It illustrates the inputs and outputs involved in training the StyleGANs and the U-Net to finally segment the test set taken from the large dataset.

We present a new strategy for augmenting segmented datasets which eliminates the differences between the segmented dataset we use for training and the dataset we wish to segment, by using texture learning. This strategy will create synthetic images which imitate dosage and scanner properties of the unsegmented dataset and maintain anatomical accuracy within the process.

As shown in Figure 1, we build on our previous work of texture learning [2] to expand our small annotated dataset with textures present in the large unlabeled dataset. In the first step, we ask an expert, a radiologist in this case, to manually segment a few CT images from the large non-segmented dataset which were chosen as "style images" for our StyleGAN training. We then use the StyleGAN architecture presented in Krishna et al. [2] to create new CT images based on the texture of a style image and the content, or segmentation maps, of another image.

We train 6 different StyleGANs with 6 separate texture images. Since these texture images came from the non-annotated dataset which contains the images we want to segment, the newly generated segmented images will make good candidates for training the segmentation network. Next, we take all the segmentation maps of the small segmented dataset and use it as "content" inputs to each of 6 trained StyleGAN networks to generate new images. This technique augments the segmented dataset by 7 folds and creates additional images with sample textures from the unsegmented dataset. We then train a U-Net [3] segmentation network with the augmented segmented dataset. The trained U-Net is then used to segment all the images of the large non-annotated dataset. Segmentation accuracy of the network is computed on a test set that consists of around 100 randomly chosen images from the non-annotated dataset. In the absence of the ground-truth for the test set, annotations are created manually with the assistance of a few radiologists.
To the best of our knowledge, we are the first to enhance segmentation accuracy by reducing texture differences between training and testing CT datasets using StyleGANs.

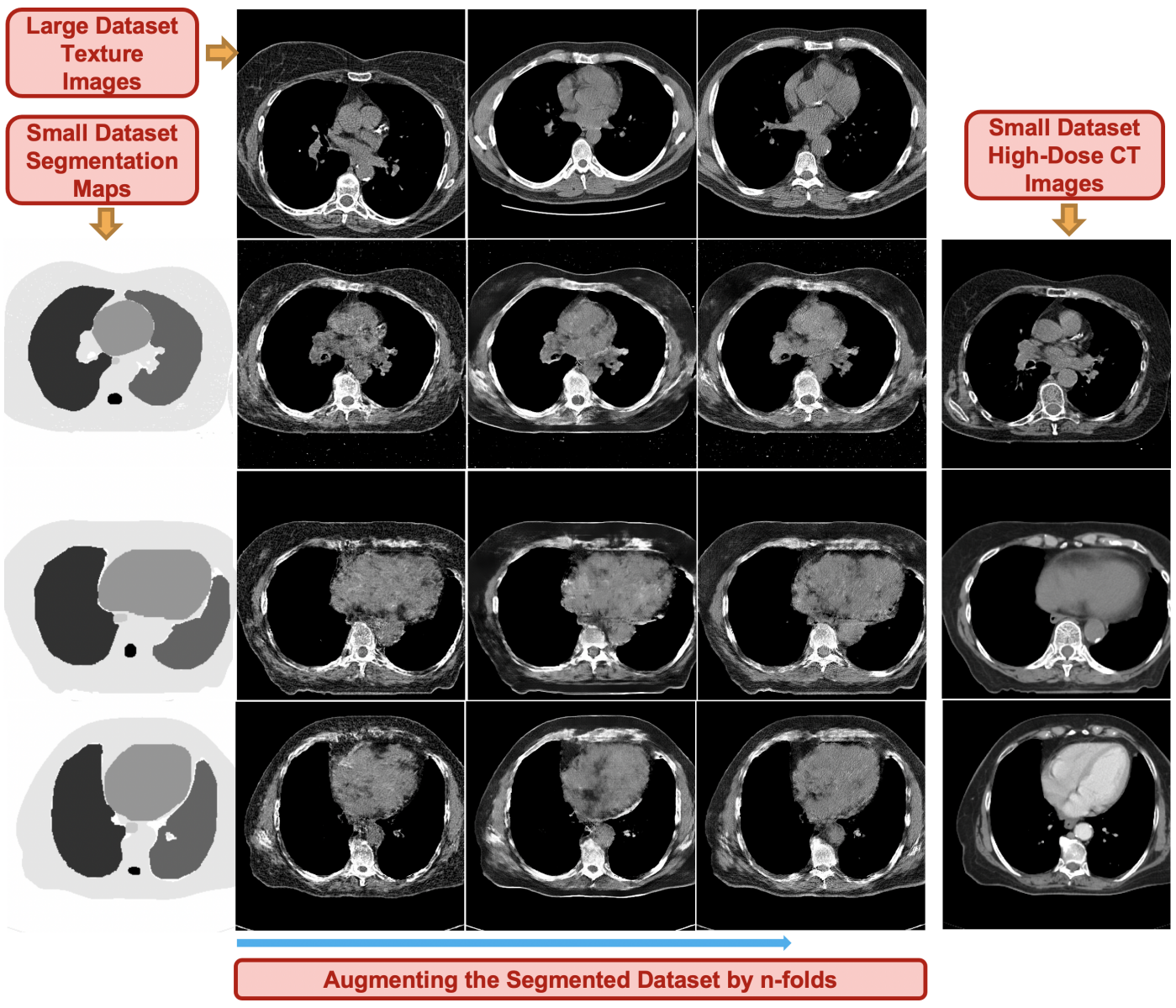

**Figure 2:** Our Expanded Dataset. For "m" images in annotated dataset and "n" style images, expanded dataset size is mx(n+1). Note, that the heart anatomy isn't perfect in the generated images, but as long as generated images could exhibit enough anatomy and texture, it could still help in improving final segmentation accuracy.

## 2 Materials and Methods

Fig. 1 highlights each step of the process. As discussed, there are three major steps: training the StyleGAN networks, augmenting the annotated dataset, and then training the segmentation U-Net with this augmented dataset. We will briefly describe each step below

### 2.1 Datasets

The top right corner of Fig. 1 highlights the two different datasets (in red colored boxes) that were involved in this work. The smaller dataset consists of 512x512 high dose chest CT images along with their segmentation maps of thirty patients depicting their lungs, heart, spinal cord, esophagus, surrounding tissue and bones within their torso. The larger dataset consists of non-annotated low dose chest CT images of similar resolution of around 14k patients taken with several different scanners at several different locations. As stated before, for using the datasets together for improving segmentation, we pre-process and augment the smaller annotated dataset multiple folds [2] using CT image textures present in the larger dataset.

### 2.2 StyleGAN

We use our encoder-decoder based styleGAN architecture and segment-wise style loss [2] for learning and generating segment-wise textures of heart, torso and surrounding tissues present in the larger dataset. For this, we first choose five or six different "style images" from the large dataset. Since the dataset contains low-dose CT scans collected over a period of three years from several different scanners, it contains a few distinct textures over all the CT-scans corresponding to distinct noise characteristics of these scanners. The "style-images" are manually selected corresponding to these distinct textures. For our large dataset six style-images seem enough to cover all the different texture features present in the dataset. Images in the top row of Figure 2 show three such style images. Note that as per our styleGAN architecture, the segmentation maps of both "style images", from the large dataset, and "content images", from the small dataset, are needed for generating CT images having textures of the large dataset's images while retaining anatomy of the small dataset's "content images". This is maintained by our segment-wise style loss, $L_s$ and content loss, $L_s$ which is as follows:

$$L_s = \sum_{sg \in SG} \sum_{l \in SL} \frac{1}{4N_l^2 M_l^2} \|G(R_{sg}^l) - G(S_{sg}^l)\|_F^2 \tag{1}$$

$$L_c = \sum_{sg \in SG} \sum_{l \in CL} \frac{1}{2N_l M_l} \|R_{sg}^l - O_{sg}^l\|_2^2 \tag{2}$$

where $O^l$, $S^l$, and $R^l$ denote the feature maps extracted from the pre-trained VGG network at layer l, for the original image $x_o$, the style image $x_s$ and the stylized image output $x_r$ respectively; $G(R^l)$ and $G(S^l)$ denote the encoded Graham Matrices [4] of those feature maps; SG is the set of all six segments including heart, torso, lungs, spinal cord and esophagus in the CT images and SL/CL is the set of all style/content layers in the VGG network. Since, the large training dataset's CT scans are not segmented, we manually build the segmentation maps of the chosen six style images with the assistance of a radiologist, as mentioned.
The StyleGAN architecture proposed by Krishna et al. [2] takes the segmentation map of one image $x_o$ and the segments' texture of another image $x_s$ as input, and outputs a new image $x_r$ with the contents (organs) of the segmentation map $x_o$ and style of image $x_s$. Similar to the original work, all the training images are resized to 512x512 pixels. Each StyleGAN is trained for 100 epochs. For the loss function, the input images, generated images, and style images along with their segmentation maps are fed into a VGG-19 network. The feature maps generated by VGG-19 are used to calculate style loss and content loss. Both content loss and style loss are calculated separately for each segmented region and then combined.

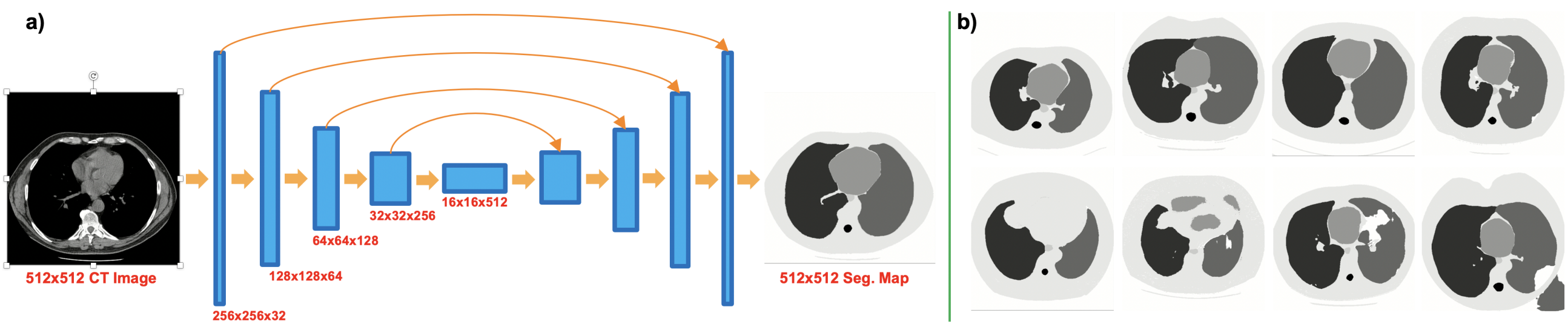

**Figure 3:** Left: A standard U-Net network which segments a low-dose soft-tissue CT image into lungs, heart, torso, spinal-cord and esophagus. Right: Top row shows some well formed segmentation maps of various ct scans, while bottom row highlights flawed results from the U-Net when trained on a smaller different dataset with high-dose CT images.

### 2.3 Augmenting Segmentation Training Dataset

Next, all the segmentation maps of the smaller CT segmentation training dataset are passed through each trained Style-GAN networks as inputs generating new CT images for each segmentation map. Note that we train six style generators corresponding to each style image hence augmenting the dataset seven fold. The original 700 images segmented lung CT dataset is now expanded to around 4900 images. In our case, the generated images have textures simulating low dosage CT images similar to the images of the large dataset we intend to segment. Also, all of the generated images have associated annotations or segmentation maps which are the same as the inputs or the annotations of the original dataset

Figure 2 shows a few results of the expanded annotated dataset. In the figure we are showing results for three such generators corresponding to three style images shown in the top row and three segmentation maps corresponding to three "content"/training dataset images shown in the right-most column. Since these generators do not serve as our final segmentation networks, the generated images do not have to be anatomically prefect as long as they can learn and exhibit enough anatomy with the large dataset's texture in generated images that could help in training a U-Net effectively in our last step. We chose the soft tissue window of CT scans for texture learning and augmenting the smaller dataset. Soft tissue images exhibit all the anatomical details necessary for learning annotations of the smaller dataset which in our case are segmentation maps having segments depicting heart, lungs, torso esophagus and spinal cord. These segments are clearly distinguishable in the soft tissue window based on the pixel values of the images. Having the ability to learn annotations or segmentation from the augmented dataset will enable the trained U-Net to recognize the CT images' segments from the larger dataset having similar textures.

### 2.4 Training the U-Net

We train a U-Net [3] for creating a segmentation map as output when given a chest CT image as input. We chose the U-Net architecture due to its excellent performance for segmenting biomedical images, especially when coupled with data augmentation techniques like elastic deformation. The architecture of the network is based on the original work done by the authors where each blue block in Figure 3a consists of two convolutional layers with a batch normalization layer in between those layers and a max-pooling layer as the last layer. In the up-sampling part, max-pooling layers are replaced by the transposed convolutions layer as the first layer followed by the two convolutional layers. As shown in the figure the results of each block in the down-sampling part are concatenated to their corresponding block in the up-sampling half. Kernels of size 3x3 are used for convolutional layers with a stride of 2. We use the augmented textured annotated dataset created in the previous step for training our U-Net. Having similar texture across the two different datasets helps training a segmentation network on the segmented dataset to segment the images of the large non-annotated dataset. We use the trained U-Net to segment randomly chosen 100 CT images of the larger dataset which forms our test set.

We analyze the improvement of the network trained on the augmented CT image dataset as compared to the network trained on the small high-dose image dataset. In the absence of the ground-truth segmentation maps for the test set, we manually corrected the erroneous predicted segmentation outputs of the network models with the help of a few radiologists. The corrected segmentation outputs of the test set images are then treated as the ground truths for the images. The first row of Figure 3b highlights some of these well formed ground truth segmentation maps, while the second row shows the erroneous predicted segmentation maps from the network trained on the smaller high-dose image dataset. We discuss the improvements and compare the two networks in the next section. Both are trained for 30 epochs with a batch size of 16.

## 3 Results, Further Work and Conclusion

Figure 4 highlights some of our results. We observe that the U-Net trained with our style learning based augmented data (U-Net 2) has a greater segmentation accuracy than the network trained with the original annotated dataset (U-Net 1). For instance, the augmented dataset-trained network classified much more of lung and torso matter correctly than the network trained with the original dataset. As for metrics,

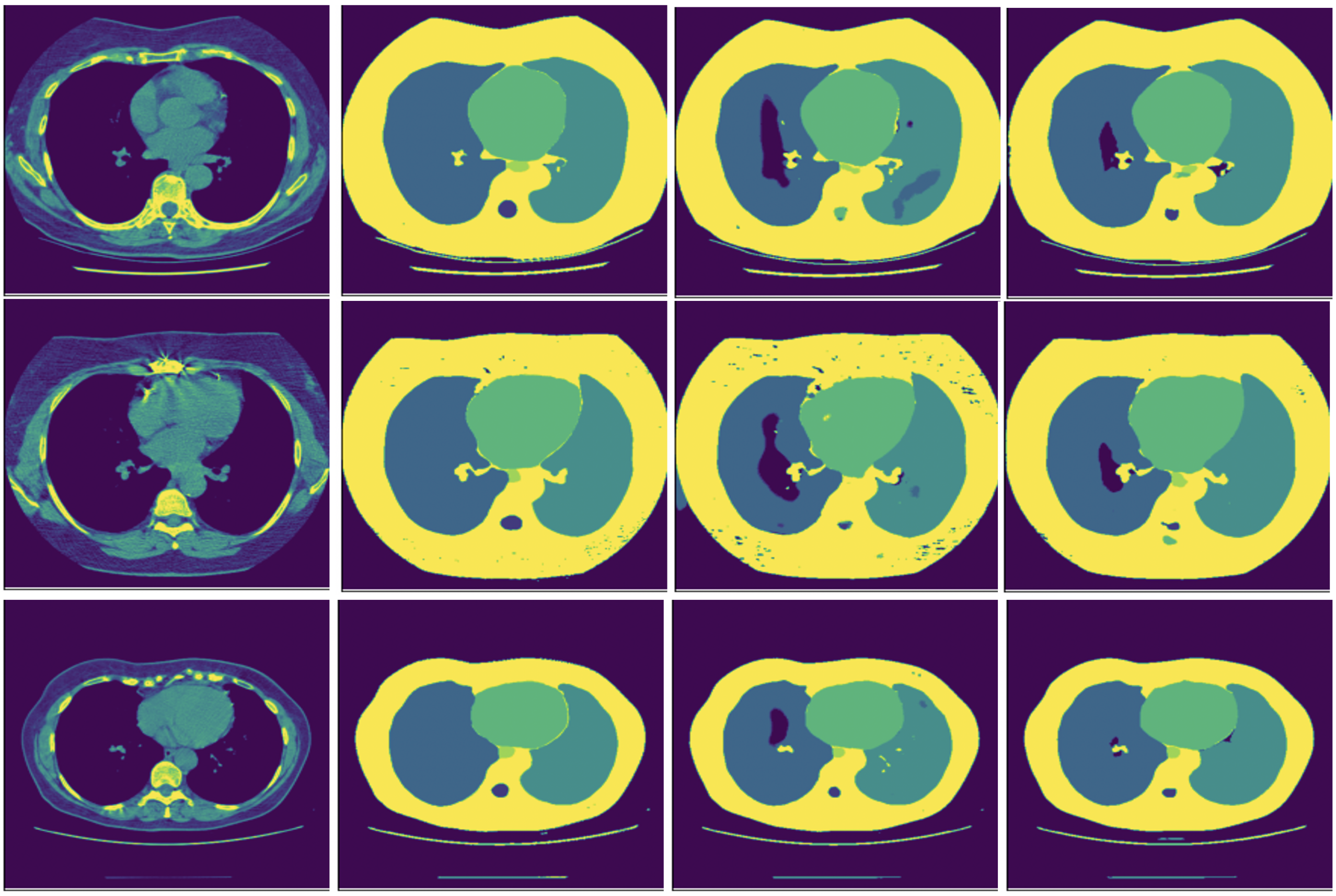

**Figure 4:** The above figures display a side-by-side comparison of the test results of the segmentation network (U-Net 1) trained on the small high-dose CT dataset versus the network (U-Net 2) trained on the augmented dataset. The segmentation maps generated by the augmented dataset show greater visual accuracy than those by the network trained on the small dataset.

U-Net 2 has a testing accuracy of 95.7% whereas the U-Net 1 has a testing accuracy of 93.1%.

We also observe that we still have segmentation flaws around smaller segments like esophagus and spinal cord. Since the small dataset represents limited anatomy of a few patients, there are errors in the segmentation outputs of the patients in the larger dataset with more varied anatomy. This becomes more evident around smaller segments since the segmentation maps of the small dataset are on an average significantly smaller, i.e., have less spatial detail, than those of the large dataset which is due to the differences in positioning of the scanner lens between the two datasets.

One way to resolve this is to explore the PCA space of segments within segmentation maps as we did in Krishna el al. [2] to generate CT images with varied anatomy through our styleGANs for augmenting the small annotated dataset. Also, we propose to further improve our segmentation network accuracy by training U-Nets and StyleGANs in a cyclic approach. We will use StyleGANs to augment a dataset, pass it through the segmentation network, and then use the segmentation maps generated by this network to provide additional data for the StyleGAN training and then repeat the process. Another option is to use a combination of all CT window images including lung and bone windows for segmenting the images. We propose to use a combination of above mentioned strategies to further improve our segmentation accuracy. According to these results, augmenting medical datasets using StyleGANs has proven to be a promising method to resolve texture differences between different medical datasets paving a way to analyze one dataset based on the annotations of another. In the segmentation task of lung CT, the StyleGANs assisted expanded dataset helped a U-Net to improve its accuracy by 2.6% on a completely different dataset. Hopefully, this new technique can mitigate the annotated medical data shortage issue that is required for training data intensive deep learning networks for various applications.

# Novel Lung CT Image Synthesis at Full Hounsfield Range With Expert Guided Visual Turing Test

Arjun Krishna[1], Shanmukha Yenneti[1], Ge Wang[2], and Klaus Mueller[1]

[1]Department of Computer Science, Stony Brook University, Stony Brook, NY, USA
[2]Biomedical Imaging Center, School of Engineering, Rensselaer Polytechnic Institute, Troy, NY, USA

**Abstract** Conventional image quality metrics are unsuitable to evaluate the realism and medical accuracy of synthetically generated CT images. We describe an approach based on the concept of Visual Turing Test that engages medical professionals to assess the generated images and provide useful feedback that can inform the generative process. We first describe our approach for synthesizing large numbers of novel and diverse CT images across the full Hounsfield range using a very small annotated dataset of around thirty patients and a large non-annotated dataset with high resolution medical images. Using an anatomy exploration interface we can generate CT images with anatomies that were non-existent within either of the datasets, without compromising accuracy and quality. Our approach works for all Hounsfield windows with minimal depreciation in anatomical plausibility. We then describe our Visual Turing Test methodology in detail and show results we have obtained.

## 1 Introduction

Deep learning in medical applications is limited due to the low availability of large labeled, annotated or segmented training datasets. The scarcity persists not only because of privacy and ownership concerns but also because of the high cost of labeling such datasets by human experts. Likewise, publicly available annotated high resolution image datasets are also often very small or even non-existent.

In this work we first present a methodology that reduces or even eliminates he problem of such small datasets by converting them into large datasets without the loss of anatomical accuracy. Our approach goes beyond simple data augmentation techniques like stretching or flipping existing images and adds new data instances with anatomies that may not even exist in these datasets. With this approach we are able to increase not only the size but the overall diversity of images in datasets significantly.

Our method uses a dataset of segmented CT images from thirty patients and a large dataset of unsegmented CT images. Our method builds on our previous work of texture learning [1] to expand the small annotated dataset with textures present in the large dataset. Subsequently we extract segmentation maps from the unsegmented large dataset via a trained U-Net. Next we train a cycleGAN on both the small segmented data and large unsegmented data in an alternate fashion to generate new images with segmentation maps as inputs. This synthesis step expands on our previous work [2] and explores the PCA space of segmentation maps in conjunction with the cycleGAN to create CT images with novel anatomies not present in either of the datasets.

Since commonly used image quality metrics are unsuitable to evaluate the realism and medical accuracy of synthetically generated CT images, we have designed a framework that engages medical professionals to assess the generated images along these qualitative figures of merit. Our evaluation interface is based on the concept of Visual Turing Test and provides several design elements to determine the degree of realism and the sources of anatomical imperfections.

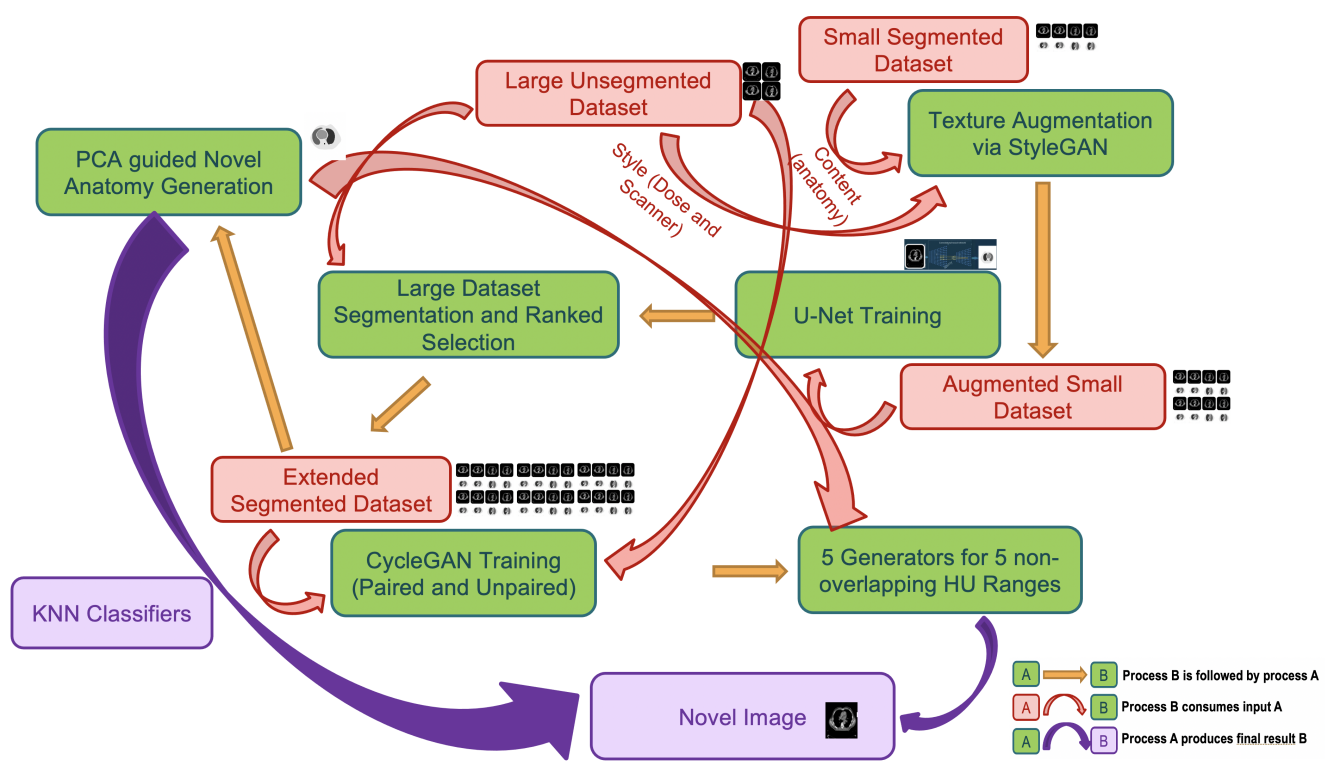

**Figure 1:** Flow starts at the top right corner with two datasets - a small segmented and a large unsegmented dataset. Three different Deep-Learning networks are used starting from a StyleGAN followed by a U-NET segmentation network and 5 CycleGANs which train generators for the final step.

## 2 Our CT Synthesis Methodology

Figure 1 highlights our sequence of steps. We will briefly summarize each step in the same sequence below.

**Texture Augmentation.** The smaller dataset consists of chest CT scans with segmentation maps (lungs, heart, etc.) of 30 patients. The larger dataset consists of non-annotated chest CT scans of ~14k patients. To use the two datasets together we modified the textures of the smaller dataset with those of the larger one, augmenting the smaller annotated dataset 3-fold. We used the network architecture of [1] for segment-wise texture learning and created new CT images with the anatomy from the small dataset and the textures from the larger dataset.

**Further Augmentation from Label Training.** We train a U-Net [3] to output a segmentation map given a chest CT image as input. We use the augmented annotated dataset created in the previous step for training our U-Net. Having similar textures across the two different datasets helps in training a segmentation network on one dataset to segment the images of another. We use the trained U-Net to segment all 14k patient images. Since the smaller dataset has limited anatomy, there are errors in the segmentation outputs of the larger dataset. k-NN classifiers are used to rank them by accuracy using certain characteristics of the segmentation images. We choose the best 1/4 of segmentation outputs and add them with their CT scans to the smaller segmented dataset. This dataset along with the larger dataset of unsegmented images is then used to train the generators for the synthesis.

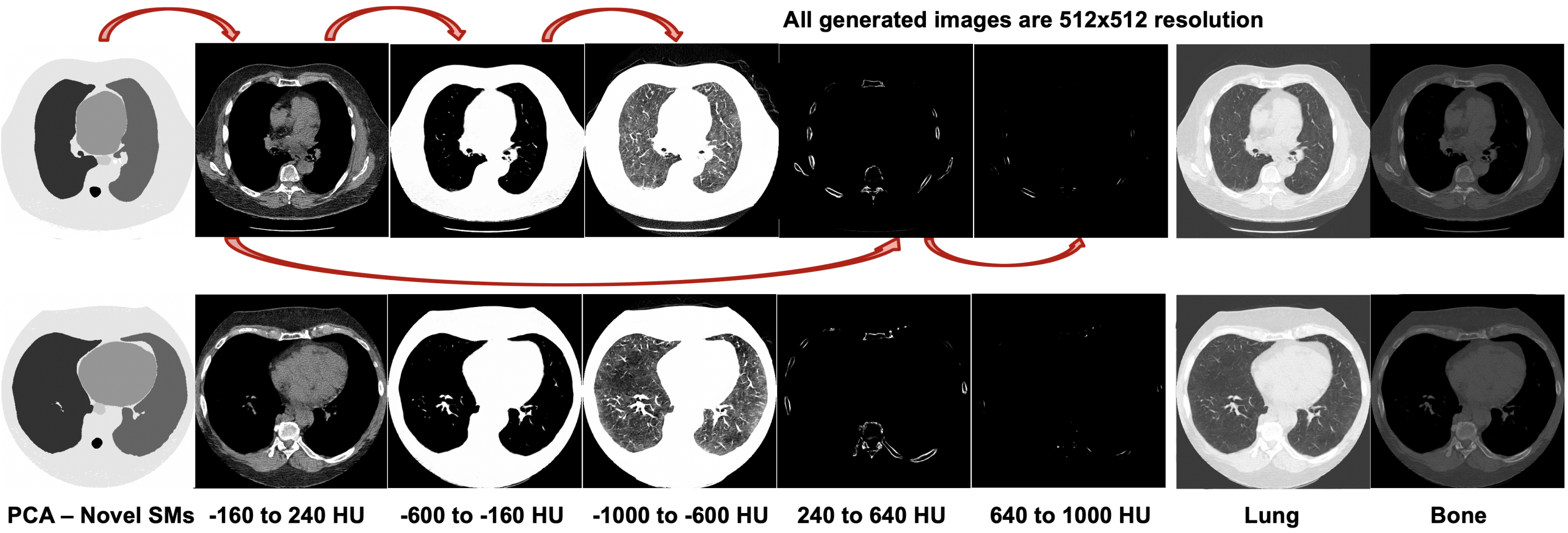

**Figure 2:** Above figure shows two examples of novel CT scans generations. The sequential training and generation learns the correlations of anatomical details and can be clearly seen within the columns as we move from left to right. The last two columns depict the anatomical consistency observed in different HU windows than in generated ones. Each red arrow represents a generator of the two generators trained in a cycleGAN setup for corresponding modalities.

**Decomposing the Hounsfield Range for Generation Steps.** Our method generates images at full Hounsfield in five separate steps. Fig. 3a shows the average distribution of pixels values of a chest CT-scan over HU values. Fig. 3b shows an image in (-160, 240) HU range while Fig. 3c shows an image in (-600, -1000) HU ranges. Two separate generators are used to generate these HU ranges thereby assisting the GANs to focus on the minute details within these ranges since discriminators within a GAN setup focus on the accuracy of the majority group of pixels within a particular HU range. Hence we use five generators to generate five distinct sets of images for five distinct HU ranges for a single CT image generation. We first generate the middle HU range image using the segmentation map as input since it details the major anatomical features such as bones and organs. We then use this generated image as input for generating the other HU range images. This is shown in detail in Fig. 2.

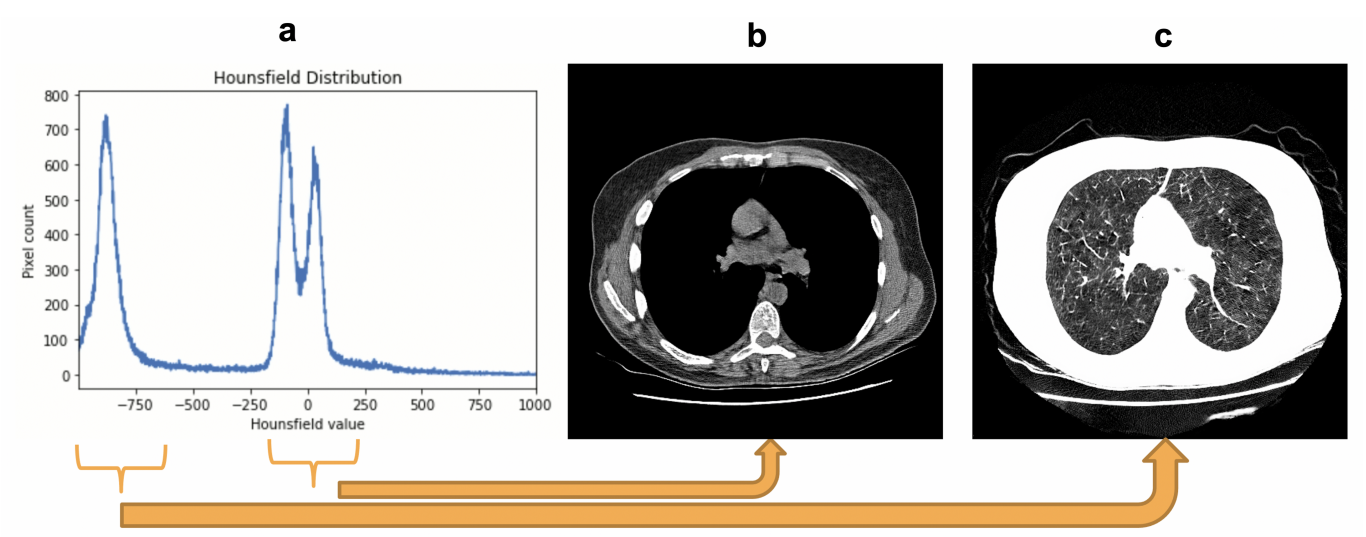

**Figure 3:** We use 5 CycleGANs to train 5 generators for 5 non-overlapping HU ranges (-1000, -600), (-600, -160), (-160, 240), (240, 640), (640, 1000)

**Paired and Unpaired Training via CycleGAN.** To generate the CT images we follow the network architecture of [4] for paired and unpaired training. We use a different algorithm and data setup for training since our paired and unpaired datasets come from different sources. We use only the large CT dataset for unpaired training while we use all the segmentation maps for both paired and unpaired training. Training was done in an alternate fashion; every iteration of paired training was followed by two iterations of unpaired training to learn the anatomical diversity present in the unsegmented dataset. As mentioned before, we have five such setups to produce five relevant generators to cover all five HU ranges. Figure 2 shows the image synthesis sequence we use to cover the full HU-range. Shown are two CT images which exhibit novel anatomy. The left two columns demonstrate their anatomical consistency in the lung and bone windows.

**Addition of Segmentation Maps via PCA.** The larger dataset contains CT scans of around 14k patients while we have segmentation maps for only 3k patients. To balance the number of segmentation maps with CT-scans for training the cycleGAN we interpolate new segmentation maps in the PCA space of existing ones. For this we used our previous methodology [2] of representing segmentation maps as a set of B-Spline curves. Since interpolations may not be perfect anatomically we use k-NN classifiers to rank the validity of segmentation maps and chose the best ones as input for training the generators in paired/unpaired training in the cycleGAN [4] setup. The creation of new segmentation maps also helps in creating CT images with novel anatomy.

## 3 Our Visual Turing Test for Evaluation

Some of the popular metrics generally used to evaluate generated medical images are Structural Similarity Index (SSIM), Peak Signal to Noise Ratio (PSNR), Fréchet inception distance (FID) and Inception Score (IS) among others. These metrics are a good representation of how much the generative model is able to mimic the training distribution and some metrics even give us a good idea of how much a model is able to diversify its outputs. When evaluating models that generate medical images like Computed Tomography (CT), Magnetic Resonance Imaging (MRI), Chest X-rays etc. a fundamental aspect to be considered is to verify the “medical accuracy” of the generated images. Currently, no metric can provide us with such evaluation of generative models used in medical imaging. Metrics such as FID and IS have a large dependence on the pre-trained networks which can be troublesome when the model fails to capture spatial relationships between various parts of the image. Other popular metrics such as PSNR and SSIM are numerical metrics that could be more reliable but they have been shown to be closely related to Mean Square Distance / Error (MSE) [5] for two images which is widely known to be poorly correlated with human perception of image quality or anatomical accuracy. This is a big drawback of these metrics in context of anatomical accuracy. So, we propose using the ability of humans having expertise in CT to assess our generated lung CT images and provide a better description of the generated images in the form of a Visual Turing Test for Medical Images.

**Introduction - Visual Turing Test.** The Visual Turing Test is a variation of the Turing Test that was first introduced by Geman et al [6] as a way to measure the level of understanding of a computer vision model. In the area of medical imaging this test was used to evaluate models based on how realistic synthetic medical images are. Chuquicusma [7] applied it to evaluate generated malignant and benign lung nodules while Han et al [8, 9] used it to evaluate brain MR images. The test is administered to human experts by showing them a randomly chosen medical image from a set of real and generated images one at a time in a random order. The expert then proceeds to give a feedback for each image shown to them without any knowledge of their actual labels. The feedback involves the experts’ opinion of whether an image is obtained from a real patient (Real) or whether it is a computer generated image (Fake). The primary idea of this test is to assess if a model is successfully able to generate medically accurate images which can be determined by measuring the number of times the model is able to fool experts into thinking that a model generated medical image is in fact a medical image obtained from a real human being. When experts are unable to separate the images into real or fake at least 50% (chance baseline) of the time, the model is said to have passed the visual Turing test.

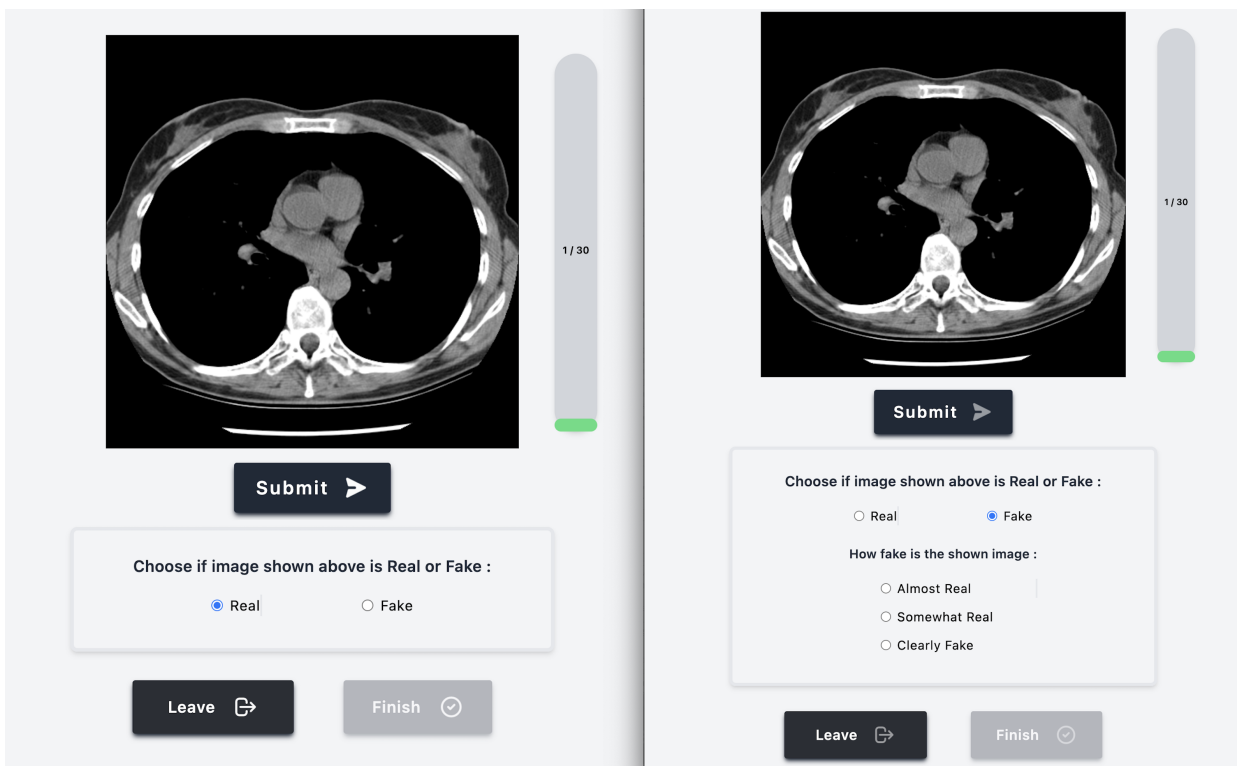

**Figure 4:** Left: Expert user evaluation interface for our Visual Turing Test. They can choose one of "Real" and "Fake" options. Right: More options pop up if they chose "Fake".

**Implementation Details.** We designed a website to carry out the Visual Turing Test with a primary focus on evaluating generative frameworks that synthesize lung CT scans. The user interface for this website was created using Next.js (a server side framework built on top of react.js), tailwindCSS, Framer motion and sanity.io (GROQ Queries) as a backend to store all the responses. All responses are stored in a state which is managed by using redux, a state container. The website is hosted using vercel and is live at https://visual-turing-test.vercel.app.
As the test begins, the study participant is presented with an image and 2 options: "Real" or "Fake". If they choose "Fake", a sub-section pops up asking them to choose another option that best represents how fake the image looks. As shown on the right side of Figure 4 they could choose one of the "Almost Real", "Somewhat Real" and "Clearly Fake" options. After choosing a "fakeness level" the participant is shown a window as in Figure 5 where they can mark the areas that look fake in the CT image. Once they are sure of their choices the participant "submits" their feedback. We designed the test to be 30 images long so as not to overwhelm the participants. It ensures their responses are well thought out and yield an accurate measure of anatomical accuracy for our synthesized CT images. The images shown are randomly chosen from one of the three windows namely bone, lung and subdural/soft-tissue.
The test provides the following functionalities:

- Evaluates a model based on human expert feedback.
- Evaluates how close the model is to generating realistic medical images, gauging medical accuracy.
- Collects the areas of an image marked as fake by the expert, which can later be used for training better models.

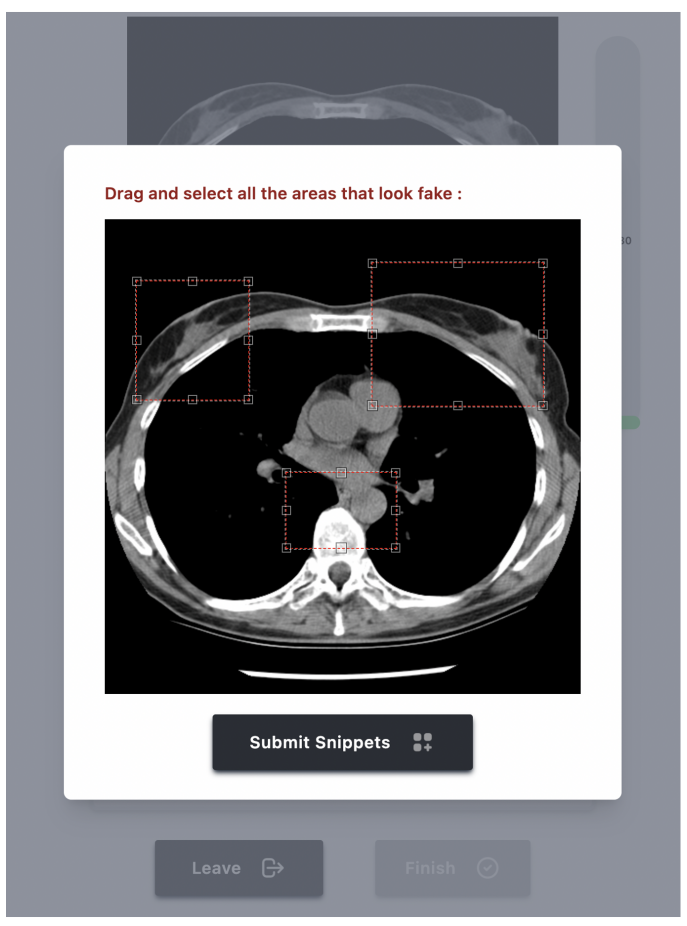

**Figure 5:** The interface that pops up after an expert user submits the "Fake" option for an image. The red boxes indicate the areas of above image annotated by the user that he/she thinks look anatomically too inconsistent for the image to be real.

| Radiologist | Accuracy | TP | TN | FP | FN | Almost Real | Somewhat Real | Clearly Fake |
|---|---|---|---|---|---|---|---|---|
| 1-Competent | 83.33% | 86.67% | 80% | 20% | 13.33% | 20% | 40% | 20% |
| 2-Competent | 73.33% | 93.33% | 53.33% | 46.67% | 6.67% | 26.67% | 26.67% | 0 |
| 3-Proficient | 96.67% | 100% | 93.33% | 6.67% | 0 | 0 | 60% | 33.33% |
| **Average** | 84.44% | 93.33% | 75.56% | 24.44% | 6.67% | 15.56% | 42.22% | 17.78% |

**Table 1:** Statistics of responses given by 3 radiologists

Experts chosen for this test consisted of doctors, radiologists and senior radiology fellows. Since every participant will have different levels of expertise, in order to measure the performance of the model across different levels of expertise, each person is asked to indicate their medical knowledge at the end of the test.

**Results**. The test was taken by 3 radiologists, 1 of whom had "proficient" expertise and the others had "competent" expertise in radiology. Each of these tests presented the radiologists with 30 images belonging to 3 different windows (soft tissue, lung and bone) comprising 15 real CT images and 15 fake CT images in a random order.

Upon analyzing the responses, it was found that senior radiologists that are proficient were able to distinguish between fake and real images better than the radiology fellows who marked themselves as "competent". This can be clearly seen in confusion matrices shown in Fig. 6 where the competent test takers had difficulties in identifying all the fake images. Also, according to the table, among the chosen fake images, very few of them seemed to be "clearly fake" to the non-experts. The statistics shown in Table 1 indicate that the generative framework in consideration has not passed the Visual Turing Test as expert radiologists can easily identify most fake lung CT images from the real ones.

**Analysis - The Heart Issue** On closer inspection of the radiologists' feedback, we found out that expert radiologists were able to identify the fake images because of the anatomical errors in the heart. Our large lung CT dataset is entirely low-dose and has CT scans from different parts of patients' chest collected over multiple scanners. Depending on the location on a patient's chest where the CT-scan was taken, a low-dose CT image could show either 2 or all 4 chambers of the heart in a blurry fashion. Since the low-dose CT images often did not clearly show these chambers, the generator could not learn these textures. This led the synthesized images to often exhibit arbitrary number of chambers. Conversely, the other parts of the synthesized CT images including bones, muscles and the surrounding tissue looked anatomically accurate, according to the participating radiologists.

**Proposed Improvements for Future Work.** One of the improvements that could correct the anatomy of a heart in low-dose CT is a more focused conditional generation of the heart in a CT image. The conditional parameter could either be size, shape or a template heart image taken from a training dataset. Denoising Diffusion Probabilistic Models (DDPMs) [10] have recently shown that they could be strong candidates for high-def. conditional generation of images. Further, conditional-DDPMs like ILVR-DDPMs [11] can also be added as a refinement / extra layer for heart generation over an existing stable diffusion-based model of lung CT, giving a user more control over the synthesis process. DDPMs are more stable as compared to GANs and unlike GANs could be easily customized over existing models eliminating the need to train new models from scratch.

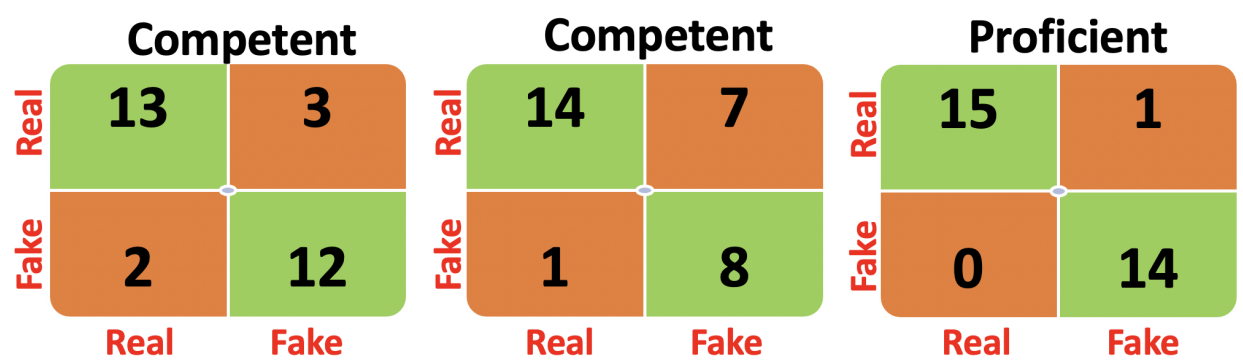

**Figure 6:** Confusion matrices for responses of 3 radiologists.

## 4 Conclusion

Our work suggests that careful implementation of texture based data augmentation combined with generative models could eliminate the "small annotated-data problem" in medical imaging domain. We also present an interactive visual turing test to evaluate these models with the help of the experts' feedback which could help develop new strategies for overcoming the shortcoming of these models.

## Acknowledgements

This research was funded in part by NIH grant R01EB032716. We thank Dr. David Yankelevitz and Dr. Yeqing Zhu from the Icahn School of Medicine at Mount Sinai, NY for their help in image evaluation. We also thank the National Cancer Institute for access to NCI's data collected by the National Lung Screening Trial. The statements contained herein are solely those of the authors and do not represent or imply concurrence or endorsement by NCI.

# Multi-class maximum likelihood expectation-maximization list-mode image reconstruction: an application to three-gamma imaging

Mehdi Latif[1,2], Jérôme Idier[1], Thomas Carlier[2], and Simon Stute[2]

[1]Nantes Université, École Centrale Nantes, LS2N, CNRS, UMR 6004, F-44000 Nantes, France
[2]Nantes Université, CHU Nantes, CRCI2NA, F-44000 Nantes, France

**Abstract** Our contribution focuses at improving the image reconstruction process for specific Compton imaging systems able to detect multiple classes of events, in the field of nuclear imaging. For each existing prototype of such systems, one or several processing methods have already been proposed to retrieve the activity map. Most of them get their inspiration from maximum likelihood expectation-maximization (MLEM), but none of them actually compute the MLEM solution. Some exploit the fully detected events only (e.g. in three-gamma imaging, the simultaneous detection of a pair of annihilation photons and of a third photon), and other combine several classes of detected events in a suboptimal way. In this paper, we first introduce a general framework for the reconstruction of a single activity map from multi-class events, and we provide the suited list-mode MLEM update equation. We then consider the case of XEMIS2, a preclinical prototype of a Compton telescope for three-gamma imaging, for which four distinct classes of partial detections coexist with the full detection class. As a preliminary step towards effective multi-class reconstruction, we generate a sensitivity map for the five classes using a dedicated Monte Carlo simulator.

## 1 Introduction

In nuclear imaging, some scanner prototypes based on Compton imaging allow the detection of different types of events. For instance,

- The MACACO camera [1] is a Compton telescope with three layers of plane detectors having different energy resolutions. Photon (or gamma) interactions in different combinations of layers lead to events with different spatial resolution properties.
- The WGI preclinical prototype [2] has a full-ring geometry with a scatterer detector close to the scanned subject and an absorber further away. In addition to Compton imaging, WGI performs positron emission tomography (PET) imaging by detecting annihilation photons in coincidence. With radio-isotopes emitting nearly simultaneously a positron and a (third) gamma, 3-gamma ($3\gamma$) imaging can be performed [3]. The emission point can be localized near the intersection of the Compton cone-of-response (COR) of the third gamma and the PET line-of-response (LOR) of the annihilation photons.
- The XEMIS2 preclinical prototype [4] is a Compton telescope using liquid xenon as a single continuous detection medium. It has been specifically designed for $3\gamma$ imaging.

All scanners described above detect distinct types of events during a single acquisition. To avoid the widespread term *type*, we call them *classes* of events hereafter.

In terms of image reconstruction, the WGI prototype separately considers Compton or PET image reconstructions [2]. With $3\gamma$, a single back-projection is performed [3]. For XEMIS2, a method has been proposed to transform $3\gamma$ data into TOF PET data in order to use conventional PET reconstruction methods [5]. Only $3\gamma$ data were used while single or double $\gamma$ events were discarded. For MACACO, all event classes were considered together in the reconstruction process. An extension of the MLEM algorithm was proposed lacking a theoretical derivation, where the update term is the sum of contributions from each event class and weighted by the sum of the sensitivities of each class [6].

Here, we first introduce a theoretical framework for the reconstruction of a single activity map from multi-class events. In particular, we derive a proper list-mode MLEM algorithm suited to the multi-class case. In a second stage, we consider $3\gamma$ imaging based on XEMIS2 as a specific case. As a preliminary step towards effective multi-class reconstruction, we generate a sensitivity map for each class using a dedicated Monte Carlo simulator.

## 2 Derivation of the multi-class list-mode MLEM

In what follows, $\boldsymbol{\lambda} := \{\lambda_j\}_{j\in[\![1,J]\!]}$ denotes a voxelized density of events, where each $\lambda_j$ represents the expected number of emitted events from the $j$th voxel. As mentioned earlier, we consider a multi-class data approach, where the available information about $\boldsymbol{\lambda}$ arises from several distinct datasets.

### 2.1 Multi-class data

Let $\boldsymbol{y} := \{\boldsymbol{y}_n\}_{n\in[\![1,N]\!]}$ be the set of $N$ collected data. It is composed of $K \in \mathbb{N}^*$ disjoint classes of physically independent events. A class refers to a specific type of event that can be detected and identified. We assume that each observed event can be attributed to one class, so that the set $\boldsymbol{y}$ can be expressed as the union of $K$ independent classes:

$$\boldsymbol{y} := \bigcup_{k=1}^{K} \{\boldsymbol{y}_n^k\}_{n\in[\![1,N^k]\!]}, \qquad N = \sum_{k=1}^{K} N^k, \tag{1}$$

where each $\boldsymbol{y}_n^k \in \mathbb{R}^{d^k}$ is a measurement vector containing photon interaction coordinates and deposited energies for the $n$th event.

### 2.2 Multi-class likelihood

We assume that within each class $k$, the events $\{\boldsymbol{y}_n^k\}_{n\in[\![1,N^k]\!]}$ are independent and identically distributed (iid) according to the probability density function of sampling $\boldsymbol{y}^k$ from any point of the $\boldsymbol{\lambda}$ distribution:

$$\boldsymbol{y}_n^k \sim p^k(\boldsymbol{y}^k|\boldsymbol{\lambda}) \quad \forall k \in [\![1,K]\!]. \tag{2}$$

The log-likelihood function then reads:

$$l(\boldsymbol{\lambda}|\boldsymbol{y}) := \sum_{k=1}^{K}\sum_{n=1}^{N^k} \ln p^k(\boldsymbol{y}_n^k|\boldsymbol{\lambda}) = \sum_{k=1}^{K} l^k(\boldsymbol{\lambda}|\boldsymbol{y}^k) \tag{3}$$

where $l^k(\boldsymbol{\lambda}|\boldsymbol{y}^k)$ denotes the log-likelihood of the $k$th class. Under standard Poisson assumptions, we can decompose $l^k(\boldsymbol{\lambda}|\boldsymbol{y}^k)$ as

$$l^k(\boldsymbol{\lambda}|\boldsymbol{y}^k) := \sum_{n=1}^{N^k} \ln\left(\sum_{j=1}^{J} a_{nj}^k \lambda_j\right) - N^k \ln\left(\sum_{j=1}^{J} \lambda_j s_j^k\right), \tag{4}$$

where $a_{nj}^k := A_j^k(\boldsymbol{y}_n^k)$ defines an element of the system matrix specific to class $k$ and voxel $j$, expressed as a continuous function of the $n$th measurement vector, and $s_j^k$ denotes the sensitivity of voxel $j$ for class $k$:

$$s_j^k := \int_{v\in\Omega^k} A_j^k(v)\,\mathrm{d}v \quad \forall j \in [\![1,J]\!],\ k \in [\![1,K]\!], \tag{5}$$

where $\Omega^k$ denotes the continuous detection domain related to the $k$th class.

### 2.3 Multi-class list-mode MLEM

To perform the estimation of $\boldsymbol{\lambda}$, we must extend the list-mode MLEM algorithm [7, 8] to simultaneously account for the $K$ classes. The result is the following update equation:

$$\widehat{\lambda}_j^{(t+1)} := \widehat{\lambda}_j^{(t)} \times \frac{1}{\sum_k s_j^k} \sum_{k=1}^{K}\sum_{n=1}^{N^k} a_{nj}^k \frac{1}{\sum_{j'} a_{nj'}^k \widehat{\lambda}_{j'}^{(t)} + \varepsilon_n^k}, \tag{6}$$

where $\varepsilon_n^k$ can be used to modelled random and scatter detections for each class. For a complete derivation of (6), we first need to express the log-likelihood of the complete dataset, along the same lines as in [7]. In the multi-class framework, a specific property is that the latter is a sum over the $K$ classes. The update equation (6) is then obtained from (3) within the model of the EM algorithm [9] and allowing the decomposition of the auxiliary function $Q(\boldsymbol{\lambda}|\boldsymbol{\lambda}^{(t)})$ into a sum of lower bound approximation on the likelihood for each class:

$$Q^k(\boldsymbol{\lambda}|\boldsymbol{\lambda}^{(t)}) := \mathbb{E}_{\boldsymbol{z}^k|\boldsymbol{y}^k,\boldsymbol{\lambda}^{(t)}}\left(\ln p^k(\boldsymbol{y}^k, \boldsymbol{z}^k \mid \boldsymbol{\lambda})\right) \forall k \in [\![1,K]\!] \tag{7}$$

where $\boldsymbol{z}^k$ denotes the latent data vector.

## 3 Application to three-gamma imaging

### 3.1 XEMIS2 camera

The above results can be applied to XEMIS2, in the context of $3\gamma$ imaging. This imaging modality makes use of an isotope that quasi-simultaneously emits a positron (leading to the annihilation in two photons of 511 keV) and a so-called third photon. In our application, scandium-44 ($^{44}$Sc) is used where the third photon is isotropically emitted with an energy of 1157 keV. The XEMIS2 camera is a preclinical prototype based on a continuous cylindrical liquid xenon (LXe) detector [4]. The axial length of the active zone is 24 cm with an internal (resp. external) radius of 7 cm (resp. 19 cm). The design of the camera only allows to operate at low count rates resulting in low random rates. The energy resolution at 511 keV is 9% (FWHM) which would allowed to remove most of scattered photon using a tight energy window.

### 3.2 Classes of detections with XEMIS2

Based on the energy of the detected photons, $K = 5$ classes of events are possible:

- $1\gamma_{\mathrm{COR}}^{511}$ events correspond to the detection of a single annihilation photon. Its emission position belongs to a COR with a half-opening angle $\beta \in [0,\pi]$ given by the Compton scattering angle formula:

  $$\beta := \operatorname{acos}\left(1 - \frac{m_e c^2 E_1}{E_0\,(E_0 - E_1)}\right) \tag{8}$$

  where $m_e c^2$ is the mass energy of an electron, $E_0$ is the total energy of the incident photon, i.e. $E_0 = 511$ keV for this class, and $E_1$ is the energy deposited in the detection medium during the scattering effect.

- $1\gamma_{\mathrm{COR}}^{1157}$ events correspond to the detection of the third photon. Akin to the former case, it is possible to define a COR using eq. (8) with $E_0 = 1157$ keV.

- $2\gamma_{\mathrm{LOR}}$ events correspond to the detection of both annihilation photons. As standard PET imaging, the source of annihilation belongs to a LOR.

- $2\gamma_{\mathrm{COR}}$ events correspond to the detection of a single annihilation photon and the third photon (combination of $1\gamma_{\mathrm{COR}}^{511}$ and $1\gamma_{\mathrm{COR}}^{1157}$). The origin of the decay belongs to the intersection of the two CORs.

- $3\gamma$ events correspond to the independent combination of a $2\gamma_{\mathrm{LOR}}$ event with a $1\gamma_{\mathrm{COR}}^{1157}$ one.

The imaging procedures currently available for XEMIS2 (*e.g.*, [5]) only exploit $3\gamma$ events. The other classes correspond to partial detections. Yet, they could bring additional information to the image reconstruction process, and thus lead to estimated activity maps $\widehat{\boldsymbol{\lambda}}$ of enhanced quality using a suited multi-class reconstruction algorithm such as eq. (6).

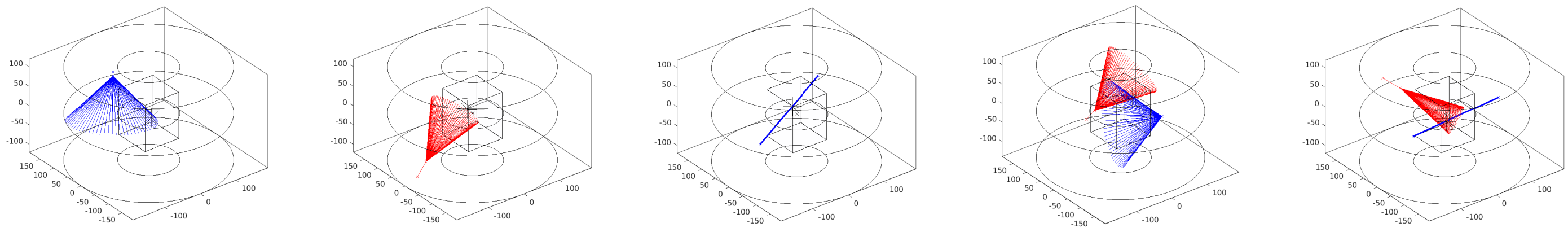

**(a)** $1\gamma^{511}_{\mathrm{COR}}$ **(b)** $1\gamma^{1157}_{\mathrm{COR}}$ **(c)** $2\gamma_{\mathrm{LOR}}$ **(d)** $2\gamma_{\mathrm{COR}}$ **(e)** $3\gamma$

**Figure 1:** Examples of detection obtained with the Monte Carlo simulator dedicated to XEMIS2. The LXe continuous detection medium is represented by the hollow cylinder, and the studied image by the box in the center. Blue LORs and CORs are obtained from annihilation photon(s), and red CORs from the third photon.

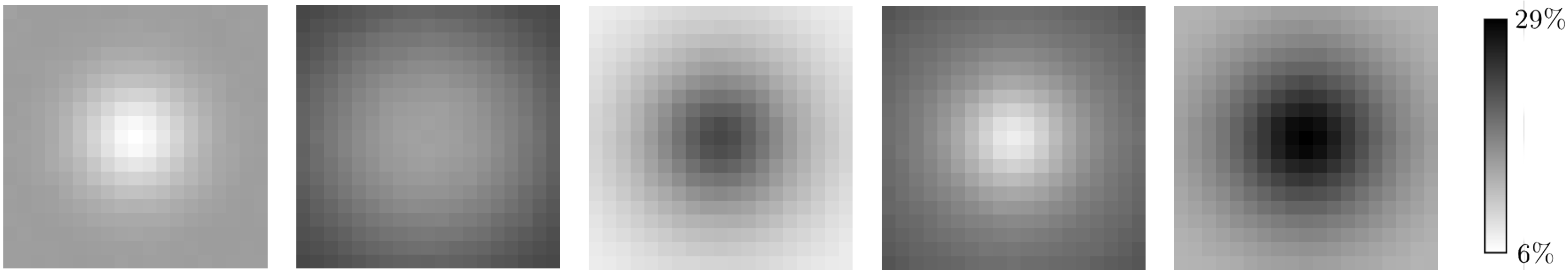

**(a)** $1\gamma^{511}_{\mathrm{COR}}$ **(b)** $1\gamma^{1157}_{\mathrm{COR}}$ **(c)** $2\gamma_{\mathrm{LOR}}$ **(d)** $2\gamma_{\mathrm{COR}}$ **(e)** $3\gamma$

**Figure 2:** A cross-section of the sensitivity maps (%) for each class of event at $z = 0$.

### 3.3 Monte Carlo simulator

Multi-class image reconstruction can be first assessed through simulation. To this end, a Monte Carlo simulator was implemented to generate $3\gamma$ emissions and their detection in a continuous medium, according to the geometry of XEMIS2. It is based on a ray-tracing method, allowing us to simulate emissions from voxels within the field-ov-view (FOV) by randomly selecting three directional vectors to define the positron range, the trajectories of the annihilation photons along the LOR, and the third photon. Once a photon reaches the LXe zone, new random draws were performed to simulate the mean free path and the radiation-matter interactions that may occur, such as Compton scattering, Rayleigh scattering and photoelectric effect. Cross-section values were extracted from the XCOM database [10]. We neglect the Rayleigh effect since its normalized cross-section is strictly lower than 10% over the entire energy range of interest i.e. for energies lower than 1157 keV. The resulting code, named Pollux, is open-source and freely available at mlatif/tep3g-pollux. Fig. 1 presents examples of detection for each event class obtained with Pollux.

### 3.4 Sensitivity computation of XEMIS2

The mathematical expression of the voxelwise sensitivity (5) involves multi-integrals that cannot be evaluated analytically. We thus used Pollux to generate approximate maps of the sensitivity for an empty FOV. To this end, the whole FOV was discretized into $19 \times 19 \times 24$ voxels of $5 \times 5 \times 10\,\mathrm{mm}^3$. $M = 2 \times 10^5$ $3\gamma$ emission points were uniformly generated in each voxel and the event class was recorded for each emission.

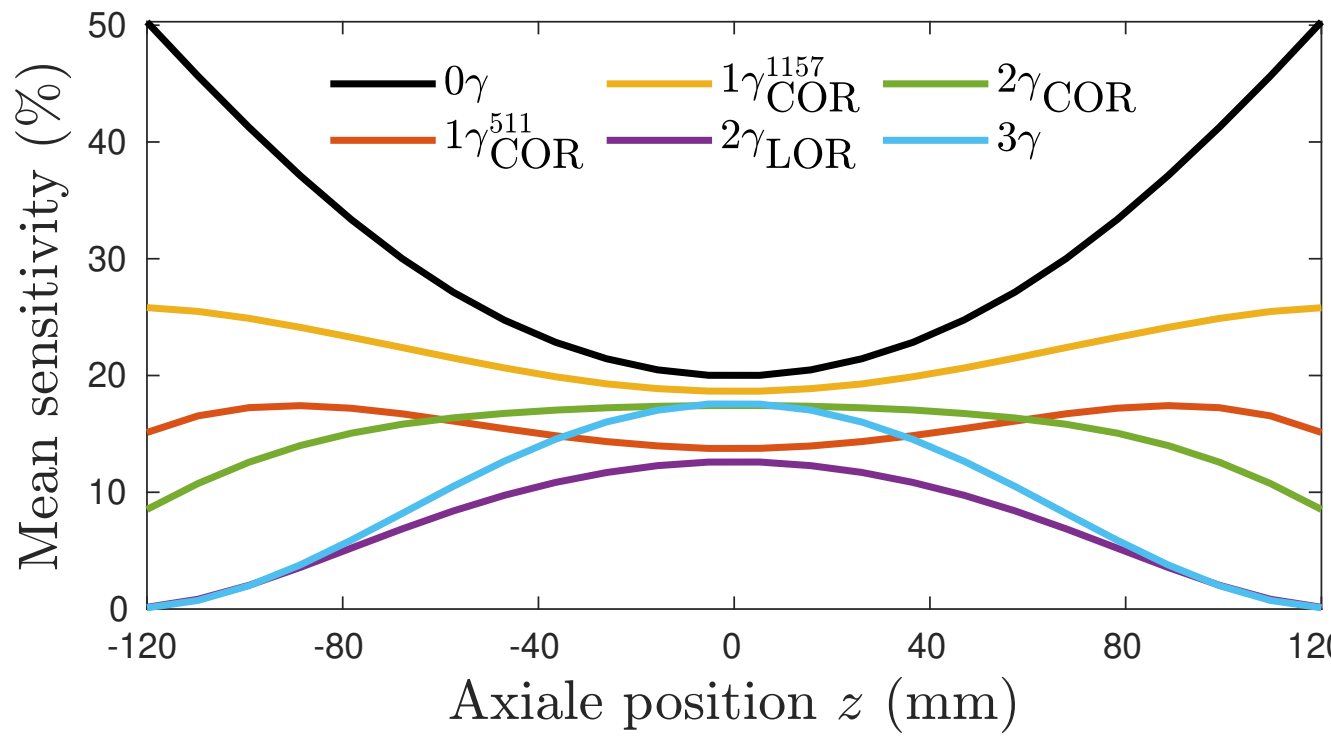

**Figure 3:** Mean sensitivity (%) along the axial direction for each class of event. The $0\gamma$ curve corresponds to the mean distribution of non-detected emissions.

Fig. 2 displays the sensitivity map obtained for each class for a transaxial cross-section at the center of the FOV, while Fig. 3 shows the same maps along the axial direction. The sensitivity distributions of the different classes are spatially heterogeneous, and tend to complement each other, which is an interesting observation in the perspective of multi-class reconstruction. For instance, the sensitivity maps of classes $3\gamma$ and $2\gamma_{\mathrm{LOR}}$ are maximal at the center of the FOV, while the trend is reversed for the sensitivity of $1\gamma^{511}_{\mathrm{COR}}$ and $1\gamma^{1157}_{\mathrm{COR}}$. Let us remark that non-detected events can be gathered in a sixth class $0\gamma$, whose "sensitivity map" (as displayed in Fig. 3) can rather be understood as a non-sensitivity one.

Table 1 displays the distribution of detected events by class, which represent 69% of the emissions that led to the detection of at least one photon. We note that approximately 13% of the emissions lead to the detection of $3\gamma$, which is small compared to the partially detected events, totalizing almost 87% of all cases. Obviously, we expect the most informative events to be in the $3\gamma$ class, and the $2\gamma_{\mathrm{LOR}}$ events to be more informative than the other events on CORs. However, $3\gamma$ and $2\gamma_{\mathrm{LOR}}$ classes represent about a quarter of the detected events.

| **Class** | $1\gamma_{\mathrm{COR}}^{511}$ | $1\gamma_{\mathrm{COR}}^{1157}$ | $2\gamma_{\mathrm{LOR}}$ | $2\gamma_{\mathrm{COR}}$ | $3\gamma$ |
|---|---|---|---|---|---|
| **% of detection** | 22.86 | 32.07 | 10.2 | 21.67 | 13.2 |

**Table 1:** Classification of detected events ($\approx 69\%$) for emissions uniformly distributed within the set voxels in the discretized FOV.

## 4 Discussion and perspectives

Our contribution focuses at designing an image reconstruction process for specific Compton imaging systems able to detect multiple classes of events, in the field of nuclear imaging.

First, we introduced a theoretical framework for the reconstruction of a single activity map from multi-class events, from which we deduced the suited list-mode MLEM update equation. The latter is potentially applicable to several existing prototypes of Compton imaging systems, such as MACACO, WGI, and XEMIS2. Let us also mention that the multi-class version of MLEM could be extended to cases where a regularization term is considered, as proposed in [11], for instance.

In a second part, we made a step further towards an application to XEMIS2, which is a preclinical prototype of a Compton telescope for $3\gamma$ imaging. In the XEMIS2 case, four distinct classes of partial detections coexist with the $3\gamma$ class. Using a dedicated Monte Carlo simulator, we determined a sensitivity map specific to each class, which is prerequisite to implement our multi-class version of MLEM. Besides, the obtained sensitivity maps clearly indicate that partial detections are far more frequent than perfect detections. Moreover, we noticed that some partial detection classes spatially complement the $3\gamma$ class. On the one hand, each partial detection is less informative than a full detection, but on the other hand, partial detections are more frequent. An interesting task will be to determine the amount of information brought by each class. This could be formally achieved by expressing the Fisher Information Matrix in the multi-class case, as an extension of the standard framework of [7]. We could then anticipate which classes are worth to be incorporated, despite the increased computational cost.

The next important step of our project will be the implementation of the proposed multi-class MLEM algorithm on the CASToR (Customizable and Advanced Software for Tomographic Reconstruction) platform [12]. Based on Pollux simulations, we are currently working at allowing continuous measurements and COR events within the CASToR framework. We also plan to implement a more realistic simulator using the GEANT4 toolkit [13].

## Acknowledgments

The authors thank Nicolas Beaupère for valuable discussions about the XEMIS2 detector and events processing. This work has been partly funded by the NExT Junior Talent project TRAC through the French "Programme d'Investissement Avenir".

# Three-dimensional maps of the tomographic incompleteness of cone-beam CT scanner geometries

Matthieu Laurendeau[1,2,3], Laurent Desbat[1], Guillaume Bernard[2], Frédéric Jolivet[2], Sébastien Gorges[2], Fanny Morin[2], and Simon Rit[3]

[1]Univ. Grenoble Alpes, CNRS, UMR 5525, VetAgro Sup, Grenoble INP, TIMC, 38000 Grenoble, France
[2]Thales AVS, Moirans, France
[3]Univ Lyon, INSA-Lyon, Université Claude Bernard Lyon 1, UJM-Saint Étienne, CNRS, Inserm, CREATIS UMR 5220, U1294, F-69373, Lyon, France

**Abstract** New generations of X-ray sources based on carbon nanotubes (CNT) enable the design of multi-sources computed tomography (CT) scanners. CT scanners with CNT often use a limited number of stationary sources and corresponding projections. Three-dimensional CT theory evaluates whether a given continuous source trajectory provides sufficient data for stable reconstruction of an imaged object. This paper extends a local incompleteness metric to derive a three-dimensional map and quantify tomographic incompleteness for a finite set of sources. We illustrate this incompleteness with a dedicated phantom. The reconstructed CT images of the phantom match the results predicted by the incompleteness map.

## 1 Introduction

Computed tomography (CT) is one of the most commonly used imaging modality for three-dimensional (3D) reconstruction in the medical and industrial fields. In the past few years, new X-ray sources have been developed based on carbon nanotube (CNT) cathodes [1]. Their small size enables the design of a new generation of CT scanners. It would benefit both industry with cheaper and motionless systems and medical applications with light-weight and mobile scanners which could be brought to emergency sites.
In a 3D context, CT scanners can be split into two categories: non-stationary architectures with mobile source(s) and stationary architectures with static source(s). Non-stationary architectures are the most common ones with source trajectories such as the conventional helix path or the circle-line path [2] which is adapted to C-arm scanners. Micro CNTs have opened new horizons for CT scanners with stationary architectures. The idea is to place several stationary sources around the scanned area. Gonzales *et al.* [3] and Vogtmeier *et al.* [4] have proposed two stationary designs for controlling luggage at airports.
3D CT theory has a few tools to help with the geometrical design of a CT scanner. Assuming non-truncated projections, Tuy [5] gave a condition to verify if a continuous source trajectory is sufficient to reconstruct an open region $\Omega$. The condition can be stated as: *every plane that intersects the imaged region* $\Omega$ *must intersect the scanning trajectory at least once*. In practice, all scanners are limited by a finite set of source locations instead of a continuous curve and Tuy's condition is never strictly satisfied. However, it is known that discrete sampling of a helical trajectory allows stable reconstruction.
Several metrics have been studied to quantify the impact of tomographic incompleteness, i.e. when Tuy's condition is not met. Metzler *et al.* [6] and Lin and Meikle [7] calculate a voxel-based percentage considering the unit sphere by locally linking Orlov's and Tuy's conditions [6]. Their metric is similar to that of Liu *et al.* [8] which was derived from the theory of the 3D Radon transform. They numerically evaluate the fraction of planes which are intersected by the source trajectory by sampling the unit sphere. This kind of measure is only possible for continuous source trajectories and the authors indicate that it is difficult to predict the reconstruction quality from this metric as, for example, 99% may lead to poorer image quality than 96%. Clackdoyle and Noo [9] quantify tomographic incompleteness to measure how far a voxel is from locally satisfying Tuy's condition in a given direction. Stopp *et al.* [10] proposed to average a similar measure in all directions. These last two criteria can be applied to both continuous and discrete source trajectories.
Our work aims at evaluating tomographic incompleteness for the design of new CT scanners assuming (first) non-truncated projections. We build on the quantification of tomographic incompleteness [9] to predict the worst direction in each spatial position of a 3D map. For four selected geometries, we demonstrate the relevance of this quantification by reconstructing a phantom made of three parallel cylinders placed at the worst location and oriented in the worst direction according to the incompleteness map.

## 2 Materials and Methods

### 2.1 Source trajectories

The incompleteness map is computed for a given source trajectory. We have chosen to compute it for four discrete geometries assuming that the imaged object is contained in the same $256^3$ mm$^3$ cube for all geometries. Fig. 1 illustrates the trajectories from the object point of view and Tab. 1 provides key parameters of each geometry. The first two architectures are conventional non-stationary architectures: circle-line and helix. The circle only satisfies Tuy's condition in the trajectory plane. We have selected the circle-line among several variants to extend the reconstructible region [2]. Here, we study the discrete form of this geometry and place the line above the circle only. In this context, Tuy's condition is not satisfied below the circle and the incompleteness should be lower above it. The helix source path is the geometry of most diagnostic CT scanners. It is composed of a source following a circle path while the imaged object translates axially through the circle which results in a helix in the coordinate system of the scanned object. For a continuous source curve, Tuy's condition is satisfied, and we anticipate a very low incompleteness for a finite set of source points.
The other two architectures are stationary: L and helix arc. The L geometry was one of the first stationary architecture commercialized for airport security [3]. It is made of two segments of sources which leaves space for detectors on the opposite side. The helix arc places sources along an arc of helix. For these two geometries, the imaged object is translated through the gantry. From the object's point of view, the pattern of sources is repeated according to the object motion in several identical cycles.

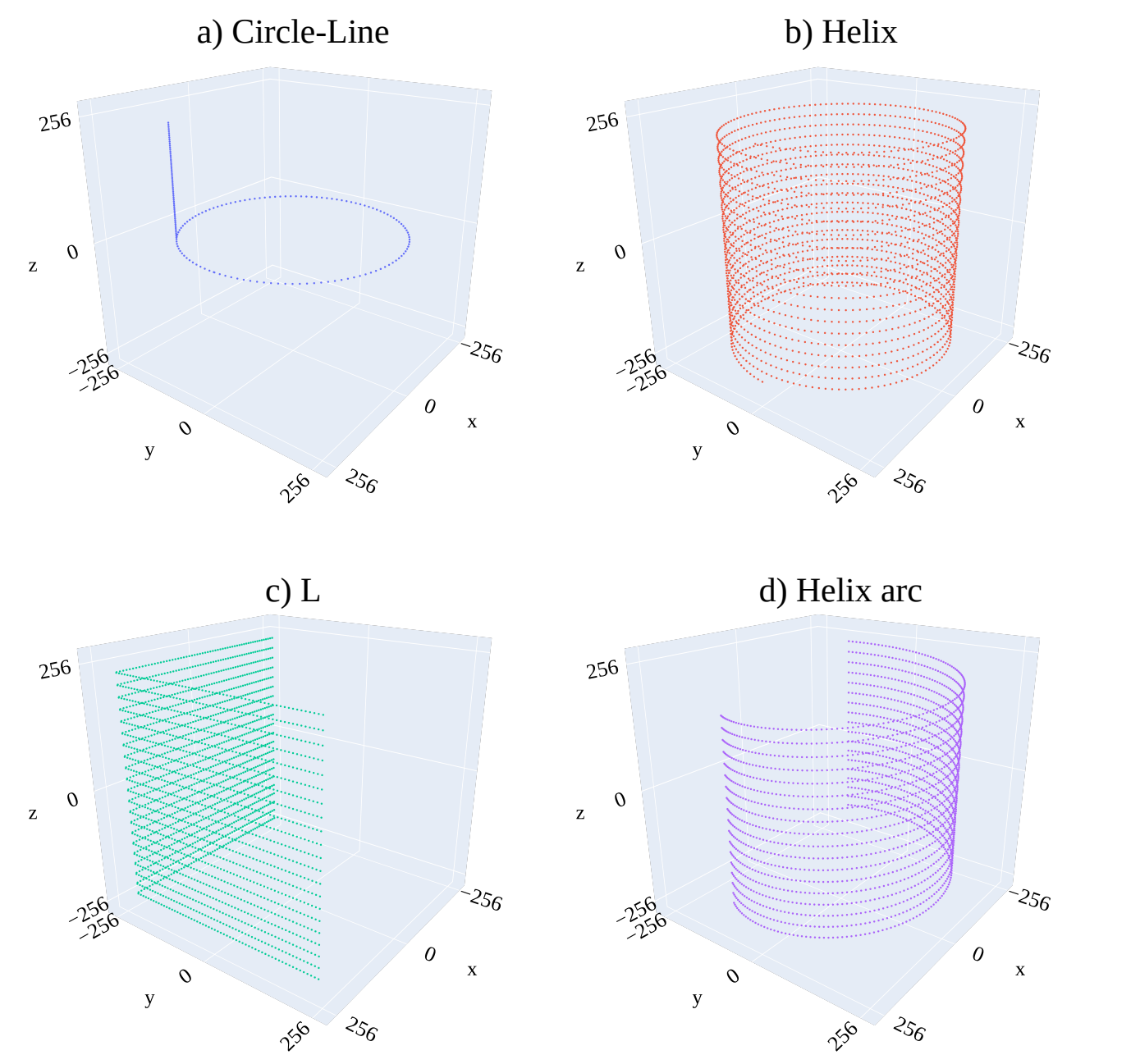

**Figure 1:** The four geometries studied in the object's coordinate system. a) combines a circle, which allows stable reconstruction only in its plane, with an orthogonal line above to complete data in the superior half of the object. The imaged object translates in the z direction for the b), c) and d) geometries. b) consists of a source with a circle path which makes it a helix when the object translates axially. c) has an L shape with sources placed on two orthogonal segments which are repeated when the imaged object moves. In d), sources are located on a helix arc and are duplicated in the axial direction.

**Table 1:** Key parameters of the four source trajectories.

| | Object translation (mm/cycle) | Sources | Radius (mm) | Special attributes (mm) |
|---|---|---|---|---|
| Circle-line | None | 120 (C) + 60 (L) | 240 (360°) | line height: 240 |
| Helix | [0,0,24] | 120 x 20 cycles | 240 (360°) | |
| L | [0,0,24] | 120 x 21 cycles | 240 (straight) | |
| Helix arc | [0,0,24] | 120 x 18 cycles | 240 (245°) | helical pitch: 120 |

### 2.2 Tomographic incompleteness

Clackdoyle and Noo [9] defined their local directional incompleteness criterion $\mathrm{I}(\mathbf{x},\ \mathbf{n}) \in \mathbb{R}^+$ at $\mathbf{x} \in \Omega \subset \mathbb{R}^3$ in the direction $\mathbf{n} \in S^2$, where $S^2$ is the unit sphere, for a source trajectory $\{\mathbf{s}_1, \mathbf{s}_2, ..., \mathbf{s}_n\} \in \mathbb{R}^{3\times n}$ as following:

$$\mathrm{I}(\mathbf{x},\mathbf{n}) = \min\left\{\frac{||\mathbf{s}_i - \mathbf{p}_i||}{||\mathbf{x} - \mathbf{p}_i||} : i = 1, 2, ..., n\right\} \tag{1}$$

where $\mathbf{p}_i \in \mathbb{R}^3$ is the projection of the source $\mathbf{s}_i$ onto the plane $\Pi_{\mathbf{x},\mathbf{n}}$ passing through the point $\mathbf{x}$ and of normal direction $\mathbf{n}$:

$$\mathbf{p}_i = \mathbf{s}_i - ((\mathbf{s}_i - \mathbf{x}) \cdot \mathbf{n})\mathbf{n}. \tag{2}$$

This criterion evaluates the minimum tangent of the angles defined by the plane $\Pi_{\mathbf{x},\mathbf{n}}$ and the X-ray lines, i.e. the lines passing through the point $\mathbf{x}$ and the source positions along the trajectory. If $\mathrm{I}(\mathbf{x},\mathbf{n}) = 0$, the plane cuts the source trajectory. Therefore, if $\mathrm{I}(\mathbf{x},\mathbf{n}) = 0$ for all $\mathbf{n} \in S^2$, the point $\mathbf{x}$ satisfies Tuy's condition and can be reconstructed if the acquired projections are not truncated. If $\mathrm{I}(\mathbf{x},\mathbf{n}) > 0$, the plane with co-direction $\mathbf{n}$ does not intersect the source trajectory and the point $\mathbf{x}$ does not satisfy Tuy's condition.
We use this criterion to evaluate the 3D spatial distribution of the tomographic incompleteness in the imaged region $\Omega$. Since we aim at identifying the less complete location and direction, we only record the maps of the worst directions $\mathbf{n}_\infty : \Omega \to S^2$

$$\mathbf{n}_\infty(\mathbf{x}) = \underset{\mathbf{n}\in S^2}{\arg\max}\{\mathrm{I}(\mathbf{x},\mathbf{n})\} \quad \forall\, \mathbf{x} \in \Omega \tag{3}$$

and the corresponding incompleteness $\mathrm{I}_\infty : \Omega \to \mathbb{R}^+$

$$\mathrm{I}_\infty(\mathbf{x}) = \mathrm{I}(\mathbf{x}, \mathbf{n}_\infty(\mathbf{x})) \quad \forall\, \mathbf{x} \in \Omega. \tag{4}$$

In practice, $\mathbf{n}_\infty$ and $\mathrm{I}_\infty$ are computed numerically by discretizing both the unit sphere $S^2$ and the object space $\Omega$. A unit hemisphere is sufficient due to the symmetry $\mathrm{I}(\mathbf{x}, -\mathbf{n}) = \mathrm{I}(\mathbf{x},\mathbf{n})$. We sampled 3000 directions using Fibonacci lattice method [11]. Fibonacci lattice arranges points along a spherical spiral homogeneously. Iteratively, each new point is placed evenly between the largest gap of the previous points.

### 2.3 Simulation phantom

To verify the incompleteness map, we simulated noiseless and untruncated projections for each geometry with a dedicated phantom. The phantom was made of three parallel cylinders with a 24 mm radius, a 6 mm height and a center-to-center distance of 6 mm. According to the incompleteness result of each geometry, the phantom's center was placed at the worst position in the subregion $\tilde{\Omega} \subset \Omega$ which can fully contain the phantom

$$\mathbf{x}^* = \underset{\mathbf{x}\,\in\,\tilde{\Omega}}{\arg\max}\{\mathrm{I}_\infty(\mathbf{x})\} \tag{5}$$

and in the worst direction $\mathbf{n}^* \in S^2$

$$\mathbf{n}^* = \mathbf{n}_\infty(\mathbf{x}^*). \tag{6}$$

CT images were reconstructed with RTK [12] using a least-squares iterative reconstruction with conjugate gradient minimization without regularization. We used 60 iterations which was visually deemed a good compromise between convergence, overfitting and image quality.

## 3 Results

### 3.1 Tomographic incompleteness map

Fig. 2 shows the incompleteness map of each studied geometry. It is a combination of the result of Eq. 3 and Eq. 4 for a given imaged region and source trajectory. In each position $\mathbf{x}$ of the imaged region, the value $\mathrm{I}_\infty(\mathbf{x})$ and the worst direction associated $\mathbf{n}_\infty(\mathbf{x})$ are calculated. The map illustrates both information using colors and 3D cones respectively.
The incompleteness maps of the circle-line and helix trajectories confirm the well known theory. For the circle-line trajectory, the incompleteness is low in the convex hull of the trajectory, $\min\{\mathrm{I}_\infty(\mathbf{x})\} \simeq 0.016$. However, below the circle plane, the imaged region is not reconstructible which translates into high incompleteness values $\max\{\mathrm{I}_\infty(\mathbf{x})\} \simeq 0.675$. The worst direction $\mathbf{n}^*$ at the bottom is almost orthogonal to the circle plane, as expected since the plane $\Pi_{\mathbf{x}^*,\mathbf{n}^*}$ is parallel to the trajectory circle and is not intersecting it nor the trajectory line. The helix's map displays a small incompleteness everywhere, $0.005 \leq \mathrm{I}_\infty(\mathbf{x}) \leq 0.016$. The residual incompleteness stems from the helix sampling.
For stationary architectures, the L trajectory is incomplete at the opposite of the two rectangles of sources with $\max\{\mathrm{I}_\infty(\mathbf{x})\} \simeq 0.456$. In this case, the worst direction $\mathbf{n}^*$ defines a plane parallel to the convex hull of the trajectory, which is consistent with Tuy's criterion. Near the sources, the incompleteness is small $\min\{\mathrm{I}_\infty(\mathbf{x})\} \simeq 0.006$, as expected, and the direction depends on the source sampling. Finally, the incompleteness map of the helix arc is similar to the non-stationary helix due to their similar trajectories: $0.005 \leq \mathrm{I}_\infty(\mathbf{x}) \leq 0.030$. It proves that a stationary design can compete with non-stationary architectures if a similar number of source locations is used.

### 3.2 Simulation & Reconstruction

The simulated 3D phantom of three parallel cylinders is placed at the worst position $\mathbf{x}^*$ (Eq. 5) and in the worst direction $\mathbf{n}^*$ (Eq. 6) in $\tilde{\Omega}$ for each geometry. The reconstructed images are shown in Fig. 3 such that the phantom is centered with $\mathbf{n}^*$ vertical.
The helix and helix arc trajectories have good image quality as predicted by their respective incompleteness maps. The circle-line trajectory has the worst image quality, and it is difficult to separate the cylinders. Finally, the L reconstruction has a slightly better image quality than the circle-line trajectory, but it is also difficult to distinguish the three cylinders.

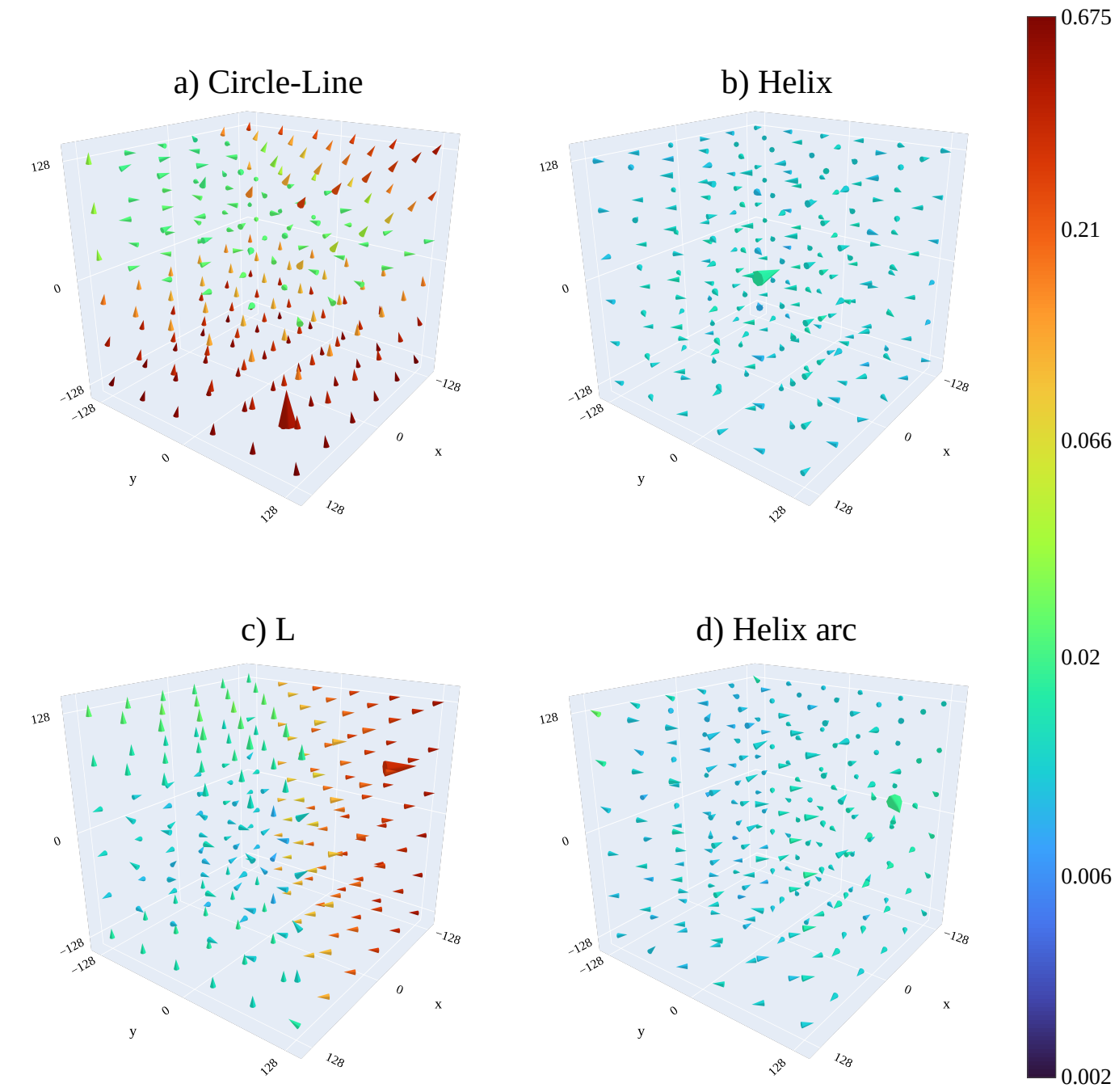

**Figure 2:** Tomographic incompleteness maps of the four studied geometries. Each map is centered on the imaged region $\Omega$ (ratio $1/2$ in each direction with respect to Fig. 1 and the same camera angle). The direction and the color of each cone represent the worst direction $\mathbf{n}_\infty(\mathbf{x})$ (Eq. 3) and the corresponding incompleteness $\mathrm{I}_\infty(\mathbf{x})$ (Eq. 4), respectively. The bigger cone shows the worst position $\mathbf{x}^*$ (Eq. 5) in the worst direction $\mathbf{n}^*$ (Eq. 6) in $\tilde{\Omega}$.

## 4 Discussion

The incompleteness map presented in this work accurately predicts image quality of a dedicated phantom. The incompleteness map is based on Tuy's theory and assumes untrucated projections. Accounting for the truncation of the projections was beyond the scope of this work.
The computation of the worst directions $\mathbf{n}_\infty$ was done numerically by sampling the unit sphere. The sampling pattern [11] was selected because it homogeneously samples the sphere. We took about 3000 points on the hemisphere which does not warrant to find the worst direction. However, the incompleteness maps $\mathrm{I}_\infty$ and $\mathbf{n}_\infty$ are quite smooth, at least for larger values which are not influenced by the trajectory sampling, and this discretization of the unit sphere may be sufficient to have a good estimate of $\mathbf{n}_\infty$. Finally, it is common for stationary architectures to use iterative reconstruction algorithms. We have chosen to use the same iterative algorithm, without regularization but a fixed number of iterations, which provides a quite fair comparison of all geometries.

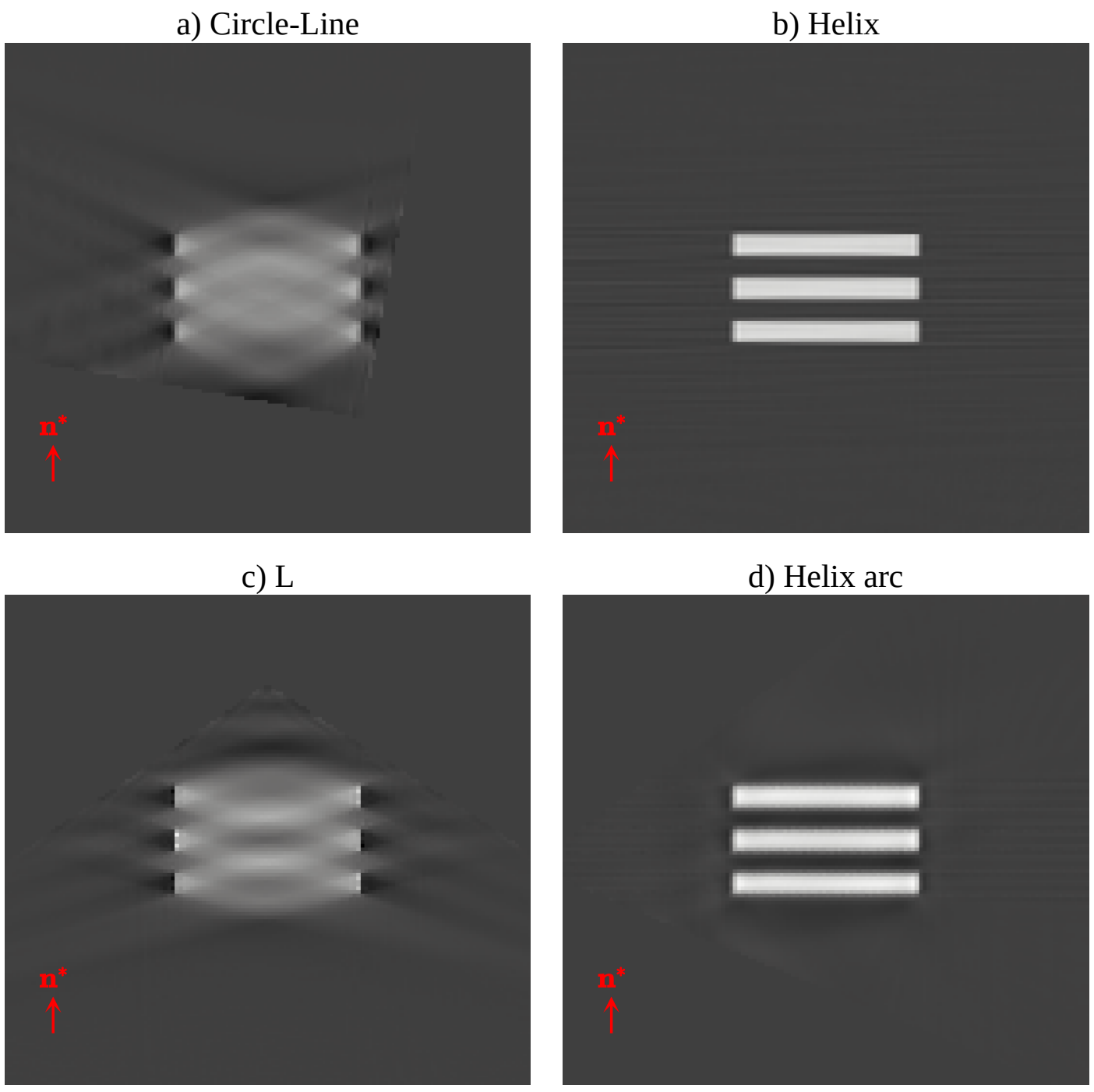

**Figure 3:** Least squares reconstruction of the same phantom for each geometry. The phantom is composed of three parallel cylinders and, for each geometry, it is placed at the worst position $\mathbf{x}^*$ (provided by Eq. 5) and in the worst direction $\mathbf{n}^*$ (provided by Eq. 6) in $\tilde{\Omega}$, as shown with a bigger cone in Fig 2. For comparison purposes, we have registered the reconstructed images to place the phantom in the center and such that $\mathbf{n}^*$ is vertical.

## 5 Conclusion

Our work defines the tomographic incompleteness map which was computed on four architectures: two non-stationary and two stationary. We have chosen a similar number of sources, radius and other characteristics to make them comparable. The results show that the incompleteness map is in agreement with Tuy's theory (on which it is based). These maps adequately predict the image quality of a dedicated disk phantom scanned at the worst location and in the worst direction provided by the incompleteness map. The incompleteness map may be used to design a compact geometry scanner with a limited number of sources by minimizing the incompleteness value in a scanned region.

# Limited-Angle CT Reconstruction using Implicit Neural Representation with Learned Initialization

Jooho Lee[1] and Jongduk Baek[1]

[1]Department of Artificial Intelligence, College of Computing, Yonsei University, Seoul, Korea

**Abstract** Limited-angle computed tomography (CT) is one of the major challenges in imaging reconstruction problems. To tackle this ill-posed inverse problem, various supervised deep learning based approaches have been proposed and shown impressive results. However, these methods have fundamental weaknesses such as the blurring effect caused by L2 loss, and difficulty in gathering a large amount of paired data in clinical practice. In this work, we propose a novel self-supervised limited-angle CT reconstruction algorithm, which effectively addresses the aforementioned limitations. We utilize the coordinate-based neural representation to obtain the missing angle data. In addition, we integrate the prior knowledge of CT image into the network via learned initialization, which dramatically enhanced the reconstruction quality. The numerical results demonstrate the superior performance of the proposed method compared to other conventional methods. We believe the presented self-supervised and patient-specific algorithm suggests a paradigm shift for limited-angle CT research based on deep learning.

## 1 Introduction

Computed Tomography (CT) is one of the most commonly used medical imaging modalities. It enables the non-destructive visualization of the internal body structures by reconstructing the image with X-ray projections acquired from different angles around the body. Mathematically, this reconstruction process is referred to as solving an inverse problem from projections to reconstruct the image of a scanned object. If a sufficient number of projections is acquired, an exact reconstruction is possible by using analytical reconstruction such as filtered-backprojection. However, when the acquired projection data is not sufficient like in limited-angle CT or sparse-view CT, the reconstructed image using filtered-backprojection suffers from severe artifacts due to the missing information.

To tackle these ill-posed inverse problems, many deep learning based approaches have been proposed and shown promising results [1,2,3]. Nevertheless, there still remain the limitations such as the blurring effect caused by L2 loss, and the difficulty in gathering a large amount of paired data for supervised learning in clinical situations. Recently, to overcome the aforementioned limitations of general deep learning (i.e., supervised learning) method, the concept of a patient-specific neural network using implicit neural representation has emerged [4,5,6]. Specifically, Kim et al. [5] proposed a streak artifact reduction algorithm using neural representation in a self-supervised fashion, and Liyue Shen et al. [6] used a similar approach with prior utilization to reconstruct sparse-view CT image.

Motivated by the previous works, we propose a novel image reconstruction algorithm for limited-angle CT. By utilizing implicit neural representation, we obtain a high-quality prior image for the imputation of missing angle data.

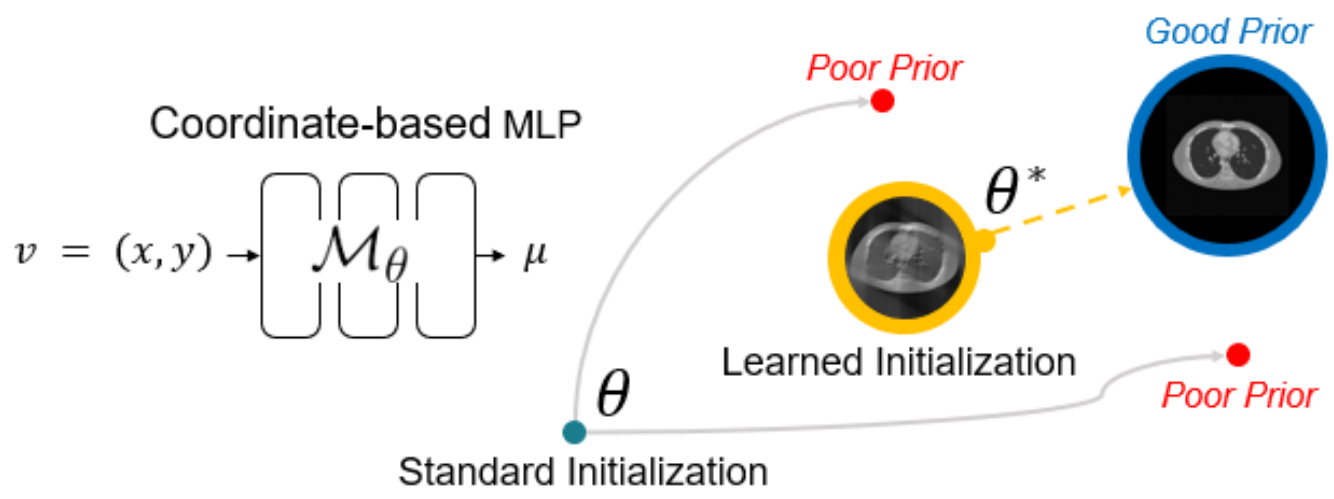

**Figure 1.** Illustration of learned initialization to find θ*

Moreover, as shown in Figure 1, CT image based weights initialization was activated to ensure the decent quality of a prior image. By using the limited-angle sinogram of a single patient, our proposed algorithm achieved an outstanding reconstruction quality. Numerical comparisons demonstrate the powerful performance of our proposed method.

To summarize, our main contributions are:

- We propose a novel image reconstruction algorithm for limited-angle CT in a self-supervised manner.
- We integrate the prior knowledge of CT image into the implicit neural representation by using a learned initialization scheme, which in fact leads to dramatic improvement of reconstruction quality.

## 2 Methods

In this section, we will illustrate the overall framework of the proposed method to reconstruct a limited-angle CT image. The main idea is to obtain a high-quality prior image (i.e., an estimate of a full-angle CT image) by using implicit neural representation with learned initialization. Then, by using a prior image, we regenerate the limited-angle artifacts, which are then subtracted from the original limited-angle FBP image to obtain the artifacts-corrected result.

The algorithm can be conceptually separated into three parts, where the first part is to initialize the weights of coordinate-based multi-layer perceptron (MLP) using the original limited-angle FBP image as an image domain prior. Second, starting from the initialized point, we optimize the MLP using a limited-angle sinogram to obtain a prior image. Lastly, we regenerate the limited-angle artifacts from the prior image to impute the missing angle data. Since this entire process can be utilized with a single limited-angle sinogram, no large-scale data is required. The schematic diagram of the proposed algorithm is illustrated in Figure 2.

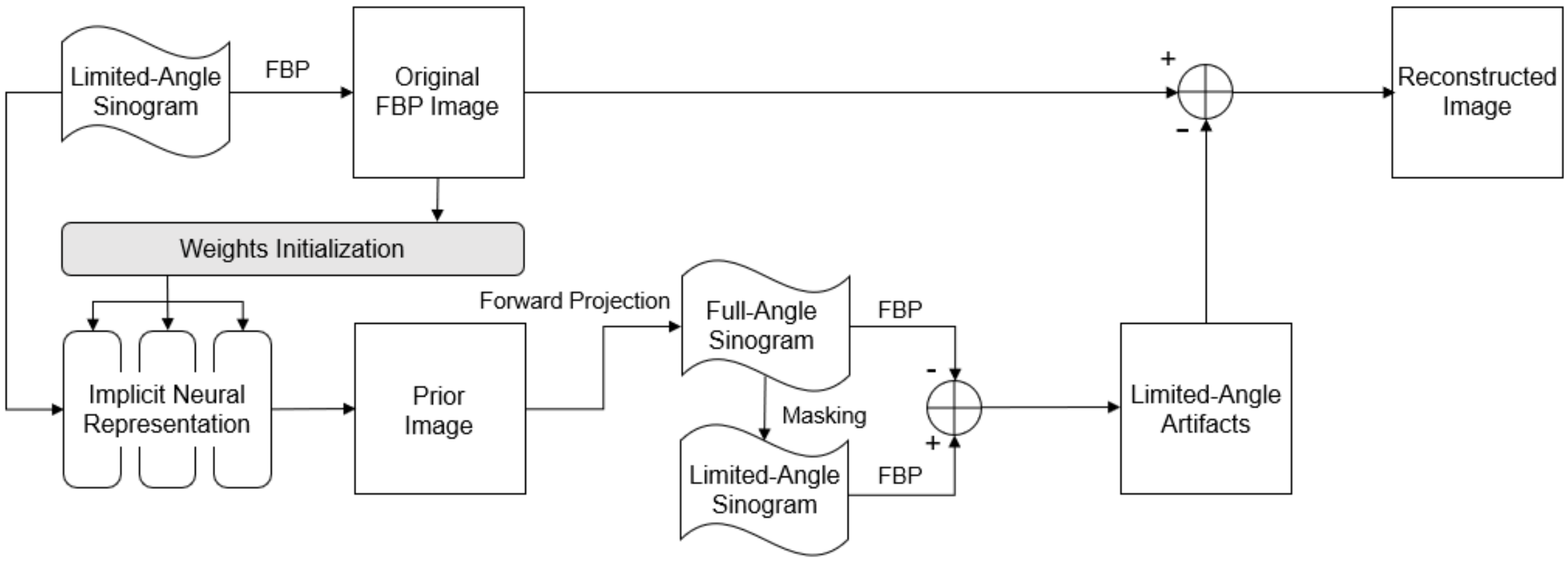

**Figure 2.** Schematic diagram of the proposed method

## 2.1 Learned Initialization of weights

In CT imaging, coordinate-based MLP can be expressed as

$$\mathcal{M}_\theta : v \to \mu \quad s.t \quad v \in \mathbb{R}^d,\ \mu \in \mathbb{R} \tag{1}$$

where $v$ represents the coordinate in the given CT image (d-dimensional space), and $\mu$ is the corresponding pixel value. A straightforward approach of using this MLP to optimize the neural representation of CT image can be an option in limited-angle CT similar to the previous work in sparse-view conditions [5]. However, when it is applied to limited-angle CT, the result fails to recover anatomical details and limited-angle artifacts remain. This is because the MLP only relies on the inductive bias when representing data that belongs to the missing angle. Furthermore, compared to sparse-view CT, limited-angle CT can be regarded as a more ill-posed inverse problem since its projection is literally limited. Therefore, additional knowledge must be activated for further improvements.

Here, we choose weights initialization scheme to feed prior information to the network. This can be easily done by first fitting the MLP to the original limited-angle FBP image with simple L2 loss described in Eq. (2). Since we can obtain the image from limited-angle projections via filtered-backprojection, there is no need for additional data.

$$L(\theta^*) = \sum_i \|\mathcal{M}_{\theta^*}(v_i) - \mu_i\|^2 \tag{2}$$

Although limited-angle FBP image has severe artifacts, we demonstrate that it can still act as an image domain prior, for instance, giving network structural shape information of CT image. In addition, as the optimization of the CT image after the weight initialization is done by minimizing the loss calculated in the sinogram domain, this operation can be viewed as integrating the knowledge from dual domain (i.e., sinogram and image domain). Once again, we make it clear that high-frequency details are not considered at this step since this process only aims to find a better starting point (Figure 1). We minimized the loss in Eq. (2) for 500 iterations using the Adam optimizer [7] with a learning rate of $5\times10^{-4}$.

## 2.2 Optimizing CT Image Representation

Using the initialized weights as a starting point, we now optimize the neural representation to obtain a high-quality prior image. Normally, this procedure can be done by minimizing the difference between an estimated value and the ground truth pixel in the image domain. However, this is impossible in practical CT scenarios where ground truth (i.e., full-angle CT image) is unavailable. Therefore, we indirectly optimize the network by utilizing a differentiable projection layer.

Let $\mathcal{P}$ denote the differentiable projection layer, then the coordinate-based MLP $\mathcal{M}$ can be supervised by updating the loss in Eq. (3) with gradient descent. Note that the entire process of optimization is in a self-supervised fashion because we calculate the loss with the acquired sinogram. For illustration, please refer to Figure 3.

$$L(\theta^*) = \sum_{image} \|\mathcal{P}(\mathcal{M}_{\theta^*}(v)) - \mathcal{P}(\mu)\|^2 \tag{3}$$

We used an MLP consisting of eight fully connected layers with a rectified linear unit (ReLU) for implicit neural representation. All hidden layers had 128 channels and the output layer had a single output channel. Furthermore, instead of putting input coordinates directly into the MLP, we used Gaussian random Fourier Feature Mapping (FFM) [8] in both the initialization and optimization process, of which the function is formulated as

$$\gamma(v) = [cos(2\pi Bv), sin(2\pi Bv)]^T \tag{4}$$

The FFM is known to improve the performance of MLPs in regard to representing high-frequency details. In our study, $B$ was sampled from Gaussian distribution $\mathcal{N}(0, \sigma^2)$, and σ was experimentally set to five. The network was optimized for 70k iterations with a learning rate of $1\times10^{-5}$.

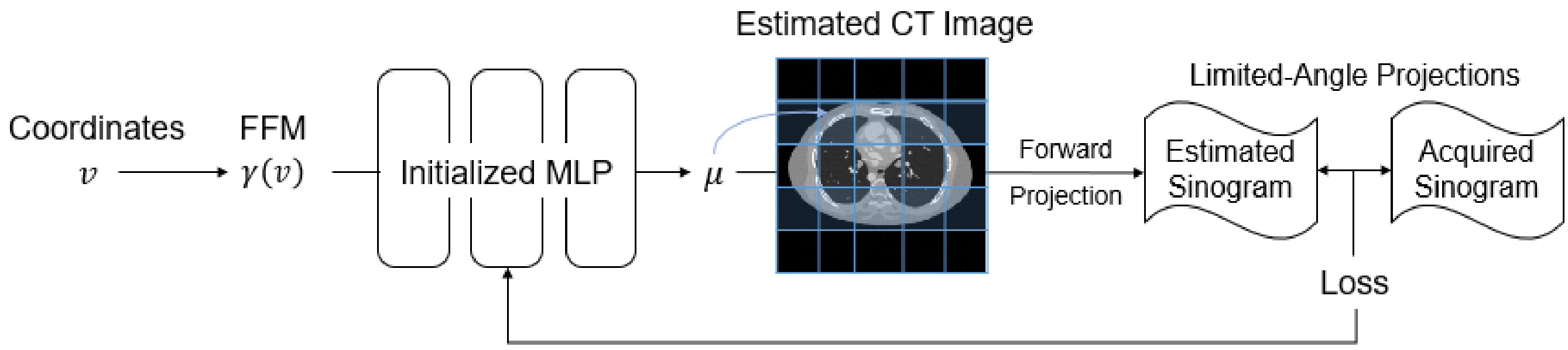

**Figure 3.** The optimization process of implicit neural representation in limited-angle CT

## 2.3 Missing Data Imputation

The output prior image of implicit neural representation can serve as a high-quality reconstruction result. Nevertheless, in order to ensure the reliability of the outcome, we decided to use the prior image for imputing the missing data from the original limited-angle FBP image. In this way, we can enhance the image quality as well as the reliability.

The severe artifacts in limited-angle CT image are induced by the missing angle. Therefore, if we impute the missing angle data with the one acquired from the prior image, we can expect to reduce the artifacts in limited-angle CT image. For imputation, we simply regenerated the limited-angle artifacts. The steps are as follows. First, we obtained an estimated full-angle sinogram by a forward projection of the prior image. Next, by masking the full-angle sinogram, we acquired the corresponding limited-angle sinogram. Then, with filtered-backprojection, we reconstructed the images of full- and limited-angle CT. After that, we calculated the difference between these two images, which becomes an estimate of limited-angle artifacts. Finally, regenerated artifacts were subtracted from the original limited-angle FBP image to present the final corrected output.

## 2.4 Data Generation

A numerical extended cardiac-torso (XCAT) phantom [9] was used for a simulation study. The fan-beam geometry for the simulation is shown in Table 1.

**Table 1.** Parameters of the fan-beam geometry

| Parameter | Value |
|---|---|
| Source to iso-center distance | 100 cm |
| Source to detector distance | 150 cm |
| Detector cell size (Size of detector array) | 0.08 cm (512 × 1) |
| Data acquisition angle | 360º (*full*) / 120º (*limited*) |
| Reconstructed image matrix size | 256 × 256 |
| Reconstructed image pixel size | 0.05 × 0.05 $cm^2$ |

The limited-angle projections were acquired using Siddon's ray-driven algorithm [10] with a 120º scanning range, and the full-angle projections were obtained with 513 views equally distributed over 360º. The angular distance between the views was set equal for both projections.

## 2.5 Compared methods

For comparison, we employed the conventional FBP reconstruction, total variation-based iterative reconstruction (TV-IR), and implicit neural representation without learned initialization. To be specific, the TV-IR was implemented with a gradient-projection-Barzilai-Borwein formulation [11] and the regularization parameter was set to 0.05 empirically. In addition, the weights of compared implicit neural representation were randomly initialized from $\mathcal{U}(-\sqrt{k},\sqrt{k})$, where $k = 1/\text{in_features}$. Note that this is the default initialization scheme for a fully connected layer in Pytorch [12]. To fairly compare the effects of initialization, we also followed the steps in section 2.3 for randomly initialized neural representation. For each method, we used peak signal-to-noise ratio (PSNR) and structural similarity index (SSIM) [13] as quantitative evaluation.

## 3 Results

Figure 4 compares the resulting images using FBP, TV-IR, randomly initialized neural representation, the proposed method, and the full-angle FBP (reference), respectively. It is shown that the proposed method effectively reduced the limited-angle artifacts and achieved superior reconstruction quality. The quantitative evaluation also demonstrates the superior performance of our proposed algorithm (Table 2).

As can be observed, it is clear that FBP and TV-IR struggle to recover anatomical details. FBP results in severe artifacts due to the incomplete input sinogram. TV-IR recovered some artifacts, but residual artifacts persist. Particularly, the image along the direction perpendicular to the missing angle is severely damaged. Compared to TV-IR, the randomly initialized neural representation suppressed the artifacts more effectively. However, there still has some structural distortions. In contrast, the proposed method reduces the artifacts of limited-angle CT very effectively, which is more evident in the regions of interest images indicated by yellow and green arrows (Figure 5). To sum up, the proposed method shows outstanding reconstruction quality in terms of limited-angle artifacts reduction and preserving anatomical details compared to all the methods above.

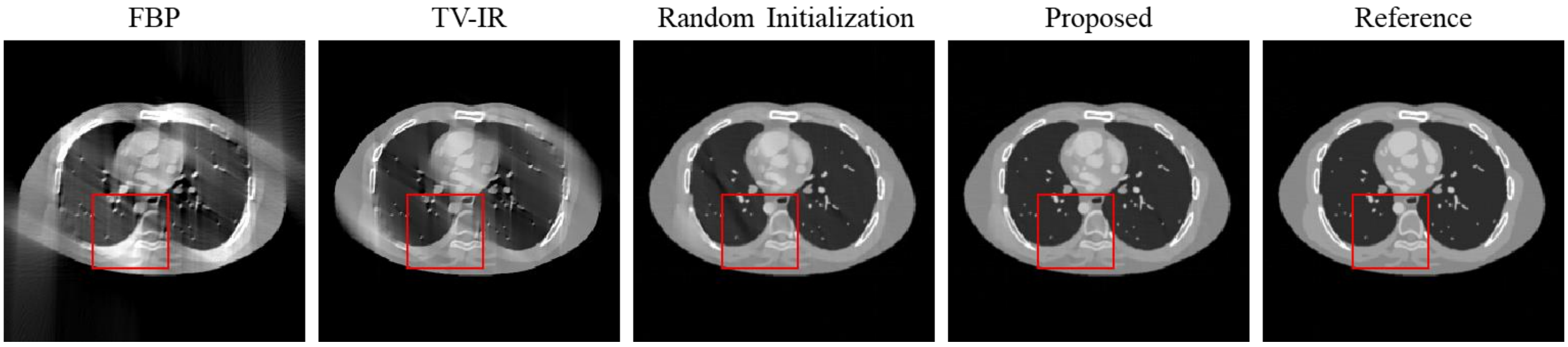

**Figure 4.** Limited-Angle CT (120°) reconstruction results. The red boxes indicate a region of interest. The display window is set to [-1000, 1000] in Hounsfield units.

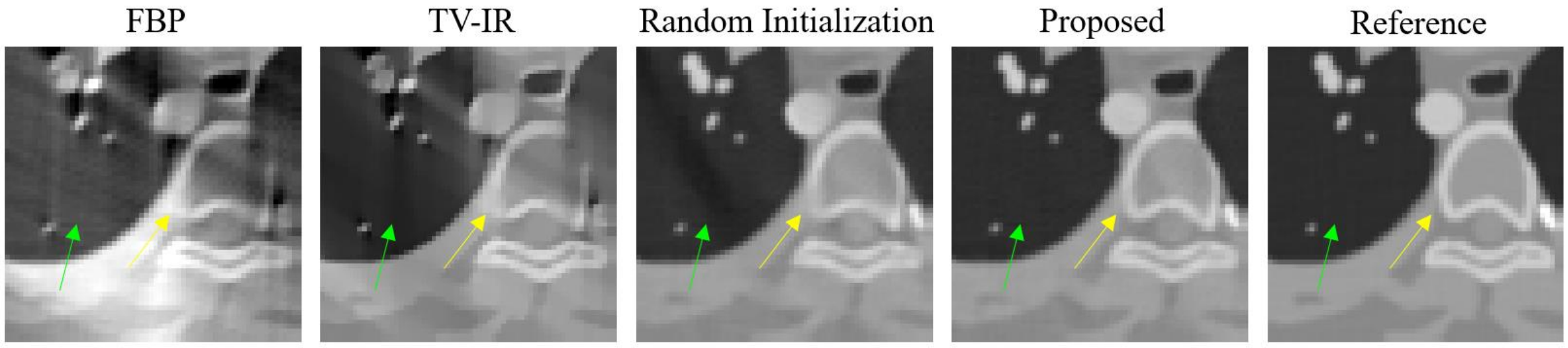

**Figure 5.** Enlarged region of interest images in Figure 4.

**Table 2.** The quantitative evaluation

| Metrics | FBP | TV-IR | Random Initialization | Proposed |
|---|---|---|---|---|
| PSNR | 18.30 | 25.71 | 36.89 | **40.03** |
| SSIM | 0.6288 | 0.8880 | 0.9721 | **0.9807** |

## 4 Conclusion

In this paper, we proposed the self-supervised implicit neural representation based algorithm for limited-angle CT reconstruction, which is one of the major challenges in image reconstruction problems. Since the naïve approach of using neural representation fails to recover anatomical structures, we integrated the image prior knowledge of CT with learned initialization. Lastly, by imputing the missing data from the acquired high-quality prior image, we effectively suppressed the artifacts.

Numerical results demonstrate that our proposed algorithm can enable limited-angle CT reconstruction of fine quality. We believe the presented self-supervised method suggests a shifting paradigm for limited-angle CT research regarding deep learning approaches.

# Computed Tomography Image Reconstruction with Different Styles of Multiple Kernels via Deep Learning

Danyang Li[1,2], Yuting Wang[3], Dong Zeng[1,2], and Jianhua Ma[1,2,*]

[1]School of Biomedical Engineering, Southern Medical University, Guangzhou 510515, China
[2]Pazhou Lab (Huangpu), Guangdong 510000, China
[3]Department of Radiation Oncology, Sun Yat-sen University Cancer Center, Guangzhou 510060, China
[*]Corresponding author

**Abstract** Computed tomography image efficiently helps diagnose potential problems or diseases before symptoms appear with the specific reconstruction kernel. Different reconstruction kernels produce CT images with different styles which exhibit various anatomical structure information. However, the standard reconstruction algorithms would reach the limit of its capacity for reconstructing the CT images with multiple image styles of different kernels, i.e., the multi-kernel style image reconstruction task. In this work, we design a deep learning network for accurate and efficient multi-kernel style image reconstruction task in low-dose CT imaging. The sorted view-by-view back-projection measurements at low-dose are fed into the deep learning network to reconstruct CT images with a bundle of various kernels. We demonstrate the feasibility of our deep learning network on Mayo CT dataset. The experimental results demonstrate that our deep learning network efficiently solves the simultaneous multi-kernel style image reconstruction issue.

## 1 Introduction

Computed tomography (CT) is a pivotal technology for clinical diagnose and radiotherapy, i.e., CT images exhibit abundant anatomical structures with high time and spatial resolution which is essential for acute disease diagnosis [1], tumor metastasis analysis [2] and image-guided intensity-modulated radiotherapy [3]. The CT image styles exhibiting different texture and structure properties are largely determined by the reconstruction kernels which filter and maintain the specific frequency information in filtered back-projection algorithm [4], indicating the selection of reconstruction is important to determine the imaging protocols and essential for specific clinical task. However, limited by the reconstruction speed and memory capability of hardware, standard reconstruction algorithms usually release the raw data after reconstructing image with one kernel, which is called multi-kernel style image reconstruction task. Therefore, it fails to reproduce the CT images at the same slice with various reconstruction kernels, which disable the retrospective analysis or longitudinal studies for the patients.

In order to handle the multi-kernel style image reconstruction task, various methods have been developed based on image post-processing technology. For example, Kenneth *et al.* developed a hybrid filter method by combining high- and low-pass kernels to simultaneously characterize different anatomical tissues [5]. Masaki *et al.* utilized the point spread functions of the system to determine the filter function to transfer the image reconstructed from one kernel to the one from another kernel [6]. Recently, inspired by the successful applications of deep learning (DL) technology in computer vision and nature language processing, DL-based methods have been developed to obtain various CT image styles of different reconstruction kernels. Such as the convolutional neural network (CNN) methods [7, 8] and a cycle-consistent generative adversarial network (cycleGAN) based method [9]. However, due to the FBP-reconstructed images lose partial information of the original anatomical structure, especially in low-dose and sparse-view imaging field, directly conducting style conversion of different kernels in image domain might produce over-smoothed results or introduce artifacts in the images.

To solve this issue, we introduce a view-by-view back-projection reconstruction network (shorten as VVBP-Net) for the multi-kernel style image reconstruction task in low-dose and sparse-view imaging situation. Specifically, the presented VVBP-Net back-projects the unfiltered noisy CT measurements into image domain view-by-view and sorts them along the view direction [10], shown in Figure 1(b). Then, the sorted back-projection of all views are fed into the followed network which employs an encoder-decoder architecture, shown in Figure 1(c). During the features being fed into following decoder, multiple reconstruction kernels are simultaneously convolved to reconstruct CT images with a bundle of various kernels in Figure 1(c). To achieve the goal, a kernel bank (bundle of various kernels) is constructed which consists of multiple reconstruction kernels, and each reconstruction kernel represents specific distribution of pixel values or noise pattern in the CT images. The novelty of this work is that our VVBP-Net enrolls all the structural characteristics of image and noise statistics of the raw measurements for CT image reconstruction, as the Figure 1(b) shows.

## 2 Materials and Methods

### 2.1 FBP Implementations

The FBP implementation generally consists of three key steps: filtering, view-by-view back-projection, and summing, as the Figure 1(a) shows. The filtering operation can remove noise and blur in the CT measurements. In the view-by-view back-projection step, the filtered CT measurements are back-projected into image domain view-by-view. Finally, the back-projection of all views are summed to obtain desired CT image. We can observe that the filtering operation might remove some critical structure details and lose measurement noise characteristics, and summing operation could miss more important structural characteristics compared with sorting back-projection of all views without summing, as the Figure 1(a) shows.

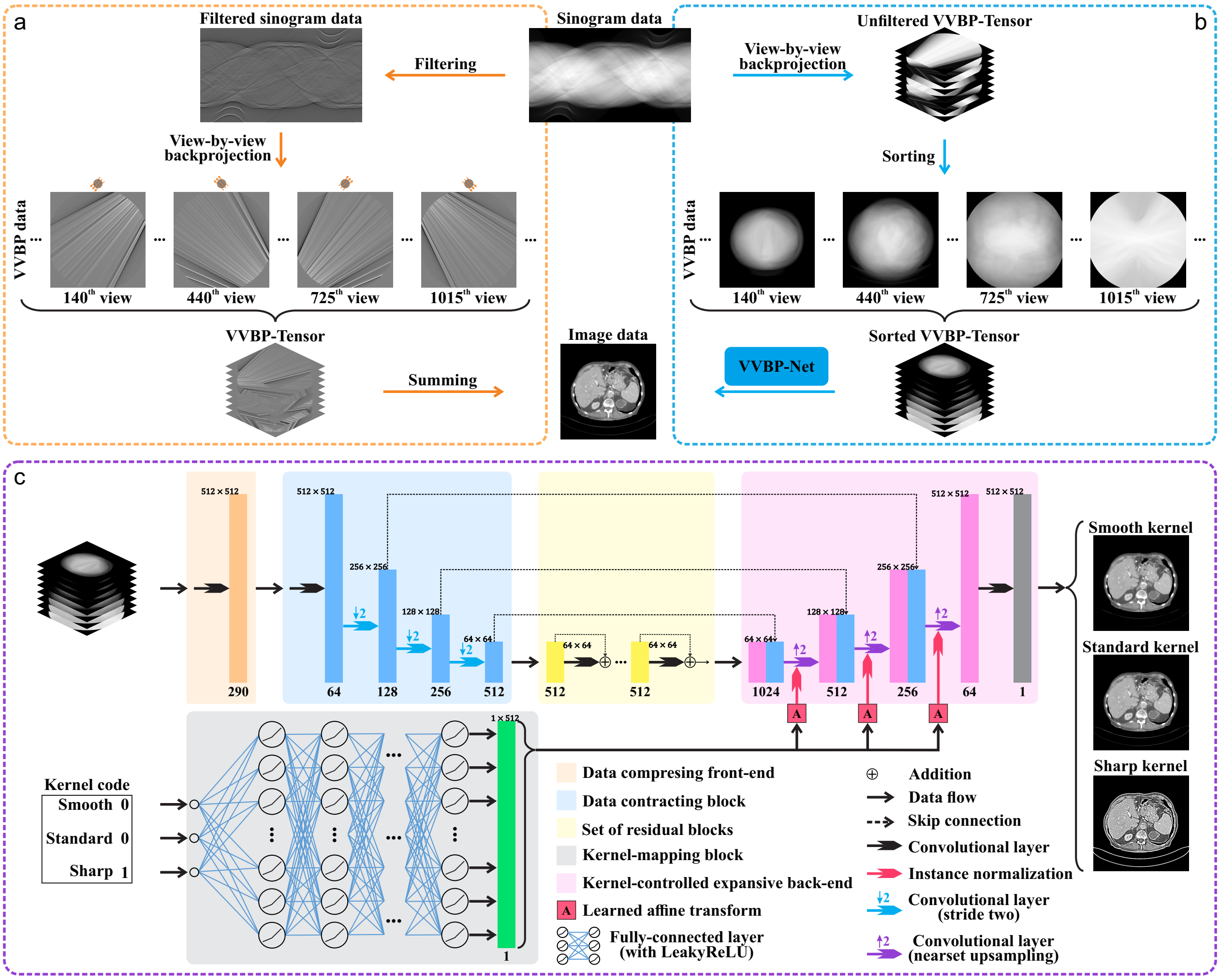

**Figure 1:** (a) The flowchart of conventional FBP reconstruction procedure. (b) The flowchart of VVBP-Net reconstruction procedure. (c) The architecture of our VVBP-Net. The VVBP-Net takes sorted backprojection of all views without unfiltering as the VVBP-Net input, and then multiple reconstruction kernels are simultaneously convolved to reconstruct CT images with a bundle of various kernels.

## 2.2 VVBP-Net framework

Figure 1(c) shows the architecture of the VVBP-Net. The VVBP-Net contains 5 components: data compressing block $\Phi^{(F)}$, data contracting block $\Phi^{(C)}$, a set of residual blocks $\Phi^{(R)}$, kernel-mapping block $\Phi^{(M)}$ and kernel-controlled expansive back-end $\Phi^{(E)}$, which can be expressed as follows:

$$\hat{I}_{N,k} = \Phi^{(E)}[\Phi^{(R)} \circ \Phi^{(C)} \circ \Phi^{(F)}, \Phi^{(M)}(c_K)], \tag{1}$$

where $\hat{I}_{N,K}$ is the estimated image with reconstruction kernel $K$. $V$ denotes the VVBP measurements. The operation $\circ$ denotes functional composition. $c_K$ is the code of kernel $K$ with the dimension $1 \times 3$. Specifically, the codes of the three different reconstruction kernels (i.e. smooth, standard and sharp) for one sample are [1, 0, 0], [0, 0, 1] and, [0, 1, 0] respectively.

## 2.3 Loss function construction

The loss function of our VVBP-Net is comprised of perceptual loss function [11] and mean-squared error (MSE) loss function. Specifically, the perceptual loss is based on VGG-16 network $\Phi^{(VGG)}$ [12]. The perceptual loss function measures the similarity of the feature representations of ground truth $I_{N,K}$ and VVBP-Net output $\hat{I}_{N,K}$, and the perceptual loss can be expressed as follows:

$$\mathscr{L}_{percep} = \sum_{j \in N} \frac{1}{x_j y_j z_j} \mathbb{E}_{(I_{N,K}, \hat{I}_{N,K})} \left\| \Phi_j^{(VGG)}(I_{N,K}) - \Phi_j^{(VGG)}(\hat{I}_{N,K}) \right\|_2^2, \tag{2}$$

where $x_j$, $y_i$ and $z_j$ are the width, height and channel of the feature maps at $j$th layer in $\Phi^{(VGG)}$, respectively. $N$ is the number of the selected layers in $\Phi^{(VGG)}$. In this work, the outputs of the 4, 9, 16, 23 and 30th layers in $\Phi^{(VGG)}$ are selected for calculating the perceptual loss. The MSE loss

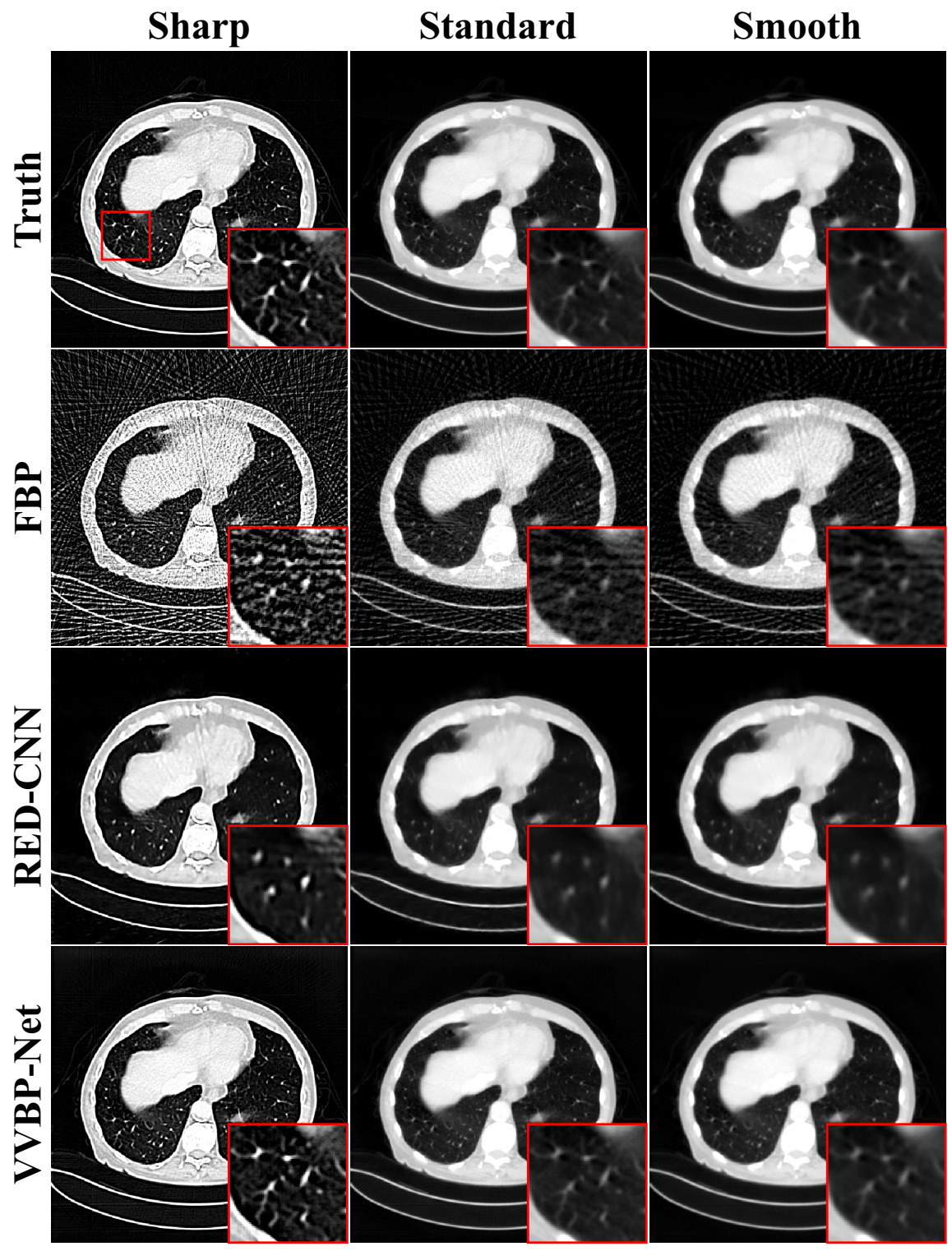

**Figure 2:** Reconstruction results of different methods from the chest CT case. The display window is [-1000, 100] HU.

function can be expressed as follows:

$$\mathscr{L}_{MSE} = \mathbb{E}_{(I_{N,K},\hat{I}_{N,K})} \left\| I_{N,K} - \hat{I}_{N,K} \right\|_2^2, \tag{3}$$

Therefore, the final loss function can be expressed as follows:

$$\mathscr{L} = \mathscr{L}_{percep} + \beta \mathscr{L}_{MSE}, \tag{4}$$

where $\beta$ is the trade-off parameter for the MSE loss and set to be 20 in this work.

### 2.4 Implementation details

In this work, we use the Mayo clinic dataset which can be obtained from the *2016 NIH-AAPM-Mayo Clinic Low-Dose CT Grand Challenge*. We simulate the low-dose with sparse-view projection data (i.e., 1/8 routine dose with 145 sampling views). Moreover, we compared our VVBP-Net with RED-CNN [13]. RED-CNN is a residual encoder decoder convolutional neural network (CNN) which combines auto-encoder, deconvolution network and shortcut connections. Both the VVBP-Net and RED-CNN are implemented in Pytorch [14] library, and their loss functions are optimized through Adam optimizer algorithm [15]. During training period, the settings of RED-CNN are determined according to the suggestions from the original paper. As for the VVBP-Net, the learning rate is set to be 1e-4 which decays at the 100th and 150th epoch by multiplying 0.5 for all 200 epochs. Limited by the hardware memory, the mini-batch size is set to be 4. All the networks are trained on two NVIDIA Tesla P40 GPUs.

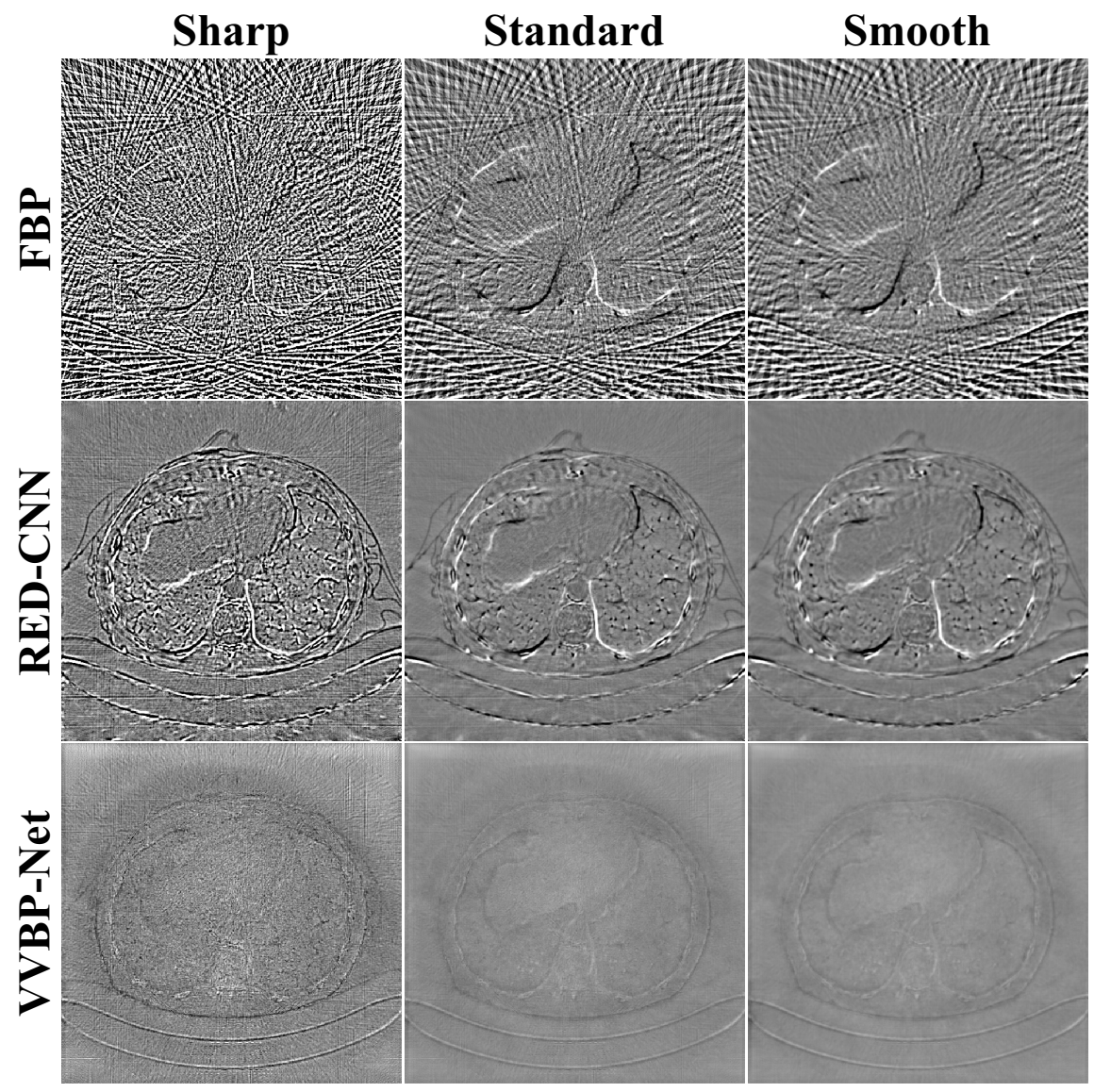

**Figure 3:** Difference images by subtracting the ground-truth and the CT images reconstructed by the FBP, RED-CNN and VVBP-Net methods, respectively. The display window is [-50, 50] HU.

## 3 Results

Figure 2 shows the low-dose with sparse-view CT images directly reconstructed by the FBP method with three different reconstruction kernels, respectively. The difference images between the FBP images and the ground-truth at the three different reconstruction kernels are also displayed. It is obviously that these images are commonly affected by the severe streak-type artifacts, which could limit the diagnostic efficacy. From the results, the RED-CNN demonstrates streak-type artifacts can be suppressed greatly compared to the network input. However, the RED-CNN results suffer from resolution loss and edge blurring compared with the VVBP-Net results. In contrast, the VVBP-Net can maintain acceptable spatial resolution at the same time suppress artifacts successfully benefiting from the VVBP operation.

Moreover, the difference images between the RED-CNN images and the ground-truth are calculated, as shown in Figure 3. As sampling views decreases and noise levels increases, we clearly observe that the VVBP-Net is superior to the RED-CNN in terms of noise-induced artifacts suppression and resolution preservation, exhibiting much better multi-kernel style reconstruction performance in comparison to the RED-CNN.

Furthermore, we use contrast-to-noise ratio (CNR), peak signal-to-noise ratio (PSNR) and structural similarity (SSIM) index measurements to evaluate the VVBP-Net performance for the multi-kernel style images reconstruction. Table 1 summarizes the quantitative results. Due to the issue of over-smooth and fine structural information loss in the RED-CNN, the RED-CNN results are obviously inferior to the VVBP-Net reconstruction results, especially in the sharp kernel cases. Therefore, the quantitative results demonstrate that the VVBP operation helps to reveal much

| | | CNR | PSNR | SSIM |
|---|---|---|---|---|
| Sharp | FBP | 10.60±4.87 | 16.23±1.33 | 0.7501±0.1477 |
| | RED-CNN | 31.32±4.68 | 22.48±1.15 | 0.8685±0.0643 |
| | VVBP-Net | 33.14±4.65 | 40.08±1.10 | 0.9909±0.0475 |
| Standard | FBP | 16.96±12.20 | 27.58±0.96 | 0.8624±0.0413 |
| | RED-CNN | 32.68±2.32 | 32.83±0.85 | 0.9402±0.0402 |
| | VVBP-Net | 38.55±2.29 | 41.99±0.70 | 0.9960±0.0224 |
| Smooth | FBP | 21.51±4.99 | 32.09±1.34 | 0.9148±0.0227 |
| | RED-CNN | 34.30±4.08 | 36.16±0.62 | 0.9659±0.0267 |
| | VVBP-Net | 39.85±3.95 | 42.47±0.35 | 0.9988±0.0221 |

**Table 1:** Quantitative measurements of the FBP, RED-CNN, and VVBP-Net reconstructed images in the testing dataset.

finer structures that might be lost in the FBP images.

## 4 Discussion

Building on recent development in deep learning, we have proposed a VVBP-Net for low-dose and sparse-view CT image reconstruction via explicit reconstruction kernel learning. The proposed VVBP-Net takes unfiltered view-by-view back-projections as network input, and then reconstructs multiple CT images with a bundle of various kernels simultaneously from an extreme low-dose scan protocol. Once the network is trained, the VVBP-Net enables flexibility to control kernels during reconstruction to obtain task-specific CT images and efficiently provides easy access to radiologists, while conventional CT reconstruction methods are regulated manually and subject to expert knowledge. All the information latent in the extreme low-dose measurements is used for VVBP-Net construction, and the intermediates from view-by-view back-projection contain rich structure details and diagnostic information similar to those of desired CT images, further improving reconstruction performance. The VVBP-Net allows for scaling to multiple CT images with multiple kernels simultaneously, either from "smooth" reconstruction kernel to "sharp" reconstruction kernel. This significantly reduces computation time compared with those methods reconstructing CT images with one reconstruction at a time, and enable more flexibilities to control reconstruction kernel switching for radiologists. Therefore, our VVBP-Net allows radiologists to efficiently reconstruct their own kernel models and conveniently diagnose disease.
However, there are some limitations in this work. First, the number of patients used in the experiments is limited and the potential bias is unknown. More clinical patients would be enrolled for evaluation. Second, the VVBP-Net is not designed for particular body region, and in the future study clinical task oriented assessment methods should be added to further evaluate and promote the VVBP-Net reconstruction performance. Third, all the clinical studies in this work are based on digital simulation due to the limited access to the raw data, but this could provide a promising way for researchers to design and construct new CT imaging system, for example, multi-source clinical CT system.

## 5 Conclusion

In conclusion, we demonstrate the VVBP-Net takes a critical step forward in accurate reconstruction of diagnosis CT images for multi-kernel style image reconstruction task, and there is great potential for designing CT systems that obtain extreme low-dose measurements reasonably with deep learning techniques.

# Realistic CT noise modeling for deep learning training data generation and application to super-resolution

Mengzhou Li[1], Peter W. Lorraine[2], Jed Pack[2], Ge Wang[1], and Bruno De Man[2*]

[1]Department of Biomedical Engineering, Rensselaer Polytechnic Institute, Troy, NY, USA
[2]GE Research, Niskayuna, NY, USA * Corresponding author (email: deman@ge.com)

**Abstract** Much progress has been made in deep learning based CT image processing, while little attention has been paid on inserted noise for network training. Additive Gaussian noise model is widely used in existing studies due to its simplicity and efficiency for large training data generation despite its distinctly uncorrelated texture compared to real CT noise. However, we find that this unmatched noise could significantly degrade the inference performance on clinical images with real CT noise and has been seriously overlooked. In this study, we investigate the impact of noise modeling on deep learning-based super-resolution (SR) in terms of noise type/level/anisotropy to emphasize the importance of realistic noise insertion. To address this challenge, we provide a step-by-step recipe for fast generation of large datasets with realistic CT noise by modulating white noise in the frequency domain with a predetermined analytical formula modeling a realistic 3D noise power spectrum (NPS). In our experiments, the generated noise patterns demonstrate almost the same textures and 3D NPS shapes as in the realistic CT noise reference. By comparing the SR performances of several models of the same network structure but trained under different noising conditions, our results suggest that (1) the Gaussian noise model is more vulnerable and a suboptimal choice compared to the model trained with CT noise, (2) the noise level and texture anisotropy can seriously affect the SR performance, and (3) covering the whole range of noise levels and noise anisotropy expected in the testing data could significantly boost the model performance and robustness.

## 1 Introduction

Deep learning has been widely used in CT imaging fields for many tasks including image denoising, low dose reconstruction, data correction, etc. Our topic of interest, CT image super-resolution (SR) with deep learning, has seen several advances in recent years with many techniques adopted from the general computer vision domain [1–3]. For better real-world performance, a certain degree of noise is often introduced into the training data to gain robustness [1]. As realistic CT noise can be affected by numerous factors during the scan, additive Gaussian noise has been widely used in most works as an efficient model, despite its distinct differences in texture and noise correlation. Inevitably, any deviation from the real degradation model could give rise to poor performance in real-world scenarios [4], just a matter of the performance drop difference. But the SR performance drop resulting from noise discrepancy has been widely overlooked, and little work has been done to quantify this impact and elaborate how to choose appropriate noise distributions for network training.

To demystify these questions, this study focuses on introducing realistic CT noise into the training and studying the impact of different noise types/level/anisotropy in the context of deep learning SR performance. We characterize the resulting models trained under different noise conditions by testing their performance both on data with the type of noise they were trained on and on the other types of noise in the study to understand whether improvements are specific to a single type of noise or are more broadly applicable.

## 2 Materials and Methods

### 2.1 Noise generation framework

There exist several ways to generate CT noise. Realistic CT simulators like CatSim [5] offer the most realistic results but have a relatively high barrier to entry and can be time-consuming to generate many large datasets. Sinogram-domain noise insertion [6] is one of the most accurate techniques, but requires access to sinogram data and can be time-consuming since every noise realization needs to undergo a full reconstruction. Image domain noise insertion [7] is another popular method that efficiently simulates CT noise in the image domain rather than the projection domain. It can generate multiple noise realizations for one CT image more efficiently compared to the simulation-based and projection-domain methods but consumes more time for a new image. Since deep learning network training needs a very large number of images and corresponding noise realizations, we are interested in developing a very fast method to generate thousands of noisy images that still produces highly realistic CT noise, and to share this recipe widely with the community.

Our approach is geared towards patch-based training architectures and our goal is to generate stationary noise patterns with realistic texture and realistic three-dimensional (3D) noise power spectrum (NPS). We propose an analytical fitting formula to depict realistic 3D NPS shapes. Realistic CT noise patterns can then quickly be computed by generating white noise and modulating its frequency-response with this predetermined analytical formula. Hence, our approach is not intended to replace realistic simulations, which have the advantage that they produce noise that is spatially-adaptive based on each specific object and scanner on a macroscopic scale. But our approach is ideal for patch-based training of networks that need to deliver robust results under a range of realistic noise conditions.

#### 2.1.1 Noise Anisotropy

Three aspects of the noise are covered in this study, i.e., type, level, and anisotropy. Specifically, we considered Gaussian noise and CT noise as two noise types. A typical range of noise levels (characterized by the standard deviation) was considered. Photon starvation-induced streaks or metal artifacts were out of scope of this study. The anisotropy refers to the dependence of the noise correlation or the noise texture on the orientation, as illustrated in Fig. 1. Representative CT noise with different anisotropy was simulated using water cylinder phantoms of various shapes in a cone beam CT geometry and using Feldkamp reconstruction with a ramp kernel. All phantoms had the same cross-section area but had different eccentricity ratios (Er), defined by the ratio between the major axis length and the minor axis length. We use Er to characterize the anisotropy of the CT noise patterns; e.g.: the noise pattern from the round shape cylinder phantom ($Er = 1.0$) demonstrates isotropic texture while the noise patterns from elliptical shape cylinder phantoms ($Er \neq 1.0$) are more non-isotropic and show more noise streaks as shown in Fig. 1. For ease of notation, we denote the former as isotropic noise and the latter as non-isotropic noise, and the corresponding NPS as isotropic NPS and non-isotropic NPS, respectively.

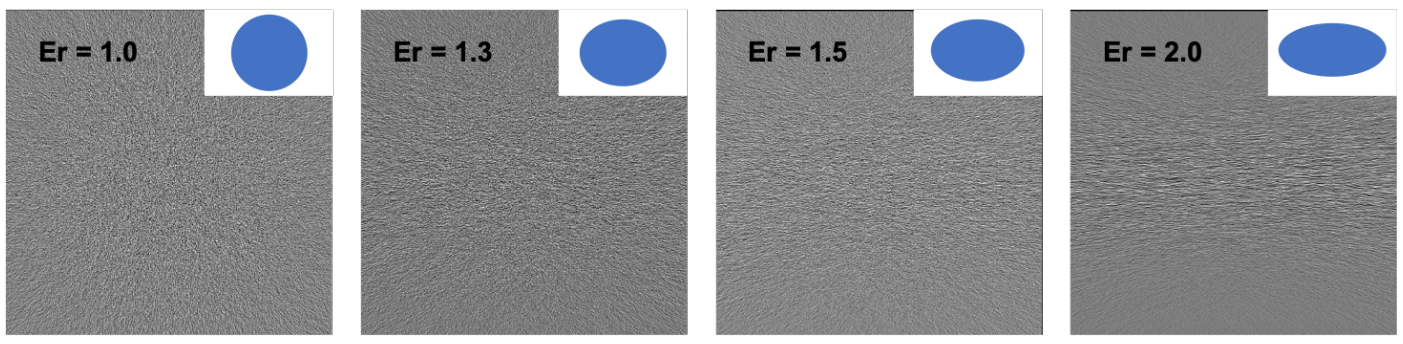

**Figure 1:** Simulated CT noise from water cylinder phantoms with same cross-section area but different shapes (illustrated in the inset figures) demonstrating textures of varying anisotropy with Er values (from left to right) of 1.0, 1.3, 1.5, and 2.0.

#### 2.1.2 3D NPS fitting formula

Our NPS model is based on the following key finding: we noticed that the dependence of the NPS on radial, azimuthal, and longitudinal frequencies is approximately separable. Therefore we factorize our NPS model into a radial component, an azimuthal component, and a longitudinal component:

$$C(f_x, f_y, f_z) = R(l)A(\theta)L(s), \tag{1}$$

where $f_x$, $f_y$ and $f_z$ are the spatial frequencies, $l = \sqrt{f_x^2 + f_y^2}$, $\tan\theta = f_y/f_x$, and $s = f_z$. For fitting the isotropic NPS, we have $A(\theta) \equiv 1$ due to the azimuthal symmetry. Fitting was performed on the average NPS of 20 simulated noise realizations. We empirically found the isotropic NPS is well modeled by the following expressions:

$$R(l) = a_3\sqrt{l}\exp\left[-(l-a_1)^2/a_2^2\right] \tag{2}$$

$$L(s) = \exp\left(-s^2/b_1^2\right), \tag{3}$$

where $a_1, a_2$ and $a_3$ are fitting parameters. For modeling the non-isotropic 3D NPS $C(l,s,\theta)$, we divide the simulated NPS by the isotropic NPS model to fit the azimuthal modulation function $A(\theta)$ to the remainder. A good fit was empirically found consisting of an exponential function and a sine function as follows,

$$A(\theta) = c_3\left[c_1\exp\left(-c_2\sin^2\theta\right) + 1\right], \tag{4}$$

where $c_1, c_2$ and $c_3$ are fitting parameters.

Combining the above formulas, we have an appropriate generation function for a general NPS. By selecting $c_1 = 0$, the model degrades to the isotropic NPS model.

#### 2.1.3 Noise generation recipe

To generate a spatially correlated CT noise volume with realistic texture, we follow the steps listed below:

(1) Generate a volume of Gaussian noise with zero mean and unit variance, e.g., using MATLAB *randn* function;

(2) Compute the Fourier spectrum of the Gaussian noise volume above with the fast Fourier transform, e.g., using MATLAB *fftn* and *fftshift* functions;

(3) Calculate the 3D NPS function of the target noise type based on Eqs. 1–4 following the values in table 1 for the coefficients;

(4) Modulate the Gaussian noise Fourier spectrum with the square root of the calculated 3D NPS via element-wise multiplication;

(5) Transform the modulated noise spectrum back to the image domain through the inverse fast Fourier transform, and keep the real part of the result as the CT noise volume, e.g., using MATLAB *ifftn*, *ifftshift*, and *real* functions;

(6) Scale the image domain CT noise obtained above to the desired noise level to obtain the final CT noise volume.

The coefficient values in table 1 are provided for the ease of replicating this study, but they can vary with different CT scanner geometries, e.g., the detector element size, scanning geometry and etc., hence, they should be adjusted accordingly for other settings. One may find the coefficients following the same procedure as detailed above.

**Table 1:** Coefficient values used for noise generation in this study with Eqs. 1–4 .

| Anisotropy | $a_1$ | $a_2$ | $b_1$ | $c_1$ | $c_2$ |
|---|---|---|---|---|---|
| $Er = 1$ | 0.3061 | 0.4707 | 0.6128 | 0 | - |
| $Er = 1.3$ | 0.3061 | 0.4707 | 0.6128 | 4.3813 | 2.6207 |
| $Er = 1.5$ | 0.3061 | 0.4707 | 0.6128 | 8.2235 | 4.3348 |
| $Er = 2.0$ | 0.3061 | 0.4707 | 0.6128 | 41.990 | 13.587 |
| $Er = 2.5$ | 0.3061 | 0.4707 | 0.6128 | 213.01 | 27.878 |
| $Er = 3.0$ | 0.3061 | 0.4707 | 0.6128 | 1047.9 | 59.425 |

### 2.2 Deep learning super-resolution

In this work we focus on studying the impact of realistic noise modeling on deep learning SR performance. Note that the goal of this study is not to find the best possible deep learning SR network architecture. To do this, we choose one popular representative network for image SR, i.e., the residual channel attention network (RCAN) [8]. RCAN is a ResNet-type network composed of improved residual blocks combined with the channel attention mechanism for SR tasks.

#### 2.2.1 Network structure and training

In our study, we used a simplified version of RCAN to speed up the training. We used one residual group with 15 residual channel attention blocks and discarded the upscaling module at the back end since we had input and output images of the same size. We changed the number of color channels to one and used 64 feature channels. Each training batch used 128 LR CT input image patches of size $64 \times 64$. The model parameters were updated with the ADAM optimizer ($\beta_1 = 0.9, \beta_2 = 0.999, \text{and} \varepsilon = 10^{-8}$), and the learning rate was initiated at $2 \times 10^{-4}$ and then exponentially decayed at a rate of 0.85 per epoch.

The loss function used for network training was defined as follows:

$$\arg\min_{\boldsymbol{\theta}} \|f(\boldsymbol{\theta};\boldsymbol{y}) - \boldsymbol{x}\|_1 + \lambda \left\| \frac{f(\boldsymbol{\theta};\boldsymbol{y}) - \boldsymbol{x}}{\boldsymbol{x} + \gamma} \right\|_2^2, \quad (5)$$

where $f$ and $\boldsymbol{\theta}$ denote the network and its trainable parameters respectively, $\boldsymbol{x}$ and $\boldsymbol{y}$ are the HR ground truth labels and the LR degraded inputs respectively (image pairs), while $\lambda$ and $\gamma$ are hyper-parameters and are set to 1 and 0.1. The L1 norm term regulates the absolute error to avoid blurring, while the L2 norm term is used to measure the relative error to emphasize the fidelity of fine details with small values [9], especially the low intensity structures in the lung regions.

In total, 16 cardiac CT exams acquired using a Revolution CT scanner (GE HealthCare, Waukesha, WI)(images courtesy of Dr Pontone and the Centro Cardiologico Monzino, University of Milan, Italy) were used for network training (12 scans) and testing (4 scans). We bicubically resampled all volumes to the same voxel size of 0.4mm and normalized the values between 0 and 1. The resultant images were defined as high-resolution (HR) ground-truth with a native resolution around 1mm in full width at half maximum. The low-resolution (LR) input images were obtained by degrading the resolution by a factor of 2 via convolving a Gaussian smoothing kernel.

A patch-based training strategy was adopted and the LR and HR patch pairs were randomly extracted within the reconstruction field-of-view of each scan. Noise was generated and added on LR patches. A variety of noise levels were used and a total number of 369k pairs of patches were obtained for network training and validation.

For final performance evaluation, 100 CT slices were randomly drawn from 4 patient scans to form a testing image set. The slices were cropped to the same size of $500 \times 500$. The different types of noise or noise with different levels/anisotropy were inserted to form the final testing data, based on the specific studies, as detailed below.

#### 2.2.2 Experimental design

To investigate the impact of noise model on network performance, we designed three experiments to study three respective aspects of realistic noise modeling: type, level, and anisotropy.

(1) Impact of noise type: We compared the performance of two networks trained from data with isotropic CT noise (**Realistic** model) and data with zero-mean Gaussian white noise (**Simplistic** model), respectively, on test images that were generated with both types of noise. The noise level was in the range of [30,40] HU in all training and test cases.

(2) Impact of noise level: We evaluated the impact of training noise level versus testing noise level. Two models were trained on data with isotropic CT noise levels uniformly sampled in the range [30, 100] HU (**Wide** model) and [30, 40] HU (**Narrow** model), respectively. Test images were generated with isotropic CT noise with noise level ranges [20,30], [30,40], [40,50], [50,60], [60,70], [70,80], [80,90], and [90,100] HU.

(3) Impact of noise anisotropy: We trained two models: one with pure isotropic noise (**Isotropic** model), and the other with a mixture of isotropic and non-isotropic images (**Non-isotropic** model). The non-isotropic noise was generated using 3 different Ers ([1.5,2.0,2.5]), in equal proportions and with random orientations. The noise levels were set in the range [30,40] HU. Noise of 4 different levels of anisotropy ([1.0,1.5,2.0,2.5]) was generated and added to the test images.

In summary, four models were trained: the baseline model, also referred as the **Reallistic/Narrow/Isotropic** model, the **Simplistic** model, the **Wide** model, and the **Non-isotropic** model. The peak signal-to-noise-ratio (PSNR) and the structural similarity index metric (SSIM) were used for quantitative performance evaluation.

## 3 Results and Discussions

### 3.1 Results for noise modeling

Due to abstract page limits, we show only a few representative results. The isotropic 3D NPS fitting results are presented in Figs. 2 (a) and (b). The fitting result (b) agrees well with the measured NPS (a) in both shape and value, and the fitting errors are overall tiny in magnitude. This indicates that our fitting model (Eqs. 2 and 3) is a good fit for the isotropic 3D NPS.

We also checked the fitting results for the modulation function in the case of a non-isotropic NPS with Ers ranging from 1.5 to 2.5. Again, the fitting curves align well with the data points, suggesting the effectiveness of the chosen function depicted as Eq. 4. One example of the generated 3D NPSs based on our proposed model is compared against the reference calculated from the simulated noise in Fig. 2(c). As seen from the cross-sections, the shape of the modeled 3D NPS is almost identical to that of the simulated NPS.

### 3.2 Results for deep learning SR

Figure 3 shows exemplary SR results from the experiment studying the impact of noise type, comparing the realistic model and the simplistic model when applied to inputs with CT noise. Both models significantly improve the image quality with cleaner appearance and finer structures. The realistic model delivers clear and sharp results under all test conditions (only test with CT noise shown here due to space limit). The simplistic model performs well on the input images with Gaussian noise but generates artifacts (grainy appearance) on the input images with CT noise, such as the false white dots/dim pores shown in the insets. Especially in the zoom-in view of the lung region, many grainy structures are generated while, in contrast, the realistic model presents a denoised version of the HR reference with a smooth noise texture.

According to the metric score distributions over the 100 test images for each experiment (not shown due to space limit),

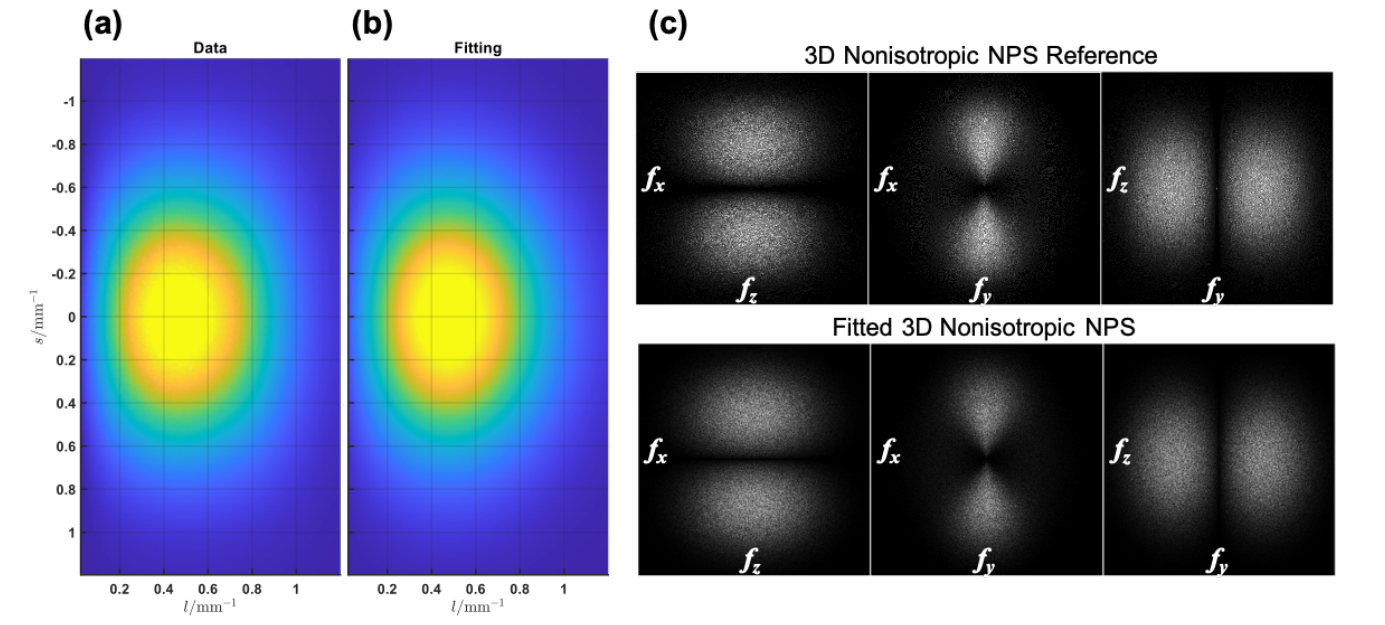

**Figure 2:** Example of a (a) 3D NPS $C_0(l,s)$ for isotropic CT noise, and the corresponding (b) fitting model result; and (c) example of a simulated 3D non-isotropic NPS (top) $C(l,\theta,s)$ with $Er = 1.5$ versus the fitted model (bottom).

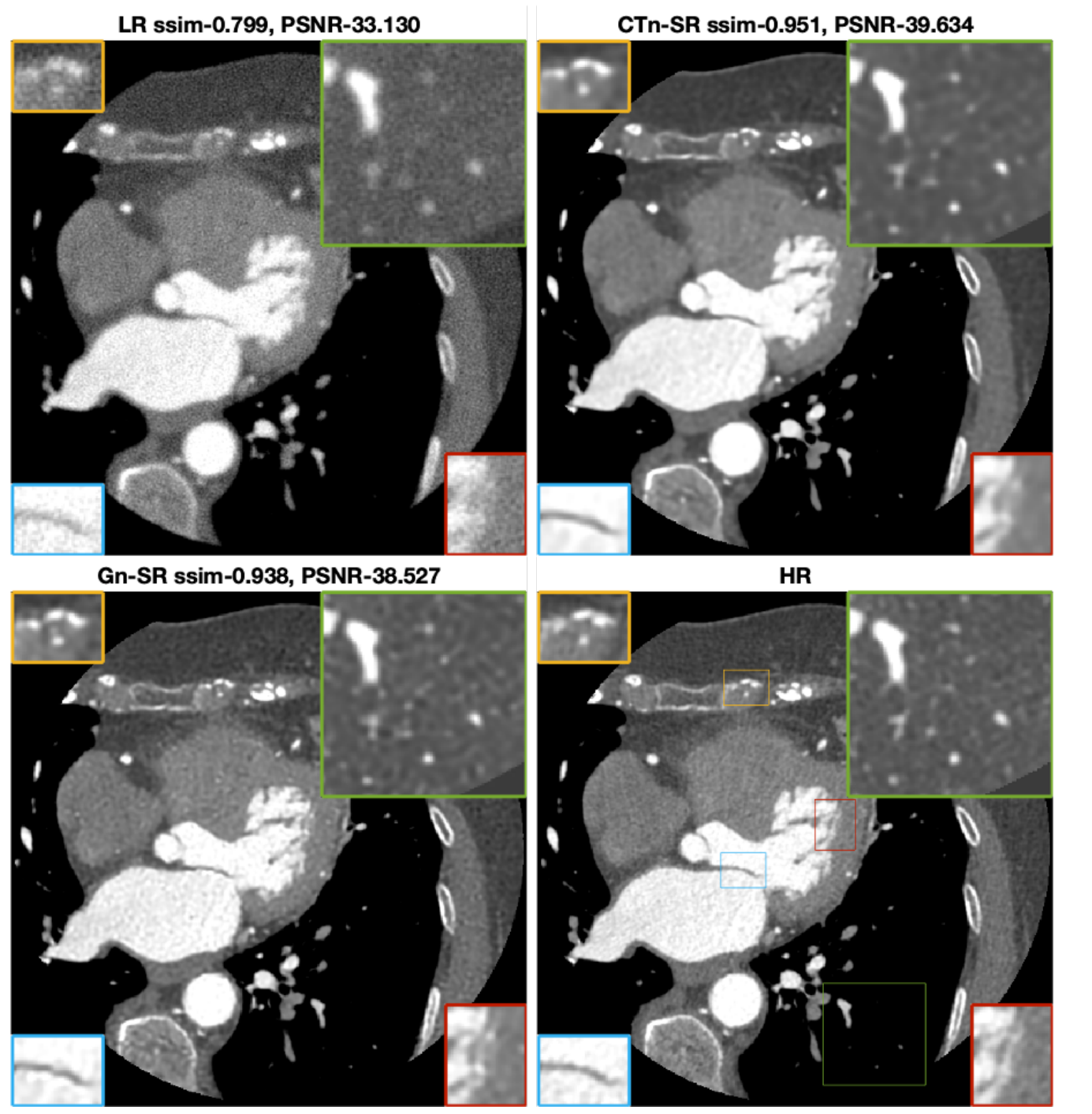

**Figure 3:** Impact of noise type on deep learning super-resolution. The images from top left to bottom right respectively show the low-resolution inputs (LR) with CT noise (CTn3040), the SR prediction with the realistic model (CTn-SR), the SR prediction with the simplistic model (Gn-SR), and the HR ground-truth, with the insets displaying a zoom-in version of a lung region (blue box) and three iodine enhanced regions (orange box, blue box, and red box). The full images, orange insets, and red insets are displayed in[-350,650] HU; the blue insets are displayed in [-350,800] HU; the green insets are displayed in [-1350, 150] HU.

the realistic model results in robust performance, whereas the simplistic model performs well in cases with Gaussian noise but performance degrades in cases with CT noise, illustrating that good performance with Gaussian noise does not necessarily imply good performance with real CT images with CT noise. Though both performance of the wide model and narrow model drops as the noise level increases, the wide model is more robust against the noise level change and always demonstrates comparable or better performance than the narrow model. Similarly, the non-isotropic model performs significantly better than the isotropic model in the conditions with non-isotropic noise. This implies that noise anisotropy is also a significant factor that impacts the SR performance.

## 4 Conclusion

In this paper, we have proposed an efficient framework for fast and realistic 3D CT noise generation, which is ideally suited for patch-based network training. A step-by-step recipe is provided and readily usable in any deep learning research in need of additive realistic CT noise. Note that our model is applicable to CT geometries with any voxel size by adjusting the coefficients. Although our fitting function is based on the noise obtained with a ramp kernel reconstruction, it is easy to adapt our result for other reconstruction kernels by multiplying a correction apodization window function as indicated by the analytical NPS formula derivation [10]. In addition, we have demonstrated the vulnerability of a deep learning model trained with the widely used Gaussian noise when testing on images with CT noise. Besides the noise type, the noise level and noise texture anisotropy also play a significant role in the final SR performance as revealed by our experiments. Our quantitative evaluation further suggests that covering noise with all cases of anisotropy and all possible noise levels expected in the testing data can improve the performance and the robustness of a network when applied to a variety of noise scenarios. Currently we used two standard data science metrics for quantitative evaluation but it is important to use criteria that are relevant to the clinical task performance for medical image evaluation, which will be part of our future work.

**Acknowledgement** This research was supported by the NIH/NHLBI grant# R01HL151561. The content is solely the responsibility of the authors and does not necessarily represent the official views of the NIH.

# Deep kernel representation learning for high-temporal resolution dynamic PET image reconstruction

Siqi Li and Guobao Wang

Department of Radiology, School of Medicine, University of California at Davis, Sacramento, CA 95817 USA

**Abstract** Dynamic positron emission tomography (PET) imaging with high temporal resolution (HTR) presents a challenge for tomographic reconstruction due to the limited count level in each short frame. The kernel methods have been demonstrated to be effective in suppressing noise for low-count dynamic PET data using data-driven spatial or spatiotemporal kernels. However, the construction of these existing kernels follows an empirical process. Our recent work has developed a trainable form for the spatial kernel representation and demonstrated image quality improvement in the spatial domain. In this paper, we further extend the concept to the temporal domain and combine it with the trained spatial kernels. The resulting deep spatiotemporal kernel method is directly applicable to single subjects in dynamic PET imaging. Results from computer simulation and patient studies indicate that the proposed deep kernel method can effectively improve image quality and surpass the performance of existing kernel methods for HTR dynamic PET image reconstruction.

## 1 Introduction

Dynamic positron emission tomography (PET) imaging with high temporal resolution (HTR) can capture rapid spatiotemporal distribution of a radiotracer in human body [1] and may allow the use of more advanced tracer kinetic models to quantify physiologically important parameters in various diseases, such as oncology and cardiology [2]. However, HTR image reconstruction is challenging because of the ill-conditioned tomographic problem and low counting statistics of dynamic PET data.

The kernel methods uniquely integrate image prior information in the forward model of iterative PET image reconstruction by using a kernel representation [3]. The resulting kernelized expectation-maximization (KEM) algorithm is easy to implement and has been demonstrated to significantly improve dynamic PET image reconstruction compared to other methods [3]. For HTR dynamic PET imaging, both spatial and temporal prior knowledge can be simultaneously included in the kernel matrix for improving image quality [1]. However, existing kernel representation is commonly built using an empirical process (e.g., analytically defined feature vectors), which may lead to unsatisfactory performance.

Our recent work [4] has shown the equivalence between the kernel representation and a trainable neural-network model in the spatial domain. A deep kernel model is then derived to enable a learned spatial kernel representation from available image prior information, rather than analytically-defined. This deep kernel method can be applied directly on single subjects without requiring a large training dataset. Compared to other unsupervised deep learning-based reconstruction methods that involve a nonlinear reconstruction problem (e.g., [5,6]), the deep spatial-kernel model has the advantage that once it is trained on the image prior, the unknown kernel coefficient image remains linear in the tomographic reconstruction, making it easier to reconstruct from the projection data.

In this paper, we propose to extend the same representation learning concept from the spatial domain to the temporal domain by making the temporal kernel representation also trainable. Different from deep spatial-kernel representation learning that is performed in the image domain, the deep temporal-kernel representation learning is pursued using the dynamic data in the projection domain for computational efficiency. Combining the learned spatial and temporal kernels together, the deep spatiotemporal kernel method is expected to outperform the kernel methods that either employ a deep spatial-kernel only [4] or an analytically derived spatiotemporal kernel [1] to improve HTR dynamic PET imaging.

## 2 Background

### 2.1 Dynamic PET image reconstruction

Dynamic PET projection data $\boldsymbol{y} = \{y_{i,m}\}$ can be well modeled as independent Poisson random variables using the log-likelihood function [7],

$$L(\boldsymbol{y}|\boldsymbol{x}) = \sum_{i=1}^{N_i}\sum_{m=1}^{N_m} y_{i,m}\log\bar{y}_{i,m} - \bar{y}_{i,m} - \log y_{i,m}!, \qquad (1)$$

where $m$ is the frame index and $N_m$ is the total number of dynamic frames. $i$ denotes the detector index and $N_i$ is the total number of detector pairs. The relationship between the expectation of the projection data $\bar{\boldsymbol{y}}$ and the unknown dynamic image $\boldsymbol{x} = \{x_{j,m}\}$ is given by

$$\bar{\boldsymbol{y}} = \boldsymbol{P}\boldsymbol{x} + \boldsymbol{r}, \qquad (2)$$

where $\boldsymbol{P}$ is the detection probability matrix for PET and includes the normalization factors for scanner sensitivity, scan duration, deadtime correction and attenuation correction. $\boldsymbol{r}$ is the expectation of random and scattered events [7].

Maximizing the Poisson log-likelihood, $\hat{\boldsymbol{x}} = \arg\max_{\boldsymbol{x}\geq 0} L(\boldsymbol{y}|\boldsymbol{x})$ gives the maximum likelihood (ML) estimate of $\boldsymbol{x}$, for example, by the expectation-maximization (EM) algorithm [8].

### 2.2 Spatiotemporal kernel method

In the spatiotemporal kernel method, the image intensity $x_{j,m}$ at the pixel $j$ in frame $m$ can be described as a linear representation of kernels [1],

$$x_{j,m} = \sum_{l\in\mathcal{N}_j}\sum_{v\in\mathcal{N}_m} \alpha_{l,v}\kappa(\boldsymbol{f}_{j,m}, \boldsymbol{f}_{l,v}) \qquad (3)$$

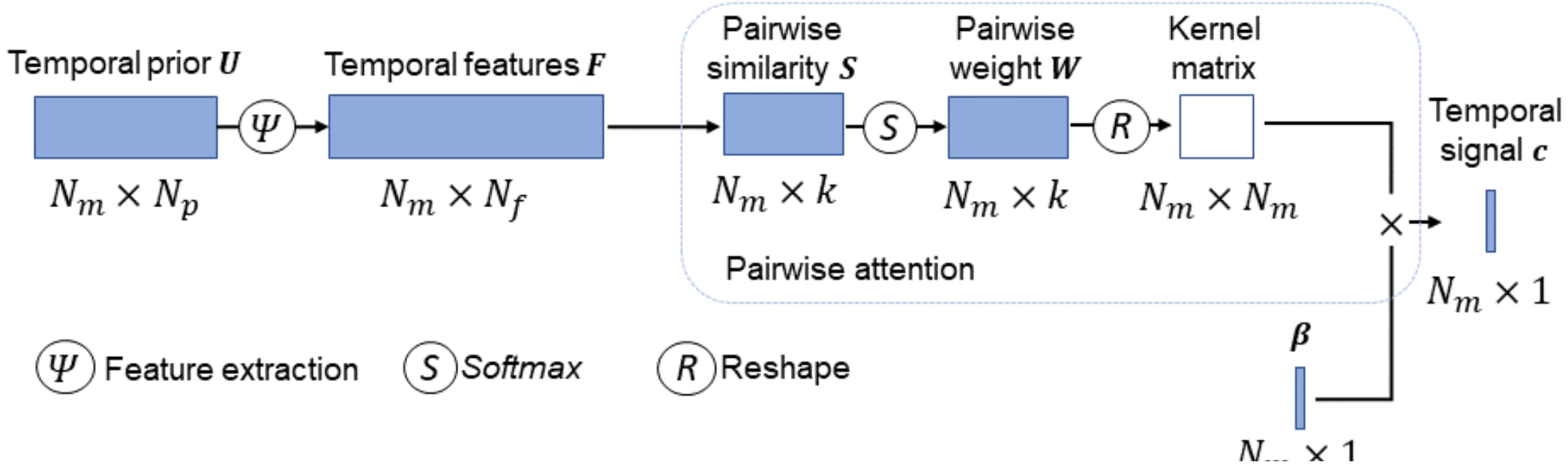

**Figure 1:** The construction of temporal kernel representation as a trainable neural network.

where $\mathcal{N}_j$ and $\mathcal{N}_m$ are neighborhoods of $j^{th}$ pixel and $m^{th}$ frame, respectively. $\alpha_{l,v}$ is the kernel coefficient at pixel $l$ in frame $v$ and $\kappa$ is the spatiotemporal kernel function. The feature vector $\boldsymbol{f}_{j,m}$ can be represented as $\boldsymbol{f}_{j,m} = [(\boldsymbol{f}_j^s)^T, (\boldsymbol{f}_m^t)^T]^T$, where $\boldsymbol{f}_j^s$ is the spatial feature vector of pixel $j$ and $\boldsymbol{f}_m^t$ is the temporal feature vector of frame $m$. The spatiotemporal kernel function can be defined separably,

$$\kappa(\boldsymbol{f}_{j,m}, \boldsymbol{f}_{l,v}) = \kappa_s(\boldsymbol{f}_j^s, \boldsymbol{f}_l^s)\kappa_t(\boldsymbol{f}_m^t, \boldsymbol{f}_v^t) \tag{4}$$

The equivalent matrix-vector form of (3) is:

$$\boldsymbol{x} = \boldsymbol{K}\boldsymbol{\alpha}, \quad with \ \boldsymbol{K} = \boldsymbol{K}_t \otimes \boldsymbol{K}_s \tag{5}$$

where the spatiotemporal kernel matrix $\boldsymbol{K}$ is decoupled into a spatial kernel matrix $\boldsymbol{K}_s$ of $N_j \times N_j$ and a temporal kernel matrix $\boldsymbol{K}_t$ of $N_m \times N_m$. $\otimes$ represents the Kronecker product. The kernel coefficient image $\boldsymbol{\alpha}$ is then estimated from the dynamic projection data $\boldsymbol{y}$ by maximizing the log-likelihood $L$: $\hat{\boldsymbol{\alpha}} = \arg\max_{\boldsymbol{\alpha}\geq 0} L(\boldsymbol{y}|\boldsymbol{K}\boldsymbol{\alpha})$, which can be solved using the kernelized EM algorithm [3],

$$\boldsymbol{\alpha}^{n+1} = \frac{\boldsymbol{\alpha}^n}{\boldsymbol{K}^T\boldsymbol{P}^T\mathbf{1}_N} \cdot \left(\boldsymbol{K}^T\boldsymbol{P}^T\frac{\boldsymbol{y}}{\boldsymbol{P}\boldsymbol{K}\boldsymbol{\alpha}^n + \boldsymbol{r}}\right), \tag{6}$$

where $n$ denotes the iteration number and the superscript "$T$" denotes matrix transpose. $\mathbf{1}_N$ is a vector of length $N = N_i \times N_m$ with all elements being 1. Once $\boldsymbol{\alpha}$ is estimated, the final dynamic PET activity image $\boldsymbol{x}$ is given by $\hat{\boldsymbol{x}} = \boldsymbol{K}\hat{\boldsymbol{\alpha}}$.

### 2.3 Deep spatial-kernel representation learning

In our recent deep spatial-kernel work [4], we describe the PET image of frame $m$, $\boldsymbol{x}_m$, as $\boldsymbol{x}_m = \boldsymbol{K}_s(\boldsymbol{\theta}_s;\boldsymbol{Z})\boldsymbol{\alpha}_m$ ,where $\boldsymbol{\alpha}_m$ denotes the kernel coefficient image of frame $m$. $\boldsymbol{K}_s(\boldsymbol{\theta}_s;\boldsymbol{Z})$ corresponds to the spatial kernel matrix in (5) but is explicitly expressed as a function of the trainable neural network parameters $\boldsymbol{\theta}_s$ and the image prior data $\boldsymbol{Z}$. The deep spatial-kernel learning problem can be formulated using:

$$\hat{\boldsymbol{\theta}}_s = \arg\min_{\boldsymbol{\theta}_s} \sum_{p=1}^{N_z} ||\boldsymbol{z}_p - \boldsymbol{K}_s(\boldsymbol{\theta}_s;\boldsymbol{Z})\tilde{\boldsymbol{z}}_p||^2, \tag{7}$$

where $\boldsymbol{Z} = \{\boldsymbol{z}_p\}_{p=1}^{N_z}$ is the image prior data that consists of $N_z$ composite images. $\tilde{\boldsymbol{z}}_p$ is a corrupted version of $\boldsymbol{z}_p$ [4]. Once $\boldsymbol{K}_s$ is trained, it is then used for frame-by-frame PET reconstruction.

## 3 Deep temporal-kernel representation learning

### 3.1 Trainable temporal-kernel model

In this work, we propose extending the same concept to the temporal domain to enable a trainable temporal kernel representation. The construction of a temporal kernel representation in [1] is equivalent to a trainable neural network model as illustrated in Fig. 1. With this model, a temporal signal $\boldsymbol{c}$ of length $n_m \times 1$ is described by

$$\boldsymbol{c} = \boldsymbol{K}_t(\boldsymbol{\theta}_t;\boldsymbol{U})\boldsymbol{\beta}, \tag{8}$$

where $\boldsymbol{\beta}$ represents the kernel coefficient vector for $\boldsymbol{c}$. $\boldsymbol{U}$ denotes the available temporal prior data which will be used to train the temporal kernel matrix $\boldsymbol{K}_t$. $\boldsymbol{\theta}_t$ includes any trainable neural network parameters. One example of $\boldsymbol{U}$ is the dynamic projection data as used in the existing spatiotemporal kernel method [1].

The trainable network model first extracts features from $\boldsymbol{U}$ to obtain the feature map, $\boldsymbol{f}_m = \Psi_m(\boldsymbol{U})$. Pairwise similarities are calculated between frame $m$ and its neighboring frames using

$$s_{mv} = -\frac{||\boldsymbol{f}_m - \boldsymbol{f}_v||^2}{2\sigma^2}, v \in \mathcal{N}_m. \tag{9}$$

The similarity measures are then converted to pairwise weights using the softmax operation,

$$w_{mv} = softmax(s_{mv}) = \frac{\exp(s_{mv})}{\sum_{v'\in\mathcal{N}_m}\exp(s_{jv'})}, \tag{10}$$

which can be explained as a pairwise attention mechanism [9]. The final step is to reshape the pairwise weight matrix $\boldsymbol{W}$ into a sparse temporal kernel matrix by using the corresponding neighborhood indices.

In this work, the feature extraction step $\Psi$ is parameterized and learned as a 1D modified U-Net model to process time series data, which shares a similar configuration with [4, 6]. In the existing temporal kernel method [1], $\Psi$ is an analytically-defined identity mapping that is equivalent to a 1D convolution with the kernel size of 1.

### 3.2 Single-subject deep temporal kernel learning

Similar to the deep spatial-kernel learning by (7), the temporal kernel respresentation learning can be also performed on single subjects. One straightforward way to train $\boldsymbol{K}_t$ would be using a set of reconstructed time activity curves,

$$\hat{\boldsymbol{\theta}}_t = \arg\min_{\boldsymbol{\theta}_t} \sum_{j=1}^{N_j} ||\boldsymbol{c}_j - \boldsymbol{K}_t(\boldsymbol{\theta}_t;\boldsymbol{U})\tilde{\boldsymbol{c}}_j||^2, \qquad (11)$$

where $\tilde{\boldsymbol{c}}_j$ denotes a corrupted version of $\boldsymbol{c}$ for pixel $j$. However, this strategy requires a pre-reconstruction of the dynamic data and is computationally expensive.

A more practical way is by utilizing the available dynamic projection data $\boldsymbol{U}$,

$$\hat{\boldsymbol{\theta}}_t = \arg\min_{\boldsymbol{\theta}_t} \sum_{i=1}^{N_i} ||\boldsymbol{u}_i - \boldsymbol{K}_t(\boldsymbol{\theta}_t;\boldsymbol{U})\tilde{\boldsymbol{u}}_i||^2, \qquad (12)$$

where $\boldsymbol{u}_i$ is the $i$th column of $\boldsymbol{U}$, denoting the time projection curve acquired by the detector pair $i$. The corrupted data $\tilde{\boldsymbol{u}}$ can be obtained by downsampling the projection data using a count reduction factor $d$ (e.g., $d$=10 in our work). Once $\boldsymbol{\theta}_t$ is trained, the temporal kernel matrix $\boldsymbol{K}_t(\boldsymbol{\theta}_t;\boldsymbol{U})$ can be combined with the trained spatial temporal kernel matrix $\boldsymbol{K}_s(\boldsymbol{\theta}_s;\boldsymbol{Z})$ to form a deep spatiotemporal kernel matrix for dynamic PET reconstruction through (6).

## 4 Validation

### 4.1 Computer simulation

We simulated a HTR dynamic PET scan in the same way as described in [1,4] for a 20-minute $^{18}$F-FDG dataset. The scanning schedule consisted of 63 time frames: 30×2s, 12×5s, 6×30s, and 15×60s. The total count level was 20 million expected events over 20 minutes.

Four types of reconstruction were compared, including (1) standard ML-EM for frame-by-frame reconstruction; (2) standard spatiotemporal kernel method (KEM-S$_A$T$_A$) [1] with both the spatial and temporal kernels analytically defined (denoted by the subscript $A$); (3) modified spatiotemporal kernel method with a deep spatial kernel (KEM-S$_D$T$_A$) as noted by the subscript $D$; and (4) proposed spatiotemporal kernel method with deep spatial and deep temporal kernels (KEM-S$_D$T$_D$).

For the deep temporal kernel learning, a 15-frame length of sliding window was used to define the temporal neighborhood $\mathscr{N}_m$. One thousand iterations were used for the training step with the learning rate set to $10^{-4}$. All reconstructions were run for 200 iterations. The mean squared error (MSE) was used to compare image quality in both spatial and temporal domains.

To demonstrate the effect of temporal kernel learning, Fig. 2 shows the different temporal basis functions for frame 2 and frame 5. Compared to the analytically defined temporal kernel construction, the deep temporal kernel learning leads to appropriate weights for each frame, especially for early frames that commonly have a sharp activity change.

Fig. 3 shows true activity image and reconstructed images by different algorithms with 100 iterations for the frame 2 and frame 5. The image MSE results in dB were included. Overall, the proposed KEM-S$_D$T$_D$ achieved the lowest MSE with good image visualization.

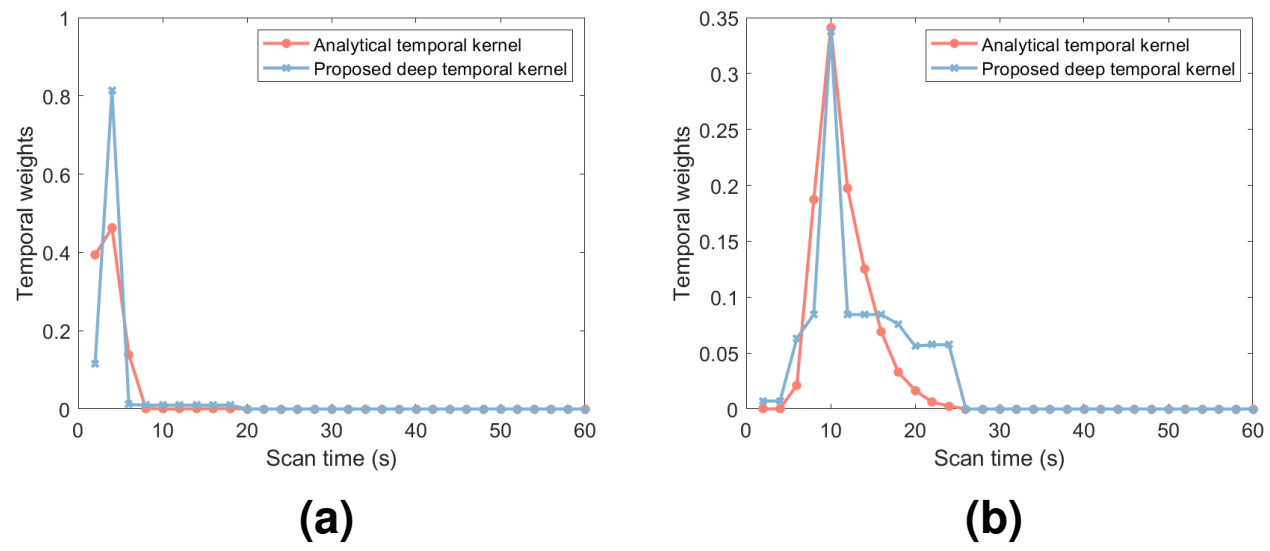

**Figure 2:** Temporal basis functions by analytically defined temporal kernel and learned temporal kernel. (a) $2^{th}$ frame and (b) $5^{th}$ frame

Fig. 4(a) shows the image MSE plots for frame 5 by varying the iteration number. The MSE results (best across iterations) of all time frames are shown in Fig. 4(b). Here the deep spatial kernel method (KEM-S$_D$) [4] was also included for comparison. KEM-S$_D$ achieved better results than KEM-S$_A$T$_A$ in some frames thanks to the learned spatial kernel matrix, but also exhibited instability in the temporal domain. When a temporal kernel was included, the performance became more stable. The proposed KEM-S$_D$T$_D$ showed further improvement for the early frames than KEM-S$_D$T$_A$ because of the learned temporal kernel matrix.

Fig. 5 shows the time activity curves (TACs) of a blood pixel (a) and a tumor pixel (b). The subfigures show the corresponding TACs of the first 60 seconds. Temporal MSE results were also included. Comparing to the analytically defined temporal kernel methods, the TACs reconstructed by the proposed KEM-S$_D$T$_D$ were closer to the ground truth.

### 4.2 Application on real patient scan

Fig. 6 shows reconstructed images using different algorithms for one HTR frame (1s/frame) of an early dynamic cardiac patient scan on a GE DST scanner operated in 2D mode with a 20 mCi $^{18}$F-FDG injection. The traditional ML-EM reconstruction was extremely noisy. The three spatiotemporal kernel methods improved the HTR image reconstruction. Specially, the proposed KEM-S$_D$T$_D$ achieved a lower background noise and had a clearer visualization of the myocardium region than the other two methods, though there is no ground truth here. Further, Fig. 7 shows the HTR time activity curves for one myocardium pixel. In comparison, the proposed KEM-S$_D$T$_D$ demonstrated substantial noise reduction in late frames without losing details in the early frames.

## 5 Conclusion

We have developed a deep kernel representation learning approach in the temporal domain to enable an improved spatiotemporal kernel method. The learning is performed on single subjects by utilizing available dynamic projection data. Computer simulation and real patient studies indicate

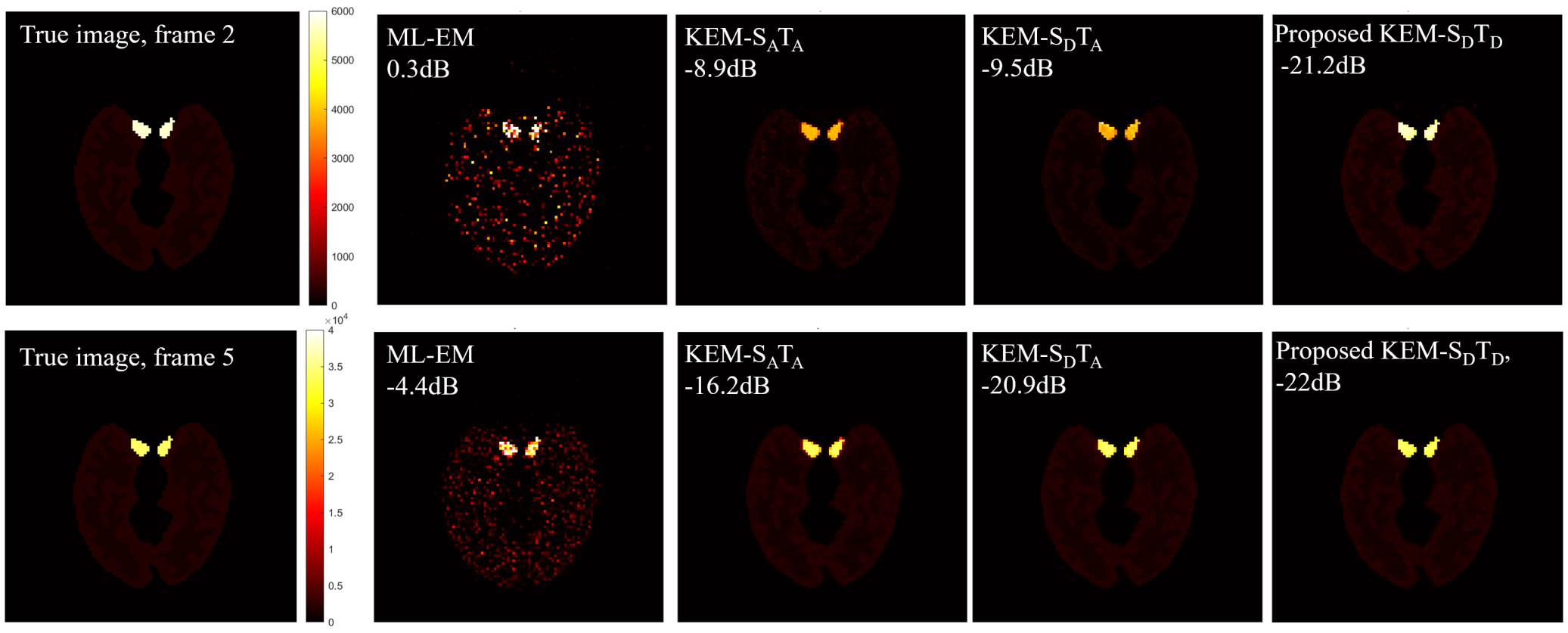

**Figure 3:** Truth activity and reconstructed images by different algorithms for frame 2 (top row) and frame 5 (bottom row).

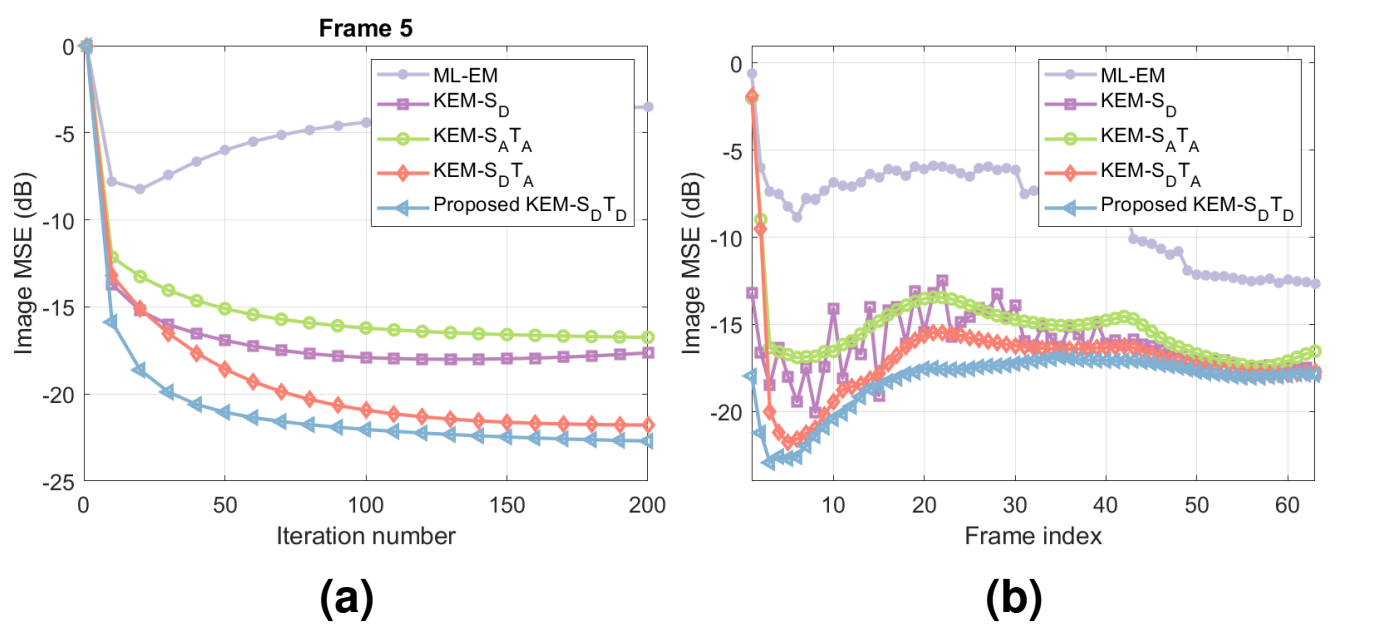

**Figure 4:** Comparison of image MSE for different reconstruction methods. (a) Plot of image MSE as a function of iteration number for frame 5, (b) image MSE of all time frames.

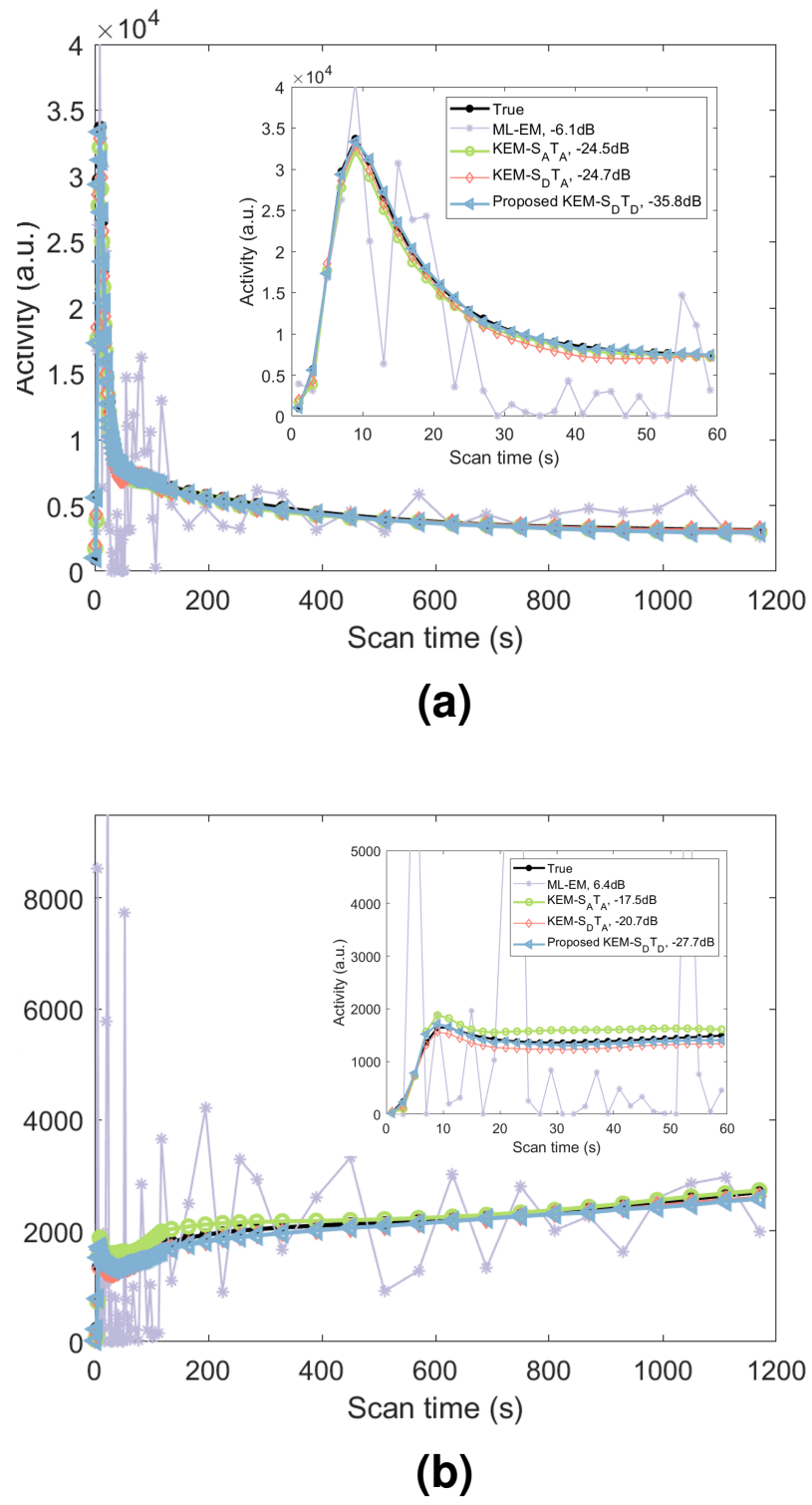

**Figure 5:** TACs reconstructed by different reconstruction methods for (a) a blood pixel and (b) a tumor pixel.

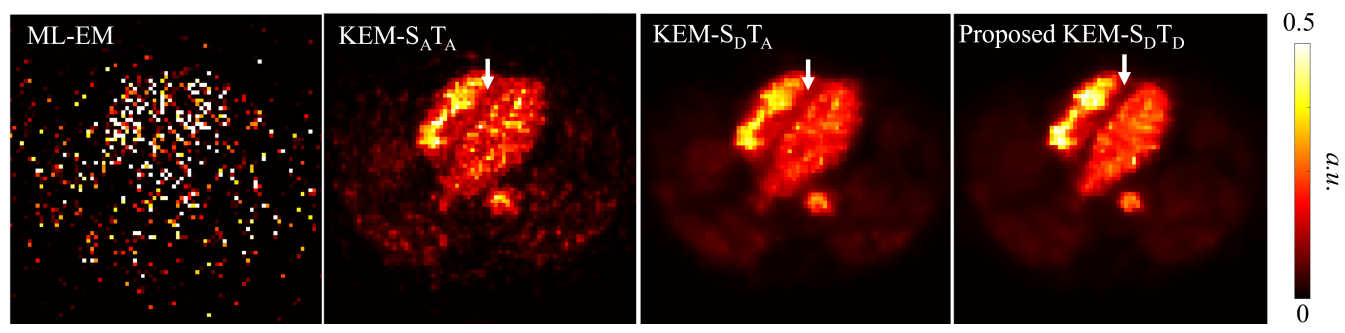

**Figure 6:** Comparison of different methods for reconstructing a HTR frame at $t = 39 - 40s$ from patient data

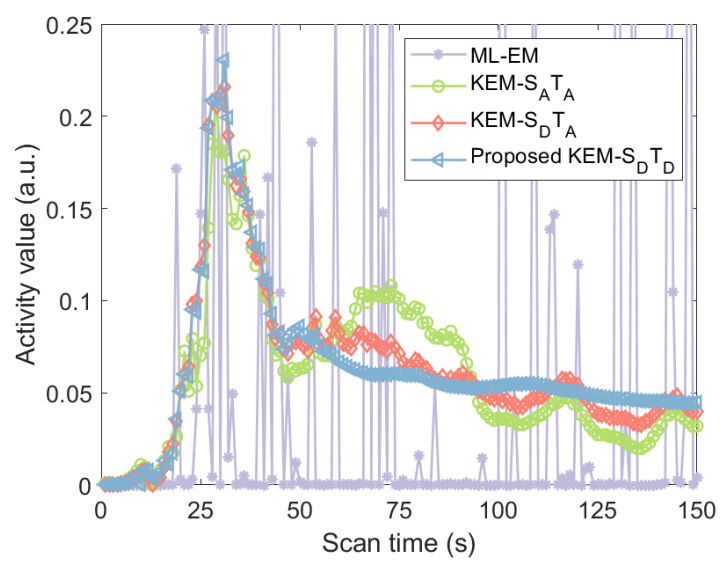

**Figure 7:** TACs reconstructed by different methods for one pixel in the myocardium from the patient scan.

the method can outperform existing kernel methods for HTR dynamic PET imaging.

# Consistency Equations in Native Coordinates and Autonomous Timing Calibration for 3D TOF PET with Depth of Interaction

Yusheng Li[1] and Hongdi Li[1]

[1]United Imaging Healthcare America, Houston, TX 77054

**Abstract** Fully 3D time-of-flight (TOF) PET scanners with depth-of-interaction (DOI) offer the potential of previously unachievable image quality in clinical PET imaging applications. Timing calibration is critical to achieve the best possible imaging performance, and maximize the benefit of TOF PET scanners. In this work, we present an autonomous timing calibration for 3D TOF PET with DOI using intrinsic TOF data consistency. The data space for 3D TOF PET data with DOI is seven-dimensional while the object space is three-dimensional. First, we derive TOF consistency equations in native coordinates to characterize the entangled redundancy and rich structure of 3D DOI TOF data. We then develop an autonomous timing calibration as an application of the TOF consistency equations. Time offsets on a per-crystal-DOI basis, can be computed by solving the two linear timing offset equations involving two TOF moments–the zeroth and first TOF moments. The proposed autonomous timing calibration can be applied to 3D DOI TOF data acquired with an arbitrary tracer distribution, which eliminates the need for a specialized data acquisition. To evaluate the method, we perform numerical simulations of a generic 2D DOI TOF PET with a brain phantom, and the time offsets were accurately computed by solving a Fredholm integral equation of the second type.

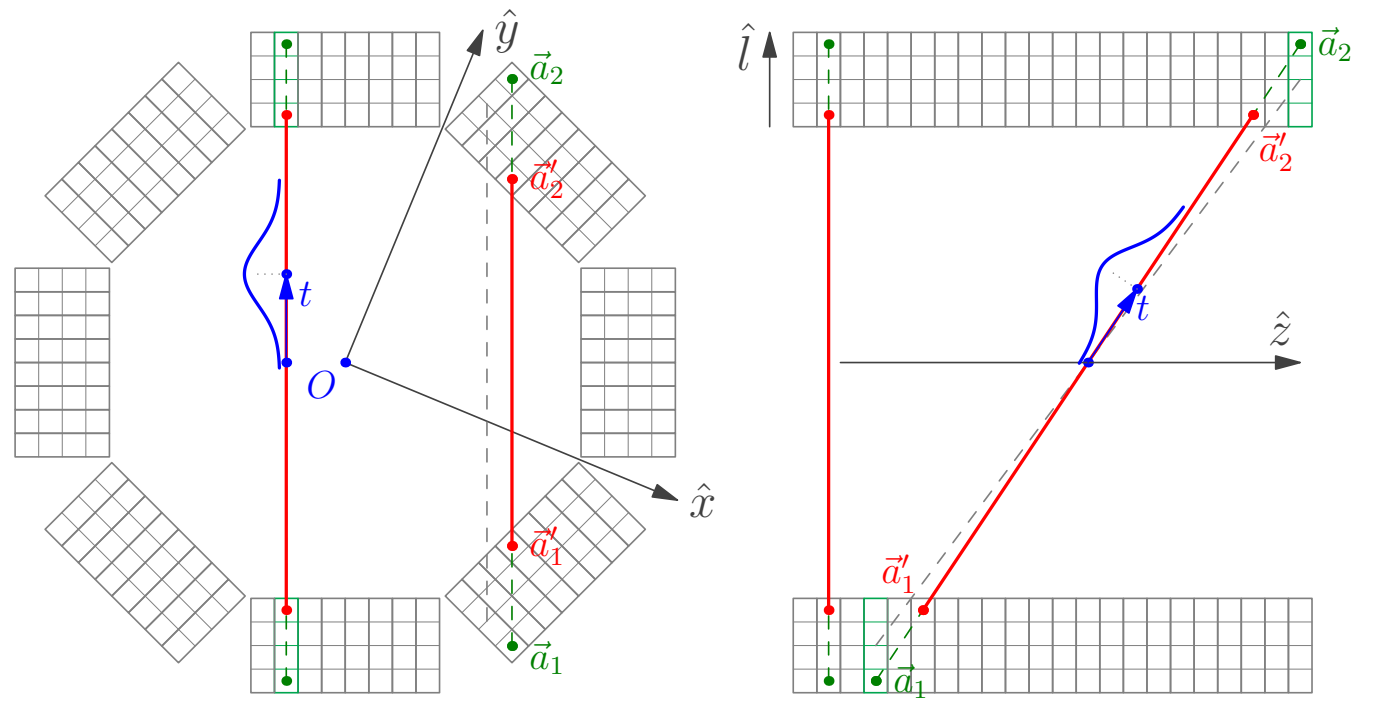

**Figure 1:** Geometry of a cylindrical TOF PET scanner with DOI using modular detectors. Each DOI TOF LOR can be represented by TOF $t$, azimuthal angles $\phi_1$ and $\phi_2$, axial coordinates $z_1$ and $z_2$, DOI layer $l_1$ and $l_2$ in native coordinates. The radius of the TOF PET scanner is a function of azimuthal angle and DOI. Note that the transverse view (left) has been rotated for illustration purposes.

## 1 Introduction

Time-of-flight PET makes use of very fast gamma-ray detectors and timing measure system to precisely measure the difference between the arrival times of the two coincident gammas. The measurements of arrival signalings have several sources of errors that can lead to biases in TOF measurements, which degrades the imaging capability of a TOF PET scanner. These factors include differences in crystal scintillation behavior, variations in detector transit times, operating temperature, electronics delays caused by, e.g., signal trace length, pulse discrimination, and triggering. Therefore, timing calibration is critical to achieve the best possible timing resolution, and maximize the benefit of TOF PET scanners [1]. The depth of interaction (DOI) can not only reduce parallax error and obtain a uniform spatial resolution across the whole field of view, but help to localize the gamma detection within a DOI segment and thus in turn improve timing resolution. In this work, we present an autonomous timing calibration for 3D TOF PET with DOI.

## 2 TOF Consistency Equations with DOI

### 2.1 3D DOI TOF Data Native Parameterization

As shown in Figure 1, we consider a TOF PET scanner with variable radius, $R(\phi,l)$, $\phi\in[0,2\pi]$, $0\le l\le L$. The two endpoints are $\vec{a}_i=[R_i\cos\phi_i, R_i\sin\phi_i, z_i]$, where $\phi_i$, $l_i$, and $z_i$ are the azimuthal angles, depth of interaction, and axial coordinates, respectively. Here, we use $R_i$ to represent $R(\phi_i,l_i), i=1,2$ for concise notations, $R_i=R+l_i$ for pure cylindrical PET scanner. The length of the LOR is given by

$$L(\phi_1,\phi_2,z_1,z_2,l_1,l_2)=\|\vec{a}_2-\vec{a}_1\| = \sqrt{R_1^2+R_2^2-2R_1R_2\cos(\phi_1-\phi_2)+(z_1-z_2)^2}. \quad (1)$$

Here $\phi_i\in[0,2\pi], i=1,2$, and $\phi_1\neq\phi_2$. For a PET scanner with a fixed transverse FOV, we have a minimum angle difference. After projecting the LOR length onto a transaxial plane, we have $L_0(\phi_1,\phi_2,l_1,l_2)=\sqrt{R_1^2+R_2^2-2R_1R_2\cos(\phi_1-\phi_2)}$. For concise parameterization of TOF data in the native coordinates, we introduce three unit vectors:

$$\hat{u}=\frac{\hat{n}\times\hat{z}}{\|\hat{n}\times\hat{z}\|}, \quad (2)$$

$$\hat{n}=\frac{\vec{a}_2-\vec{a}_1}{\|\vec{a}_2-\vec{a}_1\|}, \quad (3)$$

$$\hat{z}=[0,0,1]^{\mathsf{T}}. \quad (4)$$

These three unit vectors are more general than those for a cylindrical PET scanner, and they are the basis vectors to form a local biorthogonal system [2].

In the native detector coordinates, the 3D DOI TOF data can be parameterized as

$$q(t,\phi_1,\phi_2,z_1,z_2,l_1,l_2)=\int_{-\infty}^{+\infty}\mathrm{d}l\,h(t-l)\,f\left(\frac{\vec{a}_1+\vec{a}_2}{2}+l\hat{n}\right), \quad (5)$$

where $f \in C_0\left(\mathbb{R}^3\right)$ is a 3D tracer distribution and $h$ is a TOF profile, $t$ is the TOF parameter in a unit of length instead of time. The TOF profile is usually modeled as a Gaussian distribution with standard deviation $\sigma$,

$$h(t) = \frac{1}{\sqrt{2\pi}\sigma} \exp\left(-\frac{t^2}{2\sigma^2}\right). \tag{6}$$

The TOF data with DOI have the symmetry property, $q(t,\phi_1,\phi_2,z_1,z_2,l_1,l_2) = q(-t,\phi_2,\phi_1,z_2,z_1,l_2,l_1)$, which means that the same TOF LOR can be obtained by exchanging the two endpoints and flipping the TOF $t$. The TOF parameter $t$ is related with the time difference $T_1 - T_2$ between the two arrival times of the two gammas by $t = (T_1 - T_2) \times c/2$ with $c$ denoting the speed of light.

### 2.2 TOF Consistency Equations with DOI

The DOI TOF data $q$ are seven-dimensional (depends on 7 variables) and the object $f$ is 3D—there are four degrees of redundancy. Two degrees of redundancy can be expressed as two TOF consistency equations (CEs)[2, 3]; the other two can be expressed as two linear depth equations (DEs) discussed elsewhere [4]. The consistency equations can be used to explore the rich structure of TOF data. A Gaussian TOF profile is assumed in the derivation of the consistency equations; however, the redundancy in TOF-PET data exists beyond this assumption. We prove (omitted here due to limited space) that the TOF data $q$ in native coordinates given by (5) satisfy the following two consistency equations:

$$\begin{aligned}\Theta q = {} & \frac{L}{2}\mathscr{S}^+ q + t\mathscr{S}^- q - \frac{L}{4}\Gamma_2\frac{\partial q}{\partial t} + \frac{t}{2L_0}\Gamma_1\Gamma_2\frac{\partial q}{\partial t} \\ & + \sigma^2\mathscr{S}^-\frac{\partial q}{\partial t} - \frac{\sigma^2}{2L_0}\Gamma_1\Gamma_2\frac{\partial^2 q}{\partial t^2} = 0,\end{aligned} \tag{7}$$

$$\begin{aligned}\Xi q = {} & \frac{L}{2}\mathscr{Z}^- q + t\mathscr{Z}^+ q + t\frac{z_1 - z_2}{L}\frac{\partial q}{\partial t} \\ & + \sigma^2\mathscr{Z}^+\frac{\partial q}{\partial t} + \sigma^2\frac{z_1 - z_2}{L}\frac{\partial^2 q}{\partial t^2} = 0.\end{aligned} \tag{8}$$

Here, operators $\mathscr{S}^\pm$ and $\mathscr{Z}^\pm$ are respectively given by

$$\begin{aligned}\mathscr{S}^\pm = {} & \frac{\partial}{\partial\phi_1} \pm \frac{\partial}{\partial\phi_2} - \frac{\Gamma_1}{L_0}\left(\frac{\partial}{\partial\phi_1} \mp \frac{\partial}{\partial\phi_2}\right) \\ & - \frac{z_1 - z_2}{2L}\Gamma_2\left(\mathscr{Z}^\pm - \frac{\Gamma_1}{L_0}\mathscr{Z}^\mp\right),\end{aligned} \tag{9}$$

$$\mathscr{Z}^\pm = \frac{\partial}{\partial z_1} \pm \frac{\partial}{\partial z_2}. \tag{10}$$

and the two coefficients are given by

$$\Gamma_1 = \frac{R_1^2 - R_2^2}{L_0}, \tag{11}$$

$$\Gamma_2 = -\frac{2L}{L_0^2}R_1R_2\sin(\phi_1 - \phi_2). \tag{12}$$

It is worth noting that the coefficients $\Gamma_1$ and $\Gamma_2$, are antisymmetric, and $\Gamma_1$ is zero for the same DOI bin combinations, i.e., $(i,i), i = 0,1,\ldots,n_l - 1$. We have also derived John's equation as a linear combination of the two CEs (omitted here).

## 3 Autonomous Timing Calibration for DOI TOF PET

### 3.1 DOI TOF Data with Timing-Offset Errors

When two gamma rays are emitted from an annihilation and detected by two crystals at $\vec{a}_1$ and $\vec{a}_2$. If the true gamma arrival times at the two crystals are respectively $t_1$ and $t_2$, then the true TOF $t = t_1 - t_2$. We use $\tau(\phi,z,l)$ to denote the timing offset at crystal-DOI $(\phi,z,l)$. And the crystal timing corrections are the negative values of $\tau(\phi,z,l)$. Then the measured TOF $t'$ is given by

$$\begin{aligned}t' &= (t_1 + \tau(\phi_1,z_1,l_1)) - (t_2 + \tau(\phi_2,z_2,l_2)) \\ &= t + \tau(\phi_1,z_1,l_1) - \tau(\phi_2,z_2,l_2).\end{aligned} \tag{13}$$

The measured TOF data $m$ (after data normalization, attenuation, random and scatter corrections) are related to the ideal TOF data $q$ by

$$\begin{aligned}m&(t',\phi_1,\phi_2,z_1,z_2,l_1,l_2) = q(t,\phi_1,\phi_2,z_1,z_2,l_1,l_2) \\ &= q(t' - \tau(\phi_1,z_1,l_1) + \tau(\phi_2,z_2,l_2),\phi_1,\phi_2,z_1,z_2,l_1,l_2).\end{aligned} \tag{14}$$

### 3.2 Linear differential timing offset equations

Applying the first and second consistency equations to measured TOF data $m$, we obtain following two linear timing offsets equations (detailed proof is omitted here):

$$\begin{aligned}\mathscr{S}^- & \left[M^0\left(\tau(\phi_1,z_1,l_1) - \tau(\phi_2,z_2,l_2)\right)\right] \\ & = \frac{L}{2}\mathscr{S}^+ M^0 - \frac{1}{2}\frac{\Gamma_1\Gamma_2}{L_0}M^0 + \mathscr{S}^- M^1,\end{aligned} \tag{15}$$

$$\begin{aligned}\mathscr{Z}^+ & \left[M^0(\tau(\phi_1,z_1,l_1) - \tau(\phi_2,z_2,l_2))\right] \\ & = \frac{L}{2}\mathscr{Z}^- M^0 - \frac{z_1 - z_2}{L}M^0 + \mathscr{Z}^+ M^1.\end{aligned} \tag{16}$$

Here $M^k$ is the $k$th moment of TOF data given by

$$M^k(\phi_1,\phi_2,z_1,z_2,l_1,l_2) = \int_{-\infty}^{+\infty} \mathrm{d}t\, t^k m(t,\phi_1,\phi_2,z_1,z_2,l_1,l_2). \tag{17}$$

The zeroth moment is just the non-TOF data. Due to the symmetric property of $q$, the moments are symmetric and antisymmetric for the even and odd orders, respectively. The timing offset equations (15) and (16) for timing calibration can be solved using numerical methods [5]. The timing offsets $\tau(\phi,z,l)$ can be determined up to an additive constant by the two linear PDEs; however, this global constant does not affect timing calibration since only differences are utilized during TOF image reconstruction.

### 3.3 Linear integral equation for 2-D TOF PET

For 2D DOI TOF data, $z_1 = z_2 = z$, we can rewrite (15) as

$$\mathscr{S}_0^- \left[M^1 - M^0 \left(\tau(\phi_1, l_1) - \tau(\phi_2, l_2)\right)\right] + \frac{L_0}{2}\mathscr{S}_0^+ M^0 - \frac{1}{2}\frac{\Gamma_1\Gamma_2}{L_0} M^0 = 0, \tag{18}$$

where

$$\mathscr{S}_0^{\pm} = \left(1 - \frac{\Gamma_1}{L_0}\right)\frac{\partial}{\partial\phi_1} \pm \left(1 + \frac{\Gamma_1}{L_0}\right)\frac{\partial}{\partial\phi_2}. \tag{19}$$

In 2-D case, (20) is the generalized version of equation (7) in [6] and equation (22) in [5]. After integrating (20) over $\phi_2$ and summing over $l_2$, we obtain the following linear integral equation

$$\sum_{l_2=0}^{n_l}\int_0^{2\pi} \mathrm{d}\phi_2 \left(1 - \frac{\Gamma_1}{L_0}\right)\left[M^1 - M^0\left(\tau(\phi_1, l_1) - \tau(\phi_2, l_2)\right) + \frac{L_0}{2}M^0\right] = n_l C, \tag{20}$$

where $C$ is a constant independent of angle $\phi_1$. From (20), we can rewrite the timing offsets $\tau(\phi_1, l)$ as the solution of the following linear integral equation,

$$g(\phi_1, l_1)\tau(\phi_1, l_1) - \sum_{l_2}\int_0^{2\pi} \mathrm{d}\phi_2\, K(\cdots)\tau(\phi_2, l_2) = f(\phi_1, l_1), \tag{21}$$

where

$$K(\cdots) = K(\phi_1, \phi_2, l_1, l_2) = \left(1 - \frac{\Gamma_1}{L_0}\right)M^0, \tag{22}$$

$$g(\phi_1, l_1) = \sum_{l_2}\int_0^{2\pi} \mathrm{d}\phi_2 K(\phi_1, \phi_2), \tag{23}$$

$$f(\phi_1, l_1) = \sum_{l_2}\int_0^{2\pi} \mathrm{d}\phi_2 \left(1 - \frac{\Gamma_1}{L_0}\right)\left(M^1 + \frac{L_0}{2}M^0\right) - n_l C. \tag{24}$$

Equation (21) can be converted to a Fredholm integral equation of the second type by dividing $g(\phi_1, l_1)$.

## 4 Evaluations and Simulated Results

To evaluate the autonomous timing calibration method, we simulated a generic 2D DOI TOF PET scanner of diameter 52.35 cm. The scanner has 20 detector modules with 12 crystals in each module. The width and length of the crystals are 6 mm and 20 mm, respectively, and each crystal was equally partitioned into 4 DOI layers. The timing resolution of the TOF PET scanner is 250 ps FWHM. We used a brain phantom, and the activity and attenuation images with voxel size of 2 mm are shown in Figure 2. Figure 3 shows the generated timing offsets for the 20 × 12 crystals with 4 DOI bins.

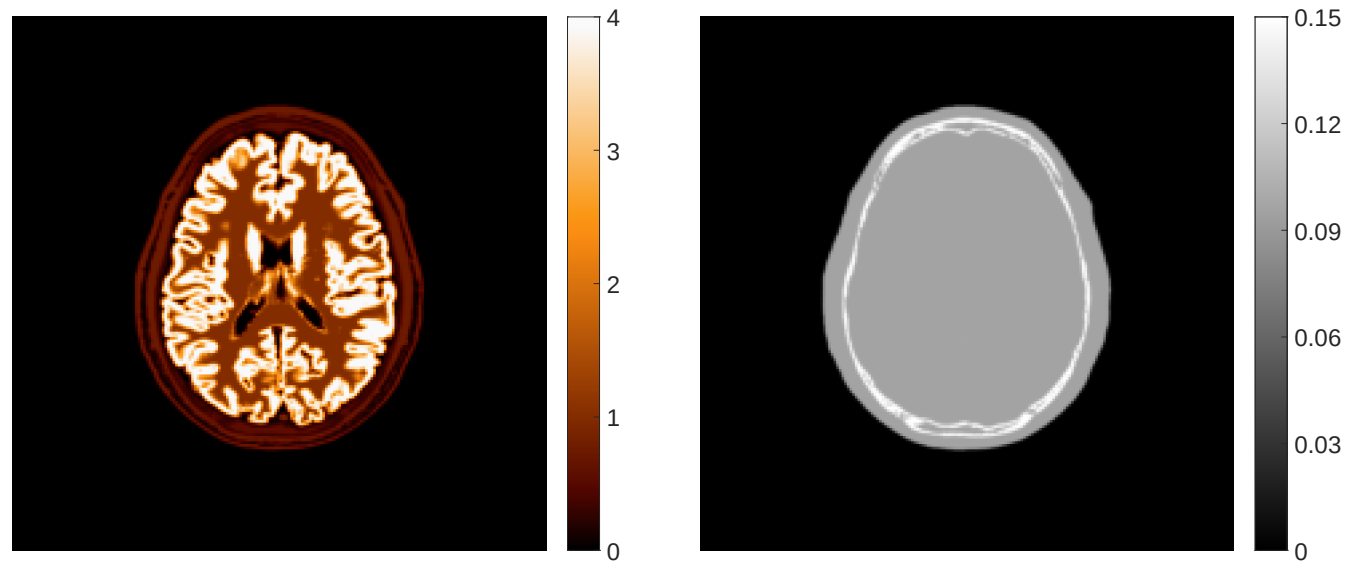

**(a)** Activity image **(b)** Attenuation image

**Figure 2:** Activity and attenuation images of the simulated brain phantom. The image size is 200 × 200 with pixel size of 2 mm.

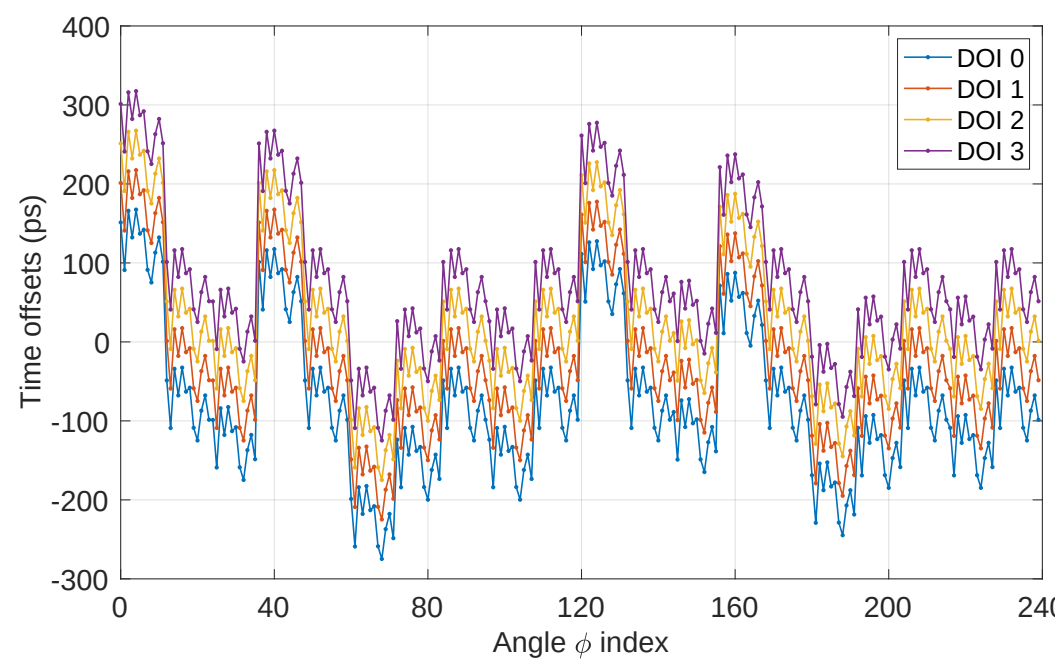

**Figure 3:** The simulated timing offsets for the 20 × 12 crystals with 4 DOI bins.

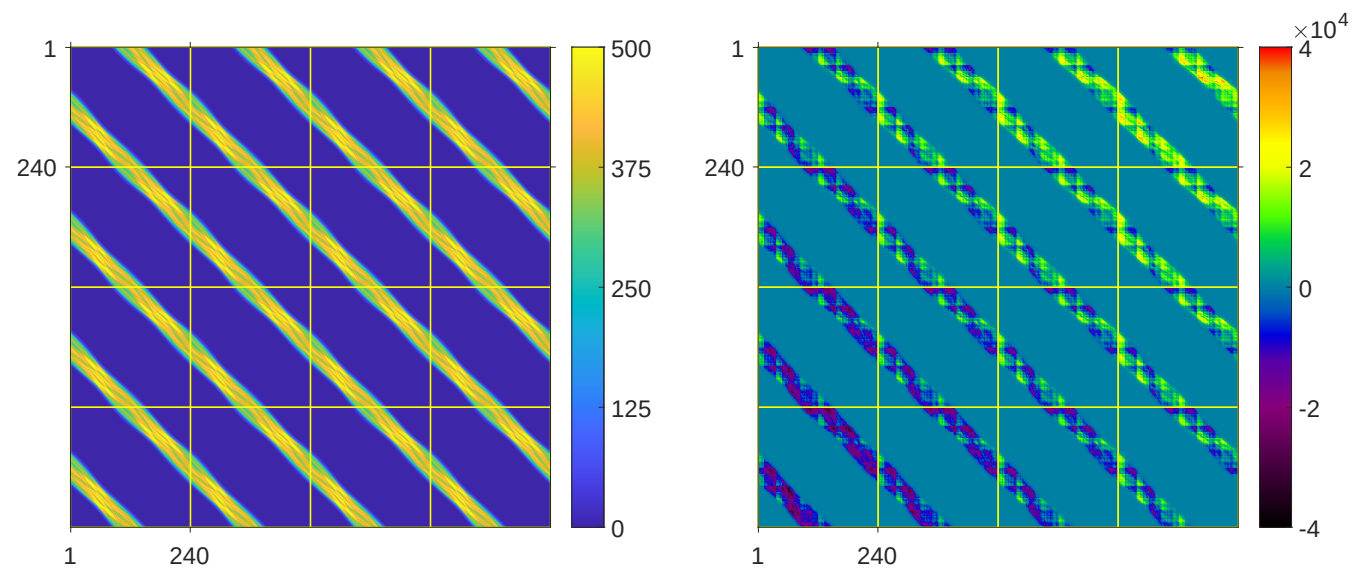

**(a)** $M^0$ **(b)** $M^1$

**Figure 4:** The zeroth (a) and first (b) DOI TOF moments computed from noise-free DOI TOF data. There are 4 × 4 DOI combinations.

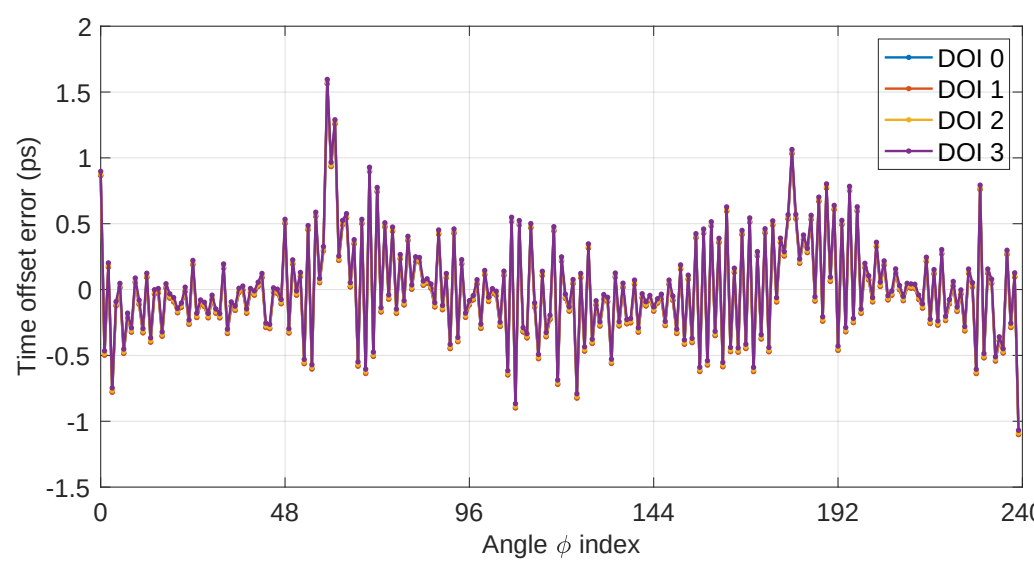

**Figure 5:** The estimated timing offset error for the 480 crystals with 4 DOI bins from noise-free DOI TOF data.

We implemented DOI TOF projector using a strip-integral model with a Gaussian TOF kernel weight. The noise-free DOI TOF projections were generated with a strip width of

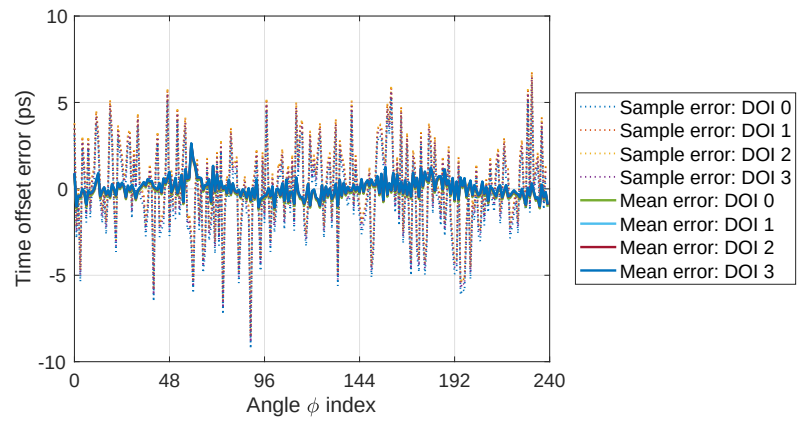

**(a)**

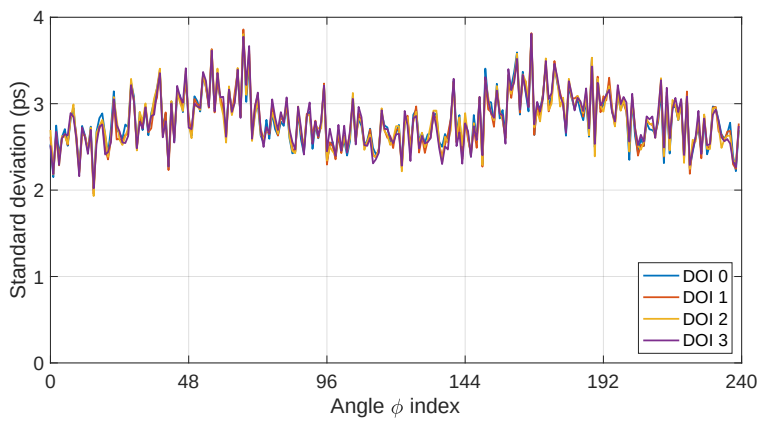

**(b)**

**Figure 6:** Timing offset error for the $240 \times 4$ crystal-DOIs estimated from noisy DOI TOF data. (a) one sample error from one noise realization and the mean error from the 60 noise realizations. (b) Standard deviation estimated from the 60 noise realizations.

3 mm and a Gaussian TOF profile of about 37.5 mm FWHM (corresponding to 250 ps). Attenuation was included based on the attenuation image; scatter and random events were not simulated. Figure 4 shows the two moments computed from noise-free DOI TOF data with attenuation correction. Figure 5 shows error between the error of the estimated time offsets from noise-free DOI TOF data.

We also estimated the timing offsets from noisy DOI TOF data. Poisson noise was added in the attenuated TOF sinograms, and attenuation was applied before estimating the timing offsets. We used 64M total counts for timing calibration. We simulated 60 independent noise realizations, and estimated the mean and variance from the 60 noisy DOI TOF data sets. Figure 6 shows the timing offset error estimated from noisy TOF data. The sample RMS error is 2.61 ps for timing offsets estimated from one noise realization. The mean RMS error is 2.85 ps from the 60 estimates. The standard deviation estimated from the 60 noise realizations is also shown, and the average standard deviation of the estimated timing offsets of the $240 \times 4$ crystal-DOIs is 2.81 ps.

Poisson statistics can affect the accuracy of the estimated timing offsets. We show in Figure 7 the estimated RMS errors at different count levels. Based on linear fitting in logarithmic scales, the RMS errors were proportional to $N^{-0.501}$, with $N$ denotes the total number of true events.

We also performed image reconstructions from both noise-free and noisy DOI TOF data to show how the timing offset errors impact reconstructed images. The images were reconstructed in an array of $200 \times 200$ with 2 mm pixel size using full DOI TOF OSEM with 20 iterations and 12 subsets. Figure 8 shows three types of reconstructions: without timing offset correction, with the known timing offsets, and with the estimated timing offsets based on TOF data consistency.

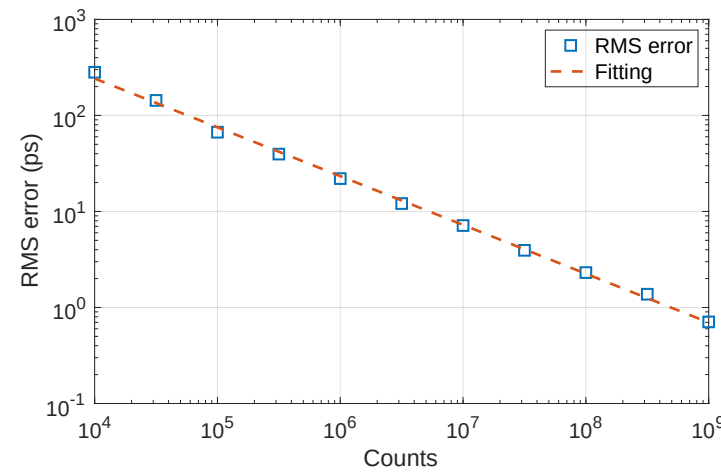

**Figure 7:** RMS errors of estimated time offsets at different count levels.

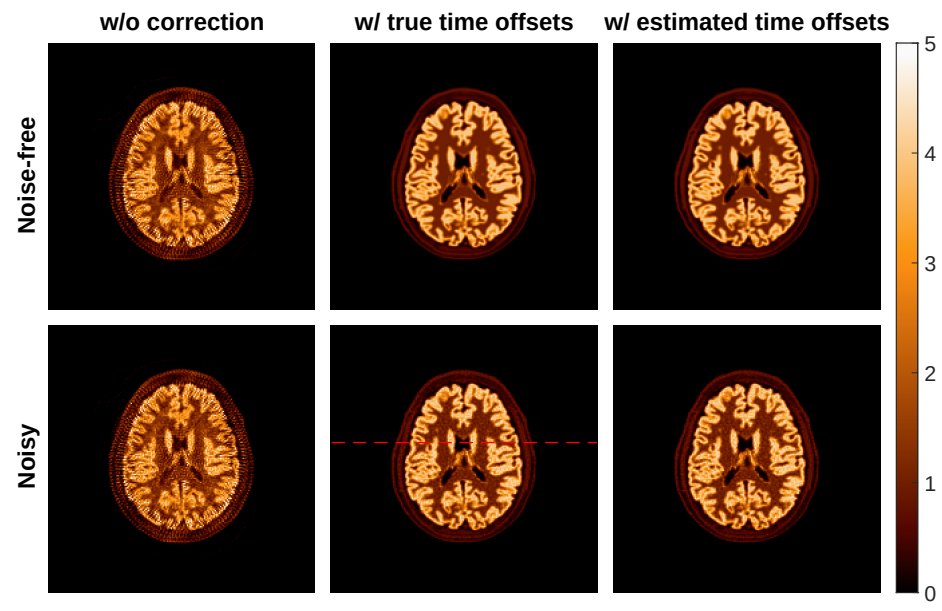

**Figure 8:** Full DOI TOF reconstructions without timing offset correction (left), with known timing offset (center) and with estimated timing offsets based on TOF data consistency (right). Top row: noise-free data; bottom row: one of the noisy data sets.

## 5 Conclusion

We presented an autonomous timing calibration for DOI TOF PET based on the TOF data consistency in native detector coordinates. The method can be applied to data acquired with an arbitrary tracer distribution, which eliminates the need for a specialized data acquistion. The timing offsets on a per-crystal-DOI basis can be determined up to a global constant which does not affect timing calibration. We showed that the algorithm can accurately estimate timing offsets from 2D DOI TOF PET simulations.

# Deep Spectrum Complex-valued Neural Network for Large-scaled Objects Super-resolution Reconstruction

Zirong Li[1] and Weiwen Wu[1]

[1]School of Biomedical Engineering, Sun Yat-sen University, Shenzhen, China

**Abstract** Computed Tomography (CT) has been widely used in industrial high-resolution non-destructive testing, but it is difficult to obtain high-resolution images for large-scale objects due to its physical limitation. To address this problem, many super-resolution deep learning networks have been proposed to map the low-resolution image to the high-resolution counterpart in the image domain. Although those methods achieve a certain effect on the main body of the image, the small structures and detail would be inevitably damaged. For restoring small structures and detail better, we find that high-frequency components (small structures and detail) are easier to be recovered in the frequency domain. Therefore, in this study, a deep spectrum complex-valued neural network has been proposed to take advantage of both global information in the image domain and high-frequency information in the frequency domain. In addition, by exploring the symmetrical property of the spectrum, we design a novel learning strategy to reduce weight parameters in the frequency domain. Besides, a novel spectrum loss has been proposed to constrain both high-frequency components and global information. The experimental results demonstrate that the proposed reconstruction method obtains better results over the state-of-the-art networks.

## 1 Introduction

Compared with conventional detection methods, super-resolution computed tomography (CT) equipping with a micro-focus X-ray source has higher sensitivity and resolution[1]. It has been widely used in many fields, such as biological, industrial, and medical applications. For example, it is employed to explore the internal plant shoots[2], the weld points of integrated circuits[3], and tiny features within capillaries[4]. It is not difficult to find that CT scanning has achieved high-resolution detection in small-scale objects.

The theoretical image spatial resolution of a CT system is determined by the equivalent beam width ($BW$). The smaller the $BW$, the higher the spatial resolution can be achieved. However, reducing the $BW$ to improve the spatial resolution of the reconstructed image has limitations. As shown in Eq. (1), the $BW$ is determined by three factors: the effective focus size of the ray source $S$, the effective pixel size of the detector $P$, and the geometric magnification $G$ of the system. With fixed $S$ and $P$ given in a certain CT system, the spatial resolution of the reconstructed CT image depends on the $G$ of the system. The $G$ of a system refers to the ratio of the source-detector distance (SDD) to the source-object distance (SOD). Since the detector has an imaging matrix of a certain size, the smaller the equivalent beam width, the smaller the detection field of view of the system. Therefore, enlarging the detection field and improving the resolution is contradictory.

$$BW = \sqrt{\frac{P^2 + (G-1)^2 S^2}{G}} \tag{1}$$

In order to obtain an image of higher resolution without changing the detection field, we propose a deep spectrum complex-valued neural network (DSCNN) to map low-resolution CT images to high-resolution ones. Most of the existing super-resolution methods focus on the image domain, but our DSCNN analyzes and explores the super-resolution task from a different spectrum perspective.

Our study demonstrates that the spectrum is critical for super-resolution (SR) reconstruction. Obviously, the low-frequency component of the image determines the main body of the image, and the high-frequency component corresponds to the details. If the high-frequency components can be well recovered in the spectrum, the SR reconstruction performance in the image domain will be better. However, as far as we know, one common drawback of existing SR methods is that they are just focusing on the image domain, which leads the high-frequency components are striking damaged in SR processing[5][6]. Those methods cannot reconstruct small structures and detail well, and the spectral shape is significantly different from that of the original image. Therefore, our DSCNN reconstructs the SR images in both frequency domains and image domains to ensure both the main body and details recovered well.

## 2 Methodology

### 2.1 Problem Statement and Motivation

Reconstructing high-resolution (HR) images from low-resolution (LR) ones can be treated as an inverse problem. $X$ and $Y$ are defined as the domains spanned by LR and HR space. The degradation can be described as $X = D(Y)$, where $X = \{x_i\}_{i=1}^{N}$ and $Y = \{y_i\}_{i=1}^{N}$, $x$ and $y$ are samples, $N$ is the sample number. SR reconstruction is to learn the inverse process of the degradation $Y = D^{-1}(X)$. The general residual learning treats the reconstruction process as $Y = D^{-1}(X) + X$.

To address these issues, we propose learning in the frequency domain to reconstruct high-quality images. Tiny details, structures, and sharp edges are high-frequency components, which means they occupy a large spectrum range and are easy to be focused on by neural networks in the frequency domain. Moreover, it is found that the shape of SR image spectrum generated by our DSCNN is radial, which is similar to the HR image spectrum. Therefore, our DSCNN synthesizes frequency domain learning and traditional image domain learning. Both the global information and details are recovered accurately.

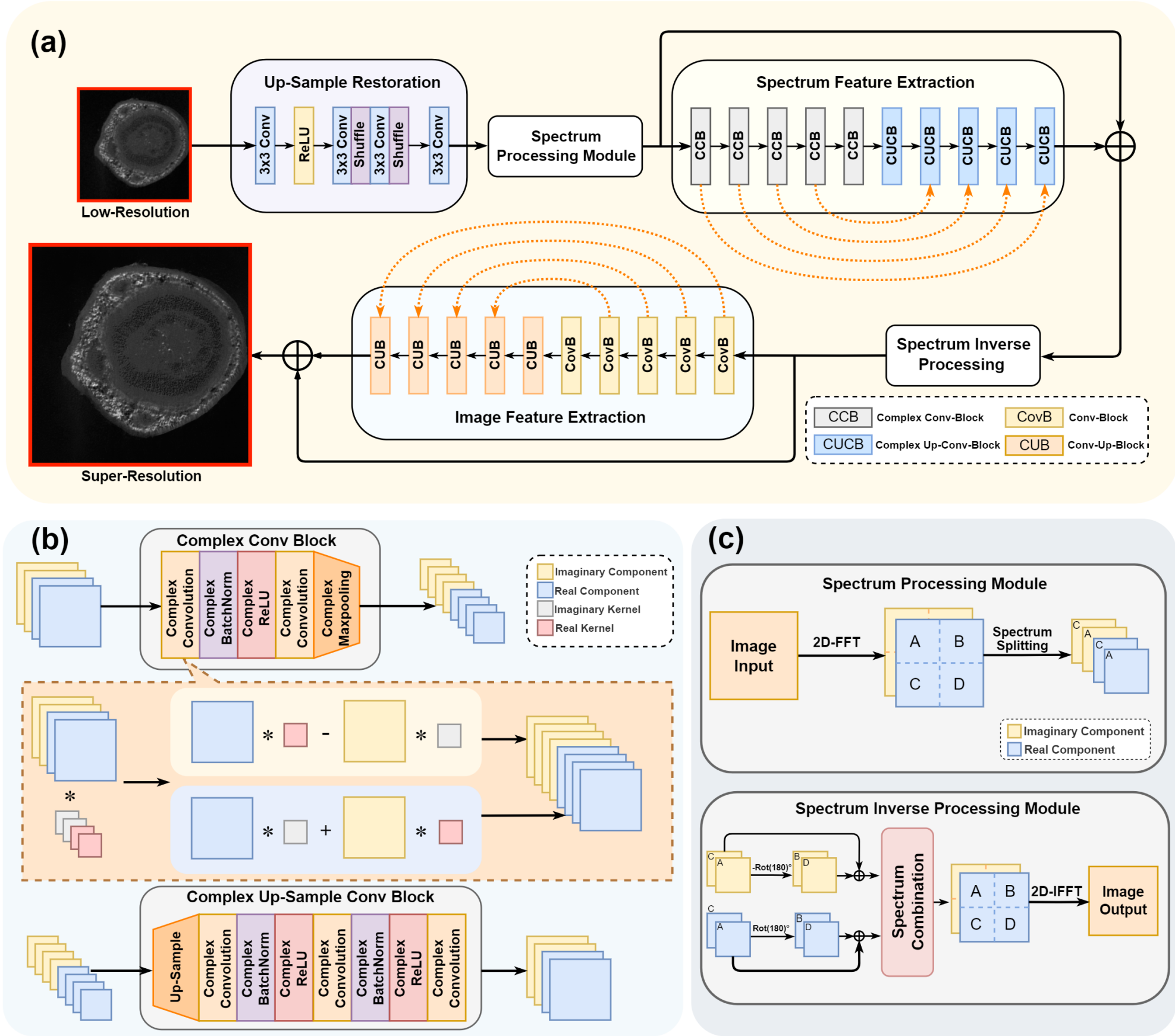

**Figure 1:** (a) shows the overall structure of our DSCNN, consisting of spectrum forward transform, spectrum feature extraction, spectrum inverse transform and space feature refinement; (b) shows a specific structure of the CCB and CUCB in spectrum feature extraction; (c) shows the detailed spectrum forward and inverse transforms modules.

## 2.2 Complex-valued methods Modeling

A pioneering related work on deep complex-valued networks is proposed in[7]. Through 2D Fourier transformation, the input image is transformed into a spectrum, which consists of real and imaginary components. The spectrum can be represented as a complex number $c = a + bi$, where $a = Re\{c\}$ is the real component and $b = Im\{c\}$ is the imaginary component. Many common methods regard $a$ and $b$ as independent parts and work with them separately. However, the amplitude and phase in the frequency domain are jointly determined by $a$ and $b$. Those methods destroy the relevance of $a$ and $b$. To solve this problem, the correlation between $a$ and $b$ is considered in the following modules.

### 2.2.1 Complex-valued Convolution

Instead of operating $a$ and $b$ separately, the entire complex number is treated as a whole. Let us define $c = a + bi$ as the vector to be convoluted. For suiting entire complex number convolution, the complex filter matrix $F = A + Bi$ is required to be utilized, where $a$ and $b$ are real vectors, $A$ and $B$ are real metrics. The convolution product of $c$ and $F$ can be given as follows:

$$F * c = (A * a - B * b) + (B * a + A * b)\, i \qquad (2)$$

It can also be represented by the matrix as follows:

$$\begin{bmatrix} Re(F*c) \\ Im(F*c) \end{bmatrix} = \begin{bmatrix} A & -B \\ B & A \end{bmatrix} * \begin{bmatrix} a \\ b \end{bmatrix} \qquad (3)$$

It is evident from Eqs. (2) and (3) that the result of spectrum convolution is jointly determined by $a$ and $b$. The correlation information is efficiently utilized.

### 2.2.2 Complex-valued Max-pooling

For a complex number supposing $a$ and $b$ are pooled separately, the corresponding relationship between the real and imaginary components will be ignored, and the spectrum phase will also be not accurate. Therefore, the top priority of modeling complex-valued max-pooling is to find an appropriate pooling basis. Here, we take the amplitude matrix $\mathbf{z} = \sqrt{\mathbf{a}^2 + \mathbf{b}^2}$ as the basis of pooling, which is determined by the real component matrix $\mathbf{a}$ and imaginary component matrix $\mathbf{b}$. Through pooling $\mathbf{z}$, the relation between $\mathbf{a}$ and $\mathbf{b}$ are preserved. We design the following strategy to get $\mathbf{a}_p$ and $\mathbf{b}_p$:

A small constant $\varepsilon$ is introduced and we further define $\mathbf{a}_\varepsilon = \mathbf{a} * \varepsilon$ satisfied $|\mathbf{a}_\varepsilon| \ll |\mathbf{z}|$. $\mathbf{a}_p$ and $\mathbf{b}_p$ are given by:

$$\begin{aligned} \mathbf{a}_p &= \frac{Pooling\,(\mathbf{a}_\varepsilon + \mathbf{z}) - Pooling\,(\mathbf{z})}{\varepsilon} \\ \mathbf{b}_p &= \sqrt{Pooling\,(\mathbf{z})^2 - \mathbf{a}_p^2} \end{aligned} \tag{4}$$

### 2.3 DSCNN Network Architecture

As shown in Fig. 1(a), first, the up-sampled input image is converted into spectrum feature maps consisting of real and imaginary components. At the same time, the spectrum feature maps are cropped within this module to reduce computational consumption. Only half of the feature maps containing complete information are preserved for the next part. The preserved feature maps learn deep-level spectral features in the Fourier domain and restore high-frequency information through the spectrum feature extraction module. Then the learned partial spectrum is restored to the complete spectrum. The spectrum inverse processing module converts the spectral feature maps to the image domain through Fourier inverse transform. The output image is finally processed through the image feature extraction module, which restores the low-frequency and comprehensive information within the image domain. Finally, we can obtain a high-quality, precise, and clear SR image.

### 2.4 Spectral Loss

In order to obtain small features and details, we develop a spectral loss for image SR reconstruction, which is based on high-frequency spectral components and image fidelity (i.e, MSE). The proposed spectral loss is beneficial to extract the high-frequency information of the SR and original HR image. The image is transformed into a spectrum by a two-dimensional discrete cosine transform. The low-frequency components are clustered in the upper left corner of the spectrum, and the high-frequency components are scattered in other regions. We set the upper left corner of the spectrum matrix to zero to extract the high-frequency details. The image will mainly contain high-frequency information through the inverse transformation. Our overall loss formula is given as follows:

$$\ell = \alpha \left\| I_{HR} - I_{SR} \right\| + \beta \left\| I_{HFHR} - I_{HFSR} \right\| \tag{5}$$

where $\alpha$, $\beta$ are constant, $I_{HR}$ is the HR image, $I_{SR}$ is the SR image, $I_{HFHR}$ is the high-frequency components within HR image and $I_{HFSR}$ is the high-frequency components within SR image. By altering $\alpha$ and $\beta$, the weights of high-frequency information can be adjusted.

## 3 Experiments and Results

### 3.1 Datasets and Implementation Details

In this study, the plant vessels of broussonetia papyrifera, fresh poplar, withered poplar, and begonia are scanned by a typical micro CT. The content of the datasets consists of 3000 tomographic images. Here they are used to verify the advantages and universality of our DSCNN in terms of details recovery and preservation.

In the first experiment, the training datasets consist of the poplar and begonia vessel images, and test datasets come from the vessel images of broussonetia papyrifer. The number of training datasets and testing datasets are 2200 and 800, respectively. In the next experiment, the poplar and broussonetia papyrifer vessel images are used to train our network, and the vessel images of begonia are employed to implement the test task.

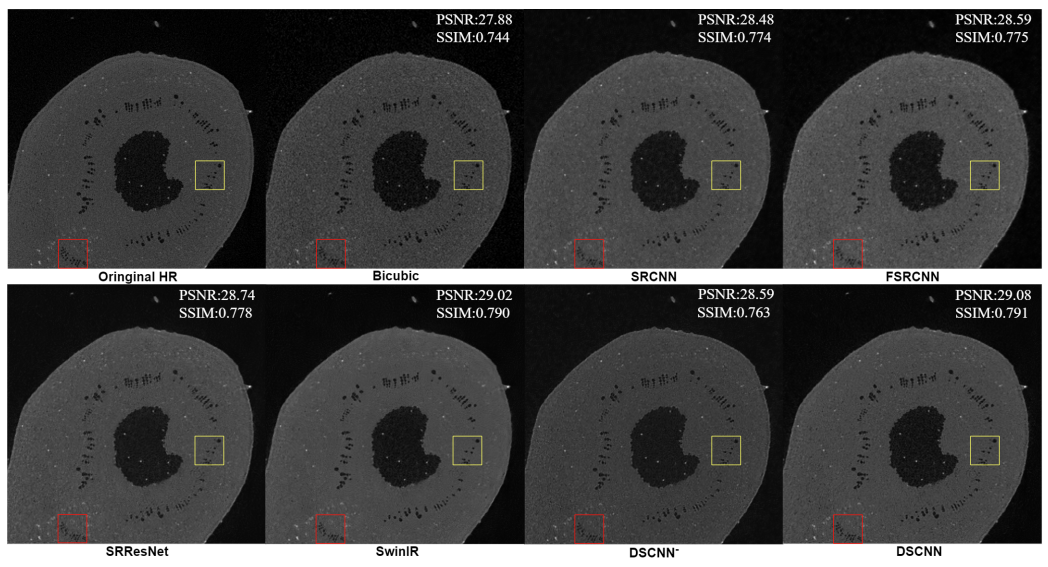

**Figure 2:** The SR reconstruction results from different algorithms in terms of the broussonetia papyrifera vessel. The results show that our proposed DSCNN achieves best performance.

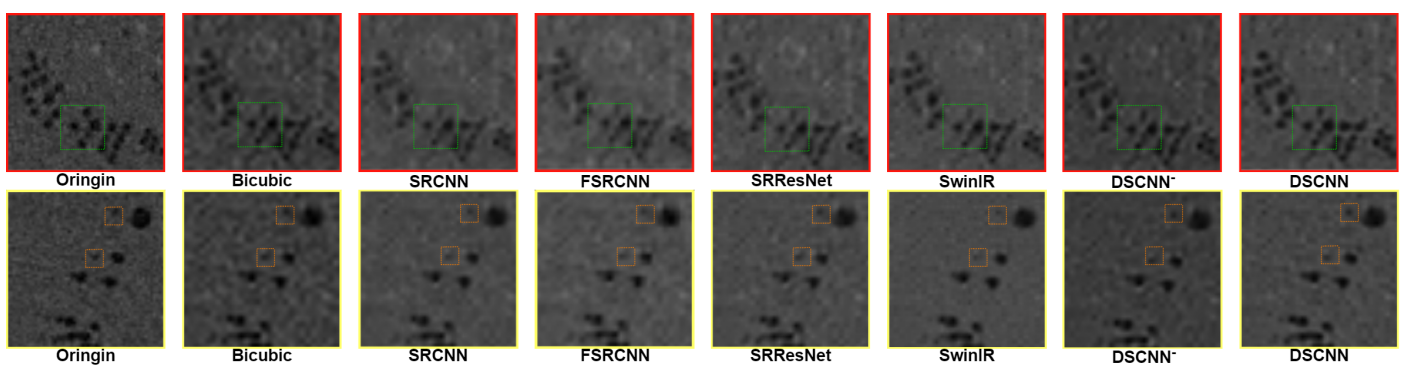

**Figure 3:** Two regions of interests (ROIs) marked in Fig. 2 are further extracted and magnified to further compare SR reconstruction performance in small features and textures.

### 3.2 SR Performance Comparison

As shown in Fig. 2, we compare the tomographic images of the broussonetia papyrifera vessel reconstructed by different SR methods. It is easy to see that our DSCNN recovers more delicate features than that obtained by other methods, and the details are closer to the original HR images. In particular, as shown in the enlarged box in Fig. 3, the ROI marked by the orange rectangle is nicely reconstructed by our DSCNN, but it almost disappears in images reconstructed by other methods.

In the next experiment, the test dataset is the vessel of begonia. The structure of the begonia vessel is more complex and challenging to be recovered than the structure of broussonetia papyrifer in the previous experiments. As shown in Fig. 4, the CNN-based SR methods, such as SRCNN[8], FSRCNN[9], and SRResNet[10], have poor performance in terms of image details recovery. SwinIR[11] with transformer has comparable results to our DSCNN, but the introduction of the transformer makes the computation several times of our DSCNN. As shown in the enlarged box in Fig. 5, the clustered dots are still well reconstructed by our DSCNN, while the images restored by other methods are relatively blurred. Expecting to analyze different deep learning networks from the spectrum perspective, in Fig. 6, we show the spectrum of SR images obtained by all methods. It can be seen that the spectrum of the original HR image is a circular spectrum emanating from the origin as the center. SwinIR, SRResNet, and our DSCNN are more similar to the HR spectrum structure. However, the spectral details of the SR images obtained by different methods are not the same. It can be seen from Fig. 6 that the spectral artifacts of our DSCNN are less, and the spectrum structure is closer to the spectrum of HR images than any others. We believe this is why our DSCNN achieves excellent results.

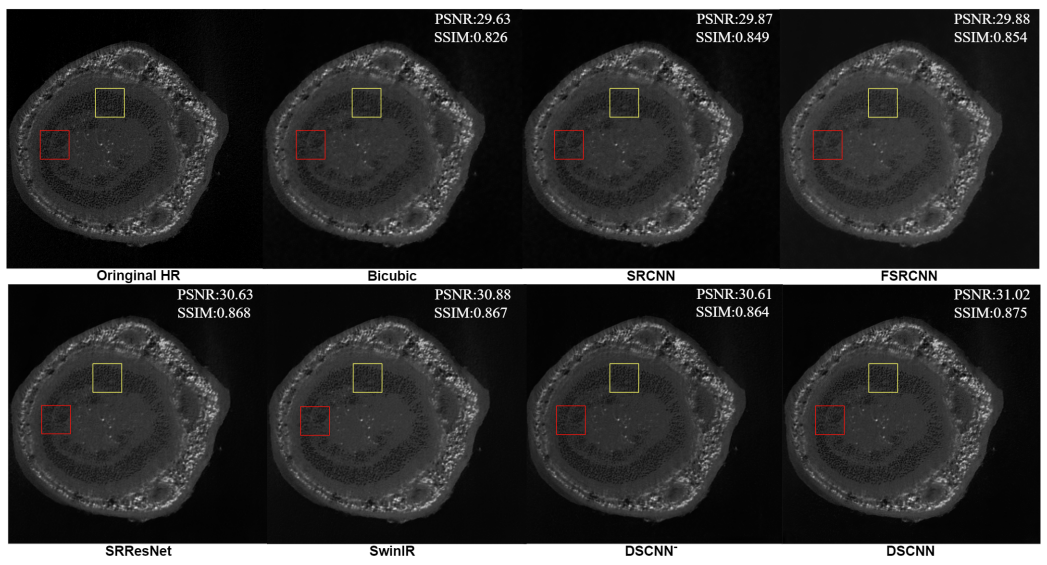

**Figure 4:** The SR reconstruction results from different algorithms in terms of begonia vessel. The SR results show that the DSCNN provides the best SR reconstruction results.

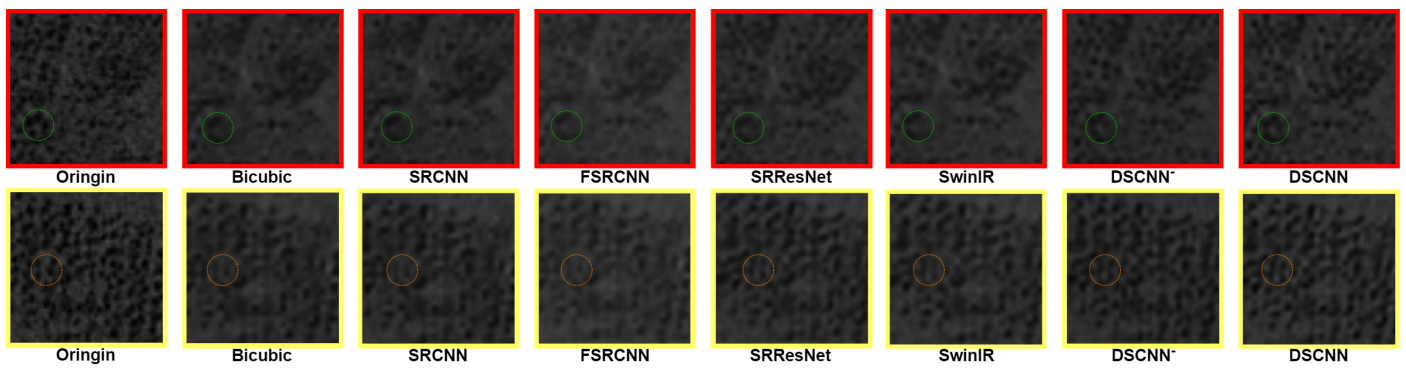

**Figure 5:** Two regions of interests (ROIs) marked in Fig. 4 are further extracted and magnified to compare SR reconstruction performance in small features and textures.

## 4 Discussion and Conclusion

The core idea of our study is to solve the image SR problem by combining both frequency and image domains. This is a pioneering work to solve the SR problem from a frequency learning perspective. Our study found that the high-frequency component in the frequency domain is positively correlated with the SR image quality. The recovery of high-frequency information becomes a key metric for evaluating network capability. We also find that deep learning based SR methods have limitations. In respect to the deep learning based SR results, the spectrum of SR images will be concentrated within a certain range. The high-frequency components beyond the region are generally small and cannot reach the original HR level.

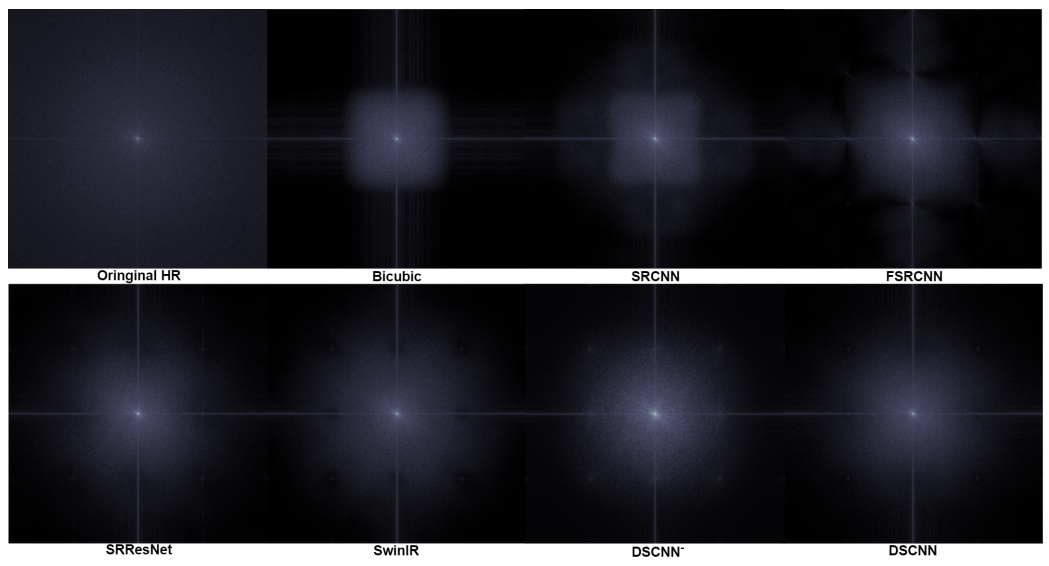

**Figure 6:** Spectrum comparison from different methods. Among the deep learning methods, the SR image spectrum structure generated by SRResNet, SwinIR, DSCNN$^{-}$ are better than others, and the spectrum produced by our DSCNN has fewer artifacts.

SSS

# Fast X-ray diffraction tomographic imaging for characterizing biological tissues

Kaichao Liang[1], Li Zhang[1], and Yuxiang Xing[1]*

[1]Department of Engineering Physics, Tsinghua University, Beijing, China
*Corresponding author: Yuxiang Xing E-mail: xingyx@mail.tsinghua.edu.cn

**Abstract** X-ray diffraction (XRD) provides material specific XRD pattern for material identification. The XRD signal characterizes light materials much better compared with traditional attenuation signal in X-ray transmission imaging. Recent researches developed various XRD tomography (XRDT) methods to acquire pixel-wise XRD patterns in a 2D cross-section for an object. The applications of XRD have been promoted in biological sample inspections. However, it takes several hours for current XRDT systems to gain high-resolution images. In this work, we combined the mechanism of coded-aperture with rotational scan to form a fast data acquisition solution and propose a system referred as a sparse-view coded-aperture XRDT. This system shortens scan time to less than 40 minutes based on our laboratory equipment that is much faster compared with several hours of pencil-beam XRDT, while maintaining similar spatial resolution. Practical experiments on breast sample inspection and kidney stone analysis verified the clinical value of the proposed system.

## 1 Introduction

X-ray diffraction (XRD) measures the Rayleigh scattering intensity distirbution of materials to form material specific XRD pattern. The XRD pattern provides a “fingerprint” property for material identification [1]. The XRD pattern refelects material intermolecular distribution law [2] which is a different physical factor compared with the attentuation signal in X-ray transmission imaging refelecting material element composition. As a result, XRD pattern provides higher material contrast and material identification accuracy for different light materials than attenuation signal [3]. Over past years, various studies on XRD have been done for biological samples, the advantages of XRD in medical applications have been acknowledged, especially for breast cancer diagmosis [4] and kidney stone component analysis. Considering the detailed structures of material distirbution in biological samples, XRD pattern acquisition with spatial resolution in a cross-section of an object is needed, which is referred as XRD tomography (XRDT). XRDT for a single 2D object slice is a problem of measuring three-dimensional XRD signals, i.e. two spatial dimensions together with another momentum transfer dimension.

Currently, the most mature XRD system is X-ray powder diffractometer [5] (single crystal diffraction is not considered here). It uses approximate monoenergetic incident X-ray from metal anode characteristic spectrum, and uses diffraction angle scanning to acquire the XRD patterns of powder samples. The XRD signal of diffractometer is from the whole sample region with no spatial imaging ability. Several spatial-resolved XRDT methods have been proposed in the field over the years. The XRDT methods can be distinguished into two types on the whole. The first type is snap-shot XRDT methods. Snap-shot type XRDT methods acquire XRD pattern images using a single measurement with no mechanical movements or scans. The most representative snap-shot XRDT is energy-dispersive XRD (EDXRD) [5] which is mostly used in security check. EDXRD uses a combination of collimator at X-ray source side (front collimator) and long pin-hole collimator at detector side (rear collimator) to track XRD signals. Each detector pixel is mapped to a specific region on object formed by the field intersection of front and rear collimators. Thus, it achieves spatial resolved imaging. The EDXRD measurements are direct with no need of signal decoupling. However, due to small diffraction angle geometry, its spatial resolution along X-ray transmission path is generally worse than 10 mm [6]. Besides, the rear collimator blocks most informative Rayleigh scattering photons, thus the detection efficiency is low. In the past decade, coded-aperture has been introduced to EDXRD to replace long pin-hole rear collimator in EDXRD [7, 8]. With a multi-hole thin coded-aperture mask, the XRD signal at each spatial location can be detected by multiple detector pixels through each aperture of the coded-aperture mask, thus the detection efficiency of XRD signal is significantly improved [7]. Signal decoupling algorithm is adopted to reconstruct XRD images. Data acquisition of this snap-shot coded-aperture XRDT (Snap-CAXRDT) takes only tens of seconds, but the spatial resolution along transmission direction is still low.

The second type XRDT methods are rotation-based XRDT [3]. These methods follow the scan mode of an X-ray transmission CT and use FBP/FDK type analytical reconstructions for XRDT projections to achieve high spatial resolution. Pencil-beam XRDT is one of the major rotation-based XRDT systems similar to a parallel CT scan. The spatial resolution can reach 1 mm. There is no rear collimator in a pencil-beam XRDT, the Rayleigh scattering photon can be fully detected in high efficiency. However, as both translation and rotation required, the total scan time is generally more than 3 hours. Fan-beam XRDT has also been proposed for XRDT [3], it adopts fan-beam illumination and adds rear grid collimator in front of detector for fan-beam XRDT projection acquisition. Although, the fan-beam illumination increases incident X-ray flux, the rear grid collimator reduces scattering photons. Thus, the total scan time cannot be significantly shortened.

Scan time is the major bottleneck of rotation-based XRDT methods for practical applications. In this work, we proposed a sparse-view rotational coded-aperture XRDT method (Sparse-CAXRDT) than can take advantages of both the fast compressed-sensing acquisition of coded-aperture and isotropic high-resolution of rotation-scan. Sparse-CAXRDT can achieve images of high spatial resolution in several minutes. We built a laboratory Sparse-CAXRDT experimental system that achieved a spatial resolution <2 mm similar to a pencil-beam XRDT, while the scan time is reduced to <40 minutes based on our existing experimental equipment. If a detector of bigger size is available, the scan time can be further reduced to less than 10 minutes. We tested the Sparse-CAXRDT for practical breast sample inspection and kidney stone component analysis, and obtained some valuable results.

## 2 Methods

### *A. Amorphous material XRD pattern*

For an amorphous material, the X-ray differential scattering cross-section of a molecule from primary Rayleigh scattering and primary Compton scattering is:

$$\frac{\partial \sigma_{\text{Scatter}}(E,\theta_S)}{\partial \theta_S} = \frac{\partial \sigma_{\text{Th}}}{\partial \theta_S} F^2(q) + \frac{\partial \sigma_{\text{KN}}}{\partial \theta_S} S(q), \text{ where } q = \frac{E \sin(\theta_S/2)}{\text{hc}} \tag{1}$$

Here, $E$ denotes the X-ray photon energy, $\theta_S$ the scattering angle, $q$ the momentum transfer measuring the photon momentum change during scattering process, $\text{h}$ the Planck constant, and $\text{c}$ the speed of light. The $\frac{\partial \sigma_{\text{Th}}}{\partial \theta_S} F^2(q)$ in Eq. (1) is differential Rayleigh scattering cross-section with $\frac{\partial \sigma_{\text{Th}}}{\partial \theta_S} = \frac{(1+\cos^2\theta_S)\text{r}_\text{e}^2}{2}$ being the Thomson scattering cross-section and $F(q)$ the molecular form factor of material including molecular interference effect. The $\frac{\partial \sigma_{\text{KN}}}{\partial \theta_S} S(q)$ in Eq. (1) is differential Compton scattering cross-section with $\frac{\partial \sigma_{\text{KN}}}{\partial \theta_S}$ being the Klein-Nishina differential cross-section, $S(q)$ the Compton scattering factor.

When $E < 160$ keV, $\theta_S < 8^\circ$, $\frac{\partial \sigma_{\text{KN}}}{\partial \theta_S} \approx \frac{\partial \sigma_{\text{Th}}}{\partial \theta_S}$ and the energy loss in Compton scattering can be ignored. Then $\frac{\partial \sigma_{\text{Scatter}}(E,\theta_S)}{\partial \theta_S} \approx \frac{\partial \sigma_{\text{Th}}}{\partial \theta_S}\left(F^2(q)+S(q)\right)$. We further consider the linear differential scattering coefficient of a material (the probability of an X-ray photon scattered toward a specific direction passing a unit thickness of a material) is:

$$\frac{\partial \mu_S(E,\theta_S)}{\partial \theta_S} = \frac{\text{N}_\text{A}\rho}{\text{M}} \frac{\partial \sigma_{\text{Th}}}{\partial \theta_S}\left(F^2(q)+S(q)\right) \tag{2}$$

Here, $\text{N}_\text{A}$ is the Avogadro constant, $\text{M}$ the relative molecular mass, and $\rho$ the material density. We use $f_{\text{XRD}}(q) \triangleq \frac{\rho}{\text{M}}\left(F^2(q)+S(q)\right)$ to include material specific physical factors and it represents material specific XRD patterns.

### *B. Sparse-view coded-aperture XRDT*

Similar to other XRDT methods, Sparse-CAXRDT measures the spatial distribution of $f_{\text{XRD}}(q)$. The targeted XRD images in Sparse-CAXRDT is defined as $f_{\text{XRDT}}(x,y,q)$, i.e. an XRD pattern $f_{\text{XRD}}(q)$ is acquired for each position $(x,y)$ on a 2D cross-section of an object.

A schematic illustration of the Sparse-CAXRDT system is in Fig. 1. An X-ray source is used to generate incident X-rays of continuous polychromatic spectrum. A slit collimator is placed before the X-ray source to form fan-beam illumination on object. There is a coded-aperture mask placed between the object and detector to code the scattering signal from the object. An energy-dispersive photon counting detector (EDPCD) is adopted to detect coded scattering signals. The beam stop in Fig.1 is used to absorb straight transmitted X-rays. During the scan, the object is rotated circularly to acquire coded scattering signal from multi-views.

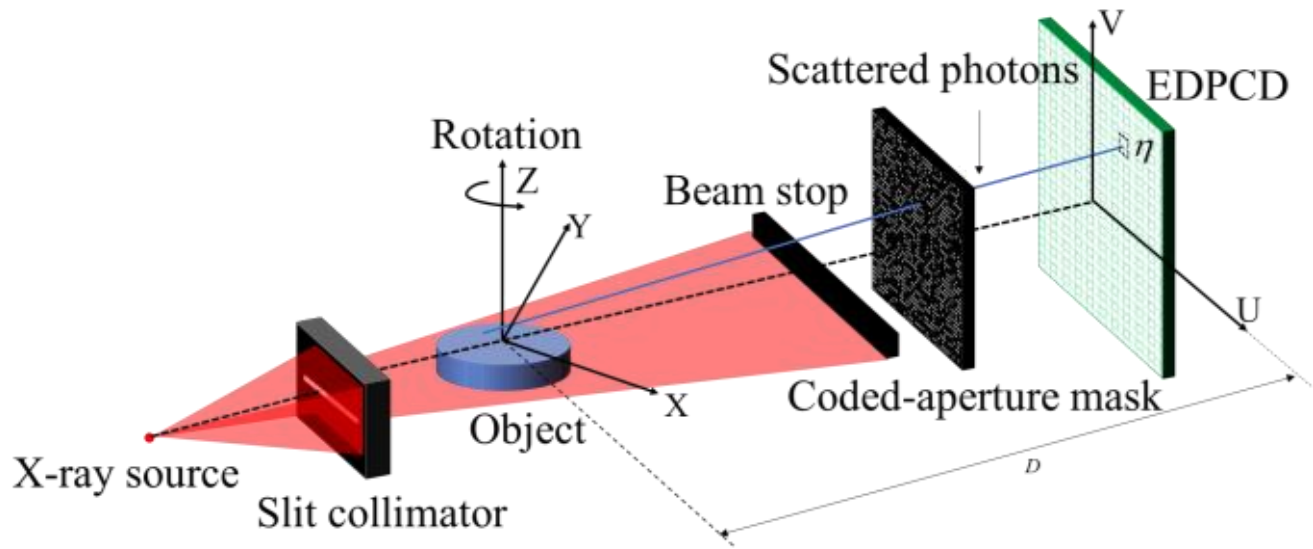

Figure 1: Illustration of the sparse-CAXRDT system.

For a specific scan angle $\varphi$, the scattering signal detected by EDPCD denoted as $I^{\text{XRDT}}(u,v,E,\varphi)$ is given by:

$$I^{\text{XRDT}}(u,v,E,\varphi) = \int_{E'} R(E,E') \int_q \iint_{x,y} I_\sigma^0(E') T(x,y,\varphi,E',u,v) C(x,y,\varphi,u,v) \times \Omega_S(x,y,\varphi,u,v) \text{N}_\text{A} \frac{\partial \sigma_{\text{Th}}}{\partial \theta_S} f_{\text{XRDT}}(x,y,q) \delta\left(q - \frac{E'\sin(\theta_S/2)}{\text{hc}}\right) \text{d}x\text{d}y\text{d}q\text{d}E' \tag{3}$$

Here, X-Y-Z denotes the object coordinate system and U-V the detector coordinate system as shown in Fig. 1. $I^{\text{XRDT}}(u,v,E,\varphi)$ is the detected signal at position $(u,v)_{\text{det}}$ on detector and at energy channel $E$. The $R(E,E')$ is the detector energy response function for practical detectors, which models the effect of X-ray at energy $E'$ on the detector signal at energy channel $E$, $R(E,E')$ can be pre-calibrated for each EDPCD. $I_\sigma^0(E')$ is the incident X-ray intensity at energy $E'$. $T(x,y,\varphi,E',u,v)$ is the attenuation factor which measures the total attenuation of the incident X-ray path and scattering X-ray path. $C(x,y,\varphi,u,v)$ is the coding function of the coded-aperture mask, it measures the ratio that scattering signal generated at $(x,y)_{\text{object}}$ on the object passes the coded-aperture mask and reaches position $(u,v)_{\text{det}}$ on the detector. Ideally, it is a 0/1 binary function. In a discrete geometrical model, its value is between [0,1]. $\Omega_S(x,y,\varphi,u,v)$ is the scattering solid angle toward a unit area on the detector plane around $(u,v)_{\text{det}}$. $\theta_S$ is the scattering angle. Based on the geometrical relation, the $\Omega_S$ and $\theta_S$ is calculated by:

$$\Omega_S = \frac{D - x\cos\varphi + y\sin\varphi}{\left((D - x\cos\varphi + y\sin\varphi)^2 + (u - y\cos\varphi - x\sin\varphi)^2 + v^2\right)^{1.5}} \quad (4)$$

$$\begin{aligned}\theta_S = &< (S + x\cos\varphi - y\sin\varphi, y\cos\varphi + x\sin\varphi, 0), \\ &(D - x\cos\varphi + y\sin\varphi, u - y\cos\varphi - x\sin\varphi, v) >\end{aligned} \quad (5)$$

with $<\mathbf{a},\mathbf{b}>$ in Eq. (5) denoting the angle between vectors $\mathbf{a}$ and $\mathbf{b}$.

In Sparse-CAXRDT, the scan angle $\varphi$ belongs to a set of discrete values $\{\varphi_1, \varphi_2, \varphi_3, \ldots, \varphi_{N_\varphi}\}$ with $N_\varphi$ the total number of views.

### C. Sparse-CAXRDT image reconstruction

We reconstruct the $f_{\text{XRDT}}(x,y,q)$ from $I^{\text{XRDT}}(u,v,E,\varphi)$ of Sparse-CAXRDT based on the physical model given by Eq. (3). To simplify the expression, we rearrange the discrete detector raw data $I^{\text{XRDT}}(u,v,E,\varphi)$ in a 4D tensor format $\mathbf{I}^{\text{XRDT}} \in \mathbb{R}^{N_u \times N_v \times N_E \times N_\varphi}$, and also rearrange the discrete XRD image $f_{\text{XRDT}}(x,y,q)$ in a 3D tensor format $\mathbf{f}_{\text{XRDT}} \in \mathbb{R}^{N_x \times N_y \times N_q}$. Here, $N_u$ is the dimension of discretized variable $u$, similar for $N_v$, $N_E$, $N_x$, $N_y$ and $N_q$. The forward physical model in Eq. (3) is denoted as a function $\Theta$, i.e.:

$$\mathbf{I}^{\text{XRDT}} = \Theta(\mathbf{f}_{\text{XRDT}}) \quad (6)$$

In practical systems, $\mathbf{I}^{\text{XRDT}}$ contains unignorable noise. Based on the physical model and noise model, the reconstruction of $\mathbf{f}_{\text{XRDT}}$ from $\mathbf{I}^{\text{XRDT}}$ can be completed by a model-based iterative reconstruction method (MBIR) with a total cost function:

$$\ell_{\text{MBIR}}(\mathbf{f}_{\text{XRDT}};\mathbf{I}^{\text{XRDT}}) = \ell_{\text{fidelity}}(\mathbf{f}_{\text{XRDT}};\mathbf{I}^{\text{XRDT}}) + \ell_{\text{prior}}(\mathbf{f}_{\text{XRDT}}) \quad (7)$$

The $\ell_{\text{fidelity}}$ is data fidelity term measuring the negative log-likelihood between $\Theta(\mathbf{f}_{\text{XRDT}})$ and $\mathbf{I}^{\text{XRDT}}$. The EDPCD detection is independent between pixels with Poisson distributed noise. Thus:

$$\ell_{\text{fidelity}}(\mathbf{f}_{\text{XRDT}};\mathbf{I}^{\text{XRDT}}) = \sum_{i_u,i_v,i_E,i_\varphi} \left[\Theta(\mathbf{f}_{\text{XRDT}}) - \mathbf{I}^{\text{XRDT}} \ln \Theta(\mathbf{f}_{\text{XRDT}})\right] \quad (8)$$

Here, subscripts $i_u, i_v, i_E, i_\varphi$ index the elements in a 4D tensor $\mathbf{I}^{\text{XRDT}}$. $\ell_{\text{prior}}(\mathbf{f}_{\text{XRDT}})$ is the prior term constraining the 3D reconstructions. In spatial dimension, we use an isotropic total-variation (iTV) constraint. While in momentum transfer dimension, we use a simple smooth constraint. Thus, the $\ell_{\text{prior}}(\mathbf{f}_{\text{XRDT}})$ is:

$$\begin{aligned}\ell_{\text{prior}} = &\lambda_1 \sum_{i_x,i_y,i_q} \sqrt{\left(\mathbf{f}_{\text{XRDT}\,i_{x+1},i_y,i_q} - \mathbf{f}_{\text{XRDT}\,i_x,i_y,i_q}\right)^2 + \left(\mathbf{f}_{\text{XRDT}\,i_x,i_{y+1},i_q} - \mathbf{f}_{\text{XRDT}\,i_x,i_y,i_q}\right)^2} \\ &+ \lambda_2 \sum_{i_x,i_y,i_q} \left(\mathbf{f}_{\text{XRDT}\,i_x,i_y,i_{q+1}} - \mathbf{f}_{\text{XRDT}\,i_x,i_y,i_q}\right)^2\end{aligned} \quad (9)$$

Here, $\lambda_1$ and $\lambda_2$ are hyper-parameters adjusting the strength of priors. The MBIR cost function in Eq. (7) is minimized with a Split-Bregman method [9] to reconstruct $\mathbf{f}_{\text{XRDT}}$.

## 3 Experiments and Results

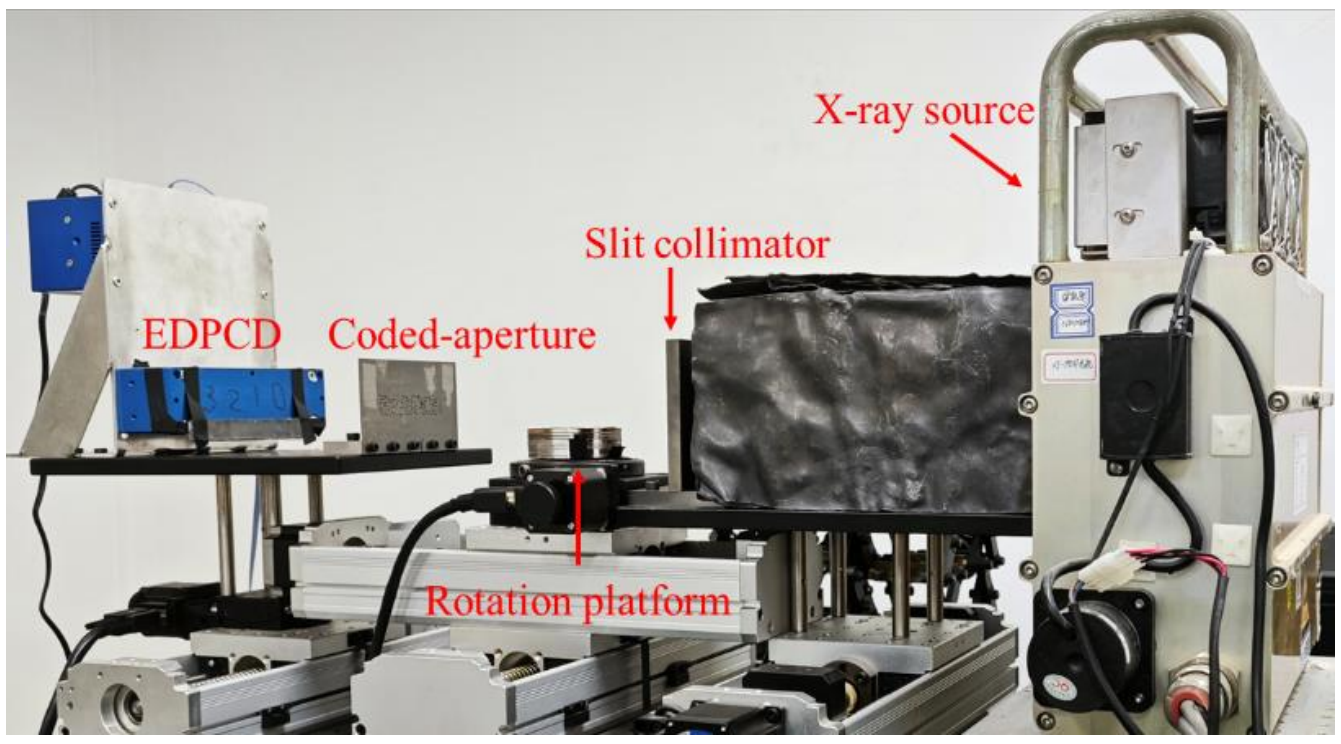

Figure 2: Laboratory sparse-CAXRDT system.

The laboratory Sparse-CAXRDT system we developed is shown in Fig. 2. In this experimental system, the X-ray source is with a tungsten anode. The tube voltage is set to 100 kV and the tube current is set to 3 mA (the maximum current of the X-ray source) in our experiment. The slit size of the front collimator is 0.5 mm in height and 50.0 mm in width. The EDPCD is of $64\times16$ pixels with pixel size 1.6 x 1.6 mm$^2$, the total active detective region is a $102.4\ \text{mm}\times25.6\ \text{mm}$ strip. Each detector pixel can output an X-ray spectrum in range [15 keV, 350 keV] by detecting photons incident to the pixel. The size of active detective region of the EDPCD is a major factor affecting imaging speed. The coded-aperture mask is made of 1.5 mm thick tungsten with 160 apertures randomly distributed in a $80.0\ \text{mm}\times16.0\ \text{mm}$ region. Each aperture is of size $1.0\ \text{mm}\times1.0\ \text{mm}$. The coded aperture mask is shown in Fig. 3. Based on the experimental system geometry, the field of view (FOV) is a circular region of 50.0 mm in diameter.

We conduct experiments for breast sample inspection and kidney stone analysis. In the breast sample experiment, a normal breast tissue sample and a breast cancer tissue sample from a collaborate hospital are imaged under research ethic approve. The two samples were both placed in test tubes for the Sparse-CAXRDT scan on our platform. There were 15 views in our experiment and the exposure time is 150 s for each view. The total exposure time is 37.5 min. The MBIR reconstruction of XRD image is of $54\times54$ pixels with pixel size $1.0\,\text{mm}\times1.0\,\text{mm}$. The range of reconstructed momentum transfer is $[0.6\,\text{nm}^{-1},4.0\,\text{nm}^{-1}]$.

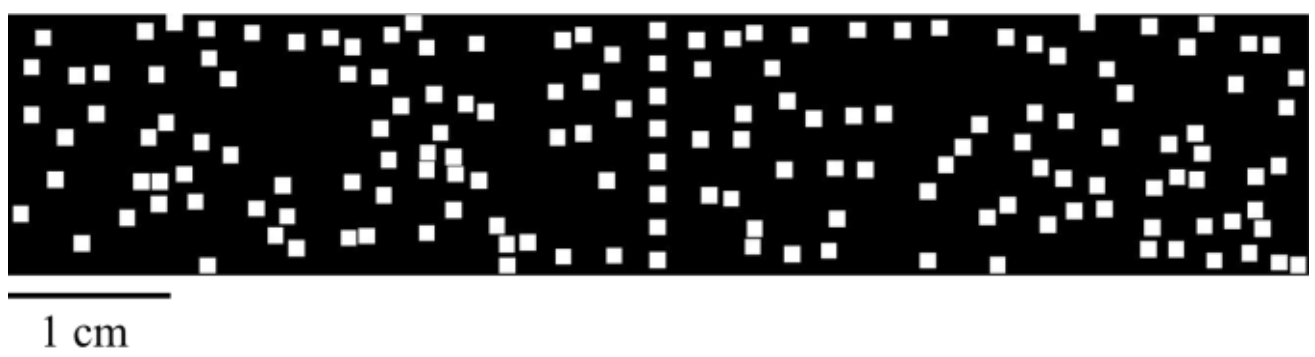

Figure 3: Sketch map of the coded-aperture mask.

The results of Sparse-CAXRDT reconstruction are displayed in Fig. 4 (1) and (2). The normal tissue and cancer tissue can be clearly distinguished as shown in Fig.4 (2). The samples are also scanned by a fan-beam photon counting CT using energy window [21.5 keV,85 keV] and attenuation map results are shown in Fig. 4(3). Obviously, the fan-beam CT image is of better resolution than the Sparse-CAXRDT result, but the difference between normal tissue and cancerous tissue is very low.

We further scanned three kidney stones on the Sparse-CAXRDT system: a uric acid stone, a stone with uric acid kernel and calcium oxalate cladding, and a cystine stone. All of them were placed in test tubes as shown in Fig. 5 (1). We use the XRDT reconstructions at the peak positions of cystine, uric acid and calcium oxalate as RGB channels to generate color image. The result is displayed in Fig. 5 (2). As we can see in Fig. 5 (2), the three type components of the kidney stones can be clearly distinguished from each other based on the results from the Sparse-CAXRDT scan.

## 4 Conclusion

In this work, we developed a Sparse-CAXRDT experimental system combining the coded-aperture fast compressed-sensing data acquisition advantage with the high spatial resolution advantage of rotational scan. The system achieves a spatial resolution better than 2 mm for XRD imaging in much shorter time than traditional pencil-beam XRDT. Our experimental results on breast sample and kidney stone imaging demonstrated the great value of this imaging modality in biological tissue inspection and analysis. Right now, our laboratory platform is with a strip detector to lower cost and the maximum tube current is limited to 3mA, thus limit the data-acquisition efficiency. If a bigger detective area and a higher X-ray source tube current can be applied, it's expectable to reduce the Sparse-CAXRDT scan time to less than 10 minutes for detailed inspection of biological samples so that make this imaging modality fairly practical.

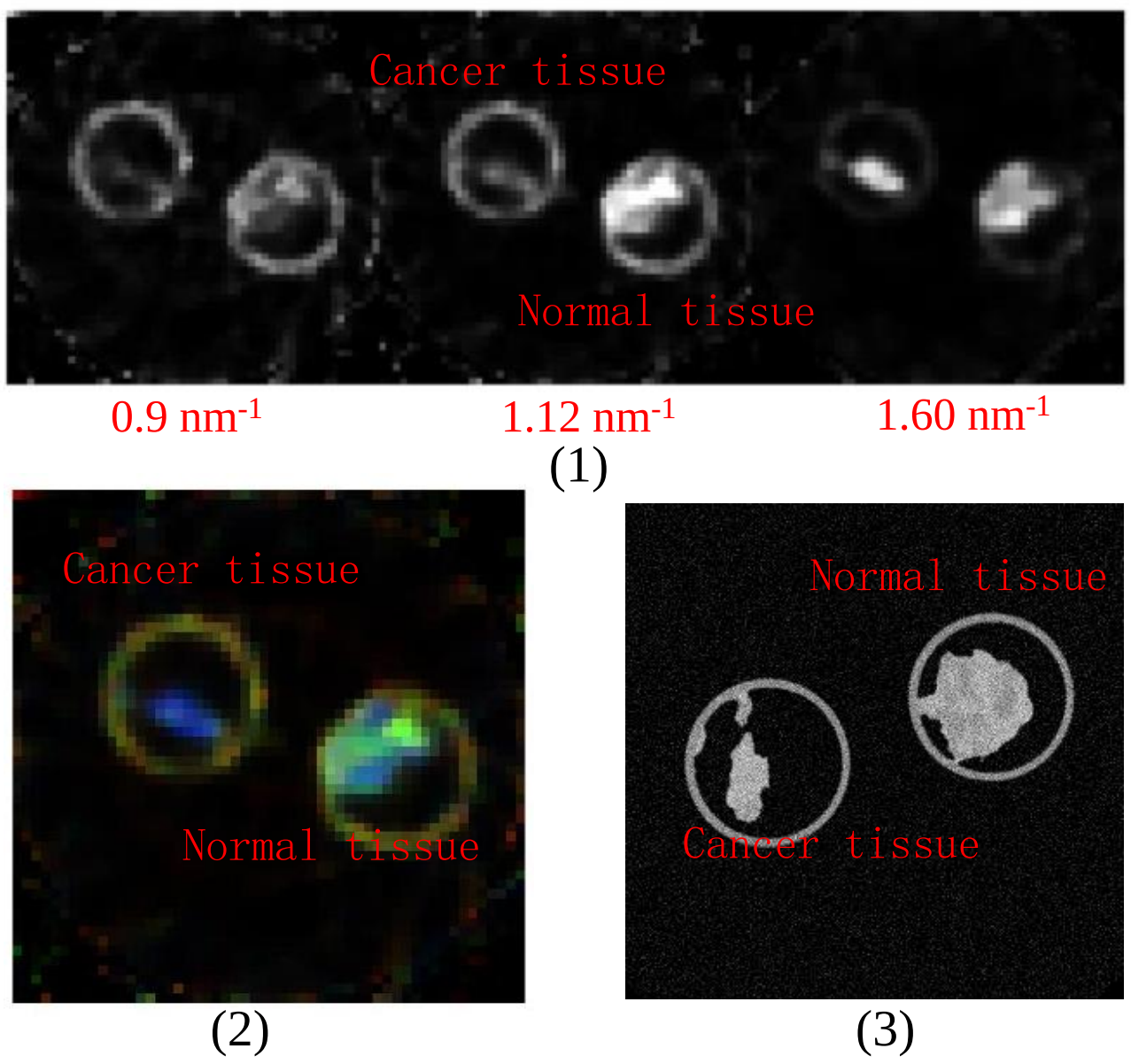

Figure 4: (1) Reconstruction of breast samples at different momentum transfers from Sparse-CAXRDT imaging. (2) Color image taking the momentum transfer channels in (1) as RGB channels. (3) Attenuation map results from Fan-beam photon-counting CT imaging using energy window [21.5 keV, 85 keV] for the breast samples for comparison.

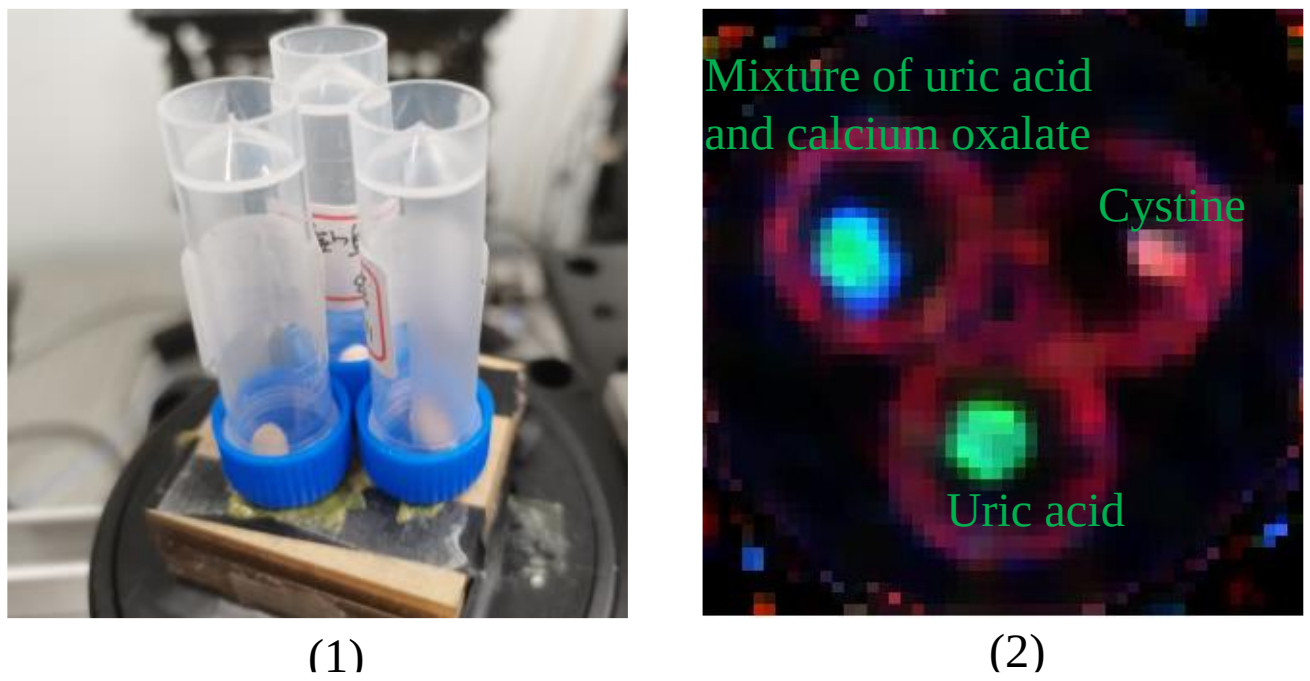

Figure 5: (1) Photo of kidney stone samples. (2) Sparse-CAXRDT reconstruction taking the peaks of cystine, uric acid and calcium oxalate as RGB channels.

## Acknoledgements

This work was supported in part by the National Natural Science Foundation of China (NNSFC) No. 12275151 and the Internal Research Funding from the Key Laboratory of Particle & Radiation Imaging (Tsinghua University), Ministry of Education, China.

# 3D Deep-learning-based image registration correction network with performance assessment using TREs estimation for dual-energy CT

Rui Liao[1], Tao Ge[1], Maria Medrano[2], Jeffrey F. Williamson[1], Bruce R. Whiting[1], David G. Politte[1], and Joseph A. O'Sullivan[1]

[1]Washington University in St. Louis, St. Louis, USA
[2]Stanford University, Stanford, USA

**Abstract** Deformable image registration (DIR) has been widely used in radiotherapy treatment planning, including proton therapy. However, quantitative assessment of the 3D DIR accuracy for real patient data is a challenging problem due to the lack of known ground-truth landmark correspondences between the source and target data. In this paper, a 3D deep-learning-based multi-modality image registration correction network (RCN) is proposed to automatically quantify the accuracy performance of a given DIR algorithm result using the target registration error (TRE) estimation. The proposed network can further refine the initial result with the dense map (same resolution and size as the input images) of the TREs acquired from the quality assessment to achieve higher accuracy and more robustness when applied to other imaging systems. The proposed convolutional neural network is a supervised approach trained using a simulated virtual human body phantom called the XCAT dataset. The proposed RCN method is evaluated using both the simulated dataset and some real patient clinical data from different CT imaging systems. The simulated test dataset with known ground truth shows an approximately 41% decrease in the TRE after the correction. Despite the fact that the RCN was trained using a digital anthropomorphic phantom without tissue heterogeneity, the validation results on real patient data indicate an accurate estimation for the TRE of our initial iterative DIR method and improvement of registration results after the correction.

## 1 Introduction

Proton therapy has the potential to treat tumors with much more conformal dose distributions compared to competing modalities, delivering little to no dose to the healthy tissue distal to the tumor [1]. Accurate CT images are required for dose planning in proton therapy. Our previous work [2, 3] introduced a DECT (dual-energy CT) stopping power ratio estimation method and demonstrated that it could reduce the uncertainty level in single-energy CT stopping-power ratio mapping techniques from $2-3.5\%$ to $1\%$ or less. Dual-energy sinograms were acquired sequentially on a single-source and –detector array commercial scanner. However, this acquisition method is vulnerable to patient motion during and between scans. Such organ motion and deformation influence the quantitative accuracy of the result and produce severe artifacts in decomposed images. In previous work[4], we integrated an iterative deformable image registration (DIR) method, SyN [5], to eliminate some of these motion artifacts in real patient data from commercial CT systems.

Registration accuracy of DIR is critical to quantitatively reconstructing the attenuation coefficients of the patient anatomy and to the overall success of dose planning. However, iterative DIR algorithms like SyN require manual adjustment of the parameters for the specific scanner and body site. For example, SyN parameters that are optimal for one scanner's data do not achieve the same accuracy level on another commercial scanner because of the difference in the imaging system. In addition, our current DIR implementation lacks tools for quantifying the geometric accuracy of the registrations. Commonly used validation metrics, DICE contour coincidence metrics are not useful since expert annotations are not available in many applications and often exhibit delineation uncertainties of comparable magnitude to DIR errors. Image metrics (e.g., mutual information, mean squared error) measure the similarity between fixed and deformed moving images based on intensity values that are not generalized and comparable for different tasks. In addition, the resulting metric value measures the similarity globally and does not indicate local mismatching. On the other hand, TREs (target registration errors) are directly defined on the DVF (displacement vector field) that does not require any intensity value in the image domain but measures the distances between corresponding landmarks in moving and fixed images [6]. In clinical practice, the TREs are often acquired by calculating the distance after registration between annotated corresponding point sets [7]. The dense TREs map proposed in this paper performs automatic TREs estimation for all locations among the image domain with same resolution of the input images.

This paper presents a novel supervised deep-learning approach (RCN: registration correction network) to estimate the registration error for sequential dual-energy CT scans without knowing the ground truth DVF or expert annotation. This approach gives an assessment of alignment errors for any given initial registration method such as SyN in our case, resulting in a dense map (same resolution as the given image) of TREs that estimate the local mismatching. The network further refines the previous transformation based on the registration error and outputs a corrected DVF with higher accuracy and robustness.

## 2 Materials and Methods

The DIR process finds an optimal coordinate transformation $\hat{T}_{M\to F}:\Omega_F\to\Omega_F$ that aligns the moving image $I_M(\mathbf{x}_M),\mathbf{x}_M\in\Omega_M$ with the fixed image $I_F(\mathbf{x}_F),\mathbf{x}_F\in\Omega_F$:

$$\hat{T}_{M\to F}(\mathbf{x}_F)=\mathbf{x}_F+\hat{\mathbf{u}}_{M\to F}(\mathbf{x}_F), \tag{1}$$

where $\hat{\mathbf{u}}_{M\to F}(\mathbf{x}_F)$ is the corresponding DVF. To evaluate the quality of this estimation $\hat{T}$, the dense TREs that measure the total displacement between all corresponding points $\mathbf{x}\in\Omega_F$ in the moving and the fixed images can be calculated as

$$\text{Dense TREs} : \Omega_F \to \mathbb{R}^+ : \mathbf{x} \mapsto \|\mathbf{u}(\mathbf{x}) - \hat{\mathbf{u}}(\mathbf{x})\|, \quad (2)$$

where $\hat{\mathbf{u}}(\mathbf{x})$ is an initial DVF generated by some other registration method, i.e SyN, and $\mathbf{u}(\mathbf{x})$ is the ground truth DVF. However, in most clinical practices, the ground truth DVF is not obtainable. In the proposed method, we introduce a convolutional neural network trained using a simulated dataset to estimate the TREs $E(\mathbf{x})$ for real patient data and later refine the initial transformation as $T^*(\mathbf{x})$.

### 2.1 Network structure

The proposed neural network is inspired by Eppenhof et al.'s work [6] using a U-net structure [8] to estimate the TREs. The proposed network estimates the dense TREs for each voxel for a pair of 3D image stacks extracted from the moving image $I_M$ and the fixed image $I_F$. The stacks have dimensions of $624\times624\times16$ $1\times1\times1\text{mm}^3$ voxels in our current implementation. We train this convolutional neural network to estimate the displacement vector for each voxel in the image domain. A supervised residual learning set-up is used to train the network to estimate the TRE $E(\mathbf{x})$ due to distorting $\bar{\mathbf{u}}(\mathbf{x})$ to the known ground truth DVF $\mathbf{u}(\mathbf{x}), \mathbf{x}\in\Omega$ by a random vector field, consisting of additive random perturbations. The trained network outputs both the refined transformation $T^*(\mathbf{x})$ (corresponding DVF as $\bar{\mathbf{u}}^*(\mathbf{x})$) and $E(\mathbf{x})$. the estimated TRE $E(\mathbf{x})$. The loss function for the network is shown in Equation 3 given by:

$$\theta = \text{argmin}_\theta \|E(\mathbf{x};\theta) - (\mathbf{u}(\mathbf{x}) - \bar{\mathbf{u}}(\mathbf{x}))\|_{L_2}, \quad (3)$$

with trainable hyper-parameters, $\theta$. The moving image is first warped with the distorted transformation $\bar{T}(\mathbf{x})$ and then stacked with the fixed image as well as the distorted DVF $\bar{\mathbf{u}}(\mathbf{x})$ before entering the U-net. The detail of the proposed network structure is shown in Figure 1.

We assume that after the initial registration, the result $\hat{T}(\mathbf{x})$ is close to the ground truth $T(\mathbf{x})$. And $\hat{T}(\mathbf{x})$ is a diffeomorphism (continuous, invertible, and one-to-one mapping) since we are using a diffeomorphic algorithm. In the training process, some random additive perturbation field is applied to the ground truth DVF $\mathbf{u}(\mathbf{x})$ to simulate the difference between $T(\mathbf{x})$ and $\hat{T}(\mathbf{x})$. The perturbation should be continuous based on the physics of the patient's organ movements and the initial diffeomorphic transformation. The details will be described in section 2.2.

### 2.2 Training set construction

The training set is derived from XCAT [9], a virtual human phantom. The 4D XCAT phantom realistically models the complex shapes of human organs and is able to model time-dependent anatomical deformations due to breathing motion. A male patient body phantom was chosen to construct our training set from the top of the head down to the lower lung area. The 4D dual-energy (80 kVp and 140 kVp) CT images ($I_M(\mathbf{x},t;$ 80 kVp$)$, $I_F(\mathbf{x},t;$ 140 kVp$)$) consists of 320 ($624\times624$) slices sampled from the human phantom every 0.5 secs for total 6 frames during a breathing cycle. The whole volume of the moving and the fixed images are further cropped into twenty $624\times624\times16$ image stacks which are dictated by the GPU memory limit. The whole dataset is randomly split into a training and validation set with the ratio 8:2. A corresponding ground truth transformation $T_{s\to t}(\mathbf{x})$ can also be generated for every pair of images ($s,t\in[0,3]$ secs) for the 4D fixed and moving images.

A 3D perturbation field $\mathbf{u}_N(\mathbf{x}), \mathbf{x}\in\Omega$ (volume with the same size of the fixed and moving images) consisting of the sum of randomly constructed 3D gaussian distributions is added to the ground truth DVF $\mathbf{u}_{s\to t}(\mathbf{x})$ in each iteration of the training process. The number $i\in[3,10]$, center $\mu_i$ (within the region near the center of the phantom), variance $\Sigma_i$, and scale $A_i$ of Gaussian distribution $\mathcal{N}(\cdot)$ are randomly sampled from uniform distributions independently for the x, y, and z dimensions yielding a $\mathbf{u}_N(\mathbf{x})$ as below:

$$\mathbf{u}_{N_{x,y,z}}(\mathbf{x}) = \sum_i A_i \cdot \mathcal{N}(\mu_i, \Sigma_i, \mathbf{x}) \quad (4)$$

An example showing one slice of the additive perturbation field in every dimension is shown in Figure 2. The proposed perturbation model randomly generate error field in every iteration of the training process to represent a complex continuous noise pattern.

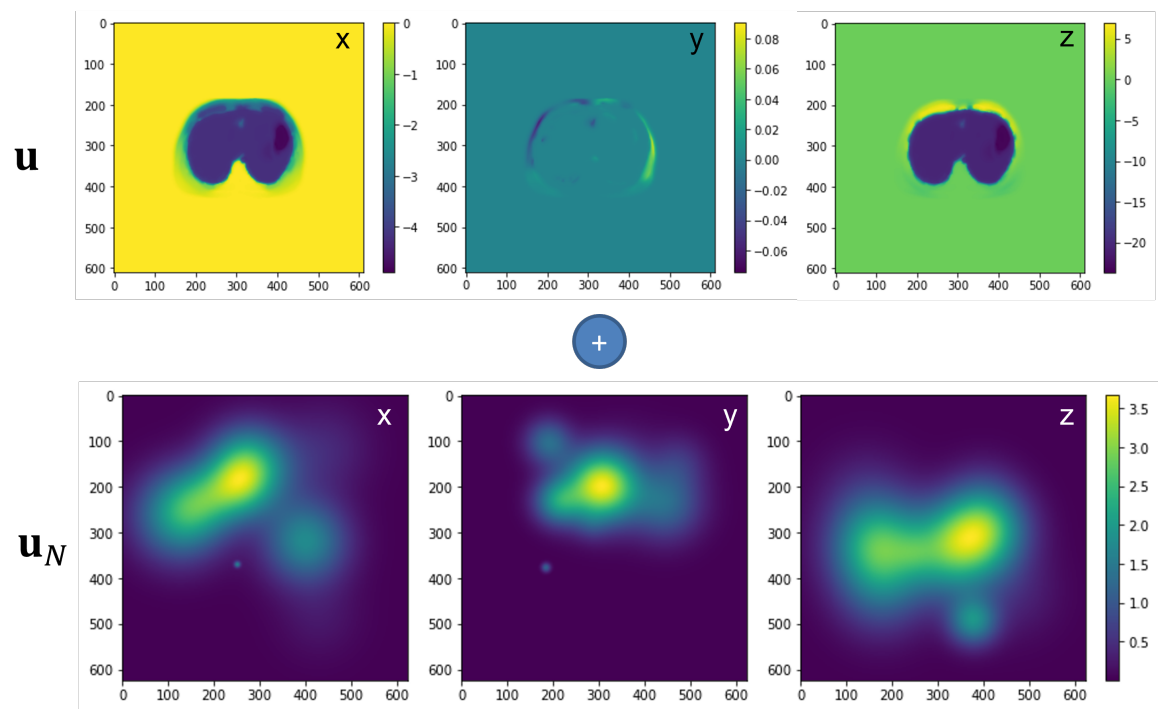

**Figure 2:** Additive random perturbation field $\mathbf{u}_N(\mathbf{x})$ as perturbation. Images show the color map of the vector length for a slice of each dimension in the ground truth DVF $\mathbf{u}(\mathbf{x})$ and the perturbation $\mathbf{u}_N(\mathbf{x})$. It represents a combination of several 3D Gaussian distributions with randomly sampled center locations, means, and deviations.

## 3 Results

Our current implementation is based on 1000 training epochs. Group normalization was applied to accelerate the training process. Figure 3 shows the total TRE ($\|\mathbf{u}^*(\mathbf{x}) - \mathbf{u}(\mathbf{x})\|$) between the refined DVF $\mathbf{u}^*(\mathbf{x})$ with the ground truth $\mathbf{u}(\mathbf{x})$ in both the training and validation sets and the perturbation level ($\|\mathbf{u}_N(\mathbf{x})\|$) for each iteration in the training process. Despite the norm of the additive random perturbation field fluctuating around $6e^{+7}$, the TREs of our estimation in both the training and the validation sets gradually reduce to $8e^{+4}$, about three levels of magnitude less than the initial error.

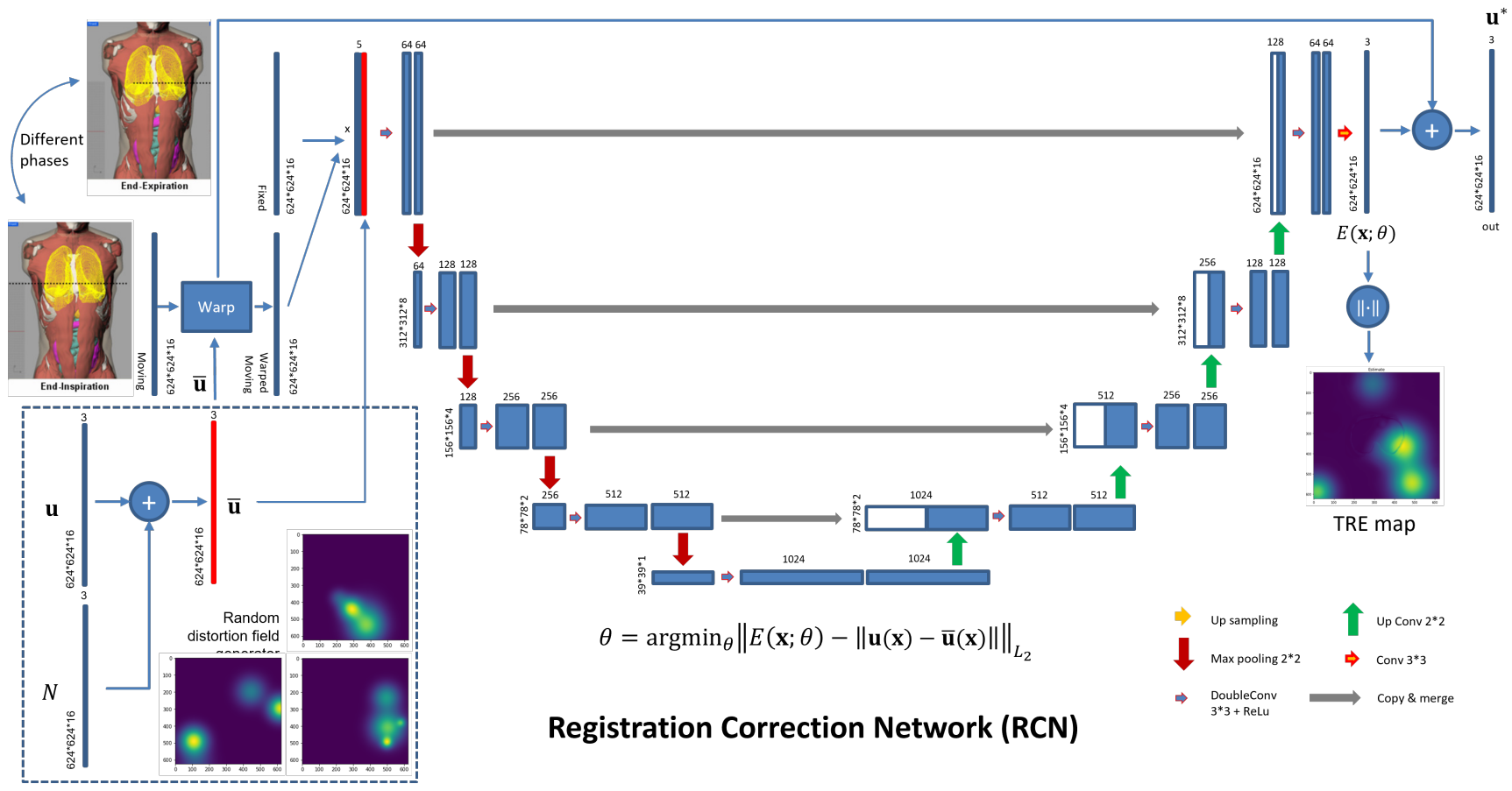

**Figure 1:** Network structure of the proposed RCN.

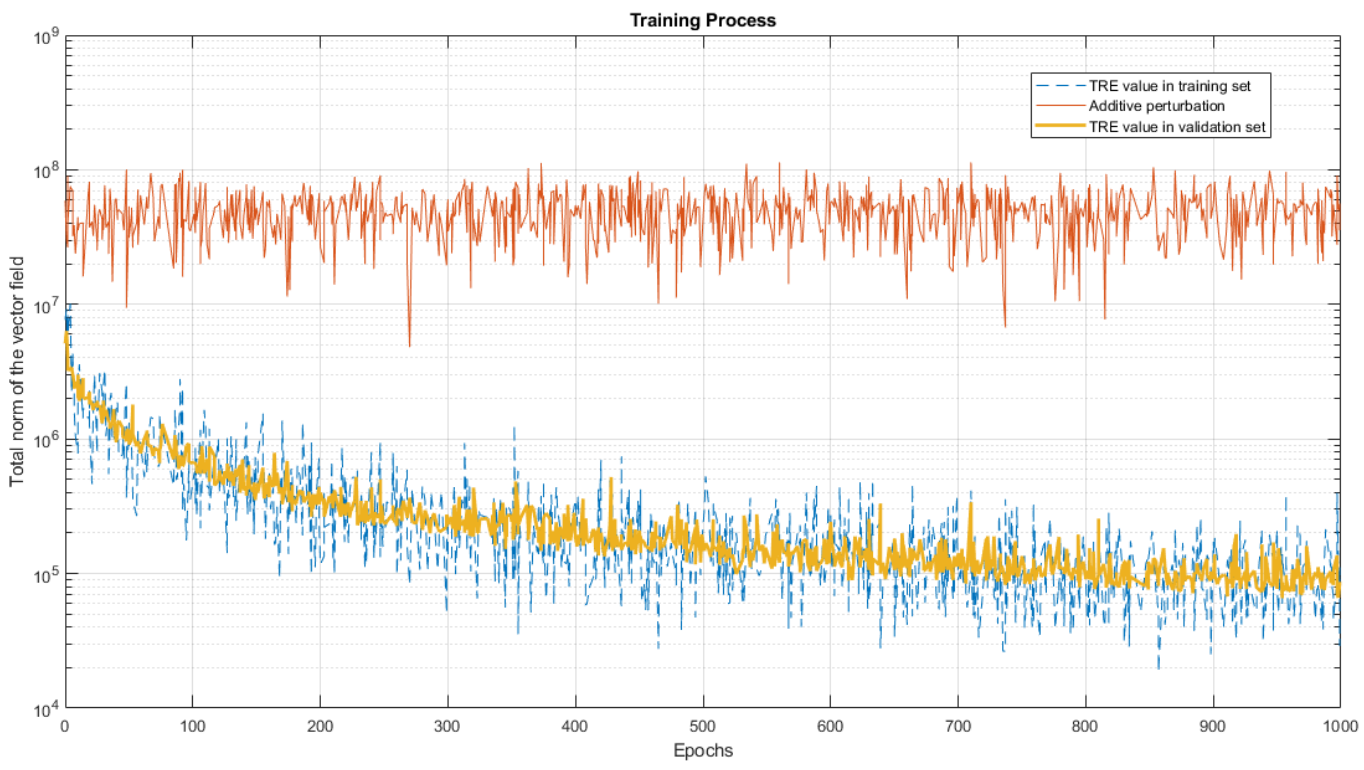

**Figure 3:** The orange line shows the norm of perturbation ($\|\mathbf{u}_N(\mathbf{x})\|$) that was randomly added to the ground truth DVF. The dashed blue line and the solid yellow line indicate the estimated TREs after the correction in the training and validation sets, respectively.

We tested RCN by applying it to another 3D simulated phantom (an XCAT female body phantom) previously unseen by the RCN during the training process. The result in Figure 4 shows an average of 41.05% decrease in the total TRE value after the RCN is applied to the initial SyN result. The TRE values outside the body in both the RCN and the SyN results have been filtered out by a mask formed by the phantom's body contour to focus on the registration accuracy assessment inside the body.

In addition, two real patient datasets were used to test the performance of the RCN. Figure 5 and 6 show a lung cancer patient scanned by a Philips Big Bore scanner and a head-and-neck patient scanned by a Siemens Confidence scanner, respectively. Mismatching was visualized by overlaying the fixed and moving images into two different color channels, and quantification was achieved using mutual information. The visualization and the image metric shown in both Philips and Siemens data indicate an increase in registration accuracy. Despite the fact that the RCN was trained using a digital anthropomorphic phantom without normal tissue density and texture heterogeneity, the promising results on real patients demonstrate its robustness.

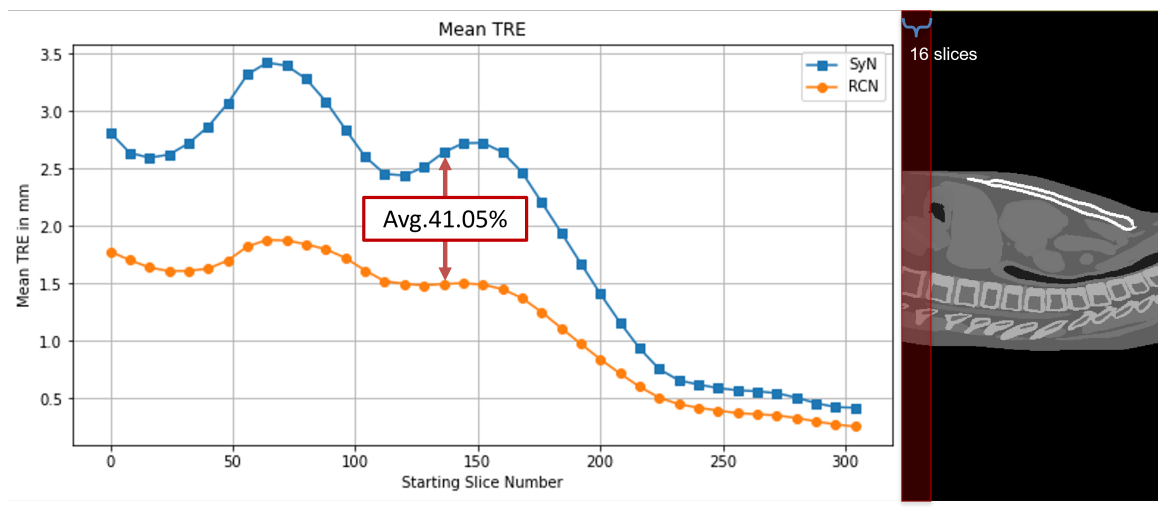

**Figure 4:** Comparison of the average TRE value between the SyN and the RCN results across the simulated 3D female body phantom shown on the right side. Each data point corresponds to a 16-slice sliding volume stack across the whole 320 slices volume.

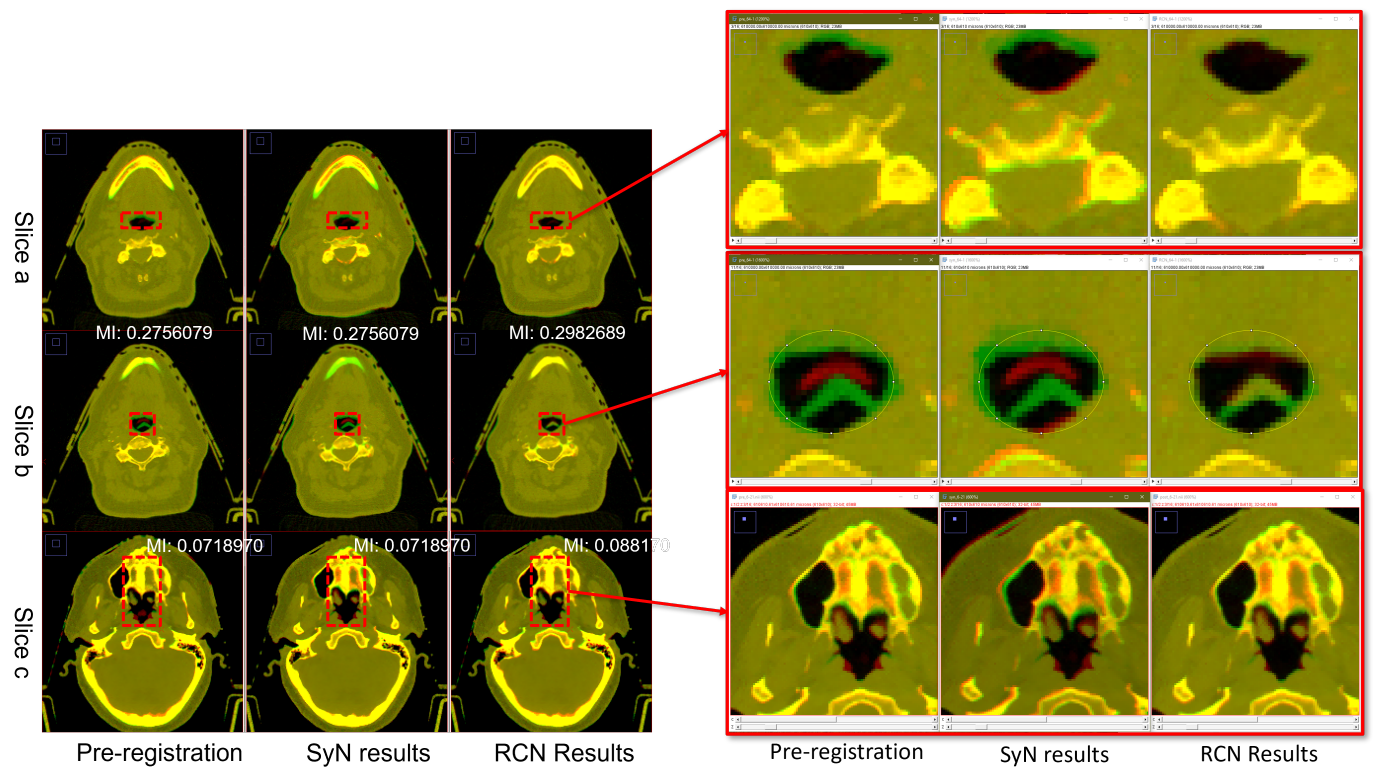

**Figure 6:** Color channel overlay for Siemens Confidence scanner dual-energy patient data (80 kVp in green and 140 kVp in red). The first column is the original patient data; the second column is the registration result using SyN, and the third column is the corrected result using our proposed RCN method.

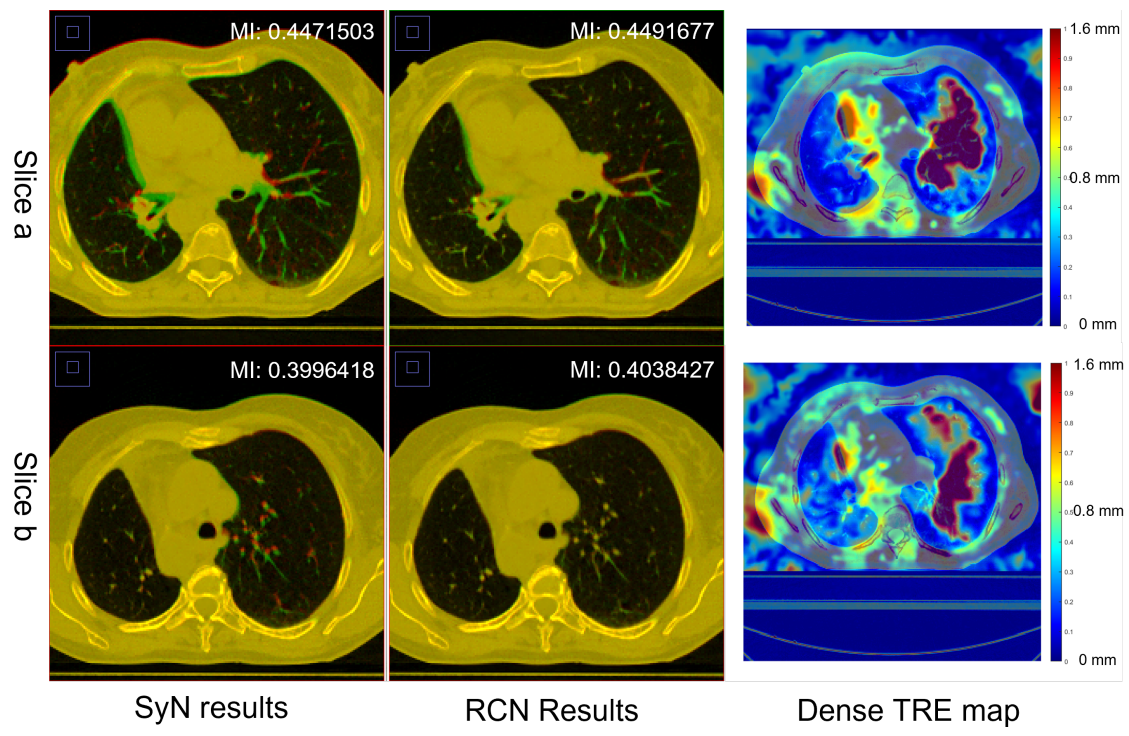

**Figure 5:** Two slices of the fixed and deformed moving images from the Philips Big Bore scanner are overlaid together in different color channels (90 kVp in green and 140 kVp in red) in the first two columns. Mismatching is highlighted as bright green or red colors. The white numbers in the upper right corner marked as 'MI' are the mutual information of the warped moving image and the fixed image calculated over 16 slice stack. The third column is the estimated corresponding dense TREs map in mm using heat-map pattern.

## 4 Discussion

Our current RCN implementation significantly improves the registration accuracy in both the sequential dual-energy CT simulated and real patient datasets. However, we have not yet tested our proposed RCN on other state-of-art deep-learning-based registration methods (e.g. VoxelMorph[10]) or some other medical imaging modality such as MRI. Ongoing work includes generalizing the RCN application to other medical imaging modalities and making thorough comparisons with the state-of-art DIR methods.

In addition, TRE estimation accuracy is compromised because the training included irrelevant TRE errors outside the patient's body contour. Those estimated errors outside the body will not affect the result in the image domain since the action is applied purely to the background air region. However, those values will be counted toward the total TRE value which brings a large gap between our estimation to the ground truth value. Inspired by [11, 12], we will add attention gates to the network that could progressively suppress feature responses in irrelevant background regions without the requirement to crop an ROI between networks.

## 5 Conclusion

To the best of our knowledge, our RCN process is the first attempt to apply a supervised deep-learning method trained on a simulated dataset to refine a given registration via dense TREs estimation for 3D dual-energy CT images. Promising test results on unseen patient data demonstrate the potential of RCN to improve DIR accuracy in clinical practice. With the potential of simulating cone-beam CT and MRI images from the XCAT dataset, the proposed RCN pipeline may also improve outcomes in other multi-modality image registration applications.

# Optimizing Reconstruction for Preservation of Perfusion Defects in Deep-Learning Denoising for Reduced-Dose Cardiac SPECT

Junchi Liu[1], Yongyi Yang[1], Hendrik Pretorius[2], and Michael A. King[2]

[1]Department of Electrical and Computer Engineering, Illinois Institute of Technology, Chicago, IL 60616
[2]Department of Radiology, University of Massachusetts Medical School, Worcester, MA 01655

**Abstract** In cardiac SPECT perfusion imaging deep learning (DL) denoising methods have been found to be highly effective for noise suppression in reduced-dose studies. However, as with conventional image filtering, DL denoising may also cause potential signal loss in the reconstructed images as a trade-off to reduced noise level. In this work, we investigate for the first time the feasibility of improving the preservation of perfusion defect signals in DL denoising by controlling the level of post-reconstruction smoothing in reduced-dose SPECT studies. In the experiments we demonstrated this approach with quarter-dose data from a set of 895 clinical acquisitions. The quantitative results indicate that use of a higher spatial resolution in the reduced-dose images than that of the standard-dose target can achieve both better preservation in defect signals and higher detection accuracy of perfusion defects after DL processing.

## 1 Introduction

Myocardial perfusion imaging with single-photon emission computed tomography (SPECT) can provide an objective assessment of the regional blood flow in the myocardium, and is widely used for diagnosis of coronary artery diseases in nuclear medicine [1]. Owing to concerns over the potential radiation risk associated with SPECT imaging studies [2,3], there is great clinical interest in reducing the dose (administered activity to patients) used for imaging [4].

With a reduced-dose study, however, the acquired data counts are lowered accordingly (unless the imaging time is increased). This will unavoidably result in elevated noise in the reconstructed images, which can adversely affect the diagnostic accuracy. To combat this issue, iterative reconstruction algorithms with resolution recovery as well as attenuation and scatter corrections have been studied for improving the image reconstruction accuracy in SPECT [5-9]. Most recently, deep learning (DL) denoising methods have been demonstrated to be highly effective for noise suppression in low-dose CT, digital breast tomosynthesis, PET, and SPECT studies [10–14]. In particular, in [13] a convolutional autoencoder (CAE) network was demonstrated to lead to improved detectability of perfusion defects in reduced-dose SPECT images.

In development of DL denoising methods for reduced-dose images, a common approach is to employ a DL network and optimize it by using a set of example input-output image pairs collected from the intended imaging application. The input is the images obtained with reduced dose and the output is typically set to be the desired (noise-free) target images. To better accommodate the variability observed in clinical subjects, a DL network is desired to be trained and optimized for clinical acquisitions. In the absence of the ground truth images with clinical subjects, images reconstructed from conventional full-dose acquisitions are often used as the learning target [10–14]. Conceptually, a DL denoising work plays a similar role to that of conventional (linear) filtering in that there can be a trade-off between noise level suppression and potential signal loss in the reconstructed images. As a consequence, a DL denoising network may lead to a loss in image resolution (over-smoothing) while suppressing the elevated imaging noise in reduced-dose studies. This effect can be problematic when the region of clinical interest in an image is small in size or has low contrast (e.g. subtle perfusion defects in cardiac SPECT).

In cardiac SPECT, a post-reconstruction filter (e.g., 3D Gaussian filter) is typically applied and optimized for the diagnostic performance in the resulting images [5]. In this study, we investigate whether optimizing the level of the post-reconstruction smoothing (which controls the spatial resolution) in the reduced-dose input can lead to a better preservation of defect signals in the resulting DL processed output. That is, we seek to optimize the reconstruction for the reduced-dose input images prior to DL processing. In particular, we demonstrate this approach by varying the smoothing parameter of the post-reconstruction filter (including no smoothing) in reduced-dose cardiac SPECT studies.

In the experiments we demonstrated this strategy with quarter-dose data obtained from a set of 895 clinical acquisitions. We investigated the DL denoising performance by measuring: 1) the similarity of perfusion defect signals in reduced-dose studies after DL processing to their counterpart in the full-dose reference, and 2) the detection accuracy of perfusion defects after DL processing.

## 2 Materials and Methods

### 2.1 Problem statement

This work is concerned with improving the accuracy of reconstructed images for perfusion defect detection by suppressing the increased noise in reduced-dose SPECT studies. Specifically, let $\boldsymbol{x}_L$ denote the image reconstructed from an acquisition with a reduced dose level. Our goal is to employ a denoising network $H(\cdot)$ (e.g., CAE) such that the output $H(\boldsymbol{x_L})$ can yield an improved detection accuracy of perfusion defects in comparison to that of $\boldsymbol{x}_L$.
For this purpose, we employ a supervised learning approach in which the denoising network is optimized with example input-output image pairs. That is, for a given reduced-dose study image $\boldsymbol{x_L}$, we define a learning target $\boldsymbol{y}$ such that the network is formulated in the following fashion:

$$\min_{H} E[||H(\boldsymbol{x_L}) - \boldsymbol{y}||^2] \quad (1)$$

where $E[\cdot]$ denotes the expectation over the ensemble of training pairs.

* This work was supported by NIH/NHLBI Grant R01-HL154687.

To ensure the applicability to clinical studies, we will employ example images acquired from clinical acquisitions for training the network $H(\cdot)$. Given the lack of ground truth images with clinical studies, for a reduced-dose input image $\boldsymbol{x}_L$, the corresponding image reconstructed with a standard dose acquisition is used as the learning target $\boldsymbol{y}$. While this may appear less accurate than using an ideal noise-free target, it can be practically desirable since the resulting output $H(\boldsymbol{x_L})$ will be optimized to be similar in appearance to that of standard dose studies, which would be more conducive to interpretations by clinicians.

## 2.2 Optimize reconstruction for DL training

In this study, our approach is to optimize the reconstruction for the input reduced-dose image $\boldsymbol{x}_L$ prior to processing by the DL network. Specifically, for cardiac SPECT images, a post-reconstruction filter such as a 3D Gaussian filter is typically applied for noise suppression and its smoothing parameter $\sigma$ is optimized for the diagnostic performance of the resulting images [15]. In this study, to demonstrate the concept, we investigate whether varying this smoothing parameter can lead to improved perfusion defect preservation in training the DL network.

Specifically, we consider reduced-dose images $\boldsymbol{x}_L$ obtained with varying degrees of post-reconstruction smoothing (i.e., spatial resolution). To be explicit, let $\boldsymbol{x}_L(\sigma)$ denote $\boldsymbol{x}_L$ obtained with smoothing parameter $\sigma$ in the Gaussian filter. The optimization problem in (1) is then re-written as

$$\min_{H} E[||H(\boldsymbol{x_L}(\sigma)) - \boldsymbol{y}||^2] \quad (2)$$

As a result, the resulting network $H(\cdot)$ depends on the choice of the σ value used. Our objective is to optimize the network performance over this parameter σ for the benefit of perfusion defect signal preservation.

Toward this objective, instead of relying on traditional image quality measures such as MSE, we consider the following two performance metrics to characterize the presence of perfusion defects in the DL output $H(\boldsymbol{x_L}(\sigma))$: 1) preservation of the defect signal, and 2) detectability of the defect signal, as described below.

*1) Perfusion defect preservation*

In clinical assessment, a perfusion defect is characterized in severity based on its extent and contrast level. Thus, it is important that the denoising network $H(\cdot)$ preserve such aspects of a defect signal while removing the imaging noise. To quantify this, we will examine the output of a known defect signal after DL processing and compute its similarity (as measured by the Pearson correlation coefficient) against its standard dose target (as detailed in Sect. 2.6).

*2) Detectability of perfusion defects*

To quantify the presence of a perfusion defect after DL denoising, we will adopt the non-prewhitening matched filter (NPWMF) as a numerical observer [15]. For this task we will make use of images reconstructed from multiple noise realizations of reduced-dose data both with and without the defect signal present. The numerical observer is then employed to determine the presence/absence of the defect signal in these images (as detailed in Sect. 2.6).

## 2.3 Clinical dataset

To demonstrate the optimization approach, we used a set of clinical acquisitions from 895 cases (453 male, 442 female), all obtained with informed consent under IRB approval [13]. These studies were acquired with standard dose in list-mode on a Philips BrightView SPECT/CT system with 64 projections (3-degree steps) and a 128x128 matrix acquired over 180 degrees. The pixel size was 0.467 cm. Among the 895 cases, 335 were clinically read as normal, 372 as having either perfusion or cardiac motion abnormalities, and the remaining 188 were read as probably normal.

From these studies we derived the reduced-dose data with 75% reduction in data counts (quarter dose). This was achieved by applying a statistical resampling procedure [5] to the standard-dose acquisitions above.

From the set of clinical acquisitions, 35 normal cases were set aside to generate 175 hybrid studies with ground truth for the optimization task on perfusion defect detection (Sect. 2.6), whereas the remaining 860 cases were randomly divided into two subsets as follows: 1) 739 cases for DL network training, and 2) 121 cases for validation (Sect. 2.5). For image reconstruction, the ordered-subsets expectation maximization (OSEM) algorithm with post-reconstruction Gaussian filtering [5] was used, in which attenuation, scatter and resolution corrections were all incorporated. For the full dose target, the σ value of the Gaussian filter was set to 1.2 voxel, which was optimized for perfusion-defect detection performance studies [5].

For the quarter dose data, the parameter σ of the Gaussian filter was varied with a set of values as σ = [0, 0.5, 0.8, 1.0, 1.2] voxels for the task of optimization.

## 2.4 Architecture of DL model

We employed for the denoising network $H(\cdot)$ a three-dimensional CAE structure, which was previously developed for reduced-dose SPECT images [13]. This network was demonstrated to achieve similar denoising performance to that of a residual CNN model, but with a smaller number of parameters [13]. The validation results in [13] indicated that the denoising performance was rather stable when the number of layers was varied from 4 to 10 in this structure. In the experiments the number of layers was set to eight (four encoding layers and four decoding layers). This was in consideration of the large variability in noise level in the reconstructed reduced images over the range of the smoothing parameter σ. As in [13], for each convolutional layer 10 feature-maps with 3×3×3 kernels were used with batch normalization and rectified linear units employed after each layer.

### 2.5 Network training

From the set of 739 training cases, about $2.2\times10^5$ input-target pairs of 3D image patches (21×21×21 voxels) were extracted from random locations within the heart region and used to form the training samples. The validation cases were used to determine the number of training epochs based on the validation error to avoid overfitting. The adaptive moment estimation (Adam) algorithm was used for network training with a batch size of 100. The CAE network was implemented with Keras on a NVIDIA GeForce GTX 1080 Ti 12G GPU.

### 2.6 Performance evaluation

To assess the detection performance of perfusion defects by the DL denoising network in clinical applications, we generated 175 hybrid studies from the set of 35 normal cases which was set aside in Sect. 2.3, in which perfusion defects were inserted randomly in size among various vascular territories according to their clinical distribution [5]. Each defect was introduced with four contrast levels (65%, 50%, 35%, and 20%) for variable defect detectability. These introduced defects were then used as the ground truth for quantifying the defect signal response in the DL output. In addition, for each of the 175 hybrid studies, we generated 50 noise realizations.

Specifically, we assessed the reconstructed images according to the two performance metrics for the preservation and detection of perfusion defects (defined earlier in Sect. 2.2) as follows:

*1) Perfusion defect preservation:* To assess the preservation of the perfusion defect signal in a reduced-dose image, we computed the Pearson correlation coefficient $\rho$ between the known defect region (in full-dose reference) and the corresponding region in the reduced dose image. In this assessment, we are mainly interested in characterizing the aspects related to the spatial extent and contrast of a defect signal. To suppress the effect of noise in reduced-dose images, we used the average image obtained from 50 noise realizations of each case to compute the defect signal.

*2) Perfusion defect detection:* To quantify the detectability of a perfusion defect, we utilized 50 noise realizations of the reduced-dose image for each test case with and without the defect present. The NPWMF detector was then applied to determine the response of the defect signal in each realization. The detection performance for the defect was summarized by using the detector signal-to-noise ratio ($\mathrm{SNR_D}$) metric as

$$\mathrm{SNR_D} = \frac{|\bar{m}_1 - \bar{m}_0|}{\sqrt{(\delta_1^2 + \delta_0^2)/2}} \quad (3)$$

where $\bar{m}_1$ and $\bar{m}_0$ denote the mean values of the detector output with and without the perfusion defect present, respectively, and $\delta_1^2$ and $\delta_0^2$ are the corresponding variances in the output.

## 3 Results and Discussions

### 3.1 Perfusion defect signal preservation

In Fig. 1 we show the similarity results of perfusion defects in DL processed quarter-dose images to that of the full dose reference, as measured by the Pearson correlation coefficient ρ, obtained with the smoothing parameter σ increased from 0 (no smoothing) to 1.2 (optimal smoothing for the full dose reference). For comparison, the results are also shown for the OSEM reconstructed images without DL processing. These results were obtained from all the 175 hybrid studies in the test set.

Interestingly, from Fig. 1 it is observed that DL achieved a higher ρ value at σ = 0.8 and 1.0 than at σ = 1.2 ($p$-value $< 10^{-6}$; paired $t$-test), with the lowest ρ value obtained at σ = 0 (i.e., no post-reconstruction filtering). The latter is likely attributed to the much higher noise level in the input images with no smoothing applied. As expected, at σ = 1.2 OSEM achieved $\rho \approx 1$ given that the same reconstruction setting was used as in the full-dose reference. However, for OSEM the ρ value is noted to decrease significantly as σ deviated from 1.2. In contrast, for DL the ρ value is maintained nearly constant as σ varied from 0.5 to 1.0. This indicates that when the low-dose input is under-smoothed (relative to the full dose reference), the DL network is able to compensate for the under-smoothing in order to match the image resolution of the full dose reference (as to be seen later in the images in Fig. 3).

### 3.2 Perfusion defect detection

In Fig. 2 we show the perfusion-defect detection results obtained from the DL processed quarter-dose images with the reconstructed parameter σ varied from 0 to 1.2. For comparison, the results are also shown for the OSEM images without DL processing. Note that σ = 0 corresponds to no post-reconstruction filtering. These results were obtained from all the 175 hybrid studies in the test set.

It is observed from Fig. 2 that DL achieved a higher $\mathrm{SNR_D}$ value at σ = 0.8 than at σ = 1.2 (smoothing level for the full dose reference) ($p$-value = 0.014; paired $t$-test). Interestingly, the results in Fig. 1 demonstrated that DL also achieved better defect signal preservation at σ = 0.8 than at σ = 1.2. Finally, it is noted that DL achieved a higher $\mathrm{SNR_D}$ at each σ value than OSEM.

### 3.3 Example images

In Fig. 3 we show the reconstructed images for an example subject in the test set (female, age=61, BMI=26.5) from the different methods. The images are shown both in short-axis slices and in polar map. For comparison, the full-dose reference is shown in Fig. 3(a); there is a small perfusion defect (50% contrast) located in the left anterior descending territories (indicated by arrows). Fig. 3(b) shows the quarter-dose OSEM images at different σ values, whereas Fig. 3(c) shows their counterpart after DL processing. As can be seen, the left ventricular (LV) wall becomes notably more noisy in quarter-dose OSEM. With DL processing, the LV wall becomes more uniform and more similar to the full-dose reference. It can also be observed that the perfusion defect region is better preserved in contrast and extent in DL with σ = 0.8 and 1.0 than with σ = 1.2.

## 3 Conclusion

In this study we investigated the feasibility of improving the preservation of perfusion defect signals in a DL denoising network by optimizing the spatial resolution in the input reduced-dose images. In the experiments we demonstrated this approach with quarter-dose imaging data from a set of 895 clinical acquisitions. The results show that using a higher resolution level in the reduced-dose reconstruction than that in the reference target can achieve a better preservation of perfusion defect signals and can lead to improved detection accuracy of perfusion defects in the DL processed images. Encouraged by these promising results, in the future we plan to further validate this approach with clinical observers.

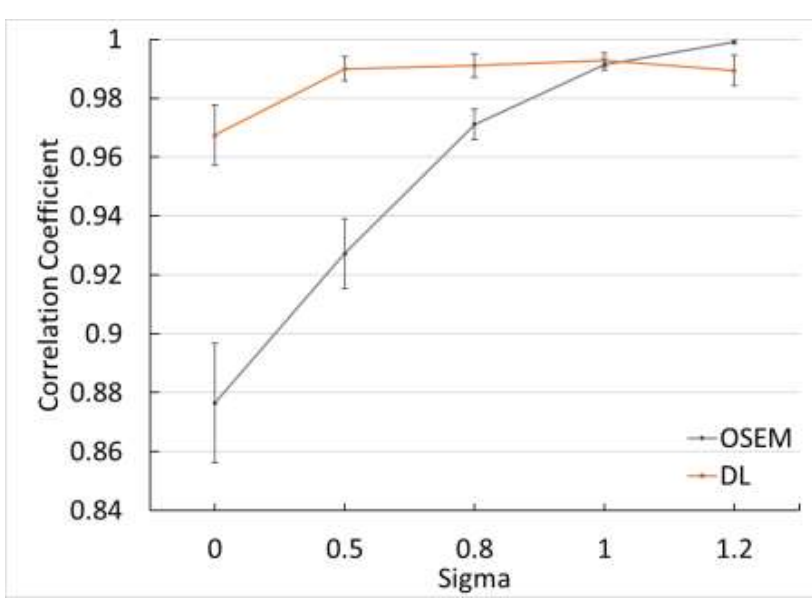

**Fig. 1.** Similarity measure ρ in perfusion defects between quarter-dose images (with and without DL processing) and their full-dose reference. Error bars indicate standard deviations.

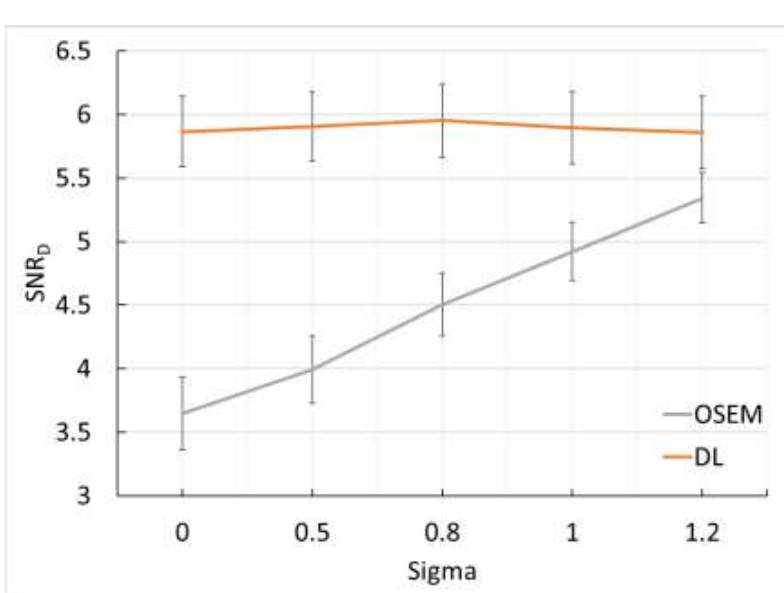

**Fig. 2.** Perfusion-defect detection results obtained on quarter-dose data (with and without DL processing). Error bars indicate standard errors.

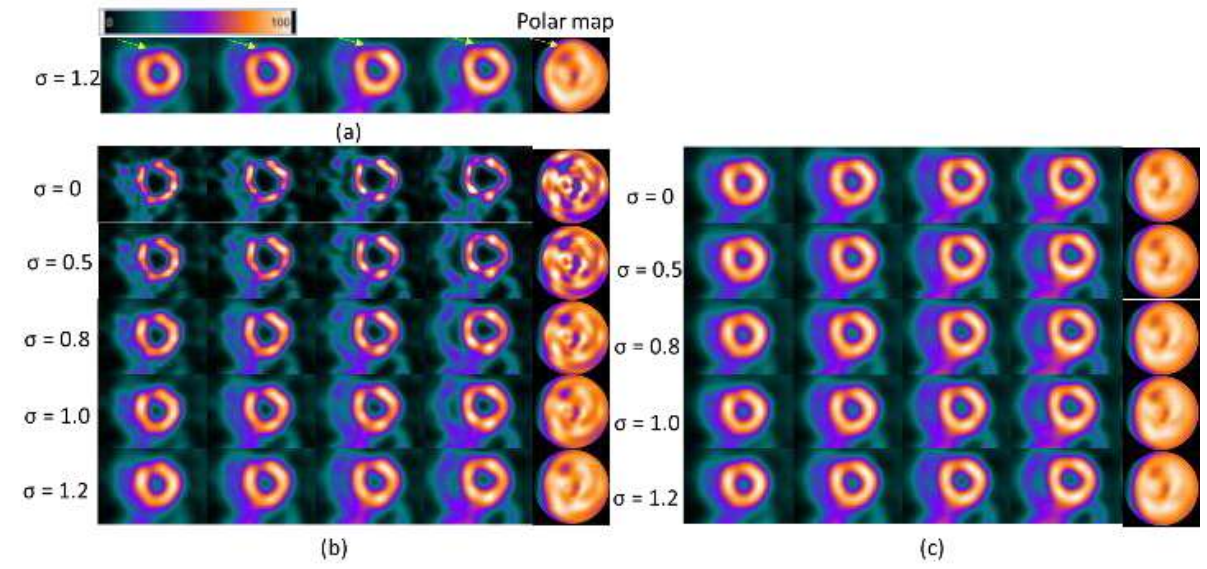

**Fig. 3.** Reconstructed images of an example subject by different methods: (a) full-dose reference (arrows indicate a perfusion defect), (b) quarter-dose OSEM, and (c) quarter-dose DL.

# Deep learning reconstruction improves image quality for Cone-Beam X-ray Luminescence Computed Tomography

Tianshuai Liu[1,2], Shien Huang[1,2], Junyan Rong[1,2], Wangyang Li[1,2] and Hongbing Lu[1,2*]

[1] Fourth Military Medical University, Biomedical Engineering Department, Xi'an, China
[2] Shaanxi Provincial key laboratory of Bioelectromagnetic Detection and Intelligent Perception, Xi'an, China
* Corresponding author. E-mail addresses: luhb@fmmu.edu.cn

**Abstract:** As an emerging hybrid imaging modality, cone-beam X-ray luminescence computed tomography (CB-XLCT) has been proposed based on the development of X-ray excitable nanoparticles. Owing to the complicated excitation process and high scattering of light propagation in biological tissues, the CB-XLCT inverse problem is inherently ill-conditioned. Here, an end-to-end three-dimensional deep encoder–decoder network(DeepCB-XLCT) is proposed to improve the quality of CB-XLCT reconstruction. It directly establishes the nonlinear mapping relationship between the inside X-ray excitable nanoparticles distribution and the boundary fluorescent signal distribution. Thus the reconstruction inaccuracy caused by the simplified linear model can be effectively reduced by the proposed network. Phantom experiments with two targets were carried out, and the results demonstrated that the DeepCB-XLCT network could improve image quality and significantly reduce reconstruction time compared with conventional methods.

## 1 Introduction

With the development of X-ray excitable nanophosphors, X-ray luminescence computed tomography (XLCT) has attracted more attention for its promising performance [1,2]. In XLCT, X-ray excitable nanophosphors are used as imaging probes and emit visible or near-infrared (NIR) light when irradiated by X-rays, which can be measured by an electron-multiplying charge-coupled device (EMCCD) camera. Then, the three-dimensional (3-D) distribution of nanophosphors in the imaged object can be resolved by solving an inverse problem using an appropriate imaging model of X-ray and photon transport. Due to the good penetrability and collimation of X-ray, XLCT can reach deeper imaging depth. In addition, the use of X-ray excitation nanoprobe can effectively avoid the interference of autofluorescence and background fluorescence, which can improve the contrast and resolution of imaging [3].

After the first demonstration of XLCT, cone-beam XLCT (CB-XLCT) is the primary imaging geometry due to its high scanning efficiency and large imaging field suited for in vivo imaging [4]. However, due to the complicated excitation process and high scattering of light propagation in biological tissues, the reconstruction of CB-XLCT is a seriously ill-posed inverse problem. To alleviate this problem and improve the reconstruction quality of CB-XLCT, a few regularization algorithm have been proposed by incorporating a priori information about the probe distribution. Although the regularization algorithm can alleviate the ill-conditioned nature of the inverse problem, it needs a lot of iterations to obtain relatively satisfactory results. In addition, the deviation between the approximate linear model of RTE and true nonlinear photon propagation cannot be avoided fundamentally based on the regularization algorithm.

In recent years, machine and deep learning have made remarkable progress in promoting the performance of multiple molecular imaging modalities such as bioluminescence tomography (BLT), photoacoustic imaging and Fluorescence molecular tomography (FMT). Some of the most popular neural network architectures used for imaging tasks offer some insight as to how these deep learning tools can solve the imaging inverse problem. A nonlinear model can be constructed, which is more consistent with the actual environment, and generally performs better than traditional analysis methods. However, deep learning technology has not been applied to CB-XLCT reconstruction up to now.

In this paper, 3D-En–Decoder network for CB-XLCT reconstruction is proposed, in which large data sets are used to learn the unknown solution to the inverse problem. The proposed 3D-En–Decoder CB-XLCT network is designed to establish a nonlinear mapping from input to output. The parameters of nonlinear mapping are studied and adjusted continuously in the process of network training. Based on this method, the inaccuracy caused by establishing the photon propagation model or solving the ill-posed inverse problem can be effectively avoided.

The remainder of this paper is organized as follows. In Section 2, the conventional forward model and inverse problem of CB-XLCT, Deep Neural Network for CB-XLCT imaging model, training date and optimization training procedure are described in detail. In Section 3, phantom experiments design and results are described for the performance evaluation of the proposed reconstruction approach. Finally, discussions and conclusions are given in Section 4 and 5.

## 2 Materials and Methods

### *2.1 Conventional Forward Model and Inverse Problem of XLCT*

For XLCT imaging, when irradiated by X-rays, nanophosphors in the object can emit visible or NIR light. Based on the previous studies, the number of optical photons emitted is proportional to the intensity distribution of the X-rays and the concentration of nanophosphor in the object, which can be expressed as[3]:

$$S(\mathbf{r}) = \Gamma X(\mathbf{r}) n(\mathbf{r}) \tag{1}$$

where *S(r)* is the light emitted, *n(r)* is the concentration of nanophosphors, Γ is the light yield of the nanophosphors, and *X(r)* is the intensity of X-rays at position *r*, which can be given by the Lambert-Beer law.

In the visible and NIR spectral window, biological tissues have the characteristics of high scattering and low absorbing. Therefore, the propagation model of the emitted light in biological tissues can be established by the diffusion equation (DE):

$$-\nabla \cdot \left[ D(\mathbf{r}) \nabla \Phi(\mathbf{r}) \right] + \mu_a(\mathbf{r}) \Phi(\mathbf{r}) = S(\mathbf{r}) \ (\mathbf{r} \in \Omega) \qquad (2)$$

where Ω is the image domain, *Φ(r)* is the photon fluence, $\mu_a(r)$ is the absorption coefficient. *D(r)* represents the diffusion coefficient that can be calculated by $D(r) = 1/(3(\mu_s'(r) + \mu_a(r)))$, in which $\mu_s'(\mathbf{r})$ is the reduced scattering coefficient.

To solve the diffusion equation (2), the Robin boundary conditions are usually applied, as shown below:

$$\Phi(\mathbf{r}) + 2\kappa D(\mathbf{r}) \left[ \mathbf{v} \nabla \Phi(\mathbf{r}) \right] = 0 \ (\mathbf{r} \in \partial\Omega) \qquad (3)$$

where $\partial\Omega$ is the boundary of $\Omega$, $\kappa$ is the boundary mismatch parameter and $\mathbf{v}$ represents the outward unit normal vector on the boundary.

Using the finite element method (FEM)[30], the total forward problem can be rewritten as follows:

$$\mathbf{\Phi} = Wx \qquad (4)$$

where $\mathbf{\Phi}$ is the projected optical data on the object surface and *W* is the system matrix, which demonstrates the weight of the unknown nanophosphor distribution $\boldsymbol{x}$ to the projection data.

To alleviate the ill-conditioned of the inverse problem, various regularization norms are usually introduced into CB-XLCT to constrain the solution space. The objective function can be written as,

$$\arg\min_{x \geq 0} \|\mathbf{W}\mathbf{x} - y\|_2^2 + \lambda \|\mathbf{x}\|_\beta \qquad (5)$$

where λ is the regularization parameter used to control the tradeoff between the regularity term and fidelity term; $\|\bullet\|_\beta$ (0≤β≤2) denotes the norm term. When β is equal to 2, it refers to L2 regularization (i.e., Tikhonov regularization). Similarly, it refers to L1 regularization when β is equal to 1.

*2.2 Deep Neural Network for CB-XLCT*

Unlike the traditional methods, CB-XLCT reconstruction based on deep neural network aims not to explicitly solve the forward and inverse problems. Instead, it establishes an end-to-end deep neural network (ETE-DNN) mapping model to directly reconstruct the distribution of fluorescence sources.

The mapping model between the surface fluorescence signals and the interior fluorescent sources is defined as follows:

$$\mathrm{y} = f_N(x) \qquad (6)$$

where $f_N$ represents the ETE-DNN framework, which can perform fast-forward inference to reconstruct fluorescent sources; y represents the measured surface fluorescence signals, which is the input of the deep network; *x* represents the unknown distribution of nanophosphors in the imaging object, which is the output of the deep network. Eventually, the inverse problem of CB-XLCT is optimized as follows:

$$\arg\min_{x \geq 0} \|f_N(y|\eta) - x_0\|_2^2 \qquad (7)$$

where $f_N(y|\eta)$ is the output prediction result of ETE-DNN with updated weight vector $\eta$; $x_0$ donates the true label (i.e., ground truth).

The 3D-En–Decoder model is composed of a 3D-Encoder network and a 3D-Decoder network with multiple layers of spatial convolution and deconvolution operators. The input is the 2D surface measurement images at 24 angles of CB-XLCT and the output is the 3D distribution of the CB-XLCT nanophosphor to be reconstructed.

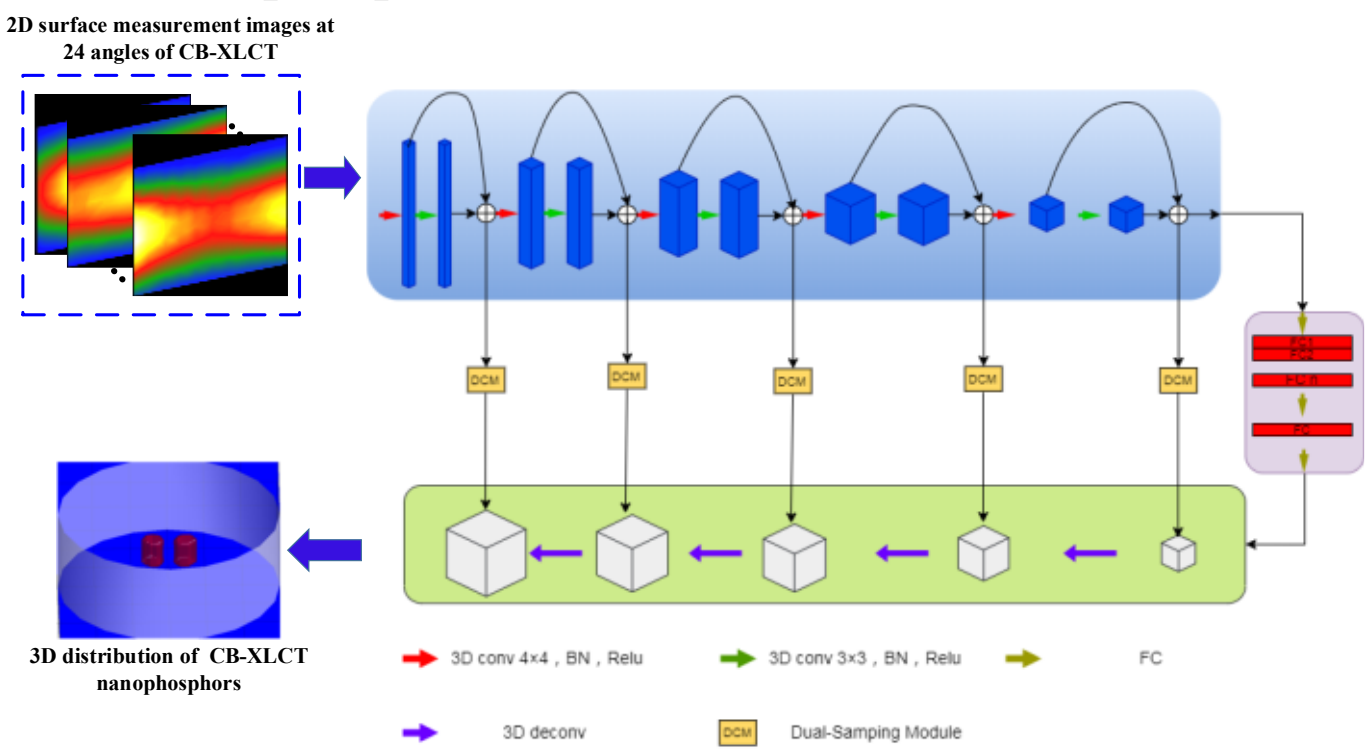

Fig. 1. Schematic illustration of the DeepCB-XLCT network architecture: the 3D deep encoder–decoder (3D-En–Decoder) network has a 3D-Encoder network and a 3D-Decoder network. The 3D encoder consists of several convolution layers, followed by batch norm, ReLU activation function, and pooling. The 3D-Decoder has several unsampling layers followed by convolution, batch norm, and ReLU activation function. There is a full connection layer between the 3D-Encoder and the 3D-Decoder.

*2.3 Training date and Optimization Training Procedure*

In order to support network training, a larger number of training datasets are needed. In CB-XLCT, because the simulation is convenient to control and evaluate, simulation is usually carried out before phantom and in vivo experiments, which means that a large number of simulation training sets can be generated through simulation programs. 6,000 simulation samples were used for training, and 2000 simulation samples were used as the validation set to determine the optimal model.

In addition, all of the projection images were resized to 128 ×128 before being fed to the network. The output of the network is a 3D image. The size of the x − y slice is also 128× 128, and the reconstruction resolution of the z axis is 1 mm. In conclusion, the shape of the input and output are, respectively, 128 × 128 ×24 and 128 × 128 × 7.

The loss function of the DeepCB-XLCT network consists of two parts:the mean square errors (MSE) and structural similarity loss (SSIM) between the reconstructed and true results. In addition, the loss of the target regions (ROI) are also considered. Therefore, the loss function of the DeepCB-XLCT network could be calculated as:

$$loss = MSE(x) + MSE(ROI) + 2*(SSIM(x) + SSIM(ROI)) \quad (8)$$

The optimizer employed in this network was the Adam algorithm implemented on Tensorflow. It was trained with epochs = 50, batch size =64.

## 3 Experimental Design and Results

In order to validate the performance of the proposed DeepCB-XLCT network with real luminescence measurements, phantom experiments were performed by using a custom-developed CB-XLCT system. The configuration of the physical phantom used in imaging experiments was shown in Fig. 2. The phantom was a transparent glass cylinder with a diameter of 3.0 cm and height of 7.0 cm, filled with 1% intralipid and water. Two small glass tubes (3mm in diameter) filled with $Y_2O_3$: $Eu^{3+}$ (50mg/ml) were symmetrically placed in the cylinder to simulate two targets. The edge-to-edge distances (EED) between the two tubes were 2.3mm and 1.7mm.

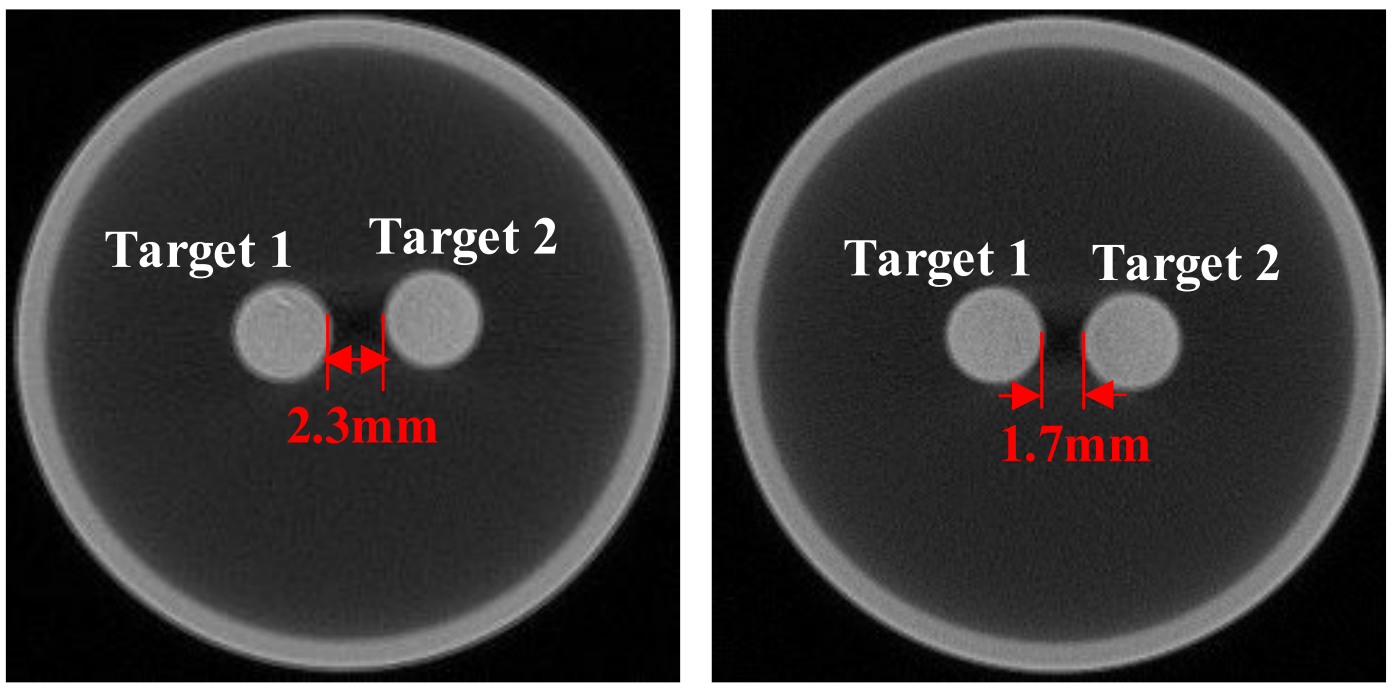

Fig. 2. The physical phantom used in imaging experiments. Transverse view of the two targets containing 50 and 50 mg/ml nanophosphors with edge-to-edge distances of 2.3mm, 1.8mm.

All phantom experiments were performed with a custom-made CB-XLCT system, which consisted of a microfocus X-ray source (Oxford Instrument, U.K.) with a maximum power of 80 W; a high precision turntable; an electron-multiplying charge-coupled device (EMCCD, iXon DU-897, Andor, U.K.), which was coupled with a 50 mm f/1.8D lens (Nikon, Melville, N.Y.) to collect luminescence signals; and a CMOS X-ray detector (2923, Dexela, U.K.) to collect X-ray signals. In the experiments, the X-ray source was set with a tube voltage of 40 kV and a tube current of 1 mA. The phantoms were placed in the same position on the turntable of the CB-XLCT system and all rotated 360° with 24 projections collected evenly by the EMCCD. More descriptions about the experiment were detailed in our published paper[5]. The exposure time, EM gain and binning of EMCCD were set to 0.5 s, 260, and 1×1, respectively. The Feldkamp-Davis-Kress (FDK) algorithmwas used for CT reconstruction.

To evaluate the performance of the proposed method, four traditional widely used methods, adaptive FISTA (ADFISTA, $L_1$ norm), MAP, T-FISTA[5], FMLEM [6], were implemented for comparison with the proposed DeepCB-XLCT network. Their hyperparameters and maximum iteration number were set according to the references published, ensuring the convergence of the reconstruction.

Fig. 3 shows the reconstruction results of two-targets positioned at different distance for phantom experiments. It can be seen that for the ADFISTA and MAP algorithms, only one target can be reconstructed, which indicate that it fail to alleviate the ill-posedness of the inverse problem. With the improved algorithms of T-FISTA and FMLEM by introducing the sparse regularization strategy, both targets could be resolved, as shown in the third and fourth columns of Fig. 3. However, the reconstruction accuracy is worse when the two targets get closer with an EED of 1.7mm. By comparison, both targets could be resolved as expected with the proposed the proposed DeepCB-XLCT method (shown in the fifth column of Fig. 3) due to the fast-forward end-to-end direct reconstruction.

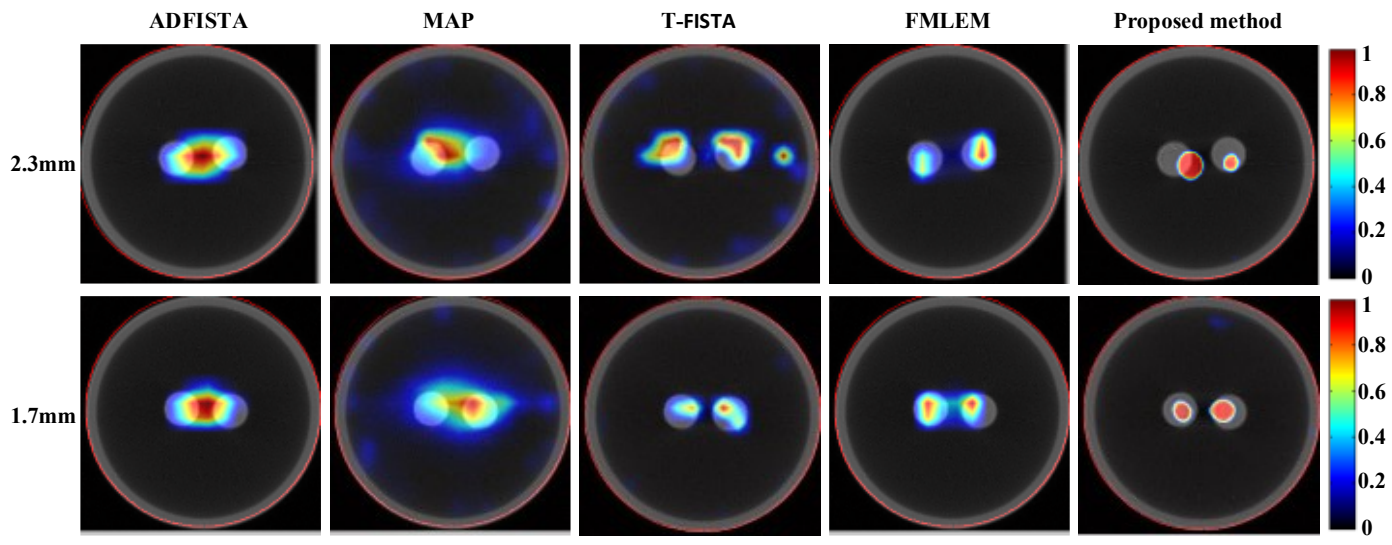

Fig. 3. Reconstruction results of two targets fused with CT (50 and 50 mg/ml) with EEDs of 2.3 and 1.7 mm. The first row shows the tomographic fused XLCT/CT images with the EED of 2.3 mm. The second row shows the tomographic fused XLCT/CT images with the EED of 1.7 mm. Reconstructions obtained by the ADFISTA, MAP, T-FISTA, FMLEM and proposed method are shown from first to fifth column, respectively.

## 4 Discussion

As an emerging hybrid imaging modality, the primary advantage of CB-XLCT is the use of X-rays, which increases the excitation depth, eliminates tissue autofluorescence, and achieves dual-mode of X-ray CT and optical molecular tomographic imaging. The performance of CB-XLCT reconstruction has a remarkably influence on the imaging results. However, the deviation between the complex imaging process and the approximate photon transmission model leads to the high ill-posed of the inverse problem, which severely limit the improvement of imaging quality.

Conventionally, researchers have proposed a variety of reconstruction methods to constrain the image and improve the reconstruction quality, including PCA method [7], wavelet theory [8], Bayesian theory [9].and $L_1$ and TV joint regularization constraints [10] to alleviate the ill-posed nature of the inverse problem. Completely different from these traditional methods, this study introduced a DeepCB-XLCT method, in which the CB-XLCT reconstruction process is completed by establishing the end-to-end nonlinear mapping model between the internal CB-XLCT nanophosphors and the surface measurement signals. Based on the DeepCB-XLCT method, it can greatly eliminate the modeling error of the forward problem and effectively avoid the artifacts caused by iterative calculation, which could greatly alleviate the ill-posed inverse problem.

Phantom experiments results confirm the superiority of the proposed method over the conventional ADFISTA, MAP, T-FISTA and FMLEM methods. Two targets could be resolved when the EED is 1.8 mm, demonstrating its advantage in improving spatial resolution. Besides, owing to the utilization of end-to-end reconstruction approach, there is no complex iterative calculation process for the DeepCB-XLCT method, which could substantially reduce the computational burden and time cost compared with conventional iterative reconstruction method.

However, there are still some limitations. Firstly, the performance of the proposed DeepCB-XLCT method has been verified through the phantom experiments with different EED, it is better to verify the performance of the proposed method through in vivo experiments in the future to fully demonstrate the superior performance of the proposed method. Secondly, as the number of reconstruction targets increase, the difficulty of network construction and training may increase. Besides, although the time cost is very low for well-trained model to reconstruct the distribution of XLCT nanophosphors, the network training takes a long time, up to several hours.

## 5 Conclusion

In summary, we have proposed an DeepCB-XLCT method to improve the quality of CB-XLCT reconstruction. It directly establishes the nonlinear mapping relationship between the inside X-ray excitable nanoparticles distribution and the boundary fluorescent signal distribution, which could effectively reduce the reconstruction inaccuracy caused by the simplified linear model and the iterative calculations. Phantom experiments results demonstrated that compared with conventional iterative method methods, the proposed DeepCB-XLCT method can improve the spatial resolution and reconstruction accuracy, which can promote the widely use of CB-XLCT in vivo.

**Funding**

National Natural Science Foundation of China (NSFC) (11805274)

National Key Research and Development Program of China (2017YFC0107400, 2017YFC0107403, 2021YFC1200104)

Key Research and Development Program of Shaanxi Province (2020SF-214, 2020SF-208)

**Disclosures**

The authors declare that there are no conflicts of interest related to this article.

# Diffusion Posterior Sampling-based Reconstruction for Stationary CT Imaging of Intracranial Hemorrhage

A Lopez-Montes[1], T. McSkimming[1,2,3], W. Zbijewski[1], A. Skeats[2], C. Delnooz[2], B. Gonzales[2], M. Maric[2], J. H. Siewerdsen[1], and A. Sisniega[1]

[1]Department of Biomedical Engineering, Johns Hopkins University, Baltimore, MD, USA
[2]Micro-X, Adelaide, SA, Australia
[3]College of Science and Engineering, Flinders University, Adelaide, SA, Australia

**Abstract** Stationary CT systems based on arrangements of multiple compact x-ray sources provide a promising platform for imaging of hemorrhagic stroke at the point-of-care. However, the reduction in mechanical complexity and increased portability comes at the cost of extremely limited and sparse volumetric sampling patterns that challenge conventional analytical and model-based iterative reconstruction methods. In this work we propose the use of learned diffusion models for image synthesis as a platform to enable accurate posterior sampling-based reconstruction, that leverages prior learning while enabling strict enforcement of consistency with the measured data.

The proposed method uses a reverse noise diffusion sampling mechanism enabled by a neural network approximation to the diffusion score function, and it was trained as a generative model following an unsupervised approach with 4000 axial slices extracted from MDCT head/brain datasets. The reverse diffusion sampling was modified to provide an approximation of the posterior sampling via a conditional score matching model that enforces data consistency. The posterior sampling approximation avoids strict data consistency iterations, offering a suitable reconstruction platform in presence of image noise.

The resulting method was evaluated for visualization of intracranial hemorrhage with a stationary multi-source scanner that features a compact configuration with 31 x-ray sources. Stationary and semi-stationary image acquisition protocols with 31 views (160 deg angular span) and 62 views (205 deg angular span), respectively, were investigated in simulation studies. Simulations were based on 11 MDCT head/brain datasets in which randomly shaped lesions with contrast pertinent to blood (60 HU) were included at random locations. Baseline performance for comparison was obtained with a conventional penalized weighted least squares (PWLS) reconstruction strategy.

Diffusion posterior sampling provided improved visualization of anatomical features and blood-like lesions across the 11 cases. Structural similarity (SSIM) measurements with the MDCT ground truth yielded median SSIM > 0.87 for the extremely limited sampling of the stationary protocol and SSIM > 0.91 for the semi-stationary protocol, and a ~10% increase in SSIM compared to PWLS. The increase in SSIM was accompanied by more consistent performance across cases, yielding ~50% reduction in SSIM interquartile range, compared to PWLS, and improved accuracy of estimated attenuation in blood regions.

The diffusion posterior sampling approach poses a new tool for accurate image reconstruction in imaging systems providing highly limited, non-homogeneous, sampling and is a promising advance towards implementation of stationary CT systems for PoC imaging of stroke.

## 1 Introduction

Stroke is a potentially life-threatening emergency in which prompt treatment drastically reduces the severity of the condition, motivating the need to start treatment directly at the point-of-care (PoC). However, triage between ischemic and hemorrhagic stroke is needed for treatment and unattainable without access to neuro-imaging at the PoC.

Recent improvements in x-ray source technologies introduced a new standard in weight and compactness with sources based on cold-cathode field emission (e.g., carbon nanotubes - CNTs) [1]. The small footprint of CNT sources enabled their arrangement into linear or curved multi-source arrays (MXAs) [2], making them perfect candidates for development of stationary CT systems with weight and footprint suitable for deployment at the PoC.

Previous work [3] proposed a mobile stationary CT concept for PoC imaging. The scanner (Fig. 1A) featured a curved array of CNT sources in combination with a curved-panel x-ray detector, yielding a compact semi-stationary system with minimal gantry mechanical requirements. However, the stationary configuration and non-circular source-detector arrangement poses challenges in the form of limited- and sparse-angular sampling and heterogeneous sampling across the field-of-view (FOV).

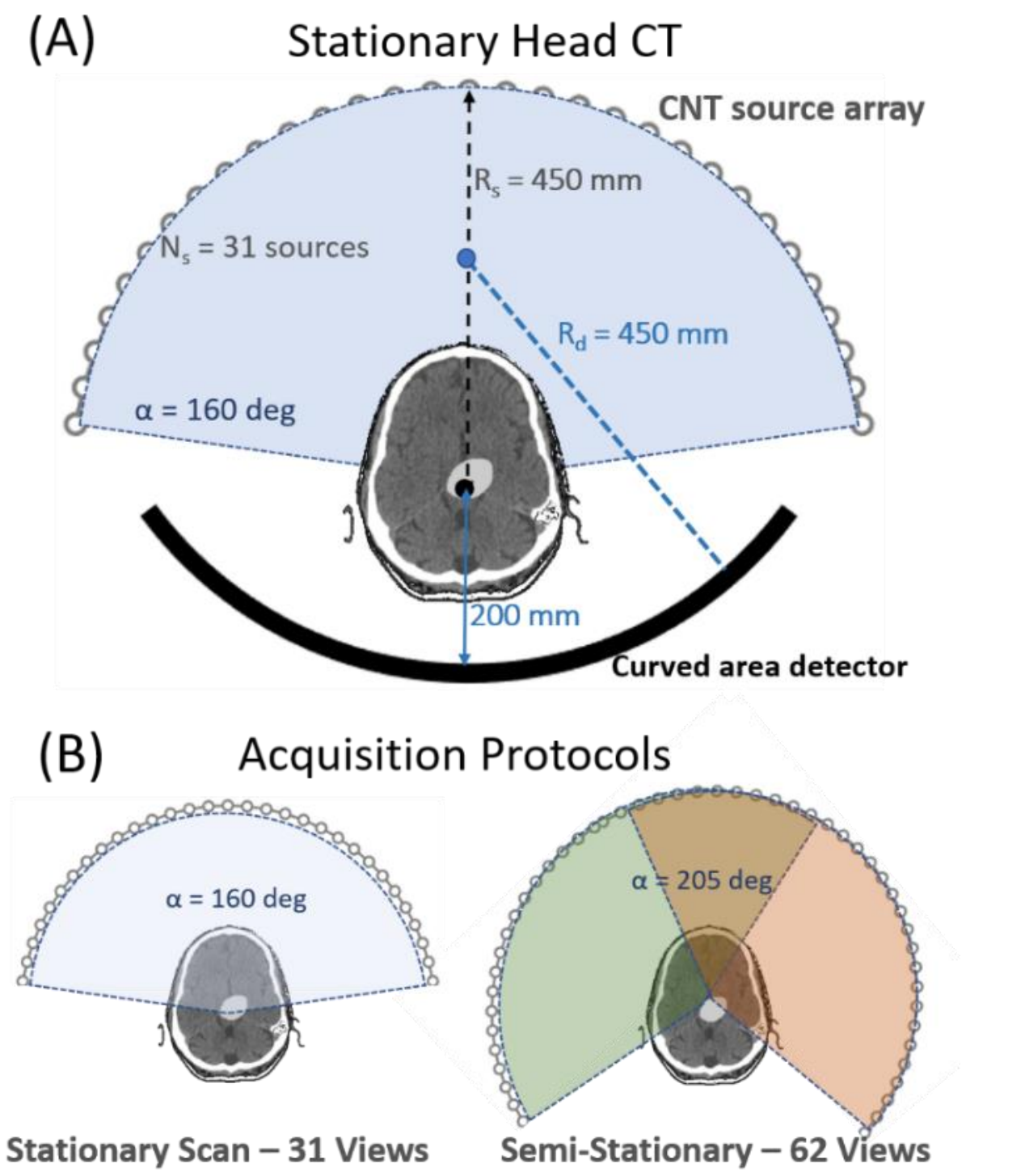

**Figure 1.** (A) Schematic representation of the stationary head CT system, consisting of 31 CNT sources distributed in a curved arc opposing a curved area detector. (B) Image acquisition protocols implemented in the scanner include a full stationary protocol spanning a 160 deg arc with 31 source positions and no motion of the gantry (posing a limited-angle and sparse sampling scenario), and a semi-stationary protocol with two sequential stationary gantry acquisitions with a single 45 deg rotation, spanning a total angular span of 205 deg and 62 source positions (posing a sparse sampling scenario).

Numerous approaches to reconstruction in undersampled scenarios have been proposed in the literature, based on model-based iterative reconstruction (MBIR) with sparsity-promoting, edge-preserving priors [4], or, more recently, on combinations of learned prior models with physically-principled reconstruction methods [5]. However, MBIR methods might fall short in heavily undersampled scenarios, while current learning-based approaches might not preserve fidelity to the measured data.

Recently introduced diffusion models for image synthesis offer a powerful platform for integration of learned features while providing a framework suitable for enforcement of fidelity to the measured data. In this work we explore the integration of diffusion learned models with posterior sampling for stationary CT reconstruction.

## 2 Materials and Methods

*2.1. Stationary CT for Head Imaging*

The stationary CT configuration is shown in Fig. 1. It consists of a MXA of 31 CNT sources arranged along a curved array with radius of $R_s$ = 450 mm. The MXA was centered at the origin of the scanner FOV and has a length of 1240 mm, resulting in total angular coverage of 160°. The curved-panel detector has the same curvature radius $R_d$ = 450 and was placed at a distance of 200 mm from the center of the FOV. The scanner features two acquisition protocols [3], shown in Fig. 1B: i) a fully-stationary protocol in which 31 views (one per source) are acquired with no motion of the gantry, and; ii) a semi-stationary protocol that featured two stationary-gantry acquisitions with a single-step rotation of 45° in between acquisitions, resulting in a total of 62 projection views. The center of rotation is placed at the center of the FOV. The stationary protocol provides a simpler acquisition with no need of gantry rotation but poses a scenario with limited- and sparse angular sampling. The semi-stationary protocol was designed to alleviate the effects of limited angular sampling (angular coverage of ~205° at the center of the FOV), still within a sparse sampling regime.

Volumetric image reconstruction from the sparse, limited angular sampling in the stationary scanner can be posed as a linear inverse problem:

$$\mathbf{y} = \boldsymbol{A}\mathbf{x} + \epsilon \tag{1}$$

where $\mathbf{y}$ is the acquired data (log converted projection images in stationary CT), $\boldsymbol{A}$ is a linear operator representing the system forward model, $\mathbf{x}$ is the image to reconstruct, and $\epsilon$ is a noise term that is assumed to follow a compound Poisson distribution in x-ray systems with area detectors. Reconstruction can then be achieved by sampling from the posterior distribution $p(\mathbf{x}|\mathbf{y})$. This problem is often tackled via MBIR methods, with sparsity assumptions on the prior distribution $p(\mathbf{x})$. For example, our previous work in stationary CT imaging used a PWLS approach with a sparsity-promoting prior acting on $\mathbf{x}$ [3]. These approaches showed limited performance with undercomplete, sparse, and non-uniform sampling.

*2.2. Diffusion Models for Posterior Sampling*

Recently, diffusion models were proposed as a new class of probabilistic generative models for general image synthesis [6]. Diffusion models for image synthesis are based on a forward process that progressively corrupts an input image $\mathbf{x}$, sampled from an unknown distribution $p(\mathbf{x})$ via a controlled sequence of noise addition stages that perturb the image making it indistinguishable from noise.

The forward noise-injection diffusion process is modeled with a stochastic differential equation (SDE) [7], of the following form:

$$d\mathbf{x}_t = f(t)\mathbf{x}_t dt + g(t)d\mathbf{w}_t \tag{2}$$

Where $f(t)$ and $g(t)$ are the drift and diffusion functions, respectively, $\mathbf{w}$ is a standard Wiener process, and $t = [0, T]$ is the stage in the diffusion. By defining the marginal probabilities of $\mathbf{x}_t$ as $p_t(\mathbf{x})$, the distributions at the initial and final stages are well defined, with $p_0(\mathbf{x}) = p(\mathbf{x})$, and $p_T(\mathbf{x})$ approximating an isotropic Gaussian distribution. Following [7], in this work we selected a Variance Exploding SDE (VESDE) in which the transition density function is given by $p_{0t}(\mathbf{x}_t|\mathbf{x}_0) = N(x_t\,|\alpha(t)\mathbf{x}_0, \beta^2(t)\boldsymbol{I})$, with $\alpha(t) = \boldsymbol{I}$, $\beta(t) = \sigma_t$, and $\sigma_t$ linearly increasing with $t$.

The generative model is then obtained via sampling of the data distribution $p(\mathbf{x})$ that is recovered from the tractable $p_T(\mathbf{x})$ by solving the following reverse SDE:

$$d\mathbf{x}_t = \left[f(t)\mathbf{x}_t dt + g(t)^2 \nabla_{\mathrm{x_t}} \log p_t(\mathbf{x}_t)\right] \mathrm{d}\,t + g(t) d\bar{\mathbf{w}}_t \tag{3}$$

Where $\bar{\mathbf{w}}_t$ is the Wiener process in the reverse direction. The term $\nabla_{x_t} log\,(p_t(\mathbf{x}_t))$ is the score function of $p_t(\mathbf{x}_t)$ that is intractable and approximated with a neural network $s_\theta(\mathbf{x}_t, t)$. This score function network is trained in an unsupervised fashion with denoising score matching [6].

To provide a sampling mechanism from the conditional score-matching model we can leverage the diffusion model as a prior, to derive the following conditional reverse SDE:

$$\begin{aligned} d\mathbf{x}_t = \big[f(t)\mathbf{x}_t dt + g(t)^2 \big(&\nabla_{\mathrm{x_t}} \log p_t(\mathbf{x}_t) \\ &+ \nabla_{\mathrm{x_t}} \log p_t(\mathbf{y}|\mathbf{x}_t)\big)\big] \mathrm{d}\,t + g(t) d\bar{\mathbf{w}}_t \end{aligned} \tag{4}$$

The added term $\nabla_{\mathrm{x_t}} \log p_t(\mathbf{y}|\mathbf{x}_t)$ results from applying the Bayes rule to the posterior distribution. It is difficult to obtain a closed-form solution for it due to its dependence with $t$ while $\mathbf{y}$ and $\mathbf{x}$ are only dependent at $t$ = 0.

A common approach to circumvent this limitation enforces data consistency by alternating sampling steps with projection into the measurement space [7]. Those approaches enforce the data fidelity as a constraint on the diffusion and might result in lower performance in presence of noise. A more stable approach can be obtained by approximating the conditional likelihood $p_t(\mathbf{y}|\mathbf{x}_t)$. We followed an approach analogous to [8] to define $p_t(\mathbf{y}|\mathbf{x}_t) \simeq p(\mathbf{y}|\hat{\mathbf{x}}_0)$, where $\hat{\mathbf{x}}_0$ is the posterior mean, approximated by the mean at each step.

Assuming a Gaussian approximation to Poisson noise model in $\mathbf{y}$, we approximate the conditional log-likelihood gradient by:

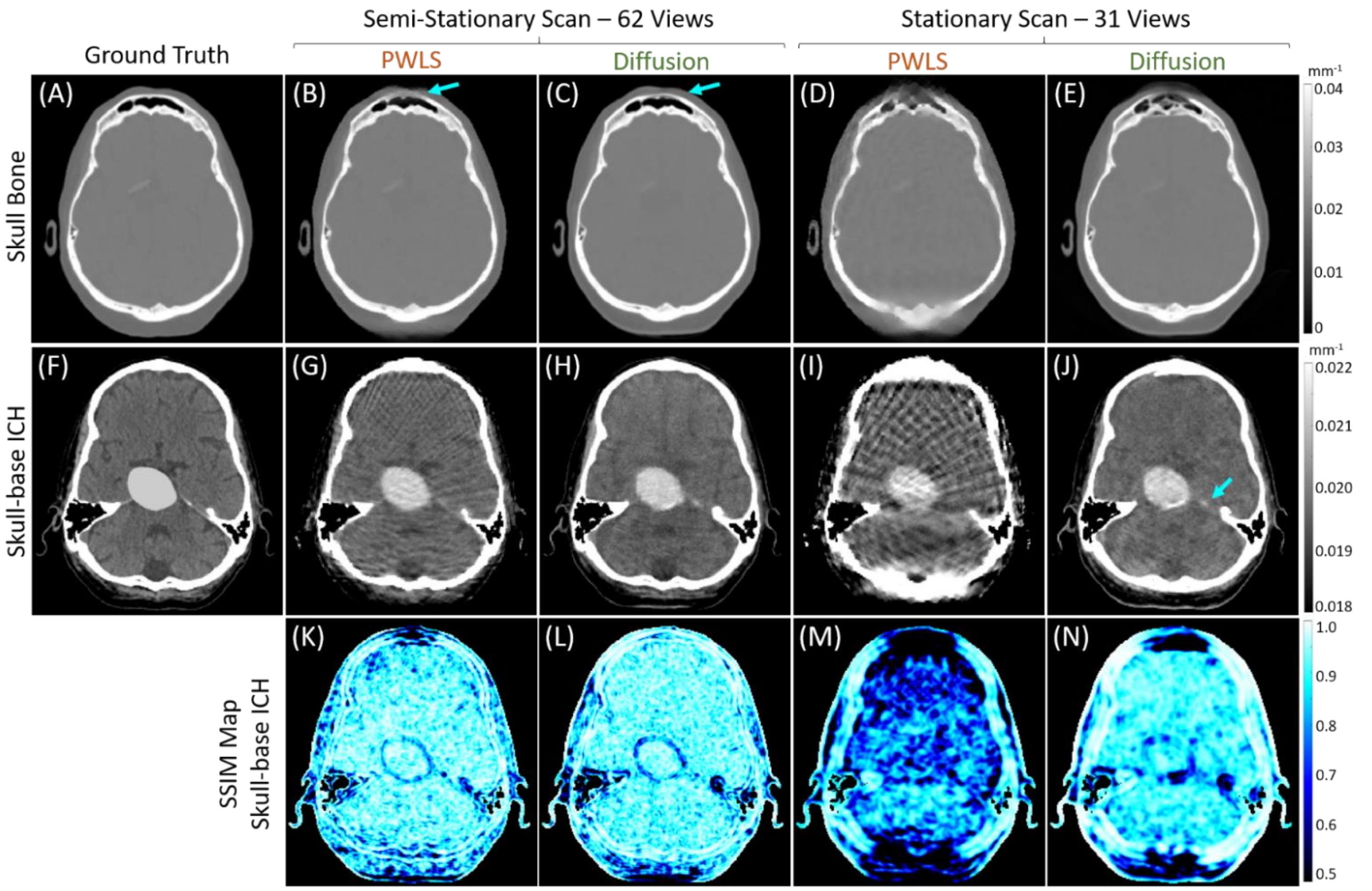

**Figure 2.** Reconstruction results for the two acquisition protocols using the proposed diffusion posterior sampling model and a conventional PWLS approach. The ground truth MDCT slices depicted detailed bone anatomy (A), and conspicuous blood-to-brain contrast (F). PWLS results for the semi-stationary protocol (B, G) showed good delineation of bone structures and intra-cranial hemorrhage, with mild limited-sampling artifacts and with streak-like sparse sampling artifacts. Diffusion posterior sampling results for the semi-stationary geometry (C, H) showed mitigated sampling artifacts, compared to PWLS, while preserving anatomical features. PWLS reconstructions for the stationary protocol (D, I) presented severe sampling artifacts that affect the visualization of the hemorrhagic lesions, soft-tissue structures and bone features. Diffusion posterior sampling reconstructions for the stationary protocol (E, J), provided improved delineation of hemorrhagic lesions, recovery of small intracranial features (see blue arrow in J), and improved bone anatomy detail. SSIM maps (K, L, M, N) illustrate the agreement between sampling artifacts and reduced similarity with the ground truth, as well as recovery of part of the underlying anatomy in heavily undersampled scenarios.

$$\nabla_{\mathbf{x}_t} \log p_t(\boldsymbol{y}|\mathbf{x}_t) = -\lambda \nabla_{x_t} \left\| \mathbf{y} - \boldsymbol{A}\left(\hat{\mathbf{x}}_0(\mathbf{x}_t)\right) \right\|_W^2 \quad (5)$$

Where $\lambda$ is a scalar step size and $\boldsymbol{W}$ is a diagonal matrix containing estimations of the variance in the measurements, approximated with $1/y$. The gradient in (5) was computed via backpropagation across the score-matching network and the forward- and back-projection operators.

## *2.3. Training Strategy and Validation Experiments*

We modeled the score matching function using the NCSNPP model [7]. Training was obtained with 4000 axial slices extracted from a collection of 22 brain MDCTs acquired as part of a previous IRB-approved study. The training set contained no images with conspicuous intracranial hemorrhage. The score-matching network was trained for 500 epochs using the Adam optimizer.

For validation of the reconstruction approach, we used 11 different MDCTs obtained analogously to the training set. To simulate hemorrhagic stroke, synthetic features with contrast pertinent to blood (60 HU) were added at random locations in the brain parenchyma. The inserts were modeled as spheres with random radius (15-25 mm). To generate realistic shapes, the spherical inserts were deformed with a random smooth, deformable vector field. Note that this kind of lesions were not present in the training set and, therefore, not included in the learning process.

Stationary CT datasets were generated by forward projection simulating the stationary (31 views) and semi-stationary (62 views) acquisition protocols. To simplify the simulation process, the curved detector was approximated as a smaller flat-panel detector (768x512 pixels, 0.5 mm pixel size) placed tangent to the intersection of the primary ray of each source and the curved detector. Polychromatic beam and quantum noise effects were introduced into the synthetic projections for an x-ray beam with 107 kV (2 mm Al, 0.2 mm Cu added filtration) and 0.5 mAs per projection. The performance of diffusion-based posterior sampling was compared to conventional MBIR in the form of PWLS with a Huber edge-preserving penalty. Axial slices with 256 x 256 x 1 voxels (1.0 mm voxel size) were reconstructed.

Note that, while 2D slices were used in this work for simplicity and computational efficiency, the 2D score matching network is applicable in 3D scenarios performing stacked 2D inferences in combination with 3D forward- and back-projection operations.

Image quality was evaluated quantitatively via structural similarity (SSIM), computed using the MDCT model as reference. The CT number in hemorrhagic lesions was measured as the mean attenuation value on an 8 x 8 mm region placed at the center of each simulated hemorrhage.

## 3 Results

Reconstruction results of two representative cases are shown in Fig. 2 for PWLS and the proposed diffusion posterior sampling model. The ground truth MDCT slices depicted detailed bone anatomy (Fig. 2A) and soft-tissue contrast (Fig. 2F). PWLS results for the semi-stationary protocol (Fig. 2B, 2G) showed good delineation of bone structures and intra-cranial hemorrhage, albeit with subtle limited-sampling artifacts at the anterior and posterior regions (see blue arrow in Fig. 2B), and with noticeable streak artifacts, visible in soft-tissue regions, particularly at anterior regions. Application of diffusion posterior sampling resulted in noticeable reduction of artifacts while preserving anatomical details in the ground truth, as illustrated in Fig. 2C and Fig. 2H. The better agreement wih the ground truth reference was confirmed by the SSIM maps (Fig. 2K, 2L) that showed larger SSIM for diffusion posterior sampling, especially in regions poised by limited sampling (anterior region of the head) and by streak artifacts (anterior and posterior regions of the brain parenchyma). In the heavy undersampling scenario posed by the stationary protocol, PWLS (Fig. 2D, 2I) yielded a noticeable drop in performance, with increased conspicuity of streak artifacts and worse delineation of the skull in the anterior and posterior regions, compared to the semi-stationary protocol. That effect is reflected in the SSIM map (Fig. 2M) that shows a consistent drop in SSIM, more noticeable in the posterior and anterior regions of the head. Application of difussion posterior sampling (Fig. 2E, 2J) resulted in a noticeable reduction of sampling artifacts and increased conspicuity of anatomical features (e.g, blue arrow in Fig. 2J) that were undistingishable from artifacts with PWLS. The SSIM map (Fig. 2N) showed agreement between the recovered anatomical features and those in the ground truth reference throughout the brain. Largely undersampled features challenged the diffusion posterior sampling approach yielding lower SSIM in the anterior and posterior regions, associated to inaccurate anatomical content.

Quantitative evaluation of SSIM and CT number accuracy for the 11 test anatomies is shown in Fig. 3. Fig. 3A shows mean SSIM computed across the complete head/brain for the two acquisition protocols and reconstruction strategies. Mean SSIM shows consistent improvement with diffusion posterior sampling, compared to PWLS, across acqisition protocols. For the stationary protocol, we observed an 8% increase in median SSIM, as well as a reduction in the SSIM interquartile range of ~50%. Similar reduction in SSIM interquartile range was obtained with the semi-stationary protocol. However, the increased image quality of PWLS with quasi-complete angular sampling yielded negligible improvement in SSIM with diffusion posterior sampling. Fig. 3B shows the accuracy of mean CT number for the hemorrhagic lesions, as a function of acquisition protocol and reconstruction metod. For both acquisition protocols, difussion posterior samplig yielded better agreement and lower range across cases, compared to PWLS. Bleed inserts in regions with highly degraded image quality resulted in outliers with inaccurate blood attenuation when using PWLS, even with semi-stationary protocols whereas diffusion posterior sampling with semi-stationary protocols yielded consistent accurate blood attenuation values.

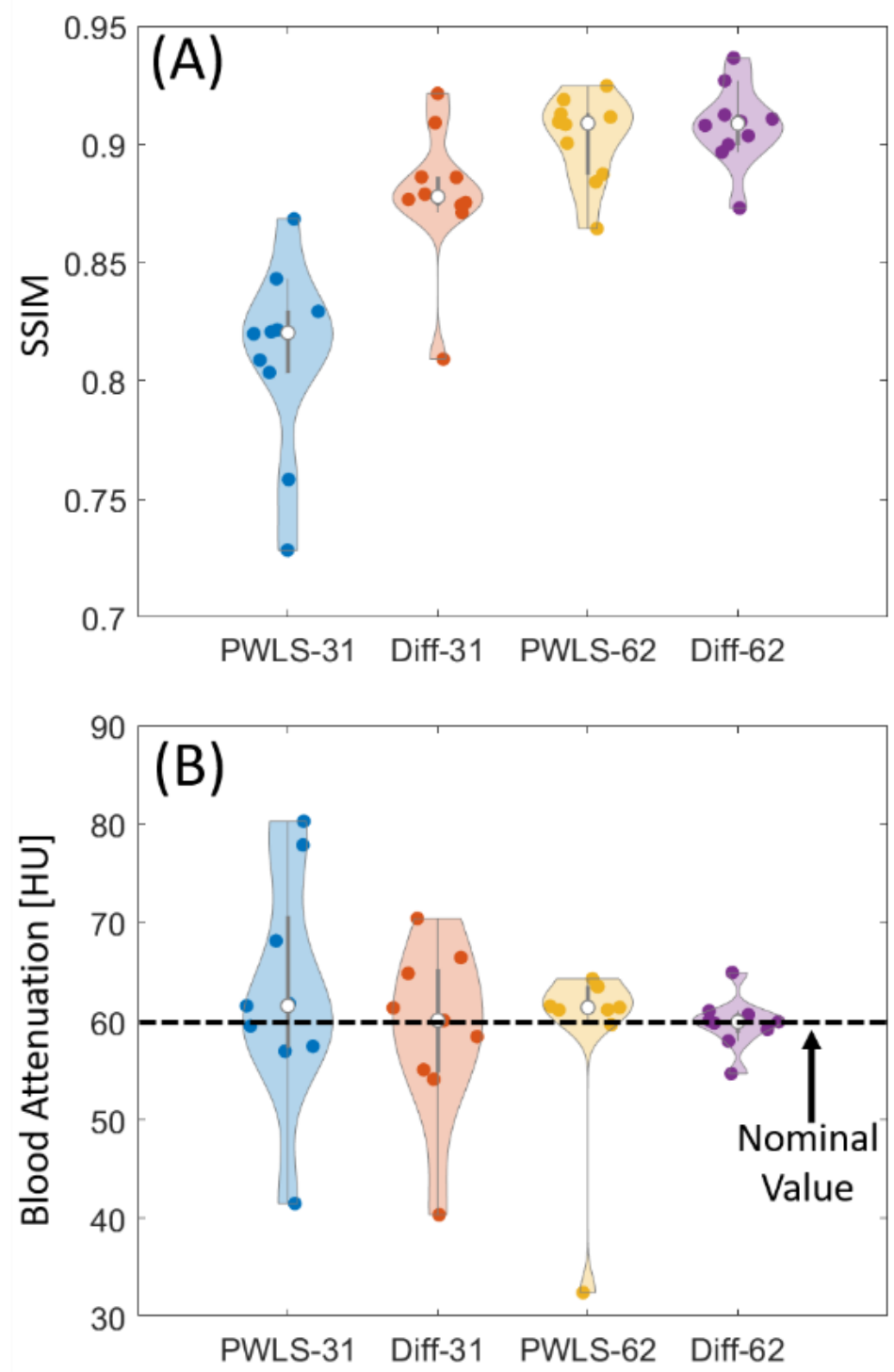

**Figure 3.** Quantification studies for the 11 cases in the validation dataset. (A) Average SSIM across the head/brain region. (B) Accuracy of blood attenuation measured in the bleed inserts vs. the nominal value (60HU).

## 4 Conclusion

We propose a diffusion-based posterior sampling model for image reconstruction in heavily undersampled scenarios, with application to stationary CT systems for PoC imaging of stroke. Diffusion posterior sampling provided promising results for accurate recovery of anatomical features and increased visibility of intracranial hemorrahgue in configurations with extreme angular undersampling, overperforming conventional MBIR approaches in the explored configurations. The results is a promising advance for implementation of fully stationary CT systems for PoC imaging of stroke.

# United Imaging PET Reconstruction Toolbox

Yihuan Lu, Yang Lv, Qing Ye, Liuchun He, Gang Yang, Yue Li, Chen Sun, Duo Zhang, Huifang Xie, Chen Xi, Yilin Liu, Yizhang Zhao, Yong Zhao, Hao Liu, Hancong Xu, Xunzhen Yu, Yu Ding, Yun Dong

United Imaging Healthcare, Shanghai, China

**Abstract:** United Imaging Healthcare (UIH) molecular imaging provides a variety of PET/CT and PET/MR solutions. Among all the systems, the total-body *uEXPLORER* and the world's first sub 200-ps time-of-flight (TOF) PET/CT scanner (*uMI Panorama*) not only provide a superior image quality to help clinicians in disease diagnosis, monitoring and treatment, but also yield unique opportunities for researchers to explore uncharted territories. To facilitate research activities of the nuclear medicine community, we have built a United Imaging PET Reconstruction Toolbox (URT). URT consists of five components: 1) Configurable iterative reconstruction platform, which provides conventional and state-of-art image reconstruction algorithms, as well as a configurable iterative reconstruction setup that allows users to implement customized algorithms; 2) Kinetic modeling and parametric imaging, which supports region of interest (ROI) analysis or parametric imaging. Automatic ROI analysis toolkits and parametric image reconstructions for multiple compartmental models as well as graphical analyses are provided; 3) Patient motion correction (MC). URT provides a comprehensive MC solution, including motion detection, estimation as well as a motion compensated image reconstruction (MCIR) platform that is capable of correcting both rigid- and non-rigid patient motion; 4) AI-based image analysis. URT provides task-specific AI models, including PET-based organ segmentation, image registration, image denoising and synthetic PET-based attenuation map generation; 5) Monte-Carlo simulation (MCS). URT provides an ultra-fast MCS tool that is capable of simulating dynamic PET with realistic patient motion for UIH scanners.

In summary, a United Imaging PET Reconstruction Toolbox (URT) has been developed for researchers to perform user customized reconstruction and to explore uncharted territories on UIH scanner.

## 1 INTRODUCTION

United Imaging Healthcare molecular imaging provides a variety of PET/CT and PET/MR solutions. Among all the systems, the world's first 194-cm-long total-body uEXPLORER and the first <200-ps time-of-flight (TOF) PET/CT scanner (uMI Panorama) not only provide a superior image quality to help clinicians in disease diagnosis, monitoring and treatment, but also yield unique opportunities for researchers to explore uncharted territories. To facilitate research activities of the nuclear medicine community, we have built a United Imaging PET Reconstruction Toolbox (URT) which consists of five components as follows:

URT provides conventional image reconstruction algorithms, e.g., OSEM, as well as various other state-of-art algorithms. Moreover, URT provides a configurable iterative reconstruction setup that allows users to implement customized algorithms. Customizable physics corrections, e.g., attenuation/scatter correction, are available in URT reconstruction.

For dynamic PET imaging protocols, region of interest (ROI) analyses or parametric imaging are used to estimate pharmacokinetic parameters, i.e., absolute PET quantification. For ROI analyses, kinetic modeling software are commercially available but often costly; the development of customized parametric reconstruction software requires specialized image reconstruction expertise. To make absolute quantification tools available to more researchers, URT provides automatic ROI analysis toolkits and parametric image reconstructions for multiple compartmental models as well as graphical analyses.

Patient motion, including both voluntary and involuntary movements, e.g., head, respiratory, limb, torso, cardiac rhythm, intestinal peristalsis and bladder filling, are commonly encountered in PET imaging and can create artifacts as well as inaccurate tracer quantification. URT provides a comprehensive motion correction (MC) solution, including motion detection, estimation as well as a motion compensated image reconstruction (MCIR) platform that is capable of correcting both rigid- and non-rigid patient motion.

Artificial intelligence (AI) technologies, especially deep learning (DL)-based algorithms, have been proven extremely useful in the medical image analysis (MIA) tasks, e.g., image denoising, segmentation and synthesis. In the foreseen future, we believe DL models will become the fundamental tools for many clinical/research applications. Unlike the conventional MIA algorithms, DL model training typically requires large amount of real clinical data, which are often publicly unavailable. To bridge the gap, URT provides users ready-to-use DL models for different MIA tasks.

Lack of ground truth using real subject data vs. lack of being physically or physiologically realistic using simulation data are often the dilemma during reconstruction algorithm development. Monte-Carlo simulation (MCS), e.g., GATE, is capable of simulating realistic PET physics, but is also computationally expensive. There is no publicly available MCS software that can simulate dynamic PET imaging with realistic patient motion. As part of the URT, an ultra-fast MCS tool that can simulate dynamic PET imaging with continuous patient motion is available to the URT users.

## 2 DESIGN OF URT SPECIFICATIONS

### A. Configurable iterative reconstruction platform

URT provides conventional OSEM list-mode iterative reconstruction algorithm with customizable parameters, e.g., number of iterations, subsets and voxel size. CT-based attenuation correction (AC), Monte-Carlo simulation-based scatter correction (SC), randoms correction, normalization and dead time correction are available for reconstruction. All the intermediate results, e.g., sensitivity map and estimated scatter sinogram, are accessible by users. Line-of-response (LOR)-based spatially-variant point-spread-function (PSF) is used to model positron range, non-collinearity, depth of interaction (if applicable) and inter-crystal scattering. Reconstruction can be performed with arbitrarily discarded list-mode data ("gating").

Besides OSEM, two advanced reconstruction algorithms are also available to the URT users, e.g., Bayesian penalized likelihood OSEM ("HYPER Iterative") [1] and deep-learning-based reconstruction ("HYPER DPR") [2]. Figure 1 shows an uMI Panorama $^{18}$F-FDG study using three different reconstruction algorithms.

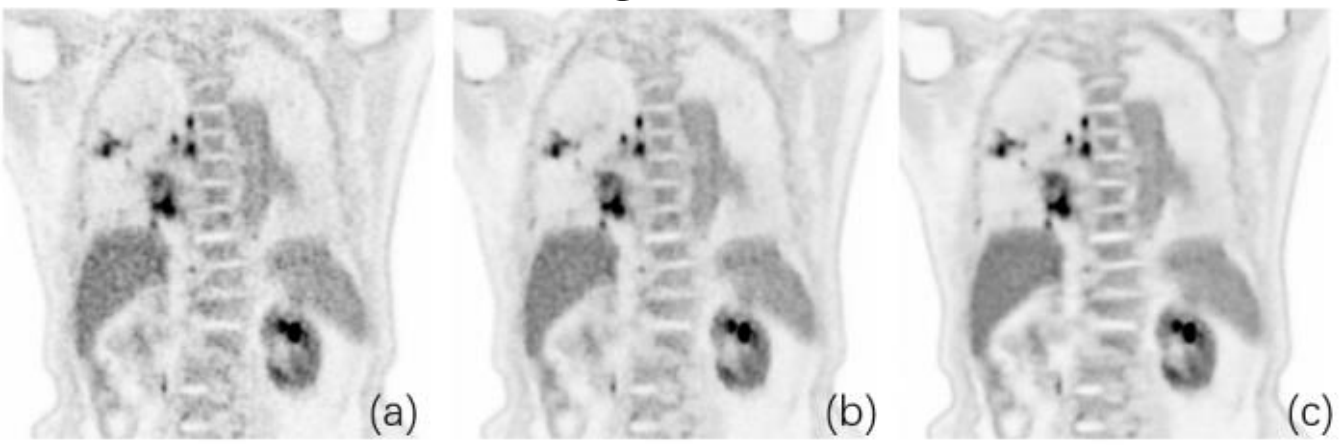

Figure 1. Reconstruction of a 1.5-min uMI Panorama $^{18}$F-FDG (9mCi, 60 min post-injection) study using (a) OSEM, (b) HYPER Iterative and (c) HYPER DPR.

In addition to providing established algorithms, URT will provide a configurable iterative reconstruction platform, which allows users to build their own algorithms using the forward- and back-projectors. Together with other correction modules, e.g., SC, AC and motion correction, URT can be used to perform new reconstruction algorithm development.

### *B. Kinetic modeling and parametric imaging*

ROI analysis and parametric imaging are supported in URT. Apart from user-defined arterial input function (IF) and whole blood time activity curve (TAC), URT can extract image-derived input function (IDIF) automatically from the reconstructed images. A tool to generate population-based input function (PBIF) template is also available. With AI-based segmentation, organ TACs can be automatically generated and can be analyzed with the specific kinetic models.

URT does not only support graphical analysis methods, e.g., Patlak, Logan, and Relative Equilibrium plots, but also common compartmental models, e.g., one-tissue, two-tissue, and two-tissue irreversible compartmental models, with or without time delay estimation. Reference tissue models, such as simplified reference tissue model (SRTM), are also provided. Indirect voxel-based analysis with above models produces the micro and macro parametric images while direct Patlak reconstruction is also provided to suppress noise in the parametric images. Careful validation of the kinetic modeling module has been performed by comparing to PMOD (PMOD Technologies LLC, Zürich, Switzerland). Figure 2 shows an uMI Panorama $^{18}$F-FDG study example of the $K_i$ and intercept images generated by URT and PMOD respectively.

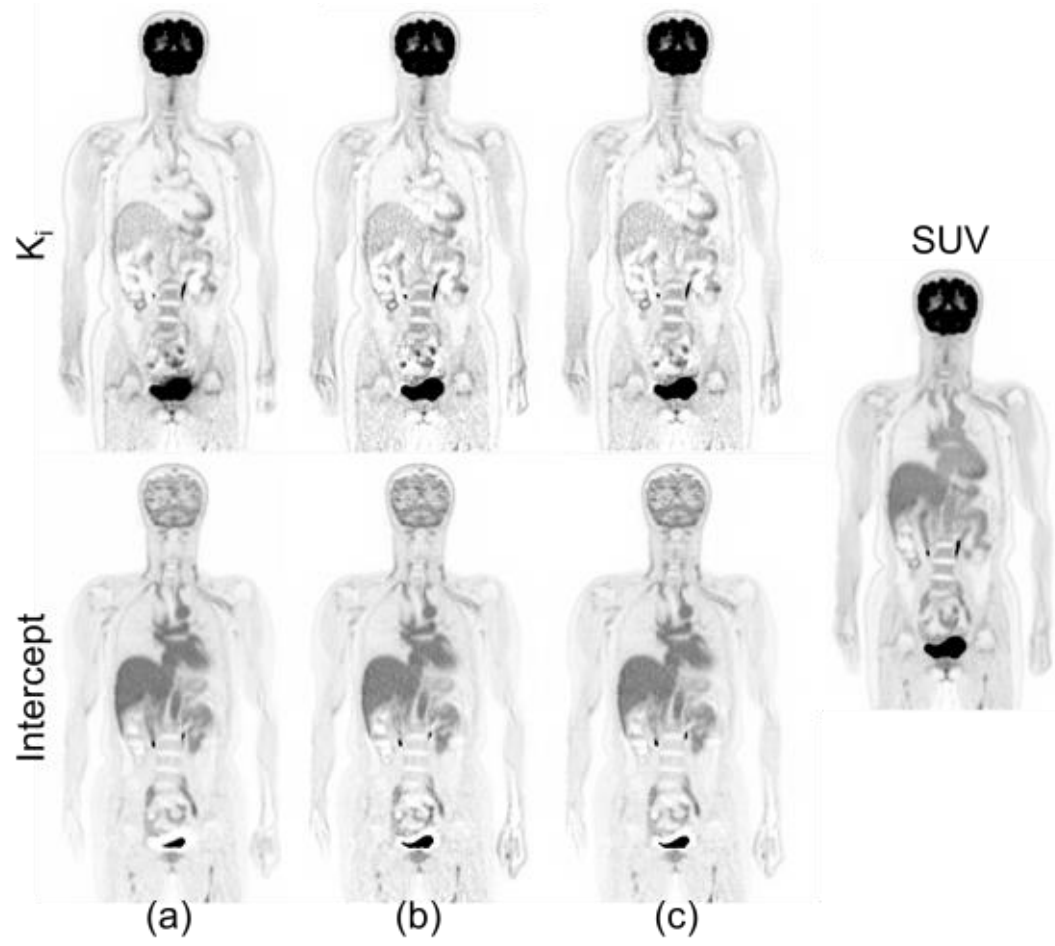

Figure 2. Comparison of the $K_i$ and intercept images from (a) direct Patlak-URT; (b) indirect Patlak-URT and (c) indirect Patlak-PMOD using a whole-bed uMI Panorama $^{18}$F-FDG study. Static SUV image is shown for reference.

### *C. Patient motion correction*

Organ-specific data-driven motion tracking and detection tools, with the aid of AI-based segmentation, are provided to extract varies types of patient motion information. GPU-based image registration modules with different similarity metrics are provided for rapid motion estimation.

Correction solution of specific motion types are provided, e.g., for respiratory motion (RM) correction (RMC), amplitude- or phase-based gated reconstruction, and subsequent RM corrected reconstruction are implemented in URT; for head motion correction (HMC), established methods are implemented. Figure 3 shows examples of HMC and RMC, respectively, for studies performed on an uMI Panorama. Users may also build their own motion models, which can be used in the URT by following a displacement field (DF) definition convention.

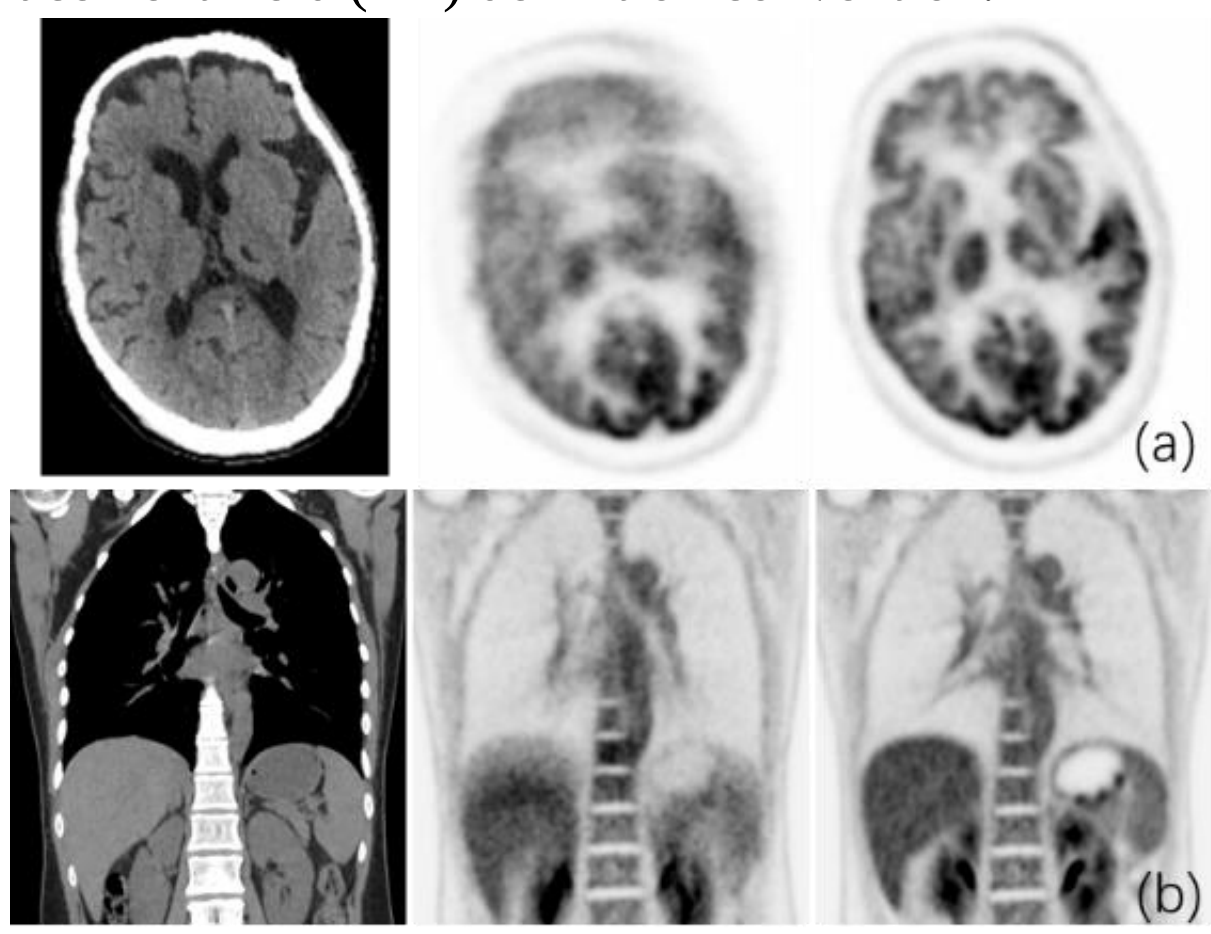

Figure 3. (a) Without and with head motion correction of an 3-min $^{18}$F-FDG brain study; (b) without and with respiratory motion correction of a 2-min single-bed $^{18}$F-FDG study on uMI Panorama. First column shows their relative CT images respectively.

To simultaneously correct multiple types of motion, e.g., non-rigid RM, bulk body motion and rigid head motion, URT will provide motion concatenation tool to generate a unified DF. In addition to correction, URT also provides quality control module to help users evaluate the MC information.

*D. AI-based image analysis*

Basic neural network-based models, e.g., convolutional neural network and generative adversarial network, have been pre-trained on the UIH datasets and shared with URT users. Specifically, task-specific AI models are provided, including PET-based organ segmentation, image registration, image denoising and synthetic PET-based attenuation map generation. Training data have undergone a rigorous data-cleaning and quality-control processes. Training and validation conditions, e.g., loss functions and training parameters, are released to URT users. Figure 4 demonstrates an example of AI-based attenuation map synthesis from a non-AC OSEM image.

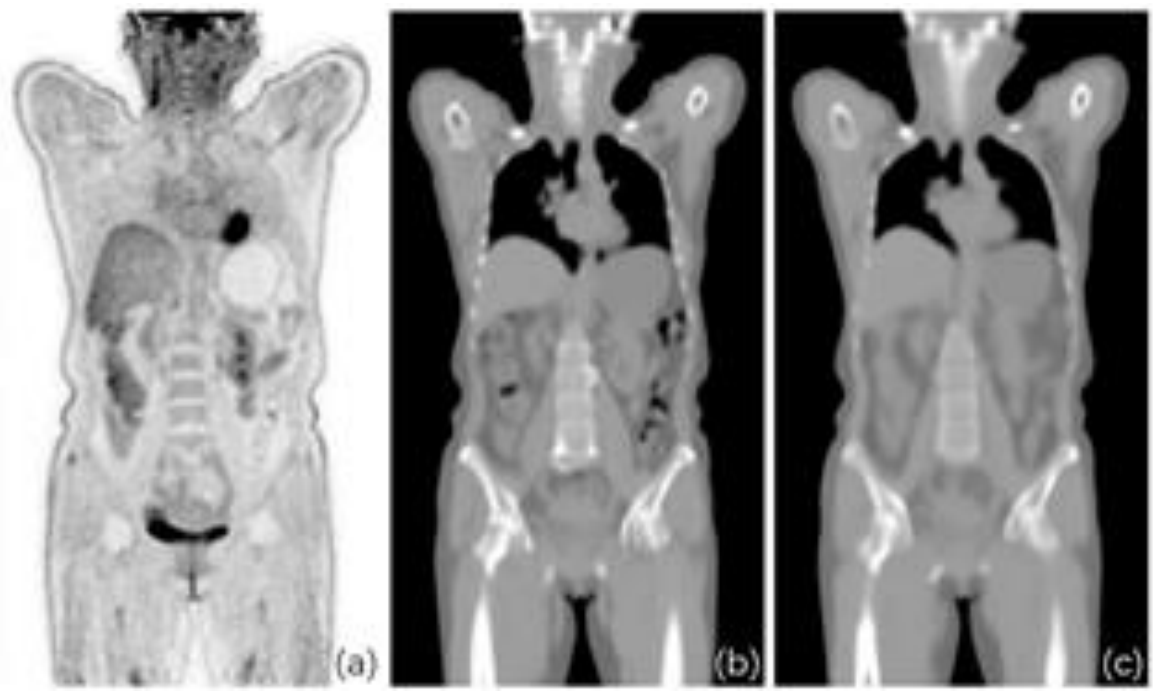

Figure 4. AI-based attenuation map synthesis: (a) non-AC OSEM (network input) of a uMI Panorama $^{18}$F-FDG study; (b) CT-based attenuation map and (c) network predicted attenuation map.

*E. Monte-Carlo simulation*

Monte Carlo simulation (MCS) in URT is a GPU-based toolkit, which is dedicated for efficient simulation of UIH PET systems. The MCS was first developed for scatter correction and was further integrated with high-precision modeling of PET detectors to perform full PET imaging chain simulation. All the PET physical processes, e.g., positron range, non-collinearity, prompt gamma, Compton scattering and inter-crystal scattering, are included in the simulation. It can also simulate UIH-specific detector features, e.g., scatter recovery in uMI Panorama. Optimized for PET simulation, MCS is ultrafast, i.e., typically three orders of magnitude faster than a CPU-based GATE simulation. Figure 5 shows an MCS example using an XCAT phantom for OSEM reconstruction with 30-min and 3-min data acquisition respectively. MCS in URT also support simulation using 3D or 4D digital phantoms, i.e., organ label maps + organ TACs or organ kinetic models, as well as rigid- and non-rigid patient motion at arbitrary temporal resolution. Figure 6 shows an MCS example using an XCAT phantom for OSEM reconstruction with or without simulating the head motion during scan.

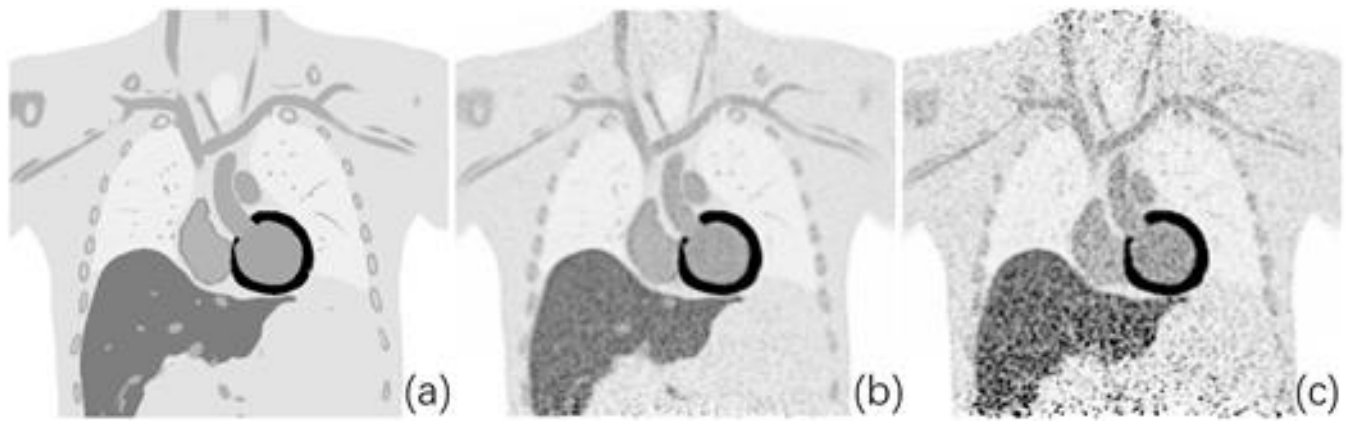

Figure 5. Monte-Carlo simulation of an uMI Panorama study. (a) XCAT emission phantom; (b) OSEM reconstruction of a 30-min and (c) 3-min $^{18}$F-FDG simulated study (10 mCi, 60-min post-injection).

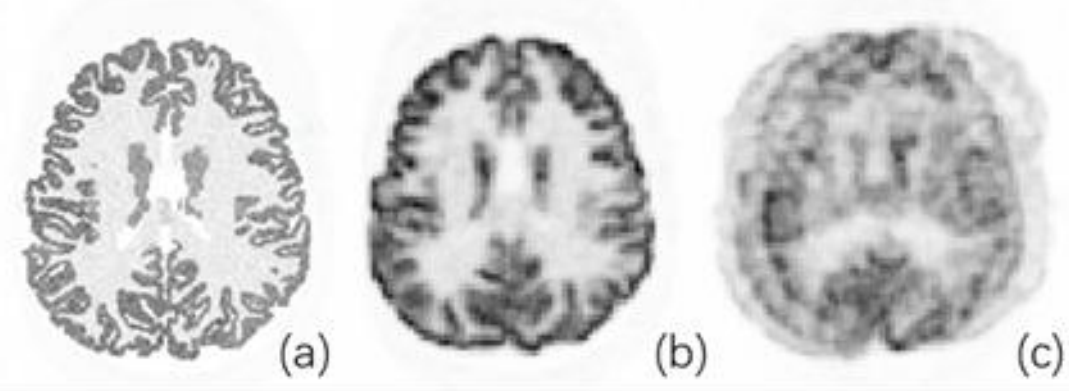

Figure 6. Monte-Carlo simulation of an uMI Panorama study. (a) XCAT emission phantom; (b) OSEM reconstruction without simulating head motion (c) OSEM reconstruction with head motion during scan.

## 3 SUMMARY

In summary, we developed URT for user customized PET reconstruction processing to facilitate research activities. URT is a platform that integrates simulation, reconstruction, AI-based image analysis and kinetic modeling for PET studies on UIH scanners. URT currently supports the uMI Panorama system, and will gradually support all the UIH scanner models in the future.

# Image reconstruction and CT-less attenuation correction in a flat panel total-body PET system

Jens Maebe[1], Meysam Dadgar[1], Maya Abi Akl[1] and Stefaan Vandenberghe[1]

[1]Department of Electronics and Information Systems, Ghent University, Ghent, Belgium

**Abstract** A new PET scanner design, the walk-through total-body PET, has recently been proposed. It consists of two flat, vertically placed, 74x106 cm$^2$ panels made up of monolithic BGO detectors. This new scanner design will offer a lower cost, high patient throughput alternative to existing total-body PET scanners. This study investigates the image reconstruction of the NEMA image quality phantom in two configurations of the scanner: fixed and rotating. In addition, two methods for CT-less attenuation correction in the system are explored: the use of a transmission source and estimation of the attenuation coefficients from the emission data itself. The fixed scanner configuration offers higher sensitivity, but streaking artifacts are observed due to the limited angle problem, which are solved in the rotating configuration. CT-less attenuation correction also proves to be feasible, with minimal discrepancies observed compared to ground-truth attenuation correction.

## 1 Introduction

Longer axial field-of-view positron emission tomography (PET) systems, advancements in detector technology and the use of deep learning for signal processing have enabled sub-one minute PET scans at reasonable dose levels and scan quality. In such a case, patient throughput is primarily constraint by patient preparation and positioning on the bed. To address this limitation, a new total-body PET scanner design has been proposed: the walk-through PET (WT-PET) [1].

It consists of two vertically placed flat detector panels, each 74 cm wide and 106 cm high. The panels are spaced 50 to 70 cm apart and adapt their position to the patient height. After radiotracer administration and uptake time, the patient walks into the scanner and stands still in an upright position for a 30 second scan. The scanner design offers simultaneous head and torso imaging at a high sensitivity for an estimated cost only slightly above a conventional (e.g. 25 cm axial field-of-view) PET/CT. This is achieved by a lower number of detectors due to the proximity of the flat panels to the patient, and the use of low-cost but high-resolution monolithic BGO detectors.

In this study, we compare the image reconstruction of two possible configurations of the WT-PET, see Figure 1.

The first is a fixed configuration where the two panels are placed 50 cm apart. This maximizes sensitivity by placing the detectors as close as possible to the patient. The gap between the panels, however, results in missing angles for image reconstruction.

In the second configuration, the two panels rotate around the central axis at a radial velocity of 180°/30s. This ensures that all emission angles are present for the reconstruction of a 30s scan. The panels do need to be placed further apart (70 cm) to make room for the patient's shoulders, thereby reducing overall sensitivity.

In addition, since the first prototype of this scanner is likely to be built without a CT, we investigate the feasibility of CT-less attenuation correction by means of either a simultaneous transmission scan, or by estimation of the attenuation coefficients from the emission data itself. The results are again compared for both configurations of the scanner.

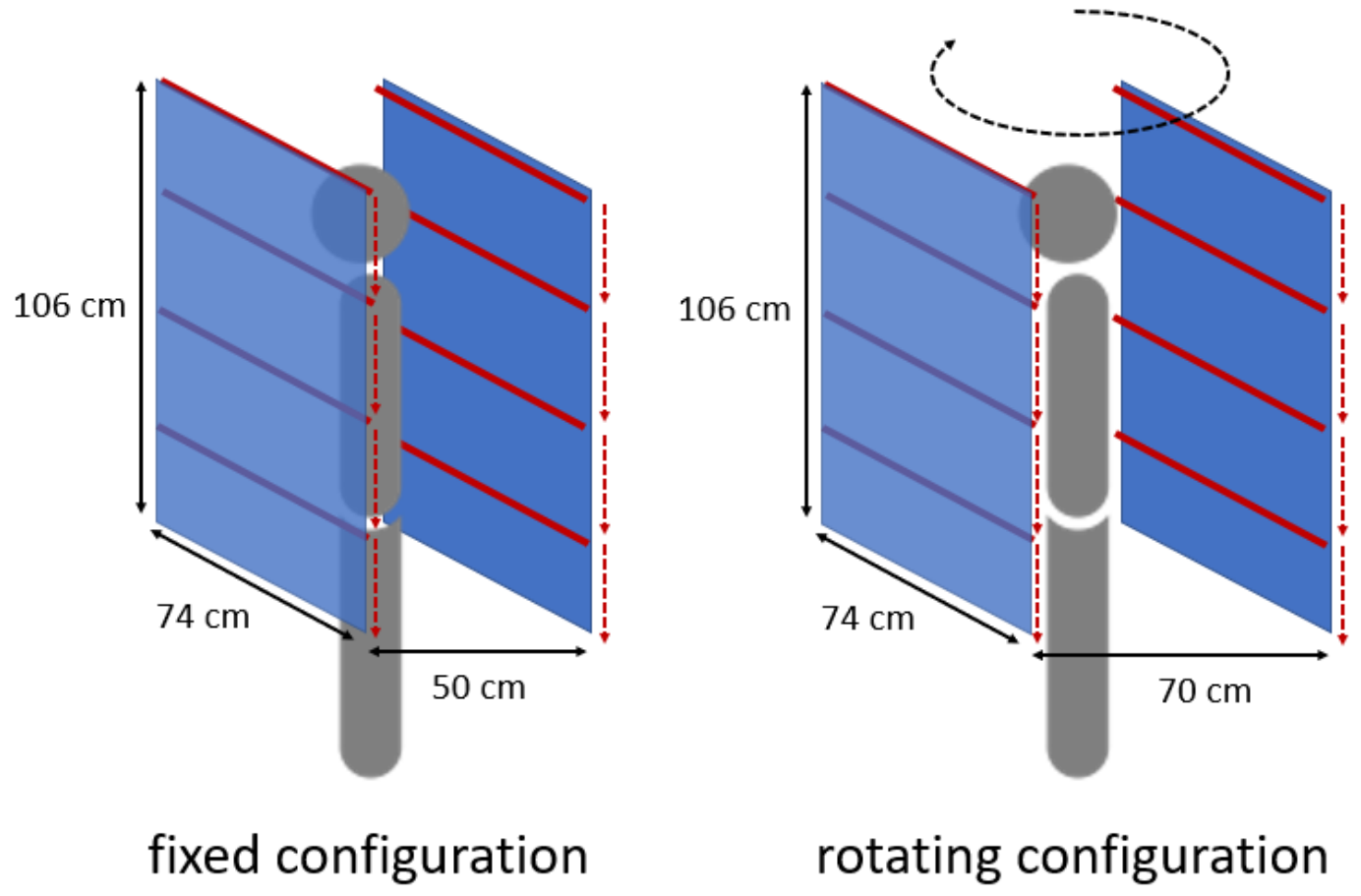

*Figure 1: Fixed and rotating configurations for the walk-through PET. The red lines represent eight optional, downwards moving line sources for a transmission scan.*

## 2 Materials and Methods

We use GATE [2] to perform 30s scan simulations of the NEMA Image Quality (IQ) phantom, with a hot to background activity ratio of 4:1 (background activity = 5.3 kBq/cc). For both scanner configurations, the geometry consists of two flat detector panels, each containing 14x20 (=280) monolithic BGO detectors. The detectors are 50x50x16 mm$^3$ in size, leaving a 3 mm gap between each other. Detector energy blurring (15%), spatial blurring (1.3 mm in 2D and 2 mm for the depth of interaction, in full width at half maximum) and time blurring (400 ps time-of-flight, or TOF, resolution) are all included. The value of 400 ps for monolithic BGO is a conservative system-level TOF value based on experimental detector results of 327 ps, obtained with a neural network for time estimation in the detector [3].

Iterative list-mode image reconstruction is performed with all images reconstructed to 2 mm voxel sizes using 10 iterations (no subsets). The activity map is reconstructed using the maximum likelihood expectation maximization (MLEM) algorithm. We include sensitivity and attenuation correction but omit any kind of regularization. Random coincidences and phantom scatter are omitted from the list-mode data prior to reconstruction, as a proper implementation of random and scatter correction is still in progress.

The attenuation map can be obtained from a scan with a transmission source. Here, we opt for 4 horizontal isotropic line sources (1 mm diameter) per panel (8 total), each with an activity of 3 MBq. This activity was chosen low enough so that no appreciable detector saturation would occur due to the transmission sources. They stretch the full width of the panel and are placed 1 cm in front of it. The line sources travel downwards during a 30s scan, so that each one sweeps over a quarter of the panel during the scan (see Figure 1).
Reconstruction of the attenuation map is done using the maximum likelihood transmission (MLTR) algorithm, operating in list-mode [4]. The attenuation image is again reconstructed to 2 mm voxel sizes using 10 iterations (no subsets).
Another option for CT-less attenuation correction is by joint reconstruction of the activity and attenuation based on emission data only, referred to as MLAA (maximum likelihood activity and attenuation) [5]. MLAA uses an interleaved updating of the activity/attenuation while keeping the attenuation/activity fixed. Given the lack of TOF for the attenuation update, convergence is slower compared to the activity reconstruction. Therefore, we perform 5 attenuation updates for each activity update, performing in total 10 iterations (no subsets) for the activity reconstruction.

## 3 Results

Figure 2 shows the iterative reconstruction of the IQ phantom for different scanner configurations, axial positions of the IQ phantom, and a non-TOF reconstruction. Note that for the fixed configuration, the panels are located at the top and bottom of the reconstructed image. The ground-truth attenuation map was used for reconstruction. We can see the result of the limited emission angles in the fixed scanner configuration, although the effect is greatly reduced by using 400 ps TOF in the image reconstruction. Placing the IQ phantom at 1/8 of the axial FOV (AFOV), the sensitivity drops (resulting in a noise increase), and the limited angle artefacts become slightly more pronounced due to fewer projection angles along the axial direction.

We also observe that the smallest contrast sphere (d = 10 mm) can still be distinguished in a 30s scan for all configurations. It is visually more easily distinguished from the background noise in the fixed configuration, due to increased sensitivity. Roughly 47% more true coincidences were detected in the fixed configuration compared to the rotating configuration.

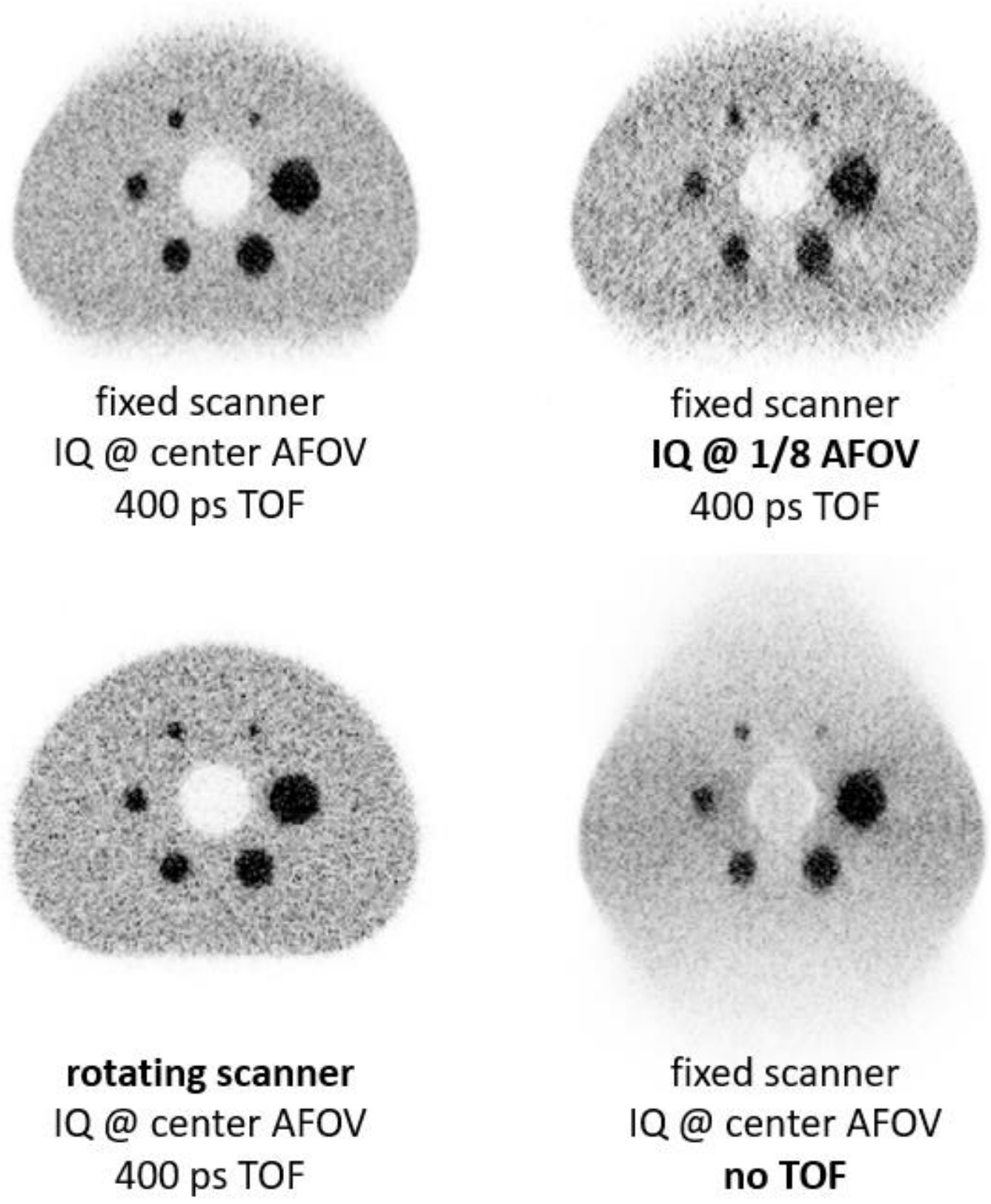

*Figure 2: MLEM reconstructions of the NEMA IQ phantom for different configurations of the WT-PET.*

Figure 3 shows, for the fixed scanner configuration, the attenuation maps obtained with MLTR and MLAA, together with the corresponding activity reconstructions. These are also compared to reconstructions without attenuation and with the ground-truth attenuation map.
The transmission source does not provide a good estimate of the attenuation map due to the limited angle problem. Nonetheless, the corresponding activity reconstruction is properly attenuation corrected. MLAA on the other hand provides a better outline of the attenuation map.

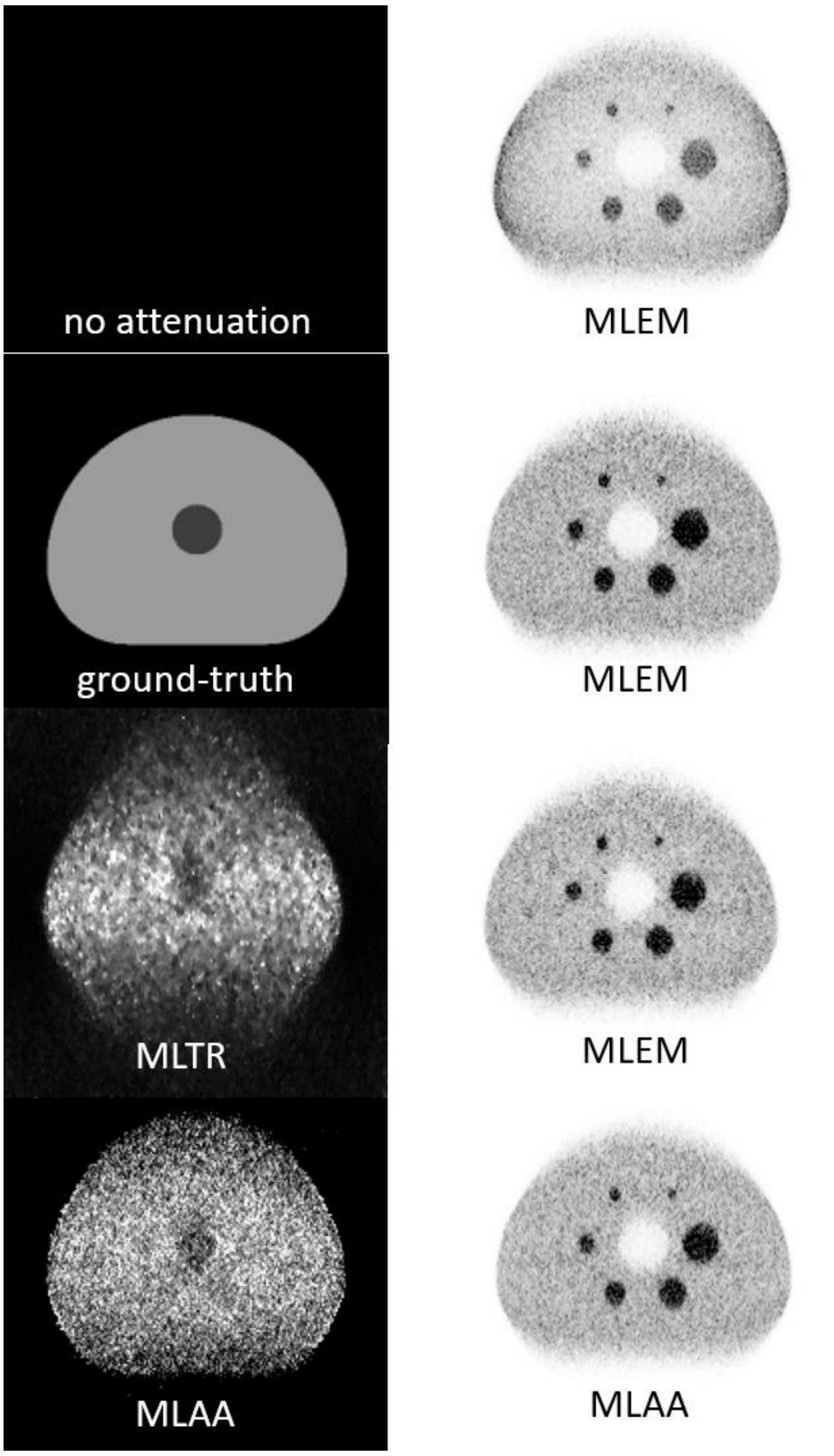

*Figure 3: Activity reconstructions based on different attenuation corrections in the* ***fixed*** *WT-TB-PET configuration.*

Figure 4 shows the same results but for the rotating scanner configuration. Here we observe that the limited angle problem is mitigated entirely, and as a result both the attenuation maps and the activity distributions are reconstructed more accurately, with the trade-off of increased noise levels due to decreased sensitivity.

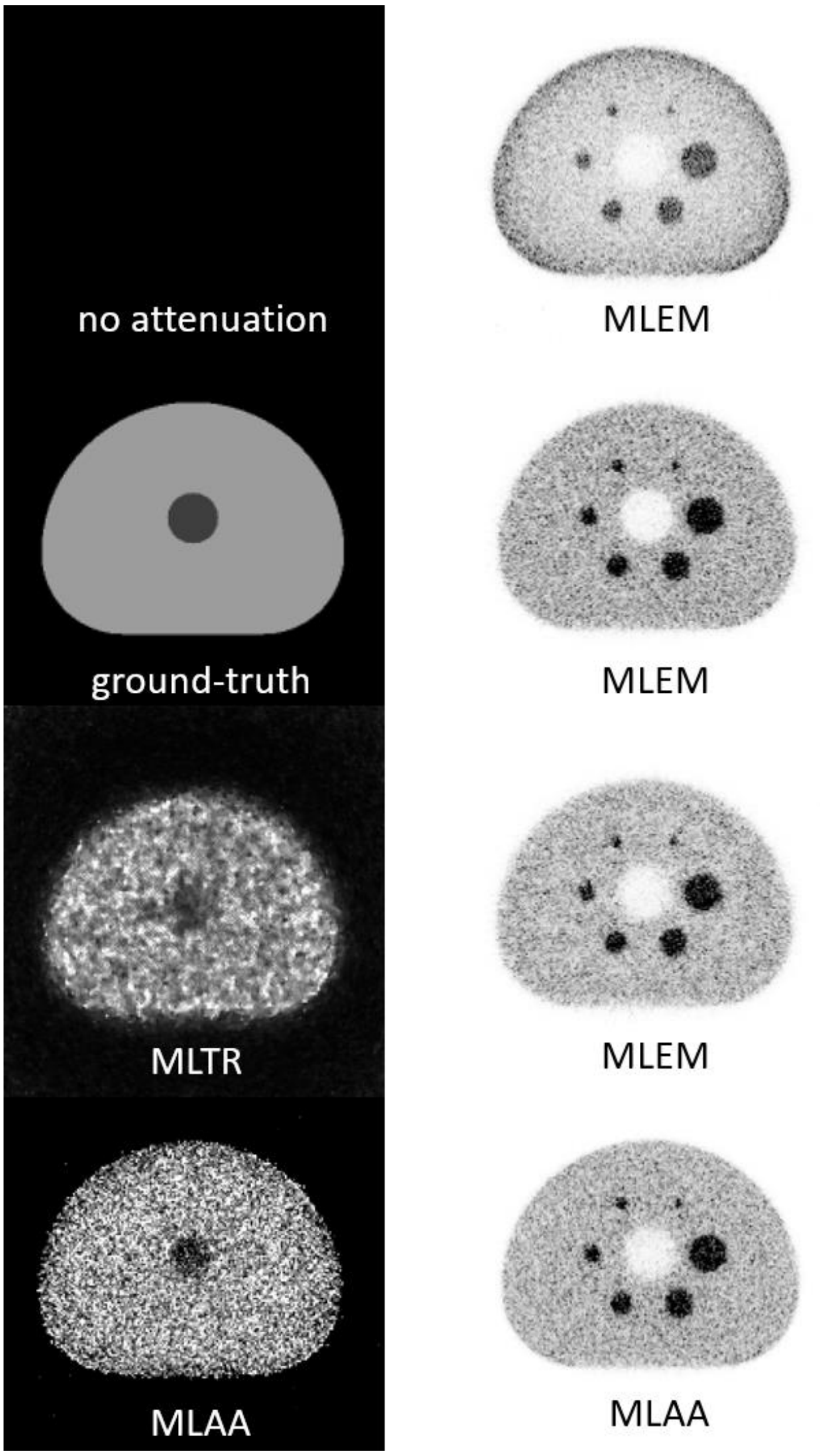

*Figure 4: Activity reconstructions based on different attenuation corrections in the* **rotating** *WT-TB-PET configuration.*

Finally, Table 1 shows the contrast recovery coefficients (CRCs) of the largest sphere in the IQ phantom for the different activity reconstructions shown in Figures 3 and 4. The rotating configuration shows better contrast recovery, with MLTR and MLAA showing slightly lower results in the fixed configuration.

| | ***no attenuation*** | ***ground-truth*** | ***MLTR*** | ***MLAA*** |
|---|---|---|---|---|
| ***fixed*** | *0.38* | *0.85* | *0.79* | *0.79* |
| ***rotating*** | *0.50* | *0.88* | *0.86* | *0.88* |

*Table 1: CRCs in different activity reconstructions for the largest sphere of the IQ phantom.*

## 4 Discussion

Although the rotating scanner configuration shows lower sensitivity due to the increased distance between both panels, overall image quality is improved by filling the missing angular gaps. While for the IQ phantom all hot-spots are still easily distinguished in the fixed configuration, this may no longer be true for more complex phantoms where the streaking artifacts could cause hot-spots to blend into each other. Note that from a practical perspective however, the rotating configuration poses additional challenges by increasing system complexity. In addition, the rotation may increase the uncertainty of the exact detector positions, which could for instance lead to a reduction of the system resolution.
The reconstructed activity images are however not regularized or post-processed, and previous studies have shown that deep learning could be used to greatly mitigate the limited angle problem in e.g. CT or PET [6] [7]. As such, it may become preferable to opt for the fixed scanner configuration in cases where sensitivity is most important.

Our results also show that reconstruction of the activity map is possible using the attenuation map generated by either MLTR or MLAA. In the fixed configuration, the transmission source does not provide a good estimate of the attenuation map, as there is no TOF information in the transmission data. However, the attenuation correction factors are still properly estimated since the emission LORs are along the same directions as the transmission LORs, resulting in a good activity reconstruction.
Note that in this study the GATE simulation of the transmission scan was kept separate from the emission scan, although in practice these could be done simultaneously in order to maintain patient throughput and limit patient movement between transmission and emission scan. Events could be assigned to the corresponding scan (transmission or emission) by TOF separation and knowledge of the transmission source positions.
MLAA on the other hand does have access to TOF information from the emission data and can therefore more accurately predict the outline of the attenuation map, although convergence can be more difficult. One major advantage of MLAA is also that no transmission source is required, reducing system complexity.

## 5 Conclusion

We compared image reconstructions for two configurations of the WT-PET system, namely a fixed and rotating configuration. While the fixed configuration shows higher sensitivity, streaking artefacts are observed due to the limited angle effect, but these are limited with 400 ps TOF. The rotating configuration on the other hand provides full angular sampling and generally better image quality.
The possibility of CT-less attenuation correction was also investigated for the system, showing its feasibility using either transmission sources or by estimating the attenuation from the emission data.

# Projection-based CBCT Motion Correction using Convolutional LSTMs

Joscha Maier[1], Timothy Herbst[1], Stefan Sawall[1], Marcel Arheit[2], Pascal Paysan[2], and Marc Kachelrieß[1]

[1]Division of X-ray imaging and CT, German Cancer Research Center (DKFZ), Heidelberg, Germany
[2]Varian Medical Systems Imaging Lab, GmbH, Baden-Daettwil, Switzerland

**Abstract** Due to the limited temporal resolution of CBCT scans, CBCT reconstructions often suffer from motion artifacts. To address this issue, existing approaches typically estimate a set of displacement vector fields in image domain that can be used to compensate for the present motion. However, since reconstruction artifacts may impair the performance of such approaches, we propose to overcome this limitation by operating in projection domain. To do so, we make use of a deep neural network which is trained to map the projections of a moving patient to projections of a static patient that is frozen in a fixed motion state. In this work, the mapping is trained using 4D CBCT simulations to account for respiratory motion. Subsequently, the network can be applied repeatedly to predict projections for different motion states of the respiratory cycle. In our experiments, these reconstructions show only minor motion artifacts and differ by less than 15 HU on average from an ideal ground truth.

## 1 Introduction

Cone-beam computed tomography (CBCT) has become increasingly important in various fields of medical imaging such as interventional radiology [1], orthopedics [2], dentistry [3], or image-guided radiation therapy [4]. However, the slow gantry rotation speed, which can be 60 s or even longer, limits the temporal resolution of CBCT scans. Therefore, anatomical regions affected by organ or patient motion often appear blurred or distorted in the corresponding CBCT reconstructions. Since these motion artifacts may severely impair the quality of the CBCT scan, motion compensation has become an active field of research.

Here, existing approaches usually distinguish between non-periodic motion such as involuntary muscle motion, twitching, or swallowing, and periodic motion such as respiratory or cardiac motion. The former is typically compensated by modeling the present motion during CT reconstruction, i.e. by incorporating some sort of motion-dependent transformation within the backprojection operation. Strategies to estimate this transformation include the use of fiducial markers [5], the use of consistency conditions in projection domain [6], 2D/3D registration of the acquired projection data to a motion-free prior volume [7, 8], as well the minimization of sharpness metrics that are sensitive to motion artifacts [9].

Applications dealing with periodic motion, on the other hand, usually rely on phase-correlated reconstructions, i.e. a time-resolved representation of the motion cycle. Here, each projection is typically assigned a motion phase based on a surrogate signal such as the displacement of a marker block on the patient's thorax (respiratory motion) or an ECG-signal (cardiac motion). In the most simple case, the projections are then sorted into bins of similar motion phase which can be reconstructed independently [10, 11]. However, since each bin is reconstructed from only a small subset of the acquired projection data, the corresponding volumes may show strong streak artifacts that degrade image quality. One option to address this issue is to use dedicated acquisition protocols, however, at the cost of longer scan times or higher patient dose [12]. Therefore, other approaches rather aim to make use of displacement vector fields (DVFs) that model inter-phase motion. Given these DVFs, a particular motion phase can be reconstructed from all available data by applying the corresponding deformation to any other phase-correlated reconstruction of the motion cycle. While first approaches estimated the DVFs from properly sampled prior CT scans [13, 14], later approaches proposed more sophisticated strategies to estimate them directly from the phase-correlated reconstructions [15–20].

However, despite recent advances, current approaches still fail if the phase-correlated reconstructions have highly irregular or sparse angular sampling due to, for example, irregular motion patterns. Other limitations are related to the need of phase-correlated reconstructions in general. Here, the phase-binning process can lead to a loss of temporal resolution as projections with slightly different motion states may be sorted into the same bin. Furthermore, the need for accurate motion surrogate signals may be time consuming as it requires additional patient preparation.

Therefore, we propose a novel deep-learning based approach that does not rely on phase-correlated reconstructions but operates directly in projection domain to overcome these limitations.

## 2 Materials and Methods

### 2.1 Projection-based Motion Correction

Considering CT imaging, a moving patient can be described by its time-dependent distribution of the attenuation coefficient $f(\boldsymbol{r},t)$. During a CBCT scan this distribution is sampled at a discrete number of time points $t \in \{t_1,\dots,t_N\}$ by acquiring x-ray projections at view angles $\theta \in \{\theta(t_1),\cdots,\theta(t_N)\}$:

$$q(\theta,u,v) = \mathsf{X}_t f(\boldsymbol{r},t), \tag{1}$$

where $u$ and $v$ denote the detector coordinates and $\mathsf{X}_t$ denotes the time-dependent X-ray transform. Given $q(\theta,u,v)$, motion correction aims to reconstruct a 3D representation of a certain motion state $f_\tau(\boldsymbol{r}) \equiv f(\boldsymbol{r},t)|_{t=\tau}$ or a sequence $S$ of successive motion states $\{f_\tau(\boldsymbol{r})\}_{\tau\in S}$. While existing approaches operate in image domain to obtain $f_\tau(\boldsymbol{r})$, we propose to predict projection images

$$p_\tau(\theta,u,v) = \mathsf{X} f_\tau(\boldsymbol{r}) \tag{2}$$

instead. To do so, we use a deep learning-based approach that is trained to learn the following mapping $M$:

$$M : q(\theta,u,v)\cdot w(\theta,u,v,c) \to p_\tau(\theta,u,v), \tag{3}$$

where $w(\theta,u,v,c)$ represents a weighting factor that encodes additional depth information according to section 2.2, and $\tau$ corresponds to the motion state of the first projection of $q$. As illustrated in figure 1, $M$ is approximated by a U-net-like neural network in which each stage is composed of a set of three 2D convolutional feature extraction layers, a layer normalization [21], and a convolutional long short-term memory (LSTM) unit [22] to learn long-term dependencies along the $\theta$-axis. Once the mapping $M$ has been learned, an estimate of $f_\tau(\boldsymbol{r})$ can be derived as:

$$\tilde{f}_\tau(\boldsymbol{r}) = \mathsf{X}^{-1} M(q\cdot w), \tag{4}$$

where $\mathsf{X}^{-1}$ is the inverse x-ray transform operator, e.g. FDK in our case.

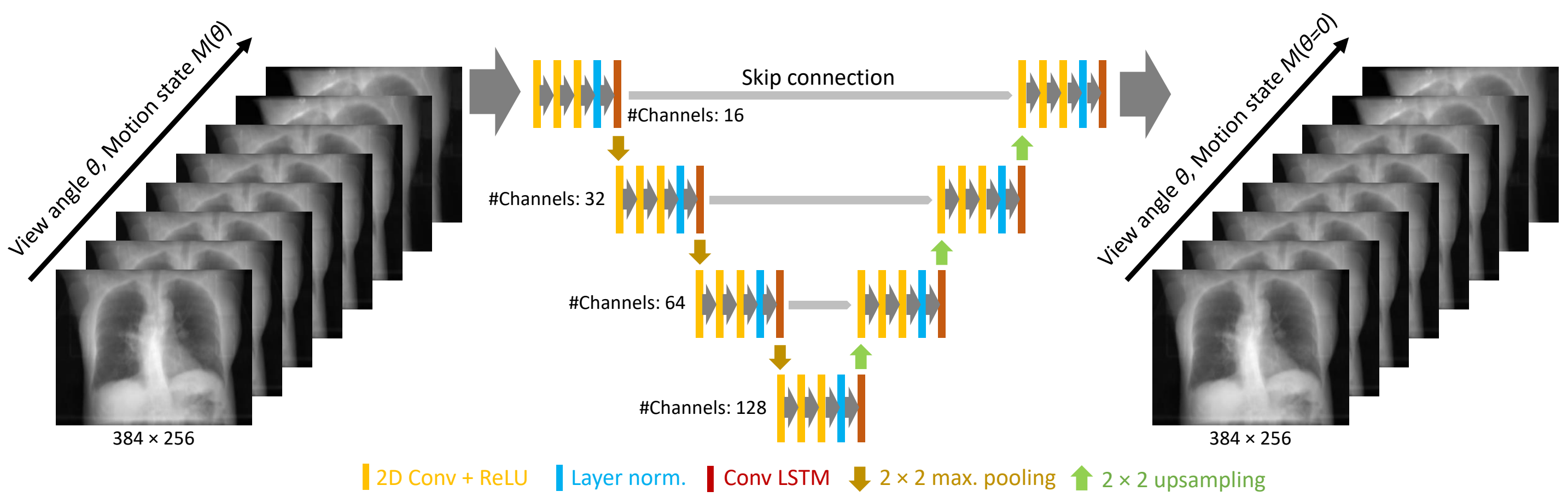

**Figure 1:** Illustration of the proposed approach. A U-net-like architecture that includes convolutional LSTMs is trained to map CBCT projections of a moving patient to CBCT projections in which the patient is frozen in the first motion state.

## 2.2 Depth Encoding

Since a 3D motion in image domain is supposed to be compensated by a neural network operating in projection domain, the performance can be increased by encoding additional depth information. Here, this information is provided in terms of a distribution function $w(\theta,u,v,c)$ that represents the interaction probability on $N_c$ successive line segments between the x-ray source $\boldsymbol{s}(\theta)$ and the detector element at $\boldsymbol{d}(\theta,u,v)$. Using an initial motion-blurred reconstruction

$$\bar{f}(\boldsymbol{r}) = \mathsf{X}^{-1} q(\theta,u,v) \tag{5}$$

as an approximation of $f(\boldsymbol{r},t)$, $w(\theta,u,v,c)$ can be calculated as:

$$w(\theta,u,v,c) = \frac{\int_0^1 \bar{f}(\boldsymbol{s}(\theta) + \frac{1}{N_c}\cdot(c+l)\cdot(\boldsymbol{d}-\boldsymbol{s})(\theta,u,v))\,dl}{\int_0^1 \bar{f}(\boldsymbol{s}(\theta) + l\cdot(\boldsymbol{d}-\boldsymbol{s})(\theta,u,v))\,dl}, \tag{6}$$

with $c \in \{0,1,\cdots,N_c-1\}$.

## 2.3 Data Generation

Training the proposed approach in a supervised manner requires paired training data, i.e. $q$'s and the corresponding $p_\tau$'s. To obtain these data, we rely on CBCT simulations that are based on 55 respiratory-gated 4D CT scans. Each of these scans consists of 10 evenly distributed respiratory phases $\{f_1(\boldsymbol{r}),\ldots,f_{10}(\boldsymbol{r})\}$ from which arbitrary motion states are generated as follows:

$$f(\boldsymbol{r},t) = \sum_{\tau=1}^{10} \Pi(a(t)-\tau+0.5)\cdot f_\tau(\boldsymbol{r}+b(t)\cdot u_\tau(\boldsymbol{r})), \tag{7}$$

where $\Pi(t)$ is the rectangular function, $a:\mathbb{R}\to[0,10]$ and $b:\mathbb{R}\to[0,1]$ are functions to define custom motion patterns, and $u_\tau(\boldsymbol{r})$ is a DVF that is calculated in advance such that $f_\tau(\boldsymbol{r}+u_\tau(\boldsymbol{r})) = f_{\tau+1}(\boldsymbol{r})$.
Projections $q$ of moving patients are then simulated according to equation (1) by inserting $f(\boldsymbol{r},t)$ as defined above. Similarly, the corresponding labels are simulated according to equation (2) by freezing the patient is a certain motion phase $\tau$. For our purpose, we simulate 5 CBCT scans per patient with a scan time of 60 s and a cyclic motion pattern with a random period between 1.5 s and 5 s. Each scan consists of 360 projections (1° angular sampling) that are acquired with a centered 384×256 pixel detector with 2×2 mm pixel spacing.

## 2.4 Training

Training data to learn the mapping given in equation (3) are generated as described in section 2.3. Here, the optimization is based on simulations of 50 patients (45 for training, 5 for validation), while the remaining 5 patients are reserved for testing. Using these data, the network is trained in a supervised manner by minimizing the mean squared error between the prediction and corresponding ground truth. Due to memory and time considerations, the gradient calculation is not performed for the entire sequence of 360 projections but each scan is divided into chunks of 16 projections which are processed one after the other. Therefore, the LSTM cells are operated in stateful mode, such that the last cell state of each chunk is used as initial state for the succeeding chunk. In that way the network was trained for 300 epochs on a NVIDIA GeForce RTX 3090 using an Adam optimizer and a learning rate of $5 \cdot 10^{-5}$. Finally, the network configuration with the lowest validation loss was used for testing.

## 3 Results

The performance of the proposed approach was evaluated for projection images as well as for the corresponding reconstructions of an independent testing data set (see section 2.4). Exemplary projections corresponding to a single motion cycle are shown in figure 2. Compared to the ground truth, which corresponds to a CBCT scan of the patient frozen in the motion state of the first projection, the input projections show strong deviations, especially in the area of the diaphragm. Applying the proposed approach clearly reduces these deviations not only around the diaphragm but also in the area of other moving structures such as the heart or the ribs. A quantitative assessment of the similarity with respect the to ground truth for all available test projections yields a mean absolute percent error of 1.9 % without correction, whereas the error with our approach is reduced to 0.7 %.
A similar evaluation was performed for reconstructions representing all motion states of the motion cycle. Since the proposed approach is trained to map all projections to the motion state of the first projection, this can be achieved by applying equation (4) $N$ times, each time with the $n^{\text{th}} \in [1, \cdots, N]$ projection being the first in the sequence. The corresponding results of two test patients in end-exhale and end-inhale are shown in figure 3. In any case the proposed approach is able to provide time-resolved reconstructions with a very reduced amount of motion artifacts. Similar trends can be observed by evaluating the error of the CT values for the entire testing data set. While reconstructions of all projections without further postprocessing show error of about 32 HU, the proposed approach reduces this error to 15 HU on average.

## 4 Discussion and Conclusions

Image-based motion compensation requires the underlying phase-correlated reconstructions to have a certain image quality. However, in case of a highly irregular or sparse angular sampling, this image quality may not be achieved, leading to poor results of the motion compensation algorithm. Here, we circumvent this problem by accounting for motion directly in projection domain. In this way the result is not impaired by reconstruction artifacts and does not need additional motion surrogate signals, since the motion state can be extracted directly from the projection image. Our initial experiments demonstrate a convincing performance in reducing motion artifacts and suggest that convolutional LSTMs are a promising architecture for such an approach. However, we realize that motion compensation is actually a 3D problem. Therefore, further research will focus on extending the current implementation to estimate 3D DVFs from 2D projection images. Finally, it has to be noted that the general concept of the proposed approach is not restricted to periodic motion patterns but can be applied in the same way to account for non-periodic motion.

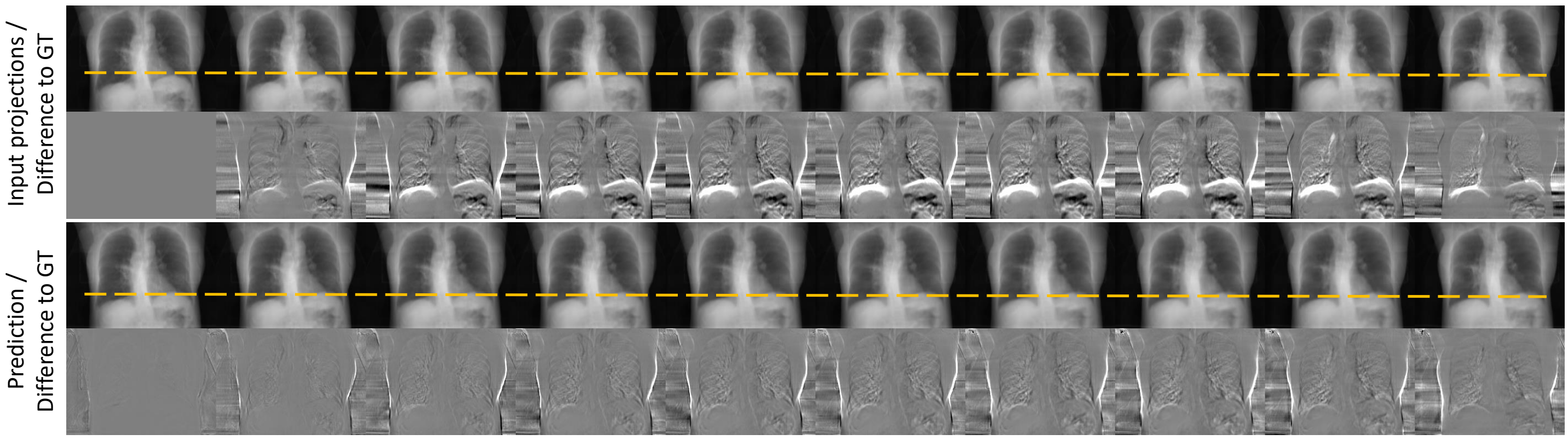

**Figure 2:** Top: sequence of consecutive projections ($\theta = 0° - 10°$) of a moving patient that are used as input to the network ($C$ = 3, $W$ = 6 HU) as well as the relative difference to the ground truth (GT) that keeps the patient frozen in the first motion phase ($C$ = 0 %, $W$ = 40 %). Bottom: prediction of the network.

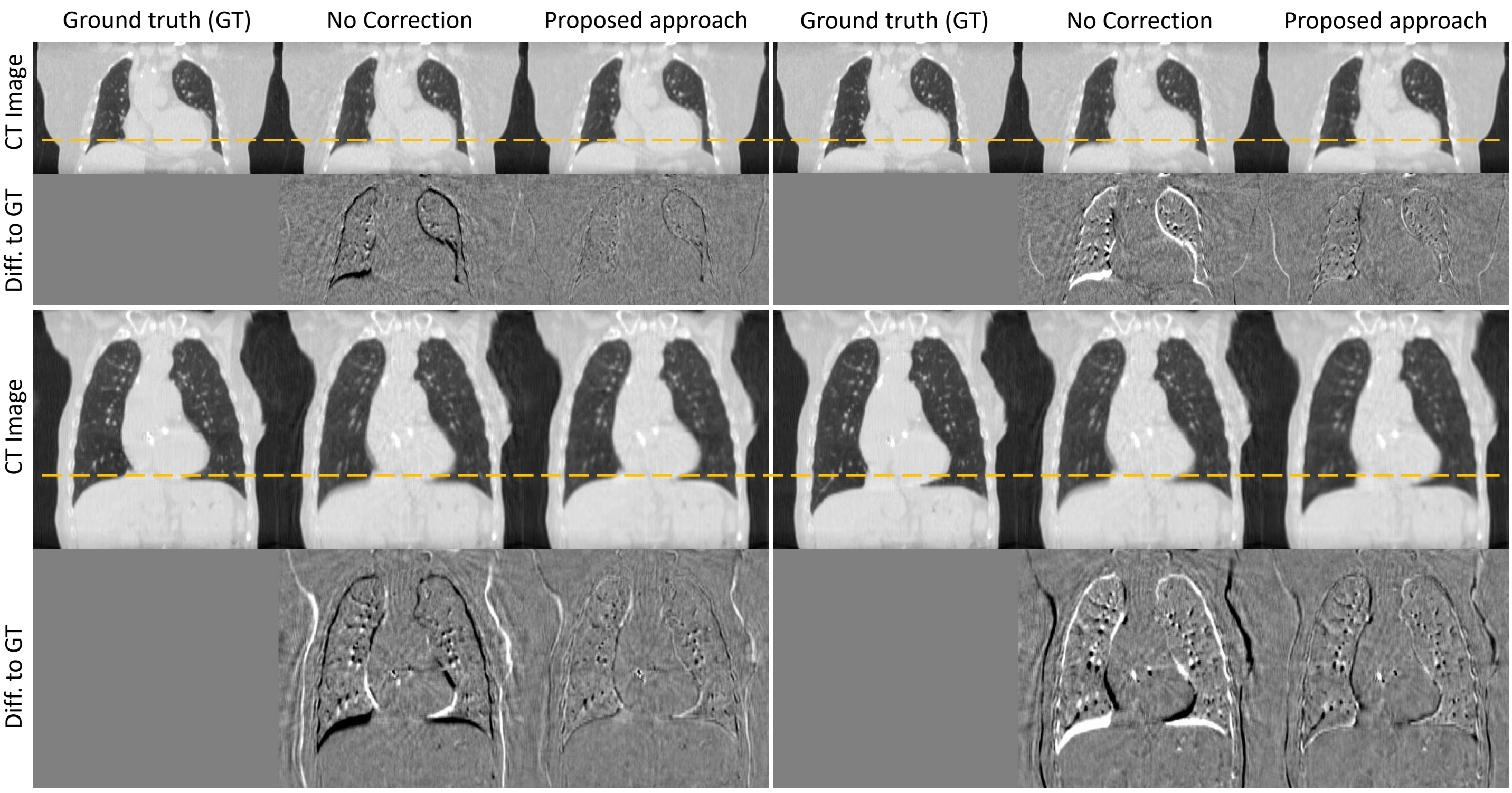

**Figure 3:** Reconstruction of two test patients (top and bottom), in two motion states corresponding to end exhale (left) and end inhale (right). The upper row ($C$ = 0 HU, $W$ = 1500 HU) shows the corresponding reconstructions with and without correction while the lower row shows the difference with respect to the ground truth ($C$ = 0 HU, $W$ = 500 HU).

# A General Solution for CT Metal Artifacts Reduction: Deep Learning, Normalized MAR, or Combined?

Yanfei Mao[1#], Mark Selles[2,3#], Michael Westmore[1], Joemini Poudel[1], Martijn F. Boomsma[2]

[1]Philips Healthcare, Cleveland, USA
[2]Department of Radiology, Isala Hospital, Zwolle, The Netherlands
[3]Department of Radiology & Nuclear Medicine, Amsterdam University Medical Centre, Amsterdam, The Netherlands
[#]Both authors contributed equally to this work

**Abstract** Many approaches have been proposed recently for reduction of metal artifacts in CT images, especially the deep learning-based metal artifact reduction (MAR) method. In this work, we evaluated three metal artifact reduction methods, namely an image-based deep learning method, a hybrid method combining deep learning and normalized MAR, as well as the traditional linear interpolation-based normalized MAR method. Each method was evaluated using clinical images with commonly encountered metal implants, including joint prosthesis, dental fillings, spine screw, and pacemaker. Both the image-based deep learning method and the hybrid method show great potential to provide a general solution to different types of metal implants. Strength and limitation of each method were discussed as well. To our knowledge, this is the first paper that comparing these three methods side by side on various metal implants.

## 1 Introduction

Reducing metal artifacts has been a challenging problem in computed tomography (CT) for decades. Numerous metal artifact reduction (MAR) techniques have been proposed, which can be classified into three main categories: physics-based correction, projection-completion, and model-based correction methods. Inspired by the great success of deep learning in image processing, the deep learning-based methods or a combination of deep learning and traditional MAR methods attract more and more attention and provide new opportunities to mitigate metal artifacts.

A popular class of deep learning-based MAR methods are image domain CNNs [1-3]. Among the traditional MAR methods, the normalized metal artifact reduction (NMAR) method [4] is considered a state-of-the-art method that has an open framework to combine with deep learning-based MAR [5-8]. Although studies of these deep learning or hybrid methods have been published, they typically only evaluate one or two clinical cases for demonstration. However, the biggest challenge for MAR is its generalization capability for different types of metal implants and clinical scenarios. Therefore, an extensive evaluation of various metal implants that are highly clinically relevant is necessary and valuable.

This study compares the performance of an image-domain deep learning-based MAR, a hybrid MAR method, and the NMAR method using four frequently encountered types of metal implants, including joint prosthesis, dental filling, spine screw, and pacemaker. In the rest of this paper, we will briefly describe the three methods in Section 2. The comparison of the results are included in Section 3. We discuss the strengths and limitations of each method for different clinical applications in Section 4 and conclude the paper in Section 5.

## 2 Materials and Methods

### 2.1 *Image-Domain Deep Learning-based MAR*

To evaluate the image-domain deep learning MAR method, a deep residual U-NET (ResUNet) architecture was developed for training. CT-images with and without simulated metal artifacts were generated using a method based on the work of Zhang, Yanbo, et al. [7], which were then used to train the network. Each 512×512 metal-corrupted image was scaled to 0-1 range using soft tissue window (L:40, W:350), bone window (L:400, W:1800) and the full dynamic window setting, respectively, which were subsequently combined to form a three-channel image. Patches of 256×256×3 were extracted from the three-channel image and used as the input to the network. The output is the artifact-free image. The network was trained for 25 epochs with a combined loss function of SSIM (weight: 0.03), L1 (weight 0.21) and L2 (weight: 0.76). For convenience, we refer to this method as AI in the rest of this paper.

### 2.2 *A Hybrid of Deep Learning-based MAR and NMAR*

The hybrid method used in this paper is similar to that proposed by [7]. In this hybrid method, the uncorrected image was first corrected by the deep learning-based MAR method described in Section 2.1. The initial corrected image was then used to generate the prior image for NMAR. In the rest of this paper, we refer to this method as AI+NMAR.

### 2.3 *NMAR*

In this paper, we used linear interpolation as the initial correction for the traditional NMAR [4], which is referred as to LI+NMAR in the rest of this paper.

Dental filling, joint prosthesis, spine screw, and pacemaker are four metal implants that are commonly observed in clinical CT images. To evaluate the performance of the above three methods, patient data with these four major metal implants were collected. Each metal category included three data sets with artifacts of varying severity. To test the reliability of the three methods, data acquired from two different systems were used for test.

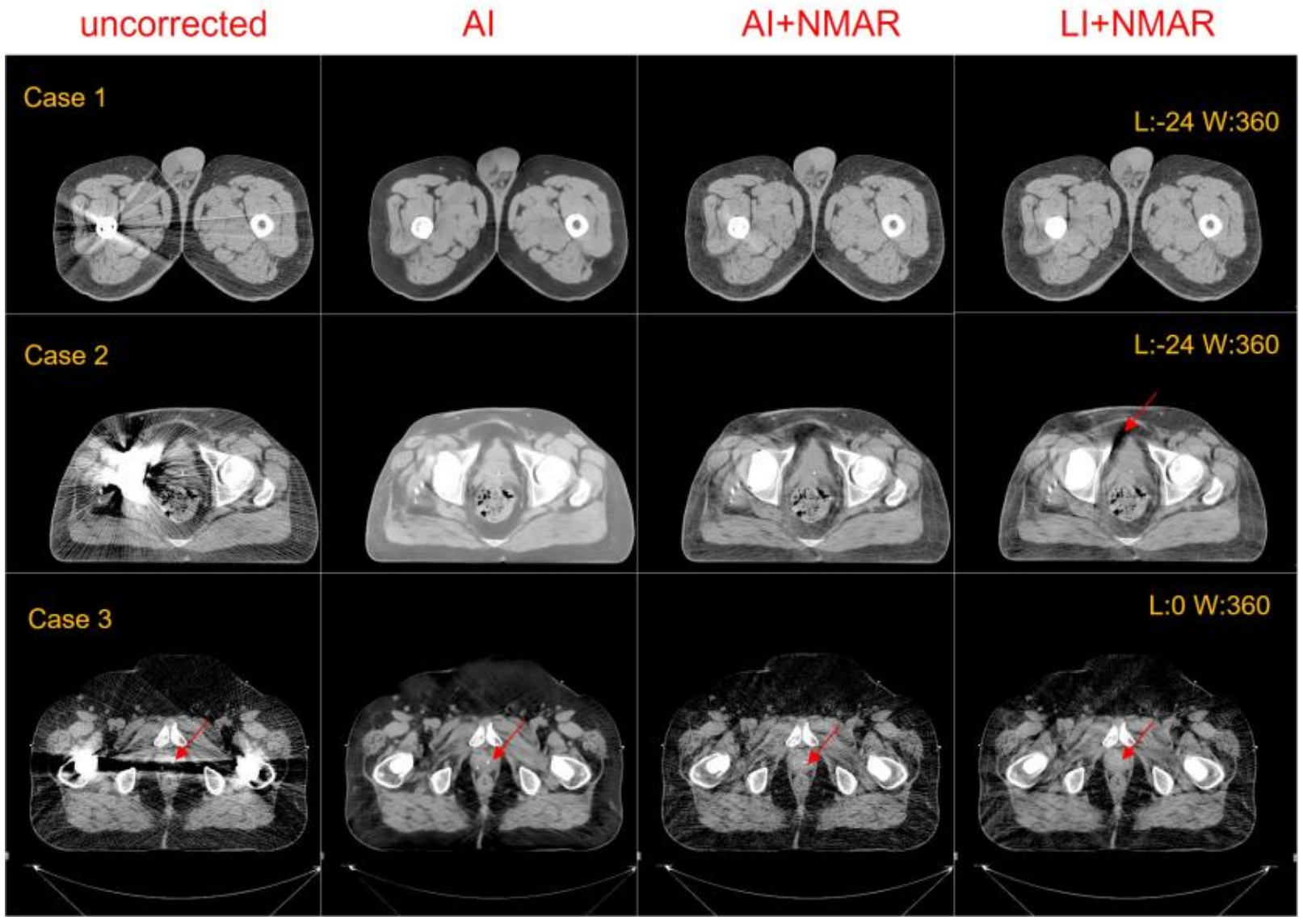

Figure 1: The hip implant images. From left to right: the original uncorrected image, the corrected results by AI, AI+NMAR, and LI+NMAR.

## 3 Results

### 3.1 *Joint Prosthesis*

Figure 1 shows the correction results of hip implant images. All three methods provided very promising results. The streak artifacts are completely suppressed and the soft tissue under dark bands are recovered. Slight under-correction is observed in LI+NMAR results of case 2 (red arrow), while resolution loss is observed in all AI results. In addition, the AI result of case 2 also shows a CT number shift, which is not seen in case 1 and 3. The difference between case 1, 2, 3 is that case 2 was collected from scanner model 1 while the other two cases were acquired from scanner model 2. In case 3, radiation beads existing in the uncorrected image (red arrows) are almost removed in the LI+NMAR image.

### 3.2 *Dental Filling*

The correction results of dental artifacts are presented in Figure 2. Compared with AI and AI+NMAR methods, the LI+NMAR method shows a big limitation on dental artifacts. Strong streak artifacts and bone shape distortion were introduced to the images after LI+NMAR correction. These artifacts are much less in the AI and AI+NMAR images. When the dental artifacts are not very strong, the results of AI and AI+NMAR are comparable, except slight resolution loss in the AI images. When the dental artifacts get more severe, more dark streaks were left over in the AI results. These dark streaks are further suppressed in the AI+NMAR results.

### 3.3 *Spine Screw*

The spine screw results are shown in Figure 3. Due to the linear interpolation operation, the bone structure between the two spine screws were completely removed in the LI+NMAR images, while both AI and AI+NMAR images showed good preservation of bone structure. Compared with the AI method, the soft tissue near spine screw (red arrows) was smeared in the AI+NMAR images.

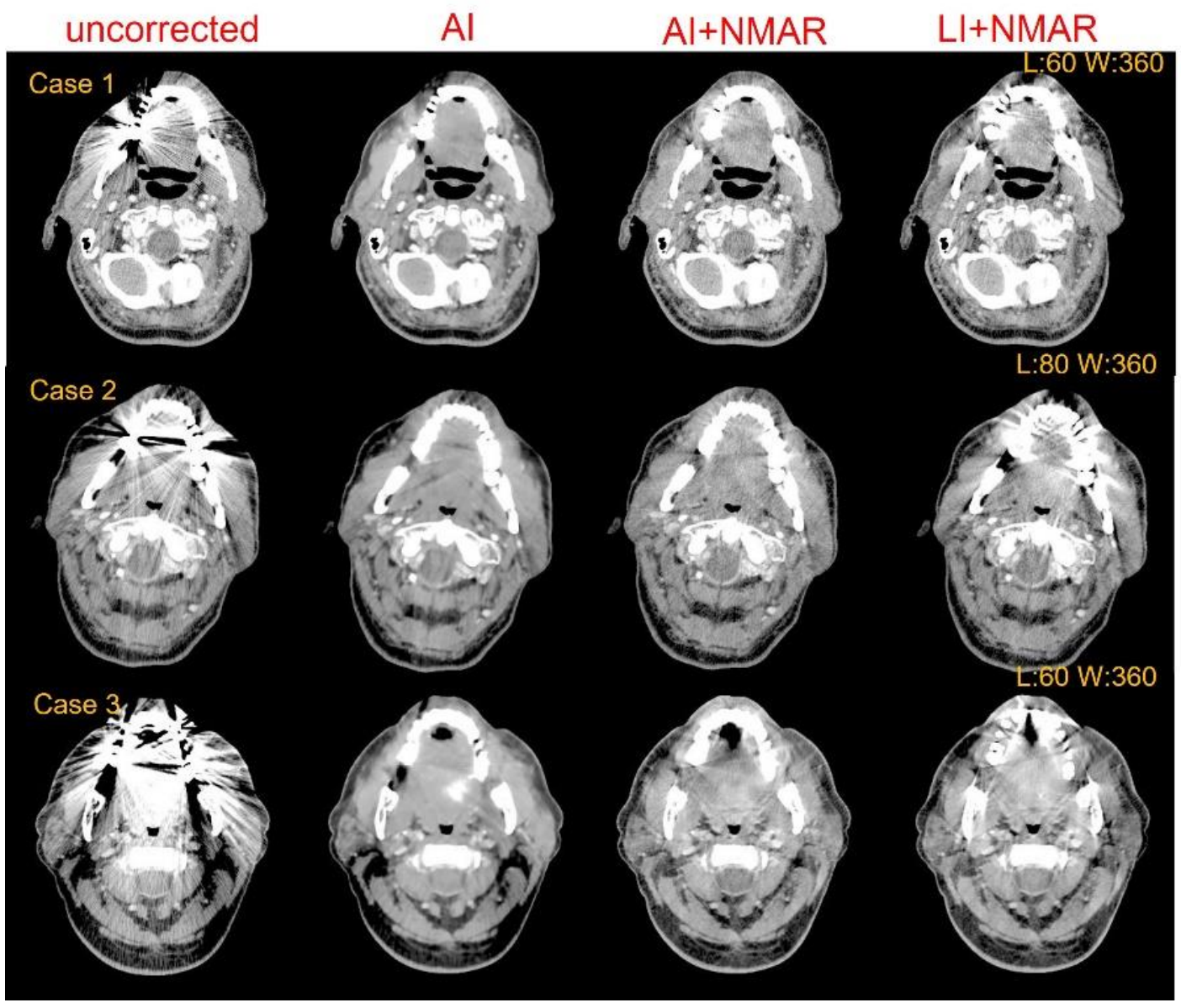

Figure 2: The head CT images with dental filling. From left to right: the original uncorrected images, the corrected results by AI, AI+NMAR, and LI+NMAR.

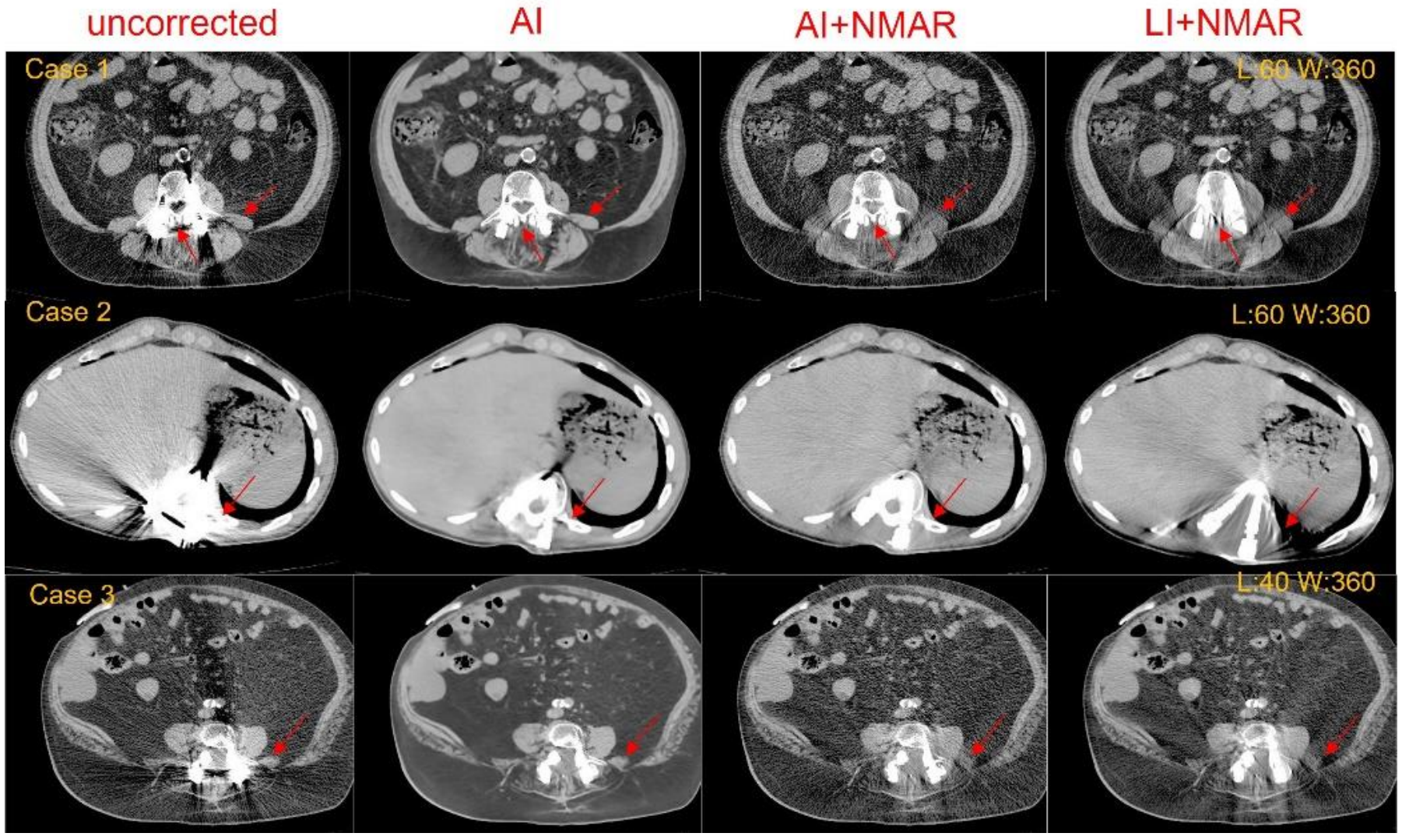

Figure 3: The abdomen CT images with spine screws. From left to right: the original uncorrected images, the corrected results by AI, AI+NMAR, and LI+NMAR.

3.4 *Pacemaker*

The artifacts from pacemaker are mild, so any introduced artifacts will be more visible in the pacemaker case, making the correction more challenging. As seen in Figure 4, the AI method works well at removing artifacts without introducing any additional streaks. In contrast, the LI+NMAR results show some additional dark streaks introduced, which greatly degrades the overall image quality. Slight resolution loss and CT number shift were also observed in the AI results of pacemaker cases. These problems were removed in the AI+NMAR images.

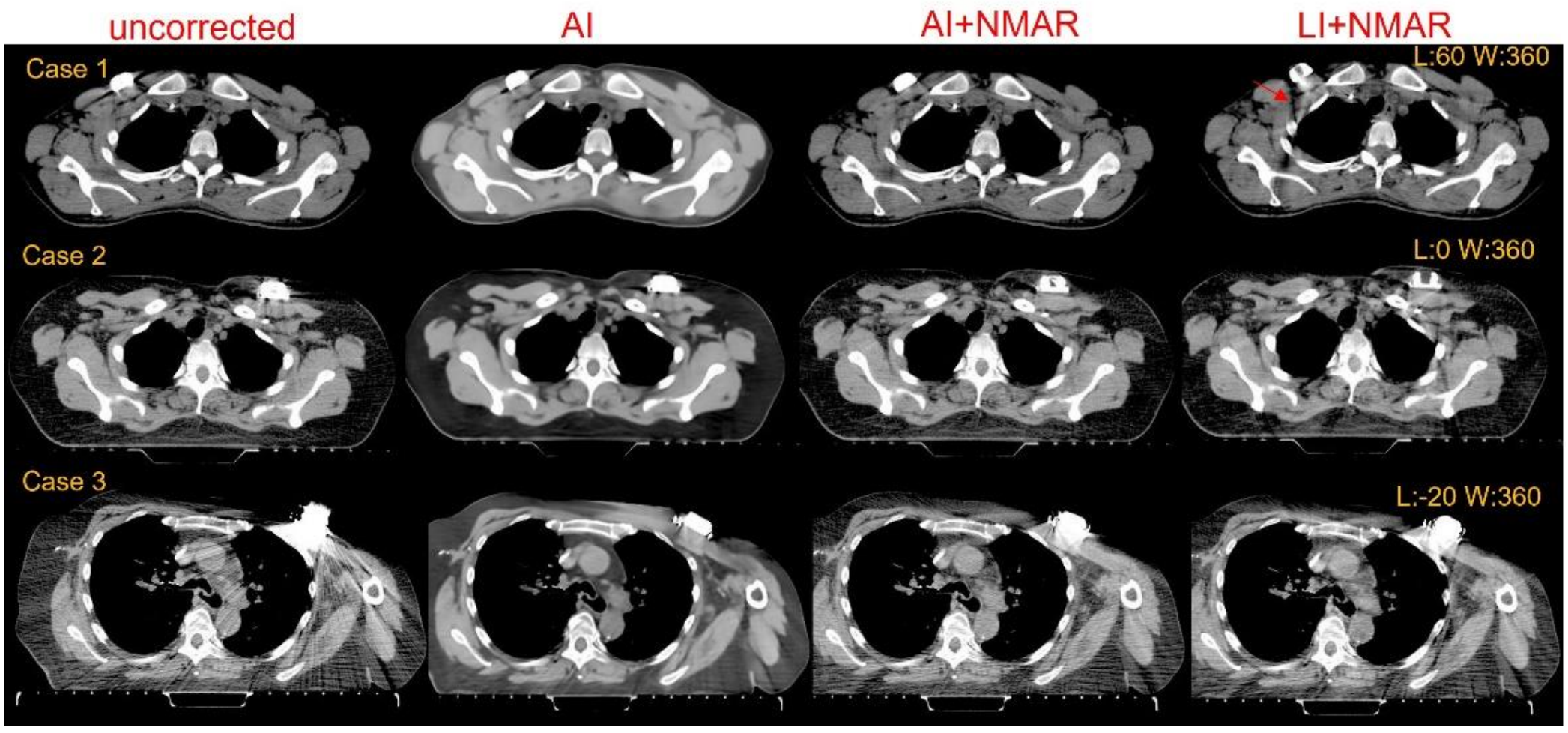

Figure 4: The abdomen CT images with pacemakers. From left to right: the original uncorrected images, the corrected results by AI, AI+NMAR, and LI+NMAR.

## 4 Discussion

The LI+NMAR method worked very well in hip implant cases, because the background doesn't have many bone structures or air pockets. The slight under-correction in case 2 of hip implant is due to some bone structure is misclassified as soft tissue. When the background is uniform, linear interpolation provides a good estimation. In contrast, in the dental cases, metals were surrounded by many bone structures, so the LI+NMAR results were not as good as those from the other two methods.

The AI method showed big advantages at preserving soft tissue details and bone structure, without requiring any parameter selection. Resolution loss was observed in all AI corrected images, which is not acceptable for diagnostic CT images. In this study, CT number shift was observed when the AI model was applied to a different scanner model from that used for training.

Training data is the bottleneck of the supervised learning approach. It requires structurally matched image pairs with and without metal artifacts, which are impossible to get in real situations. Therefore, the training data are usually generated through simulation. In clinical situations, the metal material, shape, size, and location can vary a lot. Therefore, a huge training database is required to cover the most clinical scenarios. On the other hand, due to the complexity of metal artifacts and variation of CT systems, it is hard to mimic the real metal artifacts. Most metal artifact simulators only simulate metal beam hardening effect and noise, but other physics effects, such as scatter and non-linear partial volume effect, may also play an important role [9]. It may degrade the performance of AI if the simulated metal artifacts do not fully reflect the real situation. Take dental artifacts as an example, the uncorrected image is impacted by metal beam hardening, non-linear partial volume effect, and many other physics effects. In Figure 5, AI model 1 and 2 were trained using the same network, but different training data. In addition to the metal beam hardening effect simulated in the images for model 1, the training data for model 2 also include non-linear partial volume effect. With simulated non-linear partial volume effect, the result of model 2 is much better.

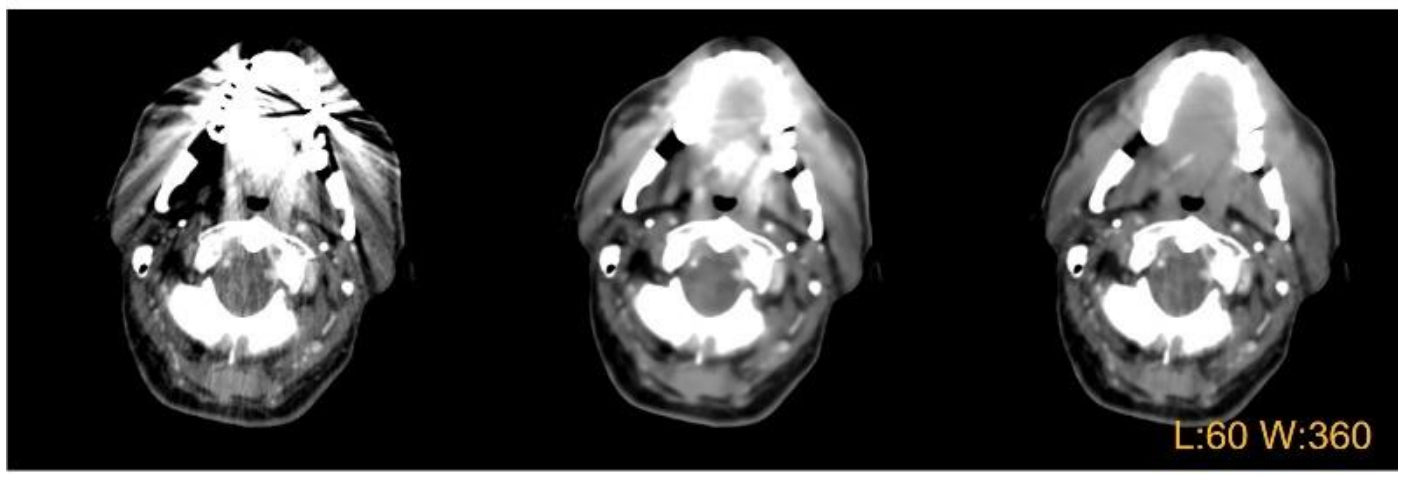

Figure 5: Impact of metal artifacts simulation on AI results. AI model 1: the metal artifacts are simulated without partial volume effects. AI model 2: the metal artifacts are simulated with partial volume effects. All other settings are the same.

With the AI image as a prior, the AI+NMAR method shows noticeable improvement over the LI+NMAR method. When AI results are under corrected, the AI+NMAR method can further reduce the metal artifacts. Unlike the AI method, the performance of AI+NMAR is more reliable when the system or scan parameters change. In some cases, such as spine screw case, new artifacts are introduced in the AI+NMAR image. This is caused by the soft tissue segmentation step in the NMAR framework, which replaces the soft tissue region with an average attenuation value of soft tissue to generate the prior image [4]. For soft tissue near spine screw, most rays are corrupted by metal, so it more relies on the information provided by the prior image. If the detail information is removed in the prior image, it will limit the recovery of details near spine screw. Without soft tissue segmentation, the soft tissue around the spine was recovered and less streak artifacts were introduced (Figure 6). To take the full benefit of AI prior, a more sophisticated classification should be used instead of the two thresholds-based classification.

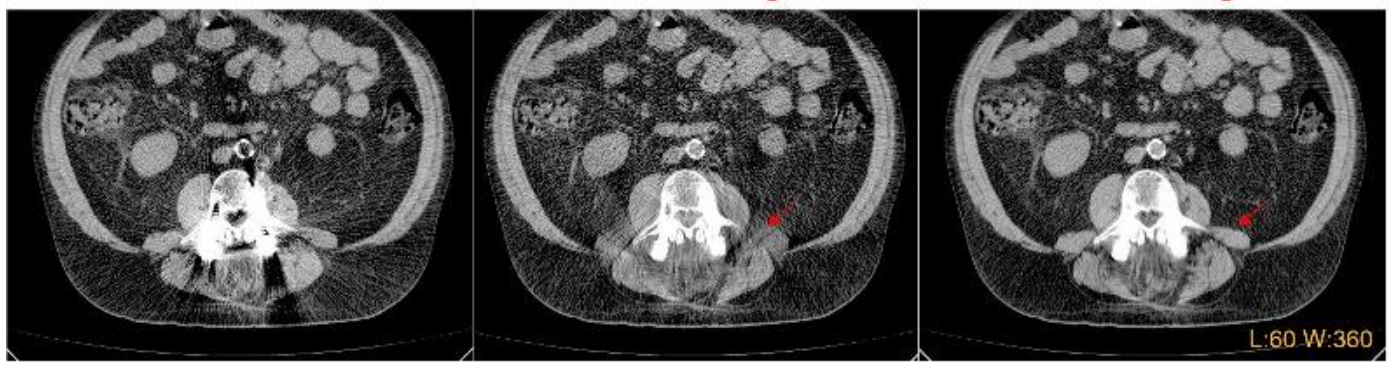

Figure 6: Impact of the segmentation of soft tissue step. Without the segmentation of soft tissue (right), more details of soft tissue were recovered, compared to with soft tissue segmentation (middle).

## 5 Conclusion

Both the AI and AI+NMAR methods outperformed the LI+NMAR method in all four metal categories. Compared with the AI+NMAR method, the AI method doesn't require parameter tuning and has a great potential to provide a general solution. The computational cost of the AI approach is much less than the AI+NMAR method. For interventional CT, the AI based MAR may be the only option that can meet the reconstruction time requirements. However, the results of AI highly rely on the acurracy of metal artifacts simulation. Due to loss of spatial resolution and potential CT number shift, image quality of the AI method still does not fully meet the requirement of diagnostic CT.

With AI results as prior, the results of AI+NMAR are promising and have less reliance on the scan parameters. The drawback of the AI+NMAR method is that new artifacts may be introduced when there is an inconsistancy in the sinogram. Adjustments to the NMAR framework is needed to take full benefit of AI.

Overall, both the AI and AI+NMAR methods have a great potental to provide a general solution for metal artifact reduction. Technical developments as well as thorough clinical evaluation are needed to allow safe and effective application of the AI in clinical practice. At current stage, the AI+NMAR appears to achieve the best overall performance.

# Fitting a 2D mesh to X-ray measurements

Jannes Merckx[1], Jan Sijbers[1], and Jan De Beenhouwer[1]

[1]imec-Visionlab, Department of Physics, University of Antwerp, Antwerp, Belgium

**Abstract** A triangular mesh is a collection of triangles connected by mutual edges that can be used to represent surfaces and interfaces between different composites of an object. This representation is potentially more efficient than a voxel grid. Furthermore, a mesh-based representation can greatly reduce partial volume effects. Manufactured objects, produced based on a CAD model, do not match this model perfectly due to production mistakes. This paper introduces two algorithms for fitting meshes to X-ray measurements, thus minimising partial volume effects. Even though the eventual goal is the extension to 3D, a 2D proof-of-concept is presented here, where vertices are placed on interfaces by using their corresponding mesh triangles as pixels for a polyethylene triangle. Possible applications of this are greatly reducing partial volume effects and adapting surface meshes to X-ray projections.

## 1 Introduction

In X-ray computed tomography (CT), a 3D image of the internal structure of an object is generated by acquiring a large number of radiographs under a variety of projection angles and applying a reconstruction algorithm on this data. Applications include medical imaging, quality control and material research [1]. Conventional algorithms distribute the scanned region into cubic voxels, each of which is attributed an attenuation value. However, dividing space in a voxel grid results in a large system of equations, meaning a large number of projection angles are required to solve it. Furthermore, when a voxel contains regions of different attenuation, partial volume effects arise [2][3][4]. The correct attenuation value can not be filled in simply because the grid does not match the structure of the sample. This problem can be partly circumvented by using octrees (in 3D), or quadtrees (in 2D), in which larger cubes in homogeneous areas reduce the number of equations that need to be solved, and smaller cubes around interfaces reduce partial volume effects [5]. Only limited research on this has been conducted in X-ray CT [6].

In conventional non-destructive testing, a Computer-aided design (CAD) model of a manufactured object is fitted to its voxel reconstruction. This requires a long processing pipeline including segmentation and surface extraction [7][8]. Recently, it has been shown that X-ray based quality control is possible in projection space by directly comparing a limited number of projections to corresponding CAD projections [9][10]. These CAD projections are accelerated X-ray simulations computed with a mesh projector. This projector requires triangular surface meshes as input and is a crucial tool in the development of X-ray-based inspection and quality control methods [11]. However, this CAD projector can not adapt its surface meshes when the material does not match the CAD model. Therefore, production mistakes and small deviations still result in incorrect simulations. Although some research has been done on adapting surface meshes to X-ray measurements [12], the main limitation of these differentiable surface meshes are their inability to adapt to unexpected smaller structures: only objects in the topology of the original mesh can be represented.

An alternative is using a volume mesh where the mesh elements (tetrahedra) are used as reconstruction elements. Like octrees, these volume elements can then be updated to match the exact structure by relocating vertices from homogeneous regions to interfaces. The advantage of volume meshes over octrees is that the tetrahedra can be oriented in more different ways. Fitting of the volume mesh to X-ray measurements can greatly reduce partial volume effects for the same memory usage. The final goal is to get reliable quality control and metrology from adaptable volume meshes starting from a volume mesh approximately matching the real object.

In this paper, a 2D adaptive volume mesh was studied as a testcase. This means triangles rather than tetrahedra were used as volume elements. Besides this, the adaptive mesh started from a regular mesh, rather than a starting condition close to the phantom. Therefore the universal potential of these adaptable surface meshes could be shown. The phantom studied was a polyethylene triangle, whose corners pose a challenge for adaptation. Two mesh refinement strategies were tested in this paper.

The first strategy replaces vertices from homogeneous regions to interface regions, changing the topology of the mesh. The second strategy moves these vertices without changing the topology, so they move onto the interface. The robustness of these methods under varying numbers of projection angles and noise was tested.

## 2 Materials and Methods

A total of $m$ triangles of a 2D mesh are used as pixels in a reconstruction. Input data for this reconstruction are intensity measurements on $p$ detector pixels under $n$ projection angles. The distance covered by incoming rays in each detector measurement through all the triangle pixels is stored in a matrix $\mathbf{A} \in \mathbb{R}^{(np)\times m}$. The attenuation value given to the triangles is stored in a vector $\boldsymbol{\mu} \in \mathbb{R}^m$. The measured intensity on different detector pixels, under different angles, is stored in a vector $\mathbf{b} \in \mathbb{R}^{np}$. After measuring the sinogram data $\mathbf{b}$ ($b_i = -\log\left(\frac{I_i}{I_0}\right)$), a system of equations is generated that can be solved to the attenuation vector $\boldsymbol{\mu}$: $\mathbf{A}\boldsymbol{\mu} = \mathbf{b}$ with an algebraic iterative reconstruction method:

$$\boldsymbol{\mu}^k \leftarrow \boldsymbol{\mu}^{k-1} + \lambda_{k-1}\frac{b_i - <\mathbf{a_i}, \boldsymbol{\mu}^{k-1}>}{|\mathbf{a_i}|^2}\mathbf{a_i}^T. \quad (1)$$

$\boldsymbol{\mu}^k$ refers to the solution $\boldsymbol{\mu}$ found in the $k$'th iteration. The step size is given by a constant $\lambda_{k-1}$. The vector $\mathbf{a_i}$ refers to the $i$'th row of $\mathbf{A}$ [13]. Even though there is not as much software available for it, solving to $\boldsymbol{\mu}$ should be equivalent to solving a reconstruction on square pixels. This solution $\boldsymbol{\mu}$ is the input needed for two mesh refinement strategies. The refinement goal is generating small triangles around interfaces, preferably with vertices on the interface. Besides matching the interface, the refinement methods push for higher quality (closer to equilateral) triangles.

### 2.1 Replacement strategy

The first strategy is based on replacing vertices. Each iteration step consists of adding a vertex, computing attenuation values in each triangle, removing a vertex, and computing attenuation values again. The addition of a vertex happens in the triangle $t$ that maximizes

$$w_t = \sum_{i=1}^{3}\left(\frac{|\mu_t - \mu_i|}{|\mathbf{z}_t - \mathbf{z}_i|}\right)^2 + \beta\left(\frac{r_t}{\min(\mathbf{l_t})}\right)^2, \quad (2)$$

where $i$ runs over all triangles neighbouring $t$. The ratio of the radius of the circumscribed circle ($r_t$) over shortest edge ($\min(\mathbf{l_t})$) is a measure for the quality of triangle $t$. The center of mass of triangle $j$ is denoted by $\mathbf{z}_j$. $\beta$ is a parameter that controls how important the two terms are in relation to each other. The vertex is added in triangle $t$ in such a way that the surrounding triangles become closer to equilateral. This is possible using Chew's algorithm[14]:

1. Calculate the center $C_t$ of the circumscribed circle of $t$.
2. If $C_t$ lies inside $t$ or one of its neighbouring triangles, add $C_t$ to the triangulation.
3. Else add the center of the first edge between $C_t$ and $t$. Remove all vertices in the diametral circle of this edge.

The circumscribed circle, diametral circle, and the vertex added in the third case, are shown in Fig. 1.
Chew's algorithm increases triangle quality every iteration step. On top of Chew's point, the added vertex can be given an extra push in such a way that it maximizes the gradient in reconstructed attenuation. This means the point $\mathbf{z}$ that is added is the point within the inscribed circle of the triangle (starting from Chew's point as center, see Fig. 2) that maximizes $|\nabla\mu(z)|^2 \approx |\nabla\mu(\mathbf{z}_{chew}) + H_\mu(\mathbf{z}_{chew})(\mathbf{z} - \mathbf{z}_{chew})|^2$, which can be computed numerically.
After the addition of a vertex, another vertex is removed. This is the vertex $v$ that minimizes the weight (2) converted to vertices.

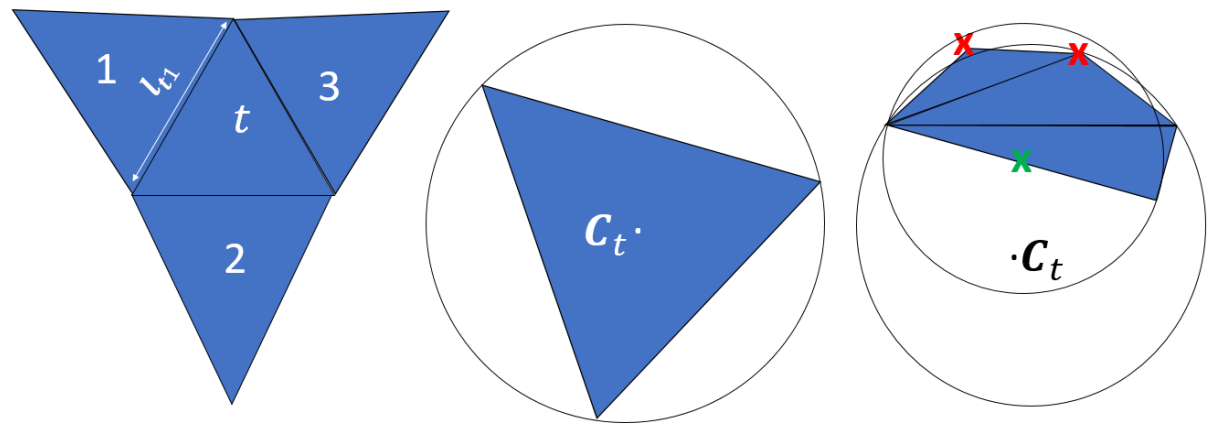

**Figure 1:** Left: triangle $t$ and surrounding edges. Middle: center circumscribed circle. Right: Point placed by Chew's algorithm in case 3.

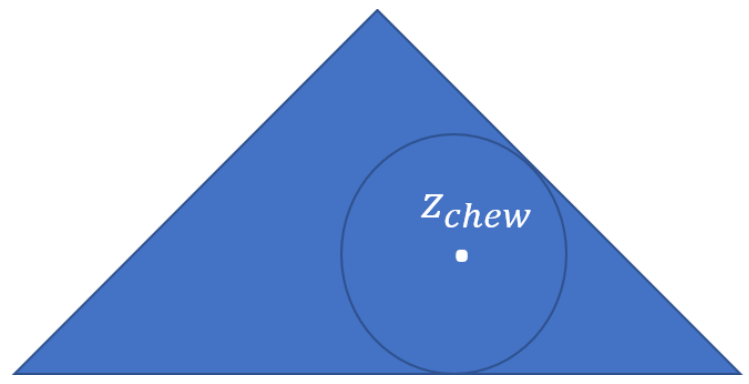

**Figure 2:** Disk that must contain $\mathbf{z}$.

### 2.2 Mobility strategy

This strategy optimizes the following objective function:

$$E = \frac{1}{2}\sum_{j=1}^{n\cdot p}\left(-\log\left(\frac{I_j}{I_0}\right) - \sum_{k=1}^{m} A_{j,k}\mu_k\right)^2 + \frac{1}{2}\gamma\sum_{j=1}^{m}\left(\frac{r_j}{\min(\mathbf{l_j})}\right)^2. \quad (3)$$

The first term minimizes the difference between the expected and measured intensity, the second term minimizes triangle quality. $\gamma$ is once again a coefficient that sets how important the two terms are in relation to each other. The gradient of (3) to vertex position can be calculated analytically in each vertex. After this, the vertices can be moved in gradient descent steps. After each step of moving all vertices, attenuation values in each triangle are updated.

### 2.3 Experiments

Experiments were conducted with a starting mesh generated from 484 regularly placed vertices, shown in Fig. 3a. The memory needed to describe this volume mesh is the sum of the memory needed for vertex coordinates, connectivity list (the vertices connected to each triangle) and the attenuation value vector ($\boldsymbol{\mu}$). In all experiments, the mesh was first refined by the replacement strategy. After convergence of this method, the resulting meshes were refined again with the mobility strategy. The experiments finished after convergence of (3). In each experiment, the parameters $\beta$ and $\gamma$ in (2) and (3) were each chosen in such a way that equal importance was given to placing

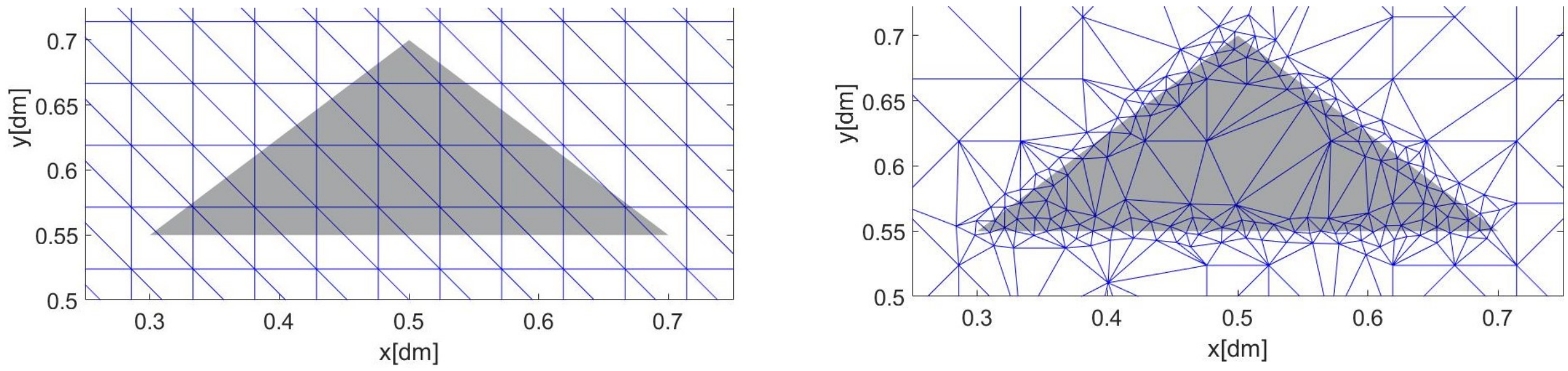

**(a)** Regular starting mesh. **(b)** Mesh refined by replacement algorithm.

**Figure 3:** Close up of attenuating triangle in mesh before and after replacement algorithm.

vertices on the interface and generating high quality triangles.

Firstly, the methods were tested with 200 noiseless projections. Both the replacement and mobility strategy require attenuation input values each iteration step. To test the performance of both strategies in an ideal scenario, input attenuation values were fixed to the ground truth values (so each triangle was attributed a weighted sum of the attenuation values of its composites).
After this, real reconstruction data, computed from simulated measurements by (1), were used as input for refinement. This was tested for several numbers projection angles, both noiseless and with 3% noise. Only a limited number of reconstruction steps were performed each iteration.

The refinement strategies were tested on a polyethylene triangle phantom, with attenuation coefficient 1.547dm$^{-1}$ for X-ray energies around 100keV. The simulated X-ray set-up was a parallel beam geometry. A measure for how good the methods perform in reducing partial volume effects is given by the partial volume fraction: $\Delta = \frac{\sum_i \min(\mathbf{s_i})}{s_p}$. $\mathbf{s}_i$ is a vector containing the areas of different homogeneous regions covered by triangle pixel $i$. The area of the polyethylene triangle phantom is denoted by $s_p$.

## 3 Results

With 200 projection angles, no noise, and when triangles are given ground truth attenuation values each iteration step, the mesh refined by the replacement strategy is shown a Fig. 3b. The mobility method refined this mesh further to the result shown in Fig. 4. The partial volume fraction $\Delta$ was as low as 0.003. A reconstruction of attenuation values on these triangle pixels is shown in Fig. 5a. Since this mesh has 882 triangles, a reconstruction on a $30 \times 30$ square pixel grid (an almost equal number of pixels) is shown in Fig. 5b, generated with the ASTRA toolbox [15]. Finally, a visualisation of how much partial volume effects remain is shown in Fig. 5c.
The partial volume fraction when the input attenuation values for the refinement strategies were computed from projection data by (1) are plotted in function of noise and varying numbers of projection angles in Fig. 6.

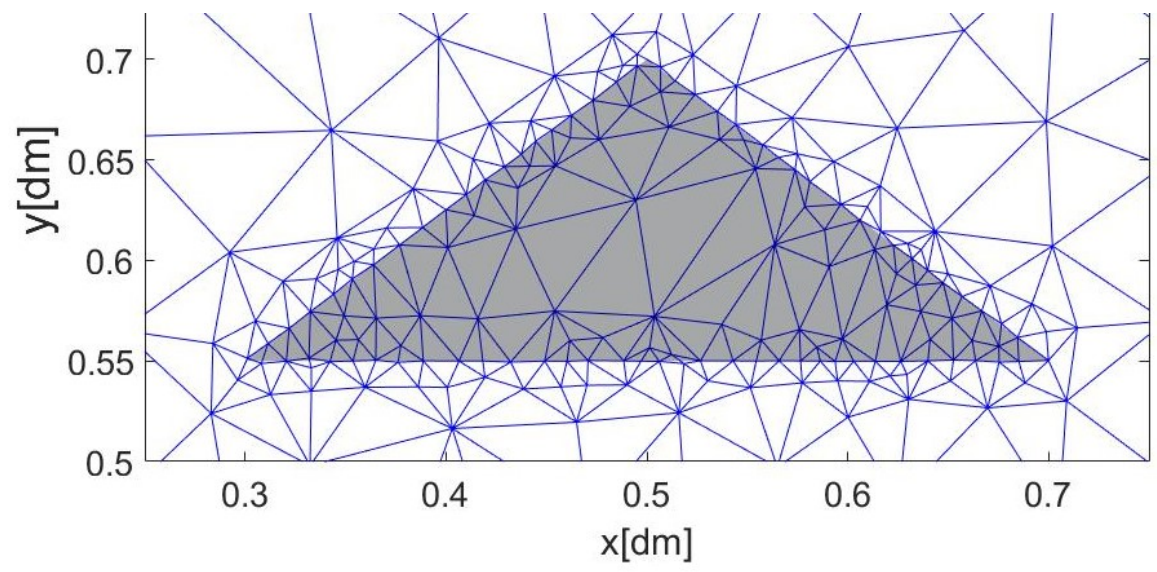

**Figure 4:** Close up of polyethylene triangle in a mesh refined by both replacement and mobility methods.

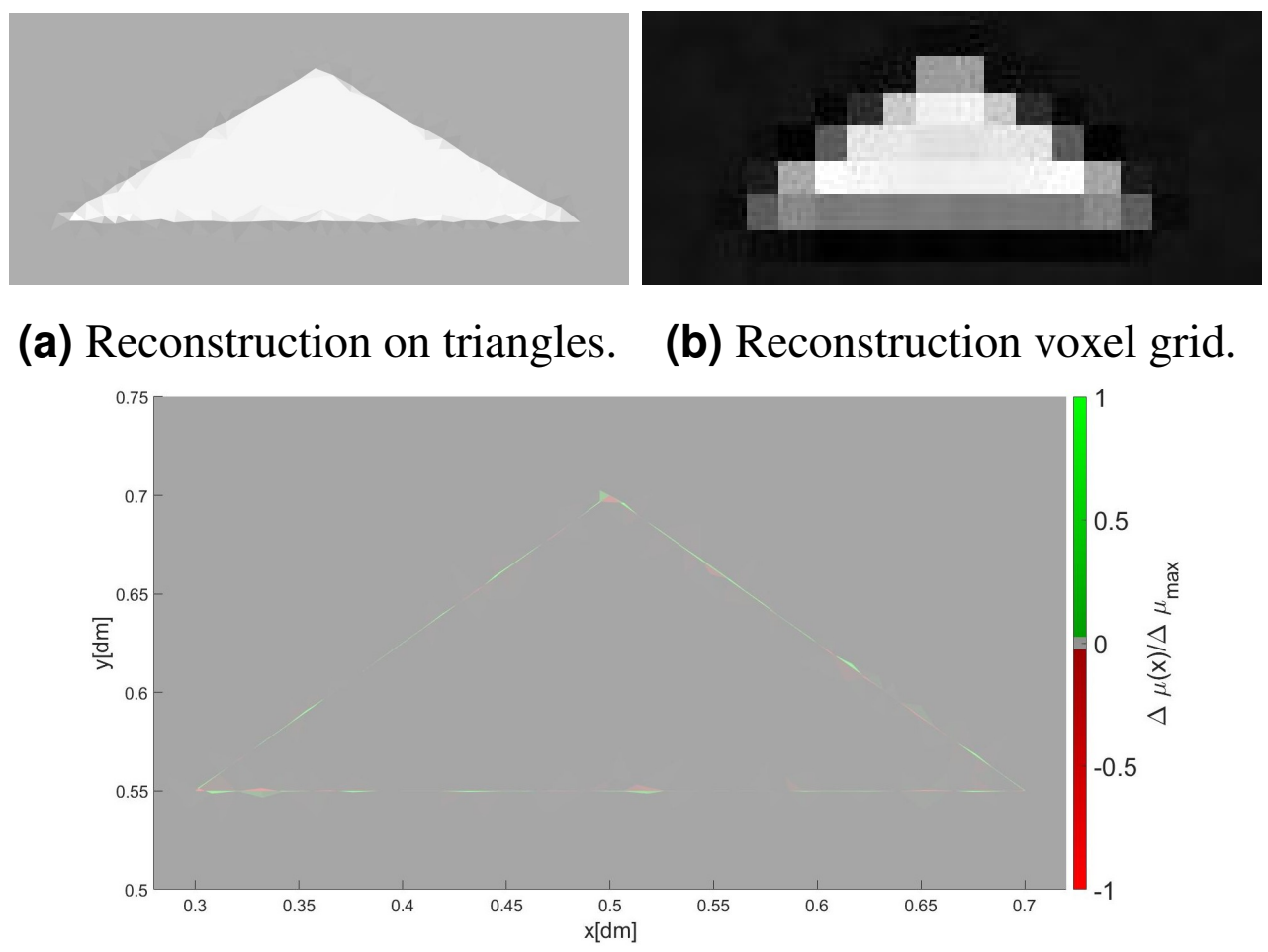

**(a)** Reconstruction on triangles. **(b)** Reconstruction voxel grid.

**(c)** Green: $\mu > \mu_0$, red: $\mu < \mu_0$.

**Figure 5:** Comparison of reconstruction results on adapted triangles to square grid reconstruction and ground truth.

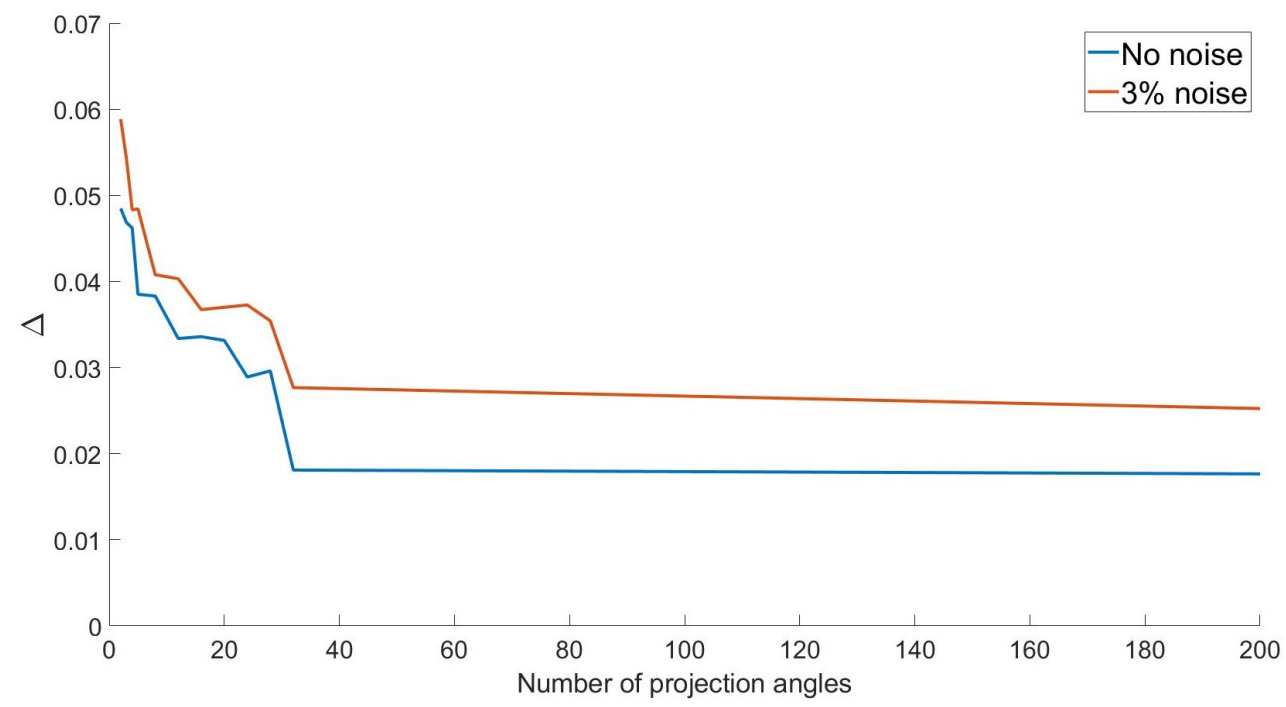

**Figure 6:** Partial volume fraction $\Delta$ for different numbers of projection angles with and without noise.

## 4 Discussion

Triangles pixels in a 2D reconstruction have the advantage over squares of allowing more flexibility and adaptability to interfaces. The mesh presented in Fig. 5a shows how triangular pixels can greatly reduce partial volume effects and Fig. 5b shows a comparison with a square grid with the same number of pixels.

Fig. 6 shows that at 32 projection angles, the partial volume fraction has not increased compared to 200 projection angles. With a high amount of noise (3%), the partial volume fraction increases from 0.018 to 0.025. With a handful projections, the partial volume fraction increases to $0.03 - 0.04$, both with and without noise. This shows the strategies are robust under noise and work well with a handful of radiographs (for this phantom). Filling in ideal reconstruction values, a result of which is shown in Fig. 4, reduces the partial volume fraction to 0.003, showing the main limitation of the methods is filling in the right attenuation values.

Even though the phantom studied in this paper is a 2D triangle, the methods can be extended to tetrahedra in 3D. Alternatively, the 3D case can be seen as a combination of 2D slices.

## 5 Conclusion

Two methods, one based on vertex replacement, one based on vertex mobility, for adapting a triangle mesh based on X-ray measurements, were introduced in this paper. These require the triangles of the mesh to be given a attenuation values, like 2D pixels. A comparison with the same number of square grid pixels showed a great potential in reducing partial volume effects. A test with different numbers of projection angles and noise showed robustness under these circumstances. With 200 projection angles, the calculations needed for getting the fit in Fig. 4 took around an hour. However, the computations were executed in sequential MATLAB code, therefore a lot of time can still be won by parallel execution. Another way to win time is by having a starting mesh closer to the real object.

## 6 Acknowledgement

The authors would like to thank Bart van Lith and Nicholas Francken (University of Antwerp) for their insights and support of this research. The authors also gratefully acknowledge support from the Research Foundation Flanders, Belgium (FWO) through projects S003421N and 1SG0523N.

# End-to-end deep learning PET reconstruction from histo-images for non-rigid motion correction

Maël Millardet[1,2], Vladimir Panin[2], Deepak Bharkhada[2], Josh Schaefferkoetter[2], Evgeny Kozyrev[1], Juhi Raj[1], Maurizio Conti[2], and Samuel Matej[1]

[1] Department of Radiology, University of Pennsylvania, Philadelphia, Pennsylvania, United States
[2]Siemens Medical Solutions USA, Inc., Knoxville, Tennessee, United States

**Abstract** End-to-end deep learning PET reconstruction is becoming increasingly important. One of its most promising applications is motion correction, where deep learning reconstruction could be used for both motion vector estimation and final reconstruction. However, several issues remain to be resolved, including whether to perform motion correction in image space after separate reconstruction of each motion gate (imageMC) or in data space prior to reconstruction (dataMC). In this study, we propose to compare these two approaches and compare the final reconstruction performed by deep learning versus OSEM. We found that the deep learning reconstruction was less noisy, at the cost of slightly reduced contrast, but had an overall higher signal-to-noise ratio. imageMC performed better than dataMC in the high-count case, but this was reversed in the low-count case, these results being the same with both of the reconstruction techniques used (OSEM and deep learning).

## 1 Introduction

End-to-end deep learning for PET image reconstruction has been steadily gaining momentum since the first publication on the topic in 2018 [1]. Recent publications [2] have achieved even better results using histo-images [3] as the network's input. One of its most promising applications is motion correction. Deep learning could improve image quality for low-statistics data and fast motion requiring very short motion gates, not only because of its ability to process low-statistics data but also because it allows high-speed processing.

PET acquisitions typically last several minutes; during this time, the patient moves with breathing, heart rate, and voluntary movements (such as head and arms). These movements blur the image, make the attenuation correction inconsistent with the emission data, and can lead to an overestimation of the lesion size and an underestimation of the uptake.

Motion correction aims to create a "frozen" image corresponding to a particular patient position. However, it is preferable to use all acquired data to preserve good image quality. Several methods exist to achieve this goal, the differences being mainly in how the motion is recorded and the activity is combined (image space versus data space [4]). The question arises especially if the data are in the histo-image format since they can be registered the same way as an image. In [4], it was shown that combining information from different gates in the data space before reconstruction can be more robust for low counts.

In this work, we study two effects:

1. We study the performance of our end-to-end network compared to OSEM + PSF (Point Spread Function).
2. We compare the motion correction in image space after a separate reconstruction of each gate to the motion correction in data space (histo-image) before reconstruction.

## 2 Materials and Methods

### *Multi-view Histo-Image*

In this work, we use multi-view histo-images as the data format as an alternative to sinograms [3]. Since they are in the same format as an image, multi-view histo-images are perfectly suitable for the input of a neural network. Contrary to the work of [2], we use multiple views in order to maintain the angular information present in the raw data, as in [5] and [6]. We form the multi-view histo-image (we used ten views in this work) according to the Most Likely Annihilation Position, where the histogrammer stores the events in the image voxels from which they were most likely emitted based on the difference in detection time of the two coincident photons. The histogrammer increments a counter for the corresponding voxel for every prompt event and decrements the counter for every delayed event. Each event is scaled on the flight during deposition by the corresponding normalization factors, including deadtime correction. After deposition, each voxel in the histo-image is multiplied by a global sensitivity factor indicating the number of lines of response ending up in that voxel. There are a few voxels located at the edge of the field of view that no line of response can reach. For the neural network to work correctly, we still assign a value to these voxels by interpolating the value of the neighboring voxels.

### *Reconstruction Pipeline*

Figure 1 shows the general reconstruction pipeline. The histo-image is first divided into smaller patches (of size 63×63×63×10, 10 being the number of views). Each patch is passed separately through the neural network, and the outputs are concatenated to form the final image volume. The patches overlap continuously to prevent the boundary between patches from creating a discontinuity in the image.

### *The U-Net Neural Network*

This work utilizes a U-Net style architecture inspired by the work of Whiteley *et al*. [2], with a contraction, bottleneck, and expansion segment. This network uses residual blocks (see Figure 2) to help the gradient propagate to the earliest layers and increase training efficiency and stability. The network contains a total number of about 15 million trainable parameters. Each layer uses 3×3×3 convolutional kernels and parametric rectified linear units (PReLU) [7] activation functions. Spatial down-sampling is performed using 3×3×3 convolutions with a stride of 2 in every dimension. The symmetrical transpose convolution performs the up-sampling.

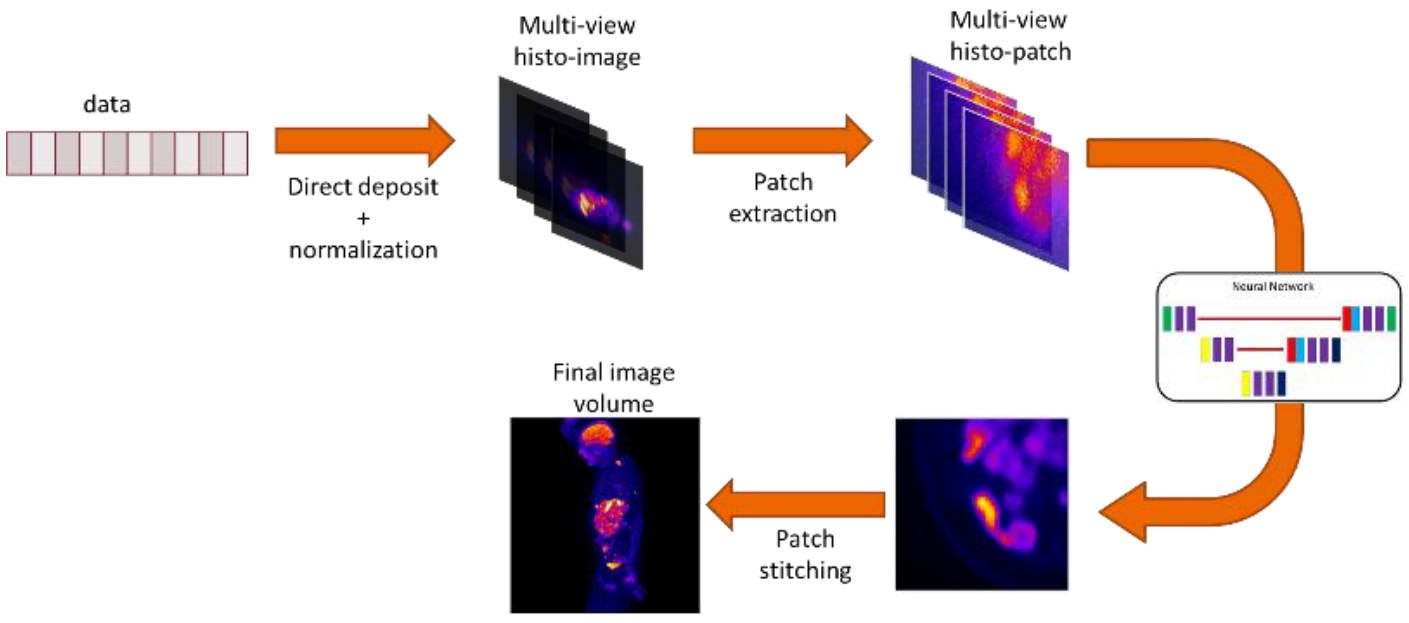

*Figure 1: Deep learning reconstruction pipeline*

We used the L1 loss function, which was shown to give sharper edges than the mean square error loss function [8]. We used the Adam optimizer with a learning rate of $10^{-4}$. The network was implemented with PyTorch. We trained three networks independently, as shown in the table below. One is trained to reconstruct non-attenuation corrected (NAC) images and is used for the estimation of the motion vectors in order to avoid the mismatch between PET end CT. The two others reconstruct fully corrected images, one being trained to reconstruct high-count data, and the other to reconstruct low-count data.

| Network | Count level | Attenuation correction | Purpose |
|---|---|---|---|
| #1 | Original (high-count) | No | Motion estimation |
| #2 | Original (high-count) | Yes | Final reconstruction |
| #3 | Subsampled ten times (low-count) | Yes | Final reconstruction |

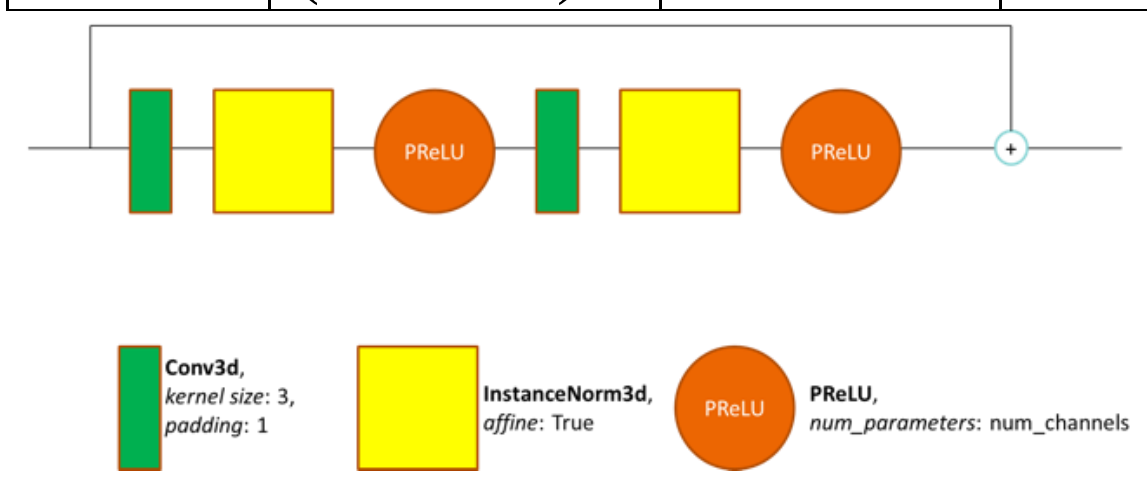

*Figure 2: Details of the residual bloc used*

***Patient Dataset***

Data from 23 patient scans were used in this study. Twenty were used for network training, and three for validation. All data were anonymized $^{18}$F-FDG whole-body studies acquired at the University Hospital of Bern with the Siemens Biograph Vision Quadra long axial field-of-view PET/CT scanner. Acquisition time ranged from 27 s to 10 min (most of the scans being 10 min scans), corresponding to 167 million prompts to 4.29 billion prompts (see table below).

The labeling data were reconstructed with MLEM + PSF (50 iterations) without post-smoothing. This number of iterations is higher than the one usually used in clinical settings and leads to noisier but sharper images. This is justified by the fact that the network reconstruction tends to be smoother than the label used for training (as reported by [2] and confirmed in the present study).

| | Minimum | Median | Maximum |
|---|---|---|---|
| Acquisition time | 27 s | 10 min | 10 min |
| Number of prompts | 167 million | 2.61 billion | 4.29 billion |
| Patient mass | 55 kg | 73.5 kg | 105 kg |

For the network trained on low-count data, we subsampled each list mode ten times but kept the high-count reconstructed images as target.

***Motion Correction Pipeline***

Here is our proposed motion correction pipeline:

Step 1: Data gating. We first compute the centroid of distribution in the axial direction over a region containing the lungs and the abdomen, based on the histo-image with the procedure described in [9]. We then use this 1D signal to sort the data into four motion gates based on amplitude gating.

Step 2: Computation of the motion fields. Following the proposal of [10], we reconstruct these images using deep learning (without attenuation correction) and then compute the motion fields between the images using the Lucas-Kanade optical flow technique [11]. See Figure 3 for an illustration. The motion vector fields are 3D images where each image voxel contains the displacement in each of the three spatial dimensions. They are first computed between adjacent gates, because optical flow performs better with small motion, and then combined so that the motion will always be relative to the reference gate 1.

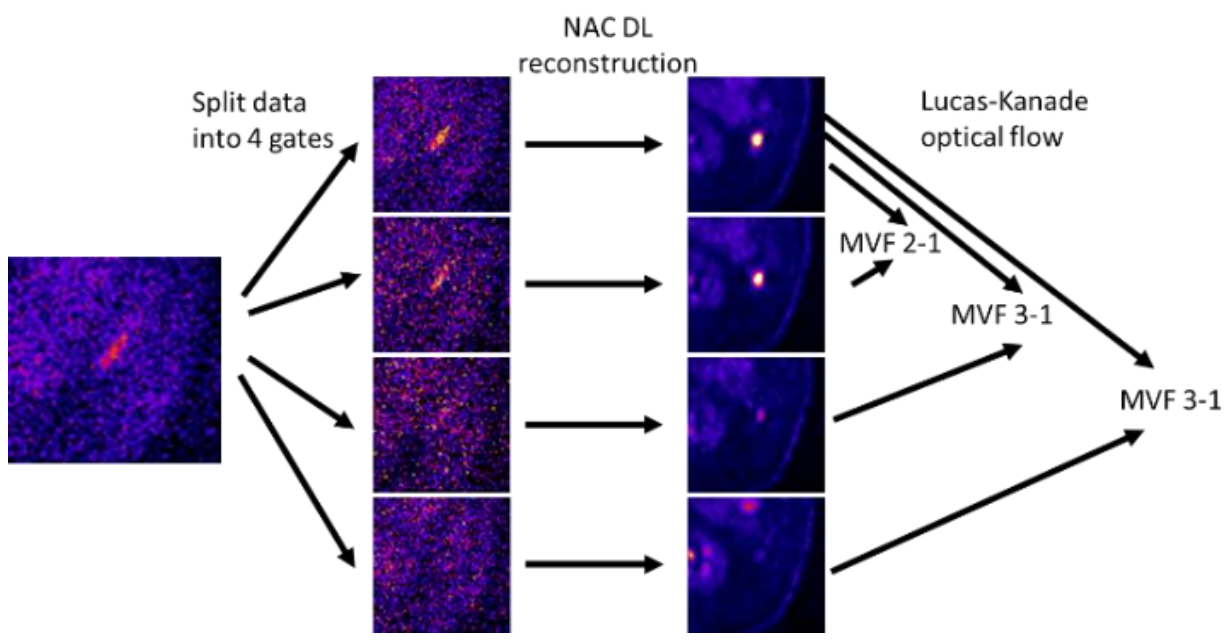

*Figure 3: Motion vector field computation pipeline*

| | Noise (std in the heart) (a.u.) | Tumor activity (mean activity in the tumor) (a.u.) | Tumor activity/noise |
|---|---|---|---|
| noMC OSEM | 0.63 | 79 | 124 |
| dataMC OSEM | 0.64 | 115 | 180 |
| imageMC OSEM | 0.61 | 116 | 191 |
| Gate 1 OSEM | 1.1 | **127** | 115 |
| noMC DL | 0.54 | 72 | 133 |
| dataMC DL | 0.52 | 104 | 200 |
| imageMC DL | **0.47** | 116 | **247** |
| Gate 1 DL | 0.71 | 118 | 166 |

*Table 1: Quantitative performance of the different motion correction techniques studied in the high-count case. The terminologies "dataMC" and "imageMC" are explained in the motion correction pipeline. "noMC" means that all data is used, but no motion correction is performed. "Gate 1" means that only data belonging to gate 1 is used for reconstruction. OSEM reconstructions are noisier than DL reconstructions.*

| **Possibility A: dataMC** | **Possibility B: imageMC** |
|---|---|
| <u>Step 3A: Motion Correction of the Data</u>. We deposit each acquired prompt event into the motion-corrected reference histo-image (dataMC DL) or sinogram (dataMC OSEM) using the MVF of step 2 | <u>Step 3B: Separate reconstruction of each gate</u>. Either by OSEM (imageMC OSEM) or deep learning (imageMC DL) |
| <u>Step 4A: Reconstruction using deep learning or OSEM from motion-corrected combined data</u> | <u>Step 4B: Motion Correction of the images</u>. We register each image to the reference gate using the MVF computed in step 2 and sum the resulting images |

The comparative OSEM + PSF reconstructions were performed using the clinical parameters, *i.e.*, four iterations and five subsets without post-smoothing. It differs from the MLEM + PSF labels used to train the network.

Robust estimation of motion deformation vectors plays a crucial role in motion correction applications. For non-rigid motion, we need to estimate motion vectors by registering reconstructed images of selected frames. This registration process is sensitive to image noise, a necessary byproduct of data subsampling into gates. The developed deep-learning reconstruction approach addresses this challenge through its ability to reduce image noise.

***Figures of Merit***

To evaluate this study, we selected one of three patients from the validation dataset with extensive liver disease and clear respiration artifacts. The scan time for this study was 4 minutes (1.19 billion prompts). We considered two regions of interest. One is in the uniform heart, with low activity and no apparent tumor. This region of interest measures the noise level (standard deviation in the region). We also study the mean activity in a small tumor subject to relatively sizeable respiratory motion (an axial displacement of about 12 mm is visible between gates 1 and 4). The top row of Figure 4 shows the deep-learning reconstruction of each gate. A cursor indicates a fixed location in the image, and the axial displacement of the tumor used for evaluation is visible. Another tumor also clearly disappears due to displacement to neighboring slices. The bottom row shows the registration of each gate to gate 1. The tumor is now at the exact same location for each gate.

## 3 Results

***Evaluation of the Network Reconstruction***

Figure 5 shows the deep learning reconstruction in the top row and the OSEM reconstruction in the bottom row. The OSEM reconstruction is noisier than the deep learning reconstruction, especially for the single gate reconstruction. This is also confirmed by Table 2, which shows that the deep learning reconstruction reduces the noise level by 15-36% for all cases compared to OSEM. This is even more dramatic in the low-count case, where the OSEM reconstructions are noisier and the deep learning reconstructions even smoother than in the high-count case. It may seem surprising that the deep learning reconstructions are smoother in low-count than in high-count, but since the noise did not match between the target and the network input, the network learned to remove the noise. On the other hand, tumor activity is also reduced by 0-9%. The loss of high-frequency details is even more dramatic in the low-count case. Overall, the signal-to-noise ratio increases by 8-45% in high-count, and 145-285% in low-count.

***Evaluation of Motion Correction***

Figure 5 and Table 2 also show that all motion correction techniques achieve tumor activity close to the single gate case and a noise level close to the reconstruction without motion correction (and even typically slightly lower) for a given reconstruction modality. In high-count, motion correction in image space gave better results than motion correction in data space, both in terms of tumor activity and noise, but those results are reversed in the low-count case.

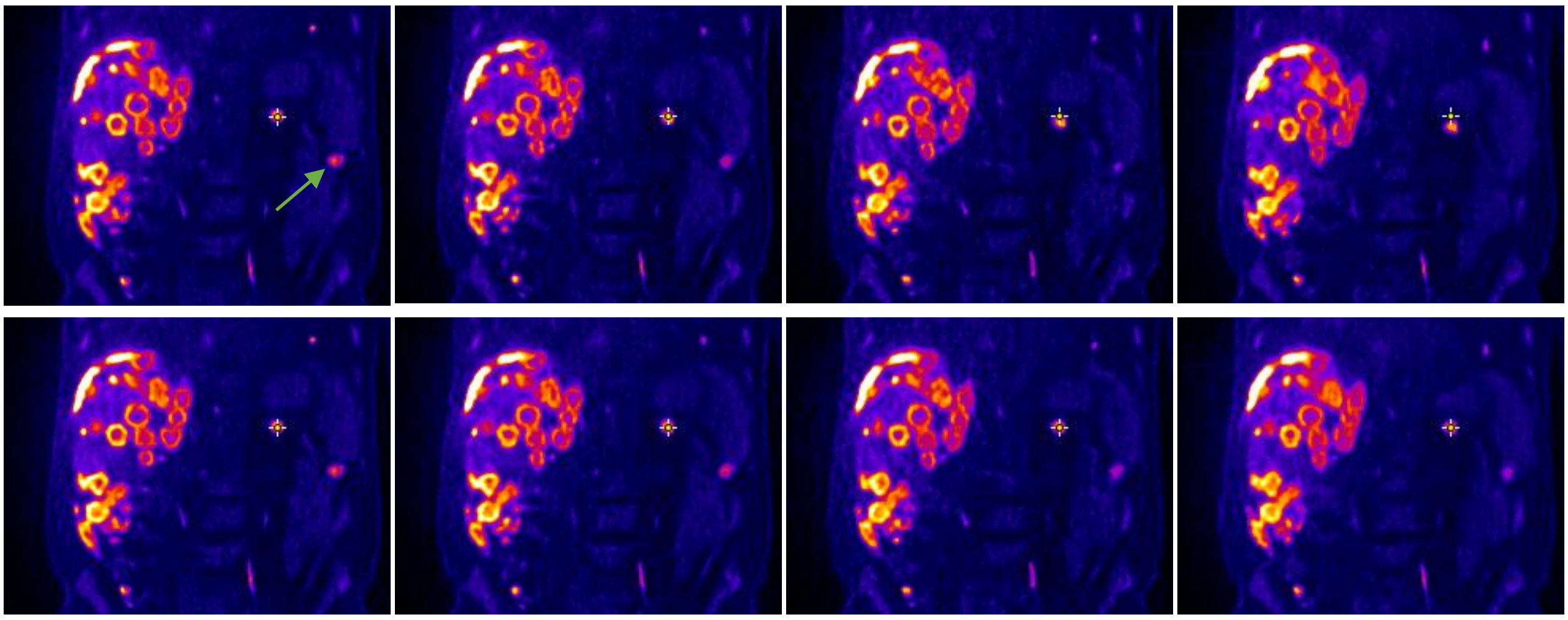

*Figure 4: patient used for evaluation. The cursor is in the same location for each gate, showing that one of the tumors (the one used for evaluation) moves about 12 mm between gates 1 and 4 due to breathing. Another tumor (indicated by the arrow) disappears into the neighboring slices. After registration, all tumors are in the same location.*

| | Noise (std in the heart) (a.u.) | Tumor activity (mean activity in the tumor) (a.u.) | Tumor activity/noise |
|---|---|---|---|
| noMC OSEM | 1.8 | 80 | 43 |
| dataMC OSEM | 1.8 | 111 | 62 |
| imageMC OSEM | 1.8 | 109 | 60 |
| Gate 1 OSEM | 3.0 | **124** | 42 |
| noMC DL | 0.50 | 77 | 155 |
| dataMC DL | **0.47** | 110 | **237** |
| imageMC DL | 0.69 | 102 | 147 |
| Gate 1 DL | 1.1 | 113 | 105 |

*Table 2: Quantitative performance of the different motion correction techniques studied in the low-count case than dataMC DL.*

## 4 Discussion and Conclusion

In this study, the motion vector fields are computed in image space and thus may be more consistent with the motion correction in image space. On the other hand, data pre-correction should give the reconstruction process more information that it can consider. This last fact is likely to be even more critical in low-count than in high-count. This may explain why we observed better results for imageMC in high-count, and for dataMC in low-count, these results being consistent with the two reconstruction techniques used (OSEM and deep learning). This is consistent with the results of [4], which showed that the motion correction in the data space was more consistent in the low-count case, especially concerning the positive bias in low-activity regions.

In our motion correction protocol, we currently use four motion gates. We plan to use a larger number in the future, especially because deep learning allows a higher signal-to-noise ratio.

This study's motion correction evaluation was based on only one patient. It would be interesting to see to which extent these results generalize to a bigger number of patients.

## Acknowledgments

The authors would like to thank all members of the Physics and Instrumentation Group of the Department of Radiology at the University of Pennsylvania and the PET Physics Team at Siemens Medical Solutions for their valuable comments and suggestions on this work. The authors would also like to thank Bern University Hospital's Inselspital Department of Nuclear Medicine, the first site of Siemens Vision Quadra, for providing anonymized patient data. This investigative work was supported by the National Institute of Biomedical Imaging and Bioengineering through grant number R01-EB031806. The content in this article is solely the responsibility of the authors and does not necessarily represent the official views of the National Institutes of Health.

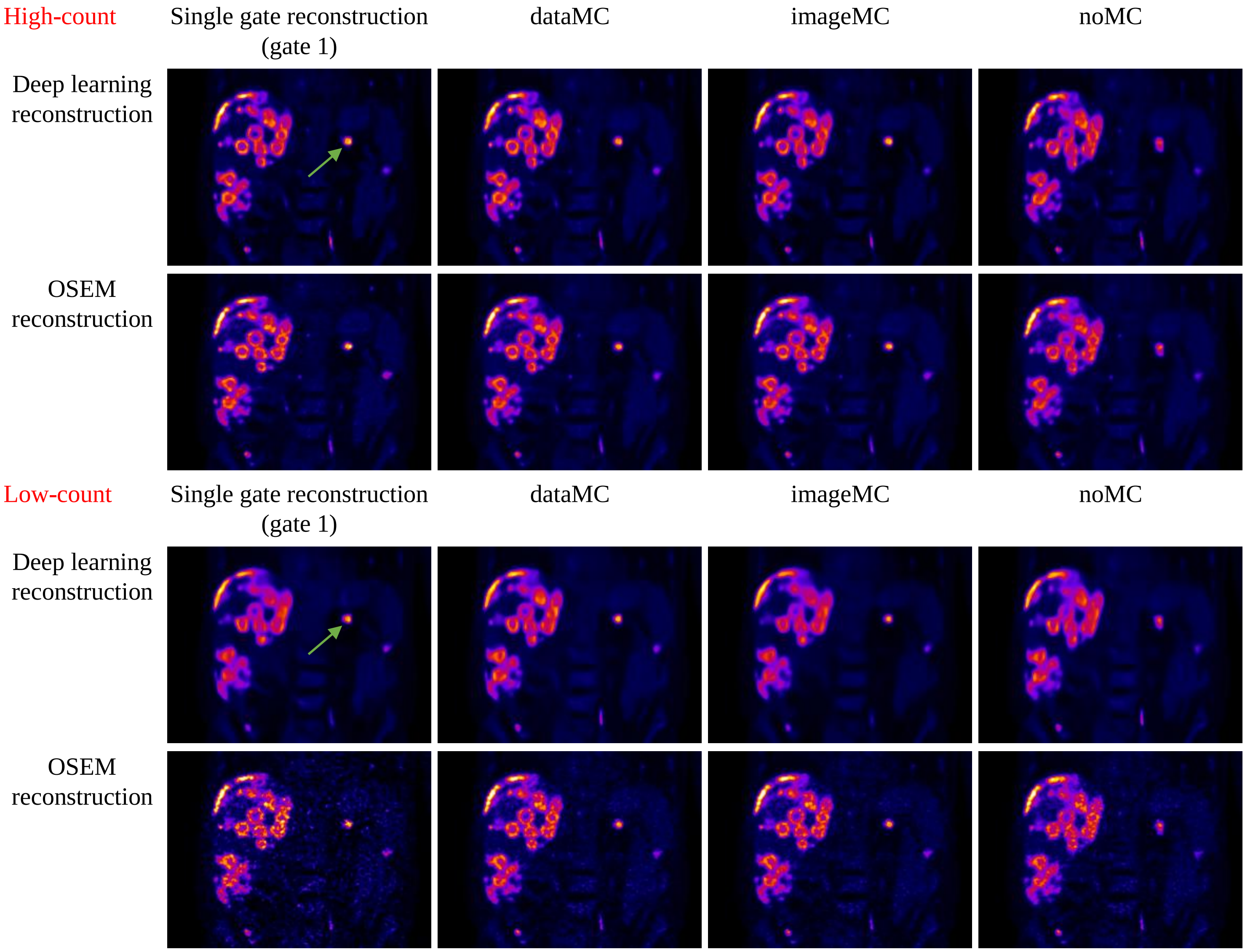

*Figure 5: Images obtained by different motion correction techniques. The OSEM reconstruction is noisier than the deep learning reconstruction, especially for the single gate reconstruction. The motion correction allows obtaining a tumor activity close to the single gate case while keeping a noise level comparable to that of the reconstruction without motion correction.*

# 3D Photon Counting CT Image Super-Resolution Using Conditional Diffusion Model

Chuang Niu[1†], Christopher Wiedeman[1†], Mengzhou Li[1], Jonathan S Maltz[2*], and Ge Wang[1*]

[1]Department of Biomedical Engineering, Center for Biotechnology & Interdisciplinary Studies, Rensselaer Polytechnic Institute, Troy, NY USA
[2]Molecular Imaging and Computed Tomography, GE HealthCare, Waukesha, WI, USA
[†]Co-first Authors
[*]Co-corresponding Authors

**Abstract** This study aims to improve photon counting CT (PCCT) image resolution using denoising diffusion probabilistic models (DDPM). Although DDPMs have shown superior performance when applied to various computer vision tasks, their effectiveness has yet to be translated to high-dimensional CT super-resolution. To train DDPMs in a conditional sampling manner, we first leverage CatSim to simulate realistic lower-resolution PCCT images from high-resolution CT scans. Since maximizing DDPM performance is time-consuming for both inference and training, especially on high-dimensional PCCT data, we explore both 2D and 3D networks for conditional DDPM and apply methods to accelerate training. In particular, we decompose the 3D task into efficient 2D DDPMs and design a joint 2D inference in the reverse diffusion process that synergizes 2D results of all three dimensions to make the final 3D prediction. Experimental results show that our DDPM achieves improved results versus baseline reference models in recovering high-frequency structures, suggesting that a framework based on realistic simulation and DDPM shows promise for improving PCCT resolution.

## 1 Introduction

Over the past two decades, CT imaging has rapidly advanced in terms of spatial, spectral and temporal resolution. However, for CT scanners based on conventional energy-integrating detectors (EIDs), further increases in spatial resolution are limited by the trade off that exists between reduced pixel size and dose-efficiency. This is because contemporary EIDs utilize segmented optical scintillators coupled to photodiode arrays. When pixel size is reduced, the reflective septa between pixels that prevent optical crosstalk occupy an increasingly fraction of the detector area, leading to losses in geometric- (and therefore dose-) efficiency.

Photon counting detectors (PCDs) are able to largely overcome this limitation. Photons incident on PCDs are directly converted into charge clouds in a semiconductor, and this charge is read out at electrodes. The nearest electrode to the point of photon interaction typically reads out the most charge, and so is identified as the pixel of interaction. The finer the spacing of the readout electrodes, the finer the potential spatial resolution of the detector.

PCDs have indeed demonstrated the combination of higher spatial resolution and dose-efficiency than EID systems. However, several physical processes prevent PCDs from realizing resolution consistent with sampling provided by electrode spacing. (1) In reality, not all charge deposited in an interaction is collected by a single electrode, but is shared among nearby electrodes. (2) In high-Z detectors (such as CdTe/CZT), detector material K-edges are present within the energy range of the diagnostic energy X-ray spectrum. K-escape fluorescence thus occurs in these detectors, in which a photon deposits only a portion its energy at the point of initial interaction, and approximately 30 keV elsewhere. (3) In low-Z PCDs, such as edge-on-irradiated Si, Compton scatter occurs, so that some of the incident photon energy can leave a pixel [1]. (4) Some common PCD designs utilize a macropixel structure, with dead space between the macropixels to accommodate circuit elements such as through-silicon-bias. This leads to non-uniform sampling of the constituent micropixels, resolution loss, and aliasing artifact. All of these 4 processes and factors broaden the detector point-spread function in complex ways that depend on the incident spectrum, the spatial frequency content of the imaged object, and the material pathlengths traversed by each ray. Approaches such as anti-coincidence processing may be used to recover resolution for (1)–(3), but these tend to fail when flux is high due to the problem of pulse pileup.

In this paper, our objective is to determine whether DL-based superresolution image postprocessing can superresolve photon counting CT images without implementing costly and potentially noise-enhancing deconvolutional methods based on physical processes that are too complex to model in practical imaging systems.

DL-based image superresolution (SR) has been extensively advanced in the computer vision field, including progress in terms of both network structure design (such as EDSR [2], SRGAN [3], RCAN [4]) and restoration frameworks (such as DPSR [5] and PULSE [6]). Despite the success of these techniques in the natural image domain, directly applying these methods to the medical image domain is challenging due to the lack of good quality low-resolution (LR) and high-resolution (HR) image pairs for network training. While downsampling techniques and Gaussian noise models have been employed to generate synthetic datasets for CT image SR and achieve promising results [7–9], the metrics used to assess performance often do not translate to the desirability of the images from a clinical perspective. This is likely due to limitations of the image degradation model, as well as the quality metrics. For example, our recent study suggests that the insertion of unrealistically-distributed noise can significantly degrade practical SR performance on images with real CT noise. Furthermore, the complex physics behind photon counting detection makes the degradation more challenging to represent realistically with simple analytic formulas [10]. To address the challenge, this study aims to leverage advanced contemporary deep learning techniques and realistic simulation tools to improve PCCT resolution.

Recently, denoising diffusion probabilistic models (DDPM) [11] have achieved great success in generative and reverse problems [12–14]. In comparison with adversarial generative models, DDPM does not suffer from mode-collapse and training instabilities, and demonstrates even better performance on various tasks. Nevertheless, effectively adapting the DDPM to improve the resolution of high-dimensional PCCT images has not yet been studied. A main obstacle we encountered in our initial application of DDPM to CT imaging is that directly training a DDPM as a conventional 3D network results in poor convergence and lengthy training times. To overcome this challenge, we decompose the 3D task into two 2D models for improving in-plane and through-plane resolution respectively. However, the 2D model trained in one dimension usually exhibits degraded performance in other dimensions. To this end, we design a joint 2D inference in the reverse diffusion process that synergizes 2D results of all three dimensions to make the final 3D prediction. We also design an alternative inference among different 2D models so that it is as efficient as a single 2D model inference.

Since it is not feasible to collect paired high and low resolution data that are perfectly registered, realistic simulation of aligned LR and HR is critical to building deep learning models. We use CatSim to generate low-resolution counterparts for CT image phantoms [15, 16]. Degradation is modulated by altering detector pitch, x-ray focal spot size, as well as noise and pixel cross-talk effects *in silico*.

## 2 Methods

### 2.1 Data and Simulation

The CatSim PCCT module is to simulate scans of 10 digital phantoms. Each phantom is a reconstructed clinical CT head scan, which is converted into a water density voxel map based on attenuation. For proprietary reasons, absolute sinogram pixel size is suppressed; we denote the simulated LR pixel side lengths as $x_{\mathrm{LR}}$ and $z_{\mathrm{LR}}$. The LR and HR scans are simulated for 1000 views, $x_{\mathrm{LR}} \times z_{\mathrm{LR}}$ pixel size and 1 mm square focal spot; and 1300 views, $0.75x_{\mathrm{LR}} \times 0.85z_{\mathrm{LR}}$ pixel size, and 0.75 mm focal spot, respectively. To achieve highest resolution and noise suppression, Poisson noise and pixel cross-talk are suppressed in HR scans. Simulating the phantoms at their true voxel size (0.293 mm in-plane, 0.625 mm axial) resulted in negligible difference between the phantom and LR reconstruction. We consequently reduced the voxel size by half in each direction to challenge the system.

All scans utilized 120 kVp tube voltage and 400 mA current for a one second rotation period and were reconstructed with filtered back projection such that there was a 1:1 voxel correspondence between the phantoms and reconstructions. One patient phantom (later used for testing) was scanned in parallel (i.e., superimposed in the sinogram domain) with in-plane and through-plane bar phantoms for quantitative evaluation.

### 2.2 Conditional DDPM

Following [17], the CT SR task is formulated as a conditional generation. Given paired LR and HR images, $\{\mathbf{x}_i, \mathbf{y}_i\}_{i=1}^N$ ($\mathbf{x}_i$ and $\mathbf{y}_i$ denote the LR and HR images, respectively, and $N$ is the number of image pairs) drawn from the conditional distribution $p(\mathbf{y}_i|\mathbf{x}_i)$, we aim to approximate $p(\mathbf{y}_i|\mathbf{x}_i)$ by learning a stochastic iterative process, where each and every iteration step is parameterized with the neural network function $f_\theta$. DDPM involves a forward Markovian process for training and a reverse Markovian diffusion process for inference. Specifically, the forward process gradually adds Gaussian noise into an HR image via a fixed Markov chain, resulting in a series of images $\mathbf{y}_0 \to \mathbf{y}_1 \to \cdots \to \mathbf{y}_T$, where the noise level gradually increases with time step $t$, $\mathbf{y}_0$ and $\mathbf{y}_T$ are the HR and pure Gaussian noise image respectively, and $T$ is the number of iteration steps. The forward Markovian diffusion process is defined by $q$:

$$q(\mathbf{y}_{1:T}|\mathbf{y}_0) = \prod_{t=1}^{T} q(\mathbf{y}_t|\mathbf{y}_{t-1}), \tag{1}$$

$$q(\mathbf{y}_{1:T}|\mathbf{y}_0) = \boldsymbol{N}(\mathbf{y}_t|\sqrt{\alpha_t}\mathbf{y}_{t-1}, (1-\alpha_t)\boldsymbol{I}), \tag{2}$$

where $\alpha_{1:T}$ are hyper-parameters that determine the variance of the Gaussian noise. Fortunately, the distribution of $\mathbf{y}_t$ conditioned on $\mathbf{y}_0$ can be derived as:

$$q(\mathbf{y}_t|\mathbf{y}_0) = \boldsymbol{N}(\mathbf{y}_t|\sqrt{\gamma_t}\,\mathbf{y}_0, (1-\gamma_t)\boldsymbol{I}), \tag{3}$$

where $\gamma_t = \prod_{i=1}^{t} \alpha_i$. Thus, any intermediate noisy image $\mathbf{y}_t$ can be calculated given $\mathbf{y}_0$ as:

$$\mathbf{y}_t = \sqrt{\gamma_t}\,\mathbf{y}_0 + (1-\gamma_t)\varepsilon, \quad \varepsilon \sim \boldsymbol{N}(\mathbf{0}, \boldsymbol{I}). \tag{4}$$

The objective function used to training the network $f_\theta$ to predict the Gaussian noise added in $\mathbf{y}_t$, conditioned on the LR image and the noise level $\gamma$ is:

$$\mathbb{E}_{\mathbf{x},\mathbf{y}}\,\mathbb{E}_{\varepsilon,\gamma}\,||f_\theta(\mathbf{x}, \sqrt{\gamma}\mathbf{y}_0 + \sqrt{1-\gamma}\varepsilon, \gamma) - \varepsilon||_l^l, \tag{5}$$

where we set $l = 1$, $(\mathbf{x}, \mathbf{y})$ is a paired training sample, $\gamma \sim p(\gamma)$ is defined as in [17], and $\varepsilon$ is normally-distributed noise. Given pure Gaussian noise $\mathbf{y}_T$ and LR image $\mathbf{x}$, the reverse Markovian process gradually removes noise from intermediate noisy images to generate the HR image, i.e., $\mathbf{y}_T \to \mathbf{y}_{T-1} \to \cdots \to \mathbf{y}_0$, which is defined by a parameterized distribution $p_\theta$:

$$p_\theta(\mathbf{y}_{0:T}) = p_\theta(\mathbf{y}_T) \prod_{t=1}^{T} p_\theta(\mathbf{y}_{t-1}\mathbf{y}_t, \mathbf{x}) \tag{6}$$

$$p(\mathbf{y}_T) = \boldsymbol{N}(\mathbf{y}_T|\mathbf{0}, \boldsymbol{I}) \tag{7}$$

$$p_\theta(\mathbf{y}_{t-1}|\mathbf{y}_t, \mathbf{x}) = \boldsymbol{N}(\mathbf{y}_{t-1}|\,\mu_\theta(\mathbf{x}, \mathbf{y}_t, \gamma_t), \sigma_t^2\boldsymbol{I}), \tag{8}$$

where the reverse process $p(\mathbf{y}_{t-1}|\mathbf{y}_t, \mathbf{x})$ is approximately Gaussian, assuming the noise variance in the forward steps is sufficiently small [18]. Based on Bayes' rule, the posterior distribution $\mathbf{y}_t$ conditioned on $(\mathbf{y}_0, \mathbf{y}_t)$ is:

$$q(\mathbf{y}_{t-1}|\,\mathbf{y}_0, \mathbf{y}_t) = \boldsymbol{N}(\mathbf{y}_{t-1}|\,\boldsymbol{\mu}, \sigma^2\boldsymbol{I}) \tag{9}$$

$$\boldsymbol{\mu} = \frac{\sqrt{\gamma_{t-1}}(1-\alpha_t)}{1-\gamma_t}\mathbf{y}_0 + \frac{\sqrt{\alpha_t}(1-\gamma_{t-1})}{1-\gamma_t}\mathbf{y}_t \tag{10}$$

$$\sigma^2 = \frac{(1-\gamma_{t-1})(1-\alpha_t)}{1-\gamma_t}. \tag{11}$$

Using Eqs. (4) and (5), $\mathbf{y}_0$ can be approximated with the trained network as:

$$\hat{\mathbf{y}}_0 = \frac{1}{\sqrt{\gamma_t}}\left[\mathbf{y}_t - \sqrt{1-\gamma_t}f_\theta(\mathbf{x}, \mathbf{y}_t, \gamma_t)\right]. \tag{12}$$

Replacing $\mathbf{y}_0$ in Eq. (10) with $\hat{\mathbf{y}}_0$ yields:

$$\mu_\theta(\mathbf{x}, \mathbf{y}_t, \gamma_t) = \frac{1}{\sqrt{\alpha_t}}\left[\mathbf{y}_t - \frac{1-\alpha_t}{\sqrt{1-\gamma_t}} f_\theta(\mathbf{x}, \mathbf{y}_t, \gamma_t)\right], \quad (13)$$

where the variance of $p_\theta(\mathbf{y}_{t-1}|\mathbf{y}_t, \mathbf{x})$ is $(1-\alpha_t)$ following [11]. Then, each iteration in inference is calculated as:

$$\mathbf{y}_{t-1} \leftarrow \frac{1}{\sqrt{\alpha_t}}\left[\mathbf{y}_t - \frac{1-\alpha_t}{\sqrt{1-\gamma_t}} f_\theta(\mathbf{x}, \mathbf{y}_t, \gamma_t)\right] + \sqrt{1-\alpha_t}\,\varepsilon_t. \quad (14)$$

Thus, once the network is trained with Eq. (5), the HR image can be iteratively generated from a pure Gaussian noise image conditioned on an LR image via Eq. (14).

### 2.3 3D Super-Resolution with Joint 2D Inference

We found that directly training a 3D version of the 2D network used in [17] on our CT datasets leads to poor convergence. In contrast, the corresponding 2D network can be easily trained. A common strategy is to process the 3D CT scan slice-by-slice using a 2D network. However, this usually leads to degraded $z$-resolution. To overcome this problem, we propose to train two 2D networks to improve in-plane and through-plane resolution respectively, where the coronal and sagittal planes share the same though-plane network. To improve resolution in three dimensions, we constrain the results of all dimensions to be consistent in the inference stage. A simple implementation is to find a consistent 3D sampling at each time step, i.e., $\tilde{\mathbf{y}}_t = \arg\min_{\mathbf{y}_t}(\lambda_c||\mathbf{y}_t - \mathbf{y}_t^h||_2^2 + \lambda_c||\mathbf{y}_t - \mathbf{y}_t^c||_2^2 + \lambda_h||\mathbf{y}_t - \mathbf{y}_t^s||_2^2) = \frac{1}{\lambda_h+\lambda_c+\lambda_s}(\lambda_h\mathbf{y}_t^h + \lambda_c\mathbf{y}_t^c + \lambda_s\mathbf{y}_t^s)$ and then use $\tilde{\mathbf{y}}_t$ as the input for next iteration, where $\mathbf{y}_t^h$, $\mathbf{y}_t^c$, and $\mathbf{y}_t^s$ are the 2D prediction results for the three dimensions. However, this implementation increases inference time threefold. Instead, we propose an efficient strategy for merging 2D results, which alternately performs a single 2D inference for one of three dimensions at each step, using each 2D result as the input for the next iteration. Computational cost is not increased, as each iteration only requires a single 2D network. In practice, these two implementations yield comparable results.

### 2.4 Implementation Details

We use 9 patient CT scans for training and 1 patient CT scan for testing. $128 \times 128$ patches are randomly cropped for training the 2D networks. We use the same network architecture as in [17], and the attention is applied to the layer with the smallest spatial dimension. The batch size is 4 and the Adam optimizer uses a learning rate of $10^{-4}$. The number of sampling steps for DDPM is set to 2000, the number of training iterations to 300000, and all other hyper-parameters are set equal to those employed in [17]. To compare DDPM models with a conventional supervised learning method, we modify the 2D network to its 3D version by converting all 2D operations to 3D ones. Since the 3D network significantly increases the memory cost, we reduce the number of inner channels to fit GPU devices, e.g., for a 24 GB GPU with a single $128 \times 128 \times 128$ sub-volume, the number of base inner channels is set to 12, and the channel multipliers are 1, 2, 4, 8, 8 for the five blocks, respectively.

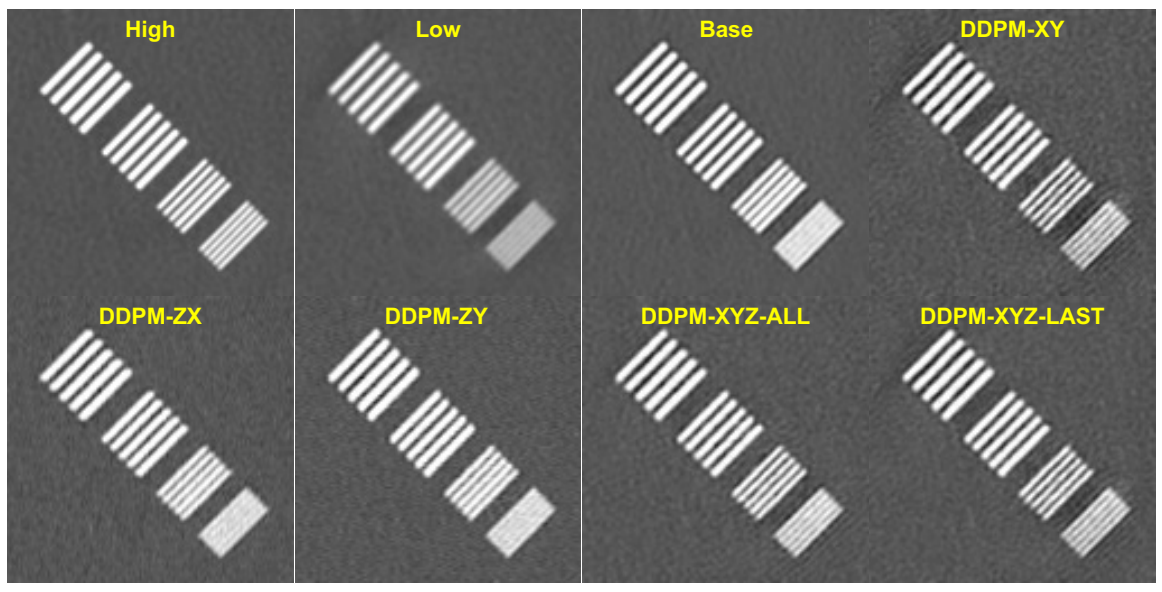

**Figure 1:** In-plane (XY) results.

All other hyper-parameters are identical to those used for 2D DDPMs. Our implementation is based on PyTorch, and it is well-known that automatic mixed precision will significantly improve the training speed. However, we find this sometimes makes the training process unstable in our experiments, so this technique is not used in this study.

## 3 Experiments and Results

In our pilot experiments, we evaluate different variants of DDPMs and the baseline 3D model trained with conventional supervised learning on the test patient CT scan with the inserted line pairs. Note that the line pair patterns are not included in the training dataset. Here we evaluate two variants of 2D joint inference: 1) DDPM-XYZ-ALL alternatively does 2D inference to merge 2D results among all inference steps; 2) DDPM-XYZ-LAST calculates the weighted sum of 2D results that are independently computed in the last step only, increasing the inference time threefold. The in-plane and through-plane results are shown in Figures 1 and 2 respectively. We observe (1) the 2D DDPM trained with in-plane slices visibly improves the resolution of in-plane slices, but its performance on through-plane dimensions is degraded. Also, some line artifacts are generated around the high-frequency line pairs. Generally, the 2D network trained on a specific dimension does not work well on other dimensions. (2) The joint inference that synergizes all-dimension results can clearly improve all-dimension performance in certain respects, e.g., artifacts present in the 2D in-plane results are reduced. (3) DDPMs appear to achieve better super-resolution results than the baseline models, recovering more detail. We also evaluated the presented DDPM and baseline models on anatomical structures. The results in Figure 3 show that, in comparison with the baseline model, DDPM achieves sharper results and the image texture better resembles that of the high-resolution phantoms for both in-plane and through-plane slices.

To quantitatively evaluate the resolution results, we calculate the corresponding modulation transfer function (MTF) of the line pairs in Figures 1 and 2, and the results are shown in Figures 4 and 5 correspondingly, where the best two DDPM results are displayed.

## 4 Discussion and conclusion

Figure 4 suggests that the in-plane resolution is improved by the DDPMs; this improvement is superior to that of the baseline model at higher frequencies, even rivaling the high-resolution reference. MTF comparison for axial resolution is slightly more ambiguous, but the select DDPMs still gen-erally outperform the baseline model (Figure 5). It should be noted that MTF measures resolution by contrast in a sinusoidal pattern, which may not comprehensively evaluate edge sharpness. Nevertheless, these quantitative results, combined with the previous qualitative observations, support the potential of DDPMs in CT super-resolution.

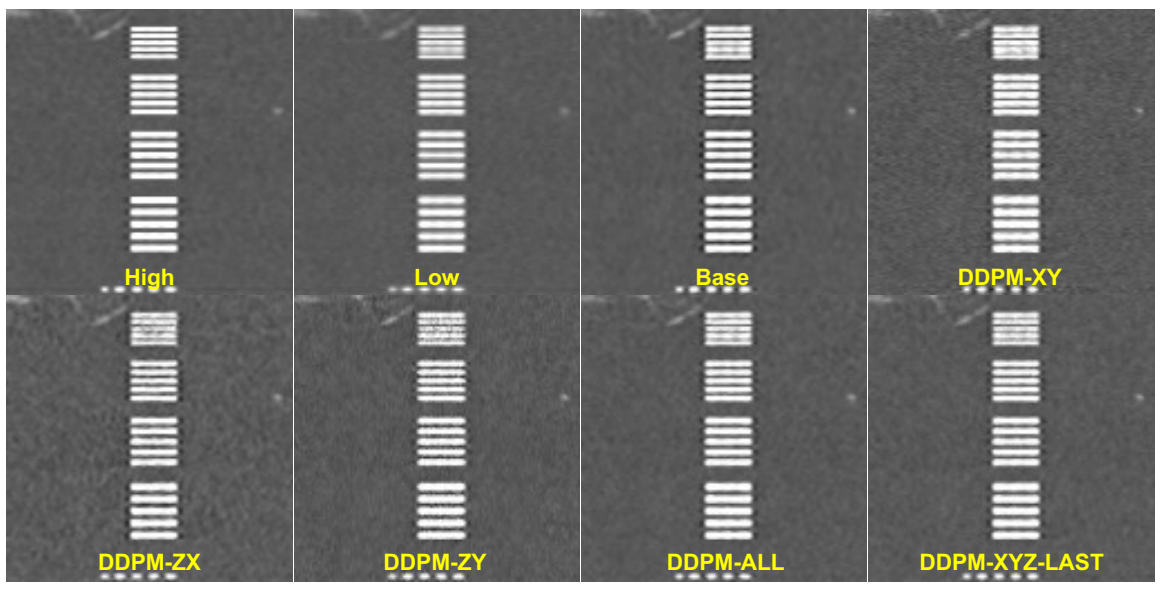

**Figure 2:** Through-plane ($z$–$x$) results.

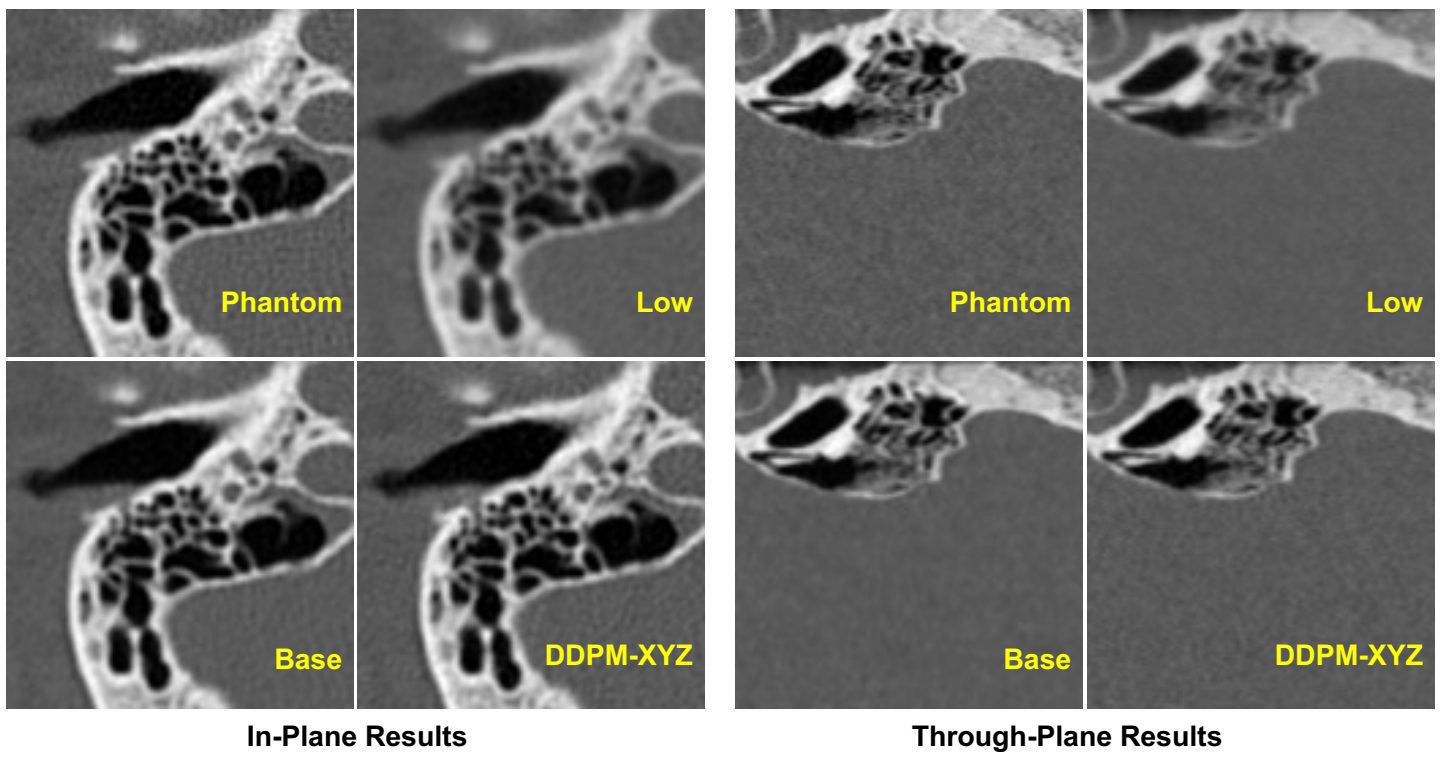

**Figure 3:** Application to CT images of the human temporal bone.

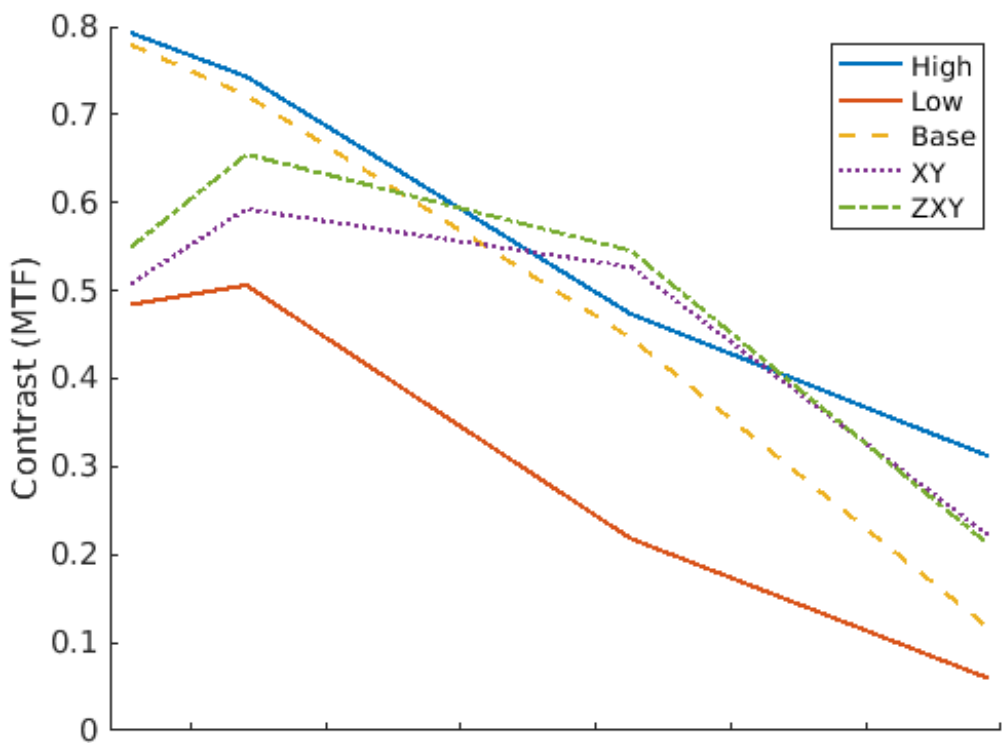

**Figure 4:** MTF Comparison (In-Plane). The frequency axis has been suppressed for proprietary reasons.

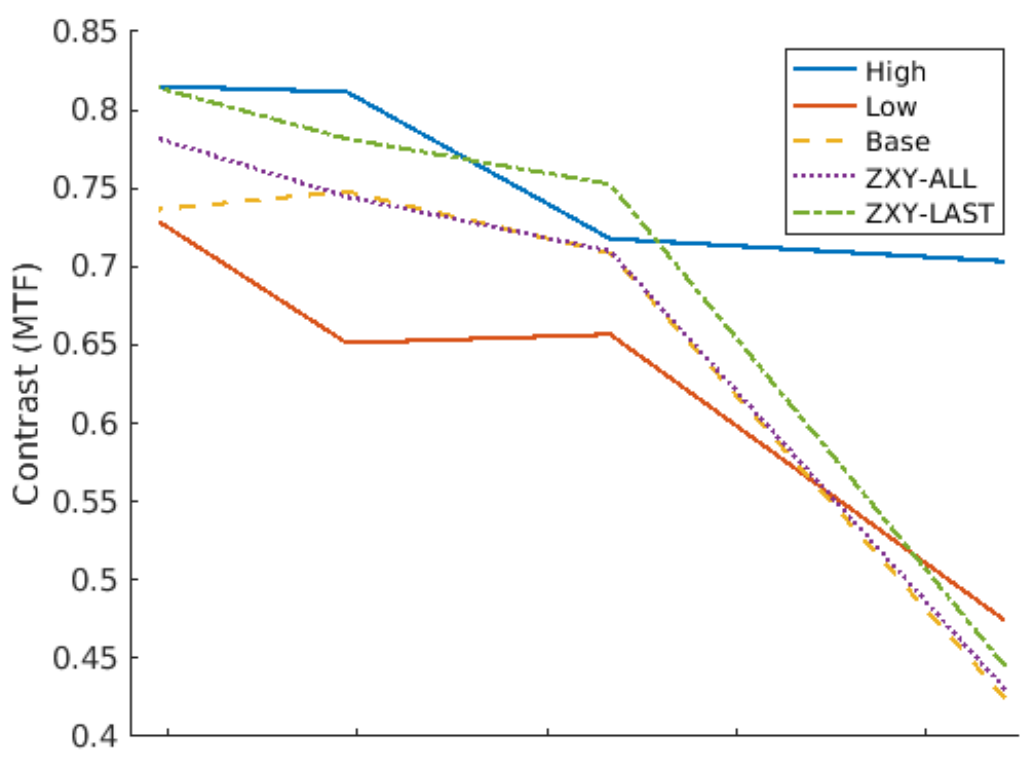

**Figure 5:** MTF Comparison (Through-Plane). The frequency axis has been suppressed for proprietary reasons.

We have demonstrated the effectiveness of conditional DDPMs in the PCCT super-resolution task. We have overcome a major challenge of training high-dimensional DDPMs by training in-plane and through-plane 2D networks, and then synergizing the 2D predictions of all dimensions. Experimental results have demonstrated the effectiveness of the presented method.

# Improving the detective quantum efficiency of detectors in cone beam computed tomography using a hybrid direct-indirect flat-panel imager

C. Orlik[1], A. F. Howansky[1], S. Léveillé[2], S. M. Arnab[2], J. Stavro[1], S. Dow[1], A. Goldan[1], S. Kasap[3], K. Tanioka[1], W. Zhao[1]

[1]Department of Radiology, State University of New York at Stony Brook, Stony Brook, USA

[2]Analogic Canada Corporation, Saint-Laurent, Canada

[3]Department of Electrical and Computer Engineering, University of Saskatchewan, Saskatoon, Canada

**Abstract** Intraoperative volumetric imaging with cone beam computed tomography (CBCT) is now widely used for vascular, neurosurgical and selective internal radiotherapy procedures to guide decision-making and for verification. As its utility grows, there is increasing demand for CBCT to provide the highest performance in spatial and contrast resolution at the lowest possible dose. Tradeoffs in these parameters are principally limited by the detective quantum efficiency (DQE) and modulation transfer function (MTF) of its active matrix flat panel imager (AMFPI). This work investigates these performance metrics for a novel direct-indirect "hybrid" AMPFI under conditions encountered in CBCT imaging. Experimental measurements from a prototype Hybrid AMFPI showed DQE(0) approaching 0.90 and 0.75 at RQA5 and RQA9, respectively. Temporal performance measurements show minimal degradation (lag and ghosting below 2%). To our knowledge, these mark the highest combined MTF and DQE for an energy-integrating imager to date under CBCT imaging conditions.

## 1 Introduction

Cone beam computed tomography (CBCT) is a widely applicable imaging modality, utilized in interventional radiology, vascular interventions, neurosurgery, and radiation therapy imaging on linear accelerations.[1] Clinical translation of CBCT was enabled by the development and commercialization of active matrix flat panel imagers (AMFPIs). Two major detector requirements for CBCT are: (1) high quantum efficiency to maximize signal from the low-dose levels employed, and (2) good temporal performance to minimize image artifacts between frames. These requirements have been met by both direct and indirect conversion AMFPIs.[2] The most developed direct conversion AMFPI consists of an amorphous selenium (a-Se) photoconductor, which has the advantage of excellent spatial resolution, but has limited x-ray quantum efficiency (XQE) at general radiographic conditions (≥ 70 kVp) due to its low atomic number. Indirect AMFPIs utilize highly attenuating scintillator materials, however blur from the generated optical photons limits spatial resolution and results in increased noise.[3,4]

Previous works have demonstrated the feasibility of direct-indirect "Hybrid" AMFPIs for digital radiography/fluoroscopy applications, including CBCT.[5] Hybrid AMFPIs comprise an a-Se layer in optical contact with a scintillator such that it serves as both an x-ray and optical sensor, as shown in Figure 1. The improvement in optical quantum efficiency (OQE) enabled by gain matching via doping a-Se with tellurium (Te) has enabled detective quantum efficiency (DQE) performance better than state-of-the-art flat panel imagers. With an increase in demand for high spatial and contrast resolution for low dose CBCT imaging modalities, we evaluate the imaging performance of Hybrid AMFPI under typical conditions encountered in CBCT.

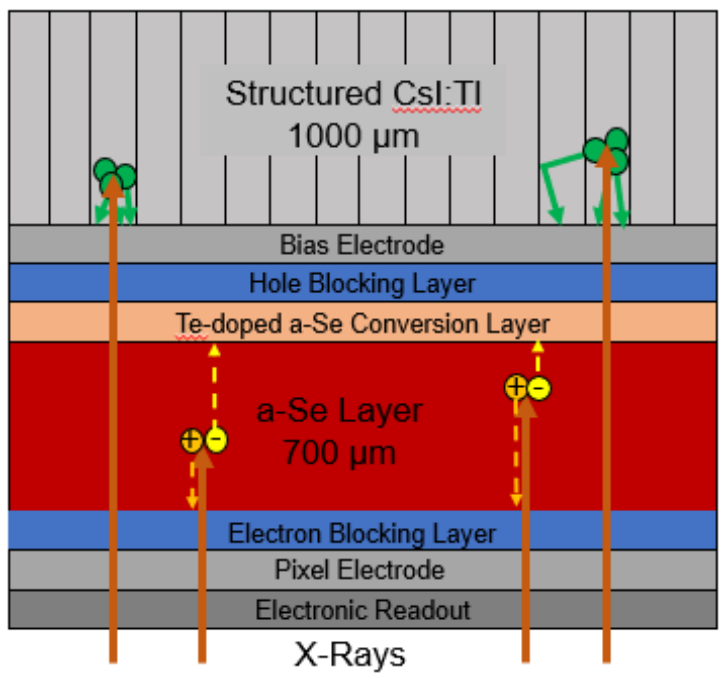

**Figure 1:** Schematic of back-irradiated (BI) Te-doped Hybrid AMFPI. Thicknesses are not drawn to scale.

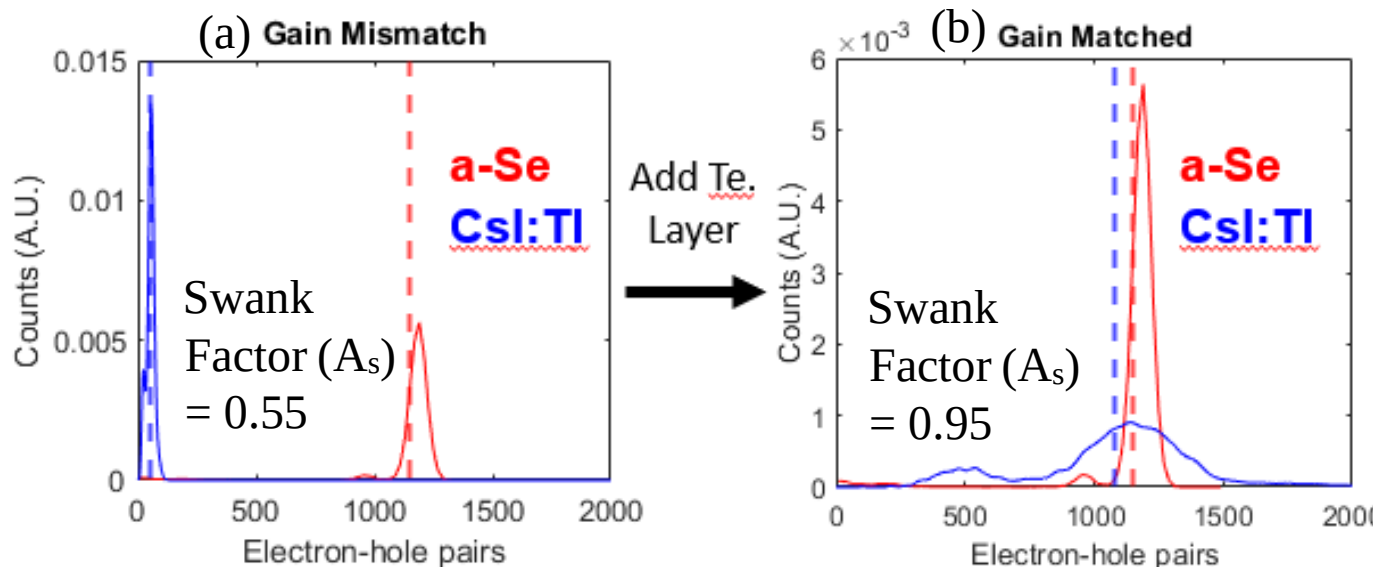

**Figure 2:** Simulated direct (red) signal pulse height spectrum (PHS) using an Am-241 beam, comprising 700 µm of a-Se ($W_{Se}$ = 50 eV) and 700 µm of substrate glass, and measured indirect (blue) signal 1000 µm CsI:Tl using an Am-241 beam PHS scaled to **(a)** OQE = 2.5% (non-doped) and **(b)** OQE = 50% (Te-doped). Dashed lines indicate averages.

## 2 Materials and Methods

A 6.5 x 6.5 cm$^2$ prototype (85 µm pixel pitch) Hybrid AMFPI was fabricated comprising 700 µm a-Se and a removable 1000 µm CsI:Tl scintillator. The x-ray sensitivity, modulation transfer function (MTF) and DQE were measured in a direct AMFPI configuration (i.e. a-Se alone) and Hybrid configuration (i.e. CsI:Tl coupled) under RQA5 and RQA9 beam qualities. Measurements were compared with representative commercial imagers. Temporal performance was investigated via lag and ghost measurements under RQA5 beam quality. To quantify lag, the detector was irradiated for a single frame, then the following 20 dark frames were recorded and the 2D mean pixel value in the same ROI was computed as a function of frame number. To quantify ghosting, a Pb slab was placed on the image receptor and an exposure was taken. The Pb slab was then removed and 9 additional frames were acquired with the same exposure. The mean pixel value of an ROI within the location of the attenuator was normalized to that of the first frame and plotted as a function of exposure number. All experiments were performed in back-irradiation (BI) geometry.

## 3 Results

| (a) | RQA5 | RQA9 |
|---|---|---|
| Direct AMFPI | 443 ADU/μGy | 367 ADU/μGy |
| Hybrid AMFPI | 632 ADU/μGy | 695 ADU/μGy |
| Hybrid Sensitivity Improvement | 43% | 89% |

**Table 1** Summary of x-ray sensitivity for Direct AMFPI (700 μm a-Se only) and Hybrid AMFPI (700 μm a-Se + 1000 μm CsI:Tl).

Table 1 shows the x-ray sensitivity in the direct and Hybrid AMFPI configurations determined from measured characteristic response curves. The direct AMFPI's x-ray sensitivity increased by 43% (RQA5) and 89% (RQA9) upon coupling with CsI:Tl in a Hybrid, owing to a ~10x improvement in the OQE with Te-doped a-Se.

Figure 3 shows the measured presampling MTFs of the direct and Hybrid AMFPI configurations at RQA5 and RQA9 beam qualities, as well as the MTF of an indirect AMFPI using the same CsI:Tl scintillator for comparison. The Hybrid MTF was expectedly lower than the direct AMFPI but markedly higher than the MTF of its indirect component (>4x at $f_{Nyquist}$).

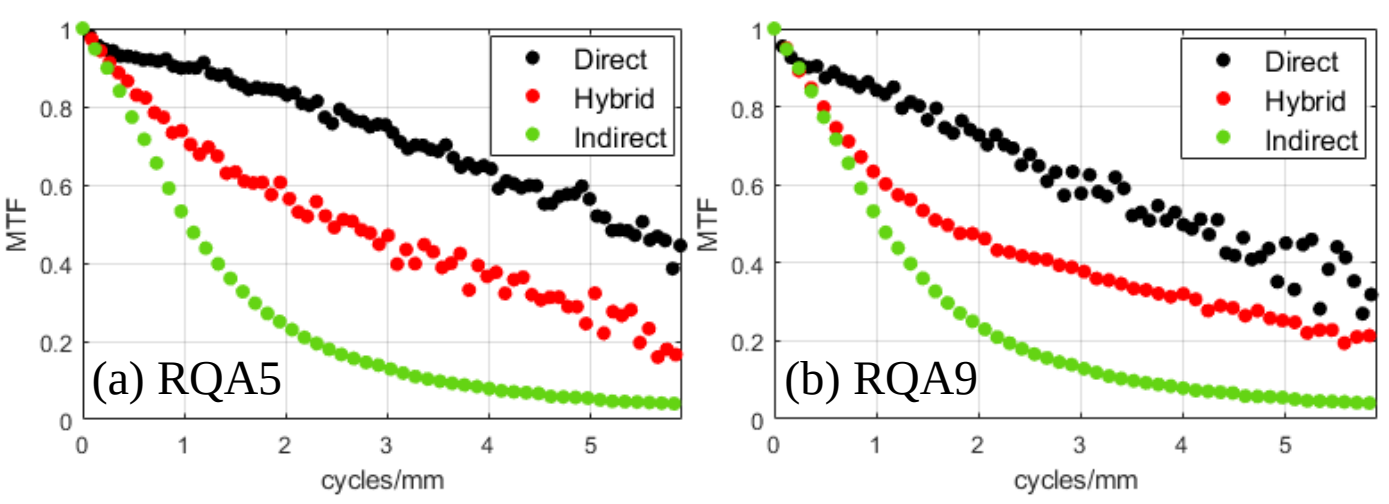

**Figure 3:** Presampling MTFs of the direct and Hybrid AMFPI under **(a)** RQA5 and **(b)** RQA9 beam conditions. Shown for comparison is the presampling MTF of an indirect AMFPI (85 μm pixel pitch) composed by the 1000 μm CsI:Tl scintillator.[6]

Figure 4 shows the DQE of both AMFPI configurations, as well as the results of the previous Hybrid prototype and a state-of-the-art commerical indirect detector (Carestream DRX-1C) for both RQA5 and RQA9 beam qualities.[7] The Te-doped Hybrid DQE was superior to both non-doped Hybrids and current commercial imagers over all spatial frequencies, with DQE(0) approaching 0.90 and 0.75 at

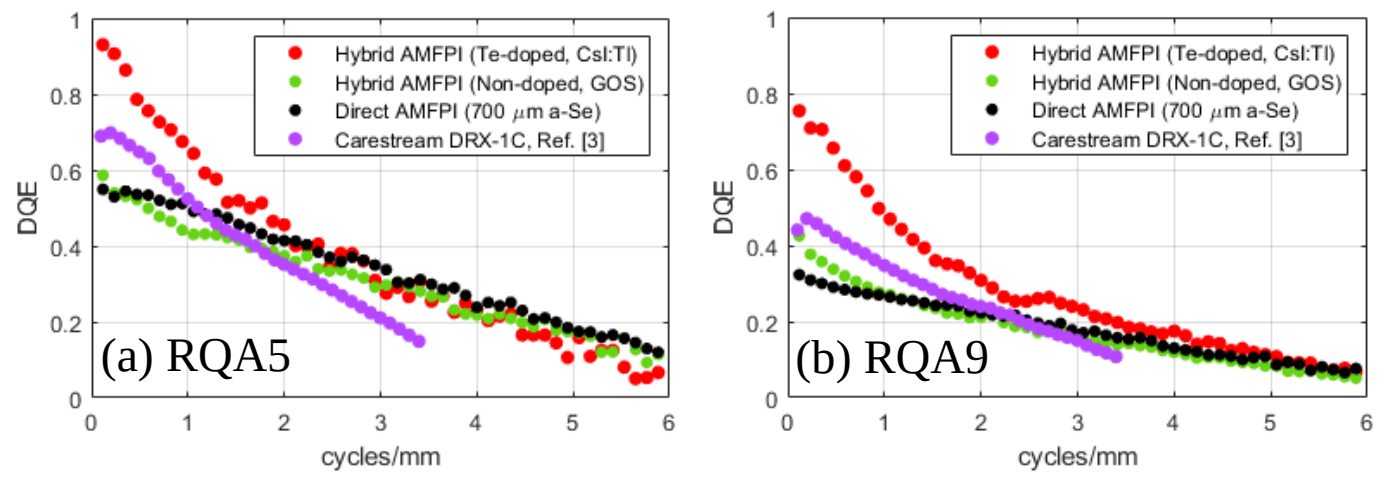

**Figure 4:** DQEs of the direct, new and previous Hybrid AMFPI configurations measured at high exposure for **(a)** RQA5 and **(b)** RQA9 beam quality. Also plotted for comparison is the DQE of the Carestream DRX-1C indirect detector.[7]

RQA5 and RQA9, respectively. To our knowledge, these mark the highest combined MTF and DQE for an energy-integrating imager to date.

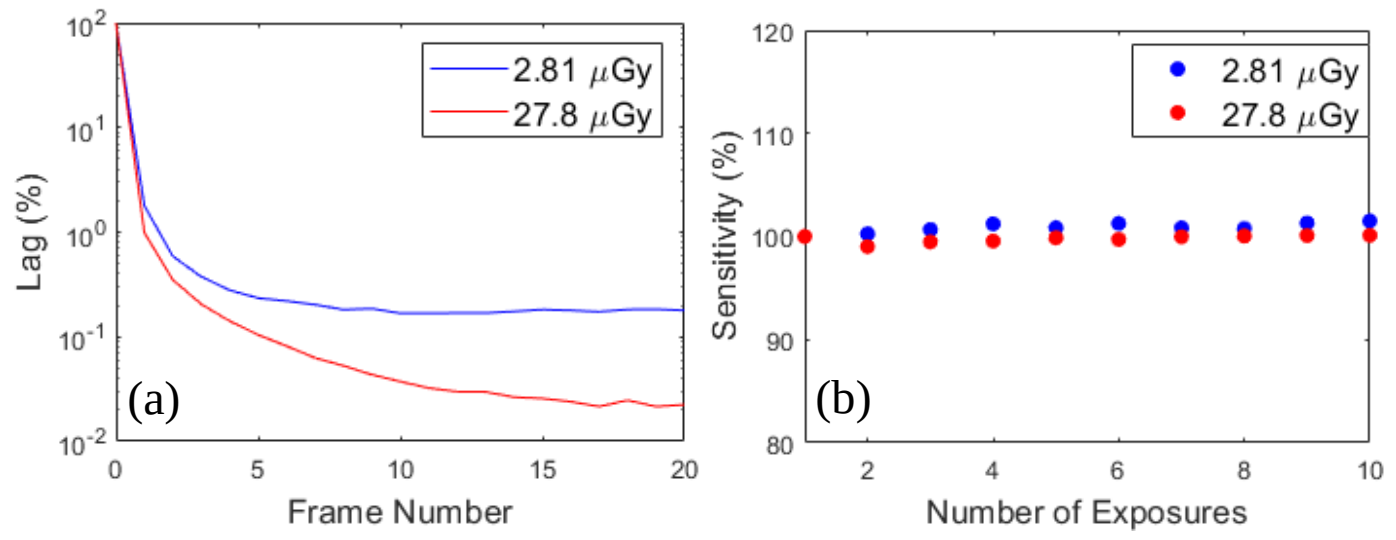

**Figure 5: (a)** Result of lag measurement for 2.81 and 27.8 μGy entrance dose. **(b)** Result of ghost measurement for 2.81 and 27.8 μGy entrance dose.

Figure 5 (a) shows the measured lag, where the zeroth frame is the irradiated frame. The entrance dose levels selected were to replicate conditions during a typical CBCT x-ray frame (2.81 μGy) and extreme conditions (27.8 μGy). The first frame lag is less than 2% for both entrance doses. Figure 5 (b) shows the x-ray sensitivity after repeated flood exposures. The observed change in sensitivity, i.e. ghosting, is negligible (less than 1%).

## 4 Discussion

Our results show improved x-ray sensitivity compared to the previous Hybrid prototype after scintillator coupling (89% increase over direct AMFPI vs 25% at RQA9).[5] This is attributed to the improved OQE provided by the Te-doped layer. This OQE improvement also allows for coupling to high-performing scintillators such as columnar CsI:Tl where the previous Hybrid was limited to blue light (~400 nm peak luminescence) emitting scintillators for better spectral-matching to non-doped a-Se.

Our results show that Hybrid MTF is reduced compared to that of the direct AMFPI, but remains substantially higher than the indirect MTF. This result is expected, as the MTF is comprised of contributions from both the direct AMFPI and the indirect AMFPI signal. In BI geometry, x-rays are preferentially absorbed in a-Se and its high intrinsic MTF dominates that of the scintillator. For higher x-ray energies (e.g., RQA9 beam quality), the MTF is degraded further as more signal contribution comes from the scintillator.

Superior DQE(f) was observed in Hybrid AMFPI for three reasons: First, Hybrid AMFPI maintains high spatial resolution compared to commercial indirect AMFPIs due to BI geometry. Second, the coupled CsI:Tl scintillator improves the x-ray quantum efficiency (XQE) compared to direct AMFPIs, and results in increased low-frequency DQE. Third, the improved OQE of the Te-doped layer prevents degradation of the Swank factor by gain matching between the direct and indirect layers of the Hybrid AMFPI.

Prelimary lag and ghost results show minimal temporal degradation and shows potential for use in real-time imaging applications like CBCT. However, the first frame lag has a dependence on frame rate,[8] which is limited to ~1 Hz with the current prototype. Future work will investigate the temporal performance of Hybrid AMFPI at higher frame rates in a future panel.

## 5 Conclusion

The results show that the Hybrid AMFPI is the highest-performing imager for digital radiography applications as quantified by MTF and DQE. Preliminary temporal performance measurements show that Hybrid AMFPI has the potential to be used in 3D imaging CBCT.

# Unrolled three-operator splitting for parameter-map learning in Low Dose X-ray CT reconstruction

Andreas Kofler[1], Fabian Altekrüger[2,3], Fatima Antarou Ba[3], Christoph Kolbitsch[1], Evangelos Papoutsellis[4,*], David Schote[1], Clemens Sirotenko[5], Felix Frederik Zimmermann[1], and Kostas Papafitsoros[6]

[1]Physikalisch-Technische Bundesanstalt (PTB), Braunschweig and Berlin, Germany
[2]Humboldt-Universität zu Berlin, Department of Mathematics, Berlin, Germany
[3]Technische Universität Berlin, Institute of Mathematics, Berlin, Germany
[4]Finden Ltd, Rutherford Appleton Laboratory, Harwell Campus, Didcot, United Kingdom
[5]Weierstrass Institute for Applied Analysis and Stochastics, Berlin, Germany
[6]School of Mathematical Sciences, Queen Mary University of London, United Kingdom

[*]Corresponding author: epapoutsellis@gmail.com

**Abstract** We propose a method for fast and automatic estimation of spatially dependent regularization maps for total variation-based (TV) tomography reconstruction. The estimation is based on two distinct sub-networks, with the first sub-network estimating the regularization parameter-map from the input data while the second one unrolling $T$ iterations of the Primal-Dual Three-Operator Splitting (PD3O) algorithm. The latter approximately solves the corresponding TV-minimization problem incorporating the previously estimated regularization parameter-map. The overall network is then trained end-to-end in a supervised learning fashion using pairs of clean-corrupted data but crucially without the need of having access to labels for the optimal regularization parameter-maps.

## 1 Introduction

Over recent years, Low Dose X-ray Computed Tomography (LDCT) has received a growing interest in the medical imaging field due to its ability to reduce the radiation dose. Patients are exposed to low levels of radiation by reducing the energy of the photons emitted from the X-ray source. Using traditional and analytic reconstruction methods such as filtered back projection (FBP), several imaging artifacts are introduced, compromising the quality of the reconstructed image and clinical diagnosis.
To overcome this problem, iterative reconstruction methods have been proposed such as algebraic reconstruction technique (ART), simultaneous algebraic reconstruction technique (SART) and projection onto convex sets (POCS). In addition, such reconstruction procedures often require the use of regularization methods in order to eliminate noise and artifacts, such as for instance, the well-known Tikhonov and Total Variation (TV) regularization [1].
The acquired measured tomography data can be described by the equation $\mathbf{z} = \mathbf{A}\mathbf{x}_{\text{true}} + \mathbf{e}$, where $\mathbf{x}_{\text{true}} \in \mathbb{R}^n$ is the ground truth image, $\mathbf{A} : \mathbb{R}^n \to \mathbb{R}^m$ is a linear operator which models the data-acquisition process, i.e. the discretized Radon transform, and $\mathbf{e} \in \mathbb{R}^m$ denotes some random noise component. Regularized iterative methods solve minimization problems of the form

$$\min_{\mathbf{x}} \mathscr{D}(\mathbf{A}\mathbf{x}, \mathbf{z}) + \mathscr{R}(\mathbf{x}), \tag{1}$$

where $\mathscr{D}(\cdot\,,\cdot)$ denotes a data-discrepancy measure and $\mathscr{R}(\cdot)$ a regularization term. A classical example is the TV tomography reconstruction problem under Gaussian noise which can be written as

$$\min_{\mathbf{x}} \frac{1}{2}\|\mathbf{A}\mathbf{x} - \mathbf{z}\|_2^2 + \lambda\|\nabla\mathbf{x}\|_1 + \mathbb{I}_{\{\mathbf{x}>0\}}(\mathbf{x}). \tag{2}$$

A key factor which impacts the quality of the reconstructed image is the careful choice of the regularization parameter $\lambda$ which balances the strength between the regularization and the data fidelity term. Underestimating $\lambda$ yields poor regularization, while overestimating it results in smooth images with an artificial "cartoon-like" appearance. Particularly in medical imaging applications, where images are at the basis of diagnostic decisions and therapy planning, a proper choice of any regularization parameter is crucial.
Employing a single scalar parameter $\lambda$ implies that the regularization is enforced with equal strength for each pixel/voxel. Depending on the application, this might be undesirable due to different features contained in the image. In this case, one can replace the scalar parameter with a spatially varying, i.e. a pixel/voxel dependent one, denoted now by $\Lambda \in \mathbb{R}_+^{qn}$. Here, $q$ denotes the number of directions for which the partial derivatives are computed. Implementation-wise, $\Lambda$ corresponds to a stack of diagonal operators which contain a different regularization parameter for every single pixel/voxel in the respective gradient domain of the image. Then, the resulting problem has the form

$$\min_{\mathbf{x}} \mathscr{D}(\mathbf{A}\mathbf{x}, \mathbf{z}) + \|\Lambda\nabla\mathbf{x}\|_1 + \mathbb{I}_{\{\mathbf{x}>0\}}(\mathbf{x}). \tag{3}$$

However, this problem requires a precise data-adaptive estimation of the spatially varying parameter-map $\Lambda$ which is a highly non-trivial task.

## 2 Methods

One approach for the automatic estimation of the spatially varying regularization parameter $\Lambda$ is employing bilevel optimization techniques. Given $M$ pairs of measured data and the corresponding ground truth $(\mathbf{z}_i, \mathbf{x}^i_{\text{true}})_{i=1}^M$, the general bilevel formulation is

$$\begin{cases} \min_{\Lambda} \sum_{i=1}^{M} l(\mathbf{x}^i(\Lambda), \mathbf{x}^i_{\text{true}}) \text{ subject to} \\ \mathbf{x}^i(\Lambda) = \operatorname*{argmin}_{\mathbf{x}} \mathscr{D}(\mathbf{Ax},\mathbf{z}) + \|\Lambda \nabla \mathbf{x}\|_1 + \mathbb{I}_{\{\mathbf{x}>0\}}(\mathbf{x}), \end{cases} \tag{4}$$

where $l$ is a suitable upper level objective. For instance, if $l(x_1, x_2) = \|x_1 - x_2\|_2^2$, the bilevel problem (4) seeks to compute the parameters $\Lambda$ which are “the best on average”, i.e. PSNR-maximizing, for the given $M$ data pairs. Hence, given some new data $\mathbf{z}_{\text{test}}$ which has been measured in a similar way as $(\mathbf{z}_i)_{i=1}^M$, solving (3) with the precomputed $\Lambda$ will yield a good reconstruction.

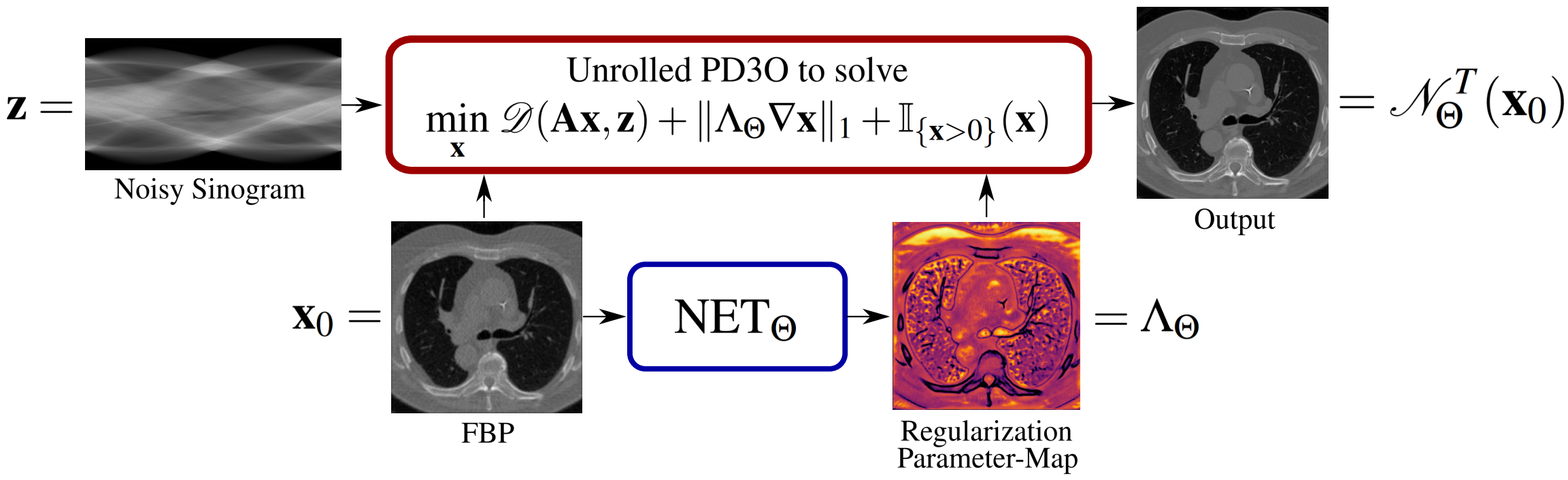

**Figure 1:** Network architecture for the LDCT reconstruction problem. It consists of a sub-network $\mathrm{NET}_\Theta$ that estimates the regularization parameter-map (blue), and a sub-network that unrolls $T$ iterations of the PD3O algorithm (red).

Although this scheme has been extensively studied both for scalar and spatially varying regularization parameters, it has been mainly applied in image denoising applications, i.e. $\mathbf{A} = \mathbf{I}_n$ with Gaussian noise, see [2–4]. Further, unsupervised approaches employing upper level energies that do not depend on the ground truth $\mathbf{x}_{\text{true}}$, i.e., $l := l(\mathbf{x}(\Lambda))$, have also been considered in a series of works [5–8]. Even though these bilevel optimization methods are typically supported by rigorous mathematical theories, they are computationally demanding which has limited their use on tomographic problems.

## 2.1 An Unrolled Neural Network Framework

Here, inspired by the recent success of unrolled neural networks (NNs) [9], we consider an unrolled neural network approach in order to learn the regularization parameter $\Lambda$. The proposed framework is summarized in Figure 1 and it is outlined next.

An unrolled NN which corresponds to an implementation of an iterative scheme of finite length is constructed to approach the solution of problem (2) assuming a *fixed* regularization parameter-map. Within the unrolled NN, the regularization parameter-map is estimated from the input data via a sub-network $\mathrm{NET}_\Theta$ and is used throughout the whole reconstruction scheme. To be more precise, given some initial estimate $\mathbf{x}_0$ we work with an iterative scheme (speficied in the next section)

$$\mathbf{x}_T = S^T(\mathbf{x}_0, \mathbf{z}, \Lambda, \mathbf{A}), \quad T = 0, 1, 2, \ldots, \tag{5}$$

that solves (3) in the limit $T \to \infty$. Then, for some fixed number of iterations $T \in \mathbb{N}$, our unrolled NN reads as follows:

$$\begin{cases} \Lambda_\Theta = \mathrm{NET}_\Theta(\mathbf{x}_0), \\ \mathbf{x}_1 = S^1(\mathbf{x}_0, \mathbf{z}, \Lambda_\Theta, \mathbf{A}), \\ \quad \vdots \\ \mathbf{x}_T = S^T(\mathbf{x}_0, \mathbf{z}, \Lambda_\Theta, \mathbf{A}). \end{cases} \tag{6}$$

Here, $\mathrm{NET}_\Theta$ denotes a U-Net [10] with learnable parameters $\Theta$. We denote by $\mathcal{N}_\Theta^T$ the overall resulting network, i.e.

$$\mathcal{N}_\Theta^T(\mathbf{x}_0) = S^T(\mathbf{x}_0, \mathbf{z}, \Lambda_\Theta, \mathbf{A}) = S^T(\mathbf{x}_0, \mathbf{z}, \mathrm{NET}_\Theta(\mathbf{x}_0), \mathbf{A}).$$

The unrolled NN can then be end-to-end trained in a supervised manner on a set of input-target image-pairs. This resulting network can be identified as a pipeline that combines in a sequential way 1) the estimation of the regularization parameter-map which is adapted to the data $\mathbf{z}$ (and hence in medical imaging to the new patient) and 2) the iterative scheme that solves the image reconstruction problem.

## 2.2 Primal-Dual Three-Operator Splitting

The iterative scheme selected here for the LDCT reconstruction problem is the Primal-Dual Three-Operator (PD3O) splitting algorithm. The PD3O was introduced in [11] and it is a generalized version of the Primal-Dual Hybrid Gradient (PDHG) algorithm [12]. It is used to minimize objectives that consist of a proximable function $g$, a composite function $f$ with the linear operator $\mathbf{K}$ and a differentiable function $h$ with a Lipschitz constant $L$:

$$\min_{\mathbf{x}} f(\mathbf{Kx}) + g(\mathbf{x}) + h(\mathbf{x}).$$

The algorithm is summarized in Algorithm 1 and explained next. Unlike the standard $\mathrm{L}^2$-squared fidelity term that is commonly used in tomography reconstruction problems with

**Algorithm 1** Unrolled PD3O algorithm

**Input:** $L = \text{Lip}(\nabla h)$, $\tau = 2/L$, $\sigma = 1/(\tau\|\mathbf{K}\mathbf{K}^\mathsf{T}\|)$, initial guess $\bar{\mathbf{x}}_0$

**Output:** reconstructed image $\mathbf{x}_{\text{TV}}$

1: $\mathbf{p}_0 = \bar{\mathbf{x}}_0$
2: $\mathbf{q}_0 = \mathbf{0}$
3: **for** $k < T$ **do**
4: $\quad \mathbf{q}_{k+1} = \text{prox}_{\sigma f^*}(\mathbf{q}_k + \sigma \mathbf{K}\bar{x}_k)$
5: $\quad \mathbf{p}_{k+1} = \text{prox}_{\tau g}(\mathbf{p}_k - \tau\nabla h(\mathbf{p}_k) - \tau \mathbf{K}^\mathsf{T}\mathbf{q}_{k+1})$
6: $\quad \bar{\mathbf{x}}_{k+1} = 2\mathbf{p}_{k+1} - \mathbf{p}_k + \tau\nabla h(\mathbf{p}_k) - \tau\nabla h(\mathbf{p}_{k+1})$
7: **end for**
8: $\mathbf{x}_{\text{TV}} = \mathbf{x}_T$

Gaussian noise, here we employ the Kullback-Leibler divergence which is more suitable to describe the noise distribution of the measured tomographic data $\mathbf{z}$. We have that $\mathbf{z} = \mathbf{A}\mathbf{x} + \mathbf{e}$, where

$$\mathbf{e} = -\mathbf{A}\mathbf{x} - \log(\tilde{\mathbf{N}}_1/N_0)\ ,\ \tilde{\mathbf{N}}_1 \sim \text{Pois}(N_0 \exp(-\mathbf{A}\mathbf{x}\mu)).$$

We denote with $\mu$ and $N_0$ the normalization constant and the mean photon count per detector bin without attenuation, respectively. The data-discrepancy in (3) can be derived from a Bayesian viewpoint and is

$$\mathscr{D}(\mathbf{A}\mathbf{x},\mathbf{z}) = \sum_{i=1}^{m} e^{-(\mathbf{A}\mathbf{x})_i\mu}N_0 - e^{-\mathbf{z}_i\mu}N_0\big(-(\mathbf{A}\mathbf{x})_i\mu + \log(N_0)\big), \tag{7}$$

see [13] for more details. To configure PD3O for (3) we define the following

$$f(\mathbf{q}) = \|\Lambda\mathbf{q}\|_1, \quad g(\mathbf{p}) = \mathbb{I}_{\{\mathbf{p}>0\}}(\mathbf{p}), \quad \mathbf{K} = \nabla,$$
$$h(\mathbf{p}) = \sum_{i=1}^{m} e^{-\mathbf{p}_i\mu}N_0 - e^{-\mathbf{z}_i\mu}N_0\big(-\mathbf{p}_i\mu + \log(N_0)\big).$$

Notice that with the standard $L^2$-squared fidelity term, it is sufficient to use the PDHG algorithm since its convex conjugate has a closed-form proximal operator, which is not the case with (7). However, the additional function in the PD3O algorithm allows to express the data discrepancy in the differentiable term $h$. Note that $\nabla h$ is not globally Lipschitz continuous but due to the non-negativity constraint, we only have to consider $\nabla h(\mathbf{p})$ for $\mathbf{p}$ with non-negative entries. Consequently, we can find an upper bound of the Lipschitz constant of $\nabla h$ by $\text{Lip}(\nabla h) \leq \|\mathbf{A}\|^2\mu^2 N_0$.

## 3 Results

To evaluate our proposed unrolled NN, we use the LoDoPaB dataset [14] for low-dose CT imaging. It is based on scans of the Lung Image Database Consortium and Image Database Resource Initiative which serve as ground truth images, while the measurements are simulated. The dataset contains 35820 training images, 3522 validation images and 3553 test images. Here the ground truth images have a resolution of $362 \times 362$ on a domain of $26\text{cm} \times 26\text{cm}$. We only use the first 300 training images and the first 10 validation images. For the forward operator we consider a normalization constant $\mu = 81.35858$, the mean photon count per detector bin $N_0 = 4096$ as well as 513 equidistant detector bins and 1000 equidistant angles between 0 and $\pi$.

In Figure 2 we compare the FBP reconstruction with the PD3O reconstructions where we use (i) a scalar parameter ($\lambda$), chosen to maximize the PSNR "on average", and (ii) our computed spatially dependent parameter map ($\Lambda_\Theta$). Using the latter, we obtain a significant improvement in the reconstruction both visually and in terms of quality measures, e.g., PSNR and SSIM. In particular, sharp edges are retained, while the constant regularizing parameter results in blurry and blocky-like reconstructions. One can observe that the network attributes higher regularization parameters to image content with smooth structures while it yields lower regularization parameters at the edges to prevent smoothing.

## 4 Discussion

We have presented a data-driven approach to automatically estimate spatially dependent parameter-maps for TV regularization for the low dose X-ray CT tomography reconstruction problem. Although only the TV regularization is considered in this paper, higher order or combinations of regularizers can be used for different CT applications, see [15, 16]. Moreover, our unrolled framework is quite flexible and can be easily used for other modalities such as qualitative and quantitative MRI reconstruction, image denoising as well as their dynamic versions, see [17]. Finally, more sophisticated network architectures than the U-Net have been proposed recently, e.g. [18, 19], which could be potentially adopted for the estimation of the regularization parameter-maps as well.

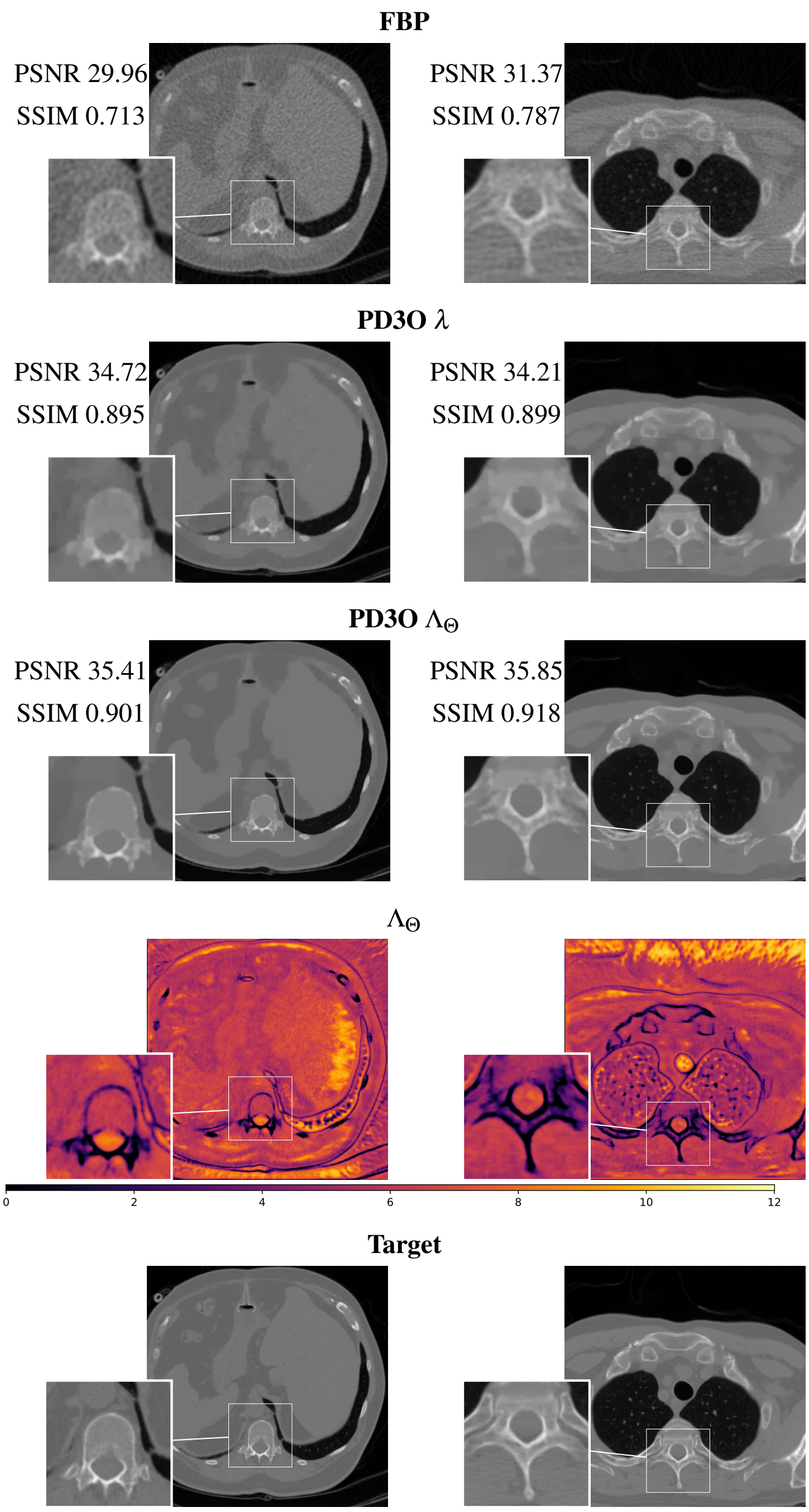

**Figure 2:** Different reconstructions obtained with PD3O employing a scalar regularization and the regularization parameter-maps obtained with the proposed CNN. From top to bottom: initial FBP-reconstruction, PD3O $\lambda$, PD3O $\Lambda_\Theta$, spatial regularization parameter-map and ground truth image. The corresponding PSNR and SSIM values are given in the top left corner of the image.

# Spherical acquisition trajectories for X-ray Computed Tomography with a robotic sample holder

Erdal Pekel[1,2], Martin Dierolf[2,3], Franz Pfeiffer[2,3], and Tobias Lasser[1,2]

[1]Department of Computer Science, School of Computation, Information and Technology, Technical University of Munich, Munich, Germany
[2]Munich Institute of Biomedical Engineering, Technical University of Munich, Munich, Germany
[3]Department of Physics, School of Natural Sciences, Technical University of Munich, Munich, Germany

**Abstract** In this work we present an X-ray computed tomography setup that integrates a seven degree of freedom robotic arm as a sample holder. The path planning and robot control algorithms are optimized for seamless execution of spherical trajectories. A precision manufactured sample holder part is attached to the robotic arm for the calibration procedure. We present experimental results with the robotic sample holder where a sample measurement on a spherical trajectory achieves improved reconstruction quality compared to a conventional circular trajectory. The proposed system is a step towards higher image reconstruction quality in flexible X-ray CT systems.

## 1 Introduction

In recent years, industrial robotic arms have become more affordable for a broader audience thanks to mass production of electrical and mechanical components. At the same time, their control and software integration has become more feasible with the emergence of open source software initiatives that aim to standardize and simplify the use of such robotic arms. The lower entry-barrier for robotic arms also offers new opportunities for X-ray computed tomography systems which benefit greatly from the improved flexibility. Capturing two-dimensional absorption images from an increased number of angles can increase reconstruction image quality and completeness.

In recent work we introduced a flexible robotic arm with seven degrees of freedom (DoF) as a sample holder within a laboratory X-ray computed tomography (CT) setup [1]. The arm adds flexibility to the setup as a sample holder by enabling arbitrary rotations and placement of the sample and hence allows non-standard trajectories that are not restricted in their sequence, in contrast to conventional circular or helical trajectories. We also introduced a suitable calibration mechanism in order to determine the exact positioning of the sample from the image. The calibration mechanism requires a sample holder part that is attached to the robotic arm, which was also introduced in [1].

In the following we present our work on the optimization of various aspects of the robotic sample holder for seamless execution of spherical trajectories. The sample holder part that is attached to the robotic arm is modified for improved coverage of spherical trajectories. We present experimental results that demonstrate improved reconstruction quality of spherical trajectories which can be executed by our system.

## 2 Methods

In this section the methods for executing spherical trajectories with the robotic arm as a sample holder in a laboratory X-ray CT setup are discussed in detail. After introducing the hardware components of the system, more specific aspects like path planning, sphere sampling, and reconstruction are described.

### 2.1 Hardware setup

The hardware components of the system are displayed in Fig. 1. The main difference to a conventional X-ray CT setup is the seven DoF robotic arm *Panda* from the manufacturer FRANKA EMIKA [2]. It has a maximum reach of 855 mm and a repeatability of 0.1 mm when repeatedly moved from a specific starting position to a goal position on a fixed trajectory. It has two fingers that can move on a linear axis and grasp objects. The maximum allowed payload is 3 kg.
Two *Intel Realsense D435* depth cameras capture the movements of the robot and provide 3d information about the surroundings as a point cloud. The cameras are connected directly to the workstation with the control software and are used for the collision detection mechanism described in [1].
The robotic arm is mounted on an optical table inside a radiation shielding enclosure for X-ray CT which houses the X-ray source and the detector (see Fig. 1a). The detector (Varex XRD 4343) has a maximum resolution of 2880x2880 and is connected to a different workstation on the network that exposes the raw 16-bit grayscale images. The robotic arm can be turned off in case of emergency from outside of the safety hutch with a power switch.

### 2.2 Sample holder part

The sample holder part is a critical component of the system as it allows the robotic arm to grasp samples of arbitrary shape and is a fundamental part of the calibration process where the position and orientation of the sample is identified. The sample holder part consists of two parts. The gripper part is where the robot's fingers can grasp the holder steadily. The cylinder part fulfills the purpose of placing a helix of fiducial markers on a cylinder next to the sample. The lower part is called *gripper part* throughout this paper and it can

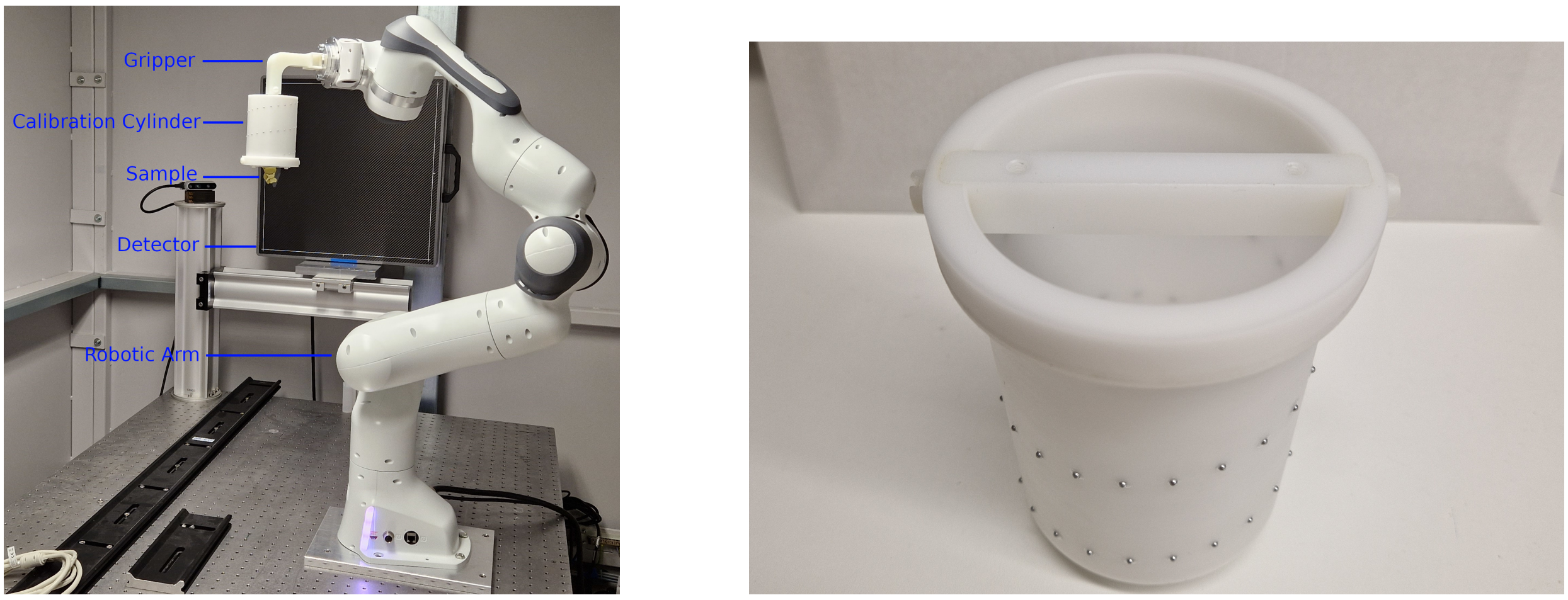

(a) Lab photo (b) Cylinder part with calibration structure (helix)

Figure 1: Hardware setup. In a the robotic arm is mounted on a table with the source and the detector inside a safety hutch. The source to robot distance is 136 cm and robot to detector distance is 79 cm. b depicts the cylinder part which houses the geometric structure for calibration.

also be directly mounted to the last link of the robotic arm when the hand is unmounted from the arm as in Fig. 1a. Prior to attaching the sample holder part to the robotic arm, the sample needs to be glued to the mounting plate which is inserted into the cylinder from the top at the intended position (see Fig. 1b).

The cylinder is 118 mm tall and 50 mm in diameter inside. The reference structure embedded in the sample holder is a helix which is made up of 50 embedded aluminium spheres of 2 mm diameter. These spheres were fixed manually in notches that were included in the design process of the holder. The spheres appear as circles on the detector images that are segmented during calibration.

Compared to our previous work, we modified the sample holder part by introducing a gripper part that connects the last link of the robotic arm with the cylindrical part of the holder. This enables the use of different gripper shapes (straight, curved) in varying lengths which enables the robotic arm to reach different areas of more specific trajectories, e.g. spherical trajectories. The gripper part can be mounted to the cylinder with a screwing mechanism.

## 2.3 Path planning and robot control

With the path planning procedure, our system exposes an abstract interface to the user for planning and executing advanced trajectories. A trajectory consists of a series of *way-points* that are approached by the robotic arm in the given order. At each way-point, the arm stops and the detector is triggered for capturing a detector image. After successful image acquisition, the robotic arm continues trajectory execution. The user specifies the parameters for sampling the way-points on the trajectory from the user interface. Given these parameters, the system first samples the way-points depending on the trajectory type. Subsequently, the underlying motion planning pipeline plans a path from each way-point to its successor. If way-point $i+1$ cannot be reached from way-point $i$, e.g. because there is no collision-free path, then a path from $i$ to way-point $i+2$ is planned and $i+1$ is marked as not reachable. This means that we know that no detector image will be captured for way-point $i+1$ before executing the trajectory. Finally, these paths are connected to each other and the output is a trajectory that starts at the robotic arm's current position and passes all way-points in the given order.

## 2.4 Sphere sampling

In order to generate a spherical trajectory we need to sample points covering the surface of a sphere. Each point represents a rotation of the sample.

Sampling a fixed number $n$ of points on the sphere is a well studied problem [3–6]. The goal is to distribute the points uniformly on the sphere's surface. Trivial approaches like sampling on each of the two polar axis independently and combining the samples to get 3d coordinates does not lead to a uniform sampling on the sphere surface. In this work we utilized *HEALPix* for this purpose [7].

With HEALPix, the surface of the sphere is partitioned into a fixed number of areas of equal size. The centers of these areas are the sampled points on the sphere. The discretization number $N_{pix}$ determines the resulting number of points on the sphere. For our experiments we chose $N_{side} = 10$, which results in $N_{pix} = 12 \cdot 10^2 = 1200$ pixels on the grid and hence 1200 potential way-points on the spherical trajectory.

## 2.5 Calibration

The calibration procedure tackles the issue that the robotic arm does not sufficiently accurately place the sample at the desired position due to inaccurate path planning and inaccurate electrical motors at its joints. Reading the sensors of the robotic arm and deducing the sample's current position is also insufficient to determine the correct position as inaccurate values are reported. However, the exact position of the sample at each view is required for the reconstruction. With the calibration procedure we are able to identify the actual positions and orientations of the sample from the detector images. For the calibration, a sample holder part with an embedded geometric reference structure that can be identified on the detector images is necessary and was introduced in section 2.2.

### 2.6 Reconstruction

The tomographic reconstruction we used around 900 equidistant X-ray projections along a circular or spherical trajectory sized $720 \times 720$ pixels with a spacing of $600\,\mu m$. For the spherical trajectory the number of projections varied by $\pm 8\%$. The reconstruction volume consisted of $720 \times 720 \times 720$ isotropic voxels with a spacing of $38\,\mu m$. Using our C++ reconstruction framework *elsa* [8], reconstruction was performed using an iterative conjugate gradient solver run for 30 iterations on a Tikhonov regularized weighted least squares problem, with the Josephs method for X-ray transform discretization and cone beam geometry. Further iterations showed no improvement on the cost function.

### 2.7 Software stack

The central part of our software stack is the *Robot Operating System* (*ROS*) [9] which is a middleware for the communication of independent processes across a network. Robot manipulation is accomplished with the *MoveIt!* framework [10, 11] and the *franka_ros* configuration package [12]. For image processing tasks and the circle segmentation we use *OpenCV* [13], for multithreading on the CPU *OpenMP* [14] and on the GPU *OpenCL* [15] and for the tomographic reconstruction *elsa* [8]. The sphere discretization in section 2.4 was implemented with the HEALPix C++ and Python interfaces [7, 16].

### 2.8 CT measurements

We conducted two experiments where the sample in Fig. 2 is measured on two different trajectories (circular and spherical) using the robotic arm. It consists of two separate parts: a bunny toy brick (Fig. 2b, left) and a solid piece of polyvinyl chloride (PVC) with a thickness of 4 mm (Fig. 2b, right). We chose this composition because these two parts differ significantly in their absorption rate, which is helpful for comparing circular and spherical trajectories in their reconstruction performance with respect to image quality. We have cut the absorber plate in a non-orthogonal shape relative to the mounting plate and arranged it next to the toy brick in order to cause beam-hardening artifacts in the reconstructions of our experiments with this sample.

For each CT measurement, the images were acquired with a source voltage of 45 kV, source power of $1445\mu$A, and exposure time of 1s. In Fig. 3 the reconstruction of our sample is shown from three different perspectives (YX, YZ and ZX) for the two trajectory types and the straight gripper part. All volumes are registered with each other with the calibration process, as the center of the helix structure serves as the coordinate system's origin.

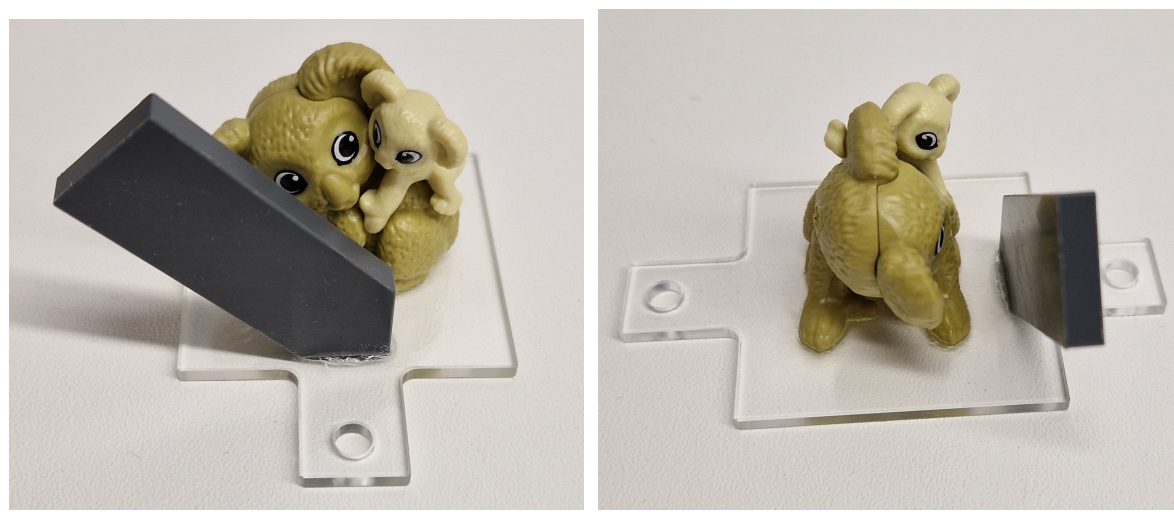

(a) Sample side (b) Sample front

Figure 2: Sample. The sample consists of the object of interest (a toy brick) and an absorber (polyvinyl chloride plate) which were both glued to a Plexiglass mounting plate by hand. The toy brick has dimensions 31 x 21 x 31 mm and the absorber has a thickness of 4 mm. The absorber plate has a significantly higher X-ray contrast absorption rate than the toy brick. We aim to introduce beam hardening artifacts with this property in the reconstructions and evaluate the performance of different trajectory types in tackling this issue.

We can see in Fig. 3 that the slices depicted in the top row (spherical trajectory, straight gripper) are sharper overall when compared to the slices in the bottom row (circular trajectory, straight gripper). When examining the region between the absorber at the top and the toy brick in the middle for the images in the left column, we can see that the slice of the experiment with the spherical trajectory does not cause artifacts, hence this area is truly black when compared to the slice on the bottom where we can spot white traces. We can also spot great differences for the slices in the center and right columns of Fig. 3, e.g. on the right column the inner structure of the toy brick is much sharper for the spherical trajectory (top right) when compared to the circular (lower right).

Another critical observation is that the absorber causes artifacts in two orthogonal directions with the two different trajectory types. In case of the circular trajectory the artifacts are parallel to x-ray beams. For the spherical trajectory the artifacts are parallel to absorber and orthogonal to the x-ray beams.

### 2.9 Future work

We plan to improve the proposed system in the future in several ways.

New gripper types could be designed that maximize the coverage of the sphere surface. The cylinder part of the sample holder with the mounting plate could also be improved for mounting bigger and heavier samples. Currently, the sample is glued to the mounting plate which could cause issues for heavier samples.

Moreover, experiments with base-scans for optimized trajectories are subject of future work. The system is expected to fully benefit from the flexibility of the robotic sample holder once it is able to determine highly-absorbing parts of the spherical trajectory. We will enable this with a scanning procedure that takes place before the actual experiment (base-scan) which is also subject of future work.

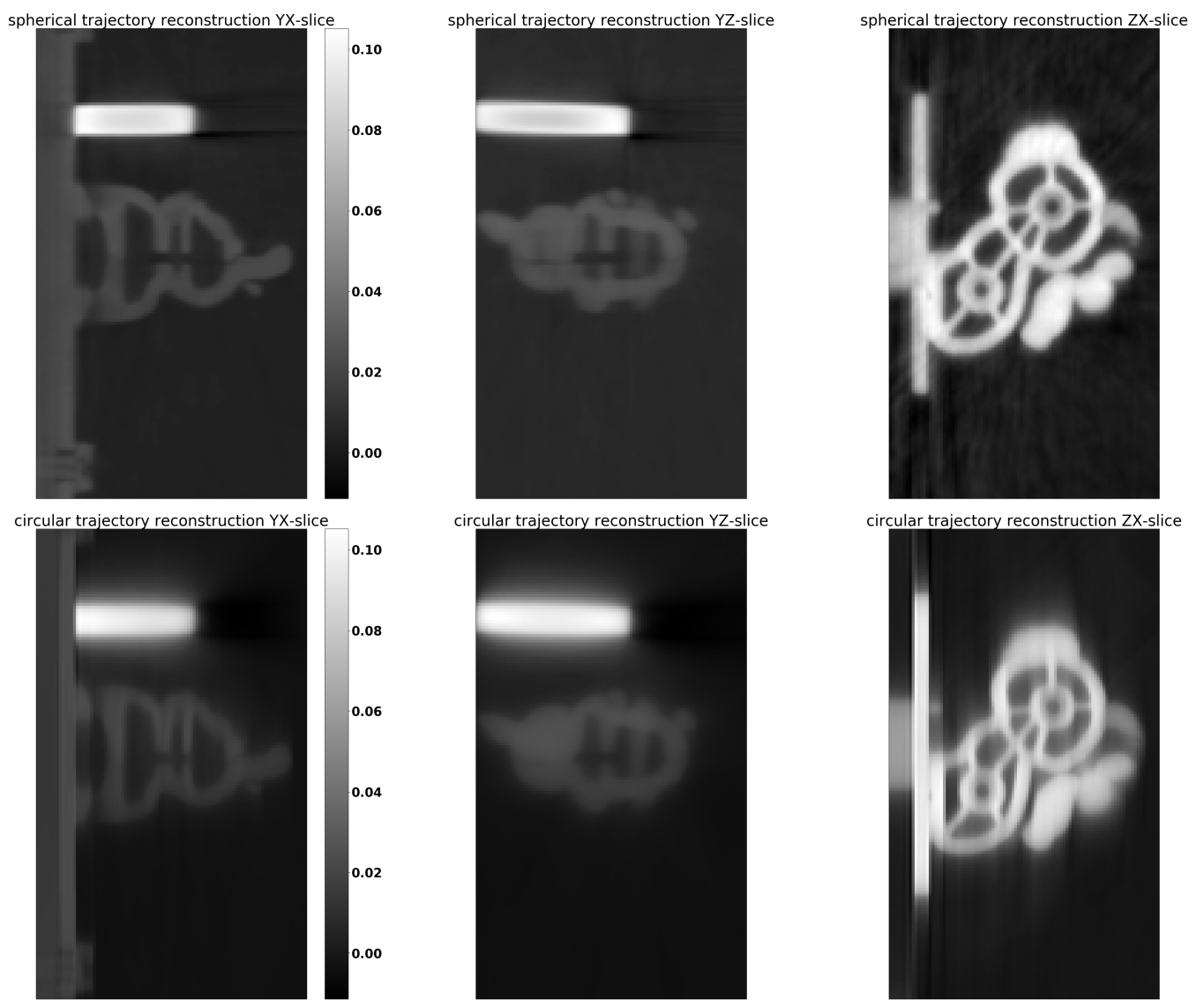

Figure 3: Experimental Results. A sample was measured and reconstructed with the robotic arm with the straight gripper part in order to compare the circular trajectory (conventional) with the spherical trajectory (advanced). The reconstruction volumes are registered and aligned with our calibration algorithm (see section 2.5). The detector images were binned with $4*4$ and the reconstruction volume has dimensions $720^3$. A zoom factor of 5x was applied to the slices to crop the region of interest. Our observation is that the reconstruction of the measurements with the spherical trajectory (top row) are superior compared to the reconstruction of the circular trajectory (bottom row). Qualitatively, there are less artifacts and the image is sharper.

## 3 Conclusion

In this work we have demonstrated the use of a seven DoF robot as a sample holder for the acquisition of spherical X-ray computed tomography trajectories using our unified software package with path planning, collision detection and calibration. We have stated at the beginning that the image quality of 3d reconstructions would benefit greatly from spherical trajectories when the sample contains highly absorbing parts. Our findings have confirmed that with a spherical trajectory the image quality is superior qualitatively when compared to a conventional circular trajectory.

# Preliminary de-multiplexing of projection images using temporal shuttering in a brain-dedicated SPECT system

Sophia Pells[1], Navid Zeraatkar[1,2], Kesava S. Kalluri[1], Stephen C. Moore[3], Micaehla May[4], Lars R. Furenlid[4,5], Phillip H. Kuo[5], and Michael A. King[1]

[1]Department of Radiology, University of Massachusetts Chan Medical School, Worcester, MA, USA
[2]Siemens Medical Solutions USA, Inc., Knoxville, TN, USA
[3]Perelman School of Medicine, University of Pennsylvania, Philadelphia, PA, USA
[4]James C. Wyant College of Optical Sciences, The University of Arizona, Tucson, AZ, USA
[5]Department of Medical Imaging, The University of Arizona, Tucson, AZ, USA

**Abstract**
Mulitplexing (mux) can increase sensitivity in SPECT imaging, but comes at the cost of increased ambiguity and can lead to image-degrading artefacts during reconstruction. Here, an algorithm was used to correct for mux in projection images. A model of the brain-dedicated multi-pinhole AdaptiSPECT-C system was modified to increase mux and used to test the performance of the de-multiplexing algorithm with digital phantoms. AdaptiSPECT-C can independently shutter each of its 120 pinholes, so the de-multiplexing algorithm was tested for acquisitions with and without a mux-free projection frame. The de-multiplexing algorithm was shown to improve the uniformity in reconstructed images of a uniform sphere of activity and uniformity was improved further with the inclusion of a multiplexing-free projection frame. In addition, the de-multiplexing algorithm improved the structural similarity and activity recovery in a brain perfusion phantom.

## 1 Introduction

Pinhole collimation can provide high spatial resolution in Single-Photon Emission Computed Tomography (SPECT) but can lead to low sensitivity, especially when small pinholes are used for high-resolution imaging. Multiplexing (mux) can be used to improve sensitivity, by permitting overlapping projection images to be acquired through multiple pinholes simultaneously. However, mux leads to ambiguity in the origin of counts and can cause artefacts to appear during image reconstruction.

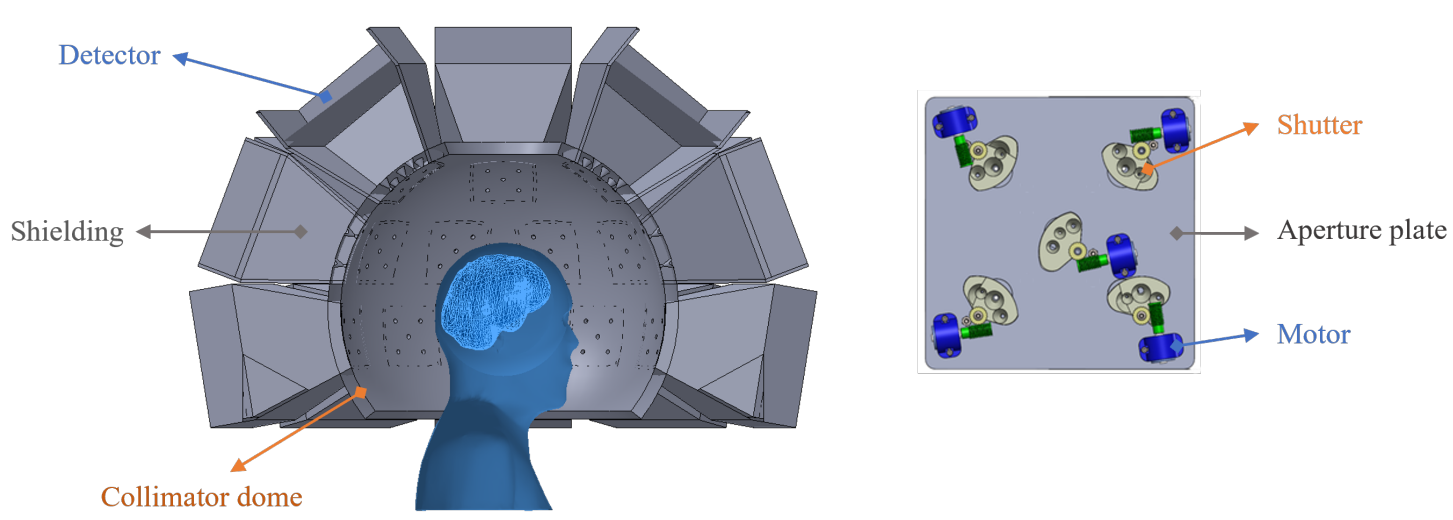

**Figure 1:** A cross-sectional diagram of the AdaptiSPECT-C system (left) with a diagram of one of the 24 aperture plates showing the shutters and motors (right), adapted from Figure 1 of [1].

Pinhole SPECT has previously been applied to brain imaging [2, 3]. AdaptiSPECT-C is a brain-dedicated SPECT system with 24 modular detectors each fitted with an aperture plate with five apertures - one central and four oblique. Adaptive shutters permit each pinhole to be set to one of three sizes or shuttered independently [4–6]. A cross-sectional diagram of the system is shown in Figure 1. Opening all five apertures on all 24 detectors simultaneously gives the highest sensitivity but leads to significant levels of mux in the projection images. Previous work proposed an algorithm for the de-multiplexing (de-mux) of projection images [7]. Additional work demonstrated that temporal shuttering of apertures to provide mux-free acquisition frames reduces mux-induced artefacts in projection images and modified the de-mux algorithm to account for temporal shuttering [4, 8]. Here, an increased-mux model of the AdaptiSPECT-C system is considered to test the algorithm and quantify its impact on reconstruction artefacts and image quantification.

## 2 Materials and methods

In-house analytic simulation and MLEM reconstruction software [9] was used to investigate de-mux in AdaptiSPECT-C. To test the robustness of the de-mux procedure in a more extreme case, the detectors were moved from the base AdaptiSPECT-C model to increase the magnification and hence the multiplexing. Two acquisition modes were compared: opening all pinholes for 100 % of the acquisition (referred to as All100), and using temporal shuttering to acquire a mux-free frame with only the central pinholes open for 20 % of the acquisition followed by a mux frame with all pinholes open for the remaining 80 % of the acquisition (referred to as Central20All80). Both modes used the same acquisition time and 2.5 mm apertures for all 120 pinholes.

A uniform spherical activity map with diameter of 21 cm to match the field-of-view of the system was used, with an attenuation map set to water for a photon of 159 keV (to mimic $^{123}$I). For a more realistic clinical image, a digital XCAT phantom [10] of a male with head size in the 99$^{th}$ percentile was also generated. A realistic activity distribution for a $^{123}$I-IMP brain perfusion scan was used, as stated in other work [11]. Both phantoms used 1 mm voxels for activity and attenuation maps. Figure 2 shows sample projection frames for both phantoms in this increased-mux model of the AdaptiSPECT-C system.

The disadvantage of acquiring a mux-free projection frame is a loss in sensitivity. Therefore both noise-free projections and ones with Poisson noise were considered to see if the increased noise in the Central20All80 acquisitions degraded the images. The Central20All80 projection images of the sphere and perfusion phantoms contained 56.7 and 21.7 million counts respectively. All projection images were reconstructed with MLEM with 2 mm voxels and attenuation correction. The noise-free images were reconstructed with 100 iterations. The iterations were stopped for the images with Poisson noise once the minimum NRMSE was reached. Five noise realisations were performed for each acquisition.

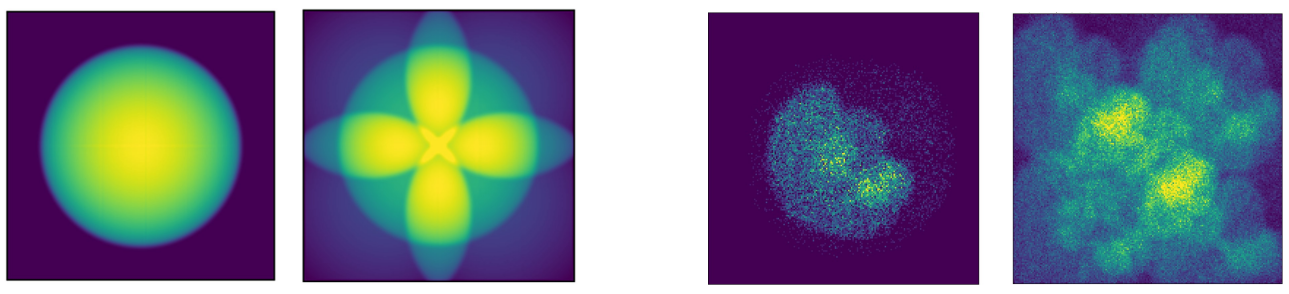

**(a)** 21 cm sphere **(b)** Perfusion phantom

**Figure 2:** Projection images acquired in the increased-mux AdaptiSPECT-C model for a 21 cm spherical source and a perfusion XCAT phantom. Each subfigure shows left: a mux-free frame with only central pinholes ($p^{mux-free}$), and right: a mux frame with all pinholes ($p^{mux}$). Each image has been normalised to the same brightness so do not reflect the relative acquisition times.

A de-mux equation was created based on previous work [7, 8]. The five pinholes on each aperture plate were numbered from the central pinhole (1) to oblique pinholes (2-5). A de-mux projection image was then estimated for the $i^{th}$ pinhole on all detectors through

$$p_i^{demux} = \frac{t_i^{mux} \sum_k a_{k,i} f_k}{\sum_{i=1}^{5} t_i^{mux} \sum_k a_{k,i} f_k} \times p^{mux} \tag{1}$$

where $t_i^{mux}$ is the relative acquisition time of the mux projection frame for the $i^{th}$ pinhole (here $t_i^{mux} = 0.8$ for Central20All80 and 1 for All100), $a_{k,i}$ are the system matrix elements corresponding to only the $i^{th}$ pinhole being open for $k$ voxels, $f_k$ are the reconstructed image voxels, and $p^{mux}$ is the mux projection image. In the two-frame acquisition mode, Central20All80, the mux-free projection frame was reconstructed separately and used for the initial activity estimation $f_k$ in Equation 1. For All100, the original mux projection reconstruction was used for the initial reconstruction. Equation 1 was applied iteratively for several rounds of de-mux.

### 2.1 Quantification metrics

Several image quality metrics were considered to quantify the differences made by the de-mux algorithm. One was the normalised root-mean-square error,

$$\text{NRMSE} = \sqrt{\frac{\sum_k (I_k - R_k)^2}{\sum_k (R_k)^2}} \tag{2}$$

where $I$ is the test image and $R$ is the ground truth image, each with $k$ voxels. The Matlab Structural Similarity Image Metric (SSIM) was also used to quantify the similarity of the images to the ground truth [12]. This metric varies between 0 and 1, where 1 indicates identical images. The global non-uniformity of the reconstructed images was calculated for the spherical phantom according to

$$\text{non-uniformity} = \frac{\max - \min}{\max + \min} \times 100\% \tag{3}$$

where max and min are the maximum and minimum pixel values in the image respectively. The differential non-uniformity was also determined by calculating the non-uniformity for groups of five adjacent pixels iteratively across the image. Horizontal and vertical differential uniformity are referred to as $\text{Diff}_\text{h}$ and $\text{Diff}_\text{v}$ respectively. Only a 19-cm-diameter volume of interest (VOI) centred on the sphere was considered in order to avoid edge effects such as Gibbs ringing.

For the perfusion brain model, VOIs were generated from the XCAT model of the true activity distribution and the percentage activity recovery (AR) was calculated with

$$\text{AR} = \frac{I_{VOI}}{R_{VOI}} \times 100\% \tag{4}$$

where $I_{VOI}$ and $R_{VOI}$ are the total VOI counts in the test and reference images respectively. To consider only the effects of multiplexing and not other artefacts due to the finite resolution of the reconstruction (such as partial volume effects), an ideal mux-free reconstruction was used as the reference image. This reconstruction used the same total projection counts, but reconstructed them as though each pinhole was acquired consecutively (this would not be possible physically without significantly increasing scan time or activity).

## 3 Results and Discussion

### 3.1 Uniform sphere

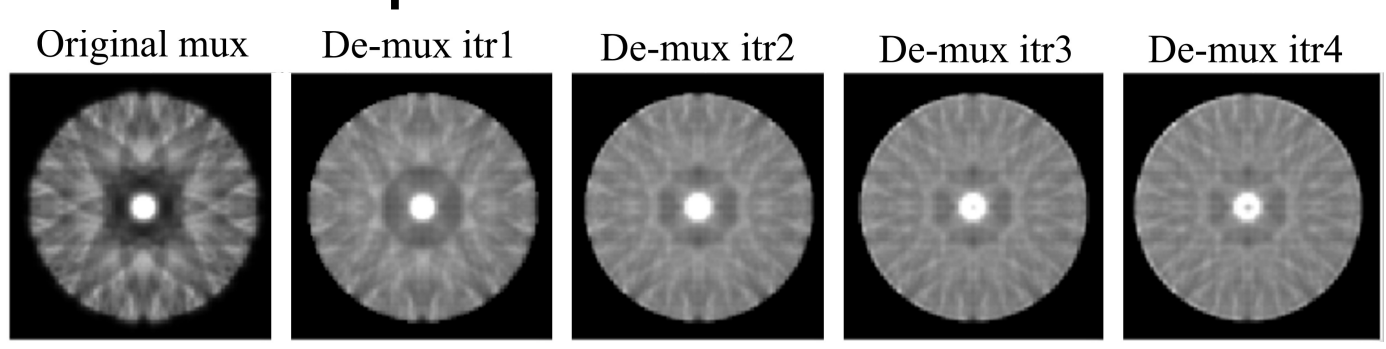

**(a)** Acquisition without mux-free frame (All100)

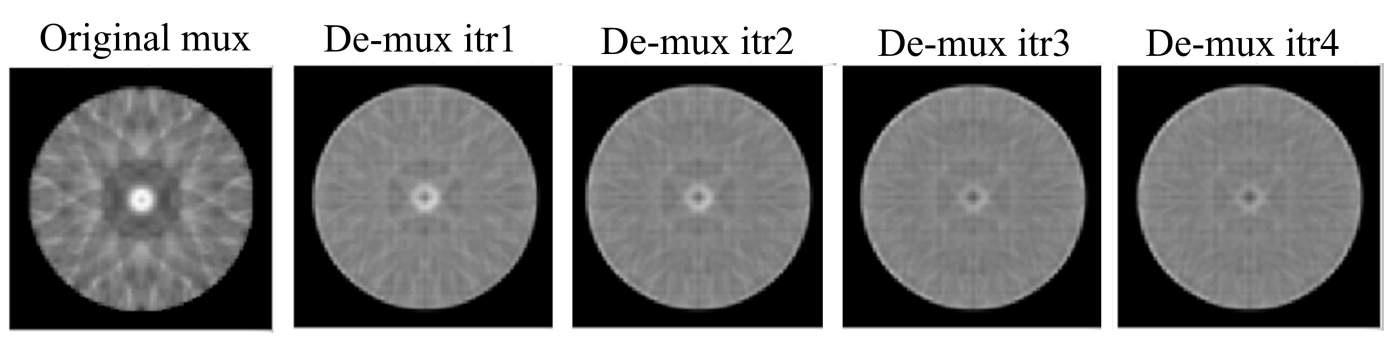

**(b)** Acquisition with mux-free frame (Central20All80)

**Figure 3:** The central trans-axial slice of reconstructions of the noise-free uniform sphere, using the two different acquisition schemes with and without a mux-free frame. The reconstruction of the original mux projection image is shown on the left, followed by the reconstruction following four de-mux iterations. 100 iterations of MLEM were used in each reconstruction and all images have been normalised to the same brightness.

Figure 3 shows the reconstructed images for four de-mux iterations for the uniform sphere, compared to the reconstruction from the original multiplexed projection image for both acquisition modes (this is $p^{mux-free} + p^{mux}$ for Central20All80). The noise-free images are shown as the mux-induced artefacts are more visible in the absence of noise. The reconstruction of the original projection for both acquisitions has clear artefacts due to the high level of multiplexing and truncation in the projection images. These artefacts decrease with de-mux iterations, but some artefacts are still present after four iterations. The Central20All80 acquisition leads to visually-smoother images. A linear profile was taken through the centre of the reconstructed images of the original mux projection and the projection after the fourth iteration of de-mux. Figure 4 shows these profiles for the noise-free data; the de-mux image from the Central20All80 acquisition gives the profile closest to the true count distribution.

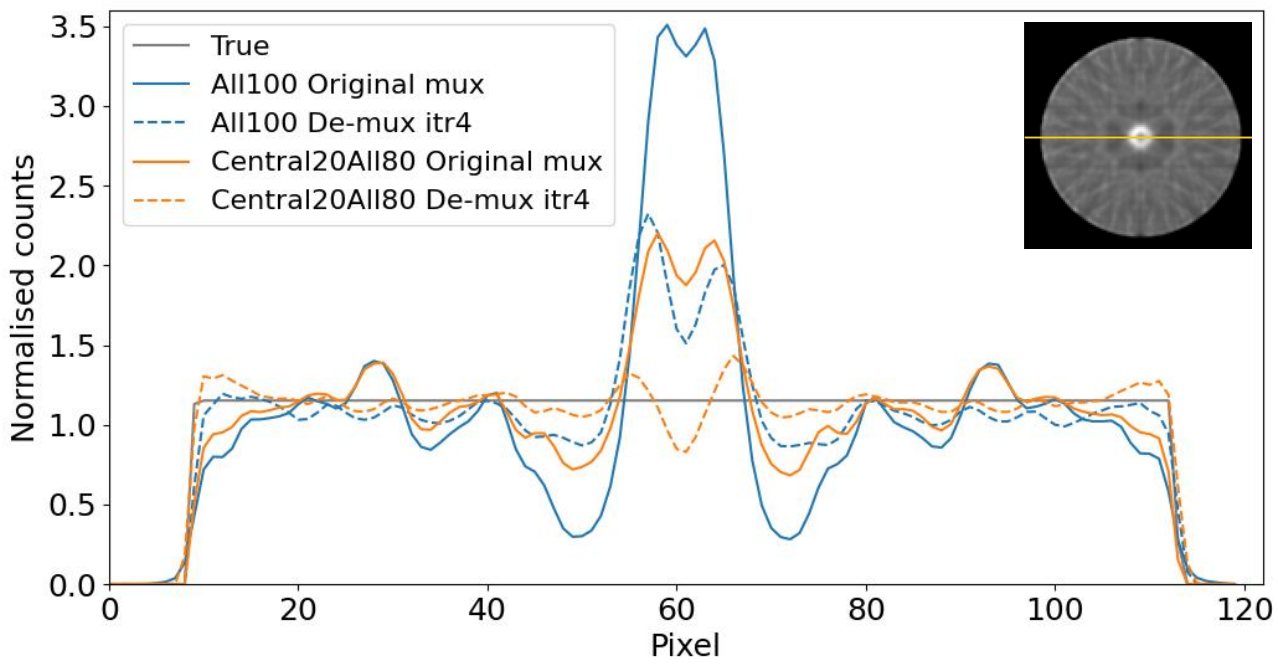

**Figure 4:** The linear profile through the centre of the noise-free reconstructed images. The inset shows the profile drawn on one of the reconstructions. All profiles have been normalised to the same average value to account for their different count levels.

The NRMSE, SSIM and global and differential non-uniformities were calculated for both acquisitions of the uniform sphere phantom with and without noise. Table 1 gives these values for the mux reconstruction and the first and fourth de-mux iterations. Regardless of noise, the Central20All80 acquisition leads to improved NRMSE, SSIM and uniformity even without any de-mux. When de-mux is applied the metrics generally improve further.

### 3.2 Perfusion phantom

Since the results for the uniform sphere showed superior performance Central20All80, this acquisition mode was considered for the perfusion phantom. Figure 5 shows the ideal (mux-free) reconstruction of the brain perfusion phantom, compared to the reconstruction of the original projection image and the reconstruction of de-mux projection image after three de-mux iterations, both with and without Poisson noise. The absolute percentage difference between the mux and de-mux reconstruction is shown on the far right; this image indicates that most of the difference is in the background region of the head rather than the brain. This is especially clear in the noise-free case where the low activity uptake in the skull bone marrow is better resolved in the de-mux image.

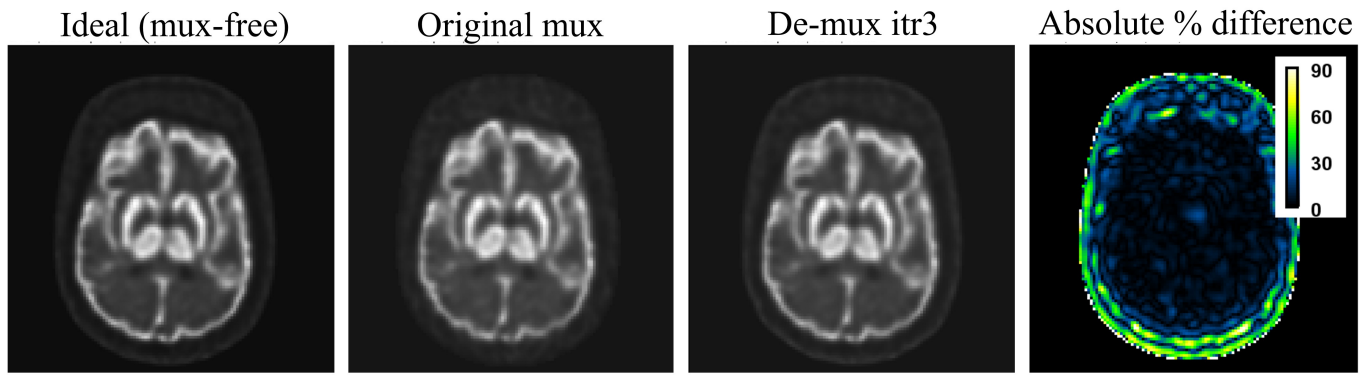

**(a)** Noise-free

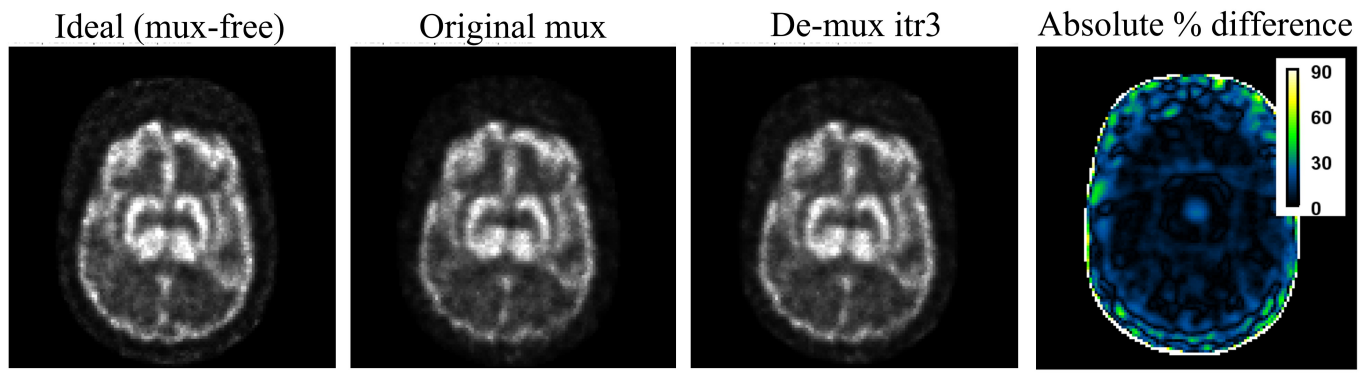

**(b)** Poisson noise

**Figure 5:** Reconstructed images of the brain perfusion phantom using the Central20All80 acquisition mode. The ideal mux-free reconstruction is shown with the mux reconstruction and the reconstruction following three de-mux iterations. The absolute percentage difference between the mux and de-mux reconstructions is given on the right with a colour bar.

A linear profile was taken through the noise-free images for the trans-axial slice shown in Figure 5 and is shown in Figure 6. The de-mux image shows a closer agreement to the ideal image than the original mux reconstruction. Table 2 gives the NRMSE and SSIM for the original and de-mux reconstructions. SSIM is improved with de-mux both with and without noise. The NRMSE is equivalent for the original mux and de-mux images within the uncertainty of the 5 noise realisations, and slightly worse after de-mux is applied in the noise-free case.

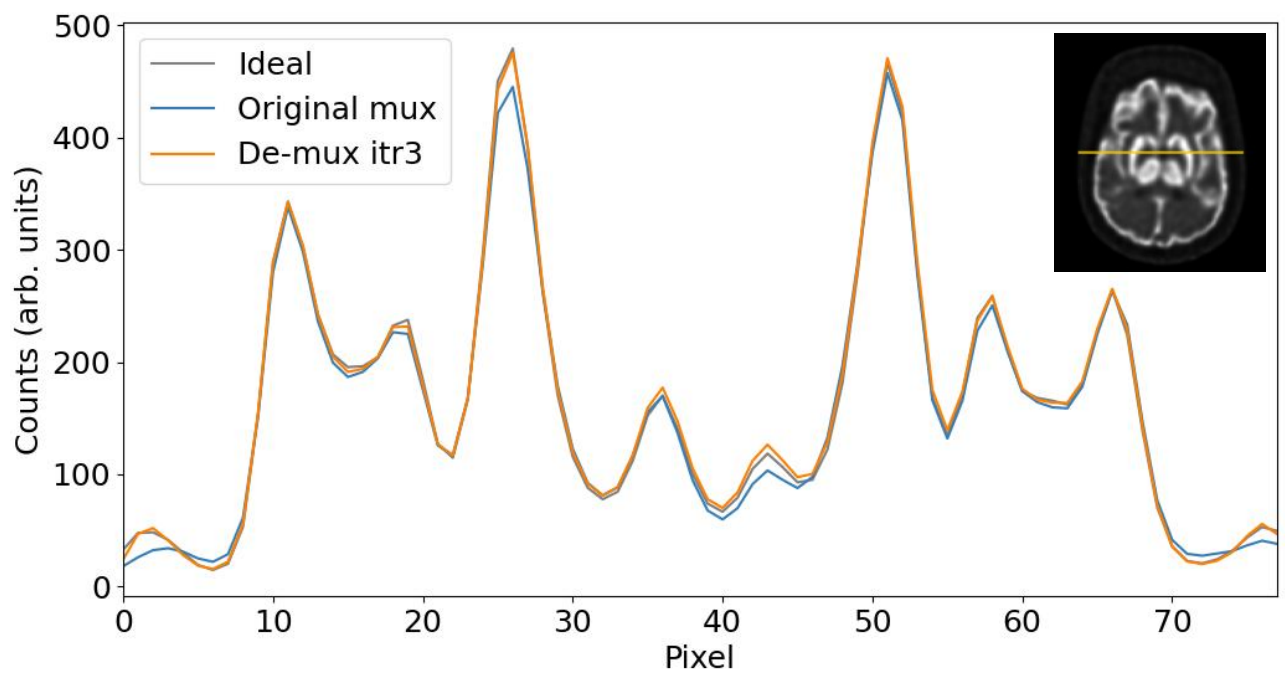

**Figure 6:** The linear profile through the centre of the reconstructed noise-free images of the perfusion phantom acquired in Central20All80. The inset shows the profile drawn on one of the reconstructions.

The percentage activity recovery was calculated for several regions of grey and white matter in the images with Poisson noise. These are shown in Figure 7. The activity recovery improves with de-mux iterations in the majority of cases but it appears that further improvement is needed for some lobes.

## 4 Future work

This work has focused on a fixed aperture shuttering scheme: 20 % mux-free followed by 80 % mux. Different acquisi-tion schemes will be investigated. Additional brain activity distributions will be considered to determine if different applications benefit from different acquisition and de-mux schemes. The convergence of the algorithm will also be investigated for different acquisition modes, as well as the impact of truncated data from the oblique pinholes. This work has focused on an increased-mux model of the AdaptiSPECT-C system; future work will verify that these findings translate to different levels of multiplexing.

| | Noise free | | | | | | Poisson noise | | | | | |
|---|---|---|---|---|---|---|---|---|---|---|---|---|
| | All100 | | | Central20All80 | | | All100 | | | Central20All80 | | |
| | | De-mux | | | De-mux | | | De-mux | | | De-mux | |
| Metric | Mux | itr1 | itr4 | Mux | itr1 | itr4 | Mux | itr1 | itr4 | Mux | itr1 | itr4 |
| NRMSE | 0.275 | 0.169 | 0.155 | 0.146 | 0.086 | 0.091 | 0.3262(2) | 0.1553(1) | 0.1706(1) | 0.2419(4) | 0.1257(5) | 0.1388(4) |
| SSIM | 0.486 | 0.669 | 0.661 | 0.538 | 0.694 | 0.667 | 0.4684(1) | 0.5263(1) | 0.5349(2) | 0.4984(2) | 0.5653(2) | 0.5777(2) |
| G(%) | 93.6 | 75.2 | 71.5 | 71.8 | 39.4 | 32.0 | 96.0(4) | 58.2(6) | 55.4(4) | 87.3(8) | 48(2) | 52.7(4) |
| $Diff_h$ (%) | 72.2 | 47.8 | 42.5 | 44.0 | 25.1 | 26.3 | 78(2) | 31(1) | 28(2) | 65(1) | 24(1) | 27(2) |
| $Diff_v$(%) | 69.6 | 45.5 | 39.5 | 42.3 | 24.0 | 26.8 | 76(1) | 30(2) | 27(2) | 65(2) | 24(2) | 26(2) |

**Table 1:** Image metrics for the reconstructions of the uniform sphere phantom. Both the noise-free images and those with Poisson noise are included. The NRMSE, SSIM, percentage global non-uniformity (G(%)) and percentage horizontal and vertical differential uniformities ($Diff_h$(%), $Diff_v$(%)) are given for the original mux reconstruction and de-mux iterations 1 and 4. The average of 5 noise realisations is given for noisy data, with the uncertainty on the last digit(s) in brackets, e.g. 96.0(4) = 96.0 ± 0.4.

| | Noise free | | Poisson noise | |
|---|---|---|---|---|
| Metric | Mux | De-mux itr3 | Mux | De-mux itr3 |
| NRMSE | 0.259 | 0.263 | 0.3240(4) | 0.3212(16) |
| SSIM | 0.770 | 0.817 | 0.7075(2) | 0.7224(44) |

**Table 2:** Imaging metrics for the original reconstruction and the reconstruction following one iteration of de-mux for the brain perfusion phantom. Average of 5 noise realisations is given for noisy data. The uncertainty on last digit(s) is given in brackets.

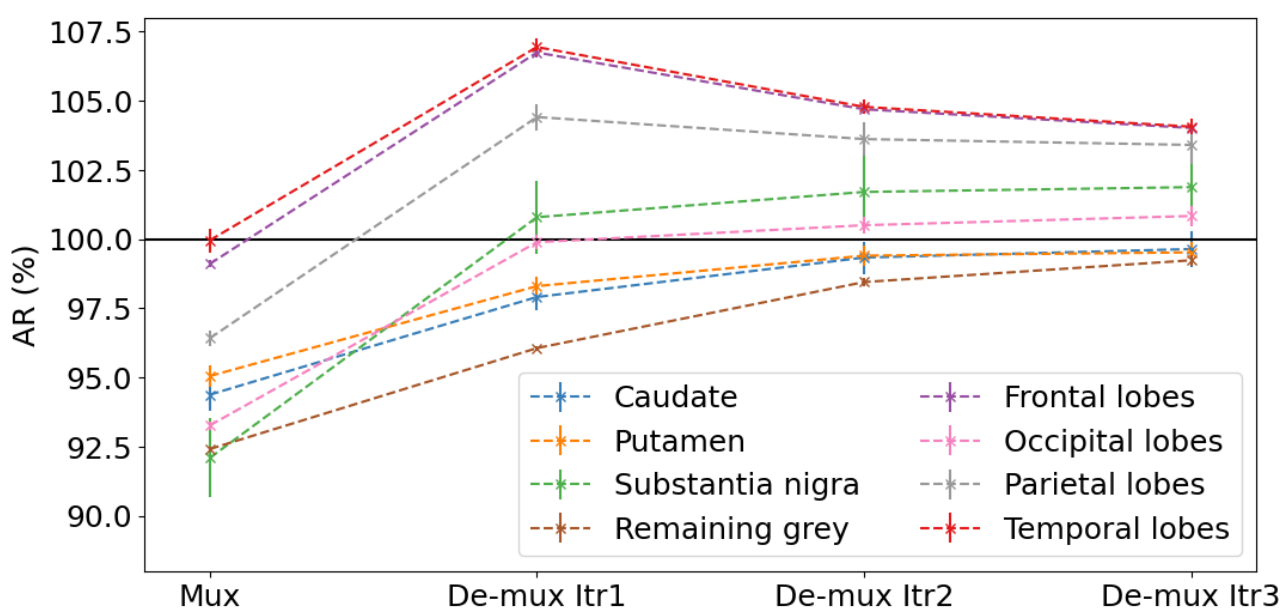

**Figure 7:** Percentage activity recovery in different regions compared to the ideal (mux-free) case for the perfusion phantom. Error bars show standard deviation of 5 Poisson noise realisations. Perfect activity recovery of 100 % is shown as a black line.

## 5 Conclusion

This work suggests that a de-mux algorithm can improve both the quality of a reconstructed image and the quantification of activity recovery in images acquired with an increased-multiplexing model of the AdaptiSPECT-C system. The images were improved further with the inclusion of a short multiplexing-free projection frame in the initial acquisition. Further investigation is needed to determine if this de-mux algorithm would mean acquisitions with multiplexing could be clinically beneficial with AdaptiSPECT-C.

**Acknowledgments and conflicts of interest**

Research reported in this publication was supported by the National Institute of Biomedical Imaging and Bioengineering of the National Institutes of Health under Award Number R01 EB022521. The content is solely the responsibility of the authors and does not necessarily represent the official views of the National Institutes of Health. N. Zeraatkar is a full-time employee of Siemens Medical Solutions USA, Inc. P.H. Kuo is a consultant and/or speaker for Amgen, Bayer, Blue Earth Diagnostics, Chimerix, Eisai, Fusion Pharma, General Electric Healthcare, Invicro, Novartis, Radionetics, and UroToday. He is a recipient of research grants from Blue Earth Diagnostics and General Electric Healthcare.

# Synergistic PET/CT Reconstruction Using a Joint Generative Model

Noel Jeffrey Pinton[1], Alexandre Bousse[1], Zhihan Wang[1], Catherine Cheze-Le-Rest[1,2], Voichita Maxim[3], Claude Comtat[4], Florent Sureau[4], and Dimitris Visvikis[1]

[1]LaTIM, INSERM UMR 1101, *Université de Bretagne Occidentale*, 29238 Brest, France.
[2]Nuclear Medicine Department, Poitiers University Hospital, F-86022, Poitiers, France.
[3]Université de Lyon, INSA-Lyon, UCBL 1, UJM-Saint Etienne, CNRS, Inserm, CREATIS UMR 5220, U1294, F-69621, LYON, France.
[4]BioMaps, Université Paris-Saclay, CEA, CNRS, Inserm, SHFJ, 91401 Orsay, France.

**Abstract** We propose in this work a framework for synergistic positron emission tomography (PET)/computed tomography (CT) reconstruction using a joint generative model as a penalty. We use a synergistic penalty function that promotes PET/CT pairs that are likely to occur together. The synergistic penalty function is based on a generative model, namely beta-variational autoencoder ($\beta$-VAE). The model generates a PET/CT image pair from the same latent variable which contains the information that is shared between the two modalities.
This sharing of inter-modal information can help reduce noise during reconstruction. Our result shows that our method was able to utilize the information between two modalities. The proposed method was able to outperform individually reconstructed images of PET (i.e., by maximum likelihood expectation maximization (MLEM)) and CT (i.e., by weighted least squares (WLS)) in terms of peak signal-to-noise ratio (PSNR). Future work will focus on optimizing the parameters of the $\beta$-VAE network and further exploration of other generative network models.

## 1 Introduction

PET/CT hybrid imaging systems have been used since the 2000's. On one hand, PET is an imaging modality used to observe and quantify molecular-level activities inside tissue through radioactive tracers, while on the other hand CT is an imaging technique that uses X-rays to produce detailed structural images of the body. In order to get clearer images, both modalities use high dose of ionizing radiation which is detrimental to the health of certain patients especially children. Therefore reducing the dosage of used ionizing radiation is crucial, but it will affect the signal-to-noise ratio (SNR) of the reconstructed images.
Current multi-modal imaging reconstruction techniques process each modality separately. However, it is possible to exploit inter-modality information which could help reduce noise during reconstruction by combining functional PET and structural CT images. This approach could ultimately lead to reduction of dosage in patient imaging. The use of deep-learning in an unrolled model-based iterative reconstruction (MBIR) algorithm for synergistic PET/magnetic resonance (MR) reconstruction [1] has shown improved performance compared to independent conventional reconstruction methods. Although this approach is promising, the training is computationally expensive and the architecture depends on the scanners system projectors/back-projectors.
In this work, we propose a synergistic penalty function which uses generated PET and CT from the same latent variable which is trained on a $\beta$-VAE. As the shared information is contained in the penalty term, the proposed approach does not depend on the system (i.e. projector/back-projectors are not part of the training).
Section 2 describes our methodology and generative model, Section 3 shows that out method performs better than traditional reconstruction techniques in terms of PSNR, and in Section 4 we discuss in what ways we can still improve the results.

## 2 Method

### 2.1 Conventional PET/CT Reconstruction

The linear X-ray attenuation and activity image are respectively denoted $\boldsymbol{\mu} = [\mu_1, \dots, \mu_J]^\top \in \mathbb{R}_+^J$ and $\boldsymbol{\lambda} = [\lambda_1, \dots, \lambda_J]^\top \in \mathbb{R}_+^J$, $J$ being be the number of pixels (or voxels) in the image. In the following we briefly introduce the MBIR settings for CT and PET.
We use a standard and monochromatic model for CT. The transmission measurement are collected along $I$ rays and stored in a vector $\boldsymbol{y} = [y_1, \dots, y_I]^\top \in \mathbb{N}^I$. Given the attenuation $\boldsymbol{\mu}$ and ignoring background events, the expected number of detected X-ray photons along each ray $i = 1, \dots, I$ is given by the Beer-Lambert law as

$$\bar{y}_i(\boldsymbol{\mu}) = h \cdot \mathrm{e}^{-[\boldsymbol{A}\boldsymbol{\mu}]_i} \tag{1}$$

where $\boldsymbol{A} \in \mathbb{R}^{I \times J}$ is a discrete line integral operator such that $[\boldsymbol{A}]_{i,j}$ denotes the contribution of pixel $j$ to ray $i$, and $h$ is the X-ray intensity. Assuming that the $I$ measurements $\{y_i\}$ are independent and follow a Poisson distribution centered on $\{\bar{y}_i(\mu)\}$, MBIR of the attenuation image $\boldsymbol{\mu}$ can be achieved by solving the following optimization problem with an iterative algorithm [2]:

$$\hat{\boldsymbol{\mu}} \in \operatorname*{arg\,min}_{\boldsymbol{\mu} \geq \boldsymbol{0}} L_{\mathrm{CT}}(\boldsymbol{\mu}) + \alpha R_{\mathrm{CT}}(\boldsymbol{\mu}) \tag{2}$$

where the loss function $L_{\mathrm{CT}}$ is defined ad $L_{\mathrm{CT}}(\boldsymbol{\mu}) = \frac{1}{2}\|\boldsymbol{A}\boldsymbol{\mu} - \boldsymbol{b}\|_{\boldsymbol{W}}^2$ with $\boldsymbol{b} = -\log \boldsymbol{y}/h \in \mathbb{R}^I$ being the vector of approximated line integrals and $\boldsymbol{W} = \mathrm{diag}\{\boldsymbol{y}\} \in \mathbb{N}^{I \times I}$ being a diagonal matrix of statistical weights, $R_{\mathrm{CT}}$ is a penalty function or regularizer, generally convex, that promotes desired image properties (in general piece-wise smoothness) while controlling the noise, and $\alpha > 0$ is a weight.

The PET data are collected along $K$ lines of response (LoRs) defined by $K$ pairs of detectors. The emission data are stored in a vector $\boldsymbol{g} \in \mathbb{N}^K$. Given the activity image $\boldsymbol{\lambda}$ and again ignoring background events, the expected number of detected $\gamma$-photon pairs along LoR $K$ is

$$\bar{g}_k(\boldsymbol{\lambda}) = \tau[\boldsymbol{P}\boldsymbol{\lambda}]_k \tag{3}$$

where $\boldsymbol{P} \in \mathbb{R}^{K\times J}$ is the PET system matrix, i.e., $[\boldsymbol{P}]_{k,j}$ is the probability that a emission at pixel $j$ is detected along LoR $k$ (accounting for attenuation), and $\tau$ is the acquisition time. Assuming that the $K$ measurements $\{g_k\}$ are independent and follow a Poisson distribution centered on $\{\bar{g}_k(\boldsymbol{\lambda})\}$, MBIR of the activity image can be achieved by solving the following penalized maximum likelihood (PML) problem iteratively:

$$\hat{\boldsymbol{\lambda}} \in \operatorname*{arg\,min}_{\boldsymbol{\lambda}\geq\boldsymbol{0}} L_{\mathrm{PET}}(\boldsymbol{\lambda}) + \delta R_{\mathrm{PET}}(\boldsymbol{\lambda}) \tag{4}$$

where the loss function $L_{\mathrm{PET}}$ is the negative Poisson log-likelihood defined as $L_{\mathrm{PET}}(\boldsymbol{\lambda}) = \sum_{k=1}^{K} -g_k \log \bar{g}_k(\boldsymbol{\lambda}) + \bar{g}_k(\boldsymbol{\lambda})$ and $R_{\mathrm{PET}}$ and $\delta$ are same as $R_{\mathrm{CT}}$ and $\alpha$ in (2).

### 2.2 Synergistic PET/CT Reconstruction Using a Joint Generative Model

$\boldsymbol{\mu}$ and $\boldsymbol{\lambda}$ can be simultaneously reconstructed by solving the following joint estimation problem:

$$(\hat{\boldsymbol{\mu}}, \hat{\boldsymbol{\lambda}}) \in \operatorname*{arg\,min}_{\boldsymbol{\mu},\boldsymbol{\lambda}\geq\boldsymbol{0}} L_{\mathrm{CT}}(\boldsymbol{\mu}) + L_{\mathrm{PET}}(\boldsymbol{\lambda}) + \gamma R_{\mathrm{syn}}(\boldsymbol{\mu}, \boldsymbol{\lambda}) \tag{5}$$

where $\gamma > 0$ and $R_{\mathrm{syn}}$ is a *synergistic* penalty function that promotes structural and/or functional correlations between the multiple images, such as parallel levelsets [3] and total nuclear variation [4].

Instead of using a handcrafted synergistic penalty function, $R_{\mathrm{syn}}$ may be trained such that $R_{\mathrm{syn}}(\boldsymbol{\mu}, \boldsymbol{\lambda}) \approx 0$ if $\boldsymbol{\mu}$ and $\boldsymbol{\lambda}$ are images that are plausible not only individually, but also *together*. In this work we propose the following penalty inspired by previous work from Duff, Campbell, and Ehrhardt [5]:

$$\begin{aligned} R_{\mathrm{syn}}(\boldsymbol{\lambda}, \boldsymbol{\mu}) = \min_{\boldsymbol{z}\in\mathbb{R}^P} \eta \frac{1}{2} \|\boldsymbol{f}_{\mathrm{PET}}(\boldsymbol{z}) - \boldsymbol{\lambda}\|^2 \\ + (1-\eta)\frac{1}{2}\|\boldsymbol{f}_{\mathrm{CT}}(\boldsymbol{z}) - \boldsymbol{\mu}\|^2 \end{aligned} \tag{6}$$

where $\boldsymbol{f}_{\mathrm{CT}}$ and $\boldsymbol{f}_{\mathrm{PET}}$ are the decoders of the trained generative model that generate CT and PET images from the same latent variable $\boldsymbol{z} \in \mathbb{R}^P$ and $\eta \in [0,1]$. To summarize (6), $R_{\mathrm{syn}}(\boldsymbol{\mu}, \boldsymbol{\lambda}) \approx 0$ if $\boldsymbol{\mu}$ and $\boldsymbol{\lambda}$ are approximately generated from the same latent variable $\boldsymbol{z}$. This approach is somehow a generalization of coupled dictionary learning for multi-modal imaging (see for example [6]) where we replaced the dictionaries by generative models. Note that a penalty can be added in (6) to control the noise. In this work we used $\beta$-VAEs for $\boldsymbol{f}_{\mathrm{CT}}$ and $\boldsymbol{f}_{\mathrm{PET}}$.

The overall reconstruction method is described in Algorithm 1. The activity and attenuation images were initialized using standard MBIR methods (MLEM and WLS). The algorithm then alternates between minimization in $\boldsymbol{z}$ (L-BFGS [7]) , $\boldsymbol{\lambda}$ (modified MLEM [8]) and $\boldsymbol{\mu}$ (penalized weighted least squares (PWLS) [2]).

**Algorithm 1** Synergistic reconstruction of PET/CT images

**Input:** Maximum iteration number `MaxIter`, sub-iteration numbers `SubIter1` and `SubIter2`, penalty parameter $\gamma$, modal parameter $\eta$

**Inputs:**
  $\gamma$, $\eta$, $\boldsymbol{g}$, $\boldsymbol{y}$
**Initialize:**
  $\boldsymbol{\lambda}^0 = \boldsymbol{\lambda}^{\mathrm{init}}$, $\boldsymbol{\mu}^0 = \boldsymbol{\mu}^{\mathrm{init}}$, $\boldsymbol{z}^0 = \boldsymbol{0}$; $\boldsymbol{p} = \boldsymbol{P}^\top \boldsymbol{1}$;
  $\boldsymbol{b} = -\log \boldsymbol{y}/h \in \mathbb{R}^I$
**for** $n = 1$ to `MaxIter` **do**
  $\boldsymbol{z}^n = \operatorname{arg\,min}_{\boldsymbol{z}\in\mathbb{R}^P} \eta \frac{1}{2}\|\boldsymbol{f}_{\mathrm{PET}}(\boldsymbol{z}) - \boldsymbol{\lambda}^{n-1}\|^2 + (1-\eta)\frac{1}{2}\|\boldsymbol{f}_{\mathrm{CT}}(\boldsymbol{z}) - \boldsymbol{\mu}^{n-1}\|^2$ ▷ *using LBFGS*
  $\tilde{\boldsymbol{\lambda}} \leftarrow \boldsymbol{\lambda}^{n-1}$
  **for** $l = 1$ to `SubIter1` **do**
    $\tilde{\boldsymbol{\lambda}}^{\mathrm{em}} \leftarrow \tilde{\boldsymbol{\lambda}} \frac{1}{\boldsymbol{p}} \boldsymbol{P}^\top \left[\frac{\boldsymbol{g}}{\boldsymbol{P}\tilde{\boldsymbol{\lambda}}}\right]$
    $\boldsymbol{c} \leftarrow \gamma\eta \boldsymbol{f}_{\mathrm{PET}}(\boldsymbol{z}^n) - \boldsymbol{p}$
    $\tilde{\boldsymbol{\lambda}} \leftarrow \frac{\boldsymbol{c} + \sqrt{\boldsymbol{c}^2 + 4\gamma\eta \boldsymbol{p}\boldsymbol{\lambda}^{\mathrm{em}}}}{2\gamma\eta}$
  **end for**
  $\boldsymbol{\lambda}^n = \tilde{\boldsymbol{\lambda}}$
  $\tilde{\boldsymbol{\mu}} \leftarrow \boldsymbol{\mu}^{n-1}$
  **for** $m = 1$ to `SubIter2` **do**
    $\tilde{\boldsymbol{\mu}}^{\mathrm{rec}} \leftarrow \tilde{\boldsymbol{\mu}} - \boldsymbol{D}^{-1}\boldsymbol{A}^\top \boldsymbol{W}(\boldsymbol{A}\tilde{\boldsymbol{\mu}} - \boldsymbol{b})$
    $\tilde{\boldsymbol{\mu}} \leftarrow \left[\frac{\boldsymbol{D}\tilde{\boldsymbol{\mu}}^{\mathrm{rec}} + \gamma(1-\eta)\boldsymbol{f}_{\mathrm{CT}}(\boldsymbol{z}^n)}{\boldsymbol{D} + \gamma(1-\eta)}\right]_+$
  **end for**
  $\boldsymbol{\mu}^n = \tilde{\boldsymbol{\mu}}$
**end for**
**return** $(\boldsymbol{\lambda}^{\texttt{MaxIter}}, \boldsymbol{\mu}^{\texttt{MaxIter}})$

### 2.3 Generative Model Network Architecture

For combining functional PET and structural CT images, we created a multi-branch $\beta$-VAE (Figure 1) architecture. The network features separate encoders and decoders for PET and CT. The encoders and decoders consist 5-layer networks with ReLU activation. We train the networks using the following loss function as described in [9]:

$$\begin{aligned} \mathrm{loss} = \mathbb{E}_{q_\phi(\boldsymbol{z}|\boldsymbol{\lambda},\boldsymbol{\mu})}[\log p_\theta(\boldsymbol{\lambda}, \boldsymbol{\mu} \mid \boldsymbol{z})] \\ - \beta D_{\mathrm{KL}}(q_\phi(\boldsymbol{z}|\boldsymbol{\lambda},\boldsymbol{\mu}) \parallel p(\boldsymbol{z})) \end{aligned} \tag{7}$$

where $\beta$ is the regularization coefficient that constrains the capacity of the latent information channel $\boldsymbol{z}$. The first term is the log-likelihood of the observed data $p_\theta(\boldsymbol{\lambda}, \boldsymbol{\mu}|\boldsymbol{z})$ averaged over the latent variable $\boldsymbol{z}$ with distribution $q_\phi(\boldsymbol{z}|\boldsymbol{\lambda},\boldsymbol{\mu})$, also referred to as the cross-entropy. The second term is the Kullback-Leibler divergence between the posterior variational approximation $q_\phi(\boldsymbol{z}|\boldsymbol{\lambda},\boldsymbol{\mu})$ and the prior distribution $p(\boldsymbol{z})$, which is selected to be a standard Gaussian distribution.

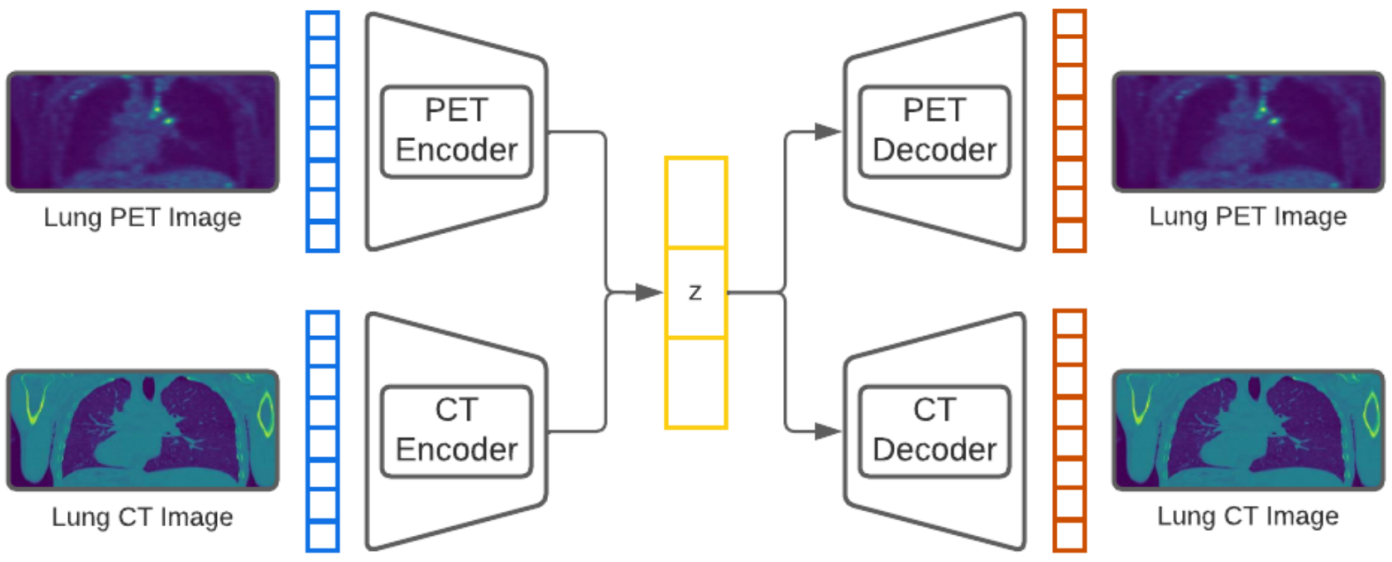

**Figure 1:** Multi-branch $\beta$-VAE architecture

### 2.4 Network Training and Implementation

The models were trained using a collection of PET/CT lung image pair $(\boldsymbol{\lambda},\boldsymbol{\mu})$ of 117 patients from CHU Poitiers, Poitiers, France. The images are $256 \times 256 \times 128$, each of the 128 slice pairs taken independently, resulting in $117 \times 128 = 14,976$ PET/CT image pairs for training. The PET images were clipped at $10^5$ Bq/mL in order to remove outlier. Both PET and CT images were scaled between 0 and 1 for training. During the final reconstruction, the PET and CT images are returned to their original domain values. A fixed value of $\beta$ at 100 of the loss function (7) was used for this study. We used Tensorflow v2.4 and Keras for the network implementation and training. The Adam Algorithm [10] was used with a learning rate of 0.0001 for 300 epochs and a batch size of 128 on a NVIDIA RTX A6000 GPU. The gradients were computed using the `GradientTape` function from Tensorflow for the L-BFGS optimization [7] used in the minimization over $\boldsymbol{z}$ in (6), which uses "automatic differentiation", breaking complex gradient calculation into simpler gradient calculations through chain rules.

### 2.5 Raw Data Generation

For the quantitative analysis of the reconstruction algorithm, we generated low-count PET/CT raw data as

$$g_k \sim \text{Poisson}(\bar{g}_k(\boldsymbol{\lambda}^\star)) \tag{8}$$

$$y_i \sim \text{Poisson}(\bar{y}_i(\boldsymbol{\mu}^\star)) \tag{9}$$

where $\boldsymbol{\lambda}^\star$ and $\boldsymbol{\mu}^\star$ are respectively PET and CT images that we used as a ground truth (GT). Attenuation was ignored for the PET data. We used SciPy [11] for the PET projectors (Radon transform) and Astra Toolbox [12] for CT projectors (for both data generation and reconstruction).

## 3 Results

### 3.1 Network-based Penalty Effect

The parameter $\gamma$ in Equation (5) determines the penalty function contribution. Higher $\gamma$ values pushes the algorithm to favor the generated image from the neural network than the data fidelity term. In order to quantitatively assess the efficacy of inter-modality information sharing in the synergistic reconstruction of PET and CT, we reconstructed the PET and CT images with different values of $\gamma$ (cf. Equation (5)), starting from $\gamma = 0$ which corresponds to standard independent reconstructions, i.e., MLEM for PET and WLS for CT.

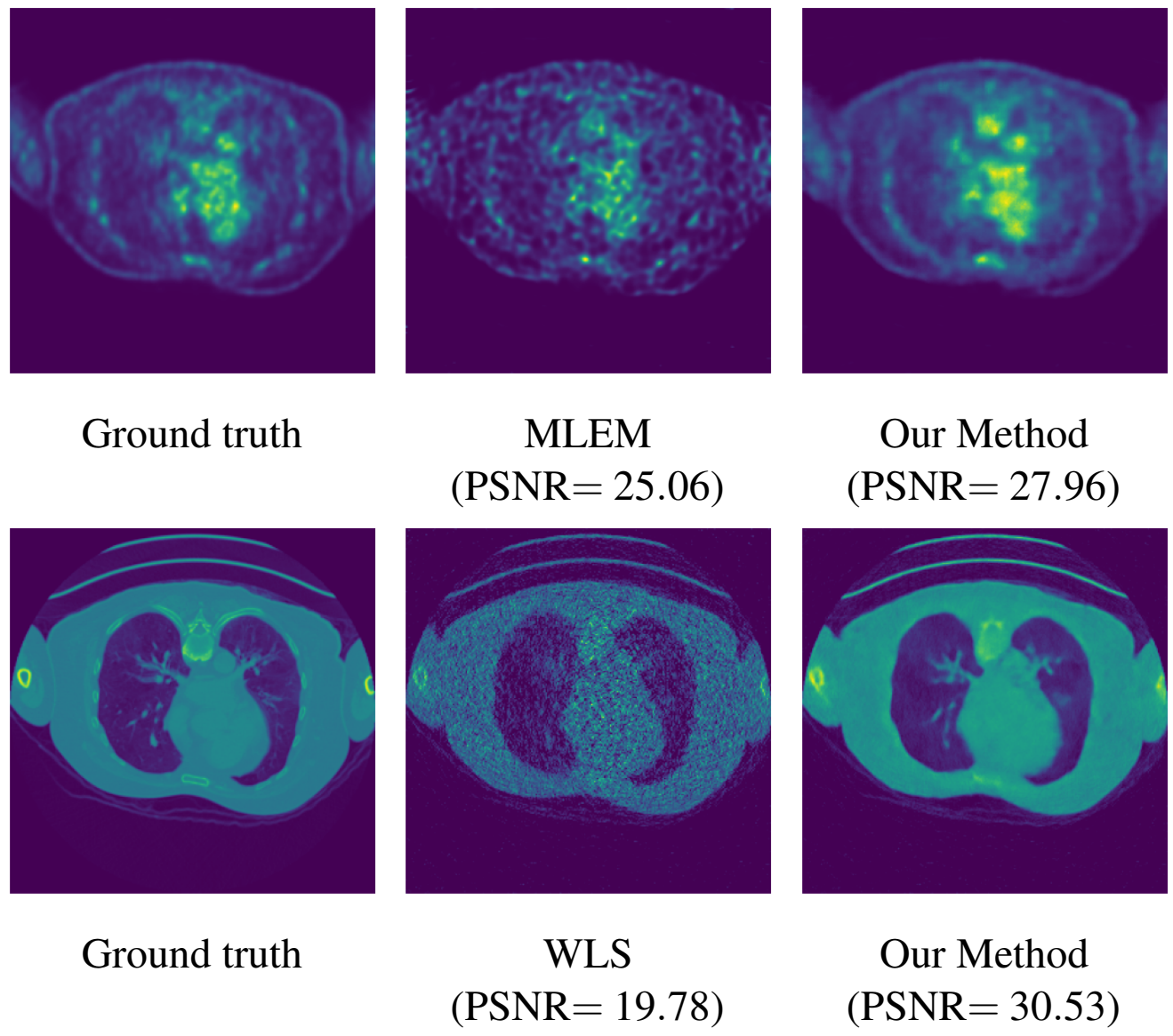

**Figure 2:** Comparison of independent PET and CT reconstructions with synergistic PET/CT reconstruction using our method at $\gamma = 5 \times 10^5$ and $\eta = 0.5$.

Figure 2 shows the reconstruction of PET (MLEM) and CT (WLS) independently (no penalties) and the synergistic reconstruction of the PET/CT image pair (our method).

The final reconstructed images by our method show that the synergistic reconstruction of PET and CT with the help of generative models as a penalty term performs better in terms of PSNR than individually reconstructed PET and CT images. We can infer from this result that both PET and CT are benefiting from the inter-modality information from each other. Figure 3 shows the graph of PSNR with respect to $\gamma$ for both PET and CT. We can see from the graph that an increase in the parameter $\gamma$ corresponds to an increase in PSNR for CT. For PET, the PSNR reaches a maximum and then starts to decline to a stable value. We observe that the optimum was not reached for the same $\gamma$. A solution is proposed on the next section.

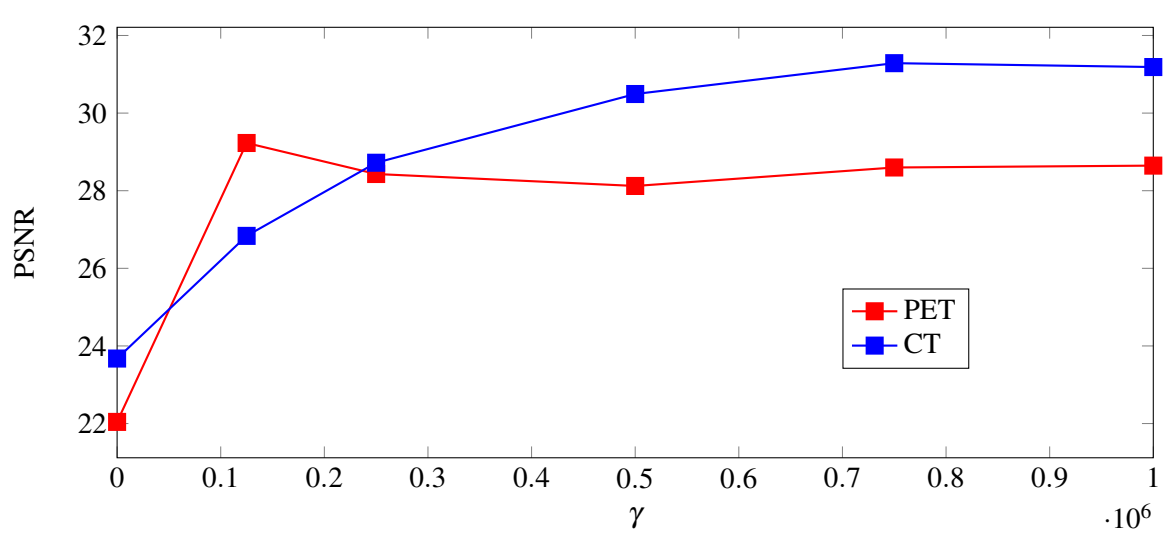

**Figure 3:** PSNR values of reconstructed PET/CT images with respect to $\gamma$ at $\eta = 0.5$.

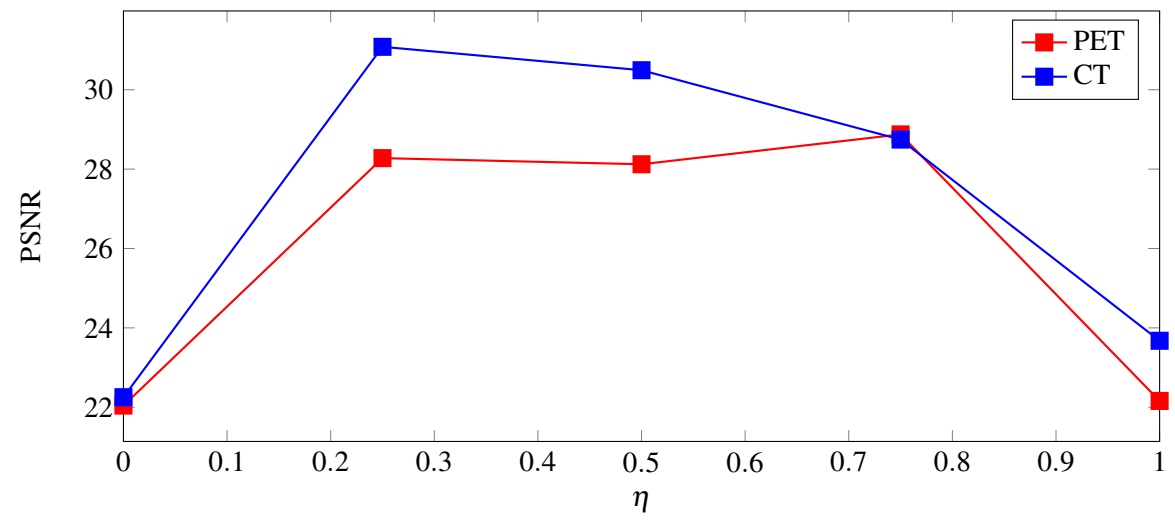

**Figure 4:** PSNR values of reconstructed PET image with respect to $\eta$ at $\gamma = 5 \times 10^5$.

### 3.2 Inter-modality Information Effect

Figure 4 shows the effect of $\eta$ from Equation (6). The variable $\eta$ is a weight that represents the contribution of PET and CT generated image into the final reconstruction. A value of $\eta = 0$ corresponds to full contribution from CT generated image and no contribution from PET in the prior, and vice versa. As shown in the graph (Figure 4), both curves have a maximum somewhere between $\eta = 0$ and $\eta = 1$. This shows that the reconstructed PET and CT images were able to exploit the inter-modal information. However, we observe that the optimum was not reached for the same values of $\eta$, the same as the previous section with $\gamma$. We conjecture that this could be fixed by weighting the data fidelity terms $L_{\text{CT}}$ and $L_{\text{PET}}$ differently as seen in Equation 10,

$$(\hat{\boldsymbol{\mu}}, \hat{\boldsymbol{\lambda}}) \in \arg\min_{\boldsymbol{\mu},\boldsymbol{\lambda} \geq \mathbf{0}} \rho L_{\text{CT}}(\boldsymbol{\mu}) + (1-\rho) L_{\text{PET}}(\boldsymbol{\lambda}) + \gamma R_{\text{syn}}(\boldsymbol{\mu}, \boldsymbol{\lambda}) \,. \quad (10)$$

## 4 Discussion

In this paper, we proposed a synergistic PET/CT reconstruction which utilizes a deep penalty method using a generative neural network $\beta$-VAE. The results showed that the technique outperforms traditional PET-only and CT-only reconstruction methods for both modalities. We have also demonstrated that the PET and CT images were able to exploit the inter-modality information between each other. The use of $\beta$-VAE generators were effective, but fine tuning the hyperparameters are needed in order to produce clearer images. Other suggestions include the tuning of the $\beta$ variable in Equation (7) and the use of attention modules in the latent space. The possibility of using other generative models such as generative adversarial network (GAN)-based models and diffusion-based models is also open. Furthermore, the addition of weights in the first two terms of Equation 5 may improve the results by optimizing the contribution of each term to the final output. Finally, we ignored attenuation factors in the PET reconstruction. In principle they should be computed from a first reconstruction of the attenuation (non synergistic). Since they correspond to line integrals, they do not suffer from noise amplification.

## 5 Conclusion

We proposed in this work a framework for synergistic PET/CT reconstruction using a generative neural network as penalty. The decoders of the trained beta-variational autoencoder (VAE) were able to effectively promote correlations between the two modalities. Evaluations using real patient dataset indicated that the proposed framework was able to exploit the inter-modal information between the two modalities which ultimately led to improved image quality.

## Acknowledgement

This work was funded by the French National Research Agency (ANR) under grant ANR-20-CE45-0020.

# Implementing FFS as a method to acquire more information at a reduced dose in CT scanners

Piotr Pluta[1], Akyl Swaby[2], and Robert Cierniak[1]

[1]Department of Intelligent Computer Systems, Czestochowa University of Technology, Czestochowa, Poland
[2]Department of Electrical and Computer Engineering, University of California, Santa Cruz, USA

**Abstract**
This paper presents an original approach to image reconstruction based on flying focal spot (FFS) technology where the X-ray source is configured to have a focal spot that is variable in position. The geometry of spiral CT scanners presents difficulties for traditional (in particular FDK type) reconstruction methods due to the non-equiangular distribution of X-rays in a given cone beam. By implementing FFS technology, we propose reducing the geometric scheme of the system to perceive X-rays in a lower level of abstraction to increase the number of projections. Improved z-sampling can increase the spatial resolution within the physical limits of a given CT scanner. This method is based on principles of statistical model-based iterative reconstruction (MBIR) where the reconstruction problem is formulated as a shift-invariant system (a continuous-to-continuous data model). Due to its unique design, it can systematically deliver selected reconstructive slices without repeating previously reconstructed sections which is advantageous during emergency CT examinations.

## 1 Introduction

Developed in the early 2000s, the progress in longitudinal spatial resolution from 4-slice to 64-slice CT has enhanced the spiral scanner to allow the use of a flying focal spot (FFS) [1], [2], [3]. The FFS feature doubles the sampling density in the channel direction and in the longitudinal direction [3]. Whereas most scanners increase the number of acquired slices by increasing the number of the detector rows, many new scanners use additional refined z-sampling techniques with a periodic motion of the focal spot in the $z$-direction ($z$-FFS) [4]. This so-called double z-sampling technique can further enhance longitudinal resolution and image quality in clinical routines [5]. Realized in cooperation with multidetector row CT (MDCT) scanners, this new technique aims at increasing the density of simultaneously acquired views in the longitudinal direction and the sampling density of the integral lines in the reconstruction planes. This is implemented with view-by-view deflections of the focal spot in the rotational $\alpha$-direction ($\alpha$FFS) and in the longitudinal $z$-direction ($z$FFS). Using this technique, the resolution of the reconstructed images may be improved, mainly by decreasing the influence of the aliasing artifacts in the reconstruction plane ($\alpha$FFS) and by reducing the windmill artifacts in the $z$-direction ($z$FFS). Due to its geometric scheme, the FFS implementation is not capable of traditional reconstruction methods. Therefore, manufacturers have primarily modified the adaptive multiple plane reconstruction (AMPR) methods for this purpose (for details see [5]).

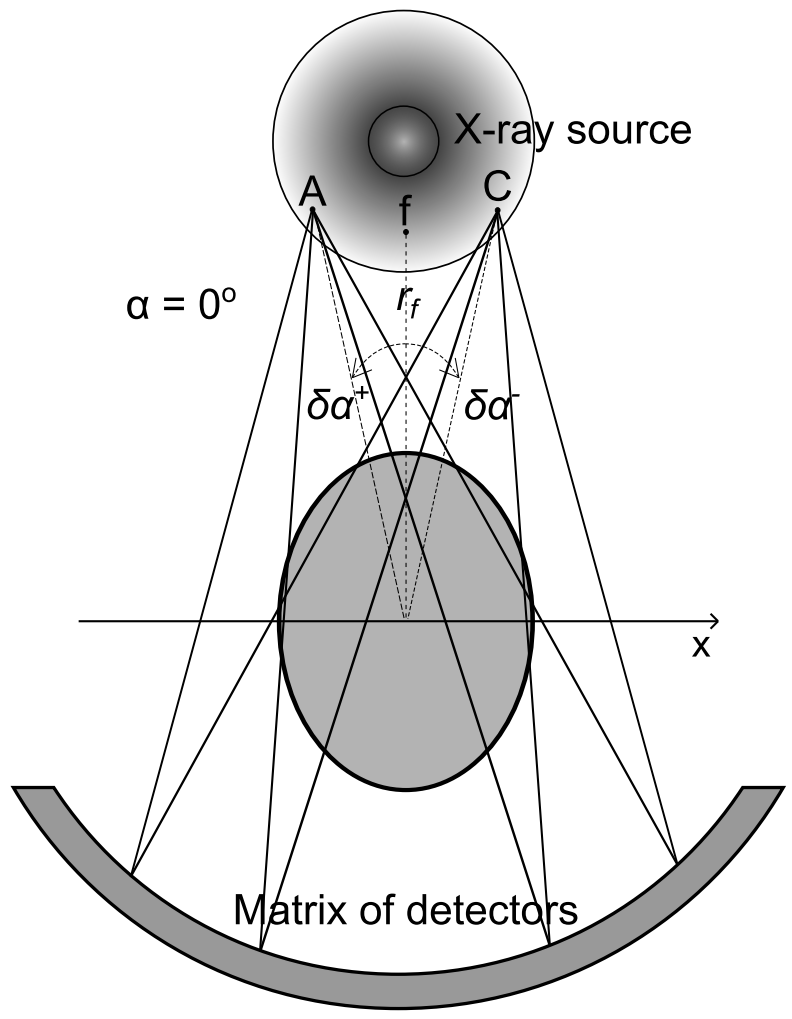

**Figure 1:** Scheme of densification of X-rays in the plane reconstruction ($\alpha$FFS).

## 2 Geometry of the Spiral CT Scanner with Flying Focal Spot

The analysis of the method, contained in this work, is based on the basic design and physical conditions that apply to the operation of a CT scanner with a multifocal X-ray tube. Illustrating these conditions is necessary to understand the subsequent reconstruction methods dedicated to this type of construction of the tomograph. The FFS technique makes use of a special construction of an X-ray tube by deflecting the electron beam (using an electric field) before it interacts with the anode of the tube. This mechanism allows for view-by-view deflections of the focal spot for X-rays emitted from that anode. As a result of the implementation of FFS, it is possible to obtain a greater density of X-rays used in the reconstruction process, both in the plane of the reconstructed image and along the z-axis around which the projection system rotates (depicted in Fig. 1 and 2).

## 3 Reconstruction Algorithm

Our proposed reconstruction method, using the FFS technique, is based on the maximum-likelihood (ML) estimation [6], [7]. Commonly, the objective of this approach to the reconstruction problem is formulated according to a discrete-to-discrete (D-D) data model, but we propose an optimization formula that is consistent with the continuous-to-continuous

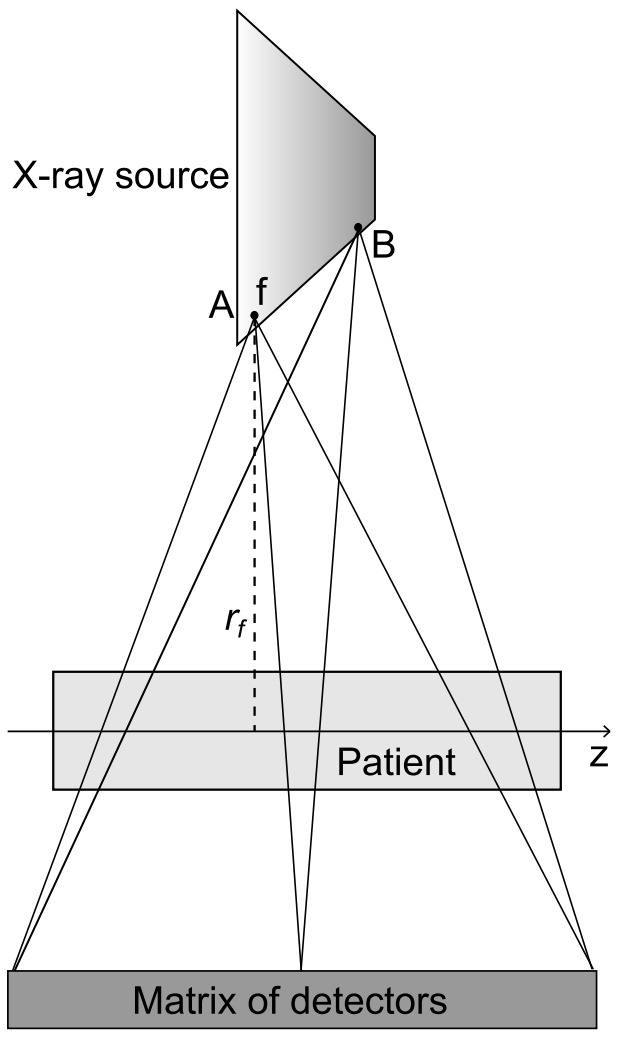

**Figure 2:** Schematic of the X-ray density along the $z$ axis ($z$FFS).

(C-C) data model [8], as follows:

$$\mu_{\min} = \arg\min_{\mu} \left( \int_x \int_y \left( \int_{\bar{x}} \int_{\bar{y}} \mu(\bar{x},\bar{y}) \cdot h_{\Delta x,\Delta y} d\bar{x} d\bar{y} - \tilde{\mu}(x,y) \right)^2 dx dy \right), \quad (1)$$

where the coefficients $h_{\Delta x,\Delta y}$ can be determined according to the following formula:

$$h_{\Delta x,\Delta y} = \int_0^{2\pi} int\left(\Delta x \cos\alpha + \Delta y \sin\alpha\right) d\alpha, \quad (2)$$

and $int(\Delta s)$ is an interpolation function with two significant parts. In this work, we implement the linear interpolation function where we begin by organizing and rebinning the raw cone beam X-ray data from the CT scanner into a parallel geometry (described in subsection 3.1). Secondly, we implement a patented (US 9.508.164 B2) statistical iterative method for the 2D reconstruction view (see subsection 3.2). Finally, the set of 2D views can be successfully used to reconstruct the 3D image from a projection. As a result of this method, we have a fully reconstructed image of the patient in 3D.

### 3.1 Rebinning part of algorithm

During the rebinning operation we decompose the X-ray into elementary parts: focus, semi-isocenter, and detector [9]. The equiangular geometry of the X-ray is then defined from a straight line value to a set of three points in 3D. Next, we compute all necessary calculations based on only those elementary components. It can be described as:

$$xray(F_x^A, F_y^A, F_z^A, F_x^B, F_y^B, F_z^B, Q_x, Q_y, Q_z, D_x, D_y, D_z, p, s_m, \alpha_\psi), \quad (3)$$

where $F$ points determine focus, $Q$ points determine semi-isocenter, $D$ points determine detector (all points are represented in 3D $(x,y,z)$), $p$ is a value of projection, $s_m$ and $\alpha_\psi$ describe the same X-ray but as parallel projections. The parameters are depicted in Fig. 3.

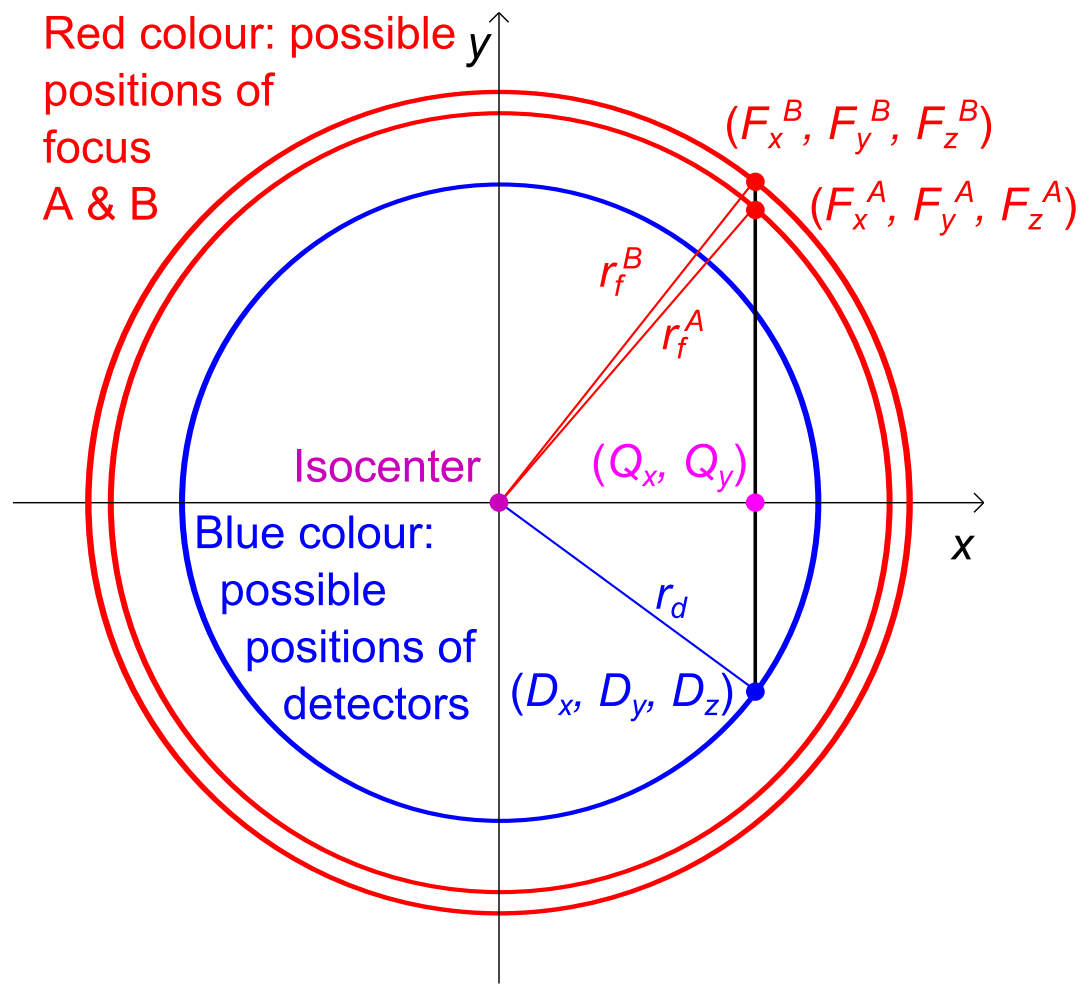

**Figure 3:** Schematic of the determined X-ray points.

After calculation, the virtual X-ray is chosen in comparison with the ideal real X-ray. First, we determine the ideal focus point with the following equations:

$$f_x = -(r_f + \Delta r_f^T) \cdot \sin(\alpha + \Delta\alpha^T); \quad (4)$$

$$f_y = (r_f + \Delta r_f^T) \cdot \cos(\alpha + \Delta\alpha^T); \quad (5)$$

$$f_z = z_0 + \Delta z^T, \quad (6)$$

where the $T$ is one of two focuses, $r_f$ is a distance between a focus and the centrum. The comparison is based on calculating of the $\zeta$ angle between real focus and virtual focus with relation to the semi-isocenter.

$$\zeta = \arccos\left( \frac{\hat{w}_x \cdot \hat{v}_x + \hat{w}_y \cdot \hat{v}_y}{\sqrt{\hat{w}_x^2 + \hat{w}_y^2} \cdot \sqrt{\hat{v}_x^2 + \hat{v}_y^2}} \right), \quad (7)$$

where:

$$\begin{aligned} \hat{v}_x &= f_x^T - Q_x; & \hat{v}_y &= f_y^T - Q_y; \\ \hat{w}_x &= F_x^T - Q_x; & \hat{w}_y &= F_y^T - Q_y. \end{aligned} \quad (8)$$

After selecting the two nearest real focuses, we must select the ideal detectors. An issue arises as we can not use the square function formula because the definition is not resolved for the line parallel to the $y$-axis. For this reason, a more complicated formula is required for the collinearity assessment of three points in 3D space. As a result of this formula, we can calculate the real detector position where we are defining for $(D_x, D_y)$:

$$\begin{cases} (D_x - f_{x0})^2 + (D_y - f_{y0})^2 = r_{fd}^2 \\ (Q_x - f_x)(D_y - f_y) - (Q_y - f_y)(D_x - f_x) = 0 \end{cases}. \quad (9)$$

After finding this detector, we can calculate the value of the virtual detector by using the three-linear interpolation function. Finally, the calculated, virtually parallel X-ray can be used for back-projection to create an image for iterative procedure (described in the next subsection). We used a FBP method to obtain a starting image for this iterative procedure.

### 3.2 Iterative part of algorithm

The iterative reconstruction procedure is schematically illustrated in the form of a diagram in Fig. 4. It is worth noting that in this diagram the block responsible for preparing the convolution kernel matrix ($h$) has been excluded from the iterative area because it is invariant through all iterations, hence it can be determined before starting the iterative reconstruction procedure. Assuming a constant image resolution, the matrix $h$ can be kept in the file to speed up the calculations. Furthermore, this diagram also excludes the block responsible for preparing the starting image outside the area of the iterative method, which is described in the previous subsection.

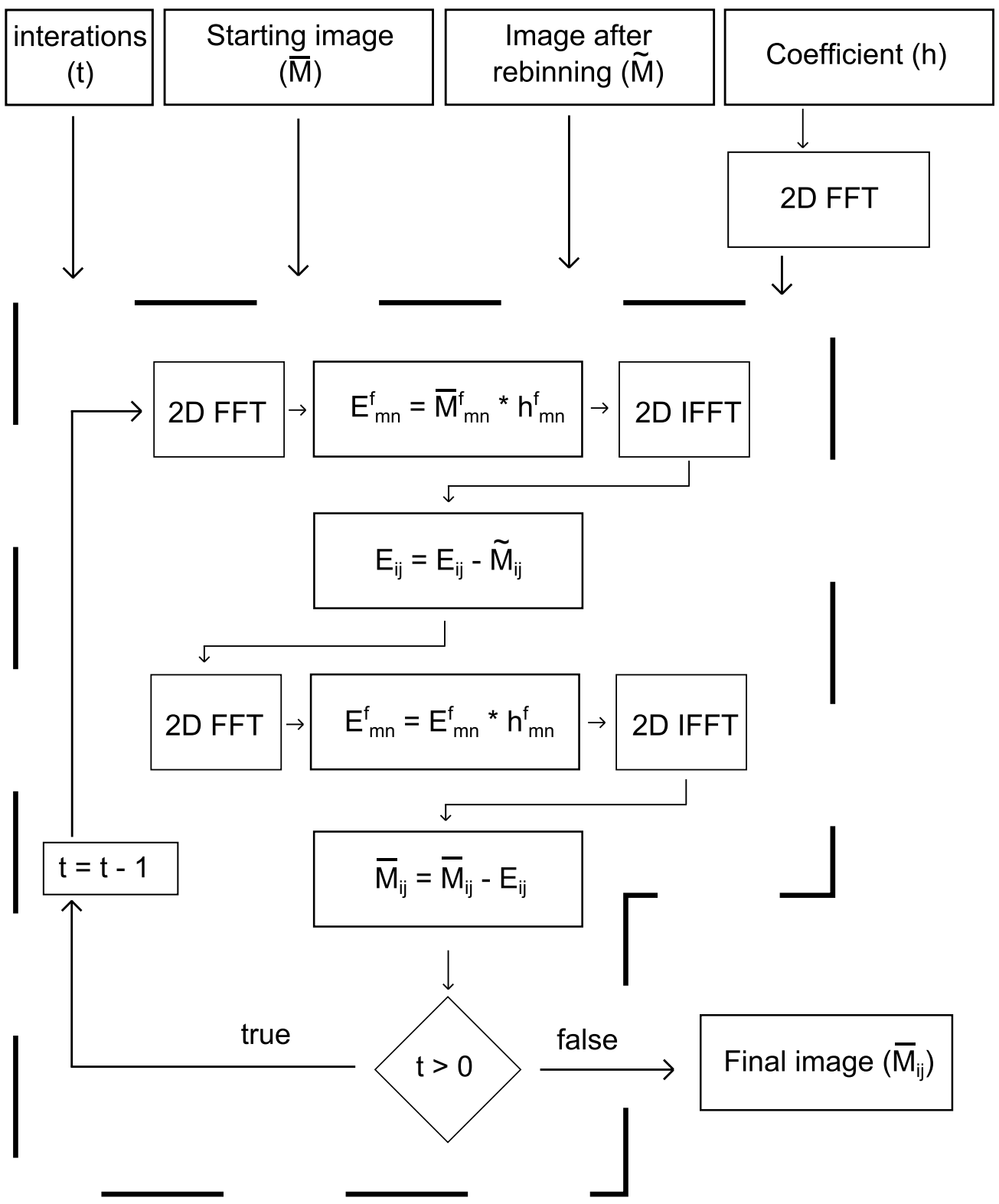

**Figure 4:** Diagram of the iterative method.

As a result of moving all calculations to the frequency domain, it is possible to reduce the computational complexity to $8log_2 4I^2$ where $I$ is the dimension of the final reconstruction image ($\overline{M}_{ij}$).

**Table 1:** Comparison of the computation times for the realizations of the rebinning and iterative reconstruction procedure on GPU.

| Iterations | GTX1080Ti | TitanV | RTX2080 | RTX3080 |
|---|---|---|---|---|
| 2 500 | 3.813 s | 1.800 s | 2.904 s | 1.513 s |
| 5 000 | 7.554 s | 3.573 s | 5.422 s | 2.874 s |
| 7 500 | 11.092 s | 5.325 s | 8.040 s | 4.239 s |

## 4 Results

In this work, we used measurements obtained from a medical scanner SOMATOM manufactured by Siemens AG, Forchheim, Germany. All projections were obtained using the helical mode, with a tube potential 120kVp and a tube current 200mAs. We carried out the calculations necessary to realize the iterative reconstruction using hardware implementations based on the NVIDIA GPU. The result of computation time of all parts of the algorithm are presented in Table 1. We also present the reconstruction images (Fig. 5, 6) of a real patient which were scanned at quarter dose. The reconstructed images were done in 5 000 iterations to recognize all of the details in comparison to the image quality of reconstructed images using the ASSR method.

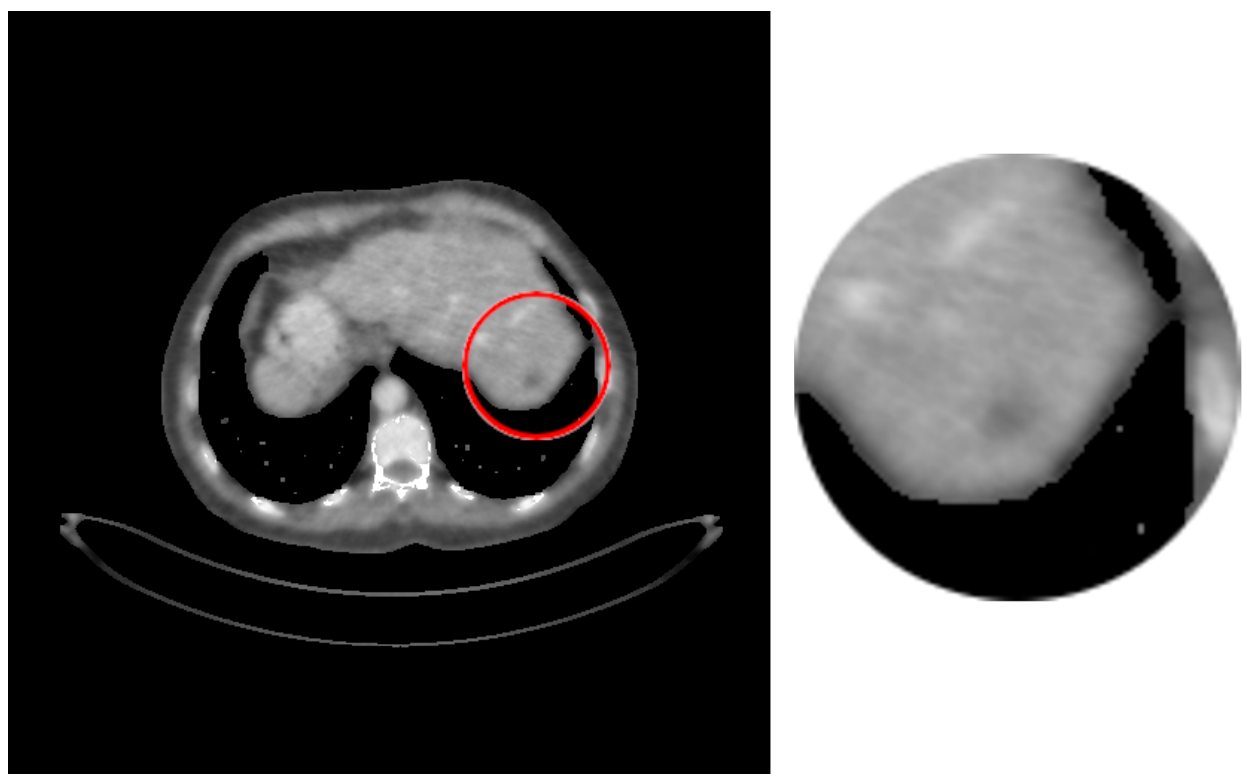

**Figure 5:** Reconstruction view of the real patient. Z 109; Iterations 5000; Dose 1/4; Proposed method.

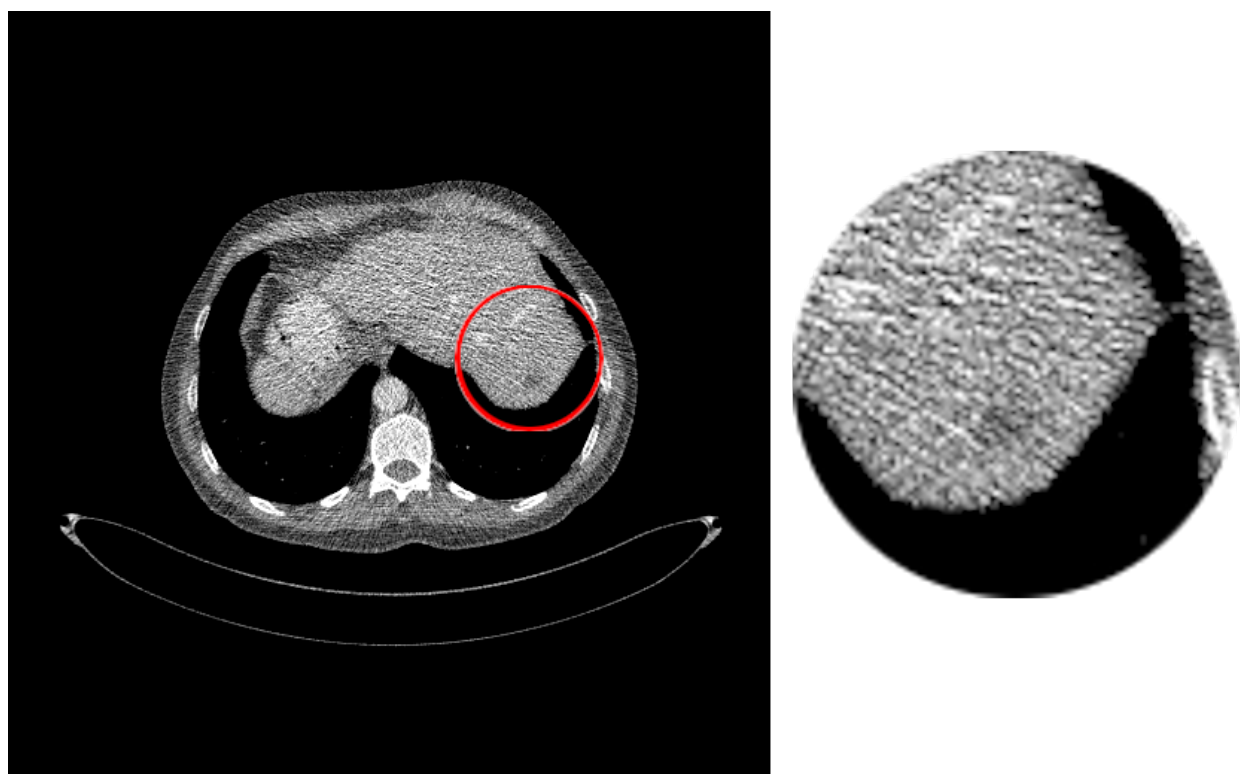

**Figure 6:** Reconstruction view of the real patient. Z 109; Dose 1/4; ASSR method.

Assessed by a radiologist, the number of iterations sufficient for an acceptable image from a diagnostic point of view is 7 000 iterations [10]. Due to the optimization and more accurate determination of the speed indicator of the iterative method, it was possible to obtain the same image quality with 5 000 iterations. This case is performed with the purpose of improving the diagnosis of liver cancer. Additionally, Fig. 7 shows the RMSE plot of the proposed method in this work in comparison to the ASSR method.

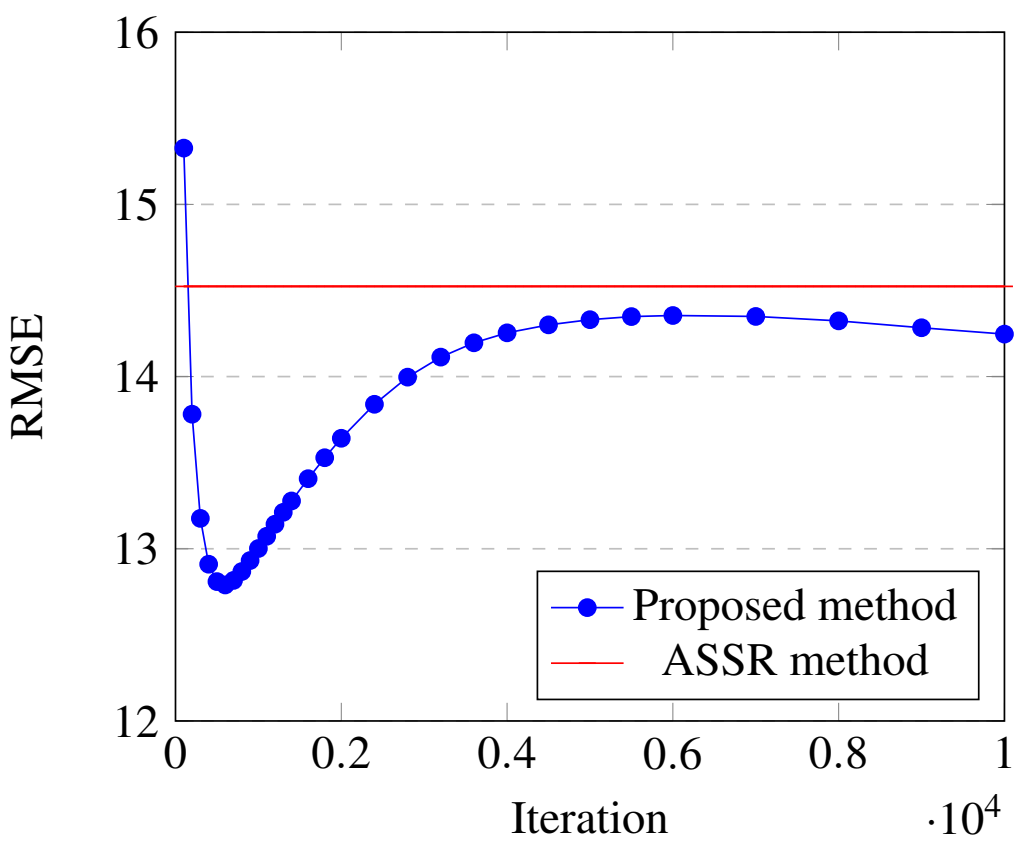

**Figure 7:** RMSE plot for the method proposed in this work & the ASSR method based on the mathematical Shepp-Logan data model, $z$: 0, Noise: Medium.

## 5 Conclusion

In this paper, we have presented a complete statistical iterative reconstruction method using FFS technology with an original rebinning method that can be used in CT scanners. Our results demonstrate improved reconstruction speeds (about 5s for all operations using a mid-range GPU) with adequate resolution to allow diagnosis of liver lesions scanned at a quarter dose. The improvement in reconstruction speed was mainly due to the use of an FFT algorithm during the most demanding calculations in the iterative reconstruction procedure as well as efficient programming techniques. For future implementation, we are interested in implementing our FFS image reconstruction approach in the context of using CT to distinguish hepatic lesions and identify metastases [11], [12]. Based on previous work [13], [14], [15], we are interested in improving the early detection of hepatic diseases, especially before patients become symptomatic, reducing mortality associated with chronic liver disease.

## Acknowledgements

The project financed under the program of the Polish Minister of Science and Higher Education under the name "Regional Initiative of Excellence" in the years 2019 - 2023 project number 020/RID/2018/19 the amount of financing PLN 12,000,000.

# Neural Network-based Single-material Beam Hardening Correction for X-ray CT in Additive Manufacturing

Obaidullah Rahman[1], Singanallur V. Venkatakrishnan[1], Zackary Snow[1], Paul Brackman[2], Thomas Feldhausen[1], Ryan Dehoff[1], Vincent Paquit[1], and Amirkoushyar Ziabari[1]

[1]Oak Ridge National Lab (ORNL), Oak Ridge, TN 37830, USA
[2]Carl Zeiss Industrial Metrology, LLC, Maple Grove, MN 55369, USA

**Abstract** Beam-hardening (BH) artifacts are ubiquitous in X-ray CT scans of additively manufactured (AM) metal components. While linearization approaches are useful for correcting beam-hardened data from single material objects, they either require a calibration scan or detailed system and material composition information. In this paper, we introduce a neural network-based, material-agnostic method to correct beam-hardening artifacts. We train a neural network to map the acquired beam-hardened projection values and the corresponding estimated thickness of the object based on an initial segmentation to beam-hardening related parameters, which can be used to compute the coefficients of a linearizing correction polynomial. A key strength of our approach is that, once the network is trained, it can be used for correcting beam hardening from a variety of materials without any calibration scans or detailed system and material composition information. Furthermore, our method is robust to errors in the estimated thickness due to the typical challenge of obtaining an accurate initial segmentation from reconstructions impacted by BH artifacts. We demonstrate the utility of our method to obtain high-quality CT reconstructions from a collection of AM components – suppressing cupping and streaking artifacts.

## 1 Introduction

X-ray CT reconstruction of AM components provides insight on defects [1, 2] introduced by the manufacturing process, allowing manufacturers to understand the impact of process parameters on part performance and, in turn, design consistent and reliable components. However, the complex attenuation of poly-chromatic X-rays as they propagate through dense metals results in beam-hardening (BH) artifacts, such as cupping and streaks, which make it challenging to detect microstructurally relevant features (e.g., cracks, porosity) in typical reconstructions. Methods to address BH can be broadly classified into hardware and algorithmic approaches [3, 4]. Hardware approaches involve filtering the X-ray beam to suppress higher energies, but this method reduces the flux, leading to poor reconstruction quality when the measurement times are kept the same. This method also requires an expert user to select the appropriate filter depending on the sample to be scanned. In contrast, software approaches include the use of a heuristic polynomial to linearize the normalized data [5] prior to the reconstruction, and more computationally expensive methods that attempt to model the non-linearities of the image formation process [6].

For industrial X-ray CT systems, linearization approaches are preferred due to their low computational complexity. These methods involve applying a polynomial correction to the normalized measurement data so that the projection vs thickness curve (p vs d) for the material is a straight line instead of the typical curve seen particularly at higher thickness values. In order to obtain this p vs d curve, one has to manufacture and measure a calibration wedge sample [3] corresponding to the same material as the component to be measured. For single material components, the linearization polynomial can be computed in theory even without a calibration sample, but this requires knowledge of source spectrum, filtering hardware specifications, and detailed knowledge of the detector spectral response to obtain the p vs d curve which is often impractical.

In this paper, we propose a new linearization approach based on the use of a neural network (NN). We first train a NN to map between the projected value and corresponding thickness to the parameters of a Van de Casteel attenuation model [7] by synthetically generating several test (p, d) data points. During inference time, the thickness values corresponding to each measurement are obtained by projecting an initial binary segmentation of the reconstruction obtained using the FDK algorithm [8]. Thus, we effectively obtain the parameters of a Van de Casteel model from the neural network, which can then be used to compute the linearization polynomial. The main advantage of our method is that it once it is trained for a collection of (p, d) data points, it can be used to correct for BH due to a range of materials. We demonstrate the value of our method by suppressing BH artifacts for numerous AM components made of different materials without any manual tuning of the algorithm - enabling a fully automated workflow for X-ray CT of metal AM components that produces high quality reconstructions. Furthermore, our method is robust against imperfect p vs d values due to an imprecise segmentation - a common occurrence for dense parts with complex geometries.

## 2 Method

Our method to correct for BH from single material scans consists of three steps: 1) use a NN to map *each* projection and estimated thickness value to the parameters of a BH model, 2) average the model parameters estimated by the

Corresponding author's email address: *rahmano@ornl.gov*. Research sponsored by the US Department of Energy, Office of Energy Efficiency and Renewable Energy, Advanced Manufacturing Office, under contract DE-AC05-00OR22725 with UT-Battelle, LLC.

NN for all the (p,d) values and 3) use the averaged model parameters to compute an $8^{th}$ order linearization polynomial. We train the proposed network solely on synthetically generated data using the bimodal energy model for BH from [7]. It was demonstrated in [7] that BH can be simplified using two dominant X-ray energies, $E_1$ and $E_2$, where $\mu_1$ and $\mu_2$ are the corresponding linear attenuation coefficients (LAC) of the material. The material's non-linear projection vs thickness can be obtained using

$$p_{bh} = \mu_2 d + \ln \frac{1+\alpha}{1+\alpha e^{-(\mu_1-\mu_2)d}} \quad (1)$$

where the left hand side is the BH-affected projection, $\alpha$ represents the ratio of the source-detector efficiency at said x-ray energies, and $d$ is the distance the x-ray beam has to traverse within the material. The ideal (BH-free) projection, which varies linearly with distance, is given by

$$p_{bhc} = \frac{\alpha\mu_1+\mu_2}{1+\alpha} d \quad (2)$$

### 2.1 Training

We used a NN, consisting of 16 fully connected layers with 512 neurons with biases and a ReLU activation, which we call beam-hardening correction network (BHCN). The input layer consists of 2 nodes for the projection value and associated thickness and the output layer consists of 3 nodes for the parameters of the model in (1). To get training data, we randomly generate vectors of $d^{tr}$, $\alpha^{tr}, \mu_1^{tr}$, and $\mu_2^{tr}$, each uniformly drawn from its realistic range.

$$\begin{aligned}
&d_i^{tr} \sim U(0\ mm,\ 20\ mm)\\
&\alpha_i^{tr} \sim U(4,\ 8)\\
&\mu_{1,i}^{tr} \sim U(0.3\ mm^{-1},\ 0.6\ mm^{-1})\\
&\mu_{2,i}^{tr} \sim U(0.03\ mm^{-1},\ 0.15\ mm^{-1})\\
&p_i^{tr} = \mu_{2,i}^{tr} d_i^{tr} + \ln \frac{1+\alpha_i^{tr}}{1+\alpha_i^{tr}\exp\{-(\mu_{1,i}^{tr}-\mu_{2,i}^{tr})d_i^{tr}\}}
\end{aligned}$$

Our claim is that the network will not need to know the material and should be able to estimate BHC parameters only from projection and distance data. Therefore we feed BHCN the pair ($p^{tr}$, $d^{tr}$) as input and train it by minimizing the weighted mean absolute error,

$$\frac{1}{N}\sum_{i=1}^{N} |\alpha_i^{tr}-\alpha_i^{out}| + 2|\mu_{1,i}^{tr}-\mu_{1,i}^{out}| + 5|\mu_{2,i}^{tr}-\mu_{2,i}^{out}|$$

The weight were empirically chosen so the losses from individual parameters are somewhat comparable, and one does not overwhelm the other.

### 2.2 Inference

In order to obtain parameters of the Van de Casteel model from the BHCN, we first reconstruct the measured data using the FDK algorithm. Next, we obtain a binary segmentation of this reconstruction using Otsu's algorithm [9] and forward project it to obtain the distance traversed corresponding to each measured projection. Then, BH-affected projection and distance vectors are fed into the BHCN to get estimates of vectors $\alpha$, $\mu_1$ and $\mu_2$. For each input data point the BHCN outputs a set of parameters. Since the input is "noisy" because of the erroneous segmentation, the output (BHC parameters) is expected to be "noisy" too. Taking respective means of the output vector estimates yields a "noise-free"/reliable version of the BHC parameters. Then, BH correction in projection using those parameters is followed by FDK reconstruction. The overall inference is outlined in Algorithm 1.

**Algorithm 1:** BHCN inference

$d_{max} \leftarrow$ largest expected object thickness;
$\varepsilon_d \leftarrow$ distance step size for polynomial fit, $n \leftarrow 8$;
$p \leftarrow$ BH-affected projection, $Im_{BH} \leftarrow FDK(p)$;
$Im_{BH}$[metal] $\leftarrow 1$, $Im_{BH}$[background] $\leftarrow 0$;
$d \leftarrow$ Forward project($Im_{BH}$);
**for** *each i* **do**
  $[\alpha_i^{est}, \mu_{1,i}^{est}, \mu_{2,i}^{est}] \leftarrow BHCN([p_i, d_i])$;
**end**
$\alpha \leftarrow \frac{1}{N}\sum_{i=1}^{N}\alpha_i^{est},\ \mu_1 \leftarrow \frac{1}{N}\sum_{i=1}^{N}\mu_{1,i}^{est},\ \mu_2 \leftarrow \frac{1}{N}\sum_{i=1}^{N}\mu_{2,i}^{est}$;
$d_{vec} \leftarrow [0,\ \varepsilon_d,\ 2\varepsilon_d, \ldots, d_{max}]$;
$proj_{bh} \leftarrow \mu_2 d_{vec} + \ln \frac{1+\alpha}{1+\alpha e^{-(\mu_1-\mu_2)d_{vec}}}$;
$proj_{bhc} \leftarrow \frac{\alpha\mu_1+\mu_2}{1+\alpha} d_{vec}$;
$f_{polyfit} \leftarrow$ poly. fit($proj_{bh}, proj_{bhc}, n, d_{vec}$);
$p_{correct} \leftarrow f_{polyfit}(p)$, $Im_{BHC} \leftarrow FDK(p_{correct})$;

## 3 Results

In Sec. 3.1, we compare our method with a baseline curve-fit method and demonstrate that the baseline method can fail in practical conditions. In Sec. 3.2, we compare our method to a more robust, CAD- and physics–based, method.

### 3.1 Comparison with a curve-fit (trivial BHC) method

One method to estimate BH model parameters from BH-affected projections is by minimizing the difference between the model, i.e. Eq. (1), and the actual projection i.e. simple curve fitting (CF) using

$$(\alpha^{CF}, \mu_1^{CF}, \mu_2^{CF}) = \underset{\alpha^{CF}, \mu_1^{CF}, \mu_2^{CF}}{\text{argmin}} \{|\, p - p_{model}\,|_2\} \quad (3)$$

where $p_{model}$ is computed from Eq. (1) using the distance and current iteration $(\alpha^{CF}, \mu_1^{CF}, \mu_2^{CF})$.

#### 3.1.1 Simulation data

We start with simulation data to compare the BHCN and the curve-fit methods by creating a synthetic metal component with BH parameters $(\alpha^{GT}, \mu_1^{GT}, \mu_2^{GT})$). Forward-projecting the metal mask provides $d^{GT}$, which, together with the BH parameters, is used to obtain $p_{bh}$ using Eq. (1) and $p_{bhc}$ using Eq. (2). We feed the vector pair $(p_{bh}, d)$ to BHCN and curve-fit methods. The estimated $(\alpha^{BHCN}, \mu_1^{BHCN}, \mu_2^{BHCN})$ and $(\alpha^{CF}, \mu_1^{CF}, \mu_2^{CF})$ will be used to get the BHC projections $p_{BHCN}$ and $p_{CF}$, respectively. In the first case, which is ideal, $d$ is exactly known, i.e. $d = d^{GT}$. This corresponds to the top left subfigure in Fig. 1. The rest of the subfigures, in clockwise direction, correspond to increasingly erroneous $d$ supplied to the two competing algorithms. $d$ is the result of incorrect segmentation (either metal declared as background, or background declared as metal) of the uncorrected FDK reconstruction using different threshold values for each of the 5 remaining scenarios. Such challenges are common when segmenting XCT scans of dense metal components with complex shapes.

It can be seen that in the ideal case, they both perform quite well, and the BHC projection vs distance from both algorithms coincide with the ground truth projection. As the segmentation starts to deteriorate, the curve-fit algorithm performance degrades more than BHCN, as seen in the departure of its curve from that of GT. After analysing the p vs d curves, we perform FDK reconstruction for each method and for each segmentation case. Fig. 2 corresponds to the reconstructions from the bottom left scenario from Fig. 1. This is the worst segmentation case among the six. The three slices shown demonstrate the robustness of BHCN over CF in case of incorrect segmentation, a common issue with complex component geometry, scattering, noise, etc.

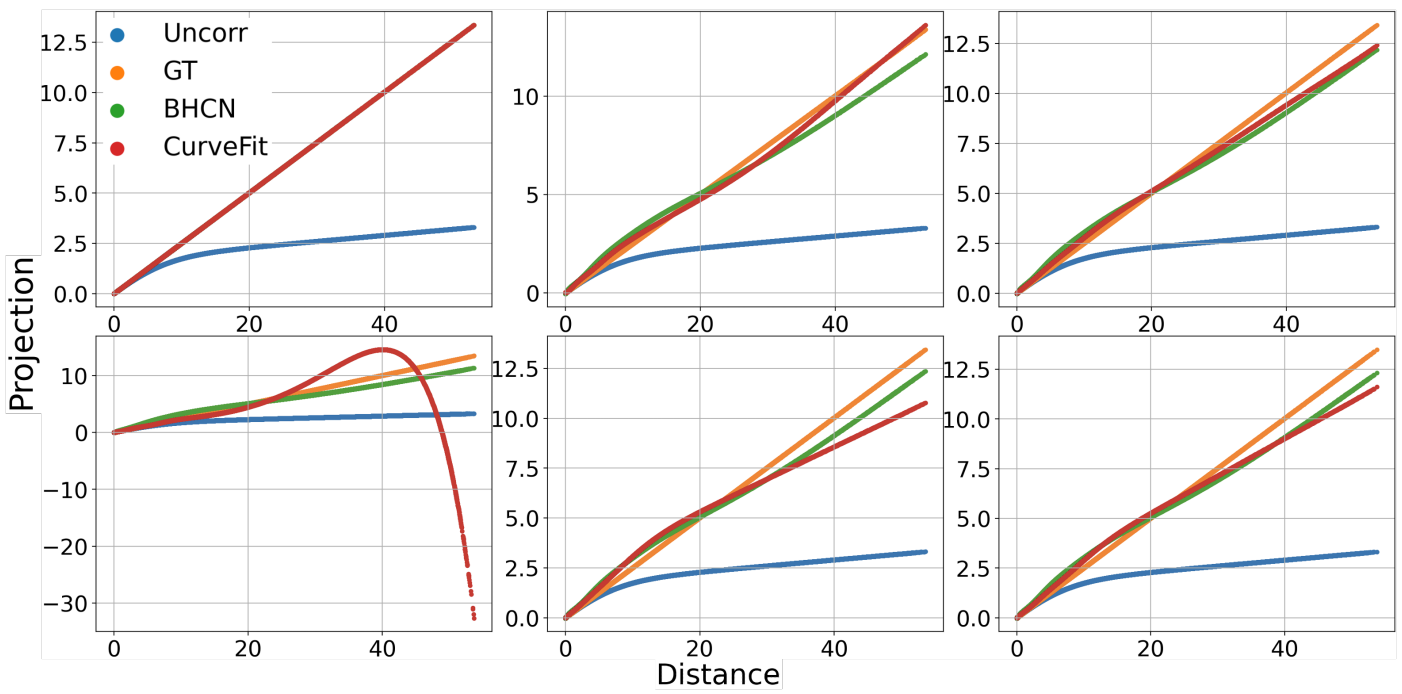

**Figure 1:** p vs d curve for increasingly worsening segmentation starting with top left in clockwise direction. Curve-fit degrades more than BHCN as segmentation becomes more inaccurate.

### 3.1.2 Experimental data

Fig. 3 demonstrates the performance of BHCN and curve-fit algorithms on an experimental data set - a scan of a steel turbine blade. Due to severe beam hardening, the binary segmentation has large errors. We observe that the proposed BHCN helps suppress cupping artifacts compared to the uncorrected image, while CF has made beam hardening worse and introduced artifacts.

## 3.2 Comparison with CAD- and physics–based BHC method

The CAD- and physics–based model proposed in [10, 11] was used to estimate the BHC parameters for alloys in our case studies and to generate a baseline reference to compare BHCN against. We refer to this method for BH artifact reduction as *reference*. When new alloys are developed, the elemental composition also changes, so the reference model would require re-estimation of the beam-hardening parameters. However, our universal BHC method is robust for different alloys without any need for re-calibration.

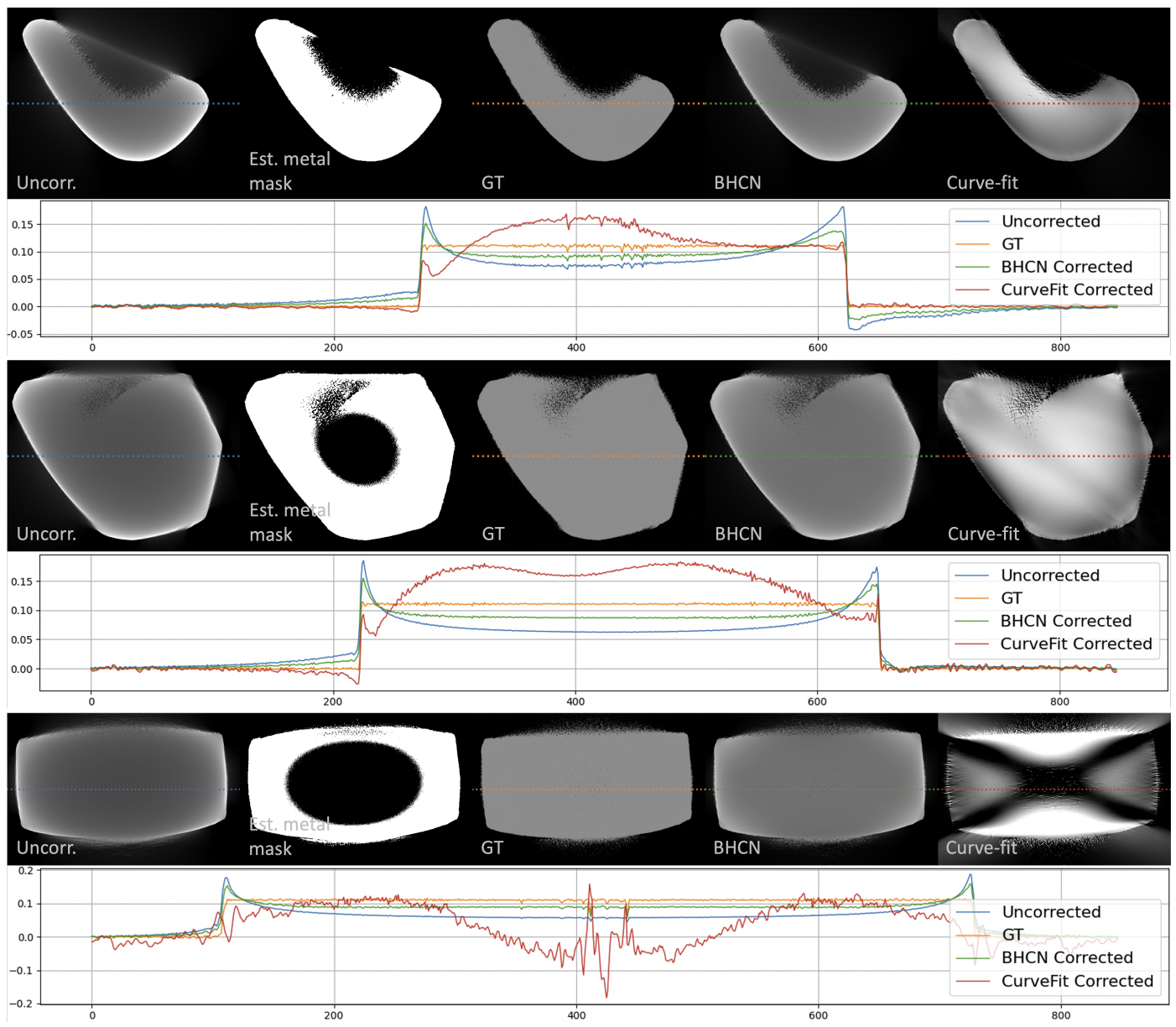

**Figure 2:** Example of incorrect segmentation of uncorrected FDK from simulated data from Fig. 1 bottom left p vs d curve. The top row of the 3 subfigures indicate different slices of the reconstruction image. (left to right): Uncorrected FDK, incorrect metal mask, ground truth, BHCN, CF. Bottom row of the subfigures: Profile plots. Curve-fit has introduced artifacts and incorrect intensities, and its profile plot, compared to BHCN's, deviates more from that of GT.

### 3.2.1 Robustness across materials

To compare BHCN images with reference images, AM components with complex geometries, including cylinders, poles, fins, and inclines, were constructed from aluminium-cerium (AlCe), stainless steel (316L), and nickel-cobalt (NiCo). In Fig. 4, the reconstruction without any BHC, displays a high degree of beam hardening. Both BHCN and the reference method have similar reduction in BH, as highlighted by the profile plots. The inset shows an expanded view of the ROI marked in each image, highlighting the better contrast of the BH corrected images near flaws in both BH corrected images.

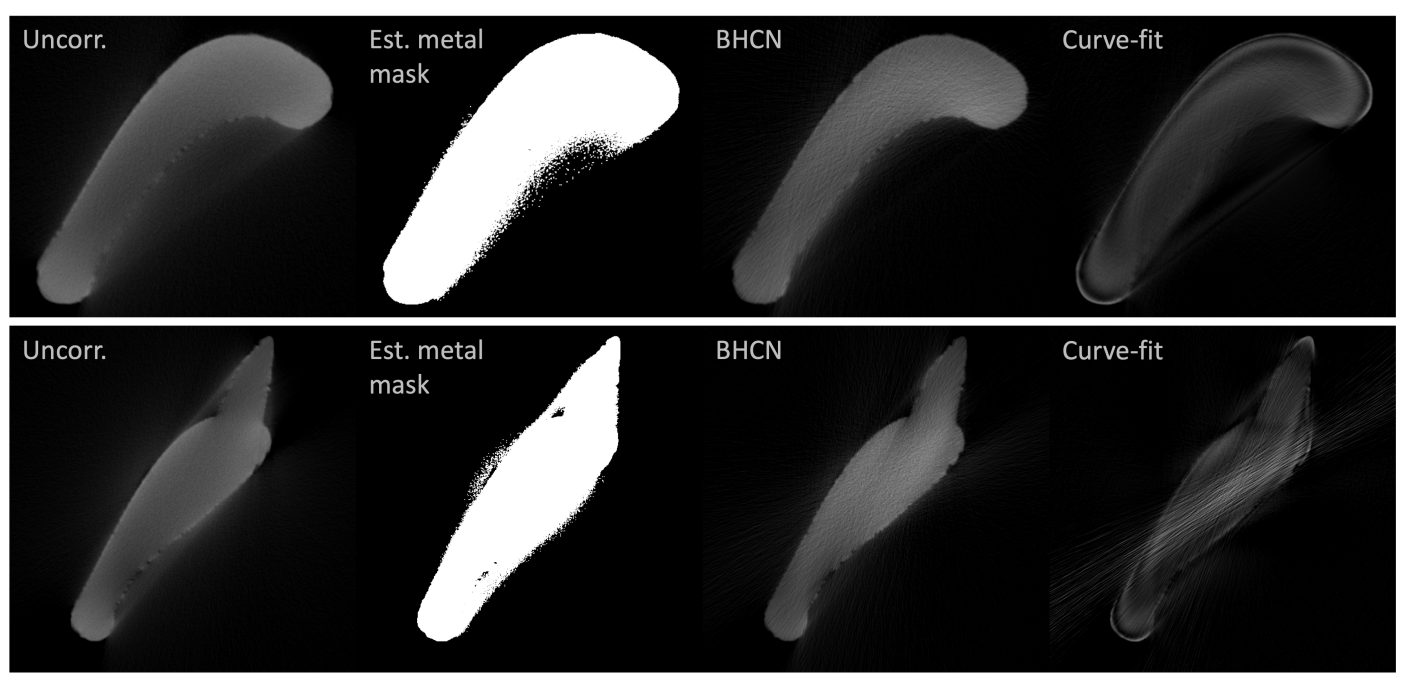

**Figure 3:** 2 slices from the reconstruction of a real turbine blade scan (left to right) Uncorrected, estimated metal mask, BHCN corrected, curve-fit corrected. BHCN makes the metal component of the image more uniform, but curve-fit seems to introduce strange intensities and distort the shape.

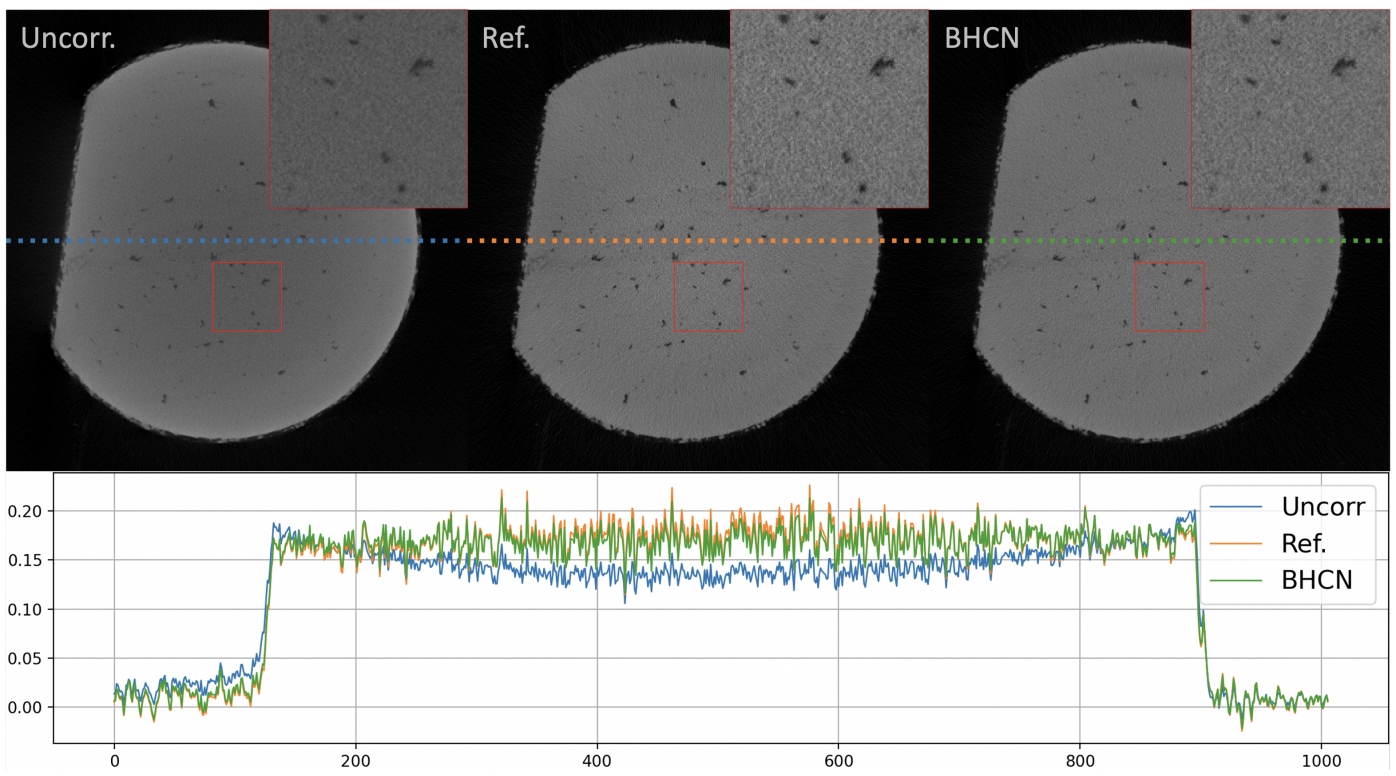

(a) AlCe

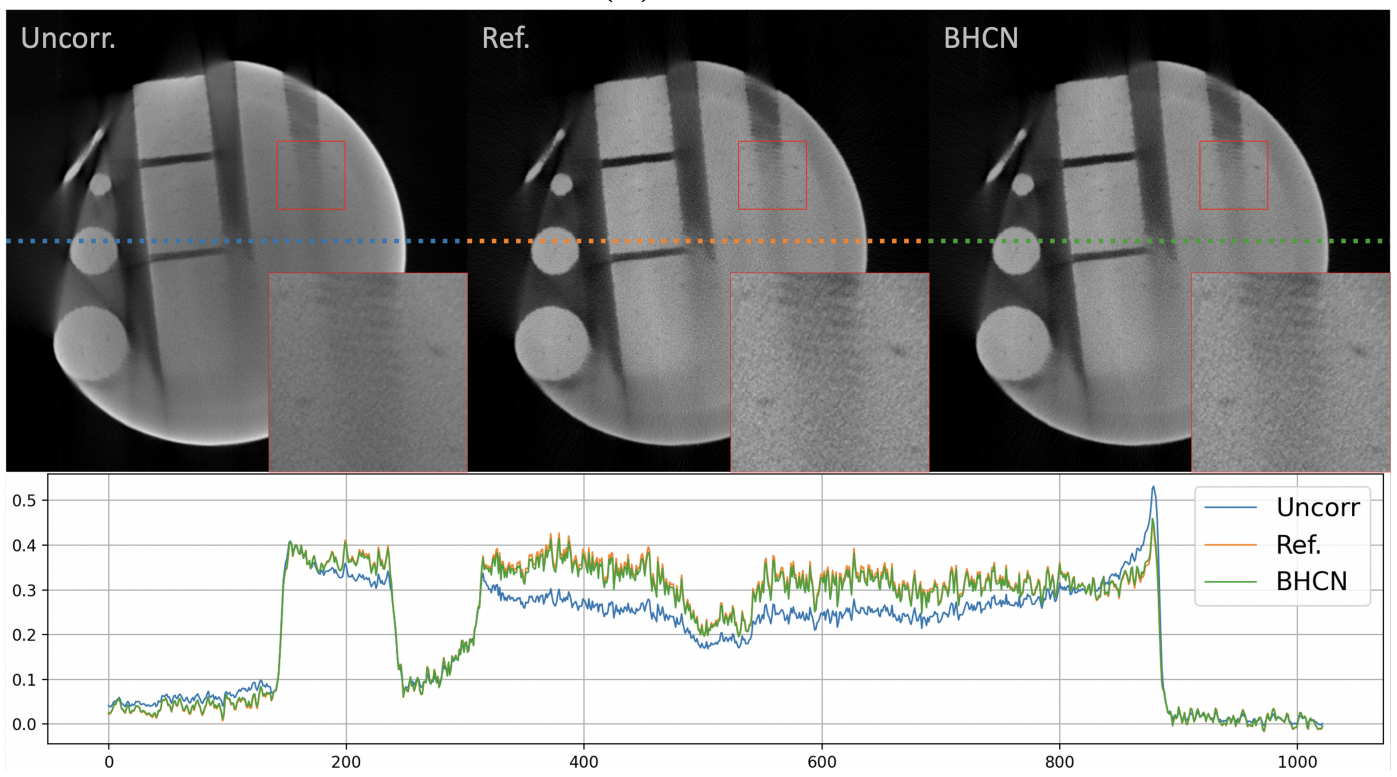

(b) NiCo

**Figure 4:** (left to right) No correction, reference, BHCN; Profile plot. BHCN reduces beam hardening artifact as much as the reference method as evident from the reduction in cupping artifact. Both reference and BHCN have better defect contrast in the RoI.

### 3.2.2 Robustness across various geometries

AM allows for printing of complex geometries, which in turn complicate beam hardening correction for XCT scans of those components. Despite that, our results suggest that the BHCN extends very well to complicated geometries. In Fig. 5 BHCN reduces BH in the top and bottom subfigures (pentagon and flower vase), and has more uniform-looking image than the reference subfigure (blade).

## 4 Conclusion

We developed a BHC network (BHCN) that is more robust than a baseline curve-fit method and compares well against the recently proposed CAD- and physics–based reference method that needs to be calibrated for each alloy. Our experiments show that the BHCN reduces BH for most alloys currently used in AM, and for different geometries. It also furnishes better defect contrast that should lead to more accurate defect characterization. Further, BHCN performs BHC by reducing cupping artifacts and producing uniform-looking images for all the alloys we tested for. We have also demonstrated that BHCN works well with simulation and real data despite the absence of accurate segmentation for the distance data needed for BHCN input. This supports the claim that in practical scenarios where perfect segmentation is not possible, BHCN can still perform beam-hardening reduction robustly without any retraining, calibration, or knowledge of the component material.

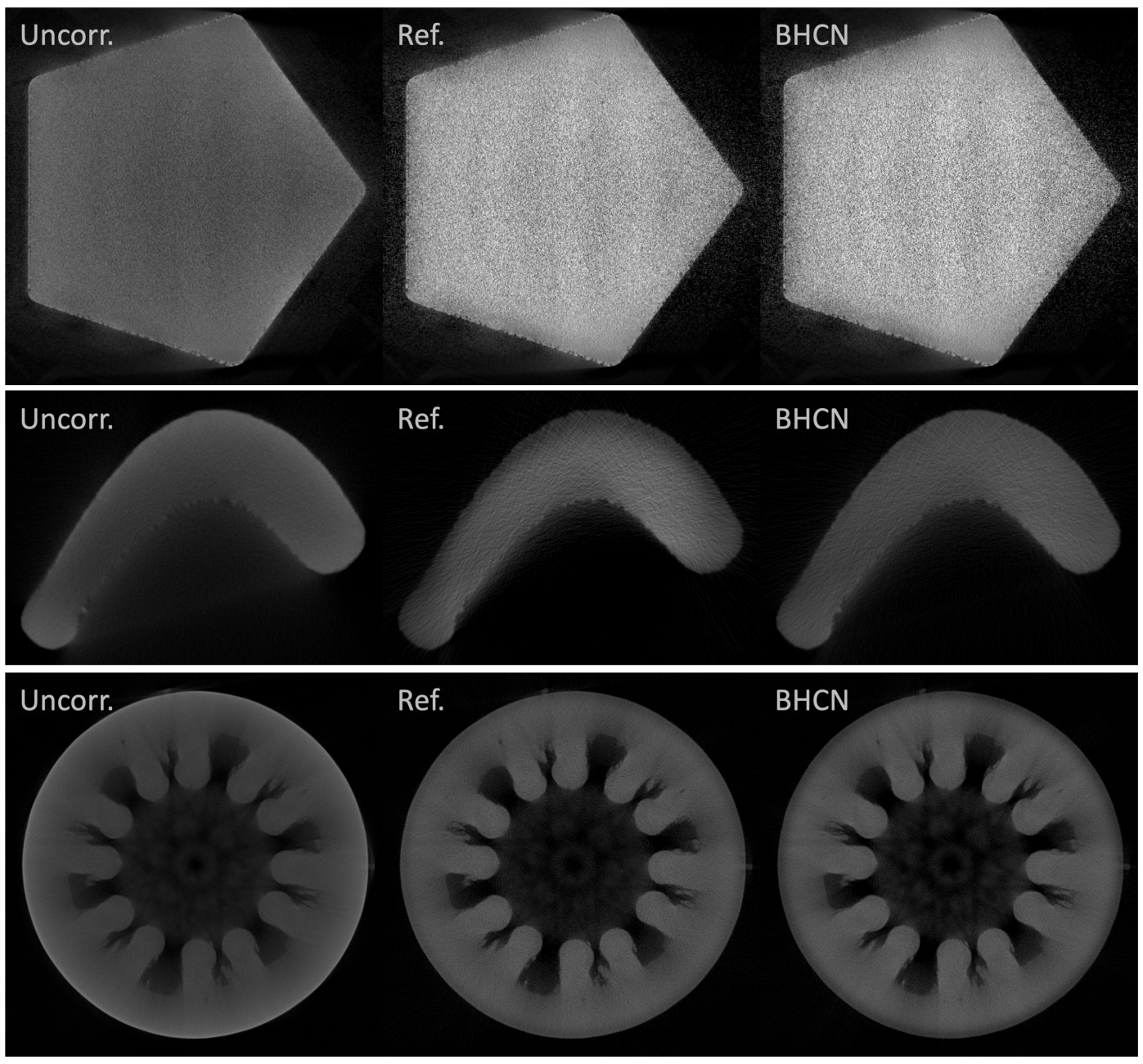

**Figure 5:** Different geometries and AM processes: (Top to bottom) Pentagon, blade and flower vase; (left to right) No correction, reference, BHCN. Both correction methods have similar reduction in BH for pentagon and vase, but BHCN blade metal part looks more uniform than that of the reference.

# Recovery of the spatially-variant deformations in dual-panel PET systems using Deep-Learning

Juhi Raj[1 *], Maël Millardet[1,2], Evgeny Kozyrev[1], Srilalan Krishnamoorthy[1], Joel S. Karp[1], Suleman Surti[1], and Samuel Matej[1]

[1]Radiology Department, University of Pennsylvania, Philadelphia, United States
[2]Siemens Medical Solutions USA, Inc., Knoxville, Tennessee, United States
[*]corresponding author, email: Juhi.Raj@pennmedicine.upenn.edu

**Abstract** Dual panel PET systems, such as Breast-PET (B-PET) scanner developed at University of Pennsylvania, exhibit asymmetric and anisotropic spatially-variant deformations in the reconstructed images due to the limited-angle data and strong depth of interaction effects for the oblique LORs inherent in such systems. In our previous work, we studied TOF effects and image-based spatially-variant PSF resolution models within dual-panel PET reconstruction to reduce these deformations. Although the application of such models led to better and more uniform quantification of small lesions across the field of view, the efficacy of such an approach is limited to small objects, such as point sources and small lesions. On the other hand, large object deformations caused by the limited-angle reconstruction cannot be corrected with the PSF modeling alone. In this work, we investigate the ability of the deep-learning networks to recover such strong spatially-variant image deformations of dual-panel systems using analytically simulated data from a dual-panel system.

## 1 Introduction

Dual panel PET systems offer certain advantages compared to the whole body PET systems, including higher sensitivity because of the detectors being placed closer to the scanned object, easier access to the object (such as for biopsy and monitoring response to therapy), and a simpler, potentially cheaper, design [1–4]. On the other hand, limited angular coverage, truncated angular data, and strong depth of interaction effects due to the steepness of LORs given by the closeness of the detectors can lead to severe artifacts and point spread function (PSF) deformations, which are strongly asymmetric and spatially variant [5].

Time-of-flight (TOF) provides additional information which reduces the limited angle artifacts; however, the problem is still quite challenging and cannot be eliminated even if a timing resolution of 200 ps or better is achieved [6, 7].
In our previous work, we investigated the application of TOF modeling [6] and image-based resolution models within statistical image reconstruction modeling the limited angle and Depth-of-Interaction (DOI) effects on the PSF deformation in the reconstructed image space [5, 7]. Although reconstructions with the optimized TOF kernel widths and with the spatially-variant PSF models led to better and more uniform quantification of small lesions across the field-of-view (FOV), the efficacy of the image-based resolution models is limited to small objects, such as point-sources and small lesions. Even though the reconstruction with PSF models showed more uniform contrast recovery of small lesions, the deformations could not be fully corrected with the PSF image-based resolution modeling alone. In [8] we proposed to use a Deep-Learning (DL) based post-reconstruction approach to reduce or eliminate the limited-angle artifacts and deformations in the images reconstructed from a dual-panel system. The potential of the DL approach lies in the fact that the topology of neural networks, defining a hypothesis functional space, can be trained to provide a mapping from the data or deformed image to expected image with an ability to reduce the limited-angle artifacts (and reject any unrealistic null-space structures).

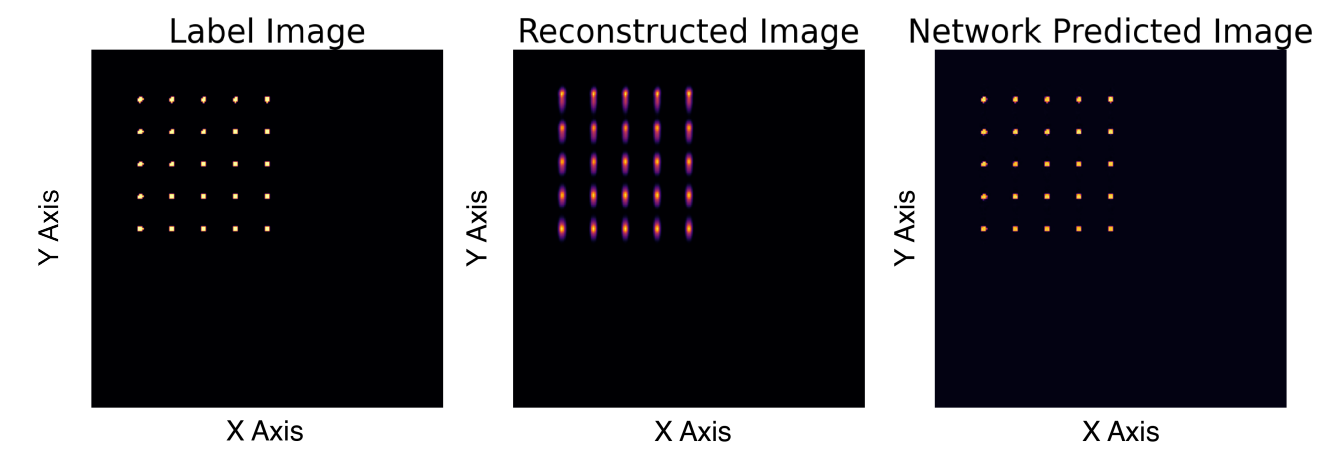

**Figure 1:** Center slice of 3D images; Spatially-variant PSF behavior of a dual-panel system, represented on the grid of point-sources placed within one quadrant of the system (horizontal detectors are placed at the top and bottom edges of illustrated images). Left: simulated point-sources, middle: deformed image affected by the spatially-variant PSF deformations (in a direction orthogonal to the detector panels), right: predicted output images of the trained network for a simplified case of the point-sources, illustrating the overall goal of the investigations in this work (but with more realistic objects).

In this study we continued with that work by further studying the ability of the DL based post-reconstruction approaches to capture and correct for the strong spatially variant deformations, as seen from the dual panel systems. It has been observed before that the neural networks with variable resolution levels (of latent feature spaces), such as U-Net [9], have the ability to capture the spatially variant image properties. However, to our best knowledge, nobody investigated so far, such a strong asymmetric (including shifts) and spatially variant deformation effects recovered on U-Net from those demonstrated in the dual panel systems.

Therefore, in this work we applied and tested DL approaches on simulated data from a dual-panel PET system with detectors placed close to the scanned object. Our future work will involve expanding the DL based post-reconstruction tool with measured data from the B-PET, a dedicated breast imager. Thereafter, the insights of this study will be applied to a direct DL reconstruction using histo-images.

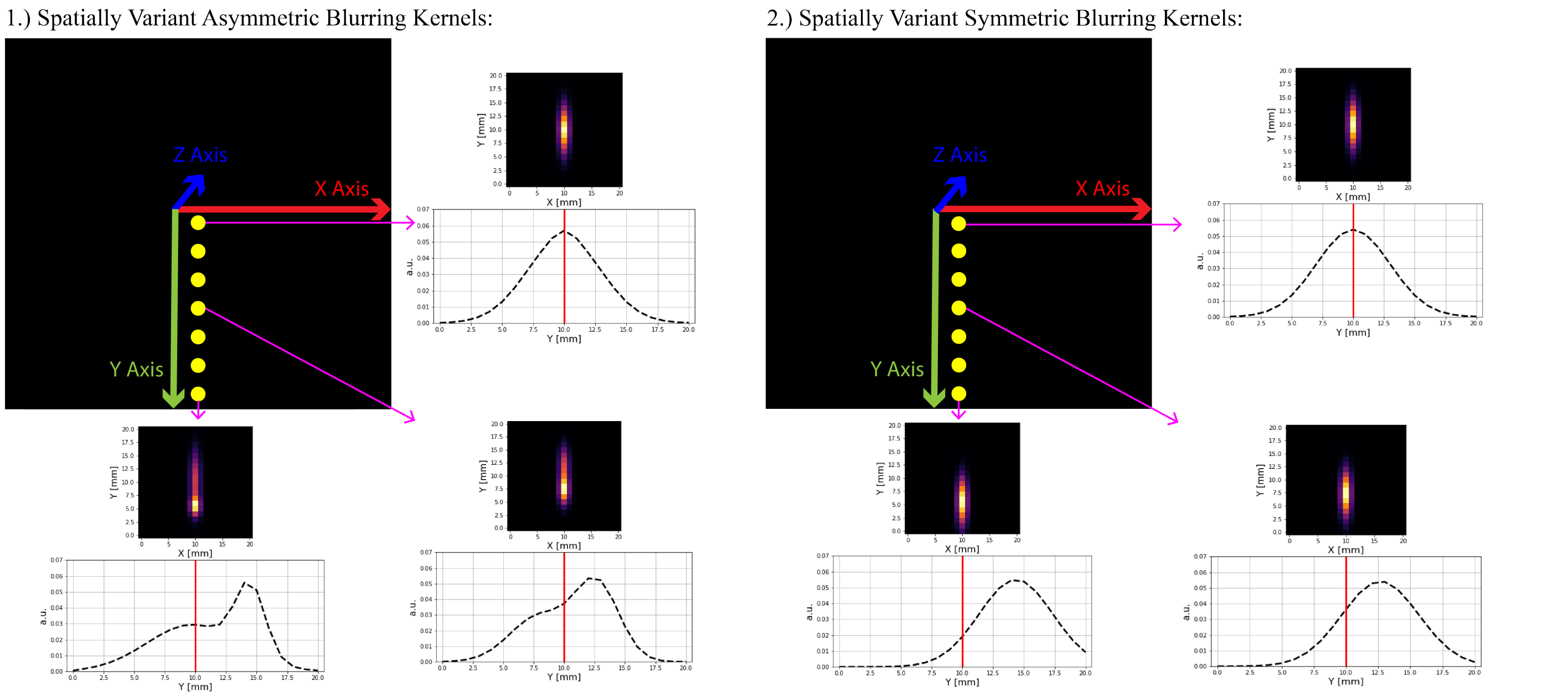

**Figure 2:** Illustration of two resolution models considered in this work to generate training and testing data sets. The left panel shows the realistic spatially-variant model (the core of this work) based on the PSF measurements in the reconstructed image space using point sources placed throughout the FOV in the XY axes. The blurring in the Z-direction is the same as in the X-direction. Each PSF kernel was parameterized in this work using a Gaussian mixture model composed of two weighted Gaussians with varying widths and shifts. The right panel shows the model using only a simple 3D Gaussian blur of constant width but with a spatial-variant shift. The point PSFs are illustrated only in one quadrant (negative Y-Axis) because the spatially variant behavior is mirrored in the positive Y-axis.

## 2 PSF Deformations

As mentioned earlier, the design of a dual panel PET system poses increased DOI effects due to the small separation between the detector heads and consequently steep incident angles for the annihilation photons with respect to the detector panel surfaces, and limited angle coverage due to the finite size of the detector panels; combination of both these effects leads to significant point spread function (PSF) deformations in the reconstructed image as shown in Figure 1. The reconstructed image in Figure 1 (center) also shows that the PSF deformations are progressively asymmetric and strongly spatially variant. The extent of this asymmetric spatial variance was employed in these simulation based on the image based resolution models that accurately capture the asymmetric PSF shape in images reconstructed from data acquired with the dual-panel scanner geometry [5].

The left panel of Figure 2 is a representation of the PSF deformations characterized in the reconstructed image space using point sources placed throughout the dual-panel FOV. The deformed PSFs become strongly anisotropic (elongated about 4-times) and asymmetric (for out of the center locations) as we progress along the Y-axis. Each PSF kernel was parameterized using a Gaussian mixture model composed of two weighted Gaussians with varying widths and shifts. The point source grid is depicted only in one quadrant (Y-Axis) because the spatially variant behavior is observed to be mirrored in the positive Y-axis.

## 3 Deformation Recovery via 3D U-Net

### 3.1 Network Architecture

As the first step we adopted a 3D U-Net architecture as shown in Figure 3. The network comprises down-sampling path, bottleneck, and an up-sampling path. Along the path, the number of convolution channels is doubled and spatial resolution is halved in the contracting path, and the reversal operations are applied during the expanding path. The total number of learnable parameters is $5\times10^6$ with input image size of 221 x 221 x 101. Each layer is configured with 3 x 3 x 3 convolution layer followed by a rectified linear unit (ReLU) as the activation functions. A kernel stride of 2 was used in spatial down- and up-sampling.

### 3.2 Network Training

Three separate data sets were generated as shown in Figure 4, i.e., point sources, spherical sources, and complex lesion-like structures (merging of three ellipsoids with random sizes and orientations) at random locations in the FOV with varying intensities. 50 training pairs and 5 validation pairs were generated and used for each of the three separate trainings and validations, respectively. The training was performed using the L1-loss function with the Adam Optimizer with the learning rate $10^{-4}$ and a batch size of 1 for 5000 epochs.

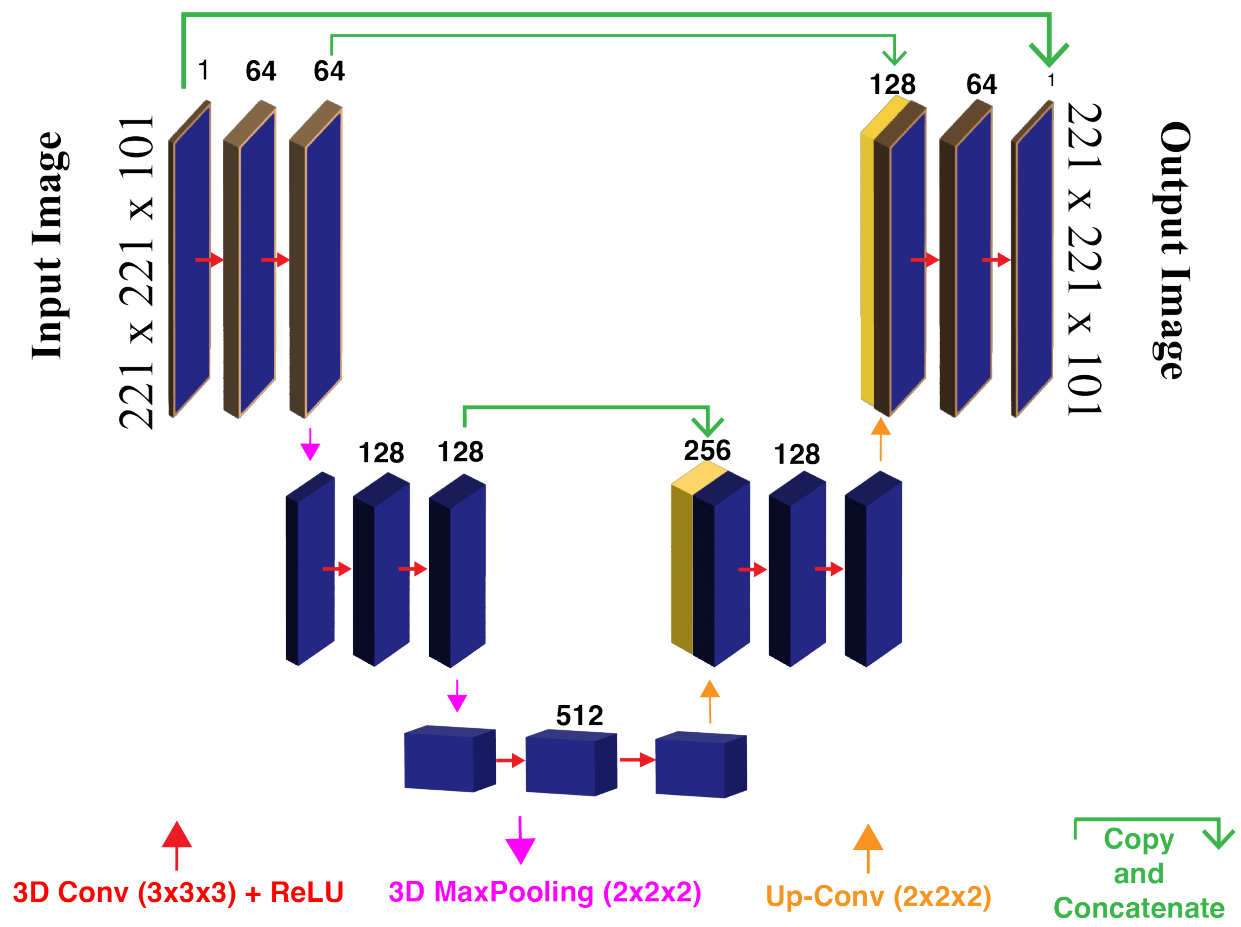

**Figure 3:** The 3D U-Net architecture adopted in this work. Each layer in the architecture is configured with 3 x 3 x 3 convolution kernels followed by a rectified linear unit (ReLU) as the activation function shown as red arrows. 3D MaxPooling function (yellow arrow) was used in the down-sampling path. The Up-Transpose is depicted as blue arrows in the up-sampling path. Skip connections are denoted as green arrows. The input and output image size is 221 x 221 x 101 with only 1 channel.

### 3.3 Test Results

The top panel of Figure 4 shows the comparison of the generated phantoms, deformed images, and the 3D U-Net predicted images. The corresponding vertical profiles of reconstructed small lesions at two extreme locations (close to the detector panels) are shown at the bottom. We additionally used the trained network on complex lesions and tested the trained network on realistic lesion shapes we obtained from a conventional whole-body PET patient scan as shown in (Figure 5). Our results demonstrate the ability of the 3D U-Net network to substantially reduce the strongly asymmetric spatially-variant deformations and recover accurate lesion locations in dual-panel PET images.

Having these very encouraging results, one important question arose: since our training data set contained only a modest number of relatively simple objects, did the network truly learn the location-dependent behavior of the dual-panel system, or did it just learn a set of particular shapes (or shape classes) and the corresponding shifts for each particular shape. We have thus designed a simple test to explore this question in the next section.

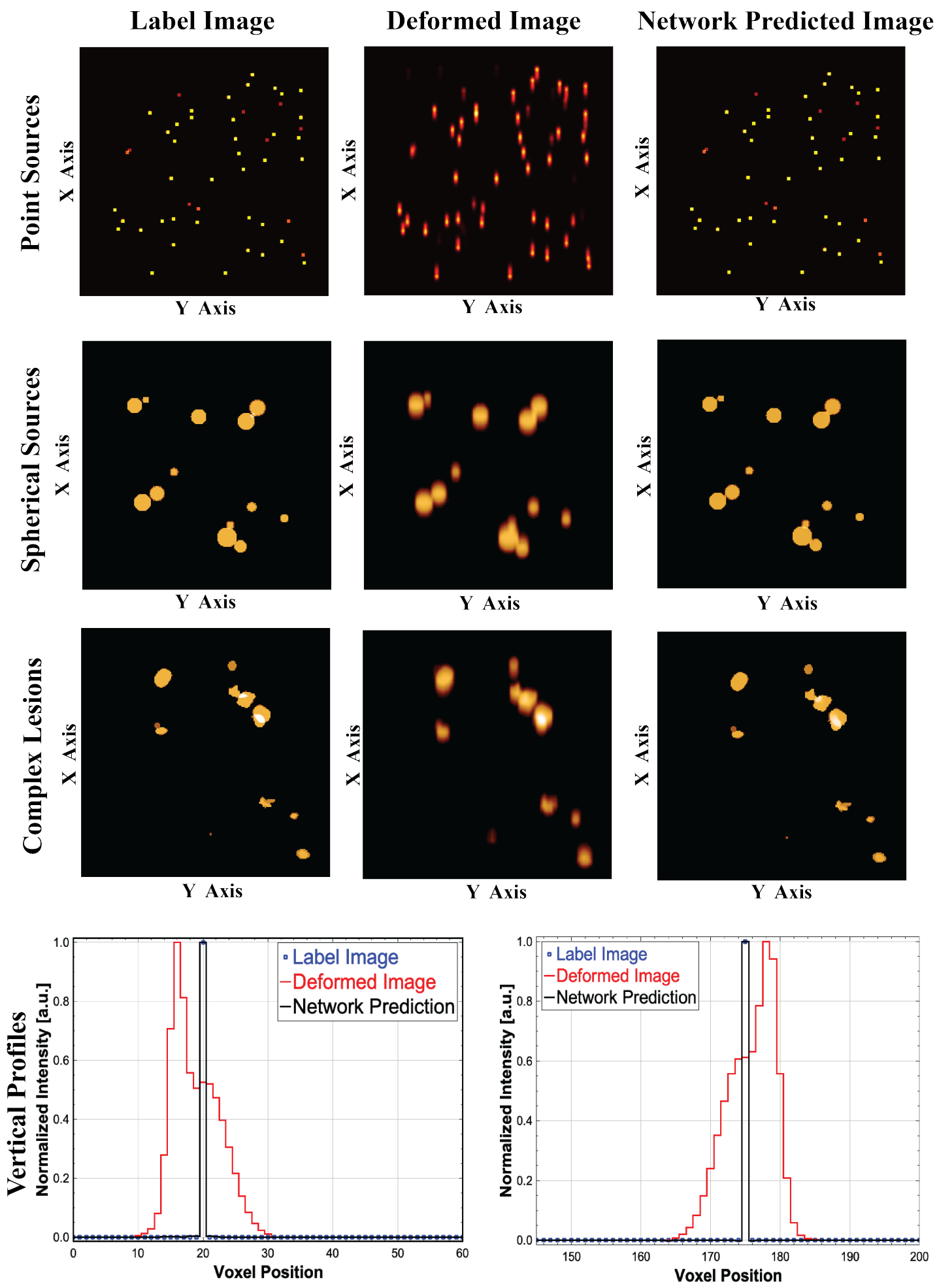

**Figure 4:** (Top Panel) Center slice of 3D label images of the randomly generated (since the network training requires learning behaviors from random locations) point sources, spherical sources, and complex lesions (given by a combination of 3 randomized ellipsoids) are shown in the first column. The corresponding deformed images and the network (3D U-Net as shown in Figure 3) predicted output is shown in the second and third columns, respectively. (Bottom Panel) Vertical line profile (along the Y-axis) of small spherical sources at different vertical positions.

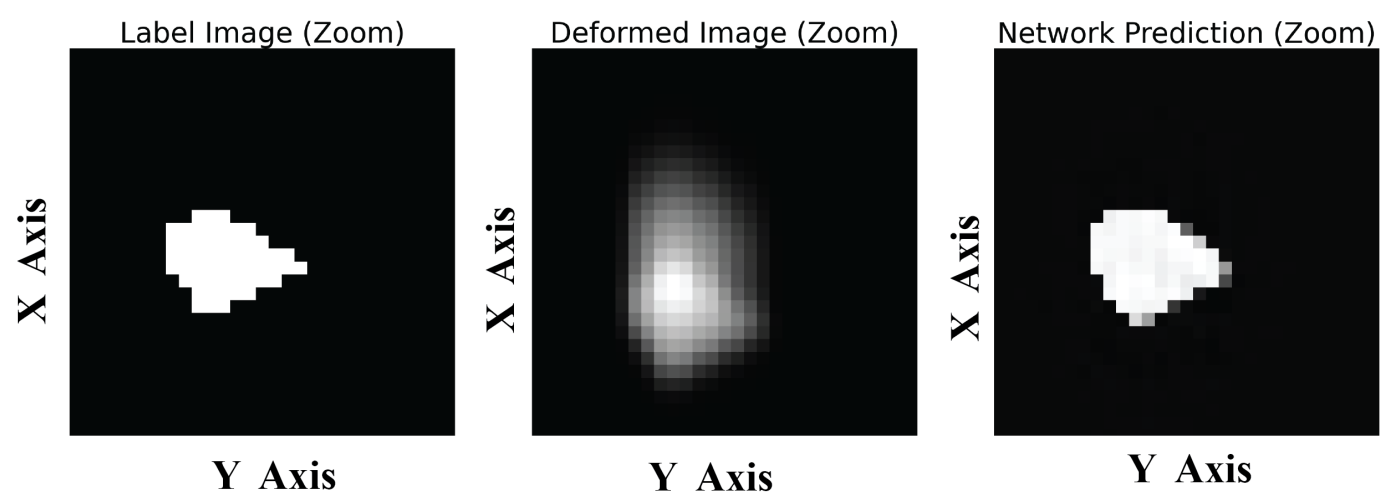

**Figure 5:** Comparison of the center slice of simulated label, deformed and 3D U-Net predicted image of a segmented lesion extracted from a conventional whole-body PET scan of a patient.

## 4 Test of Location vs. Shape Depended Recovery

### 4.1 Network Training

For this test we performed a small study using the previously described 3D U-Net. To reject the possibility that the network learned particular shapes instead of the location-dependent behaviour, we simulated a system in which the PSF shape did not change throughout the FOV - using a single 3D Gaussian with a fixed elongation in the y direction and the only change was the y-dependent shift of the 3D Gaussian, as illustrated in Figure 2-right. Using this resolution model, we generated data set composed of small spherical lesions at random locations in the FOV with varying intensities. We generated 50 training pairs and 5 validation pairs. The training was performed using the L1-loss function with the Adam Optimizer with learning rate $10^{-4}$ and a batch size of 1 for 5000 epochs.

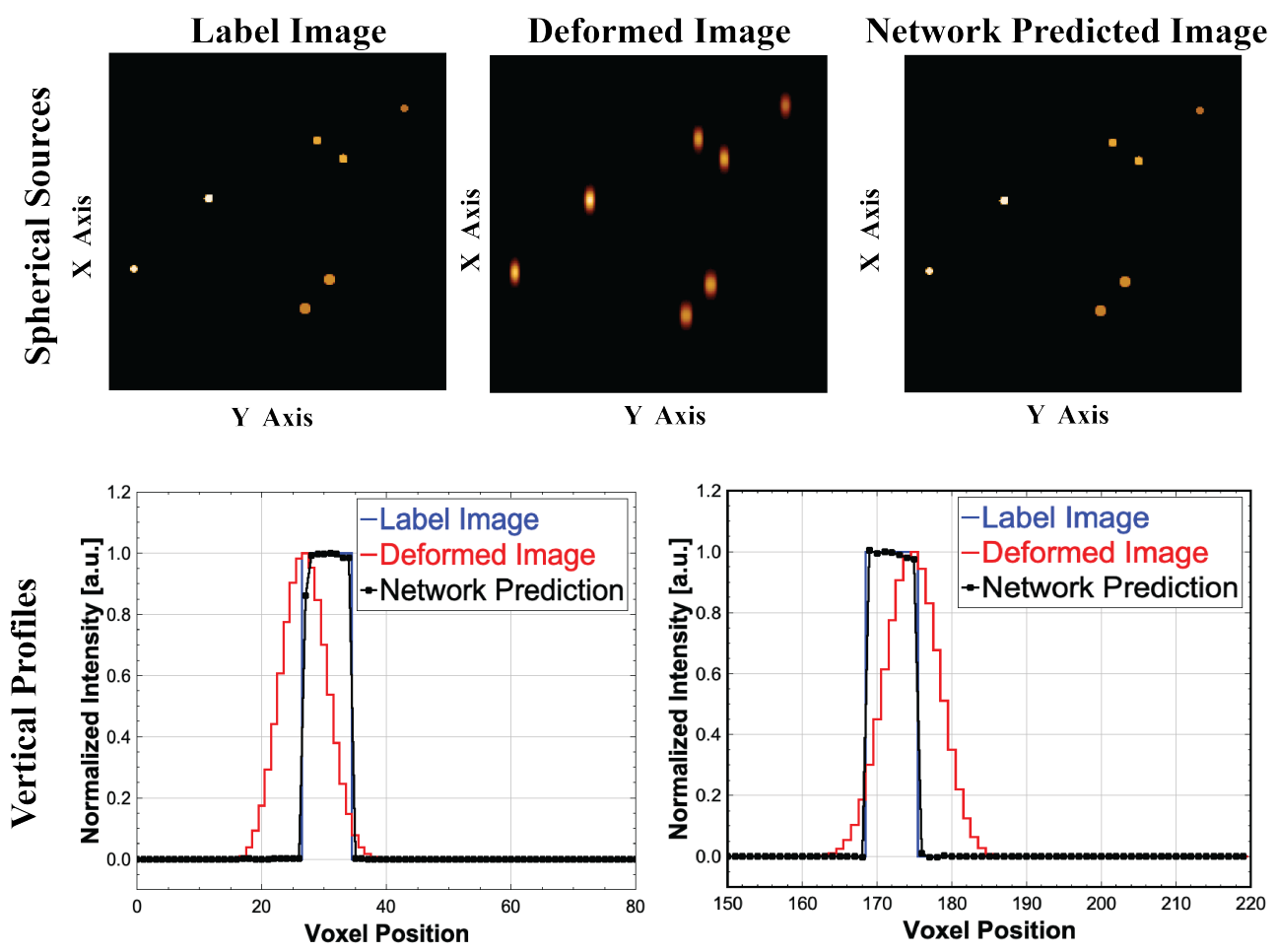

**Figure 6:** (Top Panel) Center slice of 3D label images of the randomly generated spherical sources is shown in the first column. The corresponding deformed images, and the network predicted output is shown in the second and third column, respectively. (Bottom Panel) Vertical line profiles (along Y-axis) of a circular sources at different positions.

### 4.2 Test Results

Figure 6 (top) shows the comparison of the example of the generated phantom, deformed image, and the U-Net output image. The corresponding vertical profiles of a reconstructed spherical sources at two different locations are shown in the bottom panel of the figure. These tests confirm ability of the U-Net to recover the accurate location (and shape) of the deformed lesions even if the PSF deformation shape is identical, i.e., space invariant and thus not carrying any information/indication about the lesion location.

## 5 Test for Complex Lesions in Warm Background in Noiseless Images and Noisy Images

So far, the experiments shown in the previous sections confirmed the ability of the U-Net to learn spatially variant deformations in small objects placed in the FOV without any warm background. The following test was performed with more realistic data sets where the lesion-like objects were placed in a warm background of noiseless and noisy images.

### 5.1 Network Training

Two separate data sets were generated as shown in Figure 7, i.e., complex lesion-like structures (merging of three ellipsoids with random sizes and orientations) at random locations in the FOV with varying intensities placed in a warm background with varying contrast ratio (2:1 to 10:1) and a total effective diameter of 6 mm to 30 mm. In the first dataset (Top Panel of Figure 7), the deformed images were generated with asymmetric PSF deformations as shown in the left panel of Figure 2 with no induced noise. Whereas, in the second dataset (Middle Panel of Figure 7) the deformed images were generated with asymmetric PSF deformations as shown in the left panel of Figure 2 with poisson noise. 50 training pairs and 5 validation pairs were generated and used for each of the two separate trainings and validations, respectively. The training was performed using the same hyperparameters as before.

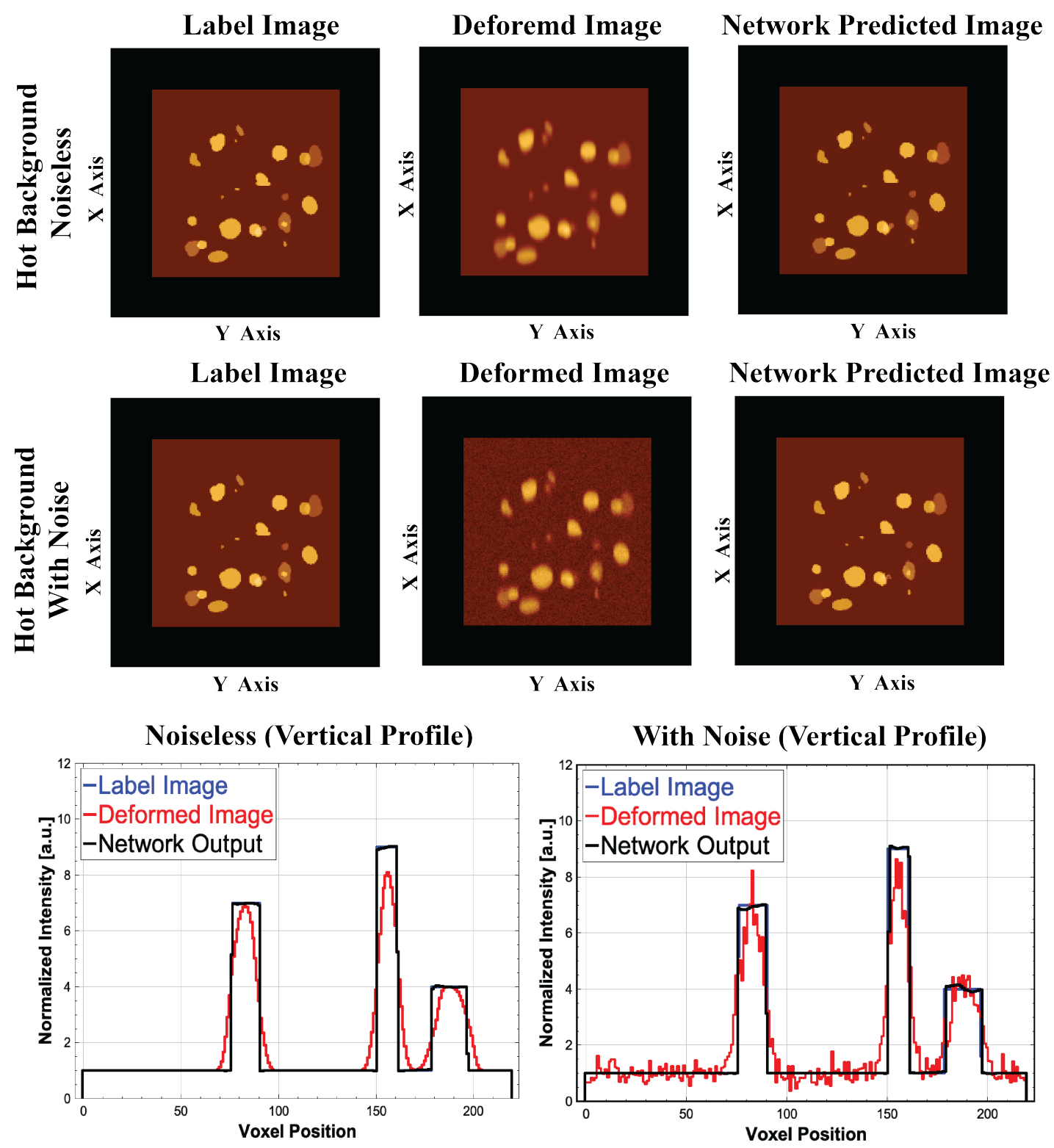

**Figure 7:** Center slice of 3D label images from the randomly generated complex lesions (given by a combination of 3 randomized ellipsoids) are shown in the first column. The corresponding deformed images without (Top panel) and with (Middle panel) induced noise and the network (3D U-Net as shown in Figure 3) predicted outputis shown in the second and third columns, respectively. (Bottom Panel) Vertical line profile (along the Y-axis) through the three images for both noiseless and noisy cases.

### 5.2 Test Results

Figure 7 (top and middle) shows the comparison of the generated phantom (with and without noise for lesions placed in a warm background), deformed image, and the U-Net output image. The corresponding vertical profiles through the label, deformed and network predicted images are shown in the bottom panel of the figure.

These results were further evaluated using the quantitative measures shown in Table 1 for the noiseless case and Table 2 for the test using noisy images. The evaluated metrics were: the average absolute lesion bias over all lesions in the testing image (with the bias in each lesion based on the mean value in the lesion VOI), the lesion Contrast Recovery Coefficient (CRC) and its average over all lesions, and the Image Roughness (IR) based on the voxel-vise standard deviation in the background regions away from the deformed lesions (i.e., background regions not affected by the blurring/deformations). These results confirm the ability and robustness of the U-Net to recover the accurate location (and shape) of the deformed lesions also in more realistic cases, with the lesions in warm background and noisy images.

| Image | Average Lesion \| Bias % \| | CRC |
|---|---|---|
| Label | 0. | 1.0 |
| Blurred | 18.5 | 0.78 |
| Network Output | 1.5 | 0.98 |

**Table 1:** Average Absolute Lesion Bias and Average Contrast Recovery Coefficient (CRC) for noiseless images shown in the top panel of Figure 7.

| Image | Average Lesion \| Bias % \| | CRC | IR% |
|---|---|---|---|
| Label | 0.0 | 1.0 | 0. |
| Blurred | 18.5 | 0.78 | 28.5 |
| Network Output | 3.85 | 0.95 | 0.23 |

**Table 2:** Average Absolute Lesion Bias, Average Contrast Recovery Coefficient (CRC) and Image Roughness (IR) percentage for noisy images shown in the middle panel of Figure 7.

## 6 Conclusion

In this work we have investigated the performance of convolution neural networks in capturing strong asymmetric and spatially variant deformation effects as those demonstrated in the dual panel systems. We constructed, trained and tested 3-dimensional U-Nets on versatile data sets with increasing complexities. We demonstrated that deep learning techniques can be used to recover strong asymmetric, anisotropic, and spatially variant deformations, as seen on the dual panel systems. In our tests we simulated PSF deformation effects for a dual-panel PET system and the results were insightful and optimistic to recover depth-of-interaction and limited angle effects. In this study we employed a relatively simple 3D U-Net with only 3 resolution layers, sufficient to represent relatively simple objects used in the presented studies. Even such a relatively shallow network was able to properly capture the spatially dependent behavior of the dual-panel system. Our future work will involve expanded studies using realistic objects simulated from B-PET and later on real B-PET data, employing potentially deeper U-Net structures. We would also like to probe the extent to which DL can correct for deformation effects with dual panel systems that may capture a smaller fraction of the data, or systems with poorer TOF resolution.

## 7 Acknowledgement

This investigative work was supported by the National Institute of Health through the grants R01-EB031806 and R01-CA196528. The content in this article is solely the responsibility of the authors and does not necessarily represent the official views of National Institutes of Health.

# Simultaneous reconstruction of activity and attenuation map with TOF-PET emission data

Zhimei Ren[1], Emil Y. Sidky[2], Rina Foygel Barber[1], Chien-Min Kao[2], and Xiaochuan Pan[2]

[1]Department of Statistics, University of Chicago, Chicago, USA
[2]Department of Radiology, University of Chicago, Chicago, USA

**Abstract** In this work, we study the simultaneous reconstruction of the activity and attenuation maps using only time-of-flight (TOF) positron emission tomography (PET). We model data in emission with a Poisson distribution, and obtain estimates of both maps via the total variation (TV)-constrained maximum likelihood estimator (MLE). We propose using the alternating direction method of multipliers (ADMM) algorithm to solve the resulting nonconvex optimization problem; the performance of our proposed estimator is demonstrated in a two-dimensional TOF-PET simulation and compared with the maximum likelihood activity and attenuation (MLAA) algorithm.

## 1 Introduction

Positron emission tomography (PET) imaging has been widely used clinically in recent years. In order to obtain a quantitatively accurate reconstruction of the activity map, it is necessary to account for the attenuation factors. While it is feasible to obtain the attenuation map with additional computed tomography (CT) scans, it is desired to simultaneously reconstruct the two maps with *only* the PET emission data since the additional CT scan could introduce additional ionizing radiation into the scan protocol, complicate work flow, and be subject to misalignment error.

The problem of simultaneous reconstruction with only the emission data is challenging since the problem is often underdetermined. [1] shows however, that with the TOF information, the problem of simultaneous reconstruction is nearly identifiable — the attenuation map can be determined up to a constant offset. Subsequently, [2] introduces the maximum likelihood activity and attenuation estimation (MLAA) algorithm for simultaneous reconstruction with TOF-PET data.

While the MLAA algorithm has shown to be effective in reducing crosstalk between the attenuation and activity maps for sufficiently narrow TOF windows, it is not clear how MLAA applies exact regularization to the images; rather, MLAA enforces regularization implicitly via early stopping of the optimization algorithm. In this work, we introduce explicit regularization to both the activity and attenuation maps by solving a total variation (TV)-constrained maximum likelihood estimation problem. To solve the resulting nonconvex optimization problem, we leverage the alternating direction method of multiplies (ADMM) algorithm following the framework in [3].

The remaining paper is organized as follows. Section 2 provides details of the model, and the optimization algorithm; in Section 3, we show numerical results of our proposed method applied to a two-dimensional TOF-PET simulation and compare its performance with MLAA. We end the paper with a discussion in Section 4.

## 2 Methods

### 2.1 The TOF-PET model

We discretize a two-dimensional image into $n_k$ pixels; for each $k \in [n_k] := \{1,2,\dots,n_k\}$, $\lambda_k$ and $\mu_k$ correspond to the activity and attenuation level, respectively. Suppose there are $n_\ell$ lines of responses (LORs) and $n_t$ TOF windows. We denote by $P \in \mathbb{R}^{n_\ell \times n_k}$ and $T \in \mathbb{R}^{n_t \times n_\ell \times n_k}$ the projection matrix and the TOF projection matrix respective, where $P_{\ell k}$ is the projection matrix element for LOR $\ell$ and pixel $k$ and $T_{i\ell k}$ is the sensitivity matrix element for TOF window $i$, LOR $\ell$ and pixel $k$. The TOF window sensitivity along the LOR is given by $w_i(t) = \exp\left[-(t-t_i)^2/(2\sigma_{\text{TOF}})\right]$, where the sampling along the LOR is half of the full-width- half-maximum (FWHM) of this Gaussian distribution $\Delta t = t_{i+1} - t_i = \text{FWHM}/2 = \sqrt{2\log 2}\cdot\sigma_{\text{TOF}}$. For each $i \in [n_t]$ and $\ell \in [n_\ell]$, we observe a measurement $C_{i\ell}$ following the Poisson distribution:

$$C_{i\ell} \sim \text{Poisson}\big(\exp(-P_\ell^\top \mu)\cdot T_{i\ell}^\top \lambda\big). \tag{1}$$

### 2.2 TV-constrained MLE

The negative log likelihood function under the above model is

$$l(\lambda,\mu) = \sum_{i\ell}\Big\{\exp(-P_i^\top \mu)\cdot T_{i\ell}^\top \lambda + C_{i\ell}\cdot\big(P_i^\top \mu - \log(T_{i\ell}^\top \lambda)\big)\Big\}.$$

Letting $N_{\text{count}}$ denote the total count of the ground-truth activity map, MLAA obtains estimated maps via minimizing $l(\lambda,\mu)$ over $(\lambda,\mu)$ with the constraint that $\sum_{k=1}^{n_k}\lambda_k = N_{\text{count}}$. In our framework, we introduce additional constraints on the images as explicit regularization. To be specific, we require that $P\mu \geq 0$ and that the total variation of both maps to be bounded by pre-specified constants: $\|\lambda\|_{\text{TV}} \leq \eta_\lambda, \|\mu\|_{\text{TV}} \leq \eta_\mu$, where $\|\cdot\|_{\text{TV}}$ denotes the isotropic TV seminorm. Collectively, our estimator is

$$\begin{aligned}(\hat\lambda,\hat\mu) = &\operatorname*{argmin}_{\lambda,\mu}\, l(\lambda,\mu)\\ &\text{s.t.}\ \sum_{k=1}^{n_k}\lambda_k = N_{\text{count}},\ P\mu \geq 0 \\ &\|\lambda\|_{\text{TV}} \leq \eta_\lambda, \|\mu\|_{\text{TV}} \leq \eta_\mu.\end{aligned} \tag{2}$$

### 2.3 The ADMM implementation

We start by noting that $l(\cdot,\cdot)$ depends on $(\lambda,\mu)$ through $(T\lambda, P\mu)$, so we can write

$$l(\lambda,\mu) = f(T\lambda, P\mu),$$

where $f(u,v) = \sum_{i\ell} u_{i\ell} \cdot e^{-v_i} - C_{i\ell}(-v_i + \log(u_{i\ell}))$. In order to implement ADMM, we shall introduce auxiliary variables $(x_\lambda, x_\mu)$ with the constraints $x_\lambda = T\lambda$ and $x_\mu = P\mu$, $(y_\lambda, y_\mu)$ such that $y_\lambda = D\lambda$ and $y_\mu = D\mu$; here $D$ is the finite difference approximation to the spatial gradient operator. The corresponding augmented Lagrangian function is

$$\mathscr{L}(x_\lambda, x_\mu, y_\lambda, y_\mu, \lambda, \mu, w_\lambda, w_\mu) = f(x_\lambda, x_\mu) \quad (3)$$
$$+\delta(\lambda^\top \mathbf{1} = N_{\text{count}}, x_\mu \geq 0, \|y_\lambda\|_1 \leq \eta_\lambda, \|y_\mu\|_1 \leq \eta_\mu) \quad (4)$$
$$+(x_\lambda - T\lambda)^\top w_\lambda + (x_\mu - P\mu)^\top w_\mu + (y_\lambda - D\lambda)^\top z_\lambda$$
$$+(y_\mu - D\mu)^\top z_\mu + \frac{\sigma_\lambda}{2}\|x_\lambda - T\lambda\|_2^2 + \frac{\sigma_\mu}{2}\|x_\mu - P\mu\|_2^2$$
$$+\frac{\rho_\lambda}{2}\|y_\lambda - D\lambda\|_2^2 + \frac{\rho_\mu}{2}\|y_\mu - D\mu\|_2^2,$$

where $\delta(\cdot)$ is the delta function; $\sigma_\lambda$, $\sigma_\mu$, $\rho_\lambda$, and $\rho_\mu$ are step-size parameters. ADMM then proceeds by iteratively updating $(x_\lambda, x_\mu, y_\lambda, y_\mu)$, $(\lambda,\mu)$ and $(w_\lambda, w_\mu, z_\lambda, z_\mu)$.

**$(x_\lambda, x_\mu, y_\lambda, y_\mu)$-update** At step $t$, ADMM updates $(x_\lambda, x_\mu, y_\lambda, y_\mu)$ via optimizing the augmented Lagrangian function; the problem can be separated into sub optimization problems over $(x_\lambda, x_\mu)$, $y_\lambda$ and $y_\mu$, respectively. The optimization problem with respect to $(x_\lambda, x_\mu)$ is nonconvex; it is however convex if we fix either $x_\lambda$ or $x_\mu$ and optimize over the other variable (this is also called a biconvex optimization problem). For the update of $(x_\lambda, x_\mu)$, we take an alternating approach: we first update $x_\lambda$ while keeping $x_\mu$ fixed, and vice versa.

$$x_\lambda^{(t+1)} = \underset{x_\lambda}{\operatorname{argmin}} \sum_{i\ell} \left\{ \exp(-x_{\mu,i}^{(t)}) \cdot x_{\lambda,i\ell} - C_{i\ell} \cdot \log(x_{\lambda,i\ell}) \right\}$$
$$+ (x_\lambda - T\lambda)^\top w_\lambda + \frac{\sigma_\lambda}{2}\|x_\lambda - T\lambda\|_2^2.$$

The above optimization problem can be analytically solved:

$$x_{\lambda,i\ell}^{(t+1)} = (-b_{i\ell} + \sqrt{b_{i\ell}^2 + 4\sigma_\lambda C_{i\ell}})/(2\sigma_\lambda),$$

where $b_{i\ell} = \exp(-x_{\mu,i}^{(t)}) + w_{\lambda,i\ell} - \sigma_\lambda \cdot (T\lambda)_{i\ell}$. Next, we update $x_\mu$ with $x_\lambda$ fixed:

$$x_\mu^{(t+1)} = \underset{x_\mu}{\operatorname{argmin}} \sum_{i\ell} \left\{ \exp(-x_{\mu,i}) \cdot x_{\lambda,i\ell}^{(t)} + C_{i\ell} x_{\mu,i} \right\}$$
$$+ (x_\mu - P\mu)^\top w_\mu + \frac{\sigma_\mu}{2}\|x_\mu - P\mu\|_2^2.$$

The above optimization problem is convex with a strictly positive second derivative, and is separable over the components of $x_\mu$. We shall use Newton's method to obtain the optimizer, and threshold the negative values in the optimizer to zero in order to satisfy the non-negativity constraint $x_\mu \geq 0$. The update of $y_\lambda$ and $y_\mu$ can be analytically written as follows:

$$y_\lambda^{(t+1)} = \operatorname{sgn}(D\lambda^{(t)} - z_\lambda^{(t)}/\rho_\lambda) \cdot \max(|D\lambda^{(t)} - z_\lambda^{(t)}| - \kappa_\lambda/\rho_\lambda, 0),$$
$$y_\mu^{(t+1)} = \operatorname{sgn}(D\mu^{(t)} - z_\mu^{(t)}/\rho_\mu) \cdot \max(|D\mu^{(t)} - z_\mu^{(t)}| - \kappa_\mu/\rho_\mu, 0).$$

Above, $\operatorname{sgn}(x)$ denotes the sign of $x$; for a vector $a$, $\max(a,0)$ means taking the maximum between each entry of $a$ and 0; the constants $\kappa_\lambda$ and $\kappa_\mu$ are determined as follows:

$$\kappa_\lambda = \inf\{\kappa : \|\max(|D\lambda^{(t)} - z_\lambda^{(t)}| - \kappa/\rho_\lambda, 0)\|_1 \leq \eta_\lambda\},$$
$$\kappa_\mu = \inf\{\kappa : \|\max(|D\mu^{(t)} - z_\mu^{(t)}| - \kappa/\rho_\mu, 0)\|_1 \leq \eta_\mu\}.$$

**$(\lambda,\mu)$-update** The update of $\lambda$ and $\mu$ is obtained via optimizing (3) with additional step-size regularization terms:

$$\text{for } \lambda : \tfrac{1}{2}(\tfrac{1}{\tau_\lambda} - \sigma_\lambda T^\top T - \rho_\lambda D^\top D) \cdot \|\lambda - \lambda^{(t)}\|_2^2,$$
$$\text{for } \mu : \tfrac{1}{2}(\tfrac{1}{\tau_\mu} - \sigma_\mu P^\top P - \rho_\mu D^\top D) \cdot \|\mu - \mu^{(t)}\|_2^2.$$

The constrained optimization problem with respect to $(\lambda,\mu)$ can be solved analytically, and the update rule for $\lambda$ is

$$\lambda^{(t+1)} = \bar{\lambda} + \frac{N_{\text{count}} - \bar{\lambda}^\top \mathbf{1}}{n_k}\mathbf{1},$$
$$\text{where } \bar{\lambda} = \lambda^{(t)} + \tau_\lambda \cdot \Big\{ T^\top (w_\lambda + \sigma_\lambda(x_\lambda - T\lambda^{(t)})) + D^\top (z_\lambda + \rho_\lambda(y_\lambda - D\lambda^{(t)})) \Big\},$$

and $\mathbf{1} \in \mathbb{R}^{n_k}$ is a vector with all the elements being 1. For $\mu$,

$$\mu^{(t+1)} = \mu^{(t)} + \tau_\mu \cdot \Big\{ P^\top (w_\mu + \sigma_\mu(x_\mu - P\mu)) + D^\top (z_\mu + \rho_\mu(y_\mu - D\mu^{(t)})) \Big\}.$$

**$(w_\lambda, w_\mu, z_\lambda, z_\mu)$-update** Finally, we update the set of variables $(w_\lambda, w_\mu, z_\lambda, z_\mu)$:

$$w_\lambda^{(t+1)} = w_\lambda^{(t)} + \sigma_\lambda \cdot (x_\lambda^{(t+1)} - T\lambda^{(t+1)}),$$
$$w_\mu^{(t+1)} = w_\mu^{(t)} + \sigma_\mu \cdot (x_\mu^{(t+1)} - P\mu^{(t+1)}),$$
$$z_\lambda^{(t+1)} = z_\lambda^{(t)} + \rho_\lambda \cdot (y_\lambda^{(t+1)} - D\lambda^{(t+1)}),$$
$$z_\mu^{(t+1)} = z_\mu^{(t)} + \rho_\mu \cdot (y_\mu^{(t+1)} - D\mu^{(t+1)}).$$

## 3 Results

We evaluate the performance of our proposed method on a simulated TOF-PET dataset, where the ground-truth activity map and attenuation map are shown in Figure 1. The data is generated from (1), and the total number of counts is approximately $10^6$.

In this numerical example, both the activity map and the attenuation map are of the size 30 cm $\times$ 30 cm, and the resolution of both maps is $n_k = 128$. The LORs are placed in a parallel-ray fashion, where for each LOR, the two detectors are 200 cm apart; in total, there are $n_\ell = 128$ LORs. The FWHM is taken to be 9 cm.

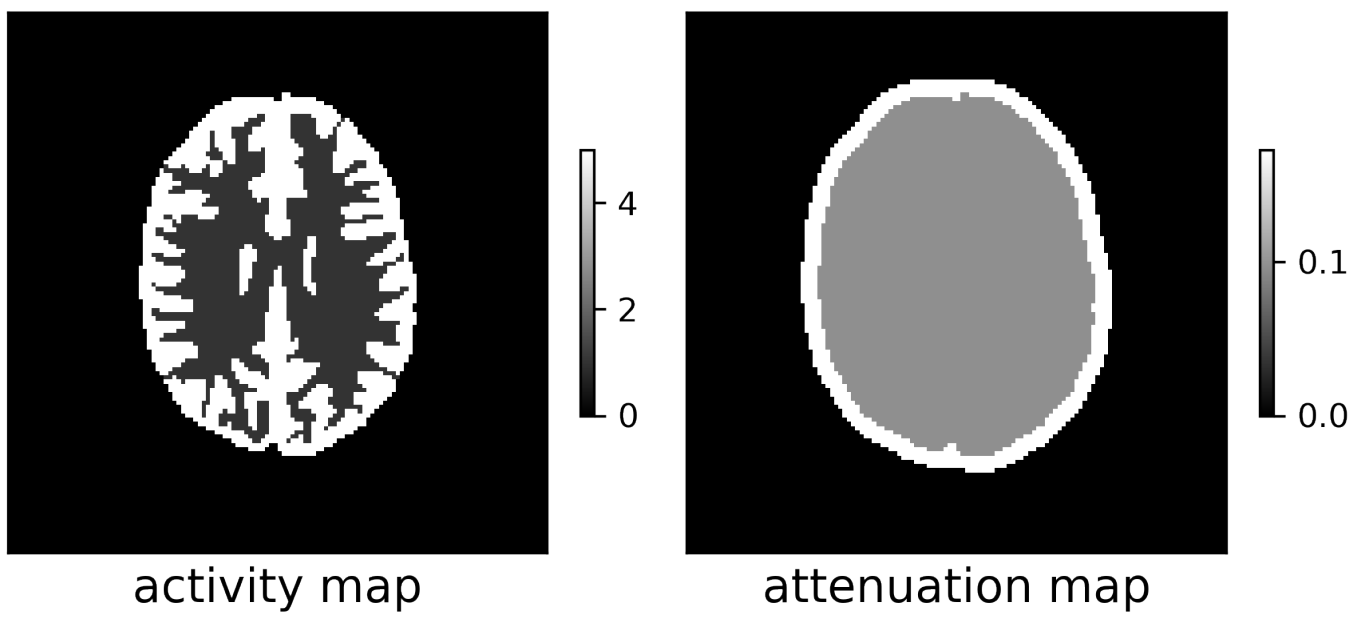

**Figure 1:** The ground-truth activity map (left) and attenuation map (right). The attenuation map contains the gray matter and the white matter, with a 5:1 activity ratio; the attenuation map includes a skull enclosing uniform water attenuation.

We run our proposed method for 200 iterations, where the parameters are chosen through a grid search. For comparison, we also implement MLAA for 200 iterations. Figure 2-5 visualize the root mean squared error (RMSE) of data, activity map, attenuation map, and the attenuation factor for the two methods, respectively.

Figure 2 shows the RMSE of the data, where we can see that both methods are able to bring down the error within a few iterations. As the number of iteration increases, both error curves converge to zero. Figure 3 concerns the RMSE of the activity map. As the number of iteration increases, the error curve of MLAA exhibits a "U" shape — it achieves the minimum RMSE at iteration 37 — while that of our method keeps decreasing as the number of iteration increases; the figure also shows that our method is able to achieve lower activity RMSE by comparing the minimum of the two curves.

Figure 4 and 5 demonstrate the RMSE of the attenuation map and the attenuation factor, respectively. Similar to the case of the activity map, by introducing the TV constraints our method is able to achieve lower attenuation map RMSE (MLAA however still has reasonable attenuation factor RMSE even though the attenuation RMSE is large).

To further compare the reconstructed maps of MLAA and our method, we present in Figure 6 the reconstructed activity map with the lowest RMSE by our method and MLAA (for our method, it is the estimate at iteration 200 and for MLAA the estimate at iteration 37). Figure 7 similarly shows the reconstructed attenuation map with the lowest RMSE for both methods (iteration 51 for MLAA and iteration 200 for our method).

In Figure 8, we also show the activity maps recovered by our method and MLAA at iteration 10, 50, and 100, respectively. As we can see again, as the number of iteration increases, the estimated map by MLAA gets noisier, while that given by our method does not suffer from overfitting.

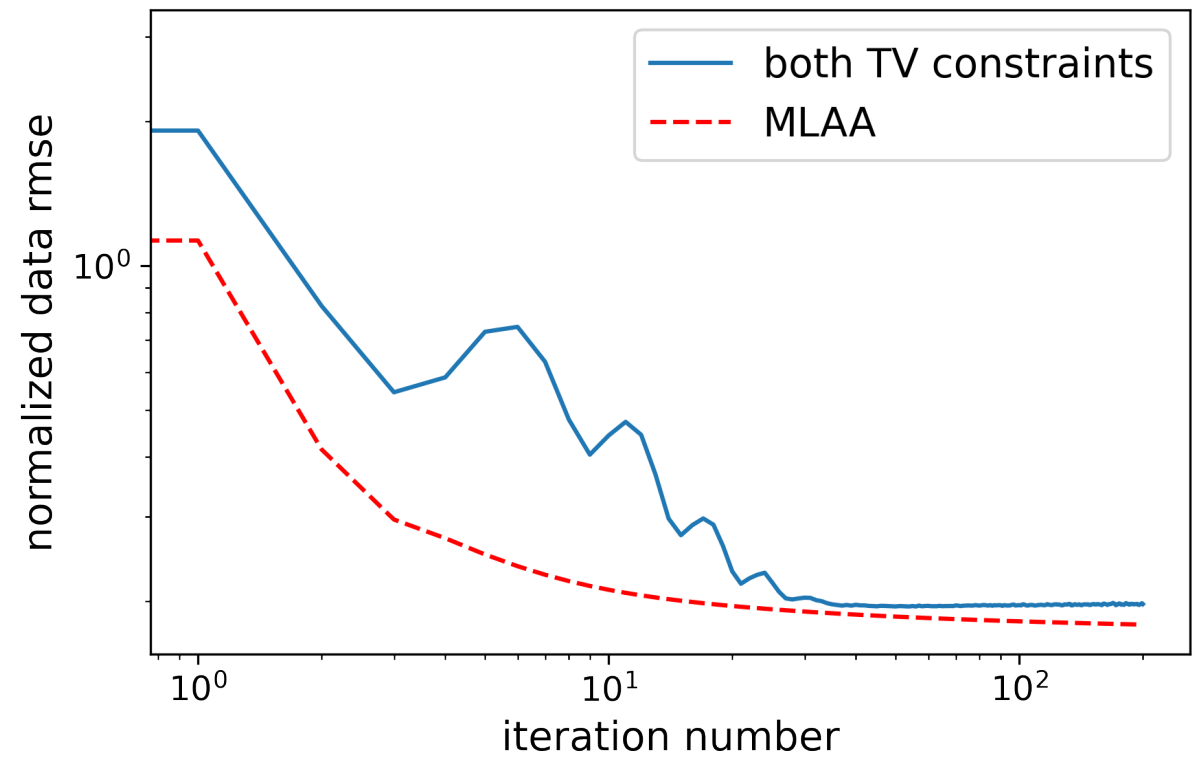

**Figure 2:** Data RMSE versus iteration numbers. The plot is on a log-log scale.

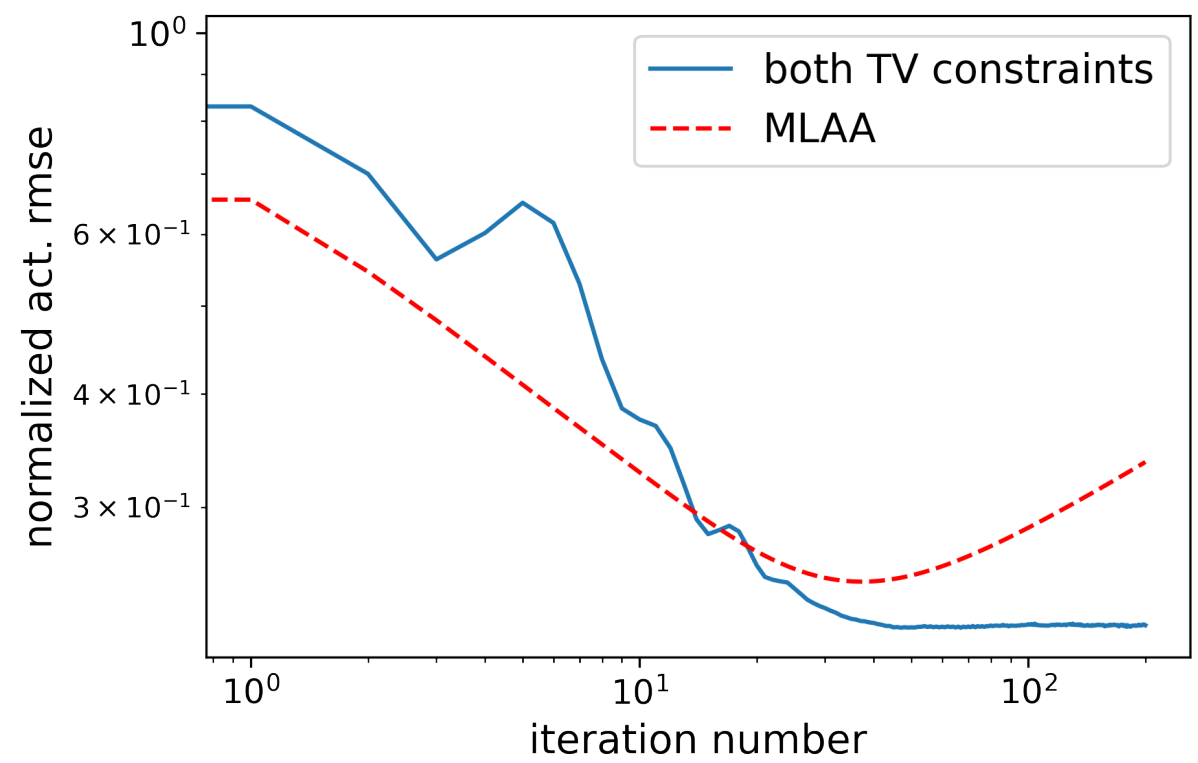

**Figure 3:** RMSE of the activity map versus iteration numbers. The plot is on a log-log scale.

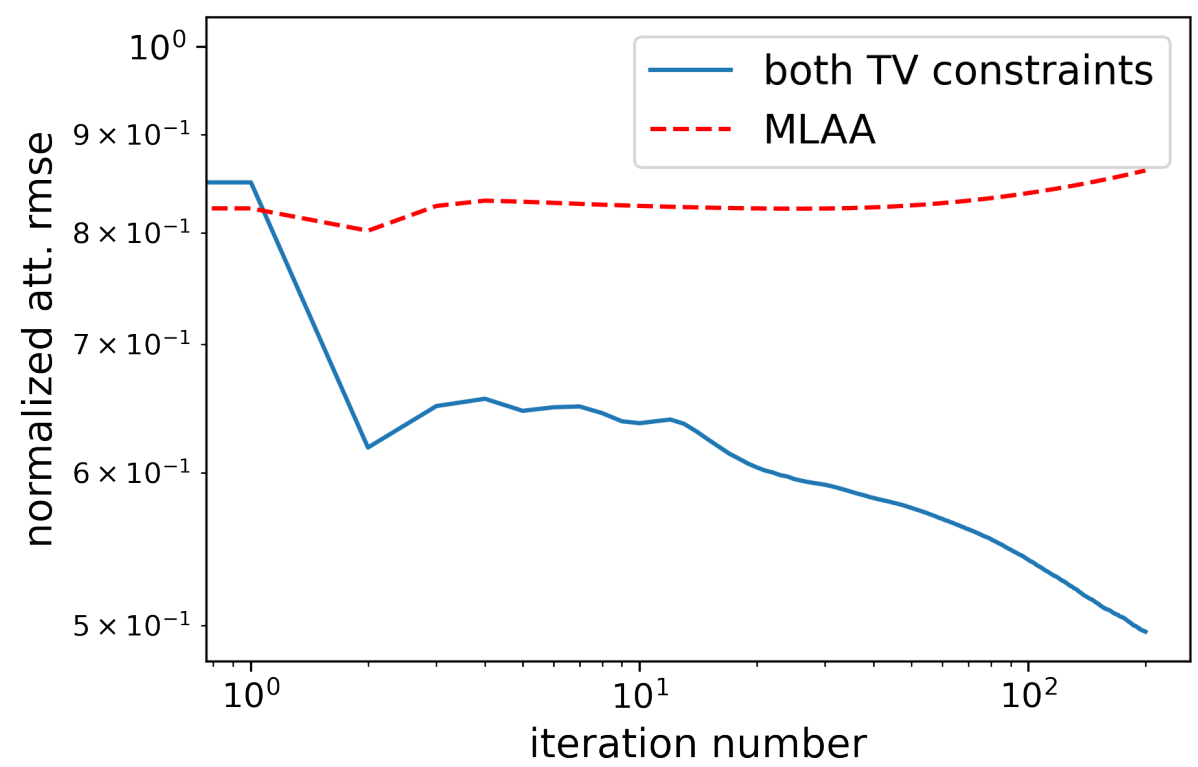

**Figure 4:** RMSE of the attenuation map versus iteration numbers. The plot is on a log-log scale.

## 4 Discussion

This paper studies the simultaneous reconstruction of the activity and attenuation map with TOF-PET data. The proposed method improves upon MLAA by introducing explicit regularization terms, i.e., the TV constraints. The ADMM procedure is then applied to solving the resulting nonconvex optimization problem, and the proposed method is evaluated on a simulated two-dimensional dataset, demonstrating improved performance when compared with the MLAA.

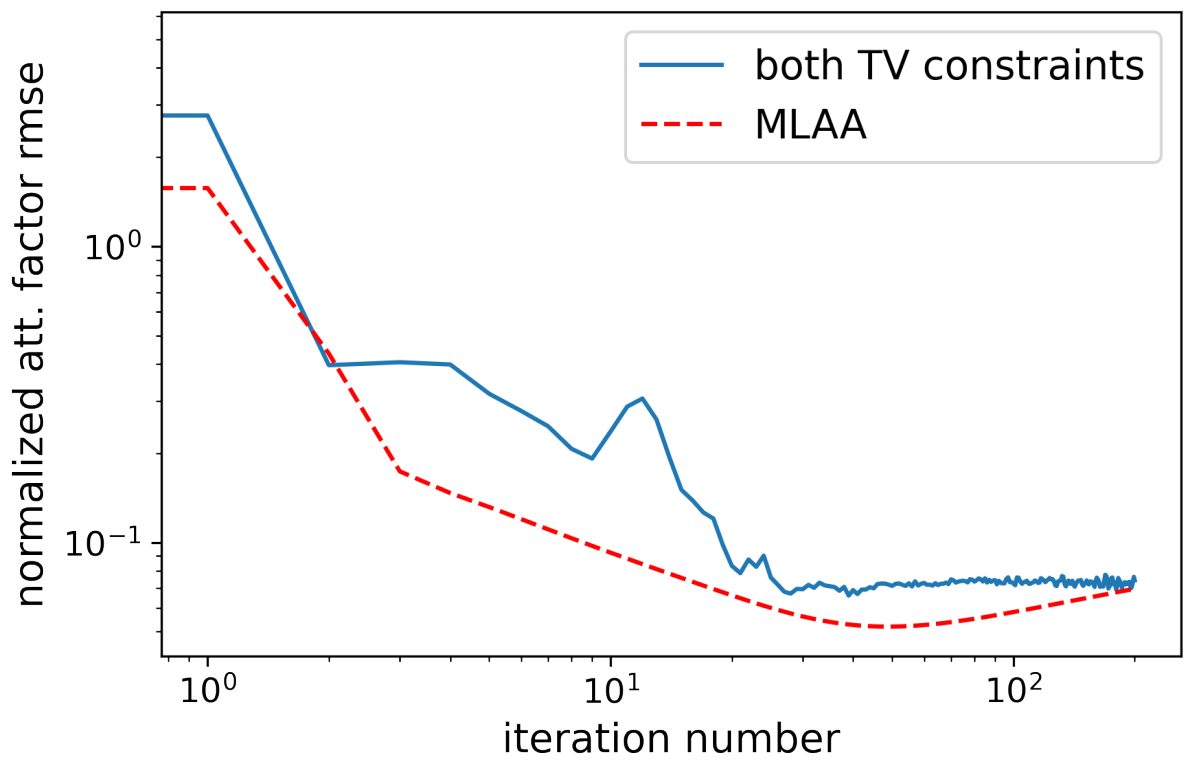

**Figure 5:** RMSE of the attenuation factor versus iteration numbers. The plot is on a log-log scale.

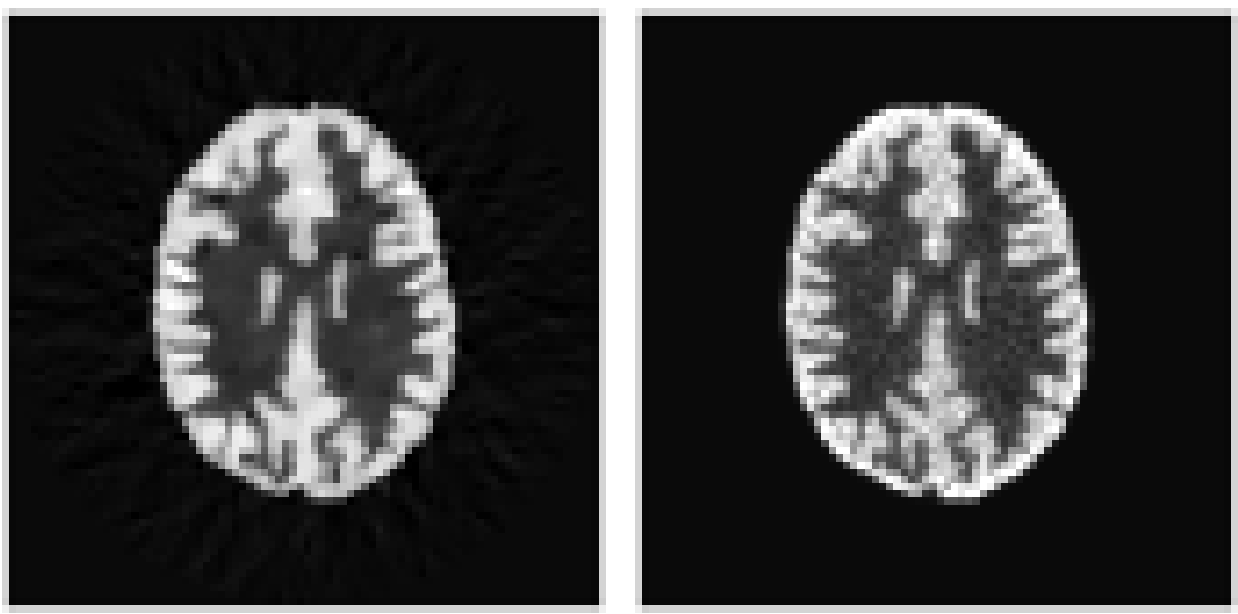

**Figure 6:** Recovered activity map with the lowest RMSE. Left: the estimated activity map of our method at iteration 200; right: the estimated activity map of MLAA at iteration 37.

## Acknowledgements

This work is supported in part by NIH Grant Nos. R01-EB026282, R01-EB023968, and R21-CA263660. The contents of this article are solely the responsibility of the authors and do not necessarily represent the official views of the National Institutes of Health.

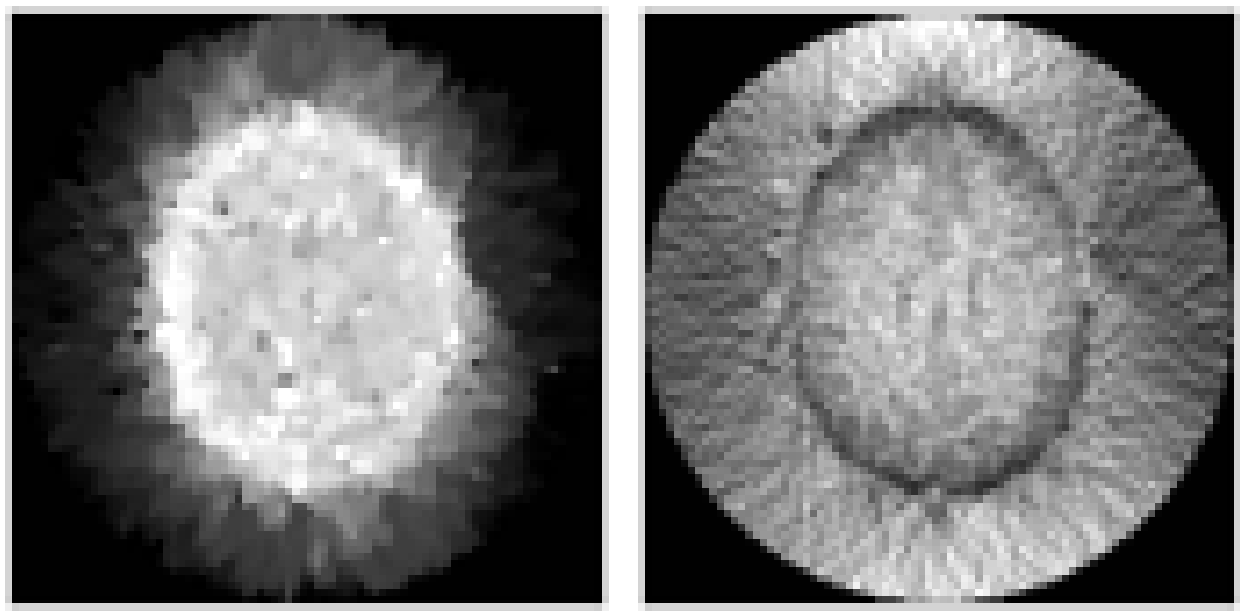

**Figure 7:** Recovered attenuation map with the lowest RMSE. Left: the estimated attenuation map of our method at iteration 200; right: the estimated attenuation map of MLAA at iteration 51.

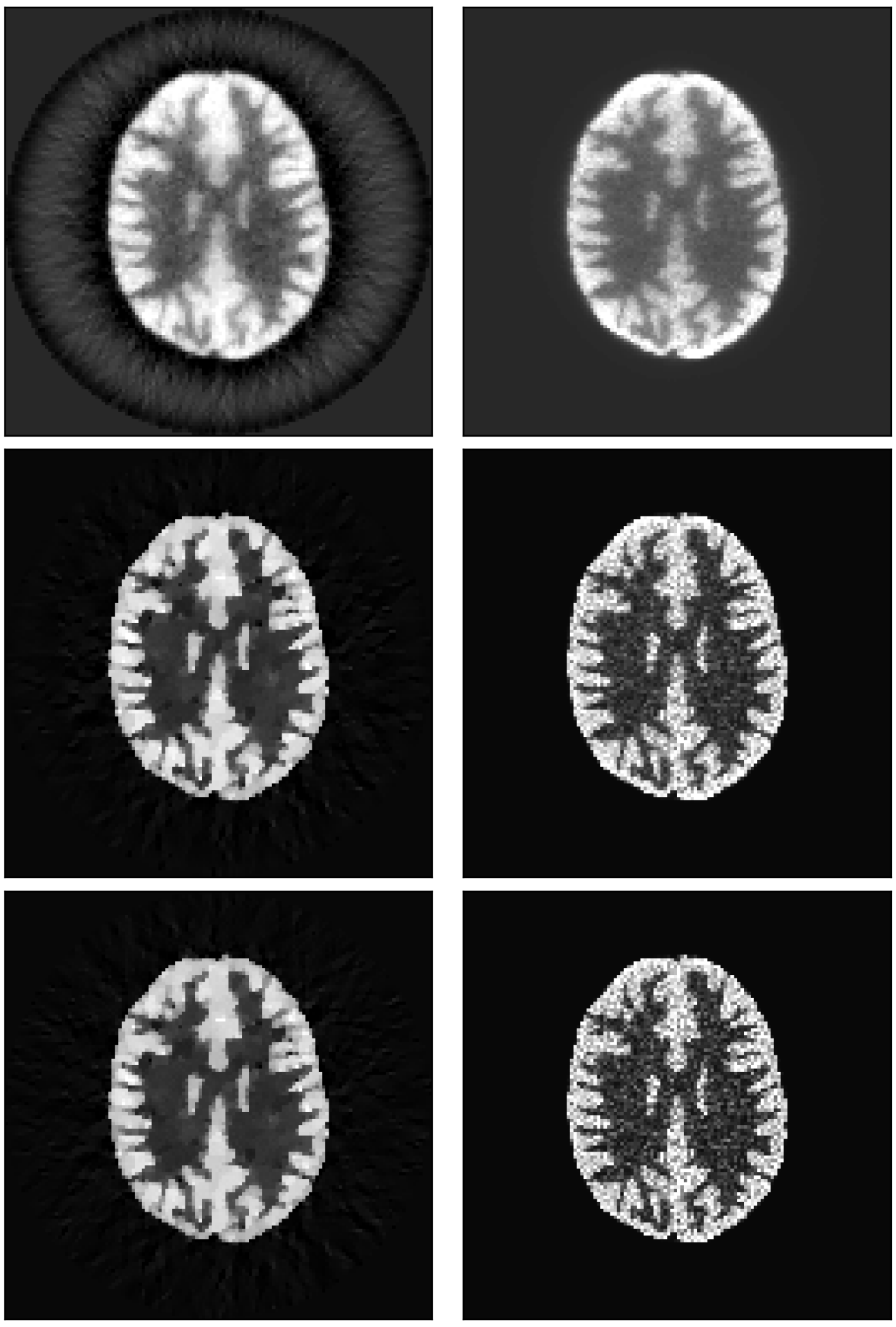

**Figure 8:** Reconstructed activity map by our method (left column) and MLAA (right) at iteration 10 (first row), iteration 50 (second row), and iteration 100 (third row).

# Ability of exponential data consistency conditions to detect motion in SPECT despite other physical effects

Antoine Robert[1], David Sarrut[1,2], Ane Etxebeste[1], Jean Michel Létang[1], and Simon Rit[1]

[1]Univ.Lyon, INSA-Lyon, Université Claude Bernard Lyon 1, UJM-Saint Etienne, CNRS, Inserm, CREATIS UMR 5220, U1206, F-69373, Lyon, France.
[2]Centre Léon Bérard, 28, rue Laennec, 69373 Lyon Cedex 08, France

**Abstract** Exponential data consistency conditions (eDCCs) are equations that express the redundancy of information between exponential projections. Exponential projections can be derived from parallel SPECT projections and be used to detect and correct for patient motion during the acquisition. However, other physical effects such as collimator resolution, scatter or noise could also introduce inconsistencies in the projections. The purpose of this work was to evaluate the impact of these effects on the eDCCs. We used ray-tracing and Monte Carlo simulations to generate different sets of projections and compared their consistency with two metrics based on eDCCs: the absolute relative difference and a noise-aware metric that takes into account the acquisition noise. The collimator resolution, the scatter and the movement increase significantly the error in the eDCCs. The noise-aware metric was more sensitive to patient motion than other effects.

## 1 Introduction

Single photon emission computed tomography (SPECT) is a key tool for diagnostic imaging which is also used for treatment planning and monitoring of radionuclide therapies. The long acquisition time makes SPECT imaging subject to blur and artifacts due to patient motion. This motion also induces a mismatch between the emission and attenuation maps that may impact the quality of the SPECT images [1].

Data consistency conditions (DCCs) are equations that express the redundancy of information between projections. DCCs have been used in PET to correct for patient motion and to align the emission and attenuation maps [2]. In SPECT, DCCs have also been used to estimate the attenuation map from the emission projections [3]. More recently, Wells *et al* [4] used exponential data consistency conditions (eDCCs) to align the attenuation map to cardiac SPECT data after rebinning pinhole data to parallel projections. eDCCs are less restrictive than SPECT DCCs in that they do not require projections taken over 360°. Wells *et al* [4] used simulated projections to evaluate eDCCs but they did not take into account the scatter or the collimator resolution, which we refer to as the point spread function (PSF) in the following. Moreover, in parallel SPECT systems, projections are acquired sequentially and may be affected by patient motion.

The purpose of this work was to assess the impact of physical effects and motion on the eDCCs. To that end, ray tracing and Monte Carlo simulations were used to generate several sets of projections and to evaluate two metrics based on eDCCs.

## 2 Materials and Methods

### 2.1 Exponential data consistency condition

In parallel SPECT, measurements can be modeled by the attenuated Radon transform. Let $f(\vec{x})$, $\vec{x} \in \mathbb{R}^2$, be the radioactivity distribution of a mono-energetic emitter and $\mu(\vec{x})$ be a known spatially varying attenuation medium. The attenuated Radon transform of $f$ is

$$g(\theta,s) = \int_{-\infty}^{+\infty} f(s\vec{u_\theta} + t\vec{v_\theta}) \exp\left(-\int_t^{\infty} \mu(s\vec{u_\theta} + t'\vec{v_\theta})\mathrm{d}t'\right)\mathrm{d}t \tag{1}$$

with $\theta \in [0, 2\pi)$ the angle of the projection, $\vec{u_\theta} = (\cos\theta, \sin\theta)^T$, $\vec{v_\theta} = (-\sin\theta, \cos\theta)^T$. We assume that we know a convex sub-region K in which the attenuation is constant ($\mu(\vec{x}) = \mu_0, \forall \vec{x} \in$ K) and out of which the activity is zero ($f(\vec{x}) = 0, \forall \vec{x} \notin$ K). Under these two assumptions, the exponential Radon transform can be computed from the attenuated Radon transform by a simple pointwise conversion [5]:

$$p(\theta,s) = \int_{-\infty}^{+\infty} f(s\vec{u_\theta} + t\vec{v_\theta})\mathrm{e}^{\mu_0 t}\mathrm{d}t = C(\theta,s)g(\theta,s) \tag{2}$$

with

$$C(\theta,s) = \exp\left(\tau_{\theta,s}\mu_0 + \int_{\tau_{\theta,s}}^{\infty} \mu(s\vec{u_\theta} + t'\vec{v_\theta})\mathrm{d}t'\right) \tag{3}$$

where $\tau_{\theta,s}$ is the location where the photons leave the region K on their way to the detector along the lines defined by the coordinates $(\theta,s)$. With this model, any pair of exponential projections $p(\theta_i,.)$ and $p(\theta_j,.)$ are consistent with each other if and only if [6]

$$P(\theta_i,\sigma_{i,j}) = P(\theta_j,\sigma_{j,i}) \quad \text{for} \quad \sigma_{i,j} = \mu_0 \tan(\frac{\theta_i - \theta_j}{2}) \tag{4}$$

where $P(\theta,\sigma) = \int_{-\infty}^{\infty} p(\theta,s)\mathrm{e}^{\sigma s}\mathrm{d}s$ is the two-sided Laplace transform of $p(\theta,.)$. There is no eDCC for opposite projections, when $\theta_i - \theta_j = \pi \,(\mathrm{mod}\, 2\pi)$.

This result can be applied independently to each line of 2D parallel projections. We define $\overline{P_{ij}}$ as the average value of $P(\theta_i,\sigma_{i,j})$ over the $N$ lines of the projections

$$\overline{P_{ij}} = \frac{1}{N}\sum_{l=1}^{N} P_l(\theta_i,\sigma_{i,j}) \tag{5}$$

where $P_l$ is the two-sided Laplace transform of the $l$-th line of the projections.

### 2.2 Data

A CT image of a thoracic patient was used as attenuation map. The emission map was a spherical tumor of 20 mm radius positioned approximately at the center of the liver (Figure 1). Another pair of attenuation and emission maps was defined by applying a 20 mm translation in the cranio-caudal direction to simulate a rigid motion occurring during the acquisition (the second emission map is shown in Figure 1). We used ray tracing with RTK [7] and Monte Carlo with Gate [8] to create several sets of attenuated projections. In RTK, one projector only models the attenuation effect (Equation 1) and another one models the attenuation and the PSF with parameters corresponding to a clinical dual-head SPECT system (General Electric Discovery NM/CT 670). The same system was modeled in the Monte Carlo simulation and a total of $51 \times 10^8$ 140 keV emission photons of $^{99m}$Tc were simulated. Several levels of scattered photons in the output projections were considered: only primary photons (no scatter), scatter correction using the double energy window [9] and no scatter correction. The projections generated with RTK were normalized against the ones obtained with Gate and Poisson noise was added. Each set of projections had 60 angles regularly sampled over 360°, each with 100 $\times$ 100 pixels and 4 mm isotropic spacing. In each case, one set of projections with no motion was generated and used as reference. A second set of projections was computed in which the first thirty projections correspond to the first patient position and the last thirty to the second one.

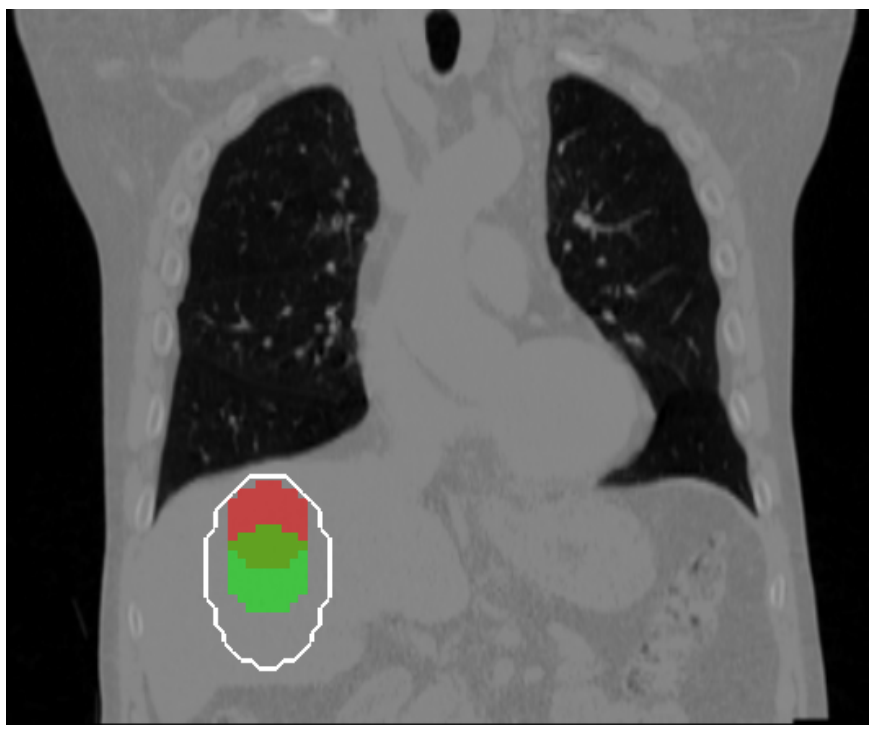

**Figure 1:** The two emission maps (in red and green) used to simulate SPECT projections overlaid over the attenuation map used with the green emission map. The white line corresponds to the region K used for the conversion of the attenuated projections to exponential projections.

### 2.3 Analysis

For each set of projections, the exponential projections were computed by choosing an elliptic region K such that it encompassed the two emission spheres while being in the liver in the two positions (Figure 1). The first attenuation map was always used for the computation of the exponential projections to mimick the clinical scenario where the CT image is acquired before the SPECT image.

Only a subset of the projection pairs was analyzed for finer analysis, those with vertical and horizontal directions, i.e. such that $\theta_1 + \theta_2 = \pi \,(\mathrm{mod}\, 2\pi)$ and $\theta_1 + \theta_2' = 0 \,(\mathrm{mod}\, 2\pi)$ respectively (blue and red segments in Figure 2). For a pair of projections, we define the signed distance $\rho = R\vec{v}_{\theta_1} \cdot (\sin(\frac{\theta_1+\theta_2}{2}), \cos(\frac{\theta_1+\theta_2}{2}))^T$ with $R = 380$ mm the detector-to-isocenter distance (Figure 2). This selection allows the evaluation of the impact of the motion on the eDCCs as the 20 mm motion occurs after the first half of the acquisition (section 2.2) so the two subsets of projections taken between 0° and 174° and between 180° and 354° are consistent. Therefore, all pairs of projections with a vertical direction are not impacted by motion whereas all the ones with a horizontal direction are. Only 28 pairs in each direction were analyzed out of the $60 \times 58 = 3480$ eDCCs.

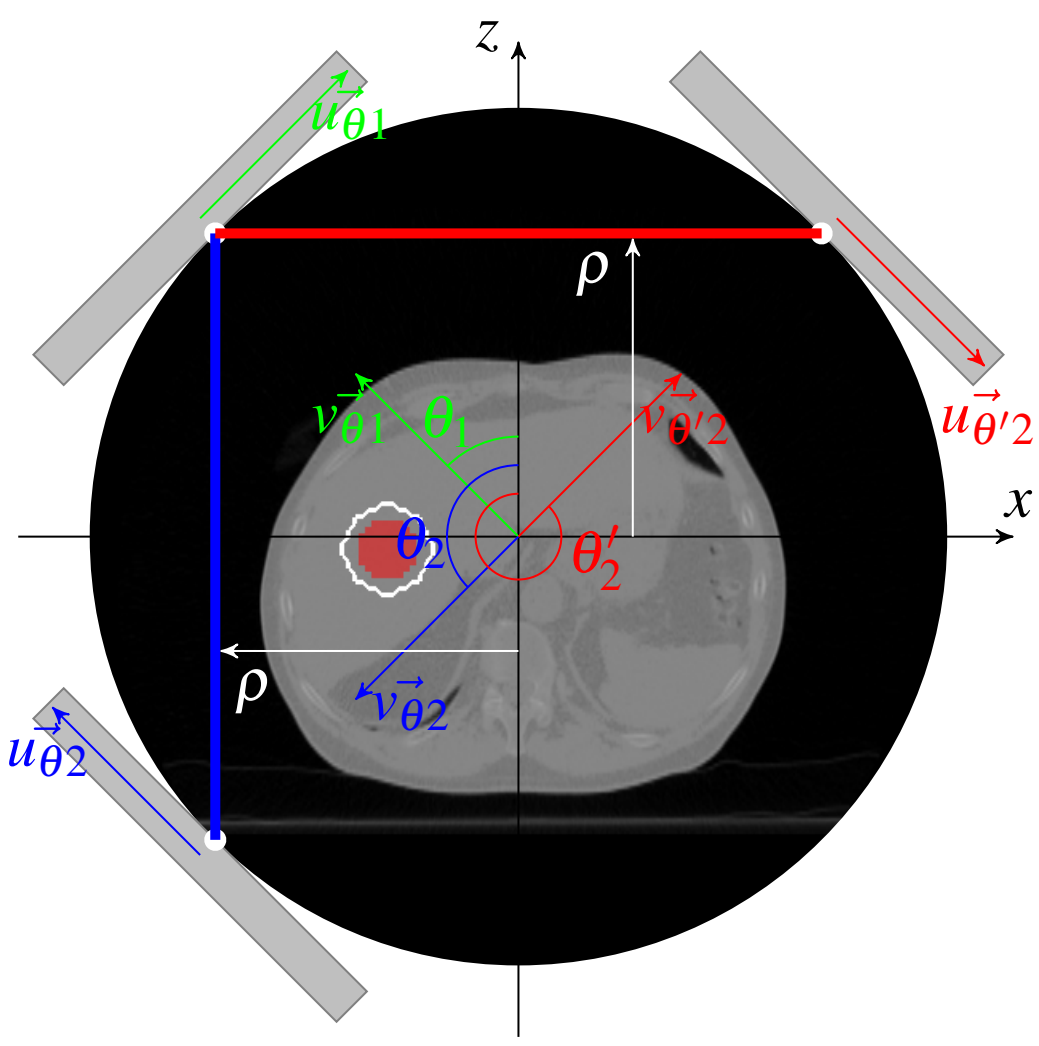

**Figure 2:** Selection of the pairs of projections for analyzing the eDCCs. The grey rectangles represent three detector positions and the white dots their centers. The blue and red segments symbolize two pairs of projections, vertical and horizontal respectively.

For each pair of projections, the eDCCs were computed and assessed with two metrics. The first one, was simply the average value of the absolute relative error:

$$\mathrm{E_{ij}} = 2 * \frac{1}{N} \sum_{l=1}^{N} \frac{|P_l(\theta_i, \sigma_{i,j}) - P_l(\theta_j, \sigma_{j,i})|}{P_l(\theta_i, \sigma_{i,j}) + P_l(\theta_j, \sigma_{j,i})}. \quad (6)$$

Following the work of Mouchet [10], we also define a noise-aware metric which corresponds to the mean absolute difference divided by the standard deviation of the mean difference:

$$\mathrm{NE_{ij}} = \frac{1}{N} \frac{\sum_{l=1}^{N} |P_l(\theta_i, \sigma_{i,j}) - P_l(\theta_j, \sigma_{j,i})|}{\sqrt{N}\sqrt{\mathrm{Var}(\overline{P_{ij}}) + \mathrm{Var}(\overline{P_{ji}})}}. \quad (7)$$

Using the properties of the variance of the weighted sum of random uncorrelated variables, we have

$$\mathrm{Var}(\overline{P_{ij}}) = \frac{1}{N^2} \sum_{l=1}^{N} \mathrm{Var}(P_l(\theta_i, \sigma_{i,j})). \quad (8)$$

After discretization and by applying the definition of the Laplace transform combined with the relation of equation 2 we can write

$$\mathrm{Var}(P_l(\theta_i, \sigma_{i,j})) = \Delta s^2 \sum_{k=1}^{M} C_l^2(\theta_i, s_k) \mathrm{Var}(g_l(\theta_i, s_k)) e^{2\sigma_{ij} s_k} \tag{9}$$

with $\Delta s$ the pixel spacing, $M$ the number of pixels per line and $C_l(\theta, s_k)$ and $g_l(\theta, s_k)$ the $k$-th pixel of the $l$-th line of $C(\theta, s)$ and $g(\theta, s)$. The photon noise follows a Poisson distribution so $\mathrm{Var}(g_l(\theta_i, s_k)) = g_l(\theta_i, s_k)$ and

$$\mathrm{Var}(\overline{P_{ij}}) = \frac{\Delta s^2}{N^2} \sum_{l=1}^{N} \sum_{k=1}^{M} C_l^2(\theta_i, s_k) g_l(\theta_i, s_k) e^{2\sigma_{ij} s_k}. \tag{10}$$

## 3 Results

The evolution of the metrics with the signed distance $\rho$ characterizing a pair of projections is shown in Figure 3 for the ray tracing simulations. In the ideal case (no noise, no PSF and no motion), the relative error $\mathrm{E_{ij}}$ had an average value of 0.2% in the vertical and horizontal directions which indicates that the projections are consistent. With motion, the relative error in the horizontal direction became much higher (108.0% on average) and a slight increase of 2.2% in the vertical direction was observed. When modeling the PSF, the relative error of eDCCs increased even without motion between the projections. In that case, the $\mathrm{E_{ij}}$ value was on average 16.3% and 54.0% in the vertical and horizontal directions, respectively. With motion, the relative error also increased in the horizontal direction compared to the case without PSF. In all cases, the increase was higher when the signed distance got close to 0 mm. Additional Poisson noise in the projections produced similar results.

When assessing the eDCCs with the noise-aware metric, the inconsistency due to the PSF seemed to be mitigated. Indeed, without motion, the value of the noise-aware metric with PSF remained close to the one without PSF. In the case of motion and PSF, the value of $\mathrm{NE_{ij}}$ in the horizontal direction was always above the one in the vertical direction but got closer when the signed distance approached 0 mm.

Similar plots are shown in Figure 4 for the eDCCs computed from the projections obtained with Monte Carlo simulations. The main difference with the ray tracing projections with PSF and noise was the addition of scattered photons. The plot with circle markers corresponds to the simulation with primary photons only, the triangles markers to the one with DEW scatter correction and the squares to the one without scatter correction. Without motion, the relative error $\mathrm{E_{ij}}$ decreased with better scatter correction for pairs in the horizontal direction but the opposite was observed in the vertical direction. In the two directions, the absolute relative error was higher when the signed distance was close to 0 mm. For all levels of scattered photons, the relative error in the horizontal direction increased with the addition of the motion but the error increased with better scatter correction.

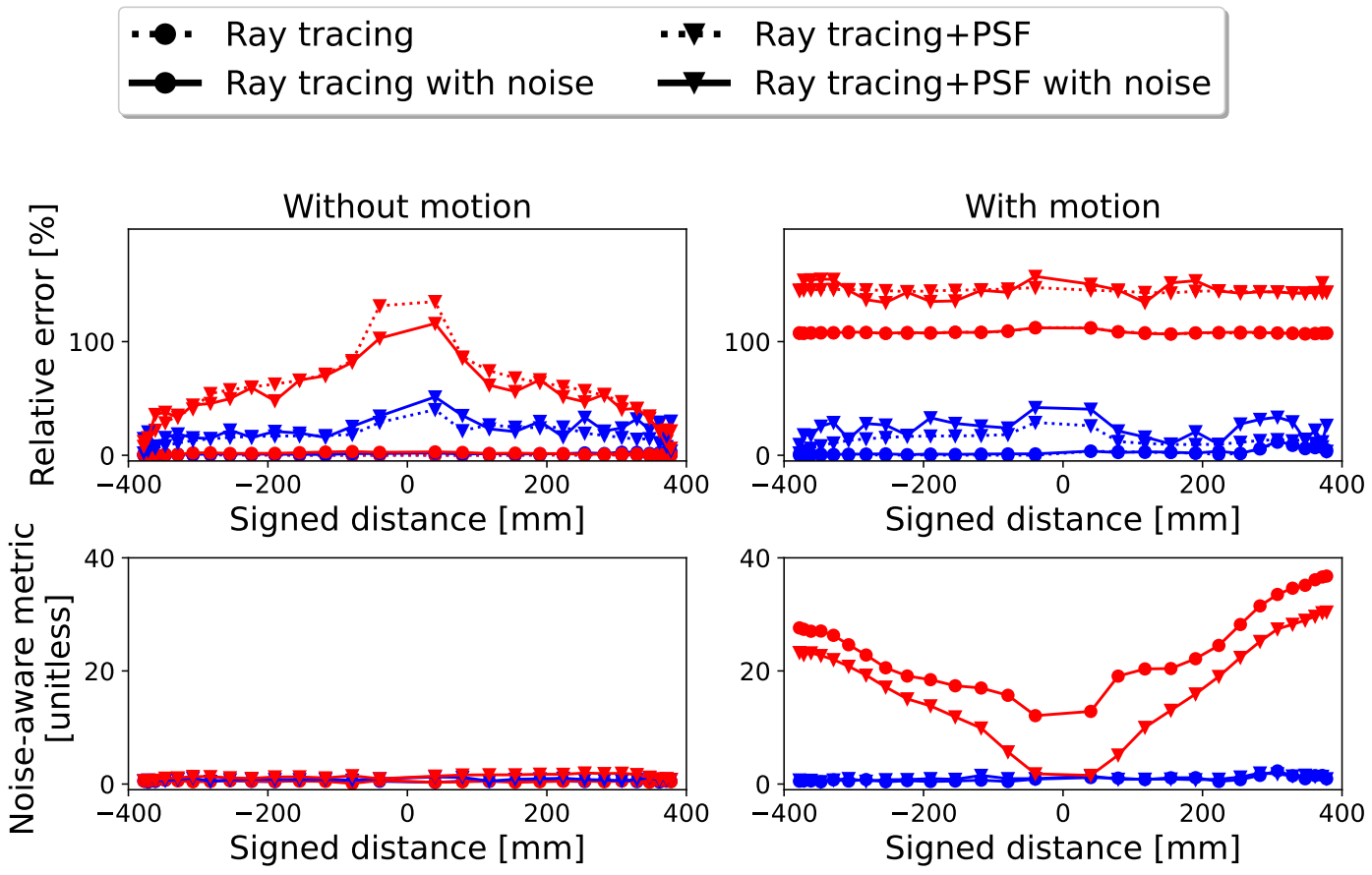

**Figure 3:** Relative error and noise-aware metric computed from the projections obtained with ray tracing. The blue and red lines correspond to the eDCCs computed in the vertical and horizontal directions, respectively (Figure 2). Each point is the metric of one pair of projections.

As for ray tracing simulations, the noise-aware metric seemed to reduce the error of the eDCCs. Without motion, it decreased with better scatter correction in both the horizontal and vertical directions. With motion, the value of the noise-aware metric in the horizontal direction increased and was similar for the three levels of scatter. The error was significantly higher in the horizontal direction than in the vertical direction except for the four pairs of projections where the signed distance was close to 0 mm.

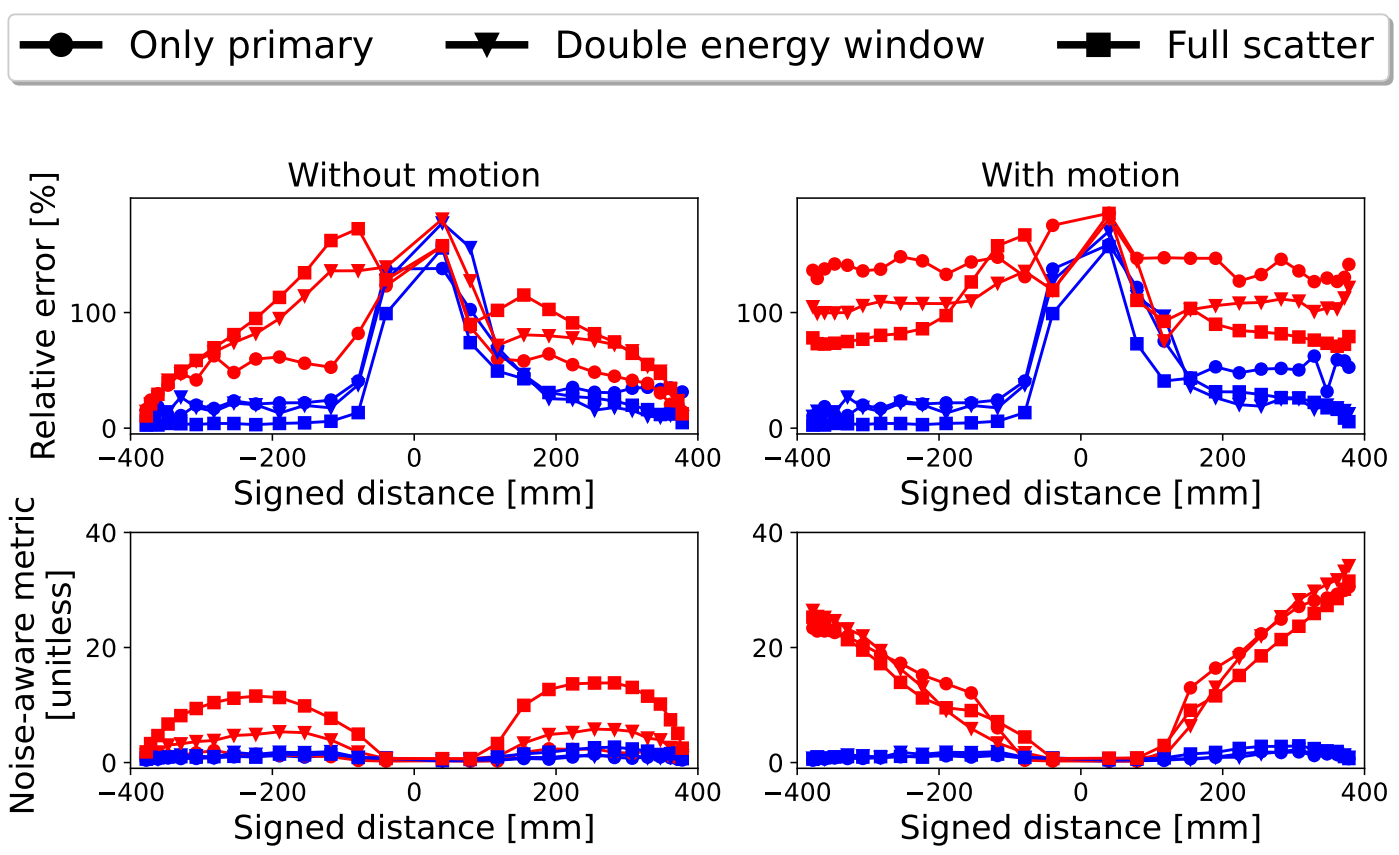

**Figure 4:** Relative error and noise-aware metric computed from the projections obtained with Monte Carlo simulations. The blue and red lines correspond to the eDCCs computed in the vertical and horizontal directions, respectively (Figure 2). Each point is the metric of one pair of projections.

## 4 Discussion

Ray-tracing and Monte Carlo simulations were used to assess the effect of noise, PSF, scatter and motion on eDCCs. They were evaluated with two metrics: the mean absolute relative error (equation 6) and a noise-aware metric (equation 7). In the closest simulations to the model of Equation 1 (ray-tracing without modelling any physical effect), the relative error of the eDCCs was close to 0% for all the pairs of projections and therefore the projections were consistent. As noticed by Wells *et al* [4], the addition of Poisson noise slightly impacts the eDCCs.

On the other hand, the PSF appeared to have a much more detrimental effect on the eDCCs. This is due to the depth-dependence of the PSF in SPECT as the source to detector distance was not the same for the two projections in our pairs. The error increases with increasing source to detector distance difference between the projections of the pair. This is why we observed a larger inconsistency in the horizontal direction than in the vertical direction in our results. If the PSF were the same for all projections, one would expect the eDCCs to be verified. This was verified with ray tracing simulations of a centered spheroid source for which the relative error of the eDCCs was close to 0% (data not shown).

The scatter also introduced inconsistencies in the projections. At first glance, it was surprising that the relative error was not always related to the level of scatter correction. This is a side effect of the normalization as the non-normalized absolute difference $\frac{1}{N}\sum_{l=1}^{N}|P_l(\theta_i,\sigma_{i,j}) - P_l(\theta_j,\sigma_{j,i})|$ led to the expected result, i.e., decreasing inconsistency with increasing scatter correction (data not shown). The inconsistency is mostly due to measuring photons outside the projection of the K-region. To overcome this issue, one could mask out the SPECT projections with the forward projections of the K-region [4]. Using this approach on our data significantly mitigates the impact of scattered photons (data not shown).

The motion was the effect that introduced the largest inconsistencies. In all cases, an important increase of the error was observed in the horizontal direction when there was motion between the projections. In the vertical direction, a slight increase was noticed for the pairs of projections with a positive signed distance. These pairs correspond to the projections obtained with the shifted emission map in both projections but the non-shifted attenuation map was used for the computation of the exponential projections (Equation 2) which introduced small inconsistencies. The error of most pairs of projections was still much higher in the horizontal direction than in the vertical one making the eDCCs a promising solution to detect patient motion during SPECT acquisitions.

In all cases, the absolute relative error was higher when the signed distance got close to 0 mm i.e. when the projections of the pair were close to be 180° apart. When taking the Laplace transform, the exponential projections are multiplied by $\exp(\sigma_{ij}s)$ and $\lim_{\theta_i-\theta_j\to\pi}\sigma_{ij} = +\infty$ so for projections close to be 180° apart, small inconsistencies could be highly magnified. By using the noise-aware metric that takes into account the variance of the measurement, we managed to reduce this effect. This metric is also more robust to the inconsistencies introduced by the PSF and the scatter and therefore more suitable for motion detection or for applying the method to real acquisitions.

## 5 Conclusion

eDCCs were evaluated on simulated datasets with increasing levels of realism. The PSF, the scatter and the movement were source of inconsistencies in the projections. We introduced a noise-aware metric which seems to be more robust to the PSF and scatter effects and which might be a promising way to detect patient motion.

## Acknowledgement

This work was supported by grant ANR-21-CE45-0026 (SPECT-Motion-eDCC) from the French National Research Agency (ANR) and was performed within the framework of the SIRIC LYriCAN INCa-INSERM-DGOS-12563 and the LABEX PRIMES (ANR-11-LABX-0063) of Université de Lyon, within the program "Investissements d'Avenir" (ANR-11-IDEX-0007) operated by the ANR.

# Combining spectral CT technologies to improve iodine quantification in pediatric imaging

Olivia F. Sandvold[1,2], Leening P. Liu[1,2], Nadav Shapira[2], Amy E. Perkins[3], J. Webster Stayman[4], Grace J. Gang[2], Roland Proksa[5], and Peter B. Noël[2]

[1]Department of Bioengineering, University of Pennsylvania, Philadelphia, PA, USA
[2]Department of Radiology, University of Pennsylvania, Philadelphia, PA, USA
[3]Philips Healthcare, Highland Heights, OH, USA
[4]Department of Biomedical Engineering, Johns Hopkins University, Baltimore, MD, USA
[5]Philips GmbH Innovative Technologies, Research Laboratories, Hamburg, Germany

**Abstract** In contrast to conventional computed tomography (CT), spectral CT provides additional information through the quantification of iodine-based contrast material for tissue perfusion. Various spectral CT designs have been introduced over the last decade with each undergoing significant improvement since initial implementation. Although current designs work for both adult and pediatric imaging, there is still room for customization to improve spectral performance and dose for pediatric imaging. It is our goal to investigate the impact of a theoretical spectral CT system using the combination of individual technologies, on improving sensitivity and decreasing noise when estimating iodinated contrast agents in pediatric imaging. To accomplish this task, we combine dedicated K-edge filter designs with a dual-layer spectral CT to estimate the spectral performance of this hybrid combination. Results show up to 17.3% decreased iodine estimation noise and increased iodine SNR by 120% with holmium prefiltration. This study provides preliminary insights into the application of combining CT technologies for improving quantification in pediatric spectral CT.

## 1 Introduction

In contrast to conventional CT, spectral CT allows enhanced tissue and material analysis through the collection of two or more measurements with distinct photon energy spectra. This quantitative ability is used for evaluation of iodinated contrast agents yielding perfusion metrics with functional information used for lesion differentiation, tumor staging, and assessment of treatment response [1].

Instrumentation of spectral CT can be achieved through spectral detectors (dual-layer, photon counting), use of multi-energy x-ray tubes (rapid kVp-switching, dual source), or use of specialized filters to produce a split spectral beam [2]. Though current spectral CT systems work for both adult and pediatric patients, there have been a limited number of studies on the optimization of spectral CT for the pediatric population [3-4]. Achieving low radiation dose is a priority for pediatric imaging. More generally, across all spectral CT applications, there is a need to improve quantification sensitivity and accuracy for low concentration of iodine. In oncologic imaging, a high sensitivity of low concentration iodine contrast is required and additional visibility can significantly increase confidence in tumor identification [5-6]. Thus, there is an opportunity to introduce spectral CT designs specifically for pediatic imaging to address current challenges.

One potential solution is to investigate implementation of "hybrid" models which combine existing techniques to enable spectral CT. There is evidence that hybrid spectral CT may rectify imperfections in quantification and increase contrast visibility at equivalent or lower doses [7].

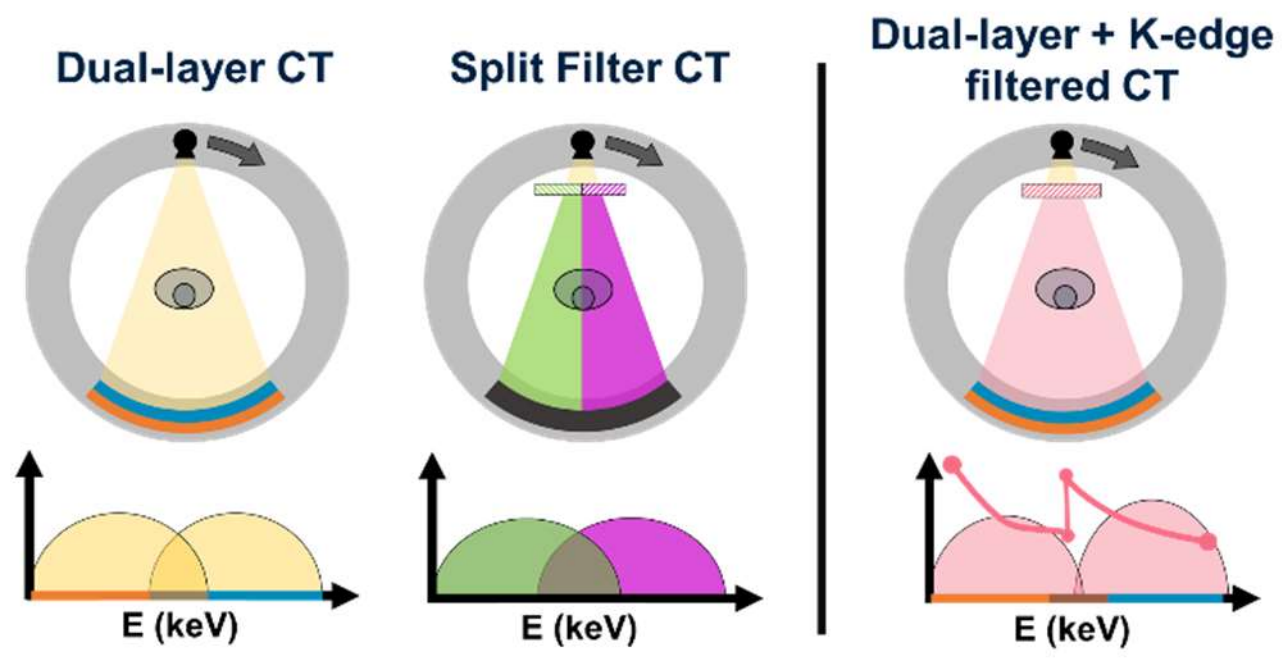

**Figure 1:** Schematic of proposed design combining dual-layer spectral CT detectors and K-edge filtration for optimal spectral performance.

We hypothesize that a configuration utilizing a clinical dual-layer detector and single K-edge filter, as depicted in **Figure 1**, may outperform non-combined systems. Spectral detectors have high dose efficiency but suffer from overlap of photon energy spectra due to hardware limitations. Implementing multiple K-edge filters to generate a number of spectral beamlets has been validated as a cost-effective technique [8-9], but it has suboptimal sampling or requires advanced reconstruction techniques [10]. Combining spectral detectors with a K-edge filter has the potential to improve the accuracy of low concentration iodine quantification in pediatric scans while also keeping radiation doses low.

In this study, we investigate whether the addition of a low-cost K-edge filter to dual-layer CT improves spectral performance for pediatric imaging. Our theoretical hybrid model utilizes the characteristics of a dual-layer spectral CT system, which includes polychromatic spectrum, detector responses, and a K-edge material filter, to determine the optimal material and thickness for pediatric imaging to increase spectral sensitivity of iodine detection.

## 2 Materials and Methods

### 2.1 Phantom design

To model patient habitus and low concentrations of iodine, phantoms were designed with varying path lengths of water and iodine as defined in Eqn. 1. For simplicity, each phantom consisted of a mixture of water [1 g/mL] and iodine with a concentration of 1 mg/mL. The water path lengths, or patient diameter, varied from 10 to 20 cm. Iodine path lengths were set to 0.5 cm and 2.5 cm, equivalent to 1 cm of 0.5 mg/mL iodine and 1 cm of 2.5 mg/mL iodine.

## 2.2 Polychromatic forward projection

Polychromatic x-ray spectra photons of a single pencil beam were simulated using the energy spectra of a clinical dual-layer detector CT (IQon Spectral CT, Philips Healthcare) at 120kVp and 200mAs. The projection value $I_i$, for layer *i,* is given by the following expression:

$$I_i = \int s_i(E) \cdot \exp\left(-\sum_\alpha \mu_\alpha(E) \cdot \rho_\alpha \cdot l_\alpha\right) dE, \quad (1)$$
$$i = 1, 2;\ \alpha = water, iodine$$

where $s_i(E)$ is the combined x-ray source spectra (in keV) and detector response (including the energy weighting) for the low and high energy layers, and $\mu_\alpha(E)$ is the mass attenuation coefficient [$cm^2/g$] at the energy E of material α, with density $\rho_\alpha$ [$g/cm^3$], with a given path length $l_\alpha$ [cm], in the phantom containing water and iodine. The central detector responses represent $s_i(E)$. $I_i$ changes with addition of a K-edge filter as follows:

$$I_i = \int s_i(E) \cdot \exp\left(-\mu_f \cdot \rho_f \cdot l_f\right) \cdot \exp\left(-\sum_\alpha \mu_\alpha(E) \cdot \rho_\alpha \cdot l_\alpha\right) dE, \quad (2)$$
$$i = 1, 2;\ \alpha = water, iodine$$

where $\mu_f(E)$ is the mass attenuation coefficient [$cm^2/g$] at the energy E for filter material *f* with thickness $l_f$ [cm] and density $\rho_f$ [$g/cm^3$].

We assume Gaussian distributed quantum noise for the energy integrating detectors. An additional assumption is no x-ray crosstalk between the two energy-integrating dual-layer detectors. Under this condition, the two signals are statistically independent theoretically and in practice [11].

## 2.3 Filter selection & dose modulation

Filter materials were selected if their respective K-edge were within the photon energy range containing the most overlap of dual-layer detectors' responses, and they were available to purchase as a thin film/sheet. Mass attenuation coefficients for each of the materials in **Table 1** were collected from XCOM NIST database [12]. Filter thicknesses ranging from 0.01 to 0.2 mm were selected based on purchase availability.

The amount of patient exposure in each simulation was set by multiplying the initial $s_i(E)$ spectra by a computed factor to yield a pre-determined number of total photons (10, 20, 100) detected by low and high energy layers after patient attenuation. We defined a single computed factor for each combination of patient size and iodine concentration. To ensure dose neutrality for simulations with pre-filtration, we scaled the sum of spectrum energy before patient attenuation, or air kerma, in filtered simulations to be equal to the air kerma of the non-filtered spectra. This is analogous to the scaling of tube current in a clinical experiment.

**Table 1:** Relevant filter materials and their properties.

| Material | Atomic number (Z) | K-edge (keV) | Physical density [$g/cm^3$] |
|---|---|---|---|
| Holmium | 67 | 55.6 | 8.80 |
| Erbium | 68 | 57.5 | 9.05 |
| Ytterbium | 70 | 61.3 | 6.98 |
| Tantalum | 73 | 67.4 | 16.60 |

## 2.4 Performance metric calculation

The negative log-likelihood ***L*** for the observation of the set of projection values $I_i$, with Gaussian quantum noise distribution with mean $\mu_i$ and variance $\sigma_i^2$, is defined in Eqn. 3. We assume $\mu_i$ and $\sigma_i^2$ depend on the line integrals for water and iodine which allows us to compute the Fisher information matrix $F_{\alpha\alpha}$ for each material α for each spectral CT configuration. A full derivation of these steps is outlined by Roessl and Herrmann [13].

$$L = -ln\left(\prod_i \left[\frac{1}{(2\pi\sigma_i^2)^{1/2}} \exp\left(-\frac{(I_i-\mu_i)^2}{2\sigma_i^2}\right)\right]\right) \quad (3)$$
$$F_{\alpha\alpha} = \sum_i \frac{1}{\sigma_j^2}\left(\frac{\partial\mu_i}{\partial l_\alpha}\right)^2 + \frac{1}{2}\sum_i \frac{1}{(\sigma_j^2)^2}\left(\frac{\partial\sigma_i}{\partial l_\alpha}\right)^2 \quad (4)$$
$$\sigma_\alpha^2 \geq F_{\alpha\alpha}^{-1} \quad (5)$$
$$SNR_\alpha = \frac{l_\alpha}{\sigma_\alpha}, \quad (6)$$
$$i = 1, 2;\ \alpha = water, iodine$$

The Cramer-Rao lower bound (CRLB) of variance, $\sigma_\alpha^2$, determines the minimum noise in the predicted line integral for water and iodine, respectively. The signal-to-noise ratio (SNR) was assessed using the CRLB variance estimate of noise in the basis decomposed projections to compare system performance of water/iodine across simulations.

Lastly, spectral separation was estimated as the difference between the weighted average photon energy (keV) of the high energy and low energy detector layers.

# 3 Results

## 3.1 Quantification noise & SNR results

Hybrid combination spectral CT with holmium, erbium, and ytterbium K-edge filters, for all filter thicknesses, patient sizes, and iodine concentrations, showed reduction in estimation noise of iodine path length and increased iodine and water SNR compared to dual-layer alone. In contrast, tantalum filtration demonstrated moderate improvement in iodine SNR and decreased estimated iodine noise in the largest phantom (20 cm) only.

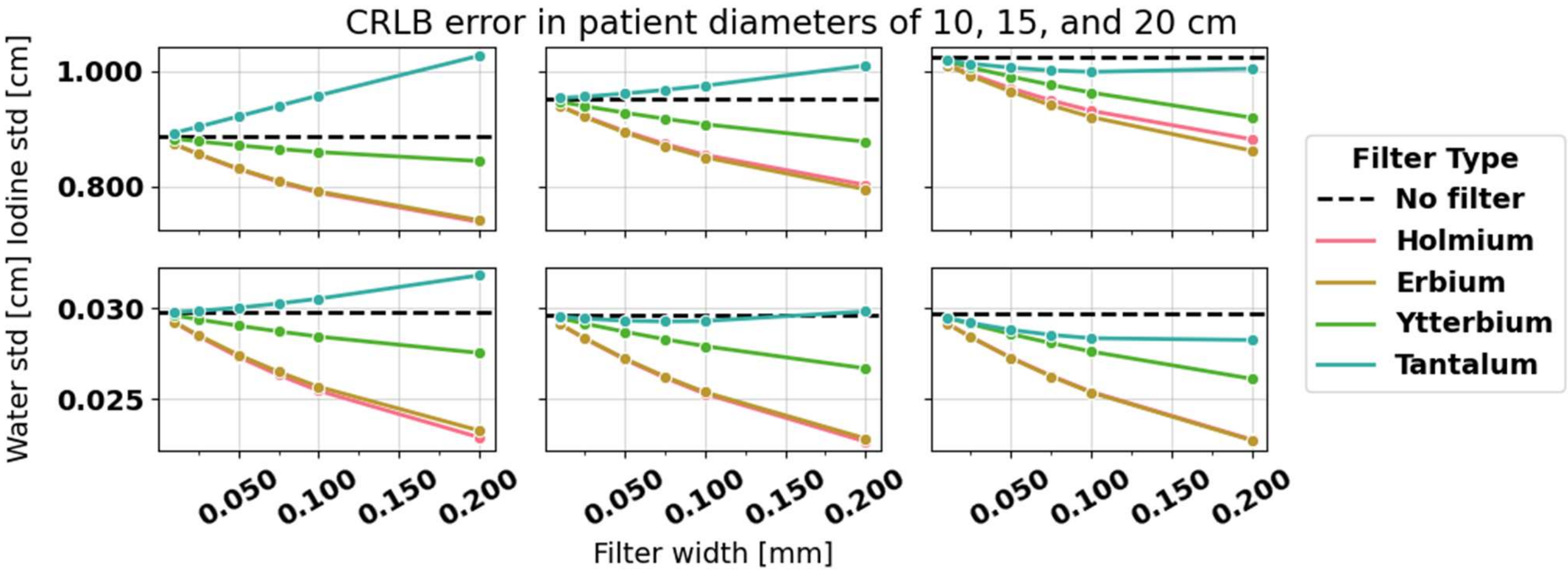

**Figure 2:** Comparison of estimated quantification error to baseline error in patient width of 10, 15, and 20 cm with 2.5 cm of iodine contrast at the lowest estimated dose level (10 total photons). The dashed black line represents quantification error in dual-layer alone.

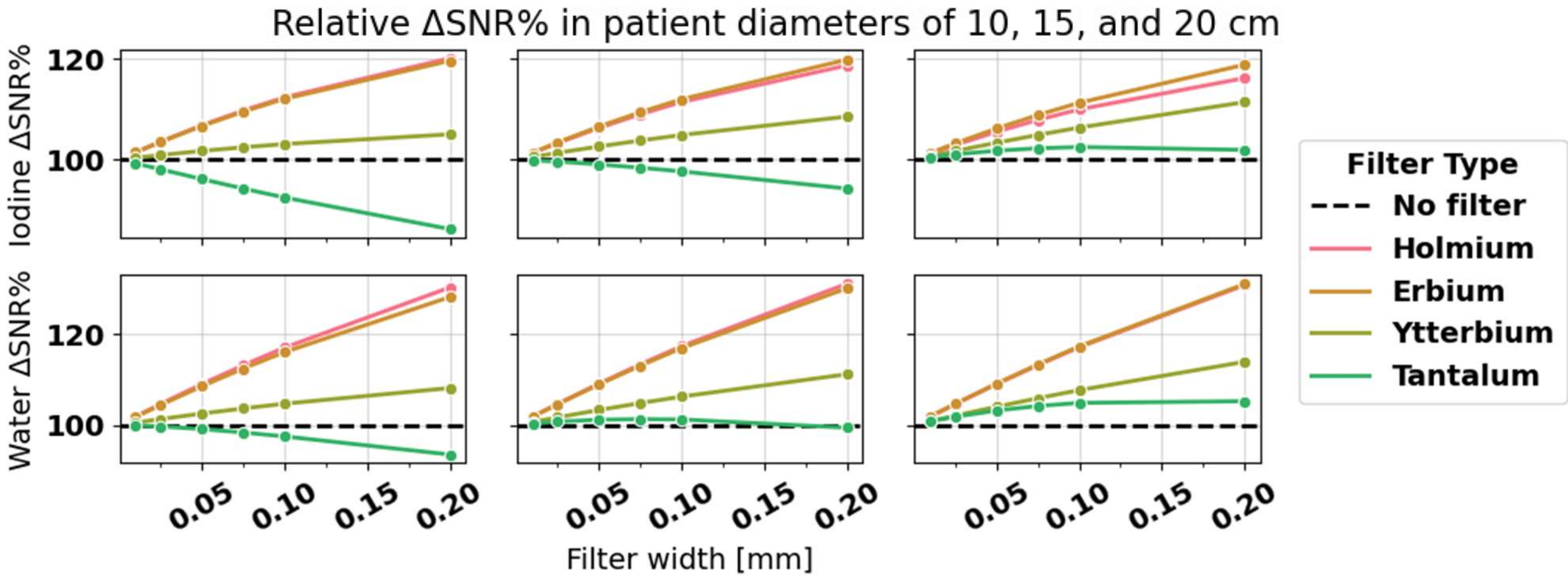

**Figure 3:** Relative changes in material domain SNR of hybrid spectral CT to dual-layer alone in patient sizes 10, 15, and 20 cm with 2.5 cm of iodine at the lowest estimated dose level (10 total photons). The dashed black line represents no change in SNR compared to the non-filtered dual-layer simulation.

**Figure 2** shows absolute improvement in iodine and water material path length noise estimation with addition of one of holmium, erbium, and ytterbium filters at all tested thicknesses, patient sizes, iodine at 2.5 cm, with the lowest estimated detector dose level (10 photons). Holmium, the K-edge filter with the largest improvement in iodine noise estimation, had relative error differences of -0.013 cm to -0.147 cm in the 10 cm sized patient. These values represent improved estimation statistics from 1.4% to 16.7% compared to non-filtered performance. Erbium and holmium were comparable for all water/iodine results across varying patient sizes and filter lengths. Ytterbium had a maximum decreased iodine noise by 11.8%. Tantalum increased iodine noise by up to 15% in the smaller patients but showed improved material estimation up to 2.8% in the 20 cm pediatric diameter. As patient sizes increased, the iodine quantification noise increased for non-filtered, holmium, erbium, and ytterbium filtered cases. Water estimation noise remained relatively constant with changing patient size in the configuration with no filter.

Filters which reduced the CRLB variance estimate also increased relative iodine SNR by up to 120% in the 10 cm phantom with 2.5 cm iodine, as seen in **Figure 3** with holmium filtration, $l_f$ = 0.02 cm. Ytterbium increased relative SNR in water and iodine, and tantalum showed improved water/iodine results in the largest phantom.

Though absolute CRLB noise estimates decreased with increased exposure, the relative SNR changes seen in **Figure 3** are the same across all dose levels (10, 20, 100 photons).

With 0.5 cm of iodine, filter performance estimation noise trends were the same as for 2.5 cm of iodine. The smallest iodine noise using the lowest detector dose was 0.87 cm in the 10 cm patient. Compare this to an estimated iodine noise of 0.73 cm using holmium at 0.2 mm, a 17.3% reduction. The average decrease in iodine noise in all phantoms was 0.1 cm for holmium with filter sizes between 0.05 to 0.2 mm. For patient sizes 10 and 15 cm, and filter thicknesses of 0.2 mm, holmium and erbium increased relative iodine SNR by 120% on average. Non-filtered simulations showed a maximum ideal iodine SNR of 0.57 for the lowest estimated dose in all patient sizes whereas holmium prefiltration increased iodine SNR up to 0.69.

### 3.2 Spectral separation

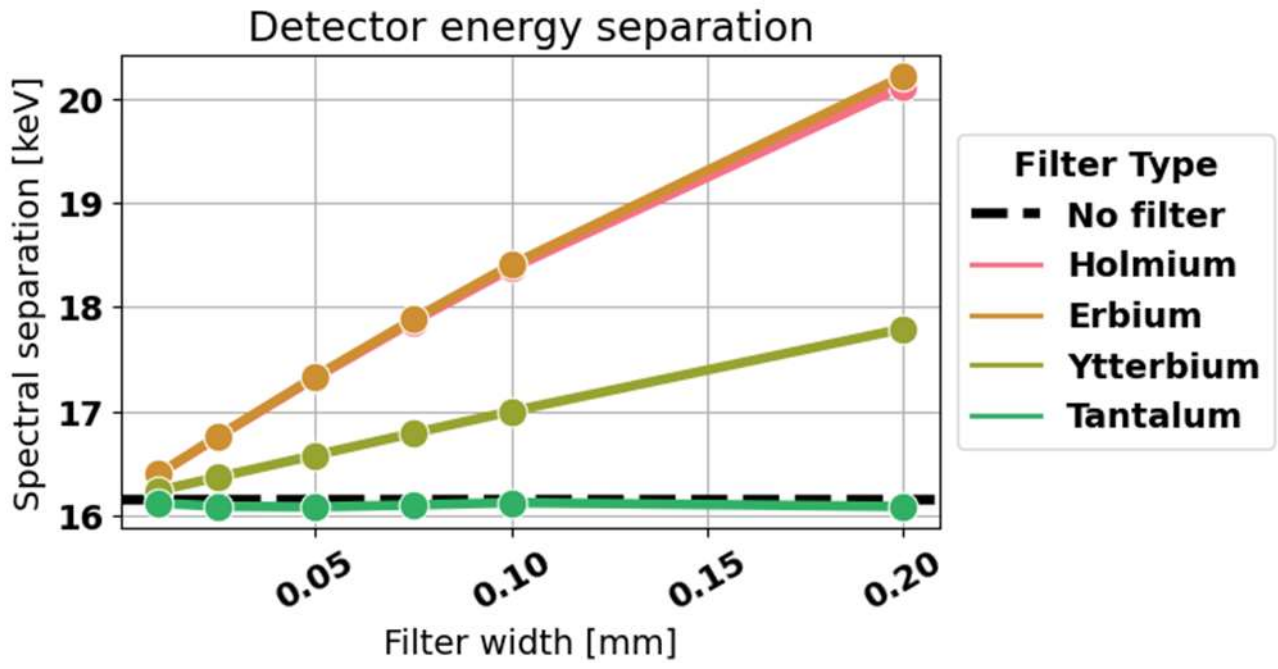

**Figure 4:** Energy separation between two detector responses after attenuation through 10 cm of water and 0.5 cm of iodine at the lowest exposure dose.

**Figure 4** shows results of estimated spectral separation for each K-edge prefilter configuration. Holmium and erbium have spectral separation up to 20.2 keV with 0.2 mm filter length compared to dual-layer CT with separation of 16.2 keV after attenuation through a 10 cm phantom with 0.5 cm of iodine. As patient sizes increase, ytterbium and tantalum increase less than 1 keV in separation for each filter thickness whereas erbium and holmium decrease separation by less than 1 keV. Tantalum prefiltration results in a wider spectral separation compared to dual-layer CT only in the 20 cm patient.

## 4 Discussion & Conclusion

In this work, we utilized our simulation framework to demonstrate that a dedicated K-edge filter design in combination with a clinical dual-layer spectral CT improves water/iodine quantification and spectral separation for pediatric imaging. Throughout this paper, it is important to note that all results are derived without the use of advanced processing techniques [14]. Filter materials holmium, erbium, and ytterbium at thicknesses of 0.01 to 0.2 mm decrease quantification noise and improve relative SNR for iodine and water in the projection domain using low concentrations (0.5 and 2.5 mg/mL) of 1 cm iodine. Holmium and erbium materials at 0.2 mm filter length showed optimal spectral performance with iodine noise reduction by 17.3% and increased iodine SNR by 120% compared to the non-filtered configuration with the lowest iodine concentration.

Spectral separation was increased for K-edge materials holmium, erbium, and ytterbium for all patient sizes. This indicates that the two detector responses benefit from additional filtration at these K-edges (55-61 keV). Though maximizing spectral separation is not a direct measure of system performance, reducing overlap is an important factor in improving material decomposition for accurate quantification [14]. An ideal dual-layer detector has a spectral separation of greater than 30 keV [14], while we were able to achieve separation around 20 keV after patient attenuation with combining technologies.

Feasibility studies will need to be conducted to determine the patient size limit where available x-ray flux becomes insufficient. In our simulations, the best performing filter materials, holmium and erbium, reduced flux by 70% at 0.2mm filter thickness. Further, use of advanced material decomposition and noise reduction methods will be applied to estimate additional improvement compared to current implementation. Lastly, simulation and measurement of scans for pediatric spectral imaging, specifically from 80 to 100kVp, will be necessary to demonstrate clinical translation.

One limitation of this study design is the dependence on a given input spectra and detector response for optimizing K-edge filter design. However, this process could be generalized and repeated for other configurations, including photon counting detectors, to identify K-edge filter designs best situated to minimize quantification error of iodine in varying patient sizes. Preliminary studies have been presented to simulate the potential for overcoming shortcomings of photon counting detector designs [15].

Given the high concern of patient delivered dose in pediatric imaging, one additional benefit of K-edge filtration may be the reduction of low energy photons that contribute to skin dose. Spectral CT may also provide advanced perfusion metrics which could mitigate the risk-benefit ratio in pediatric imaging.

Our study demonstrates the possibility of improving quantification performance by incorporating the spectral diversity of individual implementation technologies into hybrid designs. Experimental studies are necessary to translate our findings into clinical practice. While limited x-ray flux remains a concern when adding additional filtration, an opportunity arises in pediatric imaging to improve spectral performance and dose utilization due to the smaller transverse body diameter in children.

## Acknolwedgements

We acknowledge support through the National Institutes of Health (R01EB030494) and Philips Healthcare.

# Task-based Generation of Optimised Projection Sets using Differentiable Ranking

Linda-Sophie Schneider[1,2], Mareike Thies[1], Christopher Syben[2], Richard Schielein[2], Mathias Unberath[3], and Andreas Maier[1,2]

[1]Pattern Recognition Lab, Friedrich-Alexander-Universität Erlangen-Nürnberg, Germany
[2]Fraunhofer Development Center X-Ray Technology EZRT, Germany
[3]Laboratory for Computational Sensing + Robotics, Johns Hopkins University, Baltimore, MD, USA

**Abstract** We present a method for selecting valuable projections in computed tomography (CT) scans to enhance image reconstruction and diagnosis. The approach integrates two important factors, projection-based detectability and data completeness, into a single feed-forward neural network. The network evaluates the value of projections, processes them through a differentiable ranking function and makes the final selection using a straight-through estimator. Data completeness is ensured through the label provided during training. The approach eliminates the need for heuristically enforcing data completeness, which may excludes valuable projections. In addition, the calculation of the projection-based metric can be significantly accelerated by using a neural network. The method is evaluated on simulated data in a non-destructive testing scenario, where the aim is to maximize the reconstruction quality within a specified region of interest. We achieve comparable results to previous methods, laying the foundation for using reconstruction-based loss functions to learn the selection of projections.

## 1 Introduction

In the field of computed tomography (CT), a series of projections are obtained to produce a three-dimensional representation of the object of interest. However, not all projections are equally essential for image reconstruction and diagnostic purposes [2]. A projection is considered more valuable when the amount of information gain in the chosen set of projections for reconstruction is higher. Selecting the most valuable projections can enhance the detection of anomalies or defects, improve imaging efficiency, and minimize noise and artefacts in the final reconstruction. In order to determine an optimized CT trajectory, it is necessary to balance the individual value of each projection with the overall value of the set of projections used for reconstruction.
One approach to selecting valuable projections is through task-based image quality metrics. These metrics quantify the performance of a given projection with respect to a specific imaging task, such as detecting small structures or preserving low contrast details. By evaluating each projection using a task-based image quality metric, the most valuable projections can be selected for inclusion in the image reconstruction process. As a secondary criterion, data completeness should be satisfied. Simultaneously optimizing both projection-based metrics and set-based metrics is essential for improving the quality of the reconstruction. However, being effective in projection-based metrics does not necessarily result in being valuable in the context of the set-based metrics.

In our recent work [1], we investigated the trainability of the projection-dependent detectability index (PDI) for a specific class of objects and its usability for CT trajectory optimization. Because this does not ensure data variability if computed per projection, we additionally introduced a haversine distance constraint in the optimization. The aim is to optimize the quality of the reconstruction in a predefined region of interest. The optimization problem is formulated to maximize the detectability index predicted by a neural network, while ensuring that the haversine distance constraint is met. However, this approach may excluded valuable projections that enhance resolution in the region of interest.
In this work, we present an approach to address the issue of manually incorporating data completeness into the data analysis process. Our solution involves integrating this constraint into the neural network architecture. The network output directly indicates an optimized set of projections, which can be used to reconstruct the volume with high accuracy in the region of interest. To achieve this, we propose an adaptation of the ResNet-18 architecture. This outputs a hidden representation for each projection, which are processed through a differentiable ranking function to rank the projections. A straight-through estimator is used to make the final selection of projections. To connect this output to a set of projections, the integer program introduced in [1] is used to generate a label during training. Our approach is unique in that it directly balances individual projection value and overall set value, ensuring that the final selection of projections maximizes reconstruction quality in the region of interest. Furthermore, the time-consuming analytical computation of the detectability index is integrated into the neural network, resulting in a significant speedup for the computation of an optimized trajectory for the trained domain. We demonstrate the utility of our method in a non-destructive testing scenario by maximizing the reconstruction quality within a specified region of interest using a limited number of projections.
To summarize, this paper makes the following contributions:

- We introduce a network architecture that integrates both projection-based and set-based evaluation metrics for selecting a pre-defined number of projections.
- Our approach demonstrates the capability of learning a heuristic that is not limited to single projections, thereby establishing the basis for using reconstruction-based loss functions to guide the selection of projections.

## 2 Methods

We propose a combination of projection-based and set-based evaluation metrics using a three-step neural network approach. The first step involves reducing each projection to a single value using a modified ResNet-18 architecture. The regressed values are then collected in a single vector, which represents the value of each projection. The second step involves applying a differentiable ranking function (see Section 2.2) to this vector, resulting in a ranking of the projections. Lastly, the threshold function of the Straight-Through Estimator (see Section 2.3) is applied to convert the ranking into a binary vector that represents the selection of projections. An overview can be seen in Figure 1.

### 2.1 Projection-Dependent Detectability Index

The projection-dependent detectability index is a measure of the quality of a single projection and its ability to contribute to the observability of a signal in a reconstructed image. It is used to evaluate the performance of different CT projection angles. The PDI is calculated using the non-prewhitening matched filter observer model (NPWM) described in Stayman et al. [2–4]. The NPWM model defines the modulation transfer function (MTF) and noise power spectrum (NPS) as functions of the position of the target voxels in the volume, denoted by (x,y,z). The analytical equations for both the MTF and NPS in the context of iterative penalized-likelihood reconstruction were developed by Gang et al. [5]. The PDI is given by the equation

$$d^2(x,y,z) = \frac{[\iiint |MTF(x,y,z)|^2 |W_t|^2 df_x df_y df_z]^2}{\iiint |NPS(x,y,z) MTF(x,y,z)|^2 |W_t|^2 df_x df_y df_z} \quad (1)$$

where $W_t$ is the Fourier transform of the region of interest to be imaged with the highest quality. A high detectability index indicates that the required signal is effectively detected in the respective projection according to the task function $W_t$.

### 2.2 Differentiable Ranking

The ranking operator is a discontinuous function, lacking differentiability, making it unsuitable for use as a component in the training of neural networks, as the gradients necessary for optimization are not defined. To address this issue, differentiable soft ranking, a technique proposed by Blondel et al. [6] allows neural networks to learn a ranking function that can be used to order elements in a set, such as images in a dataset, based on some criterion. The algorithm, a variation of sorting networks, is more efficient in terms of computation time and memory requirements compared to other differentiable sorting algorithms.

The basic idea of the technique is to reformulate the descending ranking operation as a linear program over the permutahedron. A permutahedron $P(w)$, where $w \in \mathbb{R}^n$, is a polytope that represents the symmetries of a permutation $w$, with vertices corresponding to all possible permutations of a set of elements and edges corresponding to transpositions of adjacent elements. The linear program can be written as

$$r(\theta) = \underset{\mu \in P(\varphi)}{\arg\max} \, \langle \mu, -\theta \rangle \quad (2)$$

where $\theta \in \mathbb{R}^n$ is the vector of scores produced by the neural network and $\varphi = (n, n-1, \ldots, 1)$. An optimal solution is mostly achieved at a vertex of the permutahedron according to the fundamental theorem of linear programming [7].

To ensure differentiability, the authors introduce strongly convex regularization $\Psi$ to the linear program $P_\Psi(-\theta, \varphi)$. The regularization strength is controlled through a parameter $\varepsilon > 0$, which is multiplied by $\Psi$. The resulting $\Psi$-regularized soft ranking can be formulated as

$$r_{\varepsilon\Psi}(\theta) = P_\Psi(-\theta/\varepsilon, \varphi) \quad (3)$$

In this context, we use a quadratic regularization term $Q(\mu) = \frac{1}{2}\|\mu\|^2$. Using $\Psi = Q$, the linear program for the soft ranking over the permutahedron results in

$$r_{\varepsilon Q}(\theta) = P_Q(-\theta/\varepsilon, \varphi) = \underset{\mu \in P(\varphi)}{\arg\max} \, \langle \mu, -\theta/\varepsilon \rangle - Q(\mu) \quad (4)$$

Ascending-order soft ranking can be obtained by negating the input.

### 2.3 Straight-Through Estimator

After obtaining a ranking for the projections, the next step is to convert it into a selection of the projections. To accomplish this task, it is necessary to find a function that assigns values of 0 and 1 to the projections based on the ranking. The threshold function, as shown in Equation (5) , is one way to implement this.

$$thresh(x) = \begin{cases} 1 & x \leq k \\ 0 & x > k \end{cases} \quad (5)$$

Here, it is assumed that the ranking is in descending order and the goal is to select k projections. Thresholding, however, often causes issues during backpropagation because its derivative is zero. To address this issue, the Straight-Through Estimator (STE) is used in the backward pass as introduced in [8]. The STE allows the network to learn from the chosen projections during training by backpropagating through applying the derivative of the identity function. This allows for the gradients to be computed and updated correctly.

## 3 Experiments

### 3.1 Data

In this work, we conducted simulation experiments within the domain of non-destructive testing. The aim was to identify defects in a pre-specified region of interest. To achieve this, three test specimens of dimensions $10\,\text{cm} \times 8\,\text{cm} \times 8\,\text{cm}$, made of aluminium with a density of $2.7\,\mathrm{g/cm^3}$, were selected. Six different representations of each test specimen were generated, each with the same size and density, but with variations in the placement of the spherical-shaped defect, with a radius of 1 mm. As a result, 18 distinct test specimens were obtained.

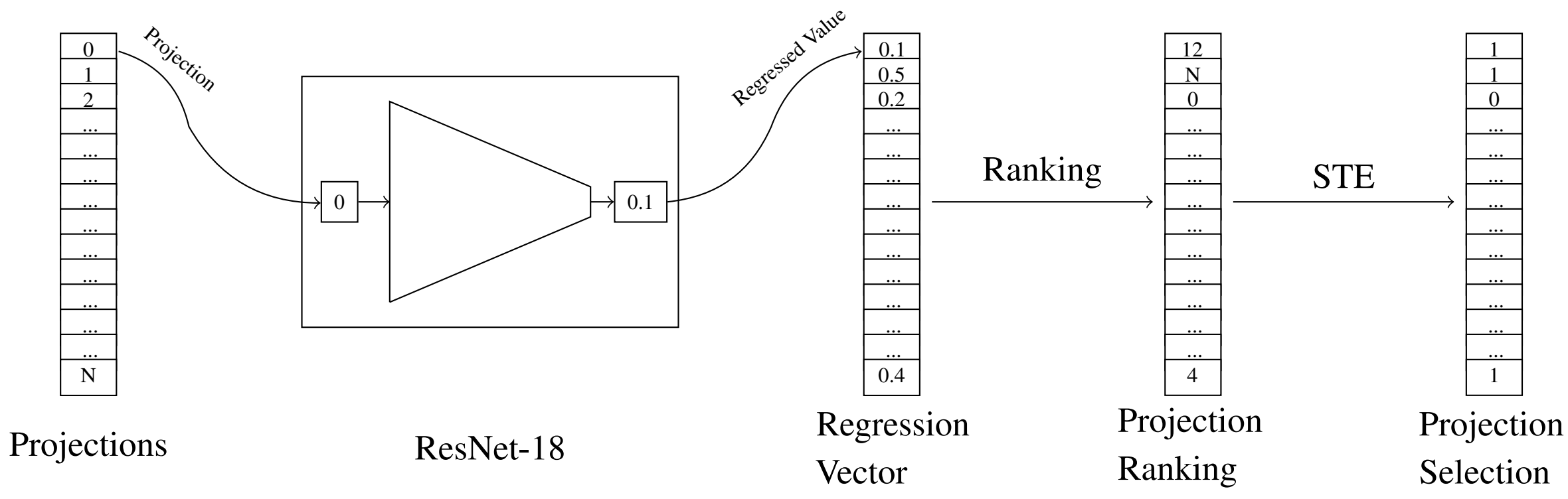

**Figure 1:** Overview of the proposed approach. First, each projection is regressed to a single value. This value is used as a basis for the ranking. To generate a selection out of the ranking, we utilize a straight trough estimator.

We defined a geometry for the CT imaging setup by fixing the detector-isocentre distance and the source-detector distance to 3 m and 4 m, respectively. The detector and source were positioned to face each other, which results in restricting the scan positions to a sphere. We parametrized the problem using azimuth angle $\varphi$ and elevation angle $\theta$ using spherical coordinates. Each CT trajectory consisted of a set of pairs $(\varphi_t, \theta_t)$ for $t \in 0,\ldots,N$, where $N$ is the total number of projection images. Initially, we sampled $N = 1000$ scan positions on a sphere using a Fibonacci-based sampling for uniform surface coverage [9].

The projection data was simulated using the Fraunhofer EZRT simulation software XSimulation with a 225 keV polychromatic spectrum. The test specimens were placed at the centre of the world coordinate system. The detector was chosen to have a size of $375 \times 375$ pixels and a pixel pitch of $400\,\mu\mathrm{m} \times 400\,\mu\mathrm{m}$. The PDI was calculated analytically using Equation (1). This allowed us to determine a set of projections through the integer optimization problem including the haversine distance constraint introduced in [1], which resulted in a binary vector $y \in \{0,1\}^N$ where 1 indicates a chosen projection. In this work, this used as a label for training. In this study, we set the number of possible projections in the optimized CT trajectory to $k = 100$.

### 3.2 Neural Network Architecture

The network was trained on 1000 projections of 15 distinct test specimens, such that 3 test specimens, one for each type of test specimen, remained to test the network performance. To predict the optimal set of projections for a specific task, we employed a Convolutional Neural Network (CNN) approach. The ResNet-18 architecture was adapted for regression by adding a single fully connected layer to perform the regression towards a scalar value. The network was trained using a Binary Cross Entropy loss function with the Adam optimizer and a learning rate of $3\mathrm{e}^{-5}$. The pre-trained ResNet-18 was initialized with weights from the ImageNet dataset and the fully connected regression layer was initialized randomly. The final selection of projections was determined using the straight-through estimator as described in Section 2.3, yielding a binary representation of the ranking where a value of 1 indicates the projection belongs to the $k$ highest-ranked projections. The training process utilized a batch size of 1 corresponding to a complete CT scan of a test specimen and was performed for a maximum of 600 epochs.

## 4 Results

To assess the performance of the proposed method, a comparison was conducted between the reconstructed volume obtained from the predicted optimal set of projections and the optimal set of projections according to the integer program in [1]. The reconstruction was performed using the Algebraic Reconstruction Technique (ART) with 3 iterations. The evaluation of the reconstructed images was based on two standard image quality metrics,the Structural Similarity Index (SSIM) and the Root Mean Squared Error (RMSE). The results of this comparison are presented in Table 1. Our analysis showed a positive correlation between a decrease in RMSE and an improvement in SSIM. Moreover, the location of the defect was crucial for both the reference method and the proposed method to perform effectively. Our proposed method was able to achieve results that were comparable to the projection set indicated by the label, especially in terms of reconstruction quality. This is shown Figure 2, where the grey values of a specific slice of our heartgear test specimen were examined.It was found that the areas of the defect between the prediction and the label are very similar. In addition, it can be seen that the structure of the defect in particular could be clearly depicted, but the same resolution could not be achieved as with the reference reconstruction. Even though there were more reconstruction artefacts in our proposed method, this did not affect the reconstruction quality in the region of interest, as shown in Table 1.

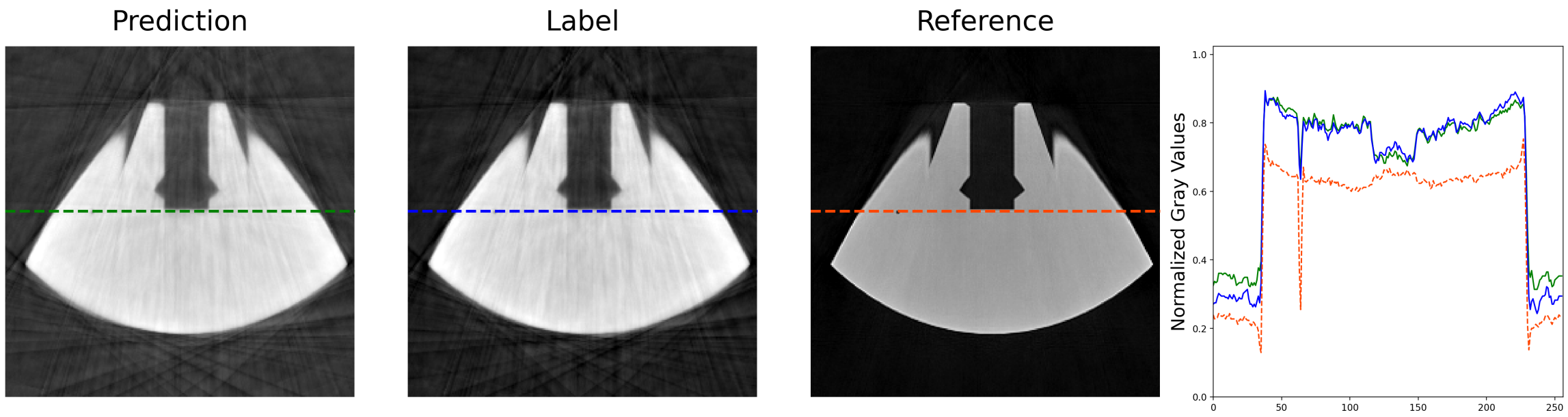

**Figure 2:** Comparison of the reconstruction quality for the heart gear between the proposed method and the method used to generate the label. Additionally, we show the reference reconstruction generated from the full projection set. One can see, that the proposed method and the labelling method do not show any significant differences, especially in the region around the error. The images are normalized.

| | RMSE | | SSIM | |
|---|---|---|---|---|
| | label | prediction | label | prediction |
| Heartgear | 0.16 | 0.13 | 0.73 | 0.79 |
| Round Heartgear | 0.21 | 0.21 | 0.45 | 0.48 |
| Dent Gear | 0.33 | 0.32 | 0.36 | 0.35 |

**Table 1:** Comparison of reconstruction quality by measuring the SSIM and RMSE between the reconstruction of the full projection set and the projections chosen by our approach or the approach in [1] at the specific region of interest $W_t$.

## 5 Discussion and Conclusion

This paper presents a new approach for selecting valuable projections in computed tomography scans, with the aim of improving reconstruction quality in a region of interest. The proposed solution uses a neural network architecture that integrates the projection-based detectability index and data completeness into a single model. The network, based on a modified ResNet-18 architecture, evaluates the value of projections, passes them through a differentiable ranking function, and the final selection is made using a straight-through estimator.

The results suggest that our proposed method is capable of selecting projections for a predefined task and a known object structure. It was found that the performance of the proposed solution is comparable to the baseline method, which is used for labelling. The fact that the performance is comparable to the baseline method demonstrates that the selection of projections can be learned by the neural network. This finding suggests that other set-based metrics, such as a reconstruction-based loss, can also be learned with our presented approach. By using the neural network to optimise the trajectory, the task can be significantly accelerated since the behaviour of the intensive analytical calculation of the detectability index is included in our proposed network.

One drawback of the proposed solution is that it is object-specific, meaning that it is trained on a specific object structure and does not generalize to structures not seen during training. Despite this, the approach is still suitable for non-destructive testing, as it is often necessary to examine similar or identical objects.

For future work, we want to add the embedding of positions, which may be useful for improvement. This can help to further guide the neural network in the direction of data completeness.

## Acknowledgements

This research was financed by the „SmartCT – Methoden der Künstlichen Intelligenz für ein autonomes Roboter-CT System“ project (project nr. DIK-2004-0009).

# PARALLELPROJ - An open-source framework for fast calculation of projections in tomography

Georg Schramm[1,2] and Fernando Boada[2]

[1]Radiological Science Laboratory, Stanford University, Stanford, US
[2]Department of Imaging and Pathology, KU Leuven, Belgium

**Abstract** In this work, we present a new open source framework, called `parallelproj`, for fast parallel calculation of projections in tomography using multiple CPUs or GPUs. This framework implements forward and back projection functions in sinogram and listmode using Joseph's method, which is also extended for time-of-flight PET projections. In a series of tests related to PET image reconstruction using data from a state-of-the-art clinical PET/CT system, we benchmark the performance of the projectors in non-TOF and TOF, sinogram and listmode. We find that the GPU mode offers acceleration factors between 20 and 60 compared to the multi CPU mode and that OSEM listmode reconstruction of real world PET data sets is possible within a couple of seconds using a single state-of-the-art consumer GPU.

## 1 Introduction

For tomographic imaging techniques used in medicine such as X-ray computed tomography (CT), positron emission tomography (PET) and single photon emission tomography (SPECT), image reconstruction results are usually expected within seconds or minutes after data acquisition, creating a severe computational challenge when reconstructing data from state-of-the-art systems. With new scanner generations, the problem size of this challenge is steadily growing, since (i) the data size is increasing due to higher resolution detectors and scanners with bigger field of view [1], and (ii) more advanced (iterative) reconstruction algorithms are being used that try to exploit more information from the acquired data, which usually necessitates the calculation of a huge amount of projections. An example of the latter is data-driven motion correction in PET [2] where instead of reconstructing a single “static frame ”, many very short time frames are reconstructed and subsequently used for motion estimation and correction. Another example for (ii) is the combination of deep learning and tomographic image reconstruction [3], using, e.g. unrolled networks, where during training also a tremendous number of projections have to be calculated to evaluate the gradient of the data fidelity during training.

For most tomographic image reconstruction algorithms, the evaluation of a linear forward model that describes the physics of the data acquisition process. is the bottleneck in terms of computation time. In CT, PET, and SPECT, the forward model includes the computation of many (weighted) line or volume integrals through an image volume, commonly called “projections” - which can be slow when executed on a single processor. Fortunately, for most reconstruction algorithms, the computation of projections can be executed in parallel on multiple processors, e.g. using multiple CPUs or one or more graphics processing units (GPUs). Note that the evaluation of the adjoint of the forward model - commonly called “back projection“‘ - is computationally more demanding, since race conditions usually occur. In recent decades, the use of GPUs for faster calculation of projections in tomographic imaging has been studied extensively; see, e.g. [4, 5]. All of these articles conclude that the time needed to calculate forward and back projections on state-of-the-art GPUs is usually much shorter compared to using multiple CPUs.

Motivated by these findings and the recent availability of very powerful low- and high-level GPU programming frameworks such as `CUDA` and `cupy`, we developed a new open source research framework, called `parallelproj`, for fast calculations of projections in tomography.

The objectives of the `parallelproj` framework are as follows:

- To provide an open source framework for fast parallel calculation of projections in tomographic imaging using multiple CPUs or GPUs.
- To provide an accessible framework that can be easily installed without the need for compilation of source code on all major operating systems (Linux, Windows, and macOS).
- To provide a framework that can be efficiently used in conjunction with `pytorch` GPU arrays to facilitate research on tomographic imaging methods, including deep learning.

## 2 Materials and Methods

### 2.1 Design principles and implementation details

The application programming interface to the `parallelproj` framework was designed such that:

- The input to the low level projector functions are as generic as possible. In practice, that means that these functions take a list of coordinates representing the start and end point of the rays to be projected as input, making the low-level functions agnostic to specific scanner geometries (or symmetries). Thus, any scanner geometry can be modeled.
- Projections can be performed in non-TOF or TOF mode.
- In the TOF mode, optimized projections for sinogram and listmode are available. In the former, the contributions to all available TOF bins along a ray are computed while traversing the image volume plane by plane, whereas in the latter only the contribution to one specific TOF bin (the TOF bin of a given listmode event) is evaluated.
- The back projections are the exact adjoint of the forward projections (matched forward and back projections).

Parallelization across multiple processors was implemented in two different ways. To enable parallelization across multiple CPUs, a first version of the `parallelproj` library was implemented using C and OpenMP (`libparalleproj_c`). Furthermore, the exact same projector functions were implemented in `CUDA` to enable parallelization on one or multiple GPUs (`libparalleproj_cuda`).

### 2.2 `parallelproj` computation modes

Using the two aforementioned projection libraries, as well as the `CUDA` projections kernels, projections can be performed in the following three different computation modes:

1. **CPU mode:** Forward and back projections of image volumes (arrays) stored on the host (CPU) can be performed using `libparallelproj_c` where parallelization across all available processors is performed using OpenMP.
2. **hybrid CPU/GPU mode:** Forward and back projections of image volumes (arrays) stored on the host canbe performed using `libparallelproj_cuda` involving data transfer from the host to all available GPUs, execution of projection kernels on the GPUs, and transfer of the results back to the host.
3. **direct GPU mode:** Forward and back projections of image volumes (arrays) stored on a GPU can be performed by direct execution of the projection kernels using a framework that supports just-in-time compilation of `CUDA` kernels, such as `cupy`. In contrast to the hybrid CPU/GPU mode, memory transfer between host and GPU is avoided.

### 2.3 Benchmark tests

To evaluate the performance of the `parallelproj` projectors using the three computation modes described above, we implemented a series of benchmarking tests. All tests are related to a PET image reconstruction task and used the geometry and properties of a state-of-the-art GE Discovery MI TOF PET/CT scanner [6] with 20 cm axial FOV. This scanner consists of 36 detector “rings”, where each “ring” has a radius of 380 mm and consists of 34 modules containing 16 detectors each such that there are 19584 detectors in total. A non-TOF emission sinogram for this scanner in span 1 has 415 radial elements, 272 views, and 1292 planes, meaning that for a full non-TOF sinogram projection, 146 million line integrals have to be evaluated. For TOF data, each line of response (LOR) is subdivided into 29 TOF bins using a TOF bin width of 169 ps. The reported TOF resolution of the scanner is 375 ps FWHM [6]. In the TOF projectors of `parallelproj`, the Gaussian TOF kernel is truncated beyond $\pm 3$ standard deviations.
To evaluate the performance of `parallelproj` for projections in sinogram mode, we measured the time needed for a forward and back projection of a span 1 subset sinogram containing 8 equally spaced views in non-TOF and TOF mode. Since it is known that the in-memory data order severely affects the computation time, especially on `CUDA` devices, we varied the order of the spatial axis of the sinogram, as well as the orientation of the image relative to the symmetry axis of the scanner. E.g. in the sinogram order mode “PVR”, the radial direction increased the fastest and the plane direction increased the slowest in memory.
In addition to the sinogram projection tests, we also evaluated the performance of `parallelproj` for non-TOF and TOF projections in listmode as a function of the number of acquired listmode events. All benchmarks were repeated 10 times and the mean and standard deviation of the results were calculated and visualized.
Finally, we also measured the time needed for a complete listmode OSEM iteration using 34 subsets as a function of the number of listmode events in the NEMA acquisition.
All tests used an image of size (215,215,71), an isotropic voxel size of 2.78 mm, and were performed on a workstation including an AMD Ryzen Threadripper PRO 3955WX 16 core 32 thread CPU with 256 GB RAM, and an NVIDIA GeForce RTX 3090 GPU with 24 GB RAM on Ubuntu 20.04 LTS using `CUDA` v11.2 and `parallelproj` v1.2.9.

## 3 Results

Figure 1 summarizes the results of the sinogram and listmode projections benchmarks in non-TOF and TOF mode in direct GPU mode. For sinogram projections, the best results in terms of the summed time needed for the forward and back projection of one subset sinogram were 0.025 s (non-TOF) and 0.189 s (TOF). In both cases, the best results were obtained when using sinogram order mode RVP and image orientation where the scanner symmetry axis corresponded to fastest increasing axis of the 3D image array (green bars). Compared to the results in CPU mode (not shown here because of the page limit), projection times in GPU mode were approximately 48-60 times faster. In listmode, the best summed projection times were 3.91 s (non-TOF) and 0.58 s (TOF) when projecting 40 million events using the GPU mode. For 1.25 million events, the corresponding times were 0.1 s, 0.018 s demonstrating an almost perfect linear relation between the number of events and the projection times. Compared to the CPU mode, TOF listmode projections were approximately 20x faster in GPU mode.
Figure 2 shows the results for the timing of a complete TOF listmode OSEM iteration, including 34 subset updates, as well as a reconstruction of the NEMA image quality phantom data. In GPU mode, the best results for 40 million and 1.25 million events were 0.63 s and 0.065 s), respectively, which was approximately 29 times faster compared to the CPU mode. For a more detailed evaluation and comparison against the hybrid CPU/GPU mode, please see our preprint [7].

## 4 Discussion

All results shown in this work demonstrate once more that parallel computation of forward and back projections using a state-of-the-art GPU is substantially faster compared to parallelization using OpenMP on a state-of-the-art multicore CPU system. Certainly, the achievable GPU acceleration factor strongly depends on the computational problem itself (e.g. sinogram or listmode reconstruction) and the problem size. In our non-TOF and TOF sinogram and listmode benchmark tests, we observed GPU acceleration factors between 20 and 60.
One important aspect that emerged from our sinogram benchmark tests - where the projection data and memory access is ordered - is the fact that the projection times varied substantially when using different memory layouts (up to a factor of 8 in the GPU mode). This can be understood by taking into

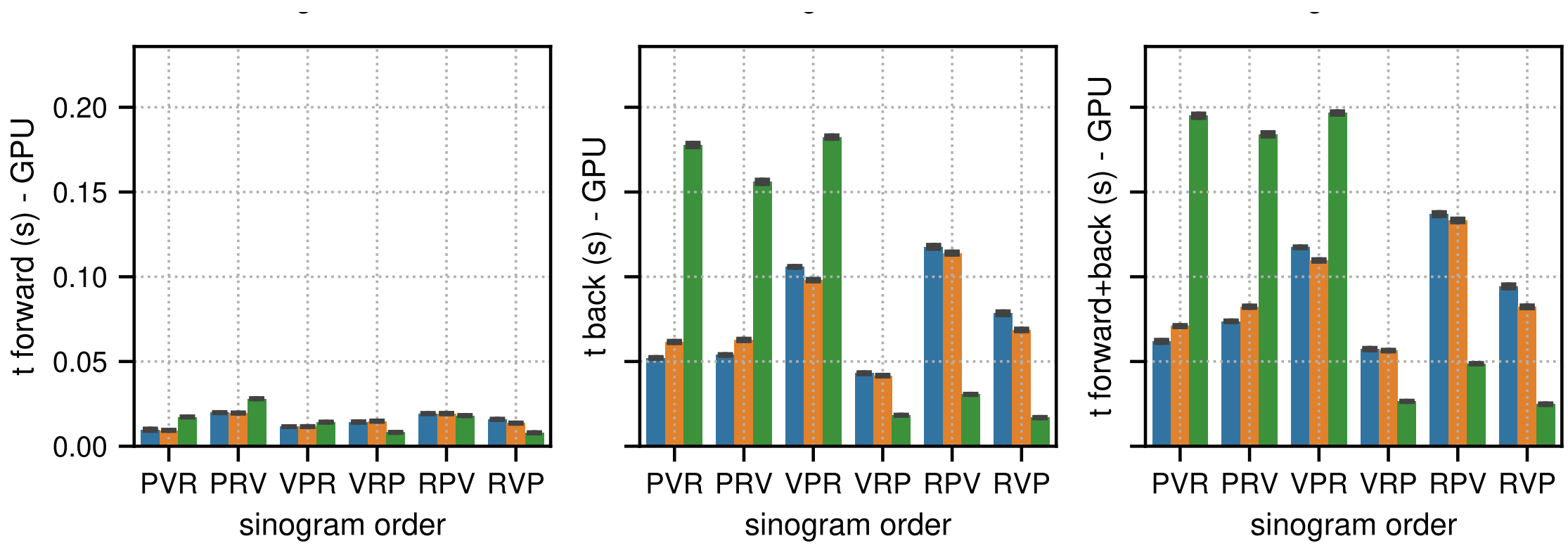

**(a)** Results for non-TOF sinogram projections.

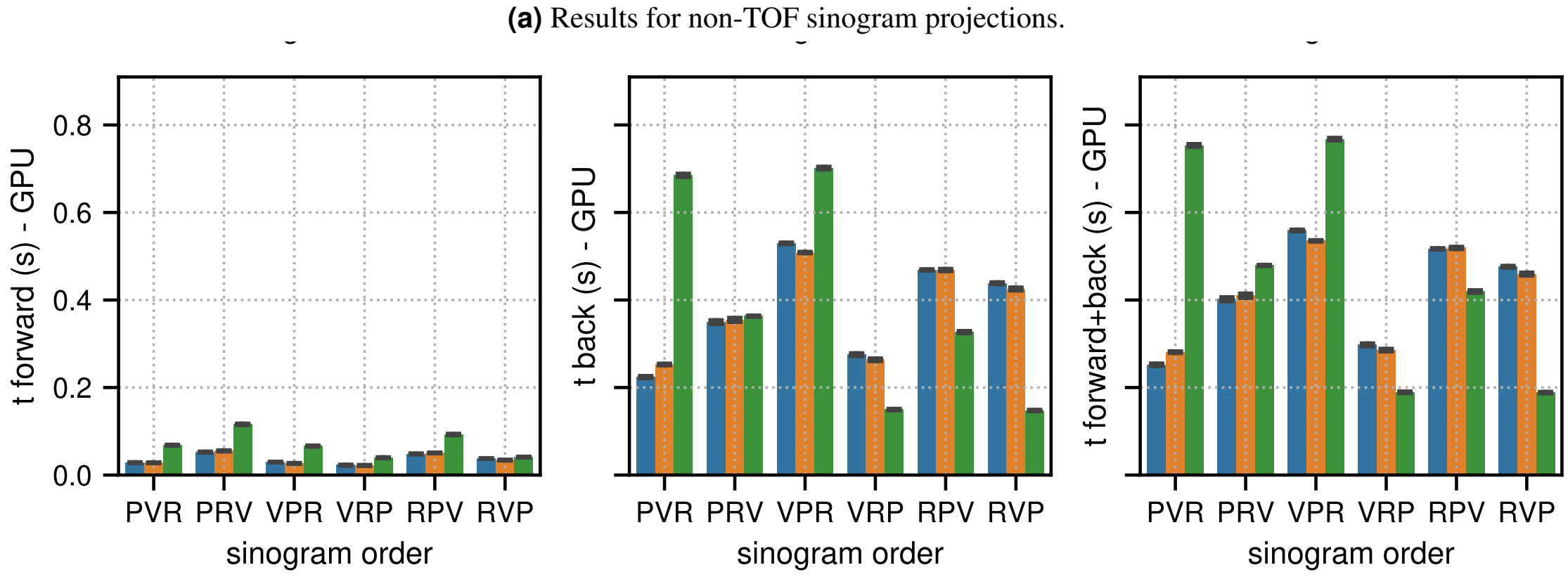

**(b)** Results for TOF sinogram projections.

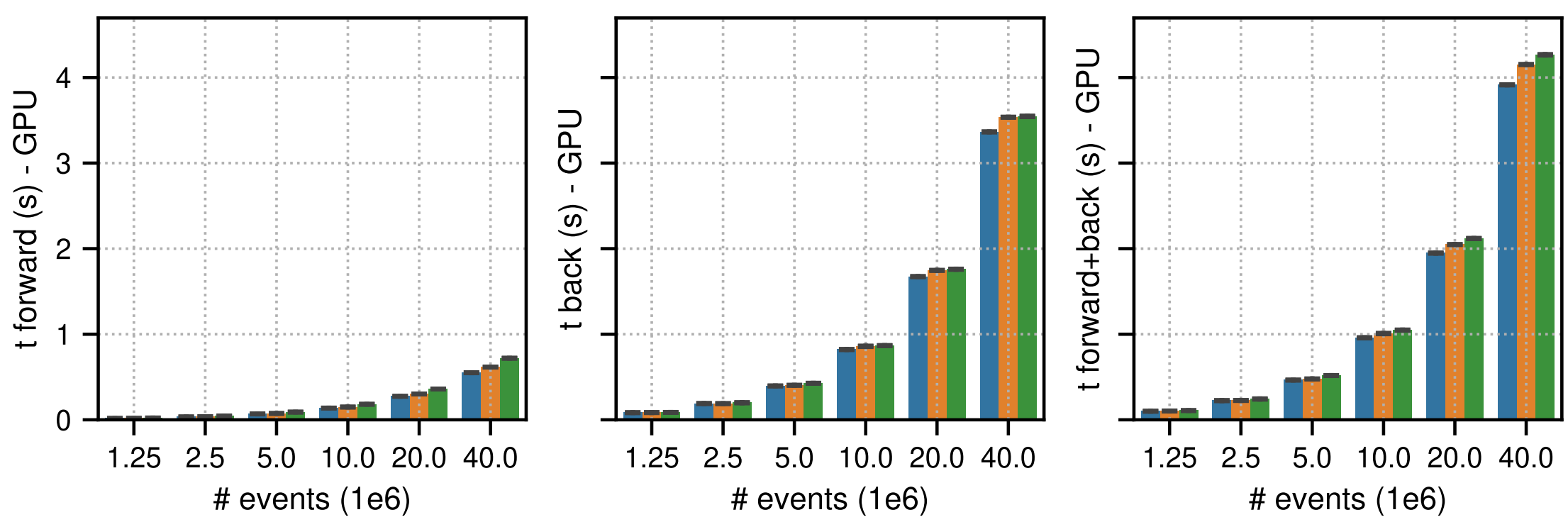

**(c)** Results for non-TOF sinogram listmode projections.

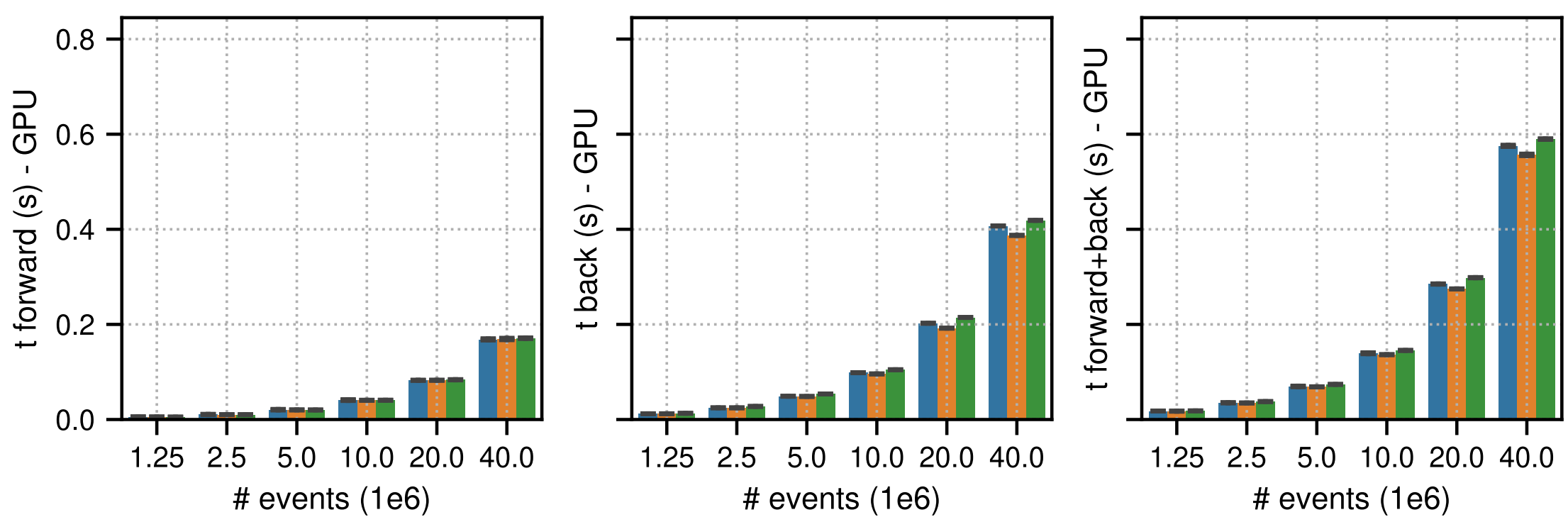

**(d)** Results for TOF sinogram listmode projections.

**Figure 1:** Results of the parallelproj benchmark tests for non-TOF/TOF sinogram/listmode projections in GPU mode. The non-TOF subset sinogram contained 415 radial elements, 8 views and 1292 planes (1 out of 34 subsets). The image used in these tests contained (215,215,71) voxels with an isotropic voxel size of 2.78 mm. The mean and the standard deviation estimated from 10 runs are represented by the colored bars and the black error bars, respectively. Note the different limits on the y axes. The colors represent different orientation of the scanner symmetry axis. (**blue**) image axis corresponding to the symmetry axis of the scanner increases the slowest. (**green**) image axis corresponding to the symmetry axis of the scanner increases the fastest. (**orange**) image axis corresponding to the symmetry axis is the "central" axis.

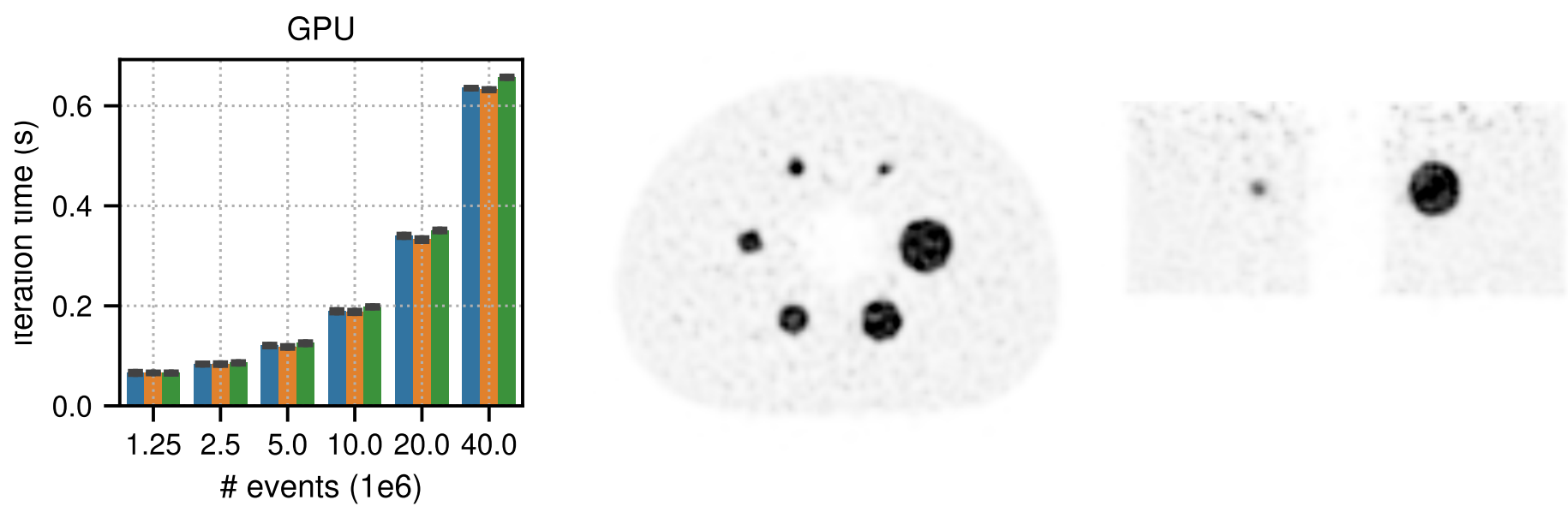

**Figure 2:** (left) Results for the timing of a complete LM OSEM iteration including 34 subset updates for the NEMA image quality phantom acquisition. The image used in these tests contained (215,215,71) voxels with an isotropic voxel size of 2.78 mm. The mean and the standard deviation estimated from 6 iterations are represented by the colored bars and the black error bars, respectively. (right) Transaxial and coronal slice of a listmode OSEM reconstruction of the NEMA image quality phantom with 40 million events after 6 iterations with 34 subsets using a standard Gaussian post filter of 4 mm FWHM. Note that for better visibility, the reconstructed image was cropped to the center portion of the transaxial FOV. See Fig. 1 for the explanations of the bar colors.

account that the amount of race conditions that are created during the back projection within a thread block heavily depends on the order and possible intersections of rays to be projected within that block. Note that in pure GPU mode, the time needed for sinogram forward projections also varied substantially across the different memory layouts, which is probably due to the way image memory is accessed and cached on `CUDA` GPUs.

Another interesting observation is the fact that the time needed to calculate TOF sinogram projections was much longer than the times needed to calculate non-TOF sinogram projections, whereas the situation was reversed in listmode. For TOF sinogram projections, more floating point operations have to be computed compared to non-TOF sinogram projections due to the evaluations of the TOF kernels between the contributing voxels and a number of TOF bins. In listmode, however, the computational work needed to project a TOF event is much lower compared to projecting a non-TOF event. This is the case because a TOF listmode event detected in a specific TOF bin is only affected by a few voxels along the complete LOR in the image, where the number of affected voxels is inversely proportional to the TOF resolution of the scanner. That in turn means that with scanner TOF resolutions becoming better and better, the gap between the TOF projection times in sinogram and listmode will become bigger and bigger, strongly favoring listmode processing. Extrapolating the timing results of one complete OSEM listmode iteration of an acquisition with 40 million counts in Figure 2, clinical listmode OSEM reconstructions of a single bed position of a standard static FDG whole-body acquisition using a PET scanner with 20-25 cm axial FOV seem to be possible in a couple of seconds and could even be faster than the acquisition time.

An important limitation of our work is the fact that we only implemented and benchmarked Joseph's projection method. Compared to other methods such as the distance-driven method, multiray models, or tube-of-response models, Joseph's method is inherently faster. Consequently, projection times are expected to be somewhat slower for more advanced projectors, but a detailed investigation of more advanced projectors is beyond the scope of this work and left for future research. [1] Note, however, that according to our experience, combining Joseph's method with an image-based and / or sinogram-based resolution model produces high-quality PET reconstructions.

Without a doubt, it is possible to further optimize the implementation of the `parallelproj` projectors, especially the `CUDA` implementation. As an example, we have decided not to use `CUDA`'s texture memory, which could substantially accelerate the image interpolations needed in the Joseph forward projections, or be also used to interpolate TOF kernel values based on a 1D lookup table. The main reason for not using texture memory is the fact that it would only accelerate the forward projections since writing into texture memory is not possible and because reconstruction times are usually dominated by the back projections. Another way to further improve the listmode projection times is to pre-sort the listmode events to minimize race conditions during back projection, as e.g. shown in [8].

[1] Since `parallelproj` is an open-source project, contributions of or discussions on more advanced projectors from the reconstruction community are more than welcome.

# Analysis of detectability index of infarct models in spectral dual-layer CBCT

Dirk Schäfer[1], Fredrik Ståhl[2,3], Artur Omar[4,5], and Gavin Poludniowski[4,6]

[1]Medical Image Acquisition, Philips Research, Hamburg, Germany
[2] Department of Neuroradiology, Karolinska University Hospital, Stockholm, Sweden
[3]Department of Clinical Neuroscience, Karolinska Institute, Stockholm, Sweden
[4]Medical Radiation Physics and Nuclear Medicine, Karolinska University Hospital, Stockholm, Sweden
[5]Department of Oncology-Pathology, Karolinska Institute, Stockholm, Sweden
[6]Department of Clinical Science, Intervention and Technology, Karolinska Institute, Huddinge, Sweden

**Abstract**
Dual-energy CBCT C-arm imaging is a new imaging modality, which may provide a variety of different spectral results in the interventional suite. We analyze model observers, i.e. the detectability index, of different infarct models, to get insight into which spectral reconstruction may be best suited for identifying infarcts prior to an intervention treating acute stroke patients. MTF and NPS measurements have been performed on a clinical dual-layer (DL) CBCT prototype system and used for model observer analysis. Infarct models with reconstruction dependent contrast amplitude are investigated using a variety of standard and spectral reconstructions. Clinical examples of subjects with small and large infarcts imaged on the same DL-CBCT system are compared visually and be means of CNR analysis of the infarcts to the results of the model observers. The model observer representing most closely a human reader prefers VMI images at higher keV values, and the same trend is obtained with simple CNR analysis. This result is partially substantiated by the CNR analysis of the clinical patient data.

## 1 Introduction

Spectral dual-layer C-arm CBCT is a new imaging modality and may be used during stroke treatment workflow to improve visualization of infarct regions in the beginning of a procedure and support blood-iodine separation for hemorrhage identification at the end of the procedure. Presenting the most relevant information to the interventionalist is a challenging task. We aim to get insight into which spectral reconstruction is best suited to support a neuroradiologist in the task of identifying infarcts in CBCT reconstructions in acute stroke treatment. Reader studies based on dual-layer, dual-energy CT have shown optimal visualization of infarcts for Virtual Monoenergetic Image (VMI) reconstructions around 70 keV [1]. Availability of clinical data for dual-energy CBCT of acute stroke patients is limited, therefore we investigate the possibility to use model observers or measures such as CNR to support the choice of most suited spectral reconstructions for the infarct identification task.

## 2 Materials and Methods

Measurements in this study were made with a non-commercial dual-layer CBCT prototype system [2]. The spectral DL-CBCT system is a modification of a commercial interventional C-arm X-ray system with CBCT imaging capability (Allura Xper FD20/15, Philips Healthcare, The Netherlands), equipped with a dual-layer 20 inch (379.4 mm x 293.2 mm) detector prototype. The detector prototype consists of two detector layers stacked on top of each other.

The prototype DL-CBCT system scans were conducted with 120 kV tube voltage, 2.5 mm aluminum-equivalent inherent tube filtering as well as 1 mm aluminum and 0.4 mm added copper filtering. A fixed 310 mA tube current was used for 622 projections (200◦ rotational scan).

Both, standard and spectral images were generated from the DL-CBCT scans. The reconstruction of the front-layer (front) corresponds to the standard reconstruction of single layer C-arm system. The addition of front- and back-layer data yields the combined reconstruction (comb). The following spectral results were derived from material-decomposed Compton scatter and photoelectric base functions [3]: Virtual monoenergetic image (VMI) and virtual non-contrast VNC images. Noise suppression in spectral images was realized by exploiting the anti-correlated noise behavior after material decomposition [4]. The denoising strength was selected by a clinical reader pilot study to yield best visualization of brain tissues for assessment of ischemic changes in stroke patients [5].

The modulation transfer function (MTF) was calculated from six scans of the head module of the multi-energy CT phantom model 662 (CIRS, Norfolk, VA, USA) using an established methodology [6], utilizing different inserts equivalent to various tissues, blood and iodine concentrations. Examples of the radially averaged MTFs for iodine blood inserts are shown in Fig. 1. Since they do not deviate appreciably, the low-contrast spatial resolution was assumed constant for a given reconstruction and the MTF curves for the 2 mg/ml I in blood was used subsequently.

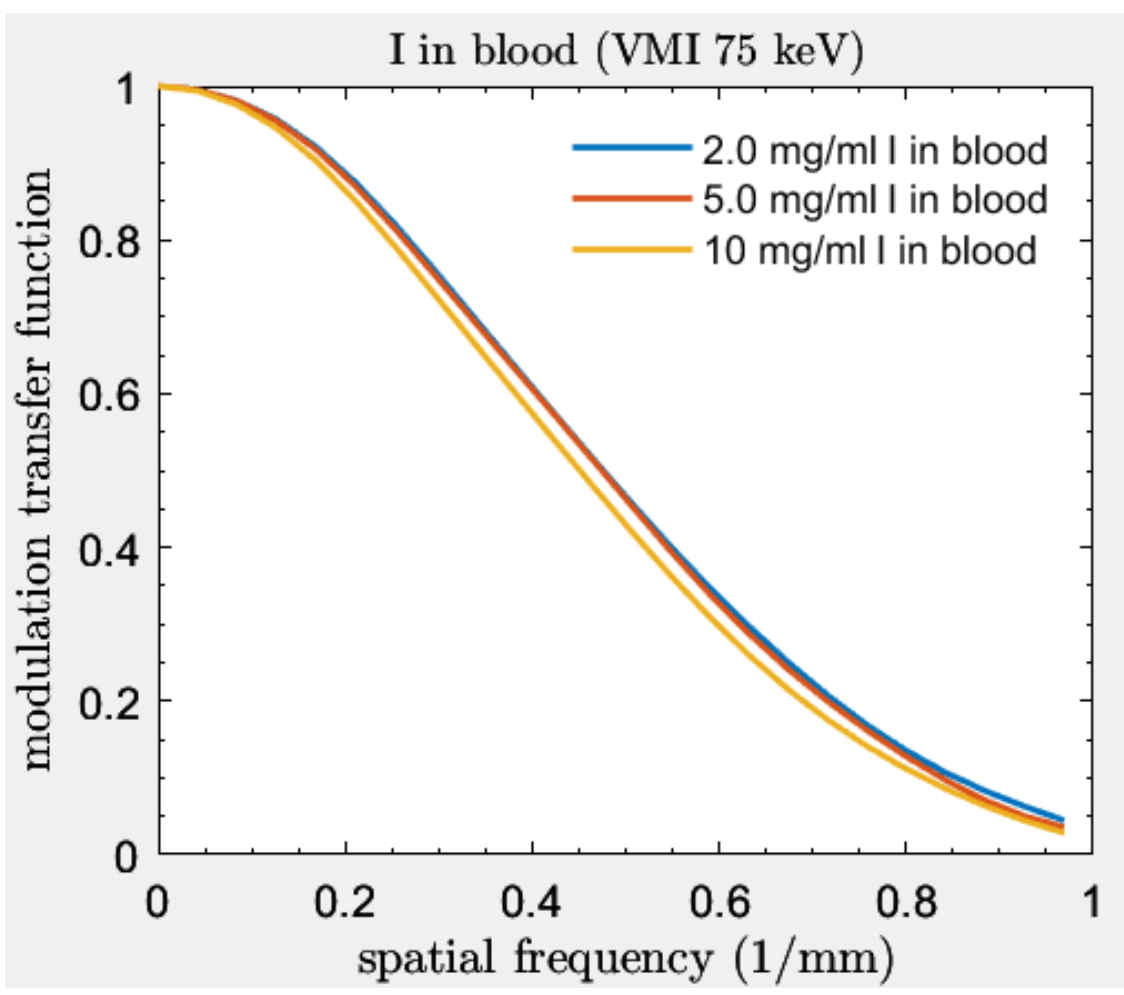

**Figure 1.** Radially averaged task-related modulation transfer functions (MTF) for Iodine blood inserts for VMI 75 keV.

The 3D noise power spectrum (NPS) was calculated from six consecutive scans made of the homogenous water phantom for all reconstructions investigated in this study, using well-established techniques [7]. The 2D NPS was then calculated. A 2D NPS of the 75keV VMI reconstructions is shown in Fig. 2.

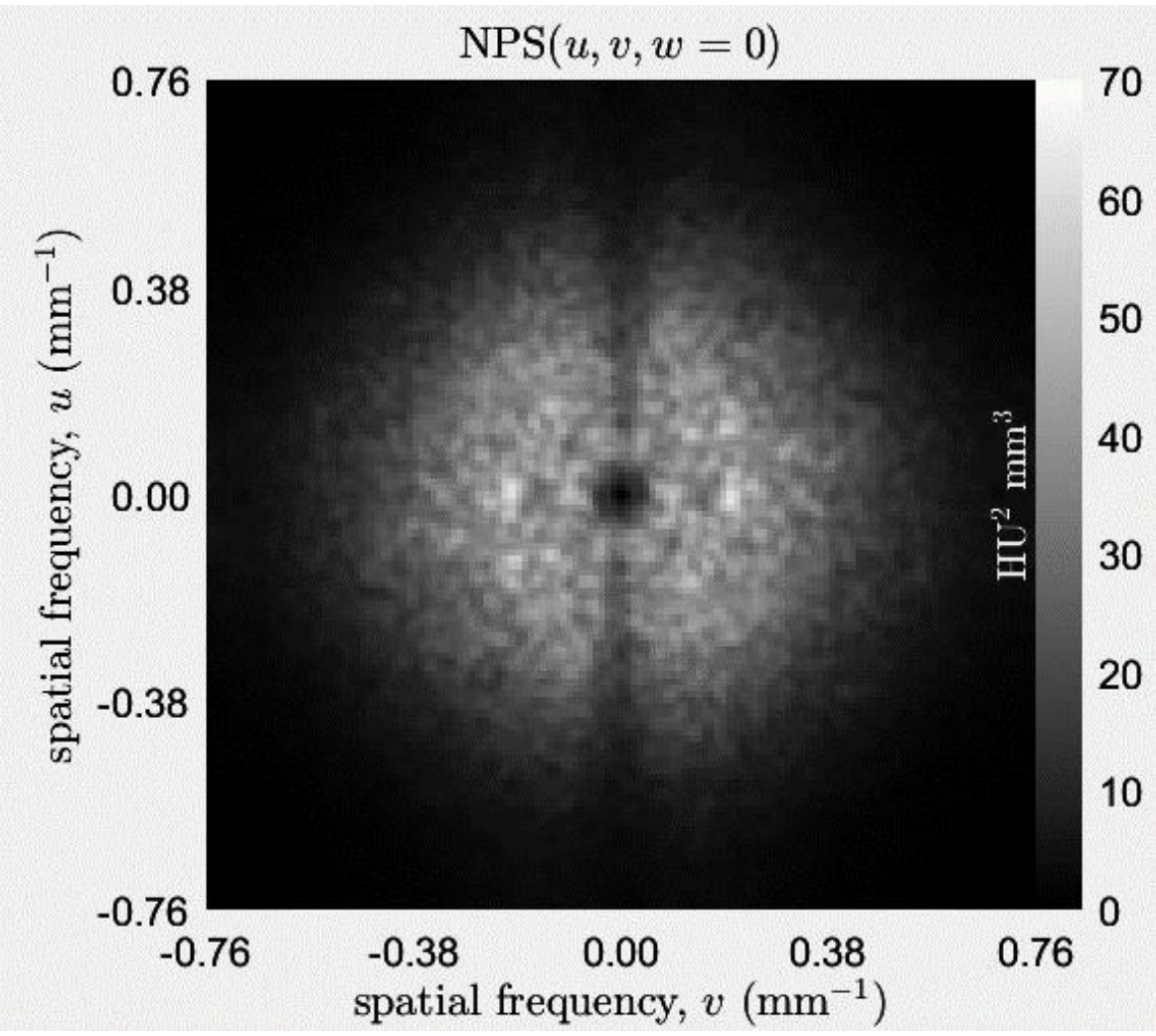

**Figure 2.** A 2D noise power spectrum corresponding to the trans-axial plane for VMI 75 keV.

The model observers are evaluated on 2D trans-axial slices, corresponding to the clinical practice of reviewing reconstructions in acute stroke treatment. The trans-axial NPS is radially averaged, as shown in Fig. 3 for selected reconstructions, and used as $\mathrm{NPS}(u,v)$ on a resampled cartesian grid in the following.

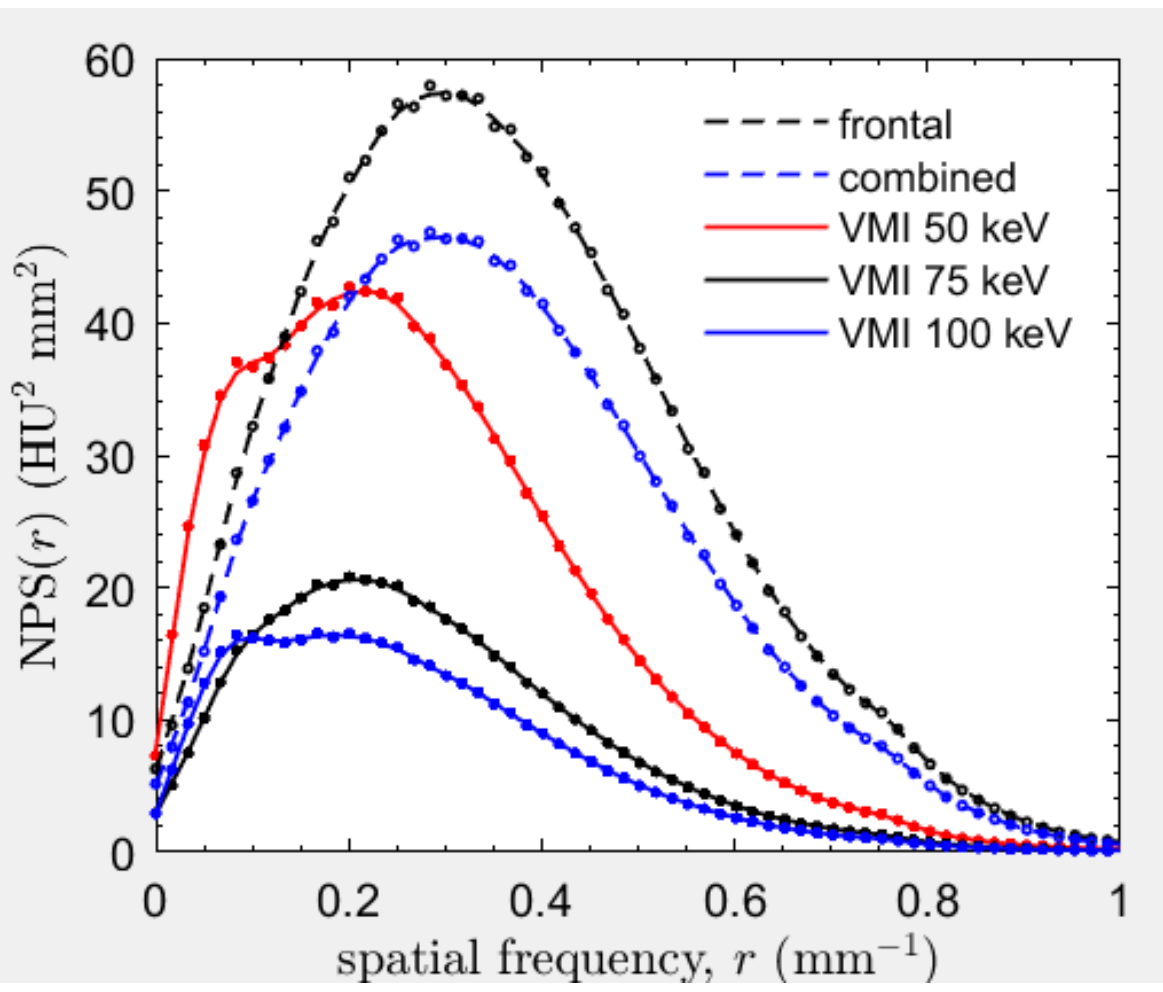

**Figure 3.** Radially averaged trans-axial 2D noise power spectrum for selected reconstructions

A general expression for the detectability index (d') in terms of the squared signal-to-noise ratio of a matched-filter model observer is given by [8]:

$$d'^2 = \frac{[\int S_O(u,v)M(u,v)dudv]^2}{\int \mathrm{NPS}_O(u,v)M(u,v)^2 dudv},$$

where $S_O(u,v)$ is the expected two-dimensional (2D) signal as perceived by the observer and $\mathrm{NPS}_O(u,v)$ is the perceived noise power spectrum. The quantity $M(u,v)$ is the "matched filter". Representations are in frequency space and assumed to be real functions.

The signal can be represented as [9]:

$$S_O(u,v) = T(u,v)\mathrm{MTF}(u,v)\mathrm{VTF}(u,v),$$

where $T(u,v)$ is the 2D Fourier transform of the contrast profile $\mathrm{C}(x,y)$ of the feature of interest, the MTF is the spatial resolution for the low-contrast task, and VTF is an assumed the visual transfer function (VTF) for a human observer, otherwise known as an "eye filter".

Different modelled contrast profiles $\mathrm{C}(x,y)$ have been investigated representing infarct regions of different size (5, 10 and 15 mm). The contrast amplitude was estimated based on different assumed percentages of water content mixed with gray matter brain tissue. Calculations of contrast were based on observed average CT numbers for water and gray matter substitutes for the different reconstructions. The resulting contrast and CNR values for standard and spectral reconstructions are shown for an infarct of 5 mm diameter with 5% water content in Fig 4.

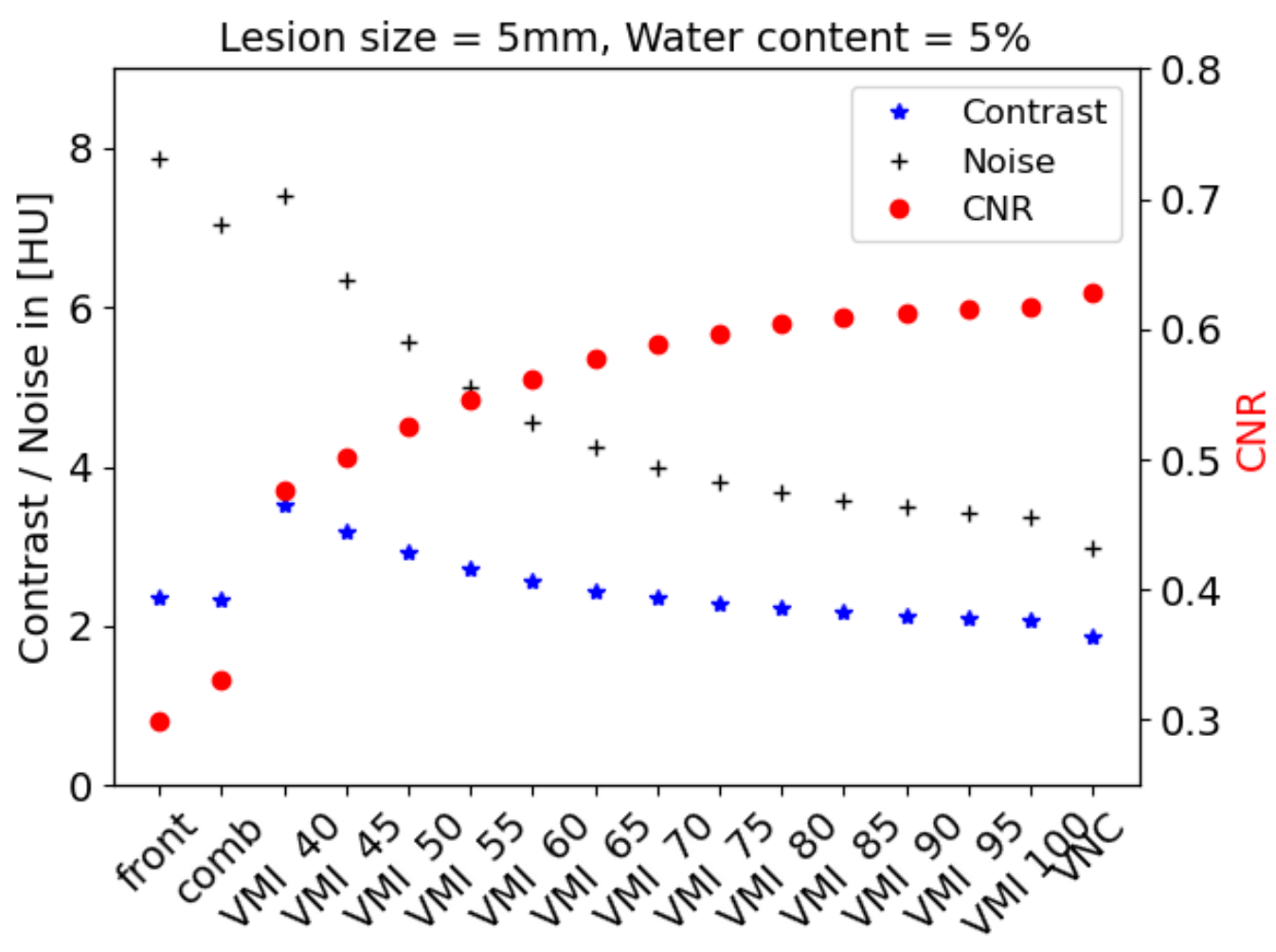

**Figure 4.** Modeled contrast, noise derived from NPS and resulting CNR used for model observers.

In this study, the VTF due to Eckstein [10] is used with a peak sensitivity at approximately 4 cycles/degree:

$$\mathrm{VTF}(u,v) = [\rho^{1.5}\exp(-0.98\rho^{0.68})]^2$$

with

$$\rho(u,v) = \sqrt{u^2+v^2}\frac{FOV\ R}{D}\frac{\pi}{180},$$

where is the reconstruction field-of-view, R is the viewing distance (assumed to be 50 cm) and D is the display size of the FOV at a viewing station (assumed to be 25 cm).

The (observed) noise power spectrum can be represented as,

$$\mathrm{NPS}_O(u,v) = \mathrm{NPS}(u,v)\mathrm{VTF}^2(u,v).$$

In this study, the matched filter will either be prewhitening or non-prewhitening [11]:

$$M(u,v) = S_O(u,v)/\mathrm{NPS}_O(u,v)$$

or

$$M(u, v) = S_O(u, v),$$

respectively. Note, that in the first case, the observer is able to decorrelate the perceived noise. This leads to simplifications of the d' formula, with the eye filter cancelling and the numerator being equal to the square of the denominator.

Three model observers will be explored:

1. PW: prewhitening matched filter. This corresponds to an ideal linear observer.
2. NPW: non-prewhitening filter. This corresponds to an observer unable to undo noise correlations.
3. NPWE: non-prewhitening filter with an eye filter. This corresponds to the NPW observer with the sensitivity of the human eye incorporated.

Moreover, visibility of infarct regions in clinical examples of spectral DL-CBCT is assessed visually and by means of CNR analysis. The patient data has been acquired in a prospective single center clinical trial (NCT04571099), which enrolled consecutive patients, 50 years or older, with ischemic or hemorrhagic stroke on initial CT [5].

## 3 Results

The detectability index d' for the modelled infarct with 5 mm diameter and 5% water content is shown for the standard and spectral reconstructions in Fig 5. The pre-whitening (pw) model observer has slightly higher d'-values compared to the non-pw (npw) observer, but the relative trend is very similar with a maximum around VMI70. The d' results of the npw observer with eye filter (npwe) are sgnificantly reduced and the trend for different VMI is more flat with the maximum of shifted towards higher VMI. The d' results for a 5 mm infarct with water content of 10% and 15% simply scale up by a factor of 2 and 3, respectively (not shown). The results for 5% water content with lesion diameter of 10 mm and 15 mm are generally higher by a factor of approximately 2.5 and 4, respectively (not shown), but similar in trend with more pronounced maxima at VMI70 for pw and npw observer and similar flat npwe–results slighlty increasing to higher energy levels of the VMI.

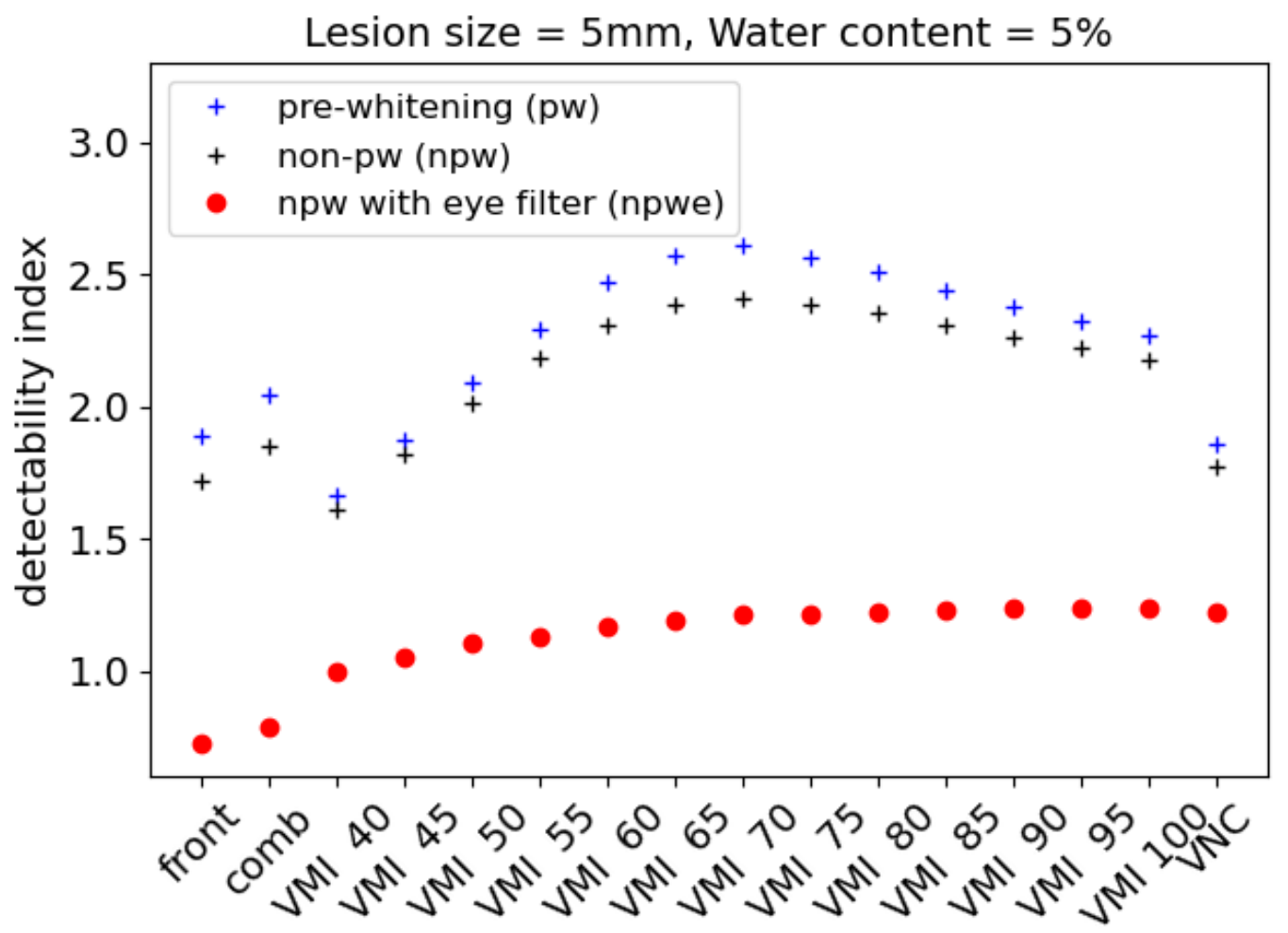

**Figure 5.** Results of different model observers for lesion size of 5 mm and water content of 5%.

Clinical examples of a small and a large infarct are shown in Fig.6. The ROIs used to quantify the CNR of the infarct region with respect to gray matter brain tissue are indicated on the front layer reconstruction. The CNR values for both infarcts are plotted in Fig. 7.

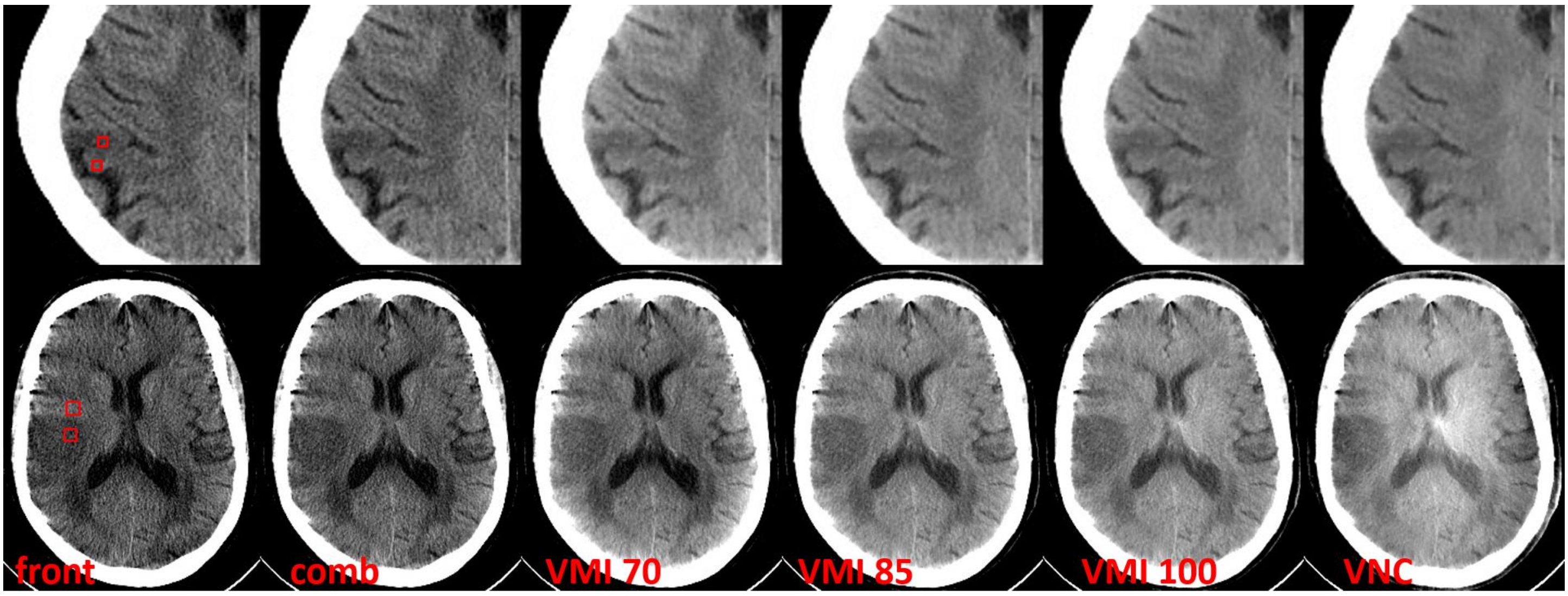

**Fig. 6.** Example of subject with a hyperacute, small infarct region (top row) and an acute large infarct region (bottom row). ROIs for measuring CNR are indicated in the front-layer reconstructions (left) and values reported in Fig. 7. All reconstructions are displayed with level / window of 25 / 70 HU and slice thickness of 5 mm.

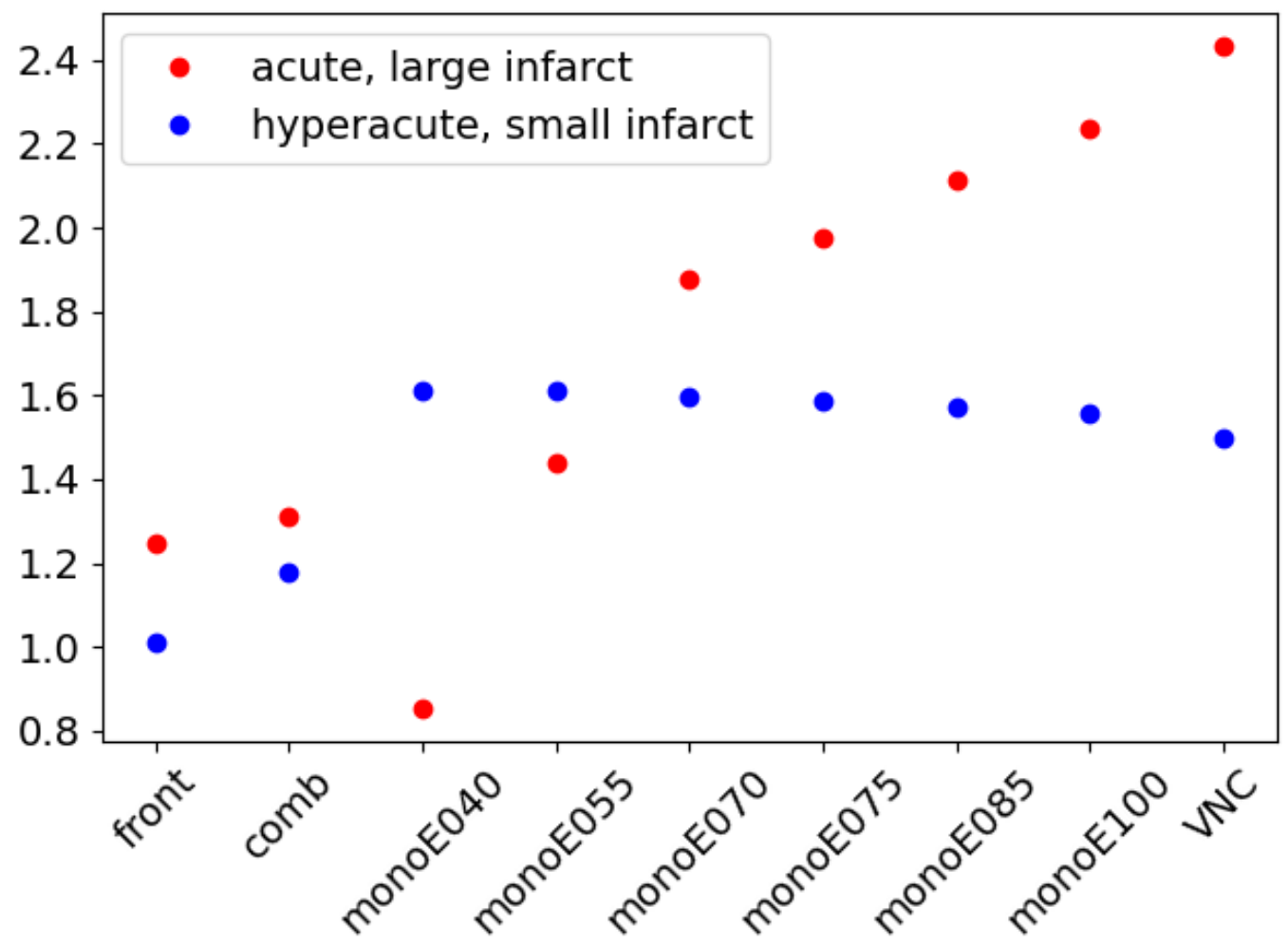

**Fig. 7.** CNR of the infarct/GM ROIs shown in Fig. 6

## 4 Discussion

There is a general observation that VMI and VNC reconstructions provide better CNR and detectability index for the task of infarct visualization compared to front-layer and combined reconstructions (see Figs. 4, 5). The CNR values of the clinical example (Fig. 7) show a similar behavior for energy levels > 50 keV. This is partly caused by the additional denoising involved in the spectral reconstructions and proper separation of these effects is not trivial and beyond the scope of this paper.

Interestingly, the optimal energy level of the VMI reconstruction for this task depends on the model observer. The non-prewhitening filter with an eye filter (npwe) is expected to match best to the performance of a human observer. The Npwe observer (Fig.5) and CNR (Fig.4) indicate higher mono-energy levels as optimal.

Optimal CNR values for the clinical example with large infarct are also obtained for higher energy levels of the VMI, whereas the small infarct shows relatively flat dependence. The different CNR trends for the different clinical cases indicate that the composition of infarcts merits further attention as well as any other potential confounding factors.

These investigations aim to identify the optimal spectral reconstruction to be presented for the task of infarct detection, which shall be cross checked in future work with reader studies on clinical data. The detectability index can be related to the area-under-the-curve (AUC) from receiver operating characteristics (ROC) studies or the percentage correct (PC) responses in multiple alternative forced choice (AFC) studies: $\mathrm{AUC} = \mathrm{PC(2\text{-}AFC)} = \Phi\left(d'/\sqrt{2}\right)$, where $\Phi(z)$ is the standard cumulative normal distribution. The inclusion of so-called *internal noise* of the observer into the model may be necessary to quantitatively predict human performance [12].

Potential limitations of the approach include the assumption of quasi-linearity (permitting image analysis in the Fourier domain) and the assumption that the clinical task can be approximated as *a signal known exactly/background known exactly* scenario.

## 5 Conclusion

This study presents the use of model observers to investigate different spectral reconstructions for infarct visualization in spectral DL-CBCT. The model observer representing most closely a human reader prefers VMI images at higher keV values, and the same trend is obtained with simple CNR analysis. This result is substantiated by the CNR analysis of a large infarct, while the clinical example of a small infarct showed less dependence on the energy level of the VMI reconstruction. These results have to be verified by clinical reader studies, but this framework may help to investigate and optimize system performance for different tasks with limited availability of clinical data.

# Real-time Liver Tumor Localization via Combined Optical Surface Imaging and Angle-agnostic X-ray Imaging

Hua-Chieh Shao, Yunxiang Li, Jing Wang, Steve Jiang, and You Zhang

Department of Radiation Oncology, University of Texas Southwestern Medical Center, Dallas, TX 75390, USA

**Abstract** The success of radiotherapy critically relies on the knowledge of tumor location for accurate dose delivery. However, intra-treatment motion, especially respiratory motion, introduces tumor localization uncertainties and decrease treatment accuracy. In image-guided radiotherapy, real-time imaging can capture patients' anatomy and motion for online target localization and treatment adaptation. However, real-time imaging is challenged by the short allowable imaging time (<500 milliseconds) to meet the temporal constraint posted by rapid patient breathing, resulting in extreme under-sampling for desired 3D imaging. In the case of livers, the diminished tumor/normal-liver-tissue x-ray imaging contrast adds another layer of difficulty to accurately localize the tumors. We addressed both challenges by developing a deep learning (DL)-based, deformable registration-driven framework to track the volumetric motion of livers and localize the tumors in real-time, by combining the information from optical body surface imaging and an x-ray projection acquired at an arbitrary gantry angle (angle-agnostic). The liver tumor localization was achieved via two steps. First, the liver boundary motion was estimated through a patient-specific surface imaging model (*Surf*), and fine-tuned using a cascaded, angle-agnostic x-ray imaging model (*X360*). Second, the volumetric motion within the liver was solved to localize the tumor via a population-based biomechanical model (*Bio*), using the liver boundary motion solved in step 1 as the boundary condition. The results show that the cascaded framework, *Surf-X360-Bio*, can achieve fast (<250 ms inference time), accurate (mean error < 2.1 mm), and robust liver tumor localizations at arbitrary x-ray angles for real-time, markerless motion tracking.

## 1. Introduction

The efficacy and safety of radiotherapy are heavily dependent on the accuracy of tumor localization to deliver conformal radiation doses and spare surrounding healthy tissues. Intra-treatment motion, particularly the respiratory motion, introduces tumor localization uncertainties for sites including the liver [1], and reduces the accuracy of radiotherapy. To deliver planned dose accurately, real-time imaging is needed for intra-treatment motion management, to provide instantaneous knowledge of the tumor motion such that the radiation beam can be adjusted simultaneously to follow the tumor [2]. However, a major challenge to real-time imaging is the time constraint. To address respiratory motion, the overall latency of real-time image guidance and simultaneous delivery adjustment should be < 500 milliseconds (ms) [3]. In such a short frame, only a single or a few onboard 2D x-ray projections can be acquired, which makes it almost impossible to use conventional image reconstruction/registration methods to track the 3D volumetric motion and deformation of livers and to localize the tumors within the liver. Furthermore, the low x-ray imaging contrast of liver tumors against their surrounding normal tissues renders 3D tumor localization even more challenging.

With recent advances in deep learning (DL)-based medical image processing, DL methods were developed to localize tumors in 3D from a single or a few onboard x-ray projections, via two general approaches: (1) image reconstruction; and (2) image registration. The reconstruction approach aims to directly reconstruct volumetric images or 3D organ shapes from one or two orthogonal x-ray projections [4, 5], which is very ill-conditioned and susceptible to multiple challenges including model robustness and generalizability. In addition, the reconstruction-based methods require further segmentation/registration to localize the tumor, which is time-consuming and error-prone for a low-contrast site like the liver. Alternatively, the registration-based approach registers previously-acquired CTs/CBCTs to new 2D x-ray projections (2D-3D registration), by correlating 3D deformable motion to image features extracted from a limited number of 2D projections. By propagating tumors contoured on prior CTs/CBCTs, the registration-based methods can directly localize the onboard tumors via solved motion fields [6, 7]. Using prior information, the registration-based methods are better conditioned and potentially more robust. Our recent study has successfully registered and localized the liver tumor to a mean error of 1.2 (±1.2) mm via a single x-ray projection, by using a graph neural network-driven approach [7]. However, one remaining challenge of the registration-based models, including our model, is that each model only works for a specific, fixed x-ray projection angle. Training separate models for all potential scan angles can be extremely time-consuming and challenging for practical applications [8], especially for rotational radiotherapy delivery techniques where the imaging gantry is continuously sweeping. Training an angle-agnostic model, however, is quite difficult due to the complex, angle-varying cone-beam projection geometry. The limited field-of-view (FOV) of the x-ray imaging systems, and the in-and-out motion of anatomies with respect to the FOV (due to gantry rotation and anatomical motion), also make it quite challenging to derive a robust, angle-agnostic model.

To address these challenges, we proposed a DL-based, deformable registration-driven framework to localize liver tumors at arbitrary x-ray projection angles. On the foundation of our previous study, we introduced two new ingredients to help train a robust, angle-agnostic model: (1). we introduced optical body surface imaging, which is widely available in radiotherapy, to continuously monitor patient surfaces under a high frame rate (10-24 Hz) with a large FOV (up to 110×140×240 $cm^3$) [9]. Optical imaging provides a stable view of the patient and offers complementary information to x-ray imaging, helping to pre-condition the x-ray model under varying x-ray angles and a limited FOV; (2). We developed a geometry-informed scheme for the x-ray model, by introducing network-inherent and angle-aware projection geometries during both training and testing stages to extract the most relevant x-ray projection features to solve the liver motion. In general, the motion of the liver and the liver tumor was solved via two steps: a) the liver boundary motion was solved by cascading a surface imaging model (*Surf*) and an angle-agnostic x-ray imaging model (*X360*), based on the input of a surface image and an x-ray projection acquired at an arbitrary scan angle (0°-360°); b) Using the liver boundary motion solved in (a), a DL-based liver biomechanical model (*Bio*) was applied to solve the intra-liver deformation for tumor localization. Combining both surface and x-ray imaging modalities helps to maximize the utilization of real-time signals to solve the liver boundary motion. Using the liver boundary motion as the boundary condition, the DL-based biomechanical model uses domain knowledge of tissue biomechanics and finite element analysis (FEA) to solve the motion of low-contrast intra-liver regions with much faster speed than conventional FEA methods to meet the real-time constraint (<250 ms inference time for the *Surf-X360-Bio* framework).

## 2. Materials and Methods

### *2.1 Overview of the Surf-X360-Bio framework*

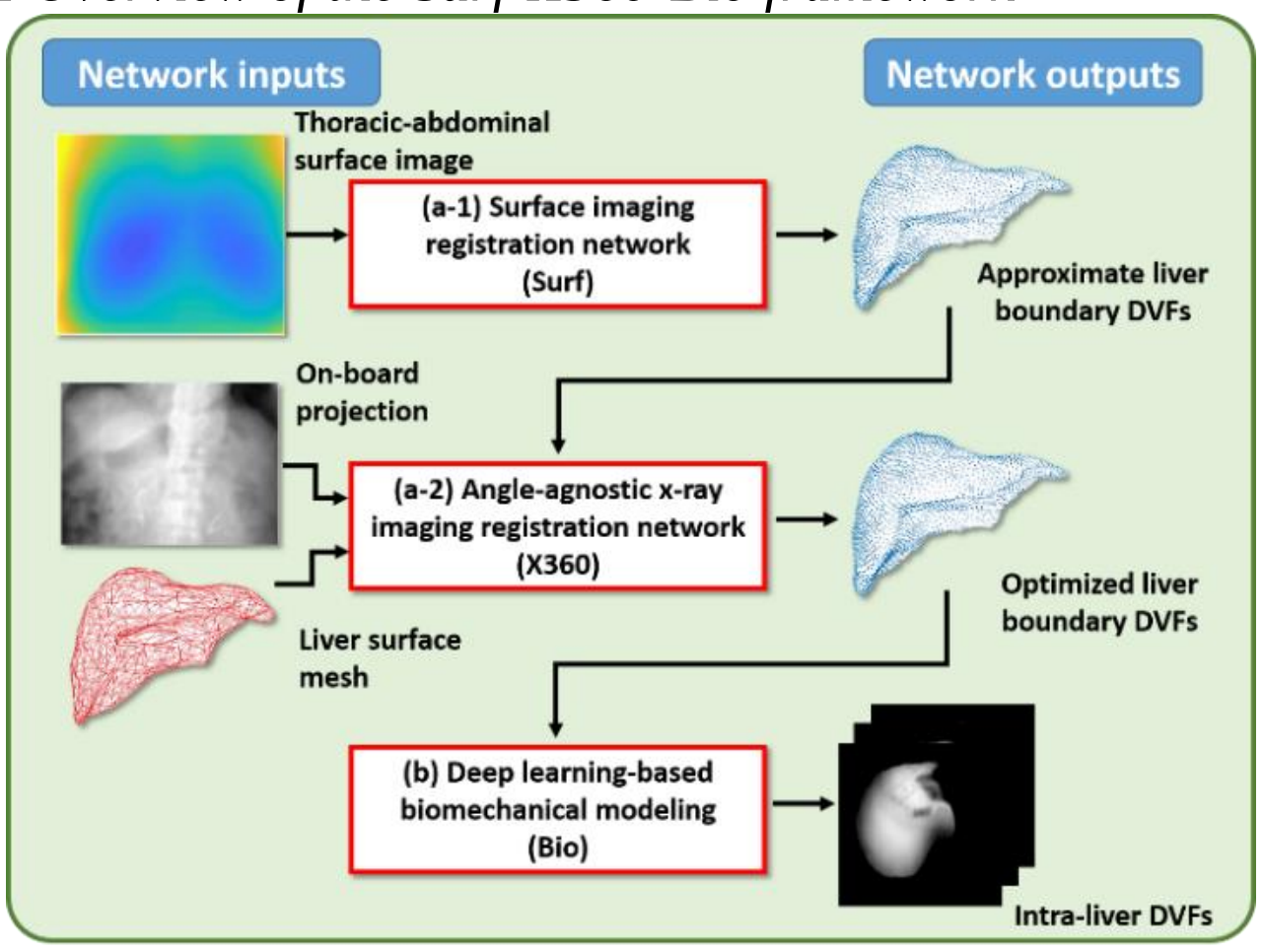

**Figure 1.** Workflow of the *Surf-X360-Bio* framework for real-time liver tumor localization. Liver tumors were localized via a deformable registration-driven approach. In sub-steps (a-1) and (a-2), liver boundary deformation vector fields (DVFs) were estimated and fine-tuned by cascaded registration networks using a surface image and an x-ray projection as inputs. Afterward, step (b) uses a DL-based biomechanical model to propagate the liver boundary DVFs into the liver to localize the liver tumors volumetrically.

Figure 1 illustrates the workflow of the *Surf-X360-Bio* framework. The 3D liver tumor was localized in real-time by three cascaded, DL-based models. The liver was converted to a mesh based on segmentations from prior CTs/CBCTs, and its deformable motion (including that of the tumor) was represented via deformation vector fields (DVFs) of the mesh nodes. The first model (*Surf*) estimated the DVFs of the liver boundary nodes using a thoracic-abdominal body surface image. However, since the internal-external motion correlation can be imperfect, the estimated liver boundary DVFs may need further correction. Therefore, the second model (*X360*) used an onboard x-ray projection acquired at an arbitrary projection angle to further fine-tune the solved DVFs of the liver surface nodes. Finally, intra-liver DVFs were solved by an efficient DL-based liver biomechanical model to propagate the boundary DVFs into the liver to localize the tumor volumetrically.

### *2.2 Surf model*

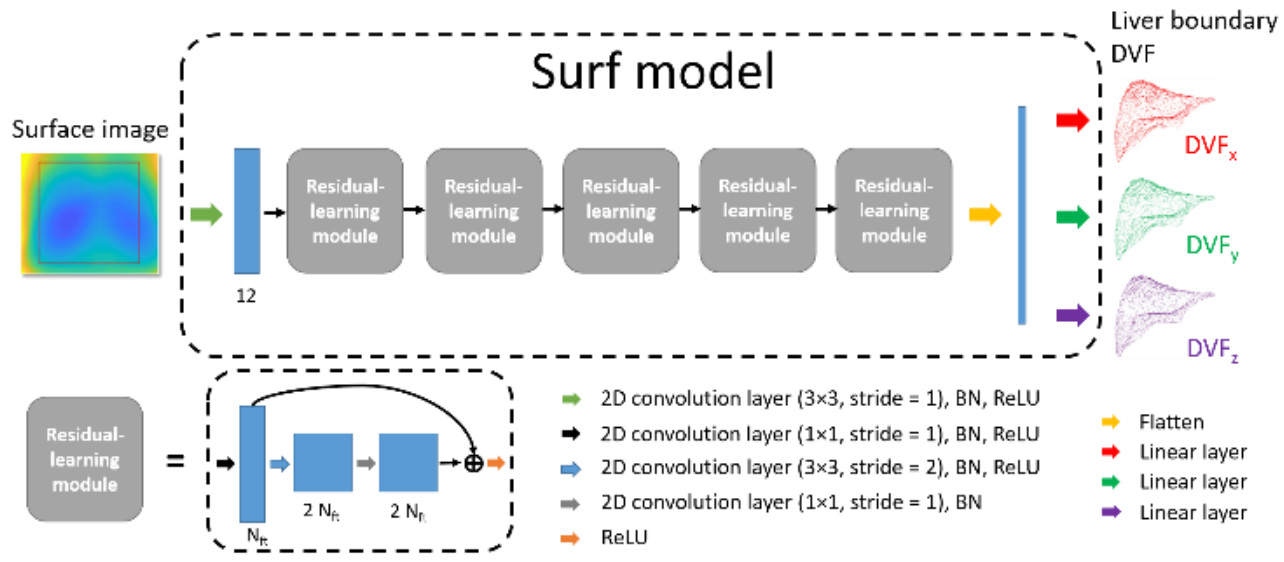

**Figure 2.** Overview of the *Surf* model. The *Surf* model estimated liver boundary DVFs from a body surface image by learning the correlations between the respiratory-induced external body surface motion and internal liver boundary motion. The network architecture was based on residual learning to predict liver boundary DVFs.

Figure 2 presents an overview of the *Surf* model. The measured 3D body surface was gridded into a Cartesian 2D image which represents the anterior-posterior patient surface coordinates. The model architecture was based on residual learning that contains repetitions of a 2D convolution layer, a batch normalization layer, and a rectified linear unit (ReLU). The feature maps from the last residual learning module were flattened and further processed by three parallel linear layers, each of which yielded a Cartesian component (x, y, z) of the liver boundary DVFs for the liver surface nodes.

### *2.3 Angle-agnostic X360 model*

The *X360* model deformed the prior liver surface mesh to match with the image features encoded in an x-ray projection, using the *Surf* model output as initialization (2.2). Figure 3 presents *X360*, which was adapted from Ref. [7] and generalized to an angle-agnostic model by incorporating geometry-informed perceptual feature pooling layers. In general, *X360* consists of a feature extraction network and a graph neural network. The feature extraction network uses ResNet-50 as the backbone to extract image features from an x-ray projection, and then the extracted feature maps were fed into the subsequent graph neural network for motion estimation. The graph neural network contains two deformation modules, each of which contains a geometry-informed perceptual pooling layer, a graph convolutional network, and a spatial transform layer. The extracted feature maps were first pooled by the geometry-informed perceptual pooling layer. As shown in Fig. 3(b), the geometry awareness was accomplished by incorporating the scan angle and the corresponding projection system matrix of each x-ray projection in the feature pooling layer, by which the model learned to adapt to angle-dependent features for corresponding motion estimation. Based on the pooled features, the following graph neural network estimates the 3D motion of the liver boundary nodes, and the spatial transform layer deforms the liver surface nodes according to the predicted motion. The two deformation modules sequentially deform the liver mesh through a rigid registration module followed by a free-form, deformable registration module. The sequential design is to improve the registration accuracy and model robustness, as an initial rigid registration allows more accurate/relevant feature pooling in the consecutive deformable registration step, especially under scenarios of large motions. The *X360* model was trained by cascading with the trained *Surf* model (*Surf-X360*) for each patient case. By randomizing the projection angle on-the-fly, an angle-agnostic model can be trained to extract only features relevant to the specific scan angle of each input x-ray projection to solve the corresponding liver surface motion.

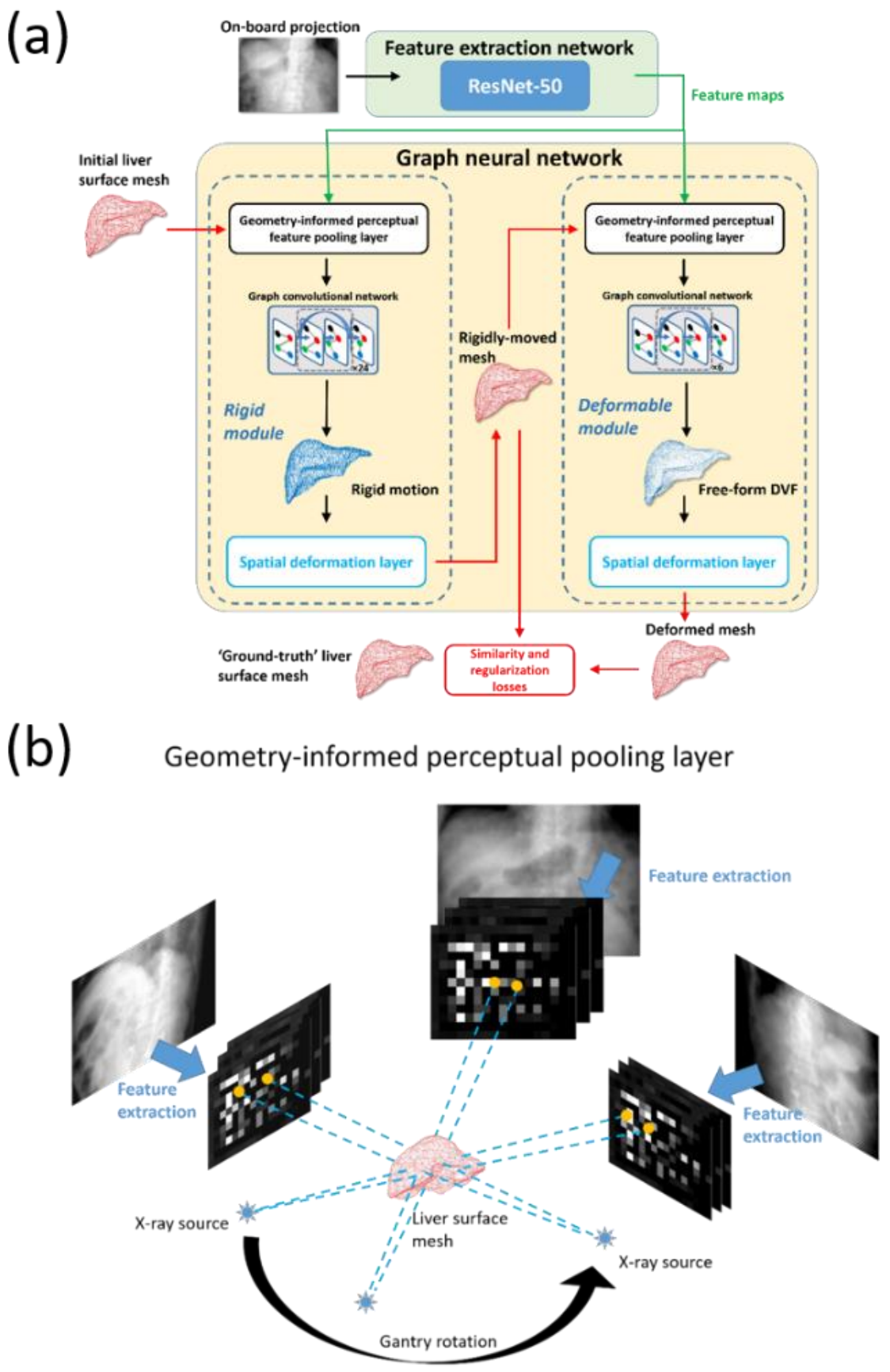

**Figure 3.** Overview of the (a) angle-agnostic *X360* model and (b) geometry-informed perceptual pooling layer. *X360* deformed a prior liver surface mesh to match the image features encoded in an onboard x-ray projection. The model architecture contains a feature extraction network and a graph neural network, which respectively extracts image features from the x-ray projection and estimates the liver boundary DVFs. The model angle awareness was achieved by the geometry-informed perceptual pooling layer that projects liver surface nodes onto the feature maps extracted from an onboard projection, according to the gantry angle and the cone-beam geometry. The liver boundary registration was sequentially solved by two modules: the first for rigid motion, and the second for free-form, deformable motion.

*2.4 Bio model*

The *Bio* model used the liver boundary DVFs solved from the cascaded *Surf-X360* models to derive the intra-liver DVFs (Fig. 4). By *Bio*, the liver boundary DVFs were decomposed into a spatially uniform component (i.e., DC component) and a residual oscillating component (i.e., AC component). This decoupling can improve model robustness/efficiency, as the DC component corresponds to a rigid translation, while the AC component reflects the residual deformation. Only the AC component needs to be propagated into the liver to generate spatially variant deformation fields. The U-Net for AC component propagation was trained in a supervised fashion, by using the Mooney-Rivlin material model and the conventional FEA to generate 'ground-truth' intra-liver DVFs for a cohort of training liver cases.

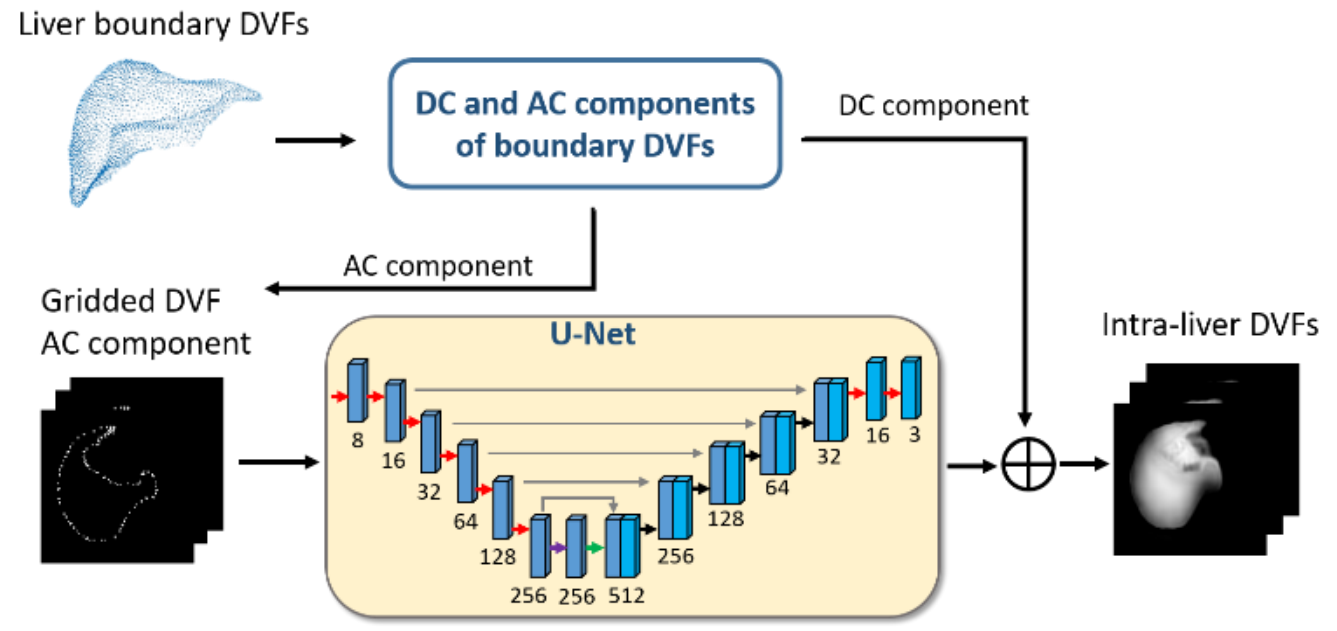

**Figure 4.** Overview of the Bio model. The input liver boundary DVFs were decomposed into a spatially homogeneous (DC) component (rigid) and a residual (AC) component (deformable). The network was trained to predict intra-liver DVFs from the AC component. The network-predicted intra-liver DVFs were re-combined with the DC component afterward to yield the overall intra-liver DVFs.

*2.5 Data curation, augmentation, and analysis*

A dataset of 4D-CTs from 34 patients with liver cancer was used to train and evaluate the proposed framework. The patients were partitioned into two subsets. The 1st subset contains 24 patients and was used to train the population-based *Bio* model, while the remaining 10 patients (2nd subset) were used to train the patient-specific *Surf-X360* models and test the overall *Surf-X360-Bio* framework.

The 4D-CTs in the 2nd subset have 10 respiratory phases (0% – 90%) each, which were insufficient to train a DL model to represent onboard motion and its variations. Thus, we applied a principal component analysis (PCA)-based statistical motion model to augment the 4D-CT motion. The details were published in Ref. [7] and are not repeated here due to length constraints. Besides the PCA-based deformable augmentations, random rigid motion augmentations were also included. The augmented dataset was partitioned into a training/validation subset and a testing subset. To prevent data leakage, the testing data were selected to correspond to the extrema of a respiratory cycle that were not seen in the training/validation subset. In addition, we amplified the PCA scaling coefficients of the testing subset to enlarge the deformable motion, and further coupled it with a larger, random rigid motion (up to 10 mm in magnitude) as compared to the training/validation subset (up to 6 mm). In total, we generated 768/384 cases for training/validation, and 174 cases for testing for each patient. For each case, the surface image was simulated, along with x-ray projections generated at random scan angles during training/validation, and 19 x-ray projections generated uniformly (across 360°) during testing.

The 'ground-truth' liver and liver tumor meshes were generated using the known augmentation DVFs for training/validation/testing. Liver boundary registration accuracy was evaluated by the root-mean-square error (RMSE) and 95-percentile Hausdorff distance (HD95). Liver tumor localization accuracy was evaluated by liver tumor center-of-mass error (COME), Dice similarity coefficient (DSC), and HD95.

*2.6 Ablation studies*

Ablation studies were performed to assess the importance of individual components in the proposed framework. The first study investigated the advantages of combining surface imaging and x-ray imaging. Two variations of the *Surf-X360-Bio* framework were generated. The first ablated the x-ray imaging (*Surf-Bio*) and the second ablated the surface imaging (*X360-Bio*). Another ablation study assessed the sequential registration modules of *X360*. The rigid registration module was ablated to form a new model, *Surf-X360-Bio-nrr*, for which the suffix '*nrr*' represents '*no rigid registration*'. We also compared *Surf-X360-Bio* to its variant without the angle-agnostic design (*Surf-X0-Bio*), for which the x-ray model was only trained for 0° projections.

## 3. Results

*3.1 Liver boundary registration accuracy*

Figure 5 compares pre- and post-registration liver surface meshes at three projection angles. The overlays between the prior and 'ground-truth' meshes (1st column) reflect the degree of motion, and *Surf-X360-Bio* deformed the meshes to match well with the 'ground-truth' (2nd column). As shown in Table 1, *Surf-X360-Bio* model outperformed the other ablated models in terms of both RMSE and HD95.

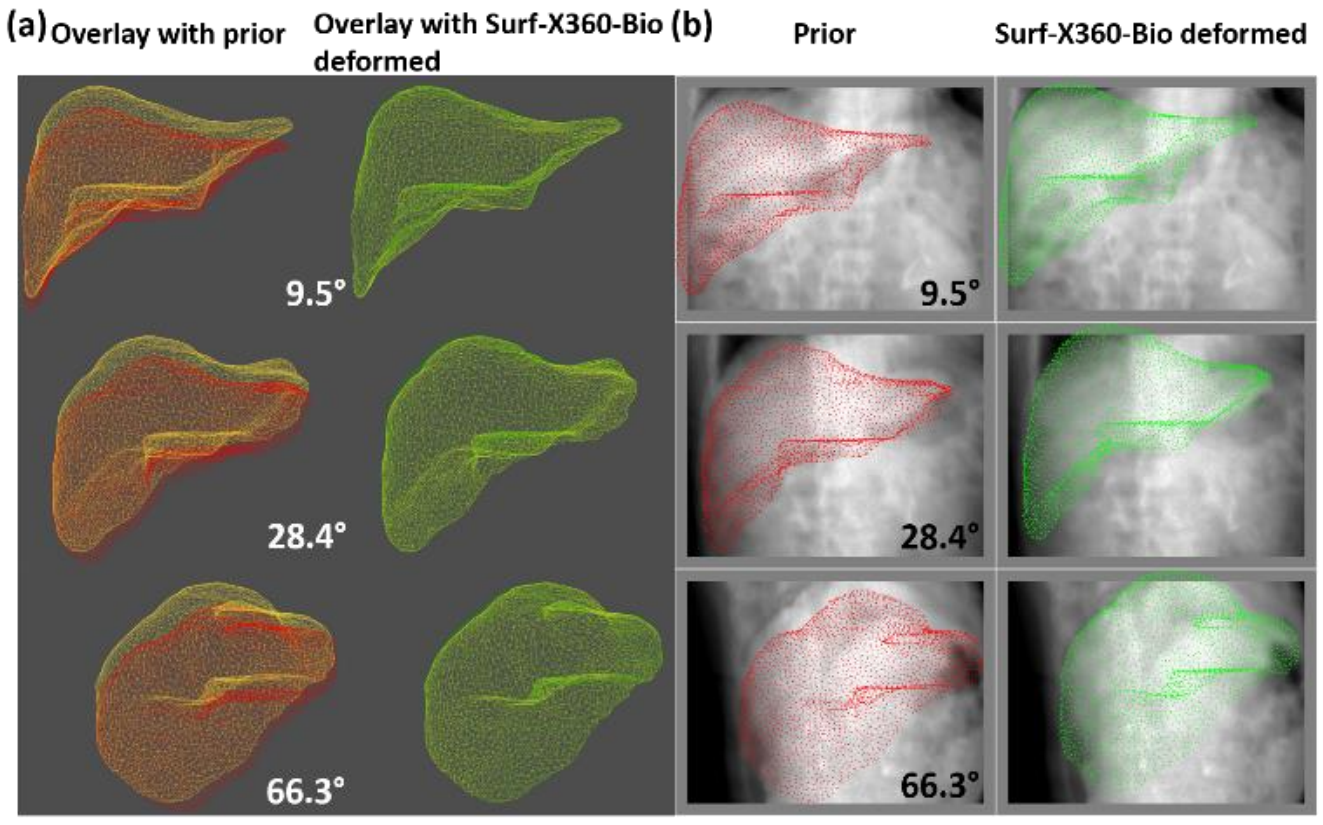

**Figure 5.** (a) Liver surface mesh overlays and (b) liver surface node projections (onto x-ray projections) at three projection angles. The prior, deformed, and 'ground-truth' liver surface meshes are in colors red, green, and yellow, respectively.

**Table 1.** Liver boundary registration accuracy (Mean±SD) in terms of root-mean-square error (RMSE) and 95-percentile Hausdorff distance (HD95) between the predicted and 'ground-truth' liver surface meshes.

| Metric | *Prior* | *Surf-Bio* | *X360-Bio* | *Surf-X360-Bio-nrr* | *Surf-X360-Bio* |
|---|---|---|---|---|---|
| RMSE (mm) | 8.0±3.7 | 2.1±1.6 | 4.2±2.0 | 1.9±1.4 | **1.6±1.2** |
| HD95 (mm) | 9.9±4.6 | 3.1±2.1 | 5.1±1.9 | 2.8±1.7 | **2.4±1.5** |

*3.2 Liver tumor localization accuracy*

Table 2 presents the liver tumor localization accuracy of different models, where *Surf-X360-Bio* performs the best.

**Table 2.** Liver tumor localization accuracy (Mean±SD).

| Metric | *Prior* | *Surf-Bio* | *X360-Bio* | *Surf-X360-Bio-nrr* | *Surf-X360-Bio* |
|---|---|---|---|---|---|
| COME (mm) | 8.5±5.2 | 2.7±2.4 | 3.7±2.5 | 2.4±2.1 | **2.1±1.8** |
| DSC | 0.42±0.29 | 0.76±0.19 | 0.68±0.21 | 0.78±0.18 | **0.81±0.16** |
| HD95 (mm) | 8.1±5.2 | 3.0±2.3 | 3.7±2.3 | 2.8±2.0 | **2.5±1.7** |

*3.3 Angular dependence*

Figure 6 compares the angle-specific and angle-agnostic models as a function of testing projection angles. *Surf-X360-Bio* showed a stable performance at arbitrary projection angles, while *Surf-X0-Bio* degraded rapidly as the angle deviates from 0°.

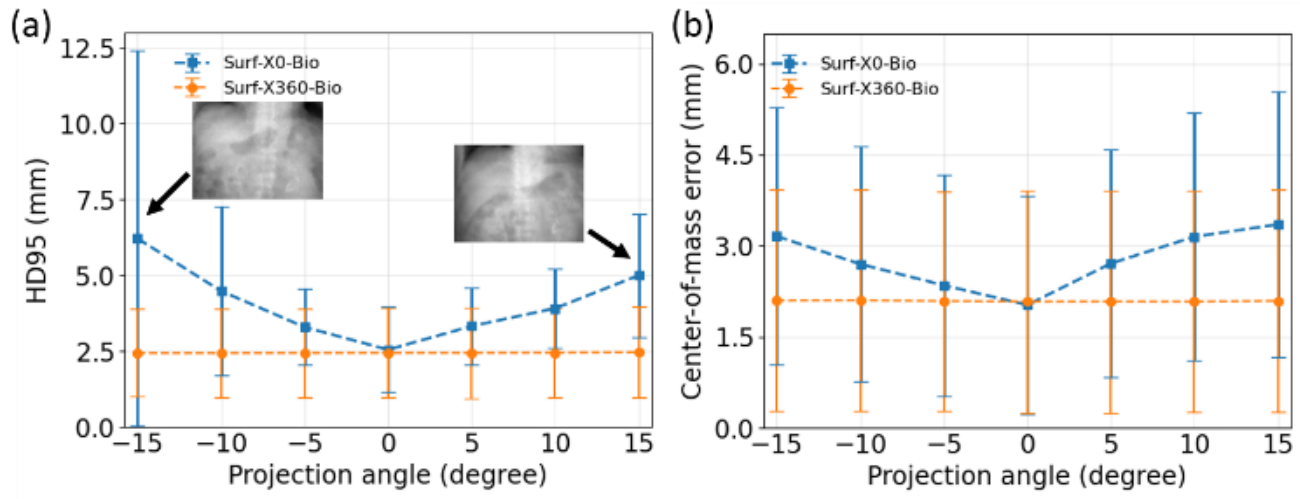

**Figure 6.** Comparison of (a) liver boundary and (b) tumor localization accuracy of angle-specific (*Surf-X0-Bio*) and angle-agnostic (*Surf-X360-Bio*) models as a function of the testing x-ray angles.

## 4. Conclusion

We presented a DL-based, deformable registration-driven approach to localize liver tumors in real-time, using a body surface image and a single x-ray projection acquired at an arbitrary scan angle. The results showed *Surf-X360-Bio* can estimate liver boundary deformation accurately, which translates to accurate liver tumor localization. The mean latency of the *Surf-X360-Bio* framework is 232 ms, which fulfills the temporal constraint of real-time imaging.

# Refine 3D Object Reconstruction from CT Projection Data via Differentiable Mesh Rendering

Le Shen[1, 2], Yuxiang Xing[1, 2], and Li Zhang[*1, 2]

[1]Key Laboratory of Particle & Radiation Imaging (Tsinghua University), Ministry of Education, China
[2]Department of Engineering Physics, Tsinghua University, Beijing 100084, China

**Abstract** 3D object reconstruction from medical images is essential in clinical medicine, such as orthopedics surgery and cardiovascular disease diagnosis. Conventional approaches to reconstruct a 3D object contains two main steps: 1) segment a target in a medical image and generate its mask by the marching cubes algorithm, 2) extract a 3D mesh from the mask. Due to the discretization error of reconstruction algorithms and the uncertainty of the segmentation, the resulting 3D objects are often inconsistent with reality. To alleviate this issue, an alternative is to reconstruct the 3D objects represented by a parametric model or geometric primitives directly from projection data. Differentiable rendering has recently been adopted in tomographic reconstruction problems and achieves superior results. In this work, we investigated the feasibility of triangle mesh object reconstruction and refinement via differentiable rendering and our preliminary results demonstrated its potential value for applications in clinical CT imaging.

## 1 Introduction

3D object reconstruction from CT images is widely used in many clinical diagnoses and treatments. For instance, coronary stenosis is regularly evaluated through the 3D reconstruction of the CT angiography images in cardiology, 3D printed orthopedic prosthesis in the joint replacement surgery is obtained from 3D models reconstructed from the CT images in orthopedics. In these scenarios, the targets in the CT images are firstly segmented manually or by some automatic segmentation algorithms. Then, the 3D masks are converted to triangle meshes by the algorithm of marching cubes and with post-processing afterwards. The quality and precision of the generated meshes depend on the spatial resolution and noise level in the reconstructed images relying on various factors including the hyper-parameters of the reconstruction algorithms, such as the filtering kernel in the analytical algorithms, the regularization strength in the iterative algorithms and the voxel size. The segmentation error and uncertainty also bring additional deviation in the target masks. Moreover, the marching cubes algorithm usually yields a staircase surface so that the smoothing is needed. Instead of this multi-pass procedure, direct reconstruction of 3D objects has attracted attention for a long time and many geometric modeling approaches have been proposed to solve this problem. For example, Fessler *et al.* proposed the single-valued generalized cylinder model to represent arterial trees by a collection of parallel ellipses and reconstructed the parametric model from two magnetic resonance angiography images[1]. Faby *et al.* proposed a parameterized bottle shape model and reconstructed bottle and liquid material densities from two dual-energy projections[2]. Jin *et al.* proposed a metal artifact and partial volume effect correction method for liquid CT scan by the projection of a hybrid of pixelized image and parameterized metallic container[3]. Besides, some non-parametric representations are also helpful in 3D object reconstruction. Among them, the polyhedral surface is a general explicit representation of homogeneous objects, and its forward projection and reconstruction in CT have been well studied[4]. In computer graphics, as a particular case of polyhedral surface, the triangle mesh is computationally efficient in ray intersection calculation and is widely adopted in ray-tracing rendering. Sawall *et al.* proposed a spatial subdivision structure to accelerate the triangle-ray intersection calculation and reconstruction of the triangle vertices from CT projections[5]. Apart from ray-tracing rendering, rasterization-based rendering is also a classical and efficient method in 3D object display, and can also be utilized for 3D surface reconstruction. Recently, differentiable rendering becomes an active research area in the computer graphics community. Many differentiable rendering methods have been developed to solve 3D reconstruction problems[6-8] and CT reconstruction problems[9]. In this work, we investigate 3D surface reconstruction from CT projection data using the differentiable mesh renderer in the PyTorch3D library[10].

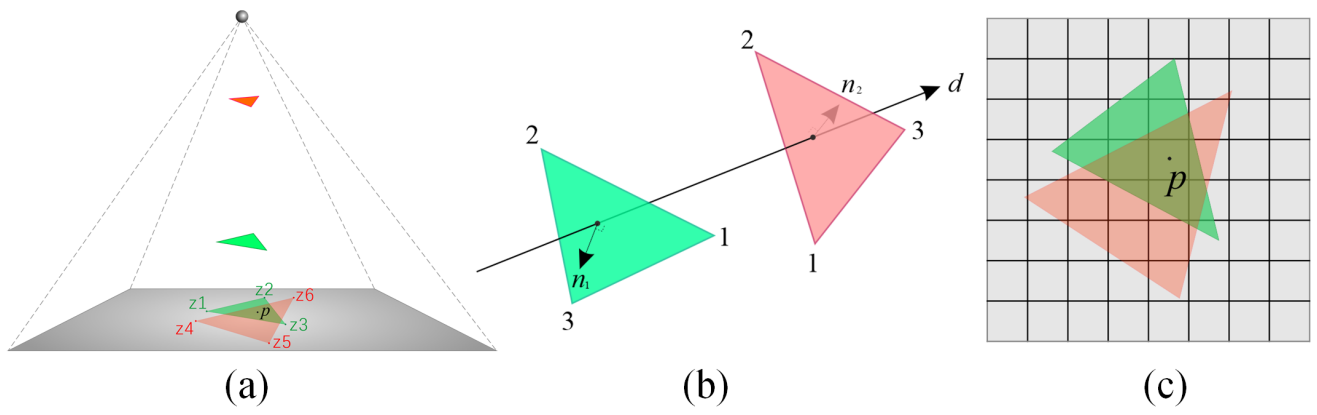

Fig. 1. Illustration of triangle mesh rendering. (a) Rendering a scene consists of a light, camera and a triangle mesh. (b) Ray-tracing based rendering via ray-triangle intersection. (c) Rasterization based rendering via discretization of the projected triangles.

## 2 Materials and Methods

Generally, we consider a 3D object consists of $N_\mathrm{m}$ triangle meshes with the $i^{\mathrm{th}}$ mesh represented by

* This work was supported by National Natural Science Foundation of China (under grant 62031020).
Corresponding author: Li Zhang (E-mail: zli@mail.tsinghua.edu.cn).

$$\mathrm{M}_i\left(\mathbf{V}_i,\mathbf{F}_i,\mathbf{E}_i\right)=\begin{Bmatrix}\mathbf{V}_i=\left(\mathbf{v}_{i1},\mathbf{v}_{i2}\cdots\mathbf{v}_{iN_{\mathrm{V}_i}}\right)\in\mathbb{R}^{3\times N_{\mathrm{V}_i}}\\ \mathbf{F}_i=\left(\mathbf{f}_{i1},\mathbf{f}_{i2}\cdots\mathbf{f}_{iN_{\mathrm{F}_i}}\right)\in\mathbb{N}^{3\times N_{\mathrm{F}_i}}\\ \mathbf{E}_i=\left(\mathbf{e}_{i1},\mathbf{e}_{i2}\cdots\mathbf{e}_{iN_{\mathrm{E}_i}}\right)\in\mathbb{N}^{2\times N_{\mathrm{E}_i}}\end{Bmatrix} \quad (1)$$

$$\mathbf{v}_*\in\mathbb{R}^{3\times1},\ \mathbf{f}_*\in\mathbb{N}^{3\times1},\mathbf{e}_*\in\mathbb{N}^{2\times1}$$

Here, $\mathrm{M}_i$ contains $N_{\mathrm{V}_i}$ vertices, $N_{\mathrm{F}_i}$ faces and $N_{\mathrm{E}_i}$ edges. We assume $\mathrm{M}_i$ is a closed mesh and the linear attenuation coefficient inside it equals $\mu_i$ homogeneously. Denoting the projection operator for a mesh as $P(\cdot)$, we have the projection data of an object

$$\hat{\mathbf{g}}=\sum_{i=1}^{N_m}\mu_i P\left(\mathrm{M}_i\right) \quad (2)$$

Intersection and inclusion between two meshes are allowed according to the addictive rule.

### 2.1 3D mesh projection via differentiable rendering

The multi-view geometrical convention can describe the cone beam CT with a flat-panel detector. The x-ray source position is determined by the camera translation matrix **T** and rotation matrix **R**. The focal length defines source-to-detector distance and the principal point defines the detector offsets in column and row directions, which can be expressed by the camera intrinsic matrix **K**. The forward projector $P(\cdot)$ consists of the following steps:

1. Project the mesh vertices from the world coordinate to the camera coordinate by the composition of **T**, **R** and **K**. Then the $x$ and $y$ coordinates are normalized to [-1, 1] while the $z$ coordinate keeps in the camera coordinate and represents the depth between the vertex and the camera (X-ray source) along the $z$-axis.
2. Rasterize the projected 2D triangle mesh. This step is similar to the fragments generation in the graphics. A fragment is the data structure that stores all the primitives overlapped with a certain pixel. For each pixel, traverse all the triangles and record those lying on the pixel center in a fragment, which stores all triangles contributing to the final line integral and their depths.
3. Render the Fragments and generate the projection image. The line integral of the $j^{\text{th}}$ detector bin is calculated as

$$\hat{g}_j=\frac{1}{\mathbf{d}_{\mathrm{prcp}}\cdot\mathbf{d}_j}\sum_t \mathrm{sgn}\left(\mathbf{n}_t\cdot\mathbf{d}_j\right)z_t \quad (3)$$

where $\mathbf{n}_t$ is the normal vector of the $t^{\text{th}}$ triangle in the Fragment, $\mathbf{d}_j$ is the ray direction and $z_t$ is the depth, $\mathbf{d}_j$ is the principle point direction and $\mathbf{d}_{\mathrm{prcp}}\cdot\mathbf{d}_j$ is the cone angle weight in cone-beam CT.

The first two steps are processed by the PyTorch3D functionality, while a customized renderer realizes the last step.

### 2.2 Objective formulation and optimization

Because the triangle mesh maintains its topology during the deformation, a proper initialization with the correct topology of the target is sufficient. Similar to an iterative CT reconstruction algorithm, the 3D object reconstruction can be formulated as a regularized least-squares problem. Given an initialization of $\mu_i$ and $\mathrm{M}_i$, we seek vertices displacement $\Delta\mathbf{V}_i$ and $\mu_i$ as a solution to the objective function

$$\begin{aligned}\Delta\mathbf{V}_i,\mu_i\in\underset{\Delta\mathbf{V}_i,\mu_i}{\operatorname{argmin}}\frac{1}{2}\left\|\sum_{i=1}^{N_m}\mu_i P\left(\mathrm{M}_i\left(\mathbf{V}_i+\Delta\mathbf{V}_i,\mathbf{F}_i,\mathbf{E}_i\right)\right)-\mathbf{g}\right\|_2^2\\+\sum_{i=1}^{N_m}R\left(\mathrm{M}_i\left(\mathbf{V}_i+\Delta\mathbf{V},\mathbf{F}_i,\mathbf{E}_i\right)\right)\end{aligned} \quad (4)$$

**g** is the measured projection data and $R$ is a group of regularization terms implemented in PyTorch3D. For example, we can use the mesh Laplace smoothing regularization with the cotangent weighting scheme[11]:

$$R_{\mathrm{L}}\left(\mathbf{V}_i\right)=\frac{1}{N_{\mathrm{V}_i}}\sum_{t=1}^{N_{\mathrm{V}_i}}\left\|\frac{\sum_{j\in\mathrm{N}_{it}}\left(\cot\alpha+\cot\beta\right)\left(\mathbf{v}_j-\mathbf{v}_{it}\right)}{\sum_{j\in\mathrm{N}_{it}}\left(\cot\alpha+\cot\beta\right)}\right\|_2 \quad (5)$$

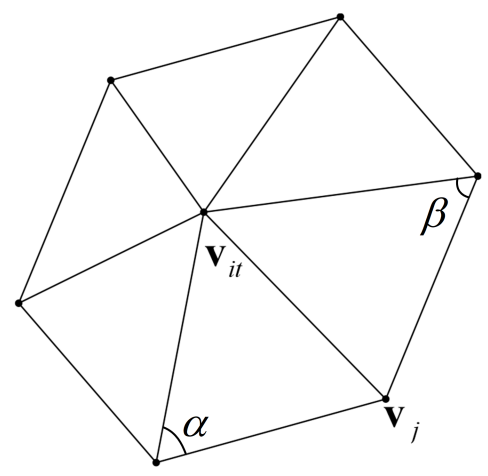

Fig. 2. Illustration of the one-ring neighborhood vertices of $\mathbf{v}_{it}$.

As shown in Fig.2, $\mathrm{N}_{it}$ is the 1-ring neighborhood vertices, $\alpha$ and $\beta$ are the angles opposite to the connecting edge of $\mathbf{v}_{it}$ and $\mathbf{v}_j$. The Laplace regularization helps to suppress the speckle noise on the surface. Another regularization to the edges is

$$R_{\mathrm{edge}}\left(\mathbf{E}_i\right)=\frac{1}{N_{\mathrm{E}_i}}\sum_{t=1}^{N_{\mathrm{E}_i}}\left\|\mathbf{v}_{ie_{it1}}-\mathbf{v}_{ie_{it2}}\right\|_2^2 \quad (6)$$

This regularization suppresses the deformation to narrow triangles. Combining (4) - (6), the objective function is

$$\begin{aligned}\Delta\mathbf{V}_i,\mu_i\in\underset{\Delta\mathbf{V}_i,\mu_i}{\operatorname{argmin}}\frac{1}{2}\left\|\sum_{i=1}^{N_m}\mu_i P\left(\mathrm{M}_i\left(\mathbf{V}_i+\Delta\mathbf{V}_i,\mathbf{F}_i,\mathbf{E}_i\right)\right)-\mathbf{g}\right\|_2^2\\+\lambda_{\mathrm{L}}\sum_{i=1}^{N_m}R_{\mathrm{L}}\left(\mathbf{V}_i+\Delta\mathbf{V}\right)+\lambda_{\mathrm{edge}}\sum_{i=1}^{N_m}R_{\mathrm{edge}}\left(\mathbf{E}_i\right)\end{aligned} \quad (7)$$

We can solve the two-block optimization problem by the alternating minimization (AM) method:

1. Fixing $\mu_i$, update the vertices displacement $\Delta\mathbf{V}_i$ using the Adam optimizer. To accelerate the convergence, the projection data are divided into ordered subsets, each containing a fixed number of projection views. We set 4 views per subset in this study. The other regularization parameters are selected empirically.
2. Fixing $\Delta\mathbf{V}_i$, update $\mu_i$ by the least-squares inversion.

## 3 Results

For simplification, we assume the projection data only contains line integrals of a finite number of homogeneous regions which can be expressed by closed meshes. The K-edge imaging of iodine contrast agent based on spectral CT satisfies this assumption. We demonstrate the differentiable rendering reconstruction method by the simulation and the real experiment of spectral CT. The iodine component is obtained by projection-domain basis material decomposition of the multi-energy data. The initialization of the meshes comes from the segmentation of the K-edge reconstruction result by thresholding, while the linear attenuation is initialized by the average of the segmented regions.

### 3.1 Simulation result

In the simulation study, we constructed a heart phantom consisting of the heart wall represented by a convex polyhedron and the left and right coronary arteries represented by two triangle meshes. The heart wall is assigned as soft tissue while the coronary arteries are filled as Iodixanol solution. As shown in Fig. 3, we simulated three distinct spectra of an ideal photon counting detector. The multi-energy cone-beam CT projection data were simulated by the ray-tracing approach at 20 views uniformly sampled in 360°. The number of incident photons for a detector bin is $2.5\times10^5$. We decompose original projection data into the projection of water and iodine density maps. The iodine image was reconstructed by iterative reconstruction with total-variation (TV) regularization. Then, the coronary arteries were segmented with a threshold of 20 mg/mL. Triangle meshes are extracted by the algorithm of marching cubes and served as initialization to (7). The average iodine density of the segmented region is 220 mg/mL while the reference value is 335 mg/mL.

Fig. 4 displays the 3D reconstruction result of the proposed method. The volume rendering image in Fig. 4 (b) indicates that the CT reconstruction yields satisfactory coronary arterie image for the few-view scan though the distal branches are almost invisible. In Fig. 4 (b), with a small threshold for segmentation, most of the arterial trees are successfully obtained from the CT images. However, the distal diameter is eroded due to the edge blur in the CT reconstruction. Besides, the surfaces suffer from the staircase artifact resulted from the segmentation. Fig. 4 (d) shows the 3D reconstruction result of the differentiable rendering method. The surfaces are as smooth as the ground truth with distal branches of accurate diameter. We also evaluated the quantitative reconstruction errors of the surfaces. Fig. 5 displays the point-to-mesh distances of the results in Fig. 4. The mean point-to-mesh distance of the segmented meshes is 0.19 mm while that of the reconstructed meshes is -0.064 mm. The iodine density estimated by the proposed method is 350 mg/mL. However, some fine vascular structures are missing in the CT reconstruction, and thus not recovered by the triangle mesh deformation.

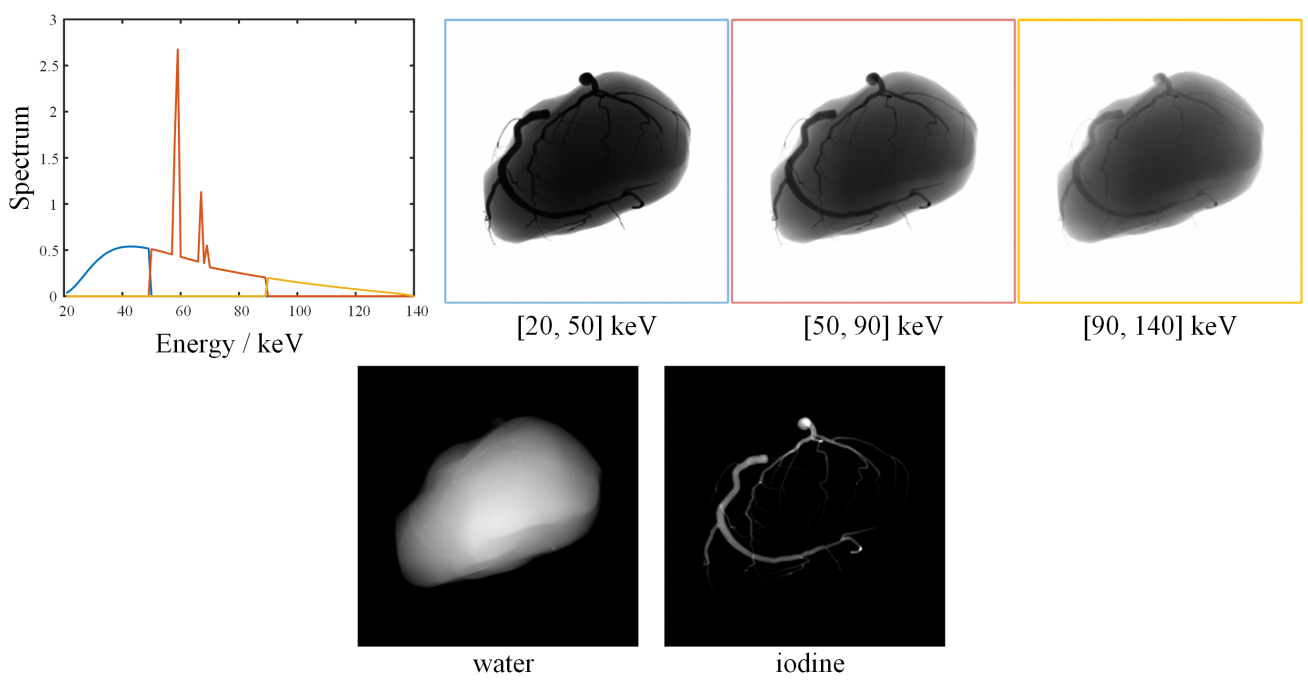

Fig. 3. Spectral CT projection simulation for a heart phantom. The heart is represented by a convex polyhedron while the coronary arteries are represented by two triangle meshes.

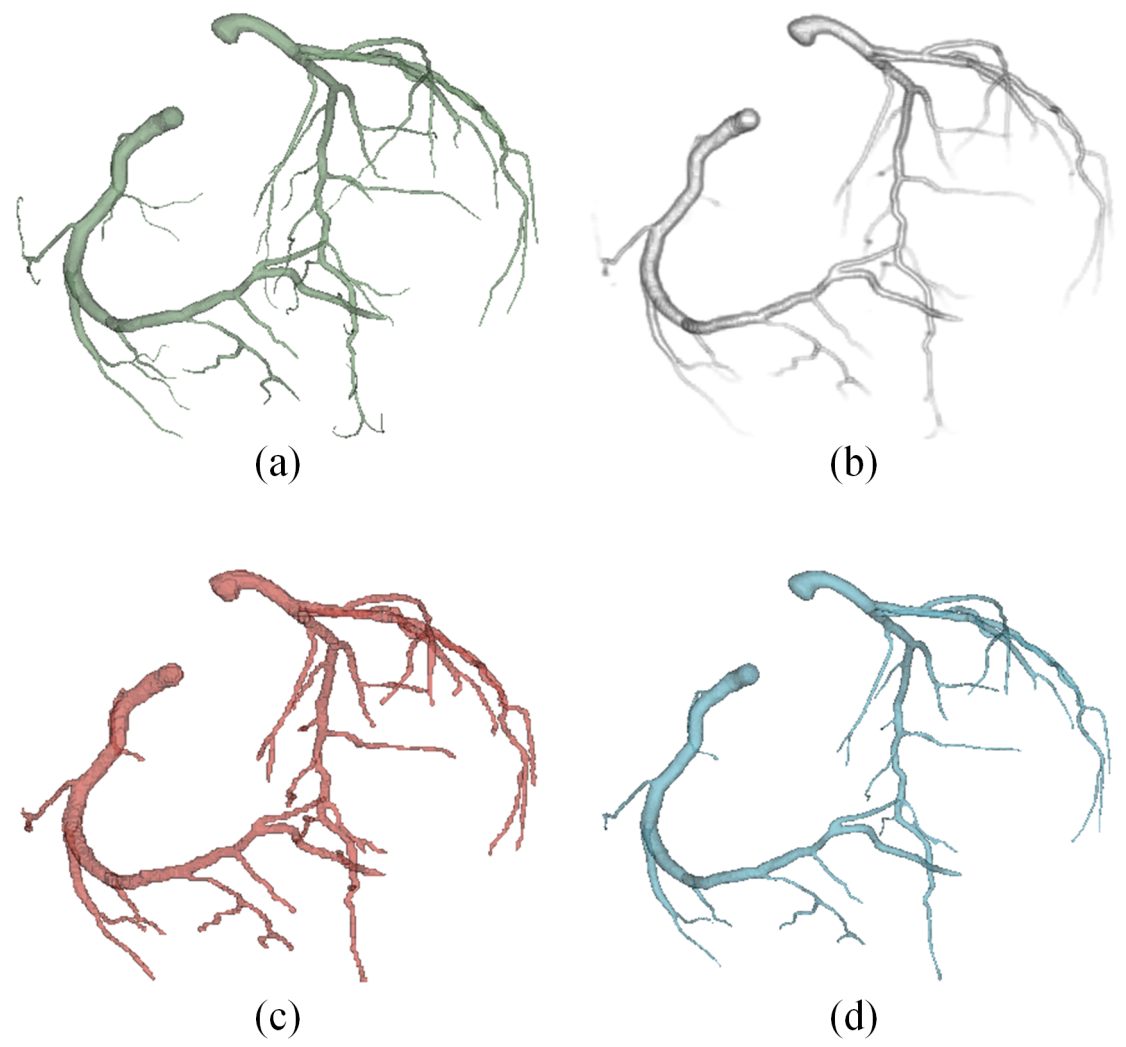

Fig. 4. 3D surfaces reconstruction of the simulation. (a) The ground truth of the coronary arteries meshes. (b) Volume rendering of the iterative reconstruction result with surface lighting. (c) Triangle meshes extracted from (b) by the marching cubes algorithm. (d 3D reconstruction result of the differentiable rendering method.

### 3.2 Real experiment result

The real experiment was conducted on a laboratory photon counting spectral CT system. We made a test phantom consisting of a water cylinder and four silicone tubes filled in 3% Iohexol solution. A metallic stent is inserted into one tube to verify the inclusion in 3D object reconstruction. In order to gain enough height of the FOV, we took three scans with the detector placed in different heights, and merged the data according to the detector positions. We set three energy thresholds and acquired 360 views of projection data over $2\pi$. The projection data were decomposed into the photoelectric effect, Compton effect and iodine component. Due to the material cross-talk and inaccurate spectra, the decomposed projection of iodine contains part of the stent. We focused on the iodine density reconstruction. The CT image was obtained by TV regularized iterative reconstruction. Since the stent density on the iodine map is about ten times of the iodine solution, we set two thresholds 10 mg/mL and 50 mg/mL for segmentation to obtain the initial triangle meshes. Then, the proposed method solved the five triangle meshes and two densities.

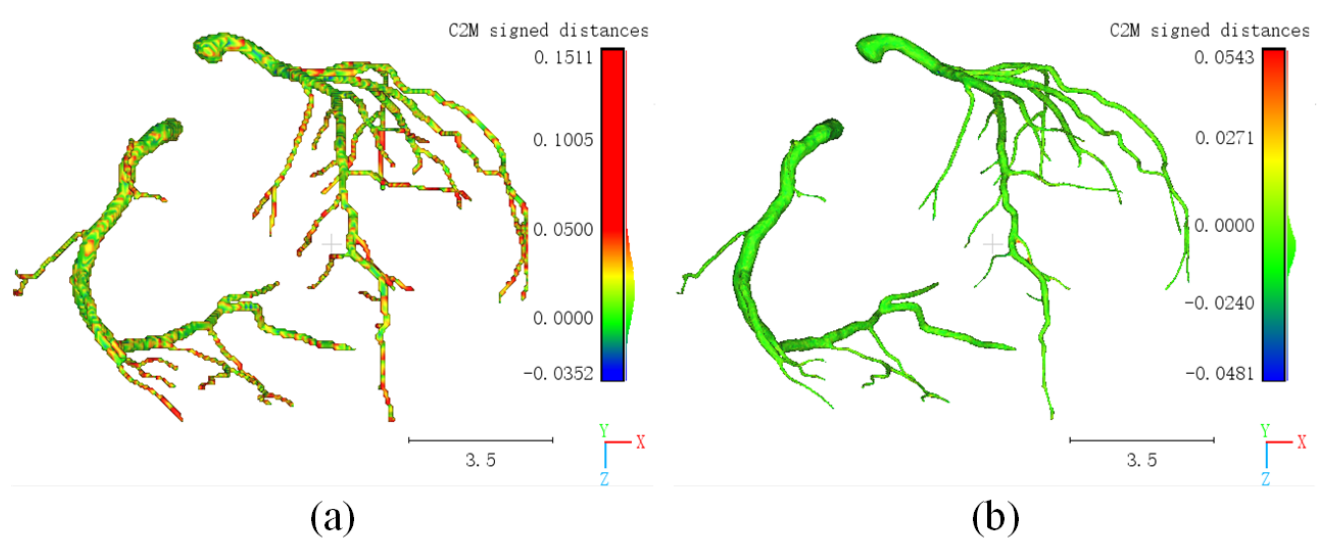

Fig. 5. Point-to-mesh distance of the surface reconstruction results. (a) Distance between Fig. 4(c) and the ground truth. (b) Distance between Fig. 4(d) and the ground truth.

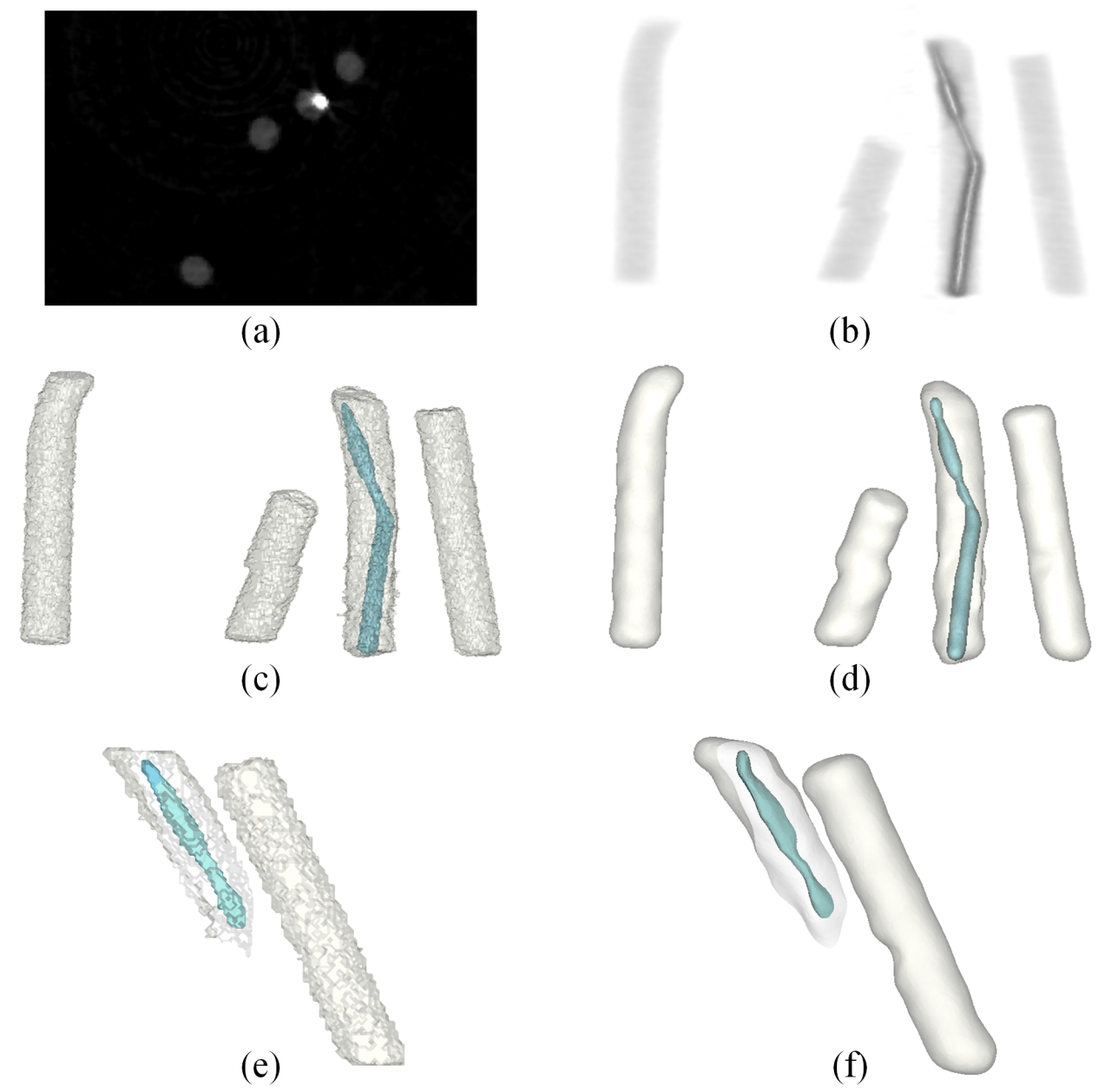

Fig. 6. 3D surfaces reconstruction of the experiment. (a) The CT image of the iodine density map. (b) The volume rendering image of the iodine density map. (c) 3D surfaces reconstructed from the two thresholds segmentation of the CT image. (d) 3D surfaces reconstructed by the differentiable rendering method. (e) Plane clipped objects in (c). (f) Plane clipped objects in (d).

Fig. 6 shows the 3D object reconstruction results of the experimental data. Fig. 6 (a) is one slice of the CT reconstruction and Fig. 5 (b) is the volume rendering image with lighting. The discontinuity of the second tube from the left is caused by the error in geometrical alignment of the three scans. Fig. 6 (c) is the 3D surfaces reconstructed from the segmentation results while Fig. 6 (d) is the 3D reconstruction results of the differentiable rendering method. The advantage in meshes quality of the proposed method is similar to the simulation study. We also show clipped surfaces of the stent in Fig. 6 (e) and Fig. 6 (f), which confirms the proposed method can adequately handle object inclusion.

## 4 Discussion

Though the results of the differentiable rendering reconstruction method achieved in the simulation and experiment are promising, we would like to address some concerns about two issues. Firstly, this method needs prior information about the 3D objects, including but not limited to their quantity, topological structures, densities or attenuation coefficients. The presented results still depend on the conventional CT reconstruction and segmentation procedure, which may be suboptimal for some cases. In future work, we will validate 3D object reconstruction starting from homeomorphic primitives. Secondly, the strong hypothesis about the constant density in the object is not always tenable and restricts the density estimation accuracy for general objects. Finally, the reconstruction of CT images is necessary for most applications. It is reasonable to combine CT image reconstruction with 3D object reconstruction.

## 5 Conclusion

In this work, we investigated the feasibility of 3D object reconstruction from CT projection via the differentiable rendering method. The 3D object reconstruction is formulated as a triangle mesh deformation problem with constraints on vertices and edges and is accomplished by optimizing a regularized least-squares objective function. The simulation and experiment demonstrate the differentiable rendering 3D object reconstruction method is effective in detailed structure restoration.

# Conversion of the Mayo LDCT Data to Synthetic Equivalent through the Diffusion Model for Training Denoising Networks with a Theoretically Perfect Privacy

Yongyi Shi, Ge Wang

Biomedical Imaging Center, Rensselaer Polytechnic Institute, Troy, NY, USA

**Abstract** Deep learning techniques are widely used in the medical imaging field; for example, low-dose CT denoising. However, all these methods usually require a large number of data samples, which are at risk of privacy leaking, expensive, and time-consuming. Because privacy and other concerns create challenges to data sharing, publicly available CT datasets are up to only a few thousand cases. Generating synthetic data provides a promising alternative to complement or replace training datasets without patient-specific information. Recently, diffusion models have gained popularity in the computer vision community with a solid theoretical foundation. In this paper, we employ latent diffusion models to generate synthetic images from a publicly available CT dataset – the Mayo Low-dose CT Challenge dataset. Then, an equivalent synthetic dataset was created. Furthermore, we use both the original Mayo CT dataset and the synthetic dataset to train the RED-CNN model respectively. The results show that the RED-CNN model achieved similar performance in the two cases, which suggests the feasibility of using synthetic data to conduct the low-dose CT research. Additionally, we use the latent diffusion model to augment the Mayo dataset. The results on the augmented dataset demonstrate an improved denoising performance.

## 1 Introduction

X-ray CT is a commonly used clinical diagnostic imaging modality for clinical tasks. However, X-ray exposure has been a public concern since ionizing radiation increases the risk of cancer and genetic diseases. Hence, low-dose CT (LDCT) has been actively studied over the past few decades. Recently, deep learning techniques have demonstrated state-of-the-art LDCT denoising performance. These deep learning-based denoising methods can be grouped into the following categories. First, projection domain preprocessing methods directly denoise the projection data using a neural network before filtered backprojection (FBP) is applied. Also, FBP unrolling methods utilize a neural network to model the FBP pipeline. Furthermore, dual domain methods conduct deep learning in the projection and image domains synergistically. Of dual domain methods, model-based iterative reconstruction methods introduce prior enforced by a neural network to reconstruct LDCT images, and model-based unrolling methods utilize a neural network to model the iteration pipeline. Finally, image post-processing methods are most convenient since it does not require access to projection data. A common limitation of all the aforementioned deep learning methods is the need of a large dataset. Although public CT datasets are available online mainly from grand challenges such as, most datasets are still limited in size and only applicable to specific medical problems.

Collecting medical data is a complex and expensive procedure. Especially, privacy and security concerns constantly challenge data sharing. As a result, publicly available CT datasets are only up to a few thousand scans. Hence, researchers are motivated to overcome this challenge using data augmentation schemes, commonly including simple modifications of images such as translation, rotation, flip and scale. Using such data augmentation to improve the training process has become a standard procedure in LDCT tasks. However, the gain in diversity is relatively small from such simple modifications of images. Thus, the deep learning-based augmentation methods are used to automatically learn the representations of images and generate plausible and realistic samples, which dramatically increase the diversity of generated images. Among them, the generative adversarial network (GAN) is one of the widely used models in the field of medical image augmentation. However, GANs suffer from inherent architectural problems such as diversity mismatch, mode collapse, and unstable training behavior. Therefore, it is highly desirable to develop new data augmentation methods for high-quality synthetic medical data.

Recently, denoising diffusion probabilistic model (DDPM) [1] and latent DDPM [2] demonstrated a superior performance over GANs in synthesis of natural images. Diffusion models were then introduced for biomedical image synthesis in various areas such as brain MRI, histopathology, chest radiographs, and eye funds. However, the diffusion model in CT image synthesis has not been performed yet, and nor has the feasibility of using synthetic data for LDCT denoising been evaluated.

In this paper, we adapt the latent diffusion model [3] to synthesize realistic CT images from an original clinical CT dataset. To demonstrate the capability of synthetic data in the LDCT denoising application, we train the residual encoder-decoder convolutional neural network (RED-CNN) [4] model using the publicly available CT dataset and the synthetic dataset respectively. The results show that the RED-CNN model has similar performance in the two cases, supporting the feasibility of using synthetic data to conduct the LDCT denoising research. Also, we use the latent diffusion model to augment the original dataset, boosting the LDCT denoising performance further.

This work was partially supported by the NIH/NCI Grant R01EB032716.

## 2 Materials and Methods

### 2.1 Variational Autoencoder

To reduce the computational demand of training diffusion models, The variational Autoencoder (VAE) [5] is employed to compress the image space of size 512×512 into a latent space of size 64×64. The training process of VAE involves two parts: encoding and decoding. In the encoding process, a distribution of real CT images $x$ is mapped to a posterior distribution through the identification model $\mathcal{E}(z|x)$ that obeys the normal distribution of mean $\mu$ and standard deviation $\sigma$ where z is the latent variable. In the decoding process, VAE generates a new CT image $\hat{x}$ from a latent variable $z$ though the generation model $\mathcal{D}(x|z)$. The goal of the VAE training is to reconstruct the image $\hat{x}$ close to the original image $x$. The VAE loss consists of an embedding loss and a reconstruction loss. The Kullback-Leiber (KL) divergence measures the embedding loss between the embedding space and white Gaussian noise. The mean squared error (MSE) captures the reconstruction loss between the input and output images. Therefore, the loss function can be defined as

$$\mathcal{L}_{VAE} = D_{KL}(\mathcal{E}(z|x)||g(z)) - \| x - \mathcal{D}(x|z)\|_2^2 \quad (1)$$

where $D_{KL}(\cdot)$ is the KL divergence. $g(z)$ is a standard Gaussian distribution.
After VAE is trained, the encoder $\mathcal{E}$ encodes a CT image $x$ into a latent representation $z = \mathcal{E}(x)$, and the decoder $\mathcal{D}$ reconstructs the image from the latent $\hat{x} = \mathcal{D}(z) = \mathcal{D}(\mathcal{E}(x))$.

### 2.2 Latent Diffusion Model

After CT images are compressed with VAE, the latent diffusion model can be used to generate a synthetic dataset. The architecture of the latent diffusion model is shown in Figure 2. It consists of the two parts: VAE and DDPM. These parts are trained subsequently. In the first training phase, VAE is trained to encode the image space into a latent space. During the VAE training phase, the latent space is directly decoded back into image space. In the second training phase, DDPM is trained in the latent space constructed using the pre-trained VAE. The weights of VAE are frozen in the DDPM training phase. DDPM is probabilistic and designed to learn a latent data distribution by gradually denoising a normally distribution variable, which corresponds to learning the reverse process of a fixed Markov Chain in length $T$. DDPM can be represented as a sequence of denoising autoencoders $\epsilon_\theta(z_t, t); t = 1, \cdots, T$, which are trained to predict a denoised variant of their input $z_t$, where $z_t$ is a noisy version of the latent variable $z$. The objective can be simplified to

$$\mathcal{L}_{LDM} = E_{\mathcal{E}(x),\ \epsilon\sim\mathcal{N}(0,1),t}[\| \epsilon - \epsilon_\theta(z_t, t)\|_2^2] \quad (2)$$

with $t$ uniformly sampled from $\{1, \cdots, T\}$. In this study, the latent space is diffused into Gaussian noise using $t =$ 1000 steps. A U-Net model is used to denoise in the latent space. Samples are then generated with a denoising diffusion implicit model (DDIM) and t=150 steps.

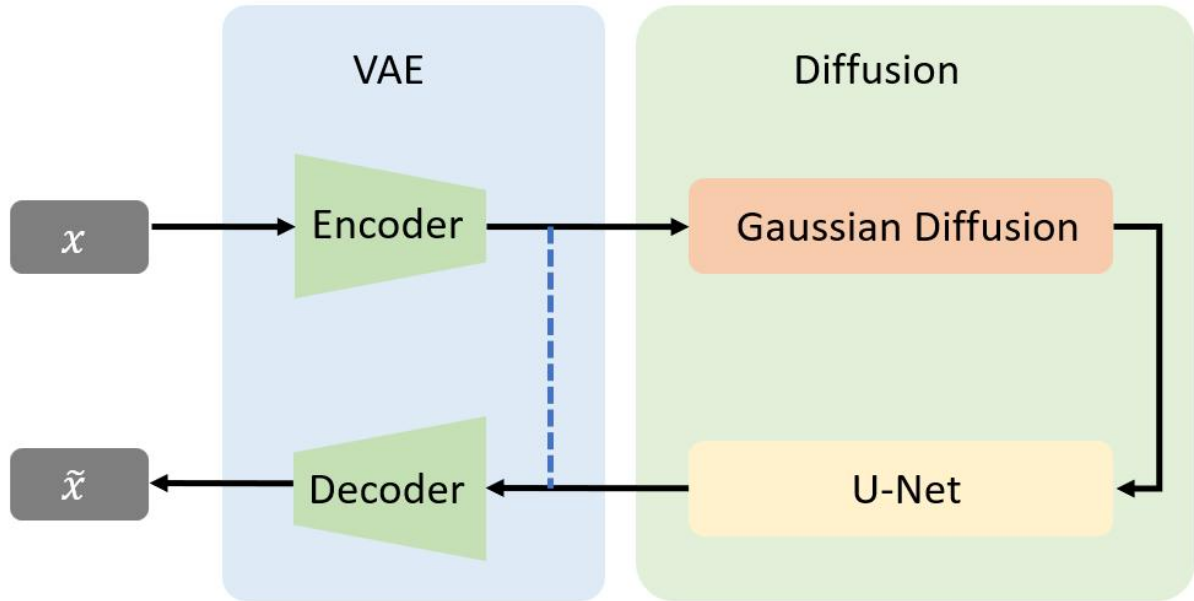

Figure 1. Architecture of the latent diffusion model coupled with the VAE.

### 2.3 Low-dose CT Denoising Network

The RED-CNN network in [4] was employed to evaluate the LDCT denoising performance using original and synthetic datasets respectively. This network consists of 10 layers, including 5 convolutional and 5 deconvolutional layers symmetrically arranged. Shortcuts are made to connect matching convolutional and deconvolutional layers. Each layer is followed by its rectified linear units (ReLU). RED-CNN is an end-to-end mapping from a LDCT image to the full-dose CT (FDCT) counterpart. MSE is utilized as the loss function.

### 2.4 Dataset

A real clinical dataset prepared and authorized by Mayo Clinics for "*the 2016 NIH-AAPM-Mayo Clinic Low Dose CT Grand Challenge*" was used to train the latent diffusion model and validate the clinical performance of RED-CNN. The dataset contains 2,588 3mm 512 × 512 FDCT images from 10 patients, which is our real dataset. In the VAE training phase, all the FDCT images were used to train the network. Then, the FDCT images were compressed into a latent space to train the DDPM model. With the trained latent diffusion model, we randomly generated 2,377 synthetic FDCT images as our synthetic dataset. The number of synthetic images is the same as that of the real data training set. For both the real and synthetic datasets, the corresponding LDCT images were produced by adding Poisson noise into the sinograms simulated from the FDCT images. With the assumed use of a monochromatic source, the projection measurements from a CT scan follow the Poisson distribution, which can be expressed as

$$y_i \sim Poisson\{b_i e^{-l_i} + r_i\}, \quad i = 1, \cdots, I \quad (3)$$

where $y_i$ is the measurement along the i-th ray path. $b_i$ is the air scan photons, $r_i$ denotes read-out noise. In Eq. (3), the noise level can be controlled by $b_i$, which was uniformly set to $10^4$ photons. The FBP algorithm was employed for image reconstruction.

To train the RED-CNN network, the images from the first 9 patients including 2,377 slices in the real dataset were selected, while the images from the remaining patient including 211 slices were used as the testing set. We also trained the RED-CNN network in the augmented dataset consisting of both the original and synthetic images to boost the performance of RED-CNN. To evaluate the feasibility of using the synthetic dataset for deep learning-based LDCT denoising, the RED-CNN networks were trained on real and synthetic datasets respectively, and then tested on the same testing set.

## 3 Results

### 3.1 Evaluation of data synthesis

Since a high quality of the images reconstructed by VAE is the prerequisite for success of the whole latent diffusion generation model, we first evaluated image quality losses due to VAE. In this phase, the FDCT images were directly encoded and decoded by VAE. The reconstructed image samples were shown in Figure 2. It can be seen that the reconstructed images have almost the same visual quality as the original FDCT images. In the FDCT images, there is still subtle noise. The noise is suppressed in the VAE images. However, the reconstructed images preserve all structural information, even tiny details. Subsequently, the RMSE, PSNR and SSIM measures between the input and reconstructed images were calculated as 0.0033, 49.6248 and 0.9931 respectively. Both our visual inspection and quantitative assessment demonstrated that the VAE architecture does not compromise image quality significantly.

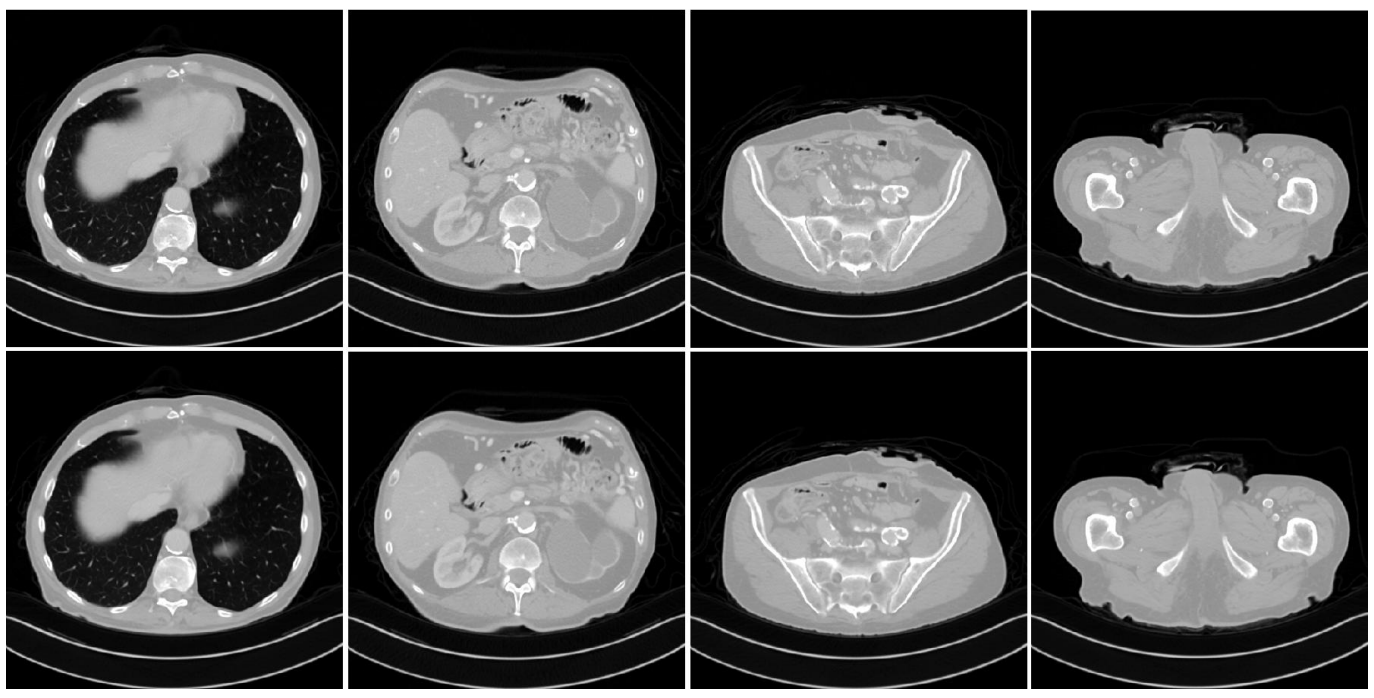

Figure 2. Images reconstructed with VAE. The top row are the input images. The bottom row are the reconstructed images.

Figure 3 shows four represented synthetic images generated using the latent diffusion model as compared to the real images. It can be observed that the latent diffusion model sampled high-quality images with sharp details and realistic textures. The synthetic and real images are visual indistinguishable in these samples.

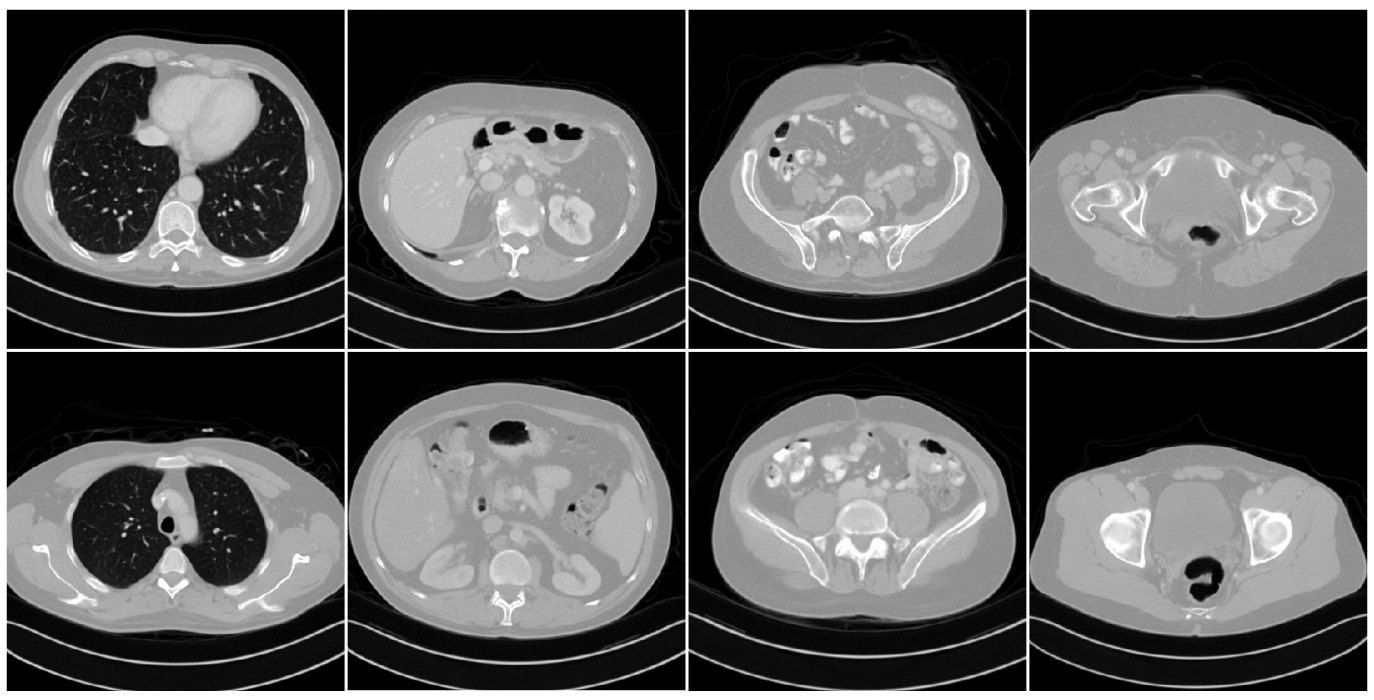

Figure 3. Images synthesized using the latent diffusion model. While the top row are the real images, the bottom row are the synthetic images.

### 3.2 Evaluation of RED-CNN

Four representative images from the testing dataset were used to demonstrate the performance of RED-CNN trained on the synthetic dataset. Figure 4 shows the LDCT denoising results using different methods. There are high image noise and streaking artifacts in LDCT images. Both the RED-CNN trained on original and synthetic dataset effectively suppressed image noise and artifacts but some details were blurred. However, the difference is visually indistinguishable between the two sets of denoised images.

For quantitative evaluation, the RED-CNN achieved high PSNR/SSIM and low RMSE compared with LDCT results in Table 1. The quantitative metrics of RED-CNN trained on the real and synthetic datasets are very close, and the denoising performance appears slightly better with synthetic data then real images.

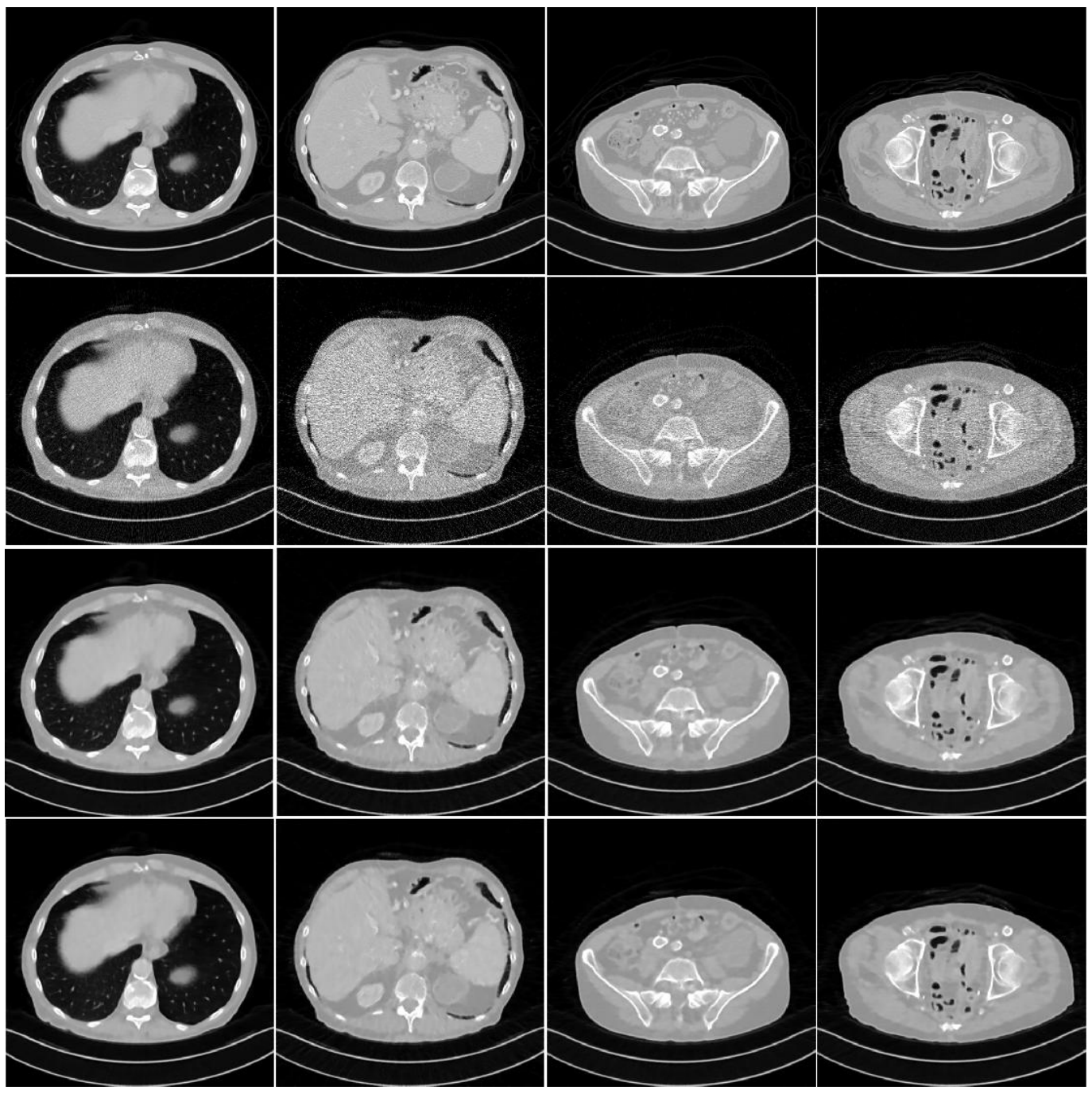

Figure 4. Four representative images from the testing dataset. From top to bottom are the reference FDCT, LDCT and VAE results trained on the original and synthetic datasets respectively.

Table 1. Quantitative results of RED-CNN using different training datasets.

| Methods | RMSE | PSNR | SSIM |
|---|---|---|---|
| Low-Dose | 0.0178 | 35.1721 | 0.8221 |
| Real | 0.0056 | 45.0165 | 0.9782 |
| Synthetic | 0.0054 | 45.3221 | 0.9790 |

Figure 5 shows the results of RED-CNN using different numbers of synthetic training images. With the increase of training data, the image quality can be gradually improved. Table 2 is the corresponding quantitative results. More training data led to better quantitative metrics.

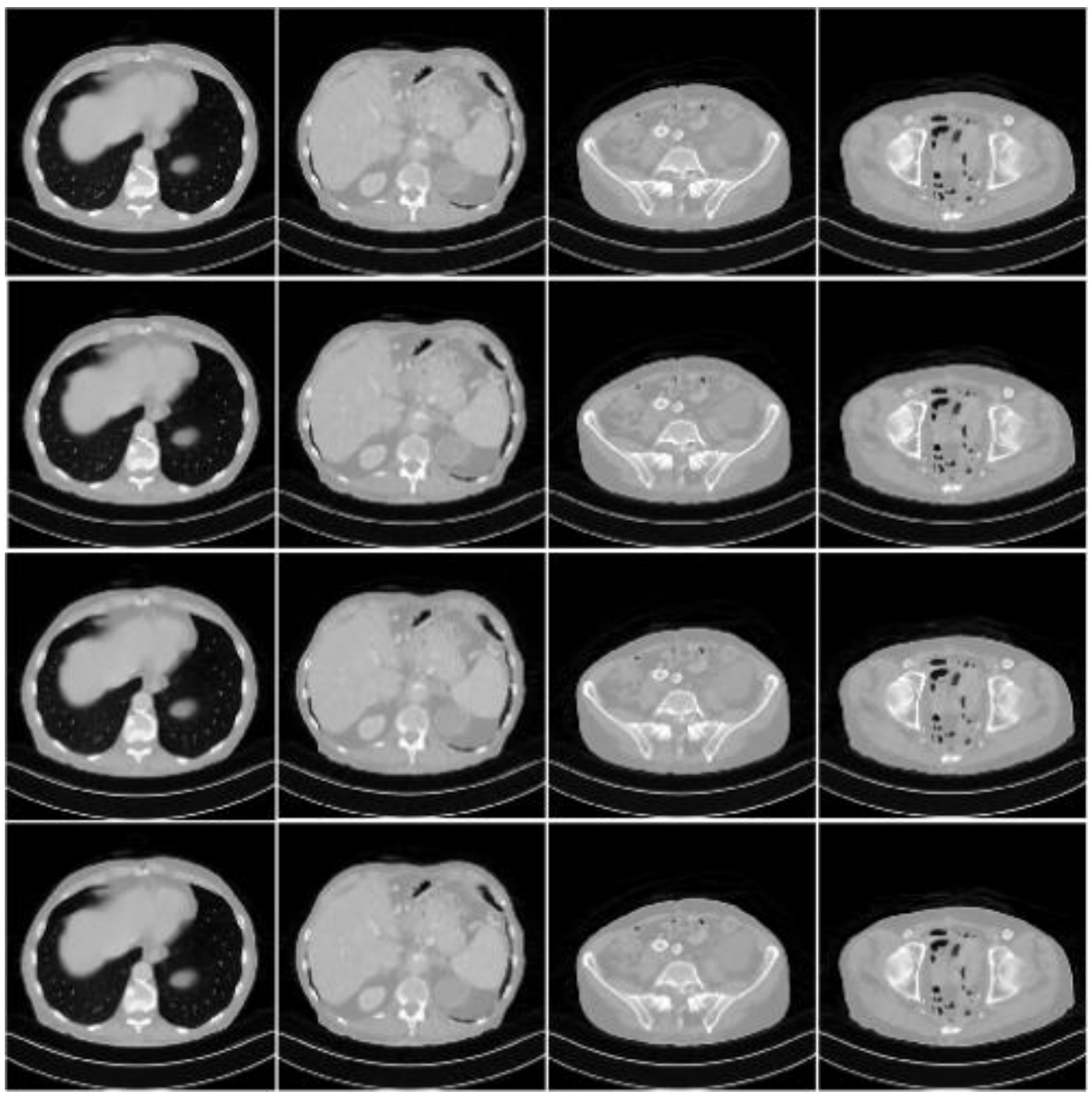

Figure 5. Four representative images reconstructed by VAE using different numbers of training images. From top to bottom are the results using the real training data, 2378, 2588, and 4000 synthetic training images respectively.

Table 2. Quantitative results of RED-CNN using different numbers of training images.

| Methods | RMSE | PSNR | SSIM |
|---|---|---|---|
| Original Data | 0.0056 | 45.0165 | 0.9782 |
| Synthetic 2378 | 0.0054 | 45.3221 | 0.9790 |
| Synthetic 2588 | 0.0053 | 45.5661 | 0.9785 |
| Synthetic 4000 | 0.0052 | 45.8011 | 0.9805 |

## 4 Discussion

In the synthetic dataset, there are some images that could be identified because they have nonideal features. Figure 6 shows four represent synthetic images with unnatural appearances. The first three images show blurred edges. The last image suppressed edge blurring but structural distributions are not consistent with real anatomy/pathology. It should be noted that although the problematic images are not globally consistent with the real situation, the local structures are the same as the real image and seem sufficient for training LDCT denoising networks. The deep learning-based LDCT denoising methods usually extract features in a limited field of view (FOV). We believe that the impact of the global nonideal distribution in Figure 6 on these LDCT reconstruction methods is quite limited. Hence, we did not exclude these suboptimal images from the training dataset.

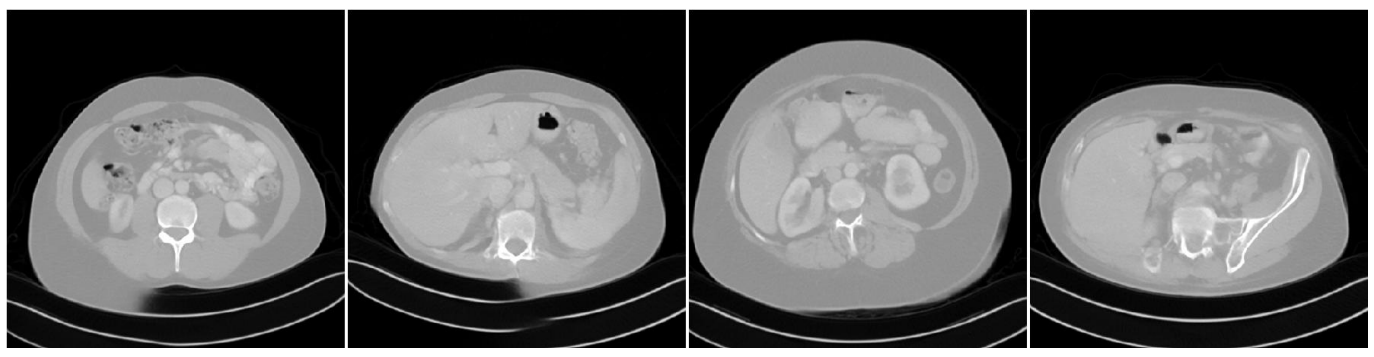

Figure 6. Examples of suboptimal samples in the synthetic image dataset.

## 5 Conclusion

In this study, we have combined the variational autoencoder and the latent diffusion model to generate a synthetic dataset in the context of deep learning-based LDCT denoising. Our experiments have demonstrated encouraging results in CT image synthesis and denoising. It seems that we can use either original or synthetic LDCT images in training deep networks for LDCT denoising without any significant difference in the denoising performance. The most important point we would like to make is that theoretically speaking, the statistical distribution of original data is the same as that of synthetic data but the latter completely blocks privacy leaking, successfully addressing the privacy/security challenge in the era of data science. With the latent diffusion model, the synthetic dataset can be infinitely large, which breaks the limitation of CT images acquisition difficulty. Our experiments have also demonstrated that when we train RED-CNN on both synthetic and real datasets, the resultant performance can be even better. In the future work, we plan to train more advanced large networks and further evaluate the equivalence between original and synthetic data and the superiority of augmented datasets.

# Task-based Assessment of Deep Networks for Sinogram Denoising with A Transformer-based Observer

Yongyi Shi[1], Ge Wang[1], and Xuanqin Mou[2]

[1]Biomedical Imaging Center, Rensselaer Polytechnic Institute, Troy, NY, USA
[2]Institute of Image Processing and Pattern Recognition, Xi'an Jiaotong University, Xi'an, Shaanxi, China

**Abstract** A variety of supervised learning methods are available for low-dose CT denoising in the sinogram domain. Traditional model observers are widely employed to evaluate these methods. However, the sinogram domain evaluation remains an open problem for deep learning-based low-dose CT denoising and other tasks. Since each lesion in medical CT images corresponds to a narrow sinusoidal strip in the sinogram domain, here we propose a transformer-based model observer to evaluate sinogram-domain-based supervised learning methods. The numerical results indicate that our transformer-based model well-approximates the Laguerre-Gauss channelized Hotelling observer (LG-CHO) for a signal-known-exactly (SKE) and background-known-statistically (BKS) task. The proposed model observer is employed to assess two classic CNN-based sinogram denoising methods. This transformer-based observer model has potential to be further developed as a guidance for deep analysis in the sinogram domain.

## 1 Introduction

Low-dose CT reconstruction is a hot topic in the CT field. One natural way is to preprocess sinogram data before image reconstruction, given the merit that the quantum noise is element-wise independent, and then the processed data can be fed to a reconstruction algorithm. Recently, convolutional neural networks (CNNs) were developed for supervised learning-based sinogram denoising [1, 2]. These networks are trained by minimizing the error between low-dose and full-dose sinograms, achieving high performance in terms of image quality metrics such as root mean square error (RMSE). However, it is well-known that such metrics may not consider the noise correlation well, which always present in CT images. To take the noise correlation in CT images into account, the modulation transfer function (MTF) and noise power spectrum (NPS) are used to evaluate image quality. However, MTF and NPS would not be proper to evaluate image quality obtained using non-linear image reconstruction methods, such as CNN-based networks.

To assess image quality from CNN-based CT reconstruction, channelized Hotelling observer (CHO) is an appropriate choice for optimizing the parameters of a CNN for signal detection tasks (e.g., detection of a lesion), where the CHO with Laguerre-Gauss channels (LG-CHO) is common in scenarios where the signal is known exactly and rotationally symmetric [3]. However, the CHO models are still linear observers, which is suboptimal compared to the ideal observer (IO) performance for detection tasks with CT images. Recent efforts have been primarily focused on training network-based model observes [4], which provide an alternative approach to CT image quality assessment. Theoretically, an ideal observer achieves the upper bound of the image quality assessment performance on the detection task. In this regard, an ideal observer that evaluates a sinogram denoise algorithm should be designed to assess the best signal detectability directly from sinogram data. By doing so, the dedicated observer would not be influenced by an image reconstruction algorithm and its parameters. Indeed, the variation in the reconstruction algorithm may affect partial volume, beam hardening, directional noise artifacts, motion artifacts, and so on. Developing a sinogram-based ideal observer involves numerous steps that bring all above-mentioned factors together. Could such a sinogram-based observer model be well established? Although existing studies suggested that it would be possible to detect a specific object from sinogram data, the observer model must be well designed to accurately simulate the human observer performance, which is rather complicated.

Our observer model focuses on the object detection task in a local region, which discriminates a local signal from its background in reference to the recognizing capability of the human vision system. The main challenge of building the model observer is that in the sinogram domain a local image structure corresponds to a narrow sinusoidal strip and all the trips of structures are overlapped to form the entire sinogram. This property of the sinogram does not match the convolutional nature of CNNs, since they specifically extract local structural information. In recent years, the transformer architecture has been applied in the field of computer vision, including image quality assessment [5]. The transformer architecture is applicable in the sinogram domain [2], since transformer designs a self-attentional mechanism to capture global interactions.

In this paper, we propose a transformer-based model observer in the sinogram domain for SKE/BKS detection, where the background is known statistically due to random background artifacts. Note that our motivation is to evaluate the influence of artifacts and noise in the supervised learning mode, where the full-dose sinogram is known exactly. Hence, the background artifacts can be obtained by subtracting the full-dose sinogram from the corresponding low-dose sinogram. We implement the transformer-based model observer in these background artifacts. The proposed transformer-based model observer is then employed to assess modern supervised CNN-based sinogram denoising methods. The canonical CNN-based denoising methods are identified for analysis, including DnCNN [1] and RED-CNN [6].

This work was partially supported by the NIH/NCI Grant R01EB032716 and NSFC No. 62071375.

## 2 Materials and Methods

### 2.1 Laguerre-Gauss Channelized Hotelling Observer

To evaluate detectability, we conduct the two alternative forced choice (2-AFC) detection task. The hypotheses for signal-absent ($H_0$) and signal-present ($H_1$) are given by:

$$H_0 : \mathbf{g} = \boldsymbol{f_b} + \boldsymbol{f_n} \quad (1)$$

$$H_1 : \mathbf{g} = \boldsymbol{f_s} + \boldsymbol{f_n} \quad (2)$$

where $\mathbf{g}$ is a column vector of a given image, $\boldsymbol{f_s}$ is a signal-present background, $\boldsymbol{f_b}$ is a signal-absent background, $\boldsymbol{f_n}$ is noise. Our goal is to assess the quality of CT images in the SKE/BKS detection task, where the background is known statistically, which is defined by random background artifacts.

For assessment of CT image quality, the HO provides the upper bound of the detection performance among all linear model observers. To avoid estimating a large covariance matrix needed in the HO, the CHO approximates the HO performance with efficient channels of much-reduced dimensionality. In the CHO, an image $\mathbf{g}$ is transformed to $\boldsymbol{v}$ by

$$\boldsymbol{v} = \boldsymbol{U}\mathbf{g} \quad (3)$$

where $\boldsymbol{U}$ is the channel matrix. LG-CHO uses a LG function for the channel matrix and can approximate the CHO performance well for a rotationally symmetric signal in a known location. The CHO template estimated from training images is computed by:

$$\boldsymbol{w} = \boldsymbol{\Delta v}^T \boldsymbol{K_v}^{-1} \quad (4)$$

where $\boldsymbol{\Delta v}$ is the mean difference between the signal-present and signal-absent transformed images, and $\boldsymbol{K_v}^{-1}$ is the image covariance matrix, $T$ denotes the transpose. The CHO decision variable can be computed by

$$d = \boldsymbol{w}^{\boldsymbol{T}}\boldsymbol{v} \quad (5)$$

### 2.2 Low-dose CT data preparation

We used the full-dose CT slice of 3 mm thickness from Mayo clinical dataset. Figure 1 shows the flowchart for generating a signal-present image in liver region. Specifically, we inserted a circularly shaped signal near the center of each region of interest (ROI) with an elevated CT value by 20 Hounsfield Unit (HU) to present a challenge of low contrast lesion detection. After forward projection, we obtained the full-dose signal-present sinogram, where the signal is a narrow sinusoidal strip in the sinogram domain. Poisson noise was superimposed to the sinogram to produce the low-dose signal-present sinogram. The filtered backprojection (FBP) algorithm was then employed for image reconstruction. The background artifacts were obtained by subtracting the full-dose signal-absent sinogram/image from the low-dose signal-present sinogram/image. After then, ROIs of 64 pixels × 64 pixels were extracted from the image to evaluate the LG-CHO method. In the sinogram domain, sinusoidal strips of 64 bins × 1160 views were extracted and reshaped to a matrix as the signal-present data for analysis by the transformer-based model observer.

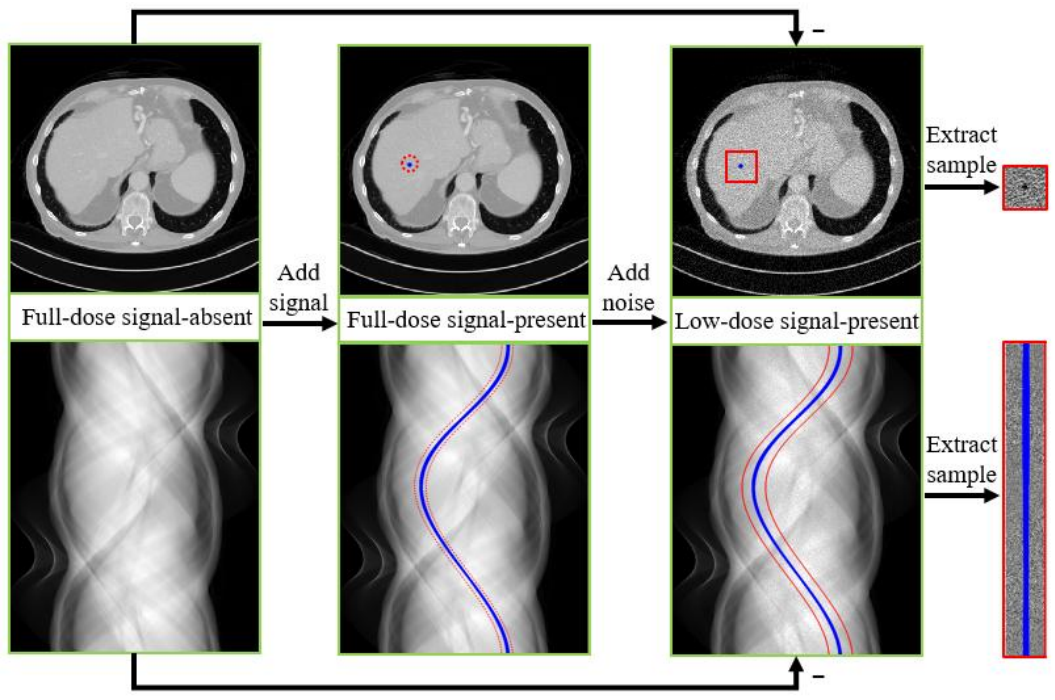

Figure 1. The flowchart for data generation.

### 2.3 Transformer-based model observer

The transformer architecture is featured by a global attention mechanism, and thus can be used to extract sinogram features from different view angles, outperforming the CNN-based approach. In traditional vision transformer, image is cut into patches as input tokens to capture their spatial relationship in nature images. Since each sinogram view represents one measurement of an imaging target, modeling relations between these views can help the network learn correlation among different views. Thus, we split each sinogram view as an input token. Figure 2 shows the top-level structure of the proposed transformer-based model observer. We modeled the detection task as a classification problem. For the 2-AFC detection task, the randomly selected signal-present and signal-absent sinogram pair serve as the input to the transformer. Note that $\dot{F}'\in\mathbb{R}^{H\times D}$ / $F'\in\mathbb{R}^{H\times D}$ randomly represents the signal-present/signal-absent or signal-absent/signal-present sinogram pairs, where $H$ and $D$ are the number of detector bins and views respectively. As often used in the vision transformer, an extra embedding $F_0$ is appended to the beginning of the input sinogram. we add the learnable position encodings $P\in\mathbb{R}^{(H+1)\times D}$ to the corresponding $F\in\mathbb{R}^{(H+1)\times D}$ to keep the position information. The output of the encoder has the same size to the input. The decoder takes another sinogram with an extra embedding as the input. The output of the encoder is used as an input of the decoder in the second MHA layer. The output of the decoder is finally obtained to feed into the following MLP head. The MLP head consists of two fully-connected (FC) layers, and the first FC layer is followed by the ReLU activation. The second FC layer is followed by the Sigmoid activation to predict the labels of the sinogram pair.

During network training, the initial learning rate was 0.00005, with the cosine learning rate decay. The entropy cross loss was used for training. The number of epochs was set to 200. To train our network, we used 5,000 signal-present sinograms and 5,000 signal-absent sinograms for the 2-AFC task. We divided the generated datasets by a 19:1 ratio for training and validation datasets.

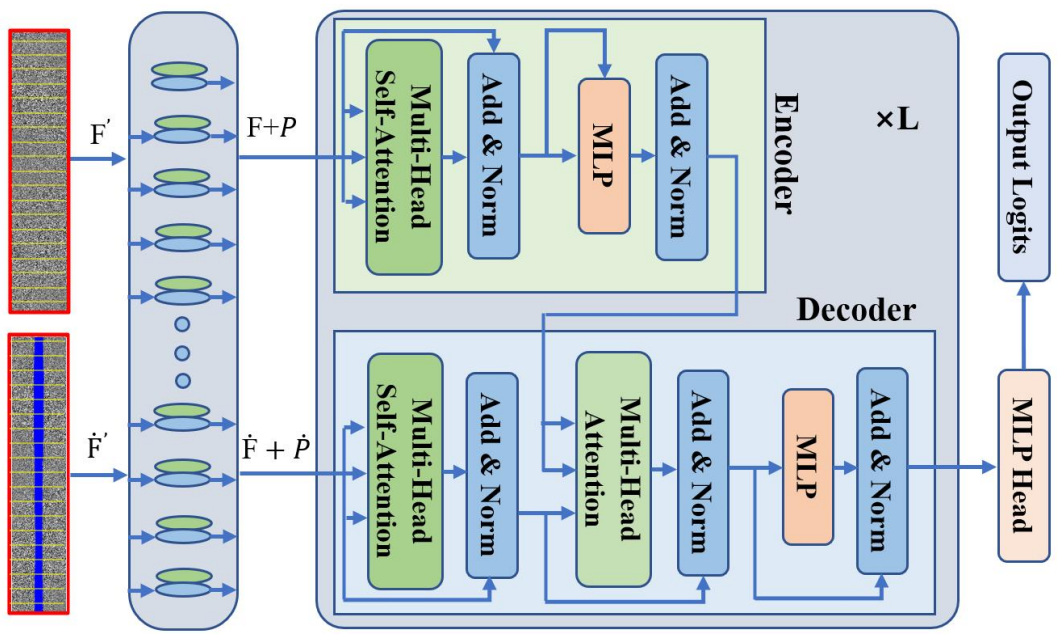

Figure 2. The architecture of the proposed transformer-based model observer.

**2.5 Evaluation of the detection performance**

To evaluate the proposed model observer applied to deep learning-based low-dose CT algorithms, the two classical networks were employed for projection domain denoising: RED-CNN and DnCNN. We used the percent correct ($P_c$) as a detection performance measure for the model observer, which is defined as

$$P_c = \frac{1}{N_t}\sum_{i=1}^{N_t} o_i \tag{6}$$

where $N_t$ is the number of test trails and $o_i$ is binary decision variables, where $o_i$ can only be 1 or 0, corresponding to correct or incorrect detection results. To obtain $o_i$, we compare the correct answer with the logit of the network output. In the case of LG-CHO, we select the image which has the largest decision variable between 2 input images and compare it with the correct answer. We use 1,000 sinogram or image pairs as the test dataset to evaluate the performance of our proposed transformer-based model observer or LG-CHO, respectively.

## 3 Results

**3.1 Detectability in different noise levels**

Figure 3 shows some samples in the training dataset. The circular signal was inserted in the center of each ROI in the artifact images to simulate a challenge of low contrast lesion detection. The signal in sinogram corresponds a narrow strip, and is difficult to distinguish visually in both artifact images and sinograms because of strong noise.

Figure 4 shows the performance curves of the signal detectability versus the dose level. It can be seen from the curves that the performance trends among the LG-CHO and transformer-based model observer are similar. This simiarity shows that our proposed transformer-based model observer can be used to evaluate the signal detection task in the sinogram domain.

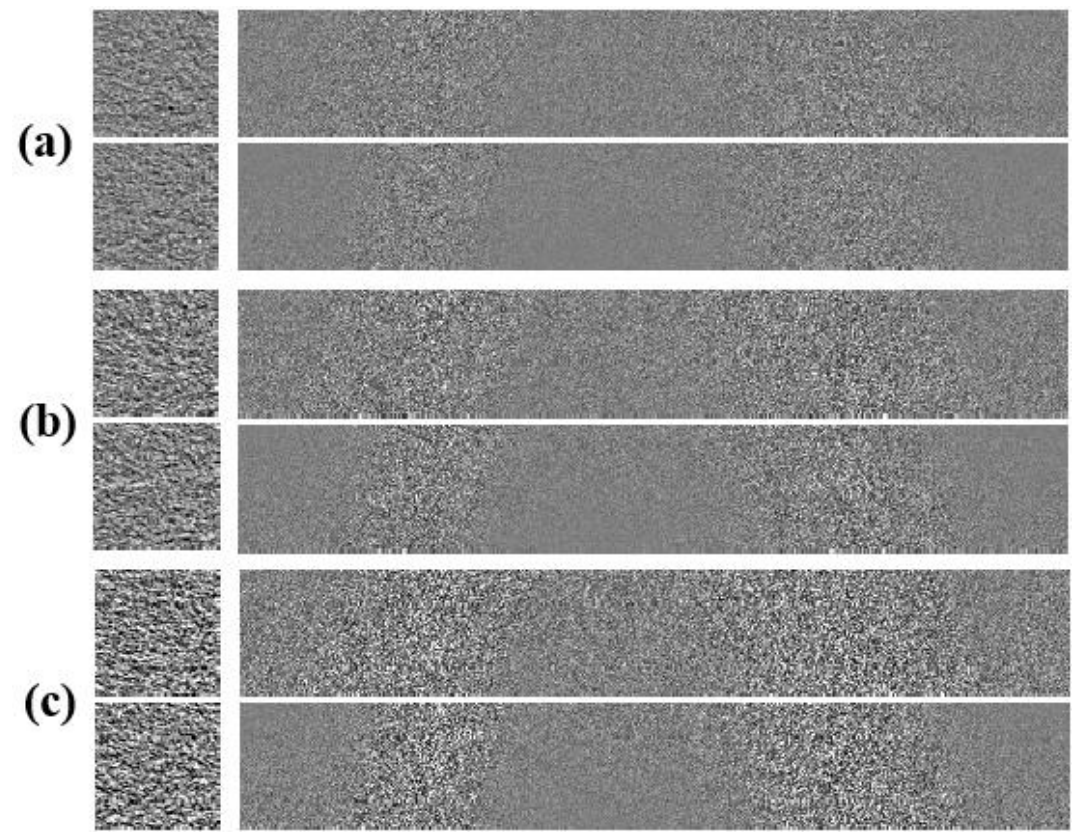

Figure 3. Representative samples in the training dataset. From (a)-(c) are the images/sinograms with 100, 50, 25 thousand incident photons per detector element. In each group of (a)-(c), the top is the signal-present images/sinograms, bottom is the signal-absent images/sinograms. The left is the ROIs from the images in a display window [-0.005, 0.005], while the right is the ROIs from the sinograms in a display window [-0.5, 0.5].

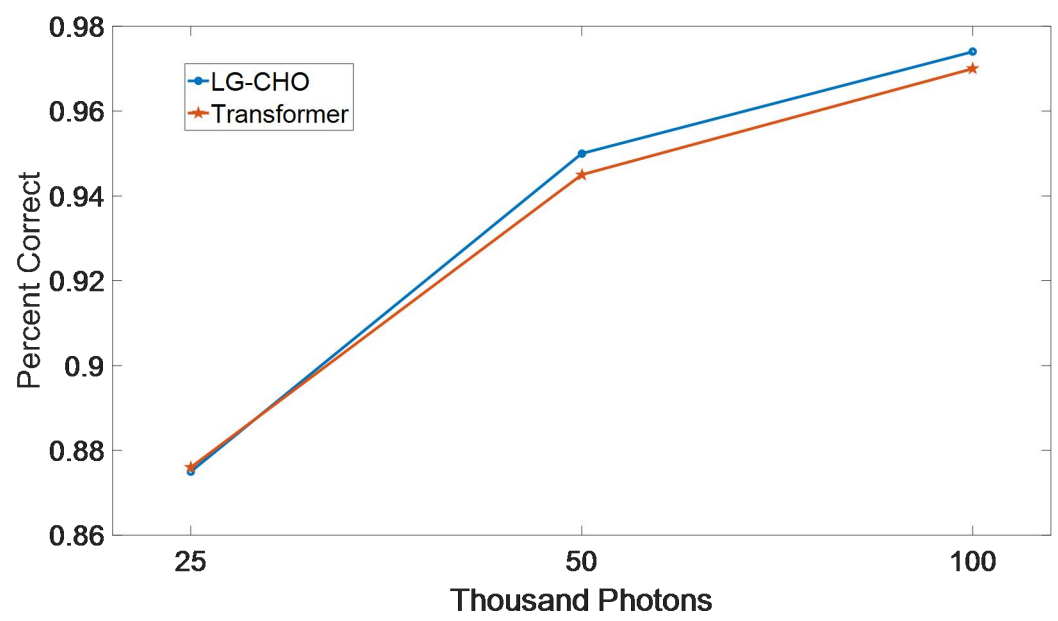

Figure 4. The lesion detectability with respect to the dose level.

**3.2 Evaluation of the Denoising networks**

Figure 5 shows the sinograms/images with 25 thousand incident photons, using RED-CNN and DnCNN respectively. The noise was suppressed by the denoising networks. Table I shows the quantitative results using different denoising networks on the whole testing set. It can be seen that the RED-CNN method produced the images with the highest peak signal-to-noise ratio (PSNR), root-mean-square error (RMSE) and structure similarity index (SSIM). DnCNN also improved on these quantitative metrics. However, some fine details were blurred. For example, the dots indicated by the red arrow can be seen in the original low-dose CT image but were blurred by the denoising networks. This is because both RED-CNN and DnCNN employ the MES loss to train the network, which does not guarantee the task-specific optimality; e. g., lesion detection. Especially for DnCNN, the introduced secondary artifacts clearly degraded the image quality.

Figure 6 shows typical samples in the training dataset. After processed by either RED-CNN or DnCNN, the noise was suppressed. However, some features were somehow distorted or blurred in the residual sinogram. Both blurred details in the image domain and stretched structures in the residual projection domain compromised the performance of the numerical observer models.

Figure 7 shows the performance curves of the signal detectability associated with different methods. It can be seen that the performance trends for the LG-CHO and transformer-based model observer are in excellent agreement. Interestingly, the proposed transformer-based model observer performed better than LG-CHO. However, the detectablitity decreased after the image was prepossed by the denoising networks, which is consistent to the visual inspection. The fine details were removed by the denosing networks, which reduced the detectability. Hence, it is desireable to propose a method that can improve the metrics of RMSE/PNSR/SSIM and the LG-CHO/Transformer-based observer simultaneously.

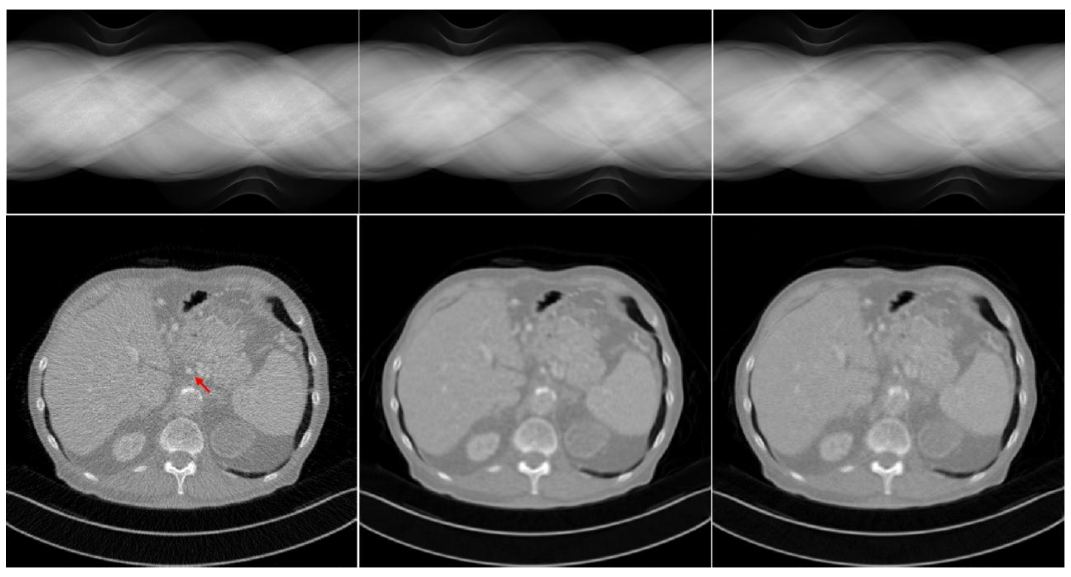

Figure 5. From left to right are the sinograms/images with 25 thousand incident photons using RED-CNN and DnCNN respectively.

Table I. Quantitative results from different denoising networks.

| Methods | Low dose | RED-CNN | Dn-CNN |
|---|---|---|---|
| RMSE | 0.0069 | 0.0025 | 0.0038 |
| PNSR | 43.4594 | 52.0860 | 48.3616 |
| SSIM | 0.9608 | 0.9981 | 0.9879 |

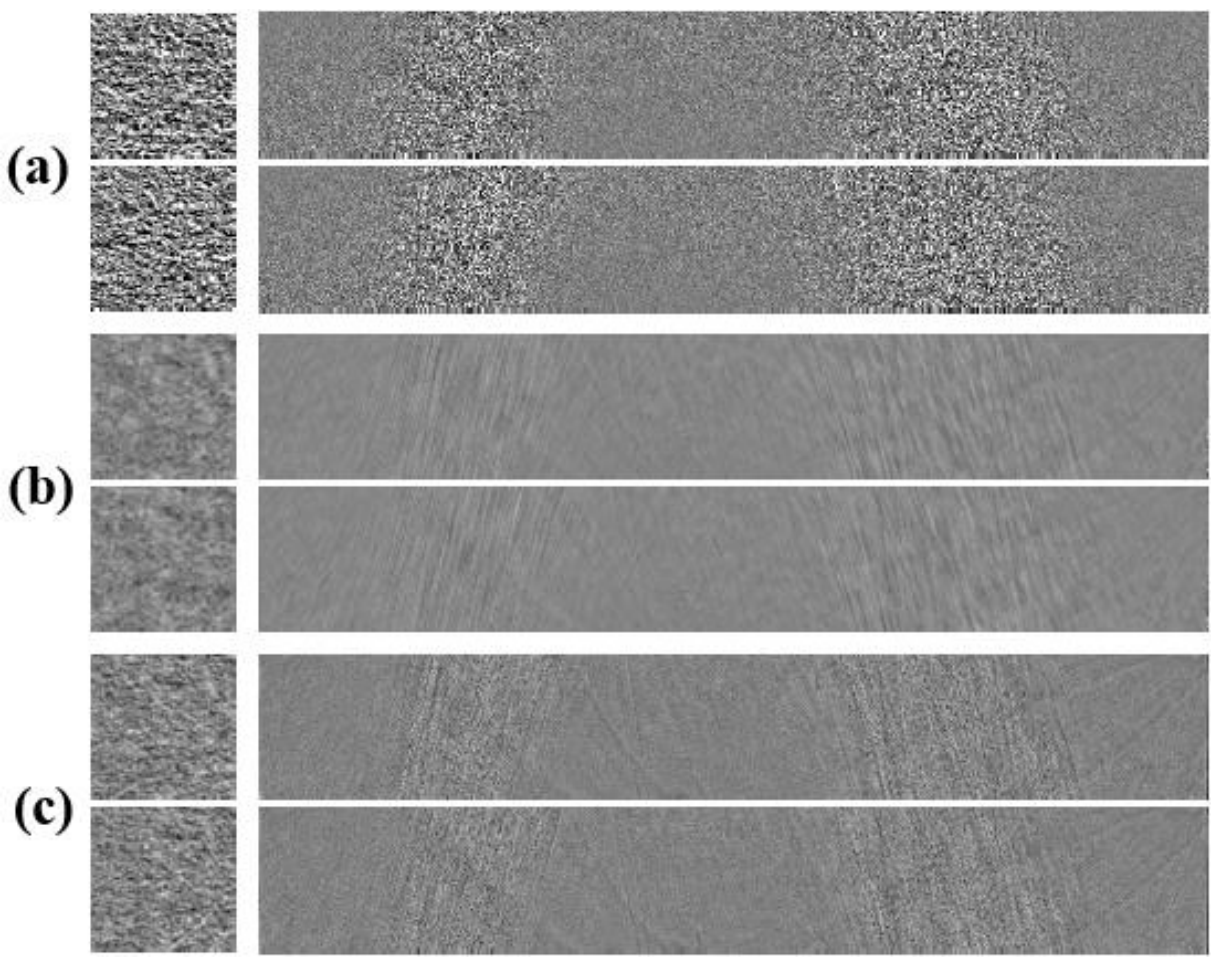

Figure 6. Typical samples in the training dataset. (a)-(c) The images/sinograms with 25 thousand incident photons per detector element, with RED-CNN and DnCNN respectively. For each group of (a)-(c), the top is the signal-present images/sinograms, while the bottom is the signal-absent images/sinograms. The left shows the ROIs from the images in a display window [-0.005, 0.005], while the right is the ROIs from the sinograms in a display window [-0.5, 0.5].

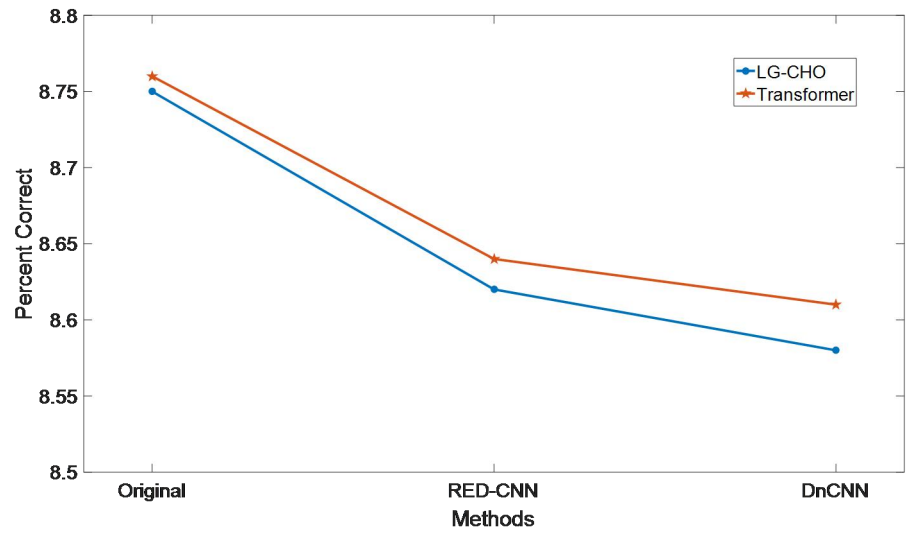

Figure 7. The lesion detectability with respect to the dose level.

## 4 Discussion and Conclusion

In this study, we have demonstrated that the transformer-based model observer yields a performance similar to the LG-CHO regarding a residual background. Hence, the reference images are required to calculate the residual background. However, in practice the reference images are usually difficult to obtain. Furthermore, the proposed transformer-based method targets the ideal observer, which is still not the same as the human observer. Implementing an anthropomorphic model observer using a transformer-based model is another interesting topic, which will not rely on reference images and bring the results closer to what the human observer achieves.

In conclusion, we have proposed a transformer-based model observer to evaluate low-dose CT sinogram denoising methods. With the transformer, the proposed observer model has yielded a performance similar to that of the LG-CHO model. However, LG-CHO has a strict assumption (i.e., a rotationally symmetric signal, a known location, and a stationary background) to approximate the ideal linear observer performance. In contrast, the transformer-based model is less restrictive. Additionally, the evaluation in the sinogram domain can avoid the image reconstruction process. As a follow-up study, we can build an ideal observer model based on sinogram data so that we could approach the upper bound evaluation of the performance of deep sinogram denoising methods.

# 3D PET-DIP Reconstruction with Relative Difference Prior Using a SIRF-Based Objective

Imraj RD. Singh[1,2], Riccardo Barbano[1], Željko Kereta[1], Bangti Jin[1], Kris Thielemans[2], and Simon Arridge[1]

[1]Department of Computer Science, UCL, UK
[2]Institute of Nuclear Medicine & Centre for Medical Image Computing, UCL, UK

**Abstract**: Deep Image Prior (DIP) is an unsupervised deep learning technique that does not require ground truth images. For the first time, 3D PET reconstruction with DIP is cast as a single optimisation via penalised maximum likelihood estimation, with a log-likelihood data-fit and an optional Relative Difference Prior term. Experimental results show that although unpenalised DIP optimisation trajectory performs well in high count data, it can fail to adequately resolve lesions in lower count settings. Introducing the Relative Difference Prior into the objective function the DIP trajectory can yield notable improvements.

## 1 Introduction

Deep Image Prior (DIP) [1] is a state-of-the-art unsupervised deep learning method for image reconstruction. It leverages the inductive bias of Convolutional Neural Networks (CNNs) to fit to natural signals faster than to noise, allowing regularisation via early stopping along the optimisation trajectory.

Gong et al. [2] were the first to apply DIP to PET reconstruction by splitting the reconstruction into Expectation Maximisation (EM) and DIP denoising. Splitting the optimisation was necessary as the PET forward model was not integrated into a deep learning framework. This was subsequently done by Hashimoto et al. [3] and DIP was implemented as a single optimisation problem, thus reducing the number of hyperparameters and the computational overhead, and simplifying implementation. But, the forward model was stored as a sparse matrix which had an excessive GPU memory overhead. Furthermore, mean-squared-error was used as the data-fidelity. Their work was recently extended to 3D PET [4] through slicing the forward operator and solving with a subset-based block iterative approach.

In this work we implement DIP for 3D PET as a single optimisation problem with log-likelihood data-fit and an optional penalisation term. Our implementation uses the wrapper developed in [5]. Projectors utilised are implicit, thus alleviating the large GPU overhead associated with explicit projection and allowing full gradient updates for 3D PET data. The implementation is tested on realistic simulated data with two count levels. Results are compared to solutions from the provably convergent Block-Sequential Regularised Expectation Maximisation (BSREM) algorithm.

## 2 Preliminaries

### 2.1 Penalised Maximum Likelihood

Penalised maximum likelihood methods for PET image reconstruction aim to solve the following optimisation problem:

$$\mathrm{argmin}_{\mathbf{x}\geq 0}\left\{\Phi(\mathbf{x}) = -L(\mathbf{y}|\mathbf{x}) + \beta R(\mathbf{x})\right\}, \qquad (1)$$

where $L(\mathbf{y}|\mathbf{x})$ is the Poisson log-likelihood describing the goodness of fit of the reconstructed image $\mathbf{x}\in\mathbb{R}^N_{\geq 0}$ to the measurements $\mathbf{y}\in\mathbb{R}^M_{\geq 0}$; $M$ and $N$ denote the number of projection bins and image voxels, respectively; $R(\mathbf{x})$ is the penalty, and $\beta > 0$ balances the data-fit and penalty. Up to an additive constant, the Poisson log-likelihood is given by: $L(\mathbf{y}|\mathbf{x}) = \sum_{i=1}^M y_i \log(\bar{y}_i(\mathbf{x})) - \bar{y}_i(\mathbf{x})$. The mean of the measurements $\bar{\mathbf{y}}$ is obtained by projecting the reconstructed image with an affine PET forward model, defined by $\bar{\mathbf{y}}(\mathbf{x}) = \mathbb{E}[\mathbf{y}] = \mathbf{A}\mathbf{x}+\bar{\mathbf{b}}$. The system matrix $\mathbf{A}$ models the PET scanner characteristics as well as physical phenomena, e.g., attenuation and positron range. The expected background events $\bar{\mathbf{b}}$ include both scatter and randoms.

In this work we consider the Relative Difference Prior (RDP) [6], defined by: $R(\mathbf{x}) = \sum_{i=1}^N \sum_{j\in\mathcal{N}_i} w_{ij} \frac{(x_i-x_j)^2}{x_i+x_j+\gamma|x_i-x_j|}$, where $\mathcal{N}_i$ is a $3\times3\times3$ neighbourhood of the $i$-th image voxel, $w_{ij}$ are the neighbourhood weights. The edge preservation parameter is set as $\gamma = 2$, as is standard in a clinical setting.

### 2.2 Block-Sequential Regularised Expectation Maximisation

BSREM [7] is a provably convergent subset algorithm for PET image reconstruction, with an iterative update given by:

$$\mathbf{x}_{k+1} = P_{\mathbf{x}\geq 0}\left[\mathbf{x}_k - \alpha_{k,n} D(\mathbf{x}_{k,n})\nabla\Phi_{m_k}(\mathbf{x}_k)\right], \quad k\geq 0. \qquad (2)$$

Here $\Phi_{m_k}$ is a subset gradient and $m_k\in\{1,\ldots,n_{\text{subsets}}\}$ is the index of a subset chosen at image update $k$, out of $n_{\text{subsets}}$ subsets. The epoch number $n\geq 0$ is incremented after every $n_{\text{subsets}}$ image updates. The step-size is given by $\alpha_n$, preconditioner by $D(\cdot)$, and $P_{\mathbf{x}\geq 0}[\cdot]$ is a non-negativity projection. The EM preconditioner is used; $D(\mathbf{x}_{k,n}) = \mathrm{diag}\left\{(\mathbf{x}_{k,n}+\delta)/\mathbf{A}^\top\mathbf{1}\right\}$, where $\delta = $ 1e-9 ensures positive definiteness, $\mathbf{A}^\top\mathbf{1}$ is the sensitivity image, and $\mathbf{x}_{k,n}$ is the reconstruction at epoch $n$. The step-size is computed with $\alpha_{k,n} = \alpha_0/(\eta n+1)$, where $\alpha_0 = 1$ is the initial step size and $\eta$ is a relaxation coefficient.

### 2.3 Deep Image Prior

DIP [1] represents $\mathbf{x}$ through learnable parameters $\theta\in\mathbb{R}^p$ of a CNN $\mathbf{f}(\mathbf{z};\theta)$ with a fixed random input $\mathbf{z}$. The optimisation problem (1) is then recast as: $\theta^\star\in\mathrm{argmin}_{\theta\in\mathbb{R}^p}\Phi(\mathbf{f}(\mathbf{z};\theta))$, where the reconstructed image is obtained from $\mathbf{x}^\star = \mathbf{f}(\mathbf{z};\theta^\star)$. In the original DIP work [1, 2] the objective function $\Phi$ consists solely of the likelihood term where the regularisation is imparted through network architecture and stopping criteria. A penalisation can be included to alleviate the lack of robust stopping criteria, which is critical to prevent overfitting to noise. The utility of additional penalisation was first investigated for CT [8] and was included for PET in [4], although the latter did not compare against traditional penalised maximum likelihood solutions.

## 3 Methods

### 3.1 Wrapping the SIRF-Objective

SIRF is a multi-modality synergistic reconstruction framework providing access to several well-established reconstruction engines. For advanced PET and SPECT reconstruction the Software for Tomographic Image Reconstruction (STIR) engine is used [9]. In this work we utilise various features of STIR through SIRF such as the parallelised C++ backend, access to 3D GPU-based projectors, and access to clinically relevant PET penalties (e.g. RDP).

The wrapper integrates SIRF into PyTorch, via exposure of `sirf.STIR.ObjectiveFunction.value` and `sirf.STIR.ObjectiveFunction.get_gradient` methods in a custom autograd function that subclasses `torch.autograd.Function`.

### 3.2 Synthetic Data Generation and System Modelling

A Monte-Carlo photon emission simulation of a voxelised XCAT torso [10] with GE Discovery 690 scanner modelled acquisition was performed using OpenGATE [11], STIR [9] and STIR-GATE-Connection [12]. The distribution of back-to-back 511 keV photon emissions is representative of activity concentrations from a $^{18}$F-FDG tracer study. Cylindrical hot lesions of dimensions 1cm diameter by 1cm length were inserted into the abdominal wall (1.6:1), liver (1.3:1), lung (2:1), and spine (1.6:1). The lesion to associated background ratio is indicated by (lesion:background).

Projection data sets containing 250 ("lower") and 1200 ("higher") million coincidence events were acquired. A true-to-background ratio of 0.93:1 was maintained for the datasets. The resulting list-mode data were re-binned into sinograms with 288 projection angles; all ring differences were used as is typical in clinical practice. The reconstruction volume had dimensions of $47 \times 128 \times 128$ with voxel-size $3.27 \times 4.0 \times 4.0$ mm. Normalisation, randoms and scatter were estimated from the Monte-Carlo data and incorporated within the forward model, see [12] for details.

### 3.3 BSREM and DIP Implementation

For BSREM the objective $\Phi(\mathbf{x})$ was split into 32 ordered subsets, accessed in accordance to the Herman-Meyer order [13]. The initial image was set as a reconstruction with ordered subset EM with 24 subsets after 1 epoch. The maximum epoch $n_{\text{epochs}}$ and relaxation coefficient $\eta$ were found through a grid search for both datasets: for higher count data $\eta = 0.02$ and $n_{\text{epochs}} = 1000$, for lower count $\eta = 0.04$ and $n_{\text{epochs}} = 500$. The grid search was assessed to ensure fast convergence and small step-sizes at $n_{\text{epochs}}$.

For DIP a three-scale 3D U-Net [14] was implemented in PyTorch (1.13.0), in-line with previous work [2, 4]. Trilinear upsampling and strided convolutions were used to change scale, with the number of features compensating for the increase/decrease of spatial dimensionality. Batch normalisation and Leaky ReLU were included after each convolution. Skip connections were also present between encoding and decoding paths of the network. ReLU was used on the network output as a non-negativity constraint. An ADAM optimiser was used for training, with initial learning rate of 1.0 and cosine annealing tending to 0 over 20,000 iterations. Two configurations of DIP were implemented; DIP with only Poisson negative log-likelihood objective (referred to as"DIP"), and with RDP in the objective ("DIP+RDP").

### 3.4 Quality Metrics

Standard metrics for quantification and detectability of lesions are used to assess image quality. Contrast Recovery Coefficient (CRC) values are calculated between the lesions and associated background Regions of Interest (ROIs) by: $\text{CRC} = (\frac{\bar{a}}{\bar{b}} - 1)/(\frac{a_t}{b_t} - 1)$, where $\bar{a}$ and $\bar{b}$ are average emissions over lesion and associated background ROIs, respectively. The subscript t denotes ground truth emission values. Standard Deviation (STDEV) was calculated on each of the background ROIs according to: $\text{STDEV} = \left(N_{\text{ROI}}^{-1} \sum (b_i - \bar{b})^2\right)^{1/2}$.

## 4 Results and Discussion

For BSREM results, a set of eleven regularisation values $\beta$ were used for each count level. For higher count $\beta \in [3.125, 31.25] \cdot$ e-3; for lower count $\beta \in [1.5, 15] \cdot$ e-2. The largest and smallest values in the range represent over-penalised and under-penalised solutions respectively. DIP results are shown at different epochs. For DIP+RDP, four regularisation values $\beta$ were used for each count level: higher count $\beta \in \{3.125, 12.5, 21.875, 31.25\} \cdot$ e-3; lower count $\beta \in \{1.5, 6.0, 10.5, 15.0\} \cdot$ e-2. DIP+RDP worked best when $\beta$ was lowest (but non-zero), results are shown in Figs. 1c and 1d. Qualitative visual comparisons of the lower count reconstructions are given in Fig. 2.

In the higher count regime, see Fig. 1a, unpenalised DIP is able to considerably out-perform BSREM across all ROIs. In the lower count data this is not the case, see Fig. 1b. Through the inclusion of RDP in the objective function, the trajectory of DIP is improved significantly such that improved image quality metrics are observed in both lower and higher count data. However, the improvement is not consistent across all lesions. From Figs. 1a and 1c, the CRC of the abdominal wall lesion decreased markedly with the inclusion of RDP for high count data. This could be due to the abdominal wall lesions' location at the edge of the axial field-of-view, where noise is higher as sensitivity is lower. These issues of lesion dependence on local sensitivity, contrast and surrounding activity have been observed and investigated with non-DIP reconstruction [15]. Extension of such work to DIP remains for the future.

A single NVIDIA RTX 3090 with 24GB of dedicated memory (VRAM) was used in this study. PARALLELPROJ [16] was used for the projection operator, both the forward and adjoint are implemented in CUDA (GPU-specific language). One full gradient 3D PET DIP iteration took $\approx 2.4$s, therefore 13.3 hours for the 20,000 iterations. This included

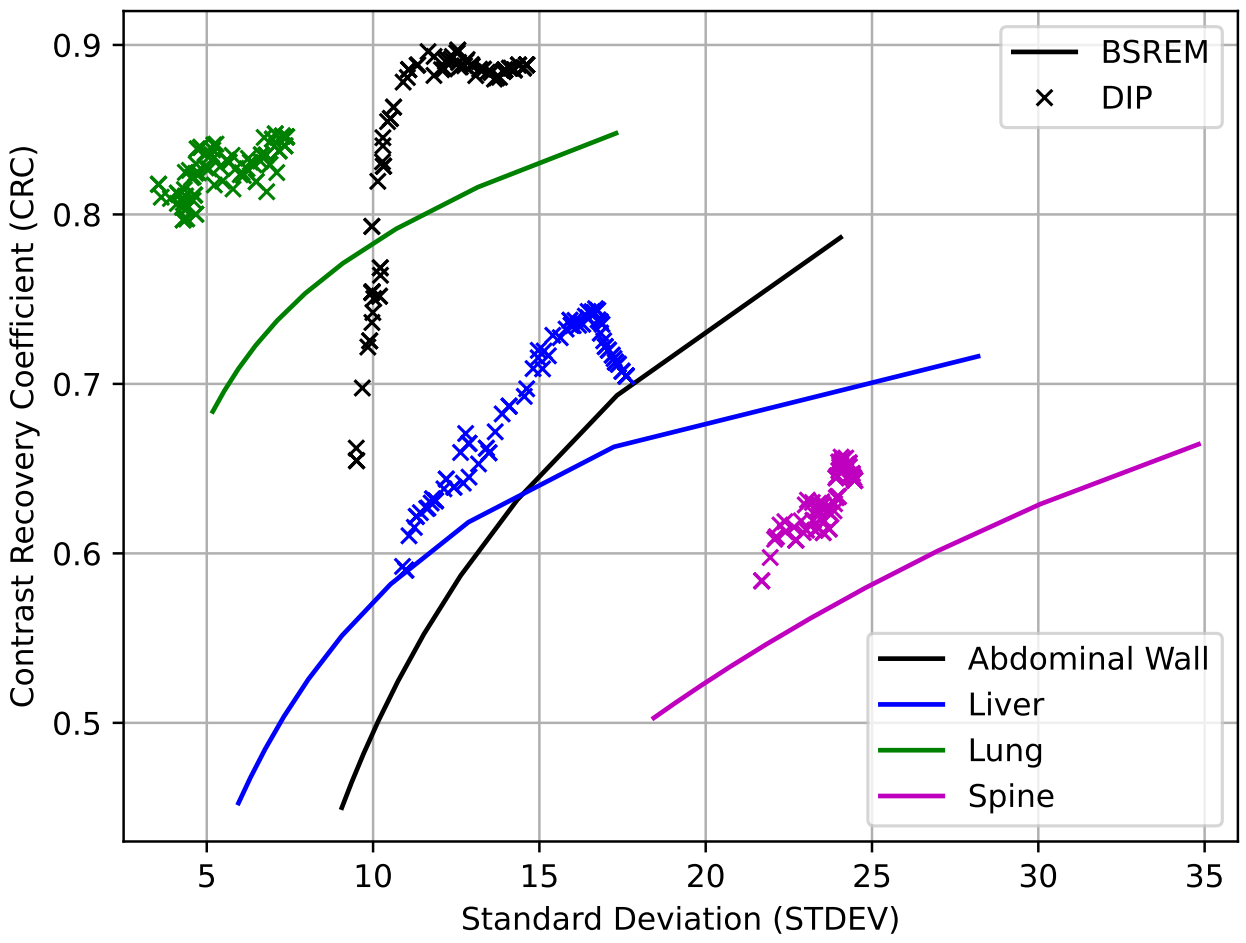

**(a)** Higher count data, DIP with no penalisation

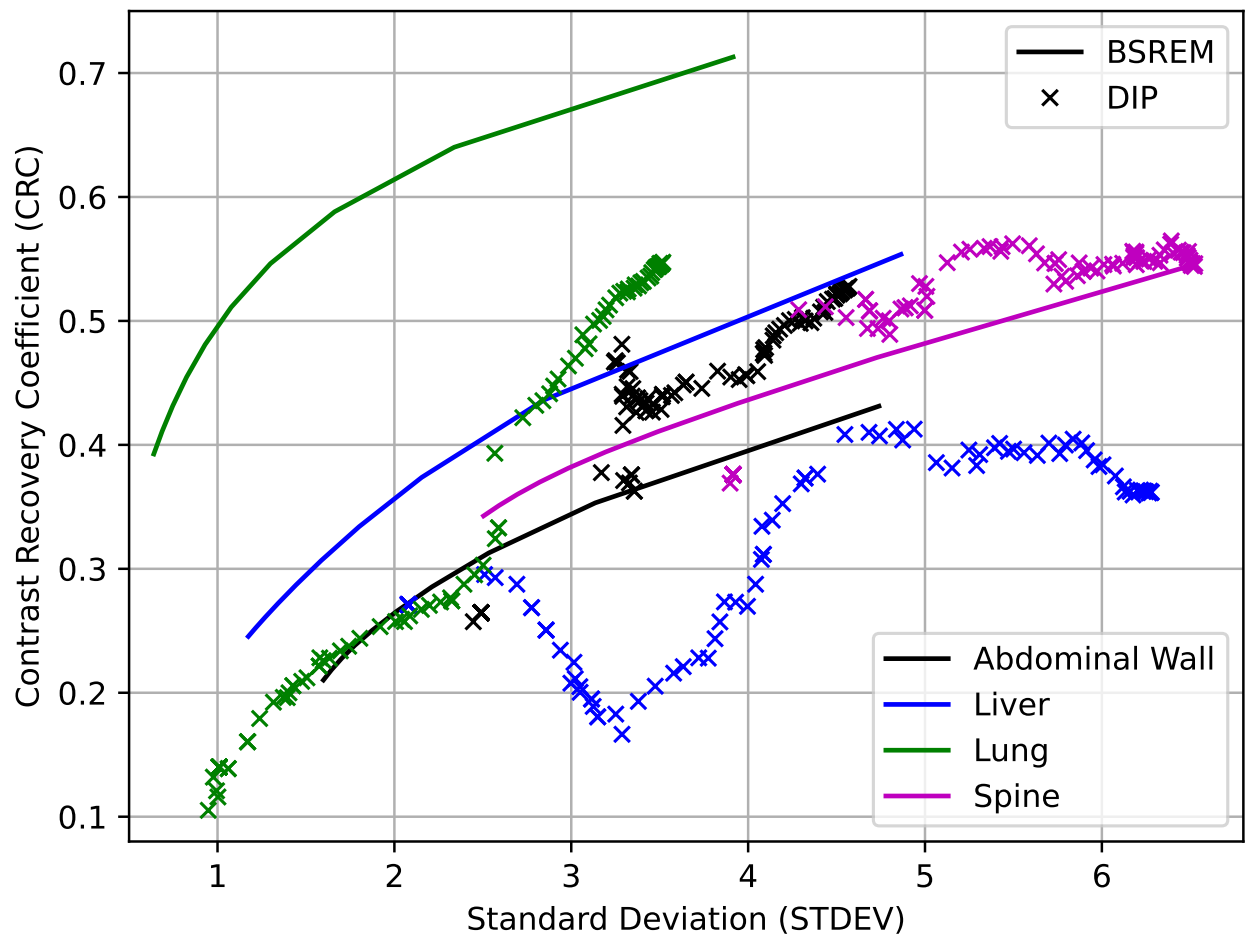

**(b)** Lower count data, DIP with no penalisation

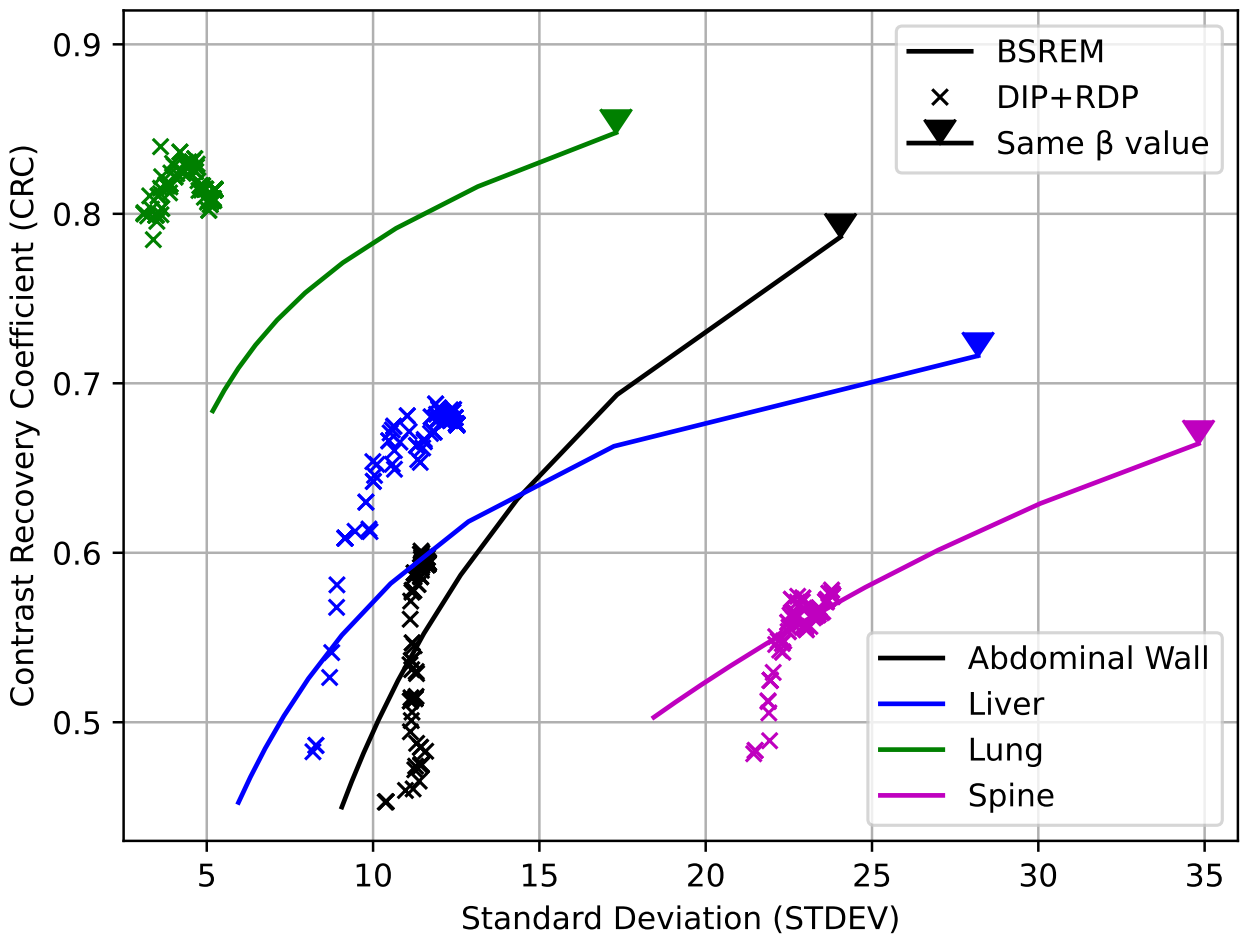

**(c)** Higher count data, DIP+RDP ($\beta = 3.125\text{e-}3$), marker indicates the same $\beta$ BSREM solution

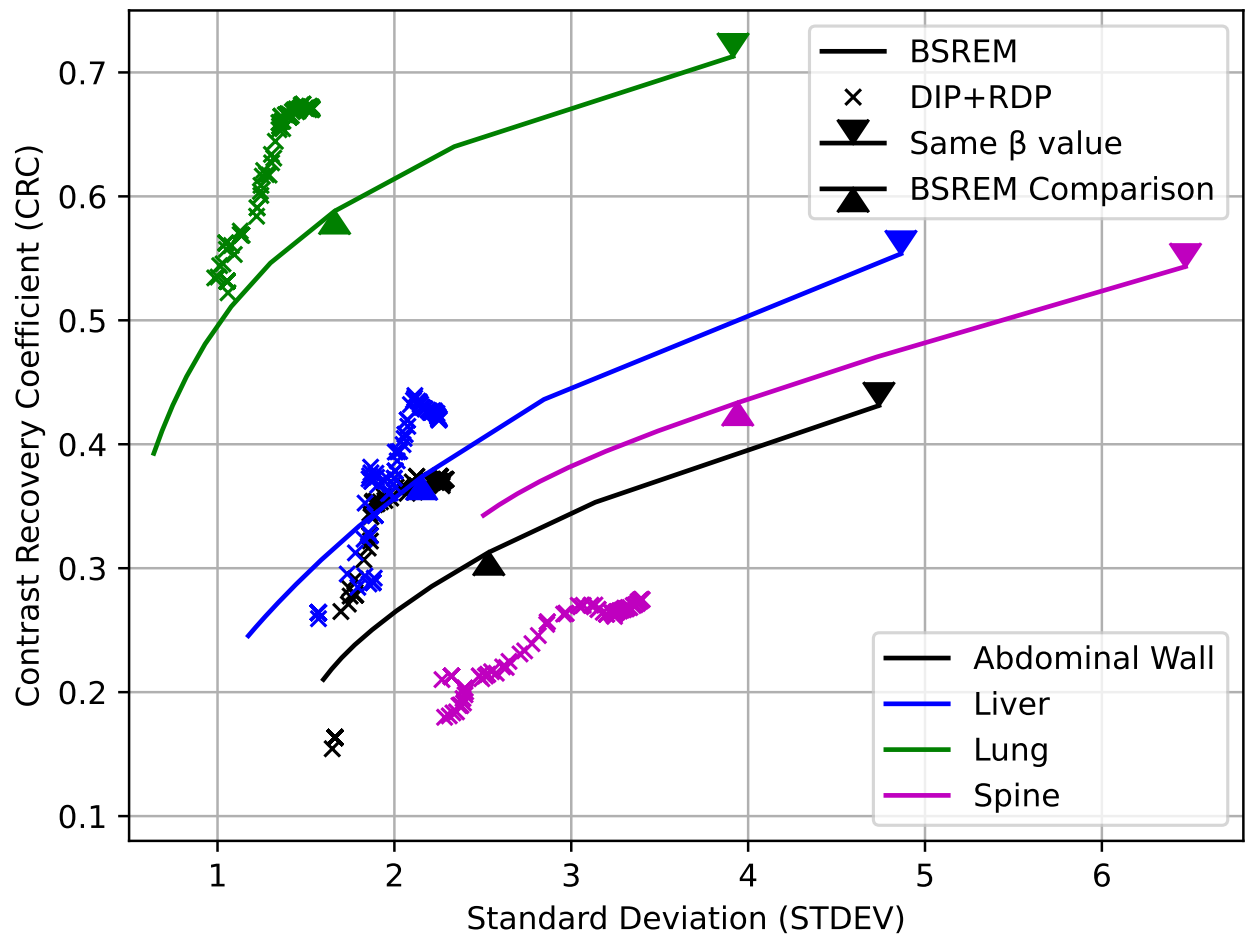

**(d)** Lower count data, DIP+RDP ($\beta = 1.5\text{e-}2$), marker indicates the same $\beta$ BSREM solution, comparison BSREM solution ($\beta = 4.5\text{e-}2$) used for qualitative assessment in Fig. 2.

**Figure 1:** Contrast Recovery Coefficient between lesion and background regions of interest against the standard deviation of the background. Results closer to the top left are best. Solid lines correspond to the BSREM solution with relative difference prior with different penalty strengths. Cross markers represent the minimum-loss DIP solution (at fixed penalty strength) taken every 100 epochs after an initial 10,000 epochs up to 20,000 epochs.

**Table 1:** GPU memory requirements on tested data for explicit vs. our implicit projector; estimated from sinogram/image sizes as well as 8-byte sparse element-size, and observing GPU memory usage respectively. Memory requirements for the 3D U-Net (see Sect. 3.3) used in forward and backward modes, and maximum image volume allowable on a 24 GB GPU.

| Projector | | 3D U-Net | | |
|---|---|---|---|---|
| Explicit matrix | Implicit (ours) | Forward | Backward | Maximum Volume |
| $>$ 100 GB | $<$ 1 GB | 0.65 GB | 0.88 GB | $300^3$ |

costly copying to-and-from the GPU which is currently necessary for integration with SIRF. The wrapper could be developed further by interfacing directly with the projector through a CUDA-based PyTorch wrapper which would keep operations on the GPU and arrays saved in VRAM; speeding up computation. Run-time could also be reduced by the use of subsets in DIP+RDP. This will be pursued in the future as it would be an important step in developing efficient deep learning techniques for PET reconstruction.

## 5 Conclusion

This is the first single optimisation implementation of 3D PET reconstruction via penalised maximum likelihood with DIP. The implementation utilises a wrapper integrating a well-established reconstruction framework (SIRF) with PyTorch. The application of DIP on high count data was able to significantly increase quality metrics, whereas on lower count data this was not observed. Introducing RDP into the objective function significantly improved the DIP trajectory for lower count data. Results indicate that further investigation is needed as conclusive and consistent improvements are not observed across count levels and lesions.

This work is supported by the EPSRC-funded UCL Centre for Doctoral Training in Intelligent, Integrated Imaging in Healthcare (i4Health) (EP/S021930/1), the UK NIHR funded Biomedical Research Centre at UCL Hospitals and the Alan Turing Institute (EPSRC EP/N510129/1). Software used in this project is partially maintained by CCP SyneRBI (EP-SRC EP/T026693/1). We thank Robert Twyman for support & supply of simulated data. Riccardo Barbano and Željko Kereta contributed equally.

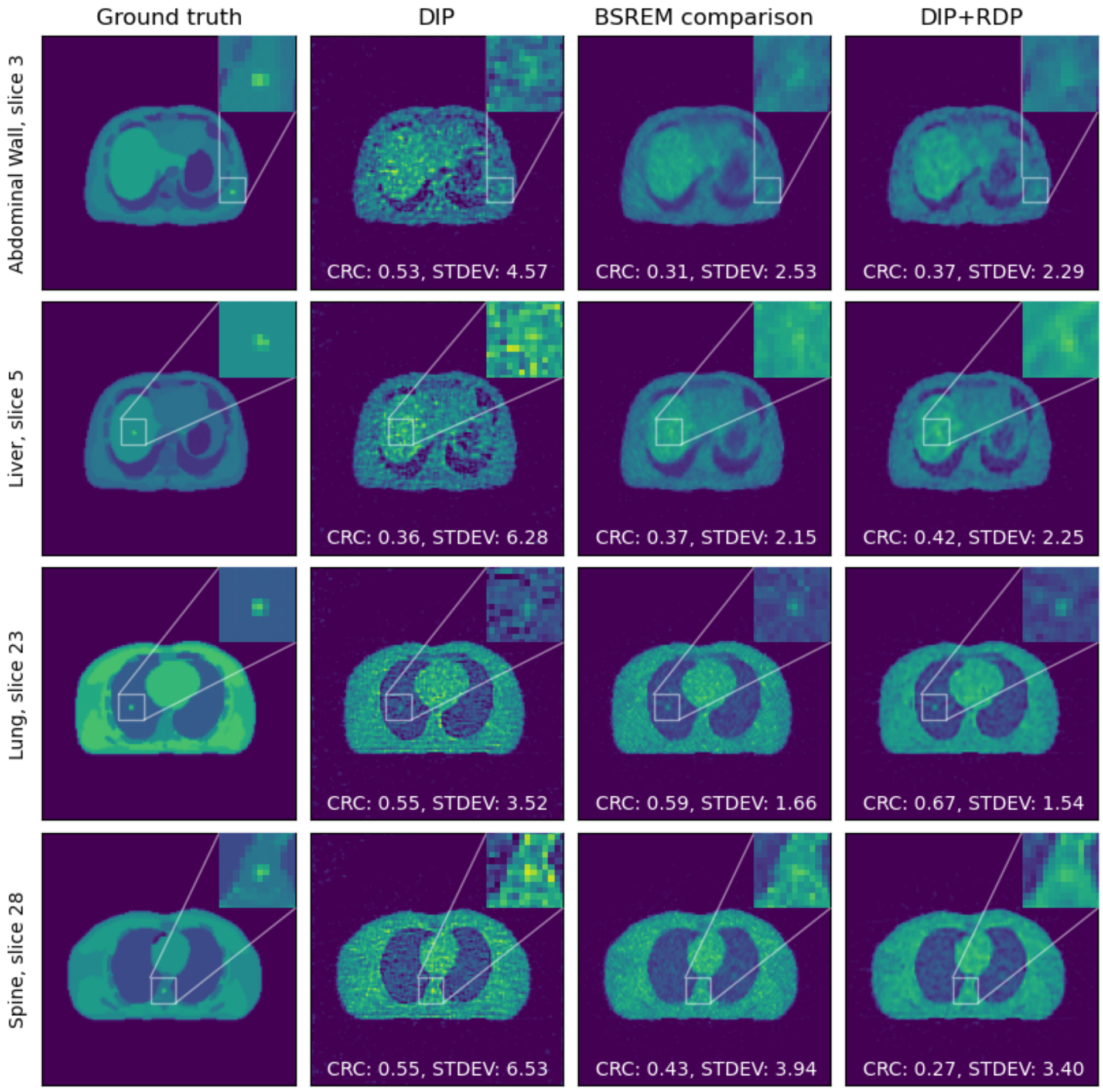

**Figure 2:** Axial slices taken through the center of lesions. Slices of ground truth emission and lower count data reconstructions with DIP, BSREM ($\beta = 4.5e-2$), and DIP+RDP ($\beta = 1.5e-2$). DIP reconstructions are the minimum-loss solutions over 20,000 epochs. CRC and STDEV values quotes are for the lesion shown in the slice. Colour-scales between reconstructed image slices are kept constant.

# Improved CT image resolution using deep learning with non-standard reconstruction kernels and CatSim training data

Somesh Srivastava[1], Mengzhou Li[3], Lin Fu[2], Ge Wang[3], Bruno De Man[2]

[1]GE Research – Healthcare, Bangalore, India
[2]GE Research – Healthcare, Niskayuna, NY, 12309 USA
[3]Department of Biomedical Engineering, Rensselaer Polytechnic Institute, Troy, NY, 12180 USA

**Abstract** Deep-learning (DL) super-resolution methods have recently shown encouraging results in improving the clarity of fine anatomical details in CT images. In this work, we explore the feasibility of using images generated with a high-frequency reconstruction kernel as the input to a DL network to boost the resolution capability of CT. Specifically, we use an "edge kernel", which is not routinely used to reconstruct clinical CT images due to its high level of noise and artifacts, but its higher-resolution image details may be exploited by a DL network. Then, we introduce the SRCAN2 network that takes both Edge and Bone kernel images as inputs to improve the super-resolution capability of DL. Also, we introduce a numerical method for generating super-resolution training data using the CatSim simulation environment. The results show that the proposed DL super-resolution method substantially improves the visual sharpness of CT images relative to the inputs. The modulation transfer function 10% threshold frequency can be increased by up to 56%. We further applied non-linear image quality metrics to characterize the contrast- and anatomy-dependent behavior of the DL network and obtained promising results.

## 1 Introduction

The importance of image resolution in x-ray CT scanning can not be overstated. Improved image resolution means that more fine detail of the human anatomy can be shown in the reconstructed images. Better resolution leads to better diagnosis and higher clinician confidence in the scan and the reconstructed image. Better image resolution could also reduce image blooming artifacts, which occur around small structures like coronary artery calcifications.

Recent literature in the field of Deep-learning based image super-resolution can be divided into 4 categories: (1) *Traditional DL networks* like EDSR [1] rely on residual blocks, while Ref [2] uses U-net [3] type architecture. (2) *GAN-based techniques*: SRGAN [4] uses a generator network with residual blocks. GAN-CIRCLE [5] was created for CT images and applies constraints like cycle-constraints. Ref [6] presents another popular GAN-based framework for microscopy images, while PULSE [7] augments GAN with exploration through the latent space of the generative model for faces and photos. (3) *Channel-attention/Spatial transformer techniques*: RCAN [8] applies channel-attention, residual blocks and image upsampling for resolution enhancement of general photographs. DFCAN [9] applies channel-attention in Fourier domain for microscopy images. Ref [10] combines spatial transformer with GANs to perform image upsampling in general images. (4) *Combination of DL and iterative techniques*: DPSR [11] applies Deep learning within ADMM iterations, and in Ref [12] the proximal regularization step within the iterative loop is replaced with a Deep neural network.

This study aims to improve DL super-resolution for CT by using non-standard input images and a new method for training data generation. Existing DL super resolution methods for CT usually directly use a standard clinical image as the network input. Although these images are optimized for human observers, they are not necessarily the optimal choice for the purpose of DL super resolution. Here, we test the idea of using higher frequency kernels like the Edge kernel to generate input images for the DL network. The Edge image is typically not used by clinicians for diagnosis due to its high noise, but it carries more high-frequency information which could be more effectively utilized by a DL network. Along with the high reconstruction kernel, we also use a very small input pixel size to preserve the fine details in the input image.

Another challenge for DL super resolution is the lack of easy ways to generate the high resolution training labels. In this work, we build high-resolution digital phantoms based on denoised and analytically sharpened clinical patient images, then apply CatSim [14] to simulate the CT imaging chain and generate the training inputs. Finally, we apply non-linear resolution metrics to characterize the contrast- and anatomy-dependent behavior of the DL network.

The organization of this paper is as follows. An SRCAN2 network for image resolution improvement is presented first (Sec 2.1). Next, we describe the method proposed to produce the training images (Sec 2.2). The details of the image evaluation methods are described in Sec 2.3. This is followed by results and conclusions.

## 2 Materials and Methods

### 2.1 Super-resolution RCAN with 2 input channels

We developed a lightweight super-resolution Fourier residual channel attention network with dual kernel inputs (SRCAN2) by adapting Ref [8]. A few modifications have been made to the RCAN network to improve the performance while reducing the total number of training parameters. First, we modified the network to have two input channels: Edge and Bone kernel filtered back projection (FBP) image reconstructions. We used an Edge kernel image as an input so that we can preserve as much high frequency information as possible from the measured sinogram. To combat the aliasing artifacts typically found in Edge kernel images, we added a Bone kernel image as the second input channel (Fig 1). The Bone kernel image is less sharp when compared to the Edge kernel image, but may help in removing the aliasing artifacts to a large extent; Second, we adopted a more advanced Fourier channel attention mechanism [9] to replace the original attention mechanism for better utilization of all frequency information from the dual kernel inputs to boost the performance; Third, we removed the upscaling module and only kept one residual group consisting of 15 Fourier channel attention blocks since our input and output image share the same size and the depth reduction of the network facilitate the network training on both speed and stability without significant performance loss in our application.

For the training loss function, we followed the settings in Ref [15, Eq 14] which consists of a L1 loss on absolute errors and a L2 loss on relative errors. The former term regulates the overall fidelity while avoid blurring, and the latter one emphasizes the fine details with small values, e.g., the anatomy with low intensity/HU in the lung regions.

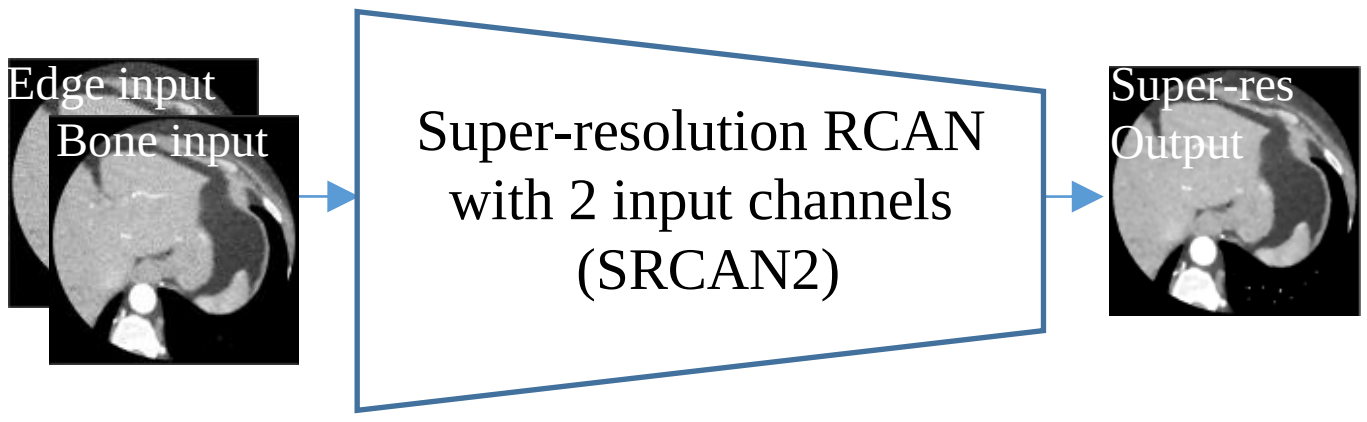

*Fig 1. Deep-learning network for image super-resolution*

The SRCAN2 was trained using 8 patient exams, each containing around 200 slices. More details about the generation of training data will be described in the next section. 240k patches of size 64×64 were generated: 80% of those were used for training and 20% for validation. For each training batch, 320×2 input CT image patches with the size of 64×64 were fed into the network. The model parameters were updated with ADAM optimizer ($\beta_1 = 0.9$; $\beta_2 = 0.999$, and $\varepsilon = 10^{-8}$), and the learning rate was set as $1\times10^{-5}$ initially then exponentially decayed at a rate of 0.95 per epoch. The network was trained for 60 epochs on 8 Tesla V100 GPUs and took 3.3 hours. The inferencing time for one 512x512 slice was 0.106 seconds on a single GPU.

### 2.2 CatSim based training data generation

We built high-resolution voxelized phantoms based on retrospective patient cardiac images acquired on a GE Revolution CT scanner (Fig 2). These original images were reconstructed with the standard kernel at 0.387-0.396mm pixel sizes. To generate clean training labels, the images were denoised using a previously developed 15-layer CNN CT denoising network . Then, to boost the image resolution, we applied a circular-symmetric Fourier-domain filter whose response is unity at zero frequency and increases quadratically to the maximum frequency:

$$H(f_x, f_y) = 1 + (f_x^2 + f_y^2)/(9.6)^2,$$

where are $f_x$ and $f_y$ are in lp/cm. Finally, we applied a non-linear gray level transform to each pixel of the image (termed "cartoonization" here), to further enhance the sharpness of the boundaries of bone and lung tissues:

$$T(x; c, r, h) = x + h\frac{x-c}{r}e^{-\left(\frac{x-c}{r}\right)^2},$$

where $x$ denotes the input gray level for each pixel, which is converted from HU by $x = 1 + HU/1000$. The transform was applied twice, with parameters $c = 0.6, r = 0.08, h = 0.113$ for lung tissues, and $c = 1.3, r = 0.08, h = 0.113$ for bone tissues, respectively.

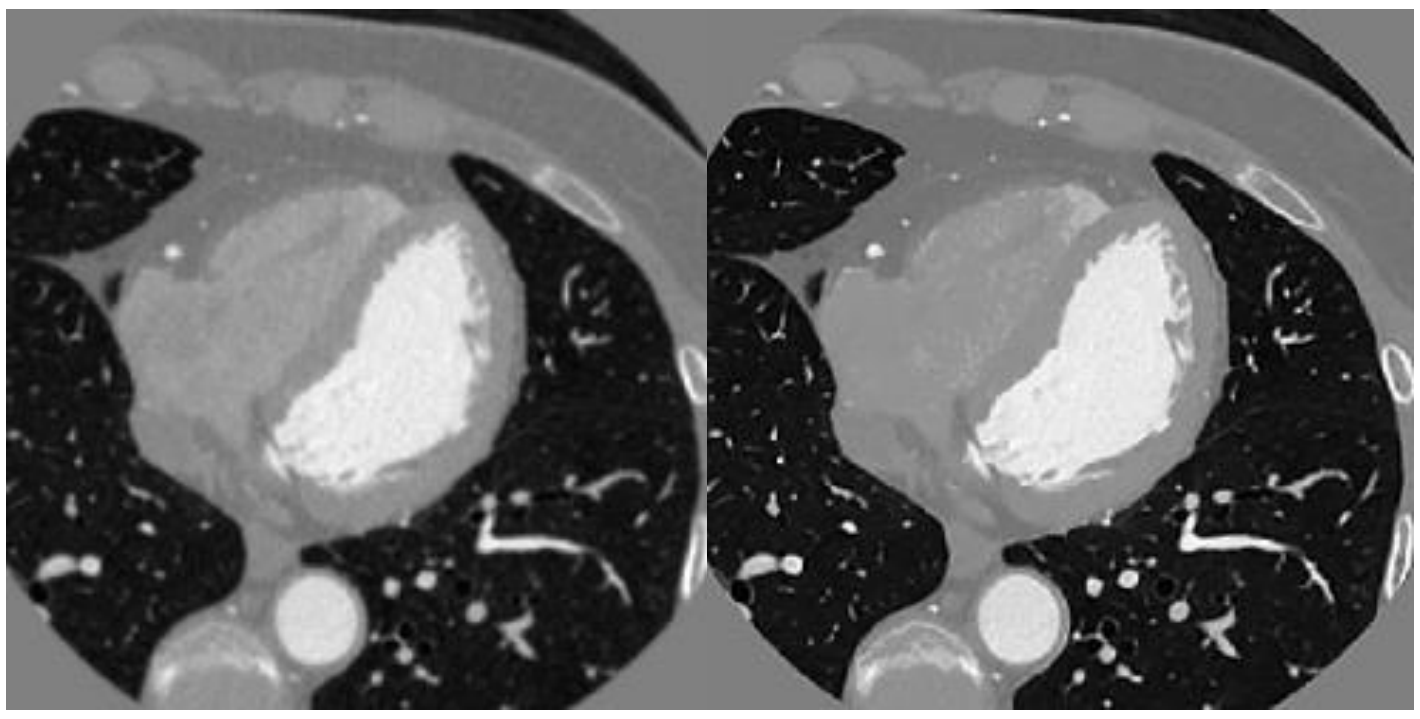

*Fig 2. High-resolution, low-noise ground-truth phantom image derived from a patient image. WW/WL=2000/1000 HU (Air=0HU).*

To generate super-resolution phantoms from these images, we also shrank the image pixel size to 0.133 mm i.e., a three-fold reduction relative to an original pixel size. Such scaling introduced miniature anatomical structures beyond the instrinsic resolution of the clinical image. This is based on the assumption of fractal self-similarity between different CT image scales.

The super-resolution phantoms were used as the ground truth training labels. CT scans of the phantoms were simulated in the CatSim environment which models the CT imaging process. The Edge and Bone kernel reconstructions from the simulation data were used as the training inputs (Fig 3).

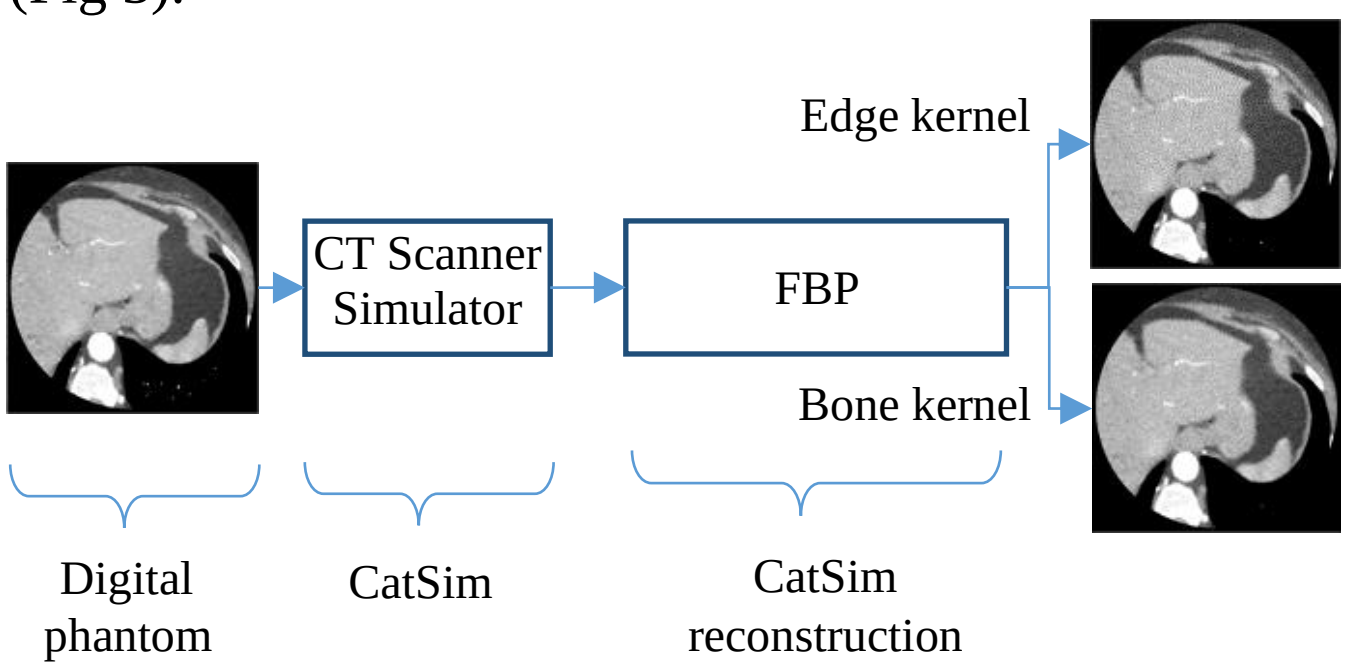

*Fig 3. The digital phantom and the simulated Edge and Bone kernel reconstructions are used as training data for SRCAN2.*

### 2.3 Image quality evaluation

We evaluated the image resolution in this study by adapting the methods in [16] and [13], where Dirac impulses (or wires) of various amplitudes were inserted at random locations in a testing image, and the difference between the output images with and without the impulses was used to quantify image resolution (Fig 4). The modulation transfer function (MTF) and contrast recovery coefficient (CRC) were computed from the wires to quantify the non-linear response of the DL network. The CRC adapted to this study was defined as the ratio between the maximum value of the output wire profile to the maximum value of the Edge kernel input wire profile.

The details of the method used to compute the resolution metrics is as follows:

Step 1. *Generation of testing wires for Edge and Bone images*. We inserted single-pixel impulses (10 g/cc) as testing wires at 65 random locations in the digital phantom, but ensuring a separation of at least 32 pixels between the wires. A CT scan of the wire-inserted phantom was simulated in CatSim. In the reconstructed image, we subtracted the background to extract the wire profiles. We then averaged the wire profiles to obtain the testing wire profiles for the Edge and Bone kernels, respectively.

Step 2. *Measurement of the wire responses of the DL network.* For each test image set (containing Edge and Bone kernel images for each image slice), we first obtained the super-resolution output images from the DL network without inserting the wires. Then, at 41 random pixel locations (but separated from each other by atleast 32 pixels) we added a scaled version of Edge and Bone wire profiles obtained in Step 1 to the test image set. The super-resolution images with the wires were obtained as the DL outputs. This was repeated with different wire contrasts in the range of 20 to 2000 HU. The contrast- and location-dependent wire responses of the DL network was obtained by taking the difference between the output images with and without the inserted wires.

Step 3. *Computation of MTF10% metric.* For each wire profile we computed the MTF curve by taking the absolute value of the 2D Fourier transform of the wire profile, and averaging the x- and y-profiles. Then, we computed the frequency (in lp/cm) where the response dropped to 10% of the DC value of the MTF curve.

Step 4. *Computation of CRC metric*. We computed CRC as the ratio between the maximum value of the output wire profile to the maximum value of the Edge kernel input wire profile.

## 3 Results

The DL network was tested on two patient exams, which were not a part of the training and validation sets. The digital phantom (i.e. ground truth) for these patient exams was generated, then the Edge and Bone kernel input images were generated from CatSim. The super-resolution images produced by the SRCAN2 network are visually substantially sharper than both the input Edge and Bone kernel reconstructions (Figs 5 and 6).

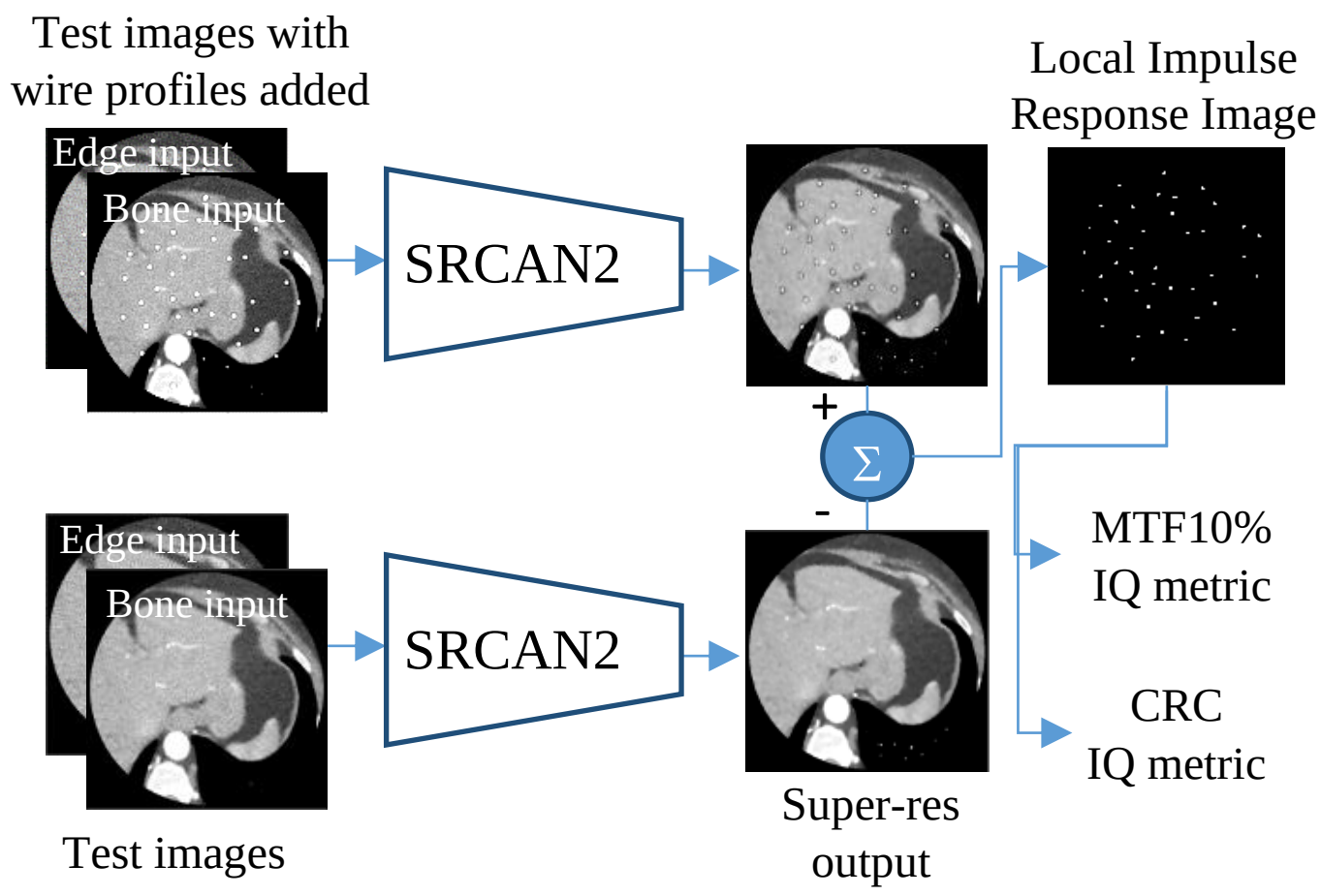

*Fig 4. Method to compute MTF10% and CRC image quality (IQ) metrics*

The wire profiles for Edge and Bone kernel input images and the network output image for an example wire is shown in Fig 7. When the wire contrast was 2000HU (in the Edge image), the MTF10% for Bone input, Edge input, and Super-res output are 10.7, 12.4 and 19.4 lp/cm respectively, i.e., an improvement of 56% over the Edge input. The peak value of the output wire was 2.27 times of the peak of the Edge input wire, i.e. CRC=2.27.

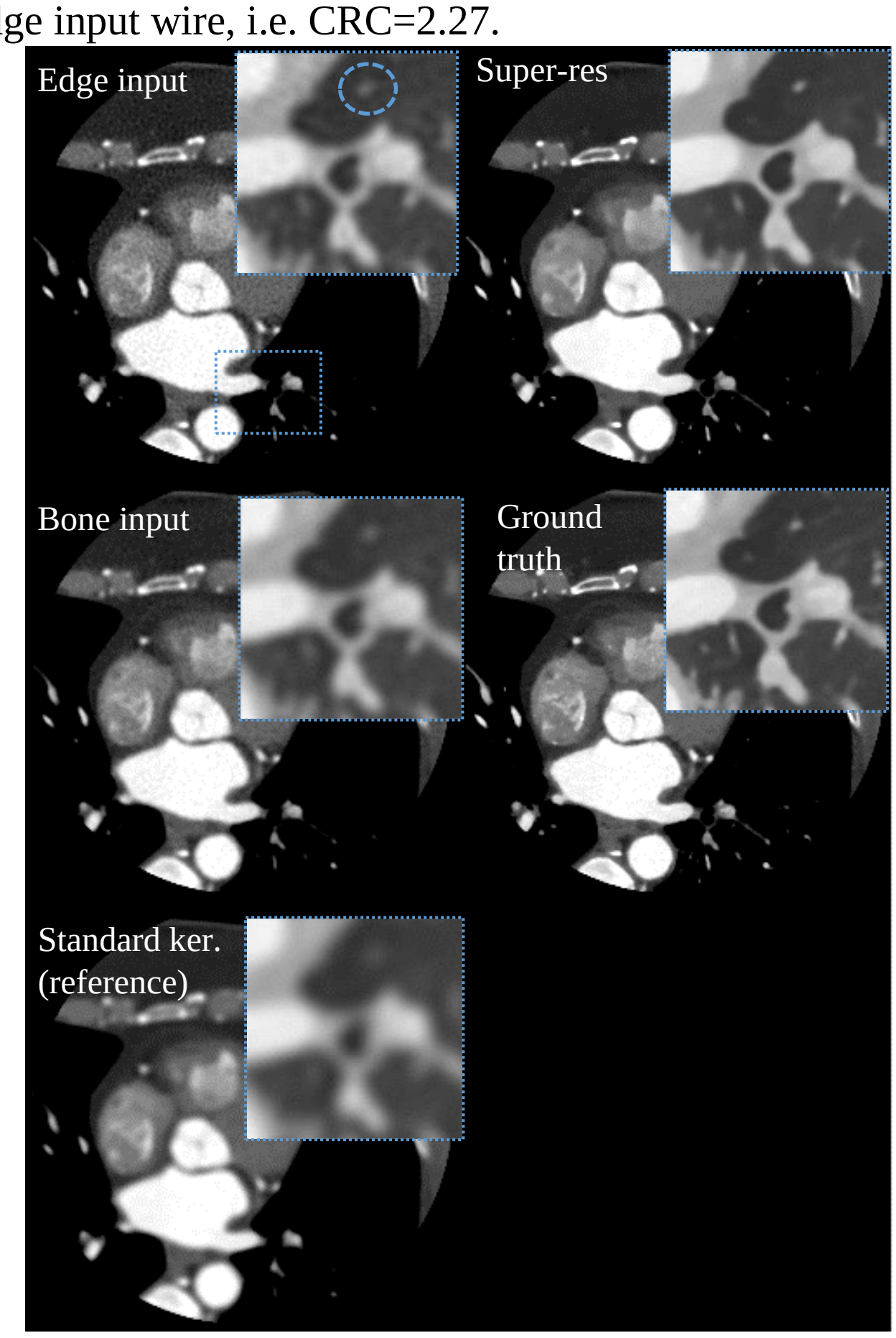

*Fig 5. DL network input images, output images, and ground truth image for a representative 2D slice. Standard kernel image is also shown for reference. WW/WL=800/1200 HU, =2000/700 HU for inlay (Air=0HU). The x- and y-profiles through the blood vessel inside the dashed circle are in Fig 6.*

Due to the non-linear nature of the DL network, the MTF10% and CRC varied with the contrast level of the test wire. The measured output image resolution reduced as the contrast level of the input test wire was reduced (Fig 8). The dependence of mean MTF10% and mean CRC on contrast level is used to characterize the non-linear behavior of the DL based image super resolution method.

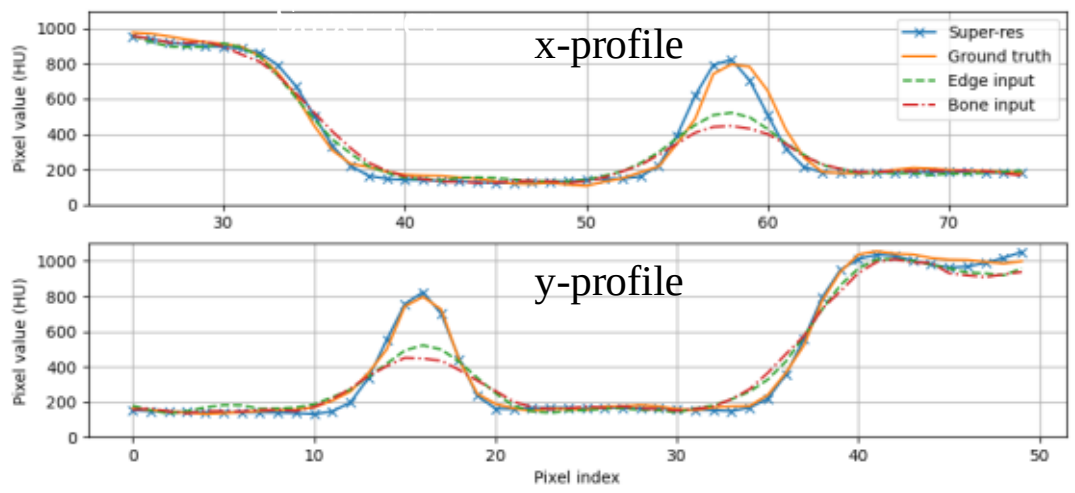

*Fig 6. The x- and y-profiles through the blood vessel inside the dashed circle in Fig 5.*

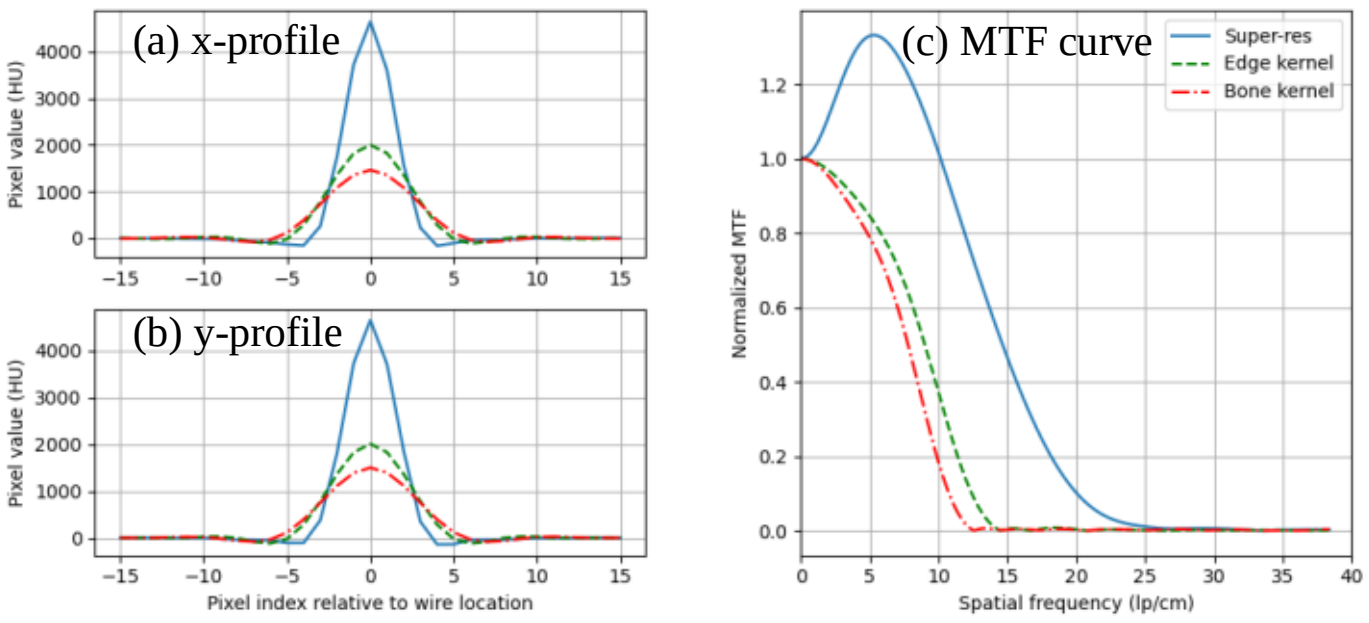

*Fig 7. (a,b) X- and y-profiles through a test wire for Edge input image, Bone input image, and Super-res image. (c) MTF curves for the same test wire for Edge input image, Bone input image, and Super-res image. Note that the contrast level of the test wire in the Edge input image is 2000HU.*

## 4 Conclusion

We explored the feasibility of using non-standard reconstructions such as images reconstructed with the Edge kernel as the input of a DL network to boost CT image resolution. Our SRCAN2 network takes multiple inputs of both Edge and Bone kernel images. To overcome the difficulty in finding super-resolution training data, we built high-resolution digital phantoms based on denoised and analytically sharpened clinical patient images, then generated the training inputs in a CatSim environment. The results showed that the proposed DL method substantially improved the visual sharpness of CT images relative to the inputs. The MTF10% increased by up to 56%. We also applied non-linear image quality metrics to characterize the dependency of the image resolution improvement on the contrast level of local features.

## Acknowledegement

Research reported in this publication was supported by the NIH/NHLBI grant# R01HL151561. The content is solely the responsibility of the authors and does not necessarily represent the official views of the NIH. We wish to thank Jed Pack for the denoising network used for training data generation. We also wish to thank Peter Lorraine for the initial implementation of the image quality metrics.

## References

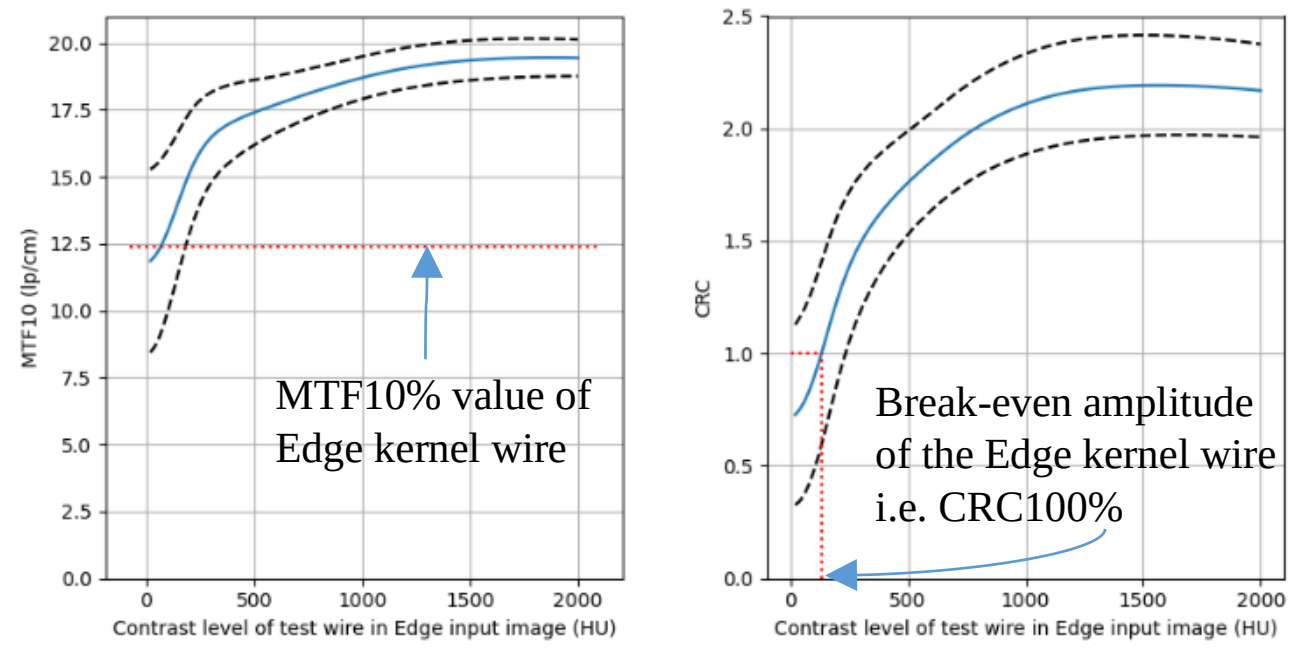

*Fig 8. The variation of the MTF10% and CRC metrics as a function of the contrast level of the test wire in the Edge input image. The dashed lines indicate +/- one standard-deviation.*

# Convergent ADMM Plug and Play PET Image Reconstruction

Florent Sureau[1], Mahdi Latreche[1,2], Marion Savanier[1], and Claude Comtat[1]

[1]BioMaps, Université Paris-Saclay, CEA, CNRS, Inserm, SHFJ, 91401 Orsay, France.
[2]Institut Denis Poisson, Université d'Orléans, UMR CNRS 7013, 45067 Orléans, France.

**Abstract** In this work, we investigate hybrid PET reconstruction algorithms based on coupling a model-based variational reconstruction and the application of a separately learnt Deep Neural Network operator (DNN) in an ADMM Plug and Play framework. Following recent results in optimization, fixed point convergence of the scheme can be achieved by enforcing an additional constraint on network parameters during learning. We propose such an ADMM algorithm and show in a realistic [$^{18}$F]-FDG synthetic brain exam that the proposed scheme indeed lead experimentally to convergence to a meaningful fixed point. When the proposed constraint is not enforced during learning of the DNN, the proposed ADMM algorithm was observed experimentally not to converge.

## 1 Introduction

In the quest for PET image reconstructions adapted to a protocol, to a specific patient or to a given task, deep learning approaches are currently a promising avenue of research. Early works have already illustrated the potential of such approaches to achieve better resolution, contrast recovery and noise propagation trade-offs compared to classical model-based variational methods for PET reconstruction [1, 2]. In particular, a reduction of the dose injected to the patient could be envisioned without sacrificing much of the reconstructed image quality compared to a standard dose exam, which would be beneficial for the patient and/or to reduce the cost of a PET scan [3].

However using deep learning in PET raises several new issues compared to the aforementioned variational approaches. In particular these specific methods can generate instabilities related to the ill-posedness of the problem that could lead to images with artefacts as already observed in biomedical image reconstruction applications [4]. Furthermore, deep learning techniques for reconstruction often use neural networks as a black box operator in the reconstruction pipeline. This leads to end estimates that lack mathematical or statistical guarantees that could make them robust, contrary to classical reconstruction techniques typically associated with a convex variational problem. Robustness of the reconstruction is also of paramount importance in PET where datasets are often small (with typically only tens of exams per protocol) leading to limited learning and validation.

In this context, hybrid techniques inspired by model-based PET reconstruction approaches in a learning framework have been proposed to avoid learning the direct model and to aim at more reliable estimates [5]. We will focus on this work on a less investigated hybrid framework for PET reconstruction than unrolled [2] or synthesis [1] approaches: the ADMM Plug and Play approach [6]. In this framework, an implicit operator related to the prior is only learned [7], making this approach flexible. Fixed-point convergence results have been investigated for this framework even though the optimization problem is non-convex and the learned operator is implicit [8, 9]. Based on these results, we propose and investigate in this work a convergent ADMM Plug and Play approach for PET image reconstruction using Deep Learning. We present in section 2 the convergent ADMM Plug and Play approach that we propose, the datasets used for training and evaluation and detail the architecture and implementation of the additional constraint needed for fixed-point convergence of the scheme. We then present and discuss our results on realistic [$^{18}$F]-FDG synthetic exams.

## 2 Materials and Methods

### 2.1 Convergent ADMM Plug and Play

We consider the reconstruction of an image denoted by $\mathbf{x} \in \mathbb{R}_+^N$, from an observed noisy sinogram $\mathbf{y} \in \mathbb{N}^M$. The ADMM plug and play algorithm is described in Algorithm 1. In our context of PET reconstruction $LL(\mathbf{y},\mathbf{x})$ is the Poisson log-likelihood and $\mathscr{D}_\theta$ is a DNN operator with inputs the reconstructed PET images, with parameters learnt in a separated step.

**Algorithm 1** ADMM Plug and Play with a DNN $\mathscr{D}_\theta$.

1: Choose $\mathbf{z}^{(0)}$, $\mathbf{u}^{(0)}$, $\rho$, $K$.
2: **for** $k = 0..K$ **do**
3: $\quad \mathbf{x}_\rho^{(k+1)} = \underset{\mathbf{x}\in\mathbb{R}_+^N}{\arg\min} -LL(\mathbf{y},\mathbf{x}) + \frac{\rho}{2}\|\mathbf{x} - (\mathbf{z}_\rho^{(k)} - \mathbf{u}_\rho^{(k)})\|_2^2$
4: $\quad \mathbf{z}_\rho^{(k+1)} = \mathscr{D}_\theta\left(\mathbf{x}_\rho^{(k+1)} + \mathbf{u}_\rho^{(k)}\right)$
5: $\quad \mathbf{u}_\rho^{(k+1)} = \mathbf{u}_\rho^{(k)} + \mathbf{x}_\rho^{(k+1)} - \mathbf{z}_\rho^{(k+1)}$
6: **end for**
7: **return** $\mathbf{x}_\rho^{(K+1)}$

Compared to unrolling techniques, this ADMM framework decouples learning the DNN and reconstructing the images, which is convenient for integrating DNN inside the reconstruction. In particular the number of ADMM iterations can be large with no impact on the GPU VRAM contrary to unrolling techniques. Besides the reconstruction problem in line 3 is a standard (convex) minimization problem, which can be solved using an efficient existing PET algorithm [10]. Compared to synthesis ADMM approaches, this Plug and Play formulation leads to a simple condition on the DNN for this scheme to converge as described below. Note in particular that no heuristic choice of an iteration-dependent $\rho$ is needed to stabilize the algorithm as often proposed as in [8], even though our final estimate still depends on $\rho$.
If the operator $\mathscr{D}_\theta$ corresponds to the proximal operator of a proper, closed and convex function, Algorithm 1 is the classical ADMM algorithm with an implicit operator and with convergence properties described in [11, 12] . However for a general $\mathscr{D}_\theta$, such algorithm may not even minimize a convex problem and there is no guarantee of convergence of such a scheme. Conditions for fixed point convergence of the ADMM Plug and Play algorithm have however been studied in [8, 9] and references therein. In particular, [9] shows that one ADMM iteration can be written equivalently as applying an operator $\mathscr{T}_\theta = \frac{1}{2}\,\mathrm{Id} + \frac{1}{2}(2\mathscr{D}_\theta - \mathrm{Id})(2\,\mathrm{Prox}_{-LL/\rho} - \mathrm{Id})$. This implies in particular that if $\mathscr{L}_\theta = (2\mathscr{D}_\theta - \mathrm{Id})$ is a non expansive operator and $\mathscr{T}_\theta$ has a fixed point, then fixed point convergence of the scheme is obtained [9]. The non-expansiveness constraint is however difficult to enforce numerically. [9] proposed to use real spectral normalization on each layer to constrain the Lispchitz constant of each layer of the DNN. In this work, we rather use the approach proposed in [13]. Taking also into account a supervised loss (Mean Squared Error - MSE), the DNN parameters are estimated in the following minimization problem:

$$\min_\theta \underbrace{\sum_{b=1}^{B} \|\mathscr{D}_\theta(\mathbf{x}_b) - \bar{\mathbf{x}}_b\|^2}_{\text{MSE}} + \beta \underbrace{\max\{\|\nabla \mathscr{L}_\theta(\tilde{\mathbf{x}}_b)\| + \varepsilon - 1, 0\}^{1+\alpha}}_{\text{Non expansiveness constraint}}, \tag{1}$$

where $b$ is the batch index, $\varepsilon$, $\alpha$ and $\beta$ are hyperparameters to balance the supervised loss and the constraint and $\tilde{\mathbf{x}}_b$ is obtained as a random convex combination of the reference image $\bar{\mathbf{x}}_b$ and the output of the neural network as follows:

$$\tilde{\mathbf{x}}_b = \kappa \bar{\mathbf{x}}_b + (1-\kappa)\mathscr{D}_\theta(\mathbf{x}_b), \kappa \sim \mathscr{U}[0,1]. \tag{2}$$

Note that the spectral norm of the Jacobian for a given entry point can be estimated using automatic differentiation but the computation is particularly intensive in terms of both GPU VRAM and execution time.

### 2.2 Datasets and Learning settings

The database used for learning and evaluation of the proposed approach was derived from 14 brain [$^{18}$F]-FDG brain exams of healthy subjects and their associated T1 weighted MR images. The T1 images were first segmented into 100 regions using FreeSurfer[1]. The PET signal was then measured in a frame between 30 minutes and 60 minutes after injection in each region using PETSurfer [14] to generate 14 distinct anatomo-functional phantoms. 3-dimensional PET simulations for a Biograph 6 TruePoint TrueV PET system were then generated using an analytical simulator [15], including normalization, attenuation, scatter and random effects. 11 phantoms were used for training and 3 for testing. Data augmentation was performed for each phantom by simulating 10 realizations of the injected dose so that the total number of counts simulated spans the range observed in the 14 exams. This results in 110 shuffled simulations used for training and 30 realizations for testing. These simulations were reconstructed with CASToR [16] using OSEM with 8 iterations of 14 subsets.
The network $\mathscr{D}_\theta$ was chosen as a U-Net [17] with 443649 parameters. We made several modifications compared to the architecture presented in [1], as a balance between performance of the network and number of parameters to learn: we use 3 levels with instance normalization, 3D average pooling and concatenation between the decoder branches and encoder branches, and we use an overall skip connection to learn on the residual image. Note that the input reconstructed images are first normalized so that the network performance is robust to dose variation. The normalization factor is then applied to the output of the network to recover the correct scale.
The DNN parameters of the U-Net were learned in two steps. In a preliminary phase, the DNN parameters are learnt only with the supervised MSE loss. The ADAM optimizer with 50 epochs and a learning rate of 0.001 was used. Batch size was 1, and the reference in the supervised loss corresponds to the noise-free images. This results on a first DNN without the constraint on the Jacobian, named "PRE" in the following. In a second phase, the total loss in Equation 1 is considered and the Power Iterative Method (with a maximum of 10 iterations) and automatic differentiation is used to compute the spectral norm of the Jacobian. In this case we use 14 additional epochs on the PRE DNN to enforce the constraint, using ADAM with a learning rate of 0.0005, batch size of 5, $\beta = 10$, $\alpha = 0.1$ and $\varepsilon = 0.05$ in Equation 1. This network is named "JAC" in the following.
Both networks are then employed in Algorithm 1 using 40 iterations and compared on the simulations. For initialization $\mathbf{z}^{(0)}$ is an OSEM reconstructed image with 8 iterations of 14 subsets and $\mathbf{u}^{(0)} = \mathbf{0}$. We first investigated the choice of $\rho$ on the convergence speed and on the solution by looking at the norm of the primal residual defined as $\mathbf{x}_\rho^{(k)} - \mathbf{z}_\rho^{(k)}$ and of the dual residual $\rho(\mathbf{z}_\rho^{(k+1)} - \mathbf{z}_\rho^{(k)})$ [12]. Both should converge to zero for ADMM to converge.

## 3 Results

MSE curves during training and testing of the PRE U-Net illustrate that 30 epochs are necessary to learn parameters in the preliminary phase (not shown). Figure 1 shows that the choice of $\beta$ leads to balanced supervised and constraint loss in the first iteration. Both MSE and Jacobian constraint components of the loss are decreasing over epochs. Performance in the testing and training datasets were comparable in terms of MSE, and the proposed implementation of the constraint leads to a Jacobian spectral norm in the testing dataset less than 1 as expected (compared to more than 3.5 for PRE).

[1] https://surfer.nmr.mgh.harvard.edu

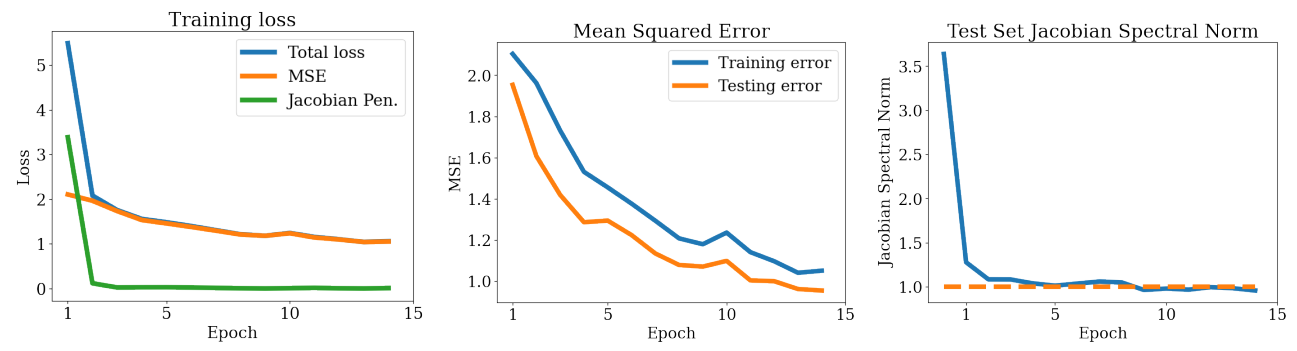

**Figure 1:** Loss functions for JAC U-Net. Left: the total loss is subdivided into its two individal contribution. Middle: MSE for training and testing phases. Right: Jacobian spectral norm for testing dataset.

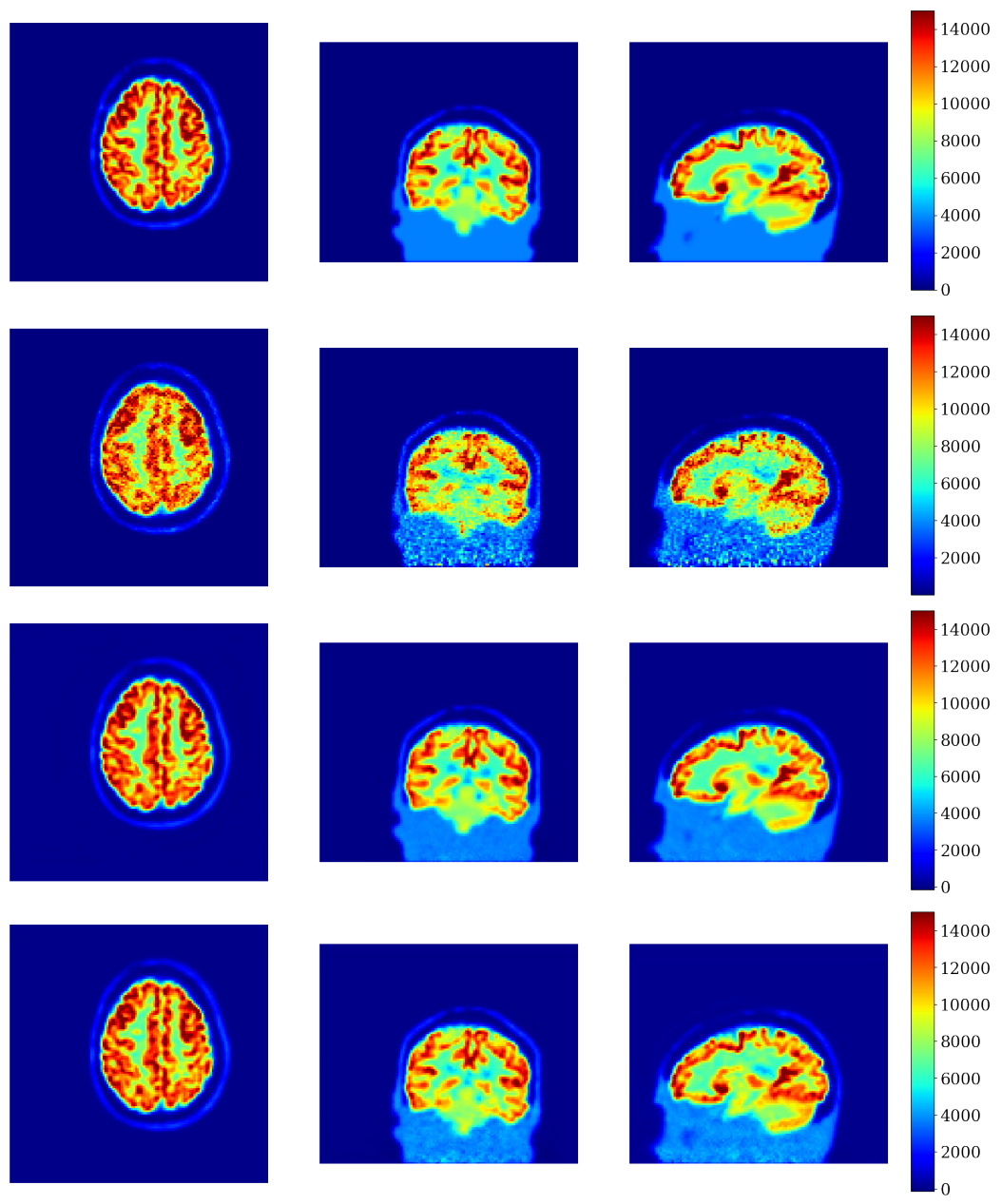

**Figure 2:** Performance of U-Nets in post-processing for a simulation in the test set. From top to bottom: noise-free reference image, OSEM image, PRE U-Net estimate, JAC U-Net estimate.

Figure 2 illustrates the performance of both U-Nets when used as simple post-processing for reconstructed PET images: propagated noise has been reduced while preserving the high frequency structures present in the original phantom. Compared to the PRE U-Net higher frequencies are observed in the background region for the JAC U-Net, indicating that the denoising performance is slightly degraded when using the Jacobian constraint.

The two networks were then employed in Algorithm 1. The impact of hyperparameter $\rho$ is illustrated on Figure 3 for the JAC U-Net. This illustrates that a carefull choice of this hyperparameter is needed to reach adequate convergence speed for the overall scheme similarly to what is observed in ADMM in convex problems. In the following, we chose $\rho = 5e-7$ which achieves fast convergence as indicated in both primal and dual residuals.

Figure 4 illustrates the performance of PRE and JAC U-Nets across ADMM iterations, for the previously selected value of $\rho$. It can be observed that JAC U-Net leads to decreasing primal and dual residuals and to a rapid stabilization of both MSE to a low value and log-likelihood to a high value. On the contrary the PRE U-Net has not converged as illustrated in primal and dual residuals, and has lower log-likelihood and higher MSE than JAC.

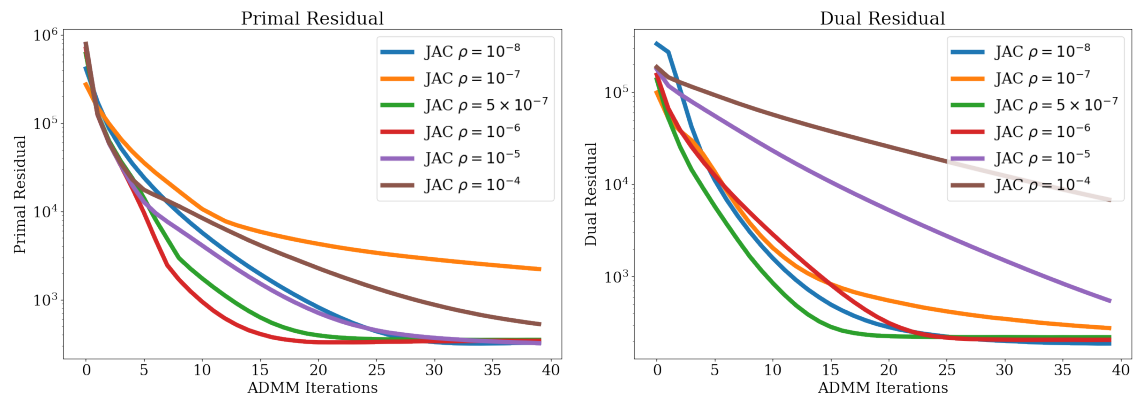

**Figure 3:** Norm of the primal (left) and dual residual (right) across ADMM Plug and Play iterations for the JAC U-Net.

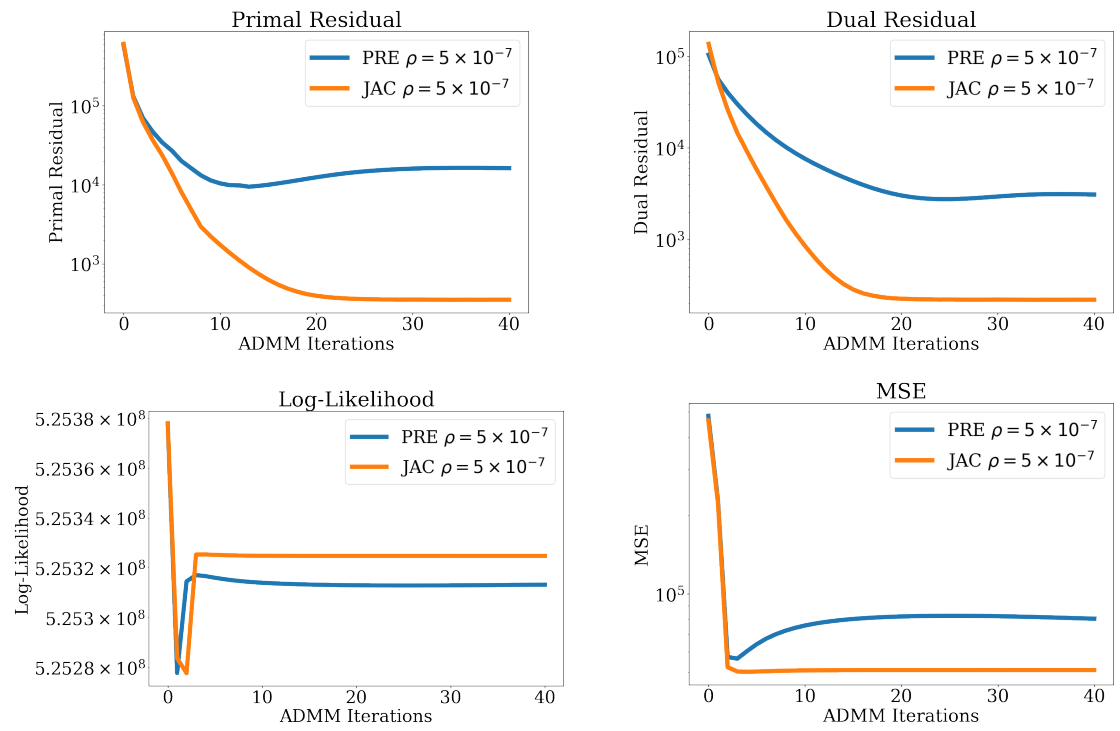

**Figure 4:** Evolution for both U-Nets of Top: primal (left) and dual (right) residuals; Bottom: log-likelihood (left) and MSE (right).

Figure 5 illustrates the recovered images using ADMM Plug and Play with PRE and JAC U-Nets, compared to the best Gaussian post-filtered OSEM image. JAC results lead to the closest image to the reference, with the lowest MSE. On the contrary, PRE U-Net leads to a not converged image further from the reference.

## 4 Discussion

In this work we have proposed a strategy to build a convergent ADMM Plug and Play algorithm by enforcing a non-expansiveness constraint during the learning of the DNN. Enforcing a strict (global) non-expansiveness constraint is actually replaced by enforcing $\|\nabla \mathscr{L}_\theta(\tilde{\mathbf{x}}_b)\|$ to be less than 1 for sampled $\tilde{\mathbf{x}}_b$ using Equation 1. More epochs are needed to ensure sufficient sampling of the space close to the solution. However, the results presented in this work illustrate that the constraint is satisfied even for the test set. It was also observed experimentally that Algorithm 1 converges as expected. Nonetheless, robustness of such a reconstruction scheme should be further assessed. We have shown that the choice of hyperparameter $\rho$ is crucial for convergence speed as in the convex case, but in this non-convex case the solution also depends on $\rho$. The choice of $\rho$ is therefore crucial, and the choice of this hyperparameter should be investigated more thoroughly. Finally we plan to investigate the performance of such algorithm in low-dose scenarios to assess the performance of such approach in a more clinically relevant setting.

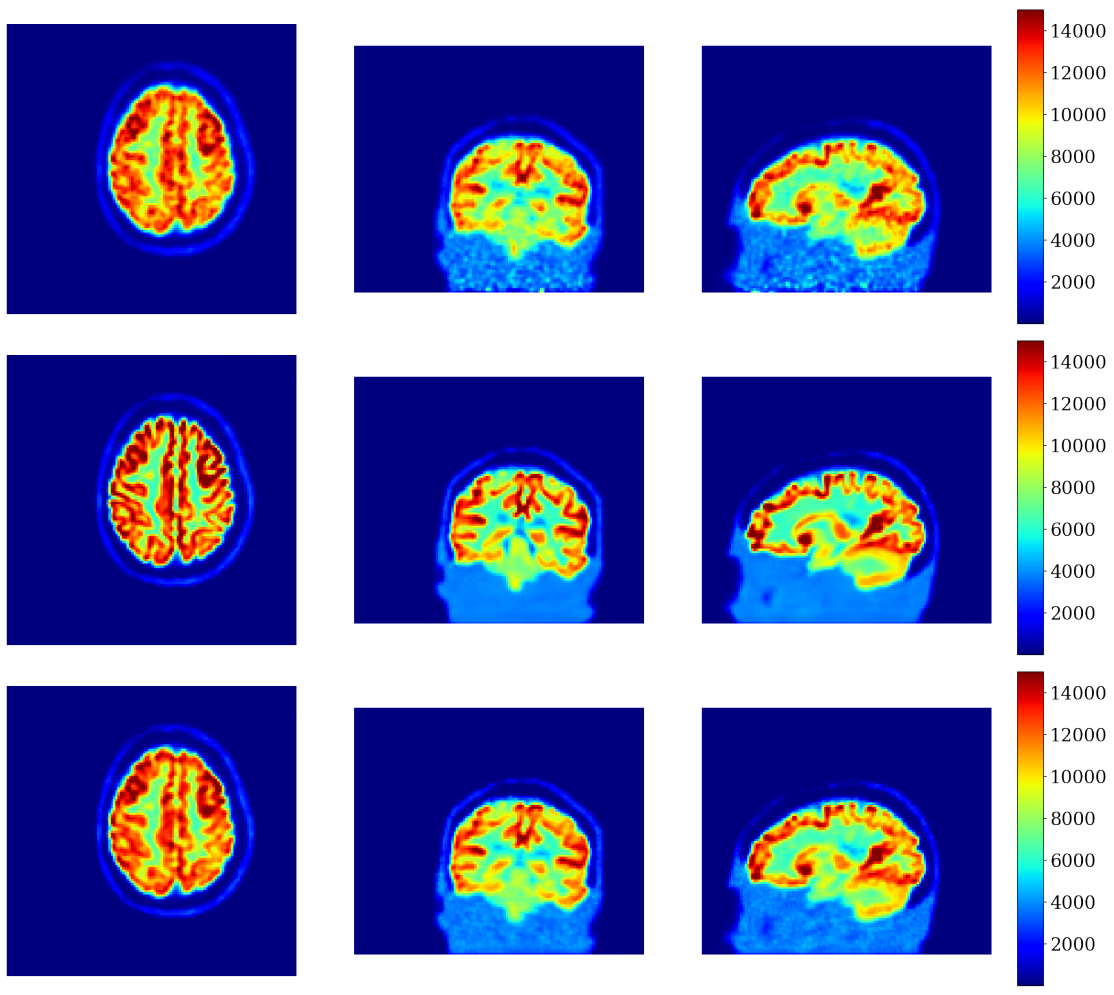

**Figure 5:** Reconstructed images for the same test set simulation as in Figure 2. On top, the OSEM reconstructed image with the best Gaussian post-filtering selected (MSE:90265). In the middle, the PRE U-Net results after 80 iterations (MSE:75227). In the last row, the result obtained with the proposed JAC U-Net after 80 iterations (MSE:50957).

## 5 Conclusion

In this work we have proposed a new approach for PET reconstruction using Deep Learning. Based on the ADMM Plug and Play framework, the proposed approach uses a constraint on the spectral norm of an operator Jacobian during learning. This promotes the non-expansiveness that leads to a convergent reconstruction scheme. We show in experimental simulations that without this constraint the ADMM does not converge. On the contrary, the proposed approach experimentally converges to a higher likelihood solution and to a lower MSE, illustrating the interest of such an approach.

## 6 Acknowledgment

We acknowledge financial support from the French National Research Agency (ANR) under grant ANR-20-CE45-0020 (ANR MULTIRECON). This work was partly funded by the France Life Imaging (ANR-11-INBS-0006 grant from the French "Investissements d'Avenir" program).

# X-Ray Small Angle Tensor Tomography

Weijie Tao[1,2], Li Lyu[1,2], Yongjin Sung[3], Grant T. Gullberg[4], Michael Fuller[5], Youngho Seo[4], and Qiu Huang[1,2]

[1]Department of Nuclear Medicine, Ruijin Hospital, School of Medicine, Shanghai Jiao Tong University, Shanghai 200240, China
[2]School of Biomedical Engineering, Shanghai Jiao Tong University, Shanghai 299240, China
[3]Department of Biomedical and Mechanical Engineering, University of Wisconsin-Milwaukee, Milwaukee, WI, USA
[4]Department of Radiology, University of California San Francisco, San Francisco, CA 94143, USA
[5]TF Instruments, Salinas 93908, CA

**Abstract** Previous approaches using X-ray dark-field imaging to obtain a tensor representation of small angle scatter in tissue involved first reconstructing the coefficients of a fixed vector field at each voxel, then fitting a Gaussian tensor representation of the small-angle scatter to the reconstructed vector space. A recent simulation study (Graetz [1]) demonstrated that small angle scatter can be represented by a linear tensor model to recover orientations with an accuracy of less than 1°. Based on this observation, we developed an iterative algorithm that reconstructs a symmetric 2nd rank tensor of small angle scatter from scalar projections of visibility measurements obtained from grating stepping acquisitions. The novelty is the design of a data acquisition method and reconstruction algorithm that provides estimates of tensor representation of small angle scatter. Its relation to medical imaging involves using the small angle scattering properties of X-ray interactions with tissue micro-structure to identify lesions with negligible density variation relative to surrounding tissues.

## 1 Introduction

There is a significant need to provide improved lesion detection using X-ray CT, especially for lesions in the lung. The goal of this work is to develop new direct tensor reconstruction algorithms that avoids having to first reconstruct the coefficients of a fixed vector field at each voxel, and then fitting a Gaussian tensor representation of the small-angle scatter to the reconstructed vector space [2]. Based on the premise that dark field scatter can be represented by a symmetric 2nd rank tensor [1], a tensor reconstruction algorithm is developed using measurements to reconstruct the six unknown elements of the symmetric tensor at each voxel representation.

In the following we first describe the tensor model for small angle X-ray scatter and then, in evaluating the tensor model, we describe how wave optics are used to simulate the projections of a phantom consisting of four layers of parallel carbon fibers.

A key aspect of this work is demonstrating that using X-ray interferometry, longitudinal directional measurements (parallel to the optical axis) and transverse directional measurements (orthogonal to the optical axis and in the direction of the gradient sensitivity) can form directional tensor projections. Using these measurements with sufficient angular projection samples, one can reconstruct 2nd-rank tensor fields of dark field scatter.

## 2 Materials and Methods

### A. *Tensor Model for Small Angle X-ray Scatter*

It has been shown [3-5] that the physical model behind directional dark-field imaging can be described by

$$p_s(x,y) = \exp\left[-\int \langle \hat{\epsilon}(x,y,z), \hat{t}\rangle^2 dz\right] \quad (1)$$

where $p_s(x,y)$ are the projections with detector coordinates $(x,y)$, $\hat{\epsilon}(x,y,z)$ is the small angle scattering coefficients with 3D coordinates $(x,y,z)$, and $\hat{t}$ is the sensitive vector of the grating (Fig. 1). In small angle X-ray scattering experiments, anisotropic scattering, represented as an ellipsoid, is related to anisotropy of the sample structure. Based on the nanostructure orientation distribution function and directional dark field imaging, a model representing the projection of small angle scatter is developed here, which establishes the relationship between the dark field projection and small angle scatter.

The relationship between the direction of the beam, the sample, and the measurement direction is defined as follows: $\hat{v}_x$ is the direction unit vector of the incoming X-ray beam; $\hat{v}_s$ is the measurement direction unit vector, which is the sensitivity direction of the grating parallel to the grating surface but perpendicular to grating lines; $\hat{v}_b$ is the unit vector of the corresponding direction of the nanostructure orientation distribution function.

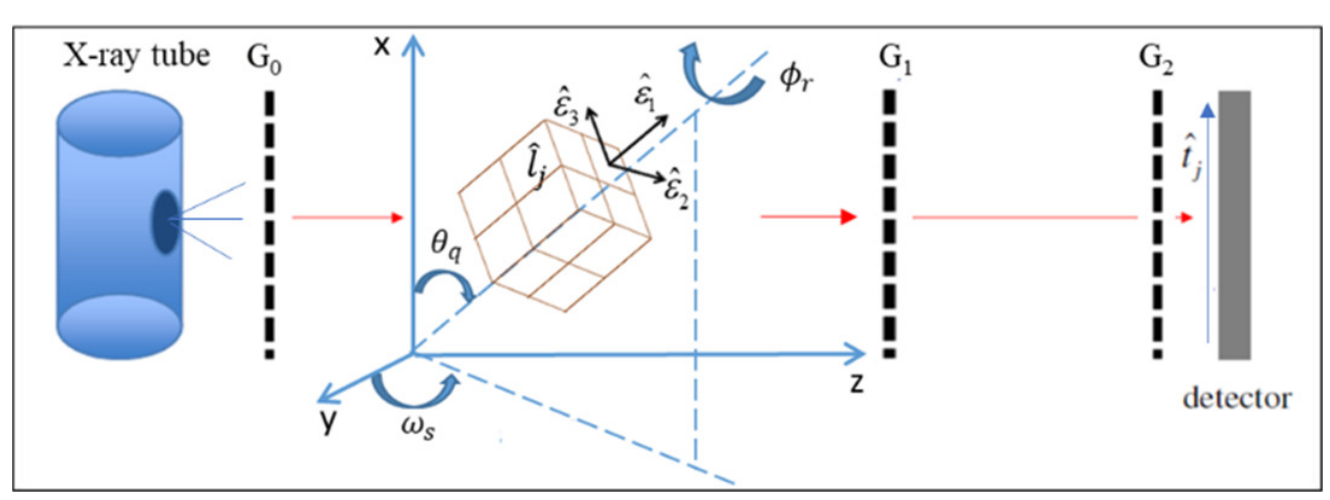

Fig. 1. Schematic diagram of the X-ray interferometry system used in our simulations.

The local small angle scatter is assumed to be a symmetric 2nd-rank tensor field $T(x,y,z)$. The measured intensity $\zeta(x,y,z)^2$ can be expressed as

$$\zeta(x,y,z)^2 = \hat{v}_s{}^T \cdot T(x,y,z) \cdot \hat{v}_s \quad (2)$$

where $\hat{v}_s$ is the sensitivity vector of the grating interferometer. Thus,

$$\hat{\epsilon}(x,y,z) = \zeta(x,y,z)\hat{v}_s = \zeta(x,y,z)\hat{t} \quad (3)$$

Substituting (3) into (1),

$$p_s(x,y) = \exp[-\int \langle \hat{\varepsilon}(x,y,z), \hat{t} \rangle^2 d\hat{v}_x] =$$
$$\exp[-\int \langle \zeta(x,y,z)\hat{t}, \hat{t} \rangle^2 d\hat{v}_x] = \exp[-\int \zeta(x,y,z)^2 d\hat{v}_x]$$
$$p_s(x,y) = exp[-\int \hat{t}^T \, T(x,y,z) \cdot \hat{t} \, d\hat{v}_x] \qquad (4)$$

where $\hat{t}$ is perpendicular to $\hat{v}_x$. The equation in (4) has the same expression as $p_{\underline{\theta}}^{\underline{\beta}\,\underline{\beta}}$ in Fig. 2, where the contribution of each voxel to the line integral in the direction of $\hat{v}_x$ is the length of the blue line in the direction of $\hat{t}$ intersecting the ellipsoid. The integration line goes through the centers of all ellipsoids.

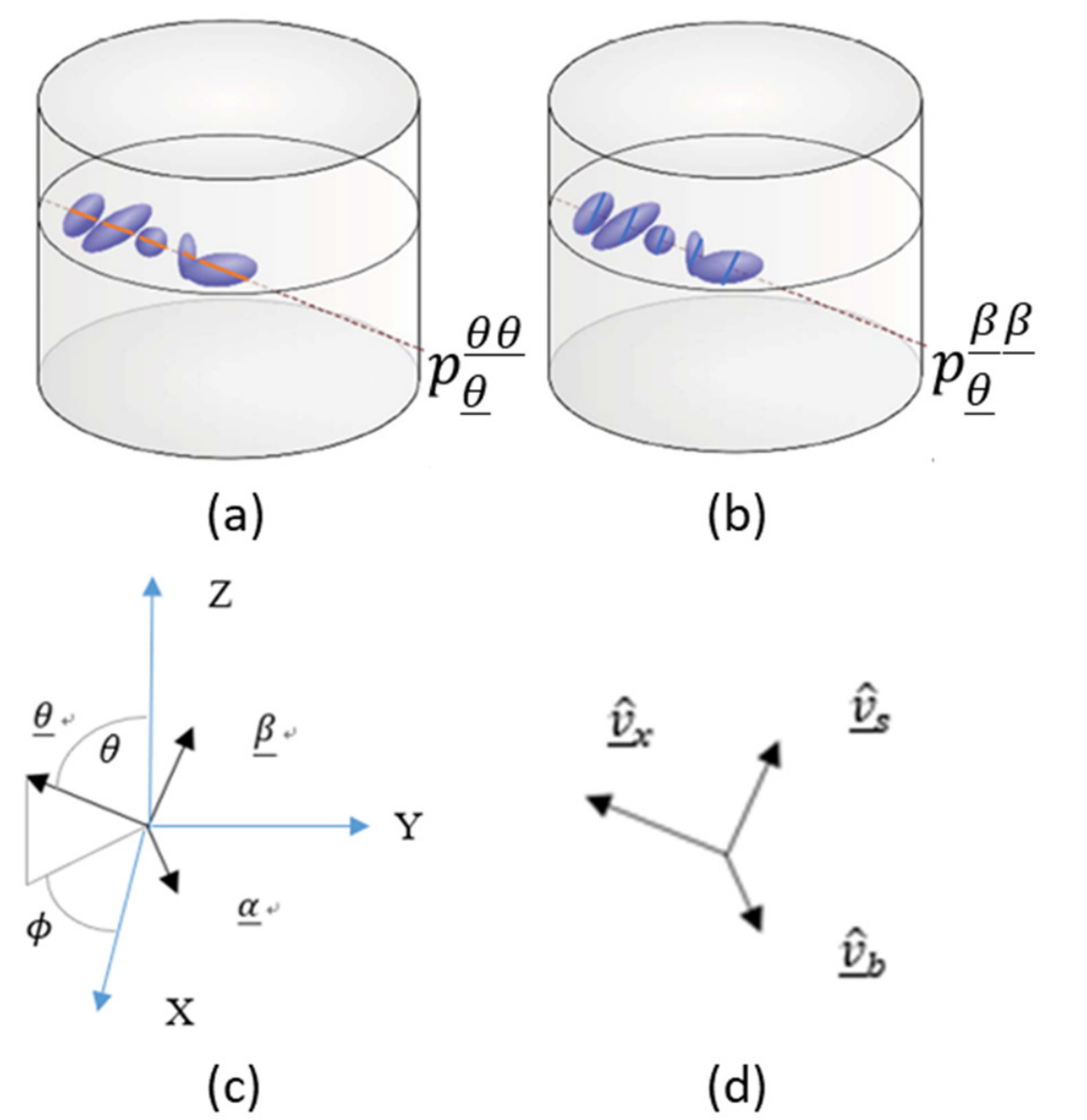

Fig. 2. Illustration of dark field projection. Here the integration line goes through the centers of all ellipsoids. (a) $p_{\underline{\theta}}^{\underline{\theta}\,\underline{\theta}}$: the integral along $\underline{\theta}$ (along $\hat{v}_x$) summing the orange intersections along $\underline{\theta}$. (b) $p_{\underline{\theta}}^{\underline{\beta}\,\underline{\beta}}$ : the integral along $\underline{\theta}$ (along $\hat{v}_x$ ) summing the blue intersections in the direction of $\underline{\beta}$. (Drawn based on Fig. 5 in [6] but modified to indicate the tensor measurements along the coordinates $\underline{\theta}$ and $\underline{\beta}$ in the lab frame.)

Due to symmetry, we can rewrite $T(x,y,z)$ as

$$T(x,y,z) = \begin{bmatrix} t_{xx} & t_{xy} & t_{xz} \\ t_{xy} & t_{yy} & t_{yz} \\ t_{xz} & t_{yz} & t_{zz} \end{bmatrix} (x,y,z) \quad . \qquad (5)$$

If $\underline{\beta} = [-cos\theta cos\phi, -cos\theta sin\phi, sin\theta]^T$, then $p_s(x,y)$ in (1), which is the projection $p_{\underline{\theta}}^{\underline{\beta}\,\underline{\beta}}$, becomes

$$-\ln[p_s(x,y)] = -\ln\left[p_{\underline{\theta}}^{\underline{\beta}\,\underline{\beta}}(x,y)\right] =$$
$$\int [(cos\theta cos\phi)^2 t_{xx} + 2(cos\theta cos\phi cos\theta sin\phi) t_{xy}$$
$$+(cos\theta sin\phi)^2 t_{yy} - 2(cos\theta cos\phi sin\theta) t_{xz}$$
$$-2(cos\theta sin\phi sin\theta) t_{yz} + (sin\theta)^2 t_{zz}] dz \qquad (6)$$

Simplifying:

$$-\ln[p_s(x,y)] = -\ln\left[p_{\underline{\theta}}^{\underline{\beta}\,\underline{\beta}}(x,y)\right] =$$
$$\int [v_{xx}t_{xx} + v_{xy}t_{xy} + v_{xz}t_{xz} + v_{yy}t_{yy} + v_{yz}t_{yz} + v_{zz}t_{zz}] dz \qquad (7)$$

where

$v_{xx} = (cos\theta cos\phi)^2$ , $v_{xy} = 2cos\theta cos\phi cos\theta sin\phi$ , $v_{xz} = -2cos\theta cos\phi sin\theta$, $v_{yy} = (cos\theta sin\phi)^2$, $v_{yz} = -2cos\theta sin\phi sin\theta$, $v_{zz} = (sin\theta)^2$ , are determined by $\underline{\beta}$, or here denoted by the sensitivity vector $\hat{t}$ of the grating interferometer.

The reconstruction problem involves solving a large system of linear equations. In this work we used a simultaneous algebraic reconstruction algorithm to reconstruct the tensor elements directly using (7). The iteration step from *k* to *k+1*:

$$t_{oj}^{(k+1)} = t_{oj}^{(k)} + \frac{\sum_{i \in P_{(\theta,\emptyset)k}} \lambda_u \frac{P_i - \sum_{\beta=1}^{6} \sum_{n=1}^{N} w_{in} v_{\beta ij} a_{\beta j}^{(k)}}{\sum_{\beta=1}^{6} \sum_{n=1}^{N} w_{in} v_{\beta ij}} w_{ij} v_{oij}}{\sum_{\beta=1}^{6} \sum_{i \in P_{(\theta,\emptyset)k}} w_{ij} v_{\beta ij}},$$
$$\forall 1 \le j \le N \qquad (8)$$

where $t_{oj}^{(k+1)}$ is *k+1* iteration for tensor matrix element *o* (for example *o=xy*) for voxel *j=1,…,N*; $w_{ij}$ are elements of the system matrix for projection *i*, voxel *j*; $v_{oij}$ are coefficients of the tensor elements, $\lambda_u$ is an update factor to improve the step size performance; sum over all projections $i\varepsilon P_{(\theta,\phi)k}$; $D_i$ is $-\ln\left(p_{\underline{\theta}}^{\underline{\beta}\,\underline{\beta}}(x,y)\right)$ in (7).

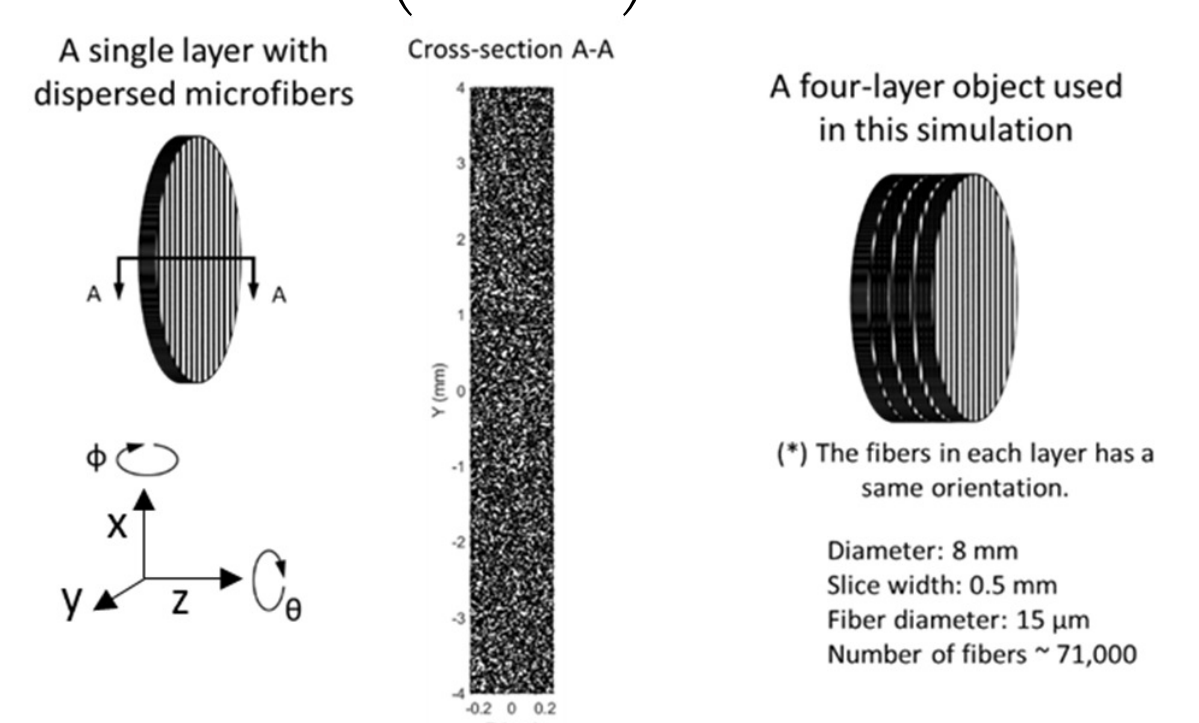

Fig. 3. Phantom used to simulate the tensor projections. The phantom consisted of four layers of parallel carbon microfibers to provide preferential scatter perpendicular to the direction of the fibers. The angles $\phi$ and $\theta$ show the rotation directions of the projections of the phantom. (This is a copy of Fig. 3 in [7].)

### B. *Phantom Simulation*

X-ray projections were simulated for an asymmetric scattering phantom in Fig. 3 consisting of four layers of parallel carbon microfibers designed to emphasize the small angle scatter in the dark field projections. Each layer width was 0.5 mm and the diameter of the disk was 8 mm. The layers each consisted of an array of 71,000 fibers of 15 $\mu$m diameter. All solid carbon fibers had the same orientation along the x-axis providing preferential scatter along the y-axis perpendicular to the fibers.

The directional projections were formed using an integrated wave optics framework [8,9] to model X-ray-matter interaction and free-space propagation. A wave optics representation of the projections involves solving the scalar wave equation for the phantom from Maxwell's equation for an electromagnet wave whose vector representation of the electric and magnetic disturbance can be represented by a scalar wave function with complex amplitude $\Psi(x,y;z)$ [8,9]. The scalar wave equation describes the interaction of X-rays with an object as a wave, and thus is appropriate to simulate the forward model for phase-sensitive X-ray imaging:

$$(\nabla^2 + k(\vec{r})^2)\Psi(\vec{r}) = 0 \quad ,$$
$$k(\vec{r}) = kn(\vec{r}), k = 2\pi/\lambda \quad ,$$

where $\lambda$ is the wavelength in a vacuum and $n(x,y;z)$ is the complex refractive index of the object. Note that the refractive index decrement $\delta$ is related to the complex refractive index $n$ : $n(x,y,z) = 1 - \delta(x,y,z) + i\beta(x,y,z)$, where $1 - \delta(x,y,z)$ and $\beta(x,y,z)$ is a measure of dispersion and of absorption, respectively.

The first-order Born or Rytov approximation [10] simplifies the solution for the scalar wave equation. The first-order Rytov approximation is more appropriate for X-ray imaging, because the imaged object is very thick (compared to the wavelength) but has a small refractive index difference (on the order of $10^{-7}$). Using the first-order Rytov approximation, the complex amplitude $\Psi(x,y;z)$ of the X-ray wave function after the interaction with the object can be written as

$$\Psi(x,y;z) = \Psi_0(x,y;z)\, exp[\phi_s(x,y;z)], \qquad (9)$$

where $\Psi_0(x,y;z)$ is the X-ray's complex amplitude assuming no object in the beam path, and $z$ is the distance from the center of the object. The complex scattered phase $\phi_s(x,y;z)$ can be related to the scattering potential of the object $Q(x,y;z)$ by

$$\tilde{\phi}_s(k_x,k_y;z) = [i4\pi(k_z + 1/\lambda)]^{-1}\, exp(i2\pi k_z z)\, \tilde{Q}(k_x,k_y,k_z), \quad (10)$$

where $\lambda$ is the wavelength in vacuum, $k_z$ is determined by $k_z = \left((1/\lambda)^2 - {k_x}^2 - {k_y}^2\right)^{1/2} - 1/\lambda$ . $\tilde{\phi}_s(k_x,k_y;z)$ is the 2D Fourier transform of $\phi_s(x,y;z)$ with respect to $x$ and $y$ . $\tilde{Q}(k_x,k_y,k_z)$ is the 3D Fourier transform of $Q(x,y,z)$. The scattering potential $Q(x,y,z)$ is given by the complex-valued refractive index $n(x,y,z)$:

$$Q(x,y,z) = (2\pi/\lambda)^2(1 - n(x,y,z)^2) \quad . \qquad (11)$$

For the phantom in Fig. 3, we derived in [7] an expression for $\tilde{Q}(k_x,k_y,k_z)$ for the entire stack of fibers as the sum $\tilde{Q}_{S_t}(k_x,k_y,0)$ of $\tilde{Q}_i(k_x,k_y,k_z)$ of individual microfibers. Relating $\tilde{Q}_{S_t}(k_x,k_y,0)$ to the scatter phase $\tilde{\phi}_{s_t}(k_x,k_y;z)$ for all layers:

$$\tilde{\phi}_{s_t}(k_x,k_y;z) = \frac{\lambda}{i4\pi} exp\left[-i\pi\lambda z\left(k_x^2 + k_y^2\right)\right] \times$$
$$\sum_{i=1}^{N_{fibers}} e^{-i2\pi(k_x x_i + k_y y_i)}\left(Q_0 R L_i/k_y\right) J_1\left(2\pi R k_y\right) sinc\left(L_i k_x\right) .$$

Therefore, the complex amplitude $\Psi_1(x,y;z)$ of X-rays after a stack of microfibers is

$$\Psi_1(x,y;z) = \Psi_0(x,y;z) exp\left[\phi_{s_t}(x,y;z)\right] \quad . \qquad (12)$$

We then showed that the propagation of $\Psi_1(x,y;z)$ through $G_1$ and transported a distance $D_2$ before $G_2$ (Fig. 4) obtained an expression $\Psi_3(x,y;z)$ (see [7]). Assuming the square modulus: $I_3(x,y;z) = |\Psi_3(x,y;z)|^2$ , is a reasonable approximation for the intensity of the irradiance distribution, we obtained an expression for $I_4(x,y;z)$. Digitizing the expression provided projection images of 16384×256. The images were oversampled along the x-scan direction to capture the small-angle scattering. The 16384×256 phase stepped images were processed as described below and then down sampled to 256×256.

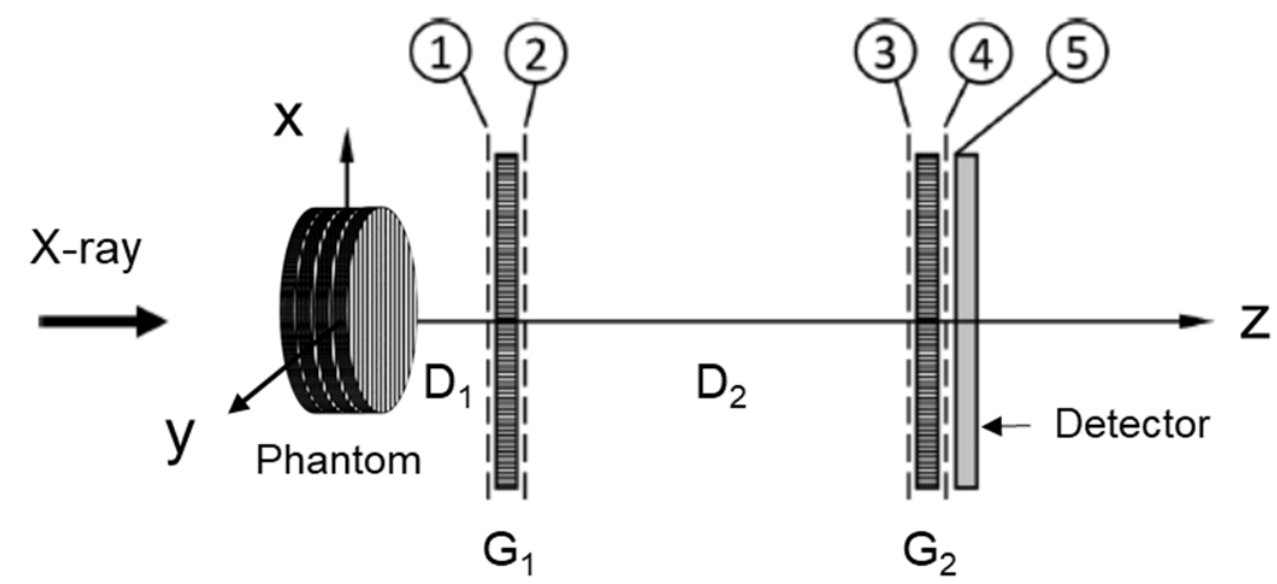

Fig. 4. Schematic diagram of the simulation geometry used in the simulation study. $G_1$ and $G_2$ are the gratings and the phantom is the four layers of fibers shown in Fig. 3. $D_1$ is the distance between the phantom and the first grating $G_1$. $D_2$ is the distance between the first grating and the second grating $G_2$. The numbers 1 through 5 in the circles refer to the planes where $\Psi_1(x,y;z) - \Psi_5(x,y;z)$ were calculated. For the phase grating simulation, the phase grating $G_2$ was shifted 8 positions in the x-direction over one period. (This is a copy of Fig. 4 in [7])

In our simulations gratings $G_1$ and $G_2$ were similar both with a grating pattern width of 10.24 mm with an 8-pixel period width of 0.005 mm and grating aperture of 0.0025 mm. From the eight phase steps, the visibility with phantom and with reference were extracted as the inverse discrete one-dimensional Fourier transform along the x coordinate to obtain the 1st harmonic of the discrete inverse of the eight phase steps / 0st harmonic of the discrete inverse of the eight phase steps: [ $V_{obj}(i,j,\theta_q,\phi_r) = a_1^{obj}(i,j,\theta_q,\phi_r)/a_0^{obj}(i,j,\theta_q,\phi_r)$ , [ $V_{ref}(i,j,\theta_q,\phi_r) = a_1^{ref}(i,j,\theta_q,\phi_r)/a_0^{ref}(i,j,\theta_q,\phi_r)$. Thus, the projections in (7) were

$$-\ln[p_s(x,y)] = -\ln\left[p_{\underline{\theta}}^{\beta\,\beta}(x,y)\right] = -\ln\left[\frac{V_{obj}(i,j)}{V_{ref}(i,j)}\right] .$$

The tomographic projections of the phantom in Fig. 3 included a total of 546 parallel projection images for θ from 0° to 90° at 18° steps (6 angles), and $\phi$ from -90° to 90° at 2° steps (91 angles). Eight phase steps were simulated for each projection. If one assumes the phantom is static, the interferometry system rotates around the phantom, so the sensitivity vector of the grating rotates. When $\theta$ is 0 and $\phi$ is 0, sensitivity vector is [1,0,0], in the direction of x axis.

Then rotating around z axis, sensitivity vectors become [1,0,0] (0°), [0.951,-0.309,0] (18°), [0.809,-0.588,0] (36°), [0.588,-0.809,0] (54°), [0.309,-0.951,0] (72°), [0,-1,0] (90°). When the phantom rotates around the x axis, it is equivalent to the sensitivity vector rotating in a plane perpendicular to the sensitivity vector, which does not change the sensitivity vector. From the six sensitivity vectors, we see the z component is always 0, causing some coefficients in (7) being 0. We cannot reconstruct the tensor elements if the coefficients are 0. In the implementation, the sensitivity vector was set to [0.99,-0.1,0], deviating from [1,0,0] a little bit when $\theta$ is 0° and $\phi$ is 0°. In this case, the z component of sensitivity vectors would not always be 0.

## 3 Results

Figure 5 shows results of the reconstruction of the six elements of the symmetric tensor for a single slice reconstructed from the simulated projections of the phantom (Fig. 3) using the algorithm described in (8).

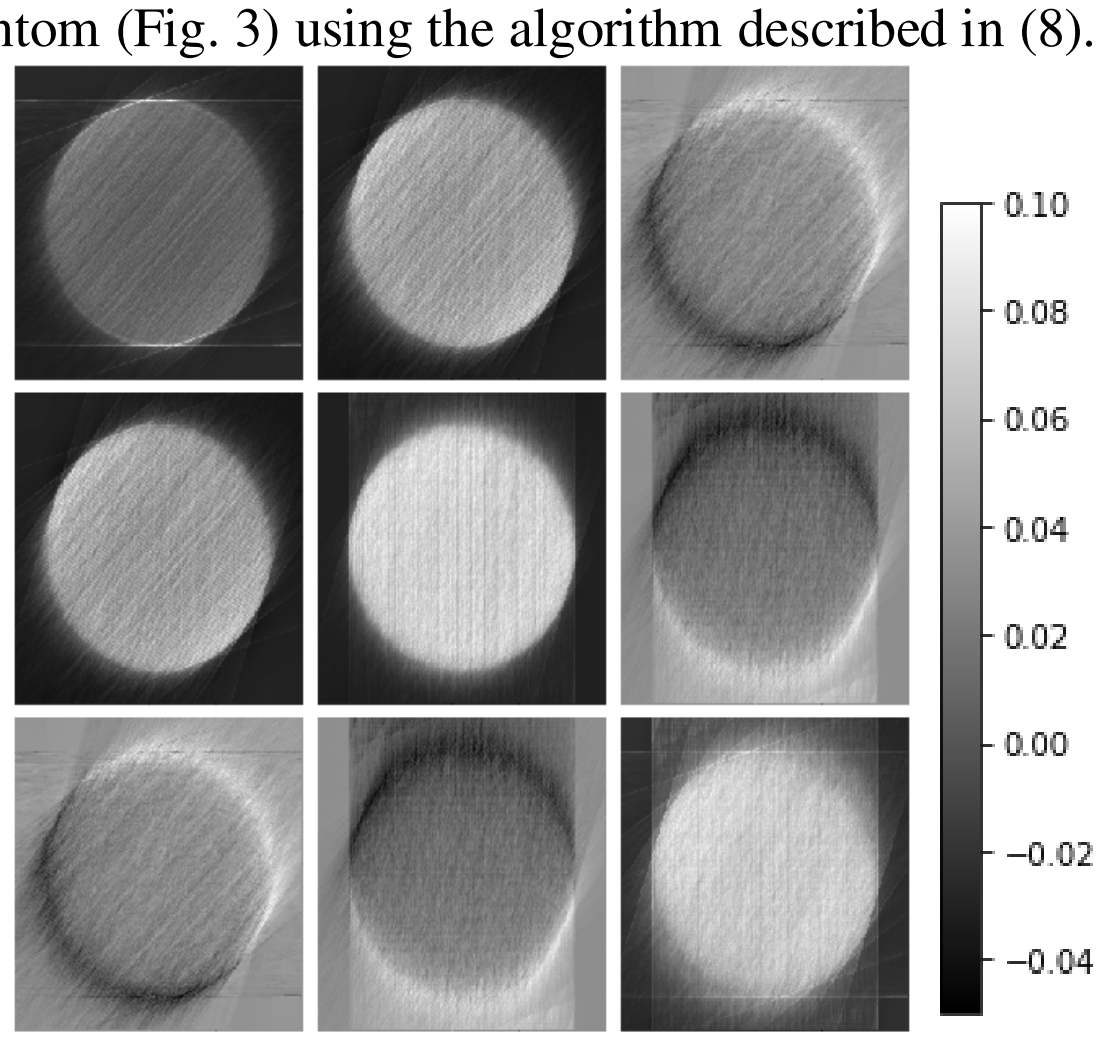

Fig. 5. Reconstructed tensor field using the direct tensor reconstruction algorithm in (8). The nine elements from slice 128 of the symmetric tensor are shown in the figure, respectively.

## 4 Discussion

A key to modeling small angle scatter as a tensor is being able to determine whether the specific orientation dependence of small-angle scatter associated with an underlying anisotropic mass distribution can be represented by a tensor model. Gratez [1] showed that X-ray dark-field scatter could be approximated as a tensor and X-ray interferometry, longitudinal and transverse directional measurements, could form scalar directional tensor projections. Using this result our work presents methodogy for the direct reconstruct of a tensor representation of X-ray small angle scatter from grating measurements of X-ray dark-field projections. Expressions in (2) and (3) are used to derive the tensor representation of the dark field projection of small angle scatter in (1). Wave optics was used for the the simulation of the grating X-ray interferometry projections. Fourier processing of the simulated projections provided scalar values of the transverse directional projection measurements $p_{\underline{\theta}}^{\underline{\beta}\,\underline{\beta}}$. From these projection measurements an iterative reconstruction algorithm was performed to directly reconstruct the tensor representation of small angle scatter for a phantom of parallel carbon fibers. In the future we need to verify whether these simulated projections were sufficient since we have shown [11] that three orthogonal orbits with two separate directional projection measurements $p_{\underline{\theta}}^{\underline{\beta}\,\underline{\beta}}$ and $p_{\underline{\theta}}^{\underline{\theta}\,\underline{\theta}}$ may be necessary to uniquely reconstruct the tensor elements using a filtered backprojection algorithm.

## 5 Conclusion

An iterive reconstruction algorithm is able to provide a minimum norm solution for the reconstruction of a symmetric 2nd rank tensor representation of small angle scatter with structure and preferential scattering directions as one would expect from a structure of parallel fibers.

# Basis image filtering enables subpixel resolution in photon-counting CT

Luca Terenzi[1], Per Lundhammar[1], and Mats Persson[1,2]

[1]Department of Physics, KTH Royal Institute of Technology, SE-106 91 Stockholm, Sweden
[2]MedTechLabs, BioClinicum, Karolinska University Hospital, Solna, Sweden

**Abstract** In this proof-of-concept work, we propose a method to further increase spatial resolution and contrast in the material decomposition approach for photon counting computed tomography (PCCT). By using different weights in the frequency domain of the two basis images obtained from a simulated phantom of water and aluminum, we improve the resolution of the reconstructed image and thereby correctly resolve the line patterns even after the first zero of the MTF. From the results, the proposed method manages to increase the modulation up to 80-200% depending on the amount of cross-talk in the detector, while correctly resolving the line pattern at low and high frequencies. To the best of our knowledge, this is the first demonstration of resolution improvement by differential basis image filtering, and this technique allows further enhancing the diagnostic quality of PCCT images.

## 1 Introduction

In the field of radiology, Computed Tomography (CT) is a well-known technique that allows to image a wide range of body structures with high resolution, from bones and muscles to blood vessels and organs.

Lately, the introduction of Photon Counting Detectors (PCD) in the CT approach, has allowed further improving the technique's performances: increased contrast and spatial resolution, improved SNR, intrinsic spectral information, reduced radiation dose and image acquisition time. [4][5][6].

Moreover, thanks to the capability of PCD to distinguish between X-rays photons of different energies, the material decomposition approach can be used to improve imaging of objects that contain material with different atomic numbers. In the case of medical imaging, the more common case is when imaging hard tissues, such as bone, together with softer tissues. Despite this, a more challenging case for conventional CT is represented when we want to tell apart two soft tissues of different materials but with similar density. In this latter case, PCD and material decomposition approach can further boost the image contrast and spatial resolution, taking advantage of the energy dependence of the linear attenuation coefficients of the different materials [7].

With this work, we simulate a phantom made of two materials (water and aluminum), and we filter with different weights in the frequency domain the two related basis images in the material decomposition approach, to further increase contrast and spatial resolution in the reconstructed image.

## 2 Materials and Methods

In the material decomposition approach, we can differentiate measurements of different incident energies $E_{inc}$ and then sort the attenuation into different energy bins. By doing so, one can obtain a sinogram for different energies, and energy becomes one further degree of freedom for the attenuation, which we can denote as $\mu(E_{inc})$. The main assumption in this approach is that all materials in the object can be decomposed into two or three basis materials. The materials we used in out simulated phantom are water and aluminum so we can write:

$$\mu(E_{inc}) = a_{Water}\mu_{water}(E_{inc}) + a_{Al}\mu_{Al}(E_{inc}) \tag{1}$$

$\mu_X$ are known functions and $a_X$ coefficients need to be determined from the measurements of the intensity. To do so we based our simulations for the material decomposition on the linearized version of the maximum-likelihood estimator obtained by linearization around the simulated ground truth of the pathlength vector $A_n$:

$$A_n^* = A_n + (J_n^T \Lambda_n^{-1} J_n)^{-1} J_n^T \Lambda_n^{-1}(Y_n - \lambda_n) \tag{2}$$

Where $J_n$ is the Jacobian of the forward model evaluated at $A_n$ and $\Lambda_n$ is the diagonal matrix of the forward model evaluated at $A_n$ (see ref [8] for derivation).

Note that the approximation used required to know the true path lengths a priori, which is not known in a realistic setting, but it remains a convenient approach when constructing a simulation.

Moreover, we made a slight modification to the equation (2), by having $Y_n - \lambda_{simp}$ where $\lambda_{simp}$ is the output of a simplified forward model built from summing over the neighbor's matrix the values in the PSF, concentrating the registered counts in the pixel of incidence. This was done because originally $Y_n = \lambda_n + noise$, and it is necessary to linearize around $\lambda_{simp}$ instead of $\lambda_n$ to model the effect of the PSF on the image spatial resolution.

The program used for simulation was MATLAB, based on the code from [8], and using the same geometry for the PCD as the one used in the reference.

The PSFs of the detector were obtained by a Monte Carlo simulation model of edge-on-irradiated silicon sensors [2][8] for two cases: standard conditions $PSF_s$ (charge cloud $\sigma = 19$ µm at 70 keV), and 4 times bigger charge cloud $PSF_{4x}$ (charge cloud $\sigma = 75$ µm at 70 keV) originated from photon interaction with the detector. This was done to study the effect of charge sharing between adjacent pixels in the detector and its effect on the reconstructed image when applying our filtering.

The total phantom used in the simulations is composed by a cylindrical water object with inserts made of parallel segment of aluminum of tunable spatial frequency disposed at the same radial distance. The parameters of the total phantom such as cylinder radius, radial distance of the aluminum inserts, their number, and their height can be all changed at will. For our simulations we mainly focused on a 4.3mm radius phantom (insert radius of 2.8mm, bar height of 1 mm and 5 bars for all the inserts), and each pixel in the simulated phantom measures 0.0063x0.0063 mm^2 to allow plotting very high lp/mm smoothly. The same results were achieved on a bigger phantom with a 2.5 cm radius.

Due to aliasing and computational limitations it was challenging to test our results on a bigger phantom, while maintaining a small pixel size in the simulated phantom to avoid being limited by it. Despite this, we do not expect to find any relevant differences in applying our method to a bigger phantom or object.

For the source parameters 1000 mAs and $2 * 10^6 \frac{photons}{mAs*mm^2}$ were used.

In the simulations we did not include noise, since our purpose here is to study the limits of attainable spatial resolution dictated by the MTF.

Moreover, we oversampled by a factor of 3 to reduce the aliasing, which is not the focus of this study and was heavily influencing the reconstructed image due to the approximation used in material decomposition at high spatial frequencies.

The main purpose behind the filter applied in the frequency domain of the two basis images is to obtain two images that exhibit the same trend of maxima and minima in attenuation values, successfully achieving 5 distinct maxima corresponding to the 5 aluminum inserts for frequencies higher than the first zero of the MTF. The general mechanism is shown in *Fig.1*. All this must be done maintaining the same attenuation coefficient values at zero frequency, while enhancing the contrast, and the resolution at high frequencies.

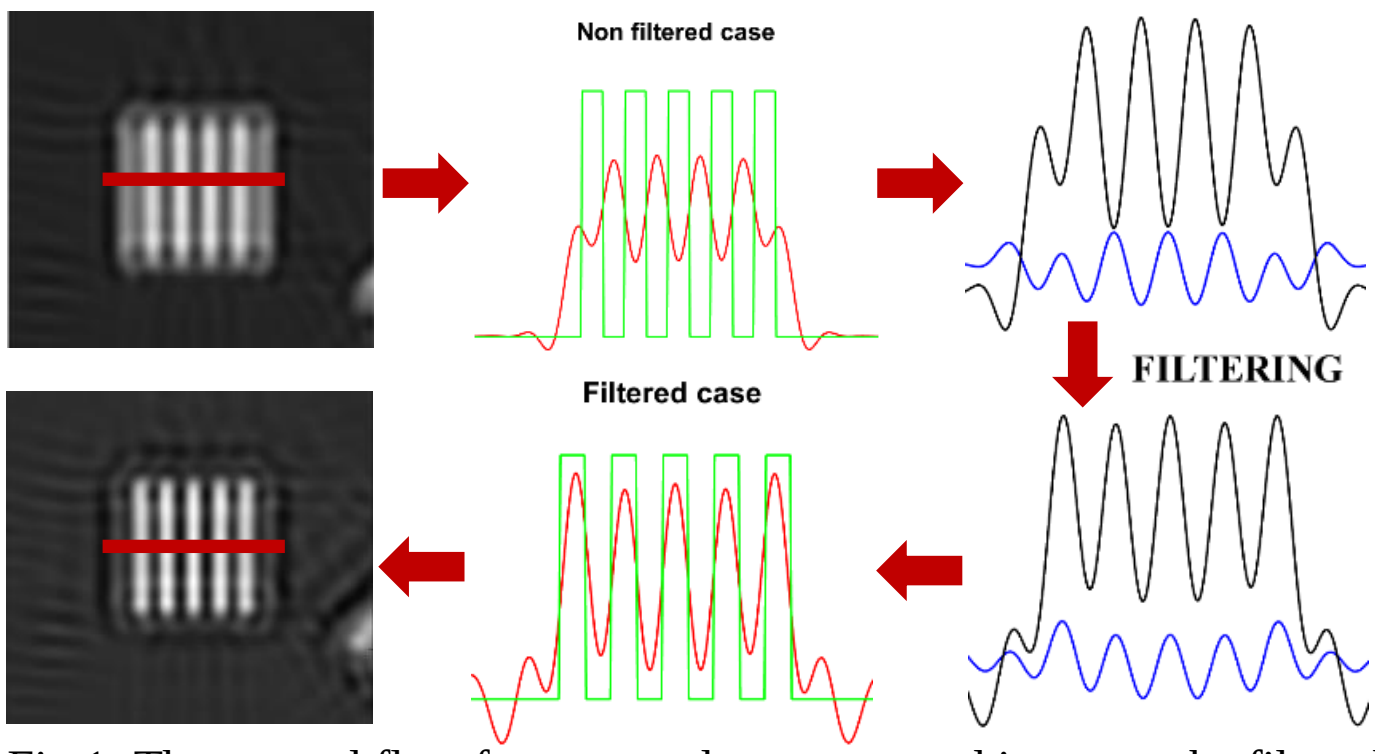

Fig.1: The general flow from normal reconstructed image to the filtered reconstructed one for a 50 lp/cm insert. The plots refer to the red line in the images on the left. In the two central images we plotted the insert section before and after filtering in red, together with the section in the same position for the insert in the phantom (square wave in green). The two plots on the right shows the water basis image (blue) and the aluminum one (black). For this example, the "overcompensating" filter was used.

In our simulations we analyzed the profile of the aluminum insert varying the frequency with integer unit steps and we noticed that the modulation approached zero for 39 lp/cm in the $PSF_s$ case and 44 lp/cm for the $PSF_{4x}$ one. After the afore-mentioned frequencies, both the water and the aluminum basis images inverted their minima and maxima position, as we would expect after having a zero in the MTF of a system. This led to obtaining 4 bright bars in the reconstructed image instead of the 5 we would expect from the aluminum inserts.

The next step was to identify potential filters that could improve contrast, also around the zero crossing of the MTFs, together with resolution at high frequencies while maintaining the same attenuation values we would expect from the materials.

Before the zero crossing, aluminum basis image exhibits maxima where the aluminum segments are, while water exhibits minima for the same positions. For this reason, we focused on two types of filters shown in *Fig.2* that could also enhance the contrast before the zero of the MTF: a first "overcompensating" filter that gives positive weights >1 to the aluminum for low frequencies, and then a "double zero crossing" filter, where we introduce a second zero crossing of the MTF for the water basis image. In both cases the weights are assigned to observe in the reconstructed image 5 brights bars at high frequencies and at the same time

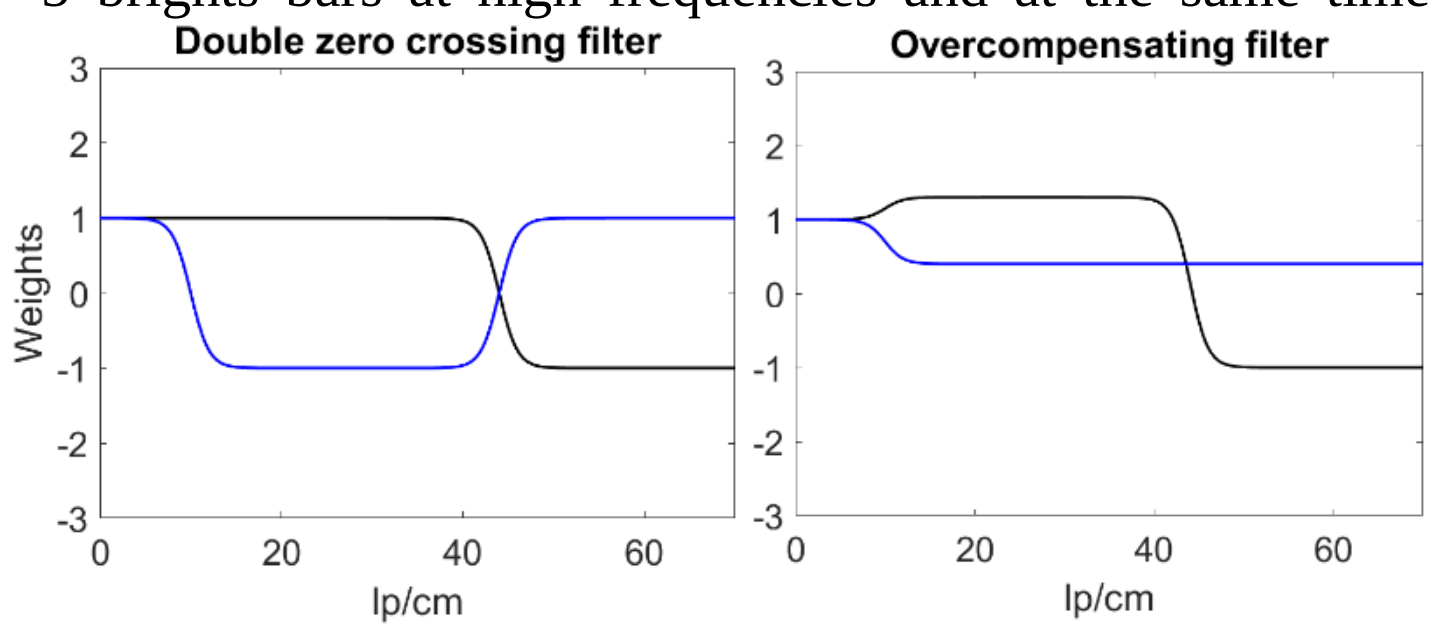

Fig.2: On the left we have the "double zero crossing" filter and on the right the "overcompensating" one for $PSF_s$. In each images black represent the filter for aluminum basis image and blue the water one. For $PSF_{4x}$ the zeros at 44 lp/cm would be at 39 lp/cm instead.

trying to increase modulation for lower and higher frequencies.

The filters were constructed following the mentioned criteria and using the sum of one or more functions with form of:

$$f(x) = \frac{y_1 - y_2}{1 + e^x} + y_2$$

This allows us to have smooth filters with two tunable asymptotes.

To compute the modulation to quantify the contrast increase we used the following formula:

$$m_f = \frac{A_{max} - A_{min}}{|A_{max} + A_{min}|}$$

where $A_{max}$ and $A_{min}$ were determined withing 2 periods of a given lp/cm from the center of the insert.

## 3 Results

Given the approximations and conditions for the simulation, in *Fig.3* we can see reconstructed images with and without filtering, for $PSF_s$ at different $E_{inc}$.

To assess the improvement of the filtering in the reconstructed image, we plotted in *Fig.3* a phantom having inserts with frequencies lower and higher than 44lp/cm, which represent the zero crossing of the MTF for $PSF_s$. We can observe how the contrast increases for all the energies. In addition, we are able in the filtered reconstructed images to correctly count 5 bars for the 50lp/cm pattern too, as *Fig.4a* shows more clearly.

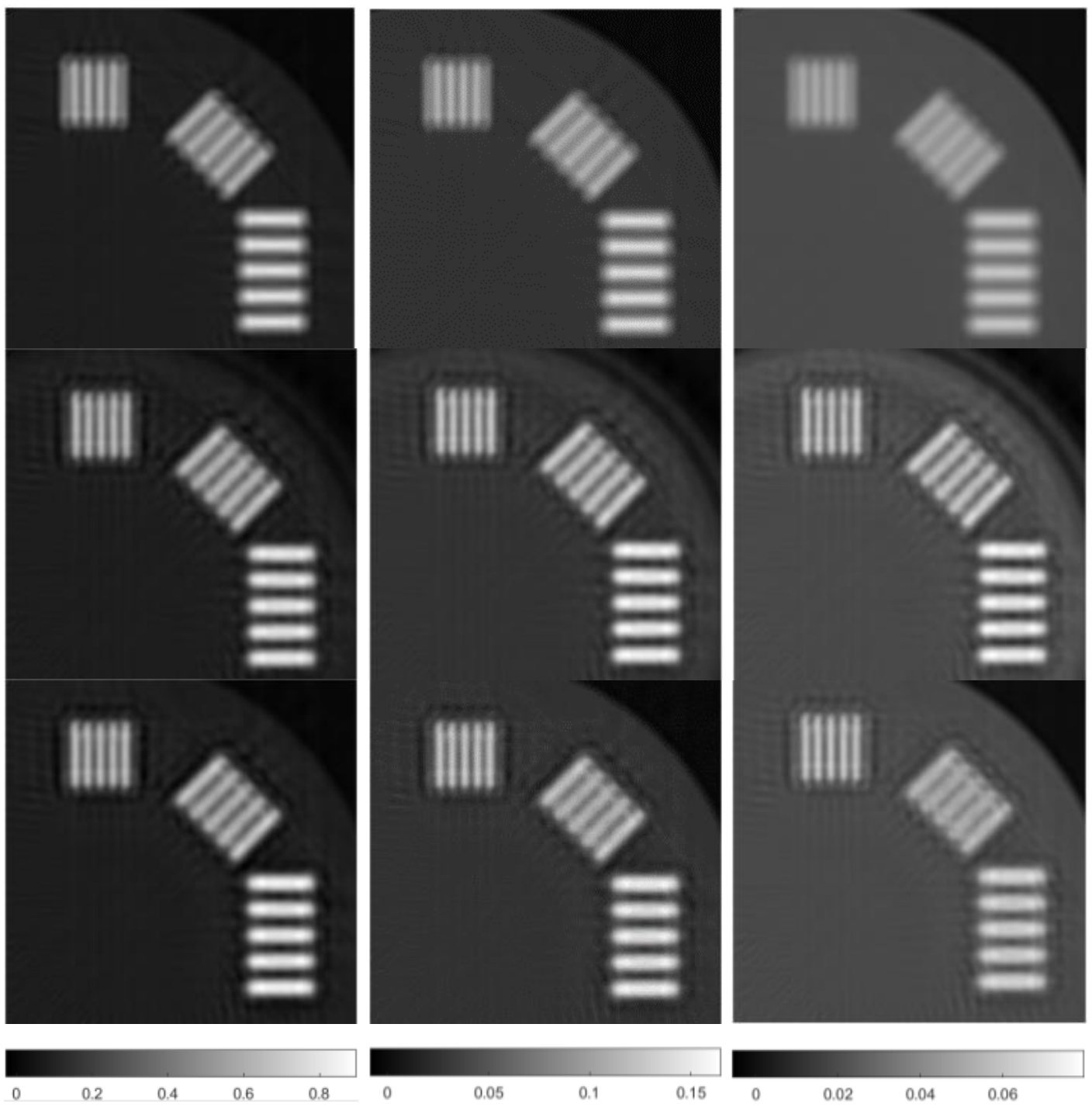

Fig.3: Reconstructed images of a phantom with 25lp/cm, 35lp/cm and 50lp/cm inserts is plotted for increasing $E_{inc}$ of 20 keV, 40 keV and 70 keV from left to right. The first row represents non filtered case, the second one deploys the "double zero crossing" filter and the third one the "overcompensating" one. The scalebar is the attenuation in $mm^{-1}$.

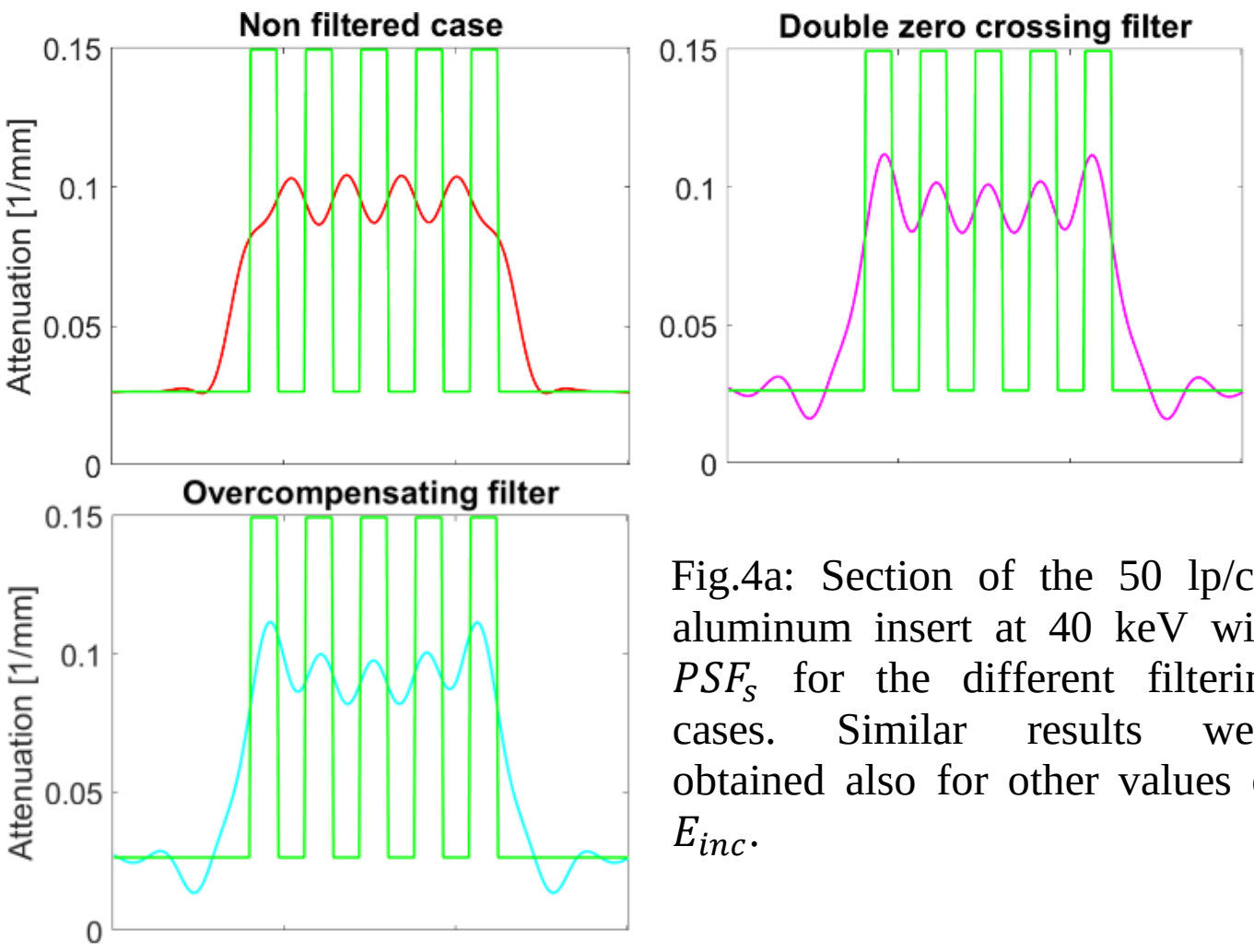

Fig.4a: Section of the 50 lp/cm aluminum insert at 40 keV with $PSF_s$ for the different filtering cases. Similar results were obtained also for other values of $E_{inc}$.

Similar images and improvements were obtained for the simulations with the $PSF_{4x}$, as we can see from *Fig.4b* and *Fig.5*.

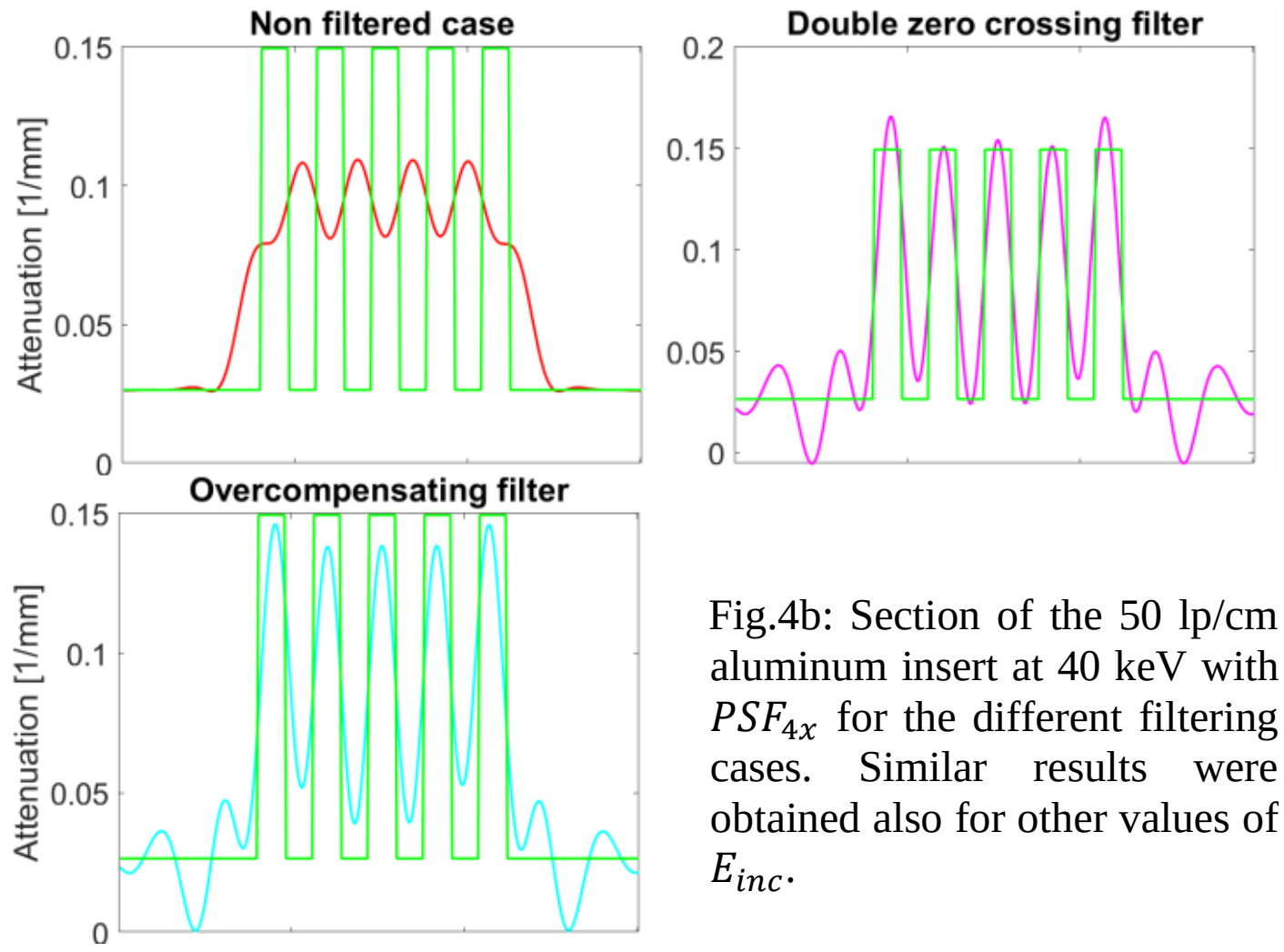

Fig.4b: Section of the 50 lp/cm aluminum insert at 40 keV with $PSF_{4x}$ for the different filtering cases. Similar results were obtained also for other values of $E_{inc}$.

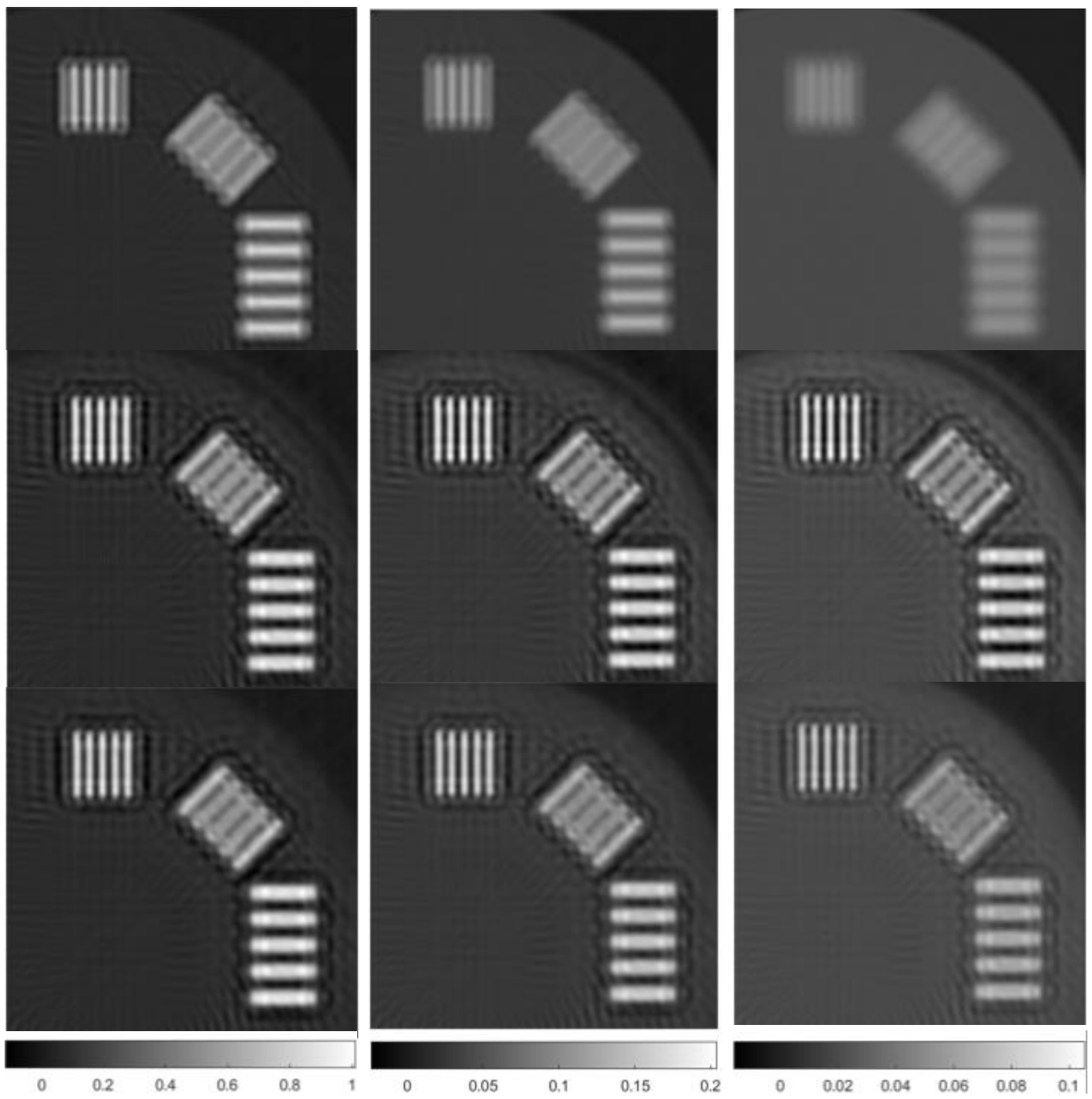

Fig. 5: Same as Fig.3 but for $PSF_{4x}$. The scalebar is the attenuation in $mm^{-1}$.

To quantify the comparison between the normal and filtered case for both PSF analyzed, we plotted in *Fig.6* the modulation at different frequencies at a given energy of $E_{inc} = 40\ keV$.

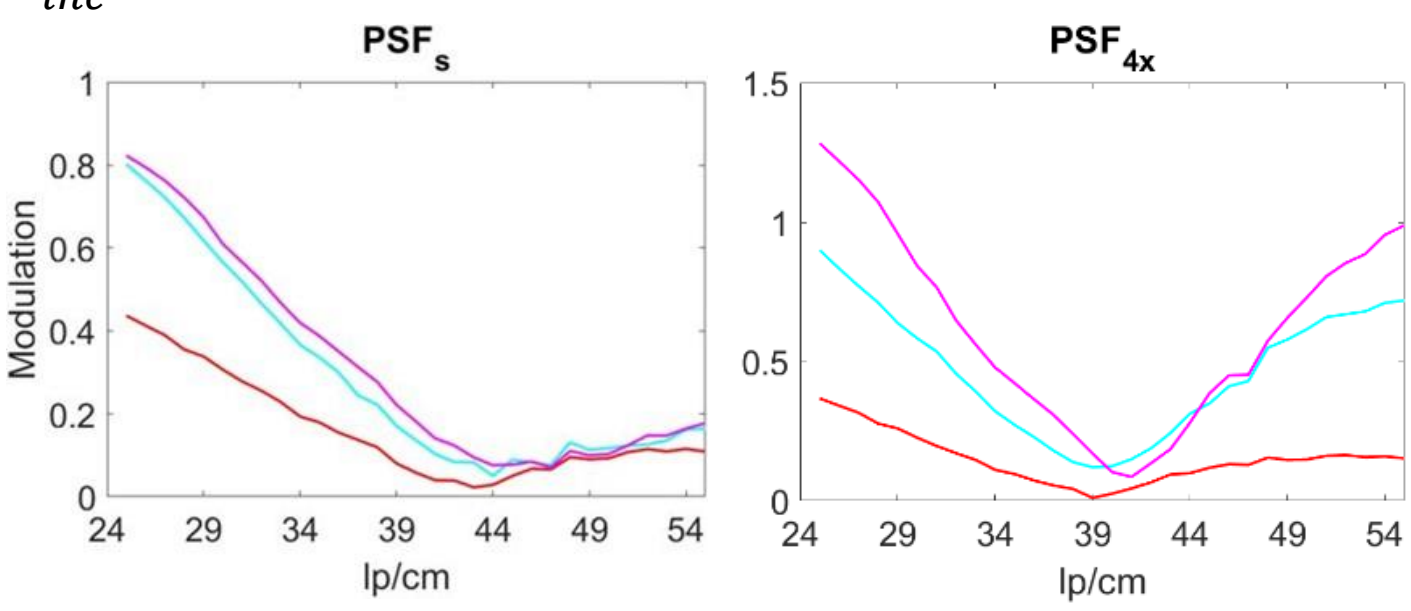

Fig.6: In red the modulation values for non-filtered reconstructed image, in magenta the filtered one with "double zero crossing" filter, and in cyan the case of the "overcompensating" filter.

For different energies, similar trends were observed, showing the consistency of the filtering method with respect to $E_{inc}$.

## 4 Discussion

As the results clearly shows, the filters proposed in this study works great for a wide range of frequencies, in both PSF cases and for different energies, managing to restore the correct five aluminum bars in the reconstructed image after the zero of the MTF that would normally lead to modulation inversion, also for frequencies higher than 50 lp/cm and before the second MTF's zero.

Despite the "double zero crossing" filter performs better in terms of modulation, we think that the "overcompensating" one would be better overall, since it introduces less artifacts in the image, such as the ringing artifact at the phantom edge in the second row (double-crossing filter) of *Fig. 3*, while still increasing contrast, and resolution at high frequencies.

Regarding the filter's choice, their shapes, zero crossing positions, and weights can be improved *ad hoc*, if the underlining idea of the proposed method is maintained as described in materials and methods section. This could be done for example for a different detector that exhibit a zero of the MTF at higher frequencies, or for different materials in the phantom that could require different weights in the filters.

By comparing the two PSFs, one interesting thing that can be noticed is the increased modulation that the filtered images with $PSF_{4x}$ show when compared to the ones obtained with $PSF_s$ before and after the zero-crossing point of the MTF (*Fig.6*). This happens even though the non-filtered images have very similar modulation for both PSFs. We hypothesize that a larger amount of charge sharing gives a larger difference between the MTFs for different energies, making it easier to avoid the zero-crossing by suitable weighting of basis images.

In addition to that, for both PSFs we can observe an increase in modulation for frequencies close to the zero crossing of the MTF.

Despite all the considerable improvements, a first drawback for this method is that one should first characterize different frequencies in a given range of interest to know where the zero of the MTF lies and what are the most appropriates weights and shapes for the filter.

Moreover, while it is true that we maintain the correct attenuation values at zero frequency for both materials, it is possible to obtain, especially at high frequencies and for $PSF_{4x}$, values that are slightly higher than the expected ones should be. In that sense, a deeper study for the characterization of the ideal weights at every frequency and energy could lead to having consistent attenuation values for all cases.

## 5 Conclusion

To conclude this proof-of-concept work, we can affirm that the proposed approach is effective in correctly resolving and increasing the visibility of high spatial frequencies inserted in the framework of material decomposition. As discussed, this may come at a price of not having exactly the exact attenuation values in the reconstructed image at high frequencies but studying in more detail the weights of the filter and its shape we believe it is possible to further reduce this problem.

In addition, the use of a bigger charge cloud PSF seemed to have increased the visibility of the aluminum inserts before and after its MTF zero crossing point.

We believe that filtering the basis images in the material decomposition approach for a PCCT, could lead to improvements in the medical imaging field, allowing doctors to correctly identify small structures and increasing the contrast around the zero of the MTF, resulting in a more clear reconstructed image.

Further studies could be made to investigate the improvement of using a bigger charge cloud PSF, as well as validating the method proposed here with an oversampling of a factor of 2 (or without), taking into account noise and testing it for different materials and with a real phantom to assess how it could perform in a real-world case.

## Acknowledgements

This study was financially supported by MedTechLabs and the Swedish research council (2021-05103). Mats Persson discloses research collaboration with GE Healthcare.

# Optimizing CT Scan Geometries With and Without Gradients

Mareike Thies[1], Fabian Wagner[1], Noah Maul[1], Laura Pfaff[1], Linda-Sophie Schneider[1,2], Christopher Syben[2], and Andreas Maier[1]

[1]Pattern Recognition Lab, Friedrich-Alexander-Universität Erlangen-Nürnberg, Erlangen, Germany
[2]Fraunhofer Development Center X-Ray Technology EZRT, Erlangen, Germany

**Abstract** In computed tomography (CT), the projection geometry used for data acquisition needs to be known precisely to obtain a clear reconstructed image. Rigid patient motion is a cause for misalignment between measured data and employed geometry. Commonly, such motion is compensated by solving an optimization problem that, e.g., maximizes the quality of the reconstructed image with respect to the projection geometry. So far, gradient-free optimization algorithms have been utilized to find the solution for this problem. Here, we show that gradient-based optimization algorithms are a possible alternative and compare the performance to their gradient-free counterparts on a benchmark motion compensation problem. Gradient-based algorithms converge substantially faster while being comparable to gradient-free algorithms in terms of capture range and robustness to the number of free parameters. Hence, gradient-based optimization is a viable alternative for the given type of problems.

## 1 Introduction

Numerous problems in computed tomography (CT) require optimizing the CT acquisition geometry based on a target function formulated on the reconstructed image. One of the most prominent examples is rigid motion compensation where the acquisition geometry is updated to compensate for involuntary patient motion occurring during the scan. This can be achieved by formulating target functions which quantify the quality of the reconstructed image via measures such as image entropy, total variation, or gradient entropy [1–4]. Minimization of the image quality criterion yields the motion parameters which produce the best reconstructed image according to the target function.

So far, these approaches have been limited to gradient-free optimization algorithms because the parameters being optimized and the target function live in different domains which are connected via the reconstruction operator. The free parameters define the geometrical relationship between scanned object, X-ray source, and detector pixels. Meanwhile, the target function is formulated in image space. The image space depends on the scan geometry in a complex manner where a change in a single geometry parameter can influence the entire reconstructed image. As a result, formulating the gradient of a target function in image space with respect to the scan geometry is not trivial and needs to incorporate the reconstruction step.

Recently, we proposed an algorithm for fan-beam CT geometries which computes all partial derivatives of the gray values in a reconstructed image with respect to the entries of the projection matrices parameterizing the scan geometry in CT [5]. These computations bridge the gap between image domain and geometry space and enable the formulation of gradients for the motion compensation problem mentioned above. Consequently, gradient-based optimization algorithms can be applied.

In this paper, we investigate the performance of different gradient-free and gradient-based optimization algorithms on the same geometry optimization problem concerning run time, capture range, and robustness to the number of free parameters. Doing so, we analyse whether gradient-free algorithms are naturally a better choice for the given type of problems or if gradient-based optimization algorithms are a comparable or even superior alternative.

## 2 Methods

### 2.1 Analytical Geometry Gradients

In our recent work [5], we presented the mathematical derivation and implementation to compute the partial derivatives of the gray values in a reconstructed CT image with respect to the entries of the $2 \times 3$-shaped CT projection matrices in fan-beam geometry. Projection matrices are a common parameterization of the CT scan geometry and describe the geometrical relationship between a point in the reconstructed image and the detector coordinate onto which this point is mapped. This fully specifies the orientation of the object (extrinsic information) as well as the detector itself (intrinsic information). We refer the reader to [5] for a detailed explanation of the mathematical steps involved in the gradient computation. For this paper, we assume that all partial derivatives of the reconstructed image $I \in \mathbb{R}^{N_x \times N_y}$ with respect to all entries of the projection matrices $P \in \mathbb{R}^{N_p \times 2 \times 3}$ can be computed analytically, i.e., the partial derivative $\frac{\partial I}{\partial P}$ is given. Here, $N_x$ and $N_y$ are the image dimensions and $N_p$ is the number of projections. Computing $\frac{\partial I}{\partial P}$ is the crucial step for formulating the gradient of a target function $f : N_x \times N_y \to \mathbb{R}$ with respect to the $N_f$ free parameters $g \in \mathbb{R}^{N_f}$ influencing the scan geometry. This is because the derivatives $\frac{\partial f}{\partial I}$ and $\frac{\partial P}{\partial g}$ are usually straight-forward to formulate or can even be computed via automatic differentiation in common deep learning libraries. Together, these variables define the gradient of the target function $f$ with respect to the free parameters $g$ via the chain rule of differentiation

$$\frac{\partial f}{\partial g} = \frac{\partial f}{\partial I} \cdot \frac{\partial I}{\partial P} \cdot \frac{\partial P}{\partial g} \quad . \tag{1}$$

## 2.2 Optimization Algorithms

We are interested in finding the optimal motion parameters $g^* \in \mathbb{R}^{N_f}$ by minimizing the target function $f$

$$g^* = \underset{g \in \mathbb{R}^{N_f}}{\arg\min} f(g) \quad . \tag{2}$$

In general, Eq. 2 is an unconstrained, non-convex problem which, however, can be locally convex depending on the parameterization and initialization of $g$. Several numerical optimization algorithms exist to solve Eq. 2 of which we compare the following ones: Covariance matrix adaptation evolution strategy (CMA-ES), Nelder-Mead Simplex, gradient descent, and Broyden-Fletcher-Goldfarb-Shanno (BFGS).

CMA-ES and Nelder-Mead Simplex are gradient-free algorithms which operate by repeatedly evaluating the target function until a minimum is met. Being an evolutionary algorithm, CMA-ES iteratively estimates a covariance matrix based on which new candidate solutions are generated. The Nelder-Mead Simplex algorithm evaluates the target function at the vertices of a $N_f + 1$ dimensional simplex followed by a number of possible operations on the simplex vertices designed to progress downhill on the target function. Both have successfully been applied to CT motion compensation via image quality criteria [2–4].

Gradient-based optimization has not been applied extensively to the considered problem even though gradient information is known to aid minimization by providing the steepest descent direction at a certain point $\hat{g}$ [6]. Gradient descent with a predefined step size updates the current solution by a step into the direction of the negative gradient multiplied by the step size. BFGS additionally uses an approximation of second order Hessian information and performs a line search along the descent direction instead of using a predefined step size.

## 2.3 Rigid Motion Compensation

The extrinsic component of each projection matrix is updated to compensate for rigid, inter-frame patient motion. The search space of this problem consists of the three rigid motion parameters rotation ($r$) and translation ($t_x$, $t_y$) for each of the $N_p$ projections leading to a total of $3N_p$ free parameters. To reduce the dimensionality of the search space and limit it to realistic motion patterns, further constraints can be enforced on each of the parameters by means of a motion model $m : \mathbb{R}^{N_f} \times \mathbb{R}^{N_p \times 2 \times 3} \rightarrow \mathbb{R}^{N_p \times 2 \times 3}$ which takes free parameters $g \in \mathbb{R}^{N_f}$ and the current estimate of projection matrices $P_n$ to yield updated projection matrices $P_{n+1}$

$$P_{n+1} = m(g, P_n) \quad . \tag{3}$$

In this case, the number of free parameters can be considerably smaller, i.e., $N_f \ll 3N_p$ [3].

# 3 Experiments

## 3.1 Data

We simulate motion-affected fan-beam sinograms from real reconstructed CT slices of publicly available cone-beam CT data of the head [7]. The images are of size $512 \times 512$ and the pixel spacing is assumed to be 1 mm. Fan-beam projection data are simulated for 360 projections over a full circle, a source-to-isocenter distance of 1000 mm, a source-to-detector distance of 2000 mm, and 1024 detector pixels with a spacing of 2 mm. Forward projections are performed using the implementation in [8]. Comparable to previous work in [2], the artificially introduced motion pattern is a step-like function for $r$, $t_x$, and $t_y$. From a start projection $p_{start}$, the perturbation increases linearly until it reaches its maximal amplitude at projection $p_{end}$ and stays constant for the rest of the scan. The maximal motion is $\pm 10$ mm for translation in x and y and $\pm 5.73°$ ($\pm 0.1$ rad) for rotation unless stated otherwise. $p_{start}$ and $p_{end}$ are chosen such that the motion pattern extends over 50 to 200 projections and is completed within the full 360 projections. The perturbation is only applied to the projection matrices, the projection data itself represent a circular motion-free trajectory.

## 3.2 Optimization Problem

Starting from the perturbed projection matrices (Sec. 3.1), we aim to find the motion parameters $t_x$, $t_y$, and $r$ which optimally annihilate the introduced step-like motion pattern and restore a circular trajectory. This is done by minimizing the mean squared error (MSE) between the motion-affected reconstruction and the ground-truth, motion-free reconstruction with respect to the motion parameters. The utilized motion model $m$ enforces a smooth curve for each of the three motion parameters by fitting a cubic spline with $N_n$ nodes which are equally distributed across the full scan range. Hence, the number of free parameters is $N_f = 3N_n$, i.e., three times the number of nodes in the respective splines. The number of nodes is $N_n = 10$ if not specified differently.

## 3.3 Optimizer Configurations

The problem described in Sec. 3.2 is optimized using each of the four optimization algorithms introduced in Sec. 2.2. All compared optimization algorithms are based on the same implementation of the target function and, if applicable, its gradient. As [5] is already implemented as a differentiable *PyTorch* operator, we further use the deep learning library to implement the spline-based motion model[1] and the MSE target function. This way, the gradient of the loss function (Eq. 1) is obtained easily by means of automatic backpropagation. Both gradient-free optimization algorithms terminate as soon as the target function value changes by less than

[1] https://github.com/patrick-kidger/torchcubicspline

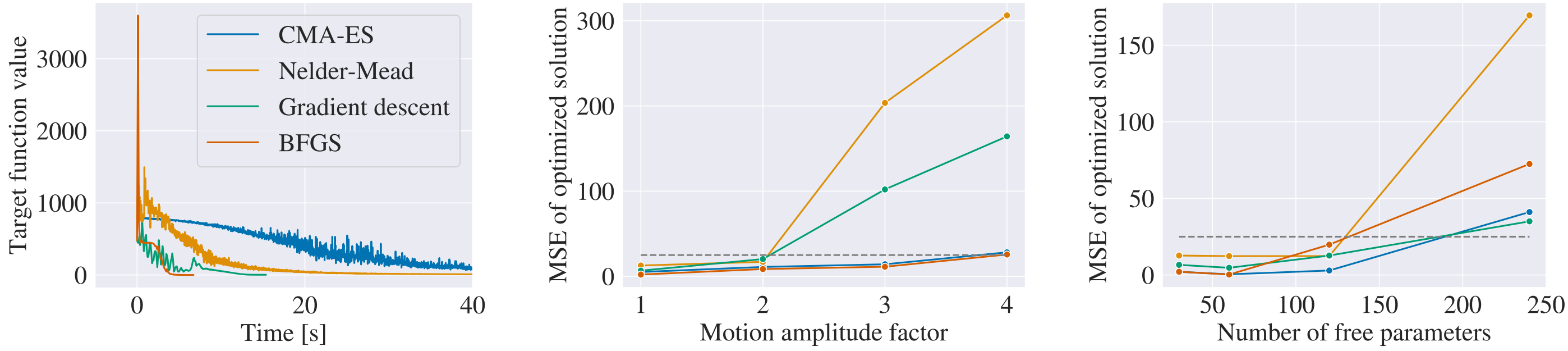

**Figure 1: Left**: Minimization of the MSE target function over time. Gradient-based optimization algorithms converge considerably faster than gradient-free algorithms. The plot does not show the gradient-free curves until full convergence. **Middle**: Evaluation of capture range. CMA-ES and BFGS converge successfully even for large motion perturbations. **Right**: Robustness against increase in free parameters. CMA-ES and gradient descent are most robust up to $\sim 200$ free parameters. The gray dashed line marks an MSE of 25 below which we consider an optimization result successful.

| | Gradient? | Source | NFEV | NJEV | MSE | Time [s] |
|---|---|---|---|---|---|---|
| CMA-ES | ✗ | pycma[1] | $7330.84 \pm 761.61$ | 0.0 | $0.99 \pm 0.52$ | $209.36 \pm 20.78$ |
| Nelder-Mead | ✗ | scipy[2] | $3481.12 \pm 1751.20$ | 0.0 | $6.28 \pm 2.95$ | $101.44 \pm 51.11$ |
| Gradient descent | ✓ | custom | $670.68 \pm 1945.85$ | $670.68 \pm 1945.85$ | $4.16 \pm 1.46$ | $36.83 \pm 52.09$ |
| BFGS | ✓ | scipy[2] | $69.56 \pm 14.46$ | $67.84 \pm 11.21$ | $1.21 \pm 0.65$ | $6.56 \pm 1.32$ |

[1] https://github.com/CMA-ES/pycma [2] https://github.com/scipy/scipy

**Table 1:** Overview of compared optimization algorithms and quantitative results. We report the number of objective function evaluations (NFEV) and the number of gradient evaluations (NJEV) needed until convergence. Further, the target function value (MSE) of the final optimized solution and the time needed for convergence is given. Standard deviations are computed across five different anatomies and five different initial motion perturbations.

0.1 or after 20000 function evaluations. The gradient-based optimization algorithms terminate once the gradient norm is less than 2 or after 500 iterations (BFGS)/ 10000 iterations (gradient descent). Initial standard deviations (CMA-ES) and step sizes (gradient descent) are adjusted to account for the different scales of translations [mm] and rotations [rad]. The step size in gradient descent decays by a factor of 0.995 after each iteration. Implementations of the optimization algorithms are summarized in Tab. 1.

## 4 Results

We first optimize the given problem with all four optimization algorithms using the standard settings described in Sec. 3. Each optimization algorithm is run 25 times using slices from five different patients with five realizations of the motion pattern each. Motion patterns vary concerning $p_{start}$ and $p_{end}$ as well as the direction of rotation and translations (positive or negative), but have the same amplitude. Each initialization and patient slice is used for all optimization algorithms identically. The resulting measurements are summarized in Tab. 1. The gradient-free optimization algorithms (CMA-ES and Nelder-Mead) have a considerably higher number of objective function evaluations (NFEV) than the gradient-based optimization algorithms (gradient descent and BFGS). For example, NFEVs for CMA-ES are two orders of magnitude higher than for BFGS. Both gradient-free optimization algorithms do not evaluate the gradient of the objective function and, hence, the number of gradient evaluations (NJEV) is zero whereas for the gradient-based algorithms NJEV $\approx$ NFEV. The MSE of the optimized solution to the ground truth is smallest for CMA-ES and BFGS. However, by visual inspection, we find that all MSE values below 25 represent successfully compensated images. The time until convergence exhibits the most notable differences. All gradient-based algorithms converge strikingly faster than the gradient-free counterparts, e.g., BFGS is over 30 times faster than CMA-ES for the investigated problem setting. This relationship is visible in the plot of the target function value over the optimization time in Fig. 1 (left) as well.

To analyse the capture range of the optimization algorithms, we measure their performance for motion patterns with increasing amplitude. On the same image slice, the amplitude of the initial motion perturbation is increased from the default value of $\pm 10$ mm to $\pm 40$ mm for translations and from $\pm 5.73°$ to $\pm 22.92°$ for rotation. The results are summarized in Fig. 1 (middle), where the motion amplitude factor describes the multiplicative increase of the motion amplitude compared to the default configuration. Hence, the higher the motion amplitude factor, the higher is the distance between initialization and true solution of the optimization problem. All four optimization algorithms exhibit a stable behavior up to motion amplitude factor of 2, and CMA-ES and BFGS even succeed for factors as high as 4. Initial and recovered reconstructions from the largest investigated motion amplitude are depicted in Fig. 2. Apparently, the initial motion degrades the image severely. Whereas CMA-ES and BFGS compensate the motion well, streaking artifacts remain for Nelder-Mead and the result optimized with gradient descent still exhibits slightly misaligned edges.

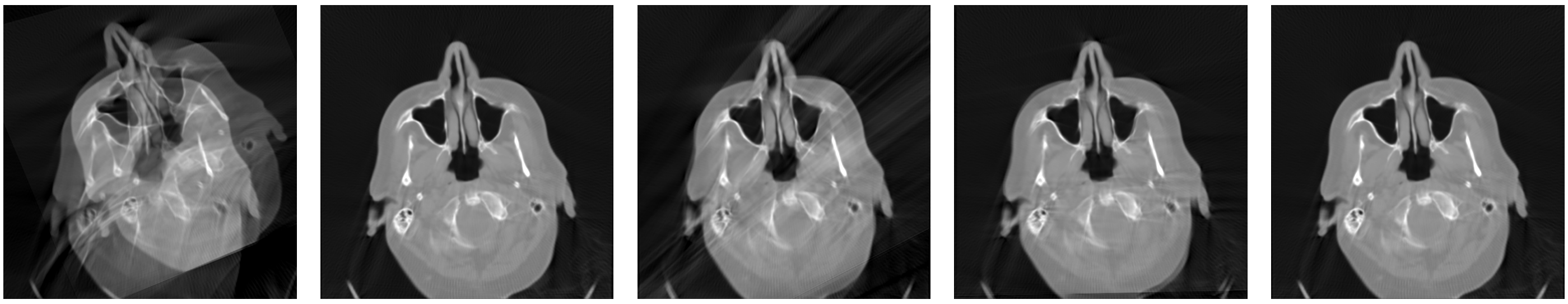

**Figure 2:** Reconstructed images recovered from the largest investigated motion amplitudes. The initial motion perturbation degrades the image severely. All optimization algorithms recover the shape of the head, but artifacts remain for Nelder-Mead and gradient descent.

Finally, we investigate the robustness of the four candidates to an increase in the number of spline nodes, and hence the number of free parameters, in the motion model. On the same image slice, we keep the motion pattern constant and vary the number of nodes per spline from 10 to 20, 40, and 80 leading to 30, 60, 120, and 240 free parameters (see Fig. 1 (right)). All algorithms succeed for up to 120 free parameters, but CMA-ES and gradient descent behave more stably for an even higher number of free parameters.

## 5 Discussion and Conclusion

We compare gradient-free and gradient-based optimization algorithms for a geometry optimization problem in fan-beam CT. Gradient-based optimization algorithms converge substantially faster mainly due to a largely reduced number of target function evaluations. In the investigated case, where each target function evaluation is expensive as it incorporates a reconstruction step, a reduction in NFEV is essential to speed up the optimization. We acknowledge that our implementation is not optimized for run time, but it is comparable and, therefore, applicable for a relative comparison between the investigated methods. While it is known that gradient-based optimization algorithms are susceptible to terminating in local minima, our investigations concerning capture range reveal a behavior comparable to the gradient-free algorithms for the given, generally non-convex problem setting. In particular, BFGS converges to a solution close to the global minimum even for largely perturbed initializations. The same holds for the robustness with respect to the number of free parameters. In this case, gradient descent performs on par with CMA-ES and succeeds even for high-dimensional optimization problems. Note that the gradient-based algorithms converge still substantially faster in both these experiments. Of course, we further acknowledge that our target function requires a ground-truth, motion-free scan at optimization time which is an unrealistic requirement for practical applications. In this work, however, the MSE target function serves as the most basic objective which lets us study the performance of different optimization algorithms without any influencing effects of a sub-optimal target function itself. In future studies, the MSE objective should still be replaced by an alternative image quality criterion. Additionally, future work might investigate the generalizability of our results to the 3D cone-beam case, but we are confident that the optimization speed up is even more pronounced in that computationally more expensive setting. Ultimately, we conclude that gradient-based optimization algorithms are a viable alternative for the studied geometry optimization problem because they accelerate the optimization without sacrificing capture range or robustness to the number of free parameters.

## Acknowledgements

The research leading to these results has received funding from the European Research Council under the European Union's Horizon 2020 research and innovation program (ERC Grant No. 810316).

# Simulations of Immuno-Contrast CT to Optimize Spectral Instrumentation and Nanoparticle Contrast Agent Materials

Matthew Tivnan[1], Grace Gang[2], and J. Webster Stayman[1]
[1]Department of Biomedical Engineering, Johns Hopkins University, Baltimore, MD
[2]Department of Radiology, Hospital of the University of Pennsylvania, Philadelphia, PA

**Abstract** Every year, new nanomedicines are developed for diagnostic and therapeutic applications. Antibody-tagged nanoparticles are particularly useful for targetting specific tissues. The current standard for preclinical evaluation of new antibody nanoparticles is positron emission tomography (PET) where the animals are divided into two groups, with one receiving the antibody nanoparticle and the other receiving a reference nanoparticle. However, there is significant mouse-to-mouse variation which means many mice must be scanned to make statistically significant claims about the effect of the antibody. With spectral CT and basis material decomposition one can potentially image the reference and antibody nanoparticles simultaneously in a single mouse. We propose a new type of functional imaging called *Immuno-Contrast CT* in which the reference nanoparticle distribution is subtracted from the antibody nanoparticle distribution to highlight the differential immunofunctional impact of the antibody. In this work we use physics simulations to jointly optimize contrast agent materials and spectral CT instrumentation for immuno-contrast CT. Our results show this optimization significantly reduces noise in the immuno-contrast images.

## 1 Introduction

Nanomedicine is a rapidly growing field that combines nanotechnology, immunology, and biomedical engineering to develop nanoparticles for diagnostic and therapeutic applications in medicine [1]. One particularly important area of research is the development of antibody-tagged nanoparticles that double as imaging contrast agents to target specific tissues or cell populations and highlight those areas in medical images [2] [3]. These antibody nanoparticles must be well-validated in a small animal model so that we understand the biological impact of the the nanoparticle before considering clinical use. Currently, one of the standard methods to determine the biodistribution of a nanoparticle with and without an antibody is to conduct a preclinical mouse imaging study with positron emission tomography (PET) [4] [5]. Typically, the mice are divided into two groups; one group would receive the antibody nanoparticle tagged with a radioisotope for PET imaging and the other group would receive a reference nanoparticle without the antibody. However, one problem with this type of PET imaging study is there can be significant mouse-to-mouse variation in the distribution of the nanoparticles. To test a hypothesis about the immunological function of a new antibody, one may need to use a large number of mice in order to have sufficient statistical power. Injecting so many mice with radioactive nanoparticles and scanning each mouse with a PET scanner can be a difficult and costly process and the final image results from PET often have low-spatial resolution and high-noise. In this work, we investigate the possibility of imaging nanoparticles with spectral x-ray computed tomography (CT). The key advantages over PET are 1) the ability to image the reference and antibody nanoparticles simultaneously in the same mouse using basis material decomposition 2) the lack of dependence on radioactive materials 3) higher spatial resolution with CT over PET.

Spectral CT uses x-ray projection measurements from multiple view angles and varied spectral sensitivity. This type of data can be used for three-dimensional tomographic image reconstruction as well as basis material decomposition. Spectral CT has already been used to image gold nanoparticles in mice [6] [7]. It has also been used to image iodine and gadolinium nanoparticles in mice simultaneously to characterize tumor vasculature [8] [9]. These studies show that imaging nanoparticles with spectral CT is a promising possibility. However, we know that material decomposition introduces noise relative to conventional CT images, so we expect that sensitivity to low concentrations of nanoparticles will be a challenging engineering problem.

We propose to investigate a new functional CT imaging method called *Immuno-Contrast CT* where a mouse is injected with both reference and antibody nanoparticles simultaneously. The two nanoparticles are labeled with two different CT contrast materials (e.g. iodine and gadolinium). Material decomposition permits reconstruction of separate image volumes for 1) water, 2) calcium, 3) reference nanoparticle concentrations, and 4) antibody nanoparticle concentrations. We define the *Immuno-Contrast Image* as the difference between the antibody nanoparticle image and the reference nanoparticle image. As a result, the immuno-contrast image will show the differential impact of the antibody on the biodistribution of the nanoparticle.

In this work, we use physical simulations of spectral CT systems to investigate different combinations of two contrast agent materials for immuno-contrast CT imaging and we simulate many spectral instrumentation designs and to answer three questions: 1) To what degree can we improve immuno-contrast CT image quality through intelligent spectral instrumentation design? 2) What are the optimal pair of contrast materials? 3) What are the optimal system designs pair of contrast materials? After this joint optimization, we evaluate our optimized design by generating a three-dimensional digital image volume with realistic mouse anatomy, simulating spectral measurements with x-ray sources and photon-counting detectors, and applying a model-based material decomposition algorithm to see the impact of our design optimization on immuno-contrast CT imaging performance.

## 2 Materials and Methods

### 2.1 Spectral CT Measurement Likelihood Model

Our mathematical model of spectral CT measurements is a random vector with a multivariate Gaussian distribution, $\mathrm{p}(\mathbf{y}|\mathbf{x})$, parameterized by the mean, $\bar{\mathbf{y}}(\mathbf{x})$, and covariance, $\Sigma_{\mathbf{y}|\mathbf{x}}$. The mean is defined as

$$\bar{\mathbf{y}}(\mathbf{x}) = \mathbf{S}\exp\left(-\mathbf{Q}\mathbf{A}\mathbf{x}\right), \tag{1}$$

where $\mathbf{x}$ is a column vector of basis material densities for each voxel, $\mathbf{A}$ is the matrix of line integrals for each x-ray projection, $\mathbf{Q}$ is the basis material mass attenuation matrix, $\mathbf{S}$ is the system spectral sensitivity matrix including all information about spectral sources and detectors, and $\mathbf{y}$ is a column vector of spectral CT measurements. Note, projection-domain material decomposition is a special case of this model where $\mathbf{A} = \mathbf{I}$. The negative log-likelihood is defined as

$$-\log \mathrm{p}(\mathbf{y}|\mathbf{x}) = \frac{1}{2}(\mathbf{y} - \bar{\mathbf{y}}(\mathbf{x}))^T \Sigma_{\mathbf{y}|\mathbf{x}}{}^{-1}(\mathbf{y} - \bar{\mathbf{y}}(\mathbf{x})) + c \quad (2)$$

where $c$ is a constant with respect to $\mathbf{x}$ and $\mathbf{y}$.

### 2.2 Spectral CT Fisher Information Matrix and the Cramer-Rao Lower-Bound on Covariance

We aim to optimize spectral CT instrumentation by quantifying imaging performance as a function of the design matrices in (1). Specifically we are interested in jointly optimizing spectral instrumentation, described by $\mathbf{S}$, and contrast materials, described by $\mathbf{Q}$. This section defines a mathematical relationship between these design matrices and the Cramer-Rao Lower Bound (CRLB).

The Cramer-Rao inequality states that the covariance of an unbiased estimator is greater than or equal to inverse of the Fisher information matrix, $\mathbf{F}$. The definition of the Fisher information matrix is the Hessian, or second derivative, of the negative log-likelihood with respect to the vector $\mathbf{x}$. Taking the gradient of (2) yields

$$-\nabla_{\mathbf{x}} \log \mathrm{p}(\mathbf{y}|\mathbf{x}) = \mathbf{A}^T \mathbf{Q}^T \mathbf{D}_{\mathbf{x}}^T \mathbf{S}^T \Sigma_{\mathbf{y}|\mathbf{x}}^{-1}(\mathbf{y} - \bar{\mathbf{y}}(\mathbf{x})) \quad (3)$$

$$\mathbf{D}_{\mathbf{x}} = D\{\exp(-\mathbf{QAx})\} \quad (4)$$

and so the Fisher information matrix is

$$\mathbf{F} = -\nabla_{\mathbf{x}}^2 \log \mathrm{p}(\mathbf{y}|\mathbf{x}) = \mathbf{A}^T \mathbf{Q}^T \mathbf{D}_{\mathbf{x}}^T \mathbf{S}^T \Sigma_{\mathbf{y}|\mathbf{x}}^{-1} \mathbf{S} \mathbf{D}_{\mathbf{x}} \mathbf{Q} \mathbf{A} \quad (5)$$

In this work, we use the CRLB, $\Sigma_{\mathbf{x}} \geq \mathbf{F}^{-1}$ to optimize spectral instrumentation design for immuno-contrast imaging. This predictive mathematical model describes the best-case-scenario for multi-material noise characteristics of unbiased material decomposition without the need to run a full material decomposition algorithm for each candidate design.

### 2.3 Simulations of Immuno-Contrast CT in Mice

We used the MOBY phantom [10] to generate three-dimensional digital image volumes of realistic mouse anatomy with 0.1 mm cubic voxels. For each voxel, there were four basis materials: water, calcium, the reference nanoparticle, and the antibody-labeled nanoparticle. To accomplish this, we ran the MOBY attenuation coefficient generator at 60 keV and 100 keV and analytically decomposed water/calcium basis image volumes. MOBY also has the ability to label activity level in specific tissues (usually for nuclear imaging simulations). We used the activity label images to define the concentration of reference and antibody nanoparticles in units of percent injected dose per gram (%ID/g). We used [5] as a rough guide for the biodistribution of reference nanoparticles and antibody nanoparticles which resulted in 15 %ID/g in the liver, 20 %ID/g in the spleen, 5 %ID/g in the kidney, and 2 %ID/g in the rest of the body for both the reference and antibody nanoparticles. We also inserted a 5 mm diameter spherical tumor into the liver which has 70 %ID/g for the antibody nanoparticle and 2 %ID/g for the reference nanoparticle. We assumed that the nanoparticle injections were 200 mL at 100 mg/mL to convert between %ID/g and mass density.

We used PYRO-NN [11], to model a cone beam forward projector, $\mathbf{A}$, with 1200 mm source-to-detector distance, 600 mm source-to-axis distance, and 0.2 mm square pixels. Our model for $\mathbf{Q}$ contains the mass attenuation spectra for water and calcium and two contrast materials. Our model for $\mathbf{S}$ is a polyenergetic x-ray source using the TASMICS model [12] with aluminium filtration as well as k-edge filtration computed using SPEKTR [13], and a photon counting detector. We do not include any model of non-ideal effects of photon counting detectors. We assume the detector has two energy bins per exposure and we have control over the energy threshold for three exposures. More details about the spectral instrumentation and contrast agent models are provided in the following section. For some select designs, we apply a projection-domain model-based material decomposition algorithm on a pixel-by-pixel basis using 1000 iterations of Newton's method to optimize the objective function in (2).

### 2.4 Optimization of Spectral CT Instrumentation and Nanoparticle Contrast Agent Materials

The goal of this work is to jointly optimize nanoparticle contrast agent materials and spectral instrumentation for immuno-contrast imaging. Therefore, our performance metric is the CRLB on the standard deviation in an immuno-contrast image formed by subtracting the reference nanoparticle image from the antibody nanoparticle image to highlight immuno-functional properties of the antibody.

For a given spectral design, $\mathbf{S}$, and set of basis materials, $\mathbf{Q}$, we consider one detector pixel in the projection domain and compute the cross-material Fisher information matrix using (5). We assume the background is 30 mm of water. Then we compute the CRLB by taking the matrix inverse and we compute the immuno-contrast variance using the formula $(\mathbf{w}^T \Sigma_{\mathbf{x}} \mathbf{w})/(\mathbf{w}^T \mathbf{w})$ where $\mathbf{w}$ is vector zero for water and calcium, positive one for the antibody nanoparticle, and negative one for the reference nanoparticle. This is equivalent to the noise variance in the immuno-contrast image. This process was repeated for the four material imaging scenario (water, calcium, X, Y) and three-material imaging scenario (water, X, Y). (We note that in the latter case, any calcium in the image volume will necessarily be modeled using the other basis materials. This model may have a potential noise advantage with fewer material bases but will generally bias material estimates.)

We evaluated the above performance metrics for 421,875 spectral designs and 10 combinations of two contrast materials for a total of over 4.2 million spectral CT imaging scenarios. The design parameters for the spectral source are: source voltage (60, 70, 80, 90, or 100 kVp), aluminium filter thickness (0.0, 0.5, 1.0, 2.0, or 4.0 mm), and k-edge filter (None, 250 $\mu$m Praseodymium, 250 $\mu$m Erbium, 125 $\mu$m Tantalum, or 125 $\mu$m Lead ). The design parameters for the photon counting detector are: energy threshold for each of the three exposures (50, 60, 70, 80, or 90% of kVp) and exposure time for each of the three exposures (1, 2, or 3 relative exposure). Exposures were normalized such that incident x-ray fluence after source filtration is $10^6$ photons per pixel per view. So, for example, if the three relative exposures are (1,1,1) then the number of incident photons is $3.33 \times 10^5$ for all three exposures, but if the three relative exposures are (1,1,3) then the first two exposures have $2.00 \times 10^5$ incident photons and the third exposure has $6.00 \times 10^5$ incident photons. Finally, we repeat the evaluation for all combinations of two contrast agent materials from among: Iron (Fe), Iodine (I), Gadolinium (Gd), Ytterbium (Yb), and Gold (Gd).

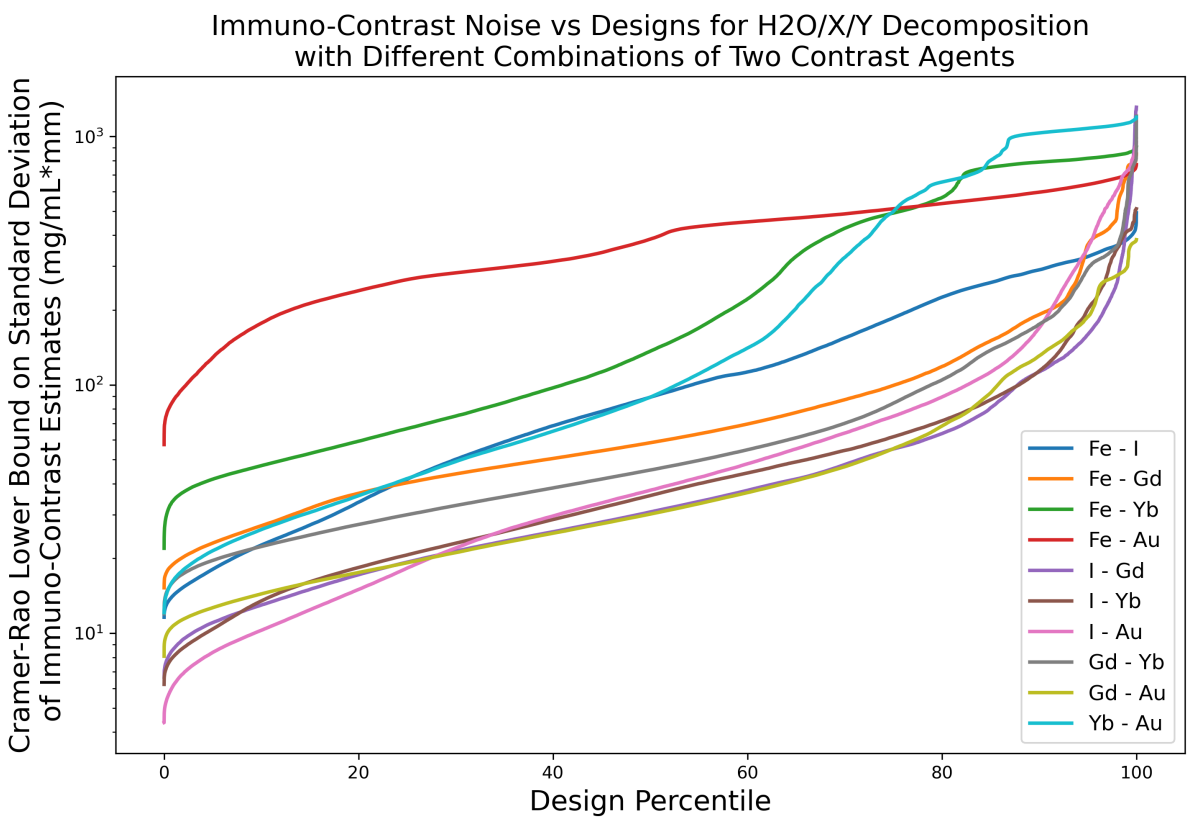

**Figure 1:** The CRLB for 4-material decomposition for all spectral designs and all contrast materials. Then the designs were sorted by immuno-contrast noise levels.

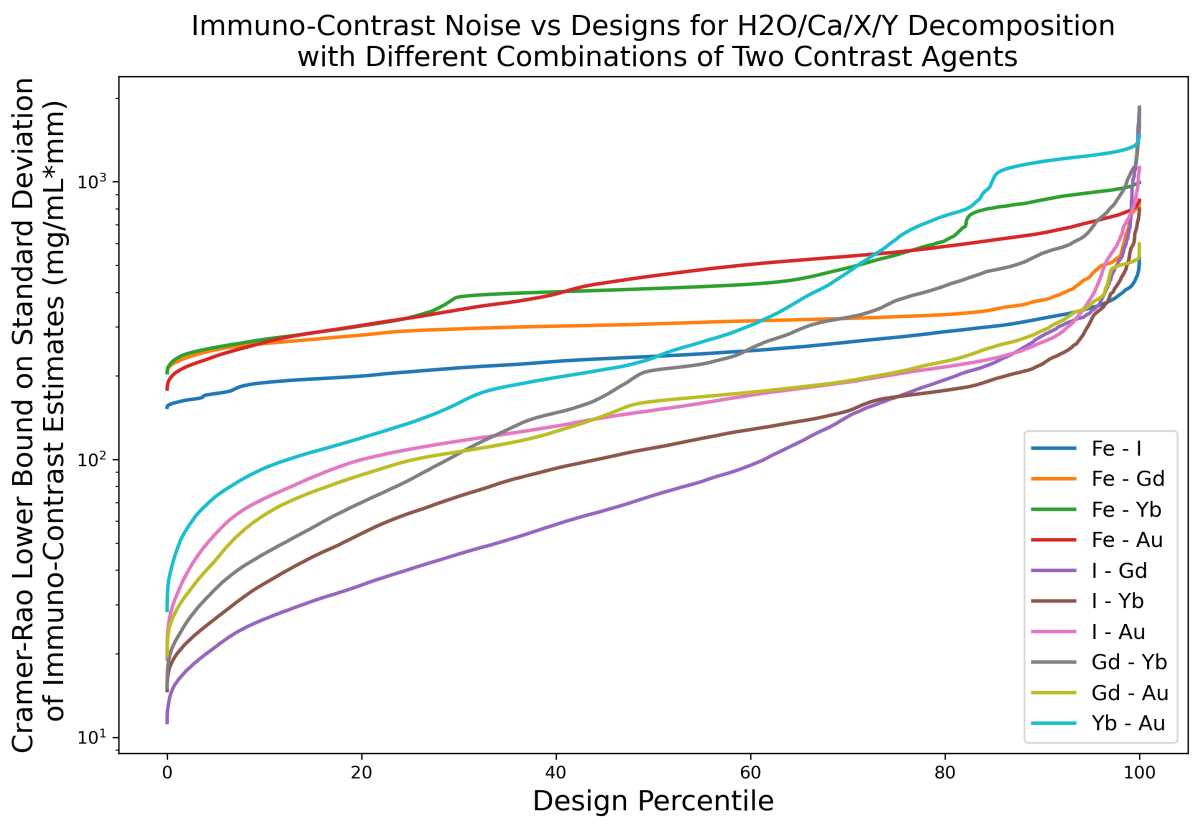

**Figure 2:** The CRLB for 3-material decomposition for all spectral designs and all contrast materials. Then the designs were sorted by immuno-contrast noise levels.

## 3 Results

The CRLB of immuno-contrast standard deviation is summarized for all designs in Figures 1 and 2. The three-material imaging case is shown in Figure 1 and the four-material case is shown in 2. For each contrast agent combination, we sorted the designs by immuno-contrast noise. The x-axis shows the design percentile and the y-axis shows our performance metric, the CRLB on immuno-contrast standard deviation. The 0th design percentile indicates the spectral designs with the best performance. We have summarized the performance of optimized spectral designs in Figure 3 and the optimized design parameters have been listed in table 1. Finally, the results of the model-based material decomposition are shown in Figures 4 5 and 6. Figure 4 shows the ground truth material density line integrals for this simulation. Figure 5 shows the material decomposition results for the 50th percentile system design and Figure 6 shows the 0th percentile top performing system design for H2O/Ca/I/Gd imaging.

## 4 Discussion

In Figures 1 and 2 we see that there is a wide range of performance levels for different spectral designs and different combinations of materials. There is a relatively steep falloff in performance relative to the best designs (near 0th percentile) which indicates that fine tuning the spectral sensitivity of the system can significantly reduce noise in the immuno-contrast images. In Figure 2 we see that all of the cases using iron as a contrast agent material have extremely high noise even for the optimized design. We believe this is due to the fact that the k-edge of iron is much lower than the x-ray spectra of these designs. Therefore, water, calcium, and iron, are approximately linearly dependent on only two basis (e.g. photoelectric effect and Compton scattering). The contrast agents which achieved the lowest immuno-contrast noise for four-material decomposition were iodine and gadolinium. These are the same contrast agents used in [8, 9] so there is already evidence it is possible to image these two materials simultaneously in mice. The material decomposition results in Figures 4 5 and 6 show visible reduction of immuno-contrast noise for the optimized design relative to the 50th percentile. These results show that a 5 mm diameter lesion with realistic nanoparticle concentrations in visible with realistic x-ray fluence levels.

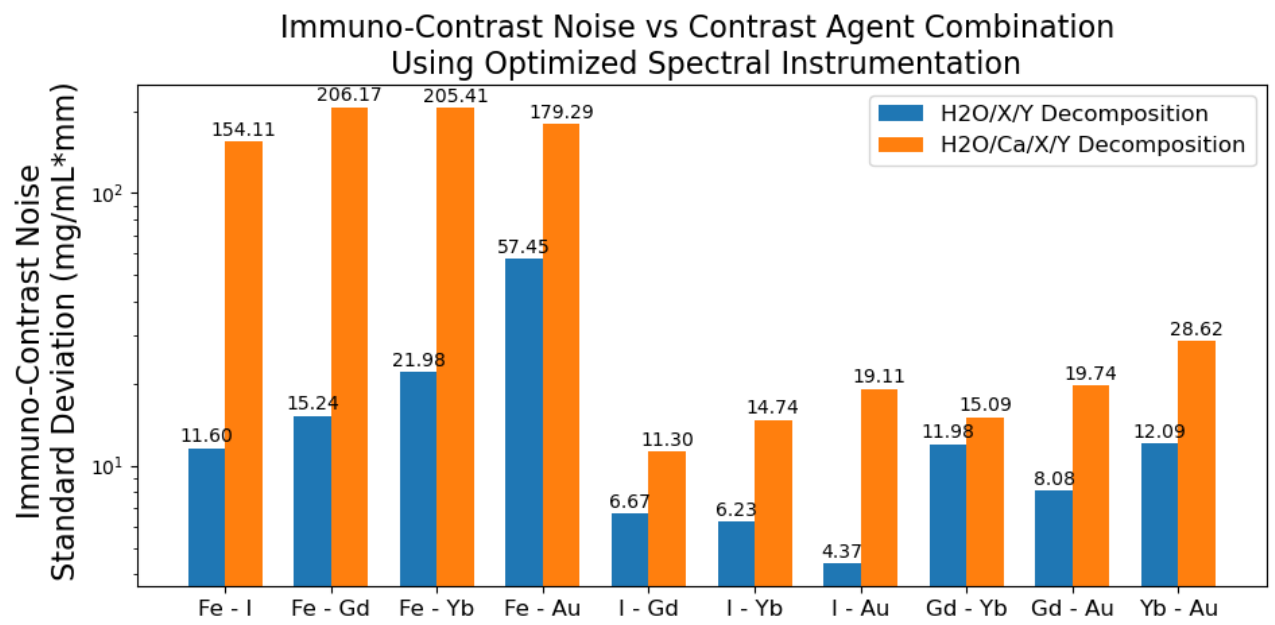

**Figure 3:** The Cramer-Rao lower bound on immuno-contrast noise standard deviation for all combinations of contrast agents using optimal spectral instrumentation.

## 5 Conclusion

This simulation study is an early stage investigation to determine the feasibility of immuno-contrast CT imaging. There are several areas where we have made assumptions and idealized approximations about both the mouse biodistribution model and spectral CT imaging simulation and material decomposition. For example, we have assumed that the reference and antibody nanoparticle biodistributions are the same with the exception of the tumor. This may not be true in practice especially when using two different contrast materials. For the material decomposition algorithm, we used a perfectly matched reconstruction model to the data generation model. In practice it is very difficult to calibrate the sensitivity of spectral CT imaging systems.

| Contrast Agents | Source kVp | Al Filter | K-Edge Filter | Exposure 1 | Threshold 1 | Exposure 2 | Threshold 2 | Exposure 3 | Threshold 3 |
|---|---|---|---|---|---|---|---|---|---|
| H2O-Ca-Fe-I | 60.0 kVp | 4.0 mm Al | 250um Pr | 2.00E+05 | 30.24 keV | 6.00E+05 | 33.6 keV | 2.00E+05 | 42.0 keV |
| H2O-Ca-Fe-Gd | 60.0 kVp | 0.5 mm Al | None | 5.00E+05 | 25.92 keV | 2.50E+05 | 28.8 keV | 2.50E+05 | 48.0 keV |
| H2O-Ca-Fe-Yb | 70.0 kVp | 0.5 mm Al | None | 5.00E+05 | 28.35 keV | 1.67E+05 | 31.5 keV | 3.33E+05 | 63.0 keV |
| H2O-Ca-Fe-Au | 60.0 kVp | 0.5 mm Al | None | 5.00E+05 | 18.9 keV | 3.33E+05 | 37.8 keV | 1.67E+05 | 42.0 keV |
| H2O-Ca-I-Gd | 70.0 kVp | 4.0 mm Al | None | 3.33E+05 | 30.24 keV | 5.00E+05 | 50.4 keV | 1.67E+05 | 63.0 keV |
| H2O-Ca-I-Yb | 80.0 kVp | 0.0 mm Al | 125um Pb | 4.29E+05 | 34.56 keV | 1.43E+05 | 57.6 keV | 4.29E+05 | 64.0 keV |
| H2O-Ca-I-Au | 100.0 kVp | 4.0 mm Al | 250um Pr | 2.00E+05 | 36.0 keV | 2.00E+05 | 40 keV | 6.00E+05 | 80.0 keV |
| H2O-Ca-Gd-Yb | 90.0 kVp | 4.0 mm Al | 125um Ta | 1.67E+05 | 45.36 keV | 3.33E+05 | 50.4 keV | 5.00E+05 | 63.0 keV |
| H2O-Ca-Gd-Au | 100.0 kVp | 4.0 mm Al | 125um Pb | 2.00E+05 | 44.8 keV | 2.00E+05 | 56 keV | 6.00E+05 | 80.0 keV |
| H2O-Ca-Yb-Au | 100.0 kVp | 4.0 mm Al | 125um Pb | 1.67E+05 | 38.4 keV | 3.33E+05 | 64 keV | 5.00E+05 | 80.0 keV |
| H2O-Fe-I | 100.0 kVp | 0.5 mm Al | None | 5.00E+05 | 27.0 keV | 3.33E+05 | 30 keV | 1.67E+05 | 50.0 keV |
| H2O-Fe-Gd | 100.0 kVp | 1.0 mm Al | None | 2.00E+05 | 40.5 keV | 2.00E+05 | 45 keV | 6.00E+05 | 50.0 keV |
| H2O-Fe-Yb | 90.0 kVp | 4.0 mm Al | 250um Pr | 2.00E+05 | 51.03 keV | 2.00E+05 | 56.7 keV | 6.00E+05 | 63.0 keV |
| H2O-Fe-Au | 100.0 kVp | 4.0 mm Al | 250um Pr | 2.00E+05 | 40.5 keV | 6.00E+05 | 81 keV | 2.00E+05 | 90.0 keV |
| H2O-I-Gd | 80.0 kVp | 0.0 mm Al | 125um Ta | 3.33E+05 | 45.36 keV | 5.00E+05 | 50.4 keV | 1.67E+05 | 56.0 keV |
| H2O-I-Yb | 70.0 kVp | 0.5 mm Al | None | 4.29E+05 | 25.2 keV | 4.29E+05 | 28 keV | 1.43E+05 | 35.0 keV |
| H2O-I-Au | 70.0 kVp | 1.0 mm Al | None | 3.33E+05 | 30.24 keV | 5.00E+05 | 33.6 keV | 1.67E+05 | 42.0 keV |
| H2O-Gd-Yb | 70.0 kVp | 4.0 mm Al | None | 1.67E+05 | 39.69 keV | 3.33E+05 | 44.1 keV | 5.00E+05 | 49.0 keV |
| H2O-Gd-Au | 70.0 kVp | 4.0 mm Al | None | 2.00E+05 | 45.36 keV | 6.00E+05 | 50.4 keV | 2.00E+05 | 56.0 keV |
| H2O-Yb-Au | 80.0 kVp | 2.0 mm Al | 125um Pb | 1.43E+05 | 51.84 keV | 4.29E+05 | 57.6 keV | 4.29E+05 | 64.0 keV |

**Table 1:** Optimized spectral instrumentation design parameters for low immuno-contrast noise for each contrast material combination.

Despite the idealized conditions, we can draw some conclusions from the results of this preliminary investigation. First, immuno-contrast CT appears to be physically possible for realistic nanoparticle concentrations and x-ray exposure levels. Second, iodine and gadolinium are a good choice of contrast agents for immuno contrast imaging. Third, optimizing the spectral design for specific combinations materials can significantly reduce noise in the immuno-contrast images.

Immuno-contrast CT imaging has the potential to accelerate the development of nanomedicines because it can simultaneously image the reference and antibody nanoparticle biodistributions, improving the statistical power of preclinical imaging studies. In the future, we look forward to addressing some of these non-ideal effects and moving on to physical experiments with spectral CT imaging systems.

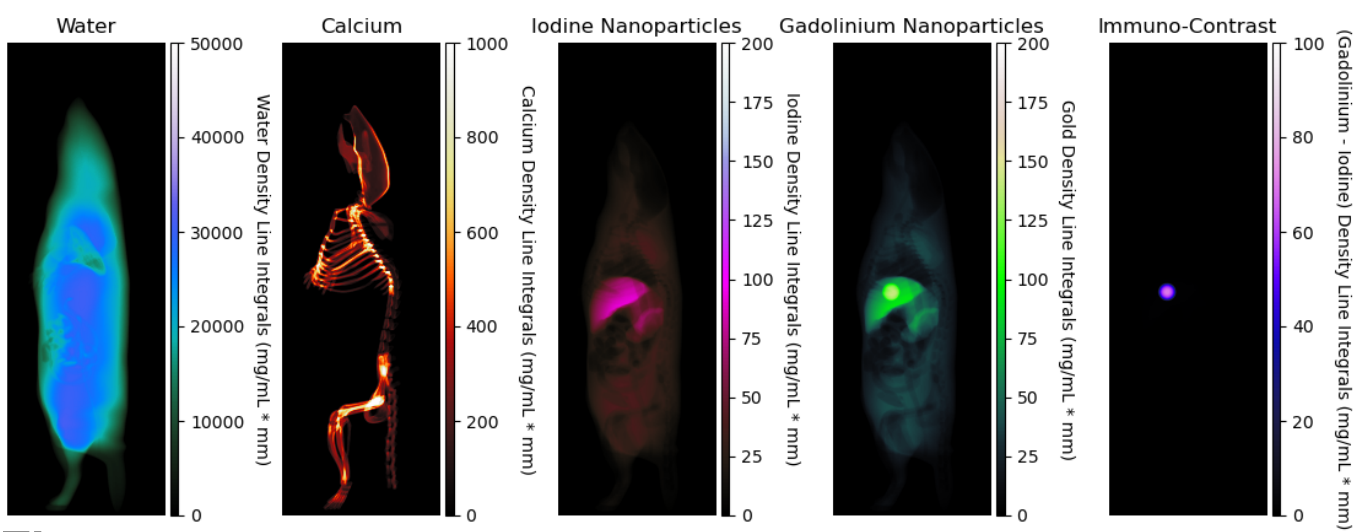

**Figure 4:** Material density line integral ground truth generated with the MOBY phantom. Immuno-contrast is Gd minus I.

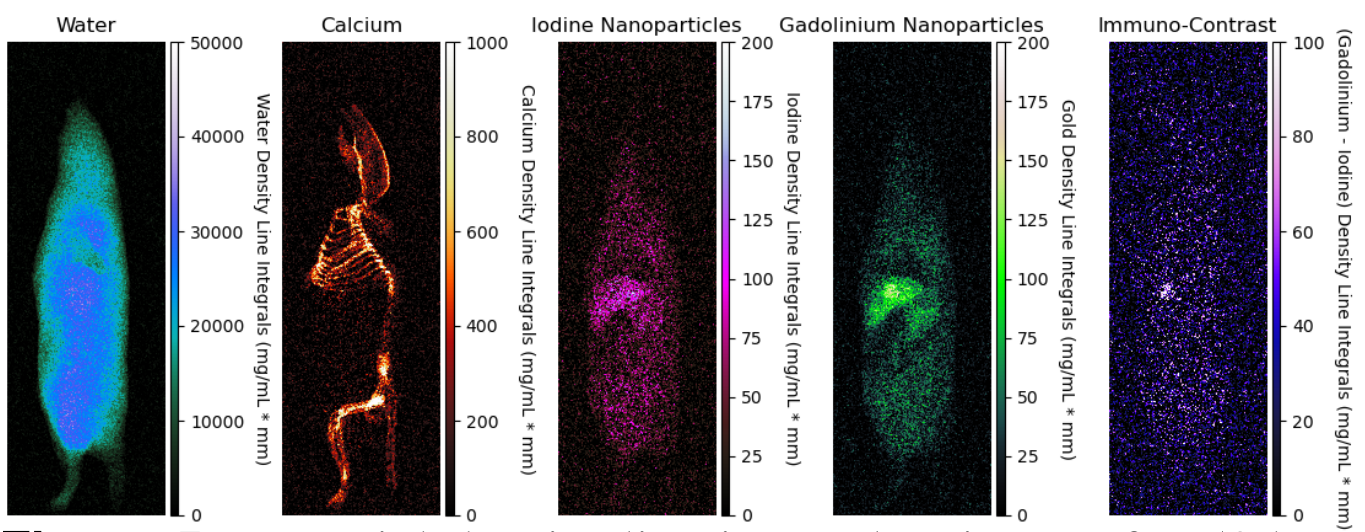

**Figure 5:** Material density line integral estimates for 50th percentile design for low immuno-contrast noise power.

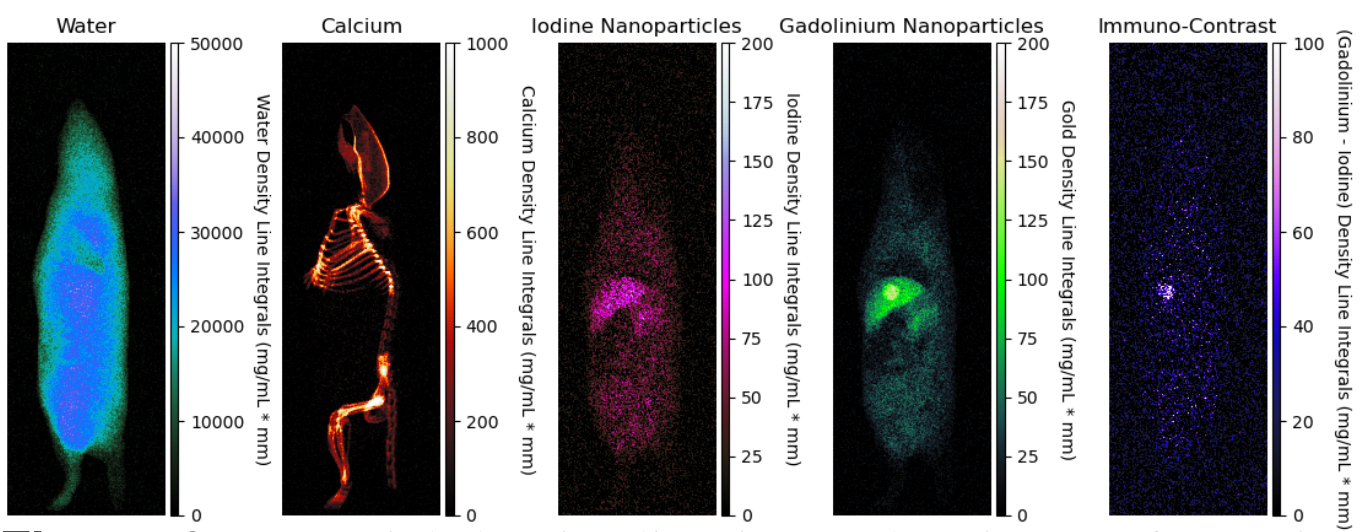

**Figure 6:** Material density line integral estimates for optimal design for low immuno-contrast noise power.

## Acknowledgements

This work is supported in part by NIH Grant R01EB030494

# 3D deep learning based cone beam artifact correction for axial CT

Artyom Tsanda[1, 3], Sebastian Wild[1], Stanislav Žabić[2], Thomas Koehler[1], Rolf Bippus[1], Kevin M. Brown[2], and Michael Grass[1]

[1]Philips Research, Hamburg, Germany
[2]Philips CT Research & Advanced Development, Cleveland, OH, USA
[3]Hamburg University of Technology, Germany

**Abstract**
Axial multi-detector computed tomography addresses multiple important clinical applications. At the same time, it suffers from cone-beam artifacts. These artifacts appear due to data insufficiency and get more pronounced with increasing axial coverage.
In this paper, we propose to train a 3D convolutional neural network to correct for cone-beam artifacts of axial multi-detector CT systems with 16 cm coverage.The approach relies on the simulation of cone-beam artifacts, patch-based training, mixed precision technology, and the 3D U-net model architecture. The method is tested using both simulated and real data. While being less accurate on average in terms of the root mean square error, we achieve better artifact suppression at the extreme compared to the two-pass approach commonly used in the field.

## 1 Introduction

Axial multi-detector computed tomography with a wide coverage scanner is a common acquisition mode for many clinical applications. It allows to quickly capture a large portion of the body within a narrow time window which makes it a favorable choice for imaging moving organs. Important application areas enabled by this mode of scan on a 16 cm wide-detector CT scanner include brain imaging (including brain perfusion) and cardiac imaging.
Larger cone angles lead to cone-beam artifacts for non-exact reconstruction methods due to the missing data problem associated with circular trajectories [1, 2]. They exhibit a low-frequency shading or streak artifacts generated by strong z-gradients in the irradiated volume. Many methods already exist for reducing cone-beam artifacts [3]. Since in this paper we use aperture weighted wedge filtered backprojection (AWW-FBP) as the reconstruction algorithm [4], we consider only the correction methods applicable to it. One of them weights redundant data based on aperture weighting. It is already integrated into the considered reconstruction algorithm. Another well-known approach to correct cone-beam artifacts is the two-pass approach [5]. It restores missing structures of high spatial gradient with soft tissue and air. After the subsequent forward- and backprojection step it provides an estimation of actual cone-beam artifacts.
Convolutional neural networks (CNNs) have become ubiquitous in computer vision ever since AlexNet [6] won the ImageNet challenge [7]. Over the last years, several approaches have been proposed to correct for cone-beam artifacts using CNNs. Han et al. [8] corrected cone-beam artifacts in coronal and sagittal views with separate 2D U-net models having 3D fusion afterwards. Training data were simulated based on a dataset of 3D CT volumes without artifacts. Maspero et al. [9] registered cone-beam CT scans to planning CT to further process it with a 2D cycle-GAN network. Minnema et al. [10] applied a 2D CNN along radial slices to achieve better training data consistency.
In this work, instead of applying 2D CNNs as used in the mentioned publications, we propose to employ a 3D CNN to directly correct for cone-beam artifacts in axial CT scans. We hypothesize that a 3D network is more suited for such correction due to the nature of the artifacts but comes with higher demands on the GPU memory for training and inference. We use helical scans to simulate axial ones with 16 cm axial coverage to generate a set of registered training pairs, patch-based training setup, mixed precision technology and the 3D U-net model architecture. We consider head scans for training and validation. The results are compared against the two-pass approach.

## 2 Materials and Methods

### 2.1 Two-Pass Approach

The basic idea of the two-pass pass approach is to use the (erroneous) reconstructed image to estimate an image containing the artifact-inducing structures. This image is then used as input to a simulation of the imaging system including forward projection and reconstruction using the same reconstruction algorithm. This is what is referred to as second pass. The reconstruction will produce similar artifacts as in the original reconstruction, but in this case the error in the images can be deduced from comparison to the forward simulation input.
The original approach [5] applied threshold operations aiming at tissue classification. In particular, soft tissue is subsequently replaced by a constant HU level which removes most cone-beam artifacts within soft tissue, while the HU values of bone are left untouched. Although soft-tissue contrast is eliminated completely, the major gradients, namely those at the bone boundaries are preserved, which are indeed the major source of artifacts. We use a modified version of this processing including additional gradient filtering, bone correction and $z$-dependent post-filtration after the second pass.
Truncation at the upper and lower boundary of the volume is a problem for this approach since artifact-inducing structures are not limited to the coverage that can be reconstructed.

Structures outside the coverage may also introduce long-range artifacts leaking into the reconstructible region. In this paper, we apply constant row extrapolation for projections.

## 2.2 Data Generation

To develop and validate our approach we simulate cone-beam artifacts using clinical CT scans acquired in helical geometry (called "helical images" in the rest of the text) . The data comes from the Philips iCT 6000 CT System (Philips Medical Systems, Cleveland, USA) with 4 cm collimation and pitch factor varying between 0.4 and 1.

Instead of directly using helical images as ground truth (GT), we process them to achieve a better match in terms of blurring. This way, the CNN is not trained to compensate for simulation-induced blurring which is not present in real scans. The starting point for obtaining a pair of artifact-prone and GT images is a helical image which is processed in two ways:

1. It gets forward projected onto an axial trajectory at a wide cone angle with a given source position along the $z$-axis, and subsequently reconstructed using aperture weighted wedge filtered backprojection. The resulting image contains axial cone-beam artifacts, and will serve as input to the CNN.

2. For each $z$-position corresponding to an image slice of the image obtained from step 1, the helical image gets forward projected onto an axial trajectory at a narrow cone angle with a source position being equal to that $z$-position. Afterwards, each of the resulting projections undergoes filtered backprojection onto a 2D image positioned at the corresponding value of $z$. Finally, all the resulting 2D images are stacked along $z$, resulting in a 3D image with identical geometry as the one from step 1. This image serves as GT for the network training.

As both images have undergone one iteration of forward- and backprojection, we achieve close levels of blurring. Note that cone-beam artifacts vanish on the image plane containing the source trajectory; hence, the GT image constructed as outlined above is free from axial cone-beam artifacts.

In addition, for a given helical image, we propose to generate several pairs of input and GT images by varying the source position along the $z$-axis used in step 1. These different offsets along $z$ lead to different cone-beam artifacts in the image, and thus serve as an efficient method of data augmentation. This approach is particularly relevant in case the number of available helical images is limited.

In our experiments we use 22 CT scans for training and 3 for testing. Each training scan is reconstructed using 3 offsets along $z$. The size of the volume for each scan is $(512, 512, 256)$ voxels with dimensions $(250, 250, 160)$ in $(x, y, z)$.

## 2.3 Training and Inference

The training setup is illustrated in Fig. 1. Before passing the data to the model during training we normalize data according to the sample mean and standard deviation calculated along all artifact-prone scans and randomly sample patches of size (128, 128, 64) corresponding to $(x, y, z)$. Since cone-beam artifacts typically cover a significant fraction of the image size, the chosen patch size should be sufficiently large. After data generation and preprocessing, the pairs of artifact-prone patches and the slice-by-slice GT deduced from helical scans can be used for training a model. We propose to train a 3D U-net model [11] from scratch (without transfer learning or pre-training) as described above, with the MSE as a loss function. In order to reduce GPU memory footprint, mixed precision training is utilized [12].

Inference also has to be done using patches, because still an image volume cannot be fit into GPU memory in our case. In our implementation patches overlap and are later fused using gaussian weights. Namely, if a point is covered by several patches, a point value from each patch will contribute with a weight proportional to its distance from the center of the patch.

For our experiments we used the MONAI library [13].

## 2.4 Metric

Since the clinical impact of cone-beam artifacts depends on where they appear in the image, we restrict the quantitative assessment of our results to the brain tissue. Specifically, the cerebrum and cerebellum as these brain parts are of interest in brain perfusion scans. In order to do that, we provide binary segmentation masks of the brain tissue for the three scans from the test set (Fig. 2h). In our experiments, we use the root mean square error (RMSE) calculated for the voxels within the mask. In addition, we also use quantiles of the absolute error within the masked region which also should better address the problem of the uneven spatial distribution of the artifacts.

# 3 Results

Table 1 shows the quantitative evaluation of the proposed approach and its comparison against the aperture weighted reconstruction and the two-pass approach using simulated 16 cm axial CT data. Metrics are calculated for the three CT scans from the test set. We explicitly position the detector so that it covers the top of the skull to eliminate the extrapolation effect for the two-pass approach.

Visual results of cone-beam correction on simulated data are demonstrated in Fig. 2. Each method is represented by a sagital slice of the head scan from the test set along with the absolute difference with the GT slice. Cone beam artifacts are visible on the images with differences as streak artifacts and low-frequent shading towards the boundaries.

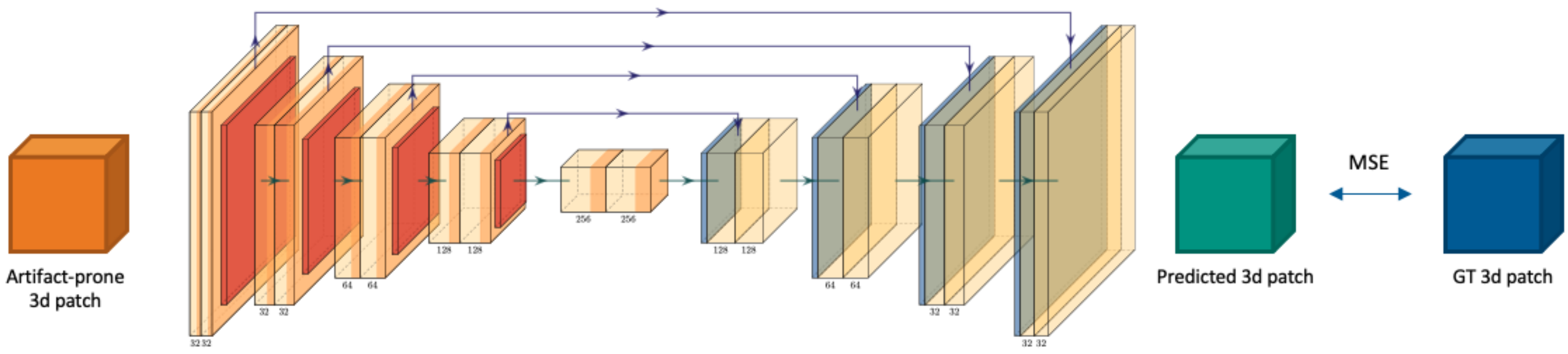

**Figure 1:** Training setup for correcting cone-beam artifacts.

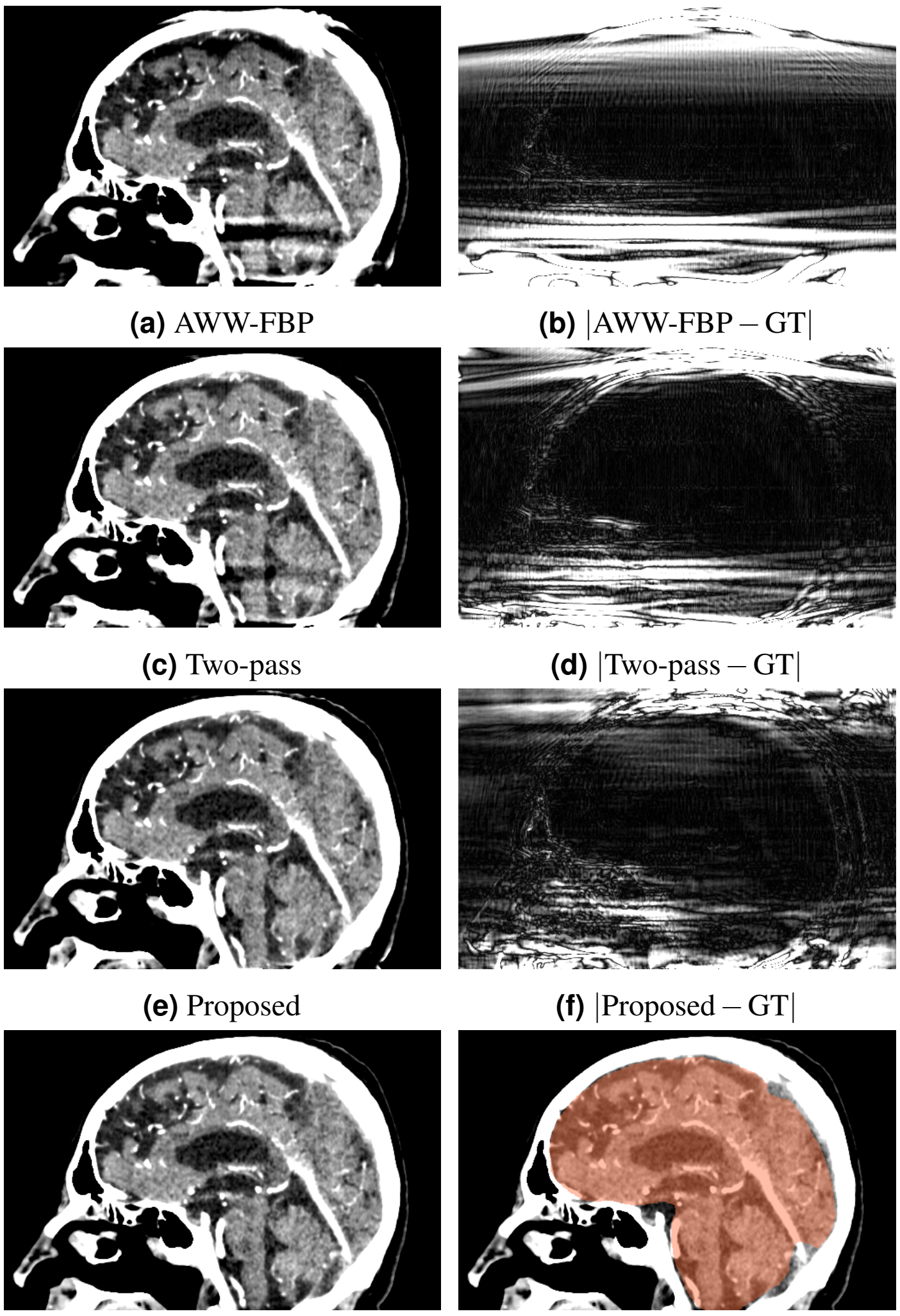

**(a)** AWW-FBP **(b)** |AWW-FBP – GT|
**(c)** Two-pass **(d)** |Two-pass – GT|
**(e)** Proposed **(f)** |Proposed – GT|
**(g)** GT **(h)** Brain mask

**Figure 2:** Comparison of cone-beam correction methods for a scan covering the top of the skull. L/W = 35 HU/70 HU is used for (a), (c), (e), (g). L/W = 10 HU/20 HU - for (b), (d), (f). (h) is the mask used for calculating the metrics.

| Method | RMSE, HU | 95% perc., HU | 99.7% perc., HU |
|---|---|---|---|
| AWW-FBP | $4.51 \pm 0.71$ | $17.97 \pm 3.75$ | $56.68 \pm 16.77$ |
| Two-pass | $\mathbf{1.97 \pm 0.32}$ | $7.62 \pm 1.89$ | $24.90 \pm 7.20$ |
| Proposed | $2.26 \pm 0.31$ | $\mathbf{6.57 \pm 0.78}$ | $\mathbf{14.73 \pm 0.39}$ |

**Table 1:** Quantitative assessment of cone-beam artifact correction methods for head imaging across three test cases. Mean and standard deviation of the corresponding metric is presented for each method.

Fig. 3 presents the results of cone-beam artifacts correction for an anthropomorphic phantom scanned on an 8 cm CT 7500 system. The data was acquired in several modes: in helical mode with pitch factor 1 at 2 cm collimation and in axial mode. The phantom was not moved between the different scans; hence we can at least visually compare the reconstructions of the helical scans to the ones from the 8 cm axial acquisitions.

## 4 Discussion

Although the two-pass approach is slightly better than the proposed one in terms of the RMSE, the quantiles of the absolute error show that it is less accurate and more divergent at extreme.

Visual analysis shows that the artifacts are quite successfully removed by the network (Fig. 2e). In particular, the horizontal streaks visible in the image reconstructed with aperture weighted wedge filtered backprojection (Fig. 2a) are almost fully removed. However, the network induces a low-frequency bias in parts of the image, visible mainly by the bright region within the skull in the absolute residuum (Fig. 2f). This is a generic finding across the different test cases studied in this work: the network-based approach in general is good at removing artifacts with a high frequency in the $z$ direction, but can introduce a more subtle low-frequency bias across a larger range in $z$. The reason for that can be due to inability of the network to accurately define the location of an incoming patch.

At the same time, the two-pass approach by design does not introduce any errors in the area of low cone angles. However,

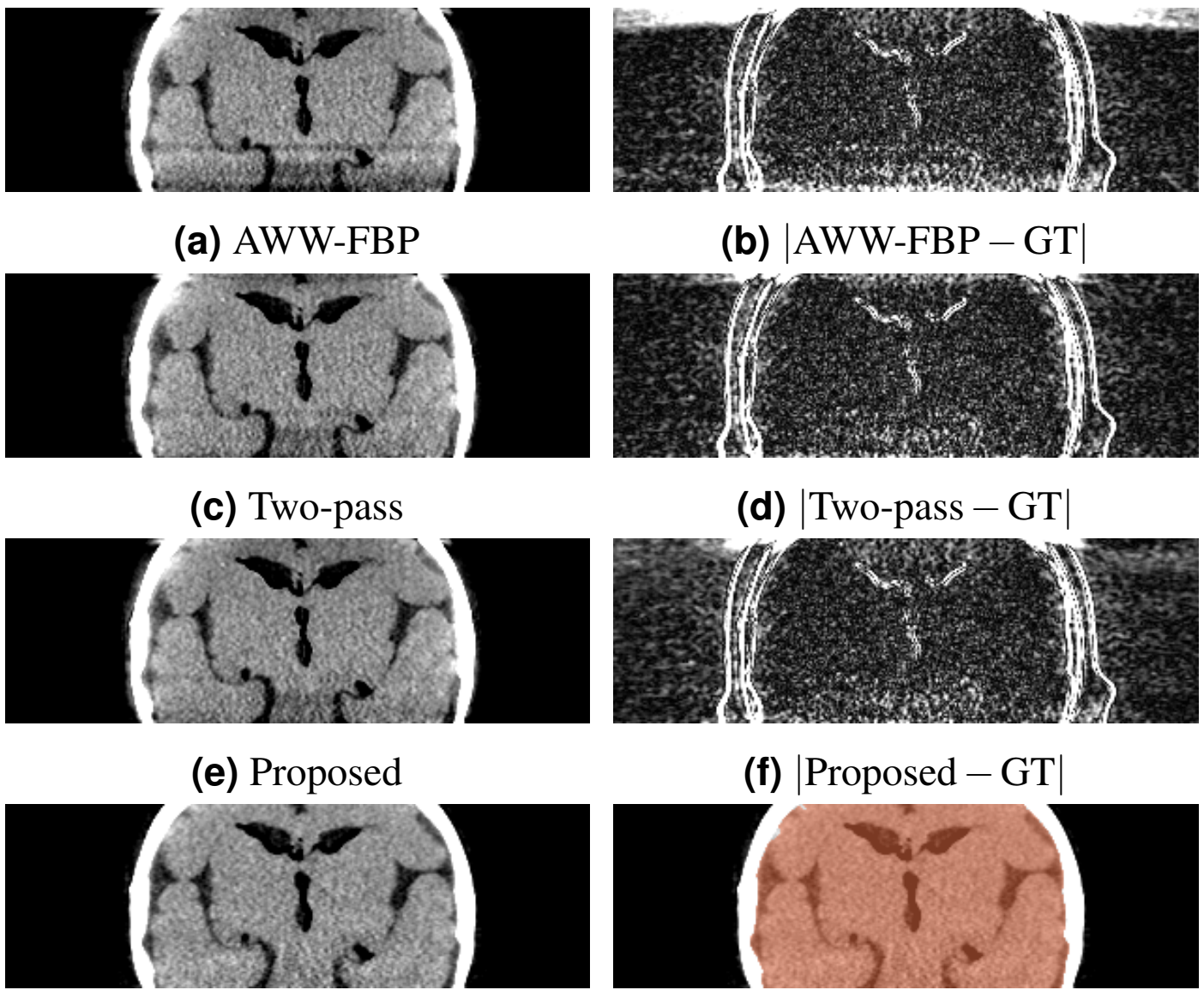

**(a)** AWW-FBP **(b)** |AWW-FBP − GT| **(c)** Two-pass **(d)** |Two-pass − GT| **(e)** Proposed **(f)** |Proposed − GT| **(g)** GT **(h)** Brain mask

**Figure 3:** Comparison of cone-beam correction methods for an anthropomorphic phantom scanned on an 8 cm CT 7500 system. The residuals are smoothed using the in-plane ($XY$) Gaussian filter in order to focus on the meaningful differences related to cone-beam artifacts rather than the different noise realizations between the helical and axial scans. L/W = 35 HU/70 HU is used for (a), (c), (e), (g). L/W = 10 HU/20 HU - for (b), (d), (f).

it still leaves visually noticeable artifacts, namely the streaks in the lower part of the brain. This visual observation correlates with the numbers presented in Table 1 where quantile values for the two-pass approach are higher.

Fig. 3e shows the result of applying the artifact-correcting network to the AWW-FBP reconstruction. Evidently, the streak-like artifacts in the lower half of the image are significantly (though not completely) reduced by the network. It is worth noting that the phantom data differs in two ways from the data used for training the network: first, it corresponds to real data and thus has a different noise texture than the forward-projected data employed for training, and second, it is based on a phantom not seen during training. Given this domain shift, it is reassuring to observe that the artifact-correcting network nevertheless successfully reduces the level of streak-like artifacts in the image.

## 5 Conclusion

While larger cone angles for axial CT acquisition open up new opportunities to improve image quality, they come at a price of inaccurate reconstruction due to missing data. In this paper, we present an approach based on a 3D CNN applied as a post-processing step to fix the artifacts introduced after reconstruction. Targeting 16 cm axial coverage, we find a way to generate a training dataset using helical scans. The results on simulated data show that the proposed approach is capable of reducing the problem. The approach can introduce a minor bias but is quite good at removing extreme streaks compared to the two-pass approach. The proposed method generalizes to 8 cm data coming from an actual CT system and demonstrates similar results. Future research directions may include designing a geometry-aware CNN-based approach for cone-beam artifacts correction preventing it from introducing bias in the central plane area.

# Geometric Constraints Enable Self-Supervised Sinogram Inpainting in Sparse-View Tomography

Fabian Wagner[1], Mareike Thies[1], Noah Maul[1], Laura Pfaff[1], Oliver Aust[2], Sabrina Pechmann[3], Christopher Syben[4], and Andreas Maier[1]

[1]Pattern Recognition Lab, Friedrich-Alexander-Universität Erlangen-Nürnberg, Germany
[2]Department of Rheumatology and Immunology, Friedrich-Alexander-Universität Erlangen-Nürnberg, Germany
[3]Fraunhofer Institute for Ceramic Technologies and Systems IKTS, Forchheim, Germany
[4]Fraunhofer Development Center X-Ray Technology EZRT, Erlangen, Germany

**Abstract** The diagnostic quality of computed tomography (CT) scans is usually restricted by the induced patient dose, scan speed, and image quality. Sparse-angle tomographic scans reduce radiation exposure and accelerate data acquisition, but suffer from image artifacts and noise. Existing image processing algorithms can restore CT reconstruction quality but often require large training data sets or can not be used for truncated objects. This work presents a self-supervised projection inpainting method that allows optimizing missing projective views via gradient-based optimization. By reconstructing independent stacks of projection data, a self-supervised loss is calculated in the CT image domain and used to directly optimize projection image intensities to match the missing tomographic views constrained by the projection geometry. Our experiments on real X-ray microscope (XRM) tomographic mouse tibia bone scans show that our method improves reconstructions by 3.1–7.4 %/7.7–17.6 % in terms of PSNR/SSIM with respect to the interpolation baseline. Our approach is applicable as a flexible self-supervised projection inpainting tool for tomographic applications.

## 1 Introduction

Computed tomography (CT) scanners allow the reconstruction of unknown 3D object density distributions from a set of acquired X-ray projection images. Most clinical applications use filtered back projection (FBP)-based reconstruction algorithms that require a dense angular sampling to meet the precondition of the analytical algorithm. Although decreasing the number of measured projection images can be beneficial to reduce patient dose, improve acquisition speed, and reduce motion effects [1], it introduces noise and image artifacts that can impair diagnostic value.

Different image processing algorithms were proposed to restore the image quality of scans acquired with reduced angular sampling and dose, intervening at different stages of the CT reconstruction pipeline. A first set of algorithms operates directly on the acquired sinogram data with the goal to upsample the number of projection images [2, 3]. Consistency conditions on CT projection data were applied to limited-angle acquisitions to improve image quality while preserving consistency with the measured data [4, 5]. However, such methods are often insensitive to in-plane artifacts and only work for non-truncated data. A second group of methods apply pure image post-processing algorithms to improve the overall quality of noisy reconstructions acquired with reduced dose [6–8]. To make CT reconstruction pipelines compatible with gradient-based data-driven training, differentiable FBP operators were presented that allow propagating a loss calculated on the reconstructed image back to the raw sinogram data [9, 10]. Such known operators [11] allow training CT pipelines employing neural networks on the sinogram and the reconstruction simultaneously [12]. Existing approaches use neural networks or other denoising operators to restore image quality [13, 14] or make use of conventional inpainting techniques to increase the angular projection sampling artificially [15].

So far most learning-based models are trained supervisedly on large-scale low and high-quality paired data sets [2]. However, multiple self-supervised training approaches exist that circumvent the need for paired training data. Noise2Noise [16] calculates a loss metric from two independent noisy image representations. Noise2Inverse [17] and other works [18] extend this principle to tomographic CT settings by splitting projection data into two independent sets. The resulting reconstructions are regarded as image representations with independent noise realizations and allow for deriving a self-supervised denoising loss. Only a few self-supervised inpainting techniques exist and even fewer are applied to CT problems [3, 19].

In this work, we present a self-supervised CT projection inpainting method. We regard missing projection images as trainable data tensors that are updated to be consistent with the measured data. Reconstructing them with a differentiable FBP operator allows deriving a self-supervised loss in the image domain with the real measured sparse-angle reconstruction. Backpropagating that loss through our fully differentiable reconstruction pipeline allows directly optimizing the missing projection data tensor to generate additional projection views. Our contributions are the following:

- We present a self-supervised projection inpainting method based on a differentiable FBP pipeline.
- We propose to directly optimize missing CT projection information through the fixed projective geometry of a differentiable FBP operator.
- We evaluate our method on high-resolution cone-beam X-ray microscope (XRM) ex-vivo mouse tibia bone data.

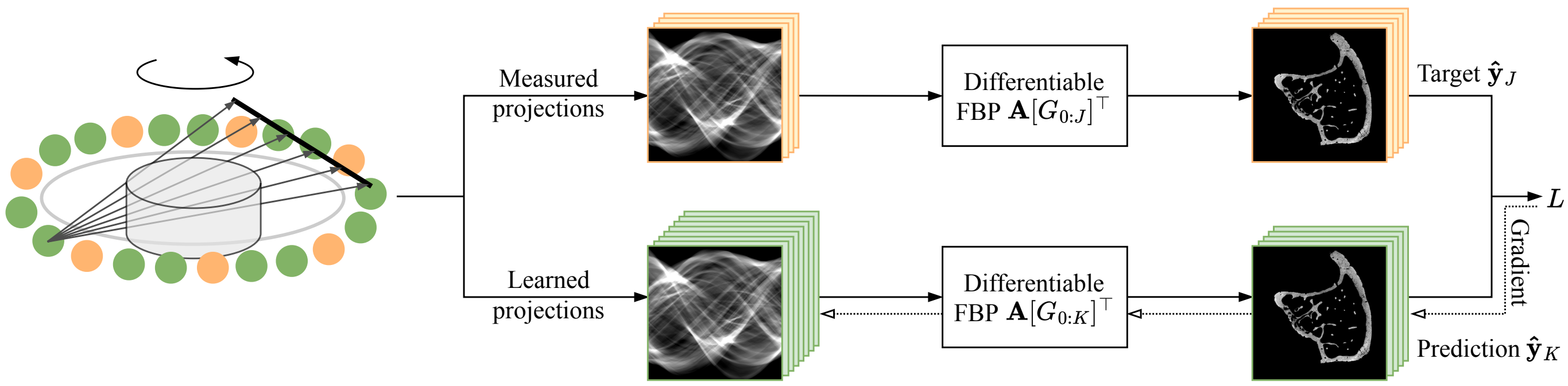

**Figure 1:** Illustration of the proposed self-supervised projection inpainting scheme. A target image is reconstructed from a sparsely sampled CT scan, e.g., one-third of the usual number of projections from a high-resolution scan (orange data). The intensities of intermediate missing projections are regarded as trainable parameters (green projections). Further, a self-supervised loss is calculated between the two reconstructed projection stacks and backpropagated through a differentiable FBP operator to the inpainted projections to update them consistently with the measured data.

## 2 Methods

### 2.1 Differentiable Reconstruction Pipeline

The acquisition of CT projection images $\mathbf{p}_j$ with $j \in \{0,\dots,J\}$ can be written as

$$\mathbf{p}_j = \mathbf{A}[G_j]\mathbf{y} + \mathbf{n}_j \tag{1}$$

with the true scanned object $\mathbf{y}$, noise vector $\mathbf{n}_j$, and the forward operator $\mathbf{A}$ conditioned to the projection geometry $G$. The inverse CT problem aims to reconstruct the unknown object from the acquired set of $J$ projections using the adjoint system matrix $\mathbf{A}^\top$ and the filter operator $\mathbf{K}$

$$\hat{\mathbf{y}}_J = \frac{1}{J}\sum_j \mathbf{A}[G_j]^\top \mathbf{K}\mathbf{p}_j \quad . \tag{2}$$

Using a differentiable backprojection operator [9] allows backpropagating a loss metric $L$ derived on the reconstructed image $\hat{\mathbf{y}}$ to the raw CT projection data using the chain rule

$$\frac{\partial L}{\partial \mathbf{p}_j} = \frac{\partial L}{\partial \hat{\mathbf{y}}}\frac{\partial \hat{\mathbf{y}}}{\partial \mathbf{p}_j} \quad . \tag{3}$$

Hence, models working in the projection domain or the projection data itself can be directly modified using gradient-based optimization.

### 2.2 Optimizing Consistent Sinogram Information

Following the Noise2Inverse [17] pipeline, a self-supervised loss $L$ can be derived from the reconstructions of two independent stacks of CT projection images $\mathbf{p}_j$ ($j \in \{0,\dots,J\}$) and $\mathbf{p}_k$ ($k \in \{0,\dots,K\}$). In the reconstruction domain, only their image content but not their noise realization is correlated which allows deriving a loss between both reconstructed images to assess the consistency of the image contents. Other works [17] minimize that loss to train the weights of denoising models $f(\cdot,w)$ in the projection and the reconstruction domain self-supervisedly

$$\underset{w}{\operatorname{argmin}}\, L(\hat{\mathbf{y}}_J, \hat{\mathbf{y}}_K) \quad . \tag{4}$$

In the sparse-angle tomographic setting where only a limited number of $J$ projection images was measured, we propose to minimize the following expression to optimize the missing set of $K$ noise-free projection images $\mathbf{p}_k$

$$\underset{\mathbf{p}_k}{\operatorname{argmin}}\, L(\hat{\mathbf{y}}_J, \hat{\mathbf{y}}_K) = \underset{\mathbf{p}_k}{\operatorname{argmin}}\, L(\hat{\mathbf{y}}_J[G_{0:J}], \hat{\mathbf{y}}_K[G_{0:K}]) \quad . \tag{5}$$

We regard the pixel intensities of the inpainted projection images $\mathbf{p}_k$ as trainable weights and optimize them in a self-supervised and data-driven way as illustrated in Fig. 1. To enable backpropagating the gradient to the projection images, we use a differentiable FBP operator [9] within our XRM reconstruction pipeline [10]. Due to the well-defined projection geometries $G_j$ and $G_k$ for the individual projection images $\mathbf{p}_j$ and $\mathbf{p}_k$, the proposed pipeline is constrained to optimize missing angular projection data in between the measured sparse-angle projections to predict realistic reconstructions $\hat{\mathbf{y}}_K$ close to $\hat{\mathbf{y}}_J$. Although the pipeline can in theory converge to an identity solution where inpainted projections equal the true measured data, this solution is not favored by the derived pixel-wise loss as both independent reconstructions would be slightly misregistered through the different sets of view geometries $G_{0:J}$ and $G_{0:K}$.

## 3 Experiments

### 3.1 Data

The used data set consists of five high-resolution X-ray microscope (XRM) cone-beam scans of ex-vivo mouse tibia bones. Investigating bone structures on the micrometer scale is instructive for understanding the cause and progression of bone-related diseases on the cellular level as well as for developing adapted therapies. Lacunar bone structures, visible in Fig. 2 as tiny holes in the bone, are in particular of interest as they contain osteocyte cells that are heavily involved in the bone-remodeling process [20]. To resolve Lacunae reasonably well, 1401 projection images are acquired, which accumulates to a total acquisition time of around 14 h. Here, neither acquisition time nor induced sample dose allow for desired in-vivo investigations [1]. Sparse-angle CT acquisitions in combination with self-supervised projection inpainting algorithms can enable faster low-dose acquisition while preserving a high reconstruction quality. The used XRM scans image the tibia bone of mice close to the knee joint and contain truncated information of the proximal fibula bone in some projective views.
The study was performed in line with the principles of the Declaration of Helsinki. Approval was granted by the Ethics Committee of FAU Erlangen-Nürnberg (license TS-10/2017).

### 3.2 Training

We evaluate the effectiveness of our method on five XRM mouse tibia scans using half (50 % dose) and one-third (33 % dose) of the available 1401 projections for all compared inpainting strategies. Projection images are reconstructed using the public cone-beam XRM reconstruction framework of Thies et al. [10]. Nearest neighbor interpolation and trilinear interpolation operators are taken from the PyTorch framework and compared to our proposed self-supervised projection inpainting method. We initialize the missing projection intensities with interpolated projections to accelerate the optimization and start from a reasonable reconstruction. Subsequently, projections $\mathbf{p}_k$ are registered in the PyTorch graph as trainable parameters and are updated using a stochastic gradient descent optimizer with learning rate 0.1 without momentum. Further, we use the mean absolute error as loss function $L$. Quantitative performance metrics are calculated with respect to the high-resolution XRM images reconstructed from all available projections.

## 4 Results and Discussion

We compute the commonly used quantitative image quality metrics peak signal-to-noise ratio (PSNR) and structural similarity index measure (SSIM) to evaluate reconstructions with respect to the high-resolution ground truth images reconstructed from all available projections. Mean ± standard deviation of the used five XRM scans is provided in Tab. 1 for both investigated sparse-angle settings. The calculated metrics indicate improved reconstruction quality of the optimized XRM projections across both sparse-angle tomographic settings and over the compared reference methods. In our experiments, optimized projection images improve the PSNR by 3.1–7.4 % and the SSIM by 7.7–17.6 % relative to the nearest neighbor interpolation baseline.
Qualitative results are compared on the reconstructions of the inpainted projection images in magnified regions of interest in Fig. 2. Biologically interesting lacunar structures are visible as small dark holes within the bone tissue. Whereas nearest neighbor interpolation only slightly improved the noise level within the bone region, trilinear interpolated and optimized projections further reduce reconstruction noise. Al-though the bone reconstructed from the optimized projections appears a bit smoother compared to trilinear interpolation, only small improvements are visible.

| 50 % dose | PSNR | SSIM |
|---|---|---|
| Nearest neigh. int. | $27.0 \pm 0.4$ | $0.620 \pm 0.008$ |
| Trilinear int. | $27.8 \pm 0.4$ | $0.656 \pm 0.008$ |
| Optimized projections | $\mathbf{29.0 \pm 0.5}$ | $\mathbf{0.729 \pm 0.009}$ |

| 33 % dose | PSNR | SSIM |
|---|---|---|
| Nearest neigh. int. | $25.8 \pm 0.5$ | $0.571 \pm 0.011$ |
| Trilinear int. | $26.2 \pm 0.5$ | $0.585 \pm 0.011$ |
| Optimized projections | $\mathbf{26.6 \pm 0.5}$ | $\mathbf{0.615 \pm 0.013}$ |

**Table 1:** Quantitative reconstruction results (mean ± std) calculated between the prediction starting from one half and one third of the full number of projections and the high-resolution reconstruction that is regarded as ground truth.

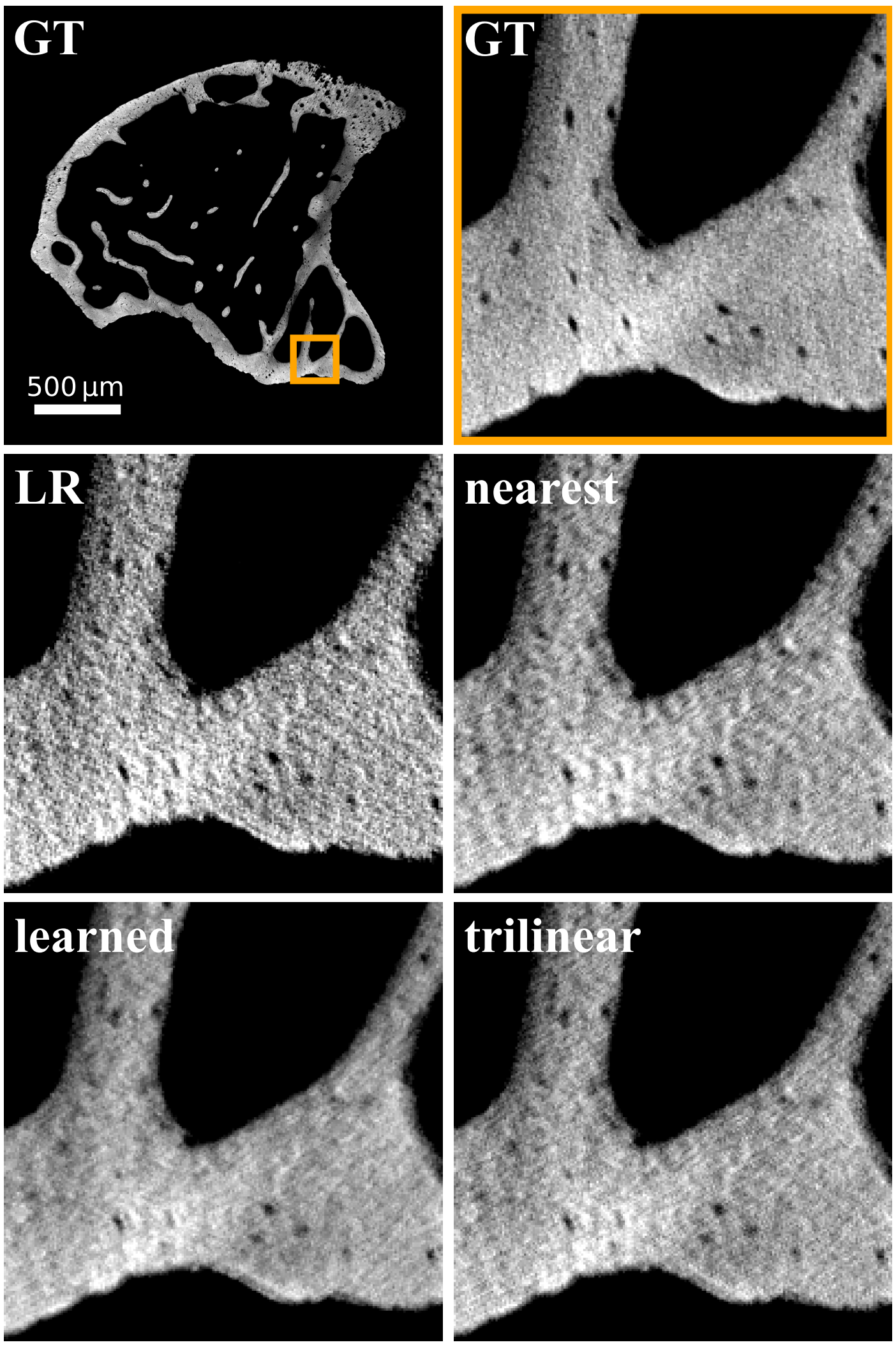

**Figure 2:** Reconstructions of mouse tibia bone XRM scans. The magnified region of interest is highlighted in the overview slice. The ground truth (GT) and low-resolution (LR) images are calculated from all and one-third of the available projection images respectively. Nearest, trilinear, and optimized denote the different reconstructions from the interpolated and optimized CT projection images.

None of the compared methods contain deep neural networks or require any form of pre-training on additional data, reference data, or a learned prior. In our proposed method, projection intensities are directly changed using gradient-based optimization within the fully differentiable reconstruction pipeline. In contrast to other techniques employing view consistency, we believe that our self-supervised pipeline is not limited to circular CT trajectories but can be applied to more difficult acquisition geometries as long as there is a differentiable reconstruction operator at hand. In addition, three of the five used bone scans contain truncated residual parts of the proximal fibula bone which are only visible in some projection views which would severely disturb most projection consistency-based algorithms. Our experiments show that the present data truncation can be handled well by our proposed projection optimization method. Further work is required to fully evaluate the generalizability of end-to-end optimization of CT projections and compare it to existing projection consistency-based methods or deep learning-based models.

## 5 Conclusion

In this work, we present a truly self-supervised projection inpainting technique to improve the reconstruction quality of sparsely acquired CT projection data. Our method allows directly optimizing projection image intensities through a differentiable FBP operator and a self-supervised loss metric calculated from two independently reconstructed projection stacks. The proposed pipeline requires further evaluation to prove its clinical effectiveness. Nevertheless, it has great potential over existing self-supervised algorithms as additional regularization can be applied and data consistency is enforced.

## 6 Acknowledgments

This work was supported by the European Research Council (ERC Grant No. 810316) and a GPU donation through the NVIDIA Hardware Grant Program.

# Circulation Federated Learning Network for Multi-site Low-dose CT Image Denoising

Hao Wang[1,2], Ruihong He[1], Jingyi Liao[1,2], Zhaoying Bian[1,2], Dong Zeng[*1,2], and Jianhua Ma[†1,2]

[1]School of Biomedical Engineering, Southern Medical University, Guangdong 510515, China
[2]Pazhou Lab (Huangpu) , Guangdong 510000, China

**Abstract** Deep learning (DL) has attracted great attention in the medical imaging field as a promising solution for high-fidelity CT image reconstruction in low-dose cases. Meanwhile, most of the existing DL-based methods are centralized machine learning frameworks with the need to centralize the data for training. These methods have poor generalization for the privacy policies. Federated learning (FL) can address privacy concerns by using local CT datasets without transferring data. The FL-based CT reconstruction performance still has room for improvement due to the limited capability of the global server. In addition, the absence of labeled data in the global server could also degrade the performance. In this work, to improve FL-based CT reconstruction performance, we first propose a novel federated learning framework with a circulation mode for low-dose CT image denoising, and the newly proposed framework is termed as Circulation FL, i.e., Ci-FL. Specifically, there is no fixed global server in the presented Ci-FL. Each local client can be regarded as the global server in each FL network training wherein the labeled data in the global server can help network training efficiently. The rest ones are regarded as the local client constructing an FL framework in each circulation node. With each site trained as a global server in the framework, we obtain the desired Ci-FL network. Experiments on multi-sites CT datasets clearly demonstrate enhanced reconstruction performance of the proposed Ci-FL against site-specific and traditional federated methods in terms of qualitative and quantitative assessments.

## 1 Introduction

Concerns have been raised about the risk of carcinogenesis from radiation doses in computed tomography (CT). Various specific CT techniques to minimize radiation dose while providing diagnostic examinations have been developed. One simple way is to lower the tube current in the examination. Meanwhile, this might suffer elevated noise if no adequate treatment in the reconstruction. To improve CT image quality, a variety of classical CT reconstruction methods have been proposed, including the sinogram-based methods, the image-based methods, and the model-based iterative reconstruction (MBIR) methods, etc. With the rapid development of deep learning (DL) techniques in recent years, DL-based CT reconstruction models have been actively explored and often obtain promising performance in the CT imaging task. A variety of DL-based CT reconstruction models have been applied, ranging from early convolutional neural network-based methods [1][2] to recent unrolling-MBIR networks [3][4]. Meanwhile, there are some intrinsic limitations in current DL-based methods. (i) Some of the DL-based methods are constructed based on datasets from one site with poor generalization ability. (ii) The other DL-based methods trained on the collected dataset from multi-site might suffer from privacy concerns [5].

Federated learning (FL) can be used to improve data privacy and efficiency in the medical field by allowing decentralized network training among multiple sites to collaborate without local data aggregation [6]. For example, Yang et al. developed a hypernetwork-based federated learning for low-dose CT imaging by learning common features from multi-sites in the global server and reconstructed CT images efficiently in local clients [7]. Meanwhile, the existing FL-based CT reconstruction methods have some disadvantages. (i) The global server without any training data only iteratively aggregates the model parameters of local clients, which might limit its capability. (ii) In each local client, the reconstruction performance highly depends on the dataset itself and cannot be comparable to the centralized DL-based CT reconstruction methods.

In this work, to promote the FL-based CT reconstruction performance, we propose a new FL strategy based on a circulation mode for low-dose CT denoising, dubbed as Circulation FL (Ci-FL). Specifically, instead of a fixed global server in the traditional FL, the proposed Ci-FL has a dynamic global server composed of each local site wherein each local site can be regarded as a global server and this global server has fully-labeled CT image pairs (i.e., normal-dose images/corresponding low-dose ones). The global server with labeled data can further improve training efficiency. In the proposed Ci-FL, except for one site as the global server, the remaining sites are regarded as local clients in each FL training in each circulation node. Moreover, in each node in the circulation, the local models are trained independently on different types of datasets in a supervised manner, and the parameters of the different local models are aggregated on the server at each round. The proposed Ci-FL selects one site as the global server in turn to aggregate parameters. Experiments on multi-site CT datasets clearly show the proposed Ci-FL substantially improves reconstruction performance compared with the other competing methods.

## 2 Materials and Methods

### 2.1 The proposed Ci-FL

Figure 1 shows the framework of the presented Ci-FL. The presented Ci-FL consists of a circulation training model and an FL framework. And, all the sites have labeled data $sitelist = [site1, site2, \ldots, siteM]$, i.e., normal-dose CT images/corresponding low-dose ones acquired with different scanner/protocols, respectively.

*Corresponding author: D. Zeng, zd1989@smu.edu.cn

†Corresponding author: J. Ma, jhma@smu.edu.cn

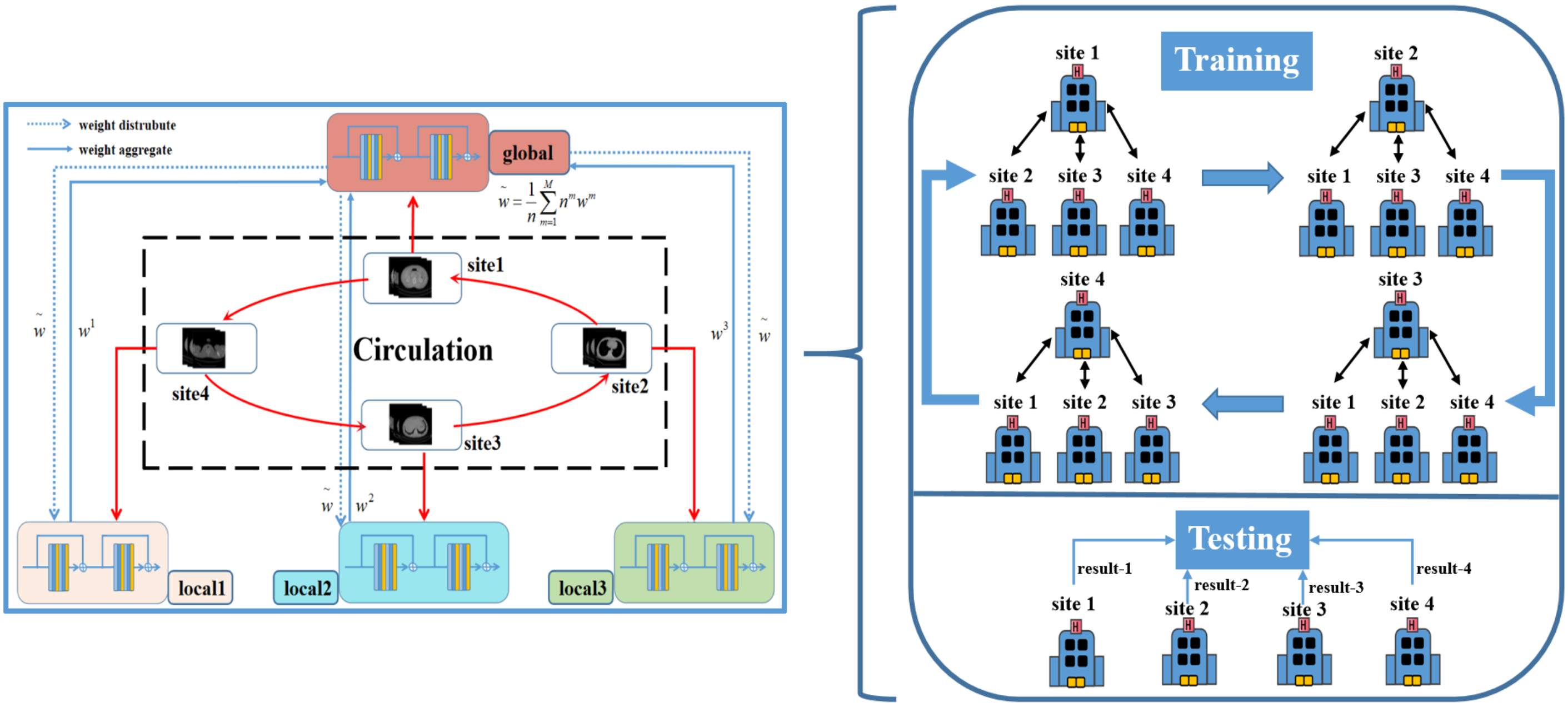

**Figure 1:** The framework of the presented Ci-FL.

At first, one of the sites, i.e., $sitem$, is chosen as the global server which is different from that in the traditional framework, and the rest are defined as local clients as the same in the traditional FL framework. Then, with the defined global server and local clients, the present Ci-FL is updated iteratively similar to the traditional FL wherein local models are trained with corresponding parameters and the global server aggregates the parameters and helps train efficiently global models with the labeled data in the global server. Next, parameters in the global model are distributed to each local site to update parameters iteratively. In particular, after a certain amount of global epochs is reached, we replace the updated datasets trained by the global server in the order in which they have been set. Similarly, each site can be selected as the global server in turn in each circulation node. In each training epoch, the dataset in each site dataset is related to model weights in the corresponding training stage. After completing multiple circulation epochs, we can obtain the final Ci-FL model that can produce optimal results.

## 2.2 The Ci-FL training

The presented Ci-FL aims to learn a generalized global model to process the CT data from multiple sites without sharing data directly and introduces the circulation mode into the FL framework to promote denoising performance. Specifically, different from the traditional FL framework, the presented Ci-FL allows each site to be selected as the global server for global model training in turn. This can help take advantage of the information among all the sites and obtain a more robust denoising model.

### 2.2.1 Local model training

In each local client, the local model is trained on local site datasets as follows:

$$\min_{w^m} \mathcal{L}_m(\Phi_{w^m}(x_i^m) - y_i^m), \quad (1)$$

where $\mathcal{L}_m$ is the loss function for the $m$th local model and in this study, it is set to be L1 norm. With $M^* = [1,2,3, \ldots, M-1]$ sites, $M^*$ local models is denoted as $\{\Phi_{w^m}\}_{m=1}^{M^*}$, where $\Phi_{w^m}$ is the $m$th local model with parameters $w^m$. $x_i^m$ is the $i$th low-dose CT image, $y_i^m$ is the normal-dose CT images in local dataset $S^m = \{x_i^m, y_i^m\}_{i=1}^{n^m}$. $n^m$ is the number of samples in the $m$th local client.

In the $t$th round, parameters at the $m$th local client can be updated as follows:

$$w_{t+1}^m = w_t^m - \sigma^m \min_{w^m} \mathcal{L}_m(\Phi_{w^m}(x_i^m) - y_i^m), \quad (2)$$

where $\sigma^m$ is the learning rate of $m$th local model.

### 2.2.2 Global model training

The parameters of local models are uploaded and aggregated in proportion to the sample number of each local dataset at a certain round $t$, and then the global model $\Phi_{\tilde{w}}$ is updated as follows:

$$\tilde{w}_{t+1} \Leftarrow \tilde{w}_t = \frac{1}{n}\sum_m^M n^m w^m, \quad (3)$$

where $n = \sum_m^M n^m$ is the total number of samples in all the local datasets.

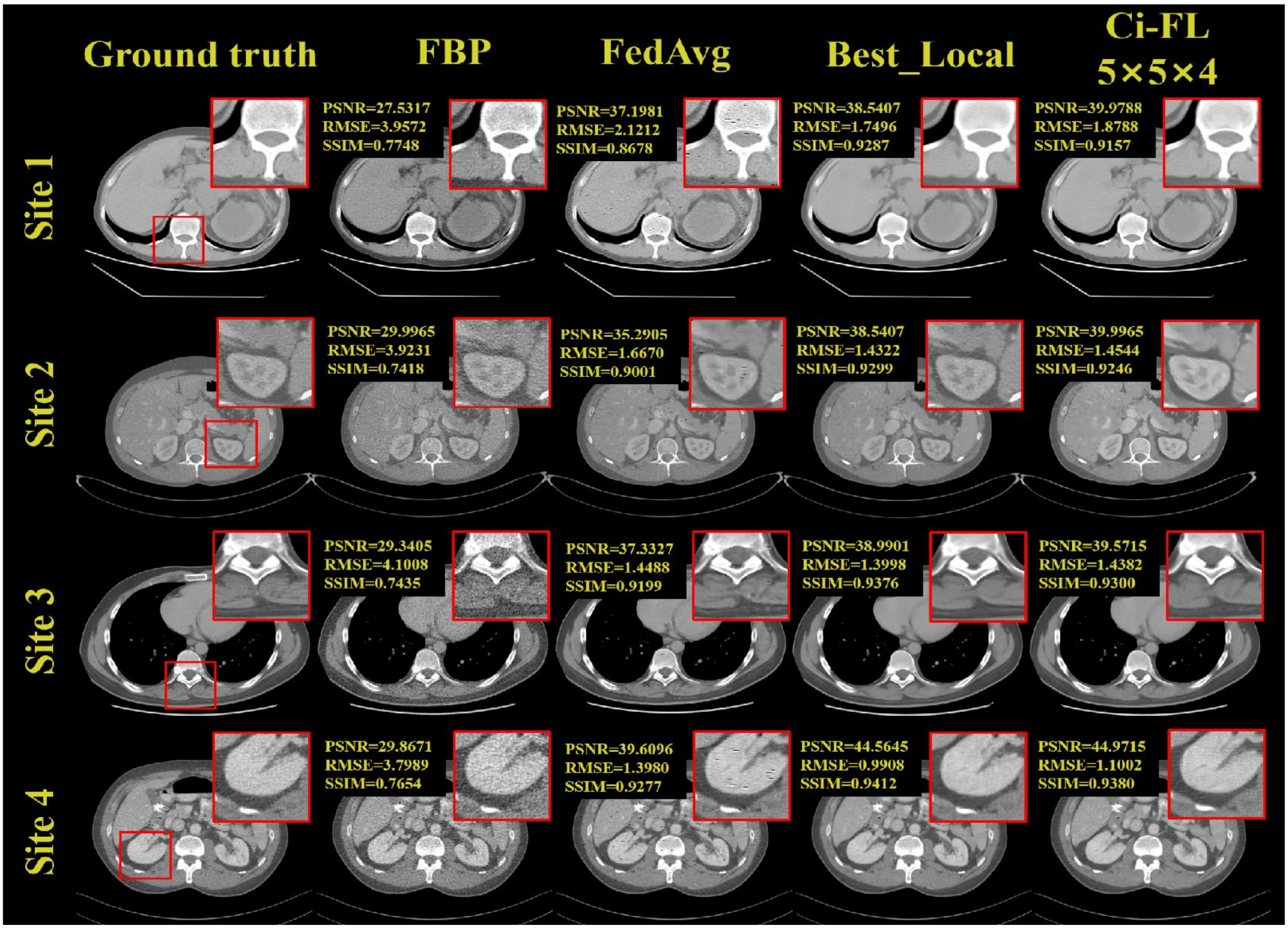

**Figure 2:** Results of Ci-FL and compared methods.
(The Ci-FL results: C × G × M, C is the circulation epoch, G is the global training epoch of the central server and M is the site number)

After updating the global model by aggregating the local parameters, the global model is further trained by the site dataset in the global server. The global model is trained on the central server similar to that in the local client. In each round, the global model is updated to obtain the trained parameters. After that, the local models receive the updated parameters of the global model at the next local training stage. This will repeat until the global model converges. In this work, the Adam algorithm is utilized to optimize both the global and local models.

## 3 Experiments

### 3.1 Dataset

To validate and evaluate the performance of the presented Ci-FL method, four different CT datasets are used in the experiment, i.e., three from local hospitals with approval from the Local Institute Research Medical Ethics Committee (i.e., site 1, site 2, and site 3), and Mayo dataset authorized by "2016 NIH-AAPM-Mayo Clinic Low Dose CT Grand Challenge" [8]. All datasets are acquired with different protocols from different scanners. Each dataset in sites 1, 2, and 3 contains 1000 normal-dose CT images, and the corresponding quarter-dose projection images are simulated based on the previous study [9]. Site 4 contains 300 normal-dose/low-dose CT image pairs. In the site 1, 2, and 3, 800 and 200 cases are assigned for training and testing datasets, respectively. In the site 4, 200 and 100 cases are assigned for training and testing datasets. The training image patches are set to 64 × 64 with a stride of 64.

### 3.2 Compared methods and implementation details

In this work, the conventional filtered back-projection (FBP) algorithm with an ideal ramp filtering kernel, FedAvg, and four local models (i.e., $Single_site\ 1, Single_site\ 2, Single_site\ 3, Single_site\ 4$) trained on each dataset are chosen as competitive methods. We empirically set the training parameters of Single_site 1, Single_site 2, Single_site 3, Single_site 4, and FedAvg to obtain optimal results by reference to the ground truth.

In the experiment, the backbone network of the presented Ci-FL method adopts a modified residual network (ResNet) with 12 residual blocks [10]. The training parameters are set as follows: (1) the number of cycle epochs and the number of global training epochs of the central server are respectively set to 5 and 5/4/3, (2) the epoch number is set to a maximum of 100 and the weight is decayed at the 100th epoch by multiplying 0.2, (3) the learning rate and batch size of the local and global models are $1.0 \times 10^{-4}$ and 32, respectively, (4) the aggregating round for the parameters of each local model is set to 1 epoch. All the networks in this work are implemented with Pytorch library, and the FBP algorithm is based on the ASTRA toolbox by utilizing one NVIDIA Tesla P40 graphics processing unit (GPU) which has 24 GB memory capacity.

## 4 Results

Figure 2 showcases the visual comparison results on the four sites from the competing methods, and the CT images at normal doses are treated as ground truth. It can be observed that the FBP results suffer from noise-induced artifacts. FedAvg can suppress noise-induced artifacts to some extent but it would lead to undesired artifacts in the final results improves the image quality and reduces the noise. But it causes over-smoothed results and decays the image resolution, as shown in the red zoomed-in regions of interest (ROIs) at site 1 and site 4. The possible reason can be that the FedAvg does not consider the information from different sites. In the Best_Local cases, the model trained with the data on a single site can process the low-dose CT images at its own site efficiently. It should be noted that these models fail to process the low-dose CT images at other sites due to the heterogeneity among different sites, as shown in Figure 3. Moreover, the presented Ci-FL successfully reduces noise-induced artifacts in the CT images from all the sites and produces promising results. The zoomed-in ROIs indicated by the red boxes also demonstrate its efficiency, indicating its generalization and robustness in processing datasets from different sites simultaneously. The main reason might be that the presented Ci-FL implements a circulation FL mode where each different site dataset takes turns acting as a central server to further train the model parameters for each round of aggregation, which has the potential to balance and improve the performance of the trained models across all medical imaging centers.

To evaluate the performance of the presented Ci-FL, three measure metrics are used, i.e., peak signal-to-noise ratio (PSNR), root mean square error (RMSE), and structural similarity index (SSIM). It can be seen that the presented Ci-FL method performs better than the FedAvg method and achieves similar performance with the single site models in terms of PSNR, RMSE, and SSIM measurements, indicating that the presented Ci-FL has stronger generalization ability than the other competing methods.

## 5 Conclusion

In this work, we propose a federated learning strategy with a circulation mode for low-dose CT image denoising, i.e., Ci-FL. Specifically, there is no fixed global server in the presented Ci-FL. Each local client can be regarded as the global server in each FL network training wherein the labeled data in the global server can help network training efficiently. Experimental results demonstrate that the presented Ci-FL strategy outperforms the other competing methods qualitatively and quantitatively. This can provide a new strategy for the FL framework in low-dose CT imaging tasks.

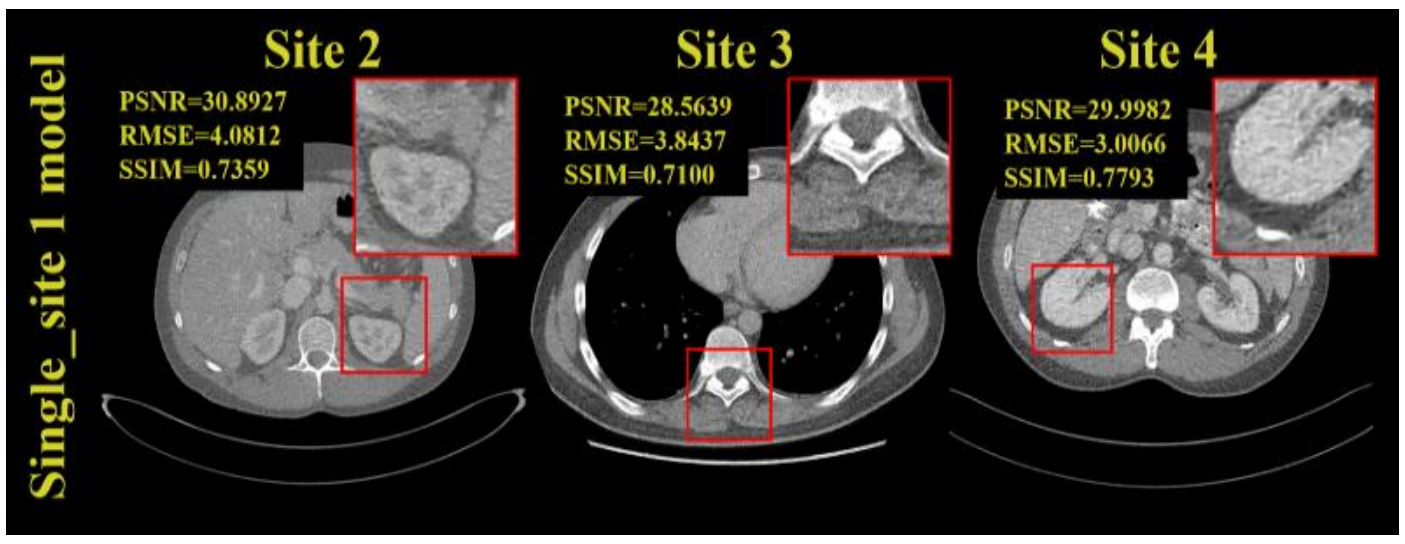

**Figure 3:** Results of site 2, site 3 and site 4 tested by the Single_site 1 model

## Acknowledge

This work was supported in part by the NSFC under Grant U21A6005, Grant 12226004, and Young Talent Support Project of Guangzhou Association for Science and Technology.

# Abdominal and pelvic CBCT-based synthetic CT generation for gas bubble motion artifact reduction and Hounsfield unit correction for radiotherapy

Kai Wang[1] and Jing Wang[1]

[1]Department of Radiation Oncology, University of Texas Southwestern Medical Center, Dallas, TX

**Abstract** The motion of gas bubbles (gastrointestinal gas) in the abdominal and pelvic region can produce significant artifacts on the cone-beam CT (CBCT) , which adversely affects the imaging quality and limits the process of image-guided radiotherapy and CBCT-based adaptive planning. In this study, we evaluated the effectiveness of cycle generative adversarial network (CycleGAN) for improving CBCT image quality and Hounsfield unit (HU) accuracy of abdominal and pelvic scans by synthesizing high-quality CT images based on the image content of CBCT images. The improved image quality of the synthetic CT (sCT) on both gas bubble artifact reduction and HU correction compared to the original CBCT demonstrated that CycleGAN is a promising tool for onboard CBCT image quality improvement, which can potentially increase the treatment precision of radiotherapy.

## 1 Introduction

Onboard cone-beam computed tomography (CBCT) is widely used in radiotherapy clinics for patient set up. It has been conventionally used for treatment target positioning in image-guided radiation therapy (IGRT), and has been increasingly used for treatment re-planning in adaptive radiation therapy (ART) more recently [1-4]. However, the application of CBCT in abdominal and pelvic cancer radiotherapy is still limited due to its low image quality. Its inherent limitation of imaging physics induces scattering and beam hardening artifacts on the image, and it is sensitive to motion of gastrointestinal gas because of the long-time onboard scanning process [5-7]. Methods for gas bubble motion artifact reduction and HU accuracy correction on CBCT are of great importance.

A lot of research has been done to reduce these artifacts and improve the CBCT image quality, especially for the scatter noise correction and beam hardening artifact correction. And much work has been done to address the periodic motion artifact caused by such as motion of heart and lung. However, the artifacts produced by isolated aperiodic motions of small structures, such as gastrointestinal gas bubble, have been rarely described, and no available method has been discussed to correct them to our knowledge [2, 8]. As the gas bubble motion artifact is frequently seen in abdominal and pelvic scan and could induce severe artifacts (Figure 1), an artifact correction method designed for gas bubble motion is desired to improve the abdominal CBCT image quality and then aid in online patient setup and adaptive radiation therapy.

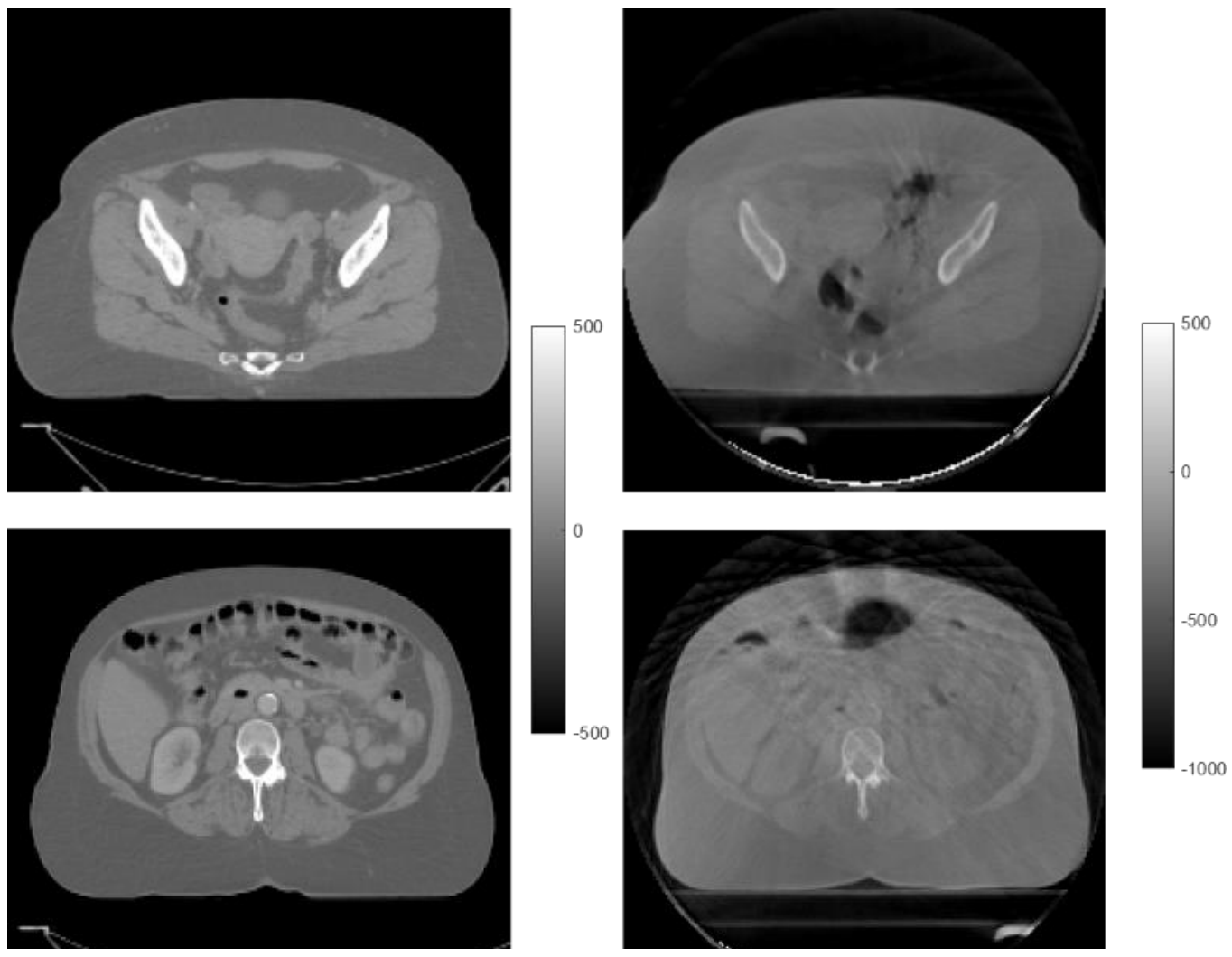

**Figure 1:** Examples to show the gas bubble artifacts and Hounsfield unit (HU) inaccuracy of onboard abdominal and pelvic CBCT. CT images are on the left, their corresponding CBCT images are on the right.

The correction of motion artifacts in medical imaging is a challenging problem, particularly in the context of real-time imaging. Traditional methods for artifact correction, such as image registration and motion compensation, update the image iteratively, which can be time-consuming and computationally intensive, making them unsuitable for real-time applications. For image guided adaptive replanning, while these methods can yield good dosimetry results for more stationary disease sites such as head and neck cancer, the performance might be poor due to the substantial patient motion and different air/water filling status in the abdominal and pelvic region.

Recently, deep learning methods, such as Cycle Generative Adversarial Networks (CycleGANs), have been applied to deal with the unpaired image data in multiple applications in medical imaging [4, 9-11]. The CycleGAN is a type of GAN that is trained to learn the mapping between two image domains, and used for cross-domain image transformation, such as CT-based synthetic MRI generation and CBCT-based synthetic CT (sCT) generation. And it showed promising performance on these tasks while preserving the anatomical well.

In this work, we propose the use of a CycleGAN for CT-quality image synthetization to reduce the artifact caused by motion of gas bubbles and HU inaccuracy caused by scattering in abdominal and pelvic onboard CBCT scans used for radiotherapy. The effectiveness of CycleGAN was evaluated using a CBCT dataset of real abdominal or pelvic cancer patients.

## 2 Materials and Methods

### 2.1 Image acquisition and processing

The dataset used for this study consisted of abdominal planning-CT and onboard CBCT scans from 70 patients who received radiotherapy in the University of Texas Southwestern Medical Center. All the planning-CTs were acquired with Philips Big Bore CT simulator (Koninklijke Philips N.V.,), and all the CBCTs were acquired on Elekta Agility XVI (Elekta, Stockholm, Sweden). The treatment areas include cervix, pancreas, uterus, vagina and vulva. Obvious gas bubble motion artifacts were identified on all these CBCT scans. All the image data were extracted retrospectively under an IRB-approved protocol.

The dataset was divided into two parts: a training set consisting of CT and CBCT scans for 60 patients, and a testing set consisting of the remaining patient scans. All the images were resample and cropped to 3D matrixes with resolution of 205×205×80 and voxel size of $2\times2\times3mm^3$. The HU range was clipped to [-1000, 1000] and then the images intensity values were rescaled to [-1, 1] for model training and validation.

### 2.2 CycleGAN architecture

As shown in Figure 2, the architecture consisted of two generator networks and two discriminator networks. The generator networks were designed to learn the mapping between the CT and CBCT iamges, where CT images are viewed as the idea image with accurate HU and without the gas bubble motion artifact. The discriminator networks were designed to distinguish between the real and generated images. Sepcifically, Generator A synthesized sCT from CBCT, and Generator B generates synthesized CBCT (sCBCT) from CT. Discriminator A discriminates between CT and sCT, while Discriminator B discriminates between CBCT and sCBCT. The two-cycled workflow includes: 1) generation of sCT based on CBCT and sCT based cycle-CBCT synthesization, 2) generation of sCBCT based on CT and sCBCT based cycle-CT synthesization. Meanwhile, the two discriminators discriminates between CT and sCT, CBCT and sCBCT, resepctively. Another set of identity images is generated by feeding CT patchs to Generator A and CBCT patches to Generator B to generate identity CT and identity CBCT, respectively.

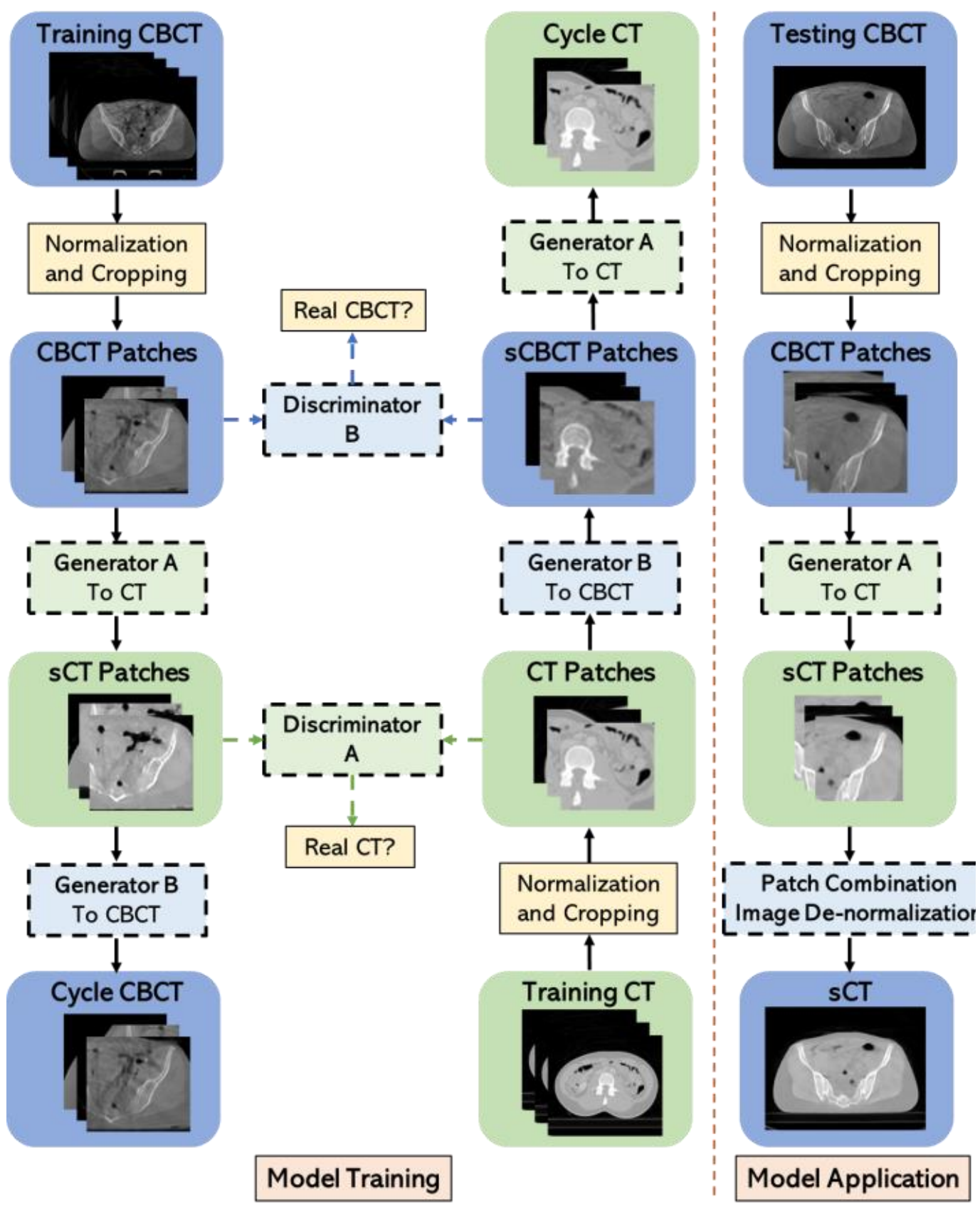

**Figure 2:** Schematic flowchart of the proposed CT image systhetic based on CBCT using cycleGAN. Two generators and discrimiators are trained in the cycleGAN structure using abdominal and pelvic CT and CBCT patches. And the trained generator for CBCT to CT conversion is used for synthetic CT (sCT) generation.

The 3D patchGAN discriminator architecture is used for our discriminators, and the a 3D CNN architecture with 9 resnet block is used for our generators. The implementation of our 3D CycleGAN architecture was based on https://github.com/davidiommi/3D-CycleGan-Pytorch-MedImaging. We use MAE loss for cycle loss and identity loss calcualtion. For calcuation of discrimitor loss and generator loss, binary cross entropy loss is used to identity systhetic images and real images.

### 2.3 Training and evaluation

The CycleGAN was trained using the Adam optimizer with a initial learning rate of 0.0001 and a batch size of 1. In the first 1000 epoch, the network was trained with the initial learning rate. In the second 1000 epoch, the network was trained with the a decreasing learning rate from 0.0001 to 0. Model weights updated after each batch. The training process was monitored using a validation set, consisting of 50 real and 50 simulated scans.

As shown in Figure 2, during training, the images are random cropped to patches of 128×128×64 and go through the generation network to produce the corresponging synthetic images. The generated patches then feed into the adverse generation network to produce the cycle images. The original patches are also feeded into different generators to generate identity images. Discriminator loss, which considers the BCE loss of identifying real image only, is used to update the discriminator. Generator loss, which considers the similarity of synthesized image to the corresding modality image measured by discrimotor loss, the similarity of original images to their cycle-images, and original images to their identity-images, measured by MAE loss, is used to update the generators.

For testing, a new CBCT image would be cropped into pathes of 3D size $128 \times 128 \times 64$ with an overlab of $16\times16\times16$ between every two neighboring patches after image normalization. Generator A will generate sCT patches based on the input CBCT patches, and all the sCT patches will be combined together to the size of the orginal CBCT image. Finally, the image denormalization step will be performed to map the sCT image intensity to HU.

The model was trained and tested on an NVIDIA GeForce RTX™ 3090 Ti GPU with 24 GB memory. The batch size was 4. The algorithm was implemented by Python 3.9 and Pytorch 1.13. The training process took around 48 h and the generation of one sCT took about 5 sec.

## 3 Results

The effectiveness of the CycleGAN for correction of the motion artifact and HU value was evaluated quantitatively and qualitatively. The qualitative evaulation included visual inspection of the generated images, and comparison of the CT, CBCT and sCT images. We selected some regions with different tissue/materials from CT, CBCT and sCT images to show the image quality improvement in HU accuracy and image uniformity for quantitative evaulation.

### 3.1 Qualitative evaluation

We generated the sCT for all the 10 testing patients based on their CBCT using the generator A in our CycleGAN model. The results of the qualitative evaluation showed that the CycleGAN was able to effectively remove the motion artifact and preserve the structural details of the abdominal images (Figure 3). Visual inspection of the generated images showed that the CycleGAN was able to correct the blurring and distortion caused by the gas bubble motion artifact, and produce images with similar visual quality to the real images.

### 3.2 Quantitative Evaluation

On each image, we identified three 7×7×7 volumes within the gas bubble artifact surrounded gas/air, fat, and muscle volumes, respectively. We comparied the HU distribtution of these volumes on each of the testing patients' CT, CBCT, and synthetic CT for HU accuracy and image uniformity evaluation (Figure 4). The results of the quantitative evaluation showed that the generated sCTs have narrower ranges of HUs within volumes of a same material/tissue than CBCTs. And the HU ranges of fat and muscle tissues in sCTs are almost the same as the ranges of CTs. The distritbution of HU values of the sCT demostrate the effectiveness of CycleGAN for HU inaccrucy correction and reduction of gas bubble motion induced streak artifacts.

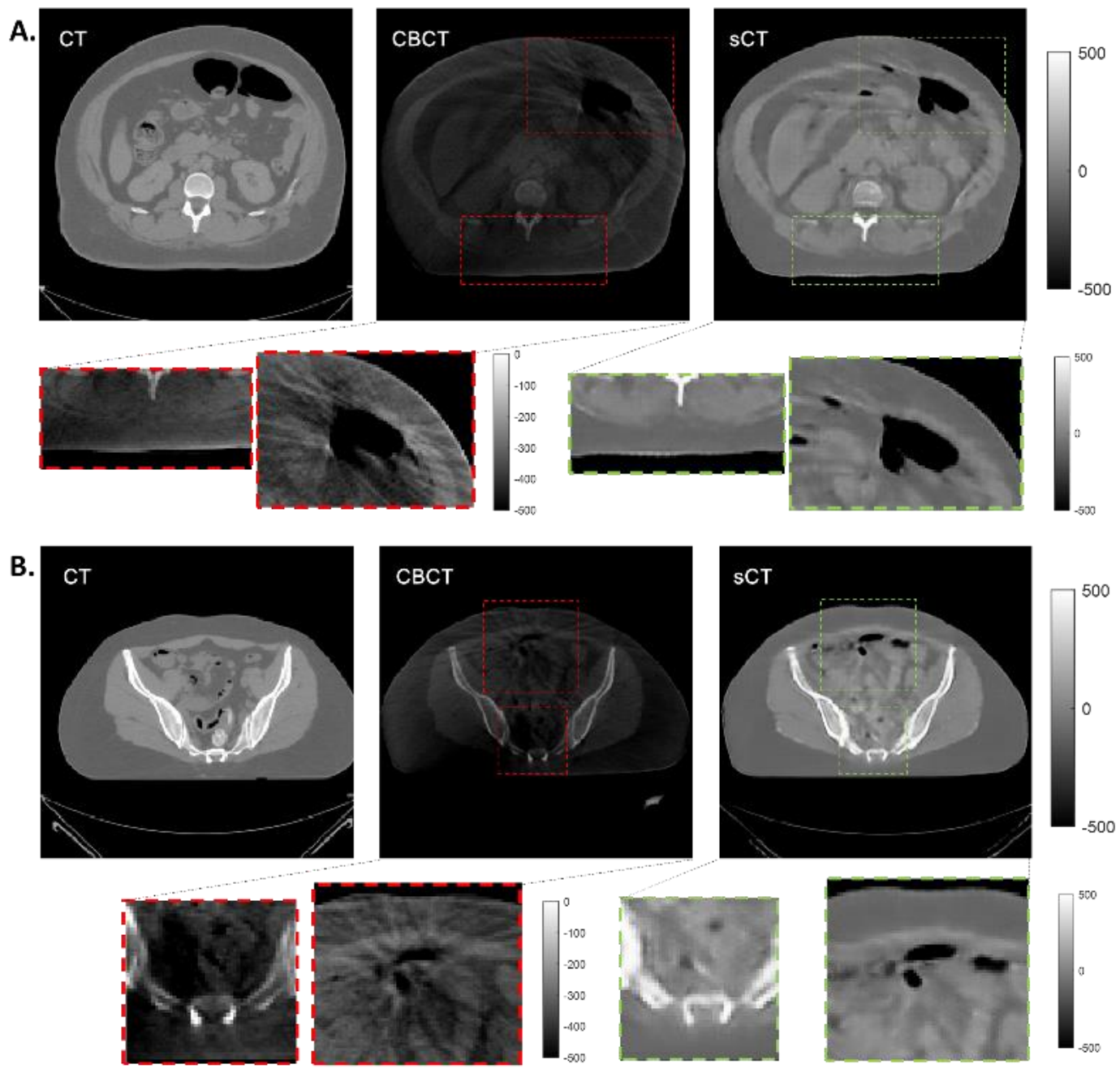

**Figure 3:** Image examples for qualitative evaluation of the gas bubble motio artifact reduction and Hounsfield Unit correction for abdominal and pelvic onboard CBCT using cycleGAN synthetic CT (sCT).

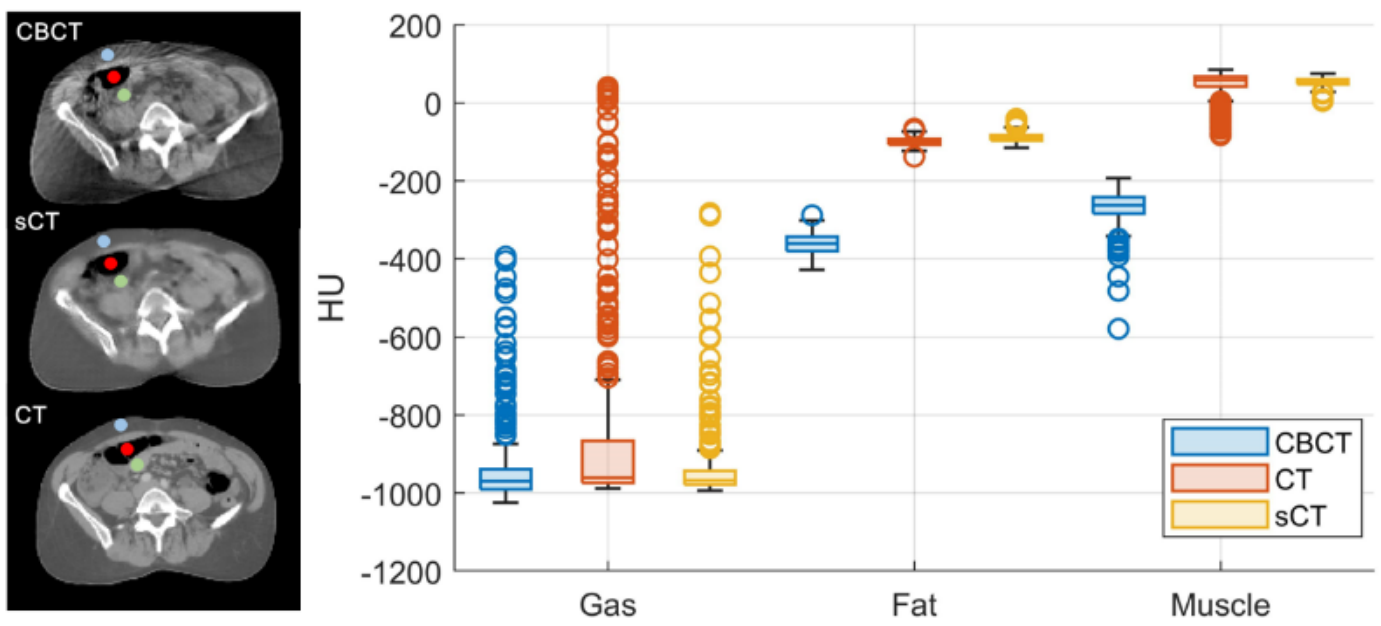

**Figure 4:** Hounsfield unit (HU) of selected gas bubble, fat, and muscle volumes around the gas bubble artifact region in different images. 7×7×7 volumes within the gas bubble, fat, and muscle were selected on all the testing patients' CT, CBCT, and synthetic CT (sCT) for comparison.

## 4 Discussion

In this study, we proposed the use of a CycleGAN for correction of HU value, and reduction of the motion artifact caused by gas bubbles in onboard abdominal/pelvical CBCT scans used for radiotherapy. Our results showed that CycleGAN was able to effectively remove the motion artifact from the images, producing scans with improved image quality and accurate HU compared to the original scans. The correction of the motion artifact was achieved in real-time, making the CycleGAN a suitable tool for use in clinical practice.

High quality onboard imaging technique is of great importance for cancer radiotherapy. For abdominal and pelvic region, serval studies have been conducted to characterize the effect of gas bubble motion artifacts through phantom and/or clinical experiment [6, 12, 13]. They suggested that the artifact can result in inaccuracy target locating and contouring, incorrect dose calcuation, and suboptimal treatment outcomes. By using a CycleGAN to correct the artifact, we can improve the accuracy of the onboard image which ensure that the better possible treatment plan can be developed for each patient. This onboard CBCT-based image genration method may help to reduce costs associated with additional devices or repeated imaging, speed up the dose calcuation workflow comparing to the tradition registration based method, and imporve the quality of treatment planning, and ultimately, has the potential to result in improved patient outcomes.

There are several limitations to our study that should be considered. First, as we don't have paired registrated CBCT and CT data, we didn't perform detailed quantitive comparision and analysis between the generated sCT and CT of a same patient in the current study. Phantom study or deformable registration of CT image to CBCT image which can genrate paired data for result analysis can conducted. Second, as the current model is not patient specific, although the pattern of gas bubble motion artifact in the image can be replaced with normal image pattern without gas bubble motion, the generated image might have fake strucuture in the previous artifact region, which is of low anatomic accuracy. To address this issue, a patient specfic model or projection image based generation model is necessary in our future work. Third, as the one important goal of this study is to improve the CBCT image quality for abdominal/pelvic cancer treatment planning. Dosimetric analysis of plan generated based on the CT, CBCT, and sCT needs to be done next to illustrate the effectiveness of the sCT based planning. Overall, even with these limitations, this approach is an important step towards the development of more advanced CBCT-based imaging methods to assist in abdominal and pelvic cancer patient radiotherapy.

## 5 Conclusion

In this study, we have trained an cycle-GAN model for abdominal and/or pelvic CBCT to sCT generation. The sCT images could effectively correct the motion artifact caused by gas bubble motion and provide more accurate HU. This approach can potentially improve the precision of IGRT, increase the accuracy of dose calculation, and facilitate the development of CBCT-based segmentation method and treatment planning method to assist in online adaptive radiotherapy.

# From coordinate system mapping to deep feature mapping: A generative deep learning approach for pixel-level alignment of images in rotation-to-rotation DECT

Peng Wang[1], Wenying Wang[2], Yuan Bao[1], and Guotao Quan[1*]
[1]Department of Reconstruction Physical Algorithm, United-Imaging Healthcare, Shanghai, China
[2]United-Imaging Healthcare America, Houston, USA

**Abstract** Dual-energy computed tomography (DECT) utilizes energy-dependent information and can improve the differentiation between tissues of different material composition that may exhibit similar attenuation in single-energy CT scans. While DECT systems are not always available in clinics, a single-source sequential rotation-to-rotation kVp-switching scan is feasible on most CT systems. However, non-negligible misalignment between the dual-energy images due to slow kVp switching limits the potential clinical application. Accurate registration between the two kVp images is required for further spectral analysis. The application of conventional registration methods is time- and computational power- consuming, making it challenging to achieve pixel-level alignment. We propose a generative approach as an alternative to coordinate registration methods to produce a perfectly aligned dual-energy set in the image domain. We discuss a statistical image mapping method based on histograms and explore the feasibility of integration with a deep learning strategy. The spectral transfer is implemented at the feature level with statistical information. Preliminary experiments using phantom and clinical data demonstrate that the proposed method can generate reliable DECT images and provide accurate material decomposition.

## 1 Introduction

DECT is a technique that uses two different spectral settings, enabling the integration of materials with different attenuations at varying photon energy levels. The DECT protocol provides several spectral image reconstruction sequences that optimize the visualization of suspicious lesions for diagnostic purposes. These sequences include virtual monochromatic images, material decompositions, virtual non-contrast images, and equivalent electron density. Among the various realizations of DECT, the single-source sequential rotation-to-rotation DECT is widely available on most CT systems. However, it suffers from patient motion-induced misalignment between the two kVp acquisitions. Such misalignment can result in significant artifacts in subsequent spectral analyses. The registration between dual-energy image pair poses a challenging task.

Image registration using a coordinate mapping strategy, encompassing effective control point searching, continuous deformation field generation, and image morphological [1]. Recent years have seen several spatial transform networks (STNs) designed for medical image registration [2]. For example, Voxel Morph, [3] a rapid deep learning-based framework for deformable medical image registration, employs a convolutional neural network to process the misaligned input image pair into a deformation field that aligns one image with the reference image. The accuracy is comparable to state-of-the-art conventional model-based methods. The registration model is adaptable to various imaging modalities, even handling cross-modality registration tasks with specific loss function designs. However, Voxel Morph struggles if any topological change of anatomy occurs. It remains challenging to interpret how Voxel Morph can translate image value information into coordinate system mapping using only convolution and normalization layers.

Alternatively, generative methods leverage the preserved structure of the original images that enables pixel-level alignment. Prior research has explored generating dual energy imaging by exploring spectral information in single kV acquisition [4], which requires precise system modeling and tuning. Other research works focus on reconstructing images from limited-angle dual-energy scans to enhance temporal resolution [5] [6]. For rotation-to-rotation DECT, pure data-driven algorithms fail to utilize the entire information effectively and risk overfitting.

Substantial studies have demonstrated the capability of deep learning technique for style and cross-modality transfer in both natural and medical images. For example, the Adaptive Instance Normalization (AdaIN) layer [7] enables real-time arbitrary style transfer. This highlights that image style transformation is as simple as data distribution modification based on feature-level statistics. While for medical images, the lack of big dataset poses sparse and noisy features with a broad dynamic range that may not achieve the required accuracy for diagnostic use.

As the attenuation properties are determined by the imaging object, a deterministic relationship between CT values at two spectral settings can be established for a known material. For homogeneous objects, one can simply apply normalization to align the mean and standard deviation between the two kVp images. This normalization can extend to histogram matching when dealing with a single-material object of varying concentrations. However, global histogram matching is unsuitable for clinical DECT images transformation due to complex material compositions. To safely apply histogram matching for precise style transfer between two kVp images, one must segment the reconstructed volumes into numerous small volumes containing only similar material, wherein local histogram matching can be applied. This work aims to integrate feature extraction, mapping, and reconstruction into one deep learning framework to generate pixel-wise aligned DECT image pairs.

*corresponding author: guotao.quan@united-imaging.com

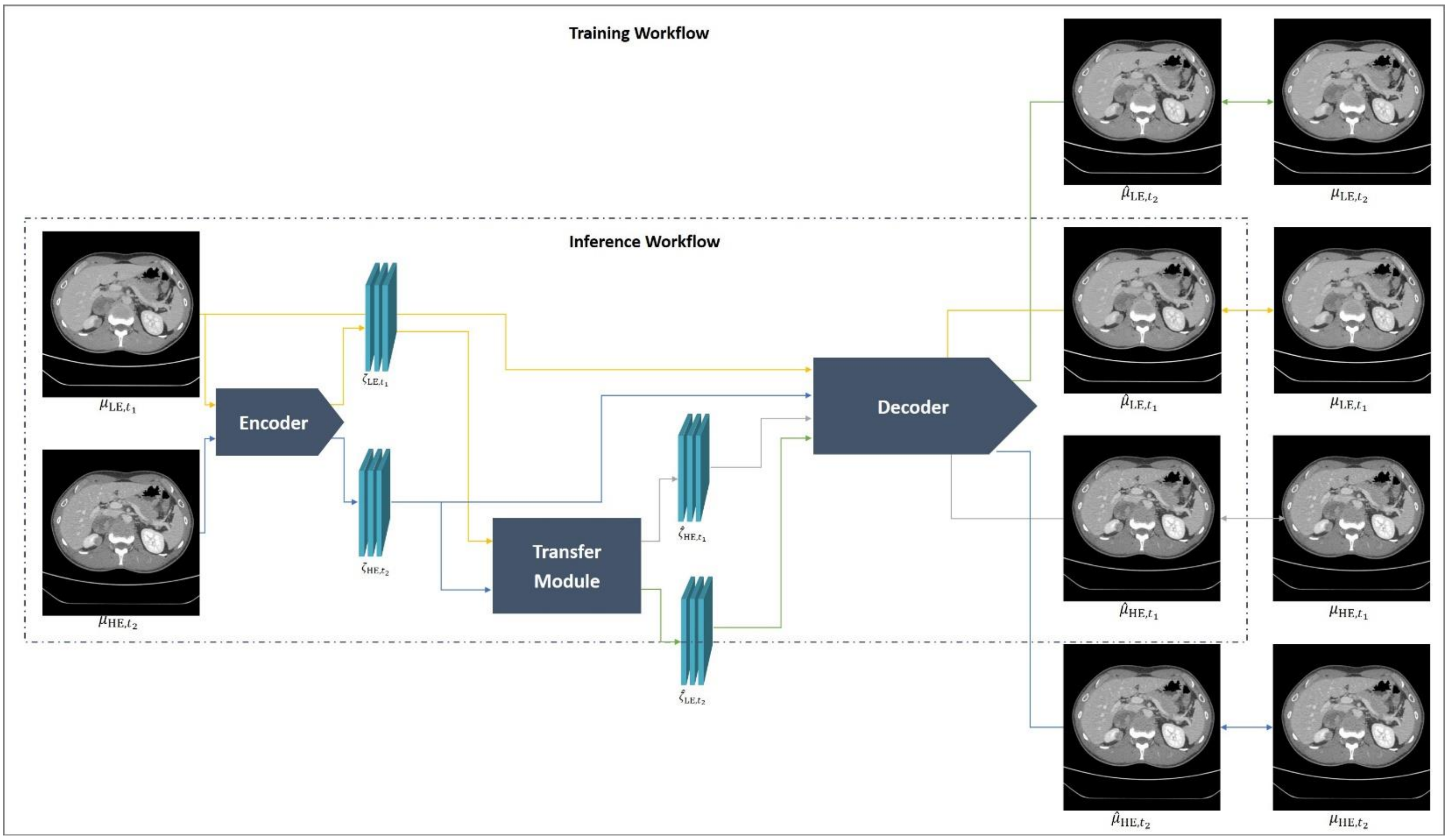

Figure 1. The architecture of the proposed Feature Transfer Network (FTN) and training/inference workflows.

## 2 Materials and Methods

Figure 1 provides an overview of the architecture of the Feature Transformation Network (FTN) and the training/inference workflow. The network takes the DECT images using filter-back projection (FBP) reconstruction as inputs. Here, $\mu_{\mathrm{LE},t_1}$ denotes the low-energy image measured at $t_1$ and $\mu_{\mathrm{HE},t_2}$ denotes the high-energy image measured at $t_2$. $\zeta$ and $\hat{\zeta}$ denote the encoded and transferred feature maps, respectively. After the decoding process, the network generates pairs of high and low energy images at both $t_1$ and $t_2$ ( $\{\hat{\mu}_{\mathrm{HE},t_1}, \hat{\mu}_{\mathrm{HE},t_2}, \hat{\mu}_{\mathrm{LE},t_1}, \hat{\mu}_{\mathrm{LE},t_2}\}$ ). In the following sections, we delve into the details of network architecture design, the dataset used for training and testing, and the training and inference procedures.

### 2.1 Network architecture

The network adopts a Variational AutoEncoder (VAE) backbone that comprises three main modules: encoder, decoder, and cross-modality transfer layer.

The encoder module includes three dense blocks with no normalization layer. Skip connections are intentionally omitted to emphasize the style transfer mapping within the down-sampled deep feature space. To address the depth limitation resulting from the absence of skip connections, we increase the number of channels to enhance the capacity of the encoder layers.

The style transfer module processes the output of the encoder layers and swaps the feature maps between the two kVp images. In this work, we select the histogram matching module (Figure 2) to find a precise pixel value on an aligned pixel grid. In preliminary investigation, we have observed that histogram matching outperforms the conventional AdaIN layer, particularly when the two kVp images share similar structural details.

The decoder module, on the other hand, reconstructs the aligned images at full resolution using either the original or transferred features as inputs.

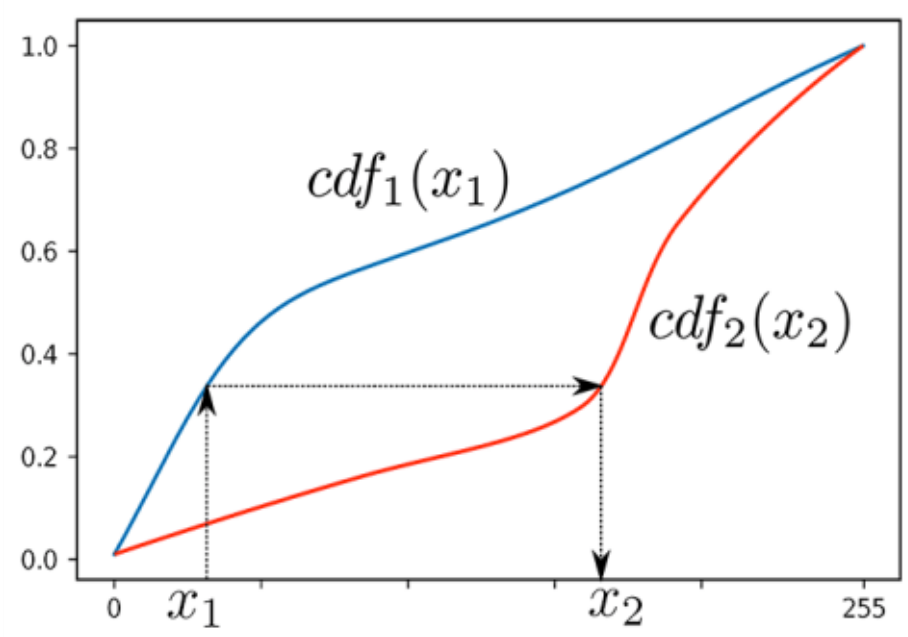

Figure 2. Example of histogram matching

| ROI Index | Ground Truth | | FBP | | FTN | |
|---|---|---|---|---|---|---|
| | Water | Iodine | Water | Iodine | Water | Iodine |
| 1 | 1000 | 0 | 997.88±56.24 | 1.81±21.95 | 1001.05±22.00 | 0.82±2.52 |
| 2 | 1000 | 20 | 999.70±62.56 | 21.7±23.33 | 1003.54±26.15 | 20.45±1.84 |
| 3 | 1000 | 50 | 1004.33±58.84 | 53.24±22.66 | 1011.62±20.19 | 51.03±3.59 |
| 4 | 1000 | 100 | 1014.50±63.12 | 99.80±26.80 | 1021.76±14.85 | 97.66±7.33 |
| 5 | 1000 | 150 | 1019.94±52.72 | 149.82±21.27 | 1023.89±18.32 | 148.59±4.84 |

Table 1. Quantitative material decomposition results of different ROIs (mean ± standard deviation) in mg/ml.

### 2.2 Dataset

The raw datasets obtained through DECT systems (United Imaging, Shanghai, China) were encrypted and granted authorization for research use. A conventional registration algorithm was applied to obtain aligned dual-energy image pairs to generate the corresponding ground truths.

For data augmentation, additional unregistered pairs were generated by applying random deformation fields to one of the registered images based on prior knowledge of human organ deformation.

### 2.3 Training procedure

We adopted a supervised learning approach to fine-tune the network parameters. Throughout the training phase, four different features were produced: two originated directly from the encoder, and two were transferred using the transfer module. The training loss features a weighted sum of value accuracy loss (mean absolute error) and the structure loss (structure similarity index measure) between the registered images and the corresponding ground truth.

$$Loss = \sum_{n=1}^{4} \left(\alpha_n \text{FedelityLoss}(\hat{\mu}_n, \mu_n) + \beta_n \text{StructLoss}(\hat{\mu}_n, \mu_n)\right)$$

### 2.4 Testing procedure

Low-energy images typically exhibit a wider dynamic range and higher noise level compared to high-energy images. Transferring from a wider dynamic range to a narrower one is more feasible through histogram matching. Therefore, during the inference process, we estimate registered low-energy images with the same distribution as the high-energy images. We evaluated our method using both GAMMEX DECT phantom data and clinical data sampling a range of different anatomical sites. Notably, none of the testing data were used during the training phase. To better visualize the registration results, we applied a standard image-domain material decomposition method to compute the water/iodine concentration maps $\rho_{\text{water}}$ and $\rho_{\text{iodine}}$:

$$\begin{bmatrix} \rho_{\text{water}} \\ \rho_{\text{iodine}} \end{bmatrix} = \begin{bmatrix} \mu^m_{\text{LE,water}} & \mu^m_{\text{LE,iodine}} \\ \mu^m_{\text{HE,water}} & \mu^m_{\text{HE,iodine}} \end{bmatrix}^{-1} \begin{bmatrix} \mu_{\text{LE}} \\ \mu_{\text{HE}} \end{bmatrix},$$

where $\{\mu^m_{\text{LE,water}}, \mu^m_{\text{LE,iodine}}, \mu^m_{\text{HE,water}}, \mu^m_{\text{HE,iodine}}\}$ are the effective mass attenuation coefficients of water and iodine at low kVp and high kVp respectively.

## 3 Results

Figure 3 summarizes the GAMMEX DECT phantom experiment where we compare the water/iodine material decomposition results using the FBP images or the FTN results. The quantitative results are summarized in Table 1. The proposed method can correct the structural misalignment, maintain the CT value accuracy, and reduce the noise.

In Figure 4, we show the clinical data results across different anatomical sites, including abdomen, thorax, and lower extremity. We compare the water/iodine material decomposition results using the original misaligned FBP images or the registered images using a conventional registration method or the proposed network. We observe improved material decomposition estimations at the edges and noise reduction in the FTN results.

## 4 Conclusions and Discussion

The proposed FTN is the first time to introduce the feature-level mapping strategy into registration tasks instead of coordinate system mapping. The trained network proves the capability of resolving the misalignment due to the low temporal resolution of rotation-to-rotation dual-energy scanning protocol, while the CT value accuracy and spectral image quality satisfy the clinical diagnostic requirement. As the noises of dual-energy images are correlated after feature mapping, the dose of scanning may be reduced as well.

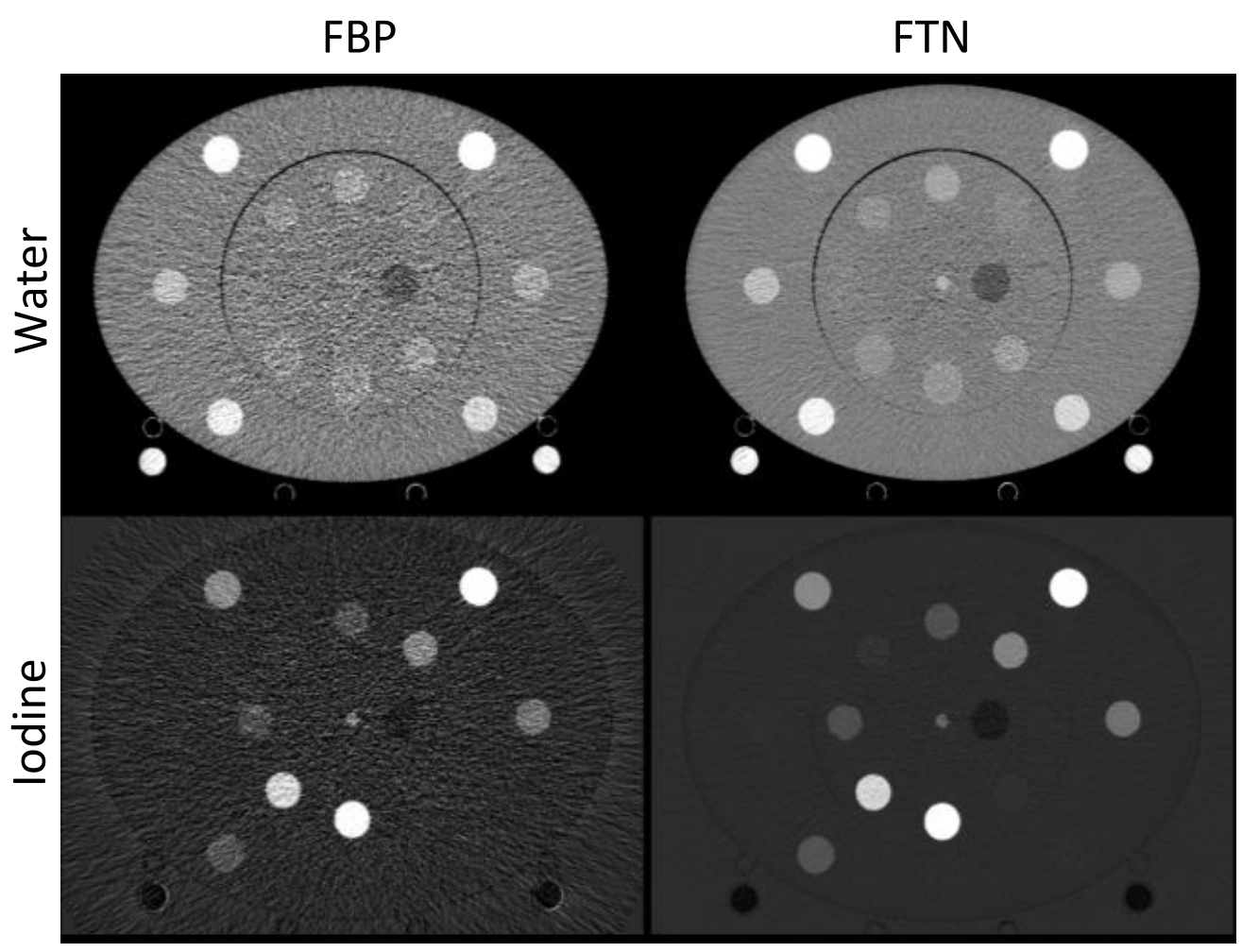

Figure 3. Quantitative comparison of water-iodine decomposition GAMMEX DECT phantom of original scan (left) and proposed method (Right). WL/WW is 1000/300 for water map (top) and 50/150 for iodine map (bottom).

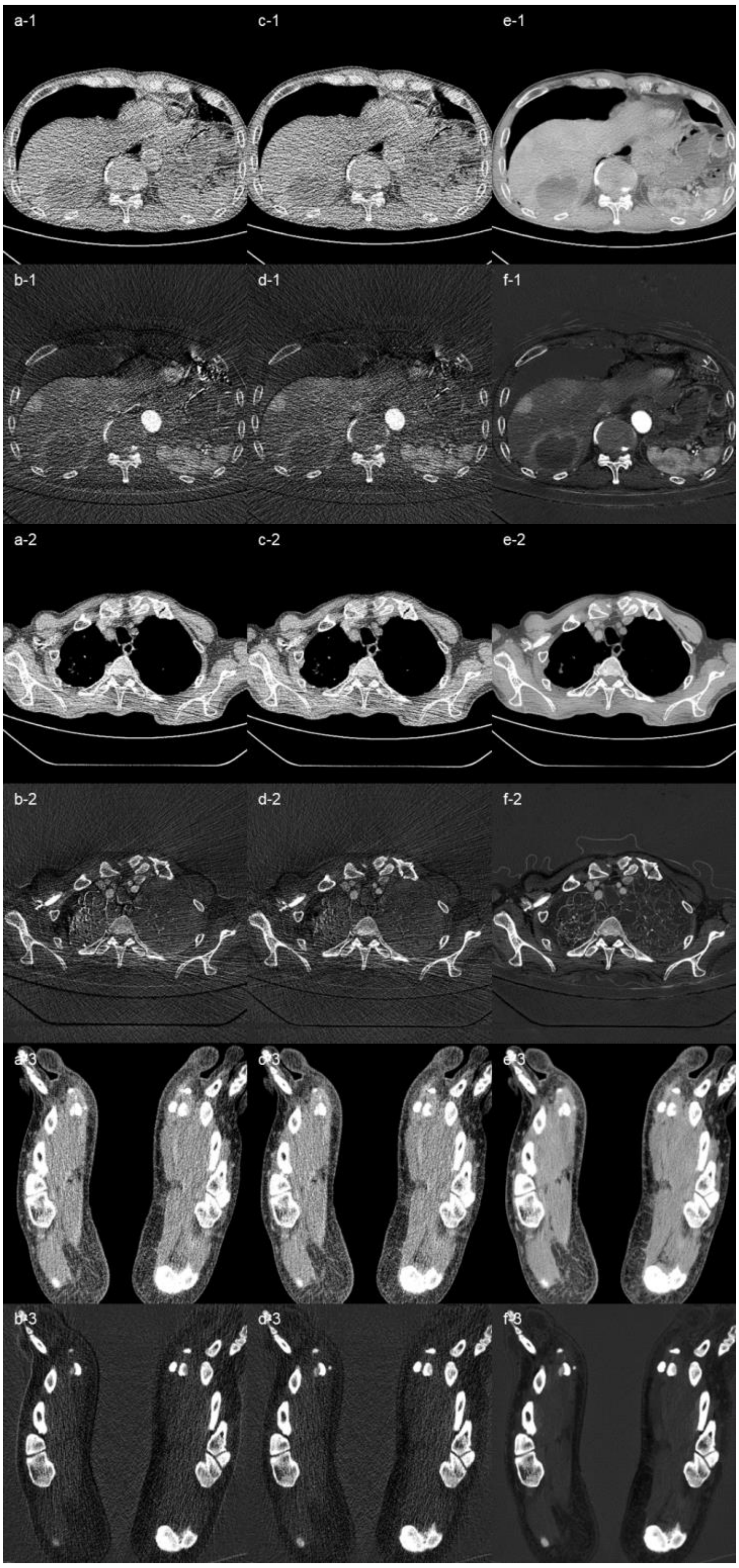

Figure 4. The testing results on different body parts. For each testing case, the above results show the water (top row) and iodine (bottom row) decomposition results. The left column is the result of original scanned dual energy images, the middle column is after non-rigid registration, the right column shows the results of the proposed method.

The deep feature transformation module was designed based on partial histogram matching. One may interpret how the CT values are transformed between the DECT images by examine the feature maps and corresponding histogram matching functions.

In this work, a coarse registration between the DECT images was applied if the mismatch is too severe. The upper limit of the resolvable displacement will be further tested to evaluate the robustness of the network. While we used a supervised learning strategy, the proposed FTN can be trained using unsupervised learning with a loss function evaluating the structure similarity between the estimated image with the corresponding image acquired at a different kVp.

The network is designed to align single-source sequential dual-energy images, where we use the same encoder and decoder for each image. This can reduce the number of parameters and enhance the generalizability. A more general implementation may adopt different encoders and decoders for other cross-modality registration tasks, including PET-CT, PET-MR, etc. This will be explored in future work. This will be investigated in future work.

# Wavelet-based stabilized score generative models

Yanyang Wang[1], Ge Wang[2], and Weiwen Wu[1*]

[1]The Department of Biomedical Engineering, Sun-Yat-sen University, Shenzhen, China
[2]The Department of Biomedical Engineering, Rensselaer Polytechnic Institute

**Abstract** Score-based generative model demonstrates strong performance in solving under-determined inverse problems. However, in the field of medical imaging, it is difficult to obtain high-quality datasets for model training. The experiment demonstrates that if the training samples are perturbed with noise, the data distribution gradient of the SGM is corrupted. It makes the reverse process of recovering images be unstable, resulting in further compromising the reconstruction performance. To address this challenge, we proposed a general unsupervised technique by incorporating the compressed sensing knowledge into the SGM for stable training. The results demonstrates our proposed method can not only reconstruct the high-quality images without clean data training, but also greatly improve the sampling stability.

## 1 Introduction

Score-matching-based generative models (SGMs) have generated tremendous impacts on various domains from super-resolution [1], adversarial interference [2], to medical image reconstruction [3]. Especially in tomographic reconstruction, SGMs have attracted great attentions. Typical SGMs work consists of the two stages, forward noising and inverse denoising. The forward diffusion process reveals a gradient distribution from training data. With the learned gradient distribution, the reverse recovery process reconstructs high-quality images through Langevin dynamics. Although the SGM can restore high-quality images, it strongly depends on clean and diverse training samples. Unfortunately, it is difficult to obtain such high-quality training data in practice. As a result, the score-matching/diffusion model could be unstable and fail to real applications.
To address this challenge, here we propose a wavelet-based stabilization technique and incorporate sparsity into the SGM for stable training. Specifically, raw training data are separated into high and low frequency components and separately purified by deep neural networks. In the reconstruction process, the wavelet-based total variation regularization is integrated into the reconstruction model for enhanced robustness and image quality. We verified on CT and MRI experiments that the results from our method in presence of noise perturbation are similar to the counterparts from clean data.
In this paper, we propose a wavelet-based technique to stabilize the SGM. First, we construct a noise suppression module using a trainable wavelet transform. The perturbed image is divided into high-frequency and low-frequency parts to filter out noise and artifacts. Second, the wavelet total variation regularization is integrated into the reconstruction model to improve the sampling stability. When the score function is subject to erroneous data, recovered images could look blurry and unrealistic. Meanwhile, the Langevin sampling dynamics are not convergent either in low-density regions, leading to fine details missing. It is well known that the wavelet total variation regularization preserves edges and details by imposing appropriate constraints. To optimize the quality of reconstructed images, we further introduce data consistency to optimize the solution. The proposed framework is verified with the DSM (Denosing Score Matching) and SDE (Stochastic Differential Equations) (two typical score-based generative models) in CT and MRI reconstruction tasks.

## 2 Materials and Methods

### 2.1 Denoising Score Matching

The target of one typical medical imaging problem is to sample high-quality images from the posterior distribution $p(x|y)$, where $x$ and $y$ represent training and testing samples. The prior distribution $p(y|x)$ of the data is usually unknown, but one can train SGMs on the clean datasets to estimate the prior distribution [3]. The denoising score matching algorithm estimates the score function from data and generate new samples with Langevin dynamics [4].
The denoising score matching algorithm can be able to approximate the gradient distribution by training the score function $\nabla_x \log p(x)$ [5]. To obtain the approximate probability of $p(x)$, multi-level Gaussian noise is injected into the clean data for training the network [6].

### 2.2 Perspective of Stochastic Differential Equations

Score-based models can be further appreciated in the perspective of Stochastic Differential Equations (SDEs) [3]. To diffuse an original image into a prior noise distribution, we start with a continuous diffusion process $\{x(t)\}_{t=0}^{T}$, where $x(t) \in R^n$ and $t \in [0,T]$. Let $P_{data}$ and $P_T$ be the data distribution and prior distribution, then $x(0) \sim P_{data}$ and $x(T) \sim P_T$. The forward process is

$$dx = f(x,t)dt + g(t)dw \tag{1}$$

where $w$ stands for a standard N-dimensional Brownian motion, $f : R^n \mapsto R^n$ and $g : R \mapsto R$ correspond to the drift and diffusion coefficients of the diffusion process respectively. Various $f$ and $g$ give different SDE functions. Typical SDEs include Variance Exploding (VE), Variance Preserving (VP), and subVP [7], where the VE-SDE is a common strategy used for medical imaging [7] [8] [9]. $f$ and $g$ in VE-SDE are formulated as

$$f = 0, g = \sqrt{\frac{d\left[\sigma^2(t)\right]}{dt}}, \tag{2}$$

where $\sigma(t) > 0$ is a monotonically increasing function [10]. The sampling process can be treated as the inverse solution of Eq. (1), which starts with samples from $x(T) \sim p_T$. The reversing procedure is expressed as

$$\begin{aligned} dx &= \left[f(x,t) - g(t)^2 \nabla_x \log p_t(x)\right] dt + g(t) d\bar{w} \\ &= \frac{d\left[\sigma^2(t)\right]}{dt} \nabla_x \log p_t(x) + \sqrt{\frac{d\left[\sigma^2(t)\right]}{dt}} d\bar{w} \end{aligned} \tag{3}$$

where $\bar{w}$ denotes another standard Wiener process in the reverse-time direction [11], and $\nabla_x \log p_t(x)$ is referred to as the score of the distribution at time $t$.

### 2.3 Wavelet Theory based Scored Generative Models

We introduce a trained module consisting of multiple CNN blocks, as well as convolution and pooling layers. Extracting multi-scale features for noise suppression, the up-sampling and down-sampling operations are replaced with the wavelet transform. Specifically, a compromised image is divided into multiple sub-bands of different frequencies, $HH_l$, $HL_l$, $LL_l$, $and$ $LH_l$. For example, one single-level wavelet transform has four sub-band filters, $f_{HH}$, $f_{HL}$, $f_{LL}$, $and f_{LH}$, and the compromised image is down-sampled through the convolution layer $x_i = (f_i \otimes (Ax + \sigma_1)) \downarrow 2$. After DWT, CNN blocks are used to learn dependencies between the sub-bands [12]. Each layer consists of 3*3 filters Conv, BN and Relu activation functions. We choose the Haar wavelet method for three-level wavelet decomposition and the number of network convolution layers is set to 24. It is known that the global information of the image is concentrated in the low frequency region, while the edge information and noise are concentrated in the high frequency region. Our target is to remove noise in the high frequency region without significantly damaging high-frequency information. According to wavelet theory, the sub-band image can be reconstructed using the inverse wavelet transform (IWT), and edge information is thus recoverable. The noise suppression process is defined as

$$W(Ax + \delta) = Ax + \varepsilon \tag{4}$$

where $\varepsilon << \delta$ represents a much-reduced noise level. Then, the forward training process is introduced in the noise-cleaned domain so that the score estimation can be greatly facilitated; i.e., $S_\theta(x, \sigma_i + \varepsilon) \approx \nabla_x \log p_{\sigma_i+\varepsilon}(x)$. this preprocessing step greatly improves the robustness of the SGM. In the reverse process, the Langevin dynamics sampling can be difficult to converge in unknown or sparsely sampled data zones. Hence, we propose the wavelet total variation regularization to stabilize Langevin dynamics sampling. At the same time, we further add data consistency to ensure the accuracy of the image reconstruction, and adjust the regularization parameters with the J-invariant calibrator. The minimization problem is defined as

$$x^{(k+1)} = \arg\min_x [\|y - Ax\|^2 + \lambda_1 \Phi_2(\Phi_1(x))], \tag{5}$$

where $\lambda_1$ is a factor to balance data consistency, $\Phi_1$ and $\Phi_2$ represent the score-matching generative model and the compressed sensing constrain respectively.
Let us introduce a variable $u$ to convert Eq. (5) into a constrained minimization problem

$$\begin{aligned} (x^{(k+1)}, u^{(k+1)}) &= \arg\min_{x,u} [\|y - Ax\|^2 + \lambda_1 \Phi_2(u)] \\ s.t., u &= \Phi_1(x), \end{aligned} \tag{6}$$

Eq. (6) can be further expressed as an unconstrained optimization problem

$$\begin{aligned} &(x^{(k+1)}, u^{(k+1)}) = \\ &\arg\min_{x,u} [\|y - Ax\|^2 + \lambda_1 \Phi_2(u) + \tfrac{\lambda_2}{2} \|\Phi_1(x) - u\|^2] \end{aligned} \tag{7}$$

Let $v$ represents a Langevin dynamic sampling based reconstructed image $\Phi_1(x)$, Eq. (15) can be then converted into a constrained linear optimization problem:

$$\begin{aligned} &(x^{(k+1)}, u^{(k+1)}, v^{(k+1)}) = \\ &\arg\min_{x,u,v} [\|y - Ax\|^2 + \lambda_1 \Phi_2(u) + \tfrac{\lambda_2}{2} \|v - u\|^2] \\ &\qquad s.t., v = \Phi_1(x) \end{aligned} \tag{8}$$

where $\lambda_2 > 0$ is a weighting factor. The above optimization problem is further converted into the following:

$$\begin{aligned} (x^{(k+1)}, u^{(k+1)}, v^{(k+1)}) &= \arg\min_{x,u,v} [\|y - Ax\|^2 \\ &+ \lambda_1 \Phi_2(u) + \tfrac{\lambda_2}{2} \|v - u\|^2 + \tfrac{\lambda_3}{2} \|v - \Phi_1(x)\|^2] \end{aligned} \tag{9}$$

where $\lambda_3 > 0$ is a coupling factor. Eq. (17) can be optimized using an alternative iterative strategy as the three sub-problems:

$$x^{(k+1)} = \arg\min_x [\|Ax - y\|^2 + \tfrac{\lambda_3}{2} \left\|\Phi_1(x) - v^{(k)}\right\|^2] \tag{10}$$

$$v^{(k+1)} = \arg\min_v [\tfrac{\lambda_2}{2} \left\|v - u^{(k)}\right\|^2 + \tfrac{\lambda_3}{2} \left\|v - \Phi_1(x^{(k+1)})\right\|^2] \tag{11}$$

$$u^{(k+1)} = \arg\min_u [\lambda_1 \Phi_2(u) + \tfrac{\lambda_2}{2} \left\|u - v^{(k+1)}\right\|^2] \tag{12}$$

## 3 Results

### 3.1 Experimental data

In the tomographic reconstruction, the clinical data of human abdominal CT scan are provided by the AAPM challenge. In tomographic experiments, the low-dose datesets of 9 patients are used for training, and 1 patient is used for evaluation. The 5480 images with 512*512 pixels and the 1mm thickness are used to train the SGM. Poisson noise are added to simulate low-dose CT, having $b_i = 5*10^3$ Poisson disturbance for training sample and $b_i = 5*10^3$ for testing sample. We add $b_i = 5*10^4$ Poisson noise to generating training sample for 60-views sparse data. Two-dimensional fan-beam geometry is employed. The tomographic datasets are collected from 720 angles and the used detector consists of 1000 pixels with 1mm. The distances starting from x-ray source to patients and detector are setting as 500mm and 1000mm.

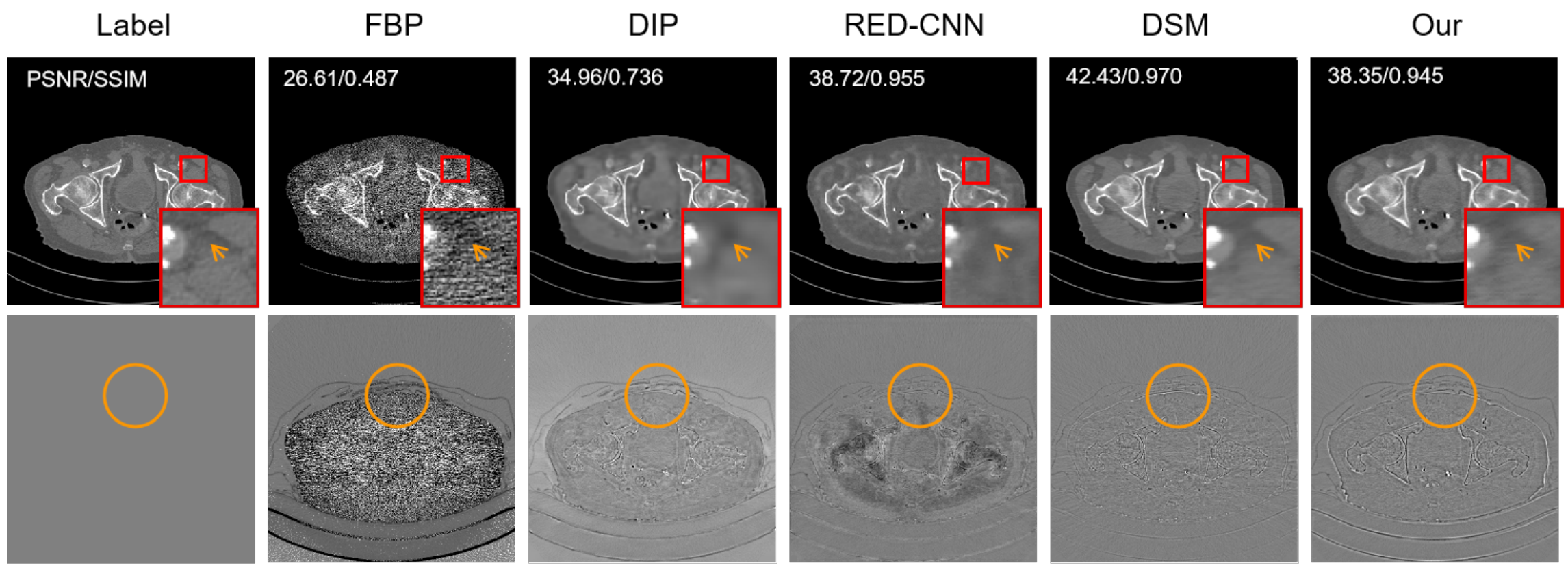

**Figure 1:** Low-dose CT reconstruction results from simulated Poisson distributed noise ($b_i = 5e3$). From left to right are ground truth, FBP, DIP, RED-CNN (supervised method), DSM, our method (SGM inbuilt).

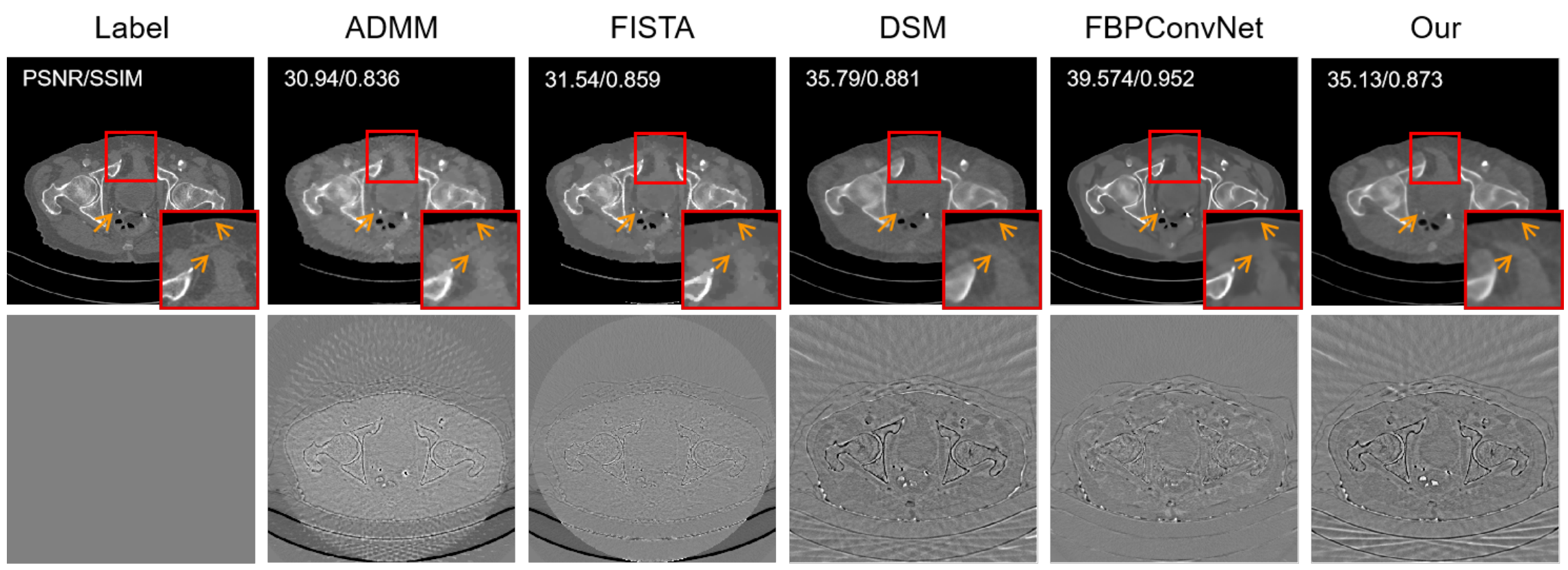

**Figure 2:** 60-views reconstruction results from AAPM low-dose CT ($b_i = 5e4$) datasets. From left to right are ground truth, ADMM, FISTA, DSM, FBPConvNet (supervised method), our method (SGM inbuilt).

In the MRI under-sampled reconstruction experiments, the fastMRI knee joint dataset is used. We train the network on real-valued part of the single-coil image with the size of 320*320 pixels. We remove the first 5 and the last 5 slices of each case. The training samples are generated by masking Gaussian 1D, 4x acceleration and center fraction 0.08. The test samples are under the Gaussian 1D masks.

### 3.2 Experiment results

In the CT denoiseing experiment, we simulated low-dose data with Poisson noise ($b_i = 5e3$) as shown in Fig. 1. Our proposed frame results is almost as great as that obtained by RED-CNN. Although there are some errors compared with the SGM in terms of detail recovery, the image quality shows the framework still achieve great reconstruction without clean image. In particular, the boundary and details of organs of our proposed frame demonstrate it has great reconstruction performance.

In the sparse-view experiment, we sampled 60 angles at equal intervals and the test samples results are shown in Fig. 2. Our method not only restores image details better than the traditional iterative method (ADMM and FISTA), but also has higher structural fidelity. Indicator performance shows that the proposed frame outperforms the baseline iterative method, and the image artifacts are almost invisible.

In the MR reconstruction, we follow the method [13] to simulate under-sampled MRI, as shown in Fig. 3. Limited by the simple network structure, the reconstruction effect of the supervised method CascadeNet and UPDNet are slightly lower than that of SDE. Our method has certain advantages in the suppression of motion artifacts. In terms of performance indicators, our method results is close to those obtained by the Score SDE.

## 4 Discussion

In this work, we propose a wavelet-based technique to stabilize SGMs, and our approach has yielded encouraging results in computed tomography (CT) and magnetic resonance imag-ing (MRI) reconstruction tasks. It is worth noting that our approach addresses the inverse problem of no clean labels in medical imaging in an unsupervised mode, which is of clinical significance. We present a series of classical medical image processing experiments – CT denoising, sparse view CT reconstruction and under-sampled MRI reconstruction, and show that our method can achieve great results without relying on clean images. Theoretical derivation also proves that our method can effectively manage the error of data gradient in training and is beneficial to data reconstruction in the sampling process.

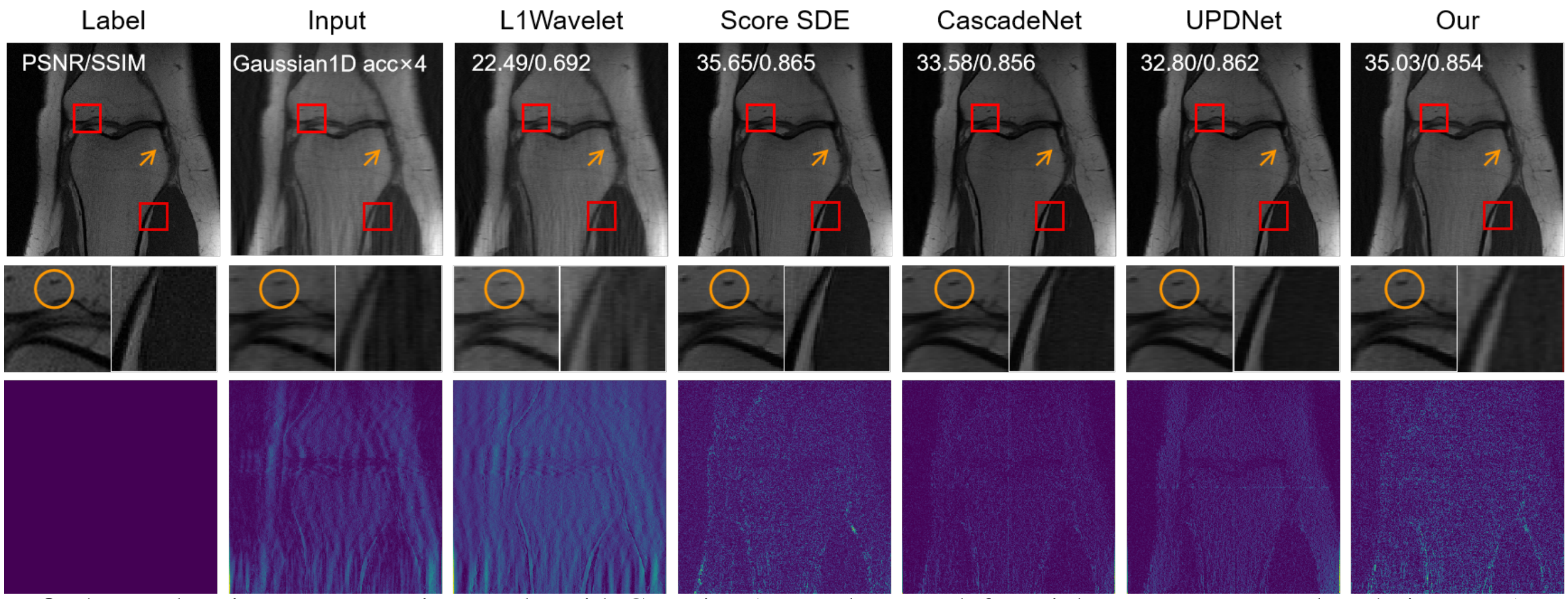

**Figure 3:** 4x acceleration reconstruction results with Gaussian 1D mask. From left to right represent ground truth, input, L1Wavelet, score SDE, CascadeNet (supervised method), UPDNet, our method (SDE inbuilt).

Although our method produced exciting reconstruction results, the results were still not as great as the results of training in clean labels. Specifically, first, SGMs are the main carrier of the algorithm, and the proposed method has inherent randomness. This method can not completely eliminate the noise interference of training samples, preserved artifacts and detail fuzzy results. We note that our approach inherits the limitations of the generated model. Unable to generate fine texture structure, image edge information distortion. Second, we train SGMs without using clean labels, and the data distribution learned by the scoring function is inaccurate.

## 5 Conclusion

In conclusion, we propose a novel approach to the unsupervised training the SGM. The implementation results show that our method can improve the robustness of the model and further optimize the sampling process. Our method opens up a new way to solve the imaging inverse problem and has a certain reference value for engineering applications.

In the future, we will actively address the limitations of the SGM. We will further reduce the error of data gradient in training, and optimize the sampling process to accelerate the convergence of Langevin dynamics. In the follow-up work, we hope that the reconstruction results can reduce the distortion of the edge structure and restore a finer texture.

# Prior information enhanced adversarial learning for kVp switching CT

Yizhong Wang[1], AiLong Cai[1], Ningning Liang[1], Shaoyu Wang[1], Junru Ren[1], Xinrui Zhang[1], Lei Li[1] and Bin Yan[1]

[1] Department of Henan Key Laboratory of Imaging and Intelligent Processing, PLA Strategic Support Force Information Engineering University, Zhengzhou, China

**Abstract** Dual energy computed tomography (DECT) can provide both structural and material information of the scanned object, and has been widely used in the medical field. However, patients may suffer from genetic damage and cancer under long-term high radiation dose of x-ray exposure. To reduce radiation dose and ensure optimal hardware cost. This work studies the switching technology based on the x-ray tube voltage (kVp). However, the kVp switching technology faces the problems of low sampling rate of each energy spectrum and the spatial misalignment of projection data of different energy spectrum. Thus, this study introduces an adversarial learning mechanism and proposes a Prior Information enhanced Projection data Inpainting Network (PINet). The experimental results show that the PINet framework is a promising approach for sparse-view angle DECT imaging.

## 1 Introduction

Different from traditional computed tomography (CT), dual energy computed tomography (DECT) can simultaneously provide the structural information and material information of the scanned object by obtaining the attenuation measurement of two different x-ray energy spectrum [1]. At present, DECT has been widely used in clinical diagnosis, such as virtual monoenergetic imaging [2],perfused blood volume imaging [3], and aortic disease diagnosis [4].

Increasing the radiation dose of x-ray is known to improve the quality of medical images. However, patients are likely to suffer from genetic damage and cancer under long-term high radiation dose of x-ray exposure. Therefore, lowering the radiation dose is also the focus of the medical imaging community. In order to reduce radiation dose and ensure optimal hardware cost. This paper study the switching technology based on x-ray tube voltage (kVp) as shown in Fig. 1. This technology only requires traditional energy integration detector and ray source, and the dose is about half of that of traditional DECT.

However, technology based on kVp switching not only faces the problem of low sampling rate of each energy spectrum, but also a common problem is that the projection of different energy spectrum is not aligned in space. Recently, deep learning has shown great potential in the field of medical image processing. Lee et al. proposed an interpolation method based on convolutional neural networks (CNN) to inpainting missing projection [5]. In 2022, Cao et al. developed a CNN framework for sparse-view projection completion and material decomposition [6]. Generative Adversarial Networks (GAN) also have great potential in the application of DECT. Kawahara et al. proposed an image synthesis framework based on GAN to material decomposition images of bone and fat scanned by DECT [7]. In 2022, Wang et al. designed a dual-way mapping GAN to mine the relationship between two different energy projection data, aiming at recovering the missing data[8].

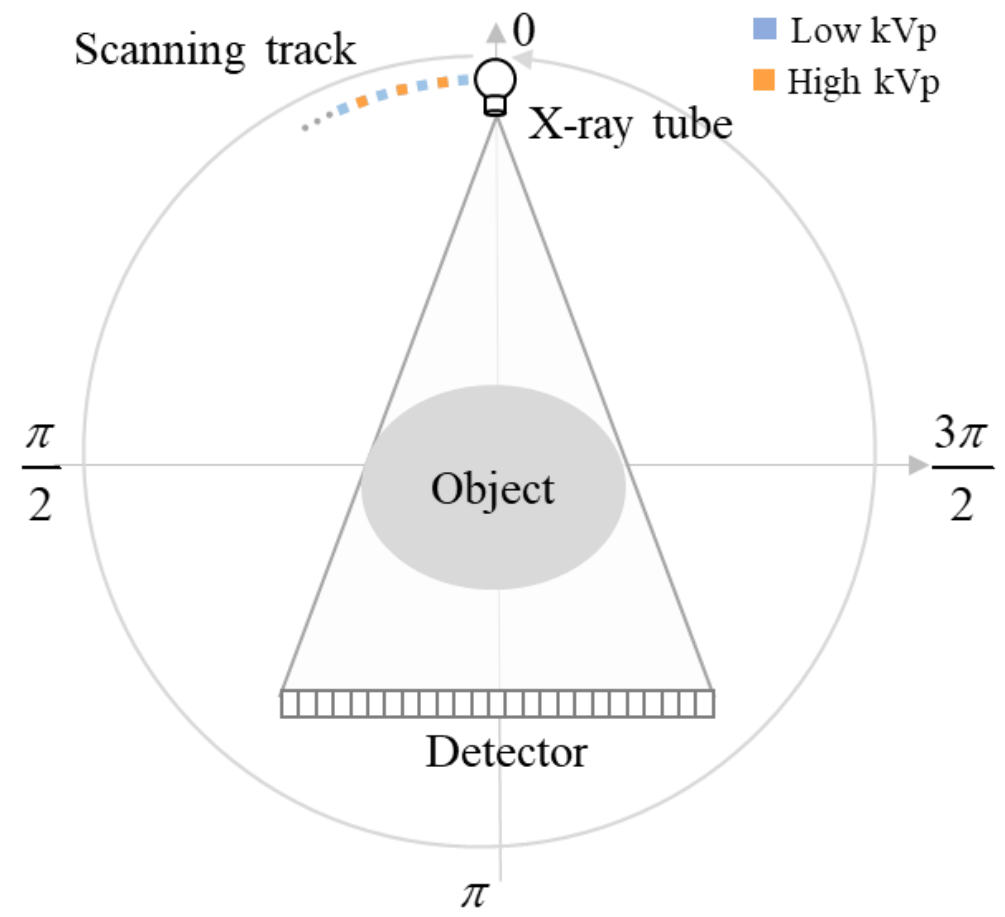

**Figure 1:** An illustration of the kVp switching technology.

There is a certain correlation between low-energy and high-energy projection data of the same object for DECT. When DECT imaging is faced with a serious shortage of projection data, researchers usually use the correlation between energy spectrum data to achieve DECT imaging [9]. Therefore, this study fuses dual energy projection data as prior information, and proposes a Prior Information enhanced Projection data Inpainting Network (PINet) for sparse-view angle DECT. In the PINet framework, this work introduces an adversarial learning mechanism to generate results close to the real projection data. In order to make full use of the prior information and original information, the prior information and sparse-view dual energy projection data are sent to two separate encoders to extract and fuse useful features. Then, the two decoders perform differential learning on the projection data in different energy channels.

## 2 Methods

The PINet framework is shown in Fig. 2. First, the kVp switching technology is used to obtain two sparse-view projection data with different energy spectrum. Then, the generator *G* takes prior information and sparse-view projection data as input, extracts and uses various features to generate low-energy and high-energy full angle (360°) projection data. The prior information is the data after fusing the sparse-view projection data of two kinds of energy. Due to the different scanning angles of different energies, the data of prior information is twice as large as that of single energy sparse-view data. At the same time, the discriminator $D$ encourages the generator $G$ to generate realistic results as much as possible. Once the PINet training is completed, the trained generator $G$ can be used to generate the completed projection data.

$$G^{*}\left(\begin{pmatrix} x_{\mathrm{L}}^{\mathrm{Sparse}} \\ x_{\mathrm{H}}^{\mathrm{Sparse}} \end{pmatrix}, x_{\mathrm{Prior}}^{\mathrm{Fusion}}\right) = \begin{pmatrix} x_{\mathrm{L}}^{\mathrm{Pred}} \\ x_{\mathrm{H}}^{\mathrm{Pred}} \end{pmatrix}, \qquad (1)$$

where $G^{*}$ denotes the trained generator. $\begin{pmatrix} x_{\mathrm{L}}^{\mathrm{Sparse}} \\ x_{\mathrm{H}}^{\mathrm{Sparse}} \end{pmatrix}$ and $x_{\mathrm{Prior}}^{\mathrm{Fusion}}$ represent sparse-view dual energy projection data and prior information respectively, which are input into the generator $G$. $\begin{pmatrix} x_{\mathrm{L}}^{\mathrm{pred}} \\ x_{\mathrm{H}}^{\mathrm{pred}} \end{pmatrix}$ represents the dual energy full angle projection data output by generator $G$. During training, the objective of PINet can be expressed as:

$$\begin{aligned} \mathcal{L}_{\mathrm{PINet}} = & \ \mathbb{E}_{\left(\begin{pmatrix} x_{\mathrm{L}}^{\mathrm{Ref}} \\ x_{\mathrm{H}}^{\mathrm{Ref}} \end{pmatrix}\right)}\left[\log D\left(\begin{pmatrix} x_{\mathrm{L}}^{\mathrm{Ref}} \\ x_{\mathrm{H}}^{\mathrm{Ref}} \end{pmatrix}\right)\right] \\ & + \mathbb{E}_{\left(\begin{pmatrix} x_{\mathrm{L}}^{\mathrm{Pred}} \\ x_{\mathrm{H}}^{\mathrm{Pred}} \end{pmatrix}\right)}\left[\log\left(1 - D\left(\begin{pmatrix} x_{\mathrm{L}}^{\mathrm{Pred}} \\ x_{\mathrm{H}}^{\mathrm{Pred}} \end{pmatrix}\right)\right)\right] \qquad (2) \\ & + \lambda \mathbb{E}_{\left(\begin{pmatrix} x_{\mathrm{L}}^{\mathrm{Ref}} \\ x_{\mathrm{H}}^{\mathrm{Ref}} \end{pmatrix}, \begin{pmatrix} x_{\mathrm{L}}^{\mathrm{Pred}} \\ x_{\mathrm{H}}^{\mathrm{Pred}} \end{pmatrix}\right)}\left[\left\| \begin{pmatrix} x_{\mathrm{L}}^{\mathrm{Ref}} \\ x_{\mathrm{H}}^{\mathrm{Ref}} \end{pmatrix} - \begin{pmatrix} x_{\mathrm{L}}^{\mathrm{Pred}} \\ x_{\mathrm{H}}^{\mathrm{Pred}} \end{pmatrix} \right\|_{1}\right] \end{aligned}$$

where $\lambda$ is the weight parameter. $\begin{pmatrix} x_{\mathrm{L}}^{\mathrm{Ref}} \\ x_{\mathrm{H}}^{\mathrm{Ref}} \end{pmatrix}$ is the label projection data of full angle. In the objective function, the mean absolute error (MAE) between the generated projection data and the label projection data is introduced to generate more realistic projection data. During network training, $G$ tries to minimize the objective function, while $D$ tries to maximize the objective function, i.e.,

$$G^{*}, D^{*} = \arg\min_{G}\max_{D} \mathcal{L}_{\mathrm{PINet}} \qquad (3)$$

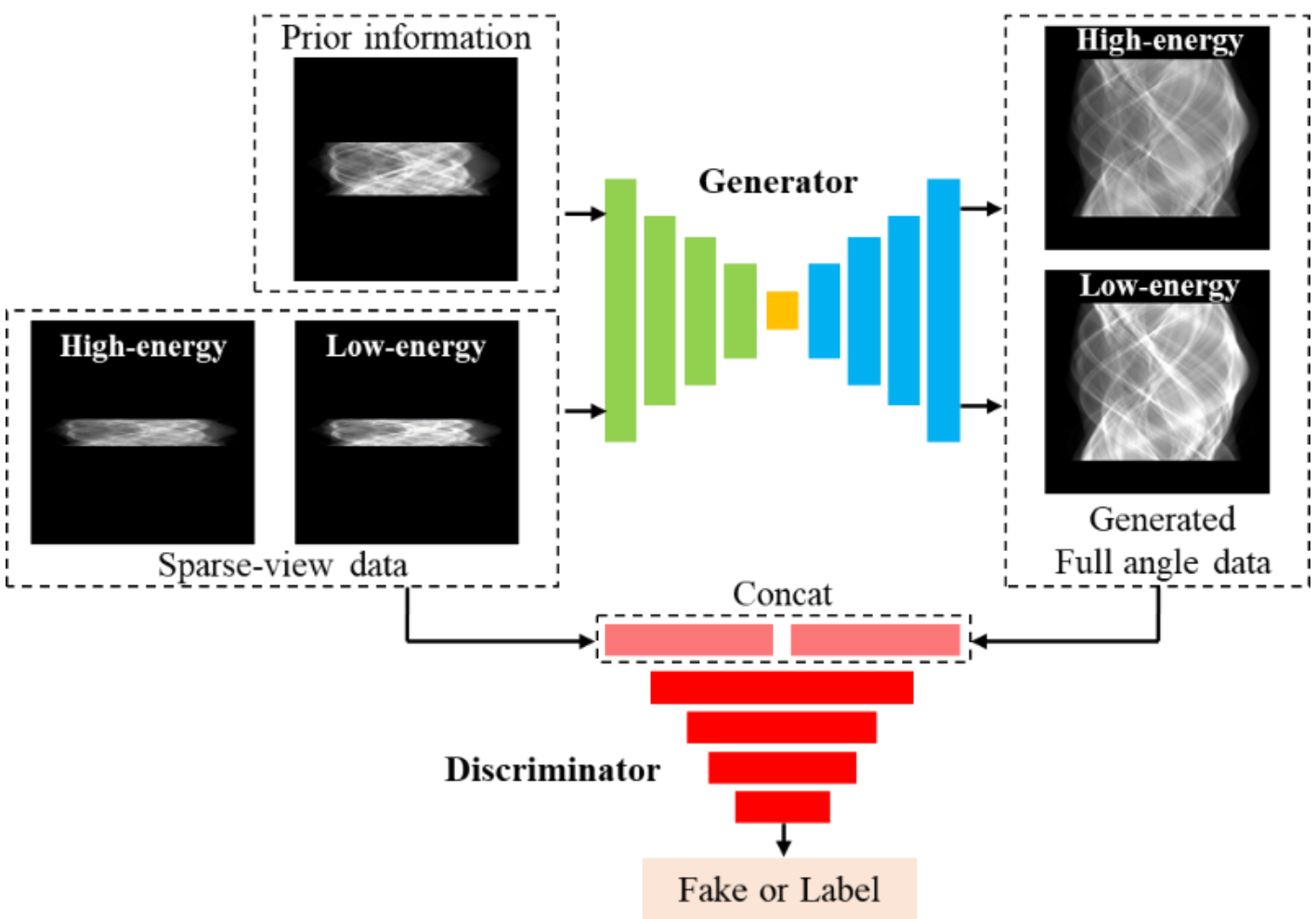

**Figure 2:** Schematic of the PINet framework.

The PINet consists of a generator $G$ and a discriminator $D$. The generator $G$ is improved from the classic U-Net [10] as shown in Fig.3. To make full use of prior information and match two energy channels, the structure of $G$ is extended to a dual input and dual output network structure. The network structure consists of three parts: encoding module, fusion module and decoding module. The encoding module contains two encoding channels, which process prior information and sparse-view projection data respectively. The sparse-view projection data channel is used as the main channel, and the initial number of channels is set to 32. The prior information channel is used as an auxiliary component, and the initial channel number is set to 8. Then, the fusion module aims to achieve the fusion of the features extracted from the two encoding channels. Considering that the difference between different energy projection data is the key to material identification, the decoding module uses two decoding channels to process the fused feature information to generate DECT data. During decoding, the shallow extracted features of the encoding module will be copied and connected to the low-energy and high-energy decoding channels.

The structure of the discriminator $D$ is a CNN, and its input is paired sparse-view angle projection data and full angle projection data (generated or label). The discriminator $D$ has five layers. The first layers contain convolution, batch norm (BN) and Rectified Linear Unit (ReLU) operations. The size of the convolution kernel is $3\times3$, the stride is 2, and the number of channels is 32. The second and third layers contain $3\times3$ convolution with stride 2, $3\times3$ convolution with stride 1, BN and ReLU operations, and the number of channels is 64 and 128 respectively. The last layer includes global average pooling, full connection and sigmoid operations The output of the discriminator $D$ is true or false to match the projection data pair, which is equivalent to 0-1 classification.

## 3 Experimental Results

The experimental dataset was established from real clinical dataset, and the DECT images were obtained using the SOMATOM Definition Flash DECT scanner (Siemens Healthcare, Germany). 80 kVp and 140 kVp spectra are used for low-energy and high-energy scanning, respectively. The dataset includes 1491 cranial cavity images of 6 patients. The size of the image is $512\times512$. The projection dataset used to train PINet is generated using 1000 images of 5 patients, while 100 images of another patient are used as the test dataset of the network.

In this paper, Sidden's ray tracing algorithm [11] is used to simulate the geometry of the fan beam. The distances from the x-ray source to the object and the detector are set to 1000 mm and 1500 mm respectively. Both low-energy and high-energy projections collect 360 frames of projection data within the 360 degree scanning range. The projection data of each frame is collected by a linear detector, which consists of 512 bins. The size of the generated full angle projection data is 360 ×512, and then two 76 ×512 size all zero matrices are added to the generated data. Finally, the size of full angle projection data is 512 × 512 obtained as the label of PINet training. In this work, low-energy and high-energy sparse-view projection data are obtained by kVp switching in every 3° rotation range. The size of each energy projection data is 60 × 512. The size of prior projection data is 120 ×512. Then, using operations similar to label data generation, the size of sparse projection data and prior projection data are 512 × 512 input into PINet. Peak Signal to Noise Ratio (PSNR), Root Mean Square Error (RMSE) and Structure Similarity Index (SSIM) are used to evaluate the reconstructed images. In order to evaluate the performance of the proposed method, it is compared with SC-CNN [6].

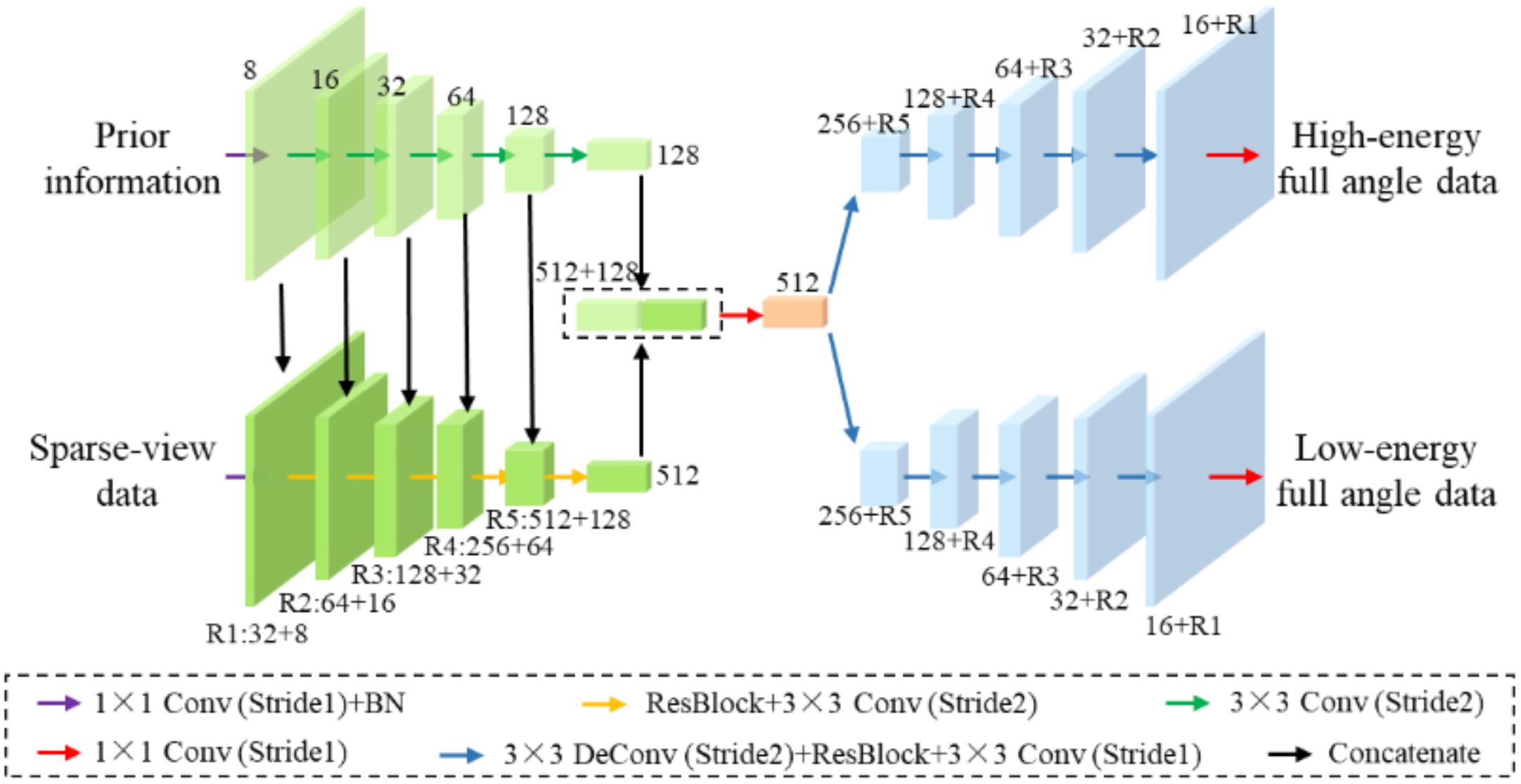

**Figure 3:** Network structure for the generator $G$.

Fig. 4 shows the results of projection data inpainting and images reconstruction by the PINet method and comparison method under sparse-view angle scanning. Then, we use the full angle projection data generated by the network to reconstruct the CT image. It can be observed that the image reconstructed by SC-CNN still has obvious artifacts, and the proposed method can effectively reduce the serious artifacts caused by the missing projection data. Furthermore, this study also compares PSNR, RMSE and SSIM of different methods as shown in Tabel 1. Compared with SC-CNN method, the PSNR of high-energy and low-energy images obtained by the proposed method is improved by 1.5413 dB and 1.2953 dB, and SSIM of proposed methods also has significant advantages. The RMSE of SC-CNN is higher than 0.0183, while the RMSE of proposed methods is lower than 0.0173. Numerical results show that the proposed method has some advantages in noise suppression and structure preservation.

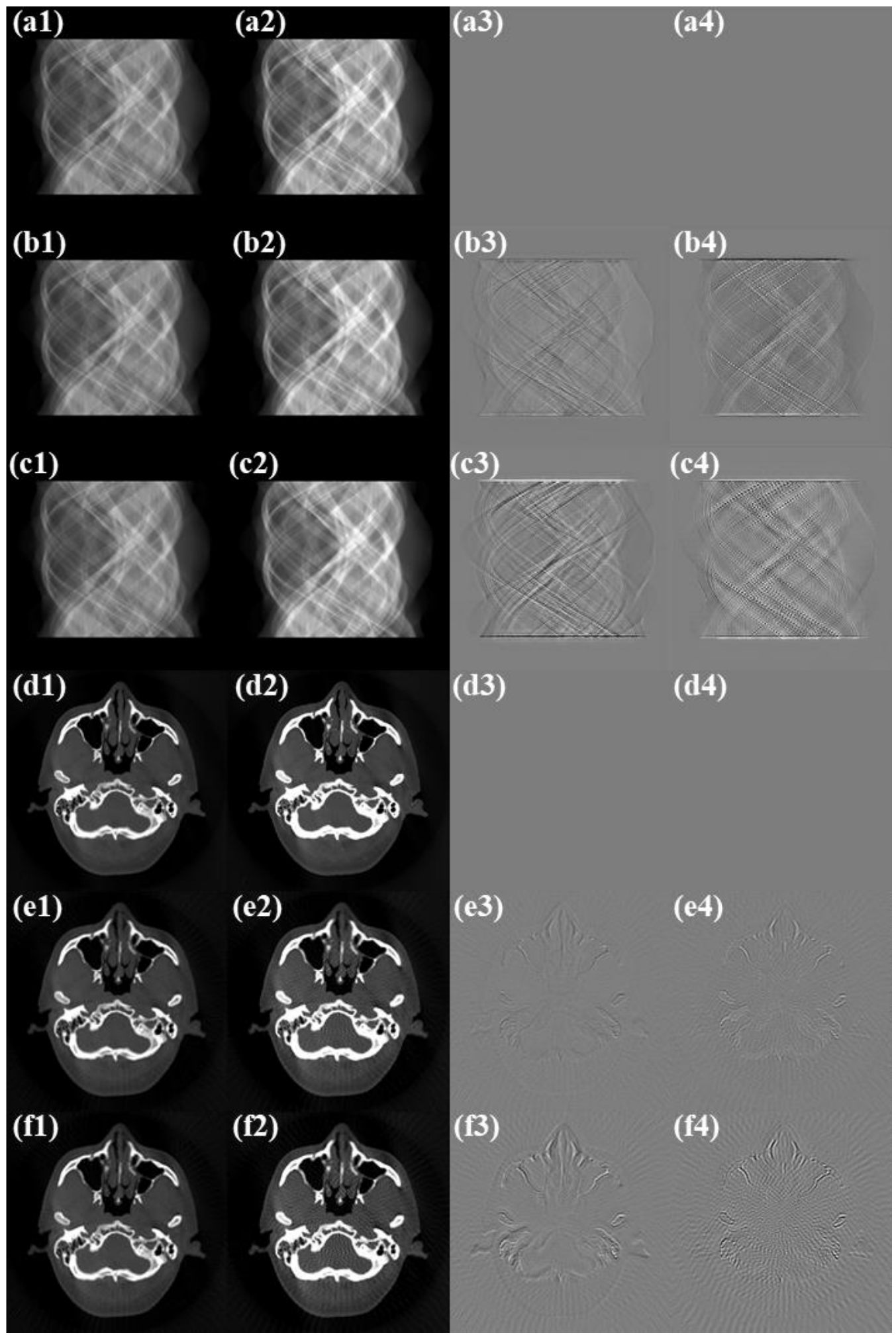

**Figure 4:** Results of inpainting projection data and reconstructed images from sparse-view angle scanning. (a) and (d) represent label projection data and reconstructed images. (b) and (e) represent projection data generated by PINet and reconstructed images. (c) and (f) represent projection data generated by SC-CNN and reconstructed images. (1) and (2) represent high-energy and low-energy, (3) and (4) represent corresponding error maps. Display windows of projection data, reconstructed images, error maps of projection data and error maps of reconstructed images are [0, 2.5], [0, 0.04], [-0.1, 0.1] and [-0.02, 0.02], respectively.

| | avg. PSNR | avg. RMSE | avg. SSIM |
|---|---|---|---|
| PINet(H) | **36.2473** | **0.0154** | **0.9030** |
| PINet(L) | **35.2881** | **0.0172** | **0.8890** |
| SC-CNN(H) | 34.7060 | 0.0184 | 0.8629 |
| SC-CNN(L) | 33.9928 | 0.0200 | 0.8535 |

**Tabel 1:** Quantitative results (PSNR: Peak Signal to Noise Ratio; RMSE: Root Mean Square Error; SSIM: Structure Similarity Index). Averaged over 100 test samples.

In order to verify the performance of the proposed method, the reconstructed images of PINet and SC-CNN are further decomposed to obtain the decomposition results of tissues and bone materials, as shown in Fig. 5. It can be seen that the basis material decomposed by PINet from the reconstructed image is closer to the ground truth, and the decomposition accuracy is higher.

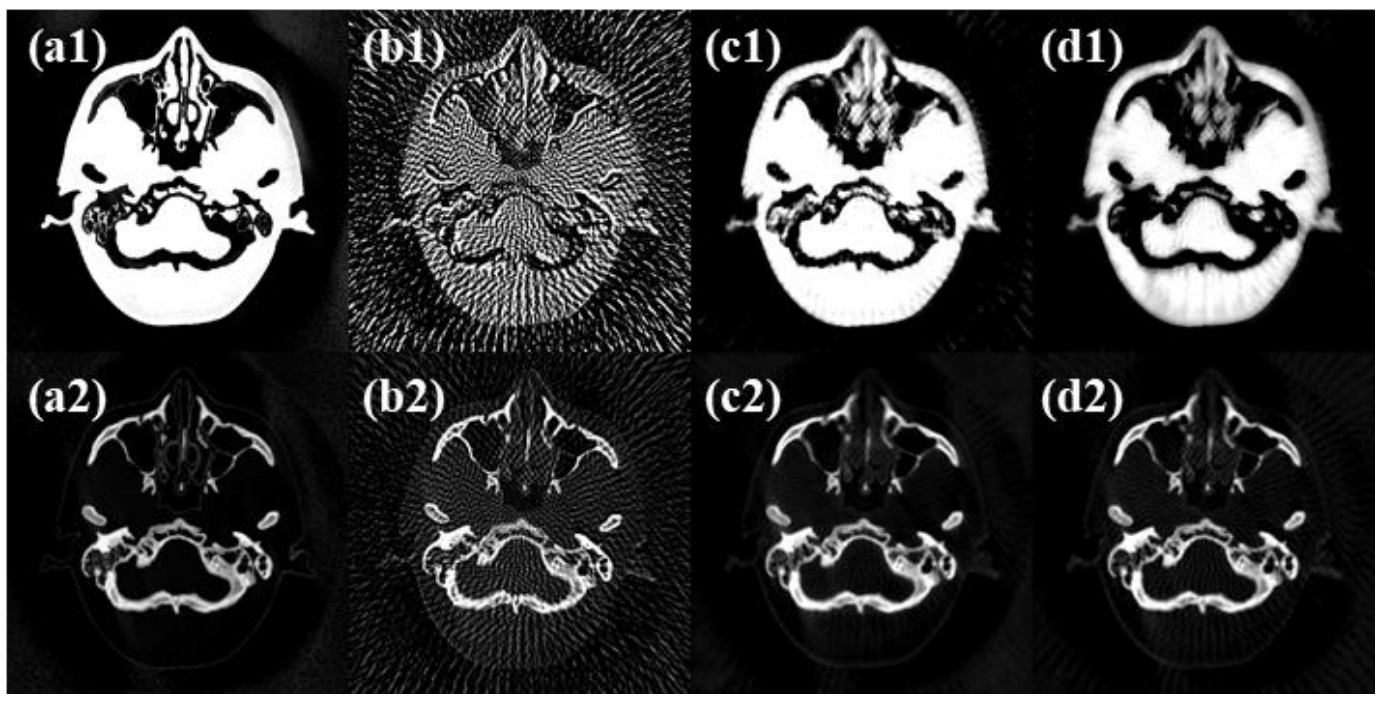

**Figure 5:** Decomposition results of different methods. (a), (b), (c) and (d) represent ground truth, raw sparse data, PINet and SC-CNN decomposition results, respectively. (1) and (2) represent tissue and bone materials. All display windows are [0, 1].

## 4 Discussion and Conclusion

DECT has great potential in medical field. To reduce the radiation dose and ensure the best hardware cost, this paper studies the kVp switching technology. In addition, aiming at the problem of missing projection data faced by this technology, this study uses the correlation between the projection data to fuse the sparse-view projection data under two different energies, and introduces the adversarial learning mechanism to propose a PINet framework with prior information. In the clinical data experiment, the feasibility of the proposed method is demonstrated. In the future work, we will extend this work to the application of spectral CT, and design corresponding image post-processing module and material decomposition module based on deep learning to achieve high-resolution spectral CT imaging.

## Acknowledgements

This work was supported by the National Natural Science Foundation of China (Grant No. 62101596), the National Key Research and Development Project of China (Grant No. 2020YFC1522002), the National Natural Science Foundation of China (Grant No. 62201616) and the China Postdoctoral Science Foundation (Grant No. 2019M663996).

# A metal artifacts reducing method using intra-oral scan data for dental cone-beam CT[1]

Yuyang Wang[1, 2], Liang Li[1,2,3]

[1] Department of Engineering Physics, Tsinghua University, Beijing, China, 100084
[2] Key Laboratory of Particle and Radiation imaging (Tsinghua University), Ministry of Education, Beijing, China, 100084
[3] Institute for Precision Medicine, Tsinghua University, Beijing, China, 100084

**Abstract** In cone-beam computed tomography (CBCT), metal implants produce severe artifacts during imaging, causing serious damage to the clinical structure and information of the teeth, reducing imaging quality, and ultimately affecting subsequent clinical treatment. However, currently, although there are many metal artifact reduction (MAR) methods, they still lack in preserving tooth structure. The proposed MAR method combines CBCT data and intra-oral scan data to better guide the segmentation of metal areas and comprehensively use projection domain and image domain data to remove metal artifacts. The experiment results are presented to demonstrate the feasibility of the proposed approach. We propose a novel MAR method that uses intraoral scan data for the first time in the analysis and processing of projection domain data, thus accurately reconstructing 3D dental images.

## 1 Introduction

Cone beam computed tomography (CBCT) has advantages of lower radiation dose, higher spatial resolution and 3D visualization images, which provide reliable imaging data for oral physicians[1, 2]. However, with the widespread use of restorative and implant materials, some oral metal materials may produce metal artifacts on CBCT images, which may affect preoperative assessment and disease diagnosis by oral physicians. Metal artifact reduction (MAR) is a challenging task as the creation of metal-induced streaks and shadowing is complexly linked to the interactions among metal, bones and tissue, with several factors such as beam hardening, scatter, non-linear partial volume effects, photon starvation and highly non-uniform attenuation[3, 4]. In order to reduce the impact of metal artifacts on image quality, scholars have been researching and exploring methods to reduce metal artifacts from various aspects, including metal artifact correction using iterative algorithms[5-7], inpainting-based correction in the projection domain[8, 9], and deep learning methods[10-13]. The above methods only used sinogram data containing metal implants, and the data itself has been severely damaged. Improving the damaged data details is a difficult task, therefore, the performance of the above methods in restoring tooth shape and structure is still unsatisfactory and has limitations in clinical application. Recently, a new method for MAR has been proposed which is based on deep learning networks[14]. In the image domain, image-enhancing network and $\alpha$ -shape-based weighted thresholding operation are used in the method to combine radiation-free intra-oral scan data with reconstruction images to extract tooth shape and provide high-quality shape information to compensate for missing or severely uncertain tooth structures, and display the parts of the tooth. Due to the addition of extra information, this method can better correct for metal artifacts.

In this study, we propose another method for metal artifact correction that also combines intra-oral scan data with reconstruction images, but differs from the above method in that we register the intra-oral scan data with the reconstructed image containing artifacts, and find the location of the metal parts in the projection data through information such as shape and threshold. This location's projection data is interpolated and replaced, and the reconstruction is done by combining the data of the original projection and the interpolated projection, and synthesizing the reconstruction images. We conduct clinical experiment to study the potential impact of the proposed method on MAR. The results of the experiment demonstrate the feasibility of the proposed method and show the benefits of using intra-oral scan data in the projection domain.

## 2 Materials and Methods

In dental CBCT, the measured sinogram data $y$ can be expressed as

$$y = T\left(-\ln\int_E \eta(E)\exp(Ax)\,dE + b\right)$$

Here, $x$ is the attenuation coefficient distribution of a patient's bone-teeth-jaw to be scanned at energy $E$, which is also the result we are looking for, $\eta$ is the energy distribution of the radiation source, $A$ is the system matrix, $b$ is the noise, and $T(\,)$ is a function of truncation caused by the size and geometric position of the detector.

The goal of the proposed method is to accurately locate metal positions by introducing additional intraoral scan information, and to repair metal-affected projection data in the projection domain, ultimately reconstructing a high-quality 3D bone-teeth-jaw model without metal artifacts.

The proposed method is based on the registration of the intra-oral scan data and the metal-affected reconstruction image, projection domain metal positioning, projection data processing and reconstruction images synthesis, as shown in Fig.1.

[1] Supported partially by the grants from Beijing Natural Science Foundation (L222001) and Tsinghua University Initiative Scientific Research Program of Precision Medicine

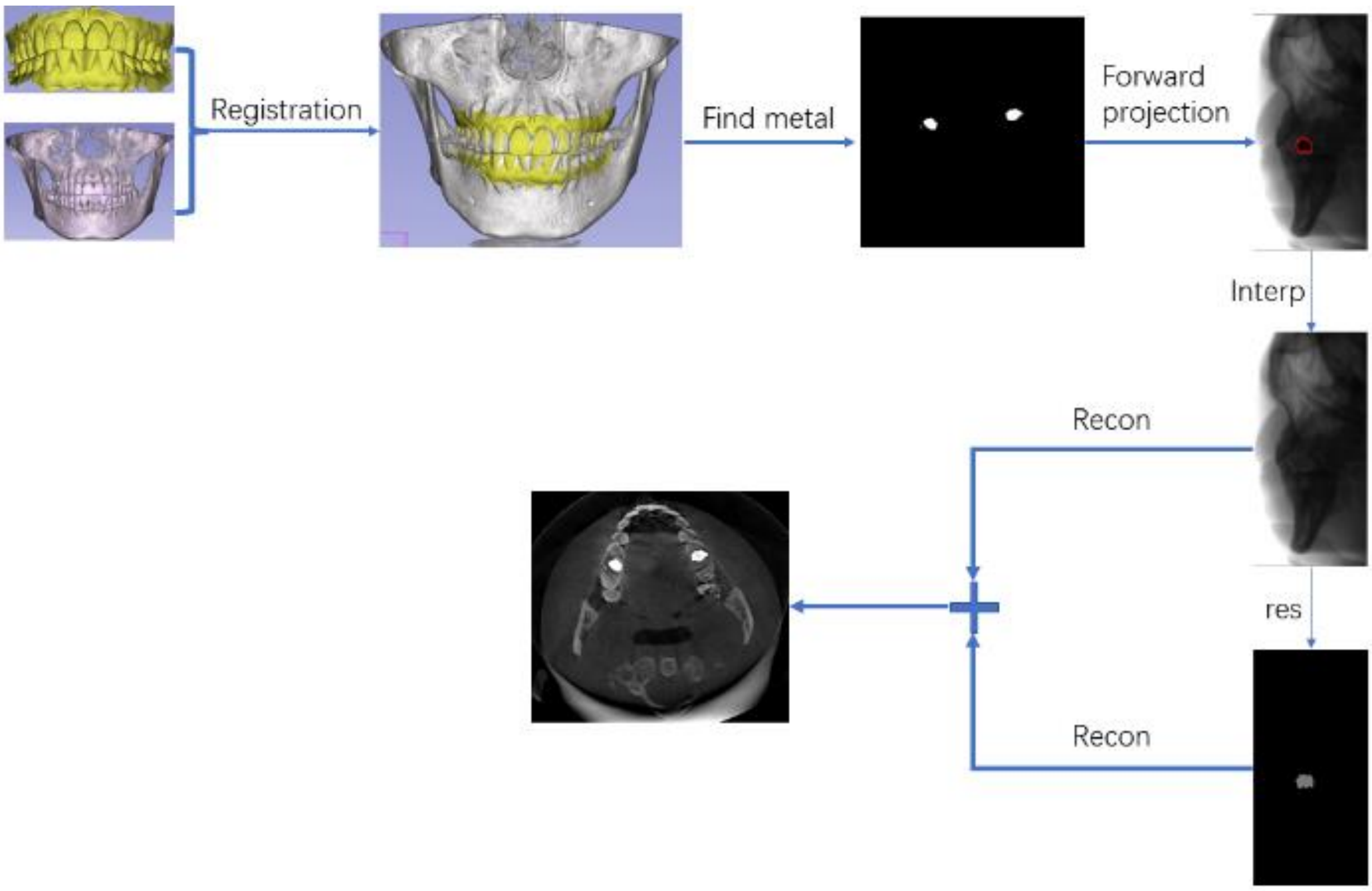

Fig.1 The proposed method

In the registration stage, we plan to use the neural network method to register the intra-oral scan data which is a point cloud image with the metal-affected CBCT reconstruction image. The so-called registration is a map of geometric position as $f_{reg}: x_{scan} \rightarrow f(x_{scan}) \approx x_{CBCT}$. Here, $x_{scan}$ is the coordinates of a point in the intra-oral scan data, $x_{CBCT}$ is the coordinates of a voxel in the reconstruction grid. At present, there have been studies on multimodal registration of CBCT[15]. However, due to the lack of clinical data, in fact only 1, we have obtained, the current research uses the manual registration method, i.e. manually find several registration points to calculate the map $f_{reg}$.

In the projection domain metal positioning stage, we first find the accurate position of the metal implant in the image domain, then project the position of the metal implant to the projection domain, thus finding the data position of the metal implant in the projection domain. We can first depict the contour of the tooth with the metal implant using the registered point cloud data. However, sometimes the implant is inside the tooth, such as root canal therapy. In this case, the tooth contour data obtained from the intra-oral scan cannot give us an accurate position of the implant. We had originally planned to use neural network methods to find the specific location of metal artifacts in the image domain by combining reconstructed images with intraoral scan data, but due to the limitation of the number of clinical data, we can only use the characteristic of the metal implant with a higher CT value to determine the position of the metal implant, so we manually adjust the CT threshold in this experiment. Then, we can obtain the implant data position in the projection domain through the projection matrix.

In the projection data processing stage, we first interpolate the data where the metal positions are located in the original projection data $y$, and replace the original metal position data with the interpolation results to obtain the interpolated projection data $y'$. After that, we can obtain the difference projection data $\Delta y = y - y'$. Since the projection data of the metal location is significantly higher than the data around the metal location, the difference projection data $\Delta y$ is non-negative. Then, We perform reconstruction of the difference projection data and interpolation projection data separately using SIRT, resulting in $x_{diff}$ and $x_{inter}$. Here, when reconstructing the difference projection data, we restrict the reconstruction area to the metal area.

In the images synthesis stage, we get $x = x_{diff} + x_{inter}$ as the final result.

## 3 Results

The projection data of a real patient were obtained from a commercial CBCT machine, while the intra-oral scan data obtained from a structrued light scanning machine. The projection data size is 645*628*546, where 645 is the number of sampled projection views, the first 640 of 645 is the number of views in $[0,2\pi)$, and 628*546 with real scal of 0.119 mm for each axis is the number of samples measured by the 2D flat detector for each projection view. The reconstruction imags were reconstructed in a voxel size of 960*960*630 with a real scale of 0.25 mm. For cone

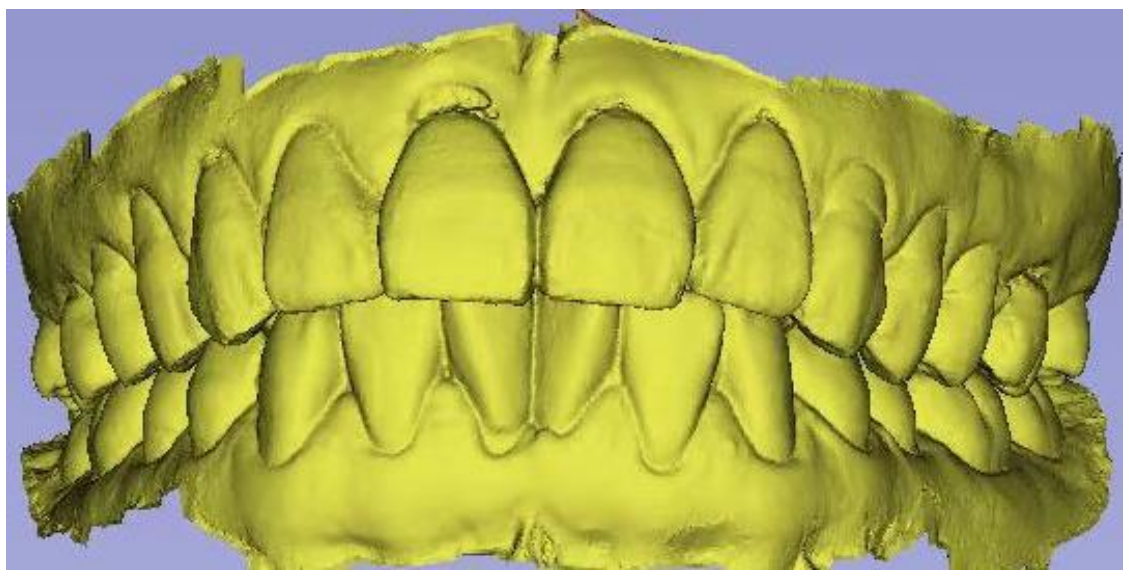

Fig.2 The intra-oral scan data

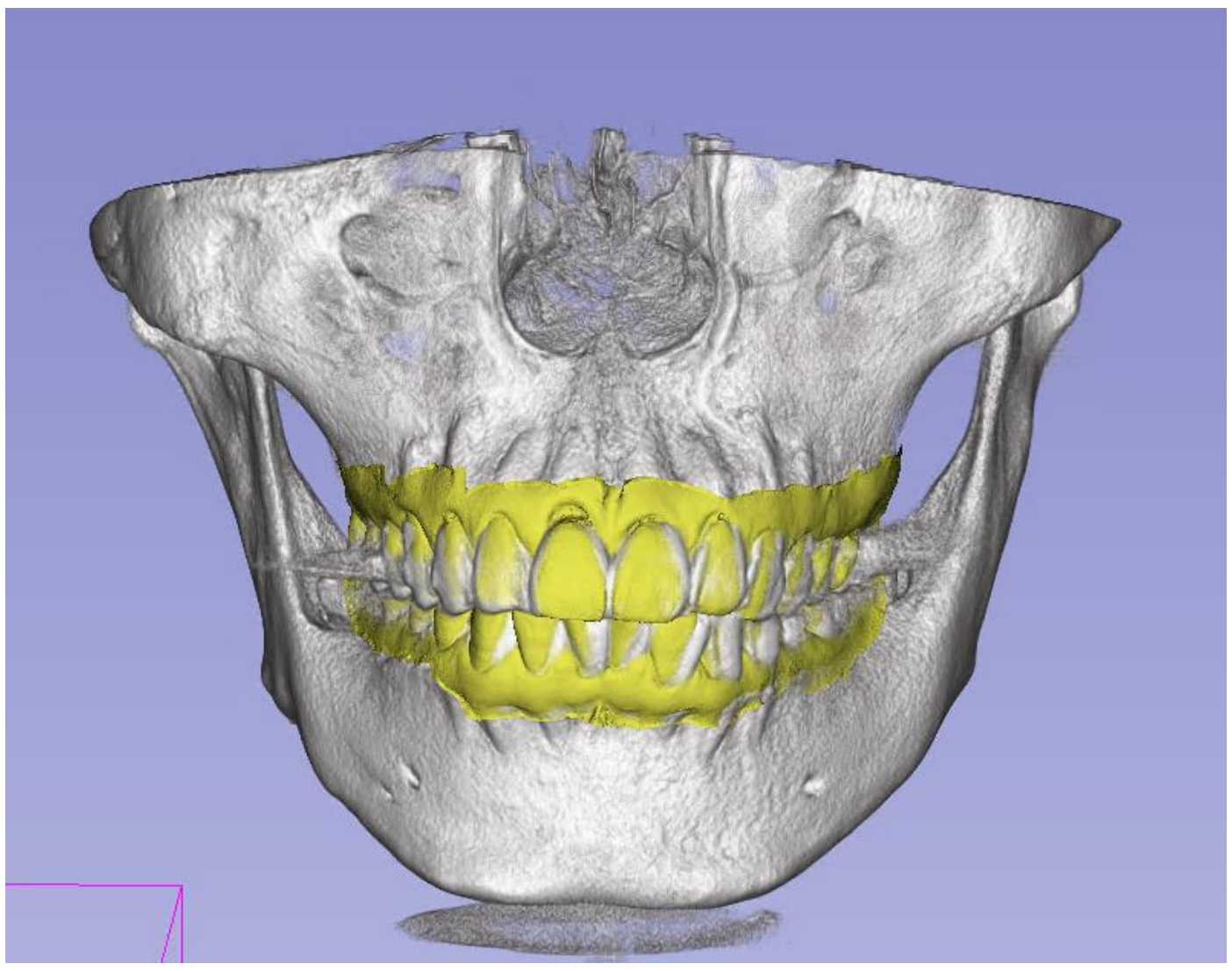

Fig.3 The result of registration of CBCT and intra-oral scan data

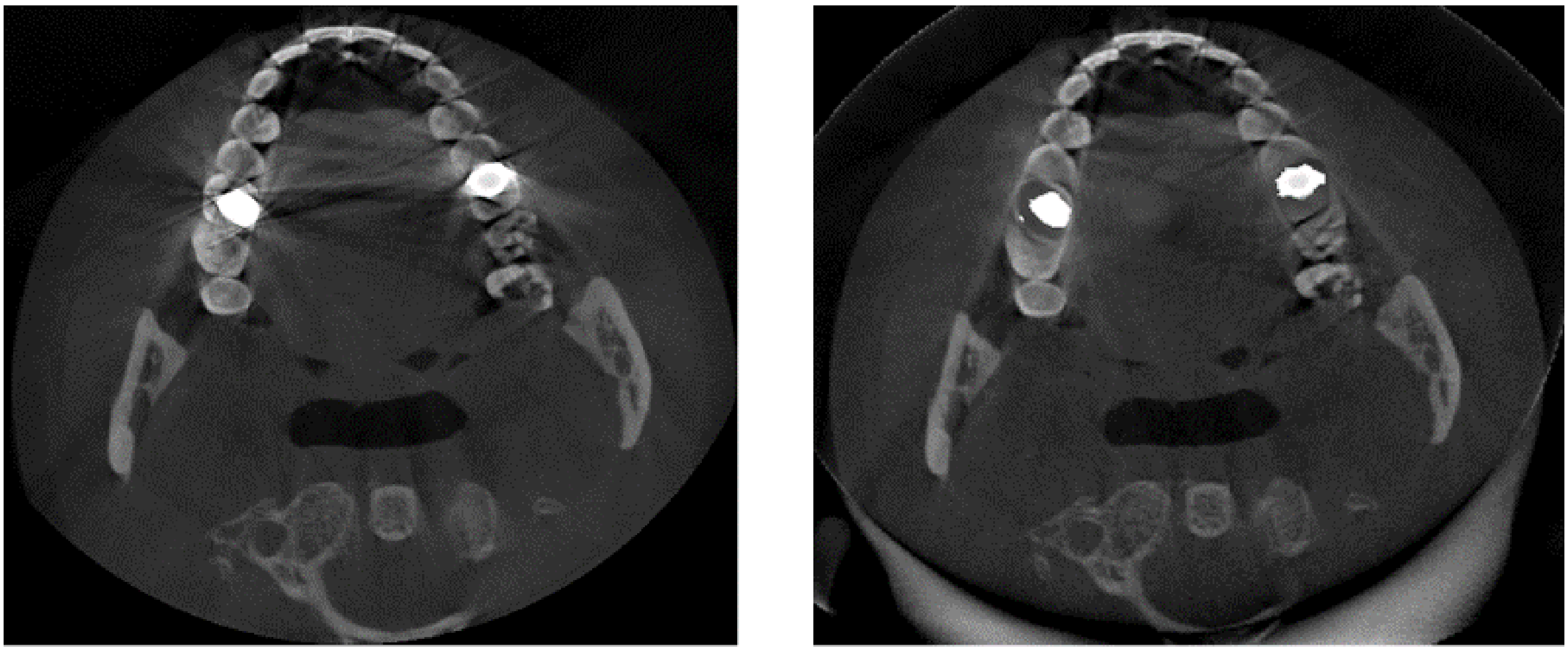

(a) The image reconstructed by FDK (b) The image reconstructed by our proposed method

Fig.4 The reconstructed images

beam projection and reconstruction, an open-source code, known as ASTRA Toolbox, is used[16, 17]. The intra-oral scan data as known as structured light scanning data is presented in the form of a point cloud, which is a 3D surface data composed of points of tooth surface, as shown in Fig.2. The result of the registration can be seen in Fig.3, it can be observed that the registration is quite good.

we use our proposed method to reconstruct the result using the astra toolbox with SIRT, as shown in Fig.4(b), while the original image reconstructed with FDK, with metal artifacts, is shown in Fig.4(a).

It can be clearly seen that the reconstructed images using the method proposed by us have significantly reduced metal artifacts, while also preserving the contour features around the metal artifacts, significantly improving the quality of the 3D images, and accurately depicting the structure of teeth and surrounding bones and tissues.

## 4 Discussion

In the current experiment, although good results have been achieved in MAR, the limitation of the number of clinical data prevents us from further adopting neural network methods for experiments, which forces us to use manual registration and manual threshold adjustment methods to find the metal region to complete our experiment. The registration stage uses the method of manually selecting registration points to calculate the registration matrix. Due to the limitations of visual recognition, the manually selected CBCT and intra-oral scan registration points may not perfectly correspond to each other, and this phenomenon did occur in the experiment. Therefore, many different registration points need to be tried, which results in certain errors in the registration process. But if neural network registration is adopted, the registration accuracy should be further improved. Manually adjusting the threshold to find metal regions also has certain limitations. We found that in the experiment, the edges of the metal regions obtained by this method are not smooth. In addition, there are also some cases of disconnection of metal regions, which leads to poor segmentation of metal regions. However, if neural network methods are used for metal region segmentation, the accuracy of region segmentation should be improved.

## 5 Conclusion

We propose a method that combines 3D CBCT data with intra-oral scan data and jointly considers the metal artifacts in both the projection and image domains. Our experiments have shown that non-radiative intraoral scan data can provide very accurate guidance for the localization of metal parts and have a very positive impact on the removal of metal artifacts while preserving the image details of teeth. However, due to the limitation of the clinical data, our experiments have not yet been able to try neural network, which will further enhance the reconstruction accuracy of our proposed method.

# On a cylindrical scanning modality in three-dimensional Compton scatter tomography

James W. Webber[1]

[1] Department of Obstetrics and Gynecology, Brigham and Women's Hospital, Boston, MA USA

**Abstract** We present injectivity and microlocal analyses of a new generalized Radon transform, $\mathscr{R}$, which has applications to a novel scanner design in 3-D Compton Scattering Tomography (CST), which we also introduce here. Using Fourier decomposition and Volterra equation theory, we prove that $\mathscr{R}$ is injective and show that the image solution is unique. Using microlocal analysis, we prove that $\mathscr{R}$ satisfies the Bolker condition (sometimes called the "Bolker assumption") [1], and we investigate the edge detection capabilities of $\mathscr{R}$. This has important implications regarding the stability of inversion and the amplification of measurement noise. This paper provides the theoretical groundwork for 3-D CST using the proposed scanner design.

## 1 Introduction

CST is an imaging technique which uses Compton scattered photons to recover an electron density, which has applications in security screening, medical and cultural heritage imaging [2–6].

We introduce a new scanning modality in 3-D CST, whereby monochromatic (e.g., gamma ray) sources and energy-sensitive detectors on a cylindrical surface scan a density passing through the cylinder on a conveyor. See figure 1, where we have illustrated $(x,y)$ and $(x,z)$ plane cross-sections of the proposed scanner geometry. The incoming photons, which are emitted from $\mathbf{s}$ with energy $E$, Compton scatter from charged particles (usually electrons) with energy $E'$, and are measured by the detector $\mathbf{d}$; meanwhile, the electron charge density, $f$ (represented by a real-valued function), passes through the cylinder in the $z$ direction on a conveyor belt. The scattered energy, $E'$, is given by the equation

$$E' = \frac{E}{1+(E/E_0)(1-\cos\omega)}, \tag{1.1}$$

where $E$ is the initial energy, $\omega$ is the scattering angle and $E_0 \approx 511\text{keV}$ denotes the electron rest energy. If the source is monochromatic (i.e., $E$ is fixed) and we can measure the scattered energy, $E'$, i.e., the detectors are energy-sensitive, then the scattering angle, $\omega$, of the interaction is fixed and determined by equation (1.1). This implies that the surface of Compton scatterers is the surface of rotation of a circular arc, which we denote as a lemon, $\mathscr{L}$. Example 2-D cross sections of a lemon are shown in figure 1. Thus, we model the Compton scattered intensity as integrals of $f$ over lemons. See, e.g., [3] for other work which models the Compton intensity in this way.

The geometry and physical modeling leads us to a new Radon transform, $\mathscr{R}$, which integrates $f$ over lemon surfaces. Using Fourier decomposition and Volterra equation theory, we prove that $\mathscr{R}$ is injective, which implies $f$ can be uniquely recovered using Compton scatter data. Using the theory of linear Fourier Integral Operators (FIO), we prove that $\mathscr{R}$ satisfies the Bolker condition [1], which gives insight into the reconstruction artifacts. In addition, we investigate the edge detection capabilities of $\mathscr{R}$ and discuss how this relates to image edge reconstruction.

The results presented here provide a novel framework for CST, and lay the theoretical foundation for 3-D density reconstruction using the proposed scanner design.

## 2 Materials and Methods

We analyze the generalized Radon transform, $\mathscr{R}$, defined by

$$\mathscr{R}f(\mathbf{s},\mathbf{d},E',z_0) = \int_{\mathscr{L}(\mathbf{s},\mathbf{d},E',z_0)} f\,\mathrm{d}S, \tag{2.1}$$

where $\mathscr{L} = \mathscr{L}(\mathbf{s},\mathbf{d},E',z_0)$ denotes a lemon surface, parameterized by $(\mathbf{s},\mathbf{d},E',z_0)$, as in figure 1, and $\mathrm{d}S$ denotes the surface measure on $\mathscr{L}$. We consider only the $\mathscr{L}$ with central axis parallel to $z$. In total, we vary four parameters, namely $\mathbf{s}$ and $\mathbf{d}$ (i.e., the positions of the tips of the lemon), the scattered energy, $E'$, which determines $\omega$, and $z_0$, which translates $\mathscr{L}$ in direction $z$. Equivalently, $z_0$ controls the position of $f$ on the conveyor.

In the following subsections, we discuss the tecniques we use to analyze injectivity and microlocal stability of $\mathscr{R}$.

### 2.1 Injectivity

In [7], Volterra type integral equations are discussed. The authors give conditions on the integral kernel for injectivity, and outline a reconstruction method using Neumann series. To prove injectivity of $\mathscr{R}$, we use the theory of [7], and derive a reconstruction formula for $f$. Showing injectivity of $\mathscr{R}$ is important, as it shows that the solution for $f$ is unique, and thus there can be no artifacts due to null space.

### 2.2 Microlocal analysis

To analyze the stability of $\mathscr{R}$ from a microlocal perspective, we apply the theory of linear FIO [8, 9]. An FIO is a special type of integral operator which propagates the edges of an image in specific ways. Through analysis of FIO we can discern important information about the recovery of image edges and image artifacts. FIO theory and microlocal analysis can also be used to predict artifact location [10] and to develop techniques for artifact suppression [11].
The analysis we present here focuses on the Bolker condition [1]. The Bolker condition relates to image artifacts in reconstructions from Radon transform data (e.g., $\mathscr{R}f$), specifically to artifacts which are additional (unwanted) edges in the reconstruction that are not in the object. If the Bolker condition is satisfied, this implies reconstruction stability, and unwanted image edges are eliminated. Conversely, if the Bolker condition fails, the capacity for artifacts is amplified.

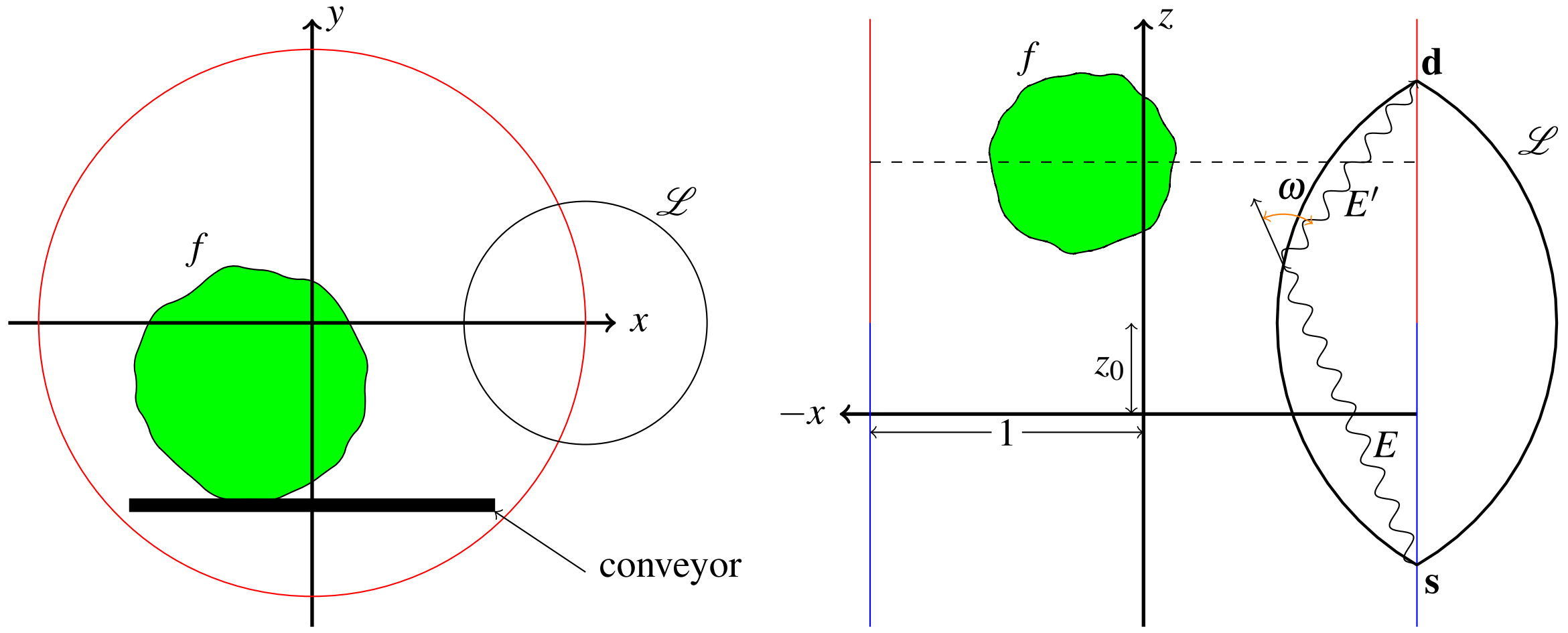

**Figure 1:** $(x,y)$ and $(x,z)$ plane cross sections of the proposed cylindrical scanning geometry. The cylinder has unit radius. The $(x,y)$ plane of the left-hand figure is highlighted as a dashed line in the $(x,z)$ plane. Cross-sections of a lemon of integration, $\mathscr{L}$, are labeled. The sources ($\mathbf{s}$) are located on the bottom half of the cylinder, highlighted in blue, and the detectors ($\mathbf{d}$), highlighted in red, are located on the upper half. The scanning target, $f$, displayed as a green, irregular disc, passes through the cylinder on a conveyor belt in the $z$ direction. The variable $z_0 \in \mathbb{R}$ is $z$ component of the center of $\mathscr{L}$.

### 2.2.1 Edge detection

We investigate the edge detection capabilities of $\mathscr{R}$ using microlocal analysis. See [12] for similar work, where the authors present a microlocal analysis of the classical straight line Radon transform, commonly applied in X-ray CT.
$\mathscr{R}$ integrates $f$ over lemon surfaces. Edges in directions normal to the lemons are detected by $\mathscr{R}$. See figure 2. The

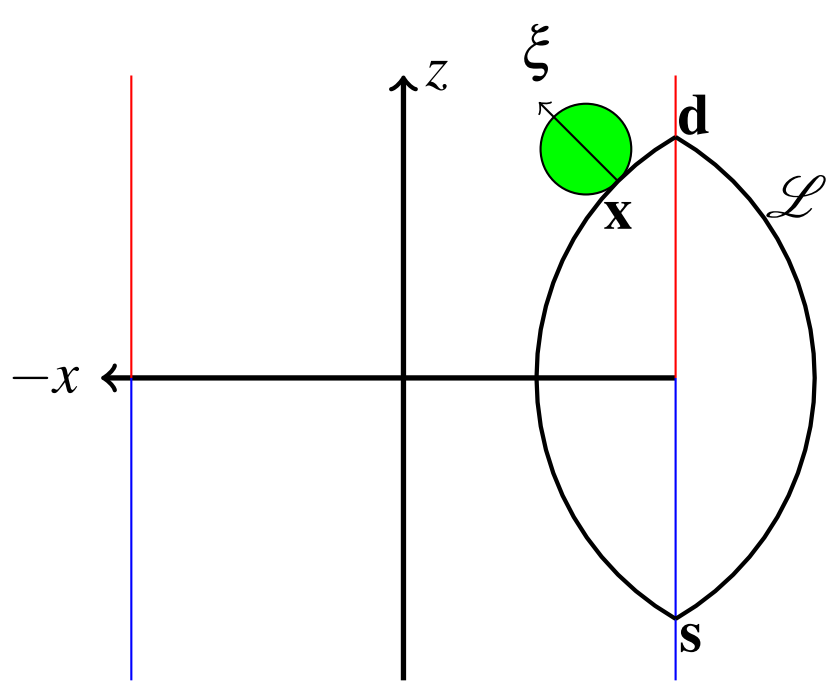

**Figure 2:** An edge, at position $\mathbf{x}$ in direction $\xi \in S^2$, on the boundary of a green disc, normal to a lemon, $\mathscr{L}$.

highlighted edge at $\mathbf{x}$ in direction $\xi \in S^2$, where $S^2$ is the unit sphere in $\mathbb{R}^3$, is normal to $\mathscr{L}$ and is thus detectable by $\mathscr{R}$. If an edge is detectable, then it can be stably reconstructed. Edges which are not detectable are invisible to the data and cannot be recovered stably without sufficient a-priori information regarding the edge map of $f$. Thus, the edge detection capabilities of $\mathscr{R}$ give great insight into the inversion stability.

## 3 Results

Here we present our theoretical results on injectivity and microlocal analysis, and show simulations to validate our results.

### 3.1 Injectivity

Let $C_\varepsilon = \{\sqrt{x^2+y^2} < 1-\varepsilon\}$, for some small offset $0 < \varepsilon < 1$, and let $L^2_c(C_\varepsilon)$ denote the set of square integrable functions with compact support in $C_\varepsilon$. Then, we have the following theorem.

**Theorem 3.1.** *Let $0 < \varepsilon < 1$ be fixed. Then, the lemon Radon transform, $\mathscr{R}$, is injective on domain $L^2_c(C_\varepsilon)$.*

Theorem 3.1 shows that any $f \in L^2_c(C_\varepsilon)$ can be recovered uniquely from $\mathscr{R}f$. To prove Theorem 3.1, we first take a Fourier decomposition of $\mathscr{R}$ on the cylinder, which reduces $\mathscr{R}$ to a set of one-dimensional Volterra operators of the first kind. After which, we apply the theory of [7] to prove injectivity of $\mathscr{R}$, and derive an inversion formula using Neumann series. The support of $f$ is required to be bounded away from the cylinder surface (i.e., by distance $\varepsilon$) to avoid singularities (division by zero) in the Volterra equation kernel.

### 3.2 Microlocal analysis

We state our main microlocal theorem below.

**Theorem 3.2.** *$\mathscr{R}$, on domain $L^2_c(C_0)$, is an FIO which satisfies the Bolker condition*

Theorem 3.2 shows that there are no added, unwanted edges in a reconstruction from $\mathscr{R}f$ data, for functions $f \in L^2_c(C_0)$. In particular, we prove that any added image edge artifacts are reflections of the true image edge map through planes tangent to the boundary of $C_0$. Thus, if $f$ is supported within $C_0$, the added artifacts must lie outside of $C_0$, and do not interfere with the scanning region. We can also use our theory to predict precisely where artifacts will occur. For example, see figure 3, where we show the artifacts predicted by our theory in reconstructions of delta functions. A delta function is supported at a single point, and has edges in all directions. The artifacts due to Bolker, in this case, are embedded in the $(x,y)$ plane and appear on cardiod type curves which lie outside of $C_0$.

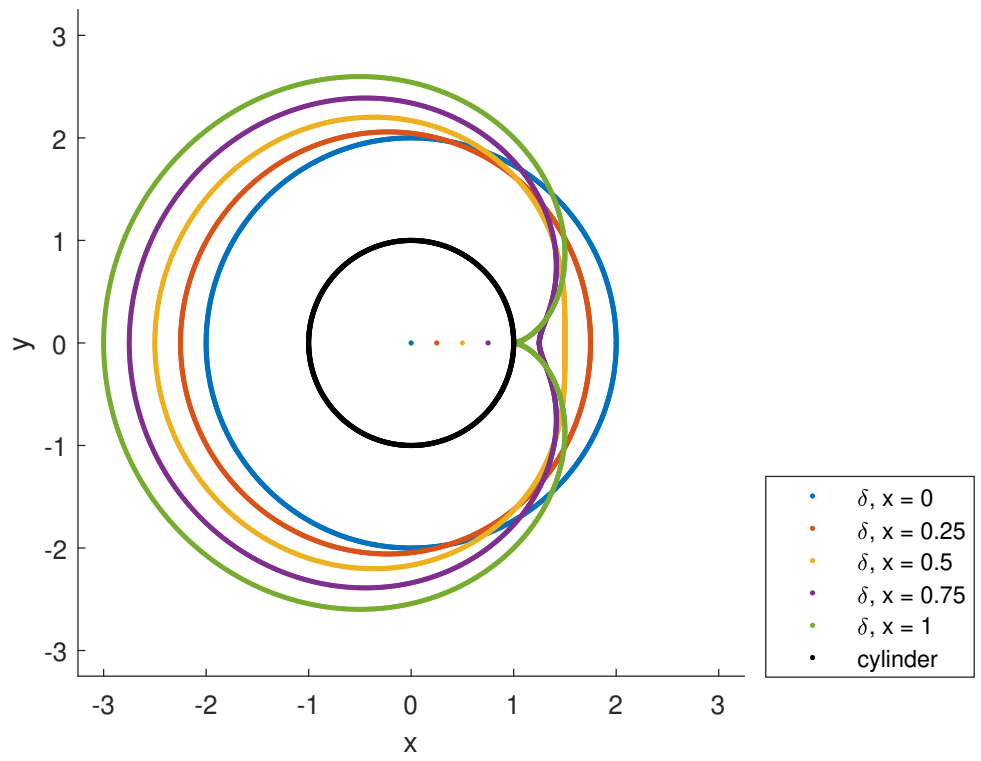

**Figure 3:** Predicted artifacts due to Bolker when reconstructing delta functions ($f = \delta$) from $\mathscr{R}f$. The boundary of $C_0$ is displayed as a black circle. The $\delta$ positions are the dots in $C_0$, and the curves, of the same color, are the corresponding locations of reconstruction artifacts.

### 3.2.1 Edge detection

Here we investigate the edge detection capabilities of $\mathscr{R}$ within $C_0$ using microlocal analysis, as discussed in subsection 2.2.1. See figure 4. For every point $\mathbf{x} \in C_0$, we calculate the proportion of directions $\xi \in S^2$ which are detectable by $\mathscr{R}$. We show our results as an image in the $(x,z)$ plane. This represents the full 3-D edge detection map due to circular symmetry. A value closer to 1 on the colorbar means greater edge detection, and conversely for values closer to 0.
We see that, as we go closer to the center of $C_0$, the edge detection ability of $\mathscr{R}$ is greatest, and this tapers off quite significantly near the boundary and top and bottom of $C_0$. Nowhere in $C_0$ are 100% of edges detectable, and thus the problem is one of limited-angle tomography, which are often severely ill-posed [12]. If the inversion of $\mathscr{R}$ is not sufficiently regularized, we will likely see a blurring effect in the reconstruction near the undetected edges, as is observed, e.g., in conventional limited angle X-ray CT [12].

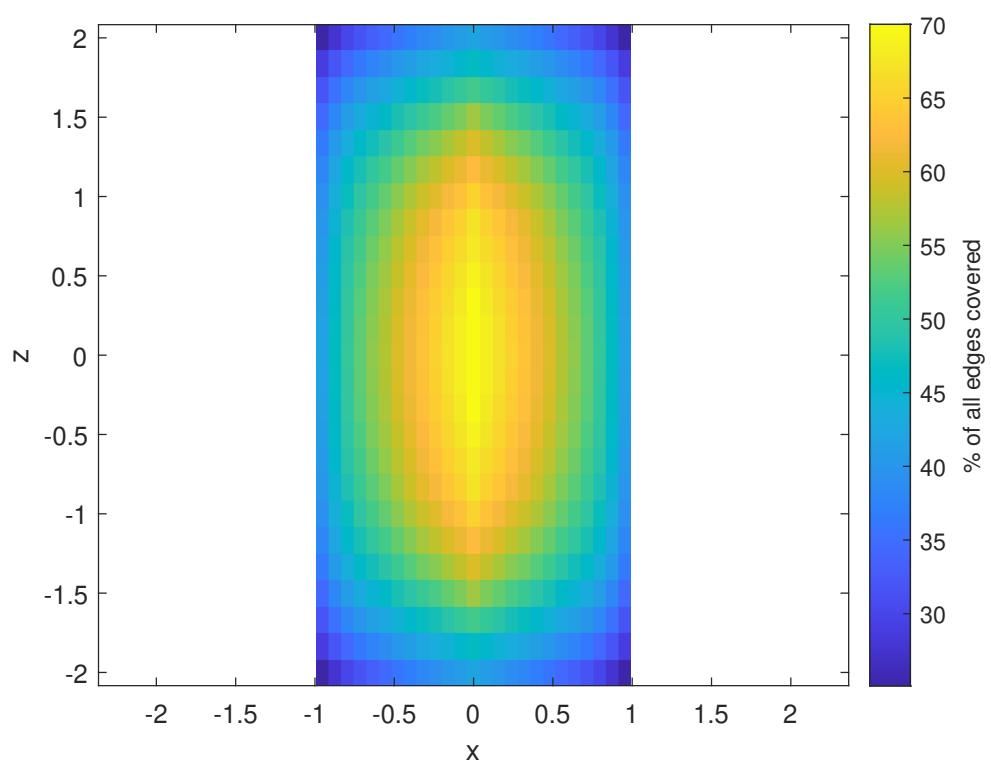

**Figure 4:** The percentage of edges detectable by $\mathscr{R}$ within $C_0$. For the purpose of this simulation, we set the height of $C_0$ as 4.

## 4 Conclusions and further work

In this paper, we introduced a novel scanning modality in 3-D CST, and a new generalized Radon transform, $\mathscr{R}$, which mathematically models the Compton signal. We showed that $\mathscr{R}$, on domain $L^2_c(C_\varepsilon)$, for some fixed $0 < \varepsilon < 1$, is injective and an FIO which satisfies the Bolker condition. We also derived a reconstruction formula using Neumann series. This work lays the theoretical foundation for 3-D Compton imaging using the proposed scanner design.
In further work, we aim to use the reconstruction formula derived here as basis for a reconstruction algorithm. The microlocal analysis results indicate the problem is severely ill-posed, and thus strong regularization (e.g., total variation, machine learning) will likely be needed to address this. We aim to pursue such regularization ideas in further work.

# Simultaneous dual-nuclide imaging based on software-based triple coincidence processing

Bo Wen[1#], Yu Shi[1#], Yirong Wang[2], Jianwei Zhou[1], Fei Kang[2*], Shouping Zhu[1*]

[1]Engineering Research Center of Molecular and Neuro Imaging of Ministry of Education, School of Life Science and Technology, Xidian University,Shaanxi 710126, China

[2]Department of Nuclear Medicine, Xijing Hospital, Fourth Military Medical University, Xi'an,Shaanxi 710032, China

Asterisk* indicates corresponding author

[#]B.Wen and Y.Shi contributed equally to this work

**Abstract** PET dual-nuclide simultaneous imaging is an advanced functional imaging technique. Most of the existing dual-nuclide imaging techniques either use extra detectors to detect prompt gammas from non-pure emitter or employ dynamic acquisition protocol to distinguish between nuclides. However, they also cause the problems of complex system structure and complicated data acquisition process. In this paper, based on the detection of prompt gammas, we used a software-based coincidence processing method to realize simultaneous dual-nuclide imaging in a conventional PET system without extra detectors. In order to evaluate the effectiveness of the dual-nuclide method used, phantom and mouse experiments of $^{18}F/^{124}I$ were performed on the in-house quad-panel PET system. In the phantom experiment, the separation images of dual-nuclide showed good separation of the different nuclides. In the mouse experiment, the separated images of $^{124}I$ and $^{18}F$ showed different activity distributions in the thyroid, heart and bladder, which were consistent with the single-nuclide reconstructed images. Comparing the separated $^{124}I$ images with the mixed dual-nuclide images using double coincidence, we found that the intensity of the signal in the thyroid was increased by more than 20 times. This work demonstrates the feasibility of dual-nuclide simultaneous imaging using a software-based triple coincidence processing method in a conventional PET system.

## 1 Introduction

Positron emission tomography (PET) is a technique that uses molecular imaging signals generated by radionuclides to detect information about human tissue. PET uses a double coincidence technique to achieve the 511-keV photon detection and then images the distribution of the nuclide. However, conventional PET imaging can only image a single-nuclide and cannot distinguish between different nuclides in one scan. With the increasing clinical application of non-pure positron emitter which emits additional gamma during β+ decay, there is a need to use dual-nuclide for simultaneous diagnosis of pathological information. For example, the application of tracers co-labeled with $^{18}F$ and $^{68}Ga$ can reflect the multiple heterogeneities of human tumors [1], allowing physicians to make a more comprehensive diagnosis of pathology. Therefore, dual-nuclide imaging research has flourished in recent years [2]. For simultaneous imaging of non-positron and pure positron nuclides, the detection of triple coincidence is important. The existing dual-nuclide imaging acquires triple-coincidence by additional prompt photon detectors [3] and hardware-based coincidence processing on field programmable gate array (FPGA). The former enhances the detection sensitivity of prompt photons, but it causes the problem of complex system as well. Another method is to separate imaging of dual-nuclide by multiple scans or 45-90 min time-sharing dynamics [4]. However, it requires a long time to scan and is difficult to apply in the clinic.

In order to solve the above problems, this paper applies 20 basic detection modules (BDMs) [5] (Raycan Technology Co., Ltd (Suzhou)) to construct a quad-panel PET system. BDM is capable of transferring the collected single-photon information, such as photon arrival time, photon energy, coordinates of the position of the photon hitting the SiPM, etc. Transfer them to the software side via User Datagram Protocol (UDP). The data format is shown in Figure 1, which provides the data basis for the implementation of the software-based triple-coincidence algorithm. In the process of coincidence processing, based on the energy characteristics and time information of prompt photons different from annihilation photons, a sorting-based time coincidence method [6] and a dual energy window-based energy coincidence method were proposed to achieve triple coincidence of dual-nuclide imaging. To evaluate the effectiveness of the software-based triple coincidence method, the simultaneous imaging and separation effects of dual-nuclide at $^{18}F/^{124}I$ were explored. The metrics were quantified on the basis of a phantom model and small animal experiments.

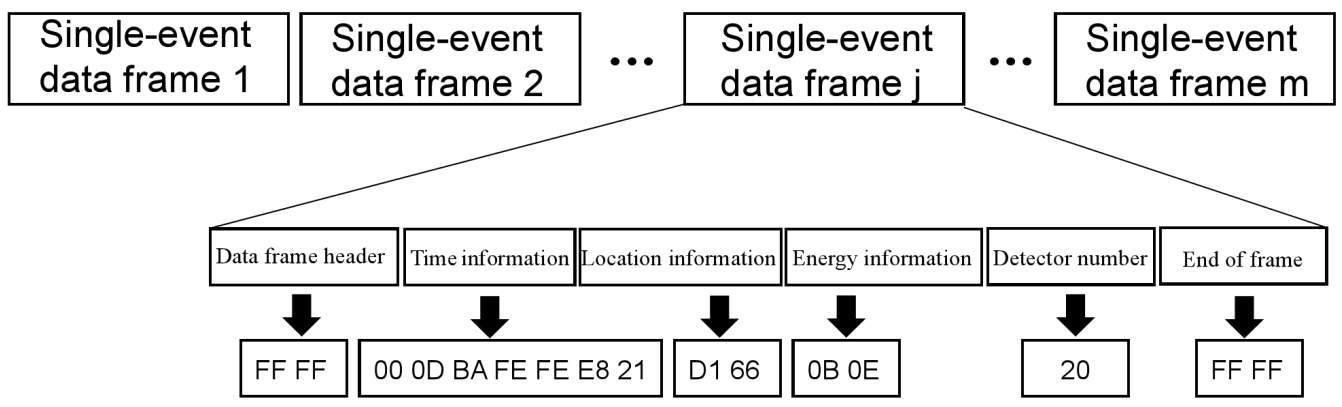

**Figure 1:** Detector data storage format

## 2 Materials and Methods

In the dual-nuclide imaging, the software-based triple coincidence method is applied to the image reconstruction of non-pure positron nuclides. Figure 2 shows the coincidence diagrams for pure and non-pure positron nuclides, respectively. The comparison reveals that the prompt photon is an important feature of the dual-nucilde separation. Therefore, in the dual-nuclide imaging process, the separation of the dual-nuclide image should focus on the detection of prompt photon. The prompt photons information of some commonly used non-pure positron nuclides is shown in Table 1. The separation process of dual-nuclide means that the lines of response (LOR) of different nuclides are counted separately by the energy-spectrum information and the time information of prompt photons.

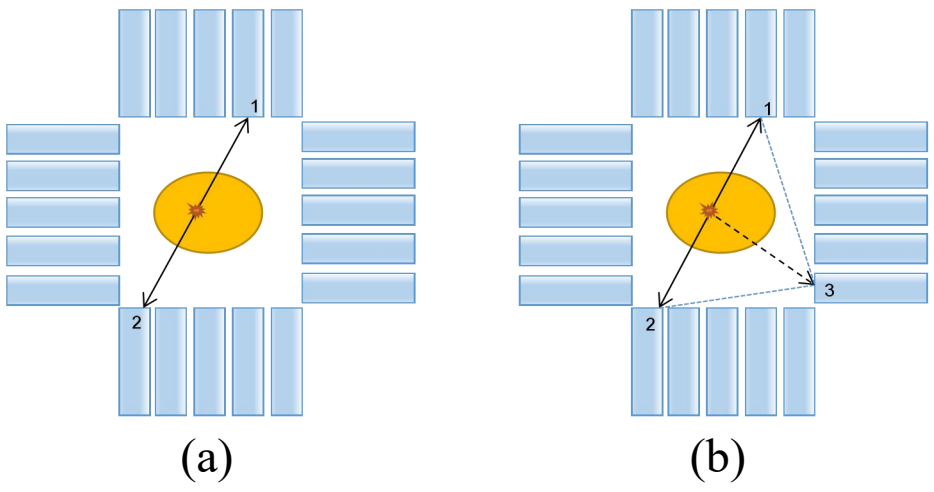

**Figure 2:** Photon coincidence diagrams for two nuclides.
(a) Schematic diagram of response lines of pure positron nuclide. (b) Schematic diagram of the response lines of non-pure positron nuclide.

| Nuclides | Half-life | annihilation photon Energy （%） | Prompt Photon Energy （keV） | Prompt photon emission ratio （%） |
|---|---|---|---|---|
| $^{22}$Na | 2.60 yr | 90.4 | 1274.5 | 99.9 |
| $^{44}$Sc | 3.97 h | 94.3 | 1157.0 | 100 |
| $^{94m}$Tc | 52 min | 70.2 | 871.1 | 96.4 |
| $^{124}$I | 4.18 days | 22.5 | 602.7 | 52.0 |
| $^{68}$Ga | 67.8 min | 88.9 | 1077 | 3.2 |

**Table 1:** Information table of non-pure positron nuclides

## 2.1 Quad-panel PET System

Because of the flexible structure of the plane PET and adjustable plate spacing, the field of view (FOV) can be adjusted according to the actual object. It is suitable for site-specific imaging, so the quad-panel PET system is built for dual-nuclide imaging, as shown in Figure 3. The BDM is made of LYSO crystals for detecting annihilation photons and prompt photons from nuclide decay. The system consists of four panels, and each panel consists of five BDMs. The detection area of each BDM is 27.5×107 mm$^2$. The imaging FOV of the system is 110×110×150 mm$^3$. The spatial resolution of the system is 1.2 mm.

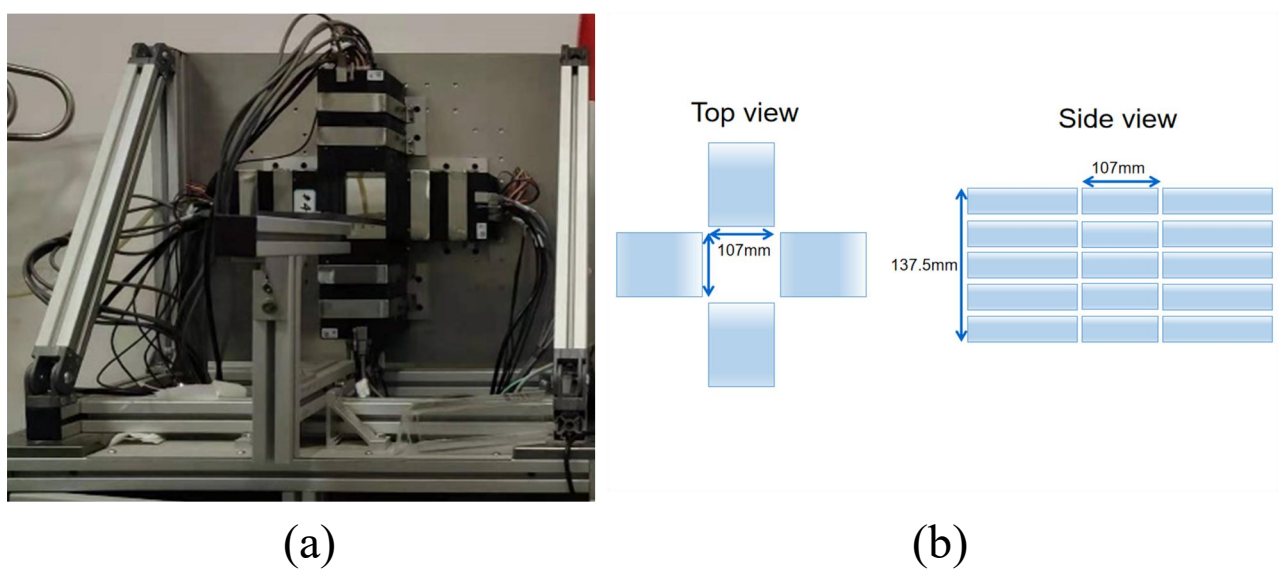

**Figure 3:** Construction of the four flat panel system

## 2.2 Software triple coincidence

The flowchart of the PET software coincidence is shown in Figure 4. The data extraction for software coincidence of the dual-nuclide is performed using a home-made quad-panel detection system. The γ-photon single-event data is obtained by synchronous clock detection of the BDM array. The coding, decoding and correction of the software coincidence data are performed at the software side. Considering the energy-spectrum difference between prompt photons and annihilation photons and the arrival time information of detectors, unlike the traditional double software coincidence, the triple software coincidence adopts a sorting-based time coincidence approach with a dual energy window-based energy coincidence to realize the double and triple coincidence LOR counting.

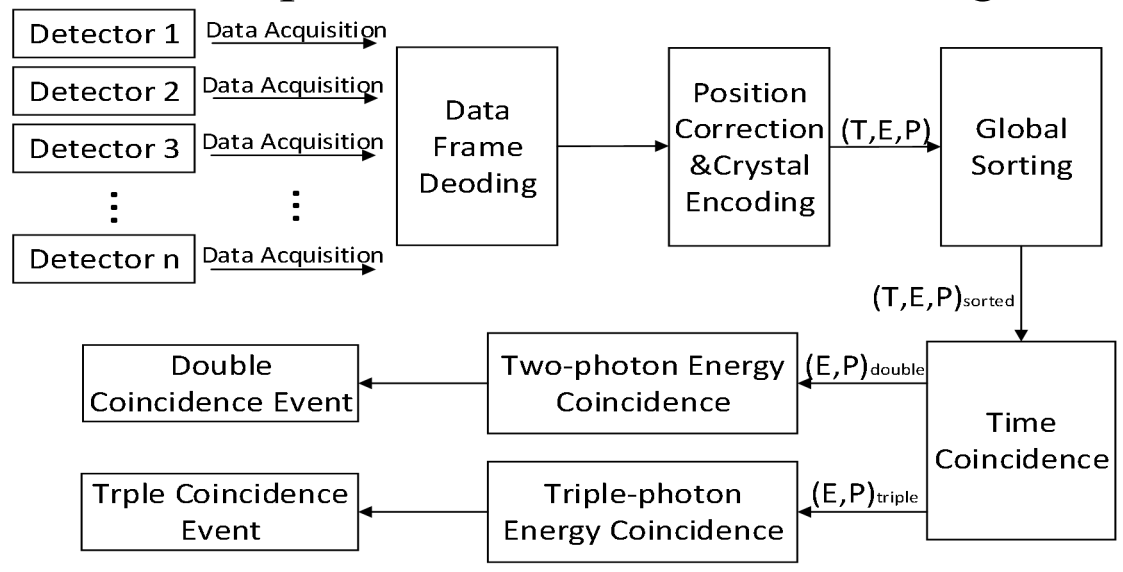

**Figure 4:** PET software coincidence flowchart

### 2.2.1 Time-coincidence methods based on sorting

Data acquisition by BDM enables to obtain the γ-photon single event data for each detector module. By decoding the storage structure of the data shown in Figure 1, each γ-photon single event is represented as $\left(t_j^i, e_j^i, \vec{p}_j^{\,i}\right)$ after decoding, where $t_j^i$ denotes the arrival time of the jth γ-photon arriving at detector i in nanoseconds (ns), $e_j^i$ denotes the energy value of the scintillation pulse caused by the jth γ-photon arriving at detector i in keV and $\vec{p}_j^{\,i}$ denotes the position coordinates of the jth γ-photon arriving at detector i. The single time stream (Ti, Ei, Pi) of the ith detector module is given by Equation 1. The time coincidence method based on ordering arranges the single events of each detector module according to the photon arrival time information from smallest to largest. Finally there can obtains a globally ordered sequence of γ-photon single events $(T,E,P)_{sorted}$, as shown in Eq. 2.

$$\left(T^i,E^i,P^i\right)=\left[\left(t_1^i,e_1^i,\vec{p}_1^{\,i}\right);\left(t_2^i,e_2^i,\vec{p}_2^{\,i}\right);\cdots\left(t_j^i,e_j^i,\vec{p}_j^{\,i}\right);\cdots;\left(t_{m_i}^i,e_{m_i}^i,\vec{p}_{m_i}^{\,i}\right)\right] \quad j=1,\cdots,m_i \qquad (1)$$

$$(T,E,P)_{sorted}=\left[\left(t_1,e_1,\vec{p}_1\right)\left(t_2,e_2,\vec{p}_2\right);\cdots;\left(t_{m_1+m_2+\cdots+m_n},e_{m_1+m_2+\cdots+m_n},\vec{p}_{m_1+m_2+\cdots+m_n}\right)\right] \quad i=1,\cdots,m_i \qquad (2)$$

After initializing the time window $\Delta t_c$, the time components of $(T,E,P)_{sorted}$ are subjected to the first-order forward difference $\Delta_1 T$ with step 1 and the first-order forward difference $\Delta_2$ T with step 2, respectively, as shown in Eqs. 3 and4. The dataset of time-double coincidence $(E,P)_{double}$ and time-triple coincidence $(E,P)_{triple}$ are obtained by comparing with the time window.

$$\Delta_1 T=\left(t_2-t_1,t_3-t_2,\cdots t_{m_1+m_2+\cdots+m_n}-t_{m_1+m_2+\cdots+m_{n-1}}\right) \qquad (3)$$

$$\Delta_2 T=\left(t_3-t_1,t_4-t_2,\cdots t_{m_1+m_2+\cdots+m_n}-t_{m_1+m_2+\cdots+m_{n-2}}\right) \qquad (4)$$

### 2.2.2 Energy coincidence based on dual-energy windows

The basic detection module (BDM) is able to achieve an energy resolution of 15%. In the pre-experimental validation, the use of the BDM was able to resolve prompt photons with an energy-spectrum similar to 511-keV, as shown in Figure 5. The energy-spectrum of the non-pure positron nuclide $^{124}$I was detected by using the quad-panel PET system. This system has 15 detection crystals at different positions of multiple detectors. The energy window of 550 keV-750 keV was selected for energy-spectrum plotting. In the figure, the 15 energy-spectrum curves can clearly distinguish the energy-spectrum peaks of 511keV and 602.7keV, indicating that dual-nuclide energy coincidence can be achieved through the energy window.

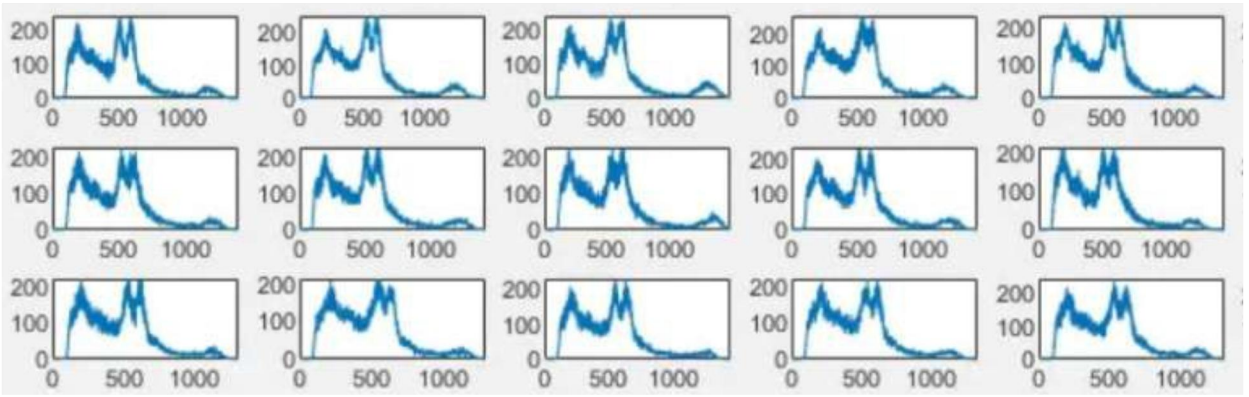

**Figure 5:** Energy-spectrum of $^{124}$I

In the process of software coincidence, by selecting the energy windows of 400-600 keV and 550-700 keV, the prompt photons and annihilation photons of $^{124}$I are energy coincided respectively. Finally the double coincidence dataset $(T,E,P)_{double}$ and triple coincidence dataset $(T,E,P)_{triple}$ can be obtained.

## 3 Results

3.1 Experimental Setups

In order to evaluate the data processing method of software-based triple coincidence and test the effectiveness of dual-nuclide separation. It is proposed to test the effect of dual-nuclide separation using phantom and small animal experiments. $^{18}$F and $^{124}$I are selected for the effect evaluation.

（1）The background-free phantom was used to evaluate the separation effect of $^{18}$F/$^{124}$I. The size of the phantom was 30×30×35 mm$^3$. A 3×3 array of 5×5×30 mm$^3$ thermal holes was arranged at the symmetric center for injection of the nuclide. The 10 µCi/ml $^{124}$I solution, 20 µCi/ml $^{124}$I solution, 10 µCi/ml $^{18}$F solution and 20 µCi/ml $^{18}$F solution were injected at the four corners of the phantom respectively. The acquisition lasted for 15 minutes, and the phantom were re-acquired for 15 minutes after 24 hours, when the $^{18}$F had completely decayed. The data of $^{124}$I single nuclide was collected.

（2）The mouse experiments were conducted using $^{18}$F/$^{124}$I for dual-nuclide imaging experiments to assess the effectiveness of the software-based triple coincidence method. The experiments were conducted by injecting 100 µCi of $^{124}$I nuclide solution, 100 µCi of $^{18}$F nuclide solution, and a mixture of $^{18}$F and $^{124}$I with activity of 200 µCi into the tail vein of mice respectively. Observing the nuclide activity changes in mice. The acquisition time of each experiment was 15 minutes.

In the experiments, twenty BDM ensure full-angle data acquisition. In the software-based triple coincidence, the energy window of annihilation photon is set to 400-600 keV. The energy window of prompt photon is set to 550-700 keV. The coincidence time window is set to 6 ns, and the image reconstruction is performed by MLEM iterative reconstruction method with 10 iterations.

3.2 Phantom experiment of $^{18}$F/$^{124}$I

The nuclide distribution of $^{18}$F/$^{124}$I is shown schematically in Figure 6. The simultaneous imaging results of $^{18}$F/$^{124}$I can be seen using the reconstructed algorithm of double coincidence as shown in Figure 7(a). The conventional reconstruction results of the dual-nuclide algorithm cannot distinguish the reconstructed images of both $^{18}$F and $^{124}$I. The reconstruction using the software-based triple coincidence algorithm separates the reconstructed images of $^{18}$F and $^{124}$I as shown in Figure 7(c)(d). Comparing the reconstructed images of the separation, the $^{18}$F and $^{124}$I separation images are able to separate the hot holes. Comparing the $^{124}$I single-nuclide image with the separation image，they have the same activity region. The comparison of the two sets of images verifies the effectiveness and accuracy of the software-based triple coincidence algorithm.

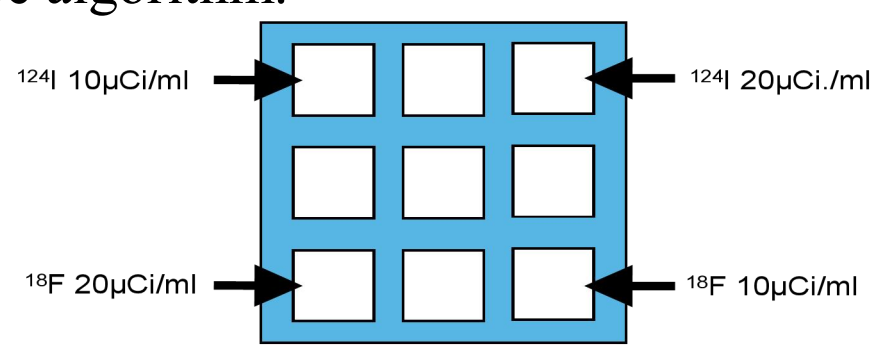

**Figure 6:** Schematic of nuclide distribution

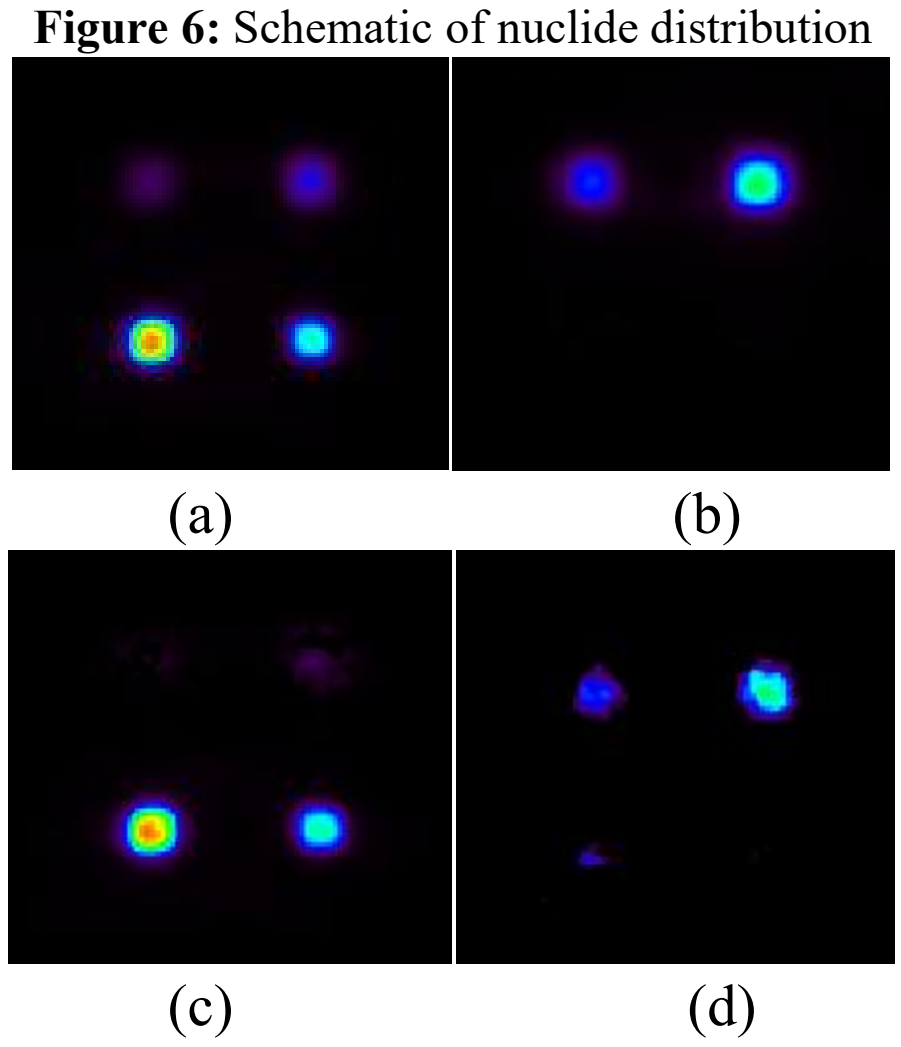

**Figure 7:** Dual-nuclide reconstructed image. (a) The dual-nuclide reconstructed image using the dual-coincidence reconstructed algorithm. (b) $^{124}$I single-nuclide reconstructed image. (c) $^{18}$F separation reconstructed image .(d) $^{124}$I separation reconstructed image.

3.3 Mouse experiment of $^{18}$F/$^{124}$I

The mice were placed in the enclosed space by means of gas hemp, as shown in Figure 8. The experimental results are shown in Figure 9. In Figure 9(a), the double coincidence reconstructed algorithm cannot distinguish the distribution of $^{18}$F and $^{124}$I in mice. Figure 9(b)(c) shows the reconstructed image of $^{124}$I injected alone and the reconstructed image of $^{124}$I separated by mixed injection, respectively. When a certain slice is selected to compare the two images, there are obvious areas of high activity in both thyroid sites, while the activity was lower in the heart and bladder. Figure 9(d)(e) shows the reconstructed images of $^{18}$F injected alone and the reconstructed images of $^{18}$F separated by mixed injection, respectively. By comparing the $^{18}$F single-nuclide reference image and selecting a certain slice to compare the two images, there were obvious high-activity regions in both the heart and bladder areas. Comparing the above results can verify that the proposed method can achieve the separation of $^{18}$F and $^{124}$I dual-nuclide image.

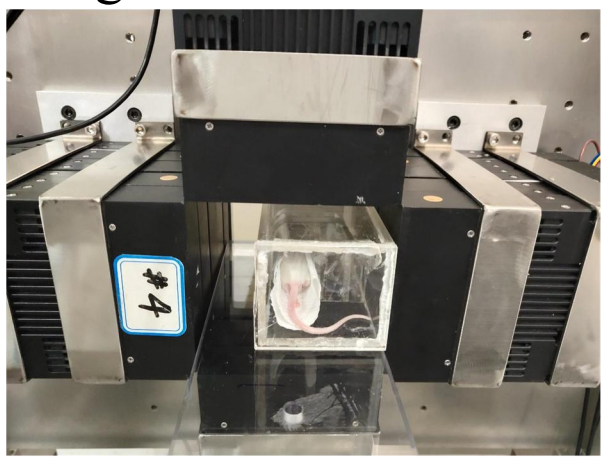

**Figure 8:** Experimental placement of mice

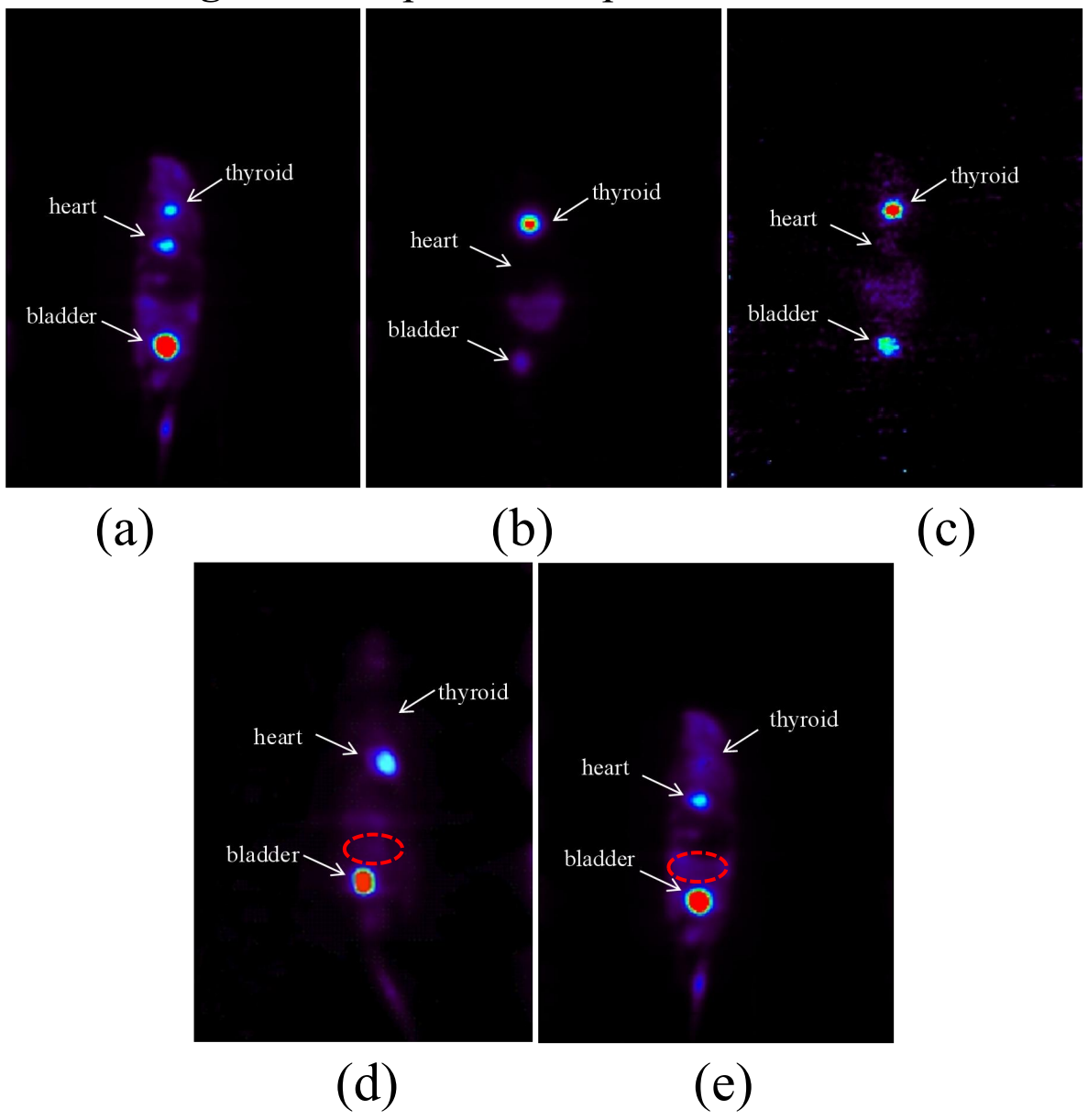

**Figure 9:** Mouse imaging reconstructed image. (a) The dual-nuclide reconstructed image using the dual-coincidence reconstructed algorithm. (b)$^{124}$I single-nuclide reconstructed image. (c) $^{124}$I separation reconstructed image .(d)$^{18}$F single-nuclide reconstructed image. (e) $^{18}$F separation reconstructed image.

The mean values of the ROIs with 3 mm, 4 mm and 5 mm in diameter were calculated in the thyroid, heart and bladder regions, respectively. As shown in Table 2, comparing the double coincidence reconstructed images with the separated images of $^{124}$I, the signal intensity at the thyroid is enhanced by more than 20-fold, but there is still signal residual at the center of the separated images of $^{124}$I.

| Signal intensity ratio | Thyroid/ Bladder | Heart/ Bladder | Thyroid / Heart |
|---|---|---|---|
| Original image | 0.205 | 0.410 | 0.501 |
| $^{124}$I separate images | 4.760 | 0.379 | 12.559 |
| $^{124}$I reference images | 16.787 | 0.034 | 491.558 |
| $^{18}$F separate images | 0.061 | 0.405 | 0.150 |
| $^{18}$F reference images | 0.018 | 0.231 | 0.078 |

**Table 2:** ROIs signal intensity ratio

Two relatively uniform regions with nuclide distribution are selected, such as the red dashed area in Figure 9(d)(e) . Their regional signal-to-noise ratios are calculated, and the results are shown in Table 3. The signal-to-noise ratio of $^{18}$F-separated images is not affected, while the signal-to-noise ratio of $^{124}$I-separated images decreases by a factor of 2-5.

| Signal to Noise Ratio | Region 1 | Region 2 |
|---|---|---|
| $^{124}$I separate images | 2.20 | 2.53 |
| $^{124}$I reference images | 4.76 | 10.02 |
| $^{18}$F separate images | 5.38 | 3.42 |
| $^{18}$F reference images | 3.77 | 3.08 |

**Table 3:** Signal-to-noise ratio of reconstructed images

## 4 Discussion And Conclusion

In this paper, a software-based triple coincidence algorithm is proposed to realize the simultaneous dual-nuclide imaging without adding extra detectors. The experimental results based on $^{18}$F/$^{124}$I showed that the proposed software-based triple coincidence algorithm could separate $^{18}$F and $^{124}$I images. Especially, in the $^{18}$F/$^{124}$I mouse test, single-nuclide separated images of different mouse organs could be distinguished based on the proposed method in current quad-panel PET system. The effectiveness of the algorithm was verified by quantitatively comparing the single-nuclide separated images with the single-nuclide reference images. However, the dual-nuclide separation in the experiment produces residual image of the other nuclide, which is still far from the ideal signal intensity. This phenomenon is caused by that the prompt photon energy of $^{124}$I used is 602.7 keV which is similar to 511 keV, considering that the detector energy resolution is usually 11%~15%, there are still some annihilation photons in the selected energy window of 550-700 keV that are conformed to the triple coincidence. Subsequently, it is necessary to introduce a random correction algorithm to eliminate the nuclide residues in the separated images. A study on denoising is still on-going. Overall, this work demonstrates the feasibility of dual-nuclide simultaneous imaging using a software-based triple coincidence processing algorithm in a conventional PET system.

# Simulated Deep CT Characterization of Liver Metastases with High-resolution FBP Reconstruction

Christopher Wiedeman[1], Peter Lorraine[2], Ge Wang[3], Richard Do[4], Amber Simpson[5], Jacob Peoples[5], Bruno De Man[2*],

[1]Department of Electrical and Computer Engineering, Rensselaer Polytechnic Institute, Troy, NY, USA
[2]GE Research - Healthcare, Niskayuna, NY, USA
[3]Department of Biomedical Engineering, Rensselaer Polytechnic Institute, Troy, NY, USA
[4]Department of Radiology, Memorial Sloan Kettering Cancer Center, NYC, NY, USA
[5]Biomedical Computing and Informatics, Queen's University, ON, Canada
*deman@ge.com

**Abstract** Computed tomography is a frontline tool for monitoring colorectal cancer and its possible progression into liver metastases, but image features predictive of metastasis (met) behavior are dependent on scanning protocols and reconstruction algorithms. We propose a simulation pipeline for studying the effects of imaging parameters on the ability to characterize liver metastasis features. This pipeline utilizes a fractal approach for generating a diverse population of virtual metastasis shapes and then superimposes these on a realistic CT liver region to perform a virtual CT scan using CatSim. We also propose filtered back projection using a high-frequency kernel, which is designed to preserve more high-resolution information than a standard reconstruction kernel. Synthetic liver metastases image patches generated with our pipeline were used to train and validate deep neural networks to recover crafted metastasis characteristics: internal heterogeneity, edge sharpness, and edge fractal dimension. In the absence of added noise, models that used high-frequency reconstruction scored significantly better compared to standard reconstruction when characterizing edge sharpness and fractal dimension ($p < 0.05$). Our novel virtual imaging framework can be used for further study of imaging feature preservation, and our results indicate the possibility of optimizing the reconstruction for enhanced AI-based anatomical feature characterization.

## 1 Introduction

Contributing to about 50,000 deaths in the United States and 900,000 deaths worldwide, colorectal cancer is considered the fourth deadliest cancer [1], [2]. Diseased patients often die from colorectal liver metastases (CRLM) rather than the primary cancer. Although many treatments exist, including resection, chemotherapy, and ablation, monitoring patient responses for the most effective therapies is a complex problem requiring further research [3], [4]. Accurate and continuous monitoring of disease progression and treatment response is critical for optimizing patient outcomes. To this end, x-ray computed tomography (CT) is among the best and most available imaging modalities for observing CRLM progression. CT image features, such as hepatic metastases texture and liver texture have been studied for predicting treatment response [5]–[8]. Fractal dimension of CT metastasis images also has been investigated for response prediction to chemoradiation therapies in patients with locally advanced rectal cancer [9], [10].

CT image features are highly sensitive to the scan and reconstruction parameters used. Despite this, these parameters are heterogeneous across studies in the current literature. The optimization of imaging parameters for the task and robustification of crucial features to various conditions are necessary for drawing consistent conclusions regarding different biomarkers and their relationship with patient outcomes [11]–[13].

Optimizing the imaging procedure for CRLM treatment requires extensive exploration of the parameter space. Simultaneously, artificial intelligence has accelerated a new paradigm of radiomics, where often inconspicuous features are learned over large datasets rather than pre-determined [14]–[16]. Satisfying these needs in practice is infeasible, requiring many high-quality patient samples, with patients subject to multiple repeated scans for comparison. As such, we propose a virtual imaging approach [17] for exploring parameters in the context of CLRM. In this paper, we:

- Define a virtual imaging pipeline for simulating CT scans of liver mets with varying scan and reconstruction parameters.
- Use this pipeline to compare radiomic performance using filtered back projection (FBP) reconstruction with a standard versus a high-frequency kernel.

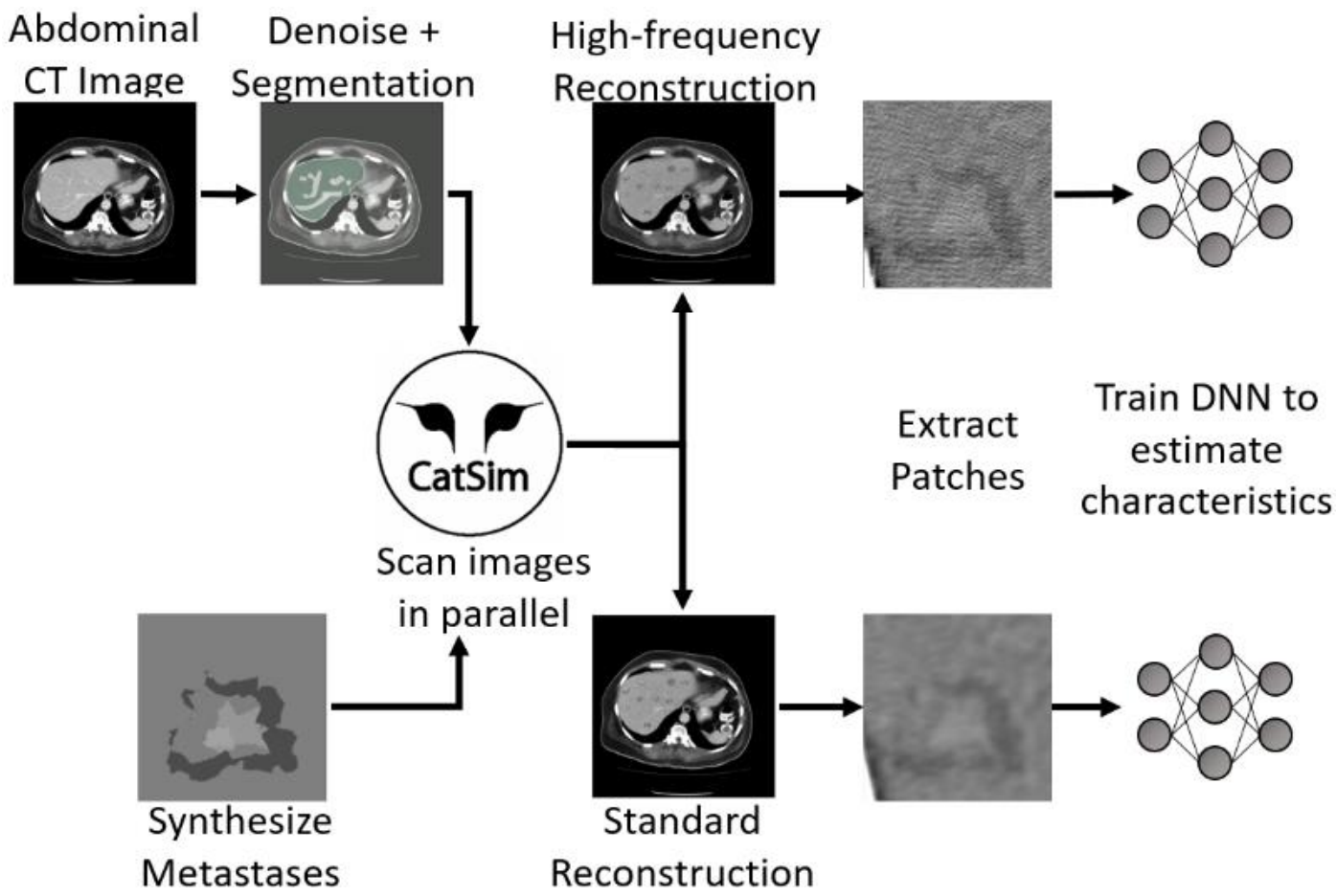

**Figure 1**: Flowchart of met simulation and radiomic analysis. Backgrounds and generated mets are scanned in parallel using CatSim. Extracted met image patches from scan reconstructions are used to train DNNs tasked with estimating the correct metastasis characteristics.

This approach is illustrated in Figure 1. Our met generation process employs a fractal generation method, which models a diverse distribution of random shapes. As these mets are synthetic, they do not have specific clinical labels, but their characteristics (such as edge fractalness) are precisely known. Consequently, they are useful for evaluating different scanning and reconstruction schemes, since the ability for an imager to preserve these characteristics of virtual mets is likely associated with its ability to preserve clinically relevant features of real mets. In our experiments, we judge a scheme's ability to preserve characteristics by a deep neural network's (DNN) ability to recover these characteristics post-reconstruction.

FBP is the basis for the reconstruction used on all commercial CT scanners and is a fast way to convert from the sensor to the image domain. Unfortunately, information – and potential prognostic detail – may be partially lost during this transformation. As such, we propose using a high-frequency kernel for data filtering, resulting in a noisier image but with closer agreement to the raw data.

## 2 Materials and Methods

*Met Synthesis:* Random initial met shapes are synthesized by generating vertices of a random fractal shape using an 'infinite detail' method inspired by [18], and then smoothing the shape with a moving average filter applied over the list of vertices. Specifically, six initial vertices defining a rough hexagon are first initialized; the midpoints along each edge are perturbed by random uniform noise scaled by the distance between the edge vertices for that mid-point and a roughness parameter, doubling the number of edges. This process is repeated until the separation between vertices is less than the resulting inter-pixel spacing. To produce diversity in edge smoothness, the list of vertex coordinates is convolved with a moving average filter of random length up to half of the number of vertices (longer kernel produces a smoother shape).

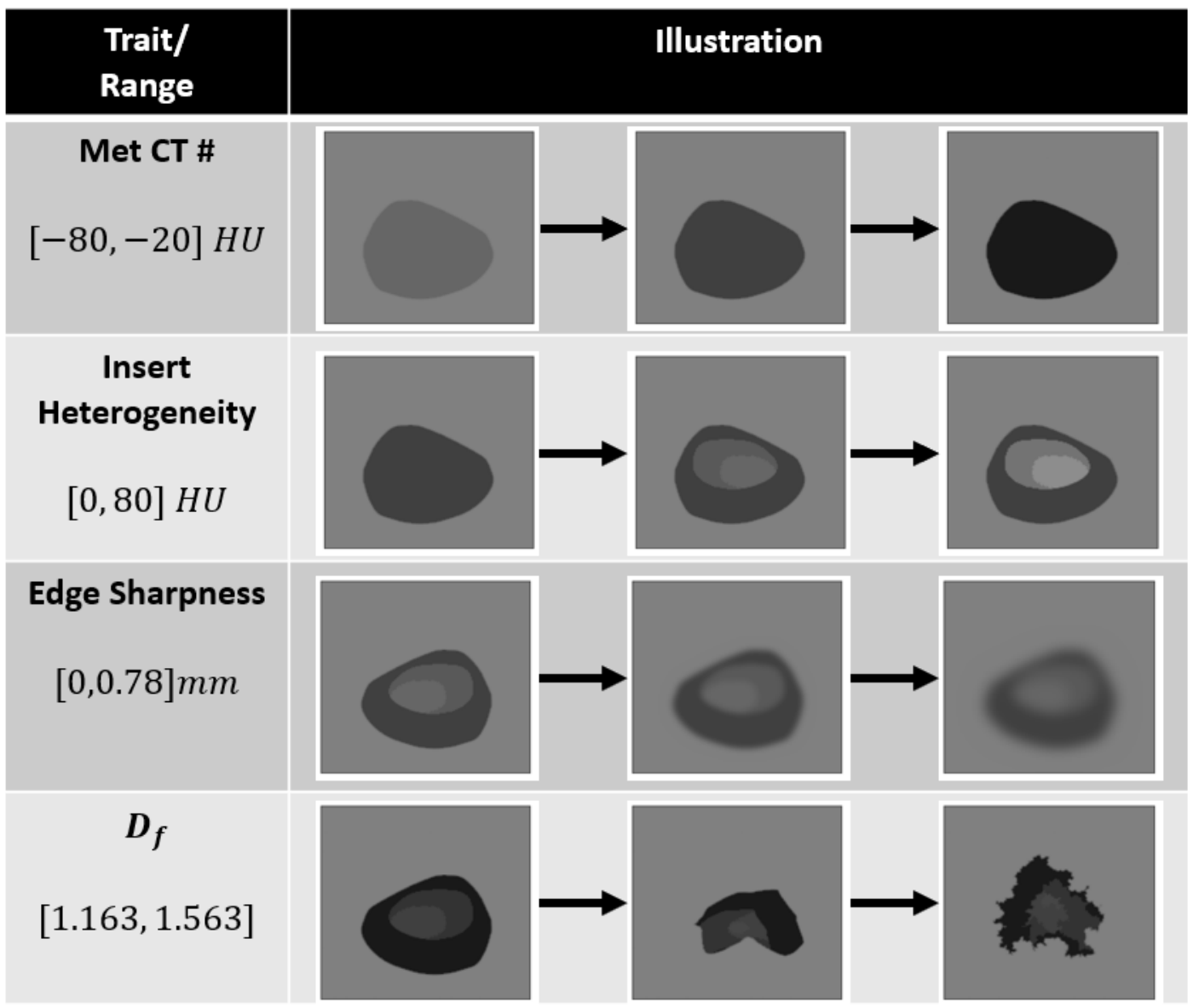

**Figure 2**: Met traits and their visual impact. Edge sharpness is measured by the standard deviation of the Gaussian kernel used to blur the shape.

The grayscale contrast (intensity relative to the background) of each met is randomly sampled from a uniform distribution between -20 HU and -80 HU (Hounsfield Units). In addition to this homogenous base, a region of heterogeneity, which we refer to as the 'insert', is superimposed over each met. These inserts each consist of 2-3 sub-shapes, which are generated using the same fractal generation and smoothing method but fit within the met boundary. This insert is scaled such that the *maximum* difference between an insert point and the met background is sampled uniformly between 0 HU and 80 HU. Edge sharpness (blur) is altered by filtering the image with a Gaussian blurring kernel with a standard deviation between 0 and 0.78mm (higher deviation creates more blurring). Figure 2 illustrates different met parameters of interest.

We also characterize the jaggedness along the outer edge of each metastasis with fractal dimension. Met shapes are quantized as a $256 \times 256$ bit map and then processed with an edge detector. The nuclear box-counting method is then used to calculate the fractal dimension of the edge images as the average slope of the log-log plot of the box scale $r_i \in \{1,2,4,8,16,32\}$ pixels and number of boxes $N(r)$ required to cover the contour [19].

*Simulation and Reconstruction:* Twenty $512 \times 512$ slices with large visible liver regions were selected as image backgrounds. The "Reduce Noise" filter from Adobe Photoshop Elements 11 was used to reduce pre-existing noise in the clinical backgrounds. The liver region of each slice was manually segmented, excluding confounding structures, to identify regions suitable for synthetic met insertion.

Scans were simulated using CatSim [20], [21]. Image backgrounds were converted to water density maps based on their CT number. The synthetic mets were randomly positioned $25\ mm \times 25\ mm$ non-overlapping patches of the liver map. Rather than superimposing the mets and clinical backgrounds in the image domain, the images were reprojected separately and superimposed in the sinogram domain, allowing the mets to be simulated at a much higher resolution (voxel size 0.156mm) than the clinical backgrounds (voxel size 0.68-0.82 mm). 10 to 12 mets were scanned with each background. Although backgrounds were reused, met placement varied between scans, resulting in diverse image patches. Scanning parameters were set to mimic the Lightspeed VCT scanner (GE HealthCare), with $140kV_p$ x-ray tube voltage. Each scanned slice was reconstructed using FBP with either a standard or high-frequency kernel. Images were reconstructed with $40\ cm$ field-of-view and $0.2\ mm$ voxel size.

Edge sharpness and fractal dimension cannot be evaluated within the same dataset, as blurring the edge of a met destroys the original fractal dimension. As such, two separate studies were simulated: a blurred study where blurring was applied to the met shapes, and a no-blur study. 10,000 mets were generated in each study.

*Deep Characterization*: The goal of characterization is to use a DNN to estimate the true met parameters from the reconstructed images. Reconstructions were cropped to $128 \times 128$ patches centered around each met. The network architecture was roughly based on ResNet V2 [22] (Figure 3). Optuna was used to optimize hyperparameters [23].

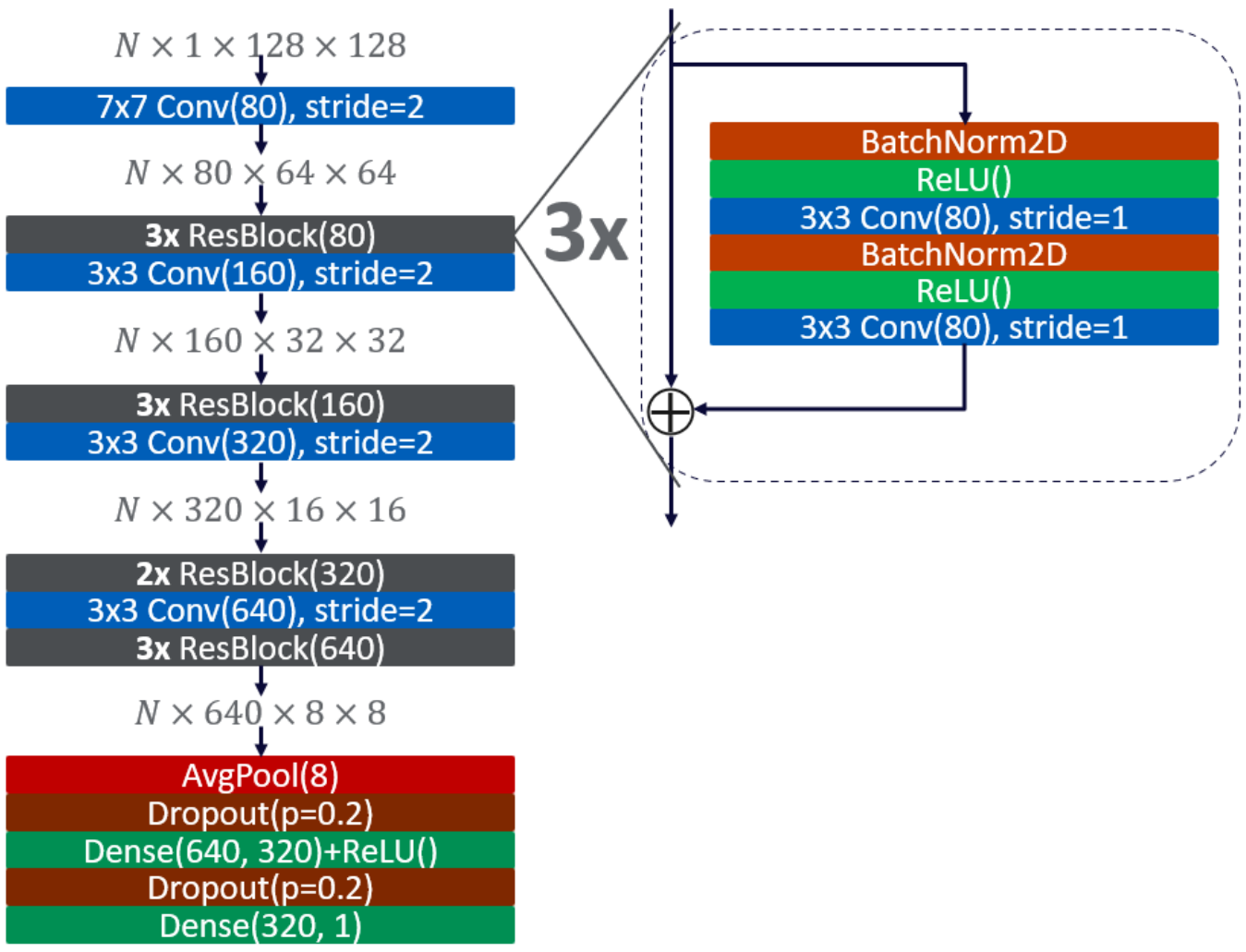

**Figure 3**: DNN architecture used for all characterization tasks. Conv(n): Convolutional layer with n output filters.

Different models were trained for each characteristic (90/10 training/validation). Fractalness and heterogeneity were evaluated using the datasets without blur, while edge sharpness was evaluated on the dataset with blur. Adam optimizer with a learning rate of $4e-5$ and a mean-squared error loss function were used.

Each network trained for 120 epochs (batch size 40). At the end of each epoch, a 'bias adjustment' was performed, where the parameters of the final dense layer were adjusted using a globally computed linear regression (across the entire training dataset) to help convergence.

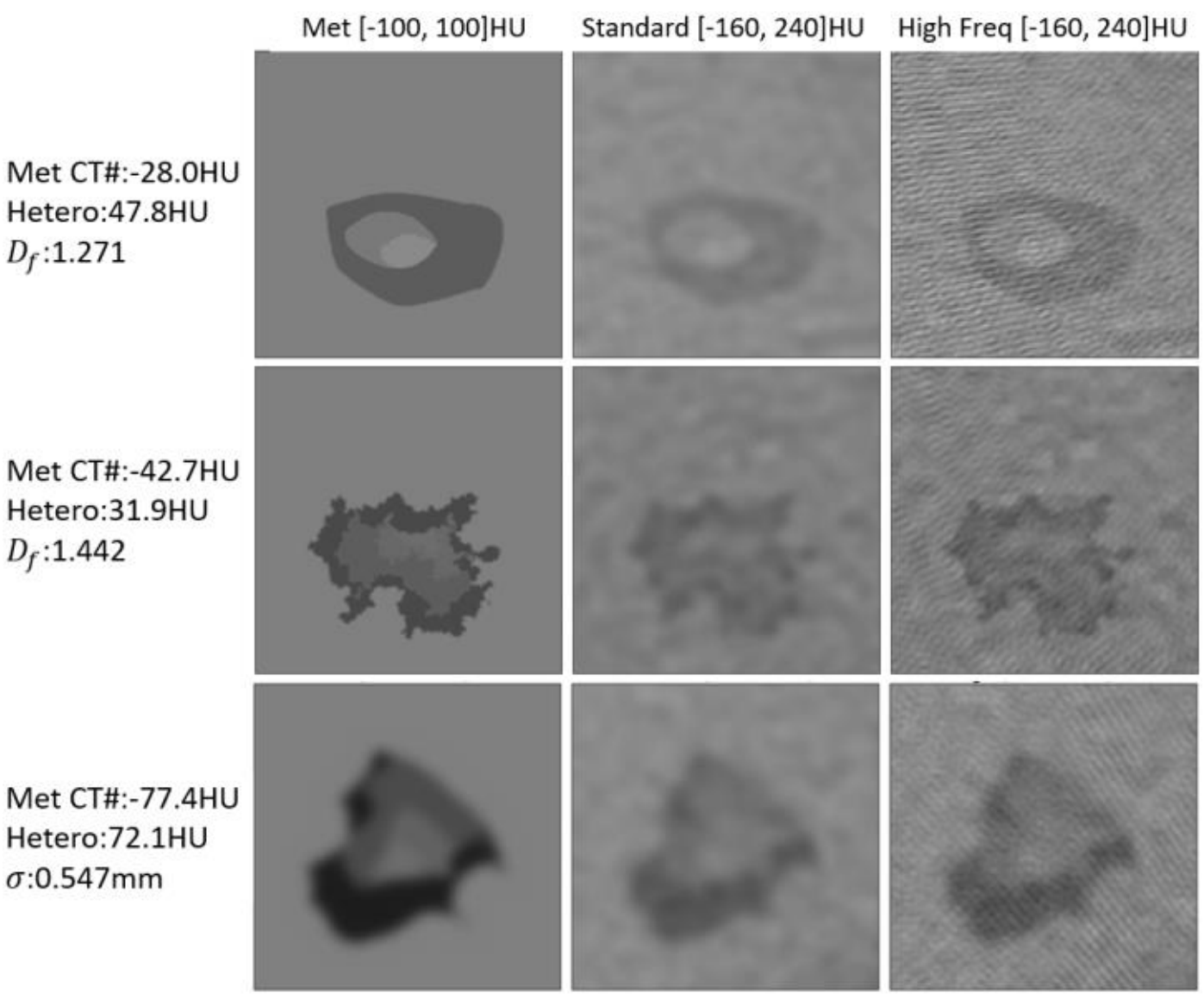

**Figure 4**: Samples of mets pre-simulation (left) and after reconstruction using standard (middle) and high frequency (right). Top two rows are from no-blur study; bottom row is from blurred study.

## 3 Results

Figure 4 illustrates example simulation image patches from standard and high-frequency reconstructions. One can notice greater noise but sharper edges in the high-frequency images. Additionally, this figure visualizes some typical met characteristics.

Characterization performance as the average squared error on validation label prediction (normalized by label variance) is reported in Figure 5. Two-tailed paired t-tests found significant differences between standard versus high-frequency reconstructions for predicting edge sharpness ($\alpha = 0.012$) and fractal dimension ($\alpha = 0.049$).

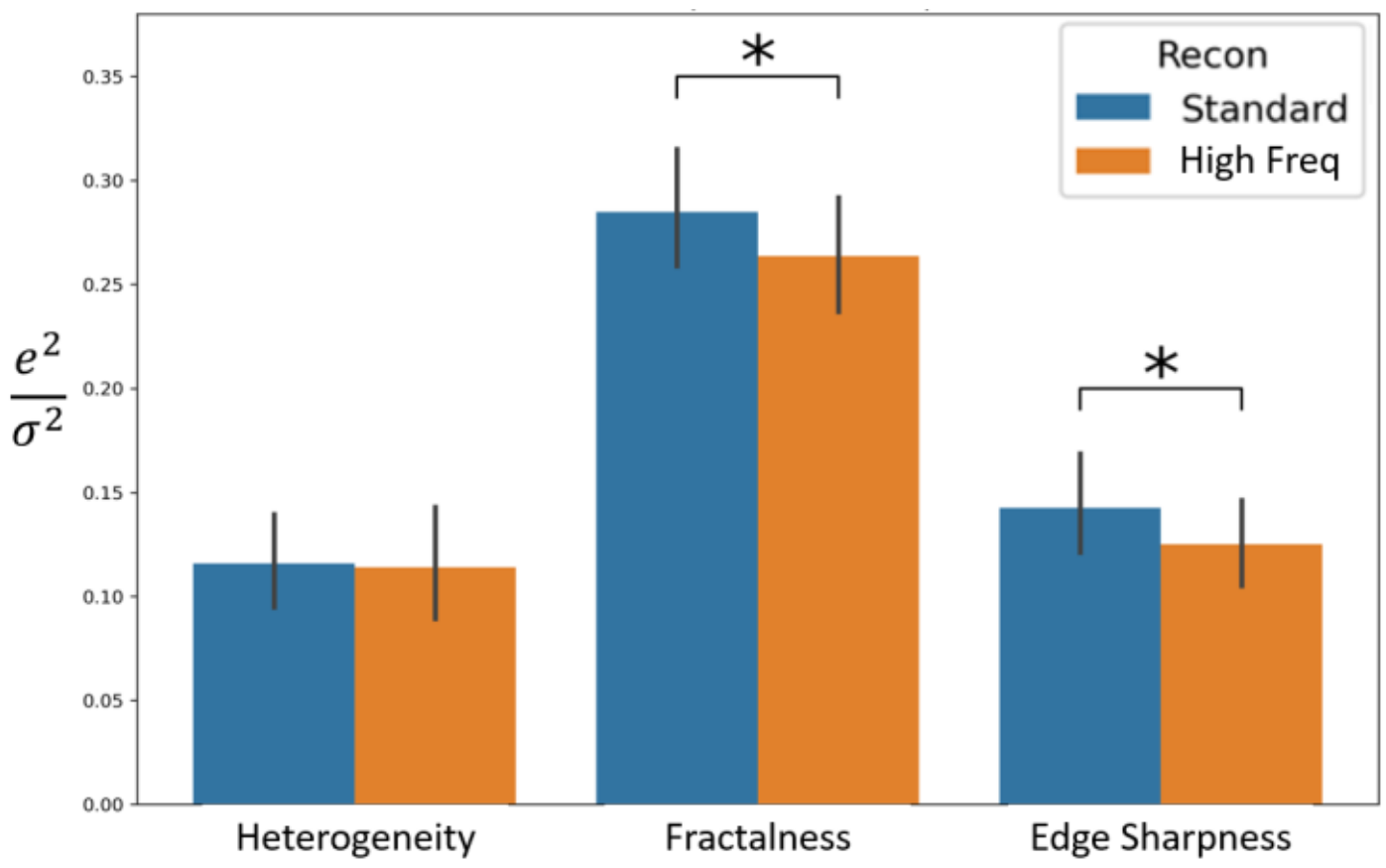

**Figure 5**: Squared error (normalized by label variance) of characterization on validation data. Error bars show 95% CI. * denotes statistically significant difference (p<0.05).

## 4 Discussion

Generating a realistic but diverse population of mets for image simulation is an outstanding problem, and the increasing complexity and variety of imaging systems has elevated the demand for virtual clinical trials [17]. Despite the relative simplicity of our approach, the output population of mets appear both plausible and diverse in terms of their shape and internal structures. Furthermore, unlike methods that superimpose shapes over reconstructed images, our approach generates shapes at high resolution *prior* to scanning and reconstruction. Shape information is known precisely, allowing one to evaluate different imaging and reconstruction settings on feature preservation.

While this simulation approach is advantageous for evaluating imaging methods, a drawback is the lack of direct clinical labels. The generated mets have well-defined anatomical characteristics, but clinical labels differentiating malignancy cannot be generated in this fashion. However, many anatomical features of mets, such as fractalness and texture, relate to clinical classification. Consequently, imaging methods that better preserve these features are assumed to have more clinical information.

Compared to the standard kernel, high-resolution reconstructions contain aliasing-like noise patterns, but also sharper edges (Figure 4). The high-frequency kernel preserves frequencies higher than what is classically permitted by the Nyquist theorem. Standard filtering reduces the noise but inevitably destroys high frequency features in this process. The results shown in Figure 5 indicate that deep learning methods can more accurately recover high frequency characteristics, such as edge sharpness and fractal dimension, from high-frequency as opposed to standard kernel reconstructions. While the noise patterns from high-frequency filtering are less appealing to humans, sufficiently trained deep models can leverage the underlying detail that is preserved by this reconstruction. Future research should investigate high-frequency reconstruction and the impact of scan noise. High-frequency reconstruction can also be used in conjunction with downstream data-driven processing, such as other analysis tasks or deep image denoising [24].

## 5 Conclusion

In this paper, we propose and assess a simulation pipeline for studying scanning and reconstruction methods for CT in CLRM imaging. Our fractal-based method for met synthesis is fast and simple yet generates a diversity of plausible shapes and variations. Our deep radiomics analysis suggests that the proposed high-frequency filter reconstruction is superior for preserving high frequency features such as edge fractalness and sharpness and might reasonably be expected to better discriminate alternative image-based metrics considered for diagnostic purposes in low noise scenarios. Future studies should expand these simulation methods to improve clinical translation and add more features such as complex texture variations. Additionally, future work should investigate using high-frequency reconstruction in low-noise, high resolution imaging applications and with data-driven image tasks, such as deep denoising and further analysis.

**Acknowledgements** Research reported in this publication was supported by the NIH/NCI grant# R01CA233888. The content is solely the responsibility of the authors and does not necessarily represent the official views of the NIH.

# Deep Learning-Based Metal Object Removal In Four-Dimensional Cardiac CT

Pengwei Wu[1], Elliot R. McVeigh[2], and Jed D. Pack[1]

[1] GE Research, One Research Circle, Niskayuna, NY, USA 12309
[2]Dept. of Bioengineering, Medicine, Radiology at University of California San Diego, CA, USA

**Abstract**: Metal objects in human hearts such as pacing leads can result in severe artifacts in reconstructed 4D CT images and hinder the development of fully automatic whole-heart functional analysis algorithms. These artifacts are difficult to remove with the conventional two-pass metal artifacts reduction (MAR) methods due to non-uniform cardiac motion over the heart cycle.

In this work, a pure image domain processing pipeline was developed to generate metal and motion artifacts free 4D CT images without accessing projection domain data. Most of the current deep learning-based MAR methods are supervised learning methods trained with emulated data and suffer from performance degradation when being translated to real patient data. To address this (domain shift) issue, we propose a new generative adversarial network that utilizes both, (i) labeled "hybrid" data (i.e., metal free clinical data augmented with emulated metal objects) and (ii) unlabeled real patient data, through a domain discriminator. The resulting metal-free images were then processed sequentially with a partial angle reconstruction-based motion estimation method.

Experimental results from real patient data confirmed the benefit of using both "hybrid" and real patient data, which resulted in a 57.6% and 43.2% improvement respectively in a uniformity metric compared to hybrid data only and real patient data only approaches. Application of the proposed processing pipeline led to significantly reduced metal and motion artifacts, resulting in images suitable for downstream automatic cardiac functional analysis tasks [e.g., left ventricle (LV) segmentation, LV function, cardiac perfusion, treatment response prediction].

## 1. Introduction

Metal artifacts present a major challenge in fully automatic functional analysis of 4D cardiac CT. Factors underlying the shading and streaks commonly termed metal artifacts include beam hardening, x-ray scatter, and photon starvation.[1] In cardiac CT, metal artifact from pacing leads is a common reason that functional analysis of the left ventricle (LV) requires time consuming human guidance and editing. While many conventional and deep learning approaches for metal artifact reduction (MAR) exist,[1,2] very few include support for dynamic objects. Success can be achieved even for moving pacing leads provided that the leads can be successfully segmented in the projection data.[3] However, doing so still requires access to raw projection data, which poses additional constraints on an otherwise purely image domain functional analysis workflow.

Motivated to address the lead artifacts issue in a fast and broadly applicable manner, in this work, we design an image domain metal artifact removal network for cardiac CT that handles *moving* pacing leads. The proposed network and training strategy properly utilize both labeled "hybrid data" (i.e., metal free clinical data augmented with emulated metal objects) and unlabeled real data (i.e., patient data with real moving pacing leads). The resulting metal-free reconstructions can then be processed sequentially with a previously proposed image domain motion compensation method.[4] Together, this processing pipeline provides metal and motion artifacts free image suitable for fully automatic functional cardiac analysis[5,6] (e.g., LV segmentation, LV function, cardiac perfusion, treatment response prediction).

## 2. Materials and Methods

As illustrated in Fig. 1: the proposed cardiac CT analysis pipeline involves three steps: (i) With raw / uncorrected images as input, the metal removal network [Fig. 1(a)] generates metal free images, which are then (ii) corrected for motion artifacts with an image domain motion compensation approach [Fig. 1(b)]. Finally, the resulting metal and motion artifact free images at different cardiac phases are used for cardiac function analysis [the key step, LV segmentation, is shown in Fig. 1(c)].

### 2.1. Deep Learning-Based Metal Object Removal

Existing deep learning-based MAR methods are mostly supervised learning methods, which requires paired metal-contaminated and metal-free reconstructions as training data. Since it's extremely difficult to acquire such paired data with real patients, most supervised MAR approaches resort to emulated data instead. However, errors in the emulation process (e.g., imperfectly modeled motion, beam hardening and scatter process) could result in severe performance degradation when translating the trained networks to real clinical datasets. This is typically referred to as the "domain shift" issue.[7] While unsupervised learning MAR approaches do not suffer from this issue[8], structural fidelity of their output are sometimes questionable due to a lack of pixel-wise similarity loss function.

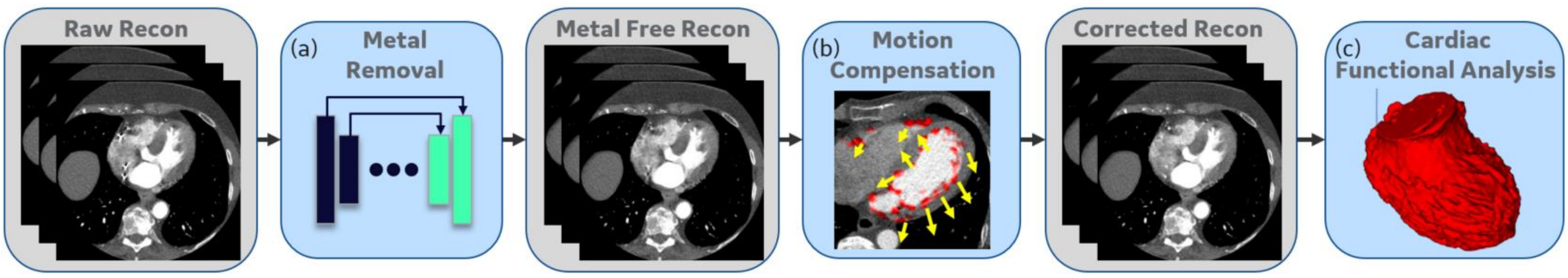

*Figure 1. Illustration of the 4D cardiac CT processing pipeline. Both (a) metal removal and (b) motion compensation steps are performed in the image domain without accessing raw data. Functional analysis is then performed on metal and motion artifacts corrected reconstructions.*

In this work, we propose a metal removal network that utilizes supervised learning from both "hybrid" data and real data domains (Fig. 2). Training input from the hybrid domain ($x_H^U$) was generated by adding emulated metal objects onto metal-free patient data (training label, $x_H^C$). As mentioned above, the fidelity of $x_H^U$ is limited by the accuracy of the emulation process (i.e., "imperfect input"). Training labels within the real domain ($x_R^C$) were generated via interpolating real metal-contaminated patient data (training input, $x_R^U$) in the projection domain. Residual artifacts from the interpolation process reduces the fidelity of $x_R^C$ compared to $x_H^C$ (i.e., "imperfect label"). Details of the training data generation procedure can be found in Sec. 2.2. Note that accessing projection domain data is only required in the network training stage.

As shown in Fig. 2, a 3D residual U-Net[9] (generator, $G$) was used to translate metal contaminated reconstructions from both hybrid and real domains into metal-free ones. Residual blocks[10] were added at each of the 4 levels of the encoding and decoding path of the U-net [Fig. 3(a)]. We utilized a domain discriminator ($D$) to differentiate between generator outputs from hybrid or real data domains. Since labels from the hybrid domain ($x_H^C$) are of higher fidelity, this adversarial framework was specifically focused on domain adaptation, where $G$ was trained to produce outputs similar to hybrid data labels with inputs from either hybrid or real domains. A three level Patch-GAN classifier was used as the domain discriminator as shown in Fig. 3(b).

The network was trained by solving the following minimax optimization problem:

$$\min_D \max_G \mathcal{L}_{GAN}(D,G) + \lambda_S \mathcal{L}_S(G) \quad (1)$$

where $\mathcal{L}_{GAN}$ and $\mathcal{L}_S$ are the GAN loss and pixel-wise similarity loss respectively. The GAN loss between the generator and the domain discriminator can be written as:

$$\mathcal{L}_{GAN}(D,G) = \mathbb{E}_{x_R^U}\left[-\log(1-D\left(G(x_R^U)\right)\right] + \mathbb{E}_{x_H^U}\left[\log\left(D(G(x_H^U))\right)\right] \quad (2)$$

Slightly different similarity losses were used for hybrid and real domains. Only the L1 loss was used for the hybrid domain thanks to its completely metal artifact free training label $x_H^C$. For the real domain, a weighted combination of L1 loss and gradient correlation loss[11] was used to account for residual artifacts in its training label $x_R^C$. Training was performed with the Adam optimizer (learning rate $5\times10^{-4}$, batch size = 32). The generator was first trained for 100 epochs with only similarity losses. Following that, both the generator and discriminator were trained following Eq. (1) for another 100 epochs.

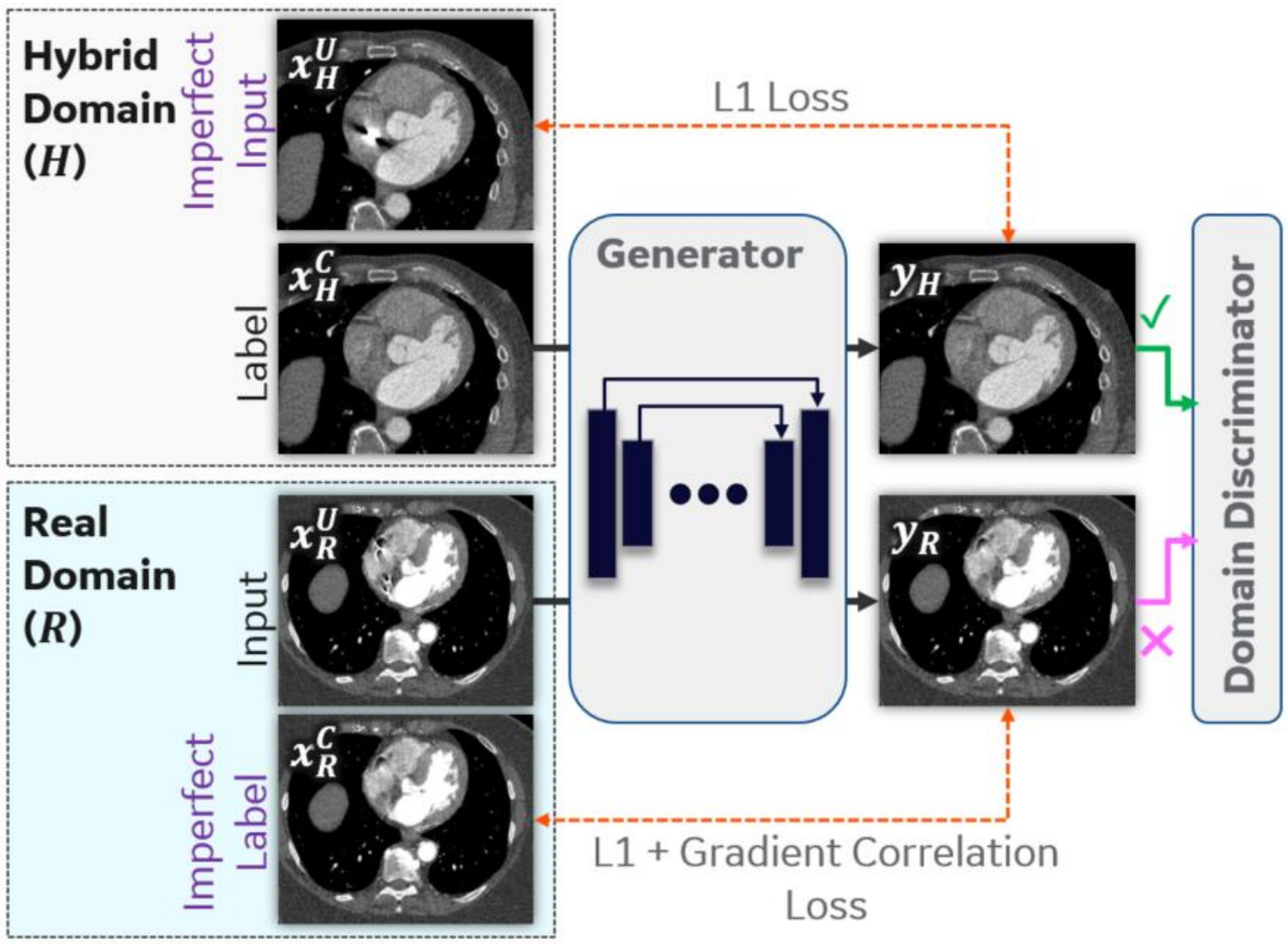

*Figure 2. Proposed framework for combining training data from hybrid (metal free patient images plus emulated metal objects) and real (real metal-contaminated patient images) domains.*

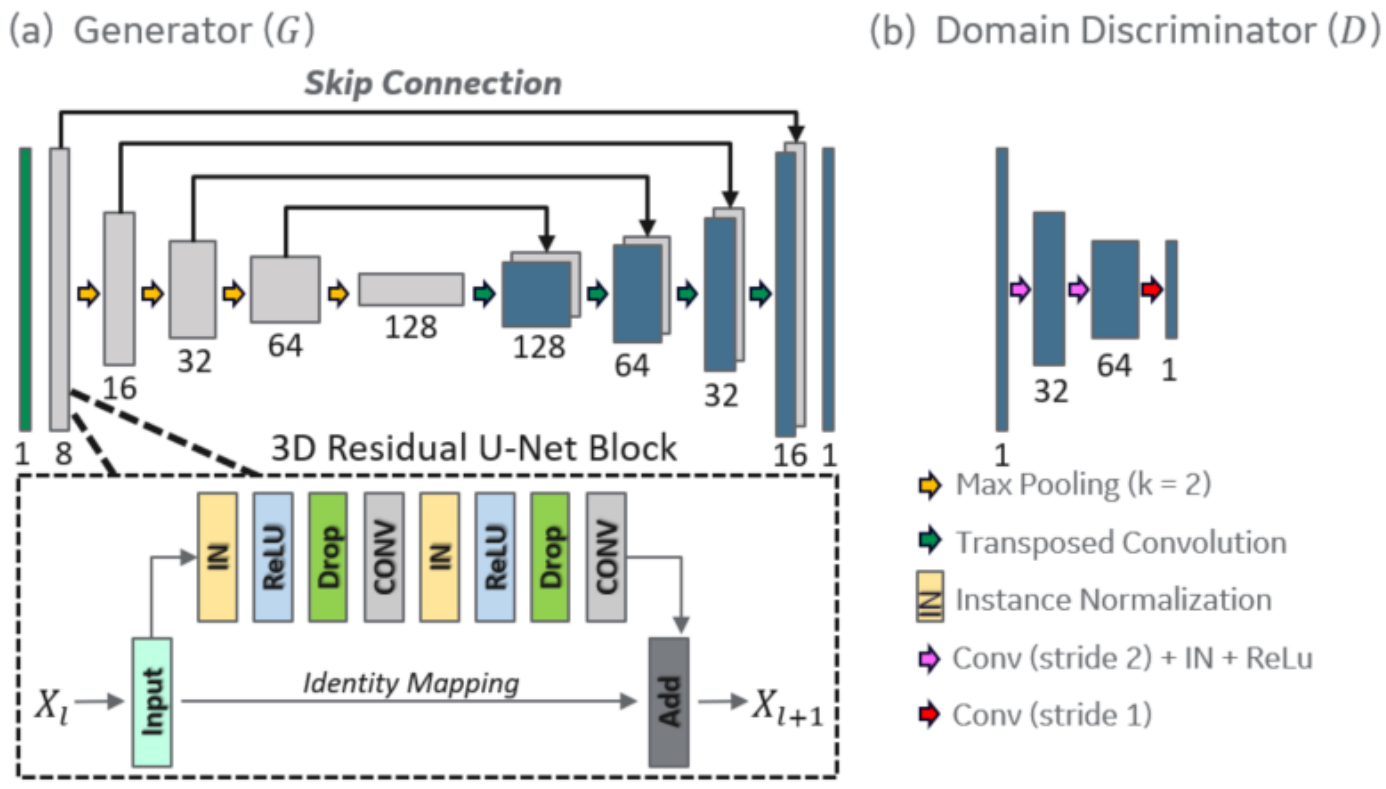

*Figure 3. Illustration of the network architectures: (a) Generator (G). (b) Domain discriminator (D).*

## 2.2. Training Data Generation

Using the CatSim toolkit[12] tuned for a 256-slice multi-detector CT system (Revolution CT™, General Electric Healthcare, Chicago, IL, US), a metal free thoracic phantom was forwarded projected with and without a library of analytical pacing lead object models at different 3D locations and orientations. Realistic metal traces (including metal attenuation, metal-induced scatter, and beam hardening effects) were obtained by subtracting metal-free forward projections from the metal-contaminated ones. These metal traces were randomly selected and combined with real metal free patient data to provide paired training data for the hybrid domain.

A projection domain interpolation type approach was used to provide the (imperfect) training labels for the real domain. Due to pacing leads motion, the conventional two-pass approach (image domain metal segmentation followed by forward projection) cannot be used to identify metal trace locations in the projection domain.[1] Instead, we propose a semi-automatic metal trace delineation approach, where 20 B-spline knots were manually placed along the pacing lead trace for one out of every 70 views. Cubic spline interpolation was then performed across and between views to create a complete metal trace delineation.

Images in both hybrid and real domains were reconstructed with short-scan 3D filtered backprojection method on a volumetric grid covering a 30×30×16 $cm^3$ FOV (axial plane voxel spacing: 0.58×0.58 $mm^2$; slice thickness: 0.625 mm). The metal artifact removal network was trained using 4200 paired 64×64×64 patches from 7 hybrid domain scans and 7 real domain scans. Two additional real domain scans were used as testing cases.

### 2.3. Motion Compensation

Following the removal of metal objects, we integrate a motion compensation approach into the processing pipeline, which is based on conjugate pairs of partial angle reconstructions (PAR).[4] Fourier domain wedge filters were used to generate conjugate PARs in image domain (separated in time by a half rotation). Motion vector fields between conjugate PARs were estimated through cross-correlation functions. These PARs were then warped by the estimated MVFs and recombined to form the motion compensated reconstructions. Note that in this work, motion compensation was performed after metal artifacts removal and in image domain.

## 3. Results

To demonstrate the benefits of our proposed network and joint hybrid-real training strategy, two additional networks were trained with (i) hybrid domain data only (using the same amount of training patches as the joint hybrid-real approach); (ii) real domain data only (using all real domain training patches in the joint hybrid-real approach, without any hybrid domain training patches). These two networks have the same structure as shown in Fig. 2 *sans* the domain discriminator, since the training data only comes from a single domain (hybrid or real).

Figure 4 shows the performance of these three metal artifacts removal networks on one example test case in the real domain, which features a large in-plane pacing lead coil near the septum, preventing reliable visualization and edge delineation of the LV myocardium from LV blood pool. The proposed joint hybrid-real approach [Fig. 4(d)] outperformed the other two approaches in terms of visual fidelity and residual artifacts [pointed by the pink arrows in Fig. 4(b-c)]. The image texture in Fig. 4(d) also appears more realistic in both axial and coronal planes, likely due to the introduced domain discriminator that promotes similarity to artifacts-free labels in the hybrid domain. As there was no ground truth image for comparison, quantitative analysis was performed by measuring the Hounsfield unit uniformity across a uniform region [i.e., 7 yellow ROIs within the myocardium in Fig. 4(a)]. The proposed method achieved a uniformity of less than 15 HU within the myocardium, which is 57.6% and 43.2% better compared to hybrid and real domain only approaches respectively.

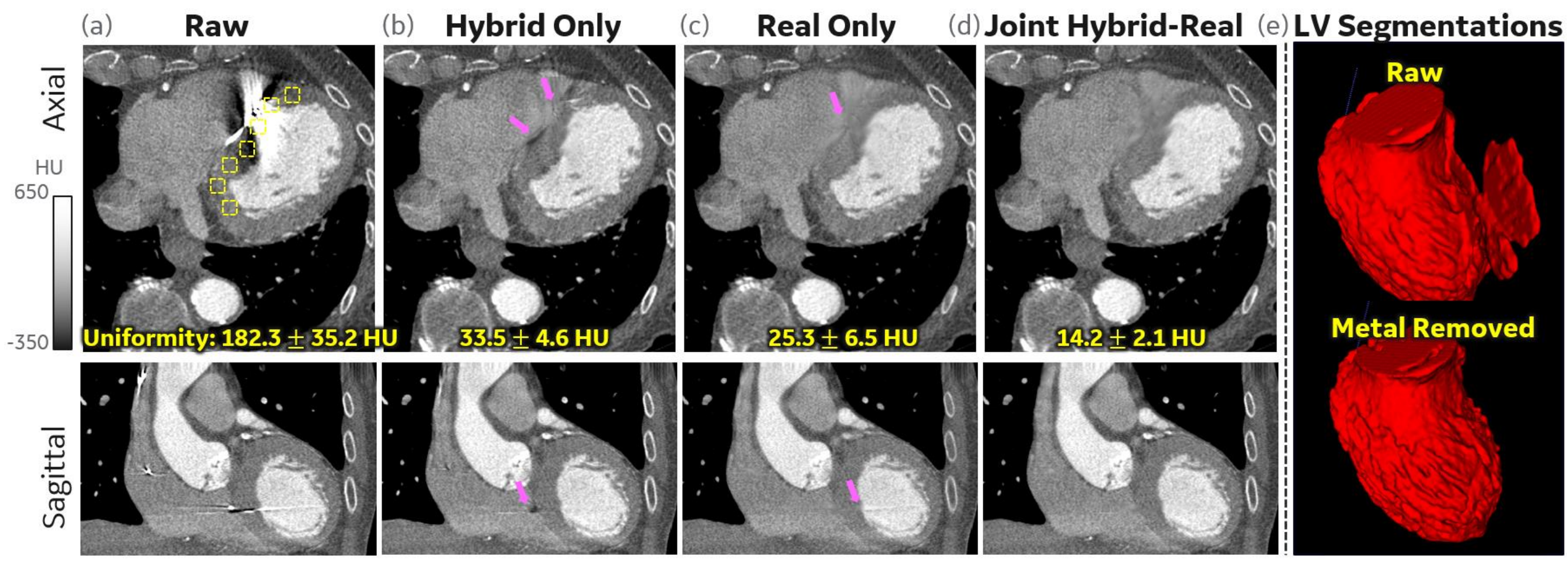

*Figure 4. Performance of the metal objects removal network on an example test case (patient with real metal objects). (a) Raw / uncorrected reconstruction. (b) Output from the network trained with hybrid domain data only. (c) Output from the network trained with real domain data only. Residual artifacts in (b) and (c) are pointed by the pink arrows. (d) Output from the network jointly trained with both hybrid and real domain data using a domain discriminator (Fig. 2). Hounsfield unit uniformity is measured from 7 yellow dashed ROIs placed inside myocardium. (e) Automatic level set-based LV blood pool segmentations on (a) (top row) and (d) (bottom row) respectively.*

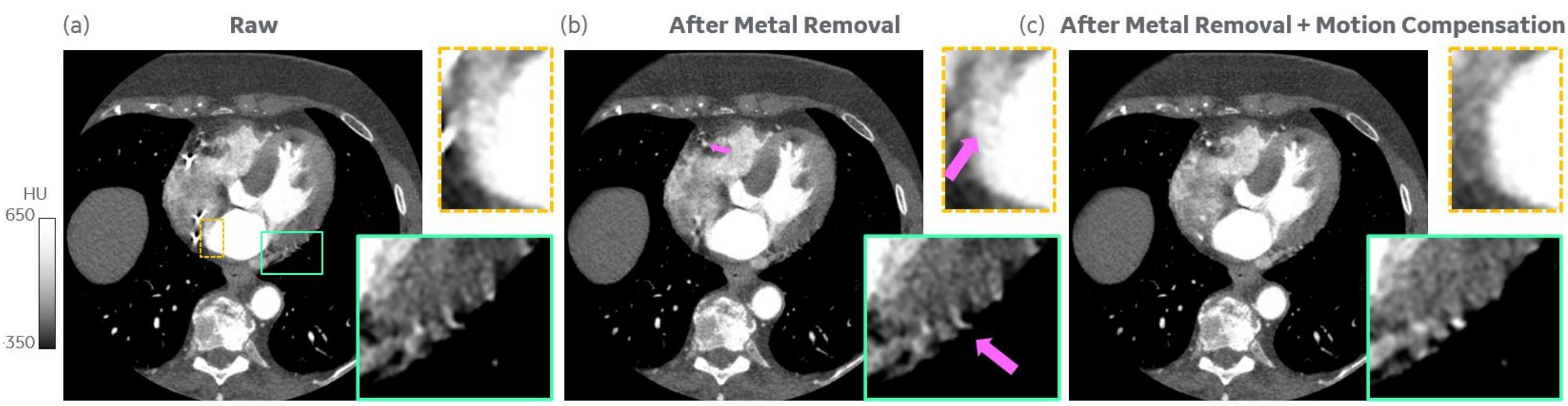

*Figure 5. Performance of the proposed 4D cardiac CT processing pipeline. (a) Raw reconstruction. (b) Reconstruction after metal artifacts removal. (c) Reconstruction after metal artifacts removal + motion compensation (i.e., input to automatic cardiac functional analysis).*

Figure 4(e) shows the dramatic improvement in level set-based automatic LV blood pool segmentation[13] with the proposed processing pipeline. After metal artifact removal, the segmentation becomes much more accurate owing to greatly reduced streaks and blooming artifacts and better-defined LV boundaries as shown in Fig. 4(d). This improvement in segmentation performance results in a high degree of reproducibility in cardiac functional analysis despite the original pacing lead artifact.

Figure 5 illustrates the feasibility of sequentially adding motion compensation immediately after metal artifact removal in the processing pipeline. The metal artifact removal network is robust to the presence of motion objects as shown in Fig. 5(b). Motion compensation resulted in a better-defined ascending aorta edge (orange zoom in Fig. 5), right coronary artery [RCA, top left pink arrow in Fig. 5(b)], and carotid arteries (cyan zoom in Fig. 5). Note again that these results were achieved purely in the image domain without accessing raw / projection domain data.

## 4. Discussion and Conclusion

This work presents a novel pure image domain metal object removal and motion compensation pipeline for 4D cardiac CT. The deep learning-based metal object removal network features a domain discriminator, which addresses the domain shift issue and allows utilization of both hybrid and real metal contaminted reconstructions as training data. Compared with hybrid and real data only approaches, the proposed joint hybrid-real training approach resulted in improved accuracy of metal removed reconstructions and uniformity in the LV muscle. Downstream tasks (e.g., LV segmentation) became much easier and more reliable after removing metal artifacts.

In this work, the network was trained to remove only metal artifacts even in the presence of metal motion, making it suitable to be chained directly with image domain motion compensation methods. Since many exisiting motion estimation method rely on registration metrics such as cross correlation or mutual information, removing high contrast metal artifacts beforehand could potentially improve the reliability of these methods. Together, the proposed rocessing pipeline provides metal and motion artifact free images for fully automatic cardiac functional analysis. Because the method is in the image domain, large clinical registries of 4D Cardiac CT studies can have lead artifacts removed retrospectively for inclusion in important studies evaluating the use of 4DCT in prognosis for many cardiac conditions (heart failure, post myocardial infarction, cardiac resynchronization therapy, for example). Ongoing work includes expanding training and testing testing datasets as well as more rigirous validation of our training data generation framework.

## Acknowledgements

Research reported in this publication was supported by NHLBI of the National Institutes of Health under award number R01 HL14678 (McVeigh). The content is solely the responsibility of the authors and does not necessarily represent the official views of the National Institutes of Health.

# Filter-independent CNN-based CT image denoising

Christian Wülker[1], Nikolas D. Schnellbächer[1], Frank Bergner[1], Kevin M. Brown[2], and Michael Grass[1]

[1]Philips Research Hamburg, Germany
[2]Philips Healthcare Cleveland, OH, USA

**Abstract** Different reconstruction filters are used in CT imaging to promote sharpness or suppress noise, for example. Designing machine-learning algorithms for CT image processing that can be used with different reconstruction filters, however, remains a challenge. In particular, it has recently been reported that CT image denoising based on convolutional neural networks (CNNs) generalizes poorly to different reconstruction filters and corresponding noise power spectra (NPS). While it is conceivable to train different CNNs for different reconstruction filters each in a dedicated manner, in this paper we argue that such a machine-learning algorithm for image denoising can and should be made fully independent of the reconstruction filter, instead. In particular, we show that it is well possible to train a single CNN-based denoising model for a standard ramp filter, and obtain the desired filter characteristics in the denoised images through an additional fast post-processing step. This is demonstrated both by visual and quantitative comparison using a clinical CT scan of the abdomen.

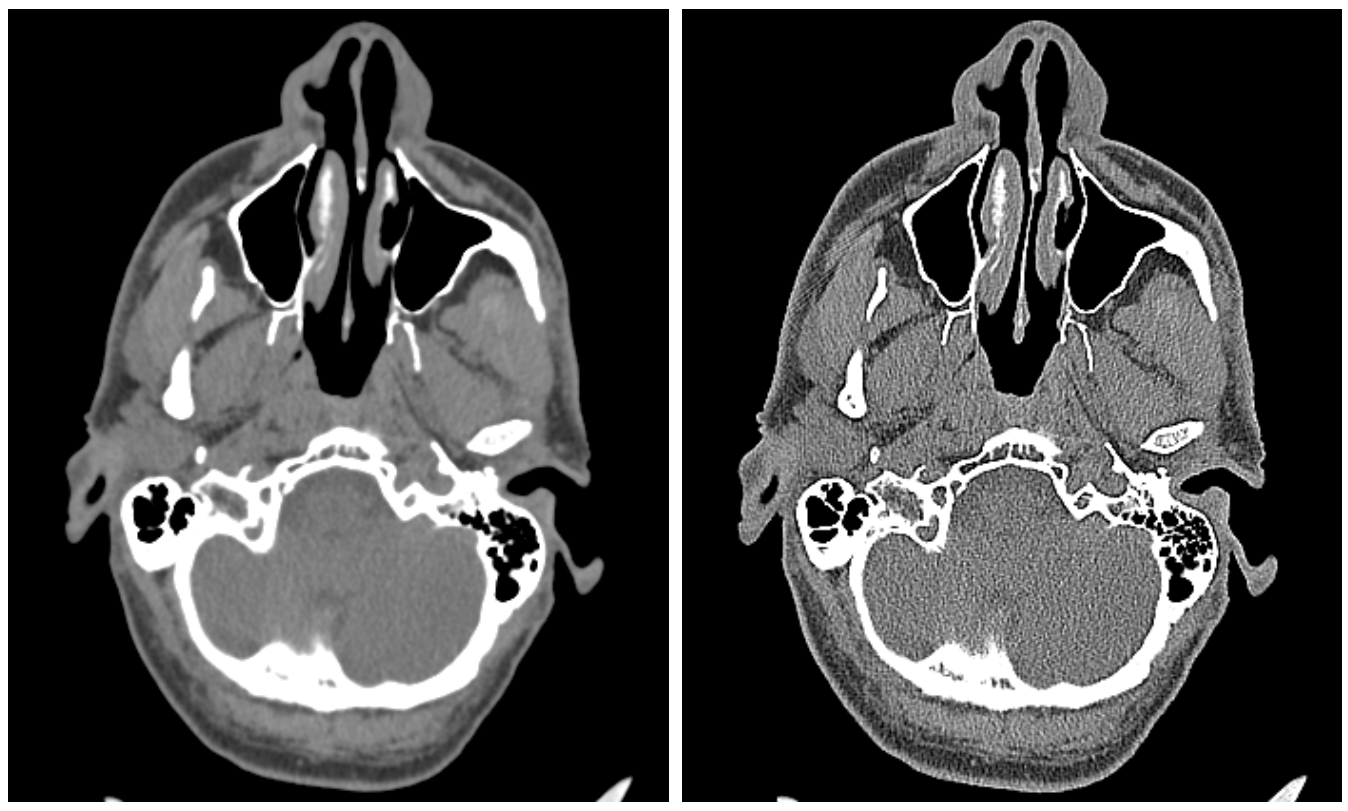

**Figure 1.** Exemplary CT images of the head (cropped FOV), reconstructed with filtered backprojection (FBP) using **(left)** a soft reconstruction filter and **(right)** a sharp reconstruction filter.

## 1 Introduction

All medical imaging modalities are susceptible to noise due to inherent statistical effects in physical signal generation and data acquisition. In X-ray computed tomography (CT), keeping the radiation exposure *as low as reasonably achievable* (ALARA) is a guideline commonly agreed upon. Low-dose CT, however, comes at the cost of higher noise levels in the reconstructed images. This explains the need for fast high-quality image denoising algorithms. Convolutional neural networks (CNNs) are a powerful machine-learning technique well suited to perform image denoising in various medical image domains.

CT images are reconstructed in practice using different reconstruction filters, thereby promoting sharpness or suppressing noise, for example (Fig. 1). It is a well-understood goal to have image processing algorithms such as for image denoising that are robust w. r. t. such changes of the reconstruction settings. Machine-learning methods, and in particular deep learning, however, are prone to overfit to the training data and thus often fail to generalize to parameter settings not sampled in training. In particular, it has recently been reported in [1] that CT image denoising based on CNNs generalizes poorly to different reconstruction filters and corresponding noise power spectra (NPS). We have made the same observation.

While it is conceivable to train multiple CNN-based denoising models for different reconstruction filters each in a dedicated manner, it is highly undesirable or may even be unfeasible to train and maintain a large number of trained CNNs as part of such a denoising solution. In this paper, we argue that CNN-based CT image denoising can and should be made fully independent of the reconstruction filter, instead, thereby reducing the number of required CNN trainings substantially. Specifically, we propose to train a single CNN-based denoising algorithm for a standard ramp filter, and always start out from images reconstructed with such a generic filter. After denoising the images with this ramp filter-trained CNN model, different image characteristics can be achieved through an additional fast filtering step in the image domain.

Our approach is motivated by the fact that standard reconstruction filters consist of two parts, namely the ramp and an additional multiplicative modulation. We refer to the latter as the modulation transfer function (MTF). The well-known Fourier slice theorem shows that the application of the reconstruction filter corresponds to a multiplication of the image spectrum with a (properly defined) filter spectrum. Notably, this holds true for the reconstruction filter parts independently, *i.e.*, we may first apply the ramp part of the filter during FBP to reconstruct an image, and subsequently filter the images with the MTF in the image domain. We have found empirically that the MTF part of the filtration commutes very well with CNN-based image denoising. This enables us to train a more generic CNN-based denoising model for the ramp filter, and obtain the desired filter characteristics in the denoised images through an additional fast post-processing step. Our proposed solution is illustrated in Fig. 2.

## 2 Materials and methods

### 2.1 Data

Raw data from five clinical body scans performed with a Philips iCT with an energy-integrating detector were used to test our strategy for reconstruction-filter agnostic CNN-based CT image denoising. Three cases were used as training data, one case as validation data (see below), and one case was reserved for testing. The tube current in these scans was 335 mA, 347 mA, and 298 mA (training cases), 384 mA (validation case), and 447 mA (testing case). The tube voltage was 120 kV in all scans. All scans had the same rotation time (500 ms) and approximately the same table speed (8 cm/s). We applied an advanced low-dose simulation technique [2] to the acquired raw data to simulate a 25% dose level from the acquired 100% dose level by artificially reducing the tube current by a factor of four. Images were reconstructed with FBP using three different reconstruction filters: a ramp filter, a sharp body filter, and a soft body filter. Reconstructions were performed with $512 \times 512$ matrix size and a field of view (FOV) of 400 mm. The slice thickness and increment were 1.0 mm and 0.5 mm, respectively. No image-based denoising was used at this stage. A total of 3,779 (1,425) slice images for each dose level (100% and simulated 25%) were reconstructed for training (validation, respectively).

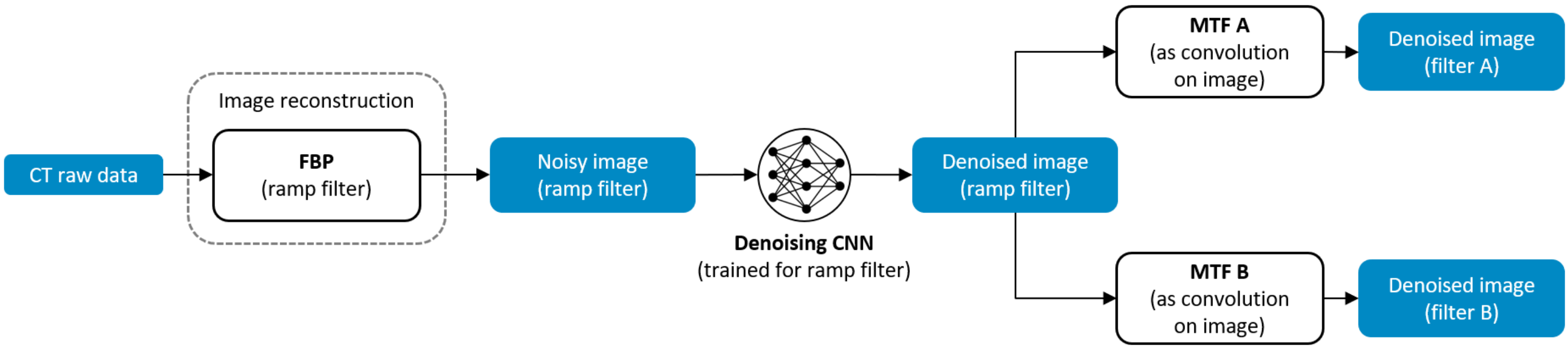

**Figure 2.** Proposed workflow for CNN-based CT image denoising. In this approach, one always reconstructs the image(s) with a ramp filter first. Then a generic denoising CNN is applied, which is trained to operate with ramp filter only. The desired filter characteristics of the images are then obtained through an additional fast post-processing step in the image domain.

## 2.2 CNN-based CT image denoising using a bias-free DnCNN

Machine-learning and in particular deep-learning approaches have recently gained popularity in medical image denoising due to substantial improvements over classical methods (see [3, 4], for example). Specifically, in *supervised learning*, models in the form of convolutional neural networks (CNNs) with dedicated topology/architecture are trained to remove noise from an input image by presenting to the model noisy images and noise-free or low-noise counterparts during training phase. Noisy images can be generated by imposing (additional) noise with the desired statistical characteristics on noise-free or low-noise images (as in our case), or by directly acquiring images with noise. The model may be trained in *residual learning mode*, which means that the network is not trained to directly produce a denoised image from the input, but instead provide as output an estimate of the noise in the input image. This noise estimate is then (partially) subtracted from the input after inference to yield a (partially) denoised image. Pixel-wise loss functions such as *mean squared error* (MSE, also referred to as $L^2$ loss) or *mean absolute error* (MAE, also referred to as $L^1$ loss) are common choices, but other types of loss function such as *structural similarity index* (SSIM) and *adversarial losses* have also been used.

An aspect of high practical relevance is whether the noise level in the input images is fixed/known *a priori*. If the noise level is not known during model inference, one speaks of *blind* image denoising. It is generally desirable to have a model that is robust w. r. t. varying noise levels (in medical imaging, often not only between different images but often also within a single image). Such a robust solution to the problem of blind image denoising has recently been introduced in [4], and we use this approach in this paper to demonstrate our approach to filter-agnostic CNN-based CT image denoising.

Specifically, we used PyTorch[1] to train the bias-free denoising CNN (BF-DnCNN) described in [4] in residual learning mode using a standard *ADAM ('adaptive moment estimation')* optimizer [5] with a learning rate of 1e-4 and with MSE as the loss function. Training was carried out three times in exactly the same manner for the ramp filter, the soft reconstruction filter, and the sharp reconstruction filter, respectively. Random initial weights (parameters) for the untrained CNNs were automatically generated by PyTorch. The BF-DnCNN architecture we used has 16 convolutional layers with $3 \times 3$ kernel size and without padding (so-called *valid convolutions*), each followed by a leaky ReLU as nonlinearity, plus one final layer with a $1 \times 1$ convolution. The number of channels between the convolutional layers was 64. Training was done in a patch-wise manner, *i.e.*, the networks were trained on image patches with a size of $64 \times 64$ pixels (the output patch size was thus $32 \times 32$). Random flipping in both directions was done on-the-fly during training as data augmentation. The target data were the corresponding patches containing only the noise from the input (that is, the difference between simulated 25% dose and acquired 100% dose images). The networks were each trained for 2,000 epochs, where one epoch was defined as one sweep over 42,513 ran-domly drawn patches using a minibatch size of 200 patches. We saved the model with the best validation error, where validation was done after each epoch using the same 3,000 patches that were drawn randomly from the validation images at the beginning of training (without augmentation).

[1] `https://pytorch.org/`

### 2.3 Image post-filtration

As explained above, the Fourier slice theorem shows that the MTF part of the reconstruction filter may be applied as a convolution operation on the reconstructed images to obtain the desired filter characteristics. This can be done in a fast way using multiplication in the frequency domain. We thus reverse-engineered the MTF part of the soft and the sharp reconstruction filter for body used in this work. This allowed for fast application of these filters to (denoised) images reconstructed with a ramp filter using the fast Fourier transform (FFT) and its inverse. If the exact MTF of the requested reconstruction kernel is known analytically, one can convert this known MTF into its corresponding 2D image domain filter representation for the described post-filtration step after image-based denoising. In absence of this knowledge one has to revert to the reverse-engineering approach as used in this work.

## 3 Experiment

We used the ramp filter-trained BF-DnCNN to denoise the 25% dose images reconstructed with the ramp filter, and applied image post-filtration to the denoised images to obtain images with the filter characteristics of the soft and sharp body reconstruction filter, respectively. For comparison, we applied the corresponding BF-DnCNN models that were dedicatedly trained for the soft and the sharp reconstruction filter to the 25% dose images reconstructed with the soft and the sharp filter, respectively.

## 4 Results and discussion

Figure 3 shows the slice-wise mean of the absolute error (AE) and structural similarity index (SSIM) computed on the test case, providing for a quantitative comparison between the dedicated filter-trained CNNs and our proposed alternative stategy with only one ramp filter-trained CNN. We also show the 95$^{th}$ percentile of the absolute error for each slice (*i.e.*, the absolute error is below this value for 95% of the pixels in each slice). It can be seen that our filter-agnostic method for CNN-based CT image denoising consistently achieves essentially the same values as the dedicatedly trained CNNs, with similar distribution of the absolute error and only slightly different SSIM (slightly lower for the soft filter and slightly higher for the sharp filter; however these particular directional differences were not always observed in other cases that we tested our approach on). Note that the absolute error was computed with the soft and sharp filter-reconstructed

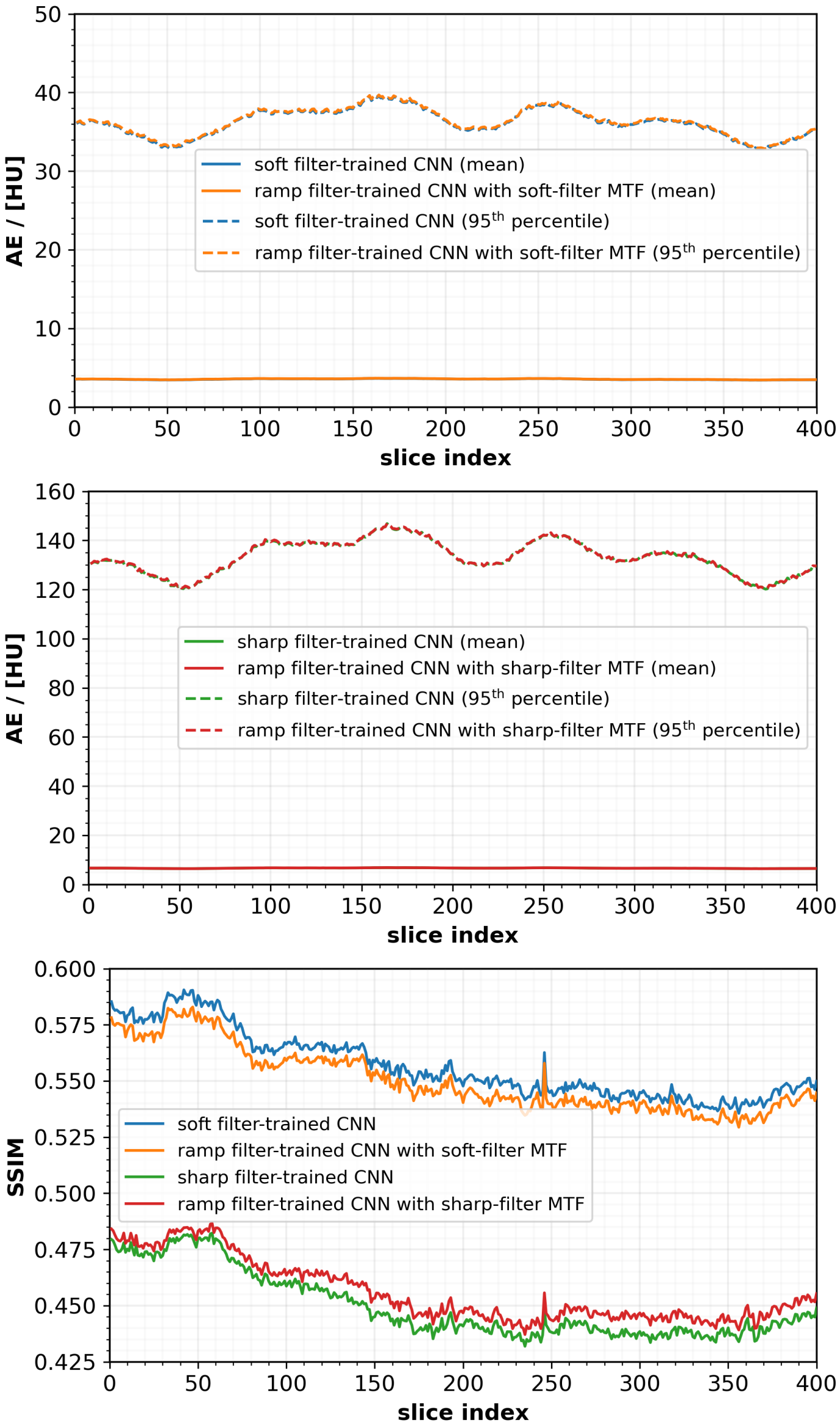

**Figure 3.** Slice-wise absolute error (AE, mean and 95$^{th}$ percentile) and structural similarity index (SSIM), computed on the test case. The horizontal axis runs through 400 slices, starting at roughly the center of the heart and ending roughly at the *ilium*. Both AE and SSIM were computed with the soft and sharp filter-reconstructed images with acquired 100% dose as the respective ground truth.

images with acquired 100% dose as the respective ground truth, which itself contains noise.

Figure 4 shows exemplary denoised images of the abdomen with different filter characteristics, as well as the corresponding ground truth images with acquired 100% dose (left column) and images with simulated 25% dose (center left column). The images in the center right column were generated from the ones in the center left column by applying the two denoising CNNs dedicatedly trained for the soft and sharp reconstruction filter, respectively. While this approach does yield high-quality denoising results, the number of CNNs in this method is unnecessarily high. The right column shows the corresponding results with the proposed alternative strategy. Here a single image with simulated 25% dose was first reconstructed using a standard ramp reconstruction filter (not shown). This image was denoised with the ramp filter-trained CNN-based denoising algorithm. Subsequent application of the MTFs for the soft and the sharp reconstruction filter in the image domain using the FFT and its inverse yielded virtually the same results as those shown in the center right column.
These results clearly demonstrate that it is well possible to train a single CNN-based denoising model for a standard ramp filter, and obtain the desired filter characteristics in the denoised images through an additional fast post-processing step. We thereby arrive at a potentially more robust and broadly applicable solution while at the same time reducing the number of required CNN trainings. Our approach may also reduce the number of reconstruction filters in general, because many of such have essentially been designed to control the SNR of the reconstructed images. This may not be necessary with improved CNN-based CT image denoising anymore.
One aspect of practical relevance is that when using FFTs to apply the MTF to the denoised images, one has to take into account the implicitly assumed periodicity of the image data in order not to introduce wrap-around artifacts with post-filtration. This amounts to using an adequate linear (*i.e.*, not circular) type of convolution.
Finally, we note that the idea presented here may also be used with other types of CNN-based CT image processing.

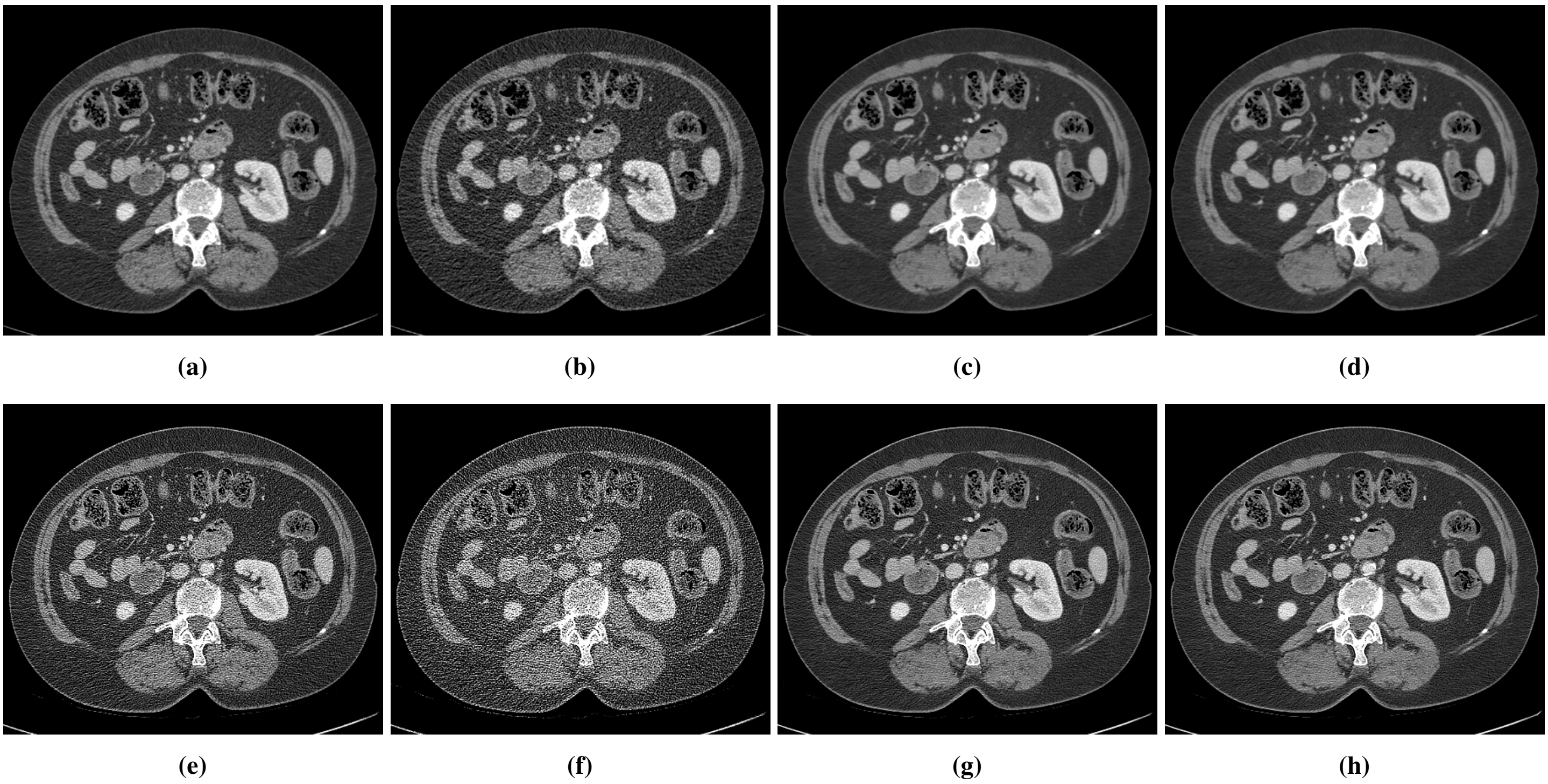

**Figure 4.** Exemplary CT images of the abdomen (cropped FOV), with **(a)** acquired 100% dose reconstructed with a soft body filter, **(b)** simulated 25% dose reconstructed with a soft body filter, **(c)** image (b) denoised with soft filter-trained CNN, **(d)** simulated 25% dose reconstructed with ramp filter and denoised with ramp filter-trained CNN with subsequent application of the soft-filter MTF, **(e)** acquired 100% dose reconstructed with a sharp body filter, **(f)** simulated 25% dose reconstructed with a sharp body filter, **(g)** image (f) denoised with sharp filter-trained CNN, and **(h)** simulated 25% dose reconstructed with ramp filter and denoised with ramp filter-trained CNN with subsequent application of the sharp-filter MTF. Note that the *same denoising CNN* was used to generate the images in the right column, yielding virtually the same results as training different CNNs for different reconstruction filters, shown in the center right column.

## Acknowledgment

The authors would like to thank Roland Proksa (Philips Research Hamburg) and Thomas Köhler (Philips Research Hamburg) for fruitful discussions.

# Patch-Based Denoising Diffusion Probabilistic Model for Sparse-View CT Reconstruction

Wenjun Xia[1], Wenxiang Cong[1], and Ge Wang[1]

[1]Department of Biomedical Engineering, Rensselaer Polytechnic Institute, Troy, NY 12180 USA

**Abstract** For sparse-view computed tomography (CT), the neural networks have limited ability to remove the artifacts with only information in the image domain. The introduction of sinogram can achieve a better anti-artifact performance, but it inevitably requires the feature maps of the whole image to be put into video memory, which makes handling large-scale or three-dimensional (3D) data a great challenge. In this paper, we propose a patch-based denoising diffusion probabilistic model (DDPM) for sparse-view CT reconstruction. A DDPM network based on patches extracted from fully sampled projection data is trained and then used to inpaint the down-sampled projection data. The network does not require paired full-sampled and down-sampled data, enabling unsupervised learning. And the processing of the data is patch-based, which can be distributedly deployed, overcoming the challenge of processing large-scale and 3D data. The experiments also show that the proposed method can achieve excellent anti-artifact performance while maintaining the texture details.

## 1 Introduction

Sparse-view CT has kept receiving wide attention from industry and academia because it can reduce the radiation risk to patients and enable healthier diagnosis. Concomitantly, a number of researchers have devoted themselves to eliminating artifacts and noise in sparse-view CT, making it practical for clinical applications.

In the last two years, the denoising diffusion probabilistic model (DDPM) [1] emerged and achieved success in the image generation field. DDPM gradually adds Gaussian noise perturbation to the image, projects the image into the latent spaces, and uses the network to learn the denoising process of the latent spaces. DDPM overcomes the mode collapse problem of GAN and also exhibits better stability than GAN in image enhancement tasks [2].

Inspired by DDPM, in this paper, we propose a projection patch-based DDPM method for sparse-view CT reconstruction. In the training stage, we use U-Net [1] to learn the reverse diffusion process for fully sampled projection patches. In the sampling process, we first implement fully sampling radon transform on the sparse-view CT images to obtain pseudo fully sampled projection data. Then pseudo fully sampled projection data is cropped into patches as the condition of reverse diffusion. The patches restored by ordinary differential equation (ODE) sampling are put together to get the final projection data, and high-quality reconstructed images are obtained using FBP. The reconstruction with our proposed method can eliminate the artifacts while preserving clinically important details. Also, two additional features of our method make it clinically friendly. First, our method does not require paired data, and both training and sampling are in an unsupervised training framework. Second, the sampling of our method is patch-based, which can split the large-scale data and 3D data into patches or cubes and process them in parallel. These features allow our proposed method to solve the challenges of deep reconstruction in the clinic.

## 2 Methodology

### 2.1 Score-Based DDPM for Patch Inpainting

We denote the fully sampled projection data as $\boldsymbol{Y} \in \mathbb{R}^{N_v \times N_d}$, where $N_v$ and $N_d$ represent the number of projection views and the detector elements, respectively. The down-sampled projection data can be obtained by a linear transform

$$\boldsymbol{Z} = P(\boldsymbol{M} \odot \boldsymbol{Y}), \tag{1}$$

where $\boldsymbol{Z} \in \mathbb{R}^{N'_v \times N_d}$ denotes the sub-sampled projection data, $\boldsymbol{M} \in \mathbb{R}^{N_v \times N_d}$ is the mask, with the entry $\boldsymbol{M}_{ij} = 1$ if the $i$-th view is sampled else $\boldsymbol{M}_{ij} = 0$, $\odot$ represents the element-wise multiplication, and $P : \mathbb{R}^{N_v \times N_d} \to \mathbb{R}^{N'_v \times N_d}$ is the operation to extract the selected view data from the projection data.

To implement the patch-based diffusion, we randomly extract a patch $\boldsymbol{y} \in \mathbb{R}^{d \times d}$ from the fully sampled projection data $\boldsymbol{Y}$. According to [1], the forward process of DDPM is a Markov chain which gradually adds Gaussian noise perturbation to the clean patch $\boldsymbol{y}_0 = \boldsymbol{y}$ with a predefined variance sequence $\boldsymbol{\beta} = \{\beta_1, \beta_2, \cdots, \beta_T\}$:

$$q(\boldsymbol{y}_{1:T}|\boldsymbol{y}_0) = \prod_{t=1}^{T} q(\boldsymbol{y}_t|\boldsymbol{y}_{t-1}), \tag{2}$$

where

$$q(\boldsymbol{y}_t|\boldsymbol{y}_{t-1}) = \boldsymbol{N}(\boldsymbol{y}_t|\sqrt{1-\beta_t}\boldsymbol{y}_{t-1}, \beta_t \boldsymbol{I}). \tag{3}$$

The iteratively perturbed patch at any time step $t$ can be directly obtained from $\boldsymbol{y}_0$:

$$q(\boldsymbol{y}_t|\boldsymbol{y}_0) = \boldsymbol{N}(\boldsymbol{y}_t|\sqrt{\bar{\alpha}_t}\boldsymbol{y}_0, (1-\bar{\alpha}_t)\boldsymbol{I}), \tag{4}$$

where $\alpha_t = 1 - \beta_t$ and $\bar{\alpha}_t = \prod_{i=1}^{t} \alpha_i$.

And a U-Net [1] can be used to learn the Gaussian perturbations:

$$\boldsymbol{L} = \mathbb{E}_{\boldsymbol{y},\boldsymbol{\varepsilon},t} \left\| \boldsymbol{\varepsilon} - \boldsymbol{\varepsilon}_{\boldsymbol{\theta}}(\sqrt{\bar{\alpha}_t}\boldsymbol{y}_0 + \sqrt{1-\bar{\alpha}_t}\boldsymbol{\varepsilon}, t) \right\|_2^2. \tag{5}$$

In the inference stage, the iterative expression of the inverse diffusion process $\boldsymbol{y}_{t-1} \sim p_\theta(\boldsymbol{y}_{t-1}|\boldsymbol{y}_t)$ can be computed as follows:

$$\boldsymbol{y}_{t-1} = \frac{1}{\sqrt{\bar{\alpha}_t}} \left( \boldsymbol{y}_t - \frac{1-\alpha_t}{\sqrt{1-\bar{\alpha}_t}} \boldsymbol{\varepsilon}_{\boldsymbol{\theta}}(\boldsymbol{y}_t, \boldsymbol{x}, t) \right) + \sigma_t \boldsymbol{\xi}. \tag{6}$$

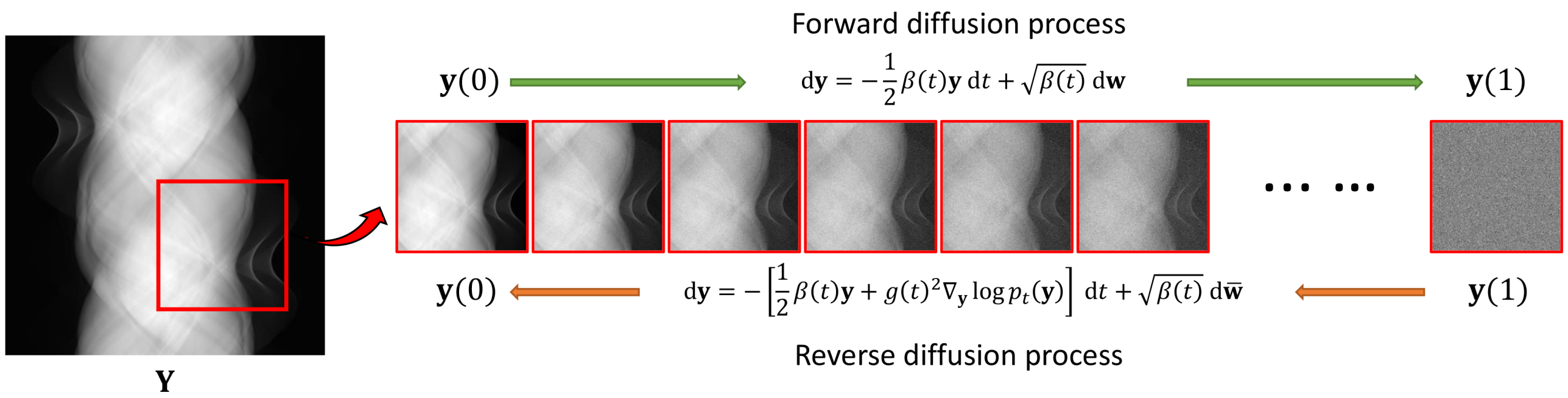

**Figure 1:** Score-based DDPM for projection patch sampling.

Song *et al.* used a stochastic differential equation (SDE) to describe the diffusion process [3]:

$$d\boldsymbol{y} = f(t)\boldsymbol{y}dt + g(t)d\boldsymbol{w}, \tag{7}$$

where $\boldsymbol{w} \in \mathbb{R}^{d\times d}$ is the standard Wiener process, $f : \mathbb{R} \to \mathbb{R}$ is the scalar function to obtain the drift coefficient, and $g : \mathbb{R} \to \mathbb{R}$ is the scalar function to obtain the diffusion coefficient. And according to [1], the reverse diffusion process can also be modeled as the solution to an SDE:

$$d\boldsymbol{y} = \left[f(t)\boldsymbol{y} - g(t)^2\nabla_{\boldsymbol{y}} \log p_t(\boldsymbol{y})\right] dt + g(t)d\bar{\boldsymbol{w}}, \tag{8}$$

where $\bar{\boldsymbol{w}} \in \mathbb{R}^{d\times d}$ is the standard Wiener process for reverse time SDE, and $\nabla_{\boldsymbol{y}} \log p_t(\boldsymbol{y})$ is referred to as the score. Instead of that DDPM network learns the noise perturbation, the score-based model uses the network to estimate the score. So we can train a time-dependent network score estimate model $\boldsymbol{s}_\theta$ with the following loss function:

$$\boldsymbol{L} = \mathbb{E}_{\boldsymbol{y},t}\lambda(t)\left\|\boldsymbol{s}_{\boldsymbol{\theta}}(\boldsymbol{y}(t),t) - \nabla_{\boldsymbol{y}(t)} \log p_t(\boldsymbol{y}(t)|\boldsymbol{y}(0))\right\|_2^2. \tag{9}$$

where $\lambda(t)$ is a positive weighting function.

In the reverse time SDE sampling, the step size is limited by the randomness of the Wiener process [4]. Song *et al.* proved that the reverse time SDE sampling shares the same marginal probability densities with an ordinary differential equation (ODE) sampling process [3]:

$$d\boldsymbol{y} = \left[f(t)\boldsymbol{y} - \frac{1}{2}g(t)^2\nabla_{\boldsymbol{y}} \log p_t(\boldsymbol{y})\right] dt, \tag{10}$$

By ODE sampling, the reverse diffusion process can reduce the noise generated by the random Wiener process and allow a larger step size to improve the sampling efficiency.

Essentially, DDPM in [1] can be seen as a special form of SDE. And The perturbation prediction network of DDPM $\boldsymbol{\varepsilon}_\theta$ can essentially be seen as estimating the scaled score $-\sigma_t\nabla_{\boldsymbol{y}} \log p_t(\boldsymbol{y}(t))$. Therefore, the trained perturbation prediction network and the score prediction network can be converted to each other. In this paper, we trained a score estimation network with patches extracted from fully sampled projection data, whose flowchart is shown in Fig. 1. And Algorithm 1 shows the training procedure of the estimation model $\boldsymbol{s}_{\boldsymbol{\theta}}$.

**Algorithm 1:** Training of the score estimation model $\boldsymbol{s}_{\boldsymbol{\theta}}$.

**Input:** Diffusion parameters $\beta_1, \beta_T$
**Output:** Trained model $\boldsymbol{s}_{\boldsymbol{\theta}}$

Initialize $\boldsymbol{s}_{\boldsymbol{\theta}}$ randomly;
**while** not converged **do**
  $\boldsymbol{Y} \sim p(\boldsymbol{Y})$
  Randomly extract patch $\boldsymbol{y}(0)$ from $\boldsymbol{Y}$
  $t \sim \text{Uniform}([0,1])$
  Update $\boldsymbol{\theta}$ with the gradient
  $\nabla_{\boldsymbol{\theta}}\left[g(t)^2\left\|\boldsymbol{s}_{\boldsymbol{\theta}}(\boldsymbol{y}(t),t) - \nabla_{\boldsymbol{y}(t)} \log p_t(\boldsymbol{y}(t)|\boldsymbol{y}(0))\right\|_2^2\right]$
**end**

### 2.2 Conditional ODE Sampling

The reverse time SDE can be solved with the Euler-Maruyama sampler and further corrected by the Langevin dynamic as demonstrated in [3]. Starting from $\boldsymbol{y}^i(t_0) \sim \boldsymbol{N}(0,\boldsymbol{I})$, such sampling will finally generate a random projection data patch. To achieve the restoration of the down-sampled projection data, we will add the down-sampled projection data as a condition to the reverse diffusion process. First, for the down-sampled projection data $\boldsymbol{Z}$, we first use the FBP algorithm to obtain the noisy image $\bar{\boldsymbol{X}}$, then implement the fully sampling radon transform on $\bar{\boldsymbol{X}}$ to obtain the noisy fully sampling projection data $\bar{\boldsymbol{Z}}$. The real projection data $\boldsymbol{Z}$ is used to rectify the noisy fully sampling projection data by inserting the real projection values in $\boldsymbol{Z}$ into $\bar{\boldsymbol{Z}}$:

$$\tilde{\boldsymbol{Z}} = P^{-1}(\boldsymbol{Z}) \odot \boldsymbol{M} + \bar{\boldsymbol{Z}} \odot (1-\boldsymbol{M}), \tag{11}$$

where $P^{-1} : \mathbb{R}^{N_v'\times N_d} \to \mathbb{R}^{N_v\times N_d}$ is the operation to reshape the down-sampled data into the fully sampling size by inserting zero into the pixels corresponding to the discarded views, $\tilde{\boldsymbol{Z}}$ is the final pseudo fully sampled projection data. Then we extract $N$ overlapped patches from pseudo fully sampled projection data and down-sampling mask with a fixed stride, and obtain two sets of patches $\{\tilde{\boldsymbol{z}}^i\}_{i=0}^N$ and $\{\boldsymbol{m}^i\}_{i=0}^N$.

For the reverse diffusion process at time moment $t_j$, we first obtain the forward diffusion distribution of $\{\tilde{\boldsymbol{z}}^i\}_{i=0}^N$ as the condition:

$$q(\tilde{\boldsymbol{z}}^i(t_j)|\tilde{\boldsymbol{z}}) = \boldsymbol{N}\left(\tilde{\boldsymbol{z}}^i(t_j)|a(t_j)\tilde{\boldsymbol{z}}, b(t_j)^2\boldsymbol{I}\right), i = 1,2,...,N. \tag{12}$$

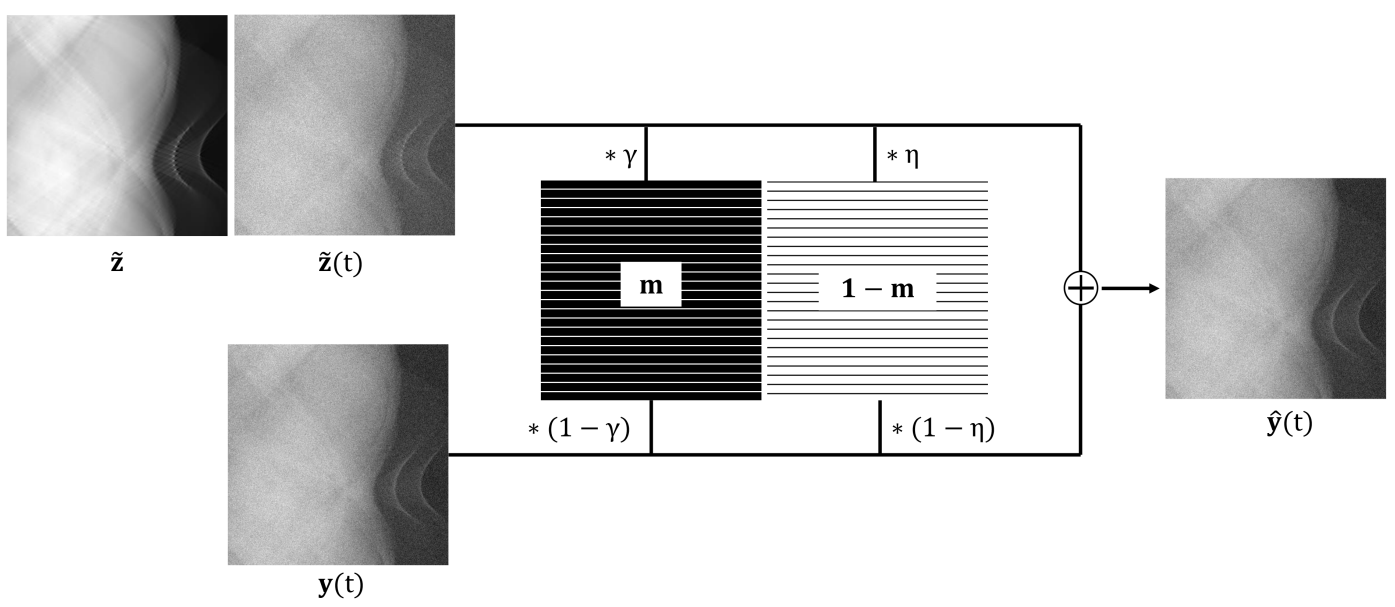

**Figure 2:** Conditioning method for projection patch sampling.

**Algorithm 2:** Inference with the trained denoising model $\boldsymbol{s_\theta}$.

**Input:** Number of time steps $T$; Diffusion parameters $\beta_1, \beta_T$; Under-sampled projection data $\boldsymbol{Z}$; Under-sampling mask $\boldsymbol{M}$

**Output:** $\boldsymbol{Y}$

Load $\boldsymbol{s_\theta}$;
$\bar{\boldsymbol{X}} = \mathrm{FBP}(\boldsymbol{Z})$; $\bar{\boldsymbol{Z}} = \mathrm{Radon}(\bar{\boldsymbol{X}})$;
$\tilde{\boldsymbol{Z}} = P^{-1}(\boldsymbol{Z}) \odot \boldsymbol{M} + \bar{\boldsymbol{Z}} \odot (1 - \boldsymbol{M})$
Extract Patches $\{\tilde{\boldsymbol{z}}^i\}_{i=0}^N$, $\{\boldsymbol{m}^i\}_{i=0}^N$
$\{t_j\}_{j=1}^T = \mathrm{linspace}(1,0)$
**for** $i = 1,2,...,N$ **do in parallel**
  $\boldsymbol{y}^i(t_0) \sim \boldsymbol{N}(0, \boldsymbol{I})$
  **for** $j = 1,2,...,T$ **do**
    $\tilde{\boldsymbol{z}}^i(t_j) = a(t_j)\tilde{\boldsymbol{z}} + b(t_j)\boldsymbol{\xi}$, $\boldsymbol{\xi} \sim \boldsymbol{N}(0,1)$
    $\hat{\boldsymbol{y}}^i(t_j) = \left[\gamma\tilde{\boldsymbol{z}}^i(t_j) + (1-\gamma)\boldsymbol{y}^i(t_j)\right] \odot \boldsymbol{m}^i + \left[\eta\tilde{\boldsymbol{z}}^i(t_j) + (1-\eta)\boldsymbol{y}^i(t_j)\right] \odot (\mathbb{1} - \boldsymbol{m}^i)$
    $\boldsymbol{y}^i(t_{j+1}) = \text{SDE Solver}(\boldsymbol{s_\theta}, \hat{\boldsymbol{y}}^i(t_j), t_j)$
  **end**
**end**
Obtain $\boldsymbol{Y}$ by assembling patches $\{\boldsymbol{y}^i(t_T)\}_{i=0}^N$

Inspired by [5], we propose a conditioning method for sparse-view CT projection data restoration. As shown in Fig. 2, before solving the SDE, the forward diffusion distribution $\{\tilde{\boldsymbol{z}}^i(t_j)\}_{i=0}^N$ is used to condition the reverse diffusion sampling as follows:

$$\begin{aligned}\hat{\boldsymbol{y}}^i(t_j) &= \left[\gamma\tilde{\boldsymbol{z}}^i(t_j) + (1-\gamma)\boldsymbol{y}^i(t_j)\right] \odot \boldsymbol{m}^i \\ &\quad + \left[\eta\tilde{\boldsymbol{z}}^i(t_j) + (1-\eta)\boldsymbol{y}^i(t_j)\right] \odot (\mathbb{1} - \boldsymbol{m}^i)\end{aligned}, \tag{13}$$

where $\mathbb{1} \in \mathbb{R}^{d \times d}$ is the matrix with all elements of one. The flowchart of the reverse diffusion process conditioned by the pseudo fully sampled projection data is shown in Algorithm 2.

However, the reverse time SDE sampling will introduce slight perturbations to the projection data due to the randomness of the Wiener process. Although these disturbances are indistinguishable to the naked eye, they will spread to the entire image domain after reconstruction with FBP algorithm, resulting in serious degradation of image quality. To avoid the perturbations caused by the Wiener process, we adopt ODE sampling in this paper. In [3], Song *et al.* used the RK45 ODE solver [6] for the ODE sampling. In this study, we introduce a more efficient ODE solver to further improve the sampling performance [7].

In [8], Kingma *et al.* defined the scalar functions in Eq. (7) as follows:

$$f(t) = \frac{\mathrm{d}\log\eta_t}{\mathrm{d}t}, \qquad g^2(t) = \frac{\mathrm{d}\sigma_t^2}{\mathrm{d}t} - 2\frac{\mathrm{d}\log\eta_t}{\mathrm{d}t}\sigma_t^2, \tag{14}$$

where $\eta_t = \sqrt{\bar{\alpha}_t}$. Substituting the perturbation prediction model into Eq. (10):

$$\mathrm{d}\boldsymbol{y} = \left[f(t)\boldsymbol{y} + \frac{g^2(t)}{2\sigma_t}\boldsymbol{\varepsilon_\theta}(\boldsymbol{y}(t),t)\right]\mathrm{d}t, \tag{15}$$

which is a semi-linear ODE [7], whose solution at time moment $t$ can be calculated with the variation of constants formula [9] as follows:

$$\begin{aligned}\boldsymbol{y}(t) &= \exp\left(\int_s^t f(\tau)\mathrm{d}\tau\right)\boldsymbol{y}(s) \\ &\quad + \int_s^t \left[\exp\left(\int_\tau^t f(r)\mathrm{d}r\right)\frac{g^2(\tau)}{2\sigma_t}\boldsymbol{\varepsilon_\theta}(\boldsymbol{y}(\tau),\tau)\right]\mathrm{d}\tau\end{aligned}. \tag{16}$$

Let $\lambda_t = \log(\eta_t/\sigma_t)$, Eq. (16) can be further simplified into

$$\boldsymbol{y}(t) = \frac{\eta_t}{\eta_s}\boldsymbol{y}(s) + \eta_t\int_s^t\left[\left(\frac{\mathrm{d}\lambda_\tau}{\mathrm{d}\tau}\right)\frac{\sigma_\tau}{\eta_\tau}\boldsymbol{\varepsilon}_\theta(\boldsymbol{y}(\tau),\tau)\right]\mathrm{d}\tau. \tag{17}$$

The predefined $\lambda_t$ can be obtained with a strictly decreasing function of $t$, denoted as $\lambda(t)$, which has a inverse function $t = t_\lambda(\lambda)$. Then, by changing the time variable $t$ into the parameter variable $\lambda$ and denoting $\hat{\boldsymbol{y}}(\lambda) := \boldsymbol{y}(t_\lambda(\lambda))$ and $\hat{\boldsymbol{\varepsilon}}_{\boldsymbol{\theta}}(\hat{\boldsymbol{y}}_\lambda,\lambda) := \boldsymbol{\varepsilon_\theta}(\boldsymbol{y}(t_\lambda(\lambda)), t_\lambda(\lambda))$, Eq. (17) can be rewritten as

$$\boldsymbol{y}_t = \frac{\eta_t}{\eta_s}\boldsymbol{y}(s) + \eta_t\int_{\lambda_s}^{\lambda_t}\left[\mathrm{e}^{-\lambda}\hat{\boldsymbol{\varepsilon}}_{\boldsymbol{\theta}}(\hat{\boldsymbol{y}}(\lambda),\lambda)\right]\mathrm{d}\lambda, \tag{18}$$

This integral can be calculated numerically by Taylor expansion. According to the order of Taylor expansion, Lu *et al.* [7] provided three solvers for the flow probability ODE, which are called *DPM-Solver-1*, *DPM-Solver-2* and *DPM-Solver-3*, respectively.

## 3 Experiments and Results

The *2016 NIH-AAPM-Mayo Clinic Low-Dose CT Grand Challenge* dataset was chosen to conduct experiments. The dataset has 2,378 paired CT images with a thickness of 3mm from 10 patients. We selected 1,923 paired images from 8 patients as the training set, and 455 paired images from the remaining 2 patients as the test set. The image size is 512x512. Then simulated projection data was obtained with the distance-driven algorithm [10]. The geometric parameters used in the simulation are consistent with those given in the dataset. In our method, the patch size is set to 64x64. And the continuous time range is $t \in [0,1]$. $\beta_1$ and $\beta_T$ are respectively set to $10^{-4}$ and 0.02 according to the recommendation in [1].

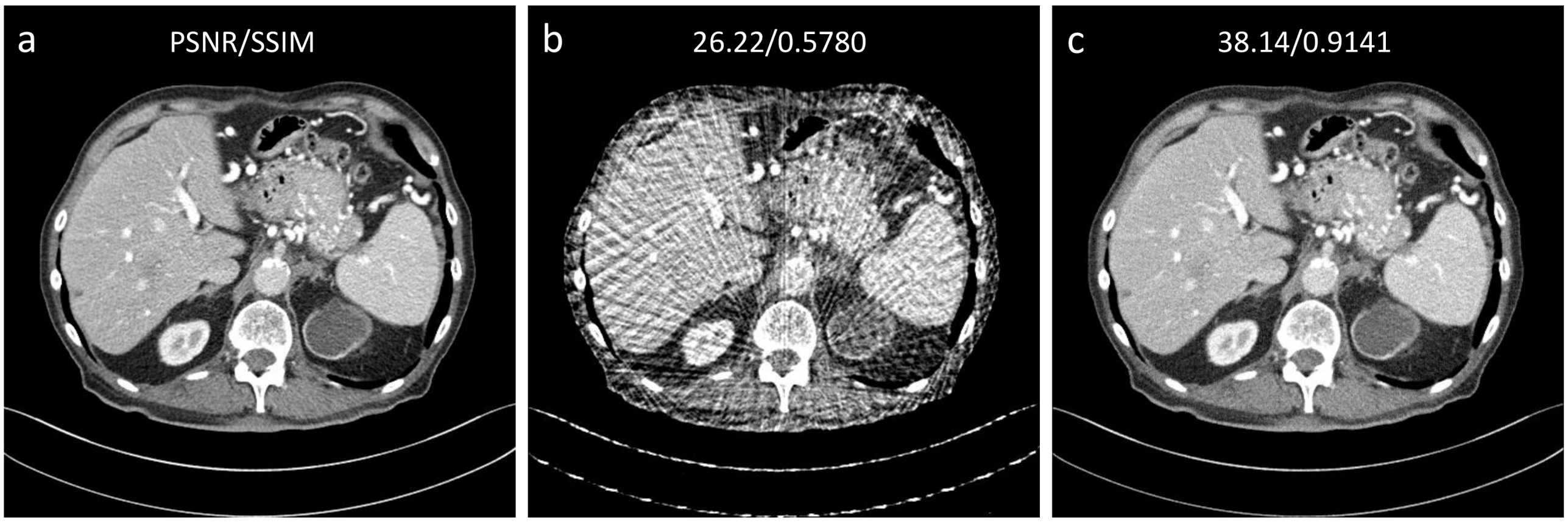

**Figure 3:** The results obtained with our proposed method. (a) Ground truth, (b) Sparse-view CT, (c) Ours. The display window is consistently set to [-160, 240] HU.

**Table 1:** Average quantitative results of our proposed method.

| | PSNR | SSIM |
|---|---|---|
| Sparse-View CT | 26.41 | 0.5608 |
| Ours | 38.69 | 0.9132 |

The model was trained with the Adam optimizer [11] with a learning rate of $1 \times 10^{-4}$. The training process converged well after $2 \times 10^5$ iterations on a computing server equipped with an Nvidia RTX A5000 GPU. When sampling, the projection data is uniformly down-sampled to 92 projection views over the full 360°. The stride for extracting patches is set to 32 for overlapping extraction. The number of functional evaluations (NFE) is set to 1000. And the condition parameters $\gamma$ and $\eta$ are set to 1.0 and 0.1, respectively.

Fig. 3 shows the results obtained with our proposed method. It can be seen that our proposed method can eliminate the artifacts contained in the sparse-view CT reconstruction with FBP. In particular, our method also preserves the structure well without any detail loss or over-smoothness, which often occur in deep learning-based methods. Observing from the perspective of texture maintenance, the reconstructed CT image with our proposed method has a very similar texture to the ground truth, which is consistent with the doctor's reading habits.

## 4 Conclusion

In this paper, we propose a patch-based DDPM model for sparse-view CT reconstruction. The method has excellent anti-artifact performance while preserving the details and textures. Our proposed method is based on unsupervised learning, which can overcome the challenge of difficulty in the acquisition of clinical paired data. At the same time, we split the entire projection data into patches so that the data can be processed in parallel, enabling the deep reconstruction of large-scale and 3D data.

## Acknowledgment

This work was partially supported by the following NIH Grants: R01EB032716, R01CA237267, and R01EB031102.

# Hybrid U-Net and Swin-Transformer Network for Limited-angle Image Reconstruction of Cardiac Computed Tomography

Yongshun Xu[1], Shuo Han[1], Dayang Wang[1], Ge Wang[2], Jonathan S. Maltz[3], Hengyong Yu[1,*]

1. Department of Electrical and Computer Engineering, University of Massachusetts Lowell, Lowell, MA, 01854, USA
2. Department of Biomedical Engineering, Rensselaer Polytechnic Institute, Troy, NY, 12180 USA
3. GE HealthCare, 3000 N Grandview Boulevard, Waukesha, WI, 53188, USA
[*]Corresponding author, Email: hengyong-yu@ieee.org

**Abstract** Cardiac computed tomography is widely used in the diagnosis of cardiovascular disease, the leading global cause of morbidity and mortality. Diagnostic confidence depends strongly on the temporal resolution of the images. To freeze heart motion, one can reduce the scanning time by acquiring limited-angle projections, but this leads to increased noise and non-motion-related artifact. The ability to reconstruct high quality images from limited-angle projections is thus highly desirable. However, this is a difficult, ill-posed problem. With the development of deep learning networks, such as U-Net and transformer networks, much success has been achieved on many image processing tasks. Here, we propose a hybrid model based on U-Net and Swin-transformer networks. U-Net has the potential to restore structural information lost due to missing projection data and related artifacts, while the Swin-transformer can gather a more detailed feature distribution. We demonstrate, through application to synthetic XCAT data, that our proposed method outperforms the state-of-the-art competing deep learning-based methods.

## 1. Introduction

Computed Tomography (CT) is one of the most highly utilized imaging tools in clinics and hospitals. In typical multislice CT imaging scanning, the detector acquires projection views over the entire range of angles required for a full-scan tomographic reconstruction. These views can be combined to reconstruct a cross-sectional image using algorithms such as filtered back projection (FBP).

The limited-angle problem typically refers to a scanning range of less than 180°, and it is a highly ill-posed problem.

By reconstructing images based on a limited angle dataset, it is possible to image moving objects within a shorter time window. Also, it may be possible to image using a lower radiation dose to the patient.

In a typical CT scan, even though the patient is typically under a breath-hold instruction and is told to keep still, it is impossible to eliminate physical movements of organs such as the beating heart. Hence, cardiac CT images exhibit motion artifact due to insufficient temporal resolution. Since temporal resolution is limited by maximum speed of the rotating gantry, short-scan data are typically used for image reconstruction and make it possible to select that cardiac phase that exhibits the least motion artifact. In many practical cases (e.g., atrial fibrillation), current temporal resolution is not adequate to freeze the beating heart. This is because the optimal phase may be different for different vessels/segments of the coronary tree, and even if the best phase is selected, motion will still occur within the corresponding temporal gating window when the heart rate is high. To further improve the temporal resolution of cardiac CT, limited-angle image reconstruction is a natural solution.

Direct application of a conventional image reconstruction algorithm (e.g., FBP) to limited-angle projections would result in images with poor quality and severe streak artifacts. Previously, researchers have assumed that an image is sparse, and that this prior knowledge can help to improve image quality. For instance, the total variation (TV) regularization model is based on the assumption of piecewise constancy of the image and the sparsity of the discrete gradient transform; TV can suppress noise while preserving edges [1].

Iterative minimization algorithms were also developed to enhance reconstruction performance. Although these iterative reconstruction algorithms can improve image quality in terms of noise-based metrics, they often suppress image detail, and reconstructed images often exhibit residual artifacts and poor texture. Additionally, these methods incur a high computational cost [2], and the performance may be even worse when applied to the limited-angle reconstruction problem.

In recent years, deep learning has been widely applied in computer vision fields such as image segmentation, as well as for denoising and super-resolution. The limited-angle reconstruction problem can be formulated as a learned mapping between low-quality and high-quality images. This requires large dataset, and the results rely on supervised learning. Convolutional neural network (CNN)-based models have been proposed to improve model representation ability by using more elaborate neural network architectures, such as generative adversarial network (GAN) based models [2][3], U-Net models [4][5], and residual blocks [6]. DD-Net is a dense and deconvolution network intended to increase the data quality by reusing the features more effectively [7]. TomoGAN adopted the adjacent noisy images and developed a GAN network with U-Net generator [8].

Transformer, a natural language processing model, has achieved great success in computer vision tasks [9]. When applied to vision problems, it divides a natural image into small patches and learns inside knowledge using a self-attention module. It also explores the global interactions between different patches. To further increase the interaction, a shift window (Swin) multi-head attention was developed by shifting the partition [10]. TransUNet applied transformer blocks on deep feature maps [11]. Swin-UNet replaced all the convolution layers with transformer blocks in U-Net [12]. SwinIR added the convolution layer and long shortcut connections to the transformer network [13].

U-Net achieves great success when applied to image segmentation tasks, which implies is may be capable of helping to restore major structural information to limited-angle images. Meanwhile, Transformer blocks perform well on the task of learning global information between patches. Those observations inspire us to develop a parallel network structure to combine U-Net and transformer blocks to jointly learn structural features and details of limited-angle cardiac CT images. Long shortcut connections and a convolution layer offer the potential to further enhance the image quality.

## 2. Materials and Methods

### 2.1 Network architecture

In this paper, we propose a Swin-Transformer-based image restoration model for limited-angle cardiac CT, which we refer to as SwinUC. The model is composed of four elements: Swin-transformer block, U-Net module, convolution layer, and skip connection. First, a convolution layer extracts the feature map, followed by a stack of residual U-Net Swin-Transformer blocks for latent feature extraction. A long skip connections are added to combine features and feed these into the last convolution layer.

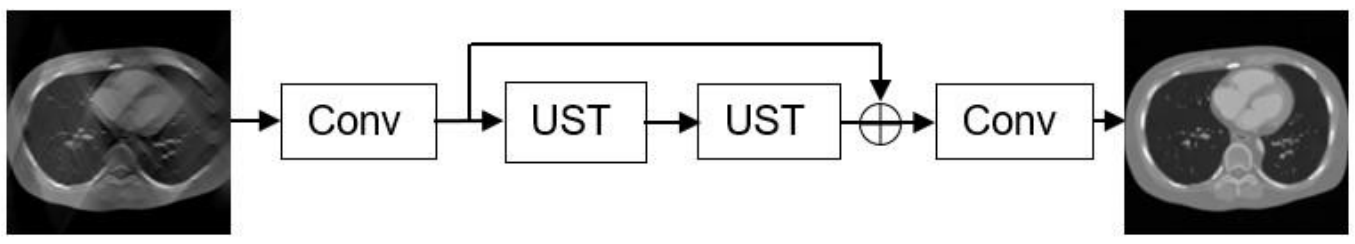

*Fig. 1.* The framework of the proposed Swin-transformer U-Net convolution network. UST represents Swin-transformer U-Net blocks. The input is a low quality FDK result from limited-angle projections, and the label is a high quality FDK result from full view projections without motion.

As shown in Fig.1, a low-quality image reconstructed from limited-angle projections serves as an input $I_{\text{LA}} \in \mathbb{R}^{H\times W}$, where H and W are image height and width. We use a convolution layer $\text{Conv}^C$ to extract to feature maps, the kernel size is $3 \times 3$, and $C$ represents the channel number. We then obtain a feature with a higher dimension $F \in \mathbb{R}^{H\times W\times C}$. UST represents Swin-transformer blocks, and two UST blocks $\text{UST}_1$ and $\text{UST}_2$ are consecutively connected after the convolution layer. Then, a residual connection combined the early-stage features and the deep features. Next, we use a convolution layer with one channel $\text{Conv}^1$ to improve the local features:

$$I_{\text{out}} = \text{Conv}^1\left((\text{UST}_2(\text{UST}_1(\text{Conv}^C(I_{\text{LA}})))) + \text{Conv}^C(I_{\text{LA}})\right). \quad (1)$$

In our image reconstruction experiment, we train the network by minimizing the mean square loss:

$$\text{Loss} = ||I_{\text{out}} - I_{\text{FV}}||_2, \quad (2)$$

where $I_{\text{FV}}$ represents the ground-truth that is reconstructed by FBP/FDK from full view motionless projections.

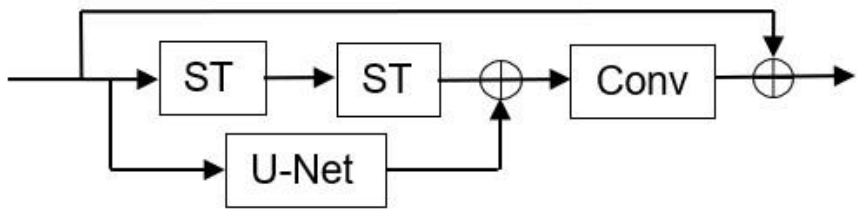

*Fig. 2.* The structure of Swin-transformer U-Net block. ST represents a Swin-transformer block.

As shown in Fig.2, the Swin-transformer U-Net block is a residual connection with parallel Swin-transformer layers and U-Net. We feed an input feature map $f_{\text{in}}$ in parallel to Swin-transformer layers $\text{ST}_1, \text{ST}_2$ and U-Net layers U. We then extract the latent features and feed these into the convolution layer Conv. Finally, we add the residual connection to get the output feature $f_{out}$:

$$f_{\text{out}} = \text{Conv}(\text{U}(f_{\text{in}}) + \text{ST}_2(\text{ST}_1(f_{\text{in}}))) + f_{\text{in}}. \quad (3)$$

In this block, U-Net gathers structural information, and the Swin-transformer layer integrates global spatial information. The skip connection layer aggregates distilled features to capture shallow features.

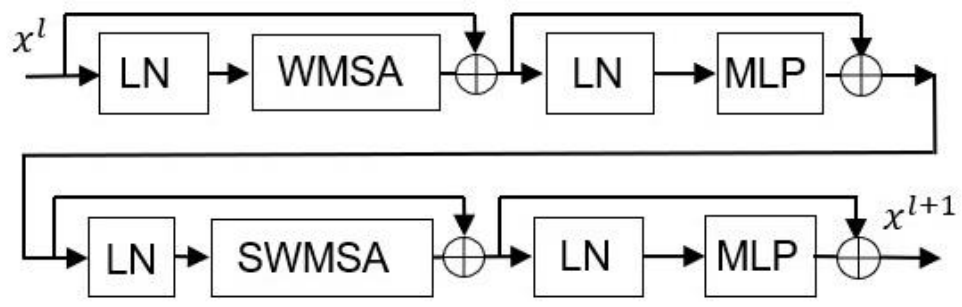

*Fig. 3.* The structure of a Swin-transformer block. LN is layer normalization, MSA is multi-head self-attention, MLP is a multilayer perceptron, and SWMSA is a shift window multi-head self-attention.

In figure 3, we show the standard Swin-Transformer architecture, which consists of two consecutive transformer blocks. The first is a window multi-head attention block, and the second , a window shift multi-head attention block. Each block includes layer normalization, skip connections, and a two-layer multilayer perceptron with ReLU layer. As introduced in [10], the shifted window partitioning approach introduces connections between non-overlapping windows in previous layer, and is effective for feature extraction. Given an input feature size of input $H \times W$, the transformer first reshapes it into $\frac{HW}{M^2} \times M^2$ features by partitioning the input into $M \times M$ windows. It computes the self-attention on each window. For a local window feature $X \in \mathbb{R}^{M^2}$, by using projection matrices $P_q, P_k, P_v$; $Q, K, V$ are computed as:

$$Q = XP_q,\ K = XP_k,\ V = XP_v. \quad (4)$$

The self-attention process is as following:

$$\text{Attention}(Q, K, V) = \text{SoftMax}(\frac{QK^T}{\sqrt{d}} + B)V, \quad (5)$$

where $Q, K, V \in \mathbb{R}^{M^2\times d}$, are the query, key, and value matrices; d is the dimension; $M^2$ is the number of features in a window; and $B$ is relative position bias $B \in \mathbb{R}^{M^2 * M^2}$.

### 2.2. Evaluation: Dataset and Experiment Setup

We use the 4D extended cardiac-torso (XCAT) phantom version 2 to generate realistic projection data to simulate a cardiac CT imaging procedure performed with injected iodinated contrast medium. This widely utilized multimodality phantom was developed at Duke University and is described in detail in [14]. XCAT can output a 4D phantom of attenuation coefficients to mimic a patient with beating heart which is close to realistic dynamic cardiac CT imaging situations. Phantoms are generated based on a set of pre-defined parameters, including spatial resolution, temporal resolution, respiration rate, and heart rate. After generating the digital phantoms, we use XCAT's CT projector to generate the cone-beam CT projections. The CT projection simulation is controlled by some key parameters, such as the distance from the object to the source, distance from the object to the detector, the detector array size, the x-ray source energy spectrum, and the beam half-fan angle. These parameters are configured to mimic a representative GE CT scanner. After generating the projection data using a circular scan, the standard FDK algorithm is employed to reconstruct the 3D volumetric image for each phase and scaled into Hounsfield Units (HU). Based on the tuned parameters, simulated images are generated for 10 patients both on dynamic and static phantoms. We reconstruct images at phases from 20% to 80%, with an interval of 3% [15].

Each simulated human phantom yields 21 different motion-blurred phases. For each phase, we apply standard FDK reconstruction to produce 32 image slices. For motion data, we select 120-degree-range views to reconstruct limited-angle images. In this way, we generate a cardiac phantom image dataset containing 6720 motion-blurred limited-angle images, and the corresponding 6720 static ground-truth images. The reconstructed images have 256 × 256 pixels. We select images from nine patients and one patient as training and test datasets, respectively.

We employ the following settings: batch size =1, number of epochs = 100, Adam is used to optimize the model, and the learning rate is set at $10^{-5}$. We implement the proposed network with PyTorch, and use one NVIDA 2080Ti GPU for training.

We compare the performance of the following methods: TomoGan, DDNet, and U-Net. Two representative slices from the test dataset that include the major vessels and bone structures are shown to enable visual comparison of the methods. Both qualitative and quantitative analysis are performed.

## 3. Results and Analysis

Limited-angle reconstructed images, full-view reconstructed images, and images reconstructed using the different models, are shown in Figures 4 and 5. Due to the limited scanning range of projections, there are severe streaking artifacts in the raw FDK reconstructions. Although major structures are recovered, much edge and texture information is lost. Figures 4 and 5 show that all the deep learning-based methods demonstrate suppression of the streaking artifacts, but there are notable differences in performance. Comparing the region of interests (ROI) indicated by the blue and red arrows (see Figure 6), U-Net restores more edge information than DDNet. It preserves piecewise smoothness better than the TomoGan. However, some of the vessel morphology is lost. The organs and bony structures contain more noise. The proposed method improves the image quality, and it is closer to the ground-truth than other methods. These results demonstrate that SwinUC can reduce artifacts at edges and reconstruct dynamic objects more precisely; blood vessel and bone structure are preserved, and organ boundaries are well-defined.

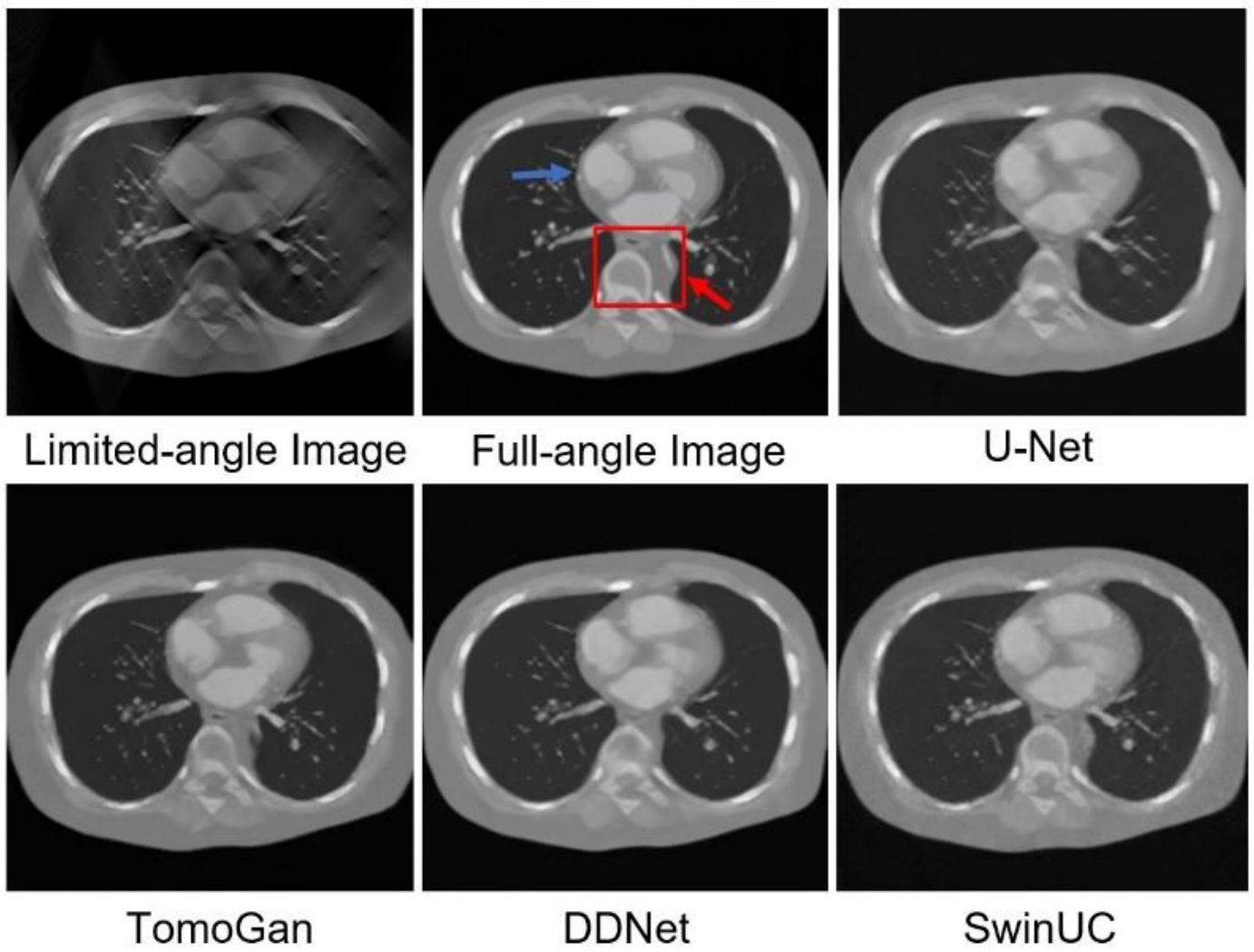

*Fig. 4.* Reconstructed images of simulated case 153 (phase 40, slice 10) using U-Net, TomoGan, DDNet, and SwinUC.

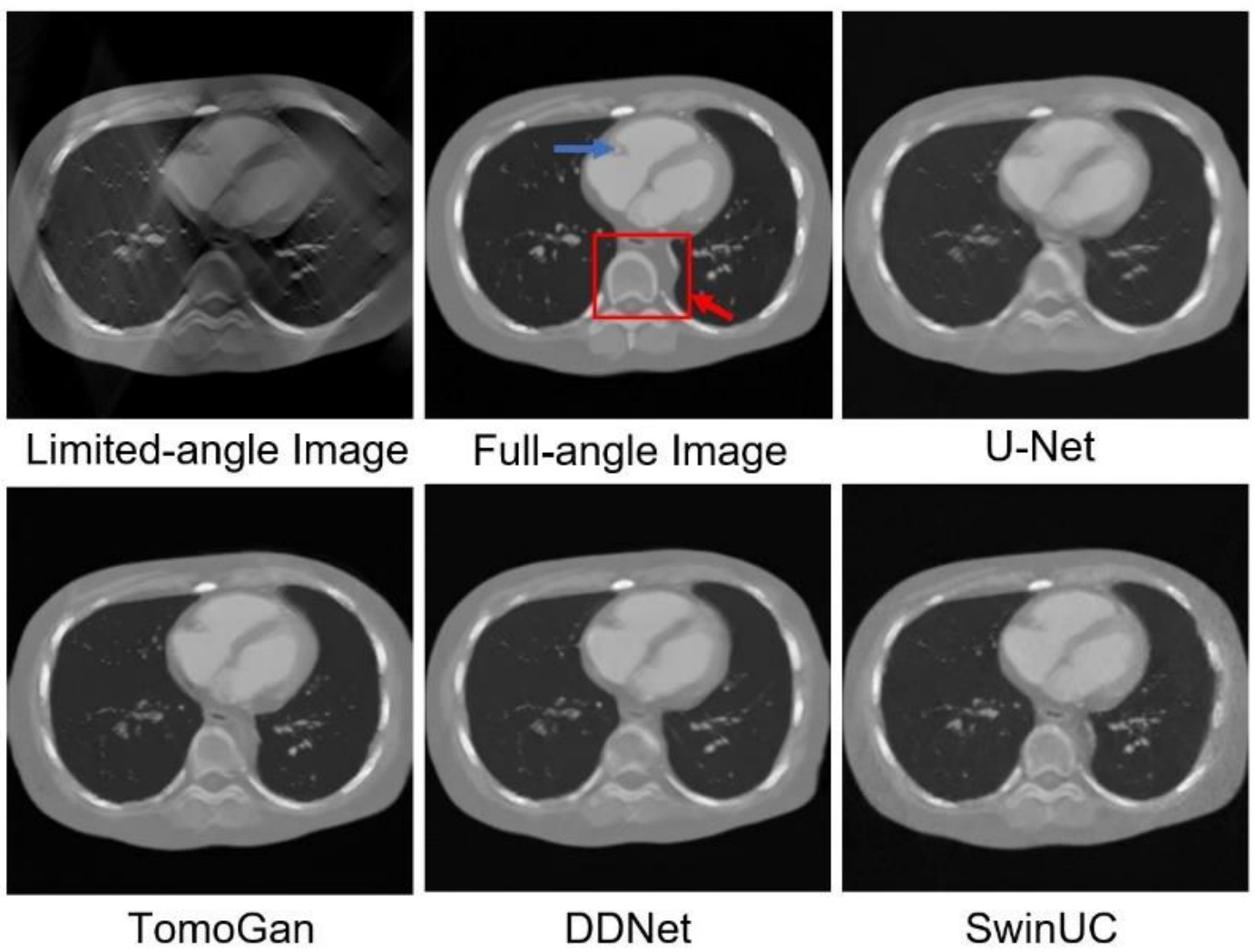

*Fig. 5.* Reconstructed images on patient 153 (phase 40, slice 21) using U-Net, TomoGan, DDNet, and SwinUC.

We compare the reconstruction results quantitatively using the structural similarity index measure (SSIM) and root mean square error (RMSE) as metrics. The average values of SSIM and RMSE results using each method appear in Table 1. Better reconstruction image quality leads to higher SSIM values and lower RMSE values. One can see that DDNet outperforms U-Net and TomoGan. Compared to DDNet, our proposed method increases the SSIM value by 0.021 and reduces RMSE by 35.7%. These quantitative measures of reconstruction confirm that our proposed method achieves better performance than competing methods.

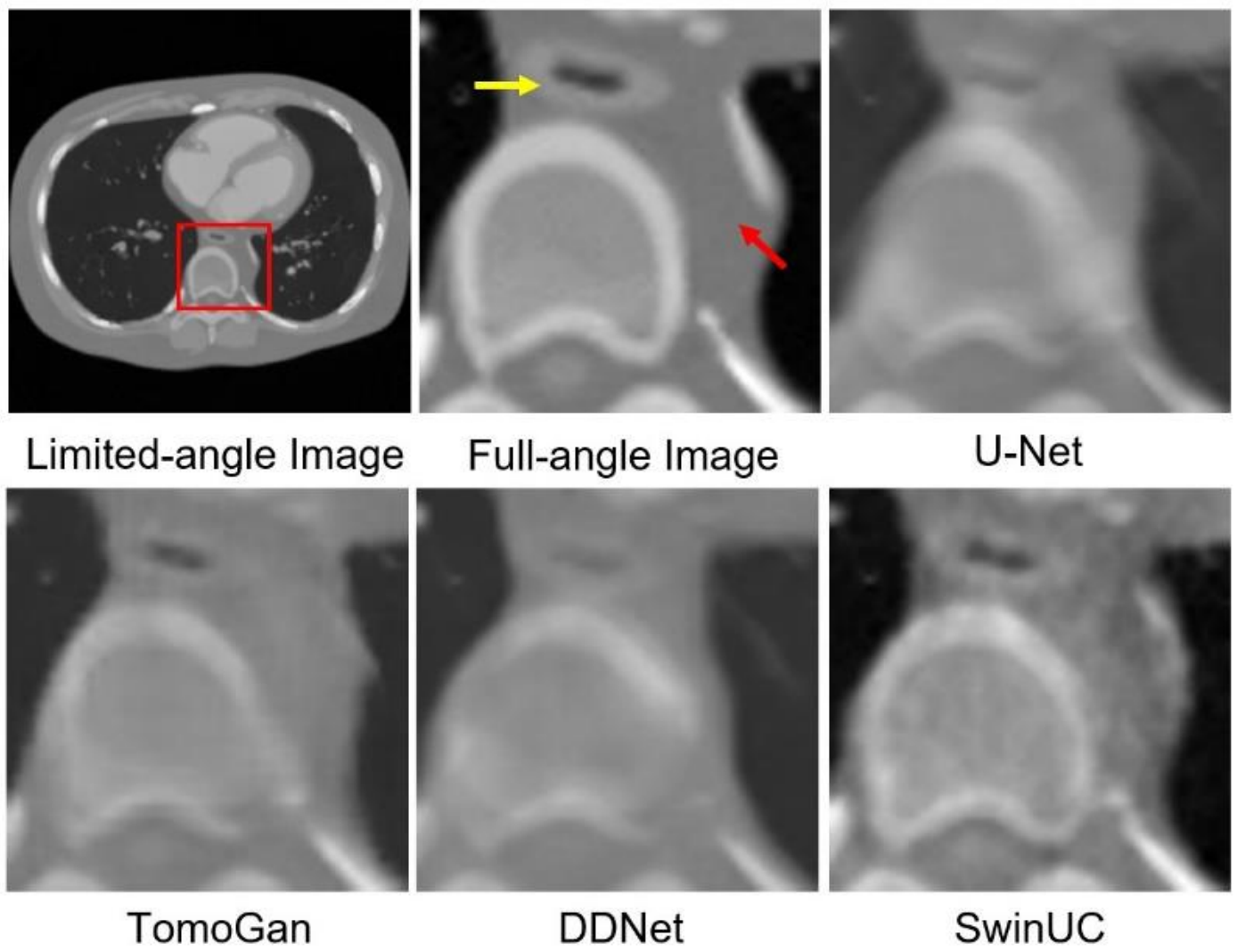

*Fig 6.* Magnified region of interest on patient 153 (phase 40, slice 21) using U-Net, TomoGan, DDNet, and SwinUC.

Table.1 Quantitative comparison (average PSNR/SSIM) on the test dataset.

| | U-Net | TomoGan | DDNet | SwinUC |
|---|---|---|---|---|
| SSIM | 0.9409 | 0.9357 | 0.9424 | **0.9631** |
| RMSE | 0.0016 | 0.0018 | 0.0014 | **0.0009** |

## 4. Discussion

Limited-angle projetion imaging can be used to enhace temporal resolution in CT, but generally at the expense of the introduction of serious artifacts. Deep learning techniques can help address this image inpainting and image denoising problem. U-Net is very effective for segmentation tasks, and can restore shape and edge information to the image. Transformer blocks gather global information by partitioning the image and using attention modules. The results of limited-angle reconstruction show that combining the U-Net and Swin-tranformer blocks creates a network that more effectively restores missing information. Our results show that edges oriented along directions within the scanning range are clear, while noise and artifacts are greatly reduced for image detail oriented along the range of missing angles. Our network outperforms other methods in the task of restoring bone and soft-tissue organ stuctures.

Our network architecture combines U-Net and the Swin-transformer to generate a UST block, and then stacks two UST blocks. In [10], stacking more transformer blocks is shown to typically improve network performance. However, the number of parameters increases dramatically and much higher computational cost is incurred. The optimal number of blocks can be found by performing further comparison experiments.

We have demonstrated both qualitative and quantitative improvement in image quality using SwinUC. However, additional work is needed to further eliminate noise and achieve higher image quality, while preserving structural information.

## 5. Conclusion

In this paper, we propose a network that combines the advantages of Swin- transformer and U-Net to solve the limited-angle tomographic reconstruciton problem. We descrubed the network structure and demonstrated its effectiveness. By comparing images reconstructed with DDnet, U-Net, and TomoGan, all trained on the same simulated dataset, we have shown that our method can effectively remove severe artifacts and improve the reconstruction image quality, as quantified by higher SSIM and smaller RMSE values.

# CT-free Total-body PET Segmentation

Song Xue[1,5], Christoph Clement[1,5], Rui Guo[2], Marco Viscione[1], Axel Rominger[1], Biao Li[2,3], Kuangyu Shi[1,4]

[1]Department of Nuclear Medicine, University of Bern, Bern, Switzerland
[2]Department of Nuclear Medicine, Ruijin Hospital, Shanghai Jiao Tong University School of Medicine, Shanghai, China
[3]Collaborative Innovation Center for Molecular Imaging of Precision Medicine, Ruijin Center, Shanghai, China
[4]Department of Informatics, Technical University of Munich, Munich, Germany
[5]These authors contributed equally: Song Xue, Christoph Clement

**Abstract** Low-dose positron emission tomography (PET) imaging is made possible with the use of high sensitivity PET/computed tomography (CT) scanners with a long axial field of view (FOV). However, using CT for anatomical localization in this process imposes a considerable radiation burden. We aim to achieve total-body PET multi-organ segmentation on non-corrected PET imaging using a deep learning (DL) approach as a step towards true CT-free PET imaging. Total-body 18F-FDG PET images of 114 patients scanned with a Siemens Biograph Vision Quadra were used for the development. The ground-truth multi-organ segmentation labels were generated using the CT images as input to the Multi-Organ Objective Segmentation (MOOSE) software. A 3D U-Net-like network was trained on the non-attenuation and non-scatter corrected PET images. Three nuclear medicine physicians independently assessed the utility of the results in a clinical setting. The trained model achieved an average Dice similarity coefficient (DSC) of 0.82 on the test dataset. These preliminary results show an accurate overlap between the MOOSE-generated labels and our predicted organ segmentations: 70% of targeted organs achieved DSCs of more than 0.80, whereas a few organs exhibited lower scores (e.g., bladder [0.70], thyroid [0.69] and pancreases [0.59]). The visual readings conducted by three nuclear medicine physicians confirmed the usability of the generated segmentations for anatomical localization. In conclusion, our study demonstrates the possibility of total-body PET multi-organ segmentation using a deep learning-based method that does not require the anatomical information from CT.

## 1 Introduction

Positron emission tomography (PET) is one of the main imaging modalities in clinical routine procedures, especially in oncology [1] and neurology [2]. PET is being widely acknowledged as an indispensable tool for diagnosis, monitoring of malignant diseases, and determination of prognosis [3]. The advent of long axial field of view (FOV) total-body PET [4] has enabled previously unachievable levels of sensitivity and quantification with reduced radiopharmaceutical dose [5]. Commercial PET/computed tomography (CT) scanners such as the Siemens Biograph Vision Quadra utilize the anatomical information provided by CT imaging mainly for localization and quantification purpose, which inevitably introduces additional ionizing radiation to patients [6] (6.4 mSv for low-dose CT [7]). As for the quantification aspect, attenuation (AC) and scatter correction (SC) are essential for precise PET quantification [8], which necessitate additional structural images from CT to calculate attenuation factors and model scatter. Deep learning (DL)-based methods have been proposed as a substitute for CT-based PET attenuation and scatter correction and have been verified on various scanners and tracers [9].

However, the localization function of CT-free PET imaging remains unadressed. Physicians continue to desire additional anatomic information from CT for localization purposes, which enhances their confidence in interpreting study results [10]. By utilizing the additional anatomic information provided by high sensitivity total-body PET [11], we aim to achieve total-body PET multi-organ segmentation on non-corrected PET imaging using a DL approach as a step towards true CT-free PET imaging.

## 2 Materials and Methods

### A. Patient Cohorts

114 subjects with 18F-FDG PET imaging were included in this study. Table 1 depicts information on the patients' demographics. The subjects were scanned with a Siemens Biography Vision Quadra total-body PET/CT scanner at Inselspital, Bern, Switzerland.

| | |
|---|---|
| **Scanner** | Siemens Biograph Vision Quadra |
| **Tracer** | 18F-FDG |
| **Number of Patients** | 114 |
| **Total dose (MBq)** | 216.7±47.6 |
| **Post-injection time (min)** | 67.3±10.8 |
| **Gender (Male/Female)** | 49/65 |
| **BMI (kg/m$^2$)** | 24.9±4.8 |

*Table 1 Information on patients' demographics*

### B. Segmentation Pipeline & Evaluation

The ground-truth multi-organ segmentation labels were generated using the CT images as input to the Multi-Organ Objective Segmentation (MOOSE) software [12]. The segmentation labels were then transformed to the PET image space and the dataset was randomly split into 91 training and 23 test cases. We set up a segmentation pipeline to generate the multi-organ labels from the non-attenuation and non-scatter corrected PET (NASC-PET) images using the nnU-Net software [13]. A 3D U-Net [14] like model with instance normalization [15] layers and leaky ReLU [16] activation functions was trained with deep supervision, where the loss is computed using multiple resolution levels. As the loss function, we used the sum of the cross-entropy and Dice [17] loss. The model was trained for 1,000 epochs with a batch size of 2, Stochastic Gradient Descent (SGD) as the optimizer, and with a decaying learning rate starting at 0.01. Common data augmentation techniques such as random rotations, scaling, and Gaussian noise were applied to the data during the training procedure. The trained model was used to predict segmentation maps of the test images with the sliding window approach. We additionally trained another model using the same procedure on attenuation and scatter corrected PET (ASC-PET) images to determine the influence of CT-based image corrections on the networks performance. Three board-certified physicians independently and blindly reviewed a mixed dataset consisting of both ground-truth and predicted segmentations.

## 3 Results

| | NASC-PET | ASC-PET |
|---|---|---|
| **Aorta** | 0.8817 | 0.8836 |
| **Bladder** | 0.6991 | 0.7084 |
| **Brain** | 0.9530 | 0.9531 |
| **Heart** | 0.8966 | 0.9100 |
| **Kidneys** | 0.8590 | 0.8767 |
| **Liver** | 0.8826 | 0.9212 |
| **Pancreas** | 0.5849 | 0.6266 |
| **Spleen** | 0.8085 | 0.8769 |
| **Thyroid** | 0.6871 | 0.6968 |
| **Lung** | 0.9306 | 0.9504 |

*Table 2 Dice similarity coefficients of the different organs in the test dataset.*

Table 2 shows the average Dice similarity coefficients (DSCs) of the different organs over the test dataset. We depict the scores for the NASC-PET as well as the ASC-PET images as a comparision. The trained model achieved an average DSC of 0.82 in the test dataset when using the NASC-PET images as input and a DSC of 0.84 when using the ASC-PET images as input. The preliminary results show an accurate overlap between the MOOSE-generated labels and our predicted organ segmentations: 70% of targeted organs achieved DSCs of more than 0.80, whereas a few organs exhibited lower scores (e.g., bladder [0.70], thyroid [0.69] and pancreases [0.59]). Figure 1 visualizes an exemplary test prediction. Visual readings of three nuclear medicine physicians shows no significant difference between ground-truth and predicted segmentations within test group.

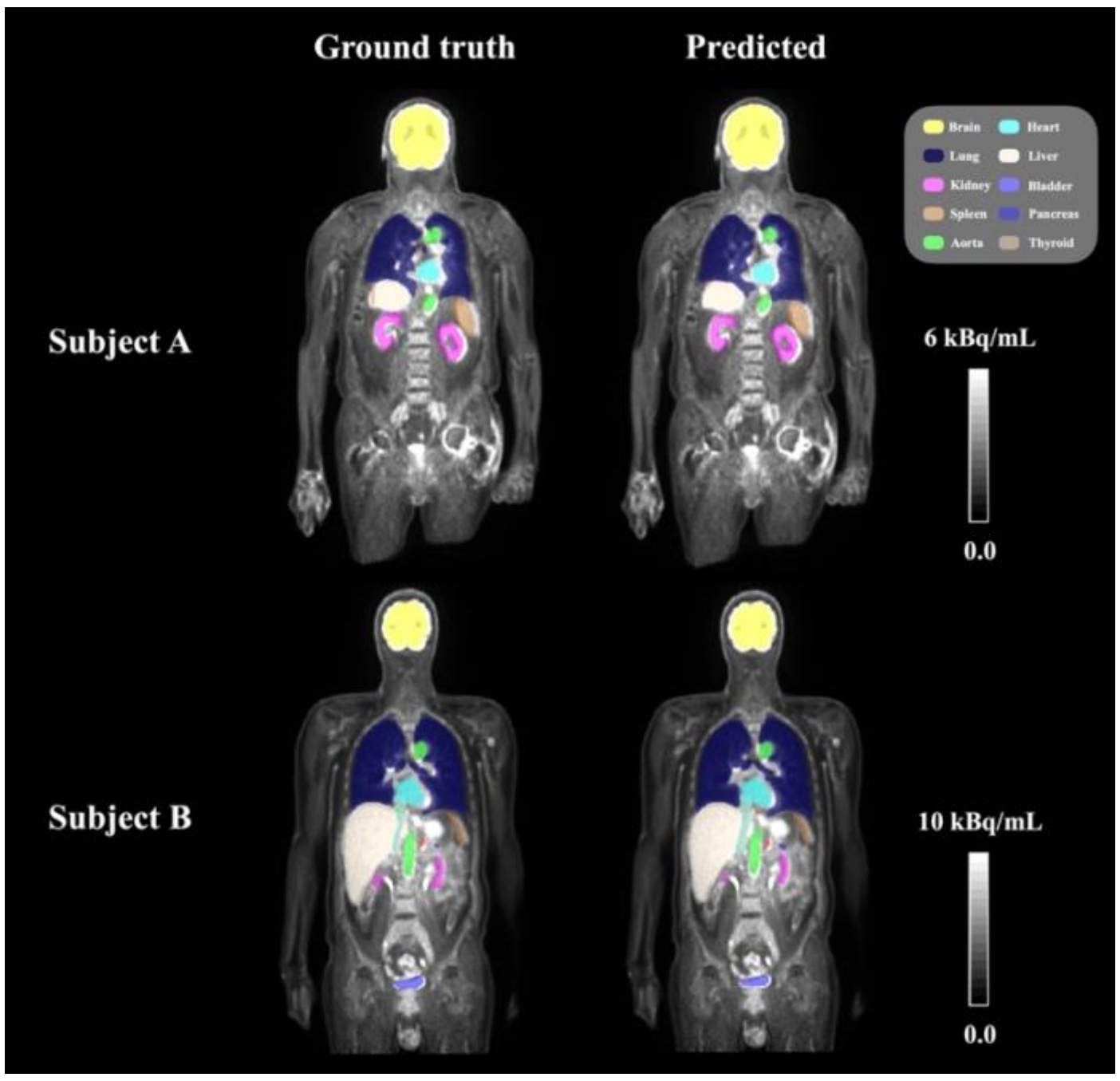

*Figure 1 Exemplary test results of 18F-FDG imaging from SIEMENS Healthineers Vision Quadra.*

## 4 Discussion

The advancement of DL-based attenuation and scatter correction technology provided a starting point for CT-free PET imaging. It enabled several application scenarios, such as multiple PET tracers examinations [18, 19], pediatric patients with previously acquired anatomic images [20], and pharmaceutical developments [21, 22]. However, for CT-free PET imaging to be considered clinically acceptable, the function of anatomical localization must be realized. The combination of PET and CT has increased the confidence of interpreting studies and the accuracy of localization and identification of non-pathologic uptake [10]. Our proposed multi-organ segmentation for total-body PET imaging can replace anatomical imaging by generating a segmentation mask from the PET image.

In our study, we trained the model on NASC-PET images and found that the results were comparable to those obtained from a model trained on attenuation and scatter corrected PET (ASC-PET) images. As the ASC-PET data already incorporates anatomical information from attenuation maps (μ-maps) after correction, we deemed it more appropriate to conduct our study using NASC-PET images to avoid using any information coming from CT images.

It should be emphasized that our preliminary study utilized a model trained on a homogeneous dataset from one scanner and one tracer, which tends to lead to the problem of overfitting of a DL model. We are currently in the process of collecting data from subjects injected with various radiopharmaceuticals on the Vision Quadra and the United Imaging uExplorer, which will be incorporated into our development in the future. For the time being, ground-truth data is generated using the MOOSE software. Our physicians are currently assisting us with manual delieaniation of multiple organs, which will serve as either the ground-truth or be used for fine-tuning. Additionally, previous studies have suggested generating μ-maps or pseudo-CTs from non-corrected PET images [23]. We are interested in comparing our results with those obtained by segmenting organs from generated pseudo-CTs.

## 5 Conclusion

Our study explored the possibility of total-body PET multi-organ segmentation using a deep learning-based method that does not require the anatomical information from CT. This represents an important step towards true CT-free PET imaging and has the potential to improve the efficiency and safety of PET scans.

# Node-Based Motion Estimation Algorithm for Cardiac CT Imaging

Seongjin Yoon[1], Alexander Katsevich[1,2], Michael Frenkel[1], Qiulin Tang[3], Liang Cai[3], Jian Zhou[3], and Zhou Yu[3]

[1]iTomography Corporation, Houston, Texas 77098, USA
[2]Department of Mathematics, University of Central Florida, Orlando, Florida 32816, USA
[3]Canon Medical Research USA, Inc., Vernon Hills, Illinois 60061, USA

**Abstract** Proposed is a semi-iterative whole heart Motion Estimation (ME) algorithm. ME is performed in an iterative fashion in small neighborhoods of motion nodes. Location of the nodes is selected according to a new scheme. Then the nodes are ordered in a tree-like structure based on spatial proximity. The motion of each node is described by a parameterized model, and the motion model at each node is estimated almost independently of the motion of the other nodes. ME at the nodes is performed sequentially according to the selected ordering. During ME, we reconstruct local patches, which are small volumes centered at the motion nodes. Reconstruction is done using short-scan data and the current motion model. Selecting the best motion model is performed by minimizing a motion artifact metric (MAM). Our MAM is the sum of two terms. The first one measures similarity between patches reconstructed from two different short-scan ranges. The second term measures image sharpness at reference phase. Once ME for all nodes is complete, a global motion model is computed by interpolating the estimated local models. Finally, the global model is used for motion compensation in an FDK algorithm. Numerical experiments show that the algorithm is robust and provides good image quality.

## 1 Introduction

Despite the increased gantry rotation speed (of $\sim 0.25$sec per rotation in most advanced scanners presently available), cardiac imaging still suffers from motion artifacts. A number of software-based approaches for improving temporal resolution and reducing motion artifacts have been proposed. The vast of majority of such approaches are based on estimating the motion of the heart. Fully iterative algorithms, which estimate both the volume at reference time and the motion model, are usually time consuming and not suitable for clinical practice [1]. Approaches, which are based on performing motion estimation (ME) first using some shortcuts, are more efficient. A group of popular approaches is based on reconstructing subphase volumes (partial angle reconstructions, or PARs) and using them for ME [2–4]. One option is to perform ME by registering pairs of PARs that have been reconstructed at pairs of points separated by 180° (e.g., as in [2]). Another option is to warp the PARs according to some motion model and add them all up to produce the full image. The optimal motion model is selected by minimizing some motion artifact metric (MAM). The choice of a MAM is usually a difficult task. Finding the optimal motion model is a difficult task too, because the cost functional is highly non-convex and multiple local minima exist [5]. One way to overcome these difficulties is to use an artificial neural network (ANN)[6, 7]. However, the use of ANNs has its own challenges, which include getting ground truth data, stability of the results with respect to fluctuations in HU values, and others. In this paper we develop an ME algorithm based on PARs and MAM, but we do not use an ANN. Instead, the difficulties inherent in the task are solved by using some novel ideas.

## 2 Algorithm description

Our cardiac Motion Estimation and Motion Compensation (ME-MC) algorithm is semi-iterative. It combines locally-iterative motion estimation (in a neighborhood of scattered motion nodes) with analytic motion compensation. The motion of each node is described by a parameterized model, and motion parameters at each node are estimated almost independently of all other nodes. The algorithm begins with the following three steps: (1) determining positions of the nodes, (2) ordering the nodes, and (3) sequential motion estimation at each node by following the selected order. Motion estimation at each node is iterative. Given the current motion model we reconstruct three versions of the same patch using three different short scan subsets of the available projection data. A patch is a small volume centered at a motion node, and the reconstructions are motion compensated. Then we evaluate the quality of the motion-compensated patches based on a cost functional. The functional measures volume similarity between the two versions of the same patch and their volume sharpness. Once motion estimation at all nodes is complete, we perform the final two steps: (4) finding a global motion model of the heart by interpolating the local motion models found in step (3), and (5) performing motion compensated reconstruction using the global motion model found in step (4).

### 2.1 Motion estimation

To evaluate the motion at each node independently, we use a set of patches, and compute the cost functional within each patch. For the purposes of local motion estimation, we assume that the motion does not change much within each patch, so the assumption of rigid body motion is made. Note that we assume rigid body motion only for local motion estimation. The global motion model used for the final motion compensation is elastic.

Our cost functional requires three patches reconstructed using three different short-scan data ranges, we call them SS-ref, SS-0 and SS-1. They are subsets of the total scan range, which is assumed to be slightly larger than one full rotation. SS-ref is the short-scan range, which is centered at the reference phase. This range is used for the final reconstruction. Note that our algorithm neither assumes nor requires that reference phase be the phase where the cardiac motion is minimal. The algorithm successfully reconstructed a motion-compensated volume with the maximum estimated speed reaching 78 mm/s at some nodes. SS-0 and SS-1 are short-scan ranges with their centers at equal distance from the reference phase and on opposite sides of it. Figure 1 illustrates the three short-scan ranges.

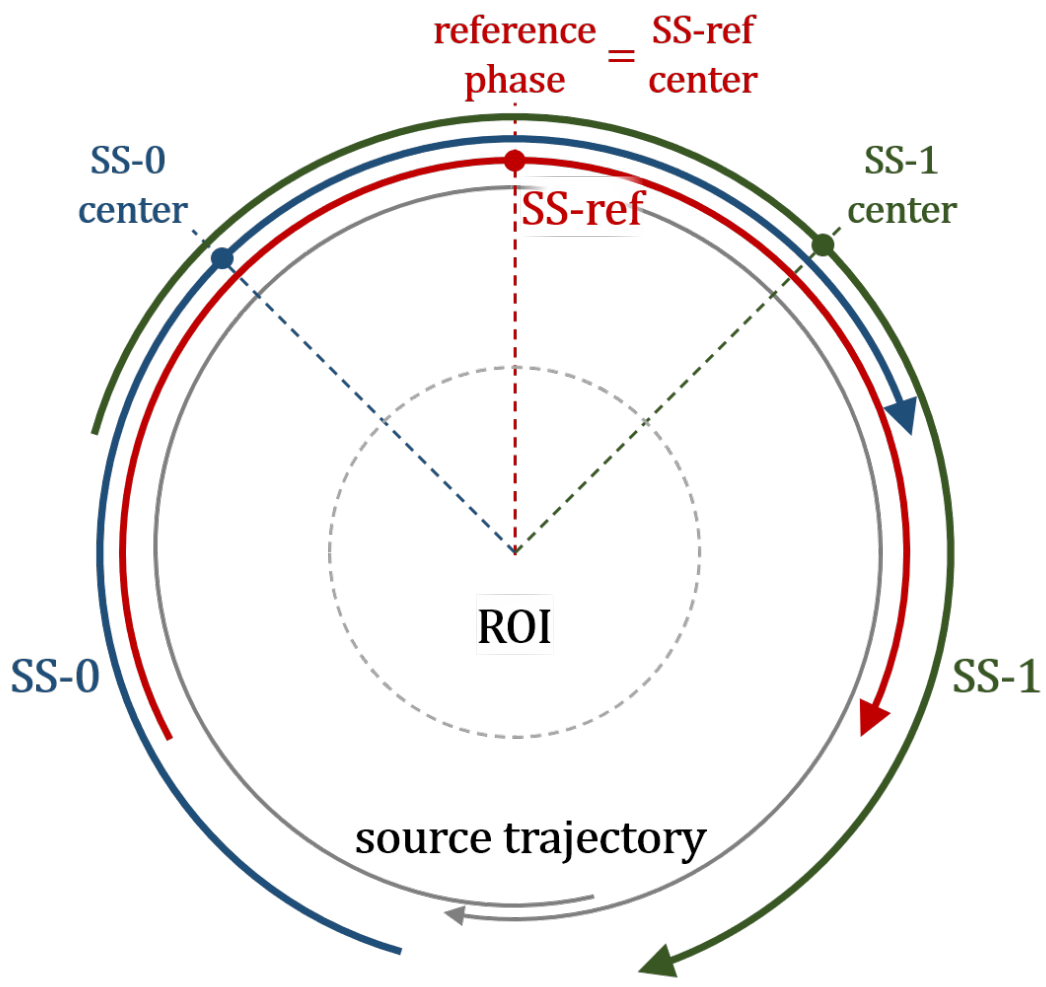

**Figure 1:** Schematic diagram of the short-scan ranges for SS-0, SS-1, and SS-ref, shown on the source trajectory. Solid dots are the short-scan center phases of the corresponding short-scans.

### 2.1.1 Cost functional and its components

For each motion node, we use the following cost functional to estimate the motion:

$$\widehat{\mathbf{M}}_j = \arg\min_{\mathbf{M}} \ \phi_j^{\text{s}}(\mathbf{M}) + \gamma_j^{\text{p}} \phi_j^{\text{p}}(\mathbf{M}) + \gamma_j^{\text{reg}} \phi^{\text{reg}}(\mathbf{M}), \quad (1)$$

where

- $\mathbf{M}$ is the set of motion parameters,
- $\phi^{\text{s}}$ is the volume similarity cost functional,
- $\phi^{\text{p}}$ is the volume sharpness cost functional,
- $\phi^{\text{reg}}$ is the motion parameter regularizer,
- $\gamma_t$ is the strength parameter of the corresponding component $\phi_t$,

### 2.1.2 Volume similarity

Let $F_j^0(\mathbf{M})$, $F_j^1(\mathbf{M})$, and $F_j^{\text{ref}}(\mathbf{M})$ denote motion compensated, reconstructed patches centered at the $j$-th node using short-scans SS-0, SS-1, and SS-ref, respectively. When the motion near the node is properly estimated and compensated, the reconstructed volumes centered at different phases should be almost identical. Therefore, we use the sum of squared differences between $F_j^0(\mathbf{M})$ and $F_j^1(\mathbf{M})$ to gauge how accurate the motion model $\mathbf{M}$ is. We call the resulting value volume similarity $\phi^{\text{s}}$, and it is the first part of our MAM.

$$\phi_j^{\text{s}}(\mathbf{M}) = \frac{1}{N_j} \| F_j^0(\mathbf{M}) - F_j^1(\mathbf{M}) \|_2^2. \quad (2)$$

Here $N_j$ is the number of voxels in the patch around the $j$-th node.

### 2.1.3 Volume sharpness

Image sharpness is used as the second part of the MAM. We estimate image sharpness by summing a vector norm of the spatial gradient over the voxels in the patch. Generally, imaging of moving objects produces blurry images. However, CT reconstruction of moving objects produces not only a blurry image, but also a depression artifact, which is characterized by image values that are too low. Frequently, these values are much lower than the average reconstructed value inside the patch. Depression artifacts tend to increase sharpness and cause bad local minima of MAM. To reject the sharpness increase due to a depression artifact, we apply a soft-thresholding function $G(\mathbf{x})$, which downweights the sharpness value around the depression artifact. Detection of depression artifacts is performed by analyzing image values. Since the final reconstruction is done for the SS-ref, we maximize the sharpness of $F_j^{\text{ref}}(\mathbf{M})$ as part of MAM to estimate motion.
Let $\mathbf{x}$ denote the vector of voxel spatial coordinates, and $f_j^{\text{ref}}(\mathbf{x}, \mathbf{M})$ denote the value of the motion compensated reconstruction $F_j^{\text{ref}}$ at the voxel $\mathbf{x}$. Then, we define the volume sharpness cost functional as follows:

$$\phi_j^{\text{p}}(\mathbf{M}) = -\frac{1}{N_j} \sum_i \left\| \nabla f_j^{\text{ref}}(\mathbf{x}_i, \mathbf{M}) \right\|_p^p G(\mathbf{x}_i), \quad (3)$$

where $p$ is the exponent in the $L_p$ vector norm.

### 2.1.4 Regularization

Minimization of the sum of the volume similarity and sharpness cost functionals with respect to $\mathbf{M}$ can be used to estimated motion. However, this sum is highly non-convex, there are many local minima, and it is often difficult to find the global minimum without an exhaustive search.
To use a gradient-based optimization technique, regularization is required to avoid undesirable local minima. We achieve this by penalizing various non-physical features, such as an excessively large motion amplitude and a non-smooth change in space.
To make motion estimation more stable, we also adjust the regularization strength $\gamma_j^{\text{reg}}$ and the patch size for each node depending on image features inside the patch.

### 2.2 Node distribution

The distribution of motion nodes inside the heart is designed in such a way so that the following two conditions are satisfied: (1) motion nodes are distributed in an optimal fashion so that the minimum number of nodes can be used to accurately represent the motion of the entire heart; and (2) motion nodes are placed only in the regions where sufficiently distinctive features (e.g., strong edges) are present. This ensures the robustness of motion estimation.

To satisfy condition (2), we distribute the nodes in regions where anatomically significant features and/or significant motion is present. Motion significance is measured in an automated fashion by computing the edge difference between uncompensated reconstructions of SS-0 and SS-1.

While the same overall algorithm is used to estimate motion at all nodes, some parameters of the cost functional may vary from node to node depending on anatomical characteristics of the image near the node.

### 2.3 Warm-start sequence

As was mentioned earlier, the motion estimation cost functional has multiple local minima, and gradient-based optimization often fails to converge to the desired minimum. Along with the regularization described in section 2.1.4, we use a predetermined warm-start sequence during motion estimation to improve robustness of the algorithm. The idea is to estimate the motion of each node one by one sequentially, so that iterative motion estimation at one node is warm-started by the motion model computed at an already processed, nearby node (called the parent node below). Creating a warm-start sequence requires sorting the nodes. Let $\mathbf{N}_k$, $k = 1, 2, \ldots$, denote the desired node sequence. The goal of the sorting is to make sure that for each node in the sequence $\mathbf{N}_k$, $k \geq 2$, there is another node earlier in the sequence $\mathbf{N}_m$, $m < k$, so that $\text{dist}(\mathbf{N}_m, \mathbf{N}_k)$ is below a threshold. The node $\mathbf{N}_{m_k}$, $m_k < k$, for which this distance is minimal is called the parent node of $\mathbf{N}_k$. This ensures that the change of motion parameters between a node and its parent node is small. The sorting algorithm works as follows. The algorithm is initiated by creating two sets of nodes: sorted and unsorted. Initially, the list of sorted nodes contains only a starting node, which is located at the start of the RCA, and the list of unsorted nodes contains all the other nodes. Then the following steps are performed. (1) Find the closest (in terms of the Euclidean distance) node pair between the sorted nodes and unsorted nodes. (2) Remove the identified unsorted node from the unsorted node list and append it to the end of the sorted node list. The other (already sorted) node from the optimal pair is marked as the parent node for the newly added node. (3) Repeat (1) and (2) until the unsorted node list is empty.

Once node sorting is over, we run motion estimation at each node as described in section 2.1 by following the order in the sorted node list. Motion estimation at the first node starts with the zero motion. Motion estimation at each subsequent node is warm-started using the estimated motion at its parent node.

## 3 Test results

We present reconstruction results for clinical datasets acquired using a Canon Aquilion ONE 320-slice CT scanner. A total of 22 clinical datasets were tested, and all results showed good improvement after ME-MC is applied. Figures 2 to 5 show four clinical data examples of our ME-MC FDK reconstructions compared with the uncorrected FDK reconstructions. The uncorrected images are on the left, and the corresponding corrected ones - on the right. Comparing the locations marked by arrows, we see that the algorithm provides good image quality.

## 4 Conclusions

In this work, we presented a computationally inexpensive algorithm for cardiac motion estimation and motion compensation. The algorithm is based on MAM minimization, and a number of ideas have been implemented in order to avoid false local minima. Numerical experiments show that the algorithm is robust and provides good image quality.

## 5 Acknowledgements

Clinical images courtesy of Dr. Marcus Chen, National Heart, Lung and Blood Institute, National Institutes of Health, USA.

A. Katsevich is a shareholder and chief technology officer of iTomography Corporation and, as such, may benefit financially as a result of the outcomes of the work reported in this publication.

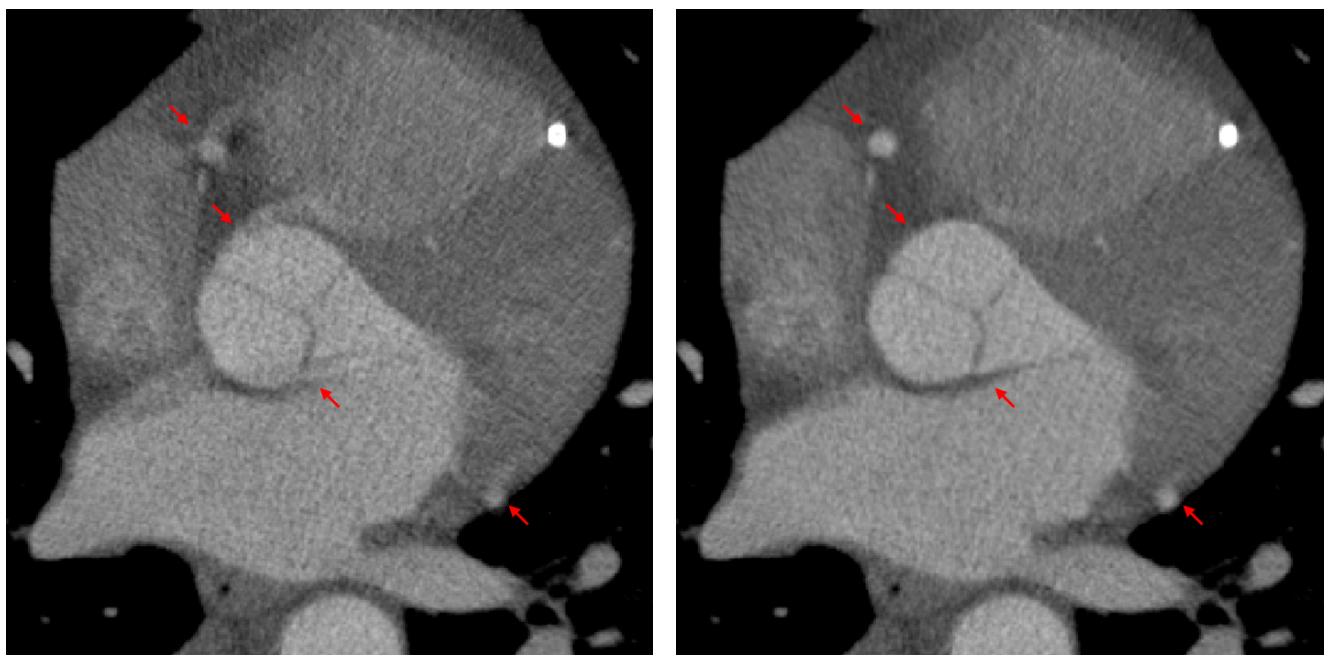

**(a)** Uncorrected FDK **(b)** ME-MC FDK

**Figure 2:** Example of ME-MC results, first dataset. Maximum estimated motion speed is 78 mm/s.

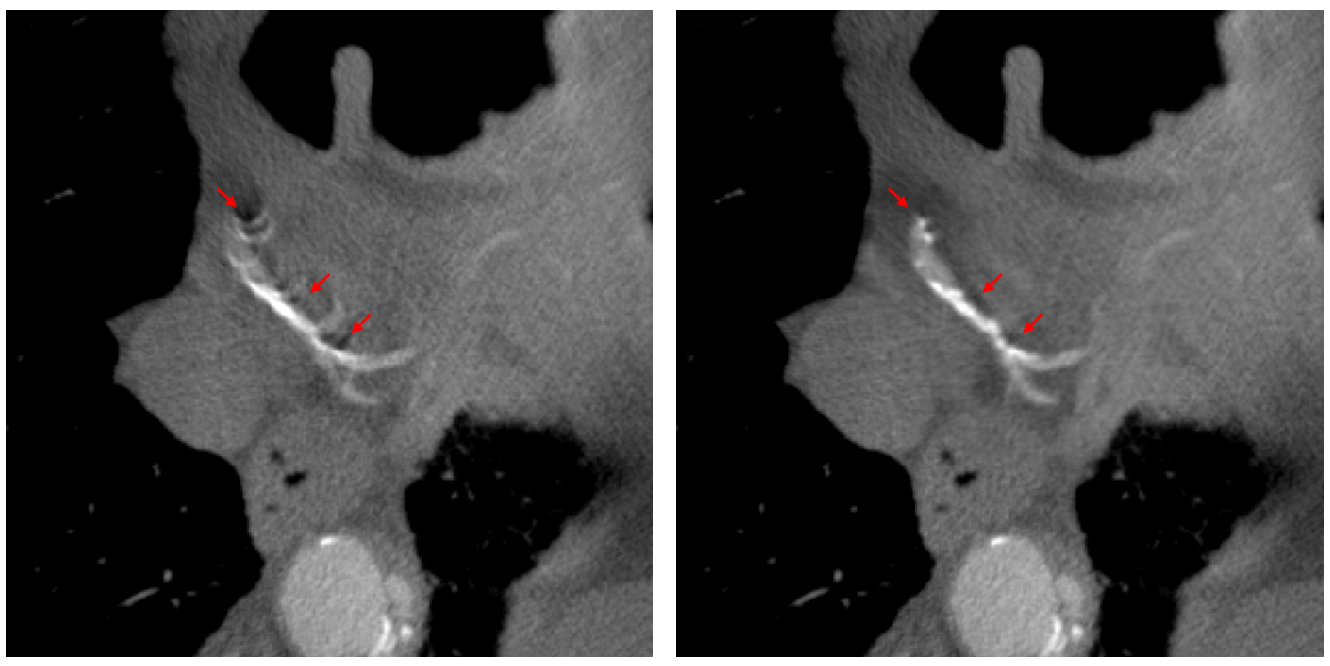

**(a)** Uncorrected FDK **(b)** ME-MC FDK

**Figure 3:** Example of ME-MC results, second dataset. Maximum estimated motion speed is 46 mm/s.

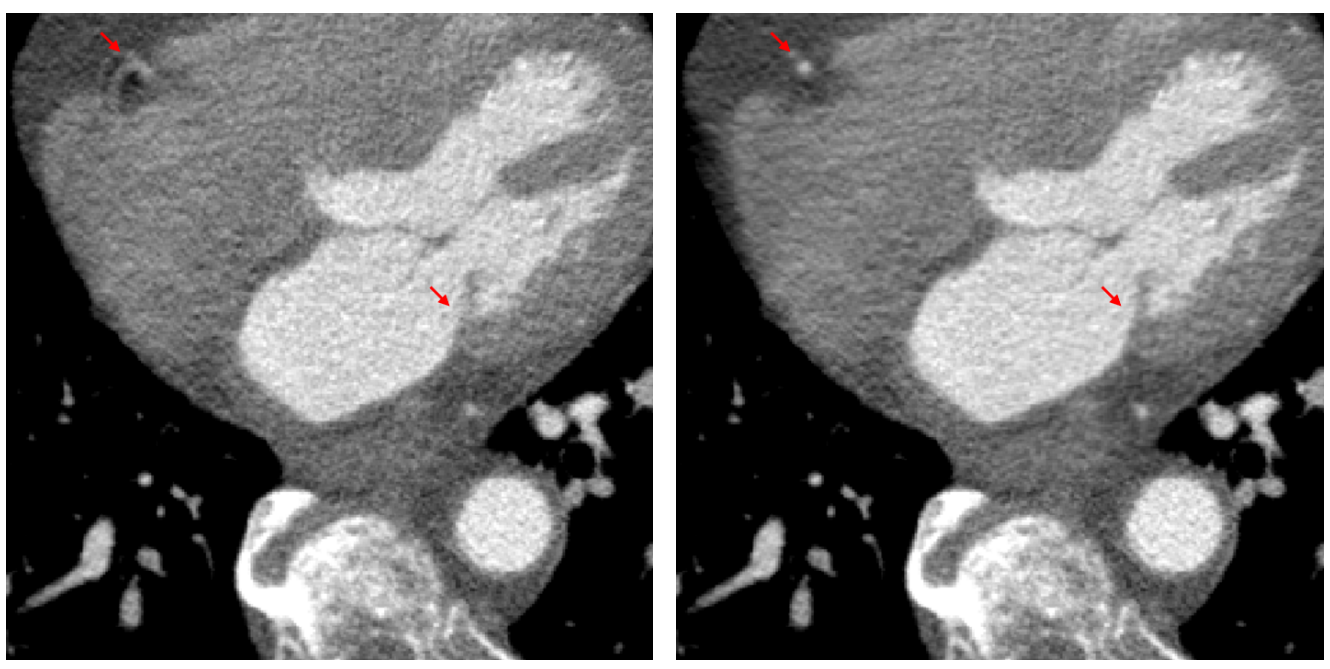

**(a)** Uncorrected FDK **(b)** ME-MC FDK

**Figure 4:** Example of ME-MC results, third dataset. Maximum estimated motion speed is 63 mm/s.

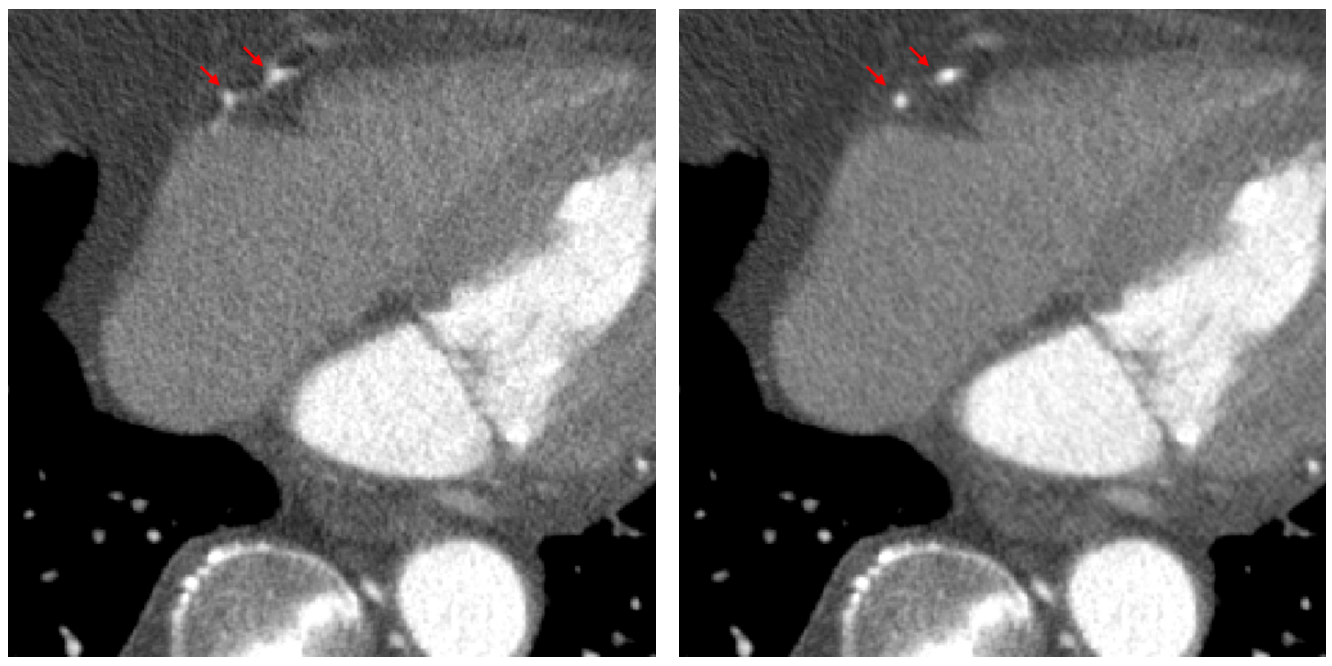

**(a)** Uncorrected FDK **(b)** ME-MC FDK

**Figure 5:** Example of ME-MC results, fourth dataset. Maximum estimated motion speed is 45 mm/s.

# Sparse-view CT Spatial Resolution Enhancement via Denoising Diffusion Probabilistic Models

Nimu Yuan[1], Jian Zhou[2], and Jinyi Qi[1]

[1] Department of Biomedical Engineering, University of California, Davis, Davis, United States
[2] Canon Medical Research USA, INC, Vernon Hills, United States

**Abstract** The temporal resolution of x-ray computed tomography (CT) is limited by the scanner rotation speed and detector readout time. One approach to reduce the readout time is to acquire fewer projections. However, reconstruction using sparse-view data could result in a loss of spatial resolution loss and reconstruction artifacts that could impact the accuracy of clinical diagnoses. Therefore, improving the spatial resolution of sparse-view CT (SVCT) is of great practical value. The aim of this study is to investigate the potential of using denoising diffusion probabilistic models (DDPMs) to eliminate noise and streaky artifacts while preserving fine details and enhancing textures in SVCT images. The DDPM was trained on a simulated dataset and its effectiveness was evaluated on both simulated and real data. The results of the study showed that the DDPM was successful in not only suppressing noise and artifacts, but also significantly enhancing the texture of the CT images.

## 1 Introduction

X-ray computed tomography (CT) plays an important role in medical imaging [1]. There are applications where high temporal resolution is important, such as reducing motion artifacts or visualizing spatial-temporal changes of contrast agents [2]–[4]. The temporal resolution of X-ray CT is limited by two factors: the rotation speed of the X-ray tube and the time it takes to read the detector for all the projection data. To reduce the detector readout time, fewer projections can be acquired per rotation. However, this can lead to a loss of spatial resolution in the reconstructed images, causing reconstruction artifacts [5] that can negatively impact the accuracy of clinical diagnoses. Improving the quality of sparse-view CT (SVCT) images is therefore of great practical significance. Compared to the traditional sparse-view problem, which is based on step-and-shoot acquisition, the SVCT investigated in this work suffers not only from angular under-sampling, but also from angular blurring due to the continuous rotation of the X-ray source.

Numerous CNN-based methods have been proposed to restore SVCT images and have shown impressive results. However, these enhanced SVCT images usually suffer from over-smoothing. The overall appearance and local textures are significantly different from the CT images reconstructed using conventional methods, such as filtered back-projection (FBP). In recent years, deep generative models (DGM), such as generative adversarial networks (GANs) [6] and denoising diffusion probabilistic models (DDPMs) [7], have shown promise in complex image enhancement tasks in the computer vision area. Compared to GANs, DDPMs have the advantage of producing more accurate distribution estimates, which can be useful in tasks such as noise and texture estimation [7], and they can effectively address the over-smoothing issue. DDPMs can also be trained using more efficient algorithms, such as maximum likelihood estimation (MLE) or maximum *a posteriori* (MAP) estimation, which are faster and more stable than the adversarial training approach used in GANs.

The purpose of this study is to examine the potentials of utilizing DDPM to enhance the spatial resolution and restore textures in SVCT images.

## 2 Materials and Methods

### 2.1 The SVCT problem

This study aims at recovering full-view CT (FVCT) images from SVCT images. We focus on the sparse-view data from helical CT scans with a continuously rotating X-ray source. In this case, each sparse-view projection covers a wider angular range than a full-view projection and is equivalent to the average of multiple full-view projections. This angular averaging leads to a radially dependent blurring effect in the reconstructed SVCT images, in addition to the streaks caused by angular under-sampling.

### 2.2 SVCT enhancement using DDPM

In this study, we applied a conditional DDPM to enhance the SVCT images, as depicted in Fig. 1. The underlying procedure of the network includes a forward diffusion process that gradually adds noise to input and a reverse denoising process that learns to generate data by denoising. DDPM works by learning to transform a standard normal distribution into an empirical data distribution through a sequence of refinement steps using a U-Net architecture network. Details can be found in references [7], [8]. In our experiments, we set the number of the root features to 64 and the number of time steps to 2000. The network was trained with SVCT images as the conditional inputs and FVCT images as the labels. Additionally, as a point of comparison, we also trained and tested a conventional U-Net utilizing the L1 loss function.

### 2.3 Experimental datasets

We obtained full-view helical CT scans of four subjects. Three scans were used to generate training data and the fourth scan was used for evaluation. The FVCT data have 1,200 projections per rotation. The corresponding SVCT scans were simulated by averaging every five projections into one, resulting in 240 projections per rotation. The tube current for the FVCT scans ranged from 103 to 461 mAs, with a tube voltage was 120kVp. Both the FVCT and SVCT images were reconstructed using a 512×512 image matrix size. To augment the training dataset, 12 image rotations were performed with 30-degree intervals.

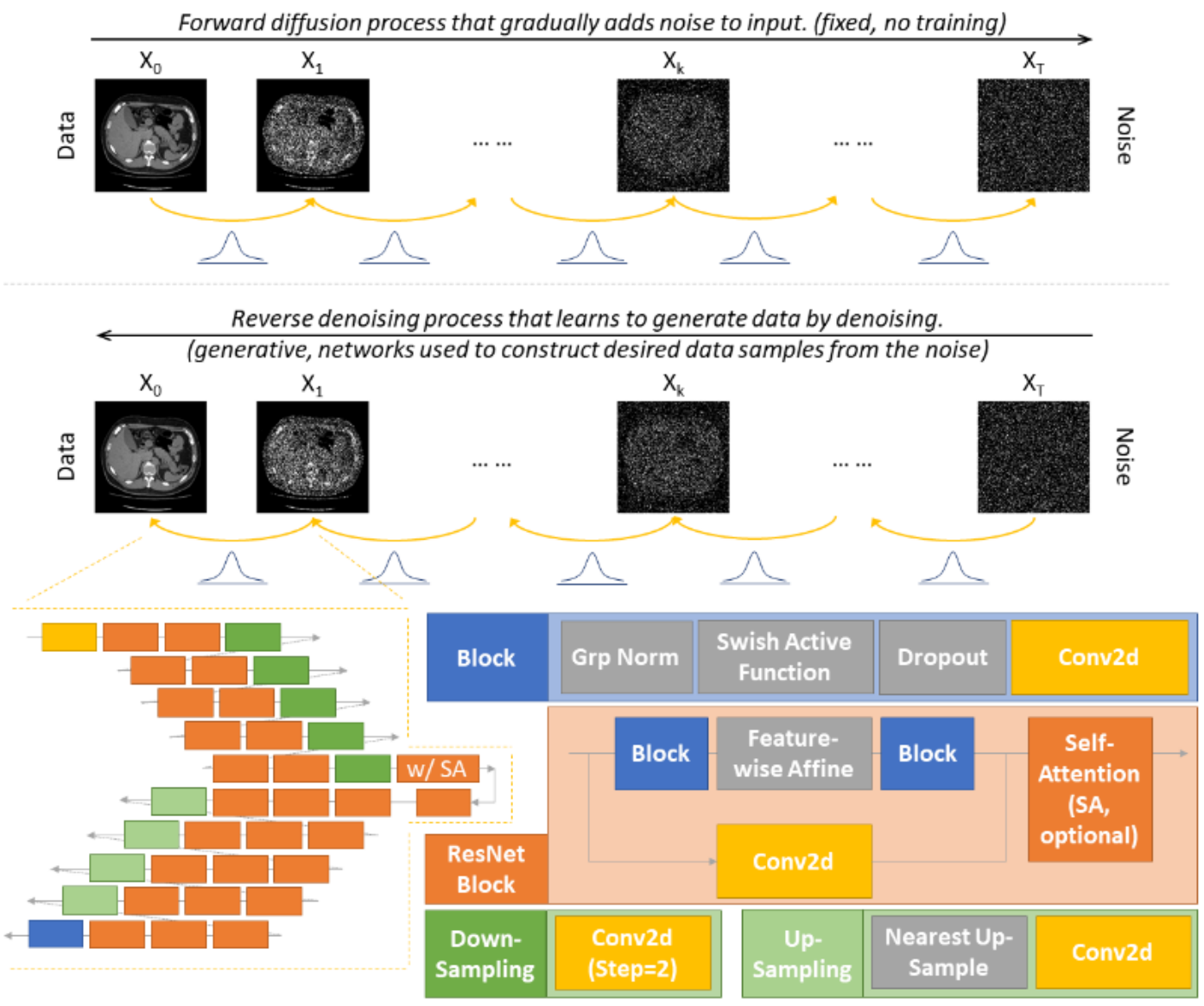

Fig. 1. The forward diffusion process ($X_0 \rightarrow X_T$) gradually adds Gaussian noise to the target image. The reverse inference process ($X_T \rightarrow X_0$) iteratively denoises the target image conditioned on a source image. The source image is not shown here.

## 2.4 Metrics for evaluation

In this study, the commonly used metrics of mean absolute error (MAE), signal-to-noise ratio (SNR), and structural similarity index (SSIM) [9] were evaluated to compare different networks. To examine the difference of textures, two addition metrics were introduced: histogram correlation (HC) and local-binary-pattern-based (LBP-based) texture similarity [10], [11]. The definitions of these two new metrics are provided in subsequent sections. In addition, we also compared the noise power spectrum (NPS) [12] of the network predictions with that of the reference FVCT images.

### 2.4.1 Histogram Correlation (HC)

HC is a macroscopic metric that measures the histogram similarity between the network predictions and labels. It is calculated as:

$$HC = \frac{\sum[H(I_{pred}) - \bar{H}(I_{pred})][H(I_{FVCT}) - \bar{H}(I_{FVCT})]}{\sqrt{\sum[H(I_{pred}) - \bar{H}(I_{pred})]^2 \sum[H(I_{FVCT}) - \bar{H}(I_{FVCT})]^2}} \quad (1)$$

where $H(I_{pred})$ and $H(I_{FVCT})$ are the Hounsfield unit (HU) histograms of the network prediction and FVCT images, respectively, and

$$\bar{H}(I_*) = \frac{1}{N} \sum H(I_*) \quad (2)$$

and $N$ is the total of histogram bins.

While HC is not a comprehensive metric and it is difficult to accurately assess image quality using HC alone, it can provide valuable information when used in conjunction with other pixel-wise metrics, such as MAE, for a more comprehensive understanding of the local texture and accuracy of the HU values in the images.

### 2.4.2 Local-binary-pattern-based (LBP-based) texture similarity

LBP [11] is a visual descriptor commonly used in computer vision to extract texture information from images. LBP-based texture similarity (LBP-TS) calculates the similarity or dissimilarity between two LBP histograms, which are generated by quantifying the frequency of different LBP patterns that appear in the images. The similarity measure used in this study was Kullback–Leibler divergence, with lower values indicating a better match between the two images. The calculation process is shown in Fig. 2.

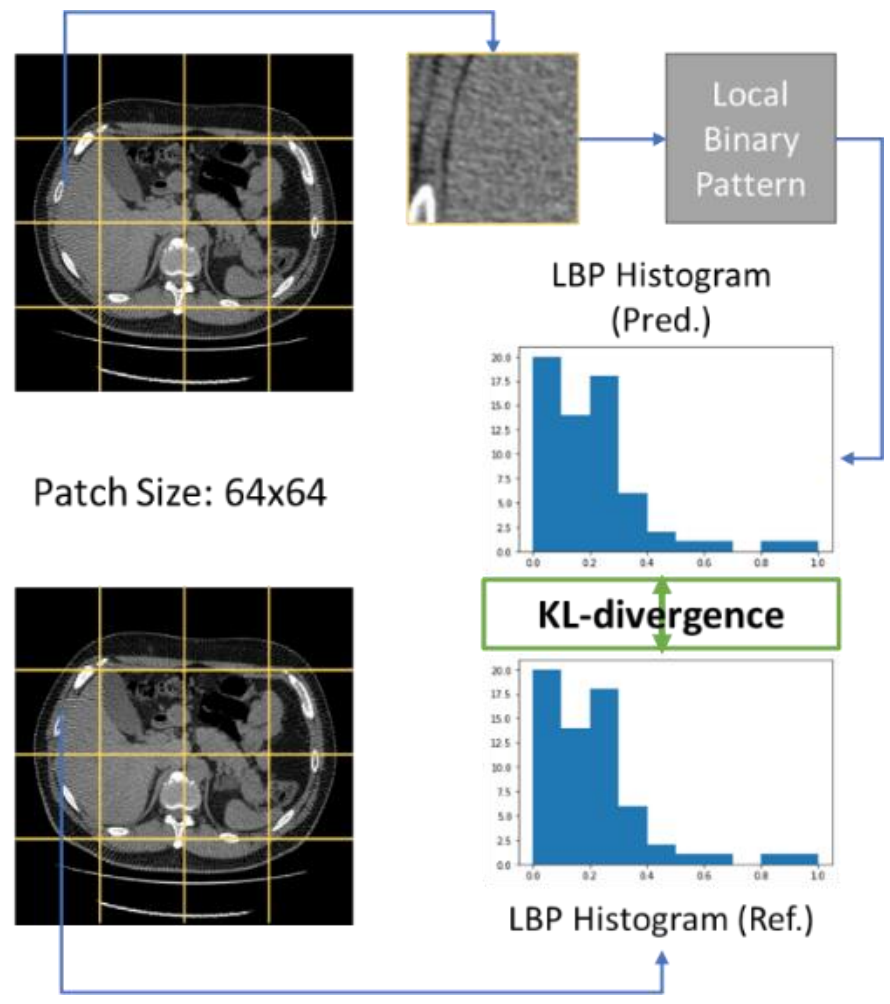

Fig. 2. The schematic diagram of the LBP-based texture similarity calculation.

## 2.5 Noise power spectrum (NPS)

NPS is employed to further compare the texture restoration abilities of the networks. Twenty image patches, each with a size of 64×64, were extracted from the liver region across 20 transaxial slices. To remove intensity variation, the averaged patch was filtered using a non-local means (NLM) method and subtracted from all the patches. The resulting patches were considered as random noise and used to calculate the NPS.

# 3 Results

## 3.1 Image comparison

Fig. 3 compares two representative image slices from the simulated test dataset. The images, from left to right, are the SVCT images, predictions from the U-Net, predictions from the proposed DDPM, and FVCT reference images, respectively. The display window for the lung slice is set to Width $(W) = 1500$ HU/Level $(L) = -600$ HU, while the display window for the abdominal slice is set to $W = 400$ HU$/L = 40$ HU. The MAE, SNR, SSIM, HC, and LBP-TS were calculated and displayed below each image. The HU histograms are also presented in the right boxes.

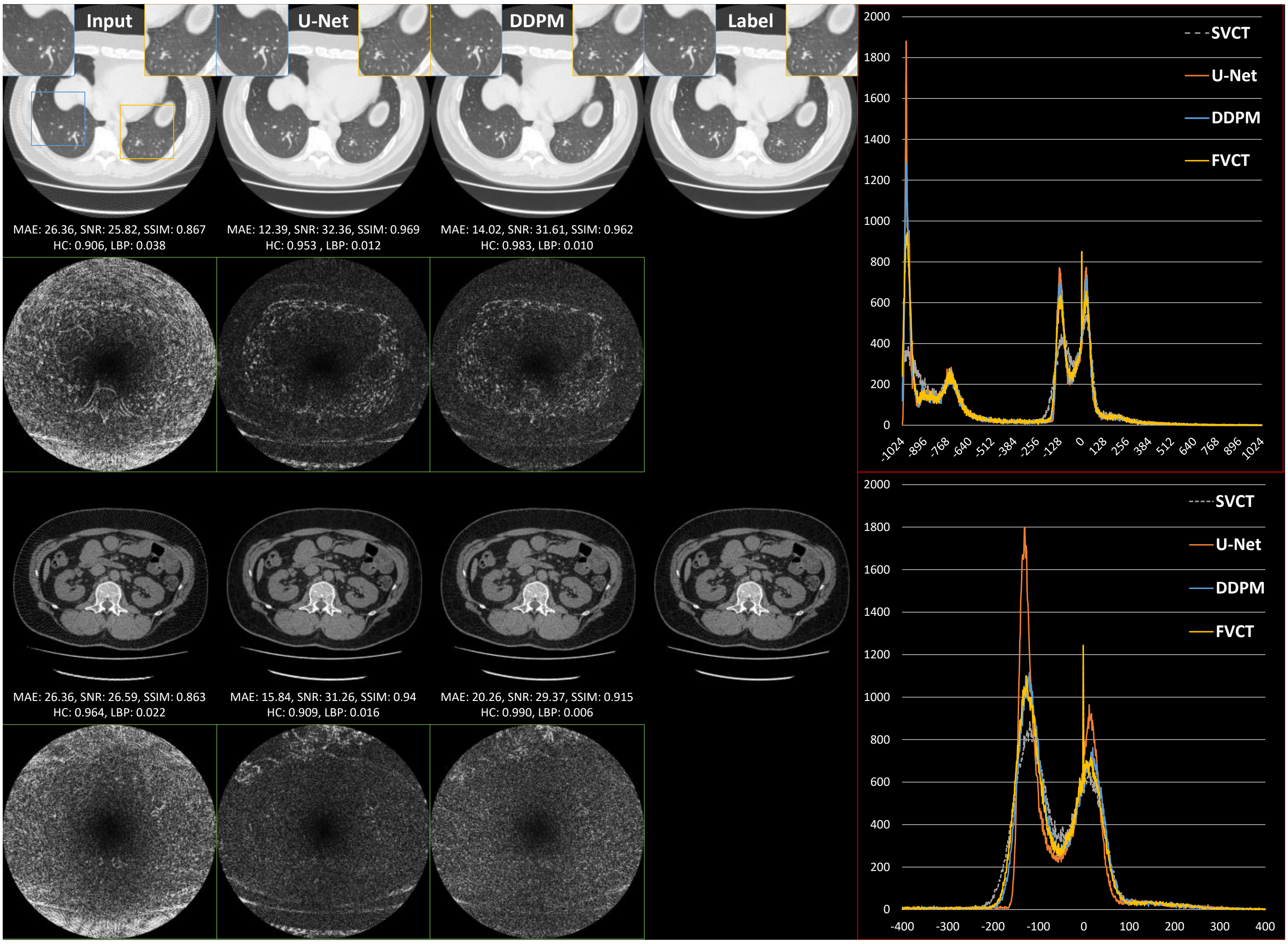

Fig. 3. Comparison of two representative slices from the simulated test dataset. From left to right, the images are SVCT images, predictions from the U-Net, predictions from the proposed DDPM and FVCT images (reference), respectively. Blue and yellow boxes are the enlarged regions, the green boxes are the corresponding absolute difference image with respect to the FVCT reference images (W=80 HU/L=40 HU), and red boxes are the HU histograms of the corresponding slices.

It is evident that both network predictions offer significant improvements over the SVCT images in terms of MAE, SNR, and SSIM, effectively reducing noise and suppressing artifacts. As shown in the magnified regions, both predictions enhance the sharpness of anatomical boundaries in the lung regions while preserving the details and fine structures of CT images in the abdominal regions. Furthermore, a comparison of the two network predictions shows that the results from the proposed DDPM exhibit a higher degree of texture restoration, a closer HU distribution to the reference, and an overall image appearance that is more similar to the FVCT reference images.

### 3.1 Quantitative results

Table 1 presents the averaged results of the MAE, SNR, SSIM, HC, and LBP-TS calculated over all tested slices for the SVCT image, the U-Net prediction, and the DDPM prediction. On average, the DDPM prediction achieved slightly lower scores in terms of MAE, SNR, and SSIM compared to the U-Net prediction, but significantly outperform the U-Net in terms of HC and LBP-TS (as determined by a paired t-test with a p-value $\ll 0.01$).

Table 1. Averaged metrics over all test CT slices (mean ± std)

| | SVCT | U-Net | DDPM |
|---|---|---|---|
| MAE | 26.40 ± 0.88 | **14.73 ± 1.10** | 18.65 ± 1.92 |
| SNR | 26.50 ± 0.53 | **31.81 ± 0.37** | 30.05 ± 0.65 |
| SSIM | 0.864 ± 0.006 | **0.951 ± 0.008** | 0.927 ± 0.015 |
| HC | 0.946 ± 0.029 | 0.897 ± 0.027 | **0.989 ± 0.006** |
| LBP-TS | 0.027 ± 0.006 | 0.015 ± 0.021 | **0.006 ± 0.002** |

### 3.2 NPS comparison

Fig. 4(a) shows the 2-D NPS and the corresponding absolute difference images with respect to the reference 2-D NPS from the FVCT. Fig 4(b) shows the averaged radial 1-D NPS. As shown in Fig. 4 (a) and (b), the U-Net not only suppressed the noise, but also removed the textures from the liver regions. In comparison, for both 2-D and 1-D NPS, the results of the DDPM predictions were much closer to those of the FVCT, providing further evidence that the proposed DDPM is capable of restoring a higher degree of textures.

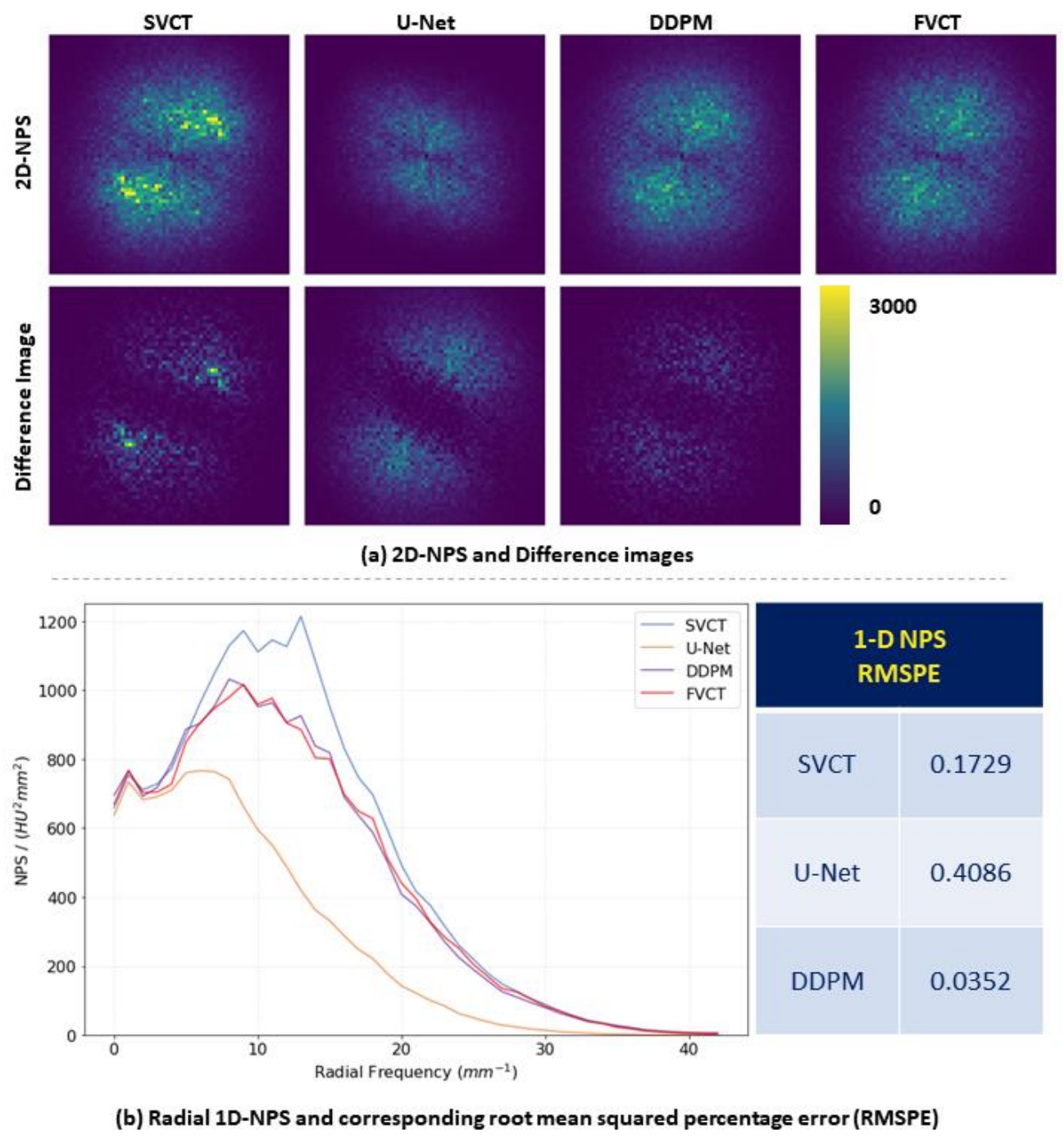

Fig 4. NPS results. (a) 2-D NPS results and the corresponding absolute difference images with respect to the FVCT reference 2-D NPS. From the left to right, results from SVCT images, U-Net predictions, DDPM predictions and FVCT images (reference), respectively. (b) 1-D radial NPS curves and corresponding mean squared perccentage error, blue: SVCT, orange: U-Net, purple: DDPM, and red:FVCT.

## 4 Discussion

The most challenging aspect of the SVCT image enhancement was the texture restoration task. Due to the angular under-sampling and blurring, the textures were obscured by streaky artifacts and rotational smoothness. U-Nets, limited by the properties of CNNs and pixel-wise loss functions, faced significant difficulties in restoring these textures, which led to over-smooth predictions, as demonstrated by the results shown in Figs. 3 and 4. In contrast, DDPMs were specifically designed to learn the underlying probability distribution of the training dataset and therefore had a more significant advantages in recovering the textures of FVCT images, as evidenced by the NPS comparison in Fig. 4.

As indicated in Table 1, U-Net outperforms DDPM in conventional metrics such as MAE and SNR. This can be attributed to the fact that U-Net tends to generate smooth images, while DDPM prioritizes matching the image style and imitating the textures. Since most of the textures are randomly distributed, DDPM aims to reproduce the random distributions, which leads to larger pixel-wise differences, instead of predicting the exact same pixels as in the label.

The main limitation of DDPM is the computational time required for deployment. Predicting a 512×512 CT slice took approximately 2 minutes on an NVidia A100 GPU. In future studies, we plan to reduce the root feature size and the number of time steps and further simplify the models to decrease the deployment time.

## 5 Conclusion

This study investigated the potential of using DDPM to enhance the SVCT images. The findings of this study suggest that DDPM has the potential to be a valuable tool to effectively suppress the noise, remove artifacts, and preserve fine structures in the SVCT images while restoring the textures. Further research is needed to validate these results using larger datasets.

# Folded-VVBP tensor network for sparse-view CT image reconstruction

Sungho Yun[1], Dain Choi[1], Seoyoung Lee[1], Jaehong Hwang[1], Jisung Hwang[1], Gyuseong Cho[1] and Seungryong Cho[1]

[1]Department of Nuclear and Quantum Engineering, KAIST, Daejeon, Republic of Korea

**Abstract** The sparse-view computed tomography (CT) is one of the effective ways to decrease the patient dose. However, the decreased sampling rate often causes aliasing artifacts in the reconstructed image. In recent, the view-by-view backprojection (VVBP) network [] was introduced for low mAs and sparse-view CT image reconstruction. The VVBP tensor has some unique characteristics that it has a structure self-similarity and a predictable artifact distribution which could helpful for reducing artifacts. However, the VVBP required a heavy computational load due to its high dimensional feature and diverse data distribution. In this study, we introduced a 'Folded-VVBP' algorithm that compresses the original signal while enhancing the structure self-similarity. The proposed folding technique could decrease the computational burden of the VVBP tensor for a network application and also it was able to increase the performance of the network. Furthermore, we introduced squeeze and excitation implemented residual-encoder-decoder convolutional neural network (SE-RED-CNN) for the network architecture to properly utilize the Folded-VVBP. Our proposed algorithm was compared with the vanilla VVBP tensor and single image-domain network with quantitative evaluation. As a result, our algorithm shows better results in terms of artifact removal and generalization performance.

## 1 Introduction

The sparse-view computed tomography (CT) is one of the effective ways for low-dose x-ray imaging. However, due to the decreased sampling rate, severe streaks occur in reconstructed images which are called aliasing artifacts. The aliasing artifact usually occurs in the reconstructed image when the sampling condition unmet the Nyquist sampling criterion. The criterion explains twice the highest spatial frequency as a minimum required sampling rate to avoid spectral overlap. But the gantry type CT scanner (3rd generation CT) cannot achieve this criterion since the measurement cannot be closer than the detector width. To combat aliasing, there have been lots of studies to alleviate the artifact. The interpolation-based method [1] has been introduced that utilizes artificially synthesized views using neighbors through linear or directional interpolation. In another way, iterative reconstruction-based approaches have been introduced with total variation terms such as SART-TV [2, 3]. The reconstructed image from the algorithm was quite good, however, the heavy computational cost that sacrifices reconstruction speed was the main issue for a practical application.

In recent, many network-based deep-learning approaches have been introduced for reducing aliasing artifacts. As a projection-domain approach, Lee *et al* introduced an interpolation network that generates view information [4]. The generative adversarial network (GAN) method which utilizes a discriminator also has been widely studied due to its realistic image representation and robust performance for artifact reduction [5]. In the image domain application, Zhang *et al* introduced a streak removal network using a dense net architecture with a deconvolution layer [6]. In further, to fully utilize co-domain knowledge for artifact removal, the hybrid network that utilizes both domains [7, 8] was introduced. Wu *et al* proposed a dual-domain residual-based optimization network (DRONE) that utilizes the additional GAN-based image-domain network to compensate for image defects that originated from the initial projection domain network. The results were superior to the compared single-domain-based network except for the additional computational cost and the network complexity. The image reconstruction network also has been introduced for low-dose CT imaging. The deep-image-prior (DIP) based method has shown promising results for the sparse acquisition and limited angle cases [10]. It makes the network predict the original image from a latent random vector by using an explicitly known forward model. However, the DIP-based method has to re-trained the network when a new input is given due to its unique training method.

Meanwhile, Tao *et al* proposed a view-by-view backprojection network (VVBP) [14], which utilizes a sorted backprojection tensor by inspiring from the intermediate process of FBP. The VVBP tensor has some unique characteristics it as structure self-similarity, predictable noise, and artifact distribution. The network shows promising results in the comparison between existing single-domain and hybrid-domain networks for noise and aliasing artifact reduction in their study. However, the VVBP required a heavy computational load due to its high dimensional feature and diverse data distribution in the view direction. Therefore, they suggested a down-sampling technique to compromise the required computational load.

In this study, we introduced a 'Folded-VVBP' algorithm that compresses the original signal twice while enhancing the structure self-similarity. The proposed folding technique could decrease the computational burden of the VVBP tensor for a network application and also it was able to increase the performance of the network. In further, we introduced squeeze and excitation implemented residual-encoder-decoder convolutional neural network (SE-RED-CNN) for the network architecture. In results section, our proposed algorithm was compared with the vanilla VVBP tensor and single image-domain approach with quantitative evaluation.

## 2 Materials and Methods

*A. View-by-view backprojection (VVBP) tensor*

The view-by-view backprojection tensor is an image-stack that composed of a respective view backprojection. If the tensor is averaged along the view-direction, we could regard the image as a FBP image. Tao *et al* [] was firstly sorted this VVBP tensor and studied its unique characteristics. In their observation, the structure self-similarity, sparsity, and predictable gaussian forms of noise statistics were observed. The structure self-similarity explain the group of similar value of backprojection signal could be visualized as an image structure (in Fig. 1). In aspects of the FBP pixel value is a mean of the total view-signal, it is explainable that the structure self-similarity is stronger near the median slice of VVBP tensor. Also, due to the sorting process, the highest and the lowest signals are located at the top and bottom of the tensor, meaning that the lowest structure self-similarity and artifact-related signal (such as noise and streaks) are observed at those indices.

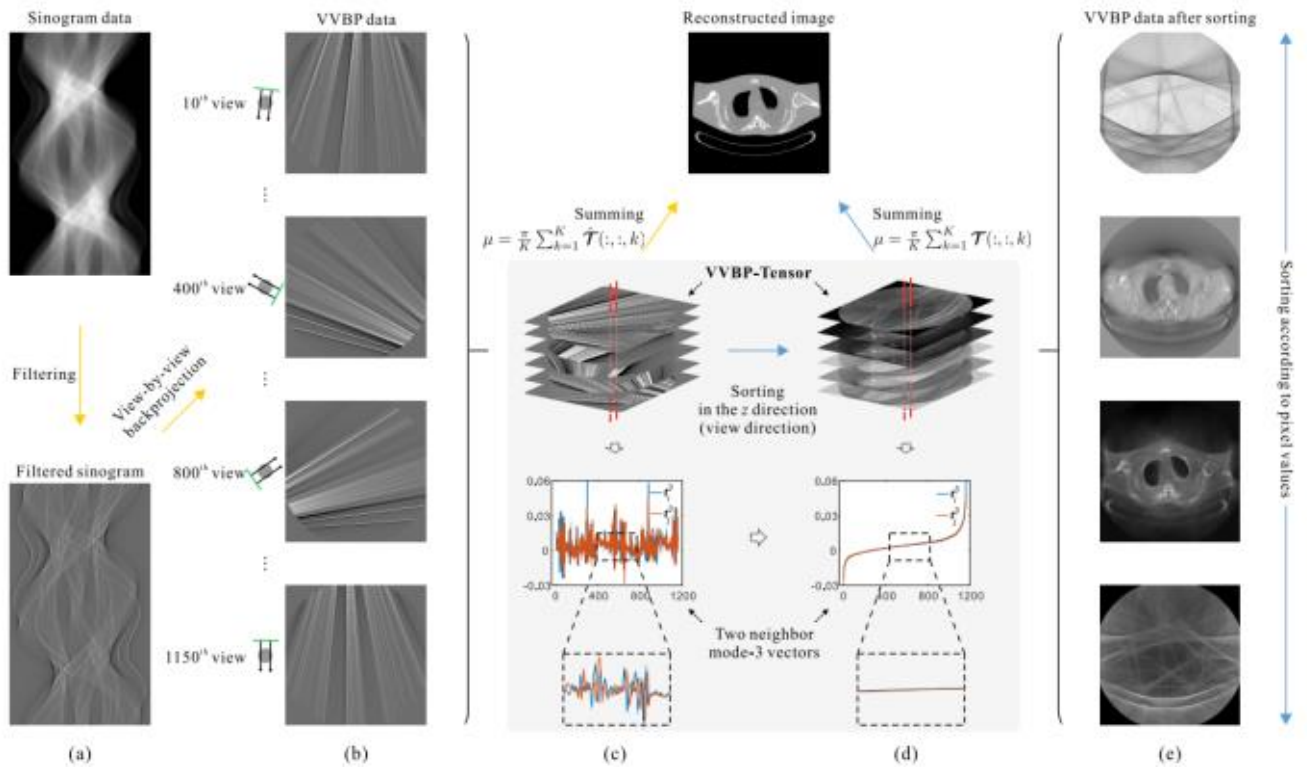

Figure 1. The illustration of the VVBP tensor. After the sorting, the tensor has a structure self-similarity near the median indices of the image stack. The image is referred to Tao et al. [14], "VVBP-Tensor in the FBP Algorithm: Its Properties and Application in Low-Dose CT Reconstruction,"

## *B. Folded-VVBP tensor*

Although the VVBP tensor has useful characteristics that are potentially helpful for artifact removal, the high dimensional data space causes a high computational burden. In addition, the useful signals as well as artifact-related signals are still located at the top and bottom of the tensor in the form of a streak or pattern. To deal with this problem, we proposed the ‘folding’ technique. The technique firstly performs sequential min-max summation to the tensor and re-scaling by considering the contributed number of pixels. Then, it re-sorts the signals again. By doing so, the signals at the view direction are compressed and have a reduced range of value which is near the mean of all signals (in Fig. 2). Therefore, it was able to make the VVBP tensor more efficient and have stronger structure self-similarity than before (in Fig. 3). Furthermore, we were still able to observe artifact-related signals at the top and bottom of the tensor while useful signals are structured due to their assembly with the counter-part signal.

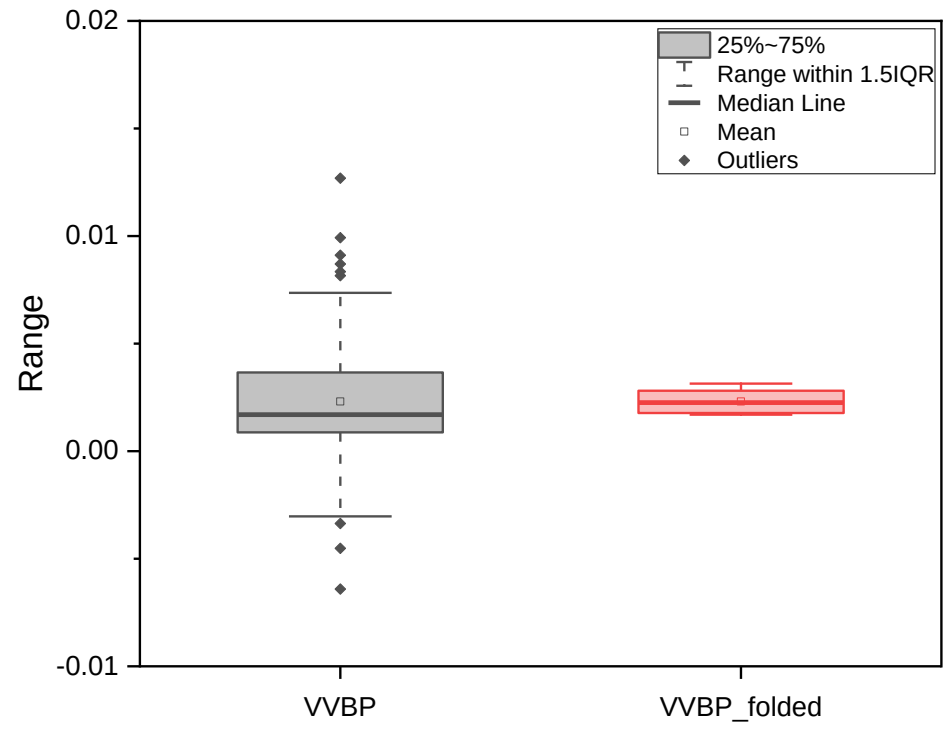

Figure 2. The box chart graph of a center pixel signal in the VVBP tensor. The left is the original VVBP and the right is the folded-VVBP. After the folding, the data range is shrinked to near the mean value of all signals

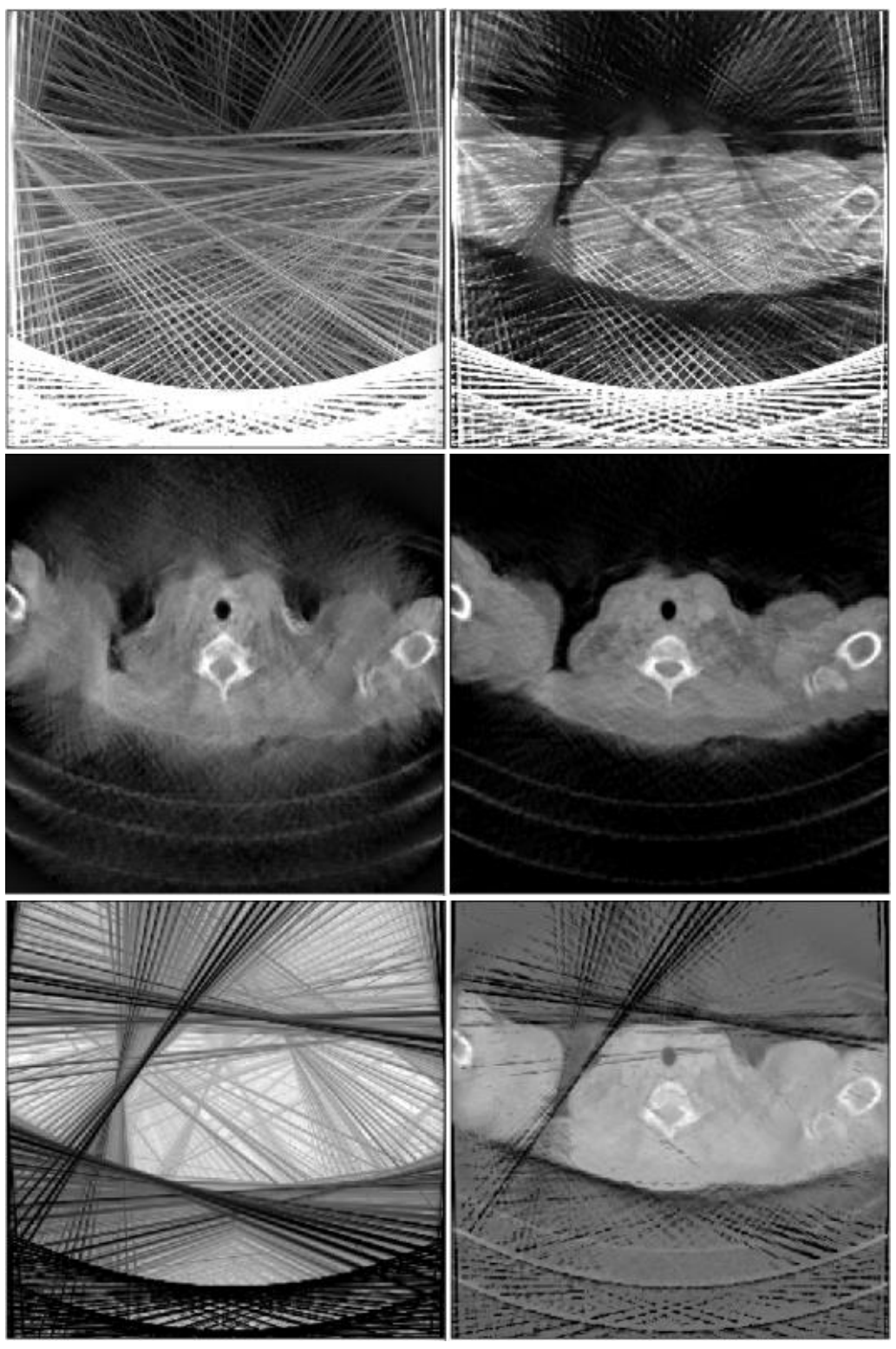

Figure 3. The image comparison between the original VVBP and the proposed folded-VVBP. The first low indicates the respective highest signal image of the tensor, the second indicates the median and the last indicates the lowest signal image

## *C. Network model*

Our proposed network is based on a residual encoder-decoder convolutional neural network (RED-CNN) [15] that is famous architecture for low-dose CT image denoising. We attached squeeze and excitation modules at each layer thereby the network may choose useful features in the VVBP tensor. The squeeze and excitation module consists of learnable weight function (ex) sigmoid) which multiplied in the channel direction of the tensor for efficient feature selection during the network training.

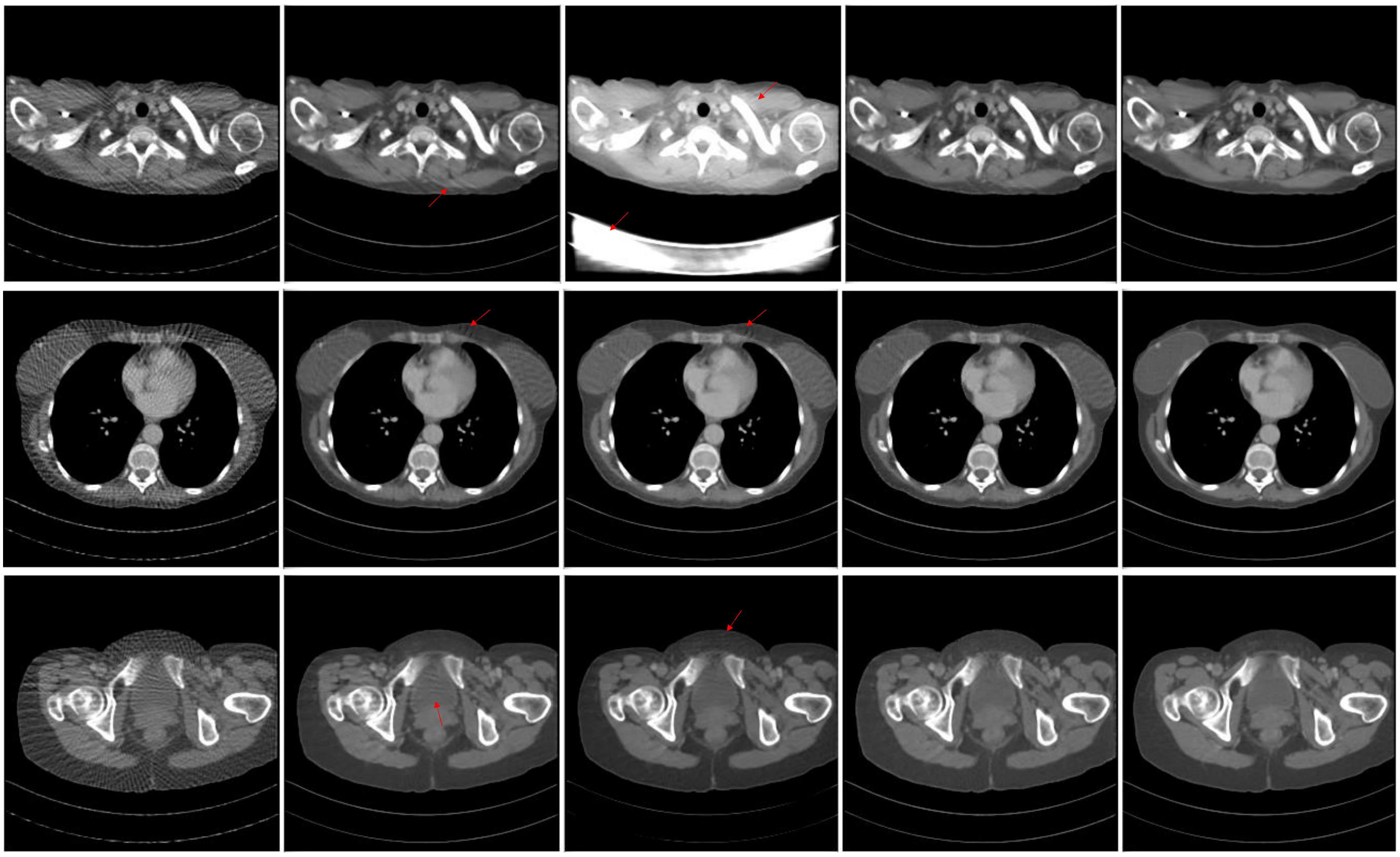

(a) (b) (c) (d) (e)

Figure 4. (a) Original FBP images, (b) output images of a single image-domain network, (c) sorted VVBP network, (d) folded-VVBP network (the proposed method), (e) Ground-truth images. The window level is [0.013 0.034] and residual artifacts are highlighted.

The Adam optimizer with step learning rate schedular was used ( lr: 0.002, step size: 10 epoch, gamma:0.86) for the network optimization. Due to the channel multiplication in encoder layers, we fixed the output channel size of the first encoder layer as 64 to consider the memory consumption.

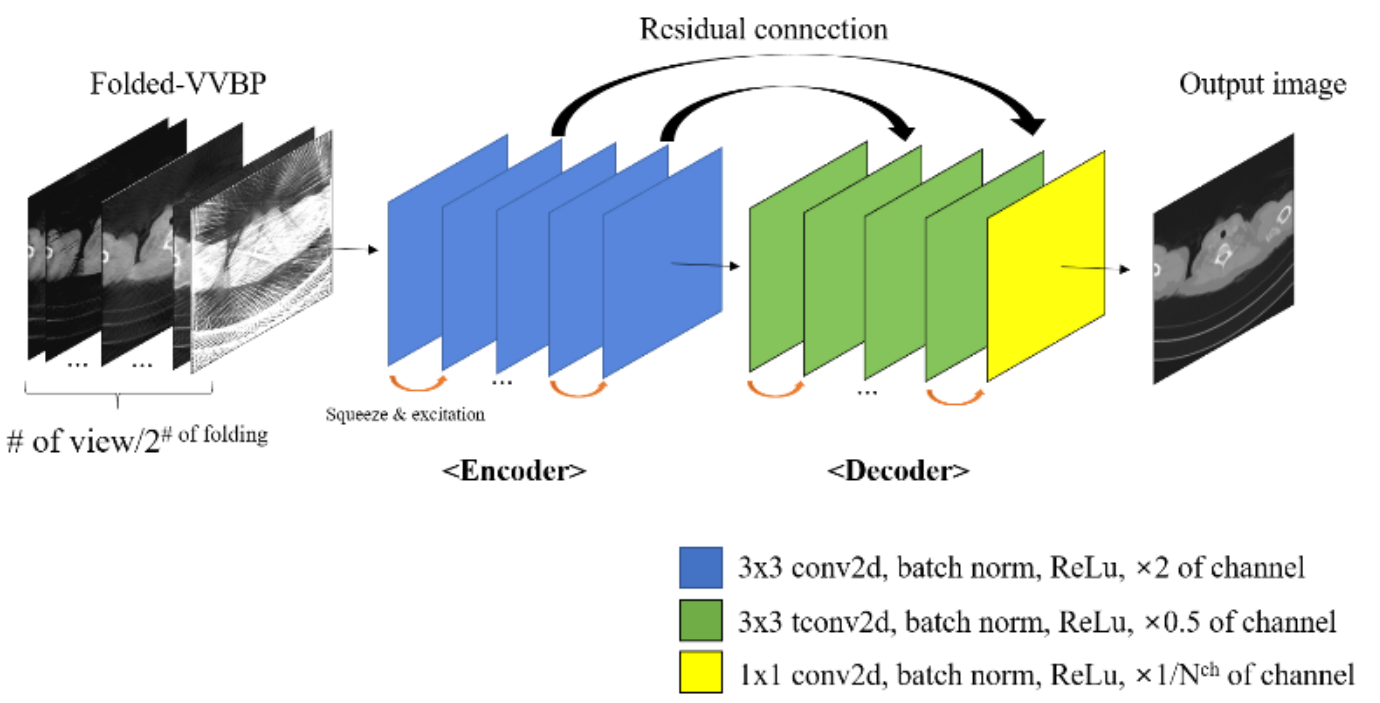

Figure 5. The proposed network architecture. A squeeze and excitation module was implemented at each layer. A total 5 encoder and decoder layers were used to consist the network and also 2 residual connections were utilized.

*D. Perceptual loss for network optimization*

In the training step, we used the VGG-16 perceptual loss in order to carry out high-contrast image recovery. This perceptual loss has been shown to significantly improve visual quality and structure details more than mean-squared error loss (MSE) because its multiple feature loss terms encourage the network carefully be aware of the spectral components of the image. The conjugated feature loss terms from four blocks of a pre-trained VGG-16 network were used in our study.

$$\text{Loss}_p = \frac{1}{R}\sum_{j=1}^{R}\sum_{i=1}^{4}\left\|\phi_i(F(\mathbf{x}_j)) - \phi_i(\mathbf{y}_j)\right\|_1$$

where $\phi_i$ is extracted feature maps from a block $i$, $F(\mathbf{x}_j) \in \mathbb{R}^{HW}$ and $\mathbf{y}_j \in \mathbb{R}^{HW}$ is the network output and target image of $j^{th}$ (1,2,...,R) training sample pair corresponding to image size $H \times W$.

*E. Materials*

The 8 clinical patient CT volume data from the Mayo Clinic (AAPM low-dose CT challenge dataset [13]) were used for the study. We used fan-beam scan geometry that has 835mm SDD and 480mm SOD. The sparse sinogram was generated as an input which consists of 128 views using a forward projection operator. Considering the memory size that was consumed during the training, we determined the voxel size as 256×256. For target data, 800 view sinogram data were used to generate an aliasing-free image. Among the total 1497 paired data, 901 pairs of data were used for the training set, 387 pairs of data were used for the validation, and 208 pairs of data were used for the test.

## 3 Results

In Fig. 4, the original FBP, network output, and ground-truth images are illustrated for comparison. As we can see, the residual streak artifacts are often observed in the single image-domain network (a). For the results of sorted VVBP net, it quite reduces artifacts well, however, sometimes it generates unreliable reconstructed images like the first-row image in (c) due to the poor generalization performance. The images of our method in (d) show the best image quality in terms of residual artifacts and recovered structure similarity. For the quantitative analysis, we evaluated root-mean-squared error (RMSE), and the structural similarity index (SSIM) value. The results of the quantitative analysis are presented in Table 1. In aspects of the computational cost, our proposed algorithm consumes 15.82GB VRAM during the network training while the original VVBP network consumes 17.24GB VRAM. Furthermore, our proposed method shows more efficient parameter utilization than the compared network (in Table 2). This gap will increase as the image size and initial number of views increase (current image size: 256×256, current number of views: 128).

TABLE 1
QUANTITATIVE ANALYSIS

| | FBP | Single image | Sorted VVBP | Folded VVBP |
|---|---|---|---|---|
| RMSE | 1.30e-4 | 4.17e-5 | 4.13e-5 | 3.58e-5 |
| SSIM | 0.36 | 0.71 | 0.69 | 0.72 |

TABLE 2
QUANTITATIVE ANALYSIS

| | Sorted VVBP | Folded VVBP |
|---|---|---|
| Consumed memory size | 17.24 GB | 15.82 GB |
| # of parameters | 12,755,761 | 12,792,625 |

## 4 Discussion

The main idea of our algorithm is to utilize folded-VVBP tensor rather than simply sorted-VVBP. By doing that, it was able to enhance the structure self-similarity of the tensor potentially beneficial to understand image context from the network point of view. Moreover, the decreased channel number of the tensor through the folding technique drastically decreases the required computational load for network utilization. Also, an important thing is that the artifact-related signals are still well presented at the top and bottom of the tensor which gives significant prior knowledge to the network. Based on the above advantages, the proposed network was able to successfully learn the CT image reconstruction without aliasing artifacts. Meanwhile, the compared sorted-VVBP network also shows quite good results than single image-domain network output, however, the poor generalization performance was observed in test sets since it can still generate good quality images for the training sets. This seems to be an overfitting issue which is often caused by weak correlations between the given data and the network model. In the future study, we will validate the proposed network and other compared networks in various data sets.

## 5 Conclusion

In this study, we proposed a folded-VVBP network to reduce the aliasing artifact in sparse-view computed tomography. The promises of the proposed network have been successfully shown. Experimental validation of the proposed method is under our research and will be presented in the near future.

## 6 Acknowledgement

This study is funded by the Korean National Research Foundation [NRF-2020R1A2C2011959]

# Neural Network Guided Sinogram-Domain Iterative Algorithm for Artifact Reduction

Gengsheng L. Zeng[1]

[1]Department of Computer Science, Utah Valley University, Orem, USA
[2]Department of Radiology and Imaging Sciences, University of Utah, Salt Lake City, USA

**Abstract** Artifact reduction or removal is a challenging task when the artifact creation physics are not well modeled mathematically. One of such situations is metal artifacts in X-ray CT when the metallic material is unknown and the X-ray spectrum is wide. In this paper, a neural network is used to act as the objective function for iterative artifact reduction when the artifact model is unknown. A hypothetical unpredictable projection data distortion model is used to illustrate the proposed approach. The model is unpredictable, because is controlled by a random variable. A convolutional neural network is trained to recognize the artifacts. The trained network is then used to compute the objective function for an iterative algorithm, which tries to reduce the artifacts in a computed tomography (CT) task. The objective function is evaluated in the image domain. The iterative algorithm for artifact reduction is in the projection domain. A gradient descent algorithm is used for the objective function optimization. The associated gradient is calculated with the chain rule. The images after the iterative treatment show the reduction of artifacts. The methodology of using a neural network as an objective function has potential value for situations where a human developed model is difficult to describe the underlying physics. Applications in medical imaging are expected to be benefit from this methodology.

## 1 Introduction

The classical image processing is frequency-component based. For example, the high frequency components beyond a certain frequency threshold are considered as noise. Then a low-pass filter is designed to remove the noise. However, the artifacts are not just random high frequency noise, but contain certain patterns. The artifacts are difficult to characterize by frequencies.

In recent years, neural networks, especially deep neural networks, are excellent tools in object classification. Therefore, neural networks are good candidates for the artifact detection task with two outcomes: artifact present and artifact absent. The main goal of this paper is to use the continuous output value of a neural network as the objective function. The final binary decision making is not used.

The algorithm to be developed in this paper is for artifact reduction. There is no restriction on the types of the artifacts. We focus on X-ray computed cosmography (CT) metal artifacts, with a twist, in this paper. The CT metal artifacts appear as bright and dark streaking, radiating from the metals.

Iterative algorithms are designed to optimize objective functions. One objective function that can indirectly measure the artifacts is the total variation (TV) norm of the reconstructed image. The TV norm minimization is able to reduce the unnecessary variation in the reconstruction, and some of the unnecessary variation is caused by the large sinogram errors [1, 2]. This method is not effective, because the TV norm tends to smooth out image details.

Another iterative reconstruction method is to treat the errors in the projections as 'noise' and to use small weights to the projections affected by metals. This method essentially tries to smooth out streaks, and the resultant images look blurry. A more successful method is referred to as sinogram inpainting. The inpainting method iteratively replaces the metal-affected sinogram by estimated metal-less projection values, as if the metals are absent. One simple way of inpainting is first to remove the metal-affected measurements and to assume that there is no metal in the object. Next, estimation methods such as interpolation, lowpass filtration, or some non-linear approaches are used to replace the original measurements by the estimated measurements. The drawback of the inpainting method is that the estimates are not accurate.

Another version of inpainting is more successful [3-5]. This version is different from the previous version in that the newer version estimate the projections with the metals in the object. It does not average the neighboring values. In [3], the objective function is the image-domain TV norm of the reconstructed image. In [4], the objective function is the energy of the negative pixel values in the reconstructed image. In [5], the objective function is a weighted combination of the objective functions in [3] and [4]. This current paper is in the category of the newer version, but its objective function is neural network generated.

In this paper, we borrow the term 'metal artifact' for a hypothetical situation where the distortion in the sinogram cannot be exactly modeled. Therefore, any model-based image reconstruction algorithm will not work for this hypothetical situation.

## 2 Materials and Methods

*A. A hypothetical unpredictable artifact model*

This paper presents an exemplary case in tomography, where some projection measurements are distorted in an unpredictable manner. Here, we distinguish distortion from noise. The magnitude of the distortion is much larger than the magnitude of noise. The noise is mainly in the high frequency range. In x-ray computed tomography (CT), the beam hardening and scattering effects make the projections through the metal deviate from the ideal line-integral model, and thus the effects are distortions. Due to the wide spectrum of the x-rays and the complicated metallic materials, the deviation (i.e., distortion) is not easy to exactly model mathematically. In this paper, some 'metals' are implanted in the computer simulated phantoms. The image values of the metal are much larger than the rest of the image, and the locations of the metals are assumed to be known. For the illustrative purposes, a hypothetical 'unpredictable' distortion is introduced by an exponential factor

$$factor = \exp(-0.05 \times projection \times rand), \quad (1)$$

where '*projection*' is the true line-integral projection value, and '*rand*' is a uniform random variable on [0, 1]. When the projection ray passes through the metals, the factor *f* defined in (1) is applied to the measured line-integral value. The random variable 'rand' in (1) makes the distorted measurement unpredictable. The value of 'rand' varies randomly from ray to ray. The exponential factor is not used if the projection ray does not pass through the metals.

The well-known filtered backprojection (FBP) algorithm is used to reconstruct the image using the distorted sinograms. When the image contains metals, the FBP reconstruction contains severe artifacts. In this paper, we refer to the artifacts caused by the hypothetical unpredictable distortion as 'metal' artifacts for convenience. We do not imply that the actual metal artifacts have a mathematical model expressed in (1).

*B. A neural network model to recognize the artifacts*

The modern neural networks have excellent capabilities to classify objects. We use a convolutional neural network to determine whether the FBP reconstruction contains 'metal artifacts.'

We do not use a neural network as a filter to remove the artifacts, because such a neural network requires a large number of image pairs to train: with and without metal artifacts. Those image pairs are difficult to obtain. It is easier to label the FBP reconstructions as 'with metal artifacts' (i.e., 1) and 'without metal artifacts' (i.e., 0). In other words, the task of 'recognizing the artifacts' is easier than the task of 'removing/reducing the artifacts.' Our strategy is to use a neural network to recognize the artifacts and then to use a sinogram-domain iterative algorithm to reduce the artifacts.

Without loss of generality, let us consider a 3-layer convolutional neural network as shown in Fig. 1. During training, the output *f* is quantified as '1' and '0.' The input layer of the neural network accepts an FBP reconstructed image, *X*. In our computer simulations in this paper, the images are two-dimensional (2D) and have the dimension of 64 × 64.

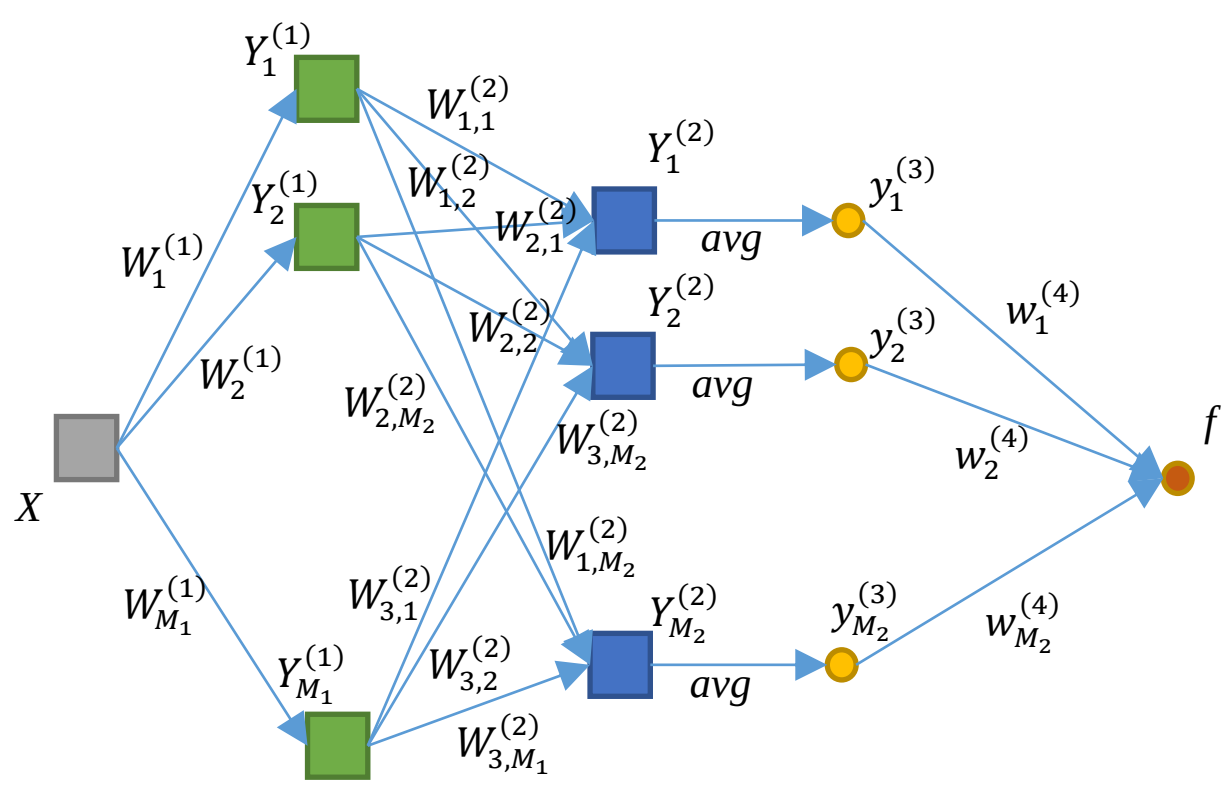

Figure 1. A neural network to recognize the 'metal' artifacts. Images as functions of (*x*, *y*): $X, Y_j^{(1)}, Y_i^{(2)}$. 2D kernels as functions of (*x*, *y*): $W_j^{(1)}, W_{j,i}^{(2)}$. Scalars: $y_i^{(3)}, f$. Scalar weights: $w_i^{(4)}$. The input is the FBP reconstructed image *X*. The output is the objective function value *f*.

This neural network assigns a numerical value, *f*, between (0, 1) to an image, *X*. If this value is closer to 1, the image *X* is likely to contain severer metal artifacts. If this value is closer to 0, the image *X* is likely to contain less severe metal artifacts. Therefore, the value *f* can be used as an objective function in an iterative algorithm that tries to reduce the artifacts.

*C. An iterative gradient descent algorithm in the sinogram domain*

An iterative gradient descent algorithm to reduce the metal artifacts in the sinogram domain is in the form of

$$P^{(k+1)} = P^{(k)} - \alpha \nabla f_P, \quad (2)$$

where $P^{(k)}$ is the estimated sinogram at the $k^{\text{th}}$ iteration of the gradient descent algorithm, and $\nabla f_P$ is the gradient of the objective function *f* in terms of the variables that are the projections in the sinogram. The function *f* is a composed function of the sinogram *P* through the following functions:

$$P \xrightarrow{yields} X \xrightarrow{yields} Y_i^{(1)} \xrightarrow{yields} Y_j^{(2)} \xrightarrow{yields} y_j^{(3)} \xrightarrow{yields} g \xrightarrow{yields} f$$

Step 1: The sinogram *P* is transformed into the image *X* by the FBP image reconstruction algorithm, which first applies the one-dimensional (1D) ramp filter in the view-by-view manner and performs the backprojection. This step can be expressed as

$$X = B^T DP, \quad (3)$$

where $B^T$ represents the backprojector in the form of a matrix and *D* is the 1D ramp filter in the form of a diagonal matrix.

Step 2: The image *X* propagates through the $1^{\text{st}}$ convolutional layer and results in $M_1$ images $Y_i^{(1)}$.

Step 3: The $M_1$ images $Y_i^{(1)}$ propagate through the $2^{\text{nd}}$ convolutional layer and results in $M_2$ images $Y_j^{(2)}$.

Step 4: The $M_2$ images $Y_j^{(2)}$ propagate through the averaging layer and results in $M_2$ scalars $y_j^{(3)}$.

Step 5: The $M_2$ scalars $y_j^{(3)}$ propagate through the perceptron layer and results in *one* scalar $f$. The output layer contains a sigmoid function.

The gradient $\nabla f_P$ can be evaluated by the chain rule going through the above five steps. Each step has a Jacobian matrix and is presented as follows.

Step 1: The partial derivative of *P* with respect to *X* is a Jacobian matrix given as

$$\frac{\partial P}{\partial X} = DB, \quad (4)$$

where *B* is the forward projection matrix.

Step 2: The partial derivative of *X* with respect $Y_i^{(1)}$ is a Jacobian matrix, whose *i*th element is given as

$$\hat{\hat{Y}}_i^{(1)}(x,y) = \frac{\partial X}{\partial Y_i^{(1)}} = W_i^{(1)} * \hat{Y}_i^{(1)}(x,y), \quad (5)$$
$$i = 1, 2, \ldots, M_1,$$

where * represents the 2D convolution, and $\hat{Y}_i^{(1)}$ is defined as

$$\hat{Y}_i^{(1)}(x,y) = \begin{cases} 1, & \text{if } Y_i^{(1)}(x,y) > 0 \\ 0, & \text{otherwise.} \end{cases} \quad (6)$$

Step 3: The partial derivative of $Y_i^{(1)}$ with respect $Y_j^{(2)}$ is a Jacobian matrix, whose ($i, j$)th element is given as

$$\hat{\hat{Y}}_{i,j}^{(2)}(x,y) = \frac{\partial Y_i^{(1)}}{\partial Y_j^{(2)}} = W_{i,j}^{(2)} * \hat{Y}_j^{(2)}(x,y), \quad (7)$$

$$i = 1, 2, \dots, M_1; \;\; j = 1, 2, \dots, M_2,$$

where $\hat{Y}_j^{(2)}$ is defined as

$$\hat{Y}_j^{(2)}(x,y) = \begin{cases} 1, & \text{if } Y_j^{(2)}(x,y) > 0 \\ 0, & \text{otherwise.} \end{cases} \quad (8)$$

Step 4: The partial derivative of $Y_j^{(2)}$ with respective to $y_j^{(3)}$ is a positive constant, which can be assumed to be 1.

Step 5: The derivative of $y_j^{(3)}$ with respective to $g$ (before the application of the sigmoid function) is

$$\frac{dy_j^{(3)}}{dg} = w_j^{(4)}. \quad (9)$$

The final output is the sigmoid function (8) with the input $g$. Its derivative is a scalar given as follows

$$S'(g) = \left(\frac{1}{1+\exp(-g)}\right)' = f(1-f). \quad (10)$$

The overall gradient for each image pixel ($x, y$), used in the gradient descent algorithm, is the combination of (4)-(10) as

$$\nabla f_X = f(1-f)\begin{bmatrix} w_1^{(4)} & \dots & w_{M_2}^{(4)} \end{bmatrix} \begin{bmatrix} \hat{\hat{Y}}_{1,1}^{(2)} & \cdots & \hat{\hat{Y}}_{1,M_1}^{(2)} \\ \vdots & \ddots & \vdots \\ \hat{\hat{Y}}_{M_2,1}^{(2)} & \cdots & \hat{\hat{Y}}_{M_2,M_1}^{(2)} \end{bmatrix} \begin{bmatrix} \hat{\hat{Y}}_1^{(1)} \\ \vdots \\ \hat{\hat{Y}}_{M_1}^{(1)} \end{bmatrix}. \quad (11)$$

Here $\nabla f_X(x,y)$ is a gradient image in the image domain as a function of ($x, y$). This gradient image can be mapped into the sinogram domain, obtaining

$$\nabla f_P(t,\theta) = DB\nabla f_X. \quad (12)$$

In other words, the gradient sinogram is obtained by forward projecting the gradient image with the operator $B$ and by applying the ramp filter $D$.

*D. Network training*

The proposed neural network (shown in Fig. 1) is, in fact, a binary classifier if appended with quantizer. The output of the classifier has two values: 0 and 1, where 1 indicates artifact present and 0 indicates artifact absent. These two values are generated by quantizing the continuous output value $f$ in the interval (0, 1) with a threshold value of 0.5.

In order to train this network, we generated 1000 random 64 × 64 images and their associated sinograms. These 1000 sinograms contained random unpredictable distortions, and the FBP reconstructed images from them were labelled as 1. We generated another set of 1000 random 64 × 64 images and their associated sinograms. These 1000 sinograms did not contain any distortions, and the FBP reconstructed images from them were labelled as 0.

The first convolutional layer had $M_1$ = 40 channels, the convolution kernel size was 3 × 3, the strides were set as (1, 1), and the activation function was the ReLU function. This layer had 1 input image and 40 output images. There were $M_1$ = 40 convolution kernels, which resulted in 40 × 3 × 3 = 360 parameters to be trained at this first convolutional layer. The biases were set to 0.

The second convolutional layer had $M_2$ = 20 channels, the convolution kernel size was 3 × 3, the strides were set as (1, 1), and the activation function was the ReLU function. This layer had 40 input images and 20 output images. There were $M_1 \times M_2$ = 40 × 20 = 800 convolution kernels, which resulted in 800 × 3 × 3 = 7200 parameters to be trained at this second convolutional layer. These two convolutional layers were implemented by 'Conv2D.'

The averaging layer was implemented by 'GlobalAveragePooling2D.' There were no parameters to train at this layer. The $M_2$ = 20 scalar outputs of this layer was then formatted as a 1D array by 'Flatten.'

The output layer was a dense layer with the 'sigmoid' activation function (8). There were $M_2$ = 20 scalar weights to be trained at this final layer. The bias parameter was not used (that is, set to 0).

The network was trained by the 'adam' optimizer, the loss function as the 'binary_crossentropy,' with the metrics set as 'accuracy.' We used epochs = 200, batch_size = 64, validation_split = 0.1, shuffle = True.

After the network was trained, the trained parameters were saved as a file on the computer.

*E. Artifact reduction computer simulations*

This part generates a new random phantom, creates its associated sinogram, incorporates random distortion (1) to the sinogram, and finds the FBP reconstruction $X$. This image $X$ contains artifacts and is the input of our proposed iterative gradient descent algorithm. This iterative algorithm uses a pre-trained neural network to calculate the objective function $f$ and to minimize this objective function by adjusting the 'metal' affected projections.

The computer simulations of the iterative gradient descent algorithm in this paper used 1000 iterations, and the step size $\alpha$ was set to 0.001. The number of output channels in the first convolutional layer was $M_1$ = 40, and the number of output channels in the first convolutional layer was $M_2$ = 20.

*F. Evaluation*

Visual inspection such as human observer studies is the most effective way to determine whether the artifacts are reduced. A quantitative method can also be used. A quantitative metric adopted in this paper is the Sum Square Difference (SSD), defined as

$$SSD = \frac{\sum_{x,y}[X_{true}(x,y) - X(x,y)]^2}{\sqrt{\sum_{x,y}[X_{true}(x,y)]^2 \sum_{i,j}[X(x,y)]^2}}, \quad (13)$$

where $X_{ture}$ is the true image, which is the originally generated noiseless and undistorted phantom, and $X$ is an FBP reconstructed image. The SSD essentially is the normalized distance between two images $X_{true}$ and $X$. The image $X$ can be the raw FBP reconstruction from the distorted sinogram and can also be the final FBP reconstruction from the processed sinogram by the proposed iterative algorithm.

## 3 Results

Computer simulation results of four random phantoms are shown in this section. The three true random phantoms are shown in Figs. 2(left), 3(left) and 4(left), respectively. The raw FBP reconstructions corresponding to the four true random phantoms are shown in Figs. 2(mid), 3(mid) and 4(mid), respectively. Severe artifacts are observed in these raw reconstructions. After the application of the proposed iterative algorithm on these raw FBP reconstructions, the corresponding final images are shown in Figs. 2(right), 3(right) and 4(right), respectively.

A 'learning curve' is defined as the objective function value at each iteration. The learning curves (not shown) indicate that at early iterations, the objective function values, $f$ , are close to 1, which implies that the neural network classifies the images as artifact presented. At later iterations, the objective function values, $f$, are close to 0, which implies that the neural network classifies the images as artifact free, even though there are still artifacts remained and visible.

All images displayed in this section are in the display window of [0, 2] using a linear gray scale. It is visually observed from the images that the severity of the artifacts has been reduced by the proposed iterative algorithm. The quantitative SSD values are provided in Table 1, showing that the final images are closer to the true images than the raw FBP images to the true images.

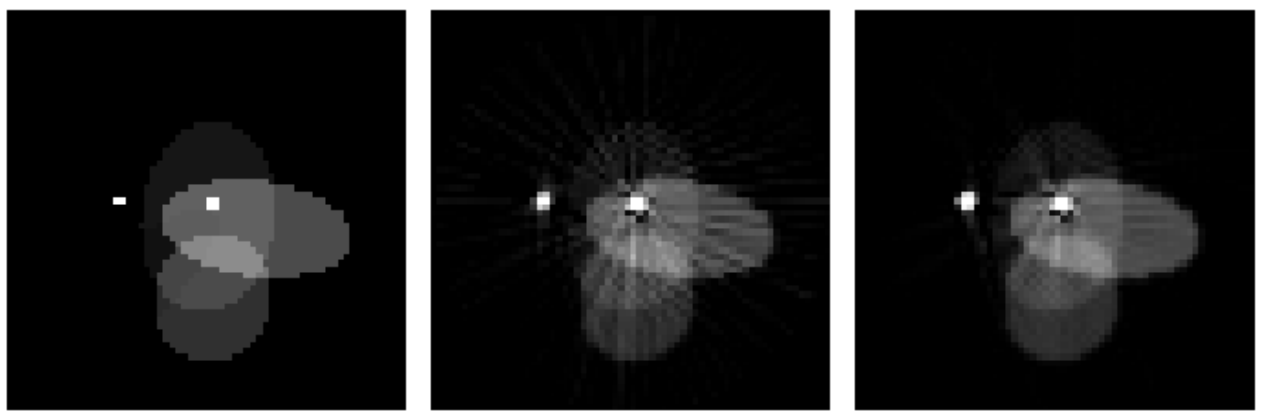

Figure 2. Study 1. Left: True image. Middle: FBP reconstruction with distorted sinogram. Right: FBP reconstruction with proposed method.

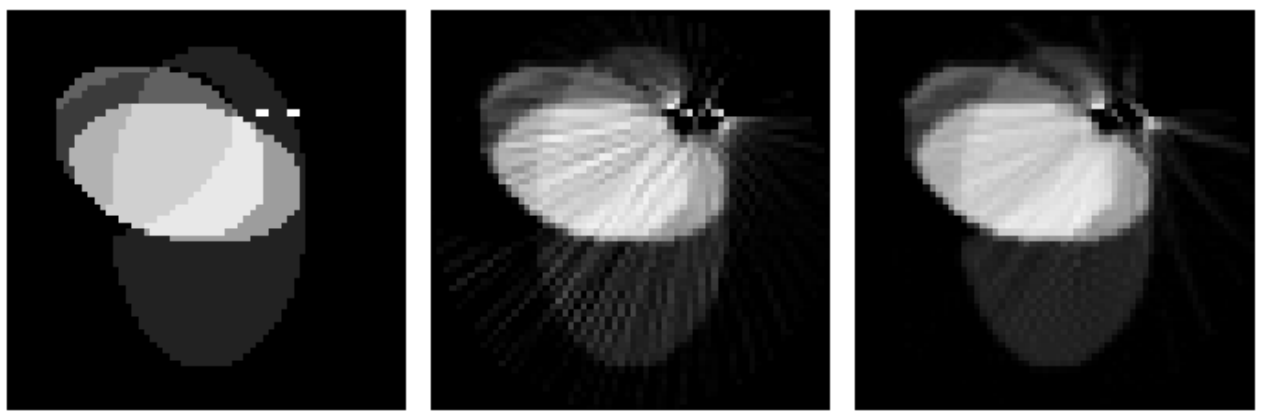

Figure 3. Study 2. Left: True image. Middle: FBP reconstruction with distorted sinogram. Right: FBP reconstruction with proposed method.

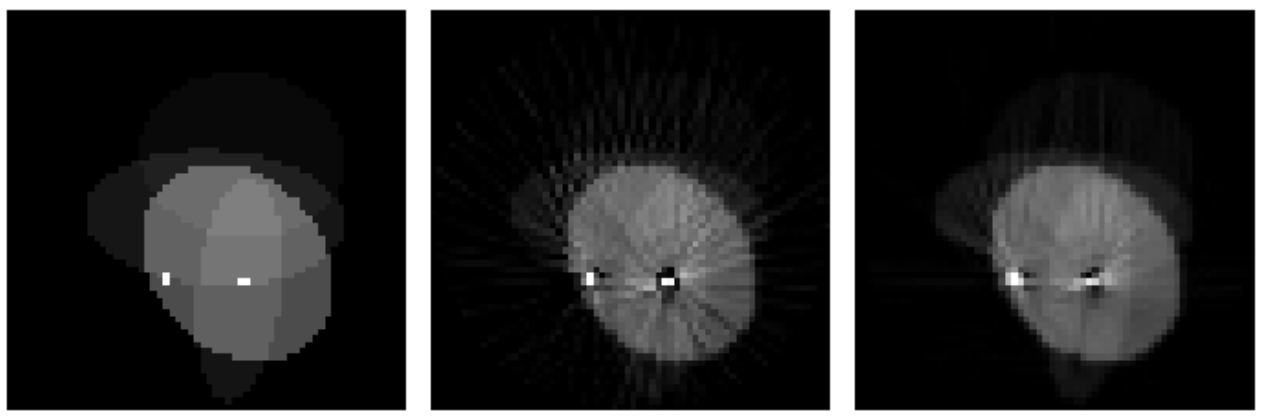

Figure 4. Study 3. Left: True image. Middle: FBP reconstruction with distorted sinogram. Right: FBP reconstruction with proposed method.

Table 1. The Sum Square Difference (SSD) between the FBP reconstruction and the true image

| Study | SSD of the raw FBP | SSD of the final FBP |
|---|---|---|
| #1 | 0.0729 | 0.0345 |
| #2 | 0.0587 | 0.0375 |
| #3 | 0.0488 | 0.0234 |

## 4 Conclusion

In many applications, the image artifacts are severe, and the measurement errors are difficult or almost impossible to derive a mathematical model for them. For this difficult situation, we suggest the use of a neural network to compute the objective function, which characterizes the severeness of the artifacts. In contract to common objective functions used in an iterative algorithm, our objective function is NOT human designed, but automatically learned by a neural network.

A large number of artifact-present and artifact-absent computer simulated images are fed to the neural network along with their binary labels. After training and removing the very last quantization layer, the sigmoid function output gives a continuous function, $f$, in the range of [0, 1]. This function, $f$, is the objective function of our proposed iterative algorithm. In each iteration, the current reconstruction is fed to the trained network, and then the trained network computes a continuous numerical value in [0, 1], which is the objective function value associated with the current reconstruction image.

In general, this objective function is not convex due to the nonlinear activation function ReLU in the neural network. Therefore, it is likely that the proposed iterative algorithm only converges to a local minimum, instead of the global minimum. It is observed from the learning curves that the objective function decreases as the iteration number increases. By comparing the images before and after the application of the proposed iterative algorithm, the artifacts in the images after the application of the iterative algorithm are less severe, although not completely removed. It is our goal in the further research investigation to improve the performance of the proposed neural network objective function methodology so that the artifacts are completely removed. After the methodology achieves a satisfactory level, it will be deployed to real-world applications.

# Development of a Solvability Map

Gengsheng L. Zeng[1,2] and Ya Li[3]

[1]Department of Computer Science, Utah Valley University, Orem, USA
[2]Department of Radiology and Imaging Sciences, University of Utah, Salt Lake City, USA
[3]Department of Mathematics, Utah Valley University, Orem, USA

**Abstract** From time to time, it is necessary to determine whether there are sufficient measurements for the image reconstruction task especially when a non-standard scanning geometry is used. When the imaging system can be approximately modeled as a system of linear equations, the condition number of the system matrix indicates whether the entire system can be stably solved as a whole. When the system as a whole cannot be stably solved, the Moore-Penrose pseudo inverse matrix can be evaluated through the singular value decomposition (SVD) and then a generalized solution can be obtained. However, these methods are not practical because they require the computer memory to store the whole system matrix, which is often too large to store. Also, we do not know if the generalized solution is good enough for the application in mind. **This paper proposes a practical image solvability map, which can be obtained for any practical image reconstruction algorithm in medical imaging.** This image solvability map measures the reconstruction errors for each location using a large number of computer-simulated random phantoms. In other words, the map is generated by a Monte Carlo approach.

## 1 Introduction

Data sufficiency conditions for continuous measurements were developed for many imaging geometries. For example, Orlov's condition uses the great-circle criterion to determine whether a positron emission tomography (PET) system measures a complete data set for analytical three-dimensional (3D) image reconstruction [1]. In Orlov's condition, the PET detector size is assumed to be infinity, and the sampling is assumed to be continuous. If the normal direction trajectory of the PET detector contains a great circle, the data set is sufficient. Orlov's condition considers the 3D parallel line integral measurements. Tuy's condition, on the other hand, considers the 3D cone-beam line integral measurements [2]. Tuy's condition is able to verify if a 3D cone-beam imaging system acquires a complete data set. Tuy's condition states that if every plane that cuts through the object intersects the cone-beam focal-point trajectory, the data set is sufficient for the reconstruction of the object. Once again, the detector is assumed to be infinite, and the sampling is continuous. A more general data sufficiency condition in the n-dimensional complex space is proposed by Kirokov [4].

For discrete sampling, the detector takes discrete finite number of positions, and the detector consists of discrete finite number of detection cells. The detector size is finite, which may lead to data truncation, where the detector does not cover the entire object. The common practice in processing discrete measurements is to use a linear model, which formulates the imaging process as a system of linear equations $AX = P$. The unknowns (i.e., the variables), $X$, of the system are the image pixels or voxels. The coefficient matrix (also known as the system matrix), $A$, is assumed to be known. The measurements, $P$, are the constant terms. The condition number analysis is a classic approach to investigate whether the normal equations $A^TAX = A^TP$ is stably solvable [5]. Singular value decomposition (SVD) analysis is able to diagnose invertibility and noise sensitivity of systems of linear equations. In [5], the condition number (i.e., the ratio of maximal and minimal singular values of matrix $A$) was calculated using the Lanczos iterative method [6] for image volumes of $65 \times 65 \times 128$. Some cone-beam imaging trajectories were analyzed and compared using the condition number analysis; a circular sine-wave trajectory was determined to be the most stable sampling scheme among the orbits investigated [5]. One drawback of the condition number analysis is that it does not work in region-of-interest (ROI) reconstruction with truncated data, because the system of equations is under-determined, and the associated condition number is essentially infinity. In this case the inverse matrix of $A^TA$ does not exist.

In situations where $A^TA$ is singular, the Moore-Penrose pseudo inverse matrix, $A^+$, can help [6]. If A is an $n \times m$ matrix, then $A^+$ is an $m \times n$ matrix. In general, $A^+A \neq I$, where $I$ is the $m \times m$ identity matrix. A method in [6] was proposed to identify the solvable subset of the unknowns. The method in [6] used the diagonal elements of $A^+A$ as a map. Each diagonal element of $A^+A$ corresponded to an image pixel. If a diagonal element is one, the corresponding image pixel can be reconstructed. A drawback of this method is that the Moore-Penrose pseudo inverse matrix $A^+$ is not easy to compute for a large imaging system, because the singular value decomposition (SVD) is required to perform on a large matrix [7], which requires a huge amount of computer memory.

This paper proposes a method to overcome the drawback in [6] so that the SVD computation is not required. This new method is Monte Carlo based and is described in Section 2 of this paper. The computer simulation results are presented in Section 3.

## 2 Materials and Methods

*A. Region-of-interest (ROI) image reconstruction*

One of the following situations can happen when an object is not completely measured. The first situation is due to the limited detector size, and only a portion of the object can be seen by the detector. The second situation is due to the lack of angular coverage. When the measurements are insufficient, it is likely that we are unable to have a stable reconstruction of the entire object. However, we may be able to have a stable reconstruction of a subset of the knowns. The aim of this paper is to determine such a subset if it exists.

*B. Proposed method*

Let us consider a generic image reconstruction algorithm, G; it can be an iterative or non-iterative algorithm; it can be a linear or nonlinear algorithm. For example, this generic image reconstruction algorithm, G, can be the iterative gradient descent (GD) algorithm, or a variate of the GD algorithm tailored for the data truncation, or a maximum-likelihood expectation-maximization (MLEM) algorithm, and so on.

We use computer simulation to create a large number of random objects, generate their projection measurements, add noise to the measurements, reconstruct the images, and compute the error between the reconstructed images and the true images. Finally, calculate the average error image for these large number of random objects. This average error image is our proposed image solvability map.

*C. Avoiding the inverse crime*

When a physical continuous system is modeled as a discrete system, modelling errors exist [8]. It is an inverse problem crime when these errors are ignored in developing and analyzing an inverse solution. For example, a typical inverse crime during computer simulations is to use the same generator to create the measurements and to be used in the reconstruction algorithm. To avoid inverse crime, in our case, if *G* is used to creating computer simulated measurements, *G* is not allowed to be used as the forward projection operator in the reconstruction algorithm.

In the computer simulations in this paper, the measurement generation uses random phantoms with the size of 384 × 384, and after projections are computed, the three adjacent projection bins are combined. Some Poisson noise is then incorporated in the combined measurements. In the image reconstruction, the image size is 128 × 128.

*D. Imaging geometry*

A hypothetical parallel-beam imaging system was simulated. The detector was asymmetric about the axis of rotation as shown in Fig. 1. The detector rotated 180° with 180 stops. In other words, the angular interval was 1°.

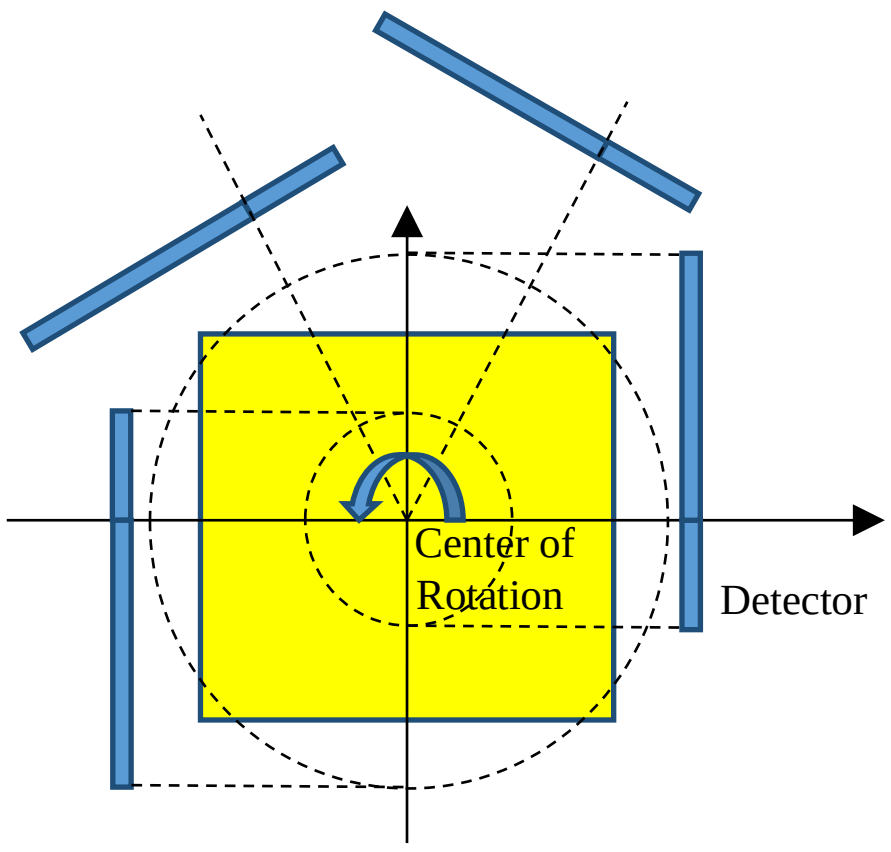

FIGURE 1. A hypothetical parallel-beam imaging system.

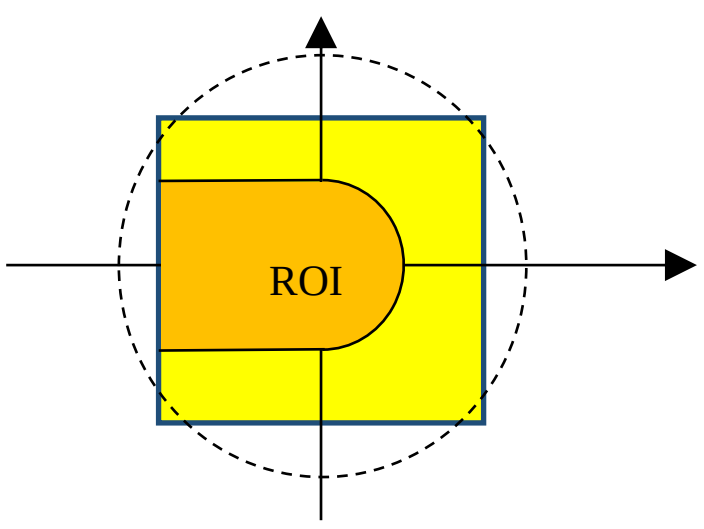

FIGURE 2. The points in the ROI have 180° angular measurements.

In 2D tomography, according to Kirokov's criterion [3], a point is fully measured if all lines passing through that point are measured. In Fig. 2, the ROI indicates the region of points that are fully measured. If an object is completely contained in the ROI, the object can be stably reconstructed, under the conditions that the projections are not truncated, and the number of views is sufficient.

*E. Image reconstruction algorithms*

In this paper, the iterative gradient descent (GD) algorithm is considered to test the feasibility of the proposed method [9]. The GD algorithm is expressed in (1).

$$GD: x_i^{(k+1)} = x_i^{(k)} - \alpha \sum_j a_{ij} \left( \sum_n a_{nj} x_n^{(k)} - p_j \right) \quad (1)$$

where $x_i^{(k)}$ is an element in image $X$ and is the $i$th image pixel value at the $k$th iteration; $a_{ij}$ is an element in the system matrix $A$ and is the contribution from the $i$th image pixel to the $j$th projection bin; $p_j$ is an element in projections $P$ and is the $j$th projection value; $k$ is the iteration number; $\alpha$ is the step size for the GD algorithm.

When the object is larger than the detector and the projections are truncated at both ends of the detector, the image reconstruction problem is referred to as the internal problem [10]. It is known that the internal problem is unsolvable [10]. A support of an object is an image, whose pixel value is non-zero (say, value one) if the corresponding object value is non-zero at the same location. If the support of the object is known, using the support information can improve the reconstruction in an internal problem [11]. An internal problem is illustrated in right part of Fig. 3.

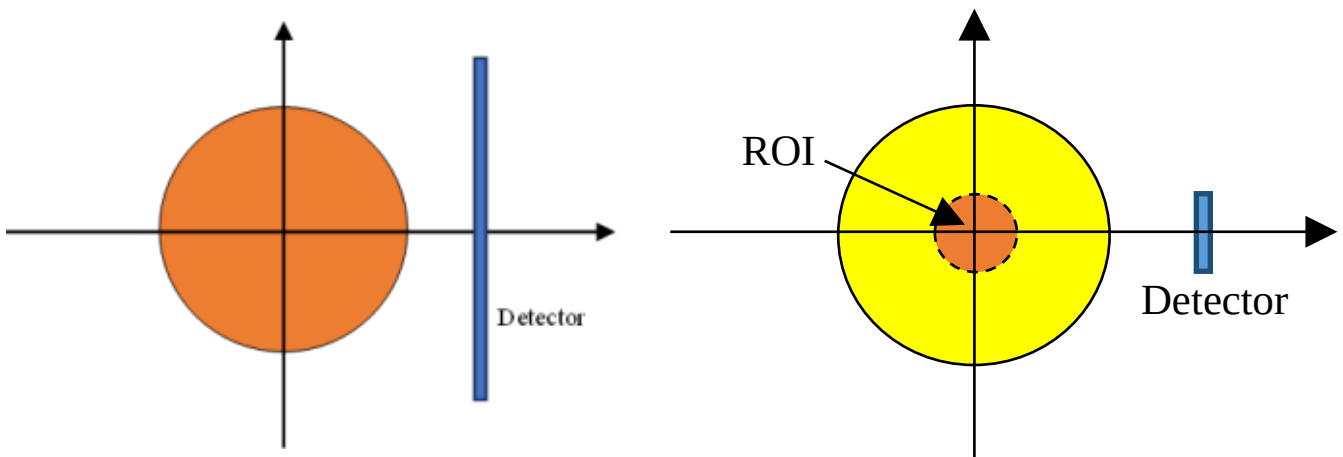

FIGURE 3. Left: Every point in the object is fully measured when the detector rotates 180°. Right: An internal problem is shown where the detector is too small to cover the entire object and truncation happens at both sides of the detector.

For the GD algorithm, we enforce the finite support at every iteration as

$$x_i^{(k+1)} = 0 \text{ if pixel } x_i \text{ is not in the support.} \quad (2)$$

It is recommended that whenever using truncated projections in an iterative algorithm, the image array be large enough to contain the entire object even though the detector is not large enough to see the entire object [11].

For the truncated data, a simple *modifie*d method can be used to reduce the artifacts [12]. This modified method assumes that if $p_j$ is not measured, $p_j$ is assigned to the forward projection value $\sum_n a_{nj} x_n^{(k)}$ at each iteration $k$, i.e.,

$$p_j = \sum_n a_{nj} x_n^{(k)}, \qquad \text{if } p_j \text{ is not measured.} \quad (3)$$

*F. Computer simulations*

In this paper, each of the computer-generated phantoms consisted of four ellipses with random sizes, locations, and intensities. The phantom size was 384 × 384. A narrow Gaussian lowpass filter with a standard deviation of one was applied to smooth out the sharp edges a little. Next, the image was normalized to the range of [0, 1].

Line integrals were calculated using the parallel-beam imaging geometry shown in Fig. 1, where the detector was asymmetric, and the number of views was 180 over 180°. After the line integrals were calculated, Poisson noise was incorporated into the simulated line-integrals. Then, the three adjacent detector bins were combined into one detector bin. In other words, the new detector's bin-size was three times larger than the original detector's bin-size. The binned-down measurements were ready for image reconstruction into an image array with the size of 128 × 128.

There were two sets of simulated measurements. The first set consisted of 1000 random phantoms and was described in the paragraphs above. The second set contained the same 1000 random phantoms as in the first set; the only thing different from the first set was that the detector was large enough to see the entire phantom as indicated in left diagram in left part of Fig. 3. The detector in the first set was asymmetric and had 107 detection bins. The detector in the second set was symmetric and had 185 detection bins. The detector bin size was the same as the image pixel size.

The following six algorithms were used to reconstruct the images and were compared:

- Iterative gradient descent (GD) algorithm (1);
- Iterative GD algorithm with the finite support enforcement (2);
- Iterative GD algorithm with the truncation modification enforcement (3);
- Iterative GD algorithm with the finite support (2) and truncation modification (3) enforcements.

*G. Image solvability map*

For each reconstructed image $X$, a squared-error image $E(X)$ is calculated as

$$e_i = (x_i - x_i^{true})^2, \quad (4)$$

where $e_i$ is the $i$th pixel in the squared-error image $E(X)$, $x_i^{true}$ is the $i$th pixel in the true image $X^{true}$, and $x_i$ is the $i$th pixel in the reconstructed image $X$.

If $n$ is the total number of random phantoms in the computer simulation (we had $n$ = 1000 in this paper), the *image solvability map* is the average image of the squared-error images, that is

$$\text{Image Solvability Map} = \frac{1}{n}\sum_{m=1}^{n} E(\text{the } m\text{th phantom's reconstruction}). \quad (5)$$

All image values in the image solvability map are non-negative. A smaller value in the map indicates that the corresponding pixel is more solvable. Due to the random noise and the determinist discrepancies introduced to fight the inverse problem crime, the minimum value in the image solvability map is not zero.

## 3 Results

Fig. 5 shows one representative of the 1000 random phantoms. The image reconstruction results from the first data set using truncated data are shown in Fig. 6 for the representative phantom shown in Fig. 5. The reconstruction algorithms are listed in the Part F of Section 2. The images in Fig. 6 are obtained from the gradient descent (GD) algorithms.

For the representative random phantom shown in Fig. 5, the squared-error images associated with the reconstructed images are shown in Fig. (7) for the GD algorithms. After finding the average of the 1000 squared-error images, an image solvability map is obtained. The image solvability maps for the four reconstruction algorithms are shown in Fig. 8.

In this paper, all phantom images (Fig. 5, (A) and (B) of Fig. 6) are displayed in the linear grayscale window of [0, 1]. All squared-error images and image solvability maps are displayed from zero to the maximum pixel value in the image. The image solvability maps are displayed with a non-linear transformation to emphasize the small values. The minimum and maximum values for the image solvability maps are listed in Table 1.

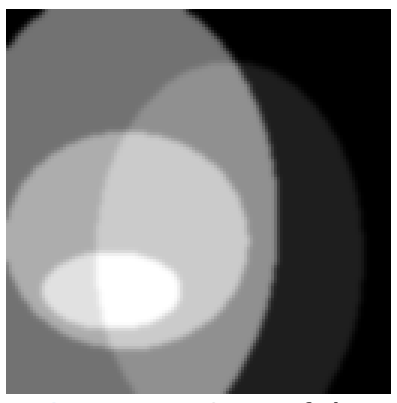

FIGURE 5. One of the 1000 random phantoms used in the computer simulations.

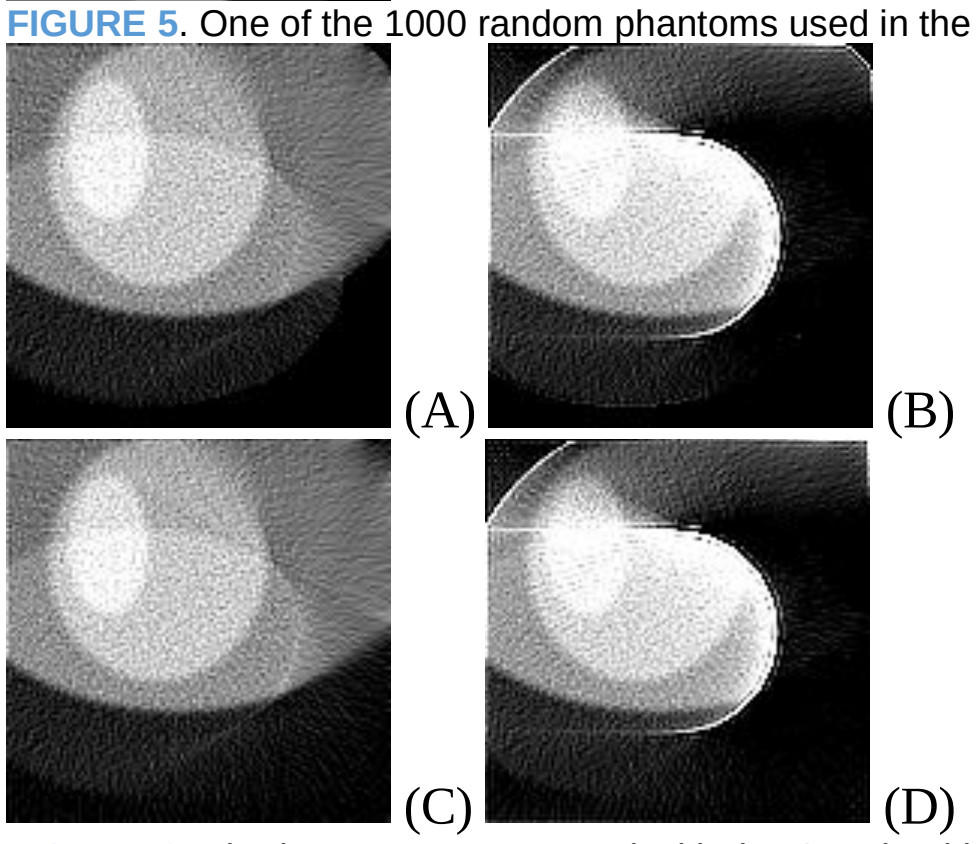

FIGURE 6. The images reconstructed with the GD algorithms using truncated data. (A) With formulas (1), (3), and (5); (B) With formulas (1) and (3); (C) With formulas (1) and (5); (D) With formula (1).

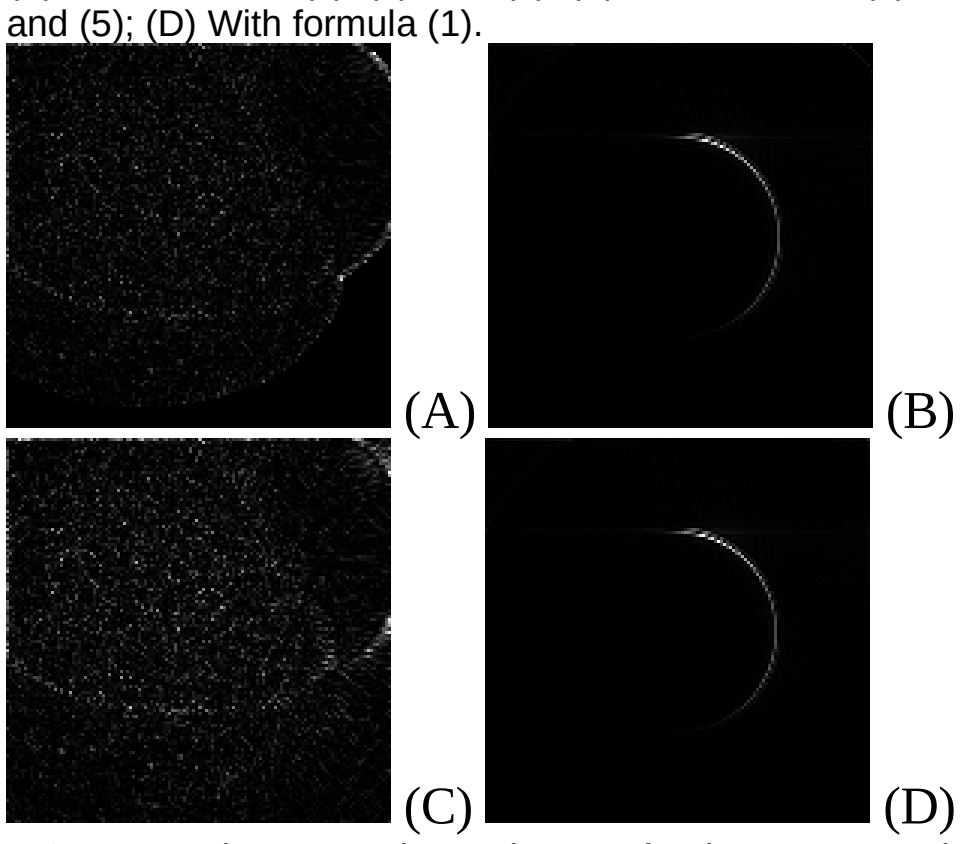

FIGURE 7. The squared-error images for the reconstructions with the GD algorithms using truncated data. (A) With formulas (1), (3), and (5); (B) With formulas (1) and (3); (C) With formulas (1) and (5); (D) With formula (1).

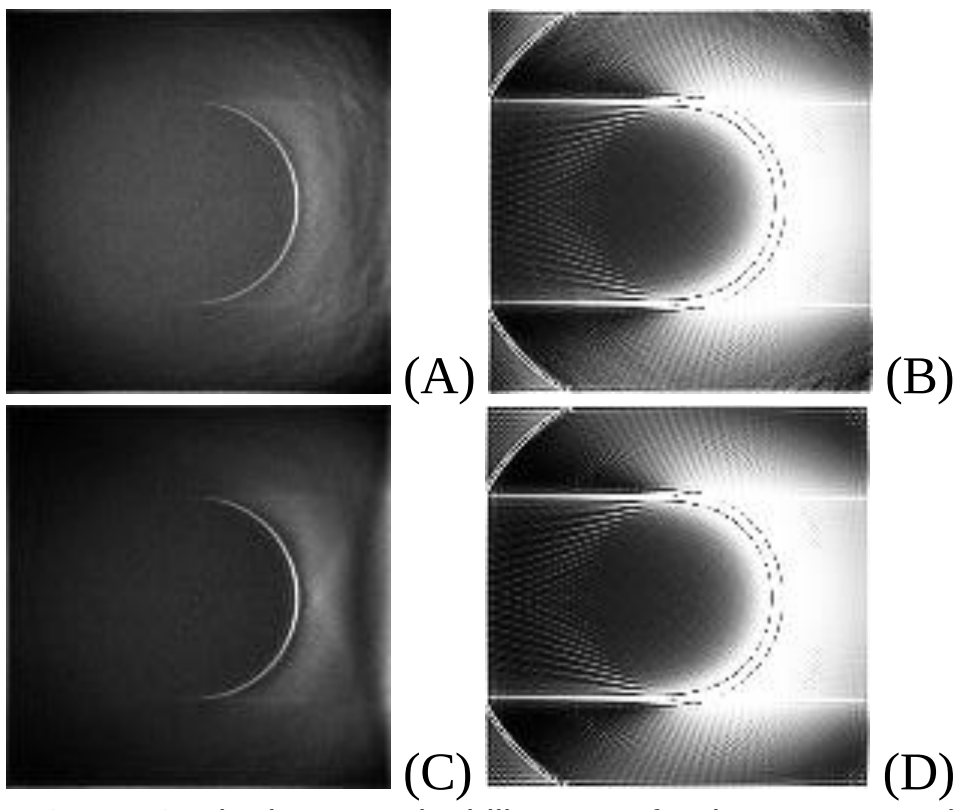

FIGURE 8. The image solvability maps for the reconstructions with the GD algorithms using truncated data. (A) With formulas (1), (3), and (5); (B) With formulas (1) and (3); (C) With formulas (1) and (5); (D) With formula (1).

TABLE 1. Maximum and minimum values in the image solvability map for the GD algorithms (see Fig. 10) using the truncated data

| Algorithm | Minimum value | Maximum value |
|---|---|---|
| GD (1) with support (3) and truncation modification (5) | $4.0035 \times 10^{-04}$ | 0.0307 |
| GD (1) with support (3) | 0.0022 | 20.7523 |
| GD (1) with truncation modification (5) | 0.0022 | 0.0537 |
| GD (1) | 0.0034 | 20.9302 |

## 4 Conclusion

It has been a desire to develop a tool that can identify which regions can be stably reconstructed if the projection measurements are not complete. It is clear that the condition number is disqualified, because the condition number only tells whether the entire system can be stably solved as a whole. Even one pixel (i.e., one unknown) is unsolvable, the condition number is extremely large or infinity. If the condition number is infinity and not all pixels can be solved, we ask a further question: "Are there any pixels that can be stably solved?"

The Moore-Penrose pseudo inverse matrix method is SVD based and is a powerful tool to use when some singular values of the system matrix are zero. However, the SVD method requires that the entire system matrix be stored in the computer memory during computation. In reality, the system matrices are too large to store. The SVD methods are not practical.

This paper proposed a practical tool that maps out the stably solvable regions in the image. The basic idea of the tool is to reconstruct a large number of random images and compute their errors with respect to their associated true images. In other words, this is a Monte Carlo based method, The errors are location dependent. The regions that have large errors are not solvable. This idea is similar to machine learning. Here we use a large number of phantoms to 'train' the image solvability map. The map is imaging-geometry dependent. If the imaging geometry is altered, we need to 're-train' a new map for the new geometry. We must point out that the image solvability map is also reconstruction algorithm dependent.

One important application of the proposed image solvability map is in C-arm cone-beam imaging trajectory design [13, 14]. Another important application is in region-of-interest (ROI) imaging system design [15, 16].

# Evaluation of CatSim's physics models for spatial resolution

Jiayong Zhang[1], Mingye Wu[1], Paul FitzGerald[1], Steve Araujo[1], and Bruno De Man[1]

[1]GE Research - Healthcare, Niskayuna, NY, USA

**Abstract** Simulation tools are crucial for efficient development of X-ray/CT imaging systems, but sophisticated and well-characterized simulators are proprietary to imaging equipment manufacturers. An open-source version of CatSim is now available as part of XCIST. We believe this is the first open-source X-ray/CT simulator that includes sophisticated modeling capability for all critical system components and physical processes. Once validated, this simulator will allow academic groups and small commercial companies to evaluate new CT technologies with minimal investment.

Detailed evaluation of CatSim is currently ongoing. Spatial resolution is one of the most important system performance characteristics to model correctly, and accurate simulation requires sophisticated models for focal spot and detector geometry and detector physics. We have now developed such models and have evaluated agreement between empirical measurements and analogous simulations in the projection and image domains. We constructed physical and analogous virtual phantoms for each domain.

We achieved good agreement in both domains, with average errors <5% and worst-case errors <10%. This represents a strong step toward full validation of CatSim's simulation performance.

## 1 Introduction

Simulation tools are crucial for efficient development of X-ray/CT imaging systems. CatSim is a proprietary simulator that has been in development for nearly two decades[1]; an open-source version is now available as the underlying simulator for the X-ray-based Cancer Imaging Simulation Toolkit (XCIST)[2].

Spatial resolution is one of the most important system performance characteristics to model correctly. In this work, we evaluate agreement between empirical results from a 64-slice scanner and simulated results when modeling the same scanner. We evaluate the spatial resolution obtained in the projection and image domains using previously developed approaches.[3] In each domain, we report the full width at half maximum (FWHM) of the point spread function (PSF), designated PSF50_proj and PSF50_image. We also report the image-domain frequency at which the modulation transfer function (MTF) reaches 50% and 10% modulation, designated MTF50 and MTF10. Importantly, we focus on agreement between empirical and simulated results, not on absolute performance of the modeled system, because that is affected by multiple factors that are controlled but not optimized in this work.

## 2 Materials and Methods

All empirical data were acquired using a Lightspeed™ VCT scanner (GE Healthcare, Chicago, IL). Data in empirical projections were converted to post-log p-values, but all standard corrections for non-ideal effects (e.g., afterglow, low signal) were disabled. All simulated data were computed using CatSim within the XCIST package (https://github.com/xcist).

We developed new models for the X-ray source focal spots using photos of the scanner's focal spot made with a pinhole camera, and we developed a model for the detector's afterglow behavior using a preiously reported method[4].

We constructed two physical phantoms, one for the projection-domain (PD) and one for the image-domain (ID) experiments (Figure 1).

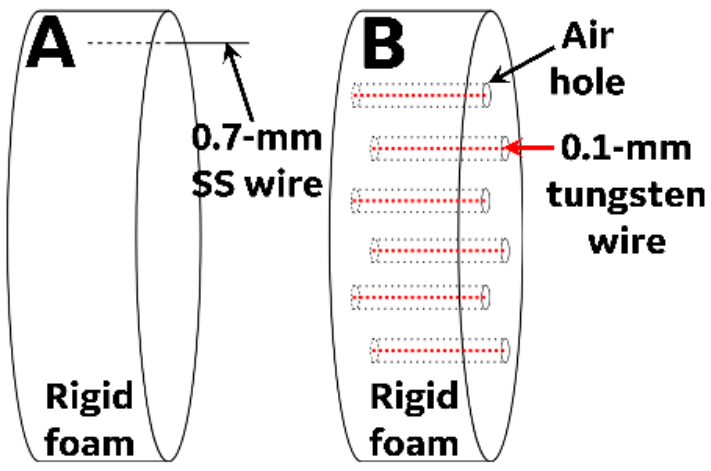

*Figure 1. Phantoms. (A) Projection-domain phantom. (B) Image-domain phantom.*

**2.1 Projection domain.** For the PD phantom we used a 0.7-mm diameter stainless steel (SS) wire, placed approximately 218 mm from the axis of rotation (AR) and approximately parallel to the AR. We then scanned the wire cantilevered in air from a 45-cm-diameter cylindrical foam phantom, using 120 kV, the small focal spot, and 300 mA. We reconstructed images only to measure the distance of the wire from the AR; we then created a virtual phantom for simulations using that wire position. We performed simulations using the same kV, focal spot, and mA as the empirical scans. We analyzed the empirical and simulated results as follows. For each of the 984 projections, we used the data from detector row 32 to measure the width of the pulse at half the peak height to determine PSF50_proj.

Note that, when scanned (Figure 2), the wire is sometimes close to the source (position S), close to the detector (position D), or tangential to the trajectory of the wire relative to the source and detector (position T). At each projection, we normalized the measured PSF50_proj by the magnification factor of the wire at that position, and converted from detector columns to mm. We next plotted the PSF50_proj values for all projections and applied a third-order spline fit to the data (Figure 3). Finally, for each projection, we calculated the difference beteen the fitted curves for empirical and simulated data.

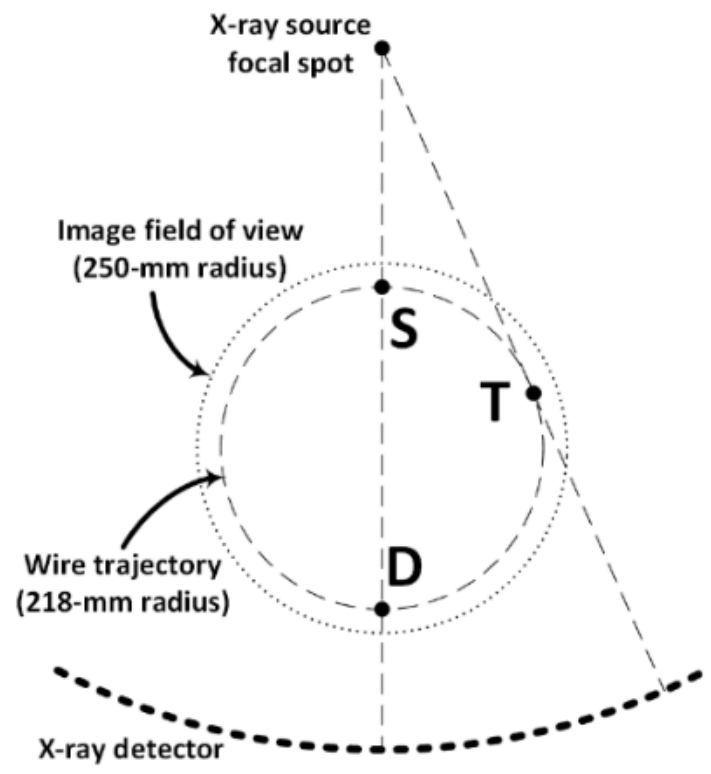

*Figure 2. Critical wire positions for the projection-domain experiments. The PSF of the wire was analyzed at 984 views angles during a complete 360° rotation, but the most interesting view angles are when the wire is closest to the source (S), the detector (D), and tangential to the wire trajectory (T).*

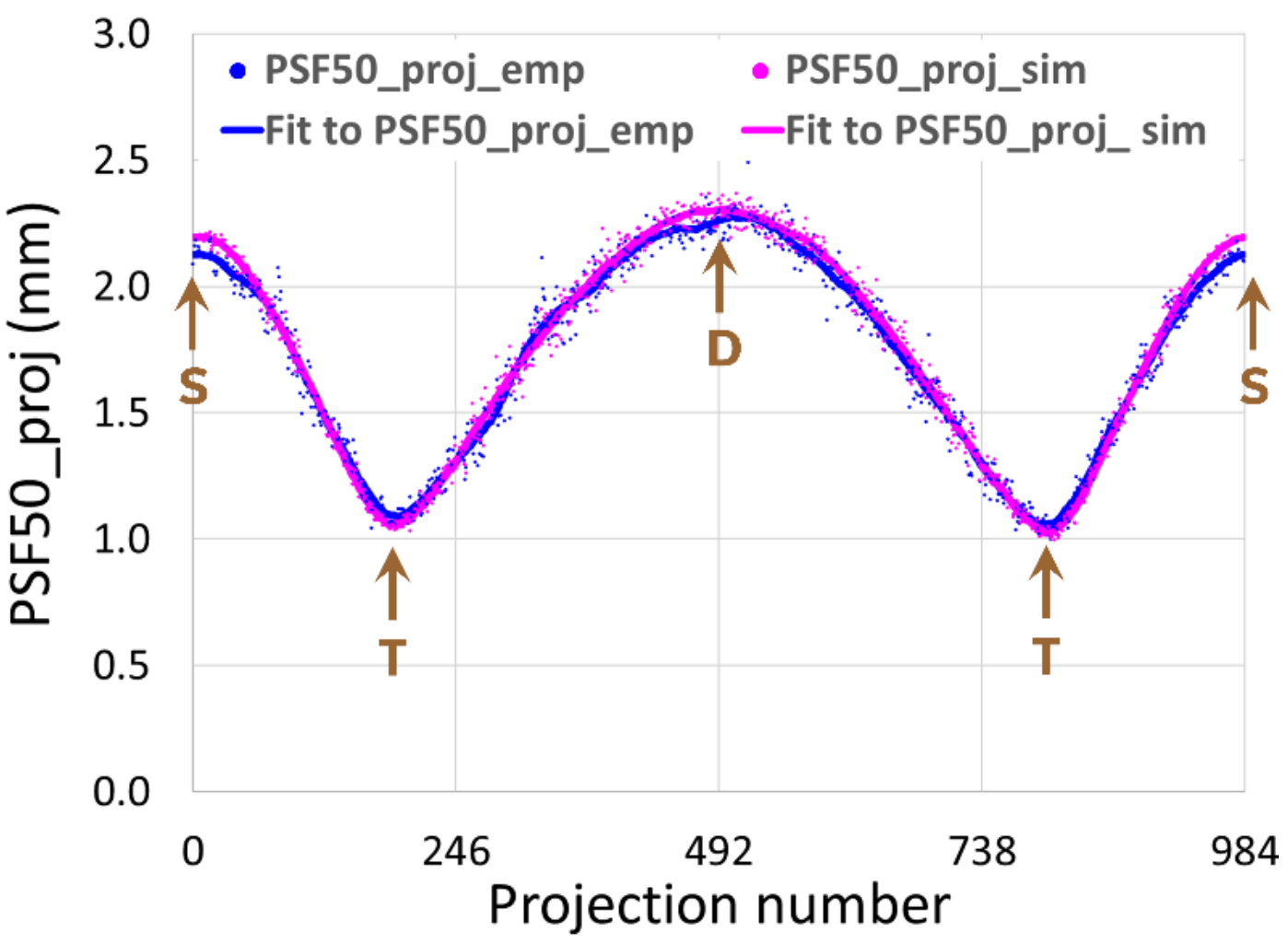

***Figure 3. PSF50_proj.*** *The projection-domain PSF is shown for 984 projections during a complete 360° rotation, for both empirical and simulated data. The approximate positions S, D, and T are indicated.*

**2.2 Image domain.** For the ID phantom, we drilled nineteen ~10-mm holes in the foam ~0 to ~225 mm from the phantom center, in ~12.5-mm increments. We fastened the ends of 0.1-mm tungsten wires at the faces of the foam, attempting to keep the wires taut and centered in the hole. We scanned the phantom (Figure 4), measured the wire positions in the images, and created a virtual phantom using the measured wire positions. When analyzing the empirical scans, we found that five wires were misaligned in the holes

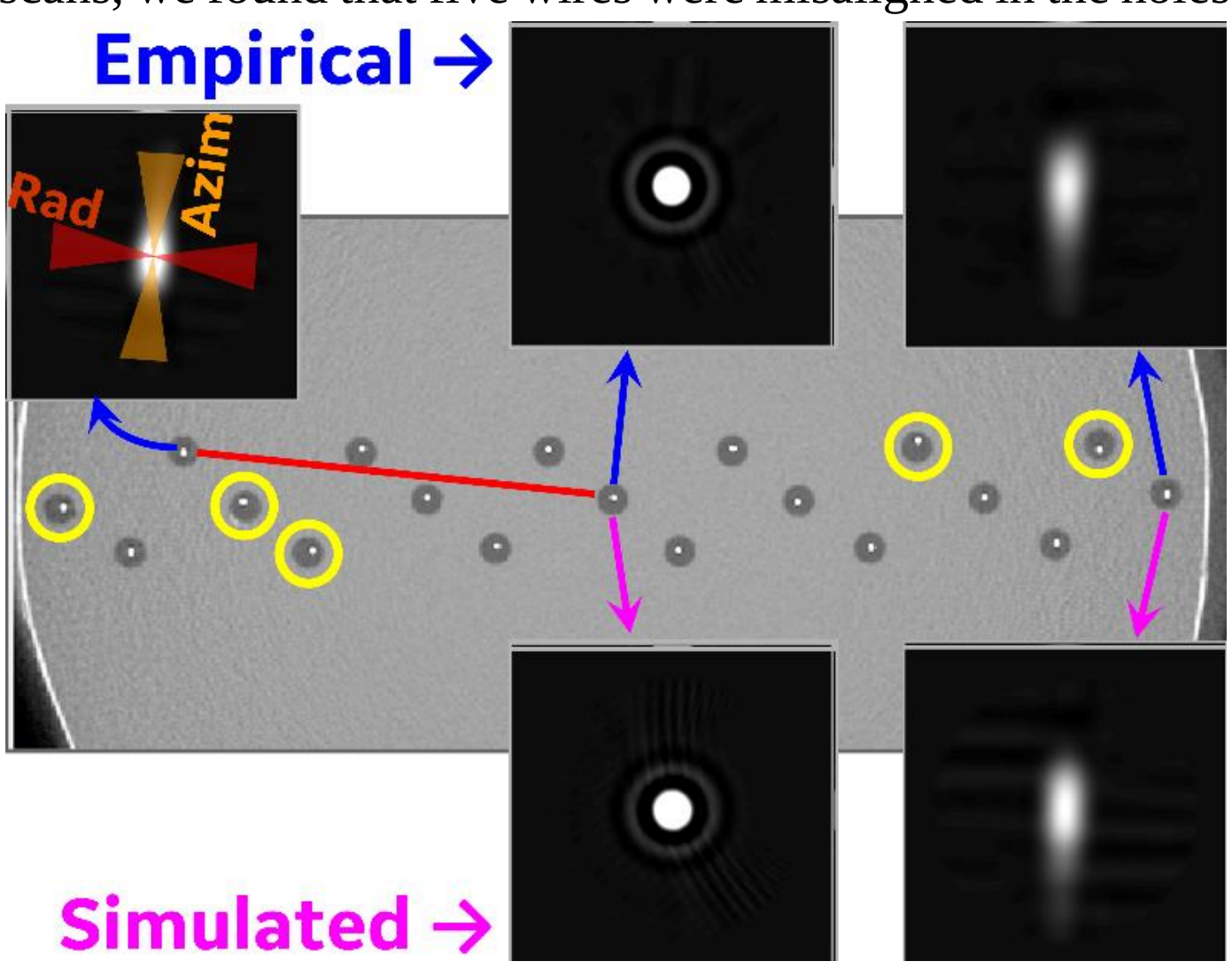

***Figure 4. Image-domain analysis.*** *One slice of a reconstructed image is shown (background) from the empirical scan of the 19-wire phantom. Five wires (yellow circles) were excluded due to alignment issues. Five examples of high-resolution region-of-interest reconstructions are shown (inset images); these are displayed at W/L = 900/-600 HU. The red line denotes the radius from the AR to a selected wire; the red and orange regions in the corresponding inset image designate the angular ranges from which the radial and azimuthal PSF curves were determined.*

and the data were unusable; these wires were excluded for both empirical and simulated experiments.

Using XCIST's Feldkamp-based reconstruction tool with the Bone kernel, we reconstructed a 10-mm ROI around each wire, then we corrected for wires potentially non-parallel to the AR by aligning the images at the maximum value, and averaged the shifted images. We applied a 2D weighting function to mask out the foam. We determined the line from the AR to each wire, and measured the PSF50 along that line for the radial response (PSF50_img_rad) and perpendicular to the line for the azimuthal response (PSF50_img_azim). We then applied a 2D FFT to the image and, for each of these orientations, we averaged pixel values over a ±15° range and measured the modulation transfer function (MTF) at 50% and 10% modulation (MTF50 and MTF10).

In total, seven parameters were reported: PSF50_proj, PSF50_img_rad, PSF50_img_azim, MTF50_rad, MTF50_azim, MTF10_rad, and MTF10_azim. As a metric for each parameter's "goodness of agreement" between empirical and simulated results, we averaged the percent errors across all samples for each metric.

## 3 Results

**3.1 Projection domain.** Selected PD results are shown in Figure 5. The general characteristics of the empirical PSFs were reproduced by the simulation, as follows. The PSFs at positions S and D showed asymmetry. The T position produced the narrowest PSF. The S position had a broad PSF and had the largest apparent AUC. The D position had the smallest apparent AUC.

PD results are summarized in Table 1. The mean error from all 984 projections was about 2%, and the maximum error was about 4%.

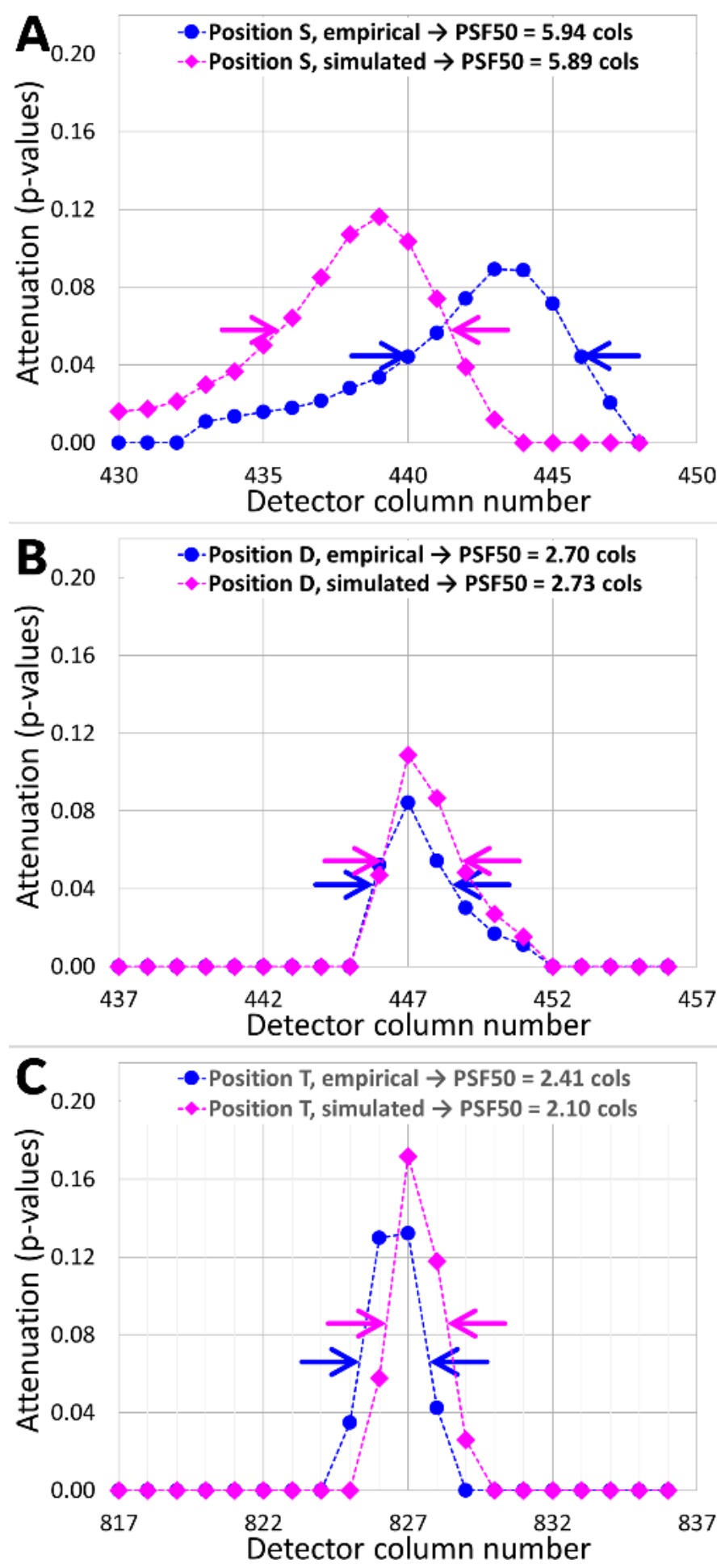

***Figure 5. PD PSFs at selected wire positions.*** *PSFs are shown at position S (A), position D (B) and position T (C). Vertical axes are equal and horizontal axis widths are equal.*

**Table 1. Projection-domain results**

| Wire | Parameter | Value | Units |
|---|---|---|---|
| S | PSF50_proj_emp | 2.08 | mm |
| | PSF50_proj_sim | 2.14 | mm |
| | Error | 3.12 | % |
| D | PSF50_proj_emp | 2.21 | mm |
| | PSF50_proj_sim | 2.25 | mm |
| | Error | 1.78 | % |
| T | PSF50_proj_emp | 1.08 | mm |
| | PSF50_proj_sim | 1.05 | mm |
| | Error | 3.02 | % |
| | Maximum error (a) | 4.27 | % |
| | Mean error (a, b) | 2.12 | % |

(a) Determined from errors at all 984 projections shown in Figure 3
(b) Mean error = RMSE/mean(all PSF50_proj_emp)

**3.2 Image domain.** The shape, asymmetry, and widths of ID PSFs (Figure 8) show excellent qualitative and quantitative agreement between empirical and simulation results. The azimuthal PSFs at the center are not shown because they were nearly identical to the radial (Figure 8A). The predominant characteristics of the positional dependence of the ID PSFs and MTFs are in good agreement (Figure 6). As the distance from center increases, the radial PSF increases slightly until a distance of about 100 mm from the AR (Figure 6A), and then remains at about 1 mm all the way to the edge (Figure 8B). The azimuthal PSF increases with the radial but continues increasing all the way to the edge (Figure 6A) to about 2.5 mm (Figure 8C), where the simulation disagreement becomes larger above 50% of maximum but PSF50 is still in close agreement. Notably, the trend as the wire position gets closer to center is not monotonic at the centermost point (Figure 6A); this characteristic in the empirical result is faithfully predicted by the simulation.

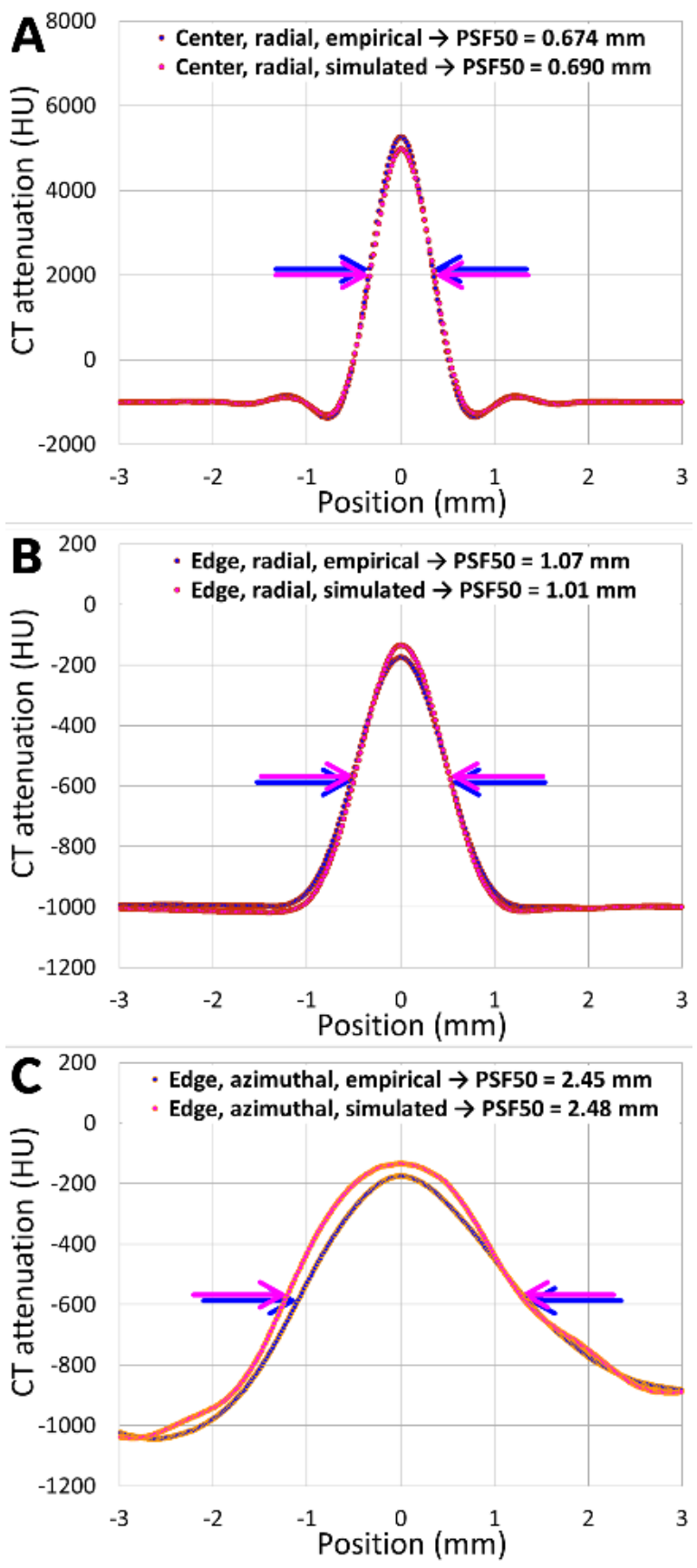

***Figure 8. Selected ID PSFs.*** *Radial PSFs are shown at the (A) center and (B) edge positions; Azimuthal PSFs are shown at the (C) edge position. Vertical axes are unequal and horizontal axes are equal.*

The positional dependence of the MTFs at 50% and 10% modulation (Figure 6B and Figure 6C) also show similar close agreement, qualitatively and quantitatively, between empirical and simulated results. Exemplary MTF curves (Figure 7) generally show good agreement but there are some notable discrepancies. The radial response at the edge (Figure 7A, blue and pink squares) is in disagreement over the range of about 1 to 3 lp/cm, but this doesn't affect the MTF50 or MTF10. The radial response at the center (Figure 7A, blue and pink triangles) and the azimuthal response at the center (Figure 7B, blue and pink circles) are in disagreement over the range of about 5 to 10 lp/cm, which affects MTF50 and MTF10. The azimuthal response at the edge (Figure 7B, blue and pink diamonds) are in disagreement over the range of about 5 to 10 lp/cm, but this doesn't affect the MTF50 or MTF10.

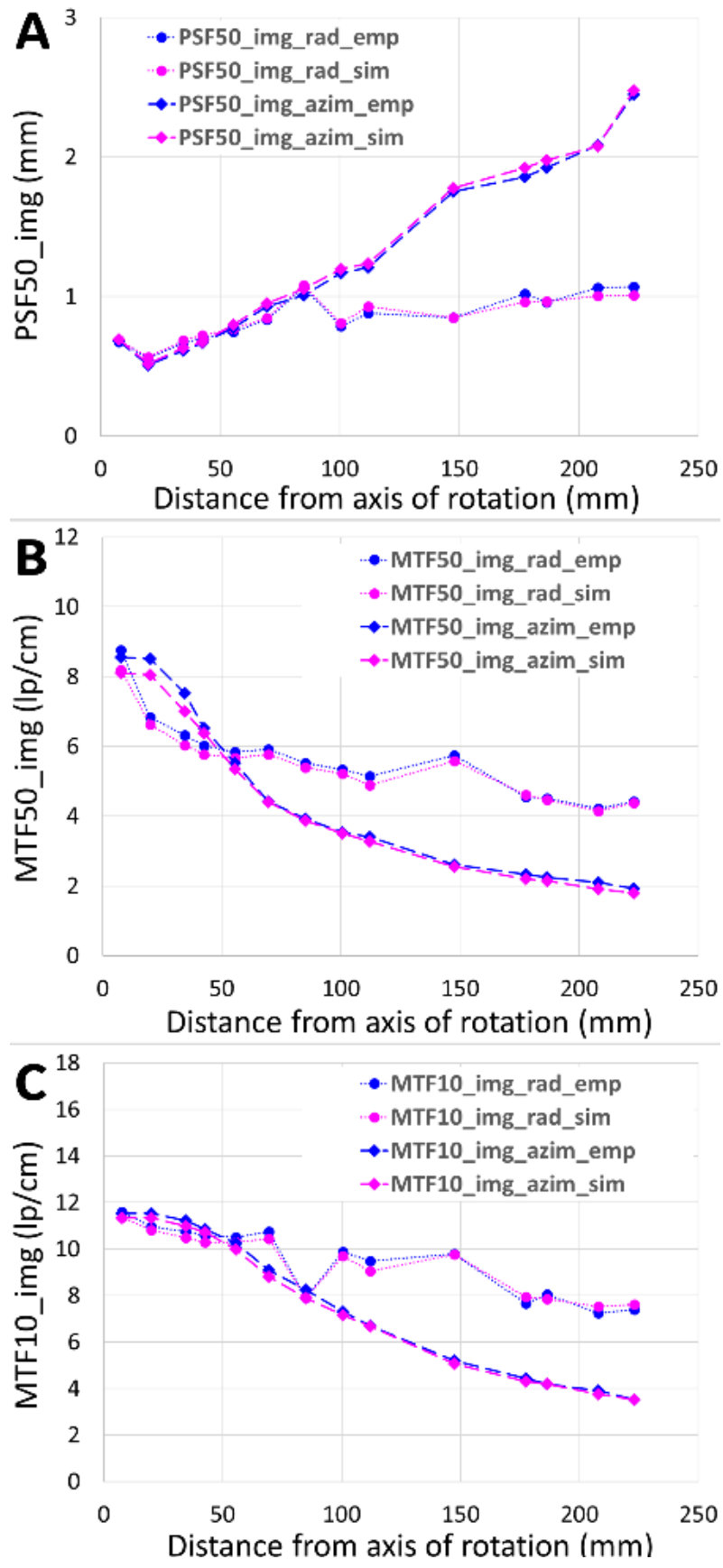

***Figure 6. ID results versus wire position.*** *(A) PSF50, (B) MTF50, and (C) MTF10.*

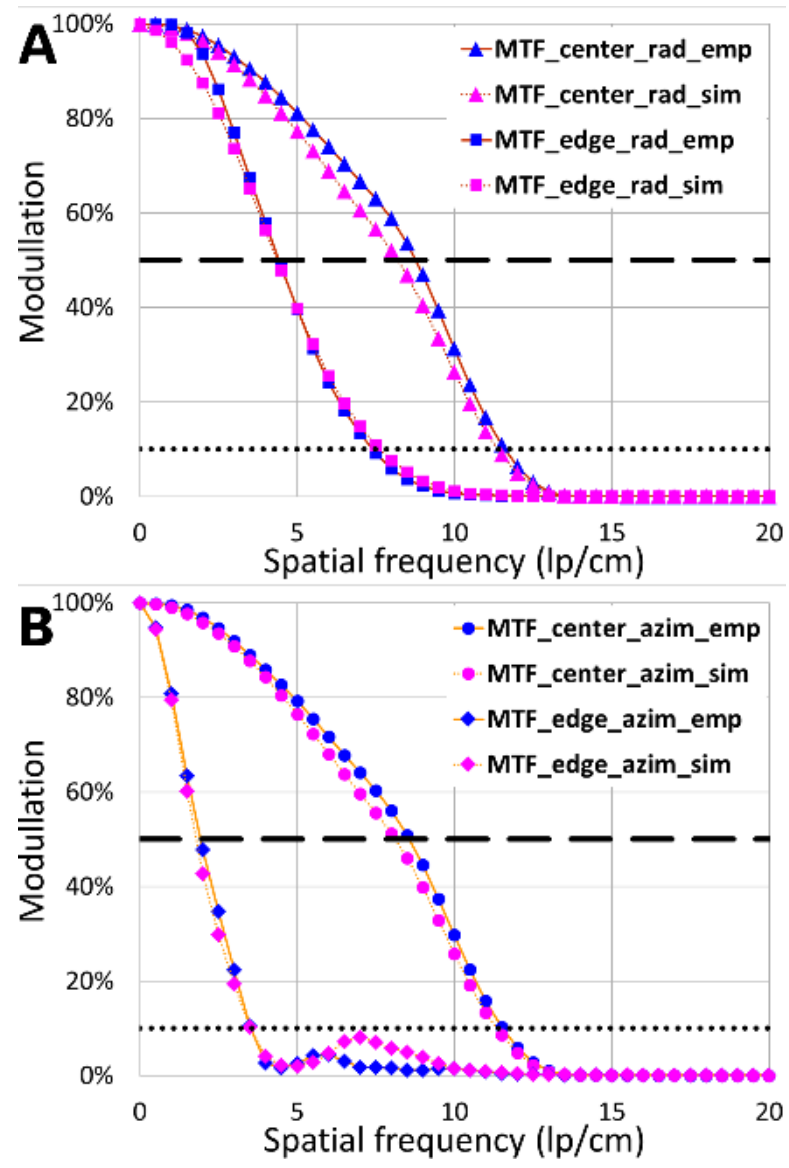

***Figure 7. Center and edge MTF curves.*** *(A) Radial (B) Azimuthal.*

ID results are summarized in Table 2. For PSF50_img, the mean(maximum) errors from all fourteen included wire positions were about 3%(6%) radially and about 2%(5%) azimuthally. For MTF50_img, the errors were about 3%(7%) radially and about 4%(10%) azimuthally. For MTF10_img, the errors were about 2%(5%) radially and about 2%(5%) azimuthally.

**Table 2. Image-domain results.**

| Wire position | Measured orientation | | PSF50_img | | MTF50_img | | MTF10_img | |
|---|---|---|---|---|---|---|---|---|
| | | | Value | Units | Value | Units | Value | Units |
| Center | Radial | Empirical | 0.67 | mm | 8.76 | lp/cm | 11.59 | lp/cm |
| | | Simulated | 0.69 | mm | 8.19 | lp/cm | 11.37 | lp/cm |
| | | Error | 2.40 | % | 6.54 | % | 1.97 | % |
| | Azimuthal | Empirical | 0.68 | mm | 8.56 | lp/cm | 11.52 | lp/cm |
| | | Simulated | 0.69 | mm | 8.10 | lp/cm | 11.34 | lp/cm |
| | | Error | 1.37 | % | 5.34 | % | 1.57 | % |
| Edge | Radial | Empirical | 1.07 | mm | 4.42 | lp/cm | 7.39 | lp/cm |
| | | Simulated | 1.01 | mm | 4.37 | lp/cm | 7.62 | lp/cm |
| | | Error | 5.61 | % | 1.05 | % | 3.14 | % |
| | Azimuthal | Empirical | 2.45 | mm | 1.93 | lp/cm | 3.53 | lp/cm |
| | | Simulated | 2.48 | mm | 1.79 | lp/cm | 3.52 | lp/cm |
| | | Error | 1.24 | % | 7.07 | % | 0.24 | % |
| | | Maximum error, radial (a) | 5.92 | % | 6.54 | % | 4.65 | % |
| | | Mean error, radial (a, b) | 2.97 | % | 2.96 | % | 2.43 | % |
| | | Maximum error, azimuthal (a) | 4.81 | % | 9.45 | % | 4.35 | % |
| | | Mean error, azimuthal (a, b) | 2.27 | % | 4.14 | % | 2.07 | % |

(a) Determined from errors at all fourteen included wires
(b) Mean error = RMSE/mean(all <parameter>_img_emp)

## 4 Discussion

The purpose of this work is to compare the results from simulation with ground-truth empirical results, and not to characterize the performance of the VCT system. Therefore, we focus on the question of agreement.

The models that determine the ability to accurately simulate the spatial resolution of a CT system include the X-ray source focal spot (FS), detector cell size, detector afterglow, and gantry rotation (the latter two resulting in azimuthal blur). Detector cell size and gantry rotation are straightforward to model, leaving the FS and afterglow as the most challenging models. At position S, PSF width is dominated by the FS convolved by azimuthal blur, and the asymmetry is produced by afterglow. At position D, PSF width is dominated by detector cell size and azimuthal blur. At position T, there is minimal effect from azimuthal blur and FS; detector cell size dominates. The close agreement in the characteristics of the PSF curves and the quantitative results suggests that our critical models are valid.

Interestingly, there is one finding in the empirical results that is unexplained but was faithfully predicted by the simulated results. Specifically, the azimuthal PSF and MTF curves show fluctuation in the range of about 60 mm to 150 mm from the AR in both empirical and simulated results.

Conversely, there were some characteristics in the results that were not in close agreement but were not captured by our metrics. For example, there is positional disagreement in the PSF_proj curves at positions S and T. Also, the MTF curves of the edge wire disagree at low frequencies in the radial direction and at higher frequencies in the azimuthal direction; these errors are undetected by our metrics.

One limitation of this study was the need to exclude some of the wires in the image-domain phantom due to alignment issues. The primary limitation is that only one focal spot model and X-ray technique was used, and only in-plane resolution was reported.

## 5 Conclusion

This work represent a strong first step toward validation of CatSim's phyics models that determine the ability to accurately simulate the spatial resolution of a typical CT system. The focal spot shape/size and detector afterglow models that have been implemented produced simulations that were in excellent agreement with empirical measurements, resulting in mean errors <5% and maximum errors <10% for all measured parameters. This study was limited to a small focal spot operated at 120 kV and 300 mA and all standard corrections for physical effects were disabled. In future work, we will extend this to the entire range of conditions used for clinical CT exams.

## Acknowledgements

Research reported in this publication was supported by the National Cancer Institute of the National Institutes of Health under Award Number U01CA231860. The content is solely the responsibility of the authors and does not necessarily represent the official views of the National Institutes of Health.

# Limited scanning arc image reconstruction with weighted anisotropic TV minimization

Leo Y. Zhang[1], Emil Y. Sidky[1], John Paul Phillips[1], Zheng Zhang[1], Buxin Chen[1], Dan Xia[1], and Xiaochuan Pan[1]

[1]Department of Radiology, The University of Chicago, Chicago, IL, USA

**Abstract** In this work, an algorithm for limited angular range scanning is developed based on data discrepancy constrained, weighted anisotropic total variation (WATV) minimization. The constraint on the data discrepancy allows for mismatch between the projection data and its estimate, and a weighting parameter controls the relative strength of ATV regularization in directions parallel and perpendicular to the scanning arc. Standard unweighted ATV is a special case of the considered optimization problem. The algorithm is demonstrated on noiseless projection data in order to investigate the accuracy of the proposed inverse problem solution as a function of the weighting parameter. The results show that the optimal recovery is found for a weighting parameter that imposes greater directional TV regularization along the direction parallel to the scan arc.

## 1 Introduction

Limited angular range scanning in CT is a classic image reconstruction problem, which is relevant for developing scanners for surgical or radiation therapy interventions. For such applications, it may not be practical to acquire transmission data over a complete scanning arc that admits accurate reconstruction by filtered backprojection (FBP). Recently, we have been investigating an optimization based image reconstruction for this problem by imposing two constraints on the directional TV of the image while minimizing the data discrepancy[1]. We have found that imposing two directional TV constraints admits exact reconstruction for angular ranges far smaller than use of a single TV constraint. Exact recovery, however, depends on the knowledge of the scanned object's directional TV values.

In this work, we seek an optimization based approach that takes advantage of different directional TV regularization strengths and that does not require prior knowledge of the object directional TV values. The use of directional weighting for TV-based regularization is a strategy that has been exploited for X-ray based tomographic imaging for under-sampled scan configurations [2, 3]. The new optimization based framework is motivated and developed in Sec. 2 and preliminary results are shown in Sec. 3.

## 2 Methods

The data model for the limited angular range problem is formulated as an under-determined linear system

$$Au = g, \tag{1}$$

where $A$ represents a discrete projection operator, applying line integrations over the attenuation coefficients of the unknown image vector $u$, with $g$ being the resultant vector of projections. The general setup for our model is shown in Fig. 1. The precise form of the fan-beam projection matrix is specified in Sec. 3. We consider convex optimization problems that impose various constraints that are used to select an unique solution out of the null space of Eq. (1). We take advantage of the convex prototyping capability of the Chambolle-Pock (CP)[4, 5] algorithm to provide an algorithm instance that can solve the problems of interest.

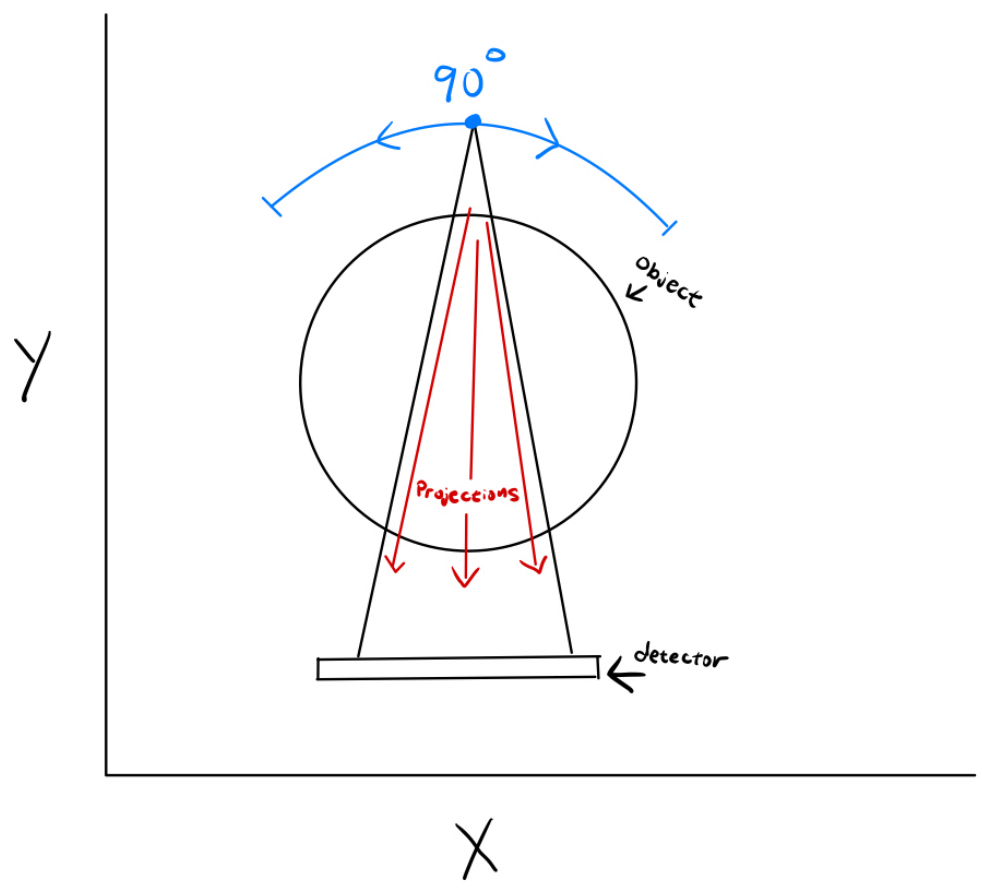

**Figure 1:** Limited Angle Setup

For investigating image reconstruction from data that has insufficient sampling, exploiting gradient sparsity, it is convenient to formulate the imaging model as a data discrepancy minimization problem with a constraint on the image ATV

$$\min_u \|Au - g\|_2^2 \text{ such that } \|\nabla u\|_1 \leq \gamma, \tag{2}$$

where $\|\nabla u\|_1$ is anisotropic total variation (ATV) and $\gamma$ is the constraint parameter. For an inverse problems type study, where exact recovery from noiseless data is being tested, $\gamma$ is set to the object ATV value, $\gamma_0$. A criticism of this approach is that it requires fore-knowledge of $\gamma_0$. This criticism is addressed by switching to a data discrepancy constraint and minimizing ATV

$$\min_u \|\nabla u\|_1 \text{ such that } Au = g, \tag{3}$$

which does not require knowledge of the object ATV ahead of time. For noiseless data, this latter optimization problem is actually equivalent to Eq. (2) through the Lagrangian formalism. For example, one could solve Eq. (3) first and use the ATV of the solution as the constraint value $\gamma$ for Eq. (2).

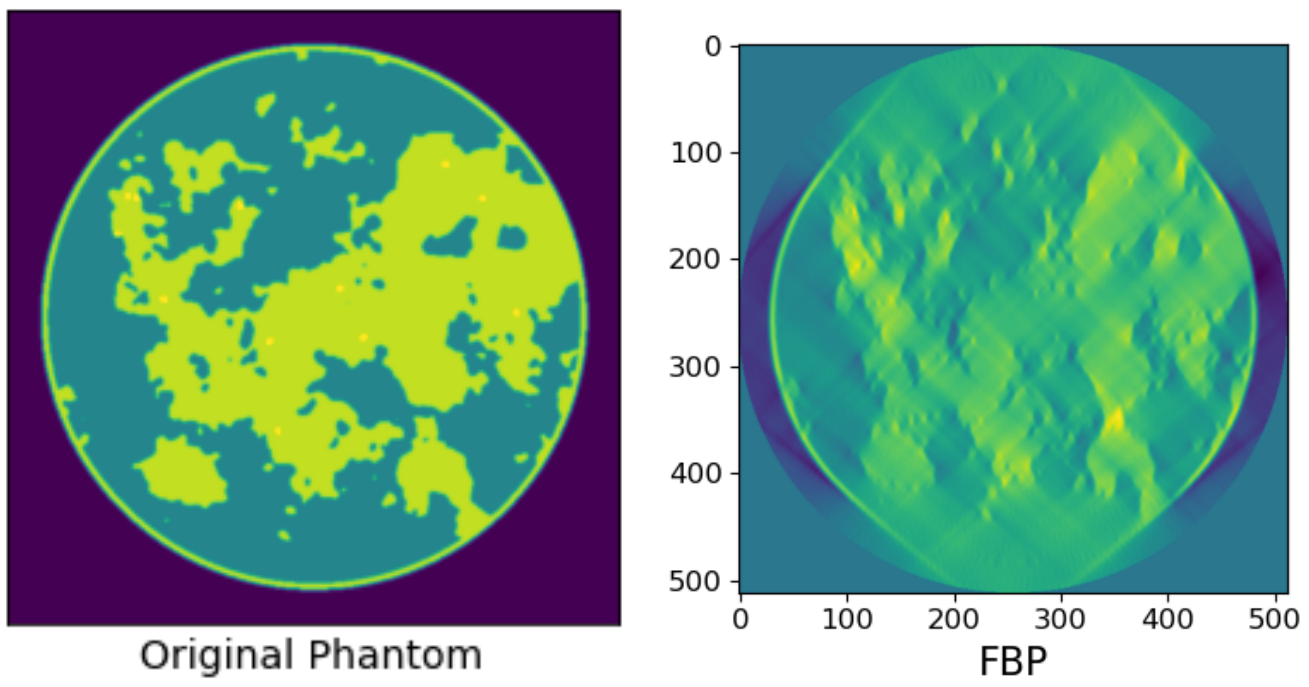

**Figure 2:** Left: ground truth image. Right: FBP reconstruction from a limited angular scan over a 90° arc.

If solving Eq. (3) yields the exact test object, then Eq. (2) will also yield the same with this procedure.

The case where Eq. (2) is generalized to having two directional TV constraints is more complicated. We consider limited angular range scanning where the scanning arc is bisected by the $y$-axis of a 2D coordinate system. The $x$-axis is then considered to be "parallel" to the scanning arc. The doubly constrained optimization problem considered in Ref. [1] is

$$\min_u \|Au-g\|_2^2 \text{ such that } \|\nabla_x u\|_1 \leq \gamma_x \text{ and } \|\nabla_y u\|_1 \leq \gamma_y, \quad (4)$$

where $\nabla_x$ and $\nabla_y$ are the $x$ and $y$-directional derivatives, respectively. For exact reconstruction from noiseless data, the directional TV values $\gamma_{0,x}$ and $\gamma_{0,y}$ of the scanned object are assumed to be known.

Because there are two directional TV constraints, it is not possible to convert this optimization problem into a parameter-free analogue to Eq. (3). Instead, the number of parameters can be reduced from two to one by converting Eq. (4) to constrained, weighted ATV minimization

$$\min_u \{\alpha\|\nabla_x u\|_1 + (2-\alpha)\|\nabla_y u\|_1\} \text{ such that } Au = g, \quad (5)$$

where the $\alpha$ weighting parameter can be tuned for optimal recovery of the scanned object. The case where $\alpha = 1$ is equivalent to Eq. (3). For image recovery from noiseless data, it is possible that exact recovery can be obtained for any value of $\alpha$ since the data equality constraint is being enforced. Furthermore, there is an optimal value $\alpha_0$ that will yield the same result as Eq. (4) with knowledge of the true directional TV values $\gamma_{0,x}$ and $\gamma_{0,y}$. The optimal combination parameter $\alpha_0$ is object dependent. Still, in principle, a “best” value of $\alpha$ can be defined as the one that yields the lowest root-mean-square-error (RMSE) for an ensemble of test objects when reconstructing images from noiseless data. Solving Eq. (5) with this value of $\alpha$ will not be as optimal as as solving Eq. (4) with knowledge of $\gamma_{0,x}$ and $\gamma_{0,y}$, but it may yield better image recovery than Eq. (3), i.e. setting $\alpha = 1$.

For this preliminary study, we consider image reconstruction from a single test object and the data equality constraint

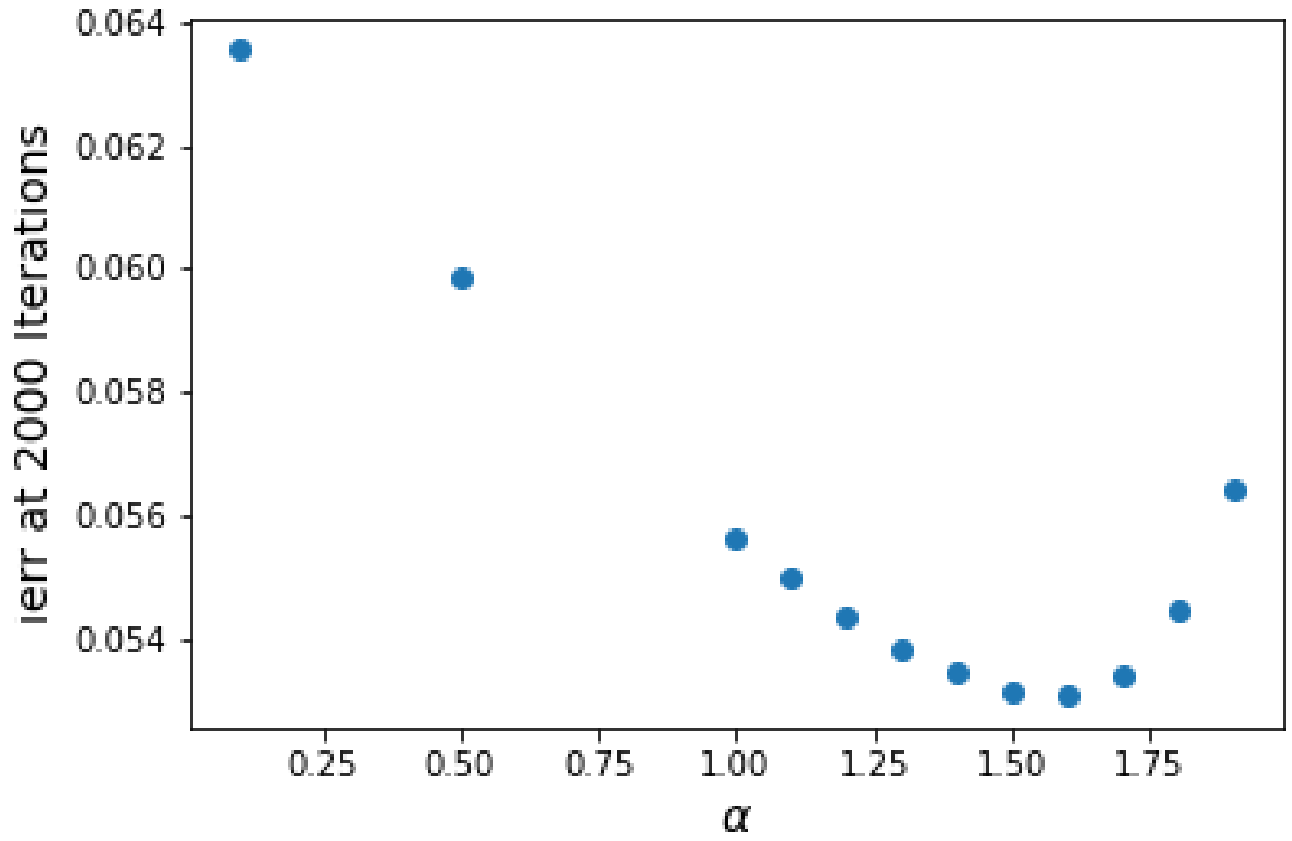

**Figure 3:** A graph of $\alpha$ values vs. image error at 2000 iterations.

is loosened to an inequality constraint governed by a data discrepancy parameter $\varepsilon$.

$$\min_u \{\alpha\|\nabla_x u\|_1 + (2-\alpha)\|\nabla_y u\|_1\} \text{ such that } \|Au-g\|_2 \leq \varepsilon. \quad (6)$$

By setting $\varepsilon$ to a small positive value, accuracy and stability of the image reconstruction model is tested.

## 3 Results

For the system of interest we consider scanning over a 90° arc with one degree spacing. As stated before, the center of the arc is bisected by the $y$-axis of a 2D coordinate system. In this way, the $x$-axis parallel to the tangent of the scanning arc at its midpoint, and the directional derivatives in the objective function of Eq. (5) are parallel and perpendicular to the the mid-arc tangent. The detector is taken to be a linear array consisting of 512 detection elements and it rotates along with the source as in a standard CT set-up. The scanning arc radius is 50 cm and the source-to-detector distance is 100 cm. The 18cm × 18cm image array consists of 512×512 pixels. Given that there are only 91 projections over a 90° arc, the reconstruction problem is challenging due to undersampling and limiting angular range scanning. The difficulty of this problem is demonstrated by Fig. 2 where image reconstruction from this scan configuration is shown for filtered backprojection (FBP).

For the results of this experiment, noiseless data is reconstructed by solving Eq. (6) varying $\alpha$ and fixing the data discrepancy to an $\varepsilon$ corresponding to an RMSE of 0.001. The CP algorithm is run for 2000 iterations for each $\alpha$ value, at which point the data discrepancy constraint value has been met. The corresponding image RMSEs are plotted in Fig. 3. As expected there is variation in the RMSE values as a function of $\alpha$, and the minimum image RMSE is obtained at approximately $\alpha = 1.6$. Accordingly, there is some potential advantage to use of WATV over the unweighted case ($\alpha = 1$). A few representative images are shown in Fig. 4 and corresponding regions of interest (ROIs) are shown in Fig. 5. All of the reconstructed images show a high degree of accuracy. We note that exact reconstruction is not expected because $\varepsilon$ is not zero. Allowing this small data discrepancy reveals which aspects of the image can be stably reconstructed. For all of the images, it is clear that the hardest parts of the phantom to recover are the top and bottom edges of the phantom. This result is expected because the scanning arc is above the test object and rays tangent to the object at the top and bottom are not measured [6].

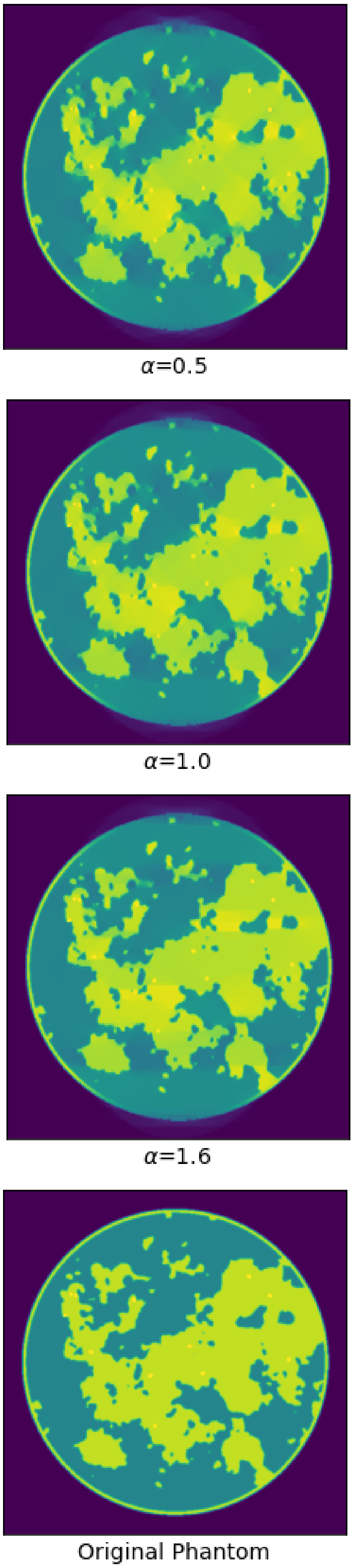

**Figure 4:** Reconstructed images for $\alpha$ values of $0.5, 1.0, 1.6$, and the original phantom

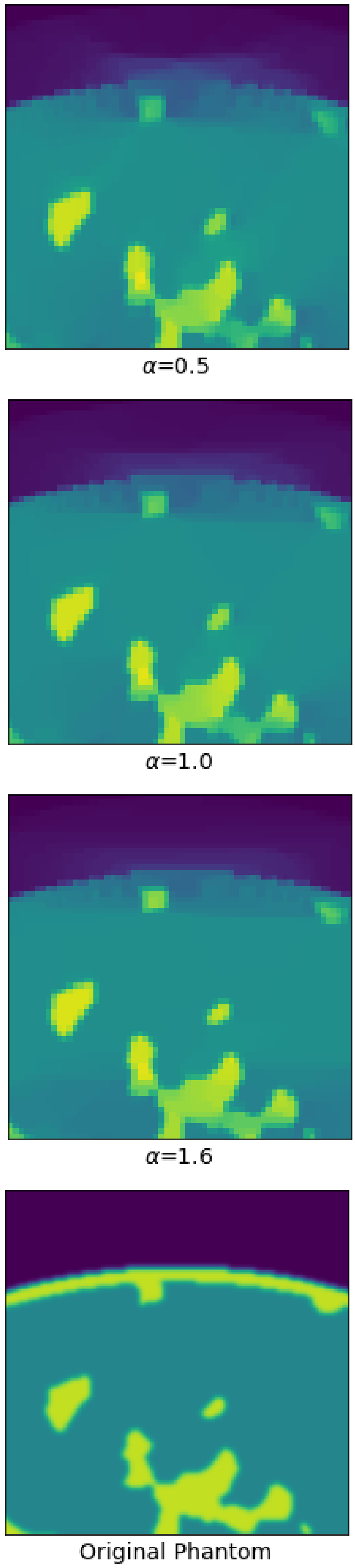

**Figure 5:** ROIs of the images shown in Fig. 4, zooming in on the top edge of the test phantom.

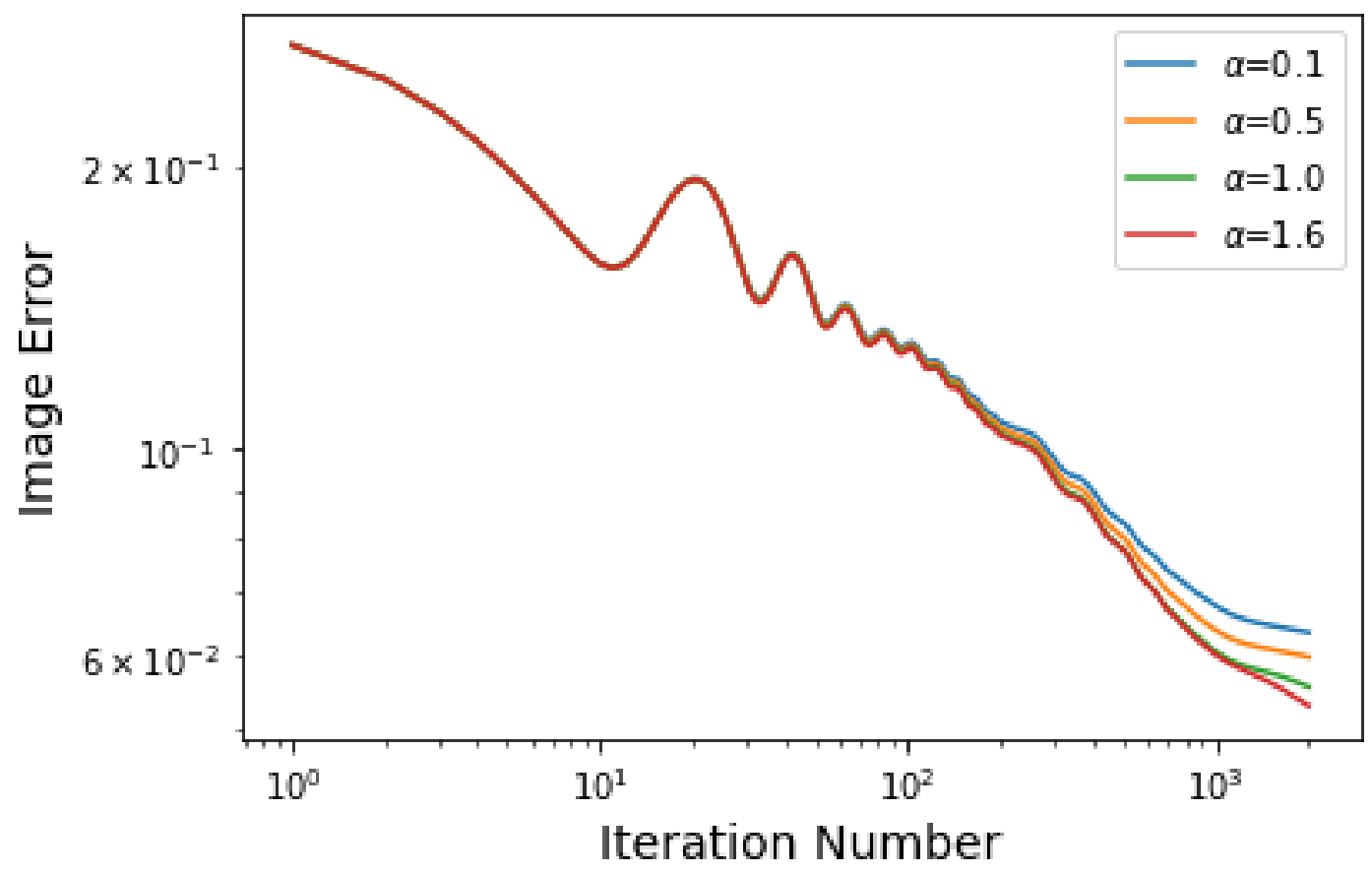

**Figure 6:** Image RSME versus iteration number for different values of the weigting parameter $\alpha$.

Use of ATV ($\alpha = 1$) has already been shown to yield higher degree of accuracy in the reconstructed images than FBP for limited angular range scanning. Optimizing the weighting parameter $\alpha$, leads to even further improvement as evidenced by the drop in image RMSE by approximately 5% when increasing $\alpha$ from 1 to 1.6. Looking at the images, the ROI corresponding to $\alpha = 1.6$ does show the best recovery of the top boundary of the phantom even though there is still obvious discrepancy when compared with the ground truth. Decreasing $\alpha$ to 0.5 results in a worsened erosion of the top edge.

Finally, in order to provide a sense of convergence of CP applied to Eq. (6) the image RMSE is shown as a function of iteration number in Fig. 6. It is clear from the figure that the image RMSE is starting to level off at 2000 iterations, but there is still a slight downward trend in all graphs, which is the largest for $\alpha = 1.6$.

## 4 Conclusion

This abstract presents an optimization-based framework for solving the inverse problem corresponding to limited angular range scanning. The proposed constrained, WATV minimization algorithm does not require prior knowledge of any aspects of the scanned object for accurate image reconstruction, and it is found that the use of weighted anisotropic TV can lead to better accuracy than the unweighted case. Further studies will include investigation into both $\alpha$ and $\varepsilon$ dependence. In conjunction with these studies, use of inconsistent data due to noise and other physical factors will also be studied.

## Acknowledgements

This work is supported in part by NIH Grant Nos. R01-EB026282, R01-EB023968, and R21-CA263660. The contents of this article are solely the responsibility of the authors and do not necessarily represent the official views of the National Institutes of Health.

# A Joint Processing Strategy for Image Quality Improvement in 3D Digital Subtraction Angiography

Xiaoxuan Zhang[1], Xiao Jiang[2], Matthew Tivnan[2], J. Webster Stayman[2], and Grace J. Gang[1]

[1] Hospital of the University of Pennsylvania Department of Radiology
[2] Johns Hopkins University Department of Biomedical Engineering

**Abstract** Three-dimensional digital subtraction angiography (3D-DSA) is a widely adopted technique for clinical evaluation of contrast-enhanced vasculatures. The distribution of a contrast agent such as iodine is often estimated via temporal subtraction. Advancements in spectral imaging technologies such as photon counting detectors offer new opportunities to improve DSA image quality. In this work, we propose a novel joint processing strategy to achieve an iodine image using two-bin spectral measurements from a photon counting detector acquired both the pre- and post-contrast injection. Simulation studies were performed using a digital phantom with iodine-enhanced vessels. The proposed method was compared with temporal subtraction and conventional spectral imaging using just the post-contrast measurements. Imaging performance was evaluated in terms of noise-resolution tradeoffs. Preliminary findings have shown measurably improved image quality given by joint processing, reducing noise by 40% and 70% compared to temporal subtraction and conventional spectral imaging using an energy-integrating detector, respectively.

## 1 Introduction

Three-dimensional digital subtraction angiography (3D-DSA) has been widely used for the diagnosis of vascular diseases in a range of clinical applications.[1,2] Conventionally, 3D-DSA relies on temporal subtraction, where cone-beam CT (CBCT) acquisitions pre- and post-contrast injection are subtracted to isolate iodine distribution in the vasculature.

With the increasing availability of spectral hardware, efforts to introduce spectral imaging for interventional applications are quickly emerging.[3] Spectral imaging has been previously proposed for 2D DSA as an alternative to temporal subtraction to mitigate patient motion. A post-contrast only acquisition using spectral measurements allow iodine distribution to be computed using material decomposition algorithms. However, significant noise amplification from material decomposition impeded clinical adoption.

Recent work[4] has proposed a joint processing strategy using spectral measurements acquired both pre- and post-contrast injection to improve the noise performance 2D DSA image estimates. Such strategy leverages both the temporal and spectral information and can yield DSA images that outperform conventional temporal-only and spectral-only processing. This work seeks to apply similar strategies to 3D-DSA imaging. We focus on spectral measurements using a photon counting detector (PCD) due to its superior spectral separability and spatial resolution compared to other spectral technologies. We envision that the low image noise and high spatial resolution enabled by this strategy can bring the most benefit to the visualization of small, low contrast targets like small vessels.

## 2 Methods

### 2.1. Theoretical methods

We propose a novel joint processing strategy for 3D-DSA which leverages both temporal and energy information for image quality improvement. Projection data are acquired using a spectral imaging system before and after contrast injection. Measurements in two spectral channels are obtained for each acquisition, yielding a total of four measurements. Under the assumption of minimal patient motion (valid for intubated patients in anatomical sites without involuntary motion like cardiac and breathing), we can formulate a general forward model that relates a three-material object (water, bone/calcium, iodine) to the four measurements as follows:

$$\bar{y}_1(l) = \mathbf{S}_1 \exp(-\mathbf{Q}^{H_2O}\mathbf{A}\rho^{H_2O} - \mathbf{Q}^{Ca}\mathbf{A}\rho^{Ca}) \quad (1)$$

$$\bar{y}_2(l) = \mathbf{S}_2 \exp(-\mathbf{Q}^{H_2O}\mathbf{A}\rho^{H_2O} - \mathbf{Q}^{Ca}\mathbf{A}\rho^{Ca}) \quad (2)$$

$$\bar{y}_3(l) = \mathbf{S}_1 \exp(-\mathbf{Q}^{H_2O}\mathbf{A}\rho^{H_2O} - \mathbf{Q}^{Ca}\mathbf{A}\rho^{Ca} - \mathbf{Q}^{I}\mathbf{A}\rho^{I}) \quad (3)$$

$$\bar{y}_4(l) = \mathbf{S}_2 \exp(-\mathbf{Q}^{H_2O}\mathbf{A}\rho^{H_2O} - \mathbf{Q}^{Ca}\mathbf{A}\rho^{Ca} - \mathbf{Q}^{I}\mathbf{A}\rho^{I}) \quad (4)$$

where $\{\bar{y}_1, \bar{y}_2\}$ denotes the pre-contrast mean measurements containing water and bone/calcium, $\{\bar{y}_3, \bar{y}_4\}$ denotes the post-contrast mean measurements containing water, calcium, and iodine, $\mathbf{S}_1$ and $\mathbf{S}_2$ models the spectral sensitivities of each spectral channel, $\mathbf{Q}$ contains the mass attenuation coefficients for each basis material, $\mathbf{A}$ denotes the system matrix, and $\rho$ represents material densities. We can alternatively combine the four equations into a single equation to facilitate joint estimation:

$$\begin{bmatrix} \bar{y}_1 \\ \bar{y}_2 \\ \bar{y}_3 \\ \bar{y}_4 \end{bmatrix} = \begin{bmatrix} \boldsymbol{S}_1 & \\ & \boldsymbol{S}_2 \\ \boldsymbol{S}_1 & \\ & \boldsymbol{S}_2 \end{bmatrix} \exp\left( - \begin{bmatrix} \mathbf{Q} & \\ & \mathbf{Q} \end{bmatrix} \begin{bmatrix} \mathbf{I} & & \\ & \mathbf{I} & \mathbf{0} \\ \mathbf{I} & & \\ & \mathbf{I} & \\ & & \mathbf{I} \end{bmatrix} \begin{bmatrix} \mathbf{A} & & \\ & \mathbf{A} & \\ & & \mathbf{A} \end{bmatrix} \begin{bmatrix} \rho^{H_2O} \\ \rho^{Ca} \\ \rho^{I} \end{bmatrix} \right) \quad (5)$$

Here, a masking matrix composed of identity matrices and zeros selectively zero out the pre-contrast measurements associated with iodine.

*2.1.1. Model-based one-step decomposition/reconstruction*

This form of forward model (linear-exponential-linear) can be solved by an optimization algorithm using a separable paraboloidal surrogate approach previously developed by the authors.[5] Assuming the measurements follow a multivariate Gaussian distribution, $y \sim \mathcal{N}(\bar{y}, \boldsymbol{\Sigma})$, the material density, $\rho$, can be solved using a nonlinear weighted least squares objective:

$$\hat{\rho} = \arg\min_{\rho} \Phi(\rho, y), \text{ where} \tag{6}$$

$$\Phi(\rho, y) = \big(y - \bar{y}(\rho)\big)^T \boldsymbol{\Sigma}_y^{-1} \big(y - \bar{y}(\rho)\big) + \beta R(\rho) \tag{7}$$

A quadratic penalty, $R$, is adopted in this work. We refer to this algorithm as the model-based one-step decpomposition (MB-OSD). The densities of water, calcium, and iodine were initialized with zeros and iteratively solved using 600 iterations of the separable quadratic surrogate algorithm with 10 ordered subsets[6] and Nesterov's acceleration.[7]

*2.1.2. Model-based projection-domain decomposition*

Alternatively, a more memory efficient and faster algorithm involves performing a projection-domain decomposition followed by an analytic or model-based reconstruction (MB-PDD). Instead of physical density $\rho$, we now estimate the material density line integral, $l = \mathbf{A}\rho$. The forward model in Eq. (5) can be modified to:

$$\begin{bmatrix}\bar{y}_1\\ \bar{y}_2\\ \bar{y}_3\\ \bar{y}_4\end{bmatrix} = \begin{bmatrix}\boldsymbol{S}_1 & \\ & \boldsymbol{S}_2\\ \boldsymbol{S}_1 & \\ & \boldsymbol{S}_2\end{bmatrix} \exp\left(-\begin{bmatrix}\mathbf{Q} & \\ & \mathbf{Q}\end{bmatrix} \begin{bmatrix}\mathbf{I} & & \\ & \mathbf{I} & \\ & & \mathbf{0}\\ \mathbf{I} & & \\ & \mathbf{I} & \\ & & \mathbf{I}\end{bmatrix} \begin{bmatrix} l^{H_2O}\\ l^{Ca}\\ l^{I}\end{bmatrix}\right) \tag{8}$$

We can write down a similar objective function as above based on $l$:

$$\Phi(l, y) = \big(y - \bar{y}(l)\big)^T \boldsymbol{\Sigma}_y^{-1} \big(y - \bar{y}(l)\big) \tag{9}$$

$$\hat{l} = \arg\min_{l} \Phi(l, y) \tag{10}$$

Note that no additional regularization is included in this objective. We instead rely on smoothing in the volumetric reconstruction (e.g., apodization filter in FDK or regularization in model-based iterative reconstruction).

We adopted a modified Newton's method to solve this objective, where a proximal regularization term was added to the diagonal of the Hessian to improve conditioning of the inversion process. We adoped zero initialization and applied 50 iterations of optimizer. Volumetric reconstruction of the estimated material density line integrals may be performed using either analytical or iterative algoirthms. In this work, we adopted filtered back-projeciton (FBP) with a Hann apodization filter with different cutoff frequencies.

2.2. Experimental methods

*2.2.1. Phantom*

A digital phantom (Figure 1) was constructed for data generation, which includes a 1000 mg/mL water background, a 4.0 mm-thick 200 mg/mL Calcium ring, and 10 mg/mL iodine-enhanced vessels of diameter ranging from 0.3 mm to 4.0 mm. The phantom is intended for evaluation of high-resolution imaging of small targets such as intracranial vessels. Pre-contrast imaging involved only the water background and the calcium ring, while post-contrast imaging included the enhanced vessels.

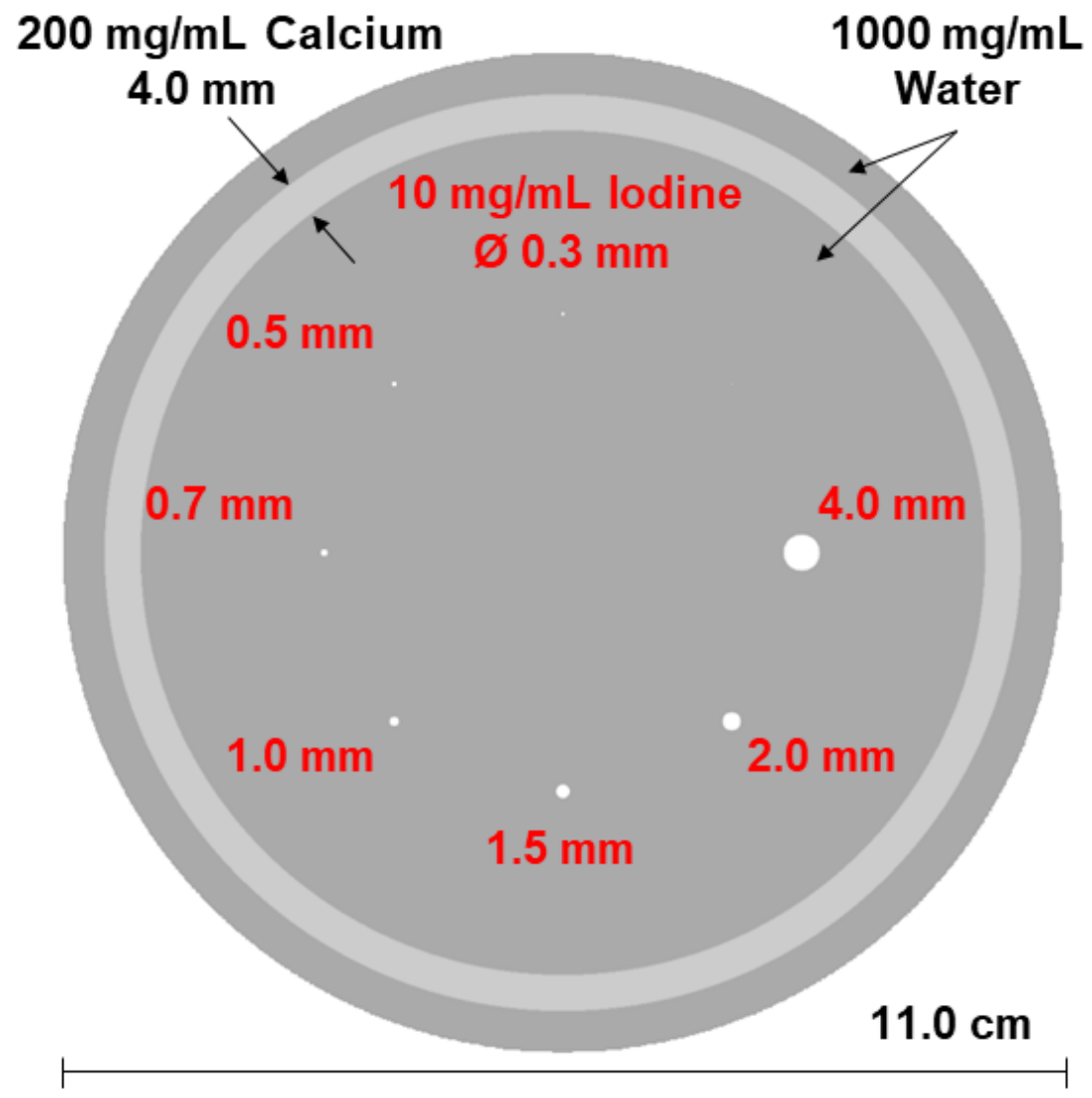

Figure 1. Digital phantom with a water background, a calcium ring, and iodine-enhanced, cylindrical vessels of varying diameters.

*2.2.2. Comparison studies*

We simulated four 3D-DSA techniques for comparison:

(1) Pre- and post-contrast measurements generated using an energy integrating detector (EID) with a 600-µm CsI scintillator. These measurements emulate data from the current standard DSA using temporal subtraction and what might be used in a 3D-DSA interventional imaging system (e.g., C-arm);

(2) Pre- and post-contrast measurements generated using a single energy bin (10–150 keV) from a photon counting detector with 750-µm CdTe. This dataset was used to evaluate performance of temporal subtraction with a PCD;

(3) Post-contrast measurements only from a PCD. Two energy thresholds were applied (40 keV and 60 keV) to obtain data from three energy bins. The PCD was modeled with "perfect" energy resolution. This data provided an example of "conventional" spectral imaging in 3D-DSA, where only energy information was leveraged for material decomposition. This technique is evaluated at both 1x and 2x the total dose for a fair comparison to techniques using both the pre- and post- measurements; and

(4) Pre- and post-contrast measurements from the PCD with two energy bins. This dataset contained both temporal

and energy information and was used to evaluate the proposed joint processing strategy. The energy threshold is set to 50 keV.

*2.2.3. Simulation*

Projection data for 3D imaging (1080 × 5 pixels, 0.154 mm pixel pitch, 360 views uniformly distributed over 360°) were generated for a system geometry with a source-to-axis distance of 800 mm and a source-to-detector distance of 1100 mm. The x-ray spectrum was generated calculated using Spektr[8] at 100 kVp with 0.25 mm Aluminum as the intrinsic filtration. The nominal tube current-time product was set to 1.0 mAs per projection. Poisson noise was added to all measurements, and detector blur measured from a physical energy integrating detector was applied to data from the EID. Blur was assumed to be negligible for the PCD. Volumetric images were reconstructed at 0.1 mm isotropic voxel size.

In temporal subtraction, the iodine distribution was obtained by subtracting two single energy reconstructions. The scale and units are therefore mismatched from physical density obtained from spectral decomposition. To enable direct quantitative comparisons, an "oracle" scaling factor computed from noiseless measurements was applied to convert attenuation coefficient differences to iodine density. In conventional spectral imaging, the density line integrals were estimated via model-based projection domain decomposition considering three energy sensitive channels and only post-contrast measurements.

For quantitative comparison of methods, the noise-resolution trade-off in the iodine basis image was evaluated by varying the cutoff frequency of the Hann filter. A similar curve can be generated by tuning the regularization strength for the one-step decomposition and will be the subject of future work. Resolution was characterized by the full width at half maximum (FWHM) of the 0.3 mm diameter vessel measured in a noiseless reconstruction. Noise was measured within a 21 × 21 × 1 region of interest at the center of the 4.0 mm diameter vessel over 10 repeated reconstructions with different noise realizations.

## 3 Results

Figure 2 shows the noise-resolution curves for conventional spectral imaging, temporal subtraction, and joint processing with model-based projection domain decomposition. Across all resolutions, conventional spectral imaging exhibits higher noise levels, even when the exposure is increased by a factor of two. Temporal subtraction using the EID provides reduced noise but also exhibits a limited improvement in spatial resolution compared to other methods – i.e., the resolution improvement stopped around 0.45 mm and increasing cutoff frequency did not improve resolution. In comparison, temporal subtraction using the PCD offers further opportunity to improve resolution, as well as ~25% reduction in noise at matched resolution. Joint processing with MB-PDD exhibited the best noise-resolution trade-off overall, permitting a ~12% noise reduction compared to temporal subtraction using the same PCD.

Figures 3(a-e) shows example decomposed iodine images of the 0.3 mm diameter vessel at a matched resolution as indicated by the black dotted vertical line in Figure 2. A high level of noise is present in Figures 3(a-b), obscuring the small vessel target. By contrast, the vessel can be easily differentiated from background noise in Figures 3(c-e), while the magnitude of noise is noticeably lower in (d) and (e).

Figure 3(f) demonstrates a preliminary model-based one-step decomposition result, which exhibits higher resolution (~0.3 mm) and markedly reduced noise compared to other methods. We conjecture that is due to the nonnegativity constraint clipping the noise at 0. While the strong noise suppression may be desirable for angiography applications, the images tend to appear "patchy" and low contrast features may be suppressed. Further investigation is underway where non-negativity constraint is turned off.

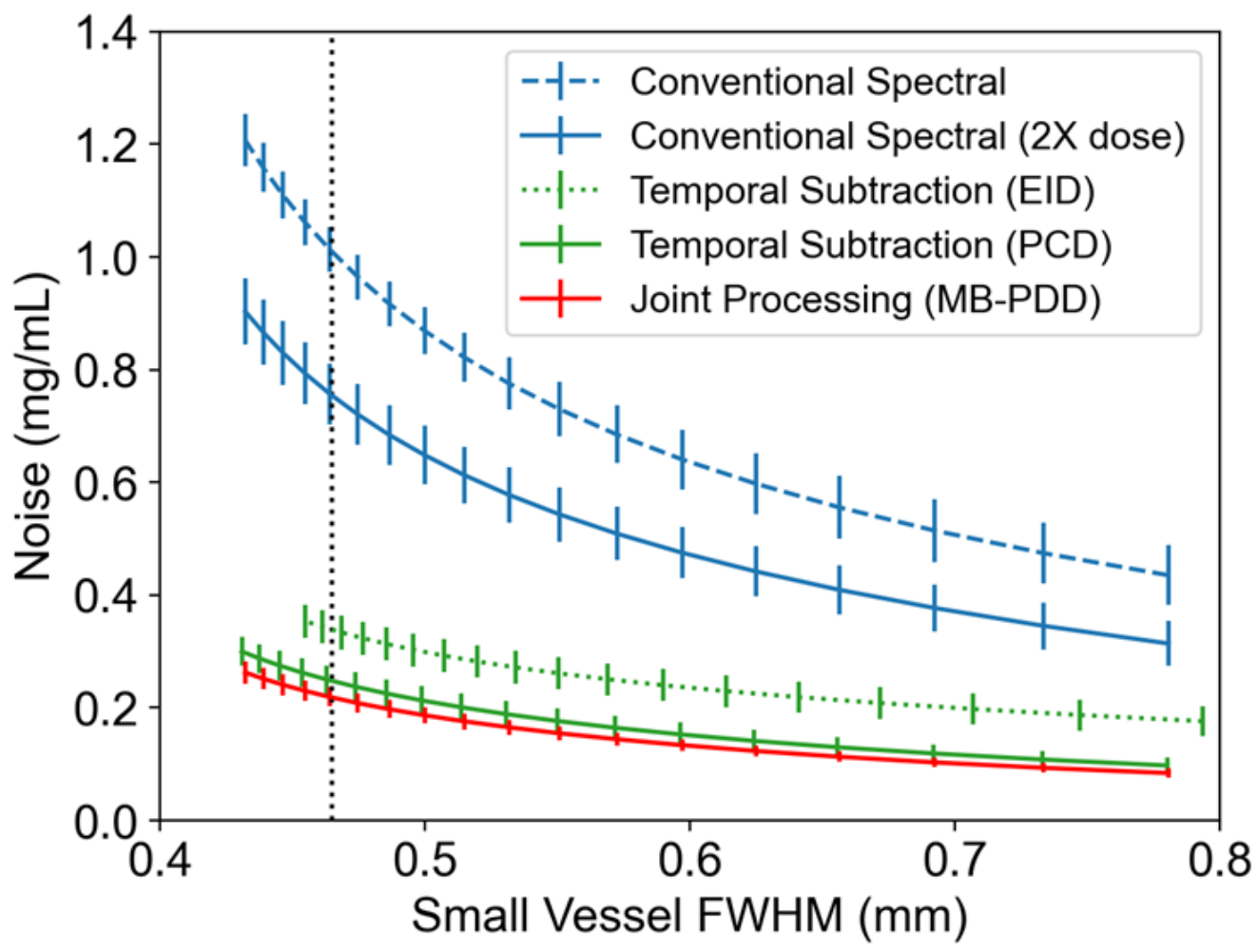

Figure 2. Noise-resolution characterization in iodine basis images estimated through conventional spectral imaging, temporal subtraction, and joint processing with varying cutoff frequencies for FBP reconstruction. Example images at matched resolution (dotted vertical line) are shown in Figure 3.

## 4 Discussion and Conclusion

This work demonstrated the capability of a joint processing strategy for improved noise and resolution performance in 3D-DSA. Leveraging both temporal and energy information, model-based decomposition approaches was shown to yield better noise-resolution tradeoffs compared to temporal subtraction or material decomposition based solely on multi-energy data.

Ongoing work includes spectral system optimization of the joint processing strategy. Material decomposition algorithms will be further optimized, including a comprehensive regularization parameter sweep of the one-step method and investigations of the behaviors of the nonnegativity constraint. Modeling of non-ideal effects in PCD including charge sharing and pulse pileup, and integration of registration algorithms to account for tissue deformation during image acquisition are also subjects of future work.

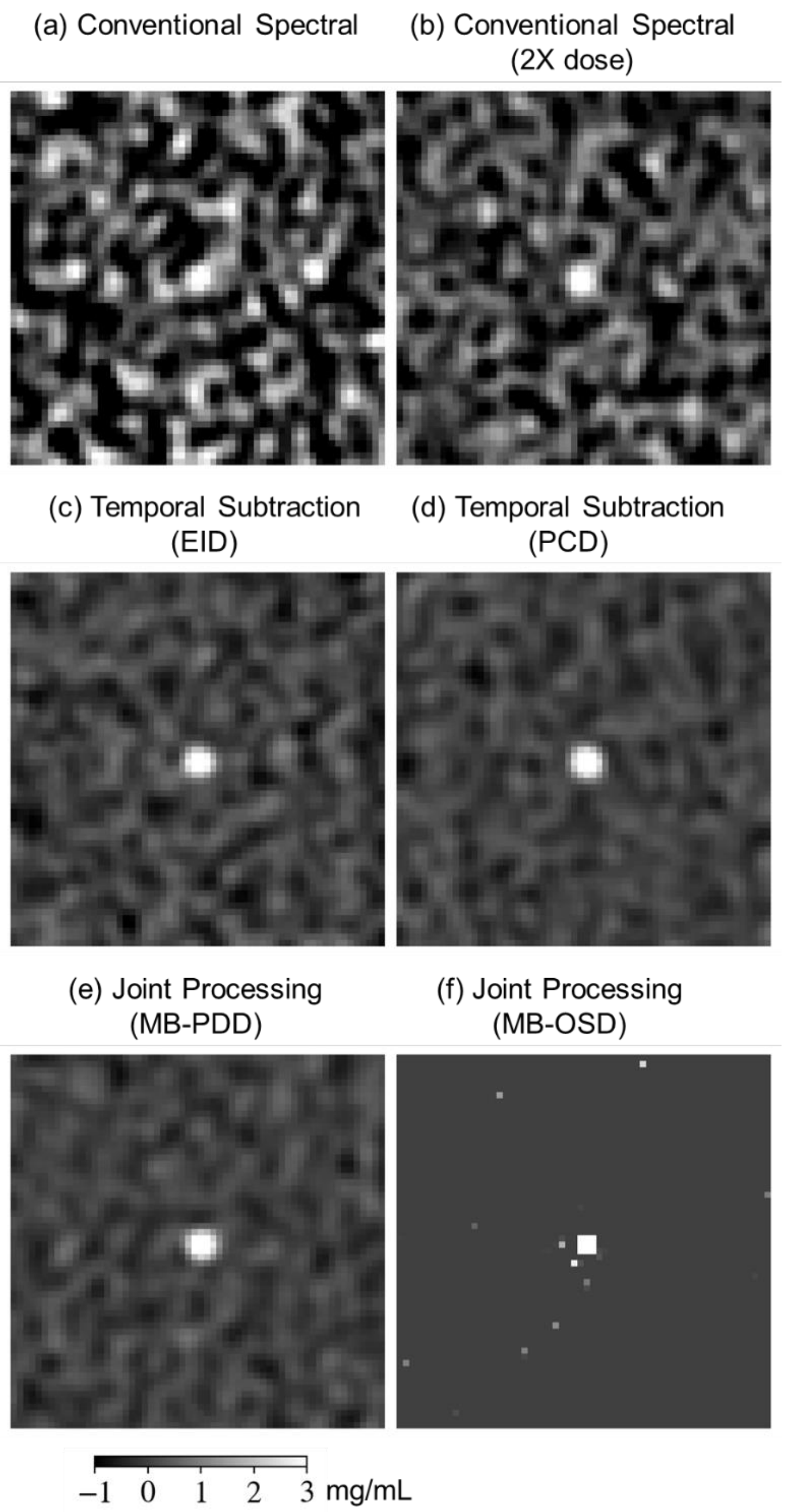

Figure 3. Example axial slice of the iodine basis images of the 0.3 mm diameter vessel estimated through (a-b) conventional spectral imaging at two dose levels, (c-d) temporal subtraction using the EID and the PCD, and (e-f) joint processing with MB-PDD and MB-OSD. The resolution was matched at 0.465 mm for (a-e) (dotted vertical line in Figure 2), and the examples were given by the mean noise level of 10 repeats. The same display window setting was applied to all images.

# Asymmetrical Dual-Cycle Adversarial Network for Material Decomposition and Synthesis of Dual-energy CT Images

Xinrui Zhang[1], AiLong Cai[1], Shaoyu Wang[1], Ningning Liang[1], Yizhong Wang[1], Junru Ren[1], Lei Li[1] and Bin Yan[1]

[1] Department of Henan Key Laboratory of Imaging and Intelligent Processing, PLA Strategic Support Force Information Engineering University, Zhengzhou, China

**Abstract** Dual-energy computed tomography (DECT) can identify the material properties with its excellent material quantitative analysis ability. However, the application of DECT is restricted by the problems of inaccuracy of energy spectrum estimation, non-linearity and inconsistency of imaging geometry, which will lead to the degradation of material distribution images. Hence, deep learning (DL)-based methods have become the state-of-the-art technique in DECT rely on its excellent feature recognition performance in the case of few spectrum prior. In this work, we propose an asymmetrical Dual-Cycle adversarial network (ADCNet) for both material decomposition and synthesis of dual-energy CT images, which has certain advantages in spectral CT multi-task parallel, improvement of image quality and radiation dose reduction. The experimental results show that the cycle framework achieves the adversarial learning of dual networks, and promotes the quality of generated images by introducing multiple mechanisms. Compared with the traditional DL-based methods, the proposed method has outstanding qualitative and quantitative indicators.

## 1 Introduction

Dual-energy computed tomography (DECT) utilizes the potential information in energy spectrum to achieve quantitative analysis of substances, which is highly promising for clinical applications. Although DECT has certain preponderance over conventional CT, a tiny disturbance in spectrum-imaging would bring out an inestimable impact on the material decomposition. Meanwhile, the radiation accumulation of DECT scanning is another question worthy of attention in clinical application. Exploring methods of reducing the radiation dose of DECT is also a key issue in the research field.

In order to effectively extract the intrinsic feature of the spectral CT images and improve the quality of decomposed images, deep learning (DL)-based methods have become the state-of-the-art technique in DECT. In 2019, Zhang *et al* [1] exploited the characteristics of DECT to optimize the traditional U-Net architecture and build a dual U-Net with butterfly structure. It shows that the dual U-Net architecture with information interaction has presented great potential in DECT material decomposition. To further improve the network performance, Shi *et al* [2] adopted the General Adversary Network (GAN) [3] to the dual U-Net structure and compared different GAN variants， creating a network called interactive Wasserstein GAN (DIWGAN). Based on this method, the effect of material decomposition has been further promoted. In 2022, Zhou *et al* [4] analysed the requirements of tradeoffs between the level of radiation and the quality of spectral CT images, and proposed a cycle adversarial network with multi-strategy to synthesize high-energy images from low- energy images. The bidirectional loop structure based on CycleGAN [5] have achieved promising results in the synthesis task of spectral CT images.

In this paper, we combine the two tasks: base material decomposition and synthesis of dual-energy CT images with an asymmetrical dual-cycle adversarial network (denoted ADCNet). In practical application, our method can use conventional CT image to synthetize dual-energy images to further reduce the radiation dose of CT scanning, and achieve accurate material decomposition at the same time, which is conducive to shorten the time of clinical diagnosis and promotes the practical application of DECT.

## 2 Materials and Methods

### 2.1 Dual-Cycle Adversarial Framework

Here, we first describe the composition of the dual-cycle generation adversarial network framework. To realize one-time conversion of multi-task in a integral framework, we design a double-entry and double-out architecture based on CycleGAN. The material decomposition module (Module I) and the image synthesis module (Module II) have been contained in the circle framework, as illustrated in Fig. 1.

Different from the traditional CycleGAN, the proposed ADCNet has been improved and innovated in network architecture, loss function and training methods.

### 2.2 Loss Function

The input high- and low-energy images $X_H^r$ , $X_L^r$ are generated as truth data to provide dual-energy spectrum information for the material decomposition. On the other side, we have prepared bone and tissue images separated from the head data of patients, which are presented as $X_B^r$, $X_S^r$. To distinguish the real material images from the fake style images converted by generator $G$, and the real high- and low-energy images from the fake style images converted by the other generator $F$ , we define the follwing loss functions:

$$\ell_{GAN}^{1}\left(G, D_{B,S}, X_{H,L}^{r}, X_{B,S}^{r}\right)=E_{x_{B,S}}\left[\log D_{B,S}(x_{B,S})\right]+E_{x_{H,L}}\left[\log\left(1-D_{B,S}\left(G(x_{H,L})\right)\right)\right] \quad (1)$$

$$\ell_{GAN}^{2}\left(F, D_{H,L}, X_{B,S}^{r}, X_{H,L}^{r}\right)=E_{x_{H,L}}\left[\log D_{H,L}(x_{H,L})\right]+E_{x_{B,S}}\left[\log\left(1-D_{H,L}\left(F(x_{B,S})\right)\right)\right] \quad (2)$$

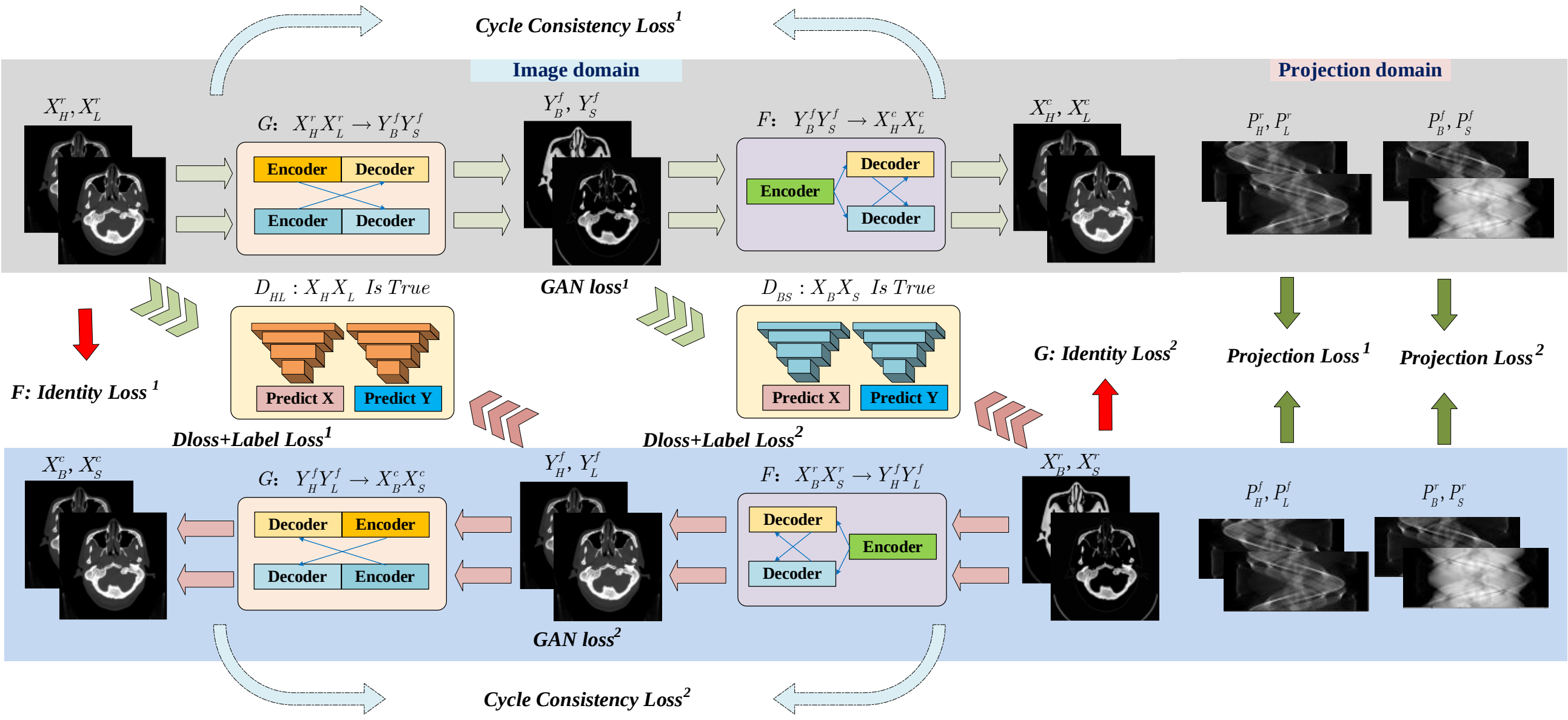

**Figure 1:** Dual-Cycle Adversarial Framework architecture. *1st line* : Module I of material decomposition. *2nd line*: Module II of synthesis of dual-energy CT images. From left to right are the loss functions in the image domain and projection domain.

$$\ell_{Cycle}(G,F)=E_{x_{H,L}}\left(\left\|F\left(G(x_{H,L})\right)-x_{H,L}\right\|_1\right)+E_{x_{B,S}}\left(\left\|G\left(F(x_{B,S})\right)-x_{B,S}\right\|_1\right) \quad (3)$$

$$\ell_{Idt}(G,F)=E_{x_{B,S}}\left[\left\|G(x_{B,S})-x_{B,S}\right\|_1\right]+E_{x_{H,L}}\left[\left\|F(x_{H,L})-x_{H,L}\right\|_1\right] \quad (4)$$

$$\ell_{edge}(G,F)=E_{x_{B,S}}\left[\left\|\nabla F(x_{B,S})-\nabla x_{B,S}\right\|_2^2\right]+E_{x_{H,L}}\left[\left\|\nabla G(x_{H,L})-\nabla x_{H,L}\right\|_2^2\right] \quad (5)$$

where $\ell_{GAN}^1$ and $\ell_{GAN}^2$ represent the loss of GAN at $D_{BS}$ and $D_{HL}$ respectively; $x_{H,L}$ and $x_{B,S}$ represent 2 spectral CT images and base materials. measuring the distance between the generated fake images and the real images and giving judgement. $\ell_{Cycle}$ aims to make the output images $X_H^c$ and $X_L^c$ (or $X_B^c, X_S^c$) consistent with $X_H^r$ and $X_L^r$ (or $X_H^c, X_L^c$) in style. The identity loss $\ell_{Idt}$ is also an indispensible link to maintain grey scale information of the materials. In addition, we introduce edge loss $\ell_{edge}$ to restore the texture and edge features of images. $Y_H^f, Y_L^f, Y_B^f$ and $Y_S^f$ represent the output of generators. Meanwhile, to further improve the quality of images, we try to excavate the projection information of the image from the projection domain as a loss function to constrain. We assume the projection of label images are $P_H^r$, $P_L^r$, $P_B^r$ and $P_S^r$ respectively. The generated images should be as close as possible to the label images. Therefore, the projection loss can be described by

$$\ell_{proj}(G,F)=E_{x_{B,S}}\left[\left\|P\left(F(x_{B,S})\right)-P(x_{B,S})\right\|_1\right]+E_{x_{H,L}}\left[\left\|P\left(G(x_{H,L})\right)-P(x_{H,L})\right\|_1\right] \quad (6)$$

The whole loss function is given by

$$\ell_{SUM}=\lambda_1\ell_{GAN}^1+\lambda_2\ell_{GAN}^2+\lambda_3\ell_{Cycle}+\lambda_4\ell_{Idt}+\lambda_5\ell_{edge}+\lambda_6\ell_{proj} \quad (7)$$

where $\{\lambda_i,\ i=1,2,\ldots,6\}$ represents the balance parameters of different loss functions.

## 2.3 ADCNet Architecture

In this paper, we proposed ADCNet which integrates multiple mechanisms to achieve the decomposition of bone and tissue from high- and low-energy CT images. The double-entry and double-out network with information interaction between the two paths can effectively acquire the internal characteristics of base materials from spectral CT images. On the contrast, to achieve the task of synthesis of DECT images, we design a single-entry and double-out network architecture to generate high- and low- energy images from fused images. Fig. 1 shows the structure of generator $G$, generator $F$ and discriminator $D$. Moreover, we also introduce DANet module [6] which combines spatial attention and channel attention to enlarge the receptive field and restore the texture details of bones and tissues. In particular, we add butterfly architecture [1] to the deepest downsampling of ADCNet in generator $G$, in order to collect high-level abstract semantic information of high- and low-energy spectral images. In generator $F$, we also introduce multi-information interaction mechanism, and merge the two kinds of materials before entering the network to assist effective feature extraction.

## 2.4 Data Preparation & Parameter Setting

The data set comes from cranial cavity slice images of 7 patients with size 512 by 512. In experiments, we prepared 1505 actual data of bone and tissue and added 120 kVp and 80 kVp spectrum to the original slice images as our simulation dual-energy data. In the training, we selected 1400 pairs of high- and low-energy images as the training dataset to train the model, and 105 pairs of images as the test dataset to validate the network performance. As for parameters of ADCNet, each downsampling path of $G$ and $F$ includes 7 convolutional layers. the number of filters is 128, 256, 512, 512, 512 and 512, respectively. Two paths are connected by 2 cascaded residual blocks. In the process of training, the initial learning rate for adam was set to 0.0002 (momentum term: 0.5, $\beta_1$ : 0.5 , $\beta_2$ : 0.999). To ensure that the training process did not produce over fitting, we set the maximum epoch to 15. The training duration of the proposed model was about 20 hours.

## 3 Results

To evaluate the performance of different modules in ADCNet, we design ablation experiments on 20 typical test data. The experiment is divided into 2 independent parts. Part I verifies the performance of information interaction, residual block, butterfly structure and DANet attention module on the original Dual-CycleGAN. As is shown in Table 1, the PSNR of decomposed Bone and Tissue increased by 5.06dB and 6.67dB respectively after adding all of the modules. We can see that ADCNet with four hybrid modules has better quantitative indicators inTo evaluate the performance of different modules in the feature extraction of bone and tissue texture. Part II verifies the improvement effect of generator *F* using fused inputs. As is shown in Fig. 2, the symmetrical dual-cycle network with 2 same generators (SDCNet) is compared with the ADCNet of 2 fused inputs in spatial dimension (ADCNet-SF) and in channel dimension (ADCNet-CF). Obviously, these two forms of ADCNet have shorter training time than SDCNet, due to the fused architecture sharing the parameters in downsampling path (the number of parameters of SDCNet and ADCNet is 387M and 329M, respectively). Especially, the quantitative indicators of PSNR and RMSE also illustrate that the ADCNet-CF has better performance far beyond than the other two networks in the synthesis of dual-energy CT images, due to the high similarity of fused material images with spectral CT images.
In addition, we select a group of head data to evaluate the qualitative performance of different networks. Fig. 3 demonstrates the comparison of four state-of-the-art networks on material decomposition including FCN, Pix2Pix [7], DIWGAN and the proposed ADCNet. The ROI (Region of Interest) of bone shows the structure of the cochlea and frontal lobe, and the ROI of tissue shows the structure of the lateral ventricle and cerebellar. Fig.4 shows the residual images of the results with the label images. Compared with the conventional FCN, the SSIM and RMSE of bone images are greatly improved by 0.3246 and 0.0188, and tissue images by 0.072 and 0.0508, which means the proposed ADCNet has marvel performance in material decomposition. Although we find that there still exists salt and pepper noise in the material images, the later operation of median filtering can also help us to further improve the quality of images. At the same time, ADCNet also achieves the synthesis of dual-energy CT images owing to its special dual-cycle architecture, even better beyond the SDCNet and ADCNet-SF. Fig.5 shows the pixel values of a certain section of 120kVp and 80kVp synthetic spectral CT images. The expected result is closer to the ground truth from the structure of the dual-energy images. In Fig.3, we also discussed the accuracy of synthetic dual-energy images from a quantitative perspective, which reached 45.85dB and 45.93dB in PSNR, and over 0.99 in SSIM. To sum up, ADCNet has achieved good results in both tasks of base material decomposition and synthesis of dual-energy CT images.

**Tabel 1:** Ablation experiments of Part I. From left to right are information interaction, residual block, butterfly structure and DANet attention module.

| ID | Interac | Res | Butterfly | DANet | PSNR-Bone | PSNR-Tissue |
|---|---|---|---|---|---|---|
| 1 | √ | | | | 37.36 | 29.17 |
| 2 | √ | √ | | | 39.58 | 31.58 |
| 3 | √ | √ | √ | | 41.10 | 32.92 |
| 4 | √ | √ | | √ | 41.43 | 34.30 |
| 5 | √ | √ | √ | √ | 42.42 | 35.84 |

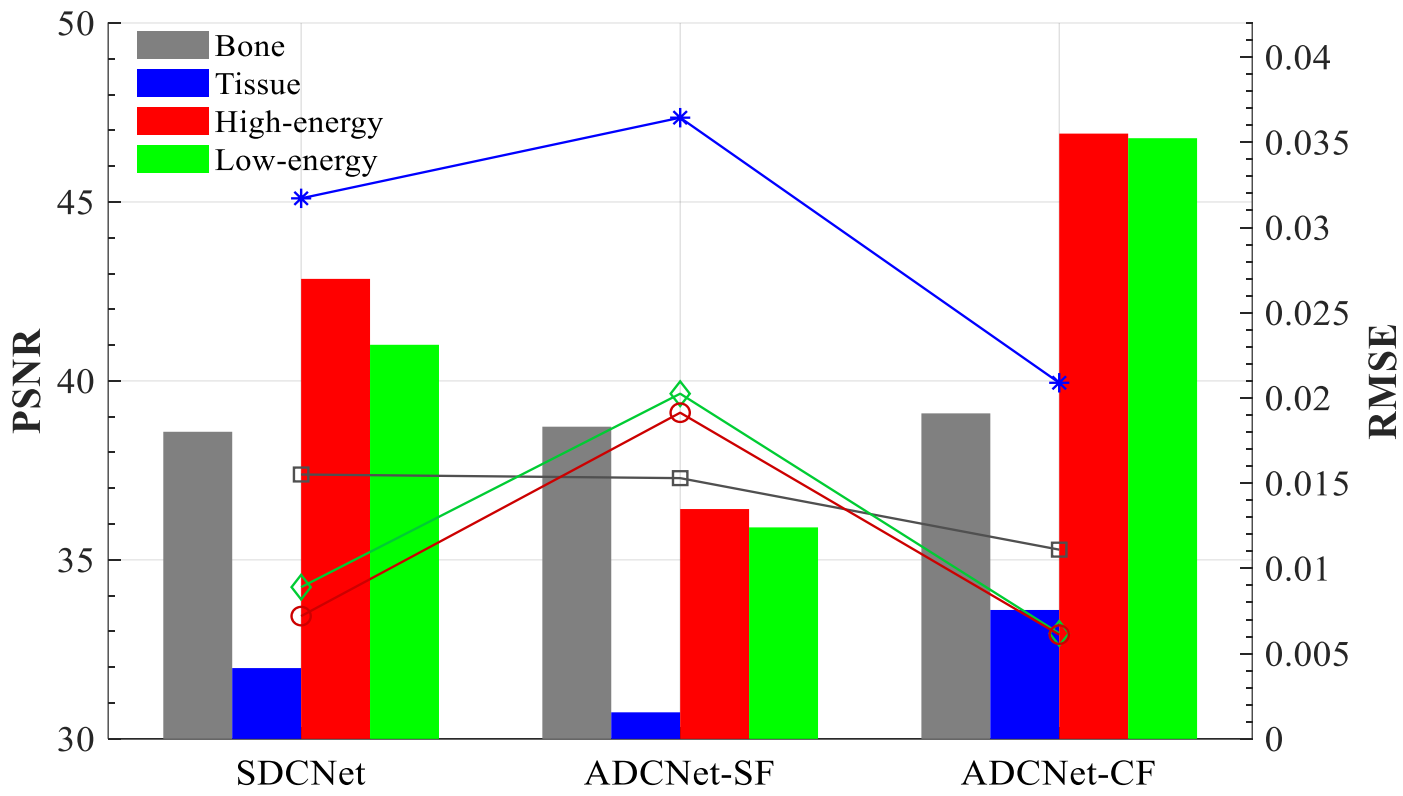

**Figure 2:** Ablation experiments of Part II. (1)SDCNet: the symmetrical dual-cycle network with 2 same generators G (2)ADCNet-SF: the asymmetrical dual-cycle network of 2 fused inputs in spatial dimension. (3)ADCNet-CF: the asymmetrical dual-cycle network of 2 fused inputs in channel dimension. *Note:* Bar chart represents PSNR, and line chart represents RMSE.

## 4 Conclusion

In this work, we proposed a new dual-cycle network architecture (ADCNet), which can achieve dual tasks in material decomposition and synthesis of dual-energy CT images. Compared with the other four typicall networks in terms of quantitative and qualitative indicators, our method demonstrated outstanding ability in material identification and accurate generation of energy spectrum images under the none-spectral prior condition, which can assist multi-functional integrated spectral CT imaging more efficiently and accurately. Additionally, Our method can achieve lower dose as the DECT images can be totally synthetic. In the future, we will further carry out relevant experiments based on the actual problems in medical diagnosis.

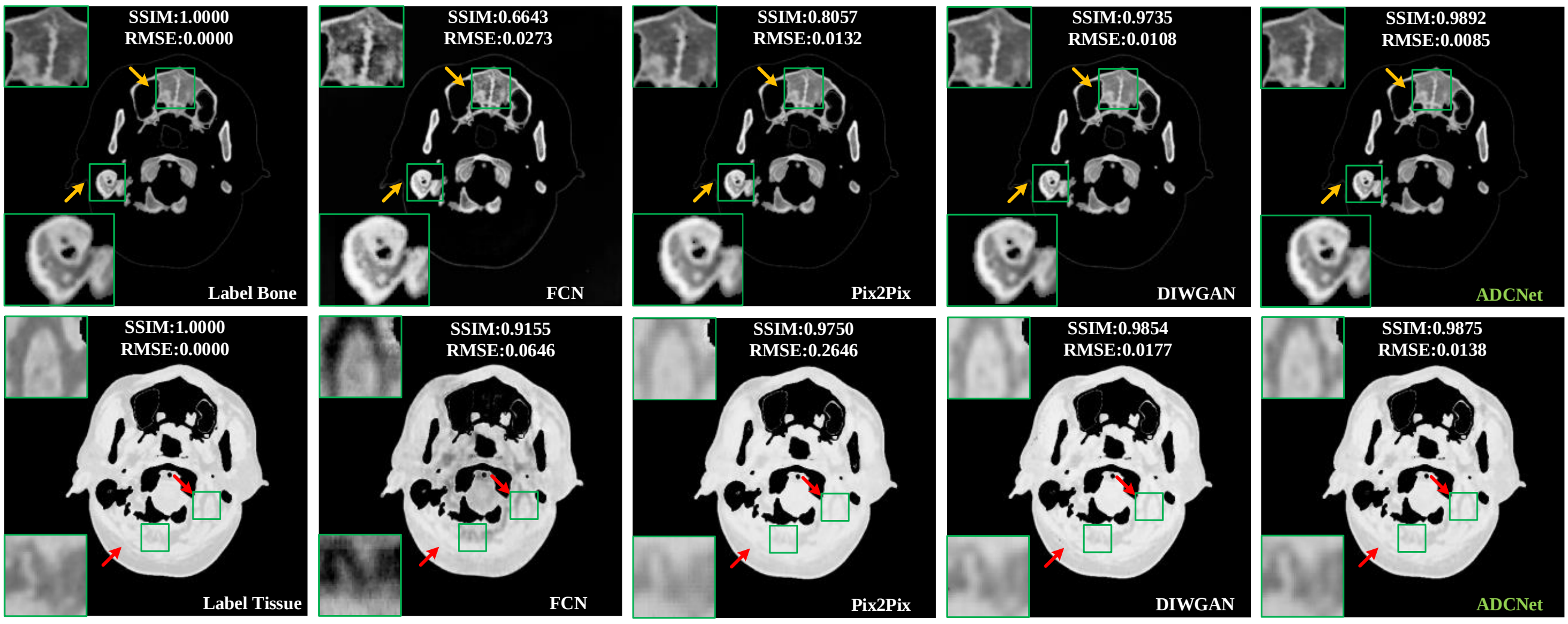

**Figure 3:** Networks on material decomposition. From left to right are FCN, Pix2Pix, DIWGAN and the proposed ADCNet. From top to bottom are bone and tissue. The 1st line represents the decomposed bone images, and the 2nd line represents the decomposed tissue images.The SSIM and RMSE values are written at the top.

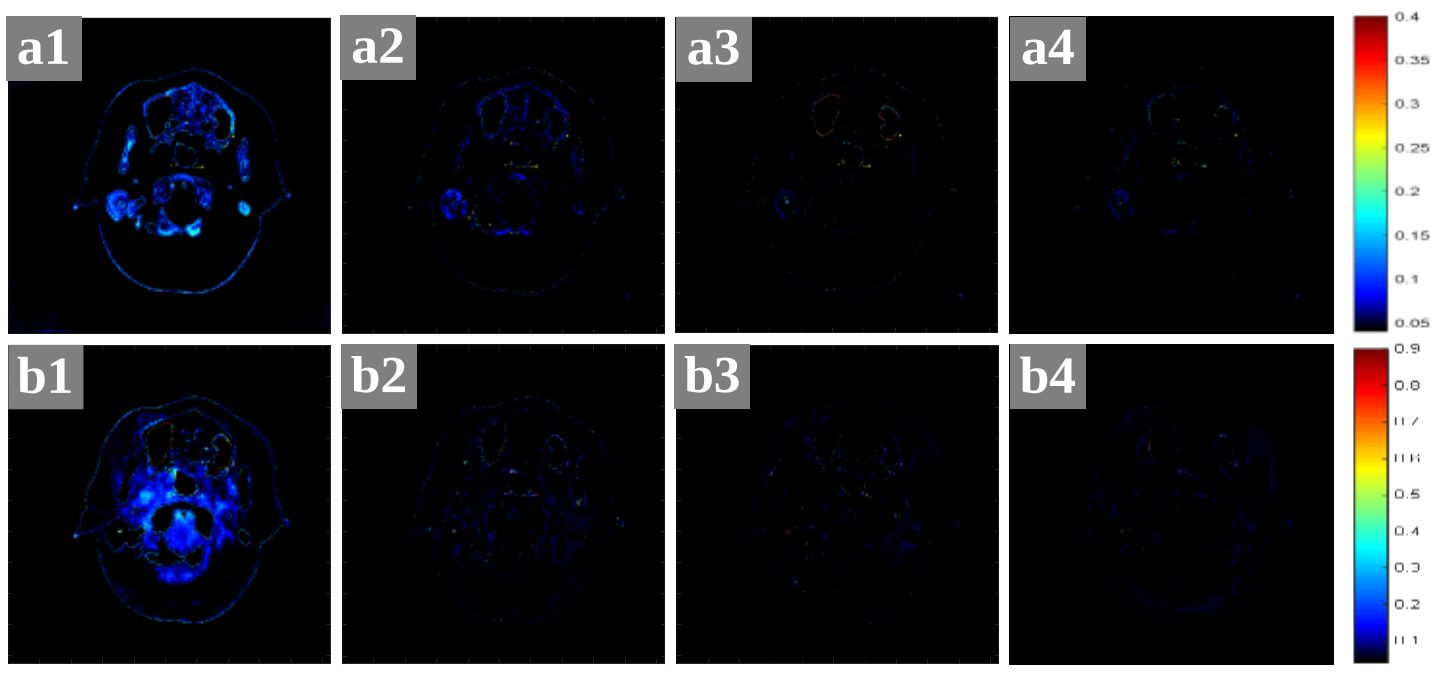

**Figure 4:** The residual images of the results of FCN, Pix2Pix, DIWGAN and ADCNet with label images. We use absolute error $\Re = \|y - y^*\|$ to express the residual images. (a1)-(a4): The residual images of bone. (b1)-(b4): The residual images of Tissue. *Note*: The range of colorbox is [0.05, 0.4] g/mm$^3$ and [0.1, 0.9] g/mm$^3$, respectively.

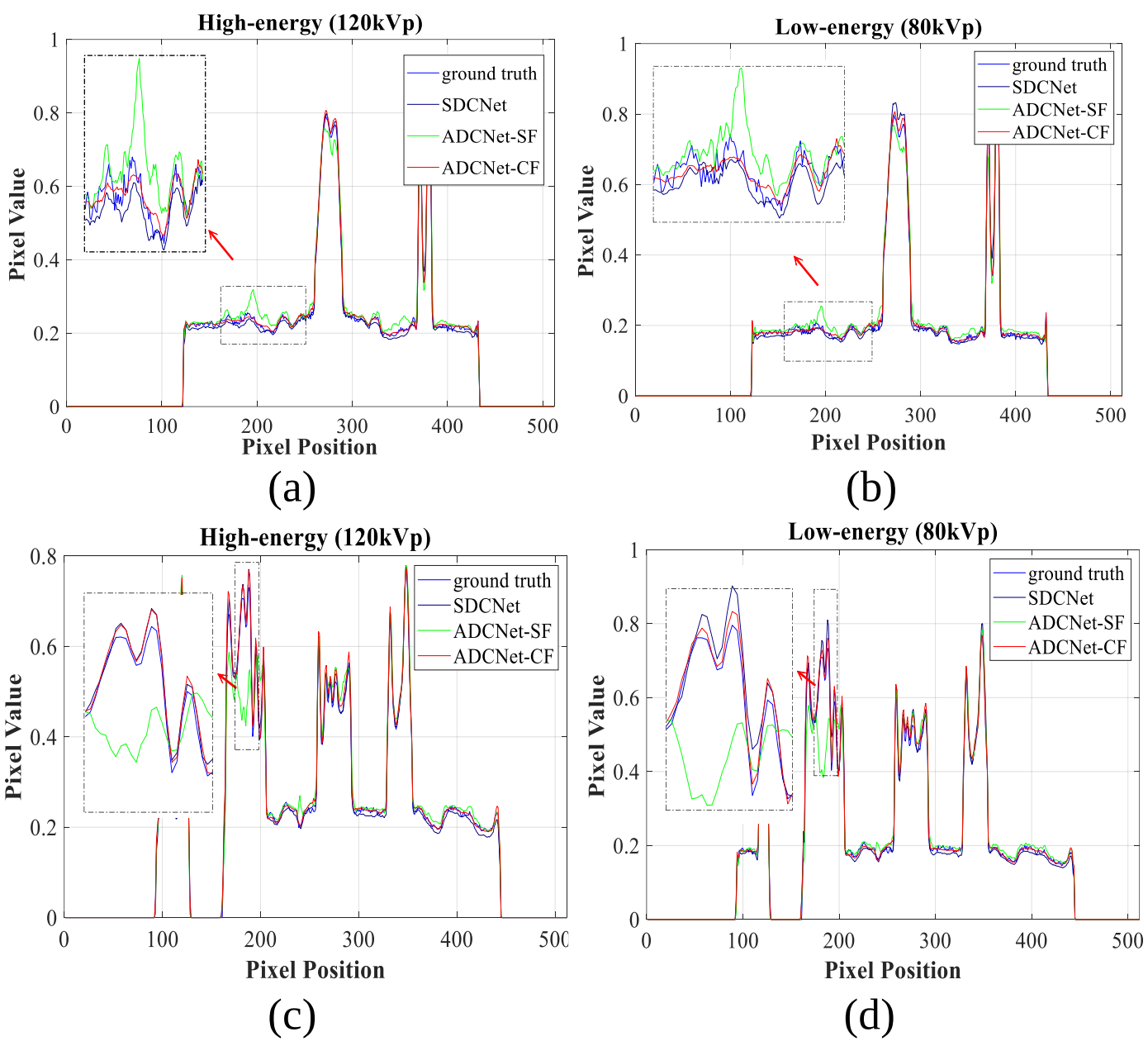

**Figure 5:** Profile plots of synthetic dual-energy CT images with different networks including SDCNet, ADCNet-SF and ADCNet-CF. (a), (b): Horizontal profile plots of 120kVp and 80kVp synthetic images. (c), (d): Vertical profile plots of 120kVp and 80kVp synthetic images.

## 5 Acknowledgments

This work was supported by the National Key Research and Development Project of China (Grant No. 2020YFC1522002), the National Natural Science Foundation of China (Grant No. 62101596), the National Natural Science Foundation of China (Grant No. 62201616) and the China Postdoctoral Science Foundation (Grant No. 2019M663996).

# Limited-angle CT imaging with a non-uniform angular sampling technique

Yinghui Zhang[1, 2], Hongwei Li[1], Xing Zhao[1], and Ke Chen[2, *]

[1]School of Mathematical Sciences, Capital Normal University, Beijing, China
[2]Department of Mathematical Sciences, University of Liverpool, Liverpool, United Kingdom

**Abstract** Non-uniform angular sampling for CT imaging has the ability to select the most informative projection angles such that better images could be reconstructed. In this paper, a non-uniform angular sampling technique is proposed for limited-angle CT reconstruction which allocates more projections close to the start and end of the angular range. Numerical experiments show that, compared with conventional uniform sampling, the proposed non-uniform sampling technique achieves better reconstruction quality consistently.

## 1 Introduction

Computed tomography (CT) is an imaging technique to discover the inner structure of scanned objects using X-ray projections. Conventionally, these projections are acquired equiangularly. However, uniform sampling cannot ensure the projections with the highest information content are acquired, especially when the number of projections is limited. It is known that the choice of the projection angles can have a crucial influence on the reconstruction quality [1, 2].

To achieve the "most informative" projection angle set, various non-uniform angular sampling methods have been proposed in the past decades. Based on the information obtained in projections that have already been measured on surrogate solutions, Batenburg et al. [3, 4] selected a new angle by maximizing the information gained by adding this projection to the set of measurements. Similarly, by applying sequential feature selection methods based on a blueprint image, Peter et al. [5] proposed several offline projection selection algorithms to sequentially find projection angles with high information content. These two methods suffer from severe computational burdens, since they have to run the reconstruction algorithms many times for determining the next best angle. Placidi et al. [6] proposed an adaptive method for selecting the projections based on the "entropy" principle with initial four projections at angles 0°, 45°, 90°, 135°. If the image is smooth or has internal symmetries, this adaptive scheme works well. Based on the spectral richness of the acquired projections and the amount of new information added by successive projections, Haque et al [7] developed two approaches that adjust the step size to adaptively select the projection angles.

Inspired by the flexibility of reinforcement learning setting, Shen et al. [8] used modern reinforcement learning methods to select projection angles and specify their doses for personalized scanning, where the CT scanning process is formulated as a Markov Decision Process. Then, a unified deep learning framework was proposed in [9] which can not only select the important projection angles but also learn a high performance reconstruction network, i.e. a pair of sampler and reconstructor is learned.

In certain CT applications, due to the restrictions on the scanning condition or the geometrical shapes of scanning objects, the projection data could be only acquired in a limited angular range which leads to the challenging limited-angle reconstruction problem. Applying classical reconstruction algorithms such as filtered backprojection (FBP) and (simultaneous) algebraic reconstruction technique ((S)ART) for limited-angle reconstruction often causes severe streak and blurring artifacts in the reconstructed images. The limited-angle reconstruction problem has been extensively studied for decades, and many methods have been proposed in the literature by utilizing priors existing in image domain or projection domain. These priors are represented either conventionally in terms of wavelet [10], total variation [11, 12], K-SVD [13] etc. or by deep learning based neural networks [14, 15].

As stated earlier, the angular sampling pattern could have a strong influence on the reconstruction quality. However, much less researcher's efforts have been put on designing the angular sampling pattern for limited-angle reconstruction. It turns out that direct translation of the non-uniform sampling methods for sparse-view CT into limited-angle CT reconstruction seems inappropriate. In [16], Zheng et al. analysed the impact of angular sampling interval on image reconstruction accuracy from limited angular range data with a scan configuration containing two orthogonal arcs. However, the non-uniform sampling strategy about how to determine non-uniform intervals is not discussed.

In this paper, we propose a non-uniform angular sampling strategy for limited-angle CT scanning and study its impact on the reconstruction quality. The remainder of this paper is organized as follows. In Section 2, a brief introduction to the directional total variation (DTV) algorithm for limited-angle reconstruction is provided, then the non-uniform sampling strategy is described. Section 3 presents experimental results on simulated phantoms and conclusions and remarks are given in Section 4.

## 2 Materials and Methods

In our tests comparing the non-uniform sampling against the uniformed ones, the DTV method is utilized as the reconstruction algorithm. So, in this section, the DTV model shall be briefly introduced, and a non-uniform angular sampling strategy is then proposed.

### 2.1 The DTV method for limited-angle CT reconstruction

Recently, an method named DTV is proposed in [12], which shows very promising reconstructions, especially for very small scanning angular ranges. The DTV model reads

$$\overrightarrow{u}^\star = \min_u \frac{1}{2}\|A\overrightarrow{u} - \overrightarrow{b}\|_2^2$$
$$s.t.\, \|\nabla_x \overrightarrow{u}\|_1 \leq t_x,\ \|\nabla_y \overrightarrow{u}\|_1 \leq t_y,\ \overrightarrow{u} \geq 0,$$

where $A \in \mathbb{R}^{I\times J}$ is the system matrix, $\overrightarrow{b}$ is a vector of length $I = V \times D$ which represents the acquired projection data, and $V$ and $D$ denote the number of projection views and the number of detector cells, respectively. The symbols $\nabla_x$ and $\nabla_y$ correspond to the discrete $x$-direction and $y$-direction gradient operators, respectively, and $t_x$ and $t_y$ control the allowed total variations along the $x$-direction and $y$-direction, respectively. In [17], the Chambolle-Pock (CP) algorithm is used to solve this convex problem optimization model with guaranteed convergence.

### 2.2 The non-uniform angular sampling strategy

Since uniform sampling scheme cannot ensure the acquired projections with highest information content, especially when dealing with limited projection data set, a non-uniform projection angle selection strategy based on the characteristics of limited-angle CT imaging problem is proposed to achieve a more informative projection angle set. For the image areas located near the rotation center, the number of received photons is relatively large, and for the projections located close to the middle of the scanning angular range, their information can be complemented by the projections on both sides, which could be sparsely sampled. However, for image areas away from the rotation center, the number of received photons is smaller, and for the projections far away from the middle of the scanning range, their information can only be complemented by one side, requiring dense sampling. Based on such a hypothesis, we employ a projection angle sampling density that decreases as it moves away from the start and end angles of limited-angle CT scan, i.e. sample more projections near the start and end of the angular range, while fewer projections are sampled near the middle of the angular range, to make the most of the limited projections. The proposed non-uniform projection angle distribution is illustrated in Fig. 1, where Fig.1 (a) illustrates the uniform angular sampling scanning while Fig. 1 (b) shows the non-uniform one.

Following the above principle, we conducted a preliminary attempt to design the non-uniform sampling angle distribution, and there are two approaches: 1. varying $\Delta\theta$ uniform sampling in different scanning range, i.e. a smaller $\Delta\theta_1$ for limited angular range of $[\theta_1, \theta_N]$ and $[\theta_{V-1-N}, \theta_V]$ and a larger $\Delta\theta_2$ for limited angular range $[\theta_{N+1}, \theta_{V-2-N}]$; 2. taking $\Delta\theta_i$ value according to geometrical series, for example, when $V$ is even,

$$\theta_i = \begin{cases} \theta_{i-1} + r^{i-2} * h & i = 2, \cdots, \frac{V}{2} \\ \theta_{i+1} - r^{V-1-i} * h & i = V-1, \cdots, \frac{V}{2}+1 \end{cases},$$

where $h = \frac{V_{mid} - V_1}{1 - r^{\frac{V}{2}-1}} * (1-r)$, $V_{mid}$ is the median of $1, 2, \cdots, V$ and $r$ is the common ratio of geometrical series. Both of them performed better than uniform distribution projection angles. To make $\Delta\theta$ vary smoothly and achieve better results, in this paper, the non-uniform angular sampling interval $\Delta\theta_i, i = 1, 2, \cdots, V-2$ are sampled from a certain normal distribution $N(0, \sigma^2)$, which are generated as follows

$$\Delta\theta_i = \frac{\int_{x_i}^{x_{i+1}} \frac{1}{\sqrt{2\pi}\sigma} exp\left(-\frac{x^2}{2\sigma^2}\right) dx}{\int_{x_1}^{x_{V-2}} \frac{1}{\sqrt{2\pi}\sigma} exp\left(-\frac{x^2}{2\sigma^2}\right) dx} * (\theta_V - \theta_1),$$

where $x$ is a vector of size $V-2$, which is $[-a, \cdots, -a + \frac{2a}{V-3} * i, \cdots, a]$, a little misuse of symbol $[\cdot]$. Both $a$ and $\sigma$ control the sampling density. Either a smaller value of $a$ or a larger value of $\sigma$ leads to a denser sampling of projection angle density close to the scanning boundary. The values of $a$ and $\sigma$ are set to 3 and 1.5 in the experiments, respectively.

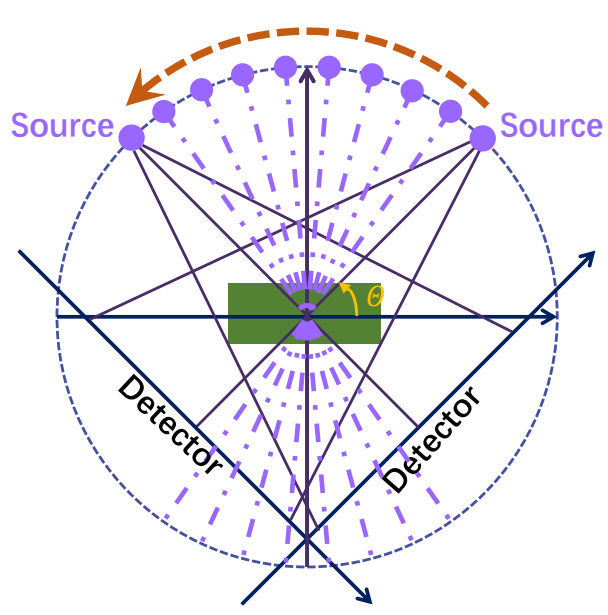

**(a)** Conventional uniform sampling

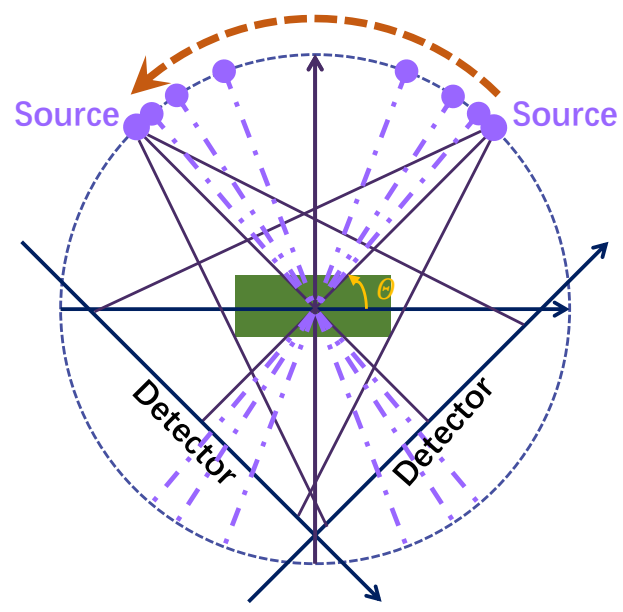

**(b)** Nonuniform sampling

**Figure 1:** The configuration of non-uniform sampling strategy for the limited-angle CT imaging ($\theta \in [45°, 135°]$).

## 3 Experiments and results

In this section, first, we verify that the choice of projection angles has a significant impact on the reconstructed image quality from limited-angle CT data. Then, to validate the effectiveness of proposed non-uniform angular sampling strategy on limited-angle CT reconstruction, numerical experiments with different scan ranges are carried out, that is, $[30°, 150°]$ and $[45°, 135°]$ and $[60°, 120°]$. Both noise-free

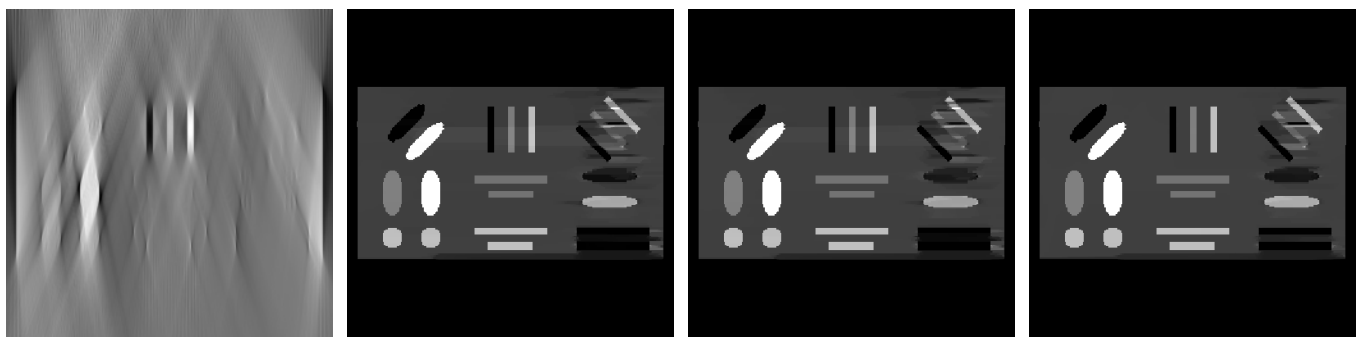

**(a)** FBP **(b)** Uniform sampling **(c)** Added angle pair $\{75^\circ + 0.05^\circ, 105^\circ - 0.05^\circ\}$ **(d)** Added angle pair $\{75^\circ + 0.4^\circ, 105^\circ - 0.4^\circ\}$ and noisy projection data are tested. For noisy projection data, Poisson noise with incidence intensity $I_0 = 1 \times 10^7$ is added to the projection data. The display window is set to [0.1, 0.5].

**Figure 2:** The reconstructions from projections of scanning angular range ($[75^\circ, 105^\circ]$) added with different projection pairs.

### 3.1 Angle-dependent reconstruction quality test

In this subsection, we will verify that the limited-angle CT image reconstruction accuracy depends on the selection of projection angles. Take noise-free limited-angle projection data from 30 degree range for example, 20 different pairs of symmetric projection angles are added to the original 30 uniform distributed projections, which are $[75^\circ + 0.05^\circ * i, 105^\circ - 0.05^\circ * i], i = 0, 1, 2, \cdots, 19$. $i = 0$ means the uniform sampling angles. The plots of PSNR and SSIM of the results reconstructed from 20 projection angle sets of size 32 of 30 degree data are shown in Fig. 3. It could be clearly notified that different 32 projection angle sets from the same angular range produce reconstructions with varying PSNR and SSIM values, i. e. the reconstruction results are projection angle-dependent. The best and worst ones of 20 reconstructions are shown in Fig. 2c and Fig. 2d, respectively. And Table 1 lists their corresponding PSNR and SSIM values.

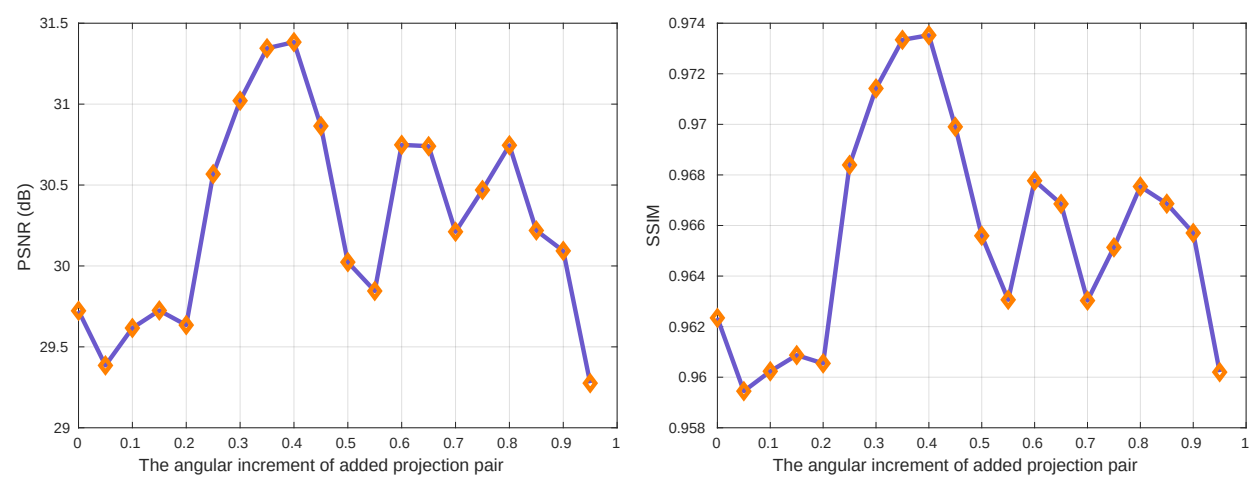

**(a)** PSNR **(b)** SSIM

**Figure 3:** The PSNR and SSIM values of different reconstructions from 30 degree angular range data ($[75^\circ, 105^\circ]$) added with 20 different projection pairs, respectively.

**Table 1:** PSNR (dB) and SSIM for reconstructions in Fig. 2.

| Added angle pair | PSNR | SSIM |
|---|---|---|
| ∅ | 29.7229 | 0.9623 |
| $\{75^\circ + 0.05^\circ, 105^\circ - 0.05^\circ\}$ | 29.3852 | 0.95945 |
| $\{75^\circ + 0.4^\circ, 105^\circ - 0.4^\circ\}$ | 31.3831 | 0.9735 |

### 3.2 Non-uniform angular sampling test

As results of Fig. 4 shown for projections of scanning angular range $[30^\circ, 150^\circ]$ and $[45^\circ, 135^\circ]$, we can clearly observe that more accurate image is achieved from non-uniform angular sampling projections, when using same number of projection angles. Streak artifacts could be recognized in reconstructed images from projection data of uniform distributing angles, as indicated by the red arrows in the third column in Fig. 4. And more steak artifacts could be identified on the left and right sides of the results than in the middle. After redistributing the angles non-uniformly with more projections near the start and end of the angular range, streak artifacts are reduced on the left and right sides of the reconstructions, while edges in the middle are well preserved.
Similarly, as shown in the last row of Fig. 4, streak and blurring artifacts could be clearly seen in the reconstruction from projection data with 110 evenly distributed angles in 60 degree angle, especially on the top right part indicated by the red frame. When reconstructing image with 110 non-uniform distributed angles selected by the proposed strategy, image distortions on the top right-hand corner part are greatly alleviated. This is because that the missing information is supplemented by densely distributed projections located near the scanning boundary. Furthermore, to produce similar high quality reconstructed images, 191 uniform distributed angles are needed for projections of scanning angular range $[60^\circ, 120^\circ]$. Table 2 lists the PSNR and SSIM values for the reconstruction results by FBP, DTV with uniform sampling projections and non-uniform sampling ones, respectively.

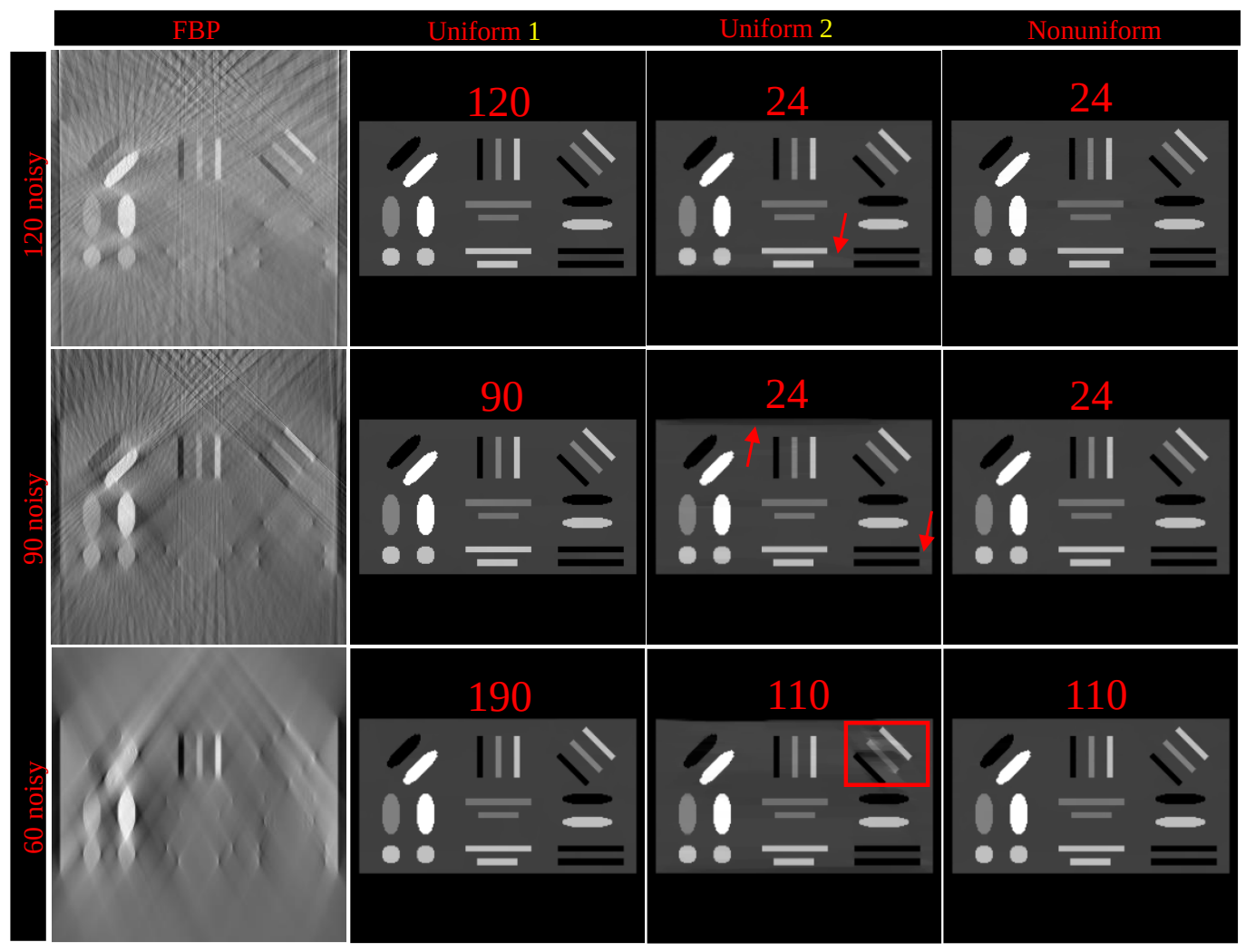

**Figure 4:** Reconstructions from projections of scanning angular ranges $[30^\circ, 150^\circ]$, $[45^\circ, 135^\circ]$ and $[60^\circ, 120^\circ]$, respectively.

## 4 Conclusion

To achieve the most informative projection angle set for limited-angle CT reconstruction, a non-uniform angular sampling strategy is proposed, which is based on the characteristics of the limited-angle CT scanning. The projection angle distribution follows the rule that more projection angles are acquired close to the start and end of the angular range. Experimental results show that, the proposed non-uniform sampling strategy could produce better reconstructions compared with the classical uniform sampling method.

**Table 2:** PSNR (dB) and SSIM for the reconstructions in Fig. 4.

| Angular range | Index | FBP | Uniform 1 | Uniform 2 | Non-uniform |
|---|---|---|---|---|---|
| $[30°, 150°]$ | PSNR | 14.1241 | 56.0174 | 45.5145 | 47.7178 |
| | SSIM | 0.2396 | 0.9999 | 0.9986 | 0.9988 |
| $[45°, 135°]$ | PSNR | 12.8975 | 56.0154 | 35.4262 | 56.0094 |
| | SSIM | 0.2199 | 0.9999 | 0.9890 | 0.9998 |
| $[60°, 120°]$ | PSNR | 11.5596 | 56.0090 | 32.4521 | 56.0126 |
| | SSIM | 0.2369 | 0.9998 | 0.9764 | 0.9998 |

# Preliminary study of image reconstruction from limited-angular-range data in spectral-spatial electron paramagnetic resonance imaging

Zheng Zhang[1], Buxin Chen[1], Dan Xia[1], Emil Y. Sidky[1], Boris Epel[2], Howard Halpern[2], and Xiaochuan Pan[1,2]

[1]Department of Radiology, The University of Chicago, Chicago, IL, USA
[2]Department of Radiation & Cellular Oncology, The University of Chicago, Chicago, IL, USA

**Abstract** Continuous-wave (CW) electron paramagnetic resonance imaging (EPRI) provides information about both the spatial distribution and spectral shape of unpaired electrons. Its signal-to-noise ratio can be enhanced by employing a Zeeman-modulation (ZM) scheme. Limited angular range (LAR) scans can decrease the strength of the magnetic field gradient or scanning time in CW-ZM EPRI, but they often lead to artifacts or biases in images reconstructed using conventional algorithms, such as filtered back projection (FBP). In this study, an optimization-based algorithm was developed to accurately reconstruct three-dimensional (3D) spatial-spectral (SS) images directly from LAR data in CW-ZM EPRI. The reconstruction was formulated as an image directional-total-variation (DTV) constrained, data-$\ell_2$-minimization problem, and a corresponding DTV algorithm was devised to solve this problem. The DTV algorithm was applied to simulated data involving various LAR scans, and its performance was evaluated using both visualization and quantitative metrics. The results indicate that SS images can be directly reconstructed from LAR data, exhibiting comparable outcomes to those obtained from standard full-angular-range scans in CW-ZM EPRI. The DTV algorithm holds the potential to enable and optimize CW-ZM EPRI with minimal imaging time and reduced artifacts through the acquisition of data in LAR scans.

## 1 Introduction

Electron paramagnetic resonance imaging (EPRI) provides information about the spatial distribution and spectral properties of unpaired electrons in an object, making it a promising tool for functional imaging, such as measuring tissue oxygen concentration [1, 2]. Commonly used for observing the distribution of unpaired electrons, continuous-wave (CW) EPRI techniques can be enhanced using the Zeeman modulation (ZM) scheme [2, 3] to further improve the signal-to-noise ratio (SNR). However, current CW-ZM EPRI faces several challenges. For example, the finite maximum strength of the magnetic field gradient results in a limited achievable angular range for image reconstruction. Additionally, current CW-ZM EPRI often suffers from long imaging time. One approach for reducing imaging time is collecting data from reduced angular ranges, known as limited-angular range (LAR). However, existing image reconstruction algorithms, such as FBP, require data collected over a full angular range, and their reconstructions from LAR data often contain streak artifacts that can lead to estimation errors of spectral parameters.

The imaging model for CW-ZM EPRI can be formulated as the Radon transform, making its image reconstruction similar to that of computed tomography (CT). Recent advances in iterative reconstruction methods with image directional total variation (DTV) constraints [4, 5] have demonstrated significant improvements in reconstruction accuracy for CT imaging with LAR configurations. Therefore, in this study, we incorporate the DTV constraints into a convex optimization program for CW-ZM EPRI image reconstruction. We employ a primal-dual-based algorithm [6, 7], referred to as the DTV algorithm, to solve the optimization program and perform image reconstruction [4]. We conduct simulated data studies and collect data using various LAR scan schemes. Subsequently, we use the DTV algorithm to reconstruct images from LAR data and evaluate the reconstruction performance through both visualization and quantitative metrics.

## 2 Materials and methods

### 2.1 Imaging model in CW-ZM EPRI

In this study, we consider a three-dimensional (3D) CW-ZM EPRI model with one spectral and two spatial dimensions. The imaging model and algorithms can readily be generalized to 4D CW-ZM EPRI model with one spectral and three spatial dimensions. Specifically, the imaging model for CW-ZM EPRI based upon continuous image and data space can be written as [2]

$$\begin{aligned} g(\xi,\hat{\alpha}) &= \cos\gamma \int d\vec{r}\,\frac{\partial f(\vec{r})}{\partial B}\,\delta(\xi-\vec{r}\cdot\hat{\alpha}) \\ &= \cos\gamma\mathbb{R}\left[\frac{\partial f(\vec{r})}{\partial B}\right], \end{aligned} \tag{1}$$

where $f(\vec{r})$ denotes a continuous 3D spatial-spectral (SS) image as 3D variable $\vec{r}=(x,y,B)^\top$, $(x,y)$ are two orthogonal axes in the spatial space and $B$ the spectral dimension; $\mathbb{R}$ denotes the 3D-Radon transform; $g(\xi,\hat{\alpha})$ is a continuous 3D-data function in a 3D space formed by scalar variable $\xi$ and 3D-unit-vector $\hat{\alpha}$, $\xi$ is the distance of a hyperplane of orientation $\hat{\alpha}$ to the origin in the 3D-image space, and $\hat{\alpha}=(\cos\phi\,\sin\gamma,\sin\phi\,\sin\gamma,\cos\gamma)^\top$; $\gamma$ depicts the angle between the $B$-axis and vector $\hat{\alpha}$, and $\phi$ the angle between $x$-axis and the projection of $\hat{\alpha}$ in the $x$-$y$ plane, as illustrated in Fig. 1.

Based upon the continuous imaging model presented in Eq. (1), we have developed a discrete-to-discrete (DD)-imaging model as

$$\mathbf{g}=\mathscr{C}\mathscr{R}\mathscr{D}_B\mathbf{f}, \tag{2}$$

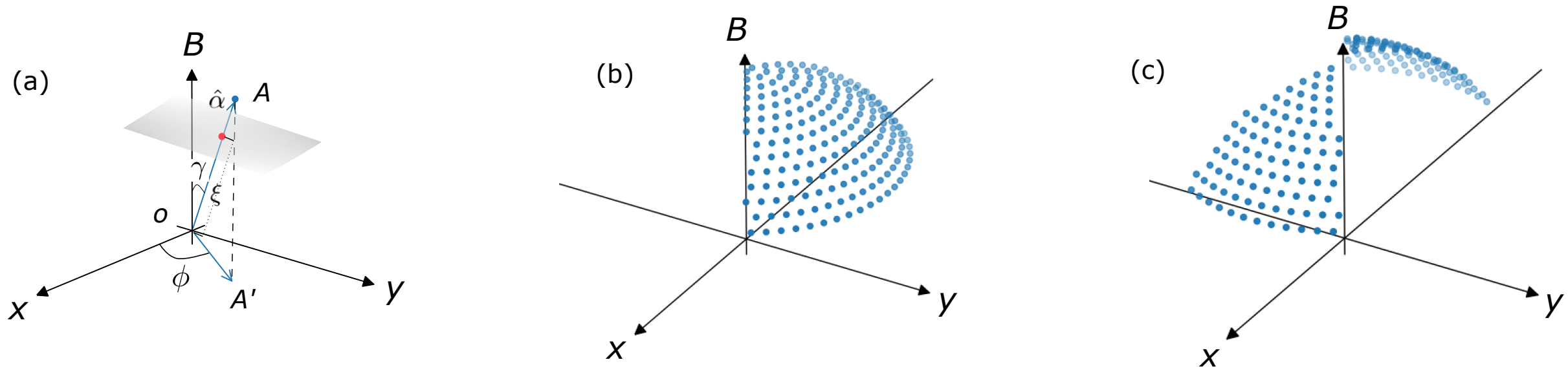

**Figure 1:** (a) The 3D spectral-spatial coordinate system with angles $\gamma$ and $\phi$ defined, along with a unit vector $\hat{\alpha}$; (b) Type I LAR scan scheme with $\gamma \in [-60°, 60°]$ and $\phi \in [30°, 90°]$; and (c) Type II LAR scan scheme with $\gamma \in [-60°, 60°]$ and $\phi \in [-30°, 30°]$.

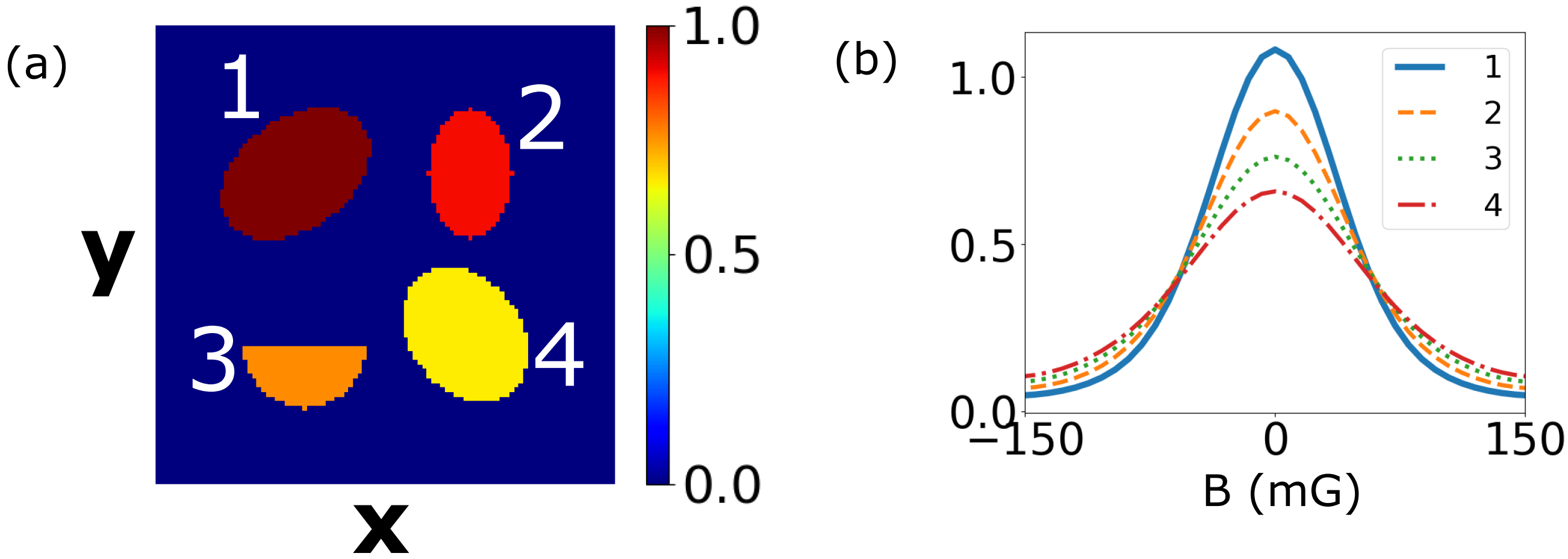

**Figure 2:** (a) Truth SS image in the slice specified by $B = 0$ G, display windows: [0.5, 1.0] arbitrary unit (AU); and (b) the truth spectral profiles at points within tubes 1-4, respectively, in the SS image in (a).

where vector $\mathbf{f}$ of size $I$ denotes the discrete 3D-SS image; $\mathbf{g}$ of size $J$ is the discrete model data; $\mathscr{C}$ a diagonal matrix of size $J \times J$ in which diagonal element $c_{jj} = \cos\gamma_{\hat{\alpha}_j}$; $\gamma_{\hat{\alpha}_j}$ the angle between the $B$-axis and vector $\hat{\alpha}_j$; matrix $\mathscr{R}$ of size $J \times I$ denotes the 3D-Radon transform in a discrete form in which element $r_{ij}$ is chosen to be the intersection hyperarea of hyperplane $j$ with voxel $i$ in the 3D-spatial-spectral space; and $\mathscr{D}_B$ denotes two-point difference along the $B$-axis. In a CW-ZM EPRI scan, $\gamma \in [-\pi/2, \pi/2]$ and $\phi \in [-\pi/2, \pi/2]$. In this study, we examine two LAR scan configurations: type I and type II. Let the values of total angular ranges along $\gamma$ and $\tau$ directions be $\gamma_\tau$ and $\phi_\tau$, which are constrained within $\gamma_\tau < \pi$ and $\phi_\tau \leq \pi$, respectively. Type I configurations result from setting $\gamma$ within $[-\gamma_\tau/2, \gamma_\tau/2]$ and $\phi$ within $[-\phi_\tau/2, \phi_\tau/2]$, while type II configurations are achieved by restricting $\gamma$ to $[-\gamma_\tau/2, \gamma_\tau/2]$ and $\phi$ to $[\pi/2 - \phi_\tau, \pi/2]$. These configurations are illustrated in Figs. 1(b)-(c) for better visualization.

## 2.2 Constrained optimization program

We formulate the reconstruction problem in CW-ZM EPRI as a convex, constrained optimization program [4] given by

$$\mathbf{f}^\star = \underset{\mathbf{f}}{\text{argmin}} \Phi(\mathbf{g}, \mathbf{g}^{[\mathscr{M}]}) \quad \text{s.t.} \quad \Psi(\mathbf{f}) \tag{3}$$

where $\Phi(\mathbf{g}, \mathbf{g}^{[\mathscr{M}]})$ denotes the $\ell_2$ data norm between measured data $\mathbf{g}^{[\mathscr{M}]}$ and model data $\mathbf{g}$ in Eq. (2); $\Psi(\mathbf{f})$ is the image-constraint term which includes two image directional total variations (DTVs) and one non-negativity of the image. Each DTV is defined as the $\ell_1$-norm of an image calculated as the two-point differences along the $x$-, $y$-, or $B$-axis. The non-negativity enforces each pixel in the reconstructed image to be greater than or equal to zero. We develop a DTV algorithm for solving the optimization program in Eq. (3) based upon a primal-dual framework [4, 6, 7]. For comparison, we also reconstruct the SS image by using the FBP algorithm.

## 2.3 Reconstruction evaluation and quantitative metrics

We first conduct visual inspection of the DTV reconstructions in the simulated-data study. Subsequently, we compare the spectral profiles in the reconstructions with those present in the truth image. In addition, we assess the images using quantitative metric Pearson-correlation-coefficient (PCC), defined as

$$\text{PCC}(\mathbf{f}^\star) = \frac{|\text{Cov}(\mathbf{f}^\star, \mathbf{f}^{[\text{ref}]})|}{\sigma(\mathbf{f}^\star)\,\sigma(\mathbf{f}^{[\text{ref}]})}, \tag{4}$$

where $\text{Cov}(\mathbf{f}^\star, \mathbf{f}^{[\text{ref}]})$ denotes the covariance between SS image $\mathbf{f}^\star$ reconstructed and reference SS image $\mathbf{f}^{[\text{ref}]}$; $\sigma^2(\mathbf{f}) =$ $\mathrm{Cov}(\mathbf{f},\mathbf{f})$ the variances of $\mathbf{f}$. In the simulation study, the reference image is the truth numerical phantom. We note that $\mathrm{PCC}(\mathbf{f}^\star) \to 1$, if $\mathbf{f}^\star \to \mathbf{f}^{[\mathrm{ref}]}$.

### 2.4 Data acquisition

In the simulated study, we design a numerical phantom with dimensions $100 \times 100 \times 100$, encompassing a 3D physical size of $1 \times 1 \times 300$ cm$^2$mG, as depicted in Fig. 2(a). The digital phantom comprises four tubes embedded within the 2D spatial space against a zero background. These four tubes contain contrast materials characterized by Voigt functions, defined as $V_k(B) = A_k, N_\sigma(B) \circledast L_{\tau_k}(B)$, where $N_\sigma(B)$ represents a Gaussian function with a zero mean and a standard deviation of $\sigma$, $L_{\tau_k}(B) = \frac{0.5\tau_k}{B^2+(0.5\tau_k)^2}$ corresponds to the Lorentzian function with width $\tau_k$, $k = 1,2,...,K$, and $K$ is the total number of material types within the imaged subject. The Voigt functions within the four tubes are specified with $\sigma = 30$ mG for $N_\sigma(B)$, and $\tau_k = 40$, 60, 80, and 100 mG for $k = 1$, 2, 3, and 4, respectively, as illustrated in Fig. 2(b). The sampling along $\gamma$ follows a uniform angular interval of 4.3°, while an equal-solid-angle (ESA) sampling scheme is applied for angles $\phi$ in the spatial domain. For the full-angular-range (FAR) scan configuration, $\gamma_\tau = 175.7°$ and $\phi_\tau = 180°$, and a total of 1128 projections are gathered. The detector encompasses 200 bins. We first generate 6 distinct noiseless datasets, each covering a different angular range, as indicated in Table 1. Subsequently, we generate noisy data by introducing Gaussian noise, yielding a signal-to-noise ratio (SNR) of 30 dB.

## 3 Results

### 3.1 Reconstructions from noiseless data

We first conduct a verification study using noiseless data obtained through the FAR configuration. In Fig. 3, we display the DTV reconstructions derived from the noiseless FAR data, accompanied by the difference image between the DTV reconstruction and the truth image. Additionally, we numerically calculate and confirm the convergence conditions, as outlined in [4], thereby demonstrating the accurate implementation of the DTV algorithm.

Having verified the correctness of the DTV algorithm, we proceed with image reconstructions using noiseless data and the LAR configurations listed in Table 1. We observe that for the considered LAR configurations, the DTV reconstructions closely resemble the true phantom image. Due to this close similarity, we omit presenting the LAR DTV reconstructions within the context of the noiseless-data study.

### 3.2 Reconstructions from noisy data

Subsequently, we employ the DTV algorithm to reconstruct SS images from noisy data acquired through LAR scans, as detailed in Table 1. We present the reconstructed SS images within the $x$-$y$ plane, determined by $B = 0$ G, as illustrated in Fig. 4. For reference, the corresponding FBP reconstructions are also displayed in the bottom row of Fig. 4. It is evident that the DTV algorithm achieves accurate SS image reconstructions from LAR scan data, as the DTV reconstructions are visually comparable to the true SS image. Conversely, the FBP images reconstructed from LAR data exhibit significant artifacts that distort the anatomical structure and introduce biases to the SS images.

Furthermore, in Fig. 5, we depict the spectral profile for a specific spatial point within tube 1 in the SS images reconstructed from LAR data. The spectral profiles in DTV reconstructions exhibit a close agreement with those present in the true SS image, while those in the FBP reconstructions differ notably from the true SS image's profile. Additionally, we quantitatively assess the reconstruction accuracy of SS images by calculating PCC defined in Eq. (4). The corresponding plots for different LAR scans listed in Table 1 are illustrated in Fig. 6. It is evident that the DTV algorithm yields more accurate reconstructions compared to FBP. Furthermore, as the angular range of the LAR scan decreases, the PCC also decreases.

## 4 Discussion

In this study, we have conducted investigations into 3D SS image reconstruction directly from LAR data in CW-ZM EPRI. We have formulated the reconstruction as an optimization problem involving data-$\ell_2$-minimization fidelity and SS-image DTV constraints. Subsequently, we developed a dedicated DTV algorithm to effectively reconstruct the SS image by solving the optimization program. Our initial focus was on noiseless-data studies using a numerical phantom to showcase the efficacy of the DTV algorithm in precisely reconstructing 3D SS images from LAR data. These studies provided insights into the upper-bound performance of the DTV algorithm for LAR CW-ZM EPRI scan setups. Subsequently, we conducted experiments with noisy data, obtained over a range of LARs. The outcomes revealed that DTV reconstructions outperform their FBP counterparts, underscoring the robustness of the DTV algorithm in practical scenarios. Furthermore, we quantitatively evaluated the quality of DTV reconstructions. Results suggest that the DTV algorithm can offer valuable guidance for devising innovative LAR scan strategies aimed at reducing scanning time in CW-ZM EPRI.

For future studies, it would be intriguing to explore alternative designs of optimization programs, involving different data fidelity terms. Additionally, evaluating image reconstruction from sparse-view data in CW-ZM EPRI holds potential for future investigations.

**Table 1:** LAR scans considered in the simulated-data study.

| Type \ $(\gamma_\tau, \phi_\tau)$ | $(150^\circ, 150^\circ)$ | $(120^\circ, 150^\circ)$ | $(120^\circ, 150^\circ)$ |
|---|---|---|---|
| I | LAR1 | LAR2 | LAR3 |
| II | LAR4 | LAR5 | LAR6 |

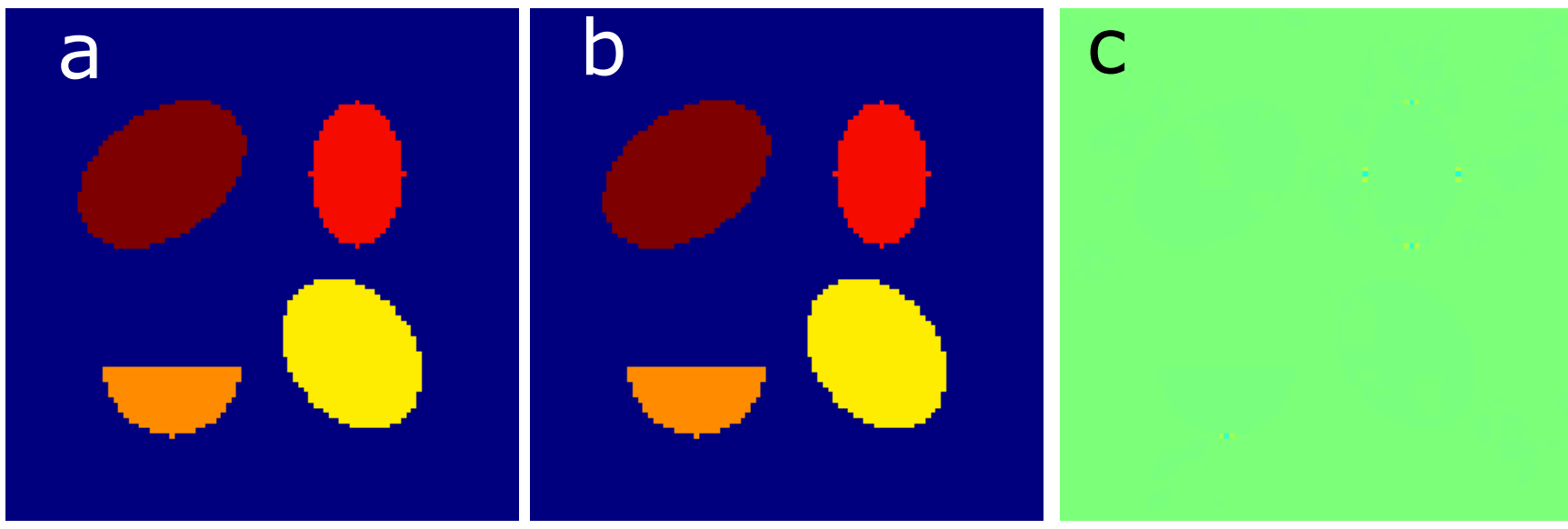

**Figure 3:** Truth numerical phantom (a) and DTV reconstruction (b) from noiseless FAR data, along with their difference image (c). Display window: [0, 1.] AU for (a) and (b), and [-0.05, 0.05] AU for (c).

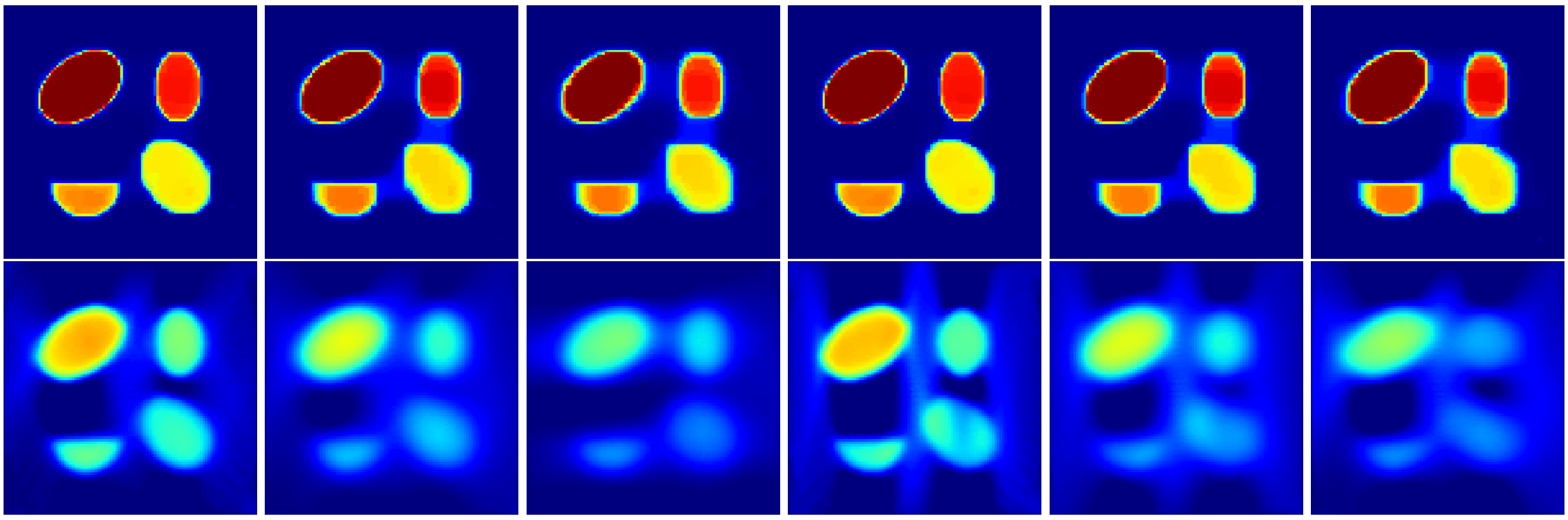

**Figure 4:** DTV (top) and FBP (bottom) reconstructions from simulated noisy data with configurations LAR1 (column 1), LAR2 (column 2), LAR3 (column 3), LAR4 (column 4), LAR5 (column 5), and LAR6 (column 6) specified in Table 1. Display window: [0, 1.] AU.

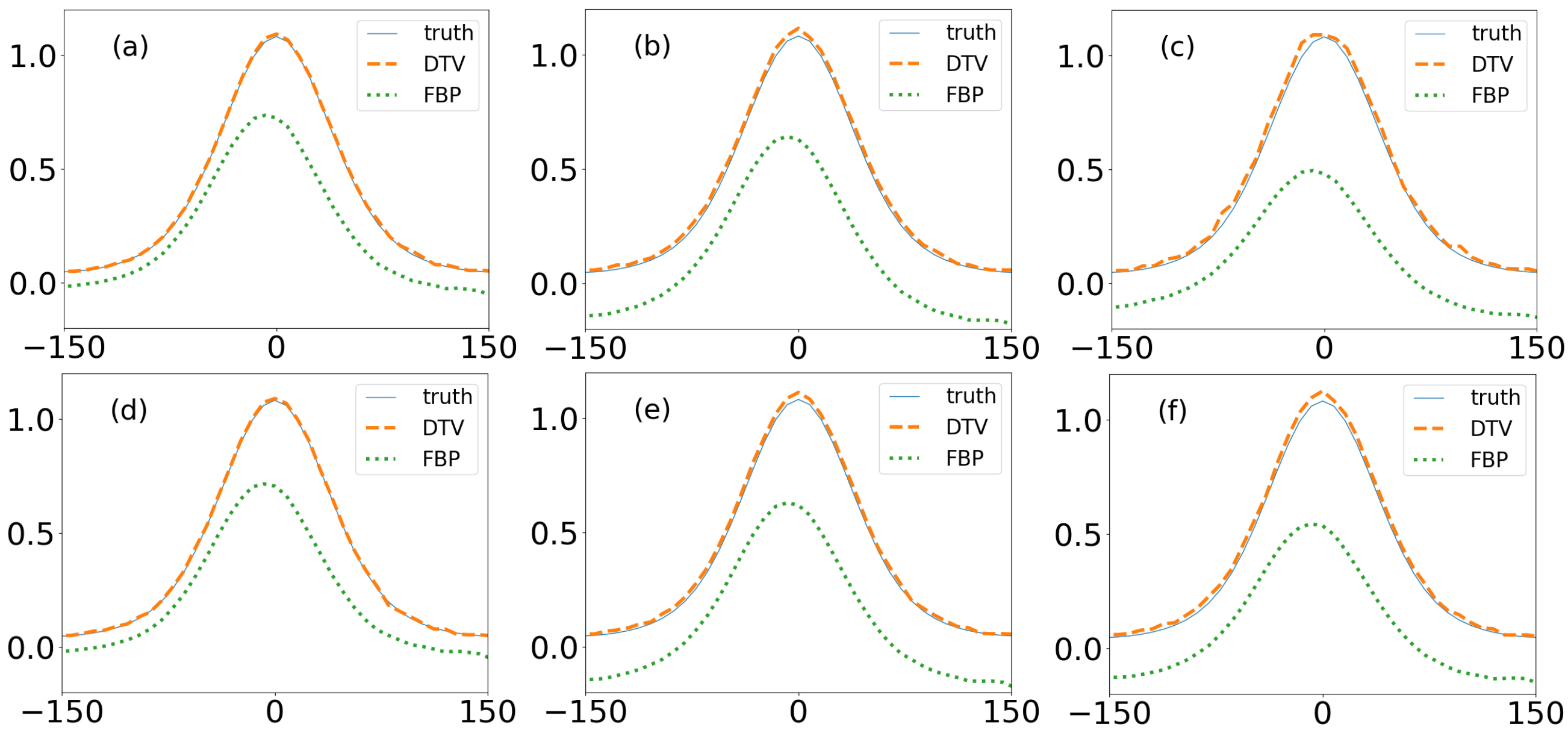

**Figure 5:** Spectral profiles (at a spatial point in tube 1) of the truth SS image (solid), and SS images reconstructed by use of the DTV (dashed) and FBP (dotted) algorithms from the simulated noisy data over LAR1 (a), LAR2 (b), LAR3 (c), LAR4 (d), LAR5 (e), and LAR6 (f) specified in Table 1.

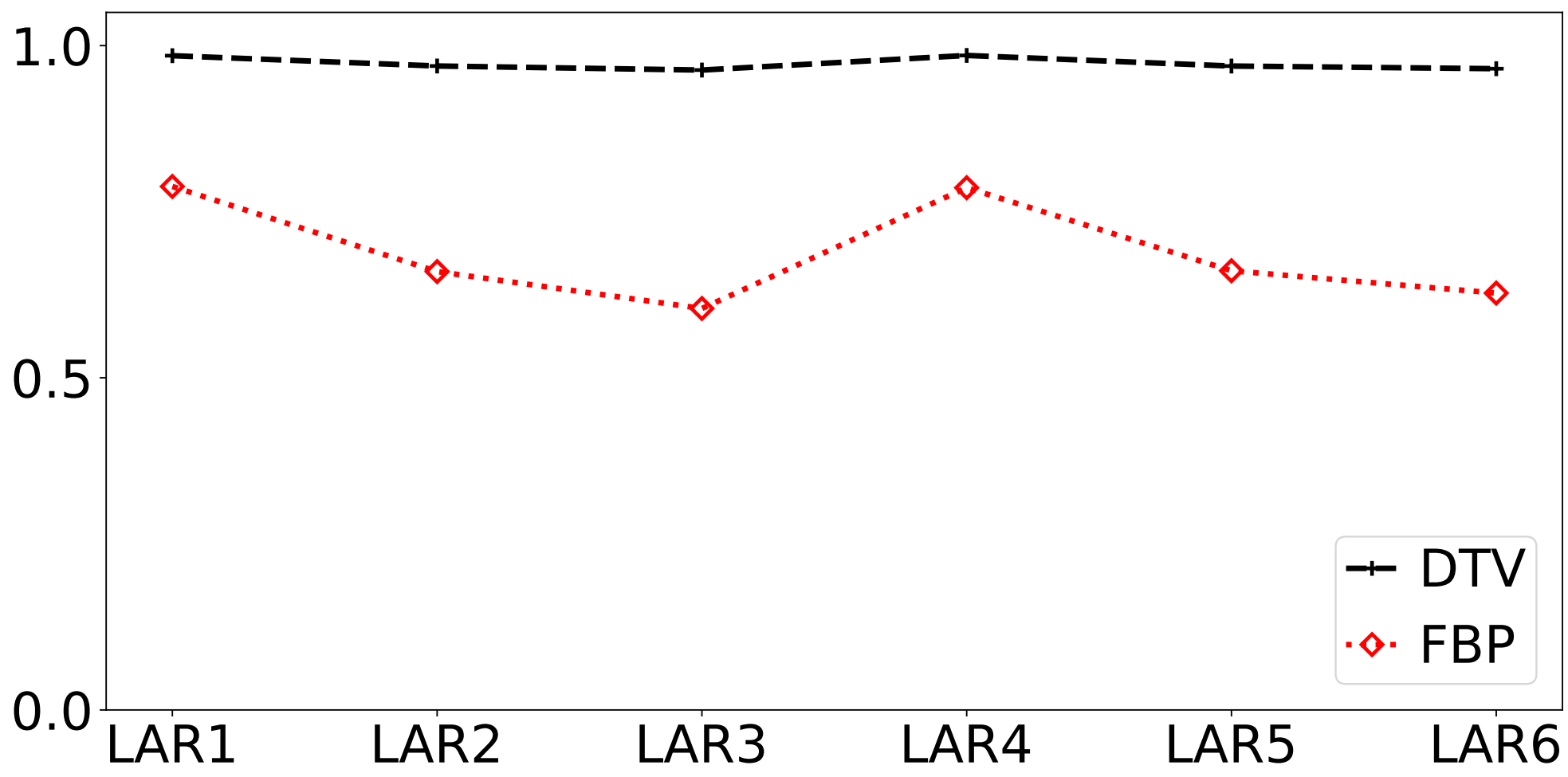

**Figure 6:** Quantitative metrics PCC of the SS images reconstructed by use of the DTV (black, dashed) and FBP (red, dotted) algorithms from simulated noisy data of LAR scans specified in Table 1.

## Acknowledgments

This work was supported in part by NIH Grants R01EB026282, R01EB023968, and R21CA263660. The computation of the work was performed in part on Computer Cluster Mel in the Department of Radiology at The University of Chicago.

# [1]A new analysis and compensation method for charge sharing in PCD

Shengzi Zhao[1,2], Katsuyuki Taguchi[3] and Yuxiang Xing[1,2]

[1]Department of Engineering Physics, Tsinghua University, Beijing, China
[2]Key Laboratory of Particle & Radiation Imaging (Tsinghua University), Ministry of Education, Beijing, China
[3]Radiological Physics Division, The Russell H. Morgan Department of Radiology and Radiological Science, Johns Hopkins University School of Medicine, Maryland, U.S.A.

**Abstract** Photon counting detectors (PCDs) offer significant advantages in capturing spectral information in computed tomography (CT). Nevertheless, their application faces limitations due to charge sharing and pile-up. In this abstract, we proposed a new analysis about how charge sharing affected the detected counts of photons pixel by pixel. By assuming uniform charge sharing probabilities across all detector pixels, our model addressed diverse charge sharing events based on their sources and destinations, and established relationships between the numbers of incident photons and detected photon counts. Our analysis reveals that the charge sharing compensation problem was always ill-posed for conventional detectors. However, MEICC detectors provide a well-determined solution to this issue by providing coincidence counts of photons in more channels. We preliminarily applied Levenberg-Marquarelt algorithm to solve the inverse problem. Utilizing data obtained from a MEICC detector, we achieved stable and physically meaningful solutions while conventional detectors could not achieve. The results demonstrated that the impact of charge sharing has been effectively mitigated.

## 1 Introduction

In recent years, Spectral CT has been studied widely for its capacity to take good use of attenuation at different energies and provide two or more material maps, which is very beneficial to distinguish object materials. In spectral CT, PCDs have great application prospects due to their ability of capturing both the number and energy information of incident photons. However, several physical effects such as charge sharing and pile up can distort recorded counts and energies, consequently limiting their applications.

In this paper, we focused on charge sharing. Dut to various factors, the charges generated by incident photons cannot be constrained in a single detector pixel and end up being detected by several adjacent pixels. As shown in Fig. 1(b), there are two types of charge sharing, spill-in and spill-out [1]. In the case of spill-in sharing, photons reach the neighboring pixels surrounding the pixel-of-interest (POI), leading to the POI collecting some charges and producing false counts. In spill-out sharing, although photons are incident on POI, the recorded counts and energies might be lower than the ground truth due to the charge escape.

In order to solve this problem and enhance the performance of PCDs, many software and hardware methods have been proposed. Lee et al introduced a correction method for spectral distortion based on a charge sharing model [2]. Some detectors were designed to establish communication between adjacent pixels and deal with their signals together [3,4]. Hsieh et al designed a special detector which could record charge sharing information and deal with them after the detection [5]. These methods addressed the charge sharing problem partially, but also had some disadvantages including extended processing time, limited accuracy and complexities in data processing. In 2020, Taguchi proposed a new MEICC detector providing additional information about charge sharing counts and directions to help correct detected spectra as well as reduce noise level [1,6].

Inspired by the MEICC detector, we proposed a new analytical model aimed at describing charge sharing process by analyzing how various types of charge sharing events affect photon counts in adjacent pixels. The model formulates the differences of conventional and MEICC detectors, which could be utilized to compensate for charge sharing. This approach gives a clear route to take advantage of MEICC data.

## 2 Methods

In this work, we simply assumed that the detection efficiency of the detectors was always 100% so that all photons including secondary photons could be fully detected. Besides, the model consider charge sharing only, not including other potential distortion factors like pile-up. Generally speaking, charge sharing happens between two, three or four adjacent detector pixels [7], denoted as double-shared, triple-shared and quadruple-shared events in this paper. Only if photons are incident on the corner of pixels, triple-shared and quadruple-shared events are considerable. Therefore, these two multiple-shared events are of only a small portion and double-shared events are predominant in all charge sharing events. We conducted our analysis by initially considering charge sharing with only double-shared events. Later, we expanded the model to incorporate triple-shared events. The investigation was carried out for conventional and MEICC detectors, respectively.

Before introducing the model, we address charge sharing utilizing a small patch as shown in Fig. 1 for demonstration. Firstly, charge sharing happens between adjacent pixels as the orange POI and blue pixels in Fig. 1(a). The black arrows denote charge sharing directions. It is necessary to gain the numbers of incident photon in the blue pixels to estimate the charge sharing events in the POI. However, a conventional detector only provides photon counts in blue ones, which are mixed with charge sharing events from the green pixels. As a result, precise compensation for counts and spectrum of the POI requires data from all the detector pixels. Secondly, all charge sharing events can be categorized into two types, spill-in ones and spill-out ones focusing on one pixel. As shown in Fig. 1(b), a spill-in event of a pixel (#5) from another pixel (#6) is equivalent to a spill-out event in the reverse direction, i. e. from pixel #6 to pixel #5 in this case. Therefore, addressing spill-out events only is sufficient to handle all charge sharing events effectively.

[1] This work is supported by National Natural Science Foundation of China (No.62031020, No. 12275151) and NIH R21 EB029739, U.S.A.
Corresponding author: Xing, Yuxiang (xingyx@mail.tsinghua.edu.cn)

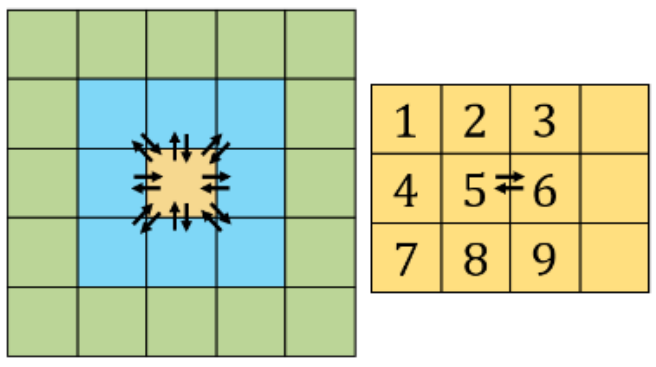

(a) (b)

Figure 1. (a) A 5 × 5 patch to show the relationship of charge sharing for its central pixel. (b) The relationship of spill-in and spill-out sharing.

## 2.1 Analytical model with double-shared events

Charge sharing happens in the detection process and causes distortions on incident spectrum. Therefore, we defined two sets of variables to characterize photon counts before and after such distortions,

$$N_{c,i} = \bar{N}_{c,i} + \epsilon_{c,i}, \tag{1}$$

$$N^0_{c,i} = \bar{N}^0_{c,i} + \epsilon^0_{c,i}. \tag{2}$$

Here, $N^0_{c,i}$ and $N_{c,i}$ are the incident and detected counts of photons in detector channel $c$ of the $i^{\text{th}}$ detector pixel. The bar on top denotes expectation. $\epsilon_{c,i}$ and $\epsilon^0_{c,i}$ are additional noise terms that are commonly Poisson. Thus, charge sharing can be regarded as functions $f_c$ to link $N_{c,i}$ and $N^0_{c,i}$,

$$N_{c,i} = N^0_{c,i} + f_c \text{ and } \bar{N}_{c,i} = \bar{N}^0_{c,i} + \bar{f}_c. \tag{3}$$

Here, we assume the principles of charge sharing i.e. the functions $f_c$ and $\bar{f}_c$ are uniform for all detector pixels.

When focusing solely on double-shared events, we listed all possible charge sharing events in Tab. 1 and defined the probabilities of charge sharing (ProbCS) in order to build our model. The subscript "$*\to*\underline{*},(i,\underline{j})$" denotes different types of double-shared event. The index "$j$" with an underline means pixel $\#j$ receives shared charges from its neighboring pixel $\#i$. Meanwhile, the shared charges are recorded as a count in one energy bin of pixel $\#j$, marked by "$\underline{*}$". For example, "$\text{L}\to\text{L}\underline{\text{L}},(i,\underline{j})$" represents a low energy photon injected on pixel $\#i$ is charge shared to be two low energy photons detected by pixel $\#i$ and its neighboring pixel $\#j$. With the assumption of uniform charge sharing response, we have $P_{*\to*\underline{*},(i,\underline{j})} = P_{*\to*\underline{*},(j,\underline{i})}$.

Considered spatial symmetry, we have much less ProbCS terms. Choosing pixel #5 in Fig. 1(b) as example, we have,

$$P_{*\to*\underline{*},(5,\underline{2})} = P_{*\to*\underline{*},(5,\underline{4})} = P_{*\to*\underline{*},(5,\underline{6})} = P_{*\to*\underline{*},(5,\underline{8})}, \tag{4}$$

$$P_{*\to*\underline{*},(5,\underline{1})} = P_{*\to*\underline{*},(5,\underline{3})} = P_{*\to*\underline{*},(5,\underline{7})} = P_{*\to*\underline{*},(5,\underline{9})}. \tag{5}$$

for direct and diagonal neighboring pixels, respectively. In total, there are ten probabilities for five types of sharing and two types of neighborhoods.

Table 1. The types of charge sharing for double-shared events. "+" "−" "None" mean count increase, decrease, and no change.

| pixel & channels | pixel #$i$ L | pixel #$i$ H | pixel #$j$ L | pixel #$j$ H |
|---|---|---|---|---|
| $P_{\text{L}\to\text{L}\underline{\text{L}},(i,\underline{j})}$ | None | None | + | None |
| $P_{\text{H}\to\text{L}\underline{\text{L}},(i,\underline{j})}$ | + | − | + | None |
| $P_{\text{H}\to\text{L}\underline{\text{H}},(i,\underline{j})}$ | + | − | None | + |
| $P_{\text{H}\to\text{H}\underline{\text{L}},(i,\underline{j})}$ | None | None | + | None |
| $P_{\text{H}\to\text{H}\underline{\text{H}},(i,\underline{j})}$ | None | None | None | + |

In the sense of mean, the detected counts $\bar{N}_{\text{L},i}$ and $\bar{N}_{\text{H},i}$ in pixel $\#i$ i.e. the POI are the sum of multiple processes,

$$\bar{N}_{\text{L},i} = \bar{N}^0_{\text{L},i} + \bar{N}_{\text{L}\to\text{L}\underline{\tilde{\text{L}}}} + \bar{N}_{\text{H}\to\tilde{\text{L}}\underline{\text{L}}} + \bar{N}_{\text{H}\to\text{L}\underline{\tilde{\text{L}}}} + \bar{N}_{\text{H}\to\tilde{\text{L}}\underline{\text{H}}} + \bar{N}_{\text{H}\to\text{H}\underline{\tilde{\text{L}}}}, \tag{6}$$

$$\bar{N}_{\text{H},i} = \bar{N}^0_{\text{H},i} - \bar{N}_{\text{H}\to\tilde{\text{L}}\underline{\text{L}}} - \bar{N}_{\text{H}\to\tilde{\text{L}}\underline{\text{H}}} + \bar{N}_{\text{H}\to\text{L}\underline{\tilde{\text{H}}}} + \bar{N}_{\text{H}\to\text{H}\underline{\tilde{\text{H}}}}, \tag{7}$$

where the processes can be expressed as,

$$\bar{N}_{\text{L}\to\text{L}\underline{\tilde{\text{L}}}} = \sum_{j\in\Omega(i)} \bar{N}^0_{\text{L},j} P_{\text{L}\to\text{L}\underline{\text{L}},(j,\underline{i})}, \tag{8}$$

$$\bar{N}_{\text{H}\to\tilde{\text{L}}\underline{\text{L}}} = \bar{N}^0_{\text{H},i} \sum_{j\in\Omega(i)} P_{\text{H}\to\text{L}\underline{\text{L}},(i,\underline{j})}, \tag{9}$$

$$\bar{N}_{\text{H}\to\text{L}\underline{\tilde{\text{L}}}} = \sum_{j\in\Omega(i)} \bar{N}^0_{\text{H},j} P_{\text{H}\to\text{L}\underline{\text{L}},(j,\underline{i})}, \tag{10}$$

$$\bar{N}_{\text{H}\to\tilde{\text{L}}\underline{\text{H}}} = \bar{N}^0_{\text{H},i} \sum_{j\in\Omega(i)} P_{\text{H}\to\text{L}\underline{\text{H}},(i,\underline{j})}, \tag{11}$$

$$\bar{N}_{\text{H}\to\text{H}\underline{\tilde{\text{L}}}} = \sum_{j\in\Omega(i)} \bar{N}^0_{\text{H},j} P_{\text{H}\to\text{H}\underline{\text{L}},(j,\underline{i})}, \tag{12}$$

$$\bar{N}_{\text{H}\to\tilde{\text{L}}\underline{\text{L}}} = \bar{N}^0_{\text{H},i} \sum_{j\in\Omega(i)} P_{\text{H}\to\text{L}\underline{\text{L}},(i,\underline{j})}, \tag{13}$$

$$\bar{N}_{\text{H}\to\text{L}\underline{\tilde{\text{H}}}} = \sum_{j\in\Omega(i)} \bar{N}^0_{\text{H},j} P_{\text{H}\to\text{L}\underline{\text{H}},(j,\underline{i})}, \tag{14}$$

$$\bar{N}_{\text{H}\to\text{H}\underline{\tilde{\text{H}}}} = \sum_{j\in\Omega(i)} \bar{N}^0_{\text{H},j} P_{\text{H}\to\text{H}\underline{\text{H}},(j,\underline{i})}. \tag{15}$$

Here, $\Omega(i)$ is the set of neighboring pixels of pixel $\#i$. The top "~" means the channel belongs to the POI. Besides, we can get equations for noisy conditions by simply adding noisy terms in Eqs. (6)–(15).

Generally speaking, charge sharing compensation is to recover numbers of incident photons $\bar{N}^0_{\text{L},i}$ and $\bar{N}^0_{\text{H},i}$ based on detected photon counts. However, in real situation, only $\bar{N}_{c,i}$ could be available and it is almost impossible to gain $\bar{N}^0_{c,j}, j \in \Omega(i)$ in Eqs. (6)–(15). Therefore, we have to estimate $\bar{N}^0_{c,i}$, $\bar{N}^0_{c,j}$ and all the ProbCS jointly. It is worth mentioning that, because all the ProbCS are independent on numbers of incident photons $\bar{N}^0_{c,i}$ and $\bar{N}^0_{c,j}$, it is possible to pre-estimate the ProbCS by some special calibration experiments and later the estimated ProbCS can be used directly to recover $\bar{N}^0_{c,i}$ in photon counting CT imaging. Unfortunately, there is no direct close-form solution in the field to estimate these probability values currently.

Now we analyze the difficulty of the problem by counting all unknown variables jointly without any other extra experiments and we start from a single pixel. The pixel #5 in Fig. 1(b) provides equations containing the numbers of incident photons of pixel # $j, j \in \Omega(5)$ as unknowns. Expanding further, we have $2\left(n_{\mathrm{p}}+2\right)^2+10$ unknowns and $2n_{\mathrm{p}}^2$ known detected counts for an $n_{\mathrm{p}} \times n_{\mathrm{p}}$ detector patch of conventional detectors. Obviously, it is always an undetermined problem no matter what $n_p$ is. However, $\lim_{n_{\mathrm{p}}\to\infty} 2n_{\mathrm{p}}^2/\left[2\left(n_{\mathrm{p}}+2\right)^2+10\right]=1$ , which means the condition will be improved by using a larger detector patch. MEICC detectors record the numbers of charge sharing events based on several extra detector channels named coincidence counters. The corresponding counts are,

$$\bar{N}_{\mathrm{LL},i}=\bar{N}^0_{\mathrm{L},i}\sum_{j\in\Omega(i)}P_{\mathrm{L}\to\mathrm{L}\underline{\mathrm{L}},(i,\underline{j})}+\bar{N}_{\mathrm{L}\to\mathrm{L}\underline{\tilde{\mathrm{L}}}}+\bar{N}_{\mathrm{H}\to\tilde{\mathrm{L}}\underline{\mathrm{L}}}+\bar{N}_{\mathrm{H}\to\mathrm{L}\underline{\tilde{\mathrm{L}}}} \tag{16}$$

$$\bar{N}_{\mathrm{LH},i}=\bar{N}_{\mathrm{H}\to\mathrm{H}\underline{\tilde{\mathrm{L}}}}+\bar{N}_{\mathrm{H}\to\tilde{\mathrm{L}}\underline{\mathrm{H}}} \tag{17}$$

$$\bar{N}_{\mathrm{HL},i}=\bar{N}^0_{\mathrm{H},i}\sum_{j\in\Omega(i)}P_{\mathrm{H}\to\mathrm{H}\underline{\mathrm{L}},(i,\underline{j})}+\sum_{j\in\Omega(i)}\bar{N}^0_{\mathrm{H},j}P_{\mathrm{H}\to\mathrm{L}\underline{\mathrm{H}},(j,\underline{i})} \tag{18}$$

$$\bar{N}_{\mathrm{HH},i}=\bar{N}^0_{\mathrm{H},i}\sum_{j\in\Omega(i)}P_{\mathrm{H}\to\mathrm{H}\underline{\mathrm{H}},(i,\underline{j})}+\bar{N}_{\mathrm{H}\to\mathrm{H}\underline{\tilde{\mathrm{H}}}} \tag{19}$$

In total, the number of equations for MEICC detectors is $6n_{\mathrm{p}}^2$. When $n_{\mathrm{p}} \geq 4$, we could have a well-determined problem by assuming that all these equations are not fully linearly dependent. By comparison, it is easier to get better performance on catching the principles of charge sharing with MEICC data than conventional ones because of extra information. If ProbCS determined, the situation of estimating $\bar{N}^0_{c,i}$ is similar that it is ill-posed for conventional detectors but well-determined for MEICC detectors.

In practice, it is crucial to ensure that all the above equations are of low co-linearity. To give a negative example, if all pixels in the $3\times 4$ patch in Fig. 1(b) receive the same photons, pixels #5 and #6 will provide the same equations. Such a patch is inefficient in importing information due to strong co-linearity of the equations. To keep low co-linearity, the received photons of $3\times 3$ patches need to be of great variety.

Despite successfully establishing the model, the direct solution of the problem remains challenging. There are no straight-forward close-form solutions for these non-linear equations generally. Hence, we settle for an alternative approach, aiming to obtain its numerical solution.

## 2.2 Taking triple-shared events into account

In this section, we consider triple-shared events. Similarly, we also define a series of $P_{*\to\underline{***},(i,\underline{j},\underline{k})}$ and because of symmetry, we have $P_{*\to\underline{***},(i,\underline{j},\underline{k})}=P_{*\to\underline{***},(j,\underline{i},\underline{k})}$. Due to the limit of photon energy, it's highly improbable for one high energy photon to be charge shared to three photons with two or more of high energy. Therefore, we ignore these "H → HHH" and "H → HHL" events to simplify the model. We also import terms in Eqs. (6)–(15) to make the equations concise. For conventional detectors, the photon counts are,

$$\bar{N}_{\mathrm{L},i,\mathrm{tri}}=\bar{N}_{\mathrm{L},i}+\bar{N}_{\mathrm{L}\to\mathrm{L}\underline{\tilde{\mathrm{L}}\mathrm{L}}}+\bar{N}_{\mathrm{H}\to\tilde{\mathrm{L}}\underline{\mathrm{LL}}}+\bar{N}_{\mathrm{H}\to\mathrm{L}\underline{\tilde{\mathrm{L}}\mathrm{L}}}+\bar{N}_{\mathrm{H}\to\mathrm{H}\underline{\tilde{\mathrm{L}}\mathrm{L}}}+\bar{N}_{\mathrm{H}\to\tilde{\mathrm{L}}\underline{\mathrm{HL}}}+\bar{N}_{\mathrm{H}\to\mathrm{L}\underline{\tilde{\mathrm{L}}\mathrm{H}}}, \tag{20}$$

$$\bar{N}_{\mathrm{H},i,\mathrm{tri}}=\bar{N}_{\mathrm{H},i}-\bar{N}_{\mathrm{H}\to\tilde{\mathrm{L}}\underline{\mathrm{LL}}}-\bar{N}_{\mathrm{H}\to\tilde{\mathrm{L}}\underline{\mathrm{HL}}}+\bar{N}_{\mathrm{H}\to\mathrm{L}\underline{\tilde{\mathrm{H}}\mathrm{L}}}. \tag{21}$$

All the processes can be expressed as,

$$\bar{N}_{\mathrm{L}\to\mathrm{L}\underline{\tilde{\mathrm{L}}\mathrm{L}}}=\sum_{j\in\Omega(i)}\bar{N}^0_{\mathrm{L},j}\sum_{k\in\Pi(i,j)}P_{\mathrm{L}\to\mathrm{L}\underline{\mathrm{LL}},(j,\underline{i},\underline{k})}, \tag{22}$$

$$\bar{N}_{\mathrm{H}\to\tilde{\mathrm{L}}\underline{\mathrm{LL}}}=\frac{1}{2}\bar{N}^0_{\mathrm{H},i}\sum_{j\in\Omega(i)}\sum_{k\in\Pi(i,j)}P_{\mathrm{H}\to\mathrm{L}\underline{\mathrm{LL}},(i,\underline{j},\underline{k})}, \tag{23}$$

$$\bar{N}_{\mathrm{H}\to\mathrm{L}\underline{\tilde{\mathrm{L}}\mathrm{L}}}=\sum_{j\in\Omega(i)}\bar{N}^0_{\mathrm{H},j}\sum_{k\in\Pi(i,j)}P_{\mathrm{H}\to\mathrm{L}\underline{\mathrm{LL}},(j,\underline{i},\underline{k})}, \tag{24}$$

$$\bar{N}_{\mathrm{H}\to\mathrm{H}\underline{\tilde{\mathrm{L}}\mathrm{L}}}=\sum_{j\in\Omega(i)}\bar{N}^0_{\mathrm{H},j}\sum_{k\in\Pi(i,j)}P_{\mathrm{H}\to\mathrm{H}\underline{\mathrm{LL}},(j,\underline{i},\underline{k})}, \tag{25}$$

$$\bar{N}_{\mathrm{H}\to\tilde{\mathrm{L}}\underline{\mathrm{HL}}}=\bar{N}^0_{\mathrm{H},i}\sum_{j\in\Omega(i)}\sum_{k\in\Pi(i,j)}P_{\mathrm{H}\to\mathrm{L}\underline{\mathrm{HL}},\left(i,\underline{j},\underline{k}\right)}, \tag{26}$$

$$\bar{N}_{\mathrm{H}\to\mathrm{L}\underline{\tilde{\mathrm{L}}\mathrm{H}}}=\sum_{j\in\Omega(i)}\bar{N}^0_{\mathrm{H},j}\sum_{k\in\Pi(i,j)}P_{\mathrm{H}\to\mathrm{L}\underline{\mathrm{LH}},(j,\underline{i},\underline{k})}, \tag{27}$$

$$\bar{N}_{\mathrm{H}\to\mathrm{L}\underline{\tilde{\mathrm{H}}\mathrm{L}}}=\sum_{j\in\Omega(i)}\bar{N}^0_{\mathrm{H},j}\sum_{k\in\Pi(i,j)}P_{\mathrm{H}\to\mathrm{L}\underline{\mathrm{HL}},(j,\underline{i},\underline{k})}. \tag{28}$$

Here, $\Pi(i,j)$ denotes the set of pixels forming L-shaped areas with pixels $\#i$ and $\#j$ as shown in Fig. 2.

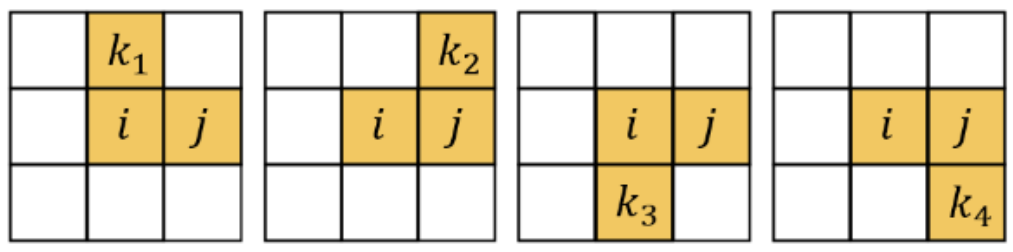

Figure 2. An example of $\Pi(i,j)$. $\Pi(i,j)=\{k_1,k_2,k_3,k_4\}$.

For MEICC detectors, we also have extra four equations,

$$\bar{N}_{\mathrm{LL},i,\mathrm{tri}}=\bar{N}_{\mathrm{LL},i}+\frac{1}{2}\bar{N}^0_{\mathrm{L},i}\sum_{k\in\Pi(i,j)}P_{\mathrm{L}\to\mathrm{L}\underline{\mathrm{LL}},(i,\underline{j},\underline{k})}+\bar{N}_{\mathrm{L}\to\mathrm{L}\underline{\tilde{\mathrm{L}}\mathrm{L}}}+\bar{N}_{\mathrm{H}\to\tilde{\mathrm{L}}\underline{\mathrm{LL}}}+\bar{N}_{\mathrm{H}\to\mathrm{L}\underline{\tilde{\mathrm{L}}\mathrm{L}}}+\bar{N}_{\mathrm{H}\to\mathrm{H}\underline{\tilde{\mathrm{L}}\mathrm{L}}}+\bar{N}_{\mathrm{H}\to\tilde{\mathrm{L}}\underline{\mathrm{HL}}}+\bar{N}_{\mathrm{H}\to\mathrm{L}\underline{\tilde{\mathrm{L}}\mathrm{H}}}, \tag{29}$$

$$\bar{N}_{\mathrm{LH},i,\mathrm{tri}}=\bar{N}_{\mathrm{LH},i}+\bar{N}_{\mathrm{H}\to\mathrm{H}\underline{\tilde{\mathrm{L}}\mathrm{L}}}+\bar{N}_{\mathrm{H}\to\tilde{\mathrm{L}}\underline{\mathrm{HL}}}+\bar{N}_{\mathrm{H}\to\mathrm{L}\underline{\tilde{\mathrm{L}}\mathrm{H}}}, \tag{30}$$

$$\bar{N}_{\mathrm{HL},i,\mathrm{tri}}=\bar{N}_{\mathrm{HL},i}+\bar{N}^0_{\mathrm{H},i}\sum_{j\in\Omega(i)}\sum_{k\in\Pi(i,j)}P_{\mathrm{H}\to\mathrm{H}\underline{\mathrm{LL}},(i,\underline{j},\underline{k})}+\sum_{j\in\Omega(i)}\bar{N}^0_{\mathrm{H},j}\sum_{k\in\Pi(i,j)}P_{\mathrm{H}\to\mathrm{L}\underline{\mathrm{HL}},(j,\underline{i},\underline{k})}, \tag{31}$$

$$\bar{N}_{\mathrm{HH},i,\mathrm{tri}}=\bar{N}_{\mathrm{HH},i}. \tag{32}$$

There are ten probabilities for triple-shared events. For an $n_{\mathrm{p}} \times n_{\mathrm{p}}$ detector patch, we have $2\left(n_{\mathrm{p}}+2\right)^2+20$ unknowns and still $2n_{\mathrm{p}}^2$ known detected counts for conventional detectors, so it's still undetermined. Larger patches can also make the condition better. For MEICC detectors, we still have a well-determined problem when $n_{\mathrm{p}} \geq 4$ with the assumption that all these equations are not fully linearly dependent.

## 3 Experimental studies

We validated our proposed method utilizing simulated data through the Monte Carlo simulation program developed by [1,6,8] that we would like to refer readers to for the details. We simulated a CdTe detector with the pixel size of $225\times 225$ μm$^2$, the thickness of 1.6mm and the density of $5.85\mathrm{g/cm^3}$. The energy windows were set to [20, 65] keV and [65, 120] keV. The X-ray source was at 120kVp with a 2mm aluminum filter. Pile-up events were avoided with impulse pulse shape.

We employed Levenberg-Marquarelt (LM) algorithm to iteratively solve the equations, focusing solely on double-shared events i.e. Eqs. (6)–(19). The patch size $n_\mathrm{p}$ was set to five, resulting in 98 unknown numbers of incident photons for each patch. Randomly generated data for 50 patches are utilized to solve the problem. With the assumption of uniform detector response, 10 probabilities are shared unknowns for all patches.

In fact, the equations for conventional detectors have a trivial solution that all the probabilities $P_{*\rightarrow\underline{**},(\underline{i},\underline{j})}$ are equal to zero and $\bar{N}_{c,i}$ is equal to $\bar{N}^0_{c,i}$, which does not make sense physically.

LM algorithm was run 30 times with random initial values for the same detected data. We randomly selected four patches to show the results in Fig. 3 focusing on the numbers of photons in the low energy bin of the central pixels compared with the ground truth of incident photon number without charge sharing. For conventional detectors, the estimated numbers of incident photons are close to detected counts, close to the trivial solution mentioned above. Moreover, they have significant standard deviations indicating poor stability. In contrast, the estimated numbers of incident photons are close to ground truths for MEICC detectors with almost negligible standard deviations, indicating strong stability.

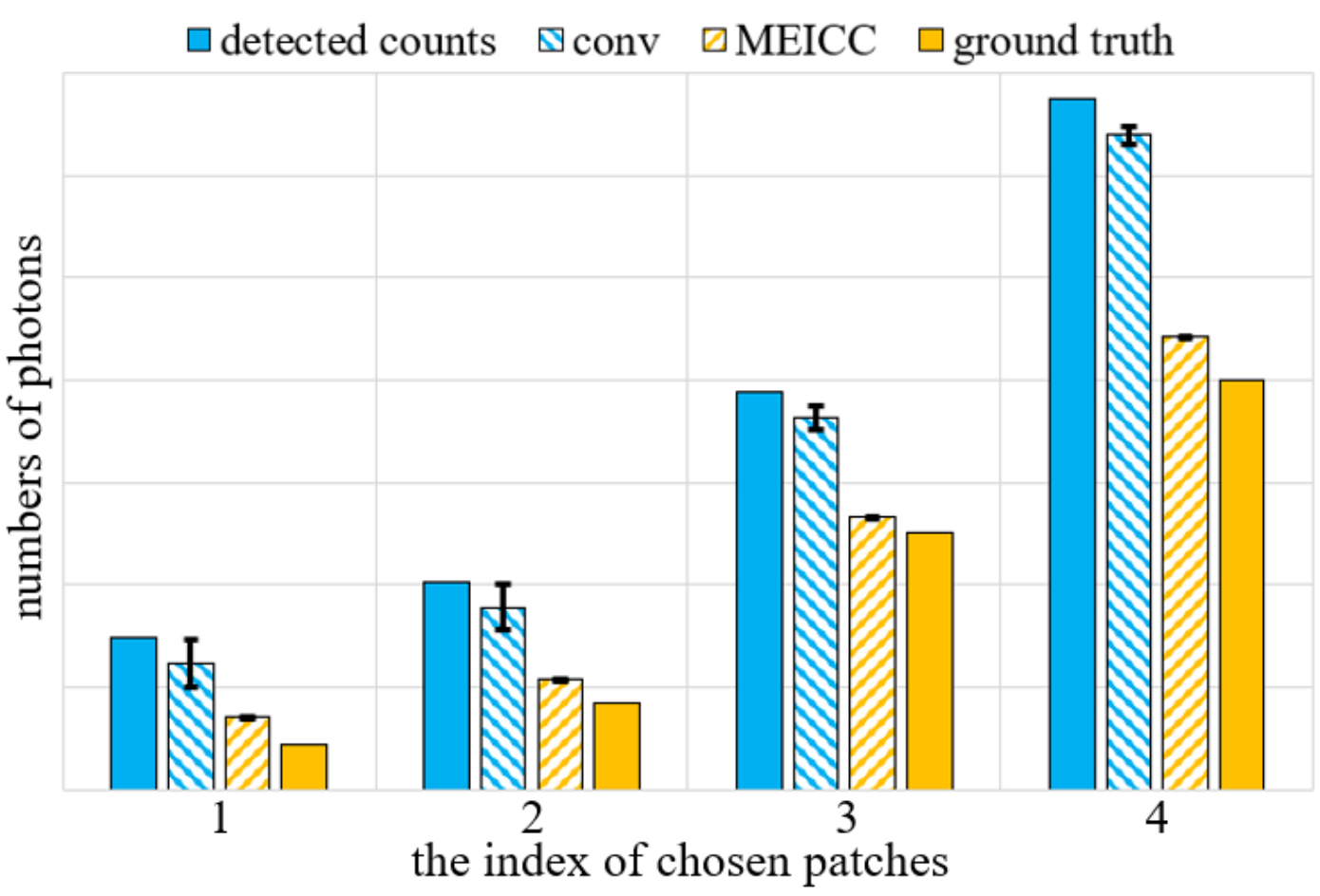

Figure 3. Four patches to demonstrate the results. The solid blue bars are detected counts in the low energy bin of their central pixels. The solid yellow bars are the numbers of incident photons without charge sharing. The striped blue bars are the results from conventional data, while striped yellow bars from MEICC data, with the error bars indicating their standard deviation.

We define a metric $\eta = \frac{N - N^*_\mathrm{ideal}}{N^*_\mathrm{ideal}}$ to measure the residual impact of charge sharing in the low energy bin. $N$ is the result obtained from our iterative LM algorithm. $N^*_\mathrm{ideal}$ is the ground truth number of incident photons. We also define $\eta_0 = \frac{N_\mathrm{det} - N^*_\mathrm{ideal}}{N^*_\mathrm{ideal}}$ as references and $N_\mathrm{det}$ is detected photon counts without any processing. For the results obtained from conventional detectors, $\eta$ is in the range of 0.3348 to 1.9489 and similar with $\eta_0$, indicating that charge sharing is only weakly compensated. For MEICC, $\eta$ is in the range of 0.014 to 0.6489, significantly smaller than $\eta_0$, indicating that charge sharing is effectively compensated.

## 4 Discussion and Conclusion

In this work, we constructed an analytical model to address charge sharing in PCDs. The model was designed to explore relationships between adjacent pixels based on charge sharing probabilities. We analyzed the conditions of charge sharing compensation problem for conventional and MEICC detectors, respectively. Notably, the MEICC data exhibited more promising results in charge shairng compensation, owing to its better conditions than conventional data. An LM iteration method was utilized to conducted preliminary validation of our model to compensate for charge sharing. It exhibits theoritical advantages of MEICC detectors and confirmed through a simulated experimental study. In future work, we are to enhance our method to obtain more valuable and acurate results for practical usage.

# Anatomical MRI-guided Deep-Learning Low-Count PET Image Recovery without the need for training data – A PET/MR study

Tianyun Zhao[1], Thomas Hagan[1], Chuan Huang[1,2,3]

[1]Department of Biomedical Engineering, Stony Brook University, Stony Brook. NY, United States
[2]Department of Radiology, Stony Brook Medicine, Stony Brook, NY, United States
[3]Department of Radiology, Emory University, Atlanta, GA, United States

**Abstract** The advent of simultaneous PET/MRI enables the possibility of using MRI to guide PET image reconstruction/recovery. Deep-learning approaches have been explored in low-count PET recovery, with current approaches focusing on supervised learning, which requires a large amount of paired low-count and full-count training data. A recently proposed unsupervised learning image-recovery approach does not require such data sets but relies on the optimal loss function. In this work, we developed an unsupervised learning-based PET image recovery approach using anatomical MRI as input and a novel loss function. Our method achieved better image recovery in both global image similarity metrics and regional standard uptake value (SUV) accuracy on FDG scans.

## 1 Introduction

Positron Emission Tomography (PET) is a sensitive and quantitative tool for clinicians and researchers to investigate a range of pathologies, such as heart disease and brain disorders [1,2]. However, conventional PET suffers from an undesirable, fundamental tradeoff between the radiation dose and signal-to-noise ratio (SNR). High SNR in PET images is beneficial for many applications, including detecting small lesions and diagnosing neurological diseases. However, obtaining a high SNR image requires a high radiation dose and/or increased scan time. Low-count PET can reduce the injected dose and/or reduce acquisition time. However, the image quality of low-count PET can be too low for clinical use if reconstructed using conventional reconstruction approaches. [3].

To solve this problem, deep learning techniques have been proposed to recover high-count-like images from low-count. Multiple deep learning models have been developed in recent years to enhance low-count PET scans; some were even approved by the FDA [4–6]. Most current techniques utilize supervised learning, where a large amount of high-count and low-count PET image pairs are needed for the network to learn. However, such data sets are hard to obtain, especially for newly developed traces, and could potentially be disease dependent.

Unsupervised learning has been proposed for this task utilizing help from anatomical MRI images. Unsupervised learning does not require low-count and high-count image pairs. However, it relies on the loss function to obtain good results without overfitting and find the best network weighting. A recent study used the L2 norm as the loss function [7]. In this work, we proposed an improved loss function for this novel approach that can be used in general studies with improved global image similarity and accurate regional uptake values.

## 2 Materials and Methods

### 2.1 PET/MRI data

A total of 28 $^{18}$F-FDG brain PET/MRI studies were gathered from our previously acquired research studies. None of these patients had brain tumors or neurological diseases. All data sets used listmode acquisition with a 20-channel PET-compatible MRI head coil on a Siemens Biograph mMR PET/MRI scanner. T1-weight MRI images were acquired using MPRAGE sequence (TE/TR/TI = 3.24/2300/900 MS, flip angle = 9, voxel size = 0.87×0.87 ×0.87mm$^3$). Attenuation maps were generated using an established MPRAGE-based algorithm incorporating bone compartment [8,9].

### 2.2 PET image reconstruction

Static PET images were reconstructed from 10-minute emission data (50-60 minutes after injection) using Siemens' E7tools with ordered subset expectation maximization (OSEM). The reconstructed image was used as the reference standard full-count PET image. Low-count PET data (10% count) were generated through Poisson thinning of the listmode data and reconstructed in the same method as the full-count PET image.

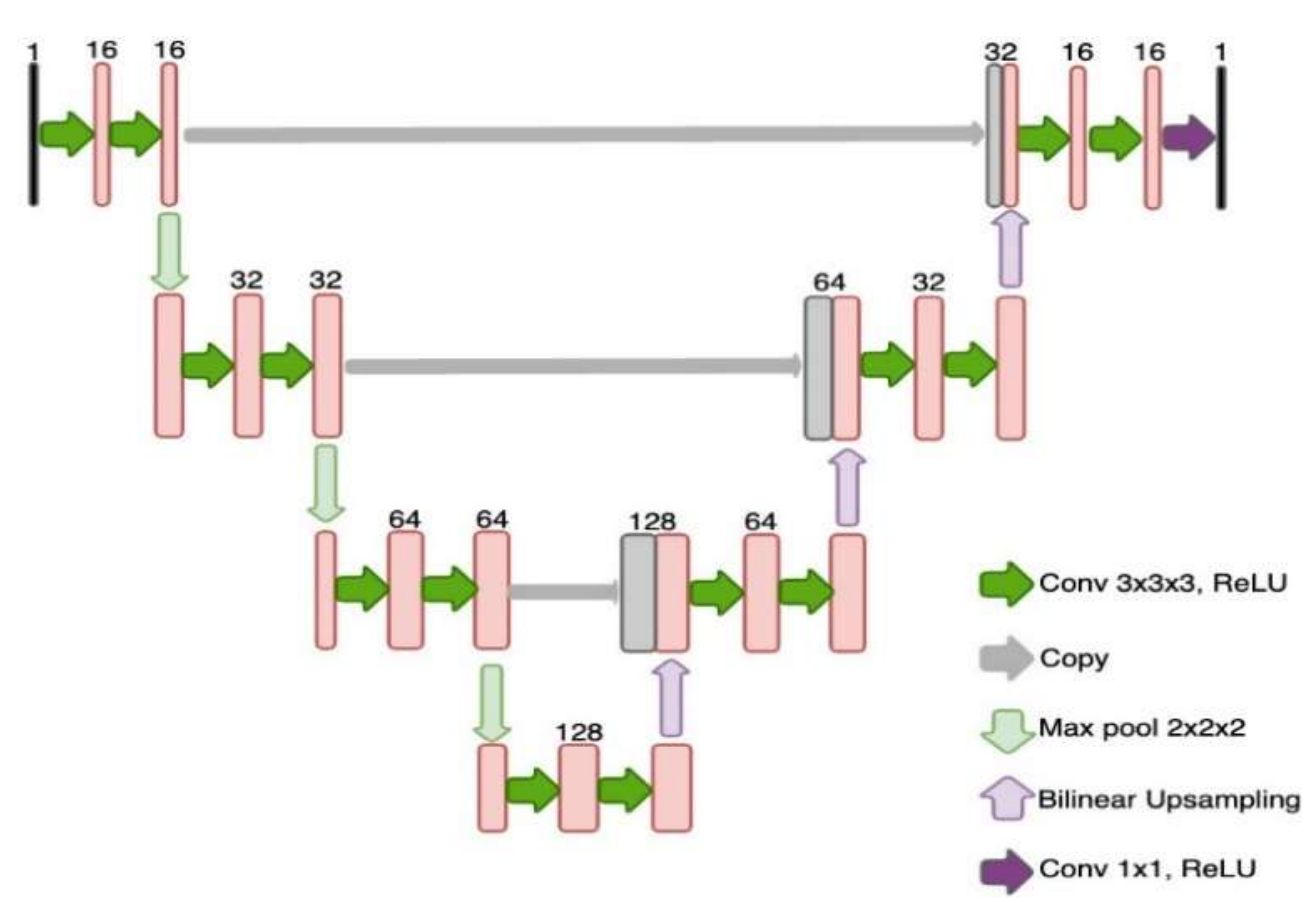

**Figure 1** U-NET architecture used.

## 2.3 Unsupervised learning image recovery

The network architecture used was a U-Net, as shown in **Figure 1.** Similar to the previous unsupervised learning framework, anatomical MPRAGE images were used as the input to the network and were trained to output low-count PET images [10].

In this work, we proposed a custom loss function (as shown below) using the structural similarity index measure (SSIM) between the recovered PET images and anatomical MRI and the mean absolute error (MAE) between the recovered PET images and the low-count PET images:

$$L(x) = \lambda \left(1 - SSIM(x, MRI)\right) + (1 - \lambda) MAE(x, PET_{lowcount}) \quad \text{[Eq. 1]}$$

where $x$ is the denoised PET image.

Each subject was trained individually. One subject was picked randomly to tune the model and optimize the loss function. The optimal iteration for the best result was chosen empirically using the training subject. Once the optimal setup was found, the same setting was applied to the rest of the data set.

### 2.4 Evaluation

Results from the unsupervised learning U-Net framework were compared to full-count data to assess the best setup using the following quantitative metrics: 1) Peak Signal to Noise Ratio (PSNR), 2) SSIM, 3) Root Mean Square Error (RMSE), all with respect to the full-count image.

The model's performance was also compared with Gaussian denoising. The optimal Gaussian denoising filter parameter was chosen using the same subject that was used to tune our model, and the optimal parameter was used on the rest of the dataset.

In addition to the quantitative image metrics, standard uptake value (SUV) was also calculated from low-count PET, unsupervised learning denoised PET, and the Gaussian denoised PET. The mean SUV ($SUV_{mean}$) values were calculated for the hippocampus and amygdala. The percentage mean SUV difference was computed between low-count to full-count PET, between the unsupervised learning denoised PET and full-count PET, and between Gaussian denoise PET and full-count PET. This comparison allows us to evaluate if the setup can be used for future quantitative PET analysis.

## 3 Results

Using the single training subject, the optimal $\lambda$ was 0.75, and the optimal iteration number was 640 with a learning rate of $3.33 \times 10^{-3}$. The network layers are shown in **Figure 1**.

The full-count PET, low-count PET, Gaussian denoised PET, and the unsupervised learning denoised PET of one representative subject are shown in **Figure 2**. Both Gaussian denoised PET and unsupervised learning denoised PET show good results compared to low-count.

The distribution of the PSNR, SSIM, and RMSE of the low-count PET, Gaussian denoised PET, and our unsupervised denoising network output for the subjects, as compared to full-count PET, are shown in **Figure 3**. The unsupervised learning denoised PET yielded the best result visible as shown in **Figure 3.** The result from the paired student t-test are also shown in **Figure 3,** indicated by asterisks, and confirms the visual finding. The unsupervised learning results achieved a significant better results with $p<0.0001$ for all three metrics when comparing to low-count PET, and it also outperformed Gaussian denoised results with statistical significance.

The calculated $SUV_{mean}$ difference percentages between each output against full-count PET for the ROIs are shown in **Tables 1**. No significant differences were found between Gaussian denoised PET and low-count PET in amygdala and hippocampus. Our unsupervised denoised PET for the FDG scan has improved SUV accuracy in terms of mean and standard deviation. The paired student t-test showed that there are significant different ($p<0.05$) between the unsupervised learning denoised PET and the Gaussian denoised PET and between the unsupervised learning denoised PET and the low-count PET. The total error was calculated and is shown in **Table 2**. It follows the observation in **Figure 3** and **Table 1** and confirms that unsupervised denoising PET did perform better when compared to the low-count PET and Gaussian denoised PET.

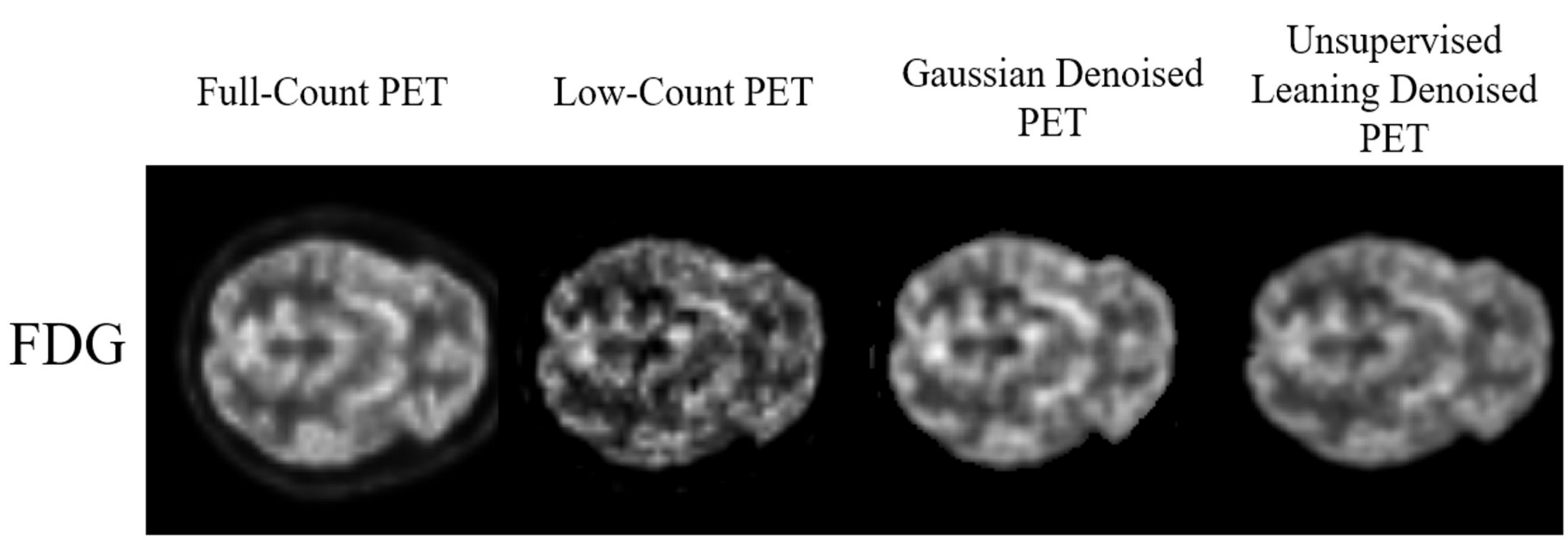

**Figure 2** Example of the full-count PET, low-count PET, Gaussian denoised PET and unsupervised-learning denoised PET.

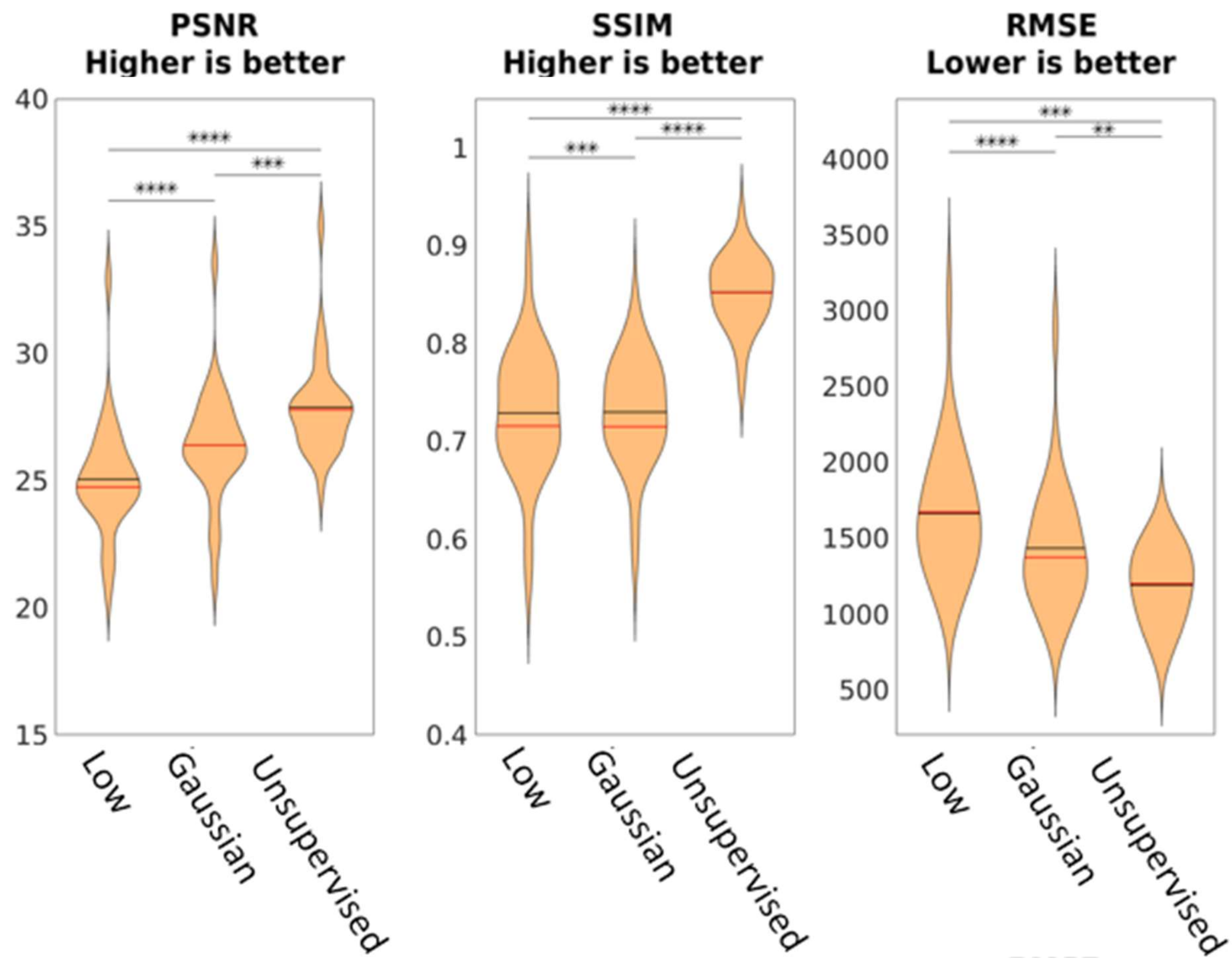

**Figure 3** The violin plot of PSNR, SSIM, and RMSE for the low-count PET, Gaussian denoised PET, and unsupervised learning denoised PET. The p-value calculated using paired student t-test was indicated by the asterisk. * $p \leq 0.05$, ** $p \leq 0.01$, *** $p \leq 0.001$, **** $p \leq 0.0001$, black line in the violine plot indicates the mean, red line indicates the median.

**Table 1** The mean and standard deviation of the percent difference between gull-count PET and the low-count PET, Gaussian denoised PET, and unsupervised learning denoised PET for FDG scans in both side of amygdala and hippocampus.

| % Difference Mean (SD) | Amygdala | | Hippocampus | |
|---|---|---|---|---|
| | Left | Right | Left | Right |
| Low-Count FDG | -7.04 (7.53) | -12.25 (12.96) | -10.69 (9.34) | -3.11 (9.55) |
| Gaussian denoised FDG | -7.42 (8.22) | -12.25 (13.51) | -10.82 (9.13) | -3.33 (8.61) |
| Unsupervised learning denoised FDG | -6.58 (7.10) | -8.10 (5.37) | -8.79 (4.88) | -1.81 (8.31) |

## 4 Discussion

Our unsupervised denoising method shows statistically significant improvement of PSNR, SSIM, and MSE when compared to low-count PET and Gaussian denoised PET. The Gaussian denoised also showed improvement over the low-count PET but is outperformed by the unsupervised learning denoised PET.

**Table 2** The total error of the percent difference between full-count

| Total Error | | Low-Count PET | Gaussian denoised PET | Unsupervised learning denoised PET |
|---|---|---|---|---|
| FDG-Amygdala | Left | 21.80 | 23.54 | 20.49 |
| | Right | 37.65 | 38.73 | 18.61 |
| FDG-Hippocampus | Left | 28.99 | 28.72 | 18.36 |
| | Right | 21.83 | 29.21 | 18.08 |

The $SUV_{mean}$ calculation shows significant differences when comparing unsupervised learning denoised PET with low-count and Gaussian denoised PET. In contrast, the Gaussian denoised PET shows no significant improvement when compared to the low-count PET. This improvement suggests that unsupervised denoised PET not only provides more visually appealing images, it also yields more quantitatively accurate results, which could be used for further kinetic modeling analysis.

In this work, only FDG results were presented. Our group is working on applying this approach to other tracers. It is expected that the loss function might need to be optimized for each tracer, depending on the nature of the tracers.

## 5 Conclusion

Our result showed that our novel loss function for this unsupervised learning denoising network outperformed conventional Gaussian denoising and improved the quality of low-count PET in all global image similarity metrics with statistical significance.

The SUV value also confirmed this conclusion and suggested that this method could be used for quantitative PET analysis. We will further improve our novel loss function together with our denoising network to allow it to optimize automatically for each subject. Further research on adjusting the loss function for other PET tracers utilizing simultaneous PET/MRI acquisition will be done.

## 6 Acknowledgment

The data used in this work was collected with the support from the National Institute of Mental Health (R01MH104512). This work was in part supported by the Parkinson Foundation.

# Author Index

# Keyword Index